\RequirePackage{./StyleFilesMacros/lhchiggs}
\documentclass[twoside]{cernyrep}
\usepackage{etoolbox}
\newtoggle{paper}
\togglefalse{paper}
\usepackage[T1]{fontenc}
\usepackage[dvipsnames]{xcolor}
\usepackage{graphicx}
\usepackage{tikz}
\usetikzlibrary{arrows,decorations.pathmorphing,backgrounds,positioning,fit,petri,automata,shadows,calendar,mindmap,decorations.markings,calc}
\tikzset{->-/.style={decoration={markings, mark=at position .5 with {\arrow{>}}},postaction={decorate}}}
\tikzset{-<-/.style={decoration={markings, mark=at position .5 with {\arrow{<}}},postaction={decorate}}}
\usepackage{cancel}
\usepackage{pdflscape}
\usepackage{float}
\usepackage{./StyleFilesMacros/axodraw}
\usepackage{enumitem}
\setlist{nolistsep} 
\usepackage{./StyleFilesMacros/lhchiggs}
\usepackage{heppennames2}
\usepackage{cernunits}
\usepackage{./StyleFilesMacros/hepparticles}

\newcommand{\met}{E\!\!\!/_T}
\newcommand{\br}{{\mathrm{Br}}}

\def\HXXHF{hXX-HF}
\def\HXXGO{hXX-GO}
\def\mirror{hidden }
\def\Mirror{Hidden }

\newcommand{\eref}[1]{Eq.~(\ref{e.#1})}

\newcommand{\ssref}[1]{Section~\ref{ss.#1}}
\newcommand{\sssref}[1]{Section~\ref{sss.#1}}
\newcommand{\cref}[1]{Chapter~\ref{c.#1}}
\newcommand{\tref}[1]{Table~\ref{t.#1}}
\newcommand{\nc}{\newcommand}
\nc{\gev}{\;\mathrm{GeV}}

\usepackage{hypernat}
\RequirePackage{./StyleFilesMacros/relsize}
\usepackage[toc,page]{appendix}
\usepackage{booktabs}
\usepackage{bigdelim}
\definecolor{gray06}{gray}{0.6}
\providecommand{\ChapterAuthor}[1]{
\hrule
\centerline{
\colorbox{gray06}{%
\begin{minipage}[t]{\textwidth}
\begin{center}
\vspace{.1cm}
\textit{#1}\\
\vspace{.1cm}
\end{center}
\end{minipage}%
}
}
\hrule
\vspace{.2cm}}

\providecommand{\SectionAuthor}[1]{\,\footnote{\textit{Author(s): #1.}}}

\providecommand{\HDECAY}{{\sc HDECAY}}

\providecommand{\Prophecy}{{\sc Prophecy4f}}

\providecommand{\FeynHiggs}{{\sc FeynHiggs}}
\providecommand\POWHEGBOX{{\sc POWHEG BOX}}  

\renewcommand{\Pl}{\ell}

\providecommand{\Powhegbox}{P\protect\scalebox{0.8}{OWHEG}\xspace B\protect\scalebox{0.8}{OX}}
\providecommand{\Powheg}{P\protect\scalebox{0.8}{OWHEG}}
\providecommand{\Sherpa}{S\protect\scalebox{0.8}{HERPA}}
\providecommand{\Powhel}{P\protect\scalebox{0.8}{OW}H\protect\scalebox{0.8}{EL}}
\providecommand{\Decayer}{D\protect\scalebox{0.8}{ECAYER}}
\providecommand{\Helacnlo}{H\protect\scalebox{0.8}{ELAC}-NLO}
\providecommand{\Mcnlo}{\protect\scalebox{0.95}{MC@NLO}}

\providecommand{\Pythia}{P\protect\scalebox{0.8}{YTHIA}}
\providecommand{\Pythiaeight}{P\protect\scalebox{0.8}{YTHIA}8}
\providecommand{\Herwig}{H\protect\scalebox{0.8}{ERWIG}}
\providecommand{\Herwigseven}{H\protect\scalebox{0.8}{ERWIG}7}
\providecommand{\ColorFull}{C\protect\scalebox{0.8}{OLOR}F\protect\scalebox{0.8}{ULL}}
\providecommand{\CVolver}{CV\protect\scalebox{0.8}{OLVER}}
\providecommand{\Openloops}{O\protect\scalebox{0.8}{PEN}L\protect\scalebox{0.8}{OOPS}}
\providecommand{\Collier}{C\protect\scalebox{0.8}{OLLIER}}
\providecommand{\Cuttools}{C\protect\scalebox{0.8}{UT}T\protect\scalebox{0.8}{OOLS}}
\providecommand{\Recola}{R\protect\scalebox{0.8}{ECOLA}}
\providecommand{\Smcnlo}{S-\protect\scalebox{0.9}{MC@NLO}}
\providecommand{\Rivet}{R-\protect\scalebox{0.8}{IVET}}
\providecommand{\Matchbox}{M\protect\scalebox{0.8}{ATCHBOX}}
\providecommand{\MadGraph}{M\protect\scalebox{0.8}{AD}G\protect\scalebox{0.8}{RAPH}}
\providecommand{\MGfiveamcnlo}{M\protect\scalebox{0.8}{AD}G\protect\scalebox{0.8}{RAPH}5\_\protect\scalebox{0.8}{A}\protect\scalebox{0.95}{MC@NLO}}
\providecommand{\MadFKS}{M\protect\scalebox{0.8}{AD}FKS}
\providecommand{\Madloop}{M\protect\scalebox{0.8}{AD}L\protect\scalebox{0.8}{OOP}}
\providecommand{\Madspin}{M\protect\scalebox{0.8}{AD}S\protect\scalebox{0.8}{PIN}}
\providecommand{\Gosam}{G\protect\scalebox{0.8}{O}S\protect\scalebox{0.8}{AM}}

\providecommand{\lsim}
{\;\raisebox{-.3em}{$\stackrel{\displaystyle <}{\sim}$}\;}
\providecommand{\gsim}
{\;\raisebox{-.3em}{$\stackrel{\displaystyle >}{\sim}$}\;}

\DeclareRobustCommand{\PA}{\HepParticle{A}{}{}\Xspace}
\DeclareRobustCommand{\PV}{\HepParticle{V}{}{}\Xspace}
\DeclareRobustCommand{\PX}{\HepParticle{X}{}{}\Xspace}
\DeclareRobustCommand{\Pf}{\HepParticle{f}{}{}\Xspace}

\DeclareRobustCommand{\PF}{\HepParticle{F}{}{}\Xspace}
\DeclareRobustCommand{\PI}{\HepParticle{I}{}{}\Xspace}

\newcommand{\myNLO}{\rm{\scriptscriptstyle{NLO}}}

\newcommand{\mySM}{\rm{\scriptscriptstyle{SM}}}

\newcommand{\ssF}{{\mathrm{F}}}
\newcommand{\ssR}{{\mathrm{R}}}
\newcommand{\ssD}{{\mathrm{D}}}

\newcommand{\ssP}{{\mathrm{P}}}

\newcommand{\ssN}{{\mathrm{N}}}

\newcommand{\bqas}{\begin{eqnarray*}}
\newcommand{\eqas}{\end{eqnarray*}}
\newcommand{\nl}{\nonumber\\}

\newcommand{\lpar}{\left(}                            
\newcommand{\rpar}{\right)}

\newcommand{\bq}{\begin{equation}}                    
\newcommand{\eq}{\end{equation}}
\newcommand{\bqa}{\arraycolsep 0.14em\begin{eqnarray}}
\newcommand{\eqa}{\end{eqnarray}}
\newcommand{\ba}[1]{\begin{array}{#1}}
\newcommand{\ea}{\end{array}}
\newcommand{\ben}{\begin{enumerate}}
\newcommand{\een}{\end{enumerate}}
\newcommand{\bei}{\begin{itemize}}
\newcommand{\eei}{\end{itemize}}
\newcommand{\eqn}[1]{Eq.(\ref{#1})}
\newcommand{\eqns}[2]{Eqs.(\ref{#1})--(\ref{#2})}

\newcommand{\ord}[1]{{\cal O}\lpar#1\rpar}

\newcommand{\mh}{\mathswitch {M_{\PH}}}
\newcommand{\mw}{\mathswitch {M_{\PW}}}

\newcommand{\mz}{\mathswitch {M_{\PZ}}}
\newcommand{\mhs}{\mathswitch {M^2_{\PH}}}

\newcommand{\mws}{\mathswitch {M^2_{\PW}}}
\newcommand{\mzs}{\mathswitch {M^2_{\PZ}}}
\newcommand{\mt}{\mathswitch {M_{\PQt}}}
\newcommand{\cph}{\mathswitch {s_{\PH}}}

\newcommand{\cpz}{\mathswitch {s_{\PZ}}}

\newcommand{\muh}{\mathswitch {\mu_{\PH}}}

\newcommand{\muhs}{\mathswitch {\mu^2_{\PH}}}
\newcommand{\gh}{\mathswitch {\gamma_{\PH}}}

\newcommand{\shat}{\mathswitch {\hat s}}
\newcommand{\muR}{\mathswitch {\mu_{\ssR}}}
\newcommand{\muF}{\mathswitch {\mu_{\ssF}}}
\newcommand{\muRs}{\mathswitch {\mu^2_{\ssR}}}

\newcommand{\tot}{{\mbox{\scriptsize tot}}}

\newcommand{\inv}{{\mbox{\scriptsize inv}}}

\newcommand{\peak}{{\mbox{\scriptsize peak}}}
\newcommand{\ren}{{\mbox{\scriptsize ren}}}

\newcommand{\ep}{\mathswitch \varepsilon}

\newcommand{\spro}[2]{{#1}\cdot{#2}}

\providecommand\POWHEG{{\sc POWHEG}}

\providecommand{\MEPSatNLO}{M\scalebox{0.8}{E}P\scalebox{0.8}{S}@N\scalebox{0.8}{LO}\xspace}
\providecommand{\MCatNLO}{M\scalebox{0.8}{C}@N\scalebox{0.8}{LO}\xspace}
\providecommand{\Sherpa}{S\scalebox{0.8}{HERPA}\xspace}
\providecommand{\Gosam}{G\scalebox{0.8}{OSAM}\xspace}

\providecommand{\OpenLoops}{O\scalebox{0.8}{PEN}L\scalebox{0.8}{OOPS}\xspace}
\providecommand{\Rivet}{R\scalebox{0.8}{IVET}\xspace}

\newcommand{\MCFM}{M\protect{CFM}\Xspace}

\DeclareRobustCommand{\Pho}{\HepParticle{\Ph}{1}{}\Xspace}
\DeclareRobustCommand{\Pht}{\HepParticle{\Ph}{2}{}\Xspace}

\providecommand{\ltbh}{``low-tb-high''}
\providecommand{\mHpm}{M_{\PSHpm}}
\providecommand{\mh}{m_{\Ph}}
\providecommand{\mhiggs}{M_{\phi}}
\providecommand{\mphi}{M_\phi}
\providecommand{\mSUSY}{M_{\scriptscriptstyle{\rm SUSY}}}
\providecommand{\mstone}{M_{\PSQtDo}}
\providecommand{\msttwo}{M_{\PSQtDt}}
\providecommand{\msbar}{\overline{\rm MS}}
\providecommand{\smallmsbar}{{\scriptscriptstyle \msbar}}
\providecommand{\mbmb}{\Mb^\smallmsbar(\Mb)}
\providecommand{\mcmc}{\Mc^\smallmsbar(\Mc)}
\providecommand{\tb}{\tan\beta}
\providecommand{\abbrev}{\relsize{0.9}}
\providecommand{\hmw}{{\abbrev HMW}}
\providecommand{\bv}{{\abbrev BV}}
\providecommand{\AR}{{\abbrev AR}}
\providecommand{\mcnlo}{{\sc MC@NLO}}
\providecommand{\powheg}{{\sc POWHEG}}
\providecommand{\prophecy}{{\sc Prophecy4f}}
\providecommand{\feynhiggs}{{\sc FeynHiggs}}
\providecommand{\hdecay}{{\sc HDECAY}}
\providecommand{\suspect}{{\sc SuSpect}}
\providecommand{\sushi}{{\sc SusHi}}
\providecommand{\moresushi}{{\sc MoRe-SusHi}}
\providecommand{\rootf}{{\sc ROOT}}
\providecommand{\thdm}{THDM}
\providecommand{\nlo}{{\abbrev NLO}}
\providecommand{\fnlo}{f\nlo}
\providecommand{\lo}{{\abbrev LO}}
\providecommand{\pt}{\ensuremath{p_T}}
\providecommand{\pth}{p^{\phi}_T}
\providecommand{\pthtwo}{p^{\phi~2}_T}
\providecommand{\dd}{\mathrm{d}}
\providecommand{\qrest}{Q_t}
\providecommand{\qresb}{Q_b}
\providecommand{\qresint}{Q_\text{int}}
\providecommand{\wrest}{w_t}
\providecommand{\wresb}{w_b}
\providecommand{\wresint}{w_\text{int}}
\providecommand{\qres}{\ensuremath{Q_{\text{res}}}}
\providecommand{\qmatch}{\ensuremath{Q_{\text{match}}}}
\providecommand{\qsh}{\ensuremath{Q_\text{sh}}}
\providecommand{\largeb}{large-$b$}
\providecommand{\largebot}{\largeb{}}
\providecommand{\largetop}{large-$t$}
\providecommand{\order}[1]{{\cal O}(#1)}
\providecommand{\sm}{{\abbrev SM}}
\providecommand{\half}{\tfrac{1}{2}}

\newcommand\sss{\scriptscriptstyle}
\newcommand\bbH{b\bar{b}\phi}
\newcommand\ttH{t\bar{t}\phi}
\newcommand{\bbh}{\bbH}
\providecommand{\tth}{\ttH}
\newcommand\mb{m_b}
\newcommand\mH{m_H}
\newcommand\yt{y_t}
\newcommand\yb{y_b}

\newcommand{\Ht}{H_{\sss T}}
\newcommand{\ptsyst}{p_{\sss T}^{\rm syst}}

\newcommand\bb{\bar{b}}
\def\beq{\begin{equation}}
\def\eeq{\end{equation}}
\newcommand\as{\alpha_{\sss S}}
\newcommand\aNLO{{\sc\small MadGraph5\_aMC@NLO}}
\newcommand\CutTools{{\sc\small CutTools}}
\newcommand\MadLoop{{\sc\small MadLoop}}

\newcommand\PYe{{\sc\small Pythia8}}

\newcommand{\Qshow}{Q_{\text{sh}}}
\newcommand{\Qshowmax}{\Qshow}
\newcommand{\llog}{{\abbrev LL}}

\def\abs#1{\left|#1\right|}
\newcommand{\mcO}{\mathcal{O}}                                                                     
\newcommand{\yvmkk}{\bar Y^2 \frac{v^2}{m_{KK}^2}}  
\nc{\bea}{\begin{eqnarray}}
\nc{\eea}{\end{eqnarray}}
\nc{\mc}{\mathcal}
\newcommand{\Br}{{\mathrm{BR}}} 
\def\ifb{{\ \rm fb}^{-1}}
\newcommand{\hsm}{h}
\newcommand{\cB}{\mathcal B}

\usepackage[hyperfootnotes=true,bookmarks=false,hypertexnames=true,pagebackref=true,linktocpage=false]{hyperref}
\iftoggle{paper}{%
	\definecolor{white}{rgb}{1.0, 1.0, 1.0}
	\renewcommand{\ChapterAuthor}[1]{
		\hrule
		\centerline{
		\colorbox{white}{%
		\begin{minipage}[t]{\textwidth}
		\begin{center}
		\vspace{.1cm}
		\textit{#1}\\
		\vspace{.1cm}
		\end{center}
		\end{minipage}%
	}}}
   \hypersetup{draft}
}{%
\hypersetup{
   unicode=true,          	
   pdftoolbar=true,        	
   pdfmenubar=true,        	
   pdffitwindow=false,     	
   pdfstartview={FitH},    	
   pdftitle={My title},    	
   pdfauthor={Author},     	
   pdfsubject={Subject},   	
   pdfcreator={Creator},   	
   pdfproducer={Producer}, 
   pdfkeywords={keyword1} {key2} {key3}, 
   pdfnewwindow=true,   	
   colorlinks=true,       	
   linkcolor=MidnightBlue,	
   citecolor=OliveGreen,	
   filecolor=Cyan,          	
   urlcolor=RedViolet       	
}
}

\renewcommand*\backref[1]{\ifx#1\relax \else (#1) \fi}

\usepackage{chngcntr}
\usepackage[subfigure]{tocloft} 
\usepackage{subfigure} 
\usepackage{wrapfig}
\usepackage{afterpage}
	\counterwithin{chapter}{part}	
	\counterwithin{section}{chapter}
	\counterwithin{equation}{chapter}
	\counterwithout{figure}{chapter}
	\counterwithout{table}{chapter}
	\counterwithout{footnote}{chapter}
\makeatletter
	\@addtoreset{chapter}{part}
	\@addtoreset{footnote}{part}
\makeatother
\begin{document}

\pagestyle{fancy}
\pagenumbering{gobble}

  

\thispagestyle{empty}
{
\setlength{\unitlength}{1mm}
\begin{picture}(0.001,0.001)
\put(-10,8){CERN Yellow Reports: Monographs}
\put(-10,2){Volume 2/2017}
\put(110,8){CERN--2017--002-M}

\put(14,-80){\LARGE\bfseries Handbook of LHC Higgs cross sections:}
\put(7,-90){\LARGE\bfseries 4. Deciphering the nature of the Higgs sector}
\put(4,-105){\Large\bfseries      Report of the LHC Higgs Cross Section Working Group}

\put(100,-150){\Large Editors:}
\put(118,-150){\Large D.~de~Florian}
\put(118,-156){\Large C.~Grojean}
\put(118,-162){\Large F.~Maltoni}
\put(118,-168){\Large C.~Mariotti}
\put(118,-174){\Large A.~Nikitenko}
\put(118,-180){\Large M.~Pieri}
\put(118,-186){\Large P.~Savard}
\put(118,-192){\Large M.~Schumacher}
\put(118,-198){\Large R.~Tanaka}
\put(58,-240){\includegraphics{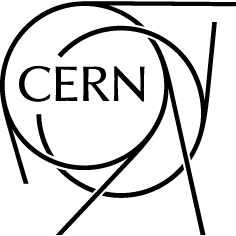}}
\end{picture}
}





\pagenumbering{roman}

\newpage

\null
\vfill

\begin{flushleft}
CERN Yellow Reports: Monographs\\
Published by CERN, CH-1211 Geneva 23, Switzerland\\

\begin{tabular}{@{}l@{~}l}
ISBN & 978--92--9083--442--7 (paperback)\\
ISBN & 978--92--9083--443--4 (PDF) \\ 
Print ISSN & 2519--8068 \\ 
Online ISSN & 2519--8076 \\ 
DOI	 & https://doi.org/10.23731/CYRM-2017-002 \\ 
\end{tabular}\\[1mm]
Accepted for publication by the CERN Report Editorial Board (CREB) on 6 April 2017\\
Available online at \url{http://publishing.cern.ch/} and \url{http://cds.cern.ch/}\\[3mm]

Copyright \copyright{} CERN, 2017\\[1mm]
\raisebox{-1mm}{\includegraphics[height=12pt]{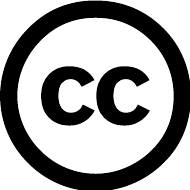}}
Creative Commons Attribution 4.0\\[1mm]
Knowledge transfer is an integral part of CERN's mission.\\[1mm]
CERN publishes this report Open Access under the Creative Commons Attribution 4.0 license
(\url{http://creativecommons.org/licenses/by/4.0/}) in order to permit its wide dissemination and use.
The submission of a contribution to a CERN Yellow Report series shall be deemed to constitute the contributor's agreement to this copyright and license statement. Contributors are requested to obtain any clearances that may be necessary for this purpose.\\[3mm]

This volume is indexed in: CERN Document Server (CDS), INSPIRE.\\[1mm]

This volume  should be cited as:\\[1mm]
LHC Higgs Cross Section Working Group, D.~de~Florian, C.~Grojean, F.~Maltoni, C.~Mariotti, A.~Nikitenko, M.~Pieri, P.~Savard, M.~Schumacher, R.~Tanaka (Eds.), Handbook of LHC Higgs Cross Sections: 4. Deciphering the nature of the Higgs sector, CERN Yellow Reports: Monographs, Vol.~2/2017, CERN-2017-002-M (CERN, Geneva, 2017), https://doi.org/10.23731/CYRM-2017-002 \\
\end{flushleft}

\newpage
\vspace*{6cm}
\begin{center} 
 {\bf Abstract}
\end{center}
\vspace{0.5cm}
This Report summarizes the results of the activities of the LHC Higgs Cross Section Working Group in the period 2014--2016.
The main goal of the working group was to present the state-of-the-art of Higgs physics at the LHC, integrating all new results that have appeared in the
last few years.
The first part compiles the most up-to-date predictions of Higgs boson production cross sections and decay branching ratios, parton distribution functions, and off-shell Higgs boson production and interference effects. 
The second part discusses the recent progress in Higgs effective field theory predictions, followed by the third part on pseudo-observables, simplified template cross section and fiducial cross section measurements, which give the baseline framework for Higgs boson property measurements. 
The fourth part deals with the beyond the Standard Model predictions of various benchmark scenarios of  Minimal Supersymmetric Standard Model, extended scalar sector, Next-to-Minimal Supersymmetric Standard Model and exotic Higgs boson decays.  
This report follows three previous working-group reports: 
{\it Handbook of LHC Higgs Cross Sections: 1.~Inclusive Observables} 
(CERN-2011-002), 
{\it Handbook of LHC Higgs Cross Sections: 2.~Differential Distributions} 
(CERN-2012-002), and 
{\it Handbook of LHC Higgs Cross Sections: 3.~Higgs properties} 
(CERN-2013-004). 
The current report serves as the baseline reference for Higgs physics in LHC Run~2 and beyond.

\newpage
\vspace*{7cm}
\begin{center}
We, the authors, would like to dedicate this Report to the memory of
\\[2.5em]
Guido Altarelli (1941 - 2015)\\[1.5em]
Thomas Kibble (1932 - 2016)\\[1.5em]
and\\[1.5em]
Yoichiro Nambu (1921 - 2015)

\end{center}

\newpage
\leftline{\bf Conveners}
\begin{flushleft}
\noindent \emph{WG1: Higgs XS\& BR:}
         B.~Mellado, \,
         P.~Musella, \,
         M.~Grazzini, \,
         R.~Harlander\,  	 
\begin{itemize}
\item \emph{BR:} A.~Denner, S.~Heinemeyer, A.~M\"uck, I.~Puljak, D.~Rebuzzi
\item \emph{ggF:} S.~Forte, D.~Gilberg, C.~Hays, A.~Lazopoulos, A.~Massironi, G.~Petrucciani, G.~Zanderighi
\item \emph{VBF and WH/ZH:} S.~Dittmaier, P.~Govoni, B.~J\"ager, J.~Nielsen, L.~Perrozzi, E.~Pianori, A.~Rizzi, F.~Tramontano
\item \emph{ttH/tH:} S.~Guindon, C.~Neu, S.~Pozzorini, L.~Reina
\end{itemize}
\vspace{0.1cm}

\noindent \emph{WG2: Higgs properties:} 	
         M.~Chen,\,
         M.~D\"uhrssen,\, 	
         A.~David,\,
         A.~Falkowski,\,
         C.~Hays,\,
         G.~Isidori
\vspace{0.1cm}

\noindent \emph{WG3: BSM Higgs:} 	
         N.~Rompotis,\, 	
         M.~Pelliccioni,\,
         I.~Low,\, 	
         M.~M\"uhlleitner,\,
         R.~Wolf
\begin{itemize}
\item \emph{MSSM neutral:} R.~Lane, S.~Liebler, A.~McCarn, P.~Slavich, M.~Spira, D.~Winterbottom
\item \emph{MSSM charged:} M.~Flechl, S.~Sekula, M.~Ubiali, M.~Zaro
\item \emph{NMSSM:} U.~Ellwanger, M.~M\"uhlleitner, F.~Staub, D.~Strom, R.~Yohay
\item \emph{Neutral extended scalars:} R.~Gerosa, H.~Logan, O.~St\aa l, R.~Santos, S.~Su, X.~Sun
\item \emph{Exotic decay:} S.~Bressler, S.~Gori, A.~Mohammadi, J.~Shelton 
\end{itemize}
\vspace{0.1cm}

\noindent \emph{Cross-group task forces:}
\begin{itemize}
\item \emph{bbH/bH:} M.~Beckingham, A.~Nikitenko, M.~Spira, M.~Wiesemann
\item \emph{HH:} S.~Dawson, C.~Englert, M.~Gouzevitch, R.~Salerno, M.~Slawinska
\item \emph{Fiducial cross-sections:} F.U.~Bernlochner, S.~Kraml, P.~Milenovic, P.~Monni
\item \emph{Offshell Higgs:} F.~Caola, Y.~Gao, N.~Kauer, L.~Soffi, J.~Wang
\item \emph{MCnet:} S.~Pl\"atzer
\item \emph{PDF:} S.~Forte, J.~Huston, R. Thorne 
\end{itemize}
\end{flushleft}

\newpage
\leftline{\bf Authors}
\begin{flushleft}
D.~de Florian$^{1}$,
C.~Grojean$^{2, 3, 4, 5}$,
F.~Maltoni$^{6}$,
C.~Mariotti$^{7}$,
A.~Nikitenko$^{8}$,
M.~Pieri$^{9}$,
P.~Savard$^{10, 11}$,
M.~Schumacher$^{12}$,
R.~Tanaka$^{13}$~(Eds.),
R.~Aggleton$^{14, 15, 16}$
% ...
\end{flushleft}
 
 \begin{itemize}
\item[$^{1}$] International Center for Advanced Studies, UNSAM, 1650 Buenos Aires, Argentina 
\item[$^{2}$] DESY, 22607 Hamburg, Germany
\item[$^{3}$] Institut f{\"u}r Physik, Humboldt-Universit{\"a}t zu Berlin, 12489 Berlin, Germany 
\item[$^{4}$] Instituci{\'o} Catalana de Recerca i Estudis Avan{\c c}ats, 08010 Barcelona, Spain  
\item[$^{5}$] Institut de F{\'\i}sica d'Altes Energies, Barcelona Institute of Science and Technology (BIST), 08193 Bellaterra, Barcelona, Spain
\item[$^{6}$] Centre for Cosmology, Particle Physics and Phenomenology (CP3), Universit{\'e} catholique de Louvain, 1348 Louvain-la-Neuve, Belgium
\item[$^{7}$] INFN Sezione di Torino, 10125 Torino, Italy
\item[$^{8}$] High Energy Physics Group, Blackett Lab., Imperial College, SW7 2AZ London, UK
\item[$^{9}$] University of California San Diego, CA 92093, USA
\item[$^{10}$] University of Toronto, Toronto, ON M5S 1A7, Canada
\item[$^{11}$] TRIUMF, Vancouver, BC V6T 2A3, Canada
\item[$^{12}$] Physikalisches Institut, Albert-Ludwigs-Universit{\"a}t Freiburg, 79104 Freiburg, Germany
\item[$^{13}$] LAL, Universit{\'e} de Paris-Sud, 91405 Orsay, France
\item[$^{14}$] University of Bristol, Bristol BS8 1TL, UK
\item[$^{15}$] School of Physics and Astronomy, University of Southampton, Highfield SO17 1BJ, UK
\item[$^{16}$] Rutherford Appleton Laboratory, Didcot OX110QX, UK
\item[$^{17}$] Institute of High Energy Physics,  Beijing 100049, China
\item[$^{18}$] DAMTP, CMS, University of Cambridge,  CB3 0WA Cambridge, UK
\item[$^{19}$] Institute for Theoretical Physics, Physics Department, ETH Z{\"u}rich, 8093 Zurich, Switzerland
\item[$^{20}$] Rudolf Peierls Centre for Theoretical Physics, University of Oxford, OX1 3NP Oxford, UK
\item[$^{21}$] Higgs Centre for Theoretical Physics, School of Physics and Astronomy, University of Edinburgh, EH9 3JZ Edinburgh, Scotland, UK
\item[$^{22}$] Institute of Theoretical Physics, Faculty of Physics, University of Warsaw, 02-093 Warsaw, Poland
\item[$^{23}$] Berkeley Center for Theoretical Physics, Department of Physics, University of California, Berkeley, CA 94720, USA
\item[$^{24}$] Lawrence Berkeley National Laboratory, Berkeley, CA 94720, USA
\item[$^{25}$] Institute for Theoretical Physics, University of T{\"u}bingen, 72076 T{\"u}bingen, Germany
\item[$^{26}$] Department of Physics and Astronomy, University of Sussex, BN1 9QH Brighton, UK
\item[$^{27}$] LAPTh, Universit{\'e} Savoie Mont Blanc, CNRS, 74941 Annecy-le-Vieux, France
\item[$^{28}$] Department of Physics, University of Warwick, CV4 7AL Warwick, UK
\item[$^{29}$] Department of Physics, New York University, New York, NY 10003, USA
\item[$^{30}$] Institute for Particle Physics Phenomenology, Department of Physics, Durham University, Durham DH1 3LE, UK
\item[$^{31}$] National Research Nuclear University MEPhI (Moscow Engineering Physics Institute), 115409 Moscow, Russia
\item[$^{32}$] Physikalisches Institut der Rheinische Friedrich-Wilhelms-Universit{\"a}t Bonn, 53115 Bonn, Germany
\item[$^{33}$] Institut f{\"u}r Experimentelle Kernphysik, Karlsruhe Institute of Technology, 76128 Karlsruhe, Germany
\item[$^{34}$] Institut f{\"u}r Theoretische Teilchenphysik und Kosmologie, RWTH Aachen University, 52056 Aachen, Germany
\item[$^{35}$] University of Agder, 4604 Kristiansand, Norway
\item[$^{36}$] Physik-Institut, Universit{\"a}t Z{\"u}rich, 8057 Zurich, Switzerland
\item[$^{37}$] Department of Physics, The Chinese University of Hong Kong, Shatin, Hong Kong
\item[$^{38}$] Department of Physics, The University of Hong Kong, Hong Kong
\item[$^{39}$] Department of Physics, The Hong Kong University of Science and Technology, Hong Kong
\item[$^{40}$] Dipartimento di Fisica, Universit{\`a} di Pavia, and INFN, Sezione di Pavia, 27100 Pavia, Italy
\item[$^{41}$] Departamento de F{\'\i}sica Te{\'o}rica and IFIC, Universitat de Val{\`e}ncia-CSIC, 46100 Burjassot, Spain
\item[$^{42}$] High Energy Physics Division, Argonne National Laboratory, Argonne, IL 60439, USA
\item[$^{43}$] Departamento de F{\'\i}sica and CFTP, Instituto Superior T{\'e}cnico, Universidade de Lisboa, 1049-001 Lisboa, Portugal
\item[$^{44}$] Institut f\"{u}r Theoretische Physik, Universit\"{a}t Heidelberg, 69120 Heidelberg, Germany
\item[$^{45}$] Nikhef, Science Park 105, 1098 XG Amsterdam, The Netherlands
\item[$^{46}$] Department of Particle Physics and Astrophysics, Weizmann Institute of Science, 7610001 Rehovot, Israel
\item[$^{47}$] Departamento de F{\'\i}sica Te{\'o}rica and Instituto de F{\'\i}sica Te{\'o}rica, IFT-UAM/CSIC, Universidad Aut{\'o}noma de Madrid, Cantoblanco, 28049 Madrid, Spain
\item[$^{48}$] Technische Universit{\"a}t M{\"u}nchen, 85748 Garching, Germany
\item[$^{49}$] Universit{\'e} Libre de Bruxelles, Service de physique des particules {\'e}l{\'e}mentaires, 1050 Bruxelles, Belgium
\item[$^{50}$] Ludwig-Maximilians-Universit{\"a}t M{\"u}nchen, Fakult{\"a}t f{\"u}r Physik, Arnold Sommerfeld Center for Theoretical Physics, 80333 M{\"u}nchen, Germany
\item[$^{51}$] CPPM, Universit{\'e} Aix-Marseille, 13288 Marseille, France
\item[$^{52}$] Department of Physics, University of Illinois at Urbana-Champaign, Urbana, IL 61801, USA
\item[$^{53}$] Institute for Nuclear Physics, Karlsruhe Institute of Technology, 76344 Eggenstein-Leopoldshafen, Germany
\item[$^{54}$] Institute for Theoretical Physics, Karlsruhe Institute of Technology, 76128 Karlsruhe, Germany
\item[$^{55}$] Theoretical Physics Department, Fermilab, Batavia, IL 60510, USA
\item[$^{56}$] CERN, 1211 Geneva 23, Switzerland
\item[$^{57}$] INFN, Sezione di Pavia, 27100 Pavia, Italy
\item[$^{58}$] Dipartimento di Fisica e Astronomia, Universit{\`a} di Padova and INFN, Sezione di Padova, 35131 Padova, Italy
\item[$^{59}$] IPHC, Universit{\'e} de Strasbourg, CNRS/IN2P3, 67037 Strasbourg, France
\item[$^{60}$] Center for High Energy Physics, Peking University, Beijing 100871, China
\item[$^{61}$] Illinois State University, Normal, IL 61790, USA
\item[$^{62}$] INFN, Sezione di Roma Tre, 00146 Roma, Italy
\item[$^{63}$] Institut de Th{\'e}orie des Ph{\'e}nom{\`e}nes Physiques, EPFL, 1015 Lausanne, Switzerland 
\item[$^{64}$] ARC CoEPP, University of Melbourne, Victoria 3010, Australia
\item[$^{65}$] Departamento de F\'\i sica da Universidade de Aveiro and CIDMA, 3810-183 Aveiro, Portugal
\item[$^{66}$] Maryland Center for Fundamental Physics, Department of Physics, University of Maryland, MD 20742, USA
\item[$^{67}$] Brookhaven National Laboratory, Upton, NY 11973, USA
\item[$^{68}$] INFN, Sezione di Roma, 00185 Roma, Italy
\item[$^{69}$] Departamento de F{\'\i}sica Te{\'o}rica I, Universidad Complutense, 28040-Madrid. Spain
\item[$^{70}$] Institut f{\"u}r Theoretische Physik und Astrophysik, 97074 W{\"u}rzburg, Germany
\item[$^{71}$] Universit{\`a} degli Studi di Roma Tre and INFN,Sezione di Roma Tre, 00146 Roma, Italy
\item[$^{72}$] Sorbonne Universit{\'e}s, Universit{\'e} Pierre et Marie Curie Paris 06, LPTHE, 75005 Paris, France
\item[$^{73}$] CNRS, UMR 7589, LPTHE, 75005, Paris France
\item[$^{74}$] SLAC National Accelerator Laboratory,  Menlo Park, CA 94025, USA
\item[$^{75}$] Max-Planck-Institut f{\"u}r Physik, 80805 M{\"u}nchen, Germany
\item[$^{76}$] LPT, UMR 8627, CNRS, Universit{\'e} de Paris-Sud, Universit{\'e} Paris-Saclay, 91405 Orsay, France
\item[$^{77}$] SUPA, School of Physics and Astronomy, University of Glasgow, G12 8QQ Glasgow, UK
\item[$^{78}$] Institute of Cosmos Sciences, Universitat de Barcelona, 08028 Barcelona, Spain
\item[$^{79}$] Dipartimento di Fisica, Universit{`a} di Milano and INFN, Sezione di Milano, 20133 Milan, Italy
\item[$^{80}$] New York City College of Technology, Brooklyn, NY 11201, USA
\item[$^{81}$] The Graduate School and University Center, The City University of New York, New York, NY 10016 USA
\item[$^{82}$] Departamento de F{\'i}sica, FCEyN, Universidad de Buenos Aires, Capital Federal, Argentina
\item[$^{83}$] Department of Mathematics, Statistics, and Physics, Wichita State University, Wichita, KS 67206, USA
\item[$^{84}$] Institute of High Energy Physics, Austrian Academy of Sciences, 1050 Wien, Austria
\item[$^{85}$] Tif Lab, Dipartimento di Fisica, Universit{\`a} di Milano, and INFN, Sezione di Milano, 20133 Milano, Italy
\item[$^{86}$] LPNHE, Universit{\'e} Pierre et Marie Curie and  Universit{\'e} Paris-Diderot, 75005 Paris, France
\item[$^{87}$] Institut Lagrange de Paris, Universit{\'e} Pierre et Marie Curie, 75005 Paris, France
\item[$^{88}$] Department of Physics and Astronomy, University of Pittsburgh, Pittsburgh, PA 15260, USA
\item[$^{89}$] INFN, Sezione di Genova, 16146 Genova, Italy
\item[$^{90}$] Department of Physics, University of Liverpool, L69 7ZE  Liverpool, UK
\item[$^{91}$] II. Institut f{\"u}r Theoretische Physik, Universit{\"a}t Hamburg, 22761 Hamburg, Germany
\item[$^{92}$] Paul Scherrer Institut, 5323 Villigen PSI, Switzerland
\item[$^{93}$] Physics Department, Carleton University, Ottawa, ON K1S 5B6 Canada
\item[$^{94}$] Department of Mathematical Sciences, University of Liverpool, L69 7ZL Liverpool, UK
\item[$^{95}$] Department of Physics, University of Cincinnati, Cincinnati, OH 45221, USA
\item[$^{96}$] Universit{\'e} de Lyon, Universit{\'e} Claude Bernard Lyon 1, CNRS-IN2P3, IPNL, 69622 Villeurbanne, France
\item[$^{97}$] Universit{\`a} Milano-Bicocca and INFN, Sezione di Milano-Bicocca, 20126 Milano, Italy
\item[$^{98}$] Physics Department, Manhattan College, Riverdale, New York, NY 10471, USA
\item[$^{99}$] Faculty of Science, University of Sarajevo, 71000 Sarajevo, Bosnia and Herzegovina
\item[$^{100}$] Institute for Theoretical Particle Physics, Karlsruhe Institute of Technology, 76128 Karlsruhe, Germany
\item[$^{101}$] Department of Physics and Astronomy, Johns Hopkins University, Baltimore, MD 21218, USA
\item[$^{102}$] Physics Department, SUNY Albany, Albany, NY 12222, USA
\item[$^{103}$] Santa Cruz Institute for Particle Physics (SCIPP) and Department of Physics, University of California, Santa Cruz, CA 95064, USA
\item[$^{104}$] Kavli IPMU (WPI), UTIAS, University of Tokyo, Kashiwa, 277-8583, Japan
\item[$^{105}$] Department of Physics, Oxford University, OX1 3RH Oxford, UK
\item[$^{106}$] Instituto de F{\'\i}sica Te{\'o}rica, IFT-UAM/CSIC, Universidad Aut{\'o}noma de Madrid, Cantoblanco, 28049 Madrid, Spain
\item[$^{107}$] Campus of International Excellence UAM+CSIC, Cantoblanco, 28049 Madrid, Spain
\item[$^{108}$] Instituto de F{\'\i}sica de Cantabria (CSIC/UC), 39005 Santander, Spain
\item[$^{109}$] Physics Department, Florida State University, Tallahassee, FL 32306, USA
\item[$^{110}$] LUPM, UMR 5299, CNRS, Universit{\'e} de Montpellier, 34095 Montpellier, France
\item[$^{111}$] Department of Physics and Astronomy, Michigan State University, East Lansing, MI 48824, USA
\item[$^{112}$] Institute for Particle Physics, Physics Department, ETH Z{\"u}rich, 8093 Zurich, Switzerland
\item[$^{113}$] PRISMA Cluster of Excellence, Johannes Gutenberg University, 55099 Mainz, Germany
\item[$^{114}$] Institut f{\"u}r Physik, Johannes Gutenberg-Universit{\"a}t, 55099 Mainz, Germany
\item[$^{115}$] University of Debrecen and MTA-DE Particle Physics Research Group, 4002 Debrecen, Hungary
\item[$^{116}$] Dipartimento di Fisica, Universit{\`a} di Torino, INFN Sezione di Torino, 10125 Torino, Italy
\item[$^{117}$] Department of Physics, Royal Holloway, University of London, Egham Hill, Egham TW20 0EX, UK
\item[$^{118}$] Bogoliubov Laboratory of Theoretical Physics, Joint Institute for Nuclear Research, 141980 Dubna, Moscow Region, Russia 
\item[$^{119}$] Department of Physics, University of Siegen, 57068 Siegen, Germany
\item[$^{120}$] Department of Physics, University of Arizona, Tucson, AZ 85721, USA
\item[$^{121}$] Institut f{\"u}r Experimentalphysik, Universit{\"a}t Hamburg, 22761 Hamburg, Germany
\item[$^{122}$] LPSC, Universit{\'e} Grenoble-Alpes, CNRS/IN2P3, 38026 Grenoble, France
\item[$^{123}$] Institute for Theoretical Physics, WWU M{\"u}nster, 48149 M{\"u}nster, Germany
\item[$^{124}$] High Energy Physics Group, Blackett Laboratory, Imperial College, SW7 2AZ London, UK
\item[$^{125}$] Physics Department, Technion, Haifa 32000, Israel
\item[$^{126}$] Universit{\`a} and INFN, Sezione di Firenze, 50019 Firenze, Italy
\item[$^{127}$] Department of Physics and Astronomy, University of Kansas, Lawrence, KS 66045, USA
\item[$^{128}$] Department of Physics and Astronomy, Northwestern University, Evanston, IL 60208, USA
\item[$^{129}$] CEA IRFU-SPP, 91191 Gif-sur-Yvette, France
\item[$^{130}$] II.Physikalisches Institut Universit{\"a}t Goettingen 37077 Germany
\item[$^{131}$] CTP, MIT, Cambridge, MA 02139, USA
\item[$^{132}$] Department of Physics, University at Buffalo, The State University of New York, Buffalo, NY 14260, USA
\item[$^{133}$] Department of Physics, Northeastern University, Boston, MA 02115, USA
\item[$^{134}$] International Center for Advanced Studies, UNSAM, 1650 Buenos Aires, Argentina
\item[$^{135}$] Department of Physics, The University of Michigan, Ann Arbor, MI 48104, USA
\item[$^{136}$] University of the Witwatersrand, School of Physics, Private Bag 3, Wits 2050, South Africa 
\item[$^{137}$] School of Physics and Astronomy, University of Manchester, Manchester, M13 9PL, UK
\item[$^{138}$] Department of Physics, University of Pennsylvania, Philadelphia, PA 19104, USA
\item[$^{139}$] Theory Center, Institute of Particle and Nuclear Studies, KEK, Tsukuba, 305-0801, Japan
\item[$^{140}$] Kansas State University, Manhattan, KS, 66506, USA
\item[$^{141}$] M. Smoluchowski Institute of Physics, Jagiellonian University, Krakow, 30-348 Poland
\item[$^{142}$] KIAS, Seoul 130-722, Republic of Korea
\item[$^{143}$] Department of Physics, Southern Methodist University, Dallas, TX 75275, USA
\item[$^{144}$] University of Virginia, Charlottesville, VA 22903, USA
\item[$^{145}$] ARC Centre of Excellence for Particle Physics at the Terascale and CSSM, Department of Physics, The University of Adelaide, Adelaide SA 5005, Australia
\item[$^{146}$] Institute for Theoretical and Experimental Physics, Moscow 117218, Russia
\item[$^{147}$] University of Birmingham, Birmingham B15 2TT, UK
\item[$^{148}$] University of Cambridge, The Cavendish Laboratory, CB3 0HE Cambridge, UK
\item[$^{149}$] Department of Physics and Astronomy, Northwestern University, Evanston, IL 60208, USA 
\item[$^{150}$] Center for High Energy Physics, University of Oregon, Eugene, OR 97403, USA
\item[$^{151}$] HKUST Jockey Club Institute for Advanced Study, Hong-Kong
\item[$^{152}$] Lomonosov Moscow State University, Skobeltsyn Institute of Nuclear, Moscow 119992, Russia
\item[$^{153}$] University of Split, 21000 Split, Croatia
\item[$^{154}$] King's College London, WC2R 2LS London, UK
\item[$^{155}$] Dipartimento di Fisica Universit{\`a} di Pisa, and INFN, Sezione di Pisa, 56100 Pisa, Italy
\item[$^{156}$] Institut f{\"u}r Kern- und Teilchenphysik, TU Dresden, 01069 Dresden, Germany
\item[$^{157}$] University of Washington, Physics Department, Seattle WA 98195, USA
\item[$^{158}$] LLR, IN2P3-CNRS, {\'E}cole Polytechnique, 91128 Palaiseau, France
\item[$^{159}$] ISEL - Instituto Superior de Engenharia de Lisboa, Instituto Polit{\'e}cnico de Lisboa, 1959-007 Lisboa, Portugal
\item[$^{160}$] Centro de F{\'\i}sica Te{\'o}rica e Computacional, Faculdade de Ci{\^e}ncias, Universidade de Lisboa, 1749-016 Lisboa, Portugal
\item[$^{161}$] Paul Scherrer Institut, 5232 Villigen PSI, Switzerland
\item[$^{162}$] Department of Astronomy and Theoretical Physics, Lund University, 22362 Lund, Sweden
\item[$^{163}$] Sorbonne Universit{\'e}s, UPMC Univ. Paris 06, LPTHE, 75005 Paris, France
\item[$^{164}$] Cornell University, Ithaca, NY 14850, USA
\item[$^{165}$] University of Florida, Gainesville, FL 32611, USA
\item[$^{166}$] The Oskar Klein Centre, Department of Physics, Stockholm University, 106 91 Stockholm, Sweden
\item[$^{167}$] Department of Physics, Harvard University, Cambridge, MA 02138, USA
\item[$^{168}$] LPC, CNRS/IN2P3 UMR 6533, 63171 Aubi{\`e}re, France
\item[$^{169}$] University College London,  WC 1E 6BT London, UK
\item[$^{170}$] Institute of Nuclear and Particle Physics, NCSR Demokritos, 15310 Agia Paraskevi, Greece
\item[$^{171}$] Dipartimento di Fisica "E. Pancini", Universit{\`a} di Napoli Federico II and INFN, Sezione di Napoli, 80126 Napoli, Italy
\item[$^{172}$] Niels Bohr International Academy, University of Copenhagen, Blegdamsvej 17, 2100 Copenhagen, Denmark
\item[$^{173}$] Enrico Fermi Institute, Kavli Institute for Cosmological Physics, University of Chicago, Chicago, IL 60637, USA
\item[$^{174}$] School of Physics and State Key Laboratory of Nuclear Physics and Technology, Peking University, Beijing 100871, China
\item[$^{175}$] Collaborative Innovation Center of Quantum Matter, Beijing 100084, China
\item[$^{176}$] Physikalisches Institut der Universit{\"a}t Bonn, 53115 Bonn, Germany
 \end{itemize}

%
%
\newpage
\cleardoublepage
\tableofcontents
%
\cleardoublepage

\newpage

\renewcommand{\thesubsection}{\Roman{part}.\arabic{chapter}.\arabic{section}.\alph{subsection}}
\renewcommand{\thesubsubsection}{\Roman{part}.\arabic{chapter}.\arabic{section}.\alph{subsection}.\roman{subsubsection}}

\pagenumbering{arabic}
\setcounter{footnote}{0}


\cleardoublepage
\phantomsection
\addcontentsline{toc}{part}{Introduction}

\chapter*{Introduction\markboth{Introduction}{Introduction}} 
\ChapterAuthor{D.~de~Florian, C.~Grojean, F.~Maltoni, C.~Mariotti, A.~Nikitenko,\\ M.~Pieri, P.~Savard, M.~Schumacher, R.~Tanaka}

\begin{figure}
\begin{center}
\includegraphics[width=\textwidth]{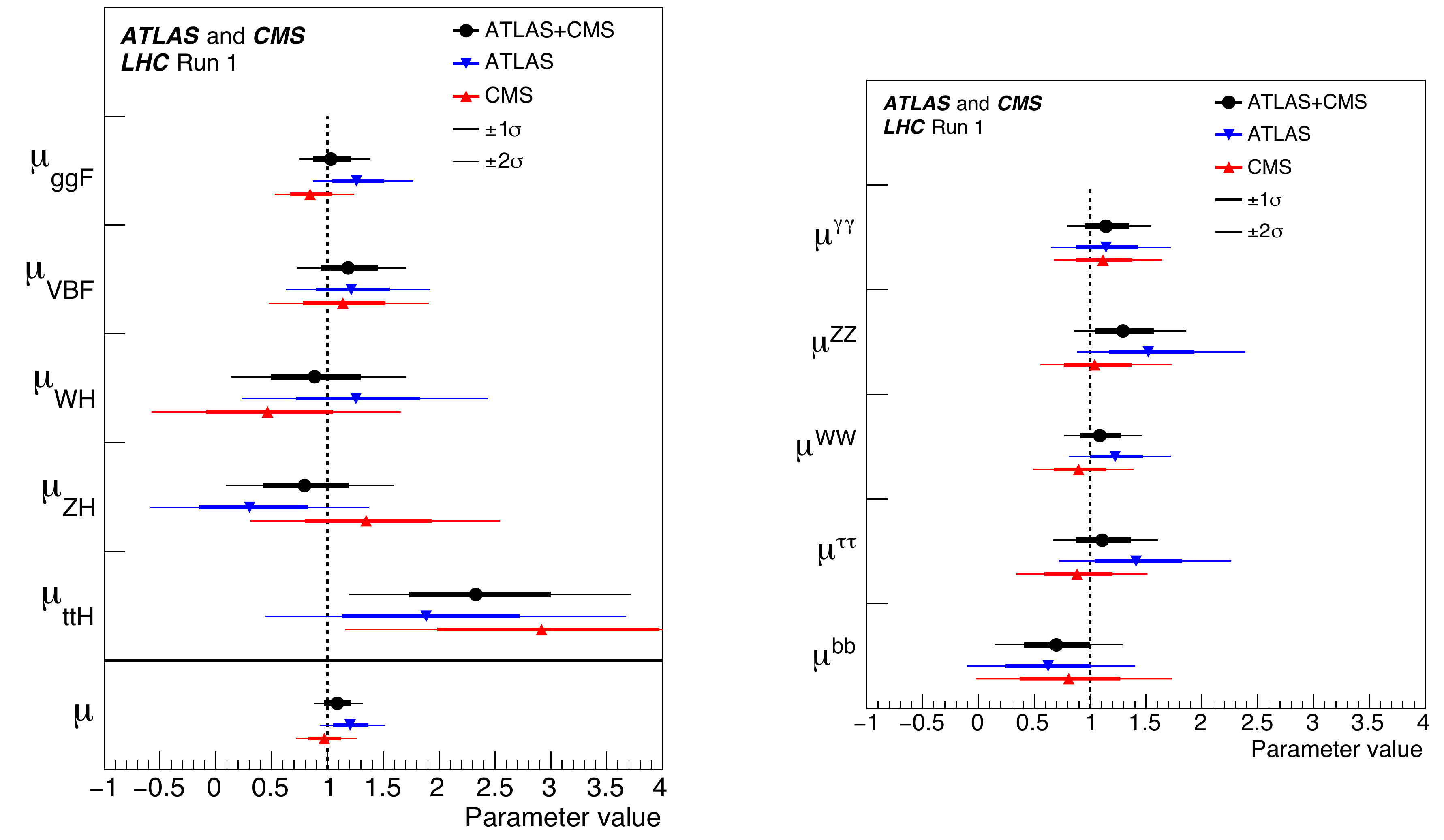}
\end{center}
\caption{Best fit results for the production (left) and decay (right) signal strengths for the combination of ATLAS and CMS data~\cite{Khachatryan:2016vau}. The error bars indicate the 1$\sigma$ (thick lines) and 2$\sigma$ (thin lines) intervals. The combined results show a remarkable agreement with the SM prediction (normalized to $\mu=1$).}
\label{fig:intro-RunI-legacy}
\end{figure}

  The observation by the ATLAS and CMS collaborations in 2012 of a new particle with properties compatible with the Higgs boson predicted by the Standard Model~\cite{Aad:2012tfa,Chatrchyan:2012xdj} was a major breakthrough in particle physics and an unprecedented advance in the understanding of the dynamics at the origin of the breaking of the electroweak symmetry.   Following the end of data-taking in 2012, a vast programme aimed at characterizing the new particle was undertaken: the full LHC Run~1 datasets collected by both experiments were reanalysed with updated detector calibrations, and improved reconstruction and analysis techniques.  The published Run~1 results cover the main production and decay channels expected of the SM Higgs boson, along with the spin and CP properties of the new particle~\cite{Aad:2013xqa,Chatrchyan:2012jja}, and precision measurements of its mass~\cite{Aad:2015zhl}.  More recently, a combination of the ATLAS and CMS measurements was performed~\cite{Khachatryan:2016vau}.  The combined results feature clear observations of the Higgs boson decay to the bosonic channels, the observation of the decay to tau leptons, and of the weak boson production mode.  Overall, the results are consistent with the predictions of the Standard Model, see \refF{fig:intro-RunI-legacy}.  In parallel, searches for an extended Higgs sector were performed covering many Beyond the Standard Model (BSM) scenarios.  No significant evidence of a signal was observed.

   In 2015, the LHC started to collide protons at the higher centre of mass energy of 13\,TeV.  The cross sections for the SM production modes increase by a factor of approximately two to four, depending on the process, and the anticipated integrated luminosity for Run~2  which is scheduled to end in 2018 is of the order of 120\,fb${}^{-1}$ per experiment.  Such a dataset will allow ATLAS and CMS to reduce the current experimental uncertainties significantly, motivating the need for improved theoretical predictions.   For BSM Higgs physics, the higher collision energy will increase substantially the reach of searches for BSM Higgs bosons in the high mass regime and the larger dataset will improve the sensitivity of searches for exotic decays of the 125\,GeV Higgs boson.  New benchmarks and updated calculations will facilitate the exploration of this newly accessible BSM parameter space.

   This report presents improved predictions for the production and decay of the Standard Model Higgs boson~\cite{Dittmaier:2011ti,Dittmaier:2012vm,Heinemeyer:2013tqa}.  These include a gluon fusion production cross section at N3LO, updated and improved Parton Distribution Functions and correlation studies following the PDF4LHC prescriptions, fully differential VBF/VH NNLO QCD and NLO electroweak calculations, NLO electroweak corrections to the ttH processes in addition to studies of the ttV backgrounds.
Differential NNLO+NNLL QCD calculations for the HH process are now available as well as the first NLO results with  full top-quark mass dependence.   
Also first 2-loop NLO calculations for $gg\to VV$ below top-threshold with a massive quark have been achieved and they will be helpful for the study of the off-shell production of the Higgs boson.

   For Higgs boson property measurements, the Interim framework for the analysis of Higgs couplings ``kappa framework"~\cite{LHCHiggsCrossSectionWorkingGroup:2012nn} was used by the ATLAS and CMS experiments to report Higgs boson coupling related results extracted from the LHC Run 1 data~\cite{Khachatryan:2016vau}. The update information for Run 2 is provided elsewhere\footnote{\url{https://twiki.cern.ch/twiki/bin/view/LHCPhysics/LHCHXSWG2KAPPA}}. With the additional statistical power of the Run~2  dataset, the experiments will be able to measure more precisely the kinematic properties of the 125 GeV Higgs and use these measurements to probe for possible deviations induced by new phenomena.  To do this, several strategies have been devised to maximize the sensitivity to BSM physics across channels and as a function of the integrated luminosity.  These strategies will allow the experiments to extract more information from the data compared to the ``kappa framework". To do this, pseudo-observables are defined as a possible alternative to more direct template, fiducial and differential cross section measurements. They will eventually be used as inputs to interpret the data in terms of Effective Field Theories (EFTs).  Several Monte Carlo tools necessary to undertake this effort are being developed and  made available, as described in this report.

  In the search for new physics in the Higgs sector, this report proposes new benchmarks for the exploration of a BSM Higgs sector along with improved and extended calculations in various scenarios including the MSSM, the NMSSM, and more generic models featuring charged and neutral Higgs bosons.  In addition, a new chapter on rare and exotic Higgs boson decays has been added and it covers rare mesonic decays, flavour-violating decays, prompt decays with and without missing energy, and decays into 
long-lived particles or with displaced vertices,  and it provides recommendations on searches for exotic decays of the 125\,GeV Higgs boson.

The updated and improved calculations of SM and BSM Higgs boson production and decay processes as reported here
provide a solid theoretical reference for experimental studies of the early Run~2  data
that are expected to decipher the properties of the 125\,GeV Higgs boson and the nature of the full Higgs sector.  
In addition, they pave the way towards the theoretical developments and improvements in precision   that will be needed in the future in order to fully exploit the potential of the complete Run~2  dataset. 

\newpage
\renewcommand*{\thefootnote}{\fnsymbol{footnote}}
\part[Standard Model Predictions]{Standard Model Predictions \footnote{M.~Grazzini, R.~Harlander, B.~Mellado, P.~Musella~(Eds.)}}
\label{chap:SM}
\renewcommand*{\thefootnote}{\Roman{part}.\arabic{footnote}}
\setcounter{footnote}{0} 
\chapter{Standard Model Parameters} 
\label{chapter:input}
\ChapterAuthor{A.~Denner, S.~Dittmaier, M.~Grazzini, R.~Harlander, R.~Thorne, M.~Spira, M.~Steinhauser}

We summarize the Standard Model input parameters 
for Higgs cross section calculations. The same parameters can be used
for other SM and BSM processes at the LHC.

\section{Lepton masses}

The lepton masses from the PDG\,\cite{Agashe:2014kda} are
\begin{align}
& m_e=0.510998928\pm 0.000000011 ~{\rm MeV}\,,\\
& m_\mu=105.6583715\pm 0.0000035 ~{\rm MeV}\,,\\
& m_\tau=1776.82 \pm 0.16 ~{\rm MeV}\,.
\end{align}

\section{Electroweak parameters}

The gauge boson masses and widths from the PDG\,\cite{Agashe:2014kda}
are
\begin{align}
m_W&=80.385\pm 0.015~{\rm GeV}\,,& \Gamma_W&=2.085\pm 0.042~{\rm GeV}\,,\\
m_Z&=91.1876\pm 0.0021~{\rm GeV}\,,&\Gamma_Z&=2.4952\pm 0.0023~{\rm GeV}\,.
\end{align}
The Fermi constant is
\begin{equation}
G_F=1.166 378 7(6)\cdot 10^{-5}~{\rm GeV}^{-2}\,.
\end{equation}
These values correspond to the physical on-shell masses. If required,
the complex pole masses can be obtained from the well-known relations
\begin{equation}
\begin{split}
m_V^\text{pole} - i\Gamma_V^\text{pole} 
= \frac{m_V-i\Gamma_V}{\sqrt{1+\Gamma^2_V/m^2_V}}\,,\qquad V\in\{W,Z\}\,.
\label{eq:ospole}
\end{split}
\end{equation}
As for the gauge boson widths, instead of $\Gamma_V^\text{pole}$ from (\ref{eq:ospole}), values consistent with the perturbative order can also be used if more appropriate.

\section{QCD parameters and parton densities}

The most important QCD parameters are the strong coupling $\alpha_s$ and
the quark masses. The default renormalization scheme for these
parameters should be the $\msbar$ scheme. In this scheme, $\alpha_s$
and the quark masses depend on a mass scale $\mu$.

$\alpha_s(\mu)$ and $m_q(\mu)$ are typically determined through their
proper renormalization group equations,
\begin{equation}
\begin{split}
\mu^2\frac{d}{d\mu^2}\alpha_s(\mu) = \alpha_s(\mu)\beta(\alpha_s)\,,\qquad
\mu^2\frac{d}{d\mu^2}m_q(\mu) = m_q(\mu)\gamma_m(\alpha_s)\,,
\end{split}
\end{equation}
combined with their numerical value at a reference mass scale, usually
$\alpha_s(m_Z)$ and $m_q(m_q)$. The perturbative expansions of the
coefficients $\beta(\alpha_s)$ and $\gamma_m(\alpha_s)$ are currently
known through
order~$\alpha_s^4$\,\cite{vanRitbergen:1997va,Czakon:2004bu}
and $\alpha_s^5$, respectively\,\cite{Baikov:2014qja}.

Also the parton density functions depend on a mass scale $\muF$,
\begin{equation}
\begin{split}
\muF^2\frac{d}{d\muF^2}\phi_i(x,\muF) =
P_{ij}(\alpha_s)\otimes\phi_j(x,\muF)\,,\qquad
\end{split}
\end{equation}
where $P_{ij}$ are the splitting functions, currently known through
order $\alpha_s^3$\,\cite{Moch:2004pa,Vogt:2004mw}, and $\otimes$
denotes the usual convolution.  Note that in principle also $\muR$
enters the PDFs implicitly through $\alpha_s$. However, the available
PDF sets assume $\muF=\muR$.

If the typical mass scale $\mu_0$ of a process is not equal to the
reference mass scale, the input quantities $\alpha_s(\mu_0)$ and
$m_q(\mu_0)$ of a perturbative calculation have to be evaluated from
their reference values by RG evolution. While 4-loop evolution may
result in the most precise currently available results for the input
parameters, consistency of the calculation often requires one to use
lower order RG evolution.

\subsection{Strong coupling constant}

The strong coupling $\alpha_s$ enters a typical theory prediction for an
LHC observable in many different ways: explicitly as expansion
parameter in the partonic calculation, or implicitly through its impact
on other input quantities. These sources may be strongly correlated, so
that inconsistencies in the input value $\alpha_s(\mu_0)$ have to be
avoided.  Specifically, the value of $\alpha_s$ used in the partonic
process should coincide with the one corresponding to the parton density
functions. This means that $\alpha_s(m_Z)$ as well as the RG running of
$\alpha_s$ for the evaluation of $\alpha_s(\mu_0)$ have to be adjusted
to the parton density functions that are used.

Concerning the default value for $\alpha_s(m_Z)$ and the estimation of
the uncertainties resulting from $\alpha_s(m_Z)$ and the PDFs, one
should follow the 2015 PDF4LHC recommendation. This amounts to choosing the central value of $\alpha_s(m_Z)$ and the ensuing uncertainty as
\begin{equation}
\alpha_s(m_Z)=0.118\pm 0.0015\, .
\end{equation}

\subsection{Quark masses}

Quark masses (in particular charm and bottom) also enter the PDFs,
albeit in a much weaker way as $\alpha_s$. This releases one from a
similar constraint as it was imposed for $\alpha_s$.

\noindent{\bf Top-quark mass.} The top quark is different from all other
known quarks in the sense that it decays before it hadronizes. To a
first approximation (i.e., neglecting soft QCD effects), the invariant
mass of its decay products may be identified with the on-shell top quark
mass. In fact, the agreement with the determination via the top quark
pair production cross section justifies this with hindsight.

In order to be consistent with existing ATLAS and CMS analyses, we
recommend to use
\begin{equation}
\begin{split}
m_t^\text{OS}=172.5\pm 1\,\text{GeV}
\end{split}
\end{equation}
as reference value for the on-shell top quark mass, corresponding to an
$\msbar$ value of
\begin{equation}
\begin{split}
m_t^{\msbar}(m_t) = 162.7\pm 1\,\text{GeV}\,,
\end{split}
\end{equation}
where we used the four-loop conversion of Ref.\,\cite{Marquard:2015qpa}.
Note that the recommended uncertainty of $\pm 1$\,GeV covers the current
world average of $m_t^\text{OS}\Big|_\text{world ave}=173.2$\,GeV
(notice also Ref.\,\cite{Khachatryan:2015hba}).

The calculation of radiative corrections may require to use the complex
pole mass for the top quark. Due to the lack of experimental data for
the top width, one should use the theoretical value for the top
quark width in the conversion formula (the analogue of
Eq.\,\ref{eq:ospole}), given by\,\cite{Jezabek:1993wk,Czarnecki:1998qc,Chetyrkin:1999ju,Blokland:2004ye,Blokland:2005vq,Gao:2012ja,Brucherseifer:2013iv}
\begin{equation}
\begin{split}
\Gamma_t = 0.89\cdot \Gamma_t^{(0)}\,,
\end{split}
\end{equation}
where
\begin{equation}
\begin{split}
\Gamma_t^{(0)} &= \frac{G_Fm_t^3}{8\sqrt{2}\pi}\left[1 -
  3\left(\frac{m_W^2}{m_t^2}\right)^2 
  + 2\left(\frac{m_W^2}{m_t^2}\right)^3\right]\,.
\end{split}
\end{equation}


\noindent{\bf Bottom-quark mass.} The situation is much different for
the bottom quark mass, because its mass can only be determined
indirectly. We recommend to use the current PDG value for the $\msbar$
bottom mass \cite{Agashe:2014kda}
\begin{equation}
m_b(m_b)=4.18\pm 0.03~{\rm GeV}
\label{eq:mbmb}
\end{equation}
as reference input value.\footnote{The dependence of this value on
  $\alpha_s$ (see Ref.\,\cite{Chetyrkin:2009fv}) can usually be neglected.}
Evolution to the characteristic scale $\mu_0$ of the process should be
done through the highest available perturbative order. The inconsistency
with the PDF value for the bottom quark mass introduced by this
procedure is expected to be small, in particular if no initial-state
$b$-quarks are involved. The uncertainty can be estimated by comparing
the results when using PDFs built on different values of the $b$ quark
mass.

If the use of the on-shell bottom-quark mass cannot be avoided, the
conversion of the $\msbar$ value should be done at the 3-loop level, and
the difference to the 4-loop result should be used as the uncertainty. Using
$\alpha_s(m_Z)=0.118$ and $m_b(m_b)=4.18$\,GeV, one
obtains\,\cite{Marquard:2015qpa}
\begin{equation}
\begin{split}
\label{eq:mb}
m_b^\text{OS}\bigg|_{4.18} = (4.18 + 0.40 + 0.20 + 0.14 + 0.13)\,\text{GeV}\,,
\end{split}
\end{equation}
where the numbers correspond to consecutive loop orders. Our
recommendation for the bottom quark pole mass is therefore obtained by
considering the sum of the first four terms in Eq.~(\ref{eq:mb}),
and assigning the last term as an uncertainty
\begin{equation}
\begin{split}
m_b^\text{OS} = 4.92\pm 0.13\,\text{GeV}\,.
\end{split}
\end{equation}

\noindent{\bf Charm-quark mass.} The charm quark mass is at the edge of
the validity range of perturbation theory. Therefore, using $m_c(m_c)$
as reference input value in order to derive the charm quark mass at a
different scale, or its perturbative on-shell mass, would force one to
apply perturbation theory at these rather low energies. We therefore
recommend to use \cite{Chetyrkin:2009fv}
\begin{equation}
m_c(3~{\rm GeV})=0.986\pm 0.026~{\rm GeV}
\end{equation}
as overall input value. To be conservative, the error quoted in
Ref.\,\cite{Chetyrkin:2009fv} has been multiplied by a factor of two here.

Concerning RG evolution and consistency with the charm quark mass in the
PDFs, the situation is analogous to the case of the bottom quark (see
above).

If the use of the on-shell charm-quark mass cannot be avoided, its mass
should be evaluated from the on-shell bottom mass (calculated as
described above) through the relation\,\cite{Bauer:2004ve}
\begin{equation}
\begin{split}
m_c^\text{OS} = m_b^\text{OS}-3.41\,\text{GeV} = 1.51\pm 0.13\,\text{GeV}\,.
\end{split}
\end{equation}

\section{Higgs boson mass}

The current combination of the measurements of the Higgs boson mass $m_H$ from ATLAS and CMS
is \cite{Aad:2015zhl}
\begin{equation}
\begin{split}
m_H=125.09\pm 0.21\text{(stat.)}\pm 0.11\text{(syst.) GeV}\, .
\end{split}
\end{equation}
As a reference value for $m_H$ in theoretical calculations we recommend to
use the rounded value
\begin{equation}
\begin{split}
m_H=125~\text{GeV}\, .
\end{split}
\end{equation}

\chapter{Parton Distribution Functions} 
\label{chap:PDFs}
\ChapterAuthor{S.~Forte, J.~Huston, R.~S.~Thorne~(Eds.); S. Carrazza, J.~Gao, Z.~Kassabov, P.~Nadolsky, J.~Rojo}
\def\lsim{\mathrel{\rlap{\lower4pt\hbox{\hskip1pt$\sim$}}
    \raise1pt\hbox{$<$}}}                

\section{The PDF4LHC recommendation}

Previous Yellow Reports~\cite{Dittmaier:2011ti,Dittmaier:2012vm,Heinemeyer:2013tqa} have provided snapshots of the state-of-the-art
for PDF determination, along with recommendations for PDF use, and for calculations of PDF uncertainties, following the guidance of the PDF4LHC group.
In a previous recommendation~\cite{Botje:2011sn}, three PDF sets were used:
CT10~\cite{Lai:2010vv}, MSTW2008~\cite{Martin:2009iq}
and NNPDF2.3~\cite{Ball:2012cx}.
These were global PDF fits involving data from a variety of experiments, including collider data from the Tevatron.
The uncertainty was provided by the envelope of all three PDF error
sets, and the  central prediction as mid-point of this envelope.
This choice is conservative but not ideal, in that it tends to be
dominated by error PDFs at 
the edge of the 
uncertainty band; it was adopted because it was felt that the degree
of agreement of the PDF sets was not sufficient to warrant their
statistical combination. Specifically, agreement was unsatisfactory for
 the gluon distribution,
particularly in the region appropriate for Higgs boson production through gluon-gluon fusion.
This disagreement prompted an intensive year-long study by the three global PDF groups,
along with HERAPDF~\cite{Abramowicz:2015mha},
but this did not uncover a clear explanation for the differences~\cite{Andersen:2014efa}.

Prior to the writing of YR4, the major PDF groups have released updates to their
PDF fits, at NLO and NNLO, including in most cases data from the LHC~\cite{Rojo:2015acz}.
The new PDF4LHC recommendation~\cite{Butterworth:2015oua} uses the updated PDFs from the three global PDF groups included in the previous recommendation:
CT14~\cite{Dulat:2015mca}, MMHT14~\cite{Harland-Lang:2014zoa} and
NNPDF3.0~\cite{Ball:2014uwa}, respectively.
Details as to why this choice was made can be found in the PDF4LHC document.
The primary requirements are that the PDFs should be based on global datasets, be carried out in a general-mass variable flavour-number scheme,
and have compatible values for the QCD coupling constant $\alpha_s(m_Z)$.
As we shall see shortly, these new PDF sets are in good agreement, not
only in the quark sector (where the agreement was satisfactory already
in  the previous generation of PDFs) but also for the gluon.
The changes can be ascribed partially to the addition of new data sets used in the PDF fits, but primarily to improvements in the fitting formalisms.
This level of agreement may change in detail with future updates, but
generally the good level of agreement should stay. An alternative recommendation~\cite{Accardi:2016ndt} is that all PDFs
(ABM12~\cite{Alekhin:2013nda}, CJ15~\cite{Accardi:2016qay},  CT14~\cite{Dulat:2015mca},
    HERAPDF2.0~\cite{Abramowicz:2015mha},
    JR14~\cite{Jimenez-Delgado:2014twa},
     HERAPDF2.0~\cite{Abramowicz:2015mha}
 MMHT14~\cite{Harland-Lang:2014zoa}, 
NNPDF3.0~\cite{Ball:2014uwa}) and
accompanying coupling and quark mass variation should
be used for precision theory predictions and any {\tt
  LHAPDF6}~\cite{Buckley:2014ana}  PDF set for 
other predictions.

Currently, there are two different representations of PDF uncertainties:
the Monte Carlo representation~\cite{DelDebbio:2004xtd,DelDebbio:2007ee} and
the Hessian representation~\cite{Pumplin:2001ct}.
Both provide compatible descriptions of the PDF uncertainties,
and recent developments have allowed for the straight-forward conversion of
one representation to the other~\cite{Watt:2012tq,Gao:2013bia,Carrazza:2015aoa}.
The use of the Monte Carlo representation makes possible a statistical combination of different PDF sets.
If different PDF sets can be assured to be equally likely representations of the underlying PDF probability distribution, they can be combined simply by taking their un-weighted average.
This can be arrived at by generating equal numbers of Monte Carlo replicas from each input PDF set, and then merging the replica sets.
The NNPDF3.0 PDF set is naturally in this format.
For the Hessian sets, CT14 and MMHT2014, the Monte Carlo replicas are generated by sampling along the eigenvector directions, assuming a Gaussian distribution.

Combinations in this manner are most appropriate when the PDF sets
that are combined are compatible with each other, as  CT14, MMHT2014
and NNPDF3.0 are.
Such a combination also allows for a direct statistical interpretation
of the resulting PDF uncertainties, unlike the envelope method.
Monte Carlo combinations of these three PDFs have been provided at both NLO and NNLO by the PDF4LHC working group.
In the following
discussion we concentrate on NNLO; similar considerations apply to NLO.  

\begin{figure}[t]
  \centering
 \includegraphics[width=.48\textwidth]{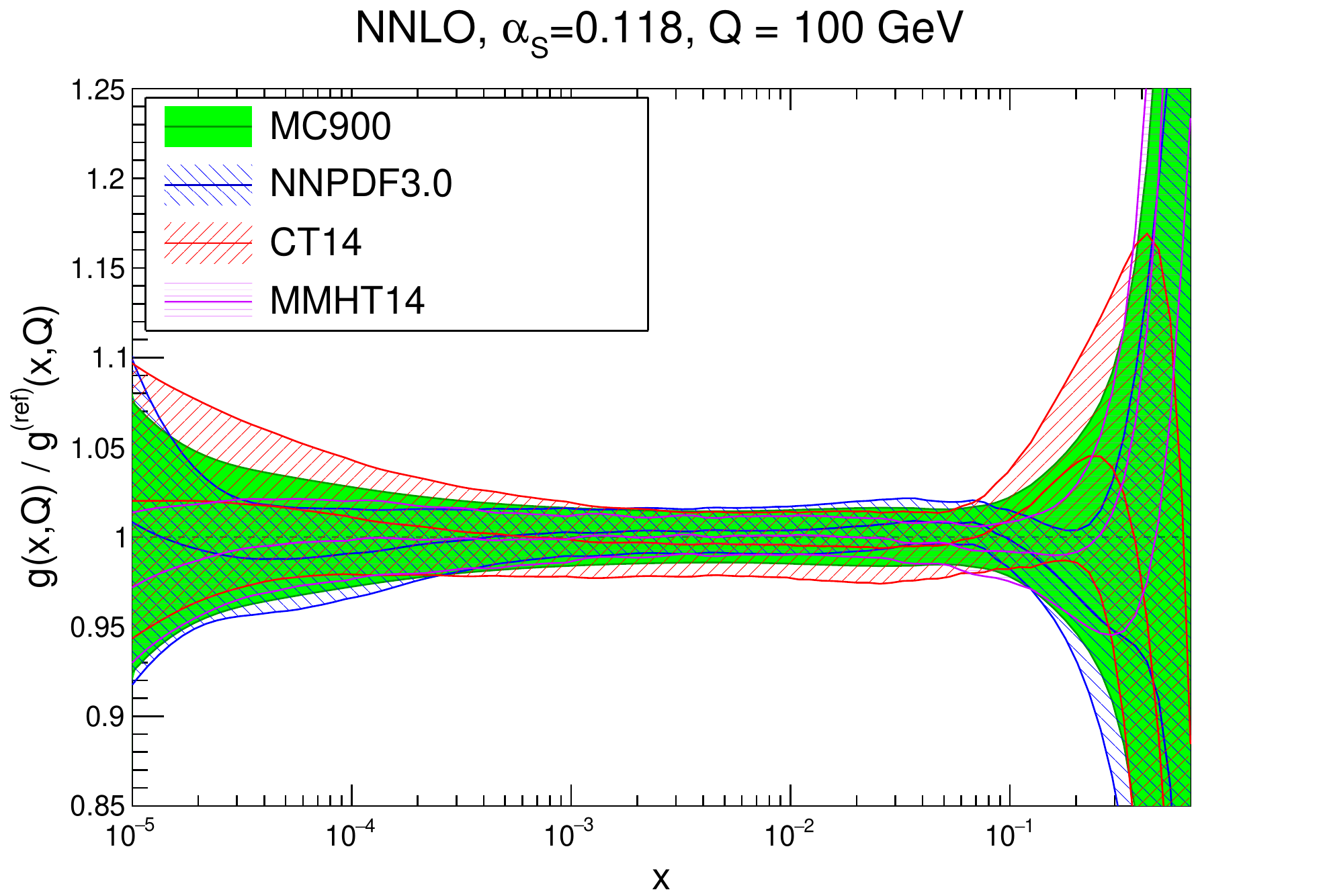}
  \includegraphics[width=.48\textwidth]{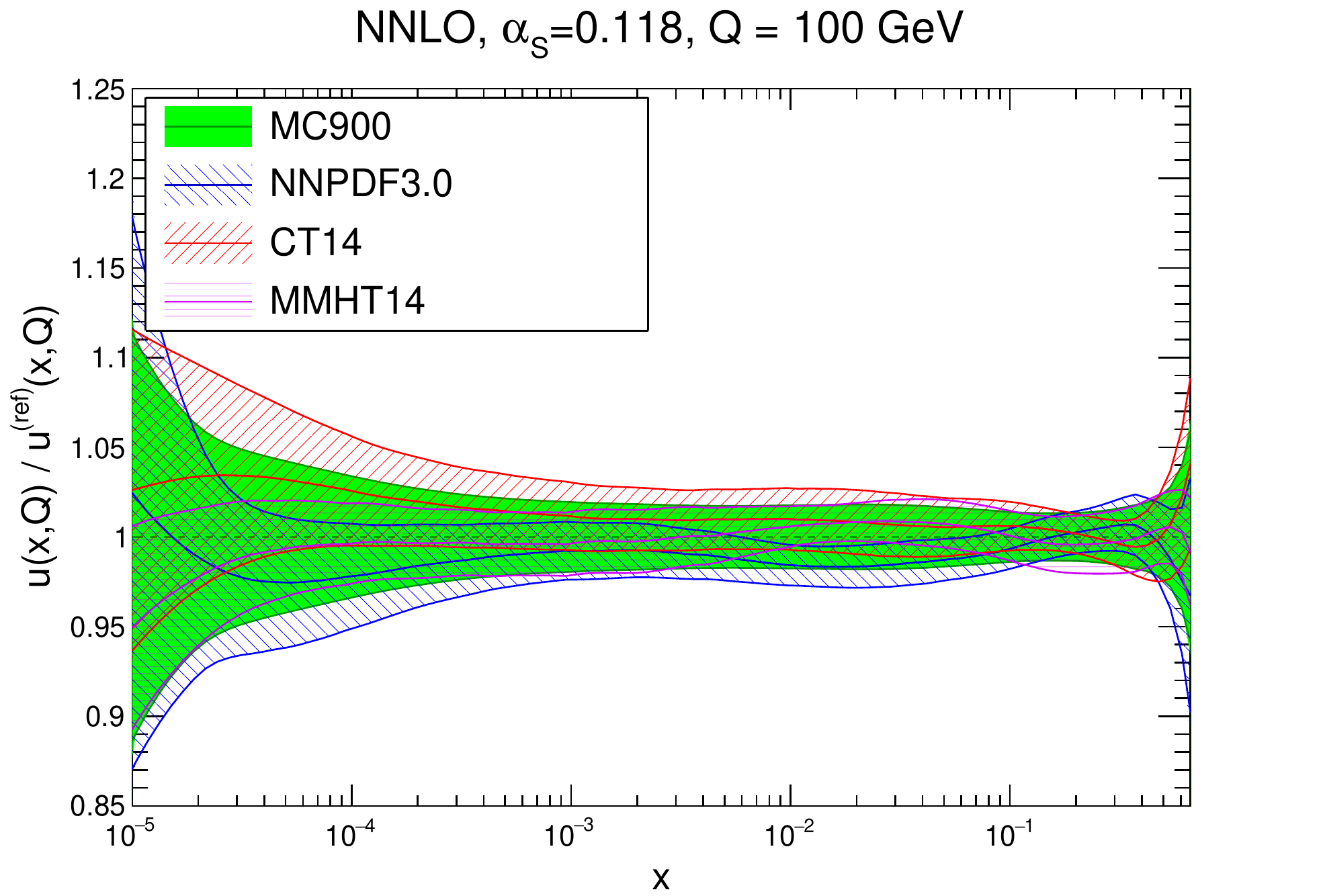}
   \caption{\small Comparison
     of the MC900 PDFs with the  sets that
     enter the combination: CT14, MMHT14 and NNPDF3.0 at
     NNLO.
     We show the gluon and the up quark
     at $Q=100$ GeV.
     Results are normalized to the central value of the prior set MC900.
}  
\label{fig:MCPDFcombV2}
\end{figure}

It was determined that $N_{\rm rep}=900$ Monte Carlo replicas,
combining $N_{\rm rep}=300$ replicas from each of the three
individual PDF sets, were sufficient to represent the combined PDF
probability distribution.
In \refF{fig:MCPDFcombV2} we show a comparison
of the combined PDF4LHC15 NNLO set (indicated by MC900 in the plot)
with the three sets that
     enter the combination, CT14, MMHT14 and NNPDF3.0.
     We show the gluon and the up quark at $Q=100$ GeV,
     normalized to the central value of the PDF4LHC15 combination.
     In these plots, as in the rest of this chapter, we use
     a fixed value $\alpha_s(m_Z)=0.118$.

It can be seen that in the ``precision physics'' region, roughly from $x\simeq 0.001$ to
     $x\simeq 0.1$,
the PDFs from the three global sets agree reasonably well with each
other, with perfect agreement for the gluon, and less good agreement
for the up quark. 
This is reflected in the combined PDF4LHC15 set, constructed from the three input PDF sets.
On the other hand, at low and high values of $x$, the uncertainty bands from the three PDF sets differ, and the uncertainty band for the 900 Monte Carlo replicas is smaller than the envelope of the three PDF uncertainty bands.
These are  regions in which PDFs are only weakly constrained by data,
as seen by the increasingly large size of the uncertainty, and the
inflated uncertainty in the combination appears to  provides a 
reasonable estimate.%

\begin{figure}[t]
  \centering
  \vspace{-3mm}
  \includegraphics[width=.48\textwidth]{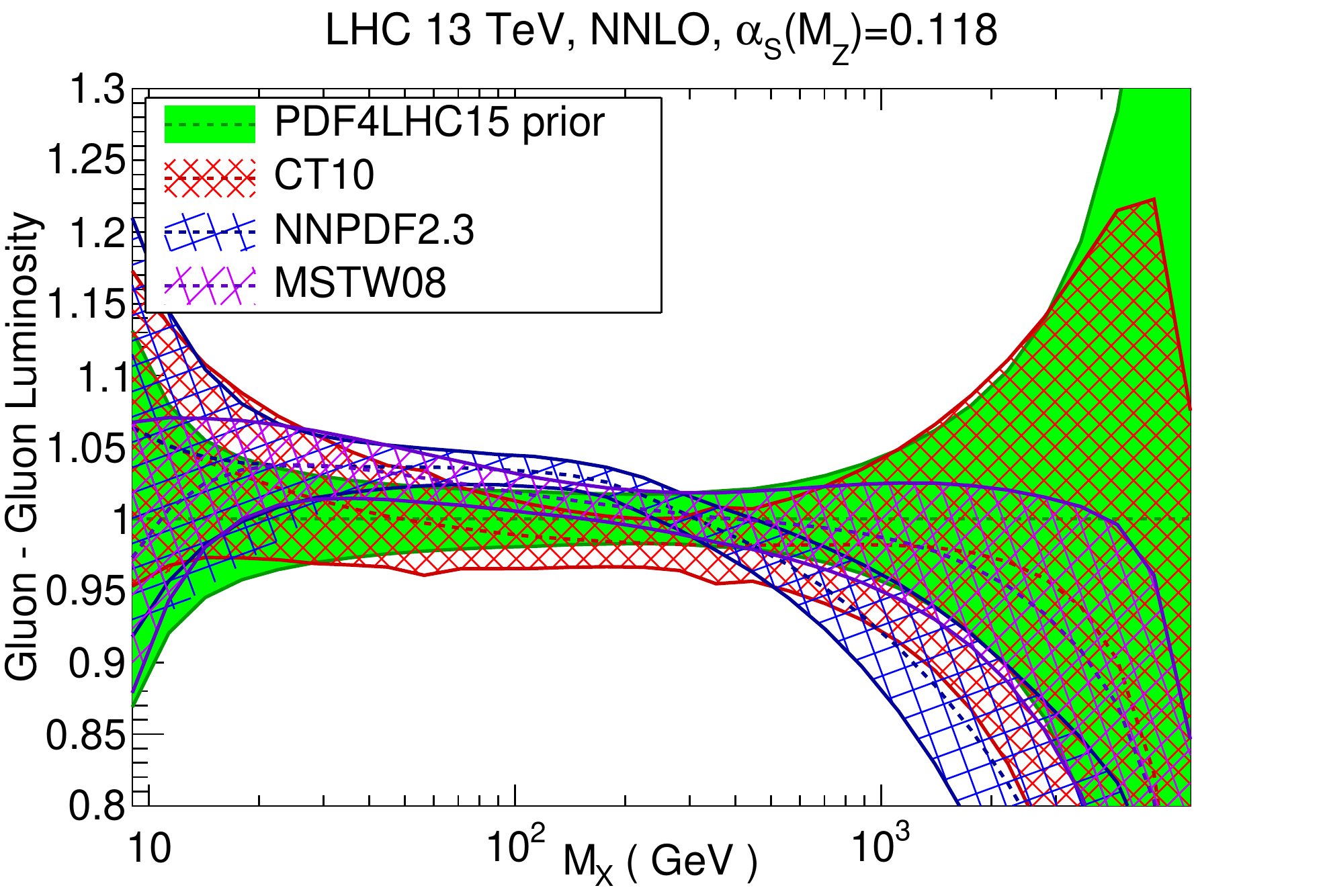}
  \includegraphics[width=.48\textwidth]{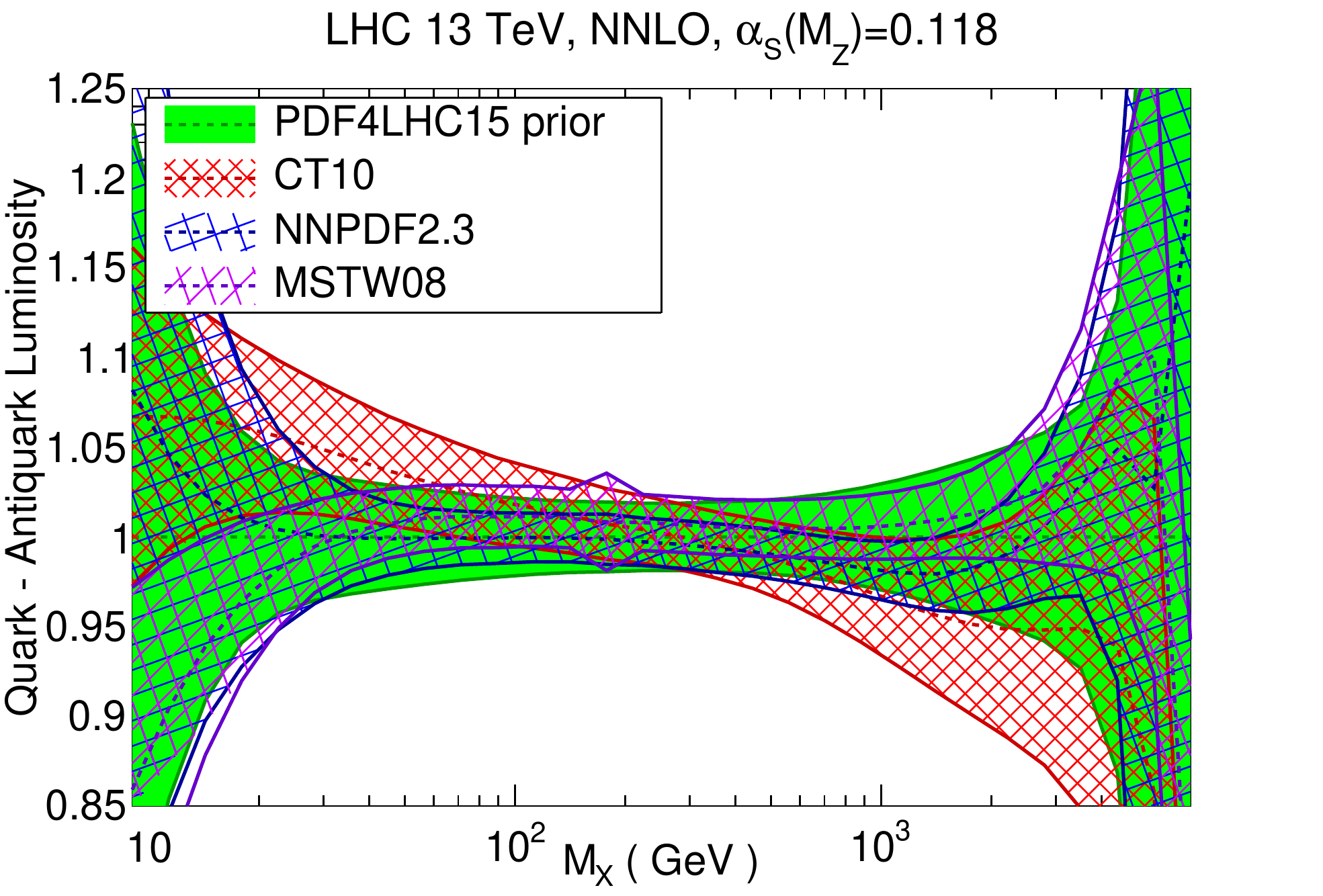}
  \includegraphics[width=.48\textwidth]{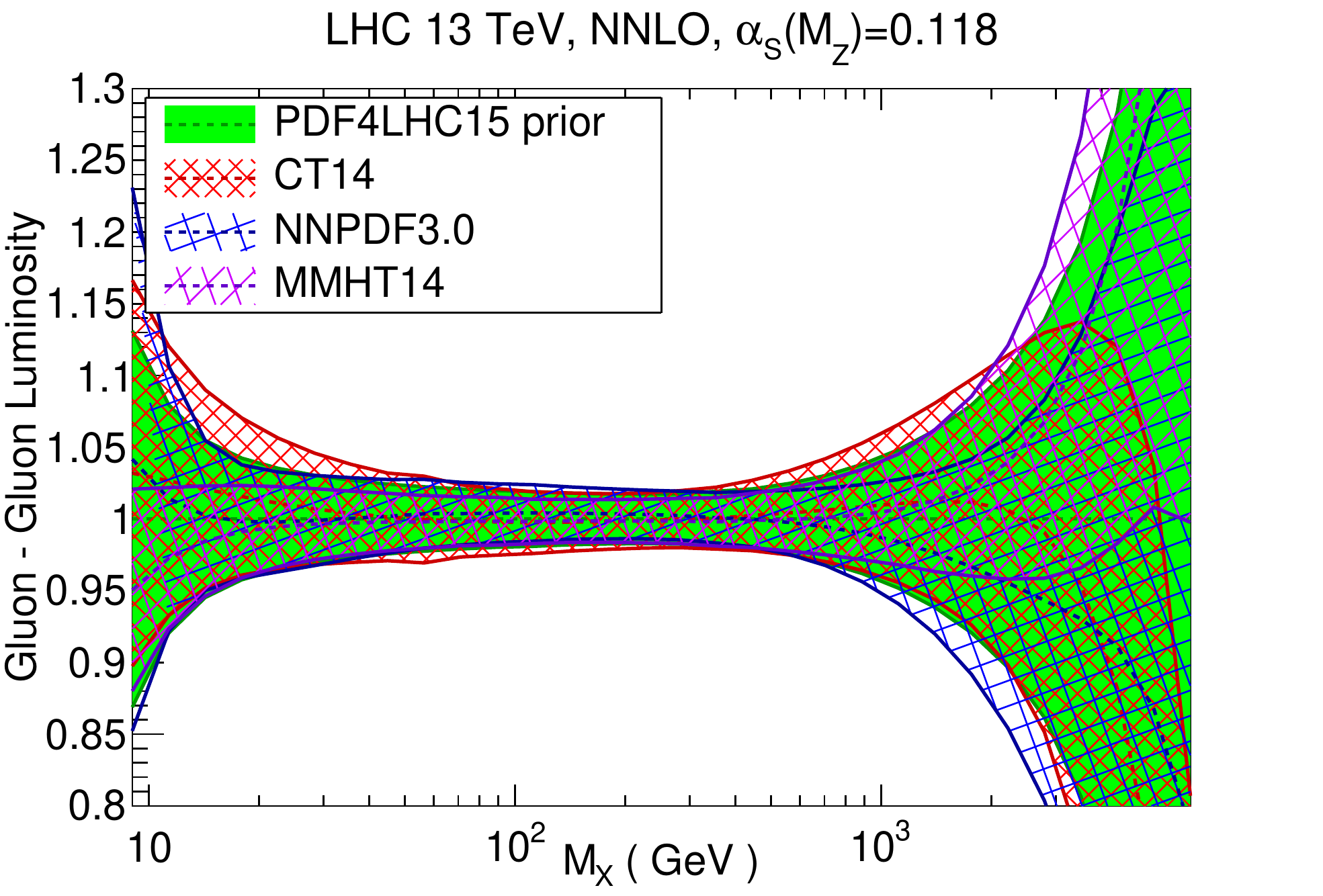}
  \includegraphics[width=.48\textwidth]{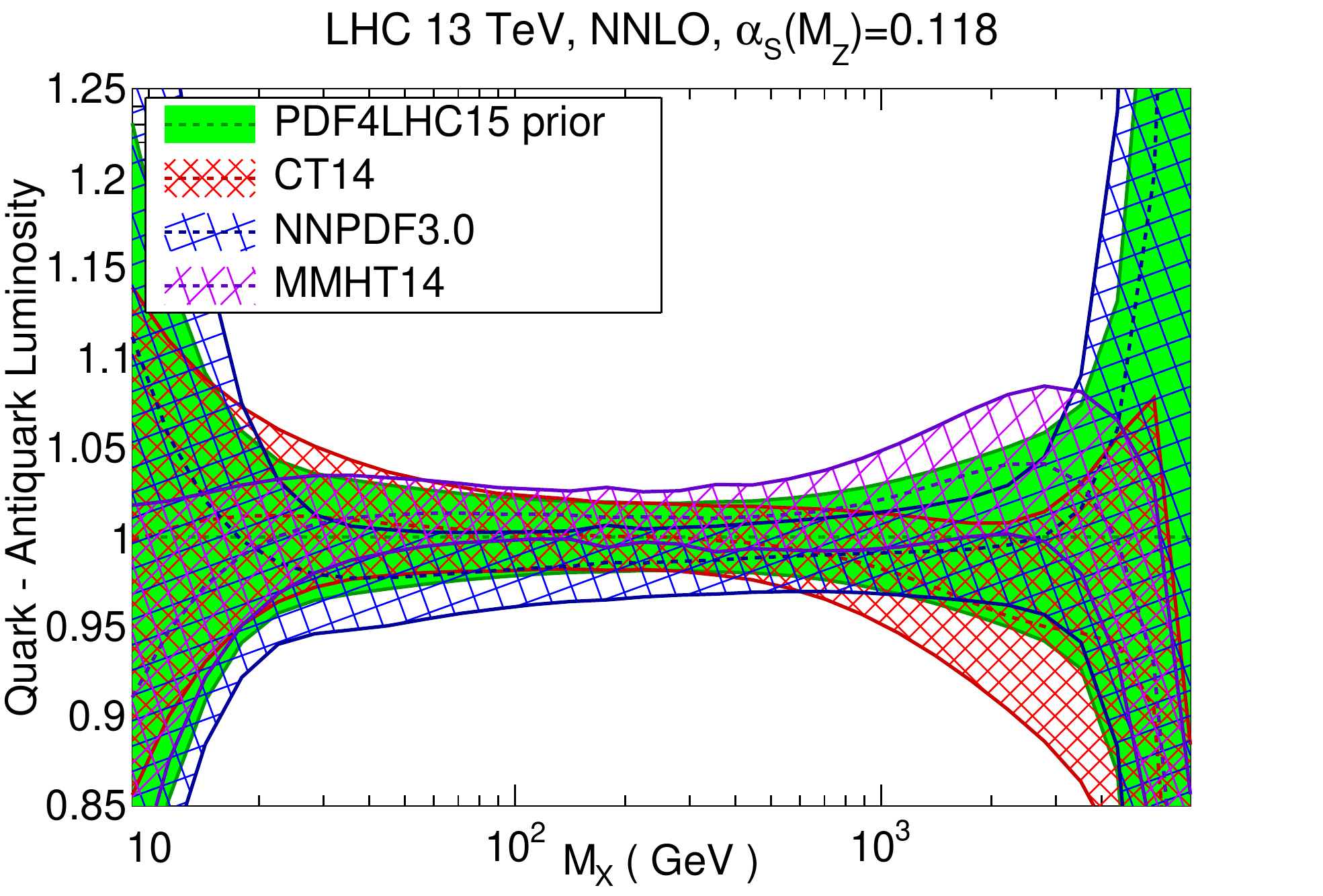}
   \caption{\small Comparison of NNLO parton luminosities at the LHC 13
     TeV. Top: the PDF4LHC15 combined set compared to the CT10,
     MSTW08, and NNPDF2.3 PDF set whose envelope was used as a
     previous PDF4LHC recommendation. Bottom: the PDF4LHC15 combined
     set compared to the three
     individual sets which enter the combination: CT14, MMHT14 and NNPDF3.0.
     The $gg$ (left) and the $q\bar{q}$ (right) luminosities
     are shown as a function of the invariant mass of the final state
     $M_X$, normalized to the central value of
     {\tt PDF4LHC15\_nnlo\_prior}.
}  
\label{fig:lumi_prior_3sets}
\end{figure}

In \refF{fig:lumi_prior_3sets} we compare
the NNLO PDF  luminosities at the LHC 13 TeV
computed using the prior set PDF4LHC15 NNLO, both to the three sets
which were used for the previous PDF4LHC recommendation (CT10, MSTW08
and NNPDF2.3), and the 
three sets which enter the current combination CT14, MMHT14 and NNPDF3.0 NNLO.
     We show the $gg$  and $q\bar{q}$ luminosities
     as a function of the invariant mass of the final state
     $M_X$, normalized to the central value of
     {\tt PDF4LHC15\_nnlo\_prior}.
The improvement in compatibility of the new sets in comparison to the old
ones, especially in the gluon sector, is apparent, particularly
in the precision mass region, say roughly 50 GeV to 1 TeV (more so for gg than for $q\bar{q}$).
Reassuringly, even though uncertainty estimates differ somewhat
between current sets,
especially for quarks, central values of
all sets in this region are in good agreement. Interestingly, they
also agree well with the mid-point of the envelope of the old sets. Hence,
in practice, in the precision region,
the central prediction with the old prescription (the
envelope of CT10, MSTW08, and NNPDF2.3) and the new prescription (the
PDF4LHC15 combined set) are actually quite close.

     There is more disagreement in the low mass region and in the high
     mass region, and the range of uncertainty for the 900 set Monte
     Carlo can be  less than the envelope of the three
     PDF groups.
     This is not surprising, as the combined uncertainty band reflects the common trend of all input PDF ensembles, while the envelope unduly emphasizes extreme behaviour of a few replicas.
While  the combined uncertainty appears to be conservative enough,
  this should be kept in mind especially  in discussions of uncertainties of
     high-mass searches.

\vspace{-2mm}

\section{The PDF4LHC15 PDF sets}

Although the $N_{\rm rep}=900$ Monte Carlo set itself could be used to
determine PDF uncertainties for any LHC process, it suffers from the
drawback of having a very large number of PDFs in the set; also, for
many applications  the non-Hessian framework may be a further drawback.
However, the most essential features of the PDF uncertainties can be
captured using three 
techniques that significantly reduce the number of error PDFs needed,
especially in view of the fact that 
there is an uncertainty in the determination of the PDF uncertainties (witness the differences between the PDF groups at low $x$ and high $x$),
and therefore very high precision is not justified in view of the
limited accuracy. 

Two of these techniques use the Hessian formalism, considering only symmetric PDF uncertainties, while the third technique uses a compressed Monte Carlo technique, which allows for asymmetric uncertainties.
Details of the derivations are provided in the PDF4LHC document.
Correspondingly, three delivery options are available for the
combined sets:
\begin{itemize}
\item {\tt PDF4LHC15\_mc}: contains 100 PDFs, including non-Gaussian features, constructed
  using the CMC method~\cite{Carrazza:2015hva}.
\item {\tt PDF4LHC15\_30}: contains 30 PDFs in a Hessian framework, determined using the META-PDF technique~\cite{Gao:2013bia}.
\item {\tt PDF4LHC15\_100}: contains 100 PDFs in a Hessian framework, determined using the MC2H approach~\cite{Carrazza:2015aoa}.
  \end{itemize}
We will henceforth refer to the starting 900 replica Monte Carlo set
as the ``prior'', from which these reduced sets are constructed: {\tt
  PDF4LHC15\_nnlo\_prior}. 

A central value of $\alpha_s(m_Z)=0.118$ is used for each of these
sets, at both NLO and NNLO, with an uncertainty of $\delta
\alpha_s(m_Z)= 0.0015$ as recommended in  the chapter on Standard Model
parameters of this document. Therefore,  for each option, individual
error sets using
$\alpha_s(m_Z)=0.1165$ and 0.1195 are provided in order to be able to compute
the uncertainty due to $\delta \alpha_s(m_Z)$ in LHC cross-sections, which should be
added in quadrature with the PDF uncertainty~\cite{Butterworth:2015oua}.
It has been verified that addition in quadrature is a good enough
approximation (in some cases exact) to the exact recipes for PDF
and $\alpha_s$ combination provided by each group~\cite{Martin:2009bu,
  Demartin:2010er,Lai:2010nw}.

\begin{figure}[t]
  \centering
  \includegraphics[width=.48\textwidth]{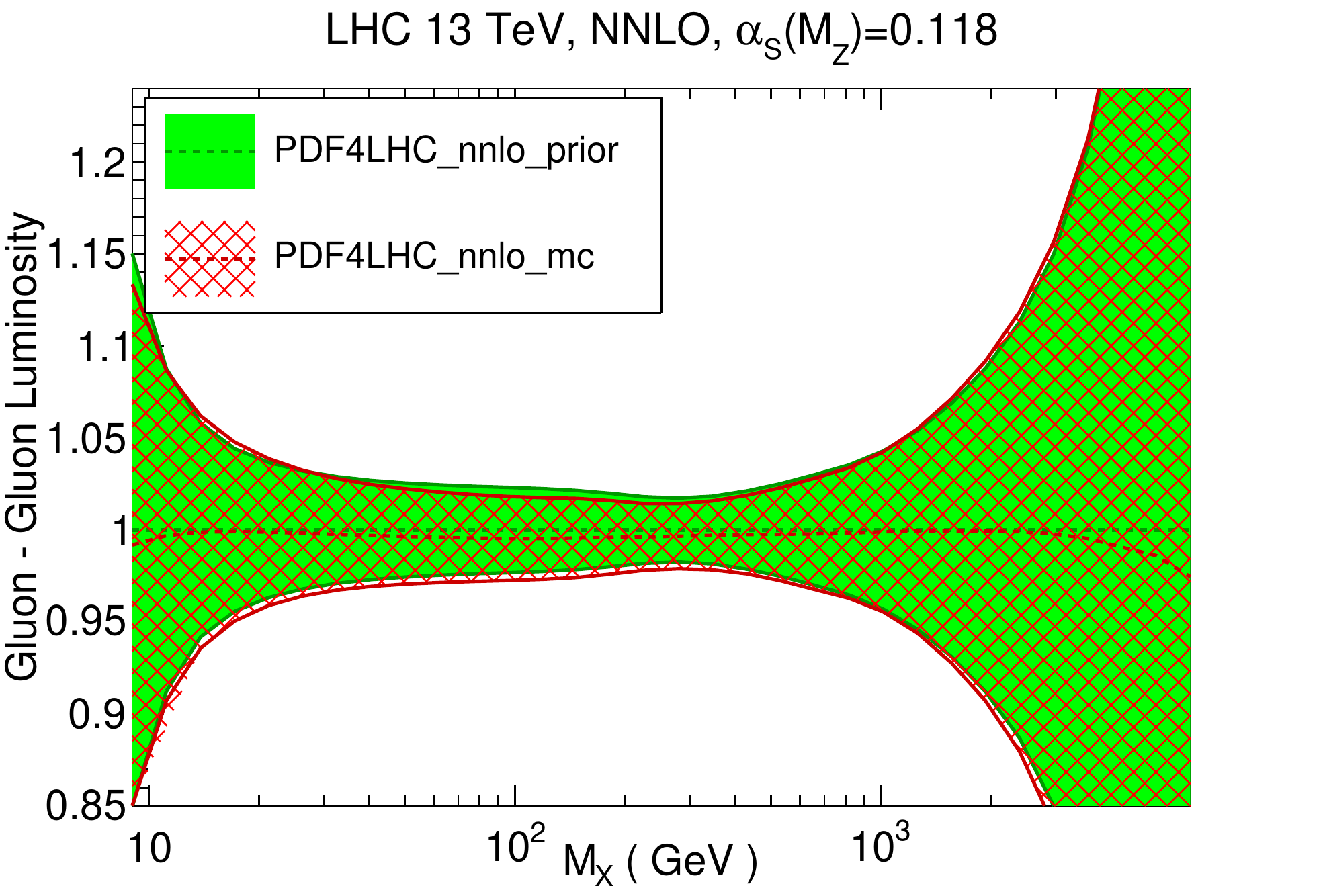}
  \includegraphics[width=.48\textwidth]{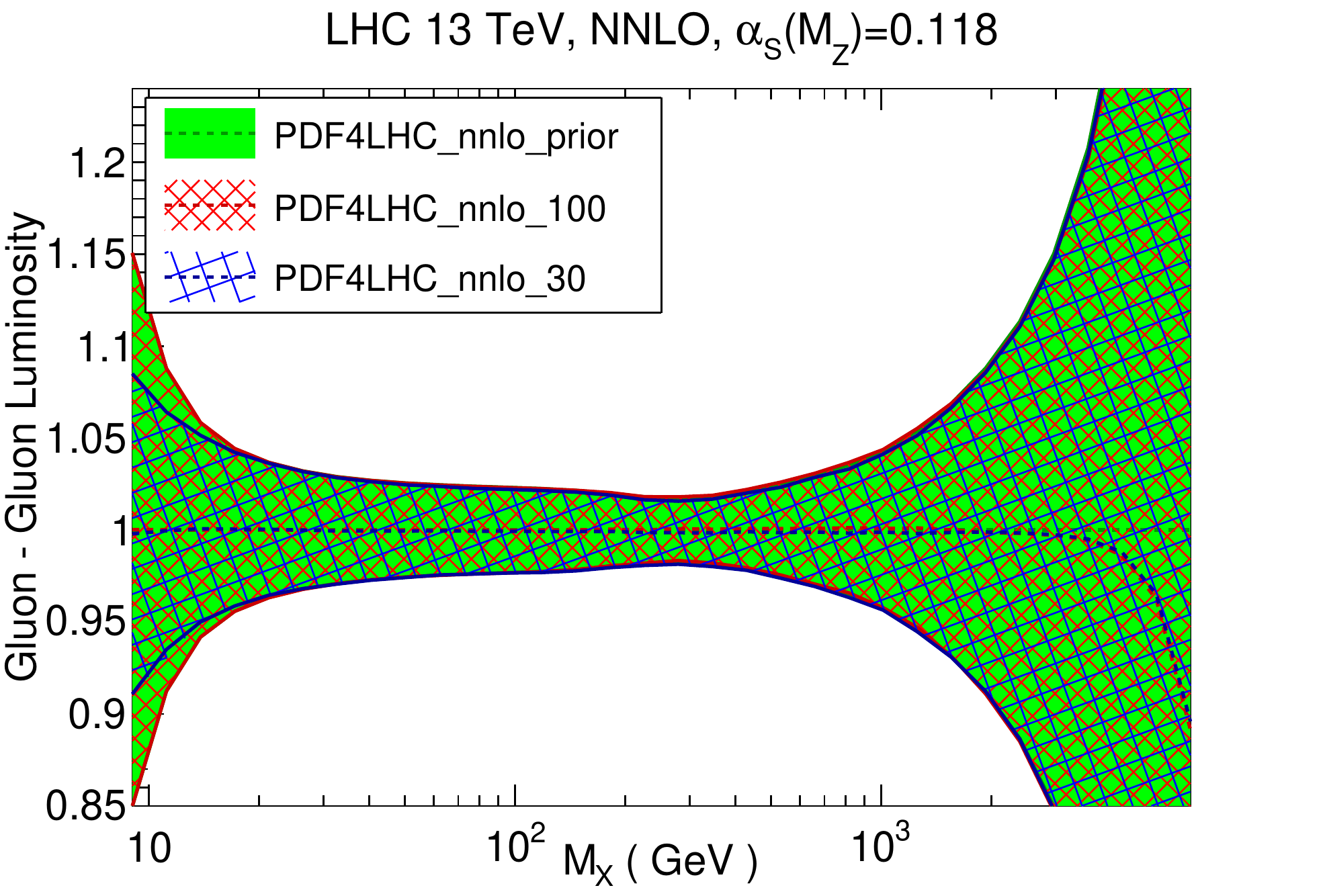}
  \includegraphics[width=.48\textwidth]{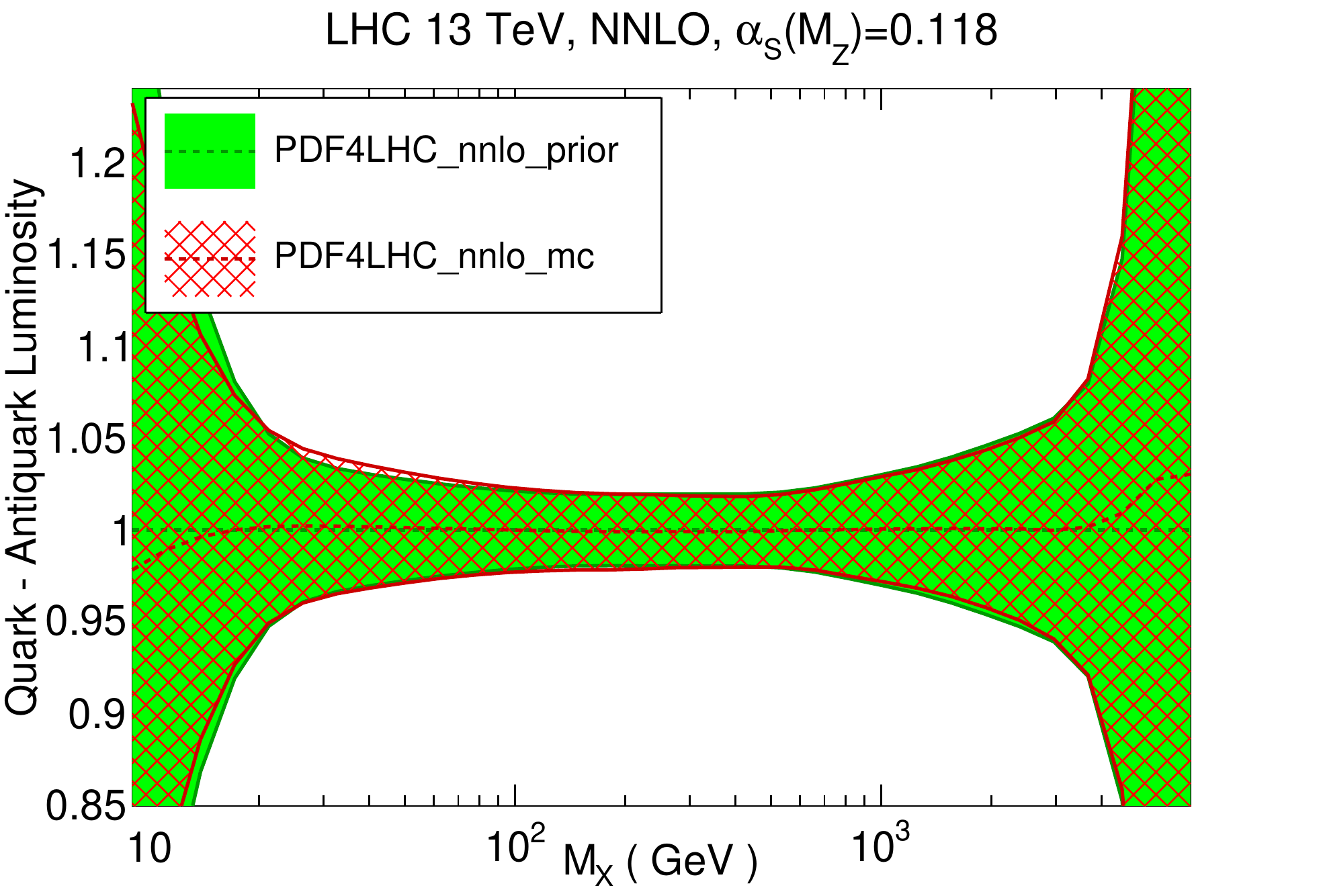}
  \includegraphics[width=.48\textwidth]{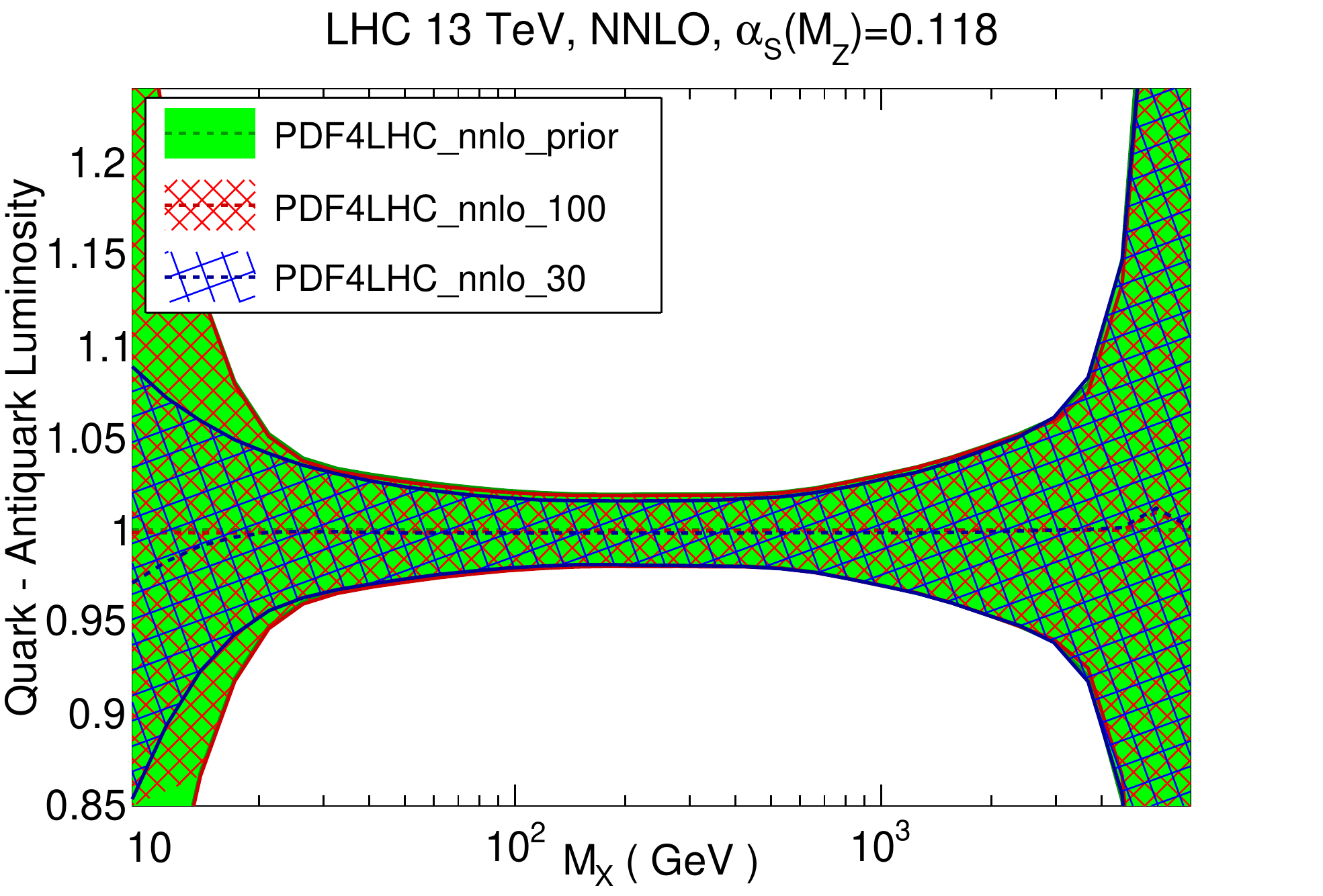}
   \caption{\small Comparison of parton luminosities at the LHC 13 TeV
     computed using the prior set {\tt PDF4LHC15\_nnlo\_prior}
     with its compressed Monte Carlo representation,
     {\tt PDF4LHC15\_nnlo\_mc} (left plots)
     and with  its two Hessian sets,
        {\tt PDF4LHC15\_nnlo\_100} and  {\tt  PDF4LHC15\_nnlo\_30}.
     We show the $gg$ (upper plots) and $q\bar{q}$ (lower plots) luminosities
     as a function of the invariant mass of the final state
     $M_X$, normalized to the central value of
     {\tt PDF4LHC15\_nnlo\_prior}.
}  
\label{fig:lumi_mc900_vs_cmc100}
\end{figure}

The gluon-gluon and quark-antiquark PDF luminosities
as a function of the final-state invariant mass $M_X$ at the LHC 13 TeV are shown
in \refF{fig:lumi_mc900_vs_cmc100}, where we compare
 the prior set {\tt PDF4LHC15\_nnlo\_prior}
with its compressed Monte Carlo representation,
     {\tt PDF4LHC15\_nnlo\_mc} 
     and with  the two Hessian reduced sets,
        {\tt PDF4LHC15\_nnlo\_100} and  {\tt  PDF4LHC15\_nnlo\_30}.
Note that by construction the central values of the two Hessian
        reduced sets coincide with the central value of the prior,
        while the central value of the Monte Carlo set reproduces it
        within the precision of the compression (which is seen to be
        quite high).
All  reduced sets
        correctly reproduce the uncertainty band for the 900 PDF Monte
        Carlo prior in the precision mass region and as the high mass
        region, while the
        {\tt PDF4LHC15\_30} shows a certain loss in precision when
        reproducing uncertainties for the very low mass region.

        The three techniques for delivering the PDF4LHC PDF uncertainties are attempts to match the uncertainty bands produced from the prior, and not the bands from the three PDF groups
        {\it per se}.
        The degree of success is a measure of the precision  of
        the three techniques for this purpose.
        It is therefore important not to confuse the precision of reproducing the prior with accuracy.
        The accuracy of the prior is not exactly known,  especially at
        high and low
mass, and the quoted PDF uncertainty represents only the best estimate by the PDF4LHC group. 

\section{Higgs boson production cross-sections}

\begin{figure}[t]
  \centering
  \includegraphics[width=.49\textwidth]{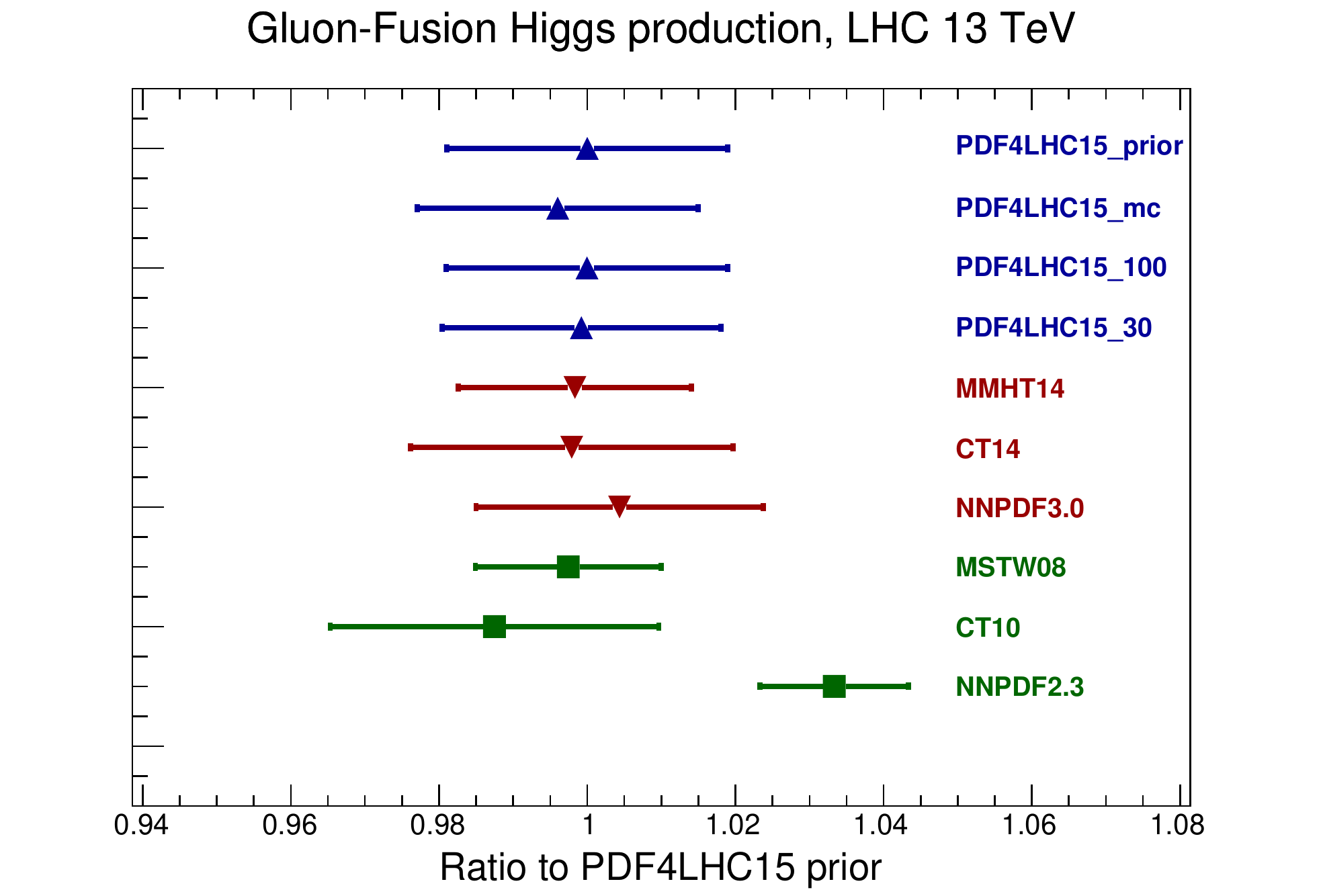}
  \includegraphics[width=.49\textwidth]{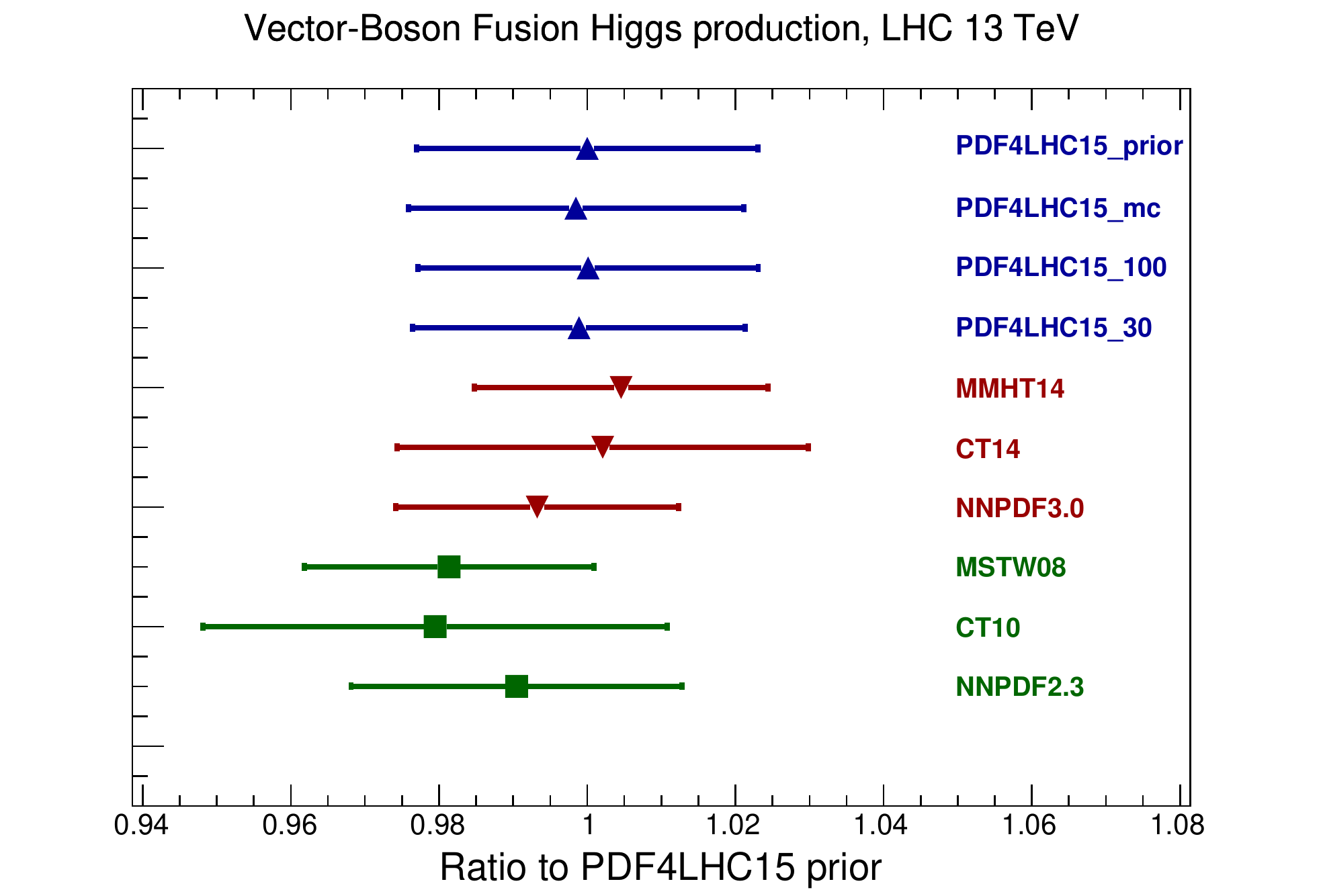}
  \includegraphics[width=.49\textwidth]{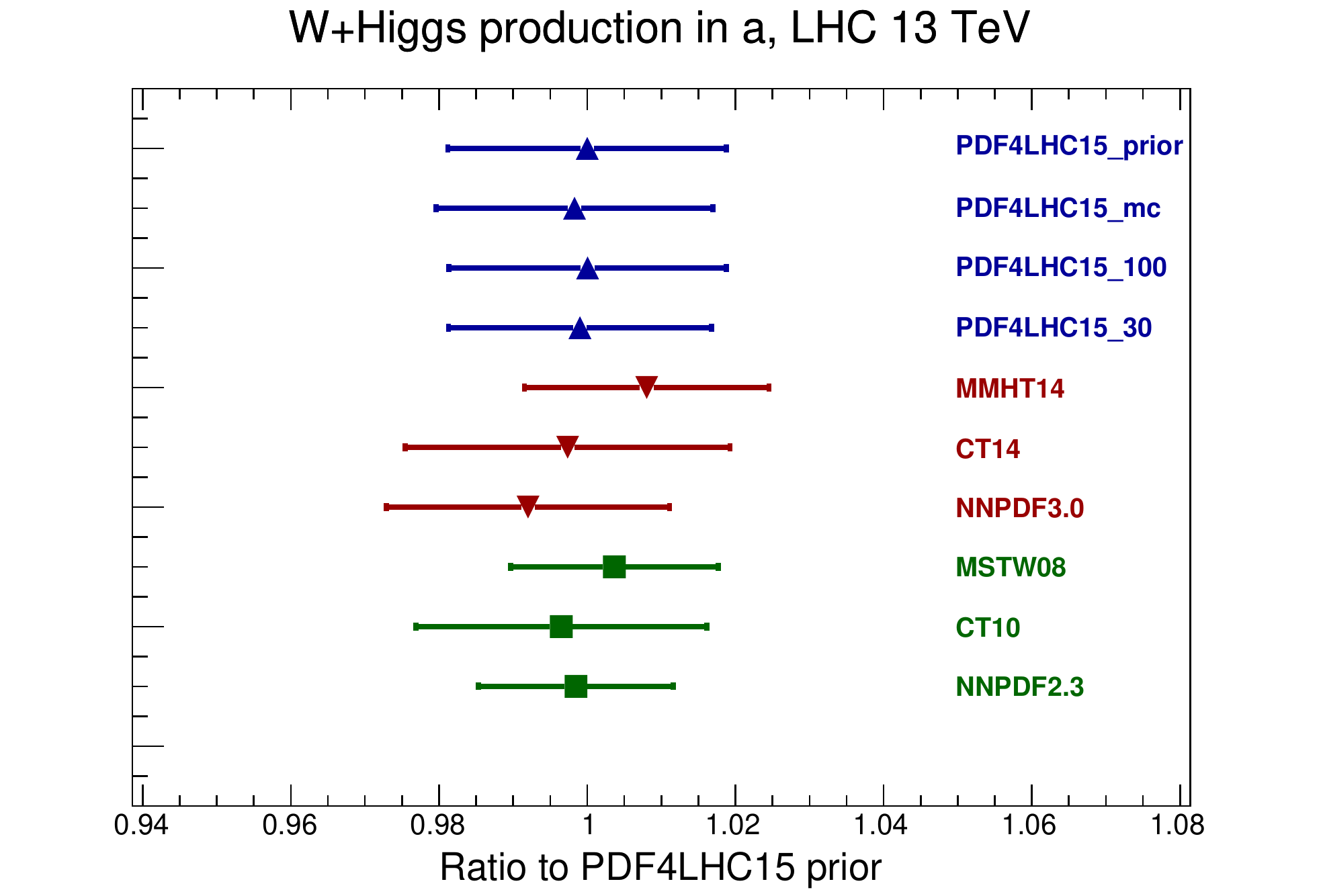}
  \includegraphics[width=.49\textwidth]{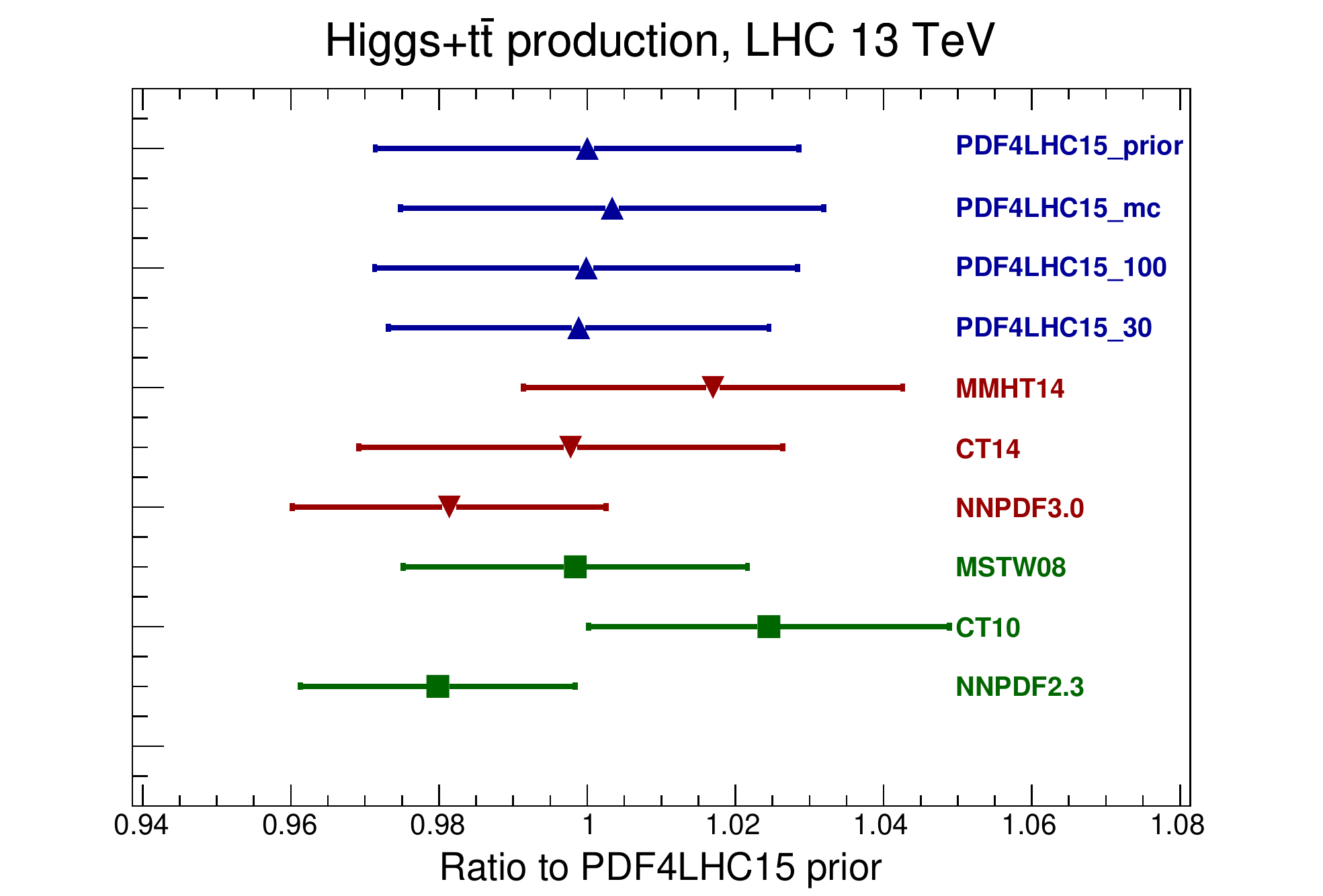}
  \caption{\small Inclusive Higgs boson production cross-sections at 13 TeV
    in the gluon-fusion, vector-boson fusion, associated production with $W$
    and associated production with a $t\bar{t}$ pair channels.
In each case  predictions of the three individual sets are shown along 
with those of the
PDF4LHC15 prior and the three reduced sets,
normalizing to the central value of the PDF4LHC15 prior set.
Predictions obtained using the three older global sets
which entered the previous PDF4LHC recommendation are also show.
All cross-sections are computed at NLO with NNLO PDFs. The value of
the strong coupling is fixed at 
 $\alpha_s(m_Z)=0.118$; the uncertainties shown are PDF uncertainties
(not including the uncertainty due to  $\alpha_s(m_Z)$).
}  
\label{fig:xsec_inclusive}
\end{figure}

        In \refF{fig:xsec_inclusive}
we show representative
inclusive Higgs boson production cross-sections in the relevant production channels
at LHC with $\sqrt{s}=13$ TeV:
gluon-fusion, vector-boson fusion, associated production with a $W$
boson,
and associated production with a $t\bar{t}$ pair.
These calculations have been performed with NLO matrix elements and NNLO PDFs,
using {\tt MG5\_aMC@NLO}~\cite{Alwall:2014hca}
interfaced to {\tt aMCfast}~\cite{Bertone:2014zva} and {\tt
  applgrid}~\cite{Carli:2010rw}, with the purpose of illustrating PDF
uncertainties and also the relative difference between PDF sets.
Indeed, since the NNLO/NLO $K$-factor is to a good approximation independent of PDFs,
it should not affect the relative differences between the predictions of
individual PDF sets.
In this study, the Higgs bosons are left undecayed.
No generation cuts are applied to Higgs bosons, jets or top quarks.
The only selection cut that is applied is given by the fact that we assume
that $W$ and $Z$ bosons decay leptonically, so the corresponding
branching fraction is included and we require $p_T^l\ge 10$ GeV and
$|\eta_l|\le 2.5$ for the charged leptons from the weak boson
decays. All uncertainties shown are pure PDF uncertainties, not
including the uncertainty due to the value of $\alpha_s$, which is
fixed at  $\alpha_s(m_Z)=0.118$ for all cross-sections shown in the plot.

%
In each case, the predictions of the
combined PDF4LHC15 prior and its three reduced versions, 
all normalized to the central value of the prior set, are shown along with
the predictions from the sets which enter the combination, MMHT14, CT14 and NNPDF3.0.
Predictions from the older  global sets,  MSTW08, CT10 and NNPDF2.3,
which entered the previous prescription~\cite{Botje:2011sn} are also
shown for comparison.                 
In particular, 
the better agreement for gluon-gluon fusion prediction using the new generation of PDFs
(CT14, MMHT2014 and NNPDF3.0) compared to the older generation (CT10,
MSTW08 and NNPDF2.3) is evident. In all cases, predictions using the
reduced sets are in excellent agreement with those obtained using the prior.

\begin{figure}[t]
  \centering
  \includegraphics[width=.49\textwidth]{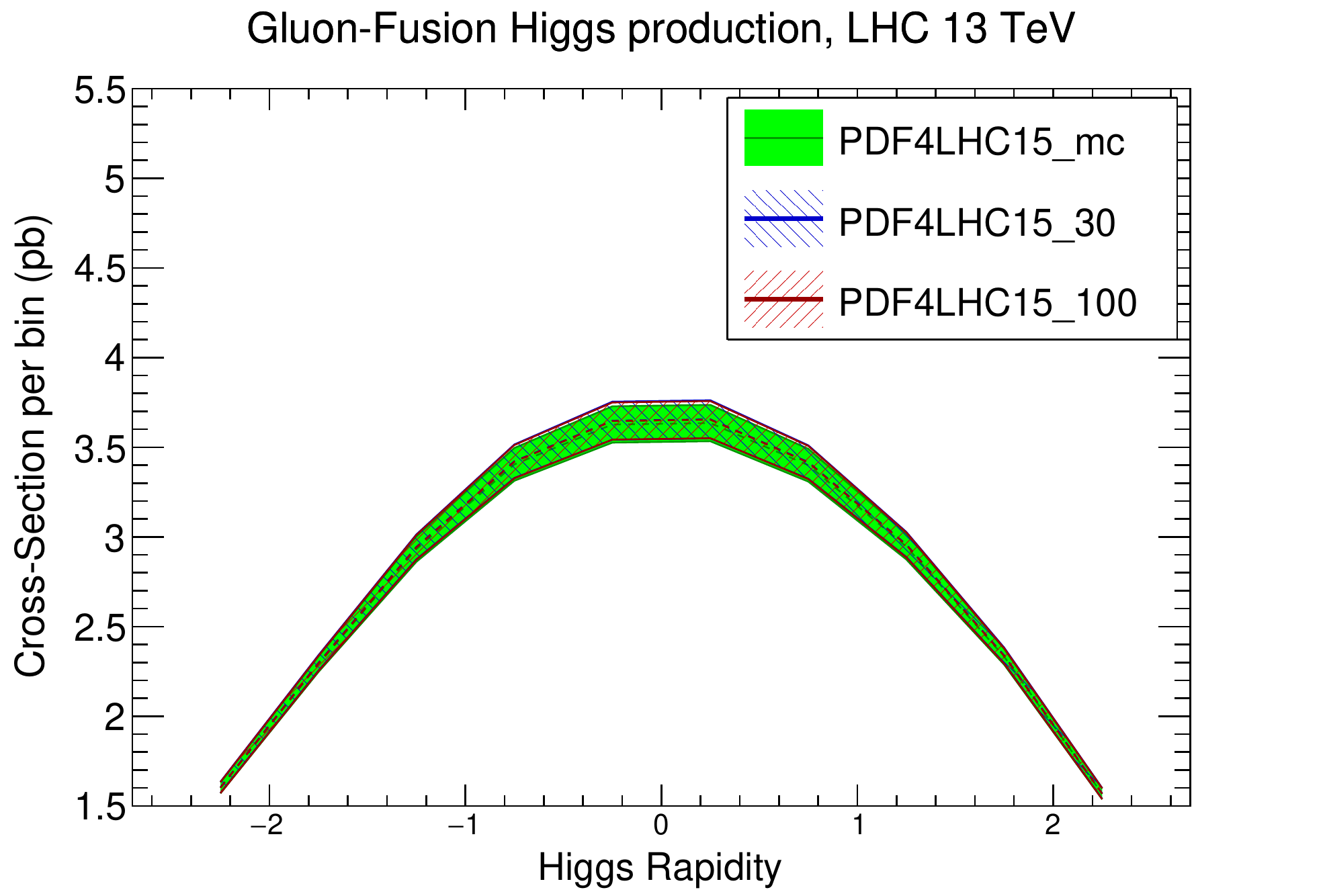}
  \includegraphics[width=.49\textwidth]{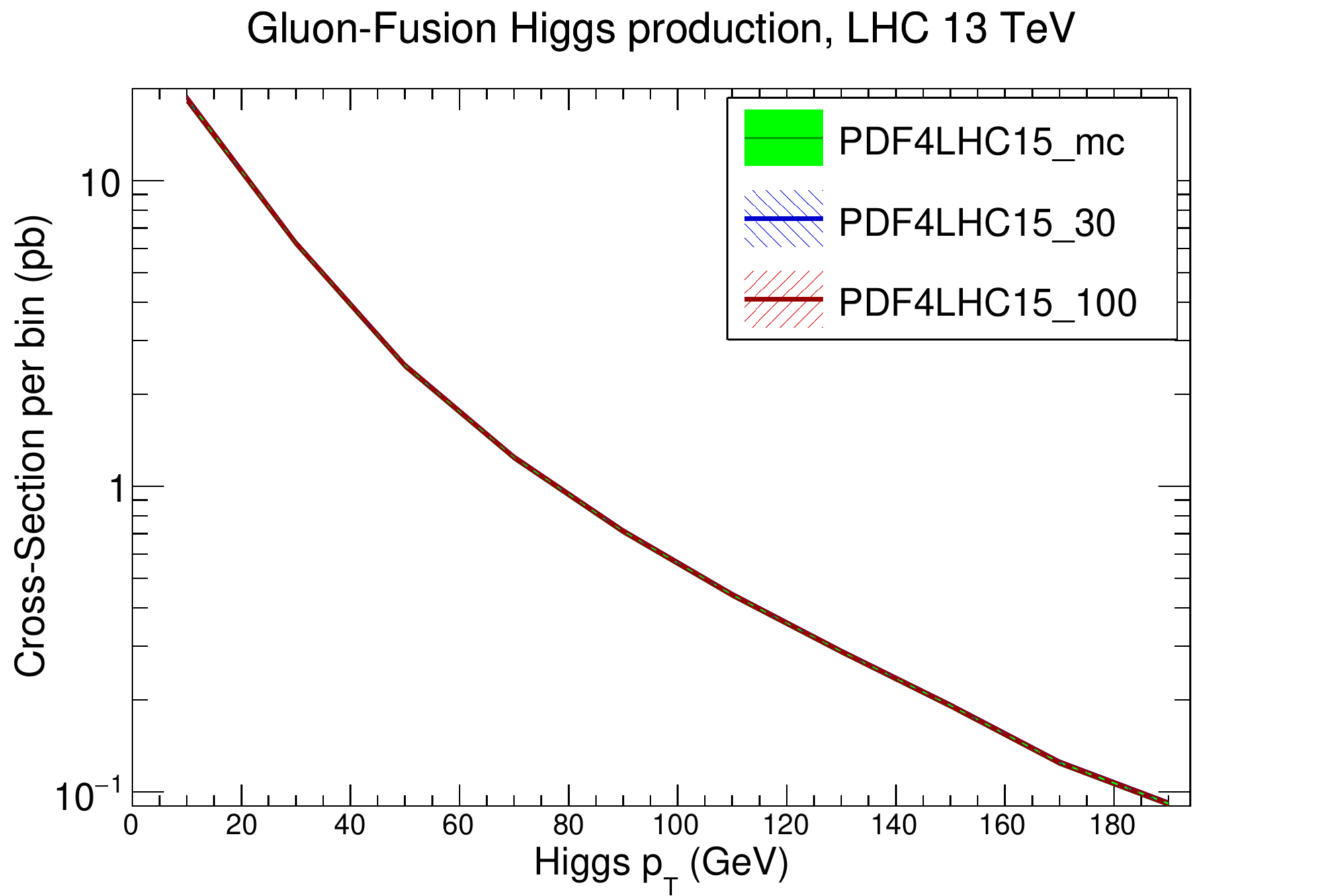}
  \includegraphics[width=.49\textwidth]{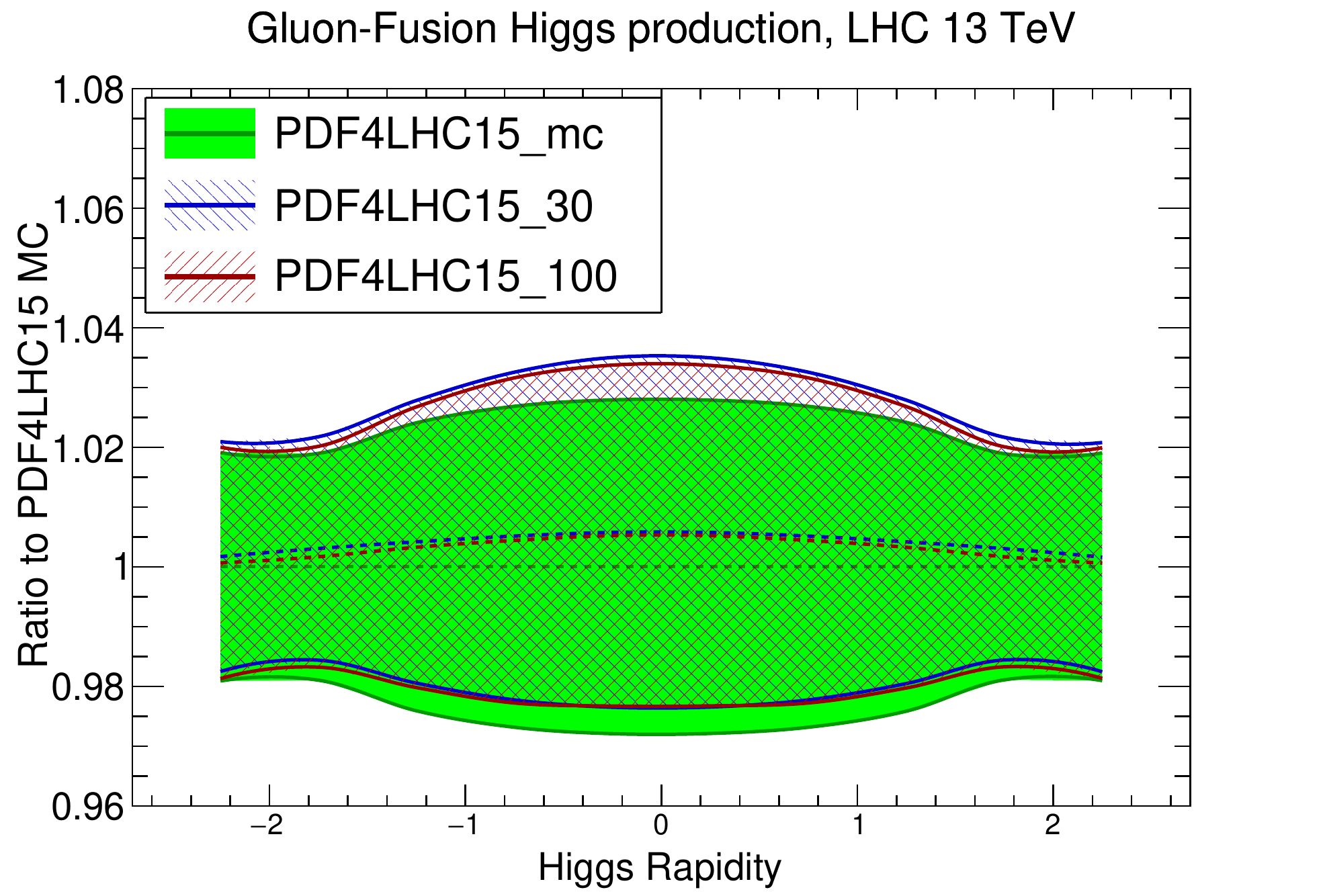}
  \includegraphics[width=.49\textwidth]{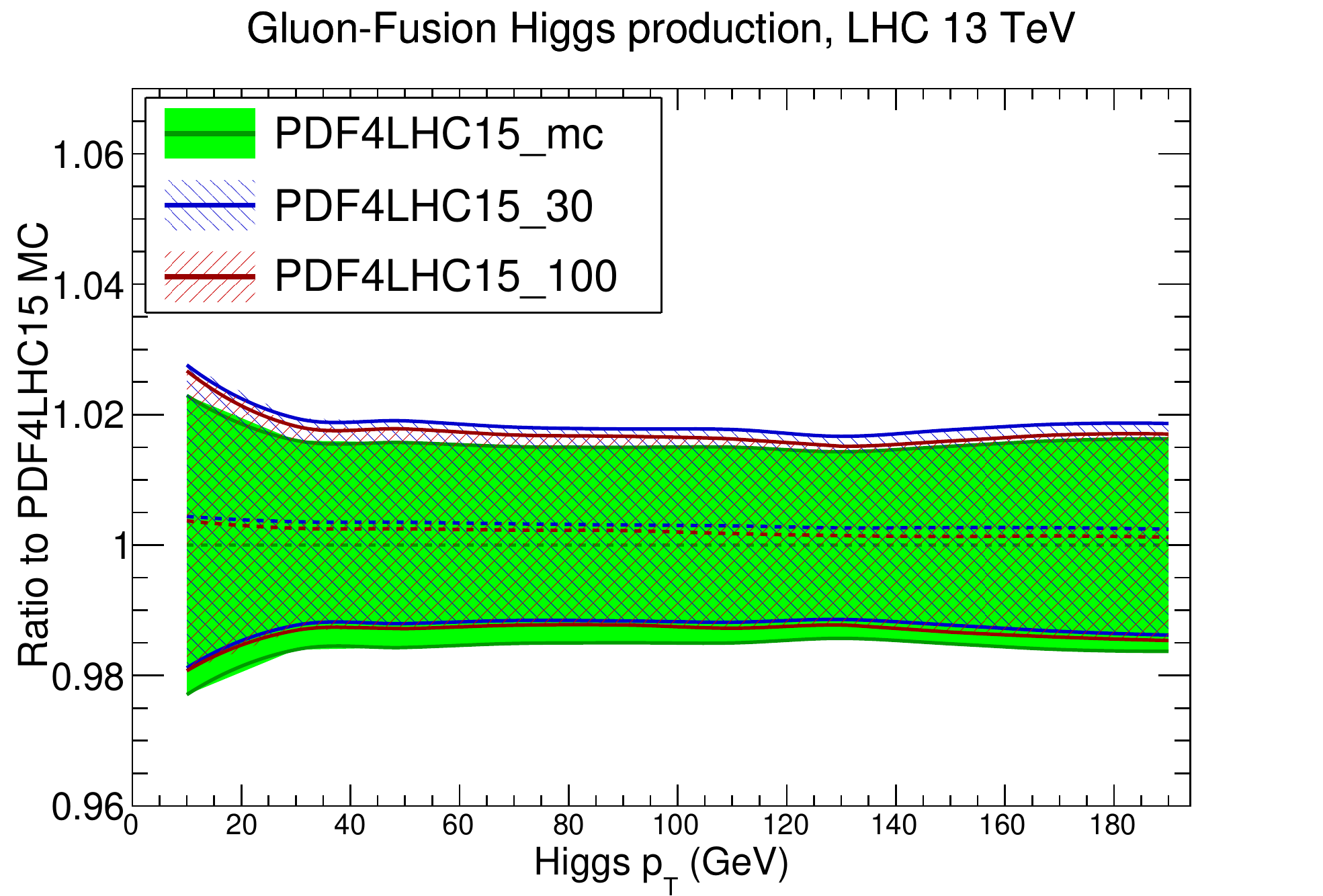}
  \caption{\small Differential distributions for Higgs boson production at gluon fusion
    at $\sqrt{s}=13$ TeV.
    The  Higgs boson rapidity 
    (left) and transverse momentum (right) distributions are shown,
    using the three different deliveries of the combined PDF4LHC15 set.
    The upper plots show the absolute distributions, while in the lower
    plot results are normalized to the central value of the
    {\tt PDF4LHC15\_mc} set.
    Cross-sections have been computed at NLO with NNLO PDFs.
}  
\label{fig:xsec_diff_higgs}
\end{figure}

In \refF{fig:xsec_diff_higgs} we show
representative differential distributions for
the Higgs boson production in gluon fusion, in
particular the transverse momentum and rapidity distributions, at the
LHC $\sqrt{s}=13$ TeV, obtained using the three different deliveries
of the combined set.
The upper plots show the cross-section per bin, in pico-barns, while the lower plots
show the corresponding results normalized to the central value of the
PDF4LHC15 NNLO prior set.
The three techniques agree well for these kinematic distributions, with a small offset 
for the predictions of the {\tt PDF4LHC15\_mc} set. Note that the
transverse momentum distribution shown is a fixed-order result, and
thus it is unreliable for $p_t\lsim30$~GeV where transverse momentum
resummation effects become important.

\section{Strong coupling and heavy quark masses}

\begin{figure}[t]
  \centering
  \includegraphics[width=.49\textwidth]{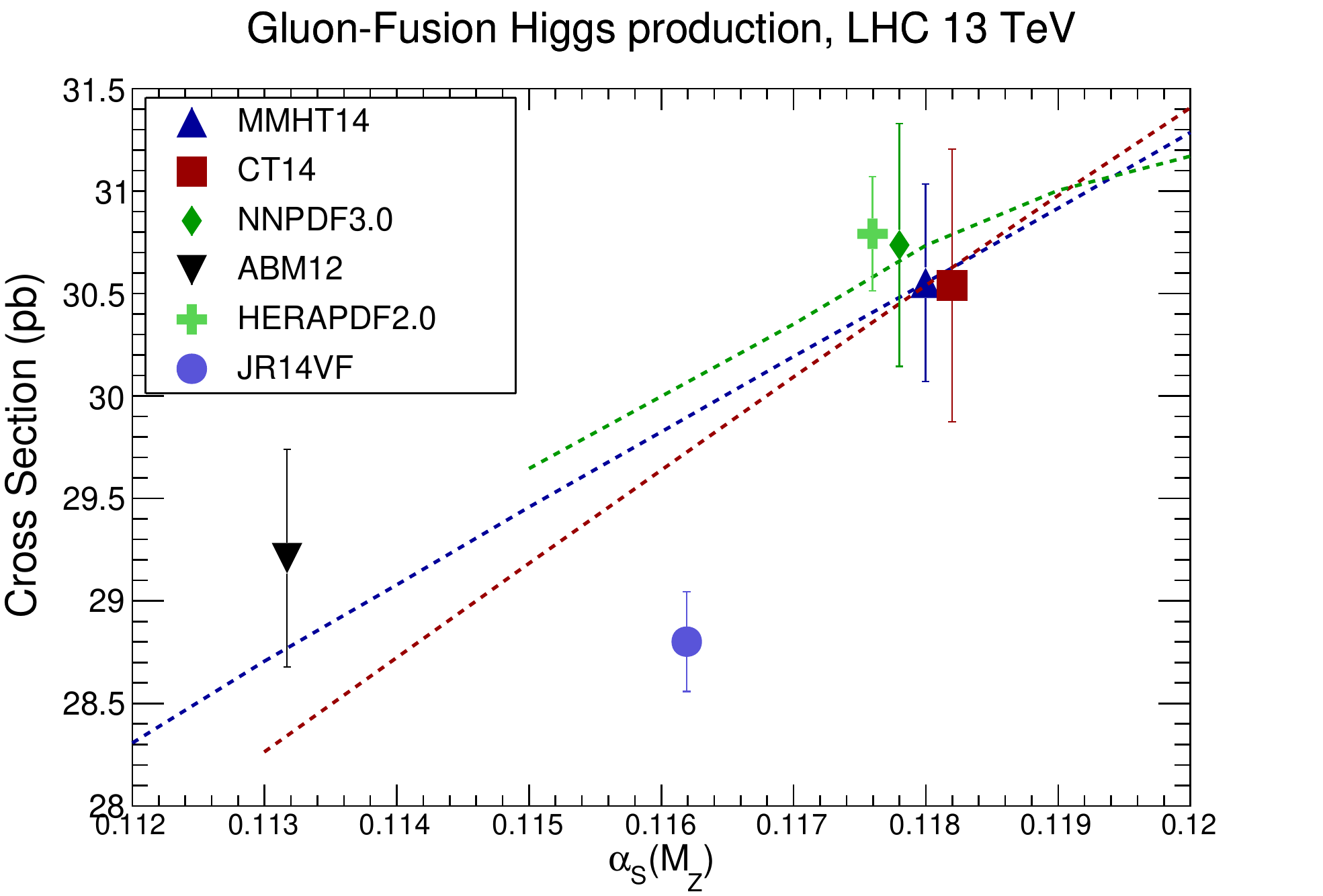}
  \includegraphics[width=.49\textwidth]{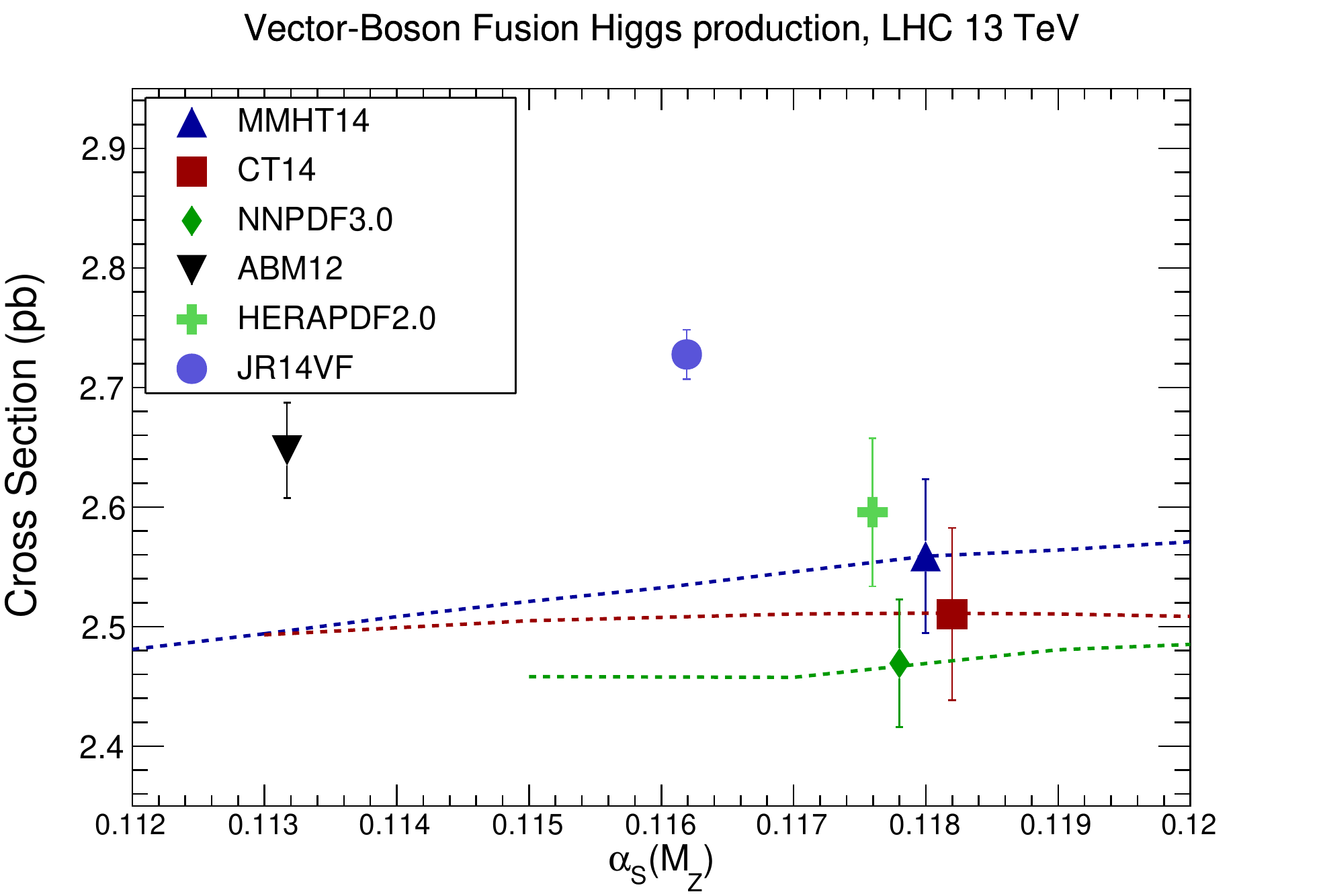}
  \includegraphics[width=.49\textwidth]{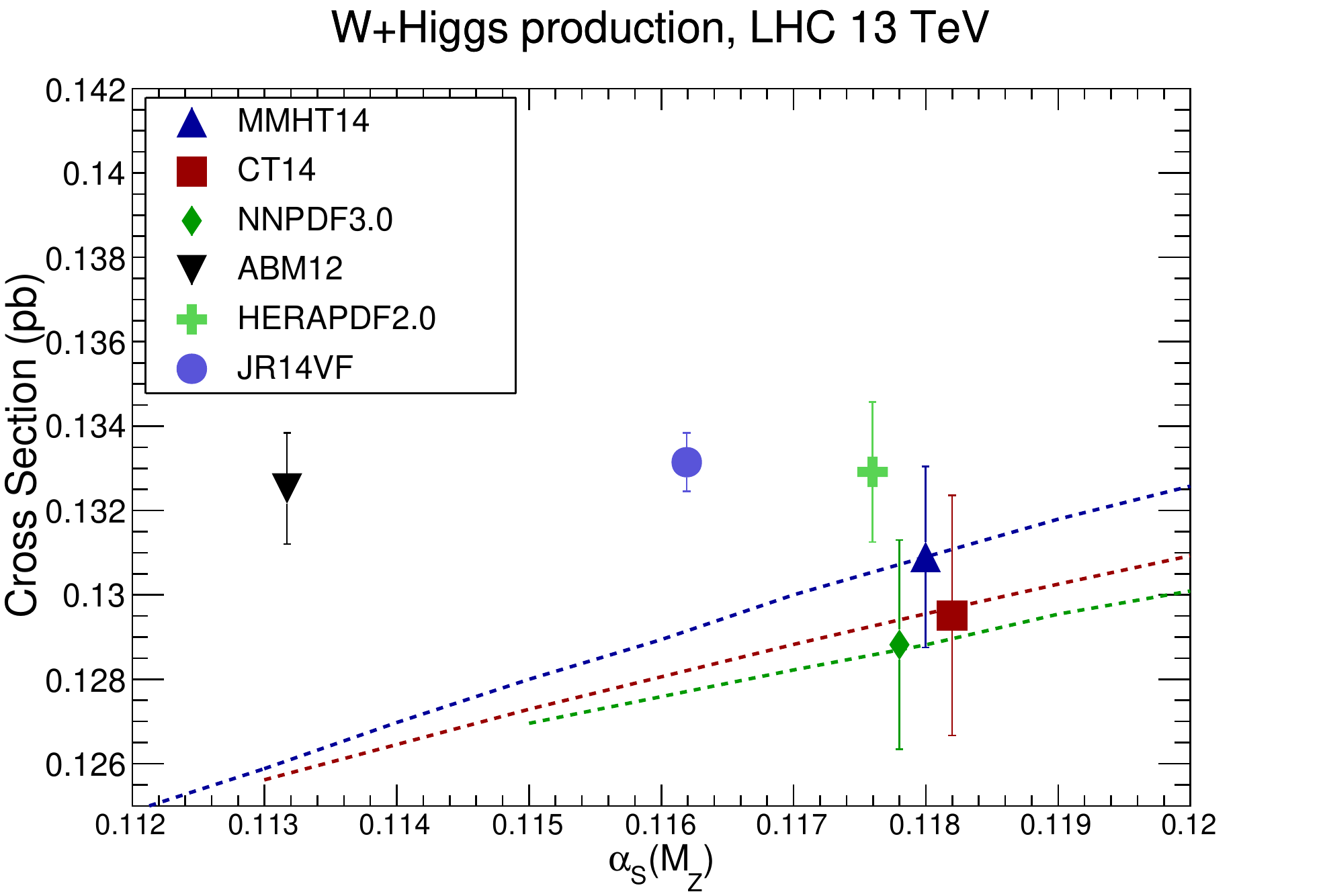}
  \includegraphics[width=.49\textwidth]{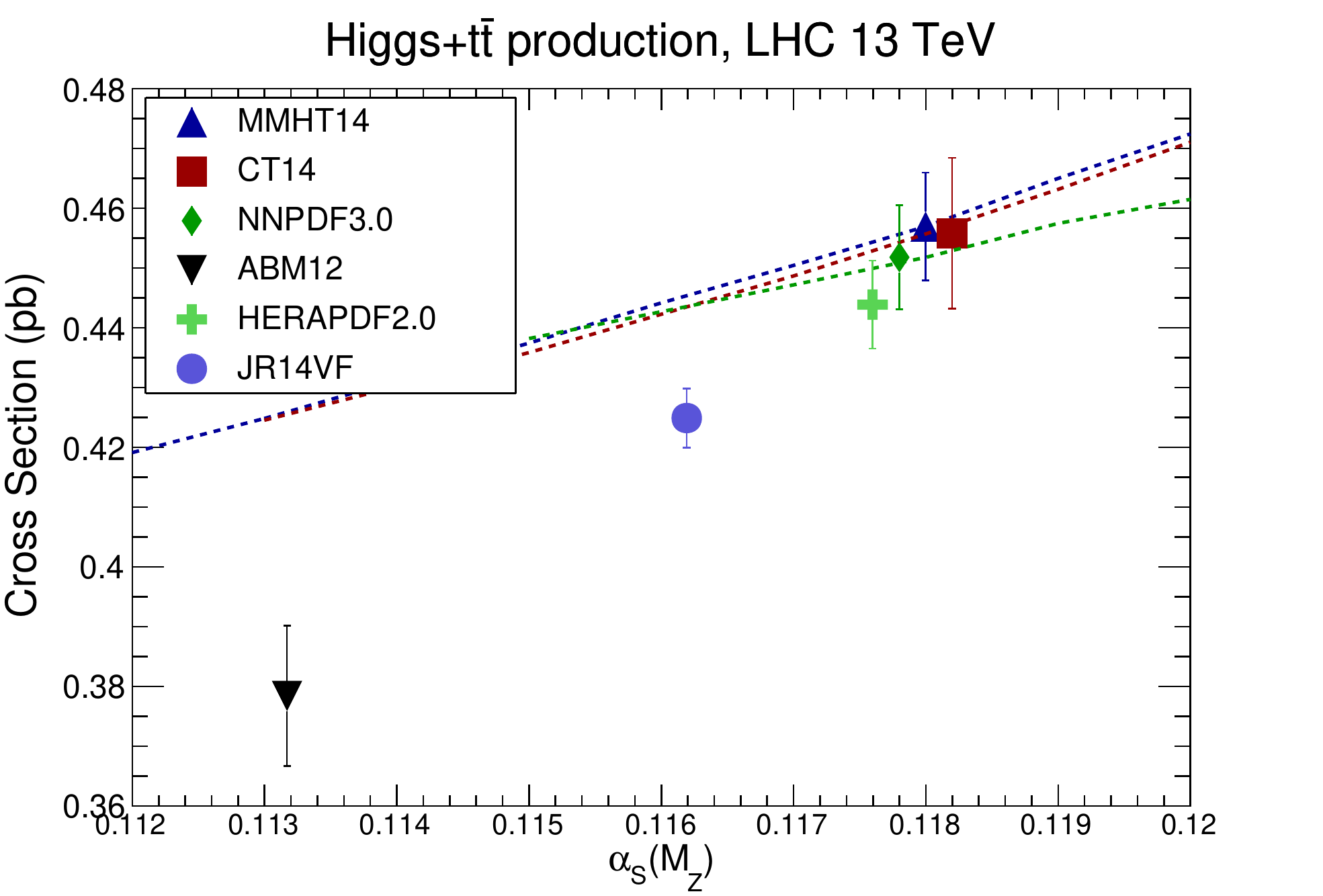}
  \caption{\small Dependence of the total inclusive cross-sections for
    Higgs boson production at $\sqrt{s}=13$ TeV in different production
    channel on the value of the strong coupling  $\alpha_s(m_Z)$ for
    the PDF sets which enter the combination: 
 MMHT14, CT14, NNPDF3.0. Predictions obtained using 
 ABM12, HERAPDF2.0, JR14VF NNLO sets are also shown at their preferred
 $\alpha_s(m_Z)$ value. The  points shown for
MMHT14, CT14, NNPDF3.0 and HERAPDF2.0 all refer to
$\alpha_s(m_Z)=0.118$ and are offset for clarity.
}
\label{fig:xsec_incl_higgs_alphas}
\end{figure}

In order to estimate the further uncertainty due to the choice of
$\alpha_s$ value it is useful to plot the cross-sections as a function
of the value of $\alpha_s(m_Z)$.
This is done in \refF{fig:xsec_incl_higgs_alphas}, where 
the total inclusive cross-sections for
    Higgs boson production at $\sqrt{s}=13$ TeV in different production
    channels are shown as a function of $\alpha_s$ for the three sets
    which enter the combination. In these plots we also include flavour
    predictions obtained using the  ABM12~\cite{Alekhin:2013nda},
    HERAPDF2.0~\cite{Abramowicz:2015mha}
    and JR14~\cite{Jimenez-Delgado:2014twa} PDF sets, each at its
    preferred value of $\alpha_s(m_Z)$.
    In the case of ABM12, we use the $N_f=5$ set.
    For HERAPDF2.0, we consider only the experimental PDF
    uncertainties.
    In the case of the JR14 set, we use the version determined
    in the variable-flavour-number scheme.
    Note that  MMHT14, CT14, NNPDF3.0 and HERAPDF2.0 all
  use the same central value of the strong coupling,
    $\alpha_s(m_Z)=0.118$; in the plot the values corresponding to
  these sets are slightly offset to improve readability.  It is
  apparent from the figures which PDF sets can produce predictions
  that may fall substantially outside of the uncertainties of the
  three PDF sets that enter the 2015 combination. 

Unlike the case for the strong coupling $\alpha_s(M_Z)$, the different PDF 
groups do not use common values of the charm and bottom masses, and also
use different definitions of a general mass variable number scheme (GM-VFN). These
are two distinct issues, particularly since each group chooses the quark 
masses at fixed default values, as opposed to trying to determine them from 
a best fit, and the values chosen have no relation to the scheme choice. 

Let us consider the issue of scheme choices first. Dependence on 
these has been very thoroughly studied in numerous articles \cite{LHhq,Guzzi:2011ew,
Thorne:2012az, Gao:2013wwa, Andersen:2014efa}. At NLO the variation in LHC 
cross section predictions for $W$ and $Z$ production
due to quite extreme differences  in  choices of GM-VFN schemes
can be of order $2-3\%$; they may be somewhat larger but still
moderate especially at high scale for processes which are directly
sensitive to charm, such as  $Z+c$ or
open charm production. However, as with other scheme choices in QCD, the 
ambiguity at fixed order is always an effect beyond the order of the 
calculation, and hence diminishes as 
one goes to higher orders. At NNLO scheme choices lead to changes in 
LHC cross section predictions of generally no more than $1\%$, and very 
often less.
This can be appreciated from Figs. 11.6, 11.7 and 11.8 in 
\cite{Andersen:2014efa}, where differences between groups for the 
total HERA cross sections calculated using the same PDFs but different schemes 
can be at most $>5\%$ at NLO, but never more than $1-2\%$ at
NNLO. Differences in $F_2^{\bar cc}(x,Q^2)$, an observable
which is directly sensitive to charm,  which is much less precisely and 
widely measured at HERA, can be $30\%$ in extreme cases at NLO but are
rarely more than $10\%$ at NNLO, and over most of the $x,Q^2$ range are 
much less than 
$5\%$.  Hence, the variation due to the choice of GM-VFN scheme
at NNLO is much less than the PDF 
uncertainty, and the variation between groups due to this can be taken as 
indicative of part on the theoretical uncertainty at NNLO. The
variation due to adopting a FFN scheme would instead be quite large,
and outside the PDF uncertainty.

The different PDFs used in the recommendation are all obtained using 
the heavy quarks defined in the pole mass 
scheme. However, the values chosen are different, with $m_c$ ranging from 
1.275--1.4~GeV and $m_b$ from 4.18--4.75~GeV. The precise determinations of 
quark masses are performed in the $\overline{\rm MS}$
 scheme, and the conversion to 
the pole masses is imprecise due to a renormalon ambiguity in the conversion 
factor. In particular, the series for the charm quark shows essentially no
convergence. Using the better behaved expression for the beauty mass, and 
the fact that $m_b^{\rm pole}-m_c^{\rm pole}=3.4~{\rm GeV}$ with a very small 
uncertainty \cite{Bauer:2004ve,Hoang:2005zw}, it was argued in 
\cite{Martin:2010db} that a reasonable estimate for pole masses is 
$m_c^{\rm pole}=1.5 \pm 0.2 ~{\rm GeV}$ and $m_b^{\rm pole}=4.9\pm 
0.2~{\rm GeV}$. Hence, the charm mass values chosen are perhaps slightly low, 
but not anomalous. The smallest $m_b^{\rm pole}$ value among the three
combined sets is somewhat low, but 
the beauty data (including the contribution to total HERA cross sections) 
has extremely little constraint on PDFs in the global fit. Moreover, it has 
been argued that at lower orders a general mass variable flavour number 
scheme is not very sensitive to the scheme in which the mass is defined
\cite{Bertone:2015gba,Harland-Lang:2015qea}. The variation in the quark 
masses between groups, i.e. the deviations from the mean values, 
is relatively small compared to the intrinsic 
uncertainty for $m_c$, but a bit larger for $m_b$. As shown in 
\cite{Ball:2011mu,Gao:2013wwa,Harland-Lang:2015qea}, the Higgs cross section 
via $gg$ fusion
can vary by about $1\%$ for $m_c$ changes of about 0.2~GeV, while variations 
with $m_b$ are much smaller than this, even for changes of 0.5~GeV. 
Hence, 
the variation in predictions between the groups due to the different quark 
masses is generally much less than the
PDF uncertainty, with the exception of cross sections directly dependent
on the $b$ quark distribution, where the mass effect on the distribution 
is more significant. 
The uncertainty due to quark masses should ideally 
be taken into account, and the current variation between groups should 
achieve this to some extent. However, in the future, it is probably preferable 
to settle on common mass values, perhaps defined in the $\overline{\rm MS}$
 scheme as 
advocated in \cite{Alekhin:2010sv}, and a common uncertainty, as now done for
$\alpha_s(M_Z)$.

For LHC calculations that are done in the $\overline{MS}$ scheme
with up to 4 active quark flavours in the running $\alpha_s $ and PDFs,
three combined  PDF4LHC sets determined in this
scheme are also provided.
 The respective PDF4LHC sets are constructed from 900 MC replicas
of CT14, MMHT14, and NNPDF3.0 PDFs for $N_f=4$ using the same 
combination techniques as for $N_f=5$. 
In this case, the initial PDF parameterizations
from the $N_f=5$ fits at initial $Q_0 \sim m_c$ are evolved to higher
$Q$ including the lightest 4 flavours only. Contributions from massive
bottom and top quarks should be then included in hard matrix
elements. The input value $\alpha_s(M_Z, N_f=4)$ 
in the $N_f = 4$ scheme  is obtained from $\alpha_s(M_Z,
N_f=5)=0.118$ by applying scheme transformation relations
\cite{Chetyrkin:2005ia} at two or three loops in QCD, 
and assuming the average $m_b=4.56$ GeV of the input PDF sets; it is
thus rather smaller than the default value 0.118.

\section{PDF correlations}

Also of importance for Higgs boson predictions and analyses are the PDF correlations,
both among Higgs boson production processes and between Higgs boson
processes and potential background processes: tables of correlations
obtained using various PDF sets were given in the previous Yellow
Report~\cite{Heinemeyer:2013tqa}. These tables can now be updated
using the more recent combined set.
In \refT{tab:correlation1}
we collect the correlation coefficients between different Higgs boson production
channels, as well as between representative Higgs boson signal and background processes.
We show the results for the PDF4LHC15 NNLO prior and for
the three reduced combined PDF sets, and
we also include the results for the three individual PDF sets.
The cross-sections have been computed at NLO with NNLO PDFs, using the same settings
as in previous plots.
All of the techniques do reasonably well reproducing the correlations of the prior, with the
{\tt PDF4LHC15\_100} PDFs reproducing the prior to within a per cent.  

It should be emphasized that 
the values of the correlation themselves, however, can be viewed as only having  a single digit (or less) accuracy in the sense that the PDF correlations for Higgs  processes and backgrounds for the 3 global PDF sets can differ by the order of 0.2 (or more).
For example, the spread in correlation coefficients for gluon-gluon fusion production and associated ($Zh$) production is 0.67.
Note that the differences in the correlation coefficients between
NNPDF3.0, CT14 and MMHT14 are large in many cases, though there
is also good agreement in other cases.

\begin{table}[t]
\caption{\small \label{tab:correlation1}
  Correlation coefficient between various Higgs boson production cross-sections
  and background cross-sections.
  In each case,
  the PDF4LHC15 NNLO prior set is compared to the Monte Carlo and with
  the two Hessian reduced sets.
  We also show the results from the three individual sets, CT14, MMHT14 and NNPDF3.0.}
  \footnotesize
  \begin{centering}
    \begin{tabular}{l|c|c|c|c|c|c}
\toprule
\multirow{2}{*}{PDF Set} & \multicolumn{5}{c}{correlation coefficient} & \tabularnewline
 & $t\bar{t},Ht\bar{t}$ & $t\bar{t},hW$ & $t\bar{t},hZ$ & $ggh,ht\bar{t}$ & $ggh,hW$ & $ggh,hZ$\tabularnewline
\midrule
{\tt PDF4LHC15\_nnlo\_prior} & 0.87 & -0.23 & -0.34 & -0.13 & -0.01 & -0.17\tabularnewline
{\tt PDF4LHC15\_nnlo\_mc} & 0.87 & -0.27 & -0.35 & -0.10 & 0.07 & -0.01\tabularnewline
{\tt PDF4LHC15\_nnlo\_100} & 0.87 & -0.24 & -0.34 & -0.13 & -0.02 & -0.17\tabularnewline
{\tt PDF4LHC15\_nnlo\_30} & 0.87 & -0.27 & -0.43 & -0.13 & -0.04 & -0.23\tabularnewline
\midrule
CT14     &  0.09  &	-0.32 &	-0.44 &	-0.26 &	-0.03  &	-0.18 \tabularnewline
MMHT14   &  0.90 &	-0.22 &	-0.52 &	0.08  &	-0.18  & 	-0.33 \tabularnewline
NNPDF3.0 &  0.90 &	-0.17 &	-0.21 &	0.18  &	0.52   & 	0.49 \tabularnewline
\bottomrule
\end{tabular}\\[0.3cm]

\begin{tabular}{l|c|c|c|c|c|c}
\toprule
\multirow{2}{*}{PDF Set} & \multicolumn{6}{c}{correlation coefficient}\tabularnewline
 & $Z,W$ & $Z,t\bar{t}$ & $Z,ggh$ & $Z,ht\bar{t}$ & $Z,hW$ & $Z,hZ$\tabularnewline
\midrule
{\tt PDF4LHC15\_nnlo\_prior} & 0.89 & -0.49 & 0.08 & -0.46 & 0.56 & 0.74\tabularnewline
{\tt PDF4LHC15\_nnlo\_mc} & 0.90 & -0.44 & 0.18 & -0.42 & 0.62 & 0.80\tabularnewline
{\tt PDF4LHC15\_nnlo\_100} & 0.91 & -0.48 & 0.09 & -0.46 & 0.59 & 0.74\tabularnewline
{\tt PDF4LHC15\_nnlo\_30} & 0.88 & -0.63 & 0.04 & -0.61 & 0.56 & 0.72\tabularnewline
\midrule
CT14     & 0.92  &	-0.69 &	0.12  &	-0.69  &	0.69 &	0.77   \tabularnewline
MMHT14   & 0.76	&  -0.70      &	0.12  &	-0.83  & 	0.15 & 0.43  \tabularnewline
NNPDF3.0 & 0.96	&  -0.13      &	0.62  &	-0.30  & 	0.84 &	0.85  \tabularnewline
\bottomrule
\end{tabular}\\[0.3cm]

\begin{tabular}{l|c|c|c|c|c|c}
\toprule
\multirow{2}{*}{PDF Set} & \multicolumn{5}{c}{correlation coefficient} & \tabularnewline
 & $W,t\bar{t}$ & $W,ggh$ & $W,ht\bar{t}$ & $W,hW$ & $W,hZ$ & $t\bar{t},ggh$\tabularnewline
\midrule
{\tt  PDF4LHC15\_nnlo\_prior} & -0.40 & 0.20 & -0.40 & 0.76 & 0.77 & 0.30\tabularnewline
{\tt  PDF4LHC15\_nnlo\_mc }& -0.44 & 0.26 & -0.42 & 0.81 & 0.82 & 0.32\tabularnewline
{\tt PDF4LHC15\_nnlo\_100} & -0.40 & 0.20 & -0.40 & 0.76 & 0.77 & 0.30\tabularnewline
{\tt PDF4LHC15\_nnlo\_30} & -0.47 & 0.19 & -0.47 & 0.77 & 0.76 & 0.31\tabularnewline
\midrule
CT14     &  -0.56  &	0.22  &	-0.56  &	0.80 &	0.77 &	0.09   \tabularnewline
MMHT14   &  -0.47  &	0.24  &	-0.53  & 	0.62 &	0.63 &	0.46  \tabularnewline
NNPDF3.0 &  -0.08  &	0.64  &	-0.26  & 	0.88 &	0.86 &	0.51  \tabularnewline
\bottomrule
\end{tabular}
\par\end{centering}
\end{table}

\begin{figure}[t]
  \centering
  \includegraphics[width=.60\textwidth]{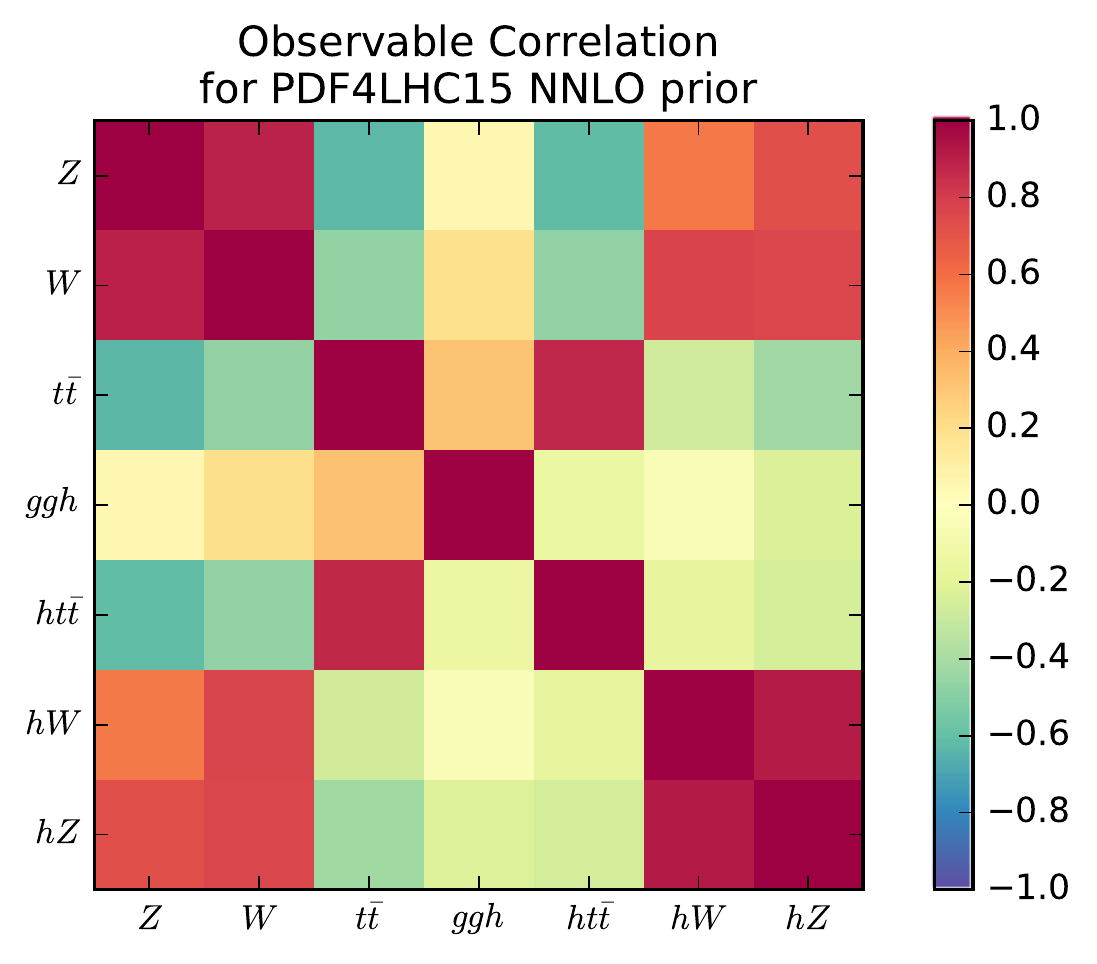}
  \caption{\small The value of the correlation coefficient between representative
    Higgs boson signal and background processes at the LHC 13 TeV, using the
    PDF4LHC15NNLO prior.
    The colour of each entry of the correlation matrix indicates the absolute
    size of the correlation coefficient.
    The processes shown in this figure are $ggh$, $ht\bar{t}$, $hW$ and $hZ$
    (for signal) and $Z$, $W$ and $t\bar{t}$ (for backgrounds).
}  
\label{fig:correlation_plot}
\end{figure}

In \refF{fig:correlation_plot} we show the absolute value of the correlation
coefficient between representative
  Higgs boson signal and background processes at the LHC 13 TeV, using the
    PDF4LHC15NNLO prior.
    The colour of each entry of the correlation matrix indicates the absolute
    size of the correlation coefficient, with a granularity of 0.2.
    The processes shown in this figure are $ggh$, $ht\bar{t}$, $hW$ and $hZ$
    (for signal) and $Z$, $W$ and $t\bar{t}$ (for backgrounds).
    Very similar results are obtained if any of the three reduced sets is used.

\section{Acceptance calculations}

An important application of the PDF4LHC15 combined sets is the
calculation of PDF uncertainties in acceptances.
These are
defined as the ratio
of cross-sections with fiducial selection cuts
to the corresponding inclusive cross-section,
$
\mathcal{A}=( \sigma|_{\rm fid}) /( \sigma|_{\rm incl})
$.
To illustrate this usage, we have 
computed the acceptances, with the corresponding PDF uncertainties,
for two Higgs boson production processes with experimentally
realistic selection cuts, namely:
\begin{itemize}
  \item Higgs boson production in the gluon fusion process,
    decaying into diphotons, $gg\to h \to \gamma\gamma$,
  \item Higgs boson production in association with
    a $W$ boson, with the latter decaying
    leptonically, $pp \to h W^{\pm} \to h l^{\pm}\nu_l$.
  \end{itemize}
As in the rest of this chapter, the calculations have been performed for
the LHC 13 TeV with {\tt MG5\_aMC@NLO} interfaced to {\tt aMCfast}, using NNLO PDFs with NLO matrix elements.

In the calculation of the fiducial cross-sections,
we use
similar acceptance requirements as those in the corresponding ATLAS and CMS
analyses.
For both processes, jets are reconstructed with the anti-$k_T$ algorithm
with $R=0.4$, and they satisfy
$p_T\ge 30$ GeV and $|\eta|\le 4.4$.
The additional selection cuts in the case of the $gg\to h \to \gamma\gamma$
are the following: 
for photons we require $p_T\ge 25$ GeV and $\eta \le 2.4$,
and the invariant mass of the diphotons should satisfy $|m_{\gamma\gamma}-125~{\rm GeV}| \le 15$ GeV.
In the case of Higgs associated production,
$pp \to h W^{\pm} \to h l^{\pm}\nu_l$, the
selection cuts on the charged leptons  are $|\eta_l|\le 2.5$ and
$p_T^l \ge 20$ GeV.

Using these kinematical cuts, we have generated {\tt applgrids}
for both fiducial and inclusive cross-sections,
and computed the resulting acceptance corrections and the corresponding
PDF uncertainties.
In Table~\ref{tab:acceptances} we collect the value of the acceptances
$\mathcal{A}$ in each case, together with the
corresponding PDF uncertainties computed with the PDF4LHC15 prior and with the
three reduced sets.
 For completeness, we also show the results for the acceptances computed
 with the individual PDF sets.
 We observe excellent agreement between the acceptances computed with
 the prior and with the three reduced sets, both
 for the central value and for the PDF
 uncertainties.

\begin{table}[h]
  \caption{\small \label{tab:acceptances} The acceptance corrections
    $\mathcal{A}$ for
    Higgs boson production at the LHC 13 TeV in two different channels with realistic
    selection cuts, and the corresponding
    PDF uncertainties.
    We compare the results of the PDF4LHC15 NNLO prior with the
    three reduced sets.
    For completeness, we also show the results for the acceptances computed
    with the individual PDF sets.
    See text for more details of the specific selection cuts in
    each case.
  }

  \centering
  \begin{tabular}{l|c|c}
    \toprule
    &  $\mathcal{A}(gg\to h\to \gamma\gamma)$ &
    $\mathcal{A}(pp\to hW\to hl\nu_l)$ \\
    \midrule
    PDF4LHC15 prior &  0.728 +- 0.006 (0.9\%)      &  0.7536 +- 0.0014 (0.18\%)  \\
\midrule
{\tt PDF4LHC15\_mc}  & 0.727 +- 0.006 (0.9\%)   &   0.7538 +- 0.0015 (0.20\%) \\
{\tt PDF4LHC15\_100}  & 0.728 +- 0.006 (0.9\%)   &
0.7536 +- 0.0013 (0.17\%)  \\
{\tt PDF4LHC15\_30} & 0.728 +- 0.006 (0.9\%)   &
0.7536 +- 0.0012 (0.15\%) \\
\midrule
MMHT14  & 0.728 +- 0.004 (0.6\%)  &
0.7532 +- 0.0012 (0.15\%)  \\
CT14   &  0.725 +- 0.007 (1.0\%)  &
0.7543 +- 0.0014 (0.18\%)  \\
NNPDF3.0  & 0.730 +- 0.005 (0.7\%)   &
0.7534 +- 0.0011 (0.14\%) \\ 
\bottomrule
  \end{tabular}
\end{table}

\section{Summary}

To summarize, in Table~\ref{tab:pdf4lhc15} we collect the
available PDF4LHC15 NNLO $N_f=5$ combined sets.
The corresponding sets are also available at NLO, as well as $N_f=4$ versions.
All the combined PDF4LHC14 sets are available
through {\tt LHAPDF6}~\cite{Buckley:2014ana}, which also includes
built-in routines for the calculation of the PDF and
PDF+$\alpha_s$ uncertainties for all relevant cases.

\begin{table}[t]
\caption{\small Summary of the combined NNLO PDF4LHC15 sets with
$N_f^{\rm max}=5$
that
  are available from {\tt LHAPDF6}.
  The corresponding NLO sets
  are also available.
 Members 0 and 1 of {\bf \tt PDF4LHC15\_nnlo\_asvar} coincide
 with members 101 and 102 (31 and 32) of
 {\bf \tt PDF4LHC15\_nnlo\_mc\_pdfas} and
 {\bf \tt PDF4LHC15\_nnlo\_100\_pdfas} ({\bf \tt PDF4LHC15\_nnlo\_30\_pdfas}).
  Recall that in {\tt LHAPDF6} there is always a zeroth member,
  so that the total number of PDF members in a given set
  is always $N_{\rm mem}+1$.
  See text for more details. \label{tab:pdf4lhc15}
}

  \centering
  \small
\begin{tabular}{l|c|c|c|l}
  \toprule
      {\tt LHAPDF6} grid  & Pert order  & {\tt ErrorType}  & $N_{\rm mem}$  & $\alpha_s(m_Z^2)$ \\
      \midrule
          {\bf \tt PDF4LHC15\_nnlo\_mc}  & NNLO  &  {\tt replicas}  &   100  & 0.118 \\
          {\bf \tt PDF4LHC15\_nnlo\_100}  & NNLO  &  {\tt symmhessian}  &   100  & 0.118 \\
          {\bf \tt PDF4LHC15\_nnlo\_30}  & NNLO  &  {\tt symmhessian}  &   30  & 0.118 \\
          {\bf \tt PDF4LHC15\_nnlo\_mc\_pdfas}  & NNLO  &  {\tt replicas+as}  &   102  & mem 0:100 $\to$ 0.118  \\
          &  &    &    & mem 101 $\to$ 0.1165  \\
          &  &    &    & mem 102 $\to$ 0.1195  \\
           {\bf \tt PDF4LHC15\_nnlo\_100\_pdfas}  & NNLO  &  {\tt symmhessian+as}  &   102  & mem 0:100 $\to$ 0.118  \\
          &  &    &    & mem 101 $\to$ 0.1165  \\
              &  &    &    & mem 102 $\to$ 0.1195  \\
          {\bf \tt PDF4LHC15\_nnlo\_30\_pdfas}  & NNLO  &  {\tt symmhessian+as}  &   32  & mem 0:30 $\to$ 0.118  \\
          &  &    &    & mem 31 $\to$ 0.1165  \\
              &  &    &    & mem 32 $\to$ 0.1195  \\
          {\bf \tt PDF4LHC15\_nnlo\_asvar}  & NNLO &  - & 1 & mem 0 $\to$ 0.1165 \\
           &   &    &  & mem 1 $\to$ 0.1195 \\
          \bottomrule
\end{tabular}
  \end{table}

Recommendations for the usage of each of these techniques are given in the
PDF4LHC 2015 document~\cite{Butterworth:2015oua},
along with explicit formulae for the calculation of PDF and
PDF+$\alpha_s(m_Z)$ uncertainties for each of the techniques.
The  recommendations can be simply summed up.
If asymmetric uncertainties are important, for example at high mass, and Hessian errors are not essential,
then {\tt PDF4LHC15\_mc} should be used.
There are two options for Hessian uncertainties.
The {\tt PDF4LHC15\_30} set provides a good estimate of the uncertainty of the prior with fewer members, sufficient in many cases, such as for the determination of nuisance parameters or acceptance calculations.
To reproduce the uncertainty of the prior exactly,
then the {\tt PDF4LHC\_100} sets should be used.

\chapter{Branching Ratios}
\label{chap:BR}
\ChapterAuthor{A.~Denner, S.~Heinemeyer, A.~M\"uck, I.~Puljak, D.~Rebuzzi~(Eds.); 
S.~Boselli, C.M. Carloni Calame, S.~Dittmaier, G.~Montagna, O.~Nicrosini, F.~Piccinini, M.~Spira, M. M\"uhlleitner}

\providecommand{\br}{{\mathrm{BR}}}
\providecommand{\Gatot}{\Gamma_{\mathrm H}}
\providecommand{\lsim}
{\;\raisebox{-.3em}{$\stackrel{\displaystyle <}{\sim}$}\;}
\providecommand{\gsim}
{\;\raisebox{-.3em}{$\stackrel{\displaystyle >}{\sim}$}\;}
\providecommand{\order}[1]{\ensuremath{{\cal O}(#1)}}
\providecommand{\Pqb}{\bar{\Pq}}
\providecommand{\Pfb}{\bar{\Pf}}
\providecommand{\PV}{\mathrm{V}}
\providecommand{\tb}{\tan\beta}
\providecommand{\Mh}{M_\mathrm{h}}
\providecommand{\MH}{M_\mathrm{H}}
\providecommand{\MA}{M_\mathrm{A}}
\providecommand{\MHp}{M_{\mathrm{H}^\pm}}
\providecommand{\mhmaxx}{\ensuremath{m_{\Ph}^{\rm max}}}
\providecommand{\mhmodp}{\ensuremath{m_{\Ph}^{\rm mod+}}}
\providecommand{\mhmodm}{\ensuremath{m_{\Ph}^{\rm mod-}}}
\providecommand{\tauphobic}{$\PGt$-phobic}
\providecommand{\hhbb}{\Ph \to \PQb \PQb}
\providecommand{\htautau}{\Ph \to \PGtp\PGtm}
\providecommand{\hmumu}{\Ph \to \PGmp\PGmm}
\providecommand{\hss}{\Ph \to \PQs \PQs}
\providecommand{\hcc}{\Ph \to \PQc \PAQc}
\providecommand{\hgaga}{\Ph \to \PGg\PGg}
\providecommand{\hZga}{\Ph \to \PZ\PGg}
\providecommand{\hWW}{\Ph \to \PW\PW}
\providecommand{\hZZ}{\Ph \to \PZ\PZ}
\providecommand{\Hbb}{\PH \to \PQb \PAQb}
\providecommand{\Htautau}{\PH \to \PGtp\PGtm}
\providecommand{\Hmumu}{\PH \to \PGmp\PGmm}
\providecommand{\Hss}{\PH \to \PQs \PQs}
\providecommand{\Hcc}{\PH \to \PQc \PAQc}
\providecommand{\Htt}{\PH \to \PQt \PAQt}
\providecommand{\Hgg}{\PH \to \Pg\Pg}
\providecommand{\Hgaga}{\PH \to \PGg\PGg}
\providecommand{\HZga}{\PH \to \PZ\PGg}
\providecommand{\HWW}{\PH \to \PW\PW}
\providecommand{\HZZ}{\PH \to \PZ\PZ}
\providecommand{\Abb}{\PA \to \PQb \PQb}
\providecommand{\Atautau}{\PA \to \PGtp\PGtm}
\providecommand{\Amumu}{\PA \to \PGmp\PGmm}
\providecommand{\Ass}{\PA \to \PQs \PQs}
\providecommand{\Acc}{\PA \to \PQc \PAQc}
\providecommand{\Att}{\PA \to \PQt \PAQt}
\providecommand{\Agg}{\PA \to \Pg\Pg}
\providecommand{\Agaga}{\PA \to \PGg\PGg}
\providecommand{\AZga}{\PA \to \PZ\PGg}
\providecommand{\AWW}{\PA \to \PW\PW}
\providecommand{\AZZ}{\PA \to \PZ\PZ}
\providecommand{\phibb}{\phi \to \PQb \PAQb}
\providecommand{\phitautau}{\phi \to \PGtp \PGtm}
\providecommand{\phimumu}{\phi \to \PGmp \PGmm}
\providecommand{\phiss}{\phi \to \PQs \PAQs}
\providecommand{\phicc}{\phi \to \PQc \PAQc}
\providecommand{\phitt}{\phi \to \PQt \PAQt}
\providecommand{\phigg}{\phi \to \Pg\Pg}
\providecommand{\phigaga}{\phi \to \PGg\PGg}
\providecommand{\phiZga}{\phi \to \PZ\PGg}
\providecommand{\phiVV}{\phi \to \PV^{(*)}\PV^{(*)}}
\providecommand{\phiWW}{\phi \to \PW^{(*)}\PW^{(*)}}
\providecommand{\phiZZ}{\phi \to \PZ^{(*)}\PZ^{(*)}}
\providecommand{\phiZh}{\phi \to \PZ\Ph}
\providecommand{\phiAh}{\phi \to \PA\Ph}
\providecommand{\phihh}{\phi \to \Ph\Ph}
\providecommand{\phiZA}{\phi \to \PZ\PA}
\providecommand{\phiAA}{\phi \to \PA\PA}
\providecommand{\phiWHp}{\phi \to \PH^\pm\PW^\mp}
\providecommand{\phiSUSY}{\phi \to \mathrm{SUSY}}
\providecommand{\Hptb}{\PH^\pm \to \PQt \PQb}
\providecommand{\Hpts}{\PH^\pm \to \PQt \PQs}
\providecommand{\Hptd}{\PH^\pm \to \PQt \PQd}
\providecommand{\Hpcb}{\PH^\pm \to \PQc \PQb}
\providecommand{\Hpcs}{\PH^\pm \to \PQc \PQs}
\providecommand{\Hpcd}{\PH^\pm \to \PQc \PQd}
\providecommand{\Hpub}{\PH^\pm \to \PQu \PQb}
\providecommand{\Hpus}{\PH^\pm \to \PQu \PQs}
\providecommand{\Hpud}{\PH^\pm \to \PQu \PQd}
\providecommand{\Hptaunu}{\PH^\pm \to \PGt\PGnGt}
\providecommand{\Hpmunu}{\PH^\pm \to \PGm\PGnGm}
\providecommand{\Hpenu}{\PH^\pm \to \Pe\PGnGe}
\providecommand{\HphW}{\PH^\pm \to \Ph\PW}
\providecommand{\HpHW}{\PH^\pm \to \PH\PW}
\providecommand{\HpAW}{\PH^\pm \to \PA\PW}
\providecommand{\HpSUSY}{\PH^\pm \to \mathrm{SUSY}}

\providecommand{\zehomi}[1]{$\cdot 10^{-#1}$}
\providecommand{\zehoze}{}
\providecommand{\zehopl}[1]{$\cdot 10^{#1}$}

\providecommand{\htofl}{{\sc Hto4l}}
\providecommand{\looptools}{{\sc LoopTools v2.10}}
\providecommand{\form}{{\sc Form}}
\providecommand{\powheg}{{\sc Powheg}}


For the accurate study of the properties of the Higgs boson, precise
predictions for the various partial decay widths and branching ratios (BRs)
along with their theoretical uncertainties are indispensable.  In
\Bref{Dittmaier:2011ti} a first precise prediction of the
BRs of the SM Higgs boson was presented.  In
\Brefs{Dittmaier:2012vm,Denner:2011mq} the BR predictions were
supplemented with an uncertainty estimate including parametric
uncertainties as well as the effects of unknown higher-order
corrections.  In \Bref{Heinemeyer:2013tqa} these predictions were
updated with a fine step size around the mass of the Higgs boson
discovered by ATLAS~\cite{Aad:2012tfa} and
CMS~\cite{Chatrchyan:2012xdj}. Moreover, error estimates were
presented in a form that is suitable for taking error correlations
into account.

In view of the updated parameter set of~\refC{chapter:input} and the
improvements in {\HDECAY} \cite{Djouadi:1997yw,Spira:1997dg,Djouadi:2006bz}
that reduce theoretical uncertainties, we provide here an update of the
predictions on SM Higgs BRs and corresponding uncertainties.

\section{Update of branching ratios and decay width for the Standard
  Model Higgs boson}

In this section we update the SM BR calculations presented in
\Brefs{Dittmaier:2011ti,Dittmaier:2012vm,Denner:2011mq,Heinemeyer:2013tqa}.
The strategy for the calculation of BRs and uncertainties is unchanged
with respect to \Bref{Dittmaier:2012vm} and \Bref{Heinemeyer:2013tqa},
respectively. However, the input parameter set has been updated (see
\refC{chapter:input}) and some improvements have been made in
{\HDECAY} that lead to a reduction of theoretical uncertainties.

\subsection{Strategy and input for branching-ratio calculations}
\label{sec:br-strategy}

We briefly summarize the strategy for the BR calculations for the
updates in this report.  A detailed description of the methods used
can be found in \Brefs{Dittmaier:2012vm,Heinemeyer:2013tqa}.
We employ {\HDECAY} \cite{Djouadi:1997yw,Spira:1997dg,Djouadi:2006bz}
and {\Prophecy}
\cite{Bredenstein:2006rh,Bredenstein:2006ha,Prophecy4f} to calculate
all the partial widths with the highest accuracy available.  The
higher-order corrections included in {\HDECAY} and {\Prophecy} have
been discussed in detail in Section 2.1.3.2 of
\Bref{Dittmaier:2012vm}. In the meantime, the following improvements
have been made in {\HDECAY}: On the one hand, for the Higgs boson decays
into fermions the complete NLO electroweak (EW) corrections have been
implemented.  For small Higgs boson masses, the theoretical
uncertainty from missing EW corrections is hence reduced to below
0.5\%.
This estimate is supported by the recent explicit calculation of the
mixed QCD-EW corrections \cite{Mihaila:2015lwa}.
On the other hand, the input parameters of HDECAY have been moved from
the bottom and charm pole masses to the masses $\Mb(\Mb)$ and
$\Mc(3\UGeV)$ in the $\overline{\mathrm{MS}}$ scheme according to the
recommendations of \refC{chapter:input}.  Following these
recommendations, the corresponding pole masses are calculated
internally in HDECAY using the three-loop relation.

The updated estimated relative theoretical uncertainties (THUs) for
the different Higgs boson partials widths resulting from missing higher-order
corrections (determined as explained in more detail in  Section
2.1.3.2 of \Bref{Dittmaier:2012vm})
are summarized in \refT{tab:uncertainty} which replaces Table 2 of
\Bref{Dittmaier:2012vm}.
\begin{table}
\caption{Estimated theoretical uncertainties from missing higher orders.}
\label{tab:uncertainty}%
\renewcommand{\arraystretch}{1.2}%
\setlength{\tabcolsep}{1.5ex}%
\centerline{
\begin{tabular}{lllll}
\toprule
\text{Partial width} & \text{QCD} & \text{electroweak} & \text{total} \\
\midrule
 $\PH \to \PQb\PAQb/\PQc\PAQc$ &    $\sim 0.2\%$
&     $\sim 0.5\%$ for $\MH <  500\UGeV$     &      $\sim 0.5 \%$\\
$\PH\to \PGtp \PGtm/\PGmp\PGmm$ & & $\sim0.5\%$  for $\MH <  500\UGeV$ &       $\sim 0.5 \%$ \\
$\PH \to \PQt\PAQt$ & $\lsim 5\%$&
      $\sim 0.5\%$   for $\MH < 500 \UGeV$     &       $\sim5\%$ \\
$\PH \to \Pg\Pg$ & ${\sim 3\%}$   &
$\sim 1\%$   &   $\sim3.2\%$\\
$\PH \to \PGg \PGg$  & ${<1\%}$ & $<1\%$    &  $\sim1\%$ \\
$\PH \to \PZ \PGg$  & ${<1\%}$ & $\sim 5\%$    &  $\sim5\%$ \\
$\PH \to \PW\PW/\PZ\PZ\to4\Pf$ & $<0.5\%$ &   $\sim 0.5\%$ for $\MH < 500\UGeV$ &  $\sim0.5\%$\\
\bottomrule
\end{tabular}
}
\end{table}
The corresponding uncertainty for the total width is obtained by
adding the uncertainties for the partial widths linearly.  In order to
determine the uncertainty for a BR, first the variations of this BR
are calculated when varying each individual partial width within the
corresponding relative error while keeping all other partial widths
fixed at their central value. Since each BR depends on all partial
widths, scaling a single partial width modifies all BRs.  Hence, there
is an individual THU of each BR due to the THU of each partial width.
We assume the THUs of all partial widths to be uncorrelated except for
all $\PH\to \PW\PW/\PZ\PZ\to4\Pf$ decays. The THUs of the latter are
assumed to be fully correlated and, hence, we only consider the
simultaneous scaling of all 4-fermion partial widths.  The derived
individual THUs for each BR are added linearly to obtain the
corresponding total THU.

Also the parametric uncertainties (PUs) have been updated in our
calculation.  The input-parameter set as defined in
\refC{chapter:input} has been used.  From the given PDG values of the
gauge-boson masses, we derive the pole masses $\MZ=91.15348\UGeV$ and
$\MW= 80.35797\UGeV$ which are used as input.  The gauge-boson widths
have been calculated at NLO from the other input parameters resulting
in $\Gamma_{\PZ}=2.49436\UGeV$ and $\Gamma_{\PW}=2.08718\UGeV$.

Concerning the PUs, we take only those
of the input parameters $\alphas$, $\Mc$, $\Mb$, and
$\Mt$ into account. The values and uncertainties are adopted
from~\refC{chapter:input} and are for convenience listed in \refT{tab:inputpu}
as well. For the masses of the light quarks, we use the masses in the
$\overline{\mathrm{MS}}$ scheme $\Mb(\Mb)$ and $\Mc(3\UGeV)$ as input,%
\footnote{Since \HDECAY\ version 6.50 it is possible to use directly
  masses in the $\overline{\mathrm{MS}}$ scheme as input for the light quarks.}
while for $\Mt$ the pole mass enters. The parametric uncertainties
resulting from the gauge-boson masses, the lepton masses, the
electromagnetic coupling and the Fermi constant are below one per
mille, and the impact of the PUs of the light quark masses on the
considered BRs is negligible.

\begin{table}\small
\renewcommand{\arraystretch}{1.2}
\setlength{\arraycolsep}{1.5ex}
\caption{Input parameters and their relative uncertainties as used for the
uncertainty estimation of the branching ratios.
}
\label{tab:inputpu}
$$\begin{array}{cccc}
\toprule
\text{Parameter} & \text{Central value}
& \text{Uncertainty}
\\
\midrule
  \alphas (\MZ)& 0.118 &  \pm0.0015\\
\Mc(3\UGeV) & 0.986\UGeV & \pm0.026\UGeV \\
\Mb(\Mb)  &  4.18\UGeV& \pm0.03\UGeV \\
\Mt^{\mathrm{pole}} & 172.5\UGeV& \pm1\UGeV  \\
\bottomrule
\end{array}$$
\end{table}

Using these PUs, for each parameter
$p=\alphas,\Mc,\Mb,\Mt$ we have calculated the Higgs BRs
for $p$, $p+\Delta p$ and $p-\Delta p$ keeping all the other parameters
fixed at their central values. The resulting shift on each BR is then given
by
\begin{eqnarray}
\Delta^p_+ \br &=&  \max \{\br(p+\Delta p),\br(p),\br(p-\Delta p)\} -
\br(p),\nonumber\\
\Delta^p_- \br &=&  \br(p) -\min \{\br(p+\Delta p),\br(p),\br(p-\Delta p)\},
\end{eqnarray}
which may lead to asymmetric uncertainties.  The total PUs have been
obtained by adding the calculated shifts due to the four parameters in
quadrature.  In analogy, the uncertainties of the partial and total
decay widths are given by
\begin{eqnarray}
\label{eqn:deltagamma}
\Delta^p_+ \Gamma &=&  \max \{\Gamma(p+\Delta
p),\Gamma(p),\Gamma(p-\Delta p)\} - \Gamma(p),\nonumber\\
\Delta^p_- \Gamma &=&  \Gamma(p) -\min \{\Gamma(p+\Delta
p),\Gamma(p),\Gamma(p-\Delta p)\},
\end{eqnarray}
where $\Gamma$ denotes the partial decay width for each considered
decay channel or the total width, respectively. The total PUs have
been calculated again by adding the individual PUs in quadrature.

The total uncertainties on the BRs, i.e.\ combinations of PUs and THUs,
are derived by adding linearly the total parametric uncertainties and
the total theoretical uncertainties, as discussed in detail in
\Bref{Dittmaier:2012vm}. To allow for taking into account
correlations in the errors of the different BRs, we provide also
the individual uncertainties for the various partial widths
in \refS{sec:br-correlations} for selected Higgs boson masses.

For completeness, we repeat that the Higgs boson total width resulting from {\HDECAY}
has been modified according to the prescription
\begin{equation}
\Gamma_{\PH} = \Gamma^{\mathrm{HD}} - \Gamma^{\mathrm{HD}}_{\PZ\PZ}
            - \Gamma^{\mathrm{HD}}_{\PW\PW} + \Gamma^{\mathrm{Proph.}}_{4\Pf}~,
\end{equation}
where $\Gamma_{\PH}$ is the total Higgs boson width, $\Gamma^{\mathrm{HD}}$
the Higgs boson width obtained from {\HDECAY},
$\Gamma^{\mathrm{HD}}_{\PZ\PZ}$ and $\Gamma^{\mathrm{HD}}_{\PW\PW}$
stand for the partial widths to $\PZ^{(*)}\PZ^{(*)}$ and $\PW^{(*)}\PW^{(*)}$
calculated with
{\HDECAY}, while $\Gamma^{\mathrm{Proph.}}_{4\Pf}$ represents the
partial width of $\PH\to 4\Pf$ calculated with {\Prophecy}.  The
latter can be split into the decays into $\PZ^{(*)}\PZ^{(*)}$,
$\PW^{(*)}\PW^{(*)}$, and the
interference,
\begin{equation}
\Gamma^{\mathrm{Proph.}}_{4\Pf}=\Gamma_{{\PH}\to \PW^{(*)}\PW^{(*)}\to 4\Pf}
+ \Gamma_{{\PH}\to \PZ^{(*)}\PZ^{(*)}\to 4\Pf}
+ \Gamma_{\mathrm{\PW\PW/\PZ\PZ-int.}}\,,
\end{equation}
where the individual contributions are defined in terms of partial widths
with specific final states according to
\begin{eqnarray}
\Ga_{{\PH}\to \PW^{(*)}\PW^{(*)}\to 4f} &=&
  9 \cdot \Ga_{{\PH}\to\PGne\Pep\PGmm\PAGnGm}
+ 12 \cdot \Ga_{{\PH}\to\PGne\Pep\PQd\PAQu}
+ 4 \cdot \Ga_{{\PH}\to\PQu\PAQd\PQs\PAQc}\, , \nonumber\\[1ex]
\Ga_{{\PH}\to \PZ^{(*)}\PZ^{(*)}\to 4f} &=&
  3 \cdot \Ga_{{\PH}\to\PGne\PAGne\PGnGm\PAGnGm}
+ 3 \cdot \Ga_{{\PH}\to\Pem\Pep\PGmm\PGmp}
+ 9 \cdot \Ga_{{\PH}\to\PGne\PAGne\PGmm\PGmp} \nonumber\\
&&{}
+ 3 \cdot \Ga_{{\PH}\to\PGne\PAGne\PGne\PAGne}
+ 3 \cdot \Ga_{{\PH}\to\Pem\Pep\Pem\Pep} \nonumber\\
&&{}+ 6 \cdot \Ga_{{\PH}\to\PGne\PAGne\PQu\PAQu}
+ 9 \cdot \Ga_{{\PH}\to\PGne\PAGne\PQd\PAQd}
+ 6 \cdot \Ga_{{\PH}\to\PQu\PAQu\Pem\Pep}
+ 9 \cdot \Ga_{{\PH}\to\PQd\PAQd\Pem\Pep} \nonumber\\
&&{}+ 1 \cdot \Ga_{{\PH}\to\PQu\PAQu\PQc\PAQc}
+ 3 \cdot \Ga_{{\PH}\to\PQd\PAQd\PQs\PAQs}
+ 6 \cdot \Ga_{{\PH}\to\PQu\PAQu\PQs\PAQs}
+ 2 \cdot \Ga_{{\PH}\to\PQu\PAQu\PQu\PAQu} \\
&&{}+ 3 \cdot \Ga_{{\PH}\to\PQd\PAQd\PQd\PAQd}\, ,\nonumber \\[1ex]
\Ga_{\mathrm{WW/ZZ-int.}} &=&
  3 \cdot \Ga_{{\PH}\to\PGne\Pep\Pem\PAGne}
- 3 \cdot \Ga_{{\PH}\to\PGne\PAGne\PGmm\PGmp}
- 3 \cdot \Ga_{{\PH}\to\PGne\Pep\PGmm\PAGnGm} \nonumber\\
&&
  {}+2 \cdot \Ga_{{\PH}\to\PQu\PAQd\PQd\PAQu}
- 2 \cdot \Ga_{{\PH}\to\PQu\PAQu\PQs\PAQs}
- 2 \cdot \Ga_{{\PH}\to\PQu\PAQd\PQs\PAQc}\, \nonumber .
\end{eqnarray}

\subsection[Partial widths and BR for Higgs boson masses close to \texorpdfstring{$125\UGeV$}{125 GeV}]{Results for partial widths and branching ratios
  including QCD and EW corrections for Higgs boson masses close to \texorpdfstring{$125\UGeV$}{125 GeV}}
\label{sec:br-125results}

We provide results for the BRs of the
SM Higgs boson using a particularly fine grid of masses
close to $\MH = 125 \UGeV$. The results
are generated and presented in analogy to the predictions and error estimates
in \Bref{Heinemeyer:2013tqa}, taking the improvements mentioned above
into account.

Here, we briefly summarize the numerical changes with respect to the
results for $\MH = 125 \UGeV$ presented in \Bref{Heinemeyer:2013tqa}.
All these changes are well within the error estimates given in
\Bref{Heinemeyer:2013tqa}. The partial width for $\PH \to \PQb\PAQb$
increases by approximately $1.5\%$, mainly due to the change in
$\alphas$. The partial widths of the other fermionic decay modes
change only at the per mille level due to the inclusion of the full
EW corrections and/or the input changes. The partial width
for $\PH \to \Pg\Pg$ decreases by approximately $4\%$. While about
half of this shift is due to the change in $\alphas$, the remaining
part comes from improvements in {\HDECAY}, in particular from the
inclusion of charm-quark-loop contributions and NLO quark-mass
effects. The partial widths for the other bosonic decay modes change
at the level of one per mille or below.  The total width increases by
approximately $0.5\%$.  Correspondingly, the relative increase for the
central value of the $\PH \to \PQb\PAQb$ BR is approximately $1\%$.
The relative decrease in the other fermionic modes is below $1\%$.
For $\PH \to \Pg\Pg$, the relative decrease of the BR is
approximately $4 \%$.  The relative decrease of the other bosonic BRs
is below $1\%$, only.

The error estimates on the BRs also change as discussed in the following: The
total error on the $\PH \to \PQb\PAQb$ BR decreases to below $2 \%$
due to the reduced errors on $\alphas$ and the bottom quark mass and
the reduced THU.  Since the error on $\PH \to \PQb\PAQb$ is a major
source of uncertainty for all the other BRs, their error is reduced by
more than $2\%$ due to this improvement alone.  In addition, the other
fermionic modes benefit from the reduced THU after the inclusion of the
full EW corrections, such that the corresponding errors are reduced
roughly by a factor of $2$ to below $2.5\%$ for the leptonic final
states and to below $7\%$ for $\PH \to \PQc\PAQc$. Also the error
estimates for the bosonic decay modes are decreased, mainly due to the
improvements in $\PH \to \PQb\PAQb$.  In particular, the error for the
decay into massive vector bosons is approximately $2\%$, i.e.\ half as
big as before. The errors on the partial widths are discussed in
\refS{sec:br-correlations}.

\begin{figure}
\begin{center}
\includegraphics[width=0.5\textwidth]{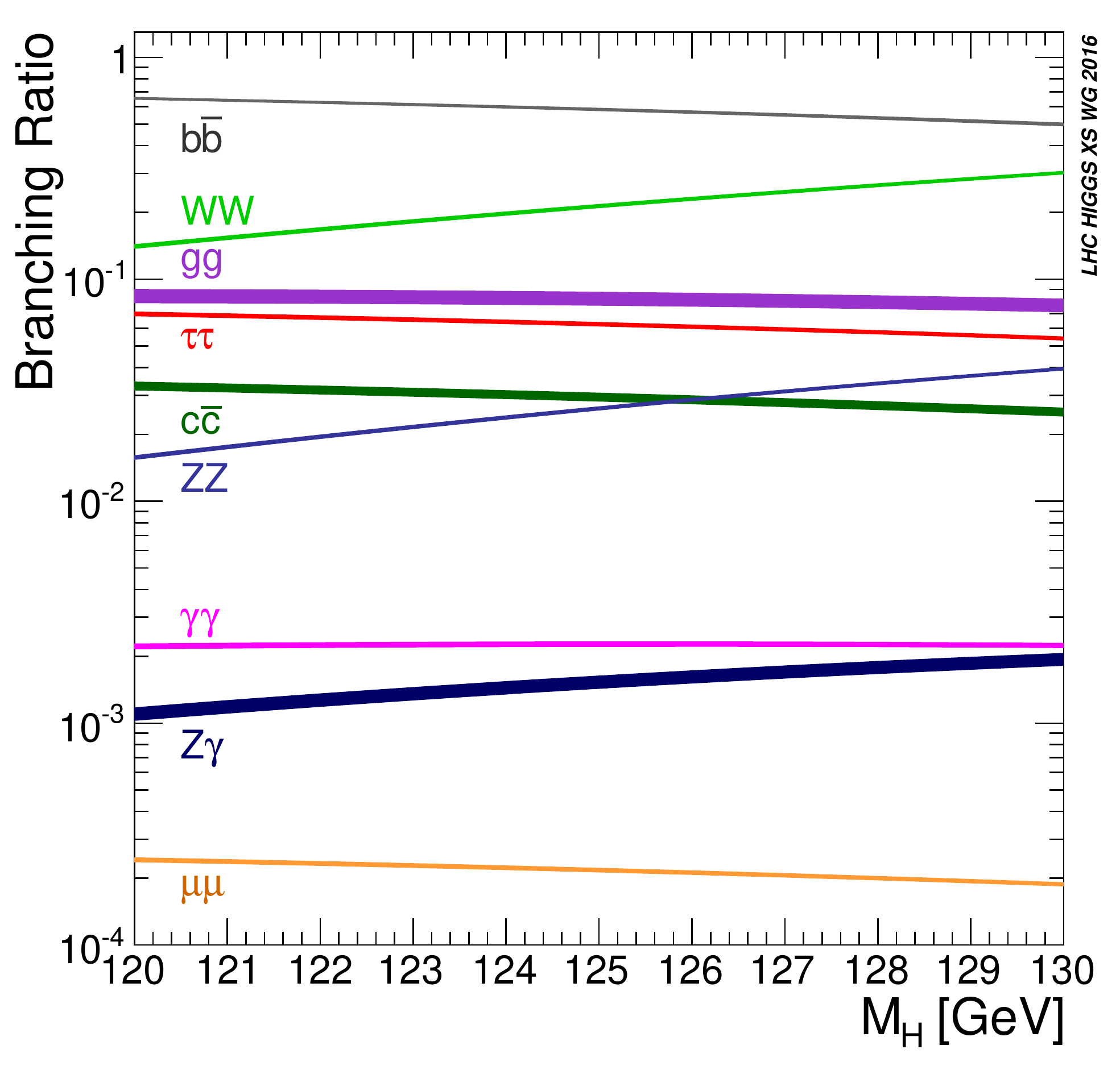}
\end{center}
\caption{Higgs boson branching ratios and their uncertainties for the
  mass range around $125\UGeV$.}
\label{fig:YRHXS4_BR_plots}
\end{figure}

The BRs for the fermionic decay modes are shown in
\refTs{tab:YRHXS4_1}--\ref{tab:YRHXS4_2}. The BRs for the
bosonic decay modes together with the total width are given in
\refTs{tab:YRHXS4_3}--\ref{tab:YRHXS4_5}.
Besides the BRs, the tables list also the corresponding
theoretical uncertainties (THU) and parametric uncertainties
resulting from the quark masses (PU($m_q$)) and the strong coupling
(PU$(\alphas)$). The PUs from the different
quark masses have been added in quadrature.
The BRs (including the full uncertainty) are also presented
graphically in \refF{fig:YRHXS4_BR_plots} for the mass region around
the Higgs boson resonance.

Finally, \refTs{tab:YRHXS4_4f_1}--\ref{tab:YRHXS4_4f_3} list the
BRs for the most relevant Higgs boson decays into four-fermion
final states. The right-most column in the tables shows the total
relative uncertainty of these BRs in per cent, obtained
by adding the PUs in quadrature and combining
them linearly with the THU. The uncertainty is
practically equal for all $\PH\to4\Pf$ BRs and the same
for those for $\PH\to\PW\PW/\PZ\PZ$.  Note that the charge-conjugate
final state is not included for $\PH \to \Plp\PGnl \PQq\PAQq$.

\subsection[Correlations and uncertainties for BR close to 125 GeV]{Correlations and uncertainties for branching ratios for Higgs boson masses close to 125\UGeV}
\label{sec:br-correlations}

As in \Bref{Heinemeyer:2013tqa}, we provide results and uncertainties
resulting from different sources for the partial widths and selected
Higgs boson masses.  These results can be used to include error
correlations for the different BRs. The reason for the correlations is
two-fold: Varying the input parameters within their error bands
induces shifts of the different partial widths and the resulting BRs
in a correlated way.  Moreover, there is a trivial correlation because
all the BRs have to add up to one. Thus, a shift in a single partial
width shifts all BRs in a correlated way.

For the partial widths the THUs are assumed to be uncorrelated (except
for the decay to massive gauge bosons). Moreover, the
correlated effect on each partial width from varying a parameter
within its errors is disentangled from the additional trivial
correlation when calculating the BRs. In Table 1 of
\Bref{Heinemeyer:2013tqa} we showed the results for the partial widths
for $\MH = 122 \UGeV$, $126 \UGeV$, and $130 \UGeV$ including for each
partial width the THU and the different PUs.

Using the updated parameter set and the updated THUs we present in
\refT{tab:br-correlation} results for the partial widths and the
corresponding uncertainties for $\MH= 124 \UGeV$, $125 \UGeV$, and $126
\UGeV$.
For each partial width, we show the THU evaluated according to
\refE{eqn:deltagamma}.  For each input parameter we show the induced
shift on each partial width for the maximal and minimal choice of the
input parameter as upper and lower entry in the table, respectively.
Hence, the table allows to read off the correlation in the variation
of the different partial widths. The associated error bands are
slightly asymmetric. However, it is a good approximation to symmetrize
the error band and assume a Gaussian probability distribution for the
corresponding prediction.

The THUs on the partial widths of all the four-fermion final-states
can be considered to be fully correlated. All other THUs are
considered to be uncorrelated. Hence, for the BRs listed in
\refTs{tab:YRHXS4_1}--\ref{tab:YRHXS4_5} only the trivial
correlation is present. However, it should be stressed again that in
contrast to the PUs, theory errors cannot be assumed to be Gaussian
errors. Assuming a Gaussian distribution and, hence, effectively
adding THUs to the PUs in quadrature will in general lead to
underestimated errors. According to the recommendations in Section 12
of \Bref{Dittmaier:2011ti}, the THUs should be considered to have a
flat probability distribution within the given range. Alternatively,
the envelope of extreme choices for the theory prediction on the
partial widths should be used as an error estimate. For all the
total errors presented on the BRs, we have added PUs and THUs of the
resulting BRs linearly, as discussed in \refS{sec:br-strategy}.
We thereby provide the envelope for each resulting BR, however, the 
correlation is lost at the level of BRs.

In total, there are four input parameters to be varied corresponding
to the PUs and one has to include eight uncorrelated THUs for the
various partial widths.  Analysing in detail the interesting
region around $\MH = 125 \UGeV$, the different uncertainties are of
different importance. Aiming for a given accuracy, some uncertainties
may be safely neglected, as can be inferred from
\refT{tab:br-correlation}.  Even sizeable uncertainties for a given
partial width can be unimportant if the decay mode has a small BR and
does not contribute significantly to combined measurements.

Concerning the PUs, the variation of $\alphas$ and the bottom quark
mass impact the BR predictions at the level of $1-2\%$.  The charm
quark mass is only relevant for $\Hcc$ and affects other BRs only at
the level of $1-2$ per mille. The THU for $\Hgaga$ amounts to 1\% and
is needed at this level of precision.  The THU for $\Hbb$, $\Hcc$,
$\Hmumu$, $\Htautau$, and $\PH\to \PW\PW/\PZ\PZ\to4\Pf$ is estimated
at 0.5\% and thus also quite small. The THU for $\Hgg$ and $\HZga$
only has sizeable effects if a measurement of the corresponding
channel is included or, in the case of $\Hgg$, errors of a few per
mille are important.

\subsection[Partial widths and BR for a wide Higgs boson mass range]{Benchmark results for partial widths and branching ratios
  including only QCD corrections for a wide Higgs boson mass range}
\label{sec:br-125results_wide}

If the minimal SM is realized, the Higgs boson is light and the best
predictions for the corresponding BRs are listed above. However, in
extended models additional Higgs bosons might show up. In order to
provide a benchmark for such additional Higgs bosons, we list the
partial widths calculated in the SM for a wide range of Higgs boson
masses. Since the EW corrections become very large and unphysical for
Higgs boson masses above $\sim600\UGeV$, we omit all EW corrections and
give results including QCD corrections only. The SM partial widths in this
scenario are listed in \refTs{tab:YRHXS4_noEW_1} and \ref{tab:YRHXS4_noEW_2}
for the 2-fermion final states and in \refTs{tab:YRHXS4_noEW_3} and \ref{tab:YRHXS4_noEW_4}
for the 2-boson final states.
Without EW corrections the results for the partial widths for
$\PH\to\PW\PW$ and $\PH\to\PZ\PZ$ calculated with {\HDECAY} and
\Prophecy\ agree within $\sim1\%$ for Higgs boson masses above $100\UGeV$.
For very light Higgs bosons the differences become bigger owing to
interference effects that are included in \Prophecy\ but not in
{\HDECAY}. Therefore, we followed our standard procedure and used \Prophecy\ to calculate the numbers
presented in \refTs{tab:YRHXS4_noEW_3} and \ref{tab:YRHXS4_noEW_4}.

\subsection[\textbf{\textsc{Hto4l}}: a generator for Higgs boson decay into
  four charged leptons]{\htofl: a generator for Higgs boson decay into
  four charged leptons at NLOPS electroweak accuracy}
\label{sec:br-hto4l}

The Monte Carlo event generator \htofl\ has been developed for the
precise simulation of the SM Higgs boson decay into four charged
leptons ($\PH\to\Pep\Pem\Pep\Pem$, $\Pep\Pem\PGmp\PGmm$, $\PGmp\PGmm\PGmp\PGmm$), including complete
NLO EW corrections and
multiple-photon effects in a matched-to-NLO Parton Shower (PS)
framework. A detailed description of the theoretical approach, the
calculation and phenomenological results can be found in
\Bref{Boselli:2015aha} and version 1.0 of the program is publicly
available \cite{Hto4l}.

\begin{figure}
  \includegraphics[width=0.48\textwidth]{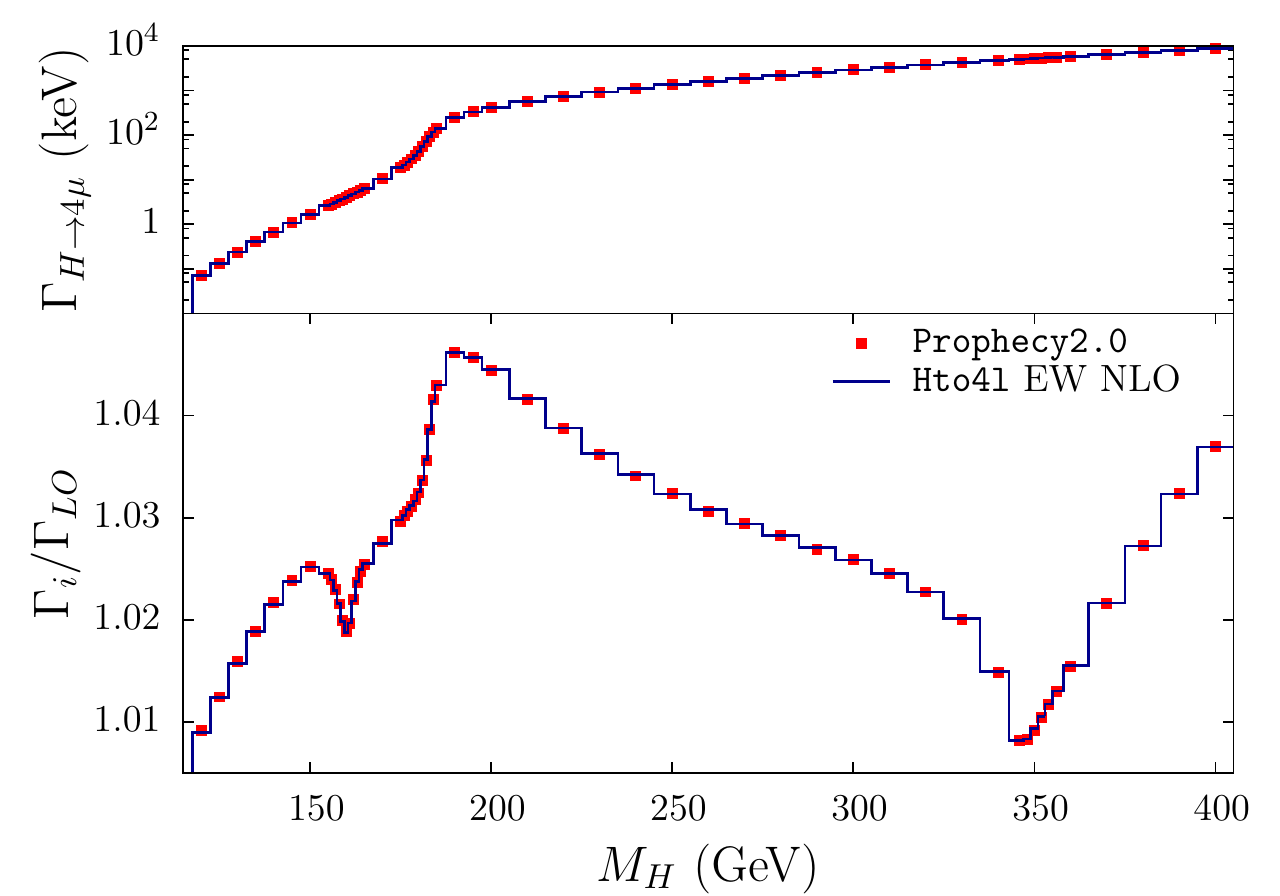}
  \includegraphics[width=0.48\textwidth]{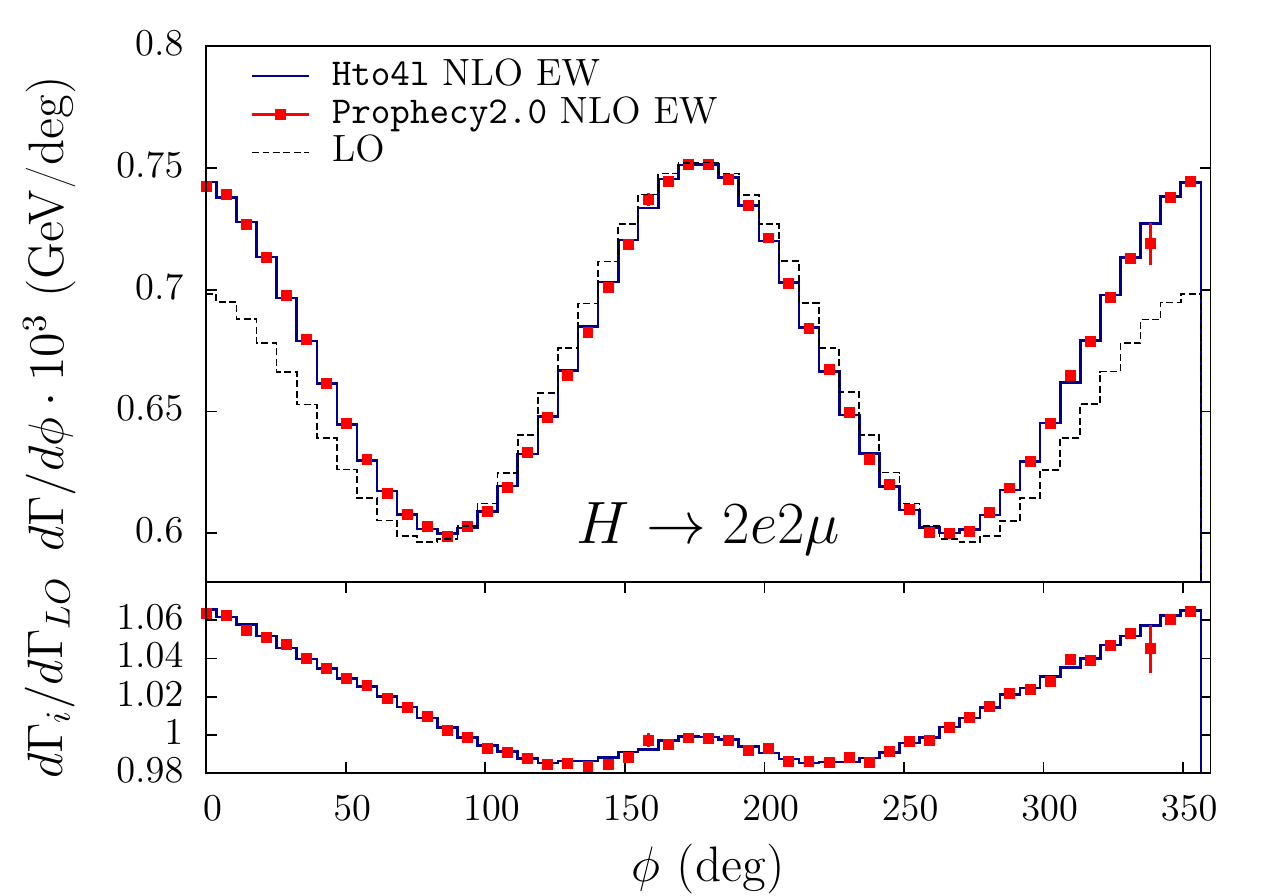}
\caption{Comparison between \htofl\ and \Prophecy. On the left, the
  $\PH\to 4\PGm$ partial width is compared as a function of $\MH$
  together with the relative effect of the NLO EW corrections. On the
  right, the $\phi$-angle distribution for $\PH\to 2\Pe 2\PGm$ is compared and the NLO EW
  effect is shown, for $\MH=125\UGeV$.}
\label{hto4lfig1}
\end{figure}

For inclusive partial widths or at NLO EW accuracy without
multi-photon emission, \htofl\ and \Prophecy\ (version
2.0)~\cite{Bredenstein:2006nk,Bredenstein:2007ec} should agree for the
leptonic final states included in \htofl. A comparison between the
codes has been carried out, both at differential and integrated level.
As examples, on the left of \refF{hto4lfig1} we show the partial
$\PH\to 4\PGm$ decay width as a function of $\MH$ (upper panel) and
the effect of the NLO EW corrections relative to the LO width (lower
panel): the predictions of \htofl\ and \Prophecy\ agree in the whole
mass range, within the negligible sub-per mille numerical integration
error. On the right of \refF{hto4lfig1}, the distribution of the angle
$\phi$ (the angle between the decay planes of the virtual $\PZ$ bosons
in the Higgs boson rest frame) in the decay $\PH\to 2\Pe 2\PGm$ is
plotted, showing both the absolute value (upper panel) and the
relative effect induced by NLO EW corrections (lower panel): again
agreement is found.

\begin{figure}
  \includegraphics[width=0.48\textwidth]{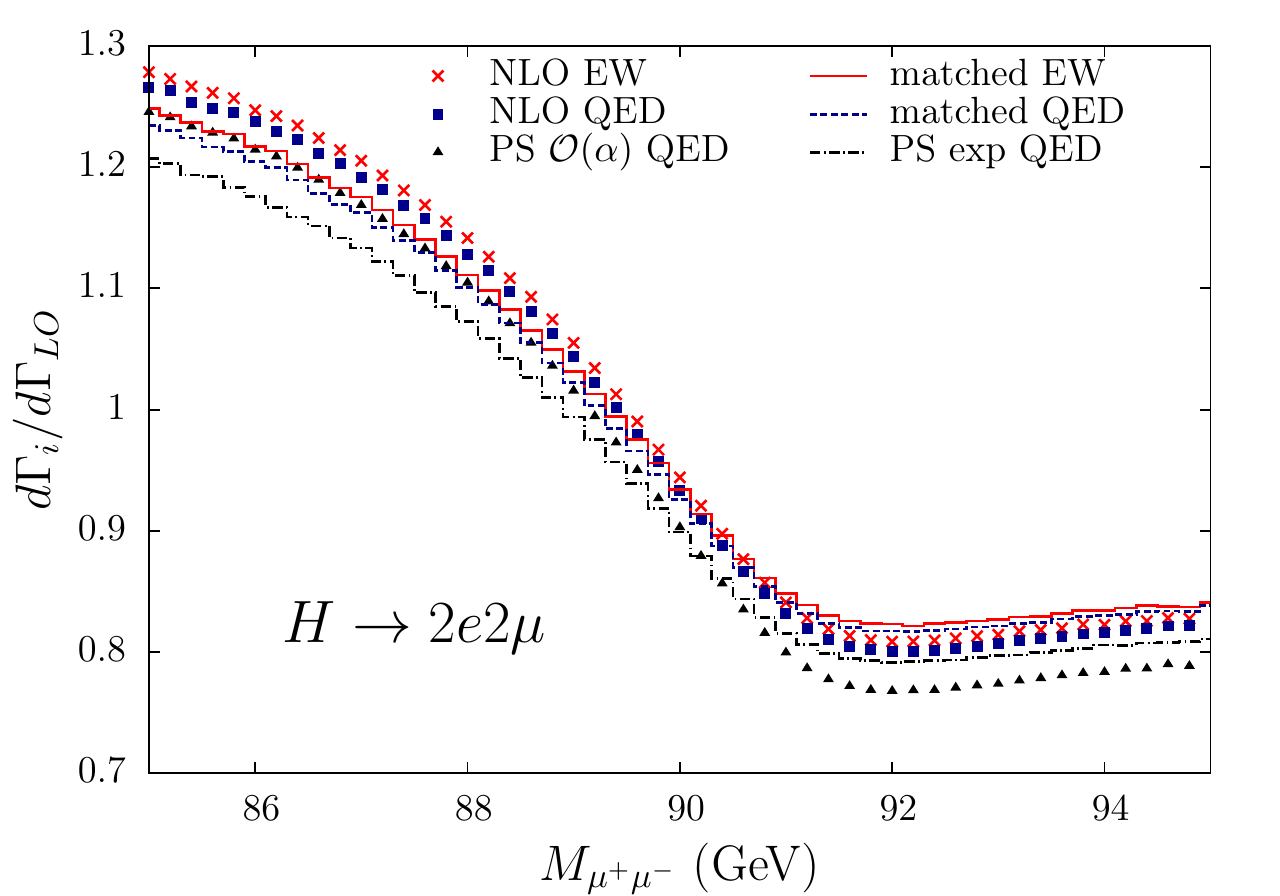}
  \includegraphics[width=0.48\textwidth]{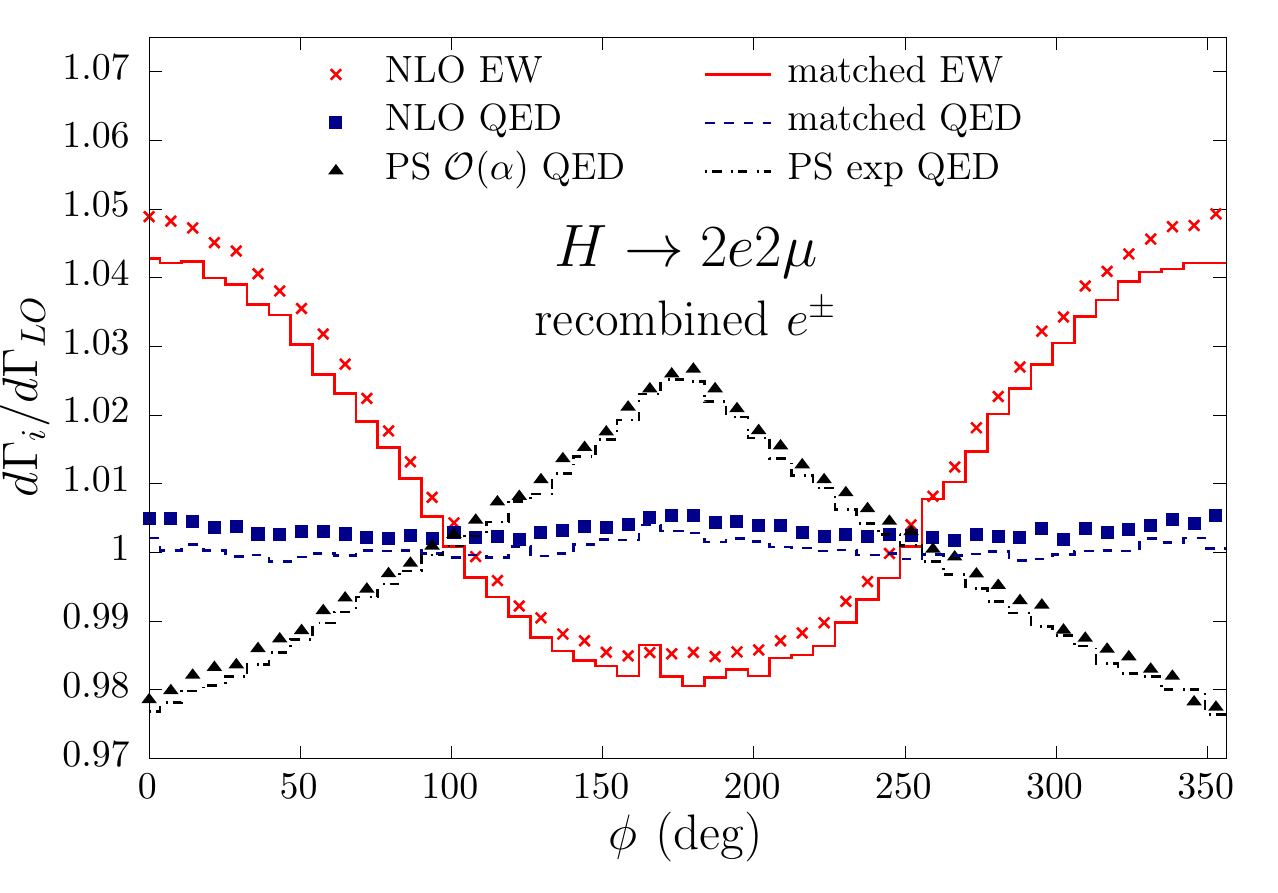}
  \caption{Effects of EW corrections at different level of accuracy on
    the $M_{\PGmp\PGmm}$ invariant mass (left) and the $\phi$-angle
    (right), with respect to the LO distribution. The decay
    $\PH\to 2\Pe 2\PGm$ with $\MH=125\UGeV$ is considered and, on the right
    plot, photons with $\Delta R_{e\gamma}< 0.1$ are recombined with
    the electron. See \Bref{Boselli:2015aha} for details.}
\label{hto4lfig2}
\end{figure}

Going beyond the NLO EW accuracy, \htofl\ includes also the
possibility to simulate multi-photon emissions in a QED PS approach,
consistently matched to the NLO calculation~\cite{Boselli:2015aha}.
Two examples of the effects induced by multi-photon radiation are
shown in \refF{hto4lfig2}, where we consider the decay $\PH\to 2\Pe
2\PGm$ with $\MH=125\UGeV$ in the Higgs boson rest frame. On the left, the
ratio between the $M_{\PGmp\PGmm}$ distribution corrected at different
levels of accuracy and the LO one is plotted, emphasizing that
multi-photon effects can reach the few per cent level on this
distribution. On the right, the size of the corrections is shown on
the $\phi$-angle distribution: while in this case the impact of QED
exponentiation is small, the inclusion of NLO QED and weak
corrections is important for studies aiming at accuracies at the level
of some per cent.
We finally remark that unweighted events are available and,
through Les-Houches-Accord (LHA) files,
\htofl\ can be used in conjunction with any
Monte Carlo event generator for (on-shell) Higgs boson production, such as, 
for instance,
\powheg~\cite{Alioli:2008tz,Alioli:2010xd}. An interface, which reads
a Higgs boson production LHA file and writes a LHA file with 
the Higgs momenta replaced by momenta of the decay products 
(leptons and photons),
is provided with the code.


\section{Update on MSSM branching ratios}

\label{sec:br-mssm}

In the MSSM the evaluation of cross sections and of branching ratios
has several common issues as outlined in Sect.\,14.1
in \Bref{Heinemeyer:2013tqa} (see also Sect.\,12.1 in \Bref{Dittmaier:2012vm}).
It was discussed that {\em before} any branching ratio calculation can be
performed, the Higgs boson masses, couplings and mixings
have to be evaluated from the underlying set of (soft SUSY-breaking)
parameters. For the case of real parameters in the MSSM, the code
\FeynHiggs~\cite{Heinemeyer:1998yj,Heinemeyer:1998np,Degrassi:2002fi,Frank:2006yh,Hahn:2009zz,Hahn:2013ria}
was selected for the evaluations in this report.
(The case with complex parameters has not been investigated so far.)
The results for Higgs boson masses and
couplings can be provided to other codes
(especially \HDECAY~\cite{Djouadi:1997yw,Spira:1997dg,Djouadi:2006bz}) via the SUSY
Les Houches Accord~\cite{Skands:2003cj,Allanach:2008qq}.

In Sect.\,2.3 of \Bref{Heinemeyer:2013tqa} it was described how the relevant
codes for the
calculation of partial decay widths, \FeynHiggs\ and \HDECAY, are
combined to give the most precise result for the Higgs boson branching
ratios in the MSSM.
The corrections included in these two codes are summarized as follows.
The full one-loop corrections in the MSSM (see the discussion in
\refS{sec:mssm-intro}) together with
resummed SUSY corrections up to the one-loop level and the $Z$~factors
up to the two-loop level have been implemented in the code
\FeynHiggs~\cite{Heinemeyer:1998yj,Hahn:2009zz,Heinemeyer:1998np,Degrassi:2002fi,Frank:2006yh,Hahn:2013ria}.
Corrections at and beyond the one-loop level in the MSSM are implemented
in the code \HDECAY~\cite{Djouadi:1997yw,Spira:1997dg,Djouadi:2006bz},
including the resummed SUSY corrections up to leading two-loop terms
and the \order{\alphas} corrections to Higgs boson decays to scalar
quarks~\cite{Accomando:2011jy}.
In \Bref{Heinemeyer:2013tqa} numerical results were shown for all MSSM Higgs
bosons (including the charged Higgs) within the updated benchmark
scenarios~\cite{Carena:2013ytb}. Here we briefly describe an update of the
branching ratio evaluation, taking into account additional final states (that
so far were included in the total width, but not evaluated as individual
BRs). The updated numbers can be found in~\cite{brsubgroupwebpage}.

After the calculation of Higgs boson masses and mixings from the
original SUSY input, the branching ratio calculation is performed as
follows.  We combine the results from \HDECAY\ and \FeynHiggs\ (where
the decays into massive gauge bosons labelled $\mathrm{FH/P4f}$ are
based on the SM evaluation taken from \Prophecy, see
\refS{sec:mssm-mssmrootfiles}) for various decay channels to obtain the
most accurate result for the branching ratios currently available. In
a first step, all partial widths have been calculated as accurately as
possible. Then the branching ratios have been derived from this full
set of partial widths.  Concretely, we used \FeynHiggs\ for the
evaluation of the Higgs boson masses and couplings from the original
input parameters, including corrections up to the two-loop level.  The
status of the various evaluations in \FeynHiggs\ and \HDECAY\ are
detailed in \Bref{Dittmaier:2012vm}.  The total decay width of the
neutral Higgs bosons, $\phi = \Ph, \PH, \PA$, is calculated as,
\begin{align}
\Gamma_\phi &=
  \Gamma^{\mathrm{FH}}_{\phitautau}
+ \Gamma^{\mathrm{FH}}_{\phimumu}
+ \Gamma^{\mathrm{FH/P4f}}_{\phiWW}
+ \Gamma^{\mathrm{FH/P4f}}_{\phiZZ} \nonumber\\
&\quad
+ \Gamma^{\mathrm{HD}}_{\phibb}
+ \Gamma^{\mathrm{HD}}_{\phitt}
+ \Gamma^{\mathrm{HD}}_{\phicc}
+ \Gamma^{\mathrm{HD}}_{\phigg}
+ \Gamma^{\mathrm{HD}}_{\phigaga}
+ \Gamma^{\mathrm{HD}}_{\phiZga} \nonumber \\
&\quad
+ \Gamma^{\mathrm{FH}}_{\phiZh}
+ \Gamma^{\mathrm{FH}}_{\phihh}
+ \Gamma^{\mathrm{FH}}_{\phiZA}
+ \Gamma^{\mathrm{FH}}_{\phiAA}
+ \Gamma^{\mathrm{HD}}_{\phiWHp}
+ \Gamma^{\mathrm{FH}}_{\phiSUSY}~,
\end{align}
followed by a corresponding evaluation of the respective branching
ratio.
With respect to previous evaluations, see \Bref{Heinemeyer:2013tqa},
$\Gamma^{\mathrm{FH}}_{\phiZh}$,
$\Gamma^{\mathrm{FH}}_{\phihh}$,
$\Gamma^{\mathrm{FH}}_{\phiZA}$,
$\Gamma^{\mathrm{FH}}_{\phiAA}$ and
$\Gamma^{\mathrm{HD}}_{\phiWHp}$ are now explicitly included as individual
channels.
The decays to SUSY particles are not included as individual branching
ratios, but of course taken into account in the total width%
\footnote{This was also done previously, but not explicitly stated
in \Brefs{Dittmaier:2012vm,Heinemeyer:2013tqa}.}%
. For completeness we also list the evaluation for the charged Higgs boson,
which has not changed wrt.\ \Bref{Heinemeyer:2013tqa}.
The total decay width of the charged Higgs boson is calculated as,
\begin{align}
\Gamma_{\PH^\pm} &=
  \Gamma^{\mathrm{FH}}_{\Hptaunu}
+ \Gamma^{\mathrm{FH}}_{\Hpmunu}
+ \Gamma^{\mathrm{FH}}_{\HphW}
+ \Gamma^{\mathrm{FH}}_{\HpHW}
+ \Gamma^{\mathrm{FH}}_{\HpAW} \nonumber\\
&\quad
+ \Gamma^{\mathrm{HD}}_{\Hptb}
+ \Gamma^{\mathrm{HD}}_{\Hpts}
+ \Gamma^{\mathrm{HD}}_{\Hptd}
+ \Gamma^{\mathrm{HD}}_{\Hpcb}
+ \Gamma^{\mathrm{HD}}_{\Hpcs}
+ \Gamma^{\mathrm{HD}}_{\Hpcd} \nonumber \\
&\quad
+ \Gamma^{\mathrm{HD}}_{\Hpub}
+ \Gamma^{\mathrm{HD}}_{\Hpus}
+ \Gamma^{\mathrm{HD}}_{\Hpud}
+ \Gamma^{\mathrm{FH}}_{\HpSUSY}~,
\end{align}
followed by a corresponding evaluation of the respective branching
ratio.


\chapter{Gluon-gluon Fusion}
\label{chap:ggF}
\ChapterAuthor{S.~Forte, D.~Gillberg, C.~Hays, A.~Lazopoulos, G.~Petrucciani, A.~Massironi, G.~Zanderighi~(Eds.);
C.~Anastasiou, A.~Banfi, M.~Bonvini, R.~Boughezal, F.~Caola, X.~Chen, B.~Di~Micco, F.A.~Dreyer, C.~Duhr, F.~Dulat, E.~Furlan, T.~Gehrmann, E.W.N.~Glover, F.~Herzog, M.~Jaquier, F.~Krauss, S.~Kuttimalai, A.~Lazopoulos, X.~Liu, P.~Maierh\"ofer, S.~Marzani, B.~Mistlberger, P.F.~Monni, F.~Petriello, E.~Re, G.P.~Salam, T.~Schmidt, M.~Sch\"onherr, M.~Spira, I.W.~Stewart, F.J.~Tackmann, K.~Tackmann, P.~Torrielli}
\providecommand{\pTcut}{p_T^{\rm cut}}
\providecommand{\cut}{{\rm cut}}

We present here an update of developments since the publication of
YR3.   First,  recent results on the inclusive
cross-section are discussed, and we provide recommendations for the
computation of its value and uncertainty. Then, jet-binned cross sections are
examined:
we provide
benchmarks for differential transverse-momentum distributions in the
effective theory. Finally the effects of heavy-quark
masses on these distributions are examined.

\section{The inclusive cross-section}
The inclusive gluon fusion Higgs boson production cross-section has a slowly
convergent perturbative expansion in QCD with large corrections at NLO
and NNLO. Therefore, uncertainties due to missing higher orders have
always been large and comparable to PDF uncertainties. Recently,
N$^3$LO QCD corrections have been computed in the effective theory as
an expansion around threshold in Ref.~\cite{Anastasiou:2016cez}, along with the evaluation of threshold resummation in
different schemes. In
this section the results of
Ref.~\cite{Anastasiou:2016cez} are reviewed; then, results on
threshold resummation
at the N$^3$LL matched  first with the NNLO (Sect.~\ref{subsec:spira}) , and then
with the N$^3$LO  (Sect.~\ref{subsec:bonvini}) fixed-order results are
presented.
fixed or level and its matching to the fixed-order result.
Finally, we present the summary for the computation of the
central value including all known corrections and the associated
total uncertainty.

\subsection[The N\texorpdfstring{$^3$}{3}LO cross section]{The N\texorpdfstring{$^3$}{3}LO cross section\SectionAuthor{C.~Anastasiou, C.~Duhr, F.~Dulat, E.~Furlan, T.~Gehrmann, F.~Herzog, A.~Lazopoulos, B.~Mistlberger}}
\label{subsec:ADDFGHLM}
\subsubsection{Ingredients of the computation}
This section summarizes our best prediction for the value of the inclusive gluon-fusion cross section and its uncertainties, following Ref.~\cite{Anastasiou:2016cez}. The main ingredient is the recent computation of gluon-fusion cross section through N$^3$LO in the effective theory where the top-quark is integrated out~\cite{Anastasiou:2014lda,Anastasiou:2014vaa,Anastasiou:2015ema}.

The master formula that summarizes all the ingredients entering our prediction for the partonic cross-sections is
\begin{equation}\label{eq:ADDmasteroriginal}
\hat{\sigma}_{ij} \simeq R_{LO}\,\left(\hat{\sigma}_{ij,EFT} + \delta_{t}\hat{\sigma}_{ij,EFT}^{NNLO} + \delta\hat{\sigma}_{ij,EW}\right) + \delta\hat{\sigma}_{ij,ex;t,b,c}^{LO}+\delta\hat{\sigma}_{ij,ex;t,b,c}^{NLO}\,.
\end{equation}
QCD corrections to the production cross-section $\hat{\sigma}_{ij,EFT}$ in the heavy-top limit have been included at NLO~\cite{Dawson:1990zj,Graudenz:1992pv,Djouadi:1991tka}, NNLO~\cite{Anastasiou:2002yz,Harlander:2002wh,Ravindran:2003um} and N$^3$LO~\cite{Anastasiou:2014lda,Anastasiou:2014vaa,Anastasiou:2015ema}. In addition, we also include effects from finite quark masses and electroweak effects, to the extent that they are available.  It was already observed at LO and NLO that the validity of the effective theory can be greatly enhanced by rescaling the effective theory by the exact LO result. We therefore rescale the cross-section $\hat{\sigma}_{ij,EFT}$ in the effective theory by the ratio
\begin{equation}
R_{LO} \equiv \frac{\sigma_{ex;t}^{LO}}{\sigma_{EFT}^{LO}}\,,
\label{eq:KLO}
\end{equation}
where $\sigma_{ex;t}^{LO}$ denotes the exact (hadronic) LO cross-section in the SM with a massive top quark and $N_f=5$ massless quarks. Moreover, at LO and NLO we know the exact result for the production cross-section in the SM, including all mass effects from top, bottom and charm quarks. We include these corrections into our prediction via the terms $\delta\hat{\sigma}_{ij,ex;t,b,c}^{(N)LO}$ in eq.~\eqref{eq:ADDmasteroriginal}, consistently matched to the contributions from the effective theory to avoid double counting. As a consequence, eq.~\eqref{eq:ADDmasteroriginal} agrees with the exact SM cross-section (with massless $u$, $d$ and $s$ quarks) through NLO in QCD. Beyond NLO, we only know the value of the cross-section in the heavy-top effective theory. We can, however, include subleading corrections at NNLO in the effective theory as an expansion in the inverse top mass~\cite{Harlander:2009mq, Pak:2009dg, Harlander:2009bw, Harlander:2009my}. These effects are taken into account through the term $\delta_{t}\hat{\sigma}_{ij,EFT}^{NNLO}$ in eq.~\eqref{eq:ADDmasteroriginal}, with the factor $R_{LO}$ scaled out.  They were originally computed with the top mass at the OS scheme, but their scheme dependence is expected to be at the sub-per mille level, following lower orders, and is hence considered negligible here.
We also include electroweak corrections to the gluon-fusion cross-section (normalized to the exact LO cross-section) through the term $\delta\hat{\sigma}_{ij,EW}$ in eq.~\eqref{eq:ADDmasteroriginal}. Unlike QCD corrections, electroweak corrections have only been computed through NLO in the electromagnetic coupling constant $\alpha$~\cite{Aglietti:2004nj,Actis:2008ug,Actis:2008ts}. Moreover, mixed QCD-electroweak corrections, i.e., corrections proportional to $\alpha\,\alpha_s^3$, are known in an effective 
theory~\cite{Anastasiou:2008tj} valid in the limit where not only the top quark but also the electroweak bosons are much heavier than the Higgs boson. In this limit the interaction of the Higgs boson with the $W$ and $Z$ bosons is described via a point-like vertex coupling the gluons to the Higgs boson. Higher-order corrections in this limit can thus be included into the Wilson coefficient in front of the dimension-five operator describing the effective interaction of the gluons with the Higgs boson.
The validity and limitations of this approximation are discussed in Section~\ref{sec:breakdown}.

\subsubsection{Summary of results}
The numerical results quoted in this section are valid for the following set of input parameters:
\begin{table}[!h]
\begin{center}
\begin{tabular}{cc}
    \toprule
    
    $\sqrt{S}$	&  13~{\textrm{TeV}}                      \\
$m_h$  		&  125~{\textrm{GeV}}                      \\
PDF 			&  {\tt PDF4LHC15\_nnlo\_100}                      \\
$\alpha_s(m_Z)$ 	&  0.118                    \\
$m_t(m_t)$	&  162.7~{\textrm{GeV}} ($\overline{\textrm{MS}}$)\\
$m_b(m_b)$	&  4.18 ~{\textrm{GeV}} ($\overline{\textrm{MS}}$)\\
$m_c(3GeV)$	&  0.986~{\textrm{GeV}}  ($\overline{\textrm{MS}}$)\\
$\mu=\mu_R=\mu_F$	&  62.5~{\textrm{GeV}}  ($=m_H/2$)\\
    \bottomrule
\end{tabular}
\end{center}
\end{table}

Using these input parameters, our current best prediction for the production cross section 
 of a Higgs boson with a mass $m_H = 125$ GeV at the LHC with a centre-of-mass energy of 13 TeV is
\begin{equation}
\label{eq:finalresult_13TeV}
\boxed{
\sigma = 48.58\,{\rm pb} {}^{+2.22\, {\rm pb}\, (+4.56\%)}_{-3.27\, {\rm pb}\, (-6.72\%)} \mbox{ (theory)} 
\pm 1.56 \,{\rm pb}\, (3.20\%)  \mbox{ (PDF+$\alpha_s$)} \,.
}
\end{equation}

The central value in eq.~\eqref{eq:finalresult_13TeV}, computed at the central scale $\mu_F=\mu_R=m_H/2$, is the combination of all the effects
considered in eq.~\eqref{eq:ADDmasteroriginal}. The
breakdown of the different effects is:
\begin{equation}\label{eq:central_value_breakdown}
\begin{array}{rrrl}
48.58\, \textrm{pb} =&  16.00 \, {\rm pb}  & \quad (+32.9\%) & \qquad (\textrm{LO, rEFT})   \\ 
                              & \,+\,  20.84 \,{\rm pb} &  \quad
                                (+42.9\%) & \qquad (\textrm{NLO, rEFT} )   \\
                              &\,-\, \phantom{a}2.05\, {\rm pb}  & \quad (-4.2\%)
                                                              &\qquad  ((t, b, c)\textrm{, exact NLO})  \\
                              & \, + \,\phantom{a} 9.56 \, {\rm pb} & \quad (+19.7\%) & \qquad (\textrm{NNLO, rEFT}) \\
                             &\,+\,\phantom{a} 0.34\, {\rm pb} &  \quad (+0.7\%) &  \qquad (\textrm{NNLO, }1/m_t)  \\
                             &\,+ \,\phantom{a} 2.40\, {\rm pb}  & \quad (+4.9\%) & \qquad
                                                         (\textrm{EW, QCD-EW}) \\
                              &\,+\,\phantom{a}  1.49\, {\rm pb}  & \quad (+3.1\%)
                                                              & \qquad (\textrm{N$^3$LO, rEFT})   
\end{array}
\end{equation}
where rEFT denotes the cross section in the effective field theory approximation rescaled by $R_{LO}$ of \eqref{eq:KLO} .
We note that  the N$^3$LO central value is completely insensitive to threshold resummation effects for $\mu_F=\mu_R=m_H/2$ and the central value obtained from a fixed-order N$^3$LO computation and a resummed computation at N$^3$LO + N$^3$LL are identical for this scale choice. We therefore conclude that threshold resummation does not provide any improvement of the central value, and it is therefore not included in our prediction.

The PDF and $\alpha_s$ uncertainties are computed following the recommendation of the PDF4LHC working group.
The remaining theory-uncertainty in eq.~\eqref{eq:finalresult_13TeV} is obtained by adding linearly various sources of theoretical uncertainty, which affect the different contributions to the cross section in eq.~\eqref{eq:ADDmasteroriginal}. The breakdown of the different theoretical uncertainties whose linear sum produces the theoretical uncertainty in eq.~\eqref{eq:finalresult_13TeV} is
\begin{center}
\begin{tabular}{cccccc}
\toprule
\begin{tabular}{c} $\delta$(scale) \end{tabular} &
\begin{tabular}{c} $\delta$(trunc) \end{tabular} &
\begin{tabular}{c} $\delta$(PDF-TH) \end{tabular} &
\begin{tabular}{c} $\delta$(EW) \end{tabular} & 
\begin{tabular}{c} $\delta(t,b,c)$  \end{tabular} & 
\begin{tabular}{c} $\delta(1/m_t)$ \end{tabular}\\ \midrule
${}^{+0.10\textrm{ pb}}_{-1.15\textrm{ pb}} $ & $\pm$0.18 pb & $\pm$0.56 pb & $\pm$0.49 pb& $\pm$0.40 pb& $\pm$0.49 pb
\\ \midrule
${}^{+0.21\%}_{-2.37\%}$ & $\pm 0.37\%$ & $\pm 1.16\%$ & $\pm 1\%$ & $\pm 0.83\%$ 
& $\pm 1\%$ \\
 \bottomrule
\end{tabular}
\end{center}
In the remainder of this note we address each of the components that enter the final theoretical uncertainty estimate in turn.

\subsubsection{Breakdown of the theoretical uncertainties}
\label{sec:breakdown}

\indent {\bf Uncertainty from missing higher orders: $\delta$(scale)}\\
The uncertainty $\delta$(scale) captures the impact of missing higher order terms in the perturbative expansion of the cross section in the rEFT. We identify this uncertainty with the scale variation when varying the renormalization and factorization scales simultaneously in the interval $\mu_F=\mu_R\in[m_H/4,m_H]$. The N$^3$LO corrections moderately increase ($\sim 3\%$) 
the cross-section for renormalization and factorization scales equal
to  $m_H/2$. In addition, they notably stabilize the scale variation,
reducing it almost by a factor of five compared to NNLO. The N$^3$LO scale-variation band 
is included entirely within the NNLO scale-variation band for scales
in the interval $[{m_H}/{4}, m_H ]$. We note that, while we vary the scales simultaneously, we have checked (see Figure 6 of \cite{Anastasiou:2016cez}) that the factorization scale dependence is flat, and the scale dependence at N$^3$LO is driven by the renormalization scale dependence. 

It is important to assess how well the scale uncertainty captures the uncertainty due to missing higher orders in the perturbative expansion, given that it failed to capture the shift in the central value due to missing perturbative orders at lower orders.  We have found 
good evidence that the N$^3$LO scale variation captures the effects of missing higher 
perturbative orders in the EFT. We base this conclusion on the following observations:
First, we observe that expanding in
$\alpha_s$ separately the Wilson coefficient and matrix-element
factors in the cross-section gives results consistent with expanding
directly their product through N$^3$LO. Second, 
a traditional threshold resummation in Mellin space up to N$^3$LL did not contribute 
significantly to the cross-section beyond N$^3$LO in the range of scales 
$\mu \in [{m_H}/{4}, m_H ]$. 
Although  the effects of threshold resummation are in general sensitive to
ambiguities due to subleading terms beyond the soft limit, we found
that within our preferred range of scales, several variants of the exponentiation formula
gave very similar phenomenological results, which are always consistent with
fixed-order perturbation theory.  Finally, a soft-gluon and
$\pi^2$-resummation using the SCET formalism also gave consistent
results with fixed-order perturbation theory at N$^3$LO. While 
ambiguities in subleading soft terms limit the use of soft-gluon
resummation as an estimator of higher-order effects, and while it is of
course possible that some variant of resummation may  
yield larger corrections, it is encouraging that this does not happen
for the mainstream prescriptions studied here. 

 We conclude this discussion by commenting on the use of resummation to estimate the uncertainty on the cross section. Based on the considerations from the previous paragraph, we are led to conclude that threshold resummation does not provide any improvement over a fixed-order calculation, and we therefore do not include it into our prediction. We base this conclusion on the following facts. First, we have already observed in the previous section that the central value at N$^3$LO for $\mu\equiv\mu_R=\mu_F=m_H/2$ is insensitive to resummation effects, excluding any need for resummation to improve the fixed order prediction. Second, the scale variation for N$^3$LO + N$^3$LL is contained inside the fixed-order N$^3$LO scale-variation band for $\mu\in[m_H/4,m_H]$ for a variety of different formalisms to perform threshold resummation, indicating that threshold resummation is more likely to underestimate the uncertainty from the variation. Finally, we point out that the resummation program itself is plagued by systematic uncertainties coming from  terms that are power suppressed in the threshold variable $(s-m_H^2)$ (or equivalently, in $1/N$ in Mellin space) and  are not controlled by the resummation. Although formally of higher order, these uncontrolled terms can have a substantial impact on the cross section, which is in our opinion not physically motivated. Any conclusion based on varying these uncontrolled power-suppressed and constant terms should therefore be discarded in our opinion, and they are not considered in our prediction.
  
{\bf The truncation uncertainty: $\delta$(trunc)}\\
The truncation uncertainty captures the uncertainty introduced by the fact the N$^3$LO corrections are currently only  known as an expansion around threshold,  i.e., as an expansion in the amount of energy available to QCD radiation, to order 37. We assign an uncertainty due to the truncation of the threshold expansion which is as large 
as 
\begin{equation}\label{eq:trunc_error}
\delta({\rm trunc}) = 10 \times \frac{\sigma^{(3)}_{EFT}(37)  -  \sigma^{(3)}_{EFT}(27)    }{\sigma^{\textrm{N$^3$LO}}_{EFT} } = 0.37\%\,.
\end{equation}
The factor $10$ is a conservative estimator of the progression of the
series beyond the first $37$
terms. Note that the complete N$^3$LO cross-section appears in the denominator of eq.~\eqref{eq:trunc_error}, i.e., the uncertainty applies to the complete  N$^3$LO result, not just the coefficient of $\alpha_s^5$.

{\bf The uncertainty from missing N$^3$LO PDFs: $\delta$(PDF-TH)}\\
So far, PDFs have only been extracted  by comparing data to theory predictions at NNLO in QCD, and so
an inconsistency may only arise due to the extraction
of the parton densities from data for which there are no N$^3$LO
predictions.
To assess this uncertainty we resort to the experience from the
previous orders and investigate the shift in the NNLO cross section
when it is computed with either NLO or NNLO PDFs.
We observe that as a function of the factorization scale in the range $\mu_F\in[m_H/4,m_H]$
(with the renormalization scale held fixed) 
scale)  
the NNLO cross-section decreases by about $2.2-2.4\%$  when NNLO
PDFs are used instead of NLO PDFs. Given that N$^3$LO
corrections are expected to be milder in general than their
counterparts at NNLO, we anticipate that they will induce a smaller
shift than at NNLO. Based on these considerations, we assign a
conservative
uncertainty estimate due to missing higher orders in the extraction of the
parton densities obtained as
\begin{equation}\label{eq:delta_PDF-TH}
\delta({\rm PDF-TH}) = \frac{1}{2}\,\left|\frac{\sigma_{EFT}^{(2),NNLO}-\sigma_{EFT}^{(2),NLO}}{\sigma_{EFT}^{(2),NNLO}}\right|=\frac{1}{2}\, 2.31\% = 1.16\%\,,
\end{equation}
where $\sigma_{EFT}^{(2),(N)NLO}$ denotes the NNLO cross-section evaluated with (N)NLO PDFs 
at the central scale $\mu_F=\mu_R=m_H/2$.
In the above, the strong coupling was set to its world-average at the $Z$ pole and evolved using three-loop renormalization group running, and we assumed conservatively that the size of the N$^3$LO corrections is
about half of the corresponding NNLO corrections.  
 This estimate is supported by the magnitude of the third-order corrections 
to the coefficient functions for deep inelastic scattering~\cite{Vermaseren:2005qc} and a 
related gluonic scattering process~\cite{Soar:2009yh}, which are the 
only two coefficient functions 
that were computed previously to this level of accuracy.

{\bf The uncertainty due to missing QCD-EW corrections: $\delta$(EW)}\\
Given the large size of the NLO QCD corrections to the Higgs cross-section, we may expect that also the EW corrections to the NLO QCD cross-section cannot be neglected. Unfortunately, these so-called mixed QCD-EW corrections are at present unknown, and only the contribution from an EFT approximation where the weak bosons are heavier than the Higgs boson are taken into account.
The effective theory method for the mixed QCD-EW corrections 
is of course not entirely satisfactory, because the computation of the EW Wilson coefficient assumes the validity of the $m_H/m_V$ expansion, $V=W,Z$ while clearly $m_H>m_V$. 
We thus need to carefully assess the uncertainty on the mixed QCD-EW corrections due to the EFT approximation. 
In the region $m_H > m_V$, we expect  effects from virtual weak bosons going on shell to be important and one should not expect that a naive application of the  EFT can give a reliable value for the cross-section. However, the EFT is only used to predict 
the relative size of QCD radiative corrections with respect to the leading 
order electroweak corrections, i.e., the dominant electroweak threshold effects from pairs of weak bosons going on shell should already be captured by the leading-order electroweak corrections. This can only vary mildly above and below threshold. For phenomenological purposes, we expect that the rescaling with the exact NLO EW corrections captures the bulk of threshold effects at 
all perturbative orders. To quantify the remaining uncertainty in this approach, we allow the EW Wilson coefficient $C_{1w}$ to vary by a factor of $3$ around its central value. We do this by introducing a rescaling factor $y_{\lambda}$ by
\begin{equation}\label{eq:y_lamda}
\lambda_{EW}\, (1+C_{1w}\,a_s +\ldots) \to \lambda_{EW} \,(1+y_\lambda\, C_{1w}\,a_s +\ldots)\,,
\end{equation}
where $a_s = \alpha_s/\pi$.
Varying $y_{\lambda}$ in the range $[1/3,3]$, we see that the cross-section varies by $-0.2\%$ to $+0.4\% $. Note that the result obtained by assuming complete factorization of EW and QCD corrections lies in the middle of the variation range, slightly higher than the $y_\lambda=1$ prediction.  Finally, we stress that the choice of the range is largely arbitrary of course. It is worth noting, however, that in order to reach uncertainties of the order of $1\%$, one needs to enlarge the range to $y_\lambda\in[-3,6]$.

An alternative way to assess the uncertainty on the mixed QCD-EW corrections is to note that the factorization of the EW corrections is exact in the soft and collinear limits of the NLO phase space. The hard contribution, however, might be badly captured. At NLO in QCD, the hard contribution amounts to $\sim 40\%$ of the $\mathcal{O}(a_s^3)$ contribution to the cross-section, where we define the \emph{hard contribution} as the NLO cross-section minus its soft-virtual contribution, i.e., the NLO contribution that does not arise from the universal exponentiation of soft gluon radiation. The \emph{hard contribution} is defined as the convolution of the parton-level quantity
\begin{equation}
\frac{\hat{\sigma}_{ij}^{(1),\textrm{hard}}}{z}\equiv\frac{\pi |C_0|^2}{8V}\,a_s^3\,\eta^{(1),\textrm{reg}}_{ij}(z)
\end{equation}
with the PDFs, which
receive contributions from the $gg$, $qg$ and $q\bar{q}$ initial state channels.
The mixed QCD-EW corrections  are $3.2\%$ of the total cross-section. Even if the 
 uncertainty of the  factorization ansatz is taken to be as large as the entire hard contribution, we will obtain
an estimate of the uncertainty equal to $0.4\times 3.2\%=1.3\%$ with respect to the total cross-section. 
  
An alternative way to define the \emph{hard contribution} is to look at the real emission cross-section regulated by a subtraction term in the FKS scheme~\cite{Frixione:1995ms}. We could then exclude the contribution of the integrated subtraction term, which is proportional to the Born matrix element, and hence of soft-collinear nature. We would then estimate the \emph{hard contribution} as $\sim 10\%$ of the $\mathcal{O}(a_s^3)$ contribution to the cross-section, which would lead to an uncertainty equal to $0.1\times 3.2\%=0.32\%$.
  
 We note that the different estimates of the uncertainty range from $0.2\%$ to $1.3\%$.
We therefore assign, conservatively, an uncertainty of $1\%$ due to mixed QCD-EW corrections for 
LHC energies.

{\bf The missing $b$ and $c$-quark mass effects: $\delta(t,b,c)$}\\
Unlike the case of the top quark, the contributions of the bottom and charm quarks at NNLO are entirely unknown. We estimate the uncertainty of the missing interference between the top and light quarks within the $\overline{ \rm MS }$-scheme as:
\begin{equation}
\label{eq:tbc_uncertainty}
\delta(t,b,c)^{\overline{\rm MS}} = \pm\, \left| \,\frac{\delta\sigma_{ex;t}^{NLO}-\delta\sigma_{ex;t+b+c}^{NLO}}{\delta\sigma_{ex;t}^{NLO}}\,\right|\,
(R_{LO}\delta\sigma_{EFT}^{NNLO}+\delta_{t}\hat{\sigma}_{gg+qg,EFT}^{NNLO}) \simeq \pm 0.31\,\textrm{pb}\,,
\end{equation}
where 
\begin{equation}
\delta\sigma_{X}^{NLO} \equiv \sigma_{X}^{NLO}-\sigma_{X}^{LO}
{\rm~~and~~}
\delta\sigma_{X}^{NNLO} \equiv \sigma_{X}^{NNLO}-\sigma_{X}^{NLO}\,.
\end{equation}

Our preferred scheme is the $\overline{\rm MS}$-scheme {\bf (with $\mu_R=m_H/2$})
  due to the bad convergence of the perturbative series
  for the conversion from an $\overline{\rm MS}$ mass to a pole
  mass for the bottom and charm
  quarks~\cite{Marquard:2016vmy,Marquard:2015qpa}. 
To account for the difference with the OS scheme, we enlarge the uncertainty on 
$\sigma_{t+b+c} - \sigma_t$, as estimated via 
eq.~\eqref{eq:tbc_uncertainty}  within the  $\overline{\rm MS}$
scheme, by multiplying it with a factor of 1.3, 
\begin{equation}
\delta(t,b,c) = 1.3  \, \delta(t,b,c)^{\overline{\rm MS}}\,.
\end{equation}

{\bf Uncertainty from top-mass effects at NNLO: $\delta(1/m_t)$}\\
The corrections due to top-mass effects at NNLO, as an expansion in $1/m_t$, are included through the term $\delta_{t}\hat{\sigma}_{ij,EFT}^{NNLO}$ in eq.~\eqref{eq:ADDmasteroriginal}.
The $1/m_t$ expansion is in fact an expansion in ${s}/m_t^2$, and consequently it needs to be matched to the high-energy limit of the cross-section, known to leading logarithmic accuracy from $k_t$-factorization. The high-energy limit corresponds to the contribution from small values of $z$ to the convolution integral with the parton luminosities. Since this region is suppressed by the luminosity, a lack of knowledge of the precise matching term is not disastrous and induces an uncertainty of roughly $1\%$, which is of the order of magnitude of the net contribution. In conclusion, following the analysis of ref.~\cite{Harlander:2009my}, whose conclusions were confirmed by ref.~\cite{Pak:2009dg}, we assign an overall uncertainty of $1\%$ due to the unknown top-quark effects at NNLO.

\subsection[N\texorpdfstring{$^3$}{3}LL resummation]{N\texorpdfstring{$^3$}{3}LL resummation\SectionAuthor{T.~Schmidt, M.~Spira}}
\label{subsec:spira}
The inclusive gluon-fusion cross section for Higgs boson production can
be improved by performing a threshold resummation of soft, virtual and
collinear gluon effects \cite{Kramer:1996iq}. This resummation is
performed in Mellin space according to the conventional formalism used
before for Higgs boson production \cite{Catani:2003zt, Moch:2005ky,
Ravindran:2005vv, Ravindran:2006cg, deFlorian:2009hc, deFlorian:2012yg,
Bonvini:2014joa, Bonvini:2016frm}.  The resummed cross section develops
a factorized kernel structure in Mellin space
\begin{eqnarray}
\tilde G_{gg}^{(res)} \left(N;\alpha_s(\mu_R^2),
\frac{M_H^2}{\mu_R^2}; \frac{M_H^2}{\mu_F^2}; \frac{M_H^2}{m_t^2}
\right)
& = & \alpha_s^2(\mu_R^2) C_{gg} \left( \alpha_s(\mu_R^2),
\frac{M_H^2}{\mu_R^2}; \frac{M_H^2}{\mu_F^2}; \frac{M_H^2}{m_t^2}
\right)
\nonumber \\
& & \times \exp \left\{ \tilde{\cal
G}_H \left( \alpha_s(\mu_R^2), \log N; 
\frac{M_H^2}{\mu_R^2}, \frac{M_H^2}{\mu_F^2} \right) \right\} \, .
\label{eq:gghres}
\end{eqnarray}
We include top mass effects up to the NLL level explicitly in the
coefficient function
\begin{eqnarray}
C_{gg} \left( \alpha_s(\mu_R^2), \frac{M_H^2}{\mu_R^2};
\frac{M_H^2}{\mu_F^2}; \frac{M_H^2}{m_t^2} \right) & = & 1 +
\sum_{n=1}^\infty
\left(\frac{\alpha_s(\mu_R^2)}{\pi}\right)^n C_{gg}^{(n)}
\left( \frac{M_H^2}{\mu_R^2}, \frac{M_H^2}{\mu_F^2}; \frac{M_H^2}{m_t^2}
\right)
\end{eqnarray}
that contains the finite parts of the virtual corrections. The NLO
contribution is given explicitly by ($\tau_Q = 4m_Q^2/M_H^2$)
\cite{deFlorian:2012yg}
\begin{eqnarray}
C_{gg}^{(1)} & = & c_H(\tau_t) + 6\zeta_2 + \frac{33-2N_F}{6}
\log\frac{\mu_R^2}{\mu_F^2} + 6(\gamma_E^2+\zeta_2) -
6\gamma_E \log\frac{M_H^2}{\mu_F^2} \, ,
\end{eqnarray}
where the function $c_H(\tau_t)$ approaches
$11/2$ in the limit of heavy top quarks. The resummed exponential
develops the expansion
\begin{eqnarray}
\tilde{\cal G}_H \left( \alpha_s(\mu_R^2), \log N;
\frac{M_H^2}{\mu_R^2}, \frac{M_H^2}{\mu_F^2} \right) & = & \log
N~g_H^{(1)}(\lambda) \nonumber \\
& + & \left. \sum_{n=2}^\infty \alpha_s^{n-2}(\mu_R^2) g_H^{(n)}\left(
\lambda,
\frac{M_H^2}{\mu_R^2}; \frac{M_H^2}{\mu_F^2} \right)
\right|_{\lambda=b_0 \alpha_s(\mu_R^2) \log N}
\label{eq:resum}
\end{eqnarray}
with $b_0$ denoting the leading order term of the QCD beta function,
\begin{equation}
b_0 = \frac{33-2N_F}{12\pi}
\end{equation} 
where $N_F$ is the number of active flavours that we choose as $N_F=5$
in the following, i.e.~the top quark has been decoupled from the strong
coupling $\alpha_s$ and the PDFs. The individual functions
$g_H^{(i)}~(i=1,\ldots,4)$ can be found e.g.~in \cite{Catani:2003zt,
Vogt:2000ci, Moch:2005ba, Laenen:2005uz}. The leading and subleading
collinear gluon effects have been included by the replacements
\cite{Kramer:1996iq, Catani:2003zt, Catani:2001ic, Schmidt:2015cea}
\begin{eqnarray}
C_{gg}^{(1)} & \to & C_{gg}^{(1)} + 6~\frac{\tilde L}{N} \nonumber \\
C_{gg}^{(2)} & \to & C_{gg}^{(2)} + (48-N_F)~\frac{\tilde L^2}{N}
\label{eq:coll}
\end{eqnarray} 
with the modified logarithm
\begin{eqnarray}
\tilde L = \log \frac{N e^{\gamma_E} \mu_F}{M_H} = \log N + \gamma_E -
\frac{1}{2} \log \frac{M_H^2}{\mu_F^2} \, .
\end{eqnarray}
These replacements reproduce the leading and subleading collinear
logarithms up to N$^3$LO.

The general expression for the inclusive cross section can be cast into
the form
\begin{eqnarray} 
\!\!\!\!\!\!\! && \sigma(s,M_H^2) = \sigma_{tt}^0
\int_{C-i\infty}^{C+i\infty} \frac{dN}{2\pi i}
\left(\frac{M_H^2}{s}\right)^{-N+1} \tilde f_g(N,\mu_F^2) \tilde
f_g(N,\mu_F^2) \nonumber \\
\!\!\!\!\!\!\! && \times \left\{ \tilde G^{(res)}_{gg}
\left(N;\alpha_s(\mu_R^2), \frac{M_H^2}{\mu_R^2};
\frac{M_H^2}{\mu_F^2};0 \right)
- \left[ \tilde G^{(res)}_{gg} \left(N;\alpha_s(\mu_R^2),
  \frac{M_H^2}{\mu_R^2}; \frac{M_H^2}{\mu_F^2};0 \right)
\right]_{(NNLO)} \right\} \nonumber \\
\!\!\!\!\!\!\! && + \sigma_{tt}^0 \int_{C-i\infty}^{C+i\infty}
\frac{dN}{2\pi i} \left(\frac{M_H^2}{s}\right)^{-N+1} \tilde
f_g(N,\mu_F^2) \tilde f_g(N,\mu_F^2) \nonumber \\
\!\!\!\!\!\!\! && \times \left\{ \tilde G^{(res)}_{gg,NLL}
\left(N;\alpha_s(\mu_R^2), \frac{M_H^2}{\mu_R^2}; \frac{M_H^2}{\mu_F^2};
\frac{M_H^2}{m_t^2} \right)
- \tilde G^{(res)}_{gg,NLL} \left(N;\alpha_s(\mu_R^2),
  \frac{M_H^2}{\mu_R^2}; \frac{M_H^2}{\mu_F^2};0 \right) \right.
\nonumber \\
\!\!\!\!\!\!\! && \left. - \left[\tilde G^{(res)}_{gg,NLL}
\left(N;\alpha_s(\mu_R^2), \frac{M_H^2}{\mu_R^2}; \frac{M_H^2}{\mu_F^2};
\frac{M_H^2}{m_t^2} \right)
- \tilde G^{(res)}_{gg,NLL} \left(N;\alpha_s(\mu_R^2),
  \frac{M_H^2}{\mu_R^2}; \frac{M_H^2}{\mu_F^2};0 \right) \right]_{(NLO)}
\right\} \nonumber \\
\!\!\!\!\!\!\! && + \sigma_{t+b+c}^{NNLO}(s,M_H^2)
\label{eq:resmatch}
\end{eqnarray}
with $\sigma_{tt}^0$ denoting top quark contribution to the LO cross
section factor
\begin{eqnarray}
\sigma_0 & = & \frac{G_F}{288\sqrt{2}\pi} \left| \sum_Q
A_Q(\tau_Q)\right|^2 \nonumber \\
A_Q(\tau) & = & \frac{3}{2}\tau[1+(1-\tau) f(\tau)] \nonumber \\
f(\tau) & = & \left\{ \begin{array}{ll}
\displaystyle \arcsin^2 \frac{1}{\sqrt{\tau}} & \tau \ge 1 \\
\displaystyle - \frac{1}{4} \left[ \log \frac{1+\sqrt{1-\tau}}
{1-\sqrt{1-\tau}} - i\pi \right]^2 & \tau < 1
\end{array} \right. 
\label{eq:ggh}
\end{eqnarray}
and $\tilde f_g$ is the Mellin moment of the gluon density. Moreover, in
order to reside to the right of all poles in the complex Mellin plane an
offset $C$ is introduced for the integration contour. The Landau
singularity at large values of $N$ on the other hand is ensured to lie
on the right side of the integration contour \cite{Catani:1996dj,
Catani:1996yz}. The index `$(NNLO)$' in the second line indicates the
fixed-order expansion of the resummed coefficient function $\tilde
G^{(res)}_{gg}$ in Mellin space up to NNLO while the index `($NLO$)'
denotes the perturbative expansion of the NLL resummed coefficient
function $\tilde G^{(res)}_{gg,NLL}$ in Mellin space up to NLO. The
first integral has been convolved with N$^3$LO $\alpha_s$ and NNLO PDFs
according to the discussion about the non-necessity of N$^3$LO PDFs of
Ref.~\cite{Forte:2013mda} and of resummed PDFs of
Ref.~\cite{Bonvini:2015ira} for the SM Higgs boson mass, while the second
integral has been evaluated with NLO $\alpha_s$ and PDFs consistently.
The fixed-order NNLO cross section \cite{Harlander:2002wh,
Anastasiou:2002yz, Ravindran:2003um} of the last term has been derived as
\begin{eqnarray}
\sigma_{t+b+c}^{NNLO}(s,M_H^2) = \sigma_{\infty}^{NNLO}(s,M_H^2) +
\sigma_{t+b+c}^{NLO}(s,M_H^2) - \sigma_{\infty}^{NLO}(s,M_H^2)
\end{eqnarray} 
with the individual parts
\begin{eqnarray}
\sigma_{\infty}^{NNLO}(s,M_H^2) & = & \sigma_{tt}^{LO} K_\infty^{NNLO}
\nonumber \\
\sigma_{\infty}^{NLO}(s,M_H^2) & = & \sigma_{tt}^{LO} K_\infty^{NLO}
\nonumber \\
\sigma_{t+b+c}^{NLO}(s,M_H^2) & = & \sigma_{t+b+c}^{LO} K_{t+b+c}^{NLO}
\end{eqnarray}
where $\sigma_{tt}^{LO}$ denotes the full LO cross section including
only top loops, $\sigma_{t+b+c}^{LO}$ the LO cross section including top
and bottom/charm loops, $K_\infty^{(N)NLO}$ the (N)NLO K-factors
obtained in the limit of heavy top quarks and $K_{t+b+c}^{NLO}$ the full
NLO K-factor including top and bottom/charm loops.  The NNLO parts have
been derived with N$^3$LO $\alpha_s$ and NNLO PDFs and the NLO terms
with NLO $\alpha_s$ and PDFs consistently as implemented in the programs
HIGLU \cite{Spira:1995mt, Spira:1996if} and SusHi
\cite{Harlander:2012pb}. This implementation guarantees that top mass
effects are treated at NLL level and bottom/charm contributions at fixed
NLO respectively.

Since the virtual coefficient of the bottom contributions behaves in the
limit $M_H^2\gg m_b^2$ as \cite{Spira:1995rr} $(C_A = 3, C_F = 4/3)$
\begin{equation}
c_H (\tau_b) \to \frac{C_A-C_F}{12} \log^2 \frac{M_H^2}{m_b^2} - C_F
\log \frac{M_H^2}{m_b^2}
\label{eq:smallmass}
\end{equation}
if the bottom mass is renormalized on-shell, it contains large
logarithms that are not resummed. The resummation of the Abelian part
proportional to $C_F$ has been performed in Ref.~\cite{Kotsky:1997rq,
Akhoury:2001mz, Melnikov:2016emg} up to
the subleading logarithmic level. These logarithms are related to the
Sudakov form factor at the virtual $Hb\bar b$ vertex that generates
these large logarithmic contributions for far off-shell bottom quarks
inside the corresponding loop contributions in the Abelian case. The
resummation of the non-Abelian part proportional to the Casimir factor
$C_A$ has not been considered so far. This type of logarithmic
contributions emerges from a different origin than the soft and
collinear gluon effects discussed so far and is the main source of the
very different size of QCD corrections to the bottom-loop contributions
\cite{Spira:1995rr, Graudenz:1992pv, Harlander:2005rq, Anastasiou:2009kn,
Anastasiou:2006hc, Aglietti:2006tp}. In order to obtain a reliable result
for the bottom contributions a resummation of these types of logarithms
is mandatory so that we do not include these contributions in our soft
and collinear gluon resummation but treat them at fixed NLO. Moreover,
according to the discussion presented about Figure~7a in
Ref.~\cite{Spira:1995rr}
we prefer to introduce quark pole masses also for the bottom and charm
quark, since the finite part of the virtual corrections is then of
moderate size due to an (accidental) cancellation of the logarithms
present in Eq.~(\ref{eq:smallmass}), while this contribution is
significantly larger if using the running $\overline{\rm MS}$ masses at
the scale $M_H/2$ so that the latter choice is disfavoured.

Following the recommendations of the LHC Higgs Cross Section Working
Group \cite{LHCHXSWG-INT-2015-006} our final results for the inclusive cross
section at N$^3$LL including NLO electroweak corrections
\cite{Djouadi:1994ge, Chetyrkin:1996wr, Chetyrkin:1996ke,
Aglietti:2004nj, Aglietti:2006yd, Degrassi:2004mx, Actis:2008ug,
Actis:2008ts} in factorized form are given in Table \ref{tb:cxn} for a
central renormalization and factorization scale choice
$\mu_R=\mu_F=M_H/2$.  Compared to our previous work
\cite{Schmidt:2015cea} the PDF+$\alpha_s$ uncertainties decreased
considerably due to the new PDF4LHC15 sets \cite{Butterworth:2015oua} of
recommended parton densities\footnote{If other PDF sets as ABM12
\cite{Alekhin:2013nda}, HERAPDF2.0 \cite{Abramowicz:2015mha} or JR14
\cite{Jimenez-Delgado:2014twa} are included the PDF+$\alpha_s$
uncertainties will increase considerably with a major part originating
from sizeable differences in the $\alpha_s$ fits at NNLO and different
data sets included in the global fits. Moreover, the proper treatment of
higher-twist effects in the global fits is an open aspect in this
context.}. The scale dependence has been obtained by an independent
variation of the renormalization and factorization scales by factors of
two up and down avoiding a splitting between these two scales by more
than a factor of two. The total uncertainties are evaluated by adding
the scale and PDF+$\alpha_s$ uncertainties linearly (to be
conservative). They range below the 10\%-level. Comparing the resummed
numbers with those of the N$^3$LO expansion of our resummed cross
section one obtains a resummation effect beyond N$^3$LO of less than two
per mille for our central scale choices so that resummation effects are
tiny. For different scale choices they can reach a level of about 2\%.
\begin{table}[hbt]
\caption{\label{tb:cxn} N$^{\,3\!}$LL Higgs boson production cross
sections via gluon fusion for different values of the Higgs boson mass
including the individual and total uncertainties due to the
renormalization and factorization scale dependence and PDF+$\alpha_s$
uncertainties including electroweak corrections using PDF4LHC15
\cite{Butterworth:2015oua} PDFs for a c.m.~energy $\sqrt{s}=13$ TeV. The
renormalization and factorization scales have been chosen as $M_H/2$.}
\renewcommand{\arraystretch}{1.5}
\begin{center}
\begin{tabular}{c|c|c|c|c}
\toprule
$M_H$ [GeV] & $\sigma(pp \to H + X)$ [pb] & scale & PDF+$\alpha_s$
& total \\ 
\midrule
124   & $47.53~pb$ & $^{+4.7\%}_{-5.4\%}$ & $\pm 3.3\%$ &
$^{+8.0\%}_{-8.7\%}$ \\ 
124.5 & $47.20~pb$ & $^{+4.6\%}_{-5.4\%}$ & $\pm 3.3\%$ &
$^{+7.9\%}_{-8.7\%}$ \\ 
125   & $46.87~pb$ & $^{+4.6\%}_{-5.4\%}$ & $\pm 3.3\%$ &
$^{+7.9\%}_{-8.7\%}$ \\ 
125.5 & $46.55~pb$ & $^{+4.6\%}_{-5.4\%}$ & $\pm 3.3\%$ &
$^{+7.8\%}_{-8.7\%}$ \\ 
126   & $46.22~pb$ & $^{+4.5\%}_{-5.4\%}$ & $\pm 3.3\%$ &
$^{+7.8\%}_{-8.7\%}$ \\ 
\bottomrule
\end{tabular}
\renewcommand{\arraystretch}{1.2}
\end{center}
\end{table}

Our numbers deviate from the explicit N$^3$LO results given in
Ref.~\cite{Anastasiou:2016cez} due to the different choice of quark mass
scheme for the top, bottom and charm contributions. If these are adopted
as running $\overline{\rm MS}$ quantities at the scale $M_H/2$ the cross
sections increase by 1.4\% compared to those with pole masses as shown
in Table~\ref{tb:cxnrun}. The numbers with $\overline{\rm MS}$ masses
agree with the full N$^3$LO results within about 2 per cent. The
differences are due to the omission of NNLO mass effects and the
virtual+soft+collinear approximation of our N$^3$LL terms in our
results.
\begin{table}[hbt]
\caption{\label{tb:cxnrun} N$^{\,3\!}$LL Higgs boson production
cross sections via gluon fusion for different values of the Higgs boson mass
for two different choices of the scheme for the top, bottom and charm 
quark
masses including electroweak corrections using PDF4LHC15
\cite{Butterworth:2015oua} PDFs for a c.m.~energy $\sqrt{s}=13$ TeV. The
last column shows the central N$^3$LO numbers of
Ref.~\cite{Anastasiou:2016cez}. The renormalization and factorization
scales have been chosen as $M_H/2$.}
\renewcommand{\arraystretch}{1.5}
\begin{center}
\begin{tabular}{c|c|c|c}
\toprule
$M_H$ [GeV] & pole masses & $\overline{m}_{Q}(M_H/2)$ &
Ref.~\cite{Anastasiou:2016cez} \\ 
\midrule
124   & $47.53~pb$ & $48.19~pb$ & $49.27~pb$ \\ 
124.5 & $47.20~pb$ & $47.83~pb$ & $48.92~pb$ \\
125   & $46.87~pb$ & $47.51~pb$ & $48.58~pb$ \\ 
125.5 & $46.55~pb$ & $47.20~pb$ & $48.23~pb$ \\
126   & $46.22~pb$ & $46.86~pb$ & $47.89~pb$ \\ 
\bottomrule
\end{tabular}
\renewcommand{\arraystretch}{1.2}
\end{center}
\end{table}


The whole framework of our resummation approach to the inclusive
gluon-fusion cross sections has also been applied to neutral Higgs boson
production within the MSSM, i.e.~providing resummed predictions at
N$^3$LL for the scalar Higgs bosons and at NNLL for the pseudoscalar
state \cite{Schmidt:2015cea, deFlorian:2007sr}.

\subsection[Combined fixed order and resummed results at N\texorpdfstring{$^3$}{3}LO+N\texorpdfstring{$^3$}{3}LL]{Combined fixed order and resummed results at N\texorpdfstring{$^3$}{3}LO+N\texorpdfstring{$^3$}{3}LL\SectionAuthor{M.~Bonvini, S.~Marzani}}
\label{subsec:bonvini}
In this contribution we briefly summarize the impact of threshold
resummation on the inclusive Higgs boson production cross-section, both in
terms of the shift in the central value, as well as a means to
faithfully estimate the theoretical uncertainty from missing higher
orders, $\delta(\text{mho})$, as detailed in
Ref.~\cite{Bonvini:2016frm}.
In this context, the proposed best estimate for the Higgs cross section is given by
\begin{equation} \label{eq:master}
\sigma_{\text{N$^3$LO+N$^3$LL}} = \sigma_{\text{N$^3$LO}} + \Delta_3\sigma_{\text{N$^3$LL}},
\end{equation}
where $ \sigma_{\text{N$^3$LO}}$ is the fixed-order cross section at N$^3$LO, as computed in
Refs.~\cite{Anastasiou:2014vaa,Anastasiou:2014lda,Anastasiou:2015ema,Anastasiou:2016cez}.
The second contribution, $\Delta_3\sigma_{\text{N$^3$LL}}$, contains the all-orders resummation of those
contributions that are enhanced in the threshold limit to N$^3$LL
accuracy~\cite{Bonvini:2014joa,Catani:2014uta,Bonvini:2014tea,Schmidt:2015cea},
minus its expansion to fixed N$^3$LO (so this contribution starts at N$^4$LO).
The computation of $\Delta_3\sigma_{\text{N$^3$LL}}$ Eq.~\eqref{eq:master} is done through the public code \texttt{TROLL}~\cite{TROLL},
formerly \texttt{ResHiggs}. \texttt{TROLL} does not compute the fixed order, but only the subtracted resummed contribution,
$\Delta_3\sigma_{\text{N$^3$LL}}$, so the fixed order has to be supplied by an external code.
In this section the code \texttt{ggHiggs}~\cite{ggHiggs} has been used.

The resummation is performed in a conjugate (Mellin) space, where the threshold limit corresponds to large $N$.
While resummation uniquely determines the coefficients of logarithmically enhanced terms and constants, there is a certain latitude in defining how the soft approximation is constructed, by making choices which differ by terms which vanish as $N\to\infty$.
In Refs.~\cite{Bonvini:2014joa, Ball:2013bra}, this freedom was
exploited to construct variants of threshold resummation that have
better perturbative properties. In particular, $\psi$-soft
resummation correctly reproduces, order by order in the strong
coupling, the analytic properties of fixed-order coefficient
functions. 
Moreover, as pointed out in Ref.~\cite{Bonvini:2014joa},
$\psi$-soft can be further improved by including in the calculation
more contributions to the soft expansion of the Altarelli-Parisi
splitting function. Two options have been considered:
\begin{itemize}
\item AP2 (default): $P_{gg}$ is retained to second order in $1-z$;
\item AP1: $P_{gg}$ is retained to first order in $1-z$.
\end{itemize}
Varying the order of this soft expansion (AP2 vs.\ AP1) can be used as an estimate of unknown contributions beyond the threshold limit.
Finally, in order to assess the impact of subleading contributions beyond N$^3$LL,
one can vary the way we deal with the constant terms, which can be treated in the default setup of Ref.~\cite{Bonvini:2016frm}
or in two extreme configurations:
\begin{itemize}
\item default: those constants coming from the Mellin transform of plus distributions~\cite{Altarelli:1981ax} are in the exponent, the others are not;
\item all constants in the exponent;
\item no constants  in the exponent.
\end{itemize}
Up to the working logarithmic accuracy, the position of the constants does not make any difference.
However, beyond the working logarithmic accuracy, moving constants produces, by interference, different subleading contributions.
Note that, since constants are known to play an important role for Higgs boson production~\cite{Ahrens:2008qu,Berger:2010xi,Stewart:2013faa},
these variations provide a robust way to estimate the perturbative uncertainty.
Combining together the different options for subleading logarithmic terms and subleading threshold terms one gets a total of $3\times2=6$
variants of the resummation.

For simplicity, and for disentangling effects coming from different sources, results are given in the clean environment of the
(rescaled) large-$m_t$ effective theory (rEFT), using the top mass $m_t=172.5$~GeV in the pole scheme,
the Higgs boson mass $m_H=125$~GeV, and the \texttt{PDF4LHC15\_nnlo\_100} PDF set~\cite
{Butterworth:2015oua,Carrazza:2015aoa,Ball:2014uwa,Harland-Lang:2014zoa,Dulat:2015mca}.
The strong coupling $\alpha_s$ is run from $\alpha_s(m_Z^2)=0.118$ (from the PDF set) to $\mu_R$ at four loops.
In order to show the stability of the resummed result, four options for the central common factorization and renormalization scale $\mu_0$ are considered:
$\mu_0 = m_H/4$, $\mu_0 = m_H/2$, $\mu_0 = m_H$, $\mu_0 = 2m_H$.
Then, the scales $\mu_R$ and $\mu_F$ are varied about $\mu_0$ by a factor of 2 up and down,
keeping the ratio $\mu_R/\mu_F$ never larger than 2 or smaller than $1/2$ (canonical 7-point scale variation).

In Ref.~\cite{Bonvini:2016frm}, $\psi$-soft with AP2 and with the natural choice for the constants is considered as the best option for threshold resummation. 
However, the other variants of $\psi$-soft have the same formal accuracy and are used to estimate the uncertainty from $1/N$ terms and subleading logarithmic corrections.
Thus, combining these variations with the 7-point scale variation a robust estimate of the uncertainty coming from unknown missing higher orders is proposed:
\begin{equation*}
\boxed{\delta(\text{mho}): \;\text{envelope of the canonical 7-point scale variations and the 6 variants of $\psi$-soft resummation}}
\end{equation*}
This corresponds to a total of $7\times 6=42$ cross section points from which one takes the highest and the lowest
cross sections as the maximum and minimum of the uncertainty band, and provides the most conservative way of including these uncertainties.

\begin{table}
  \caption{Fixed-order results and their scale uncertainty together with resummed results and their uncertainty
    (as given by the envelope of prescription and scale variations)
        for four choices of the central scale. Results are given in
        the (rescaled) large-$m_t$ effective theory with pole top mass
        (see text).}
\begin{center}
  \begin{tabular}{l|llll}
  \toprule
    & $\mu_0=m_H/4$ & $\mu_0=m_H/2$ & $\mu_0=m_H$ & $\mu_0=2m_H$ \\
    \midrule
             LO & $ 18.6^{+5.8}_{-3.9} $ & $ 16.0^{+4.3}_{-3.1} $ & $ 13.8^{+3.2}_{-2.4} $ & $ 11.9^{+2.5}_{-1.9} $ \\[1ex]
            NLO & $ 44.2^{+12.0}_{-8.5} $ & $ 36.9^{+8.4}_{-6.2} $ & $ 31.6^{+6.3}_{-4.8} $ & $ 27.5^{+4.9}_{-3.9} $ \\[1ex]
           NNLO & $ 50.7^{+3.4}_{-4.6} $ & $ 46.5^{+4.2}_{-4.7} $ & $ 42.4^{+4.6}_{-4.4} $ & $ 38.6^{+4.4}_{-4.0} $ \\[1ex]
        N$^3$LO & $ 48.1^{+0.0}_{-7.5} $ & $ 48.1^{+0.1}_{-1.8} $ & $ 46.5^{+1.6}_{-2.6} $ & $ 44.3^{+2.5}_{-2.9} $ \\[1ex]
\midrule
          LO+LL & $ 24.0^{+8.9}_{-6.8} $ & $ 20.1^{+6.2}_{-5.0} $ & $ 16.9^{+4.5}_{-3.7} $ & $ 14.3^{+3.3}_{-2.8} $ \\[1ex]
        NLO+NLL & $ 46.9^{+15.1}_{-12.6} $ & $ 46.2^{+15.0}_{-13.2} $ & $ 46.7^{+20.8}_{-13.8} $ & $ 47.3^{+26.1}_{-15.8} $ \\[1ex]
      NNLO+NNLL & $ 50.2^{+5.5}_{-5.3} $ & $ 50.1^{+3.0}_{-7.1} $ & $ 51.9^{+9.6}_{-8.9} $ & $ 54.9^{+17.6}_{-11.5} $ \\[1ex]
N$^3$LO+N$^3$LL & $ 47.7^{+1.0}_{-6.8} $ & $ 48.5^{+1.5}_{-1.9} $ & $ 50.1^{+5.9}_{-3.5} $ & $ 52.9^{+13.1}_{-5.3} $ \\[1ex]
\bottomrule
  \end{tabular}
  \label{tab:bonviniresults}
  \end{center}
\end{table}

\begin{figure}
  \centering
  \includegraphics[width=0.495\textwidth,page=2]{./WG1/ggF/figs/hxswg_hadr_xsec_res_eff_125_13_PDF4LHC15_uncertainty_summary3.pdf}
  \includegraphics[width=0.495\textwidth,page=1]{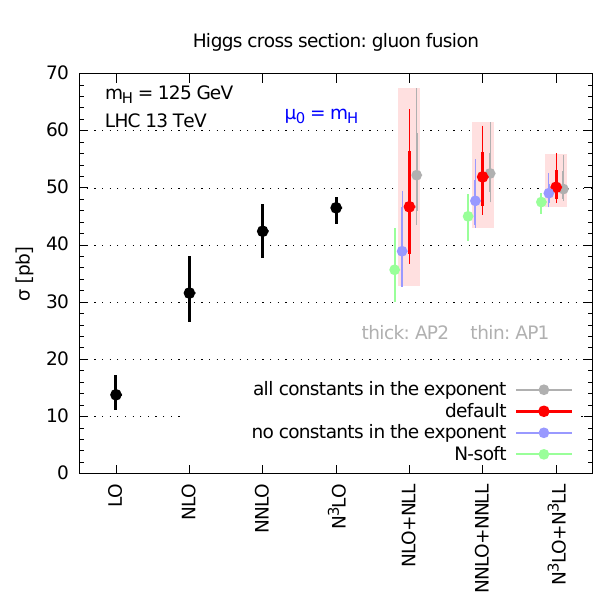}\\
  \includegraphics[width=0.495\textwidth,page=3]{./WG1/ggF/figs/hxswg_hadr_xsec_res_eff_125_13_PDF4LHC15_uncertainty_summary3.pdf}
  \includegraphics[width=0.495\textwidth,page=4]{./WG1/ggF/figs/hxswg_hadr_xsec_res_eff_125_13_PDF4LHC15_uncertainty_summary3.pdf}
  \caption{Higgs cross section at 13 TeV in the rescaled effective theory (rEFT), for four different choices of the central scale $\mu_F=\mu_R$:
    at the top we show $m_H/2$ and $m_H$, while at the bottom  $m_H/4$ and $2m_H$.
    The uncertainties on the fixed-order predictions and on $N$-soft come solely from scale variation;
    for the $\psi$-soft AP2 results the scale variation is shown as the thick uncertainty band. The 
    thinner bands correspond to the 7-point scale variation
    envelope on the $\psi$-soft AP1 instead, whose central value is not shown.
    The light-red rectangles are the envelope of all $\psi$-soft variants,
    corresponding to the 42-point uncertainty described in the text.}
  \label{fig:13TeV}
\end{figure}

Results for the cross section at fixed LO, NLO, NNLO and N$^3$LO accuracy, and its resummed counterpart at
LO+LL, NLO+NLL, NNLO+NNLL and N$^3$LO+N$^3$LL accuracy, are reported in \refT{tab:bonviniresults},
for the four central scale choices. The same results are shown as plots in \refF{fig:13TeV}.
For comparison, results for the ``standard'' threshold resummation (which we call $N$-soft) are also shown in the plots.
For $N$-soft one only keeps non-vanishing contributions in the large $N$ limit:
all the logarithmically enhanced contributions are in the resummed exponent, while the constant terms are not exponentiated.
For completeness, we also provide in \refT{tab:res} the various results at N$^3$LO+N$^3$LL for individual scales and resummation prescription.

\begin{table}[t]
 \caption{Resummed cross sections at N$^3$LO+N$^3$LL for the different prescriptions as a function of 
  the scales $\mu_F$ and $\mu_R$ over a wide range, for $m_H=125$~GeV and $\sqrt{s}=13$~TeV in the rEFT.}
  \centering
  \begin{tabular}{cc|rrrrrrr}
  \toprule
    & & \multicolumn{6}{c|}{$\psi$-soft} & \\
    & & \multicolumn{2}{c|}{default} & \multicolumn{2}{c|}{all constants in exp} & \multicolumn{2}{c|}{no constants in exp} & \\
    $\mu_F/m_H$ & $\mu_R/m_H$ & AP2 & AP1 & AP2 & AP1 & AP2 & AP1 & \multicolumn{1}{|c}{$N$-soft}\\
    \midrule
$   4$ & $   4$ & $56.8$ & $66.0$ & $56.8$ & $66.0$ & $51.2$ & $58.7$ & $49.4$ \\
$   4$ & $   2$ & $55.1$ & $62.3$ & $54.9$ & $62.0$ & $52.2$ & $58.6$ & $50.5$ \\
$   2$ & $   4$ & $53.2$ & $57.2$ & $53.7$ & $57.9$ & $48.2$ & $51.4$ & $46.0$ \\
$   2$ & $   2$ & $52.9$ & $56.0$ & $52.7$ & $55.8$ & $49.9$ & $52.5$ & $47.9$ \\
$   2$ & $   1$ & $51.2$ & $53.0$ & $50.9$ & $52.6$ & $50.5$ & $52.1$ & $48.9$ \\
$   1$ & $   2$ & $50.2$ & $50.4$ & $50.6$ & $50.9$ & $47.6$ & $47.7$ & $45.6$ \\
$   1$ & $   1$ & $50.1$ & $50.1$ & $49.8$ & $49.8$ & $49.1$ & $49.0$ & $47.5$ \\
$   1$ & $ 1/2$ & $48.5$ & $48.3$ & $48.3$ & $48.0$ & $49.1$ & $48.8$ & $48.3$ \\
$ 1/2$ & $   1$ & $48.4$ & $47.4$ & $48.8$ & $47.7$ & $47.6$ & $46.6$ & $46.3$ \\
$ 1/2$ & $ 1/2$ & $48.5$ & $48.0$ & $48.3$ & $47.8$ & $48.6$ & $48.1$ & $48.0$ \\
$ 1/2$ & $ 1/4$ & $47.0$ & $47.1$ & $47.1$ & $47.2$ & $47.7$ & $47.7$ & $47.9$ \\
$ 1/4$ & $ 1/2$ & $47.8$ & $47.4$ & $48.2$ & $47.7$ & $48.0$ & $47.6$ & $47.6$ \\
$ 1/4$ & $ 1/4$ & $47.7$ & $48.0$ & $47.6$ & $47.9$ & $48.0$ & $48.2$ & $48.2$ \\
$ 1/4$ & $ 1/8$ & $44.7$ & $45.1$ & $45.4$ & $45.7$ & $44.6$ & $45.0$ & $44.9$ \\
$ 1/8$ & $ 1/4$ & $45.5$ & $46.1$ & $46.1$ & $46.6$ & $46.2$ & $46.6$ & $46.5$ \\
$ 1/8$ & $ 1/8$ & $41.0$ & $40.9$ & $41.4$ & $41.2$ & $40.9$ & $40.8$ & $40.9$ \\
\bottomrule
  \end{tabular}
   \label{tab:res}
\end{table}

Several comments are in order:
\begin{itemize}
  
\item The uncertainty on the fixed order reduces to the canonical 7-point variation,
  while at resummed level we have the 42-point variation $\delta(\text{mho})$ detailed above.
  The fixed-order 7-point variation gives an uncertainty similar to $\delta(\text{scale})$ of Ref.~\cite{Anastasiou:2016cez},
  which is based on a 3-point variation of $\mu_0/2<\mu_R=\mu_F<2\mu_0$.

\item Even ignoring the LO (which contains too few information for being predictive),
  one can consider the convergence pattern of the fixed-order perturbative expansion when going from NLO to NNLO and to N$^3$LO,
  relative to the scale uncertainty.
  The pattern is worse at larger central scales and improves at smaller scales.
  For instance, at the largest central scale considered ($\mu_0=2m_H$), the central result at each order lies outside the band of the previous order;
  conversely, at the smallest central scale considered ($\mu_0=m_H/4$), NNLO is contained in the NLO band, and central N$^3$LO in the NNLO band
  (although the scale error at N$^3$LO is large and very asymmetric).
  A similar convergence pattern is observed at $\mu_0=m_H/2$; however, we note that the N$^3$LO band does not overlap with the NLO band.  
  Additionally, the N$^3$LO results at the four central scales shown in Table~\ref{tab:bonviniresults}
  are barely compatible.
  This analysis shows that an estimate of the uncertainty from missing higher orders  that solely relies on $\delta(\text{scale})$,
  i.e.\ a canonical 7-point scale variation, is not reliable at fixed order.
  
\item As far as the resummed results and their uncertainty $\delta(\text{mho})$ are concerned,
 one can note that, for each choice of the central scale $\mu_0$, the uncertainty of the resummed
  results from NLO+NLL onwards covers the central value and at least a portion of the band of the next (logarithmic) order.
  In fact, with the exception of the choice $\mu_0=m_H/4$ (the pathological behaviour of which seems to be driven by the N$^3$LO contribution),
  the NNLO+NNLL band is fully contained in the NLO+NLL band, and the N$^3$LO+N$^3$LL band is fully contained in the NNLO+NNLL band.
  This, together with the observation of the systematic reduction of the scale uncertainty when going from one logarithmic order to the next,
  shows that the proposed $\delta(\text{mho})$ provides one with a very reliable estimate of the uncertainty from missing higher orders.
  This is further supported by the observation that resummed results at each order are all compatible among the different
  choices of the central scale $\mu_0$.
\item
  Note that the different options for the position of the constants, while giving
  a large spread at NLO+NLL, is of little importance at higher orders, especially at N$^3$LO+N$^3$LL, suggesting a good convergence of the resummed series.
\item
  In many respects, the choice $\mu_0=m_H/2$ seems optimal,
  in full agreement with previous analyses, e.g.~\cite{Anastasiou:2016cez}.
  The convergence of the fixed-order is already good, and the convergence of the resummed result
  is very good. The error band at N$^3$LO+N$^3$LL is smaller than for other central scales,
  but compatible with the results computed at different values of $\mu_0$.
  Moreover, given that the way of estimating the uncertainty is very conservative, and successful at previous orders,
  the uncertainty on the N$^3$LO+N$^3$LL seems reasonably trustful.
\end{itemize}

The result advocated as best result in Ref.~\cite{Bonvini:2016frm} within the rEFT setup considered there is hence
\begin{equation}
\text{rEFT}:\; \sigma_{\text{N$^3$LO+N$^3$LL}} = \sigma_{\text{N$^3$LO}} + \Delta_3\sigma_{\text{N$^3$LL}} = 48.5 \pm1.9\, (4 \%)\, \text{pb},
\end{equation}
where, to be even more conservative, the error has been symmetrized.
Note that the effect of adding the resummation to the N$^3$LO on the central value
is rather small, $+0.4$~pb, corresponding to $+0.8\%$, which, however, is \emph{not} covered by the very asymmetric fixed-order scale uncertainty.
The authors of Ref.~\cite{Bonvini:2016frm} firmly believe that the uncertainty estimate derived from resummation
is much more reliable and trustful than that obtained by simple (asymmetric) scale variation at fixed order.\footnote
{If scale variation error at fixed-order is symmetrized the resulting uncertainty becomes more reasonable.
However, given that the small uncertainty comes from the vicinity of a stationary point in the scale dependence,
it might still underestimate the missing higher order uncertainty.}

In order to go beyond the rEFT approach, we have to discuss the role of quark mass effects on the resummed contribution.
The approach of Ref.~\cite{Schmidt:2015cea} consists of including finite $m_t$ effects to NLL, while treating $m_b$ and $m_c$ at fixed-order. 
Because of the accuracy of the rEFT approach for the top contribution, this leads to a resummed contribution very close to the rEFT one considered so far.
In Ref.~\cite{Bonvini:2014joa} a more aggressive approach was considered and bottom and charm were included to NLL and the top contribution to NNLL,
albeit in the usual $1/m_t$ expansion. However, the interplay between soft logarithms and logarithms of $m_b$ is still to be fully understood
(see for instance Ref.~\cite{Schmidt:2015cea}).
In any case, we believe that the uncertainty $\delta(t,b,c)$ of Ref.~\cite{Anastasiou:2016cez} likely covers the differences between the two approaches. 

In Ref.~\cite{Bonvini:2016frm}, results from resummation have been compared to different methods for estimating the uncertainty from missing higher orders.
First, the Cacciari-Houdeau Bayesian approach~\cite{Cacciari:2011ze} has been considered, which employs
the known perturbative orders to construct a probability distribution for the subsequent unknown order.
In its modified incarnation~\cite{Forte:2013mda,Bagnaschi:2014wea},
the method gives an uncertainty of $\pm2$~pb at $95\%$ degree of belief, fully compatible with the estimate obtained from resummation.
Second, following an idea by David and Passarino~\cite{David:2013gaa}, several algorithms to accelerate
the convergence of the perturbative series, based on non-linear sequence transformations, have been considered.
By performing a survey of different algorithms, it was possible to show that both the fixed-order and resummed series
exhibit good convergence properties at $m_H/2$ (and also at $m_H/4$).
Noticeably, the mean of each distribution is very close to the N$^3$LO+N$^3$LL prediction.

In conclusion, these tests  further support the claim that the N$^3$LO+N$^3$LL calculation,
together with its uncertainty $\delta(\text{mho})$, provides the most reliable estimate for the Higgs cross-section in gluon gluon fusion.
In terms of relative contributions, which are likely to remain unchanged when quark-mass and electroweak effects are included in the fixed order,
the results from this section can be summarized as
\begin{equation}
\boxed{\sigma_{\text{N$^3$LO+N$^3$LL}} - \sigma_{\text{N$^3$LO}} = +0.4\,\text{pb},
\qquad
\delta(\text{mho}) = \pm 4\%,}
\end{equation}
which are the recommended shift with respect to the N$^3$LO and recommended uncertainty from missing higher orders
by the Authors of Ref.~\cite{Bonvini:2016frm} for the SM Higgs boson at LHC 13~TeV.


\subsection[Summary for the total cross-section]{Summary for the total cross-section\SectionAuthor{S.~Forte, D.~Gillberg, C.~Hays, A.~Lazopoulos, G.~Petrucciani, A.~Massironi, G.~Zanderighi}}
\label{subsec:recommendation}
We now summarize the working group recommendation for the total cross-section 
and associated uncertainty for the LHC at~13~TeV.

The recommendation for the central value is to take the pure N$^3$LO
result, evaluated at $\mu_R=\mu_f=m_H/2$. This choice of scale is
motivated by the observation that the perturbative expansion is more
stable both at fixed order and at the resummed level. With this
choice of scale, the effect of the resummation is (at LHC energies)
much smaller than
the uncertainty related to the choice of resummation prescription, and
much smaller than the residual scale uncertainty. Furthermore, the
N$^3$LO EFT result should be rescaled by the $R_{LO}$
Eq.~(\ref{eq:KLO}); charm, bottom and top contributions should be
included exactly up to NLO, and finite top mass effects at NNLO using the
expansion in $1/m_t$ from
Ref.~\cite{Harlander:2009bw,Harlander:2009mq,Harlander:2009my}. Finally,
electroweak corrections~\cite{Actis:2008ug,Actis:2008ts} should be
included multiplicatively using 
complete factorization. The value of the $\overline{\textrm{MS}}$
heavy quark masses and of $\alpha_s$ given in Chapter~\ref{chapter:input} should be used. This leads to the result of Ref.~\cite{Anastasiou:2016cez}, 
given in Eq.~(\ref{eq:finalresult_13TeV}) above. Note that changing
from pole (previous recommendation~\cite{Dittmaier:2011ti}) 
to  $\overline{\textrm{MS}}$ masses leads to an increase of the
cross-section of order 2\%~\cite{Anastasiou:2016cez},

For the treatment of uncertainties, we distinguish parametric
uncertainties and theoretical uncertainties.

{\bf PDFs and $\alpha_s$} give the parametric uncertainty for this
process. For these, we recommend to follow the PDF4LHC15
  recommendation~\cite{Butterworth:2015oua} summarized in
  Chapters~\ref{chapter:input}-\ref{chap:PDFs}. This leads to an
  absolute 
  uncertainty $\Delta_{{\rm PDF}+\alpha_s}=1.56$~pb, i.e. a relative
  uncertainty of $3.2$~\%. This is in agreement with
  Ref.~\cite{Anastasiou:2016cez}, see also Eq.~(\ref{eq:finalresult_13TeV})
  of Section~\ref{subsec:ADDFGHLM} above.
This is currently the dominant source of uncertainty. It should be
considered Gaussianly distributed, and the above interval provides the
standard deviation of the Gaussian, corresponding to a symmetric 68\% confidence
level. 

The estimation of theoretical uncertainties inevitably involves a
somewhat subjective component. For these we provide two different
procedures, based on two possible different interpretations of
theoretical uncertainties.\begin{itemize}
\item In {\bf Procedure F} the final theoretical uncertainty is
  interpreted as a flat 100\% confidence level. This means that if the
  missing information which provides all the given sources of
  theoretical uncertainty was supplied, then the result is expected to
  lie with certainty within the given interval, but not more likely to
  be in any region within this interval. The interval is constructed
  as the linear sum of individual sources of theoretical uncertainty,
  each estimated as explained below.  Each individual source of
  uncertainty should not be endowed with a statistical interpretation
  and merely concurs to the  determination of  the final overall range. 
\item In {\bf Procedure G} each source of theoretical uncertainty is
  interpreted as a one-sigma range. The final uncertainty is thus
  obtained by combining in quadrature individual sources of
  uncertainty. Because there are many sources of uncertainty, the
  final distribution is expected to be approximately Gaussian, and the
  final combined uncertainty should thus be interpreted as a symmetric
  68\% confidence level.
\end{itemize}

We now provide a list of sources of uncertainty. For each source, we
briefly describe and provide the uncertainty estimate to be used in
either procedure. It is important to stress that individual
uncertainties have a different interpretation in the two procedures:
specifically F-uncertainties are merely components of the final
uncertainty range, while G-uncertainties are one-sigma ranges. Even
when taken to be numerically equal in the two procedures they thus
have a different meaning. 

The various sources of uncertainty and the corresponding
F-uncertainties and G-uncertainties are:

\begin{itemize}
\item{\bf missing higher QCD orders}
In order to estimate the uncertainty related to missing higher order
QCD corrections beyond N$^3$LO various options have been considered:
\begin{itemize}

\item In  Ref.~\cite{Anastasiou:2016cez} it is suggested to  perform a
   scale variation scan of the N$^3$LO result including mass
   effects, in the range  $m_H/4\le \mu_R=\mu_F\le m_H$, (see
   Section~\ref{subsec:ADDFGHLM} above).\footnote{The inclusion of
   mass effects reduces the whole cross section and thus also the
   uncertainty by a factor of about 0.7 in comparison to what one
   would obtain performing a scale variation scan with mass effects
   not included.} 
This procedure gives
   $\Delta_{\rm MHOU}=[-1.2;+0.1]$~pb, i.e. a relative
   uncertainty of $[-2.4;+0.2]$~\%. Note that a three-point scale
   variation, rather than a scan, would give a vanishing upper uncertainty.
\item A 7-point scale variation can be performed  with $m_H/4\le
   \mu_R, \mu_F\le m_H$ keeping $\frac12 \le \frac{\mu_R}{\mu_F} \le
   2$. This is a standard procedure  
 used to estimate uncertainties
   in fixed-order calculations. It turns out to give a result which is
   similar to the previous one, namely $\Delta_{\rm
       MHOU}=[-1.4;+0.1]$~pb, i.e. a relative uncertainty of
     $[-3.0;+0.2]$~\%. Note that the lower variation is somewhat
   larger than that found in the scan, as the latter was performed
   with fixed ratio $\mu_F/\mu_R=1$.
\item Since scale variation probes only the size of higher-order
    terms, but not their sign, and the scale-variation
    uncertainties quoted above are very asymmetric, one can argue that
    they should be symmetrized while keeping the central
    value fixed.~\footnote{Yet another alternative would be to keep
      the scale-uncertainty band and quote as a central value the
      midpoint.} In the case of the 7-point scale variation this 
    gives $\Delta_{\rm MHOU}=\pm1.4$~pb,
      i.e. a relative uncertainty of $[-3.0;+3.0]$~\%. 
\item While resummation has a minimal impact at central scale, it
  provides a more stable perturbative expansion at all previous
  orders. In Ref.~\cite{Anastasiou:2016cez} a variety of resummation
  schemes were examined and it was found that, within  an equal scale
  variation scan, resummation contributions lie within the fixed order
  scale uncertainty interval in all cases considered. However,  one may argue 
 that seven-point scale variation of the
  resummed result provides a more reliable estimate
  of the perturbative uncertainty. Taking the default N$^3$LL+N$^3$LO
  resummation scheme of Ref.~\cite{Bonvini:2016frm} (see also
  Section~\ref{subsec:bonvini} above) one gets $\Delta_{\rm MHOU}=[-1.6;+1.5]$~pb,
      i.e. a relative uncertainty of $[-3.2;+3.2]$~\%. Note that the
  scale variations reported in \refT{tb:cxn} are larger, but the
  results in that table are at NNLO+N$^3$LL, rather than
  N$^3$LO+N$^3$LL. 
This is very
      close to the symmetrized seven-point scale variation of the
      fixed order result.\footnote{In Ref.~\cite{Bonvini:2016frm} it
      is instead recommended to take the envelope of the seven-point
      scale variations for a variety of different resummation
      prescriptions; this leads to a marginally more conservative
      uncertainty of about $\pm4\%$.}
\end{itemize}
 
The {\bf F-uncertainty} is estimated by performing a
   scale variation scan in the range  $m_H/4\le \mu_R=\mu_F\le m_H$
   following
 the suggestion of
  Ref.~\cite{Anastasiou:2016cez},
   which gives    $\Delta_{\rm MHOU}=[-1.2;+0.1]$~pb, corresponding to a relative
   uncertainty of $[-2.4;+0.2]$~\%.\\
\noindent The {\bf G-uncertainty} is estimated by performing
a  symmetrized
  seven-point scale uncertainty, which leads to a conservative result
  that agrees with the uncertainty based on resummation arguments,
  i.e.
$\Delta_{\rm MHOU}=\pm1.4$~pb, corresponding to
       a relative uncertainty of $[-3.0;+3.0]$\%. 

\item{\bf missing electroweak corrections}
In our recommendation, electroweak (EW) corrections are included assuming
complete factorization. This gives rise to a positive $5\%$ correction
to the pure QCD result. If instead QCD and EW corrections are
combined additively, one gets an enhancement of $1.7\%$. Finally,
mixed QCD-EW corrections computed using an effective field theory
(EFT) with
$m_W,\,m_Z\gg m_H$~\cite{Anastasiou:2008tj} give an enhancement of
$3.2\%$.  

The {\bf F-uncertainty} is estimated as $\Delta_{\rm ew}=0.5$~pb
    corresponding to  $\pm 1\%$ uncertainty, 
which is deemed to be conservative enough and it corresponds to an
intermediate value between various possible estimates, as seen in
Sect.~\ref{sec:breakdown} and 
  Ref.~\cite{Anastasiou:2016cez}.\\
\noindent
The {\bf G-uncertainty} is conservatively
estimated as  the average of
  the difference between our chosen  complete factorization ($5\%$)
  and either of the alternative possibilities (additive:  $1.7\%$, or
  EFT: $3.2\%$). This corresponds to an uncertainty of $\pm 2.5\%$ or
  $\pm1.2$~pb.
\item{\bf bottom and charm interference with top}
Bottom and charm interference with top quark loops are known exactly
only up to NLO. At LO and NLO, bottom-top interference leads to a
correction which is about the same as the finite top mass correction,
but with the opposite sign. At NNLO, including rEFT and $1/m_t$
effects leads to a correction of about 1~pb. Furthermore, the NLO
top-bottom and charm interference correction changes by 0.7~pb if
$\overline{\rm MS}$ or pole heavy quark masses are used. 

The {\bf F-uncertainty} is estimated as $\Delta_{\rm t,bc}=\pm0.4$~pb,
      i.e. a relative uncertainty of $\pm0.8$\%, following
Sect.~\ref{sec:breakdown} and Ref.~\cite{Anastasiou:2016cez}.\\
\noindent
The {\bf G-uncertainty} is estimated taking the scheme dependence of the NLO
  interference as a reasonably conservative estimate. This leads to
$\Delta_{\rm t,bc}=\pm0.7$~pb,
      i.e. a relative uncertainty of $\pm1.5$\%.

\item{\bf finite top mass} 
Both the {\bf F-uncertainty} and 
the {\bf G-uncertainty} are estimated  
as $\Delta_{\rm 1/m_t}=\pm0.49$~pb,
      i.e. a relative uncertainty of $\pm1$\%,   following
      Sect.~\ref{sec:breakdown} and
      Ref.~\cite{Anastasiou:2016cez}. 
\item{\bf missing N$^3$LO PDFs} 
Both the {\bf F-uncertainty} and 
the {\bf G-uncertainty} are estimated  
assigning to lack of knowledge of
  the N$^3$LO PDFs
  an uncertainty of $\Delta_{\rm PDF-TH}=\pm0.56$~pb,
      i.e. a relative uncertainty of $\pm1.2$\%,  following  Sect.~\ref{sec:breakdown} and
      Ref.~\cite{Anastasiou:2016cez}. 
\item{\bf truncation of the soft expansion} Both the {\bf F-uncertainty} and 
the {\bf G-uncertainty} are estimated  assigning to the truncation of the
  soft expansion used to derive the N$^3$LO QCD corrections 
$\Delta_{\rm soft}=\pm0.18$~pb,
      i.e. a relative uncertainty of $\pm0.4$\%, following  Sect.~\ref{sec:breakdown} and
      Ref.~\cite{Anastasiou:2016cez}. 
\end{itemize}

The total uncertainty is thus:
\begin{itemize}
\item {\bf F-uncertainty:} $[-6.7,+4.6]\%$ corresponding to $[-3.3,+2.2]$~pb.
\item {\bf G-uncertainty:}  $\pm4.5$\% corresponding to  $\pm2.2$~pb
\end{itemize}
We recall that the F-uncertainty is a 100\% flat confidence interval,
while the G-uncertainty is a one-sigma Gaussian uncertainty. They can
be compared by noting that a symmetric flat interval has a variance
equal to its half-width divided by $\sqrt3$. The symmetrized
$F$-uncertainty hence corresponds to a variance of $6.7/\sqrt{3}=3.9$,
which is not far from the G-uncertainty. The two estimates are thus
roughly compatible.

The final recommendation for gluon fusion cross-section at the LHC is presented in Chapter~\ref{chap:WG1Summary}.


\section{Differential and jet-binned cross sections}
\providecommand{\ptjv}{p_{\rm t,veto}}

In some decay channels
it is common to perform different analyses depending on the number of jets accompanying
the Higgs boson.
This is because the Higgs boson signal in different jet multiplicities is affected
by different backgrounds.
Most notably, when the Higgs boson decays to $WW$, the dominant top background is significantly suppressed by requiring zero jets in the final state. 
Jet veto transverse momentum thresholds used by ATLAS and CMS are of order $25-30$ GeV, and thus substantially smaller than the Higgs boson mass $m_H$. In this case real radiation is suppressed and the imbalance between virtual and real corrections produces logarithms of the form $\ln(\ptjv/m_H)$ which may invalidate the fixed order perturbative expansion. Resummed predictions for the cross section in the 0-jet bin
have been obtained both in full QCD \cite{Banfi:2012jm} and in the framework of
Soft Collinear Effective field Theory (SCET) \cite{Becher:2012qa,Stewart:2013faa}.
Several methods for determining the uncertainties and their correlations across jet-bins were proposed in the past, and used in Run~1 measurements. At the end of Run~1 further improvements were proposed. 
In this section we review different approaches to the treatment of correlated uncertainties in jet bins.

In Section~\ref{subsec:binuncertainties} a general approach to theory uncertainties in kinematic bins is presented. It reduces to the commonly used ST \cite{Stewart:2011cf} or JVE \cite{Banfi:2012jm} methods in particular cases
and is applicable also to treat simplified template cross sections.
In Section~\ref{subsec:run1_method} the ST and JVE methods are compared up to NNLO and results for 13 TeV are presented.
In Section~\ref{subsec:stwz_blptw} an updated calculation of the cross section in the 0-jet, 1-jet and $\geq 2$-jet bins is presented. This calculation, based on the work of Ref.~\cite{Stewart:2013faa}, includes the direct resummation of the logarithmically enhanced terms in the 0-jet bin, and the indirect resummation of the corresponding terms in the 1-jet bin. Heavy quark mass effects are accounted for through an overall rescaling factor.

Recently, the N$^3$LO result for the inclusive cross section (see Section~\ref{subsec:ADDFGHLM}), and the H+jet cross section at NNLO (see Section~\ref{sec:antenna}),  became available.
Both these calculation refer to perturbative corrections at ${\cal O}(\as^5)$ and can be used to improve
the computation of the 0-jet cross section.
In Section~\ref{sec:jve} an updated calculation of the jet vetoed cross section is presented, which consistently includes the above information. The calculation includes the resummation of the $\ln(\ptjv/m_H)$ terms at NNLL and a resummation of the logarithmically enhanced contribution of the jet radius $R$. Heavy-quark mass effects are included according to the approach of Ref.\cite{Banfi:2013eda}.

When the recoiling QCD radiation is integrated over, the NNLO calculation of H+jet provides an NNLO prediction for the $p_T$ spectrum. In Section~\ref{subsec:MRT} an NNLO calculation of the $p_T$ spectrum is combined with an NNLL resummed computation in momentum space \cite{Monni:2016ktx} and NNLL+NNLO predictions for the cumulative distribution are presented, together with a comparison of NNLL+NLO results with available reference predictions for this observable.

\subsection[General treatment of theory uncertainties in kinematic bins]{General treatment of theory uncertainties in kinematic bins\SectionAuthor{F.J.~Tackmann, K.~Tackmann}}
\label{subsec:binuncertainties}
Whenever the experimental measurements are separately performed in different kinematic regions (or bins),
the theoretical predictions and their uncertainties must also be evaluated separately for each
kinematic region. This is necessary also when the information from all measured bins is eventually combined in the interpretation,
since different bins can in general have different sensitivities and therefore contribute with different relative weights
to the final result. In this context, the correlations of the theoretical uncertainties for
different bins must be taken into account. This is particularly important whenever a binning cut induces an important
additional source of perturbative uncertainties that affects each bin but should cancel in their sum. This is precisely what happens in the context of jet binning, and it requires one to treat the uncertainties induced by the binning as anti-correlated between the bins~\cite{Berger:2010xi, Stewart:2011cf}. In general, to properly treat the theoretical uncertainties one should thus try to identify and distinguish different sources of uncertainties and take into account the correlation implied by each source.

\subsubsection{Single bin boundary}

We first review the case where the cross section is split into two bins by a single perturbatively nontrivial binning cut. To be concrete, we use the $0$-jet cross section as an important example.

In this case, the total inclusive cross section, $\sigma_\tot \equiv \sigma_{\geq 0}$, is divided into an exclusive $0$-jet bin, $\sigma_0(\pTcut)$, where the $p_T$ of the leading jet is restricted to $p_T^{\rm jet} < \pTcut$ and the remaining inclusive $1$-jet bin, where the leading jet is required to have $p_T^{\rm jet} \geq \pTcut$,
\begin{equation}
\sigma_{\geq 0} = \sigma_0(\pTcut) + \sigma_{\geq 1}(\pTcut)
\,.\end{equation}
Typically, the experimentally used $\pTcut$ values are smaller than the hard-interaction scale $\sim m_H$. In this case, the $\pTcut$ restriction induces Sudakov double logarithms~\cite{Tackmann:2012bt} $\ln(p_T^{\rm cut}/m_H)$ at each order in $\alpha_s$, which grow as $\pTcut$ is lowered. When $\pTcut$ is small enough for $\sigma_{\geq1}(\pTcut)$ to contain a nonnegligible fraction of the total cross section, this implies that the cut-induced perturbative corrections have a nontrivial influence on the perturbative series of $\sigma_0$, corresponding to an additional and a priori nonnegligible source of uncertainty that is not present in $\sigma_{\geq 0}$. (Note that this is irrespective of whether the cut effects are computable in fixed-order or resummed perturbation theory.)

The uncertainties involved in the jet binning can be described in general in terms of fully correlated and fully anticorrelated components, which amounts to parameterizing the uncertainty matrix for $\{\sigma_0, \sigma_{\geq 1}\}$ as
\begin{equation} \label{eq:Cgeneral}
C(\{\sigma_0, \sigma_{\geq 1}\}) =\!
\begin{pmatrix}
(\Delta^{\rm y}_0)^2 &  \Delta^{\rm y}_0\,\Delta^{\rm y}_{\geq 1}  \\
\Delta^{\rm y}_0\,\Delta^{\rm y}_{\geq 1} & (\Delta^{\rm y}_{\geq 1})^2
\end{pmatrix}
\!+
\begin{pmatrix}
 \Delta_\cut^2 &  - \Delta_\cut^2 \\
-\Delta_\cut^2 & \Delta_\cut^2
\end{pmatrix}
\!.\end{equation}
While this form was originally utilized in the context of the ST method~\cite{Stewart:2011cf, Gangal:2013nxa}, it
is simply a general parameterization of a $2\times2$ symmetric matrix, and not specific to a particular calculation or
framework for determining theory uncertainties. That is, the uncertainties obtained with any prescription can always be written in this form, provided sufficient information or assumptions on the correlations are available.

The parameterization in \eqref{eq:Cgeneral} proves convenient for two reasons:
First, the separation into independent components that are 100\% correlated or anticorrelated between the different observables allows for a straightforward implementation in terms of independent nuisance parameters for each component. That is, one has two nuisance parameters $\kappa^{\rm y}$ and $\kappa_\cut$ whose uncertainty amplitudes for $\{\sigma_{\geq 0}, \sigma_0, \sigma_{\geq 1}\}$ are
\begin{equation}
\kappa^{\rm y}: \quad \{ \Delta^{\rm y}_{\geq 0},\, \Delta^{\rm y}_{0},\, \Delta^{\rm y}_{\geq 1} \}
\qquad\qquad
\kappa_\cut: \quad \{ 0,\, \Delta_\cut, -\Delta_\cut \}
\,,\end{equation}
where $\Delta_{\geq 0}^{\rm y} = \Delta^{\rm y}_{0} + \Delta^{\rm y}_{\geq 1}$.
Hence, this provides a baseline for the experimental implementation, independent of a particular theoretical prediction.
Second, this parameterization admits a simple physical interpretation, which is very useful to identify and estimate each component for a given theory calculation: The first correlated component, denoted with a superscript ``y'', can be interpreted as an overall yield uncertainty of a common source for all bins. The second anticorrelated component can be interpreted as a migration uncertainty between the two bins, which is introduced by the binning and drops out in their sum.

The existing prescriptions for estimating perturbative uncertainties in jet binning and their justifications have been documented extensively before~\cite{Stewart:2011cf, Banfi:2012yh, Gangal:2013nxa, Dittmaier:2012vm, Heinemeyer:2013tqa}. Here, we only give a brief summary in order to illustrate the above for the case of the $0/1$-jet boundary. In fixed-order predictions, there is no way to unambiguously identify different sources for $\Delta^{\rm y}$ and $\Delta_{\rm cut}$, so one has to make some assumptions. Using a naive correlated scale variation for all jet bins amounts to setting $\Delta_i^{\rm y} \equiv \Delta_i^{\rm FO}$, where $\Delta_{i}^{\rm FO}$ are the default perturbative uncertainties estimated by the usual scale variations in the fixed-order predictions, while $\Delta_\cut \equiv 0$ is neglected. As mentioned above, once the binning effects become important, the associated migration uncertainty should not be neglected, otherwise this can easily lead to an underestimate. In the ST method~\cite{Stewart:2011cf}, this is avoided by taking
\begin{equation}  \label{eq:ST}
\text{ST}:\qquad
\Delta_0^{\rm y} = \Delta_{\geq 0}^{\rm y} = \Delta_{\geq 0}^{\rm FO}
\,,\quad
\Delta_{\geq 1}^{\rm y} = 0
\,,\qquad\qquad
\Delta_\cut = \Delta_{\geq 1}^{\rm FO}
\,.\end{equation}
That is, the migration uncertainty is approximated by the perturbative uncertainty of $\sigma_{\geq 1}(\pTcut)$, which is motivated by the structure of the perturbative series at small $\pTcut$. Maintaining as the total uncertainty for $\sigma_{\geq 1}$ its usual fixed-order uncertainty then requires setting $\Delta_{\geq 1}^{\rm y} = 0$. As a result, one effectively treats the usual fixed-order perturbative uncertainties in $\sigma_{\geq 0}$ and $\sigma_{\geq 1}$ as independent sources. This can also be generalized~\cite{Gangal:2013nxa}, by using an additional parameter $\rho$ to separate $\Delta_{\geq 1}^{\rm FO}$ into yield and migration parts, which then effectively determines the correlation between $\Delta_{\geq 0}^{\rm FO}$ and $\Delta_{\geq 1}^{\rm FO}$,
\begin{equation}  \label{eq:STrho}
\text{ST($\rho$)}:\qquad
\Delta_0^{\rm y} = \Delta_{\geq 0}^{\rm FO}
\,,\quad
\Delta_{\geq 1}^{\rm y} = \rho\, \Delta_{\geq 1}^{\rm FO}
\,,\qquad\qquad
\Delta_\cut = \sqrt{1 - \rho^2}\, \Delta_{\geq 1}^{\rm FO}
\,.\end{equation}

Another prescription is the JVE method~\cite{Banfi:2012yh}, which typically yields similar uncertainties for the same perturbative inputs. It relies on the assumption that the perturbative uncertainty in the $0$-jet fraction $\epsilon_0 = \sigma_0(\pTcut)/\sigma_{\geq 0}$ is uncorrelated with that of the inclusive cross section $\sigma_{\geq 0}$, i.e., the perturbative uncertainties in $\epsilon_0$ and $\sigma_{\geq 0}$ are treated as the independent sources. This amounts to taking
\begin{equation}  \label{eq:JVE}
\text{JVE}:\qquad
\Delta_{\geq 0}^{\rm y} = \Delta_{\geq 0}^{\rm FO}
\,,\quad
\Delta_0^{\rm y} = \epsilon_0\, \Delta_{\geq 0}^{\rm FO}
\,,\quad
\Delta_{\geq 1}^{\rm y} = (1 - \epsilon_0)\, \Delta_{\geq 0}^{\rm FO}
\,,\qquad
\Delta_\cut = \sigma_{\geq 0}\, \Delta(\epsilon_0)
\,.\end{equation}
This means that the relative yield uncertainties for all bins are equal to the relative uncertainty of the total cross section, $\Delta_i^{\rm y}/\sigma_i = \Delta_{\geq 0}^{\rm FO}/\sigma_{\geq 0}$.
In the migration uncertainty, $\Delta(\epsilon_0)$ is the perturbative uncertainty of $\epsilon_0$, which is obtained by considering its scale variation together with several variations of how to write its perturbative series that differ by higher-order terms. This also means that the total uncertainty for $\sigma_{\geq 1}$ is not just given by the usual $\Delta_{\geq 1}^{\rm FO}$. When used together with a resummed calculation, the fixed-order estimate of $\Delta(\epsilon_0)$ is replaced by its resummed counterpart, see Section~\ref{subsec:stwz_blptw}.

\subsubsection{Multiple bin boundaries}

We now move on to discuss the more general case where the total cross
section is split into multiple mutually exclusive bins, as is the case
in most experimental analyses. Specifically, this applies to the case
of
the simplified template cross section framework discussed in
Section~\ref{STCS}. A full implementation of the
theory-independent parameterization discussed here in the simplified
template cross sections would allow utilizing theoretical predictions
in a very flexible manner. In particular, it would enable easily
switching between different theory predictions in the interpretation
of the experimental measurements. 

With multiple bins, each bin can have more than one boundary and vice versa any given boundary can be shared by different bins.
To make this tractable in a systematic fashion, we first note that Eq.~\eqref{eq:Cgeneral} applies in an obvious way to any single bin boundary and binning cut when all additional subdivisions are removed. That is, a given binning cut, labelled ``$a/b$'', separates the cross section as $\sigma_{ab} = \sigma_a + \sigma_b$ with an associated migration uncertainty $\Delta_\cut^{a/b}$, which is anticorrelated between $\sigma_a$ and $\sigma_b$. (Note that $\sigma_{ab}$ is not necessarily the total cross section but can itself correspond to a bin; what is relevant is that the $a/b$ cut does not act outside of $\sigma_{ab}$.)
Including additional cuts that further subdivide $\sigma_a$ and $\sigma_b$, we have $\sigma_a = \sum_i \sigma_a^i$ and $\sigma_b = \sum_j \sigma_b^j$, where we have labelled the individual sub-bins according to whether they are part of $\sigma_a$ or $\sigma_b$. Since we interpret the $a/b$ boundary as a common uncertainty source, we can consider it as fully correlated among each set of sub-bins and implement it via a single nuisance parameter $\kappa_\cut^{a/b}$. The corresponding uncertainty amplitudes for all the bins are given as
\begin{equation}
\kappa_\cut^{a/b}: \quad \Delta_\cut^{a/b} \times \bigl\{\{ x_a^i \}, - \{ x_b^j \} \bigr\}
\qquad\text{with}\qquad
\sum_i x_a^i = \sum_j x_b^j = 1
\,,\end{equation}
where the parameters $x_a^i$ and $x_b^j$ specify how $\Delta_\cut^{a/b}$ gets distributed among the sub-bins. (With this information it is also straightforward to construct a corresponding uncertainty matrix, but this is not actually necessary.)

In this way, we can consider each binning cut (or bin boundary) as a potential source of an uncertainty with an associated nuisance parameter.
(Of course, in practice with sufficiently complicated bin boundaries, one has to apply some theoretical judgement in how one chooses the relevant independent binning cuts.) In addition, we can have an overall yield uncertainty correlated among all bins.

To illustrate this for a simple example, consider the case of 3 mutually exclusive jet bins $\{\sigma_0, \sigma_1, \sigma_{\geq 2}\}$. In this case, we can easily identify two bin boundaries as the two relevant sources of migration uncertainties, namely the cut on the leading jet separating $\sigma_{\geq 0} = \sigma_0 + \sigma_{\geq 1}$ and the cut on the 2nd jet separating $\sigma_{\geq 1} = \sigma_1 + \sigma_{\geq 2}$. In addition, we have an overall yield uncertainty. Considering the five (interdependent) observables $\{\sigma_{\geq 0}, \sigma_0, \sigma_{\geq 1}, \sigma_1, \sigma_{\geq 2} \}$, the three nuisance parameters and their respective uncertainty amplitudes are
\begin{align} \label{eq:3binnuisances}
\kappa^{\rm y}: &\quad \{ \Delta^{\rm y}_{\geq 0},\, \Delta^{\rm y}_{0},\, \Delta^{\rm y}_{\geq 1},\, \Delta^{\rm y}_{1},\, \Delta^{\rm y}_{\geq 2} \}
\quad\text{with}\quad
\Delta^{\rm y}_{\geq 0} = \Delta^{\rm y}_{0} + \Delta^{\rm y}_{\geq 1}
\,,\qquad
\Delta^{\rm y}_{\geq 1} = \Delta^{\rm y}_{1} + \Delta^{\rm y}_{\geq 2}
\,,\nn\\
\kappa_\cut^{0/1}: &\quad \Delta^{0/1}_\cut \times \{ 0,\, 1,\, -1,\, -(1 - x_1),\, -x_1 \}
\,,\nn\\
\kappa_\cut^{1/2}: &\quad \Delta^{1/2}_\cut \times \{ 0,\, x_2,\, -x_2,\, 1 - x_2,\, -1 \}
\,,\end{align}
Here, $x_1$ determines how $\Delta^{0/1}_\cut$ is split between $\sigma_1$ and $\sigma_{\geq 2}$, and $x_2$ determines how $\Delta_\cut^{1/2}$ is split between $\sigma_0$ and $\sigma_1$.
For $x_1\neq 0$, the $0/1$ migration into $\sigma_{\geq 1}$ can affect both $\sigma_1$ and $\sigma_{\geq 2}$. For
$x_1 = 0$, the $0/1$ migration happens entirely between $\sigma_0$ and $\sigma_1$, so the binning is effectively treated as $\sigma_{0+1} = \sigma_0 + \sigma_1$. Similarly, for $x_2 = 0$ the $1/2$ migration is contained within $\sigma_{\geq 1} = \sigma_1 + \sigma_{\geq 2}$. On the other hand, allowing for $x_2 \neq 0$ corresponds to the more general case $\sigma_{\geq 0} = \sigma_{<2} + \sigma_{\geq 2}$.

In terms of the uncertainty matrices for the 3 quantities $\{\sigma_0, \sigma_1, \sigma_{\ge 2}\}$, Eq.~\eqref{eq:3binnuisances} corresponds to
\begin{align}
C(\{\sigma_0, \sigma_1, \sigma_{\ge 2}\})
&= C^{\rm y}(\{\sigma_0, \sigma_1, \sigma_{\ge 2}\}) + C_\cut^{0/1}(\{\sigma_0, \sigma_1, \sigma_{\ge 2}\}) + C_\cut^{1/2}(\{\sigma_0, \sigma_1, \sigma_{\ge 2}\})
\,, \\
C^{\rm y}(\{\sigma_0, \sigma_1, \sigma_{\ge 2}\})
&= \begin{pmatrix}
(\Delta_0^{\rm y})^2 \;&\; \Delta_0^{\rm y} \Delta_1^{\rm y} \;&\; \Delta_0^{\rm y} \Delta_{\ge2}^{\rm y} \\
\Delta_0^{\rm y} \Delta_1^{\rm y} \;&\; (\Delta_1^{\rm y})^2 \;&\; \Delta_1^{\rm y} \Delta_{\ge2}^{\rm y} \\
\Delta_0^{\rm y} \Delta_{\ge2}^{\rm y} \;&\; \Delta_1^{\rm y} \Delta_{\ge2}^{\rm y} \;&\; (\Delta_{\ge2}^{\rm y})^2
\end{pmatrix}
\,, \\
C_\cut^{0/1}(\{\sigma_0, \sigma_1, \sigma_{\ge 2}\})
&= \bigl(\Delta_\cut^{0/1} \bigr)^2\begin{pmatrix}
1 \;&\; -(1-x_1) \;&\; -x_1 \\
-(1 - x_1) \;&\; (1 - x_1)^2 \;&\; x_1(1 - x_1) \\
-x_1  \;&\; x_1(1 - x_1) \;&\; x_1^2
\end{pmatrix}
\,, \\
C_\cut^{1/2}(\{\sigma_0, \sigma_1, \sigma_{\ge 2}\})
&= \bigl(\Delta_\cut^{1/2} \bigr)^2\begin{pmatrix}
x_2^2 \;&\; x_2(1 - x_2) \;&\; -x_2 \\
x_2(1 - x_2) \;&\; (1 - x_2)^2\;&\; -(1 - x_2) \\
-x_2 \;&\; -(1 - x_2) \;&\; 1
\end{pmatrix}
\,.\end{align}
Note that for the most generic case with 3 bins, one could in principle add an analogous third migration component $0/2$ for $\sigma_{\neq 1} = \sigma_0 + \sigma_{\geq 2}$. In this jet-binning example, this is clearly artificial, since $\sigma_0$ and $\sigma_{\geq 2}$ do not actually share a boundary and the migration between them is only indirect and already captured by the other two migration components.

The ST method applied to the case of 3 jet bins is equivalent to using
$\Delta_0^{\rm y} = \Delta_{\geq 0}^{\rm FO}$,  $\Delta_{1}^{\rm y} = \Delta_{\geq 2}^{\rm y} = 0$, 
$\Delta_\cut^{0/1} = \Delta_{\geq 1}^{\rm FO}$, $\Delta_\cut^{1/2} = \Delta_{\geq 2}^{\rm FO}$, and taking $x_1 = x_2 = 0$. This effectively considers $\sigma_{\geq 0}$, $\sigma_{\geq 1}$, and $\sigma_{\geq 2}$ as the independent sources, and in particular one can directly identify the corresponding nuisance parameters $\kappa_{\geq i}$ in existing implementations of the ST method as $\kappa^{\rm y} \equiv \kappa_{\geq 0}$, $\kappa_\cut^{0/1} \equiv \kappa_{\geq 1}$, and $\kappa_\cut^{1/2} \equiv \kappa_{\geq 2}$.
Section~\ref{subsec:stwz_blptw} discusses how estimates for the above parameters can be obtained when both jet boundaries are treated in resummed perturbation theory.

Mathematically speaking, the above generic parameterization has some redundancy, as it has more than the minimal number of six parameters that would be required to describe a general 3x3 symmetric matrix.
This is desired and makes it flexible enough to accommodate different scenarios while maintaining the simple physical interpretation in terms of the underlying sources. In contrast, using a mathematically minimal parameterization one would inevitably be forced to reexpress the contributions from several physically independent sources in terms of a minimal number of mathematically independent components and thus lose their physical meaning. In general, there can be several (independent) sources of theory uncertainties that give rise to each type of component. As for any other systematic uncertainty, these should then be implemented via separate nuisance parameters, which preserves their physical origin and in particular allows for the possibility to correlate them with other observables if necessary. For example, when electroweak and QCD corrections are treated as factorized, one could consider separately estimating the perturbative uncertainties of each, with each having their own set of nuisance parameters. (In practice, when the electroweak corrections are applied as an overall correction factor, they give rise to a yield uncertainty and a single nuisance parameter.)

\subsection[Exclusive fixed-order cross sections and jet-veto efficiencies at NNLO]{Exclusive fixed-order cross sections and jet-veto efficiencies at NNLO\SectionAuthor{B.~Di~Micco}}
\label{subsec:run1_method}

During Run-1 of LHC data taking at 7 and 8 TeV the two methods reviewed
in Sect.~\ref{subsec:binuncertainties}  were used to compute uncertainties on the jet bin 
acceptance for the Higgs boson signal: the so called Stewart-Tackmann method (ST) \cite{Stewart:2011cf} and the Jet Veto Efficiency method  (JVE) \cite{Banfi:2012jm}.
The first method has been used from the Higgs boson discovery paper
\cite{Aad:2012tfa}, \cite{Chatrchyan:2012xdj} while the second method
has been used in the  Run-1 ATLAS paper on the $h \to WW^*$ channel
\cite{ATLAS:2014aga}.
 In the present section we will present results obtained using both
 methods at 13~TeV, providing all details
 on the inputs used for the computation of the uncertainties.

\subsubsection{The Stewart-Tackmann method}


The ST method can be applied to all jet bins and assumes that inclusive jet bin cross sections, that is the cross section to have at least N jets,
($\sigma_{\ge N}$) are uncorrelated for all values of $N$. In this case the $N-$jet cross section can be written as:
\[
\sigma_N = \sigma_{\ge N} - \sigma_{\ge N+1}
\]
and the uncertainty on $\sigma_N$ is computed using the relation:
\[
\Delta^2 \sigma_{N} = \Delta^2 \sigma_{\ge N} + \Delta^2 \sigma_{\ge N+1}
\]
 The values of $\Delta^2\sigma_{\ge N, \ge N+1}$ are evaluated as the envelope
 of all cross sections obtained by changing the renormalization
 ($\mu_r$) and factorization ($\mu_f$) scales
by  a factor two around the central scale of $\mu = m_H/2$, excluding the values $\mu_f/\mu_r = 4$ and  $\mu_f/\mu_r = 1/4$. The values of $\sigma_{\ge 0}$
are computed at NNLO using the HNNLO program~\cite{Catani:2007vq}, and
 the process $pp \to h \to WW^{*} \to  e^+ \nu e^- \bar{\nu}$ is used
 for the computation of the cross section. The branching fraction to $W$ pairs used in HNNLO is Br$(h \to W^+W^-)
=0.2054$, and it is used, together with the PDG~\cite{Agashe:2014kda} value of Br$(W\to
e\nu) = 0.1070$, to correct the HNNLO cross section by the total branching
fraction, in order to extract the $h$ production cross section. The
value used is:
\[
{\rm Br}(h \to W^+ W^-){\rm Br}^2(W\to e \nu) = 0.002356
\]
The  obtained values are tabulated in Table~\ref{tab:inclusive}.

The $H+1$ jet and $H+2$ jet cross sections have been computed
at LO and NLO for the  $pp \to h \to WW^{*} \to  e^+ \nu e^-
\bar{\nu}$ process using the MCFM program~\cite{Campbell:2006xx},
the branching fraction to $W$ pairs used in MCFM is Br$(h \to W^+W^-)
=0.214$, and it is used to correct the MCFM cross section by the total branching
fraction:
\[
{\rm Br}(h \to W^+ W^-){\rm Br}^2(W\to e \nu) = 0.00245
\]
In Table \ref{tab:Hplus1jet} and Table
\ref{tab:Hplus2jet} we show the Higgs boson production cross section after
having corrected the MCFM results for the branching fraction
above. The jet $p_T$ thresholds of 25 GeV and 30 GeV are used.

\begin{table}
\caption{\label{tab:inclusive} Inclusive cross sections
($\sigma_{\ge0}$) in pb for the process
  $pp \to H$ for several renormalization and
  factorization scale values,  the cross section is evaluated with
  HNNLO at $\sqrt{s}=13$ TeV, the central value scale is set to 
$m_H/2$. The Higgs boson mass is set at $m_H = 125.09$ GeV. PDF4LHC15 NNLO MC PDFs are
used. The error quoted is the statistical error of the computation. The
computation is performed in the infinite top quark mass approximation.
The cross section is reported at LO, NLO and NNLO.}
\centering
\begin{tabular}{cc|ccc}
\toprule
 & & \multicolumn{3}{c}{Inclusive cross section (HNNLO)} \\ 
\cmidrule{3-5}
$\mu_{F}/m_H$ & $\mu_{R}/m_H$ & LO & NLO & NNLO \\
\midrule
1&1&12.697$\pm$0.003&30.30$\pm$0.15&41.50$\pm$0.15\\
1&1/2&15.519$\pm$0.003&36.51$\pm$0.26&46.14$\pm$0.26\\
1/4&1/4&16.691$\pm$0.003&42.82$\pm$0.34&50.16$\pm$0.34\\
1/4&1/2&13.354$\pm$0.003&34.38$\pm$0.20&45.33$\pm$0.20\\
1/2&1&11.958$\pm$0.003&29.43$\pm$0.16&40.99$\pm$0.16\\
1/2&1/4&18.267$\pm$0.004&43.99$\pm$0.35&49.65$\pm$0.35\\
1/2&1/2&14.615$\pm$0.003&35.61$\pm$0.25&45.64$\pm$0.25\\
\bottomrule
\end{tabular}
\end{table}

\begin{table}
\caption{\label{tab:Hplus1jet} H+1 jet inclusive cross sections
($\sigma_{\ge1}$) in pb for  the process
  $pp \to H$ for several renormalization and
  factorization scale values,  the cross section is evaluated with
  MCFM at $\sqrt{s}=13$ TeV, the central value scale is set to $m_H/2$. 
The Higgs boson mass is set at $m_H = 125.09$ GeV. PDF4LHC15 NNLO MC PDFs are
used. }
\centering
\begin{tabular}{cc|cc|cc}
\toprule
 & & \multicolumn{4}{c}{ H+1jet inclusive cross section (MCFM)} \\ 
\cmidrule{3-6}
\multicolumn{2}{c}{}&\multicolumn{2}{|c|}{$p_T > 25$
                      GeV}&\multicolumn{2}{c}{$p_T > 30$ GeV} \\
$\mu_{F}/m_H$ & $\mu_{R}/m_H$ & LO & NLO & LO & NLO \\
\midrule
1 & 1 & 9.303$\pm$0.002&16.48$\pm$0.05&7.947$\pm$0.002&14.11$\pm$0.02 \\
1 & 1/2 & 12.764$\pm$0.002&19.80$\pm$0.05&10.900$\pm$0.002&16.87$\pm$0.03 \\
1/4 & 1/4 & 18.373$\pm$0.004&22.04$\pm$0.08&15.836$\pm$0.003&18.96$\pm$0.06 \\
1/4 & 1/2 & 12.874$\pm$0.002&19.18$\pm$0.05&11.097$\pm$0.002&16.47$\pm$0.05 \\
1/2 & 1 & 9.398$\pm$0.002&16.42$\pm$0.02&8.060$\pm$0.0016&14.04$\pm$0.03 \\
1/2 & 1/4 & 18.396$\pm$0.003&22.36$\pm$0.10&15.777$\pm$0.003&19.29$\pm$0.06 \\
1/2 & 1/2 & 12.893$\pm$0.002&19.42$\pm$0.10&11.056$\pm$0.002&16.77$\pm$0.05 \\
\bottomrule
\end{tabular}
\end{table}

\begin{table}
\caption{\label{tab:Hplus2jet} H+2 jet cross-section ($\sigma_{\ge2}$)
in pb for
 the process
  $pp \to H$ for several renormalization and
  factorization scale values.
  The values are computed with MCFM  at $\sqrt{s}=13$ TeV, R=0.4
  for jet thresholds of  $p_{T} > 25$ and $p_{T} > 30$  GeV. The central value scale is chosen to be
  $m_H/2$. PDF4LHC15 NNLO MC  pdfs are used.}
\centering
\begin{tabular}{cc|cc}
\toprule
 \multicolumn{4}{c}{gg$\to$H+2jets cross section (MCFM)} \\ 
\midrule
$\mu_{F}/m_H$ & $\mu_{R}/m_H$ & LO & NLO \\
\multicolumn{2}{c}{} & \multicolumn{2}{|c}{$p_T > 25$ GeV} \\
\midrule
1 & 1 & 5.250$\pm$	0.002&	6.96$\pm$
0.03 \\
1 & 1/2 & 8.003	$\pm$0.003	&6.90$\pm$	0.06\\
1/4 & 1/4 & 14.565$\pm$	0.005	&2.26$\pm$	0.14\\
1/4 & 1/2 &  9.068$\pm$	0.003	&4.73$\pm$	0.09\\
1/2 & 1 & 5.586	$\pm$0.002	&5.67	$\pm$0.05\\
1/2 & 1/4 & 13.679$\pm$	0.005	&4.10	$\pm$0.11\\
1/2 & 1/2 & 8.514	$\pm$0.003	&5.55	$\pm$0.07\\
\midrule
\multicolumn{2}{c}{} & \multicolumn{2}{|c}{$p_T > 30$ GeV} \\
\midrule
1 & 1 & 3.980	$\pm$0.001&	5.20	$\pm$0.03\\
1 & 1/2 & 6.064	$\pm$0.002&	5.27	$\pm$0.04\\
1/4 & 1/4 & 11.192	$\pm$0.004&	-1.6	$\pm$0.5\\
1/4 & 1/2 & 6.966	$\pm$0.002&	3.52	$\pm$0.05\\
1/2 & 1 & 4.262	$\pm$0.002&	4.28	$\pm$0.04\\
1/2 & 1/4 & 10.434	$\pm$0.004&	2.81	$\pm$0.10\\
1/2 & 1/2 & 6.496	$\pm$0.002&	4.12	$\pm$0.05\\
\bottomrule
\end{tabular}
\end{table}

From Table \ref{tab:inclusive} and Table \ref{tab:Hplus1jet} we can
compute $\Delta \sigma_{\ge 0}$, $\Delta \sigma_{\ge 1}$,
$\Delta \sigma_{\ge 2}$ and $\sigma_0$, $\sigma_1$ using the
definitions:\footnote{Note that $\sigma_{\ge 2}$ and its uncertainty
are computed here at leading order in order to match the power of
$\alpha_s$ with   the NLO computation of $\sigma_{\ge
1}$. Alternatively, one could choose to evaluate $\sigma_{\ge 2}$ at
the highest known order, namely, NLO. }
\[
\sigma_0 = \sigma_{\ge 0} - \sigma_{\ge 1} ,
\, \sigma_1 = \sigma_{\ge 1} - \sigma_{\ge 2}^{\rm LO}.
\]
where the cross section corresponding to the scale choice  $\mu_{F} =
\mu_{R} = m_H/2$ is used as central value and the LO value of
$\sigma_{\ge 2}$ is used in order to preserve the $\alpha_s$ power
counting in $\sigma_1$, being $\sigma_{\ge 1}$ computed up to NLO.
The central values of the exclusive cross sections and their
uncertainties are summarized in Table \ref{tab:ST_unc} together with
the fractional error on $\sigma_0$ and $\sigma_1$. 
The upward and downward fractional errors are obtained using the following
formulae:
\[
\frac{\Delta^+ \sigma_0}{\sigma_0} = \frac{\sqrt{\Delta^{+2} \sigma_{\ge 0} + \Delta^{-2} \sigma_{\ge 1}}}{\sigma_0} \quad
\frac{\Delta^- \sigma_0}{\sigma_0} = \frac{\sqrt{\Delta^{-2} \sigma_{\ge 0} + \Delta^{+2} \sigma_{\ge 1}}}{\sigma_0} 
\]

\[
\frac{\Delta^+ \sigma_1}{\sigma_1} = \frac{\sqrt{\Delta^{+2} \sigma_{\ge 1} + \Delta^{-2} \sigma_{\ge 2}}}{\sigma_1} \quad
\frac{\Delta^- \sigma_1}{\sigma_1} = \frac{\sqrt{\Delta^{-2} \sigma_{\ge 1} + \Delta^{+2} \sigma_{\ge 2}}}{\sigma_1} 
\]
where the ${}^+$ sign indicates the upward uncertainty and the ${}^-$
sign the downward uncertainty. In the same table, the symmetrized
number are reported, using the maximum of the upward and downward
errors. We recommend to use the symmetrized  values as final uncertainties.

\begin{table}
\caption{\label{tab:ST_unc} Summary of jet-bin uncertainties on the $0$ and
  $1$ jet exclusive cross sections obtained using the ST method. The
  last two lines show symmetrized uncertainty intervals from the $7^{th}$ and $8^{th}$ row.}
\centering
\begin{tabular}{c|c|c}
\toprule
$\Delta \sigma_{\ge 0}$ &\multicolumn{2}{c}{ $[-4.6,+4.6]  $ pb} \\
\midrule
& $p_T > 25$ GeV & $p_T > 30$ GeV \\
\midrule
$\Delta \sigma_{\ge 1}$ & $[-3,+2.9]  $ pb & $[-2.7,+2.5 $ pb\\
$\Delta \sigma_{\ge 2}^{\rm LO}$ & $[- 3.3, +6.0] $ pb &  $[-2.5,+4.7] $ pb\\
$\sigma_{0}$ & 26.2 pb & 28.9 pb  \\
$\sigma_{1}$ &10.9 pb & 10.3 pb \\
\midrule
$\Delta \sigma_0 / \sigma_0$ S.T &[-0.22, +0.22]  & [-0.18,+0.18]  \\
$\Delta \sigma_1 / \sigma_1$ S.T &[-0.62, +0.40]  & [-0.53,+0.34]  \\
\midrule
$\Delta \sigma_0 / \sigma_0$ S.T &[-0.22, +0.22]  & [-0.18,+0.18]  \\
$\Delta \sigma_1 / \sigma_1$ S.T &[-0.62, +0.62]  & [-0.53,+0.53]  \\
\bottomrule
\end{tabular}
\end{table}

\subsubsection{The Jet Veto Efficiency method.}

The first version of Jet Veto Efficiency method, presented in
Ref.~\cite{Banfi:2012jm}, 
computes the jet veto acceptance uncertainties using three different definitions of the jet veto efficiency.
Such definitions differ among each other for terms beyond $\alpha_s^2$ and the related efficiencies show different behaviour as a function of the vetoed jet $p_T$. 
The uncertainty is computed, in this case, by doing the envelope of the
naive scale uncertainty of the reference method and the central values
obtained with the three jet veto efficiency definitions. The definitions are commonly referred as ``schemes'' $a,b,c$ and are reported below:
\begin{equation}
 \epsilon_N^{a}= 1-\frac{\sigma_{\ge N+1}^{\rm NLO}}{\sigma_{\ge N}^{\rm NNLO}}     \quad \epsilon_N^{b} = 1 -
\frac{\sigma_{\ge N+1}^{\rm NLO}}{\sigma^{\rm NLO}_{\ge N}} \quad \epsilon_N^{c} = 1 - \frac{\sigma_{\ge
    N+1}^{\rm NLO}}{\sigma_{\ge N}^{\rm NLO}} + \left ( \frac{\sigma_{\ge N}^{\rm
      NLO}}{\sigma_{\ge N}^{\rm LO}} - 1 \right ) \frac{\sigma_{\ge
    N+1}^{\rm LO}}{\sigma_{\ge N}^{\rm LO}}  \label{eq:schemes}
\end{equation}
The $N$ value represents the number of jets of the exclusive
selection.

\subsubsubsection{The JVE method for zero jet}
In the 0-jet case the large logs that are produced by the
 introduction of the jet $p_T$ threshold can be resummed using the
 JetVHeto  program.
Inputs to the JetVHeto computation are the LO, NLO and NNLO inclusive
cross sections, that are shown in Table \ref{tab:inclusive} and the
$\sigma_{\ge 1}$ cross section shown in Table \ref{tab:Hplus1jet}. The
resummed computation can be matched to the fixed order result using
three different schemes  that can be
considered equivalent to the three different Jet Veto Efficiency
schemes listed above. In the resummed case, all scales, including the resummation scale, are varied by a factor two,
and the envelope built using such uncertainty band together with the central value
obtained with the three different matching schemes is quoted as final uncertainty.
We report results using both the fixed-order computation and  the
resummed one.

In Table \ref{table:JVE0} we show the 0-jet JetVHeto efficiencies obtained for the factorization, renormalization and resummation scale considered in the envelope,
and the central value of each  scheme. The envelope is built using
the scale variations of the ``scheme a'' only, therefore only for this
scheme  are detailed values for each scale choice  shown.
The efficiencies are reported for the 25 and 30 GeV $p_T$ threshold
using fixed order, resummed only and resummed results matched with the
fixed order computation at NNLO+NNLL.

\begin{table}
\centering
\caption{\label{tab:JVE0} $H+0$ jet efficiencies of the process
  $pp \to H \to e^+ \nu e^- \bar{\nu}$ for several renormalization and
  factorization scale values, matching schemes and resummation scales.
  The values are computed with JetVHeto  at $\sqrt{s}=13$ TeV, R=0.4
  and $p_{T} >25, 30$ GeV. The central value scale is chosen to be
  $m_H/2$. The Higgs boson mass is set at $m_H = 125.09$ GeV. PDF4LHC15 NNLO MC  are
used. } \label{table:JVE0}
\footnotesize
\begin{tabular}{cc|c|c|ccc|ccc}
\toprule
\multicolumn{10}{c}{0-jet JetVHeto efficiencies for different $p_T$ thresholds} \\
\midrule
\multicolumn{4}{c|}{} & \multicolumn{3}{c|}{$p_T > 25$ GeV} & \multicolumn{3}{c}{$p_T > 30$ GeV} \\ 
\cmidrule{5-10}
$\mu_F/m_H$ & $\mu_R/m_H$ & $Q_{\rm res}/m_H$ & Scheme &  NNLO+NNLL &
                                                                      NNLL&
                                                                            NNLO & NNLO+NNLL & NNLL & NNLO \\
\midrule
1 & 1& 1/2 & a & 0.5841 & 0.5727 & 0.6028 & 0.6509 & 0.6289 &0.6600\\
1 & 1/2& 1/2 & a & 0.5810 & 0.5513 & 0.5708  & 0.6472 & 0.6025 & 0.6344\\
1/4 & 1/4& 1/2 & a & 0.5792 & 0.5176 & 0.5606  & 0.6431 & 0.5642 & 0.6221 \\
1/4 & 1/2& 1/2 & a & 0.5549 & 0.5290 & 0.5768 & 0.6248 & 0.5845 & 0.6366\\
1/2 & 1& 1/2 & a & 0.56790 & 0.5586 & 0.5995& 0.6379 & 0.6169 & 0.6575\\
1/2 & 1/4& 1/2 & a & 0.5866 & 0.5207 & 0.5495  & 0.6462 & 0.5656 & 0.6114 \\
1/2 & 1/2& 1/2 & a & 0.5726 & 0.5413 & 0.5744  & 0.6367 & 0.5940 & 0.6325 \\
1/2 & 1/2& 1/4 & a & 0.5650 & 0.5295 & 0.5744 & 0.6273 & 0.5686 & 0.6325\\
1/2 & 1/2& 1 & a & 0.6336 & 0.6387 & 0.5744  & 0.6987 & 0.6938 & 0.6325\\
1/2 & 1/2& 1/2 & b & 0.5147 & 0.5413 & 0.4544  & 0.5760 & 0.5940 & 0.5289\\
1/2 & 1/2& 1/2 & c & 0.7557 & 0.5413 & 0.9379 & 0.8207 & 0.5940 & 0.9389\\
\bottomrule
\end{tabular}
\end{table}

\subsubsubsection{The JVE method for one jet}
The $JVE$ method in his fixed order form can be easily extended to the
1-jet bin. Such approach has been used by the ATLAS collaboration for
the publication of the  Run-1 paper on the $h \to WW$ channel~\cite{ATLAS:2014aga}. This channels is, at the
moment, the decay channel that provides the most accurate
measurements of the Higgs boson production cross section and its couplings. 
Equations~(\ref{eq:schemes}) can be used also for  the 1-jet bin where $\epsilon_1$ represents the ratio of events with exactly one
jet over the number of events with at least one jet.
In scheme $a$  the NNLO $H+1$ jet cross section
is used. Such value has been nowadays evaluated~\cite{Caola:2015wna} but
software tools are still not publicly available, therefore the same
approach of~\cite{ATLAS:2014aga} will be followed in the following, assuming  that scheme $a$  lies between scheme $b$ and scheme $c$. Such assumption has been tested up to NLO,
and at NNLO for the $gg$ induced process at the $\sqrt{s}$ of 8 TeV.
Therefore, the average of schemes $b$ and $c$ is used instead of
the scheme $a$. Using the cross sections reported in Table \ref{tab:inclusive}, Table \ref{tab:Hplus1jet} and
Table \ref{tab:Hplus2jet} , the $\epsilon_1$ values are computed for schemes
$b$, $c$ and their average and tabulated in Table \ref{tab:i}. The uncertainty is evaluated as the envelope
of the average of schemes $b$ and $c$, computed for all renormalization
and factorization scales,  and the central values of the schemes  $b$
and $c$.

\begin{table}
\caption{\label{tab:i} H+1 jet efficiency of the process
  $pp \to H \to e^+ \nu e^- \bar{\nu}$ for several renormalization and
  factorization scale values.
  The values are computed with MCFM  at $\sqrt{s}=13$ TeV, R=0.4
  and and $p_{T} >25, 30$ GeV. The central value scale is chosen to be
  $m_H/2$. The Higgs boson mass is set at $m_H = 125.09$ GeV. PDF4LHC15 NNLO MC  are
used. }
\centering
\begin{tabular}{cc|ccc|ccc}
\toprule
\multicolumn{8}{c}{$\epsilon_1$ NLO} \\
\midrule
\multicolumn{2}{c}{} & \multicolumn{3}{|c|}{$p_T > 25$ GeV} &\multicolumn{3}{c}{$p_T > 30$ GeV} \\
\cmidrule{3-8}
$\mu_F/m_H$ & $\mu_R/m_H$ & $\epsilon_1^{b}$ & $\epsilon_1^{c}$ &
                                                                  $(\epsilon_1^b+\epsilon_1^c)/2$
                                             & $\epsilon_1^{b}$ & $\epsilon_1^{c}$ & $(\epsilon_1^b+\epsilon_1^c)/2$\\
\midrule
1 & 1 & 0.5777 & 0.6874 & 0.6326 &  0.6311 & 0.7335 & 0.6823\\
1 & 1/2 & 0.6517 & 0.8054 & 0.7285 & 0.6878 & 0.8215 & 0.7546\\
1/4 & 1/4 & 0.8976 & 1.035 & 0.9665 & 1.086 & 1.242 & 1.164
  \\
1/4 & 1/2 & 0.7533 & 0.9776 & 0.8654 & 0.7860 & 0.9865 & 0.8862\\
1/2 & 1 & 0.6549 & 0.8411 & 0.7480  & 0.6952 & 0.8615 &
                                                                  0.778337 \\
1/2 & 1/4 & 0.8167 & 0.9375 & 0.8771 & 0.8543 & 0.9692 &
                                                                   0.9117 \\
1/2 & 1/2 & 0.7142 & 0.9040 & 0.8091 & 0.7541 & 0.9308 & 0.8424\\

\bottomrule
\end{tabular}
\end{table}

\subsubsubsection{The JVE method final results}
 In Table \ref{tab:JVE_Final} we
summarize the jet veto efficiencies and their uncertainties for the 0-jet and
1-jet bin indicating,  in the 0-jet case,  both the resummed and  the fixed order results. 
We recommend to use the resummed result as reference values. 

\begin{table}
\caption{\label{tab:JVE_Final} Jet veto efficiency, exclusive jet bin
  cross sections and their  uncertainties for the $0$ and
  $1$ jet bin using the JVE method. The labels F.O. and RES. refer to
  the usage of fixed order and resummed inputs, respectively. The last four lines of the table
  show symmetrized uncertainty intervals.}
\centering
\begin{tabular}{c|c|c}
\toprule
\multicolumn{3}{c}{Summary of JVE related uncertainties.}\\
\midrule
& $p_T > 25$ GeV & $p_T > 30$ GeV\\
\midrule
$\epsilon_0$ RES.   &  0.57  & 0.64 \\
$\epsilon_0$ FO.   &  0.57  & 0.63 \\
$\Delta \epsilon_0$ RES. & [- 0.06,+0.18]  &[-0.06,  +0.18 ]  \\
$\Delta \epsilon_0$ F.O. & [-0.12, +0.36]  &[-0.10, +0.31]    \\
$\Delta \epsilon_0/\epsilon_0  $ RES. &[-0.11 , +0.32]  &[-0.09,
                                                          +0.28]    \\
$\Delta \epsilon_0/\epsilon_0  $ F.O. &[-0.21 , +0.63]  &[-0.16, +0.49]    \\

\midrule
$\epsilon_1$ F.O. & 0.81 & 0.84 \\
$\Delta \epsilon_1$ F.O.  &[ -0.18, +0.16] &[-0.16, +0.32] \\
$\Delta \epsilon_1/\epsilon_1$ F.O.  &[-0.22, +0.20]  &[-0.19, +0.38] \\
\midrule
$\Delta \sigma_0/\sigma_0$ F.O. &[-0.18, +0.37]  &[-0.12, +0.33]\\
$\Delta \sigma_1/\sigma_1$ F.O. &[-0.87, +0.36]  &[-0.86, +0.48]\\
$\Delta \sigma_0/\sigma_0$ RES. &[-0.13, +0.32]  &[-0.12, +0.29]\\
$\Delta \sigma_1/\sigma_1$ RES. &[-0.48, +0.25]  &[-0.54, +0.42]\\
\midrule
$\Delta \sigma_0/\sigma_0$ F.O. &[-0.37, +0.37]  &[-0.33, +0.33]\\
$\Delta \sigma_1/\sigma_1$ F.O. &[-0.87, +0.87]  &[-0.86, +0.86]\\
$\Delta \sigma_0/\sigma_0$ RES. &[-0.32, +0.32]  &[-0.29, +0.29]\\
$\Delta \sigma_1/\sigma_1$ RES. &[-0.48, +0.48]  &[-0.54, +0.54]\\
\bottomrule
\end{tabular}
\end{table}

\noindent In order to compare the JVE results with the ST ones, in the same
table we report the
0 ($\sigma_0$) and 1 ($\sigma_1$)  jet cross sections with the Jet Veto method.
They are computed from the Higgs  total cross section and the
jet veto efficiencies $\epsilon_0$ and $\epsilon_1$ according to the
following formulae:
\[
\sigma_0  = \sigma_{\rm tot} \cdot \epsilon_0 \quad \sigma_1 = \sigma_{\rm tot} \cdot (1 - \epsilon_0) \cdot \epsilon_1
\]

\noindent In this expression, the best available computation of the Higgs total cross section can be used because full factorization between total cross section and
 acceptance is assumed  in this approach. The cross section value computed using the De Florian-Grazzini method for resummation and
including the complex pole scheme is:
\[
 \sigma_{\rm tot} = 43.92 \Upb {}^{+7.4 \%}_{- 7.9 \%} {\rm (scale)}
 {}^{+7.1 \%}_{-6.0 \%} {\rm (PDF)}
\]
the PDF error is usually accounted independently in experimental analyses,
therefore only the scale variation related  error is accounted
for. Using simple error propagation, the central values and their
uncertainties have been computed and  summarized in Table \ref{tab:JVE_Final}.

\subsection[Combined resummed predictions for the \texorpdfstring{$0$-jet, $1$-jet, and $\geq 2$-jet}{0-jet, 1-jet, and 2-or-more-jet} bins]{Combined resummed predictions for the \texorpdfstring{$0$-jet, $1$-jet, and $\geq 2$-jet}{0-jet, 1-jet, and 2-or-more-jet} bins\SectionAuthor{R.~Boughezal, X.~Liu, F.~Petriello, I.W.~Stewart, F.J.~Tackmann}}
\label{subsec:stwz_blptw}
\providecommand{\pTjet}{p_T^{\rm jet}}
\providecommand{\pTjetone}{p_T^{\rm jet1}}
\providecommand{\pTjettwo}{p_T^{\rm jet2}}
\providecommand{\pToff}{p_T^{\rm off}}
\providecommand{\resum}{\mathrm{resum}}
\providecommand{\nons}{\mathrm{nons}}
\providecommand{\Rnons}{{R\mathrm{sub}}}
\providecommand{\muns}{\mu_\mathrm{ns}}
Experimental analyses require a consistent treatment of cross sections and their uncertainties for several jet bins.
In the following we discuss predictions for the $0$-jet,
$1$-jet, and $\ge 2$-jet bins with a resummation of jet-veto logarithms,
and provide updated results for $13\,{\rm TeV}$.
We utilize a theoretical approach that provides flexible control over uncertainties
allowing for the identification of different sources of yield and migration uncertainties.
It is thus well-suited to provide a theoretical description of jet binning, including multiple
jet-bin boundaries.

\subsubsection{Jet \texorpdfstring{$p_T$}{pT} resummation at NNLL\texorpdfstring{$'+$}{prime+}NNLO\SectionAuthor{I.~W.~Stewart, F.~J. Tackmann}}

We discuss the resummed predictions for the $H+0$-jet cross section from gluon fusion with a $p_T$ veto on jets and with the resummation of jet-veto logarithms at NNLL$'+$NNLO order~\cite{Stewart:2013faa}.
We place a particular emphasis on a careful estimate of the perturbative uncertainties and include a detailed discussion of how yield and migration uncertainties are determined. The different contributions to the uncertainty are estimated by appropriate variations of the different scales in virtuality and rapidity space appearing in a factorization theorem.
This allows us to distinguish between and account for the uncertainties due to higher fixed-order corrections as well as higher-order towers of jet-$p_T$ logarithms.

\begin{figure}[t!]
\includegraphics[width=0.3\textwidth]{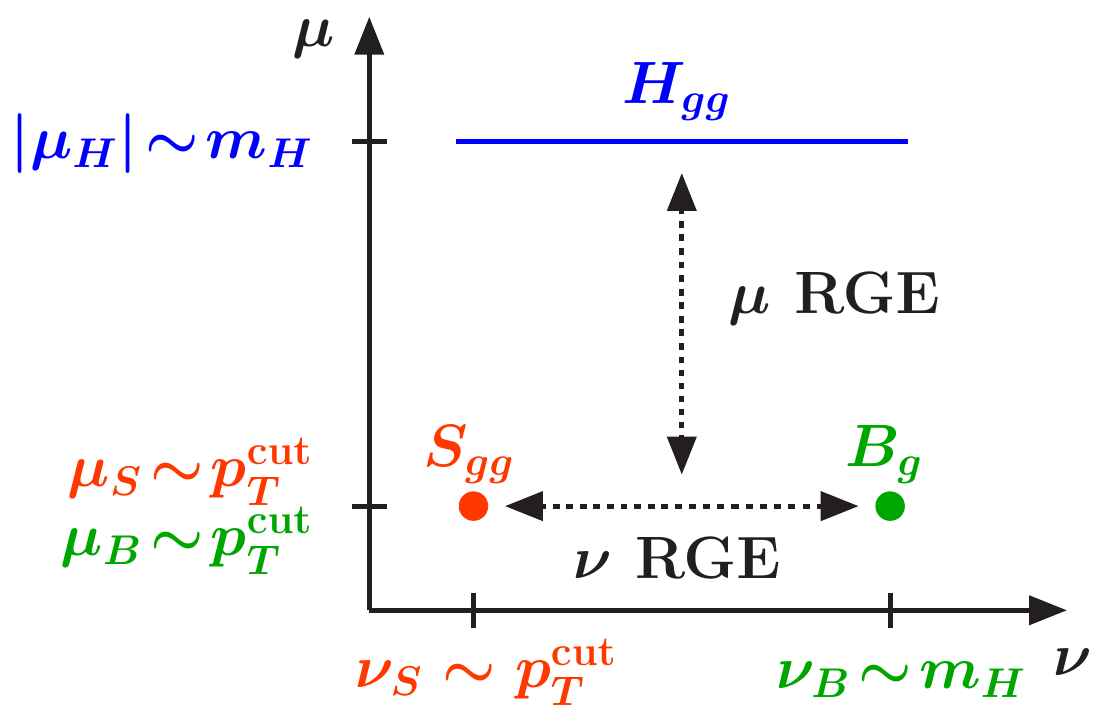}%
\includegraphics[width=0.35\textwidth]{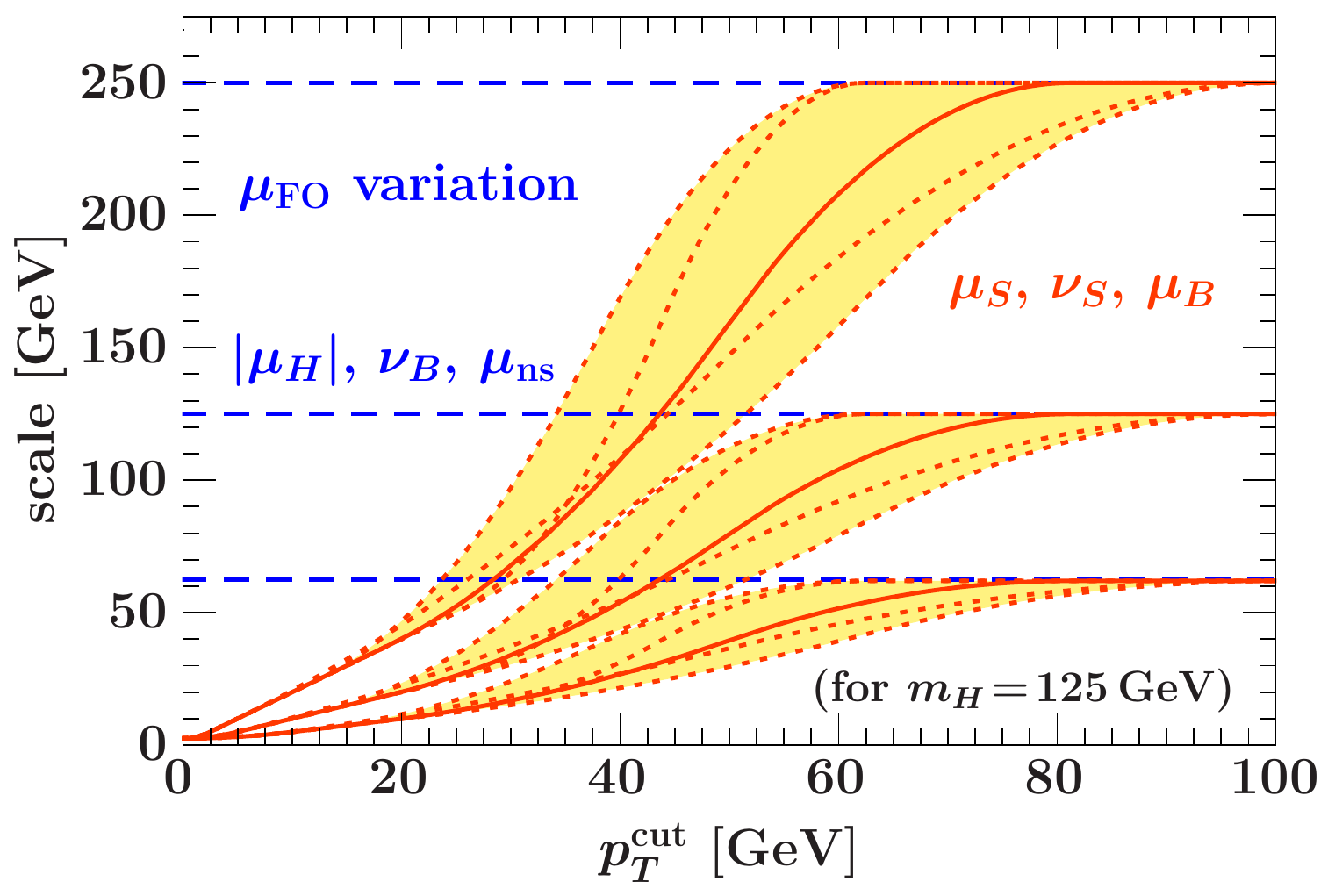}%
\includegraphics[width=0.35\textwidth]{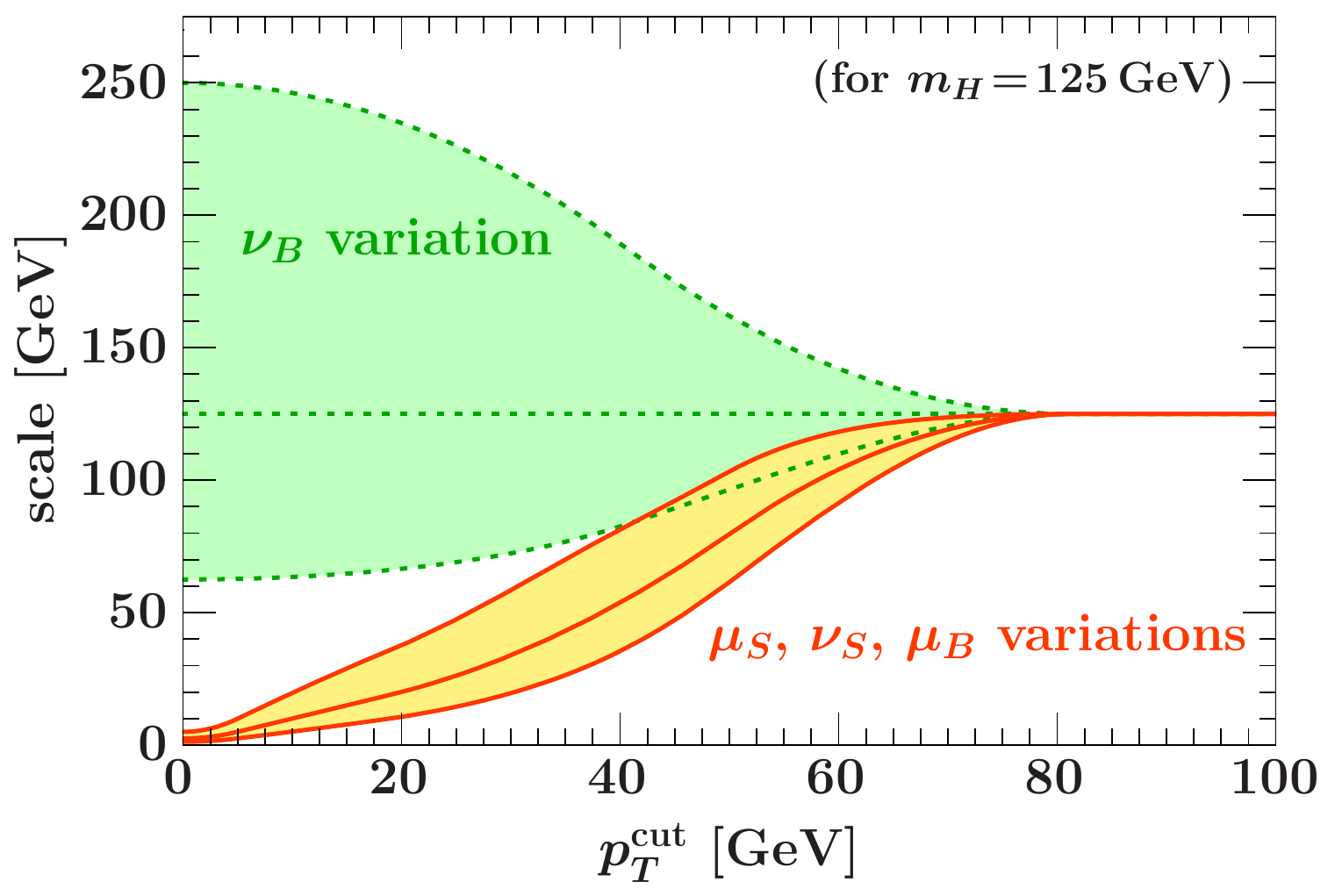}%
\caption{Left panel: Illustration of the RGE in virtuality and rapidity space to resum jet-$p_T$ logarithms.
Middle panel: Profile scale variations contributing to the overall fixed-order (yield) uncertainty.
Right panel: Profile scale variations whose combinations are used to assess the resummation (migration) uncertainty.}
\label{fig:RGscales}
\end{figure}

We utilize the framework of soft-collinear effective theory (SCET)~\cite{Bauer:2000ew, Bauer:2000yr, Bauer:2001ct, Bauer:2001yt}
for jet-veto resummation at hadron colliders~\cite{Stewart:2009yx,Berger:2010xi,Becher:2012qa,Tackmann:2012bt,Stewart:2013faa}.
The factorized $pp\to H+0$-jet cross section with a cut on $\pTjet < \pTcut$ is given by
\begin{align} \label{eq:fact}
\sigma_0(\pTcut)
&= \frac{\sqrt{2} G_F\, m_H^2}{576 \pi \Ecm^2}\, H_{gg} (m_H^2, \mu_H) \int\!d Y
B_g\Bigl(m_H, \pTcut, R, \frac{m_H}{\Ecm}\,e^{Y}, \mu_B, \nu_B \Bigr)
\nn \\ & \qquad
B_g\Bigl(m_H, \pTcut, R, \frac{m_H}{\Ecm}\,e^{-Y}, \mu_B, \nu_B \Bigr)\, S_{gg}(\pTcut, R, \mu_S, \nu_S)
\,U_0(\pTcut, R; \mu_H, \mu_B, \mu_S, \nu_B, \nu_S)
\nn \\ & \qquad
+ \sigma_0^\Rnons(\pTcut, R) + \sigma_0^\nons(\pTcut, R, \muns)
\,.\end{align}
The first term is the leading contribution containing all the singular logarithmic terms $\alpha_s^i\ln^j(\pTcut/m_H)$. The resummation of the logarithms is performed by renormalization group evolution (RGE) in both virtuality ($\mu$) and rapidity ($\nu$) space, illustrated on the left of \refF{fig:RGscales}. The factorized hard ($H_{gg}$), beam ($B_g$), and soft ($S_g$) functions are evaluated at their own natural virtuality scales $\mu_i$ and rapidity scales $\nu_i$, where they contain no large logarithms and are calculable at fixed order in $\alpha_s$. From there they are evolved to a common (arbitrary) scale, yielding the combined evolution factor $U_0$ which resums the logarithms of the virtuality ratios $\mu_i/\mu_j$ and rapidity ratios $\nu_i/\nu_j$.
The resummation is performed to NNLL$'$ order, which in addition to the NNLL resummation includes the full $\ord{\as^2}$ corrections to $H_{gg}$, $B_g$, and $S_{gg}$ (which includes the $\ord{\as^2}$ effects from jet clustering). These are formally part of the N$^3$LL resummation for which they provide the correct RGE boundary conditions, and incorporate all dominant (singular) NNLO corrections into the resummed result.

The second term in Eq.~\eqref{eq:fact}, $\sigma_0^\Rnons(\pTcut, R)$, contains $\ord{R^2}$ contributions. For $R = 0.4$ they are numerically very small and are treated as subleading power corrections. The last term in Eq.~\eqref{eq:fact}, $\sigma_0^\nons(\pTcut, R, \muns)$, contains $\ord{\pTcut/m_H}$ ``nonsingular'' corrections, which vanish for $\pTcut\to0$ but become important at large $\pTcut$.
These terms are added to achieve the full NNLL$'$+NNLO accuracy, which incorporates the complete NNLO cross section for all values of $\pTcut$, including the inclusive NNLO cross section.

The RGE scales $\mu_H, \mu_B, \mu_S, \nu_B$, and $\nu_S$ are chosen as functions of $\pTcut$, which are referred to as profile scales~\cite{Ligeti:2008ac, Abbate:2010xh}. They have to satisfy several constraints and their construction is discussed in detail in Ref.~\cite{Stewart:2013faa}. Essentially, in the resummation region at small $\pTcut$ they have to parametrically follow the canonical scaling dictated by the RGE, while at large $\pTcut \gtrsim m_H/2$ they must approach a common fixed-order scale $\mu_{\rm FO}$ in order to turn off the resummation and avoid unphysical behaviour. The remaining freedom in the choice of the profile scales provides a flexible and powerful way to assess the perturbative uncertainties in the resummed predictions. For this reason, profile scales have been applied by now in a large variety of different contexts and have become an established and reliable method for assessing perturbative uncertainty in resummed predictions.
As detailed in Ref.~\cite{Stewart:2013faa}, in the context of jet binning the profile scale variations allow us to identify the different uncertainty sources contributing to the yield and migration uncertainties discussed in Section~\ref{subsec:binuncertainties}.

Before discussing the variations, we stress that the scales are unphysical parameters, and their variations simply provide a convenient way to probe the ``typical'' size of the associated missing higher-order terms. The observed variations in the results must be interpreted as such. In particular, we do not assign any meaning to accidentally small one-sided scale variations that yield asymmetric
uncertainties, which are just the result of nonlinear scale dependence, which is frequently encountered at higher orders and including resummation. Hence, we always consider the maximum absolute deviation from the chosen central scale as the (symmetric) uncertainty. To be explicit, an observed variation of $+|x|$ and $-|y|$ in the cross section is interpreted as uncertainty of $\pm \max\{|x|, |y|\}$.

The first type of variation is a collective variation of all scales by a factor of $1/2$ and $2$. This keeps all scale ratios and thus all logarithms fixed, and at large $\pTcut$ reproduces the uncertainty in the fixed-order cross section. Hence, this corresponds to an overall fixed-order uncertainty (within the resummed prediction), and is naturally identified as a common source for all $\sigma_i$, giving rise to a yield uncertainty $\Delta_{\mu i}$. A second type of variation included in $\Delta_{\mu i}$ is to the profile shape controlling the transition points where the resummation is turned off.
The total of 14 profile variations $V_\mu$ contributing to $\Delta_{\mu 0}$ are displayed in the middle panel of \refF{fig:RGscales}.
For each profile $v_i$ in $V_\mu$ we obtain $\Delta_{\mu i}$ as the maximum absolute deviation from the central value,
\begin{align} \label{eq:Deltamui}
\Delta_{\mu 0} (\pTcut) = \max_{v_i \in V_\mu} \bigl\lvert \sigma_0^{v_i} (\pTcut) - \sigma_0^{\rm central} (\pTcut) \bigr\rvert
\,,\qquad
\Delta_{\mu \geq 0} = \max_{v_i \in V_\mu} \bigl\lvert \sigma_{\geq 0}^{v_i} - \sigma_{\geq 0}^{\rm central} \bigr\rvert
\,.\end{align}
For $\Delta_{\mu \geq 0}$, only the variation of $\mu_{\rm FO}$ by a factor $2$ contributes, and $\Delta_{\mu \geq 1}(\pTcut) = \Delta_{\mu \geq 0} - \Delta_{\mu 0} (\pTcut)$.

The profile scale variations $V_\resum$ contributing to the resummation uncertainty, $\Delta^{0/1}_\resum$, are shown in the right panel of \refF{fig:RGscales}. They separately vary each of the beam and soft scales up and down but keep $\mu_{\rm FO}$ fixed. They thus directly probe the intrinsic uncertainty in the resummed logarithmic series. The variation is chosen to approach the conventional factor of 2 for $\pTcut\to 0$. Out of the $80$ possible combinations of all variations, all combinations leading to arguments of logarithms which are more than a factor of 2 different from their central values are not considered. This leaves a total of 35 profile scale variations in $V_\resum$ that are used to estimate
\begin{equation} \label{eq:Delta01resum}
\Delta^{0/1}_\resum(\pTcut) = \max_{v_i \in V_\resum} \bigl\lvert \sigma_0^{v_i} (\pTcut) - \sigma_0^{\rm central} (\pTcut) \bigr\rvert
\,.\end{equation}
The $\pTcut$ logarithms are the primary source of uncertainty caused by the jet binning at small $\pTcut$ and we can therefore identify $\Delta_\resum$ as the corresponding migration uncertainty. Furthermore, $\Delta^{0/1}_\resum$ smoothly turns off at large $\pTcut$ where the logarithms become unimportant and the resummation is turned off. This is consistent with the fact that in this limit migration effects become irrelevant since $\sigma_{\geq 1}(\pTcut)$ becomes numerically much smaller than $\sigma_0(\pTcut)$.

Our predictions use a complex hard scale $\mu_H = -i \mu_{\rm FO}$ with $\mu_{\rm FO} = m_H$.
This is the canonical scale at which the hard function
contains no large logarithms and shows a significantly improved perturbative stability
compared to $\mu_H = \mu_{\rm FO}$ (for any value of $\mu_{\rm FO}$).
In the transition from small to large $\pTcut$ we keep the hard scale at its complex value.
In principle, one could contemplate rotating it to the real axis as a function of $\pTcut$ to turn off the resulting
resummation of logarithms of $\ln(\mu_H/\abs{\mu_H})$. However, this would inevitably lead to
an unphysical behaviour of a decreasing cross section with increasing $\pTcut$. Instead, the improved convergence observed in the small $p_T$ region, where the factorization of the hard virtual corrections into the overall factor
$H_{gg}$ is manifest, also translates into the fixed-order cross section at large $\pTcut$, because the majority of the total cross section comes from the small $p_T$ region. Consequently, as one can see from Table~\ref{tab:incl}, the inclusive cross section for $\mu_H = -i m_H$ shows a much better perturbative convergence and as a result is already at NNLO in close agreement with the N$^3$LO result at the default scale $\mu_H = m_H/2$.
The uncertainty related to this resummation is estimated by varying the phase of $\mu_H = \mu_{\rm FO} \exp(-i \varphi)$ as $\varphi = \pi/2\pm\pi/4$. This phase variation is roughly equivalent to the usual factor of 2 since $\pi/4 \simeq \ln 2$. We have
\begin{equation}
\Delta_{\varphi i} = \max_{\varphi \in \{\pi/4, \pi/2, 3\pi/4\}} \bigl\lvert \sigma_i^{\varphi} - \sigma_i^{\rm central} \bigr\rvert
\,.\end{equation}
We consider this as an additional independent uncertainty source, and since it affects all cross sections via an overall multiplicative factor, it is treated as a yield uncertainty.

In summary, we have the following three uncertainty amplitudes for $\{\sigma_{\geq 0}, \sigma_0, \sigma_{\geq 1}\}$,
\begin{align} \label{eq:01nuisance}
\kappa_\mu^{\rm y}: &\quad \{ \Delta_{\mu \geq 0},\, \Delta_{\mu 0},\, \Delta_{\mu \geq 1} \}
\,,\qquad\qquad
\kappa_\varphi^{\rm y}: \quad \{ \Delta_{\varphi \geq 0},\, \Delta_{\varphi0},\, \Delta_{\varphi\geq 1} \}
\,,\nn\\
\kappa^{0/1}_\cut: &\quad \Delta^{0/1}_\resum \times \{ 0,\, 1, -1\}
\,.\end{align}

\begin{figure}[t!]
\includegraphics[width=0.5\textwidth]{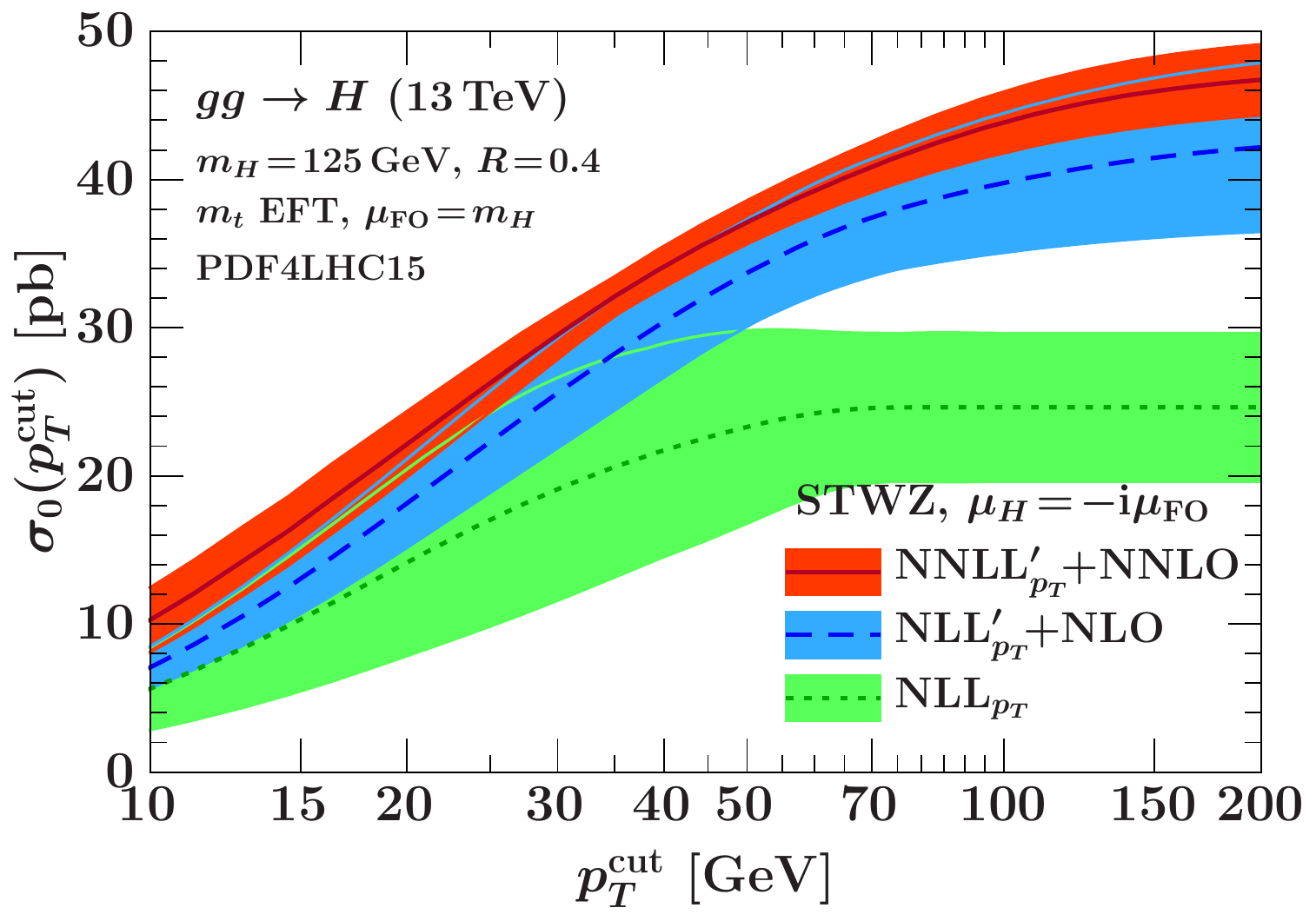}%
\includegraphics[width=0.5\textwidth]{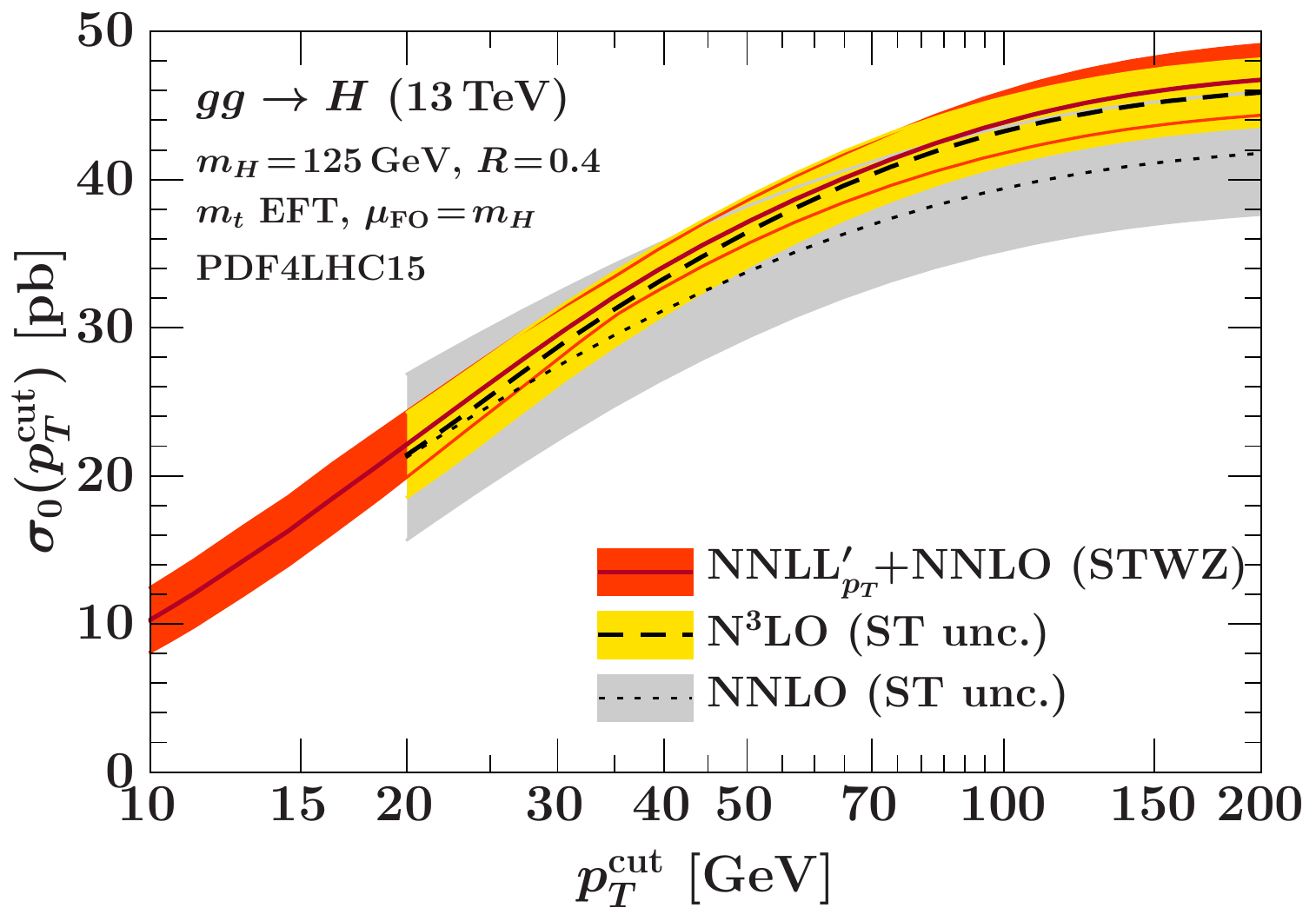}
\caption{The $0$-jet cross section $\sigma_0(\pTcut)$, comparing different resummation orders (left) and resummed and fixed-order predictions (right). The N$^3$LO result on the right utilizes the results of Refs.~\cite{Anastasiou:2015ema, Caola:2015wna}.}
\label{fig:STWZ0jet}
\end{figure}

\begin{figure}[t!]
\includegraphics[width=0.5\textwidth]{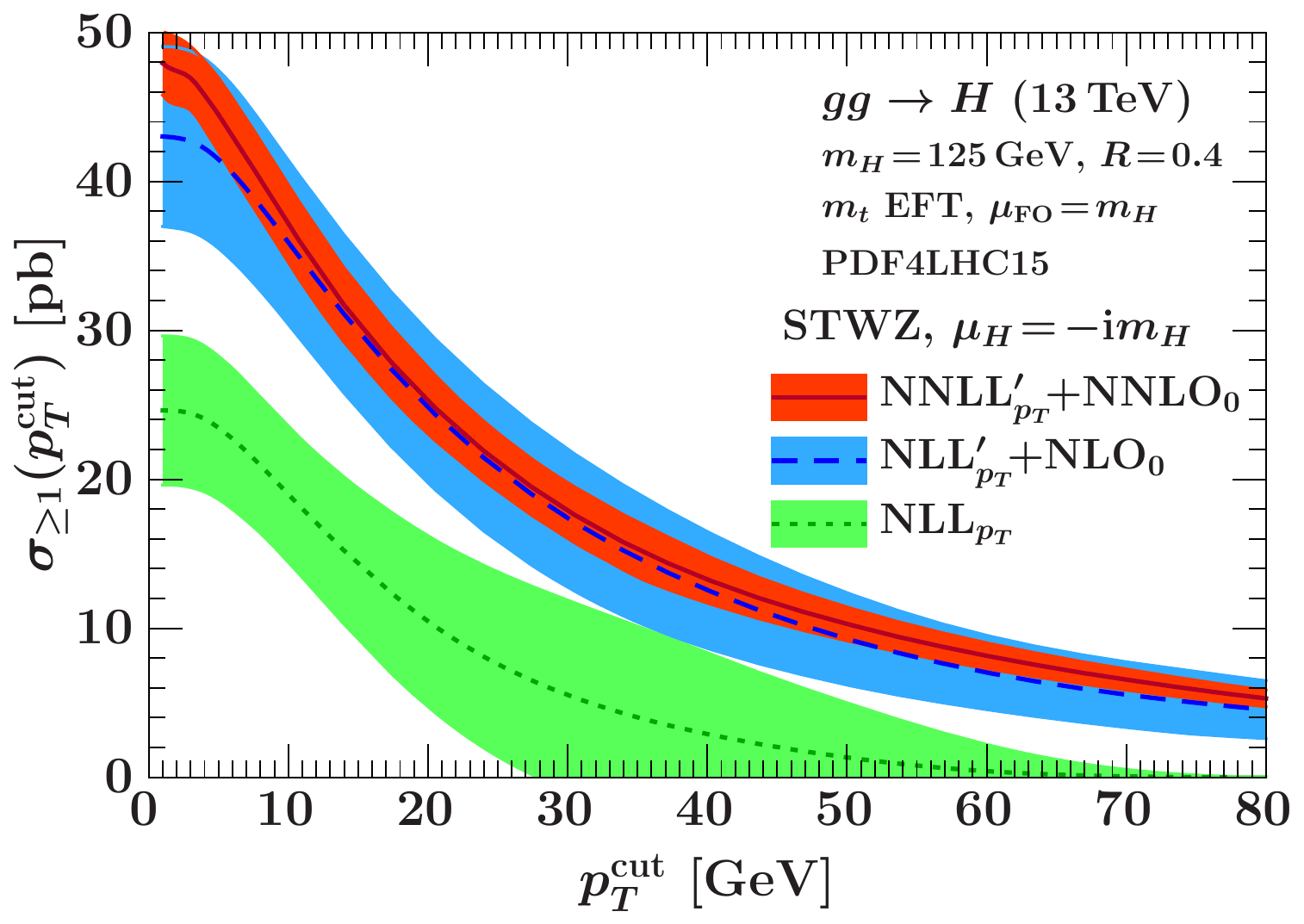}%
\includegraphics[width=0.5\textwidth]{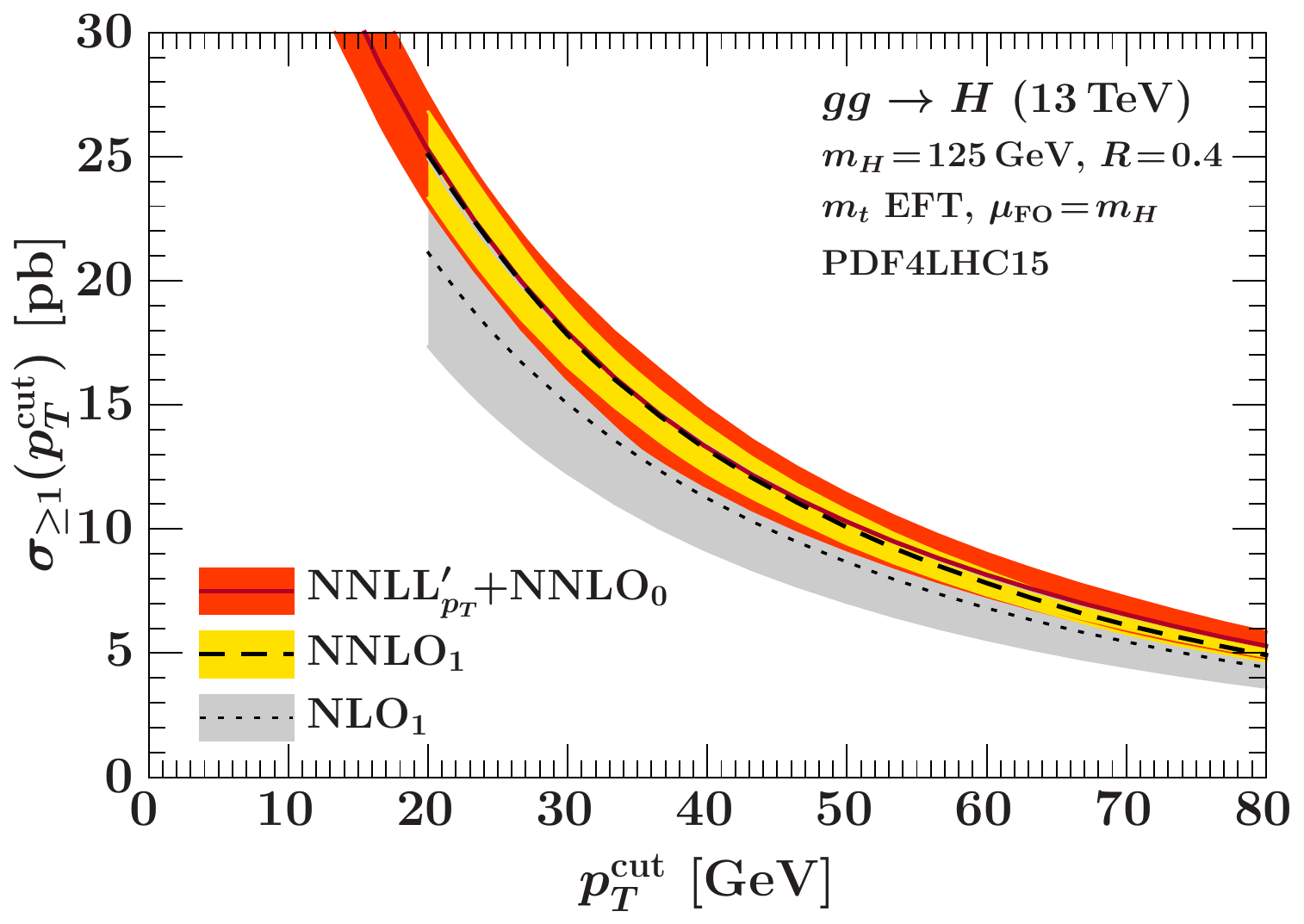}
\caption{The $\geq 1$-jet cross section $\sigma_{\geq1}(\pTcut)$, comparing different resummation orders (left) and resummed and fixed-order predictions (right). The NNLO$_1$ result on the right utilizes the results of Ref.~\cite{Caola:2015wna}.}
\label{fig:STWZ1jet}
\end{figure}

\begin{table}
\caption{The inclusive cross section $\sigma_{\geq0}$, comparing different orders and the conventional scale choices (columns 2 and 3) and with complex hard scale (columns 4 and 5). The uncertainties are the perturbative uncertainties. The last column corresponds
to the default scale choice for $\mu_H$ used for the resummed predictions discussed here.}
\label{tab:incl}
\renewcommand{\arraystretch}{1.1}
\begin{tabular}{l||l|l||l|l}
\toprule
$\sigma_{\geq 0}/\mathrm{pb}$ & $\mu_H = m_H/2$ & $\mu_H = m_H$ & $\mu_H = -i m_H/2$ & $\mu_H = -i m_H$
\\
\midrule
LO & $16.04 \!\pm\! 15.8\%$ & $13.80 \!\pm\! 16.3\%$ & $26.70 \!\pm\! 12.8\%_\mu \!\pm\! 22.9\%_\varphi$ & $23.29 \!\pm\!  14.6\%_\mu \!\pm\! 14.7\%_\varphi $
\\
NLO & $36.94 \!\pm\! 19.7\%$ & $31.61 \!\pm\! 16.9\%$ & $47.40 \!\pm\! 14.1\%_\mu \!\pm\! 10.5\%_\varphi$ & $42.18 \!\pm\! 12.4\%_\mu \!\pm\! 6.8\%_\varphi $
\\
NNLO & $46.55 \!\pm\! 9.1\%$ & $42.42 \!\pm\! 9.7\%$ & $48.76 \!\pm\! 2.8\%_\mu \!\pm\! 1.5\%_\varphi$ & $47.41 \!\pm\! 4.6\%_\mu \!\pm\! 2.0\%_\varphi$
\\
N$^3$LO & $48.03 \!\pm\! 3.2\%$ & $46.51 \!\pm\! 4.9\%$ & $47.85 \!\pm\! 0.6\%_\mu \!\pm\! 0.8 \%_\varphi$ & $47.96 \!\pm\!  1.5\%_\mu \!\pm\! 0.5\%_\varphi $
\\
\bottomrule
\end{tabular}
\end{table}

The numerical results we present are obtained in the top EFT limit rescaled with the exact LO $m_t$ dependence. Bottom quark and EW effects are not yet included. Their numerical effects go in opposite directions and largely cancel each other, so they should be included together.
Jets are defined with a jet radius of $R = 0.4$ and without any cut on the jet rapidity. We use the PDF4LHC15 PDFs with $\alpha_s(m_Z) = 0.118$. In Table~\ref{tab:incl} we compare the corresponding results for the inclusive cross section at different hard scales. We also give the corresponding N$^3$LO results, obtained utilizing the results of Ref.~\cite{Anastasiou:2015ema}.

In \refF{fig:STWZ0jet} we show our results for the resummed and matched $0$-jet cross section, comparing the different consistently matched resummation orders on the left. On the right, we compare our best prediction at NNLL$'+$NNLO with the pure fixed-order results at the corresponding scale with ST uncertainties at NNLO and N$^3$LO (obtained from the results of Refs.~\cite{Anastasiou:2015ema, Caola:2015wna}). The NNLL$'+$NNLO agrees well with the fixed N$^3$LO with ST uncertainties. (The same is also observed for NLL$'+$NLO compared to NNLO.) Since the typical values of $\pTcut \simeq 30\UGeV$ lie in the transition region, it is not unexpected that the resummed result is roughly predicting the next-higher fixed order. This also confirms that the ST method produces reasonably sized uncertainties here. Our resummed predictions still have the advantage of allowing one to separately estimate all types of yield and migration uncertainties, in contrast to the fixed-order predictions where some explicit assumptions on the correlations between cross sections are needed.
The corresponding results for the $\geq 1$-jet cross section are shown in \refF{fig:STWZ1jet}. Here the lowest NLL result is not very meaningful as it does not even contain the correct LO$_1$ result. In the right panel we see that the highest order resummed result agrees well with the corresponding fixed NNLO$_1$~\cite{Boughezal:2015dra, Boughezal:2015aha, Caola:2015wna} result albeit with larger uncertainties toward larger $\pTcut$.

\begin{table}
\caption{Predictions for the $0$-jet cross section at $\pTcut = 25\UGeV$ (top) and $\pTcut = 30\UGeV$ (bottom) corresponding to \refF{fig:STWZ0jet} with a breakdown of the perturbative uncertainties.}
\label{tab:STWZ0jet}
\centering
\begin{tabular}{l|l|ccc|c}
\toprule
& $\sigma_0(25\UGeV)/\mathrm{pb}$ & $\Delta_{\mu0}$ & $\Delta_{\varphi0}$ & $\Delta^{0/1}_\resum$ & total pert. unc.
\\ \midrule
NLL & $17.04 \!\pm\! 7.21$ & $18.7\%$ & $12.8\%$ & $35.7\%$ & $42.3\%$
\\
NLL$'+$NLO$_0$ & $22.29 \!\pm\! 3.43$ & $7.7\%$ & $5.1\%$ & $12.3\%$ & $15.4\%$
\\
NNLL$'+$NNLO$_0$ & $26.25 \!\pm\! 1.97$ & $4.7\%$ & $0.6\%$ & $5.8\%$ & $7.5\%$
\\ \bottomrule
& $\sigma_0(30\UGeV)/\mathrm{pb}$ & $\Delta_{\mu0}$ & $\Delta_{\varphi0}$ & $\Delta^{0/1}_\resum$ & total pert. unc.
\\ \midrule
NLL & $19.10 \!\pm\! 7.52$ & $17.4\%$ & $12.8\%$ & $32.9\%$ & $39.4\%$
\\
NLL$'+$NLO$_0$ & $25.59 \!\pm\! 3.72$ & $7.9\%$ & $5.3\%$ & $11.0\%$ & $14.5\%$
\\
NNLL$'+$NNLO$_0$ & $29.51 \!\pm\! 1.65$ & $3.8\%$ & $0.1\%$ & $4.1\%$ & $5.6\%$
\\ \bottomrule
\end{tabular}
\end{table}

\begin{table}
\caption{Predictions for the $\geq 1$-jet cross section at $\pTcut = 25\UGeV$ (top) and $\pTcut = 30\UGeV$ (bottom) corresponding to \refF{fig:STWZ1jet} with a breakdown of the perturbative uncertainties.}
\label{tab:STWZ1jet}
\centering
\begin{tabular}{l|l|ccc|c}
\toprule
& $\sigma_{\geq 1}(25\UGeV)/\mathrm{pb}$ & $\Delta_{\mu\geq 1}$ & $\Delta_{\varphi\geq1}$ & $\Delta^{0/1}_\resum$ & total pert. unc.
\\ \midrule
NLL$'+$NLO$_0$ & $20.69 \!\pm\! 4.88$ & $17.4\%$ & $8.7\%$ & $13.3\%$ & $23.6\%$
\\
NNLL$'+$NNLO$_0$ & $21.16 \!\pm\! 1.96$ & $4.5\%$ & $3.8\%$ & $7.1\%$ & $9.3\%$
\\ \bottomrule
& $\sigma_{\geq 1}(30\UGeV)/\mathrm{pb}$ & $\Delta_{\mu\geq 1}$ & $\Delta_{\varphi\geq1}$ & $\Delta^{0/1}_\resum$ & total pert. unc.
\\ \midrule
NLL$'+$NLO$_0$ & $17.39 \!\pm\! 4.63$ & $19.1\%$ & $9.1\%$ & $16.2\%$ & $26.6\%$
\\
NNLL$'+$NNLO$_0$ & $17.90 \!\pm\! 1.88$ & $6.0\%$ & $5.2\%$ & $6.8\%$ & $10.5\%$
\\ \bottomrule
\end{tabular}
\end{table}

At this point we should note that matching to the N$^3$LO$_0$ and NNLO$_1$ results without also going to the corresponding N$^3$LL$^{(\prime)}$ resummation order amounts to including unresummed singular logarithms in the nonsingular matching corrections. Doing so would reduce the overall scale dependence but not necessarily improve the accuracy of the final prediction, because in the resummation region there is no guarantee that this would move the result in the right direction. (Thrust in $e^+e^-$ is a known example where it would not.) It would also come at the cost of reentangling the uncertainties sources, since on the one hand it should not reduce the resummation uncertainties while on the other hand it would impact the dominant $\pTcut$ dependence. Given these potential subtleties and given that our results with a complex hard scale already agree very well with the N$^3$LO$_0$ and NNLO$_1$ results, we do not see sufficient reason to add these terms until the corresponding resummation order is available as well.

The numerical results for $\sigma_0(\pTcut)$ and $\sigma_{\geq 1}(\pTcut)$ for $\pTcut = 25\UGeV$ and $\pTcut = 30\UGeV$ with a full breakdown of the uncertainties are given in Tables~\ref{tab:STWZ0jet} and \ref{tab:STWZ1jet}.
For $\pTcut = 30\UGeV$, $\Delta_\mu$ and $\Delta^{0/1}_\resum$ contribute roughly equally at each order, while $\Delta_\varphi$ is subdominant, except at the highest order in $\sigma_{\geq 1}$ where it contributes almost equally.
For $\pTcut = 25\UGeV$, the picture is roughly similar, except that as expected the uncertainties in $\sigma_0$ are somewhat increased compared to $\pTcut = 30\UGeV$. Note that for $\sigma_{\geq1}$ the total uncertainty actually slightly decreases from $\pTcut = 30\UGeV$ to $\pTcut = 25\UGeV$. The reason is that the cross section increases and the logarithms are resummed which results in the relative size of the remaining overall yield uncertainties to decrease.

\subsubsection{Combining resummed predictions for \texorpdfstring{$0/1/2$}{0/1/2}-jet bins \SectionAuthor{R.~Boughezal, X.~Liu, F.~Petriello, F.~J. Tackmann}}

Here, we discuss resummation-improved predictions for $0/1/2$-jet bins. We provide updated numerical results for 13 TeV and PDF4LHC15 PDFs and include the full breakdown of the uncertainties in terms of the parameterization of Eq.~\eqref{eq:3binnuisances}. We follow Ref.~\cite{Boughezal:2013oha} for combining the $0$-jet resummation of Ref.~\cite{Stewart:2013faa}, discussed above, with the $1$-jet resummation of Refs.~\cite{Liu:2012sz, Liu:2013hba}, which was documented in Ref.~\cite{Heinemeyer:2013tqa}, and for the estimation of yield and migration uncertainties.

\begin{figure}[t!]
\centering
\includegraphics[width=0.4\textwidth]{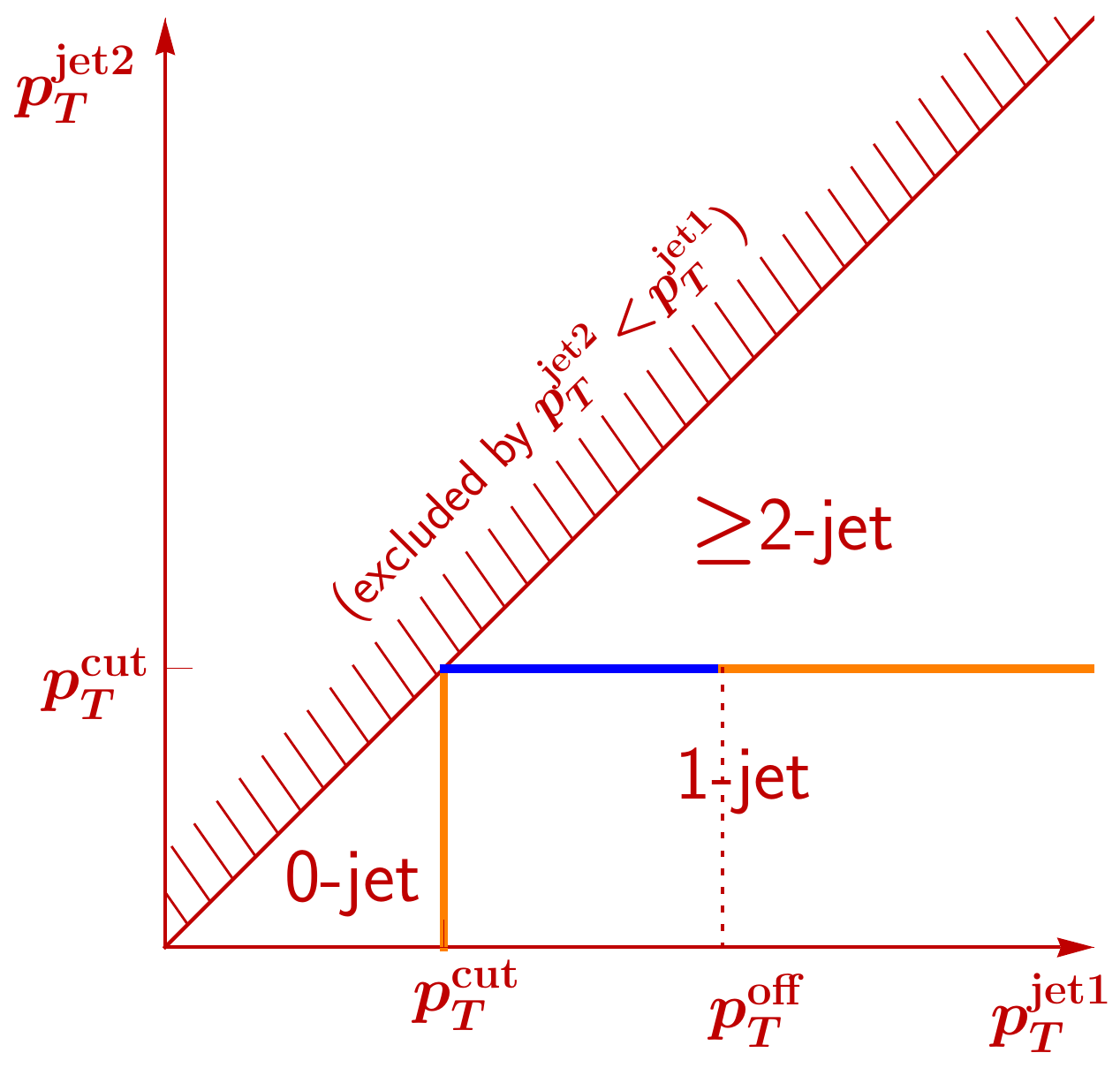}%
\caption{Illustration of the $\pTjetone$-$\pTjettwo$ plane relevant for the $0/1/2$-jet binning. The orange jet boundaries are treated in resummed perturbation theory, while the blue boundary is treated at fixed order. The dependence of the final results on the arbitrary $\pToff$ parameter is negligible compared to the perturbative uncertainties.}
\label{fig:012jetbins}
\end{figure}

We denote the $p_T$ of the leading and 2nd leading jet by $\pTjetone$ and $\pTjettwo$ and by definition $\pTjettwo<\pTjetone$. The different jet bins are illustrated in \refF{fig:012jetbins}. We define
\begin{align}
\sigma_{\geq 0} &: \text{ the inclusive cross section}
\,, \\
\sigma_0 (\pTcut) &: \text{ the 0-jet cross section, with } \pTjetone\! < \pTcut
\,, \nn \\
\sigma_{\ge 1} (\pTcut) &: \text{ the inclusive 1-jet cross section, with } \pTjetone\! \geq \pTcut
\,, \nn \\
\sigma_1 ([p_{Ta}, p_{Tb}]; \pTcut) &: \text{ the exclusive 1-jet cross section, with } p_{Ta} \leq \pTjetone\! < p_{Tb}
\,,\quad \pTjettwo\! < \pTcut
\,, \nn \\
\sigma_{\ge 2} ([p_{Ta}, p_{Tb}]; \pTcut) &: \text{ the inclusive 2-jet cross section, with } p_{Ta} \leq \pTjetone\! < p_{Tb}
\,,\quad \pTjettwo\! \geq \pTcut
\,. \nn
\end{align}

The exclusive 1-jet bin is theoretically quite nontrivial as it is affected by both the $0/1$-jet and $1/2$-jet boundaries. To construct a resummation-improved expression for it, we introduce a parameter $\pToff > \pTcut$ to separate the low and high $\pTjetone$ regions, shown by the dotted line in \refF{fig:012jetbins},
\begin{equation}
\sigma_1(\pTcut) \equiv \sigma_1 ([\pTcut, \infty]; \pTcut) = \sigma_1 ([\pTcut, \pToff]; \pTcut) + \sigma_1 ([\pToff, \infty]; \pTcut)
\,.\end{equation}
In practice, $\pToff$ is taken to be around $m_H/2$. The second term above contains logarithms of $\pTcut/Q$ with $Q \sim \pToff \sim m_H$.
These can be resummed to NLL$'+$NLO using the direct resummation of Refs.~\cite{Liu:2012sz, Liu:2013hba}, which is valid for $\pTjetone \sim m_H$ much larger than the $\pTcut$ on the 2nd jet.
The first term above contains the $0/1$ jet boundary. To improve it by the corresponding resummation, we can write it as
\begin{equation}
\label{eq:sigma1lowpTJ}
\sigma_1 ([\pTcut, \pToff]; \pTcut) = \bigl[ \sigma_0 (\pToff) - \sigma_0 (\pTcut) \bigr] - \sigma_{\ge 2} ([\pTcut, \pToff], \pTcut)
\,,\end{equation}
where the first term in brackets is equivalent to $\sigma_0 (\pToff) - \sigma_0 (\pTcut) = \sigma_{\ge 1} (\pTcut) - \sigma_{\ge 1} (\pToff)$.
We can then use the NNLL$'+$NNLO resummation of Ref.~\cite{Stewart:2013faa} for resumming the logarithms of $\pTcut / m_H$ associated with the $0/1$-jet boundary. The additional 2-jet terms in Eq.~\eqref{eq:sigma1lowpTJ} are calculated at fixed NLO$_2$. The validity of this indirect resummation approach is discussed and studied in Ref.~\cite{Boughezal:2013oha}. It essentially relies on the fact that in the region of interest, the $0$-jet terms dominate while the 2-jet terms are numerically small.

The combination of the indirect resummation for $\pTjetone < \pToff$ and the direct resummation for $\pTjetone \geq \pToff$ thus allows for a resummation-improved description of the complete 1-jet bin, i.e.,
\begin{equation}
\label{eq:1jetmaster}
\sigma_1 ([\pTcut, \infty]; \pTcut) = \sigma_1^{\rm indirect} ([\pTcut, \pToff]; \pTcut) + \sigma_1^{\rm direct} ([\pToff, \infty]; \pTcut) \,.\end{equation}
An important consistency check of this method is that the final result should be largely independent of the precise choice of $\pToff$. More precisely, the residual dependence on $\pToff$ should be much smaller than the estimated perturbative uncertainties. As shown in Ref.~\cite{Boughezal:2013oha}, by using a complex scale hard scale in the $0$-jet resummation as well as including the $H+1$ jet NNLO$_1$ virtual corrections in the direct resummation, the results become practically independent of $\pToff$.

To estimate the perturbative uncertainties, we use profile scale variations in both the $0$-jet and $1$-jet resummations, using the same physical interpretations to identify the different uncertainties sources.
The first is the overall fixed-order uncertainty, $\Delta_{\mu i}$, as discussed above Eq.~\eqref{eq:Deltamui}, which is treated as a yield uncertainty. It is estimated by collectively varying all scales that appear in the resummed predictions and reduces to the usual fixed-order scale variations in the respective limits of large $\pTcut$. The uncertainties for $\{\sigma_{\geq 0}, \sigma_0, \sigma_{\geq 1}, \sigma_1, \sigma_{\geq 2}\}$ are
\begin{equation}
\kappa_\mu^{\rm y}: \quad \{ \Delta_{\mu \geq 0},\, \Delta_{\mu 0},\, \Delta_{\mu \geq1},\, \Delta_{\mu 1},\, \Delta_{\mu \geq 2} \}
\,,\end{equation}
where the first three are as in Eqs.~\eqref{eq:Deltamui} and \eqref{eq:01nuisance} and
\begin{align} \label{eq:Delta1mu}
\Delta_{\mu 1} &\equiv \Delta_{\mu 1}([\pTcut, \infty]; \pTcut)
= \Delta_{\mu 0}(\pToff) - \Delta_{\mu 0}(\pTcut) + \Delta_{\mu1}([\pToff, \infty]; \pTcut)
\,,\nn\\
\Delta_{\mu \geq 2} &\equiv \Delta_{\mu \geq0} - \Delta_{\mu0}(\pTcut) - \Delta_{\mu 1}([\pTcut, \infty]; \pTcut)
\,.\end{align}
Since the yield uncertainties are fully correlated, the yield uncertainty in the 1-jet bin is the linear sum from the two regions. For the region below $\pToff$ we use $\Delta_{\mu0}(\pToff) - \Delta_{0\mu}(\pTcut)$ from the $0$-jet resummation in Eq.~\eqref{eq:Deltamui}. For the region above $\pToff$, $\Delta_{\mu1} ([\pToff, \infty]; \pTcut)$ is determined by the overall profile scale variations in the $1$-jet resummation~\cite{Liu:2013hba}. The result for $\Delta_{\mu \geq 2}$ then follows from consistency and e.g. ensures that $\Delta_{\mu\geq1} = \Delta_{\mu 1} + \Delta_{\mu \geq2}$.

Next, the resummation uncertainties induced by the $0/1$-jet and $1/2$-jet boundaries are used to estimate the respective migration uncertainties. Parameterizing the migration uncertainties by two independent nuisance parameters as in Eq.~\eqref{eq:3binnuisances}, the uncertainties for $\{\sigma_{\geq 0}, \sigma_0, \sigma_{\geq 1}, \sigma_1, \sigma_{\geq 2}\}$ are
\begin{align}
\kappa_\cut^{0/1}: &\quad \Delta^{0/1}_\cut \times \{ 0,\, 1,\, -1,\, -(1 - x_1),\, -x_1 \}
\,,\nn\\
\kappa_\cut^{1/2}: &\quad \Delta^{1/2}_\cut \times \{ 0,\, 0,\, 0,\, 1,\, -1 \}
\,,\end{align}
where we set $x_2 = 0$ and
\begin{align}
\Delta^{0/1}_{\cut} &= \Delta^{0/1}_\resum(\pTcut)
\,,\qquad
x_1 = \Delta^{0/1}_\resum (\pToff) / \Delta^{0/1}_\resum (\pTcut)
\,,\nn\\
\Delta^{1/2}_\cut
&= \Bigl\{ \bigl[\Delta^{\rm FO}_{\ge 2} ([\pTcut, \pToff], \pTcut) \bigr]^2 + \bigl[ \Delta^{1/2}_\resum ([\pToff, \infty]; \pTcut) \bigr]^2
\Bigr\}^{1/2}
\,.\end{align}
Here, $\Delta^{0/1}_\resum$ is the $0$-jet resummation uncertainty from Eq.~\eqref{eq:Delta01resum} which accounts for the $0/1$-jet boundary (vertical orange line in \refF{fig:012jetbins}).
The nonzero value for $x_1$ arises from the region below $\pToff$, which is computed from the difference of the 0-jet cross sections at $\pTcut$ and $\pToff$ as shown in Eq.~\eqref{eq:sigma1lowpTJ}. Hence, the contribution of the $0$-jet resummation uncertainty to $\sigma_1$ is $\Delta^{0/1}_\resum (\pToff) - \Delta^{0/1}_\resum(\pTcut)$ (where the two are treated as correlated). Comparing this to the parameterization above, this should be equal to $-\Delta^{0/1}_\cut(1 - x_1)$, which determines $x_1$.

The $1/2$-jet boundary above $\pToff$ (the horizontal orange line in \refF{fig:012jetbins}) is accounted for by the $1$-jet resummation uncertainty $\Delta_\resum^{1/2}([\pToff, \infty]; \pTcut)$, which is determined by resummation profile scale variations in Ref.~\cite{Liu:2013hba}. For the $1/2$-jet boundary below $\pToff$ (blue line in \refF{fig:012jetbins}), we follow the original ST method and use the inclusive 2-jet uncertainty $\Delta^{\rm FO}_{\ge 2}$, given by the usual scale variation of the 2-jet fixed-order contributions in Eq.~\eqref{eq:sigma1lowpTJ}. The two are considered independent and added in quadrature.

\begin{table}
\caption{Predictions for the $0/1/2$-jet bins for $\pTcut = 25\UGeV$ (top) and $\pTcut = 30\UGeV$ (bottom).}
\label{tab:BLPTW12jet}
\centering
\begin{tabular}{l||c|cccc|c}
\toprule
$\pTcut = 25\UGeV$ & $\sigma/\mathrm{pb}$ & $\Delta_\mu$ & $\Delta_\varphi$ & $\Delta^{0/1}_\cut$ & $\Delta^{1/2}_\cut$ & total pert. unc.
\\ \midrule
$\sigma_{\geq 0}$ &  $47.41 \!\pm\! 2.40 $ &  $4.6\%$ & $2.0\%$ & - & - & $5.1\%$
\\
$\sigma_0$ & $26.25 \!\pm\! 1.97$ & $4.7\%$ & $0.6\%$ & $5.8\%$ & - & $7.5\%$
\\
$\sigma_{\geq 1}$ & $21.16 \!\pm\! 1.96$ & $4.5\%$ & $3.8\%$ & $7.1\%$ & - & $9.3\%$
\\
$\sigma_{1}$ & $13.28 \!\pm\! 1.76$ & $4.2\%$ & $3.3\%$ & $9.8\%$ & $7.2\%$ & $13.3\%$
\\
$\sigma_{\geq 2}$ & $\,\,\,7.88 \!\pm\! 1.12$ & $5.1\%$ & $4.6\%$ & $2.7\%$ & $12.2\%$ & $14.3\%$
\\ \bottomrule
$\pTcut = 30\UGeV$ & $\sigma/\mathrm{pb}$ & $\Delta_\mu$ & $\Delta_\varphi$ & $\Delta^{0/1}_\cut$ & $\Delta^{1/2}_\cut$ & total pert. unc.
\\ \midrule
$\sigma_{\geq 0}$ &  $47.41 \!\pm\! 2.40 $ &  $4.6\%$ & $2.0\%$ & - & - & $5.1\%$
\\
$\sigma_0$ & $29.51 \!\pm\! 1.65$ & $3.8\%$ & $0.1\%$ & $4.1\%$ & - & $5.6\%$
\\
$\sigma_{\geq 1}$ & $17.90 \!\pm\! 1.88$ & $6.0\%$ & $5.2\%$ & $6.8\%$ & - & $10.5\%$
\\
$\sigma_{1}$ & $11.94 \!\pm\! 1.58$ & $5.5\%$ & $4.8\%$ & $8.4\%$ & $7.2\%$ & $13.2\%$
\\
$\sigma_{\geq 2}$ & $\,\,\,5.96 \!\pm\! 1.05$ & $7.1\%$ & $6.1\%$ & $3.6\%$ & $14.5\%$ & $17.6\%$
\\ \bottomrule
\end{tabular}
\end{table}

For our numerical results we use the same input parameters as for the $0$-jet resummation in the previous subsubsection. The results are presented in Table~\ref{tab:BLPTW12jet} for $\pTcut = 25\UGeV$ and $\pTcut = 30\UGeV$ with a complete breakdown of all uncertainty contributions. The results present a very consistent picture. The underlying parameters entering the migration uncertainties are given by $\Delta^{0/1}_\cut(25\UGeV) = 1.51\,\mathrm{pb}$, $x_1(25\UGeV) = 0.141$, $\Delta^{1/2}_\cut(25\UGeV) = 0.96\,\mathrm{pb}$, and
$\Delta^{0/1}_\cut(30\UGeV) = 1.21\,\mathrm{pb}$, $x_1(30\UGeV) = 0.175$, $\Delta^{1/2}_\cut(30\UGeV) = 0.86\,\mathrm{pb}$. As one might expect, $x_1$ is small and most of the $0/1$-migration uncertainty enters in $\sigma_1$. We also see that $\sigma_1$ is dominated by the migration uncertainties with similar contributions from each boundary, while the uncertainty in $\sigma_{\geq 2}$ is dominated by the $1/2$ boundary. However, due to the resummation improvement the total uncertainties in all bins are substantially smaller than in the pure fixed-order predictions.

\subsection[Jet-vetoed Higgs cross section in gluon fusion at N\texorpdfstring{$^3$}{3}LO+NNLL]{Jet-vetoed Higgs cross section in gluon fusion at N\texorpdfstring{$^3$}{3}LO+NNLL with small-\texorpdfstring{$R$}{R} resummation\SectionAuthor{A.~Banfi, F.~Caola, F.A.~Dreyer, P.F.~Monni, G.P.~Salam, G.~Zanderighi, F.~Dulat}}
\label{sec:jve}




\providecommand{\LLR}{\text{LL$_R$}\xspace}
\providecommand{\NNLLNNLO}{\text{NNLO+NNLL}\xspace}
\providecommand{\NNLLNNNLO}{N$^3$LO+NNLL\xspace}
\providecommand{\NNNLO}{\text{N$^3$LO}\xspace}
\providecommand{\NNNLL}{\text{N$^3$LL}\xspace}
In some Higgs boson decay modes (most notably $WW^*$ and $\tau\tau$),
it is standard to perform different analyses depending on the number
of accompanying jets.
This is because different jet multiplicities have different dominant
backgrounds.
Of particular importance for the $WW$ decay is the zero-jet case, where
the dominant top-antitop background is dramatically reduced.
For precision studies it is important to predict accurately the
fraction of signal events that survive the zero-jet constraint, and to
assess the associated theory uncertainty.
Jet-veto transverse momentum thresholds used by ATLAS and CMS are
relatively soft ($\ptjv\sim 25-30$ GeV), hence QCD real radiation is
constrained by the cut and the imbalance between virtual and real
corrections results in logarithms of the form $\ln(\ptjv/m_H)$ that
should be resummed to all orders in the coupling constant.
This resummation has been carried out to next-to-next-to-leading
logarithmic accuracy (\text{NNLL}, i.e.\ including all
terms $\as^n \ln^k (\ptjv/m_H)$ with $k \ge n-1$ in the logarithm of the
cross section) and matched to next-to-next-to-leading order (\text{NNLO}) in
refs.~\cite{Banfi:2012jm,Stewart:2013faa,Becher:2013xia}
(some of the calculations also included partial \NNNLL{} contributions).
At this order one finds that the effect of the resummation is to shift
central predictions only moderately, and to reduce somewhat the
theoretical uncertainties.
Yet, the residual theoretical uncertainty remains sizeable, roughly
$10\%$~\cite{Banfi:2012jm}, and the impact of higher-order effects
could therefore be significant.

Since the first \NNLLNNLO{} predictions for the jet-veto, three
important theoretical advances happened: firstly, the \NNNLO{}
calculation of the total gluon-fusion cross
section~\cite{Anastasiou:2015ema}; secondly the calculation of the
\text{NNLO} corrections to the Higgs plus one-jet
cross-section~\cite{Boughezal:2015dra,Boughezal:2015aha,Caola:2015wna};
and finally the LL resummation of logarithms of the jet-radius
$R$~\cite{Dasgupta:2014yra}.

All these recent results were merged together in
ref.~\cite{Banfi:2015pju} by extending the matching of the jet-veto
cross-section to N$^3$LO+NNLL+\LLR. The effect of heavy-quark masses
has been considered following the approach of
ref.~\cite{Banfi:2013eda}. The code used to produce the following
results can be downloaded from~\cite{JetVHeto}.

\subsubsection{\texorpdfstring{N$^3$LO+NNLL+LL$_{R}$}{N3LO+NNLL+LL_R} cross section and 0-jet efficiency at 13 TeV}
\label{sec:all}

In this section we report predictions for the jet-veto efficiency and
cross section in Higgs boson production in gluon fusion at the
LHC. The results are based on the calculation obtained in
ref.~\cite{Banfi:2015pju}. The reader should refer to the latter work
for additional information.
The various ingredients that we use are summarized below:
\begin{itemize}
\item The total N$^3$LO cross section for Higgs boson production in gluon
  fusion~\cite{Anastasiou:2015ema}, obtained in the heavy-top
  limit. The Wilson coefficient is expanded out consistently both in
  the computation of the total and the inclusive one-jet cross
  section.
\item The inclusive one-jet cross section at NNLO taken from the code
  of ref.~\cite{Caola:2015wna}, in the heavy-top limit.
  In this computation the $qq$ channel is included only up to NLO, and
  missing NNLO effects are estimated to be well below scale variation
  uncertainties~\cite{Boughezal:2015aha}.
\item Exact top- and bottom-mass effects up to NLO in the jet-veto
  efficiency and cross section~\cite{Spira:1995rr}. Beyond NLO, we use
  the heavy-top result, without any modifications.
\item Large logarithms $\ln(m_H/\ptjv)$ resummed to NNLL accuracy
  following the procedure of~\cite{Banfi:2012jm}, with the treatment of
  quark-mass effects as described in ref.~\cite{Banfi:2013eda}.
\item Logarithms of the jet radius resummed to LL accuracy,
  following the approach of ref.~\cite{Dasgupta:2014yra}.
\end{itemize}

We consider 13 TeV LHC collisions with a Higgs boson mass of
$m_H = 125$ GeV.
For the top and bottom pole quark masses, we use $m_t\,=\,172.5$ GeV and
$m_b\,=\,4.75$~GeV.
Jets are defined using the anti-$k_t$
algorithm~\cite{Cacciari:2008gp}, as implemented in \texttt{FastJet
  v3.1.2}~\cite{Cacciari:2011ma}, with radius parameter $R=0.4$, and
perform the momentum recombination in the standard $E$ scheme (i.e.\
summing the four-momenta of the pseudo-particles).
We use PDF4LHC15 parton distribution functions at NNLO with
$\alpha_s(m_Z) = 0.118$
(\texttt{PDF4LHC15\_nnlo\_mc})~\cite{Butterworth:2015oua}.
The impact of higher-order logarithmic corrections is probed by
introducing a resummation scale $Q$ as shown in ref.~\cite{Banfi:2012jm}.
The central prediction for the jet-veto efficiency is obtained by
using the matching scheme $(a)$~\cite{Banfi:2015pju}, setting the
renormalization and factorization scales to $\mu_R=\mu_F = m_H/2$, and
the resummation scale relative to both top and bottom contributions to
$Q=Q_0=m_H/2$.
To determine the perturbative uncertainties for the jet-veto
efficiency we follow the Jet-Veto efficiency (JVE) method as outlined
in ref.~\cite{Banfi:2015pju}. This differs from the original method of
refs.~\cite{Banfi:2012yh,Banfi:2012jm} which has been modified to take
into account the excellent convergence observed at the perturbative
order considered here ~\cite{Banfi:2015pju}. The uncertainty band is
determined as described below.

We vary $\mu_R$, $\mu_F$ by a factor of 2 in either direction,
requiring $1/2 \le \mu_R/\mu_F \le 2$.
Maintaining central $\mu_{R,F}$ values, we also vary $Q$ in the
range $\frac{2}{3}\le Q/Q_0\le\frac{3}{2}$.
As far as the small-$R$ effects are concerned, subleading logarithmic
effects are estimated by means of a second resummation scale $R_0$,
which acts as the initial radius for the evolution of the gluon-jet
fragmentation which gives rise to the $\ln R$ terms.  We choose the
default value $R_0=1.0$,\footnote{The initial radius for the small-$R$
  evolution differs from the jet radius used in the definition of
  jets, which is $R=0.4$.} and vary it conservatively by a factor of
two in either direction.
Finally, keeping all scales at their respective central values, we
replace the default matching scheme $(a)$ with scheme $(b)$, as
defined in ref.~\cite{Banfi:2015pju}.
The final uncertainty band is obtained as the envelope of all the
above variations.
We do not consider here the uncertainties associated with the parton
distributions (which mostly affect the cross section, but not the
jet-veto efficiency), the value of the strong coupling or the impact
of finite quark masses on terms beyond NLO. Moreover, our results do
not include electro-weak effects.

\begin{figure}
  \centering
  \includegraphics[width=0.49\textwidth]{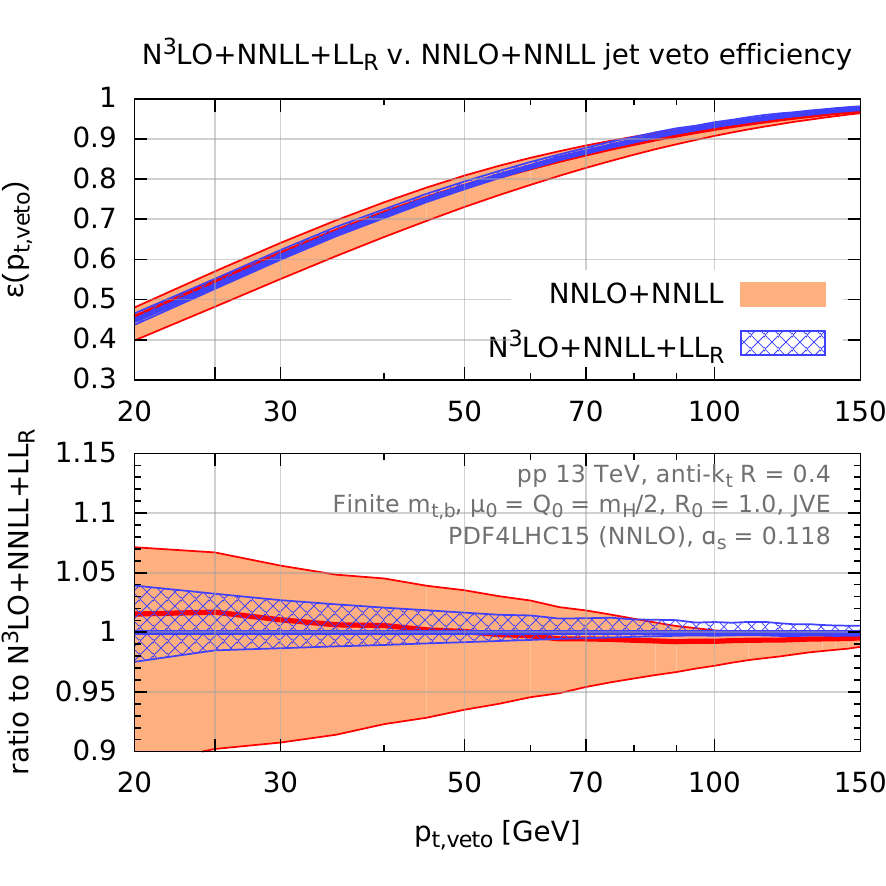}\hfill
  \includegraphics[width=0.49\textwidth]{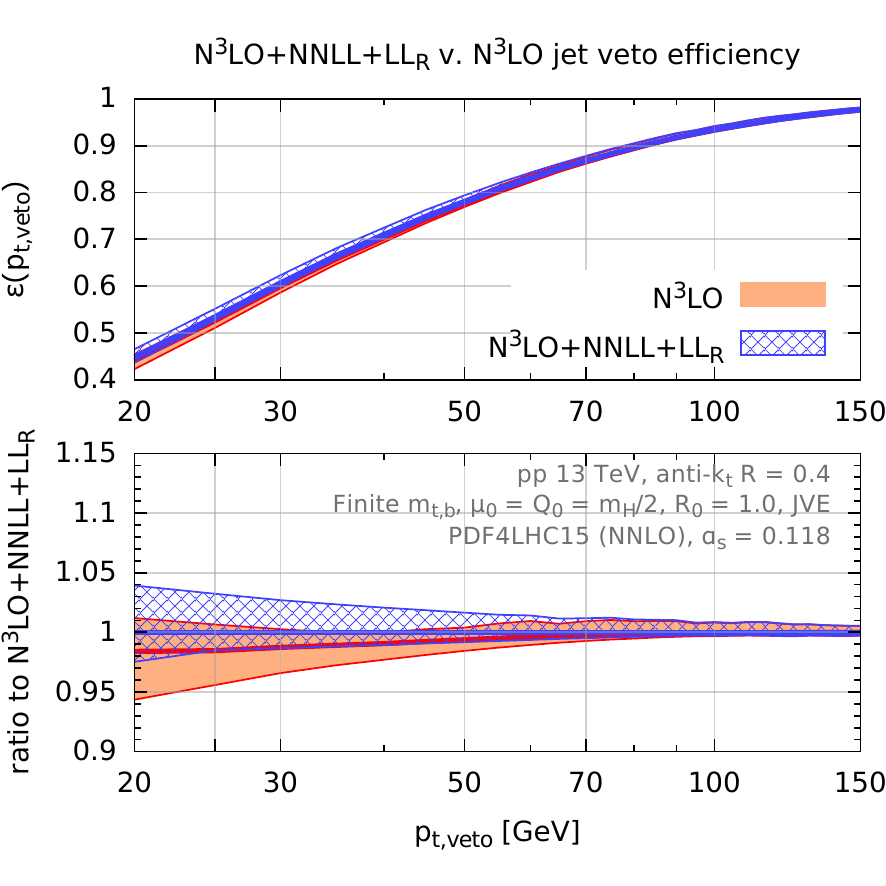}
  \caption{N$^3$LO+NNLL+LL$_R$ best prediction for the jet-veto efficiency (blue/hatched)
    compared to NNLO+NNLL (left) and fixed-order at N$^3$LO
    (right).
  }
  \label{fig:bestprediction-efficiency}
\end{figure}
In the left plot of Figure~\ref{fig:bestprediction-efficiency} we show
the comparison between our best prediction for the jet-veto efficiency
(N$^3$LO+NNLL+LL$_R$) and the previous NNLO+NNLL accurate prediction,
both including mass effects.
 We see that the impact of the N$^3$LO correction on the central value
 is in the range 1-2\% at relevant jet-veto scales. The uncertainty
 band is significantly reduced when the N$^3$LO corrections are
 included, going from about 10\% at NNLO down to a few per cent at
 N$^3$LO.
Figure~\ref{fig:bestprediction-efficiency} (right) shows the
comparison between the N$^3$LO+NNLL+LL$_R$ prediction and the pure
N$^3$LO result. We observe a shift of the central value of the order
of 2\% for $\ptjv > 25\UGeV$ when the resummation is turned on.
In that same $\ptjv$ region, the uncertainty associated with the
N$^3$LO prediction is at the 3\% level, comparable with that of the
N$^3$LO+NNLL+LL$_R$ prediction.
The fact that resummation effects are nearly of the same order as the
uncertainties of the fixed order calculation suggests that the latter
might be accidentally small.
This situation is peculiar to our central renormalization and
factorization scale choice, $\mu_R = \mu_F = m_H/2$, and does not
occur at, for instance, $\mu_R=\mu_F=m_H$ (see
ref.~\cite{Banfi:2015pju} for details).

\begin{figure}
  \centering
  \includegraphics[width=0.49\textwidth]{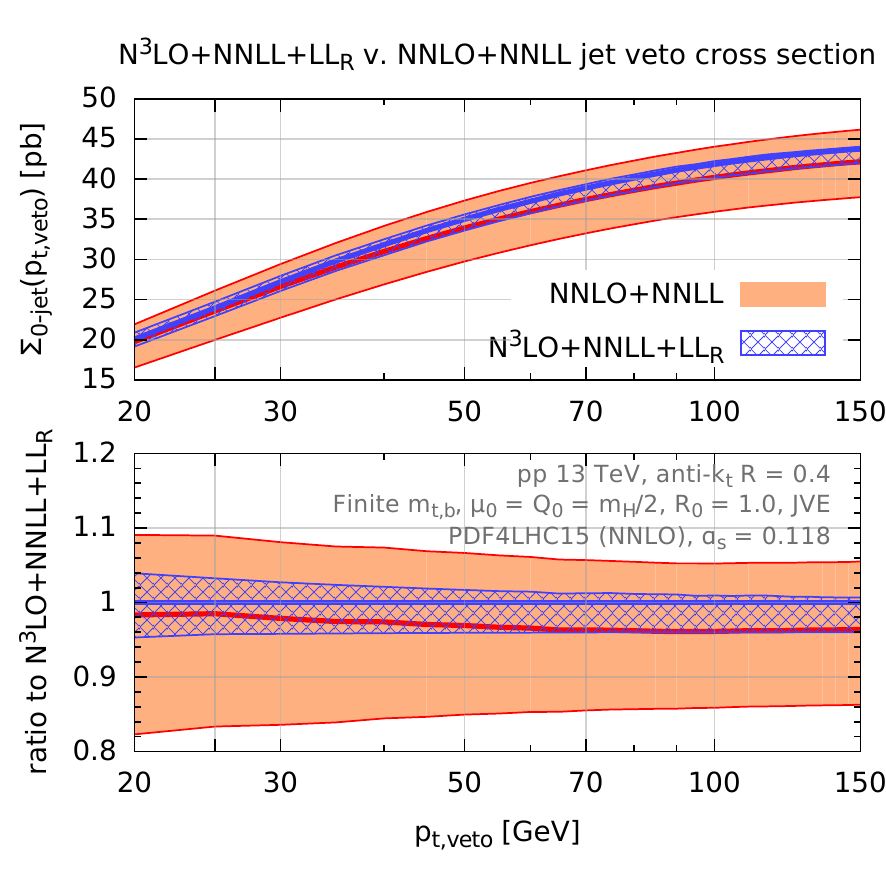}\hfill
  \includegraphics[width=0.49\textwidth]{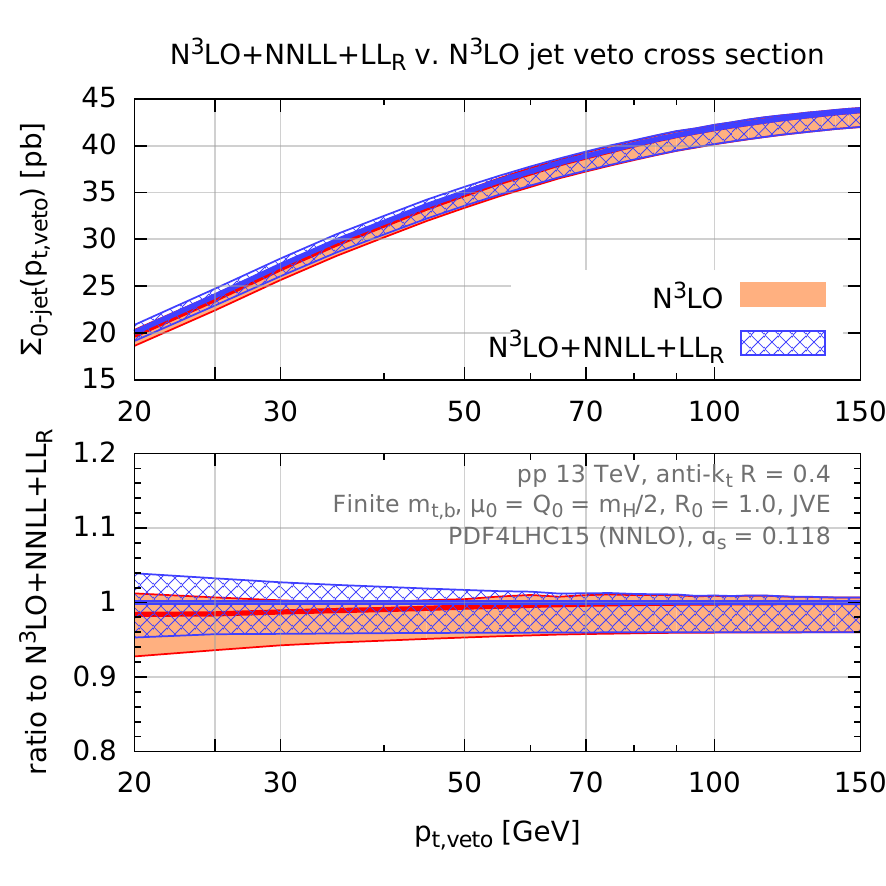}
  \caption{N$^3$LO+NNLL+LL$_R$ best prediction for the jet-veto cross section (blue/hatched)
    compared to NNLO+NNLL (left) and fixed-order at N$^3$LO
    (right).  }
  \label{fig:bestprediction-sigma}
\end{figure}

The zero-jet cross section is obtained as
$\Sigma_\text{0-jet}(\ptjv)=\sigma_{\rm tot}\,\epsilon(\ptjv)$, and
the inclusive one-jet cross section is obtained as
$\Sigma_{\ge\text{1-jet}}(p_{t,\rm{min}})=\sigma_{\rm
  tot}\,\left(1-\epsilon(p_{t,\rm{min}})\right)$.
The associated uncertainties are obtained by combining in quadrature
the uncertainty on the efficiency obtained as explained above and that
on the total cross section, for which we use plain scale
variations. The corresponding results are shown in
\refF{fig:bestprediction-sigma}.
For this scale choice, we observe that the effect of including
higher-order corrections in the zero-jet cross section is quite
moderate at relevant $\ptjv$ scales. This is because the small $K$
factor in the total cross section compensates for the suppression in
the jet-veto efficiency. The corresponding theoretical uncertainty is
reduced by more than a factor of two.

The predictions for jet-veto efficiency and the zero-jet cross section
are summarized in Table~\ref{tab:13-TeV-0jet}, for two experimentally
relevant $\ptjv$ choices.
Results are reported both at fixed-order, and including the various
resummation effects.

\begin{table}
 \caption{Predictions for the jet-veto efficiency and cross
   section at N$^3$LO+NNLL+LL$_{R}$, compared to the N$^3$LO and NNLO+NNLL cross
   sections.
   The uncertainty in the fixed-order prediction is obtained using the
   JVE method. All numbers include the effect of top and bottom quark
   masses, treated as described in the text, and are for a central scale $\mu_0=m_H/2$.}
 \label{tab:13-TeV-0jet}
\begin{center}
  \begin{tabular}{c|c|c|c|c}
    \toprule
    LHC 13 TeV
    & $\epsilon^{{\rm N^3LO+NNLL+LL_R}}$ & \,$\Sigma^{{\rm
                                N^3LO+NNLL+LL_R}}_\text{0-jet}\,{\rm [pb]}$\,  &
                                                          \,$\Sigma^{{\rm
                                                          N^3LO}}_\text{0-jet}$ &
                                                          \,$\Sigma^{{\rm
                                                          NNLO+NNLL}}_\text{0-jet}\,$
    \\[0.2em] \midrule
    $\ptjv=25\,{\rm GeV}$ & $0.534^{+0.017}_{-0.008}$ & $24.0^{+0.8}_{-1.0}$ & $23.6^{+0.5}_{-1.2}$ & $23.6^{+2.5}_{-3.6}$ \\
    $\ptjv=30\,{\rm GeV}$ & $0.607^{+0.016}_{-0.008}$ & $27.2^{+0.7}_{-1.1}$ & $26.9^{+0.4}_{-1.2}$ & $26.6^{+2.8}_{-3.9}$ \\
   \bottomrule
  \end{tabular}
\end{center}
\end{table}

\begin{figure}
  \centering
  \includegraphics[width=0.49\textwidth]{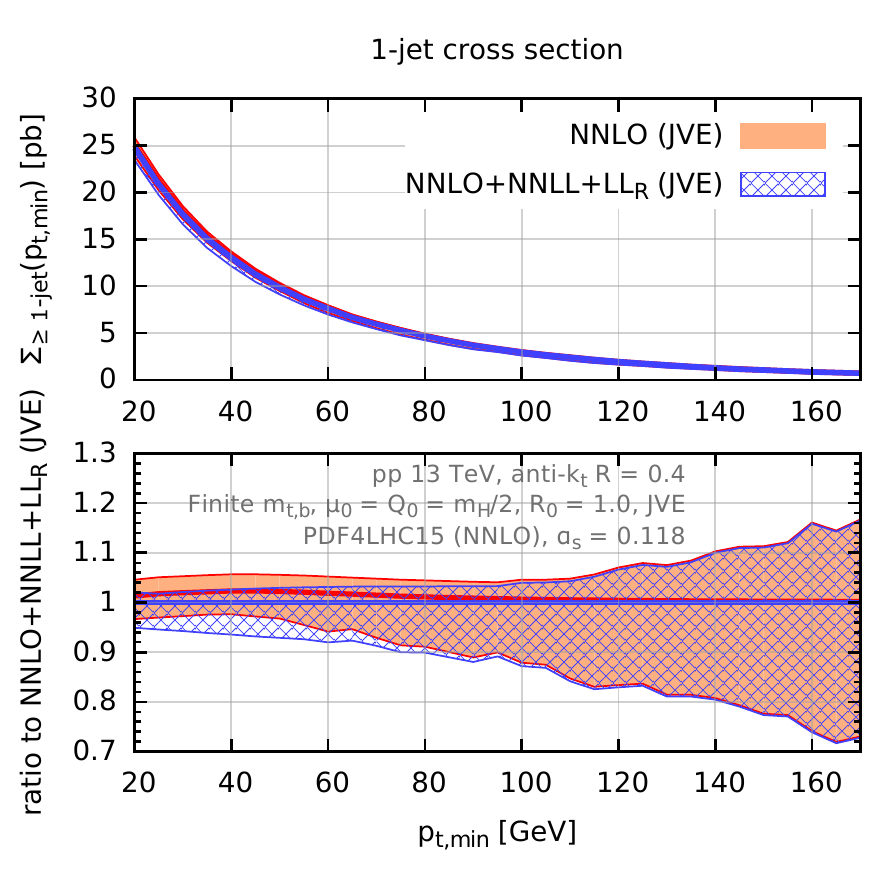}\hfill
  \includegraphics[width=0.49\textwidth]{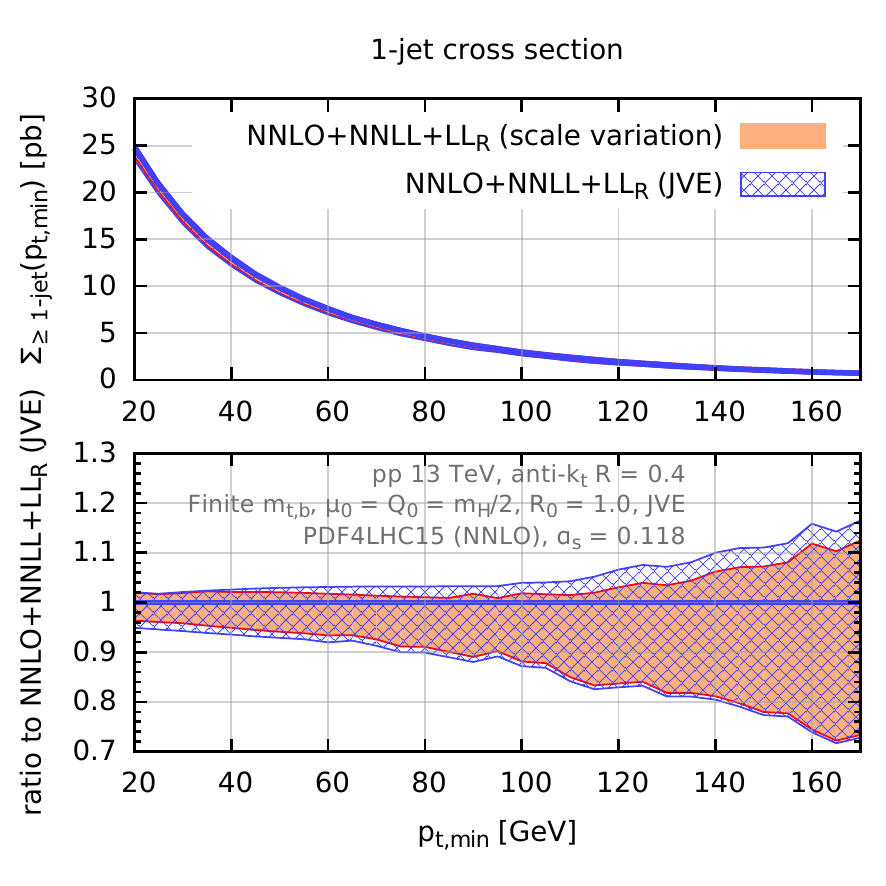}
  \caption{Matched NNLO+NNLL+\LLR prediction for the inclusive one-jet
    cross section (blue/hatched) compared to fixed-order at NNLO
    (left) and to the matched result with direct scale variation for
    the uncertainty (right), as explained in the text.}
  \label{fig:1jetxs}
\end{figure}

Figure~\ref{fig:1jetxs} shows the inclusive one-jet cross section
$\Sigma_{\ge \text{1-jet}}$, for which the state-of-the-art
fixed-order prediction is
NNLO~\cite{Boughezal:2015dra,Boughezal:2015aha,Caola:2015wna}. The
left-hand plot shows the comparison between the best prediction at
NNLO+NNLL+LL$_R$, and the fixed-order at NNLO. Both uncertainty bands
are obtained with the JVE method outlined in
ref.~\cite{Banfi:2015pju}. We observe that the effect of the
resummation on the central value at moderately small values of $\ptjv$
is at the per cent level. Moreover, the inclusion of the resummation
leads to a slight increase of the theory uncertainty in the small
transverse momentum region.

The right-hand plot of \refF{fig:1jetxs} shows our best prediction
for the one-jet cross section with uncertainty obtained with the JVE
method, compared to the case of just scale (i.e.\ $\mu_R$, $\mu_F$,
$Q$) variations. We observe a comparable uncertainty both at small and
at large transverse momentum, which indicates that the JVE method is
not overly conservative in the tail of the distribution. We have
observed that the same features persist for the corresponding
differential distribution.
Table~\ref{tab:13-TeV-1jet} contains the predictions for the
inclusive one-jet cross section for two characteristic $p_{t,\text{min}}$ choices.

\begin{table}
 \caption{Predictions for the inclusive one-jet cross section at
   NNLO+NNLL+LL$_{R}$ and NNLO. The uncertainty in the fixed-order
   prediction is obtained using the JVE method. All numbers include
   the effect of top and bottom quark masses, treated as described in
   the text, and are for a central scale $\mu_0=m_H/2$. }
 \label{tab:13-TeV-1jet}
  \begin{center}
      \begin{tabular}{c|c|c}
        \toprule
        LHC 13 TeV
        &\, $\Sigma^{{\rm
          NNLO+NNLL+LL_R}}_{\ge \text{1-jet}}\,{\rm [pb]}$\,  &
                                                               \,$\Sigma^{{\rm
                                                               NNLO}}_{\ge\text{
                                                               1-jet}}\,{\rm
                                                               [pb]}$
        \\[0.2em] \midrule
    $p_{\rm t, min}=25\,{\rm GeV}$ & $20.9^{+0.4}_{-1.1}$ & $21.2^{+0.7}_{-1.0}$\\
    $p_{\rm t,min}=30\,{\rm GeV}$ & $17.6^{+0.4}_{-1.0}$ & $17.9^{+0.6}_{-0.8}$ \\
   \bottomrule
  \end{tabular}
\end{center}
\end{table}

%

\subsection[Higgs-\texorpdfstring{$\pt$}{pT} resummation in momentum space at NNLL+NNLO in gluon fusion]{Higgs-\texorpdfstring{$\pt$}{pT} resummation in momentum space at NNLL+NNLO in gluon fusion\SectionAuthor{P.F.~Monni, E.~Re, P.~Torrielli}}
\label{subsec:MRT}
\newcommand{\asCMW}{\alpha_s^{\rm CMW}}
\newcommand{\pb}{\;\mathrm{pb}}
\newcommand{\nb}{\;\mathrm{nb}}
\newcommand{\TeV}{\;\mathrm{TeV}}
\newcommand{\mrtpt}{{p_t^{\tiny{\mbox{H}}}}}
\newcommand{\mur}{\mu_{\tiny{\mbox{R}}}}
\newcommand{\muf}{\mu_{\tiny{\mbox{F}}}}
\newcommand{\cF}{{\cal F}}
\newcommand{\dZ}{d{\cal Z}[\{\hat{R}'(\kto), k_i\}]}
\newcommand{\kto}{k_{t,1}}
\newcommand{\kti}{k_{t,i}}
\newcommand{\dki}{\langle dk_{i}\rangle}
\newcommand{\dko}{\langle dk_{1}\rangle}

In the gluon-fusion production mode, the Higgs boson transverse momentum $\mrtpt$
is defined as the inclusive vectorial sum of the transverse momenta of the
recoiling QCD partons radiated off the incoming gluons.
The fixed-order perturbative description of its differential
spectrum features large logarithms in the form
$\as^n \ln^m(\mh/\mrtpt)/\mrtpt$, with $m\leq 2n-1$, which spoil the
convergence of the series at small $\mrtpt$.
In order to obtain meaningful predictions in that phase-space region,
such terms must be resummed to all orders in $\as$, so that the
perturbative series can be recast in terms of dominant all-order
towers of logarithms. The logarithmic accuracy is commonly defined at
the level of the {\it logarithm} of the cumulative cross section, henceforth
referred to as $\Sigma(\mrtpt)$, where one refers to the dominant terms
$\alpha_s^n \ln^{n+1}(\mh/\mrtpt)$ as leading logarithms (LL), to terms
$\alpha_s^n \ln^{n}(\mh/\mrtpt)$ as next-to-leading logarithms (NLL), to
$\alpha_s^n \ln^{n-1}(\mh/\mrtpt)$ as next-to-next-to-leading logarithms
(NNLL), and so on.

The all-order computation of the logarithms of the ratio
$\mh/\mrtpt$ has been performed up to NNLL order in
refs.~\cite{Bozzi:2003jy,Bozzi:2005wk} using the formalism developed
in~\cite{Collins:1984kg,Catani:2000vq}, and in
ref.~\cite{Becher:2010tm} using an effective-field-theory approach.
These resummed results are usually matched to fixed-order predictions
in order to obtain a description of $\mrtpt$ which gives a reliable
coverage of the whole phase space.
The recent computations of the differential $\mrtpt$ distribution at next-to-next-to-leading
order (NNLO)~\cite{Boughezal:2015dra,Boughezal:2015aha,
Caola:2015wna,Chen:2016zka}, and of the inclusive gluon-fusion cross section
at next-to-next-to-next-to-leading order (N$^3$LO) in~\cite{Anastasiou:2015ema,
Anastasiou:2016cez}, once combined with state-of-the-art resummation, allow
to obtain a formal NNLL+NNLO accuracy for $d\sigma/d\mrtpt$.

All of the resummation approaches mentioned so far rely on an
impact-parameter-space formulation~\cite{Dokshitzer:1978yd,Parisi:1979se},
which is motivated by the fact that the observable naturally factorizes in this
space as a product of the contributions of each individual emission.
Conversely, in $\mrtpt$ space one is unable to find, at a given order
beyond LL, a closed analytic expression for the resummed distribution
which is simultaneously free of logarithmically subleading
corrections and of singularities at finite $\mrtpt$
values~\cite{Frixione:1998dw}.
This fact has a simple physical origin: the region of small $\mrtpt$
receives contributions both from configurations in which each of the
transverse momenta of the radiated partons is equally small (Sudakov
limit), and from configurations where $\mrtpt$ tends to zero owing to
cancellations among non-zero transverse momenta of the emissions. The
latter mechanism is in fact the dominant one at small $\mrtpt$ and, as a
result, the cumulative cross section in that region vanishes as
${\cal O}(\mrtpt^2)$ rather than being exponentially
suppressed~\cite{Parisi:1979se}.
If these effects are neglected in a resummation performed in
transverse-momentum space, the latter would feature a geometric
singularity at some finite value of $\mrtpt$. The same issue is present
in an impact-parameter-space formulation whenever one tries to obtain a
result in $\mrtpt$ space free of any contamination from subleading
logarithmic terms.

However, it has recently been shown~\cite{Monni:2016ktx} that the
problem can be solved also in transverse-momentum space, upon
extending the formalism developed in~\cite{Banfi:2004yd,Banfi:2014sua}
to treat observables that feature such kinematic cancellations.

The method of~\cite{Monni:2016ktx} organizes the computation of the
cumulative cross section $\Sigma(\mrtpt)$ as an ensemble of emissions off
the incoming gluons; the amplitude for each emission, characterized by
a certain transverse momentum $\kti$, is then expanded around the
largest transverse momentum in the ensemble (denoted as $\kto$), and
one just retains the terms in this expansion that contribute to a
given logarithmic accuracy, discarding subleading contributions.
The latter expansion is always justified since, by construction, all
emissions are softer than $\kto$ and, owing to the recursive infrared
and collinear (rIRC) safety of the observable, emissions with $\kti\muchless \kto$
that would invalidate the expansion do not contribute to the
logarithmic structure of the resummed cross
section~\cite{Banfi:2004yd}. The hardest emission is integrated over
all of its natural phase space, including the region $\kto\gg\mrtpt$,
which is regularized by means of the Sudakov exponential.

This formulation ensures that all kinematic effects that contribute to
the $\mrtpt\to0$ limit are properly taken into account, and not only is
$\Sigma(\mrtpt)$ free of singularities at finite $\mrtpt$, but it also
features the correct $\mrtpt^2$ scaling \cite{Parisi:1979se} at low
$\mrtpt$. This procedure can be effectively interpreted as a resummation
of the large logarithms $\ln(\mh/\kto)$, and the logarithmic order is
defined in terms of the latter; the corresponding formal accuracy in
terms of the logarithms $\ln(\mh/\mrtpt)$ is the same as in terms of
$\ln(\mh/\kto)$, and the difference amounts to
logarithmically-subleading corrections.

In~\cite{Monni:2016ktx} the cumulative cross section is efficiently
computed with a fast Monte Carlo method where, for each event, the
kinematics of the ensemble of emissions is stochastically generated
with weights that take into account all of the physical effects that
occur at a given logarithmic accuracy.  In particular, at NLL
accuracy, all emissions after the hardest can be generated with equal
weights obtained in the soft-collinear approximation; at NNLL,
one emission weight per event is modified to take into account the
corrections which arise when a single parton in the ensemble is either
emitted collinearly to the beam with a significant fraction of the emitter's
momentum (hard-collinear), or close in rapidity to another parton, and
therefore it is sensitive to its correct rapidity bounds.
The Sudakov-exponential weight associated with the hardest
emission is correspondingly evaluated including terms up to
$\mathcal O(\as^n\ln^{n}(\mh/\kto))$ at NLL, and up to
$\mathcal O(\as^n\ln^{n-1}(\mh/\kto))$ at NNLL, and the overall parton
luminosity incorporates coefficient functions at the relevant order in
the strong coupling -- i.e.~${\cal O}(\as^3)$ and ${\cal O}(\as^4)$
for a NLL and NNLL matching, respectively -- and is evaluated at a
factorization scale of the order of $\kto$.

The expansion of the cumulative rate, necessary for the matching with
fixed-order predictions, is performed in a semi-analytic way, where the
contributions to the different orders in $\as$ are reduced
analytically to linear combinations of $\mrtpt$-dependent
master-integral grids that are evaluated numerically with high
accuracy once and for all, and interpolated dynamically at runtime,
ensuring optimal speed performances.

The computation is entirely carried out in momentum space, without the
need to transform the observable and the parton luminosities into
impact-parameter space, with benefits in terms of speed.
Moreover, in transverse-momentum space a clear physical picture emerges of the
effects that determine the low-$\mrtpt$ region, where all contributions to a given logarithmic
accuracy can be systematically tracked down and accounted for.

The approach used to perform the resummation is fully general and it
can be straightforwardly extended to the entire class of
rIRC-safe~\cite{Banfi:2004yd} observables that feature kinematic
cancellations in the infrared.

In the following we report predictions for the Higgs boson transverse
momentum distribution in the heavy-top effective theory at
the $13$\,TeV LHC, with $\mh=125$\,GeV, and {\tt PDF4LHC15}
\cite{Butterworth:2015oua} parton densities at NNLO. The central
prediction uses $\mur=\muf=\mh$, and $Q=\mh/2$, where $Q$ represents
the resummation scale, introduced as usual to estimate the impact on
physical results of the neglected higher-order logarithmic
contributions. The perturbative uncertainty for all predictions is
estimated by varying both $\mur$ and $\muf$ by a factor of two in
either direction while keeping $1/2\leq \mur/\muf \leq 2$.  Moreover,
for central $\mur$ and $\muf$ scales, the resummation scale $Q$ is
varied by a factor of two in either direction.
%
\begin{figure}
 \begin{minipage}[b]{8.0cm}
   \centering
   \includegraphics[width=8.0cm]{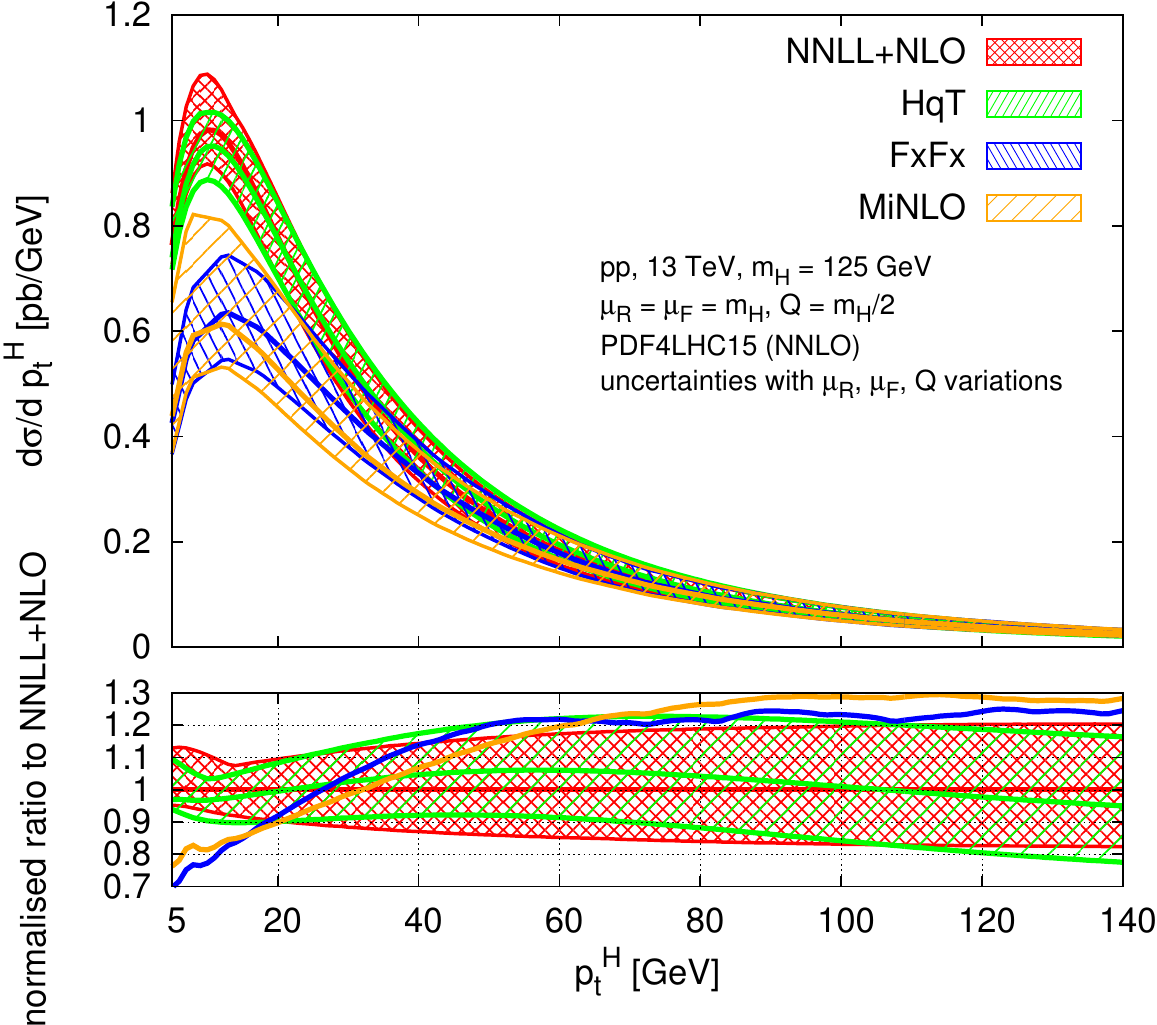}
   \caption{\footnotesize{Comparison of the Higgs $\mrtpt$ NNLL+NLO
      prediction as obtained in~\cite{Monni:2016ktx} (red) to \texttt{HqT}
      (green). For reference, the predictions obtained with MiNLO at
      NLO (orange), and FxFx (blue) are shown. Lower panel: ratio of
      the various distributions, normalized to their respective
      central-scale inclusive cross sections, to the central NNLL+NLO
      prediction~\cite{Monni:2016ktx}. Uncertainty bands are shown only for the
      resummed results.}}
\label{fig:results1}
 \end{minipage}
 \ \hspace{2mm} \hspace{0mm} \
 \begin{minipage}[b]{8.0cm}
  \centering
   \includegraphics[width=8.0cm]{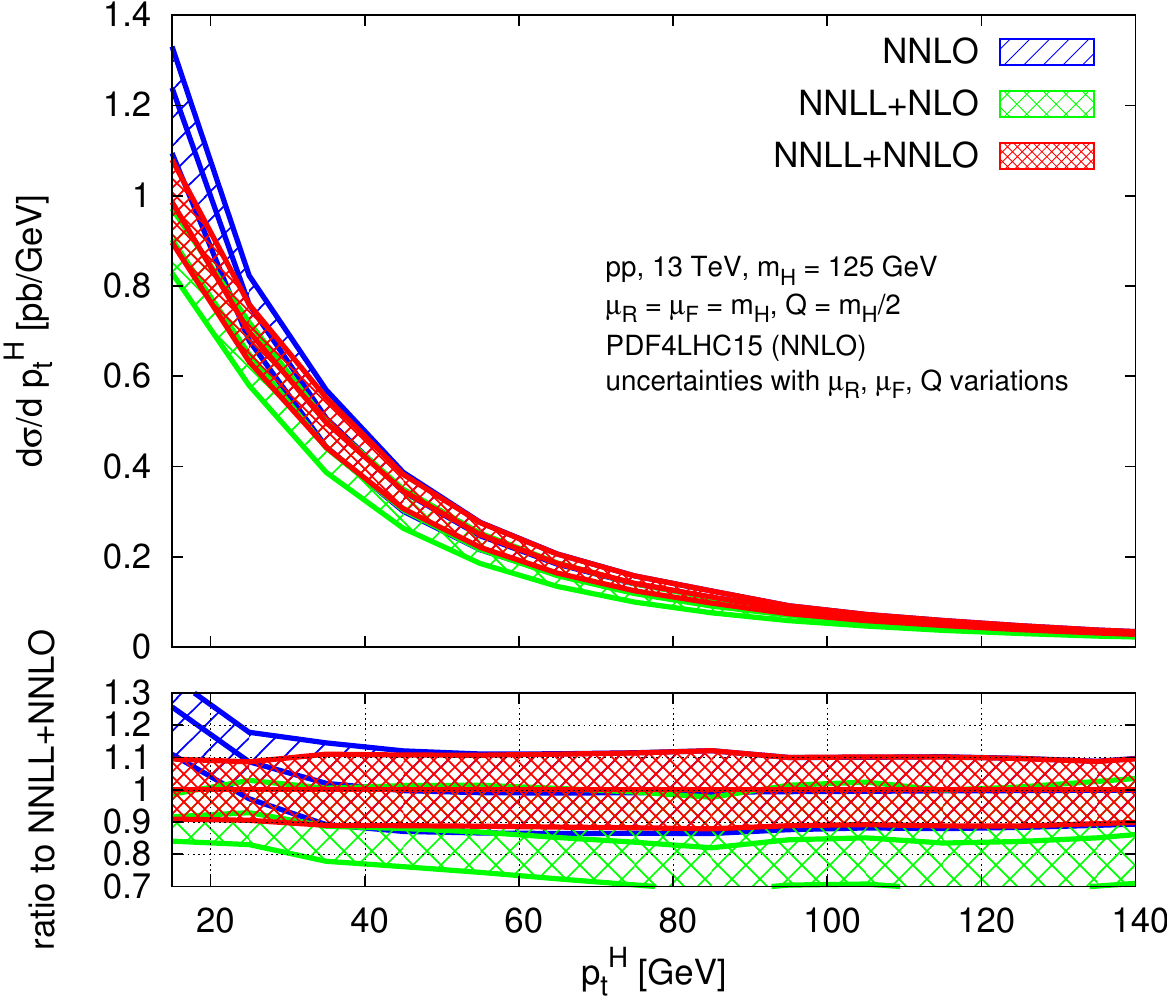}
   \caption{\footnotesize{Higgs $\mrtpt$ at NNLL+NNLO (red), NNLL+NLO
    (green), and NNLO (blue). Lower panel: ratio of the
    three predictions to the NNLL+NNLO one.\vspace{22mm}}}
  \label{fig:results2}
 \end{minipage}
\end{figure}

The matching to the fixed-order prediction is obtained with an
additive scheme, according to the formula
\begin{equation}
\label{eq:ptH_NNLL+NNLO_matching}
\Sigma^{\tiny{\mbox{NNLL+(N)NLO}}}(\mrtpt)=\Sigma^{\tiny{\mbox{(N)NLO}}}(\mrtpt)+
\Sigma^{\tiny{\mbox{NNLL}}}(\mrtpt)-\Sigma^{\tiny{\mbox{EXP}}}(\mrtpt),
\end{equation}
where $\Sigma^{\tiny{\mbox{EXP}}}(\mrtpt)$ represents the expansion of
the resummed formula to $\mathcal O(\as^4)$ for a matching to the NLO-accurate
differential distribution, or to $\mathcal O(\as^5)$ for a matching to the NNLO-accurate
differential distribution. The introduction of modified logarithms of the form
$\ln(Q/\kto)\to \ln\big[(Q/\kto)^p+1\big]/p$ ensures that the matched
cumulative cross section on the left-hand side
of~\eqref{eq:ptH_NNLL+NNLO_matching} reduces to the fixed-order one at
large transverse momentum.

In the main panel of \refF{fig:results1} a comparison is shown of the
prediction of ref.~\cite{Monni:2016ktx} at NNLL+NLO to that obtained with
  \texttt{HqT}~\cite{Bozzi:2005wk,deFlorian:2011xf}. As expected, very good
  agreement over the entire $\mrtpt$ range is observed between these two results,
  which have the same perturbative accuracy. The \texttt{HqT} prediction is
  moderately lower in the peak of the distribution, and higher at intermediate
  $\mrtpt$ values, although this pattern may slightly change with different central-scale
  choices. These small differences have to do with the different treatment of
  subleading effects in the two resummation methods.
 The agreement of the two results, both for the central scale and for the
  uncertainty band, is even more evident in the lower inset of \refF{fig:results1},
  which displays the ratio of the various distributions, each normalized to its central-scale 
  inclusive rate, to the normalized central NNLL+NLO curve of~\cite{Monni:2016ktx}.

  \refF{fig:results1} also reports the $\mrtpt$ distribution obtained
  with the NLO version of \texttt{POWHEG}+MiNLO
  \cite{Alioli:2010xd,Hamilton:2012rf}, and with the
  \texttt{MG5\_aMC@NLO}+FxFx
  \cite{Alwall:2014hca,Frederix:2012ps} event generators, using
  default renormalization and factorization scales for the two methods
  (in FxFx a merging scale $\mu_Q=\mh/2$ has been employed). Both
  generators are interfaced to \texttt{Pythia\,8.2}
  \cite{Sjostrand:2014zea}, without including hadronization,
  underlying event, and primordial $k_{\perp}$ (whose impact has been
  checked to be fully negligible for this observable), and use {\tt
    PDF4LHC15} parton densities at NLO. By inspecting the normalized
  ratios shown in the lower panel, one observes that the shape of the
  Monte-Carlo predictions deviates significantly from the NNLL+NLO
  results at $\mrtpt \lesssim 60$\,GeV. In order to avoid possible
  misunderstandings between the \texttt{POWHEG}+MiNLO result shown in
  \refF{fig:results1} and the so called ``NNLOPS'' approach, we
  recall that the \texttt{POWHEG}+MiNLO generator was further improved
  to achieve NNLO accuracy for fully inclusive
  observables in ref.~\cite{Hamilton:2013fea}. Although the logarithmic
  accuracy of the two ``\texttt{POWHEG}-based'' results is the same,
  the $\mrtpt$ spectrum obtained with the NNLOPS approach was found to
  be, numerically, in very good agreement with the \texttt{HqT}
  NNLL+NLO result, certainly more than what the \texttt{POWHEG}+MiNLO
  result does. More details can be found in~\cite{Hamilton:2013fea}.

  \refF{fig:results2} shows the comparison of the matched
  NNLL+NNLO result to the NNLL+NLO and the fixed-order NNLO
  predictions.  The inclusion of the NNLO corrections leads to a
  $10-15\%$ increase in the matched spectrum for $\mrtpt > 15$\,GeV, and
  to a consistent reduction in the perturbative uncertainty.
  The impact of the NNLL resummation on the fixed order becomes
  increasingly important for $\mrtpt \lesssim 40$\,GeV, leading to a
  suppression of the differential spectrum in this phase-space region
  which reaches about $25\%$ at $\mrtpt = 15$\,GeV.  For
  $\mrtpt \gtrsim 40$\,GeV, the matched prediction reduces to the NNLO
  one. The final theory uncertainty is at the $\pm10\%$-level in the
  phenomenologically relevant $\mrtpt$ range.

\subsection[NNLOJET: \texorpdfstring{$H + j$}{H+j} at NNLO using Antenna subtraction]{NNLOJET:: \texorpdfstring{$H + j$}{H+j} at NNLO using Antenna subtraction\SectionAuthor{X.~Chen, E.W.N.~Glover, T.~Gehrmann, M.~Jaquier}}
\label{sec:antenna}
The NNLO corrections to $H + j$ receive contributions from processes which, with respect to the leading order, feature two additional external legs (double real emission, \cite{DelDuca:2004wt,Dixon:2004za,Badger:2004ty}), one external leg and one internal loop (real virtual, \cite{Dixon:2009uk,Badger:2009hw,Badger:2009vh}) or two additional internal loops (double virtual, \cite{Gehrmann:2011aa}). Each of these is separately infrared (IR) divergent, with the divergences cancelling upon integration over the phase space. Since this integration is  performed numerically, a procedure is needed to extract the singularities from the various contributions and to achieve their cancellation prior to the phase space integration.

In the integration of the real radiation contributions over their phase space, IR divergences appear in configurations where the momentum of a parton becomes unresolved. In such configurations, a QCD matrix element factorizes into products of universal unresolved factors, which contain the IR singularities, and reduced matrix elements, which depend only on the resolved kinematics. The antenna subtraction formalism \cite{GehrmannDeRidder:2005cm,Daleo:2006xa,Currie:2013vh} makes use of this
factorization in describing the divergent behaviour of matrix elements in terms of simpler normalized matrix elements involving an adequate set of partons. Using these antenna functions, one can construct subtraction terms which, when added to the corresponding matrix elements, make them well-defined and integrable over the whole phase space, thus allowing for numerical integration.

With the reduced matrix element $M_n$ (possibly with loops) and the jet function $J_n$, a typical subtraction term has the form:
\begin{equation}
 X_m^l(p_a,p_2,\ldots,p_{m-1},p_b)|M_n(\ldots,\tilde{p}_a,\tilde{p}_b,\ldots)|^2J_n(\ldots,\tilde{p}_a,\tilde{p}_b,\ldots),
\end{equation}
where $X_m^l(p_a,p_2,\ldots,p_{m-1},p_b)$ is the $m$-parton $l$-loop antenna function, which features the full IR divergent behaviour of the partons $p_2,\ldots,p_{m-1}$ in the colour ordering $\{p_a,p_2,\ldots,p_{m-1},p_b\}$. At NNLO, one is required to consider values of $(m,l)$ up to $(4,0)$ and $(3,1)$. The momenta $\tilde{p}_a$ and $\tilde{p}_b$ are given through mappings $\{p_a,p_2,\ldots,p_{m-1},p_b\}\rightarrow\{\tilde{p}_a,\tilde{p}_b\}$ which interpolate between all unresolved configurations of the partons $p_2,\ldots,p_{m-1}$ with the hard radiators $p_a$ and $p_b$.
The subtraction terms can then be integrated analytically over the phase spaces of the unresolved partons to obtain the integrated antenna functions $\mathcal{X}_m^l(p_A,p_B)$, which feature explicit poles up to order $2(m-2+l)$ in $D-4$. Together with the mass factorization counterterms, which originate from the renormalization of the PDFs, they can be shown to fully cancel the pole structure of the virtual loop matrix elements.

We have implemented all subprocesses for $H+j$ at NNLO into the
parton-level event generator NNLOJET \cite{Chen:2014gva,Chen:2016zka}, relying on the antenna subtraction method to cancel the implicit and explicit IR divergences appearing in the matrix elements of the various contributions. This program allows for the computation of all IR-safe observables related to $H+j$ final states to NNLO accuracy. Renormalization and factorization scale can be chosen on an event-by-event basis. In order to stabilize the numerical integration, and to arrive at reliable error estimates on it, we divide the sample of Monte Carlo integration points into sub-samples, on which a weighted average is performed (see \cite{Ridder:2015dxa,Ridder:2016rzm,Ridder:2016nkl} for details on this procedure). The decay of the Higgs boson to two photons is included, and other decay modes will be added in the future. The program allows us to make predictions for fiducial cross sections taking into account appropriate event selection criteria and Higgs boson decay matrix elements.

The NNLOJET code is used for the fixed order predictions of the
observables discussed in Section~\ref{sec:jbenchmarks}. Besides this,
it also provides an independent validation of the fixed-order results
described in Section~\ref{subsec:stwz_blptw} (JVE method), which use
the code of Ref.~\cite{Caola:2015wna}. For example, the NNLOJET result for the 
inclusive cross section with scale choice $\mu_F=\mu_R=1/2*m_H$ and under infinite top mass assumption is 17.7~pb for $p_{T,{\rm min}} = 30$~GeV. By removing the NNLO contribution from the quark-quark initiated channels and including the finite top quark mass effect at LO (which is the exact setup used for the fixed order results),
the NNLOJET result fully agrees with the corresponding NNLO inclusive
cross sections quoted in Table~\ref{tab:STWZ0jet}.
 

\section[Benchmarks for cross sections and differential distributions]{Benchmarks for cross sections and differential distributions\SectionAuthor{S.~Forte, D.~Gillberg, C.~Hays, A.~Lazopoulos, G.~Petrucciani, A.~Massironi, G.~Zanderighi}}
\label{sec:jbenchmarks}
An accurate modelling of the differential distributions in gluon
fusion production is important since the experimental analyses
typically combine measurements in different phase space regions,
either to improve the sensitivity to a signal or to target other 
Higgs boson production modes to which gluon fusion is a background.  In this
section, comparisons are performed between the predictions of
different parton-level computations and hadron-level event generators,
to assess their compatibilities and the accuracy of the modelling.

Unless otherwise specified, all the predictions correspond to a Higgs boson 
mass of $125$ \UGeV, $\sqrt{s} = 13$ \UTeV, and the choice of SM
input parameters and PDFs in
Sects.~\ref{chapter:input}--~\ref{chap:PDFs}. We will first list the
various codes and calculations used in the benchmarking, and then
discuss predictions for several observables in turn.

\begin{figure}\centering
\begin{center}
  \includegraphics[width=0.8\textwidth]{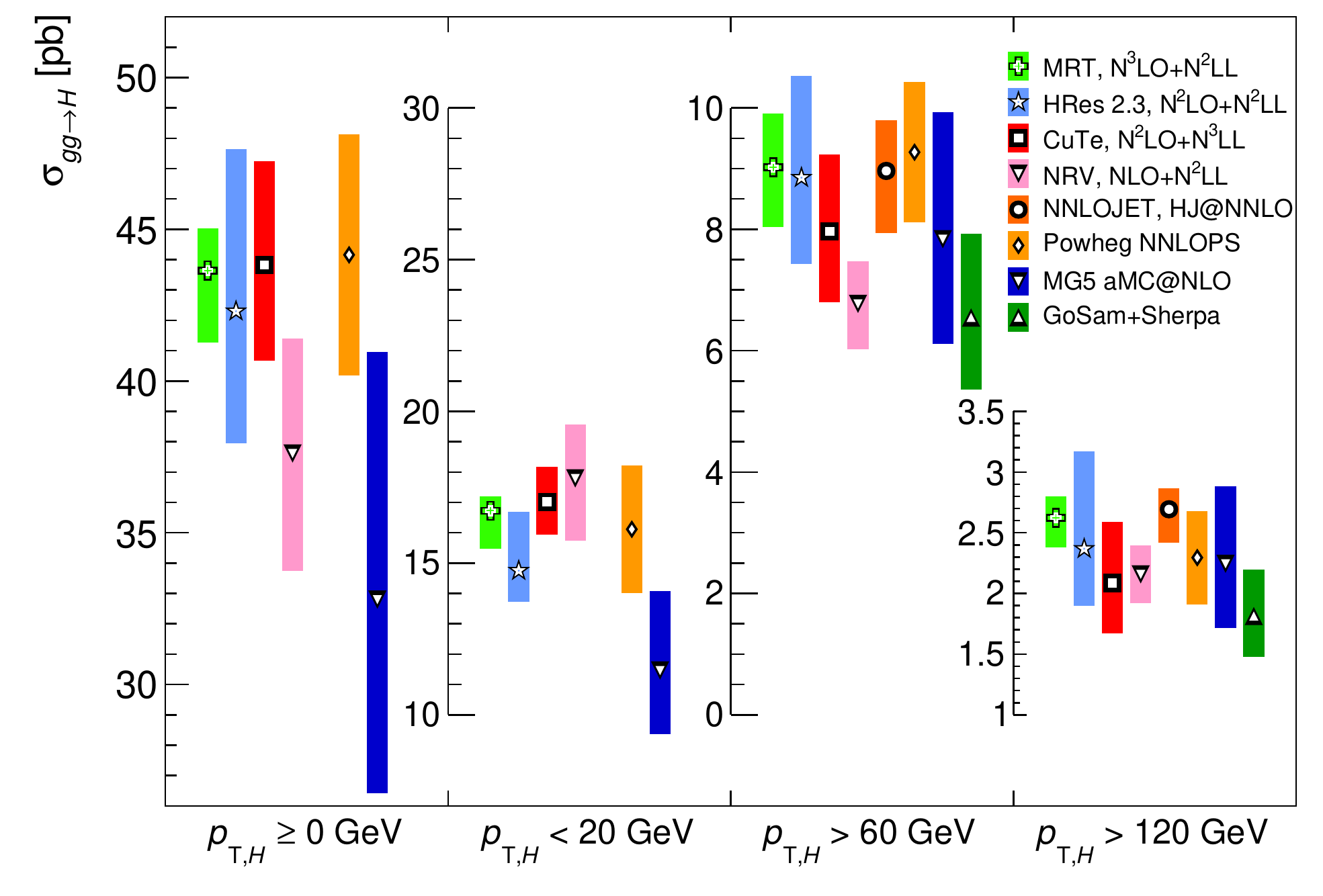}
\end{center}
\vspace*{-0.3cm}
\caption{Comparison of cross section in different regions of $p_{T,\,H}$.}
\label{fig:WG1-ggF-differential-parton-pt}
\end{figure}

\subsection{Calculations and codes}

We list here the calculations and codes used in the following
benchmarking, and provide information about settings whenever they
differ from our default.

\subsubsection{Parton level codes }
\label{sec:partonlevelcodes}
\noindent
{\sc \bf HRes}: A parton-level
code~\cite{deFlorian:2012mx,Grazzini:2013mca} to compute differential distributions
in gluon fusion production at NNLO QCD accuracy, with NNLL QCD resummation for small $\pT(\PH)$
and matching to NLO QCD $\PH+1\text{jet}$ at large $\pT(\PH)$.
Finite top, bottom, and charm quark masses are included at NLO QCD accuracy.
In this comparison, predictions were computed with a dynamical scale  
$\mu_R=\mu_F=(\sqrt{m_H^2+p_t^2})/2$ and with resummation scales
$Q_1=m_H/2$, $Q_2=2 m_b$. The overall uncertainties are estimated by
taking the envelope of  a seven-point scale variation of $\mu_R$ and $\mu_F$ for
central $Q_1$ and $Q_2$ and a  
variation of $Q_1$ and $Q_2$ one at a time by a factor two about their
central value, with central  $\mu_R$ and $\mu_F$.\\ 
\noindent
{\sc \bf CuTe}: A parton-level code~\cite{Becher:2012yn}, for the $\pT(\PH)$ differential
distribution, with NNLL QCD resummation at small $\pT(\PH)$ and matching to NLO QCD
$\PH+1\text{jet}$ at large $\pT(\PH)$. The resummation is based on
Soft Collinear Effective Theory (SCET) and NNLL accuracy is
accordingly defined.
The finite top quark mass is included at LO QCD accuracy. 
The scale is chosen as $\mu= p_T + q^*\exp(p_t/q^*)$
with $q^*=7.83$~GeV. Uncertainties are estimated by varying $\mu$ by a
factor two about its central value and 
by varying the unknown coefficient $F_{(3,0)}^g$ of the
anomaly exponent $F^g$ in the range $\pm 2 (4 \pi) F_{(2,0)}^g$, and
also by including one-sigma PDF uncertainties and varying $\alpha_s$
in the range $[0.1165,0.1195]$.
\\
\noindent
{\sc \bf NRV}: A parton-level computation to NNLL resummed accuracy
matched to the fixed-order $O(\alpha_s^4)$
computation~\cite{Neill:2015roa}. Resummation is
performed using a SCET approach and the logarithmic order is defined
accordingly. All quarks but top are assumed massless.  Uncertainties
are estimated as described in Ref.~\cite{Neill:2015roa}.
The 
individual NNLO PDF sets MMHT and NNPDF3.0 entering the PDF4LHC
recommendations have been used, and uncertainties within and between
the sets are 
included in the error estimation.\\
\noindent{\sc \bf MRT}: This code was described in
Sect.~\ref{subsec:MRT}. It is based on a parton-level computation at
NNLL for the Higgs boson $p_T$ distribution matched to the NNLO
($O(\alpha_s^5)$) fixed-order prediction for Higgs+jet. The central
prediction uses $\mu_R = \mu_F = m_H$, and $Q = m_H/2$. The perturbative 
uncertainty for all predictions is estimated by seven-point scale
variation of  $μ_R$ and $μ_F$ with fixed resummation scale, and
varying the resummation 
scale by a factor two for fixed $\mu_R$ and $\mu_F$. 
\\\noindent
{\sc \bf BFGLP}: A parton-level prediction for $\PH+1\text{jet}$ at NNLO
QCD using jettiness  
subtraction~\cite{Boughezal:2015aha}.   The prediction is not matched to inclusive gluon fusion production, and thus the resulting
 $\pT(\PH)$ distribution can be directly compared to the other predictions only well above the
jet $\pT$ threshold used in the computation ($30\UGeV$), but has a higher accuracy because 
of the extra order in QCD.  Predictions shown use the settings of
Ref.~\cite{Heinemeyer:2013tqa} but with NNPDF2.3 parton distributions.
\\
\noindent
{\sc \bf NNLOJET}: This code~\cite{Chen:2016zka} was described in Sect.~\ref{sec:antenna}. It has
the same perturbative 
accuracy as BFGLP. A threshold of $p_T=30$~GeV is adopted for
jet counting. Uncertainties are estimated by three-point scale
variation about $\mu_r=\mu_F=m_H$. The NNLOJET code was validated extensively
    against the H+j NNLO calculation of~\cite{Caola:2015wna}
    (discussed in Section~\ref{sec:jve}), yielding excellent agreement at below one per cent for all
    distributions.
\\
\noindent
{\sc \bf STWZ-BLPTW} This code was described in
Sect.~\ref{subsec:stwz_blptw}. It is a SCET-based resummation for the
jet veto at NNLL$^\prime$+NNLO. The
prescription  used for uncertainty estimation was described in detail
in Sect.~\ref{subsec:stwz_blptw}.
\\
\noindent{\sc \bf JVE} This code was described in Sect.~\ref{sec:jve}. It
performs a N$^3$LO+NNLL+LLR accurate resummation for the jet veto,
including heavy quark mass effects up to NLO. Beyond NLO, the heavy top
result is used as explained in Ref.~\cite{Banfi:2015pju}. The
prescription  used for uncertainty estimation was described in detail
in Sect.~\ref{sec:jve}.
\\
{\sc \bf Gosam + Sherpa}: The predictions used in this comparison
include up to three additional  
jets at NLO QCD
accuracy~\cite{Cullen:2011ac,Cullen:2014yla,Gleisberg:2008ta} in the approximation of an infinitely 
heavy top quark.
The predictions presented in this report were computed using sets of
Ntuples with generation cuts set to 
%
$ p_T\;>\;25~\UGeV~, |\eta|\;<\;4.5$.
%
The cut on the pseudorapidity can not be removed, but
an explicit computation at LO shows that the effect of these cuts is
at the per cent level.
%
The renormalization and factorization scales were set to
$\frac{\hat{H}^\prime_T}{2}\;=\;\frac{1}{2}\left(\sqrt{m_{\mathrm{H}}^{2}+p_{T,\mathrm{H}}^{2}}+\sum_{i}|p_{T,i}|\right)$, 
where the sum runs over all partons accompanying the Higgs boson in
the event. The theoretical uncertainties were estimated by varying
both of them by factors of $0.5$ and $2$ around this central
value. Further details are provided in Ref.~\cite{Greiner:2015jha}.
\\

\begin{figure}\centering
\begin{center}
  \includegraphics[width=0.7\textwidth]{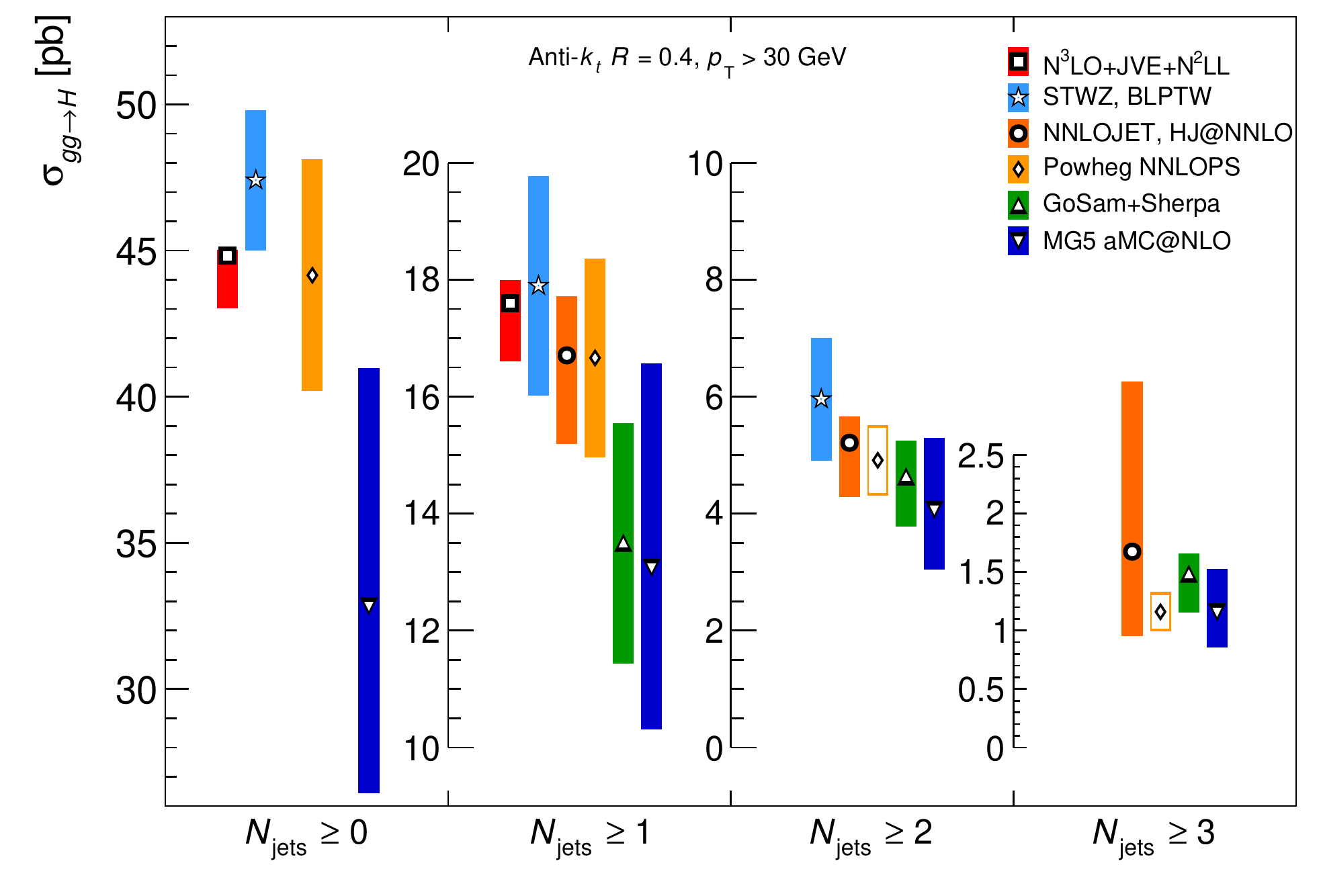}
  \includegraphics[width=0.7\textwidth]{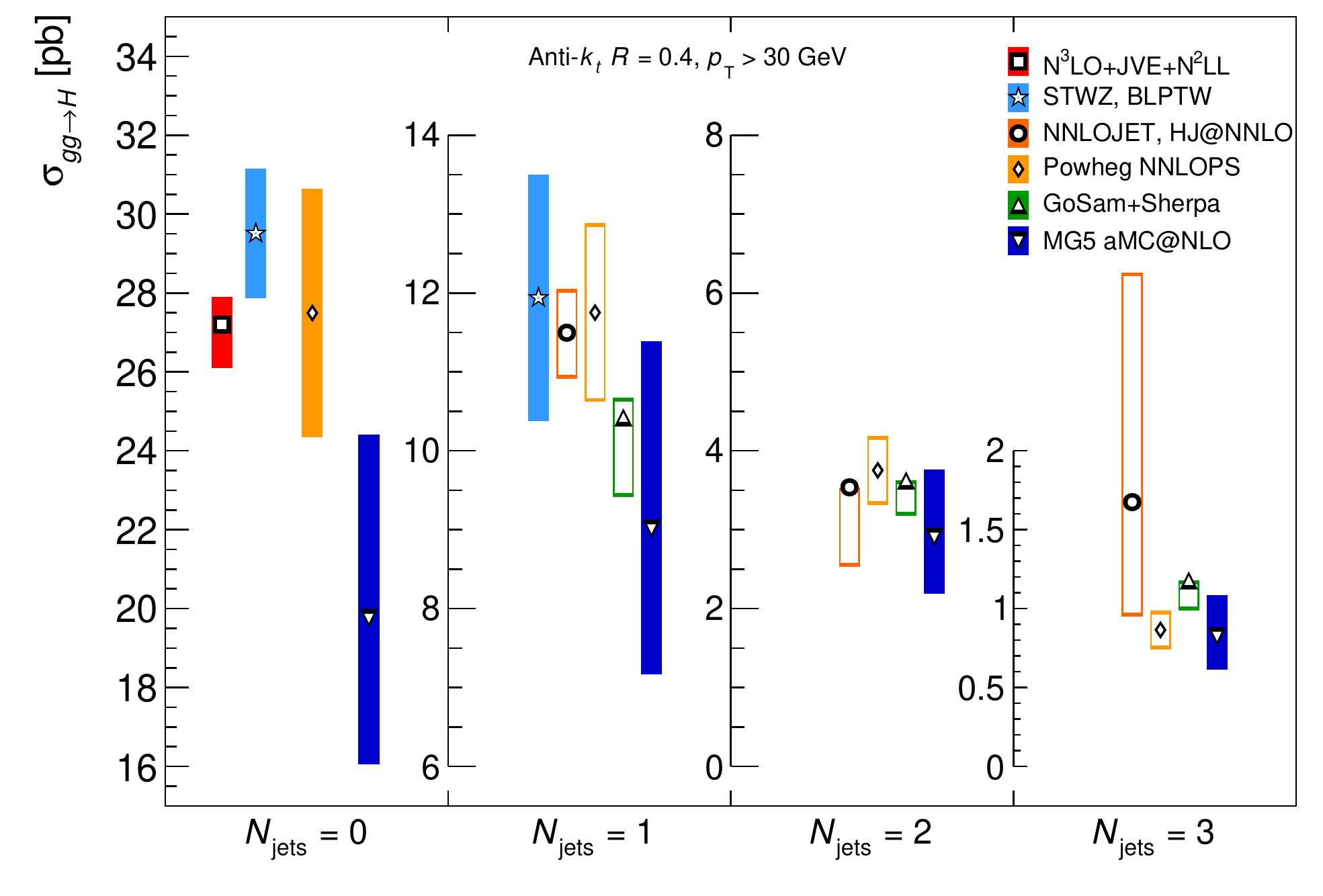}
\end{center}
\vspace*{-0.3cm}
\caption{
  Various predictions for the  inclusive and exclusive jet cross
  section for gluon fusion Higgs boson production. 
  } \label{fig:jetbins}
\end{figure}

\subsubsection{Hadron-level event generators}
\noindent
\noindent
{\sc \bf MG5\_ aMC@NLO}: 
The predictions used in this comparison include up to two additional jets at NLO QCD accuracy
with the {\sc FxFx} merging scheme, in the Effective Field Theory approach ($\Mt\to\infty$).
The top quark mass is included via reweighting of the events.  The full mass effects are included
in all multiplicities in all contributions apart from the two-loop
matrix element in the virtual for $\PH+1\text{jet}$ and $\PH+2\text{jets}$ NLO matrix elements, for which
the correction is approximated using that of the Born contribution.
The bottom quark mass and the top-bottom interference are included in $\PH+0\text{jet}$ at NLO
using the resummation scales suggested in \cite{Harlander:2014uea}.
The central value of merging scale is set to 30~GeV, and the deviations from using merging scales of 20~GeV and 50~GeV is taken as an uncertainty. Further details are provided in Ref.~\cite{Frederix:2016cnl}. \\
\noindent
{\sc \bf Powheg NNLOPS}: Predictions have  NNLO QCD accuracy for inclusive
events, and NLO+PS for Higgs+one jet.
Top and bottom quark mass effects are included up to NLO according to
Ref.~\cite{Hamilton:2015nsa}. The central scale choice is
$\mu_F=\mu_R=m_H/2$. Uncertainties are obtained by preforming a
three-point scale variation of $\mu_R=\mu_F$ by a factor two about the
central value for the NNLO part, and using a 9-point variation for the Powheg scale.
These two uncertainties are taken as uncorrelated and are added in quadrature to obtain the total QCD scale variation. Uncertainties are also provided for switching off bottom and top mass effects.

\begin{figure}\centering
\includegraphics[width=0.49\textwidth]{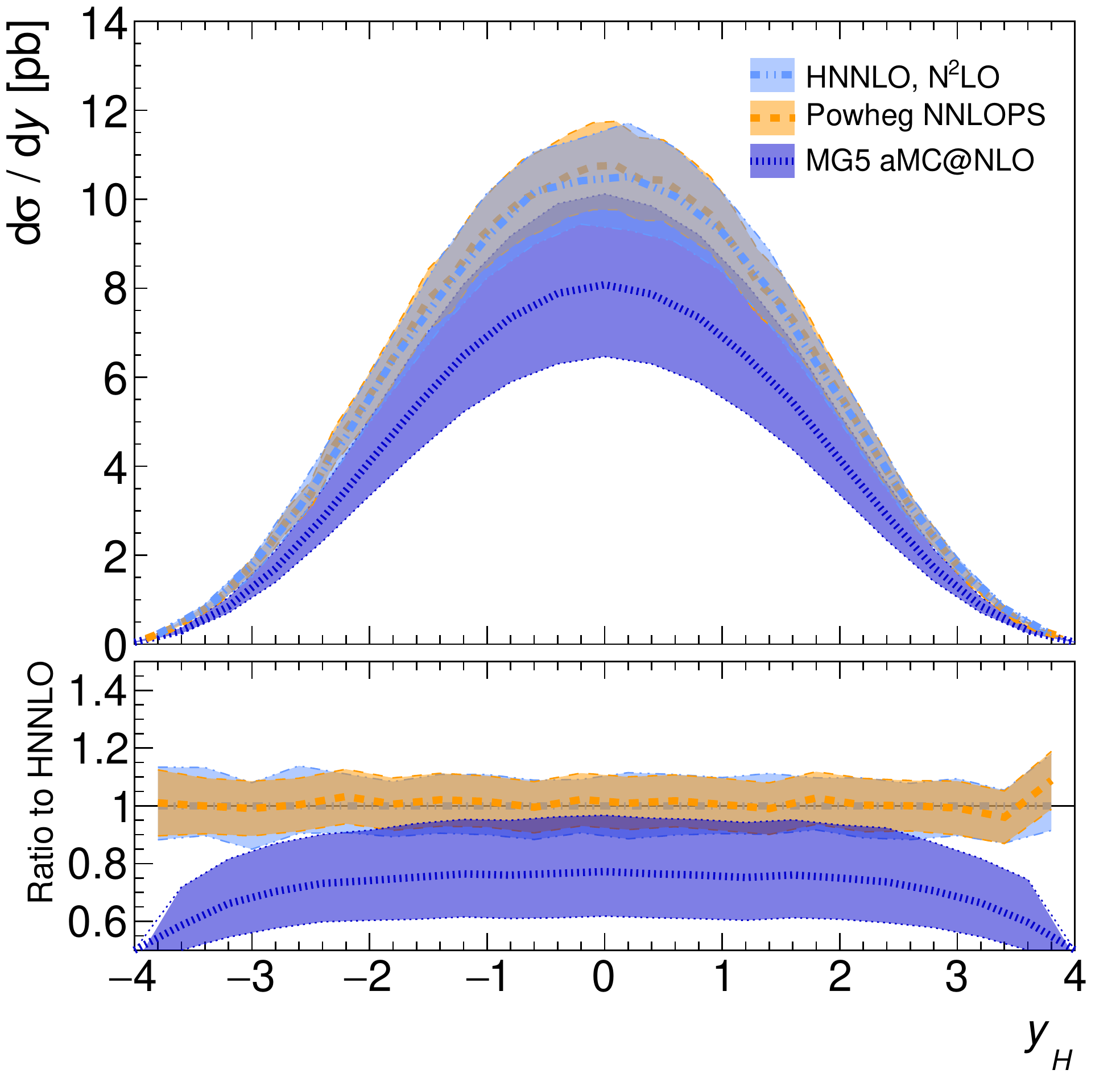}
\includegraphics[width=0.49\textwidth]{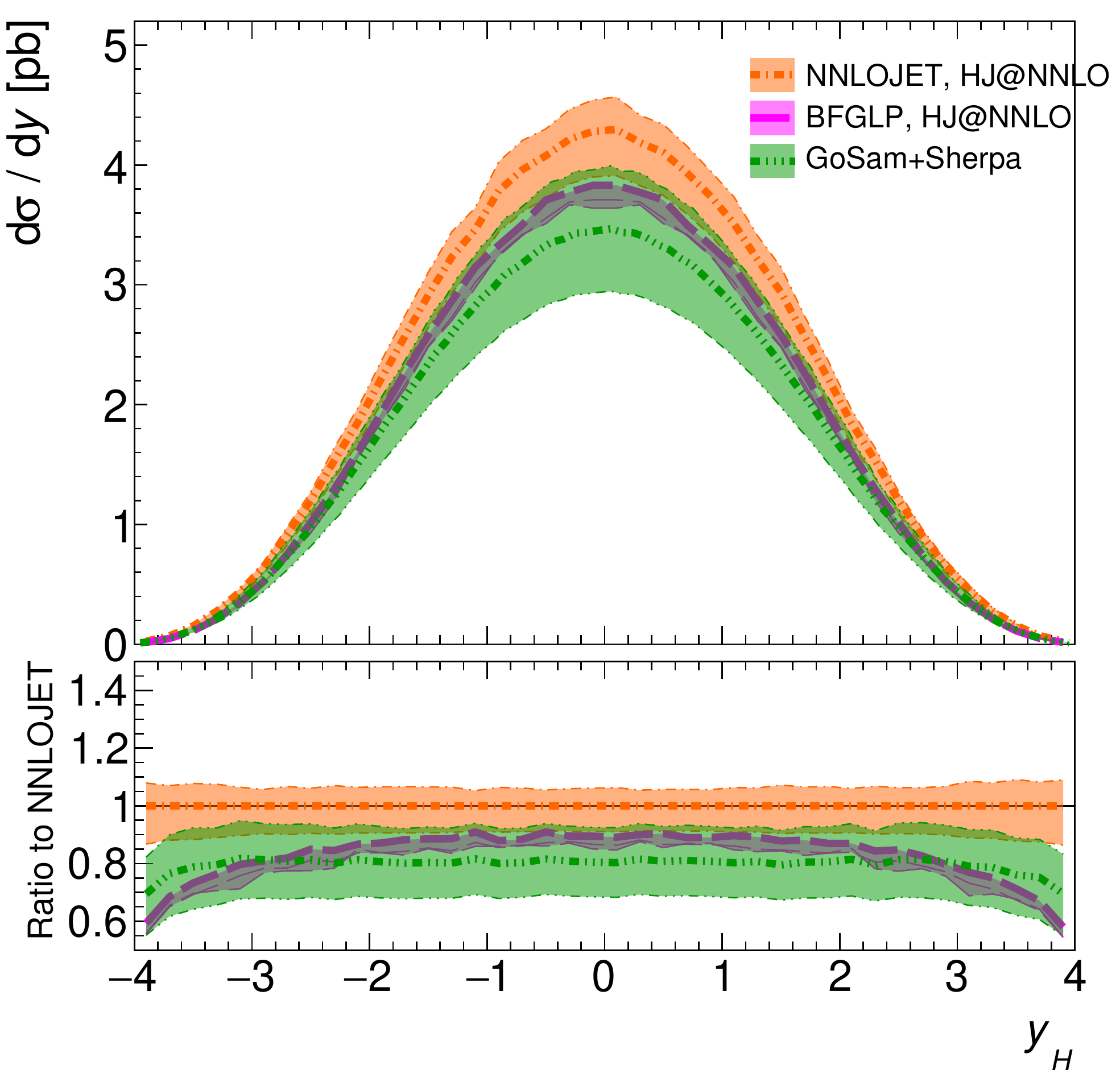}

\vspace*{-0.3cm}
\caption{Predictions of the Higgs boson rapidity distribution
for gluon fusion production, including both parton-level calculations
and hadron-level MC predictions. The spectra are shown at the
inclusive level (left) and with a jet cut (right).}
\label{fig:Higgs-rapidity}
\end{figure}

\subsection{Observables}
We will first discuss results for integrated cross-sections, then
consider various differential distributions. 

\begin{table}
\caption{Fraction of $gg\to H$ events in different kinematic regions from various predictions
  (central values only). The numbers in italic font have been obtained by using
  the N$^3$LO cross section with the EW component subtracted $48.58-2.4 = 46.18$~pb.
  {\sc GoSam+Sherpa} use separate predictions for each jet bin: ggF +
  1, 2 and 3 jets at NLO, respectively. In all other cases numbers are
normalized to their respective predictions.}
\label{tab:ggbenchmarkTable}
\renewcommand{\arraystretch}{1.2}
\setlength{\tabcolsep}{1.5ex}
\begin{center}
\begin{tabular}{l|ccc|cccc}
\toprule
 & \multicolumn{3}{c|}{$p_{T,H}/\UGeV$} &
   \multicolumn{4}{c}{$N_{\rm jets}$, ~$p_{T,j} > 30\UGeV$} \\ 
Prediction & $<20$ & $>60$ & $>120$ & $=0$ & $=1$ & $\geq 2$ & $\geq 3$ \\ 
\midrule
  HRes                      & {34.9\%}& {20.9\%}& {5.60\%}&   $-$ &   $-$ &   $-$ &   $-$ \\ 
  {\sc CuTe}                & {38.8\%}& {18.2\%}& {4.76\%}&   $-$ &   $-$ &   $-$ &   $-$ \\ 
  {\sc MRT}                 & {38.3\%}& {20.7\%}& {6.00\%}&   $-$ &   $-$ &   $-$ &   $-$ \\ 
  NRV                       & {47.3\%}& {18.0\%}& {5.74\%}&   $-$ &   $-$ &   $-$ &   $-$ \\ 
  BLPTW                     &   $-$ &   $-$ &   $-$ & {62.2\%}& {25.2\%}& {12.6\%}&   $-$ \\ 
  JVE                       &   $-$ &   $-$ &   $-$ & {60.7\%}&   $-$ &   $-$ &   $-$ \\ 
  {\sc NNLOJET}                &   $-$ & {\it19.4\%}& {\it5.83\%}& {\it63.8\%}& {\it24.9\%}& {\it11.3\%}& {\it3.63\%}\\ 
  {\sc GoSam+Sherpa}        &   $-$ &   $-$ &   $-$ &   $-$ & {\it22.6\%}& {\it10.0\%}& {\it3.22\%}\\ 
  {\sc Powheg Nnlops}       & {36.5\%}& {21.0\%}& {5.20\%}& {62.3\%}& {26.6\%}& {11.1\%}& {2.63\%}\\ 
  aMC\@NLO MG5              & {34.9\%}& {23.9\%}& {6.84\%}& {60.2\%}& {27.5\%}& {12.4\%}& {3.52\%}\\ 
\bottomrule
\end{tabular}
\end{center}
\end{table}

\subsubsection{Cross-sections}
In \refF{fig:WG1-ggF-differential-parton-pt} we compare
predictions for cross-sections in different ranges of the Higgs boson $p_T$. 
Cross-sections are obtained integrating differential
$p_{T,\,H}$ distributions separately for each uncertainty variation provided.
The uncertainties about the central value are obtained using the
prescription associated with each prediction, most commonly quadratic
addition of the difference from the nominal for each effect
considered.

The QCD scale uncertainty is obtained by integrating the spectrum
for each scale choice and taking the envelope around the nominal integral.
The exception is NNLOPS, for which two QCD scale uncertainties are considered:
one uncertainty from the three-point envelope of the NNLO part, and a separate
uncertainty from the nine-point envelope of the Powheg scale. These
uncertainties 
are treated as independent sources and are hence added in quadrature. The
former affects the normalization ("yield") while the latter affects the
migration between low and high Higgs boson (and jet) $p_T$.

It should be pointed out that not all predictions include the same
set of uncertainties. All include QCD scale variations, which is always the
dominant uncertainty. Some also
include other uncertainties, for example from choices of PDF set and
resummation scale.

For the following discussion we find it useful to take the MRT
prediction as a reference, since it has the nominally highest accuracy
both at low and high $p_{T,\,H}$. At low $p_{T,\,H}<20$~GeV we see
that MRT is in good agreement with all codes which include N$^2$LL
resummation (HRes, CuTe and NRV). At high $p_{T,\,H}$ MRT agrees well
with NNLOJET both in central value and (small) uncertainty. We note that
POWHEG NNLOPS agrees well with MRT both at low $p_{T,\,H}$  (even though it
does not have formally NNLL accuracy) and at high $p_{T,\,H}$ (even if
it does not have NNLO corrections to Higgs+1 jet). On the other hand,
we observe that MG5\_aMC@NLO is lower and with rather larger
uncertainty at low $p_T$ since it does not include the NNLO correction
to Higgs boson production. The GoSam+Sherpa prediction at high $p_T$ is
lower.
Finally we remark that
the NRV prediction tends to be lower at medium and large $p_{T,\,H}$,
which also reflect on their total cross section.  This might be
related to the shape function used in the intermediate region by this
group in order to switch to the fixed order result at high  $p_{T,\,H}$.

Next, in \refF{fig:jetbins} we compare predictions for
inclusive (top) and exclusive (bottom) jet-binned cross sections.
The 0-jet inclusive cross-sections obtained integrating the
N$^3$LO+JVE+N$^3$LL construction is by definition the recent N$^3$LO
result. 
Furthermore, for the 0-jet inclusive cross-section Powheg NNLOPS is accurate at
NNLO; STWZ is formally also NNLO accurate, but it is higher (and with
smaller uncertainty) as it
includes $\pi^2$ resummation; MG5\_aMC@NLO is on the other hand lower
and with larger uncertainties as
it is only accurate to NLO. In the inclusive  one-jet bit all
predictions which are NNLO accurate (for Higgs+1 jet) are in good
agreement with each other, while GoSam+Sherpa and  MG5\_aMC@NLO are
lower. Note that even though Powheg NNLOPS does not include NNLO
corrections, it is in good agreement with the NNLO-accurate
predictions.  In the higher inclusive jet bins all results are in
reasonable agreement, with differences most likely due to choices of
scale and the treatment of heavy quarks. We note that for the 3-jet
cross-section the NNLOJET prediction has a very large uncertainty due to
its LO nature. For Powheg NNLOPS the third jet is only provided by
parton showering; it is however known that the uncertainty is
underestimated in this case.

Exclusive cross sections follow a similar pattern. Note that in this
case the uncertainties shown for MG5\_aMC@NLO and Powheg NNLOPS are
unreliable as jet veto effects are not properly accounted for. 

The fractions of events in the various inclusive and exclusive bins
are tabulated in \refT{tab:ggbenchmarkTable}.
\begin{figure}\centering
\includegraphics[width=0.49\textwidth]{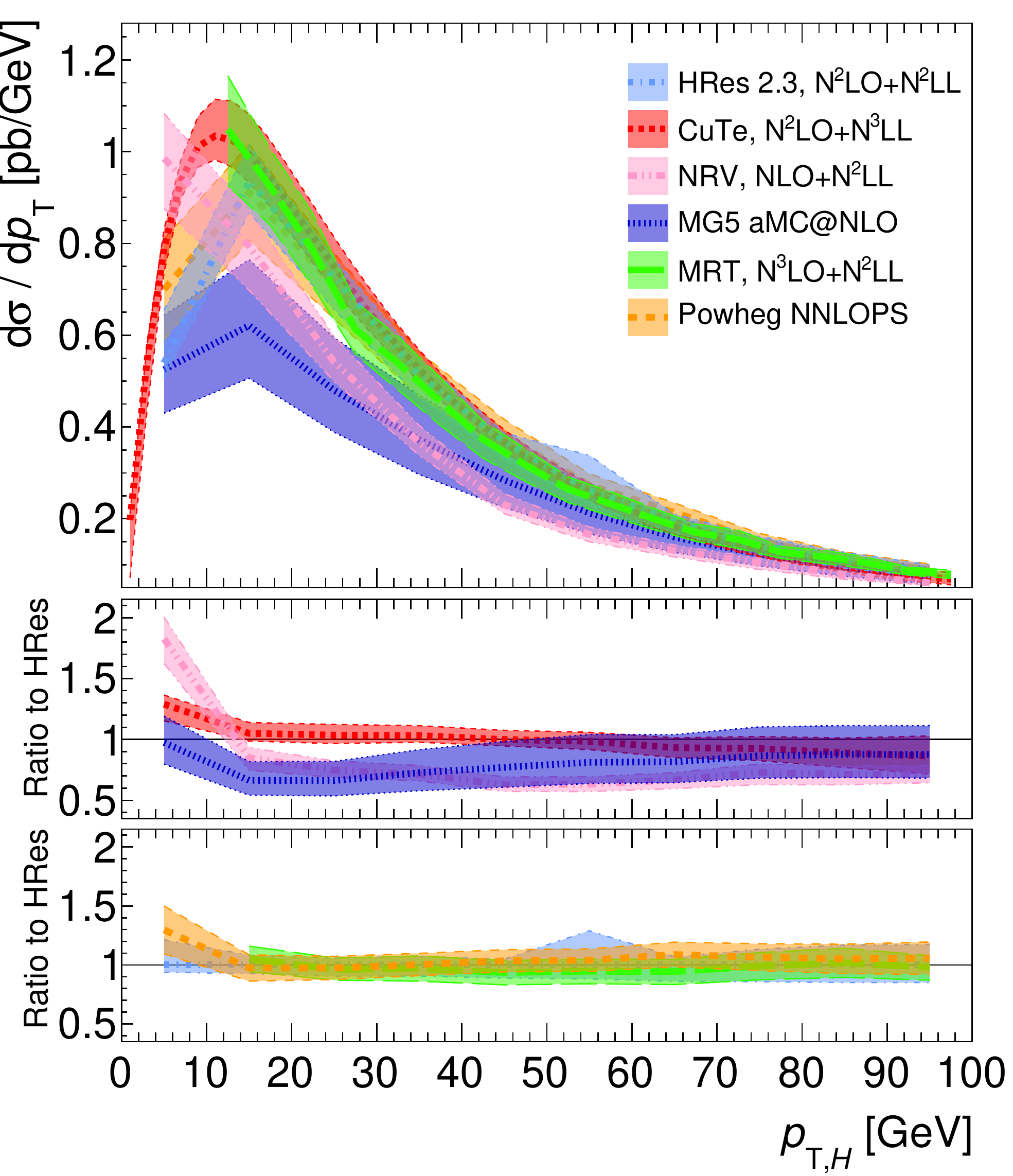}

\vspace*{-0.3cm}
\caption{ Same as \refF{fig:Higgs-rapidity}, but for the Higgs boson transverse 
momentum distribution at low $p_T$.
}
\label{fig:Higgs-pT}
\end{figure}


\subsubsection{Differential distributions}

In \refF{fig:Higgs-rapidity} we compare Higgs boson rapidity
distributions, both at the inclusive level (left) and with a jet cut
(right) defined as above. At the inclusive level, as before, the NLO
MG5\_aMC@NLO result undershoots the predictions from HNNLO and Powheg
NNLOPS which by construction agree with each other. Note that the
$K$-factor is not flat: NNLO corrections are more important for large rapidity.
Also in the presence of a jet the NLO result from GoSam+Sherpa is
somewhat low. The BFGLP and NNLOJET results substantially differ in shape,
especially at high rapidity, and the BFGLP result seems to have
very small scale  uncertainties.  However, it should be noted that the BFGLP 
setup is different to the default (see Section~\ref{sec:partonlevelcodes}), and 
in particular different PDFs are used.

\begin{figure}\centering
\includegraphics[width=0.49\textwidth]{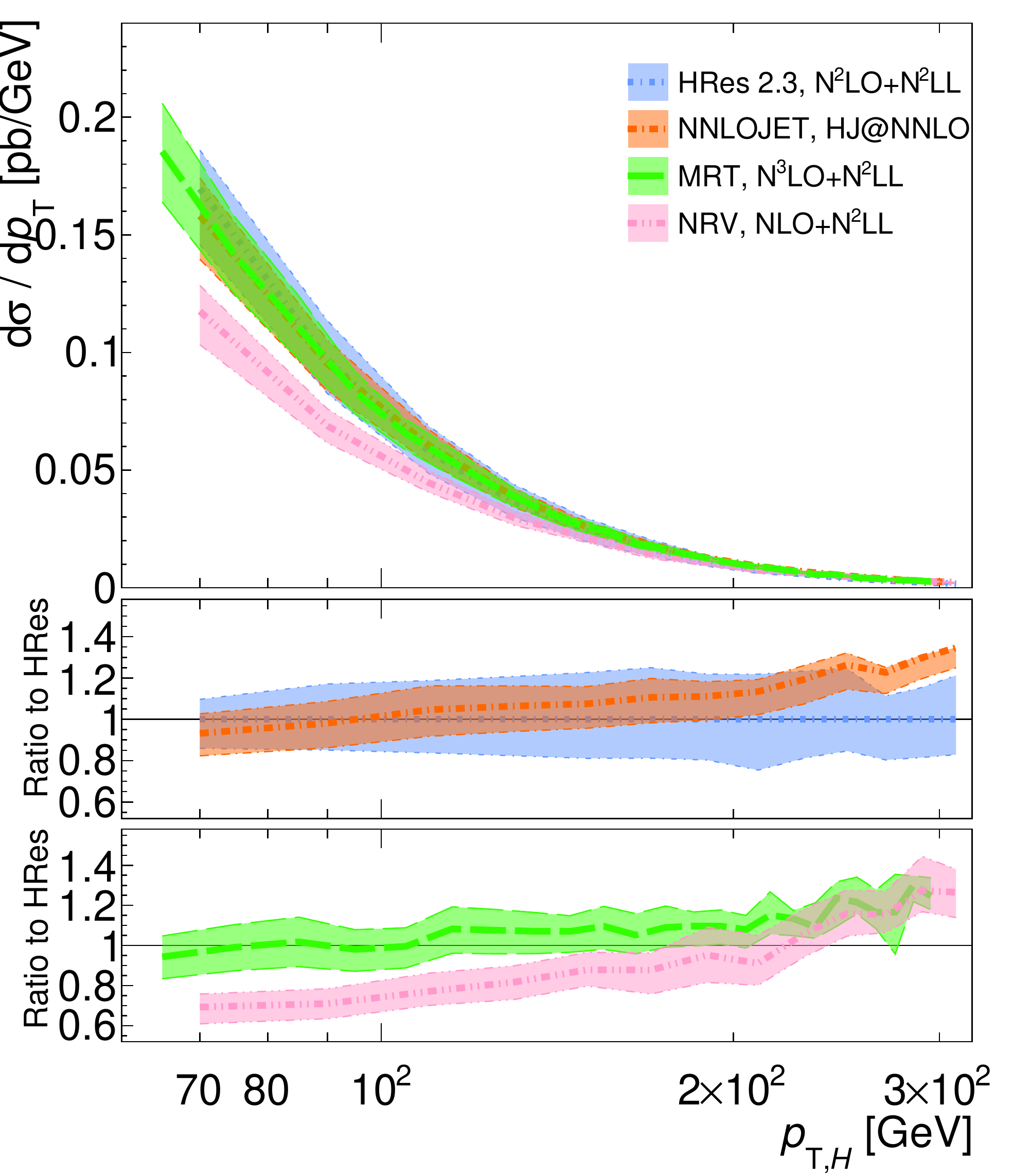}
\includegraphics[width=0.49\textwidth]{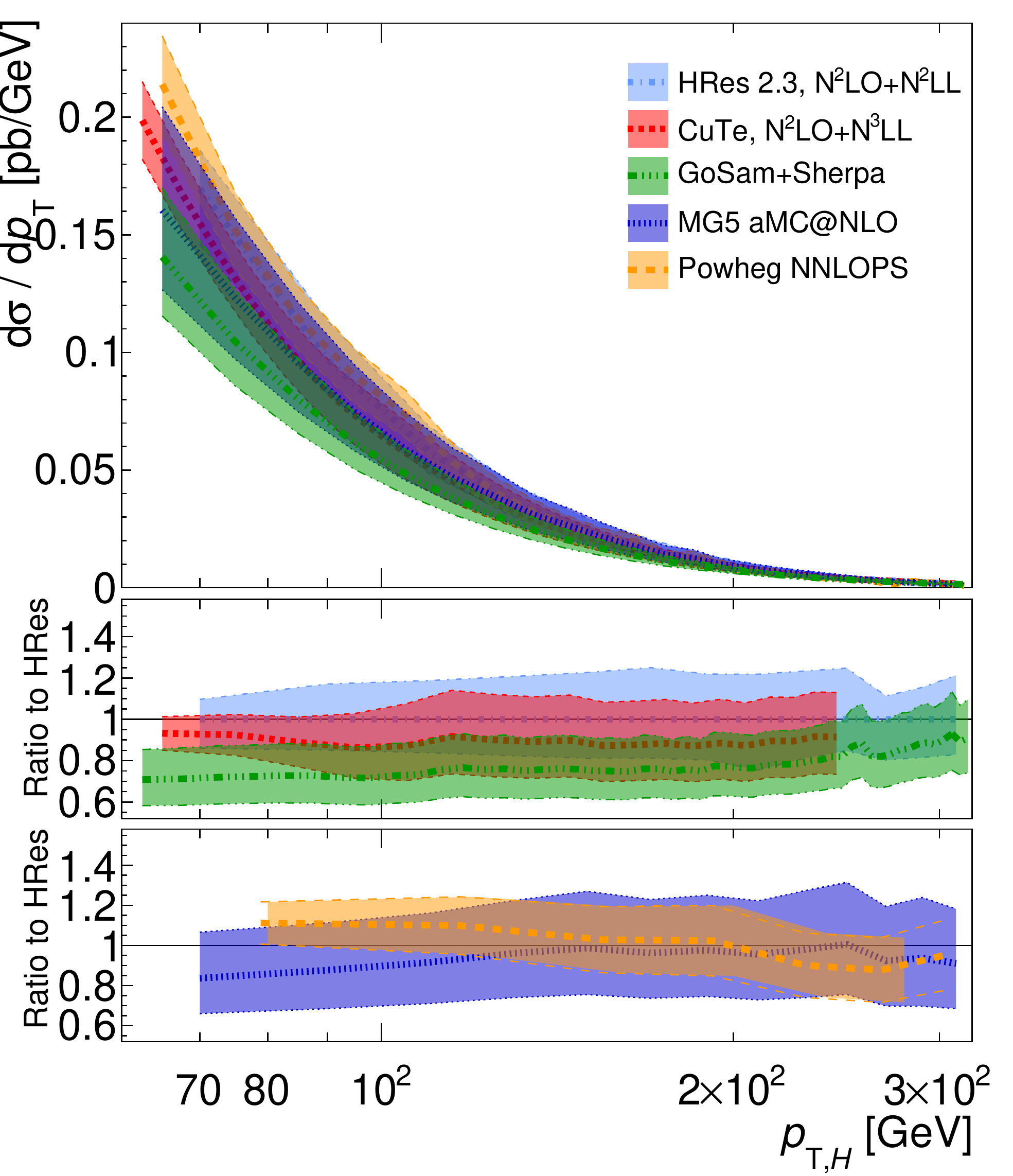}
\vspace*{-0.3cm}
\caption{Same as \refF{fig:Higgs-pT} for  $p_{T,H}>60\UGeV$.}
\label{fig:WG1-ggF-differential-parton-ptlarge}
\end{figure}

Next, in \refF{fig:Higgs-pT}
we show the Higgs boson transverse momentum distribution at
low  $p_T$. 
Note that all predictions but CuTe have bins with
width of 10~GeV. The  CuTe, MRT and HRes results are in good
agreement throughout the $p_T$ range, with some differences appearing
for $p_T\lsim 10$~GeV. Again the NLO prediction from MG5\_aMC@NLO is
lower and with a somewhat different shape; similar considerations
apply to NRV at high $p_T$ which has the same fixed-order accuracy. At
low $p_T$ NRV does not appear to have a Sudakov peak at the same $p_T$
value as other results. NNLOPS follows closely the HRes result, with
only minor differences in the smallest $p_T$ bin.

The high $p_T$ region for the same distributions is shown in
\refF{fig:WG1-ggF-differential-parton-ptlarge}. Note that 
predictions should  be taken with care for $p_T\gsim m_T$, as they are
all obtained in the infinite top mass approximation. As above, HRes and
Powheg NNLOPS agree well within uncertainties for all $p_T$ values. At
high $p_T$ MRT and NNLOJET display a somewhat harder spectrum because
they include NNLO corrections  to the one-jet configuration. The CuTe
prediction agrees well with HRes for all $p_T$ values.
The NRV prediction
appears to have a somewhat different shape, and it overshoots the HRes
prediction at the largest $p_T$ despite not including NNLO corrections.  
Uncertainties are all comparable, with the MRT uncertainty smallest as
expected since it includes the N$^3$LO correction to the inclusive
result.

\begin{figure}\centering
\includegraphics[width=0.49\textwidth]{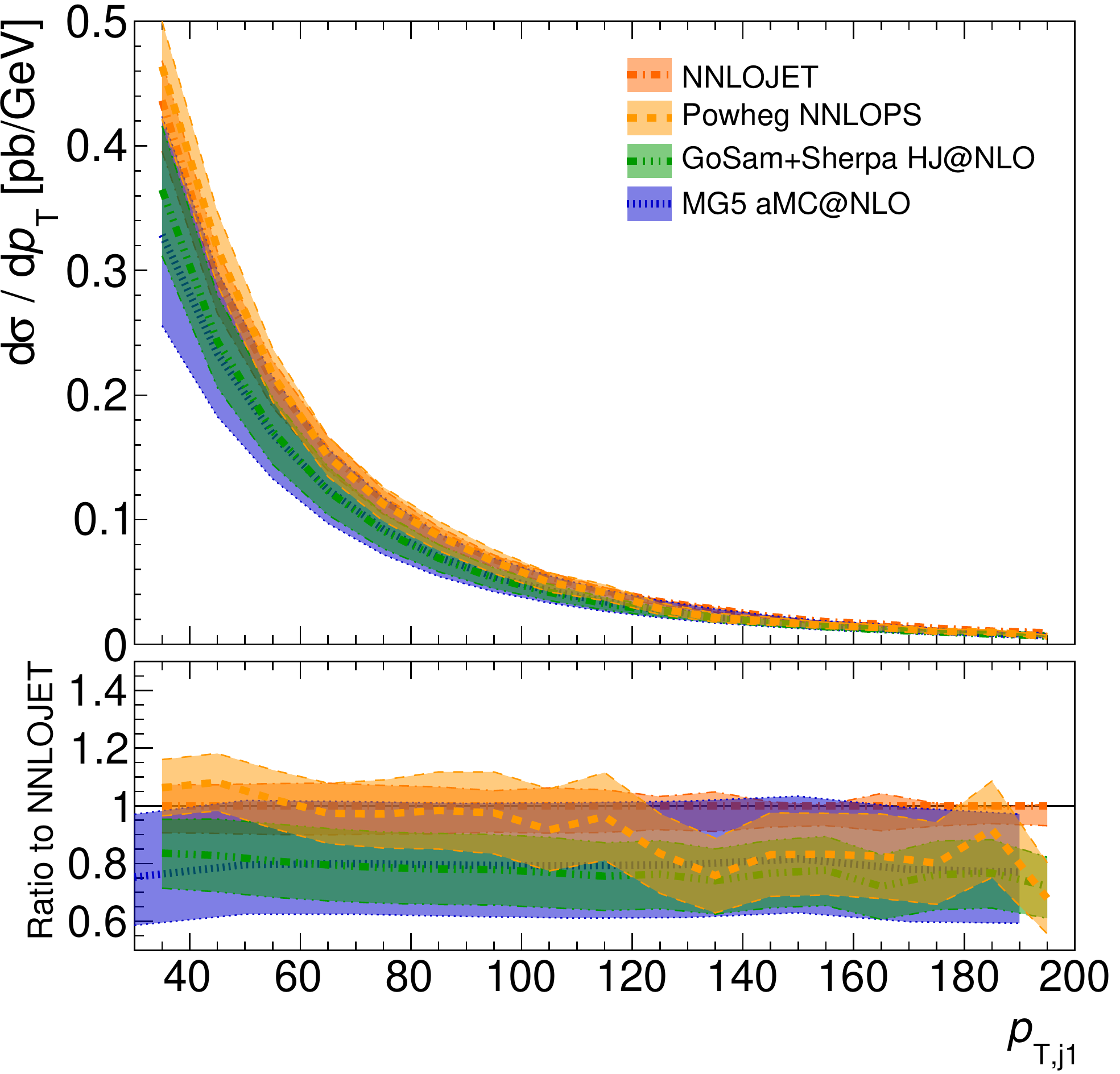}
\includegraphics[width=0.49\textwidth]{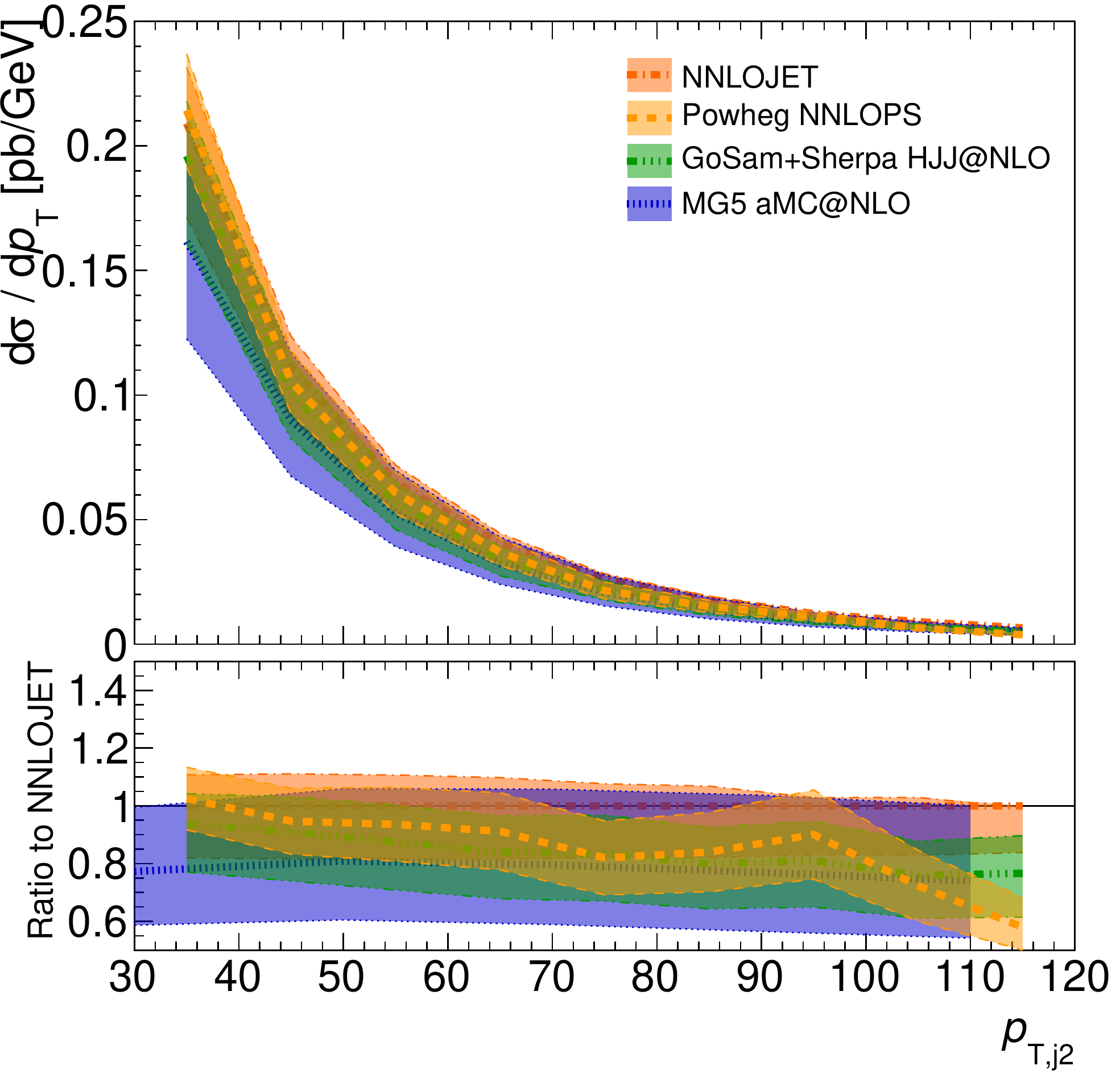}
\vspace*{-0.3cm}
\caption{
  Leading (left) and subleading (right jet $p_T$ distributions.
}
\label{fig:WG1-ggF-differential-hadron-pt1}
\end{figure}

We now turn to the leading and subleading jet $p_T$ distributions,
shown in \refF{fig:WG1-ggF-differential-hadron-pt1}. For the
leading jet (left plot) the NNLOJET result, which is NNLO, is higher and
with smaller uncertainty than GoSam+Sherpa and MG5\_aMC@NLO, which
agree well with each other. The Powheg prediction is affected by large
statistical fluctuations, but the shape can be understood by noting
that at low $p_T$ it agrees with NNLOJET as it includes the NNLO
correction to Higgs boson production, while at high $p_T$ it reproduces the
behaviour of the other NLO Monte Carlos. For the subleading jet
(right plot) all predictions but Powheg NNLOPS have the same NLO
accuracy and agree  within uncertainties. Powheg NNLOPS on the other
hand is leading-order only and its uncertainty is known to be somewhat
underestimated, yet it is in reasonable agreement with the other
results.

\begin{figure}\centering
\includegraphics[width=0.49\textwidth]{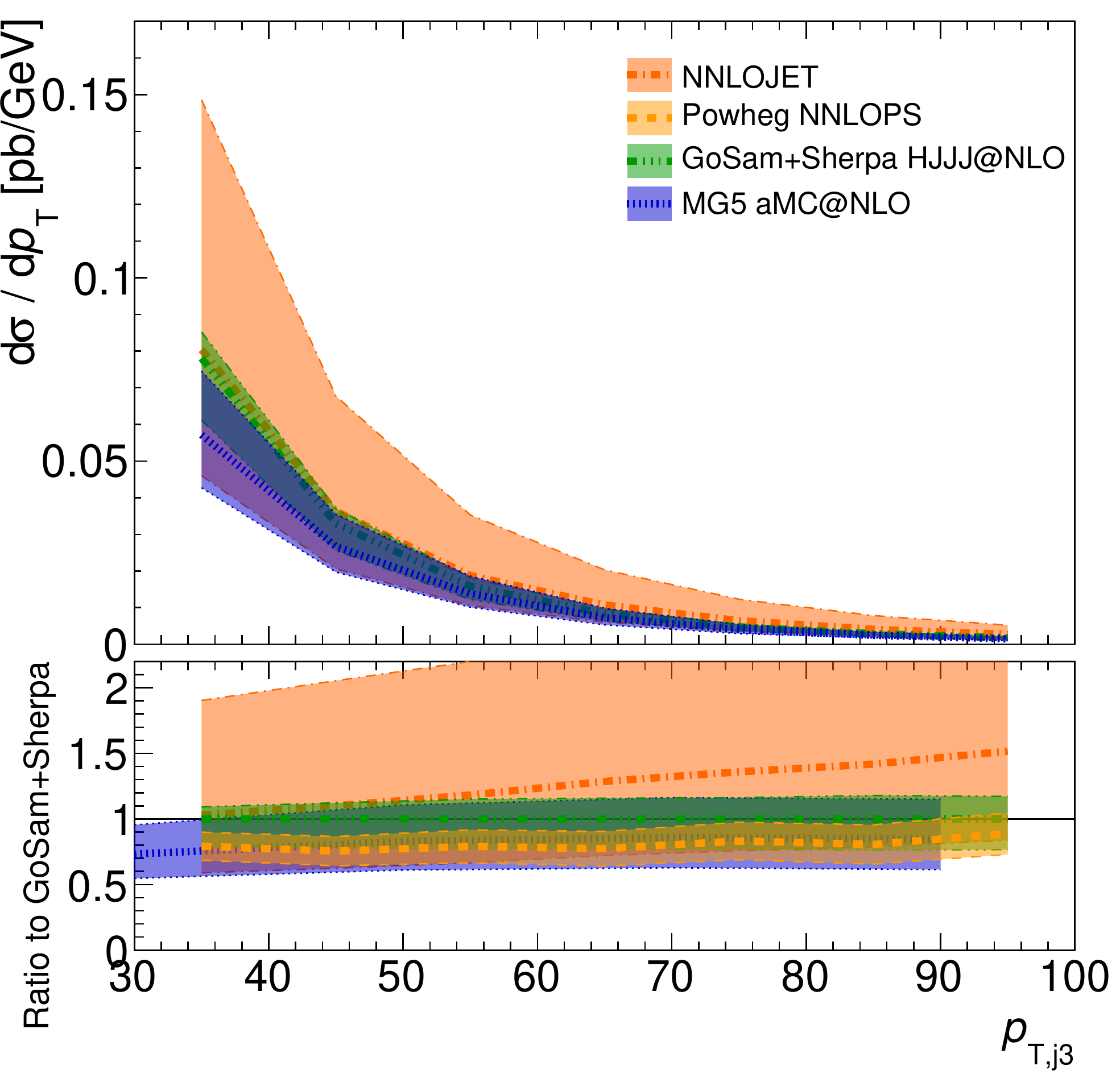}
\includegraphics[width=0.49\textwidth]{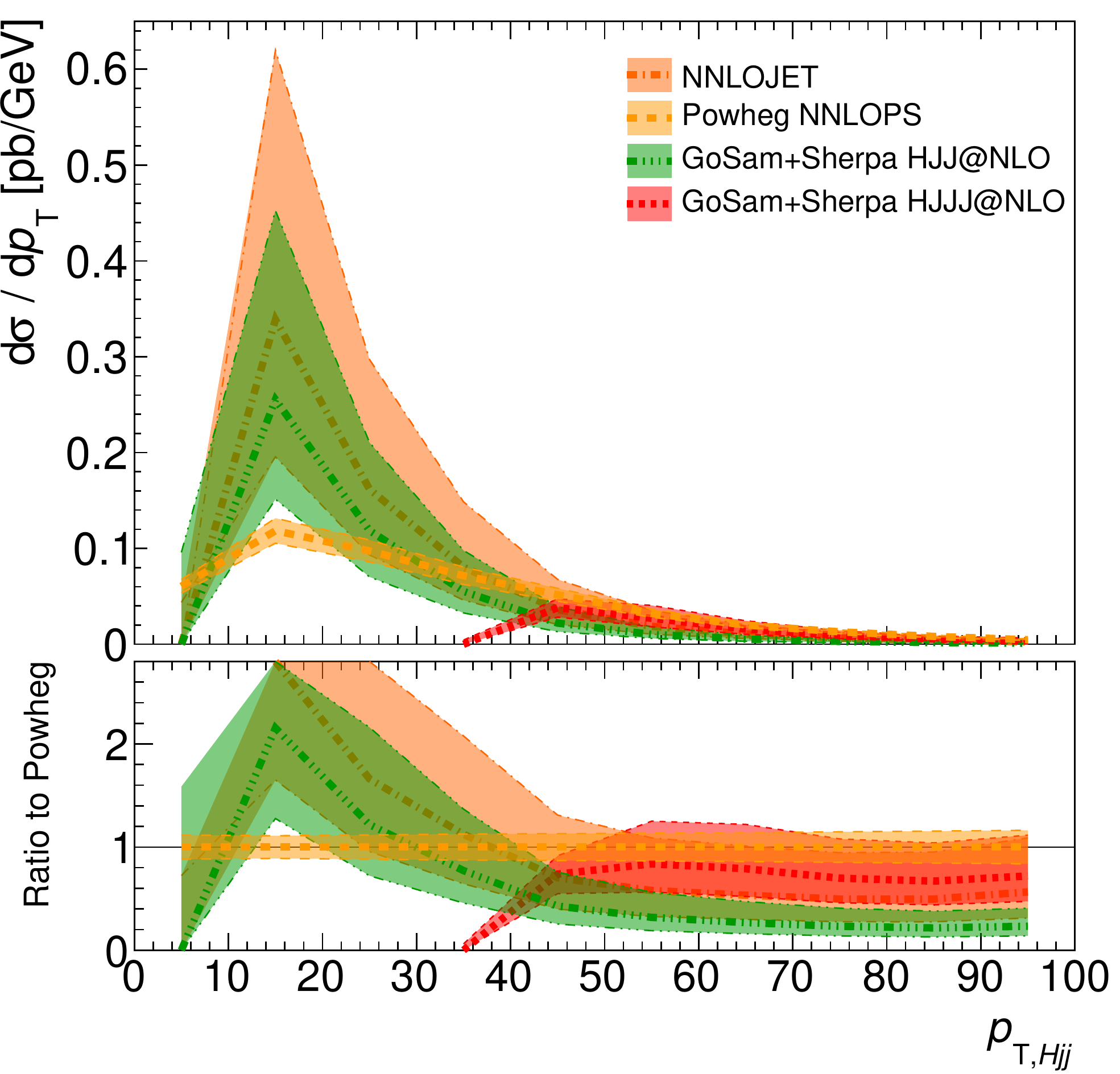}
\vspace*{-0.3cm}
\caption{ The transverse momentum of the 
  third jet  (left) and of the Hjj system (right). }
\label{fig:WG1-ggF-differential-hadron-pt2}
\end{figure}

In \refF{fig:WG1-ggF-differential-hadron-pt2} we show the
 transverse momentum of the 
  third jet  (left) and of the Hjj system (right). These two
 distributions start at $O(\alpha_s^5)$  and thus they coincide in the
 NNLOJET computation, which provides a purely leading-order description
 of these quantities and is thus affected by a large uncertainty. On
 the other hand, the
 $p_T$ of the 
 third jet (left plot) is described by 
 GoSam+Sherpa   at NLO, and 
in this case the
 uncertainty is reduced. The Powheg NNLOPS result for this
 distribution
 agrees well with these computations
 despite the fact that the third jet is only given by the parton
 shower. As far as the Hjj $p_T$ is concerned now GoSam+Sherpa HJJ@NLO
 and NNLOJET both provide a leading-order description while 
GoSam+Sherpa HJJJ@NLO provides  a NLO
 description at high $p_T$. The leading order
 predictions agree well within their large uncertainties. The NLO
 correction is positive at large $p_T$ but the uncertainty is not
 significantly reduced~\cite{Greiner:2015jha}. At lower $p_T$ the HJJJ@NLO becomes unreliable as
 it misses part of the NLO correction since the transverse momentum
 of all jets, including the third one,  is by 
 construction $p_T>30$~GeV.
  The Powheg NNLOPS follows, and in fact exceeds HJJJ@NLO  at high
 $p_T$ but it lies below the LO result  at lower $p_T$.

\begin{figure}\centering
\includegraphics[width=0.32\textwidth]{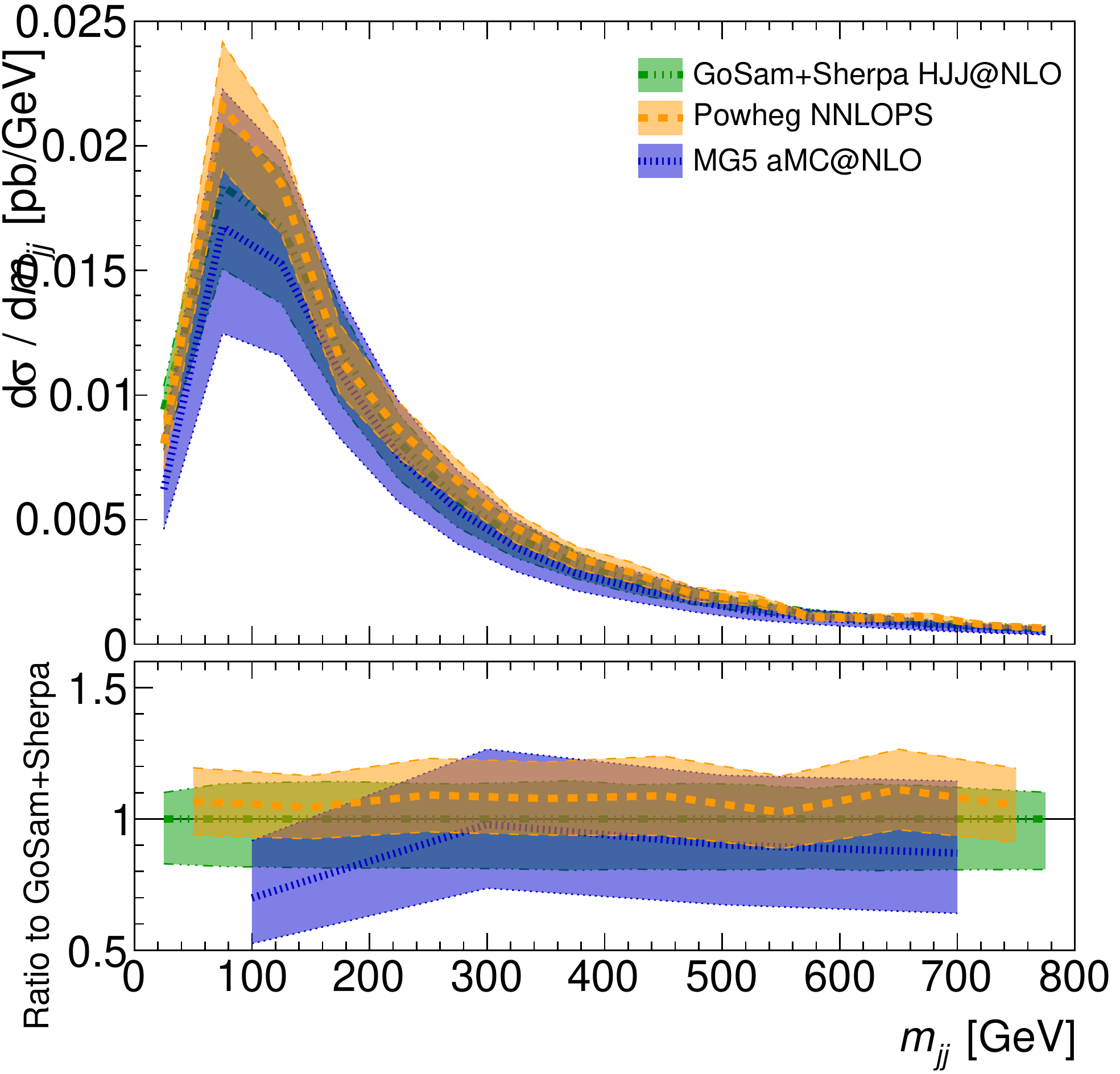}
\includegraphics[width=0.32\textwidth]{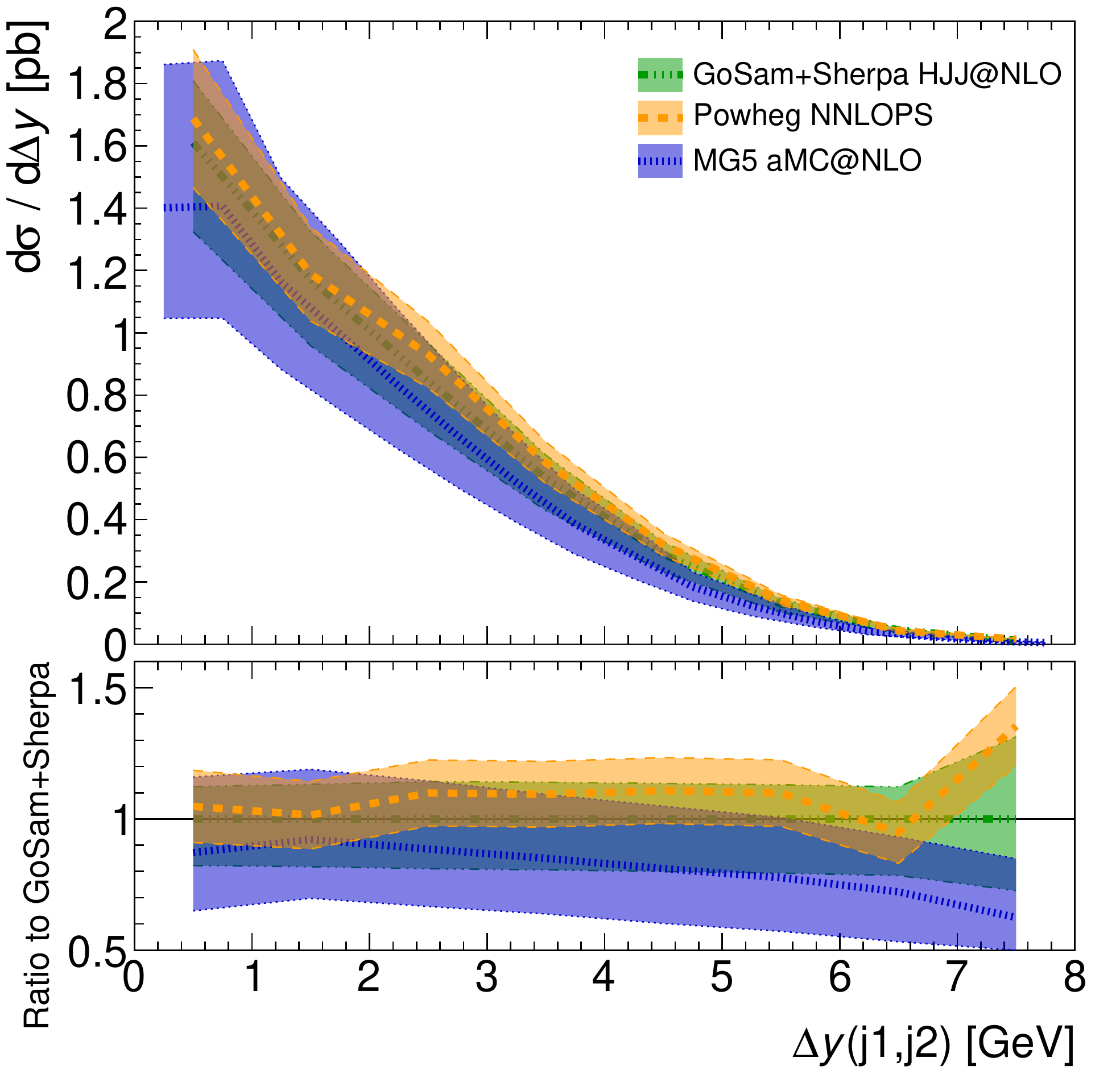}
\includegraphics[width=0.32\textwidth]{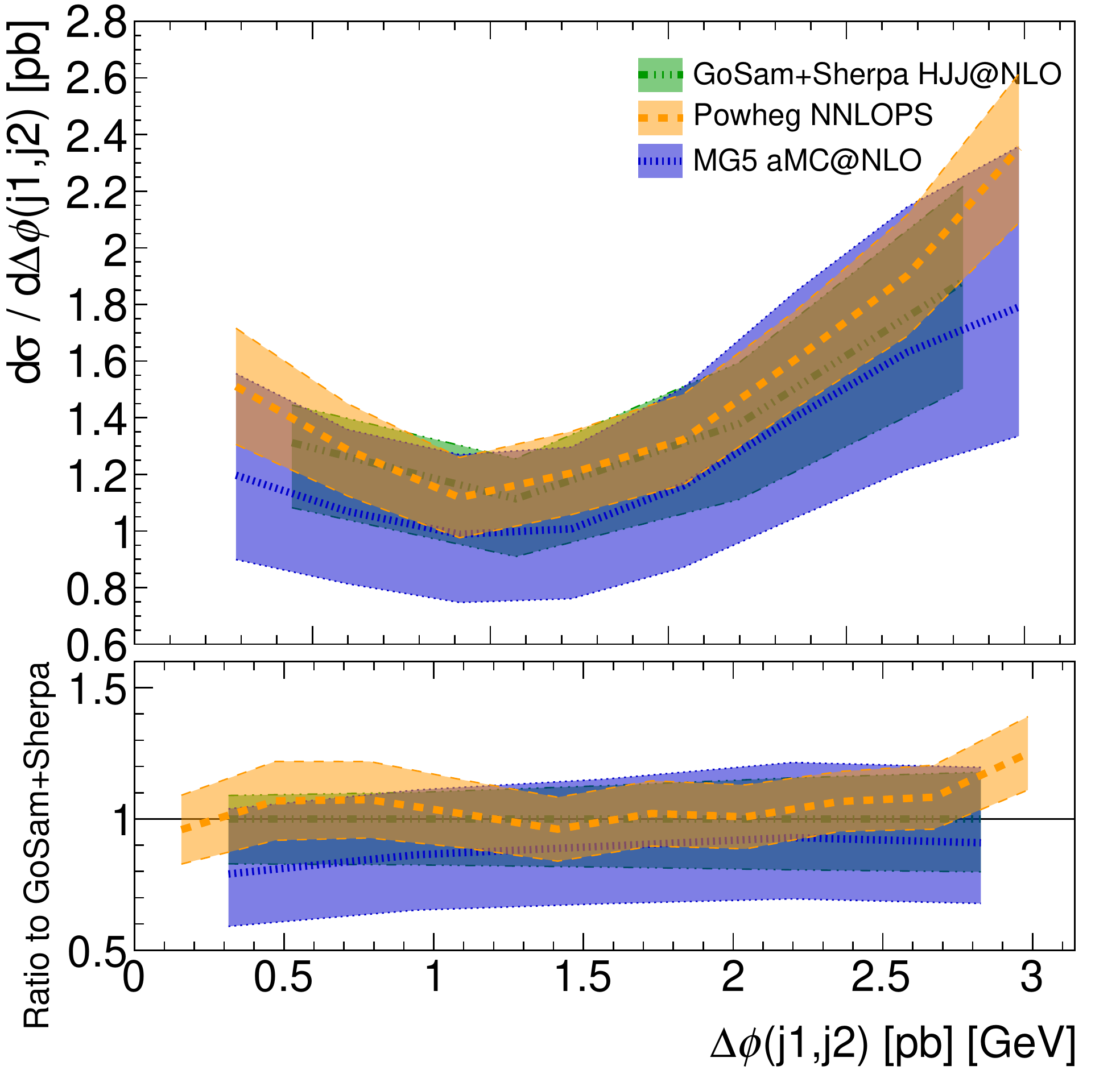}
\vspace*{-0.3cm}
\caption{Two-jet variables typically used in VBF studies: from left to
 right, mass of the dijet system, distance in rapidity, and azimuthal angle between the two
 jets.
}
\label{fig:WG1-ggF-differential-hadron-pt3}
\end{figure}

Finally in \refF{fig:WG1-ggF-differential-hadron-pt3} we show
variables relevant for VBF studies. Cross-sections are also tabulated
in \refT{tab:benchmarkVBF}. All results agree reasonably well
within the large uncertainties. 

\begin{table}
\caption{Predicted cross sections for $gg\to H$ with VBF topology. The QCD uncertainties shown for {\sc Powheg Nnlops} are not valid (the third jet is from the showering).
Fixed order NLO predictions with two and three jets are provided by {\sc GoSam+Sherpa}.
The last two rows show results from normalizing the inclusive cross section to 46.18~pb.}\label{tab:benchmarkVBF}
\renewcommand{\arraystretch}{1.2}
\setlength{\tabcolsep}{1.5ex}
\begin{center}
\begin{tabular}{l|rrr|rrr}
\toprule
 & \multicolumn{3}{c|}{$m_{jj}>400\UGeV$, $\Delta y_{jj}>2.8$} &
   \multicolumn{3}{c}{$m_{jj}>600\UGeV$, $\Delta y_{jj}>4.0$} \\
&              & \multicolumn{2}{c|}{$p_{T,j3}$ / GeV}
&              & \multicolumn{2}{c}{$p_{T,j3} $ / GeV} \\

Prediction
& no jet veto & \multicolumn{1}{c}{$<30$} & \multicolumn{1}{c|}{$>30$}
& no jet veto & \multicolumn{1}{c}{$<30$} & \multicolumn{1}{c}{$>30$} \\
\midrule

{\sc Powheg Nnlops}
& $653^{+86}_{-86}$~fb  & $435^{+54}_{-54}$~fb & $218^{+32}_{-32}$~fb
& $283^{+36}_{-36}$~fb  & $198^{+24}_{-24}$~fb & $85^{+12}_{-12}$~fb \\

MG5\_ aMC@NLO
& $512^{+152}_{-133}$~fb  & $329^{+92}_{-84}$~fb& $183^{+67}_{-54}$~fb
& $214^{+62}_{-57}$~fb    & $142^{+39}_{-37}$~fb& $72^{+29}_{-21}$~fb\\
  {\sc hjj@nlo}
&  $610^{+74}_{-120}$~fb  & $435^{+0}_{-70}$~fb  & \multicolumn{1}{c|}{$-$}
& $268^{+32}_{-55}$~fb& $195^{+0}_{-31}$~fb & \multicolumn{1}{c}{$-$}\\

{\sc hjjj@nlo}
& \multicolumn{1}{c}{$-$} & \multicolumn{1}{c}{$-$} & $240^{+17}_{-54}$~fb
& \multicolumn{1}{c}{$-$} & \multicolumn{1}{c}{$-$} &  $97^{+5}_{-22}$~fb\\
\midrule

{\sc  Nnlops}, $k=1.05$
& $683^{+90}_{-90}$~fb& $455^{+57}_{-57}$~fb & $228^{+33}_{-33}$~fb
& $296^{+38}_{-38}$~fb& $207^{+25}_{-25}$~fb & $89^{+13}_{-13}$~fb\\
  MG5, $k=1.41$
& $721^{+214}_{-188}$~fb & $463^{+129}_{-118}$~fb & $258^{+94}_{-76}$~fb
& $302^{+87}_{-80}$~fb& $200^{+55}_{-52}$~fb & $101^{+41}_{-29}$~fb\\

\bottomrule
\end{tabular}
\end{center}
\end{table}

\section[Effects of heavy-quark masses]{Effects of heavy-quark masses\SectionAuthor{F.~Krauss, S.~Kuttimalai, P.~Maierh\"ofer, M.~Sch\"onherr}}
The Higgs Effective Field Theory (HEFT) framework for perturbative
calculations of the gluon fusion Higgs boson production process is a well
established tool that allows a significant reduction of complexity
in higher-order QCD calculations.  In this approach, the
heavy-quark-loop induced Higgs-gluon coupling of the Standard Model
(SM) is approximated by taking into account only the top quark
contribution and by calculating production amplitudes in the limit of
an infinite top quark mass. This is typically achieved by deriving the
relevant amplitudes from the effective Lagrangian
\begin{align}
\mathcal{L}_\mathrm{HEFT} = -\frac{C_1}{4v} H G_{\mu\nu}G^{\mu\nu}\, ,
\end{align}
with the gluon field strength tensor $G_{\mu\nu}$, the Higgs field
$H$, and a perturbatively calculable Wilson coefficient $C_1$. This
Lagrangian gives rise to tree-level couplings that replace the
loop-induced SM couplings between gluons and the Higgs boson, effectively
reducing the number of loops in any calculation by one.

When considering the total inclusive Higgs boson production cross section,
finite top quark mass effects remain very moderate even at higher
orders in QCD~\cite{Marzani:2008az,Pak:2009bx,Pak:2009dg,
  Harlander:2009my,Harlander:2009mq,Harlander:2009bw}.
In the tail of the transverse momentum spectrum of the Higgs boson
or for heavy Higgs boson (virtual) masses, however, the corrections
can become very large, indicating a complete breakdown of the HEFT
approximation~\cite{Baur:1989cm,Ellis:1987xu}. It has also
been known for a long time that the bottom quark loops, which are
entirely neglected in the HEFT, affect the spectrum in the
small-$p_\perp^H$ region \cite{Ellis:1987xu,Keung:2009bs}. In this
region, an all-order resummation of QCD corrections is
required. Standard techniques need to be adapted in order to achieve
this due to the bottom quark mass that introduces an additional scale
into the calculation \cite{Grazzini:2013mca}.

Several fully differential Monte Carlo codes have therefore been
developed that take into account the full heavy quark mass dependence
at NLO~\cite{Langenegger:2006wu,Anastasiou:2009kn,Bagnaschi:2011tu,
  Grazzini:2013mca}.
NLO results for Higgs boson production in association with a jet are not
available for finite heavy quark masses due to missing two-loop
amplitudes for this process.

In this note, we present an approximate treatment of finite top mass
effects at NLO based on one-loop amplitudes only. This allows us to
calculate Higgs plus $n$-jet processes at NLO, while retaining finite
top mass effects in an approximate way. Using this approximation, we
employ multijet merging techniques~\cite{Hoeche:2012yf} to
merge higher-multiplicity NLO processes matched to a parton shower
into one exclusive event sample, extending similar
approaches~\cite{Catani:2001cc,Krauss:2002up,Lonnblad:2001iq,Alwall:2011cy}
in terms of jet multiplicity and $\alpha_s$ accuracy.
Based on leading order merging, we also suggest a method to address
the issues raised in \cite{Grazzini:2013mca} concerning the inclusion
of bottom quark contributions in the low-$p_\perp^H$ region.

\subsection{Implementation of quark mass corrections}

In order to take into account the full heavy quark mass effects in the
hard scattering at leading order, we replace the approximate HEFT
tree-level matrix elements provided by Sherpa's matrix element
generator Amegic++~\cite{Krauss:2001iv} with the exact one-loop matrix
elements provided by OpenLoops~\cite{Cascioli:2011va} in combination
with Collier~\cite{Denner:2014gla}. This allows the calculation of
processes with up to three additional jets in the final state at
leading order, with the full top and bottom quark mass dependency
taken into account.

At NLO, the cross section for the production of a Higgs boson accompanied by
a certain number $m-1$ of jets receives contributions from two
integrals of different phase space dimensionality.
\begin{align}
  \sigma = \int (B+V+I)d\phi_m + \int (R-S)d\phi_{m+1} \label{gghmq_nloxs}
\end{align}
The born term $B$ and the real emission term $R$ are present already
at leading order for processes of the respective jet multiplicity and
can be corrected as in the leading order calculation. $I(\phi_m)$ and
$S(\phi_{m+1})$ denote the integrated and differential Catani-Seymour
subtraction terms, respectively~\cite{Catani:1996vz}. They render both
integrals separately finite and are built up from leading-order
$m$-particle matrix elements dressed with appropriate splitting
kernels and can henceforth be corrected by using the full one-loop
matrix elements instead of the tree-level HEFT approximation. Note
that because we correct $R$ and $S$ with matrix elements of different
final state multiplicity, the mere convergence of the corresponding
integral already provides a crucial test for our implementation.

The IR-subtracted virtual correction $V$ receives contributions from
two-loop diagrams when taking into account the full heavy quark mass
dependencies. Since these amplitudes are available only for the Higgs boson 
plus zero-jet final state, we employ an ad-hoc approximation that only
involves one-loop matrix elements (even for the Higgs boson plus zero-jet
final state). We assume a factorization of the $\alpha_s$-corrections
from the quark mass corrections and set
\begin{align}
V = V_\mathrm{HEFT}\frac{B}{B_\mathrm{HEFT}}\,.\label{gghmq_approx}
\end{align}
In this approximation, we can straightforwardly apply finite top mass
corrections in simulations employing CKKW multi jet merging at NLO in
the MEPS@NLO scheme~\cite{Hoeche:2012yf}.

We expect the approximation~\eqref{gghmq_approx} to give reasonable
results only if the HEFT-approximation is valid. For any contribution
involving the bottom Yukawa coupling $y_b$ , it cannot be used due to
the small bottom quark mass. This applies to the interference terms
proportional to $y_ty_b$ as well as the squared bottom contributions
proportional to $y_b^2$. We therefore calculate terms that involve
$y_b$ as separate processes at leading order. The NLO corrections to
the total inclusive cross sections for the $y_ty_b$ contributions and
the $y_b^2$ contributions are only ${\cal{O}}{1}\%$ and ${\cal{O}}{20}\%$, 
respectively~\cite{Bagnaschi:2015bop}.  Furthermore, the $y_b^2$ 
contributions featuring the slightly larger NLO K-factor are 
significantly suppressed compared to the $y_ty_b$ 
terms~\cite{Bagnaschi:2015bop}. We therefore consider a treatment at
leading order sufficiently accurate. Any terms proportional to $y_t^2$
will however be calculated at NLO in the approximation described
above.

\subsection{Finite top mass effects}

\begin{figure}
  \centering
    \includegraphics[width=0.45\textwidth]{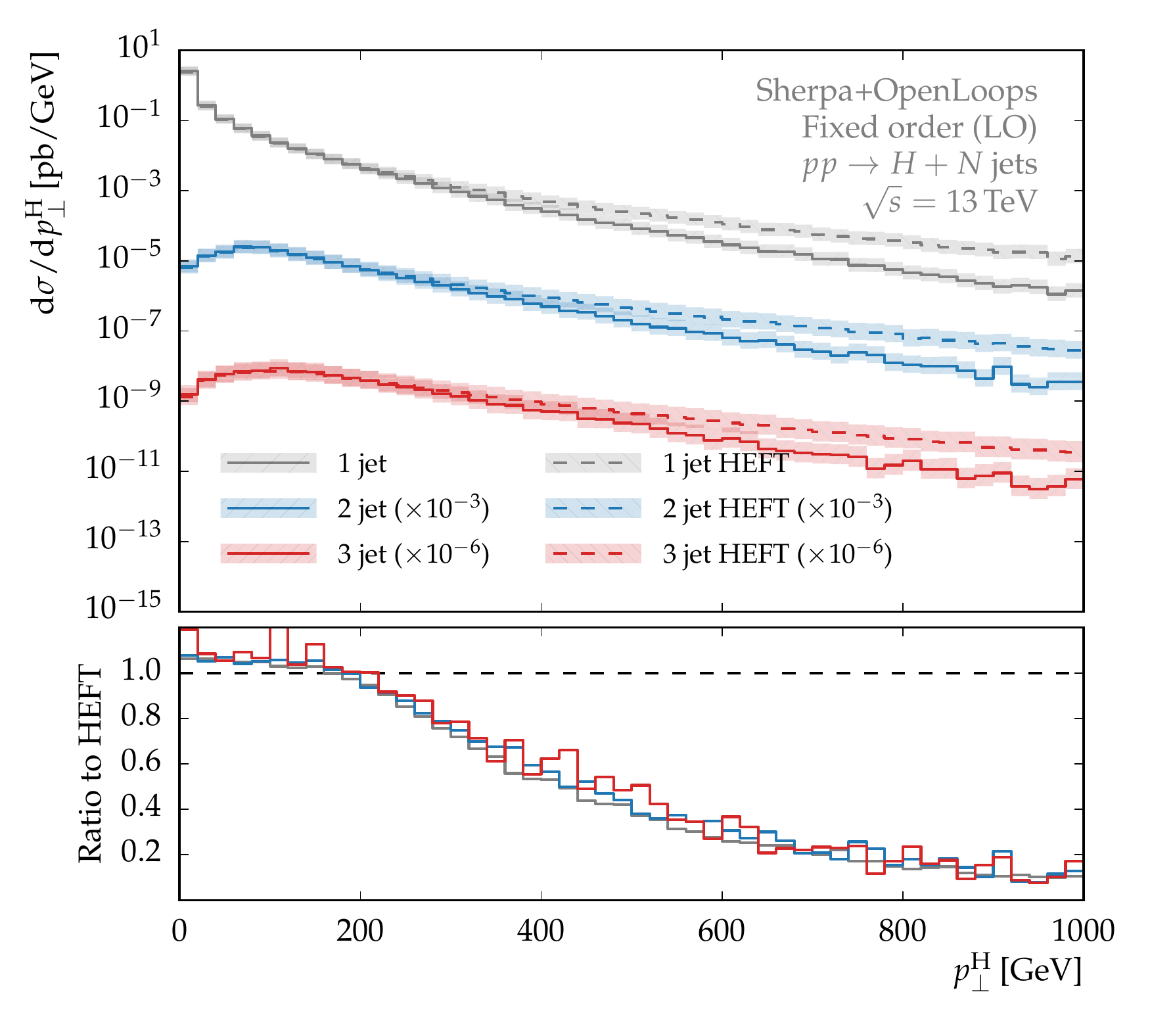}
  \hfill
    \includegraphics[width=0.45\textwidth]{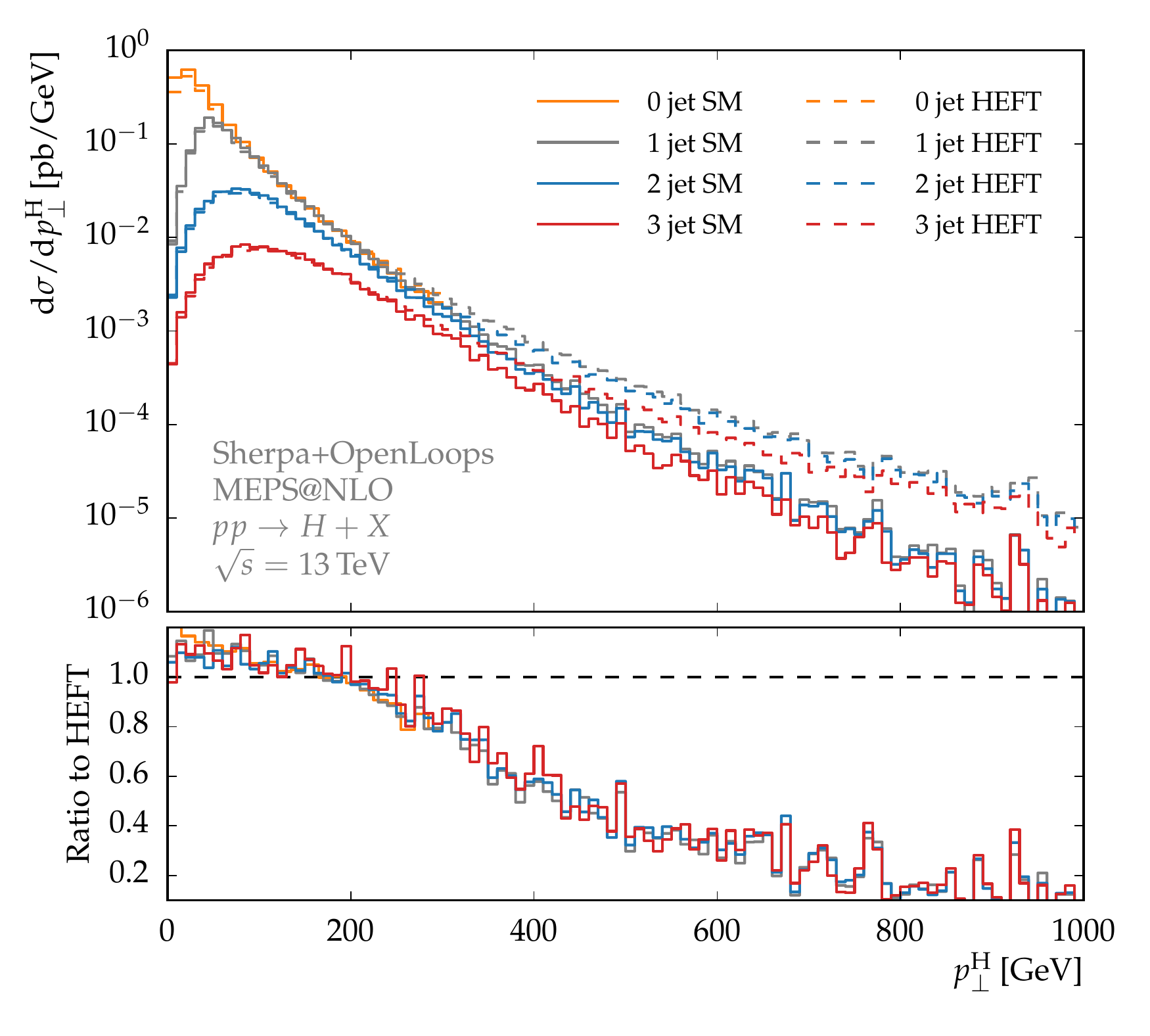}
    \caption{  The Higgs boson transverse momentum spectrum in gluon fusion. We
    show individual curves for the HEFT approximation (dashed) and the
    full SM result taking into account the mass dependence in the top
    quark loops. The lower panel shows the ratio of the SM results to
    the HEFT approximation.
    Left: LO fixed order calculation for up to three jets. The
    error bands indicate the uncertainties obtained from variations
    of the factorization and renormalization scales.  
    Right: Multijet merged calculation. We include the zero and one jet 
    final states at NLO as well as the two jet final state at leading order. 
    The individual curves show inclusive $N$-jet contributions.}
    \label{gghmq_meps}
\end{figure}

As mentioned in the introduction, the total inclusive cross section is
only mildly affected by finite top mass effects. The low-$p^H_\perp$
region, where the bulk of the cross section is located, can therefore
be expected to exhibit only a moderate dependence on the top quark
mass. In kinematic regimes where any invariant significantly exceeds
$m_t$, however, we expect the HEFT approximation to break down. The
$p_\perp^H$ distributions in \refF{gghmq_meps} (left) exemplify
this picture. We show Higgs boson transverse momentum distributions for
final states with one, two, and three jets calculated at fixed leading
order. Jets are reconstructed using the anti-$k_T$ algorithm with a
radius parameter of $R=0.4$ and a minimum jet $p_\perp$ of
$\sim{30}$~GeV except in the 1-jet case, where we apply
only a small minimum $p_\perp$-cut of $\sim{1}$~GeV. The
distributions for all three jet multiplicities exhibit a very similar
pattern when comparing the full SM result to the HEFT approximation.
Below $p_\perp^H\approx m_H$, we observe a flat excess of around
$\sim{6}\%$ that recovers the correction factor to the total
inclusive Higgs boson production cross section at leading order. The
deviations become very large when $p_\perp^H$ significantly exceeds
$m_t$, as expected. The similarity of the top mass dependency of the
$p_\perp^H$ spectrum for all jet final multiplicities confirms similar
findings for one- and two-jet configurations in
\cite{Buschmann:2014twa}.

In \refF{gghmq_meps} (right), we show analogous results obtained 
from the MEPS@NLO simulation. We included NLO matrix elements for the 
zero- and one-jet final states as well as leading order matrix elements 
for the two-jet final state in the merged setup and set $Q_\mathrm{cut}$
to $\sim{30}$~GeV. From the ratio plot in \refF{gghmq_meps} it is evident that in our approximation we recover
the same suppression patterns as in the respective fixed leading order
calculations for all jet multiplicities. This is a nontrivial
observation as an $m$-jet configuration receives corrections from
$m$-jet matrix elements as well as from $m+1$-jet matrix elements
through the real emission corrections $R$ in \eqref{gghmq_approx}.

\subsection{Nonzero bottom mass effects}

\begin{figure}
  \centering
  \includegraphics[width=.48\textwidth]{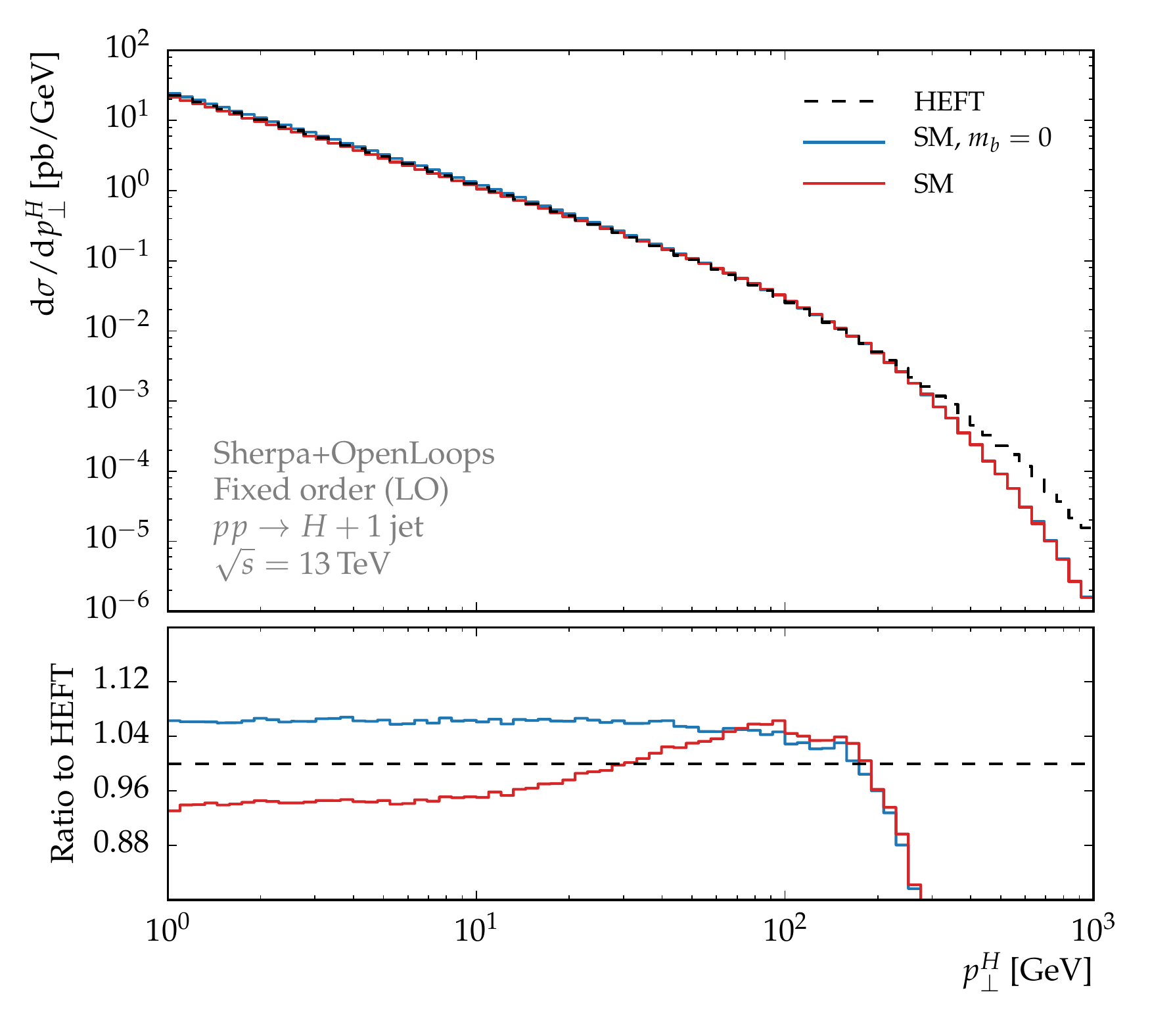}
  \caption{Bottom quark mass effects at fixed leading order. The
      minimum jet $p_\perp$ is set to $\sim{1}$~GeV in
      order to display map out the low $p_\perp^H$ region as well.}
    \label{gghmq_tb_fo}
  \end{figure}
  
As pointed out already in~\cite{Ellis:1987xu,Keung:2009bs}, the
inclusion of the bottom quark in the loops affects the $p_\perp^H$
distribution only at small values of $p_\perp^H$ around $m_b$. In
\refF{gghmq_tb_fo} we reproduce these findings for the process
$pp\rightarrow H+j$ at fixed order. In the $p_\perp$ range around
$m_b$ where the bottom effects are large, a fixed order prediction is
of course unreliable due to the large hierarchy of scales between
$m_H$ and the transverse momentum. This large separation of scales
induces Sudakov logarithms $\ln(m_H/p_\perp)$ that spoil any fixed
order expansion and require resummation.

It was argued in~\cite{Grazzini:2013mca} that the resummation of these
logarithms is complicated by the presence of the bottom quark in loops
that couple to the Higgs boson. The bottom quark introduces $m_b$ as
an additional scale above which the matrix elements for additional QCD
emissions do not factorize. Since a factorization is essential for the
applicability of resummation techniques, it was proposed to use a
separate resummation scale of the order of $m_b$ for the contributions
involving $y_b$, thereby restricting the range of transverse momenta
where resummation is applied to the phase space where factorization is
guaranteed. Two quantitative prescriptions have been proposed for the
determination of a specific numerical value for the resummation scale
of the bottom
contributions~\cite{Bagnaschi:2015qta,Harlander:2014uea}. These two
methods yield numerical values of $\sim{9}$~GeV and
$\sim{31}$~GeV~\cite{Bagnaschi:2015bop} for the dominant
top-bottom interference terms. In addition to $m_b$, we will therefore
consider these values for our numerical studies. The pure top quark
contributions proportional to $y_t^2$ will be treated as usual, with
the resummation scale set to $m_H$.

\begin{figure}
  \centering
    \includegraphics[width=0.45\textwidth]{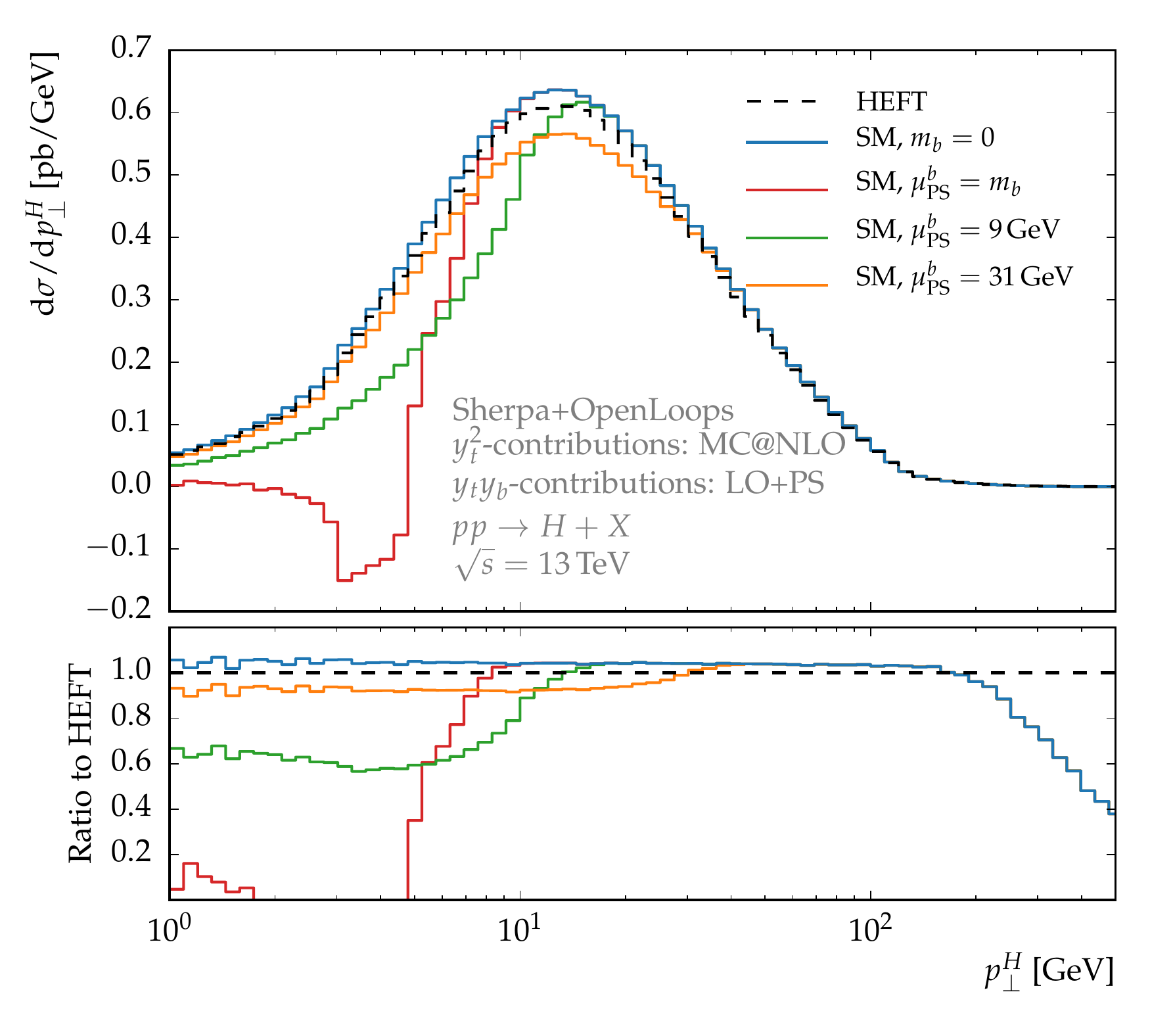}
  \hfill
    \includegraphics[width=0.45\textwidth]{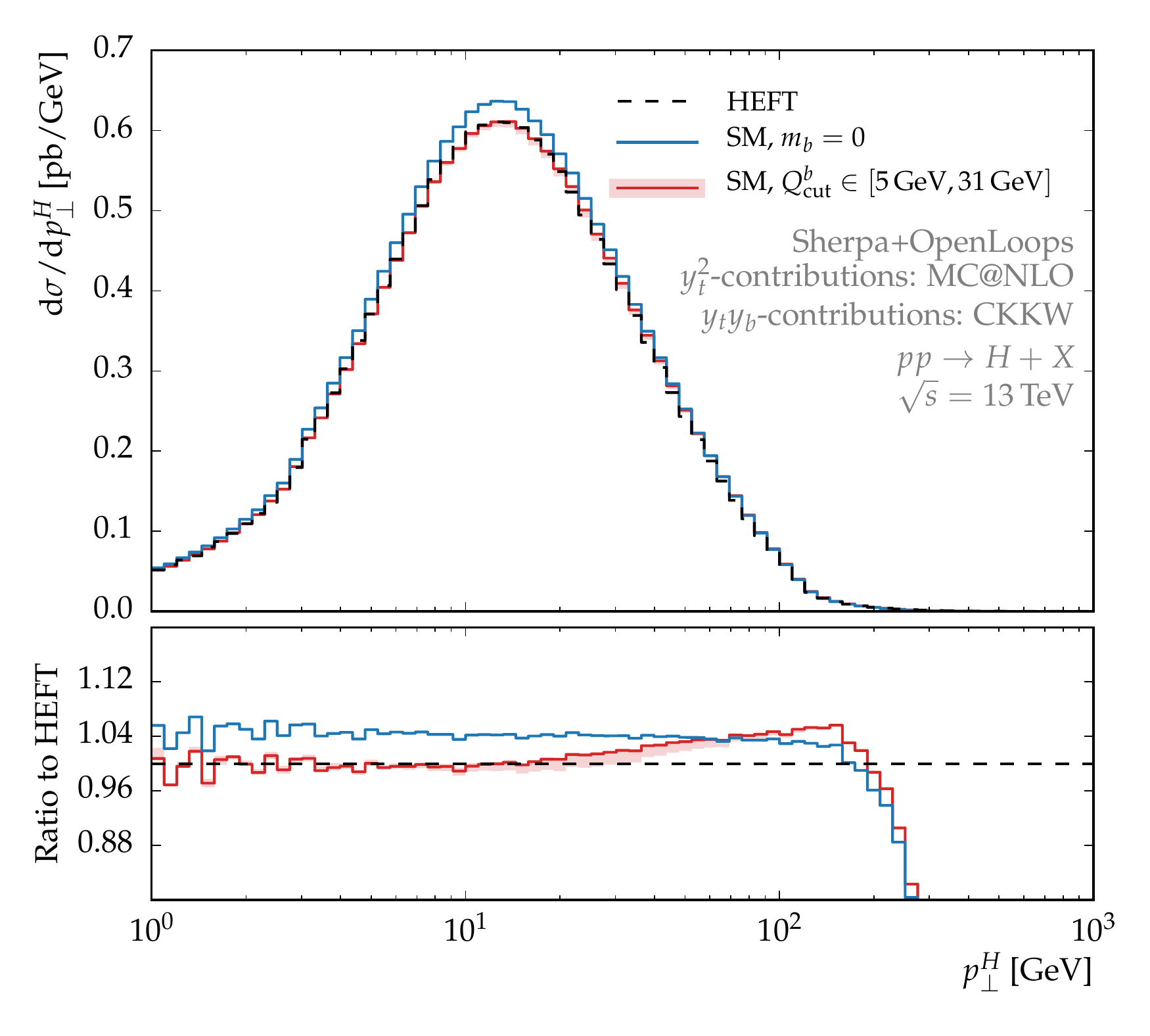}
    \caption{Left: Bottom quark mass effects at LO+PS accuracy with small
      parton shower starting scales $\mu_\mathrm{PS}^b$. The specific
      values chosen for $\mu_\mathrm{PS}^b$ are motivated in the
      text.
      Right: Bottom quark mass effects taken into account by means of
      CKKW merging with a small merging scale
      $Q^b_\mathrm{cut}=\mathcal{O}(m_b)$. The red error band shows
      variations of this scale as indicated.}
    \label{gghmq_tb_meps}
\end{figure}

While reference~\cite{Grazzini:2013mca} was concerned with analytical
resummation techniques, similar approaches were studied in the context
of NLO-matched parton shower Monte Carlos
\cite{Bagnaschi:2015qta,Harlander:2014uea,Bagnaschi:2015bop}. Our
discussion will be restricted to the leading order as the
approximation used for the NLO calculation of the top quark
contributions \eqref{gghmq_approx} is invalid for the bottom quark
terms. The equivalent of the resummation scale in analytic
calculations is the parton shower starting scale $\mu_\mathrm{PS}$
because it restricts parton shower emissions to the phase space below
this scale and because this scale enters as the argument in the
Sudakov form factors. Using separate parton shower starting scales for
the top and the bottom contributions, respectively, requires to
generate and shower them separately as well. A corresponding
separation of terms in the one-loop matrix elements can be achieved
with OpenLoops. By means of this separation into terms proportional to
$y_t^2$ and the remainder, we can generate an MC@NLO sample for the
top quark contributions while calculating the terms involving $y_b$ at
leading order. \refF{gghmq_tb_meps} (left) shows the $p_\perp^H$
spectrum obtained this way. We show results with $\mu_\mathrm{PS}^b$
set to $m_b$, $\sim{9}$~GeV, and
$\sim{30}$~GeV as motivated above. The parton shower
starting scale for the top quark contributions will be
$\mu_\mathrm{PS}^t=m_H$ throughout. For small $\mu_\mathrm{PS}^b$, the
suppression in the low $p_\perp$ region below $m_b$ is much more
pronounced than in the fixed order result in \refF{gghmq_tb_fo}.
This is because, pictorially speaking, without changing the cross
section of the individual contributions, the parton shower simulation
spreads the $y_t^2$ part over a much wider range, up to
$\mathcal{O}(m_H)$, than for the negative $y_ty_b$, up to
$\mathcal{O}(m_b)$ only. The spectrum in this region is therefore
extremely sensitive to variations of $\mu_\mathrm{PS}^b$. When varying
$\mu_\mathrm{PS}^b$ to sufficiently low values, the differential cross
section may even become negative, clearly an unphysical result. We
stress, again, that this is not a physical effect but an artefact of
the unitary nature of the parton shower. Setting the value of
$\mu_\mathrm{PS}^b$ to a small value, the entire leading order bottom
cross section contributions will be distributed in a phase space with
Higgs boson transverse momenta not significantly exceeding
$\mu_\mathrm{PS}^b$. Since this cross section is negative, the
spectrum must become negative at some point when lowering
$\mu_\mathrm{PS}^b$.

We therefore suggest another approach at taking into account the
bottom quark contributions in a parton shower Monte Carlo simulation.
We account for the non-factorization of the real emission matrix
elements above some scale $Q^b_\mathrm{cut}$ by correcting parton
shower emissions harder than this scale with the appropriate fixed
order matrix elements. This can be done consistently in the CKKW
merging scheme~\cite{Catani:2001cc,Hoeche:2009rj}.  Setting the
merging scale for the bottom contributions $Q_\mathrm{cut}^b$ to $m_b$
allows the correction of the parton shower in the regime where the
matrix elements involving $m_b$ do not factorize (without restricting
all emissions to the phase space below).  Above $Q_\mathrm{cut}$, the
fixed-order accuracy of the real emission matrix
elements is thereby retained. Since any NLO prediction of the
inclusive process describes the $p_\perp$ spectrum only at leading
order, our approach retains the same parametric fixed order accuracy
when considering the $p_\perp^H$ distribution. Beyond fixed order, the
differences should be small since the NLO corrections to the inclusive
cross section are at the per cent level for the $y_ty_b$ interference
terms.

In \refF{gghmq_tb_meps} (right) we show the bottom quark effects on 
the $p_\perp^H$ spectrum in this approach. We include matrix elements with
up to one jet in the merging such that a leading order accuracy in
$\alpha_s$ is guaranteed for both the top and the bottom contributions
to the $p_\perp^H$ spectrum. This allows a comparison to 
\refF{gghmq_tb_fo}. The effects of the bottom quarks lead to a very
similar suppression pattern over the entire displayed range of
$p_\perp^H$. The large NLO K-factor that appears in the MC@NLO
calculations of the top contributions however affects the overall
relative normalization of the bottom quark effects. They are
correspondingly smaller by roughly $\sim{50}\%$ in 
\refF{gghmq_tb_meps} when compared to \refF{gghmq_tb_fo}. The
sensitivity to variations of the scale in the calculation that
effectively accounts for the presence of the bottom mass in the
problem is drastically reduced. \refF{gghmq_tb_meps} includes an
error band corresponding to variations of $Q_\mathrm{cut}^b$ in the
large interval between $m_b$ and $\sim{31}$~GeV. On the
displayed scale, these variations are hardly visible.

\subsection{Conclusions}

We presented in this section an implementation of heavy quark mass
effects in gluon fusion Higgs boson production that allows to systematically
include finite top mass effects in an approximate way at NLO for in
principle arbitrary jet multiplicities in the final state. Based on
this approximation, we presented results for the Higgs boson transverse
momentum distributions obtained from NLO matched and merged samples.
When comparing the top quark mass dependence in one-, two-, and
three-jet final states, we observed a universal suppression pattern
that agrees very well with the corresponding leading order results.

Our treatment of contributions involving the bottom Yukawa coupling is
based on $m_b$- and $m_t$-exact leading order matrix elements in
combination with tree-level multijet merging techniques. We argued
that this approximation is appropriate since it allows to retain
leading order accuracy for the corresponding contributions in the
$p_\perp^H$-spectrum and it also allows to account for the
non-factorization of real emission amplitudes at scales above $m_b$.
In this approach, the uncertainty associated with the appearance of
$m_b$ as an additional scale in the calculation is drastically
reduced.



\vspace{0.5cm}
\noindent
{\bf Acknowledgements}

The editors are grateful to the following people who have provided  numbers used for
the benchmarks for cross sections and differential distributions presented in Section~\ref{sec:jbenchmarks}: Thomas Becher, Massimiliano Grazzini, 
Nicolas Greiner, Thomas L\"ubbert, Gionata Luisoni, Duff Neill, Matthias Neubert, Hayk Sargsian, Ira Rothstein, Varun Vaidya, Daniel Wilhelm and Jan Winter.

\chapter{VBF and VH}
\label{chap:VBF+VH}
\ChapterAuthor{S.~Dittmaier, P.~Govoni, B.~J\"ager, J.~Nielsen, L.~Perrozzi, E.~Pianori, A.~Rizzi, F.~Tramontano~(Eds.);
W.A.~Astill, J.~Bellm, W.~Bizon, F.~Campanario, J.~Campbell, A.~Denner, F.A.~Dreyer, R.K.~Ellis, G.~Ferrera, T.~Figy, P.~Francavilla, R.~Frederix, S.~Frixione,
M.~Grazzini, R.~Harlander, B.~Hespel, J.~Huston, S.~Kallweit, A.~Karlberg, F.~Krauss, A.~Kulesza, G.~Luisoni, S.-O.~Moch, A.~M\"uck, C.~Oleari, C.E.~Pandini, A.~Papaefstathiou, S.~Pl\"atzer, E.~Re, G.P.~Salam, P.~Schichtel, M.~Sj\"odahl, V.~Theeuwes, P.~Torrielli, E.~Vryonidou, C.~Williams, G.~Zanderighi, M.~Zaro, T.~Zirke}
\providecommand{\muR}{\mathswitch {\mu_{\mathrm{R}}}}
\providecommand{\muF}{\mathswitch {\mu_{\mathrm{F}}}}
\providecommand\DY{\mathrm{DY}}
\providecommand\DIS{\mathrm{DIS}}
\providecommand\VBF{\mathrm{VBF}}
\providecommand\NLL{\mathrm{NLL}}
\providecommand\VH{\mathrm{VH}}
\providecommand\WH{\mathrm{WH}}
\providecommand\ZH{\mathrm{ZH}}
\providecommand\ELWK{\mathrm{EW}}
\providecommand\HAWK{{\tt HAWK}}
\providecommand\MCFM{{\tt MCFM}}
\providecommand\VBFNLO{{\tt VBFNLO}}
\providecommand\VHNNLO{{\tt VH@NNLO}}
\providecommand\vhnnlo{{\tt VHNNLO}}
\providecommand\POWHEG{{\tt POWHEG}}
\providecommand\POWHEGBOX{{\tt POWHEG BOX}}
\providecommand\HERWIG{{\tt HERWIG}}
\providecommand\PYTHIA{{\tt PYTHIA}}
\providecommand\HDECAY{{\tt HDECAY}}
\providecommand{\kT}{\ensuremath{k\sb{\scriptstyle\mathrm{T}}}}
\providecommand\NNLOPS{{\tt NNLOPS}}
\providecommand\HVNNLO{{\tt HVNNLO}}
\providecommand{\HVNNLOPS}{{\tt HVNNLOPS}}
\providecommand{\HWJMINLOPS}{{\tt HWJ-MiNLO (Pythia8+hadr)}}
\providecommand{\HWNNLOPS}{{\tt HW-NNLOPS (Pythia8+hadr)}}
\providecommand{\HWNNLOPSshort}{{\tt HW-NNLOPS}}
\providecommand\DYNNLOPS{{\tt DYNNLOPS}}
\providecommand\MINLO{{\tt MiNLO}}
\providecommand\HWJMINLO{{\tt HWJ-MiNLO}}
\providecommand\FASTJET{{\tt FastJet}}
\providecommand\HNNLOPS{{\tt HNNLOPS}}
\providecommand\HW{{\tt HW}}
\providecommand\PhiHW{\Phi_{\scriptscriptstyle HW^*}}
\providecommand\PhiHWsimp{\Phi_{\scriptscriptstyle HW}}
\providecommand\thetacs{\theta^*}
\providecommand\phics{\phi^*}

\providecommand\ptH{p_{\scriptscriptstyle \mathrm{T,H}}}
\providecommand\ptw{p_{\scriptscriptstyle \mathrm{T,W}}}
\providecommand\ptwh{p_{\scriptscriptstyle \mathrm{T,HW}}}
\providecommand\ptjone{p_{\scriptscriptstyle
    \mathrm{T,j_{1}}}}
\providecommand\yjone{y_{\scriptscriptstyle
    \mathrm{j_{1}}}}

\providecommand\yh{y_{\scriptscriptstyle \mathrm{H}}}
\providecommand\ptw{p_{\scriptscriptstyle \mathrm{T,W}}}

The production of a Standard Model Higgs boson in association with two hard
jets in the forward and backward regions of the detector, frequently quoted
as the ``vector-boson fusion''~(VBF) channel,
and the production of a Higgs boson in association with a W or Z~boson,
known as ``VH production'' or ``Higgs-strahlung'',
represent cornerstones in a comprehensive study of
Higgs boson couplings at the LHC.
These production channels do not only provide valuable information on
the couplings of Higgs bosons to the massive gauge bosons by themselves, but
also allow for the isolation of the Higgs boson decays into
$\tau$-lepton or bottom-quark pairs, whose investigation is essential in
the Higgs boson couplings analysis.

In the previous reports~\cite{Dittmaier:2011ti,Dittmaier:2012vm,Heinemeyer:2013tqa},
state-of-the-art predictions and error estimates for the total and differential
cross sections for $\Pp\Pp\to\PH+ 2\,\mathrm{jets}$ and
$\Pp\Pp\to\PH V \to \PH+2\,$leptons were compiled (with $V=\PW,\PZ$), but the process of improving and
refining predictions is still ongoing, even within the Standard Model.
In this contribution we update the cross-section predictions for
VBF and VH production, covering integrated total and fiducial cross sections
as well as differential distributions.
In detail, the presented state-of-the-art predictions include QCD corrections
up to next-to-next-to-leading order (NNLO), electroweak (EW) corrections
up to next-to-leading order (NLO), and contributions from specific partonic channels
that open at higher perturbative orders, such as photon-induced collisions or
gluon-fusion contributions.
Apart from collecting numerical results, we give recommendations as to how to
combine the individual components and to assess conservative estimates of
remaining theoretical uncertainties.
Moreover, issues connected to the matching and the impact of parton
showers (PS) are discussed.

\section{VBF cross-section predictions}
\label{sec:VBF-XS}

\subsection{Programs and tools for VBF}

\subsubsection{HAWK}
\label{sec:HAWK-VBF-sub-sub}

\HAWK{}~\cite{Denner:2014cla,HAWK}
is a parton-level event generator for Higgs boson production in
vector-boson fusion~\cite{Ciccolini:2007jr, Ciccolini:2007ec},
$\Pp\Pp\to\PH+2\,\mathrm{jets}$, and Higgs-strahlung \cite{Denner:2011id},
$\Pp\Pp\to\PH V \to \PH+2\,$leptons (with $V=\PW,\PZ$).
Here we summarize its most important features for the VBF channel.

\HAWK{}
includes the complete NLO QCD and EW corrections and all weak-boson
fusion and quark--antiquark annihilation diagrams, i.e.~$t$-channel
and $u$-channel diagrams with VBF-like vector-boson exchange and
$s$-channel Higgs-strahlung diagrams with hadronic weak-boson decay,
as well as all interferences.
\HAWK{} allows for an on-shell Higgs boson or for an off-shell Higgs
boson (with optional decay into a pair of gauge singlets).
The EW corrections include also the contributions from photon-induced
channels, but contributions from effective Higgs--gluon couplings,
which are part of the QCD corrections to Higgs boson production via
gluon fusion,
are not taken into account.
External fermion masses are neglected
and the renormalization and factorization scales are set to $\MW$ by
default. Since version 2.0, \HAWK{} includes
anomalous Higgs-boson--vector-boson couplings.

\subsubsection{MG5\_aMC@NLO}
\label{sec:aMCNLO-sub-sub-VBF}

Higgs boson production through VBF, possibly in association with extra jets,
can be generated automatically in {\tt MG5\_aMC@NLO}, and is thus
exactly on the same footing as any other generic process.
A phenomenology study of H+2jet VBF production has been presented in
\Bref{Frixione:2013mta}, where NLO QCD results matched to different parton
showers (\HERWIG6, \HERWIG++ and virtuality-ordered \PYTHIA6) have been compared
to fixed-order NLO predictions and the corresponding \POWHEG-matched
ones. Predictions for VBF matched to \PYTHIA8 have been successively presented
in \Bref{Alwall:2014hca}.  The code for simulating VBF Higgs boson production
at the NLO(+PS) accuracy can be generated and run via the commands
\begin{verbatim}
generate p p > h j j $$ w+ w- z [QCD]
output VBF-MG5_aMC
launch
\end{verbatim}
where the \$\$ syntax forbids $s$-channel $\PW$ and $\PZ$ bosons.
Virtual corrections featuring electroweak bosons in the loop (pentagons) are not included when using the above command lines. Note that diagrams of this class are either zero, or negligible for all practical purposes.%
\footnote{For this reason, the internal check of pole cancellation fails.
In order to disable these checks, the parameters \texttt{IRPoleCheckThreshold}
and \texttt{PrecisionVirtualAtRunTime} inside \texttt{Cards/FKS\_params.dat} must be set to -1.}
As the default
Standard Model in {\tt MG5\_aMC@NLO} assumes a non-vanishing bottom mass, no
$\PQb$~quark is included in the definition of the \Pp\ and j multiparticles. In
order to include $\PQb$ quarks, it is sufficient to load the `loop\_sm-no\_b\_mass`
model before generating the code, with the command
\begin{verbatim}
import model loop_sm-no_b_mass
\end{verbatim}
In both cases, a $V_{CKM}=1$ is assumed and the Higgs boson is kept on its
mass shell.\\
For what concerns Higgs plus three jets production in VBF,
predictions for the third and the veto jet at NLO+PS accuracy have been
presented in \Bref{Alwall:2014hca}, considering $t$-channel modes
only. The relevant code can be generated and executed with the commands
\begin{verbatim}
generate p p > h j j j $$ w+ w- z [QCD]
output VBF-MG5_aMC
launch
\end{verbatim}

\subsubsection{POWHEG}
\label{sec:POWHEG-sub-sub}

The \POWHEGBOX\ is a program package that allows for the matching of
NLO QCD calculations with parton shower generators using the \POWHEG\
method. VBF-induced Higgs boson production has been implemented in the
\POWHEGBOX\ in the factorized approximation, where cross-talk between
the fermion lines is neglected, in Ref.~\cite{Nason:2009ai}.  More
recently, also an implementation of Higgs boson production in association
with three jets via VBF, based on the NLO QCD calculation of
Ref.~\cite{Figy:2007kv}, has become available~\cite{Jager:2014vna}.

Both implementations are based on the respective NLO QCD calculations for genuine weak-boson fusion topologies, i.e.\ the VBF approximation.
Quark--antiquark annihilation and interference contributions between $t$- and $u$-channel contributions are disregarded.
The CKM matrix elements can be assigned by the user. External fermion masses are neglected throughout.
For the choice of renormalization and factorization scales, various options are available.
For this report, fixed scales, $\muF=\muR=\MW$, are used, and contributions from external bottom and top quarks are entirely disregarded.

\subsubsection{VBF@NNLO}
\label{sec:VBFNNLO-sub-sub}

{\tt VBF@NNLO}~\cite{Bolzoni:2010xr,Bolzoni:2011cu} computes VBF total Higgs cross sections
at LO, NLO, and NNLO in QCD via the structure-function approach.  This
approach~\cite{Han:1992hr} consists  in considering VBF process as a
double deep-inelastic scattering (DIS) attached to the colourless pure
electroweak vector-boson fusion into a Higgs boson.  According to this
approach one can include NLO QCD corrections to the VBF process employing the
standard DIS structure functions $F_i(x,Q^2);\,i=1,2,3$ at
NLO~\cite{Bardeen:1978yd} or similarly the corresponding structure functions
at NNLO~\cite{Kazakov:1990fu,Zijlstra:1992kj,Zijlstra:1992qd,Moch:1999eb}.

The effective factorization underlying the structure-function approach does not include all types of contributions.
At LO an additional contribution arises from the
interferences between identical final-state quarks (e.g.,\
$\PQu\PQu\rightarrow \PH\PQu\PQu$) or between processes where either a $\PW$
or a $\PZ$ can be exchanged (e.g.,\ $\PQu\PQd\rightarrow \PH\PQu\PQd$).  These
LO contributions have been added to the NNLO results presented here, even if they are very small.
Apart from such contributions, the structure-function approach is exact also at NLO.
At NNLO, however, several types of diagrams violate the underlying factorization.
Their impact on the total rate has been computed or estimated in~\Bref{Bolzoni:2011cu} and found to be negligible.
Some of them are colour suppressed and kinematically
suppressed~\cite{vanNeerven:1984ak,Blumlein:1992eh,Figy:2007kv}, others have
been shown in \Bref{Harlander:2008xn} to be small enough not to produce
a significant deterioration of the VBF signal.

At NNLO QCD, the theoretical QCD uncertainty is reduced to less than
$2\%$. Electroweak corrections, which are at the level of $5\%$, are
not included in {\tt VBF@NNLO}.
The Higgs boson can either be produced on its mass-shell, or off-shell effects can be included in the
complex-pole scheme.

\subsubsection{proVBFH}
{\tt proVBFH} is a parton-level Monte Carlo program for the
calculation of differential distributions for VBF Higgs boson production to
NNLO QCD accuracy.
It is based on \POWHEG's fully differential NLO QCD calculation for
Higgs boson production in association with three jets via
VBF~\cite{Figy:2007kv,Jager:2014vna}, and an inclusive NNLO QCD
calculation~\cite{Bolzoni:2010xr},
the latter being taken in the
structure-function approximation, which are combined using the
projection-to-Born method described in Ref.~\cite{Cacciari:2015jma}.

{\tt proVBFH} uses a diagonal CKM matrix, Breit--Wigner distributions
for the W and Z bosons, and neglects fermion masses.
It assumes that there is no cross-talk between the upper and lower hadronic sectors.
For this report, the factorization and renormalization scales are set
to the W-boson mass, $\muR = \muF = \MW$.

\subsubsection{VBFNLO}
\label{sec:VBFNLO-sub-sub}

\VBFNLO~\cite{Arnold:2008rz} is a parton-level Monte Carlo generator for the simulation of various processes
with weak bosons at NLO QCD accuracy. In particular, Higgs boson production in association with two~\cite{Figy:2003nv}
or three jets~\cite{Figy:2007kv} via VBF is implemented with different options for the decays of the Higgs boson.
For VBF Higgs boson production in association with two jets, in addition to the default SM implementation,
options are available for the inclusion of anomalous coupling effects~\cite{Hankele:2006ma} and
VBF Higgs boson production in the context of the MSSM~\cite{Figy:2010ct}. NLO EW corrections to VBF can also be computed~\cite{Figy:2010ct}.
Quark--antiquark annihilation and interference contributions between $t$- and $u$-channel contributions are
not taken into account. In the following we will refer to this setup as ``VBF approximation''.

\subsection{VBF parameters and cuts}
\label{sec:VBFcuts-sub}

The numerical results presented in the next section have been
computed using the values of the EW parameters given in Chapter~\ref{chapter:input}.
The electromagnetic coupling is fixed in the
$\GF$ scheme,
\beq
 \alpha_{\GF} = \sqrt{2}\GF\MW^2(1-\MW^2/\MZ^2)/\pi,
\eeq
and the weak mixing angle, $\theta_{\mathrm{w}}$, is defined in the on-shell scheme,
\begin{equation}
\sw^2\equiv\sin^2\theta_{\mathrm{w}}=1-\MW^2/\MZ^2.
\end{equation}
The renormalization and factorization scales are set equal to the
$\PW$-boson mass,
\begin{equation}
\label{eq:VBF_ren_fac_scales}
\mu = \muR = \muF= \MW,
\end{equation}
and both scales are varied in the range $\MW/2 < \mu < 2\MW$ keeping $\muF=\muR$,
which catches the full scale uncertainty of integrated cross sections
(and of differential distributions in the essential regions).

In the calculation of the QCD-based cross sections, we have used the
PDF4LHC15\_nnlo\_100 PDFs~\cite{Butterworth:2015oua}, for the calculation of the EW corrections we have employed
the NNPDF2.3QED PDF set~\cite{Ball:2013hta}, which includes a photon PDF.
%
Note, however, that the relative EW correction factor, which is used in the following,
hardly depends on the PDF set, so that the uncertainty due to the
mismatch in the PDF selection is easily covered by the other remaining
theoretical uncertainties.

For the fiducial cross section and for
differential distributions the following reconstruction scheme
and cuts have been applied.
Jets are constructed according to the anti-$\kT$ algorithm~\cite{Cacciari:2008gp} with
$D=0.4$, using the default
recombination scheme ($E$ scheme).  Jets are constructed from partons $j$ with
\begin{equation}
\label{eq:VBF_cuts1}
|\eta_j| < 5\,,
\end{equation}
where $\eta_j$ denotes the pseudo-rapidity.  Real photons, which
appear as part of the EW corrections, are an input to the jet
clustering in the same way as partons.
Thus, in real photon radiation events, final states may consist of jets
only or jets plus a real identifiable photon, depending on whether
the photon was merged into a jet or not, respectively.
Both events with and without isolated photons are kept.

Jets are ordered according to their $\pT$ in decreasing
progression. The jet with highest $\pT$ is called leading jet $(j_1)$, the
one with next highest $\pT$ subleading jet $(j_2)$, and both are the tagging
jets.  Only events with at least two jets are kept.  They must satisfy
the additional constraints
\begin{equation}
\label{eq:VBF_cuts2}
{\pT}_j > 20\UGeV, \qquad
|y_j| < 5, \qquad
|y_{j_1} - y_{j_2}| > 3\,, \qquad M_{jj} > 130\UGeV,
\end{equation}
where $y_{j_{1,2}}$ are the rapidities of the two leading jets.
The cut on the 2-jet invariant mass $M_{jj}$ is sufficient to suppress
the contribution of $s$-channel diagrams to the VBF cross section
to the level of $1{-}2\%$, so that the DIS approximation of
taking into account only $t$- and $u$-channel contributions is justified.
In the cross sections given below, the $s$-channel contributions will be given
for reference, although they are not included in the final VBF cross sections
by default.

While the VBF cross sections in the DIS approximation are independent
of the CKM matrix, quark mixing has some effect on $s$-channel contributions.
For the calculation of the latter we employed a Cabbibo-like CKM matrix
(i.e.\ without mixing to the third quark generation) with Cabbibo angle,
$\theta_{\mathrm{C}}$, fixed by $\sin\theta_{\mathrm{C}}=0.225$.
Moreover, we note that we employ complex W- and Z-boson masses in the
calculation of $s$-channel and EW corrections in the standard \HAWK{} approach,
as described in
\Brefs{Ciccolini:2007jr, Ciccolini:2007ec}.

The Higgs boson is treated as on-shell particle in the following consistently,
since its finite-width and off-shell effects in the signal region are
suppressed in the SM.

\subsection{Integrated VBF cross sections}
\label{subsec:VBF-XS}

The final VBF cross section $\sigma^{\VBF}$ is calculated according to:
\begin{equation}
\sigma^{\VBF} = \sigma_{\NNLO \QCD}^{\DIS} (1+\delta_{\ELWK}) + \sigma_{\gamma},
\label{eq:sigmaVBF}
\end{equation}
where $\sigma_{\NNLO \QCD}^{\DIS}$ is the NNLO QCD prediction for the
VBF cross section in DIS approximation, based on the calculation
of \Bref{Cacciari:2015jma} with PDF4LHC15\_nnlo\_100 PDFs.  The
relative NLO EW correction $\delta_{\ELWK}$ is calculated with
\HAWK{}, but taking into account only $t$- and $u$-channel diagrams
corresponding to the DIS approximation.  The contributions from
photon-induced channels, $\sigma_{\gamma}$, and from $s$-channel
diagrams, $\sigma_{\mbox{\scriptsize $s$-channel}}$ are obtained from
\HAWK{} as well, where the latter includes NLO QCD and EW corrections.
To obtain $\sigma^{\VBF}$, the photon-induced contribution is added
linearly, but $\sigma_{\mbox{\scriptsize $s$-channel}}$ is left out
and only shown for reference, since it is not of true VBF origin (like
other contributions such as H+2jet production via gluon fusion).

Tables~\ref{tab:vbf_XStot} and \ref{tab:vbf_XSfiducial} summarize the total and
fiducial Standard Model VBF cross sections and the corresponding uncertainties
for the different proton--proton collision energies
for a Higgs boson mass $\MH=125\UGeV$.

\begin{table}
\caption{Total VBF cross sections including QCD and EW corrections
and their uncertainties for different proton--proton collision energies
$\sqrt{s}$ for a Higgs boson mass $\MH=125\UGeV$.}
\label{tab:vbf_XStot}
\begin{center}%
\begin{small}%
\tabcolsep5pt
\renewcommand{\arraystretch}{1.2}
\begin{tabular}{ccccccc|c}%
\toprule
$\sqrt{s}$[GeV] & $\sigma^{\VBF}$[fb] & $\Delta_{\mathrm{scale}}$[\%] &
$\Delta_{\mathrm{PDF}/\alphas/\mathrm{PDF\oplus\alphas}}$[\%] &
$\sigma_{\NNLO \QCD}^{\DIS}$[fb] & $\delta_{\ELWK}$[\%] & $\sigma_{\gamma}$[fb] & $\sigma_{\mbox{\scriptsize $s$-channel}}$[fb]
\\
\midrule
$7$  & $1241.4(1)$ &$^{+0.19}_{-0.21}$ &$\pm 2.1/\pm 0.4/\pm2.2$ &$1281.1(1)$ & $-4.4$ & $17.1$ & $584.5(3)$
\\
$8$  & $1601.2(1)$ &$^{+0.25}_{-0.24}$ &$\pm 2.1/\pm 0.4/\pm2.2$ &$1655.8(1)$ & $-4.6$ & $22.1$ & $710.4(3)$
\\
$13$ & $3781.7(1)$ &$^{+0.43}_{-0.33}$ &$\pm 2.1/\pm 0.5/\pm2.1$ &$3939.2(1)$ & $-5.3$ & $51.9$ & $1378.1(6)$
\\
$14$ & $4277.7(2)$ &$^{+0.45}_{-0.34}$ &$\pm 2.1/\pm 0.5/\pm2.1$ &$4460.9(2)$ & $-5.4$ & $58.5$ & $1515.9(6)$
\\
\bottomrule
\end{tabular}%
\end{small}%
\end{center}%
\vspace{2em}
\caption{Fiducial VBF cross sections including QCD and EW corrections
and their uncertainties for different proton--proton collision energies
$\sqrt{s}$ for a Higgs boson mass $\MH=125\UGeV$.}
\label{tab:vbf_XSfiducial}
\begin{center}%
\begin{small}%
\tabcolsep5pt
\renewcommand{\arraystretch}{1.2}
\begin{tabular}{ccccccc|c}%
\toprule
$\sqrt{s}$[GeV] & $\sigma^{\VBF}$[fb] & $\Delta_{\mathrm{scale}}$[\%] &
$\Delta_{\mathrm{PDF}/\alphas/\mathrm{PDF\oplus\alphas}}$[\%] &
$\sigma_{\NNLO \QCD}^{\DIS}$[fb] & $\delta_{\ELWK}$[\%] & $\sigma_{\gamma}$[fb] & $\sigma_{\mbox{\scriptsize $s$-channel}}$[fb]
\\
\midrule
$7$  & $602.4(5)$ &$^{+1.3}_{-1.6}$ &$\pm 2.3/\pm 0.3/\pm2.3$ & $630.8(5)$ & $-6.1$ &  $9.9$ & $8.2$
\\
$8$  & $795.9(6)$ &$^{+1.3}_{-1.5}$ &$\pm 2.3/\pm 0.3/\pm2.3$ & $834.8(7)$ & $-6.2$ & $13.1$ & $11.1$
\\
$13$ & $1975.4(9)$ &$^{+1.3}_{-1.2}$ &$\pm 2.1/\pm 0.4/\pm2.2$ & $2084.2(10)$ & $-6.8$ & $32.3$ & $29.0$
\\
$14$ & $2236.6(26)$ &$^{+1.5}_{-1.3}$ &$\pm 2.1/\pm 0.4/\pm2.1$ & $2362.2(28)$ & $-6.9$ & $36.7$ & $33.1$
\\
\bottomrule
\end{tabular}%
\end{small}%
\end{center}%
\end{table}

The scale uncertainty, $\Delta_{\mathrm{scale}}$, results from a variation
of the factorization and renormalization scales
\eqref{eq:VBF_ren_fac_scales} by a factor of $2$ keeping $\muF=\muR$,
as indicated above, and the combined PDF${\oplus}\alphas$ uncertainty
$\Delta_{\mathrm{PDF\oplus\alphas}}$ is obtained following the PDF4LHC
recipe~\cite{Butterworth:2015oua}. Both $\Delta_{\mathrm{scale}}$ and
$\Delta_{\mathrm{PDF\oplus\alphas}}$ are actually obtained from
$\sigma_{\NNLO \QCD}^{\DIS}$, but this QCD-driven uncertainties can be
taken over as uncertainty estimates for $\sigma^{\VBF}$ as well.  The
theoretical uncertainties of integrated cross sections originating
from unknown higher-order EW effects can be estimated by
\begin{equation}
\Delta_\ELWK = \max\{0.5\%,\delta_{\ELWK}^2,\sigma_\gamma/\sigma^{\VBF}\}.
\end{equation}
The first entry represents the generic size of NNLO EW corrections, while the second accounts for
potential enhancement effects.
Note that the whole photon-induced cross-section contribution $\sigma_\gamma$ is treated
as uncertainty here, because the PDF uncertainty of $\sigma_\gamma$ is estimated to be $100\%$
with the NNPDF2.3QED PDF set. At present, this source, which is about $1.5\%$,
dominates the EW uncertainty of the integrated VBF cross section

Results for the VBF cross sections from a scan over the SM Higgs boson mass $\MH$
can be found in \refA{VBFappendix}.
In detail the total cross sections are collected in
\refTs{tab:vbf_XStot_7}--\ref{tab:vbf_XStot_14}
and the fiducial cross sections in
\refTs{tab:vbf_XSfiducial_7}--\ref{tab:vbf_XSfiducial_14}.
Inclusive cross sections at NNLO QCD computed with the VBF@NNLO code are shown in
\refTs{tab:vbfnnlo_XStot_7}--\ref{tab:vbfnnlo_XStot_14} for the SM
\MH scan, and in \refTs{tab:table_vbf_nnlo_escan_125}--\ref{tab:table_vbf_nnlo_escan_12509} for the
energy scan. No interpolation of results has been performed.

\subsection{Differential VBF cross sections}

\begin{figure}
\includegraphics[width=.47\textwidth]{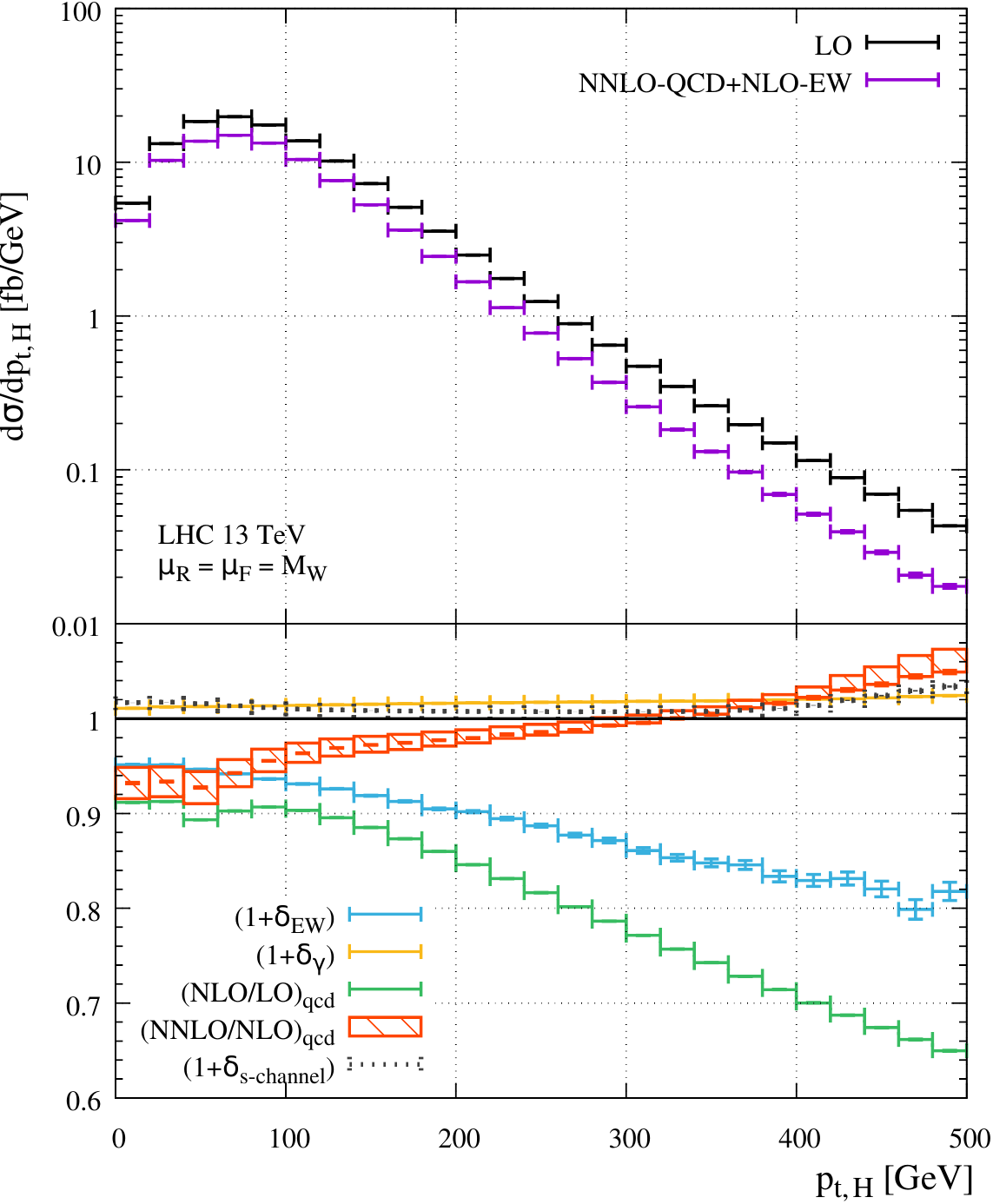}
\hfill
\includegraphics[width=.47\textwidth]{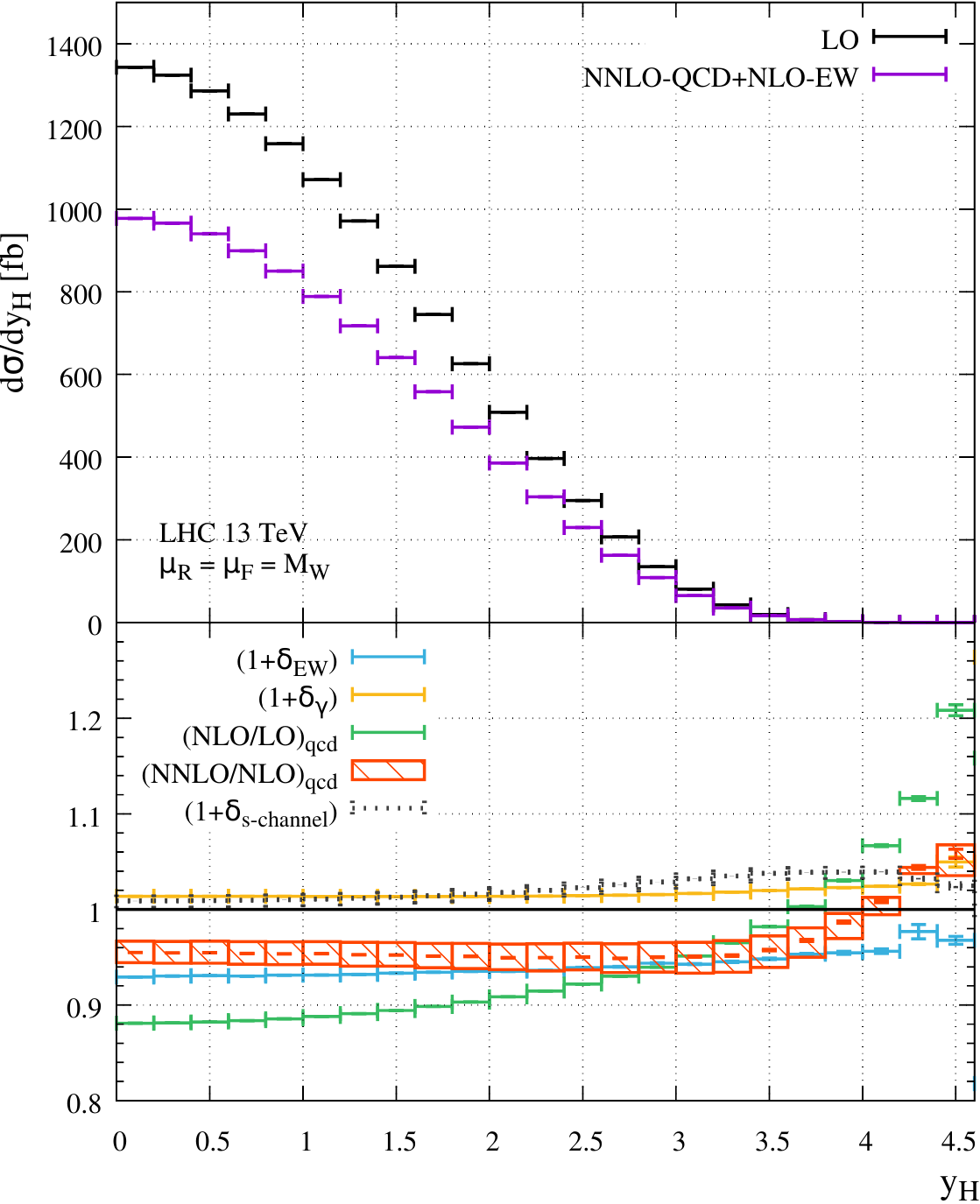}
\caption{Transverse-momentum and rapidity distributions of the Higgs boson in VBF
at LO and including NNLO QCD and NLO EW corrections (upper plots)
and various relative contributions (lower plots) for $\sqrt{s}=13\UTeV$ and $\MH=125\UGeV$.}
\label{fig:SM-VBF-ptH-yH}
\end{figure}

\begin{figure}
\includegraphics[width=.47\textwidth]{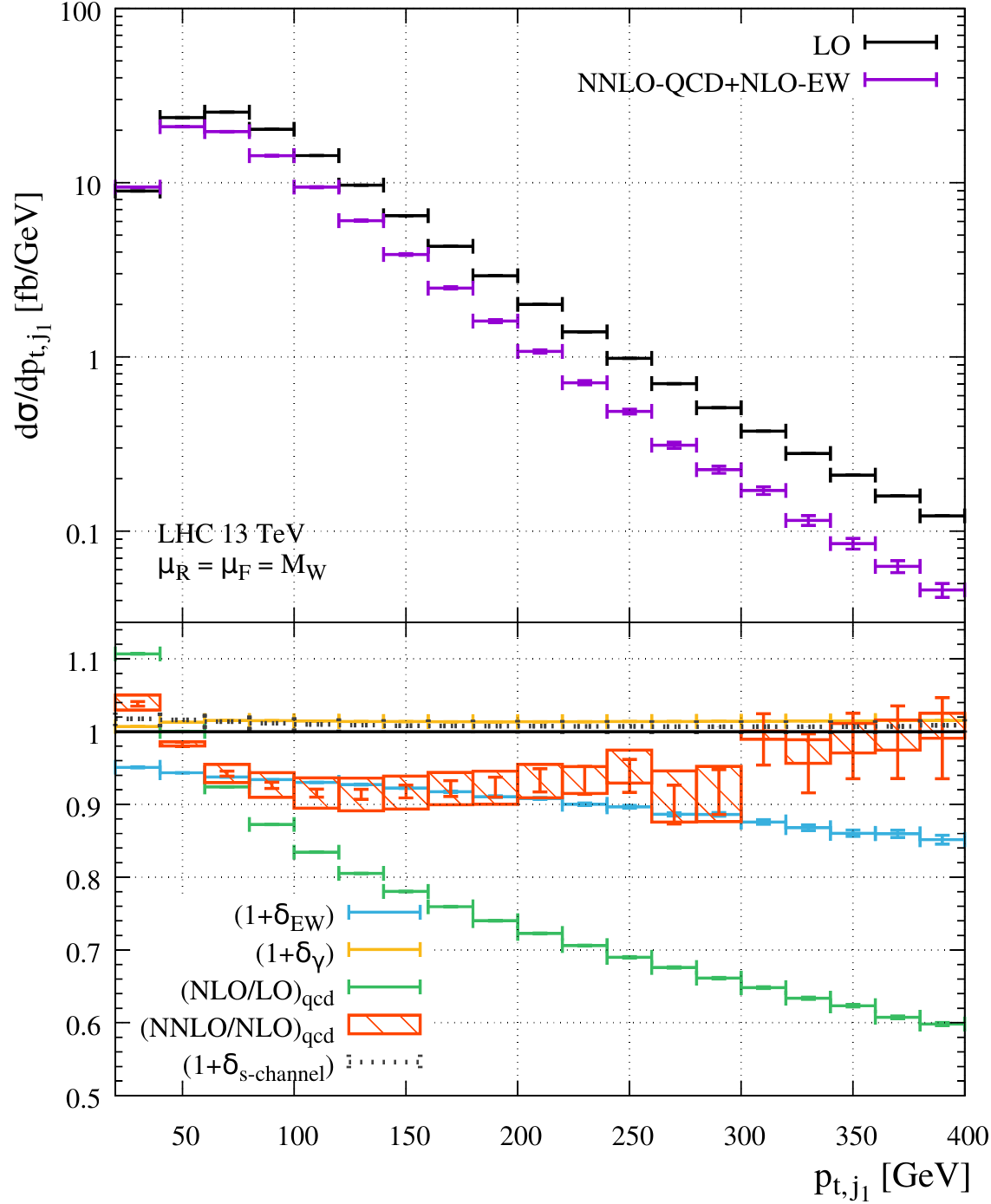}
\hfill
\includegraphics[width=.47\textwidth]{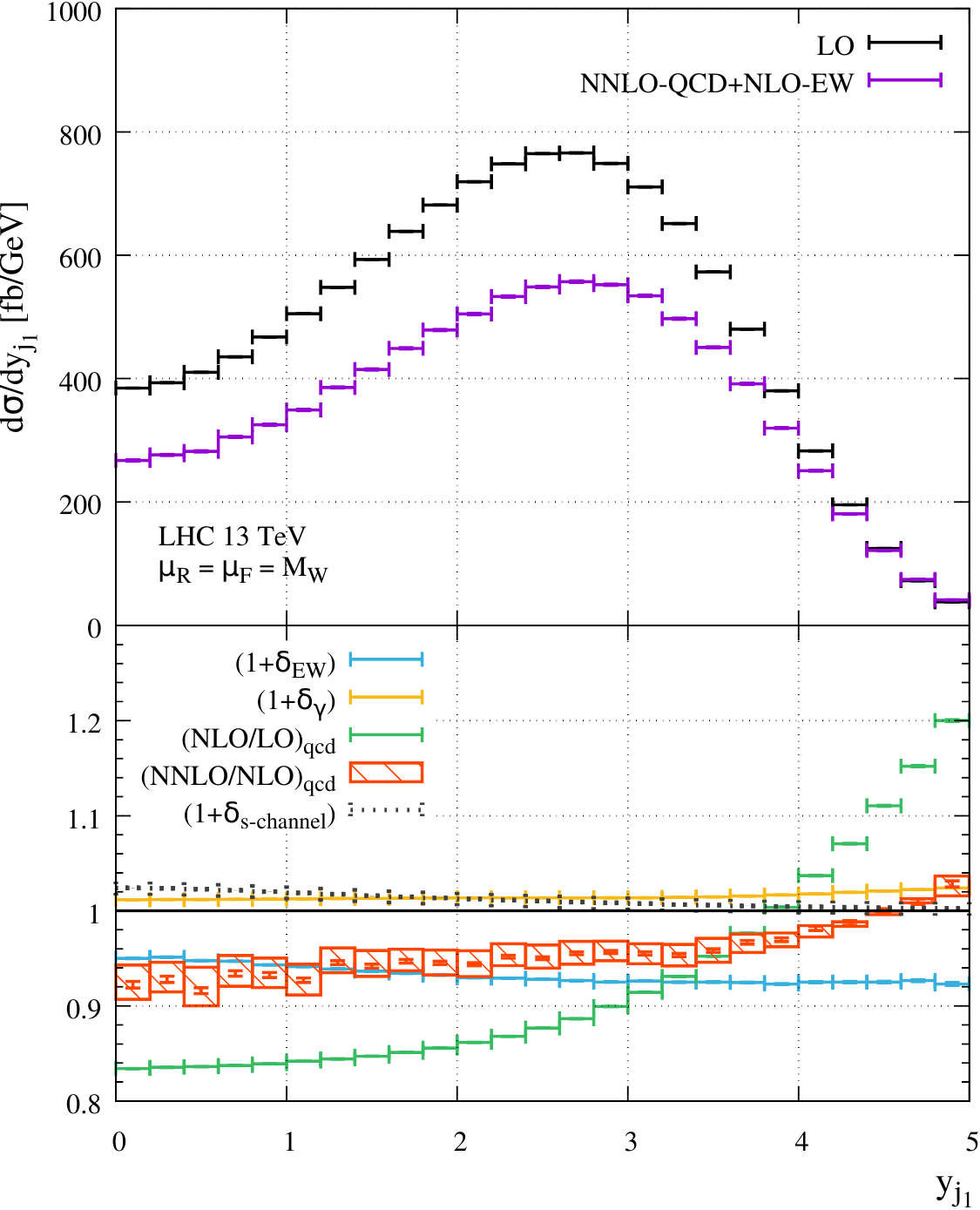}
\caption{Transverse-momentum and rapidity distributions of the leading jet in VBF
at LO and including NNLO QCD and NLO EW corrections (upper plots)
and various relative contributions (lower plots) for $\sqrt{s}=13\UTeV$ and $\MH=125\UGeV$.}
\label{fig:SM-VBF-ptj1-yj1}
\end{figure}

\begin{figure}
\includegraphics[width=.47\textwidth]{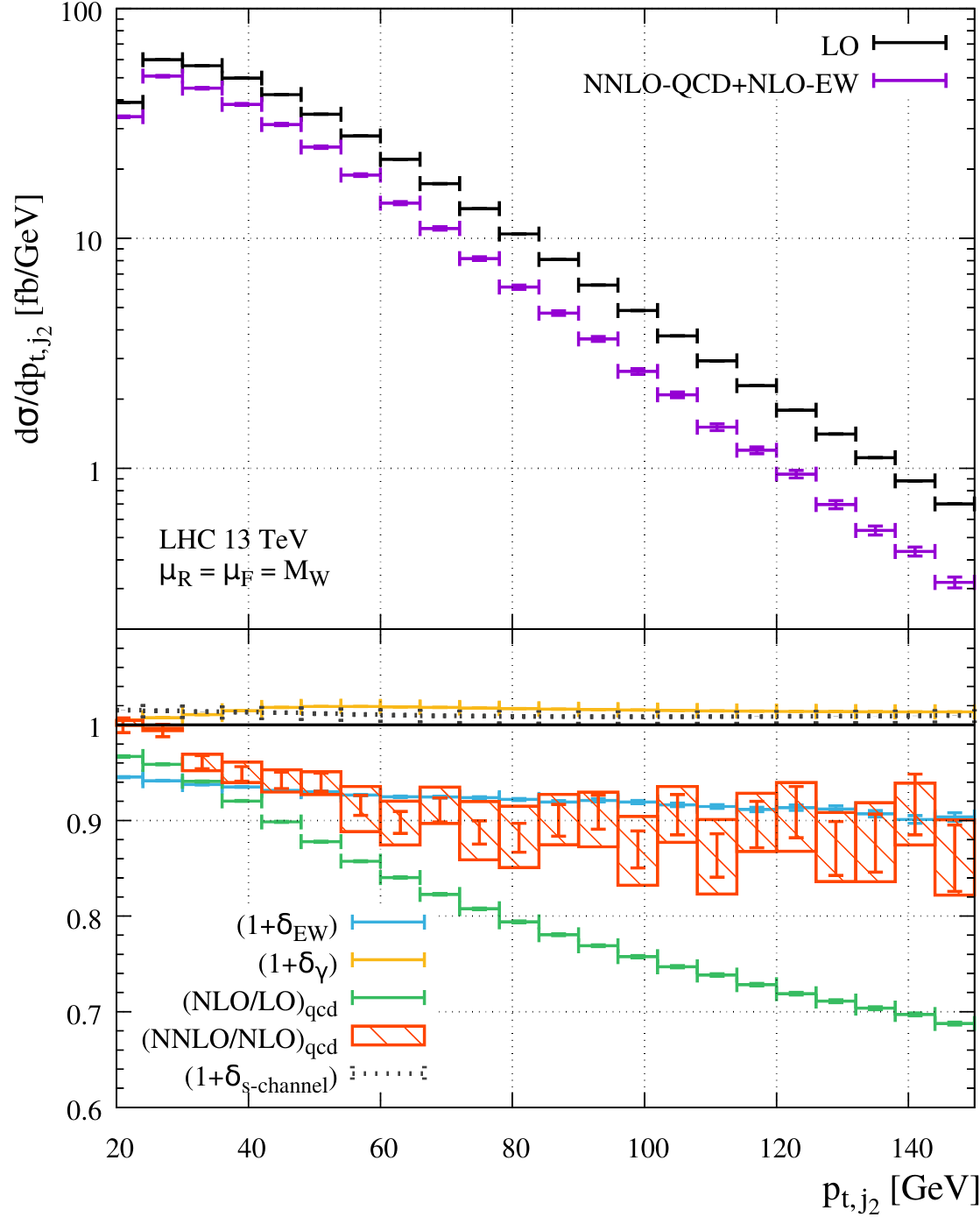}
\hfill
\includegraphics[width=.47\textwidth]{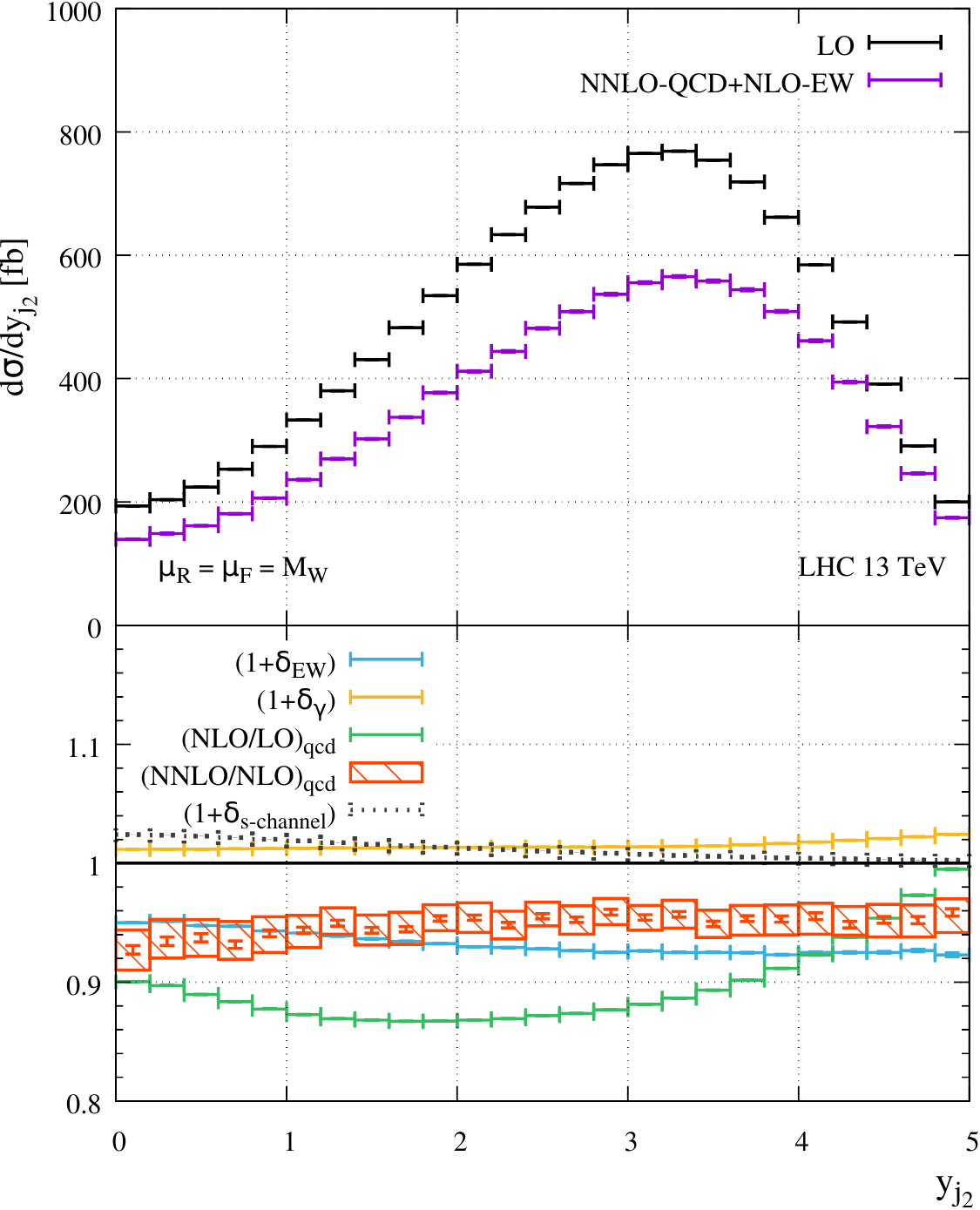}
\caption{Transverse-momentum and rapidity distributions of the subleading jet in VBF
at LO and including NNLO QCD and NLO EW corrections (upper plots)
and various relative contributions (lower plots) for $\sqrt{s}=13\UTeV$ and $\MH=125\UGeV$.}
\label{fig:SM-VBF-ptj2-yj2}
\end{figure}

\begin{figure}
\includegraphics[width=.47\textwidth]{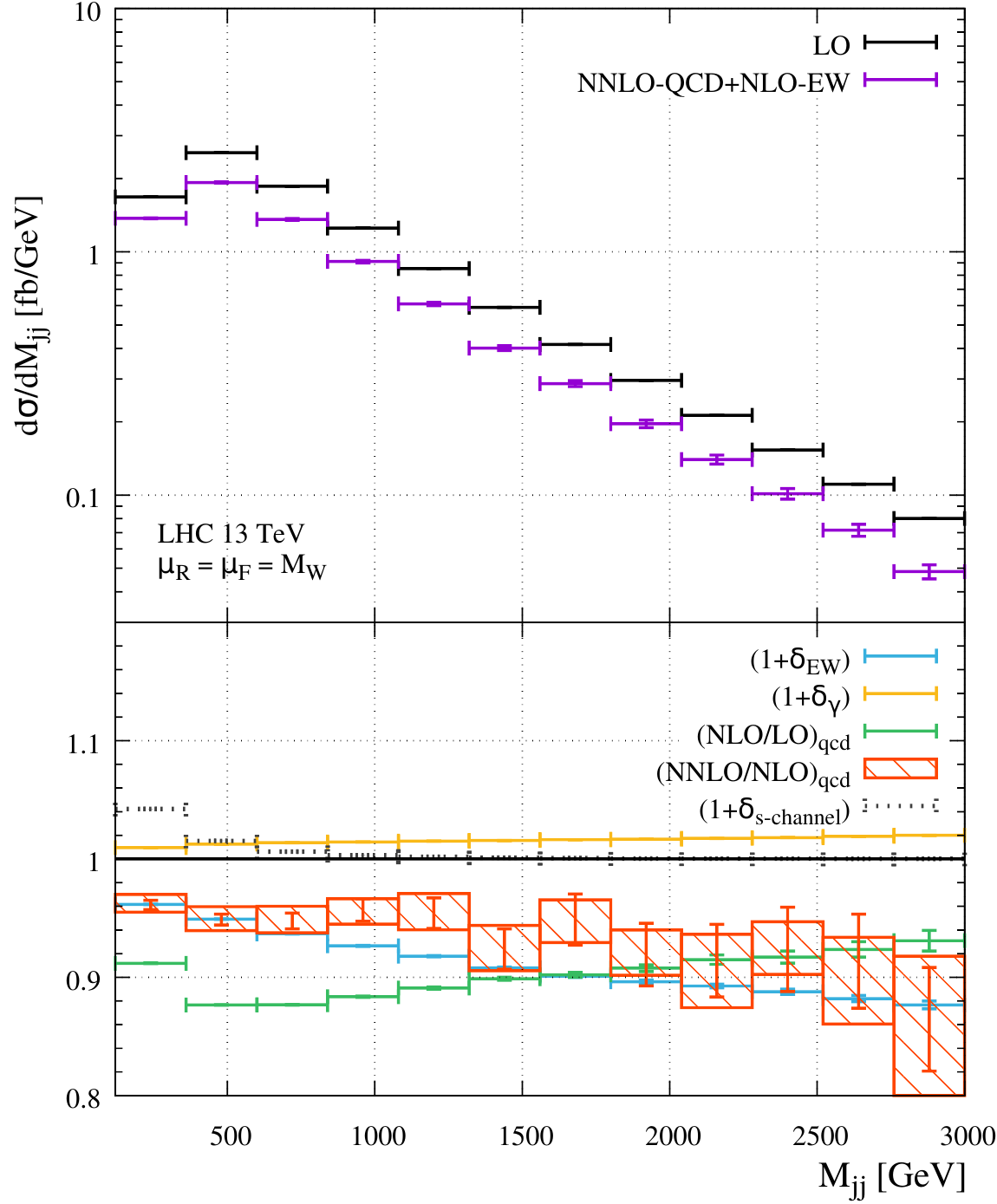}
\hfill
\includegraphics[width=.47\textwidth]{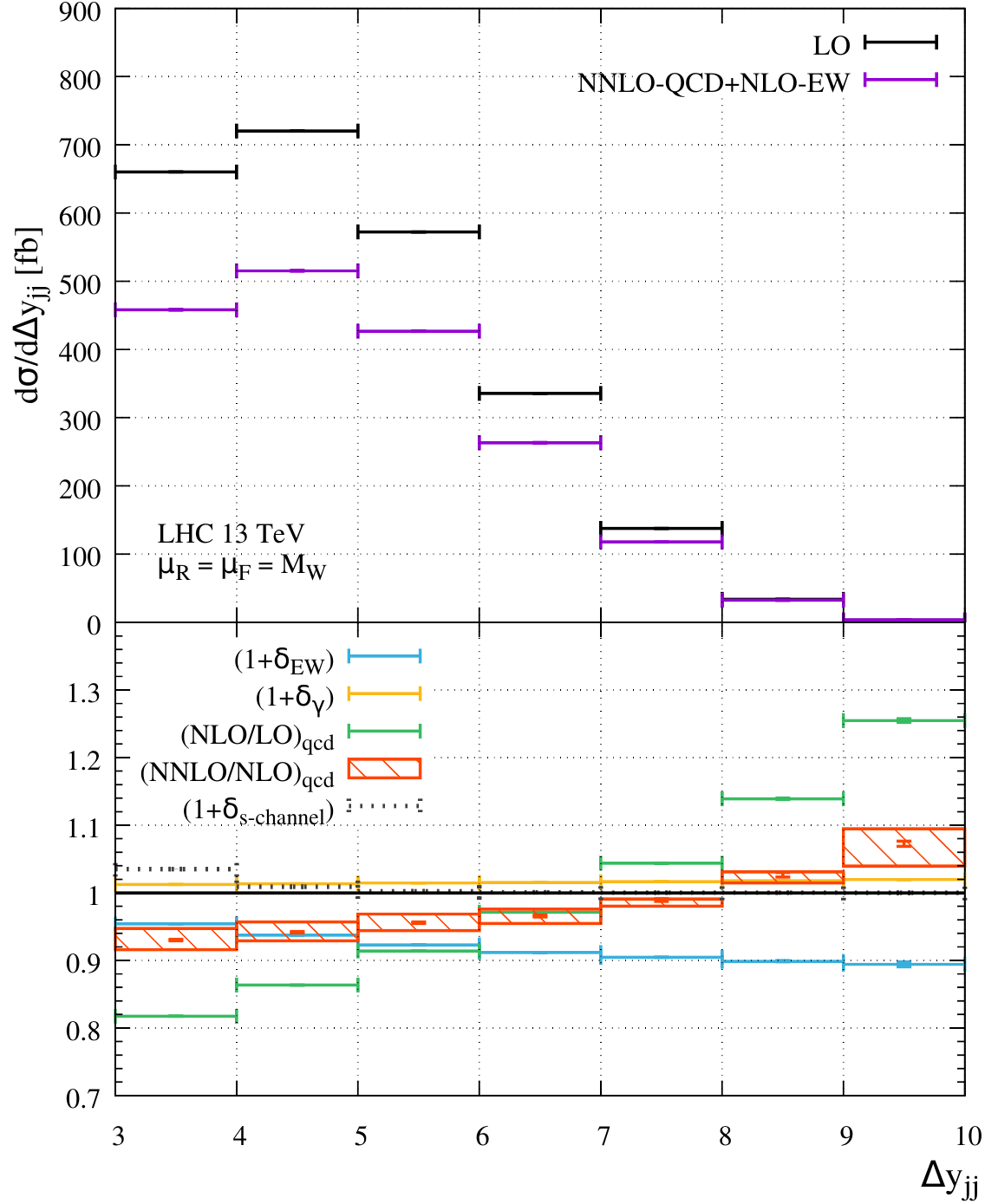}
\caption{Distributions in the invariant mass and in the rapidity difference of the first two
leading jets in VBF
at LO and including NNLO QCD and NLO EW corrections (upper plots)
and various relative contributions (lower plots) for $\sqrt{s}=13\UTeV$ and $\MH=125\UGeV$.}
\label{fig:SM-VBF-Mjj-yjj}
\end{figure}

\begin{figure}
\centerline{
\includegraphics[width=.47\textwidth]{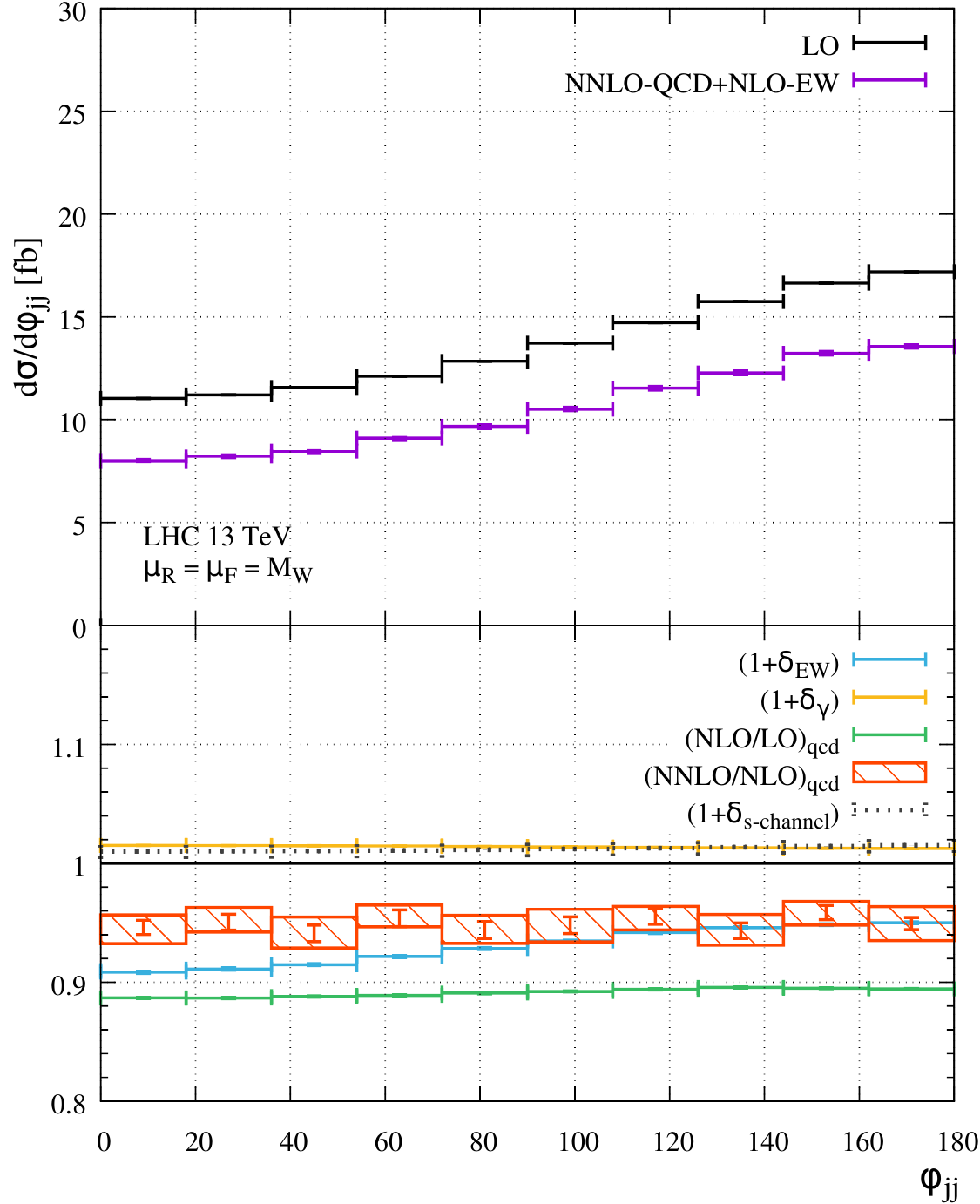}
}
\caption{Distribution in the azimuthal-angle difference of the first two
leading jets in VBF
at LO and including NNLO QCD and NLO EW corrections (upper plots)
and various relative contributions (lower plots) for $\sqrt{s}=13\UTeV$ and $\MH=125\UGeV$.}
\label{fig:SM-VBF-phijj}
\end{figure}

Figures~\ref{fig:SM-VBF-ptH-yH}--\ref{fig:SM-VBF-phijj} show the most important
differential cross sections for Higgs boson production via VBF in the SM.
The upper panels show the LO cross section as well as the best fixed-order prediction,
based on the analogue of Eq.~\eqref{eq:sigmaVBF} for differential
cross sections.
The lower panels illustrate relative contributions and the ratios
$(\NLO/\LO)_{\mathrm{qcd}}$ and $(\NNLO/\NLO)_{\mathrm{qcd}}$
of QCD predictions when going from LO to NLO QCD to
NNLO QCD. Moreover, the relative EW correction to the (anti)quark--(anti)quark
channels ($\delta_{\ELWK}=\sigma_{\ELWK}/\sigma_{\LO}$) and the relative correction induced by initial-state
photons ($\delta_\gamma=\sigma_\gamma/\sigma_\LO$) are shown.
Finally, the relative size of the $s$-channel contribution for
Higgs+2jet production
($\delta_{\mbox{\scriptsize $s$-channel}}=\sigma_{\mbox{\scriptsize $s$-channel}}/\sigma_\LO$) is
depicted as well, although it is not included in the definition of the VBF cross section.
Integrating the differential cross sections shown in the following, and all its
individual contributions, results in the fiducial cross sections discussed in the
previous section.

The ratio $(\NLO/\LO)_{\mathrm{qcd}}$ shows a quite large impact of NLO QCD
corrections, an effect that can be traced back to the scale choice $\mu=\MW$,
which is on the low side if mass scales such as $p_{\mathrm{T}}$ and $M_{jj}$
get large in some distributions. The moderate ratio
$(\NNLO/\NLO)_{\mathrm{qcd}}$, however, indicates nice convergence of perturbation
theory at NNLO QCD. The band around the ratio $(\NNLO/\NLO)_{\mathrm{qcd}}$ illustrates the
scale uncertainty of the NNLO QCD cross section, which also applies to $\sigma^{\VBF}$.

The EW corrections $\delta_{\ELWK}$ to (pseudo)rapidity and angular distributions are rather flat,
resembling the correction to the integrated (fiducial) cross section.
In the high-energy tails of the $p_{\mathrm{T}}$ and $M_{jj}$ distributions,
$\delta_{\ELWK}$ increases in size to $10$--$20\%$, showing the onset of the well-known large
negative EW corrections that are enhanced by logarithms of the form
$(\alpha/\sw^2)\ln^2(p_{\mathrm{T}}/\MW)$.
The impact of the photon-induced channels uniformly stays at the generic level of $1$--$2\%$, i.e.\
they cannot be further suppressed by cuts acting on the variables shown in the distributions.

The contribution of $s$-channel (i.e.\ VH-like) production uniformly shows the relative size
of about $1.5\%$ observed in the fiducial cross section, with the exception of the
$M_{jj}$ and $\Delta y_{jj}$ distributions, where this contribution
is enhanced at the lower ends of the spectra. Tightening the VBF cuts at theses ends,
would further suppress the impact of $\sigma_{s-\mathrm{channel}}$, but reduce the signal
at the same time.
As an alternative to decreasing $\sigma_{s-\mathrm{channel}}$, a veto on subleading jet pairs with
invariant masses around $\MW$ or $\MZ$ may be promising. Such a veto, most likely, would
reduce the photon-induced contribution $\delta_\gamma$, and thus the corresponding uncertainty, as well.

The theoretical uncertainties of differential cross sections originating from unknown
higher-order EW effects can be estimated by
\begin{equation}
\Delta_\ELWK = \max\{1\%,\delta_{\ELWK}^2,\sigma_\gamma/\sigma^{\VBF}\},
\end{equation}
i.e.\ $\Delta_\ELWK$ is taken somewhat more conservative than for integrated cross sections, accounting for possible
enhancements of higher-order effects due to a kinematical migration of events in distributions.
Note that $\delta_{\ELWK}^2$, in particular, covers the known effect of enhanced EW corrections at high
momentum transfer (EW Sudakov logarithms, etc.).
As discussed for integrated cross sections in the previous section, the large uncertainty of the
current photon PDF forces us to include the full contribution $\sigma_\gamma$ in the EW
uncertainties. While the photon-initiated contribution obtained with NNPDF2.3QED or MRST2004QED is
 currently affected by large PDF uncertainties, it should be noted that those uncertainties are considerably 
 reduced when the more recent LUXqed\_plus\_PDF4LHC15\_nnlo\_100 PDF set is instead employed~\cite{Manohar:2016nzj}.


\section{VH cross-section predictions}
\label{sec:VH-XS}

\subsection{Programs and tools for VH}

\subsubsection{HAWK}
\label{sec:HAWK-VH-sub-sub}

\HAWK{}~\cite{Denner:2014cla,HAWK}
is a parton-level event generator for Higgs boson production in
vector-boson fusion~\cite{Ciccolini:2007jr, Ciccolini:2007ec},
$\Pp\Pp\to\PH jj$, and Higgs-strahlung \cite{Denner:2011id},
$\Pp\Pp\to\PH V \to \PH+2\,$leptons (with $V=\PW,\PZ$).
Here we summarize its most important features for the VH channel.

\HAWK{} calculates the complete NLO QCD and EW corrections
to the processes
$\Pp\Pp\to\PW\PH\to\PGn_{l}\,l\,\PH$ and
$\Pp\Pp\to\PZ\PH\to l^-l^+\PH/\PGn_{l}\PAGn_{l}\PH$,
i.e.\ the leptonic decays and all off-shell effects
of the W/Z bosons are included.
The Higgs boson can be taken as on-shell or off-shell
(with optional decay into a pair of gauge singlets).
The EW corrections include also the contributions from photon-induced
channels, but gluon-fusion contributions ($\Pg\Pg\to\PZ\PH$)
are not taken into account.
External fermion masses are neglected,
and the renormalization and factorization scales are set to $M_V+\MH$
($V=\PW,\PZ$) by default. Since version 2.0, \HAWK{} includes
anomalous Higgs-boson--vector-boson couplings.

\subsubsection{MG5\_aMC@NLO}
\label{sec:aMCNLO-sub-sub-VH}

Similar to the generation of any generic process, also Higgs boson production
in association with a vector boson can be generated automatically with
{\tt MG5\_aMC@NLO}. At the NLO QCD accuracy, multiple jets can
also be included using the FxFx merging technique~\cite{Frederix:2012ps}. In
\Bref{Alwall:2014hca} the example of $\PH\,\Pe^+ \nu_{\Pe} + 0,1 \textrm{ jets}$ has
been presented, and adding a further jet at the NLO is feasible with
a small-scale cluster. The situation is entirely similar for the case
of $\PZ\PH$ (possibly plus jet) production. The commands to generate
the corresponding codes are
\begin{verbatim}
import model loop_sm-no_b_mass
define l+ = e+ mu+ ta+
define l- = e- mu- ta-
define vl~ = ve~ vm~ vt~
define vl = ve vm vt
generate p p > h l+ l- [QCD] @0
add process p p > h l+ vl [QCD] @0
add process p p > h l- vl [QCD] @0
add process p p > h l+ l- j [QCD] @1
add process p p > h l+ vl j [QCD] @1
add process p p > h l- vl~ j [QCD] @1
add process p p > h l+ l- j j [QCD] @2
add process p p > h l+ vl j j [QCD] @2
add process p p > h l- vl~ j j [QCD] @2
output VH-MG5_aMC
\end{verbatim}
The first command loads a five-flavour scheme model (the {\tt MG5\_aMC} default
uses a four-flavour scheme model), which sets the $\PQb$-quark mass to zero and
includes it in the definition of the \texttt{p} and \texttt{j} multi-particle
labels. The next four commands define the multi-particle labels for the
leptons and neutrinos used in the \texttt{generate} and \texttt{add process}
commands. After writing the code to disk (with the \texttt{output} command)
the event generation can be started by executing the command
\texttt{launch}. The FxFx merging algorithm is available when matching to
\HERWIG6, \PYTHIA8, or \HERWIG++ partons
showers~\cite{Frederix:2012ps,Frederix:2015eii}, and can be turned on by
setting the \texttt{ickkw} parameter to \texttt{3} in the file run\_card.dat.

\subsubsection{MCFM}

The calculation is performed at NNLO QCD
and includes the decays of the unstable
Higgs and vector bosons. We also include all
$\mathcal{O}(\alpha_s^2)$ contributions that occur in production for
these processes: those mediated by the exchange of a single
off-shell vector boson in the $s$-channel, and those which arise
from the coupling of the Higgs boson to a closed loop of
fermions.

\begin{figure}
\begin{center}
\includegraphics[width=14cm]{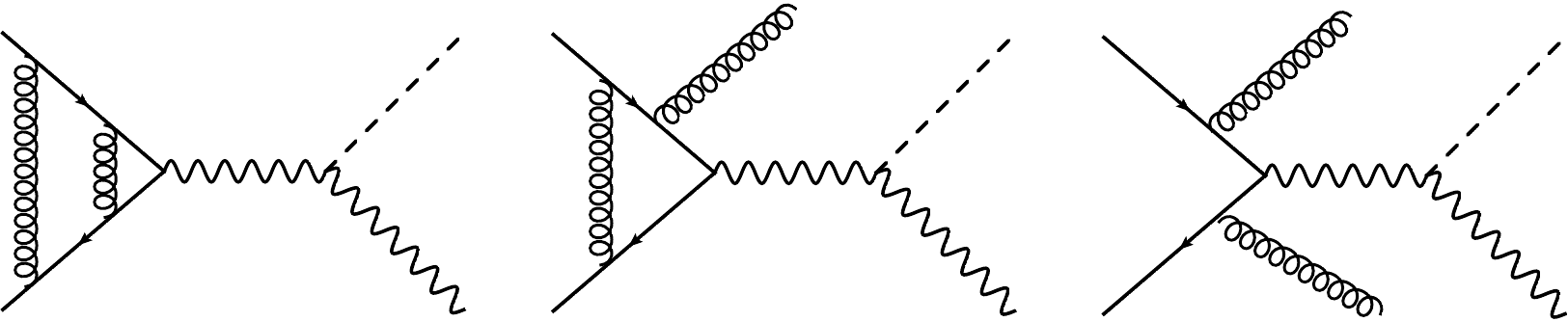}
\caption{Drell--Yan-like production modes for the associated production of a Higgs boson. Shown are representative Feynman diagrams needed
to compute the $\mathcal{O}(\alphas^2)$ corrections to the process. Examples are shown for each of the 0-, 1-, and 2-parton phase-space configurations.}
\label{fig:DYNNLO}
\end{center}
\end{figure}
\begin{figure}
\begin{center}
\includegraphics[width=8cm]{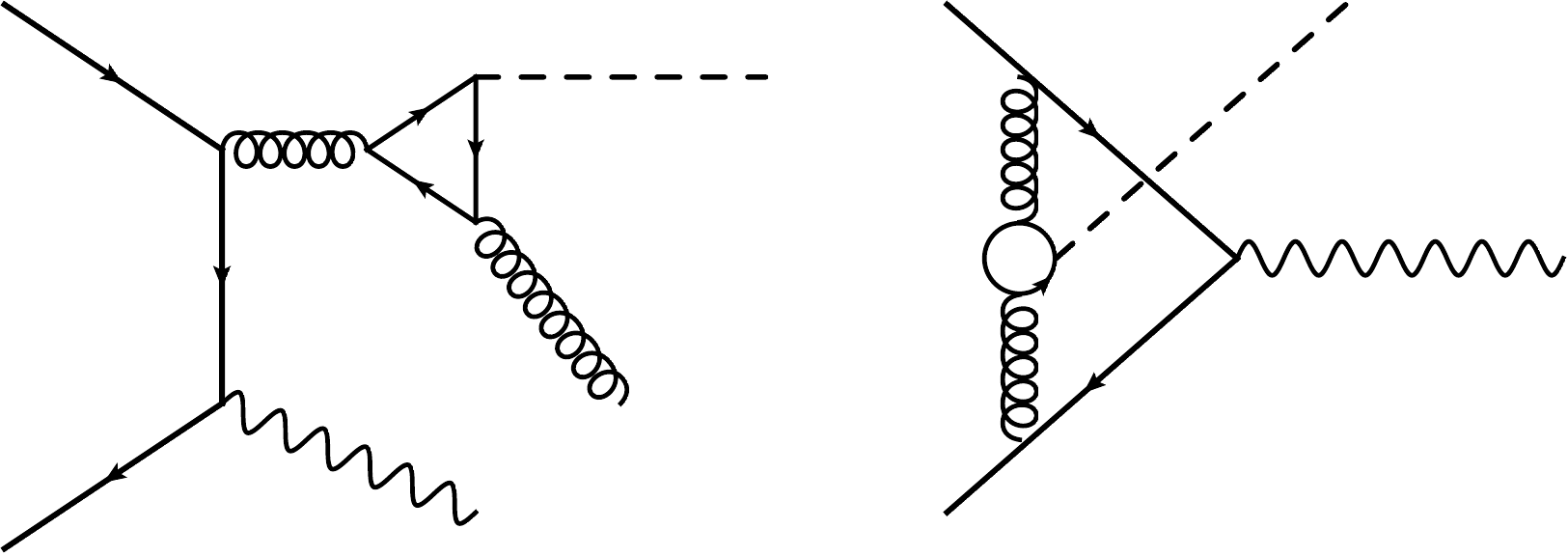}
\includegraphics[width=12cm]{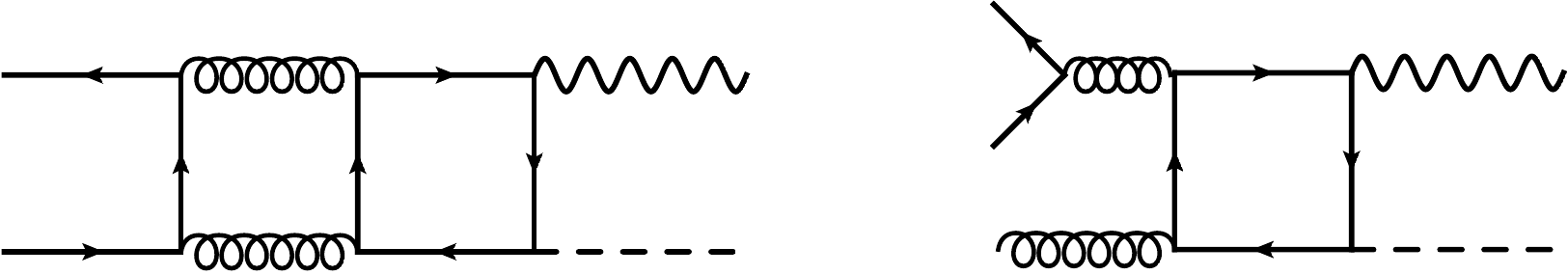} \\ \vspace*{4mm}
\includegraphics[width=12cm]{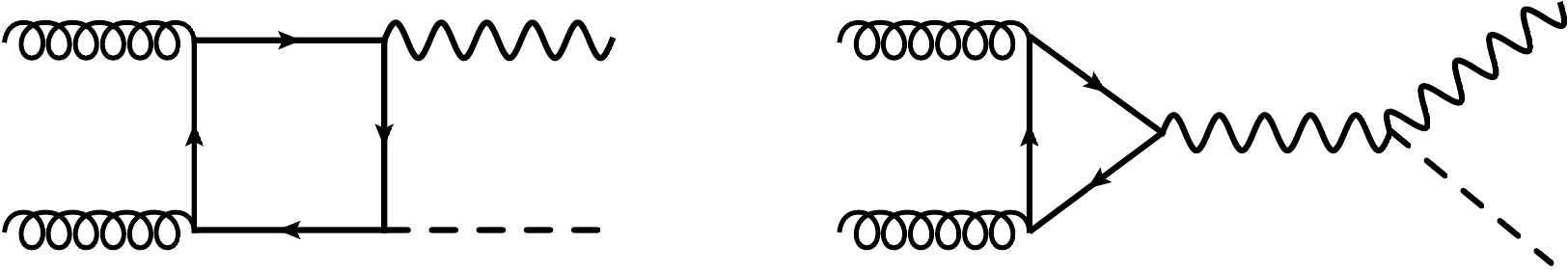}
\caption{Diagrams representing associated production of a Higgs boson that
are sensitive to the top Yukawa coupling $y_{\PQt}$.
The topologies indicated in the top line occur for either $\PW\PH$ or $\PZ\PH$ production
and interfere with the LO amplitude. The remaining topologies
only occur for $\PZ\PH$ production.  The $\Pg\Pg\rightarrow \PZ\PH$ contribution represented
in the bottom line is not proportional to $y_{\PQt}$, as can be seen from
the examples on the left ($y_{\PQt}$-dependent) and right (no~$y_{\PQt}$).
}
\label{fig:Yt}
\end{center}
\end{figure}
Examples of diagrams that contribute at NNLO QCD are
shown in Figs.~\ref{fig:DYNNLO} and~\ref{fig:Yt}.
The first type of contributions has the same structure
as single vector-boson production, c.f.\ \refF{fig:DYNNLO}.
Diagrams of the second type, shown in \refF{fig:Yt}, all contain
a closed loop of fermions and, in general, represent
the Higgs boson coupling directly to a heavy quark (predominantly a top-quark).
Note that some of the contributions shown in \refF{fig:Yt} only
occur for the case of $\PZ\PH$ production and, for the $\Pg\Pg \to \PZ\PH$ contributions,
not all diagrams are proportional to the top-quark Yukawa coupling.
Each of these contributions results in NNLO QCD corrections at the few per cent level
for typical cuts, so that inclusion of them all is necessary in order
to obtain sufficient theoretical control of this process.
A further complication is the inclusion of decays of the Higgs boson into
bottom quarks, which we consider.  Our calculation also includes the
significant impact of NLO QCD corrections in this decay, using the factorized approach
described in \Brefs{Ferrera:2013yga,Ferrera:2014lca}.  This method takes advantage
of improved descriptions of the total decay rate that are available in the \HDECAY\ code~\cite{Djouadi:1997yw}.
The assembly of a complete calculation at NNLO QCD requires the regularization of infrared singularities,
which we handle using the recently-developed ``jettiness subtraction''
procedure~\cite{Gaunt:2015pea,Boughezal:2015dva,Boughezal:2015aha,Boughezal:2015ded}
that has been implemented in the Monte Carlo program \MCFM~\cite{Campbell:1999ah,Campbell:2011bn,Campbell:2015qma}.
A detailed description of our calculation can be found in \Bref{Campbell:2016jau}.

\subsubsection{VHNNLO}

\vhnnlo~\cite{Ferrera:2011bk,Ferrera:2013yga,Ferrera:2014lca} is a parton level
program for the calculation of fully differential cross sections for $\Pp\Pp\to\PW\PH$
and $\Pp\Pp\to\PZ\PH$ including up to second order QCD corrections and the decays
of the weak bosons to leptons and of the Higgs boson to bottom quarks.

\subsubsection{VH@NNLO}

\VHNNLO~\cite{Brein:2012ne,Harlander:2013mla} calculates the total
inclusive cross section for $\Pp\Pp\to\PW\PH$ and $\Pp\Pp\to\PZ\PH$
production, including all available QCD corrections through ${\cal
O}(\alphas^2)$, i.e.\ NNLO.\footnote{Large parts of \VHNNLO\ are taken
over from {\tt ZWPROD} by W.~van Neerven~\cite{Hamberg:1990np}.}
Specifically, these are the Drell--Yan-like terms (see \refF{fig:DYNNLO}), given by the
process $\PQq\bar\PQq\to\PH V$ (with $V=\PW,\PZ$) plus radiative corrections due to
virtual and real gluon and/or quark
radiation~\cite{Hamberg:1990np,Brein:2003wg}, as well as terms involving
closed top or bottom loops (see \refF{fig:Yt}). For the latter, we distinguish those that
interfere with the lowest-order $\PQq\bar\PQq\to\PH V (\Pg)$ (with $V=\PW,\PZ$) amplitude
(plus crossings) and which we simply denote as ``top-loop'' terms
$\sigma_{\Pt\mbox{\scriptsize -loop}}$~\cite{Brein:2011vx}, and the
$\Pg\Pg\to\PH\PZ$
process~\cite{Kniehl:1990iv,Dicus:1988yh,Brein:2003wg,Kniehl:2011aa}.
\VHNNLO\
also includes the NLO corrections for this latter process, which are of
order $\alphas^3$~\cite{Altenkamp:2012sx}. The NLO+NLL corrections for
that process quoted below are not yet included
in \VHNNLO~\cite{Harlander:2014wda}.

\subsection{VH parameters and cuts}
\label{sec:VHcuts-sub}

The numerical results presented in the next section have been
computed using the values of the EW parameters given in Chapter~\ref{chapter:input}.
The electromagnetic coupling is fixed in the
$\GF$ scheme,
\beq
 \alpha_{\GF} = \sqrt{2}\GF\MW^2(1-\MW^2/\MZ^2)/\pi,
\eeq
and the weak mixing angle is defined in the on-shell scheme,
\begin{equation}
\sin^2\theta_{\mathrm{w}}=1-\MW^2/\MZ^2.
\end{equation}

In the calculation of the QCD-based cross sections,
the renormalization and factorization scales are set equal to the
invariant mass of the $\VH$ system,
\begin{equation}
\label{eq:VH_ren_fac_scales}
\mu=\muR = \muF= M_{\PV\PH}, \quad M_{V\PH}^2\equiv(p_V+p_{\PH})^2,
\end{equation}
and both scales are varied independently in the range $M_{V\PH}/3 < \mu < 3M_{V\PH}$.
The PDFs are taken from the set PDF4LHC15\_nnlo\_mc PDFs.

For the calculation of the EW corrections we employed
the NNPDF2.3QED PDF set~\cite{Ball:2013hta}, which includes
EW corrections and a photon PDF.
For the calculation of photon-induced contributions to the cross sections
with a realistic error estimate we took into account the photon PDF
of the MRST2004QED PDF set~\cite{Martin:2004dh} as well. 
A considerable reduction in the photon PDF uncertainty can be achieved
by using the more recent LUXqed\_plus\_PDF4LHC15\_nnlo\_100 PDF set ~\cite{Manohar:2016nzj}.

Note, however, that the relative EW correction factor, which is used in the following,
hardly depends on the PDF set, so that the uncertainty due to the
mismatch in the PDF selection is easily covered by the other remaining
theoretical uncertainties. Moreover, the EW corrections show a very small
dependence on the factorization scale, so that the use of
$\muF=M_V+\MH$ is acceptable,%
\footnote{In its present version, \HAWK{} does not support dynamical scales.}
although full consistency would require to
use equal QCD and QED factorization scales.

For the fiducial cross section and for
differential distributions the following reconstruction scheme
and cuts have been applied.
Jets are constructed according to the anti-$\kT$ algorithm~\cite{Cacciari:2008gp} with
$D=0.4$, using the default recombination scheme ($E$ scheme).
Jets are constructed from partons $j$ with
\begin{equation}
\label{eq:VH_cuts1}
|\eta_j| < 5\,,
\end{equation}
where $y_j$ denotes the rapidity of the (massive) jet.
In the presence of phase-space cuts and in the generation of
differential distributions, the treatment of real photons,
which appear as part of the NLO EW corrections, has to be specified.
In the following we assume perfect isolation of photons from leptons.%
\footnote{Perfect isolation to some extent applies to
muons going out into the muon chamber. A simulation of radiation off
electrons requires some recombination of collinear electron--photon pairs,
mimicking the inclusive treatment of electrons within electromagnetic showers
in the detector.
The two different treatments were compared in \Bref{Denner:2011id}, revealing
differences at the $1\%$ level for the relevant physical observables.}
The charged leptons $l$ have to pass the following acceptance cuts,
\begin{equation}
\label{eq:VH_cuts2}
{\pT}_{l} > 15\UGeV, \qquad
|y_{l}| <2.5\,.
\end{equation}
For $\PZ\PH$ production with $\PZ\to \ell^+\ell^-$ the invariant mass of the two leptons
should further concentrate around the Z~pole,
\begin{equation}
75\UGeV < M_{ll} < 105\UGeV.
\end{equation}

While the $\PZ\PH$ cross sections are independent
from the CKM matrix, quark mixing has some effect on WH production.
For the calculation of the latter we employed a Cabbibo-like CKM matrix
(i.e.\ without mixing to the third quark generation) with Cabbibo angle
$\theta_{\mathrm{C}}$ fixed by $\sin\theta_{\mathrm{C}}=0.225$.
Moreover, we note that we employ complex W- and Z-boson masses in the
calculation of the EW corrections in the standard \HAWK{} approach,
as described in \Bref{Denner:2011id}.

The Higgs boson is treated as on-shell particle in the following consistently,
since its finite-width and off-shell effects in the signal region are
suppressed in the Standard Model.

\subsection{Total VH cross sections}

\begin{table}
\caption{Total $\PWp({\to}l^+\nu_{l})$H cross sections including QCD and EW corrections
and their uncertainties for different proton--proton collision energies
$\sqrt{s}$ for a Higgs boson mass $\MH=125\UGeV$.}
\label{tab:wph_XStot}
\begin{center}%
\begin{small}%
\tabcolsep5pt
\renewcommand{\arraystretch}{1.2}
\begin{tabular}{cccccccc}%
\toprule
$\sqrt{s}$[GeV] & $\sigma$[fb] & $\Delta_{\mathrm{scale}}$[\%] &
$\Delta_{\mathrm{PDF}/\alphas/\mathrm{PDF\oplus\alphas}}$[\%] &
$\sigma_{\NNLO \QCD}^{\DY}$[fb] & $\sigma_{\Pt\mbox{\scriptsize -loop}}$[fb] &
$\delta_{\ELWK}$[\%] & $\sigma_{\gamma}$[fb]
\\
\midrule
$7$ & $  40.99$ & ${}_{-0.9}^{+ 0.7}$ & $\pm1.9/\pm0.7/\pm 2.0$ & $  42.78$ & $   0.42$ & $-7.2$ & $ 0.88^{+1.10}_{-0.10}$ \\
$8$ & $  49.52$ & ${}_{-0.9}^{+ 0.6}$ & $\pm1.8/\pm0.8/\pm 2.0$ & $  51.56$ & $   0.53$ & $-7.3$ & $ 1.18^{+1.38}_{-0.14}$ \\
$13$ & $  94.26$ & ${}_{-0.7}^{+ 0.5}$ & $\pm1.6/\pm0.9/\pm 1.8$ & $  97.18$ & $   1.20$ & $-7.4$ & $3.09^{+3.33}_{-0.37}$ \\
$14$ & $ 103.63$ & ${}_{-0.8}^{+ 0.3}$ & $\pm1.5/\pm0.9/\pm 1.8$ & $ 106.65$ & $   1.36$ & $-7.4$ & $3.55^{+3.72}_{-0.43}$ \\
\bottomrule
\end{tabular}%
\end{small}%
\end{center}%
\vspace{2em}
\caption{Total $\PWm({\to}l^-\bar\nu_{l})$H cross sections including QCD and EW corrections
and their uncertainties for different proton--proton collision energies
$\sqrt{s}$ for a Higgs boson mass $\MH=125\UGeV$.}
\label{tab:wmh_XStot}
\begin{center}%
\begin{small}%
\tabcolsep5pt
\renewcommand{\arraystretch}{1.2}
\begin{tabular}{cccccccc}%
\toprule
$\sqrt{s}$[GeV] & $\sigma$[fb] & $\Delta_{\mathrm{scale}}$[\%] &
$\Delta_{\mathrm{PDF}/\alphas/\mathrm{PDF\oplus\alphas}}$[\%] &
$\sigma_{\NNLO \QCD}^{\DY}$[fb] & $\sigma_{\Pt\mbox{\scriptsize -loop}}$[fb] &
$\delta_{\ELWK}$[\%] & $\sigma_{\gamma}$[fb]
\\
\midrule
$7$ & $  23.04$ & ${}_{-0.8}^{+ 0.6}$ & $\pm2.2/\pm0.6/\pm 2.3$ & $  23.98$ & $   0.24$ & $-7.0$ & $  0.51_{ -0.05}^{+  0.69}$ \\
$8$ & $  28.62$ & ${}_{-0.8}^{+ 0.6}$ & $\pm2.1/\pm0.6/\pm 2.1$ & $  29.71$ & $   0.31$ & $-7.1$ & $  0.70_{ -0.07}^{+  0.94}$ \\
$13$ & $  59.83$ & ${}_{-0.7}^{+ 0.4}$ & $\pm1.8/\pm0.8/\pm 2.0$ & $  61.51$ & $   0.78$ & $-7.3$ & $  2.00_{ -0.22}^{+  2.34}$ \\
$14$ & $  66.49$ & ${}_{-0.6}^{+ 0.5}$ & $\pm1.7/\pm0.9/\pm 1.9$ & $  68.24$ & $   0.89$ & $-7.3$ & $  2.32_{ -0.26}^{+  2.65}$ \\
\bottomrule
\end{tabular}%
\end{small}%
\end{center}%
\end{table}
Tables~\ref{tab:wph_XStot} and \ref{tab:wmh_XStot} summarize the total
Standard Model $\PW^\pm$H cross sections with $\PWp{\to}l^+\nu_{l}$
and $\PWm{\to}l^-\bar\nu_{l}$
as well as the corresponding uncertainties
for different proton--proton collision energies
for a Higgs boson mass $\MH=125\UGeV$.
Tables~\ref{tab:zllh_XStot} and~\ref{tab:znnh_XStot} likewise show
the respective results on the total
Standard Model $\PZ$H cross sections with $\PZ\to \ell^+\ell^-$
and $\PZ\to\nu\bar\nu$ (summed over three neutrino generations).

\begin{table}
\caption{Total $\PZ$H cross sections with $\PZ\to \ell^+\ell^-$ including QCD and EW corrections
and their uncertainties for different proton--proton collision energies
$\sqrt{s}$ for a Higgs boson mass $\MH=125\UGeV$.}
\label{tab:zllh_XStot}
\begin{center}%
\footnotesize
\tabcolsep5pt
\renewcommand{\arraystretch}{1.2}
\begin{tabular}{ccccccccc}%
\toprule
$\sqrt{s}$[GeV] & $\sigma$[fb] & $\Delta_{\mathrm{scale}}$[\%] &
$\Delta_{\mathrm{PDF}/\alphas/\mathrm{PDF\oplus\alphas}}$[\%] &
$\sigma_{\NNLO \QCD}^{\DY}$[fb] & $\sigma^{\Pg\Pg\PZ\PH}_{\NLO+\NLL}$[fb] &
$\sigma_{\Pt\mbox{\scriptsize -loop}}$[fb] &
$\delta_{\ELWK}$[\%] & $\sigma_{\gamma}$[fb]
\\
\midrule
$7$ & $  11.43$ & ${}_{-2.4}^{+ 2.6}$ & $\pm1.6/\pm0.7/\pm 1.7$ & $  10.91$ & $   0.94$ & $   0.11$ & $-5.2$ & $  0.03_{ -0.00}^{+  0.04}$ \\
$8$ & $  14.18$ & ${}_{-2.4}^{+ 2.9}$ & $\pm1.5/\pm0.8/\pm 1.7$ & $  13.36$ & $   1.33$ & $   0.14$ & $-5.2$ & $  0.04_{ -0.00}^{+  0.05}$ \\
$13$ & $  29.82$ & ${}_{-3.1}^{+ 3.8}$ & $\pm1.3/\pm0.9/\pm 1.6$ & $  26.66$ & $   4.14$ & $   0.31$ & $-5.3$ & $  0.11_{ -0.01}^{+  0.12}$ \\
$14$ & $  33.27$ & ${}_{-3.3}^{+ 3.8}$ & $\pm1.3/\pm1.0/\pm 1.6$ & $  29.47$ & $   4.87$ & $   0.36$ & $-5.3$ & $  0.12_{ -0.01}^{+  0.13}$ \\
\bottomrule
\end{tabular}%
\footnotesize
\end{center}%
\vspace{2em}
\caption{Total $\PZ$H cross sections with $\PZ\to\nu\bar\nu$ (summed over three neutrino generations)
including QCD and EW corrections
and their uncertainties for different proton--proton collision energies
$\sqrt{s}$ for a Higgs boson mass $\MH=125\UGeV$.}
\label{tab:znnh_XStot}
\begin{center}%
\footnotesize
\tabcolsep5pt
\renewcommand{\arraystretch}{1.2}
\begin{tabular}{ccccccccc}%
\toprule
$\sqrt{s}$[GeV] & $\sigma$[fb] & $\Delta_{\mathrm{scale}}$[\%] &
$\Delta_{\mathrm{PDF}/\alphas/\mathrm{PDF\oplus\alphas}}$[\%] &
$\sigma_{\NNLO \QCD}^{\DY}$[fb] & $\sigma^{\Pg\Pg\PZ\PH}_{\NLO+\NLL}$[fb] &
$\sigma_{\Pt\mbox{\scriptsize -loop}}$[fb] &
$\delta_{\ELWK}$[\%] & $\sigma_{\gamma}$[fb]
\\
\midrule
$7 $ & $  68.18$ & ${}_{-2.4}^{+ 2.6}$ & $\pm1.6/\pm0.7/\pm 1.7$ & $  64.70$ & $   5.59$ & $   0.64$ & $-4.3$ & $  -0.00$ \\
$8 $ & $  84.56$ & ${}_{-2.4}^{+ 2.9}$ & $\pm1.5/\pm0.8/\pm 1.7$ & $  79.25$ & $   7.89$ & $   0.81$ & $-4.3$ & $  -0.00$ \\
$13$ & $ 177.62$ & ${}_{-3.1}^{+ 3.8}$ & $\pm1.3/\pm0.9/\pm 1.6$ & $ 158.10$ & $  24.57$ & $   1.85$ & $-4.4$ & $  -0.00$ \\
$14$ & $ 198.12$ & ${}_{-3.3}^{+ 3.8}$ & $\pm1.3/\pm1.0/\pm 1.6$ & $ 174.77$ & $  28.88$ & $   2.11$ & $-4.4$ & $  -0.00$ \\
\bottomrule
\end{tabular}%
\end{center}%
\end{table}

The total VH cross sections $\sigma^{\VH}$ are calculated according to
\begin{eqnarray}
\sigma^{\PW\PH} &=& \sigma_{\NNLO \QCD}^{\PW\PH,\DY} (1+\delta_{\ELWK}) + \sigma_{\Pt\mbox{\scriptsize -loop}} + \sigma_{\gamma},
\label{eq:sigmaWH}
\\
\sigma^{\PZ\PH} &=& \sigma_{\NNLO \QCD}^{\PZ\PH,\DY} (1+\delta_{\ELWK})
+ \sigma_{\Pt\mbox{\scriptsize -loop}} + \sigma_{\gamma} + \sigma^{\Pg\Pg\PZ\PH},
\label{eq:sigmaZH}
\end{eqnarray}
where $\sigma_{\NNLO \QCD}^{\VH,\DY}$ is the Drell--Yan-like part of the
NNLO QCD prediction for the VH cross section, based on the calculation
of \Bref{Brein:2003wg} with NNLO PDFs.
Since we include the leptonic decays of the W/Z~bosons, we multiply the
cross sections from \VHNNLO\ with the branching ratios
\begin{equation}
\mathrm{BR}_{\LO}(\PW\to \ell\nu_{\ell}) = 0.108894, \quad
\mathrm{BR}_{\LO}(\PZ\to \ell^+\ell^-) = 0.0335950, \quad
\mathrm{BR}_{\LO}(\PZ\to\nu\bar\nu) = 0.199218,
\end{equation}
which are the ratios of the LO partial widths and the total widths as defined in Chapter~\ref{chapter:input}.
With these branching ratios our combination of QCD and EW corrections results in
NNLO QCD + NLO EW accuracy.
The relative NLO EW correction $\delta_{\ELWK}$ is calculated with \HAWK{}.
Note that there is no issue with photon isolation in the calculation of the total cross section,
where all mass singularities from collinear photon emission off leptons vanish owing to the
KLN theorem.
The contributions from photon-induced channels, $\sigma_{\gamma}$
are obtained from \HAWK{} as well and added linearly to the cross section.
It is important to notice that $\sigma_{\gamma}$ is based on the average of the
median of the cross sections obtained with PDF replicas of NNPDF2.3QED PDFs
and the cross section obtained with MRST2004QED PDFs ``set~1''.
The lower error corresponds to the lower limit of all NNPDF2.3QED PDFs,
the upper error to the maximum of the 68\% smallest cross sections from
the NNPDF2.3QED set and the cross section obtained with MRST2004QED ``set~0''.
Since the photon PDF is constrained by data rather loosely, the error on
$\sigma_{\gamma}$ is large and non-Gaussian. In fact the mean value of
$\sigma_{\gamma}$ calculated with NNPDF2.3QED PDF replicas is
larger than the shown median by factors $\sim2{-}2.5$.

The contribution $\sigma^{\Pg\Pg\PZ\PH}$ of the gluon-fusion channel is
calculated through NLO
using \VHNNLO \cite{Brein:2012ne,Harlander:2013mla,Altenkamp:2012sx};
the NLL effects are added on top of that, following \Bref{Harlander:2014wda}.
The scale uncertainty
$\Delta_{\mathrm{scale}}$ results from a variation of the factorization
and renormalization scales \eqref{eq:VH_ren_fac_scales} by a factor of
$3$, as indicated above.
The errors $\Delta_{\mathrm{PDF}}$ and $\Delta_{\alphas}$ induced by
uncertainties in the PDFs and $\alphas$, respectively, are given separately
together with the combined version
$\Delta_{\mathrm{PDF\oplus\alphas}}$, which is calculated
from the 68\%~CL interval using the PDF4LHC15\_nnlo\_mc PDF
set. The
$\Delta_{\mathrm{scale}}$ and $\Delta_{\mathrm{PDF\oplus\alphas}}$ are
evaluated without taking into account EW effects.

The theoretical uncertainties of integrated cross sections originating from unknown higher-order
EW effects can be estimated by
\begin{equation}
\Delta_\ELWK = \max\{0.5\%,\delta_{\ELWK}^2,\Delta_\gamma\}.
\label{eq:VH_EW_THU}
\end{equation}
This estimate is based on the maximum of the generic size $\sim0.5\%$ of the neglected
NNLO EW effects, taking into account a possible systematic enhancement $\sim\delta_{\ELWK}^2$,
and the potentially large relative uncertainty $\Delta_\gamma=\Delta\sigma_\gamma/\sigma$ of the photon-induced
contribution $\sigma_\gamma$, whose absolute uncertainty $\Delta\sigma_\gamma$
can be read from the tables.

In order to extract the total VH production cross sections without
leptonic W/Z decays in NNLO QCD + NLO EW accuracy (neglecting off-shell effects),
one should divide the results on total cross sections by the respective
leptonic W/Z branching ratios in NLO EW accuracy.
These are given by
\begin{equation}
\mathrm{BR}_{\NLO}(\PW\to \ell\nu_{\ell}) = 0.108535, \quad
\mathrm{BR}_{\NLO}(\PZ\to \ell^+\ell^-) = 0.0335962, \quad
\mathrm{BR}_{\NLO}(\PZ\to\nu\bar\nu) = 0.201030,
\end{equation}
calculated from the ratios of the NLO partial widths
(calculated in the \HAWK{} setup) and the total widths
as defined in Chapter~\ref{chapter:input}.
In this extraction, one should subtract the photon-induced contributions $\sigma_\gamma$ from the
cross sections before dividing through
the branching ratio, since $\sigma_\gamma$ receives a significant contribution
from incoming photons coupling to the charged leptons of the W or Z~decays.
Thus, $\sigma_\gamma/\mathrm{BR}$ is an uncertainty of the resulting VH cross section, which
is quite significant in the WH case.

Results for the total VH cross sections from a scan over the SM Higgs boson mass $\MH$
can be found in \refA{VHappendix}. 
In detail the total cross sections for the production of
$\PW^+({\to} \ell^+\nu_{\ell})\PH$,
$\PW^-({\to} \ell^-\bar\nu_{\ell})\PH$,
$\PZ({\to} \ell^+\ell^-)\PH$, and
$\PZ({\to}\nu\bar\nu)\PH$ final states are summarized in
\refTs{tab:wph_XStot_7}--\ref{tab:znnh_XStot_14}.
The energy scan is presented in \refTs{tab:WH_XStot_125}--\ref{tab:ZH_XStot_12509}.

\subsection{Fiducial and differential VH cross sections}

\begin{table}
\caption{Fiducial $\PWp({\to}l^+\nu_{l})$H cross sections including QCD and EW corrections
and their uncertainties for proton--proton collisions at
$\sqrt{s}=13\rm{TeV}$ for a Higgs boson mass $\MH=125\UGeV$.}
\label{tab:wph_XSfiducial}
\begin{center}%
\begin{small}%
\tabcolsep5pt
\begin{tabular}{ccccccc}%
\toprule
$\sqrt{s}$[GeV] & $\sigma$[fb] & $\Delta_{\mathrm{scale}}$[\%] & $\Delta_{\mathrm{PDF}}$[\%] &
$\sigma_{\NNLO \QCD}^{\DY}$[fb] & $\delta_{\ELWK}$[\%] & $\sigma_{\gamma}$[fb]
\\
\midrule
$13$ & $73.90$ & ${}_{-0.3}^{+ 0.3}$ & $\pm 1.4$ & $78.61$ & $-8.3$ & $1.81^{+1.10}_{-0.23}$
\\
\bottomrule
\end{tabular}%
\end{small}%
\end{center}%
\vspace{2em}
\caption{Fiducial $\PWm({\to}l^-\bar\nu_{l})$H cross sections including QCD and EW corrections
and their uncertainties for proton--proton collisions at
$\sqrt{s}=13\rm{TeV}$ for a Higgs boson mass $\MH=125\UGeV$.}
\label{tab:wmh_XSfiducial}
\begin{center}%
\begin{small}%
\tabcolsep5pt
\begin{tabular}{ccccccc}%
\toprule
$\sqrt{s}$[GeV] & $\sigma$[fb] & $\Delta_{\mathrm{scale}}$[\%] & $\Delta_{\mathrm{PDF}}$[\%] &
$\sigma_{\NNLO \QCD}^{\DY}$[fb] & $\delta_{\ELWK}$[\%] & $\sigma_{\gamma}$[fb]
\\
\midrule
$13$ & $ 42.77 $ & ${}_{-0.3}^{+ 0.2}$ & $\pm 1.8$ & $45.29$ & $-8.0$ & $1.11^{+0.65}_{-0.12}$
\\
\bottomrule
\end{tabular}%
\end{small}%
\end{center}%
\vspace{2em}
\caption{Fiducial $\PZ$H cross sections with $\PZ\to \ell^+\ell^-$ including QCD and EW corrections
and their uncertainties for proton--proton collisions at
$\sqrt{s}=13\rm{TeV}$ for a Higgs boson mass $\MH=125\UGeV$.}
\label{tab:zllh_XSfiducial}
\begin{center}%
\begin{small}%
\tabcolsep5pt
\begin{tabular}{cccccccc}%
\toprule
$\sqrt{s}$[GeV] & $\sigma$[fb] & $\Delta_{\mathrm{scale}}$[\%] & $\Delta_{\mathrm{PDF}}$[\%] & $\sigma_{\NNLO \QCD}^{\DY}$[fb] &
$\sigma^{\Pg\Pg\PZ\PH}$[fb] & $\delta_{\ELWK}$[\%] & $\sigma_{\gamma}$[fb]
\\
\midrule
$13$ & $16.08$ & ${}_{-1.4}^{+ 2.2}$ &  $\pm 1.2$ & $16.21$ & $1.36$ & $-9.2$ & $0.00$
\\
\bottomrule
\end{tabular}%
\end{small}%
\end{center}%
\end{table}
Tables~\ref{tab:wph_XSfiducial} and \ref{tab:wmh_XSfiducial} summarize the fiducial
Standard Model $\PW^\pm$H cross sections with $\PWp{\to}l^+\nu_{l}$
and $\PWm{\to}l^-\bar\nu_{l}$
as well as the corresponding uncertainties
for the proton--proton collision at $\sqrt{s}=13\UTeV$
for a Higgs boson mass $\MH=125\UGeV$.
Table~\ref{tab:zllh_XSfiducial} likewise shows
the respective results on the total
Standard Model $\PZ$H cross sections with $\PZ\to \ell^+\ell^-$.
The fiducial cross sections are calculated as follows: For QCD corrections we have used
\vhnnlo{} with the $\rm{NNPDF3.0\_nnlo\_as\_0118}$ PDF set. Renormalization and factorization scales
are varied independently by factors of $2$ and $1/2$ including $7$ combinations, avoiding the
cases $(2,1/2)$ and $(1/2,2)$. The envelope is taken as a scale uncertainty to parameterize
missing higher-order QCD corrections.
A representation of the PDF error for the QCD part in VH production has been obtained
with SM-PDF~\cite{Carrazza:2016htc} starting from a prior of NNPDF3.0 at NNLO. The SM-PDF derived
in~\cite{Carrazza:2016htc} for $\rm{VH}$ processes adopted an analysis very close to the one used
here and contains in total five symmetric eigenvectors.
The EW corrections are again calculated with \HAWK{} in the same way as described for the
total cross section. Moreover, the recipe \eqref{eq:VH_EW_THU} for estimating the
EW uncertainty $\Delta_\ELWK$ applies for the fiducial cross section as well.

The combination of QCD and EW corrections
has been done following the same procedure as described
in Eqs.~\eqref{eq:sigmaWH} and \eqref{eq:sigmaZH}, as far as the
corresponding contributions are available (see tables).
The $\Delta_{\mathrm{scale}}$ and $\Delta_{\mathrm{PDF}}$ are
evaluated without taking into account EW effects;
the uncertainties of the latter can be estimated again following Eq.~\eqref{eq:VH_EW_THU}.

Differential cross section results in NNLO QCD + NLO EW accuracy have been computed
following the same procedure as outlined above for the fiducial cross section.
QCD corrections are calculated with \vhnnlo{} using
the settings reported above for the computation of the fiducial cross sections.
The EW corrections are again calculated with \HAWK{} as in the previous section,
with the only difference in the calculation of the photon-induced contribution.
Instead of
working with many PDF replicas we
have calculated $\sigma_\gamma$ with the central PDF of NNPDF2.3QED.
In order to obtain $\sigma_\gamma$ in the same setup as for the integrated
cross sections of the previous section (for $\sqrt{s}=13\UTeV$),
the shown results on $\sigma_\gamma$ in $\PW\PH$ production should be rescaled
by a factor of $0.7$. This rescaling is based on the corresponding
integrated results for $\sigma_\gamma$. Taking over the relative uncertainty
from the integrated cross section as well, we get the estimate
$\Delta_\gamma\sim1.5\%$.
For $\PZ\PH$ production $\sigma_\gamma$ and $\Delta_\gamma$ have a
phenomenologically negligible impact.

The theoretical uncertainties of differential cross sections originating from unknown
higher-order EW effects can be estimated by
\begin{equation}
\Delta_\ELWK = \max\{1\%,\delta_{\ELWK}^2,\Delta_\gamma\},
\end{equation}
i.e.\ $\Delta_\ELWK$ is taken somewhat more conservative than for integrated cross sections,
accounting for possible enhancements of higher-order effects due to a kinematical migration
of events in distributions.
Note that $\delta_{\ELWK}^2$, in particular, covers the known effect of enhanced EW corrections at high
momentum transfer (EW Sudakov logarithms, etc.).

\begin{figure}
\includegraphics[width=.47\textwidth]{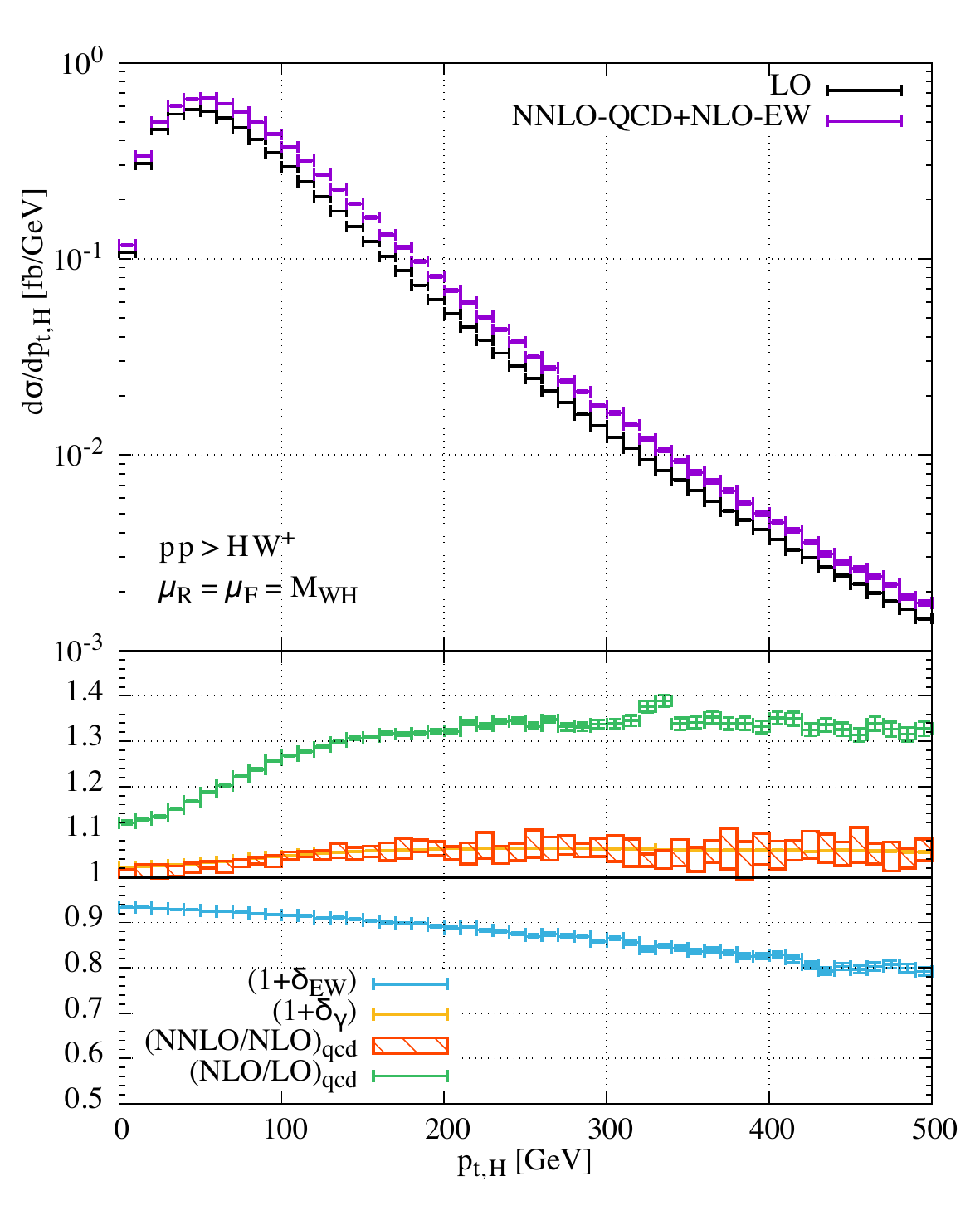}
\hfill
\includegraphics[width=.47\textwidth]{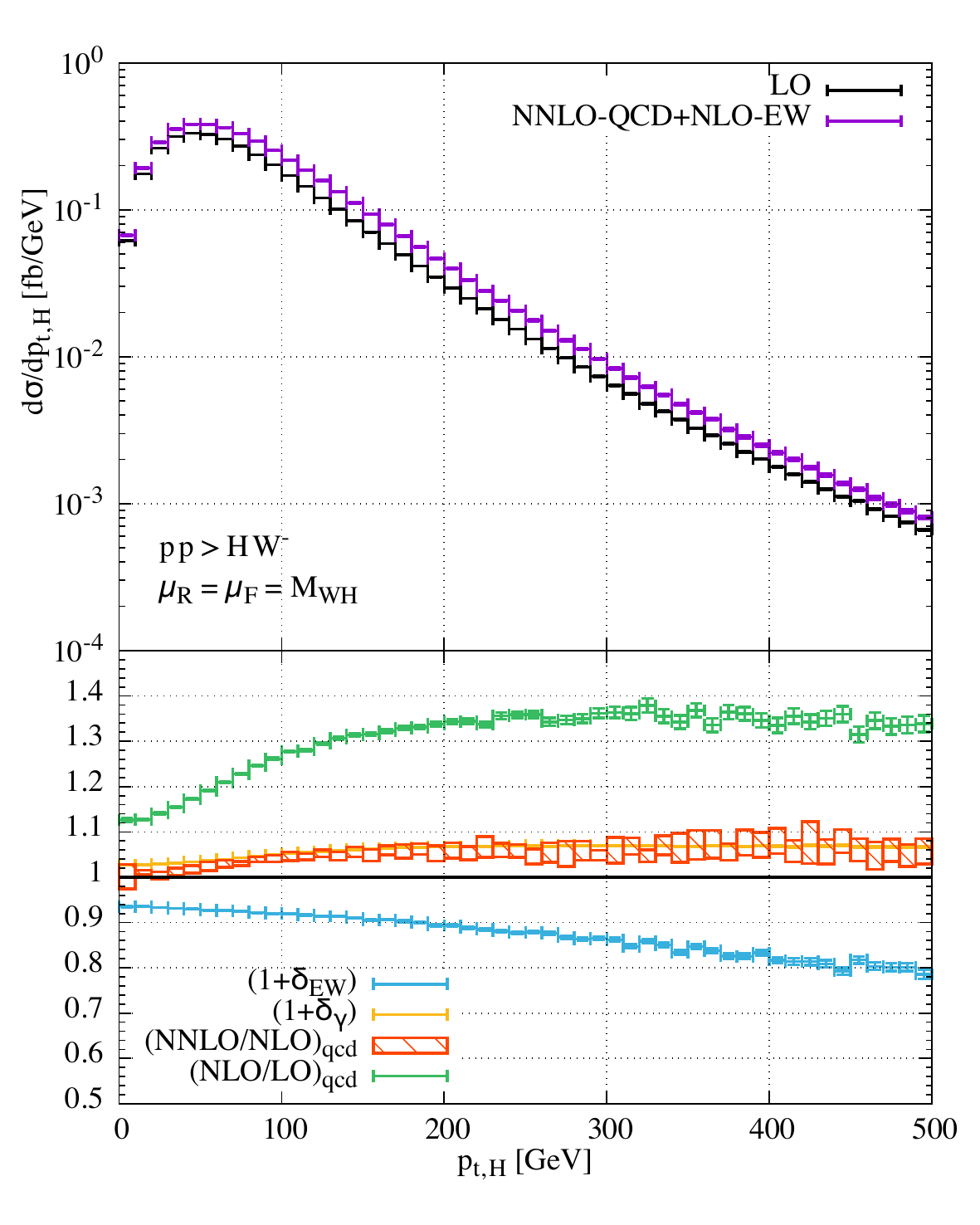}
\caption{Left: transverse-momentum distributions of the Higgs boson in $\PW^+\PH$
production at LO and including NNLO QCD and NLO EW corrections (upper plots)
and relative higher-order contributions (lower plots) for $\sqrt{s}=13\UTeV$ and $\MH=125\UGeV$.
Right: the same for $\PW^-\PH$ production.
Note that $\delta_\gamma$ is based on the central value
of the photon PDF of NNPDF2.3QED, while $\sigma_\gamma$ in \Trefs{tab:wph_XStot}--\ref{tab:zllh_XSfiducial}
is based on combined results using the median and the photon PDF of MRST2004QED (and smaller by a factor 0.7), see text.
}
\label{fig:SM-WH-ptH}
\end{figure}
\begin{figure}
\includegraphics[width=.47\textwidth]{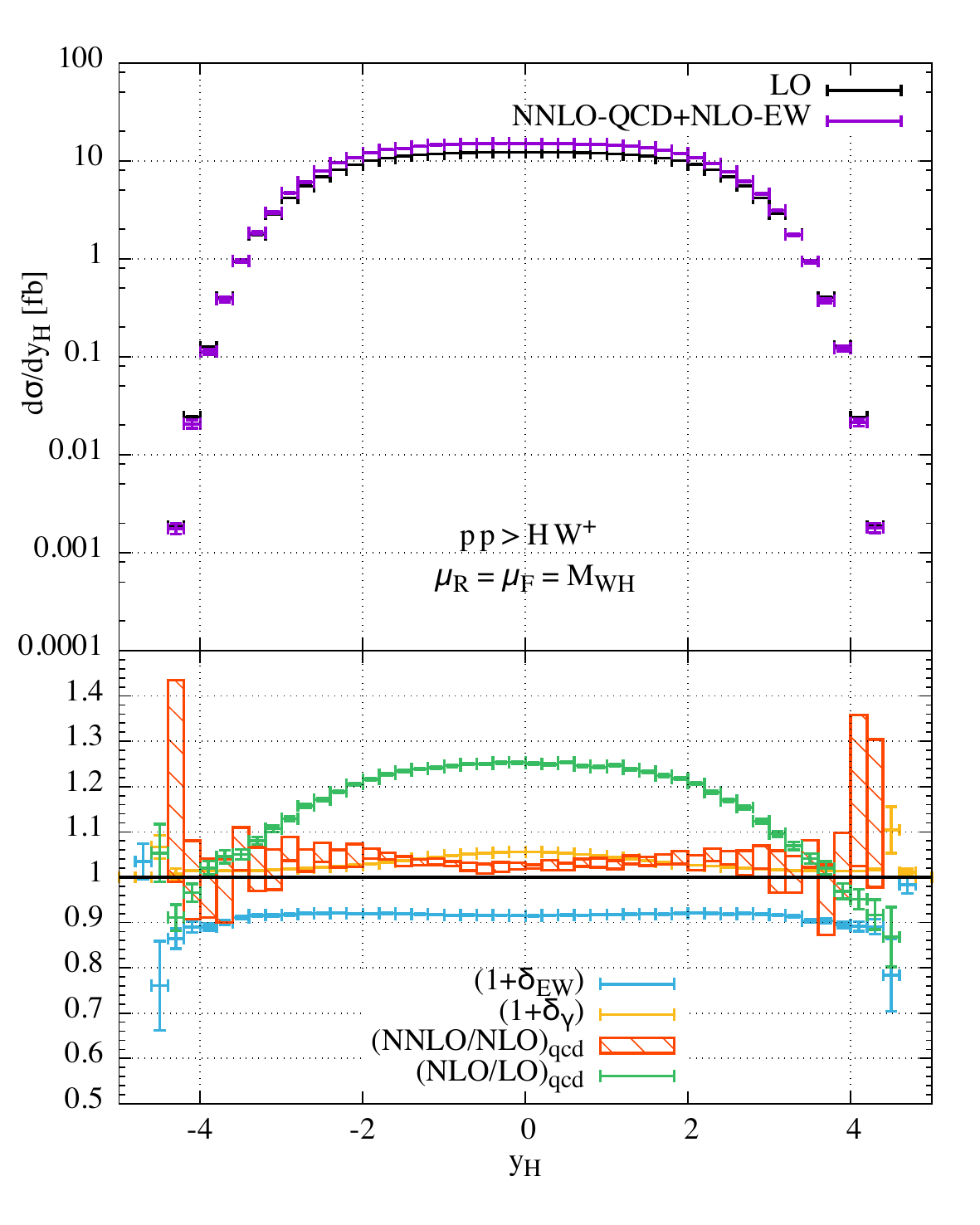}
\hfill
\includegraphics[width=.47\textwidth]{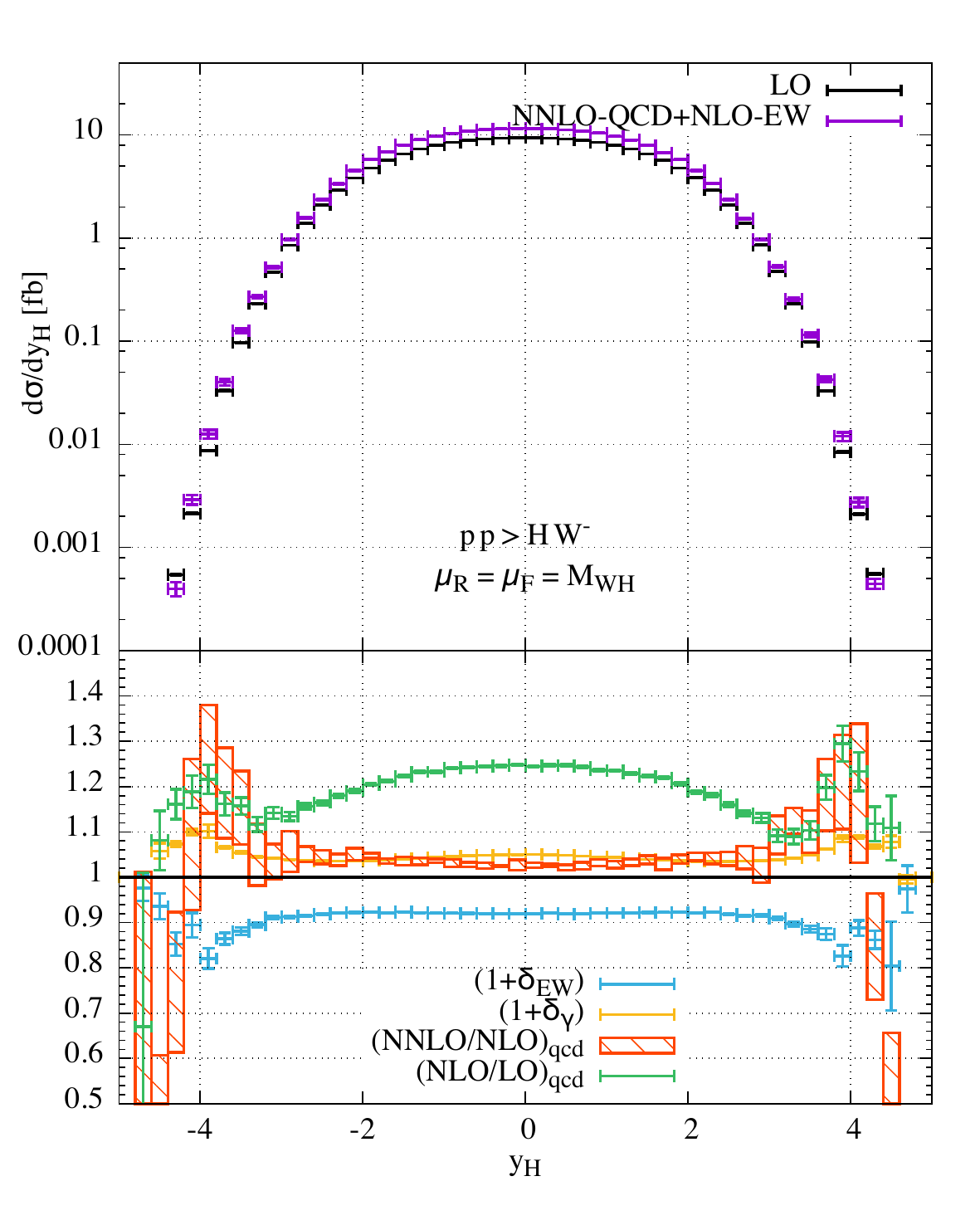}
\caption{Left: rapidity of the Higgs boson in $\PW^+\PH$
production at LO and including NNLO QCD and NLO EW corrections (upper plots)
and relative higher-order contributions (lower plots) for $\sqrt{s}=13\UTeV$ and $\MH=125\UGeV$.
Right: the same for $\PW^-\PH$ production.
Note that $\delta_\gamma$ is based on the central value
of the photon PDF of NNPDF2.3QED, while $\sigma_\gamma$ in \Trefs{tab:wph_XStot}--\ref{tab:zllh_XSfiducial}
is based on combined results using the median and the photon PDF of MRST2004QED (and smaller by a factor 0.7), see text.
}
\label{fig:SM-WH-yH}
\end{figure}
\begin{figure}
\includegraphics[width=.47\textwidth]{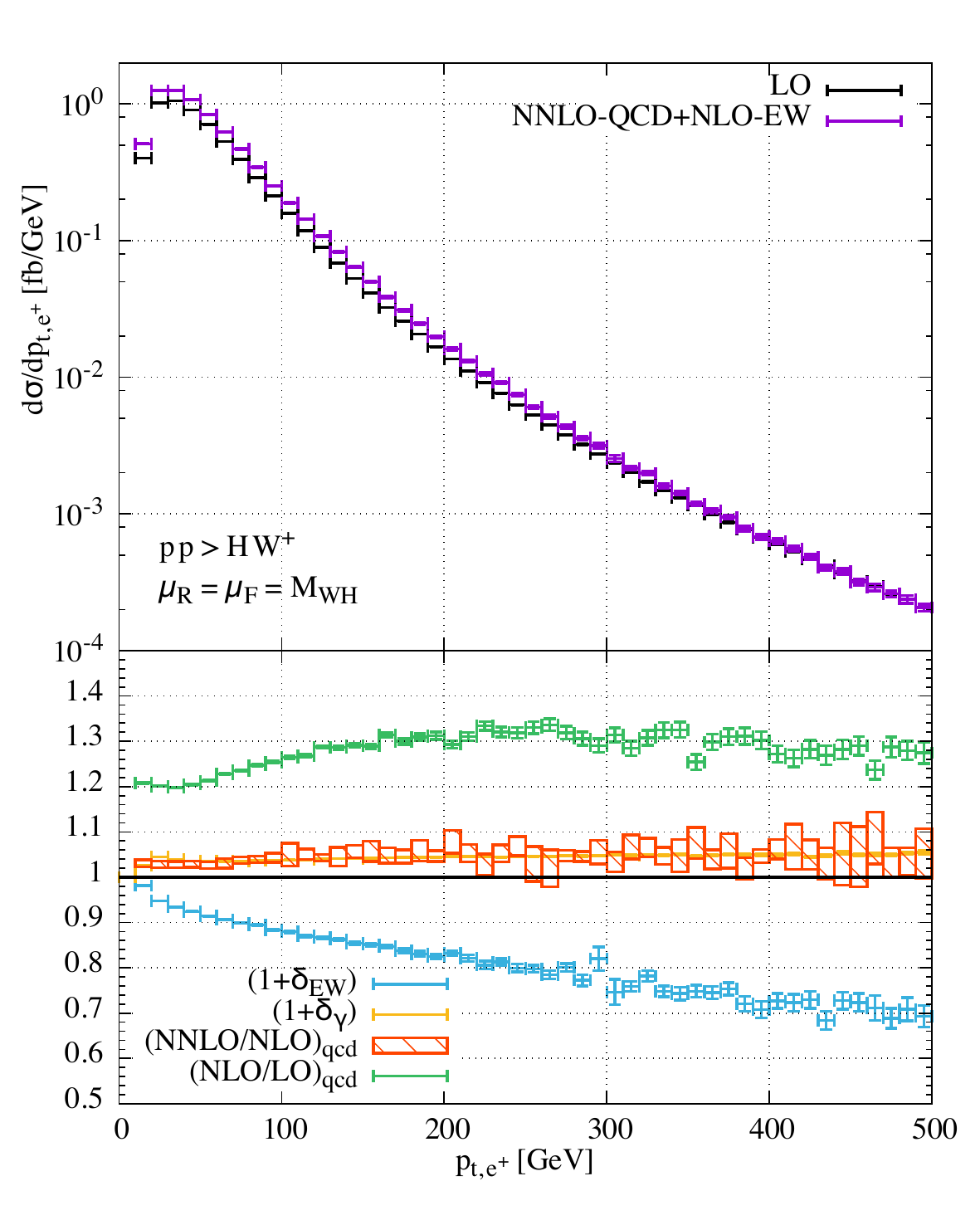}
\hfill
\includegraphics[width=.47\textwidth]{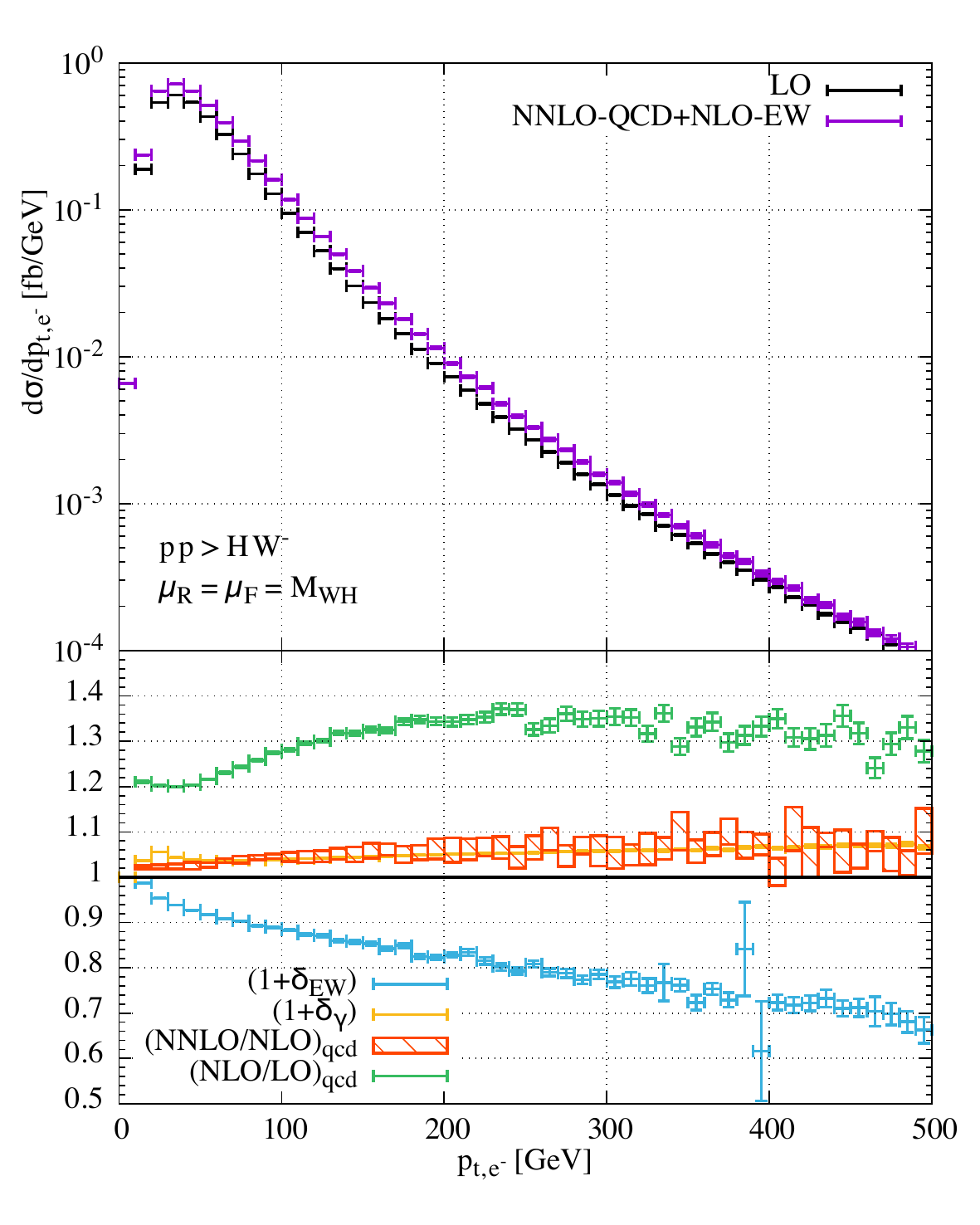}
\caption{Left: transverse-momentum distribution of the charged lepton in $\PW^+\PH$
production at LO and including NNLO QCD and NLO EW corrections (upper plots)
and relative higher-order contributions (lower plots) for $\sqrt{s}=13\UTeV$ and $\MH=125\UGeV$.
Right: the same for $\PW^-\PH$ production.
Note that $\delta_\gamma$ is based on the central value
of the photon PDF of NNPDF2.3QED, while $\sigma_\gamma$ in \Trefs{tab:wph_XStot}--\ref{tab:zllh_XSfiducial}
is based on combined results using the median and the photon PDF of MRST2004QED (and smaller by a factor 0.7), see text.
}
\label{fig:SM-WH-ptlep}
\end{figure}
\begin{figure}
\includegraphics[width=.47\textwidth]{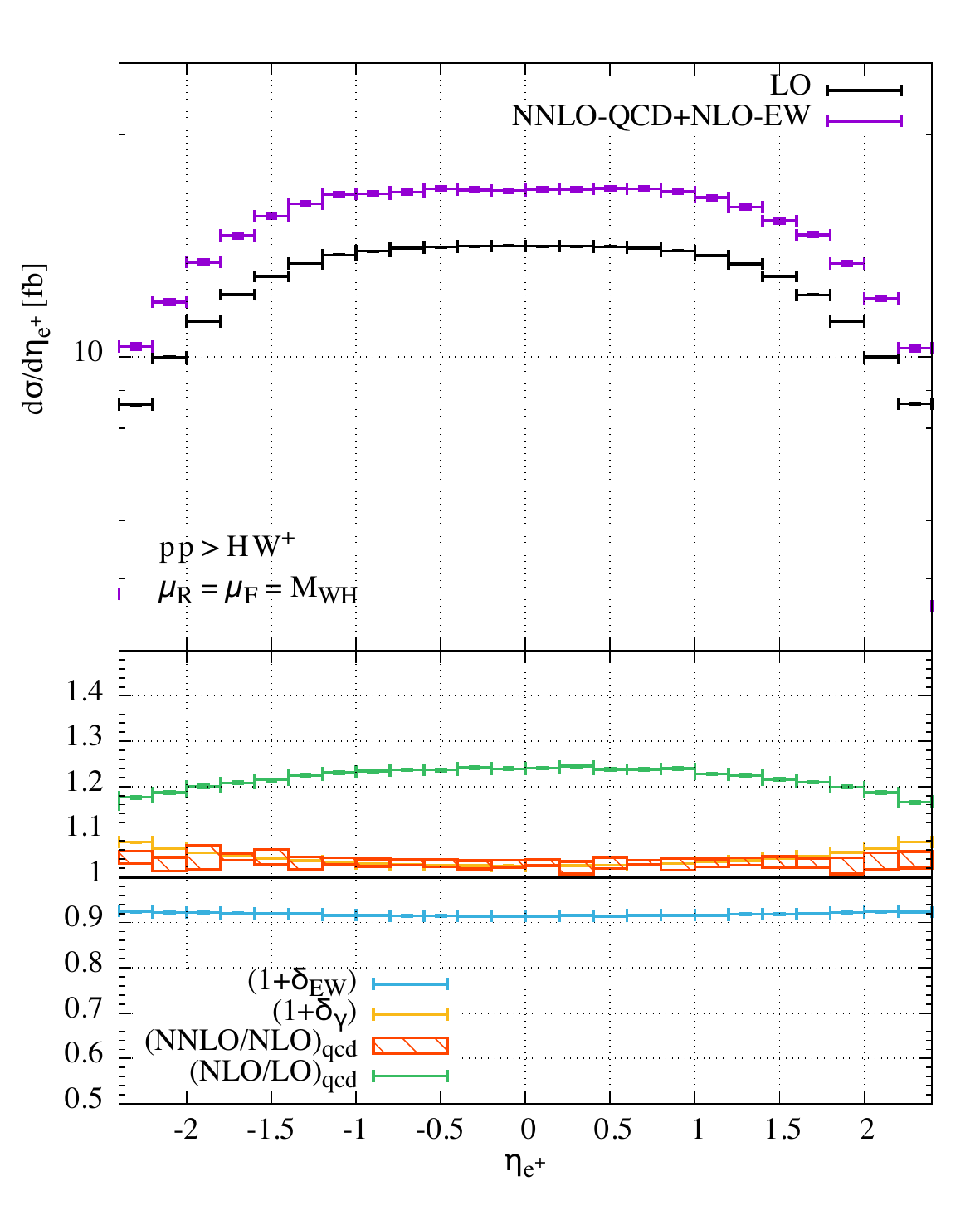}
\hfill
\includegraphics[width=.47\textwidth]{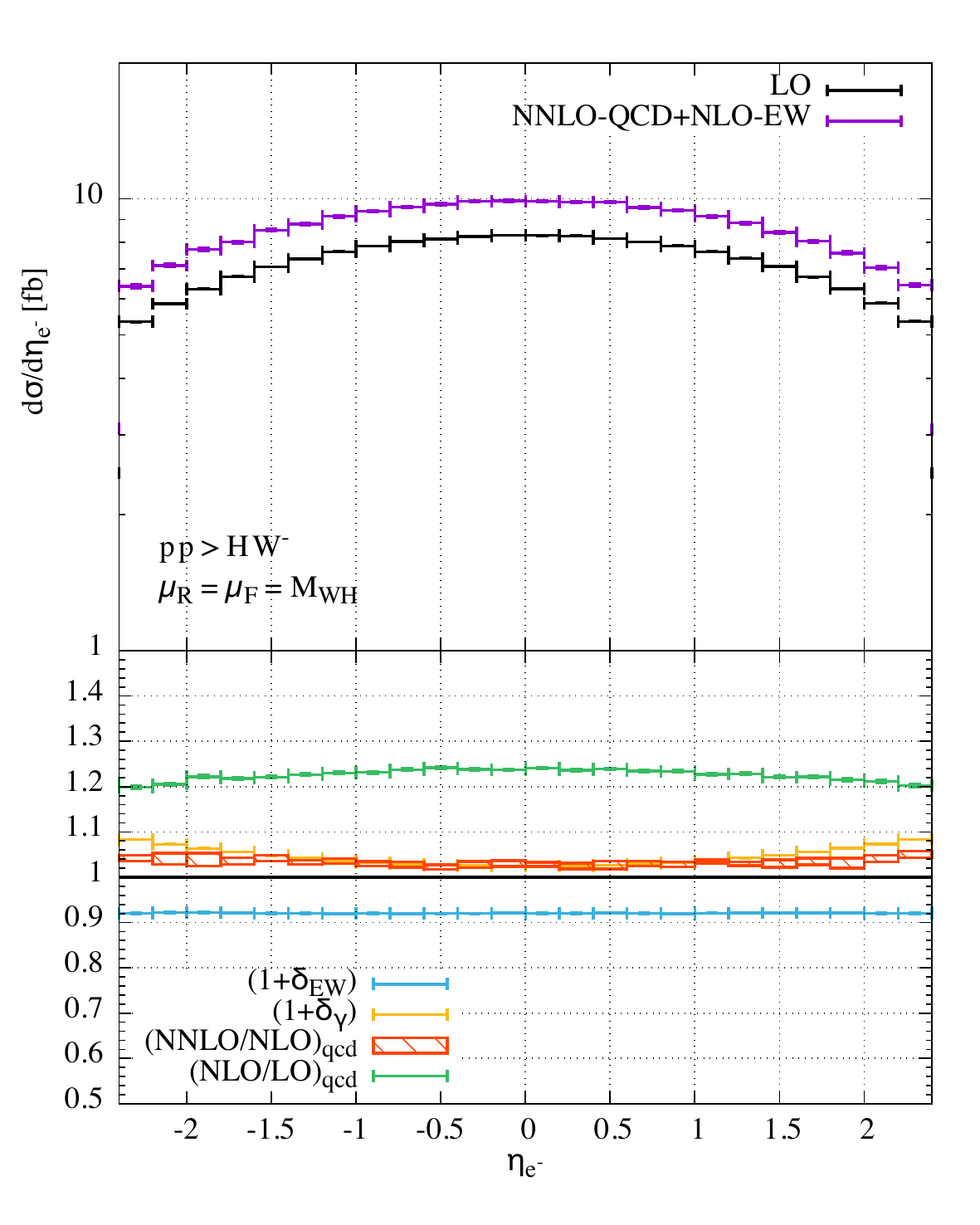}
\caption{Left: pseudorapidity distribution of the charged lepton in $\PW^+\PH$
production at LO and including NNLO QCD and NLO EW corrections (upper plots)
and relative higher-order contributions (lower plots) for $\sqrt{s}=13\UTeV$ and $\MH=125\UGeV$.
Right: the same for $\PW^-\PH$ production.
Note that $\delta_\gamma$ is based on the central value
of the photon PDF of NNPDF2.3QED, while $\sigma_\gamma$ in \Trefs{tab:wph_XStot}--\ref{tab:zllh_XSfiducial}
is based on combined results using the median and the photon PDF of MRST2004QED (and smaller by a factor 0.7), see text.
}
\label{fig:SM-WH-pseudorapidity}
\end{figure}
\begin{figure}
\includegraphics[width=.47\textwidth]{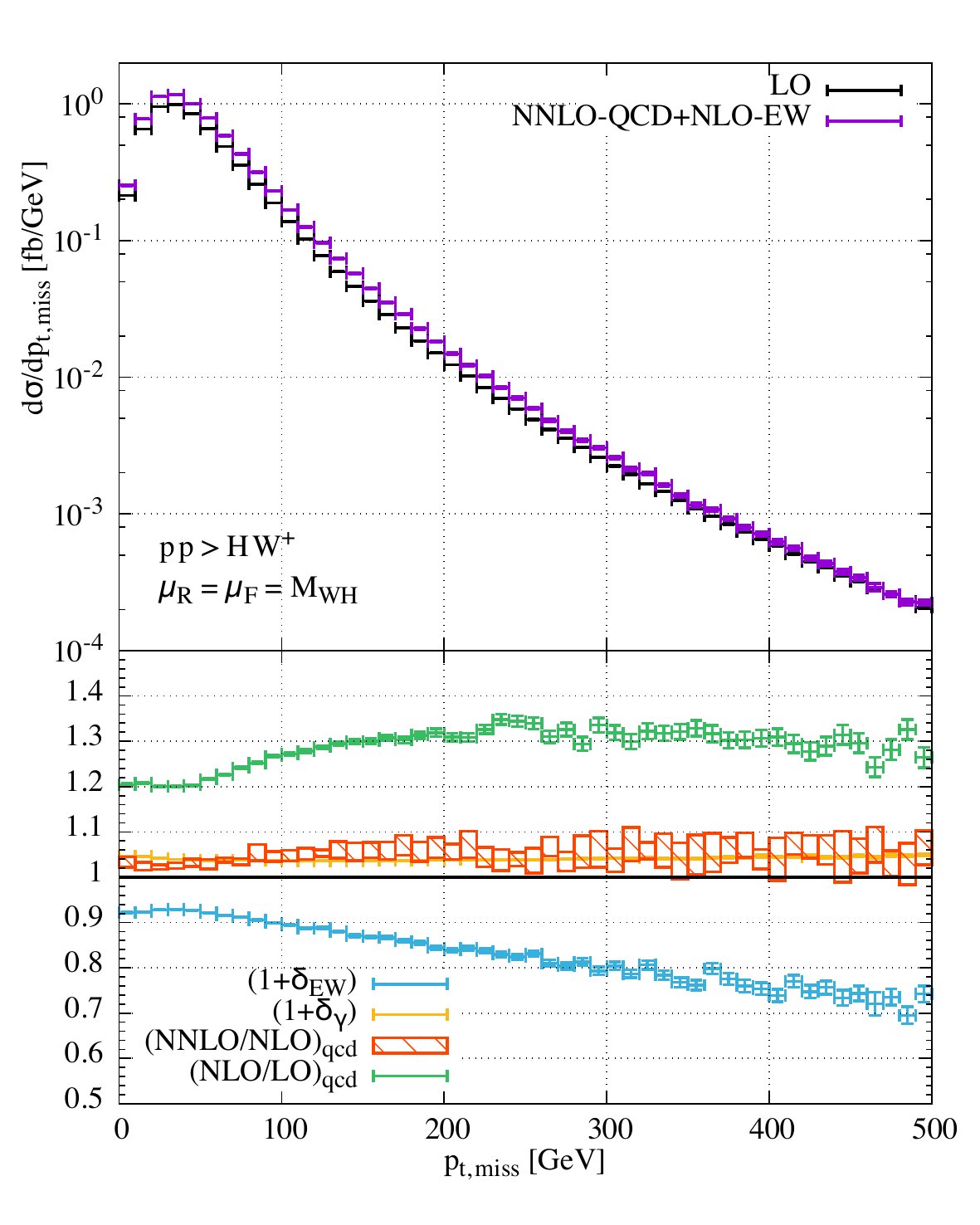}
\hfill
\includegraphics[width=.47\textwidth]{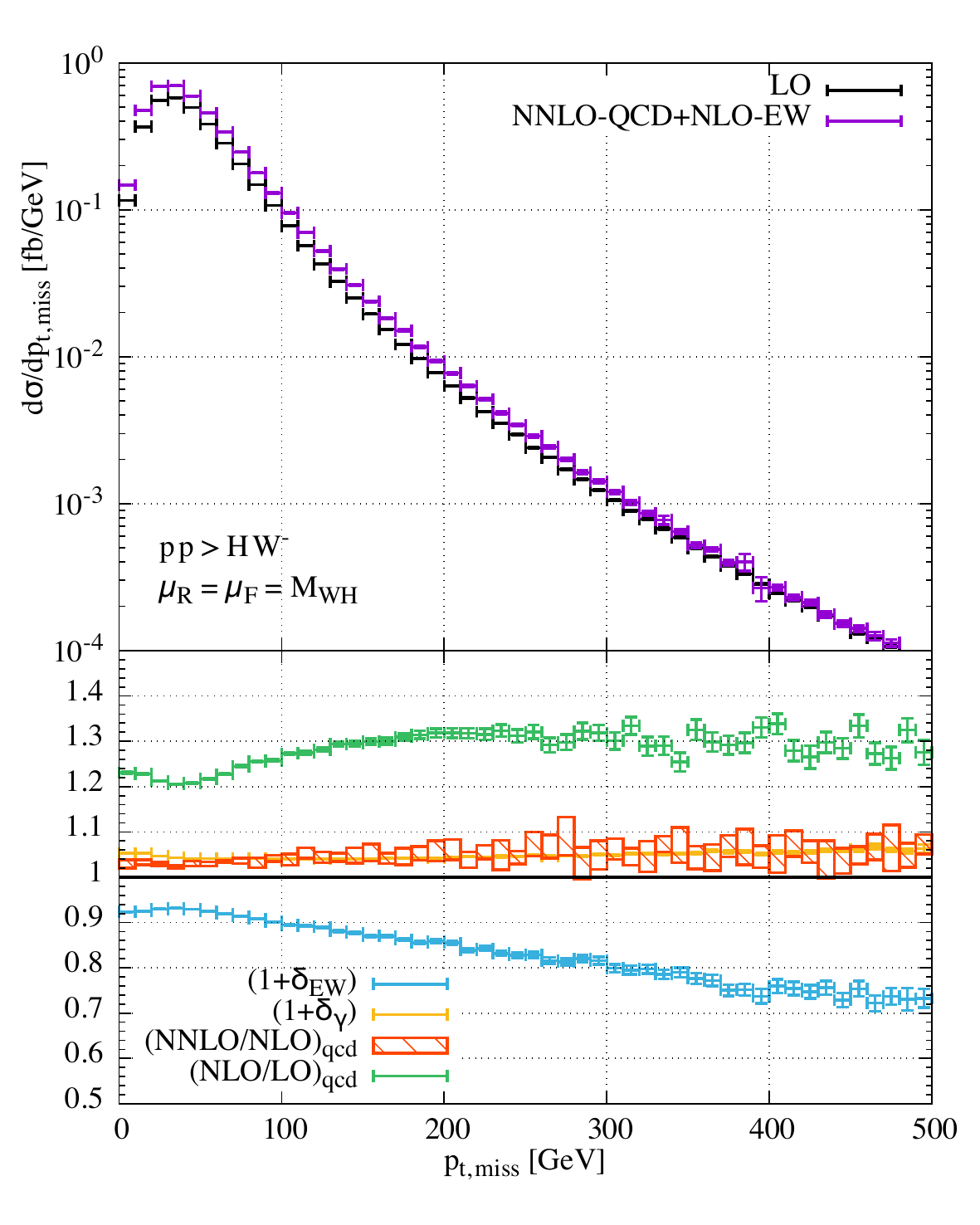}
\caption{Left: missing transverse momentum in $\PW^+\PH$
production at LO and including NNLO QCD and NLO EW corrections (upper plots)
and relative higher-order contributions (lower plots) for $\sqrt{s}=13\UTeV$ and $\MH=125\UGeV$.
Right: the same for $\PW^-\PH$ production.
Note that $\delta_\gamma$ is based on the central value
of the photon PDF of NNPDF2.3QED, while $\sigma_\gamma$ in \Trefs{tab:wph_XStot}--\ref{tab:zllh_XSfiducial}
is based on combined results using the median and the photon PDF of MRST2004QED (and smaller by a factor 0.7), see text.
}
\label{fig:SM-WH-ptmiss}
\end{figure}
\begin{figure}
\includegraphics[width=.47\textwidth]{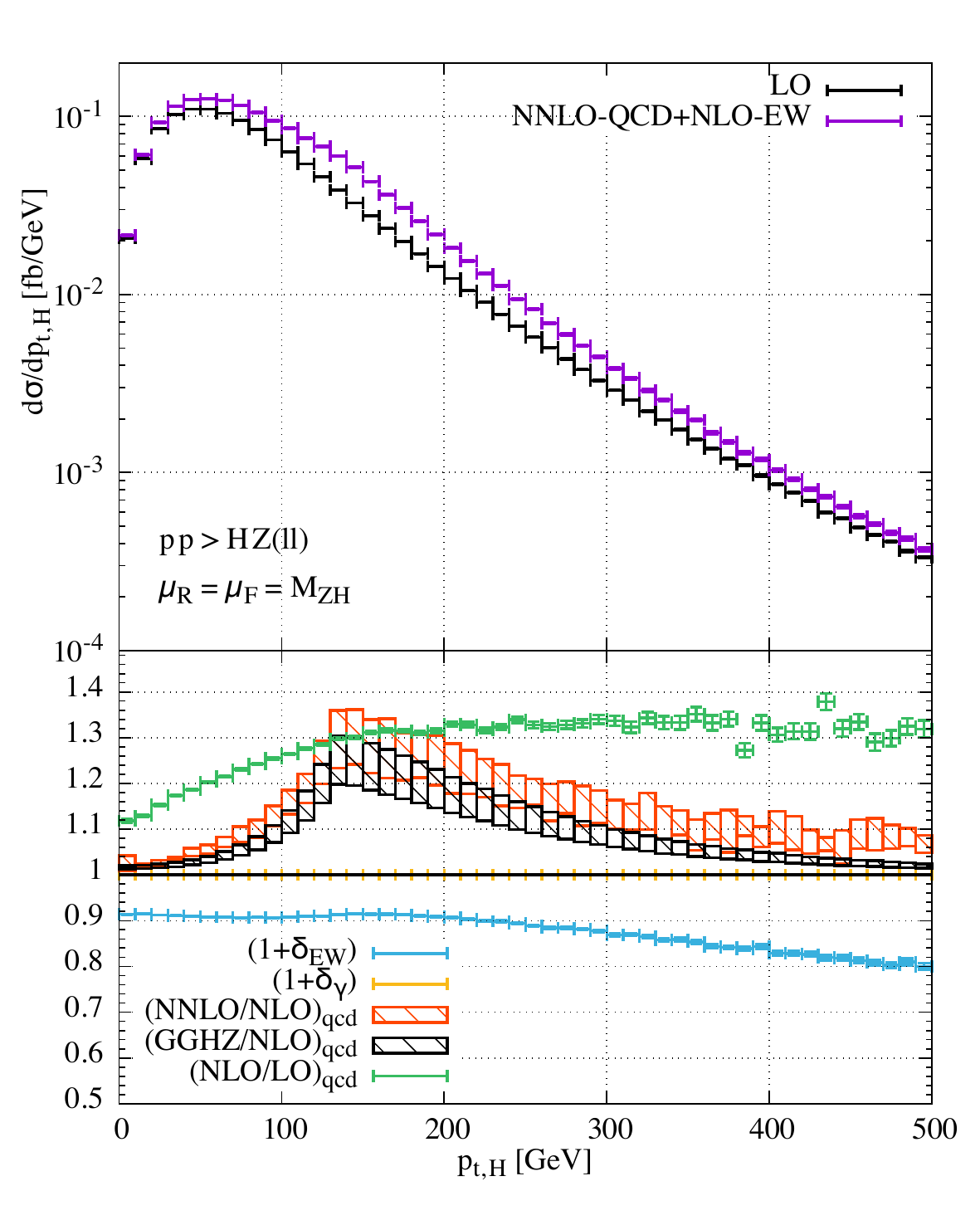}
\hfill
\includegraphics[width=.47\textwidth]{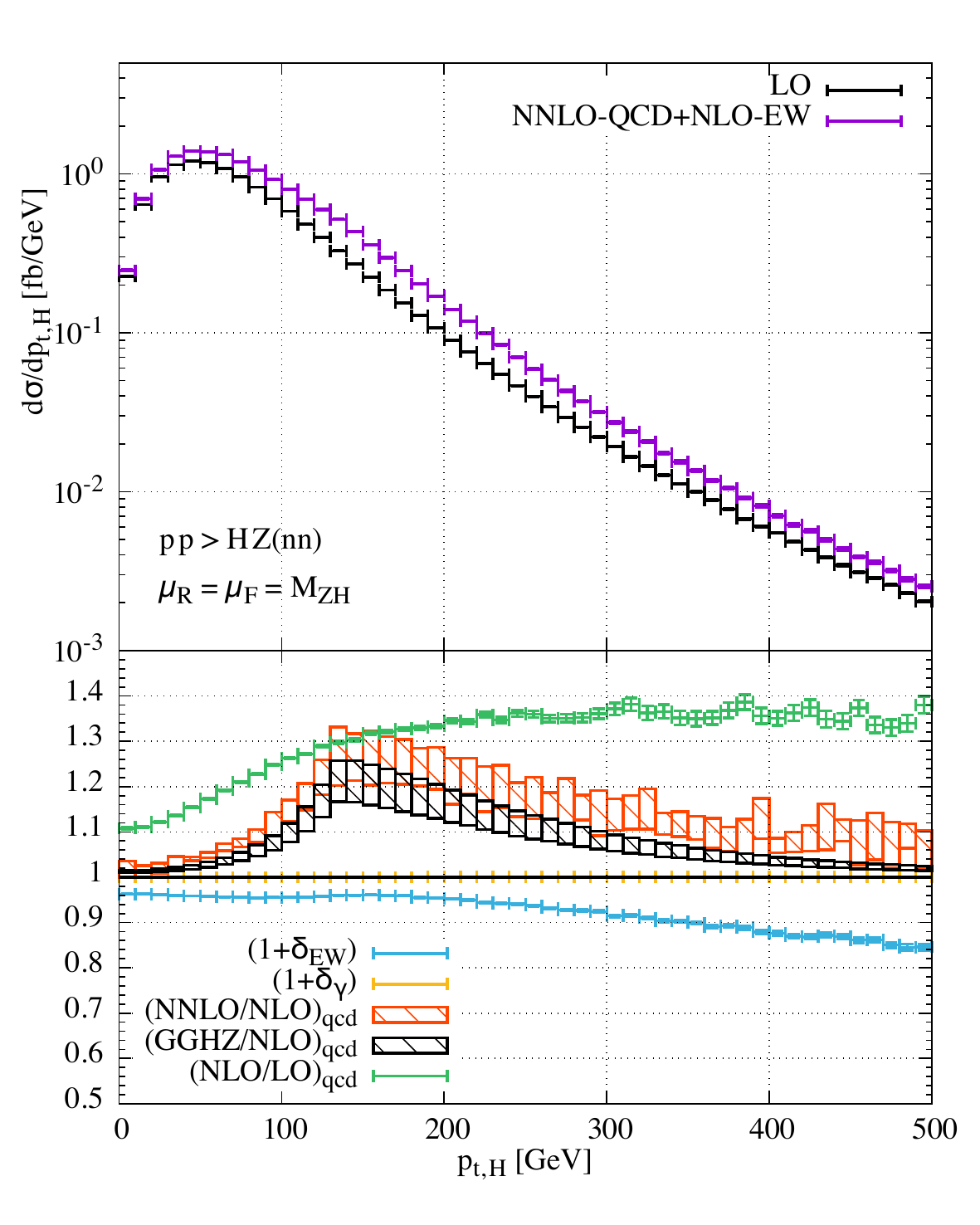}
\caption{Left: transverse-momentum distributions of the Higgs boson in $\PZ(\rightarrow \ell^+\ell^-)\PH$
production at LO and including NNLO QCD and NLO EW corrections (upper plots)
and relative higher-order contributions (lower plots) for $\sqrt{s}=13\UTeV$ and $\MH=125\UGeV$.
Right: the same for $\PZ(\rightarrow \nu \bar\nu)\PH$ production.}
\label{fig:SM-ZH-ptH}
\end{figure}
\begin{figure}
\includegraphics[width=.47\textwidth]{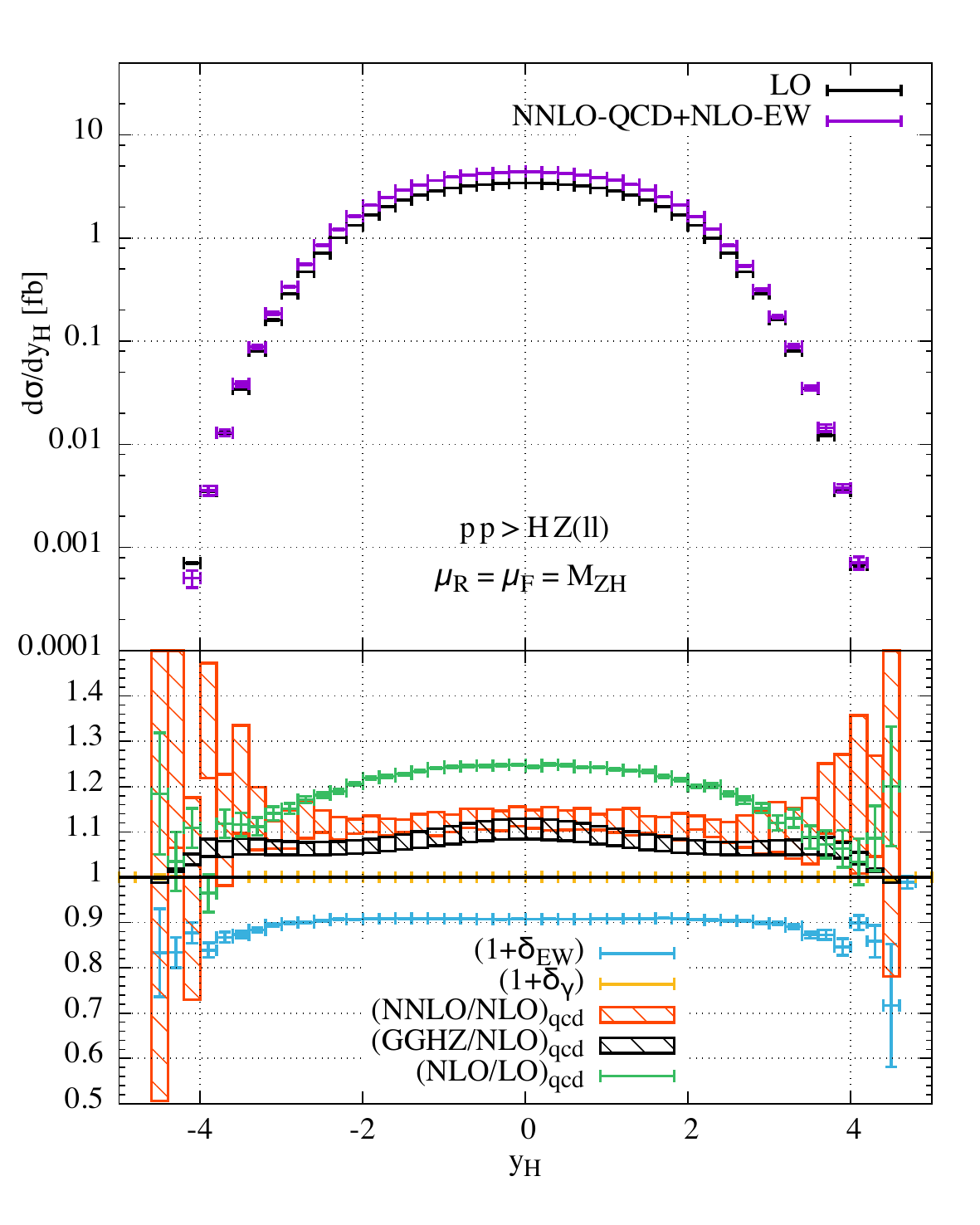}
\hfill
\includegraphics[width=.47\textwidth]{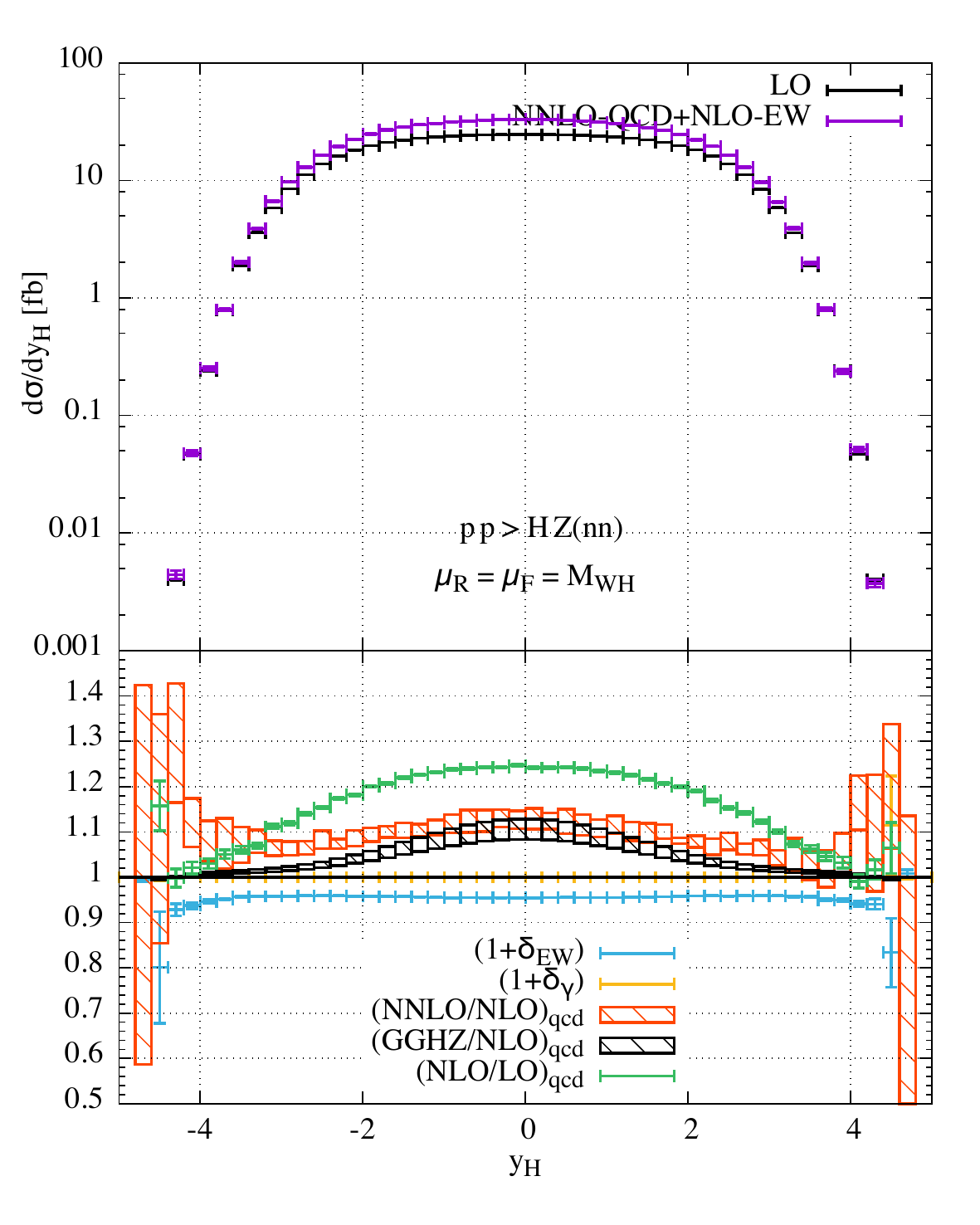}
\caption{Left: rapidity distributions of the Higgs boson in $\PZ(\rightarrow l^+l^-)\PH$
production at LO and including NNLO QCD and NLO EW corrections (upper plots)
and relative higher-order contributions (lower plots) for $\sqrt{s}=13\UTeV$ and $\MH=125\UGeV$.
Right: the same for $\PZ(\rightarrow \nu \bar\nu)\PH$ production.}
\label{fig:SM-ZH-yH}
\end{figure}
\begin{figure}
\includegraphics[width=.47\textwidth]{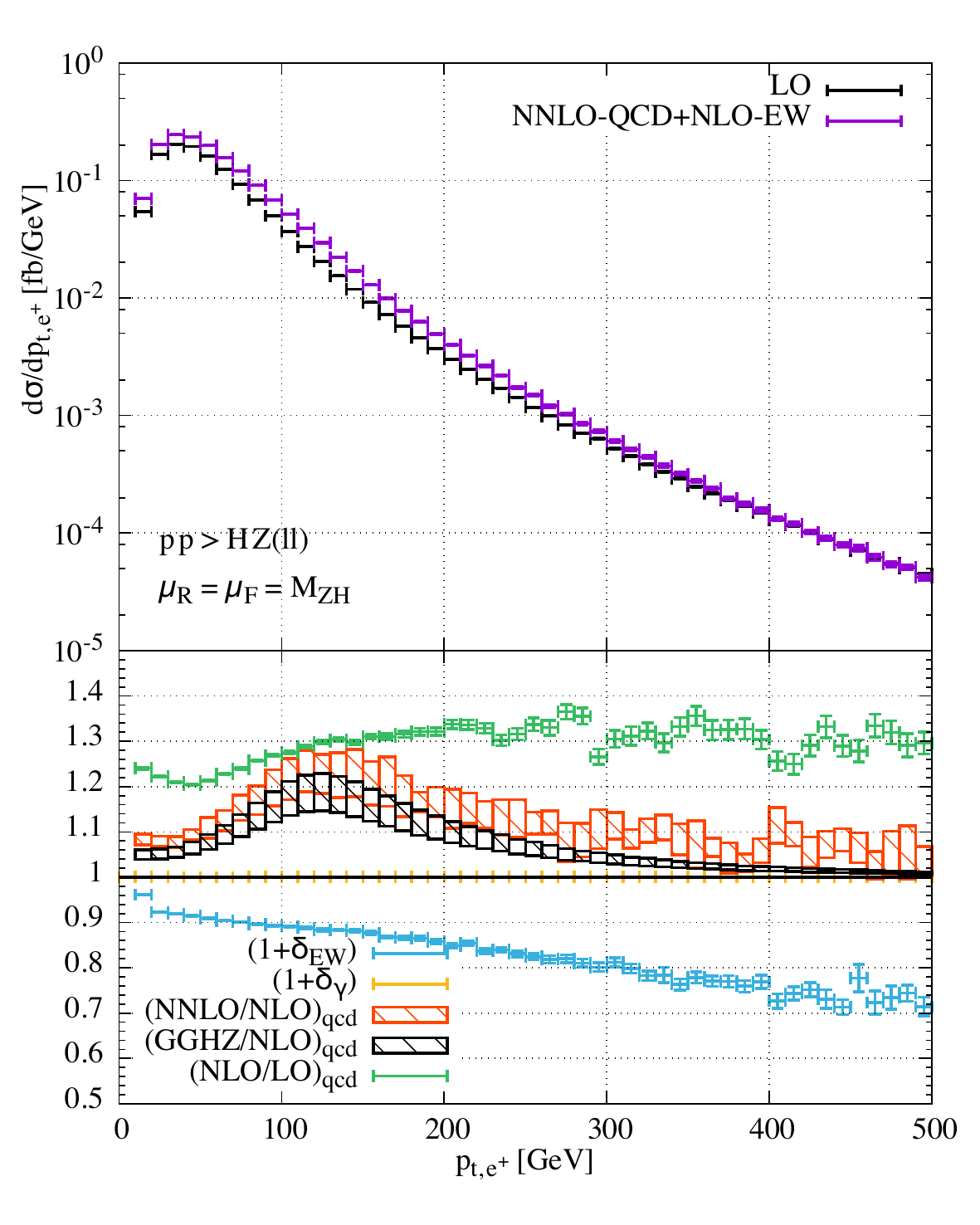}
\hfill
\includegraphics[width=.47\textwidth]{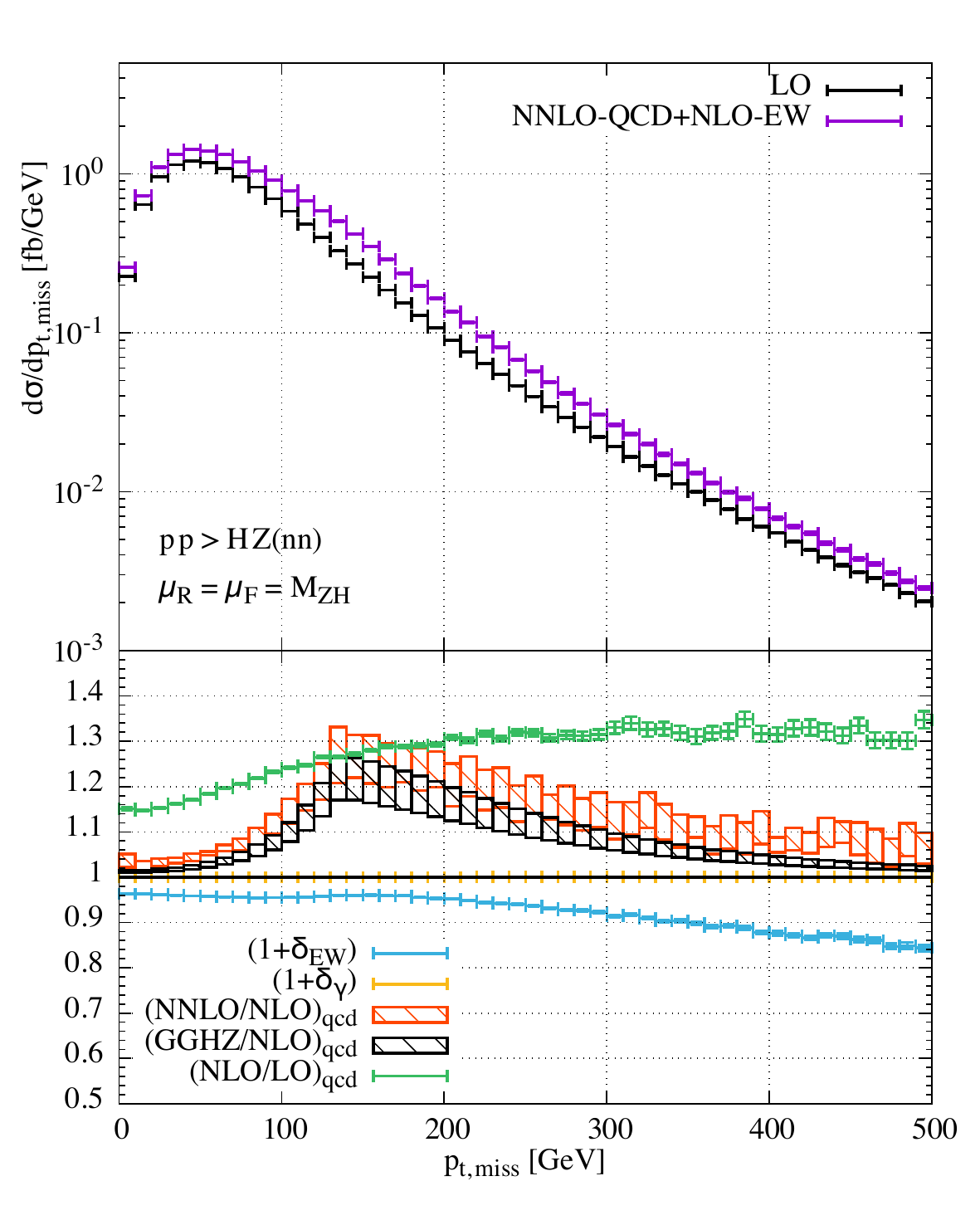}
\caption{Left: transverse-momentum distributions of the positive charged lepton in $\PZ(\rightarrow \ell^+\ell^-)\PH$
production at LO and including NNLO QCD and NLO EW corrections (upper plots)
and relative higher-order contributions (lower plots) for $\sqrt{s}=13\UTeV$ and $\MH=125\UGeV$.
Right: the same for the missing-transverse-momentum distribution in $\PZ(\rightarrow \nu \bar\nu)\PH$ production.}
\label{fig:SM-ZH-ptlpmiss}
\end{figure}
Figures~\ref{fig:SM-WH-ptH}--\ref{fig:SM-WH-ptmiss} show the impact of radiative corrections
of the most important differential distributions for Higgs boson production via WH mode in the SM,
while in \refFs{fig:SM-ZH-ptH}--\ref{fig:SM-ZH-ptlpmiss} the same effects are shown for
the Higgs boson production in association with a Z boson.
The figures generically show the known size of the NLO QCD corrections at the
level of $\sim20{-}30\%$ in the most important phase-space regions.
At NNLO, the QCD corrections amount to some per cent in the dominating regions, but
can grow to $10{-}20\%$ in the tails of distributions.
In those regions the QCD scale uncertainty accordingly grows to $\sim5\%$ or more.
The EW corrections are generically flat in (pseudo)rapidity distributions,
where they resemble the EW corrections to the fiducial cross sections.
In transverse-momentum distributions the EW correction grow further negative
to $-(10{-}20)\%$ for $p_{\mathrm{T}}$ of some $100\UGeV$.
The photon-induced corrections turn out to be only significant for WH
production. They have a tendency to grow in the tails of distributions as well,
but do hardly exceed the $5\%$ level there.

Finally, we emphasize that the contributions $\sigma_{\Pt\mbox{\scriptsize -loop}}$ are not
included in the discussion of fiducial cross sections and differential distributions
presented here, while the contribution $\sigma^{\Pg\Pg\PZ\PH}$ are included at leading
order ($\alphas^2$).

\subsection{Cross-section predictions including the decay \texorpdfstring{$H\to b\bar b$}{H to bb}}

We use the Standard Model parameters as recommended by the LHCHXSWG,
supplemented by CKM matrix elements $V_{ud} = 0.975$ and
$V_{cs} = 0.222$.  For the
parton distribution functions we use the NNLO CT14 set and associated
strong coupling, $\alphas(\MZ)=0.118$ with 3-loop running.  Central
predictions correspond to the scale choice $\muR = \muF = \mu_0$ where
$\mu_0 = M_V + \MH$ and
we consider an envelope of variations around this choice to define
the scale uncertainty.  For $\PW^\pm \PH$ production the extreme choices
correspond to $\muR = 2\mu_0$, $\muF=\mu_0/2$ and
$\muR = \mu_0/2$, $\muF=2\mu_0$.  For $\PZ\PH$ production the extrema
are instead represented by $\muF = \muR = 2\mu_0$ and
$\muF = \muR = \mu_0/2$.  Our results are obtained for the LHC operating
at $\sqrt s=13\UTeV$.

We cluster all jets according to the anti-$k_{\mathrm{T}}$ jet algorithm
with distance parameter $R=0.4$.  We subsequently require that two of
the jets contain the $\PQb$ and $\bar \PQb$ quarks from the Higgs boson decay
and that these jets satisfy,
\begin{equation}
p_{\mathrm{T}}(\mbox{\rm b-jet}) > 25\UGeV, \qquad
|\eta(\mbox{\rm b-jet})| < 2.5 \,.
\end{equation}
Note that the calculation of the $\PH \to \PQb\bar \PQb$ decay is performed at NLO QCD.

We begin by considering the $\PW^\pm \PH$ process, with the $\PW$~boson
decaying to a lepton and a neutrino.  The acceptance cuts for the
decay products are,
\begin{equation}
p_{\mathrm{T}}(\mbox{lepton}) > 15~\UGeV, \qquad
|\eta(\mbox{lepton})| < 2.5, \qquad
p_{\mathrm{T}}(\mbox{neutrino}) > 15~\UGeV \,.
\end{equation}

The cross sections under these cuts, using NNLO PDFs, are found to be
\begin{eqnarray}
&&
\sigma^{\NLO \QCD}(\PW^+\PH) = 23.56~\mathrm{fb} \,, \qquad
\sigma^{\NNLO \QCD}(\PW^+\PH) = 24.18^{+0.36}_{-0.64}~\mathrm{fb} \,, \nonumber \\
&&
\sigma^{\NLO \QCD}(\PW^-\PH) = 15.49~\mathrm{fb} \,, \qquad
\sigma^{\NNLO \QCD}(\PW^-\PH) = 15.87^{+0.26}_{-0.46}~\mathrm{fb} \,.
\end{eqnarray}
The NNLO QCD corrections under these cuts are small and positive, increasing
the NLO QCD cross sections by less than 1\%.  The scale uncertainty at NNLO QCD
is at the 3\% level.

For the $\PZ\PH$ process we consider the decay of the $\PZ$ boson into a single
family of leptons and apply the cuts,
\begin{eqnarray}
&&
p_{\mathrm{T}}(\mbox{lepton}) > 15\UGeV, \qquad
|\eta(\mbox{lepton})| < 2.5, \qquad \nonumber \\
&&
75\UGeV < M_{ll} < 105\UGeV \,.
\end{eqnarray}
These result in the following cross sections,
\begin{equation}
\sigma^{\NLO \QCD}(\PZ\PH) = 6.041~\mathrm{fb} \,, \qquad
\sigma^{\NNLO \QCD}(\PZ\PH) = 6.891^{+0.101}_{-0.162}~\mathrm{fb} \,.
\end{equation}
In this case the NNLO corrections increase the NLO cross section by
about 15\% and the scale uncertainty at NNLO QCD is around 2\%.

\begin{figure}
\begin{center}
\includegraphics[width=0.48\textwidth]{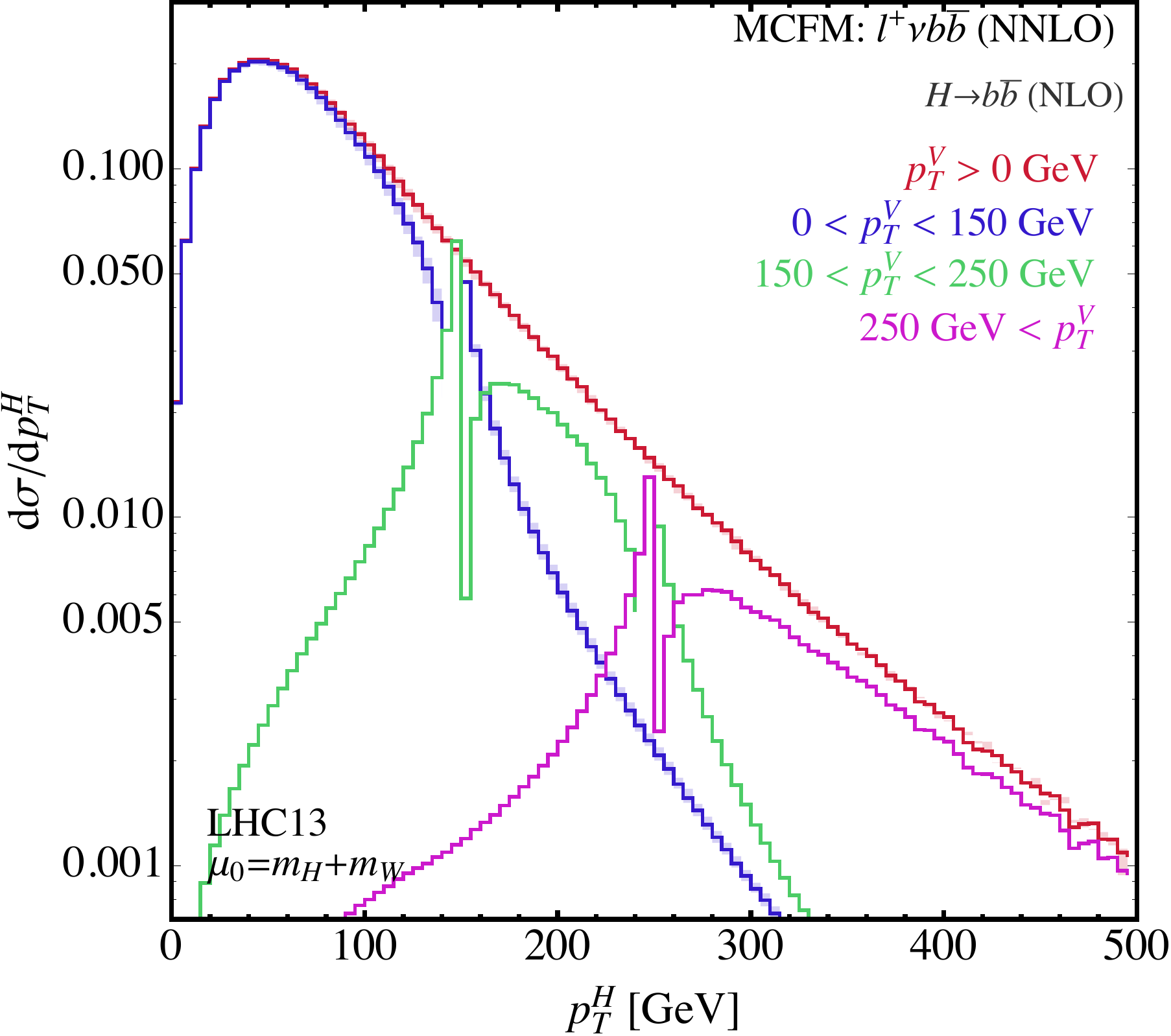}
\includegraphics[width=0.48\textwidth]{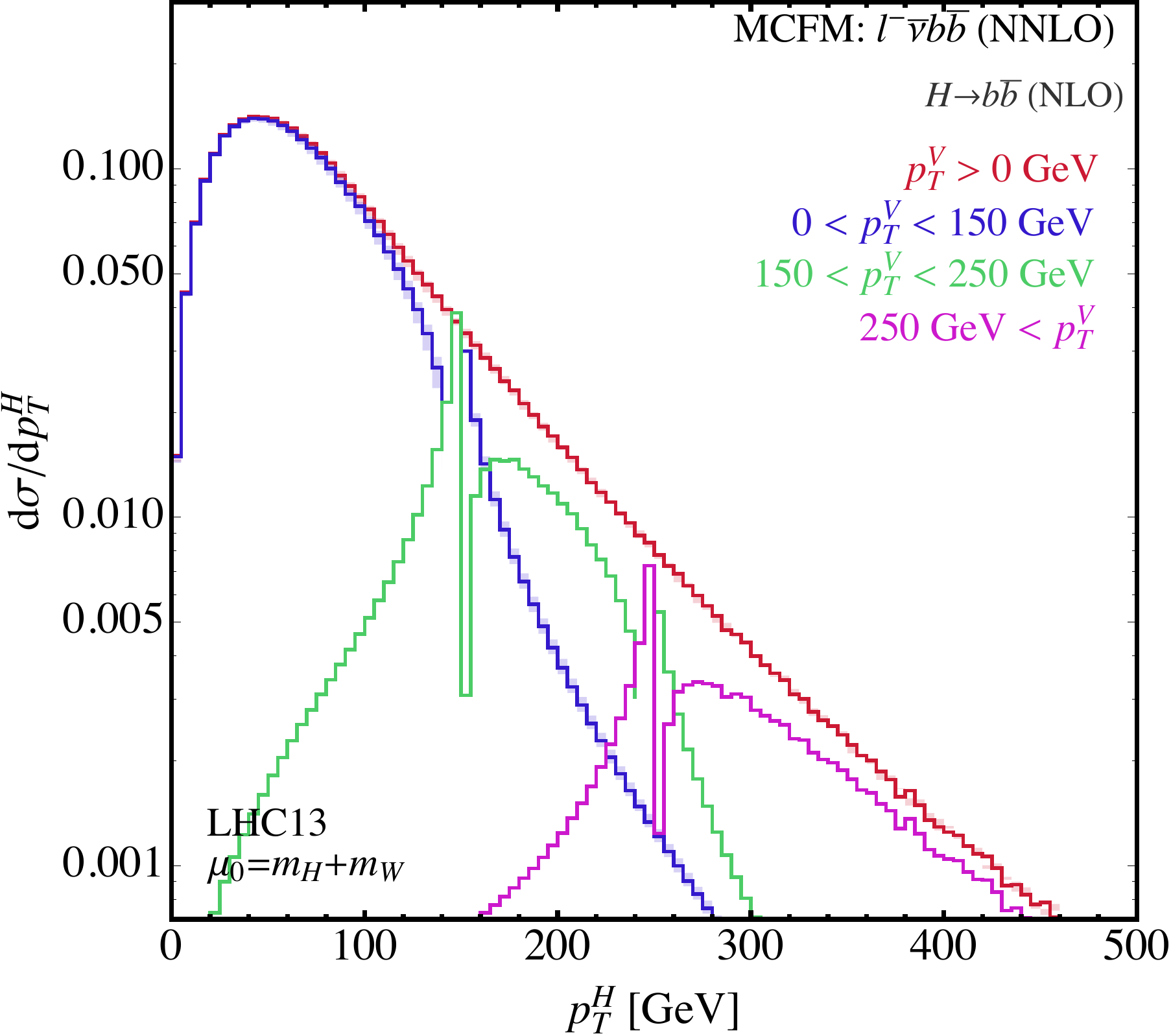}
\end{center}
\caption{The transverse momentum $p_{\mathrm{T}}^{\PQb\bar\PQb}$ for $\PW^+\PH$ (left) and $\PW^-\PH$ (right) at the 13 TeV LHC, phase space cuts are described in the text.}
\label{fig:plots1}
\end{figure}

\begin{figure}
\begin{center}
\includegraphics[width=0.45\textwidth]{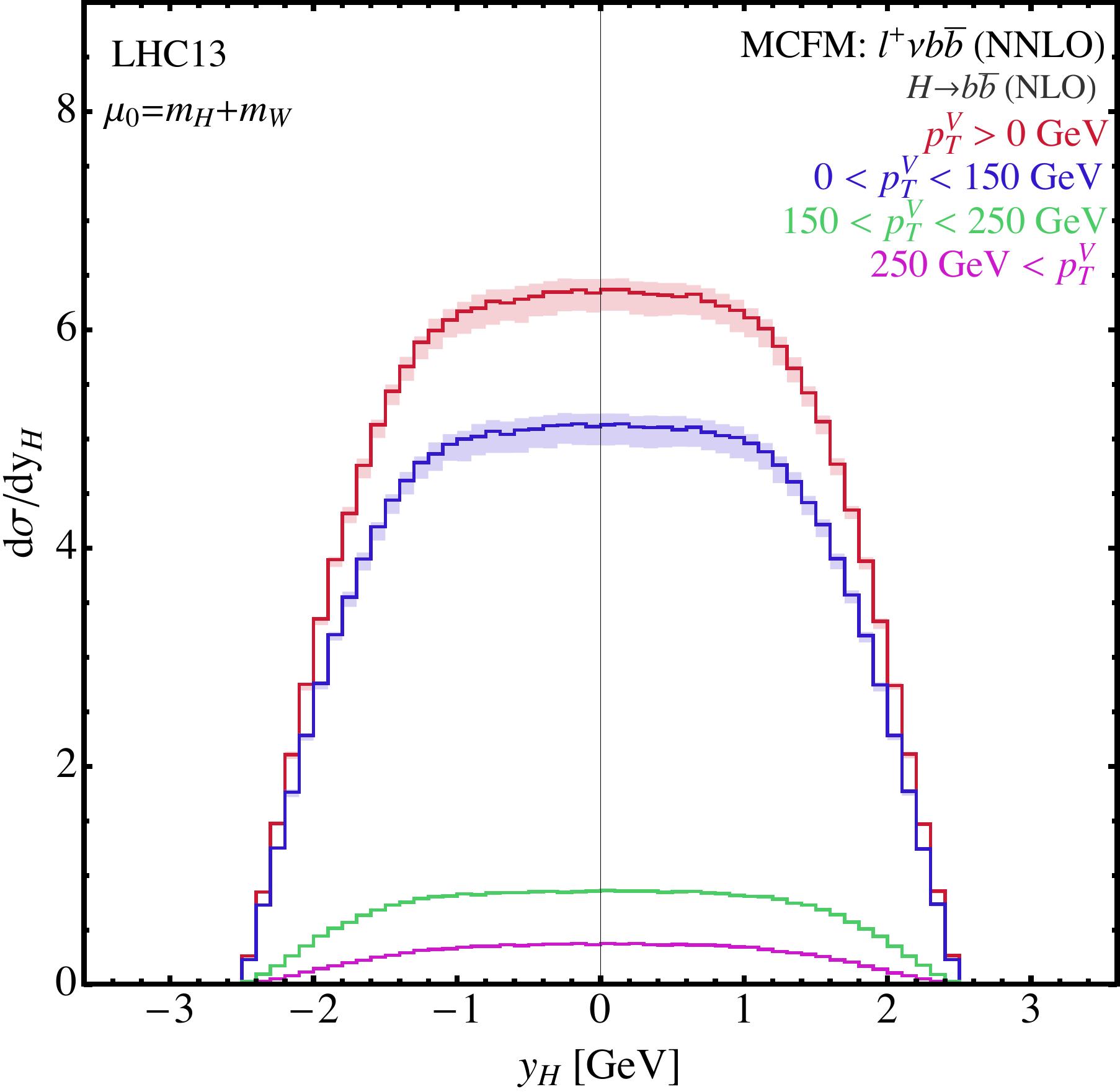}
\includegraphics[width=0.45\textwidth]{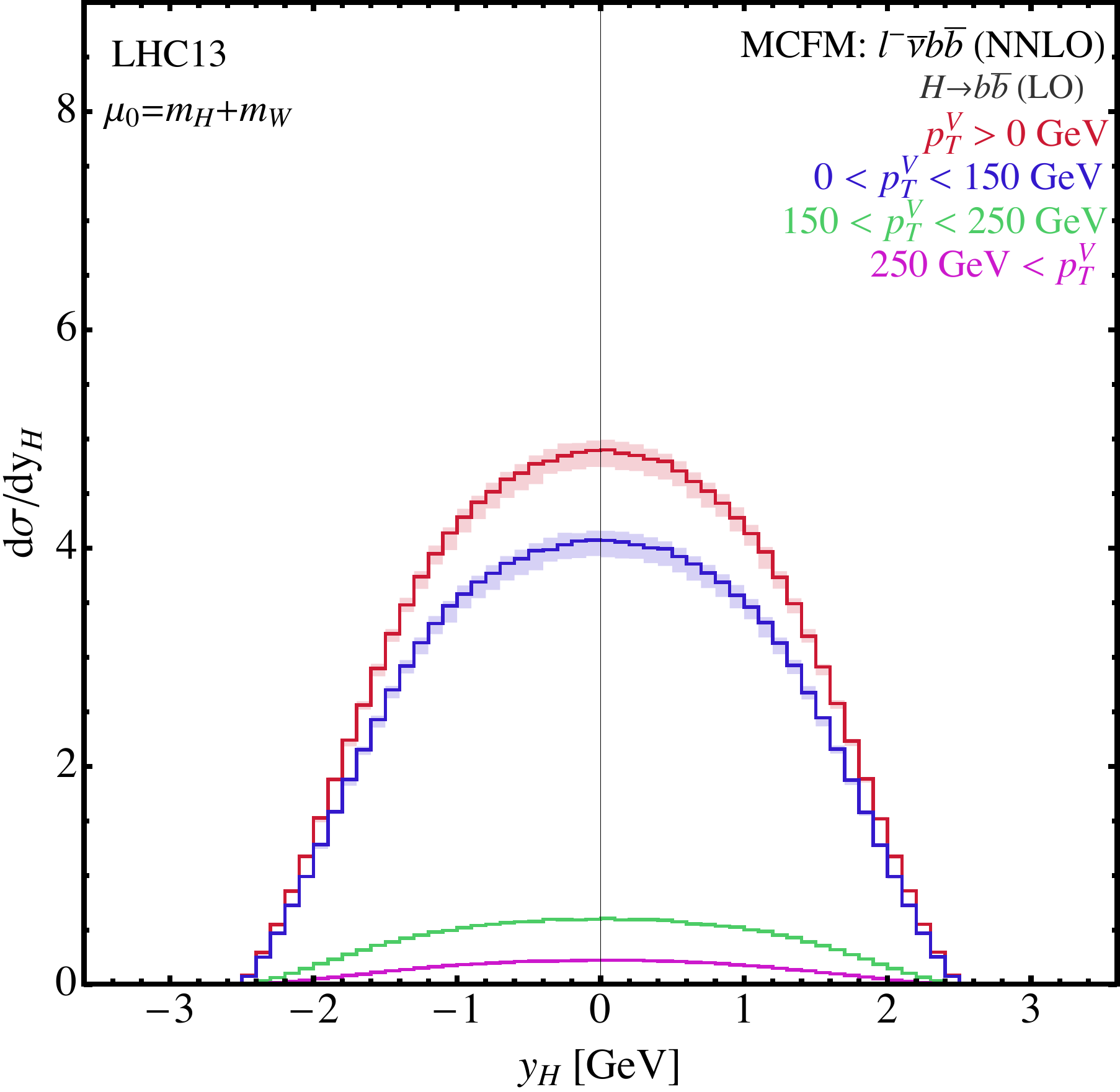}
\end{center}
\caption{The rapidity of the $\PQb\bar\PQb$ pair for $\PW^+\PH$ (left) and $\PW^-\PH$ (right) at the 13 TeV LHC, phase space cuts are described in the text.}
\label{fig:plots2}
\end{figure}

\begin{figure}
\begin{center}
\includegraphics[width=0.48\textwidth]{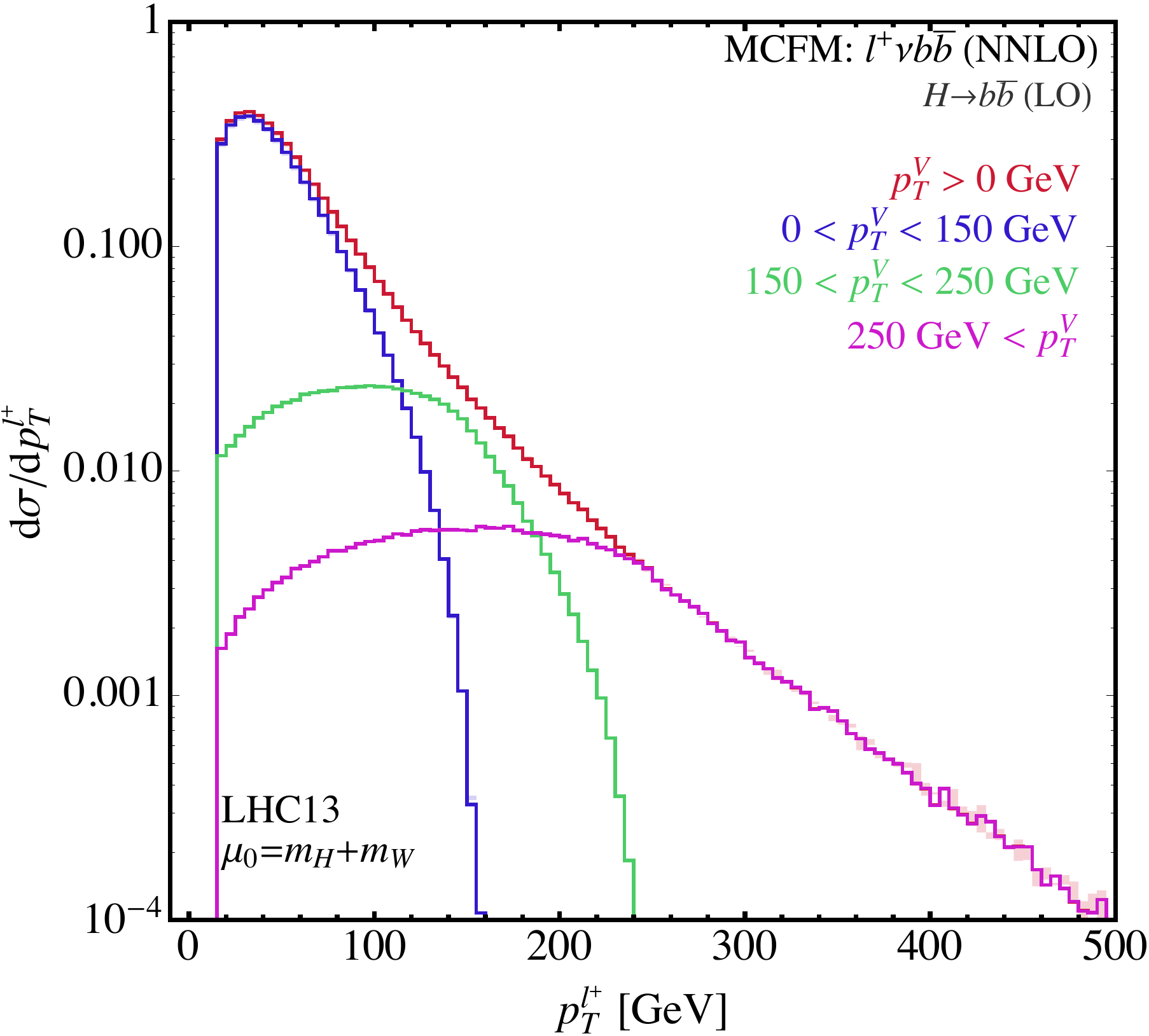}
\includegraphics[width=0.48\textwidth]{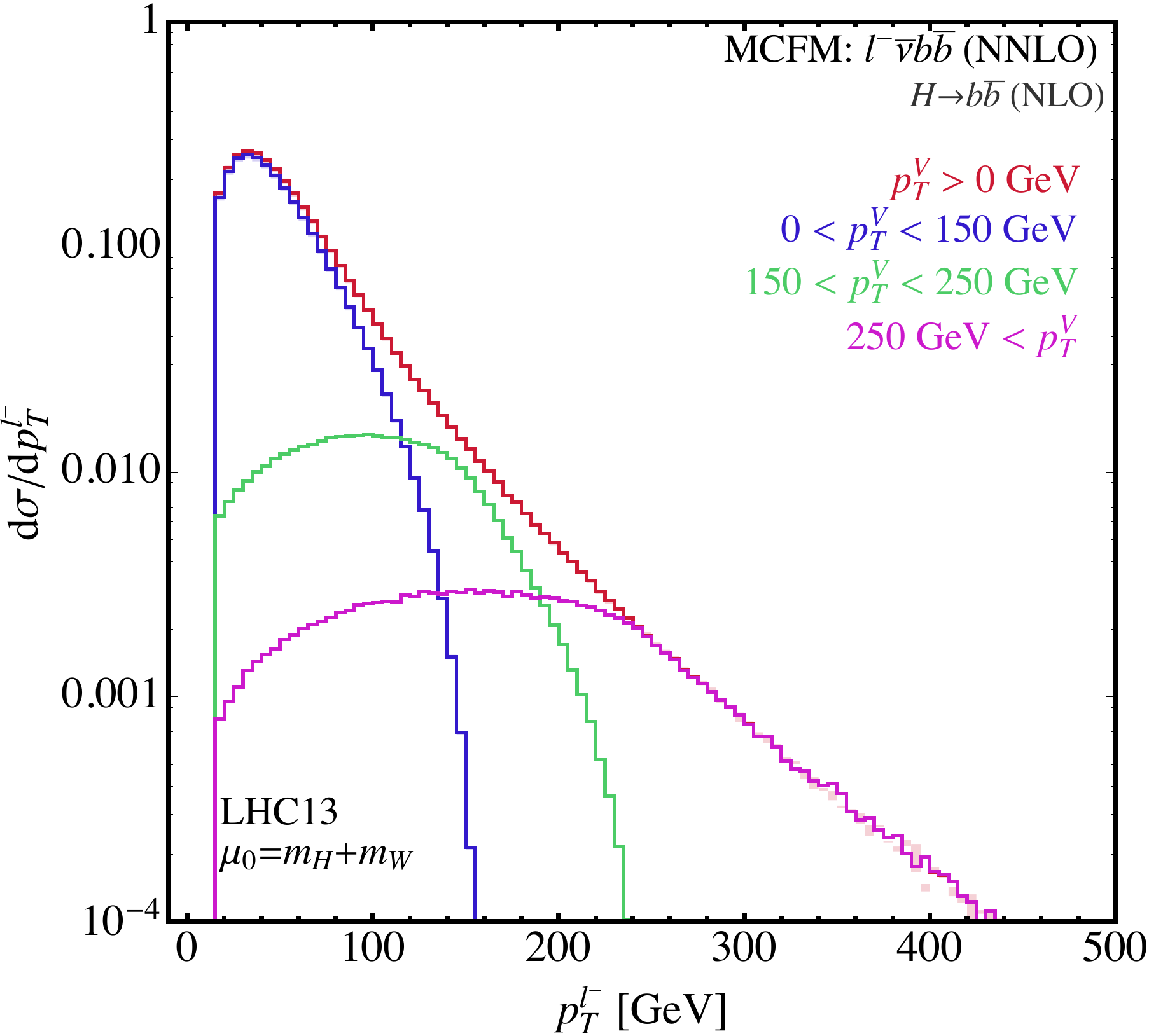}
\includegraphics[width=0.45\textwidth]{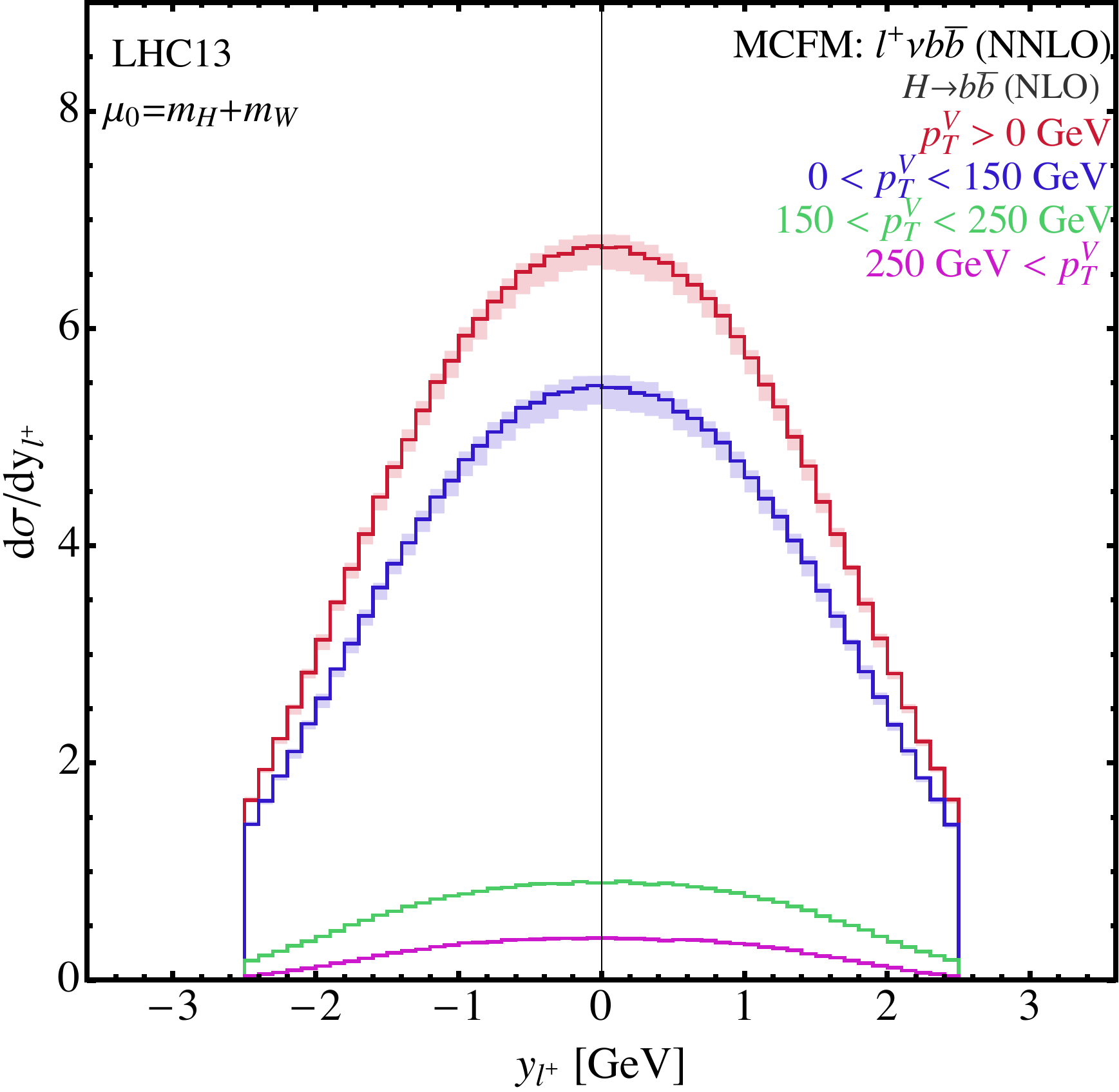} \; \;
\includegraphics[width=0.45\textwidth]{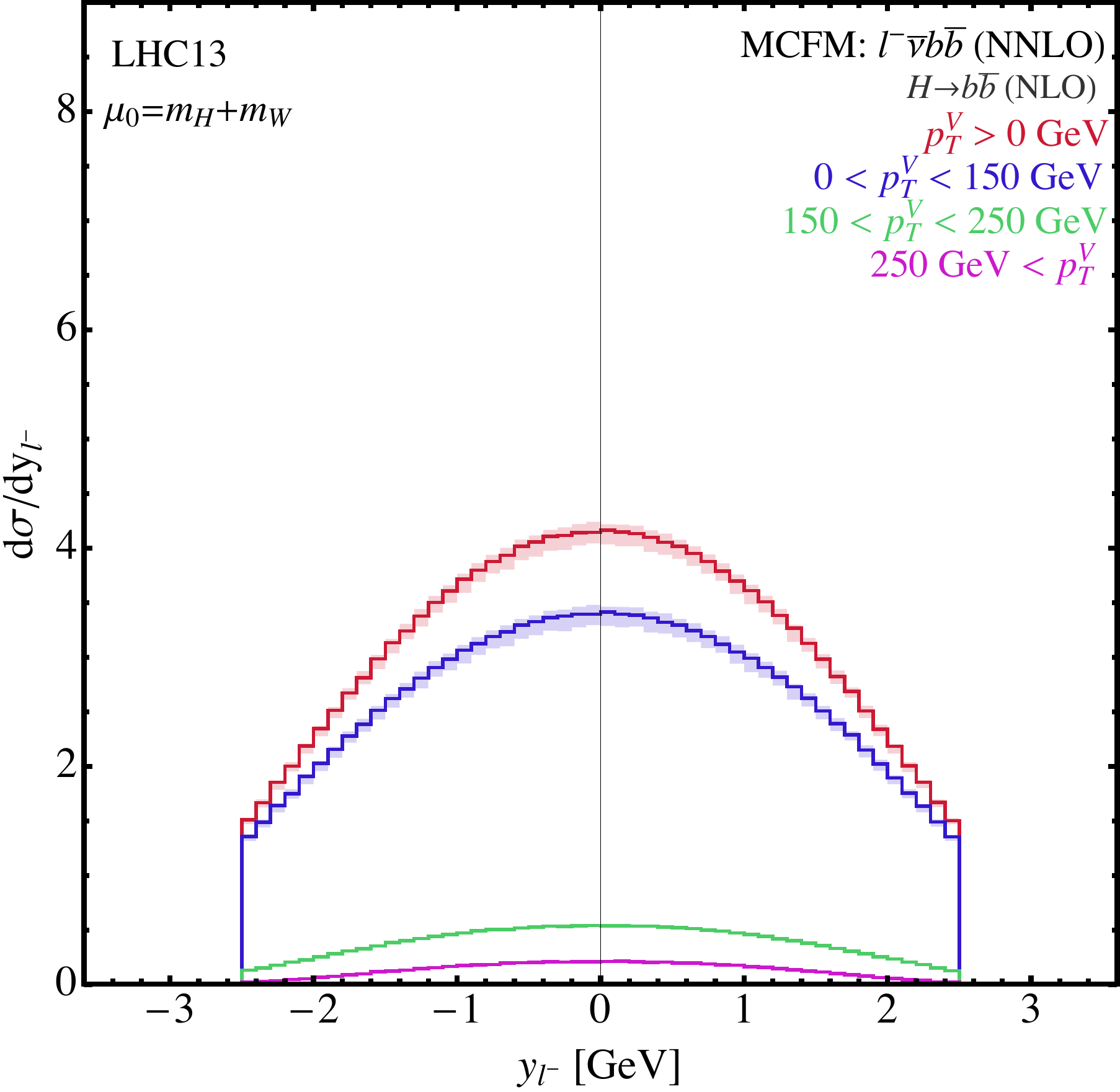}
\end{center}
\caption{The transverse momentum $p_{\mathrm{T}}^{l}$ for $\PW^+\PH$ (left, upper) and $\PW^-\PH$ (right, upper) and
lepton rapidity $\PW^+\PH$ (left, lower) and $\PW^-\PH$ (right, lower)  at the 13 TeV LHC, phase space cuts are described in the text.}
\label{fig:plots3}
\end{figure}

\begin{figure}
\begin{center}
\includegraphics[width=0.487\textwidth]{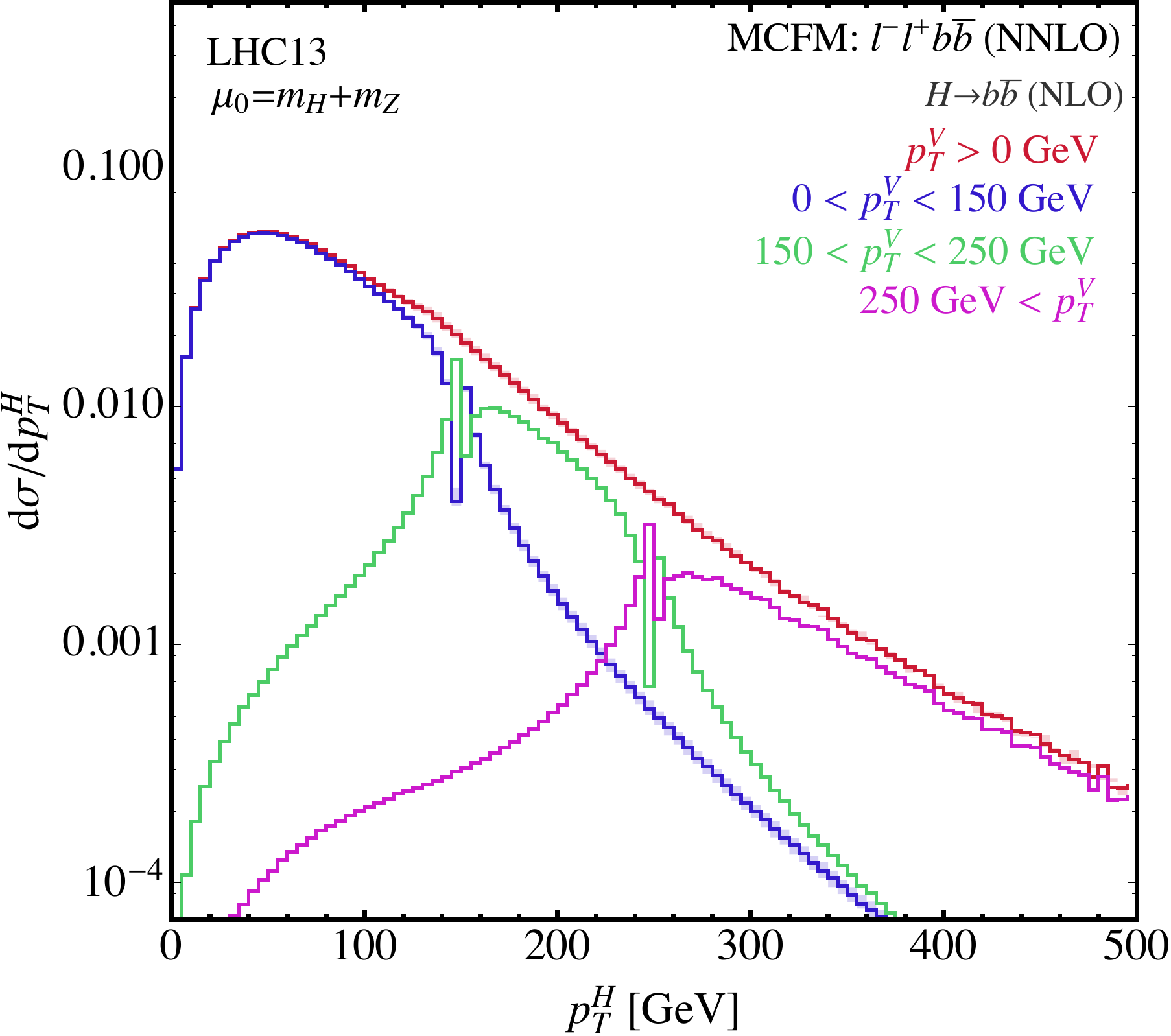}
\includegraphics[width=0.45\textwidth]{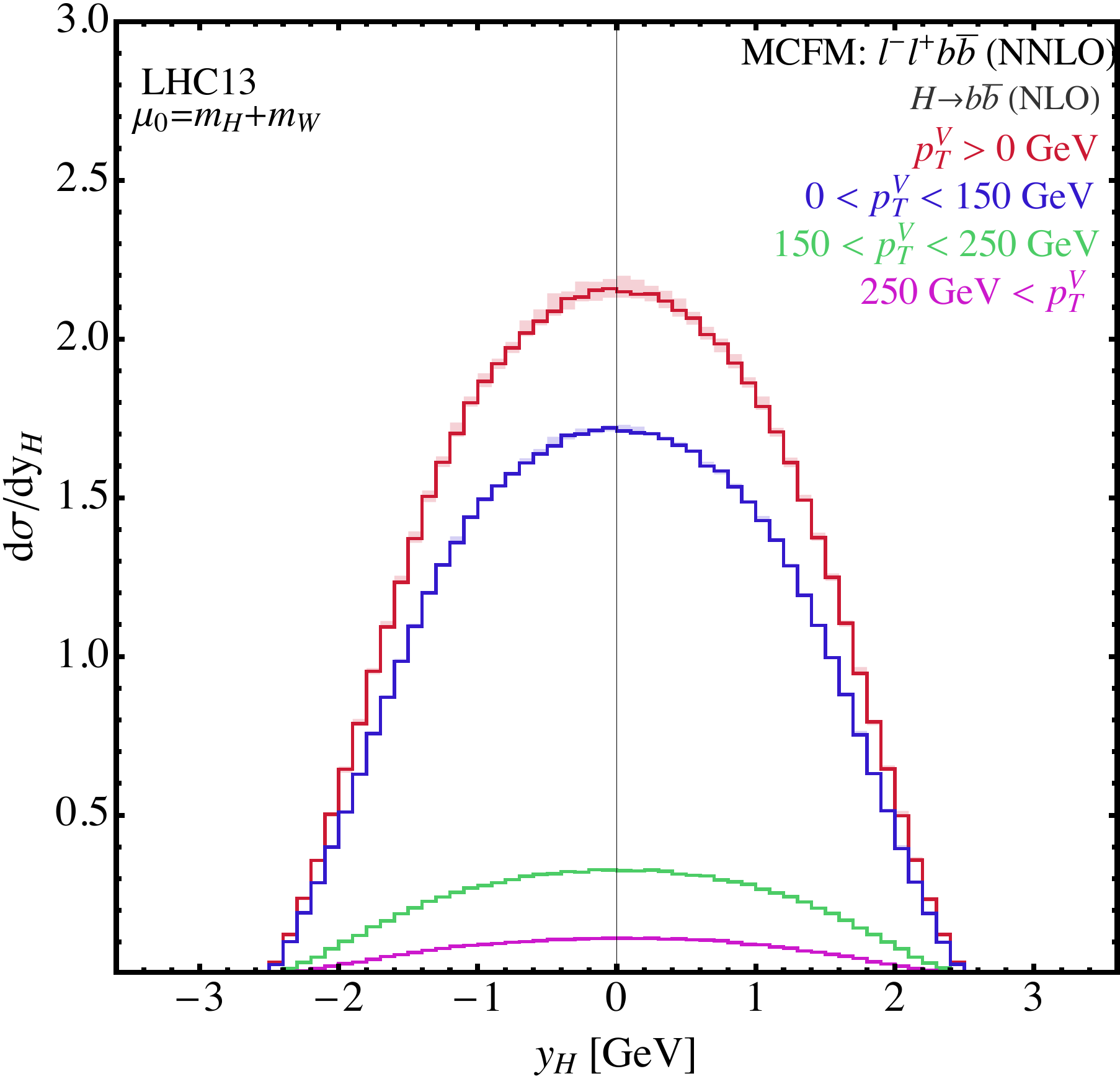}
\end{center}
\caption{The transverse momentum and rapidity of the $\PQb\bar\PQb$ pair for $\PZ\PH$  at the 13 TeV LHC, phase space cuts are described in the text.}
\label{fig:plots4}
\end{figure}

\begin{figure}
\begin{center}
\includegraphics[width=0.482\textwidth]{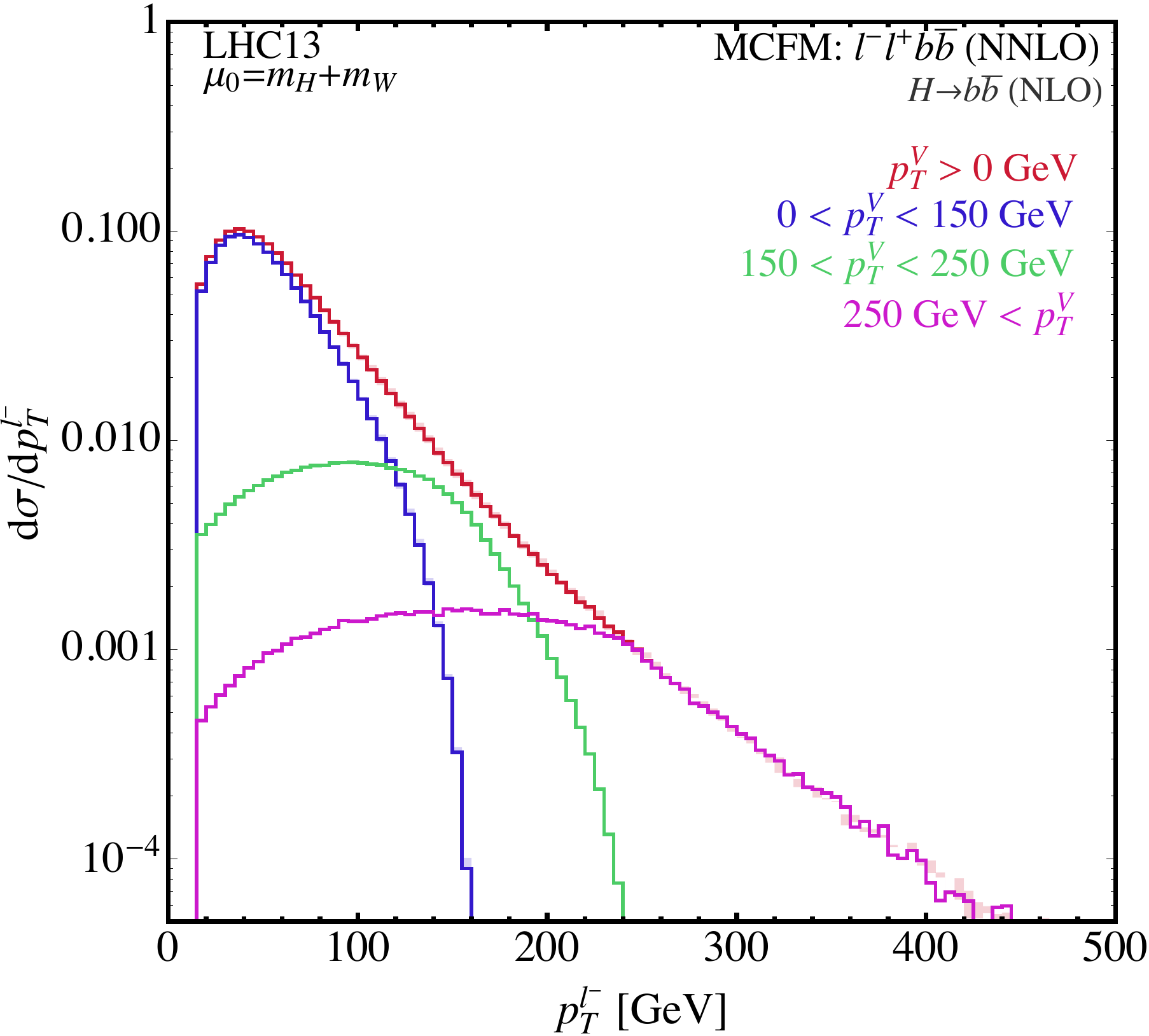}
\includegraphics[width=0.45\textwidth]{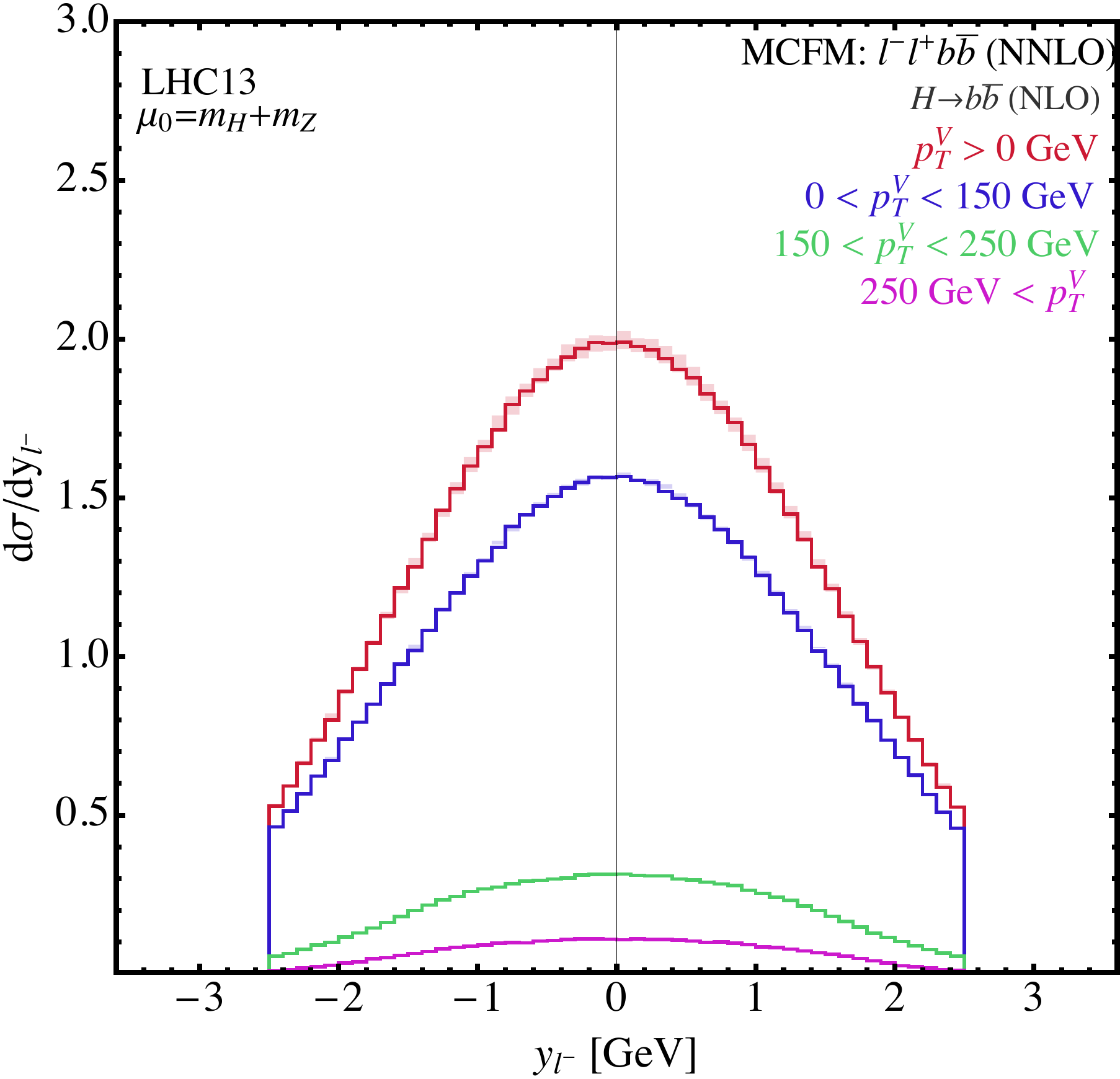}
\end{center}
\caption{The lepton transverse momentum and rapidity for $\PZ\PH$  at the 13 TeV LHC, phase space cuts are described in the text.}
\label{fig:plots5}
\end{figure}

Differential predictions for the final state $V(\rightarrow l_1l_2) \PH(\rightarrow \PQb\bar{\PQb})$ are
presented in Figs.~\ref{fig:plots1}-\ref{fig:plots5}. For $\PW^{\pm}\PH$ production we present the differential
observables side-by-side on the same scale, so that the relative suppression of $\PW^-$ compared to $\PW^+$ is
readily apparent. Figure~\ref{fig:plots1} presents the $p_{\mathrm{T}}$ of the Higgs boson candidate, whose four-momentum
is defined as the sum of those of the identified $\PQb$-jets. We present
predictions for a variety of selection cuts. The red curve corresponds to an ``inclusive'' $p_{\mathrm{T}}^V$ selection
while the remaining curves slice the phase space into various $p_{\mathrm{T}}^V$ bins:
\begin{eqnarray}
\mbox{red:} && p_{\mathrm{T}}^V~\mbox{inclusive} \,, \nonumber\\
\mbox{blue:} && 0 < p_{\mathrm{T}}^V < 150~\mbox{GeV} \,, \nonumber\\
\mbox{green:} && 150 < p_{\mathrm{T}}^V < 250~\mbox{GeV} \,, \nonumber\\
\mbox{magenta:} && p_{\mathrm{T}}^V > 250~\mbox{GeV} \,.
\end{eqnarray}
The inclusive curve is thus recovered by summing over all of the remaining curves.  At leading order $p_{\mathrm{T}}^V
\equiv p_{\mathrm{T}}^{\PH}$, with departures from this equality the result of additional radiation that is inherent in the
higher-order corrections. The discontinuities that are apparent in the regions of phase space around $p_{\mathrm{T}}^V =
p_{\mathrm{T}}^{\PH}$, namely at $p_{\mathrm{T}}^{\PH} = 150, 250$~GeV, are indicators of the fact that  perturbation theory is unreliable at
such boundaries. The situation is exacerbated by the inclusion of higher-order corrections in the Higgs boson
decay, since boundary logarithms also appear due to radiation in the
decay~\cite{Ferrera:2013yga,Ferrera:2014lca,Campbell:2016jau}. The differences between the $\PW^+\PH$ and $\PW^-\PH$
predictions are most clear in the $y_{\PH}$ observable (Figure~\ref{fig:plots2}). This observable is sensitive to the
valence/sea quark distribution inside the proton. The valence $u$ distribution is more favored for $\PW^+\PH$, and
stiffens the $y_{\PH}$ distribution by favoring more forward regions of phase space. On the other hand $\PW^-\PH$
production is associated with the production of more central Higgs bosons.  We present the leptonic observables
in Figure~\ref{fig:plots3}. As the $p_{\mathrm{T}}^V$ cut is increased the $p_{\mathrm{T}}^{l}$ distribution flattens out.  Finally
in Figures~\ref{fig:plots4}-\ref{fig:plots5} we present the differential predictions for $\PZ\PH$. The conclusions
are broadly similar, although the phase-space boundary effects are somewhat damped for this process.
This is due to the presence of $\Pg\Pg\rightarrow \PZ\PH$ contributions that provide a sizeable correction to the cross
section at NNLO. Since this switches-on in the LO phase space, the large negative bin is partially compensated
by the inclusion of these pieces. There is a noticeable inflection in the $p_{\mathrm{T}}^{\PH}$ spectrum at around
$p_{\mathrm{T}}^{\PH} \sim m_t$, which is where these pieces begin to become important.


\section{Electroweak production of H+3jets at NLO+PS}
Electroweak production of a Higgs boson in association with three jets has first been considered at NLO-QCD accuracy in Ref.~\cite{Figy:2007kv} in the VBF approximation. A matching of this calculation to parton shower programs in the framework of the \POWHEGBOX\ has been presented in \cite{Jager:2014vna}.  In Ref.~\cite{Campanario:2013fsa}, NLO-QCD corrections have been provided without resorting to the VBF approximation.
This latter calculation is based on spinor helicity techniques in
combination with the methods developed in the context of
\cite{Campanario:2011cs}. For its implementation, a module has been developed:
\textsf{HJets++}
 is a plugin to \HERWIG7's
\textsf{Matchbox}~\cite{Bellm:2015jjp,Platzer:2011bc} module, providing
amplitudes for calculating the production of a Higgs boson in
association with $n_{{\rm jet}}=2,3$ jets at next-to-leading order in QCD,
{\it i.e.} at ${\cal O}(\alpha^3 \alpha_s^{n_{{\rm jet}}-1})$\footnote{In this approach, Yukawa couplings are counted as a separate expansion parameter; thus finite heavy  quark loop contributions are to be considered separately.}.
The plugin nature of this module enables the amplitudes to be
directly used in an NLO-plus-parton shower matched simulation, with both
subtractive (MC@NLO-type) and multiplicative (\POWHEG{}-type) matchings being
available. Either of the two parton showers available in \HERWIG7  can be used in
the matching.

\begin{figure}
\includegraphics[width=.47\textwidth]{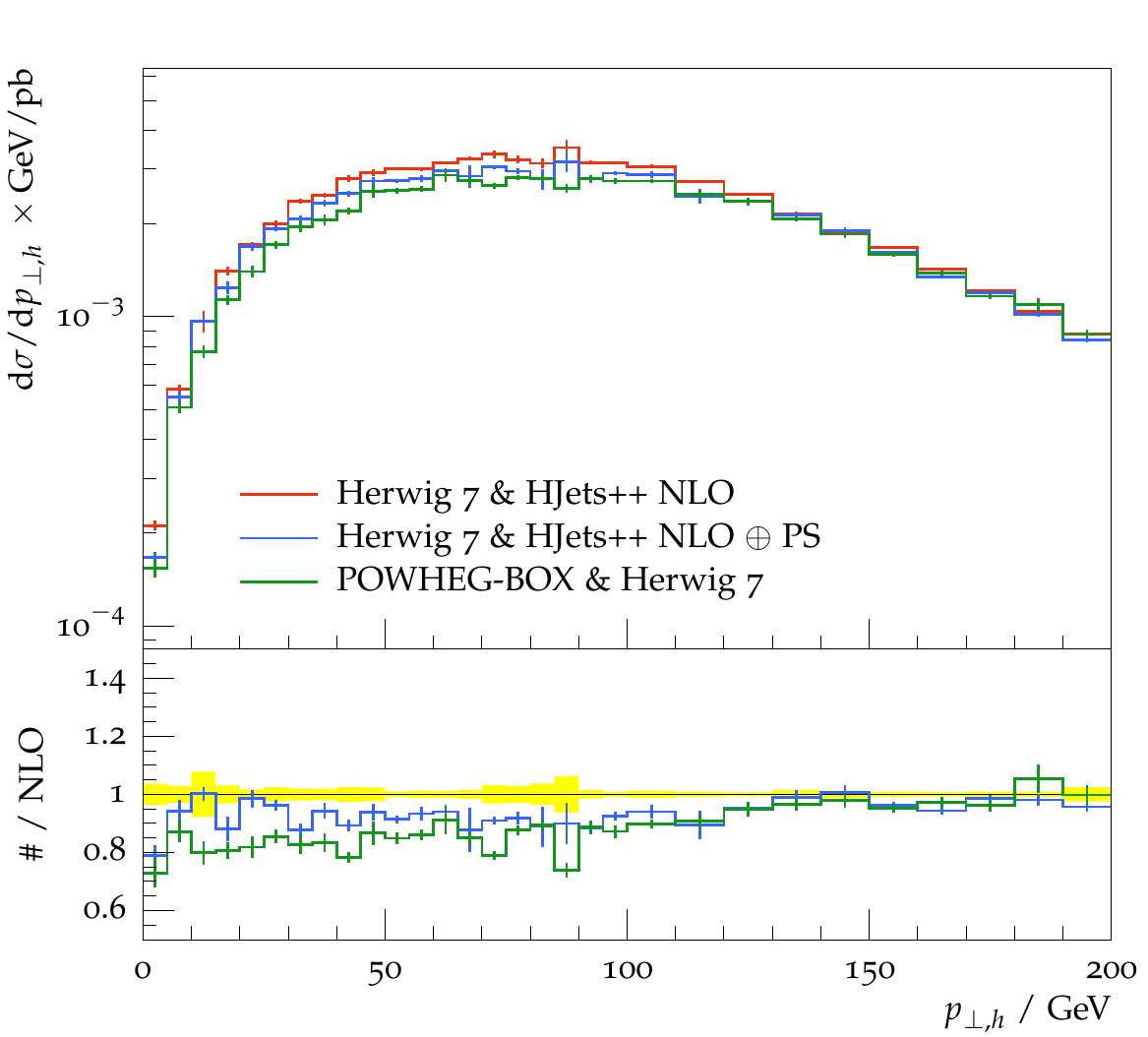}
\hfill
\includegraphics[width=.47\textwidth]{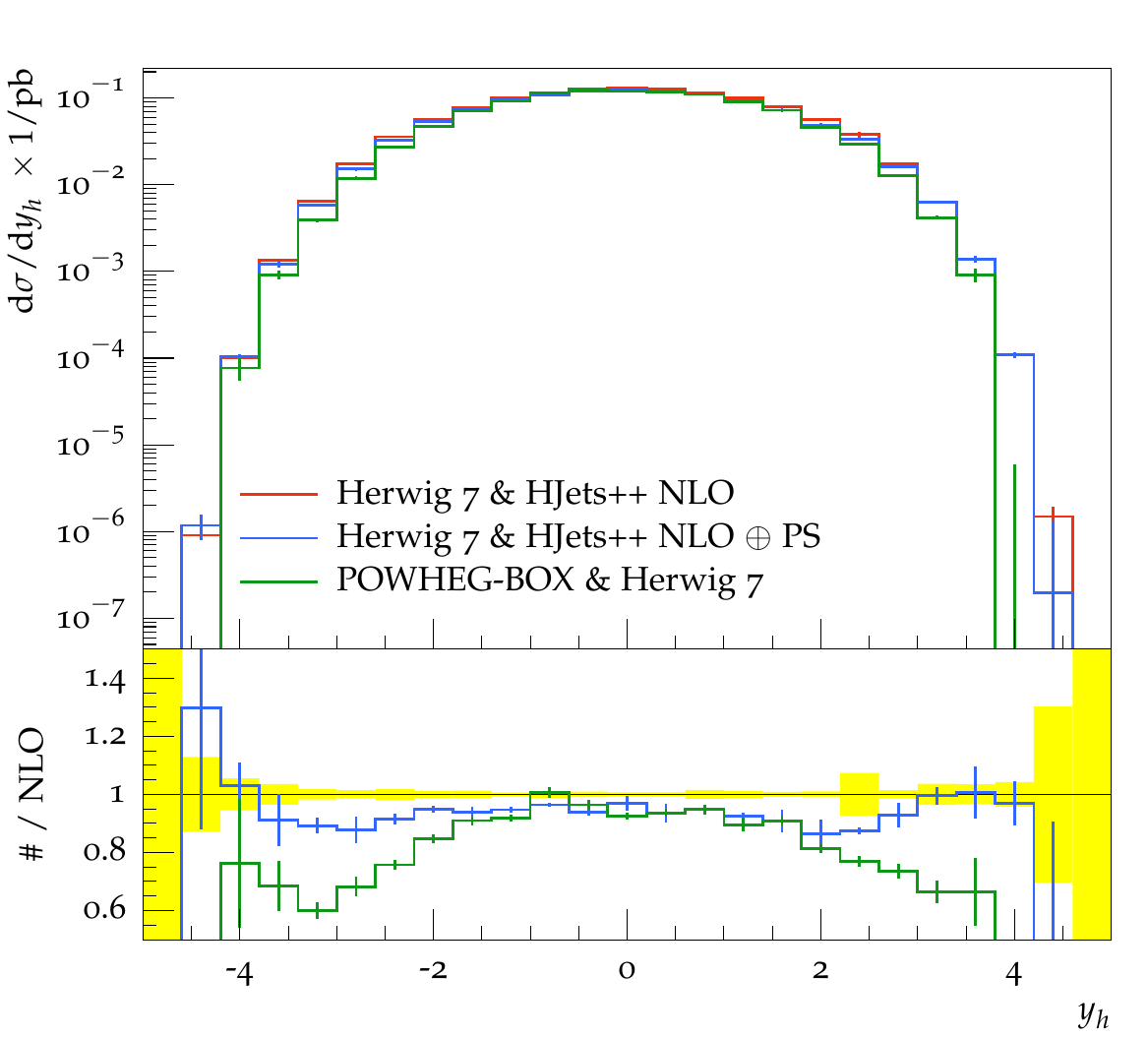}
\caption{Transverse-momentum and rapidity distributions of the Higgs boson in EW H+3~jet production at NLO QCD (red line) as obtained from using the \textsf{Matchbox} framework
of \HERWIG7\ with the \textsf{HJets++} plugin, and at NLO QCD matched with the
\HERWIG7\ angular ordered parton shower in the same framework (blue line),
and with the \POWHEGBOX\ (green line), respectively. The lower panels show the respective ratios of the NLO+PS to the fixed-order NLO QCD result for $\sqrt{s}=13\UTeV$ and $\MH=125\UGeV$. The yellow bands indicate the statistical uncertainty of the NLO result.}
\label{fig:SM-VBF-H3j-H}
\end{figure}

\begin{figure}
\includegraphics[width=.47\textwidth]{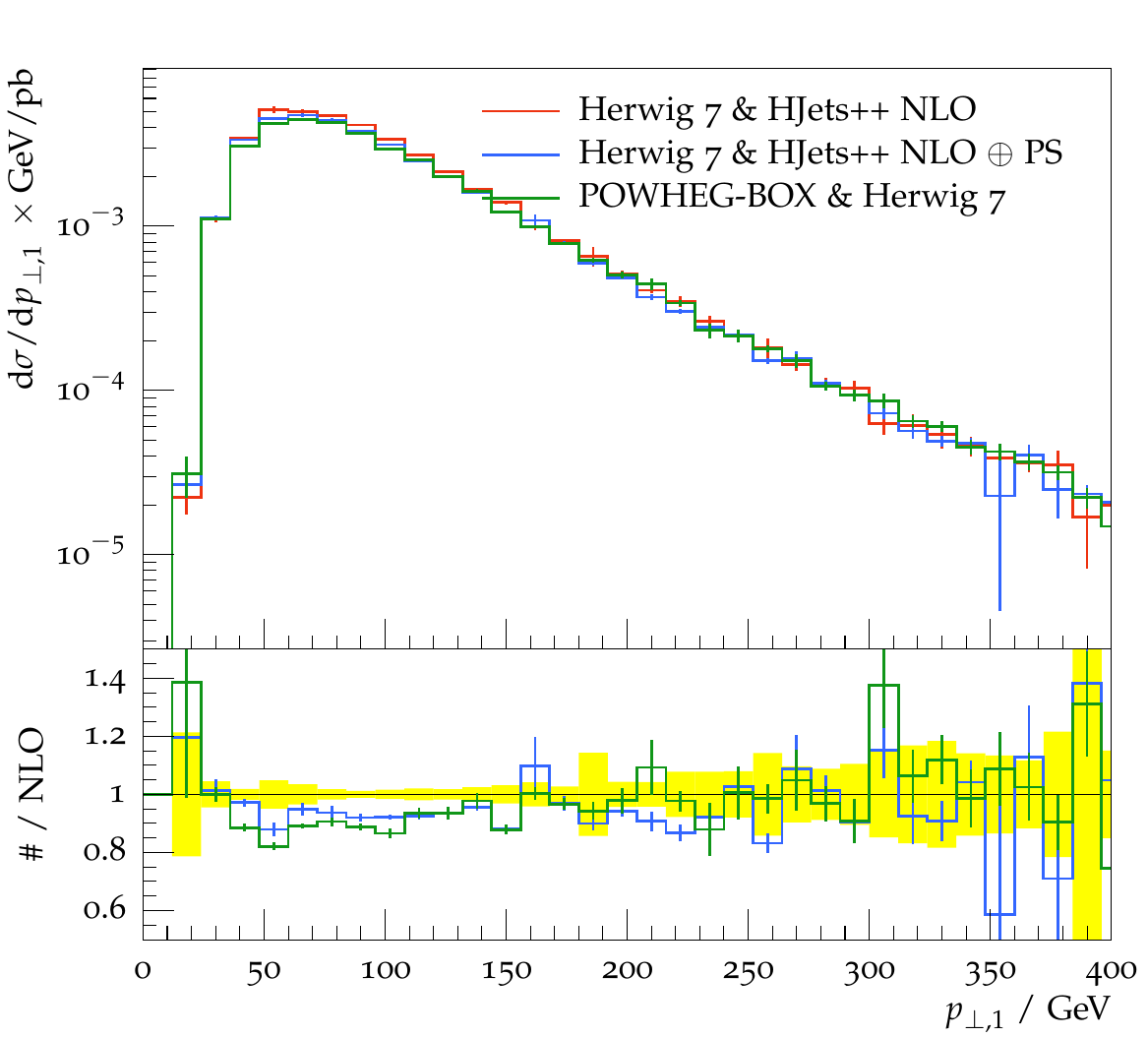}
\hfill
\includegraphics[width=.47\textwidth]{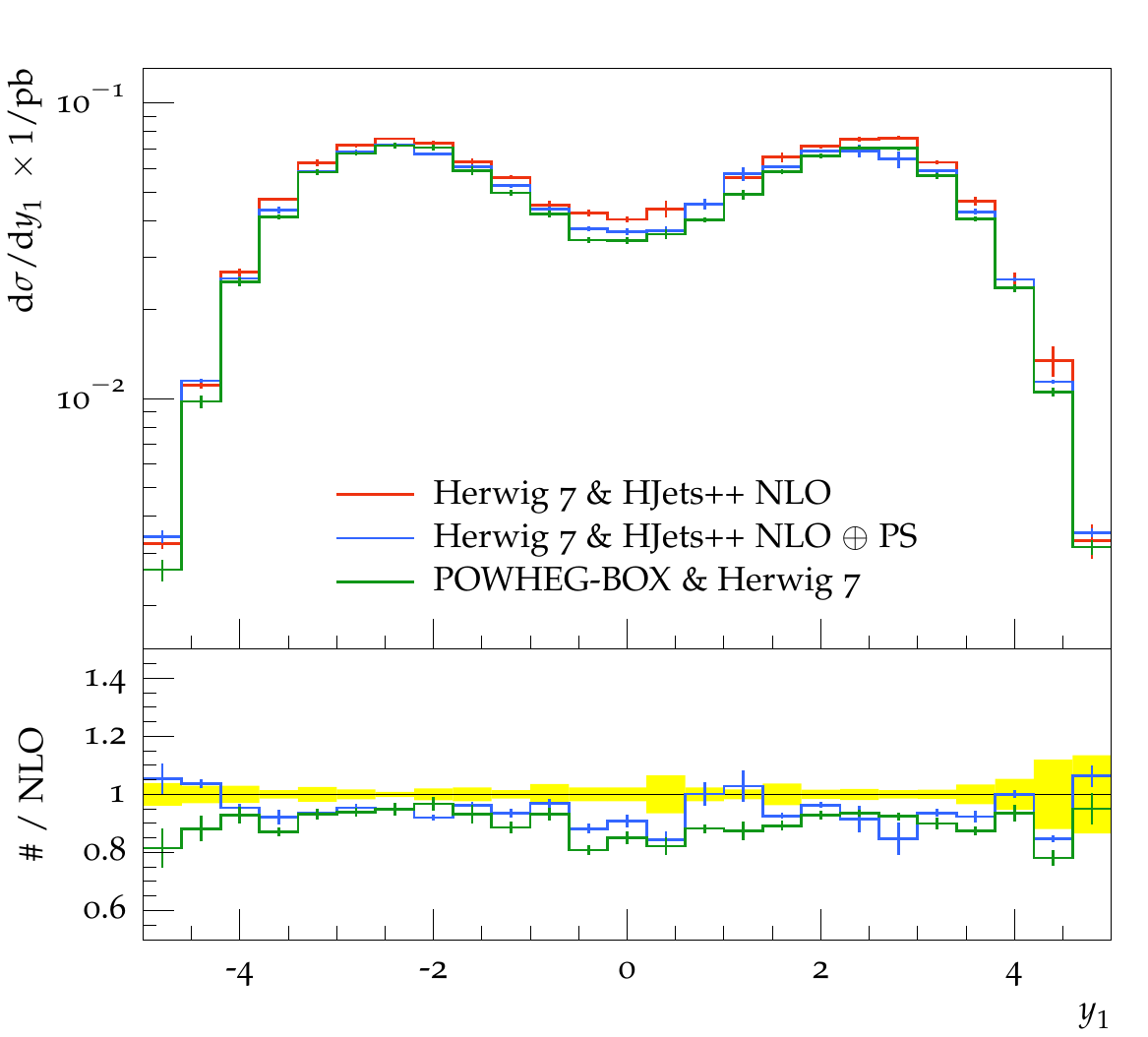}
\caption{Transverse-momentum and rapidity distributions of the hardest tagging jet in EW H+3~jet production at NLO QCD (red line)  as obtained from using the \textsf{Matchbox} framework
of \HERWIG7\ with the \textsf{HJets++} plugin, and at NLO QCD matched with the
\HERWIG7\ angular ordered parton shower in the same framework (blue line),
and with the \POWHEGBOX\ (green line), respectively.
The lower panels show the respective ratios of the NLO+PS to the fixed-order NLO QCD result for $\sqrt{s}=13\UTeV$ and $\MH=125\UGeV$. The yellow bands indicate the statistical uncertainty of the NLO result. }
\label{fig:SM-VBF-H3j-jet1}
\end{figure}

\begin{figure}
\includegraphics[width=.47\textwidth]{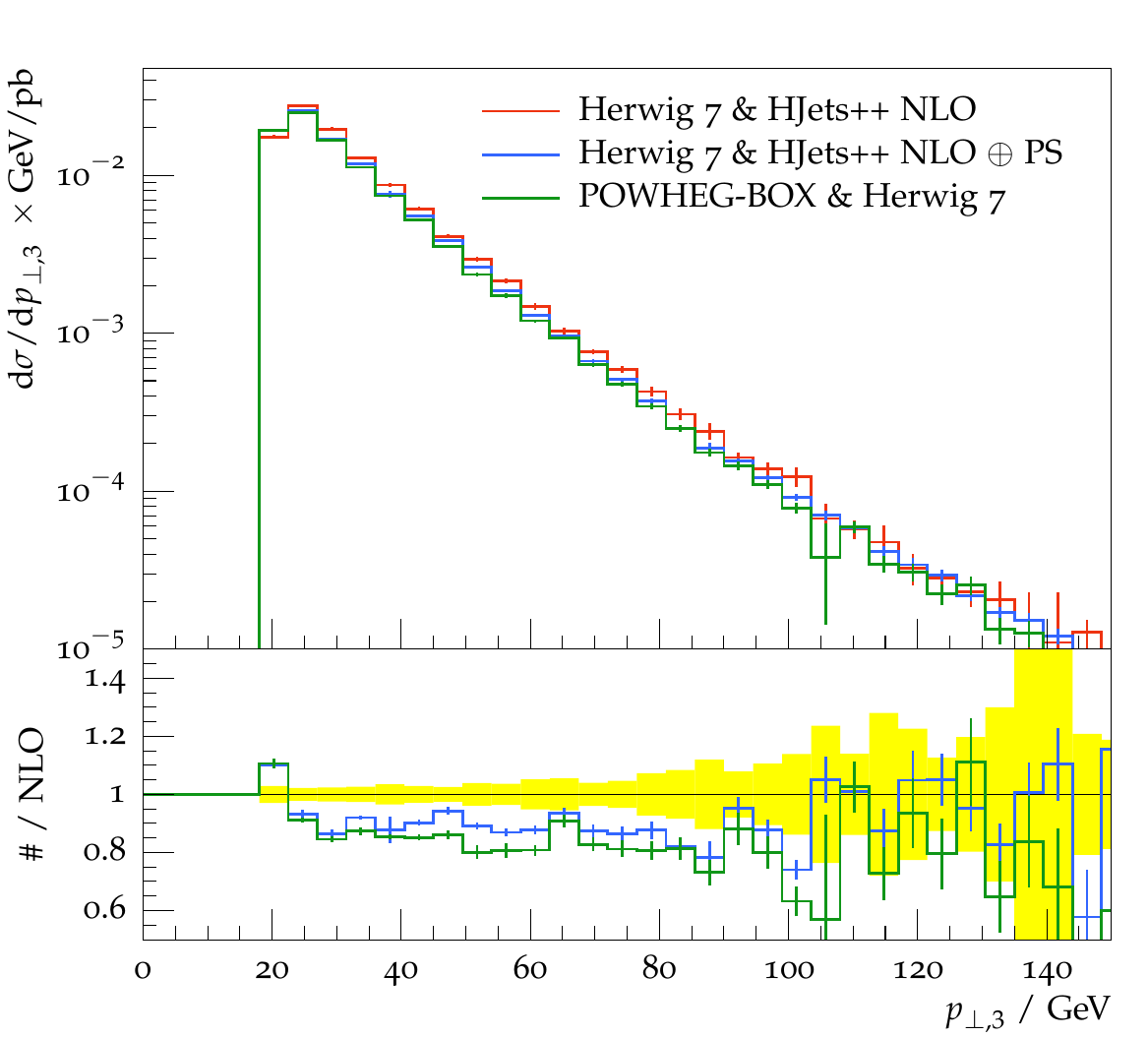}
\hfill
\includegraphics[width=.47\textwidth]{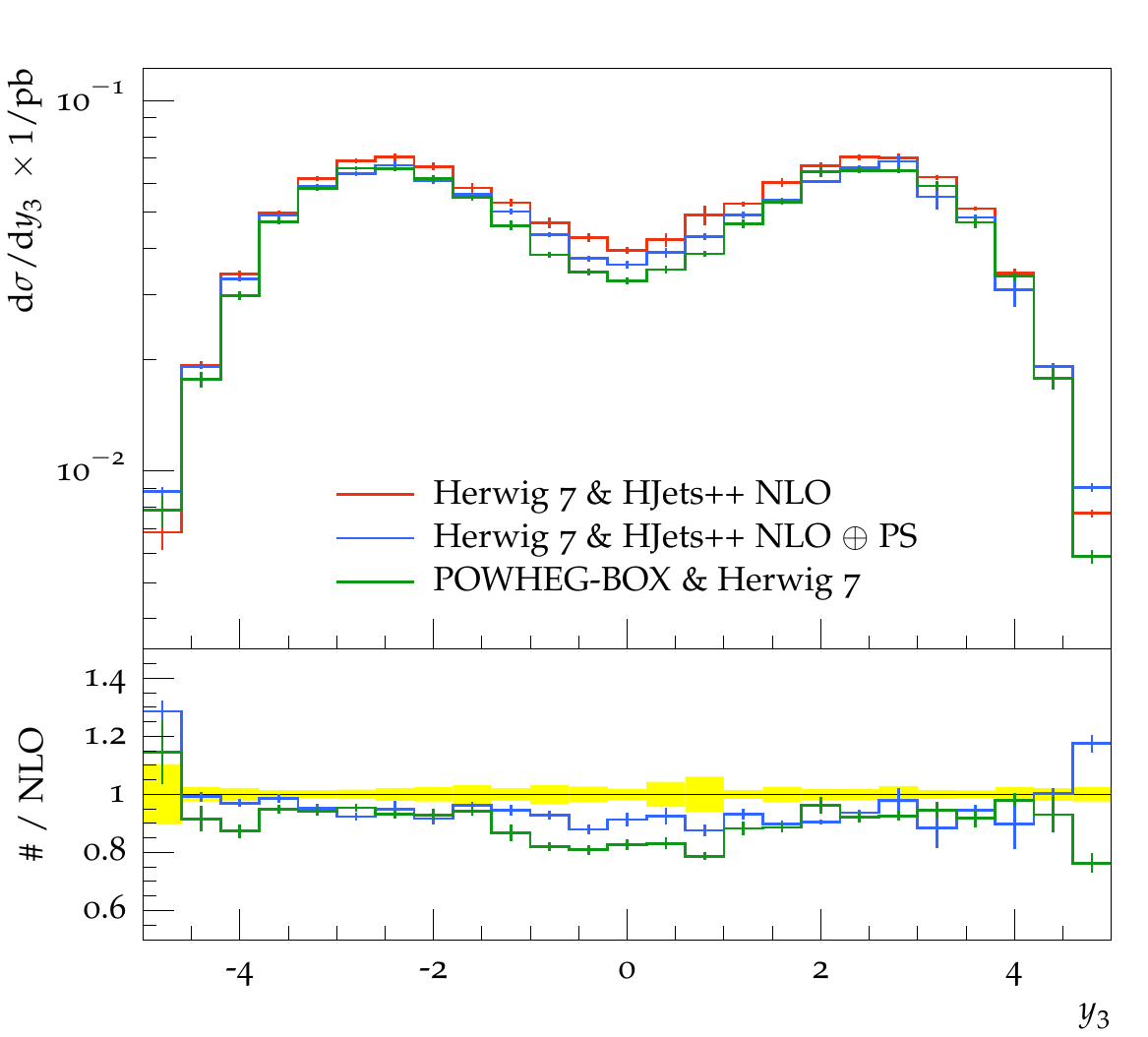}
\caption{Transverse-momentum and rapidity distributions of the third jet in EW H+3~jet production at NLO QCD (red line)  as obtained from using the \textsf{Matchbox} framework
of \HERWIG7\ with the \textsf{HJets++} plugin, and at NLO QCD matched with the
\HERWIG7\ angular ordered parton shower in the same framework (blue line),
and with the \POWHEGBOX\ (green line), respectively. The lower panels show the respective ratios of the NLO+PS to the fixed-order NLO QCD result for $\sqrt{s}=13\UTeV$ and $\MH=125\UGeV$. The yellow bands indicate the statistical uncertainty of the NLO result. }
\label{fig:SM-VBF-H3j-jet3}
\end{figure}

Here, results obtained with the \textsf{HJets++} module matched  through the subtractive matching algorithm with the default \HERWIG7 angular ordered shower are presented and compared to those obtained with the \POWHEGBOX\   implementation. Multiple partonic interactions and hadronization are disregarded throughout. Contributions from external top- and bottom quarks are neglected and, consistently, the CT10 four-flavour PDF set is used \cite{Lai:2010vv}. In addition to the selection cuts of Eqs.~(\ref{eq:VBF_cuts1})--(\ref{eq:VBF_cuts2}) a third jet is required with
\begin{equation}
\label{eq:VBF_jet3_cut}
{\pT}_j > 20\UGeV, \qquad
|y_j| < 5\,.
\end{equation}
Results for the transverse-momentum and rapidity distributions of the Higgs boson and the hardest tagging jet are shown in \refFs{fig:SM-VBF-H3j-H} and \ref{fig:SM-VBF-H3j-jet1}, respectively. In addition, the respective distributions of the third-hardest jet are illustrated in \refF{fig:SM-VBF-H3j-jet3}.
For the given setup results obtained with the \POWHEGBOX\  code that resorts to the VBF approximation are in good agreement with the full calculation of the \textsf{HJets++} implementation. A comparison of the \textsf{HJets++} at NLO QCD  and at NLO QCD matched with parton shower reveals that tune effects are moderate for the considered observables.
%


\section{VH production at NLO+PS}
Calculations for the VH process matched to parton shower programs are
available at NLO accuracy with the \POWHEGBOX~\cite{Luisoni:2013kna} and
{\tt MG5\_aMC}~\cite{Alwall:2014hca} with FxFx
merging~\cite{Frederix:2012ps,Frederix:2015eii}, for ZH and WH, and at
LO for ggZH. Although not used in this report, the gluon gluon fusion contribution
can be generated with additional jets that can be merged with MLM merging~\cite{Hespel:2015zea}.

Within the \POWHEGBOX{} the computation is carried out using the
improved \MINLO{} prescription~\cite{Hamilton:2012rf} applied to {\tt
  HZJ} ({\tt HWJ-MiNLO}) and {\tt HWJ} (\HWJMINLO). The event
generation was performed in a similar way as described in
ref.~\cite{Luisoni:2013kna}, but using the
NNPDF30\_nlo\_as\_118~\cite{Ball:2014uwa} PDF set.

A systematic comparison of these calculation for 13 TeV LHC collisions
has been carried out in several regions of the phase space
making use of several Rivet~\cite{rivet} analyses, differing for the vector boson and Higgs boson candidate selection.
The $Z(ll)H(bb)$ process is studied in the Z pT bins: inclusive, [0-100]~GeV, (100-200]~GeV, >200~GeV.
The Z leptons are selected with the cuts $|\eta| <2.5$, $p_{T} > 15$ GeV.
The dilepton invariant mass $m_{ll}$ in required to be in the range [75-105]~GeV.
The $Z(\nu\nu)H(bb)$ process is studied in the Z $p_{T}$ bins: inclusive, [0-150]~GeV, (150-250]~GeV, >250~GeV.
The Z $p_{T}$ is evaluated through the missing transverse energy of the event.
The $W(l\nu)H(bb)$ process is studied in the W $p_{T}$ bins: inclusive, [0-150]~GeV, (150-250]~GeV, >250~GeV.
The W lepton is required to have $|\eta|<2.5$, $p_{T} > 15$ GeV.
The neutrino $p_{T}$, evaluated through the missing transverse energy of the event, is required to be above 15 GeV.

The processes are studied as a function of the number of additional jets,
reconstructed with fastjet~\cite{Cacciari:2011ma} with the anti-$k_T$ algorithm and a cone of 0.5,
and selected to have $p_T > 20$ GeV and $|\eta| < 4.5$.
The jet counting is used to define the exclusive VH+0-jet and VH+1-jet regions and the VH+$\geq$1-jet one,
used in the experimental analyses~\cite{Aad:2014xzb,Chatrchyan:2013zna}.

For each process, the Higgs boson $p_T$ and rapidity, the lepton $p_T$ and rapidity, and the neutrino $p_T$
are compared in each of the boson $p_T$ bins, for different bins of additional jets
after normalizing the inclusive cross section to unity.

The results obtained with \POWHEGBOX\ matched with the default \PYTHIA6 (\POWHEG{}+PY6) shower are presented and compared
to those obtained with the {\tt MG5\_aMC} implementation matched with both default \PYTHIA8 ({\tt MG5\_aMC}+PY8) and default \HERWIG7~\cite{Bellm:2015jjp,Bahr:2008pv} ({\tt MG5\_aMC}+HW7) tune.
%
Comparisons are made keeping the Higgs boson stable.
The plots are shown for the ZH case but similar conclusions apply as well to the WH process.
%
%

\begin{figure}[hptb]
\centering
\includegraphics[width=.47\textwidth]{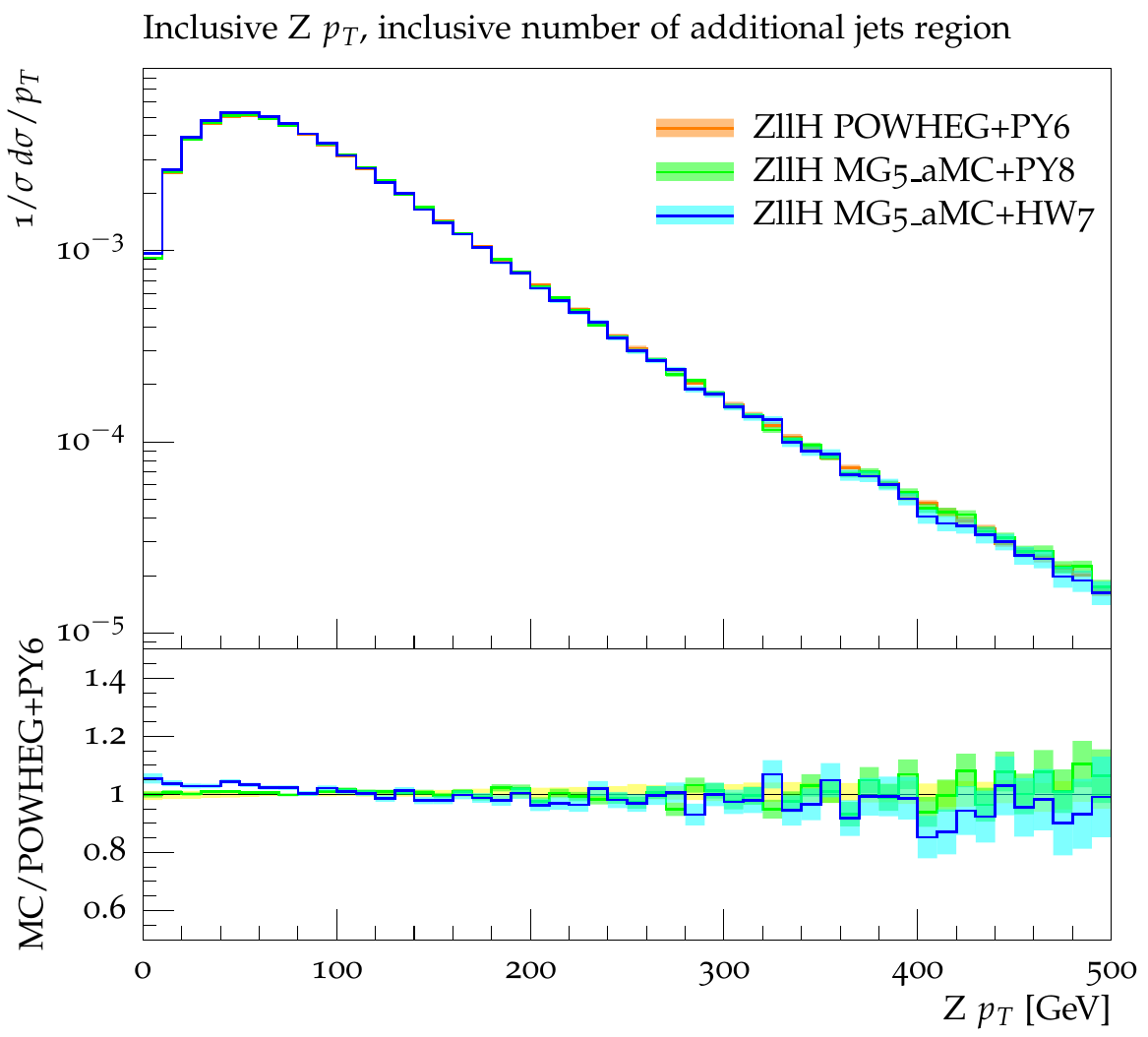}
\includegraphics[width=.47\textwidth]{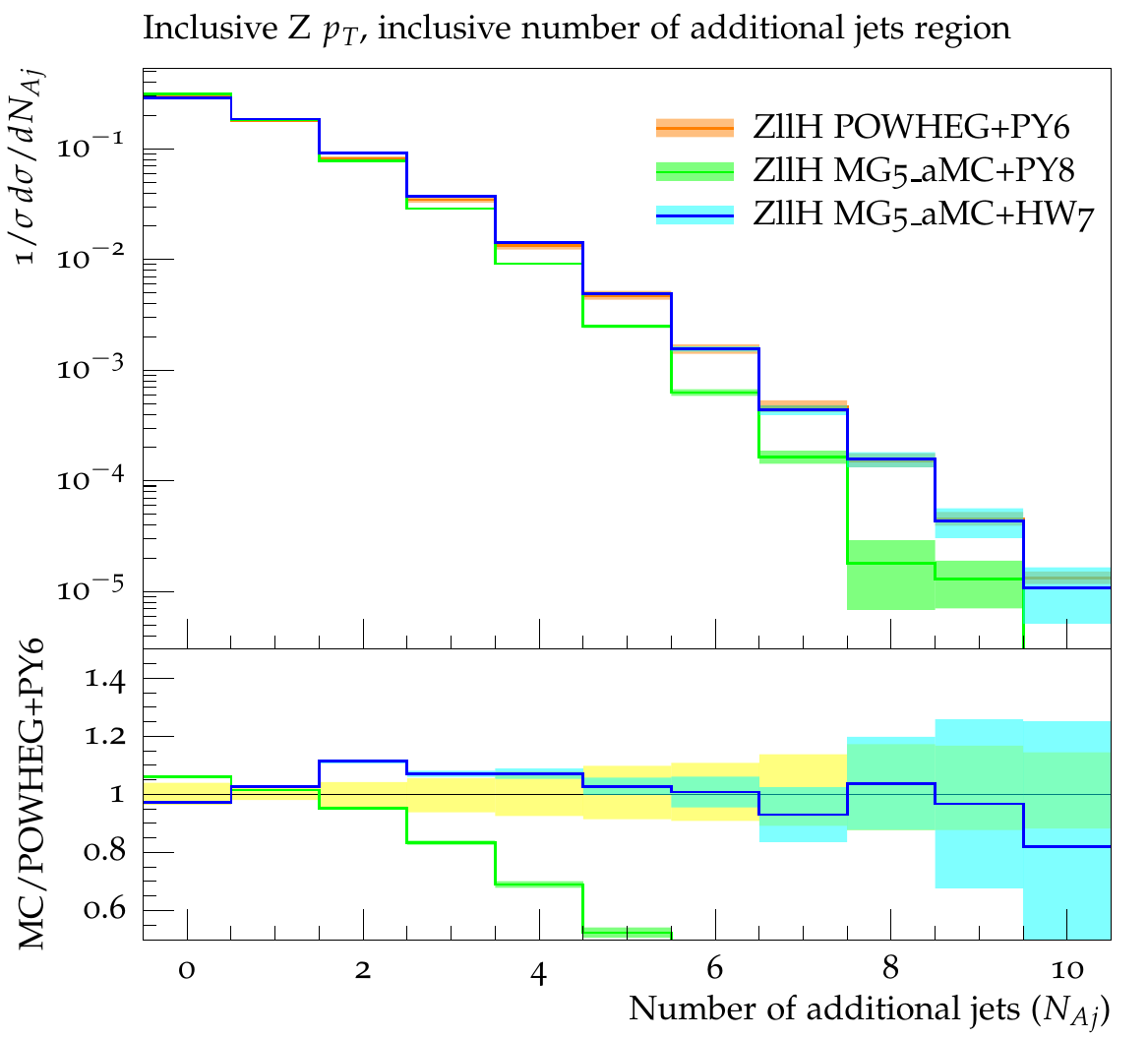}
\caption{Comparison of the boson \pt\ (left) and number of additional jets (right) in the inclusive case for $Z(ll)H$.}
\label{fig:stable__incl_vpt_jets}
\end{figure}

The boson \pt\ and additional jet multiplicity distributions in the inclusive case are shown in \refF{fig:stable__incl_vpt_jets}.
A very small trend is visible in the boson \pt\ for {\tt MG5\_aMC}+HW7 when compared with \POWHEG{}+PY6 and {\tt MG5\_aMC}+PY8,
while the distribution of additional jets for {\tt MG5\_aMC}+PY8 deviates at high multiplicity when compared with the other two cases.
The discrepancies highlighted in the comparison of {\tt MG5\_aMC}+PY8 and {\tt MG5\_aMC}+HW7 clearly indicate the need
for a careful choice of the parton shower and underlying event tune when performing analyses
which require categories with exclusive number of jets and boson \pt\ binning.

\begin{figure}[hptb]
\centering
\includegraphics[width=.47\textwidth]{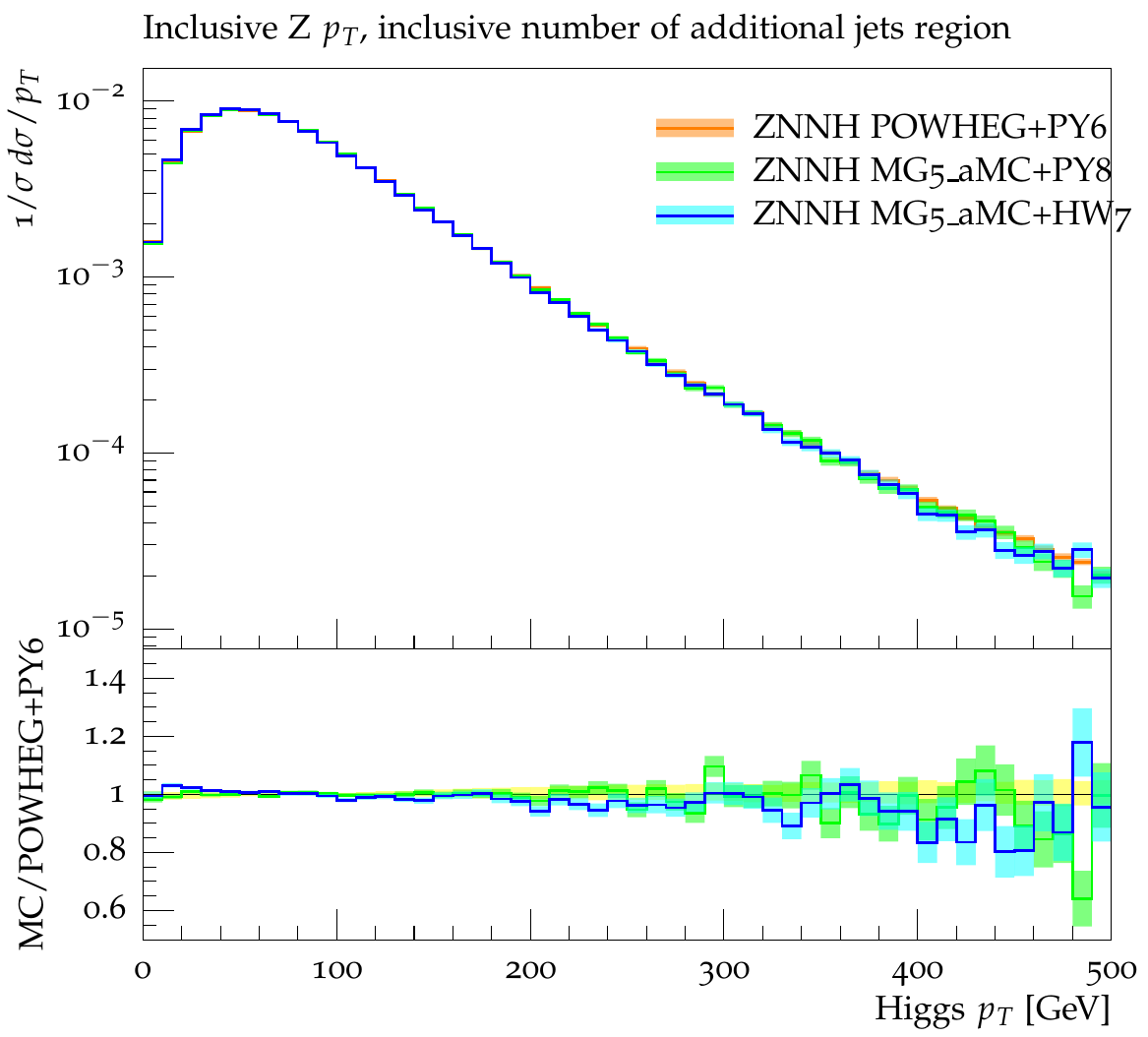}
\includegraphics[width=.47\textwidth]{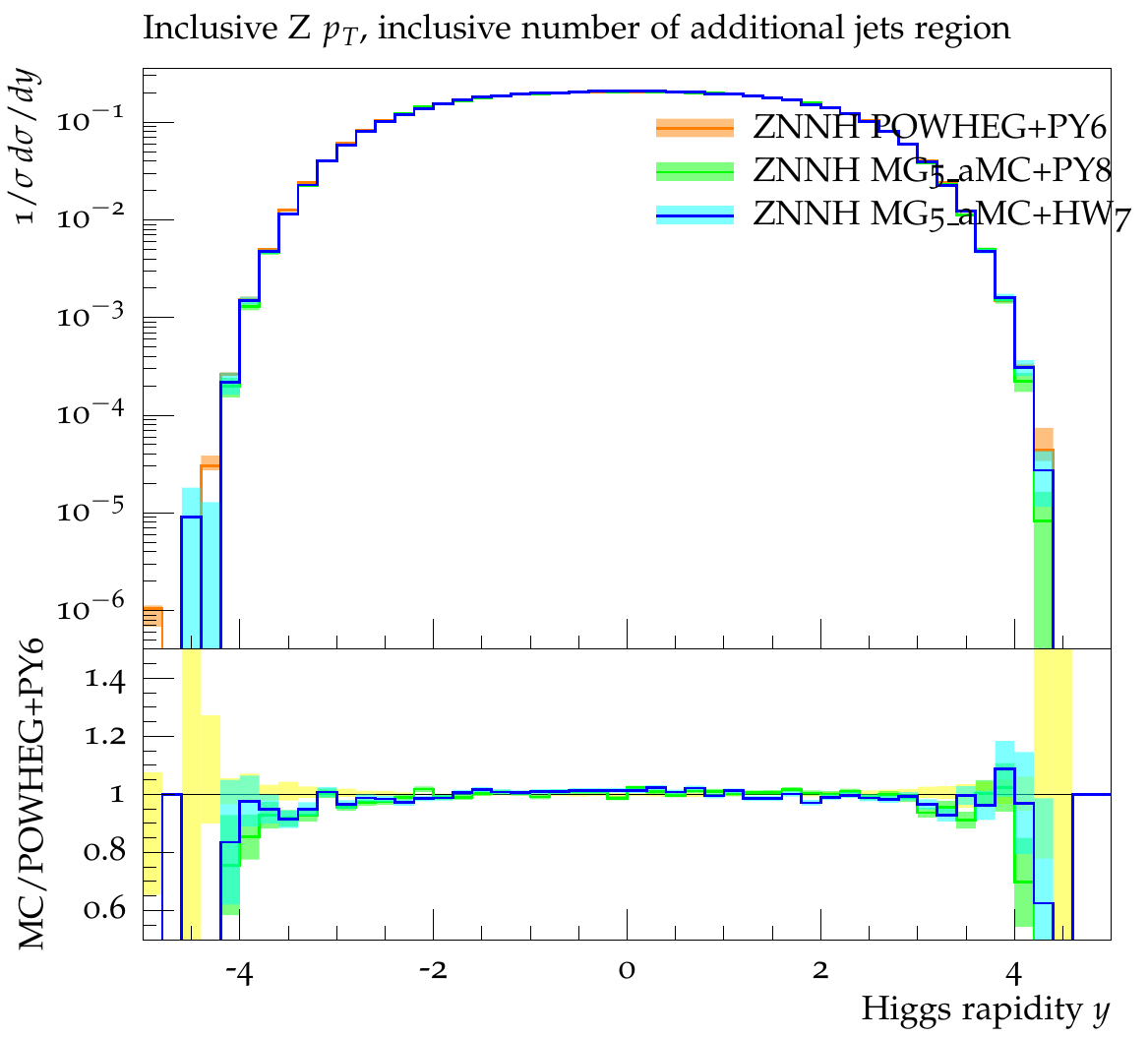}
\includegraphics[width=.47\textwidth]{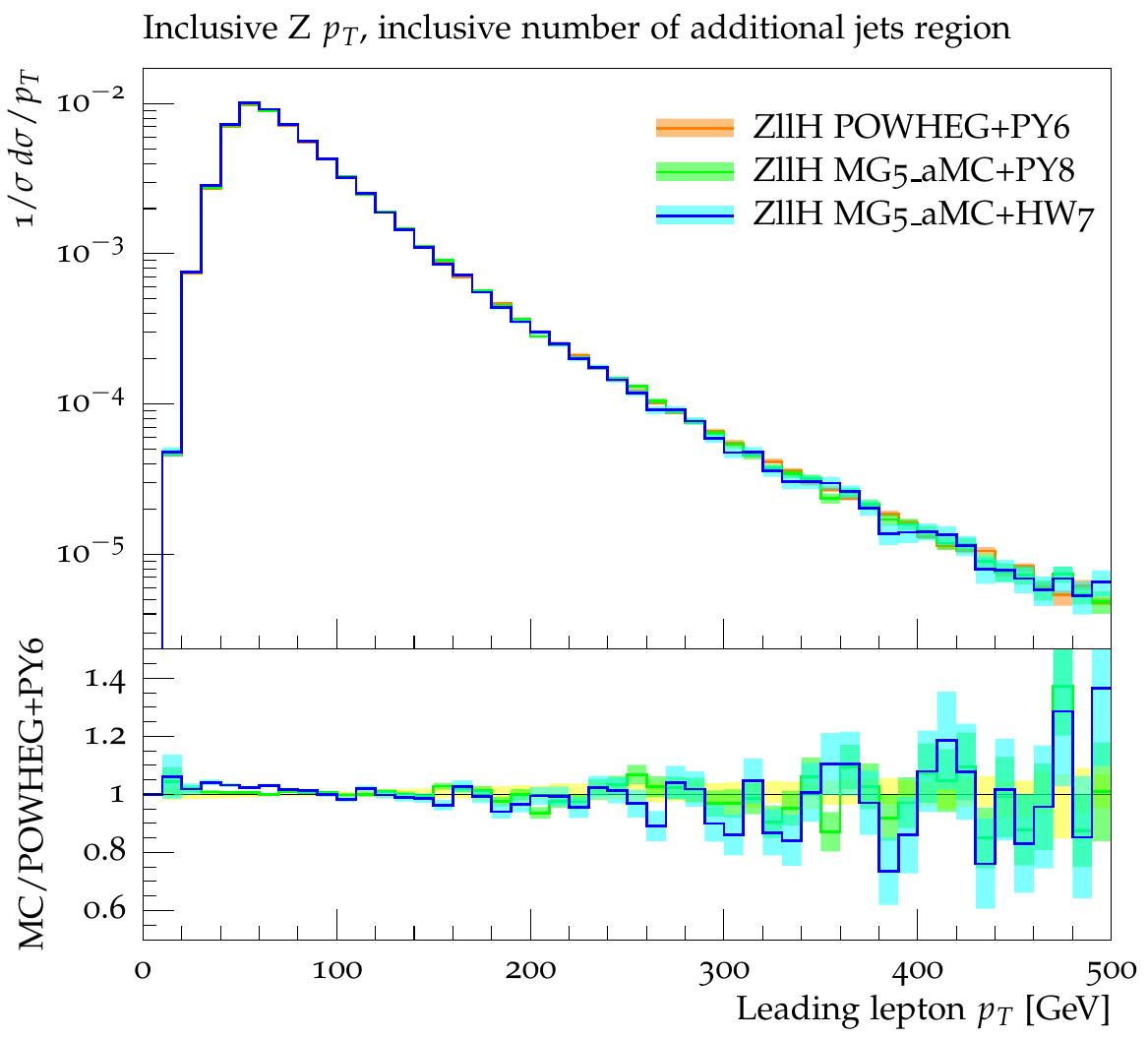}
\includegraphics[width=.47\textwidth]{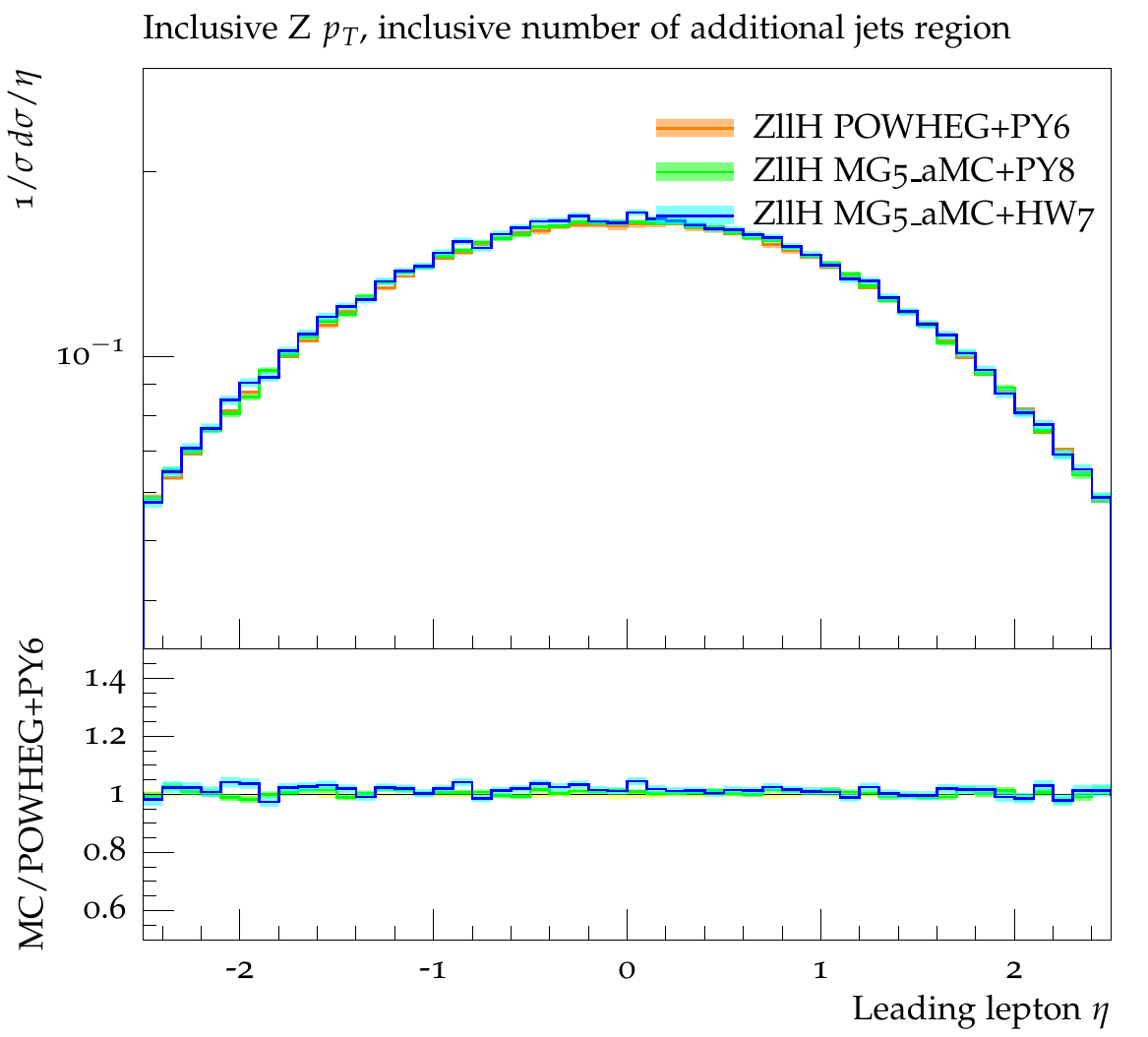}
\caption{Comparison of the boson \pt\ and rapidity in $Z(\nu\nu)H$ events, and leading lepton \pt\ and $\eta$ in $Z(ll)H$ events in the inclusive jet region.}
\label{fig:stable__incl_hig}
\end{figure}

In the same phase space, characterized by inclusive boson \pt\ and additional jet selection, the Higgs boson \t\ for {\tt MG5\_aMC}+HW7 exhibits
the same trend visible for the boson \pt, while the rapidity shapes are well compatible
as shown in \refF{fig:stable__incl_hig}.

\begin{figure}[hptb]
\centering
\includegraphics[width=.47\textwidth]{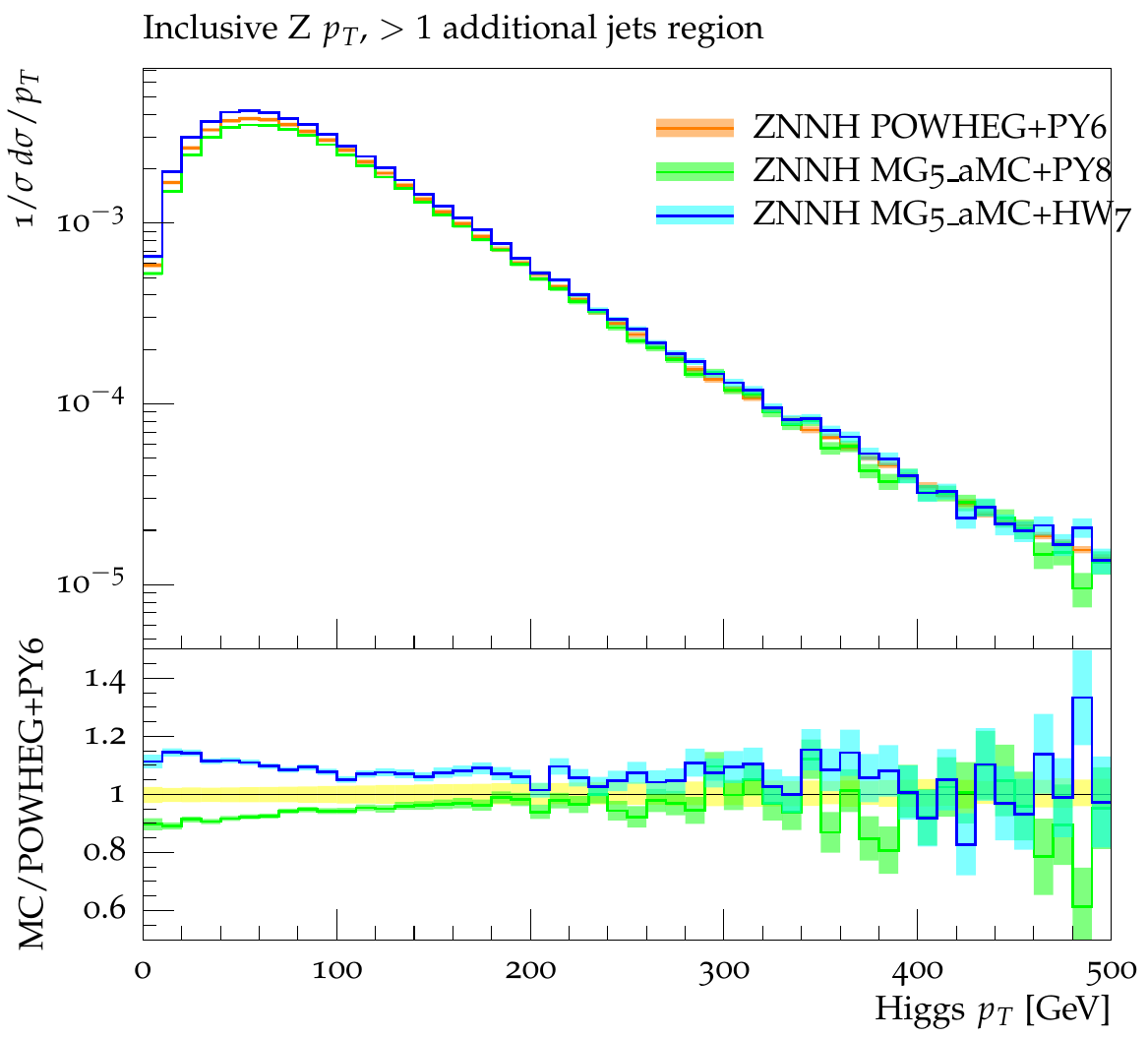}
\includegraphics[width=.47\textwidth]{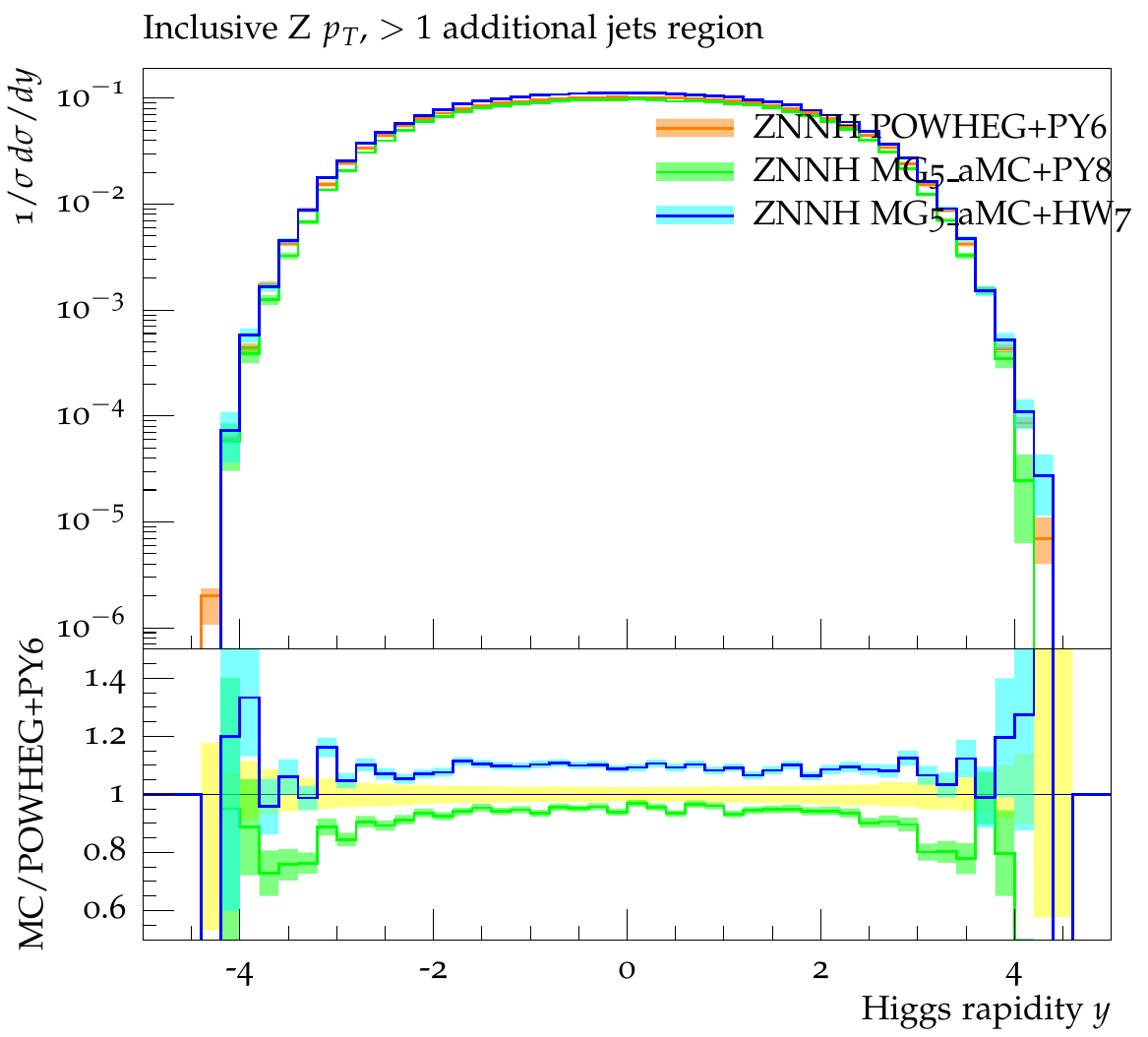}
\includegraphics[width=.47\textwidth]{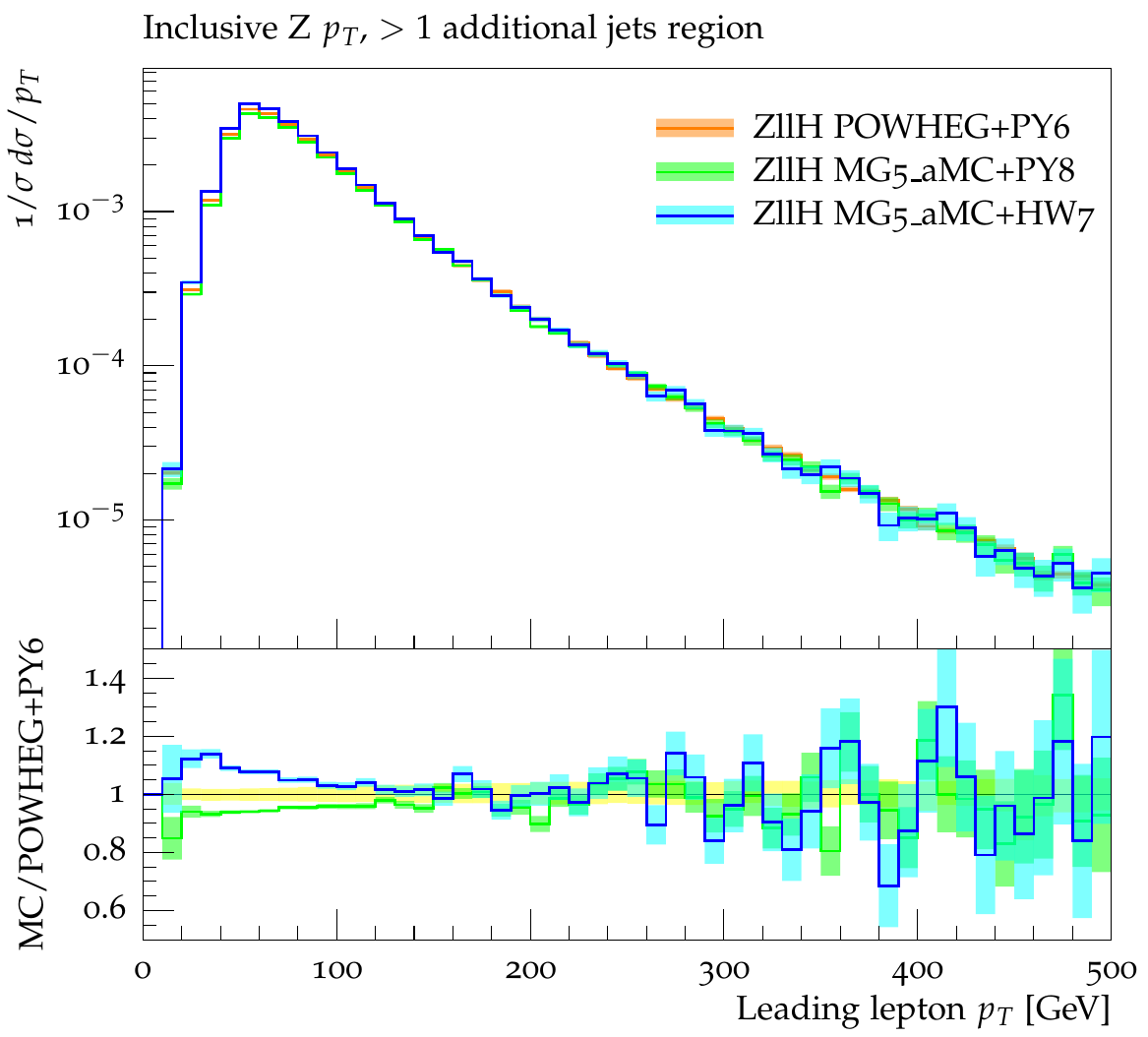}
\includegraphics[width=.47\textwidth]{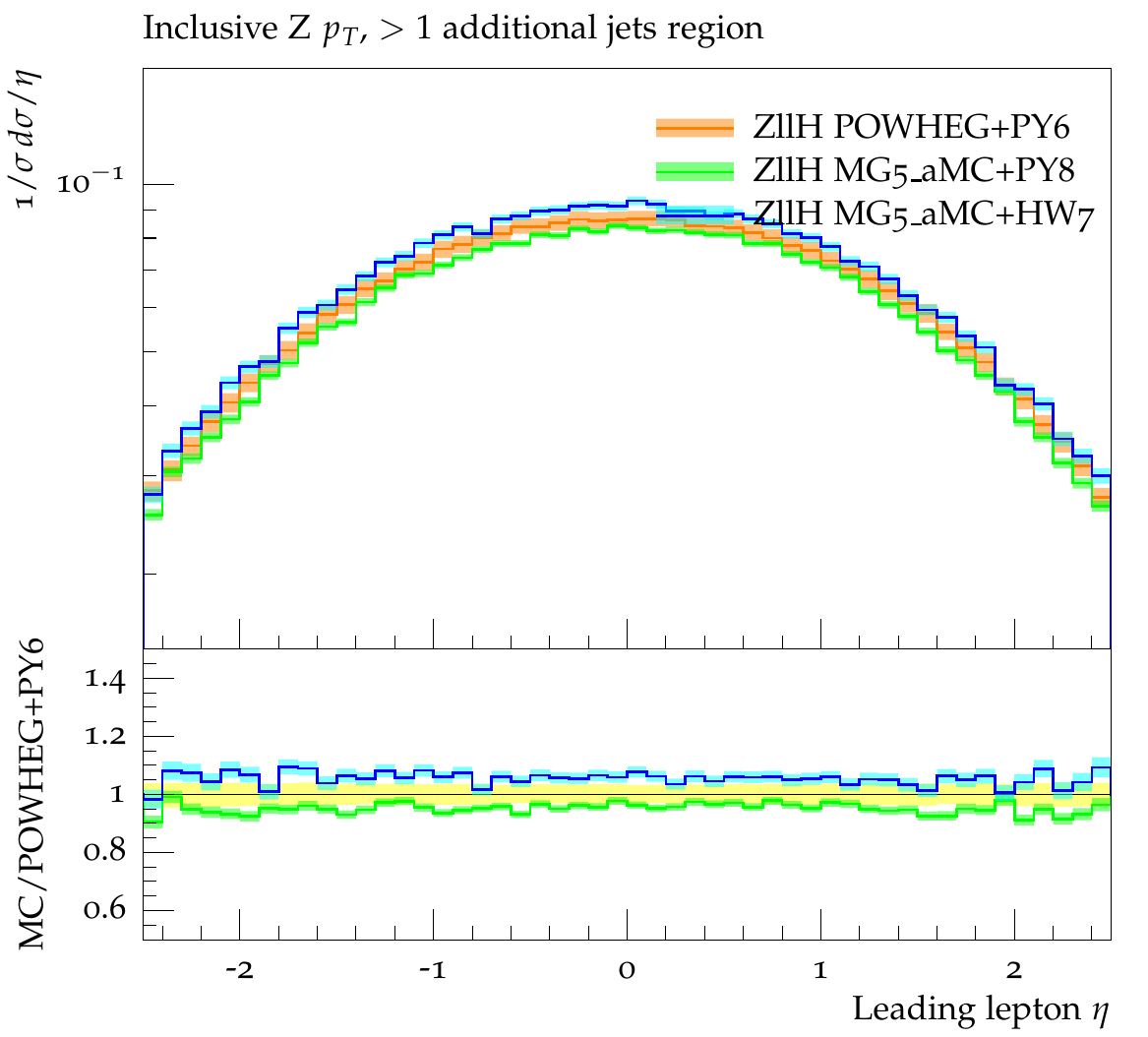}
\caption{Comparison of the boson \pt\ and rapidity in $Z(\nu\nu)H$ events, and leading lepton \pt\ and $\eta$ in $Z(ll)H$ events in the inclusive
the inclusive boson \pt\ region requiring at least 1 additional jet.}
\label{fig:stable__gt1j_hig}
\end{figure}

In the phase space characterized by an inclusive boson \pt\ and the explicit request for 0 additional jets,
the Higgs boson $p_T$ and rapidity, the lepton $p_T$ and rapidity, and the neutrino $p_T$
distribution shapes remain well compatible, but a different normalization can be observed as a reflection of the
different distribution in the additional jet multiplicity.
While requiring exactly 1 additional jet,
the lepton and Higgs boson \t shapes modeled by HW7 tend to deviate slightly,
further increasing their discrepancy
when requiring the inclusive boson \pt\ and at least 1 additional jet,
as well as the overall normalization due to the aforementioned differences in the additional jet multiplicity.
The the boson \pt\ and rapidity in $Z(\nu\nu)H$ events, and leading lepton \pt\ and $\eta$ in $Z(ll)H$ events
for the latter case are shown in \refF{fig:stable__gt1j_hig}.

\begin{figure}[hptb]
\centering
\includegraphics[width=.47\textwidth]{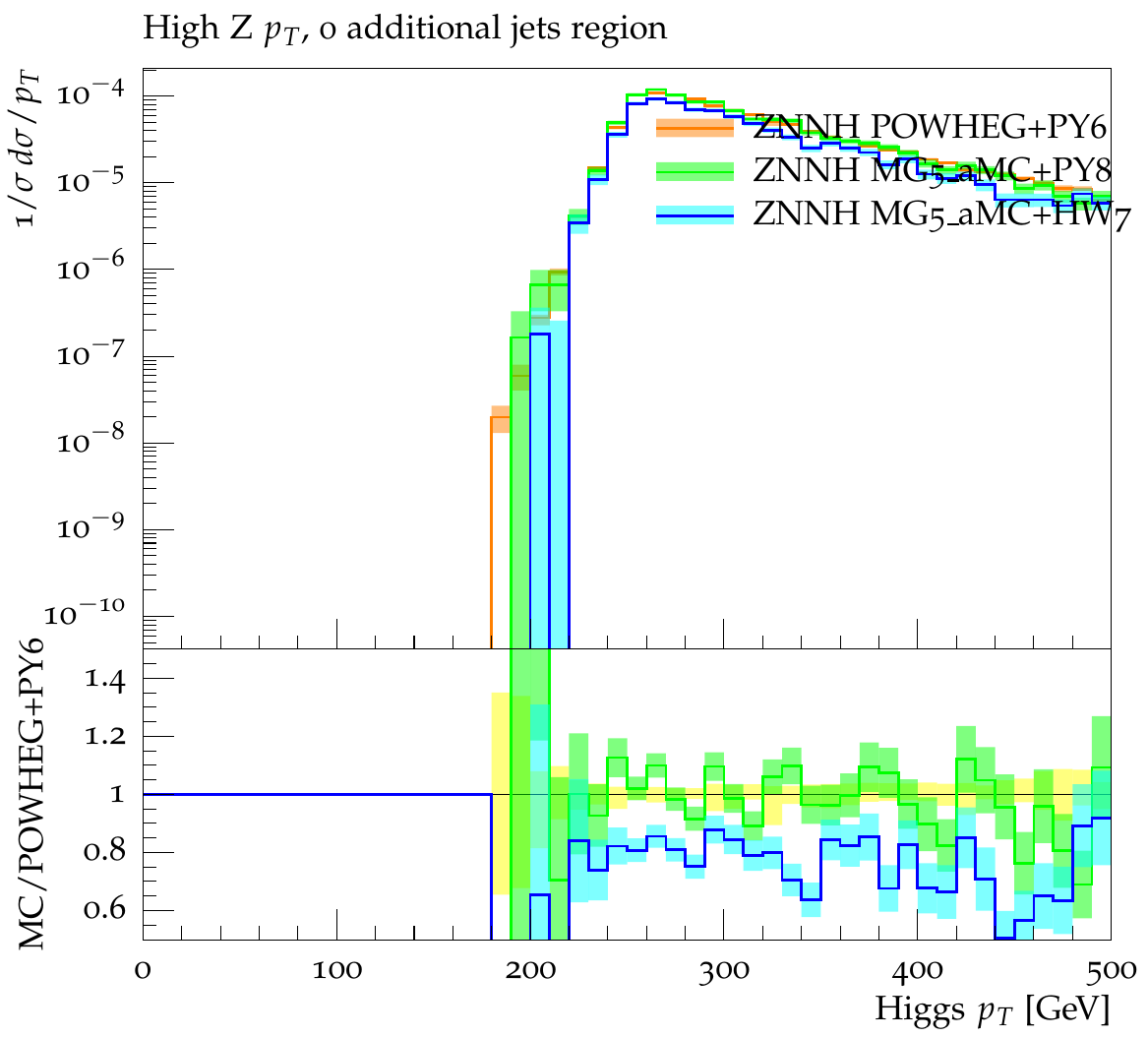}
\includegraphics[width=.47\textwidth]{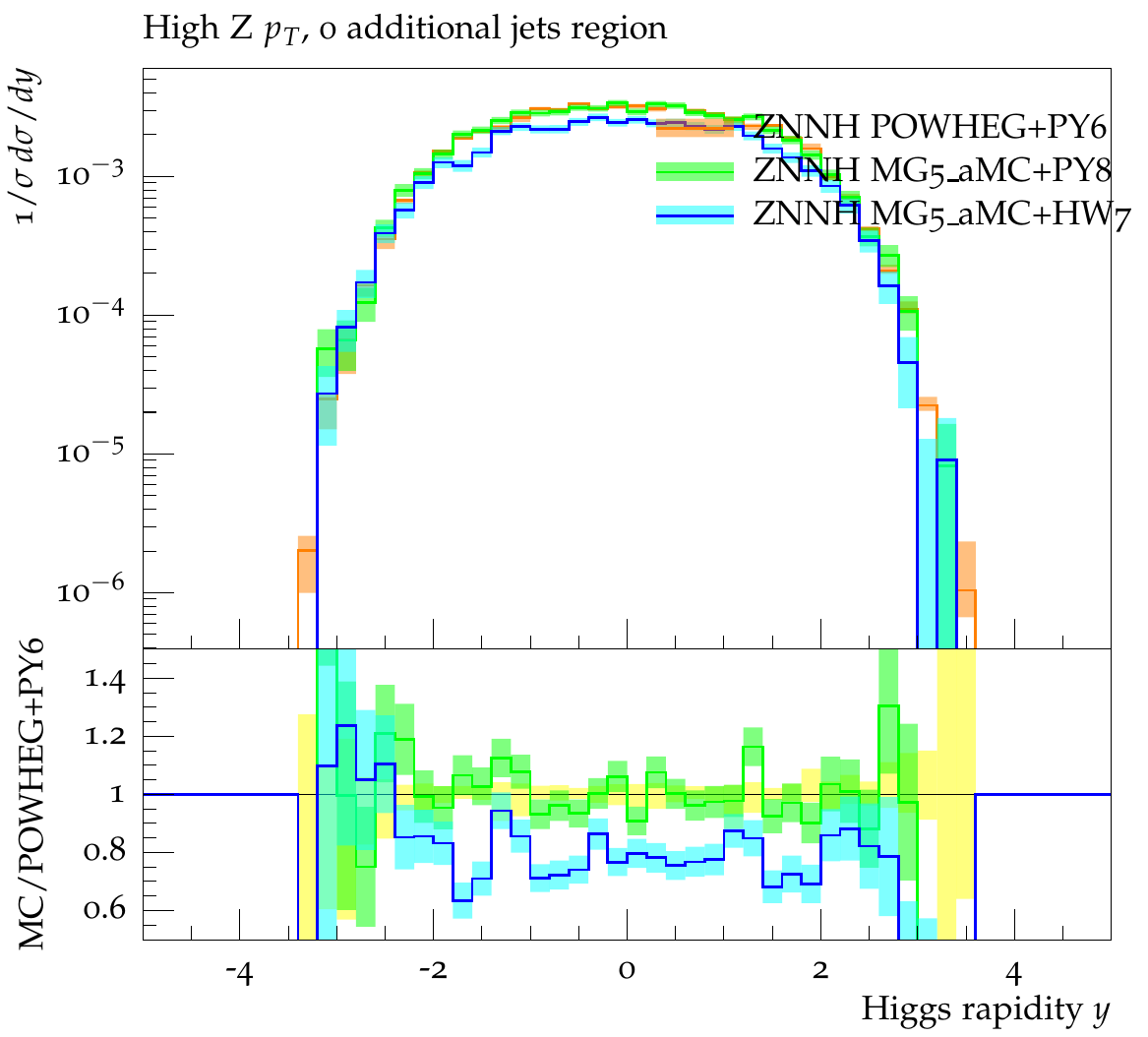}
\includegraphics[width=.47\textwidth]{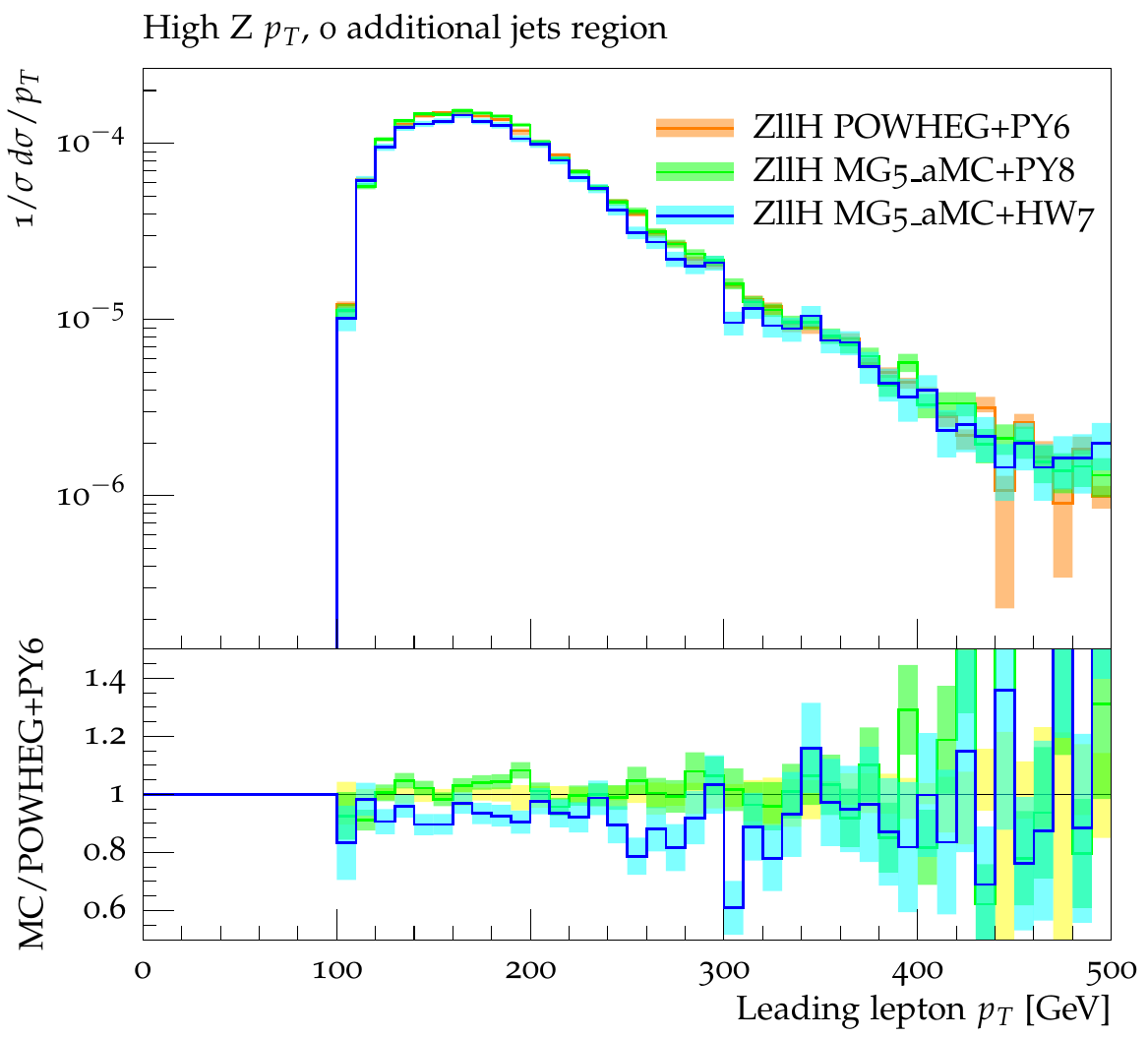}
\includegraphics[width=.47\textwidth]{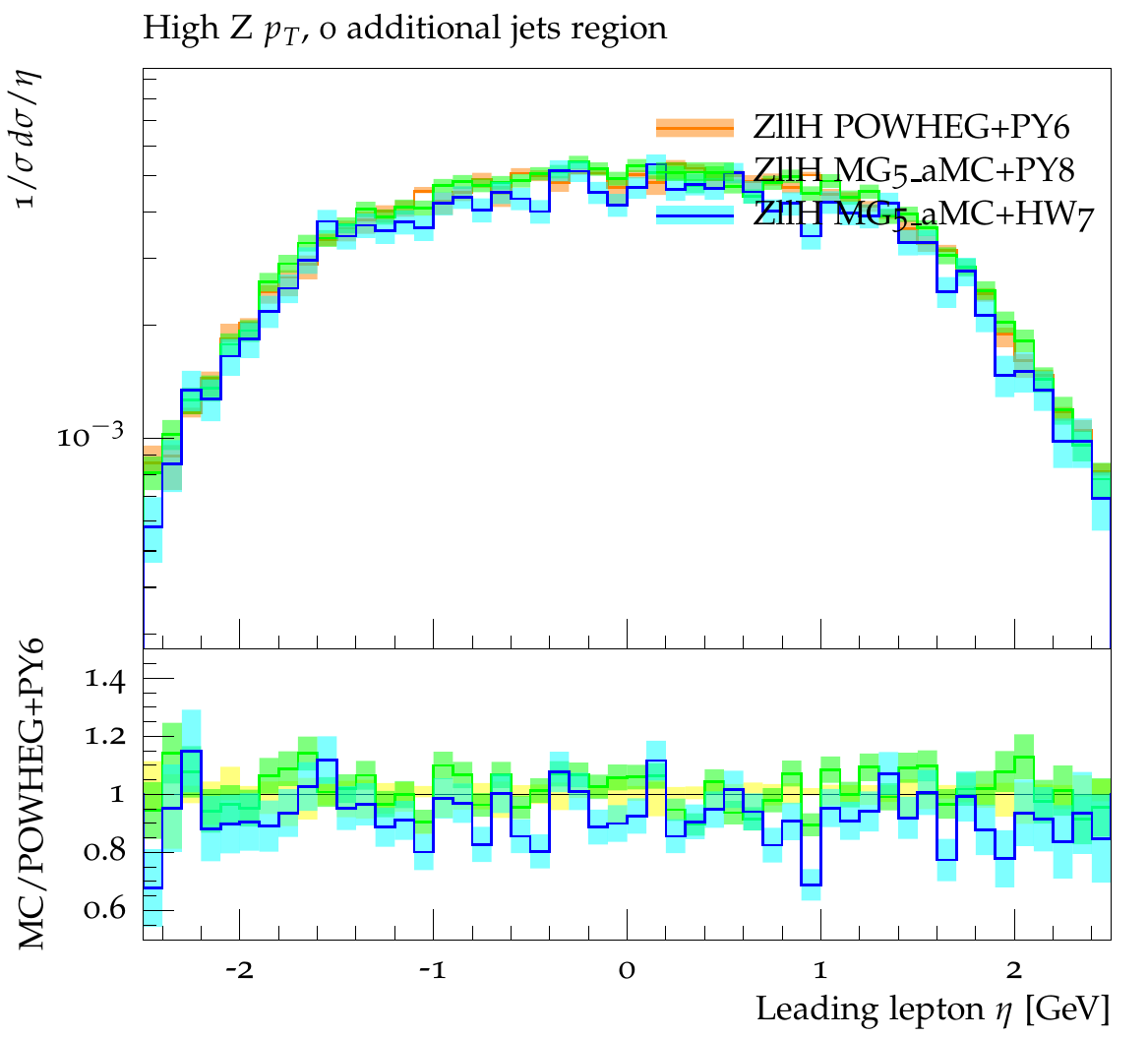}
\caption{Comparison of the boson \pt\ and rapidity in $Z(\nu\nu)H$ events, and leading lepton \pt\ and $\eta$ in $Z(ll)H$ events in the inclusive
the inclusive boson \pt\ region requiring 0 additional jet.}
\label{fig:stable__gt1j_highvpt}
\end{figure}

Extending the comparison to the low, medium and high boson \pt\ regions defined above,
consistent results are observed.
In particular, well compatible shapes are observed for the inclusive jet selection,
apart from some minor trend in the low Higgs boson \t\ region, especially for HW7.
The same level of agreement is also observed for the 0- and 1-jet phase spaces, as well as when requiring at least 1 jet,
with the normalization offset discussed previously.
The the boson \pt\ and rapidity in $Z(\nu\nu)H$ events, and leading lepton \pt\ and $\eta$ in $Z(ll)H$ events
for high boson \pt\ case when requiring exactly 0 additional jets are shown in \refF{fig:stable__gt1j_highvpt}.

Finally, a comparison of the quark-quark ($ZH$) and gluon-gluon ($ggZH$) initiated processes is performed using
{\tt MG5\_aMC} interfaced with a common parton shower, namely \PYTHIA8. The plots are shown for the leptonic decay of the Z boson
but apply as well to the decay into neutrinos.

\begin{figure}[hptb]
\centering
\includegraphics[width=.47\textwidth]{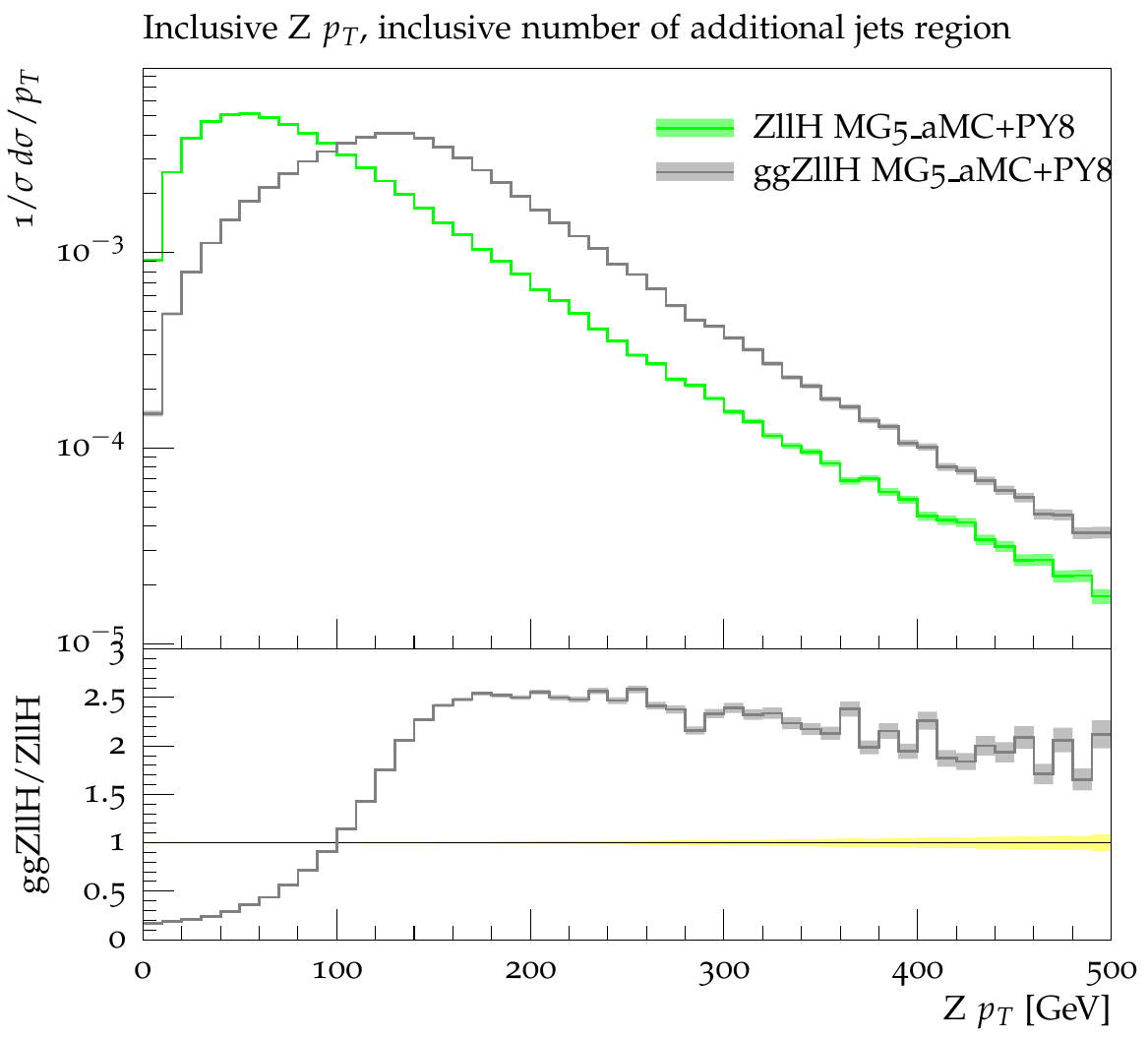}
\includegraphics[width=.47\textwidth]{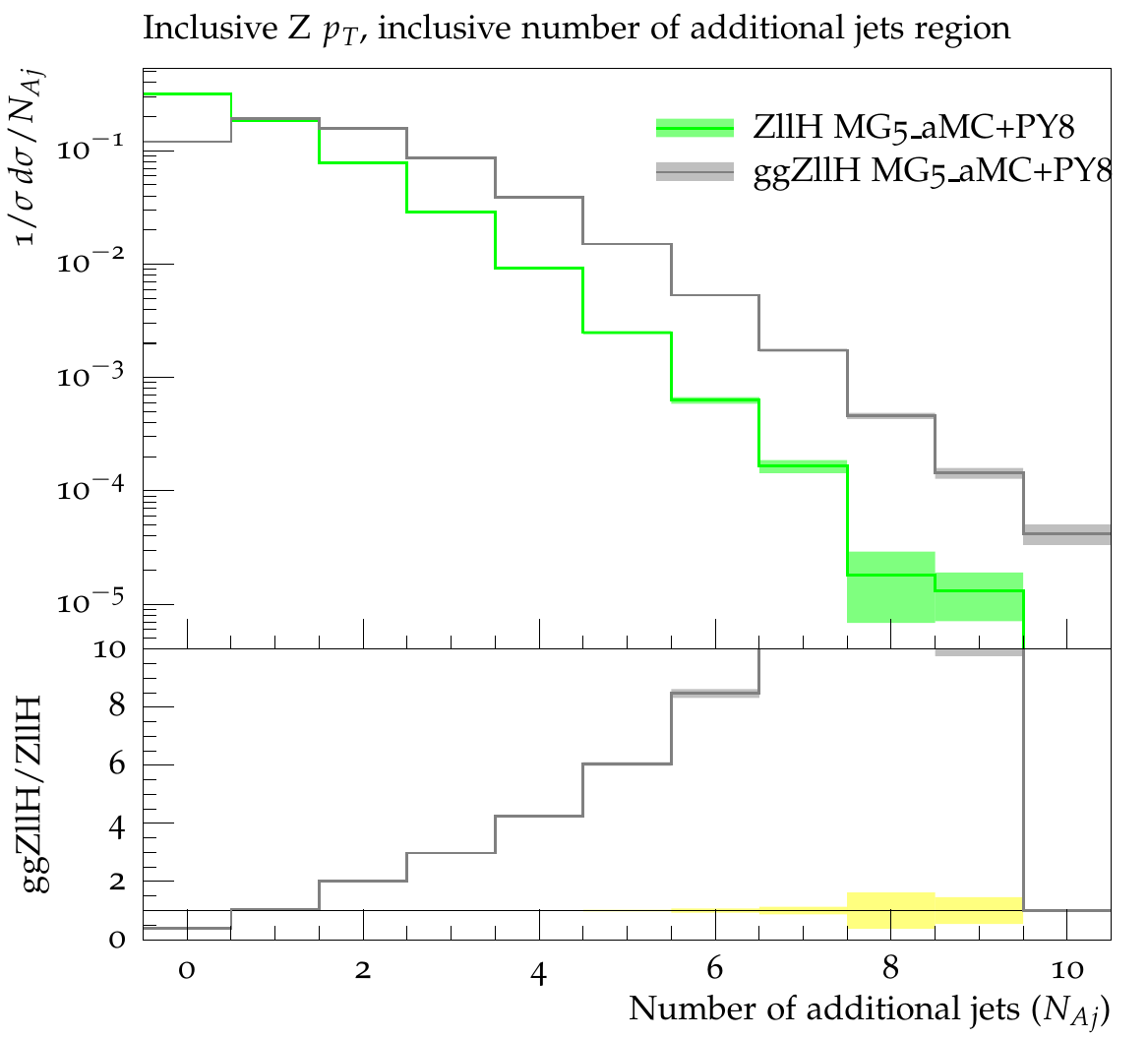}
\caption{Comparison of the boson \pt\ (left) and number of additional jets (right) in the inclusive case for $Z(ll)H$.}
\label{fig:stable__incl_vpt_jets_ggzh}
\end{figure}

\begin{figure}[hptb]
\centering
\includegraphics[width=.47\textwidth]{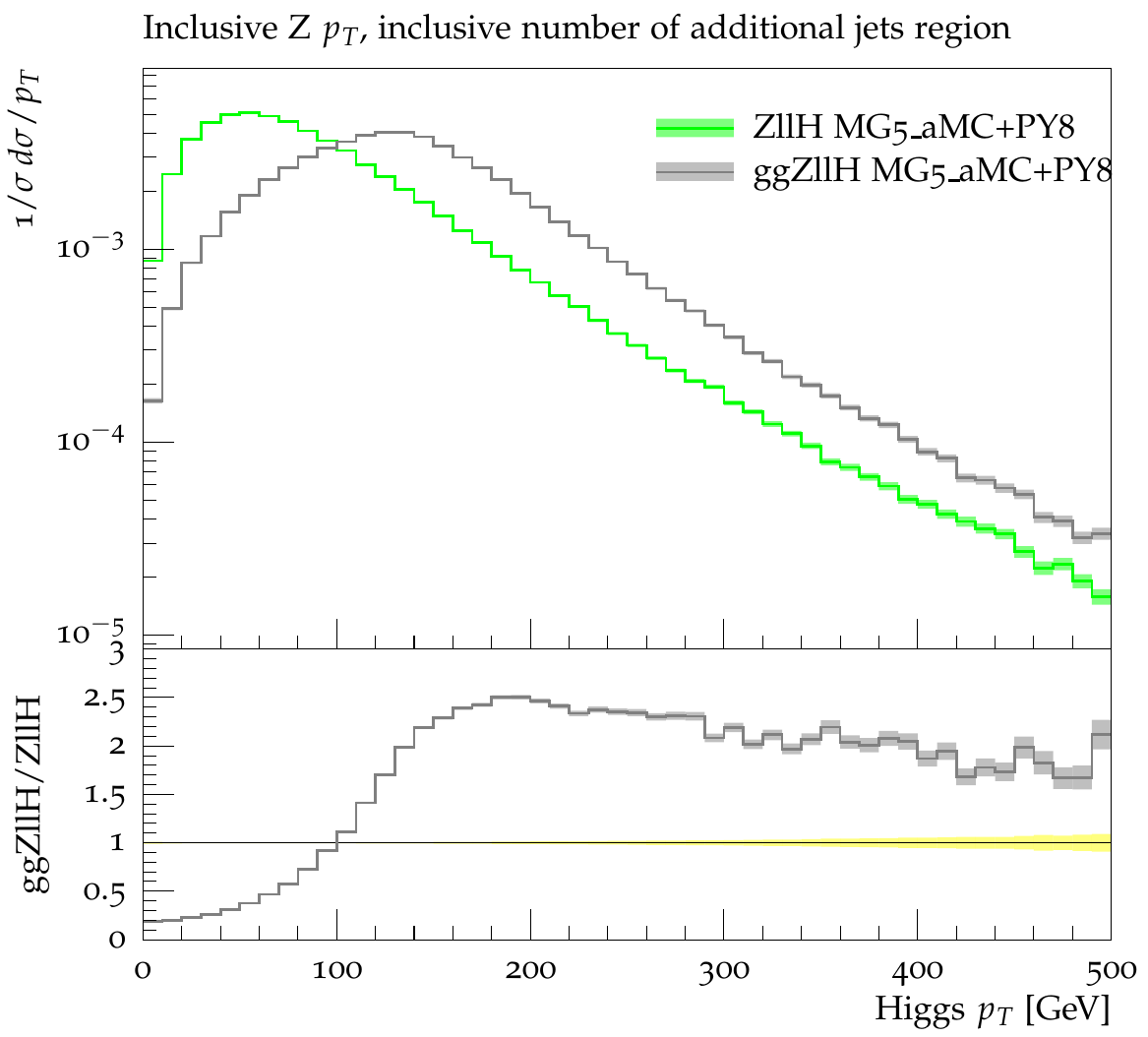}
\includegraphics[width=.47\textwidth]{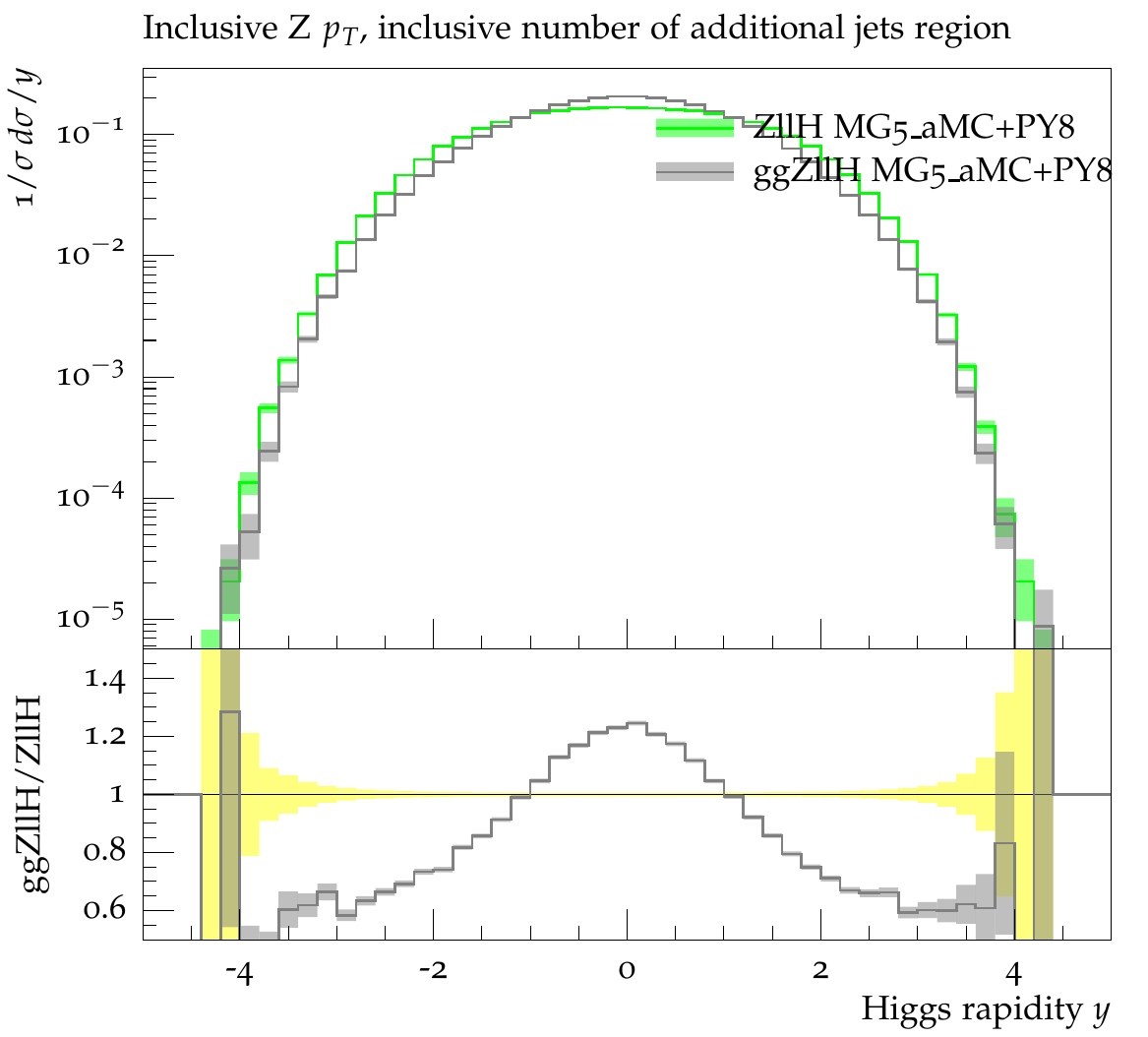}
\includegraphics[width=.47\textwidth]{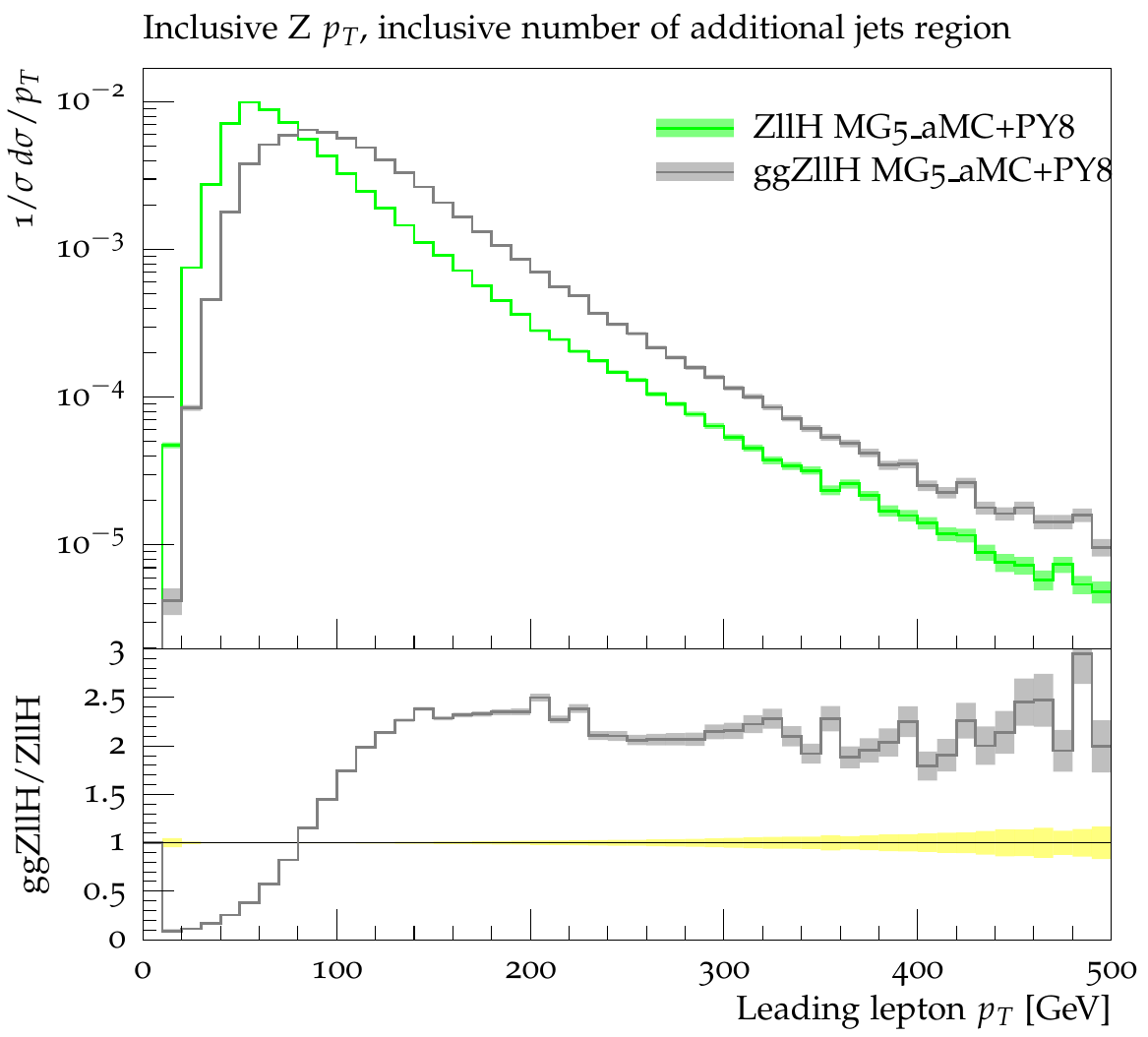}
\includegraphics[width=.47\textwidth]{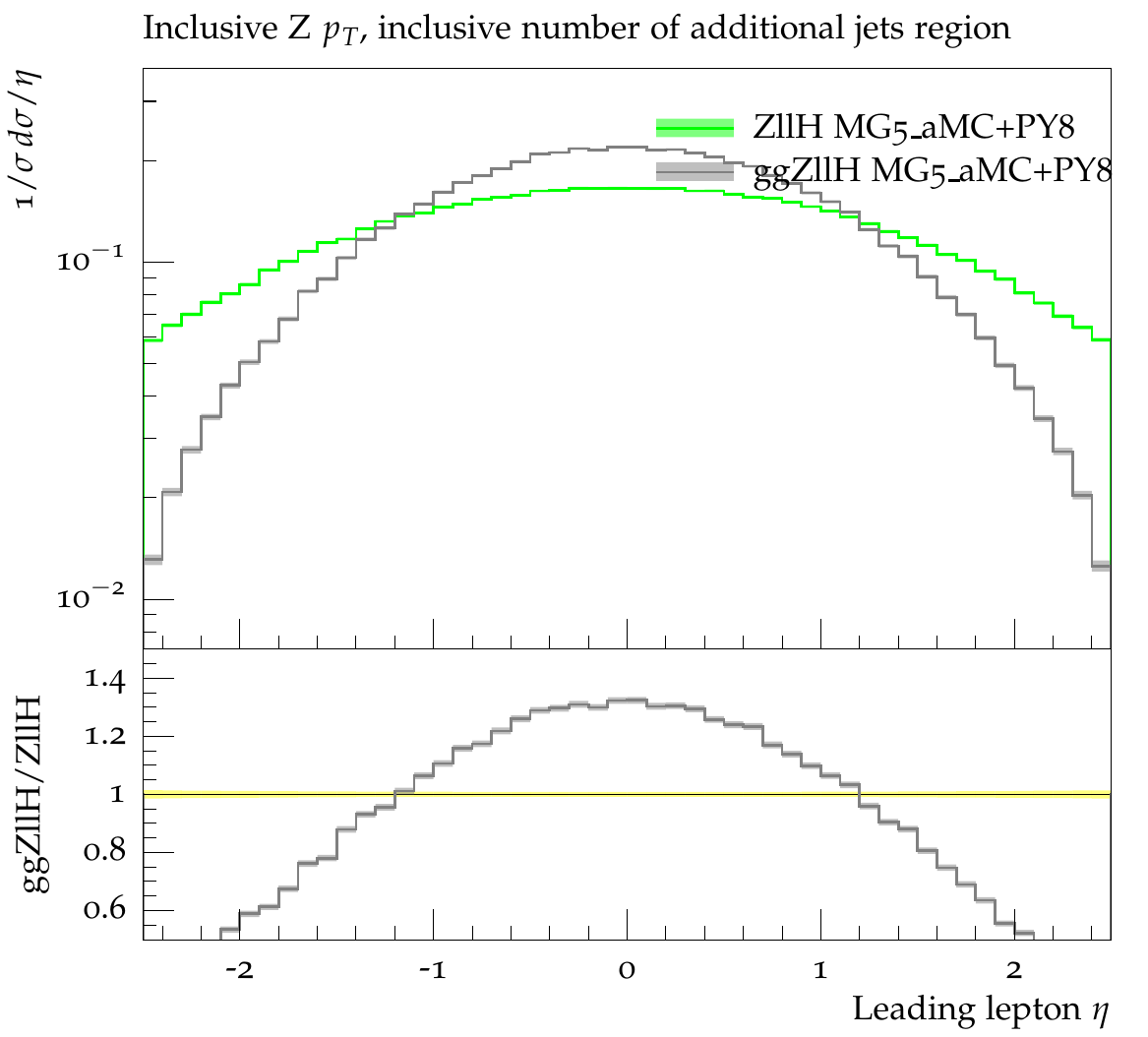}
\caption{Comparison of the boson \pt\ and rapidity, and leading lepton \pt\ and $\eta$ in $ggZ(ll)H$ $Z(ll)H$ events in the inclusive jet region.}
\label{fig:stable__incl_vpt_jets_ggzh2}
\end{figure}

The relative cross section of the gluon initiated process is $\sim15\%$, but a common normalization to unitary cross section is used,
to better underline the shape differences.
The boson \pt\ and additional jet multiplicity distributions in the inclusive case are shown in \refF{fig:stable__incl_vpt_jets_ggzh}
Two correlated features can be observed: both the boson \pt\ and the multiplicity of additional jets are much harder for the gluon initiated contribution.
Therefore in the high boson \pt\ region, usually regarded as the most sensitive, and in presence of at least 1 jet, the relative contribution of $ggZ(ll)H$
is much higher.
The the boson \pt\ and rapidity, and leading lepton \pt\ and $\eta$ in $ggZ(ll)H$ events
for high boson \pt\ case when requiring exactly 0 additional jets are shown in \refF{fig:stable__incl_vpt_jets_ggzh2} compared to $Z(ll)H$.

\section{NNLOPS for VH}

We report about a study of the Higgs boson production in association with a $W^{+}$
boson at next-to-next-to-leading order accuracy including parton
shower effects (\NNLOPS{})~\cite{Astill:2016hpa}
\begin{equation}
  pp\rightarrow HW^{+} \rightarrow H l^{+}\nu_{l}\,,
  \label{eq:def-process}
\end{equation}
where $l = \{e,\mu\}$.
To achieve \texttt{NNLOPS} accuracy we have implemented a reweighting
method similar to the one introduced in \HNNLOPS{}
\cite{Hamilton:2013fea} and \DYNNLOPS{} \cite{Karlberg:2014qua}. We
reweight events obtained with the \POWHEG{} NLO+PS accurate
calculation of HW in association with a jet, and upgraded with the
\MINLO{} procedure (\HWJMINLO{}) \cite{Luisoni:2013kna}, by a factor:
\begin{eqnarray}
  \mathcal{W}\left(\PhiHWsimp,\, p_{{\scriptscriptstyle
      \mathrm{T}}}\right)&=&h\left(\pt\right)\,\frac{\smallint
    d\sigma^{{\scriptscriptstyle
        \mathrm{NNLO\phantom{i}}}}\,\delta\left(\PhiHWsimp-\PhiHWsimp\left(\Phi\right)\right)-\smallint
    d\sigma_{B}^{{\scriptscriptstyle
        \mathrm{MINLO}}}\,\delta\left(\PhiHWsimp-\PhiHWsimp\left(\Phi\right)\right)}{\smallint
    d\sigma_{A}^{{\scriptscriptstyle
        \mathrm{MINLO}}}\,\delta\left(\PhiHWsimp-\PhiHWsimp\left(\Phi\right)\right)}\nonumber\\ &+&\left(1-h\left(\pt\right)\right)\,,\label{eq:W}
\end{eqnarray}
where $d\sigma^{\mathrm{NNLO}}$ and $d\sigma_{A/B}^{\mathrm{MINLO}}$
are multi-differential distributions obtained at pure NNLO level and
by analysing produced \HWJMINLO{} events, respectively. The function
$h(p_{\scriptscriptstyle\mathrm{T}})$ is defined as
\begin{equation}
h(p_\mathrm{T}) = \frac{(M_{H}+M_W)^2}{(M_{H}+M_W)^2+p_\mathrm{T}^2}\,, \label{eq:h_pt}
\end{equation}
where $p_{\scriptscriptstyle\mathrm{T}}$ is the transverse momentum of
the leading jet, and it is used to split the \MINLO{} cross section
into
\begin{equation} d\sigma_A^{\scriptscriptstyle\mathrm{MINLO}} =d
\sigma^{\scriptscriptstyle\mathrm{MINLO}}\,
h(p_{\scriptscriptstyle\mathrm{T}})\,,\qquad
d\sigma_B^{\scriptscriptstyle\mathrm{MINLO}} =
d\sigma^{\scriptscriptstyle\mathrm{MINLO}}\,
(1-h(p_{\scriptscriptstyle\mathrm{T}})\,.
\end{equation}
Therefore the function $h(p_{\scriptscriptstyle\mathrm{T}})$ ensures that the
reweighting is smoothly turned off when the leading jet is hard since
in that region the \HWJMINLO{} generator is already NLO accurate, as
is the NNLO calculation of HW.

For the process in eq.~(\ref{eq:def-process}) the Born kinematics is
fully specified by 6 independent variables. We have chosen them to be:
the transverse momentum of Higgs boson
($p_{\scriptscriptstyle\mathrm{T,H}}$); the rapidity of \HW{} system
($y_{\scriptscriptstyle HW}$); the difference of Higgs boson rapidity and
the $W^{+}$ rapidity ($\Delta y_{\scriptscriptstyle HW}$); the
invariant mass of $e^{+}\nu_{e}$ system $(m_{e\nu})$; and the two
Collins-Soper angles $(\theta^{*},\phi^{*})$ \cite{Collins:1977iv}:
\begin{equation}
  \Phi_{B} = \lbrace p_{\scriptscriptstyle\mathrm{T,H}}, y_{\scriptscriptstyle HW}, \Delta y_{\scriptscriptstyle HW}, m_{e\nu},
      \theta^{*}, \phi^{*}  \rbrace\,.
\end{equation}
In this setup the multi-differential cross-section can be written in the form:
\begin{eqnarray}
\label{eq:sigma}
\frac{d\sigma}{d\Phi_B}  &=&
\frac{d^6\sigma}{dp_{\scriptscriptstyle\mathrm{t},H}\,dy_{\scriptscriptstyle HW}\, d\Delta y_{\scriptscriptstyle HW}\, dm_{e\nu}\, d\thetacs d\phics} \nonumber \\
&=&  \frac{3}{16\pi}  \left (
\frac{d \sigma}{d\PhiHW}(1+\cos^2\thetacs) + \sum_{i=0}^{7} A_i(\PhiHW ) f_i(\thetacs, \phics)
\right)\,,
\end{eqnarray}
where $\PhiHW =\lbrace p_{\scriptscriptstyle\mathrm{T,H}}, y_{\scriptscriptstyle HW}, \Delta y_{\scriptscriptstyle HW},
m_{e\nu}\rbrace$, and the angular dependence is encoded in
the coefficients $A_{i}(\PhiHW)$ and the functions:
\begin{eqnarray}
\label{eq:f}
f_0(\thetacs,\phics) = \left(1-3\cos^2\thetacs\right)/2\,,\qquad &
f_1(\thetacs,\phics) = \sin2\thetacs \cos\phics\,, \nonumber\\
f_2(\thetacs,\phics) = (\sin^2\thetacs \cos2\phics)/2\,, \qquad&
f_3(\thetacs,\phics) = \sin\thetacs \cos\phics\,, \nonumber\\
f_4(\thetacs,\phics) = \cos\thetacs\,, \qquad &
f_5(\thetacs,\phics) = \sin\thetacs \sin \phics\,, \nonumber\\
f_6(\thetacs,\phics) = \sin 2\thetacs \sin \phics\,, \qquad&
f_7(\thetacs,\phics) = \sin^2\thetacs \sin 2\phics\,.
\end{eqnarray}
Since the angular dependence is fully expressed in terms of the
$f_i(\thetacs,\phics)$ functions, the coefficients of the expansion
$A_i(\PhiHW)$ depend only on the remaining kinematic variables. Using
orthogonality properties of spherical harmonics we can extract these
coefficients.

In our work we have simplified our procedure by noting that the
$m_{e\nu}$ invariant mass distribution has a flat K-factor. This is
true even when examining the $d\sigma/d m_{e\nu}$ distribution in
different bins of $\PhiHWsimp = \{p_{\scriptscriptstyle \mathrm{T},H},
y_{\scriptscriptstyle HW}, \Delta y_{\scriptscriptstyle HW}
\}$. Therefore, in eq.~(\ref{eq:sigma}) we replace the 4-dimensional
$\PhiHW$ with the 3-dimensional $\PhiHWsimp$. This is an
approximation, however we believe that it works extremely well as
discussed in ref.~\cite{Astill:2016hpa}. In our work we obtain
$\frac{d\sigma}{d\PhiHWsimp}$ and $A_i(\PhiHWsimp)\:(i=0,7)$ at pure
NNLO level by running the \HVNNLO{} code
\cite{Ferrera:2011bk,Ferrera:2013yga}, and we obtain the
results at \MINLO{} level by running \HWJMINLO{}
\cite{Luisoni:2013kna}. We store the results in 9 three-dimensional
tables. Following this step, we use these tables along with
eq.~(\ref{eq:sigma}) to obtain the function eq.~(\ref{eq:W}) to
reweight each produced event. The final ensemble of events is NNLO
accurate for all observables at Born level and a parton shower can now
be applied without affecting the NNLO accuracy.

In the following we show results for 13 TeV LHC collisions applying the
lepton cuts reported in Eq.~\eqref{eq:VH_cuts2}.
Jets have been clustered using the anti-$k_t$
algorithm with $R=0.4$~\cite{Cacciari:2008gp} as implemented in
\FASTJET{}~\cite{Cacciari:2005hq,Cacciari:2011ma} and count if they fulfil
the following conditions:
\begin{equation}
p_{\mathrm{T}}(\mbox{\rm jet}) > 20\UGeV, \qquad
|\eta(\mbox{\rm jet})| < 4.5 \,.
\end{equation}
As for the PDF, we have used the MMHT2014nnlo68cl
set~\cite{Harland-Lang:2014zoa}, corresponding to a value of
$\alpha_s(M_{\scriptscriptstyle Z}) = 0.118$.
For \HWJMINLO{} events the scale choice is dictated by the \MINLO{}
procedure, while for the NNLO we have used for the central renormalization
and factorization scales $\mu_0 = M_{\scriptscriptstyle H}+M_{\scriptscriptstyle W}$.
To estimate uncertainties we
calculate both the fixed order NNLO and \HWJMINLO{} results at 7
scales, each with renormalization and factorization scale varied
independently up and down by a factor of 2. When these results are
then used in eq.~(\ref{eq:W}) this gives 49 combinations for the
\NNLOPS{} results. We define our perturbative uncertainty as the envelope
of these 49 variations.

To shower partonic events, we have used
\PYTHIA{8}~\cite{Sjostrand:2007gs} (version 8.185) with the ``Monash
2013'' \cite{Skands:2014pea} tune. We consider events after parton
showering and hadronization effects, unless otherwise
stated. Underlying event and multiple parton interactions were kept
switched off. To define leptons from the boson decays we use the Monte 
Carlo truth, \emph{i.e.}~we assume that if other leptons are present,
the ones coming from the $W$ decay can be identified correctly.  To
obtain the results shown in the following, we have switched on the
``doublefsr'' option introduced in ref.~\cite{Nason:2013uba}.
The plots shown throughout this study have been obtained keeping the
veto scale equal to the default \POWHEG{} prescription.
In some figures we also compare our results against \HVNNLO{}, run
with $\mu_0 = M_{\scriptscriptstyle H}+M_{\scriptscriptstyle W}$ as
central scale choice, and with the same PDF set used for \HWJMINLO{}
and \HVNNLOPS{}.

\begin{figure}
  \centering
  \includegraphics[width=0.47\textwidth,page=3]{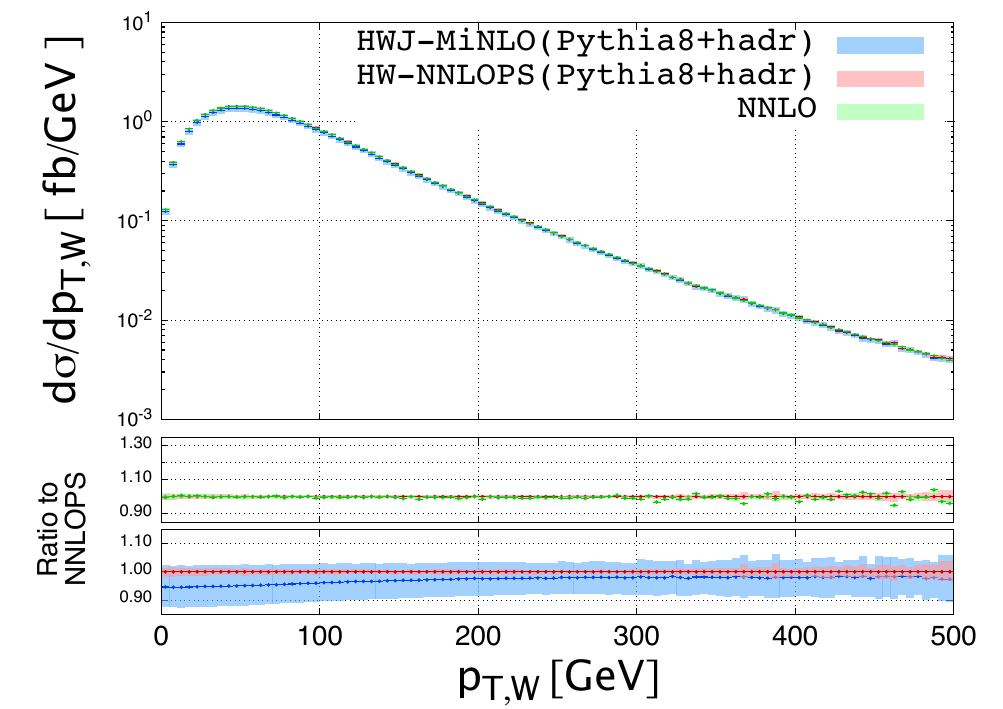}
  \includegraphics[width=0.47\textwidth,page=2]{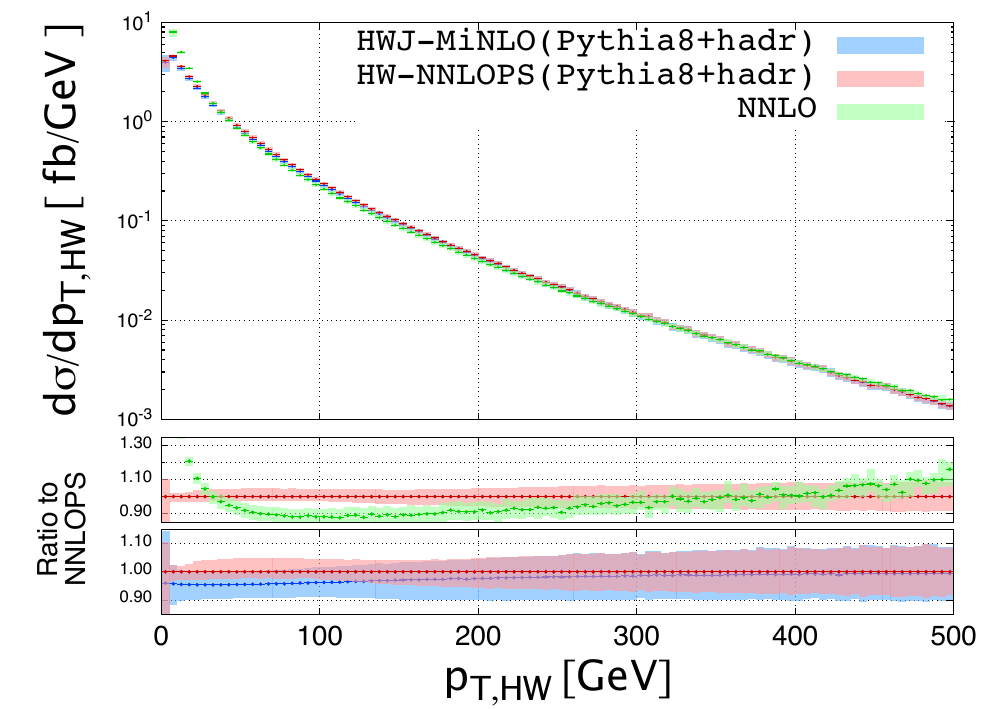}
  \caption{Comparison of \HWJMINLOPS{} (blue), NNLO (green), and \HWNNLOPS{} (red) for $\ptw$ (left) and $\ptwh$ (right). }
  \label{fig:extra_ptw_ptwh}
\end{figure}

In \refF{fig:extra_ptw_ptwh} we show distributions for the transverse momenta
of the $W$ boson and the $WH$ system, respectively.
NNLO results (from \HVNNLO{}) are compared against those obtained with
\HWJMINLO{} and \HVNNLOPS{}. For observables that are fully inclusive
over QCD radiation, as $\ptw$, the agreement among the \HVNNLO{} and
\NNLOPS{} predictions is perfect, as expected. One also notices the
sizeable reduction of the uncertainty band when \HWJMINLO{} results
are upgraded to \NNLOPS{}. As no particularly tight cuts are imposed,
the NNLO/NLO K-factor is almost exactly flat.  The right panel shows
instead the effects due to the Sudakov resummation. At small
transverse momenta, the NNLO cross section becomes larger and larger
due to the singular behaviour of the matrix elements for $HW$
production in association with arbitrarily soft-collinear
emissions. The \MINLO{} method resums the logarithms associated to
these emissions, thereby producing the typical Sudakov peak, which for
this process is located at $1$ GeV $ \lesssim \ptwh \lesssim 4$ GeV,
as expected from the fact that the LO process is Drell-Yan like.  It
is also interesting to notice here two other features that occur away
from the collinear singularity, and which are useful to understand
plots to be shown in the following. Firstly, the $\pt$-dependence of
the NNLO reweighting can be explicitly seen in the bottom panel, where
one can also appreciate that at very large values not only the
\NNLOPS{} and \MINLO{} results approach each other, but also that the
uncertainty band of \HVNNLOPS{} becomes progressively larger (in fact,
in this region, the nominal accuracy is NLO). Secondly, in the region
$50$ GeV $\lesssim \ptwh \lesssim 300$ GeV, the NNLO and \NNLOPS{}
lines show deviations of up to about 10 \%: these are due both to the
compensation that needs taking place in order for the two results to
integrate to the same total cross section, as well as to the fact that
the scale choices are different (fixed for the NNLO line, dynamic and
set to $\ptwh$ in \MINLO{}). When $\ptwh\gtrsim 250$ GeV the two
predictions start to approach, as this is the region of phase space
where the \MINLO{} scale is similar to that used at NNLO ($\mu =
M_{\scriptscriptstyle H}+M_{\scriptscriptstyle W}$).

\begin{figure}
  \centering
  \includegraphics[width=0.47\textwidth,page=10]{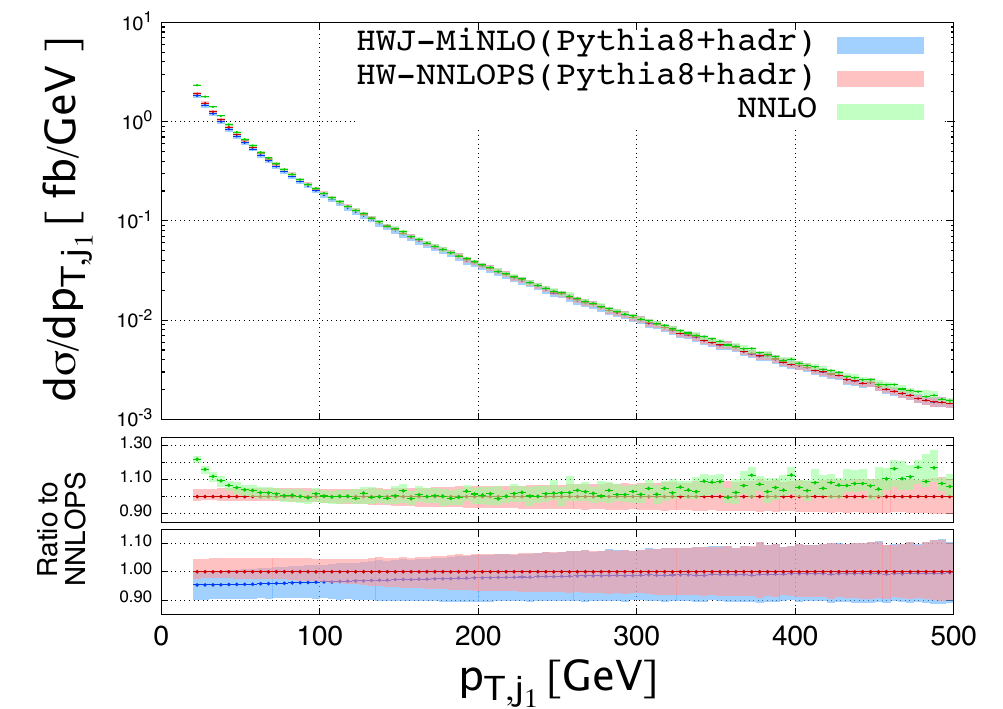}
  \includegraphics[width=0.47\textwidth,page=11]{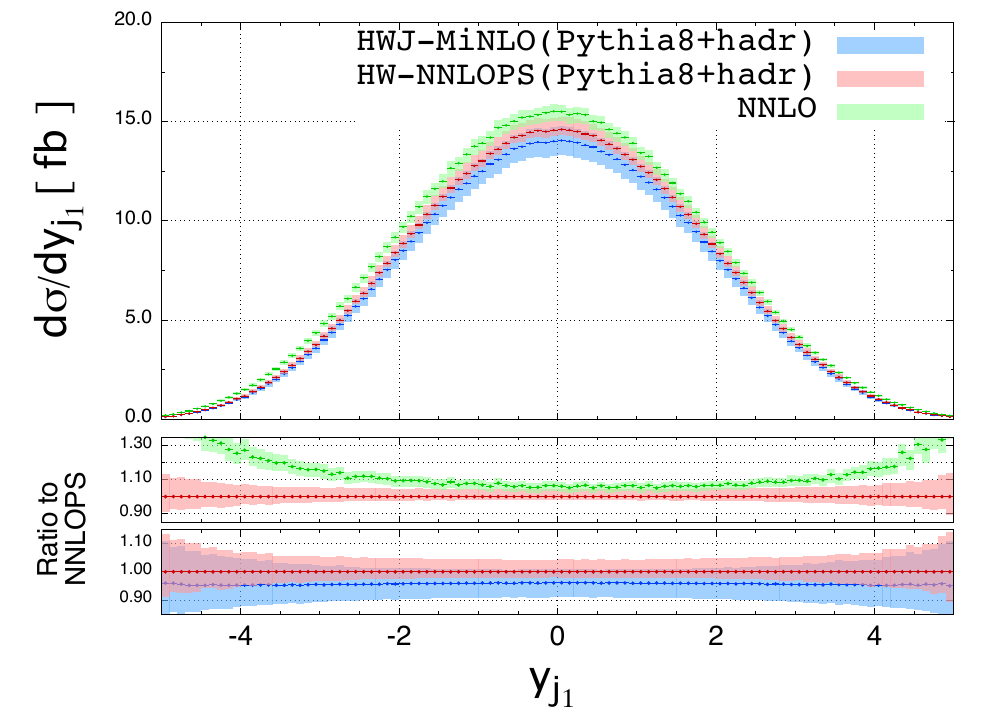}
  \caption{Comparison of \HWJMINLOPS{} (blue), NNLO (green), and \HWNNLOPS{} (red) for $\ptjone$ (left) and $\yjone$ (right).}
  \label{fig:extra_ptj1_yj1}
\end{figure}

In \refF{fig:extra_ptj1_yj1} we show the transverse momentum and the
rapidity of the hardest jet.
Most of the differences among these three predictions can be easily
explained by the considerations made above on the $\ptwh$ spectrum,
although here effects due to multiple radiation as well as
hadronization are bound to play some role too.  In
\refF{fig:extra_ptj1_yj1} we notice that, for large values of
$|y_{j_1}|$, there are large differences among the NNLO result and
those containing Sudakov resummation: this is expected, since a
large-rapidity jet has on average a smaller transverse momentum, hence
the singular nature of the NNLO result is more evident in these
kinematics configurations.

Next we find it interesting to examine the size of
non-perturbative effects.
\begin{figure}
  \centering
  \includegraphics[width=0.47\textwidth,page=1]{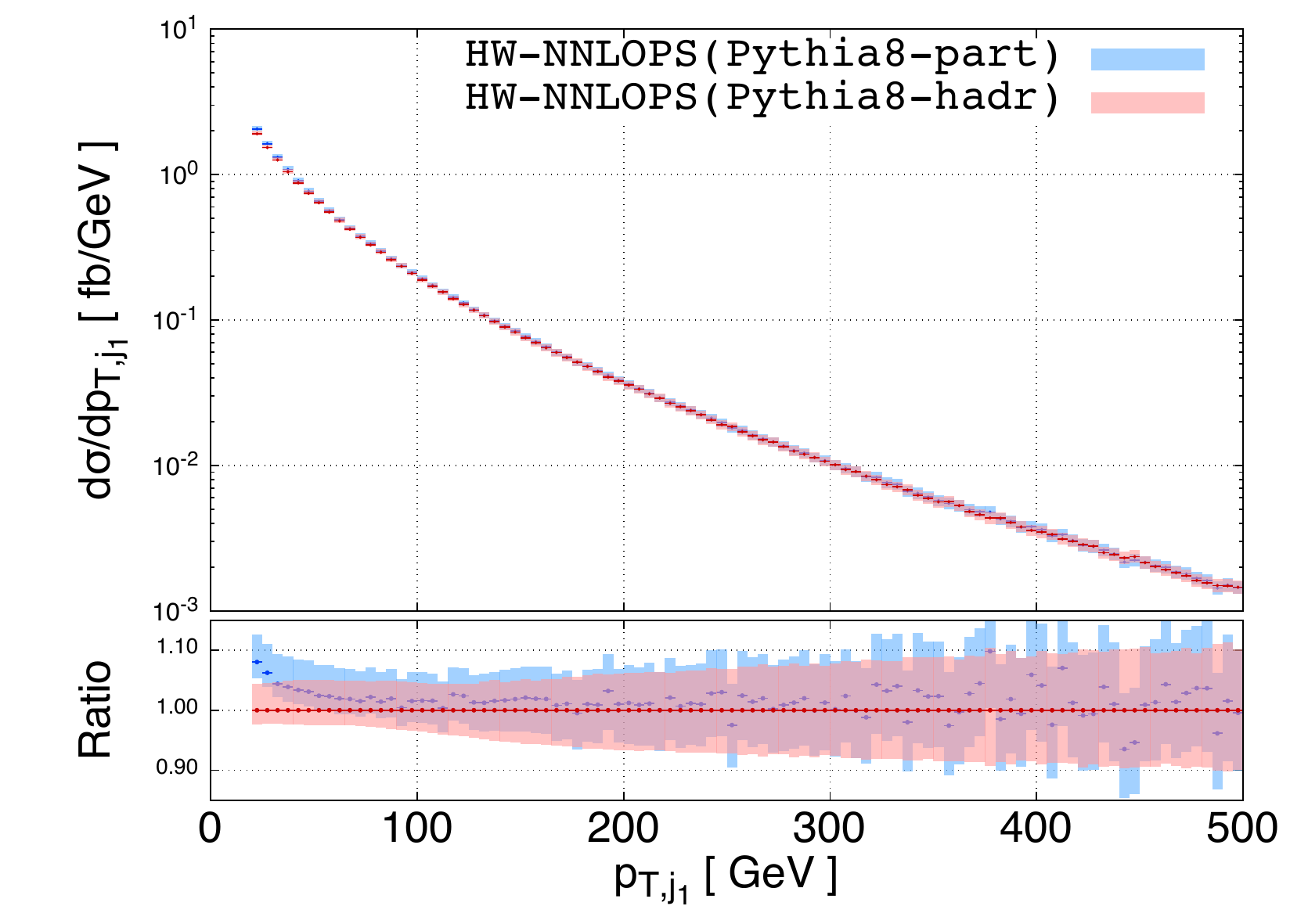}
  \includegraphics[width=0.47\textwidth,page=2]{./WG1/VBFplusVH/plots/final-all-pheno-plots-2.pdf}
  \caption{Comparison of \HWNNLOPSshort{} with (red) and without (blue) hadronization for $\ptjone$ (left) and $\yjone$ (right).}
  \label{fig:extra_nohad_ptj1_yj1}
\end{figure}
As shown in \refF{fig:extra_nohad_ptj1_yj1}, hadronization has a
sizeable impact on the shapes of jet distributions: differences up to
$7\hspace{-0.05cm}-\hspace{-0.05cm}8~\%$ can be seen in the jet $\pt$
spectrum at small values, and are still visible at a few per cent
level till when relatively hard jets are required ($\ptjone > 100$
GeV). Even larger effects can be seen in the rapidity distribution (right panel) at large rapidities.
The \HVNNLOPS{} generator allows us to simulate these features in a
fully-exclusive way, retaining at the same time all the virtues of an
NNLO computation for fully inclusive observables, as well as
resummation effects, thanks to the interplay among \POWHEG{}, \MINLO{}
and parton showering.

\begin{figure}
  \centering
  \includegraphics[width=0.47\textwidth,page=1]{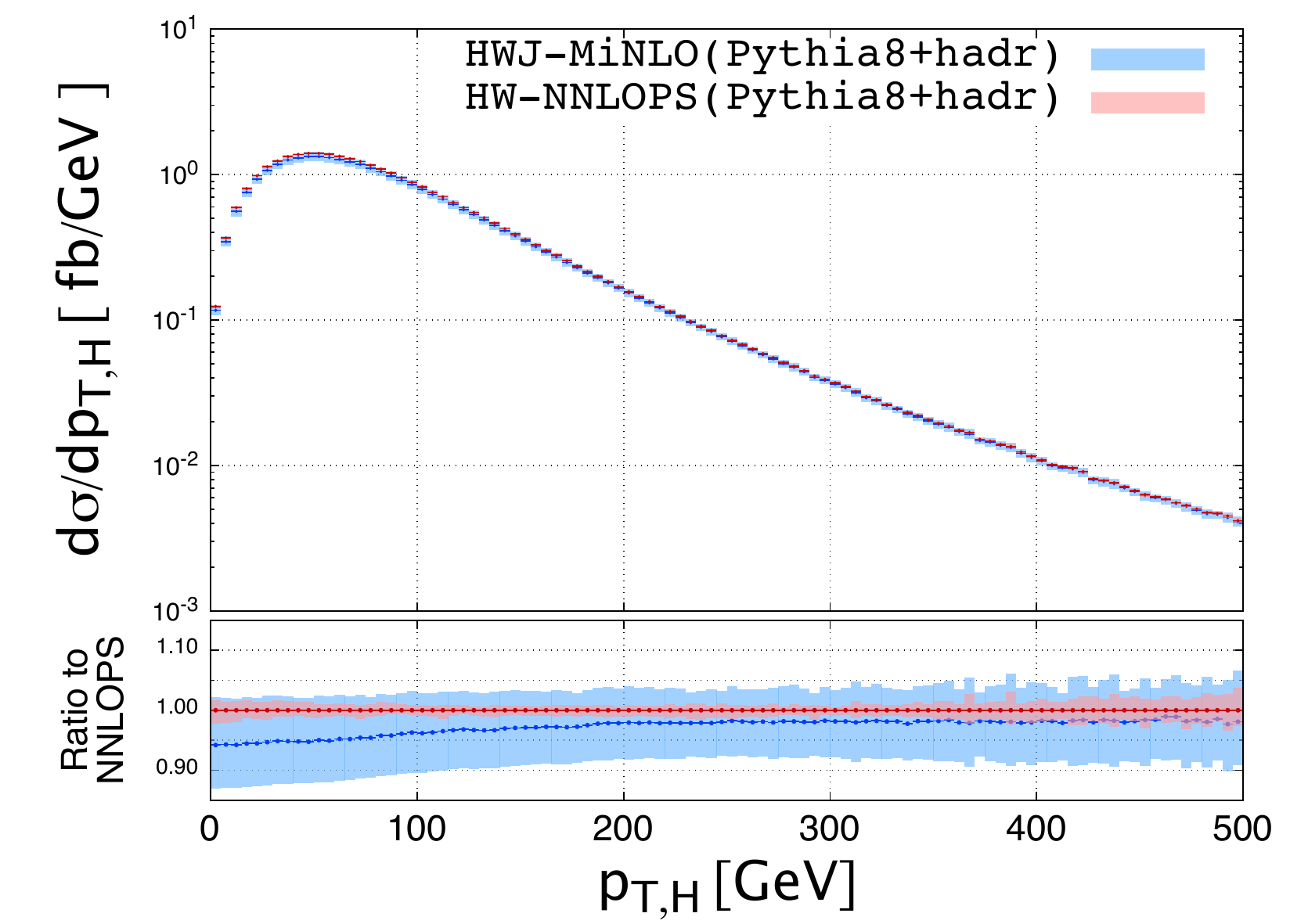}
  \includegraphics[width=0.47\textwidth,page=2]{./WG1/VBFplusVH/plots/final-all-yr4.pdf}\\
  \includegraphics[width=0.47\textwidth,page=4]{./WG1/VBFplusVH/plots/final-all-yr4.pdf}
  \includegraphics[width=0.47\textwidth,page=3]{./WG1/VBFplusVH/plots/final-all-yr4.pdf}
  \caption{Comparison of \HWJMINLOPS{} (blue) and \HWNNLOPS{} (red) for $\pt$ (left) and rapidity (right) for Higgs (upper) and lepton (lower).}
  \label{fig:yr4_pth_yh_ptlep_ylep}
\end{figure}

\begin{figure}
  \centering
  \includegraphics[width=0.47\textwidth,page=5]{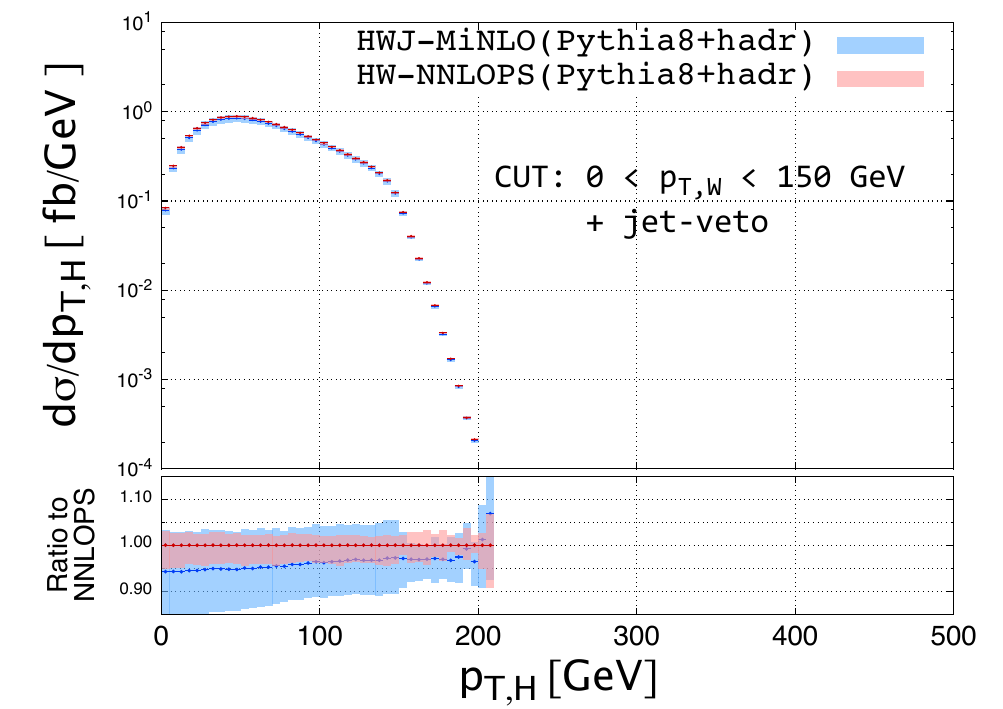}
  \includegraphics[width=0.47\textwidth,page=6]{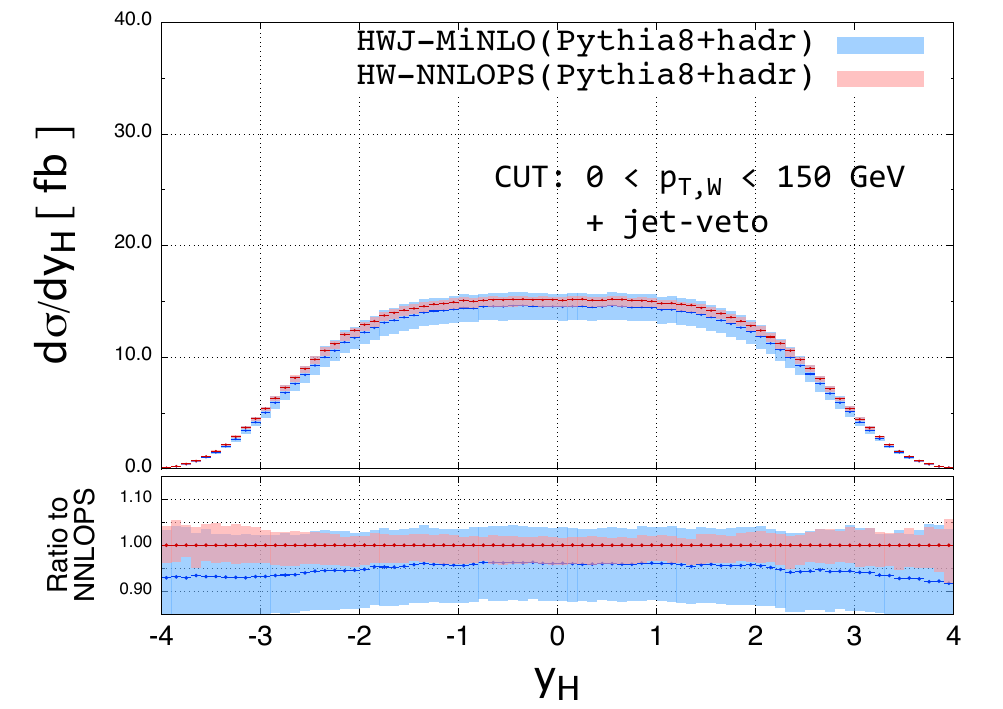}
  \caption{Comparison of \HWJMINLOPS{} (blue) and \HWNNLOPS{} (red) for $\ptH$ (left) and $\yh$ (right) for $\ptw<150$ GeV and no jet.}
  \label{fig:yr4_0j_w1_pth_yh}
\end{figure}

\begin{figure}
  \centering
  \includegraphics[width=0.47\textwidth,page=7]{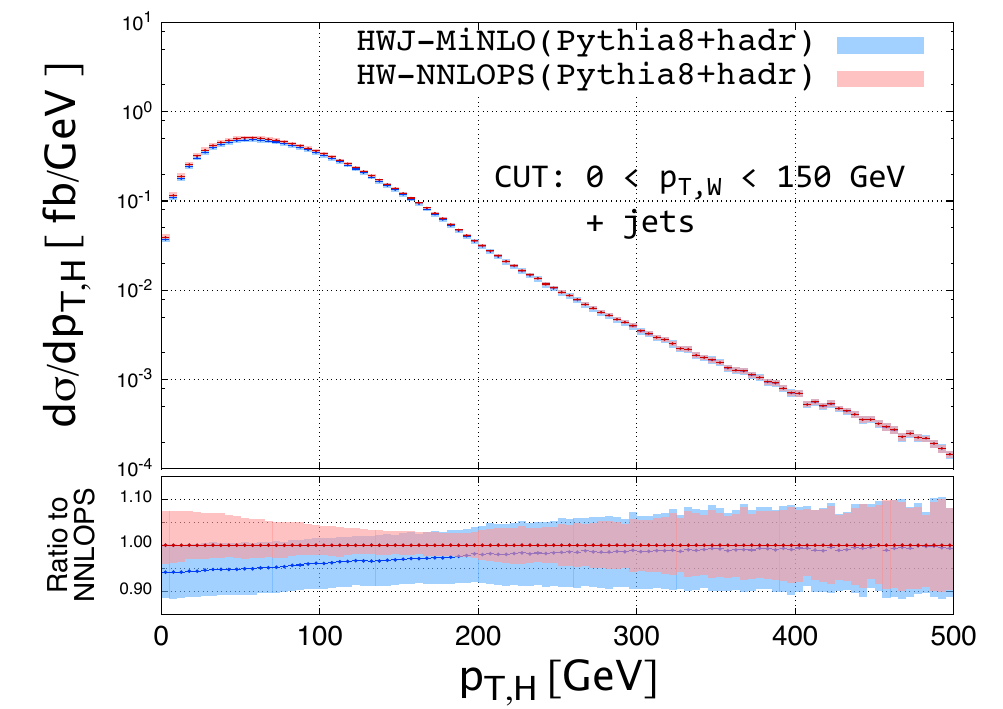}
  \includegraphics[width=0.47\textwidth,page=8]{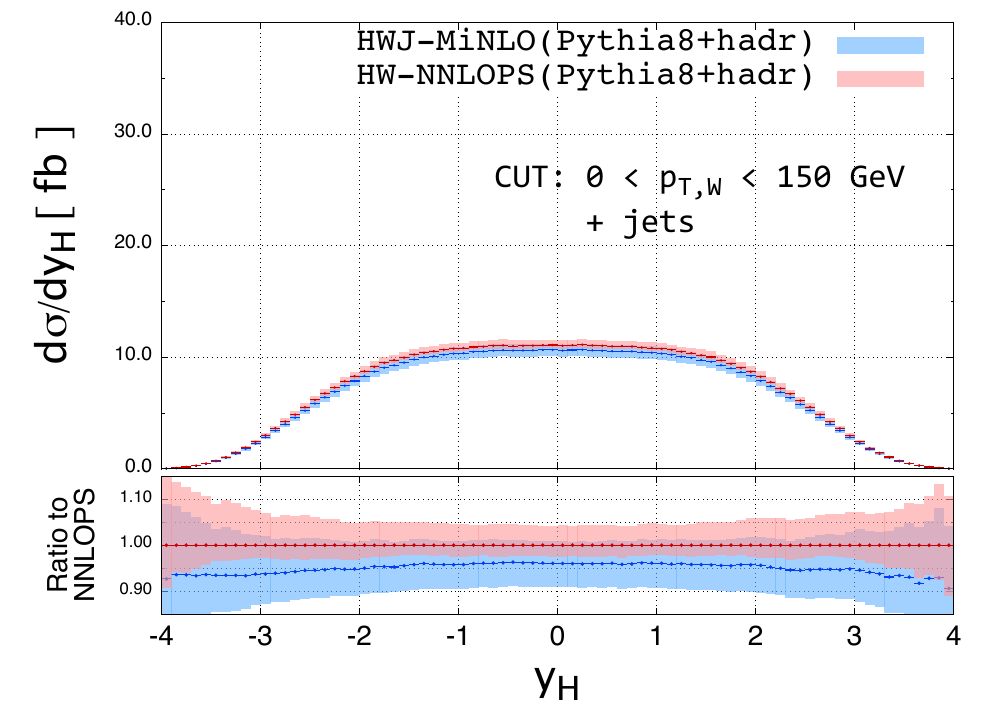}
  \caption{Comparison of \HWJMINLOPS{} (blue) and \HWNNLOPS{} (red) for $\ptH$ (left) and $\yh$ (right) for $\ptw<150$ GeV and at least 1 jet.}
  \label{fig:yr4_1j_w1_pth_yh}
\end{figure}

\begin{figure}
  \centering
  \includegraphics[width=0.47\textwidth,page=9]{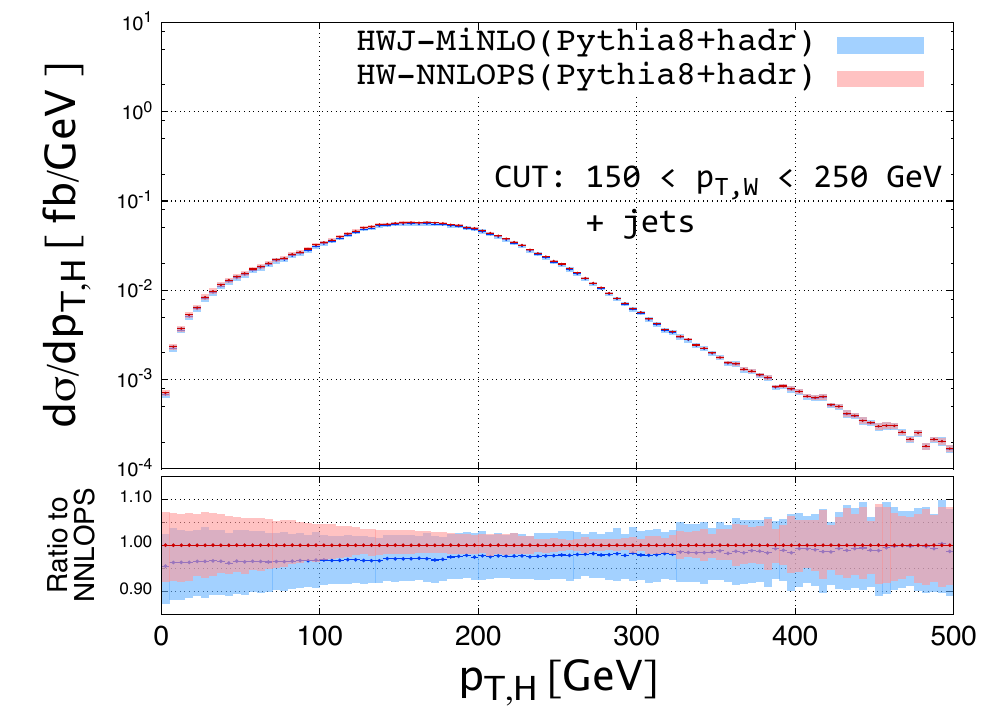}
  \includegraphics[width=0.47\textwidth,page=11]{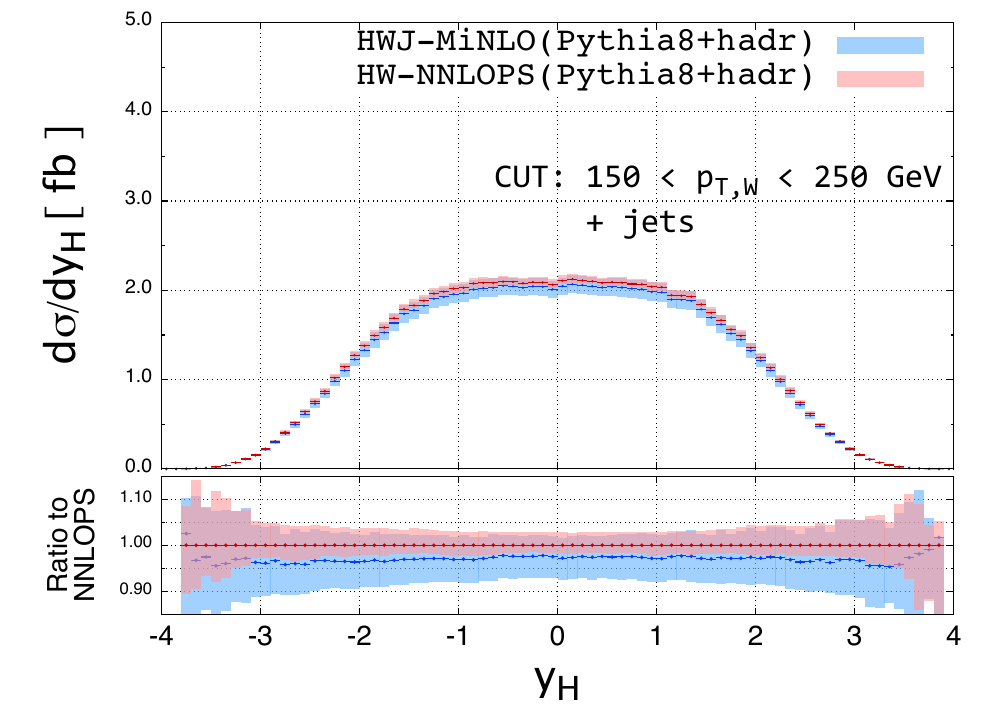}
  \caption{Comparison of \HWJMINLOPS{} (blue) and \HWNNLOPS{} (red) for $\ptH$ (left) and $\yh$ (right) for $150\mbox{ GeV}<\ptw<250$ GeV and at least 1 jet}
  \label{fig:yr4_1j_w2_pth_yh}
\end{figure}

In \refF{fig:yr4_pth_yh_ptlep_ylep} we show the transverse
momentum and rapidity distributions of the Higgs boson and the charged
lepton, as predicted by the \HVNNLOPS{} code and by the underlying
\HWJMINLO{} simulation.
No particular feature needs be commented in these plots: since no cuts
are applied on extra radiation, the inclusion of higher order
corrections just makes the \HVNNLO{} predictions more accurate, as
expected. On the other hand it is interesting to see how these
distributions are affected by requiring further cuts, like imposing a
jet veto or requiring the presence of at least one jet, whilst restricting at
the same time the phase space to different windows for
$\ptw$. Figs.~\ref{fig:yr4_0j_w1_pth_yh}, \ref{fig:yr4_1j_w1_pth_yh}
and~\ref{fig:yr4_1j_w2_pth_yh} display the Higgs boson transverse
momentum and rapidity in the three following cases:
\begin{itemize}
\item no jet (``jet veto''), $\ptw<150$ GeV
\item at least 1 jet, $\ptw<150$ GeV
\item at least 1 jet, $150\mbox{ GeV}<\ptw<250$ GeV
\end{itemize}
The first thing to notice is that, in general, the uncertainty band of
the \NNLOPS{}-accurate prediction is not as narrow as in
\refF{fig:yr4_pth_yh_ptlep_ylep}: this is expected and physically
sound, because the phase space is not fully inclusive with respect to
the QCD activity, due to the requirements on jets. In the jet-veto
case, however, the results show that the inclusion of NNLO corrections
within a \MINLO{}-based simulation is important, since the uncertainty
band of \HVNNLOPS{}, although larger than in
\refF{fig:yr4_pth_yh_ptlep_ylep}, is still narrower than the
\HWJMINLO{} one.

The second thing to notice is that, when jets are required, the
\HVNNLOPS{} predictions display larger uncertainties, a bit smaller
but in general similar to those obtained with \HWJMINLO{}. This is
expected, since this is exactly the phase space region where both
computations are formally NLO accurate. The effect of the NNLO/NLO
reweighting is still quite visible (both in the overall normalization
and in the slightly smaller bands) though, due to the fact that the
cut on the jet transverse momentum is relatively small. This also
means that the \HWJMINLO{} and \HVNNLOPS{} results are likely to be
different from fixed order computations, since the use of dynamic
scales in \MINLO{} and its interplay with resummation has an impact in
this phase space region, as shown in \refF{fig:extra_ptj1_yj1} for
the associated jet distributions.

The final thing to notice, and the one exception to the general trend
in the previous observations, is the shrinking of the uncertainty band
at intermediate values of $\ptH$ in Figs.~\ref{fig:yr4_1j_w1_pth_yh}
and~\ref{fig:yr4_1j_w2_pth_yh}, which is even more noticeable in the
$y_H$ distributions, the latter being dominated by the kinematics
where $\ptH$ peaks. This feature is due to the requirement on $\ptw$,
and can be explained as follows. For a fully inclusive kinematics, the
transverse momenta of the $W$ and $H$ boson are typically balanced,
with a value of about $40$ GeV (see \emph{e.g.} the peak in
Figs.~\ref{fig:extra_ptw_ptwh}
and~\ref{fig:yr4_pth_yh_ptlep_ylep}). When jets are required, at least
the hardest jet $\pt$ will play a role in the momentum conservation in
the transverse plane: its typical value, however, depends on the
requirements on the massive bosons kinematics. From this observation
the band shrinking in the $\ptH$ spectrum can be understood. For
instance, in \refF{fig:yr4_1j_w1_pth_yh}, when $\ptH$ approaches
values close to the larger values available for $\ptw$, one enters a
region where the jet has to be just above its minimum allowed value:
this is the region where the uncertainty band in the jet $\pt$
spectrum is minimal, as shown in \refF{fig:extra_ptj1_yj1}. As
soon as larger $\ptH$ values are probed whilst keeping $\ptw<150$ GeV,
harder jets are required by momentum conservation, hence the
uncertainty band from \HVNNLOPS{} rapidly approaches the one from
\HWJMINLO{}.  This effect is even more evident in
\refF{fig:yr4_1j_w2_pth_yh}: if $\ptH$ is relatively small, then
momentum conservation doesn't constrain $\ptjone$ very strongly,
yielding a standard uncertainty band, relatively similar to
\HWJMINLO{}. In the region where cuts push $\ptw$ and $\ptH$ to
similar values, once more the jet must be close to its threshold
region, and hence the uncertainty band is reduced.


\chapter{ttH and tH}
\label{chap:ttH}
\ChapterAuthor{S.~Guindon, C.~Neu, S.~Pozzorini, L.~Reina~(Eds.);
A.~Broggio, M.~Casolino, F.~Demartin, A.~Denner, R.~Feger, A.~Ferroglia, R.~Frederix, S.~Frixione, M.~V.~Garzelli, 
S.~Gieseke, M.~Harrendorf, H.~B.~Hartanto, V.~Hirschi, S.~H\"oche, S.~Honeywell, B.~J\"ager, A.~Juste, A.~Kardos, 
A.~Kulesza, J.~Lindert, M.~Mantoani, K.~Mawatari, M.~Moreno Llacer,  N.~Moretti, L.~Motyka, D.~Pagani, 
B.~D.~Pecjak, S.~Pl\"atzer, R.~Podskubka, C.~Reuschle, E.~Shabalina, H.~Shao, M.~Sch\"onherr, F.~Siegert, A.~Signer, 
T.~Stebel, V.~Theeuwes, Z.~Trocsanyi, I.~Tsinikos, D.~Wackeroth,  L.~L.~Yang, M.~Zaro}
\renewcommand{\ttbar}{t\bar{t}}
\renewcommand{\bbbar}{b\bar{b}}
\providecommand{\ttbb}{t\bar{t}b\bar{b}}
\renewcommand{\tth}{t\bar{t} H}
\renewcommand{\al}{\alpha}
\providecommand{\als}{\alpha_{\mathrm{s}}}
\renewcommand{\shat}{\hat s}
\newcommand{\sigh}{\hat \sigma}
\renewcommand{\si}{\sigma}
\renewcommand{\nn}{\nonumber}
\newcommand{\tosv}{{\scriptscriptstyle \to}}
\newcommand{\CF}{C_{\mathrm{F}}}
\newcommand{\CA}{C_{\mathrm{A}}}

\providecommand{\deltaEW}{\delta_{\mathrm{EW}}}
\providecommand{\sigmaLOqcd}{\sigma^{\mathrm{LO}}_{\mathrm{QCD}}}
\providecommand{\sigmaNLOqcd}{\sigma^{\mathrm{NLO}}_{\mathrm{QCD}}}
\providecommand{\sigmaNLOqcdew}{\sigma^{\mathrm{NLO}}_{\mathrm{QCD+EW}}}
\providecommand{\dsigmaew}{\delta\sigma_{\mathrm{EW}}}
\providecommand{\ewpdf}{\mathrm{EW, PDF}}

\newcommand{\thxs}[8]{$#1$ &$#2$ & $#3$ & $#4$ & #5 & #6 & $#7$ & $#8$ \\}
\newcommand{\tthxs}[8]{$#1$ &$#2$ & $#3$ &$#4$ & #5 & #6 & $#7$ & $#8$ \\}
\newcommand{\var}[2]{$+#1$\;\;$-#2$}

\newcommand{\exthxs}[8]{$#1$ &$#2$ & $#3$ & $#4$ & #5 & #6 & $#7$ & $#8$ \\}
\newcommand{\extthxs}[8]{$#1$ &$#2$ & $#3$ &$#4$ & #5 & #6 & $#7$ & #8  \\}
\newcommand{\varvar}[6]{$\pm#1$  & $\pm#3$ & $\pm#5$}


\def\ttHpath{WG1/ttH}

\def\ttHintropath{\ttHpath/ttH-introduction}
\section{Introduction}
\label{sec:ttH-introduction}

The production of a Higgs boson in association with a top-quark pair ($\tth$) or a single top quark ($tH$) is going to play a very important role in the
Higgs boson physics program of Run~2 of the LHC since it can provide
a direct measurement of the top-quark Yukawa coupling.

In this context, the $\tth/tH$ working group has discussed the
current status and future plans of $\tth$ and $tH$ experimental
analyses and has reviewed the status of theoretical predictions, for
both signals and backgrounds. The emphasis has been on identifying and
characterizing state-of-the art theoretical predictions and tools for
signals and backgrounds in all relevant $\tth$ and $tH$ searches,
as well as on identifying various theory-related sources of
uncertainties and prioritizing them according to both their impact on
experimental analyses and the likelihood of theory improvements in the
near future.

On top of providing tables of cross sections for both $\tth$ and
$tH$ production, which include all most up-to-date calculations of QCD
and electroweak (EW) corrections, as well as the estimated theoretical
uncertainty from scale variations, $\als$, and parton distribution
functions (PDFs), we review in \Sref{sec:ttH-XS} the
next-to-leading-order (NLO) QCD+EW predictions for $\tth$
inclusive production, and in \Sref{sec:tH_NLO_QCD} the NLO QCD
predictions for $tH$ production. Recent developments in improving the
prediction for $\tth$ production by either including off-shell
effects or subsets of next-to-next-to-leading-order (NNLO) QCD
corrections are discussed in \Srefs{sec:ttH-off-shell} and~\ref{sec:ttH-resummation}, respectively.
Dedicated studies of some of the main background processes are
presented in \Sref{se:ttH-ttV} ($t\bar{t}V$ and $t\bar{t} VV'$ with $V,V'=W^\pm,Z$)
and \Sref{sec:ttH-ttbb} ($t\bar{t}b\bar{b}$).

One major activity of this working group has been the
comparison and validation of state-of-the-art theoretical tools
available to calculate both signal and backgrounds including the
proper matching of fixed-order NLO QCD corrections and parton-shower evolution.
\Sref{sec:ttH-nlo-qcd+ps} presents such comparison for
$\tth$ production, while \Srefs{sec:ttH-ttV-tools} and~\ref{sec:ttH-ttbb} present
analogous studies for the $t\bar{t}V$ ($V=W^\pm,Z$) and $t\bar{t}b\bar{b}$ backgrounds.

\label{se:tth_tthxs}

\def\ttHNLOpath{\ttHpath/ttH-NLO-QCD+EW}
\def\tthxstabhead{
\midrule
$\sqrt{s}$&
$\mh$ &
$\sigmaNLOqcd$&
$\sigmaNLOqcdew$ &
$K_{\mathrm{QCD}}$ &
$\deltaEW$[\%] &
Scale[\%] &
$\als$[\%] &
PDF[\%] &
PDF+$\als$[\%]
\\   \midrule}

\section{NLO QCD+EW predictions for \texorpdfstring{$t\bar{t}H$}{ttH} production}
\label{sec:ttH-XS}

Predictions for inclusive and differential $\tth$ production at NLO
QCD are available since more than a
decade~\cite{Reina:2001sf,Dawson:2002tg,Dawson:2003zu,Beenakker:2001rj,Beenakker:2002nc},
while EW corrections have been calculated only
recently~\cite{Yu:2014cka,Frixione:2014qaa,Frixione:2015zaa}. Although
their effect on total rates is usually suppressed with respect to NLO
QCD corrections by a factor of order $\alpha/\als$, when hard
scales are probed they can be enhanced by electroweak Sudakov
logarithms~\cite{Ciafaloni:1998xg,Ciafaloni:2000df,Denner:2000jv,Denner:2001gw}.
For what concerns $\tth$ production, in particular for a precise
extraction of the top quark Yukawa coupling $y_{\mathrm{t}}$, EW corrections
should be accounted for because of at least two reasons. First, EW
corrections, unlike QCD corrections, spoil the trivial dependence of
the total cross section on $\sim y_{\mathrm{t}}^2$ , introducing also (small)
terms where the Higgs couples to $W^\pm$ and $Z$ bosons, or to
itself. Second, EW corrections show Sudakov effects: in order to
suppress backgrounds, many $\tth$ searches are performed in a boosted
regime~\cite{Butterworth:2008iy,Plehn:2009rk,Buckley:2013auc}, where
Sudakov logarithms can be important.

This section presents NLO QCD+EW predictions for inclusive $\tth$
production.  All input parameters are chosen according
to~\cite{LHCHXSWG-INT-2015-006}, and the hadronic cross section is obtained
using the PDF4LHC15~\cite{Butterworth:2015oua} and
NNPDF2.3QED~\cite{Ball:2013hta} parton distribution functions as explained in
detail below.  For the top-quark and Higgs boson masses the on-shell
scheme is used, and the top-quark Yukawa
coupling\footnote{In the adopted convention the Feynman rule of
  the $t\bar{t}H$ vertex is $(-iy_{\mathrm{t}})$.} is related to the
top-quark mass and the Fermi constant ($G_\mu$) by
\begin{equation}
y_{\mathrm{t}}=(\sqrt{2}G_{\mu})^{1/2}\mt\,.
\end{equation}
The central value for renormalization and factorization scales is set to
\begin{equation}
            \mu = \mt + \mh/2\,,
\end{equation}
and the scale uncertainty is estimated by independent variations of
renormalization ($\muR$) and
factorization ($\muF$) scales in the range $ \mu/2 \le \muR, \muF \le
2\mu$, with $1/2 \le \muR/ \muF \le 2$.

The NLO QCD+EW predictions presented in the following,
\begin{equation}
\sigmaNLOqcdew=\sigmaNLOqcd+\dsigmaew\,,
\end{equation}
result from the combination of various contributions.  The usual NLO QCD cross section,
\begin{equation}
\sigmaNLOqcd=\sigmaLOqcd+\delta\sigmaNLOqcd,
\end{equation}
comprises LO terms of $\mathcal{O}(\als^2\alpha)$ and NLO terms of
$\mathcal{O}(\als^3\alpha)$, which involve $gg$, $q\bar q$, and $gq$
partonic channels.  The remaining EW corrections, denoted as
$\dsigmaew$, include three types of terms:
\begin{enumerate}
\item LO EW terms of $\mathcal{O}(\alpha^3)$ that result from squared
  EW tree amplitudes in the $q\bar q$ and $\gamma\gamma$ channels;
\item LO mixed terms of $\mathcal{O}(\als\alpha^2)$ that result
  from the interference of EW and QCD tree diagrams in the $b\bar b$
  and $\gamma g$ channels (other $q\bar q$ channels do not contribute
  at this order due to the vanishing interference of the related
  colour structures);
\item NLO EW corrections of $\mathcal{O}(\als^2\alpha^2)$ in the
  $q\bar q$, $gg$ and $\gamma g$ channels.  Subleading NLO terms of
  $\mathcal{O}(\als\alpha^3)$ and $\mathcal{O}(\alpha^4)$ are not
  included as they are expected to be strongly suppressed.
\end{enumerate}
For $\sqrt{s}=7$--$14\,\UTeV$ and $\mh=125\,\UGeV$, the corrections
resulting from LO EW, LO mixed and NLO EW effects are all positive and
amount, respectively, to $0.5\%$, $0.8$--$1.5\%$ and $1.1$--$1.9\%$ of
the NLO QCD cross section.\footnote{Here $\mathcal{O}(\alpha)$ effects
  related to QED evolution of PDF (see below) are not included.}
Photon-induced partonic channels dominate the LO mixed terms, while
their contribution to LO EW and NLO EW terms is almost negligible.

A fully consistent treatment of NLO QCD+EW corrections requires
corresponding precision in the employed PDF.  In particular, parton
distributions should include QED evolution effects and, consequently,
a photon density.  In order to circumvent the absence of QED effects
in the PDF4LHC15 distributions the following approach is adopted:
\begin{enumerate}
\item NLO QCD contributions are computed using the PDF4LHC15 set: more
  precisely, the PDF4LHC15 set with 30+2 members is used for PDF and
  $\als$ uncertainty estimates;
\item all EW correction effects resulting form partonic channels with
  initial-state quarks and/or gluons are computed with the same
  PDF4LHC15 set;
\item for all $\gamma$-induced EW correction effects the NNPDF2.3QED set is used;
\item the missing $\mathcal{O}(\alpha)$ effect due to the QED
  evolution of quark PDF is estimated from the difference between
  NNPDF2.3QED parton densities and their NLO QCD counterpart without
  QED evolution, NNPDF2.3, both with $\als(\mz)=0.118$. The
  relevant $\mathcal{O}(\alpha)$ correction factor is determined as
  follows by means of a LO QCD calculation,
\begin{equation}
\label{eq:tthpdfqed}
\delta\sigma_{\ewpdf}
= \sigmaNLOqcd \left( 1 + \delta_{\ewpdf} \right)\,,\qquad\mbox{where}\qquad
\delta_{\ewpdf} = \frac{\sigma_{\rm LO~QCD}^{\rm NNPDF\,QED}}{\sigma_{\rm LO~QCD}^{\rm NNPDF}} -1 \,.
\end{equation}
For $\sqrt{s}=7$--$14\,\UTeV$ and $\mh=125\,\UGeV$, the effect of QED
PDF evolution ranges from $-0.7\%$ to $-0.9\%$. Being negative it
compensates in part the effect of LO and NLO EW corrections to the
partonic cross sections.
\end{enumerate}

At the same order of $\sigma^{\rm NLO}_{\rm EW}$ also the real
emission of an extra heavy weak gauge boson can in principle
contribute to the cross section for inclusive $\tth$ production.  Such
a contribution from heavy-boson radiation (HBR) is generally not
considered as part of EW corrections, owing to the fact that the
emission of an extra heavy boson can be distinguished from the
corresponding non emission.  However, it can contribute to the cross
section when the decay products of the heavy boson escape from, {\it
  e.g.}, the detector acceptance or the experimental selection cuts.
Furthermore, these contributions might compensate the Sudakov
logarithms which enhance the NLO EW corrections at large scales.  We
will not include HBR contributions in the following results. Their
impact has been computed for the total cross section and differential
observables in~\cite{Frixione:2014qaa, Frixione:2015zaa}, where it has
been found to be small (less than $1\%$) on total rates and, unlike
NLO EW corrections, only marginally enhanced when large energy scales
are probed.

\begin{figure}
    \centering
    \includegraphics[width=0.49\textwidth]{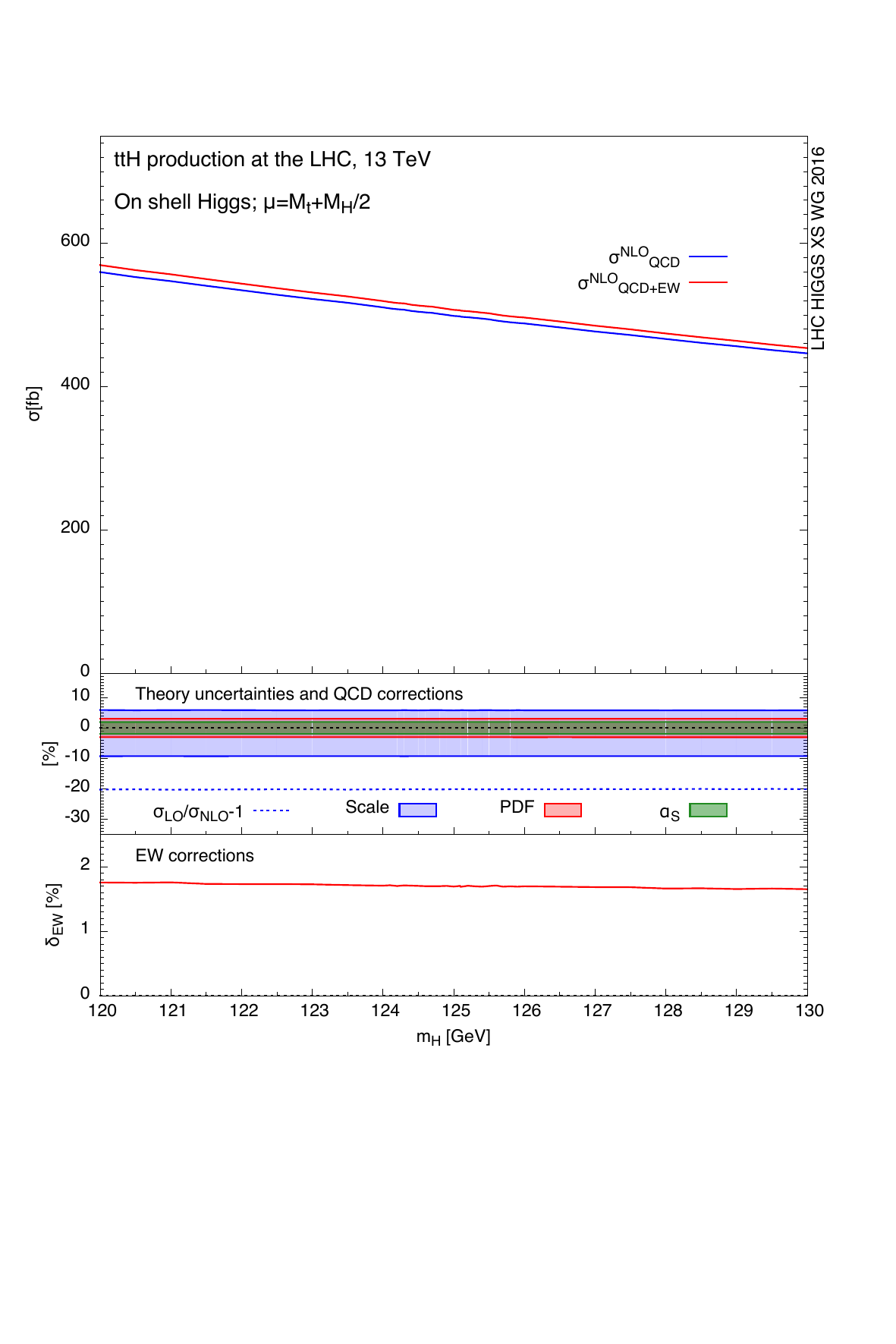}
    \includegraphics[width=0.49\textwidth]{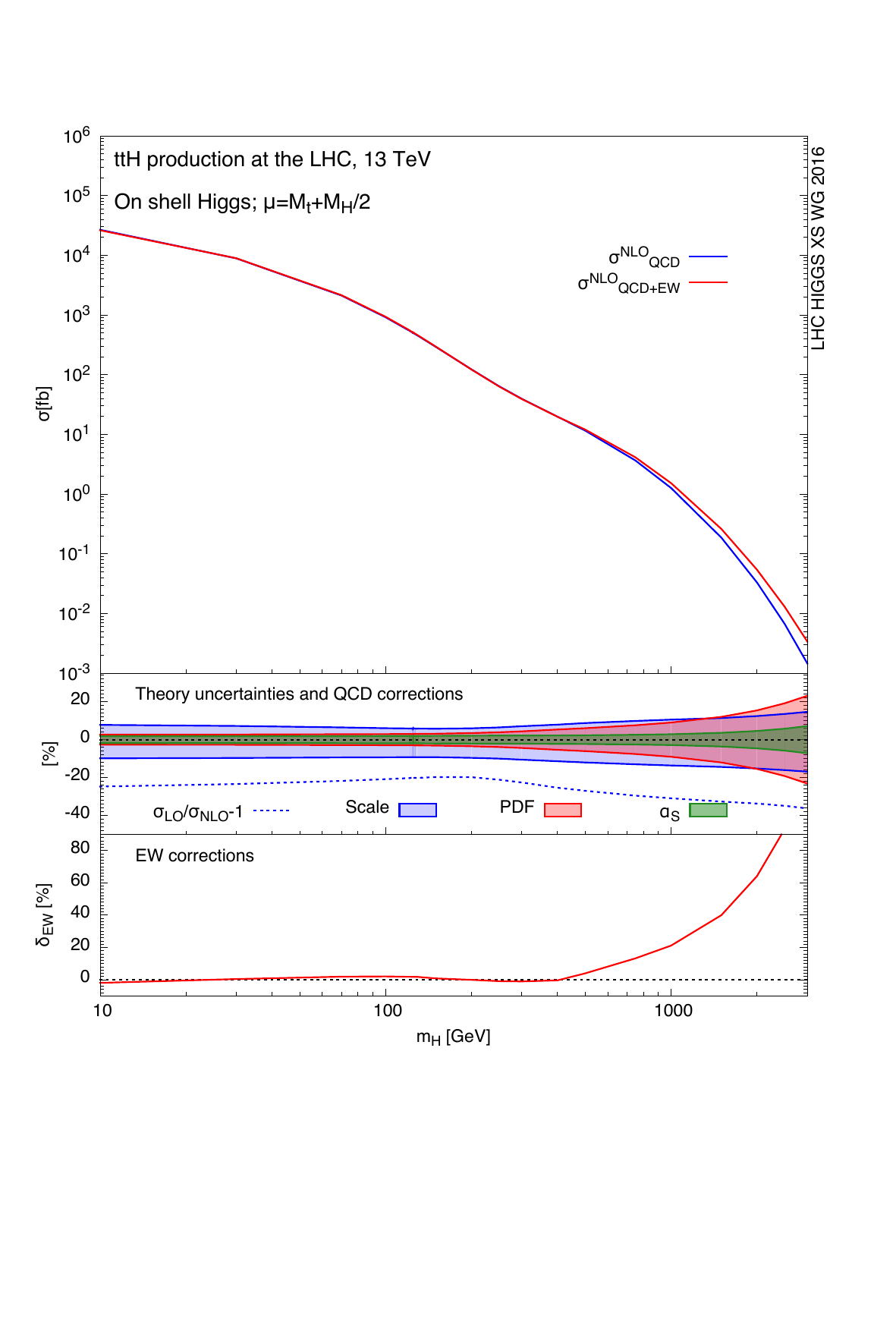}
    \caption{\label{fig:ttHEWplot13} The upper panel shows the $\tth$
      total cross section as a function of $\mh$, at $13~\UTeV$,
      including only NLO QCD corrections (blue curve) and both NLO
      QCD+EW corrections (red curve). The intermediate panel
      illustrates the estimated theoretical uncertainties from scale,
      PDF, and $\als$ variation over the same $\mh$ ranges. The
      lower panel shows the size of the electroweak corrections as a
      function of $\mh$.}
\end{figure}

Tables~\ref{tab:sm7tev}--\ref{tab:smescan12509} present NLO QCD+EW
predictions for different collider energies and Higgs boson mass
values.  The relative impact of QCD and EW corrections is illustrated
in the form of a QCD correction factor
\begin{equation}
K_{\mathrm{QCD}}=\frac{\sigmaNLOqcd}{\sigmaLOqcd}\,,
\end{equation}
and a relative EW correction factor
\begin{equation}
\delta_{\mathrm{EW}}=\dsigmaew/\sigmaNLOqcd\,.
\end{equation}
All result in Tables~\ref{tab:sm7tev}--\ref{tab:smescan12509} are based on
\MGfiveamcnlo~\cite{Alwall:2014hca}, similarly as
in~\cite{Frixione:2015zaa}.  A cross check against an independent
calculation based on \Sherpa+\Openloops~\cite{Kallweit:2014xda} has
confirmed the correctness of NLO QCD+EW predictions for
$\sqrt{s}=7,8,13,14~\UTeV$ and $\mh=125~\UGeV$ at the per mille level.
Predictions for the production of a Higgs boson in the mass range
$\mh=120$--$130~\UGeV$ are reported in
Tables~\ref{tab:sm7tev}--\ref{tab:sm14tev} for
$\sqrt{s}= 7,8,13,14~\UTeV$ respectively.  The relative scale and
PDF+$\als$ uncertainties are the same for NLO QCD as for NLO
QCD+EW cross sections, therefore they can be computed for the former
and applied to the latter.
Tables~\ref{tab:smescan125},~\ref{tab:smescan12509} list numbers at
different $\sqrt s$ for $\mh=125.00~\UGeV$ and $\mh=125.09~\UGeV$,
respectively.  The integration uncertainty affecting results is at
0.1\% level for $\sigma^{\rm NLO}_{\rm QCD+EW}$.
The
left and right plots in \refF{fig:ttHEWplot13} show the $\tth$
cross section as a function of the Higgs boson mass at $13~\UTeV$, for
the SM and BSM range respectively. The scale, PDF, and $\als$
uncertainties are also shown, together with the QCD and EW correction
factors.

\def\ttHnlopspath{\ttHpath/ttH-NLO-QCD+PS}
\section{Comparison of NLO QCD+Parton Shower simulations for \texorpdfstring{$t\bar{t}H(b\bar{b})$}{ttH(bb)}}
\label{sec:ttH-nlo-qcd+ps}

In recent years fixed-order NLO QCD calculations of $t\bar{t}H$ have been interfaced
with parton-shower (PS) Monte-Carlo generators
(\Herwig~\cite{Marchesini:1991ch,Corcella:2000bw,Bahr:2008pv},
\Pythia~\cite{Sjostrand:2006za,Sjostrand:2007gs,Skands:2014pea,Sjostrand:2014zea},
and \Sherpa~\cite{Gleisberg:2008ta}) using one of the methods proposed in
the literature, namely \Mcnlo~\cite{Frixione:2002ik,Frixione:2003ei},
\Powheg~\cite{Nason:2004rx,Frixione:2007vw,Frixione:2007nw}, and
\Smcnlo~\cite{Hoeche:2011fd,Hoeche:2012ft}, and are nowadays
implemented in a variety of tools, from
\MGfiveamcnlo \cite{Frederix:2011zi,Alwall:2014hca,Hirschi:2011pa},
to \Powhel~\cite{Bevilacqua:2011xh,Garzelli:2011vp},
\Powhegbox~\cite{Alioli:2010xd,Hartanto:2015uka},
\Sherpa+\Openloops~\cite{Gleisberg:2008ta,Cascioli:2011va,Cascioli:2013gfa},
and \Herwigseven~\cite{Bellm:2015jjp,Bahr:2008pv}.

The accurate description of the $\tth$ signal, from the energy
scale of the hard scattering to the hadronization energy scale,
crucially relies on these tools and their use in experimental analyses
is highly recommended. Due to the prominent role that $\tth$
production will play in the Higgs-physics program of Run~2 of the
LHC, it is crucial to validate different implementations against each
other and verify their compatibility. Given the multiplicity and
diversity of NLO QCD parton-shower Monte-Carlo generators available to
calculate $\tth$ observables (total cross section and
distributions), a systematic comparison requires to define a common
set-up that takes into account the technical aspects of different
matching schemes between fixed-order NLO QCD calculation and parton
shower.  It is the first necessary step towards a better control of
the theoretical accuracy of $\tth$ predictions, and has been the
purpose of a dedicated study in the context of the $\tth$
working group. Previous
studies~\cite{Dittmaier:2012vm,Andersen:2014efa} have shown
compatibility among different subsets of these tools, but different
choices made in each existing study prevent to derive from them a
more uniform comparison.

In this section we present details and outcomes of a new comprehensive
comparison of the most up-to-date tools currently available for Run~2 
studies, and compare them using a common choice of input parameters
for the fixed-order NLO QCD calculation and the PS. Some arbitrariness
in the choice of PS-specific parameters can still be present, as will
be manifest in the comparison of observables that are more sensitive
to regions of phase space that are dominated by the PS. We recommend
that the comparison presented in this section serves as the main
reference to anybody interested in using any of the NLO QCD+PS tools
that will be discussed in the following for the production of official
samples of $\tth$ showered events.

We have compared five NLO QCD calculations of $\tth$ consistently
interfaced with either \Sherpa, \Pythiaeight, or \Herwigseven. Namely,
we have compared results from:
\begin{itemize}
\item \Smcnlo\, using \Openloops~1.2.3~+~\Sherpa~2.2.0,
\item \MGfiveamcnlo~2.3.2~+~\Pythiaeight~2.1.0,
\item \Powhel~+~\Pythiaeight~2.1.0,
\item \Powhegbox~+~\Pythiaeight~2.1.0,
\item  \Herwigseven\, using \Openloops~1.2.4+~\MGfiveamcnlo~2.3.0+~\Herwigseven.
\end{itemize}
\Sherpa+\Openloops\, uses \Openloops~\cite{Cascioli:2011va} as a
one-loop generator, and relies on the \Cuttools\,
library ~\cite{Ossola:2007ax} for the numerically stable
evaluation of tensor integrals. Real-emission
contributions, infrared subtractions based on the Catani-Seymour
technique~\cite{Catani:1996vz,Catani:2002hc}, and phase-space
integration are handled by \Sherpa. The NLO corrections are matched to
the \Sherpa\, PS generator~\cite{Schumann:2007mg} using
the \Sherpa\, formulation~\cite{Hoeche:2011fd,Hoche:2012wh} of the
\Mcnlo~\cite{Frixione:2002ik,Frixione:2003ei} method, also dubbed
\Smcnlo.

Within \MGfiveamcnlo~\cite{Hirschi:2011pa,Alwall:2014hca}, fixed-order
NLO QCD results are obtained by adopting the FKS
method~\cite{Frixione:1995ms,Frixione:1997np} for the subtraction of
the infrared divergences of the real-emission matrix elements
(automated in the module \MadFKS~\cite{Frederix:2009yq}), and the
OPP integral-reduction procedure~\cite{Ossola:2006us} for the
computation of the one-loop matrix elements (automated in the module
\Madloop~\cite{Hirschi:2011pa}). Matching with parton showers is
achieved by means of the \Mcnlo\,
formalism~\cite{Frixione:2002ik,Frixione:2003ei}.

The \Powhegbox\,
framework~\cite{Nason:2004rx,Frixione:2007vw,Alioli:2010xd} adopts the
FKS subtraction scheme~\cite{Frixione:1995ms,Frixione:1997np} to
factor out the infrared singularities of the real-emission cross
section, while the the virtual one-loop matrix elements can be
provided with different methods. In the public \Powhegbox~(V2)
distribution of $\tth$, the NLO QCD virtual corrections are
implemented using the one-loop routines from the NLO QCD calculation
of Ref.~\cite{Reina:2001sf,Dawson:2002tg,Dawson:2003zu}. The
corresponding results are labelled as \Powhegbox\, in this section.  On
the other hand, the \Powhel\, generator~\cite{Garzelli:2011vp} uses the \Helacnlo\,
package~\cite{Bevilacqua:2011xh} for the computation of all matrix
elements provided as input to the \Powhegbox. Within the \Powhegbox\,
the matching with parton showers is obtained implementing the \Powheg\,
matching scheme~\cite{Nason:2004rx,Frixione:2007vw,Frixione:2007nw}.
The matched results from the NLO computations are interfaced to
\Pythiaeight\, via Les-Houches event (LHE) files.

\Herwigseven, based on extensions of the previously developed \Matchbox\,
module~\cite{Platzer:2011bc,Bellm:2015jjp}, implements the
Catani-Seymour dipole subtraction
method~\cite{Catani:1996vz,Catani:2002hc} for the infrared divergences
of the real-emission matrix elements, provides for the final-state phase-space
integration, and can interface to a variety
of LO and NLO matrix elements providers, either at the level of
squared matrix elements, based on extensions of the BLHA
standard~\cite{Binoth:2010xt,Alioli:2013nda,Andersen:2014efa}, or at
the level of colour--ordered subamplitudes, where the colour bases are
provided by an interface to the \ColorFull~\cite{Sjodahl:2014opa} and
\CVolver~\cite{Platzer:2013fha} libraries. For this study the relevant
tree-level matrix elements are provided by
\MGfiveamcnlo~\cite{Alwall:2011uj,Alwall:2014hca} (at the level of
colour-ordered subamplitudes), whereas the relevant
tree-level/one-loop interference terms are provided by
\Openloops~\cite{Cascioli:2011va,Cascioli:2014wya} (at the level of
squared matrix elements).  Fully automated NLO matching algorithms are
available, henceforth referred to as subtractive (NLO$\oplus$) and
multiplicative (NLO$\otimes$) matching -- based on the
\Mcnlo~\cite{Frixione:2002ik} and \Powheg~\cite{Nason:2004rx}
formalism respectively -- for the systematic and consistent
combination of NLO QCD calculations with both shower variants (an
angular--ordered parton shower~\cite{Gieseke:2003rz} and a dipole
shower~\cite{Platzer:2009jq}) in \Herwigseven. For this study the
subtractive matching in combination with the angular--ordered parton
shower has been chosen.

For the purpose of the comparison presented in this section, the NLO
QCD calculation has been performed using $N_{\mathrm{F}}=5$ light flavours,
$\sqrt{s}=13$~\UTeV\, for the centre-of-mass energy, $\mh=125$~\UGeV\, for the
Higgs boson mass, and $\mt=172.5$~\UGeV\, for the top-quark mass. The
top-quark Yukawa coupling has been defined in terms of the Fermi
constant as $y_{\mathrm{t}}=(\sqrt{2}G_F)^{1/2}\mt$. We have followed the
recommendation of the Higgs Cross-Section Working
Group~\cite{LHCHXSWG-INT-2015-006} for all other parameters that are not
explicitly given here. We have used a dynamical renormalization
($\muR$) and factorization ($\muF$) scale defined as the geometric
mean of the transverse energies ($E_T$) of the final-state particles
($t$, $\bar{t}$, and $H$). The central value of both scales is then
set to $\mu_0=(E_{\mathrm{T}}(t)E_{\mathrm{T}}(\bar{t})E_{\mathrm{T}}(H))^{1/3}$, where
$E_{\mathrm{T}}=\sqrt{M^2+p_{\mathrm{T}}^2}$ for $M$ the mass of a given particle and $p_{\mathrm{T}}$
its corresponding transverse momentum. Finally, following the 
Higgs Cross Section Working Group recommendation, we have used the PDF4LHC15
parton distribution functions (PDF)~\cite{Butterworth:2015oua}, and
more specifically the central set of PDF4LHC15\_nlo\_30, with NLO
$\als(\mu)$ and $\als(\mz)=0.118$.

The different parton-shower generators have all been set up not to
include hadronization, underlying events, and QED effects in the shower.  
The shower resummation scale has been chosen as consistently as possible among different generators. This corresponds to setting $\mu_Q=H_T/2$ in \Smcnlo\, where $H_{\mathrm{T}}$ is defined as the sum of the $\tth$ final-state transverse energies ($H_{\mathrm{T}}=E_{\mathrm{T}}(t)+E_{\mathrm{T}}(\bar{t})+E_{\mathrm{T}}(H)$),
while using $h=H_{\mathrm{T}}/2$ in the
definition of $h_{\mathrm{damp}}=h^2/(h^2+p_{\mathrm{T}}^2)$ in \Powhegbox\,
where $p_{\mathrm{T}}$ is the transverse momentum of the hardest parton in the $O(\als)$ QCD real emission. In the case of \MGfiveamcnlo\ a different choice had to be adopted since only resummetion scales of the form $\mu_Q=\xi\sqrt{\hat{s}}$ are supported, where the prefactor $\xi$ is randomly distributed in a freely adjustable
$(\xi_{\mathrm{min}},\xi_{\mathrm{max}})$ range. In particular
we have adopted the default recommendation and used
$(\xi_{\mathrm{min}},\xi_{\mathrm{max}})=(0.1,1)$. 
In the case of \Herwigseven\, the hard shower
scale, similarly to the renormalization and factorization scale, has been
set to be the geometric mean of the transverse energies (see
above). Internal studies have shown that a different scale choice for
the hard shower scale results in only small differences in the distributions.

The theoretical uncertainty bands have been calculated purely from the
renormalization and factorization scale dependence, estimated by
varying these scales independently by a factor of two about their
central value ($\muR=\mathswitch{\xi_{\ssR}}\mu_0$ and
$\muF=\mathswitch{\xi_{\ssF}}\mu_0$, with
$\mathswitch{\xi_{\ssR,\ssF}}=1/2,1,2$). For the purpose of this
comparison we have used a common set of PDF and therefore we have not
included in the theoretical error any uncertainty from PDF variation (since it
would have been the same for all results).
No uncertainty from the parton shower has been included. It has been
our goal to investigate if, under physically equivalent choices of the
parton-shower setup, and having eliminated the differences that can
come from different treatment of hadronization and underlying events,
all the tools considered in this comparison give results that are
compatible within the scale uncertainty for all observables
that are not directly affected by parton-shower effects (see discussion
of Figs.~\ref{fig:ttH_stable_NLOPS_1}-\ref{fig:ttH_decays_NLOPS_4}).
Differences that are
observed in regions of phase space dominated by the PS
should be resolved by properly including parton-shower uncertainties,
and this study should serve as solid ground to investigate these
parton-shower specific effects.

Finally, we have considered two scenarios: without and with decays of
the $\tth$ final-state particles. In the second case, we have let
the Higgs boson decay to $b$ quarks, $H\rightarrow b\bar{b}$, while the
top and antitop quarks decay leptonically to $t\rightarrow b e^+\nu_{\mathrm{e}}$
and $\bar{t}\rightarrow \bar{b}\mu^-\bar{\nu}_\mu$, respectively.
Notice that, for the purpose of this comparison, the results presented
include only this specific decay chain, i.e. correspond to the process
$t\bar{t}H\rightarrow e^+\mu^-\nu_{\mathrm{e}}\bar{\nu}_\mu b\bar{b}b\bar{b}$.
Events are required to contain one $e^+$ and one $\mu^-$ with
transverse momentum $p_{\mathrm{T}}^l>20$~\UGeV\, and pseudorapidity $|\eta^l|<2.5$
($l$=lepton), as well as missing transverse energy
$E_{\mathrm{T}}\!\!\!\!\!\!/>30$~\UGeV, and four $b$ jets.  $b$ jets are defined
using the anti-$k_T$ jet algorithm with $R=0.4$, and requiring that
the jet contains at least one $b$ or $\bar{b}$ quark and has
transverse momentum $p_{\mathrm{T}}^{\mathrm{b}}>25$~\UGeV\, and pseudorapidity $|\eta^{\mathrm{b}}|<2.5$.
Finally, spin-correlation effects have been taken into account using
the built-in implementations of each package, like \Madspin~\cite{Artoisenet:2012st} for
\MGfiveamcnlo, \Decayer~\cite{Garzelli:2014dka} for \Powhel, and analogous modules for
\Powhegbox, \Smcnlo, and \Herwigseven, all based on the approach
originally proposed in Ref.~\cite{Frixione:2007zp}.  The results from the five NLO
QCD+PS simulations listed above have been processed through a common
\Rivet\, analysis that implements the selection cuts described above.

In Figs.~\ref{fig:ttH_stable_NLOPS_1}-\ref{fig:ttH_stable_NLOPS_2} we
present results for the comparison of the on-shell case (no decay of
$t$, $\bar{t}$, and $H$ included), while in
Figs.~\ref{fig:ttH_decays_NLOPS_1} -\ref{fig:ttH_decays_NLOPS_4} we
present results for the case of
$t\bar{t}H \rightarrow e^+\mu^-\nu_{\mathrm{e}}\bar{\nu}_\mu b\bar{b}b\bar{b}$.
In order to minimize the effect of treating decays at LO the
corresponding branching ratios have been normalized to
$\mathrm{Br}(H\rightarrow b\bar{b})=57.7\%$ (from
Ref.~\cite{Heinemeyer:2013tqa}, for $\mh=125$~\UGeV), and
$\mathrm{Br}(t\rightarrow b e^+ \nu_{\mathrm{e}})=\mathrm{Br}(W^+\rightarrow
e^+\nu_{\mathrm{e}})=10.83\%$ (from~\cite{Agashe:2014kda}), and similarly for
$\bar{t}\rightarrow\bar{b}\mu^-\bar{\nu}_\mu$.

In each case we compare results for various standard differential
observables. Each plot shows in the upper window the comparison of
results obtained using the five different NLO QCD+PS tools used in our
study, as well as the pure NLO QCD fixed-order results (see
Table~\ref{tab:sm13tev}) which have been used for validation. The
lower windows of each plot illustrate the theoretical uncertainty from
renormalization- and factorization-scale dependence calculated as
previously explained. More specifically, each lower window shows all
results normalized to a particular one, together with the uncertainty
band of the latter. For NLO distributions this uncertainty is of the
order of 10-15\%, but can grow to 20\% or more in the tails of
distributions. On the other hand, a much larger uncertainty affects
distributions like the $p_{\mathrm{T}}$ of the hardest light jet since the
underlying hard process is LO in nature.

The comparison of both total cross sections and distributions shows in
general full compatibility among all sets of results within the
theoretical uncertainty considered in this study.  This is in
particular true for the on-shell case, when decays of the final-state
particles are not considered. This validates the set-up chosen for the
comparison, both at the level of the fixed-order NLO QCD calculation
and at the level of the matching with the PS.

In the case in which the decays of both top-quarks and Higgs boson are
implemented we still see overall very good agreement. We notice some
moderate discrepancies in the distribution in the number of $b$ jets
($d\sigma/dN_{\mathrm{b-jets}}$). In the case of
\Powhel+\Pythiaeight\, the excess in the low $N_{\mathrm{b-jets}}$
bins and the deficit in the high $N_{\mathrm{b-jets}}$ bins are mainly due
to having considered the bottom quarks as massless in the decays of
the top quarks, matched to a parton shower that uses massive bottom 
quarks. As all other distributions in
Figs.~\ref{fig:ttH_decays_NLOPS_1}-\ref{fig:ttH_decays_NLOPS_4} are
obtained from events with exactly four $b$ jets, the difference in the
exclusive $b$-jet multiplicity distribution at $N_{{\rm b-jets}} = 4$
affects the normalization of the distributions, but leaves their
shapes intact.

On the other hand, the overall discrepancy between most
implementations considered for $N_{\mathrm{b-jets}}>4$ is mainly of
parton-shower origin.  Indeed, since the $N_{\mathrm{b-jets}}>4$ bins
are mainly populated by $b$ jets originating in the parton shower,
these effects depend on the specific set up of the parton-shower
algorithm used in each case and should be considered as part of the
theoretical uncertainty coming from the parton shower, which we have
not explicitly quantified in this study. A dedicated study of
parton-shower effects acquires more meaning in the context of specific
experimental analyses, if, for instance, observables like
$d\sigma/dN_{\mathrm{b-jets}}$ for large numbers of $b$ jets had to
become relevant. Having provided a sound comparison of a broad variety
of main NLO QCD+PS frameworks, we have laid the foundation for further
dedicated studies that will likely happen in the context of specific
experimental analyses.

\providecommand{\higgsstableplot}[2]{
\includegraphics[width=0.75\textwidth, trim=0 #1 0 0,clip]{\ttHnlopspath//#2}}

\providecommand{\higgsstablefig}[1]{
\begin{minipage}{0.5\textwidth}
\higgsstableplot{0.115\textwidth}{figs_stable/main/#1}\\
\higgsstableplot{0.115\textwidth}{figs_stable/sherpaopenloops/#1}\\
\higgsstableplot{0.115\textwidth}{figs_stable/mg5amc/#1}\\
\higgsstableplot{0.115\textwidth}{figs_stable/powhel/#1}\\
\higgsstableplot{0.115\textwidth}{figs_stable/powheg/#1}\\
\higgsstableplot{0}{figs_stable/herwig/#1}
\end{minipage}}

\begin{figure}
\higgsstablefig{PT_TOP}
\higgsstablefig{ETA_TOP}\\[2mm]
\higgsstablefig{PT_H}
\higgsstablefig{ETA_H}
\caption{\label{fig:ttH_stable_NLOPS_1} NLO QCD+PS and fixed-order NLO
  QCD predictions for differential $\tth$ observables at 13~\UTeV.
  Each ratio plot shows all results normalized to one particular NLO
  QCD+PS prediction and the scale variation band of the latter.  }
\end{figure}

\begin{figure}
\begin{tabular}{cc}
	\higgsstablefig{PT_TT}	&	\higgsstablefig{PT_TTH}\\[2mm]
	\multicolumn{2}{c}{\higgsstablefig{PT_J1}} \\
\end{tabular}
\caption{\label{fig:ttH_stable_NLOPS_2} NLO QCD+PS and fixed-order NLO
  QCD predictions for differential $\tth$ observables at 13~\UTeV.
  The ratio plots are defined as in \refF{fig:ttH_stable_NLOPS_1}.  }
\end{figure}

\providecommand{\higgsdecaysplot}[2]{
  \includegraphics[width=0.75\textwidth, trim=0 #1 0 0,clip]{\ttHnlopspath/#2}}

\providecommand{\higgsdecaysfig}[1]{
\begin{minipage}{0.49\textwidth}
\higgsdecaysplot{0.115\textwidth}{figs_decays/main/#1}\\
\higgsdecaysplot{0.115\textwidth}{figs_decays/sherpaopenloops/#1}\\
\higgsdecaysplot{0.115\textwidth}{figs_decays/mg5amc/#1}\\
\higgsdecaysplot{0.115\textwidth}{figs_decays/powhel/#1}\\
\higgsdecaysplot{0.115\textwidth}{figs_decays/powheg/#1}\\
\higgsdecaysplot{0}{figs_decays/herwig/#1}
\end{minipage}}

\begin{figure}
\centering
\higgsdecaysfig{PT_EL}
\higgsdecaysfig{ETA_EL}\\[2mm]
\higgsdecaysfig{PT_MU}
\higgsdecaysfig{ETA_MU}
\caption{\label{fig:ttH_decays_NLOPS_1}
NLO QCD+PS predictions for differential $\tth$ observables with
$t\bar{t}H\to e^+\mu^-\nu_{\mathrm{e}}\bar{\nu}_\mu+b\bar{b}b\bar{b}$
at 13~\UTeV. The ratio plots are defined as in \refF{fig:ttH_stable_NLOPS_1}.}
\end{figure}

\begin{figure}
\centering
\higgsdecaysfig{MET}
\higgsdecaysfig{DPHI_LL}\\[2mm]
\higgsdecaysfig{NBj_Exc_XS}
\higgsdecaysfig{NBj_Inc_XS}
\caption{\label{fig:ttH_decays_NLOPS_2}
NLO QCD+PS predictions for differential $\tth$ observables with
$t\bar{t}H\to e^+\mu^-\nu_{\mathrm{e}}\bar{\nu}_\mu+b\bar{b}b\bar{b}$ at 13~\UTeV.
The ratio plots are defined as in \refF{fig:ttH_stable_NLOPS_1}.}
\end{figure}

\begin{figure}
\centering
\higgsdecaysfig{PT_B1}
\higgsdecaysfig{PT_B2}\\[2mm]
\higgsdecaysfig{PT_B3}
\higgsdecaysfig{PT_B4}
\caption{\label{fig:ttH_decays_NLOPS_3}
NLO QCD+PS predictions for differential $\tth$ observables with
$t\bar{t}H\to e^+\mu^-\nu_{\mathrm{e}}\bar{\nu}_\mu+b\bar{b}b\bar{b}$ at 13~\UTeV.
The ratio plots are defined as in \refF{fig:ttH_stable_NLOPS_1}.}
\end{figure}

\begin{figure}
\centering
\higgsdecaysfig{ETA_B1}
\higgsdecaysfig{ETA_B2}\\[2mm]
\higgsdecaysfig{ETA_B3}
\higgsdecaysfig{ETA_B4}
\caption{\label{fig:ttH_decays_NLOPS_4}
NLO QCD+PS predictions for differential $\tth$ observables with
$t\bar{t}H\to e^+\mu^-\nu_{\mathrm{e}}\bar{\nu}_\mu+b\bar{b}b\bar{b}$ at 13~\UTeV.
The ratio plots are defined as in \refF{fig:ttH_stable_NLOPS_1}.}
\end{figure}


\def\ttHoffpath{\ttHpath/ttH-decays-offshell}

{
\let\text\textrm

\newcommand{\fb}{{\ensuremath\unskip\,\text{fb}}\xspace}

\newcommand{\Pj}{\ensuremath{\mathrm{j}}\xspace}

\newcommand{\ptsub}[1]{\ensuremath{p_{T,#1}}\xspace}

\newcommand{\qqb}{\ensuremath{q\bar{q}}\xspace}

\newcommand{\fullProcessH  }{\ensuremath{\Pp\Pp\to\Pe^+\nu_{\Pe} \mu^-\bar{\nu}_\mu\PQb\bar{\PQb}\PH}\xspace}

\newlength{\width}
\newlength{\height}
\newcommand{\brabar}[1]{%
    \settoheight{\height}{\ensuremath{#1}}%
    \settowidth{\width}{\ensuremath{#1}}%
    \makebox[0pt][l]{\ensuremath{#1}}%
    \raisebox{1.26ex}{\scalebox{.3}{\textbf{(}}}%
    \rule[1.41\height]{0.7\width}{0.35pt}%
    \raisebox{1.26ex}{\scalebox{.3}{\textbf{)}}}%
    }

\newcommand{\ttbarbbbar}{\ensuremath{\PQt\bar{\PQt}\PQb\bar{\PQb}}\xspace}
\newcommand{\ttbarh}{\ensuremath{\PQt\bar{\PQt}\PH}\xspace}
\newcommand{\ttbarhProcess}{\ensuremath{\Pp\Pp\to\PQt\bar{\PQt}\PH\to\Pl^+\nu_{\Pl} \Pj\Pj\PQb\bar{\PQb}\PQb\bar{\PQb}}\xspace}
\newcommand{\ttbarProcess }{\ensuremath{\Pp\Pp\to\PQt\bar{\PQt}\PQb\bar{\PQb}\to\Pl^+\nu_{\Pl} \Pj\Pj\PQb\bar{\PQb}\PQb\bar{\PQb}}\xspace}
\newcommand{\fullProcess}{\ensuremath{\Pp\Pp\to\Pl^+\nu_{\Pl} \Pj\Pj\PQb\bar{\PQb}\PQb\bar{\PQb}}\xspace}
\newcommand{\ggProcess    }{\ensuremath{\Pg\Pg\to\Pl^+\nu_{\Pl} q'\bar{q}''\PQb\bar{\PQb}\PQb\bar{\PQb}}\xspace}
\newcommand{\qqbProcess    }{\ensuremath{q\bar{q}\to\Pl^+\nu_{\Pl} q'\bar{q}''\PQb\bar{\PQb}\PQb\bar{\PQb}}\xspace}
\newcommand{\ttbbb}{\ttbarbbbar}

\section{Off-shell effects in \texorpdfstring{$t\bar{t}H$}{ttH} production}
\label{sec:ttH-off-shell}

\subsection{\texorpdfstring{$t\bar{t}H$}{ttH} with off-shell top decays: \texorpdfstring{$W^+W^-b\bar{b}H$}{WWbbH} production at NLO QCD}

In this section predictions for the hadronic production of
top--antitop pairs in association with a Higgs boson at
next-to-leading-order QCD, including the decay of the top and antitop
quark into bottom quarks and leptons, are presented.  The computation
is based on full leading and next-to-leading-order matrix elements for
$\Pe^+\nu_{\Pe} \mu^-\bar{\nu}_\mu\PQb\bar{\PQb}\PH(\Pj)$ and includes
all non-resonant contributions, off-shell effects and interferences
(for more details see \Bref{Denner:2015yca}).  Besides off-shell
effects also NLO corrections to top-quark decays are included, which
is not the case in many NLO and NLO$\,+\,$parton shower (PS) $\ttbarh$
calculations on the market.

\subsubsection{Method of calculation}

The study is based on the tree-level amplitudes at
$\order{\als\alpha^{5/2}}$ for gluon-induced and
quark--antiquark-induced processes and the corresponding NLO
corrections of order $\als$. The bottom quark is considered
massless.  The corresponding real corrections receive
also contributions of quark--gluon- and antiquark--gluon-initiated
processes.  The Catani--Seymour subtraction formalism
\cite{Catani:1996vz,Catani:2002hc} is applied for the regularization
and analytical cancellation of IR singularities.  For the computation
of all matrix elements as well as colour- and spin-correlated squared
matrix elements needed for the evaluation of subtraction terms, the
recursive amplitude generator \Recola~\cite{Actis:2012qn} is employed.
A consistent description of all resonances is achieved using the
complex-mass scheme \cite{Denner:1999gp,Denner:2005fg,Denner:2006ic}.
The top-quark Yukawa coupling is defined in the on-shell scheme.

The matrix elements for the virtual corrections are calculated with
\Recola, which uses the
\Collier \cite{Denner:2014gla,Denner:2016kdg} library for the numerical evaluation of
one-loop scalar
\cite{'tHooft:1978xw,Beenakker:1988jr,Dittmaier:2003bc,Denner:2010tr}
and tensor integrals
\cite{Passarino:1978jh,Denner:2002ii,Denner:2005nn}.  The results for
the virtual NLO contribution to the squared amplitude,
$2\Re\mathcal{M}^*_0\mathcal{M}_1$, have been successfully compared
with \MGfiveamcnlo~\cite{Alwall:2014hca}. In addition the Ward
identity for the matrix elements of the gluon-initiated process at
tree and one-loop level has been checked.

\subsubsection{Setup of the analysis}
\label{sec:tthnlo_off_setup}

Cross section and differential distributions for the LHC operating at
$13\UTeV$ are investigated.  LHAPDF 6.05 with CT10NLO parton
distributions are employed for LO and NLO cross sections and
contributions from the suppressed bottom-quark parton density and
flavour mixing are neglected. The value of the strong coupling
constant $\als$ as provided by LHAPDF based on a two-loop accuracy
with $N_\text{F}=5$ active flavours is used. The electromagnetic
coupling $\alpha$ is derived from the Fermi constant in the $G_\mu$
scheme. The width of the top quark $\Gt$ is calculated at LO and NLO
QCD including effects of off-shell W bosons according to
\Bref{Jezabek:1988iv}.  The top-quark width is calculated at the
scale $\mt$, which is kept fixed when studying scale uncertainties.

For the jet reconstruction the anti-$k_\text{T}$ algorithm
\cite{Cacciari:2008gp} is used with a jet-resolution parameter
$R=0.4$. Only final-state quarks and gluons with rapidity $|y|<5$ are
clustered into infrared-safe jets.
After recombination standard selection cuts are imposed on transverse
momenta and rapidities of charged leptons and b~jets, missing
transverse momentum and rapidity--azimuthal-angle distance between b
jets. Two b jets and two charged leptons in the final
state are required, with bottom quarks in jets leading to b jets, and
\begin{equation}\label{eqn:cuts-topdecays-nlo}
        \begin{aligned}
                \text{b jets:}                      && \ptsub{{\PQb}}         &>  25\UGeV,  & |y_{\PQb}|   &< 2.5, &\qquad\\
                \text{charged lepton:}              && \ptsub{\Pl}         &>  20\UGeV,  & |y_{\Pl}| &< 2.5, &\\
                \text{missing transverse momentum:} && \ptsub{\text{miss}} &>  20\UGeV,                      &\\
                \text{b-jet--b-jet distance:}       && \Delta R_{\PQb\PQb}   &> 0.4.                          &\\
        \end{aligned}
\end{equation}

As default, a dynamical scale
\begin{equation}\label{eqn:DynamicalScale}
        \mu_\text{dyn} = \muR = \muF = \left(M_{\text{T},\Pt}\,M_{\text{T},\bar{\Pt}}\, M_{\text{T},\PH}\right)^\frac{1}{3}\quad\text{with}\quad M_{\text{T}}=\sqrt{M^2+p^2_\text{T}}
\end{equation}
is used for the renormalization $\muR$ and factorization scale
$\muF$ following \Bref{Frederix:2011zi}.  Alternatively, a
fixed scale according to \Bref{Beenakker:2002nc} is chosen:
\begin{equation}\label{eqn:FixedScale-topdecays-nlo}
        \mu_\text{fix} = \muR = \muF = \frac{1}{2}\left(2\mt + M_{\PH} \right) = 236\UGeV.
\end{equation}
Scale uncertainties are determined by computing integrated and
differential cross sections at seven scale pairs,
$(\muR/\mu_0$,
$\muF/\mu_0)=(0.5,0.5),(0.5,1),(1,0.5),(1,1),(1,2),(2,1),(2,2)$.
The central value corresponds to $(\muR/\mu_0$,
$\muF/\mu_0)=(1,1)$, and the error band is constructed from the
envelope of these seven calculations.

\subsubsection{Results for integrated cross sections}

\begin{table}
        \centering
        \renewcommand\arraystretch{1.3}
        \caption[Composition of the integrated cross section]{\label{table:results_summary}
                Composition of the integrated cross section for $\Pp\Pp \to
                \Pe^+\nu_{\Pe} \mu^- \bar{\nu}_\mu \PQb \bar{\PQb}
                \PH(\Pj)$
                at the $13\UTeV$ LHC
                with the dynamical  scale. In column one the partonic
                initial states are listed, where $q=\PQu,\PQd,\PQc,\PQs$ and $\brabar{q}=q,\bar{q}$. The second
                and third column give the integrated cross sections in fb for LO and
                NLO, resp., including scale uncertainties. The last column provides the $K$
                factor with $K=\sigma_\text{NLO}/\sigma_\text{LO}$.
        }
        \begin{tabular}{llll}
                \toprule
                  \multicolumn{1}{l}{\text{pp}}                
                & \multicolumn{1}{c}{\text{$\sigma_\text{LO}$ [fb]}}%
                & \multicolumn{1}{c}{\text{$\sigma_\text{NLO}$ [fb]}}%
                & \multicolumn{1}{c}{$K$}\\                
                \midrule
                $\Pg \Pg$        &  $1.5906(1) ^{+33.7\%}_{-23.6\%}$ & $2.024(3) ^{+8.4\%  }_{-16.2\%}$   & 1.273(2)\\
                $q \bar{q}$      &  $0.67498(9)^{+24.1\%}_{-18.1\%}$ & $0.495(1) ^{+17.2\%  }_{-39.5\%}$  & 0.733(2)\\
                $\Pg \brabar{q}$ &                                   & $0.136(1) ^{+295\%   }_{-166\%}$ &     \\ 
                \midrule
                $\sum$         &  $2.2656(1) ^{+30.8\%}_{-22.0\%}$ & $2.656(3) ^{+0.9\%   }_{-4.6\%}$   & 1.172(1)\\

                \bottomrule
        \end{tabular}
\end{table}
In \refT{table:results_summary} the integrated cross
sections for the dynamical scale \eqref{eqn:DynamicalScale} is presented.
The cross sections for the fixed scale
\eqref{eqn:FixedScale-topdecays-nlo} are lower by only about 1\,\%, and the
$K$~factor for the fixed scale is 1.176(1).
The contribution of the dominating gluon-fusion channel increases from
about 70\,\% at LO to 76\,\% at NLO.  The contribution of the quark--antiquark annihilation drops
from about 30\,\% at LO to 19\,\% at NLO. The gluon--(anti)quark
induced real-radiation subprocesses contribute about 5\,\% at NLO. The
inclusion of NLO QCD corrections reduces the scale dependence from
31\,\% to 5\,\%.
Note that the NLO scale uncertainty band in  \refT{table:results_summary}
is by a factor 3 smaller than for on-shell $\ttbarh$ production at
$13\UTeV$ with the PDF4LHC15 prescription.
This might be due to the acceptance cuts, the PDFs, or to the fact
that scale variations are not applied to the top width.

\begin{figure}
        \begin{minipage}[t]{.490\textwidth}
                \includegraphics[width=\linewidth]{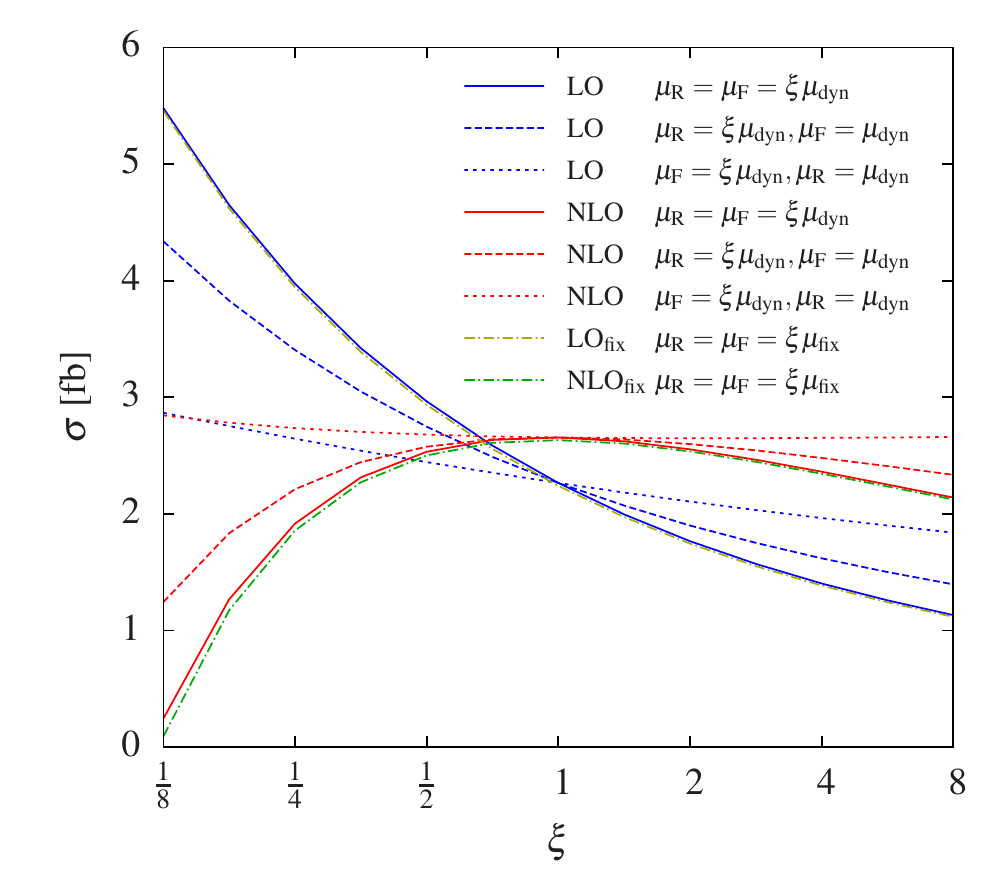}
                \caption{\label{plot:scale_dependence}%
                  Scale dependence of the LO and NLO integrated cross
                  section at the $13\UTeV$ LHC. The
                    renormalization and factorization scales are
                    varied around the central values of the fixed
                    ($\mu_0=\mu_\text{fix}$, dash-dotted lines) and
                    dynamical scale ($\mu_0=\mu_\text{dyn}$, solid
                    lines). For the dynamical scale the variation with
                    $\mu_\text{R}$ while keeping
                    $\mu_\text{F}=\mu_\text{dyn}$ fixed and vice versa
                    is shown with dashed lines.}
        \end{minipage}
        \hfill
        \begin{minipage}[t]{.490\textwidth}
                \includegraphics[width=\linewidth]{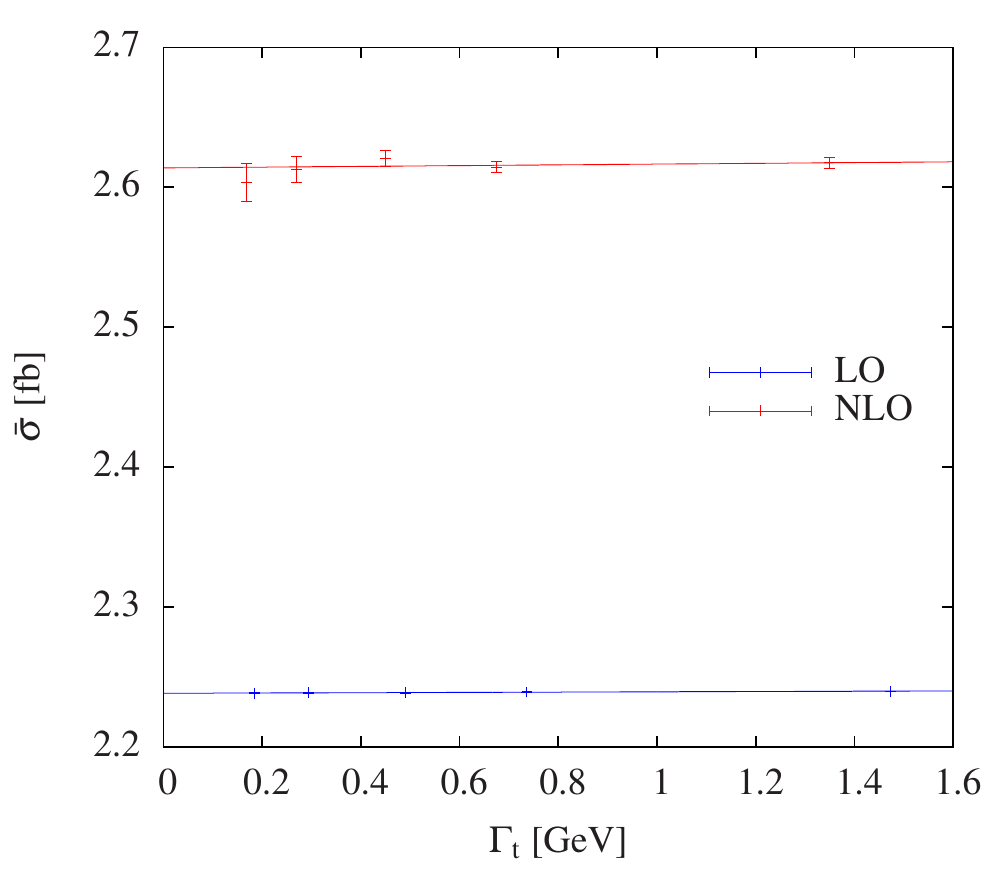}
                \caption{\label{plot:top_width_extrapolation}%
                  Zero-top-width extrapolation of the LO and NLO cross
                  section at the LHC at $\sqrt{s}=13\UTeV$ for fixed
                  scale $\mu_0=\mu_\text{fix}$.}
        \end{minipage}
\vspace{-5mm}        
\end{figure}
The dependence of the integrated LO (blue) and NLO (red) cross
sections on the values of the fixed and dynamical scale is displayed
in \refF{plot:scale_dependence}. Solid lines for the dynamical scale
and dash-dotted lines for the fixed scale show the scale dependence
for a simultaneous variation of the renormalization and factorization
scales and dashed lines the individual variation, where one of the
scales is kept fix at the central value, for the dynamical scale only.
While the largest scale variation is obtained when both scales are
changed simultaneously, the smallest effect results if only the
factorization scale is varied. The cross sections for the fixed and
dynamical scale choices are uniformly shifted relative to each other
by about 1\,\% as for the central scale $\mu_0$ both for LO and NLO
except for $\mu<\mu_0/2$, where the fixed scale leads to a faster
decrease of the cross section with $\mu$ as the dynamical scale. For
the fixed and dynamical scale the maximum of the NLO cross section is
near $\mu\simeq\mu_0$, justifying the use of both scale choices to be
stable against scale variations. The $K$~factor equals one at the
slightly lower scale of about $\mu\simeq 0.7\mu_0$

The effects of the finite top-quark width have been determined via a
numerical extrapolation to the zero-top-width limit, $\Gamma_{\Pt} \to
0$ (see \refF{plot:top_width_extrapolation}).  For fixed scale $\mu_{\mathrm{fix}}$
finite-top-width effects shift the LO and NLO cross section by
$-0.07\pm0.01\,\%$ and $-0.14\pm0.22\,\%$, respectively, which are
within the expected order of $\Gamma_{\Pt}/M_{\Pt}$.

\subsubsection{Results for distributions}

\begin{figure}
                \includegraphics[width=0.49\textwidth]{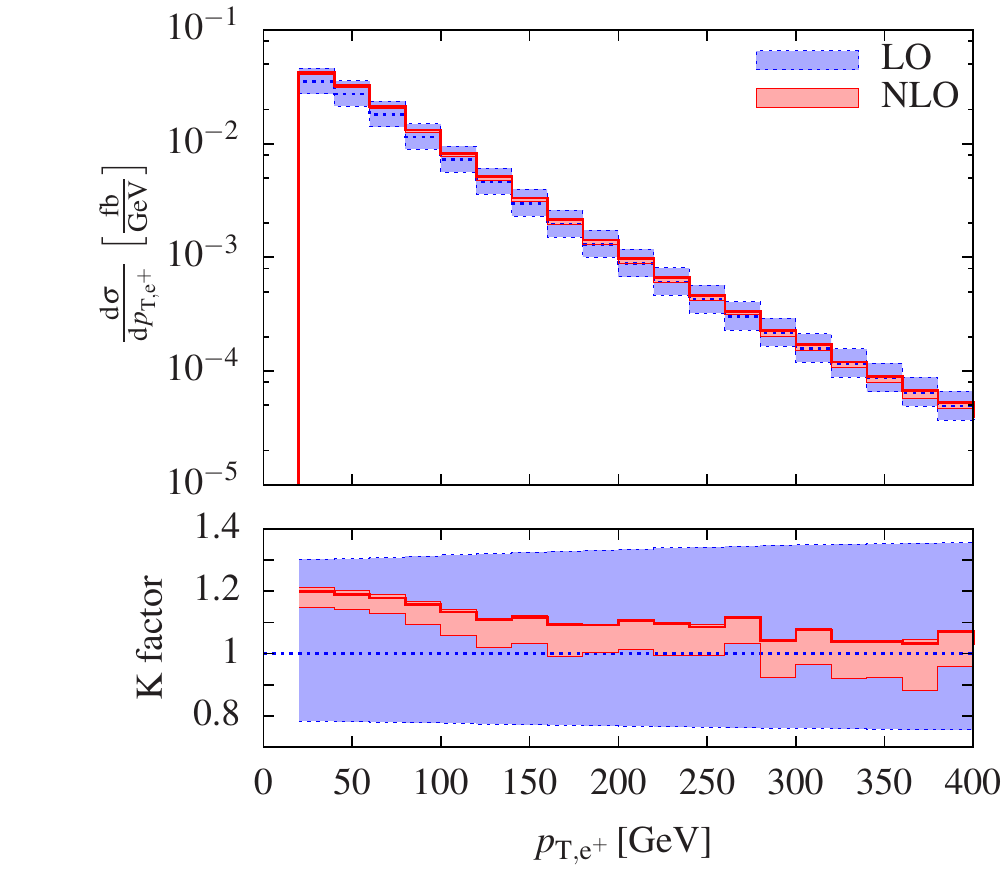}%
                \includegraphics[width=0.49\textwidth]{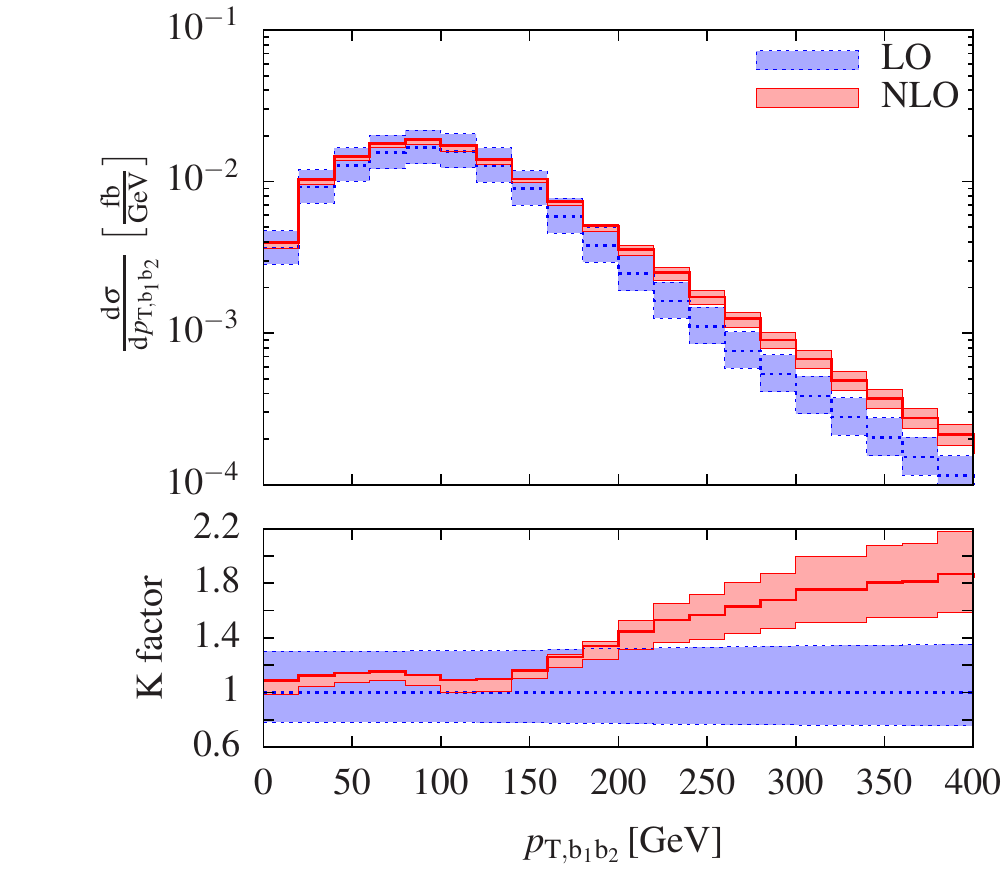}%

                \includegraphics[width=0.49\textwidth]{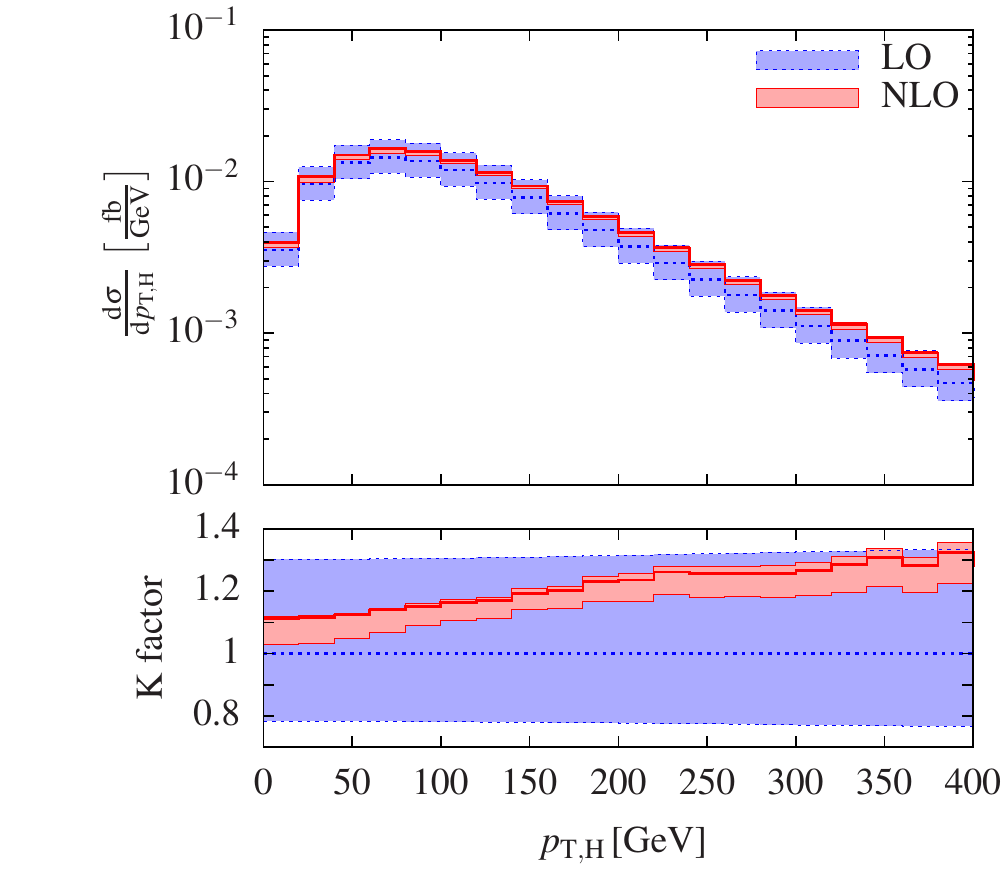}%
                \includegraphics[width=0.49\textwidth]{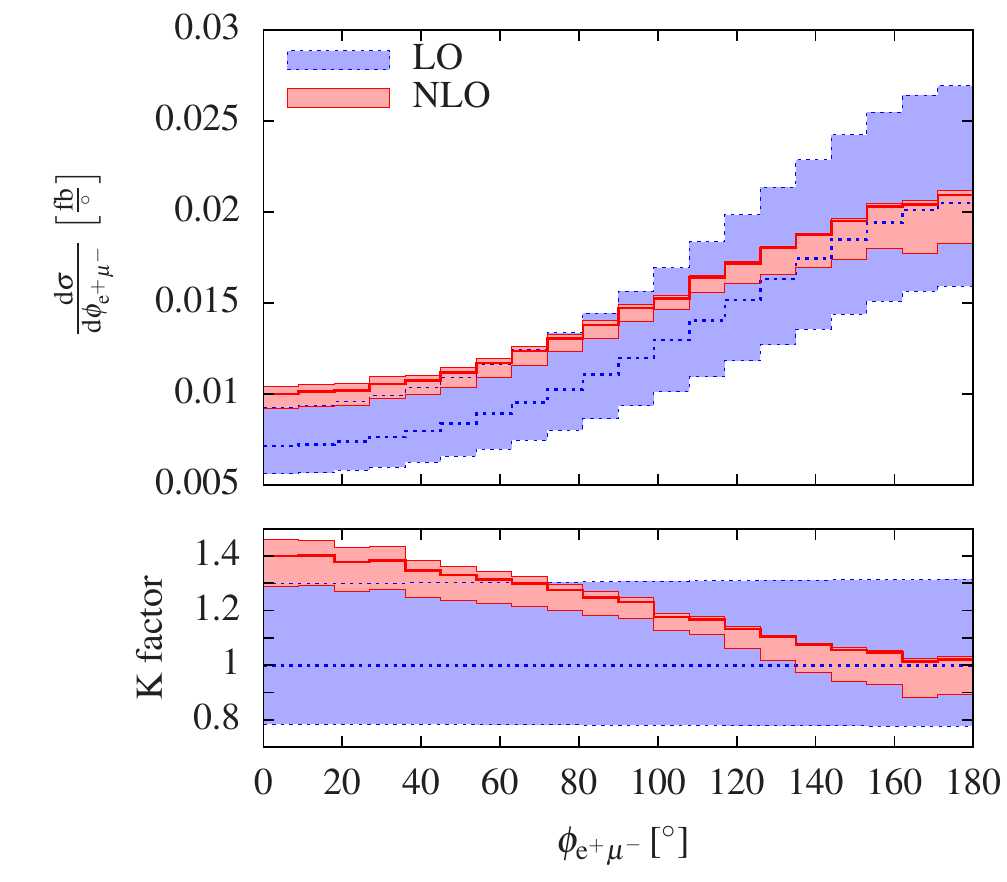}%

        \caption{\label{fig:WWbbH-distributions}
                Distributions at the LHC at $\sqrt{s}=13\UTeV$ for dynamical scale $\mu_0=\mu_\text{dyn}$:
                for the transverse momentum of the positron~(upper left), %
                for the  transverse momentum of the \PQb-jet
                pair~(upper right), %
                for the  transverse momentum of the Higgs boson~(lower
                left) and %
                the azimuthal angle between the positron and the muon in the transverse plane~(lower right), %
                The lower panels show the $K$~factor.}
\end{figure}%
Four differential distributions are shown in
\refF{fig:WWbbH-distributions} for the dynamical scale choice
\eqref{eqn:DynamicalScale}.  The upper panels show the LO (blue,
dashed) and NLO (red, solid) predictions with uncertainty bands from
scale variations. The lower panels display the LO (blue) and NLO (red)
predictions with scale uncertainties normalized to the LO results at
the central scale, i.e.  $K_\text{LO}=\text{d}\sigma_\text{LO}(\mu)/
\text{d}\sigma_\text{LO}(\mu_0)$ and
$K_\text{NLO}=\text{d}\sigma_\text{NLO}(\mu)/
\text{d}\sigma_\text{LO}(\mu_0)$.  Thus, the central red curve
corresponds to the usual NLO correction factor ($K$ factor), defined
as $K=\sigma_\text{NLO}(\mu_0)/\sigma_\text{LO}(\mu_0)$. The blue band
shows the relative scale uncertainty of the LO differential cross
section. The scale uncertainties are determined as explained at the
end of Section~\ref{sec:tthnlo_off_setup}.

The upper left plot of \refF{fig:WWbbH-distributions} shows the
transverse-momentum distribution of the positron. Using the dynamical
scale, the $K$~factor changes only slightly (within 20\,\%) over the
displayed range, and the NLO band lies within the LO band.  The
residual scale variation is at the level of $10\,\%$ at NLO.  This
behaviour is typical for most other distributions (see
\Bref{Denner:2015yca}).
A notable exception is the distribution in the transverse momentum of
the \PQb-jet pair (upper right in \refF{fig:WWbbH-distributions}), where we
observe an increase of the $K$ factor for high transverse momentum to
a value of 1.8 at $\pt\simeq400\UGeV$. This originates from the fact
that this region is suppressed for on-shell top quarks, an effect
known already from $\PQt\bar\PQt$ production, where it is even more pronounced
\cite{Denner:2012yc}.
Using the fixed scale \eqn{eqn:FixedScale-topdecays-nlo} instead leads to a much
larger variation of the relative corrections and the NLO prediction
moves outside the LO band in the  high-$\pt$ tails.

The lower left panel in \refF{fig:WWbbH-distributions} displays the
transverse-momentum distribution of the Higgs boson. The average $\pt$
of the Higgs boson is around 70\UGeV. The cross section decreases more
moderately with $\pt$ in the plotted range than for other
transverse-momentum distributions.
In the lower right panel of \refF{fig:WWbbH-distributions} the
distributions in the azimuthal angle in the transverse plane between
the two charged leptons are presented. It exhibits sizeable NLO effects for
small angles and the $K$ factor varies by $40\,\%$. NLO corrections of
similar size are found for the distribution in the cosine of the angle
between the two charged leptons.
Such large NLO effects in the kinematics
of top-decay products can be attributed to order $\als$
radiative corrections to top decays. It would be interesting to compare them
against conventional NLO+PS simulations, where such effects are modelled
by the parton shower.

\vskip 5mm

To summarize, this study, which includes NLO correction effects to the
top--antitop--Higgs-boson production and the top decay processes,
showed that for the inclusive investigated set-up, non-resonant and
off-shell top-quark effects are below one per cent. Including NLO
corrections, the scale
uncertainty is reduced to $5\,\%$ for the integrated cross section and to
the level of $10\,\%$ for distributions.

\subsection[Background and interference effects:
  \texorpdfstring{$\Pl\nu+\Pj\Pj+\PQb\protect\bar{\PQb}\PQb\protect\bar{\PQb}$}{lnu+jj+bbbb}
  production at LO QCD]{Irreducible background and interference effects:
  \texorpdfstring{$\Pl\nu+\Pj\Pj+\PQb\protect\bar{\PQb}\PQb\protect\bar{\PQb}$}{lnu+jj+bbbb}
  production at LO QCD}

A LO analysis of Higgs boson production in association with a
top-quark pair at the LHC investigating the semi-leptonic final state
consisting of four b jets, two jets, one identified charged lepton and
missing energy is presented.  All the various contributions of order
$\als^k\alpha^{4-k}$ with $k=0,1,2,3$ to the LO matrix elements and
their interferences are taken into account.  The Standard Model
predictions in three scenarios are considered: the resonant Higgs boson plus
top-quark-pair production, the resonant production of a top-quark pair
in association with a b-jet pair and the full process including all
non-resonant and interference contributions.  By comparing these
scenarios the irreducible background for the production
rate and several kinematical distributions is examined.
More details of this study can be found in \Bref{Denner:2014wka}.

\subsubsection{Setup of the analysis}
\label{sec:ttH_background_setup}

The full LO process $\fullProcess$ involves partonic channels with up
to 78,000 diagrams. All matrix elements are calculated with
\Recola~\cite{Actis:2012qn} which provides a fast and numerically
stable computation. \Recola\, uses recursive methods and allows to
specify intermediate particles for a given process.
The complex-mass scheme
\cite{Denner:1999gp,Denner:2005fg,Denner:2006ic} is used for the
consistent description of all resonances that are not treated in the
pole approximation. If resonant particles are required as intermediate
states, these are treated in the pole approximation
\cite{Kleiss:1988xr,Stuart:1991xk,Aeppli:1993rs}, i.e.\ nonresonant
contributions are neglected and the momenta are projected such that
the corresponding production and decay matrix elements are on shell
and gauge invariant.

Cross section and differential distributions are investigated for the
LHC operating at $13\UTeV$.  LHAPDF 6.05 with CT10 parton
distributions (which is an NLO PDF set) is employed and contributions from the suppressed
bottom-quark parton density and flavour mixing are neglected. The
strong coupling $\als$ is taken from the PDF set and the
electromagnetic coupling $\alpha$ is derived from the Fermi constant
in the $G_\mu$ scheme. The width of the top quark $\Gt$ is calculated
at LO QCD including effects of off-shell W bosons according to
\Bref{Jezabek:1988iv}.

Three scenarios to calculate the process $\fullProcess$ are considered:
\begin{itemize}
\item In the first scenario, the \textit{full process}, all SM
  contributions to the process $\fullProcess$ are included.
  Matrix elements involving external gluons receive contributions of
  $\order{\als\alpha^3}$, $\order{\als^2\alpha^2}$ and
  $\order{\als^3\alpha}$, whereas amplitudes without external
  gluons receive an additional $\order{\alpha^4}$ term of pure
  electroweak origin.  In this scenario many different resonant
  subprocesses contribute, such as $\PQt\PQt\PH$,
  $\PQt\PQt\PZ/\gamma^*$, or $\PQt\PQt\PQb\PQb$ production, as well as
  single $\PW$ or
  $\PW+{}$multiboson production in association with light and/or b~jets.
\item In the second scenario only those diagrams are taken into account
  that contain an intermediate top--antitop-quark pair. The resulting
  amplitude, labelled \textit{$\ttbarbbbar$ production} in the
  following, corresponds to the production of a bottom--antibottom
  pair and an intermediate top--antitop pair followed by its
  semileptonic decay, i.e.\ $\ttbarProcess$.
  Note that the pole approximation is used
 for the top quarks only,
  hence all off-shell effects of the remaining
  unstable particles are taken into account.
  As a consequence of the
  required top--antitop-quark pair the amplitudes receive no
  contribution of $\order{\als^3\alpha}$. In this scenario only the resonant
  subprocesses $\ttbarh$,
  $\PQt\bar\PQt\PZ/\gamma^*$, and $\ttbarbbbar$ production contribute.
\item Finally, the signal process $\ttbarhProcess$ is considered and
  labelled $\ttbarh$ \textit{production}. In addition to the
  intermediate top--antitop-quark pair an intermediate Higgs boson
  decaying into a bottom--antibottom-quark pair is required and the
  pole approximation is used for the top-quark pair and the Higgs boson.  
  The requirement of the Higgs boson eliminates contributions
  of $\order{\als^2\alpha^2}$ and all resonant subprocesses other
  than $\ttbarh$
  from the amplitude.
\end{itemize}

The bottom quarks are taken massive, and a fixed
renormalization and factorization scale is used according to
\Bref{Beenakker:2002nc},
\begin{equation}\label{eqn:FixedScale-irrback-lo}
        \mu_\text{fix} = \mu_\text{R} = \mu_\text{F} = \frac{1}{2}\left(2\mt + M_{\PH}\right) = 236\UGeV.
\end{equation}
While this hard scale choice is appropriate for $\ttbarh$ production,
it tends to underestimate $\ttbarbbbar$ production and other QCD contributions
by a factor 2 or more \cite{Bredenstein:2010rs}, leading to a low
$\ttbarbbbar/\ttbarh$ ratio.  In the future one should consider an improved scale
choice for the QCD contributions.

The following analysis requires  4~b~jets, 2~jets and one charged
lepton within the following acceptance cuts:
\begin{equation}\label{eqn:cuts-irrback-lo}
        \begin{aligned}
                \text{non-b jets:}                        && p_{\text{T},\Pj}         &>  25\UGeV,  & |y_{\Pj}| &< 2.5, \hspace{15ex}\\
                \text{b jets:}                      && p_{\text{T},\PQb}         &>  25\UGeV,  & |y_{\PQb}| &< 2.5,              \\
                \text{charged lepton:}              && p_{\text{T},\Pl^+  }         &>  20\UGeV,  & |y_{\Pl^+}|   &< 2.5,              \\
                \text{missing transverse momentum:} && p_{\text{T},\text{miss}} &>  20\UGeV,                                        \\
                \text{jet--jet distance:}           && \Delta R_{\Pj\Pj}        &> 0.4,                                            \\
                \text{b-jet--b-jet distance:}       && \Delta R_{\PQb\PQb}        &> 0.4,                                            \\
                \text{jet--b-jet distance:}         && \Delta R_{\Pj\PQb}        &> 0.4.
        \end{aligned}
\end{equation}

\subsubsection{Background and interference contributions in integrated cross sections}
\label{sec:ttH_background_crosssections}

In \refT{table:results_full_matrixelement_summary} individual contributions to the
integrated cross section for the three scenarios are presented.  While
the first column specifies the scenario, the following columns contain
the contributions resulting from the square of matrix elements of
specific orders in the strong and electroweak coupling.  The column
labelled ``Sum'' represents the sum of the preceding columns, whereas
the column labelled ``Total'' provides the integrated cross section
including in addition all interference effects between different
orders in the couplings, and the last column labelled ``Int'' gives
the interference contributions in per cent of the ``Total''.
\begin{table}\small
\setlength{\tabcolsep}{1.0ex}
        \centering
        \renewcommand\arraystretch{1.3}
        \setlength{\tabcolsep}{5pt}
        \caption{\label{table:results_full_matrixelement_summary}
                Composition of the cross section in fb for \ttbarh production,
                $\ttbarbbbar$ production and the full process at the
                LHC at $13\UTeV$.}
        \begin{tabular}{lccccccc}
                \toprule
                  \text{scenario}
                & \multicolumn{6}{l}{\text{Cross section [fb]}}\\
                & \multicolumn{1}{c}{$\order{(\alpha^4)^2}$}%
                & \multicolumn{1}{c}{$\order{(\als\alpha^3)^2}$}%
                & \multicolumn{1}{c}{$\order{(\als^2\alpha^2)^2}$}%
                & \multicolumn{1}{c}{$\order{(\als^3\alpha)^2}$}%
                & \multicolumn{1}{c}{Sum}%
                & \multicolumn{1}{c}{Total}%
                & \multicolumn{1}{c}{Int}\\

                \midrule
                $\ttbarh$      & 0.014887(2)
                & 7.377(1)  & --- & ---    &  7.3920(9) &  7.3920(9) &
                --- \\
                $\ttbarbbbar$      &  0.018134(6)
                & 10.311(4)  & 17.570(9)  & --- &  27.90(1)   &
                26.446(7) & -5.2(3)\% \\
                full process           &  0.02120(3)
                &  10.87(2)   &  18.69(6)   & 0.516(2)
                &  30.10  (6)   &  28.60 (6) & -5.50(5)\% \\
                \bottomrule
        \end{tabular}
\end{table}

For \ttbarh production (2nd row of
\refT{table:results_full_matrixelement_summary}), where about $70\,\%$
of the events originate from the gluon-fusion process, the bulk of the
contributions results from matrix elements of order
$\order{\als\alpha^3}$, quark--antiquark annihilation receives an
additional tiny contribution from pure electroweak interactions. Note
that there are no interferences between diagrams of $\order{\alpha^4}$
and $\order{\als\alpha^3}$ in this scenario.

For $\ttbarbbbar$ production (3rd row of
\refT{table:results_full_matrixelement_summary}), the production rate is
significantly enhanced compared to \ttbarh production, and thus the
irreducible background $\sigma^{\rm Irred.}_{\ttbarbbbar} =
\sigma^{\rm Total}_{\ttbarbbbar} - \sigma^{\rm Total}_{\ttbarh} =
19.06\fb$ exceeds the \ttbarh signal by a factor of 2.6. The major
contribution to the irreducible background is due to QCD production of
$\order{(\als^2\alpha^2)^2}$.  The additional contributions of
$\order{(\als\alpha^3)^2}$ in the $\ttbarbbbar$~scenario result
from Feynman diagrams involving electroweak interactions with
$\PZ$~bosons, $\PW$~bosons and photons, where in particular
$\Pt\bar{\Pt}\PZ$ production contributes $1.01\fb$.
The difference between the sixth (Sum) and seventh (Total) column in
\refT{table:results_full_matrixelement_summary} is due to interference contributions
between matrix elements of different orders in the coupling constants.
These cause a reduction of the cross sections by about $5\,\%$ with
respect to the full cross section. With respect to \ttbarh production
or $\ttbarbbbar$ production at order $(\als^2\alpha^2)^2$ this
amounts to $20\,\%$ and $8\,\%$, respectively.  The dominant effect is
due to interferences of diagrams of $\order{\als\alpha^3}$ where a
W~boson is exchanged in the $t$-channel (see
\refF{fig:ttH_feynman_diagram} for an example) with diagrams of
$\order{\als^2\alpha^2}$ that yield the dominant irreducible
background.
\begin{figure}
\centerline{
\includegraphics[width=.3\linewidth]{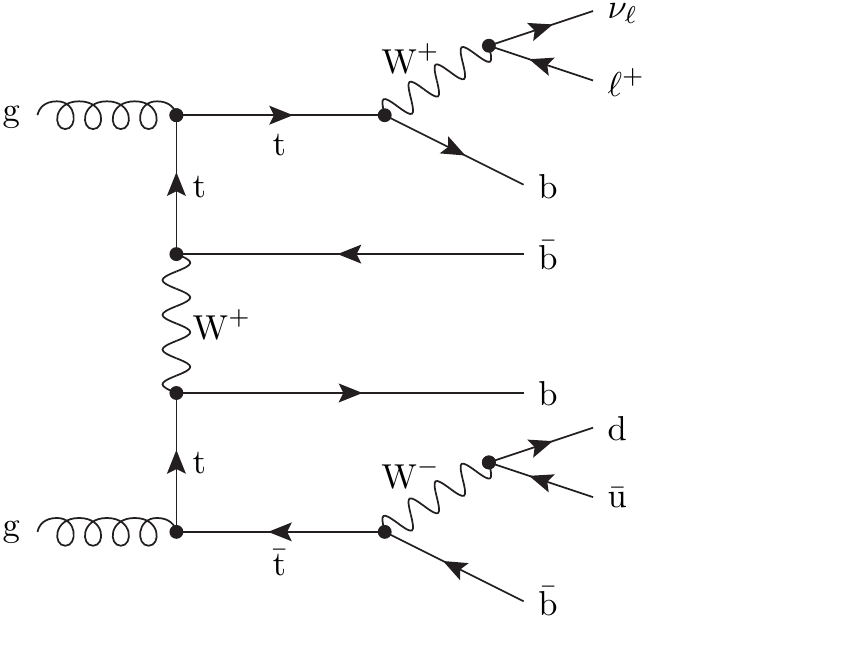}
}
        \caption{\label{fig:ttH_feynman_diagram}%
                Representative Feynman diagram that gives rise to
                large interferences with the $\ttbarbbbar$ production
                diagrams of order $\order{\als^2\alpha^2}$.}
\end{figure}
 These kinds of
interferences are absent in the $\qqb$ channel.  On the other hand,
the interference of the $\Pt\bar\Pt\PH$ signal process with the
dominant irreducible background of order $\order{\als^2\alpha^2}$
is below one per cent.

The results for the full process are listed in the last row of
\refT{table:results_full_matrixelement_summary}.  Here, additional
partonic channels ($\Pg q$, $\Pg \bar{q}$, $q q^{(\prime)}$)
contribute about $5\,\%$.  Nevertheless, the cross section increases
by merely $8\,\%$ relative to $\ttbarbbbar$ production.  With respect
to \ttbarh production or $\ttbarbbbar$ production at order
$(\als^2\alpha^2)^2$ this, however, amounts to $30\,\%$ and $12\,\%$,
respectively.  The contributions of order
$\order{(\als^3\alpha)^2}$ are below $2\,\%$ and the interference
pattern is similar to the case of $\ttbarbbbar$ production.

\subsubsection{Background contributions to differential distributions}
\label{sec:ttH_background_distributions}

Turning to differential distributions for the three
scenarios, results for the full process are compared with $\ttbarbbbar$
production and $\ttbarh$ production to assess the irreducible
background to $\ttbarh$ production.  The upper panels
in each plot of \refF{fig:ttH_background_distributions}
 show the differential distribution of the full process
with a black solid line, \ttbarbbbar production with a dashed blue
line and \ttbarh production with a dotted red line. The lower panels
provide the ratio of \ttbarbbbar production to the full process with a
dashed blue line and the ratio of \ttbarh production to the full
process with a dotted red line.

Motivated by \Bref{ATLAS:2012cpa} the two b jets that most likely
originate from the decay of the top quark ($\PQt\to\PW^+
\PQb\to\Pl^+\nu_{\Pl}\PQb$) and antitop quark
($\bar{\Pt}\to\PW^-\bar{\PQb}\to\bar{u}d\bar{\PQb}$, with $u=\PQu,
\PQc$, $d=\PQd, \PQs$) are determined and the invariant mass of the
remaining b-jet pair is plotted.  Since in most events the top quark
and antitop quark in \ttbarh production are nearly on-shell, the two b
jets maximizing the corresponding propagator contributions are most
likely to originate from the top-quark and antitop-quark decay. To
determine the maximizing b-jet combination a top-momentum
candidate is computed with the charged lepton, neutrino and a b-jet momentum
($p_{\text{b}_i}$),
\begin{equation}
        p_{\Pl^+\nu_{\Pl} \text{b}_i} = p_{\Pl^+} + p_{\nu_{\Pl}} + p_{\PQb_i}
\end{equation}
and an antitop-momentum candidate with the two momenta of
the non-b jets and a
different b-jet momentum ($p_{\PQb_j}$),
\begin{equation}
        p_{\text{j}_1\text{j}_2\text{b}_j} = p_{\text{j}_1} + p_{\text{j}_2} + p_{\text{b}_j}.
\end{equation}
As b jets originating from the top-quark and antitop-quark decay we
select those that maximize the likelihood function $\mathcal{L}$
defined as a product of two Breit--Wigner distributions corresponding
to the top-quark and antitop-quark propagators:
\begin{equation}
        \mathcal{L} \propto \left[\Bigl(p_{\Pl^+\nu_{\Pl} \PQb_i}^2 \!- \mt^2\Bigr)^2+\left(\mt\Gt\right)^2\right]^{-1} \left[\Bigl(p_{\text{j}_1\text{j}_2\text{b}_j}^2\!- \mt^2\Bigr)^2+\left(\mt\Gt\right)^2\right]^{-1}.
\end{equation}

\refF{fig:ttH_background_distributions} upper left presents the
b-jet-pair invariant mass that has been identified to originate from
the Higgs boson decay by the maximum-likelihood method described
above.  In the off-shell region the ratio of \ttbarh production to the
full process drops considerably below the corresponding ratio for the
total cross section of about a fourth.  Since this method tags the
b~jets resulting from the top and antitop quarks, any resonance in the
invariant mass of the remaining b-jet pair is resolved, and thus the
Z~resonance is clearly visible in the plot.  Note, however, that this
analysis uses an idealized $\ttbarbbbar$ reconstruction. In
practice, multi-jet emissions, finite jet-energy resolution, and
light-jet mistagging lead to a severe dilution of $\PQb\bar\PQb$
resonances.

\begin{figure}
  \includegraphics[width=.49\linewidth]{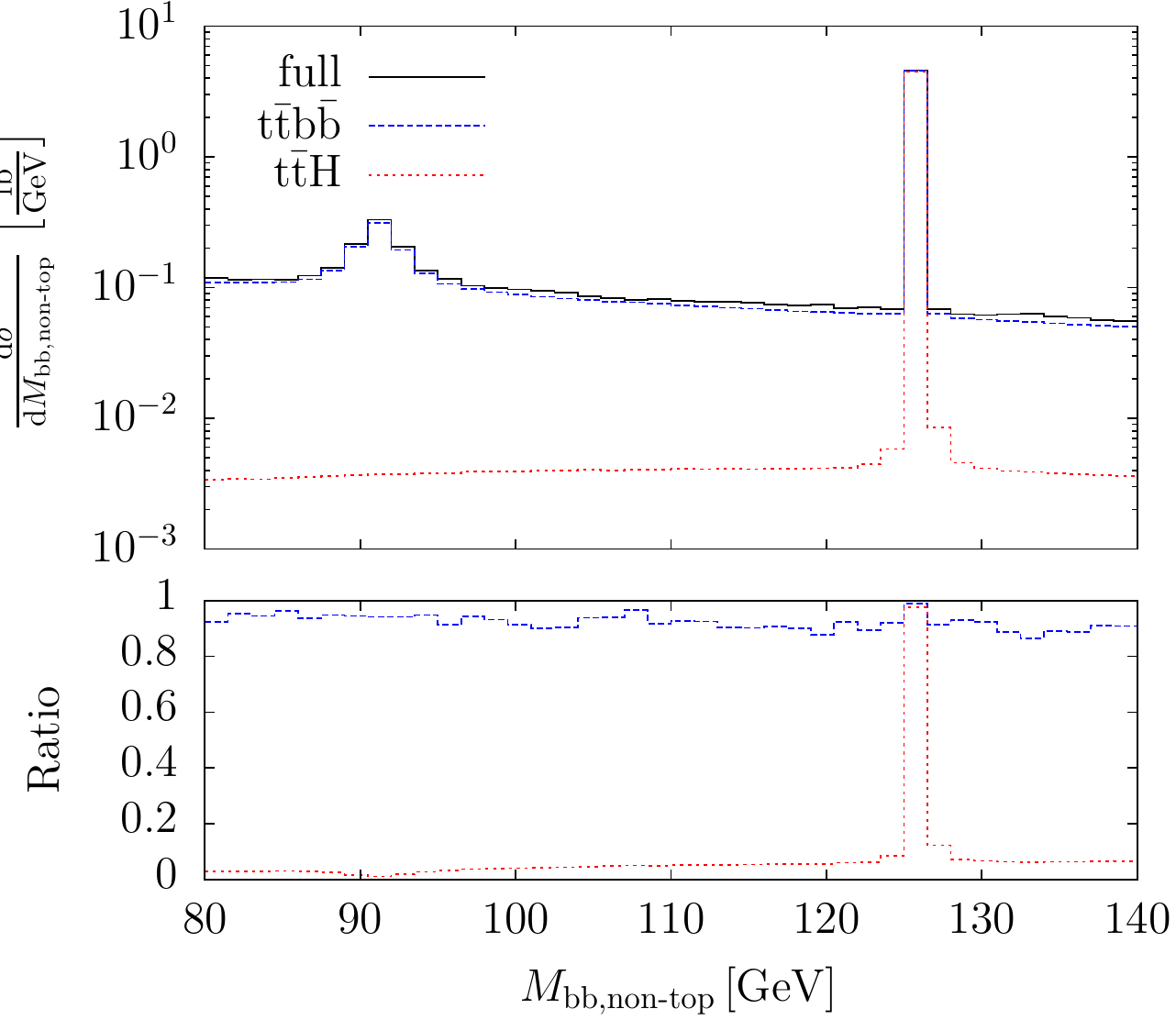}
        \hfill
  \includegraphics[width=.49\linewidth]{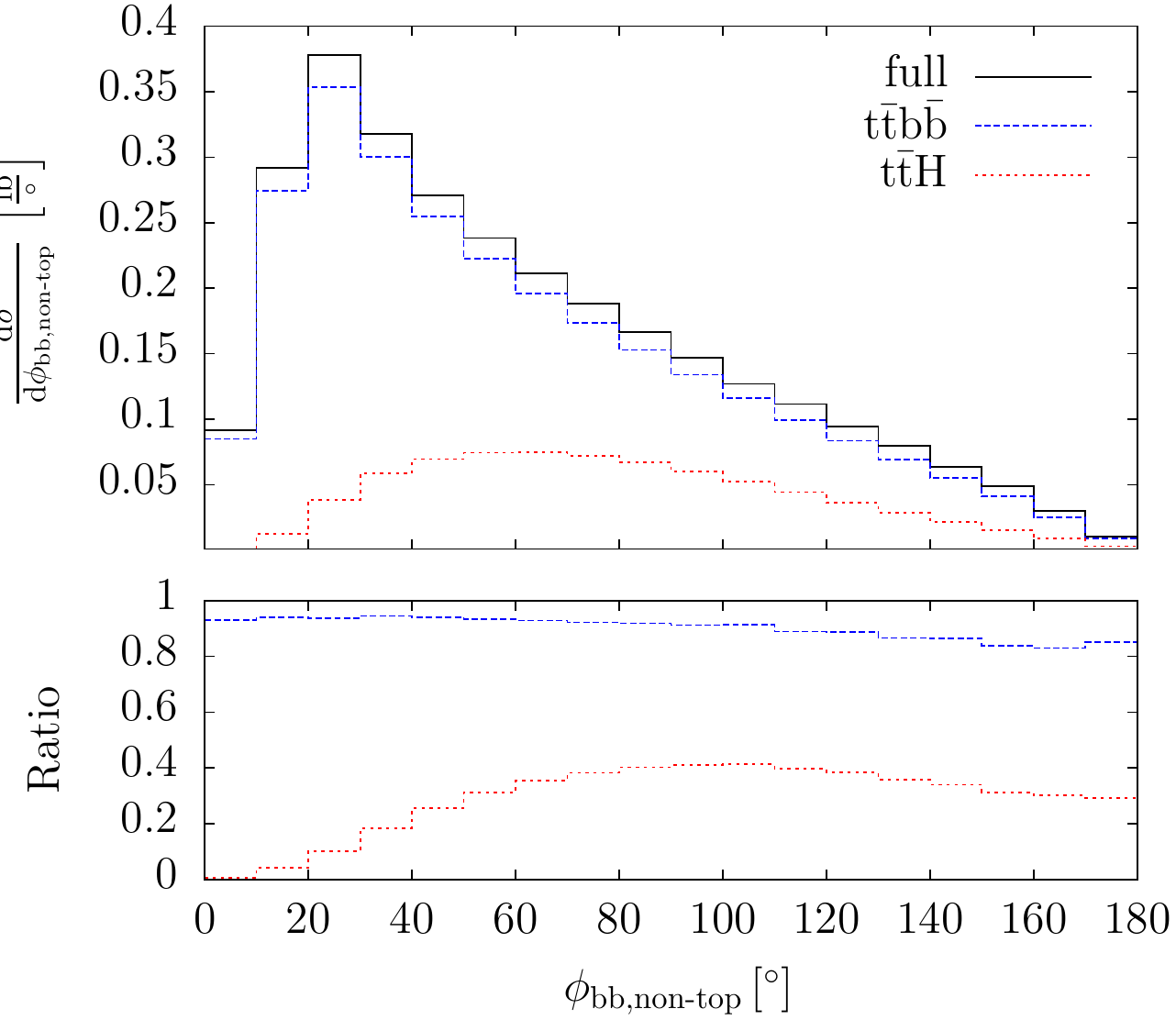}

  \includegraphics[width=.49\linewidth]{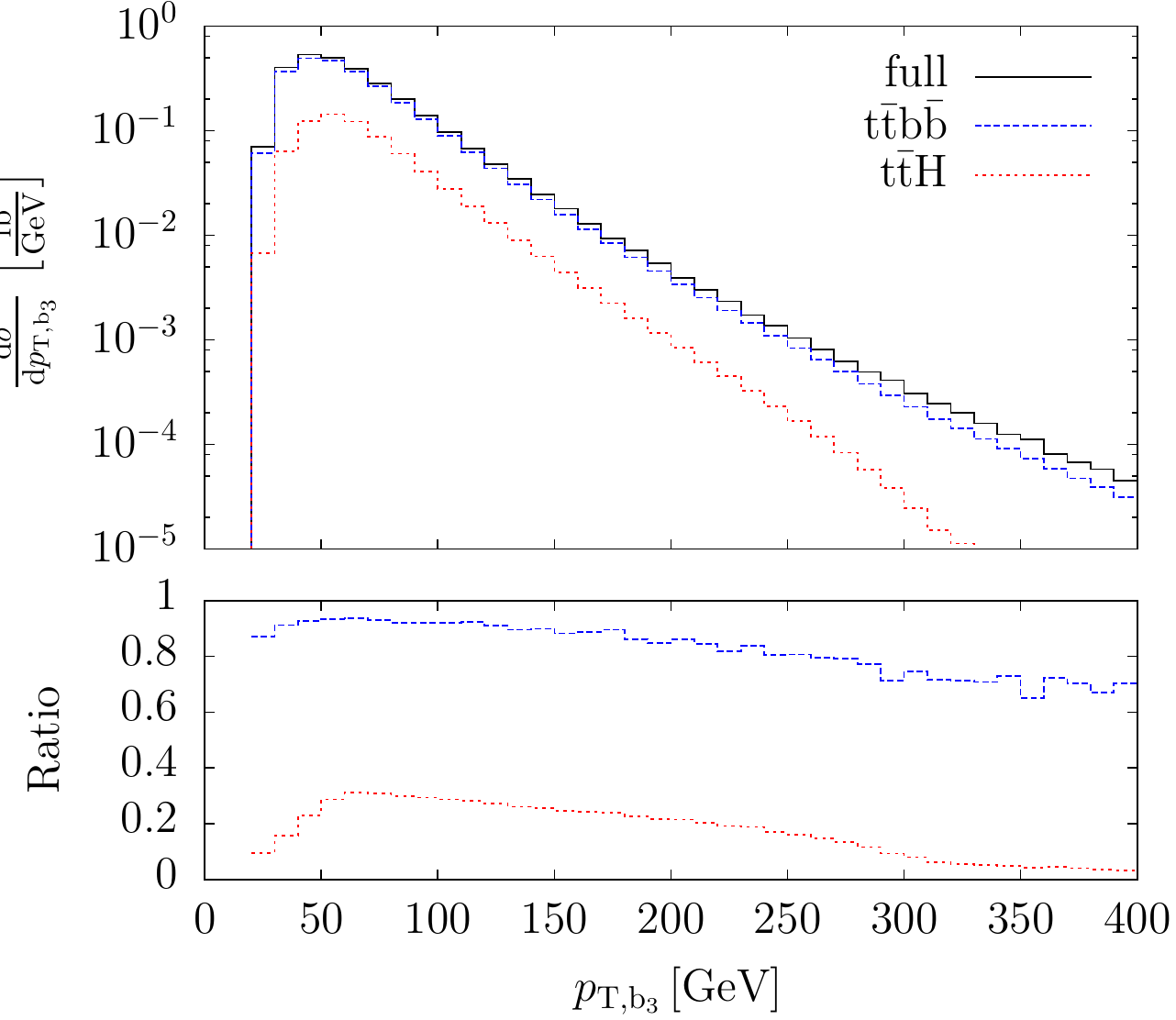}
        \hfill
  \includegraphics[width=.49\linewidth]{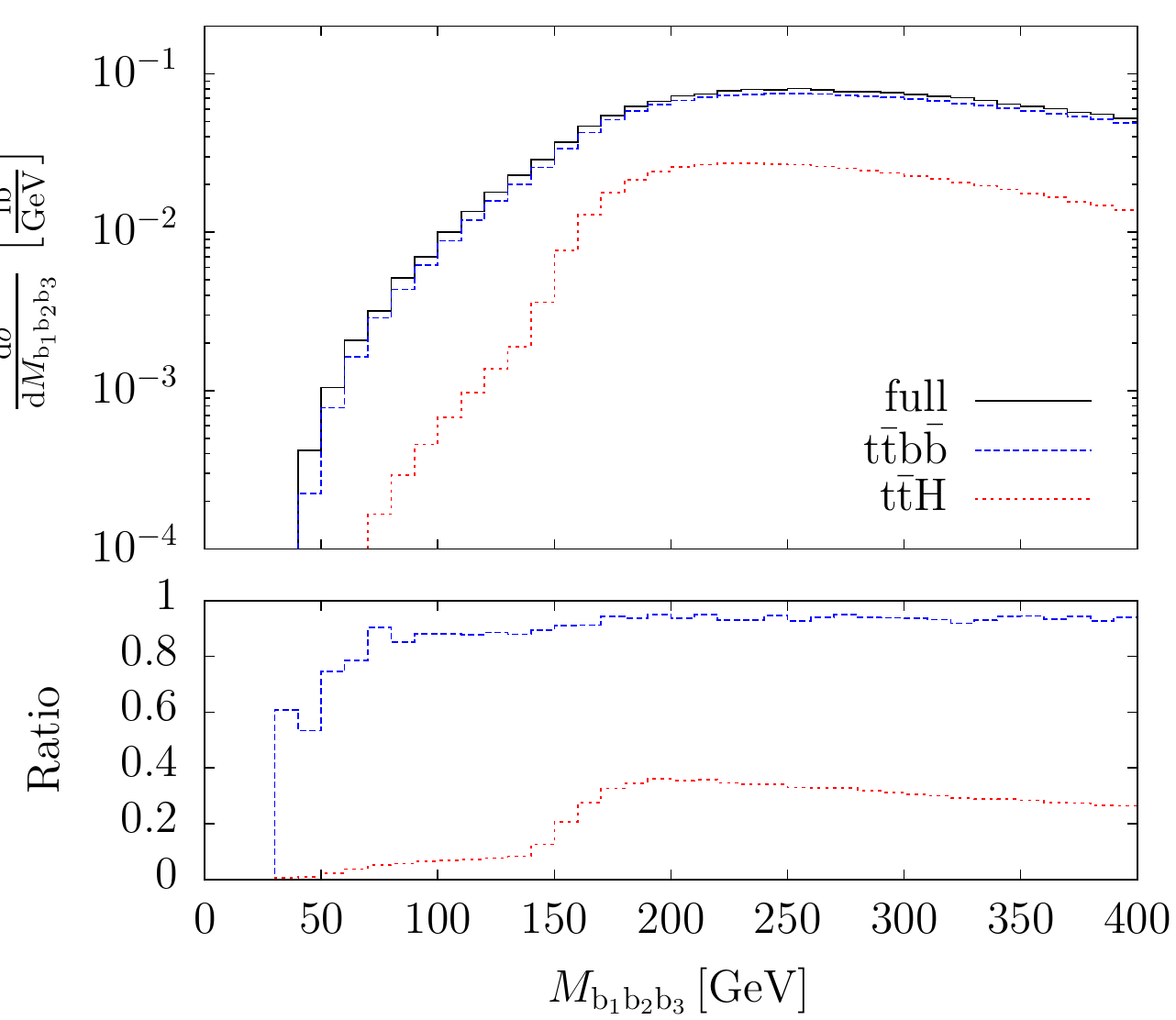}
        \caption{\label{fig:ttH_background_distributions}
                Differential distributions at the LHC at $13\UTeV$ for
                the three different scenarios: invariant-mass
                distribution of the b-jet pair %
                determined by top--antitop Breit--Wigner maximum
                likelihood (upper left),
                azimuthal separation of this b-jet pair (upper right), %
                transverse momentum of the 3rd-hardest \PQb~jet (lower left), %
                and invariant mass of the three hardest \PQb~jets (lower right). %
                The lower panels show the relative size of $\ttbbb$ and $\ttbarh$ production normalized to the full process.}
\end{figure}

In the following distributions that show deviations in the shape
between the full process and the $\ttbbb$ and $\ttbarh$ sub-processes
are discussed.
\refF{fig:ttH_background_distributions} upper right shows the
azimuthal separation of the b-jet pair determined by top--antitop
Breit--Wigner maximum likelihood. While \ttbarbbbar production and the full
process yield a very similar shape, \ttbarh production exhibits
clearly a different shape.  This behaviour can be explained by the
dominant production mechanisms of bottom--antibottom pairs. In the
signal process these result from the Higgs boson and owing to the
finite Higgs boson mass tend to have a finite opening angle. In the
background processes the bottom--antibottom pairs result mainly from
gluons and thus tend to be collinear leading to a peak at small
$\phi_{\PQb\PQb}$ that is cut off by the acceptance function.  Thus,
this distribution can help to separate bottom--antibottom pairs
resulting from Higgs bosons from those of other origin.

\refF{fig:ttH_background_distributions} lower left displays the
transverse-momentum distribution of the third-hardest b jet. Here all
three approximations are similar in shape for $p_\text{T}$~values
below $150\UGeV$. For higher transverse momenta the distribution for
\ttbarh production diverges from those of the full process and
\ttbarbbbar production. This behaviour is not found in the transverse
momentum distributions of the two harder b~jets but to some extent in
the one of the fourth-hardest b jet. This results from the fact that
in the $\ttbarh$ signal all b~jets originate from heavy-particle
decays, while in the full process some are directly produced yielding
more b~jets with high transverse momenta.

Finally, \refF{fig:ttH_background_distributions} lower right presents the
invariant mass of the three hardest b jets. Below the threshold
$\MH+p_{\mathrm{T},\PQb,\mathrm{cut}}\approx150\UGeV$ the signal process
is strongly suppressed, above its ratio to the full process rises to
$36\,\%$ at $M_{\mathrm{b_1b_2b_3}}\sim 195\UGeV$ and then drops slowly
to $26\,\%$ at $M_{\mathrm{b_1b_2b_3}}\sim 400\UGeV$. The ratio of
\ttbarbbbar production and the full process on the other hand is
roughly constant for $M_{\mathrm{b_1b_2b_3}}\gsim 70\UGeV$.

\subsubsection{Interference effects in differential distributions}
\label{sec:ttH_interference_distributions}

\begin{figure}
  \includegraphics[width=0.49\linewidth]{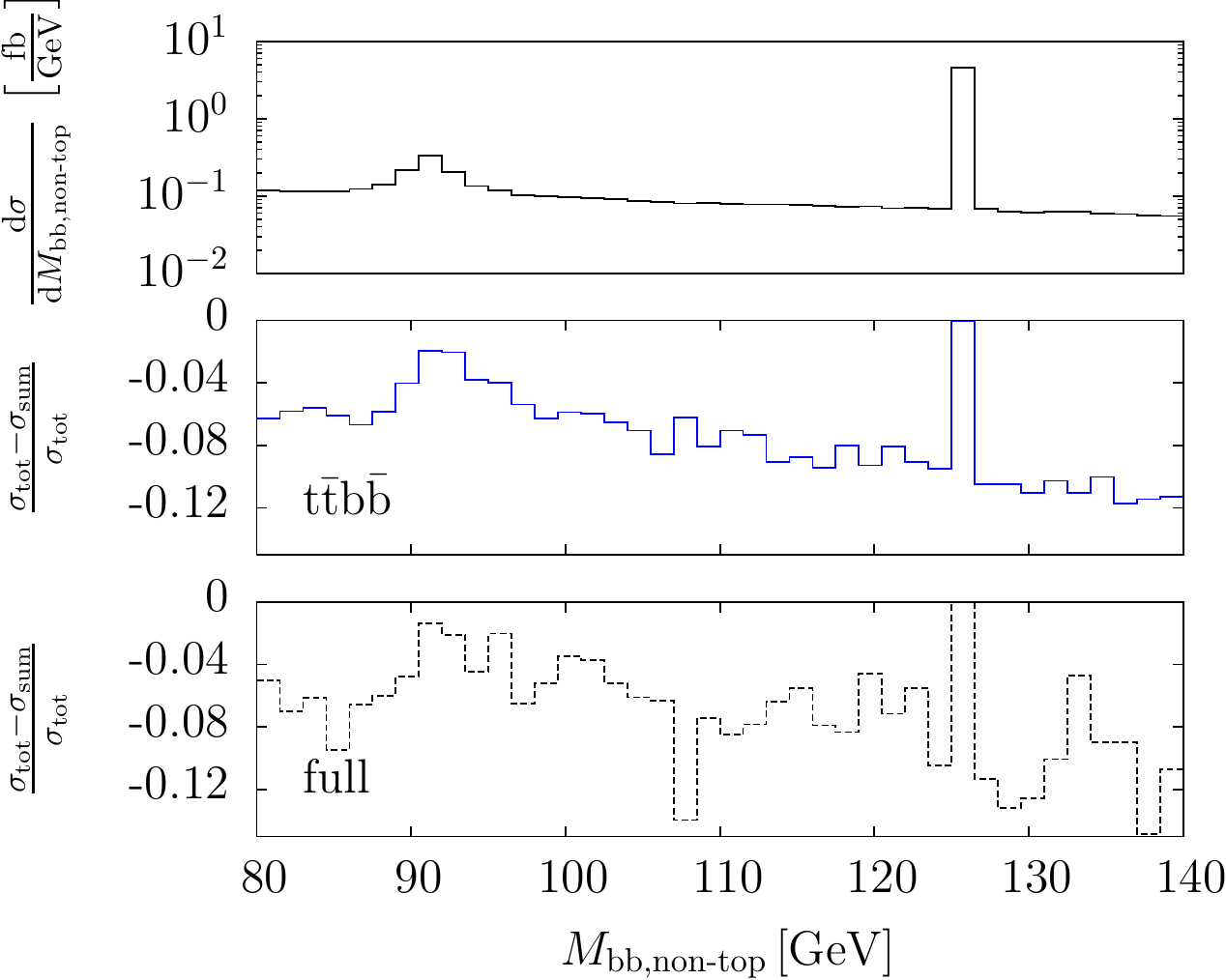}
  \hfill
  \includegraphics[width=0.49\linewidth]{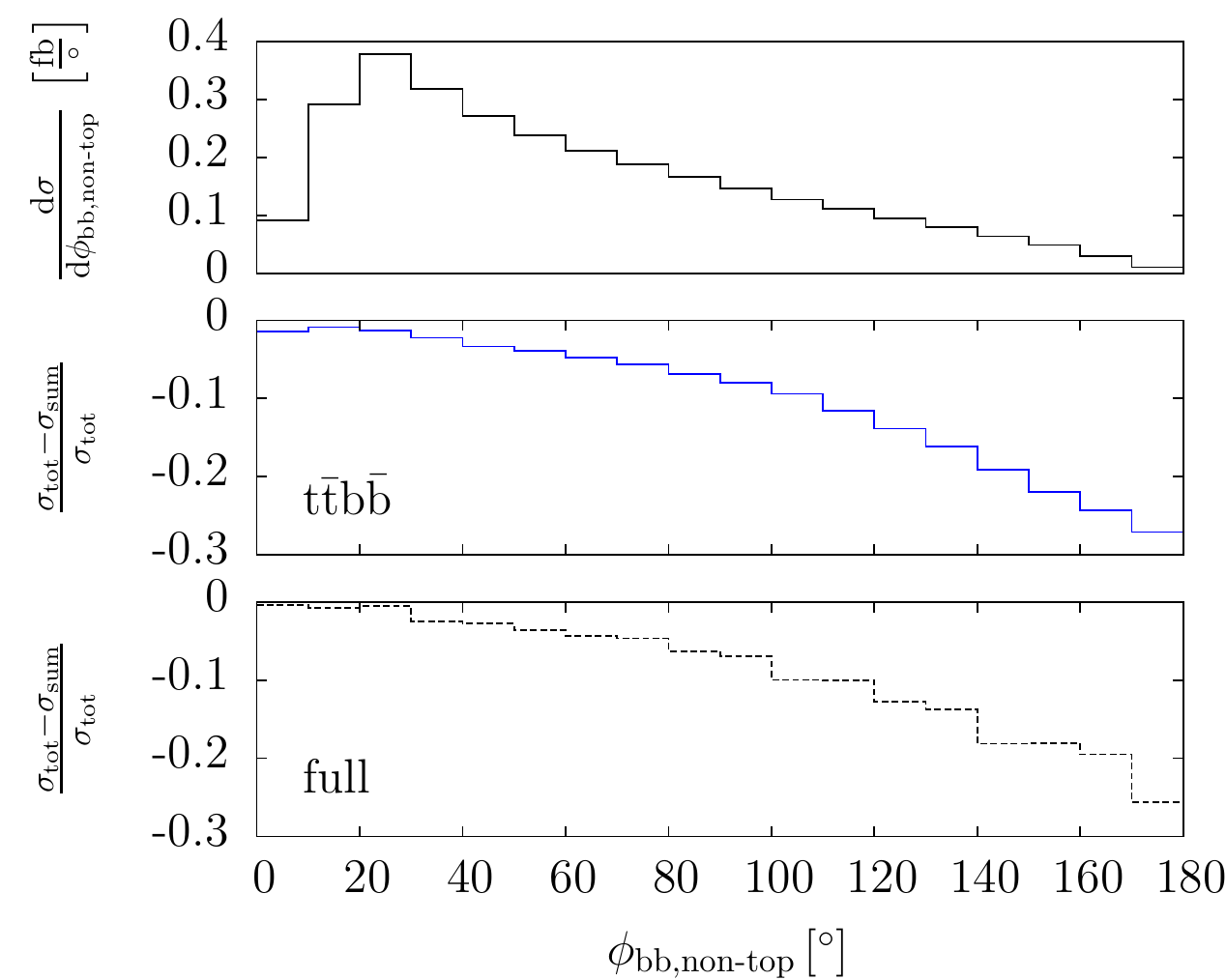}
  \caption{\label{plot:ttH_interference}
    Interference effects versus invariant mass of the b-jet
    pair determined by top--antitop Breit--Wigner maximum likelihood
    (left) and azimuthal separation of this b-jet pair  (right).  The lower
    panels show the relative interference effects of $\ttbbb$
    production and the full process, respectively.  The upper panel
    shows the corresponding differential distribution of the full
    process as reference.}
\end{figure}
In the following the effects of the interference contributions between
matrix elements of different orders in $\als$ are investigated.  For
most distributions a uniform shift by roughly the same amount as for
the total cross section, i.e.\ about $5\,\%$ for \ttbarbbbar
production and the full process (interference effects are absent in
\ttbarh production), is observed. For both scenarios a few kinematical
distributions are found that are sensitive to these interference
effects.  The upper panels of \refF{plot:ttH_interference} show the
results for the full process and the central and lower panels
highlight the interference effects.  Specifically, the central panels
show the relative difference
${(\sigma_\text{tot}-\sigma_\text{sum})}/{\sigma_\text{tot}}$ for
\ttbarbbbar production with a solid blue line and the lower panels the
same relative difference for the full process with a dashed line.

\refF{plot:ttH_interference} left shows the interference effects
on the distribution of the invariant mass of the b-jet pair determined
by top--antitop Breit--Wigner maximum likelihood. The suppression of
interference in the regions of the Higgs- and Z-boson resonances is
clearly visible. For invariant masses above the Higgs threshold the
interference effect exceeds $-10\,\%$.
As shown in \refF{plot:ttH_interference} right, the relative
interference effects grow with increasing azimuthal-angle separation
of the b-jet pair determined by top--antitop Breit--Wigner maximum
likelihood from almost zero at small angles to $-25\,\%$ for
$\phi_{\mathrm{bb,non\textnormal{-}top}}=180^\circ$, while the cross
section drops with increasing azimuthal-angle separation.

Finally it should be noted that in \Bref{Denner:2014wka} the
irreducible background and interference effects for various other
distributions have been investigated.

\vskip 5mm

To summarize, the cross section for  $\ttbarbbbar$ production agrees
with the one for the full process within $8\,\%$ and interferences
between matrix elements of different orders in the coupling constants
contribute about $5\,\%$ of the full process. Note, however, that these
effects amount to $30\,\%$ and $20\,\%$ of the Higgs-signal process.

}


\def\ttHrespath{\ttHpath/ttH-NLO-QCD+resummation}
\section{\texorpdfstring{$t\bar{t}H$}{ttH} production beyond NLO}
\label{sec:ttH-resummation}

A fixed-order computation of the NNLO QCD corrections to hadronic $\tth$
production is still beyond the technical reach of current higher-order
perturbative calculations. Nevertheless, their impact on the
theoretical predictions for both total cross sections and
distributions can be larger than the first order of EW corrections, and
their inclusion could substantially reduce the systematic dependence
on renormalization and factorization scales. It is therefore important
to investigate the possibility of gaining information on the NNLO QCD
corrections to $\tth$ production by studying particular sets of
higher-order corrections that can be calculated analytically from
first principles. Recently two
studies~\cite{Kulesza:2015vda,Broggio:2015lya} have explored the
effect of soft-gluon emission beyond NLO. A brief summary of the
methods used as well as the results obtained for the LHC at both 13
and 14~\UTeV\, are presented in Section~\ref{subsec:NLO+NLL} for
Ref.~\cite{Kulesza:2015vda} and in Section~\ref{subsec:nNLO} for
Ref.~\cite{Broggio:2015lya}.

In spite of the fact that both studies target the same kind of
radiative corrections, the two approaches are quite different. In
Ref.~\cite{Kulesza:2015vda} the soft-gluon corrections are calculated
in the absolute threshold limit \ie in the limit when the partonic
centre-of-mass energy approaches the square of the production
threshold energy ($2\mt+\mh$), while in Ref.~\cite{Broggio:2015lya} the
soft-gluon corrections are calculated in the final-state-invariant-mass
threshold limit \ie in the limit when the partonic centre-of-mass
energy approaches the square of the invariant mass of the $\tth$
state.

Moreover, while Ref.~\cite{Kulesza:2015vda} relies on the
classical Mellin resummation technique and performs an all-order
resummation of Next-to-Leading-Logarithms (NLL), Ref.~\cite{Broggio:2015lya}
uses techniques of Soft-Collinear Effective Theory (SCET) to obtain
approximate NNLO corrections from a truncated expansion of a
Next-to-Next-to-Leading-Logarithms (NNLL) resummed formula matched
with the full NLO fixed-order calculation.

Given these differences, a comparison of results between the two
studies is not obvious. The NLO+NLL results of
Section~\ref{subsec:NLO+NLL} and Table~\ref{tab:results}, and the NLO+approximate NNLO
results of Section~\ref{subsec:nNLO} and
Tables~\ref{tab:CSat620}-\ref{tab:LBCS} correspond to different
extensions of the NLO QCD results and cannot be directly compared,
even when they are evaluated at the same renormalization and
factorization scale, because they include different orders of
differently-defined soft-gluon corrections. What is however
interesting is that within the uncertainty from scale variation and
from corrections formally subleading in the soft limit, they both
overlap with the NLO QCD fixed-order calculation within its
uncertainty range.

It will be interesting to explore directions in which these studies can be used to further improve the theoretical understanding and accuracy of both Mellin and momentum-space resummations, and eventually use them to more systematically control the theoretical uncertainty on $\tth$ production at the LHC.

\subsection[NLO+NLL soft-gluon resummation in the partonic
  centre-of-mass theshold limit]{\texorpdfstring{$t\bar{t}H$}{ttH} production including NLO+NLL soft-gluon resummation in the partonic
  centre-of-mass theshold limit}
\label{subsec:NLO+NLL}

\subsubsection{Introduction}

In this section we discuss how to improve the NLO QCD calculation of $pp \to \tth$ at the LHC by adding the resummation of soft-gluon corrections in the partonic centre-of-mass threshold limit, performed using the Mellin space formalism~\cite{Kulesza:2015vda}.  This particular type of corrections arises due to soft-gluon emission in the threshold limit \ie when the production is considered close to the absolute threshold $\shat \sim M^2=(2\mt +\mh)^2$, where $\hat{s}$ is the partonic centre-of-mass energy, $\mt$ the top-quark mass, and $\mh$ the Higgs boson mass. In this region, the cross section receives enhancements in the form of logarithmic corrections in $\beta = \sqrt {1-M^2/\shat}$. The quantity $\beta$ measures the distance from the absolute production threshold and can be related to the maximal velocity of the $t\bar{t}$ system.

Threshold resummation in Mellin space has been so far well developed and copiously applied only to $2 \to 2$ processes. Its application to the $pp \to \tth$ process requires developing the formalism for the case of a $2 \to 3$ process with two coloured massive particles in the final state. While the universality of resummation concepts warrants their applications to scattering processes with many partons in the final state~\cite{Bonciani:2003nt, Aybat:2006wq}, the specifics of the colour structure and kinematics need to be taken into account in the process-dependent terms. In particular, the non-trivial colour flow between four coloured partons at the Born level influences the contributions from wide-angle soft-gluon emissions which have to be included at the next-to-leading-logarithmic (NLL) accuracy. The evolution of the colour exchange at NLL is governed by the one-loop soft anomalous dimension. Correspondingly, the application of the threshold resummation to the $pp \to \tth$ process requires the calculation of the one-loop soft anomalous dimension  for a $2 \to 3$ process with two coloured massive particles in the final state. An additional improvement of the calculation at the NLL accuracy is the inclusion of the ${\cal O}(\als)$ non-logarithmic threshold corrections originating from hard off-shell dynamics.

\subsubsection{Resummation at production threshold}

At the partonic level, the Mellin moments for the process $ ij \to kl B$, where $i$ and $j$ denote massless coloured partons, $k$ and $l$ two massive quarks, and $B$ a massive colour-singlet particle, are given by
\begin{equation}
\sigh_{ij \to kl B, N} (m_k, m_l, m_B, \mu_F^2, \mu_R^2) = \int_0^1 d \hat\rho \, \hat\rho^{N-1} \sigh_{ij \to kl B} (\hat \rho, m_k, m_l, m_B, \mu_F^2, \mu_R^2)\,,
\end{equation}
where $\hat \rho = 1- \beta^2=M^2/\shat$, and $\muR$ and $\muF$ denote the renormalization and factorization scales, respectively.

At LO, the $\tth$ production process receives contributions from the $q\bar{q}$ and $gg$ channels. Analysing the colour structure of the underlying processes in the $s$-channel colour bases, $\{ c_I^q\}$ and $\{c_I^g\}$, with \mbox{$c_{\bf 1}^q =  \delta^{\al_{i}\al_{j}} \delta^{\al_{k}\al_{l}},$}
\mbox{$c_{\bf 8}^{q} = T^a_{\al_{i}\al_{j}} T^a_{\al_{k}\al_{l}},$}
\mbox{$c_{\bf 1}^{g} = \delta^{a_i a_j} \, \delta^{\al_k
  \al_l}, $}
\mbox{$c_{\bf 8S}^{g}=  T^b _{\alpha_l \alpha_k} d^{b a_i a_j} ,$}
\mbox{$c^{g}_{\bf 8A} = i T^b _{\alpha_l \alpha_k} f^{b a_i a_j} $} ($\alpha_{i,j,k,l}=1,\ldots,N$, and $a,b,a_{i,j}=1,\ldots, N^2-1$, with $N=3$),
the soft anomalous-dimension matrix becomes diagonal in the production threshold limit~\cite{Kidonakis:1996aq}, and the NLL resummed cross section in the
$N$-space has the form~\cite{Kidonakis:1996aq,Bonciani:1998vc}
\begin{equation}
\label{eq:res:fact}
\sigh^{{\rm (res)}}_{ij\tosv kl B,N}\!\! =\!\! \sum_{I} \!\sigh^{(0)}_{ij\tosv
  klB,I,N}\, {C}_{ij\tosv kl B,I}\,
\Delta^i_{N+1} \Delta^j_{N+1}
\Delta^{\rm (int)}_{ij\tosv kl B,I,N+1}\,,
\end{equation}
where the explicit dependence on the scales is suppressed.  The index $I$ in Eq.~(\ref{eq:res:fact}) distinguishes between contributions from different colour channels. The colour-channel-dependent contributions to the LO partonic cross sections in Mellin-moment space are denoted by $\sigh^{(0)}_{ij\tosv klB,I,N}$. The radiative factors $\Delta^i_{N}$ describe the effect of the soft-gluon radiation collinear to the initial-state partons and are universal, see e.g.~\cite{Bonciani:1998vc} . Large-angle soft-gluon emission is accounted for by the factors $\Delta^{\rm (int)}_{ ij\tosv kl B,I,N}$ which are directly related to the soft-gluon anomalous dimension calculated in~\cite{Kulesza:2015vda}. As indicated by the lower indices, the wide-angle soft emission depends on the partonic process under consideration and the colour configuration of the participating particles. In the limit of absolute threshold production, $\beta \to 0$, the factors $\Delta^{\rm (int)}_{ ij\tosv kl B,I,N}$ coincide with the corresponding factors for the $2 \to 2$ process $ij \to kl$~\cite{Kulesza:2015vda}. All perturbative functions governing the radiative factors up to the terms needed to obtain NLL accuracy in the resummed expressions are considered.

The coefficients ${C}_{ij\tosv klB,I}= 1 + \frac{\als}{\pi} {C}^{(1)}_{ij\tosv klB,I}+ \dots$ contain all non-logarithmic contributions to the NLO cross section taken in the threshold limit. More specifically, these consist of Coulomb corrections, $N$-independent hard contributions from virtual corrections, and $N$-independent non-logarithmic contributions from soft emissions.  Although formally the coefficients $C_{ij\tosv kl B,I}$ begin to contribute at NNLL accuracy, in the numerical studies of the $pp \to \tth$ process presented in the following we consider both the case of $C_{ij\tosv kl B,I}=1$, i.e. with the first-order corrections to the coefficients neglected, as well as the case with these corrections included. In the latter case the Coulomb corrections and the hard contributions are treated additively, i.e.  ${C}_{ij\tosv klB,I}^{(1)}={C}_{ij\tosv klB,I}^{(1, \rm hard)}+{C}_{ij\tosv klB,I}^{(1, \rm Coul)}$.
For $k,l$ denoting massive quarks, and more specifically for $kl=t\bar{t}$,
the Coulomb corrections are ${C}_{ij\tosv klB,{\bf 1}}^{(1, \rm Coul)} = \CF \pi^2 /(2 \beta_{kl})=\CF \pi^2 /(2 \beta_{t\bar{t}})$ and ${C}_{ij\tosv klB,{\bf 8}}^{(1, \rm Coul)} = (\CF -\CA/2) \pi^2 /(2 \beta_{kl})=(\CF -\CA/2) \pi^2 /(2 \beta_{t\bar{t}})$, with $\beta_{t\bar{t}}=\sqrt{1- 4\mt^2/\shat_{t\bar{t}}}$, and $\shat_{t\bar{t}}=(p_t+p_{\bar{t}})^2$. As the $N$-independent non-logarithmic contributions from soft emission are accounted for using the techniques developed for the $2\to2$ case~\cite{Beenakker:2011sf,Beenakker:2013mva}, the problem of calculating the $C_{ij\tosv \tth,I}^{(1)}$ coefficients reduces to the calculation of the $O(\als)$ virtual corrections to the $\tth$ process. We extract them numerically using the publicly available \Powhegbox\,  implementation of the $\tth$ process~\cite{Hartanto:2015uka}, based on the calculations developed in~\cite{Reina:2001sf,Dawson:2002tg,Dawson:2003zu}. The results are then cross-checked using the standalone \Madloop~\cite{Hirschi:2011pa} implementation in MG5\_aMC@NLO. Since the $q\bar{q}$ channel receives only colour-octet contributions, the extracted value contributing to $C^{(1, {\rm hard})}_{q\bar{q} \tosv t \bar{t} H, {\bf 8}}$ is exact. In the $gg$ channel, however, both the singlet and octet production modes contribute. We extract the value which contributes to the coefficient $\bar{C}^{(1, {\rm hard})}_{gg \tosv t \bar{t} H}$ averaged over colour channels and use the same value to further calculate $C_{gg \tosv t \bar{t} H, {\bf 1}}$ and $C_{gg \tosv t \bar{t} H, {\bf 8}}$.

The resummation-improved NLO+NLL cross sections for the $pp \to \tth$ process are then
obtained through matching the NLL resummed expressions with the full NLO cross sections
\begin{eqnarray}
\label{eq:hires}
&& \si^{\rm (NLO+NLL)}_{pp \tosv \tth}(\rho, \muF^2, \muR^2)\! =\!
\si^{\rm (NLO)}_{pp \tosv \tth}(\rho,\muF^2, \muR^2) +
\si^{\rm  (res-exp)}_{pp \tosv \tth}(\rho, \muF^2, \muR^2) \nn \\
&&\!\!\!\!\!\!\!\!\!{\rm with} \nn \\
&& \si^{\rm  (res-exp)}_{pp \tosv \tth}  \! =   \sum_{i,j}\,
\int_{\cal C}\,\frac{dN}{2\pi  i} \; \rho^{-N}
 f^{(N+1)} _{i/p} (\muF^2) \, f^{(N+1)} _{j/p} (\muF^2) \nn \\
&& \! \times\! \left[
\sigh^{\rm (res)}_{ij\tosv \tth ,N} (\muF^2, \muR^2)
-  \sigh^{\rm (res)}_{ij\tosv \tth ,N} (\muF^2, \muR^2)
{ \left. \right|}_{\scriptscriptstyle({NLO})}\! \right],
\end{eqnarray}
where $\sigh^{\rm (res)}_{ij\tosv  \tth ,N}$ is given in Eq.~(\ref{eq:res:fact}) and  $ \sigh^{\rm
  (res)}_{ij\tosv \tth ,N} \left. \right|_{\scriptscriptstyle({NLO})}$ represents its perturbative expansion truncated at NLO.  The moments of the parton distribution functions (PDF) $f_{i/p}(x, \muF^2)$ are defined in the standard way $f^{(N)}_{i/p} (\muF^2) \equiv \int_0^1 dx \, x^{N-1} f_{i/p}(x, \muF^2)$.  The inverse Mellin transform (\ref{eq:hires}) is evaluated numerically using a contour ${\cal C}$ in the complex $N$-space according to the ``Minimal Prescription'' method developed in Ref.~\cite{Catani:1996yz}.
\begin{figure}[htb]
\centering
\includegraphics[width=0.45\textwidth]{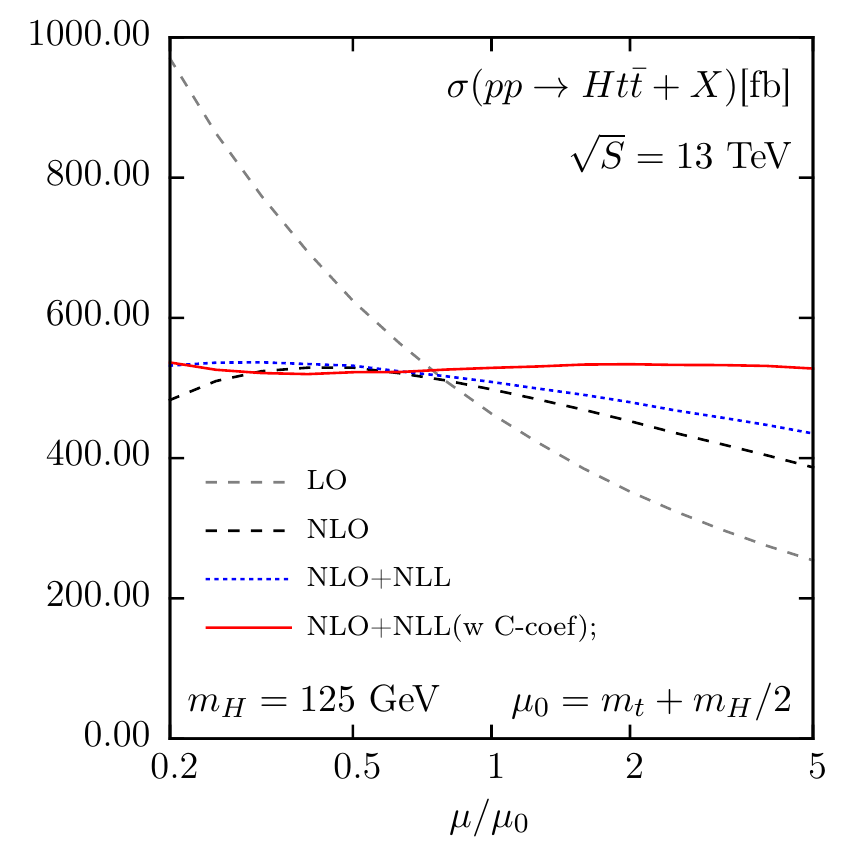}
\includegraphics[width=0.45\textwidth]{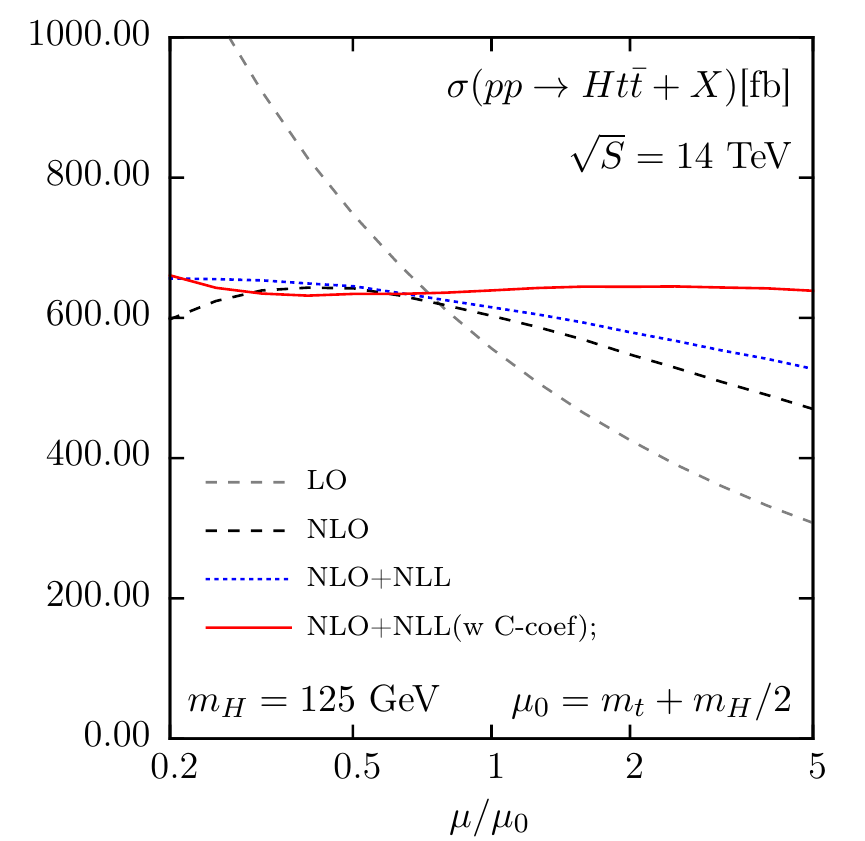}
\caption{Scale dependence of the LO, NLO, and NLO+NLL cross sections
  at $\sqrt{s}=13$ and $\sqrt{s}=14$~\UTeV\, LHC collision energy. The
  results are obtained by simultaneously varying $\muF$ and $\muR$,
  with $\mu=\muF=\muR$. See Section~\ref{subsec:NLO+NLL}.}
\label{f:scaledependence:sim}
\end{figure}
\subsubsection{Numerical predictions}
The numerical results presented in this section are obtained using the parameter values listed in~\cite{LHCHXSWG-INT-2015-006}, in particular using the PDF4LHC15\_100 PDF sets~\cite{Butterworth:2015oua} for the NLO and NLO+NLL predictions. The LO results, shown here only for illustration, are obtained with the MMHT14 PDF sets~\cite{Harland-Lang:2014zoa}. The NLO cross section is calculated using the MG5\_aMC@NLO code~\cite{Alwall:2014hca}.

In~\refF{f:scaledependence:sim} we analyse the scale dependence of the resummed total cross section for $pp \to \tth$ at $\sqrt{s}=13$ and 14~\UTeV, varying simultaneously the factorization and renormalization scales as  $\mu=\muF=\muR$. As demonstrated in \refF{f:scaledependence:sim}, adding the soft-gluon corrections stabilizes the dependence on $\mu=\muF=\muR$ of the NLO+NLL predictions with respect to NLO.  The central values, calculated at $\mu=\mu_0= \mt +\mh/2$, and the scale error at $\sqrt{s}=13$~\UTeV\, changes from $499^{+5.8\%}_{-9.3\%}$~fb at NLO to $530^{+0.8\%}_{-1.6\%}$~fb at NLO+NLL (with $C^{(1)}_{ij \tosv \tth,I}$ coefficients included) and correspondingly, from $604^{+6.0\%}_{-9.2\%}$~fb to $641^{+0.8\%}_{-1.3\%}$~fb at $\sqrt{s}=14$~\UTeV. It is also clear from \refF{f:scaledependence:sim} that the coefficients $C_{ij\tosv \tth}^{(1)}$ strongly impact the predictions, especially at higher scales. In fact, their effect is more important than the effect of the logarithmic corrections alone, in correspondence to the strong suppression $\sim \beta^4$ for the real emission in the $2\to 3$ process due to the massive three-particle phase space. This observation also indicates the relevance of the contributions originating from the region away from the absolute threshold which need to be known in order to further improve theoretical predictions.

The effect of including NLL corrections is summarized in Table~\ref{tab:results} for the LHC collision energy of 13 and 14~\UTeV. Here we choose to estimate the theoretical uncertainty due to scale variation by varying $\muF$ and $\muR$ independently via the 7-point method \ie by considering the minimum and maximum values obtained with
 $(\muF/\mu_0, \muR/\mu_0) = (0.5,0.5), (0.5,1), (1,0.5), (1,1), (1,2), (2,1), (2,2)$. The NLO+NLL predictions show a significant reduction of the scale uncertainty, compared to NLO results. The reduction of the positive and negative scale errors amounts to around 25-30\% of the NLO error for $\sqrt{s}=13, 14$~\UTeV. This general reduction trend is not sustained for the positive error after including the  $C^{(1)}_{ij \tosv t \bar{t} H,I}$ coefficients. More specifically, the negative error is further slightly reduced, while the positive error is increased.  The origin of this increase can be traced back to the substantial dependence on $\mu_F$ of the resummed predictions with non-zero $C^{(1)}_{ij \tosv t \bar{t} H,I}$ coefficients, manifesting itself at larger scales. However, even after the redistribution of the error between the positive and negative parts, the overall size of the scale error, corresponding to the size of the error bar, is reduced after resummation by around  12\% with respect to the NLO uncertainties.  The scale error of the predictions is still larger than the PDF error of the NLO predictions ($\sim$3 \%) which is not expected to be significantly influenced by the soft-gluon corrections.

\begin{table}
\caption{NLO+NLL and NLO total cross sections for $pp \to \tth$ at $\sqrt{s}=13$ and 14~\UTeV. Results have been obtained using $\mu_0=235$~\UGeV, $\mt=172.5$~\UGeV, $\mh=125$~\UGeV, and the PDF4LHC15 set of PDF. The error ranges given together with the NLO and NLO+NLL results indicate the scale uncertainty. See Section~\ref{subsec:NLO+NLL}.}
\label{tab:results}
\begin{center}
\renewcommand{\arraystretch}{1.3}
\begin{tabular}{c |c c c c c c}
\toprule
$\sqrt{s}$ {[}\!\UTeV{]} & NLO {[}fb{]} & \multicolumn{2}{c}{NLO+NLL} & \multicolumn{2}{c}{NLO+NLL with $C$} & \tabularnewline
 & & Value {[}fb{]} & K-factor & Value {[}fb{]}  & K-factor & \tabularnewline
\midrule
13 & $499_{-9.3\%}^{+5.8\%}$ & $509_{-6.5\%}^{+4.2\%}$ & 1.02 &$530_{-5.5\%}^{+7.8\%}$ &1.06 & \tabularnewline
14 & $604_{-9.2\%}^{+6.0\%}$ & $616_{-6.5\%}^{+4.5\%}$ & 1.02 & $641_{-5.5\%}^{+7.9\%}$& 1.06 & \tabularnewline
\bottomrule
\end{tabular}
\end{center}
\end{table}

\subsection[Approximate NNLO via soft-gluon resummation in the
  ``PIM'' limit]{\texorpdfstring{$t\bar{t}H$}{ttH} production at approximate NNLO via soft-gluon resummation in the
  ``PIM'' limit}
\label{subsec:nNLO}
\subsubsection{Method}

A different approach to the estimate of the approximate NNLO QCD corrections
to the total and differential $\tth$ cross sections was considered in Ref.~\cite{Broggio:2015lya}.
In this case, the approximate formulas were obtained by studying soft-gluon corrections in the limit where the partonic centre-of-mass energy approaches the invariant mass of the $\tth$ final state, where the latter can be arbitrarily large.  The soft limit employed is the exact analogue of the so-called pair-invariant mass (PIM) threshold limit used to study top-quark pair production at NNLL and approximate NNLO in \cite{Ahrens:2010zv}. The approximate NNLO corrections are extracted from the perturbative information contained in a soft-gluon resummation formula valid to NNLL accuracy.  The derivation of this formula is based on SCET methods (for a recent review, see \cite{Becher:2014oda}). The soft-gluon resummation formula for this process contains three essential ingredients, all of which are matrices in the colour space needed to describe four-parton scattering. These ingredients are 1) a hard function, related to virtual corrections; 2) a soft function, related to real emission corrections in the soft limit; and 3) a soft anomalous dimension, which governs the structure of the all-order soft-gluon corrections through the renormalization group (RG).
Of these three ingredients, both the NLO soft function \cite{Ahrens:2010zv, Li:2014ula} and the NLO soft anomalous dimension \cite{Ferroglia:2009ep, Ferroglia:2009ii} needed for NNLL resummation in processes involving two massless and two massive partons can be adapted directly to $\tth$ production. The NLO hard function is instead process dependent and it was evaluated in \cite{Broggio:2015lya} by using a modified version of the one-loop providers \Gosam~\cite{Cullen:2011ac,Cullen:2014yla}, \Madloop~\cite{Hirschi:2011pa} and \Openloops~\cite{Cascioli:2011va} in combination with \Collier~\cite{Denner:2002ii, Denner:2005nn, Denner:2010tr, Denner:2014gla}.

Due to the mechanism of dynamical threshold enhancement \cite{Becher:2007ty,Ahrens:2010zv}, it is often the case that observables receive their dominant contributions from soft-gluon corrections derived in the PIM threshold limit. Obvious examples are the cross section differential in the invariant mass of the $\tth$ final state, or the total cross section obtained by integrating this distribution.
Furthermore, given that results in the  PIM threshold limit are fully differential in the Mandelstam variables characterizing the Born process, it is possible to use them in order to estimate the NNLO corrections to any differential distribution which is non-vanishing at Born level. In \cite{Broggio:2015lya} an in-house parton-level Monte Carlo was written and employed to study approximate NNLO corrections to the differential cross sections with respect to the $p_T$ of the Higgs boson, the $p_T$ of the top quark, the invariant mass of the $\tth$ pair, and the rapidities of the top quark or Higgs boson, in addition to the total cross
 section and differential cross section with respect to the $\tth$ final state.  The  NNLO approximation was then matched to the complete NLO calculation \cite{Reina:2001bc, Reina:2001sf, Beenakker:2001rj, Dawson:2002tg, Beenakker:2002nc, Dawson:2003zu} obtained from \MGfiveamcnlo~\cite{Alwall:2014hca}.
 This procedure can in principle be extended to include the decays of the final-state particles and by retaining information on the spins of the final-state top quarks, as it was done in \cite{Broggio:2014yca} for the top-pair production process.

\subsubsection{Approximate NNLO formulas}
\label{sec:approxNNLO}
The fixed-order expansion of the cross section and the resummation of
soft-emission effects are two complementary approaches to the precise
determination of physical observables. For this reason, one typically
wants to match resummed and fixed-order calculations in order to
account for all of the known effects when obtaining phenomenological
predictions. However, there are situations in which the perturbative
expansion in $\als$ is still justified, but soft-gluon emission
effects provide the bulk of the corrections at a given perturbative
order. In those cases, one can re-expand the resummed hard scattering
kernels in order to obtain approximate formulas which include all of
the terms proportional to plus distributions up to a given power of
$\als$ in fixed-order perturbation theory. In PIM kinematics, the
hard scattering kernels $C$ depend, among other arguments, on
$z \equiv M^2/\hat{s}$, where $M$ indicates the invariant mass of the
three heavy particles in the final state (notice that this is
different from the $M$ used in Section~\ref{subsec:NLO+NLL}) and
$\hat{s}$ is the partonic centre-of-mass energy. The PIM soft-emission
kinematics corresponds to the limit $z \to 1$. Schematically, the NNLO
contribution to the hard-scattering kernels has the following
structure
 \begin{equation}
 C^{(2)} \left(z,\mu \right) = \sum_{n=0}^3 D_n(\mu) \left[\frac{\ln^n(1-z)}{1-z} \right]_+  + C_0(\mu) \delta(1-z) + R(z,\mu)\, \label{eq:C2struct} \, ,
 \end{equation}
where several arguments of the functions $D_n$, $C_0$, and $R$ in (\ref{eq:C2struct}), such as masses and scattering angles have been dropped, and only the $\mu$ and $z$ dependence has been kept.
 The approximate NNLO formulas for
 the partonic cross sections derived in \cite{Broggio:2015lya} include the
 complete set of functions $D_i$, some of the scale-dependent terms in the
 function $C_0$ as well as partial information on the function $R(z)$
 which is non singular in the $z \to 1$ limit\footnote{For a detailed
 	description of what is included in the functions $C_0$ and
 	$R$ in the SCET approach, and how this information improves
 	the agreement between different kinematic schemes in top pair
 	production and stop pair production, the reader
 	can refer to \cite{Ahrens:2010zv, Ahrens:2011mw, Broggio:2013uba}.}.
 The information obtained from approximate NNLO formulas can be matched to the complete NLO calculation of a given observable in order to obtain precise predictions for a physical quantity, details can be found in \cite{Broggio:2015lya}. As an example, for the total cross section  the matching operation is straightforward and can be schematically written as,
\begin{equation}
\label{eq:approx-NNLO}
\sigma^{\mathrm{NLO+approx NNLO}}=
\sigma^{\mathrm{NLO}}+
\sigma^{\mathrm{approx. NNLO}}-\sigma^{\mathrm{approx. NLO}}\,,
\end{equation}
where the subtraction of the last term avoids double counting of NLO terms proportional to plus distributions and delta functions, and all of the terms on the r.h.s. of Eq.~(\ref{eq:approx-NNLO}) are evaluated with NNLO PDF.
 To avoid lengthy superscripts, in
 the following the matched NLO + approximate NNLO calculation is denoted as ``nNLO''.

 \subsubsection{Numerical analysis \label{sec:NumAnalyis}}
 \label{sec:pheno}

 This section includes results for the total cross section and the final-state invariant mass distribution obtained from the numerical
 evaluation of the nNLO formulas.
 In order for corrections in the soft limit to be dominant
 in observables which are also sensitive to regions of phase space far away from $z\to 1$, the mechanism of dynamical threshold
 enhancement \cite{Becher:2007ty, Ahrens:2010zv} must occur. This
 means that the parton luminosities  should drop off quickly enough away from the
 integration region where $z\to 1$ that an expansion under the
 integral of the partonic cross section in the soft limit is
 justified.
 Approximate NLO calculations are obtained by re-expanding the NNLL resummed partonic cross section to NLO; consequently they reproduce completely all of the terms singular in the $z\to 1$ limit in the NLO partonic cross section, but they miss terms which are subleading in the soft limit.  In \cite{Broggio:2015lya} it was checked that the soft approximation works quite well for the NLO total cross section and differential distributions analysed in that work. This fact proves that dynamical threshold enhancement at NLO does take place.  This does not immediately imply that the same holds at higher orders, but is an important sanity check nonetheless.

 \begin{figure} \begin{center} \begin{tabular}{cc}
                                 \includegraphics[width=6.6cm]{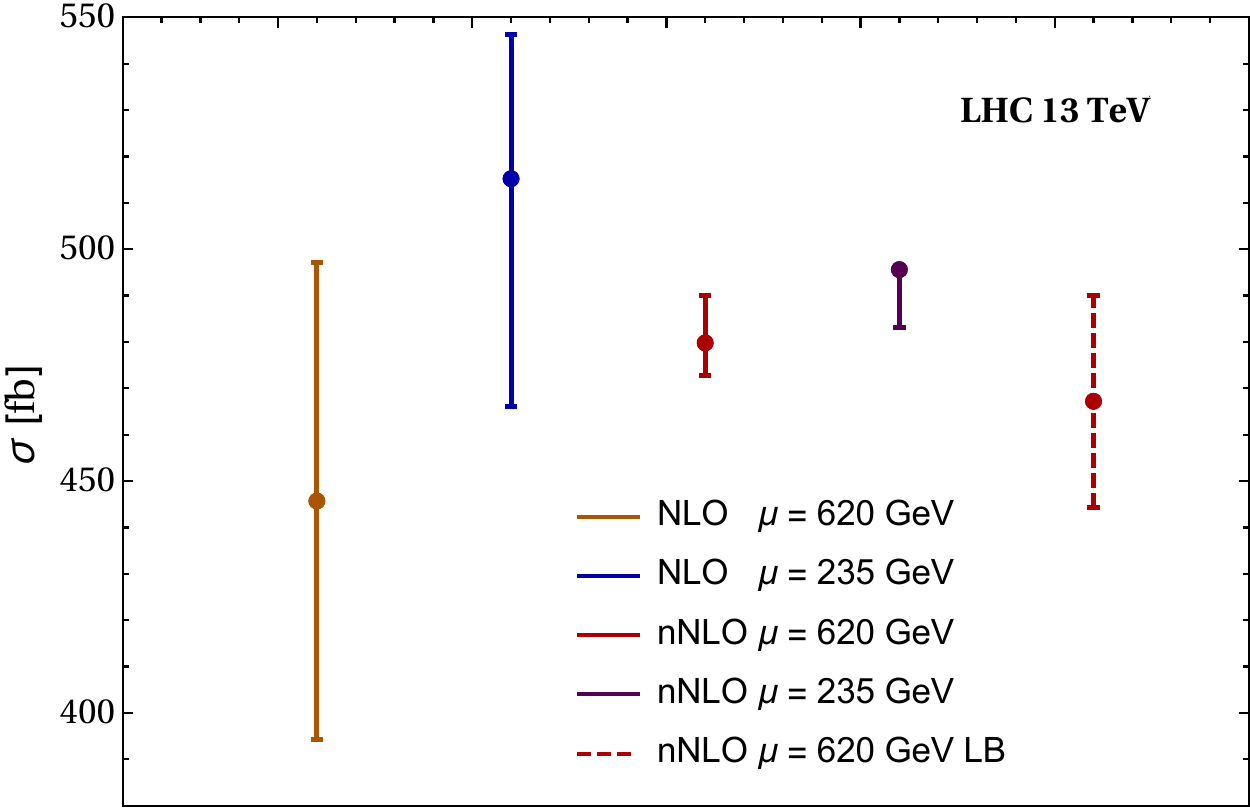} &        \includegraphics[width=8.5cm]{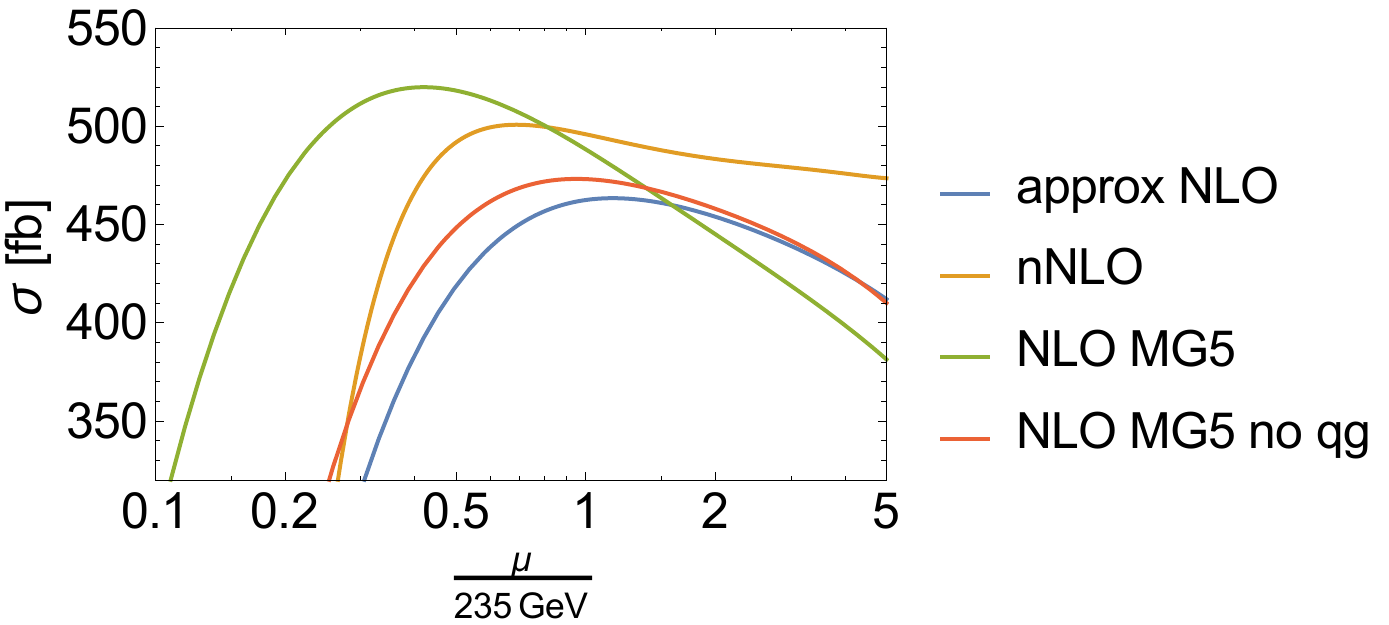} \\
                               \end{tabular} \end{center}
   \caption{Left panel: Comparison among the NLO and nNLO calculations
   of the total cross section reported in Table~\ref{tab:CSat620} for
   the LHC at $\sqrt{s} =13$~\UTeV. The cross section labelled LB refers
   to the conservative estimate of the nNLO uncertainty reported in
   Table~\ref{tab:LBCS}. Right panel: Scale dependence of the total
   cross section (from \cite{Broggio:2015lya}). The curves represent
   the NLO cross section evaluated with {\tt MG5} by excluding the
   quark-gluon channel contribution (red line), the complete NLO cross
   section evaluated with {\tt MG5} (green line), the nNLO cross
   section (orange line), and the approximate NLO cross section (light
   blue line). In the right panel, all perturbative orders are
   evaluated with NNLO MSTW2008 PDF. See Section~\ref{subsec:nNLO}.\label{fig:ScaleDep}}
\end{figure}

The traditional choice of the factorization and renormalization scale employed in
\cite{Beenakker:2001rj, Beenakker:2002nc, Dawson:2003zu}, namely $\mu_0 = (2 \mt + \mh)/2 \sim 235~\UGeV$
is not an ideal choice for the evaluation of approximate formulas obtained from the soft emission limit. While the approximate NLO
calculations  reproduce very well the exact contribution of the NLO gluon-fusion and quark-annihilation channels also at this scale,  the NLO cross section receives a sizeable contribution from the quark gluon channel. In particular, this channel dominates the uncertainty derived from scale variation in the range $[
\mu_0/2, 2 \mu_0]$. While the contribution of the quark-gluon channel is included in nNLO predictions via NLO matching, the soft gluon emission corrections alone do not provide any information on this channel since it is subleading in the soft limit $z \to 1$. Finally, a steep drop of the gluon-fusion and quark-annihilation channel cross sections occurs  for $\mu \sim 100$~\UGeV\, (see right panel of \refF{fig:ScaleDep}; all of the curves in that panel are evaluated using NNLO PDF). If one includes the contribution of the quark-gluon channel, the steep decrease of the cross section occurs at smaller values  of $\mu$. Consequently, while the choice $\mu_0 = 235$~\UGeV\, works well for complete fixed-order calculations at NLO, the same choice makes an evaluation of the theoretical uncertainty affecting nNLO calculations based on scale variation unreliable.
For these reasons, the main findings of \cite{Broggio:2015lya}  were obtained for a factorization and renormalization scale fixed at a value $\mu_0 = 620$~\UGeV, which is close to the peak of the final state invariant mass distribution.
The location of this peak is not very sensitive to the LHC energy.

In the calculations reported in the following and in \cite{Broggio:2015lya}, no cuts are applied on the momenta of the final state particles and all of the calculations are carried out with MSTW 2008 PDF. The complete list of input parameters can be found in Table~1 in \cite{Broggio:2015lya}.  The NLO results needed for the matching and for comparison are obtained from the code \MGfiveamcnlo~\cite{Alwall:2014hca}, which for convenience is indicated by the acronym {\tt MG5}.

 \paragraph{Total cross section}
 \label{sec:xs}


 \begin{table}
 	\caption{Total cross section at $\sqrt{s} =
 		13$~\UTeV\, and $\sqrt{s} =
 		14$~\UTeV. Each order is evaluated with the MSTW2008
 		PDF at the corresponding perturbative order
 		(meaning, e.g. NNLO PDF for the nNLO
 		calculation). The uncertainties reflect scale
 		variation only. The top quark mass is $\mt = 172.5$~GeV, the Higgs boson mass is $\mh = 125$~GeV. See Section~\ref{subsec:nNLO}.\label{tab:CSat620}}
 	\centering
 	\def\arraystretch{1.3}
 	\begin{tabular}{ccccc}
		\toprule
 		$\sqrt{s}$~[\!\UTeV] & $\mu_0$~[\UGeV] & LO   [fb] & NLO  {\tt MG5} [fb] &nNLO [fb] \\
 		\midrule 
		$13$ & $620$ & $317.2^{+30.7~\%}_{-21.8~\%}$ & $445.7^{+11.5~\%}_{-11.5~\%}$ &$479.8^{+2.1~\%}_{-1.5~\%}$\\
 		$13$ & $235$ & $464.2^{+35.4~\%}_{-24.1~\%}$ & $515.2^{+6.0~\%}_{-9.5~\%}$ &$495.6^{+0.0~\%}_{-2.5~\%}$\\
 		$14$ & $620$ & $383.2^{+30.6~\%}_{-21.6~\%}$ & $539.7^{+11.4~\%}_{-11.4~\%}$ &$580.7^{+2.1~\%}_{-1.4~\%}$\\
 		$14$ & $235$ & $558.2^{+34.9~\%}_{-24.0~\%}$ & $623.3^{+6.3~\%}_{-9.4~\%}$ &$599.5^{+0.0~\%}_{-2.5~\%}$\\
 		\bottomrule
 	\end{tabular}
 \end{table}

 The total cross section at LO, NLO, and nNLO calculated at $\mu_0=620$~GeV can be found in Table~\ref{tab:CSat620}. If one accounts for the
 approximate NNLO corrections, the central value of the cross section
 increases with respect to the NLO calculation carried out at $\mu_0 = 620$~GeV, while the
 scale uncertainty is significantly reduced. If one compares instead the nNLO prediction at $\mu_0 = 620$~GeV with the NLO cross section evaluated at the standard scale choice $\mu_0 = 235$~GeV (which is also shown in Table~\ref{tab:CSat620}), one sees that the central value of the nNLO cross section is smaller than the NLO one. However, as shown in the left panel of \refF{fig:ScaleDep} for the LHC at $\sqrt{s} =13$~\UTeV, the nNLO scale uncertainty interval at $\mu_0 = 620$~GeV is completely included in the  NLO uncertainty interval evaluated with the standard choice $\mu_0 = 235$~GeV. For completeness we report also the numbers obtained by evaluating the nNLO cross section with the traditional scale choice
 $\mu_0 = 235$~GeV. In this case the NNLO soft emission corrections are positive and small, however the use of NNLO PDF in the nNLO calculation leads to a result which is smaller than the one found with a NLO calculation employing NLO PDF.

The uncertainty of the nNLO cross sections quoted in Table~\ref{tab:CSat620}, based on scale variation alone, is likely to underestimate the residual  perturbative uncertainty of these results. A more conservative estimate of the residual uncertainty affecting the approximate formulas, which accounts also for the numerical effect of terms which are
formally subleading in the soft limit and cannot be determined starting from soft emission resummation formulas, was considered in \cite{Broggio:2015lya}.
 %
 %
 \begin{table}
 	\caption{Total cross section  with an estimate of the error
          associated to the scale variation and to the formally
          subleading terms, as explained in the text and in
          \cite{Broggio:2015lya}.  Each order is evaluated with the
          MSTW2008 PDF at the corresponding perturbative order. See Section~\ref{subsec:nNLO}.
 		\label{tab:LBCS}}
 	\centering
 	\def\arraystretch{1.3}
 	\begin{tabular}{ccccc}
		\toprule
 		$\sqrt{s}$~[\!\UTeV] & $\mu_0$~[\UGeV] &  NLO  {\tt MG5} [fb] & approx. NLO [fb] & nNLO [fb]  \\
 		\midrule
		$13$ &$620$ & $445.7^{+11.5~\%}_{-11.5~\%}$ & $442.4^{+10.0~\%}_{-10.0~\%}$ &$467.2^{+4.9~\%}_{-4.9~\%}$  \\
 		$14$ &$620$ & $539.7^{+11.4~\%}_{-11.4~\%}$ & $534.8^{+9.8~\%}_{-9.8~\%}$ & $565.2^{+4.9~\%}_{-4.9~\%}$  \\
 		\bottomrule
 	\end{tabular}
 \end{table}

This more conservative estimate of the uncertainty leads to the approximate NLO predictions found in the fourth column of Table~\ref{tab:LBCS}. The central value of the approximate NLO cross section is placed in the middle of the uncertainty interval. The central value and the uncertainty interval obtained in this way are quite close to the complete NLO results evaluated at $\mu_0 = 620$~GeV, which are  shown in the third column of Table~\ref{tab:LBCS}.  While this can be somewhat accidental, it shows that, at least for this choice of the scale $\mu_0$, the terms subleading in the soft limit are numerically of the same size of the quark-gluon channel contributions, which is neglected in approximate NLO calculation.  The last column in Table~\ref{tab:LBCS} shows the nNLO total cross section calculated by estimating the residual perturbative uncertainty according to the more conservative procedure, as it was done in the next-to-last column for the approximate NLO case. We stress that nNLO results are obtained by matching the NNLO corrections in the soft limit to the complete NLO results. As such, they include the same quark-gluon channel contribution included in the NLO result. By looking at Table~\ref{tab:LBCS} one sees that the central value of the nNLO total cross section evaluated at $\mu_0 = 620$~GeV is larger than the NLO one (evaluated at the same scale) and the residual perturbative uncertainty is roughly half the one found at NLO. For both $\sqrt{s} = 13$~\UTeV\, and $\sqrt{s} = 14$~\UTeV, the nNLO results reported in Table~\ref{tab:LBCS} are included in the NLO uncertainty interval evaluated at $\mu_0 = 620$~GeV, and their central values are included in the NLO uncertainty interval evaluated with the standard scale choice $\mu_0 = 235$~GeV. For the $\sqrt{s} = 13$~\UTeV\, case, this situation is illustrated in the left panel of \refF{fig:ScaleDep}.

 \paragraph{Differential distributions}
 \label{sec:dxs}

 \begin{figure}[ht]
 	\begin{center}
 		\begin{tabular}{cc}
 			\includegraphics[width=7cm]{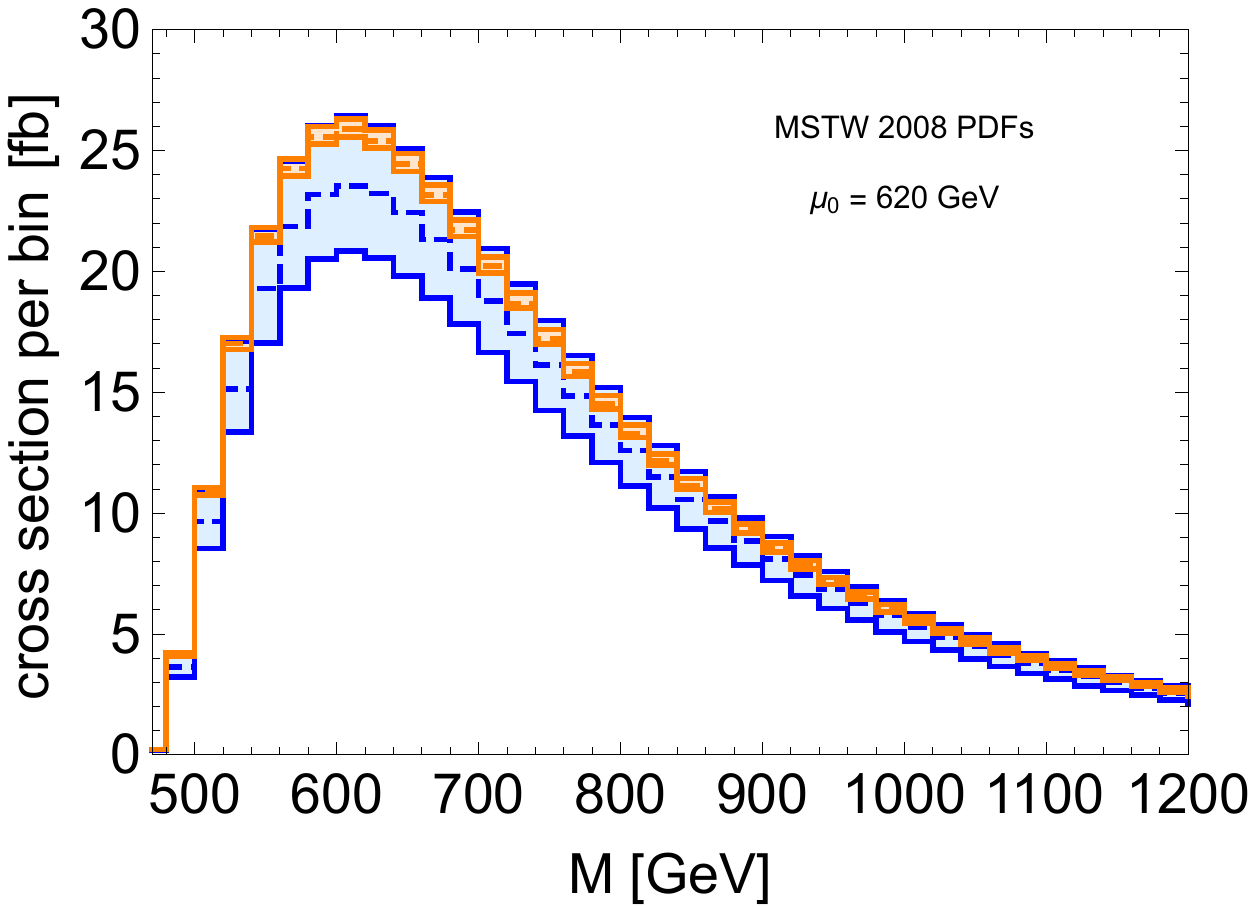} &
 			\includegraphics[width=7cm]{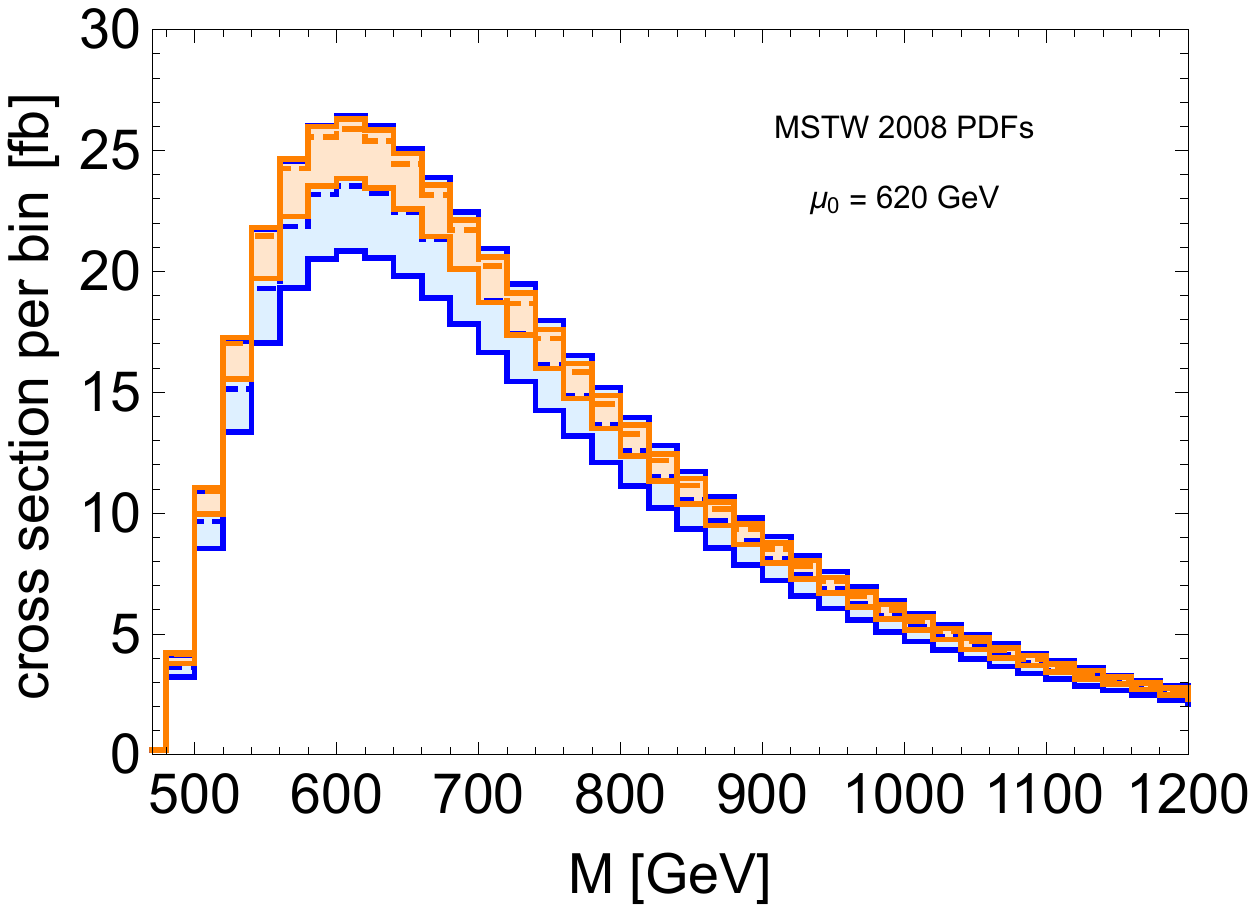} \\
 		\end{tabular}
 	\end{center}
 	\caption{Differential distributions at nNLO (orange band) compared to the NLO calculation carried out with \texttt{MG5}  (blue band) for the LHC at $\sqrt{s} = 13$~\UTeV. NLO distributions are evaluated with NLO PDF, nNLO distributions with NNLO PDF.
 		The scale is set to $\mu_0 = 620$~GeV in both NLO and nNLO distributions,  and it is varied in the range $[\mu_0 /2 ,2 \mu_0]$. The nNLO band in the left panel reflects exclusively the scale variation, while in the right panel it includes also the uncertainty associated with the treatment of subleading terms.  \label{fig:MdistnnLOvsNLO}
 	}
 \end{figure}

An advantage of the approach followed in \cite{Broggio:2015lya} is that it can be employed  to calculate arbitrary differential cross sections.  This can be done  by employing standard Monte-Carlo methods. In \cite{Broggio:2015lya} several differential distributions for the LHC operating at 13~\UTeV\, were considered.
It was shown in \cite{Broggio:2015lya} that \emph{i)} the approximate NLO distributions reproduce very well the corresponding NLO calculations carried out by excluding the contribution of the quark-gluon channel, and \emph{ii)} the distributions and their uncertainty, evaluated with the conservative method described above, reproduce well the complete NLO distributions and their scale-variation uncertainty bands obtained from {\tt MG5}, at least for $\mu_0 = 620$~GeV. A calculation of the differential distributions at nNLO is therefore justified.

\refF{fig:MdistnnLOvsNLO} shows the invariant-mass distribution at nNLO in comparison with the corresponding complete NLO calculation. In both panels, the NLO perturbative uncertainty band is obtained by varying the scale in the usual range. In the left panel, the nNLO band is obtained by scale variation alone. In the right panel, the band reflecting the residual perturbative uncertainty was obtained by considering both the effect of subleading terms and scale variation, as described above for the total cross section calculation.
 In \cite{Broggio:2015lya} it was observed that for all of the distributions considered in that work, including the $M$ distributions, the nNLO predictions at $\mu_0 = 620$~GeV are slightly larger than the NLO ones when the latter are evaluated at the same $\mu_0$ value. The nNLO bands for the differential distribution obtained with the method which considers both the effect of subleading terms and the scale variation are roughly half as large as the ones obtained by evaluating these quantities at NLO.


\def\ttHtHpath{\ttHpath/ttH-tH}
\section{\texorpdfstring{$tH$}{tH} production at NLO in QCD}
\label{sec:tH_NLO_QCD}

The production of a Higgs boson in association with a single top quark ($tH$)
at hadron colliders shares important analogies with the
electroweak production of a single top quark.
At LO one can organize the Feynman diagrams into three
independent (non-interfering) sets, based on the virtuality
of the $W$ boson coupled to the heavy-quark $b{-}t$ current:
$t$-channel production features a virtual space-like $W$,
$s$-channel production a virtual time-like $W$,
and $W$-associated production an on-shell $W$ boson in the final state.
This classification is useful for event generation, but is valid
only at LO and in the five-flavour scheme (5FS); when adding higher-order
corrections, or when employing the four-flavour scheme (4FS),
the picture becomes fuzzy because some amplitudes can interfere.

In the 5FS the separation between $t$-channel, $s$-channel
and $W$-associated production is exact up to NLO in QCD
(channels start to interfere at NNLO), and calculations are
typically simpler than in the 4FS.
In the 4FS, instead, the $t$-channel process
can interfere already at NLO in QCD both with (NNLO) $s$-channel
and with $W$-associated production (only if the $W$ boson decays hadronically);
nevertheless, such interference is tiny at NLO and can be neglected if
the aim is to compute the dominant $t$-channel process.
Therefore, the separate simulation of $t$-channel and $s$-channel $tH$
processes is not a problem up to NLO accuracy in QCD.
A detailed study of these two channels at the LHC
has been presented recently in \Bref{Demartin:2015uha}, including NLO QCD
corrections and addressing many sources of uncertainty
(notably the flavour-scheme dependence of the $t$-channel process).
We will review $t$-channel $tH$ production in \Section~\ref{sec:tH_tcha},
and for $s$-channel $tH$ production in \Section~\ref{sec:tH_scha}.

Associated $tWH$ production, on the other hand, interferes also
with the much larger $\tth$ process starting from NLO in QCD in the 5FS,
and this happens already at LO in the 4FS.
In general, such interference is large and cannot be neglected,
thus NLO simulations of the $tWH$ channel are not straightforward to carry out. For a recent NLO computation of this process in the 5FS, see \Ref~\cite{Demartin:2016axk}.


\subsection{\texorpdfstring{$t$}{t}-channel \texorpdfstring{$tH$}{tH} production}
\label{sec:tH_tcha}

\begin{figure}
\center
\includegraphics[width=0.6\textwidth]{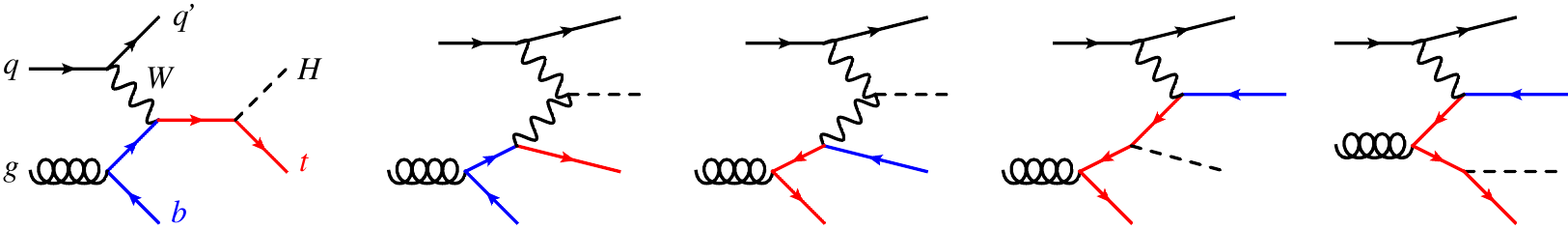}
\hspace{4em}
\includegraphics[width=0.24\textwidth]{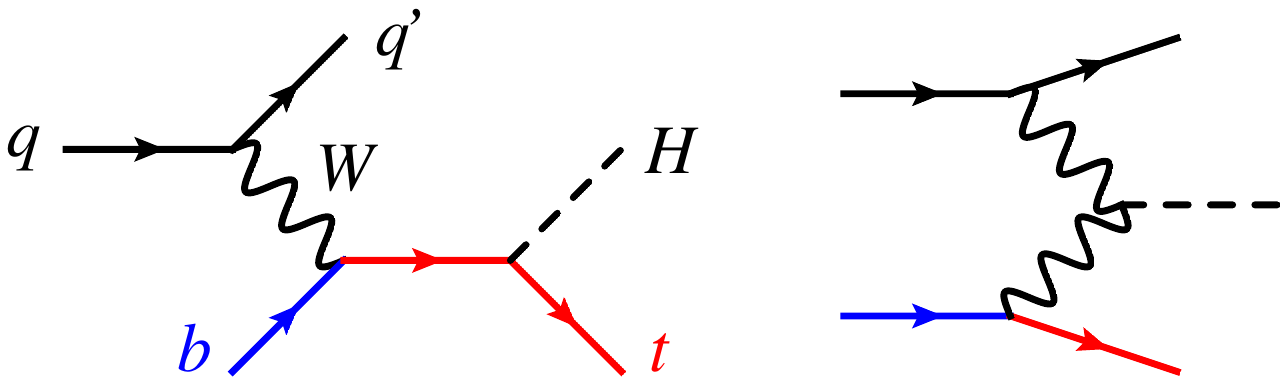}\\[1ex]
\caption{LO Feynman diagrams for $t$-channel $tH$ production in the 4FS
 (left) and in the 5FS (right).}
\label{fig:tH_tcha_diagrams}
\end{figure}

Being a process initiated by bottom quarks,
the $t$-channel $tH$ production can be computed either with the 4FS
or with the 5FS approach.
In the 4FS the hard matrix element and the phase space are computed taking
into account all the bottom-quark (pole) mass effects, but there is no
$b$ parton distribution function (PDF), thus $b$ quarks are generated
perturbatively in the hard matrix element via initial-state gluon splitting into
a $b\bar b$ pair (see \refF{fig:tH_tcha_diagrams}, left).
In the 5FS potentially large logarithms associated with
such $g \to b \bar b$ splitting are resummed
to all orders into an initial-state bottom-quark PDF
(see \refF{fig:tH_tcha_diagrams}, right),
while all other bottom-quark mass effects are neglected
($M_{\mathrm{b}}=0$), and a correct description of the process transverse dynamics is included only at NLO.
Both approaches feature advantages and shortcomings at the LHC.
Their difference mainly consists of what kind of terms are kept in the
perturbative expansion or pushed into
missing higher-order corrections, thus it becomes milder when including
more terms in perturbative QCD.
Therefore, NLO accuracy is mandatory to reduce the flavour-scheme dependence
of theoretical predictions. Moreover, a comparison of 4FS and 5FS results
can offer some insights on the relevance of missing higher-order
corrections for processes that are more sensitive to large
initial-state logarithmic corrections, and should therefore be taken
into account in estimating the theoretical uncertainty for a given process.

The results presented in this section have been obtained
in the \MGfiveamcnlo\, framework~\cite{Alwall:2014hca},
where the NLO simulation can be generated
automatically in both the 4FS and 5FS~\cite{Demartin:2015uha}.
The top quarks can then be decayed with \Madspin~\cite{Artoisenet:2012st},
thereby keeping spin correlations.

\subsubsection{Total cross section at NLO}

In this section we address the inclusive cross section at NLO accuracy in QCD.
The numerical results presented here have been obtained using
the input parameters listed below, which follow the prescriptions given
in \Brefs{LHCHXSWG-INT-2015-006,Butterworth:2015oua}.
The pole mass of the top quark and its Yukawa coupling\footnote{In the adopted convention the Feynman rule of
  the $q\bar{q}H$ vertex is $(-iy_{\mathrm{q}})$ for a generic quark $q$.}
(renormalized on shell) are
\begin{equation}
\label{eq:mt_yt}
\mt = 172.5\UGeV \,, \hspace{4em}
y_{\mathrm{t}}=\frac{\mt}{v} = \big( \sqrt{2} G_\mu \big)^{1/2} \mt \,,
\end{equation}
where $v \simeq 246\UGeV$ is the EW vacuum expectation value.
The bottom-quark pole mass in the 4FS (left) and 5FS (right) is set to
\begin{equation}
M_{\mathrm{b}}^\mathrm{(4FS)} = 4.92\UGeV \,,
\hspace{4em}
M_{\mathrm{b}}^\mathrm{(5FS)} = 0 \,,
\end{equation}
while the bottom-quark Yukawa coupling is always set to zero,
$y_{\mathrm{b}}=0 $\,,
because its impact on the
total cross section amounts to less than $0.1\%$.
Actually, to speed up the 4FS code, the corresponding diagrams
are not even generated.
The EW parameters are
\begin{equation}
G_\mu = 1.166379 \cdot 10^{-5}\UGeV^{-2} \,, \hspace*{3em}
\mz= 91.1876\UGeV \,, \hspace*{3em}
\mw = 80.385\UGeV \,,
\end{equation}
which in turn fix the electromagnetic coupling (no running) and the on-shell
weak mixing angle to
\begin{equation}
\alpha = \sqrt{2} G_\mu \mw^2 (1 - \mw^2/\mz^2) / \pi \simeq 1/132.233  \,,
\hspace*{3em}
\sin^2 \theta_{\mathrm{W}} = 1 - \mw^2/\mz^2 \simeq 0.2229 \,.
\end{equation}
We assume $V_{\mathrm{tb}}=1$ and, for simplicity, the whole CKM matrix to be diagonal%
\footnote{The only important assumption here is $V_{\mathrm{tb}}=1$;
once the third generation is decoupled from the first two, and if one is
inclusive over the first two generations, then the result doesn't depend on the
mixing between the first two generations (i.e. the Cabibbo angle) due to unitarity.}
\begin{equation}
 V_{\mathrm{CKM}} = \mathrm{diag} \big\{ V_{\mathrm{ud}}, V_{\mathrm{cs}}, V_{\mathrm{tb}} \big\} = \mathrm{diag} \big\{ 1,1,1 \big\} \,.
\end{equation}
The proton content in terms of parton distribution functions (PDFs)
is evaluated by using the NLO \textsc{PDF4LHC15} sets
in the corresponding flavour-number scheme.
The PDFs also determine the reference value of the strong coupling
used in the simulation, which then is automatically run at 2-loop accuracy.
In the 5FS this value and its uncertainty are
\begin{equation}
\label{eq:tH_tcha_alphas}
\als^\mathrm{(5FS)}(\mz) = 0.1180 \pm 0.0015 \,,
\end{equation}
while in the 4FS $\als(\mz)$ is slightly smaller and consistent with
a four-flavour running~\cite{Butterworth:2015oua}.
The combined PDF+$\als$ uncertainty is computed from the
Hessian set with 30 (PDF) + 2 ($\als$) members, accordingly to
\Eq~(28) in \Ref~\cite{Butterworth:2015oua}.
The renormalization $\muR$ and factorization $\muF$ scales
are both set equal to the reference value
\begin{equation}
\label{eq:tH_tcha_scale}
\mu_0^{(t-\mathrm{channel})} = (\mh+\mt)/4 \,,
\end{equation}
while the scale dependence in each flavour scheme is estimated from the
maximum and minumum variations of the cross section
among six scale points with
\begin{equation}
\label{eq:tH_tcha_scalevariations}
1/2 < \mu_{\mathrm{R,F}}/\mu_0 < 2 \,, \qquad 1/2 < \muR/\muF < 2 \,.
\end{equation}
The reference scale choice in \Eq~\eqref{eq:tH_tcha_scale} is motivated
by physical arguments in the 4FS description~\cite{Maltoni:2012pa}. In particular,
it ensures that the discrepancy between the 4FS and 5FS results
is not unreasonably large, and that the 5FS uncertainty is not underestimated,
which might happen when using a very high scale
(see Figure~3 in \Ref~\cite{Demartin:2015uha}).

In \Trefs{tab:tH_tcha_SM_lhc7}, \ref{tab:tH_tcha_SM_lhc8},
\ref{tab:tH_tcha_SM_lhc13} and \ref{tab:tH_tcha_SM_lhc14}
we collect the results for the combined $t$-channel $pp \to tH + \bar tH$ production
at the LHC, at centre-of-mass energies of
$\sqrt{s} = 7,8,13,$ and $14\UTeV$, respectively, and for various Higgs boson masses
in the range $120{-}130\UGeV$.
In the third column we report the reference cross section (in\Ufb),
$\sigma_{tH+\bar tH}$, computed at NLO and in the 5FS, while in the fourth
column we report the NLO $K_{\mathrm{QCD}}$ factor, defined as
\begin{equation}
K_{\mathrm{QCD}}= \sigma^{\mathrm{NLO~QCD}}_{tH + \bar tH} / \sigma^{\mathrm{LO}}_{tH + \bar tH} \,,
\end{equation}
where both the LO and NLO cross sections are computed with the same inputs.
In the fifth column we report the combined scale plus flavour-scheme (FS)
uncertainty, expressed as upper and lower per cent variations
with respect to the reference 5FS prediction.
The combined scale+FS uncertainty band is the largest source of theoretical
uncertainty, and it is computed from the maximum
and minimum variations of the cross section among the 6+6 scale points
according to \Erefs{eq:tH_tcha_scalevariations}
in the two flavour schemes.
This translates into the following equations
\begin{equation}
 \sigma^{+} = \max \limits_{
 \substack{(\mu_R,\mu_F)~\mathrm{points} \\[0.2ex] \mathrm{4FS,~5FS}} }
 \, \sigma^{(\mathrm{FS})}_{tH + \bar tH}(\muR,\muF)  \,,
 \hspace*{4em}
 \sigma^{-} = \min \limits_{
 \substack{(\muR,\muF)~\mathrm{points} \\[0.2ex] \mathrm{4FS,~5FS}} }
 \, \sigma^{(\mathrm{FS})}_{tH + \bar tH}(\mu_R,\mu_F)  \,,
\end{equation}
\begin{equation}
 \mathrm{Scale+FS~[\%]} =
100\left( \sigma^{+} / \sigma^{(\mathrm{5FS})}_{tH+\bar tH} -1 \right)
 \hspace*{2em}
 100 \left( \sigma^{-} / \sigma^{(\mathrm{5FS})}_{tH+\bar tH} -1 \right) \,.
\end{equation}
In the sixth, seventh, and eighth columns we report the $\als$, PDF, and
combined PDF+$\als$ uncertainty in the 5FS,
which is the second-largest source of theoretical uncertainty.
We recall that it is computed employing the \textsc{PDF4LHC15}
Hessian set with 30 (PDF) + 2 ($\als$) members, with the $\als$
uncertainty given in \Eq~\eqref{eq:tH_tcha_alphas}, and combining
the two uncertainties in quadrature accordingly to the
\textsc{PDF4LHC15} prescription.
Finally, in the last two columns we report the separate top ($tH$)
and anti-top ($\bar tH$) contributions to the 5FS cross section (in\Ufb).
In \Tref{tab:tH_tcha_SM_lhc6to15} we repeat the exercise, this time
keeping the Higgs boson mass fixed to $\mh=125\UGeV$ and varying instead
the collider energy in the range $\sqrt{s} = 6{-}15\UTeV$, to show
the gain in the cross section and the reduction of uncertainties.
The numbers in \Trefs{tab:tH_tcha_SM_lhc7} to \ref{tab:tH_tcha_SM_lhc6to15},
relevant for the SM Higgs boson, are summarized in the plots of
\refF{fig:tH_tcha_SM_plots} where the blue uncertainty band
is computed summing the scale+FS and PDF+$\als$ uncertainties.

We conclude the discussion of results relevant for the SM Higgs boson
by commenting on two minor uncertainties, namely the ones associated
with the bottom-quark and top-quark masses.
According to \Ref~\cite{LHCHXSWG-INT-2015-006}, we take the uncertainty
on the bottom-quark mass to be $M_{\mathrm{b}} = 4.92 \pm 0.13\UGeV$. At $13\UTeV$,
and for a $125\UGeV$ Higgs boson mass, this translates into a 4FS cross section
of $\sigma^\mathrm{(4FS)}_{tH + \bar tH} = 67.4^{+0.7}_{-0.5}\Ufb$,
which corresponds to an uncertainty of about $1\%$.
Since no \textsc{PDF4LHC15} set with heavy-quark mass variations
has been published yet, we estimate the impact on the 5FS cross section
using the numbers in \Ref~\cite{Demartin:2015uha},
where previous-generation PDF sets have been used.

The $\pm 0.25\UGeV$ bottom-mass uncertainty quoted in \Ref~\cite{Demartin:2015uha} returned an uncertainty in the 5FS cross section of about $2\%$.
A crude rescaling to $\pm 0.13\UGeV$ results in an uncertainty of roughly  $1\%$, comparable to the one in the 4FS.


Similarly, we consider a top-quark mass uncertainty of
$\mt = 172.5 \pm 1.0\UGeV$, which returns a 5FS cross section of
$\sigma^\mathrm{(5FS)}_{tH + \bar tH} = 74.3^{+0.4}_{-0.3}\Ufb$ at
$13\UTeV$.  Thus, the $\mt$ uncertainty in the total cross section is
below $1\%$, since increasing the top-quark mass causes a reduction of
the available phase space which is however partly compensated by the
larger top-quark Yukawa coupling.

Associated $tH$  production in the $t$-channel  is known for having
maximal destructive interference in the SM between $H-W$ interactions
on the one hand, and $H-t$ interactions on the other hand:
deviations from the SM top-quark Yukawa coupling can result in a large enhancement
of the cross section.
This has prompted the LHC experiments to perform
searches for the $125\UGeV$ Higgs boson in this process~\cite{Aad:2014lma,Khachatryan:2015ota}
assuming that the sign of the top-quark Yukawa coupling is opposite to the SM coupling
in \Eq~\eqref{eq:mt_yt},
\begin{equation}
\label{eq:-yt}
y_{\mathrm{t}}=-y_{\mathrm{t}}^{\mathrm{(SM)}}=-\mt/v \,,
\end{equation}
which results in maximally constructive interference between the two subsets of diagrams.
Given the interest in experimental searches, in \Tref{tab:tH_tcha_-ytSM}
we provide reference cross sections and uncertainties for this scenario at 13 and $14\UTeV$.
For further applications of this process to constrain deviations from the SM
interactions of the $125\UGeV$ particle,
see also \refS{s.efttools}.

Finally, we extend our investigation to Higgs boson masses in the range
$\mh=10{-}3000\UGeV$ keeping the Higgs boson as stable particle and neglecting Higgs boson width effects,
which might provide a useful reference for BSM Higgs searches.
The results at 13 and $14\UTeV$ are
plotted in \refF{fig:tH_tcha_BSM_plots}.
These results should be taken with care, since an hypothetical BSM
Higgs boson may contribute to the same $tH$ final state through
different interactions than the ones described by SM-like diagrams.

\subsubsection{Differential distributions at NLO+PS}

\begin{figure}
\center
\includegraphics[width=0.32\textwidth]{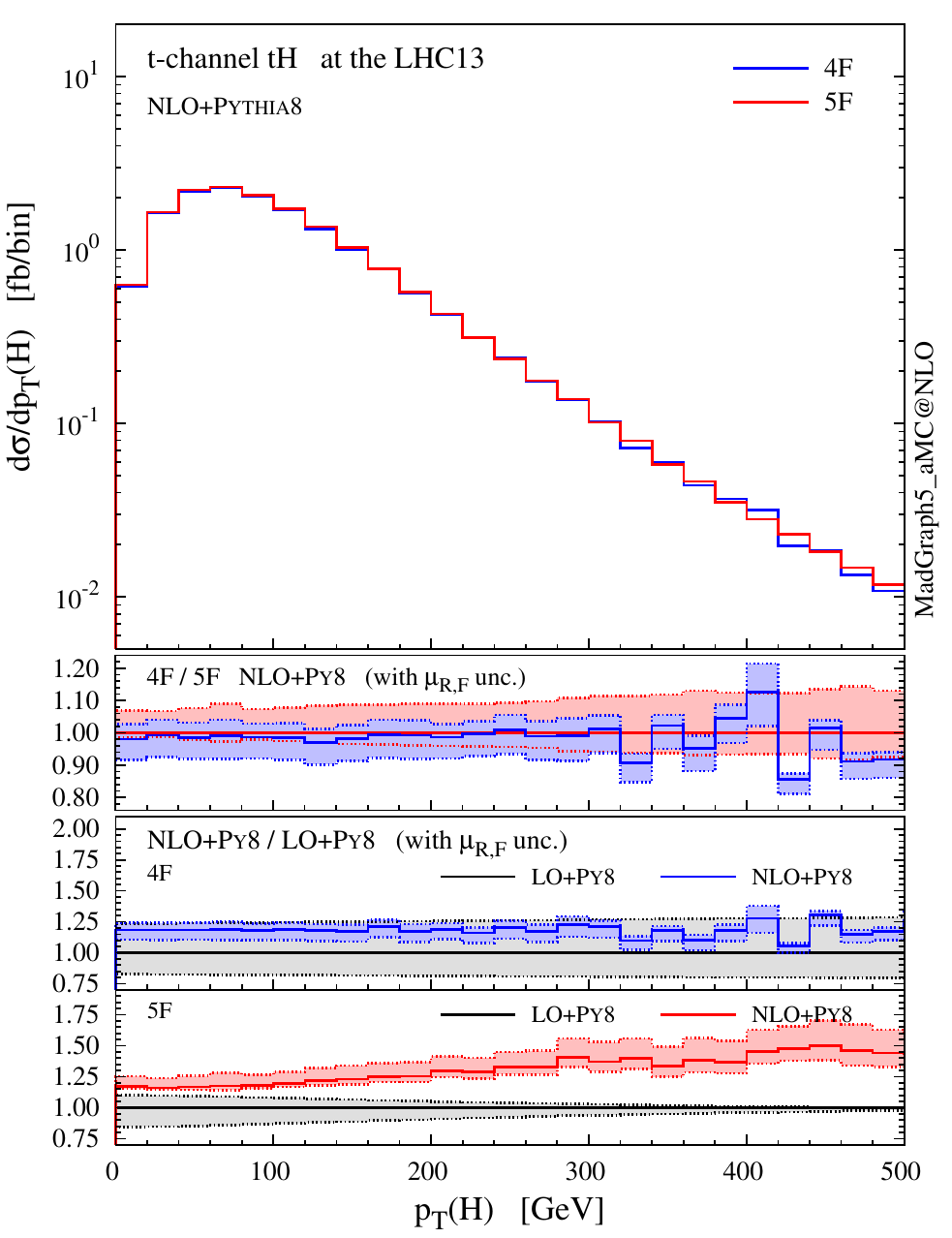}
\hspace*{0.2em}
\includegraphics[width=0.32\textwidth]{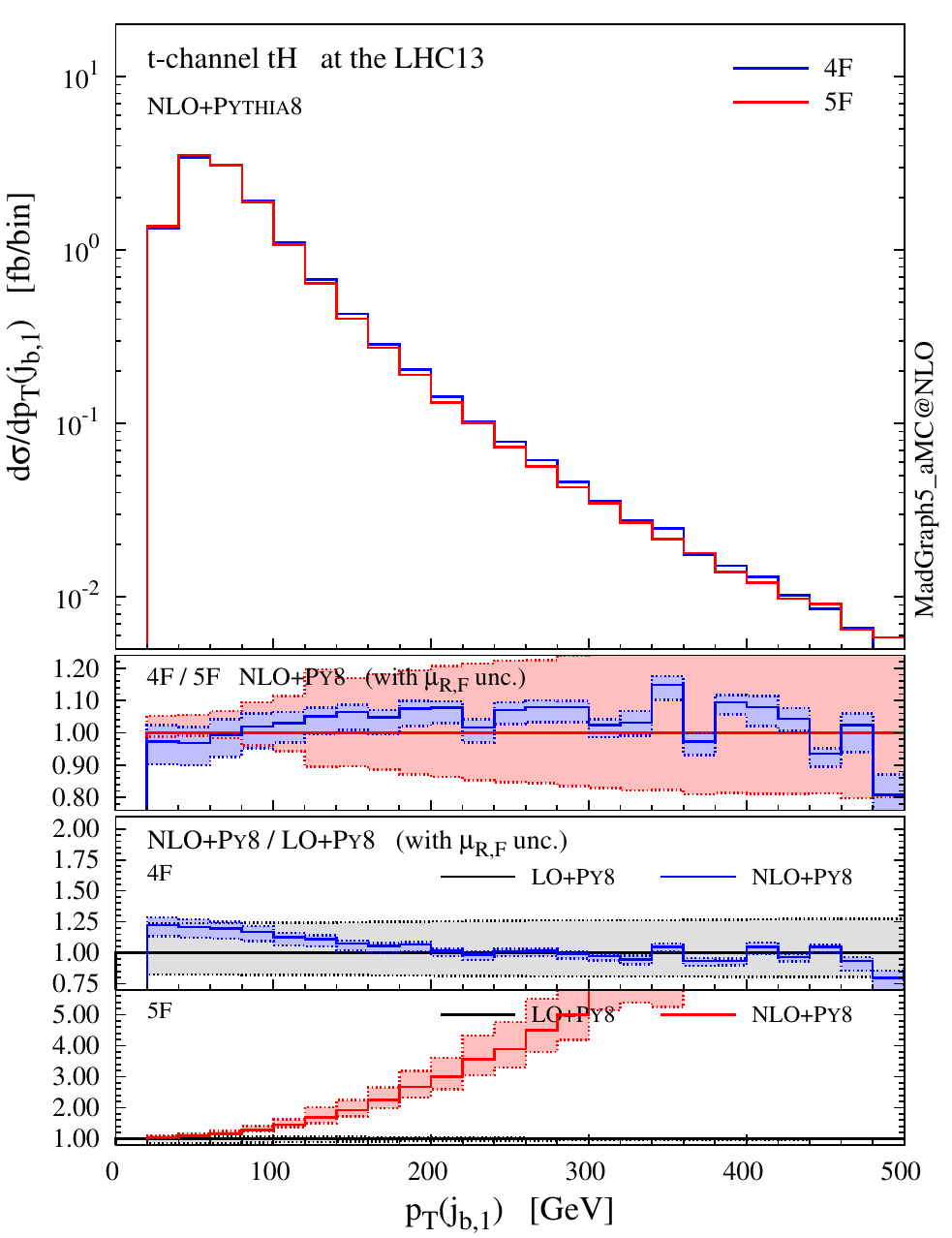}
\hspace*{0.2em}
\includegraphics[width=0.32\textwidth]{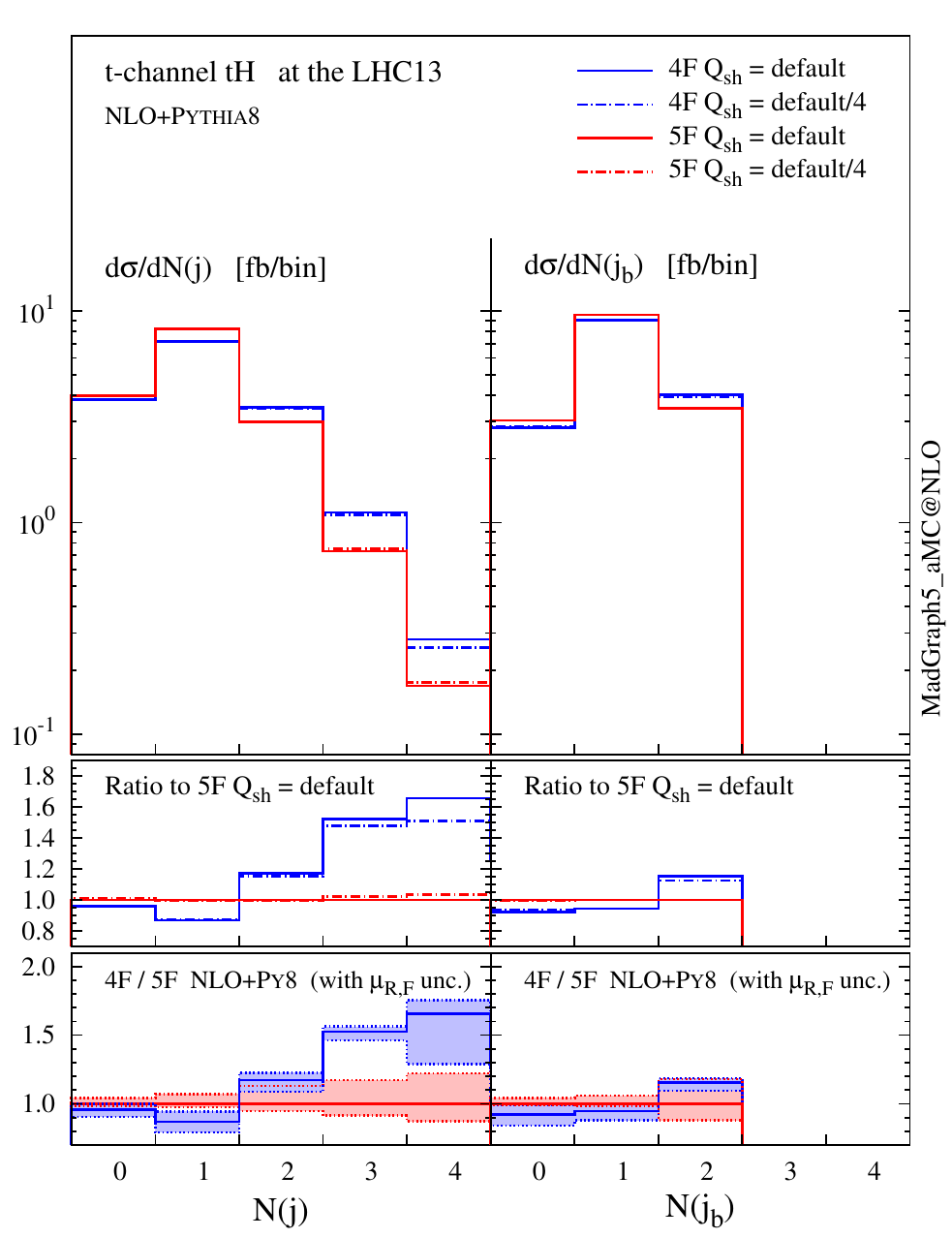}
\caption{Differential distributions in $t$-channel $pp \to tH + \bar tH$
production at the $13\UTeV$ LHC, computed at NLO matched to \Pythiaeight,
in the 4FS (blue) and in the 5FS (red).
We show the transverse momentum of the Higgs boson $p_{\mathrm{T}}(H)$ on the left,
of the hardest $b$-tagged jet $p_{\mathrm{T}}(j_{\mathrm{b},1})$ in the centre,
and the light-jet and $b$-jet multiplicities on the right.
Plots are taken from \Ref~\cite{Demartin:2015uha}.}
\label{fig:tH_tcha_distributions}
\end{figure}

In this section we briefly address NLO differential distributions
matched to a parton shower (NLO+PS).
To generate events for distributions, we recommend to use a dynamic
event-by-event scale instead of the static one in \Eq~\eqref{eq:tH_tcha_scale}.
In \Ref~\cite{Demartin:2015uha} we have employed the fraction of transverse energy
(restricted to the set of $H$, $t$, and $b$ particles) given by the formula
\begin{equation}
\label{eq:tH_tcha_dynscale}
\mu_0^{(t-\mathrm{channel,~dynamic})} = \sum_{\mathrm{i=H,t,b}} M_{\mathrm{T}}(i)/6 \,,
\hspace*{4em}
\mathrm{where} \quad M_{\mathrm{T}} = \sqrt{M^2+p_{\mathrm{T}}^2} \,,
\end{equation}
is the transverse mass of a particle of mass $M$.
We have found that this choice of dynamic scale returns a total cross section very close
to the one computed with the static scale in \Eq~\eqref{eq:tH_tcha_scale}.
On top of that, there is a remarkable agreement at NLO+PS between the 4FS and 5FS
predictions for many observables, such as those related to the
Higgs boson (see left plot in \refF{fig:tH_tcha_distributions}),
the top quark, and the forward jet.
This is non trivial, especially in the light of the inadequacy of
the 5FS predictions at LO, which can suffer of large differential $K$ factors
after the inclusion of bottom-quark transverse dynamics,
see central plot in \refF{fig:tH_tcha_distributions}.
On the other hand, the 4FS gives in general more stable results
(flatter $K$ factors, smaller scale dependence in the tails)
and is able to provide accurate predictions for a wider set of observables,
including those related to the spectator $b$-quark and the extra jets.
Finally, we find the choice of shower starting scale, $Q_{\mathrm{sh}}$, to have a tiny impact on NLO+PS results,
see right-hand-side plot in \refF{fig:tH_tcha_distributions}.


\subsection{\texorpdfstring{$s$}{s}-channel \texorpdfstring{$tH$}{tH} production}
\label{sec:tH_scha}

\begin{figure}
\center
\vspace*{0.08\textwidth}
\includegraphics[width=0.33\textwidth]{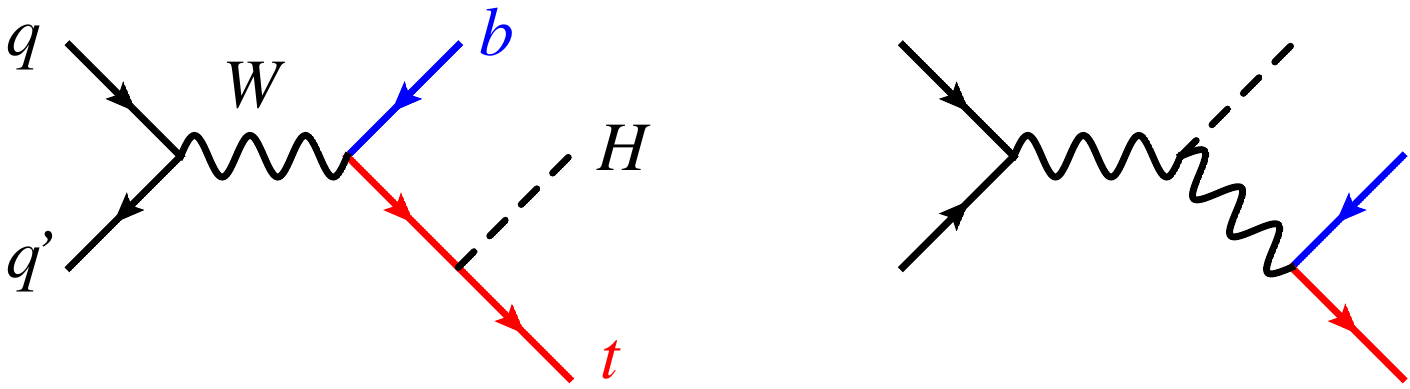}
\hspace*{0.63\textwidth} \\
\vspace*{-0.18\textwidth} \hspace*{0.38\textwidth}
\includegraphics[width=0.28\textwidth]{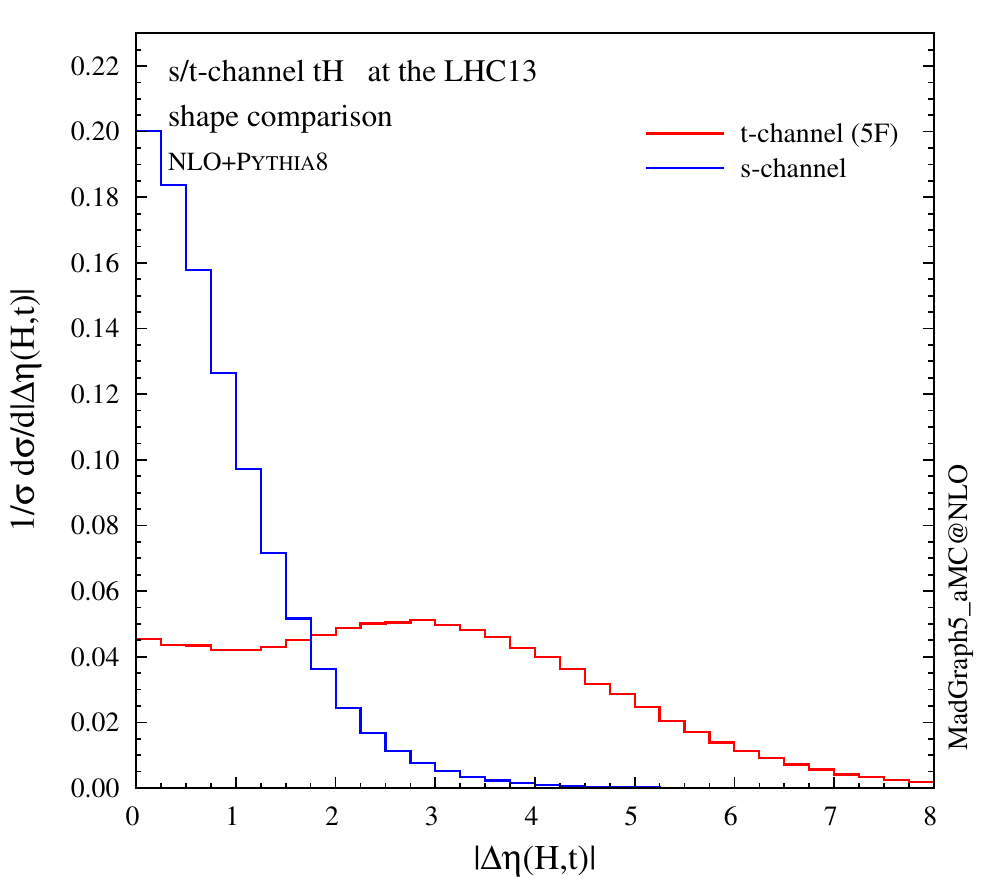}
\hspace*{0.2em}
\includegraphics[width=0.28\textwidth]{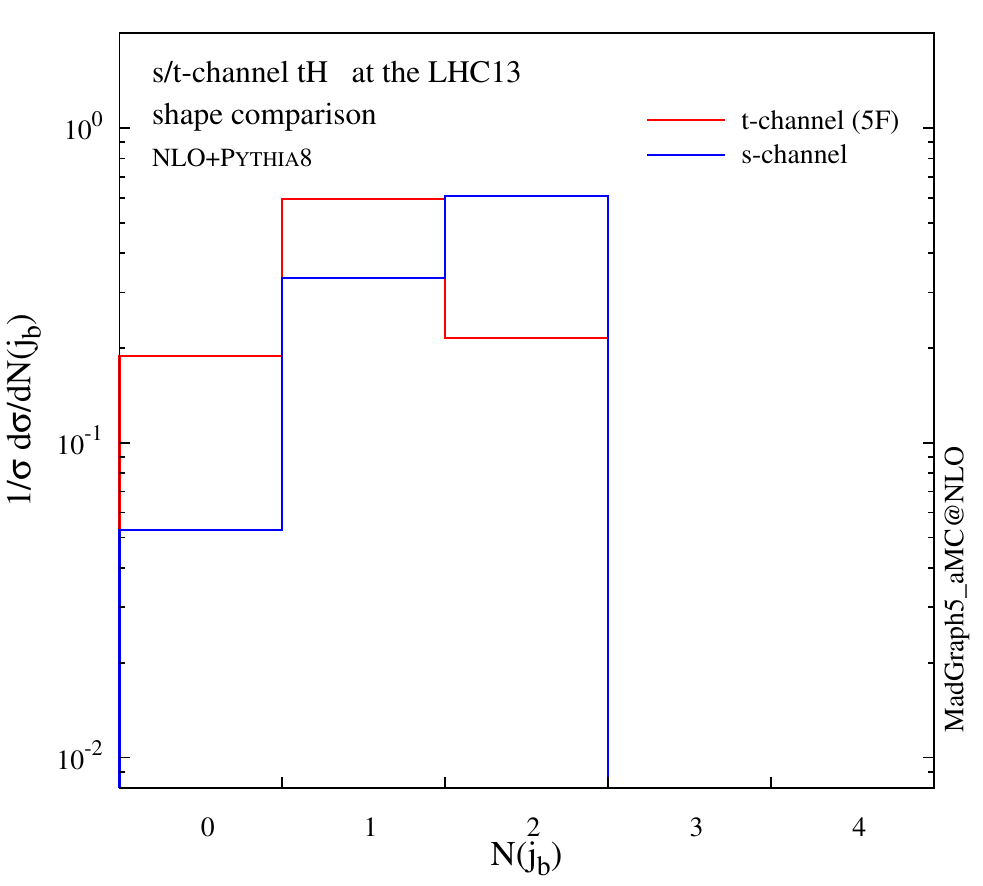}
\caption{On the left: LO Feynman diagrams for $s$-channel $tH$ production.
At the centre: pseudorapidity distance between the Higgs boson and the top quark
$\Delta \eta (H,t)$ in the $s$-channel process (blue)
compared to the $t$-channel one (red).
On the right: multiplicity of $b$-tagged jets in the $s$-channel process (blue)
compared to the $t$-channel one (red).
Plots are taken from \Ref~\cite{Demartin:2015uha}.}
\label{fig:tH_scha_diagrams}
\end{figure}

Unlike the $t$-channel process, $s$-channel $tH$ production is not
affected by flavour-scheme ambiguities, since at LO it is initiated by
quark-antiquark annihilation (see left-hand-side of
\refF{fig:tH_scha_diagrams}).  Therefore, one can simply employ
the 5FS for simulating this process.  Results presented in this section have been obtained
in the \MGfiveamcnlo\, framework. Once again, spin correlations can be kept
by decaying the top quarks with \Madspin.
The same input parameters as for the $t$-channel process in the 5FS
have been used, with the exception of the reference scale choice,
which in this case is
\begin{equation}
\mu_0^{(s-\mathrm{channel})} = (\mh+\mt)/2 \,.
\end{equation}

In \Trefs{tab:tH_scha_SM_lhc7}, \ref{tab:tH_scha_SM_lhc8},
\ref{tab:tH_scha_SM_lhc13} and \ref{tab:tH_scha_SM_lhc14}
we collect the results for the combined $s$-channel $pp \to tH + \bar tH$ production
at the LHC, at centre-of-mass energies of
$\sqrt{s} = 7,8,13$, and $14\UTeV$ respectively, and for various Higgs boson masses
in the range $120{-}130\UGeV$.
These tables are analogous to the ones presented in the previous section for the
$t$-channel process:
in the third column we report the reference cross section;
in the fourth the QCD $K$ factor, defined in \Eq~\eqref{eq:tH_tcha_scalevariations};
in the fifth the scale dependence, computed from the maximum and minimum variations
of the cross section among the 6 points listed in \Eq~\eqref{eq:tH_tcha_scalevariations};
in the sixth, seventh, and eight the $\als$, PDF, and combined PDF+$\als$ uncertainty,
computed employing the 30+2 \textsc{PDF4LHC15} Hessian set;
and finally, in the last two columns we report the separate top and anti-top contributions
to the cross section.
In \refT{tab:tH_scha_SM_lhc6to15} we show the cross-section results
obtained varying the
LHC energy in the range $\sqrt{s} = 6{-}15\UTeV$ and keeping the Higgs boson mass fixed
to $\mh=125\UGeV$.
All these numbers are summarized in the plots
of \refF{fig:tH_scha_SM_plots}, where the the blue uncertainty band
is produced summing the scale and PDF+$\als$ uncertainties.


We also plot in
\refF{fig:tH_scha_BSM_plots} the analogous cross-section results in the extended Higgs boson mass range
$10{-}3000\UGeV$, at 13 and $14\UTeV$.
Finally, we remark that even though the $s$-channel cross section is very tiny
(around $25{-}30$ times smaller than the $t$-channel cross section at $13{-}14\UTeV$),
this process features kinematical distributions with shapes rather different than the
$t$-channel ones, see central and right plots in \refF{fig:tH_scha_diagrams}.

\begin{figure}
\centering
\includegraphics[width=.4\textwidth]{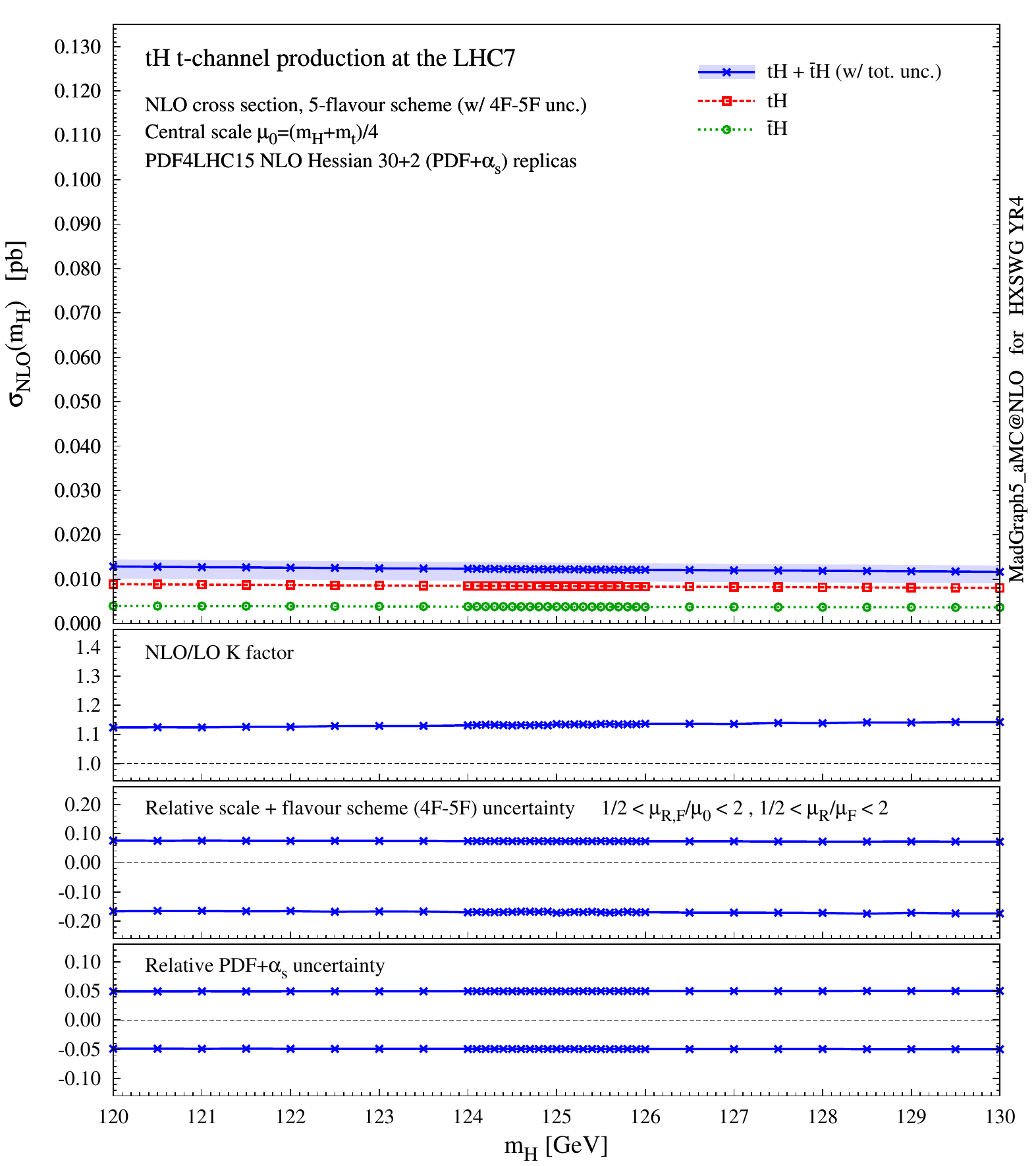}
\includegraphics[width=.4\textwidth]{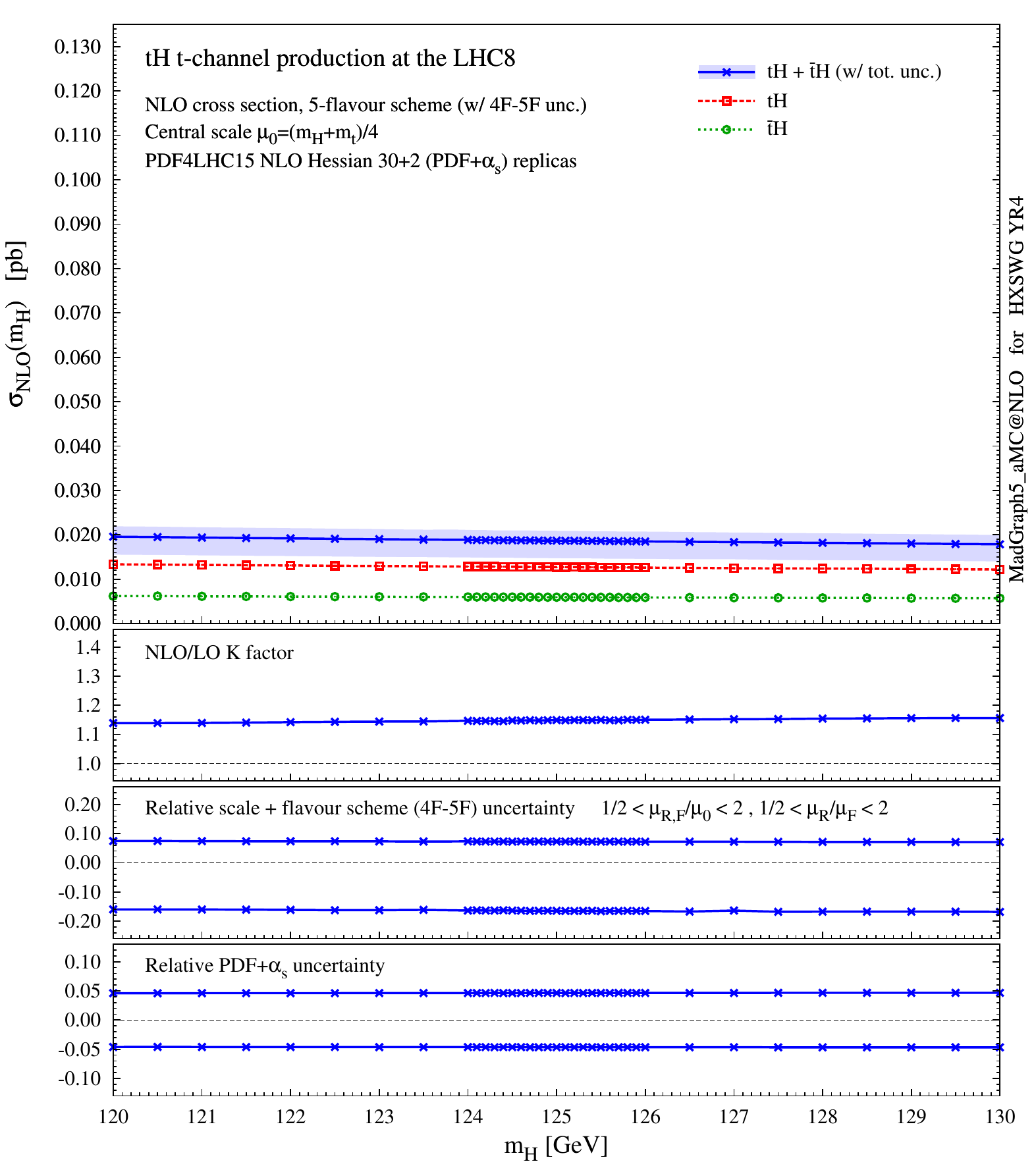}\\
\includegraphics[width=.4\textwidth]{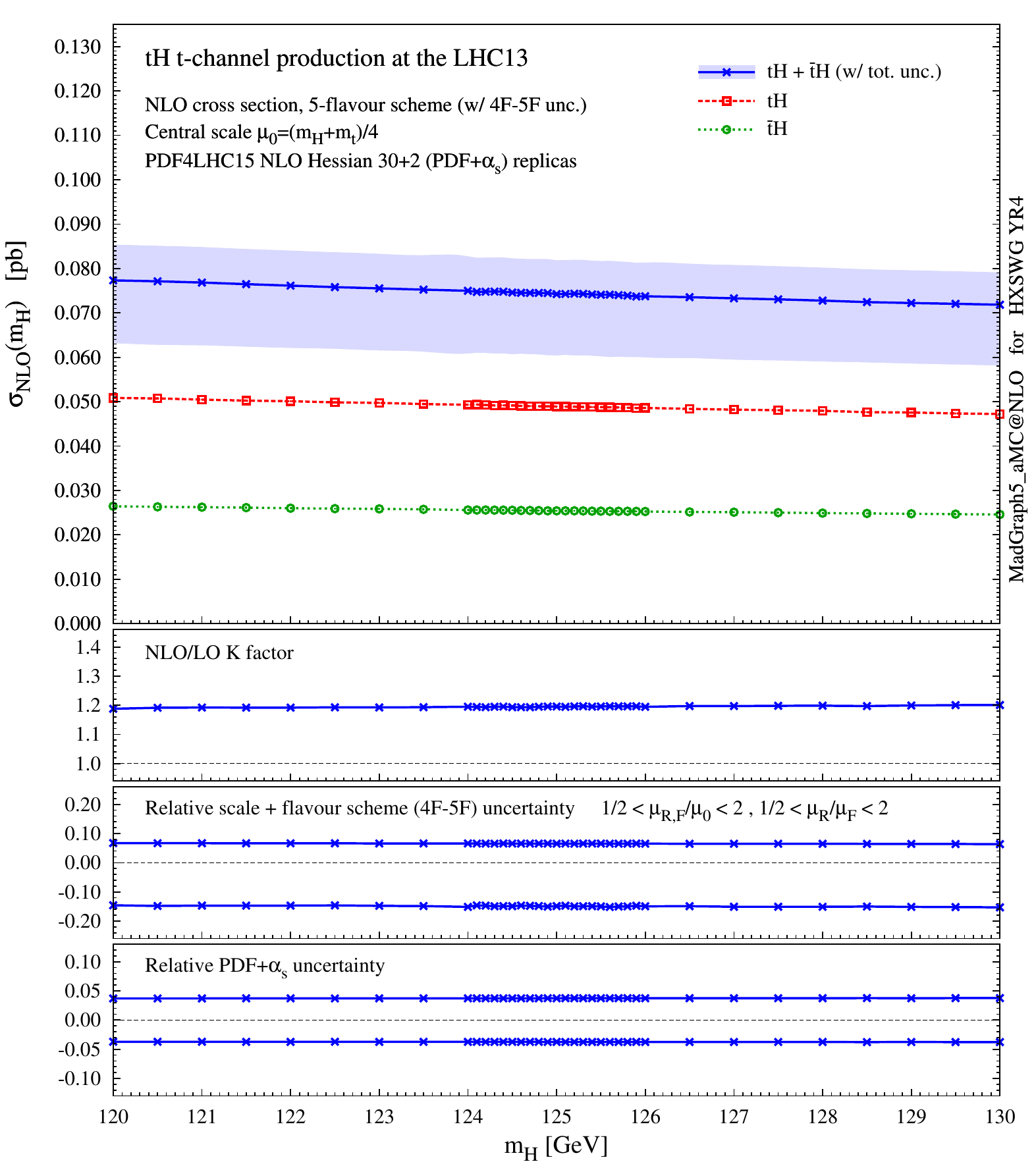}
\includegraphics[width=.4\textwidth]{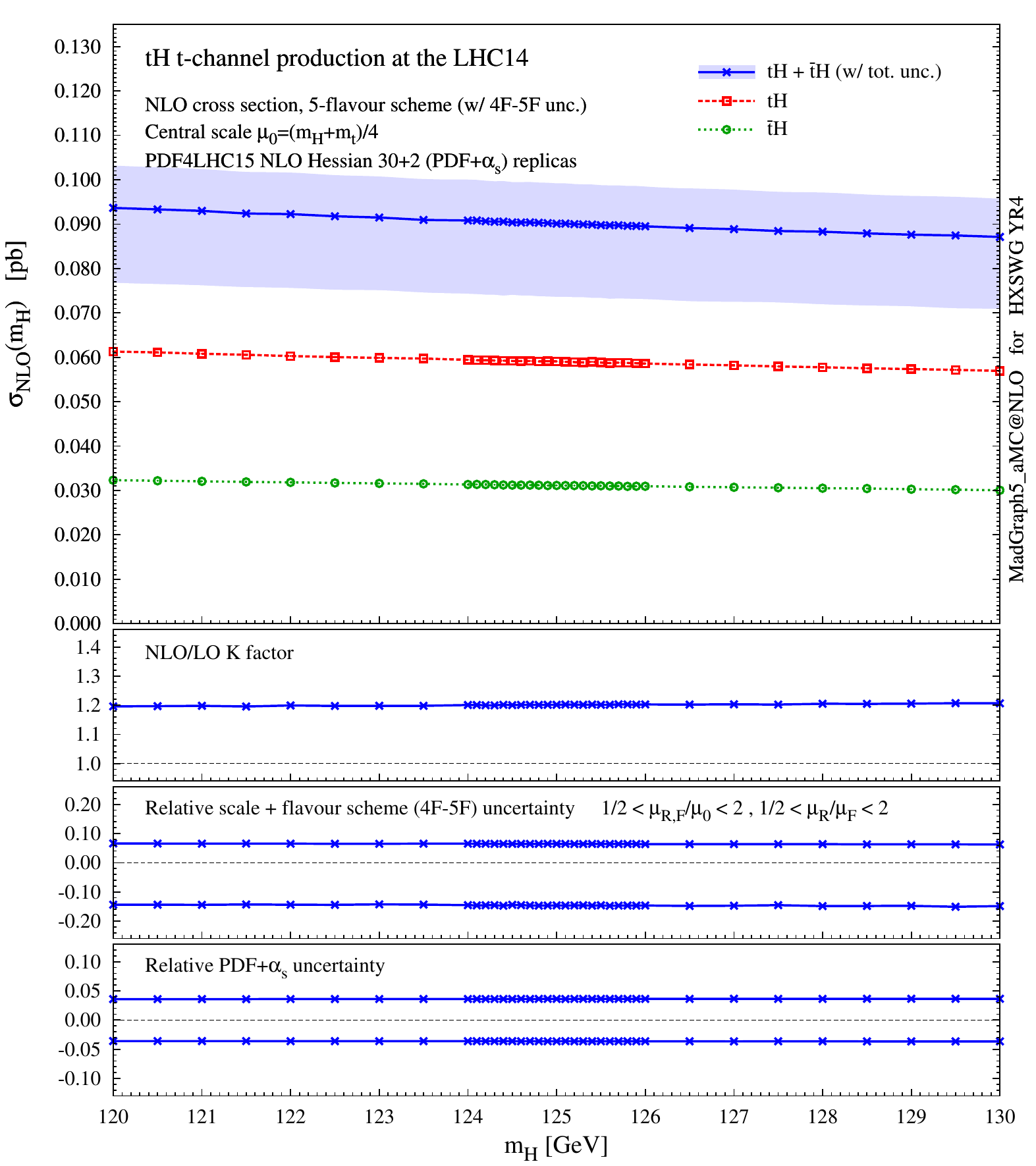}\\
\includegraphics[width=.4\textwidth]{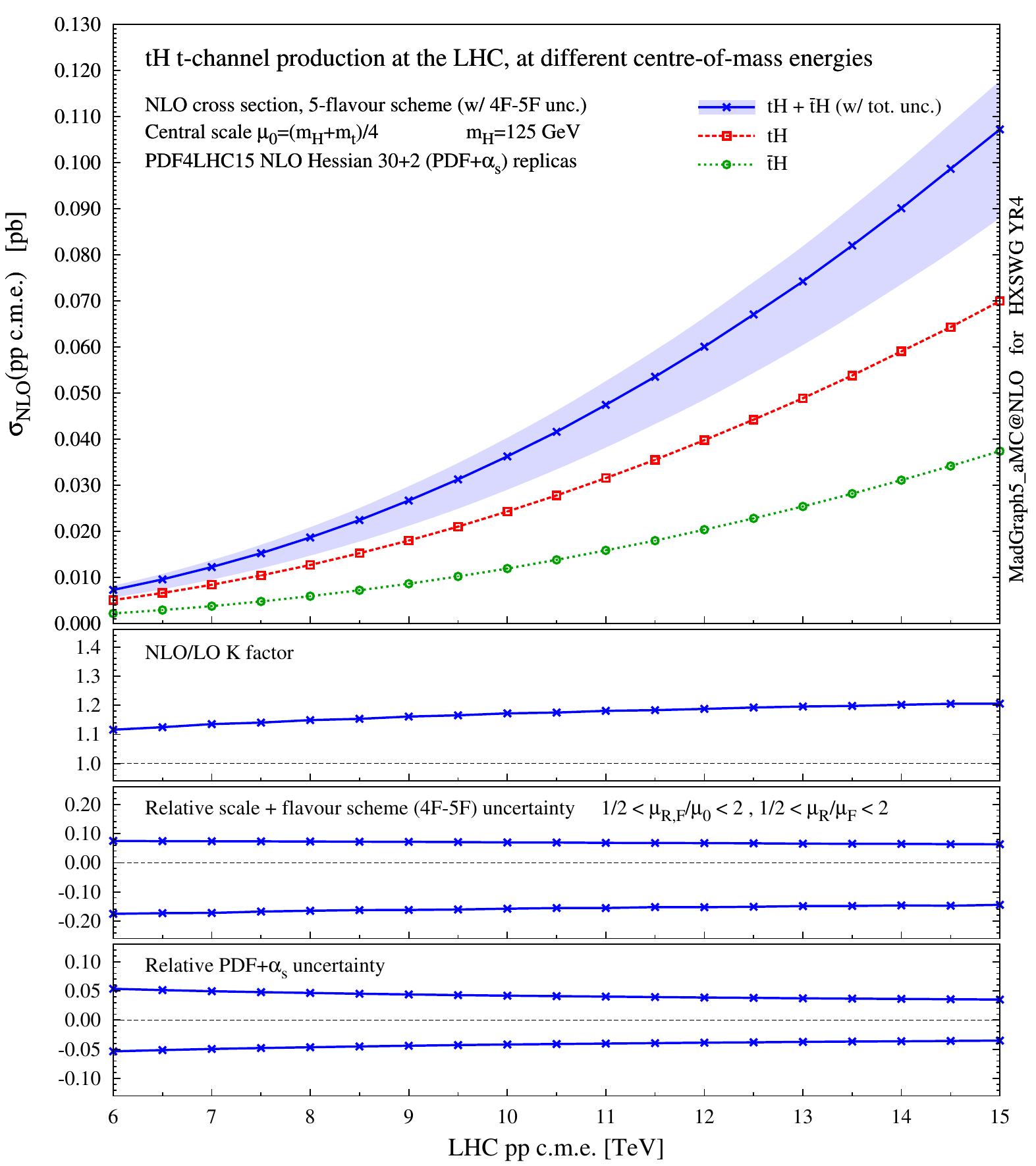}
\caption{Cross sections for $t$-channel $tH$ and $\bar t H$ production.}
\label{fig:tH_tcha_SM_plots}
\end{figure}

\begin{table}\footnotesize
\caption{Cross section for $t$-channel $tH$ and $\bar t H$ production at the 13 and 14\UTeV~LHC, for $y_t=-y_t^{\mathrm{(SM)}}$.}
\label{tab:tH_tcha_-ytSM}
\setlength{\tabcolsep}{1.0ex}
\centering
\begin{tabular}{cccccccccc}
\toprule
$\sqrt{s}$[\UTeV] &
$\mh$[\UGeV] &
$\sigma_{tH+\bar tH}$[\Ufb] &
$K_{\mathrm{QCD}}$ &
Scale+FS  [\%] &
$\alpha_S$ [\%] &
PDF [\%] &
PDF+$\alpha_S$ [\%] &
$\sigma_{tH}$[\Ufb] &
$\sigma_{\bar tH}$[\Ufb]  \\   \midrule
\input{\ttHtHpath/table_-ytSM_tchannel.dat} 
\bottomrule
\end{tabular}
\end{table}


\begin{figure}
\centering
\includegraphics[width=.4\textwidth]{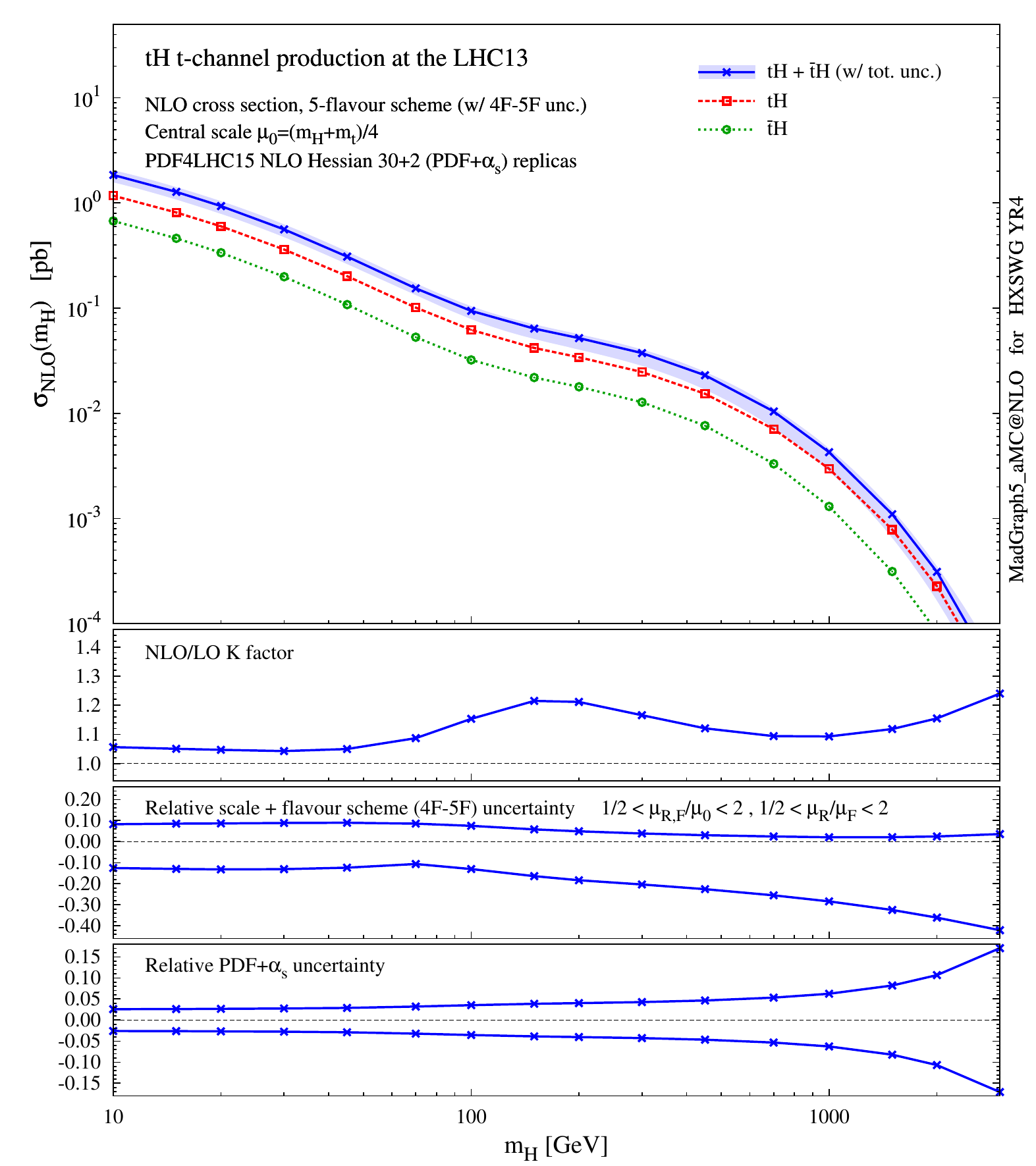}
\includegraphics[width=.4\textwidth]{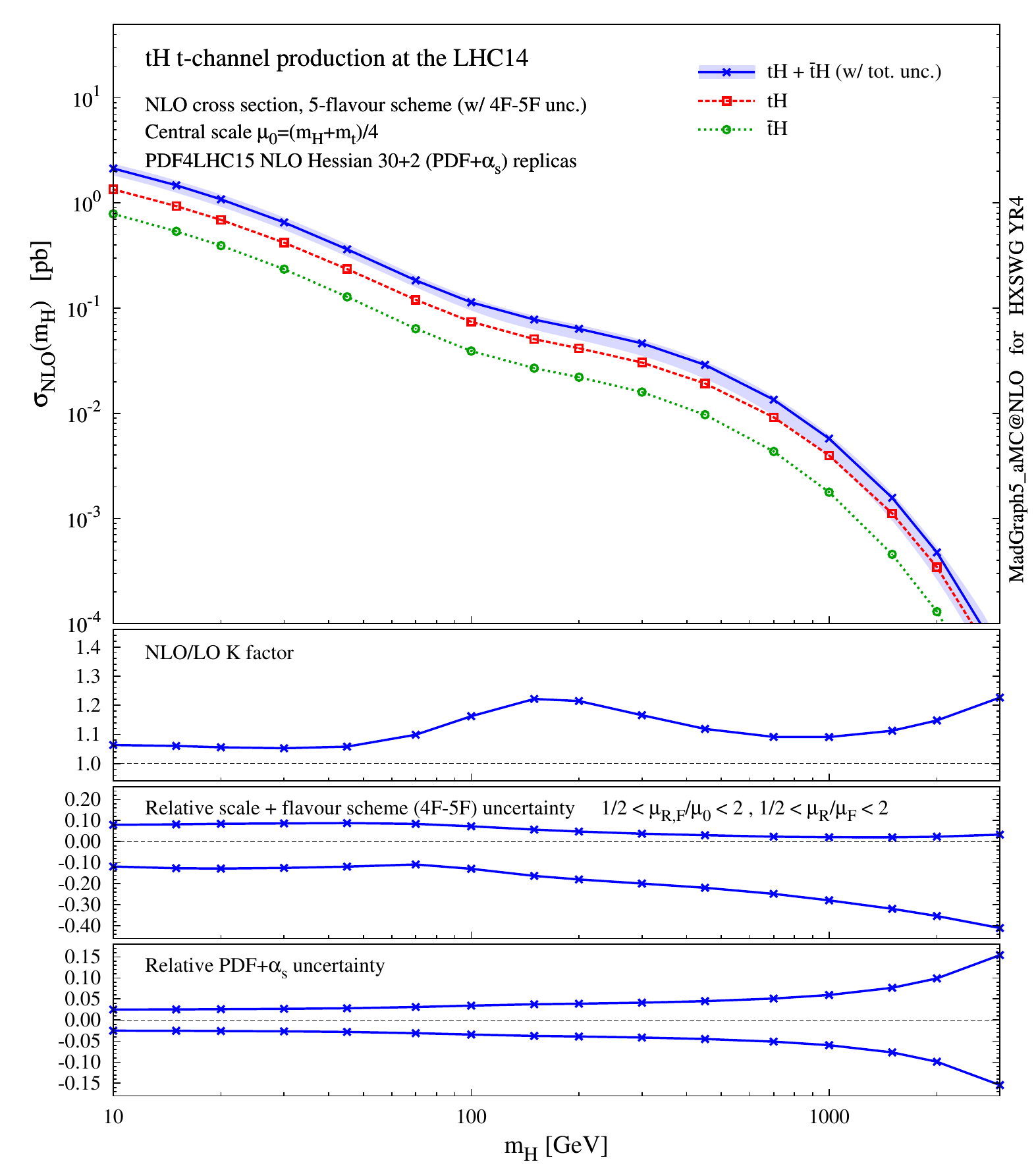}
\caption{Cross sections for $t$-channel $tH$ and $\bar t H$ production
in the extended Higgs boson mass range.}
\label{fig:tH_tcha_BSM_plots}
\end{figure}

\begin{figure}
\centering
\includegraphics[width=.4\textwidth]{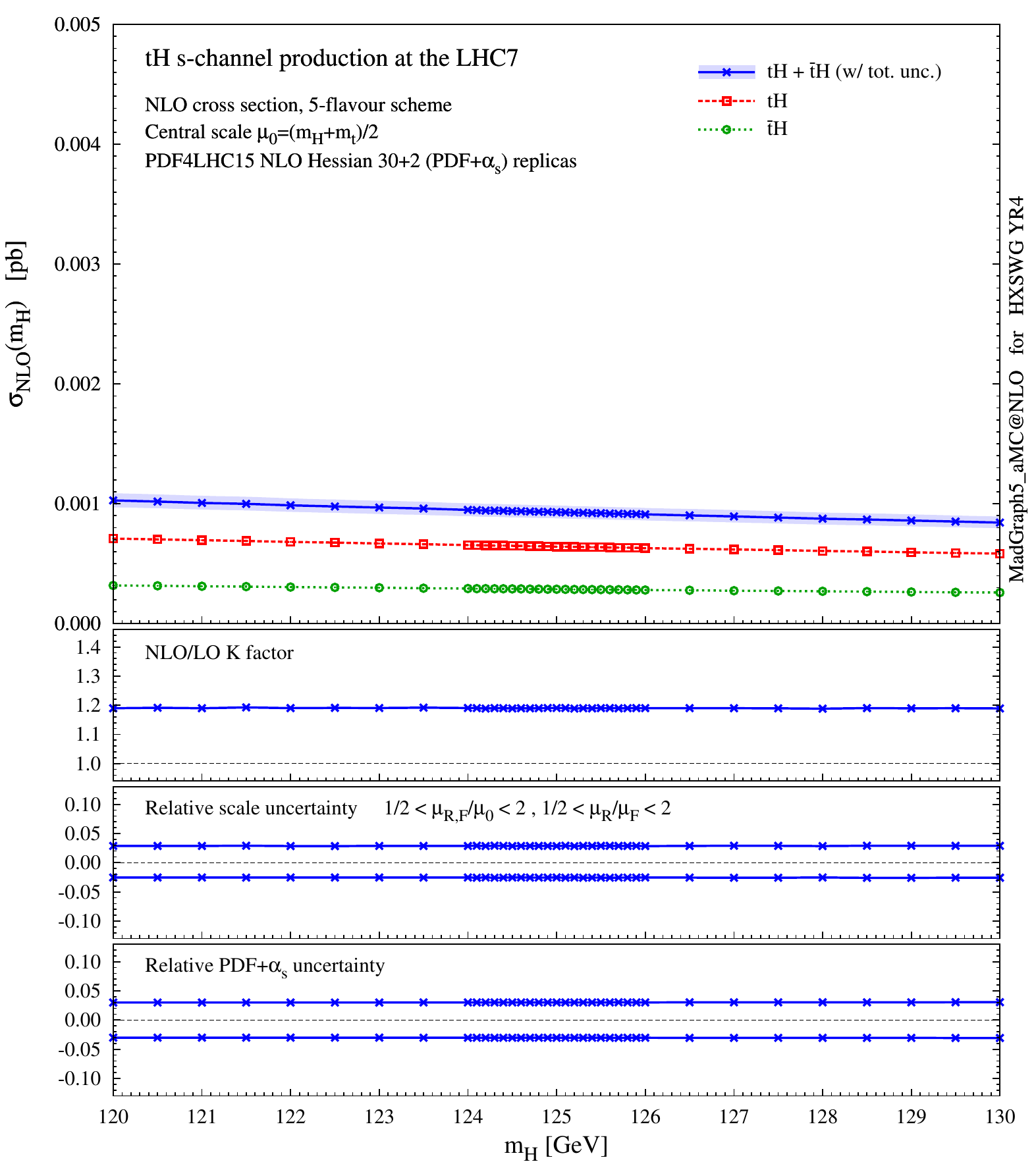}
\includegraphics[width=.4\textwidth]{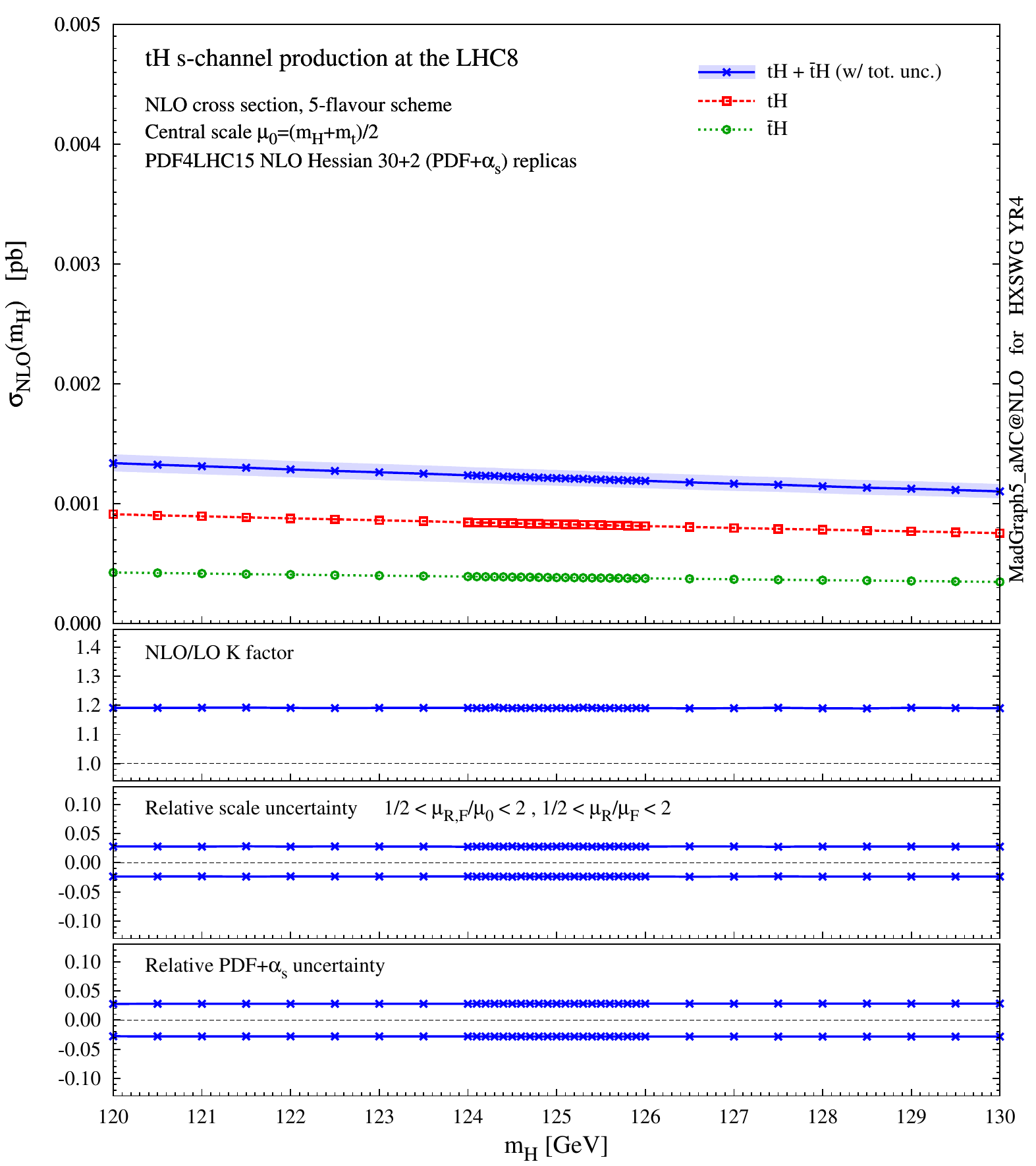}\\
\includegraphics[width=.4\textwidth]{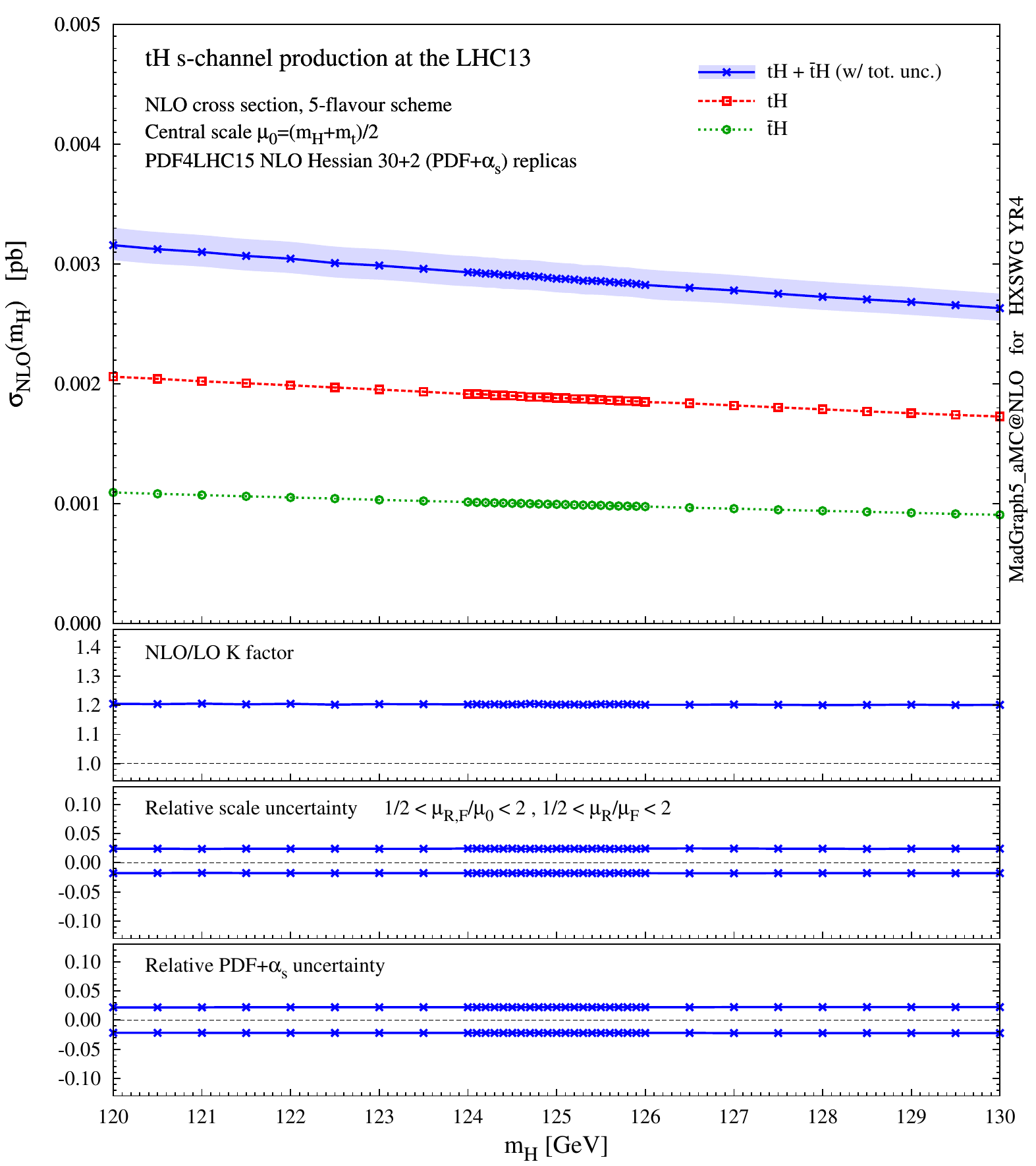}
\includegraphics[width=.4\textwidth]{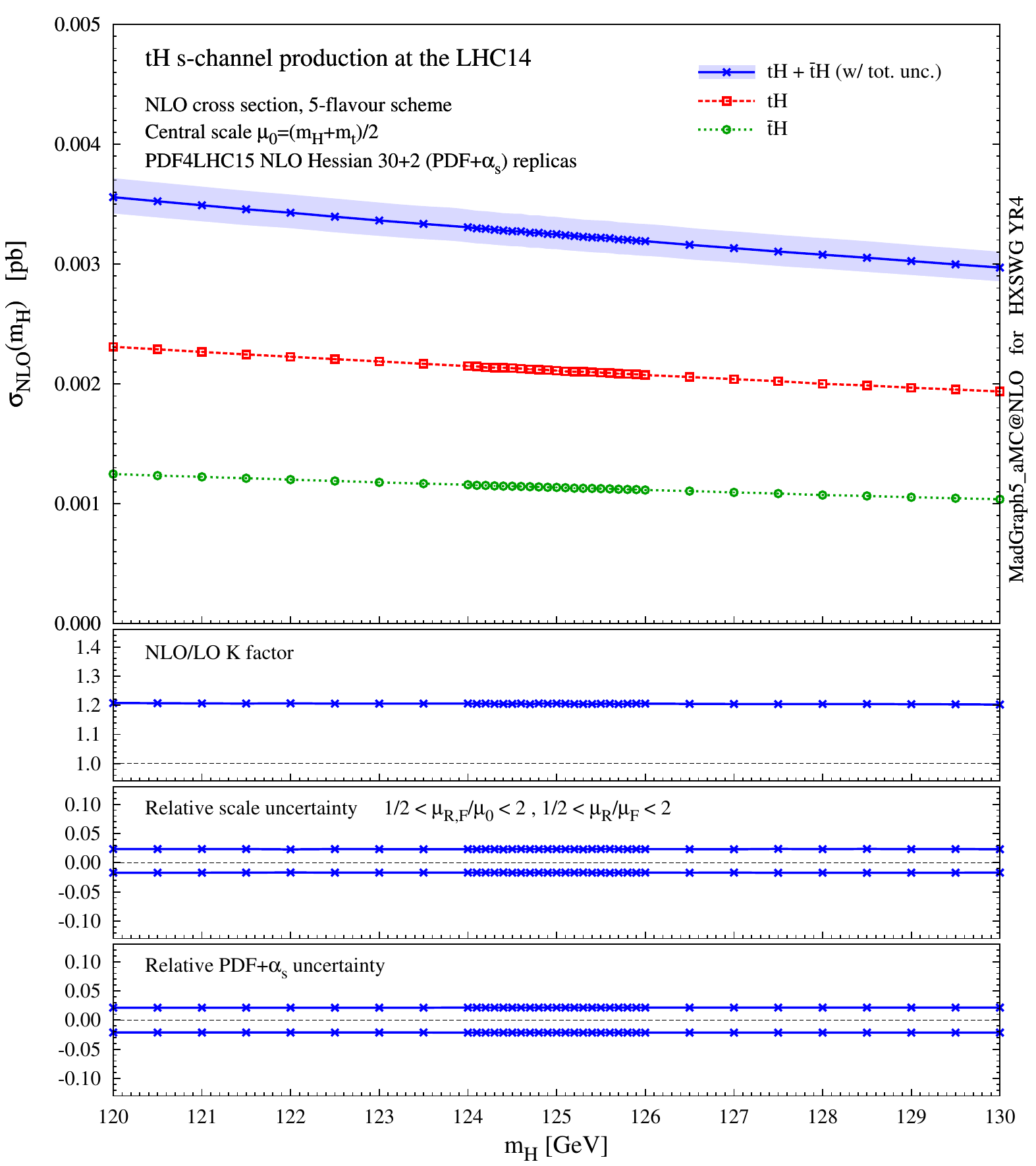}\\
\includegraphics[width=.4\textwidth]{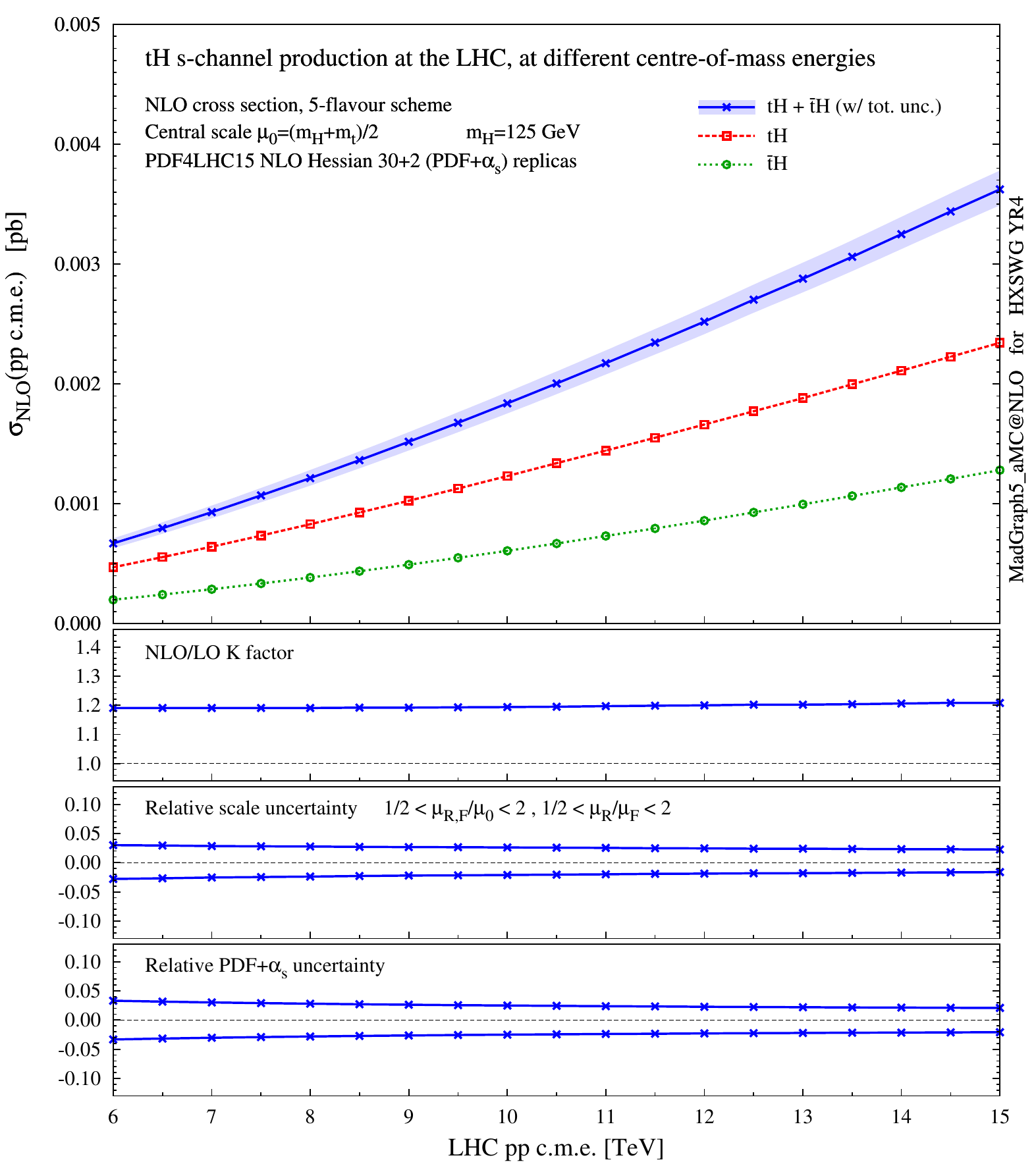}
\caption{Cross sections for $s$-channel $tH$ and $\bar t H$ production.}
\label{fig:tH_scha_SM_plots}
\end{figure}

\begin{figure}
\centering
\includegraphics[width=.4\textwidth]{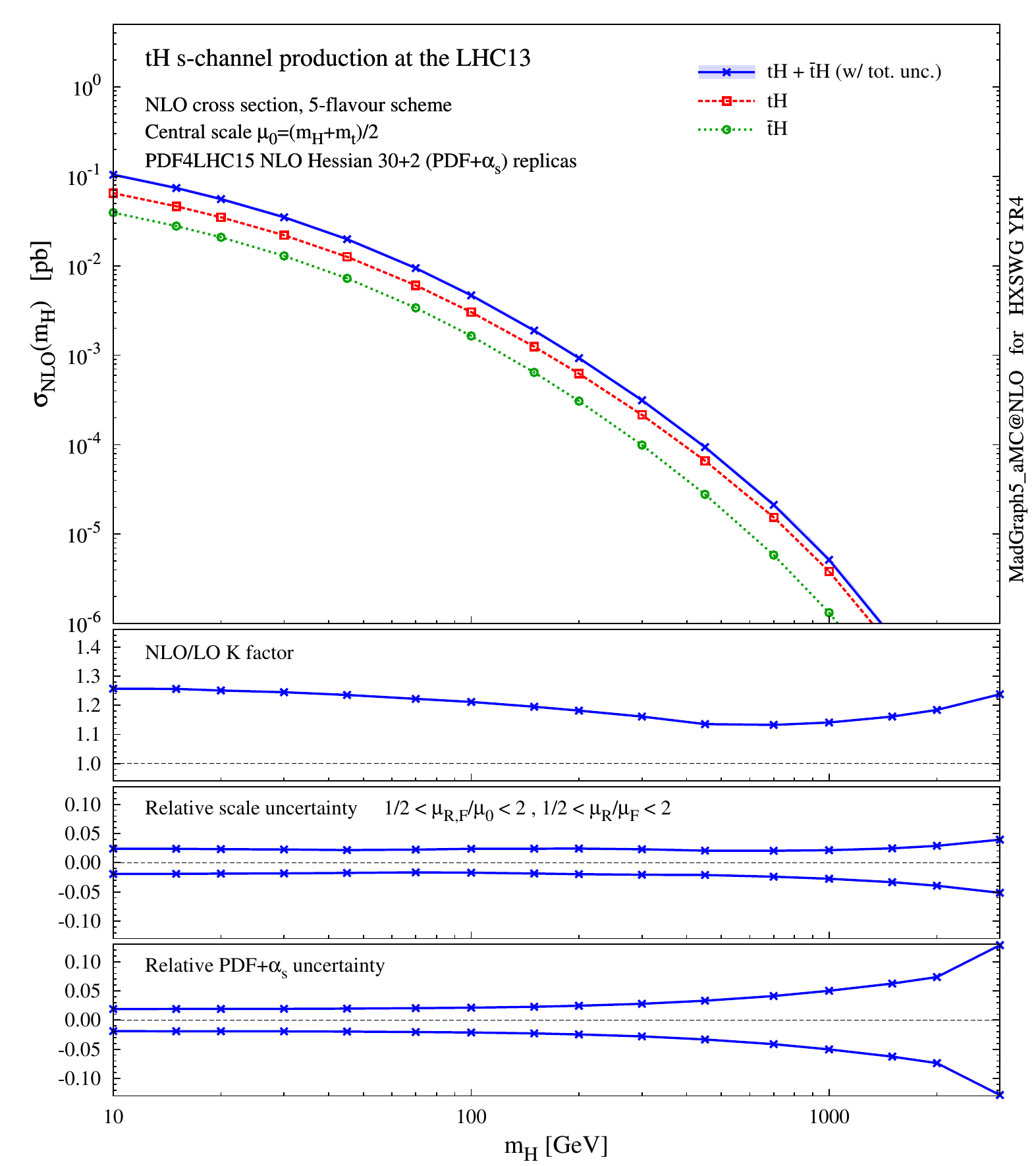}
\includegraphics[width=.4\textwidth]{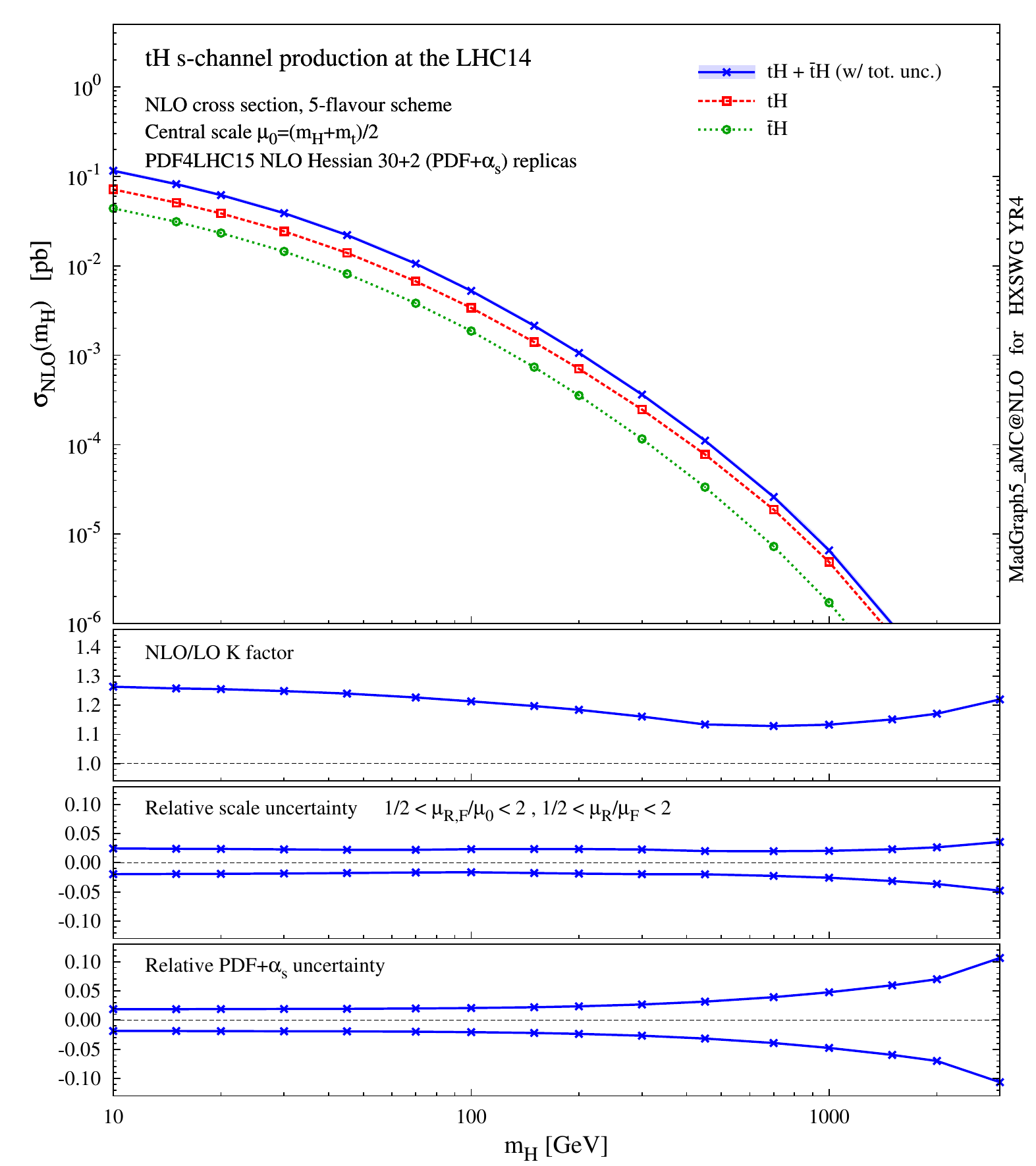}
\caption{Cross sections for $s$-channel $tH$ and $\bar t H$ production
in the extended Higgs boson mass range.}
\label{fig:tH_scha_BSM_plots}
\end{figure}


\section{\texorpdfstring{$t\bar{t}Z$}{ttZ} and \texorpdfstring{$t\bar tW^\pm$}{ttW} production}
\label{se:ttH-ttV}
\def\ttHttvqcdewpath{\ttHpath/ttH-ttV-QCD+EW}
\providecommand{\ttv}{t\bar t V}
\providecommand{\ttvv}{t\bar t VV}
\providecommand{\ttz}{t\bar t Z}
\providecommand{\ttwpm}{t\bar t W^{\pm}}
\providecommand{\ttwp}{t\bar t W^{+}}
\providecommand{\ttwm}{t\bar t W^{-}}

The production of a $\ttbar$ pair in association with electroweak vector bosons represent an important source of background
to $\tth$ production in the $H\to \bbbar$, $H\to WW^*$ and $H\to \tau\tau$ channels.
In this section we present NLO QCD+EW predictions for inclusive and differential $\ttz$ and $\ttwpm$
cross sections, a comparison of NLO QCD distributions obtained with various automated tools,
as well as NLO QCD predictions for $\ttbar$ production in association with two vector bosons.

\subsection{NLO QCD+EW predictions for \texorpdfstring{$t\bar{t}Z$}{ttZ} and \texorpdfstring{$t\bar{t}W^\pm$}{ttW} production}
\label{sec:ttH-ttV-QCD+EW}

Predictions for $\ttbar$ production in association with a vector boson
$V=Z,W^\pm$ at NLO QCD have been presented in~\cite{Lazopoulos:2008de,Kardos:2011na,Campbell:2012dh} and matched to the parton shower
in~\cite{Garzelli:2011is,Garzelli:2012bn}, while the first calculation of electroweak corrections
for this class of processes has been completed more
recently~\cite{Frixione:2015zaa}.  In the following we present NLO QCD+EW
predictions for inclusive $\ttz$ and $\ttwpm$ production
at $\sqrt{s}=13\,\UTeV$ and $14\,\UTeV$.
All input parameters  are chosen according to the HXSWG recommendations~\cite{LHCHXSWG-INT-2015-006}, and the
hadronic cross section is obtained using the PDF4LHC15~\cite{Butterworth:2015oua}
and NNPDF2.3QED~\cite{Ball:2013hta} parton distributions as explained below.
The top-quark and the electroweak vector bosons are treated as stable particles,
and for the renormalization of the respective mass paramaters the on-shell scheme is used.
The Higgs boson mass is set to $\mh=125\UGeV$.
The electroweak couplings and mixing angle are evaluated in the $G_\mu$-scheme using the
Fermi constant and the vector-boson masses as input parameters.
The central value for renormalization and factorization
scales is set to
\begin{equation}
\mu_0 = \mt + M_{\mathrm{V}}/2\,.
\end{equation}
For the NLO QCD part of the calculation scale uncertainties are
estimated by independent variations of renormalization and factorization
scales in the range $ \mu/2 \le \muR, \muF \le 2\mu$, with $1/2 \le \muR
/ \muF \le 2$, while for PDF and $\als$
uncertainties the  PDF4LHC15 prescription is used.
The resulting uncertainties are applied also to NLO EW correction effects.

The NLO QCD+EW predictions presented in the following,
\begin{equation}
\sigmaNLOqcdew=\sigmaNLOqcd+\dsigmaew\,,
\label{eq:ttvqcdew}
\end{equation}
result from the combination of various contributions.  The usual NLO QCD cross section,
\begin{equation}
\sigmaNLOqcd=\sigmaLOqcd+\delta\sigmaNLOqcd,
\end{equation}
comprises LO terms of $\mathcal{O}(\als^2\alpha)$ and NLO QCD corrections of
$\mathcal{O}(\als^3\alpha)$, which involve $gg$, $q\bar q$ and $gq$
partonic channels. Note that the $gg$ channel
starts contributing only at NNLO QCD in the case of
$\ttwpm$ production.
The remaining EW corrections, denoted as $\dsigmaew$,
include three types of terms:
\begin{enumerate}

\item LO EW terms of
$\mathcal{O}(\alpha^3)$ that result from squared EW tree amplitudes in the
$q\bar q$ and $\gamma\gamma$ channels.

\item LO mixed terms  of
$\mathcal{O}(\als\alpha^2)$ that result from the interference of EW and
QCD tree diagrams in the $b\bar b$ and $\gamma g$ channels. Other $q\bar q$
channels do not contribute at this order due to the vanishing interference
of the related colour structures. Thus $\ttwpm$ production does not receive any
$\mathcal{O}(\als\alpha^2)$ contribution.

\item NLO EW corrections of
$\mathcal{O}(\als^2\alpha^2)$ in the $q\bar q$, $gg$ and $\gamma g$
channels.  Subleading NLO terms of $\mathcal{O}(\als\alpha^3)$ and
$\mathcal{O}(\alpha^4)$ are not included as they are expected to be strongly
suppressed.
\end{enumerate}
At $\sqrt{s}=13$--$14\,\UTeV$, LO EW contributions
to $\ttz$ and $\ttwpm$ production
correspond to $+1.2\%$ and $+0.6\%$ of the respective NLO QCD cross sections,
while LO mixed effects are at the sub-per mille level.
Order $\als^2\alpha^2$  NLO EW corrections to the
$\ttz$, $\ttwp$ and $\ttwm$ cross sections
amount to $-0.9\%$, $-3.3\%$ and $-2.6\%$, respectively.\footnote{Here $\mathcal{O}(\alpha)$ effects related
to the QED evolution of PDF (see below) are not included.}
Photon-induced channels have a non-negligible impact ($+0.8\%$) only
in the $\ttz$ cross section at $\mathcal{O}(\als\alpha^2)$. However
their effect is almost completely cancelled by the contribution of the $b\bar b$ channel
at the same order.

Effects of $\mathcal{O}(\alpha)$ related to the QED evolution of PDF and initial-state photons
are included in the same way as discussed in Sect.~\ref{sec:ttH-XS}
for $\tth$ production: (i) all NLO QCD+EW
predictions for partonic channels with initial-state gluons and quarks are
computed using the PDF4LHC15 set with 30+2 members; (ii) for $\gamma$-induced
channels NNPDF2.3QED set (with $\als(\mz)=0.118$) is used;
(iii) the effect of the QED evolution of quark PDF is estimated from the
difference between NNPDF2.3QED and NNPDF2.3 parton densities
as indicated in~\refE{eq:tthpdfqed}, and
for a more detailed discussion we refer to Sect.~\ref{sec:ttH-XS}.
Similarly as for $\tth$ production, the effect of QED PDF evolution is
negative and ranges from $-0.6\%$ to $-0.8\%$.

Corrections of order $\alpha^2\als^2$ resulting from the emission of and
extra heavy boson have been discussed in~\cite{Frixione:2015zaa}.  Although
they enter at the same perturbative order as the NLO EW corrections to
$\ttv$ production, such contribution represent
separate physics processes, namely $\ttbar$ production in associated with
two vector bosons. Corresponding
predictions at NLO QCD are presented in Sect.~\ref{sec:ttH-ttVV}.

Predictions for inclusive $\ttv$ cross sections at NLO QCD+EW for $\sqrt{s}=13$ and $14\,\UTeV$ are listed
in \refTs{tab:ttvXSa}--\ref{tab:ttvXSb}. The
impact of QCD and EW corrections is shown in the form of a QCD correction
factor
\begin{equation}
\label{eq:ttvqcd}
K_{\mathrm{QCD}}=\frac{\sigmaNLOqcd}{\sigmaLOqcd}\,,
\end{equation}
and a relative EW correction factor
\begin{equation}
\delta_{\mathrm{EW}}=\dsigmaew/\sigmaNLOqcd\,.
\label{eq:ttvdsiew}
\end{equation}
Electroweak corrections include all LO and NLO EW
effects discussed above, i.e.~also those arising form the QED evolution of PDF.
All quantities in \refE{eq:ttvqcd}--\refE{eq:ttvdsiew}
are obtained using NLO QCD PDF.
Results in \refTs{tab:ttvXSa}--\ref{tab:ttvXSb} have been obtained with
\MGfiveamcnlo~\cite{Alwall:2014hca}.  A cross check against an independent
calculation based on \Sherpa+\Openloops~\cite{Kallweit:2014xda} has
confirmed the correctness of NLO QCD+EW predictions at the level of the
quoted statistical accuracy (0.1--0.2\%).


\begin{table}\footnotesize
\caption{Inclusive $\ttv$ cross sections at NLO QCD and NLO QCD+EW accuracy for $\sqrt{s}=13$~\UTeV.
NLO QCD+EW results represent the best predictions and should be used in experimental analyses.
Scale, PDF, and $\als$ uncertainties are quoted in per cent. Absolute statistical uncertainties
are indicated in parenthesis. We also quote the NLO QCD+EW $\ttwm +
\ttwp$  combined cross
sections where correlation effects have been
consistently included in the estimate of the corresponding uncertainties.  Collider energy and cross sections are in \UTeV\, and femtobarn, respectively.}
\label{tab:ttvXSa}
\setlength{\tabcolsep}{1.0ex}
\centering
\renewcommand{\arraystretch}{1.2}
\begin{tabular}{lcccccccc}
\toprule
Process &  $\sqrt{s}$ &  $\sigmaNLOqcd$   &  $\sigmaNLOqcdew$  & $K_{\mathrm{QCD}}$ & $\deltaEW$[\%]&  Scale[\%]              &  PDF[\%]   &  $\alpha_S$[\%]   \\   
\midrule
$\ttz$  &  $13$       &  $841.3( 1.6)$ &  $839.3( 1.6)$  & $1.39$             & $-0.2$        &  $+9.6\%\;\;-11.3\%$  &  $+2.8\%\;\;-2.8\%$   &  $+2.8\%\;\;-2.8\%$   \\
$\ttwp$ &  $13$       &  $412.0(0.32)$ &  $397.6( 0.32)$ & $1.49$             & $-3.5$        &  $+12.7\%\;\;-11.4\%$ &  $+2.0\%\;\;-2.0\%$   &  $+2.6\%\;\;-2.6\%$   \\
$\ttwm$ &  $13$       &  $208.6(0.16)$ &  $203.2( 0.16)$ & $1.51$             & $-2.6$        &  $+13.3\%\;\;-11.7\%$ &  $+2.1\%\;\;-2.1\%$   &  $+2.9\%\;\;-2.9\%$   \\
$\ttwm + \ttwp$ &  $13$       &  $620.6(0.36)$ &  $600.8(0.36)$ &  $1.50$     & $-3.2$      &  $+12.9\%\;\;-11.5\%$ &  $+2.0\%\;\;-2.0\%$   &  $+2.7\%\;\;-2.7\%$   \\
\bottomrule
\end{tabular}
\end{table}


\begin{table}
\footnotesize
\caption{Inclusive $\ttv$ cross sections at NLO QCD and NLO QCD+EW accuracy for $\sqrt{s}=14$~\UTeV.
NLO QCD+EW results represent the best predictions and should be used in experimental analyses.
Scale, PDF, and $\als$ uncertainties are quoted in per cent. Absolute statistical uncertainties
are indicated in parenthesis. Collider energy and cross sections are in \UTeV\, and femtobarn, respectively.}
\label{tab:ttvXSb}
\setlength{\tabcolsep}{1.0ex}
\centering
\renewcommand{\arraystretch}{1.2}
\begin{tabular}{lcccccccc}
\toprule
Process &  $\sqrt{s}$ &  $\sigmaNLOqcd$ &  $\sigmaNLOqcdew$ & $K_{\mathrm{QCD}}$ & $\deltaEW$[\%]&  Scale[\%]               &  PDF[\%] &  $\alpha_S$[\%] \\   
\midrule
$\ttz$  &  $14$       &  $1018(2.2)$   &  $1015(2.2)$   & $1.40$             & $-0.3$        &  $+9.6\%\;\;-11.2\%$   &  $+2.7\%\;\;-2.7\%$ &  $+2.8\%\;\;-2.8\%$   \\
$\ttwp$ &  $14$       &  $474.9(0.36)$ &  $458.2(0.36)$ & $1.51$             & $-3.5$        &  $+13.2\%\;\;-11.6\%$  &  $+1.9\%\;\;-1.9\%$ &  $+2.6\%\;\;-2.6\%$   \\
$\ttwm$ &  $14$       &   $244.5(0.17)$ &  $238.0(0.17)$ & $1.54$             & $-2.7$        &  $+13.8\%\;\;-11.8\%$ &  $+2.0\%\;\;-2.0\%$ &  $+2.9\%\;\;-2.9\%$   \\
\bottomrule
\end{tabular}
\end{table}

The size of the corrections as well as the scale and PDF+$\als$
uncertainties vary very little form 13 to 14~\UTeV. For $\ttwpm$
production the QCD and EW corrections as well as the NLO scale
uncertainties turn out to be slightly more pronounced as compared to
$\ttz$ production.  Scale variations range from 10 to 13\% and
represent the dominant source of uncertainty.  It was checked that
replacing the fixed scale $\mu=\mt+M_{\mathrm{V}}/2$ by a dynamic
scale $\mu=H_{\mathrm{T}}/2$ shifts all $\ttv$ cross sections by
$-7\%$, which is consistent with the scale uncertainties quoted
in~\refTs{tab:ttvXSa}--\ref{tab:ttvXSb}.  The combined PDF+$\als$
uncertainty amounts to $2$--$3\%$, and EW correction effects turn out
to be similarly small (between $-0.2\%$ and $-3.5\%$).

\begin{figure}
	\centering
	\includegraphics[page=1, trim=80 100 40 80, clip,scale=0.32]{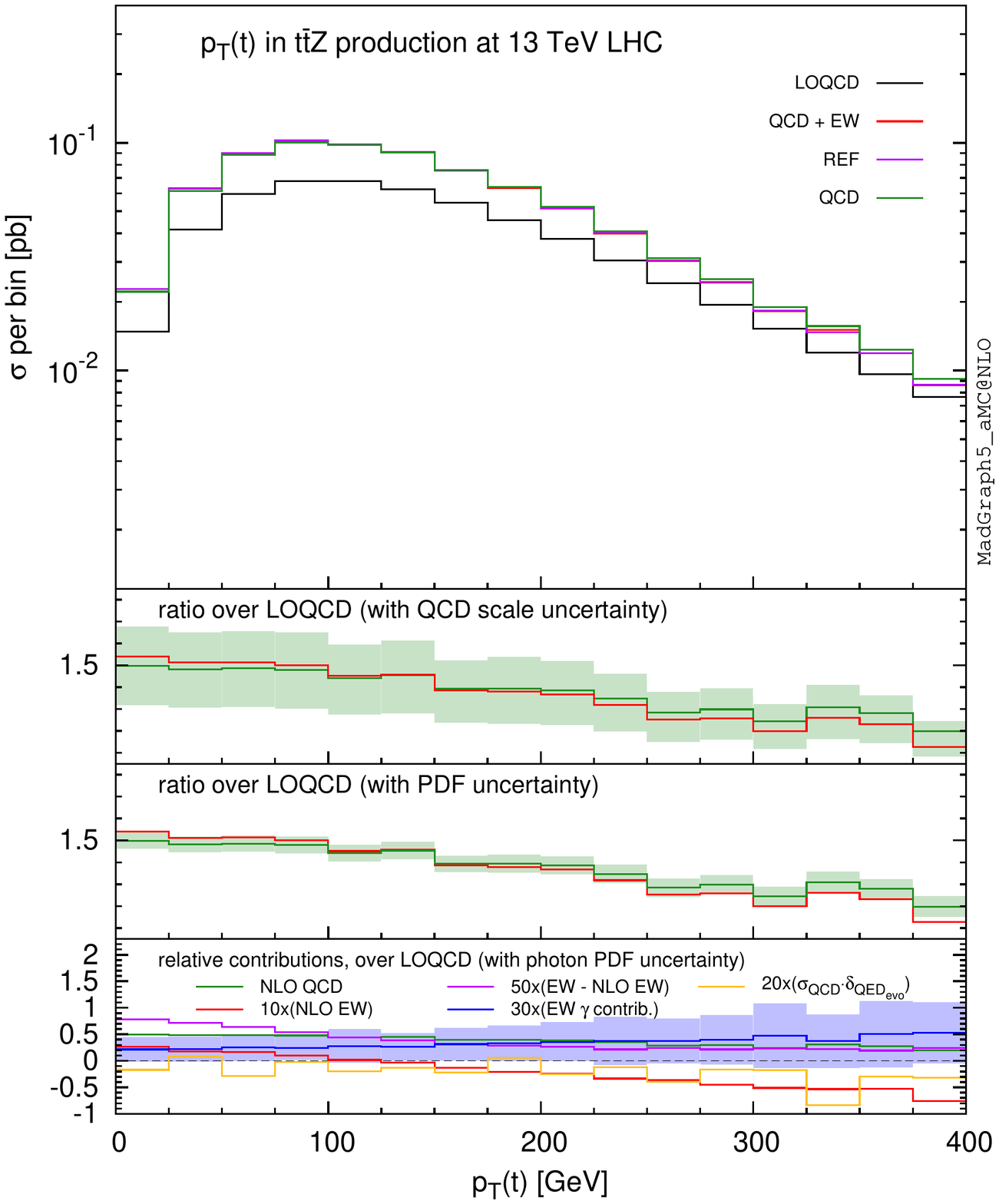}
	\includegraphics[page=2, trim=80 100 40 80, clip,scale=0.32]{\ttHttvqcdewpath/Figures/ttZ.pdf}
	\includegraphics[page=3, trim=80 100 40 80, clip,scale=0.32]{\ttHttvqcdewpath/Figures/ttZ.pdf}
	\caption{Transverse momentum ($p_{\mathrm{T}}$) distribution of the top quark, anti-top quark, and $Z$ boson.}
\label{fig:ttZEW}
\end{figure}

\begin{figure}
	\centering
	\includegraphics[page=1, trim=80 100 40 80, clip,scale=0.32]{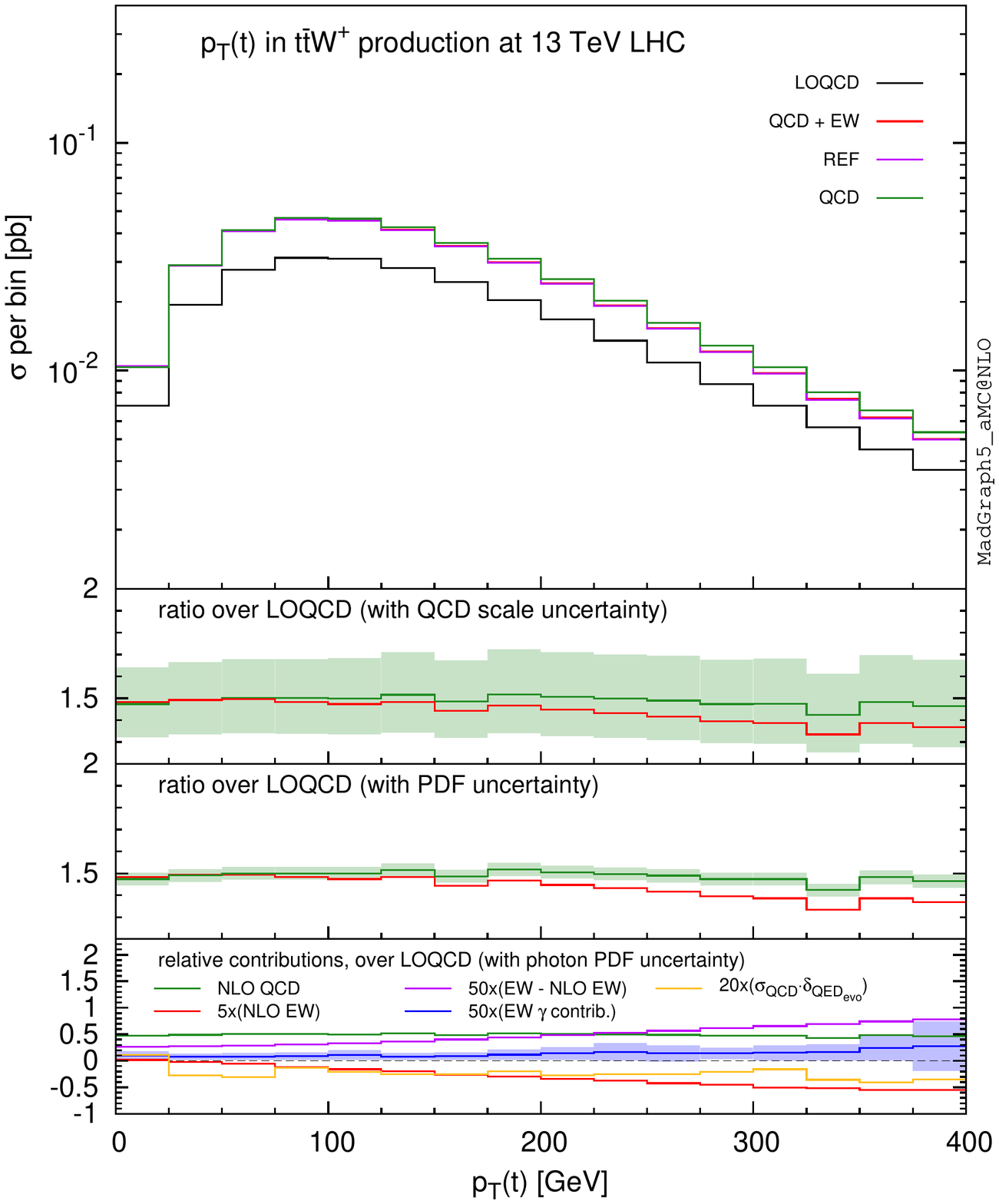}
	\includegraphics[page=2, trim=80 100 40 80, clip,scale=0.32]{\ttHttvqcdewpath/Figures/ttWp.pdf}
	\includegraphics[page=3, trim=80 100 40 80, clip,scale=0.32]{\ttHttvqcdewpath/Figures/ttWp.pdf}
	\caption{Transverse momentum ($p_{\mathrm{T}}$) distribution of the top quark, anti-top quark, and $W^+$ boson.}
\label{fig:ttWpEW}
\end{figure}

\begin{figure}
	\centering
	\includegraphics[page=1, trim=80 100 40 80, clip,scale=0.32]{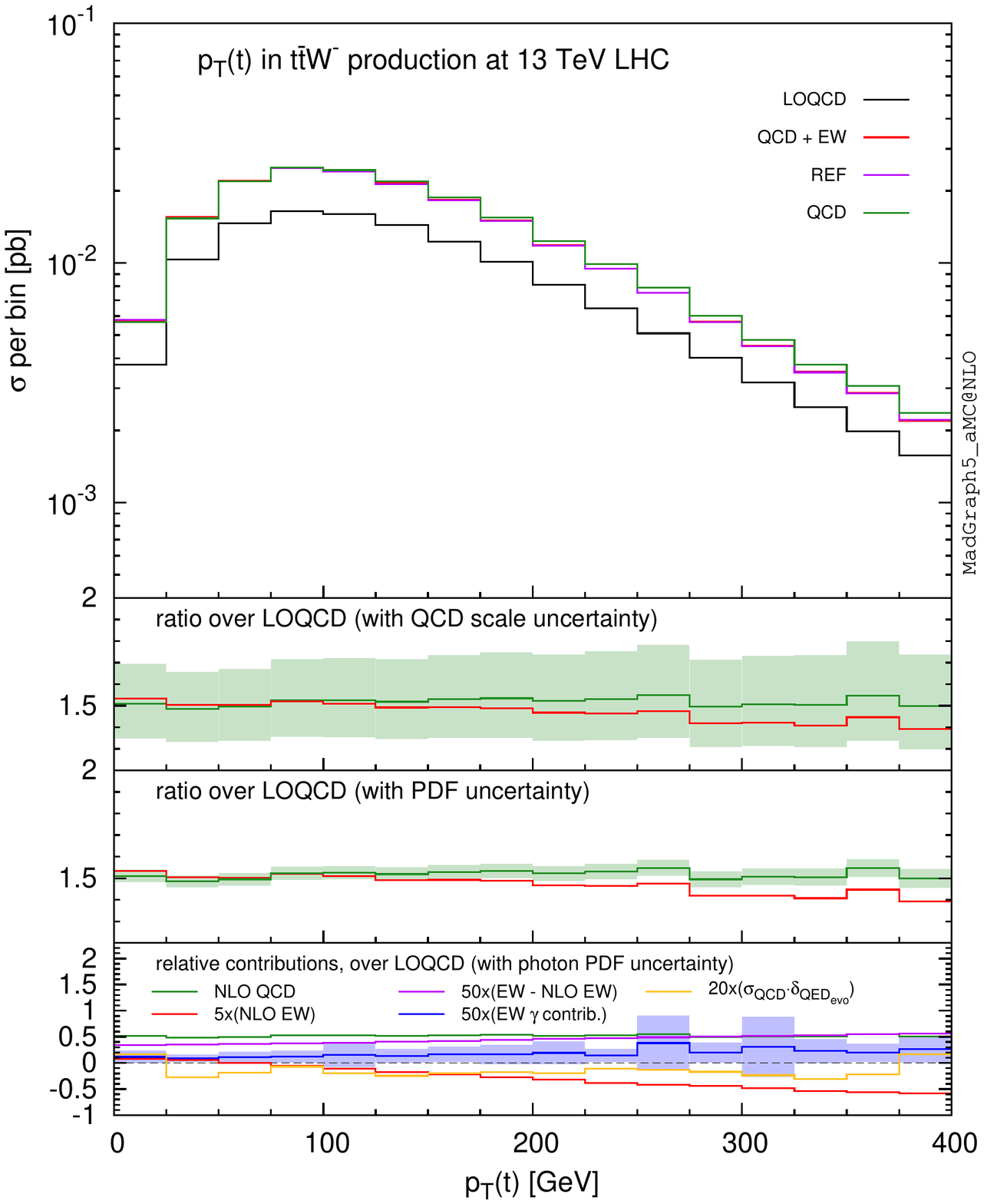}
	\includegraphics[page=2, trim=80 100 40 80, clip,scale=0.32]{\ttHttvqcdewpath/Figures/ttWm.pdf}
	\includegraphics[page=3, trim=80 100 40 80, clip,scale=0.32]{\ttHttvqcdewpath/Figures/ttWm.pdf}
	\caption{Transverse momentum ($p_{\mathrm{T}}$) distribution of the top quark, anti-top quark, and $W^-$ boson.}
\label{fig:ttWmEW}
\end{figure}

In the tails of transverse-momentum distributions NLO electroweak
effects can become more sizeable due to the appearance of Sudakov
logarithms. This is illustrated in
\refFs{fig:ttZEW}--\ref{fig:ttWmEW}, which display differential NLO
QCD+EW predictions obtained with \MGfiveamcnlo\, for $\ttz$ , $\ttwp$
and $\ttwm$ production, respectively, using the same setup and fixed
scale choice as described above.


\def\ttHttvtoolspath{\ttHpath/ttH-ttV-tools}
\subsection{Comparison of NLO QCD predictions for differential distributions}
\label{sec:ttH-ttV-tools}

In this section we provide a comparison among fixed-order NLO QCD predictions 
of differential distributions for $\ttv$ ($V=Z,W^\pm$) production at
the LHC with $\sqrt{s}=13\UTeV$, obtained using the following tools:
\begin{itemize}
\item \Sherpa~2.2.0 ~+~\Openloops~1.2.3,
\item \MGfiveamcnlo~2.3.2,
\item \Powhel.
\end{itemize}
The impact of EW corrections has been discussed in
Section~\ref{sec:ttH-ttV-QCD+EW}. Here the main goal is only to assess
the agreement and compatibility of existing fixed-order NLO QCD
calculations at the level of differential cross sections. Future
studies will test the compatibility of different implementations of
these calculations which also include parton-shower effects. In this
context, all the tools listed above have been used as fixed-order QCD
Monte Carlo generators.

\Sherpa+\Openloops\, uses \Openloops~\cite{Cascioli:2011va} as a
one-loop generator, and relies on the \Cuttools\,
library ~\cite{Ossola:2007ax} for the numerically stable
evaluation of tensor integrals. Real-emission
contributions, infrared subtractions based on the Catani-Seymour
technique~\cite{Catani:1996vz,Catani:2002hc}, and phase-space
integration are handled by
\Sherpa~\cite{Schumann:2007mg,Hoeche:2011fd,Hoche:2012wh}.

In \MGfiveamcnlo~\cite{Hirschi:2011pa,Alwall:2014hca}, fixed-order
NLO QCD results are obtained by adopting the FKS
method~\cite{Frixione:1995ms,Frixione:1997np} for the subtraction of
the infrared divergences of the real-emission matrix elements
(automated in the module \MadFKS~\cite{Frederix:2009yq}), and the OPP
integral-reduction procedure~\cite{Ossola:2006us} for the computation
of the one-loop matrix elements (automated in the module
\Madloop~\cite{Hirschi:2011pa}).

Finally, the \Powhel\, generator~\cite{Garzelli:2012bn} uses the \Helacnlo\,
package~\cite{Bevilacqua:2011xh} for the computation of all matrix
elements provided as input to the \Powhegbox.  The \Powhegbox\,
framework~\cite{Nason:2004rx,Frixione:2007vw,Alioli:2010xd} adopts the
FKS subtraction scheme~\cite{Frixione:1995ms,Frixione:1997np} to
factor out the infrared singularities of the real-emission cross
section, while the the virtual one-loop matrix elements can be
provided with different methods.

For this comparison all input parameters are chosen according to the
HXSWG recommendations~\cite{LHCHXSWG-INT-2015-006}, and the hadronic cross
section is obtained using the PDF4LHC15~\cite{Butterworth:2015oua}
and a five-flavour scheme (5FS).  Renormalization ($\muR$) and
factorization ($\muF$)
scales have been fixed to a dynamical central value
$\mu_0=H_{\mathrm{T}}$, where $H_{\mathrm{T}}$ is the sum of the
transverse energies of the $\ttv$ final state ignoring extra jet
emission. The  theoretical uncertainty has been
estimated by varying both $\muR$ and $\muF$ by a factor of 2 about $\mu_0$.

Results for the following distributions:
\begin{itemize}
\item transverse momentum ($p_{\mathrm{T}}$)  of the top quark,
  vector boson, $t\bar{t}$ system, and $\ttv$ system;
\item pseudorapidity ($\eta$) of the top quark and
  vector boson,
\end{itemize}
are presented in Figs.~\ref{fig:ttWp_stable_NLOPS_1}-\ref{fig:ttz_stable_NLOPS_2}, for the case of $V=W^+,W^-$, and
$Z$, respectively.


\providecommand{\ttwpplot}[2]{
\includegraphics[width=0.85\textwidth, trim=0 #1 0 0,clip]{\ttHttvtoolspath/ttwp/#2}}

\providecommand{\ttwpplotfig}[1]{
\begin{minipage}{0.5\textwidth}
\ttwpplot{0.115\textwidth}{main/#1}\\
\ttwpplot{0.115\textwidth}{sherpaopenloops/#1}\\
\ttwpplot{0.115\textwidth}{mg5amc/#1}\\
\ttwpplot{0\textwidth}{powhel/#1}
\end{minipage}}

\begin{figure}
\ttwpplotfig{ETA_TOP}
\ttwpplotfig{PT_TOP}\\[2mm]
\ttwpplotfig{ETA_V}
\ttwpplotfig{PT_V}\\
\caption{\label{fig:ttWp_stable_NLOPS_1} 
Fixed-order NLO predictions for differential $\ttwp$ observables at 13\,TeV.
Each ratio plot shows all results normalized to one particular 
NLO+PS prediction and the scale variation band of the reference prediction.
}
\end{figure}

\begin{figure}
\ttwpplotfig{PT_TT}
\ttwpplotfig{PT_TTV}
\caption{\label{fig:ttWp_stable_NLOPS_2} 
Fixed-order NLO predictions for differential $\ttwp$ observables at 13\,TeV.
Ratio plots as in \refF{fig:ttWp_stable_NLOPS_1}. 
}
\end{figure}



\providecommand{\ttwmplot}[2]{
\includegraphics[width=0.85\textwidth, trim=0 #1 0 0,clip]{\ttHttvtoolspath/ttwm/#2}}

\providecommand{\ttwmplotfig}[1]{
\begin{minipage}{0.5\textwidth}
\ttwmplot{0.115\textwidth}{main/#1}\\
\ttwmplot{0.115\textwidth}{sherpaopenloops/#1}\\
\ttwmplot{0.115\textwidth}{mg5amc/#1}\\
\ttwmplot{0\textwidth}{powhel/#1}
\end{minipage}}

\begin{figure}
\ttwmplotfig{ETA_TOP}
\ttwmplotfig{PT_TOP}\\[2mm]
\ttwmplotfig{ETA_V}
\ttwmplotfig{PT_V}\\
\caption{\label{fig:ttWm_stable_NLOPS_1} 
Fixed-order NLO predictions for differential $\ttwm$ observables at 13\,TeV.
Each ratio plot shows all results normalized to one particular 
NLO+PS prediction and the scale variation band of the reference prediction.
}
\end{figure}

\begin{figure}
\ttwmplotfig{PT_TT}
\ttwmplotfig{PT_TTV}
\caption{\label{fig:ttWm_stable_NLOPS_2} 
Fixed-order NLO predictions for differential $\ttwm$ observables at 13\,TeV.
Ratio plots as in \refF{fig:ttWm_stable_NLOPS_1}. 
}
\end{figure}



\providecommand{\ttzplot}[2]{
\includegraphics[width=0.85\textwidth, trim=0 #1 0 0,clip]{\ttHttvtoolspath/ttz/#2}}

\providecommand{\ttzplotfig}[1]{
\begin{minipage}{0.5\textwidth}
\ttzplot{0.115\textwidth}{main/#1}\\
\ttzplot{0.115\textwidth}{sherpaopenloops/#1}\\
\ttzplot{0.115\textwidth}{mg5amc/#1}\\
\ttzplot{0\textwidth}{powhel/#1}
\end{minipage}}

\begin{figure}
\ttzplotfig{ETA_TOP}
\ttzplotfig{PT_TOP}\\[2mm]
\ttzplotfig{ETA_V}
\ttzplotfig{PT_V}\\
\caption{\label{fig:ttz_stable_NLOPS_1} 
Fixed-order NLO predictions for differential $\ttz$ observables at 13\,TeV.
Each ratio plot shows all results normalized to one particular 
NLO+PS prediction and the scale variation  of the reference prediction.
}
\end{figure}

\begin{figure}
\ttzplotfig{PT_TT}
\ttzplotfig{PT_TTV}
\caption{\label{fig:ttz_stable_NLOPS_2} 
Fixed-order NLO predictions for differential $\ttz$ observables at 13\,TeV.
Ratio plots as in \refF{fig:ttz_stable_NLOPS_1}. 
}
\end{figure}


\def\ttHttvvpath{\ttHpath/ttH-ttVV}
\label{se:ttH-ttVV}
\subsection{\texorpdfstring{$t\bar{t}VV$}{ttVV} production (\texorpdfstring{$V=Z,W^\pm,H$}{V=Z,W,H}) at NLO QCD}
\label{sec:ttH-ttVV}

\begin{table}
\caption{NLO and LO cross sections for $\ttvv$ processes ($V=Z,W^\pm,H$) at 13~\UTeV. The renormalization and factorization scales are set equal to half of the sum of the masses of the final-state particles.  \label{table:ttvv}}
\small
\renewcommand{\arraystretch}{1.3}
\begin{center}
\begin{tabular}{  c | c c c c c }
\toprule
13 TeV $ \sigma$ [ab]  & $t \bar t W^+Z$ &  $t \bar t W^-Z$ & $t\bar t ZZ$ \\
\midrule
  NLO QCD &  $2705(3)^{+9.9 \%}_{-10.6 \%}~^{+2.7 \%}_{-2.7 \%}$ & $1179(2)^{+11.2 \%}_{-11.2 \%}~^{+3.7 \%}_{-3.7 \%}$ & $1982(2)^{+5.2 \%}_{-9.0 \%}~^{+2.6 \%}_{-2.6 \%}$ \\
 LO & $1982(2)^{+28.4 \%}_{-20.6 \%}~^{+3.3 \%}_{-3.3 \%}$ & $839.4(6)^{+28.2 \%}_{-20.5 \%}~^{+4.2 \%}_{-4.2 \%}$ & $1611(1)^{+31.4 \%}_{-22.1 \%}~^{+2.7 \%}_{-2.7 \%}$ \\
  $K$-factor & 1.36 & 1.40 & 1.23 \\

\toprule
13 TeV $ \sigma$ [ab]  & $t \bar t W^+H$ &  $t \bar t W^-H$ & $t\bar t ZH$ \\
\midrule
  NLO QCD & $1089(1)^{+1.8 \%}_{-5.9 \%}~^{+2.6 \%}_{-2.6 \%}$  & $493.0(5)^{+2.6 \%}_{-6.4 \%}~^{+3.4 \%}_{-3.4 \%}$ & $1535(2)^{+1.9 \%}_{-6.8 \%}~^{+3.0 \%}_{-3.0 \%}$ \\
 LO & $997.0(9)^{+26.9 \%}_{-19.8 \%}~^{+3.0 \%}_{-3.0 \%}$ & $440.0(4)^{+26.9 \%}_{-19.8 \%}~^{+3.8 \%}_{-3.8 \%}$ & $1391(1)^{+32.2 \%}_{-22.6 \%}~^{+2.8 \%}_{-2.8 \%}$ \\
  $K$-factor & 1.09 & 1.12 & 1.10 \\

\toprule
13 TeV $ \sigma$ [ab]  & $t \bar t W^+W^-$ &  $t \bar t W^+W^- $ (4f) & $t\bar t HH$ \\
\midrule
  NLO QCD&                                       --                                                  & $11500(10)^{+8.1 \%}_{-10.9 \%}~^{+3.0 \%}_{-3.0 \%}$ & $756.5(7)^{+1.1 \%}_{-4.4 \%}~^{+3.3 \%}_{-3.3 \%}$ \\
 LO & $8380(5)^{+33.2 \%}_{-23.1 \%}~^{+3.0 \%}_{-3.0 \%}$ & $8357(5)^{+33.3 \%}_{-23.1 \%}~^{+3.0 \%}_{-3.0 \%}$   & $765.4(5)^{+31.8 \%}_{-22.4 \%}~^{+2.9 \%}_{-2.9 \%}$ \\
 $K$-factor & -- & 1.38 & 0.99 \\
\bottomrule
\end{tabular}
\end{center}
\end{table}

Besides the dominant contribution from $\ttz$ and $\ttwpm$, the production of a top-quark pair ($t\bar{t}$) in association with two heavy bosons $V$ with $V=Z,W^{\pm},H$ can  also  be a non-negligible component of the background to leptonic signatures emerging from $\tth$. In this section we provide LO and NLO QCD results for $\ttvv$~\footnote{In this section the symbol $V$ indicates also the Higgs boson.} total cross sections at 13~\UTeV.  We set $\mh=125~\UGeV$ and all the other input parameters  according to the prescription in \Ref~\cite{LHCHXSWG-INT-2015-006}. In order to be consistent with the results at NLO QCD+EW accuracy in Section~\ref{sec:ttH-XS} ($\tth$) and Section~\ref{sec:ttH-ttV-QCD+EW} ($\ttz$ and $\ttwpm$), we used as renormalization and factorization scale $\muF=\muR=\mu:=\mt+\left(M(V_1)+M(V_2)\right)/2$, where $V_1$ and $V_2$ are the two heavy bosons in the final states. A detailed study of the dependence on the definition of the scales both for total cross sections and differential distributions for $\ttvv$ production can be found in \Ref~\cite{Maltoni:2015ena}. All the predictions in \Ref~\cite{Maltoni:2015ena} and those reported here have been calculated and can be reproduced with the public version of \MGfiveamcnlo~\cite{Alwall:2014hca}.

LO and NLO QCD cross sections for all the $\ttvv$ processes are listed in  Table~\ref{table:ttvv}. All the results include scale uncertainties, obtained by varying independently $\muF$ and $\muR$ in the range $\mu/2<\muF,\muR<2\mu$, and PDF errors evaluated with the
LHAPDF set \cite{Butterworth:2015oua}. All the calculations are performed in the five-flavour scheme (5FS), with the exception of $t \bar t W^+W^-$ production where we used the four-flavour scheme (4FS) in order to avoid additional resonant top quarks in the matrix elements of the real $b$-quark radiation. However, also in this case we used the same PDF set with the associate $\als$ in the 5FS. As can be seen in  Table~\ref{table:ttvv}, the difference between LO results in the two different schemes is of the order of 20 ab, which is negligible for LHC phenomenology, especially below 300 fb$^{-1}$ luminosity.

We want to stress that we did not include in Table~\ref{table:ttvv} results for $t \bar{t}V \gamma$ production, which may be also relevant. However, their LO contribution, without any cut on the photon, is already included in the NLO EW corrections to $\tth$, $\ttz$ and $\ttwpm$ production and consequentially in the best predictions $\sigma^{\rm NLO}_{\rm QCD+EW}$  that are reported  in Sections~\ref{sec:ttH-XS} and \ref{sec:ttH-ttV-QCD+EW}. Thus, LO contributions from $t \bar{t}V \gamma$ production must not be reevaluated when $\sigma^{\rm NLO}_{\rm QCD+EW}$ predictions for $\tth$, $\ttz$ and $\ttwpm$ are used. Equivalently, for the same reason, specific cuts on additional photon emission require $t \bar{t}V \gamma$ simulation with the corresponding cuts and the best predictions for $\tth$, $\ttz$ and $\ttwpm$ cannot be used. \\

We want also to point out that, similarly to the case of $t \bar{t}V \gamma$, the LO predictions for $t \bar{t}V_{1} V_{2}$ processes may also be classified as part of the NLO EW corrections to  $t \bar{t}V_{1}$ or $t \bar{t}V_{2}$ production. Indeed, $\ttvv$ LO cross sections  are of $\mathcal{O}(\als^2 \alpha^2)$, {\it i.e.}, of the same perturbative order of NLO EW corrections to $\tth$, $\ttz$ and $\ttwpm$ production.
The contribution that we denote as Heavy Boson Radiation (HBR) and define as
\begin{equation}
\sigma_{\rm HBR}(t \bar{t} V_1)=\sum_{V_2=H, Z, W^{\pm}} \sigma_{\rm LO}(t \bar{t}V_{1}V_{2})
\end{equation}
can enter the inclusive $\sigma(t \bar{t} V_1)$ at $\mathcal{O}(\als^2 \alpha^2)$.
We stress that the HBR contribution has not been included in the best predictions $\sigma^{\rm NLO}_{\rm QCD+EW}$ for $\tth$, $\ttz$ and $\ttwpm$ production and should not be included whenever  the $t \bar{t}V_{1} V_{2}$ processes are treated separately. It is nevertheless interesting to evaluate its size,  compare it with the corresponding  $\sigma^{\rm NLO}_{\rm QCD+EW}$ and verify possible cancellations, which are induced by the partial compensations of the Sudakov logarithms in the $\sigma^{\rm NLO}_{\rm EW}$ component.

\begin{table}[h]
\caption{$\delta_{\rm HBR}$ and $\delta_{\rm EW}$ for all the $\ttv$ processes. \label{table:dHBR}}
\small
\renewcommand{\arraystretch}{1.3}
\begin{center}
\begin{tabular}{  c | c c c c c }
\toprule
$\ttv$  & $\tth$ & $\ttz$ & $\ttwp$ &  $\ttwm$  \\
\midrule
$\delta_{\rm HBR}$ [\%]             &  1.0  &       0.9           &      4.1             &    7.0  \\
$\delta_{\rm EW}$ [\%]             &  1.7  &      -0.2          &      -3.5             &    -2.6  \\
\bottomrule
\end{tabular}
\end{center}
\end{table}

In Table~\ref{table:dHBR} we list for all the $\ttv$ processes the value of $\delta_{\rm HBR}$ defined as
\begin{equation}
\delta_{\rm HBR}=\frac{\sigma_{\rm HBR}}{\sigma_{\rm LO}}\, ,
\end{equation}
together with the values of $\delta_{\rm EW}$ defined in sections ~\ref{sec:ttH-XS} and \ref{sec:ttH-ttV-QCD+EW}.
As can be seen in Table~\ref{table:dHBR} the HBR contributions are, as expected, of the same order of the electroweak corrections. Here we report only results at the level of the total cross section, in \Ref~\cite{Frixione:2015zaa} the impact of HBR contributions has been studied also at differential level.


\def\ttHttbbpath{\ttHpath/ttH-ttbb}
\section{NLO+PS simulations of \texorpdfstring{$t\bar t b\bar b$}{ttbb} production}
\label{sec:ttH-ttbb}

The production of $\ttbar$ pairs  in association with two $b$-jets constitutes
a large irreducible background to $\tth$ production in the $H\to\bbbar$ channel,
and the rather large uncertainty of
Monte Carlo simulations of $\ttbar+b$-jets  production
is one of the main limitations of current $\tth(\bbbar)$ searches at the LHC.
A reliable theoretical description of $\ttbar$ production
in association with two $b$-jets requires
hard-scattering cross sections for the relevant partonic
processes $q\bar q/gg\to \ttbb$ at
NLO QCD~\cite{Bredenstein:2009aj,Bevilacqua:2009zn,Bredenstein:2010rs}.
The inclusion of NLO QCD effects reduces scale uncertainties from the 70--80\% level at LO
to about 20--30\%.
To become applicable in the context of experimental analyses,
NLO calculations need to be matched to parton showers.
A NLO+PS simulation of $\ttbb$ production
based on the five flavour number scheme (5FNS), where b-quarks are treated as massless partons,
was presented in~\cite{TROCSANYI:2014lha,Garzelli:2014aba},
while an alternative NLO+PS simulation that includes b-quark mass effects
in the four flavour number scheme (4FNS)
was published in~\cite{Cascioli:2013era}.

Finite b-quark masses permit to extend $\ttbb$ matrix elements to the full
phase space, including regions where b-quark pairs become collinear and
matrix elements with $m_b=0$ would be divergent.  Thus, using $\ttbb$ matrix
elements with $m_b>0$ in the 4FNS it is possible to simulate $\ttbar+b$-jets
production in a fully inclusive way, including also signatures where a b-quark remains
unresolved and a single b-jet is observed.\footnote{Here and in the
following we consistently exclude top-decay products from the counting of
$b$ jets.}
In contrast, the applicability of
5FNS calculations is limited to phase-space regions where the two
$b$-quarks in $\ttbb$ matrix elements have sufficient transverse momentum
and angular separation in order to avoid the breakdown of the $m_b=0$ approximation
in the collinear regions.
Such a requirement needs to be imposed at the
level of generation cuts, i.e.~before matching matrix elements to the parton
shower, and one might expect that the resulting NLO+PS predictions for observables with
two or more hard b-jets should be insensitive to generation cuts.
However, this is not the case, since events with multiple hard $b$-jets
can result from collinear
$g\to \bbbar$ splittings in $\ttbb$
matrix elements combined with the conversion of hard gluons into
$b$-jets via $g\to \bbbar$ parton shower splittings.
In fact, as pointed out in~\cite{Cascioli:2013era}, this so-called double-splitting mechanism
can lead to a sizeable enhancement of the
$\ttbb$ background in the Higgs-signal region, $M_{bb}\sim M_H$.

Given the importance of (quasi) collinear $g\to \bbbar$ splittings,
the choice of the flavour number scheme and the inclusion of b-quark mass effects
play a critical role.
For what concerns 5FNS calculations,
while it is clear that setting $m_b=0$ and omitting the
singular phase-space regions leads to
a logarithmic sensitivity to the unphysical
generation cuts for observables with a single hard b-jet,
double-splitting (or multiple-splitting)
contributions imply such a sensitivity also
for observables with two or more hard b-jets.
Such a logarithmic dependence can naturally be avoided in
the framework of NLO merging~\cite{Hoeche:2012yf,Frederix:2012ps,Lonnblad:2012ix},
where the singular phase space regions,
defined in terms of an appropriate merging cut,
are populated by the parton shower combined with matrix elements
for $\ttbar+0,1$\,jet production.
However, applying NLO merging to $\ttbar+0,1,2$\,jet
production~\cite{Hoeche:2014qda} is technically much more challenging as compared to
NLO+PS simulations of $\ttbb$ production.  Moreover, in the merging
approach all $b$-quarks produced via double-splitting contributions
would tend to arise from the parton shower,
which implies a strong dependence on parton-shower modelling.
In contrast, 4FNS simulations with $m_b>0$ have the advantage that one of
the $g\to \bbbar$ splittings is entirely described in terms of $\ttbb$
matrix elements at NLO,
while the additional hard $b$-jet arises from $\ttbb g$ tree amplitudes
matched to the parton shower via $g\to \bbbar$ splittings that can take
place at any stage of the shower evolution.
In conventional 4FNS calculations as the ones presented in this study, the
number of active quark flavours is limited to four both in the evolution of
the PDFs and $\alpha_s$.  Thus, renormalization group logarithms associated
with $b$- and $t$-quark loops are included only at fixed-order NLO.\footnote{In the 4FNS,
$b$-quark contributions to the running of $\alpha_s$ are consistently restored at NLO
accuracy by  including $b$-quark loops in the matrix elements
and renormalizing them via zero-momentum subtraction.  Also top-quark loop
contributions to $\alpha_s$ have been renormalized via zero-momentum
subtraction in the 4FNS.}
As
discussed in~\cite{Cascioli:2013era}, such logarithms can be easily resummed
by using modified 4FNS PDF sets that include all relevant quark flavours in
the running of $\alpha_s$.  At $\sqrt{s}=8$\,TeV it was found that
higher-order contributions of this type increase the NLO 4FNS $\ttbb$ cross
section by about $9\%$~\cite{Cascioli:2013era}, while in the 5FNS they are
naturally included.

Finally, we observe that simulations of $\ttbb$ production in the 4FNS can be
combined with fully inclusive $\ttbar+$jets samples based on the 5FNS in a
rather straightforward way~\cite{Moretti:2015vaa}.  In fact, in order to
avoid the double counting of $b$-quark production in the $\ttbb$ and
$\ttbar+$jets sample it is sufficient to veto events that involve $b$
quarks in the $\ttbar+$jets sample.  This prescription has to be applied after
parton showering, and the $b$-quark veto should be restricted to showered
$\ttbar+$jets matrix elements before top decays, i.e.~it should not be
applied to $b$ quarks that arise from (showered) top decays or from the
underlying event.

From the above discussion it is clear that the parton shower and the
choice of the flavour number scheme
play a critical role
in the description of $\ttbb$ production, and
the thorough understanding of the related uncertainties is of prime importance for the success of
$\ttbar H(\bbbar)$ searches. As a first step in this direction, in the following we present a systematic
comparison of various NLO+PS simulations of $\ttbb$ production
based on different parton showers,
matching schemes,
and flavour number schemes.

\subsection{NLO+PS tools and simulations}
Three different NLO+PS simulations of $\ttbb$ production at $\sqrt{s}=13$\,TeV based on \Sherpa+\Openloops \cite{Gleisberg:2008ta,Cascioli:2011va,Cascioli:2013gfa},
\MGfiveamcnlo+\Pythiaeight~\cite{Alwall:2014hca,Hirschi:2011pa,Skands:2014pea,Sjostrand:2014zea}
and
\Powhel+\Pythiaeight~\cite{Kardos:2013vxa,Bevilacqua:2011xh,Skands:2014pea,Sjostrand:2014zea}
have been compared.
The various NLO matching methods, parton showers and flavour number
schemes employed in the three simulations are summarized
in~\refT{tab:ttbb_tools}.
The \Sherpa+\Openloops\, and \MGfiveamcnlo\, simulations employ the
4FNS and the \Mcnlo\, matching
method\footnote{More precisely, \MGfiveamcnlo\, employs the original
  formulation of \Mcnlo\,
  matching~\cite{Frixione:2002ik,Frixione:2003ei}, while
  \Sherpa+\Openloops\, implements an alternative
formulation~\cite{Hoeche:2011fd,Hoche:2012wh}
denoted as \Smcnlo, which
  is characterized by an improved treatment of colour correlations but
  is otherwise equivalent to the method
  of~\cite{Frixione:2002ik,Frixione:2003ei}.} to combine matrix
elements with the \Sherpa\, and \Pythiaeight\, parton showers,
respectively.  Therefore possible differences between
\Sherpa+\Openloops\, and \MGfiveamcnlo\, predictions
can be attributed to parton shower effects or to differences in the two
implementations of the MC@NLO approach and related technical parameters.
Instead, the \Powhel\, simulation differs
from the other two simulations in at least two aspects: it is
performed in the 5FNS and it employs the
\Powheg\, matching method~\cite{Nason:2004rx,Frixione:2007vw}.  Moreover, when comparing
\Sherpa+\Openloops\, and \Powhel\, predictions one should keep in mind
that also the respective parton showers (\Sherpa\, and \Pythiaeight) are
different.

Since the \Powhel\, simulation is performed with massless
$b$ quarks,\footnote{To be precise, in the \Powhel\, simulation
$b$-mass effects are neglected at the matrix element level but are taken
into account in the parton shower.}
in order to avoid collinear $g\to\bbbar$ singularities,
the hard matrix elements need to be restricted to
phase space regions where both $b$ quarks remain resolved.  This is achieved
through the generation cuts
\begin{equation}
\label{ttbb:5Fgencuts}
m_{bb}> 2 \xi_m m_b,\qquad
p_{T,b}>\xi_T m_b.
\end{equation}
Technically, in order to guarantee infrared
safety, in the case of real emission events the cuts in \Eref{ttbb:5Fgencuts}
are applied after a projection onto the Born phase space.
The above cuts are chosen in a way that
mimics, at LO in $\alpha_S$,  the  $\log(m_b)$ dependence
that arises when one $b$-quark is integrated out.
Thus they can be regarded as an
heuristic approach in order to obtain reasonable 5FNS predictions also for
$\ttbar+b$-jet observables with a single resolved $b$ jet.
However, as discussed above,
NLO+PS $\ttbb$ predictions in the 5FNS are sensitive to the
generation cuts in \Eref{ttbb:5Fgencuts}. In particular they depend on the
unphysical parameters $\xi_m$ and $\xi_T$,
which have been set to one in the present study.\footnote{Variations of these parameters
will allow to quantify up to which extent,
in practice, predictions in conditions typically met in experimental
analyses depend on the choice of these technical cuts.}
Moreover, one should keep in mind that $\log(m_b)$ terms beyond LO and finite terms of order $m_b$
are not consistently included in the 5FNS.

\begin{table}\footnotesize
\caption{Employed tools, matching methods, parton showers, flavour number scheme (FNS) and
generation cuts in the NLO+PS simulations of $\ttbb$ production.
The \Powhel~generator implements matrix elements with massless $b$-quarks but
includes $b$-mass effects through \Pythia.}
\label{tab:ttbb_tools}
\setlength{\tabcolsep}{1.0ex}
\begin{center}
\renewcommand{\arraystretch}{1.2}
\begin{tabular}{llllll}
\toprule
Tools                                    & Matching method     & Shower   & FNS   & $m_b$\,[GeV] & Generation cuts \\
\midrule
\Sherpa~2.2.1+\Openloops~1.2.3              & \Smcnlo           & \Sherpa   & 4FNS & 4.75         & fully inclusive  \\
\MGfiveamcnlo~2.3.2+\Pythiaeight~2.1.0  & \Mcnlo              & \Pythiaeight   & 4FNS & 4.75         & fully inclusive  \\
\Powhel+\Pythiaeight~2.1.0                     & \Powheg              & \Pythiaeight   & 5FNS & 0            & $m_{bb}> 2 m_b$,\; $p_{T,b}>m_b$ \\
\bottomrule
\end{tabular}
\end{center}
\end{table}

\subsection{Parton showers, PDF, and \texorpdfstring{$\alpha_s$}{alpha_s}}

Full Monte Carlo simulations of $\ttbb$ production
involve hard $\ttbb$ cross section at NLO,
top-quark decays, parton showering,
hadronization, hadron decays, and the underlying event.  The main
source of theoretical uncertainty in this involved simulation
framework is given by the mechanism that governs $b$-quark production
in association with top-quark pairs.  Thus, in order to obtain a
sufficiently transparent picture of the nontrivial QCD dynamics
of $b$-quark production, it was decided to reduce the complexity
that results from the presence of the additional b-quarks that arise from
top-quark decays via well-understood weak interactions.
To this end, top quarks have been treated as stable particles in the simulations.
Moreover, all NLO+PS simulations have been performed at the parton level,
including only the perturbative phase of parton shower evolution, and
neglecting hadronization as well as any other non-perturbative aspect.
The quantitative importance of hadronization and the
  possible bias that can result from switching off hadronization in
  the comparison of two $\ttbar+b$-jet simulations based on different
  parton showers was assessed by comparing \Sherpa~2.1 and \Pythia~8.2
  LO+PS simulations of $pp\to H+$jets (including
  $b$ jets) at 14\,TeV.  Thanks to the colour neutral nature of the
  Higgs boson, this process allows one to assess the impact of
  hadronization by turning it on and off.  The effects of
  hadronization increase with decreasing jet transverse momenta. Thus
  they predominantly arise in the vicinity of the jet-$p_T$
  threshold. For the production of $b$-jets with $p_T>25\UGeV$ and
  $|\eta|<2.5$ they amount to about $-2\%(-4\%)$ per $b$ jet in
  \Pythiaeight\, (\Sherpa).  This suggests that the bias that results from
  turning off hadronization should be well below the typical NLO+PS
  uncertainties in $\ttbb$ production.

In order to reduce uncontrolled sources of bias related to shower modelling
in the comparison of NLO+PS simulations based on \Sherpa\, and \Pythiaeight, those
free parton-shower parameters related to the strong coupling have
been chosen in a uniform way. Specifically, the rescaling factors $x$ that are
applied to the strong coupling terms $\alpha_S(x\,k_T^2)$
for each shower emission have been set to $x=1$ both for initial- and final-state radiation.
Furthermore the option of resumming
subleading logarithms
of Catani-Marchesini-Webber kind~\cite{Catani:1990rr}
was deactivated.
Note that these choices neither correspond to the \Sherpa\, default nor to the
\Pythiaeight\, default settings.  Moreover they are not meant to provide an
optimal description of data.  They are only aimed at a consistent comparison of
the two showers, where simple parametric differences are avoided, and the
remaining deviations can be attributed to intrinsic shower features, such as
the different definition of the shower evolution variables.

Since parton-shower tunes and PDFs are intimately connected, it is not
trivial to identify a common PDF set that is optimal for all parton
showers.  For the present study the NNPDF3.0 NLO PDF set was adopted,
keeping in mind that this choice might bias the comparison of
\Sherpa\, against \Pythiaeight. The specific PDF set was chosen according to
the employed flavour number scheme (4FNS or 5FNS), while the value of
$\alpha_s(M_Z)$ in NLO matrix elements and for the first shower
emission was chosen consistently with the PDF. The same holds for the
running of $\alpha_s$, whose evolution is implemented at 2-loops both
in matrix elements and parton showers.  For subsequent shower
emissions the 4FNS (5FNS) together with the corresponding value of
$\alpha_s(M_Z)$ was used in \Sherpa\,~(\Pythiaeight).

\subsection{Input parameters and scale choices}
To simulate $\ttbb$ production at 13\,TeV the input parameters
$m_t=172.5$~GeV, $m_b=4.75$~GeV and $\alpha_s^{(5F)}(M_Z)=0.118$
have been used together with NNPDF3.0 parton distributions at NLO,
as discussed above.\footnote{Note that the employed NNPDFs
and related $\alpha_s(M_Z)$ value in the 4FNS are
derived from variable-flavour-number NNPDFs with $\alpha_s^{(5F)}(M_Z)=0.118$
via appropriate backward and forward evolution with five and four active flavours,
respectively.}
The central values of the renormalization and factorization scales have been chosen as
\begin{equation}
\label{ttbb:scales}
\mu_{R,0}=\left(\prod_{i=t,\bar t,b, \bar b} E_{T,i}\right)^{1/4},
\qquad
\mu_{F,0}=\frac{H_T}{2} =\frac{1}{2}\sum_{i=t, \bar t,b, \bar b,j} E_{T,i},
\end{equation}
where $E_{T,i}=\sqrt{M_i^2+p^2_{T,i}}$ denotes the transverse
energy of top and bottom quarks, defined at parton level. Note that also
extra parton emissions contribute to the total transverse energy $H_T$ in
\Eref{ttbb:scales}.
Theoretical uncertainties have been assessed by means of standard
variations $\mu_R=\xi_R \mu_{R,0}$, $\mu_F=\xi_F \mu_{F,0}$ with
$0.5< \xi_R, \xi_F <2$ and $0.5< \xi_R/\xi_F <2$.

The CKKW inspired renormalization scale choice in \Eref{ttbb:scales}
is based on~\cite{Cascioli:2013era} and takes into account the fact
that top and bottom quarks are produced at widely different scales
$E_{T,b}\muchless E_{T,t}$. This turns out to improve the perturbative
convergence as compared to a hard global scale of order $m_t$.
In particular, in the 4FNS it was checked that using $\mu_{R}=H_T/2$ instead
of $\mu_{R}=\mu_{R,0}$ increases the $K$-factor by 0.25 and reduces the NLO
cross section by about 40\%, which is only barely consistent with the level
of uncertainty expected from factor-two scale variations.  Moreover,
computing LO and NLO cross sections using PDFs and $\alpha_s$ values at NLO
throughout\footnote{With this approach $K$-factors are much less dependent
on the employed PDF sets and reflect the convergence of the perturbative
expansion in a more realistic way as compared to using LO inputs for the LO
cross section.} yields $K$-factors around 2 with $\mu_R=\mu_{R,0}$ and about
0.25 higher with $\mu_R=H_T/2$.  Thus both scale choices seem to be
suboptimal, and in order to improve the convergence of the perturbative
expansion, a scale even softer than~\Eref{ttbb:scales} should be considered
in the future.  In any case a hard scale of type $\mu_R=H_T/2$ is not
recommended.

\def\ximin{\xi_{\mathrm{min}}}
\def\ximax{\xi_{\mathrm{max}}}

In the context of the \Mcnlo\, matching approach, where the
resummation scale $\mu_Q$, i.e.~the parton shower starting scale, is a
free parameter, it is natural to identify this scale with the
factorization scale.  Thus $\mu_Q=\mu_{F,0}=H_T/2$ was used in the
\Sherpa+\Openloops\, simulation.  In the case of \MGfiveamcnlo\, a
different choice had to be adopted since only resummation scales of
the form $\mu_Q=\xi \sqrt{\hat{s}}$ are supported, where the prefactor $\xi$
is randomly distributed in the freely adjustable range
$[\ximin,\ximax]$ with a distribution that is strongly peaked at
$(\ximin+\ximax)/2$~\cite{Alwall:2014hca}. Comparing the $H_T/2$ and $\mu_Q=\xi \sqrt{\hat{s}}$
distributions it was observed that the
respective peaks lie around $200\UGeV$ and $400\UGeV$
when the default \MGfiveamcnlo\, settings
$(\ximin,\ximax)=(0.1,1)$ are used, i.e.~the default $\mu_Q$
in \MGfiveamcnlo\, is much harder.

Given that \Mcnlo\,
predictions for $\ttbb$ production are quite sensitive to $\mu_Q$,
it was decided to lower the $\xi$ upper bound to $\ximax=0.25$, which
brings the $\mu_Q$ reasonably close to $H_T/2$.  We note that this
choice is also supported by the study of an \MGfiveamcnlo\, simulation
of $H\bbbar$ production in the 4FNS~\cite{Wiesemann:2014ioa}, where it
was found that reducing $\ximax$ from 1 to 0.25 strongly improves the
convergence of NLO+PS and NLO distributions at large transverse
momenta.

\def\hdamp{h_{\mathrm{damp}}}

In the \Powheg\, matching method, the resummation scale is not a freely
adjustable parameter, since the first emission on top of $\ttbb$ events is
entirely described by matrix elements, and the corresponding transverse momentum
scale sets the upper bound for subsequent shower emissions.
Nevertheless, \Powheg\, simulations involve a parameter $\hdamp$ that
separates the first-emission phase space into a singular region, where the
first emission is resummed and corrected with a local $K$-factor, and a
remnant region, where it is handled as at fixed-order NLO.
Given the analogy with the separation of
soft and hard events in the \Mcnlo\, approach,
and given that $\mu_Q$ represents the upper bound for
emissions off soft events,
it is natural to chose
$\hdamp$ of the same order of $\mu_Q$.  Thus the choice $\hdamp=H_T/2$ was
adopted in the \Powhel\, simulation.

Variations of the resummation scale and of the $\hdamp$ parameter
have not been considered in this study.

\subsection{NLO+PS predictions for \texorpdfstring{$t\bar t+b$}{tt+b}-jets cross sections in \texorpdfstring{$b$}{b}-jet bins}

In the following we compare integrated and differential NLO+PS
predictions for $\ttbar+b$-jets production with a certain minimum
number of $b$ jets, $n_b>N_b$.  In particular we focus on the
bins with $n_b\ge 1$ or $n_b\ge 2$, which are the most relevant ones
for $\ttbar H(\bbbar)$ analyses.  For the jet definition the anti-$k_T$
algorithm with $R=0.4$ is adopted, and jets that involve one or more
$b$-quark constituents are classified as $b$-jets.  Note that also
jets that result from collinear $g\to\bbbar$ splittings are handled as
$b$ jets.  Moreover no requirement is imposed on the minimum transverse
momentum of $b$ quarks inside $b$ jets.  Events are categorized
according to the number $n_b$ of resolved $b$ jets within the
acceptance region,
\begin{equation}
p_{T,b}>25\,\UGeV\,,
\qquad
|\eta_{b}|<2.5\,.
\end{equation}
Let us recall that top quarks are treated as stable particles, thus
the two $b$ quarks that arise from top decays as well as possible
extra $b$ quarks from the showering of top-decay products are not
included in $n_b$.  Apart from the requirement $n_b\ge N_b$ no additional cut
will be applied.\footnote{To be more precise, the \Sherpa+\Openloops\,
  and \MGfiveamcnlo\, samples are fully inclusive, while in the case
  of \Powhel\, the technical cuts \Eref{ttbb:5Fgencuts} are applied as
  discussed above.} In order to illustrate the importance of parton
shower effects, the various NLO+PS predictions presented in the
following are also compared to fixed-order NLO predictions.  The
latter are based on \Sherpa+\OpenLoops and are obviously independent
of the employed parton shower and matching scheme.

All quoted theoretical uncertainties correspond to factor-two
variations of the renormalization and factorization scales. In
\refFs{fig:ttbb_1}--\ref{fig:ttbb_9} they are shown as bands, and, to
improve readability, three different ratio plots are shown, where all results are normalized to
one particular NLO QCD+PS prediction and the corresponding scale
variation band is shown.

\newcommand{\xs}[2]{#1}
\def\TTB{\mathrm{ttb}}
\def\TTBB{\mathrm{ttbb}}
\def\TTBBB{\mathrm{tt+3b}}
\def\TTBBBB{\mathrm{tt+4b}}

\begin{table}
\caption{Fixed-order NLO and NLO+PS predictions for integrated $\ttbar+b$-jets cross sections
at 13\,TeV in bins with $n_b\ge 1$ and $n_b\ge 2$ $b$ jets.}
\label{tab:XS}
\vspace*{0.3ex}
\begin{center}
\renewcommand{\arraystretch}{1.2}
\begin{tabular}{llll@{\hspace{12mm}}l@{\hspace{12mm}}l}
\toprule
Selection
&  Tool
&  {$\sigma_{\mathrm{NLO}}\,[\mathrm{fb}]$} 
&  {$\sigma_{\mathrm{NLO+PS}}\,[\mathrm{fb}]$} 
&  {$\sigma_{\mathrm{NLO+PS}}/\sigma_{\mathrm{NLO}}$}  
\\ 
\midrule
$n_b\ge 1$
& \Sherpa+\OpenLoops
& $12820^{+35\%}_{-28\%}$ 
& $12939^{+30\%}_{-27\%}$ 
& $1.01$ 
\\
& \MGfiveamcnlo
& 
& $13833^{+37\%}_{-29\%}$ 
& $1.08$ 
\\
& \Powhel
& 
& $10073^{+45\%}_{-29\%}$ 
& $0.79$ 
\\
\midrule
$n_b\ge 2$
& \Sherpa+\OpenLoops
& $2268^{+30\%}_{-27\%}$ 
& $2413^{+21\%}_{-24\%}$ 
& $1.06$ 
\\
& \MGfiveamcnlo
& 
& $3192^{+38\%}_{-29\%}$ 
& $1.41$ 
\\
& \Powhel
& 
& $2570^{+35\%}_{-28\%}$ 
& $1.13$ 
\\
\midrule
\end{tabular}
\end{center}
\end{table}

Results for the $\ttbar+b$-jets cross sections with $n_b\ge N_b$ b
jets for various values of $N_b$ are presented in \refT{tab:XS} and
\refF{fig:ttbb_1}.  In the following we will refer to the results for
$N_b=1,2,3,4$ as $\TTB$, $\TTBB$, $\TTBBB$ and $\TTBBBB$ cross
sections, respectively.  For the $\TTB$ and $\TTBB$ cross sections,
which are described at NLO accuracy, the various NLO+PS predictions
turn out to be in decent mutual agreement.  More precisely, $\TTB$
predictions based on the 4FNS (\Sherpa+\Openloops\, and \MGfiveamcnlo)
agree very well with each other and with fixed-order NLO, and the 5FNS
$\TTB$ simulation (\Powhel) lies only 20\% lower, despite that
it was not designed to describe final states with a single
b-jet (due to the generation cuts).

For the $\TTBB$ cross section one finds excellent agreement between
fixed-order NLO, \linebreak
\Sherpa+\Openloops\, and \Powhel.  This seems to
suggest that this observable has little sensitivity to parton shower
effects and to the choice of the flavour number scheme.  However this
interpretation is challenged by the fact that the \MGfiveamcnlo\,
$\TTBB$ result lies more than $30\%$ above the other predictions.
The only significant differences between \MGfiveamcnlo\, and
\Sherpa+\Openloops\, simulations lie in the employed parton showers and details
of MC@NLO matching, thus the origin of the observed discrepancy is likely to lie in the choice
of shower starting scale in \MGfiveamcnlo\, combined with the higher
intensity of QCD radiation in \Pythiaeight\, with respect to \Sherpa.
This is confirmed by the further enhancement of the \MGfiveamcnlo\, cross section in
the bins with $n_b\ge 3$ and $n_b\ge 4$ $b$-jets (see
\refF{fig:ttbb_1}), where the additional b quarks arise from
$g\to \bbbar$ parton-shower splittings, which results in a much
stronger sensitivity to shower effects.  Note that this kind of
uncertainty for $N_b=3,4$ is not included in the quoted scale
variations.
In the \Sherpa+\Openloops\, simulation, the size of scale uncertainties and
the difference between NLO and NLO+PS predictions are fairly similar to what
observed at $\sqrt{s}=8$\,TeV in~\cite{Cascioli:2013era}.  In particular,
NLO+PS scale uncertainties range between 20--30\% in all $b$-jet bins and
are smaller as compared to the case of fixed-order NLO.  Scale variations in
\MGfiveamcnlo\, and \Powhel\, tend to be larger and agree well with each other
for $N_b=2$, while \Powhel\, features a larger scale dependence in the
other bins, especially for $N_b=3,4$.  These various differences can be
attributed to the employed flavour-number schemes and
to technical aspects of the implementation of scale variations in the
three different NLO+PS Monte Carlo tools.

\subsection{\texorpdfstring{$\TTB$}{ttb} differential analysis}

Various differential observables for an inclusive $\TTB$ analysis with
$n_b\ge 1$ $b$-jets are presented in \refFs{fig:ttbb_2}--\ref{fig:ttbb_4}.
For all distributions that are inclusive with respect to extra light-jet emissions
one observes a rather similar behaviour as for the $\TTB$ cross section,
i.e.~\Sherpa+\OpenLoops, \MGfiveamcnlo\, and fixed-order NLO predictions
agree well, while \Powhel\, lies about 20\% lower. Only \Powhel\, features significant shape
distortions with respect to fixed-order NLO in the region of low rapidity and/or
low $p_T$ for the leading top and bottom quarks and for the $\ttbar$ system
(\refFs{fig:ttbb_2}--\ref{fig:ttbb_3}).
Observables that explicitly involve the first light-jet emission
(\refF{fig:ttbb_4}) turn out to behave differently. While
for \Sherpa+\OpenLoops, \Powhel\, and fixed-order NLO
there is mutual agreement within scale variations,
the \MGfiveamcnlo\, prediction turns out to lie up to
50\% higher at $p_{T,j}\sim 50$\,\UGeV.
This enhancement of QCD radiation
in \MGfiveamcnlo+\Pythiaeight\,
disappears at $p_{T,j}\sim 150$\,\UGeV.
It is most likely related to what was observed above
in $b$-jet bin cross sections with $N_b\ge 2$.

\subsection{\texorpdfstring{$\TTBB$}{ttbb} differential analysis}

Various differential observables for an inclusive $\TTBB$ analysis with
$n_b\ge 2$ $b$-jets are presented in \refFs{fig:ttbb_5}--\ref{fig:ttbb_9}
Observables that depend on the top-quark and b-jet kinematics
but are inclusive with respect to extra jet emission are presented
in \refFs{fig:ttbb_5}--\ref{fig:ttbb_7}. For all such distributions
a fairly good agreement between \Sherpa+\OpenLoops, \Powhel\, and fixed-order NLO is
observed, both at the level of shapes and normalization.
The most significant shape differences show up in the $p_T$
of the 2nd b-jet and do not exceed 20\%.
In \MGfiveamcnlo\ the matching to the  \Pythiaeight\ shower increases the $\TTBB$
rates by about 35\% with respect to \Sherpa+\OpenLoops,
and turns out to have an non-trivial dependence on the
top and b-jet kinematics.  In particular it tends to enhance distributions in
the the regions with small top-quark and b-jet $p_T$ and at large
$\Delta R$ separation between the two $b$-jets.

For the distribution in the invariant mass of the $b$-jet pairs, which corresponds to the
mass of the $H\to \bbbar$ candidate, it turns out that all NLO+PS results
are in reasonably good mutual agreement. The results also confirm the presence of an NLO+PS
distortion of the invariant-mass distribution, which was
attributed to double-splitting effects in~\cite{Cascioli:2013era}.
More precisely, in the vicinity of the Higgs boson resonance
the NLO+PS enhancement w.r.t.~NLO is close to 20\% and thus less pronounced
to what was observed in~\cite{Cascioli:2013era} at $\sqrt{s}=8$\,TeV,\footnote{This can be due to the
different collider energy and to the different scale choices in this study and in~\cite{Cascioli:2013era}.}
while the \Powhel\,and \MGfiveamcnlo\, distributions
feature an additional enhancement of about 10\% and 35\%, respectively,
w.r.t.~\Sherpa+\OpenLoops in the Higgs boson signal region.

Various observables that are directly sensitive to the emission of
an additional jet are shown in \refFs{fig:ttbb_8}--\ref{fig:ttbb_9}.
Despite the intrinsic LO nature and stronger shower dependence of such distributions,
\Sherpa+\OpenLoops\, and \Powhel\, remain in good
agreement: the most important deviations, which show up in the $p_T$
tail of the first light jet, do not exceed 40\%.
In contrast, the excess of \MGfiveamcnlo\, w.r.t.~the other predictions
grows by about a factor two, reaching about 70\% in average, and gives rise
to more pronounced shape distortions as compared to the case of
inclusive $\TTBB$ observables. Similarly as for the $\TTBB$ analysis, the enhancement is
concentrated at light-jet momenta between 50--150\UGeV,
where it reaches up to 100\%. A similarly strong
increase shows up also in the region of central light-jet rapidity,
as well as in angular and mass distributions that involve light and $b$-jets.

\subsection{Summary and conclusions}
In summary, we have presented a systematic study of Monte Carlo simulations of $pp\to
\ttbar+b$-jets at 13\,TeV that compares various NLO+PS
predictions based on different matching methods, parton showers and matching
schemes.  While the inclusion of $b$-mass effects is the only fully
consistent way of describing inclusive $\ttbar+b$-jets production in terms
of $\ttbb$ matrix elements, the observed agreement between
\Sherpa+\OpenLoops and \Powhel\, predictions indicates that also
simulations with massless b-quarks and appropriate generation cuts
provide predictions in agreement well within the scale uncertainties.

The various NLO+PS simulations considered in this study confirm
that the invariant mass of the b-jet pair receives significant NLO+PS corrections
that can reach 20-30\% in the $H\to \bbbar$ signal region~\cite{Cascioli:2013era}.
Based on standard variations of the renormalization and factorization scales,
the expected accuracy of NLO predictions should be at the 25--35\% level.
However in various phase-space regions the differences between the various NLO+PS simulations
tend to be larger.

In particular, some of the distributions generated with
\MGfiveamcnlo+\Pythiaeight\,
have significantly different shapes, resulting in larger predictions for up
to 100\%, compared to the other NLO+PS simulations.  These are probably
related to the high intensity of the QCD radiation
in \Pythiaeight\, and are quite sensitive to the choice of the
shower starting scale in the MC@NLO matching framework.
These findings should be regarded as a first step towards a thorough
investigation of NLO matching and parton shower effects,
including all relevant sources
of uncertainty, in the Monte Carlo modelling of $\ttbar+b$-jets production.
In the future also top-quark decays should be investigated.

\providecommand{\ttbbtableplot}[2]{
\includegraphics[width=0.85\textwidth, trim=0 #1 0 0,clip]{\ttHttbbpath/#2}}

\providecommand{\ttbbfig}[1]{
\begin{minipage}{0.5\textwidth}
\ttbbtableplot{0.115\textwidth}{plots/main/#1}\\
\ttbbtableplot{0.115\textwidth}{plots/sherpaopenloops/#1}\\
\ttbbtableplot{0.115\textwidth}{plots/mg5amc/#1}\\
\ttbbtableplot{0}{plots/powhel/#1}
\end{minipage}}

\begin{figure} 
\bce
\ttbbfig{NBj_XS}
\ece
\caption{\label{fig:ttbb_1} 
  Fixed-order NLO and NLO+PS predictions for integrated
  $\ttbar+$b-jets cross sections at 13\,TeV in inclusive bins with
  $n_b>N_b$ $b$ jets.  Each ratio plot shows all results normalized to
  one particular NLO QCD+PS prediction and the corresponding scale
  variation band.  }
\end{figure}

\begin{figure}
\ttbbfig{1_PT_T1}
\ttbbfig{1_PT_T2}
\\[2mm]
\ttbbfig{1_ETA_T1}
\ttbbfig{1_ETA_T2}
\caption{\label{fig:ttbb_2} Fixed-order NLO and NLO+PS top-quark
 distributions for $pp\to\ttbar+\ge 1\,b$ jets at 13\,TeV.  Ratio
  plots like in \refF{fig:ttbb_1}.  }
\end{figure}

\begin{figure}
\ttbbfig{1_PT_T1T2}
\ttbbfig{1_PT_B1}
\\[2mm]
\bce
\ttbbfig{1_ETA_B1}
\ece
\caption{\label{fig:ttbb_3} Fixed-order NLO and NLO+PS
top-quark and b-jet
distributions for $pp\to\ttbar+\ge 1\,b$ jets at 13\,TeV.
  Ratio plots like in \refF{fig:ttbb_1}.  }
\end{figure}

\begin{figure}
\ttbbfig{1_ETA_J1}
\ttbbfig{1_PT_J1}
\\[2mm]
\bce
\ttbbfig{1_M_J1B1}
\ece
\caption{\label{fig:ttbb_4}
Fixed-order NLO and NLO+PS light-jet distributions
for $pp\to\ttbar+\ge 1\,b$ jets
at 13\,TeV.
Ratio plots like in \refF{fig:ttbb_1}.
}
\end{figure}

\begin{figure} 
\ttbbfig{2_PT_T1}
\ttbbfig{2_PT_T2}
\\[2mm]
\ttbbfig{2_ETA_T1}
\ttbbfig{2_ETA_T2}
\caption{\label{fig:ttbb_5} Fixed-order NLO and NLO+PS top-quark
distributions for $pp\to\ttbar+\ge 2\,b$ jets at 13\,TeV.  Ratio
  plots like in \refF{fig:ttbb_1}.  }
\end{figure}

\begin{figure} 
\ttbbfig{2_ETA_B1}
\ttbbfig{2_ETA_B2}
\\[2mm]
\ttbbfig{2_PT_B1}
\ttbbfig{2_PT_B2}
\caption{\label{fig:ttbb_6} Fixed-order NLO and NLO+PS b-jet
distributions for $pp\to\ttbar+\ge 2\,b$ jets at
13\,TeV.  Ratio plots like in \refF{fig:ttbb_1}.  }
\end{figure}

\begin{figure} 
\ttbbfig{2_M_B1B2}
\ttbbfig{2_DR_B1B2}
\\[2mm]
\ttbbfig{2_PT_T1T2}
\ttbbfig{2_PT_B1B2}
\caption{\label{fig:ttbb_7} (7) Fixed-order NLO and NLO+PS
 distributions of the $\bbbar$ and $\ttbar$ systems for $pp\to\ttbar+\ge 2\,b$ jets at
  13\,TeV.  Ratio plots like in \refF{fig:ttbb_1}.  }
\end{figure}

\begin{figure} 
\ttbbfig{2_M_J1B1}
\ttbbfig{2_M_J1B2}
\\[2mm]
\bce
\ttbbfig{2_PT_J1}
\ece
\caption{\label{fig:ttbb_8} Fixed-order NLO and NLO+PS
  light-jet mass and transverse momentum distributions for $pp\to\ttbar+\ge 2\,b$ jets at
  13\,TeV.  Ratio plots like in \refF{fig:ttbb_1}.  }
\end{figure}

\begin{figure} 
\ttbbfig{2_DR_J1B1}
\ttbbfig{2_DR_J1B2}
\\[2mm]
\bce
\ttbbfig{2_ETA_J1}
\ece
\caption{\label{fig:ttbb_9} Fixed-order NLO and NLO+PS
  light-jet angular distributions for $pp\to\ttbar+\ge 2\,b$ jets at
  13\,TeV.  Ratio plots like in \refF{fig:ttbb_1}.  }
\end{figure}


\chapter{Higgs Boson Pair Production}
\label{chap:HH}
\ChapterAuthor{S.~Dawson, C.~Englert, M.~Gouzevitch, R.~Salerno, M.~Slawinska~(Eds.);
J.~Baglio, S.~Borowka, A.~Carvalho,  M.~Dall'Osso, P.~de Castro Manzano, D.~de Florian, T.~Dorigo, F.~Goertz, C.A.~Gottardo, N.~Greiner, J.~Grigo, R.~Gr\"ober, G.~Heinrich, B.~Hespel, S.~Jones, M.~Kerner, I.M.~Lewis, J.~Mazzitelli, M.~M\"uhlleitner, A.~Papaefstathiou, T.~Robens, J.~Rojo, J.~Schlenk, U.~Schubert, M.~Spannowsky, M.~Spira, M.~Tosi, E.~Vryonidou, M.~Zaro, T.~Zirke}
\def\fb{fb}
\def\ab{ab}
\def\hsm{h}
\def\mhsm{m_h}
\def\mt{m_t}
\def\mz{m_Z}
\def\tev{\hbox{TeV}}
\def\gev{\hbox{GeV}}
\def\sqrts{\sqrt{s}}
\def\mhh{M_{hh}}
\def\mhhh{M_{hhh}}
\def\FTapp{FT$_{\text{approx}}$}


\def\mXsz{m_X^{\rm spin\text{-}0}}
\def\Xsz{X^{\rm spin\text{-}0}}


\section{Introduction}
\label{sec:hhintro}
In the SM, the Higgs self-couplings are uniquely determined by the structure of the scalar
potential,
\begin{equation}
V={\mhsm^2\over 2}\hsm^2+\lambda_3 v \hsm^3 +{\lambda_4\over 4}\hsm^4\, ,
\end{equation}
where $\lambda_3=\lambda_4={\mhsm^2/(2 v^2)}$.  Experimentally measuring $\lambda_3$ and
$\lambda_4$ is thus a crucial test of the mechanism of electroweak symmetry breaking.  A measurement of $\lambda_3$
requires double Higgs boson production while $\lambda_4$ is first probed in the production of $3$ Higgs bosons.

The phenomenology of multi-Higgs boson final states will provide
complementary information to that found from single Higgs physics at the LHC. Due to generically
small inclusive cross sections and a difficult signal vs. background
discrimination, the best motivated multi-Higgs final states at the Large
Hadron Collider are Higgs boson pair final states, of which gluon fusion
$gg\to \hsm\hsm$ is the dominant production mode.

\begin{figure}[!b]
  \begin{center}
    \includegraphics[scale=1.1]{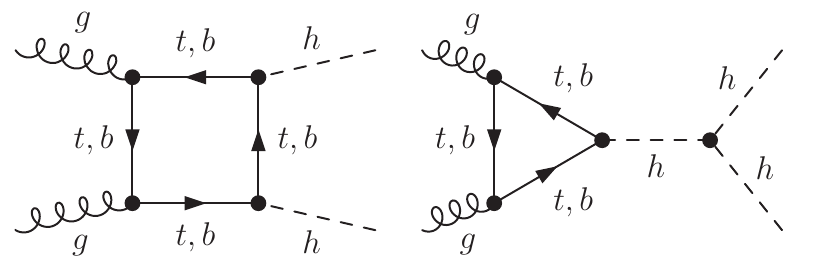}
    \caption{\label{fig:feyndiag} Feynman diagrams contributing to Higgs boson pair
      production via gluon fusion at leading order.}
  \end{center}
\end{figure}

Many models of physics beyond the Standard Model with SM-compatible single
Higgs boson signal strengths can exhibit a di-Higgs phenomenology vastly different from
the SM expectation. In this sense, a successful discovery of Higgs boson pair
production at the LHC and the subsequent measurement of potential
deviations from the SM constitute an important avenue in the search for
physics beyond the SM. In particular, modifications of the Higgs trilinear 
couplings (e.g. via a modified Higgs self interaction) can
only be directly observed in Higgs boson pair production.  In the
gluon fusion process this occurs via the
interference of the box and triangle diagrams shown in
\refF{fig:feyndiag}~\cite{Plehn:1996wb,Djouadi:1999rca,Glover:1987nx}.

To facilitate such a measurement, it is crucial to establish the Higgs boson pair 
production cross section in the SM to the best theoretical
accuracy possible and to provide BSM benchmarks that reflect the phenomenology of Higgs boson pairs at the LHC in a consistent and concise fashion. 

This report summarizes the results of the HH cross section group of the 2014-2015 LHC Higgs Cross Section
working group that aims to establish SM predictions for a range of dominant and subdominant Higgs boson pair production modes at the LHC at the highest available theoretical precision. In addition, we provide benchmarks for resonant and non-resonant extensions of the SM phenomenology of Higgs boson pairs in light of current single Higgs property measurements which are aligned with the efforts of other subgroups.
This note is structured as follows: In \refS{sec:hhxsec} we provide an update on the dominant Higgs boson pair production modes at the LHC; special care is devoted to the dominant gluon fusion production mode in \refS{sec:hhgf}.  Section~\ref{sec:hhdif} contains representative distributions at NLO for the SM gluon fusion pair production channel. Section~\ref{sec:hhbsm} discusses benchmarks of motivated BSM scenarios. The BSM phenomenology of multi-Higgs final states can be divided into resonant and non-resonant extensions of the SM. The latter is discussed in \refS{sec:hheft} using the language of Effective Field Theory. Benchmarks for resonant di-Higgs final state searches are discussed using the singlet-extended SM and the 2 Higgs doublet model in \refS{sec:hhsinglet}, which provide theoretically clean avenues to introduce new resonant physics into Higgs boson pair production.

\section{Total rates in the SM}
\label{sec:hhxsec}

\subsection{Gluon fusion}
\label{sec:hhgf}

The NLO~\cite{Dawson:1998py} and NNLO~\cite{deFlorian:2013jea} fixed order corrections to $gg\rightarrow\hsm\hsm$ are
known in the large top mass limit.  Recently, the complete NLO fixed order corrections, including
all top quark mass effects, have become available\cite{Borowka:2016ehy,Borowka:2016ypz}.
 The QCD corrections are large, typically doubling the cross section
from LO to NLO, with another $\sim 20\%$ increase going from NLO to NNLO.  The threshold resummation
corrections for Higgs boson pair production  at NNLL~\cite{Shao:2013bz, deFlorian:2015moa} further increase the rate.
The  NNLL threshold resummed cross sections are combined consistently with the fixed order NNLO results in 
 \refT{tab:hhsignnllb}, with the rate being weighted by the exact LO finite $\mt$ result normalized
 to the $\mt\rightarrow \infty$ LO result.  This is the HEFT approximation. 
For all the predictions we use  the {\tt PDF4LHC15\_nnlo\_mc} proton PDF set that is recommended by the PDF4LHC group \cite{Butterworth:2015oua}.

The scale choice $\mu_0=M_{hh}/2$ is shown, 
with  the scale variation  taken to be  ${\mu_0/ 2} < \mu_R, \mu_F<2 \mu_0$, with the restriction ${1/2} < {\mu_R/ \mu_F} < 2$.    The effect of choosing the central
scale to be $\mu_0=M_{hh}$ is shown in \refT{tab:hhsignnll}.  The numerical importance of
the threshold resummation is minimized for $\mu_0=M_{hh}/2$, and so we recommend this as our
preferred choice. The scale uncertainties are $\sim 4-6\%$ and the PDF uncertainties are $\sim 2-3\%$
at $\sqrts=13~\tev$.

\begin{table}[!t]
\begin{center}
\caption{\label{tab:hhsignnllb} NNLL  matched to NNLO cross sections for $gg\rightarrow \hsm\hsm$ with a central scale $\mu_0=M_{hh}/2$ with $\mhsm=124.5~\gev$, $\mhsm=125~\gev$, $\mhsm=125.09~\gev$~and $\mhsm=125.5~\gev$~\cite{deFlorian:2015moa} computed in the HEFT approximation.
The uncertainties from top quark mass effects are not included in this table. 
 Uncertainties are evaluated using the PDF4LHC recommendation and are based on the {\tt PDF4LHC15\_nnlo\_mc} set.} 
\begin{tabular}{l| c c  c c}
\toprule
$\mhsm=124.5~\gev$ & $\sigma_{\text{NNLL}}(\fb)$ & Scale Unc. $(\%)$ & PDF Unc. $(\%)$ & 
$\alpha_s$ Unc. $(\%)$ \\
\midrule
$\sqrts=7$~\tev & $7.772$ & $+4.0-5.7$ & $\pm 3.4$ & $\pm 2.8$ \\
$\sqrts=8$~\tev & $11.26$ & $+4.1-5.7$ & $\pm 3.0$ &  $\pm 2.6 $ \\
$\sqrts=13$~\tev & $38.20$ & $+4.3-6.0$ & $\pm 2.1$ &  $\pm 2.3$ \\
$\sqrts=14$~\tev & $45.34$ & $+4.4-6.0$ & $\pm 2.1$ &  $\pm 2.2$ \\
$\sqrt{s}=100$~\tev & $1760$ & $+5.0-6.7$ & $\pm 1.7$ & $\pm 2.1$ \\
\toprule
$\mhsm=125~\gev$ & $\sigma_{\text{NNLL}}(\fb)$ & Scale Unc. $(\%)$ & PDF Unc. $(\%)$ & 
$\alpha_s$ Unc. $(\%$)  \\
\midrule
$\sqrts=7$~\tev & $7.718$ & $+4.0-5.7$ & $\pm 3.4$ &  $\pm 2.8$ \\
$\sqrts=8$~\tev & $11.18$ & $+4.1-5.7$ & $\pm 3.1$ & $\pm 2.6 $ \\
$\sqrts=13$~\tev & $37.95$ & $+4.3-6.0$ & $\pm 2.1$ & $\pm 2.3$ \\
$\sqrts=14$~\tev & $45.05$ & $+4.4-6.0$ & $\pm 2.1$ &  $\pm 2.2$ \\
$\sqrts=100$~\tev & $1749$ & $+5.1-6.6$ & $\pm 1.7$ & $\pm 2.1$\\
\toprule
$\mhsm=125.09~\gev$ & $\sigma_{\text{NNLL}}(\fb)$ & Scale Unc. $(\%)$ & PDF Unc. $(\%)$ & 
$\alpha_s$ Unc. $(\%)$ \\
\midrule
$\sqrts=7$~\tev & $7.708$ & $+4.0-5.7$ & $\pm 3.4$ & $\pm 2.8$ \\
$\sqrts=8$~\tev & $11.17$ & $+4.1-5.7$ & $\pm 3.1$ & $\pm 2.6$ \\
$\sqrts=13$~\tev & $37.91$ & $+4.3-6.0$ & $\pm 2.1$ & $\pm 2.3$ \\
$\sqrts=14$~\tev & $45.00$ & $+4.4-6.0$ & $\pm 2.1$ & $\pm 2.2$ \\
$\sqrts=100$~\tev & $1748$ & $+5.0-6.5$ & $\pm 1.7$ & $\pm 2.0 $ \\
\toprule
$\mhsm=125.5~\gev$ & $\sigma_{\text{NNLL}}(\fb)$ & Scale Unc. $(\%)$ & PDF Unc. $(\%)$ & 
$\alpha_s$ Unc. $(\%)$ \\
\midrule
$\sqrts=7$~\tev & $7.663$ & $+4.0-5.7$ & $\pm 3.4$ & $\pm 2.8$ \\
$\sqrts=8$~\tev & $11.11$ & $+4.1-5.7$ & $\pm 3.1$ & $\pm 2.6$ \\
$\sqrts=13$~\tev & $37.71$ & $+4.3-6.0$ & $\pm 2.1$ & $\pm 2.3$ \\
$\sqrts=14$~\tev & $44.76$ & $+4.4-5.9$ & $\pm 2.1$ & $\pm 2.2$ \\
$\sqrts=100$~\tev & $1738$ & $+5.2-6.4$ & $\pm 1.7$ & $\pm 2.1$ \\
\bottomrule
\end{tabular}
\end{center}
\end{table}

\begin{table}[t]
\begin{center}
\caption{\label{tab:hhsignnll} NNLL cross sections for $gg\rightarrow \hsm \hsm$ with a central scale $\mu_0=M_{hh}$
with $\mhsm=125~\gev$. The uncertainties from top quark mass effects are not included in this table. 
Uncertainties are evaluated using the PDF4LHC recommendation and are based on the {\tt PDF4LHC15\_nnlo\_mc set.} } 
\begin{tabular}{l| c c c c}
\toprule
$\mu_0=M_{hh}$ & $\sigma_{NNLL}(\fb)$ & Scale Unc. $(\%)$ & PDF Unc. $(\%)$ &  $\alpha_s$ Unc.$(\%)$ \\
\midrule
$\sqrts=7$~\tev & $7.61$ & $+5.6-6.0$ & $\pm 3.3$ &$\pm 2.8$\\
$\sqrts=8$~\tev & $11.0$ & $+5.5-6.0$ & $\pm 3.0$ & $\pm 2.6$ \\
$\sqrts=13$~\tev & $37.3$ & $+5.1-6.1$ & $\pm 2.1$ & $\pm 2.3 $ \\
$\sqrts=14$~\tev & $44.2$ & $+5.2-6.1$ & $\pm 2.0$ & $\pm 2.2 $ \\
$\sqrts=100$~\tev & $1712$ & $+5.2-6.2$ & $\pm 1.7$ & $ \pm 2.0$ \\
\bottomrule
\end{tabular}
\end{center}
\end{table}

For convenience, we define two $K$ factors, computed in  $\mt\rightarrow
\infty$ limit, for the total cross sections, where $\sigma_{\text{NNLL}}$ is
the fixed order NNLO rate matched to the NNLL rate,
\begin{eqnarray}
K&\equiv&{\sigma_{\text{NNLL}}\over \sigma_{\text{NLO}}}\nonumber \\
K^\prime&\equiv& {\sigma_{\text{NNLL}}\over \sigma_{\text{LO}}}\, .
\label{eq:kdef}
\end{eqnarray}
The $K$ factors for the scale choices $\mu_0=M_{hh}$ and $\mu_0=M_{hh}/2$ are shown in
Tabs.~\ref{tab:kdef1} and \ref{tab:kdef2}, respectively. 

\begin{table}[!t]
\begin{center}
\caption{\label{tab:kdef1} $K$ factors as defined in Eq.~\eqref{eq:kdef} for
 $gg\rightarrow \hsm\hsm$ with a central scale $\mu_0=M_{hh}/2$ and $\mhsm=125~\gev$~\cite{deFlorian:2015moa}.}
\begin{tabular}{c| c c c c c}
\toprule
& $\sqrts=7~\tev$ &  $\sqrts=8~\tev$ &  $\sqrts=13~\tev$ &  $\sqrts=14~\tev$ &  $\sqrts=100~\tev$\\
\midrule
$K$& 1.203 & 1.200 & 1.193 & 1.192 & 1.195 \\
$K^\prime$ & 2.299& 2.296 & 2.301 & 2.304 & 2.472 \\
\bottomrule
\end{tabular}
\end{center}
\end{table}
\begin{table}[!t]
\caption{\label{tab:kdef2} $K$ factors as defined in Eq.~\eqref{eq:kdef} for
 $gg\rightarrow \hsm\hsm$ with a central scale $\mu_0=M_{hh}$ and $\mhsm=125~\gev$~\cite{deFlorian:2015moa}.}
\begin{centering}
\begin{tabular}{c| c c c c c}
\toprule
& $\sqrts=7~\tev$ &  $\sqrts=8~\tev$ &  $\sqrts=13~\tev$ &  $\sqrts=14~\tev$ &  $\sqrts=100~\tev$\\
\midrule
$K$& 1.426 & 1.413 & 1.378 & 1.373 & 1.305 \\
$K^\prime$ & 2.987& 2.949 & 2.847 & 2.835 & 2.699 \\
\bottomrule
\end{tabular}
\end{centering}
\end{table}

\subsubsection{Top Quark Mass Uncertainties}
Recently, the complete $2$-loop reducible contributions have been 
calculated~\cite{Degrassi:2016vss}, followed 
shortly  by the full NLO calculation including all top quark mass effects\cite{Borowka:2016ehy,Borowka:2016ypz}, 
and we can compare with various approximations in the literature,
\begin{itemize}
\item An $\mt$ expansion to NNLO estimates uncertainties from top mass effects 
at $\mathcal{O}(5\%)$~\cite{Grigo:2015dia,Grigo:2014jma,Grigo:2013rya}.
\item The exact inclusion of the $\mt$ dependence in the real contributions to the NLO corrections and reweighting the virtual corrections evaluated in the $\mt\to \infty$ approximation with the full LO finite $\mt$ dependence
estimates uncertainties to be $\mathcal{O}(-10\%)$. This is the  $FT_{approx}$ of 
Refs.~\cite{Frederix:2014hta,Maltoni:2014eza}.
\end{itemize}

 We provide approximate NLO
rates for $\hsm\hsm$ production in \refT{table:xsec21} in the limit in which the top mass effects are retained 
in the real contributions, the  $FT_{approx}$ of Refs.~\cite{Frederix:2014hta} and~\cite{Maltoni:2014eza}.
   These results  can be compared with the NLO results obtained by computing a $K$ factor in the 
   $\mt\rightarrow\infty$ limit and re-weighting by the exact LO result (HEFT approximation)  as shown
in \refT{table:xsec33}~\cite{deFlorian:2013jea,Dawson:1998py}.
 There is good, but not exact, agreement between the two approximations.

The complete  NLO results for $\mhsm=125~\gev$ and $\mt=173~\gev$, with all top mass effects,
 are shown in Table \ref{tab:hhfinnlo}.  The inclusion of the mass effects consistently decreases the NLO
 cross section from the HEFT result, while the $FT_{approx}$ is a better approximation to the total rate.
\begin{table}
\begin{center}
\caption{\label{tab:hhfinnlo}NLO results for $gg\rightarrow \hsm\hsm$. The uncertainty in per cent is from scale variations only.
The central scale is $\mu_0=\mhh/2$. We use $\mt=173~\gev$ and  $\mhsm=125~\gev$.
The PDF set is {\sc PDF4LHC15}\_nlo\_100\_pdfas.\cite{Borowka:2016ehy}}
 \begin{tabular}{r|ccc}
\toprule
$\sqrt{s}$\ \ \ &NLO HEFT&NLO FT$_{approx}$& NLO$^{exact}$\\
\midrule
7 TeV&6.44$^{+20.1\%}_{-16.9\%}$&	5.95$^{+17.3\%}_{-15.7\%}$&	5.80$^{+16.3\%}_{-15.2\%}$\\
&&&\\
8 TeV&9.37$^{+19.8\%}_{-16.5\%}$&	8.61$^{+16.7\%}_{-15.1\%}$&	8.35$^{+15.7\%}_{-14.6\%}$\\
&&&\\
13 TeV & 32.22$^{+18.2\%}_{-15.1\%}$&	28.90$^{+15.0\%}_{-13.4\%}$&	27.80$^{+13.8\%}_{-12.8\%}$\\
&&&\\
14 TeV & 38.32$^{+18.1\%}_{-14.9\%}$&	34.26$^{+14.7\%}_{-13.2\%}$&	32.91$^{+13.6\%}_{-12.6\%}$\\
\bottomrule
\end{tabular}
\end{center}
\end{table}
The inclusion of the mass effects reduces the NLO rate by an energy dependent factor
which can be parameterized as 
\begin{equation}
\sigma(gg\rightarrow hh)_{NLO}^{exact}=\sigma (gg\rightarrow hh)_{NLO}^{HEFT}(1+\delta_t)\, ,
\end{equation}
where, using $\mt=173~\gev$, 
\begin{eqnarray}
\delta_t  (7~\tev) & =&  -9.94\%\\
\delta_t (8~\tev) & =&  -10.88\%\\
\delta_t (13~\tev) & = & -13.72\%\\
\delta_t (14~\tev) &=& -14.11\%\, . 
\end{eqnarray}
The top mass effects can be included consistently by writing our final result as
\begin{equation}
\sigma_{NNLL}^\prime=\sigma_{NNLL}+\delta_t\sigma_{NLO}^{HEFT}\, ,
\label{sigf}
\end{equation}
where $\sigma_{NNLL}$ is given in Table \ref{tab:hhsignnllb}.
This prescription amounts to subtracting $5.49~\fb$ from the $14~\tev$,  $4.50~\fb$ from the $13
~\tev$, $1.02~\fb$ from the $8~\tev$ and $0.64~\fb$ from the $7~\tev$
 numbers of Table \ref{tab:hhsignnllb} . (We note that Table  \ref{tab:hhsignnllb} uses $\mt=172.5~\gev$,
so there is a slight mis-match in the $\mt$ values.)
 Furthermore, we neglect  any possible $\mhsm$ dependence
of $\delta_t$.  Our recommended results are given in Table \ref{tab:hhfin} and correspond to
the convention of Eq. \ref{sigf}.  We arbitrarily assume 
a top mass uncertainty of $\pm 5\%$ from unknown top quark mass effects at NNLO, and do not include
a theoretical error on $\delta_t$.




\begin{table}
 \renewcommand{\arraystretch}{1.6}
\begin{center}
 \caption{Signal cross section (in \fb) for $g g \to \hsm\hsm$ at NLO QCD  in the { \FTapp}~approximation of Ref.~\cite{Maltoni:2014eza,Frederix:2014hta}, with a central
 scale  $\mu_0=M_{hh}/2$. The first uncertainty is the scale uncertainty and
 the second is the PDF 
 uncertainty based on the {\tt{PDF4LHC15\_nlo\_mc}} set.  
 There is an additional  uncertainty from 
 top quark mass effects, along with an $\alpha_s$ uncertainty. \label{table:xsec21}}
\scriptsize
 \begin{tabular}{l|ccccc}
\toprule
$\mhsm~(\gev)$ & $\sqrts=7$ \tev & $\sqrts=8$ \tev & $\sqrts=13$ \tev & $\sqrts=14$ \tev & $\sqrts=100$ \tev \\ 
 \midrule
124.5 & $6.08^{+17.3\% }_{-15.7\% }\pm 4.0\%$ & $8.74^{+16.8\% }_{-15.1\% }\pm 3.6\%$ & $29.43^{+15.1\% }_{-13.5\% }\pm 2.7\%$& $35.08^{+14.8\% }_{-13.2\%}\pm 2.6 \% $& $1254^{+14.5\%}_{-14.2\% }\pm 2.1 \%$ \\ 
  125 & $6.01^{+17.2\%}_{-15.6\%}\pm 4.0\%$ & $8.62^{+16.8\%}_{-15.2\%}\pm 3.7\%$ & $29.26^{+15.0\%}_{-13.4\%}\pm 2.7\%$& $34.59^{+14.6\%}_{-13.1\%}\pm 2.6\%$& $1237^{+14.3\% }_{-14.1\% }\pm 2.1\%$ \\ 
    125.09 & $6.03^{+17.2\%}_{-15.6\%}\pm 4.0 \%$ & $8.70^{+16.7\%}_{-15.1\%}\pm 3.7 \%$ & $29.27^{+15.1\%}_{-13.5\%}\pm 2.7\%$& $34.59^{+14.7\%}_{-13.2\%}\pm 2.6\%$& $1229^{+14.6\%}_{-14.2\%}\pm 2.1 \%$ \\ 
      125.5 & $5.99^{+17.4\%}_{-15.7\%}\pm 4.0\%$ & $8.64^{+16.9\%}_{-15.2\% }\pm 3.6 \%$ & $29.28^{+14.9\%}_{-13.4\% }\pm 2.7 \%$& $34.41^{+14.9\% }_{-13.2\%}\pm 2.6\%$& $1227^{+14.3\%}_{-14.1\%}\pm 2.1 \%$ \\ 
\bottomrule      
\end{tabular}
\end{center} 
\begin{center}
 \caption{Signal cross section (in \fb) for $g g \to \hsm\hsm$ at NLO QCD  in the 
 $\mt\rightarrow\infty$ limit, reweighted by the exact LO result,
 of Ref.~\cite{deFlorian:2013jea,Dawson:1998py}, with a central
 scale  $\mu_0=M_{hh}/2$.  
 Only the uncertainties due to scale variation are shown. There is an additional uncertainty from 
 top quark mass effects, along with PDF and $\alpha_s $ uncertainties.\label{table:xsec33}}  
\begin{tabular}{l|ccccc}
\toprule
$\mhsm~(\gev)$ & $\sqrts=7$ \tev & $\sqrts=8$ \tev & $\sqrts=13$ \tev & $\sqrts=14$ \tev & $\sqrts=100$ \tev \\ 
 \midrule
  125 & $6.415^{+20\%}_{-16.8\%}$ & $9.318^{+19.5\%}_{-16.4\%}$ & $31.81^{+18.2\%}_{-15.0\%}$& 
  $37.79^{+18\%}_{-14.8\%}$& $1464^{+16.1\% }_{-13.8\% }$ \\ 
  \bottomrule
 \end{tabular}
\end{center}  
\end{table}

\begin{table}[!t]
\begin{center}
\caption{\label{tab:hhfin} NNLL matched to NNLO cross sections for $gg\rightarrow \hsm\hsm$
including top quark mass effects to NLO\cite{Borowka:2016ehy}, as in Eq.\ref{sigf},
with a central scale $\mu_0=M_{hh}/2$ with $\mhsm=124.5~\gev$, $\mhsm=125~\gev$, $\mhsm=125.09~\gev$~and $\mhsm=125.5~\gev$~\cite{deFlorian:2015moa}.
Uncertainties are evaluated using the PDF4LHC recommendation and are based on the {\tt PDF4LHC15\_nnlo\_mc} set.
These are our recommended numbers. }
\begin{tabular}{l| c c c c}
\toprule
$\mhsm=124.5~\gev$ & $\sigma_{\text{NNLL}}^\prime(\fb)$ & Scale Unc. $(\%)$ & PDF Unc. $(\%)$ &
$\alpha_s$ Unc. $(\%)$ \\
\midrule
$\sqrts=7$~\tev & $7.132$ & $+4.0-5.7$ & $\pm 3.4$ & $\pm 2.8$ \\
$\sqrts=8$~\tev & $10.24$ & $+4.1-5.7$ & $\pm 3.0$ & $\pm 2.6 $ \\
$\sqrts=13$~\tev & $33.78$ & $+4.3-6.0$ & $\pm 2.1$ & $\pm 2.3$ \\
$\sqrts=14$~\tev & $39.93$ & $+4.4-6.0$ & $\pm 2.1$ & $\pm 2.2$ \\
\toprule
$\mhsm=125~\gev$ & $\sigma_{\text{NNLL}}^\prime(\fb)$ & Scale Unc. $(\%)$ & PDF Unc. $(\%)$ &
$\alpha_s$ Unc. $(\%$) \\
\midrule
$\sqrts=7$~\tev & $7.078$ & $+4.0-5.7$ & $\pm 3.4$ & $\pm 2.8$ \\
$\sqrts=8$~\tev & $10.16$ & $+4.1-5.7$ & $\pm 3.1$ & $\pm 2.6 $ \\
$\sqrts=13$~\tev & $33.53$ & $+4.3-6.0$ & $\pm 2.1$ & $\pm 2.3$ \\
$\sqrts=14$~\tev & $39.64$ & $+4.4-6.0$ & $\pm 2.1$ & $\pm 2.2$ \\
\toprule
$\mhsm=125.09~\gev$ & $\sigma_{\text{NNLL}}^\prime(\fb)$ & Scale Unc. $(\%)$ & PDF Unc. $(\%)$ &
$\alpha_s$ Unc. $(\%)$ \\
\midrule
$\sqrts=7$~\tev & $7.068$ & $+4.0-5.7$ & $\pm 3.4$ & $\pm 2.8$ \\
$\sqrts=8$~\tev & $10.15$ & $+4.1-5.7$ & $\pm 3.1$ & $\pm 2.6$ \\
$\sqrts=13$~\tev & $33.49$ & $+4.3-6.0$ & $\pm 2.1$ & $\pm 2.3$ \\
$\sqrts=14$~\tev & $39.59$ & $+4.4-6.0$ & $\pm 2.1$ & $\pm 2.2$ \\
\toprule
$\mhsm=125.5~\gev$ & $\sigma_{\text{NNLL}}^\prime(\fb)$ & Scale Unc. $(\%)$ & PDF Unc. $(\%)$ &
$\alpha_s$ Unc. $(\%)$ \\
\midrule
$\sqrts=7$~\tev & $7.023$ & $+4.0-5.7$ & $\pm 3.4$ & $\pm 2.8$ \\
$\sqrts=8$~\tev & $10.09$ & $+4.1-5.7$ & $\pm 3.1$ & $\pm 2.6$ \\
$\sqrts=13$~\tev & $33.29$ & $+4.3-6.0$ & $\pm 2.1$ & $\pm 2.3$ \\
$\sqrts=14$~\tev & $39.35$ & $+4.4-5.9$ & $\pm 2.1$ & $\pm 2.2$ \\
\bottomrule
\end{tabular}
\end{center}
\end{table}

\subsection{Other production channels}
There are a number of additional subdominant Higgs boson pair production modes at the LHC~\cite{Baglio:2012np,Frederix:2014hta}. In particular, Higgs boson pairs can be produced in association with electroweak bosons ($W^\pm \hsm\hsm$, $Z\hsm\hsm$),  top quarks ($t {\overline t} \hsm\hsm$ and $tj\hsm\hsm$,) or jets ($\hsm\hsm j j$). The associated production of $hh$ with vector bosons is known at NNLO QCD \cite{Baglio:2012np},  where the NNLO corrections to the NLO rate are of order $10\%$.
The NNLO rates are given in Tables \ref{table:xsec22}- \ref{table:xsec25}.  The scale choice $\mu_0=M_{hhV}$ ($V=W,Z$) is shown,
with  the scale variation  taken to be  ${\mu_0/ 2} < \mu_R, \mu_F<2 \mu_0$, with the restriction ${1/ 2} < {\mu_R/ \mu_F} < 2$.    

While the weak boson fusion configuration (WBF) is reliably known at NLO precision~\cite{Baglio:2012np,Frederix:2014hta} (with NNLO corrections negligible \cite{Liu-Sheng:2014gxa}), a LO estimate of the gluon fusion contribution to the WBF configuration~\cite{Dolan:2015zja} suggests that it is non-negligible even for tight WBF selections~\cite{Dolan:2013rja,Englert:2014uqa}. 

The WBF NLO rates are given in \refT{table:xsec26}. 
The WBF cross section is highly stabilized at NLO QCD; electroweak corrections are not available.
A recent study suggests that angular correlations could be useful to separate this gluon fusion contamination from the WBF configuration~\cite{Nakamura:2016agl}.
The $t {\overline t} \hsm \hsm $ and $t{\overline t} \hsm j$   cross sections are given in Tabs.~\ref{table:xsec23} and \ref{table:xsec27}. The $t {\overline t} \hsm\hsm$ 
rate is particularly interesting in composite models, where it may be enhanced over the SM rate. The
scale variation for $t{\overline t} h h$ and  $tj\hsm\hsm$  is chosen as ${\mu_0/ 2} < \mu_R, \mu_F<2 \mu_0$ with $\mu_0=\mhh/2$.

In \refT{table:xsec28} we also provide numbers for $pp\to \hsm\hsm\hsm$ at NLO QCD in the \FTapp~approximation of~\cite{Frederix:2014hta,Maltoni:2014eza}.

\begin{table}[h!]
 \renewcommand{\arraystretch}{1.6}
  \centering
   \caption{\label{table:xsec22}Signal cross section (in fb) for $\hsm \hsm Z$  production at NNLO QCD with
     the central scale $\mu_0 = \mu_R = \mu_F = M_{hhZ}$ \cite{Baglio:2012np}. The
     first uncertainty is the scale uncertainty and the second is the
     PDF + $\alpha_s$ uncertainty based on the {\tt PDF4LHC15\_nnlo\_mc} set.}
  \scriptsize   
  \begin{tabular}{l|ccccc}
    \toprule
    $\mhsm~(\gev)$ & \;\;\;$\sqrt{s} = 7$ \tev\;\;\; &
    \;\;\;$\sqrt{s} = 8$ \tev\;\;\; & \;\;\;$\sqrt{s} = 13$ \tev\;\;\; &
    \;\;\;$\sqrt{s} = 14$ \tev\;\;\; & \;\;\;$\sqrt{s} = 100$ \tev\;\;\; \\
    \midrule
    124.5 & $0.109^{+2.8\%}_{-2.2\%} \pm 2.9\%$
    & $0.145^{+2.8\%}_{-2.3\%} \pm 2.6\%$
    & $0.368^{+3.5\%}_{-2.6\%} \pm 1.9\%$
    & $0.420^{+3.6\%}_{-2.7\%} \pm 1.8\%$
    & $8.33^{+5.9\%}_{-4.6\%} \pm 1.7\%$\\ 

    125 & $0.108^{+2.6\%}_{-2.2\%} \pm 2.9\%$
    & $0.143^{+2.8\%}_{-2.2\%} \pm 2.6\%$
    & $0.363^{+3.4\%}_{-2.7\%} \pm 1.9\%$
    & $0.415^{+3.5\%}_{-2.7\%} \pm 1.8\%$
    & $8.23^{+5.9\%}_{-4.6\%} \pm 1.7\%$\\

    125.09 & $0.108^{+2.6\%}_{-2.2\%} \pm 2.9\%$
    & $0.143^{+2.7\%}_{-2.3\%} \pm 2.6\%$
    & $0.362^{+3.4\%}_{-2.6\%} \pm 1.9\%$
    & $0.414^{+3.5\%}_{-2.7\%} \pm 1.8\%$
    & $8.22^{+5.9\%}_{-4.6\%} \pm 1.7\%$\\

    125.5 & $0.106^{+2.6\%}_{-2.2\%} \pm 2.9\%$
    & $0.141^{+2.8\%}_{-2.2\%} \pm 2.6\%$
    & $0.359^{+3.5\%}_{-2.7\%} \pm 1.9\%$
    & $0.409^{+3.5\%}_{-2.7\%} \pm 1.9\%$
    & $8.13^{+5.9\%}_{-4.6\%} \pm 1.7\%$\\
    \bottomrule
  \end{tabular}
 \end{table}
 \begin{table}[!ht]
 \renewcommand{\arraystretch}{1.6}
  \centering
   \caption{Signal cross section (in fb) for $\hsm \hsm W^-$ production  at NNLO QCD with
     the central scale $\mu_0 = \mu_R = \mu_F = M_{hhW}$ \cite{Baglio:2012np}. The
     first uncertainty is the scale uncertainty and the second is the
     PDF + $\alpha_s$ uncertainty based on the {\tt PDF4LHC15\_nnlo\_mc} set.}
   \label{table:xsec24}
  \scriptsize      
  \begin{tabular}{l|ccccc}
    \toprule
    $\mhsm~(\gev)$ & \;\;\;$\sqrt{s} = 7$ \tev\;\;\; &
    \;\;\;$\sqrt{s} = 8$ \tev\;\;\; & \;\;\;$\sqrt{s} = 13$ \tev\;\;\; &
    \;\;\;$\sqrt{s} = 14$ \tev\;\;\; & \;\;\;$\sqrt{s} = 100$ \tev\;\;\; \\
    \midrule
    124.5 & $0.0516^{+0.98\%}_{-1.2\%} \pm 4.0\%$
    & $0.0688^{+1.0\%}_{-1.2\%} \pm 3.7\%$
    & $0.176^{+1.2\%}_{-1.3\%} \pm 2.8\%$
    & $0.200^{+1.2\%}_{-1.3\%} \pm 2.7\%$
    & $3.34^{+3.6\%}_{-4.3\%} \pm 1.9\%$\\ 

     125 & $0.0509^{+0.98\%}_{-1.2\%} \pm 4.0\%$
    & $0.0679^{+1.0\%}_{-1.2\%} \pm 3.7\%$
    & $0.173^{+1.2\%}_{-1.3\%} \pm 2.8\%$
    & $0.198^{+1.2\%}_{-1.3\%} \pm 2.7\%$
    & $3.30^{+3.5\%}_{-4.3\%} \pm 1.9\%$\\ 

   125.09& $0.0508^{+0.98\%}_{-1.2\%} \pm 4.0\%$
    & $0.0677^{+1.0\%}_{-1.2\%} \pm 3.7\%$
    & $0.173^{+1.2\%}_{-1.3\%} \pm 2.8\%$
    & $0.197^{+1.2\%}_{-1.3\%} \pm 2.7\%$
    & $3.30^{+3.5\%}_{-4.3\%} \pm 1.9\%$\\ 

    125.5 & $0.0502^{+0.98\%}_{-1.2\%} \pm 4.0\%$
    & $0.0670^{+1.0\%}_{-1.2\%} \pm 3.7\%$
    & $0.171^{+1.2\%}_{-1.3\%} \pm 2.8\%$
    & $0.195^{+1.2\%}_{-1.3\%} \pm 2.7\%$
    & $3.27^{+3.5\%}_{-4.3\%} \pm 1.9\%$\\ 
    \bottomrule
  \end{tabular}
 \end{table}
\begin{table}[h!]
 \renewcommand{\arraystretch}{1.6}
  \centering
   \caption{  \label{table:xsec25}Signal cross section (in fb) for $\hsm \hsm W^+$ production  at NNLO QCD with
     the central scale $\mu_0 = \mu_R = \mu_F = M_{hhW}$ \cite{Baglio:2012np}. The
     first uncertainty is the scale uncertainty and the second is the
     PDF + $\alpha_s$ uncertainty based on the {\tt PDF4LHC15\_nnlo\_mc} set.}
  \scriptsize        
  \begin{tabular}{l|ccccc}
    \toprule
    $\mhsm~(\gev)$ & \;\;\;$\sqrt{s} = 7$ \tev\;\;\; &
    \;\;\;$\sqrt{s} = 8$ \tev\;\;\; & \;\;\;$\sqrt{s} = 13$ \tev\;\;\; &
    \;\;\;$\sqrt{s} = 14$ \tev\;\;\; & \;\;\;$\sqrt{s} = 100$ \tev\;\;\; \\
    \midrule
    124.5 & $0.114^{+0.47\%}_{-0.59\%} \pm 3.0\%$
    & $0.147^{+0.43\%}_{-0.52\%} \pm 2.8\%$
    & $0.333^{+0.32\%}_{-0.41\%} \pm 2.2\%$
    & $0.373^{+0.33\%}_{-0.39\%} \pm 2.1\%$
    & $4.74^{+0.90\%}_{-0.96\%} \pm 1.8\%$\\ 

     125 & $0.113^{+0.47\%}_{-0.59\%} \pm 3.0\%$
    & $0.145^{+0.43\%}_{-0.52\%} \pm 2.8 \%$
    & $0.329^{+0.32\%}_{-0.41\%} \pm 2.2\%$
    & $0.369^{+0.33\%}_{-0.39\%} \pm 2.1\%$
    & $4.70^{+0.90\%}_{-0.96\%} \pm 1.8\%$\\ 

   125.09& $0.113^{+0.47\%}_{-0.59\%} \pm 3.0\%$
    & $0.145^{+0.43\%}_{-0.52\%} \pm 2.8\%$
    & $0.329^{+0.32\%}_{-0.41\%} \pm 2.2\%$
    & $0.368^{+0.33\%}_{-0.39\%} \pm 2.1\%$
    & $4.69^{+0.90\%}_{-0.96\%} \pm 1.8\%$\\ 

    125.5 & $0.111^{+0.47\%}_{-0.59\%} \pm 3.0\%$
    & $0.143^{+0.43\%}_{-0.52\%} \pm 2.8\%$
    & $0.326^{+0.32\%}_{-0.41\%} \pm 2.2\%$
    & $0.365^{+0.33\%}_{-0.39\%} \pm 2.1 \%$
    & $4.65^{+0.90\%}_{-0.96\%} \pm 1.8\%$\\ 
    \bottomrule
  \end{tabular}
 \end{table}
\begin{table}[h!]
\renewcommand{\arraystretch}{1.6}
\begin{center}
 \caption{Cross section (in \fb) for weak boson fusion $\hsm\hsm jj$ at NLO QCD with the central
 scale $\mu_0=M_{hh}/2$~\cite{Frederix:2014hta}. The first uncertainty is the scale uncertainty and
 the second is the PDF
 uncertainty based on the  {\tt{PDF4LHC15\_nlo\_mc}} set. \label{table:xsec26}}  
  \scriptsize    
\begin{tabular}{l|ccccc}
        \toprule
$\mhsm~(\gev)$ & $\sqrts=7$ \tev & $\sqrts=8$ \tev & $\sqrts=13$ \tev & $\sqrts=14$ \tev & $\sqrts=100$ \tev \\ 
 \midrule
124.5 &   $0.320^{+3.2\%}_{-3.7\%}\pm 2.7\%$  &$0.470^{+2.4\%}_{-3.1\%}\pm 2.6\%$ & $1.65^{+2.4\%}_{-2.7\%}\pm 2.3\%$ & $1.97^{+2.3\%}_{-2.6\%}\pm 2.3\%$ & $81.9^{+0.2\%}_{-0.2\%}\pm 1.8\%$ \\ 

  125 & $0.316^{+3.7\%}_{-4.1\%}\pm 2.7\%$   & $0.468^{+2.8\%}_{-3.3\%}\pm 2.6\%$ & $1.64^{+2.0\%}_{-2.5\%}\pm 2.3\%$ & $1.94^{+2.3\%}_{-2.6\%}\pm 2.3\%$ &  $80.3^{+0.5\%}_{-0.4\%}\pm 1.7\%$   \\ 

    125.09 & $0.313^{+3.2\%}_{-3.8\%}\pm 2.6\%$ & $0.459^{+3.2\%}_{-3.6\%}\pm 2.6\%$  & $1.62^{+2.3\%}_{-2.7\%}\pm 2.3\%$ & $1.95^{+1.8\%}_{-2.3\%}\pm 2.4\%$ & $80.8^{+0.8\%}_{-0.8\%}\pm 1.8\%$  \\ 

      125.5 & $0.312^{+3.6\%}_{-4.0\%}\pm 2.7\%$  & $0.458^{+2.9\%}_{-3.4\%}\pm 2.6\%$ & $1.63^{+2.0\%}_{-2.5\%}\pm 2.3\%$ & $1.94^{+1.3\%}_{-1.9\%}\pm 2.3\%$  &  $80.7^{+0.7\%}_{-0.7\%}\pm 1.8\%$ \\ 
 \bottomrule     
\end{tabular}
\end{center} 
\end{table}
%
\begin{table}[h!]
\renewcommand{\arraystretch}{1.6}
\begin{center}
 \caption{Cross section (in \fb) for $t\bar{t}\hsm\hsm$  production at NLO QCD with the central
 scale $\mu_0=M_{hh}/2$~\cite{Frederix:2014hta}.  The first uncertainty is the scale uncertainty and
 the second is the PDF 
uncertainty based on the  {\tt{PDF4LHC15\_nlo\_mc}} set. \label{table:xsec23}}
  \scriptsize   
\begin{tabular}{l|ccccc}
        \toprule
$\mhsm~(\gev)$ & $\sqrts=7$ \tev & $\sqrts=8$ \tev & $\sqrts=13$ \tev & $\sqrts=14$ \tev & $\sqrts=100$ \tev \\ 
 \midrule
124.5 & $0.112^{+3.5\%}_{-12.5\%}\pm 4.2\%$  & $0.176^{+2.9\%}_{-10.7\%}\pm 3.9\%$ & $0.786^{+1.3\%}_{-4.5\%}\pm 3.2\%$& $0.968^{+1.7\%}_{-4.6\%}\pm 3.1\%$ & $87.2^{+7.9\%}_{-7.3\%}\pm 1.6\%$ \\ 

  125 & $0.110^{+3.5\%}_{-12.5\%}\pm 4.2\%$ & $0.174^{+2.9\%}_{-10.6\%}\pm 3.9\%$ & $0.775^{+1.5\%}_{-4.3\%}\pm 3.2\%$& $0.949^{+1.7\%}_{-4.5\%}\pm 3.1\%$  &  $82.1^{+7.9\%}_{-7.4\%}\pm 1.6\%$  \\ 

    125.09 & $0.109^{+3.5\%}_{-12.8\%}\pm 4.2\%$  & $0.174^{+2.8\%}_{-10.6\%}\pm 3.9\%$  & $0.772^{+1.7\%}_{-4.5\%}\pm 3.2\%$& $0.949^{+1.8\%}_{-4.8\%}\pm 3.2\%$ & $82.1^{+8.3\%}_{-7.6\%}\pm 1.6\%$  \\ 

      125.5 & $0.107^{+3.3\%}_{-12.9\%}\pm 4.2\%$  & $0.172^{+2.9\%}_{-10.4\%}\pm 4.0\%$  & $0.762^{+1.3\%}_{-4.5\%}\pm 3.2\%$ & $0.937^{+1.5\%}_{-4.5\%}\pm 3.1\%$& $81.9^{+8.2\%}_{-7.6\%}\pm 1.6\%$   \\ 
\bottomrule      
\end{tabular}
 \end{center} 
\end{table}
\begin{table}[h!]
\renewcommand{\arraystretch}{1.6}
\begin{center}
 \caption{Signal cross section (in \fb) for $\hsm\hsm tj$  production at NLO QCD with the central
 scale $\mu_0=M_{hh}/2$~\cite{Frederix:2014hta}.  The first uncertainty is the scale uncertainty and
 the second is the PDF 
uncertainty based on the  {\tt{PDF4LHC15\_nlo\_mc}} set. \label{table:xsec27}}  
  \scriptsize   
\begin{tabular}{l|ccccc}
        \toprule
$\mhsm~(\gev)$ & $\sqrts=7$ \tev & $\sqrts=8$ \tev & $\sqrts=13$ \tev & $\sqrts=14$ \tev & $\sqrts=100$ \tev \\ 
 \midrule
124.5 & $0.00335^{+3.9\%}_{-1.7\%}\pm 6.2\%$    & $0.00551^{+5.6\%}_{-3.2\%}\pm 5.8\%$ & $0.0289^{+5.4\%}_{-3.4\%}\pm 4.6\%$ & $0.0365^{+4.4\%}_{-1.6\%}\pm 4.7\%$ & $4.44^{+5.2\%}_{-5.6\%}\pm 2.3\%$  \\ 

  125 & $0.00331^{+3.9\%}_{-1.8\%}\pm 6.1\%$  &  $0.00538^{+5.3\%}_{-3.0\%}\pm 5.6\%$ & $0.0289^{+5.5\%}_{-3.6\%}\pm 4.7\%$ & $0.0367^{+4.2\%}_{-1.8\%}\pm 4.6\%$ & $4.44^{+2.2\%}_{-2.8\%}\pm 2.4\%$ \\ 

    125.09 & $0.00331^{+4.3\%}_{-2.1\%}\pm 6.3\%$  & $0.00540^{+5.4\%}_{-3.1\%}\pm 5.6\% $ & $0.0281^{+5.2\%}_{-3.2\%}\pm 4.5\%$ & $0.0364^{+3.7\%}_{-1.3\%}\pm 4.7\%$ & $4.43^{+2.0\%}_{-2.6\%}\pm 2.4\%$ \\ 

      125.5 &  $0.00326^{+3.9\%}_{-1.6\%}\pm 6.1\%$ & $0.00521^{+5.5\%}_{-3.4\%}\pm 5.8\% $ & $0.0279^{+6.1\%}_{-4.6\%}\pm 6.4\%$ & $0.0359^{+3.8\%}_{-1.6\%}\pm 4.7\%$&  $4.43^{+2.1\%}_{-2.6\%}\pm 2.4\%$ \\ 
\bottomrule      
\end{tabular}
\end{center} 
\end{table}
\begin{table}[h!]
 \renewcommand{\arraystretch}{1.6}
\begin{center}
 \caption{Signal cross section (in \ab) for $g g \to \hsm\hsm\hsm$ at NLO QCD for $\mhsm = 125~\gev$ with $\mu_R=\mu_F=\mu_0$~\cite{Maltoni:2014eza}.  The first uncertainty is the scale uncertainty and
 the second is the PDF 
uncertainty based on the  {\tt{PDF4LHC15\_nlo\_mc}} set.  \label{table:xsec28} }  
  \scriptsize   
\begin{tabular}{l|ccccc}
        \toprule
$\mu_0$ & $\sqrt{s}=7$ \tev & $\sqrt{s}=8$ \tev & $\sqrt{s}=13$ \tev & $\sqrt{s}=14$ \tev & $\sqrt{s}=100$ \tev \\ 
 \midrule
 $M_{hhh}/2$ & $12.03^{+17.8\% }_{-16.3\% }\pm 5.2\%$ & $17.99^{+16.5\%}_{-15.4\% }\pm 4.8\%$ & $73.43^{+14.7\% }_{-13.7\% }\pm 3.3\%$& $86.84^{+14.0\%}_{-13.2\%}\pm 3.2\%$& $4732^{+11.9\% }_{-11.6\% }\pm 1.8\%$ \\ 

  $M_{hhh}$ & $9.91^{+19.3\% }_{-16.6\% }\pm 5.3\%$ & $15.14^{+18.4\%}_{-16.0\% }\pm 4.7\%$ & $63.32^{+16.1\% }_{-14.1\% }\pm 3.4\%$& $76.15^{+15.9\%}_{-14.0\%}\pm 3.2\%$& $4306^{+14.0\% }_{-12.3\% }\pm 1.8\%$ \\ 
  \bottomrule
\end{tabular}
\end{center} 
\end{table}

\clearpage

\begin{figure}[!b]
  \centering
  \includegraphics[width=0.49\textwidth]{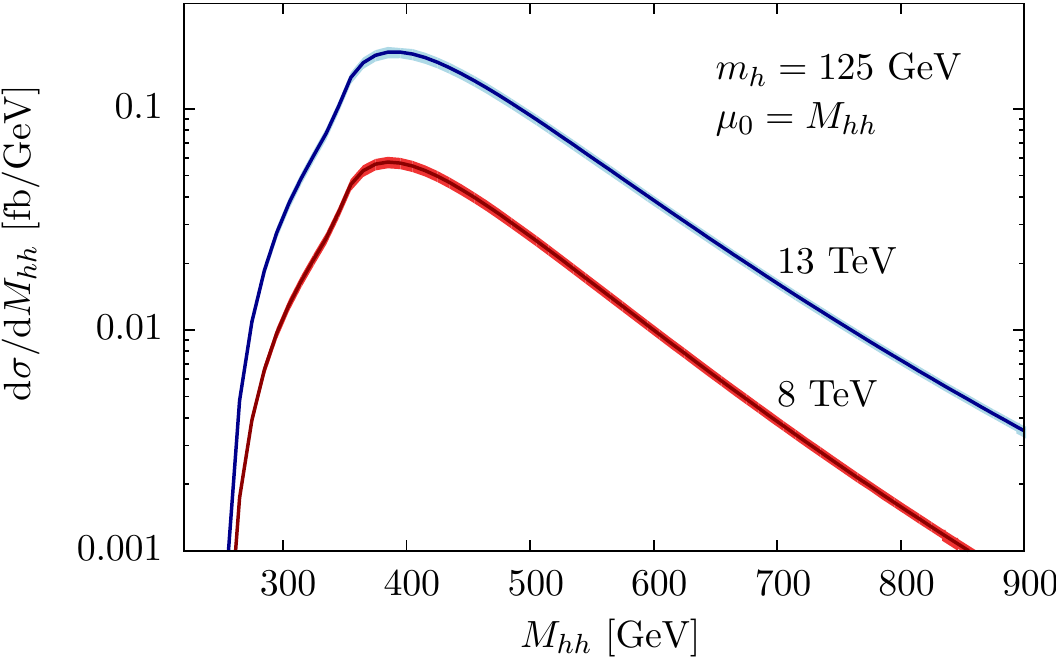}
  \hfill
  \includegraphics[width=0.49\textwidth]{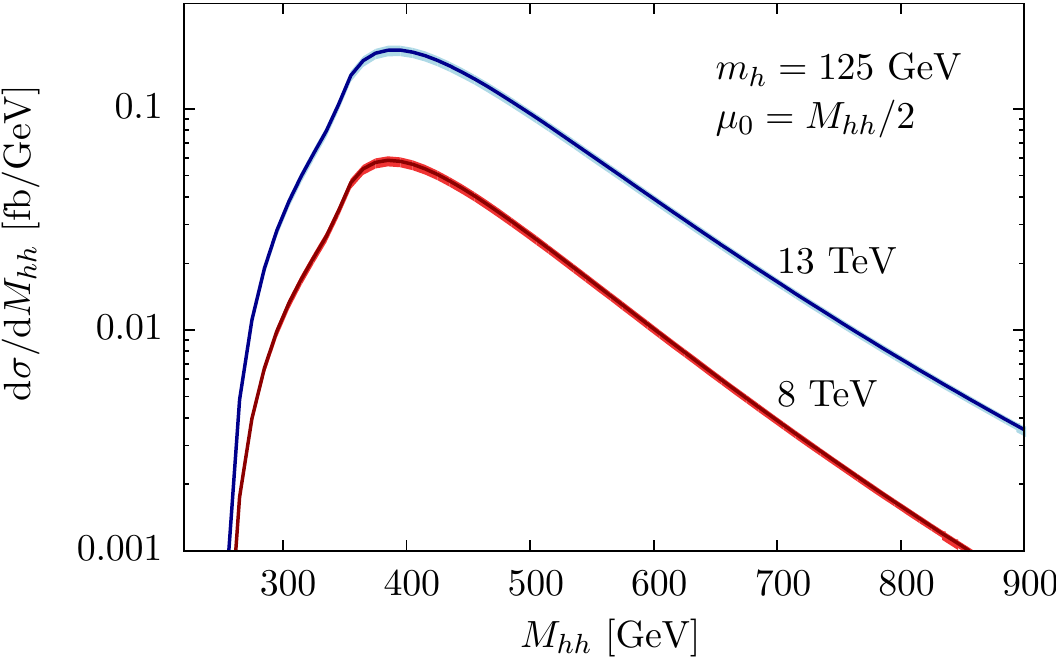}
  \caption{\label{fig:nnlodiff} $\sqrts=8~\tev$ and $13~\tev$
    NNLO+NNLL cross section distribution for $\mt\to \infty$
    calculation detailed in the previous section, also showing the
    scale uncertainty\cite{deFlorian:2015moa}. For details see text.}
\end{figure}

\section{Differential distributions}
\label{sec:hhdif}
The di-Higgs differential mass distribution of the NNLO+NNLL
calculation detailed in the previous section is shown in
\refF{fig:nnlodiff}.  This figure includes the higher order corrections
in the $\mt\rightarrow\infty$ limit.
Due to the intricate destructive interplay of the trilinear and box
contributions depicted in \refF{fig:feyndiag}, however,  the top mass
threshold significantly impacts the differential distributions for the
gluon fusion process, and the invariant di-Higgs boson mass differential cross section in particular.  
On the one hand, the momentum-dependent distributions of the di-Higgs system 
are exploited in phenomenological analyses (either implicitly
or explicitly), as they exhibit a highly sensitive response to
BSM-induced modifications of the SM coupling pattern (see below).  On
the other hand, experimental characteristics of a particular set of
selection cuts motivated from the desire to enhance signal over
background strongly depend on the transverse Higgs momentum (and
therefore on $\mhh$) selection thresholds; 
boosted Higgs kinematics~\cite{Dolan:2012rv,Gouzevitch:2013qca,Papaefstathiou:2012qe,Behr:2015oqq} are a
particularly drastic example of this. Both the theoretical appeal and
the experimental necessity of studying non-inclusive fiducial cross
sections have far-reaching consequences for di-Higgs analyses when we
extrapolate the findings of the previous section to realistic
selection criteria.

%
  \begin{figure}[!p]
    \centering
    	\includegraphics[width=0.50\textwidth]{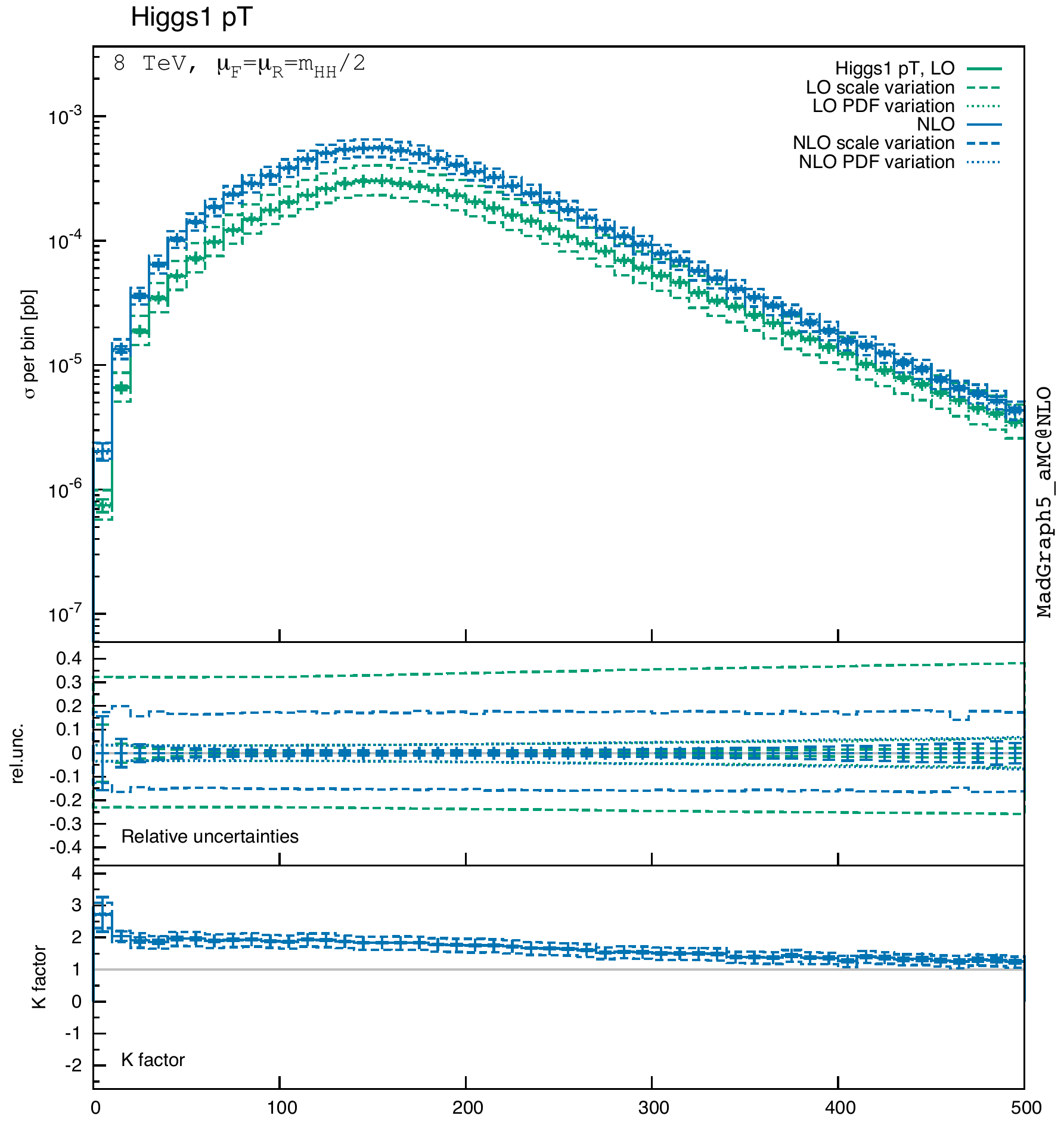}
	\caption{\label{fig:diffxsec1a}    
        Transverse  momentum distribution of the leading Higgs boson in \gev\ for $pp\to \hsm \hsm$
        using {\tt MG5\_aMC@NLO + Pythia8} at NLO with the {{\FTapp}}~approximation, for $\mhsm=125~\gev$, $\mt=172.5~\gev$, and
        $\sqrts =8~\tev$. The scales are chosen to be
        $\mu_R=\mu_F=\mhh/2$.  Scale and PDF uncertainties are added
        linearly in the distribution. The $K-$factor is defined as the ratio between NLO and LO cross sections. } 
    	\includegraphics[width=0.50\textwidth]{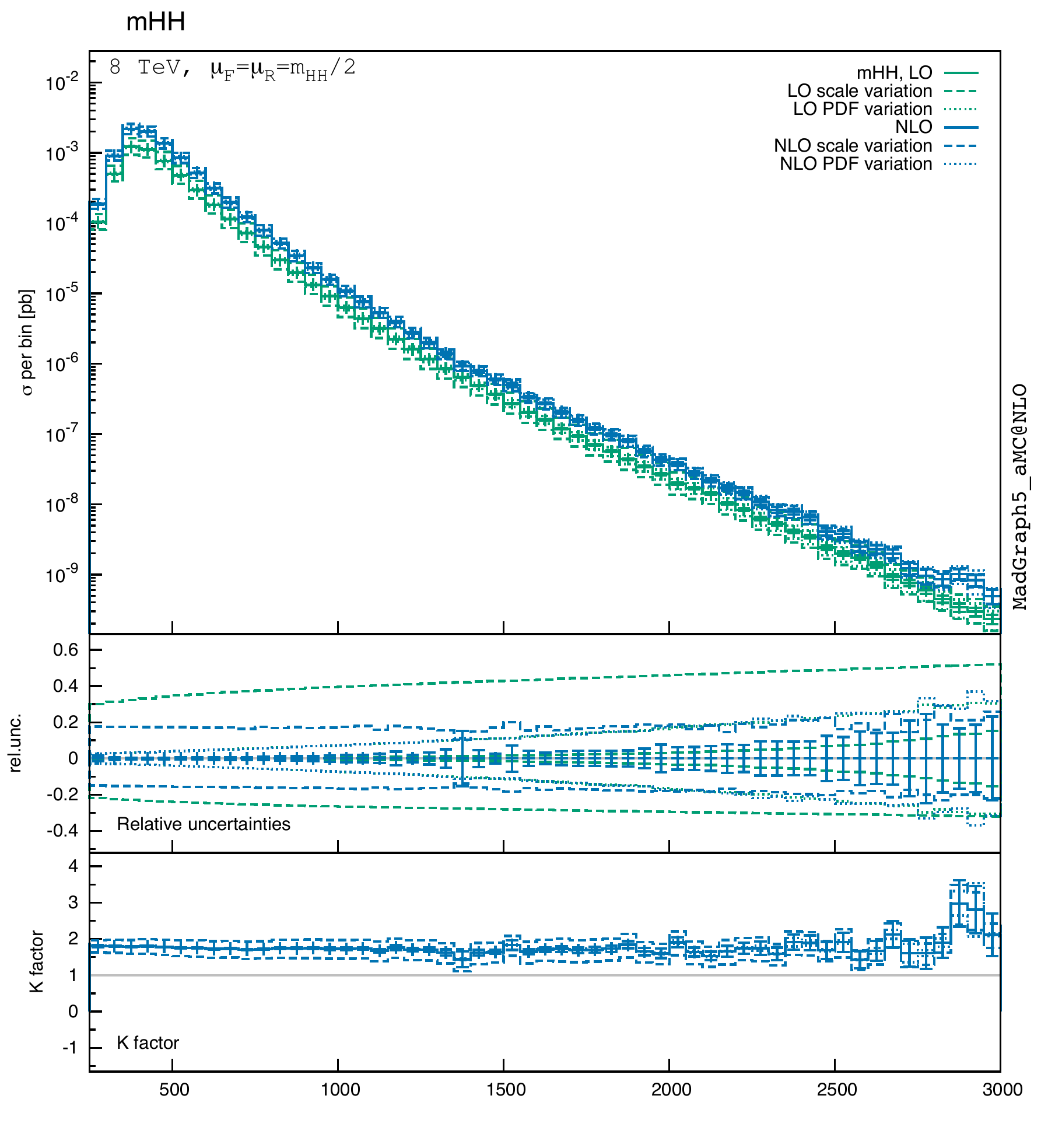} 
	\caption{\label{fig:diffxsec1b}    
        Invariant di-Higgs boson mass differential distribution in \gev\ for $pp\to
        \hsm \hsm$ using {\tt MG5\_aMC@NLO + Pythia8} at NLO
        with the {{\FTapp}}~approximation, for $\mhsm=125~\gev$,
        $\mt=172.5~\gev$, and $\sqrts =8~\tev$. The scales are chosen
        to be $\mu_R=\mu_F=\mhh/2$.  Scale and PDF uncertainties are
        added linearly in the distribution. The $K-$factor is defined as the ratio between NLO and LO cross sections. } 
  \end{figure}
  \begin{figure}[!p]
    \centering
	\includegraphics[width=0.50\textwidth]{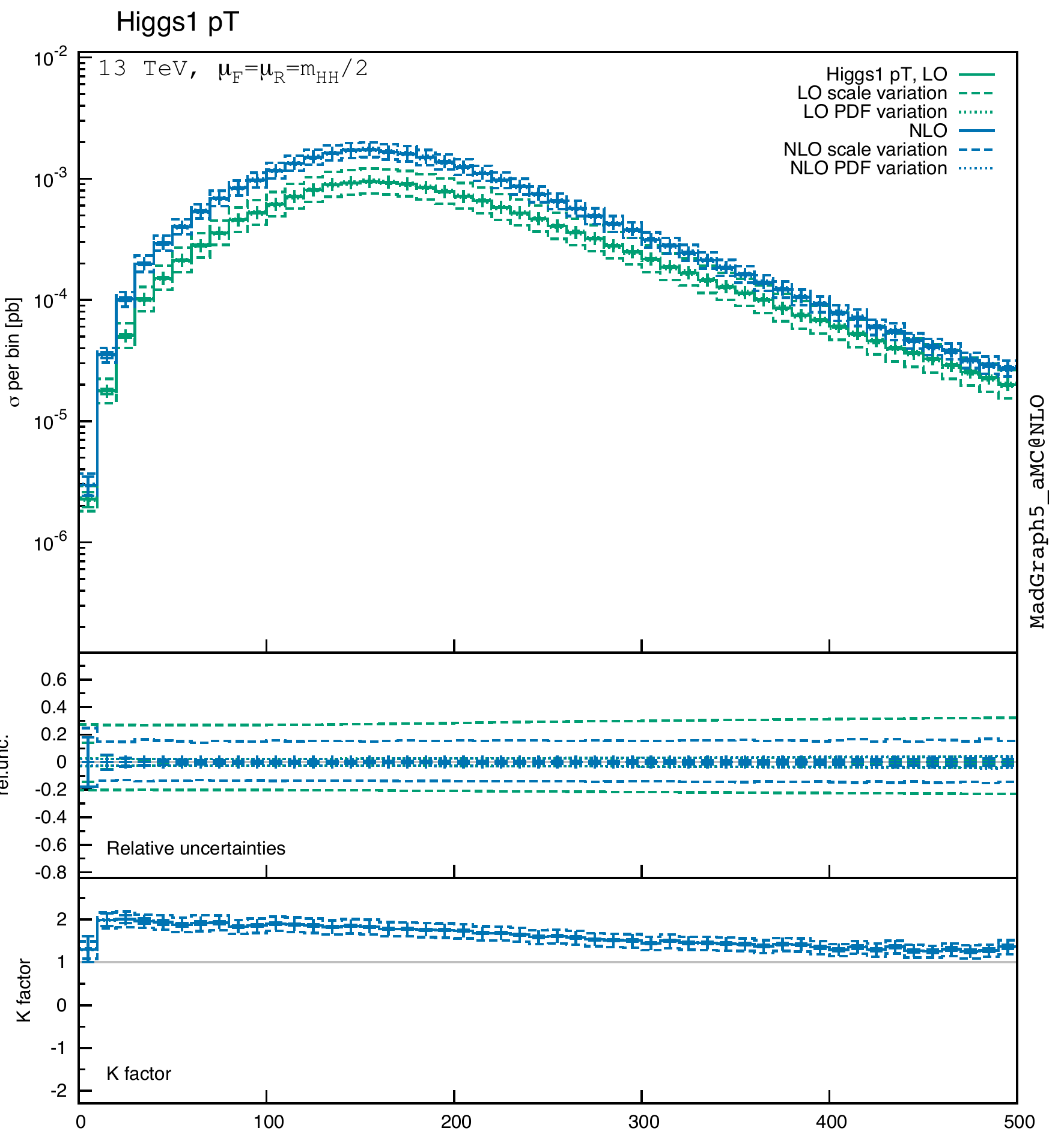}
	\caption{\label{fig:diffxsec2a}
        Transverse momentum  distribution of the leading Higgs boson in \gev\ for $pp\to \hsm \hsm$
        using {\tt MG5\_aMC@NLO + Pythia8} at NLO with the {{\FTapp}}~approximation, for $\mhsm=125~\gev$, $\mt=172.5~\gev$, and
        $\sqrts =13~\tev$. The scales are chosen to be
        $\mu_R=\mu_F=\mhh/2$.  Scale and PDF uncertainties are added
        linearly in the distribution.  The $K-$factor is defined as the ratio between NLO and LO cross sections.} 
	\includegraphics[width=0.50\textwidth]{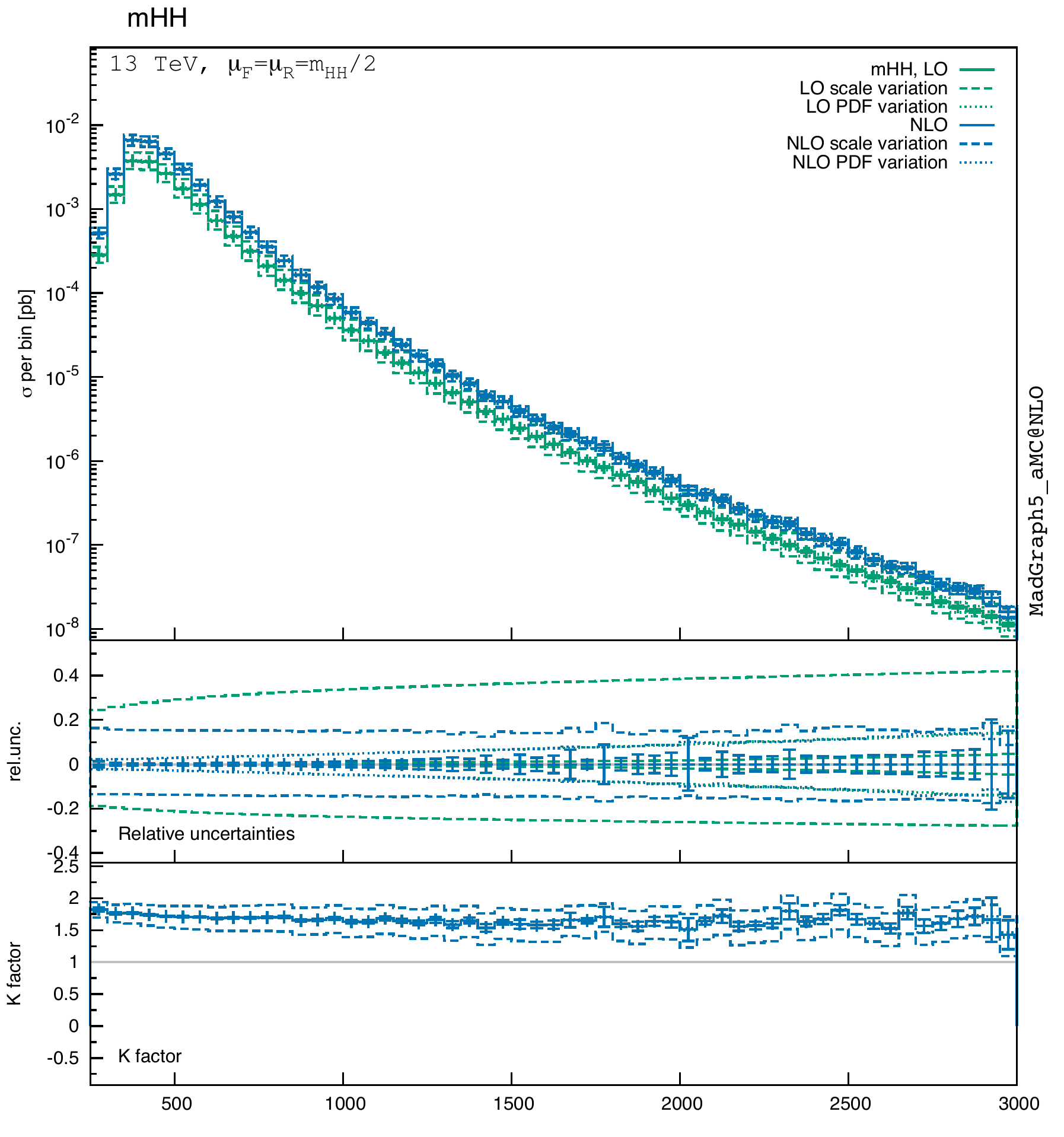}
	\caption{\label{fig:diffxsec2b}
        Invariant di-Higgs boson mass differential distribution in \gev\ for $pp\to
        \hsm \hsm$ using {\tt MG5\_aMC@NLO + Pythia8} at NLO
        with the {{\FTapp}}~approximation, for $\mhsm=125~\gev$,
        $\mt=172.5~\gev$, and $\sqrts =13~\tev$. The scales are chosen
        to be $\mu_R=\mu_F=\mhh/2$.  Scale and PDF uncertainties are
        added linearly in the distribution.  The $K-$factor is defined as the ratio between NLO and LO cross sections.} 
  \end{figure}

In this section we present some distributions obtained using the {{\FTapp}}~approximation for the NLO results to establish to which extent
approximate NLO event generators can be used, and to which extent this
calculation provides a guideline to relate fiducial cross sections to
inclusively modeled quantities. We also comment on the impact of
higher order corrections beyond the \FTapp~approximation that have
become available \cite{Borowka:2016ehy,Borowka:2016ypz} while this report was
completed.  Technically, the  {{\FTapp}} calculation is
performed using matched {\tt MG5\_aMC@NLO + Pythia8}
simulations~\cite{Alwall:2014hca,Frederix:2014hta}. As already mentioned, this
approximation contains exact, full $\mt$-dependent real emission
contributions combined with a finite $\mt$ Born-reweighted
$\mt\to\infty$ calculation of the virtual loop corrections to obtain
an approximation of the fully differential NLO cross section.

\begin{figure}[t]
\centering
\includegraphics[width=0.50\textwidth]{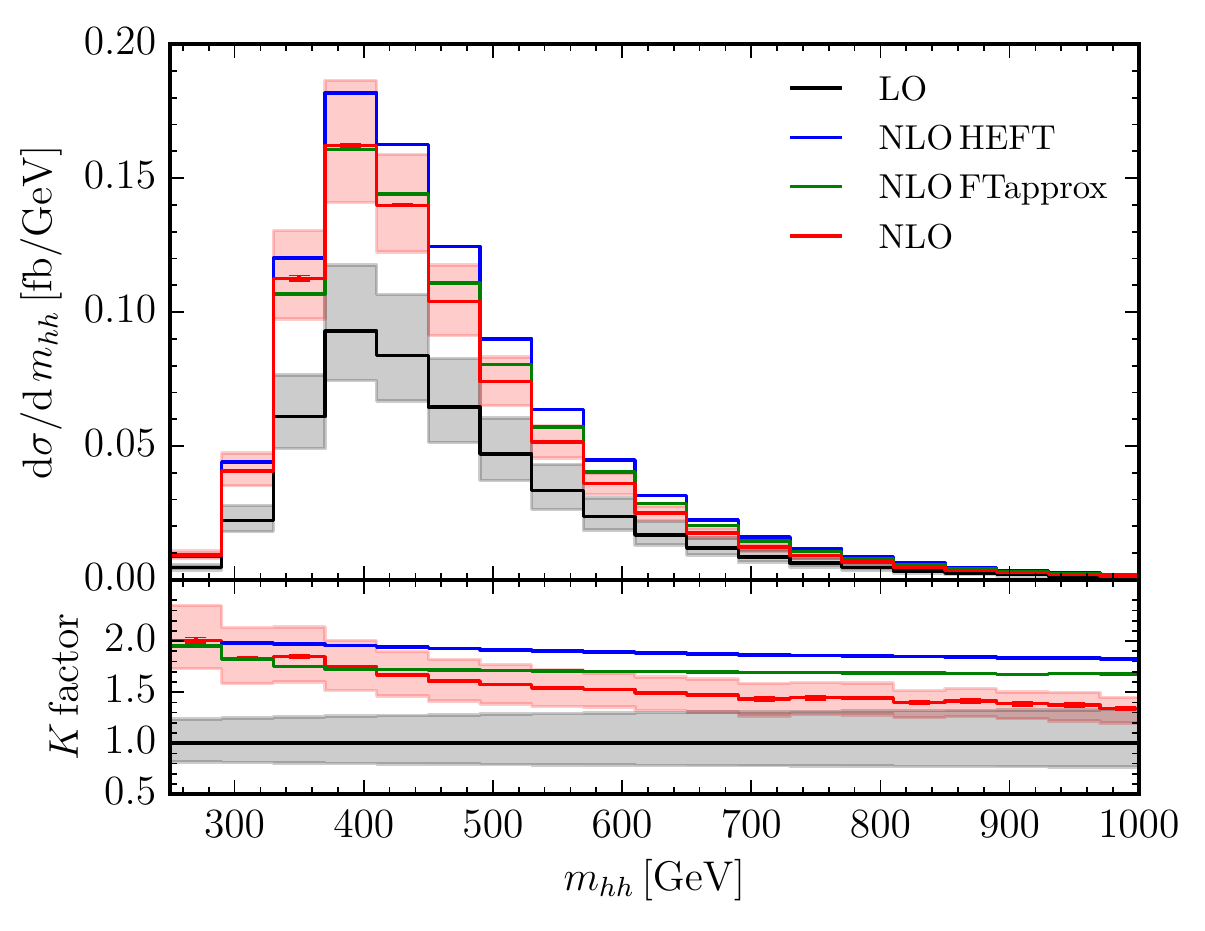}
\caption{\label{fig:mhhfull} Invariant di-Higgs boson mass distribution for
  various approximations, taken from~\cite{Borowka:2016ehy}. The red curves include the complete
  $\mt$ dependent NLO calculation. The uncertainty is computed by varying the scales by a factor 2 around $\mhh/2$. The Higgs boson mass is chosen to be $\mhsm=125~\gev$ and $\mt=173~\gev$, the centre of mass energy is $\sqrts=14~\tev$.}
\end{figure}

\begin{figure}[!p]
    \centering
     	\includegraphics[width=0.50\textwidth]{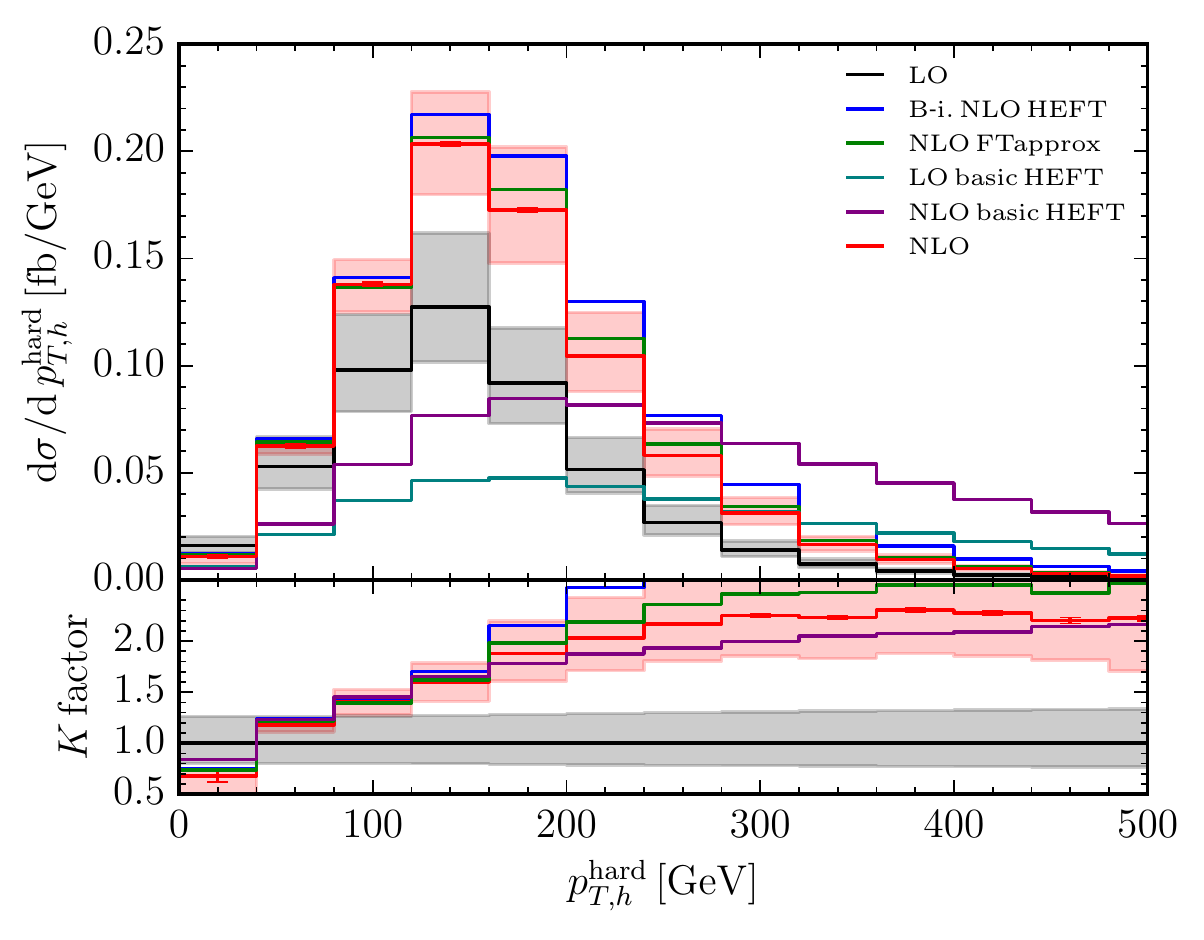}
     	\caption{\label{fig:pthardfull14}   
        Transverse momentum  distribution of the leading Higgs boson in \gev\ for $pp\to \hsm \hsm$ 
         at NLO with the full top mass dependence, taken from~\cite{Borowka:2016ypz}, for $\mhsm=125~\gev$, $\mt=173~\gev$, and
        $\sqrts =14~\tev$. The scales are chosen to be
        $\mu_R=\mu_F=\mhh/2$.} 
   	\includegraphics[width=0.50\textwidth]{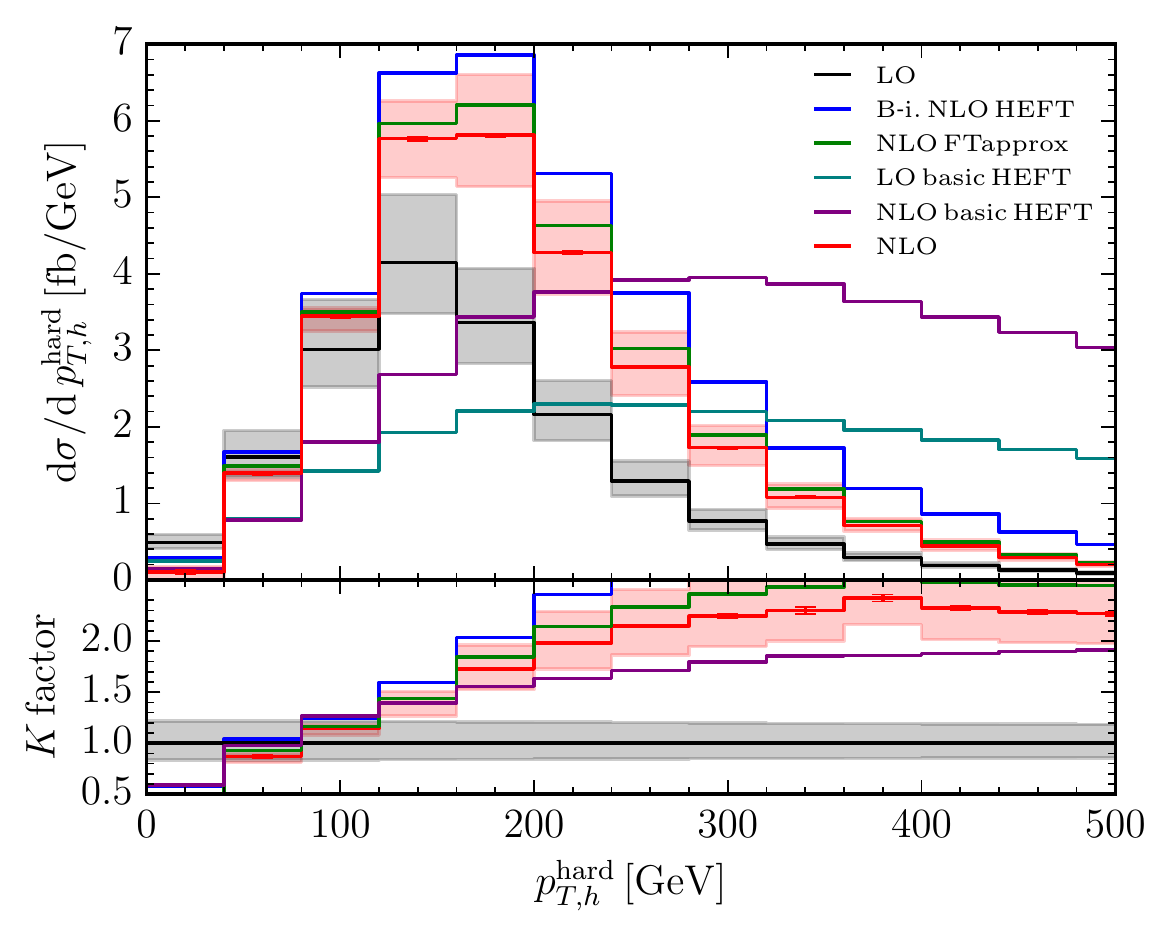}
	\caption{\label{fig:pthardfull100}    
         Transverse momentum  distribution of the leading Higgs boson in \gev\ for $pp\to \hsm \hsm$ 
         at NLO with the full top mass dependence, taken from~\cite{Borowka:2016ypz}, for $\mhsm=125~\gev$, $\mt=173~\gev$, and
        $\sqrts =100~\tev$. The scales are chosen to be
        $\mu_R=\mu_F=\mhh/2$.} 
  \end{figure}

We show in Figs.~\ref{fig:diffxsec1a} - \ref{fig:diffxsec2b} the \FTapp~distributions for the $\hsm\hsm$ invariant mass and the leading
Higgs boson $p_T$, including PDF and scale uncertainties for centre-of-mass energies of
$8~\tev$ and $13~\tev$. These distributions also rely on the
{\tt PDF4LHC15\_nlo\_mc} sets with 30 replicas. The scale variation dominates over the PDF
uncertainty leading to a rather flat $+30\%,~ -25\%$ uncertainty over a
broad, phenomenologically interesting energy regime, calculated from a
scale variation that is again $\mu_0/2<\mu<2 \mu_0$, for
$\mu_0=\mhh/2$. 
The scale uncertainty is similar when considering the full NLO QCD corrections, see Figs.~\ref{fig:mhhfull} - \ref{fig:pthardfull100}. 

For some observables, for example all rapidity distributions, the differential QCD
corrections are simple rescalings of the LO distribution with the
total K factor in the \FTapp~scheme. 
For most distributions, however, the discrepancy between  the \FTapp~and the full result moves out of the scale variation uncertainty band in the tail of the distributions.
This is shown for the di-Higgs boson invariant mass $\mhh$-distribution in \refF{fig:mhhfull} and for the
leading Higgs boson transverse momentum $p_{T,h_1}$ in Figures~\ref{fig:pthardfull14} and \ref{fig:pthardfull100}.

\begin{figure}[t]
\centering
\includegraphics[width=0.49\textwidth]{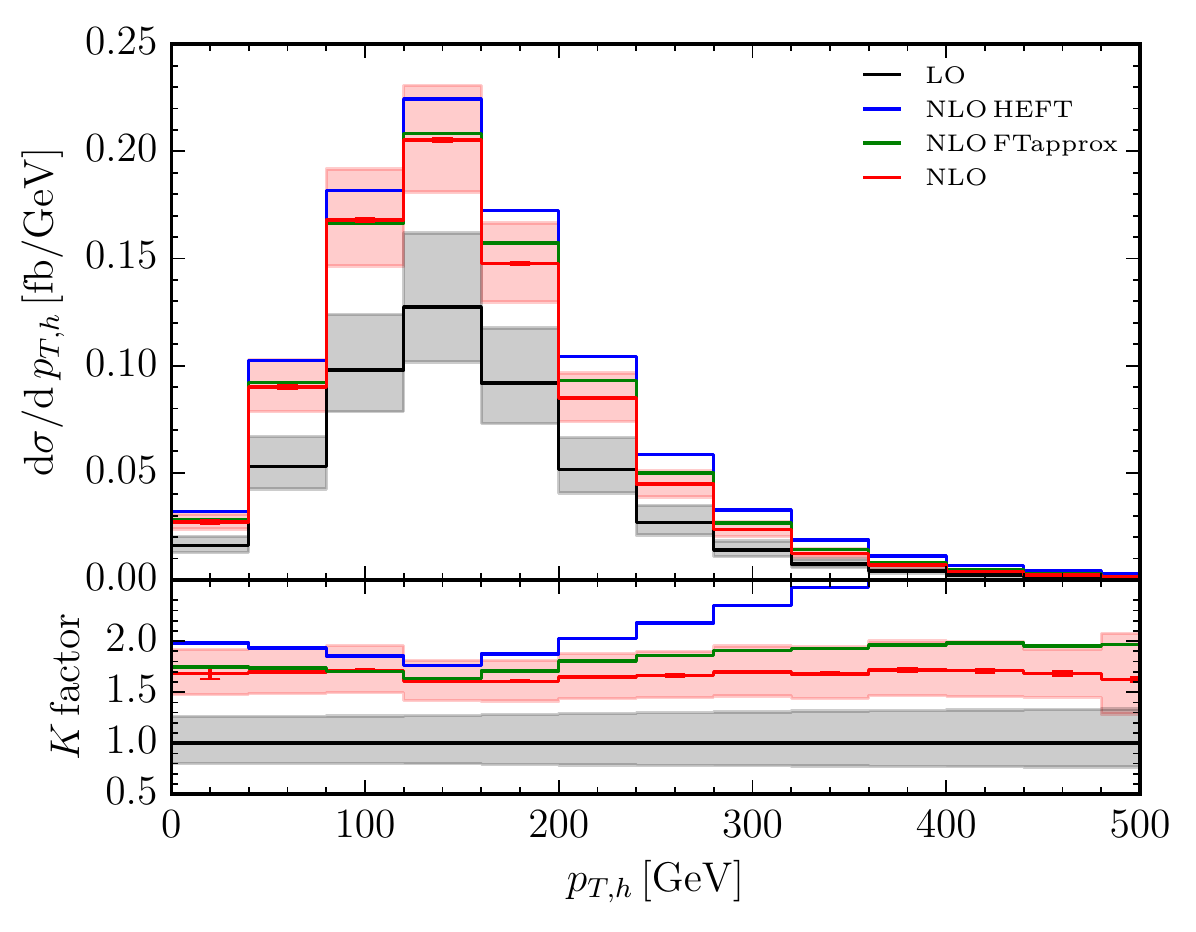}
\caption{\label{fig:pthfull} $p_{T,h}$  distribution for
  various approximations, taken from~\cite{Borowka:2016ehy}. The red curves include the complete
  $\mt$ dependent NLO calculation. The uncertainty is computed by varying the scales by a factor 2 around $\mhh/2$. The Higgs boson mass is chosen to be $\mhsm=125~\gev$ and $\mt=173~\gev$, the centre of mass energy is $\sqrts=14~\tev$.}
\end{figure}

While the \FTapp~scheme provides a reasonable approximation in the low energy regime within uncertainties, the virtual corrections in the finite $m_t$ limit significantly soften the high energy events compared to the \FTapp~approximation, see \refF{fig:mhhfull}. These corrections can be as big as 30\% and a corresponding reweighting or associated uncertainty needs to be included in analyses that particularly focus on the phase space region $m_{hh}\gtrsim 400~\text{GeV}$.
The full NLO QCD corrections have a significant effect on the $p_T$ distribution (both the distribution of the leading-$p_T$ Higgs boson, $p_{T,h_1}$,  
and the one of any Higgs boson, $p_{T,h}$), as can be seen in 
Figs.~\ref{fig:pthardfull14} - \ref{fig:pthfull}.  
Both the HEFT and \FTapp  approximation fail to reproduce the distribution above the top mass. 

The leading jet transverse momentum and the transverse momentum of
the di-Higgs system are also  not well approximated
by a rescaling of the LO results.  It should be noted that the leading jet transverse momentum distribution is only
known to leading order precision and should be considered with care by
including LO uncertainties. For the $p_{T}(hh)$ distribution, resummed results at NLL+NLO, including the full NLO mass dependence in the $p_T(hh)\to 0$ region, 
have been obtained recently~\cite{Ferrera:2016prr}.

Other uncertainties of equal importance become particularly apparent from investigating different
shower and matching/merging approaches that we detail in the
following. The comparison we detail in the following is based on the MC@NLO
calculation that appears in~\cite{Maltoni:2014eza} generated using
\texttt{MG5\_aMC@NLO}~\cite{Alwall:2011uj, Alwall:2014hca}
and the merged calculation of~\cite{Maierhofer:2013sha} using
\texttt{OpenLoops} matrix elements~\cite{Cascioli:2011va}. All the
results use the \texttt{HERWIG++} general-purpose Monte Carlo for the
parton shower~\cite{Bahr:2008pv, Gieseke:2011na,
  Arnold:2012fq,Bellm:2013hwb,Bellm:2015jjp}. The Higgs bosons are
taken to be stable and hadronization and the underlying event
simulation is turned off. The central
factorization/renormalization scale for both calculations is chosen
to be $\mu_0 = \mhh/2$. This scale is varied between $\mu = \mu_0/2$ and
$\mu = 2 \mu_0$ in the merged calculation. Furthermore, the merging
scales used for the MLM procedure were varied in the range 40-90~GeV in steps of 10~GeV and a
uniform smooth function, as described in~\cite{Maierhofer:2013sha}, of
widths $\epsilon = 10, 20, 30$~GeV is used.

\begin{figure}[t]
  \centering
  \parbox{0.49\textwidth}{\includegraphics[width=0.49\textwidth]{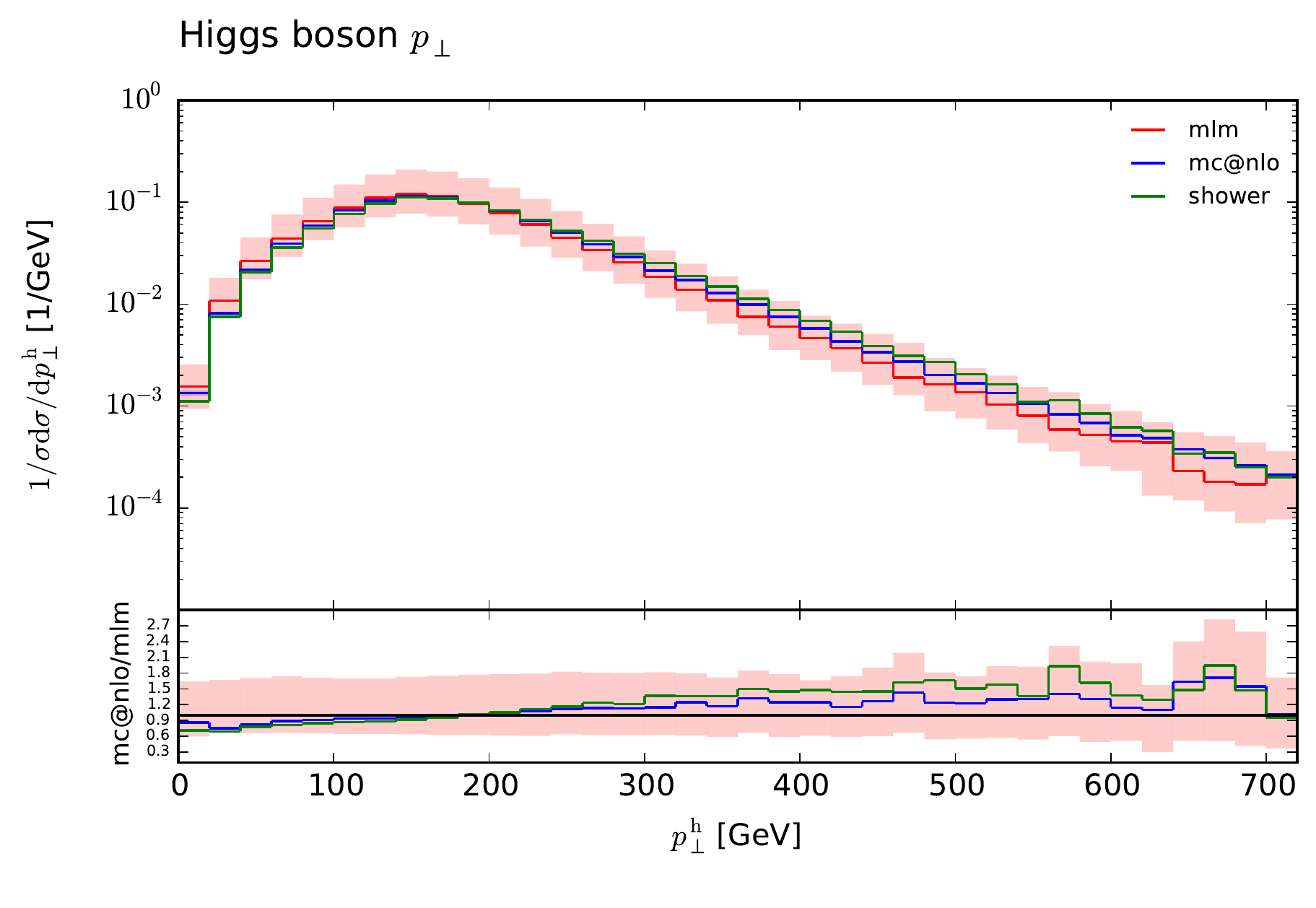}}\hfill
  \parbox{0.49\textwidth}{\includegraphics[width=0.49\textwidth]{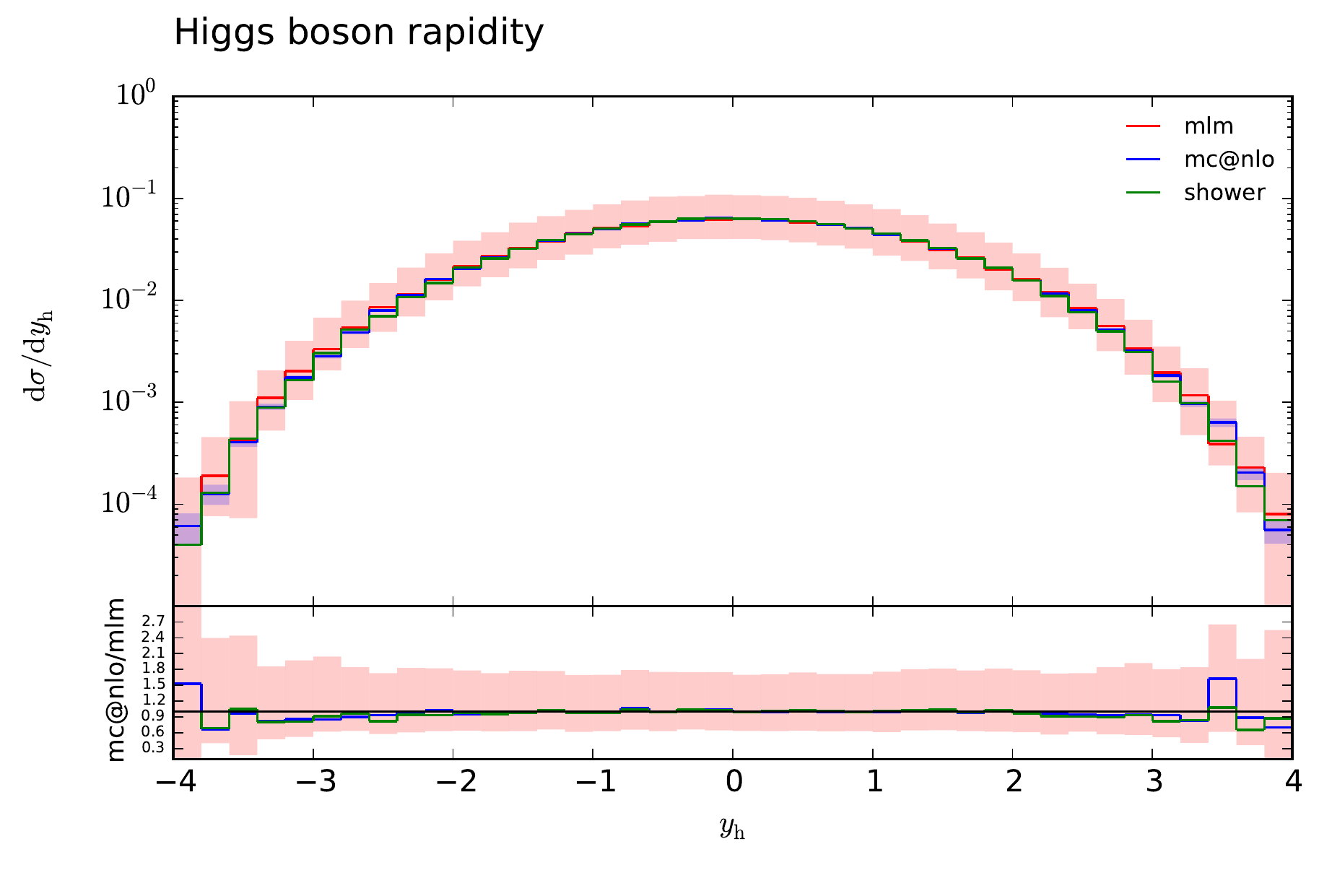}}\\
  \parbox{0.49\textwidth}{\includegraphics[width=0.49\textwidth]{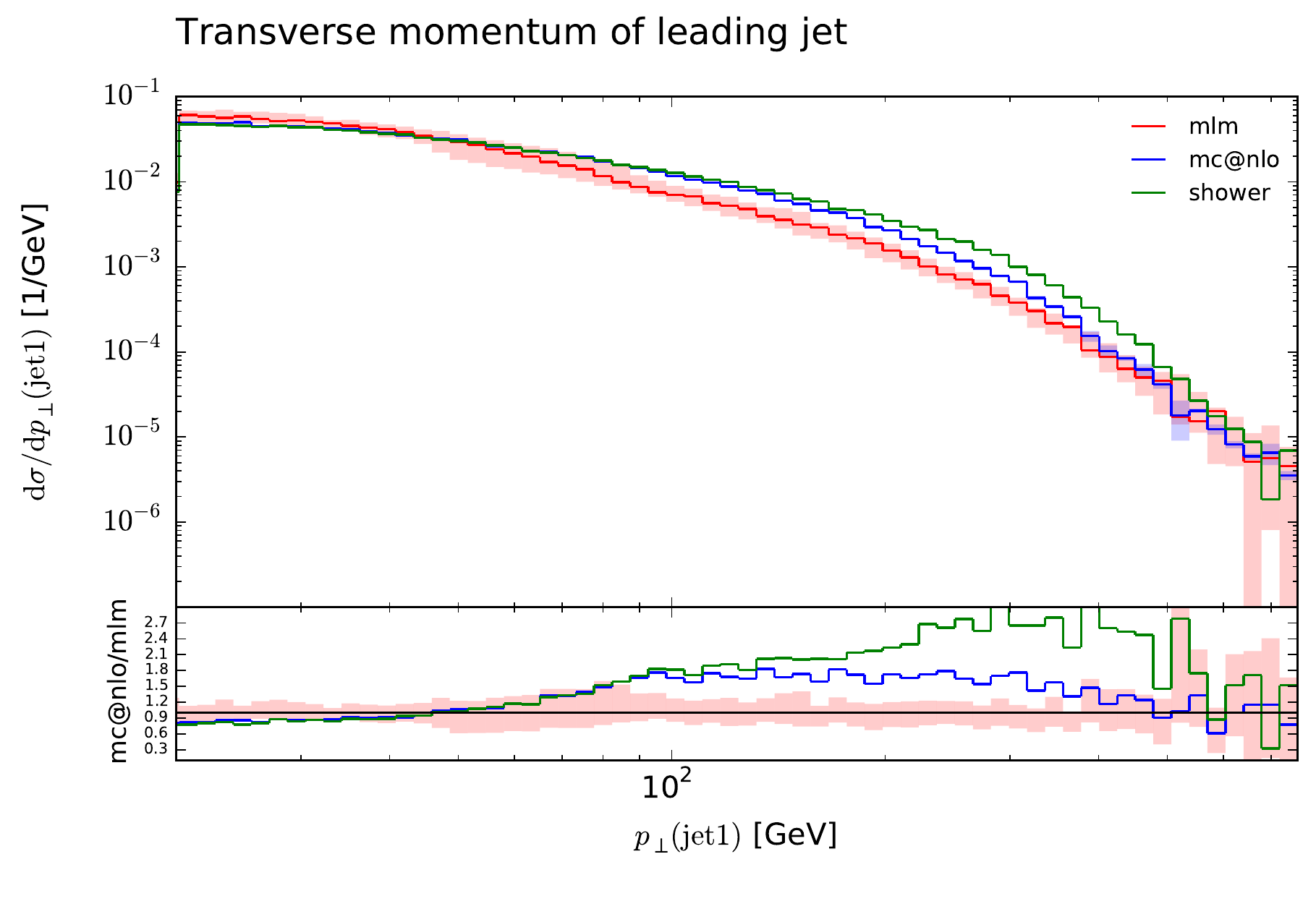}}\hfill
  \parbox{0.45\textwidth}{\vspace{0.6cm}\caption{{{Comparisons of
      distributions of observables in di-Higgs boson production. We
      show the transverse momentum of (any) Higgs boson, $p_{T,h}$, in
      the top left figure, the rapidity of (any) Higgs boson, $y_h$,
      in the top right figure and the transverse momentum of the hardest
      jet $p_T(\text{jet}_1)$, in the
      figure at the bottom. The uncertainty band corresponds to the scale uncertainties described above.}}\label{fig:dihiggs1}}}\\[0.2cm]
\end{figure}
\begin{figure}[htb]
  \centering
  \parbox{0.49\textwidth}{\includegraphics[width=0.49\textwidth]{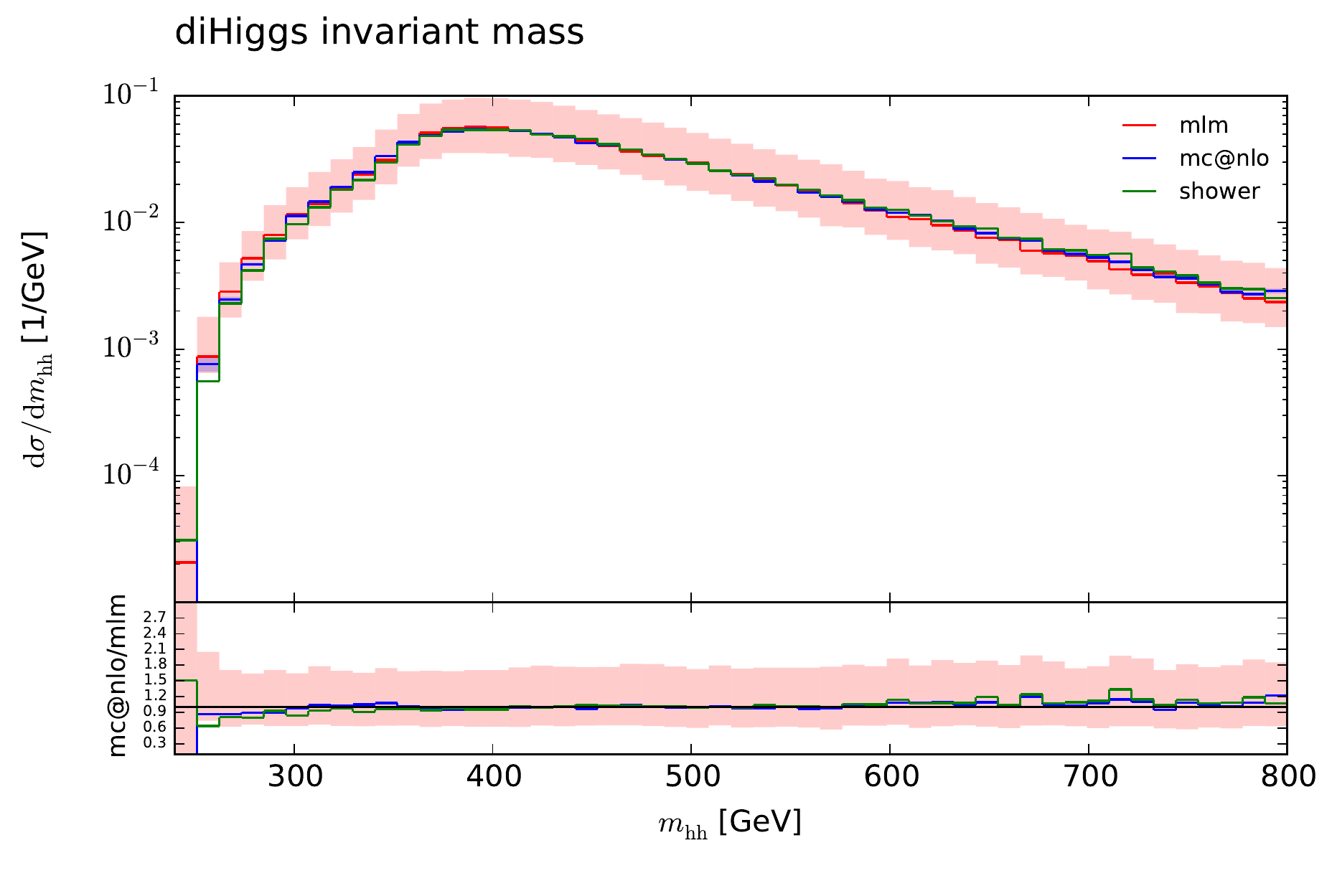}}\hfill
  \parbox{0.49\textwidth}{\includegraphics[width=0.49\textwidth]{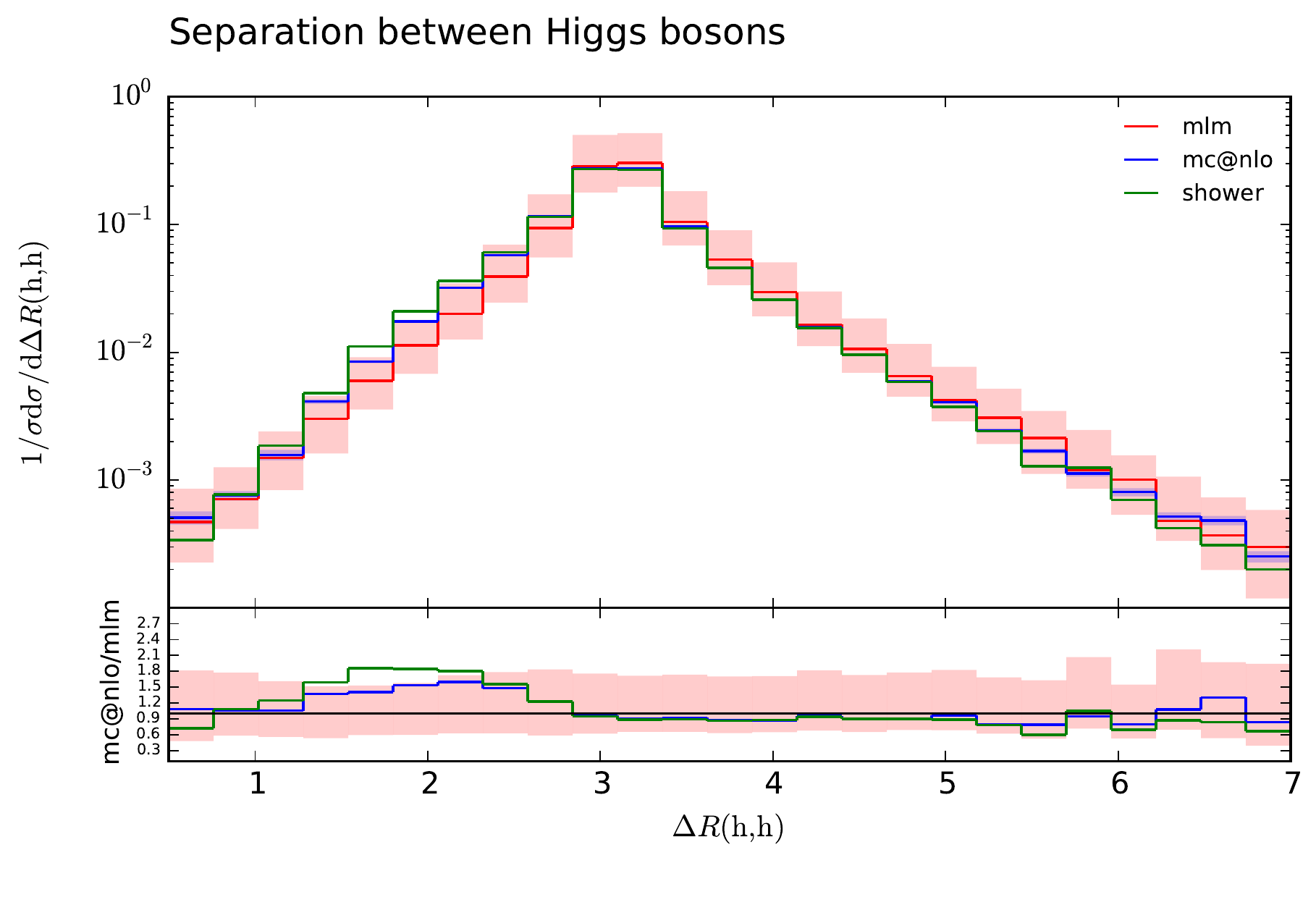}}
  \parbox{0.49\textwidth}{\includegraphics[width=0.49\textwidth]{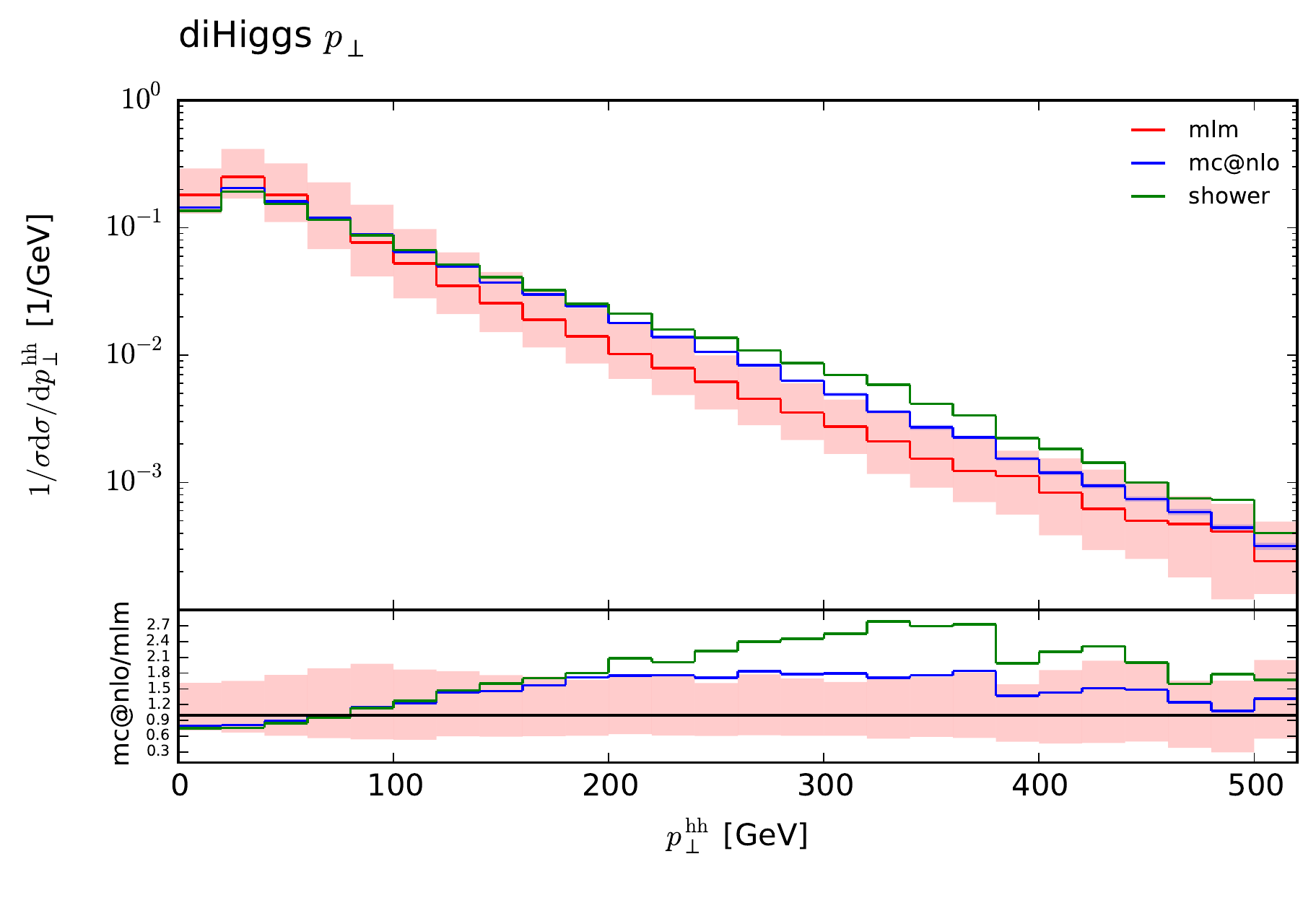}}\hfill
  \parbox{0.45\textwidth}{\vspace{0.6cm}\caption{{{Comparisons of
      distributions of di-Higgs observables. We show the invariant
      mass of the di-Higgs system, $\mhh$, in the top left figure, the
      separation between the two Higgs bosons, $\Delta R(h,h)$, in the
      top right figure and the transverse momentum of the di-Higgs system, 
      $p_T({hh})$ in the lower
      figure. The uncertainty band corresponds to the scale uncertainties described above.}}\label{fig:dihiggs2}}}\\[0.6cm]
\end{figure} 

\begin{figure}[h!tb]
\centering
 \centering
  \includegraphics[width=0.6\textwidth]{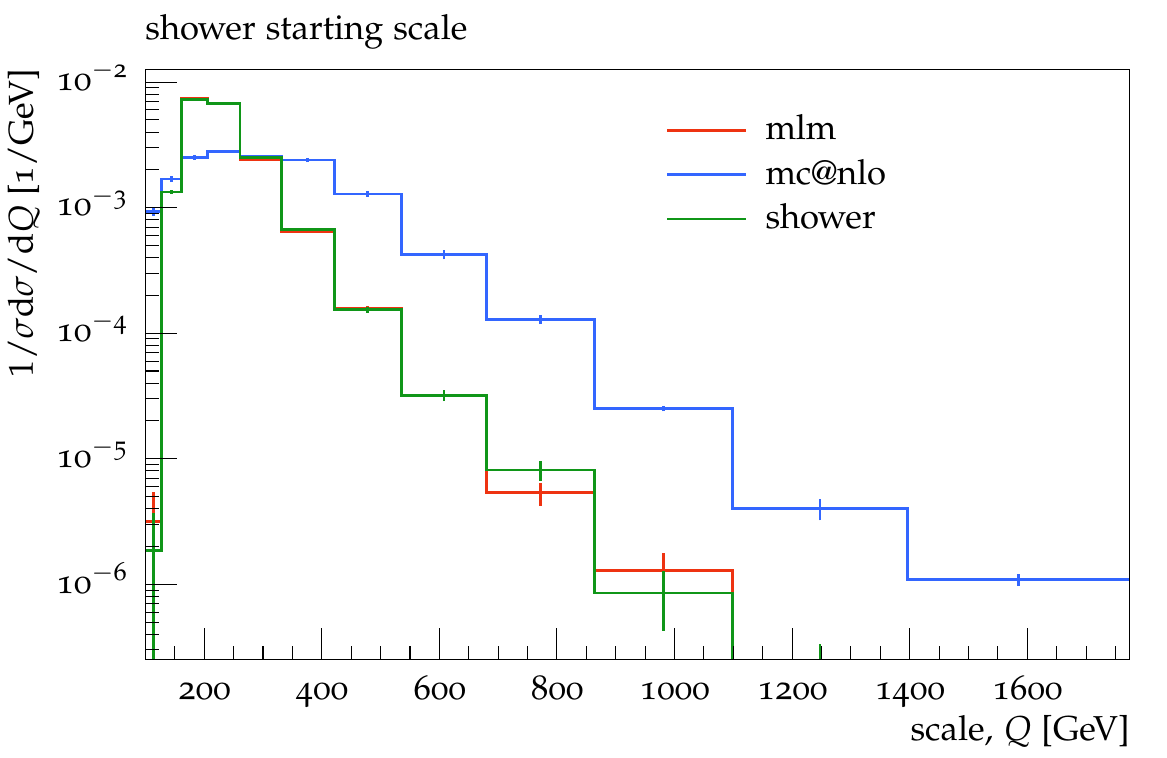}
  \caption{The dynamical starting scale for the \texttt{HERWIG++} shower used in
    each of the calculations. }\label{fig:scales}
\end{figure}

We present comparisons of the MC@NLO samples, labelled `mc@nlo', the
merged calculation, labelled `mlm' and the pure leading-order
calculation, labelled `shower', all showered with \texttt{HERWIG++} in
Figs.~\ref{fig:dihiggs1} and~\ref{fig:dihiggs2}. We show,
respectively, distributions for the transverse momentum and rapidity
of (any) single Higgs boson, the transverse momentum of the hardest
jet, the invariant mass of the di-Higgs boson pair, the di-Higgs boson transverse
momentum and the separation between the Higgs bosons. All
distributions are normalized to unity. The di-Higgs boson invariant
mass and separation between the Higgs bosons, as well as the single
Higgs observables, are in good agreement within the uncertainties of
the merged calculation {{indicated by the band}}. 
There exists a discrepancy in the di-Higgs
transverse momentum and the transverse momentum of the hardest jet.
One of the reasons that could explain this effect is the difference in the choice
of the dynamical starting scale of the shower $Q$.
\refF{fig:scales} provides the differential distribution of the $\hsm\hsm$ production cross section
as function of $Q$. We can clearly see that the choice of the scale in the MC@NLO sample
differs from MLM sample. 
This observation suggests that further assessment of the
systematics owing to the parton shower should be performed, a task left
to future studies.

\section{Benchmark BSM scenarios}
\label{sec:hhbsm}

In this section we propose BSM models recommended for study in
di-Higgs boson production, assuming the two final state Higgs bosons are
SM-like. We do not consider the production of two different Higgs particles, 
such as $AH$, since these cross sections are highly
model-dependent~\cite{Dolan:2012ac,Baglio:2014nea,Hespel:2014sla} and typically
suppressed compared to the production of the SM-like Higgs, making generic predictions
difficult.

The benchmarks have been chosen based on the following criteria:
\begin{itemize}
\item They can by directly related to other (e.g. single Higgs)
  measurements.
\item Their signatures cover resonant and non-resonant production modes and
  in the former case are distinguishable from each other and from the
  SM.
\end{itemize}

We will adopt the Effective Field Theory framework for non-resonant
production and advocate the singlet-extended SM  and 2HDM for initial studies of resonant
di-Higgs final states.

\subsection{Effective Field Theory}
\label{sec:hheft}

If no new light particles are observed at the LHC and there are no
observable resonances in di-Higgs boson production, then an effective field
theory approach is useful and well-motivated.  The phenomenologically
relevant terms we consider for double Higgs boson production in $gg$ fusion  are (written
in terms of the physical Higgs boson field, $h$)\cite{Falkowski:2015fla}
\begin{eqnarray}
  L& \overset{LO}{=} &L_{SM}+\biggl(c_{g}{\hsm\over v}+c_{gg}{\hsm^2
    \over 2 v^2}\biggr) {g_s^2\over4}G_{\mu\nu}^AG^{A,\mu\nu}
  -{\hsm\over v}
  \Sigma_f\Sigma_{ij}\sqrt{m_{f_i}m_{f_j}}[\delta y_f]_{ij}{\overline f_i} f_j
  \nonumber \\&&-
  {\hsm^2\over 2 v^2}\Sigma_f \Sigma_{ij}\sqrt{m_{f_i}m_{f_j}}[ y_f^{(2)}]_{ij}\biggl[ {\overline f}_{iR} f_{jL}+hc\biggr]+\delta \lambda_3 \hsm^3
  +L_{CPV}\nonumber \\
  L_{CPV}&=&\biggl({\tilde c}_{g}{\hsm\over v}+{\tilde c}_{gg}{h^2\over 2 v^2}\biggr) {g_s^2\over4}G_{\mu\nu}^A{\tilde G}^{A,\mu\nu}
  \nonumber \\ &&
  -{\hsm\over v}
  \Sigma_f\Sigma_{ij}\sqrt{m_{f_i}m_{f_j}}[\delta y_f]_{ij}\biggl\{\biggl(\cos \phi^f_{ij}-1\biggr){\overline f_i} f_j
  -i\sin \phi^f_{ij} {\overline f_i}\gamma_5 f_j\biggr\}\, ,
  \label{eftlag}
\end{eqnarray}
where $i,j$ are generation indices, and the sum over $f$ is over all
charged fermion species. A hierarchy between CP-violating and CP-conserving
interactions has already been established~\cite{Contino:2013kra,Dumont:2013wma} and
for simplicity we assume CP conservation and flavour diagonal Higgs boson couplings, 
leading to the Lagrangian relevant for di-Higgs boson production, 
\begin{eqnarray}
  L&=&L_{SM}+\biggl(c_{g}{h\over v}+c_{gg}{h^2\over 2 v^2}\biggr) {g_s^2\over4}G_{\mu\nu}^AG^{A,\mu\nu}-{h\over v}
  \Sigma_f\Sigma_i m_{f_i}[\delta y_f]_{i}{\overline f_i} f_i
  \nonumber \\&&-
  {h^2\over 2 v^2}\Sigma_f \Sigma_{i}m_{f_i}[ y_f^{(2)}]_{i} {\overline f}_{i} f_i+\delta \lambda_3 h^3\, .
  \label{eq:eftwg2}
\end{eqnarray}
We do not consider enhanced $b$ couplings, since in order to be
relevant for double Higgs boson production, the enhancement
must be extremely large\footnote{See for example, Figure 6 in
  Ref.~\cite{Dawson:2012mk}.}.

The inputs in this realization of the EFT can thus be taken in general as,
\begin{itemize}
\item $c_g$, $c_{gg}$, $\delta y_t$, $y_t^{(2)}$, $\delta
  \lambda_3$\hfill{\hbox{Non-Linear EFT}}
\end{itemize}
which can be reduced to
\begin{itemize}
\item $c_g$, $\delta y_t$, $\delta \lambda_3$, $y_t^{(2)}$\hfill{\hbox{Linear EFT}}
\end{itemize}
in the linear realization of EWSB, where the Higgs boson is part of a weak doublet,
leading to correlations between the couplings.

Note that a combination of $c_{g}$ and $\delta y_t$ is fixed by the
requirement that single Higgs boson production has the experimentally
observed value,
\begin{equation}
  R_\hsm\equiv {\sigma(gg\rightarrow \hsm)\over\sigma(gg\rightarrow \hsm)_{SM}}\sim \left| 1+ 12 \pi^2 c_{g}+\delta y_t\right|^2
  \, .
\end{equation} 
The couplings $c_{gg}$, $y_t^{(2)}$ and $\delta\lambda_3$ cannot be
probed in single Higgs boson production, but require measurement of the
di-Higgs rate and distributions\cite{Goertz:2014qta,Azatov:2015oxa,Gillioz:2012se,Chen:2014xwa}.

A fit to the total cross section in terms of the EFT coefficients shown here has
been given in \refT{tab:coefs} and is
detailed in a separate note~\cite{Carvalho:2016rys}, (a similar procedure is performed in Ref.~\cite{Azatov:2015oxa}). Those references construct
a cross section fit in terms of effective Higgs boson couplings, combining the relevant terms
in~\ref{eq:eftwg2} as:
\begin{multline}
L^{hh}={1\over 2}
\partial_\mu \hsm \partial^\mu \hsm -
{\mhsm^2\over 2}\hsm^2
-\kappa_\lambda \lambda_{SM}v \hsm^3 \\ 
-{\mt\over v} \biggl(v+\kappa_t\hsm+{c_2\over v}\hsm\hsm
\biggr)\biggl({\overline t}_Lt_R+h.c.\biggr)+{\alpha_s\over 12 \pi v} 
\biggl(c_{1g} \hsm-{c_{2g}\over 2 v} \hsm\hsm\biggr)G^{A}_{\mu\nu}G^{A,\mu,\nu}\, . 
\label{eq:hhcluster}
\end{multline}
The SM limit is $\kappa_2=\kappa_\lambda=1$ and $c_2=c_{1g}=c_{2g}=0$.
This fit can be straightforwardly mapped onto the EFT parameters of
Eq.~\eqref{eftlag} via the identities
\begin{equation}
c_g  =  {c_{1g}\over 12\pi^2} \,,\quad
c_{gg}  =  -{c_{2g}\over 12\pi^2} \,,\quad
y_t^{(2)} = 2 c_2 \,,\quad
\delta y_t =( \kappa_t -1)\,,\quad
\delta \lambda_3  =  -v(\kappa_{\lambda}-1) \lambda_{SM}\,.
\end{equation}

Further information on the EFT coefficients can be found from $\hsm\hsm$
production by noting that different EFT operators have different
kinematic dependences. The LO box and triangle diagram exactly cancel
each other at threshold in the SM. This implies that $d\sigma/d\mhh$ is most
sensitive to variations in $\kappa_{t}$ and $\kappa_{\lambda}$ at
threshold, while the dependence on $\kappa_{\lambda}$ is suppressed at
high partonic energies. The NLO corrections to the EFT predictions for
double Higgs boson production have been investigated in the large $\mt$ limit 
Ref.~\cite{Grober:2015cwa}, with the conclusion that the $K$ factor of
the EFT shows little kinematic dependence and little dependence on the effective couplings, 
however with the same caveats as mentioned in Secs.~\ref{sec:hhxsec} and {\ref{sec:hhdif}.

We can take advantage of this property of the K-factors, approximating the ratio
between the cross sections obtained for different EFT parameters and the SM cross
section with the corresponding LO ratio:

\begin{eqnarray}
R_{hh} \equiv {\sigma_{\hsm\hsm}\over \sigma_{\hsm\hsm}^{SM}}&\overset{LO}{=}& A_1\kappa_t^4+A_2c_2^2
+(A_3\kappa_t^2+A_4 c_g^2)\kappa_\lambda^2
+A_5 c_{2g}^2
+(A_6 c_2+A_7\kappa_t\kappa_\lambda)\kappa_t^2\nonumber \\ &&
+(A_8\kappa_t\kappa_\lambda+A_9 c_g \kappa_\lambda)c_2 +A_{10}c_2 c_{2g}
+(A_{11}c_g\kappa_\lambda+A_{12}c_{2g})\kappa_t^2\nonumber \\
&&+(A_{13}\kappa_\lambda c_g+A_{14}c_{2g})\kappa_t\kappa_\lambda+
A_{15}c_g c_{2g}\kappa_\lambda\, .
\label{eq:interpolation}
\end{eqnarray}

The $A_i$ coefficients are extracted from a simultaneous fit, based on the maximization of a likelihood, to the cross sections obtained from a LO simulation and provided in \refT{tab:coefs}. A detailed study of theoretical uncertainties was performed in Ref.~\cite{Carvalho:2016rys}. The uncertainties related to PDF and $\alpha_S$  variations induces less than  a 2\% variation in  the $A_i$ values.
\begin{table}[!t]
\caption{\label{tab:coefs} Values of $A_i$ parameters for Eq. \eqref{eq:interpolation}\cite{Carvalho:2016rys}.}
\centering
\begin{tabular}{rccccc}
\toprule
$\sqrt{s} $ & 7 TeV & 8 TeV & 13 TeV & 14 TeV & 100 TeV\\[1mm]
\midrule
$A_1$  &  2.21  &  2.18  &  2.09  &  2.08  &  1.90  \\
$A_2$  &  9.82  &  9.88  &  10.15  &  10.20  &  11.57  \\
$A_3$  &  0.33  &  0.32  &  0.28  &  0.28  &  0.21  \\
$A_4$  &  0.12  &  0.12  &  0.10  &  0.10  &  0.07  \\
$A_5$  &  1.14  &  1.17  &  1.33  &  1.37  &  3.28  \\
$A_6$  &  -8.77  &  -8.70  &  -8.51  &  -8.49  &  -8.23  \\
$A_7$  &  -1.54  &  -1.50  &  -1.37  &  -1.36  &  -1.11  \\
$A_8$  &  3.09  &  3.02  &  2.83  &  2.80  &  2.43  \\
$A_9$  &  1.65  &  1.60  &  1.46  &  1.44  &  3.65  \\
$A_{10}$  &  -5.15  &  -5.09  &  -4.92  &  -4.90  &  -1.65  \\
$A_{11}$  &  -0.79  &  -0.76  &  -0.68  &  -0.66  &  -0.50  \\
$A_{12}$  &  2.13  &  2.06  &  1.86  &  1.84  &  1.30  \\
$A_{13}$  &  0.39  &  0.37  &  0.32  &  0.32  &  0.23  \\
$A_{14}$  &  -0.95  &  -0.92  &  -0.84  &  -0.83  &  -0.66  \\
$A_{15}$  &  -0.62  &  -0.60  &  -0.57  &  -0.56  &  -0.53  \\
\bottomrule
\end{tabular}
\end{table}

In general, $\mhh$ is a sensitive variable for differentiating the
effects of the different EFT coefficients. 
In the vicinity of the SM limit the utility of this
approach, however, is limited by the small di-Higgs rate, although
it can nevertheless provide essential additional information on the various effective couplings,
in particular in the case of an enhanced cross section.
There are large correlations, typically requiring a global fit rather than a fit
to a single EFT coefficient~\cite{Azatov:2015oxa,Dall'Osso:2015aia}.
To this end, the analysis of~\cite{Dall'Osso:2015aia} divides kinematic points into $12$
clusters, scanning over a range of EFT coefficients, where within the clusters the dependence on the kinematic
parameters is similar.  The clusters have clear kinematic differences
between them, particularly in the peak structure of the $\mhh$
distributions.  Scanning over a range of EFT coefficients this
procedure allows us to formulate a set of benchmark choices that
exhibit a particularly interesting di-Higgs phenomenology and to exploit the kinematical particularities of a large 
portion of the parameter space with a limited number of analyses. These
recommendations are given in \refT{tab:benchmarkseft}. For the particularly 
interesting case for which the only non-zero EFT
coefficient is $\kappa_{\lambda}$, we have the results in
\refT{tab:nnlomodlam}, that give the relative modification
compared to the SM at NNLL~\cite{deFlorian:2015moa}. 
The difference between the $R_{\hsm\hsm}^{\text{NNLL}}$ found in \refT{tab:nnlomodlam}  
and the $R_{\hsm\hsm}$ calculated by the LO interpolation and the coefficients of \refT{tab:coefs} 
does not exceed 5\%~\cite{Carvalho:2016rys}. This  variation  is inside the theoretical uncertainty of the 
cross section normalization (\refT{tab:hhsignnllb}), 
and slightly larger that the NLO K-factor uncertainty~\cite{Grober:2015cwa}.

In summary, to obtain $\sigma_{\hsm\hsm}$ in the EFT,  we recommend the use of the 
relation $ \sigma_{\hsm\hsm} = R_{\hsm\hsm} \cdot \sigma_{\hsm\hsm}^{SM}$. Here $\sigma_{\hsm\hsm}^{SM}$ is the most up to date calculation based on NNLO+NNLL, with the associated uncertainties, and $R_{\hsm\hsm}$ is found from Eq.~\ref{eq:interpolation}. The  PDF and $\alpha_s$ uncertainties of $A_i$ was found to be well below 1\% and can safely be neglected~\cite{Carvalho:2016rys}. The missing order uncertainties on $R_{\hsm\hsm}$ are covered by the ones assigned to  $\sigma_{\hsm\hsm}^{SM}$.

\begin{table}[!t]
\caption{Coefficient choices of EFT benchmarks. \label{tab:benchmarkseft}}
\centering
\begin{tabular}{c | ccccc}
\toprule
Benchmark & $\kappa_{\lambda}$ & $\kappa_{t}$ & $c_{2}$	& $c_{g}$ & $c_{2g}$ \\[1mm]
\midrule
1 &	7.5	 & 1.0	 &	-1.0	& 0.0	& 0.0 \\
2 &	1.0	 & 1.0	 &	0.5		& -0.8	& 0.6 \\ 	
3 &	1.0	 & 1.0	 &	-1.5	& 0.0	& -0.8 \\  	
4 &	-3.5 & 1.5  &	-3.0	& 0.0	& 0.0 \\ 
5 &	1.0	 & 1.0	 &	0.0		& 0.8	& -1 \\ 
6 &	2.4	 & 1.0	 &	0.0		& 0.2	& -0.2 \\ 
7 &	5.0	 & 1.0	 &	0.0		& 0.2	& -0.2 \\ 
8 &	15.0 & 1.0	 &	0.0		& -1	& 1 \\ 
9 &	1.0	 & 1.0	 &	1.0		& -0.6	& 0.6 \\ 
10 &	10.0 & 1.5   &	-1.0	& 0.0	& 0.0 \\ 
11 &	2.4	 & 1.0	 &	0.0		& 1		& -1 \\ 
12 &	15.0 & 1.0	 &	1.0		& 0.0	& 0.0 \\ 
SM &	1.0 & 1.0	 &	0.0		& 0.0	& 0.0 \\ 
\bottomrule
\end{tabular} 
\end{table} 


\begin{table}[!b]
\caption{\label{tab:nnlomodlam} Values for $\sigma_{\text{NNLL}}/\sigma_{\text{NNLL},SM}$ for non-SM values
of the trilinear Higgs boson coupling, with all other EFT couplings set to their SM values~\cite{deFlorian:2015moa}.}
\centering
\begin{tabular}{c  | c c c c  c}
\toprule
\multicolumn{6}{c}{ $R_{\hsm\hsm}^{\text{NNLL}} \equiv \sigma_{\text{NNLL}}/\sigma_{\text{NNLL},SM}(\kappa_{\lambda})$ } \\[1mm] \midrule
$\kappa_{\lambda}$ & $-1$ & $-0.5$ &
$0$ & $0.5$ & $2 $\\
\midrule
$\sqrt{s}=7$ \tev & 4.17 & 3.12 & 2.24 & 1.53 & 0.452  \\
$\sqrt{s}=8$ \tev & 4.09 & 3.06 & 2.21 & 1.52 & 0.455  \\
$\sqrt{s}=13$ \tev & 3.85 & 2.92 & 2.13 & 1.49 & 0.466  \\
$\sqrt{s}=14$ \tev & 3.82 & 2.90 & 2.12 & 1.49 & 0.467  \\
$\sqrt{s}=100$ \tev & 3.39 & 2.62 & 1.97 & 1.43 & 0.492  \\
\bottomrule
\end{tabular}
\end{table}

\subsection{Higgs Singlet Model}
\label{sec:hhsinglet}
The Higgs singlet model~\cite{Binoth:1996au,Schabinger:2005ei,Patt:2006fw} is a simple example where double Higgs boson production can receive large
contributions from a resonance. The model contains a Higgs doublet, 
$\Phi^T=(\phi^+, {\tilde{\phi_0}}={\phi_0+v\over\sqrt{2}})$, 
and Higgs singlet, $S={s+\langle S\rangle \over\sqrt{2}},$
and  is described by $5$ parameters in the potential:
\begin{equation} 
V=-m^2\Phi ^\dagger \Phi -\mu^2 S^2
+\lambda_1(\Phi^\dagger \Phi)^2+\lambda_2S^4+\lambda_3 \Phi^\dagger \Phi S^2 \, ,
\end{equation} 
where a $Z_2$ symmetry $S\rightarrow -S$ and $\Phi\rightarrow \Phi$ has been imposed for simplicity.
After electroweak symmetry breaking, both ${\tilde \phi}_0$ and $S$ get vacuum expectation values and
the physical fields  $h,H$ are mixtures of the original fields
\begin{alignat}{5}
h&=\,& \cos\alpha &\, {\phi_0}-\sin\alpha &\,s  \, \phantom{.}\nonumber \\
H&=\,&\sin\alpha &\, {\phi_0}+\cos\alpha &\,s \, ,
\end{alignat}
and we assume $M_H>m_h$ in the following. 
The LO trilinear Higgs boson couplings are,
\begin{eqnarray}
\lambda_{hhh}&=&-{3m_h^2\over v}\biggl(\cos^3\alpha-\tan\beta \sin^3\alpha\biggr)\\
\lambda_{Hhh}&=&-{m_h^2\over v}\sin(2\alpha)(\cos\alpha+\sin\alpha\tan\beta)\biggl(1+{M_H^2\over 2 m_h^2}\biggr).
\end{eqnarray}
The NLO relations for the trilinear couplings are in Ref. ~\cite{Bojarski:2015kra}.

The input parameters can be taken as (see e.g. \cite{Choi:2013qra}),
\begin{itemize}
\item
$\mhsm=125~\gev$, $M_H$, $\cos\alpha$, $v$, $\tan\beta=v/\langle s\rangle $\, ,
\end{itemize}
and the Higgs boson branching ratios to SM particles, $X_{SM}$, are:
\begin{eqnarray}
\Gamma(\hsm\rightarrow X_{SM} X_{SM})&=&\cos^2\alpha \, \Gamma(\hsm\rightarrow X_{SM} X_{SM})_{SM} \nonumber \\
\Gamma(H\rightarrow X_{SM} X_{SM})&=&\sin^2\alpha \,\Gamma(H\rightarrow X_{SM} X_{SM})_{SM}\nonumber \\
\Gamma_H&=& \sin^2\alpha\, \Gamma_{H,SM}(M_H)+\Gamma(H\rightarrow \hsm \hsm)\nonumber \\
\Gamma_\hsm&=&\cos^2\alpha \,\Gamma_{\hsm,SM}(\mhsm)\, ,
\end{eqnarray}
where $\Gamma_{H,SM}(M_H) $ are the Standard Model Higgs boson widths evaluated at $M_H$ which are
completely fixed in terms of $\tan\beta, M_H$, and $\cos\alpha$.  ATLAS~\cite{Aad:2015pla}  considered the restrictions
from Higgs boson coupling measurements on the parameters of the singlet model and found $|\cos\alpha|>0.94$, where we omit
the possibility of the $H$ decaying to some new invisible particles.   The heavier Higgs boson contributes to the $W$ mass, which imposes a 
further limit on $\cos\alpha$ as
a function of $M_H$~\cite{Robens:2015gla,Lopez-Val:2014jva}. 
The branching
ratio, $H\rightarrow \hsm \hsm$,  can be quite large, ${\cal O}\sim 20-30\%$, leading to large effects
in di-Higgs boson production.  Values of the LO branching ratios and widths for $H\rightarrow hh$ for representative
values of the parameters are shown in \refFs{fig:singlet_br} and \ref{fig:singlet_gam2}.  
The maximum and minimum allowed branching ratios, consistent with experimental restrictions, are shown in \refT{tab:highm2HH} as a function of $M_H$.

\begin{table}[!t]
\caption{\label{tab:highm2HH} Maximal and minimal allowed branching ratios, consistent with experimental restrictions,
in the singlet model, taken at the maximal allowed value of $|\sin\alpha|$ for $H\rightarrow hh$. Note that the  minimal values for the branching ratio stem from $\sin\alpha\,\leq\,0$. Decay branching ratios correspond to the branching ratios of a SM Higgs of the same mass, rescaled by $1-\text{BR}({H\,\rightarrow\,h\,h})$\cite{Robens:2015gla,Robens:2016xkb}.}
\centering
\begin{tabular}{c|c|c|c}
\toprule
$M_H (\gev)$&$|\sin\alpha|_{max}$&$BR({H\rightarrow\,h\,h})_{min}$&$BR({H\rightarrow\,h\,h})_{max}$ \\ 
\midrule
255 &0.31	& 0.09 &0.27 \\
260&	0.34 & 0.11 & 0.33 \\
265&	0.33&0.13  &0.36 \\
280&	0.32& 0.17&0.40 \\
290&	0.31&0.18&0.40 \\
305&	0.30&0.20&0.40 \\
325&	0.29& 0.21 & 0.40 \\
345&	0.28&0.22&0.39 \\
365&	0.27& 0.21 & 0.36 \\
395&	0.26& 0.20 &0.32 \\
430&	0.25&0.19&0.30 \\
470&	0.24&0.19 & 0.28 \\
520&	0.23&0.19 & 0.26 \\
590&	0.22& 0.19 &0.25 \\
665&	0.21& 0.19&0.23 \\
770&	0.20& 0.19&0.23 \\
875&	0.19&0.19&0.22\\
920&0.18&0.19&0.22\\
$\geq$ 975&0.17&0.19&0.21\\
\bottomrule
\end{tabular}
\end{table}

\begin{figure}[!t]
\begin{center}
\includegraphics[width=.62\textwidth]{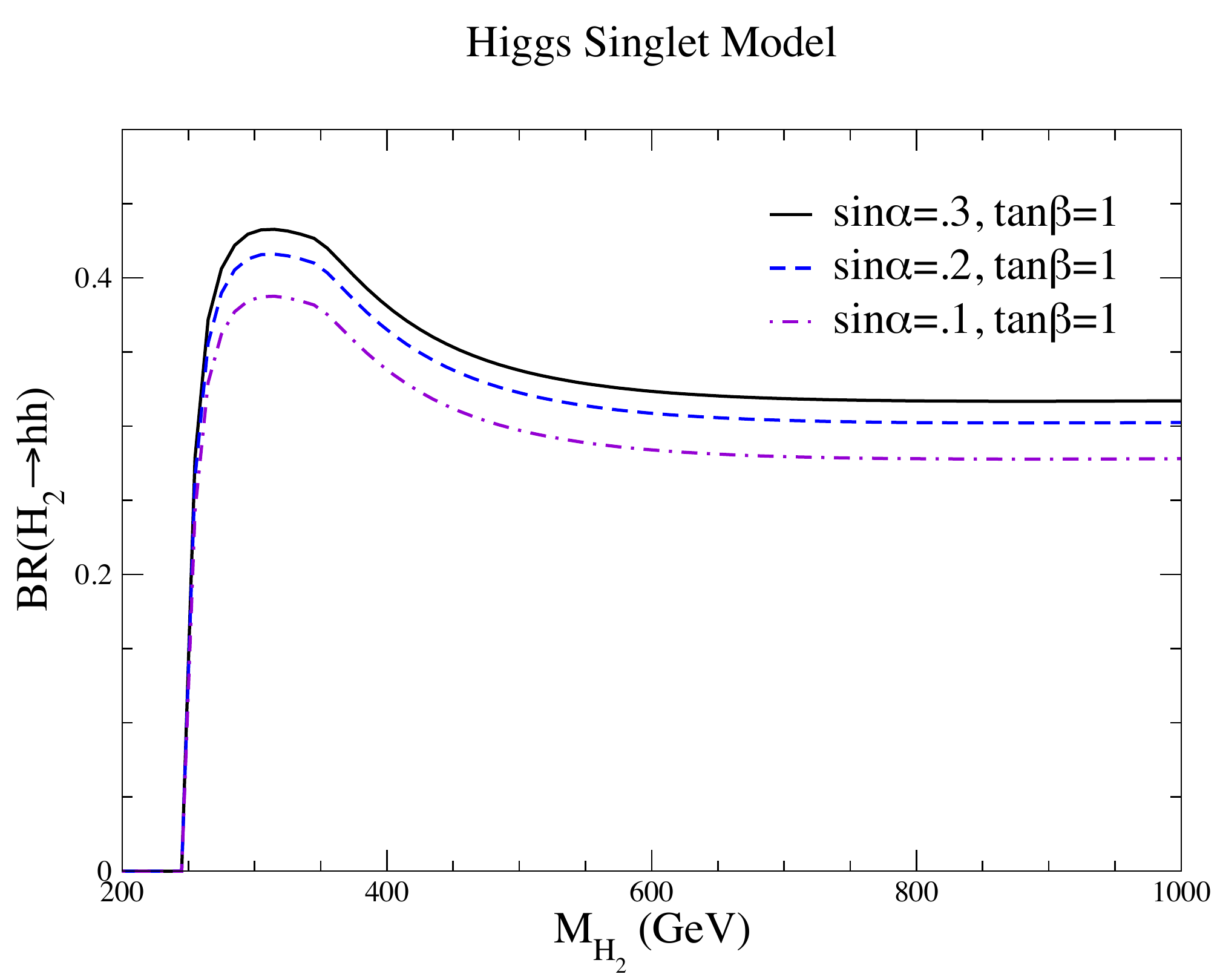}
\caption{\label{fig:singlet_br} Leading order branching ratio of $H\rightarrow hh$  in the singlet model for representative 
values of the parameters.}
\end{center}
\end{figure}

\begin{figure}[!t]
\vspace{-7cm}
\begin{center}
\includegraphics[width=.75\textwidth]{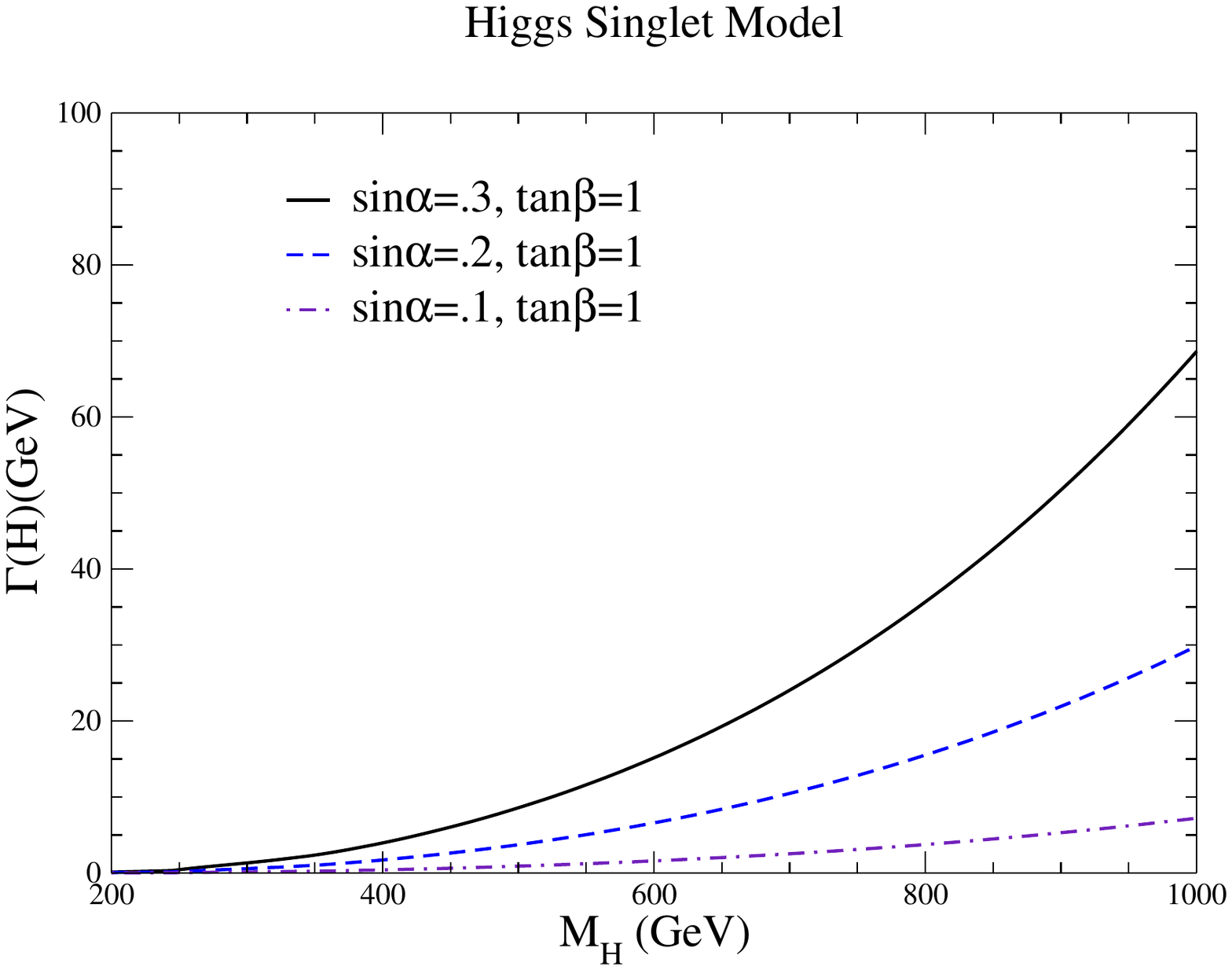}
\caption{\label{fig:singlet_gam2} Total leading order width of $H\rightarrow hh$  in the singlet model for representative 
values of the parameters.}
\end{center}
\end{figure}

The $\mhh$
distributions in the singlet model show clear resonance peaks as illustrated in \refF{fig:singlet}.
\begin{figure}[!t]
\begin{center}
\includegraphics[width=.62\textwidth]{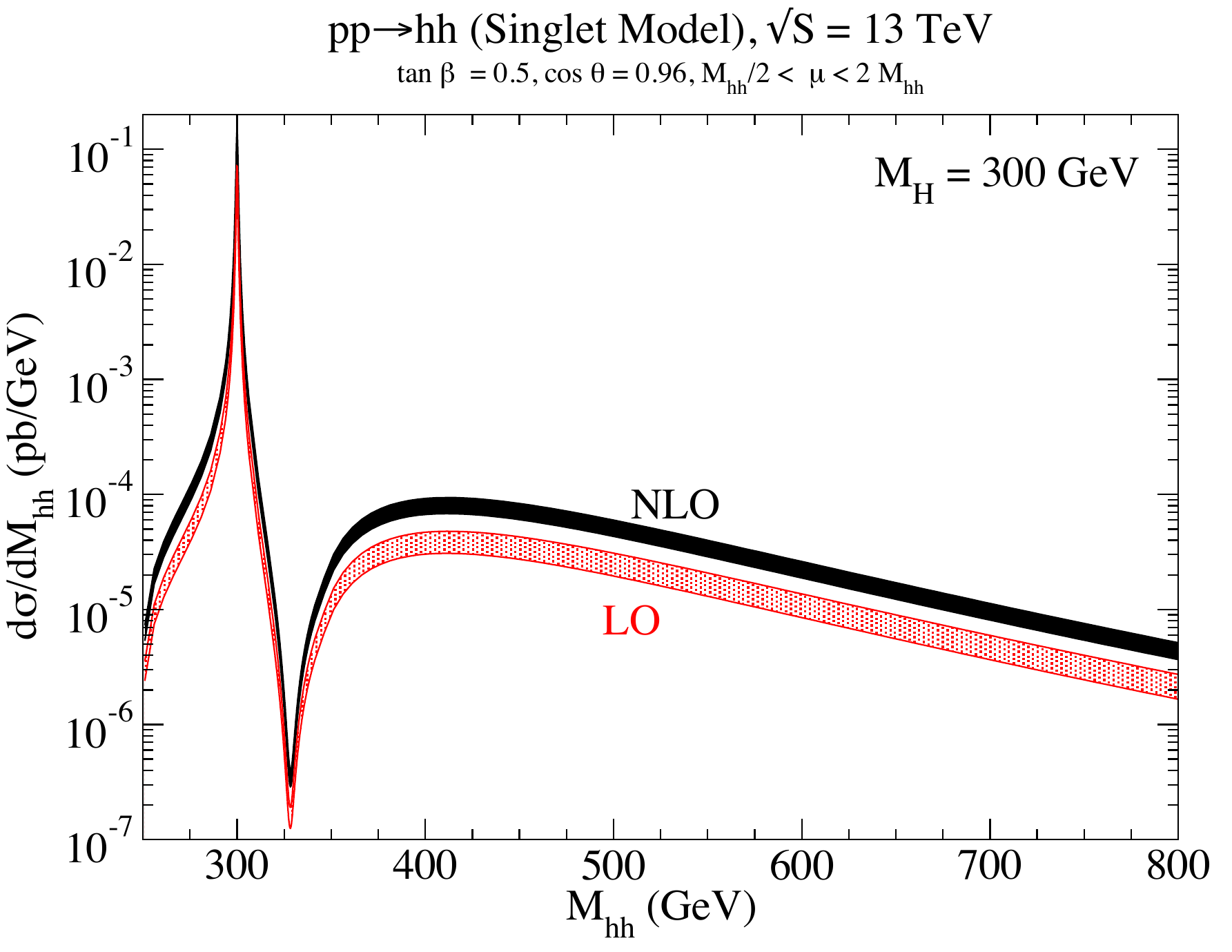}
\caption{\label{fig:singlet} Invariant di-Higgs boson mass differential distribution for $pp\to \hsm \hsm$ 
in the singlet model with a heavy Higgs with $M_H=300~\gev$.}
\end{center}
\end{figure}
The NLO QCD corrections  to double Higgs boson production can be found in the large $\mt$ 
limit~\cite{Dawson:2015haa} and give a $K$ factor which is approximately the same as in
the Standard Model.  For $M_H\sim 2\mhsm$, the rate is dominated by the resonance contribution
which is implemented in the code sHDECAY\cite{Costa:2015llh}. For fixed
$\sin\alpha=0.28$ and $\tan\beta=0.5$, the predictions for a range of heavy Higgs boson masses are given in Tabs.
\ref{tab:singhh14}-\ref{tab:singhh7}. 
 
 \begin{table}
  \caption{\label{tab:singhh14}NLO cross sections in the singlet model  for fixed $\sin\theta=0.28$,
$\tan\beta=0.50$
 and $\sqrts=14~\tev$, with $\mu=M_{hh}/2$~\cite{Dawson:2015haa}.}
 \centering
 \begin{tabular}{c|c|c|c|c}
 \toprule
 $M_H$ (\gev) & Cross Section (\fb) & PDF (\%) &$\alpha_s$ (\%)& scale (\%)  \\
 \midrule
  260 &   278.06&    2.2 &    2.0 & +    18.9   -14.8 \\
  275 &    311.39&    2.2 &    2.0 & +    18.8   -14.9 \\
  300 &     303.35&    2.2 &    2.0 & +    18.9   -14.9 \\
  325 &     290.68&    2.2 &    2.0 & +    18.7   -14.9 \\
  350 &     307.86&    2.3 &    1.9 & +    18.7   -15.0 \\
  400 &     286.17&    2.4 &    1.9 & +    18.6   -15.0 \\
  450 &     217.24&    2.5 &    1.9 & +    18.4   -15.1 \\
  500 &     163.98&    2.7 &    1.8 & +    18.4   -15.1 \\
  600 &     103.53&    2.7 &    1.8 & +    18.3   -15.1 \\
  700 &      76.07&    2.8 &    1.8 & +    18.2   -15.1 \\
  750 &      68.32&    2.8 &    1.8 & +    18.2   -15.1 \\
  800 &      62.86&    2.8 &    1.8 & +    18.2   -15.1 \\
  900 &      56.04&    2.7 &    1.9 & +    18.3   -15.1 \\
 1000 &     52.28&    2.6 &    1.9 & +    18.3   -15.0 \\
 1100 &      50.06&    2.7 &    1.9 & +    18.3   -15.1 \\
 1200 &     48.71&    2.7 &    1.9 & +    18.3   -15.0 \\
 1300 &     47.84&    2.6 &    1.9 & +    18.3   -15.0 \\
 1400 &      47.22&    2.6 &    1.9 & +    18.3   -15.0 \\
 1500 &      46.81&    2.6 &    1.9 & +    18.3   -15.0 \\
 1600 &      46.48&    2.6 &    1.9 & +    18.3   -15.0 \\
 1800 &     46.03&    2.6 &    1.9 & +    18.3   -15.0 \\
 2000 &     45.71&    2.6 &    1.9 & +    18.3   -15.0 \\
 2250 &     45.37&    2.6 &    1.9 & +    18.3   -15.0 \\
 2500 &     45.03&    2.6 &    1.9 & +    18.3   -15.0 \\
 2750 &    44.67&    2.6 &    1.9 & +    18.3   -15.0 \\
 3000 &      44.25&    2.6 &    1.9 & +    18.3   -15.0 \\
 \bottomrule
 \end{tabular}
 \end{table}


 \begin{table}
 \caption{\label{tab:singhh13}NLO cross sections in the singlet model  for fixed $\sin\theta=0.28$,
 $\tan\beta=0.50$ and
 $\sqrts=13~\tev$, with $\mu=M_{hh}/2$~\cite{Dawson:2015haa}.}
 \centering
 \begin{tabular}{c|c|c|c|c}
 \toprule
 $M_H$ (\gev)& Cross Section (\fb) & PDF (\%) &$\alpha_s$ (\%)& scale (\%)  \\
 \midrule
  260 &     240.06&    2.2 &    2.0 & +    19.2   -15.0 \\
  275 &    268.80&    2.2 &    2.0 & +    19.1   -15.1 \\
  300 &    260.78&    2.3 &    2.0 & +    19.1   -15.1 \\
  325 &     248.87&    2.3 &    1.9 & +    18.9   -15.1 \\
  350 &    262.72&    2.4 &    1.9 & +    18.9   -15.2 \\
  400 &     242.67&    2.5 &    1.9 & +    18.8   -15.2 \\
  450 &     183.37&    2.7 &    1.9 & +    18.6   -15.3 \\
  500 &     137.85&    2.7 &    1.9 & +    18.6   -15.3 \\
  600 &      86.49&    2.9 &    1.9 & +    18.4   -15.3 \\
  700 &      63.49&    2.9 &    1.9 & +    18.4   -15.3 \\
  750 &     57.05&    2.9 &    1.8 & +    18.4   -15.3 \\
  800 &     52.49&    2.9 &    1.9 & +    18.4   -15.3 \\
  900 &      46.86&    2.9 &    1.9 & +    18.5   -15.3 \\
 1000 &     43.81&    2.8 &    1.9 & +    18.4   -15.3 \\
 1100 &     42.02&    2.8 &    1.9 & +    18.5   -15.3 \\
 1200 &     40.91&    2.7 &    1.9 & +    18.5   -15.2 \\
 1300 &     40.18&    2.6 &    1.9 & +    18.5   -15.2 \\
 1400 &     39.70&    2.6 &    1.9 & +    18.5   -15.2 \\
 1500 &      39.34&    2.6 &    1.9 & +    18.5   -15.2 \\
 1600 &     39.08&    2.6 &    1.9 & +    18.5   -15.2 \\
 1800 &      38.70&    2.6 &    1.9 & +    18.5   -15.2 \\
 2000 &      38.44&    2.7 &    1.9 & +    18.5   -15.2 \\
 2250 &     38.16&    2.7 &    1.9 & +    18.6   -15.2 \\
 2500 &      37.88&    2.7 &    1.9 & +    18.5   -15.2 \\
 2750 &     37.57&    2.7 &    1.9 & +    18.5   -15.2 \\
 3000 &     37.22&    2.7 &    1.9 & +    18.5   -15.2 \\
 \bottomrule
  \end{tabular}
 \end{table}

 \begin{table}
 \caption{\label{tab:singhh8}NLO cross sections in the singlet model for fixed $\sin\theta=0.28$,  
$\tan\beta=0.50$ and
  $\sqrts=8~\tev$, for $\mu=M_{hh}/2$~\cite{Dawson:2015haa}.}
 \centering
  \begin{tabular}{c|c|c|c|c}
  \toprule
 $M_H$ (\gev) & Cross Section (\fb) & PDF (\%) &$\alpha_s$ (\%)& scale (\%)  \\
 \midrule
  260 &    85.64&    2.7 &    2.1 & +    20.7   -16.5 \\
  275 &      94.39&    2.8 &    2.1 & +    20.6   -16.5 \\
  300 &      89.08&    3.0 &    2.1 & +    20.6   -16.5 \\
  325 &     82.78&    3.1 &    2.0 & +    20.5   -16.6 \\
  350 &     85.29&    3.2 &    2.0 & +    20.4   -16.6 \\
  400 &      75.37&    3.3 &    2.0 & +    20.2   -16.6 \\
  450 &      54.71&    3.6 &    2.0 & +    20.1   -16.7 \\
  500 &      39.82&    3.8 &    2.0 & +    20.0   -16.7 \\
  600 &      24.07&    3.9 &    2.0 & +    19.9   -16.8 \\
  700 &      17.54&    4.0 &    2.0 & +    19.9   -16.8 \\
  750 &     15.82&    3.9 &    2.0 & +    19.9   -16.7 \\
  800 &      14.64&    3.9 &    2.0 & +    19.9   -16.8 \\
  900 &      13.23&    3.8 &    2.0 & +    20.0   -16.7 \\
 1000 &      12.49&    3.8 &    2.1 & +    19.9   -16.7 \\
 1100 &     12.07&    3.7 &    2.0 & +    20.0   -16.7 \\
 1200 &      11.82&    3.7 &    2.0 & +    20.0   -16.7 \\
 1300 &      11.66&    3.7 &    2.0 & +    20.0   -16.7 \\
 1400 &      11.55&    3.7 &    2.0 & +    20.0   -16.6 \\
 1500 &     11.47&    3.7 &    2.0 & +    19.9   -16.7 \\
 1600 &     11.41&    3.7 &    2.0 & +    20.0   -16.7 \\
 1800 &      11.32&    3.7 &    2.0 & +    20.0   -16.6 \\
 2000 &     11.25&    3.7 &    2.0 & +    20.0   -16.7 \\
 2250 &    11.18&    3.7 &    2.0 & +    20.0   -16.6 \\
 2500 &      11.10&    3.7 &    2.0 & +    19.9   -16.6 \\
 2750 &     11.01&    3.6 &    2.0 & +    20.0   -16.7 \\
 3000 &      10.91&    3.6 &    2.0 & +    20.0   -16.7 \\
 \bottomrule
 \end{tabular}
 \end{table}

 \begin{table}
 \caption{\label{tab:singhh7}NLO cross sections in the singlet model for fixed $\sin\theta=0.28$, 
$\tan\beta=0.50$, and
  $\sqrts=7~\tev$ with $\mu=M_{hh}/2$~\cite{Dawson:2015haa}.}
 \centering
 \begin{tabular}{c|c|c|c|c}
 \toprule
 $M_H$ (\gev)& Cross Section (\fb) & PDF (\%) &$\alpha_s$ (\%)& scale (\%)  \\
 \midrule
  260 &    62.89&    3.0 &    2.1 & +    21.2   -16.9 \\
  275 &     68.99&    3.0 &    2.1 & +    21.1   -16.9 \\
  300 &     64.40&    3.2 &    2.1 & +    21.0   -16.9 \\
  325 &      59.45&    3.2 &    2.1 & +    20.9   -17.0 \\
  350 &    60.78&    3.3 &    2.1 & +    20.8   -17.0 \\
  400 &     52.85&    3.7 &    2.1 & +    20.7   -17.1 \\
  450 &     37.81&    4.0 &    2.1 & +    20.5   -17.1 \\
  500 &     27.23&    4.1 &    2.1 & +    20.5   -17.2 \\
  600 &      16.28&    4.2 &    2.1 & +    20.3   -17.2 \\
  700 &     11.85&    4.2 &    2.1 & +    20.4   -17.2 \\
  750 &     10.71&    4.2 &    2.1 & +    20.4   -17.2 \\
  800 &      9.93&    4.1 &    2.1 & +    20.3   -17.2 \\
  900 &      9.01&    4.1 &    2.1 & +    20.4   -17.1 \\
 1000 &      8.54&    4.0 &    2.1 & +    20.3   -17.1 \\
 1100 &       8.27&    4.0 &    2.1 & +    20.4   -17.1 \\
 1200 &      8.11&    4.0 &    2.1 & +    20.5   -17.1 \\
 1300 &      8.01&    4.0 &    2.1 & +    20.4   -17.1 \\
 1400 &      7.94&    4.0 &    2.1 & +    20.5   -17.1 \\
 1500 &     7.89&    4.0 &    2.1 & +    20.4   -17.1 \\
 1600 &       7.85&    4.0 &    2.1 & +    20.4   -17.1 \\
 1800 &     7.79&    4.0 &    2.1 & +    20.4   -17.1 \\
 2000 &      7.75&    4.0 &    2.1 & +    20.4   -17.1 \\
 2250 &     7.70&    4.0 &    2.1 & +    20.5   -17.1 \\
 2500 &      7.64&    4.0 &    2.1 & +    20.4   -17.1 \\
 2750 &      7.58&    4.0 &    2.1 & +    20.4   -17.1 \\
 3000 &     7.51&    4.0 &    2.1 & +    20.4   -17.1 \\
 \bottomrule
  \end{tabular}
 \end{table}

The enhancements of the di-Higgs cross section in the singlet model  can be as large as factors of 
${\cal{O}}(10-20)$ and are typical of those which can
be obtained in models with a heavy Higgs particle with a mass near $2\mhsm$ and a large branching ratio to $\hsm\hsm$,
such as the 2HDM, the MSSM, or the NMSSM.  It is interesting to tabulate the largest allowed values of the 
di-Higgs cross section in the singlet model using the restrictions of \refT{tab:highm2HH}.  These cross
sections are shown in \refT{tab:hhmax14}-\ref{tab:hhmax7}.

\begin{table}
 \caption{\label{tab:hhmax14}$\sqrts=14~ \tev$ NLO cross sections in the singlet model with parameters chosen to maximize the
 cross section, with $\mu=M_{hh}/2$~\cite{Robens:2016xkb}.}
\centering
\begin{tabular}{c|c|c|c|c|c|c}
\toprule
$M_H$ (\gev)& $\sin\theta$ & $\tan\beta$ & Cross Section (\fb) & PDF (\%) &$\alpha_s$ (\%)& scale (\%)  \\
\midrule
260 &   0.31 &   0.80 &  365.71&    2.1 &    2.0 & +    19.0   -14.8 \\
  275 &   0.31 &   0.80 &  407.56&    2.2 &    2.0 & +    18.9   -14.9 \\
  300 &   0.31 &   0.80 &  395.31&    2.2 &    2.0 & +    18.8   -14.9 \\
  325 &   0.27 &   0.58 &  279.16&    2.2 &    1.9 & +    18.7   -14.9 \\
  350 &   0.27 &   0.58 &  295.73&    2.3 &    1.9 & +    18.7   -14.9 \\
  400 &   0.27 &   0.58 &  275.47&    2.5 &    1.9 & +    18.5   -15.1 \\
  450 &   0.24 &   0.46 &  169.33&    2.5 &    1.9 & +    18.4   -15.1 \\
  500 &   0.24 &   0.46 &  130.40&    2.7 &    1.8 & +    18.3   -15.1 \\
  600 &   0.23 &   0.37 &   81.05&    2.7 &    1.8 & +    18.2   -15.1 \\
  700 &   0.21 &   0.31 &   58.65&    2.7 &    1.8 & +    18.2   -15.1 \\
  750 &   0.21 &   0.25 &   54.15&    2.6 &    1.8 & +    18.3   -15.1 \\
  800 &   0.21 &   0.25 &   51.23&    2.6 &    1.9 & +    18.2   -15.1 \\
  900 &   0.19 &   0.25 &   45.96&    2.6 &    1.8 & +    18.3   -15.0 \\
 1000 &   0.17 &   0.23 &   43.13&    2.7 &    1.9 & +    18.3   -15.0 \\
 1100 &   0.17 &   0.23 &   42.39&    2.6 &    1.9 & +    18.3   -15.0 \\
 1200 &   0.17 &   0.23 &   41.90&    2.6 &    1.9 & +    18.3   -15.0 \\
 1300 &   0.17 &   0.23 &   41.59&    2.6 &    1.9 & +    18.3   -15.0 \\
 1400 &   0.17 &   0.23 &   41.38&    2.6 &    1.9 & +    18.3   -15.0 \\
 1500 &   0.17 &   0.23 &   41.24&    2.6 &    1.9 & +    18.3   -15.0 \\
 1600 &   0.17 &   0.23 &   41.14&    2.6 &    1.9 & +    18.3   -15.0 \\
 1800 &   0.17 &   0.23 &   41.00&    2.6 &    1.8 & +    18.3   -15.0 \\
 2000 &   0.17 &   0.23 &   40.92&    2.6 &    1.9 & +    18.3   -15.0 \\
 2250 &   0.17 &   0.23 &   40.85&    2.6 &    1.9 & +    18.3   -15.0 \\
 2500 &   0.17 &   0.23 &   40.81&    2.6 &    1.9 & +    18.3   -15.0 \\
 2750 &   0.17 &   0.23 &   40.78&    2.6 &    1.9 & +    18.3   -15.0 \\
 3000 &   0.17 &   0.23 &   40.75&    2.6 &    1.9 & +    18.3   -15.0 \\  
 \bottomrule
 \end{tabular}
 \end{table}

 \begin{table}
  \caption{\label{tab:hhmax13}$\sqrts=13~ \tev$ NLO cross sections in the singlet model with parameters chosen to maximize the
 cross section, with $\mu=M_{hh}/2$~\cite{Robens:2016xkb}.}
  \centering
 \begin{tabular}{c|c|c|c|c|c|c}
 \toprule
 $M_H$ (\gev)& $\sin\theta$ & $\tan\beta$ & Cross Section (\fb) & PDF (\%) &$\alpha_s$ (\%)& scale (\%)  \\
 \midrule
  260 &   0.31 &   0.80 &  315.92&    2.2 &    2.0 & +    19.2   -15.0 \\
  275 &   0.31 &   0.80 &  351.78&    2.2 &    2.0 & +    19.2   -15.1 \\
  300 &   0.31 &   0.80 &  340.05&    2.3 &    2.0 & +    19.1   -15.1 \\
  325 &   0.27 &   0.58 &  239.07&    2.3 &    1.9 & +    18.9   -15.1 \\
  350 &   0.27 &   0.58 &  252.31&    2.4 &    1.9 & +    18.9   -15.1 \\
  400 &   0.27 &   0.58 &  233.65&    2.5 &    1.9 & +    18.7   -15.3 \\
  450 &   0.24 &   0.46 &  142.91&    2.7 &    1.9 & +    18.6   -15.3 \\
  500 &   0.24 &   0.46 &  109.61&    2.7 &    1.9 & +    18.5   -15.3 \\
  600 &   0.23 &   0.37 &   67.81&    2.9 &    1.8 & +    18.4   -15.3 \\
  700 &   0.21 &   0.31 &   49.05&    2.9 &    1.9 & +    18.4   -15.3 \\
  750 &   0.21 &   0.25 &   45.31&    2.9 &    1.9 & +    18.5   -15.3 \\
  800 &   0.21 &   0.25 &   42.90&    2.9 &    1.9 & +    18.4   -15.3 \\
  900 &   0.19 &   0.25 &   38.57&    2.7 &    1.9 & +    18.5   -15.3 \\
 1000 &   0.17 &   0.23 &   36.24&    2.7 &    1.9 & +    18.4   -15.2 \\
 1100 &   0.17 &   0.23 &   35.62&    2.7 &    1.9 & +    18.5   -15.2 \\
 1200 &   0.17 &   0.23 &   35.22&    2.7 &    1.9 & +    18.5   -15.2 \\
 1300 &   0.17 &   0.23 &   34.96&    2.7 &    1.9 & +    18.5   -15.2 \\
 1400 &   0.17 &   0.23 &   34.80&    2.7 &    1.9 & +    18.5   -15.2 \\
 1500 &   0.17 &   0.23 &   34.68&    2.7 &    1.9 & +    18.5   -15.2 \\
 1600 &   0.17 &   0.23 &   34.60&    2.7 &    1.9 & +    18.5   -15.2 \\
 1800 &   0.17 &   0.23 &   34.49&    2.7 &    1.9 & +    18.5   -15.2 \\
 2000 &   0.17 &   0.23 &   34.42&    2.7 &    1.9 & +    18.5   -15.2 \\
 2250 &   0.17 &   0.23 &   34.37&    2.7 &    1.9 & +    18.5   -15.2 \\
 2500 &   0.17 &   0.23 &   34.33&    2.7 &    1.9 & +    18.5   -15.2 \\
 2750 &   0.17 &   0.23 &   34.31&    2.7 &    1.9 & +    18.5   -15.2 \\
 3000 &   0.17 &   0.23 &   34.29&    2.7 &    1.9 & +    18.5   -15.2 \\
 \bottomrule
  \end{tabular}
 \end{table}

 \begin{table}
 \caption{\label{tab:hhmax8}$\sqrts=8~ \tev$ NLO cross sections in the singlet model with parameters chosen to maximize the
 cross section, with $\mu=M_{hh}/2$~\cite{Robens:2016xkb}.}
 \centering
  \begin{tabular}{c|c|c|c|c|c|c}
  \toprule
 $M_H$ (\gev)& $\sin\theta$ & $\tan\beta$ & Cross Section (\fb) & PDF (\%) &$\alpha_s$ (\%)& scale (\%)  \\
 \midrule
 260 &   0.31 &   0.80 &  113.38&    2.7 &    2.1 & +    20.8   -16.5 \\
  275 &   0.31 &   0.80 &  124.18&    2.8 &    2.1 & +    20.7   -16.5 \\
  300 &   0.31 &   0.80 &  116.69&    3.0 &    2.0 & +    20.5   -16.5 \\
  325 &   0.27 &   0.58 &   79.42&    3.1 &    2.0 & +    20.4   -16.6 \\
  350 &   0.27 &   0.58 &   81.89&    3.2 &    2.0 & +    20.4   -16.6 \\
  400 &   0.27 &   0.58 &   72.54&    3.3 &    2.0 & +    20.2   -16.7 \\
  450 &   0.24 &   0.46 &   42.59&    3.6 &    2.0 & +    20.1   -16.7 \\
  500 &   0.24 &   0.46 &   31.68&    3.8 &    2.0 & +    20.0   -16.7 \\
  600 &   0.23 &   0.37 &   19.01&    3.9 &    2.0 & +    19.9   -16.7 \\
  700 &   0.21 &   0.31 &   13.79&    3.9 &    2.0 & +    19.9   -16.7 \\
  750 &   0.21 &   0.25 &   12.79&    3.9 &    2.0 & +    19.9   -16.7 \\
  800 &   0.21 &   0.25 &   12.16&    3.8 &    2.0 & +    19.9   -16.7 \\
  900 &   0.19 &   0.25 &   11.07&    3.7 &    2.0 & +    20.0   -16.7 \\
 1000 &   0.17 &   0.23 &   10.49&    3.7 &    2.0 & +    19.9   -16.7 \\
 1100 &   0.17 &   0.23 &   10.34&    3.6 &    2.0 & +    20.0   -16.6 \\
 1200 &   0.17 &   0.23 &   10.26&    3.7 &    2.0 & +    20.0   -16.7 \\
 1300 &   0.17 &   0.23 &   10.20&    3.7 &    2.0 & +    19.9   -16.7 \\
 1400 &   0.17 &   0.23 &   10.16&    3.7 &    2.0 & +    20.0   -16.6 \\
 1500 &   0.17 &   0.23 &   10.14&    3.7 &    2.0 & +    19.9   -16.7 \\
 1600 &   0.17 &   0.23 &   10.12&    3.7 &    2.0 & +    20.0   -16.7 \\
 1800 &   0.17 &   0.23 &   10.10&    3.7 &    2.0 & +    20.0   -16.7 \\
 2000 &   0.17 &   0.23 &   10.08&    3.6 &    2.0 & +    19.9   -16.7 \\
 2250 &   0.17 &   0.23 &   10.07&    3.7 &    2.0 & +    20.0   -16.6 \\
 2500 &   0.17 &   0.23 &   10.06&    3.7 &    2.0 & +    19.9   -16.7 \\
 2750 &   0.17 &   0.23 &   10.05&    3.7 &    2.0 & +    20.0   -16.6 \\
 3000 &   0.17 &   0.23 &   10.05&    3.7 &    2.0 & +    20.0   -16.7 \\  
 \bottomrule
 \end{tabular}
 \end{table}

 \begin{table}
 \caption{\label{tab:hhmax7}$\sqrts=7~ \tev$ NLO cross sections in the singlet model with parameters chosen to maximize the
 cross section, with $\mu=M_{hh}/2$~\cite{Robens:2016xkb}. }
 \centering
\begin{tabular}{c|c|c|c|c|c|c}
\toprule
 $M_H$ (\gev)& $\sin\theta$ & $\tan\beta$ & Cross Section (\fb) & PDF (\%) &$\alpha_s$ (\%)& scale (\%)  \\
 \midrule
  260 &   0.31 &   0.80 &   83.40&    3.0 &    2.1 & +    21.2   -16.9 \\
  275 &   0.31 &   0.80 &   90.93&    3.0 &    2.1 & +    21.1   -16.9 \\
  300 &   0.31 &   0.80 &   84.51&    3.2 &    2.1 & +    21.0   -16.9 \\
  325 &   0.27 &   0.58 &   57.05&    3.2 &    2.1 & +    20.9   -17.0 \\
  350 &   0.27 &   0.58 &   58.34&    3.3 &    2.1 & +    20.8   -17.0 \\
  400 &   0.27 &   0.58 &   50.84&    3.7 &    2.1 & +    20.7   -17.1 \\
  450 &   0.24 &   0.46 &   29.42&    4.0 &    2.1 & +    20.5   -17.1 \\
  500 &   0.24 &   0.46 &   21.68&    4.1 &    2.1 & +    20.4   -17.2 \\
  600 &   0.23 &   0.37 &   12.90&    4.2 &    2.1 & +    20.3   -17.2 \\
  700 &   0.21 &   0.31 &    9.37&    4.1 &    2.1 & +    20.4   -17.2 \\
  750 &   0.21 &   0.25 &    8.71&    4.1 &    2.1 & +    20.4   -17.1 \\
  800 &   0.21 &   0.25 &    8.29&    4.1 &    2.1 & +    20.4   -17.1 \\
  900 &   0.19 &   0.25 &    7.58&    4.0 &    2.1 & +    20.4   -17.1 \\
 1000 &   0.17 &   0.23 &    7.20&    4.0 &    2.1 & +    20.3   -17.1 \\
 1100 &   0.17 &   0.23 &    7.11&    4.0 &    2.1 & +    20.4   -17.1 \\
 1200 &   0.17 &   0.23 &    7.05&    4.0 &    2.1 & +    20.5   -17.1 \\
 1300 &   0.17 &   0.23 &    7.02&    4.0 &    2.1 & +    20.4   -17.1 \\
 1400 &   0.17 &   0.23 &    6.99&    4.0 &    2.1 & +    20.4   -17.1 \\
 1500 &   0.17 &   0.23 &    6.98&    4.0 &    2.1 & +    20.4   -17.1 \\
 1600 &   0.17 &   0.23 &    6.96&    4.0 &    2.1 & +    20.4   -17.1 \\
 1800 &   0.17 &   0.23 &    6.95&    4.0 &    2.1 & +    20.4   -17.1 \\
 2000 &   0.17 &   0.23 &    6.94&    4.0 &    2.1 & +    20.4   -17.1 \\
 2250 &   0.17 &   0.23 &    6.93&    4.0 &    2.1 & +    20.4   -17.1 \\
 2500 &   0.17 &   0.23 &    6.92&    4.0 &    2.1 & +    20.4   -17.1 \\
 2750 &   0.17 &   0.23 &    6.92&    4.0 &    2.1 & +    20.4   -17.1 \\
 3000 &   0.17 &   0.23 &    6.92&    4.0 &    2.1 & +    20.4   -17.1 \\
\bottomrule
 \end{tabular}
 \end{table}
 
 \subsection{2 Higgs Doublet Model}
 The 2 Higgs doublet model (2HDM) is a simple extension of the SM which can exhibit large resonance effects.
The 2HDM has $5$ physical Higgs bosons: 2 neutral scalars, $h^0,H^0$, a pseudo-scalar, $A$, and 
a charged Higgs boson pair $H^\pm$, 
 In general, 2HDMs have Higgs mediated tree level flavour changing neutral currents (FCNCs), 
which must be suppressed.   Most 2HDMs eliminate FCNCs by imposing a discrete $Z_2$ symmetry in which the 
fermions of a given charge only couple to one of the Higgs doublets.    The two most familiar versions are the type I model, in which all of the fermions couple to the same Higgs doublet, and the type II model, in which the $Q=2/3$ quarks couple to one doublet and the $Q=-1/3$ quarks and leptons couple to the other.    The type II model is the Higgs sector of the MSSM. 
Two additional versions interchange the lepton assignments.  In the ``lepton-specific" model, all of the quarks couple to one doublet while the leptons couple to the other, and in the ``flipped" model, the $Q=2/3$ quarks and leptons couple to one doublet and the $Q=-1/3$ quarks couple to the other.    All four of these models have been extensively studied \cite{Branco:2011iw}.   

The couplings of the Higgs bosons to fermions are described by two free parameters.   
The ratio of vacuum expectation values of the two Higgs doublets is  $\tan\beta \equiv \frac{v_2}{v_1}$, 
and the mixing angle which diagonalizes the neutral scalar mass matrix is $\alpha$.   The couplings of the light (heavy) CP even Higgs boson, 
$h^0$ ($H^0$), to fermions and gauge bosons
relative to the Standard Model couplings
 are given for all four 2HDMs considered here in \refT{table:coups} (\refT{table:coupsH}). 

2HDMs are significantly limited by experimental data.  Higgs boson coupling measurements restrict $\cos (\alpha-\beta)$  to be close to
the SM limit, $\cos(\alpha-\beta)\sim 0$, while heavy Higgs searches restrict $M_{H^0}$ as 
a function of $\cos(\alpha-\beta)$\cite{Aad:2015pla}. 
A set of benchmarks which respect all experimental limits was found in Refs. \cite{Hespel:2014sla,Haber:2015pua}, and representative benchmarks are given
in \refT{tab:benchmarks}.  The benchmarks were further chosen such that the total rate for di-Higgs boson production is similar 
to that of the SM.  These benchmarks can exhibit significant resonance effects (B1 and B2), while B7 is almost
indistinguishable from the SM.  The NLO di-Higgs boson invariant mass distributions for these benchmarks are shown in \refF{fig:2hdmfig}.  
Note that other benchmarks have been proposed in the literature, such as in Ref. \cite{Baglio:2014nea}, where resonant effects
can also be important already at the level of the inclusive rate as in their benchmark H-1.


\begin{table}
\caption{Light Neutral Higgs ($h^0$) Couplings in the 2HDM.}
\label{table:coups}
\centering
\renewcommand{\arraystretch}{1.3}
\begin{tabular}[t]{c|c|c|c|c}
\toprule
& I& II& Lepton Specific& Flipped\\
\midrule
$g_{hVV}$ & $\sin(\beta-\alpha)$ & $\sin (\beta-\alpha)$ &$\sin (\beta-\alpha)$&$\sin (\beta-\alpha)$\\
$g_{ht\overline{t}}$&${\cos\alpha\over\sin\beta}$&${\cos\alpha\over\sin\beta}$&${\cos\alpha\over\sin\beta}$&${\cos\alpha\over\sin\beta}$\\
$g_{hb{\overline b}}$ &${\cos\alpha\over\sin\beta}$&$-{\sin\alpha\over\cos\beta}$&${\cos\alpha\over\sin\beta}$&$-{\sin\alpha\over \cos\beta}$\\
$g_{h\tau^+\tau^-}$&${\cos\alpha\over \sin\beta}$&$-{\sin\alpha\over \cos\beta}$&$-{\sin\alpha\over \cos\beta}$&${\cos\alpha\over\sin\beta}$\\
\bottomrule
\end{tabular}
\end{table}

\begin{table}
\caption{Heavy Neutral CP Even Higgs ($H^0$) Couplings in the 2HDM.}
\label{table:coupsH}
\centering
\renewcommand{\arraystretch}{1.3}
\begin{tabular}[t]{c|c|c|c|c}
\toprule
& I& II& Lepton Specific& Flipped\\
\midrule
$g_{HVV}$ & $\cos(\beta-\alpha)$ & $\cos (\beta-\alpha)$ &$\cos (\beta-\alpha)$&$\cos (\beta-\alpha)$\\
$g_{Ht\overline{t}}$&${\sin\alpha\over\sin\beta}$&${\sin\alpha\over\sin\beta}$&${\sin\alpha\over\sin\beta}$&${\sin\alpha\over\sin\beta}$\\
$g_{Hb{\overline b}}$ &${\sin\alpha\over\sin\beta}$&${\cos\alpha\over\cos\beta}$&${\sin\alpha\over\sin\beta}$&${\cos\alpha\over \cos\beta}$\\
$g_{H\tau^+\tau^-}$&${\sin\alpha\over \sin\beta}$&${\cos\alpha\over \cos\beta}$&${\cos\alpha\over \cos\beta}$&${\sin\alpha\over\sin\beta}$\\
\bottomrule
\end{tabular}
\end{table}

\begin{table}
 \caption{Parameter choices for the 2HDM benchmarks. All
 masses are given in GeV\cite{Hespel:2014sla}.}
 \label{tab:benchmarks}
\begin{center}
 \begin{tabular}{l||cc|cccc} 
 \toprule
 & $\tan\beta$ & $\alpha$ & $m_{H^0} $ &  $m_{A^0} $  & $m_{H^{\pm}} $   & $m^2_{12} $  \\ 
 \midrule
B1 & 1.75 & -0.5881 & 300 & 441 & 442 & 38300  \\
B2 & 1.50 & -0.6792 & 700 & 701 & 670 & 180000 \\
B7 & 10.00 & 0.1015 &  500 & 500 & 500 & 24746 \\ 
\bottomrule
 \end{tabular}
 \end{center}
\end{table}

\begin{figure}[!t]
\begin{center}
\includegraphics[width=.62\textwidth]{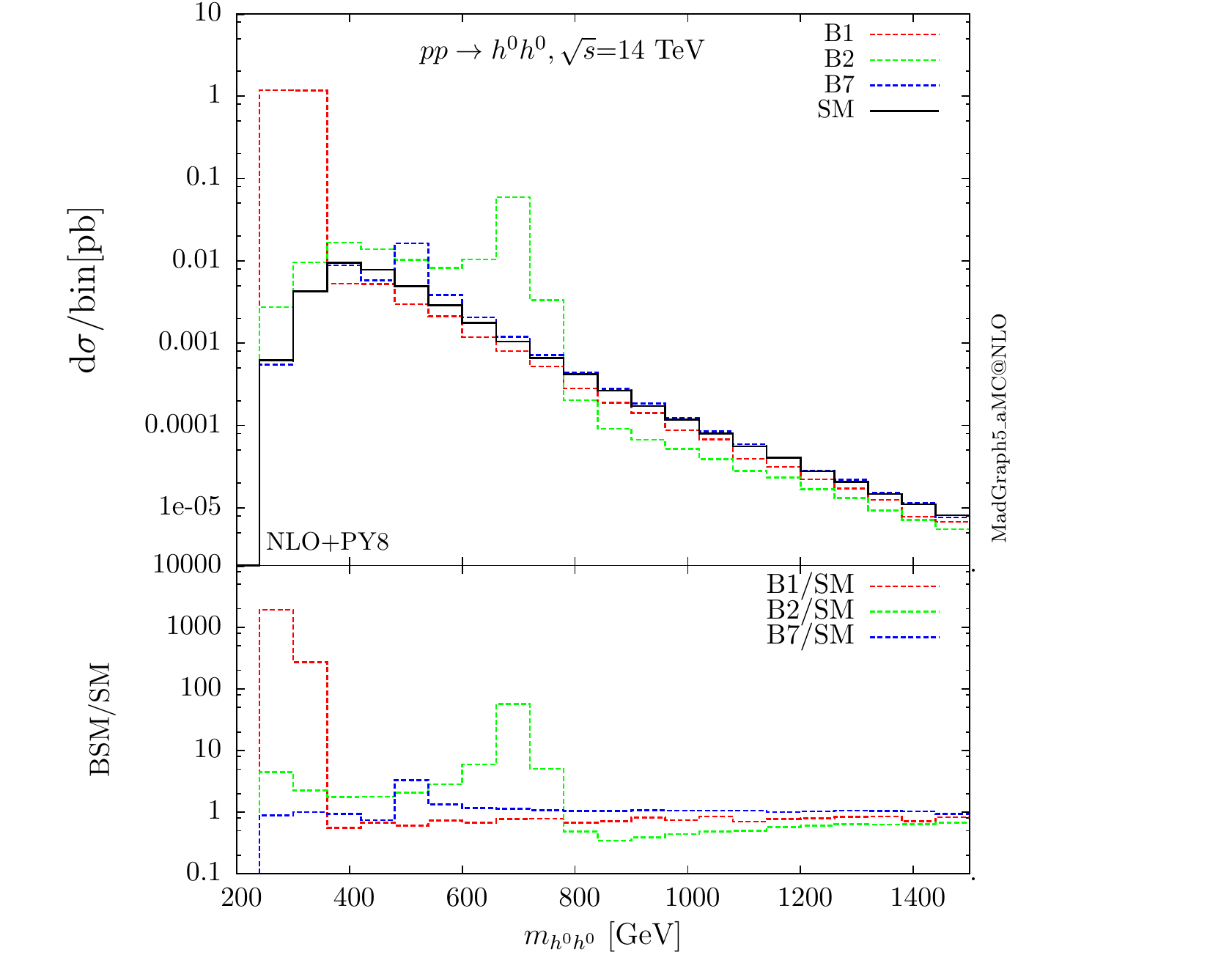}
\caption{\label{fig:2hdmfig}Invariant mass for the $pp\rightarrow h^0h^0$ process at NLO using the benchmarks
from \refT{tab:benchmarks}\cite{Hespel:2014sla} .}
\end{center}
\end{figure}

\clearpage
\section{Experimental results}

In Run~1 ATLAS and CMS performed searches for BSM di-Higgs boson production in gluon-gluon fusion process assuming resonant and nonresonant hypotheses. Taking into account the Higgs bosons decays, four different final states were explored.
One search requires both Higgs bosons to decay to $b\bar{b}$, 
that is the largest decay branching fraction within the SM.
In other two the second Higgs boson decays to $\gamma\gamma$ or $\tau\tau$ final states that help to reduce the SM background. 
The fourth channel, explored by ATLAS, features one  Higgs boson decaying to $WW^*$ with a subsequent leptonic decay and the other to $\gamma\gamma$.
A summary of the searches, obtained assuming a di-Higgs boson production through a spin-0 resonance in s-channel with a negligible natural width, is shown in \refF{fig:combined_plot}\footnote{One may notice than for some of the analyses a spin-2 interpretation is also available as well as an interpretation assuming a significant natural width~\cite{Aad:2015uka, Khachatryan:2015yea}.}. To compare different final states the decays branching fractions of the Higgs boson are assumed to be those of the SM.
Limits are provided from $\mXsz = 260$~GeV to $\mXsz = 3$~TeV and span over 3 orders of magnitude from typically 1-10 pb around the lowest edge and 1-10 fb around the highest edge. They are interpreted in the context of two simplified scenarios of the Minimal Supersymmetric Standard Model, 2 Higgs Doublet Model and Warped Extra Dimensions.

\begin{figure}[h!tb]
\centering
 \centering
  \includegraphics[width=0.9\textwidth]{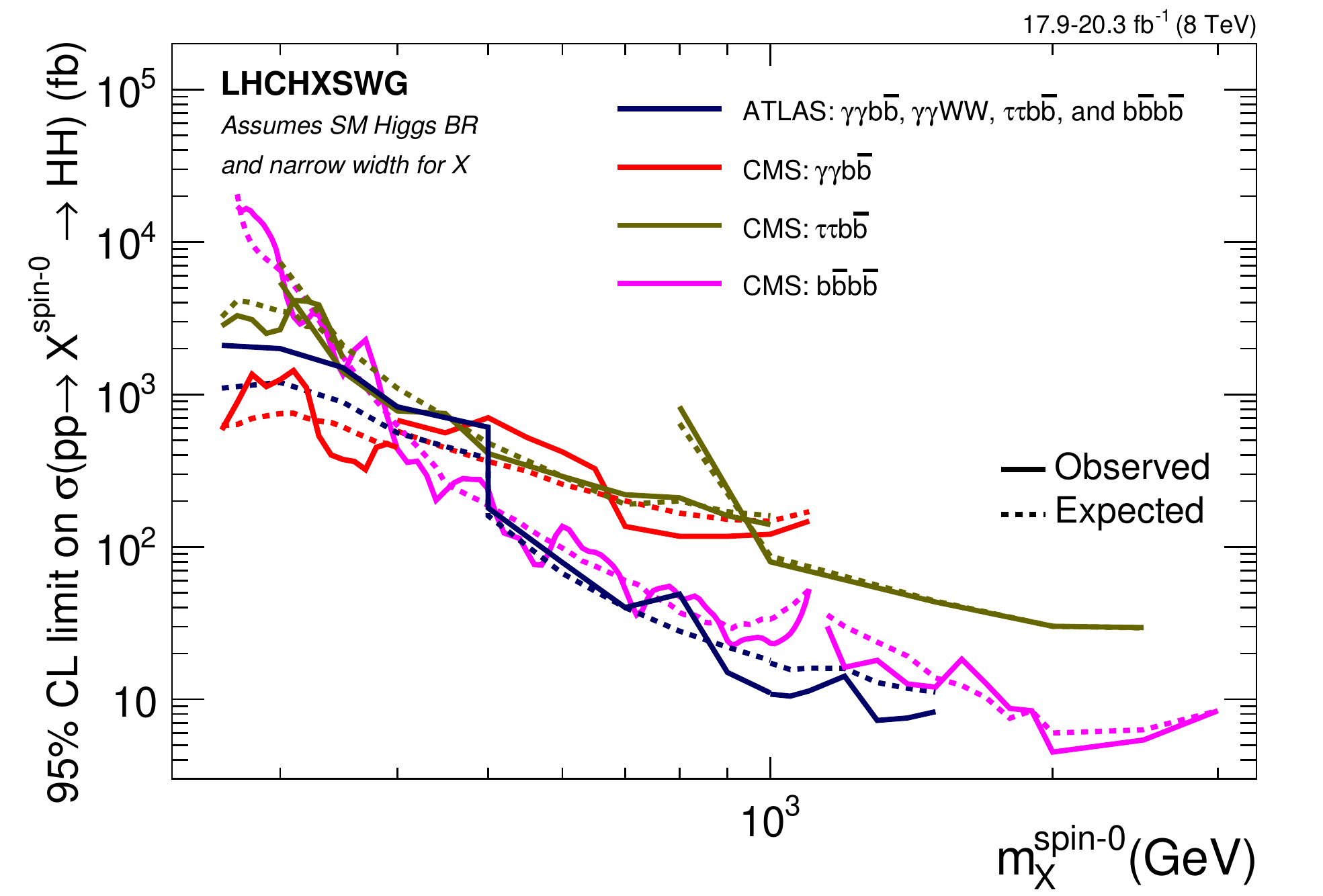}
  \caption{Comparison of the observed and expected 95\% confidence level (CL) upper limits on the product of cross section and the
  branching fraction $\sigma(pp \rightarrow \Xsz) \times \mathcal{B}(\Xsz \rightarrow \hsm\hsm)$. We assuming a narrow-width approximation for $\Xsz$ and SM branching fractions for Higgs boson decays. Results are provided by ATLAS and CMS collaborations based on results from Run~1 data taking period.\label{fig:combined_plot}}
\end{figure} 

The ATLAS collaboration performed searches at $\sqrt{s} = 8$~TeV
using an integrated luminosity of $20.3$~fb$^{-1}$~\cite{Aad:2014yja, Aad:2015uka, Aad:2015xja} and subsequently combined them for $\mXsz < 1$~TeV hypothesis in Ref.~\cite{Aad:2015xja}. The latter result is shown on \refF{fig:combined_plot} complemented with  $b\bar{b}b\bar{b}$ results for $\mXsz > 1$~TeV hypothesis.

Similar searches were performed by the CMS collaboration in $\gamma\gamma b\bar{b}$~\cite{Khachatryan:2016sey}, $\tau\tau b\bar{b}$~\cite{Khachatryan:2015tha, CMS:2016zxv, CMS:2015zug}, $b\bar{b} b\bar{b}$ \cite{Khachatryan:2015yea, Khachatryan:2016cfa} using a data sample of $17.9$ to $20.3$~fb$^{-1}$ depending on the analysis. The results obtained by different analyses looking at an identical final state are shown in \refF{fig:combined_plot} with the same colour. In particular in the case of $\tau\tau b\bar{b}$ final state, Ref.~\cite{Khachatryan:2015tha} was optimized to look for low-mass region $\mXsz < 350$~GeV; Ref.~\cite{CMS:2016zxv} concentrates its efforts on middle-mass region $300 < \mXsz < 1000$~GeV using boosted taus; Ref.~\cite{CMS:2015zug} extends the search up to 2.5 TeV looking at boosted Higgs bosons decaying to nearby taus and jets. Similar logic divides the low-mass~\cite{Khachatryan:2015yea}, $\mXsz < 1100$~GeV, and high-mass~\cite{Khachatryan:2016cfa}, $\mXsz > 1150$~GeV, searches in $b\bar{b} b\bar{b}$ final state.

The properties discussed below are valid for the results from both ATLAS and CMS collaborations.
In general, we observe that at low mass, $\mXsz \lesssim 350$~GeV, $\gamma\gamma b\bar{b}$ channel is the most sensitive one. It benefits from a good trigger efficiency looking online for a pair of photons, a good reconstruction efficiency of this pair and a low SM background. In contrast, above 500 GeV the most sensitive channel is $b\bar{b} b\bar{b}$. At high mass the trigger efficiency looking for 3 or 4 b-tagged jets with high $p_T$ improves compared to the low mass. Therefore the branching fraction of this channel provides a decisive advantage. The properties of the $\tau\tau b\bar{b}$ channel are intermediate between these two. The sensitivity of different channels crosses around $\mXsz \approx 400-500$~GeV depending on the exact details of each analysis. Finally the $\gamma\gamma WW^*$ channel is the less sensitive one since it benefits from all the advantages of $\gamma\gamma b\bar{b}$ channel, but suffers from a significantly lower branching fraction and reconstruction efficiency of $h \rightarrow WW^*$ compared to $h \rightarrow b\bar{b}$.

The searches for nonresonant Higgs boson pair production assuming SM like kinematics was performed by ATLAS in all four channels followed by a subsequent combination~\cite{Aad:2015xja} and by CMS in  $\gamma\gamma b\bar{b}$~\cite{Khachatryan:2016sey} and $\tau\tau b\bar{b}$~\cite{CMS:2016zxv} channels. The results are shown in Table \ref{tab:nonresonant}. The observed limits exceeds by at least a factor of 40 the total SM cross section provided in Section \ref{sec:hhxsec}. 
The CMS collaboration performed also a first generic nonresonant search within the framework of EFT using the $\gamma\gamma b\bar{b}$ final state~\cite{Khachatryan:2016sey}. This search was designed to exploit the shape properties of the nonresonant $m_{hh}$ spectrum already discussed in Section \ref{sec:hhbsm}. The limits were interpreted in term of parameters $\kappa_\lambda$, $\kappa_t$, and $c_2$ excluding $|c_2| > 3$ and $\kappa_\lambda < - 17.5$ or $\kappa_\lambda > 22.5$.

Although the Run~1 results are still far away from being sensitive to the SM $hh$ production we can already drive interesting conclusions for future analyses. The sensitivity of $\gamma\gamma b\bar{b}$, $\tau\tau b\bar{b}$ and $b\bar{b} b\bar{b}$ is similar for the SM like nonresonant search and for the resonant search with $\mXsz \approx 400$~GeV. This is not a coincidence, indeed the SM $m_{hh}$ spectrum exhibits a broad peak around 400 GeV. This means that a measurement of the SM $hh$ production would equally benefit from a combination of those three channels. This observation confirms the prospect from the ATLAS~\cite{ATL-PHYS-PUB-2014-019, ATL-PHYS-PUB-2015-046} and CMS~\cite{Butler:2020886} collaborations for the HL-LHC program. One shall notice that those prospects includes also $WW^*b\bar{b}$ as a promising channel~\cite{Butler:2020886} .

\begin{table}
\caption{\label{tab:nonresonant} Comparison of the observed and expected 95\% CL upper limits on the nonresonant cross section $\sigma(pp \rightarrow \hsm\hsm)$ assuming SM-like kinematics.}
\begin{center}
\def\arraystretch{1.3}
\begin{tabular}{cc|ll}
\toprule
Channel &  Experiment &  Observed (pb) &  Expected (pb) \\
\midrule
$\gamma\gamma b\bar{b}$ & ATLAS & 2.2 &  1.0 \\
$\gamma\gamma b\bar{b}$ & CMS & 0.71 & 0.60 \\
$\tau\tau b\bar{b}$  & ATLAS & 1.6 & 1.3 \\
$\tau\tau b\bar{b}$  & CMS & 0.59 & 0.94\\
$b\bar{b} b\bar{b}$  & ATLAS & 0.62 & 0.62 \\
$\gamma\gamma WW^*$  & ATLAS & 11.0 & 6.7 \\
\midrule
Combination & ATLAS & 0.69 & 0.47 \\
\bottomrule
\end{tabular}
\end{center}
\end{table}


\clearpage


\chapter{Off-shell Higgs Production and Higgs Interference}
\label{chap:offshell_interf}
\ChapterAuthor{F.~Caola, Y.~Gao, N.~Kauer, L.~Soffi, J.~Wang~(Eds.);
A.~Ballestrero, C.~Becot, F.~Bernlochner, H.~Brun, A.~Calandri, F.~Campanario, F.~Cerutti, D.~de Florian, R.~Di Nardo, L.~Fayard, N.~Fidanza, N.~Greiner, A.~V.~Gritsan, G.~Heinrich, B.~Hespel, S.~H{\"o}che, F.~Krauss, Y.~Li, S.~Liebler, E.~Maina, B.~Mansouli\'{e}, C.~O'Brien, S.~Pozzorini, M.~Rauch, J.~Roskes, U.~Sarica, M.~Schulze, F.~Siegert, P.~Vanlaer, E.~Vryonidou, G.~Weiglein, M.~Xiao, S.~Yuen}


\section{Introduction}

The Higgs boson measurements in the resonant region (on-peak) are broadly consistent
with Standard Model expectations.
The observed Higgs boson cross-sections are primarily measured via decays
into two electroweak bosons ($WW$, $ZZ$ and $\gamma\gamma$).
However, the measured on-peak cross-sections are affected by an intrinsic scaling 
ambiguity between the Higgs boson couplings and the total Higgs boson width: 
$\sigma_{i\to H\to f}\sim g_i^2 g_f^2/\Gamma_H$.
Disentangling this ambiguity would make it possible to constrain or even measure the 
total Higgs boson width at the LHC, which would be highly desirable.
The total width of the SM Higgs boson is about 4 MeV, and hence much smaller than the 
experimental resolution of the Higgs boson mass measurements in the two high-resolution 
channels $H\to 4\ell$ and $H\to \gamma\gamma$, which is of the order of 1 GeV.
For this reason, a direct measurement of the Higgs boson width is not feasible at the LHC.

A novel method has recently been proposed to constrain the Higgs boson
width using events away from the on-peak region in the decays into $ZZ$ and $WW$ 
\cite{Kauer:2012hd,Caola:2013yja,Campbell:2013una}.
The off-shell cross-section of $gg\to H^\ast\to VV$ contributes $\mathcal{O}(15\%)$ due to
two threshold effects, near $2 M_{V}$ from the Higgs boson decay and $2 m_{t}$ from the $gg\to H$ production.
The electroweak diboson continuum $gg\to VV$ plays an important role in this off-shell region,
mainly due to the large destructive interference with the $gg\to H^\ast\to VV$ signal.
At leading order, $gg\to VV$ proceeds through a box diagram, which makes higher
order calculations difficult.
In this off-shell region, where $M_{VV}\gg M_H$, the cross-section dependence on the total Higgs boson width is negligible, providing a unique opportunity to measure the absolute Higgs boson couplings.
The off-shell Higgs boson couplings can then be correlated with the on-shell cross-sections
to provide a novel indirect constraint on the total Higgs boson width.
It has been pointed out \cite{Englert:2014aca,Logan:2014ppa} that BSM physics that alters 
the relation between Higgs cross-sections in the on-peak and off-shell regions could invalidate the method as applied in \cite{Caola:2013yja,Campbell:2013una}.  Using future LHC 
data to constrain New Physics affecting the off-shell Higgs boson couplings is therefore
important \cite{Gainer:2014hha,Azatov:2014jga}.

The method has been promptly adopted by the CMS and ATLAS collaborations.
The analyses \cite{Khachatryan:2014iha,Aad:2015xua,Khachatryan:2016ctc} present constraints on the off-shell Higgs boson event
yields normalized to the Standard Model prediction (signal strength) in the
$ZZ$$\rightarrow$4$\ell$, $ZZ$$\rightarrow$2$\ell$2$\nu$ and $WW$$\rightarrow\ell\nu\ell\nu$ channels.
In the ATLAS analysis \cite{Aad:2015xua}, using the CLs method, the observed 95\% confidence level (CL)
upper limit on the off-shell signal strength is in the range 5.1--8.6, with an expected range of 6.7--11.0.
This range is determined by varying the unknown  \footnote{cf.\ \refS{sec:offshell_interf_ggVV_NLO}}
 $gg$$\rightarrow$ $ZZ$ and $gg$$\rightarrow$$ WW$ background K-factor
from higher-order QCD corrections between half and twice the value of the evaluated signal K-factor.
Under the assumption that the Higgs boson couplings are independent of the energy scale of the Higgs boson production,
a combination of the off-shell constraint with the on-shell Higgs peak measurement yields an observed (expected) 95\% CL
upper limit on the Higgs boson total width normalized to the one predicted by the Standard Model, i.e. $\Gamma_{\mathrm{H}}/\Gamma_{\mathrm{SM}}$,
in the range of 4.5--7.5 (6.5--11.2) employing the same variation of the background K-factor.
Assuming that the unknown gg$\rightarrow VV$ background K-factor is equal to the signal K-factor,
this translates into an observed (expected) 95\% CL upper limit on the Higgs boson total width of 22.7 (33.0) MeV.

In the CMS analysis of the $ZZ$ and $WW$ channels combined \cite{Khachatryan:2016ctc}, an observed (expected) upper limit on the off-shell Higgs boson event yield normalized to the Standard Model prediction of 2.4 (6.2) is obtained at the 95\% CL for the gluon fusion process and of 19.3 (34.4) for the VBF process. The observed and expected constraints on the Higgs boson total width are 13 MeV and 26 MeV, respectively, at the 95\% CL. Concerning the $gg\rightarrow$$VV$ background K-factor, the central values and uncertainties are assumed to be equal to those of the signal K-factor, with an additional 10\% uncertainty.

In addition to the off-shell $H^*\to VV$ channels, the $H\to\gamma\gamma$
channel also provides a very clean signature for probing Higgs boson properties, including its mass.
However, there is also a large continuum background $gg\to \gamma\gamma$
to its detection in this channel.
It is important to study how much the coherent interference between the Higgs boson signal
and the background could affect distributions in
diphoton observables, and possibly use it to constrain Higgs boson properties.
An interesting study\cite{Martin:2012xc,Dixon:2013haa} showed that this interference
can lead to a shift in the Higgs boson mass, which has a
strong dependence on the $p_T$ of the diphoton system and the total Higgs boson width.
This provides another way to constrain the Higgs boson width.



\section{Overview}

This chapter contains selected studies and benchmark results for
off-shell Higgs boson production and Higgs interference.
In \refS{sec:offshell_interf_VV}, theoretical and experimental studies
of the SM Higgs boson signal in the off-shell/high-mass region for 
the gluon-fusion and VBF $\PH\to \PV\PV$ channels ($\PV=\PW,\PZ$) 
including the interference with the background are presented.
More specifically, \refS{sec:offshell_interf_vv_settings} details the 
used input parameters and gives our recommendations for the QCD scale and 
the order of the gluon PDF and illustrates the corresponding cross section 
dependence.
Benchmark cross sections and distributions are collected
in \refS{sec:offshell_interf_vv_bench_sm} for the Standard Model, including recommended experimental selections for use in $\Pg\Pg\to \PV\PV$ calculations,
and for the Higgs Singlet Model in \refS{sec:offshell_interf_vv_bench_1hsm}.
Multi-jet merging and parton shower effects are discussed in 
Sections~\ref{sec:offshell_interf_vv_sherpa} and \ref{sec:offshell_interf_vv_atlas_mc_gg}.
Interference effects for heavy Higgs bosons or Higgs-like
resonances in SM extensions are illustrated in Sections~\ref{sec:offshell_interf_vv_jhugen_mcfm} and \ref{sec:offshell_interf_vv_gosam}.
In \refS{sec:offshell_interf_ggVV_NLO}, the status of NLO $\Pg\Pg\to \PV\PV$ calculations is reviewed, and $\Pg\Pg \to 4\Pl$ benchmark results and our recommendation for the 
treatment of the $\Pg\Pg (\to \PH) \to \PZ\PZ$ interference $K$-factor are 
given.
In \refS{sec:interf_2gamma}, the interference in the $\PH\to \PGg\PGg$ channel is discussed.  A theory overview is given and Monte Carlo interference implementations and related experimental studies are described.



\section{\texorpdfstring{$H\to VV$}{H to VV} modes (\texorpdfstring{$V=W,Z$}{V=W,Z})}
\label{sec:offshell_interf_VV}

\providecommand{\MVV}{\mathswitch {M_{\PV\PV}}}
\providecommand{\Mtltl}{\ensuremath{M_{2\Pl2\Pl}}}
\providecommand{\M}[1]{\mathswitch {M_{#1}}}
\providecommand{\Mtt}[1]{\ensuremath{M_{\mathrm{T}#1}}}
\providecommand{\ptt}[1]{\ensuremath{p_{{\mathrm{T}{#1}}}}}
\providecommand{\MpT}{\ensuremath{p_{\mathrm{T}}^{\mathrm{miss}}}}
\providecommand{\aaa}[1]{\HepAntiParticle{{#1}}{}{}\Xspace}


\subsection[Input parameters and recommendations for input parameters and PDF]{Input parameters and recommendations for the QCD scale and the 
order of the gluon PDF}
\label{sec:offshell_interf_vv_settings}

The SM input parameters for Higgs physics given in 
\Bref{LHCHXSWG-INT-2015-006} 
are adopted with the $G_\mu$ scheme: 
$\MW = 80.35797$\UGeV, $\MZ = 91.15348$\UGeV, $\GW = 2.08430$\UGeV, 
$\GZ = 2.49427$\UGeV, $\Mt = 172.5$\UGeV, $\Mb(\Mb) =
4.18$\UGeV\ and $\GF = 1.1663787\cdot 10^{-5}$\UGeV$^{-2}$. The CKM matrix
is approximated by the identity matrix.
Finite top and bottom quark mass effects 
are included.  Lepton and light quark masses are neglected.
Results are given for $\Pp\Pp$ collisions at $\sqrt{s}=13$\UTeV\ unless
otherwise noted. The PDF set \texttt{PDF4LHC15\_nlo\_100} \cite{Butterworth:2015oua} is used by default. All PDF sets are used with the default $\alpha_s$ 
of the set.
A fixed-width Breit-Wigner propagator $D(p) \sim (p^2 - M^2 + i M \Gamma)^{-1}$
is employed for $W,Z$ and Higgs bosons, where $M$ and $\Gamma$ are determined 
by the complex pole of the amplitude due to unstable particle propagation.%
\footnote{ In agreement with \textsc{HDECAY}, the $W$ and $Z$ masses and widths 
have been changed from physical on-shell masses to the pole values, see 
\Eq{} (7) in \Bref{LHCHXSWG-INT-2015-006}.  The relative deviation is at the 
$3\cdot 10^{-4}$ level.}
The SM Higgs boson mass is set to 125\UGeV.  The SM Higgs boson width parameter 
is calculated using \textsc{HDECAY} v6.50 \cite{Djouadi:1997yw}. For $\MH=125$\UGeV\ one obtains $\GH= 4.097\cdot 10^{-3}$\UGeV.

For off-shell and high-mass $\PH\to \PV\PV$ cross-section and interference
calculations, we recommend and employ the QCD scale $\mu_R=\mu_F=\MVV/2$
unless otherwise noted.  Next, we elucidate the choice of the PDF order
for the $\Pg\Pg\to \PV\PV$ continuum background and the corresponding 
Higgs-continuum interference.
Combining any $n$-order PDF fit with a $m$-order 
parton-level calculation is theoretically consistent 
as long as $n\ge m$.  Deviations are expected to be
of higher order if the same $\alphas(\MZ)$ is used.
But, using a LO gluon PDF with $\alphas(\MZ)$ 
obtained in the LO fit is not recommended: The
gluon PDF is mostly determined by DIS data, especially in the 
SM Higgs region.  
At LO, DIS does not have a gluon channel. It only enters at 
NLO, with a large $K$-factor. A LO fit cannot properly account for 
this $\Ord(50\%)$ contribution, but incorrectly adjusts the gluon
evolution to compensate, which results in an overestimated value of 
$\alphas(\MZ)$ of approximately $0.13$.
We therefore recommend using a NLO PDF set when computing 
the $\Pg\Pg\ (\to \PH)\to \PV\PV$ interference 
and the $\Pg\Pg$ continuum background at LO as well as NLO. 
For consistency, we also use the NLO PDF set for the 
corresponding signal process.\footnote{We note that the LO and NLO VBF results
have also been obtained with the NLO PDF set.}

The variation induced by different PDF and QCD scale 
choices is illustrated in 
Tables\ \ref{tab:ofs:pdfmin}, \ref{tab:ofs:pdfcms}, \ref{tab:ofs:scalemin}
and \ref{tab:ofs:scalecms} 
using the process $\Pg\Pg\ (\to \PH)\to \Pl\PAl\Pl'\PAl'$.
The Higgs boson signal ($S$), $\Pg\Pg$ background ($B$) and
the signal-background interference ($I$) are displayed
at LO for four Higgs boson invariant mass regions:
\begin{itemize}
\item \textit{off-shell} (OFS): $\MVV > 140$\UGeV 
\item \textit{off-shell high-mass (interference)} (HM1): $220 < \MVV < 300$\UGeV 
\item \textit{off-shell high-mass (signal enriched)} (HM2): $\MVV > 300$\UGeV 
\item \textit{resonance} (RES): $110 < \MVV < 140$\UGeV 
\end{itemize}
Motivated by the Higgs boson width constraints of \Brefs{Khachatryan:2014iha,Aad:2015xua}, the off-shell high-mass region is divided into the 
interference-sensitive (HM1) and signal-enriched (HM2) regions.
Two sets of selection cuts are considered:
\begin{itemize}
\item \textit{minimal cuts} (MIN): $\M{\Pl\PAl}>10$\UGeV, $\M{\Pl'\PAl'}>10$\UGeV
\item \textit{CMS} $\PH\to 4\Pl$ \textit{cuts} (CMS): $\ptt{1}>20$\UGeV, $\ptt{2}>10$\UGeV, $\ptt{3,4}>5$\UGeV, $|\eta_{\Pe}|<2.5$, $|\eta_{\PGm}|< 2.4$, $\M{\Pe\aaa{\Pe}}>4$\UGeV, $\M{\PGm\aaa{\PGm}}>4$\UGeV
\end{itemize}
The PDF4LHC15 \cite{Butterworth:2015oua} NLO and NNLO sets ($\alphas(\MZ)=0.118$) and the CT14 \cite{Dulat:2015mca} LO sets with $\alphas(\MZ)=0.130$ and $\alphas(\MZ)=0.118$ (and $1$- and $2$-loop evolution, respectively) are compared in Tables\ \ref{tab:ofs:pdfmin} and \ref{tab:ofs:pdfcms}.  As expected, the deviations
for PDF sets with $\alphas(\MZ)=0.118$ are of order $10\%$ or less while 
the LO set with $\alphas(\MZ)=0.130$ yields results that differ by up to 
$30\%$.  The deviation between the NLO and NNLO sets is at the per cent level.
Furthermore, different choices for the QCD scale $\mu=\mu_R=\mu_F$ are compared
in Tables\ \ref{tab:ofs:scalemin} and \ref{tab:ofs:scalecms}.
As central scale choices, the dynamic scale $\mu_0=\Mtltl/2$ and the fixed 
scales $\MH/2$ and $\MZ$ are considered.  The LO scale variation is 
estimated for $\mu_0$ using the scales $\mu_0/2$ and $2\mu_0$.  
The results illustrate that using a fixed scale appropriate for
resonant signal or background will significantly overestimate the 
signal, background and interference cross sections in the far off-shell 
and high-mass regions.  With the recommended central scale $\Mtltl/2$,
a factor-two scale variation yields a LO scale uncertainty of 
$20\%{-}25\%$ for the off-shell signal and signal plus background interference.
The results of these comparisons were calculated using 
\textsc{gg2VV} \cite{Kauer:2012hd}.

\begin{table}
\caption{PDF dependence of off-shell $\Pg\Pg\ (\to \PH)\to \Pl\PAl\Pl'\PAl'$ 
cross sections at LO in \Ufb\ for one lepton flavour combination. 
MIN cuts are applied.  $R$ is the ratio of NNLO, LO result to NLO 
result.  The bottom rows show the ratio of OFS, HM1, HM2 to RES result
for $S$ and $S+I$. The recommended QCD scale $\mu_R=\mu_F=\Mtltl/2$ is used.
The MC error is given in brackets.
See main text for other details.}
\label{tab:ofs:pdfmin}%
\renewcommand{\arraystretch}{1.2}%
\setlength{\tabcolsep}{1.5ex}%
\begin{center}
\footnotesize
\begin{tabular}{llccccccc}
\toprule
\multicolumn{2}{c}{} & \multicolumn{7}{c}{PDF set order} \\
\cmidrule(lr){3-9}
Reg. & Amp. & \textbf{NLO} & NNLO & $R$ & LO($0.118$) & $R$ & LO($0.130$) & $R$ \\
\midrule
    & $S$  & $0.1266(1)$  & $0.1255(1)$ & $0.991(2)$  & $0.1255(1)$  & $0.992(2)$  & $0.1414(2)$  & $1.116(2)$ \\
OFS & $S+I$  & $-0.1313(2)$ & $-0.1298(2)$ & $0.988(2)$   & $-0.1307(2)$  & $0.995(2)$  & $-0.149(1)$ & $1.138(8)$  \\
    & $B$  & $2.988(4)$  & $2.945(5)$ & $0.986(2)$  & $2.960(4)$  & $0.991(2)$  & $3.448(5)$  & $1.154(3)$ \\
\midrule
    & $S$  & $0.01933(4)$ & $0.01906(4)$ & $0.986(3)$ & $0.01899(4)$ & $0.982(3)$ & $0.02210(5)$  & $1.143(4)$ \\
HM1 & $S+I$  & $-0.04550(8)$  & $-0.04475(8)$  & $0.984(3)$  & $-0.04486(7)$  & $0.986(3)$  & $-0.0516(6)$  & $1.13(2)$ \\
    & $B$  & $1.182(3)$ & $1.165(3)$ & $0.985(3)$ & $1.166(3)$ & $0.986(3)$ & $1.354(3)$ & $1.145(4)$ \\
\midrule
    & $S$  & $0.0981(1)$  & $0.0974(1)$  & $0.993(2)$  & $0.0973(1)$  & $0.992(2)$  & $0.1084(2)$  & $1.105(2)$ \\
HM2 & $S+I$  & $-0.0465(1)$ & $-0.04622(9)$ & $0.994(3)$ & $-0.04637(9)$ & $0.997(3)$ & $-0.0522(6)$ & $1.12(2)$ \\
    & $B$  & $0.611(2)$  & $0.605(2)$  & $0.990(4)$  & $0.598(2)$  & $0.980(4)$  & $0.676(2)$  & $1.107(5)$ \\
\midrule
    & $S$  & $0.800(1)$ & $0.780(1)$ & $0.976(2)$ & $0.843(1)$ & $1.054(2)$ & $1.021(2)$ & $1.276(3)$ \\
RES & $S+I$  & $0.803(2)$  & $0.784(2)$  & $0.976(4)$  & $0.845(4)$  & $1.052(6)$  & $1.023(3)$  & $1.274(5)$ \\
    & $B$  & $0.1092(2)$ & $0.1063(2)$ & $0.974(2)$ & $0.1150(2)$ & $1.053(3)$ & $0.1389(2)$ & $1.272(3)$ \\
\toprule
OFS/ & $S$   & $0.1583(3)$  & $0.1609(3)$  &  & $0.1490(3)$  &  & $0.1385(3)$  &  \\
RES  & $S+I$ & $-0.1635(4)$  & $-0.1655(5)$  &  & $-0.1547(7)$ &  & $-0.146(2)$  &  \\
HM1/ & $S$   & $0.02418(6)$  & $0.02443(6)$  &  & $0.02253(6)$  &  & $0.02165(5)$  &  \\
RES  & $S+I$ & $-0.0566(2)$ & $-0.0571(2)$ &  & $-0.0531(3)$  &  & $-0.0504(6)$  &  \\
HM2/ & $S$   & $0.1227(2)$  &  $0.1249(3)$   &  & $0.1155(2)$  &  & $0.1062(2)$  &  \\
RES  & $S+I$ & $-0.0579(2)$  & $-0.0589(2)$  &  & $-0.0549(3)$  &  & $-0.0510(6)$  &  \\
\bottomrule
\end{tabular}
\end{center}
\end{table}

\begin{table}
\caption{PDF dependence of off-shell 
$\Pg\Pg\ (\to \PH)\to \Pem\Pep\PGmm\PGmp$ 
cross sections at LO in \Ufb. 
CMS cuts are applied.  Other details as in \refT{tab:ofs:pdfmin}.}
\label{tab:ofs:pdfcms}%
\renewcommand{\arraystretch}{0.8}%
\setlength{\tabcolsep}{1.5ex}%
\begin{center}
\footnotesize
\begin{tabular}{llccccccc}
\toprule
\multicolumn{2}{c}{} & \multicolumn{7}{c}{PDF set order} \\
\cmidrule(lr){3-9}
Reg. & Amp. & \textbf{NLO} & NNLO & $R$ & LO($0.118$) & $R$ & LO($0.130$) & $R$ \\
\midrule
    & $S$  & $0.0952(3)$  & $0.09396(8)$  & $0.986(3)$  & $0.09034(7)$  & $0.949(3)$  & $0.10191(8)$  & $1.070(3)$ \\
OFS & $S+I$  & $-0.0893(3)$  & $-0.0883(1)$  & $0.989(3)$  & $-0.08436(9)$  & $0.944(3)$  & $-0.0973(1)$  & $ 1.089(4)$ \\
    & $B$  & $1.869(3)$  & $1.841(3)$  & $0.985(2)$  & $1.736(2)$  & $0.928(2)$  & $2.033(3)$  & $1.088(2)$ \\
\midrule
        & $S$  & $0.01303(9)$  & $0.01278(3)$  & $0.981(7)$  & $0.01200(3)$  & $0.921(7)$  & $0.01402(3)$  & $1.076(8)$ \\
HM1 & $S+I$  & $-0.0298(2)$  & $-0.02942(6)$  & $0.986(6)$  & $-0.02759(5)$  & $0.925(5)$  & $-0.03227(6)$  & $1.082(6)$ \\
    & $B$  & $0.738(2)$  & $0.727(2)$  & $ 0.986(3)$  & $0.679(2)$  & $0.920(3)$  & $0.795(2)$  & $1.079(4)$ \\
\midrule
    & $S$  & $0.0761(3)$  & $0.07531(8)$  & $ 0.990(4)$  & $0.07271(7)$  & $0.956(3)$  & $0.08123(8)$  & $1.067(4)$ \\
HM2 & $S+I$  & $-0.0349(2)$  & $-0.03471(7)$  & $ 0.994(6)$  & $-0.03376(6)$  & $0.967(6)$  & $-0.03757(7)$  & $1.076(7)$ \\
    & $B$  & $0.382(2)$  & $0.377(2)$  & $0.987(5)$  & $0.353(1)$  & $ 0.925(5)$  & $0.403(2)$  & $1.055(5)$ \\

\midrule
    & $S$  & $0.4392(7)$  & $0.4284(7)$  & $0.975(3)$  & $0.4343(7)$  & $0.989(3)$  & $0.5267(8)$  & $1.199(3)$ \\
RES & $S+I$  & $0.439(2)$  & $0.428(2)$  & $0.975(4)
$  & $0.433(2)$  & $0.988(4)$  & $0.527(2)$  & $1.200(5)$ \\
    & $B$  & $0.06294(8)$  & $0.06155(8)$  & $0.978(2)$  & $0.06243(9)$  & $ 0.992(2)$  & $0.0755(1)$  & $1.200(3)$ \\
\toprule
OFS/ & $S$   & $0.2169(7)$  & $0.2193(4)$  &  & $0.2080(4)$  &  & $0.1935(4)$  &  \\
RES  & $S+I$ & $-0.2036(8)$  & $-0.2065(6)$  &  & $-0.1946(6)$  &  & $ -0.1847(6)$  & \\
HM1/ & $S$   & $0.0297(2)$  & $0.02984(8)$  &  & $0.02762(8)$  &  & $0.02662(7)$  &  \\
RES  & $S+I$ & $-0.0680(4)$  & $-0.0688(3)$  &  & $-0.0637(3)$  &  & $-0.0613(2)$  &  \\
HM2/ & $S$   & $0.1733(6)$  & $0.1758(4)$  &  & $0.1674(4)$  &  & $0.1542(3)$  &  \\
RES  & $S+I$ & $-0.0796(5)$  & $-0.0811(3)$  &  & $-0.0779(3)$  &  & $-0.0714(3)$  &  \\
\bottomrule
\end{tabular}
\end{center}
\end{table}


\begin{table}

\vspace{-0.9cm}

\caption{QCD scale $\mu=\mu_R=\mu_F$ dependence and symmetric scale uncertainty of 
off-shell $\Pg\Pg\ (\to \PH)\to \Pl\PAl\Pl'\PAl'$ 
cross sections at LO in \Ufb\ for one lepton-flavour combination. 
MIN cuts are applied.  $R$ is the ratio of the result to the cross section with 
the recommended scale choice $\mu=\Mtltl/2$.    
As recommended, the NLO PDF set is used. Other details as in \refT{tab:ofs:pdfmin}.}
\label{tab:ofs:scalemin}%
\renewcommand{\arraystretch}{1}%
\setlength{\tabcolsep}{1.5ex}%
\begin{center}
\footnotesize
\begin{tabular}{llccccccc}
\toprule
\multicolumn{2}{c}{} & \multicolumn{3}{c}{Dynamic scale} & \multicolumn{4}{c}{Fixed scales} \\
\cmidrule(lr){3-9}
     &      &                   & $\Delta(\Mtltl)$ & $R$ &  &  &  & \\
Reg. & Amp. & \textbf{$\Mtltl/2$} & $\Delta(\Mtltl/4)$ & $R$ &  $\MH/2$ & $R$ & $\MZ$ & $R$ \\
     &      &                   & symmetr.\  $\Delta$  & $R$ &  &  &  & \\
\midrule
    & $S$  &   & $-0.0258(2)$  & $-0.204(2)$  &   &   &   &  \\
    & $S$  & $0.1266(1)$ &  $0.0349(2)$ &  $0.276(2)$ & $0.2038(2)$ & $1.610(2)$ & $0.1760(2)$& $1.390(2)$\\
    & $S$  &   & $\pm0.0303(2)$  & $\pm0.240(1)$  &   &   &   &  \\
    & $S+I$  &   & $0.0251(2)$  & $0.182(2)$  &   &   &   &  \\
OFS & $S+I$  & $-0.1313(2)$ & $-0.0328(2)$ & $-0.250(2)$& $-0.1831(2)$ & $1.394(2)$ & $-0.1604(2)$ & $1.221(2)$ \\
    & $S+I$  &   & $\pm0.0290(2)$  &$\pm0.221(1)$ &   &   &   &  \\
    & $B$  &   &  $-0.545(5)$ & $-0.182(2)$  &   &   &   &  \\
    & $B$  & $2.988(4)$ & $0.699(7)$ &  $0.234(3)$ & $3.751(4)$ & $1.255(3)$ & $3.327(4)$ & $1.114(2)$ \\
    & $B$  &   &  $\pm0.6225(4)$ & $\pm0.209(2)$    &   &   &   &  \\
\midrule
    & $S$  &   &  $-0.00355(4)$ &$-0.184(3)$  &   &   &   &  \\
    & $S$  & $0.01928(3)$ &$0.00455(6)$ &  $0.236(3)$ & $0.02406(6)$ & $1.248(4)$ & $0.02150(5)$ & $1.115(3)$ \\
    & $S$  &   &  $\pm0.00405(4)$ & $\pm0.210(2)$  &   &   &   &  \\
    & $S+I$  &   & $0.0085(1)$ & $0.187(3)$  &   &   &   &  \\
HM1 & $S+I$  & $-0.04553(8)$ &  $-0.0106(2)$& $-0.233(3)$ & $-0.0561(1)$ & $1.233(3)$ & $-0.05002(9)$ & $1.099(3)$ \\
    & $S+I$  &   & $\pm0.0096(1)$  & $\pm0.2095(2)$  &   &   &   &  \\
    & $B$  &   & $-0.223(4)$  & $-0.188(3)$ &   &   &   &  \\
    & $B$  & $1.186(3)$ &  $0.273(5)$ & $0.230(4)$ & $1.462(3)$ & $1.232(4)$ & $1.302(3)$ & $1.098(4)$ \\
    & $B$  &   & $\pm0.248(2)$  & $\pm0.209(3)$  &   &   &   &  \\
\midrule
    & $S$  &   &$-0.0207(2)$  & $-0.211(2)$  &   &   &   &  \\
    & $S$  & $0.0982(2)$& $0.0284(2)$ & $0.289(2)$ & $0.1693(2)$ & $1.724(3)$ & $0.1451(2)$ & $1.478(3)$ \\
    & $S$  &   & $\pm0.0246(2)$  & $\pm0.250(2)$  &   &   &   &  \\
    & $S+I$  &   & $0.0099(2)$  & $0.212(3)$  &   &   &   &  \\
HM2 & $S+I$  & $-0.04651(8)$ & $-0.0136(2)$ & $-0.293(3)$ & $-0.0818(2)$ & $1.760(5)$ & $-0.0700(2)$ & $1.505(4)$ \\
    & $S+I$  &   & $\pm0.0118(1)$  & $\pm0.253(2)$  &   &   &   &  \\
    & $B$  &   & $-0.123(2)$ & $-0.201(3)$  &   &   &   &  \\
    & $B$  & $0.610(1)$ &  $0.167(3)$ & $0.275(5)$ & $0.929(3)$ & $1.524(5)$ & $0.807(2)$ & $1.323(4)$ \\
    & $B$  &   & $\pm0.145(2)$  & $\pm0.238(3)$  &   &   &   &  \\
\midrule
    & $S$  &   & $-0.115(2)$  & $-0.143(2)$  &   &   &   &  \\
    & $S$  & $0.800(1)$ & $0.131(2)$ & $0.164(2)$ & $0.801(2)$ & $1.001(2)$ & $0.737(1)$ & $0.921(2)$ \\
    & $S$  &   & $\pm0.123(2)$  & $\pm0.154(2)$ &   &   &   &  \\
    & $S+I$  &   & $-0.116(3)$  & $-0.145(2)$  &   &   &   &  \\
RES & $S+I$  & $0.803(2)$ & $0.130(3)$ & $0.162(3)$ & $0.803(2)$ & $1.000(3)$ & $0.739(2)$ & $0.920(3)$ \\
    & $S+I$  &   & $\pm0.123(2)$  & $\pm0.153(2)$ &   &   &   &  \\
    & $B$  &   & $-0.0158(3)$ & $-0.145(3)$  &   &   &   &  \\
    & $B$  & $0.1092(2)$ & $0.0176(3)$ &  $0.162(3)$ & $0.1089(2)$ & $0.998(2)$ & $0.1002(2)$ & $0.917(2)$ \\
    & $B$  &   & $\pm0.0167(2)$  & $\pm0.153(2)$ &   &   &   &  \\
\toprule
OFS/ & $S$   & $0.1583(3)$  &   &  & $0.2545(5)$  &  &$0.2389(4)$ &  \\
RES  & $S+I$ & $-0.1635(4)$  &   &  & $-0.2279(5)$  &  & $-0.2172(5)$  &  \\
HM1/ & $S$   & $0.02411(5)$  &   &  & $0.03005(8)$ &  & $0.02918(8)$ &  \\
RES  & $S+I$ & $-0.0567(2)$  &   &  & $-0.0699(2)$  &  & $-0.0677(2)$  &  \\
HM2/ & $S$   & $0.1228(3)$  &   &  & $0.2114(4)$ &  & $0.1970(4)$ &  \\
RES  & $S+I$ & $-0.0579(2)$  &   &  & $-0.1019(3)$  &  & $-0.0948(3)$  &  \\
\bottomrule
\end{tabular}
\end{center}
\end{table}


\begin{table}

\vspace{-0.1cm}

\caption{QCD scale $\mu=\mu_R=\mu_F$ dependence and symmetric scale uncertainty of 
off-shell $\Pg\Pg\ (\to \PH)\to \Pl\PAl\Pl'\PAl'$ 
cross sections at LO in \Ufb\ for one lepton-flavour combination. 
CMS cuts are applied.  Other details as in \refT{tab:ofs:scalemin}.}
\label{tab:ofs:scalecms}%
\renewcommand{\arraystretch}{1.2}%
\setlength{\tabcolsep}{1.5ex}%
\begin{center}
\footnotesize
\begin{tabular}{llccccccc}
\toprule
\multicolumn{2}{c}{} & \multicolumn{3}{c}{Dynamic scale} & \multicolumn{4}{c}{Fixed scales} \\
\cmidrule(lr){3-9}
     &      &                   & $\Delta(\Mtltl)$ & $R$ &  &  &  & \\
Reg. & Amp. & \textbf{$\Mtltl/2$} & $\Delta(\Mtltl/4)$ & $R$ &  $\MH/2$ & $R$ & $\MZ$ & $R$ \\
     &      &                   & symmetr.\  $\Delta$  & $R$ &  &  &  & \\
\midrule
    & $S$  &   & $-0.0196(3)$  & $ -0.206(4)$  &   &   &   &  \\
    & $S$  & $0.0952(3)$ & $0.0257(4)$ & $0.270(4)$ & $0.1545(4)$ & $1.622(6)$ & $0.1338(4)$ & $1.405(5)$ \\
    & $S$  &   & $\pm0.0227(3)$  & $\pm0.238(3)$  &   &   &   &  \\
    & $S+I$  &   & $ 0.0164(4)$  & $0.184(4)$  &   &   &   &  \\
OFS & $S+I$  & $-0.0893(3)$ & $-0.0223(4)$ & $-0.250(5)$ & $-0.1282(4)$ & $1.435(6)$ & $-0.1119(3)$ & $1.253(5)$ \\
    & $S+I$  &   & $\pm0.0194(3)$  & $\pm0.217(3)$  &   &   &   &  \\
    & $B$  &   &  $-0.331(4)$ & $-0.177(2)$  &   &   &   &  \\
    & $B$  & $1.869(3)$ & $0.430(4)$ & $ 0.230(2)$ & $2.341(3)$ & $1.252(3)$ & $2.084(3)$ & $1.115(2)$ \\
    & $B$  &   &  $\pm0.381(3)$ & $\pm0.204(2)$  &   &   &   &  \\
\midrule
    & $S$  &   &  $-0.00235(3)$ & $-0.181(2)$  &   &   &   &  \\
    & $S$  & $0.01302(2)$ & $0.00303(3)$ & $0.233(3)$ & $0.0163(2)$ & $1.25(1)$ & $0.0145(1)$ & $1.115(8)$ \\
    & $S$  &   &  $\pm0.00269(2)$ & $\pm0.207(2)$  &   &   &   &  \\
    & $S+I$  &   & $0.00536(6)$  & $0.179(2)$  &   &   &   &  \\
HM1 & $S+I$  & $-0.02986(5)$ & $-0.00682(7)$ & $-0.228(3)$ & $-0.0370(2)$ & $1.241(7)$ & $-0.0326(2)$ & $1.092(6)$ \\
    & $S+I$  &   & $\pm0.00609(5)$  & $\pm0.204(2)$  &   &   &   &  \\
    & $B$  &   & $-0.132(2)$  & $-0.178(2)$ &   &   &   &  \\
    & $B$  & $0.739(1)$ & $0.168(2)$ & $0.227(3)$ & $0.908(2)$ & $1.229(3)$ & $0.811(2)$ & $1.097(3)$ \\
    & $B$  &   & $\pm0.150(1)$  & $\pm0.203(2)$  &   &   &   &  \\
\midrule
    & $S$  &   & $-0.0160(2)$  & $-0.210(2)$  &   &   &   &  \\
    & $S$  & $0.0761(1)$ & $0.0218(2)$ & $0.286(3)$ & $0.1315(4)$ & $1.727(6)$ & $0.1131(4)$ & $1.485(5)$ \\
    & $S$  &   & $\pm0.0189(1)$  & $\pm0.248(2)$  &   &   &   &  \\
    & $S+I$  &   & $0.00740(7)$  & $0.211(2)$  &   &   &   &  \\
HM2 & $S+I$  & $-0.03505(6)$ & $-0.01006(9)$ & $-0.287(3)$ & $-0.0630(3)$ & $1.798(9)$ & $-0.0537(3)$ & $1.533(8)$ \\
    & $S+I$  &   & $\pm0.0088(1)$  & $\pm0.249(2)$  &   &   &   &  \\
    & $B$  &   & $-0.0768(8)$ & $-0.201(2)$  &   &   &   &  \\
    & $B$  & $0.3822(6)$ & $0.1019(9)$ & $0.267(3)$ & $0.582(2)$ & $ 1.522(5)$ & $0.506(2)$ & $1.324(4)$ \\
    & $B$  &   & $\pm0.090(1)$  & $\pm0.234(2) $  &   &   &   &  \\
\midrule
    & $S$  &   & $-0.0603(9)$  & $-0.137(2)$  &   &   &   &  \\
    & $S$  & $ 0.4392(7)$ & $0.066(1)$ & $0.151(3)$ & $0.4389(7)$ & $0.999(3)$ & $0.4044(6)$ & $0.921(2)$ \\
    & $S$  &   & $\pm0.064(2)$  & $\pm0.145(2)$ &   &   &   &  \\
    & $S+I$  &   & $-0.060(2)$  & $-0.136(4)$  &   &   &   &  \\
RES & $S+I$  & $0.439(2)$ & $0.067(2)$ & $ 0.154(5)$ & $0.438(2)$ & $0.999(4)$ & $ 0.406(2)$ & $ 0.925(4)$ \\
    & $S+I$  &   & $\pm0.064(2)$  & $\pm0.145(3)$ &   &   &   &  \\
    & $B$  &   & $-0.0086(2)$  & $-0.136(2)$  &   &   &   &  \\
    & $B$  & $0.06294(8)$ & $0.0097(2)$ & $0.155(2)$ & $0.06302(9)$ & $1.001(2)$ & $0.05816(8)$ & $0.924(2)$ \\
    & $B$  &   & $\pm0.0092(1)$  & $\pm0.146(2)$ &   &   &   &  \\
\toprule
OFS/ & $S$   & $0.2169(7)$  &   &  & $ 0.352(1)$  &  & $0.331(1)$  &  \\
RES  & $S+I$ & $-0.2036(8)$  &   &  & $-0.292(2)$  &  & $-0.276(2)$  &  \\
HM1/ & $S$   & $0.02964(6)$  &   &  & $0.0371(3)$  &  & $0.0359(3)$  &  \\
RES  & $S+I$ & $-0.0681(3)$  &   &  & $-0.0845(5)$  &  & $-0.0804(5)$  &  \\
HM2/ & $S$   & $0.1734(4)$  &   &  & $0.300(1)$  &  & $0.280(1)$  &  \\
RES  & $S+I$ & $-0.0799(3)$  &   &  & $-0.1437(8)$  &  & $-0.1325(7)$  &  \\
\bottomrule
\end{tabular}
\end{center}
\end{table}


\subsection[Off-shell and interference benchmarks: Standard Model]{Off-shell and interference benchmark cross sections and distributions: Standard Model}

\label{sec:offshell_interf_vv_bench_sm}

Gluon-fusion SM benchmark results were computed with \texttt{gg2VV} \cite{Kauer:2012hd} (see also \Brefs{Binoth:2005ua,Binoth:2006mf,Kauer:2013qba,Kauer:2015dma}) and 
\texttt{MG5\_aMC@NLO} \cite{Alwall:2014hca,Hirschi:2015iia} (see also \Bref{Kauer:2015dma}).  The \texttt{gg2VV} and \texttt{MG5\_aMC} 
results were found to be in good agreement.
Benchmark cross sections for $\Pg\Pg\ (\to \PH)\to \PV\PV \to 4$ leptons 
processes in $\Pp\Pp$ collisions at $\sqrt{s}=13\UTeV$ in the SM
are given in \refT{tab:ofs:leptonicxs}.
Results for the Higgs boson signal, the signal including signal-background
interference as well as the interfering background without Higgs contribution 
are displayed for minimal cuts 
$M_{\Pl\PAl} > 10\UGeV, M_{\Pl'\PAl'} > 10\UGeV$.  
The cross sections are calculated at loop-induced leading order.  
The recommended next-to-leading 
order $K$-factor is discussed in \refS{sec:offshell_interf_ggVV_NLO}.
Similarly, in \refT{tab:ofs:semileptxs} benchmark cross sections for $\Pg\Pg\ (\to \PH)\to \PV\PV \to$ semileptonic final states
in $\Pp\Pp$ collisions at $\sqrt{s}=13\UTeV$ in the SM are given.
As above, the signal amplitude is calculated at loop-induced leading order.
But, for the semileptonic decay modes, the $\Ord(g_{\mathrm{s}}^2e^2)$ tree-level as 
well as the important loop-induced $\Ord(g_{\mathrm{s}}^2e^4)$ amplitude contributions 
to the interfering background are taken into 
account \cite{Kauer:2015dma}.
The following minimal cuts are applied: $M_{\Pl\PAl} > 10\UGeV, M_{\PQq\PAQq} > 10\UGeV,
\ptt{j}>25\UGeV$.
Higgs boson invariant mass distributions corresponding to the cross sections given 
in \refTs{tab:ofs:leptonicxs} and \ref{tab:ofs:semileptxs} are displayed in 
\refFs{fig:ofs:ZAZA2l2l_ZAZA4l}, \ref{fig:ofs:ZAZ_2l2v_WW2l2v}
and \ref{fig:ofs:WWZAZ_2l2v_WWlvqq}.

\begin{table}
\caption{\label{tab:ofs:leptonicxs}
Cross sections (\UfbZ) for $\Pg\Pg\ (\to \PH)\to \PV\PV \to 4$ leptons processes in 
$\Pp\Pp$ collisions at $\sqrt{s}=13\UTeV$ in the SM. 
Results for the Higgs boson signal (S), the signal including signal-background
interference (S+I) as well as the interfering background without 
Higgs contribution ($\Pg\Pg$ bkg.) are given.
Minimal cuts are applied: $M_{\Pl\PAl} > 10\UGeV, M_{\Pl'\PAl'} > 10\UGeV$.
Cross sections are given at loop-induced leading order and for a single
lepton flavour ($\Pl$) or single different-flavour combination ($\Pl$,$\Pl'$).
$\PGg^\ast$ contributions are included in $\PZ\PZ$.
The integration error is displayed in brackets.
} 
\renewcommand{\arraystretch}{1.2}%
\setlength{\tabcolsep}{1.5ex}%
\begin{center}
\begin{tabular}{cccc}
\toprule
final state & S & S+I & $\Pg\Pg$ bkg. \\ 
\midrule
$\Pl\PAl \Pl^{\prime}\PAl^{\prime}$ 	& $0.9284(7)$ & $0.6707(8)$ & $4.264(2)$ \\
$\Pl\PAl \Pl\PAl$	& $0.4739(8)$	& $0.3467(8)$	& $1.723(3)$ \\
$\Pl\PAl \PGn_{\Pl'}\PAGn_{\Pl'}$ & $1.896(2)$ & $1.386(2)$ & $5.730(5)$ \\
$\PAl\PGnl\PAGn_{\Pl'}\Pl'$ 	& $37.95(4)$ 	& $33.60(4)$	& $45.31(4)$ \\
$\PAl\PGnl \PAGnl\Pl$   & 36.01(3)	& $31.19(3)$	& $50.52(4)$	\\
\bottomrule
\end{tabular}
\end{center}
\end{table}

\begin{table}
\caption{\label{tab:ofs:semileptxs}
Cross sections  (\UfbZ) for $\Pg\Pg\ (\to \PH)\to \PV\PV \to$ semileptonic final states
in $\Pp\Pp$ collisions at $\sqrt{s}=13\UTeV$ in the SM. 
Results for the Higgs boson signal (S), the signal including signal-background
interference (S+I) as well as the interfering background without 
Higgs contribution ($\Pg\Pg$ bkg.) are given.
The signal amplitude is calculated at loop-induced leading order.
For the semileptonic decay modes, the $\Ord(g_{\mathrm{s}}^2e^2)$ tree-level as 
well as the important loop-induced $\Ord(g_{\mathrm{s}}^2e^4)$ amplitude contributions 
to the interfering background are taken into 
account \cite{Kauer:2015dma}.
Minimal cuts are applied: $M_{\Pl\PAl} > 10\UGeV, M_{\PQq\PAQq} > 10\UGeV,
\ptt{j}>25\UGeV$.
Cross sections are given for a single lepton flavour.
Other details as in \refT{tab:ofs:leptonicxs}.
}
\renewcommand{\arraystretch}{1.2}%
\setlength{\tabcolsep}{1.5ex}%
\begin{center}
\begin{tabular}{cccc}
\toprule
final state & S & S+I & $\Pg\Pg$ bkg. \\ 
\midrule
$\Pl \PAl  \PQd \PAQd$ 	& $1.711(3)$	 & $0.96(1)$  	& $1.575(6){\cdot}10^3$	\\
$\Pl \PAl  \PQu \PAQu$	& $1.334(3)$	& $0.750(5)$	& $2.30(5){\cdot}10^3$	\\
$\Pl \PAGnl  \PQu \PAQd$ & $38.66(5)$ 	& $30.58(8)$	& $1.111(3){\cdot}10^4$	\\
$\PAl \PGnl  \PAQu \PQd$&  $38.68(5)$	 & $30.59(8)$ 	& $1.112(3){\cdot}10^4$	\\
\bottomrule
\end{tabular}
\end{center}
\end{table}

\begin{figure}
\includegraphics[trim=3cm 0cm 0cm 0cm, clip, width=.49\textwidth]{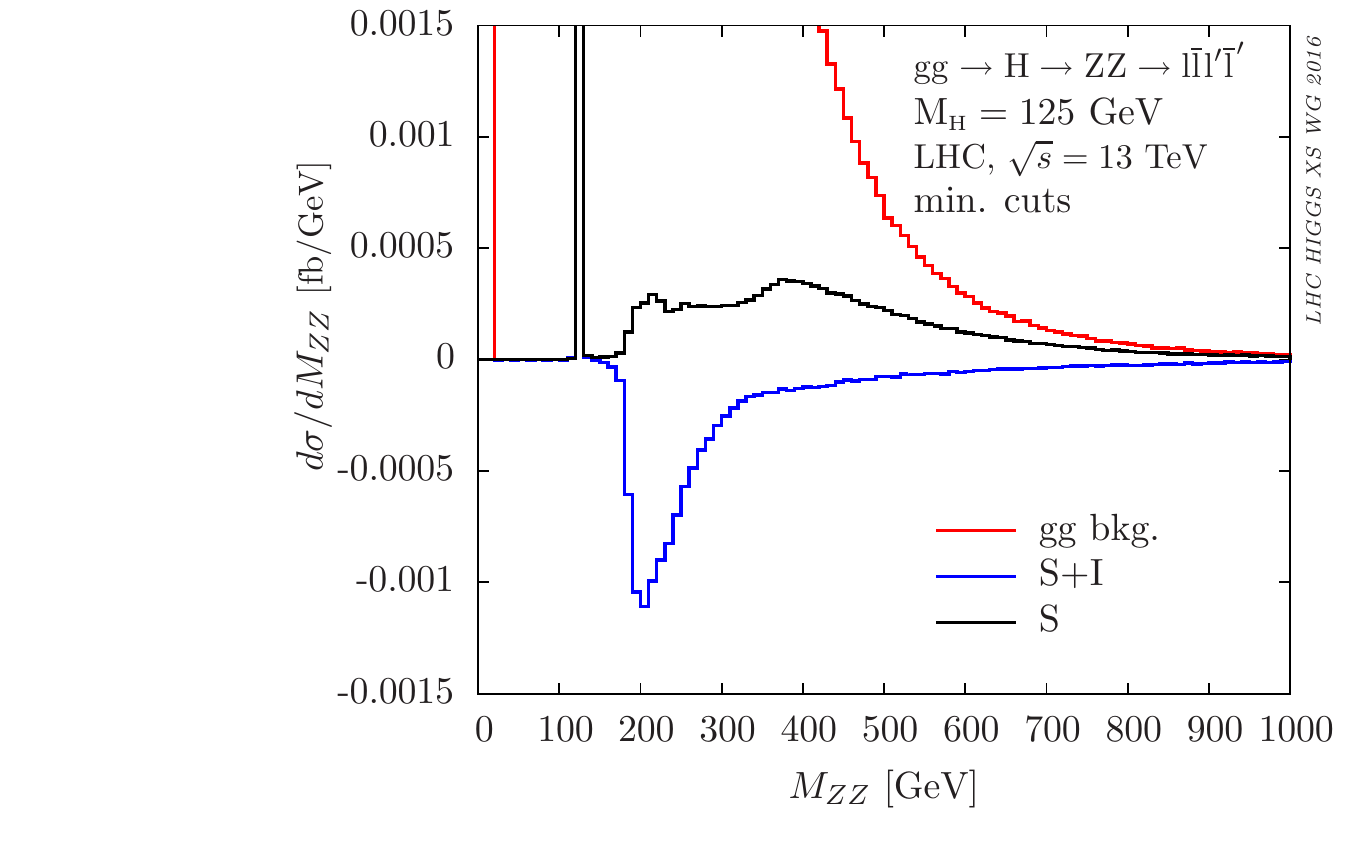}\hfil
\includegraphics[trim=3cm 0cm 0cm 0cm, clip, width=.49\textwidth]{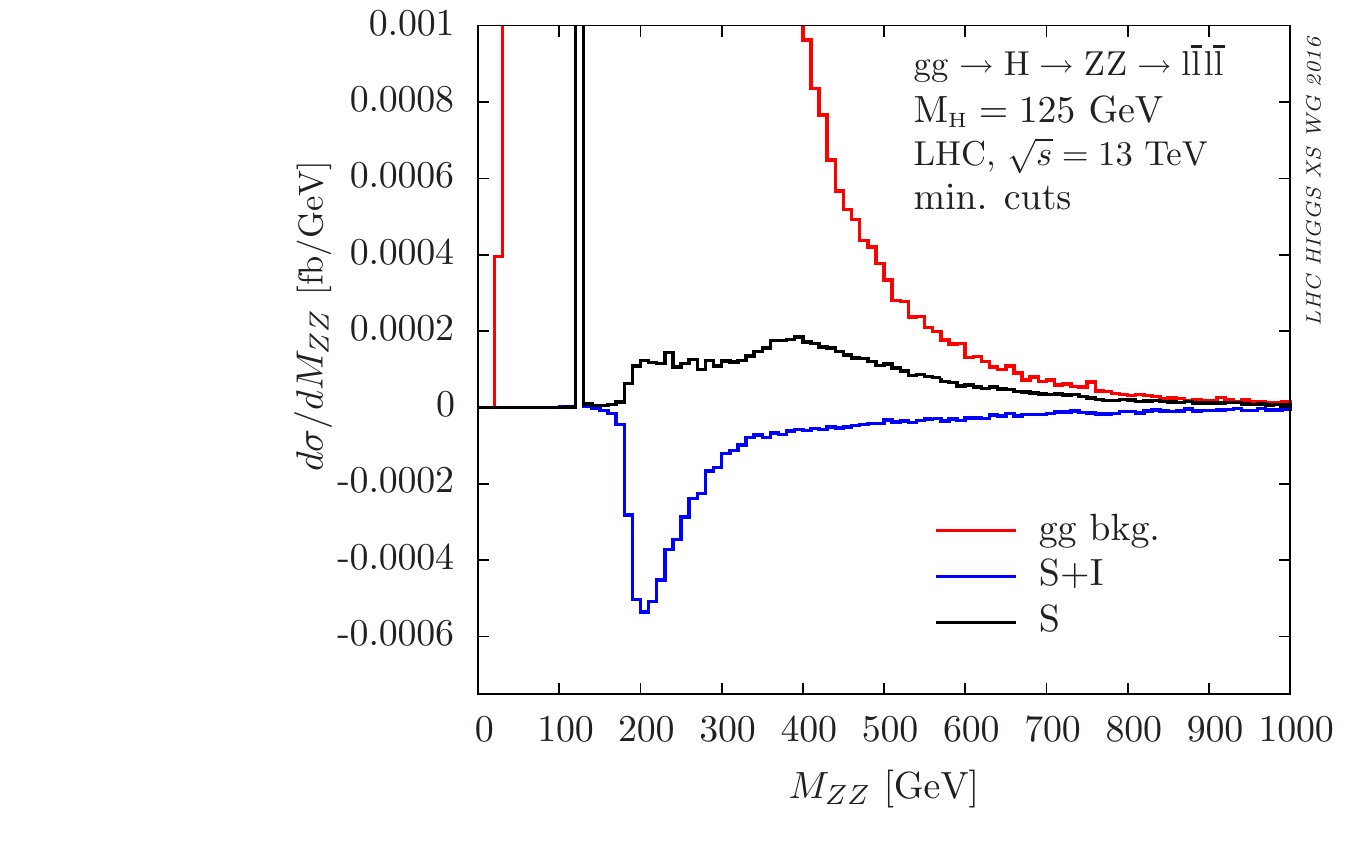}
\caption{\label{fig:ofs:ZAZA2l2l_ZAZA4l}Invariant mass distributions for 
$\Pg \Pg\ (\to \PH) \to \PZ\PZ \to \Pl \PAl  \Pl' \PAl'$
and 
$\Pg \Pg\ (\to \PH)  \to \PZ\PZ \to  \Pl \PAl  \Pl \PAl$.
Other details as in \refT{tab:ofs:leptonicxs}.}
\end{figure}

\begin{figure}
\includegraphics[trim=3cm 0cm 0cm 0cm, clip, width=.49\textwidth]{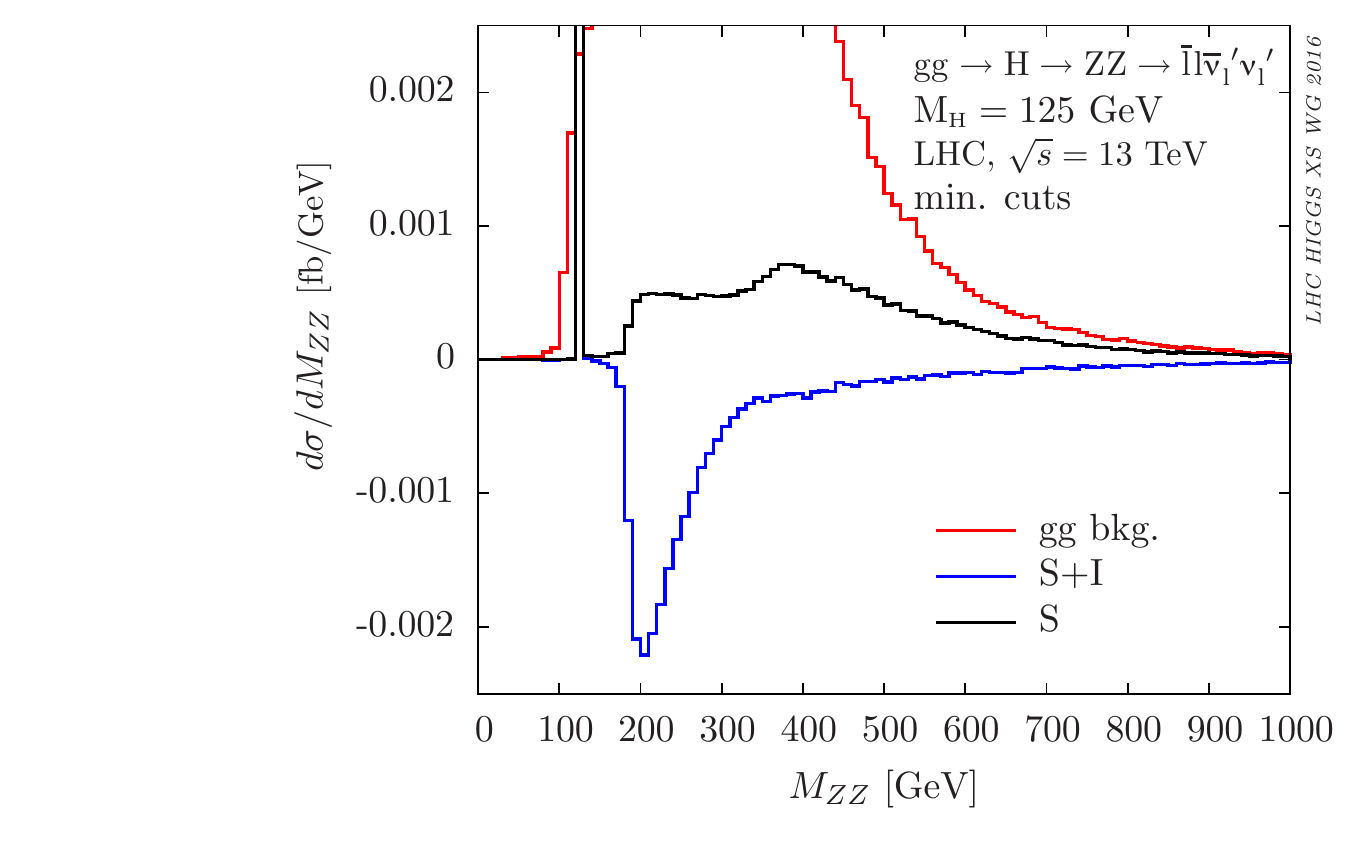}\hfil
\includegraphics[trim=3cm 0cm 0cm 0cm, clip, width=.49\textwidth]{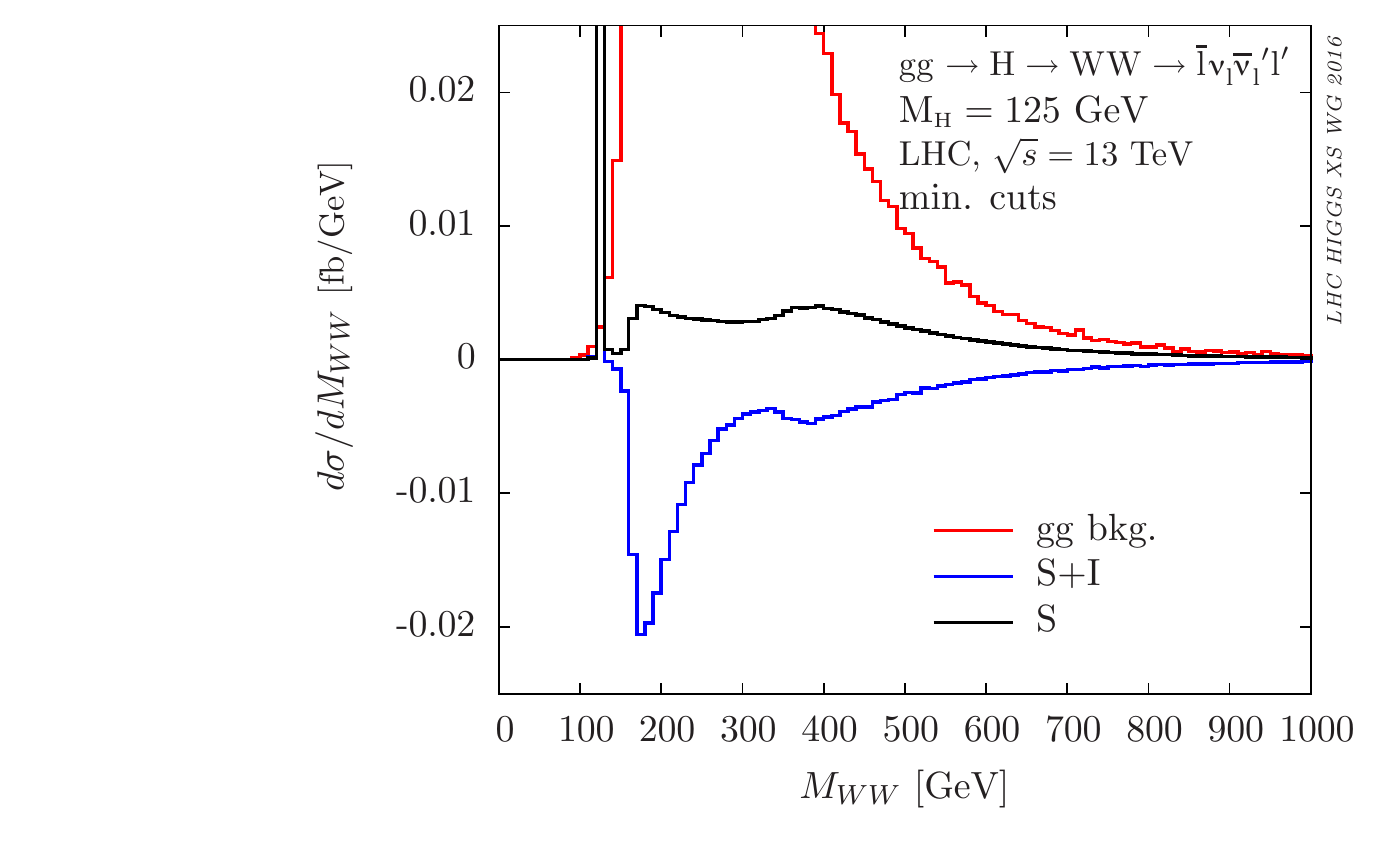}
\caption{\label{fig:ofs:ZAZ_2l2v_WW2l2v}Invariant mass distributions for 
$ \Pg \Pg\ (\to \PH) \to \PZ\PZ \to \Pl\PAl \PGn_{\Pl'}\PAGn_{\Pl'}$
and 
$\Pg \Pg\ (\to \PH) \to \PW\PW \to \PAl\PGnl\PAGn_{\Pl'}\Pl'$.
Other details as in \refT{tab:ofs:leptonicxs}.}
\end{figure}

\begin{figure}
\includegraphics[trim=3cm 0cm 0cm 0cm, clip, width=.49\textwidth]{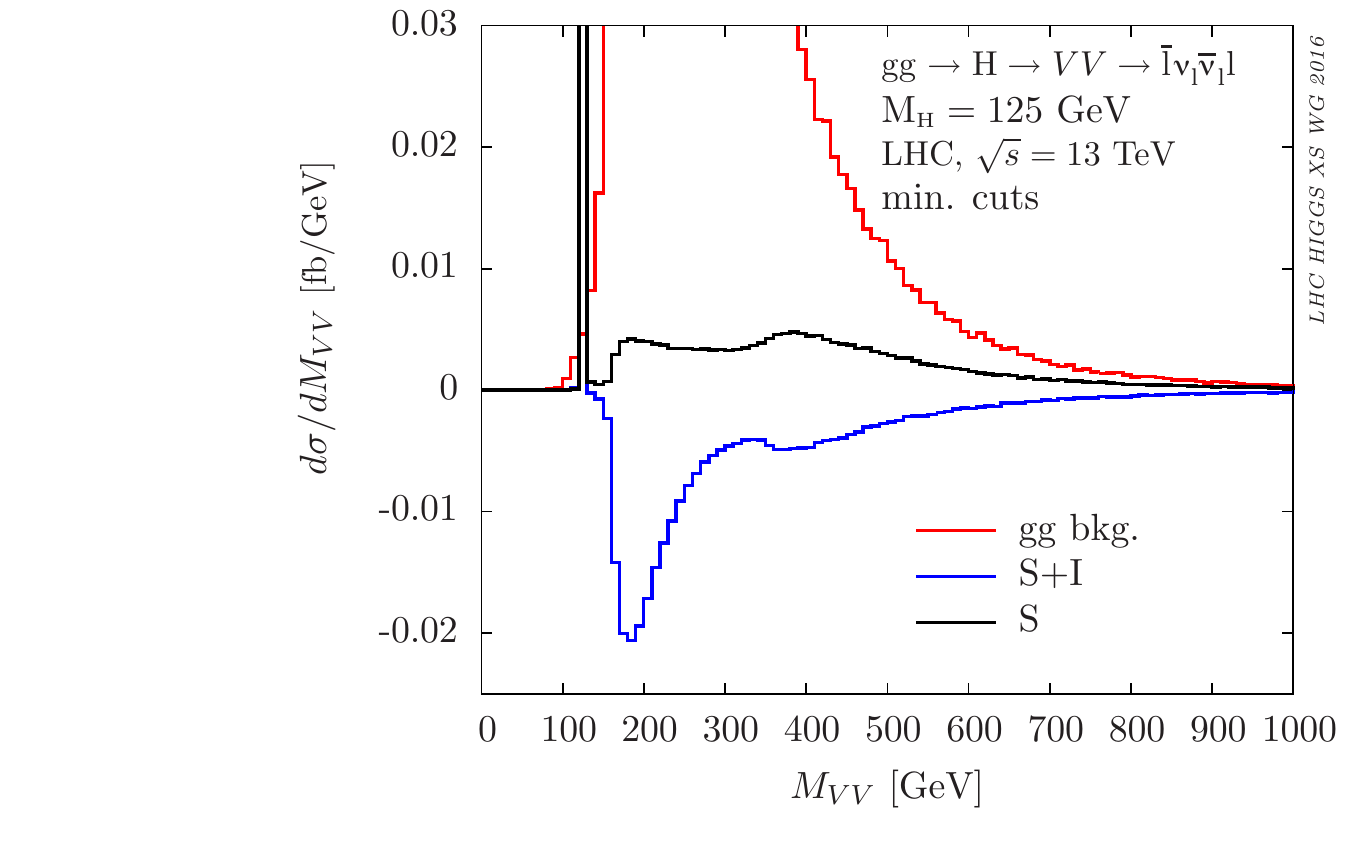}\hfil
\includegraphics[trim=3cm 0cm 0cm 0cm, clip, width=.49\textwidth]{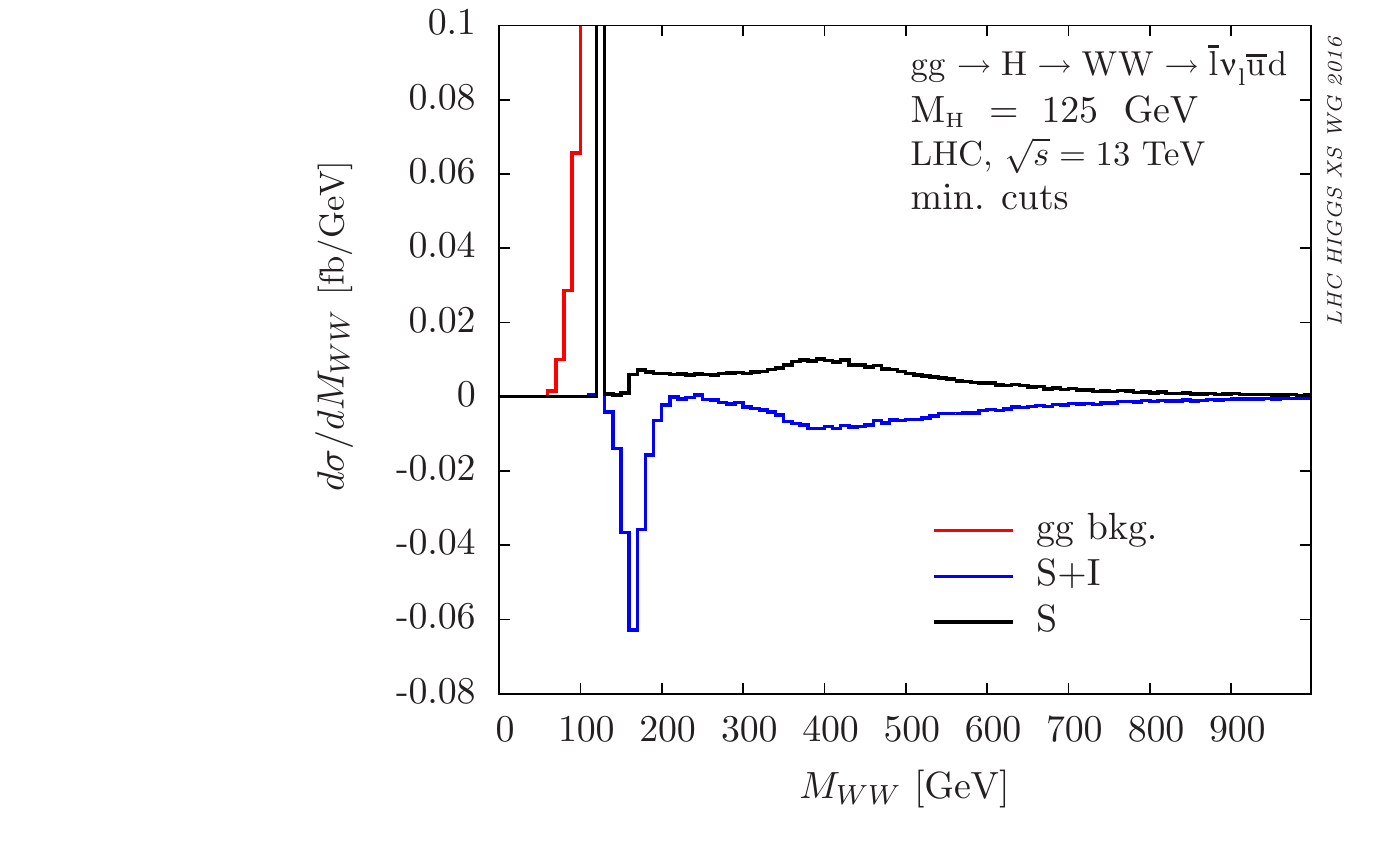}
\caption{\label{fig:ofs:WWZAZ_2l2v_WWlvqq}Invariant mass distributions for 
$ \Pg \Pg\ (\to \PH)  \to \PW\PW/\PZ\PZ \to  \PAl\PGnl \PAGnl\Pl$
and 
$ \Pg \Pg\ (\to \PH) \to \PW\PW \to \PAl \PGnl  \PAQu \PQd$.
Other details as in \refTs{tab:ofs:leptonicxs} and \ref{tab:ofs:semileptxs}.}
\end{figure}


The following experimental Higgs off-shell search selections 
are recommended for use in $\Pg\Pg\to \PV\PV$ calculations:\\
Jets: ATLAS: $\ptt{j} > 25$\UGeV{} for $|\eta_j|<2.4$, $\ptt{j}>30$\UGeV{} for 
$2.4<|\eta_j|<4.5$\\
Jets: CMS: $\ptt{j} > 30$\UGeV{} for $|\eta_j|<4.7$\\
$\PH\to \PZ\PZ\to 4\Pl$ channel: ATLAS:\\
$\ptt{\Pl,1}> 20\UGeV$, 
$\ptt{\Pl,2}> 15\UGeV$,
$\ptt{\Pl,3}> 10\UGeV$,
$\ptt{\Pe,4}> 7\UGeV$,
$\ptt{\PGm,4}> 6\UGeV$,
$|\eta_{\Pe}|<2.47$,
$|\eta_{\PGm}|<2.7$,
$M_{4\Pl} > 220\UGeV$\\
$\PH\to \PZ\PZ\to 4\Pl$ channel: CMS:\\
$\ptt{\Pl,1}> 20$\UGeV,
$\ptt{\Pl,2}> 10$\UGeV,
$\ptt{\Pe,3,4}> 7$\UGeV,
$\ptt{\PGm,3,4}> 5$\UGeV,
$|\eta_{\Pe}|<2.5$,
$|\eta_{\PGm}|<2.4$,
$M_{4\Pl} > 220$\UGeV\\
$\PH\to \PZ\PZ\to 2\Pl2\PGn$ channel: ATLAS:\\
$\ptt{\Pl}> 20$\UGeV\ (electron, muon),
$|\eta_{\Pe}|<2.47$,
$|\eta_{\PGm}|<2.5$,
$E_{T,miss} > 180\UGeV$,
$\Delta\phi_{\Pl\Pl} < 1.4$,
$\Mtt{,\PZ\PZ} > 380\UGeV$\\
ATLAS transverse mass definition (recommended for $M_{\PV\PV}> 2 M_{\PZ}$):
\begin{equation}
\label{eq:ofs:MTZZ}
\Mtt{,\PZ\PZ}=\sqrt{(\Mtt{,\ell\ell}+\Mtt{,miss})^2-({\bf{p}}_{T,\ell\ell}+{\bf{p}}_{T,miss})^2}\,,\
{\rm where}\ \ \ \Mtt{,X}=\sqrt{\ptt{,X}^2+M_Z^2}
\end{equation}
$\PH\to \PZ\PZ\to 2\Pl2\PGn$ channel: CMS:\\
$\ptt{\Pl}> 20$\UGeV\ (electron, muon),
$E_{T,miss} > 80\UGeV$\\
$\Mtt{,\PZ\PZ}$ used by CMS: Eq.~(\ref{eq:ofs:MTZZ}) with $M_Z$ replaced by
$M_{\ell\ell}$\\
$\PH \to \PW\PW\to 2\Pl2\PGn$ channel: ATLAS:\\
$\ptt{\Pl,1}> 22$\UGeV, 
$\ptt{\Pl,2}> 10$\UGeV, 
$|\eta_{\Pe}|<2.47$, 
$|\eta_{\PGm}|<2.5$, 
$M_{\Pl\Pl} > 10\UGeV$, 
$\ptt{,miss} > 20\UGeV$,
$\Mtt{,\PW\PW}>200\UGeV$\\
ATLAS transverse mass definition (recommended):
\begin{equation}
\label{eq:ofs:MTWW}
\Mtt{,\PW\PW}=\sqrt{(\Mtt{,\ell\ell}+\ptt{,miss})^2-({\bf{p}}_{T,\ell\ell}+{\bf{p}}_{T,miss})^2}\,,\
{\rm where}\ \ \ \Mtt{,\ell\ell}=\sqrt{\ptt{,\ell\ell}^2+M_{\ell\ell}^2}
\end{equation}


Vector-boson-fusion SM benchmark results were computed with 
\texttt{PHANTOM} \cite{Ballestrero:2007xq} (see also \Bref{Ballestrero:2015jca})
and 
\texttt{VBFNLO} \cite{Arnold:2008rz,Baglio:2014uba} (see also \Brefs{Jager:2006zc,Jager:2006cp,Bozzi:2007ur,Jager:2009xx,Feigl:2013naa}).
Good agreement was achieved for all fully-leptonic Higgs boson decay modes.
For VBF, two selection cut sets are applied which have the 
following selection in common:
\begin{itemize}
\item $\ptt{j} > 20\UGeV$, $|\eta_j| < 5.0$, $M_{jj} > 60\UGeV$ for all jets, anti-$k_\mathrm{T}$ jet clustering with $R=0.4$
\item $\ptt{\Pl} > 20\UGeV$, $|\eta_{\Pl}| < 2.5$, $M_{\Pl\PAl}>20\UGeV$ (same flavour only), $E_{\mathrm{T}}^{\mathrm{miss}} > 40\UGeV$
\item tagging jets: $j_1,j_2$, ordered by decreasing $|\eta_j|$
\end{itemize}
Exception: for the \textit{resonance} (RES) region (see \refS{sec:offshell_interf_vv_settings}) 
$M_{\Pl\PAl}>10\UGeV$ is applied instead of $M_{\Pl\PAl}>20\UGeV$.
With this common selection, we define:
\begin{itemize}
\item \textit{loose VBF cuts}: common selection and $M_{j_1j_2} > 130\UGeV$
\item \textit{tight VBF cuts}: common selection and $M_{j_1j_2} > 600\UGeV$, $\Delta y_{j_1j_2} > 3.6$, $y_{j_1} y_{j_2} < 0$ (opposite hemispheres)
\end{itemize}

Benchmark cross sections for $\Pq\Pq'(\to \Pq\Pq'\,\PH) \to \Pq\Pq'\,\PZ(\PGg^\ast)\PZ(\PGg^\ast) \to \Pq\Pq'\,\Pl \PAl  \Pl' \PAl'$ and $\Pq\Pq'(\to \Pq\Pq'\,\PH) \to \Pq\Pq'\,\PW\PW \to \Pq\Pq'\,\PAl\PGnl\PAGn_{\Pl'}\Pl'$ in $\Pp\Pp$ collisions at $\sqrt{s}=13\UTeV$ in the SM with tight and loose VBF cuts are given 
in \refTs{tab:ofs:SM_VBF_ZZ_tight}, \ref{tab:ofs:SM_VBF_WW_tight}, 
\ref{tab:ofs:SM_VBF_ZZ_loose} and \ref{tab:ofs:SM_VBF_WW_loose}.
Leading order and next-to-leading order results for the Higgs boson signal, 
the signal including signal-background interference as well as the interfering background without Higgs contribution are displayed.
Corresponding Higgs boson invariant mass (for $\PZ\PZ$) and transverse mass 
(for $\PW\PW$) distributions are shown in 
\refFs{fig:ofs:VBFNLO_MZZ_loose_tight} and \ref{fig:ofs:VBFNLO_MTWW_loose_tight},
respectively, for loose and tight VBF cuts.

The full set of SM benchmark cross sections and distributions is available
at \url{https://twiki.cern.ch/twiki/bin/view/LHCPhysics/LHCHXSWGOFFSHELL}.

\begin{table}
\caption{\label{tab:ofs:SM_VBF_ZZ_tight}
Cross sections for $\Pq\Pq'(\to \Pq\Pq'\,\PH) \to \Pq\Pq'\,\PZ(\PGg^\ast)\PZ(\PGg^\ast) \to \Pq\Pq'\,\Pl \PAl  \Pl' \PAl'$ in $\Pp\Pp$ collisions at $\sqrt{s}=13\UTeV$ in the SM. 
Leading order (LO) and next-to-leading order (NLO) results for 
the Higgs boson signal (S), the signal including 
signal-background interference (S+I) as well as the interfering 
background without Higgs contribution (B) are given.
Tight VBF cuts are applied (see main text).
Cross sections are given for a single lepton flavour combination. 
The integration error is displayed in brackets.
}
\renewcommand{\arraystretch}{1.2}%
\setlength{\tabcolsep}{1.5ex}%
\begin{center}
\scriptsize 
\begin{tabular}{c c c c c c}
\toprule
$\sigma[\UfbZ]$ & & $110\UGeV <M_{\PZ\PZ}< 140\UGeV$ & $M_{\PZ\PZ}>140\UGeV$ & $220\UGeV<M_{\PZ\PZ}<300\UGeV$ & $M_{\PZ\PZ}>300\UGeV$ \\ 
\midrule
S & LO & $6.88(2) {\cdot} 10^{-3}$ & $1.2501(9) {\cdot} 10^{-2}$ &  $1.316(3) {\cdot} 10^{-3} $   &  $1.0644(9) {\cdot} 10^{-2}$ \\ 
S+I & LO &  $6.92(4) {\cdot} 10^{-3}$ & $- 1.398(6) {\cdot} 10^{-2}$   & $- 1.85(3) {\cdot} 10^{-3}$  & $- 1.126(5) {\cdot} 10^{-2}$ \\ 
B & LO & $1.0(2) {\cdot} 10^{-4}$ & $6.554(4) {\cdot} 10^{-2} $  & $1.672(2) {\cdot} 10^{-2}$   & $4.126(3) {\cdot} 10^{-2}$ \\ 
\midrule
S & NLO & $5.67(4) {\cdot} 10^{-3}$ & $1.371(3) {\cdot} 10^{-2}$ &  $1.234(8) {\cdot} 10^{-3} $   &  $1.198(3) {\cdot} 10^{-2}$ \\  
S+I & NLO &  $5.2(6) {\cdot} 10^{-3}$ & $- 1.55(2) {\cdot} 10^{-2}$   & $- 1.75(6) {\cdot} 10^{-3}$  & $- 1.288(9) {\cdot} 10^{-2}$ \\  
B & NLO & $5(2) {\cdot} 10^{-5}$ & $6.749(9) {\cdot} 10^{-2} $  & $1.627(5) {\cdot} 10^{-2}$   & $4.400(7) {\cdot} 10^{-2}$ \\ 
\bottomrule
\end{tabular}
\end{center}
\end{table}

\begin{table}
\caption{\label{tab:ofs:SM_VBF_WW_tight}
Cross sections for $\Pq\Pq'(\to \Pq\Pq'\,\PH) \to \Pq\Pq'\,\PW\PW \to \Pq\Pq'\,\PAl\PGnl\PAGn_{\Pl'}\Pl'$ in $\Pp\Pp$ collisions at $\sqrt{s}=13\UTeV$ in the SM. 
Tight VBF cuts are applied (see main text).
Cross sections are given for a single lepton flavour combination, but taking
into account both charge assignments, e.g.\ $(\Pl,\Pl')=(\Pe,\PGm)$ or $(\PGm,\Pe)$.
Other details as in \refT{tab:ofs:SM_VBF_ZZ_tight}.
}
\renewcommand{\arraystretch}{1.2}%
\setlength{\tabcolsep}{1.5ex}%
\begin{center}
\scriptsize 
\begin{tabular}{c c c c c c}
\toprule
$\sigma[\UfbZ]$ & & $110\UGeV <M_{\PW\PW}< 140\UGeV$ & $M_{\PW\PW}>140\UGeV$ & $220\UGeV<M_{\PW\PW}<300\UGeV$ & $M_{\PW\PW}>300\UGeV$ \\ 
\midrule
 S & LO & $1.7411(9)$ & $2.370(6) {\cdot} 10^{-1}$ &  $3.08(2) {\cdot} 10^{-2} $   &  $1.783(5) {\cdot} 10^{-1}$ \\  
S+I & LO & $1.740(3)$ & $- 3.00(4) {\cdot} 10^{-1}$   & $- 4.9(2) {\cdot} 10^{-2}$  & $- 0.197(3)$ \\  
B & LO & $8(2) {\cdot} 10^{-4}$ & $3.387(2) $  & $0.8642(6)$                 & $1.856(2)$ \\ 
\midrule
 S & NLO & $1.453(4)$ & $2.51(2) {\cdot} 10^{-1}$ &  $2.96(6) {\cdot} 10^{-2} $   &  $1.95(2) {\cdot} 10^{-1}$ \\  
S+I & NLO & $1.45(1)$ & $- 3.0(2) {\cdot} 10^{-1}$   & $- 3(2) {\cdot} 10^{-2}$  & $- 0.234(9)$ \\  
B & NLO & $6.7(7) {\cdot} 10^{-4}$ & $3.381(6) $  & $0.825(4)$                 & $1.933(4)$ \\ 
\bottomrule
\end{tabular}
\end{center}
\end{table}

\begin{table}
\caption{\label{tab:ofs:SM_VBF_ZZ_loose}
Cross sections for $\Pq\Pq'(\to \Pq\Pq'\,\PH) \to \Pq\Pq'\,\PZ(\PGg^\ast)\PZ(\PGg^\ast) \to \Pq\Pq'\,\Pl \PAl  \Pl' \PAl'$ in $\Pp\Pp$ collisions at $\sqrt{s}=13\UTeV$ in the SM. 
Loose VBF cuts are applied (see main text).
Other details as in \refT{tab:ofs:SM_VBF_ZZ_tight}.
}
\renewcommand{\arraystretch}{1.2}%
\setlength{\tabcolsep}{1.5ex}%
\begin{center}
\scriptsize 
\begin{tabular}{c c c c c c}
\toprule
$\sigma[\UfbZ]$ & & $110\UGeV <M_{\PZ\PZ}< 140\UGeV$ & $M_{\PZ\PZ}>140\UGeV$ & $220\UGeV<M_{\PZ\PZ}<300\UGeV$ & $M_{\PZ\PZ}>300\UGeV$ \\ 
\midrule
S & LO & $1.202(2) {\cdot} 10^{-2}$  & $1.662(2) {\cdot} 10^{-2}$   & $2.153(5) {\cdot} 10^{-3}$  & $1.351(2) {\cdot} 10^{-2}$ \\ 
 S+I & LO & $1.197(7) {\cdot} 10^{-2}$ & $- 1.95(2) {\cdot} 10^{-2}$ &  $- 3.34(5) {\cdot} 10^{-3} $   &  $- 1.441(8) {\cdot} 10^{-2}$ \\  
B & LO &  $2.2(2) {\cdot} 10^{-4}$ & $1.3535(7) {\cdot} 10^{-1} $  & $3.821(3) {\cdot} 10^{-2}$   & $7.909(5) {\cdot} 10^{-2}$ \\ \midrule
S & NLO & $1.035(4) {\cdot} 10^{-2}$  & $1.781(3) {\cdot} 10^{-2}$   & $1.993(9) {\cdot} 10^{-3}$  & $1.495(3) {\cdot} 10^{-2}$ \\ 
 S+I & NLO & $1.02(2) {\cdot} 10^{-2}$ & $- 2.04(2) {\cdot} 10^{-2}$ &  $- 3.1(1) {\cdot} 10^{-3} $   &  $- 1.58(2) {\cdot} 10^{-2}$ \\  
B & NLO &  $2.0(4) {\cdot} 10^{-4}$ & $1.346(2) {\cdot} 10^{-1} $  & $3.651(5) {\cdot} 10^{-2}$   & $8.108(9) {\cdot} 10^{-2}$ \\ \bottomrule
\end{tabular}
\end{center}
\end{table}

\begin{table}
\caption{\label{tab:ofs:SM_VBF_WW_loose}
Cross sections for $\Pq\Pq'(\to \Pq\Pq'\,\PH) \to \Pq\Pq'\,\PW\PW \to \Pq\Pq'\,\PAl\PGnl\PAGn_{\Pl'}\Pl'$ in $\Pp\Pp$ collisions at $\sqrt{s}=13\UTeV$ in the SM. 
Loose VBF cuts are applied (see main text).
Other details as in \refT{tab:ofs:SM_VBF_WW_tight}.
}
\renewcommand{\arraystretch}{1.2}%
\setlength{\tabcolsep}{1.5ex}%
\begin{center}
\scriptsize 
\begin{tabular}{c c c c c c}
\toprule
$\sigma[\UfbZ]$ & & $110\UGeV <M_{\PW\PW}< 140\UGeV$ & $M_{\PW\PW}>140\UGeV$ & $220\UGeV<M_{\PW\PW}<300\UGeV$ & $M_{\PW\PW}>300\UGeV$ \\ 
\midrule
 S & LO & $3.271(2)$ & $3.325(9) {\cdot} 10^{-1}$ &  $5.10(3) {\cdot} 10^{-2} $   &  $2.301(8) {\cdot} 10^{-1}$ \\  
S+I & LO & $3.278(6)$  & $- 4.79(9) {\cdot} 10^{-1}$   & $- 9.7(3) {\cdot} 10^{-2}$  & $- 2.61(7) {\cdot} 10^{-1}$ \\  
B & LO & $1.8(3) {\cdot} 10^{-3}$ & $7.449(5) $  & $2.004(2)$   & $3.830(3)$ \\ 
\midrule
 S & NLO & $2.836(7)$ & $3.46(3) {\cdot} 10^{-1}$ &  $4.75(7) {\cdot} 10^{-2} $   &  $2.50(3) {\cdot} 10^{-1}$ \\  
S+I & NLO & $2.85(5)$  & $- 4.4(2) {\cdot} 10^{-1}$   & $- 7.6(9) {\cdot} 10^{-2}$  & $- 2.7(2) {\cdot} 10^{-1}$ \\  
B & NLO & $1.8(2) {\cdot} 10^{-3}$ & $7.402(9) $  & $1.928(4)$   & $3.949(7)$ \\ 
\bottomrule
\end{tabular}
\end{center}
\end{table}

\begin{figure}
\includegraphics[trim=0cm 0cm 0cm 0cm, clip, width=.49\textwidth]{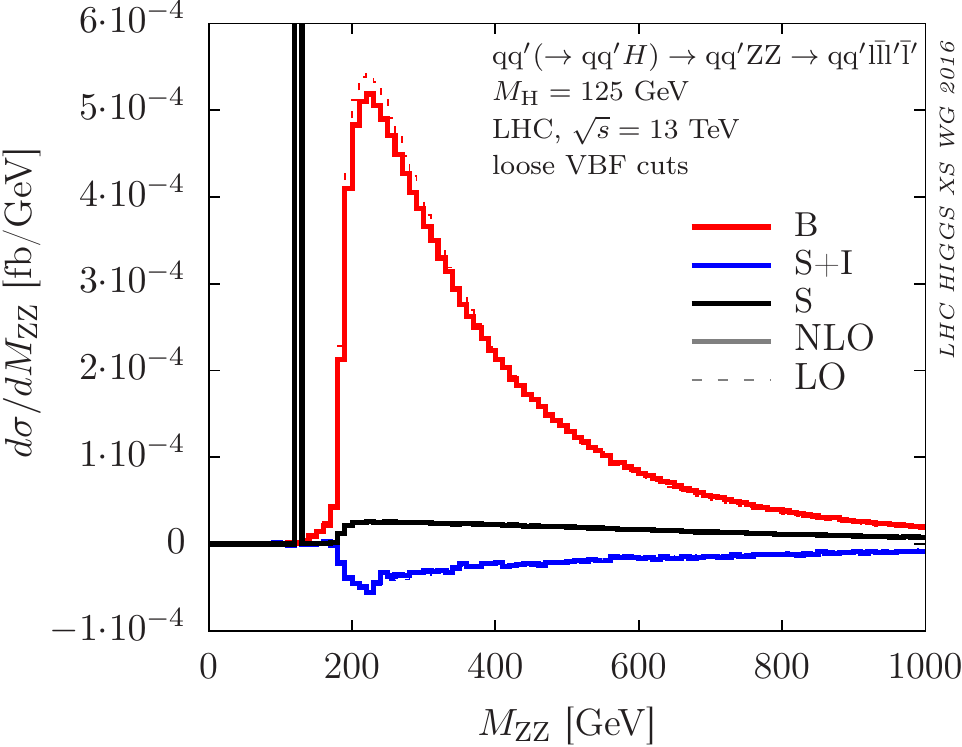}\hfil
\includegraphics[trim=0cm 0cm 0cm 0cm, clip,width=.49\textwidth]{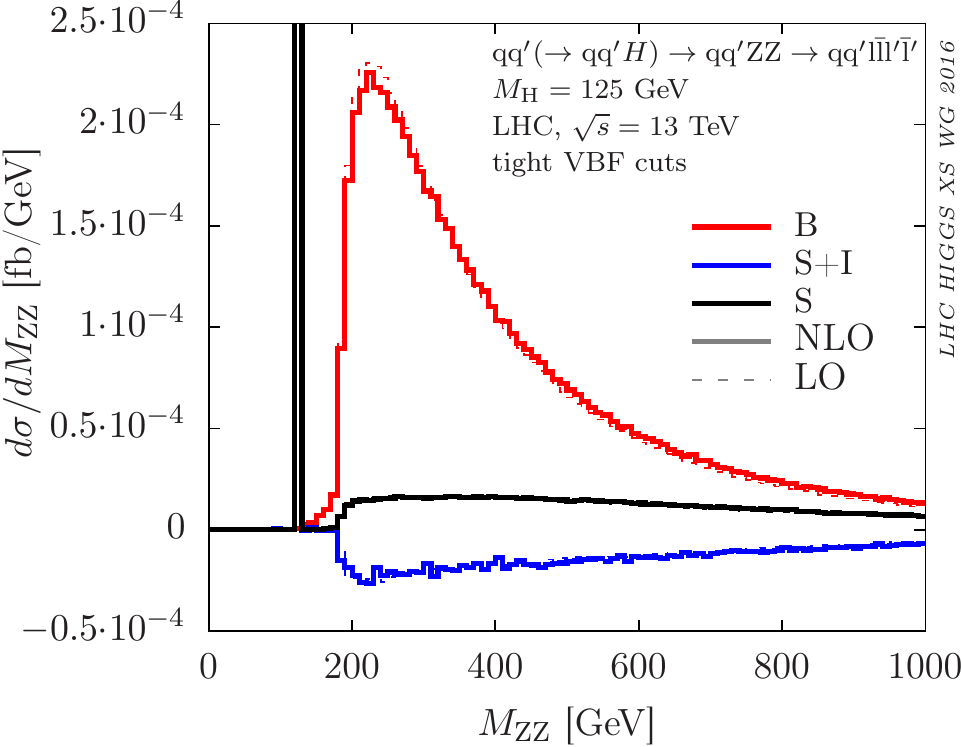}
\caption{\label{fig:ofs:VBFNLO_MZZ_loose_tight} Invariant mass distributions for 
$\Pq\Pq'(\to \Pq\Pq'\,\PH) \to \Pq\Pq'\,\PZ(\PGg^\ast)\PZ(\PGg^\ast) \to \Pq\Pq'\,\Pl \PAl  \Pl' \PAl'$ in $\Pp\Pp$ collisions at $\sqrt{s}=13\UTeV$ in the SM. Loose and tight VBF cuts are applied in the left and right graphs, respectively.
Leading order (dashed) and next-to-leading order (solid) results for 
the Higgs boson signal (S), the signal including signal-background 
interference (S+I) as well as the interfering background without 
Higgs contribution (B) are given.
Cross sections are given for a single lepton flavour combination.
}
\end{figure}

\begin{figure}
\includegraphics[trim=0cm 0cm 0cm 0cm, clip, width=.49\textwidth]{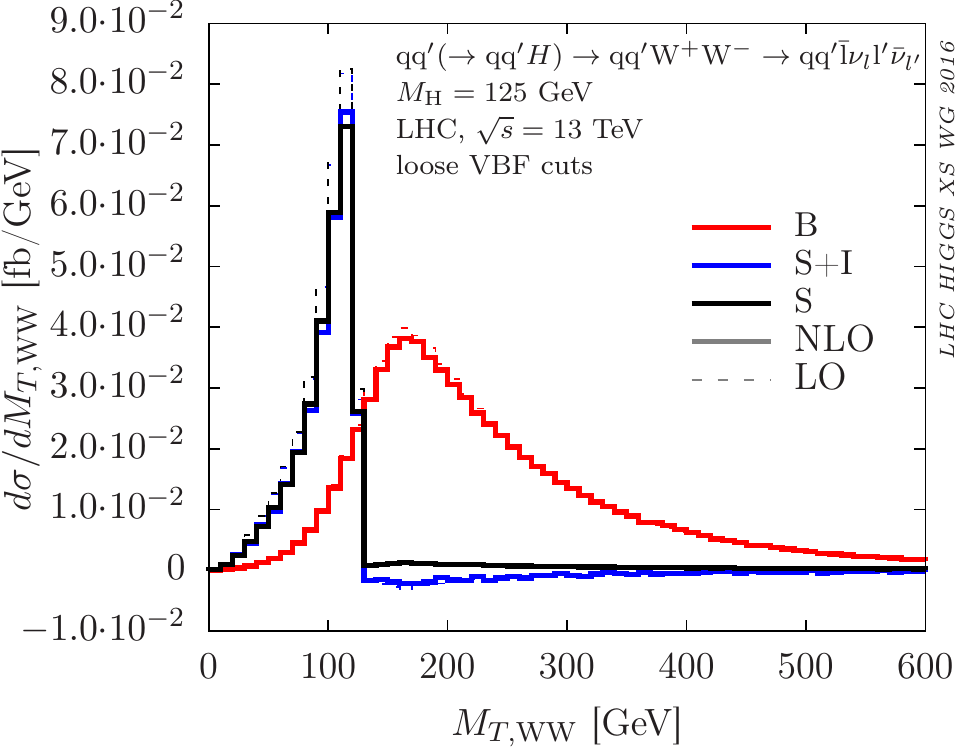}\hfil
\includegraphics[trim=0cm 0cm 0cm 0cm, clip,width=.49\textwidth]{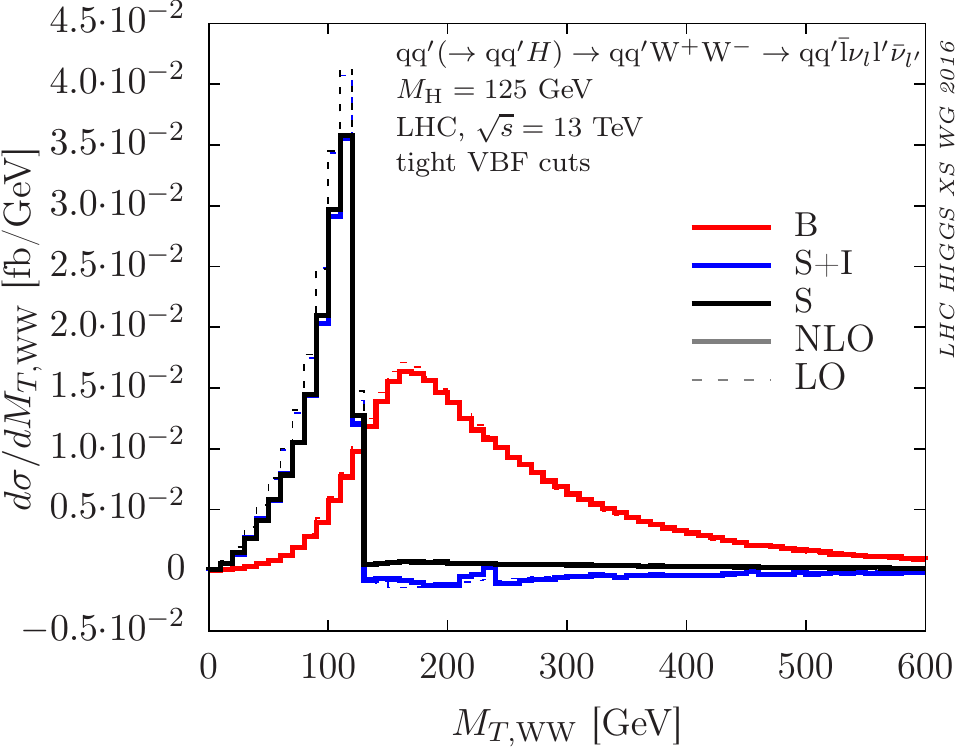}
\caption{\label{fig:ofs:VBFNLO_MTWW_loose_tight}
Transverse mass $\Mtt{,\PW\PW}$ (see Eq.~(\ref{eq:ofs:MTWW})) distributions for 
$\Pq\Pq'(\to \Pq\Pq'\,\PH) \to \Pq\Pq'\,\PW\PW \to \Pq\Pq'\,\PAl\PGnl\PAGn_{\Pl'}\Pl'$ in $\Pp\Pp$ collisions at $\sqrt{s}=13\UTeV$ in the SM. 
Cross sections are given for a single lepton flavour combination, but taking
into account both charge assignments, e.g.\ $(\Pl,\Pl')=(\Pe,\PGm)$ or $(\PGm,\Pe)$.
Other details as in \refF{fig:ofs:VBFNLO_MZZ_loose_tight}.}
\end{figure}


\subsection{Off-shell and interference benchmarks: 1-Higgs Singlet Model}

\label{sec:offshell_interf_vv_bench_1hsm}

\providecommand{\Ph}{\HepParticle{h}{}{}\Xspace}
\providecommand{\Pho}{\Ph_1}
\providecommand{\Pht}{\Ph_2}
\providecommand{\Mho}{\mathswitch {M_{\Ph 1}}}
\providecommand{\Mht}{\mathswitch {M_{\Ph 2}}}

The simplest extension of the SM Higgs sector is given by the addition
of a singlet field which is neutral under the SM gauge groups.  
We adopt the definition of the 1-Higgs Singlet Model (1HSM), a.k.a.\ EW Singlet Model,
which is given in Section 13.3 of \Bref{Heinemeyer:2013tqa}.
Here, interference benchmark cross sections and distributions in 
the 1HSM are presented.  We employ basis (335) of 
\Bref{Heinemeyer:2013tqa} and specify four benchmark points:
\begin{enumerate}
\item $\Mht = 400$\UGeV, $\sin\theta = 0.2$,
\item $\Mht = 600$\UGeV, $\sin\theta = 0.2$,
\item $\Mht = 600$\UGeV, $\sin\theta = 0.4$,
\item $\Mht = 900$\UGeV, $\sin\theta = 0.2$,
\end{enumerate}
where $\Mho = 125$\UGeV\ and $\mu_1 = \lambda_2 = \lambda_1 = 0$ for all 
points.  The corresponding Higgs boson widths are given in \refT{tab:ofs:widths}.  
They have been calculated using \textsc{FeynRules} \cite{Alloul:2013bka}.
\begin{table}
\caption{\label{tab:ofs:widths}Widths of the physical Higgs bosons $\Pho$ and $\Pht$ in the 1-Higgs-Singlet Extension of the SM with mixing angles $\sin\theta=0.2$ and $\sin\theta=0.4$ as well as $\mu_1= \lambda_1=\lambda_2=0$.}
\renewcommand{\arraystretch}{1.2}%
\setlength{\tabcolsep}{1.5ex}%
\begin{center}
\begin{tabular}{c|c|c|ccc}
\toprule
\multicolumn{2}{c}{} & $\Pho$ & \multicolumn{3}{c}{$\Pht$}  \\ 
\cmidrule(lr){3-6}
$\sin\theta$ & $M$ [\UGeV] & 125 & 400 & 600 & 900 \\ 
\midrule
$0.2$ & $\Gamma$ [\UGeV] & $4.34901{\cdot} 10^{-3}$ & 1.52206 & 5.95419 & 19.8529 \\ 
$0.4$ &  $\Gamma$ [\UGeV] & $3.80539{\cdot} 10^{-3}$ & & 22.5016 	&  	\\ 
\bottomrule
\end{tabular}
\end{center}
\end{table}

Gluon-fusion 1HSM benchmark results were computed with \textsc{gg2VV} \cite{Kauer:2012hd} 
(see also \Bref{Kauer:2015hia}).  More specifically,
cross sections for $\Pg\Pg\ (\to \{\Pho,\Pht\}) \to \PZ(\PGg^\ast)\PZ(\PGg^\ast) \to 
\Pl\PAl\Pl'\PAl'$ for the 13\UTeV\ LHC are given in \refTs{tab:ofs:sigplusint} and 
\ref{tab:ofs:allcontribs}.  The corresponding distributions are shown in 
\refFs{fig:ofs:sigplusint} and \ref{fig:ofs:allcontribs}, respectively.
Results for the heavy Higgs boson signal and its interference with the 
light Higgs and continuum background and the combined interference are given in 
\refT{tab:ofs:sigplusint} and \refF{fig:ofs:sigplusint}. In \refT{tab:ofs:sigplusint}, 
the ratio $R_i = (S+I_i)/S$ is used to illustrate the 
relative change of the heavy Higgs boson signal due to interference with the 
light Higgs and continuum background amplitude contributions.
Heavy-Higgs-light-Higgs interference effects and the coherent sum of all interfering 
contributions is shown in \refT{tab:ofs:allcontribs} and \refF{fig:ofs:allcontribs}.

Vector-boson-fusion 1HSM benchmark results were computed with 
\textsc{PHANTOM} \cite{Ballestrero:2007xq} (see also \Brefs{Maina:2015ela,Ballestrero:2015jca})
and 
\textsc{VBFNLO} \cite{Arnold:2008rz,Baglio:2014uba} (see also \Brefs{Jager:2006zc,Jager:2006cp,Bozzi:2007ur,Jager:2009xx,Feigl:2013naa}).
Good agreement was achieved for all fully-leptonic Higgs boson decay modes.
Cross sections for $\Pq\Pq'(\to \Pq\Pq'\{\Pho, \Pht\}) \to \Pq\Pq'\,\PZ(\PGg^\ast)\PZ(\PGg^\ast) \to \Pq\Pq'\,\Pl \PAl  \Pl' \PAl'$ 
and 
$\Pq\Pq'(\to \Pq\Pq'\{\Pho, \Pht\}) \to \Pq\Pq'\,\PW\PW \to \Pq\Pq'\,\Pl \PAGnl  \PAl'\PGn_{\Pl'}$ 
for the 13\UTeV\ LHC are given in 
\refTs{tab:ofs:singlet_vbf_ZZ_tight} and \ref{tab:ofs:singlet_vbf_ZZ_loose}
and \refTs{tab:ofs:singlet_vbf_WW_tight} and \ref{tab:ofs:singlet_vbf_WW_loose} for tight and loose VBF cuts (see \refS{sec:offshell_interf_vv_bench_sm}), respectively.  More specifically, the sum of the light and heavy Higgs contributions including light-heavy interference, the interfering background without Higgs contributions and the sum of the Higgs boson signal and its interference with 
the background are given.
VBF Higgs boson invariant mass distributions in the 1HSM are shown in \refF{fig:ofs:singlet_vbf}.

The full set of 1HSM benchmark cross sections and distributions is available
at \url{https://twiki.cern.ch/twiki/bin/view/LHCPhysics/LHCHXSWGOFFSHELL}.

\begin{table}
\caption{\label{tab:ofs:sigplusint}
Cross sections (\UfbZ) for $\Pg\Pg\ (\to \{\Pho,\Pht\}) \to \PZ(\PGg^\ast)\PZ(\PGg^\ast) \to \Pl\PAl\Pl'\PAl'$ in $\Pp\Pp$ collisions at $\sqrt{s}=13\UTeV$\ at loop-induced leading order 
in the 1-Higgs-Singlet Extension of the SM (1HSM) with $\Mho = 125 \UGeV$, $\Mht = 400, 600, 900\UGeV$ and mixing angle $\sin\theta=0.2$ or $0.4$ as indicated. 
Results for the heavy Higgs ($\Pht$) signal ($S$) and its interference with the 
light Higgs ($I_{\Ph 1}$) and the continuum background ($I_{bkg}$) and the full 
interference ($I_{full}$) are given.
The ratio $R_i = (S+I_i)/S$ illustrates the relative change of 
the heavy Higgs boson signal due to interference with 
the light Higgs and continuum background amplitude contributions.
Cross sections are given for a single lepton flavour combination.  
Minimal cuts are applied: $M_{\Pl\PAl} > 4\UGeV, M_{\Pl'\PAl'} > 4\UGeV, 
p_{\mathrm{T}\PZ} > 1\UGeV$.
The integration error is displayed in brackets.
} 
\renewcommand{\arraystretch}{1.3}%
\setlength{\tabcolsep}{1.5ex}%
\begin{center}
\footnotesize
\begin{tabular}{ccccccccc}
\toprule
\multicolumn{3}{c}{} & \multicolumn{3}{c|}{interference} & \multicolumn{3}{c}{ratio} \\ 
\cmidrule(lr){4-9}
 $\sin\theta$ & $ M_{\Ph2}$  [\UGeV]  & $S (\Pht)$& $I_{\Ph1}$  & $I_{bkg}$ & $I_{full}$  &$R_{\Ph1}$ & $R_{bkg}$ & $R_{full}$ \\ 
 \midrule
0.2 & 400	& 0.07412(6)		& 0.00682(6)		& -0.00171(2)		& 0.00511(6)		& 1.092(2)		& 0.977(1)		& 1.069(2)		\\ 
0.2 & 600	& 0.01710(2)		& -0.00369(3) 		& 0.00384(3) 		& 0.00015(4) 		& 0.784(2)		& 1.225(2)		& 1.009(3)		\\ 
0.2 & 900	& 0.002219(2)		& -0.003369(9)		& 0.003058(8) 		& -0.00031(2)		& -0.518(4) 	& 2.378(4)		& 0.860(6)			\\ 
0.4 & 600	& 0.07065(6)		& -0.01191(6)		& 0.01465(6)		& -0.00274(9)		& 0.831(2)		& 1.207(2)		& 1.039(2)		\\ 
\bottomrule
\end{tabular}
\end{center}
\end{table}

\begin{figure}
\includegraphics[trim=3cm 0cm 0cm 0cm, clip, width=.49\textwidth]{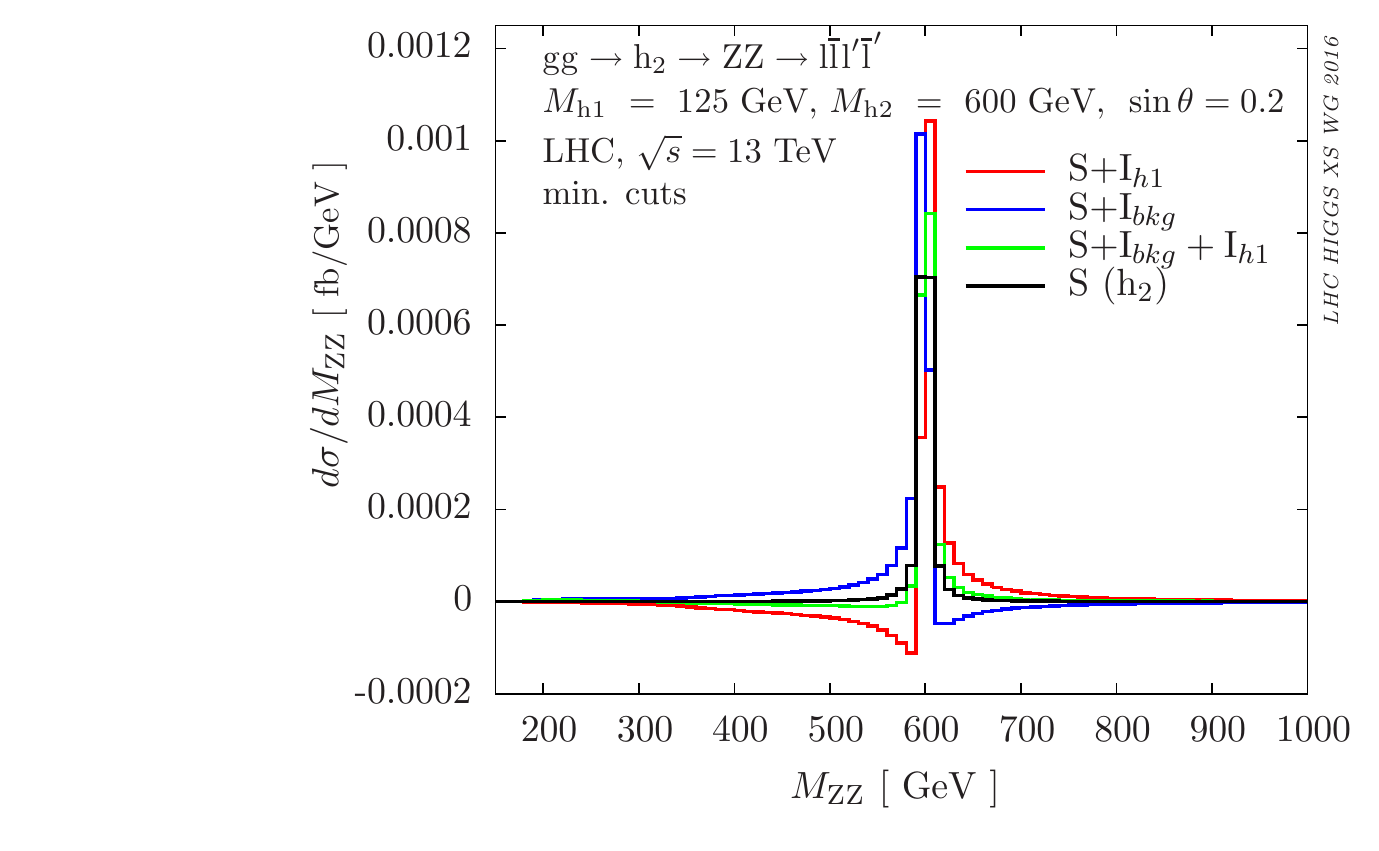}\hfil
\includegraphics[trim=3cm 0cm 0cm 0cm, clip,width=.49\textwidth]{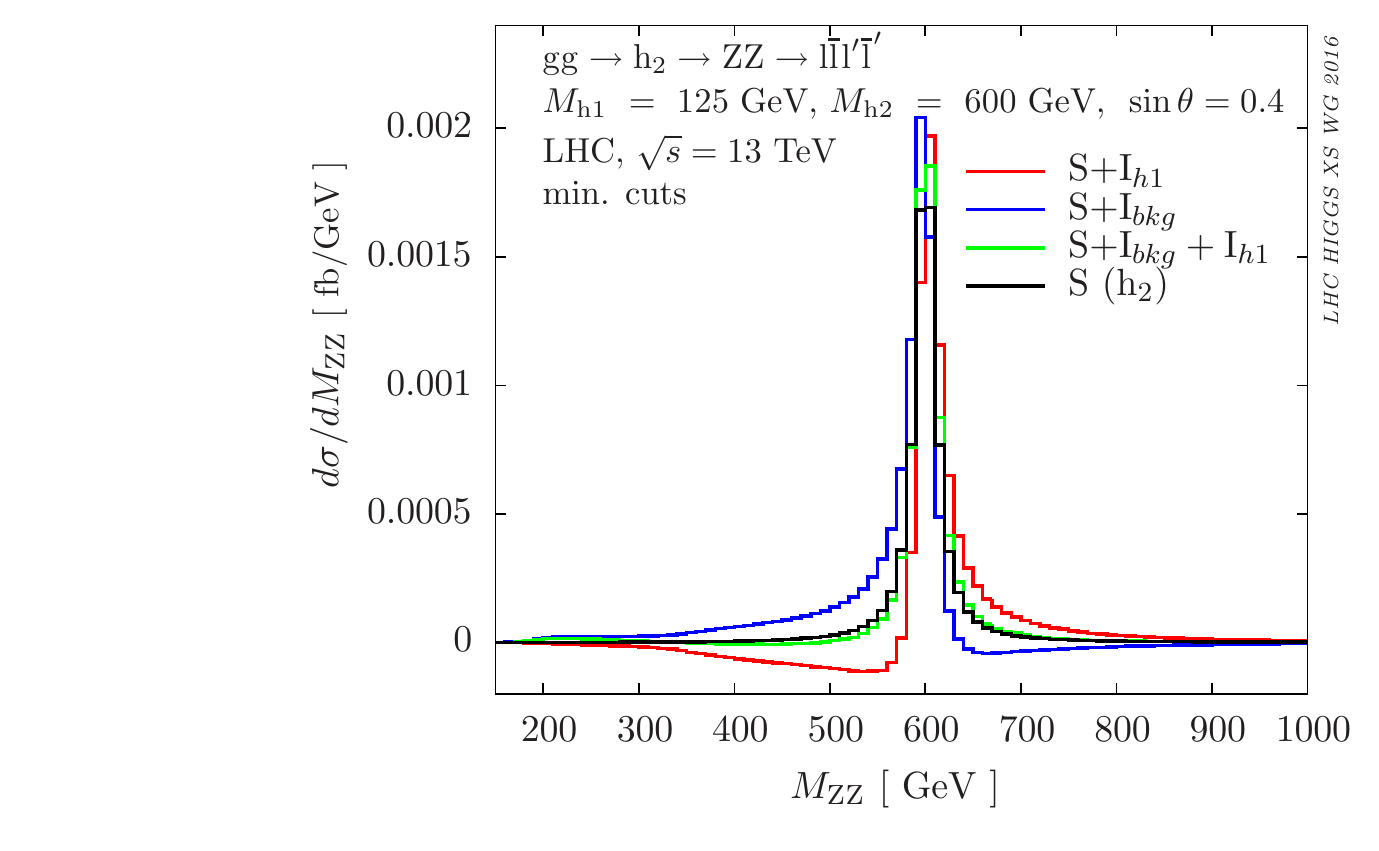}
\caption{\label{fig:ofs:sigplusint}Invariant mass distributions for $ \Pg \Pg\ (\to \{\Pho, \Pht\}) \to \PZ(\PGg^\ast)\PZ(\PGg^\ast) \to \Pl \PAl  \Pl' \PAl'$, other details as in \refT{tab:ofs:sigplusint}. }
\end{figure}

\begin{table}
\caption{\label{tab:ofs:allcontribs}
Cross sections (\UfbZ) for $\Pg \Pg\ (\to \{\Pho,\Pht\}) \to \PZ(\PGg^\ast)\PZ(\PGg^\ast) \to \Pl \PAl  \Pl' \PAl'$ in $\Pp\Pp$ collisions at $\sqrt{s}=13$ \UTeV\ in the 1HSM with $\Mho = 125$ \UGeV, $\Mht= 400, 600, 900$ \UGeV\ and mixing angle $\sin\theta=0.2$ or $0.4$ as indicated.
Results for the heavy Higgs ($\Pht$) signal ($S$), light Higgs background ($\Pho$) and continuum background (gg bkg.) are given.  Where more than one contribution is 
included, all interferences are taken into account.
Other details as in \refT{tab:ofs:sigplusint}.
} 
\renewcommand{\arraystretch}{1.2}%
\setlength{\tabcolsep}{1.5ex}%
\begin{center}
\begin{tabular}{ccccccc}
\toprule
 $\sin\theta$ & $ M_{\Ph2}$  [\UGeV]  & $S (\Pht)$ & $ \Pho$  & gg bkg. & $S$ + $\Pho$ + $I_{\Ph1}$  & all  \\ 
 \midrule
0.2 & 400	& 0.07412(6)		& 0.854(2)		& 21.18(7)	 	& 0.934(2)		& 21.86(7)		\\
0.2 & 600	& 0.01710(2)		& 0.854(2)		& 21.18(7)		& 0.867(2) 	& 21.80(7)		\\
0.2 & 900	& 0.002219(2)		& 0.854(2)		& 21.18(7)		& 0.852(2)		& 21.79(7)		\\
0.4 & 600	& 0.07065(6)		& 0.734(2)		& 21.18(7)		& 0.793(2)		& 21.77(7)		\\
\bottomrule
\end{tabular}
\end{center}
\end{table}

\begin{figure}
\includegraphics[trim=3cm 0cm 0cm 0cm, clip, width=.49\textwidth]{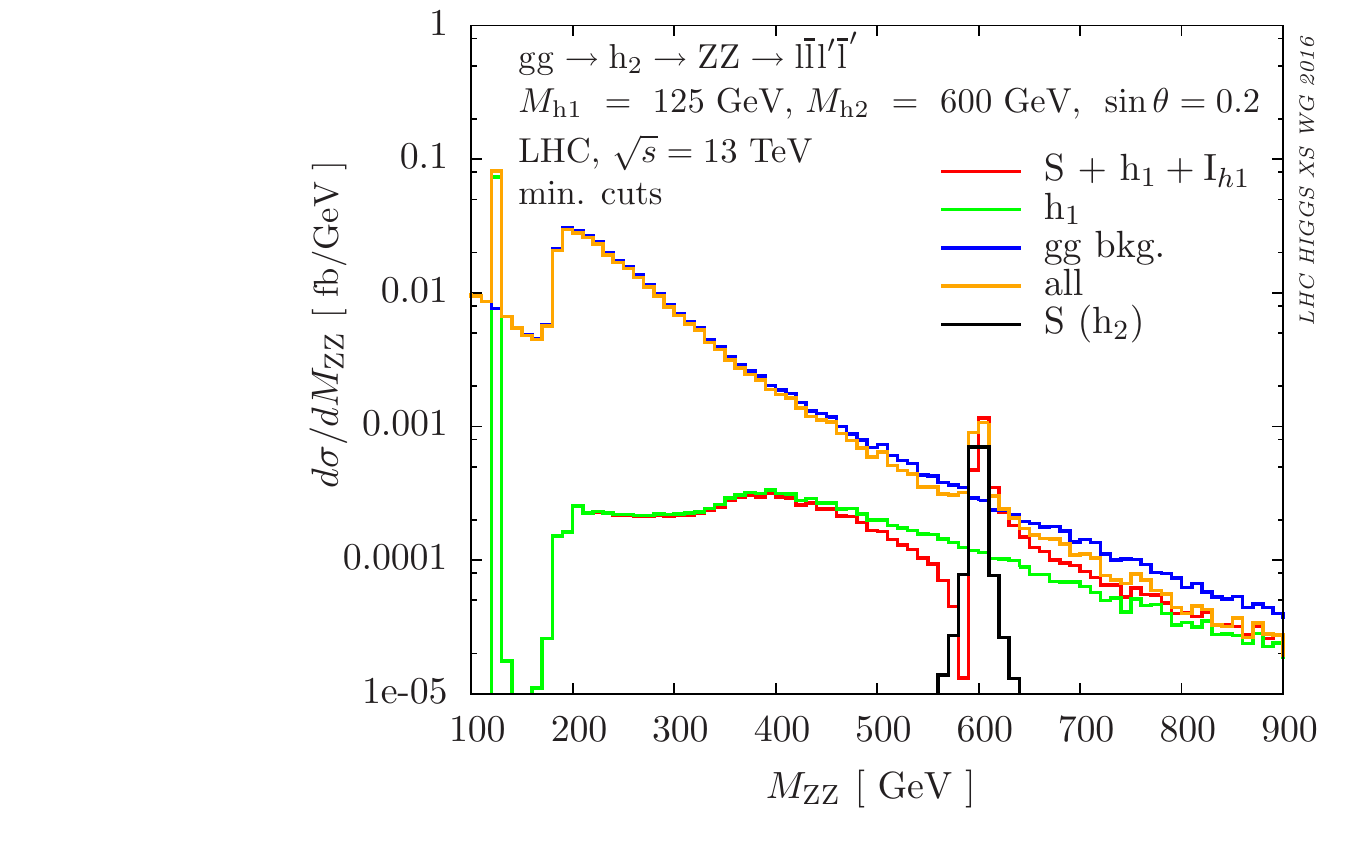}\hfil
\includegraphics[trim=3cm 0cm 0cm 0cm, clip,width=.49\textwidth]{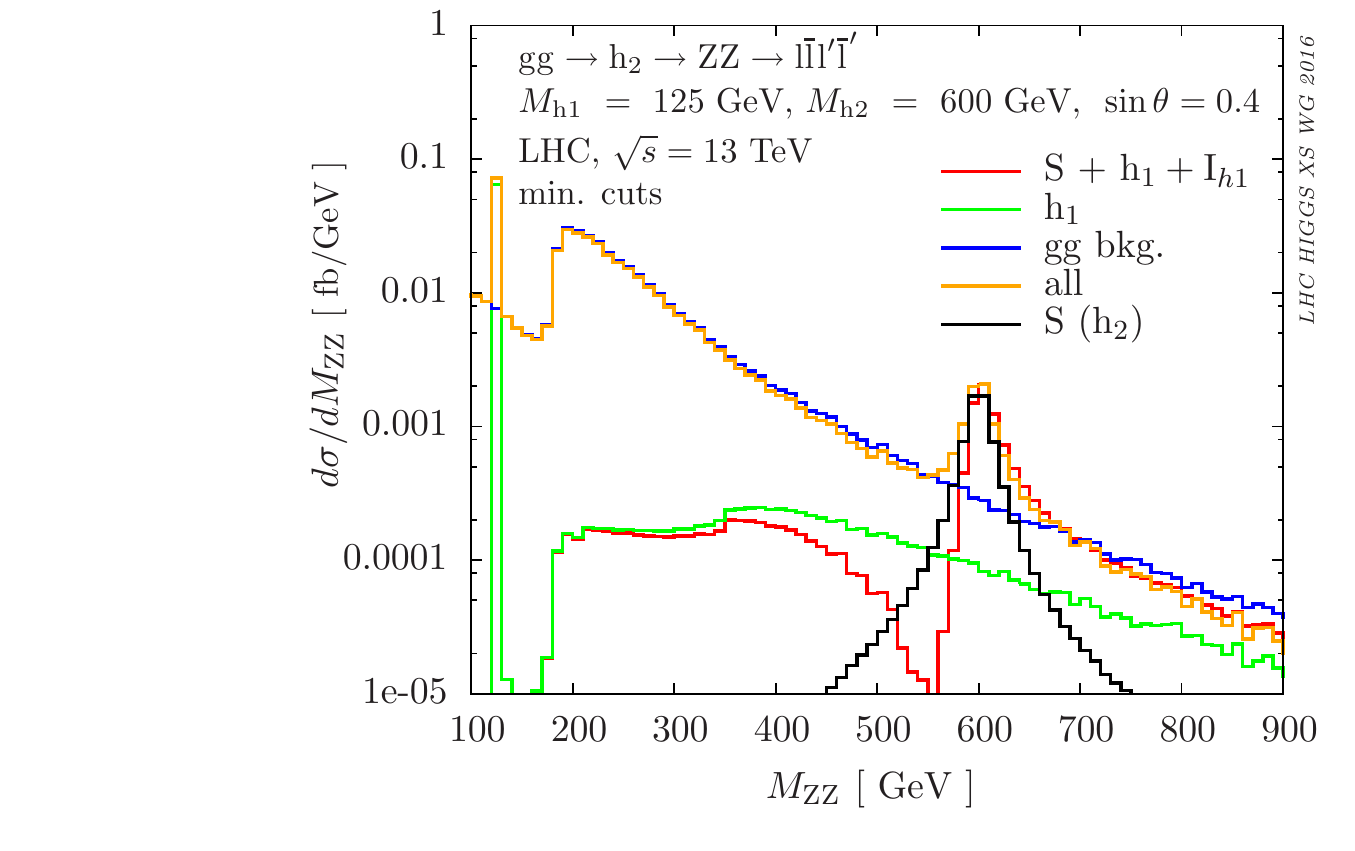}
\caption{\label{fig:ofs:allcontribs} Invariant mass distributions for $ \Pg \Pg\ (\to \{\Pho, \Pht\}) \to \PZ(\PGg^\ast)\PZ(\PGg^\ast) \to \Pl \PAl  \Pl' \PAl'$, other details as in \refT{tab:ofs:allcontribs}.}
\end{figure}

\begin{table}
\caption{\label{tab:ofs:singlet_vbf_ZZ_tight}
Cross sections for $\Pq\Pq'(\to \Pq\Pq'\{\Pho, \Pht\}) \to \Pq\Pq'\,\PZ(\PGg^\ast)\PZ(\PGg^\ast) \to \Pq\Pq'\,\Pl \PAl  \Pl' \PAl'$ in $\Pp\Pp$ collisions at $\sqrt{s}=13\UTeV$ in the 1-Higgs-Singlet Extension of the SM (1HSM). Tight VBF cuts (see \refS{sec:offshell_interf_vv_bench_sm}) are applied.  Results are given for the first, second, third and fourth 1HSM benchmark points with $\Mho = 125$\UGeV,\ $\mu_1 = \lambda_2 = \lambda_1 = 0$ and $(\Mht[\UGeVZ],\sin\theta)=(400,0.2),(600,0.2),(600,0.4),(900,0.2)$, respectively.  The sum of the light and heavy Higgs contributions including light-heavy interference (S), the interfering background without Higgs contributions (B) and the sum of the Higgs boson signal and its interference with 
the background (S+I) are given.
Cross sections are given at leading order and for a single lepton flavour combination.  
The integration error is displayed in brackets.
} 
\renewcommand{\arraystretch}{1.2}%
\setlength{\tabcolsep}{1.5ex}%
\begin{center}
\begin{tabular}{c c c c c}
\toprule
$\sigma[\UfbZ]$ & 1HSM point & $M_{\PZ\PZ}>140\UGeV$ &   $220\UGeV<M_{\PZ\PZ}< 300\UGeV$ & $M_{\PZ\PZ}> 300\UGeV$ \\ 
\midrule
 S & 1 & $1.686(2){\cdot}10^{-2}$    &  $1.185(4){\cdot}10^{-3} $   &  $1.514(2){\cdot}10^{-2}$ \\  
S+I & 1 & $- 9.69(3){\cdot}10^{-3}$   & $- 1.85(2){\cdot}10^{-3}$    & $- 6.90(2){\cdot}10^{-3}$ \\  
B & 1  & $6.725(2){\cdot}10^{-2} $   & $1.750(1){\cdot}10^{-2}$     & $4.148 (2){\cdot}10^{-2}$ \\ 
\midrule
 S & 2 & $1.436(1){\cdot}10^{-2}$    &  $1.232(4){\cdot}10^{-3} $   &  $1.259(1){\cdot}10^{-2}$ \\  
S+I & 2 & $- 1.180(3){\cdot}10^{-2}$  & $- 1.88(2){\cdot}10^{-3}$    & $- 9.00(2){\cdot}10^{-3}$ \\  
B & 2 & $6.725(2){\cdot}10^{-2} $   & $1.750(1){\cdot}10^{-2}$     & $4.148 (2){\cdot}10^{-2}$ \\ 
\midrule
 S & 3 & $2.025(2){\cdot}10^{-2}$    &  $8.90(4){\cdot}10^{-4} $    &  $1.895(2){\cdot}10^{-2}$ \\  
S+I & 3 & $- 4.34(3){\cdot}10^{-3}$  & $- 1.74(2){\cdot}10^{-3}$    & $- 1.72(3){\cdot}10^{-3}$ \\  
B & 3 & $6.725(2){\cdot}10^{-2} $   & $1.750(1){\cdot}10^{-2}$     & $4.148 (2){\cdot}10^{-2}$ \\ 
\midrule
 S & 4 & $1.263(1){\cdot}10^{-2}$    &  $1.238(4){\cdot}10^{-3} $   &  $1.085(1){\cdot}10^{-2}$ \\  
S+I & 4 & $- 1.309(3){\cdot}10^{-2}$  & $- 1.86(2){\cdot}10^{-3}$    & $- 1.029(2){\cdot}10^{-2}$ \\  
B & 4 & $6.725(2){\cdot}10^{-2} $   & $1.750(1){\cdot}10^{-2}$     & $4.148 (2){\cdot}10^{-2}$ \\ 
\bottomrule
\end{tabular}
\end{center}
\end{table}

\begin{table}
\caption{\label{tab:ofs:singlet_vbf_WW_tight}
Cross sections for $\Pq\Pq'(\to \Pq\Pq'\{\Pho, \Pht\}) \to \Pq\Pq'\,\PW\PW \to \Pq\Pq'\,\Pl \PAGnl  \PAl'\PGn_{\Pl'}$ in $\Pp\Pp$ collisions at $\sqrt{s}=13\UTeV$ in the 1-Higgs-Singlet Extension of the SM. Tight VBF cuts are applied.  
Cross sections are given for a single lepton flavour combination, but taking
into account both charge assignments, e.g.\ $(\Pl,\Pl')=(\Pe,\PGm)$ or $(\PGm,\Pe)$.
Other details as in \refT{tab:ofs:singlet_vbf_ZZ_tight}.
} 
\renewcommand{\arraystretch}{1.2}%
\setlength{\tabcolsep}{1.5ex}%
\begin{center}
\begin{tabular}{c c c c c}
\toprule
$\sigma[\UfbZ]$ & 1HSM point & $M_{\PW\PW}>140\UGeV$ &   $220\UGeV<M_{\PW\PW}< 300\UGeV$ & $M_{\PW\PW}> 300\UGeV$ \\ 
\midrule
 S & 1 & $3.283(3){\cdot}10^{-1}$  &  $2.68(1){\cdot}10^{-2} $    &  $2.758(3){\cdot}10^{-1}$ \\  
S+I & 1 & $- 1.98(2){\cdot}10^{-1}$ & $- 4.9(1){\cdot}10^{-2}$     & $- 9.8(1){\cdot}10^{-2}  $ \\  
B & 1 & $3.382(2) $           & $8.63(1){\cdot}10^{-1}$      & $1.854(1)$ \\ 
\midrule
 S & 2 & $2.727(3){\cdot}10^{-1}$  &  $2.80(1){\cdot}10^{-2} $    &  $2.189(2){\cdot}10^{-1}$ \\  
S+I & 2 & $- 2.48(2){\cdot}10^{-1}$ & $- 4.9(1){\cdot}10^{-2}$     & $- 1.48(1){\cdot}10^{-1}  $ \\  
B & 2 & $3.382(2) $            & $0.863(1)$                & $1.854(1)$ \\ 
\midrule
 S & 3 & $3.937(4){\cdot}10^{-1}$  &  $2.01(1){\cdot}10^{-2} $    &  $3.541(4){\cdot}10^{-1}$ \\  
S+I & 3 & $- 8.4(2){\cdot}10^{-2}$ & $- 4.6(1){\cdot}10^{-2}$      & $ 9(1){\cdot}10^{-3}  $ \\  
B & 3 & $3.382(2) $           &  $0.863(1)$               & $1.854(1)$ \\ 
\midrule
 S & 4 & $2.377(2){\cdot}10^{-1}$  &  $2.81(1){\cdot}10^{-2} $    &  $1.836(2){\cdot}10^{-1}$ \\  
S+I & 4 & $- 2.75(1){\cdot}10^{-1}$ & $- 4.88(1){\cdot}10^{-2}$    & $- 1.74(1){\cdot}10^{-1}  $ \\  
B & 4 & $3.382(2) $            & $0.863(1)$               & $1.854(1)$ \\ 
\bottomrule
\end{tabular}
\end{center}
\end{table}

\begin{table}
\caption{\label{tab:ofs:singlet_vbf_ZZ_loose}
Cross sections for $\Pq\Pq'(\to \Pq\Pq'\{\Pho, \Pht\}) \to \Pq\Pq'\,\PZ(\PGg^\ast)\PZ(\PGg^\ast) \to \Pq\Pq'\,\Pl \PAl  \Pl' \PAl'$ in $\Pp\Pp$ collisions at $\sqrt{s}=13\UTeV$ in the 1-Higgs-Singlet Extension of the SM. Loose VBF cuts are applied.  Other details as in \refT{tab:ofs:singlet_vbf_ZZ_tight}.
} 
\renewcommand{\arraystretch}{1.2}%
\setlength{\tabcolsep}{1.5ex}%
\begin{center}
\begin{tabular}{c c c c c}
\toprule
$\sigma[\UfbZ]$ & 1HSM point & $M_{\PZ\PZ}>140\UGeV$ &   $220\UGeV<M_{\PZ\PZ}< 300\UGeV$ & $M_{\PZ\PZ}> 300\UGeV$ \\ 
\midrule
S & 1 & $2.272(2){\cdot}10^{-2}$   &  $1.94(1){\cdot}10^{-3}$       & $1.983(2){\cdot}10^{-2}$ \\ 
 S+I & 1 & $- 1.34(1){\cdot}10^{-2}$  &  $- 3.33(4){\cdot}10^{-3} $    &  $- 8.17(5) {\cdot}10^{-3}$ \\  
B & 1 & $1.3950(5){\cdot}10^{-1} $ & $4.005(3){\cdot}10^{-2}$       & $7.964(4){\cdot}10^{-2}$ \\ 
\midrule
S & 2 & $1.889(2){\cdot}10^{-2}$   &  $2.00(1){\cdot}10^{-3}$       & $1.592(2){\cdot}10^{-2}$  \\ 
 S+I & 2 & $- 1.68(1){\cdot}10^{-2}$  &  $- 3.40(4){\cdot}10^{-3} $    &  $- 1.154(5) {\cdot}10^{-2}$ \\  
B & 2 & $1.3950(5){\cdot}10^{-1} $ & $4.005(3){\cdot}10^{-2}$       & $7.964(4){\cdot}10^{-2}$ \\ 
\midrule
S & 3 & $2.590(3){\cdot}10^{-2}$   &  $1.45(1){\cdot}10^{-3}$       & $2.372(2){\cdot}10^{-2}$ \\ 
 S+I & 3 & $- 6.9(1){\cdot}10^{-3}$  &  $- 3.12(4){\cdot}10^{-3} $    &  $- 2.1(1) {\cdot}10^{-3}$ \\  
B & 3 & $1.3950(5){\cdot}10^{-1} $ & $4.005(3){\cdot}10^{-2}$       & $7.964(4){\cdot}10^{-2}$ \\ 
\midrule
S & 4 & $1.658(2){\cdot}10^{-2}$   &  $2.02(1){\cdot}10^{-3}$       & $1.359(2){\cdot}10^{-2}$ \\ 
 S+I & 4 & $- 1.85(1){\cdot}10^{-2}$  &  $- 3.36(4){\cdot}10^{-3} $    &  $- 1.329(5) {\cdot}10^{-2}$ \\ 
B & 4 & $1.3950(5){\cdot}10^{-1} $ & $4.005(3){\cdot}10^{-2}$       & $7.964(4){\cdot}10^{-2}$ \\ 
\bottomrule
\end{tabular}
\end{center}
\end{table}

\begin{table}
\caption{\label{tab:ofs:singlet_vbf_WW_loose}
Cross sections for $\Pq\Pq'(\to \Pq\Pq'\{\Pho, \Pht\}) \to \Pq\Pq'\,\PW\PW \to \Pq\Pq'\,\Pl \PAGnl  \PAl'\PGn_{\Pl'}$ in $\Pp\Pp$ collisions at $\sqrt{s}=13\UTeV$ in the 1-Higgs-Singlet Extension of the SM. Loose VBF cuts are applied.  Other details as in \refT{tab:ofs:singlet_vbf_WW_tight}.
} 
\renewcommand{\arraystretch}{1.2}%
\setlength{\tabcolsep}{1.5ex}%
\begin{center}
\begin{tabular}{c c c c c}
\toprule
$\sigma[\UfbZ]$ & 1HSM point & $M_{\PW\PW}>140\UGeV$ &   $220\UGeV<M_{\PW\PW}< 300\UGeV$ & $M_{\PW\PW}> 300\UGeV$ \\ 
\midrule
 S & 1 & $4.600(5){\cdot}10^{-1}$      &  $4.46(1){\cdot}10^{-2} $   &  $3.692(4){\cdot}10^{-1}$ \\  
S+I & 1 & $- 3.25(4){\cdot}10^{-1}$     & $- 9.3(2){\cdot}10^{-2}$    & $- 1.23(3){\cdot}10^{-1}$ \\  
B & 1 & $7.424(3) $               & $2.001(1)$               & $3.815(2)$ \\ 
\midrule
 S & 2 & $3.733(3){\cdot}10^{-1}$      &  $4.59(1){\cdot}10^{-2} $   &  $2.805(3){\cdot}10^{-1}$ \\  
S+I & 2 & $- 4.05(4){\cdot}10^{-1}$     & $- 9.2(2){\cdot}10^{-2}$    & $- 2.00(3){\cdot}10^{-1}$ \\  
B & 2 & $7.424(3) $               & $2.001(1)$               & $3.815(2)$ \\ 
\midrule
 S & 3 & $5.17(1){\cdot}10^{-1}$      &  $3.33(1){\cdot}10^{-2} $    &  $4.482(5){\cdot}10^{-1}$ \\  
S+I & 3 & $- 1.88(4){\cdot}10^{-1}$     & $- 8.5(2){\cdot}10^{-2}$    & $+ 1(3){\cdot}10^{-3}$ \\  
B & 3 & $7.424(3) $               & $2.001(1)$               & $3.815(2)$ \\ 
\midrule
 S & 4 & $3.274(3){\cdot}10^{-1}$      &  $4.65(1){\cdot}10^{-2} $   &  $2.339(3){\cdot}10^{-1}$ \\  
S+I & 4 & $- 4.43(4){\cdot}10^{-1}$     & $- 9.6(2){\cdot}10^{-2}$    & $- 2.38(3){\cdot}10^{-1}$ \\  
B & 4 & $7.424(3) $               & $2.001(1)$               & $3.815(2)$ \\ 
\bottomrule
\end{tabular}
\end{center}
\end{table}

\begin{figure}
\includegraphics[clip, width=.49\textwidth]{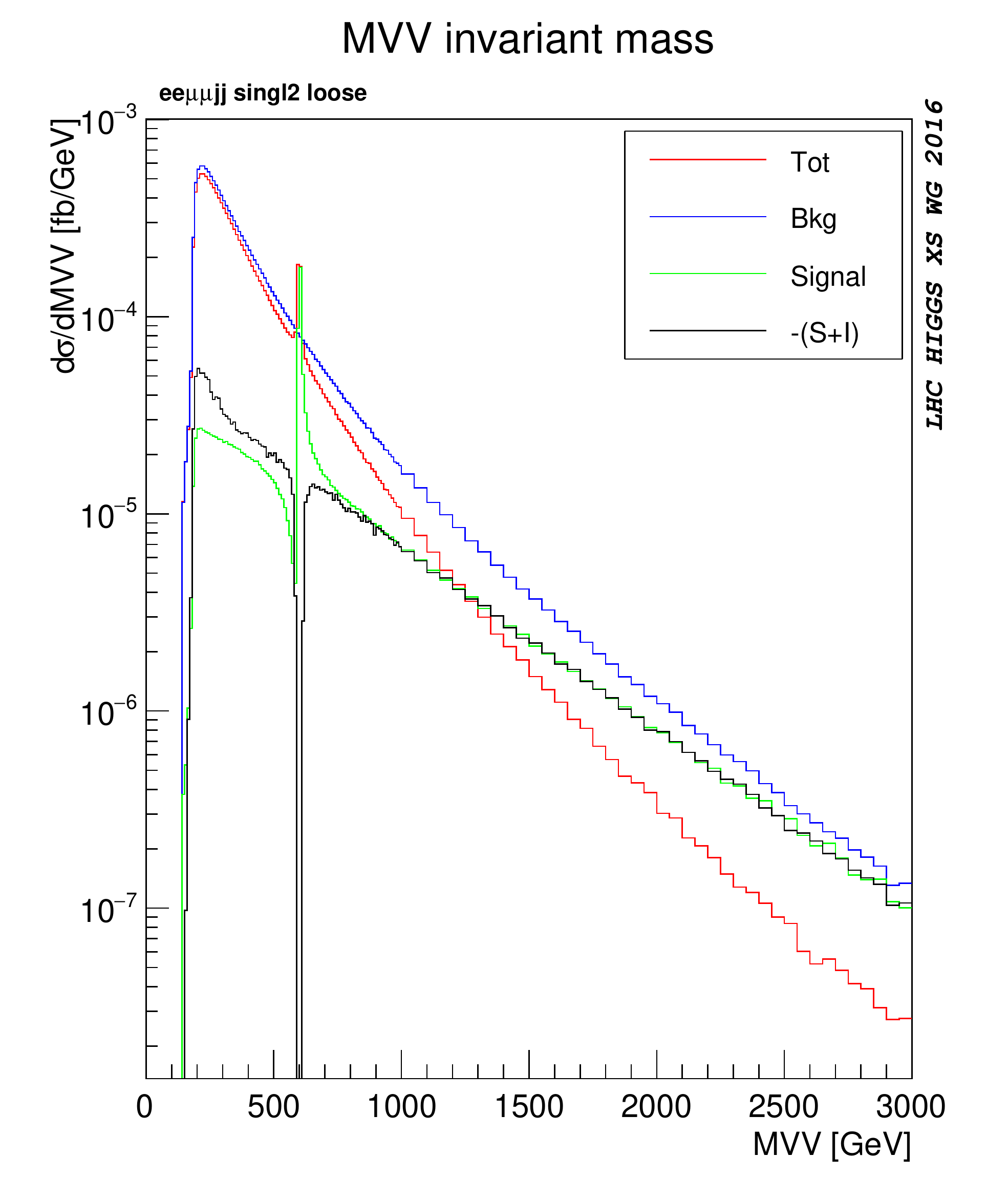}\hfil
 \includegraphics[clip,width=.49\textwidth]{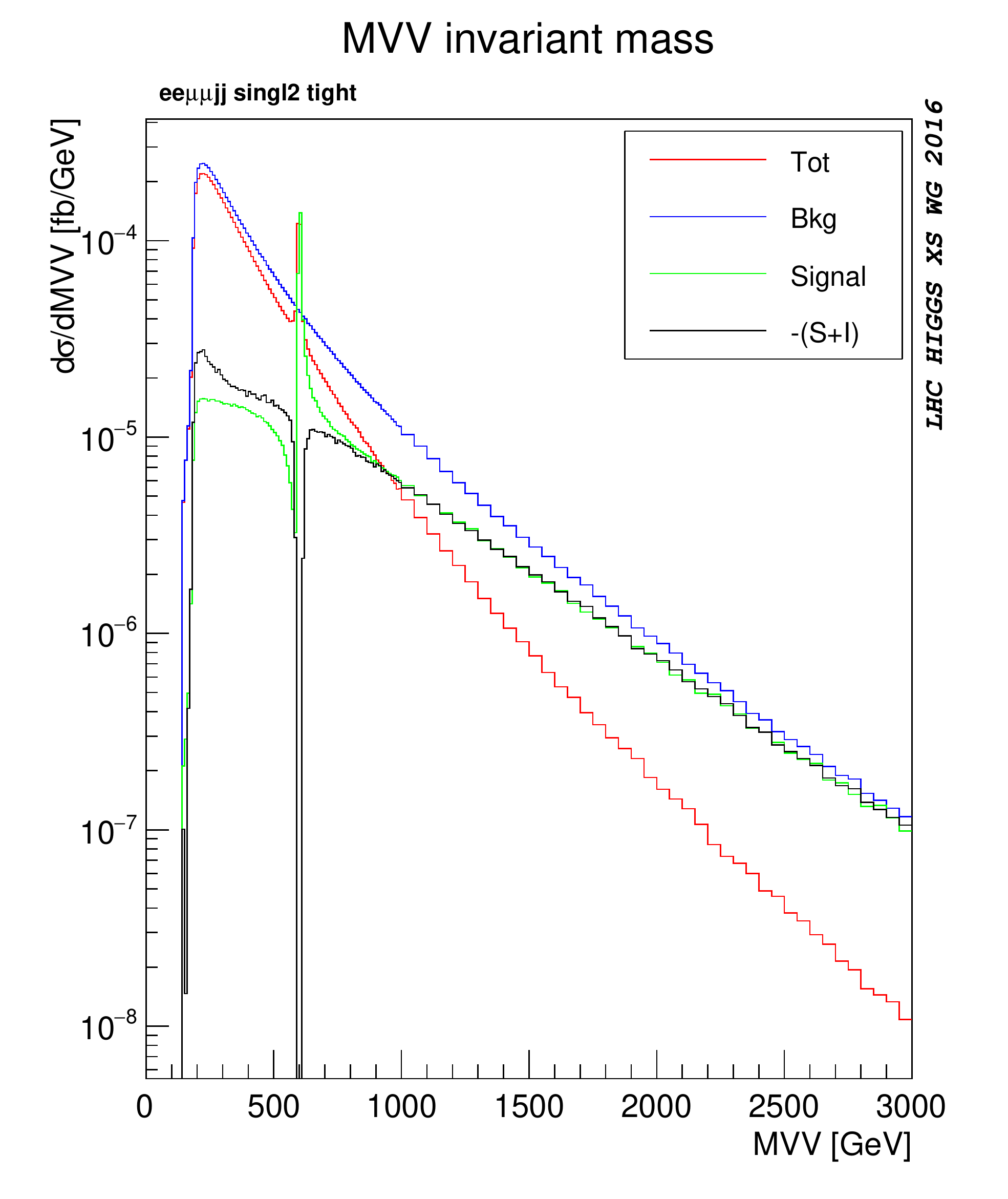}
\caption{\label{fig:ofs:singlet_vbf} Invariant mass distributions for 
$\Pq\Pq'(\to \Pq\Pq'\{\Pho, \Pht\}) \to \Pq\Pq'\,\PZ(\PGg^\ast)\PZ(\PGg^\ast) \to \Pq\Pq'\,\Pl \PAl  \Pl' \PAl'$ in $\Pp\Pp$ collisions at $\sqrt{s}=13\UTeV$. Loose and tight VBF cuts are applied in the left and right graphs, respectively.  Results for the second 1HSM benchmark point ($\Mho = 125 \UGeV$, $\Mht = 600\UGeV$, $\sin\theta=0.2$) are shown: the sum of the light and heavy Higgs contributions including light-heavy interference (Signal), the interfering background without Higgs contributions (Bkg), the sum of Signal and Bkg including interference (Tot), and the 
negative of the sum of Signal and its interference with Bkg (-(S+I)).
Other details as in \refT{tab:ofs:singlet_vbf_ZZ_tight}.}
\end{figure}




\subsection{Multijet merging effects in \texorpdfstring{$gg\to\ell\bar\nu_\ell\bar{\ell'}\nu_{\ell'}$}{gg to lnulnu} using \texorpdfstring{\protect\Sherpa}{SHERPA}}
\label{sec:offshell_interf_vv_sherpa}
\subsubsection{Set-up}
\noindent
In this section, results for the loop--\-induced process
$gg\to\ell\bar\nu_\ell\bar{\ell'}\nu_{\ell'}$ obtained with the {\Sherpa}  event
generation framework~\cite{Gleisberg:2008ta} will be presented, with the goal
to highlight the effect of multijet merging~\cite{Catani:2001cc} on some
critical observables.  This is accomplished by directly comparing the results
where the leading order processes depicted in Figure~\ref{Fig::LO_feymans}
have been supplemented with the parton shower (labelled LOOP2+PS) with a sample
where an additional jet has been produced, {\it i.e.}\xspace the quark-loop
induced processes $gg\to\ell\bar\nu_\ell\bar{\ell'}\nu_{\ell'}g$ and
$qg\to\ell\bar\nu_\ell\bar{\ell'}\nu_{\ell'}q$ (labelled MEPS@LOOP2) as shown in
Figure~\ref{Fig::NLO_feynmans}.  In addition, these two samples are further
subdivided into those including a Higgs boson of $m_H=125$ GeV and those where
the Higgs boson has been decoupled with $m_H\to\infty$.
\begin{figure}
  \begin{center}
    \begin{tabular}{ccc}
      \includegraphics[width=0.3\textwidth]{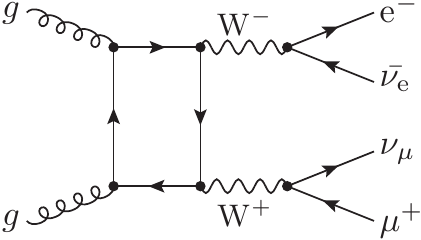} &
      \includegraphics[width=0.3\textwidth]{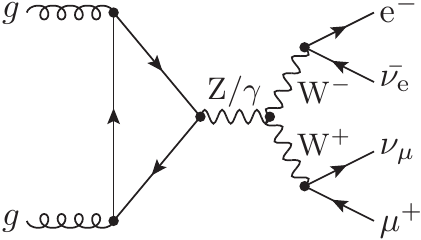} &
      \includegraphics[width=0.3\textwidth]{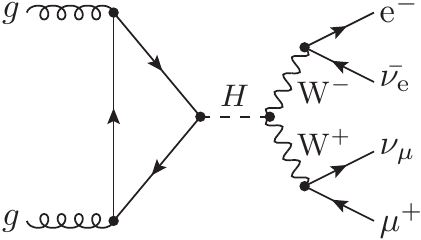}
    \end{tabular}
    \caption{Leading order Feynman diagrams contributing to
      $gg\to\ell\bar\nu_\ell\bar{\ell'}\nu_{\ell'}$.}
    \label{Fig::LO_feymans}
  \end{center}
\end{figure}
\begin{figure}
  \begin{center}
    \begin{tabular}{ccc}
      \includegraphics[width=0.3\textwidth]{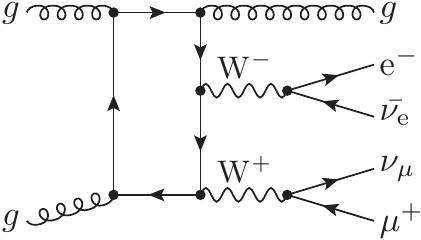} &
      \includegraphics[width=0.3\textwidth]{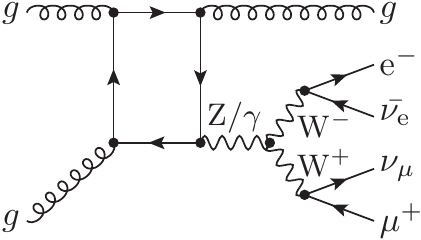} &
      \includegraphics[width=0.3\textwidth]{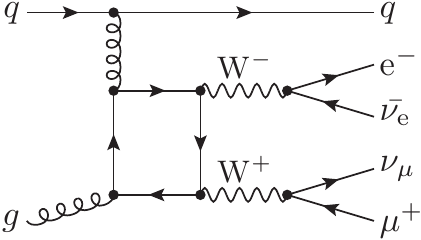}
    \end{tabular}
    \caption{Leading order Feynman diagrams contributing to the background
      production of final states $\ell\bar\nu_\ell\bar{\ell'}\nu_{\ell'}+$jet
      through a quark loop.}
    \label{Fig::NLO_feynmans}
  \end{center}
\end{figure}
Here, the matrix elements are provided from the \OpenLoops+{\Collier}  
package~\cite{Cascioli:2011va,Denner:2014gla} are being used.  For parton
showering, the implementation of~\cite{Schumann:2007mg} is employed, with a
starting scale
\begin{equation}
  \mu_Q^2=p_{\perp,\ell\bar\nu_\ell\bar{\ell'}\nu_{\ell'}}^2+m^2_{\ell\bar\nu_\ell\bar{\ell'}\nu_{\ell'}}\,.
\end{equation}
A similar analysis, although for centre-of-mass energies of 8 TeV has already
been presented in~\cite{Cascioli:2013gfa}.  Here, in addition, the effect of
including a Higgs boson with mass $m_H=125$ GeV is investigated, which was not
the case in the previous analysis.  Results without the Higgs boson are
obtained by effectively decoupling it, pushing its mass to very high values
in the calculation, $m_H\to\infty$.

\subsubsection{Results}
\noindent
In this investigation the following cuts have been applied:
\begin{eqnarray}
  \begin{array}{lclclcl}
    p_{\perp,\,\ell}\,&\ge&\,25\,\mbox{\rm GeV}\,,&\;\;\;&
    |\eta_\ell|\,&\le\,& 2.5\nonumber\\
    p_{\perp,\,j}\,&\ge&\,30\,\mbox{\rm GeV}\,,&\;\;\;&
    |\eta_j|\,&\le\,& 5\,,
  \end{array}
\end{eqnarray}
where jets are defined by the anti-$k_T$ algorithm with $R=0.4$.  In addition
a cut on the missing transverse momentum has been applied,
\begin{equation}
  E\!\!\!/_T\,\ge\,25\,\mbox{\rm GeV}\,,
\end{equation}
which of course is practically given by the combined neutrino momenta.

\noindent
In Figure~\ref{Fig::jetmultis} inclusive and exclusive jet multiplicities
as obtained from the samples described above are displayed.  They clearly
show that especially for jet multiplicities $N_{\mathrm jet}\ge 1$ the impact
of multijet merging is sizeable and important.  Furthermore, there is a visible
difference in the overall rate of about a factor of 2 between the results
with and without the Higgs boson.
\begin{figure}
  \begin{center}
    \begin{tabular}{cc}
      \includegraphics[width=0.45\textwidth]{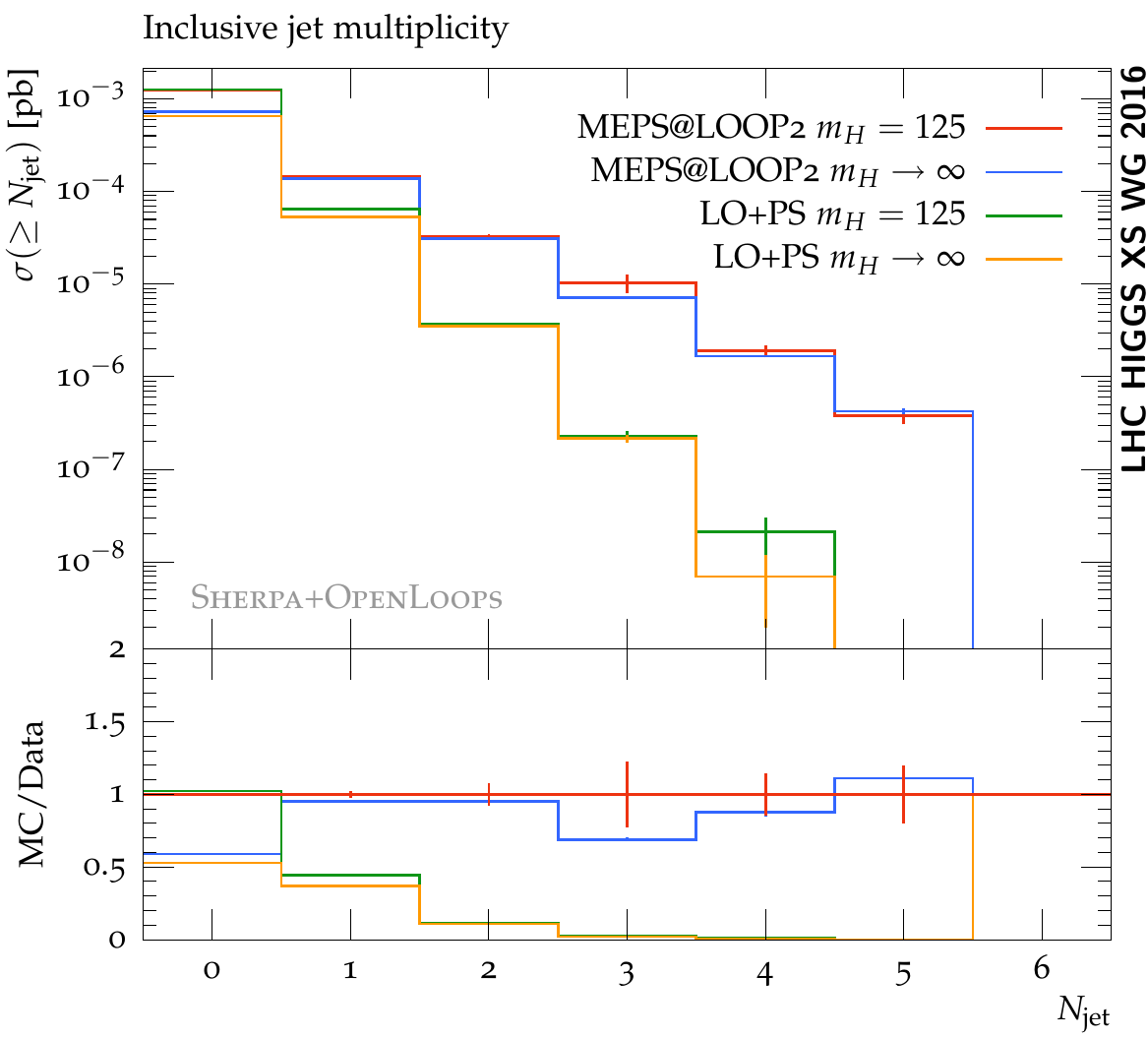} &
      \includegraphics[width=0.45\textwidth]{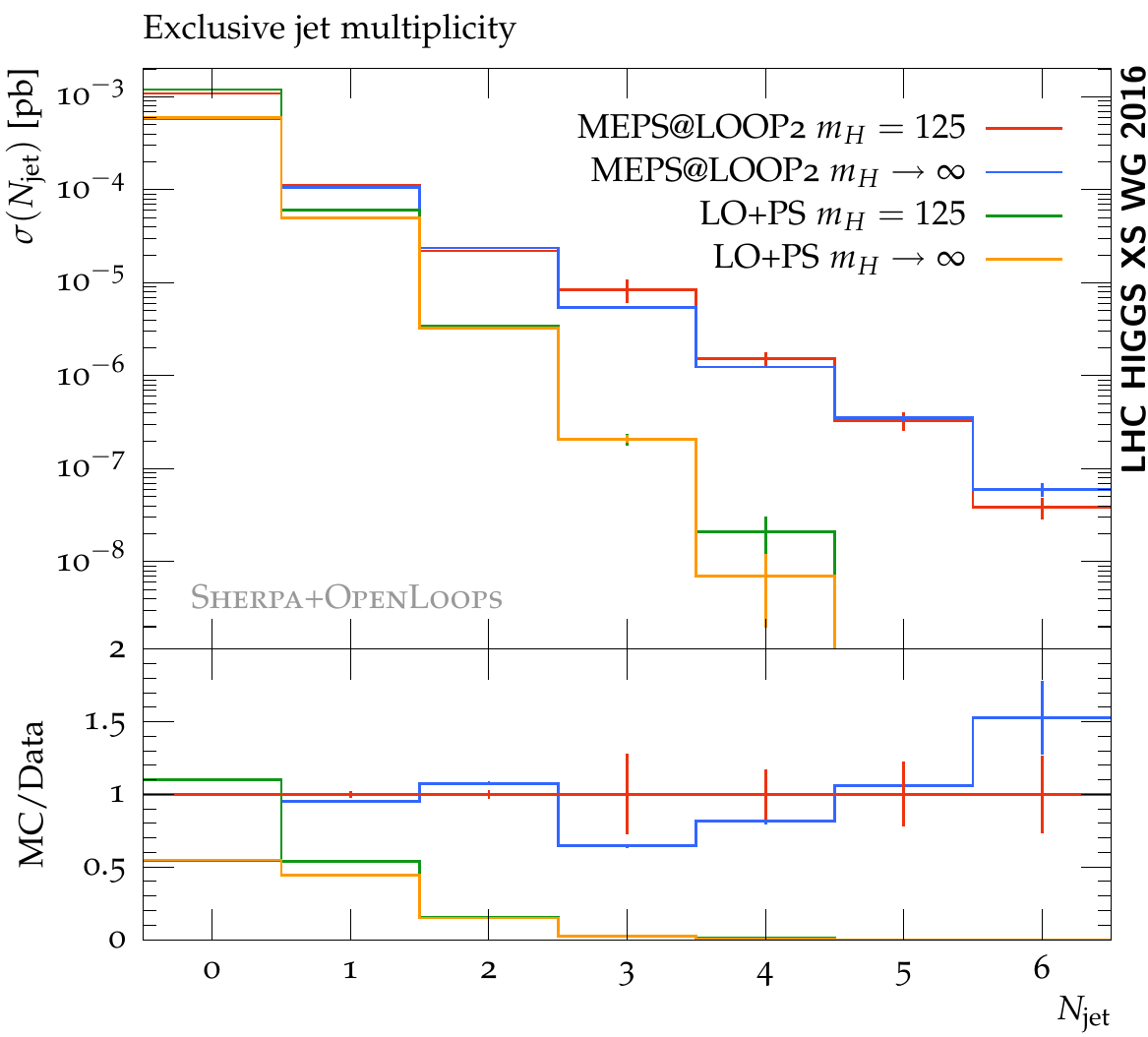}
    \end{tabular}
    \caption{Inclusive (left) and exclusive (right) jet cross sections
      with and without multijet merging and with ($m_H=125$ GeV) and without
      ($m_H\to\infty$) including a Higgs boson, including multijet merging or
      merely relying on the parton shower to simulate all QCD emissions.}
    \label{Fig::jetmultis}
  \end{center}
\end{figure}
This becomes even more visible when considering cross sections after the
application of a jet veto, cf.~the right panel of Figure~\ref{Fig::jetvetoes}.
\begin{figure}
  \begin{center}
    \begin{tabular}{cc}
      \includegraphics[width=0.45\textwidth]{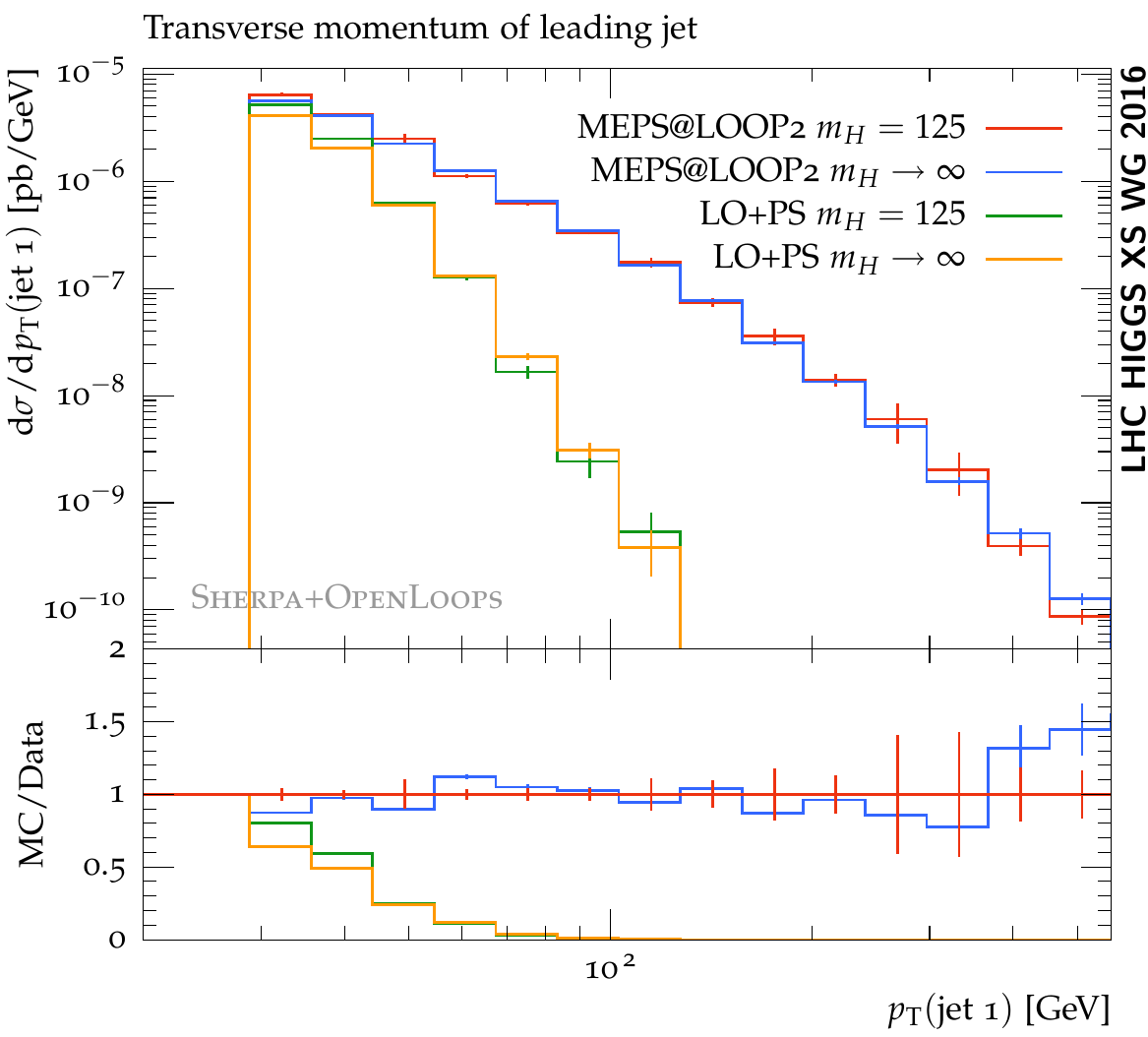}&
      \includegraphics[width=0.45\textwidth]{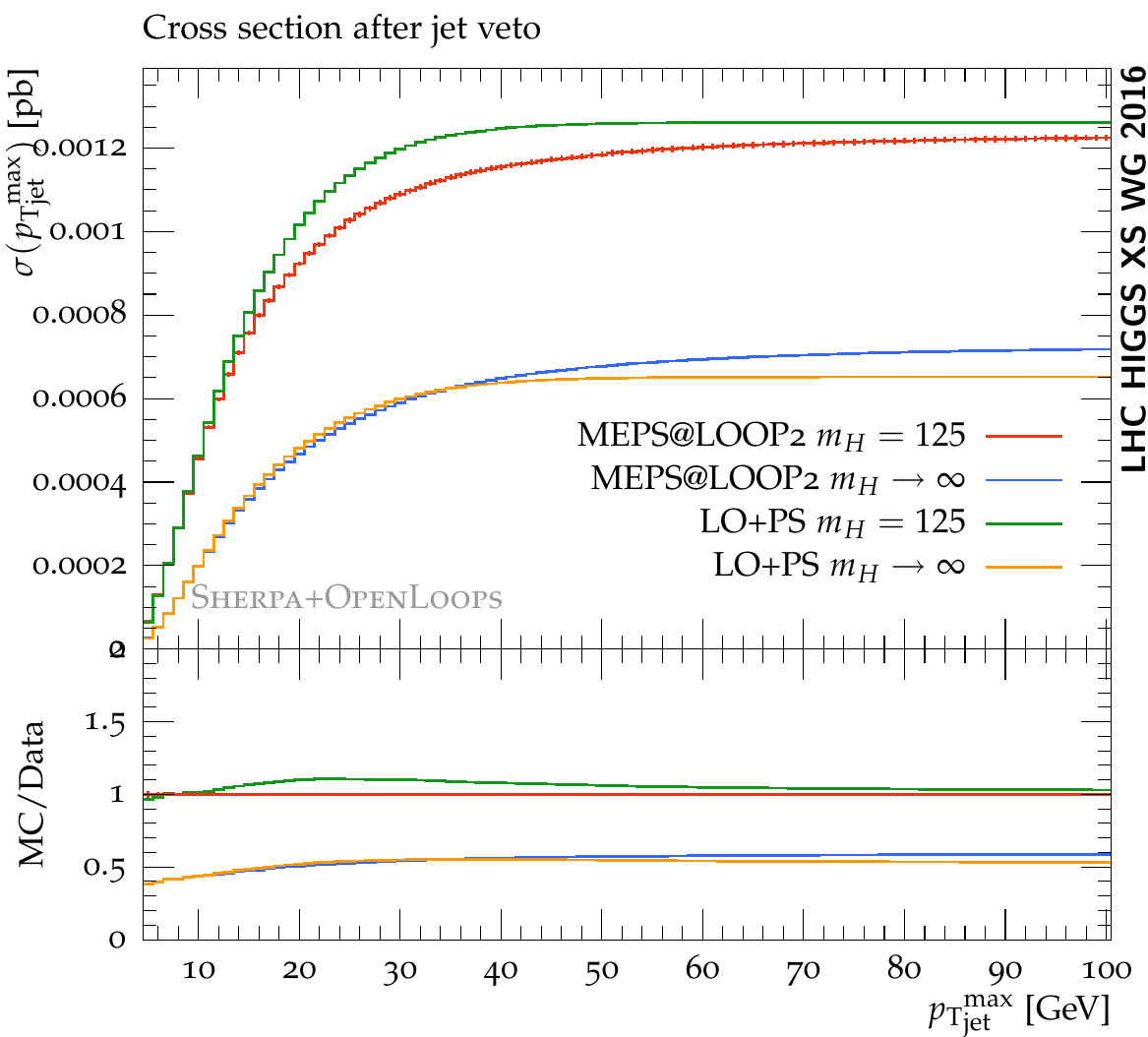}
    \end{tabular}
    \caption{Differential cross section in dependence of the transverse
      momentum of the leading jet (left) and the cross section after
      application of a jet veto in dependence of the transverse momentum
      cut on jets (right).}
    \label{Fig::jetvetoes}
  \end{center}
\end{figure}
Multijet merging leads to jets that are visibly harder -- the LOOP2+PS
results fall of very quickly with respect to the merged result, see the left
panel of \refF{Fig::jetmultis}.  However, since the bulk of the inclusive
cross section is related to jet transverse momenta below about 30 GeV, the
jet-vetoed cross section saturates relatively quickly and is thus
correspondingly independent of the hard tails in transverse momentum.  This
ultimately leads to effects of the order of about 10\% or so from multijet
merging.  At the same time, in the linear plot of the jet-vetoed cross section
the rate difference due to the inclusion of the Higgs boson becomes visible.
As expected, these differences manifest themselves in the usual kinematic
regions stemming from spin effects in the decay of the $W$ bosons,
illustrated in Figure~\ref{Fig::llcorrels}.
\begin{figure}
  \begin{center}
    \begin{tabular}{cc}
      \includegraphics[width=0.45\textwidth]{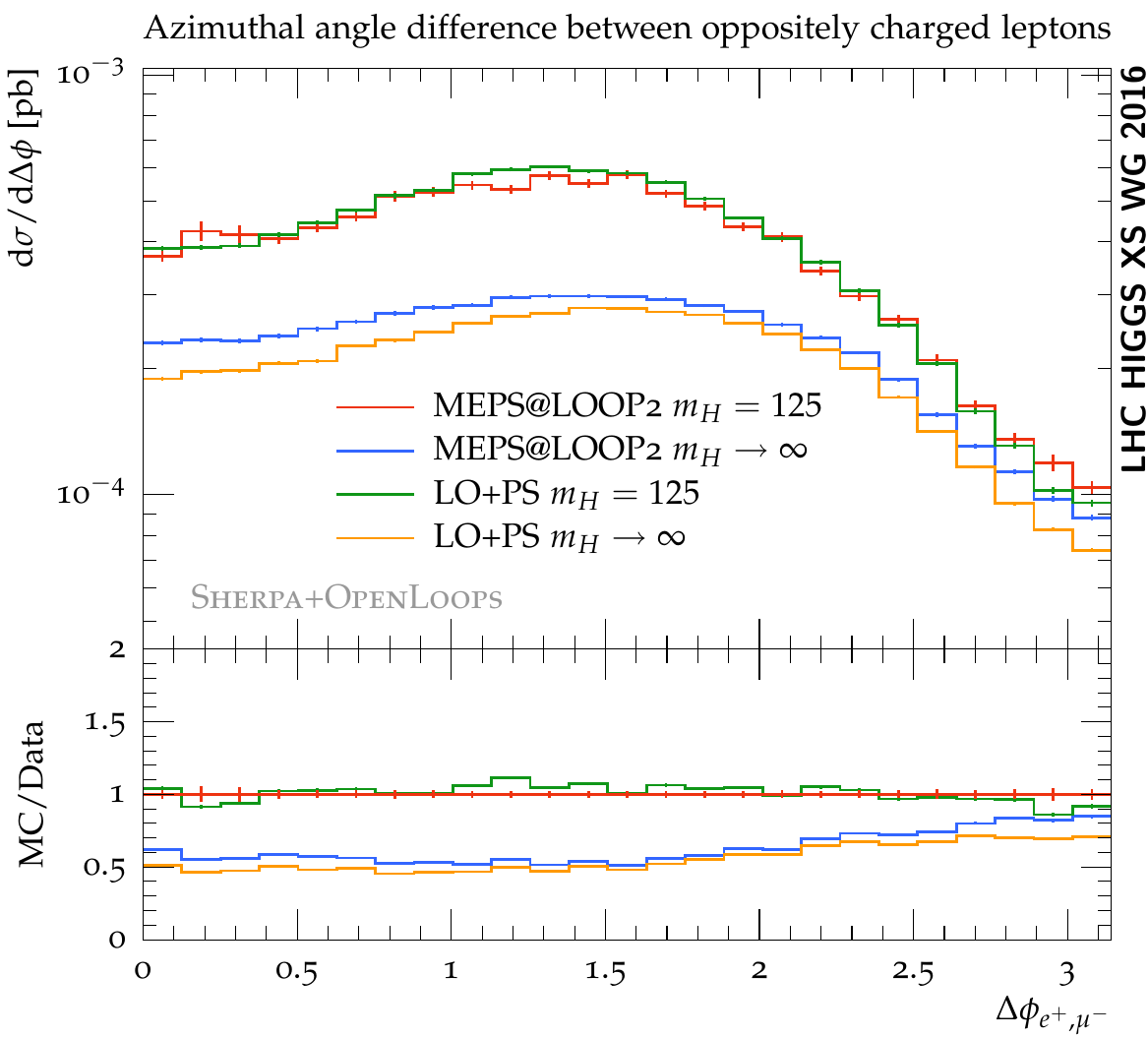}
      \includegraphics[width=0.45\textwidth]{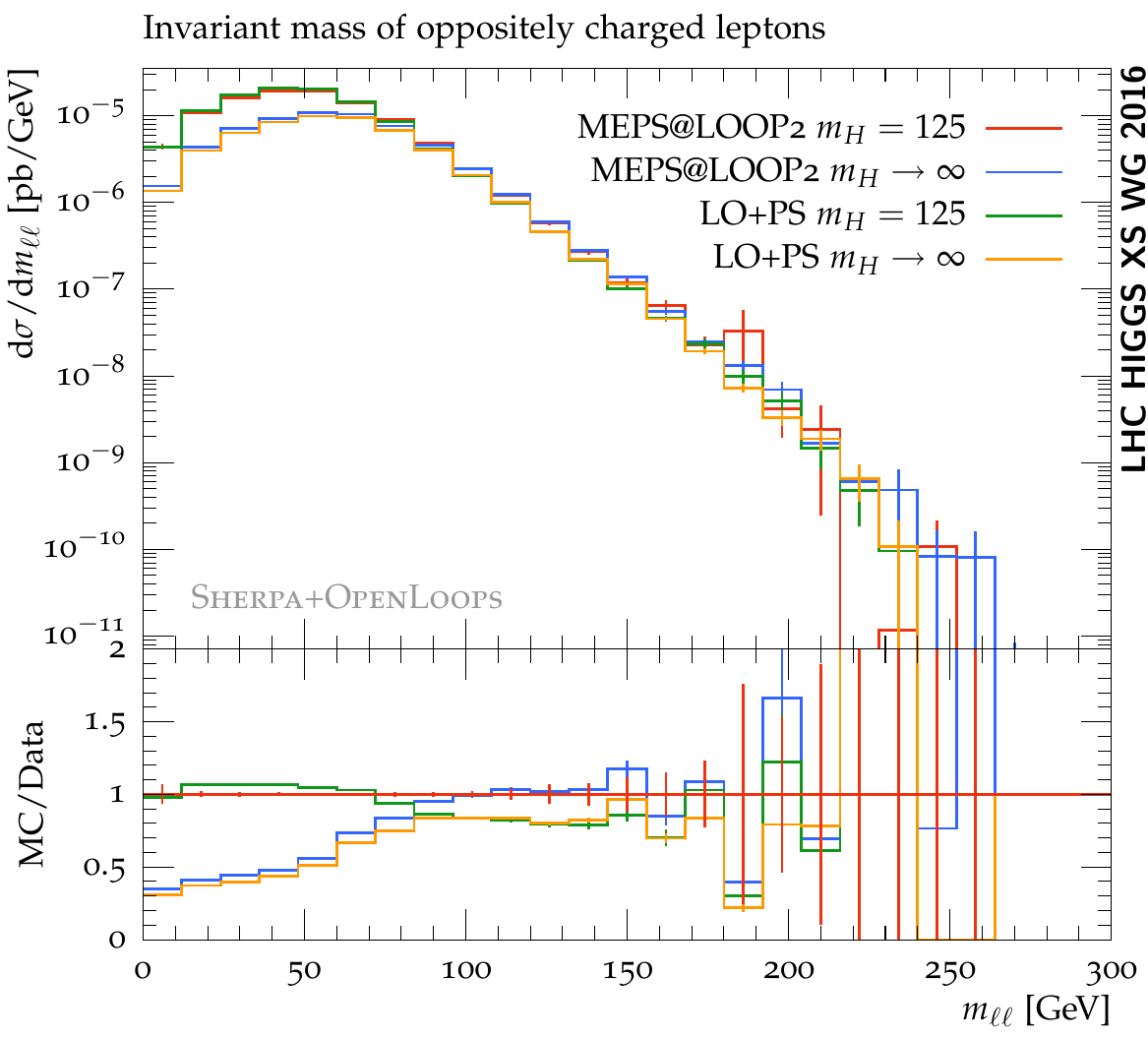}
    \end{tabular}
    \caption{Differential cross section in dependence of the transverse
      separation of the two leptons (left) and of their invariant mass
      (right).}
    \label{Fig::llcorrels}
  \end{center}
\end{figure}
Clearly, the presence of a Higgs boson pushes the leptons closer in phase
space.  Since the overall rate is dominated by the 0-jet bin, the
differences between merged and LO samples are again relatively small, of
the order of 10\% or below.

\noindent
To summarize: the application of multijet merging to loop--\-induced processes
$gg\to VV^{(*)}$ leads to visibly harder jet spectra and significantly larger
jet multiplicities, irrespective of whether this process is mediated by a
Higgs boson or not.  It is clearly the overall scale of the process and the
fact that the initial states are identical that is responsible here.  The
effect on jet-vetoed cross sections in the 0-jet bin is small, 10\% or
below, since these cross sections essentially appear after integration over
the jet-cross section up to the veto scale.  Clearly, though, this would be
different when asking for exactly one jet and vetoing further jets.  The
impact of the merging is small on the lepton correlations in the regions 
that are important for the definition of signal and background regions.


\subsection{Study of higher-order QCD corrections in the \texorpdfstring{gg$\rightarrow$H$\rightarrow$VV}{gg to H to VV} process}
\label{sec:offshell_interf_vv_atlas_mc_gg}

\subsubsection{Introduction}
\label{sec:offshell_gghvv_shower_intro}

The analysis \cite{Aad:2015xua} employed to extract the off-shell signal strength in the high mass ($m_{\mathrm{4\ell}}>$220 GeV) ZZ$\rightarrow$4$\ell$, ZZ$\rightarrow$2$\ell$2$\nu$ and WW$\rightarrow\ell\nu\ell\nu$ final states, is based on two Monte Carlo simulations for gg-initiated processes,
namely gg2VV \cite{Kauer:2013qba} and MCFM \cite{Campbell:2013una}. The dominant gg-initiated processes used in the analysis \cite{Aad:2015xua} are listed below:
\begin{enumerate}
\item gg$\rightarrow H \rightarrow$ ZZ, the signal (S) comprising both the on-shell peak at $m_{\mathrm{H}}=$125.5 GeV and the off-shell region where the Higgs boson acts as a propagator;
\item gg$\rightarrow$ ZZ, the continuum background (B);
\item gg$\rightarrow (H^{*}) \rightarrow$ ZZ, the signal, continuum background and interference contribution, labelled as SBI in what follows.
\end{enumerate}
However, only Lowest-Order (LO) in QCD Monte Carlo simulations are available, namely gg2VV and MCFM with Pythia8 \cite{Sjostrand:2007gs} showering. For this reason, mass-dependent K-factors to higher order accuracy are needed to achieve a better precision.
\begin{itemize}
\item For the signal process, higher order QCD corrections are computed: LO to Next-to-Next-to-Leading-Order (NNLO) K-factors are calculated as a function of the diboson invariant mass $m_{\mathrm{ZZ}}$.
\item  For the background process, the full K-factor from LO to NNLO accuracy is currently not available.
\end{itemize}

Different approaches exploited in order to take into account the absence of higher order QCD corrections in gg$\rightarrow (H^{*}) \rightarrow$VV final states (it is to note that Next-to-Leading Order, NLO, gg$\rightarrow$ZZ QCD calculation has been recently performed \cite{Caola:2015psa}) and the systematic uncertainties associated to these processes will be detailed in the following sections.

\subsubsection{Parton Shower Scheme Dependence}
\label{sec:offshell_gghvv_shower_psscheme}

Given that no higher order matrix element calculations are available for the $gg$-initiated processes, the only way to simulate QCD radiation is through the parton shower. However, as the generation is done at LO in QCD, there is no clear prescription to evaluate the systematic uncertainties on the QCD scale. According to the maximum jet $p_{\mathrm{T}}$ scale emission characterizing the parton showers, two different configurations \cite{ATL-PHYS-PUB-2016-006} are exploited, the \textit{power shower} (the emission is allowed up to the kinematical limit) and the \textit{wimpy shower} (the shower is started at the value of the factorization or the renormalization scale). Pythia8 is tuned as default with the power shower option. The comparison is carried out involving the following parton shower schemes at generator level:

\begin{itemize}
\item Pythia8 power shower including a matrix element correction on the first jet emission such that information coming from the exact matrix element calculation is exploited for the hardest jet in the shower \cite{Sjostrand:2007gs};
\item Pythia8 power shower  without a matrix element correction;
\item Pythia8 wimpy shower without a matrix element correction;
\item Herwig6.5 \cite{Corcella:2000bw} in combination with Jimmy.
\end{itemize}
The items above are finally compared to high-mass Powheg-Box \cite{Alioli:2008tz} NLO gg$\rightarrow H \rightarrow$ ZZ event sample with a Higgs boson mass generated with $m_{\mathrm{H}}$=380 GeV, chosen around the most sensitive off-shell invariant mass region for the analysis. The normalized $p_{\mathrm{T}}$(ZZ) distributions, detailed in \refF{fig:ps} as reported in Ref. \cite{ATL-PHYS-PUB-2016-006} for the sample above in the text are plotted in the same high ZZ mass range (345$<m_{\mathrm{4\ell}}<$415) GeV in order to ensure a compatible mass of the hard interaction system. As the default samples are generated with the LO gg$\rightarrow (H^{*})\rightarrow$ZZ matrix element with Pythia8 using the power shower parton shower option and this sample shows the largest deviation from Powheg, the full difference of the order of 10\% is taken as a systematic uncertainty in the ATLAS analysis as in \cite{Aad:2015xua}.

\begin{figure*}
\begin{center}
\includegraphics[width=0.6\linewidth]{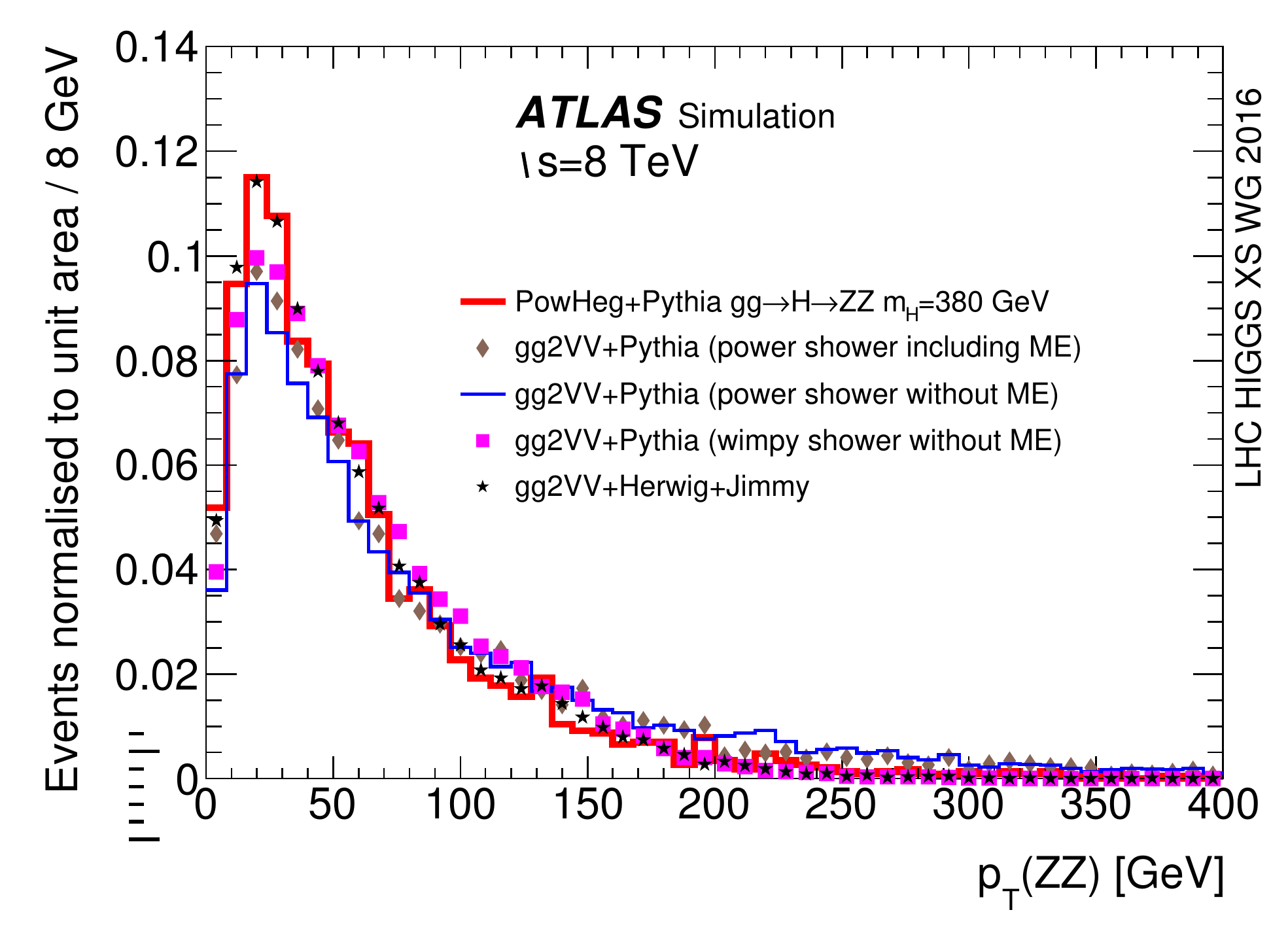}
\end{center}
\caption{Distribution of $p_{\mathrm{T}}$(ZZ) comparing the NLO generator Powheg showered with Pythia8, the LO generator gg2VV + Pythia (power or wimpy shower), the LO generator gg2VV showered with Jimmy+Herwig. All samples are restricted to the range (345$<m_{\mathrm{4\ell}}<$415) GeV.}
\label{fig:ps}
\end{figure*}

\subsubsection{Higher order QCD corrections to the transverse momentum and the rapidity of the ZZ system}
\label{sec:qcdPS}
Higher order QCD corrections for the gg$\rightarrow$ ZZ processes are studied using the Sherpa+OpenLoops \cite{Gleisberg:2008ta,Cascioli:2011va} generator that contains the LO gg$\rightarrow$ZZ+1-jet matrix element and merges this with the LO gg$\rightarrow$ZZ+ 0-jet matrix element. For the gg$\rightarrow H\rightarrow$ ZZ signal contribution with $m_{\mathrm{H}}$=380 GeV (on-shell signal), the Powheg generator reweighted (as a function of $p_{\mathrm{T}}$) to the HRes2.1 prediction \cite{deFlorian:2012mx} to reach NNLO+NNLL accuracy is also used. Figures \ref{fig:allPLOT}, \ref{fig:allPLOT2} and \ref{fig:allPLOT3} include validation distributions of various comparisons of the variables of interest, namely the transverse momentum, $p_{\mathrm{T}}$(ZZ),  and the rapidity, Y(ZZ), of the ZZ system in both on-shell and off-shell mass regions using Powheg+Pythia8, Sherpa+OpenLoops and gg2VV+Pythia8 generators using kinematic variables computed at truth level. The list of cuts applied in the generation level can be found below ($p_{\mathrm{T}}^{\mathrm{\ell}}$ is the transverse momentum of each lepton in the final state, $|\eta^{\mathrm{\ell}}|$ represents its rapidity \footnote{ATLAS uses a right-handed coordinate system with its origin at the nominal interaction point (IP) in the centre of the detector, and the z-axis along the beam line. The x-axis points from the IP to the centre of the LHC ring, and the y-axis points upwards. Cylindrical coordinates ($r,\phi$) are used in the transverse plane, $\phi$ being the azimuthal angle around the beam line. Observables labelled \textit{transverse} are projected into the x-y plane. The pseudorapidity is defined in terms of the polar angle $\theta$ as $\eta$=-$\ln \tan(\frac{\theta}{2})$.} while $m_{\textrm{Z1}}$ is the Z boson mass closest to the Z peak, being $m_{\mathrm{Z2}}$ the mass of the second lepton pair):

\begin{itemize}
\item $m_{\mathrm{4\ell}}>$100 GeV;
\item $p_{\mathrm{T}}^{\mathrm{\ell}}>$3 GeV;
\item $|\eta^{\mathrm{\ell}}|<$2.8;
\item $m_{\mathrm{Z1,Z2}}>$ 4 GeV.
\end{itemize}

Additional selection criteria are applied on the final state quadruplet (the leptons in the quadruplet are ordered in transverse momentum and denoted with the superscript $\mathrm{\ell}$ in what follows) in the Monte Carlo samples in such a way to mimic the standard selection reported in Ref. \cite{Aad:2015xua}, namely:
\begin{itemize}
\item  $p_{\mathrm{T}}^{\mathrm{\ell 1}}>$20 GeV, $p_{\mathrm{T}}^{\mathrm{\ell 2}}>$15 GeV, $p_{\mathrm{T}}^{\mathrm{\ell 3}}>$10 GeV, $p_{\mathrm{T}}^{\mathrm{\ell 4}}>$5 (6) GeV for muons (electrons);
\item $|\eta^{\mathrm{\ell}}|<$2.5;
\item (50$<m_{\mathrm{Z1}}<$106) GeV;
\item if $m_{\mathrm{4\ell}}<$140 GeV $\rightarrow$ $m_{\mathrm{Z2}}>$12 GeV, if 140$<m_{\mathrm{4\ell}}<$190 GeV $\rightarrow$ $m_{\mathrm{Z2}}>$0.76$\cdot$($m_{\mathrm{4\ell}}$-140)+12 GeV, if $m_{\mathrm{4\ell}}>$190 GeV $\rightarrow$ $m_{\mathrm{Z2}}>$50 GeV.
\end{itemize}

The errors bars in \refFs{fig:allPLOT}, \ref{fig:allPLOT2} and \ref{fig:allPLOT3} indicate the statistical uncertainty related to the finite Monte Carlo statistics only. The systematic uncertainties, when applicable, are drawn as shaded boxes, extracted from scale variations on Sherpa+OpenLoops and HRes2.1 as described in the following Section \ref{sec:sysBDT}. The systematic uncertainties from the HRes2.1 are applicable here as the Powheg generator is directly reweighted to the HRes2.1 prediction. The results and the distributions reported in the following figures refer to Monte Carlo samples generated  at a collision energy $\sqrt{s}$=8 TeV.
\\
\\
As highlighted in \refF{fig:allPLOT} (a) for what concerns the on-shell and \refF{fig:allPLOT} (b) for the off-shell, the lack of higher QCD calculations in gg2VV results in different $p_{\mathrm{T}}$ spectra (order of 20\% in the relevant kinematic region) compared to the higher order Powheg and Sherpa+OpenLoops Monte Carlo. In the high mass region, the off-shell (generated with $m_{\mathrm{H}}$=125.5 GeV) and on-shell (produced with $m_{\mathrm{H}}$=380 GeV) Higgs boson productions with gg2VV match fairly well as shown in \refF{fig:allPLOT} (b).
\\
\refF{fig:allPLOT2} (a) shows that the differences in $p_{\mathrm{T}}$ between Sherpa and gg2VV in the off-shell high mass region are not fully covered by the uncertainties assigned to Sherpa. Since the Sherpa generator has a better treatment of the first hard jet emission, in the $H\rightarrow$ZZ$\rightarrow$4$\ell$ analysis, gg2VV is reweighted to the Sherpa prediction in the ATLAS analysis \cite{Aad:2015xua}. As for the rapidity distribution reported in \refF{fig:allPLOT2} (b), no significant difference between gg2VV and Sherpa is present in the high mass region; hence, the reweighting procedure on Y is not necessary.
\\
Figures \ref{fig:allPLOT2} (c) and (d) stress the fact that the ZZ-transverse momentum and the rapidity of the signal process gg$\rightarrow (H^{*})\rightarrow$ ZZ differ from the gg$\rightarrow$ ZZ background process and the SBI unlike the gg2VV generator as noted in Figures \ref{fig:allPLOT3} (a) and (b). This is caused by the presence of the additional matrix element correction to the first jet emission included in Sherpa that generates a different treatment of signal and background components. This statement has been explicitly validated by removing the 1-jet matrix element computation in Sherpa: full compatibility is found between signal and background once the 1-jet ME treatment is removed in Sherpa.
\\
\\
In the analysis deployed by ATLAS \cite{Aad:2015xua}, the LO gg2VV generator, whose $p_{\mathrm{T}}$ and y distributions are displayed in \refF{fig:allPLOT3}, is reweighted to Sherpa+OpenLoops in the $p_{\mathrm{T}}$ of the VV system to achieve a better description of the $p_{\mathrm{T}}$ spectrum: the impact of the reweighting on the acceptance is calculated to be below 1\% for the signal and at the level of 4-6\% for the background. In the ZZ$\rightarrow$4$\ell$ channel, the reweighting procedure is only used to account for the acceptance effects, as the matrix-element discriminant employed to disentangle signal and background components is insensitive to the $p_{\mathrm{T}}$ of the ZZ system. For the ZZ$\rightarrow$2$\ell$2$\nu$ channel, the reweighting is applied in both the transverse mass shape and acceptance as the $m_{\mathrm{T}}$ holds dependence on the transverse momentum of the ZZ system.

\begin{figure}
\begin{center}
\subfigure[]{\includegraphics[width=0.47\linewidth]{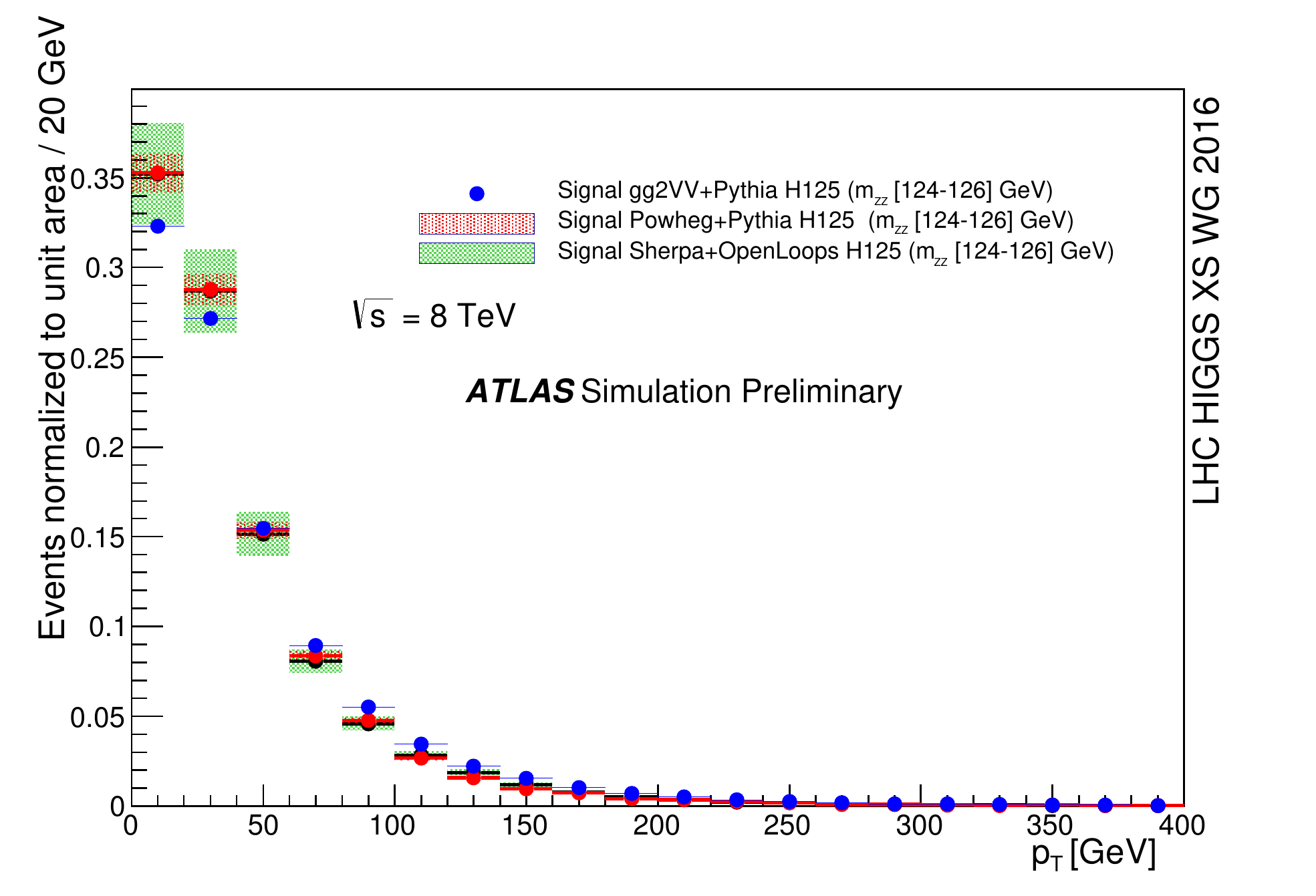}}
\subfigure[]{\includegraphics[width=0.47\linewidth]{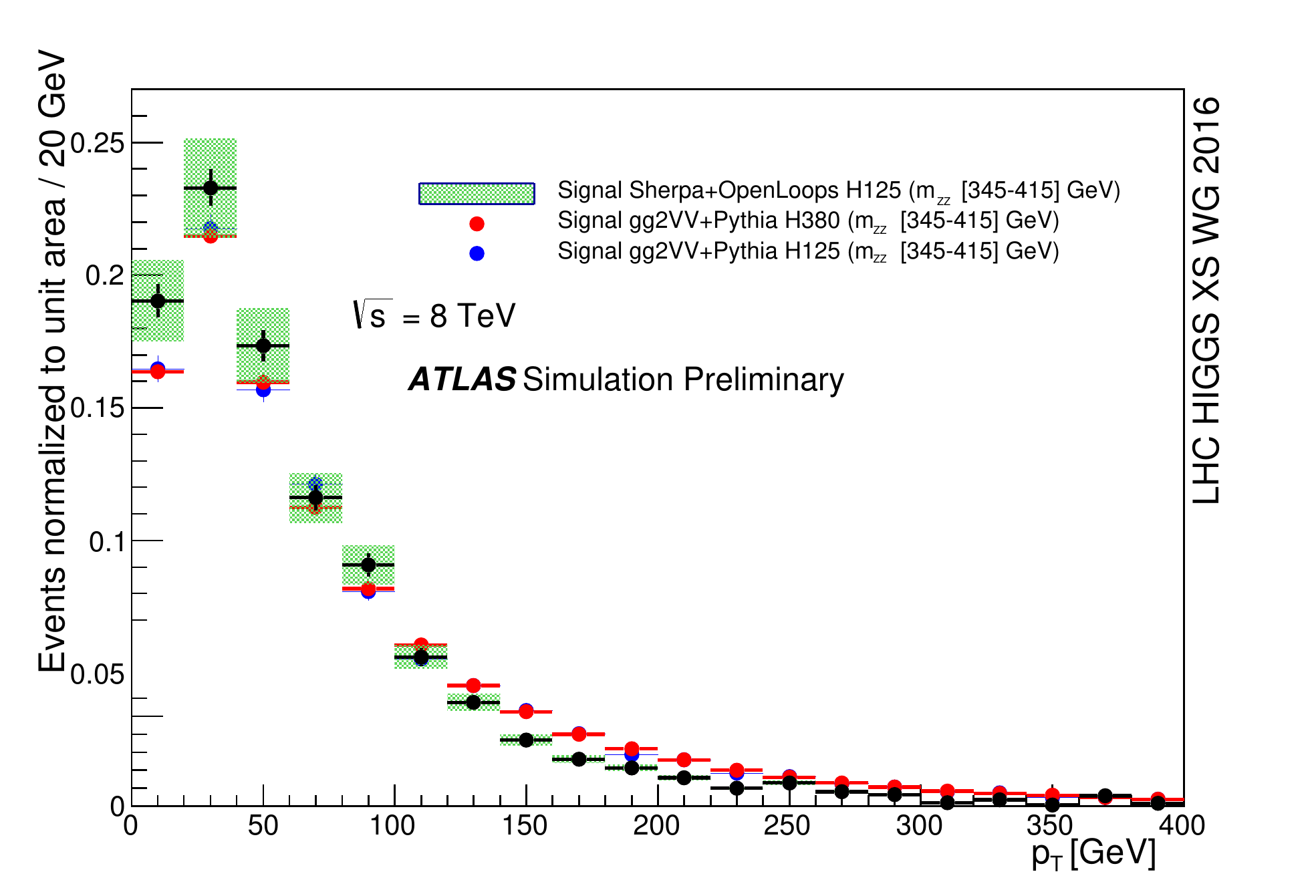}}
\caption{Comparison of the on-shell gg$\rightarrow (H^{*})\rightarrow$ ZZ signal process in $p_{\mathrm{T}}$ (a) generated with $m_{\mathrm{H}}$=125.5 GeV in the mass range $m_{\mathrm{ZZ}} \in$ [124,126] GeV for Powheg, Sherpa and gg2VV. Comparison of the gg$\rightarrow (H^{*})\rightarrow$ ZZ off-shell signal process in $p_{\mathrm{T}}$ (b) with $m_{\mathrm{H}}$=125.5 GeV produced with gg2VV and Sherpa and gg$\rightarrow (H^{*})\rightarrow$ ZZ signal process with $m_{\mathrm{H}}$=380 GeV using gg2VV (on-shell) in the region $m_{\mathrm{ZZ}} \in$ [345,415] GeV.}
\label{fig:allPLOT}
\end{center}
\end{figure}

\begin{figure}
\begin{center}
\subfigure[]{\includegraphics[width=0.47\linewidth]{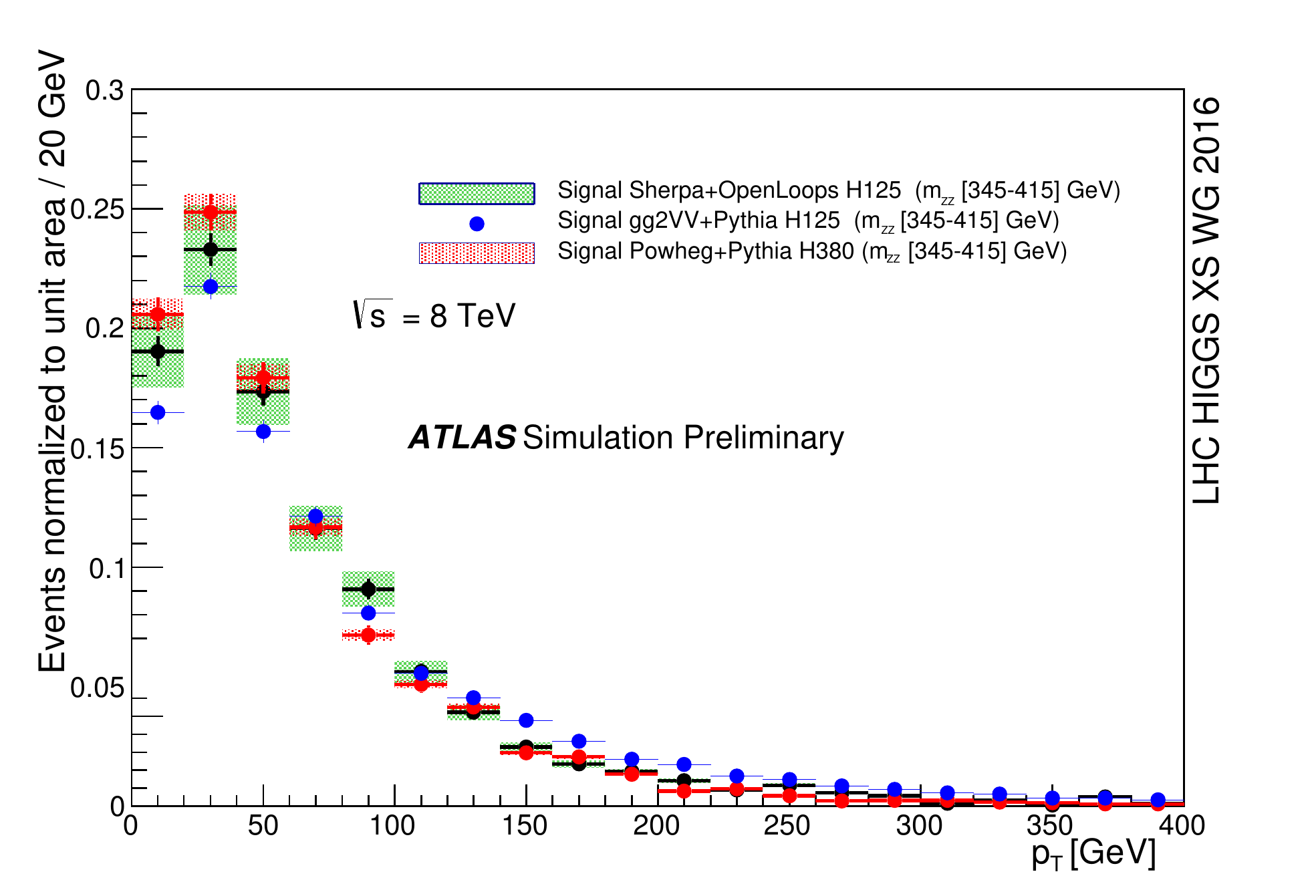}}
\subfigure[]{\includegraphics[width=0.47\linewidth]{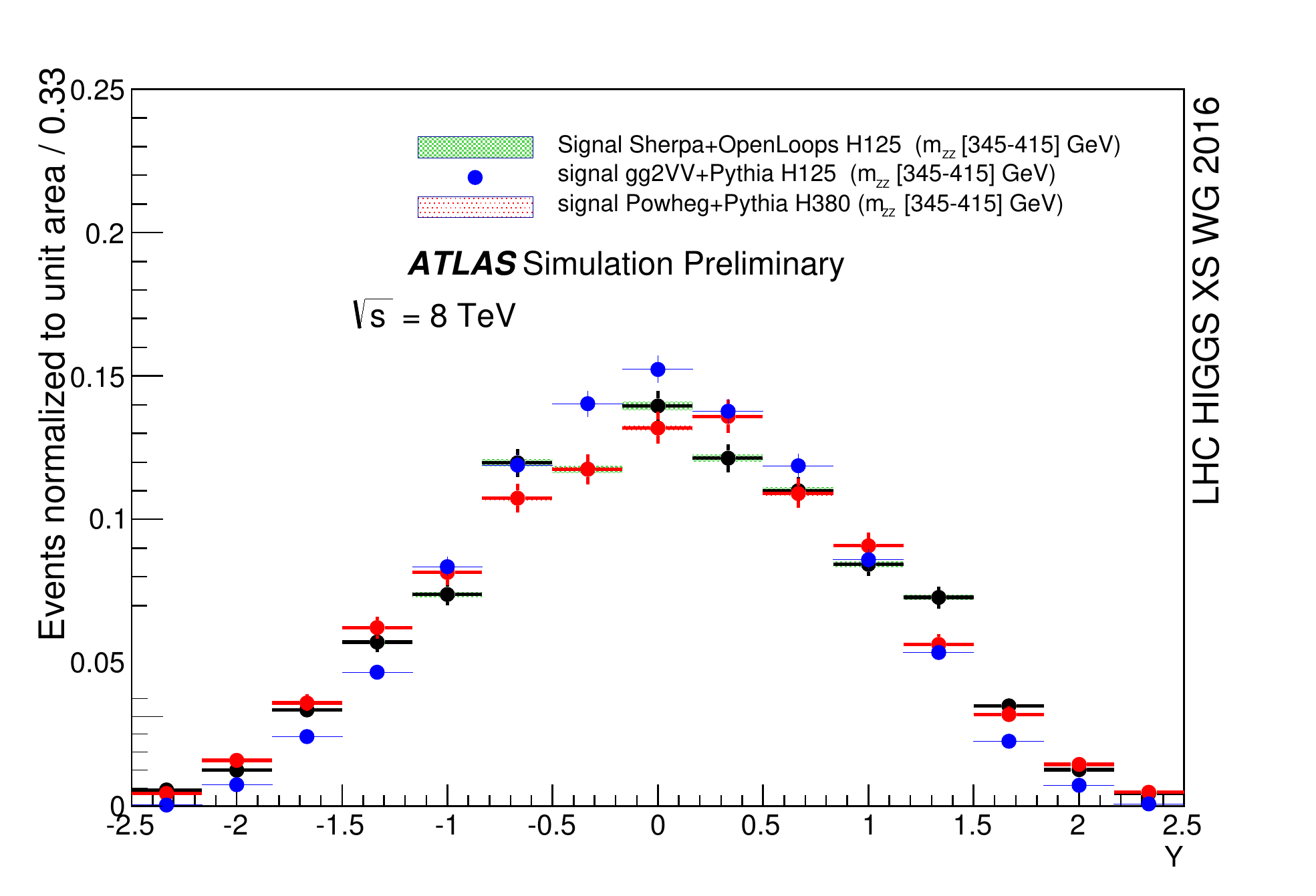}}\\
\subfigure[]{\includegraphics[width=0.47\linewidth]{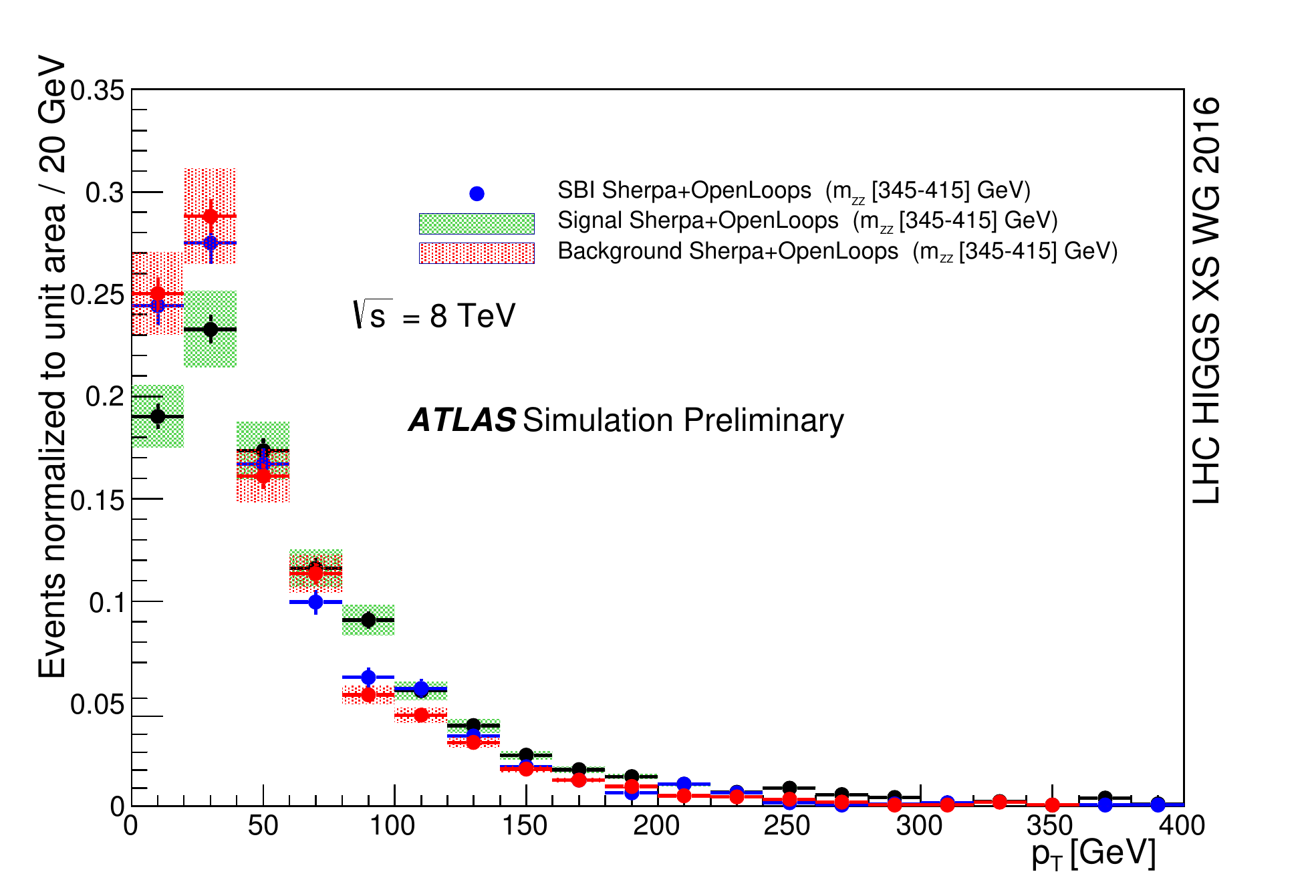}}
\subfigure[]{\includegraphics[width=0.47\linewidth]{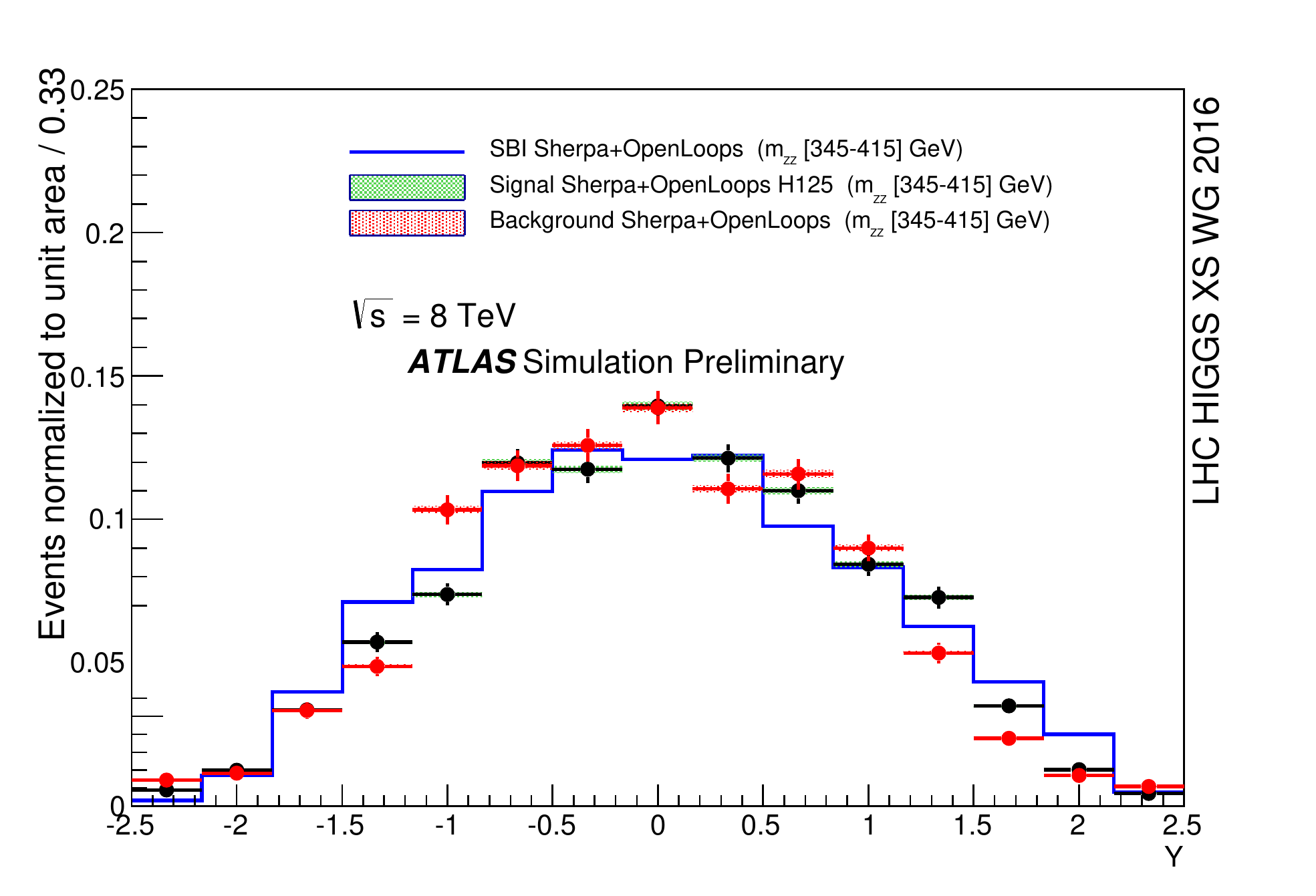}}
\caption{Comparison of the gg$\rightarrow (H^{*})\rightarrow$ ZZ off-shell signal process in $p_{\mathrm{T}}$ (a) and rapidity (b) generated with $m_{\mathrm{H}}$=125.5 GeV produced with gg2VV and Sherpa and gg$\rightarrow (H^{*})\rightarrow$ ZZ signal process with $m_{\mathrm{H}}$=380 GeV using Powheg (on-shell) in the region $m_{\mathrm{ZZ}} \in$ [345,415] GeV. Off-shell comparison in $p_{\mathrm{T}}$ (c) and rapidity (d) of the gg$\rightarrow (H^{*})\rightarrow$ ZZ signal sample generated with $m_{\mathrm{H}}$=125.5 GeV, the gg$\rightarrow$ ZZ background and the SBI contribution using Sherpa in the mass range $m_{\mathrm{ZZ}} \in$ [345,415] GeV.}
\label{fig:allPLOT2}
\end{center}
\end{figure}

\begin{figure}
\begin{center}
\subfigure[]{\includegraphics[width=0.47\linewidth]{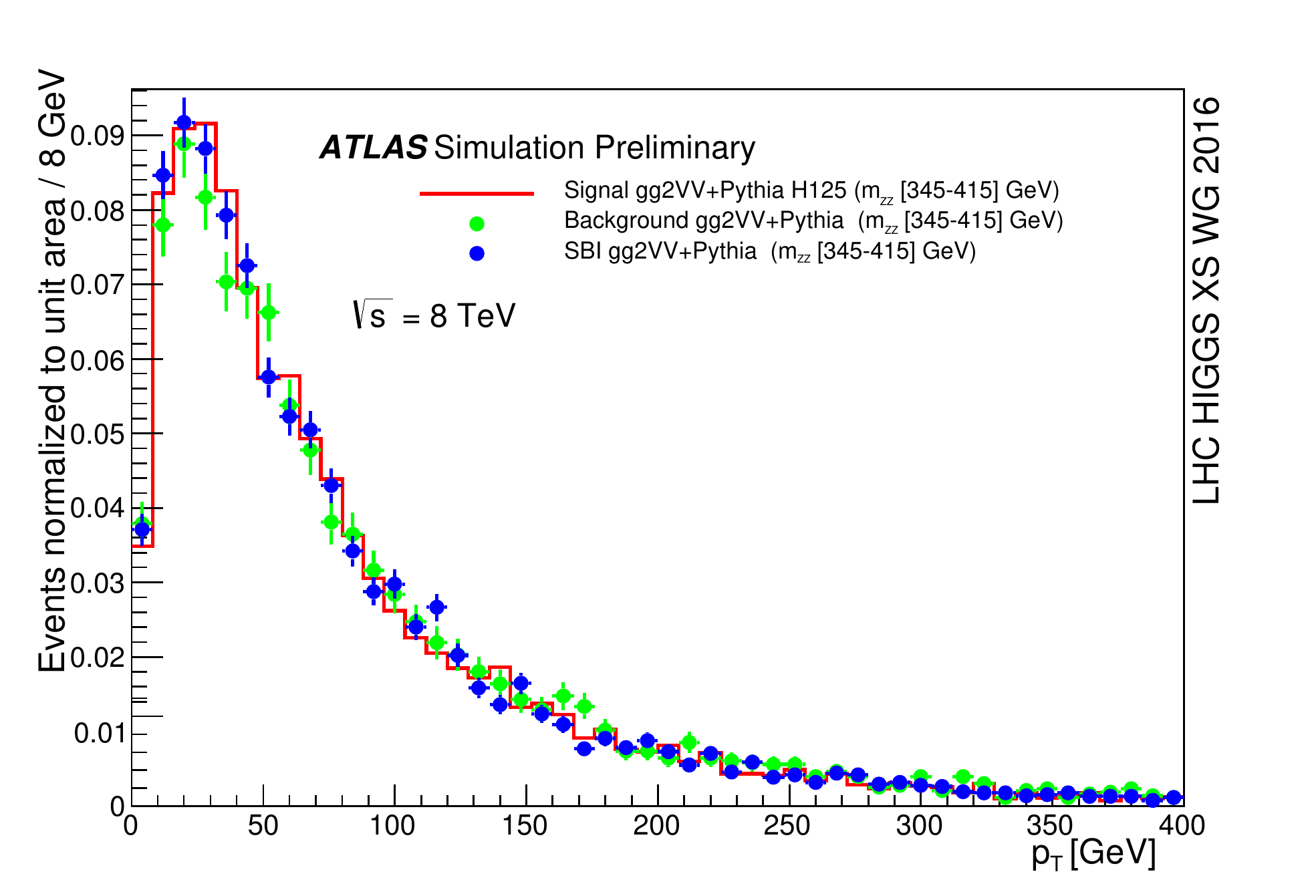}}
\subfigure[]{\includegraphics[width=0.47\linewidth]{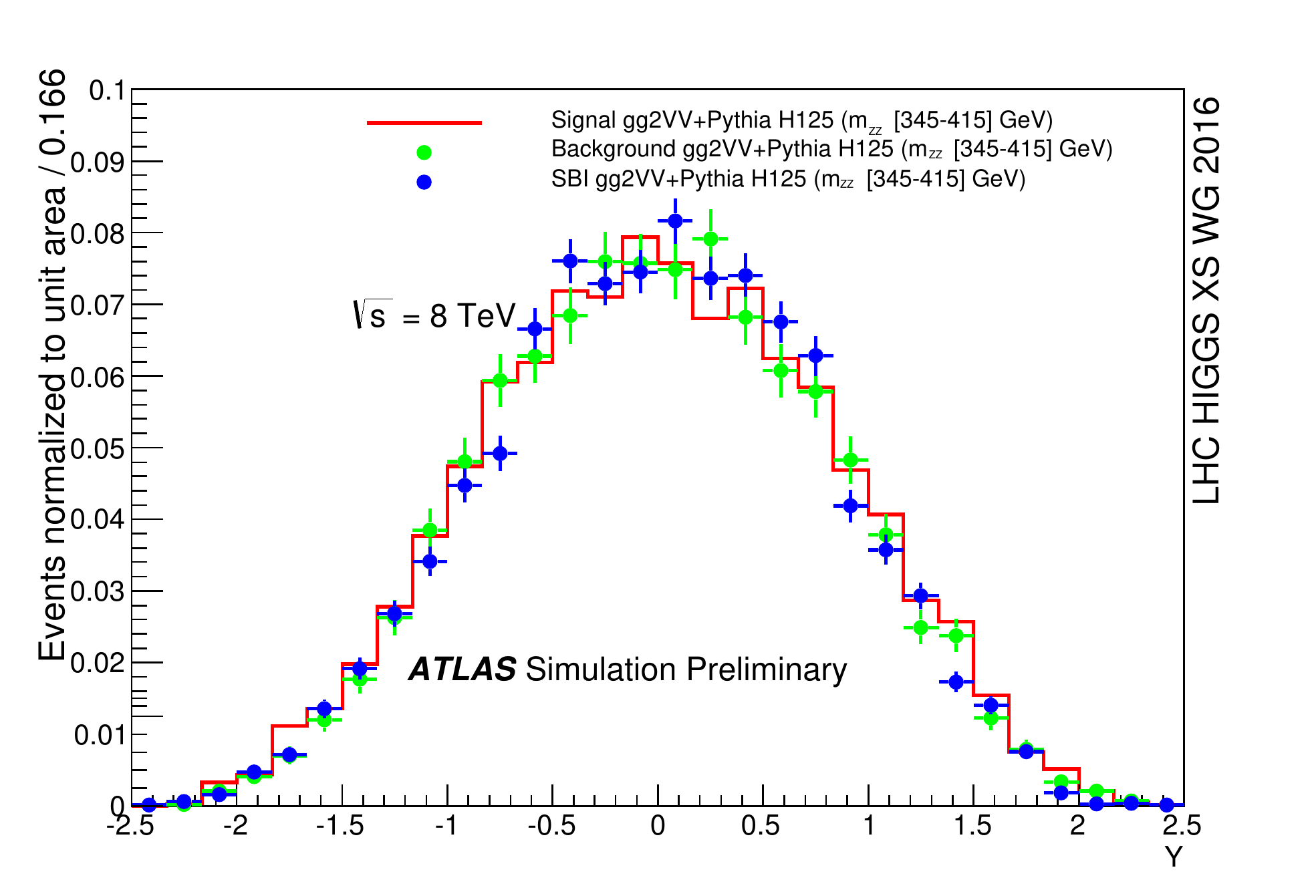}}
\caption{Comparison in $p_{\mathrm{T}}$ (a) and rapidity (b) of the three gg2VV contributions (signal generated with $m_{\mathrm{H}}$=125.5 GeV, background and SBI) in the mass region $m_{\mathrm{ZZ}} \in$ [345,415] GeV.}
\label{fig:allPLOT3}
\end{center}
\end{figure}

\subsubsection{Scale variations on the gg-initiated samples}
\label{sec:sysBDT}
In order to evaluate the systematic effects on the uncertainties on $p_{\mathrm{T}}$ and $\eta$ in the ZZ frame, the procedure is applied by varying the renormalization scale ($\mu_{\mathrm{R}}$), the factorization scale ($\mu_{\mathrm{F}}$), the resummation scale ($\mu_{\mathrm{Q}}$) and the resummation scale related to the bottom quark mass ($\mu_{\mathrm{B}}$).
\\
\\
The impact of the PDF uncertainties is also evaluated: the nominal PDF set, CT10 \cite{Gao:2013xoa}, applied on the Powheg signal sample at $m_{\mathrm{H}}$=125.5 GeV are compared with MSTW2008 \cite{Martin:2009iq} and with NNPDF2.3 \cite{Ball:2012cx} in bins of ZZ-transverse momentum and rapidity. Its impact is found to be below 3\%.

The Monte Carlo simulations employed for these studies and the full scheme of scale variations applied to these samples are listed in Table \ref{tab:scaleVAR}. Assuming that the resummation scales ($\mu_{\mathrm{Q}}$ and $\mu_{\mathrm{B}}$) variations are independent of the normalization and factorization scales ($\mu_{\mathrm{R}}$ and $\mu_{\mathrm{F}}$), we fix the vector pair ($\mu_{\mathrm{R}}$, $\mu_{\mathrm{F}}$) while varying $\mu_{\mathrm{Q}}$ or $\mu_{\mathrm{B}}$. Similarly we fix the resummation scales, $\mu_{\mathrm{Q}}$ and $\mu_{\mathrm{B}}$, while varying $\mu_{\mathrm{R}}$ and $\mu_{\mathrm{F}}$. Following the usual prescriptions, the nominal scale of the process is set to $m_{\mathrm{ZZ}}/2$ while the nominal value for the resummation scale related to the bottom mass is set to $m_{\mathrm{b}}$ and the Powheg nominal values for renormalization and factorization scales are set to $m_{\mathrm{ZZ}}$.

\begin{table}
\caption{Scale variations considered in the evaluation of the theoretical uncertainties related to the $p_{\mathrm{T}}(\mathrm{ZZ})$ and Y(ZZ) for the $gg\to H\rightarrow$ZZ and $q\bar{q}\rightarrow$ZZ processes. The scale variations on Sherpa signal detailed in the second row are also applied on the Sherpa gg$\rightarrow$ZZ continuum background as stated in the text. The merging scale for Sherpa has not been modified for this study.}
\label{tab:scaleVAR}
\centering 
\tabcolsep=0.11cm
\begin{tabular}{c |  c |  c | c | c}
  \toprule
Process & MC & Nominal Scales & Scale variations & \# Variations \\ [0.5ex]
\midrule
 {$gg\to H\to ZZ$} & {HRes} & $\mu_{\mathrm{R}}=\mu_{\mathrm{F}}={m_{\mathrm{ZZ}}\over2}$
& $({1\over2}\mu_{\mathrm{R/F}},2\mu_{\mathrm{R/F}})$, ${1\over2}\le\mu_{\mathrm{F}}/\mu_{\mathrm{R}}\ge2$ & 6 \\ [0.8ex]
&  &  $\mu_{\mathrm{Q}}=m_{\mathrm{ZZ}}/2$, $\mu_{\mathrm{B}}=m_{\mathrm{b}}$ & $({1\over2}\mu_{\mathrm{Q}}, 2\mu_{\mathrm{Q}})$, $({1\over4}\mu_{\mathrm{B}}, 4\mu_{\mathrm{B}})$ & 8 \\ [0.8ex]
\midrule
 {$gg\to H\to ZZ$}  &{Sherpa}
& $\mu_{\mathrm{R}}=\mu_{\mathrm{F}}={m_{\mathrm{ZZ}}\over2}$ & $({1\over2}\mu_{\mathrm{R/F}},2\mu_{\mathrm{R/F}})$, ${1\over2}\le\mu_{\mathrm{F}}/\mu_{\mathrm{R}}\ge2$ & 6 \\ [0.8ex]
& &  $\mu_{\mathrm{Q}}=m_{\mathrm{ZZ}}/2$, $\mu_{\mathrm{B}}=m_{\mathrm{b}}$ & $({1\over\sqrt{2}}\mu_{\mathrm{Q}}, \sqrt{2}\mu_{\mathrm{Q}})$ & 2 \\ [0.8ex]
\midrule
$q\bar{q}\to ZZ$ & Powheg & $\mu_{\mathrm{R}}=\mu_{\mathrm{F}}=m_{\mathrm{ZZ}}$ & $({1\over2}\mu_{\mathrm{R/F}},2\mu_{\mathrm{R/F}})$  &  6\\ [0.5ex]
\bottomrule
 \end{tabular}
\end{table}

\refF{fig:Hresplot} shows the shape-only variations on $p_{\mathrm{T}}$(ZZ) and Y(ZZ) for a high mass $m_{\mathrm{H}}$=380 GeV gg$\rightarrow H \rightarrow$ ZZ signal process, produced by QCD scale variations evaluated with the HRes2.1 Monte Carlo generator. The scale variations on the rapidity in \refF{fig:Hresplot} (b) can be neglected since they are much smaller than those of the transverse momentum, \refF{fig:Hresplot} (a). \refF{fig:sherpaPL} shows the variation of the signal process (a) and the background processes on $p_{\mathrm{T}}$(ZZ) created with the Sherpa+OpenLoops Monte Carlo sample. The envelope of these independent variations on $p_{\mathrm{T}}$(ZZ) is calculated as the maximal up and down contribution for each $p_{\mathrm{T}}$ bin for the HRes2.1 case as well as for Sherpa signal and background. Since the contribution of the resummation scale is dominant, a first envelope encompassing renormalization and factorization scales summed it in quadrature with the envelope extracted from the resummation scale provides enough accuracy for this study. Note that the Sherpa variations enclose the variations of HRes2.1 because Sherpa does not contain the full NLO calculations, hence its variations are larger than the typical scales of HRes2.1.The systematic uncertainties reported in Ref. \cite{Aad:2015xua} associated with the Sherpa-based reweighting in $p_{\mathrm{T}}$ of the VV system are assessed by varying the relevant scales in Sherpa: the larger in value between the scale variations in Sherpa and 50\% of the difference between Sherpa and gg2VV+Pythia  is assigned as the systematic uncertainty. This conservative approach is chosen to consider potential uncertainties not accounted for by the scale variations. The impact of the PDF uncertainties is found to be negligible.

\begin{figure*}
\subfigure[]
{
\includegraphics[width=0.47\linewidth]{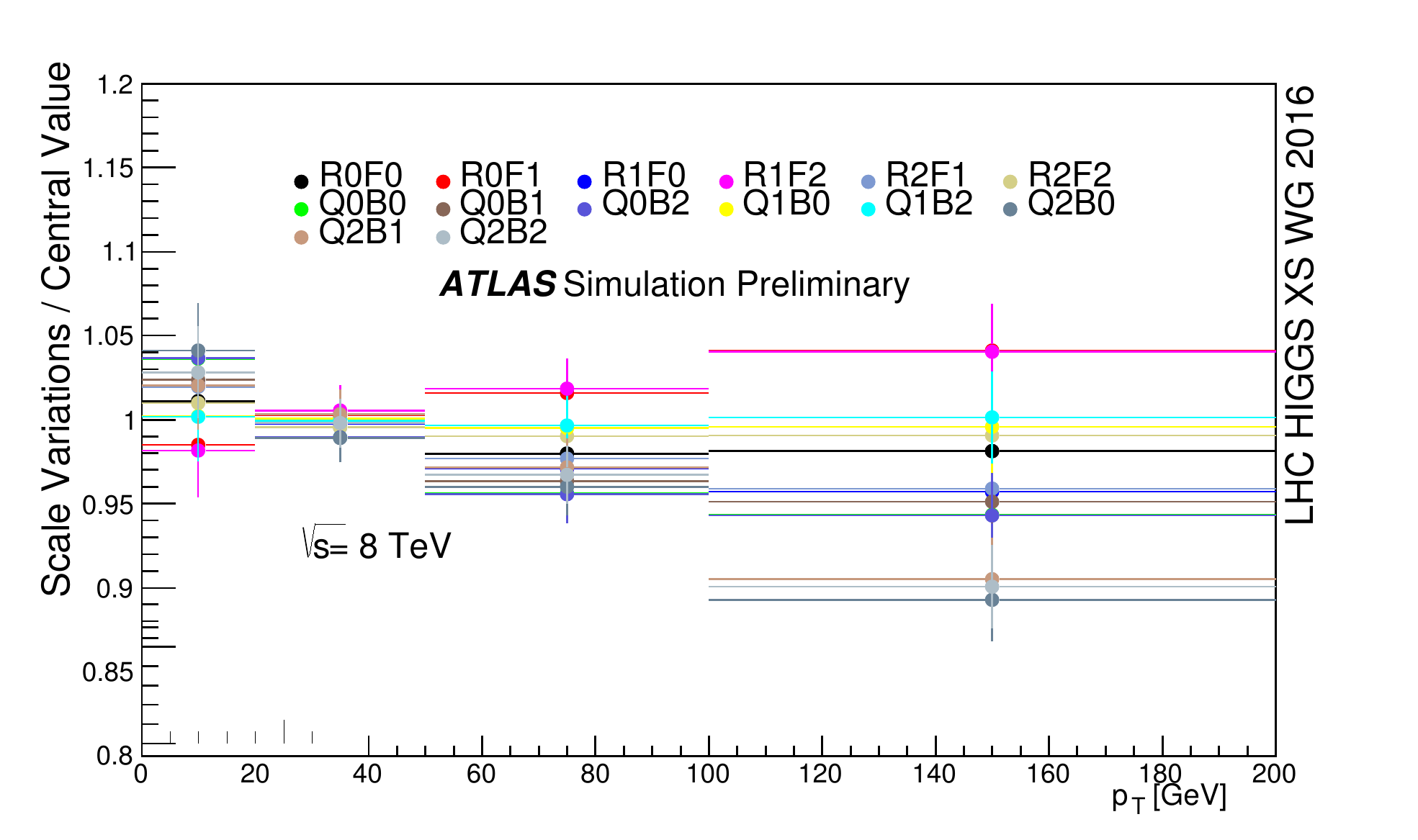}
}
\subfigure[]
{
\includegraphics[width=0.47\linewidth]{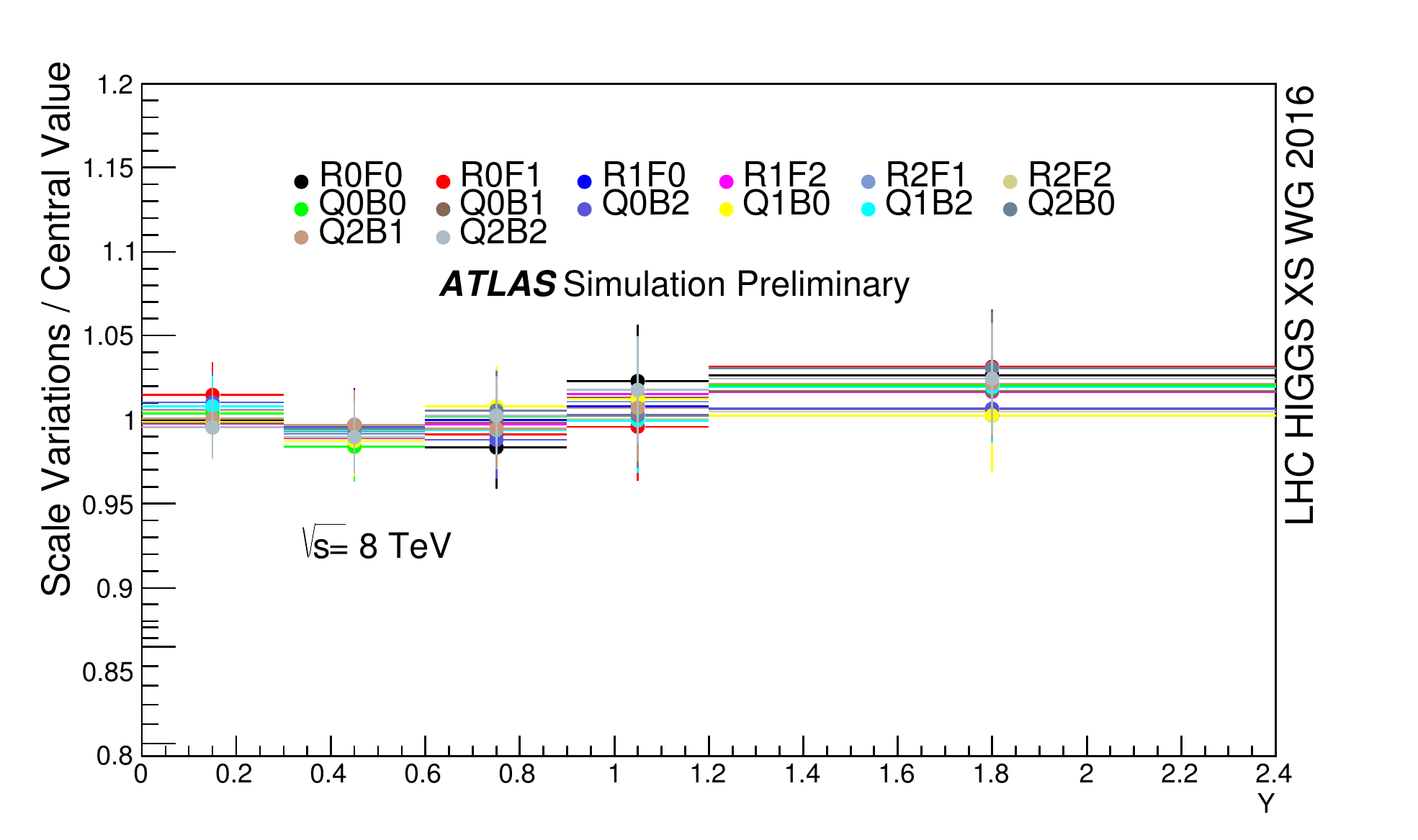}
}
\caption{Relative change of the $p_{\mathrm{T}}$ and Y spectra due to the QCD scale variations produced with HRes2.1 signal generated at $m_{\mathrm{H}}$=380 GeV: ratio of the up or down variations $p_{\mathrm{T}}$ or rapidity with respect to the nominal distribution. Q labels the resummation scale, B the resummation scale related to the bottom quark mass, R the renormalization scale, F the factorization scale. The numbers coupled with each variation characterize the nominal value (1), the down variation (0) and the up variation (2).}
\label{fig:Hresplot}
\end{figure*}

\begin{figure*}
\subfigure[]
{
\includegraphics[width=0.47\linewidth]{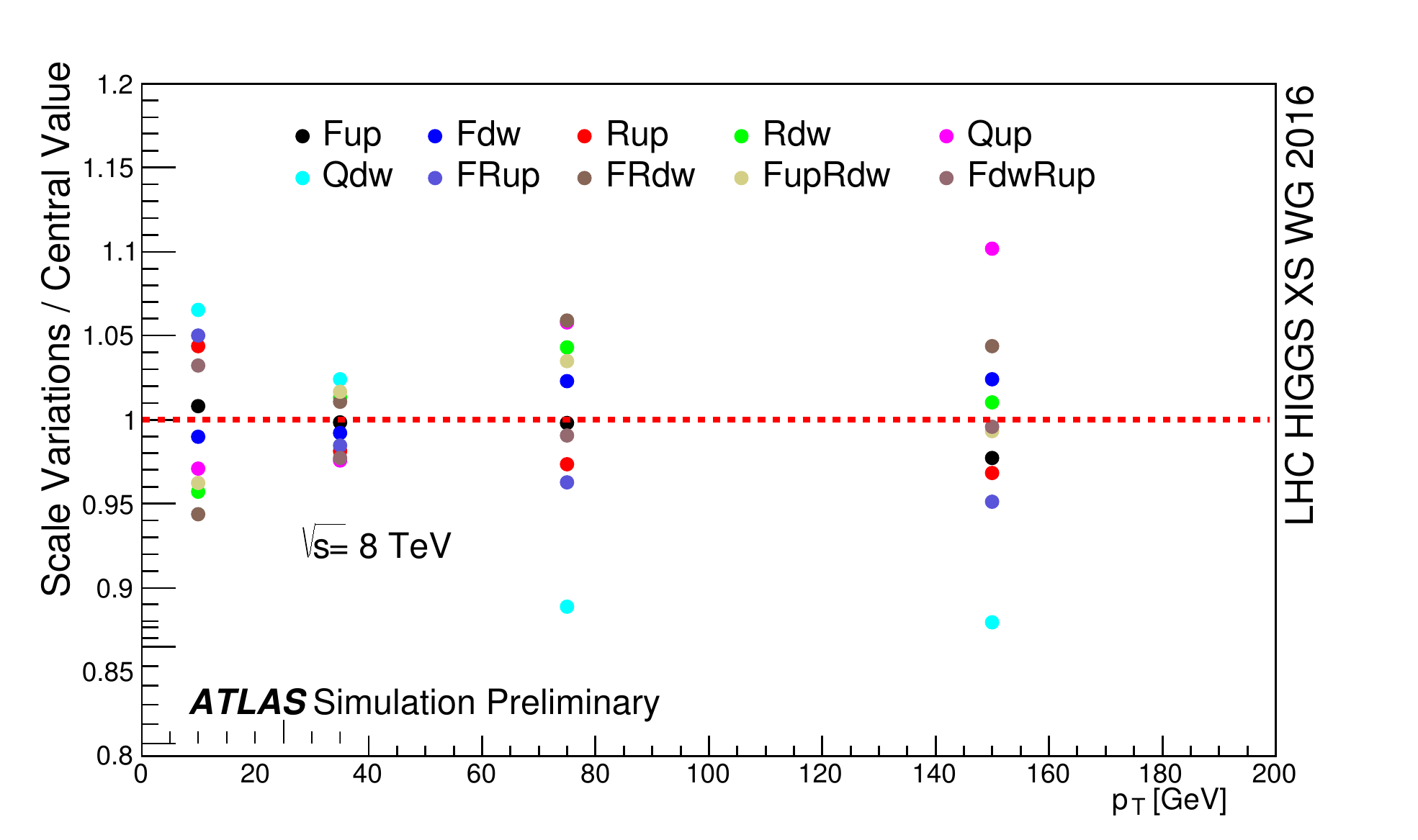}
}
\subfigure[]
{
\includegraphics[width=0.47\linewidth]{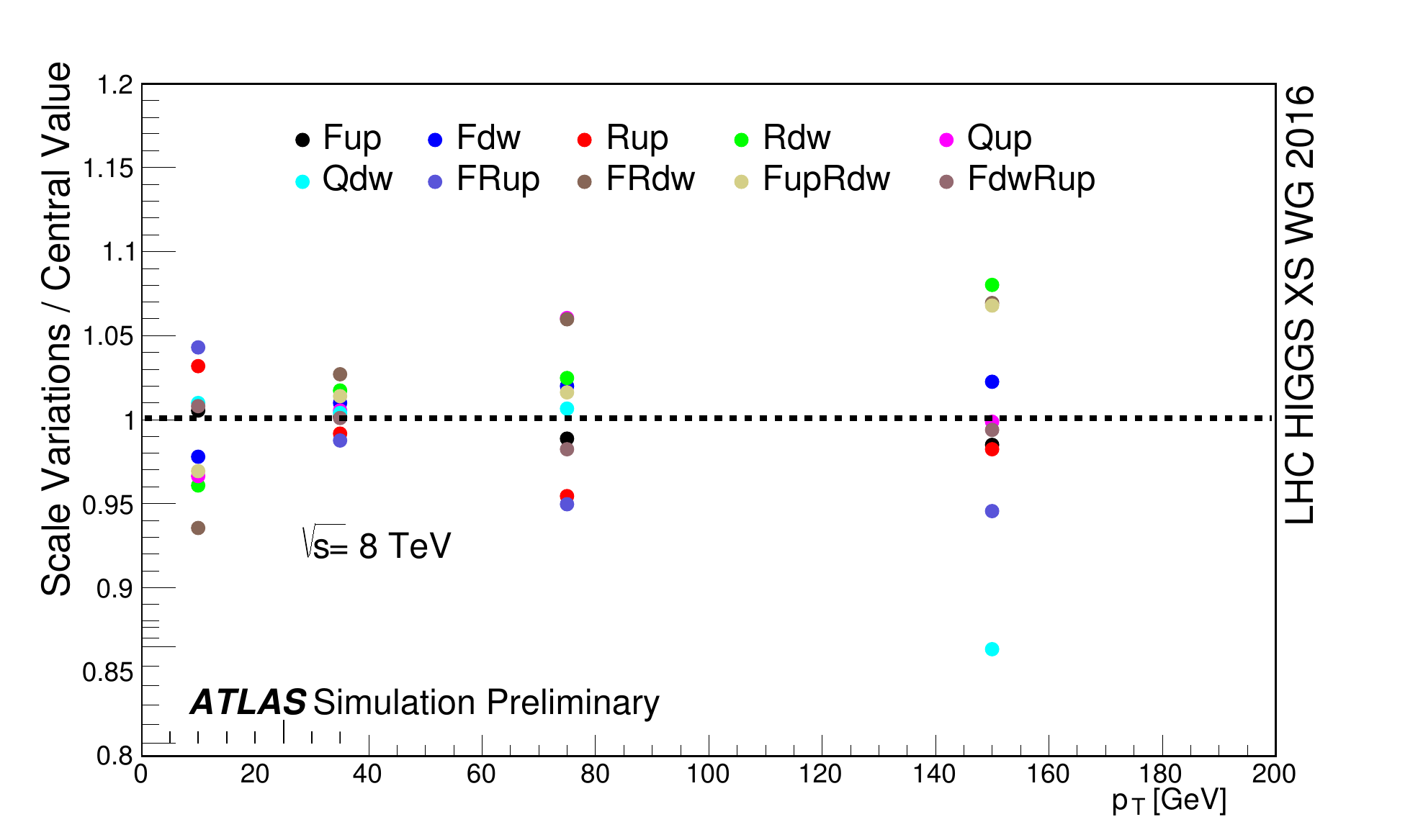}
}
\caption{Relative uncertainties on the $p_{\mathrm{T}}$ spectrum for the Sherpa+OpenLoops signal (a) and background (b) samples induced by the QCD scale variations: ratio of the up or down variations with respect to the nominal distribution. Q labels the resummation scale, R the renormalization scale, F the factorization scale.}
\label{fig:sherpaPL}
\end{figure*}


\newcommand{\fr}[2]{\frac{#1}{#2}} 
\newcommand{\non}{\nonumber}
\newcommand{\ar}{\mbox{$\rightarrow$}}
\def\ra{\rightarrow}
\newcommand{\re}{{\Re}e}
\newcommand{\im}{{\Im}m}
\newcommand{\Et}{\ensuremath{E_\mathrm{T}}}                                     
\newcommand{\Z}{Z}
\newcommand{\F}{F}
\newcommand{\cPg}{g}
\newcommand{\GHSM}{\Gamma_H^{\rm SM}}
\newcommand{\V}{V}   
\providecommand{\ssst}{\scriptscriptstyle}

\subsection{Higgs boson off-shell simulation with the MCFM and JHU generator frameworks}
\label{sec:offshell_interf_vv_jhugen_mcfm}

The JHU Generator and MELA framework~\cite{Gao:2010qx,Bolognesi:2012mm,Anderson:2013afp}  
is designed for the study of anomalous couplings of a resonance to vector bosons and fermions 
in various decay and production processes on LHC,
and is applicable to either the already discovered boson $H(125)$ or a new resonance $X(m_X)$. 
In addition to stand-alone generation, the framework is also integrated with the MCFM Monte Carlo 
package~\cite{Campbell:2010ff,Campbell:2011bn,Campbell:2013una} 
for modelling of the background processes and allows simulation of anomalous couplings 
in off-shell $H(125)^*$ boson production including interference with continuum diboson production. 
The simulation  of an additional broad resonance $X$ is also included, 
allowing for the study of a new Higgs-like resonance with arbitrary couplings interfering with the SM processes. 
The MELA framework allows various likelihood functions either for construction of kinematic 
discriminants or re-weighting of MC simulation. 

The formalism in the JHUGen / MELA framework follows the convention for the tensor structure of $HVV$ couplings
\begin{equation}
A(\PH\V\V) \propto
\Biggl[ a_{1}
 - e^{i\phi_{\Lambda{Q}}} \frac{\left(q_{{\V}1} + q_{{\V}2}\right)^{2}}{\left(\Lambda_{Q}\right)^{2}}\\
 - e^{i\phi_{\Lambda{1}}} \frac{\left(q_{{\V}1}^2 + q_{{\V}2}^2\right)}{\left(\Lambda_{1}\right)^{2}}
\Biggr]m_{\V}^2 \epsilon_{{\V}1}^* \epsilon_{{\V}2}^*
\\
+ a_{2}^{}  f_{\mu \nu}^{*(1)}f^{*(2),\mu\nu}
+ a_{3}^{}   f^{*(1)}_{\mu \nu} {\tilde f}^{*(2),\mu\nu},
\label{eq:ampl-spin0-offshell}
\end{equation}
where $f^{(i){\mu \nu}} = \epsilon_{{\V}i}^{\mu}q_{{\V}i}^{\nu} - \epsilon_{{\V}i}^\nu q_{{\V}i}^{\mu} $
is the field strength tensor of a gauge boson with momentum $q_{{\V}i}$ and polarization
vector $\epsilon_{{\V}i}$, ${\tilde f}^{(i)}_{\mu \nu} = \frac{1}{2} \epsilon_{\mu\nu\rho\sigma} f^{(i),\rho\sigma}$
is the dual field strength tensor.
Spin-one and spin-two resonance couplings, higher-order terms in $q^2$ expansion, 
and terms asymmetric in $q_{V1}^2$ and $q_{V2}^2$ are supported by the generator but are not shown here,
see Refs.~\cite{Gao:2010qx,Bolognesi:2012mm,Anderson:2013afp} and generator manual for details. 
The above $q^2$ expansion is equivalent to the effective Lagrangian  notation with operators up to dimension 
five~\cite{Khachatryan:2014kca,Khachatryan:2015mma}
\begin{eqnarray}
 {L}(\PH\V\V) \propto &&
a_{1}\frac{m_{\ssst\Z}^2}{2} \PH \Z^{\mu}\Z_{\mu} 
- \frac{\kappa_1}{\left(\Lambda_{1}\right)^{2}} m_{\ssst\Z}^2 \PH  \Z^{\mu} \Box \Z_{\mu}
- \frac{\kappa_3}{2\left(\Lambda_{Q}\right)^{2}} m_{\ssst\Z}^2 \Box \PH  \Z^{\mu} \Z_{\mu} 
 \nonumber \\
&& - \frac{1}{2}a_{2} \PH  \Z^{\mu\nu}\Z_{\mu\nu} 
- \frac{1}{2}a_{3} \PH  \Z^{\mu\nu}{\tilde \Z}_{\mu\nu} 
  \nonumber \\
&& + a_{1}^{\PW\PW}{m_{\ssst\PW}^2} \PH \PW^{+\mu} \PW^{-}_{\mu}
- \frac{1}{\left(\Lambda_{1}^{\PW\PW}\right)^{2}} m_{\ssst\PW}^2 \PH
  \left(  \kappa_1^{\PW\PW} \PW^{-}_{\mu} \Box \PW^{+\mu} + \kappa_2^{\PW\PW} \PW^{+}_{\mu} \Box \PW^{-\mu} \right)
    \nonumber \\
&& 
 - \frac{\kappa_3^{\PW\PW}}{\left(\Lambda_{Q}\right)^{2}} m_{\ssst\PW}^2 \Box \PH  \PW^{+\mu} \PW^{-}_{\mu}
- a_{2}^{\PW\PW} \PH \PW^{+\mu\nu}\PW^{-}_{\mu\nu}
- a_{3}^{\PW\PW} \PH \PW^{+\mu\nu}{\tilde \PW}^{-}_{\mu\nu} 
\nonumber \\
&&                          + \frac{\kappa_2^{\Z\gamma}}{\left(\Lambda_{1}^{\Z\gamma} \right)^{2}} m_{\ssst\Z}^2 \PH  \Z_\mu \partial_\nu \F^{\mu\nu}
                          - a_{2}^{\Z\gamma} \PH \F^{\mu\nu} \Z_{\mu\nu} - a_{3}^{\Z\gamma} \PH  \F^{\mu\nu}{\tilde \Z}_{\mu\nu}
\nonumber \\
&&
                           - \frac{1}{2}a_{2}^{\gamma\gamma} \PH  \F^{\mu\nu}\F_{\mu\nu} - \frac{1}{2}a_{3}^{\gamma\gamma}\PH  \F^{\mu\nu}{\tilde F}_{\mu\nu}
 %
- \frac{1}{2}a_{2}^{\Pg\Pg} \PH  G^{\mu\nu}_aG^a_{\mu\nu} 
- \frac{1}{2}a_{3}^{\Pg\Pg}\PH  G^{\mu\nu}_a{\tilde G}^a_{\mu\nu},
\label{eq:lagrangian-offshell}
\end{eqnarray}
where $\V_{\mu\nu} = \partial_\mu V_\nu - \partial_\nu V_\mu $,
$G^a_{\mu\nu} = \partial_\mu A^a_\nu - \partial_\nu A^a_\mu + g f^{abc}A^b_\mu A^c_\nu$,
${\tilde V}^{\mu \nu} = 1/2 \epsilon^{\mu \nu \alpha \beta} V_{\alpha \beta}$,
$Z$ is the $Z$ field, $W$ is the $W$ field, $F$ is the $\gamma$ field, and $G$ is the $g$ field. 

Both on-shell $H$ production and off-shell $H^*$ production are considered. 
There are no kinematic constraints on either $q^2_{\ssst Vi}$ or $(q_{\ssst V1}+q_{\ssst V2})^2$, 
other than the relevant parton luminosities. 
Since the scale of validity of the nonrenormalizable higher-dimensional operators is {\it a priori}
unknown, effective cut-off scales $\Lambda_{V1,i}, \Lambda_{V2,i}, \Lambda_{H,i}$ are introduced
for each term in Eq.~(\ref{eq:ampl-spin0-offshell}) with the form factor scaling the anomalous
contribution $g_i^{\rm BSM}$ as
\begin{eqnarray}
g_i = g_i^{\rm SM} \times \delta_{i1} +
g_i^{\rm BSM} \times \frac{\Lambda_{V1,i}^2 \Lambda_{V2,i}^2 \Lambda_{H,i}^2}
{(\Lambda_{V1,i}^2+|q^2_{\ssst V1}|)(\Lambda_{V2,i}^2+|q^2_{\ssst V2}|)(\Lambda_{H,i}^2+|(q_{\ssst V1}+q_{\ssst V2})^2|)} 
\,.
\label{eq:formfact-spin0-offshell}
\end{eqnarray}

The $gg \to ZZ/Z\gamma^*/\gamma^*\gamma^*\to 4f$ process is  generated at LO in QCD.
In simulation shown in \refF{fig:MCFM_BSM_GenLevel}, the QCD factorization and renormalization 
scales are chosen to be running as $m_{4\ell}/2$ and NNPDF30 parton structure functions are adopted. 
In order to include higher-order QCD corrections, LO, NLO, and NNLO signal cross section calculation 
is performed using the MCFM and HNNLO programs~\cite{Catani:2007vq,Grazzini:2008tf,Grazzini:2013mca} 
for a wide range of masses using narrow width approximation. 
The ratio between the NNLO and LO, or between the NLO and LO, values is used as a weight ({\it k}\,-factor).
The NNLO {\it k}\,-factors are applied to simulation as shown in \refF{fig:MCFM_BSM_GenLevel}. 
While this calculation is directly applicable for signal, it is approximate for background. However, the 
NLO calculation is available~\cite{Caola:2015psa,Melnikov:2015laa} for background for the mass range
$2m_Z<m_{4\ell}<2m_t$. There is a good agreement between the NLO {\it k}\,-factors calculated for signal
and background, and any differences set the scale of systematic uncertainties from this procedure. 

Two applications of off-shell $H(125)$ simulation are shown in \refF{fig:MCFM_BSM_GenLevel}.
In one case, anomalous $HVV$ couplings introduce distinct kinematics in the mass range $m_{4\ell}>2m_Z$.
In the other case, a hypothetical $X(m_X)$ resonance interferes with both $H(125)$ off-shell tail and 
the $gg\to4\ell$ background. In all cases, most general $HVV$ and $XVV$ couplings discussed above are possible. 
Anomalous coupling parameterization in terms 
effective fractions of events follows LHC convention~\cite{Khachatryan:2014kca,Khachatryan:2015mma} 
and is equivalent to parameterization in Eq.~(\ref{eq:ampl-spin0-offshell}) 
with  $f_{ai}= |a_i|^2\sigma_i / \Sigma_j |a_j|^2\sigma_j$.

\begin{figure}
\centering
\centerline{
\includegraphics[width=0.4\textwidth]{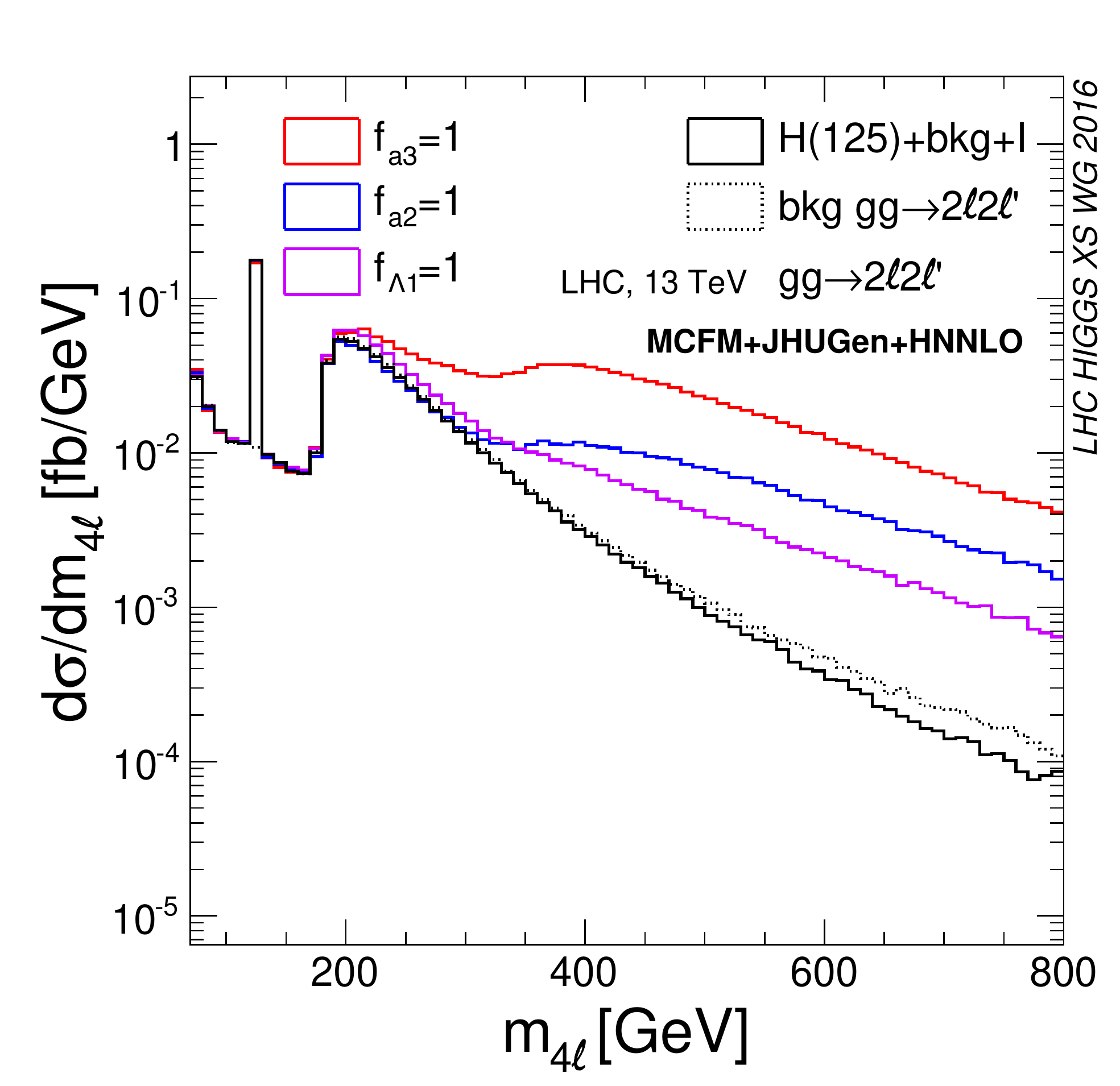}
\includegraphics[width=0.4\textwidth]{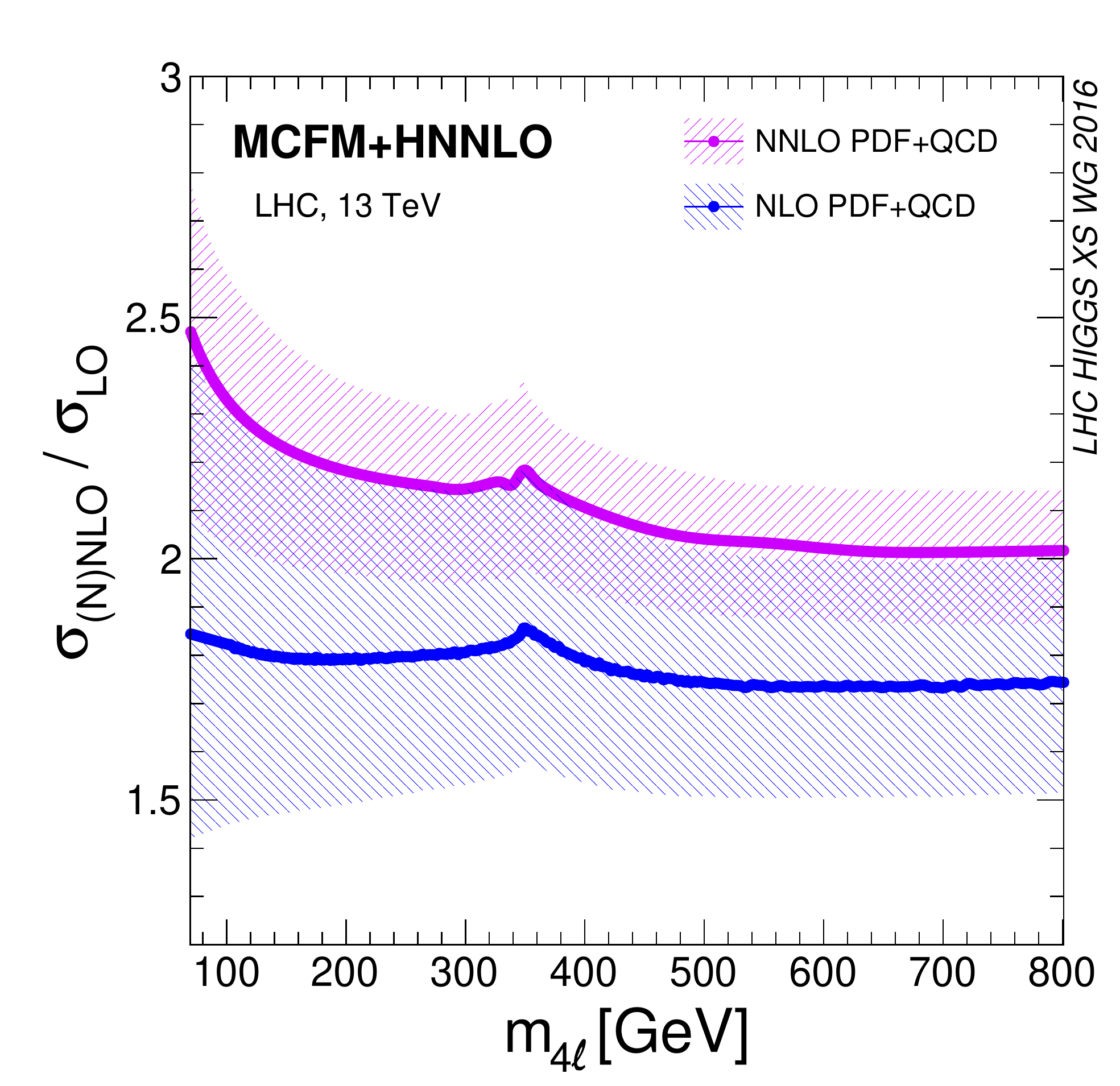}
}
\centerline{
\includegraphics[width=0.85\textwidth]{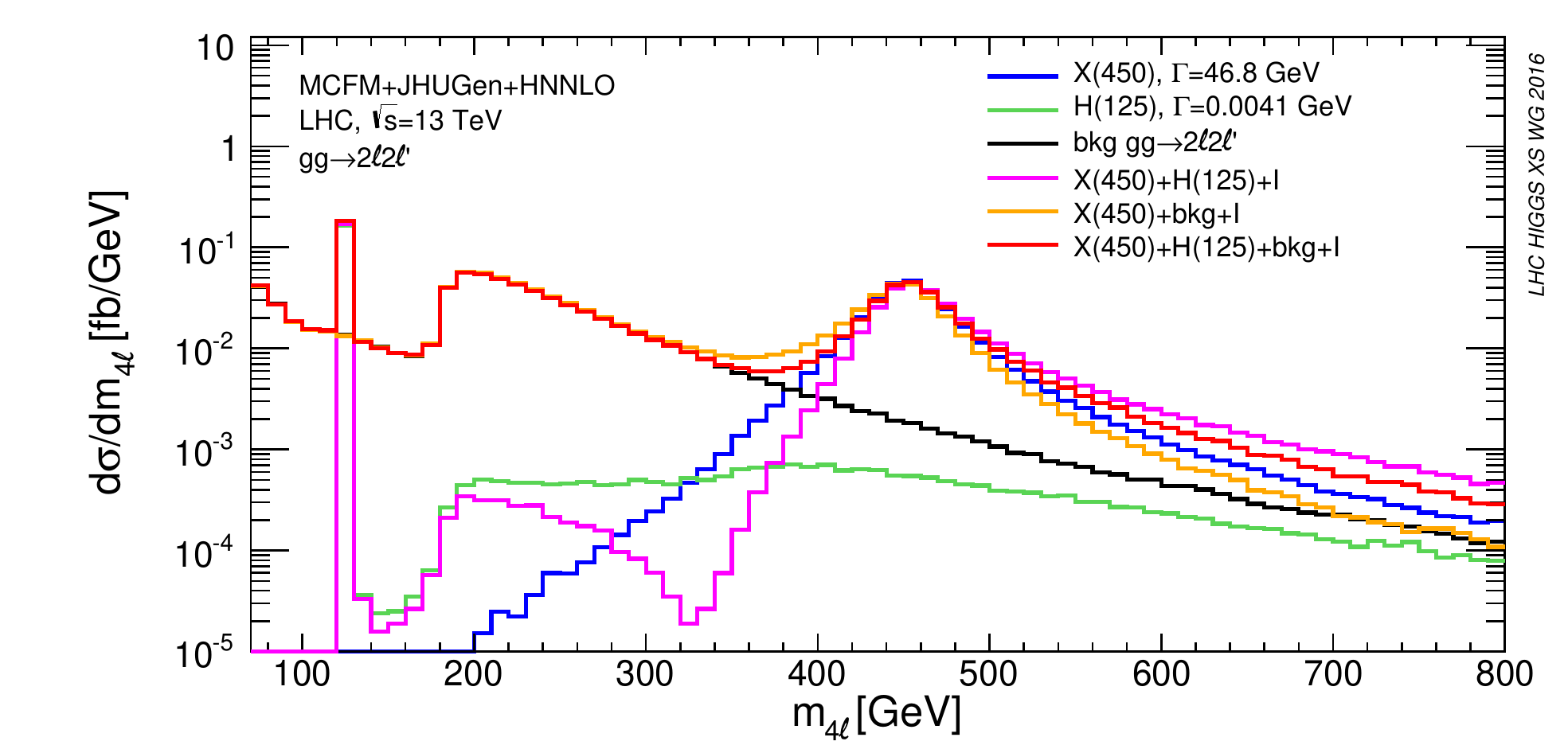}
}
\caption{
Differential cross section of the process $gg \to ZZ/Z\gamma^*/\gamma^*\gamma^*\to 2\ell 2\ell^\prime$ (where $\ell$, $\ell^\prime= e$, $\mu,$ or $\tau$) as a function of invariant mass $m_{4\ell}$ generated with the MCFM+JHUGen framework, including the NNLO in QCD weights calculated with MCFM+HNNLO. The NNLO and NLO weights ({\it k}\,-factors) as a function of $m_{4\ell}$ are shown on the top-right plot. The top-left plot shows several scenarios of $H(125)$ anomalous couplings to two weak vector bosons with enhancement in the off-shell region with the $a_3$, $a_2$, and $\Lambda_{1}$ terms, as coloured histograms, as well as the $a_1$ term (SM), as the solid black histogram, from Eq.~(\ref{eq:ampl-spin0-offshell}) in decreasing order of enhancement at high mass. The bottom plot shows distributions in the presence of a hypothetical $X(450)$ resonance with several components either isolated or combined. In all cases interference (I) of all contributing amplitudes is included.}
\label{fig:MCFM_BSM_GenLevel}
\end{figure}



\providecommand{\cba}{c_{\beta-\alpha}}
\providecommand{\sba}{s_{\beta-\alpha}}
\providecommand{\tb}{t_{\beta}}
\providecommand{\smallz}{{\scriptscriptstyle Z}}
\providecommand{\mzz}{m_{\smallz\smallz}}
\providecommand{\gosam}{{\sc GoSam}}

\subsection[Interference contributions to heavy Higgs boson production 
in the 2HDM]{Interference contributions to gluon-initiated heavy Higgs boson production 
in the 2HDM using \gosam}
\label{sec:offshell_interf_vv_gosam}

\subsubsection{\gosam}

\gosam~\cite{Cullen:2011ac,Cullen:2014yla} is a package for the
automated calculation of one-loop (and tree-level) amplitudes. 
It can be used either in standalone mode or as a {\it One Loop
Provider} (OLP) in combination with a Monte Carlo program, where the
interface is automated, based on the standards defined
in \Brefs{Binoth:2010xt,Alioli:2013nda}.  
\gosam{} is not a library of pre-computed processes, but calculates 
the amplitude for the process specified by the user in a 
{\it run card} on the fly. 
In the OLP version, the information for the code generation is taken
from the order file generated by the Monte Carlo 
program. 
The amplitudes are evaluated using $D$-dimensional reduction at
integrand level~\cite{Ossola:2006us,Ellis:2008ir,Mastrolia:2008jb},
which is available through the reduction procedures and
libraries
{\sc Samurai}~\cite{Mastrolia:2010nb,vanDeurzen:2013pja} or {\sc Ninja}~\cite{vanDeurzen:2013saa,Peraro:2014cba}. 
Alternatively, tensorial reconstruction~\cite{Heinrich:2010ax} is also
available, based on the
library {\tt golem95C}~\cite{Binoth:2008uq,Cullen:2011kv,Guillet:2013msa}.
The scalar master integrals can be taken from  {\sc
OneLOop}~\cite{vanHameren:2010cp} or {\sc QcdLoop}~\cite{Ellis:2007qk}. 

The \gosam{} package comes with the built-in model files 
{\tt sm}, {\tt smdiag}, {\tt smehc}, 
{\tt sm\_complex}, {\tt smdiag\_complex}, 
where the latter two should be used if complex masses and
couplings are present in the amplitude.
Complex masses, stemming from the consistent inclusion of  decay
widths  for unstable particles at NLO~\cite{Denner:2005fg}, are particularly important for the inclusion of
electroweak corrections, which also can be calculated with \gosam~\cite{Chiesa:2015mya}.
The model files {\tt smehc}  contain the effective Higgs-gluon
couplings. It has been used for example in the calculation of the NLO
corrections to H+3\,jet production in gluon fusion~\cite{Cullen:2013saa,Greiner:2015jha} and in the
calculation of $HH$+2\,jet production in both the gluon fusion and the
vector boson fusion channel~\cite{Dolan:2015zja}.

Other models can be imported easily, using the {\tt UFO} (Universal
FeynRules Output)~\cite{Degrande:2011ua,Degrande:2014vpa} format. 
This feature has been exploited for example in \Brefs{Greiner:2013gca,Cullen:2012eh}.

Therefore, \gosam{} comprises all the features which are needed to
calculate interference effects, both within and beyond the Standard
Model. 
An example for interference effects within the 2-Higgs-Doublet Model will be given below.

\subsubsection{Interference contributions to gluon-initiated heavy Higgs boson production in the 2HDM}

In this section
we discuss the loop-induced processes
$gg\to ZZ$ and $gg\rightarrow VV(\rightarrow e^+e^-\mu^+\mu^-/e^+e^-\nu_l\bar\nu_l)$
at LO QCD in the context of a CP-conserving
Two-Higgs-Doublet-Model (2HDM).
In particular, we study the effect of the interference between light
and heavy Higgs bosons, and with the background.
The 2HDM contains two Higgs doublets, which we name $H_1$ and $H_2$.
The models can be
classified into type I and type II, if we demand no tree-level
flavour-changing neutral currents and CP conservation. By
convention~\cite{Branco:2011iw}, 
the up-type quarks couple to $H_2$. 
In models of type I, the  down-type quarks also couple to $H_2$, 
while in type II models, they couple to $H_1$. 
The coupling to the leptons can either be through
$H_1$ or $H_2$,  but as our studies are not sensitive to the coupling of 
the Higgs bosons to leptons, we  do not need a further type distinction.
The two Higgs doublets form one CP-odd field $A$ and two
CP-even Higgs fields $h$ and $H$ due to CP conservation,
as well as two charged Higgs bosons $H^\pm$.
The 2HDM can be described in different basis representations. We make use
of the ``physical basis'', in which the masses of all physical Higgs bosons,
the ratio of the vacuum expectation values $\tb:=\tan\beta=v_2/v_1$
and the Higgs mixing angle in the CP-even sector $\alpha$, or alternatively
$\sba:=\sin(\beta-\alpha)$, are taken as input parameters.
We choose $\beta-\alpha$ in between $-\pi/2\leq \beta-\alpha\leq \pi/2$, such that $-1\leq \sba \leq 1$
and $0\leq \cba\leq 1$.
Our scenarios are thus specified by the two angles $\alpha$ and $\beta$,
which completely determine the relative couplings (with respect to the couplings
of a SM Higgs boson) of the light and the heavy
Higgs boson to quarks and the heavy gauge bosons. They are provided in \Eq(\ref{eq:gVV})
and \Tref{tab:couplingsGoSam} (together with \Eq(\ref{eq:gtbH}) for a decomposition in
terms of $\beta-\alpha$ and $\beta$). Moreover, our analysis is sensitive to $m_h$ and $m_H$,
whereas it is rather insensitive to the mass of
the pseudoscalar $m_A$ and the heavy charged Higgs boson mass $m_{H^\pm}$,
as long as they are heavy enough not to open decay modes of the
heavy Higgs $H$ into them and as long as the decay mode $H\rightarrow hh$ is sub-dominant.
\begin{table}
\caption{Relative couplings $g_f^\phi$ (with respect to the SM coupling) for the two 2HDM types.}
\begin{center}
\begin{tabular}{ c c c c c}
\toprule
Model      & $g_u^h$ & $g_d^h$  & $g_u^H$ & $g_d^H$  \\ 
\midrule
Type I     & $\cos\alpha/\sin\beta$ & $\cos\alpha/\sin\beta$  & $\sin\alpha/\sin\beta$ & $\sin\alpha/\sin\beta$ \\
Type II    & $\cos\alpha/\sin\beta$ & $-\sin\alpha/\cos\beta$ & $\sin\alpha/\sin\beta$ & $\cos\alpha/\cos\beta$ \\
\bottomrule
\end{tabular}
\end{center}
\label{tab:couplingsGoSam}
\end{table}
The strengths of the Higgs boson couplings to the gauge bosons $V\in\lbrace W,Z\rbrace$
are given by
\begin{align}
 g_{V}^h = \sin(\beta-\alpha)=:\sba,\qquad g_{V}^{H}=\cos(\beta-\alpha)=:\cba\quad.
\label{eq:gVV}
\end{align}
The pseudoscalar has no lowest-order couplings to a pair of gauge bosons.
It can in principle contribute to the considered processes with four
fermions in the final state. Because of the suppression of the 
Yukawa couplings to leptons, however, these contributions are very small, 
and thus diagrams involving the pseudoscalar are
not of relevance for our discussion.
In case of $|\sba|=1$ the light Higgs boson $h$ couples to the gauge bosons
with same strength as the SM Higgs boson. In contrast the coupling of the heavy
Higgs boson~$g_V^H$ vanishes according to the sum rule $(g_V^h)^2+(g_V^H)^2=1$.
Of large relevance for our discussion are the relative couplings of the heavy
Higgs boson to bottom-quarks and top-quarks, which are given by
\begin{align}
\nonumber
g_t^H&=\frac{\sin\alpha}{\sin\beta}=-\sba\frac{1}{\tb}+\cba,\\
\text{Type I: } g_b^H&=\frac{\sin\alpha}{\sin\beta}=-\sba\frac{1}{\tb}+\cba,\quad
\text{Type II: } g_b^H=\frac{\cos\alpha}{\cos\beta}=\sba\tb+\cba\quad.
\label{eq:gtbH}
\end{align}

\subsubsubsection{Details of the calculation}

We make use of \gosam{}~\cite{Cullen:2011ac,Cullen:2014yla}
to discuss the processes $gg\rightarrow e^+e^-\mu^+\mu^-$ and
$e^+e^-\nu_l\bar\nu_{l}$ (including all three neutrino flavours).
For a study of the relevance of interference contributions
we also consider the process $gg\rightarrow ZZ$, which we
generated with the help of {\tt FeynArts}~\cite{Hahn:2000kx}
and {\tt FormCalc}~\cite{Hahn:1998yk} and linked to
{\tt LoopTools}~\cite{Hahn:1998yk} for the calculation of the employed one-loop Feynman diagrams.
We added its amplitudes
to a modified version~\cite{Harlander:2013mla}
of {\tt vh@nnlo}~\cite{Brein:2012ne}.
It allows to be linked to {\tt 2HDMC}~\cite{Eriksson:2009ws} which we need for the
calculation of the Higgs boson widths $\Gamma_h$ and $\Gamma_H$.
In the case of the four lepton final state we have to sum over all possible intermediate configurations leading to the given final state.
This particularly means that depending on the sub-process, also intermediate $W$-bosons as well as non-resonant
contributions and photon exchange have to be taken into account. 
For the numerical integration over the four particle phase space we have combined the \gosam{} amplitudes
with the integration routines provided by {\tt MadEvent}~\cite{Maltoni:2002qb,Alwall:2007st}.\\
It is well-known that the calculation of processes including internal 
Higgs bosons, in particular if one includes higher orders,
needs a gauge invariant formulation of the Higgs boson
propagator. Since we are working at LO QCD only,
a simplistic Breit-Wigner propagator is sufficient for all our purposes.
We checked our modified {\tt vh@nnlo} and our \gosam{} implementations
against each other for $gg\rightarrow ZZ$ at the amplitude level 
and reproduced parts of the results presented in \Bref{Kauer:2015hia} for
the four leptonic final state within the numerical uncertainties.\\
We consider four benchmark scenarios to cover different aspects of a heavy Higgs boson
in the phenomenology of a 2HDM, given in \Tref{tab:2hdm}. 
All scenarios include a light Higgs boson with mass $m_h=125$\,GeV.
We keep the couplings of the light Higgs close to the ones of 
the SM Higgs by a proper choice of $\tb$ and $\sba$. 
The masses (and widths) of quarks
and gauge bosons are set to $m_t=172.3\,\text{GeV}, m_b(m_b)=4.16\,\text{GeV},
 m_Z=91.1876\,\text{GeV}, m_W=80.398\,\text{GeV},\Gamma_Z = 2.4952\,\text{GeV}, \Gamma_W= 2.085\,\text{GeV}$.

\begin{table}
\caption{2HDM scenarios considered in our analysis.}
\begin{center}
\begin{tabular}{ c  c  c  c  c  c }
\toprule
Scenario & 2HDM type &$\tb$ & $\sba$ & $m_H$ & $\Gamma_H$ \\
\midrule
S1 & II & $2$   & $-0.995$  & $200$\,GeV & $0.0277$\,GeV \\
S2 & II & $1$   & $0.990$   & $400$\,GeV & $3.605$\,GeV \\
S3 & I  & $5$   & $0.950$   & $400$\,GeV & $2.541$\,GeV \\
S4 & II & $20$  & $0.990$   & $400$\,GeV & $5.120$\,GeV \\
\bottomrule
\end{tabular}
\end{center}
\label{tab:2hdm}
\end{table}

Our studies presented here are carried out for the LHC with a centre-of-mass energy of $\sqrt{s}=13$\,TeV.
The role of interference effects is a bit less pronounced at $7/8$\, TeV compared to  $13$\,TeV. 
We make use of {\tt CT10nnlo}~\cite{Gao:2013xoa} as PDF set for the gluon luminosities.
Since our calculations are purely performed at LO the renormalization scale dependence enters
through the strong coupling~$\alpha_s$ only, which we take from the employed PDF set.
We choose the renormalization and factorization scale to be dynamical, namely
half of the invariant mass of the gauge boson system $\mu_R=\mu_F=m_{VV}/2$, i.e. $\mu_R=\mu_F=m_{4l}/2$
in case of the four leptonic final states.
It is known to have a small effect on the cross section~\cite{Kauer:2012ma,Campbell:2013una},
which we numerically confirm for the processes under consideration.
In case of the four lepton or the two lepton and two neutrino final states, we additionally cut on
the transverse momentum and the pseudorapidity of each lepton~$l$, $p_T^l>10$\,GeV and $|\eta_l|<2.7$,
the $R$-separation between individual leptons $R^{ll'}>0.1$ as well as $m_{ll}>5$\,GeV, where
$ll$ is an oppositely charged same-flavour dilepton pair.
For the neutrinos we ask for a total missing transverse momentum of $E_T^{\text{miss}}>70$\,GeV.
The cuts are inspired by the recent ATLAS analysis carried out in \Bref{Aad:2015kna}.
One of the most important observables is certainly the invariant mass distribution of the four leptons as the two Higgs bosons manifest themselves 
in Breit-Wigner peaks in this distribution.
For the process $gg\rightarrow e^+e^-\mu^+\mu^-$ this observable $m_{4l}$ is also experimentally
easily accessible due to two electrons and two muons in the final state. In the cases with neutrinos in the final state 
the situation is more involved. The invariant mass is no longer an observable that is experimentally accessible but only
a transverse component can be measured. If one is interested in a heavy Higgs boson that will decay into the four leptons via 
two intermediate electroweak gauge bosons a sensible choice is to consider the transverse mass of the underlying two boson
system. In our case the two boson system can be $ZZ$ as well as $WW$. 
We therefore define a general transverse mass via \cite{Aad:2015agg}
\begin{equation}
 m_{VV,T}^2= \left(E_{T,ll} +E_{T,\nu\nu}\right)^2 - \left|\vec{p}_{T,ll} + \vec{p}_{T,\nu\nu}\right|^2\;,
 \label{eq:mT}
\end{equation}
with
\begin{equation}
 E_{T,ll}=\sqrt{p_{ll}^2+|\vec{p}_{T,ll}|^2}\;,\quad \text{and}\;\; E_T^{\text{miss}}=E_{T,\nu\nu}=\left|\vec{p}_{T,\nu\nu}\right|\;.
 \label{eq:ET}
\end{equation}
\subsubsubsection{Discussion of four fermionic final states}
\begin{figure}
\begin{center}
\begin{tabular}{cc}
\includegraphics[width=0.5\textwidth]{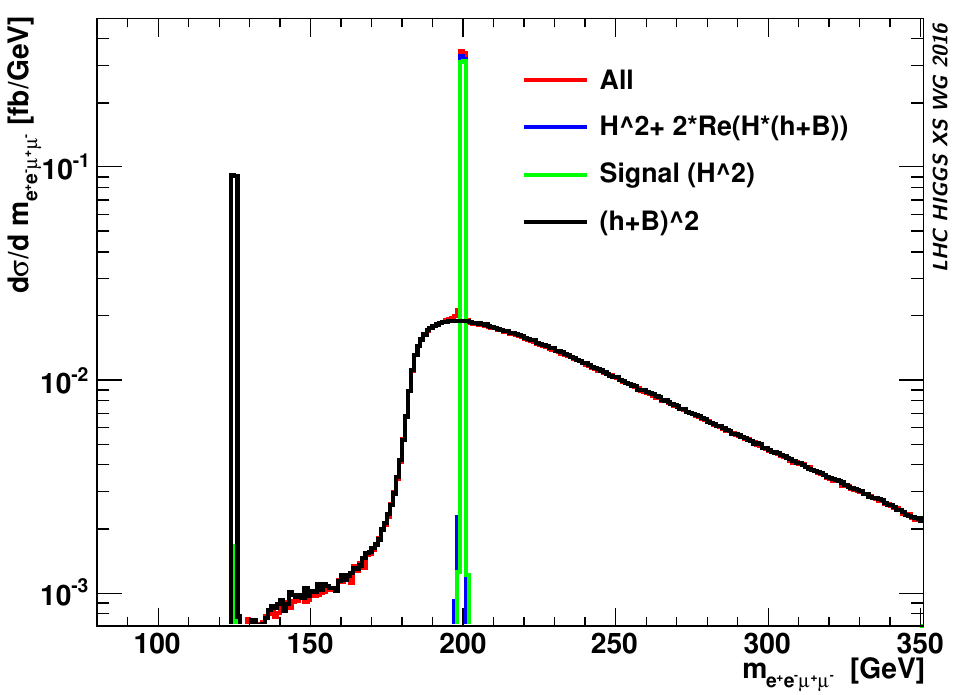} &
\includegraphics[width=0.5\textwidth]{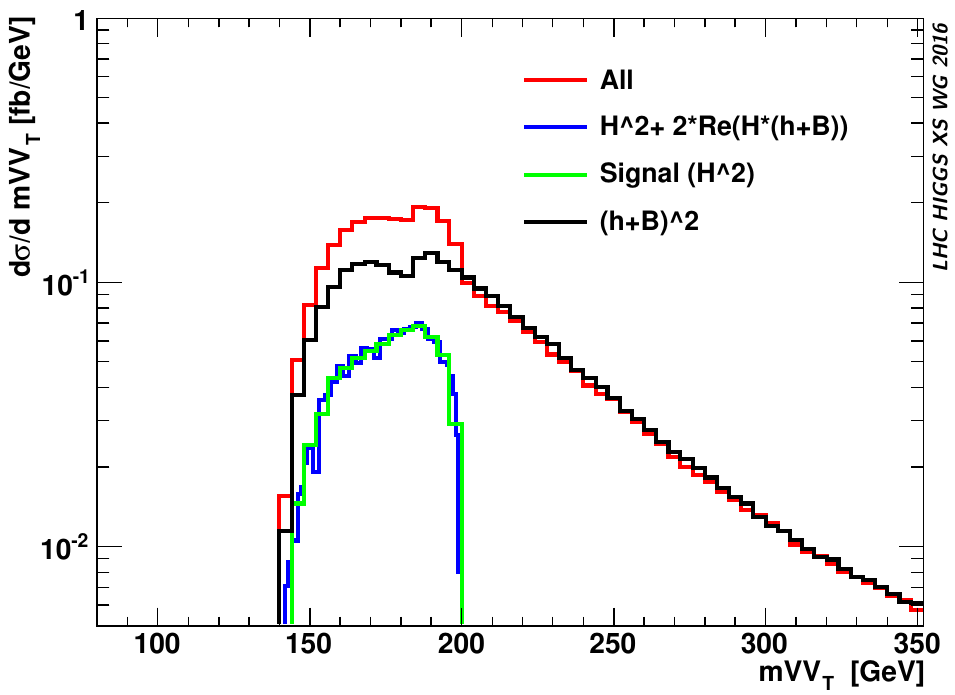}\\[-0.2cm]
 (a) & (b)
\end{tabular}
\end{center}
\caption{(a) Invariant mass distribution for $gg\rightarrow e^{+}e^{-}\mu^{+}\mu^{-}$ and (b) transverse mass
distribution for $gg\rightarrow e^{+}e^{-}\nu_l\bar\nu_l$
for scenario~S1 at $\sqrt{s}=13$\,TeV.}
\label{fig:S1_mass} 
\end{figure}

We exemplify the results for the four fermionic final state by discussing  the results of scenario~S1.
\refF{fig:S1_mass} shows the invariant mass distribution of the four leptons for $gg\rightarrow e^{+}e^{-}\mu^{+}\mu^{-}$ and 
the transverse mass distribution using the definition in \Eq(\ref{eq:mT}) for the processes involving final state neutrinos. 
We distinguish four different contributions. In red, denoted with 'All', we plot all
contributions that lead to the given final state in the considered scenario. In green, we only plot the contribution from
the heavy Higgs boson, whereas in blue we also add the interference of the heavy Higgs boson with the background and the light Higgs boson.
The contribution $|h+B|^2$, plotted in black, contains besides the contributions without any Higgs also contributions of the light Higgs as well as the interference
contributions of the light Higgs boson with non-Higgs diagrams.\\
In the invariant mass plot of $gg\rightarrow e^{+}e^{-}\mu^{+}\mu^{-}$, see \refF{fig:S1_mass}~(a),
the two Higgs boson peaks at $m_{4l}=125$ and $200$\,GeV can be clearly seen.
Due to the very small width of the heavy Higgs boson there is no distortion of the Breit-Wigner shape visible, and also the impact
of the interference contribution to the total height of the peak is rather small. The transverse mass distribution 
for $gg\rightarrow e^{+}e^{-}\nu_l\bar\nu_l$ shows a quite different pattern. First of all there is no peak from the light Higgs boson.
The reason for this are the different cuts compared to the process without neutrinos. The requirement of $E_T^{\text{miss}}>70$\,GeV
excludes this region of phase space. 
Due to the fact that the four momenta of the neutrinos are experimentally not accessible one sets
$E_{T,\nu\nu}=\left|\vec{p}_{T,\nu\nu}\right|$, which ignores the invariant mass of the neutrino system.
This removes the sharp peak of the heavy Higgs boson, 
which is visible in the invariant mass distribution of the muon process.
Instead of a distinguished peak one obtains a broad distribution.
But also here the contribution of the interference remains small. 
A second difference compared to the muon process is the occurrence of a small dip at around $m_{VV,T}=180$\,GeV in both signal and background.
This specific shape is due to the fact that the total contribution to the process with neutrino final state consists
of the sum of two different sub-processes, namely the one with the electron neutrino and the ones with muon- and tau neutrino
in the final state.
Whereas the first sub-process also has contributions from intermediate $W$-bosons, this is not the case for the latter sub-processes.
The two sub-processes therefore show a different kinematical behaviour and the sum of the two contributions leads to the given
distribution.\\
For a more detailed discussion of the other scenarios and different observables we refer to \Bref{Greiner:2015ixr}.

\subsubsubsection{Relevance of interference contributions}
The interference contributions of the heavy Higgs boson with
the light Higgs boson and the background are significantly
enhanced in two cases:
Naturally small couplings involved in the signal process increase
the mentioned interferences. This is either of relevance in the
decoupling limit of the 2HDM where $\sba\to 1$
and thus the coupling of the heavy Higgs boson to gauge bosons
vanishes or through a small coupling of the heavy Higgs boson
to top- and/or bottom-quarks. According to Eq. (2) the top-quark
coupling vanishes for a specific value of $\sba$
for fixed $\tb$. In a 2HDM type~I the bottom-quark coupling
vanishes for the same value, such that the cross section
$\sigma(gg\rightarrow H\rightarrow VV)$ gets zero, whereas
in a 2HDM type~II the cross section is minimal.
Moreover the interferences are found to be large for an enhanced
bottom-quark Yukawa coupling, i.e. large $\tb$. Again, for further details
we refer to \Bref{Greiner:2015ixr}. Interferences in the mentioned
two cases can help to lift the signal cross section by more
than a factor of $2$ and thus
enhance the sensitivity of heavy Higgs boson searches. 

\subsubsubsection{Interferences at high invariant masses}
\begin{figure}
\begin{center}
\begin{tabular}{cc}
\includegraphics[width=0.47\textwidth]{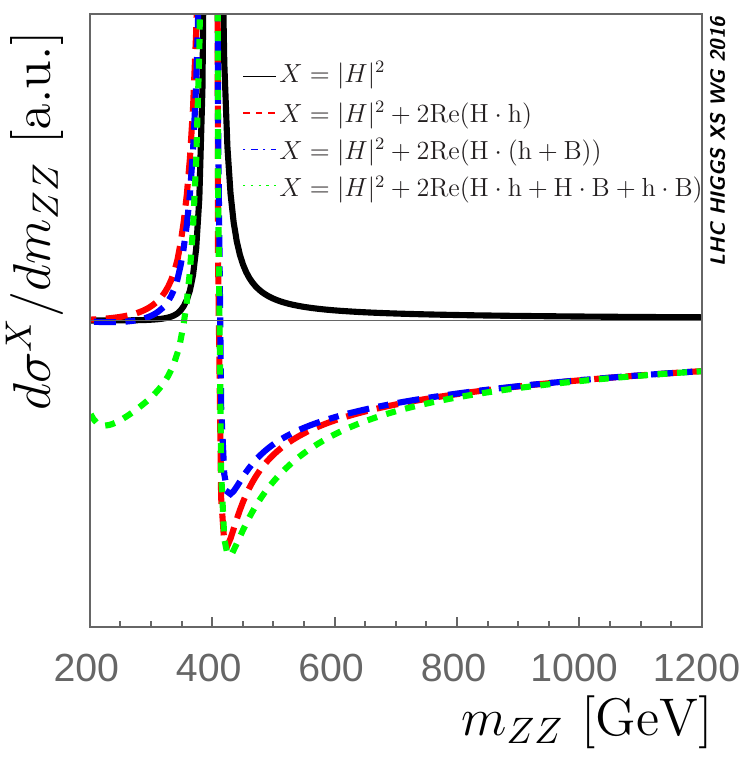} & 
\includegraphics[width=0.47\textwidth]{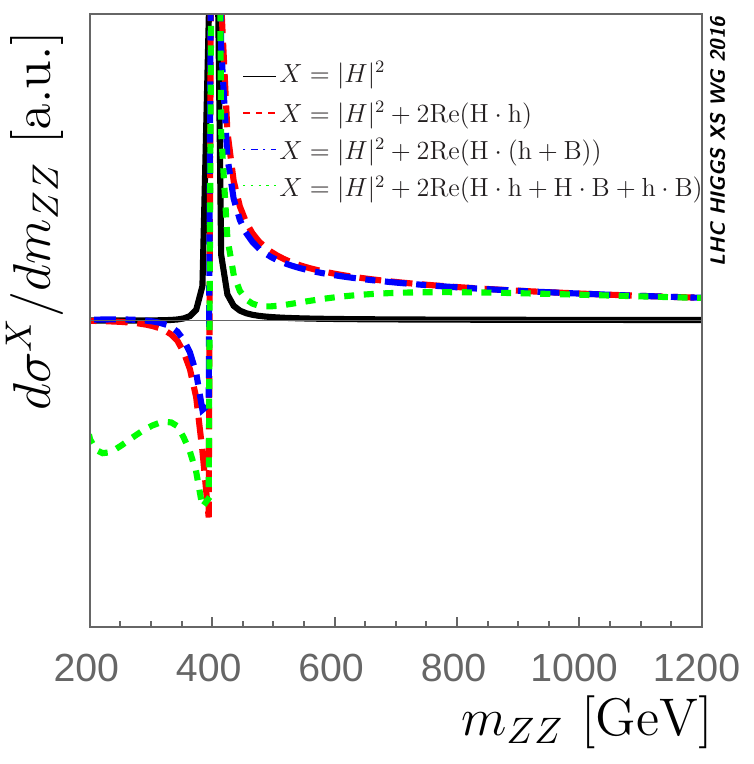} \\
 (a) & (b) \\
\multicolumn{2}{c}{\includegraphics[width=0.47\textwidth]{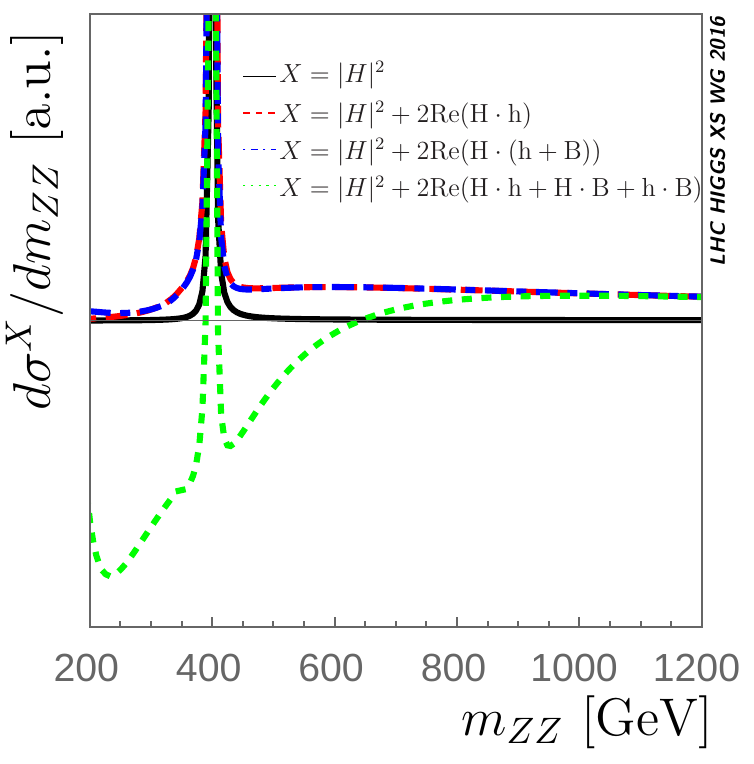}} \\
\multicolumn{2}{c}{(c)} 
\end{tabular}
\end{center}
\caption{Partonic cross sections $d\sigma^X/d\mzz$ for $gg\rightarrow ZZ$ in arbitrary units as a function of the invariant mass $\mzz$ in GeV
for scenario (a) S2, (b) S3 and (c) S4
(black: $X=|H|^2$; red, dashed: $X=|H|^2+2\text{Re}(H\cdot h)$;
blue, dot-dashed: $X=|H|^2+2\text{Re}(H\cdot h)+2\text{Re}(H\cdot B)$; green, dotted: $X=|H|^2+2\text{Re}(H\cdot h)+2\text{Re}(H\cdot B)
+2\text{Re}(h\cdot B)$).}
\label{fig:S246hb} 
\end{figure}
So far we focused on the interference effects between the heavy Higgs and the background 
as well as the heavy Higgs and the light Higgs in the vicinity of the heavy Higgs boson resonance,
since the interference between the light Higgs boson and the background can be considered
constant in this region.
However, at high invariant masses of the diboson system
the interplay between all three contributions $h$ and $H$ and the background $B$
is of relevance, to a certain extent related to the unitarization of the cross section.
In \refF{fig:S246hb} we plot the differential cross 
section $gg\rightarrow ZZ$ as a function of the invariant mass of the diboson system $\mzz$ up to high masses beyond the heavy
Higgs boson resonance. We exemplify the discussion for the three scenarios S2, S3 and S4. The differences
between the coloured curves display the importance of the different interference terms.
Since the figures are obtained for the partonic cross section and we are interested in
the relative effects of the interferences among each other, we do not display units for
$d\sigma/d\mzz$. At high invariant masses the interference between the heavy Higgs boson
and the background is negligible, in contrast to the interference of the light Higgs and the
heavy Higgs boson, which remains large and can have either sign.
Moreover the smoothly falling interference of the light Higgs boson and the background
comes into the game within a certain window of invariant masses below $1$\,TeV.
\refF{fig:S246hb} depicts different cases, where the interference $h\cdot H$ 
is either negative similar to the interference $h\cdot B$ or leads to a positive contribution to
the differential cross section
in a region $\mzz\in\left[450\,\text{GeV},1000\,\text{GeV}\right]$.
The latter case is true for scenarios~S3 or S4, where a sign change of
the total depicted contribution leads to a dip and a subsequent ``peak''-like
structure when added to the background. This structure also appears in the total four particle final state,
where the gluon luminosities further suppress the cross section at high invariant masses.
Thus all interferences need to be taken into account in order to correctly describe the cross section
at high invariant masses.




\section{\texorpdfstring{$gg\to VV$}{gg to VV} at NLO QCD}
\label{sec:offshell_interf_ggVV_NLO}

\subsection{The status of theoretical predictions}
A good theoretical control of the off-shell region requires the
knowledge of higher order QCD correction for both the signal $pp\to H
\to 4l$ and the SM background $pp\to 4l$ processes.  At high invariant
masses, the signal $gg\to H \to 4l$ and the background $gg\to 4l$
processes individually grow with energy, eventually leading to
unitarity violations. In the SM, a strong destructive interference
between signal and background restores unitarity in the high energy regime,
and its proper modelling is important for reliable predictions in
the off-shell tail. At invariant masses larger than the top threshold
$m_{4l} > 2 \mt$ the effect of virtual top quarks running in the
loops is non negligible and must be taken into account.

The state of the art for theoretical predictions of signal, background
and interference is very different. For an exhaustive description of
the signal cross section we refer the reader to the relevant sections
of this report. As far as perturbative QCD is
concerned, the signal is known through NLO with exact quark mass
dependence~\cite{Djouadi:1991tka,Spira:1995rr}. NNLO corrections are
known as an expansion around the $\mt \to \infty$
limit~\cite{Harlander:2003ai,Anastasiou:2002yz,Ravindran:2003um},
matched to the exact high-energy limit~\cite{Marzani:2008az} to avoid
a spurious growth at high energies~\cite{Harlander:2009my,Pak:2009dg}.
Very recently, the N$^3$LO corrections became
available~\cite{Anastasiou:2015ema} in the infinite top mass
approximation. They turned out to be moderate, with a best stability
of the perturbative expansion reached for central scale
$\mu=\mh/2$. So far, results are known as an expansion around
threshold, which is expected to reproduce the exact result to better
than a per cent.

We now briefly discuss the status of theoretical description of the
background. In the SM, four-lepton production is dominated by quark
fusion processes $q\bar q \to VV \to 4l$.  Recently, NNLO QCD
corrections were computed for both the $ZZ$~\cite{Cascioli:2014yka} and the
$WW$~\cite{Gehrmann:2014fva} processes, leading to a theoretical uncertainty coming
from scale variation of a few per cent. In these prediction, the
formally NNLO gluon fusion channel $gg\to 4l$ enters for the first
time, i.e. effectively as a LO process. At the LHC, it is enhanced
by the large gluon flux and corresponds to roughly 60\%(35\%) of the
total NNLO corrections to the $ZZ$($WW$) process.
Despite being sub-dominant for $pp\to4l$ production, the $gg\to 4l$
sub-channel is of great importance for off-shell studies. First of
all, as we already mentioned there is a strong negative interference
between $gg\to 4l$ and $gg\to H\to 4l$. Second, the gluon fusion SM
background is harder to separate from the Higgs boson signal.

Computing NLO corrections to $gg\to 4l$ is highly non trivial as it
involves the knowledge of complicated two-loop amplitudes with both
external and internal massive particles. Parton shower studies based
on merged $gg\to 4l + 0,1~{\rm jet}$ have been performed for
example in~\cite{Cascioli:2013gfa}.  Very recently, NLO QCD corrections
for $gg\to VV\to 4l$ process were computed in the case of massless
quark running in the loop~\cite{Caola:2015psa,Caola:2015rqy}.
This approximation is expected to hold very well below threshold,
$m_{4l} < 2 \mt \sim 300~\rm{GeV}$. As in the Higgs case, finite top
quark effects are known as an expansion in
$1/\mt$~\cite{Melnikov:2015laa}.  Going beyond that would require
computing two-loop amplitudes which are currently beyond our
technological reach, so the exact result is not expected in the near
future.

\subsection{Brief description of the NLO computation for \texorpdfstring{$gg\to 4l$}{gg to 4l}}
\subsubsection{Massless quark contribution}
In this section, we briefly report the main details of the $gg\to VV
\to 4l$ NLO QCD
computations~\cite{Caola:2015psa,Caola:2015rqy}. Despite being NLO
calculations, they pose significant technical challenges.
First, complicated two-loop amplitude are required, see
\refF{offshell_interf_ggVV_NLO:twoloop} for a representative sample. These amplitudes were
recently computed in~\cite{Caola:2015ila,vonManteuffel:2015msa}. They
include decay of the vector bosons and account for full off-shell
effects. For the results in~\cite{Caola:2015psa,Caola:2015rqy}, the
public C++ implementation of Ref.~\cite{vonManteuffel:2015msa} was
used. To ensure the result is stable, the code code compares numerical
evaluations obtained with different (double, quadruple and, if
required, arbitrary) precision settings until the desired accuracy is
obtained.  For a typical phase space point, the evaluation of all
two-loop amplitudes requires about two seconds.

\begin{figure}
\centering
\includegraphics{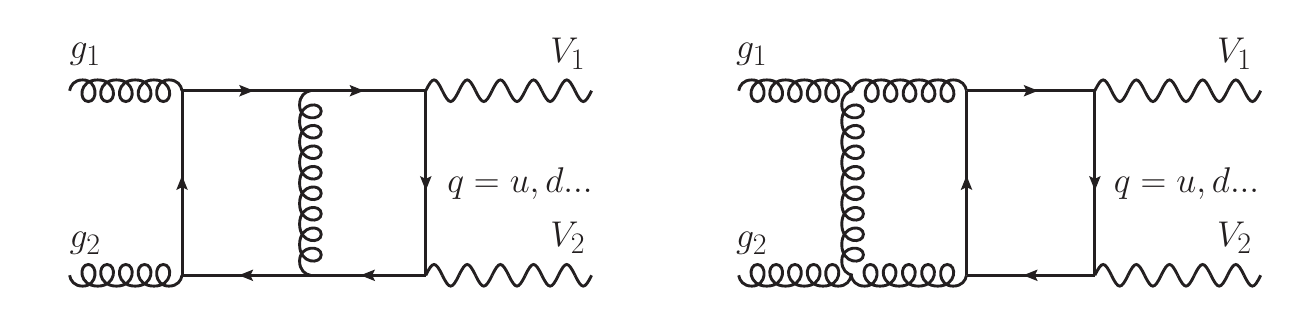}
\caption{Representative two-loop diagrams for the $gg\to 4l$ process.
Leptonic decays of the vector bosons is assumed.}\label{offshell_interf_ggVV_NLO:twoloop}
\end{figure}

\begin{figure}
\centering
\includegraphics{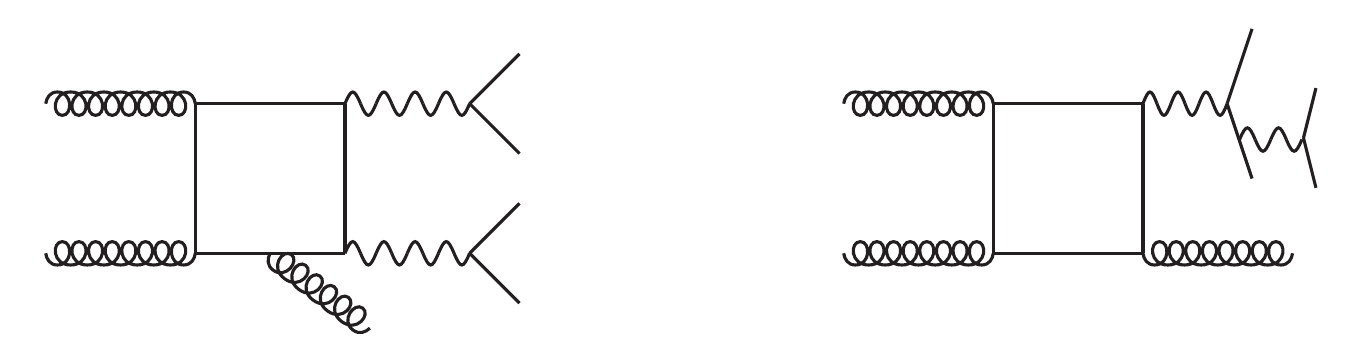}
\caption{Representative double (left) and single (right) resonant one-loop diagrams
for the $gg\to4l+g$ process.}\label{offshell_interf_ggVV_NLO:oneloop}
\end{figure}

Second, one-loop real emission amplitudes are required, see
\refF{offshell_interf_ggVV_NLO:oneloop}.  Despite being only one-loop amplitudes, they must
be evaluated in degenerate soft/collinear kinematics, so they must be
numerically stable.  For the computations
in~\cite{Caola:2015psa,Caola:2015rqy}, these amplitudes were computed
using a mixture of numerical~\cite{Ellis:2007br} and
analytical~\cite{Badger:2008cm} unitarity. As a cross-check, the
obtained amplitudes were compared against
OpenLoops~\cite{Cascioli:2011va} for several different kinematic
points.  Possible numerical instabilities are cured by increasing the
precision of the computation. The typical evaluation time for a phase
space point, summed over colour and helicities, is about 0.1
seconds. Also in this case, full decay of the vector bosons into
leptons/neutrinos and off-shell effects are understood. Note that the
latter involve single-resonant diagrams, see 
\refF{offshell_interf_ggVV_NLO:oneloop}(right).
Arbitrary cuts on the final state
leptons/neutrinos (and additional jet) are possible.  In the
computations~\cite{Caola:2015psa,Caola:2015rqy}, interference between
$WW$ and $ZZ$ mediated processes for $2l2\nu$ final states are
neglected. They are expected to be irrelevant in the
experimental fiducial regions. Full $ZZ$/$\gamma\gamma$ interference
effects are included.

In~\cite{Caola:2015psa,Caola:2015rqy}, contributions coming from
$qb\to VV q$ mediated by closed fermion loops were not included. This
is because at $\mathcal O(\alpha_s^3)$ there are several other
contributions to the $qg$ channel other than one-loop squared
amplitudes, which in principle are not sub-dominant. Neglecting these
channels is fully justified in the large gluon approximation
of~\cite{Caola:2015psa,Caola:2015rqy}. Residual factorization scale
uncertainties are expected to give an estimate of the impact of neglected
channels.

In the $ZZ$ computation~\cite{Caola:2015psa}, the top quark
contribution is neglected and the bottom quark is considered massless
(see~\cite{Caola:2015psa} for more details).  This approximation is
expected to work at the 1\% level for the total $gg\to ZZ$
cross-section, but it is not reliable in the high invariant mass
regime. To quantify this, in \refF{offshell_interf_ggVV_NLO:mass_fig} we compare at LO the full massive
computation with the approximation~\cite{Caola:2015psa}.
\begin{figure}
\centering
\includegraphics[width=0.45\textwidth]{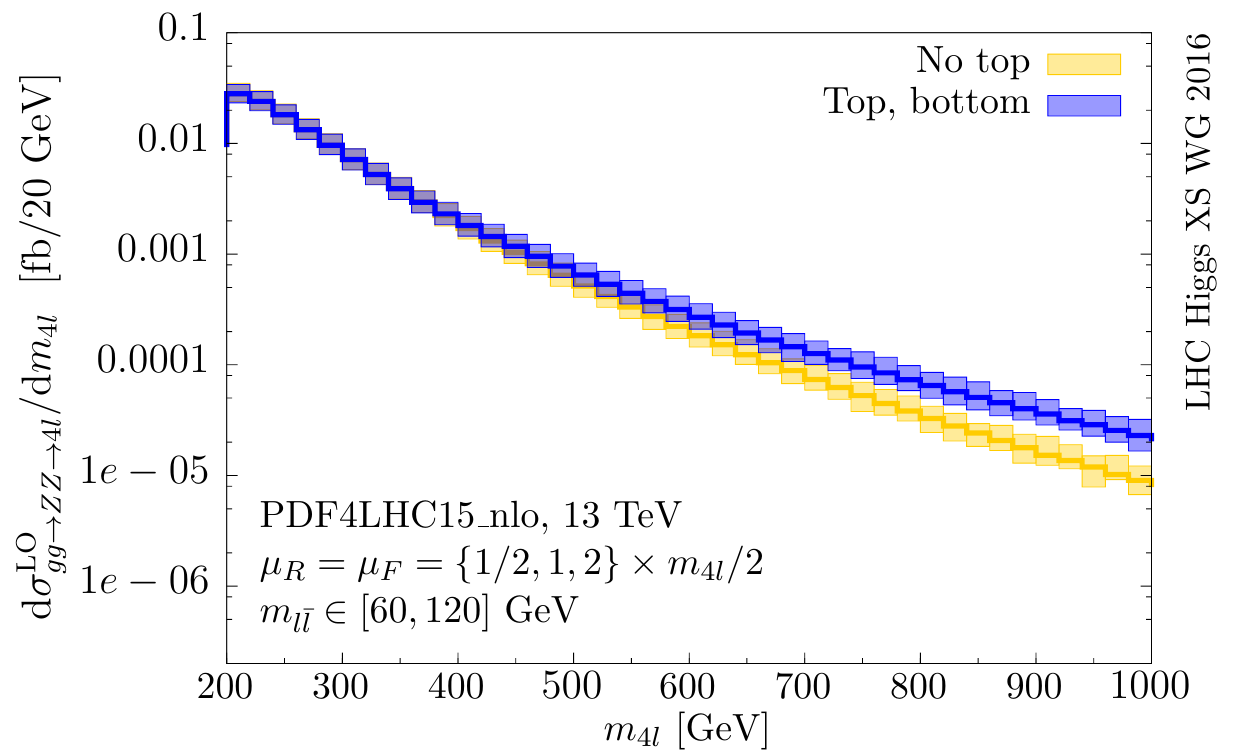}
\includegraphics[width=0.45\textwidth]{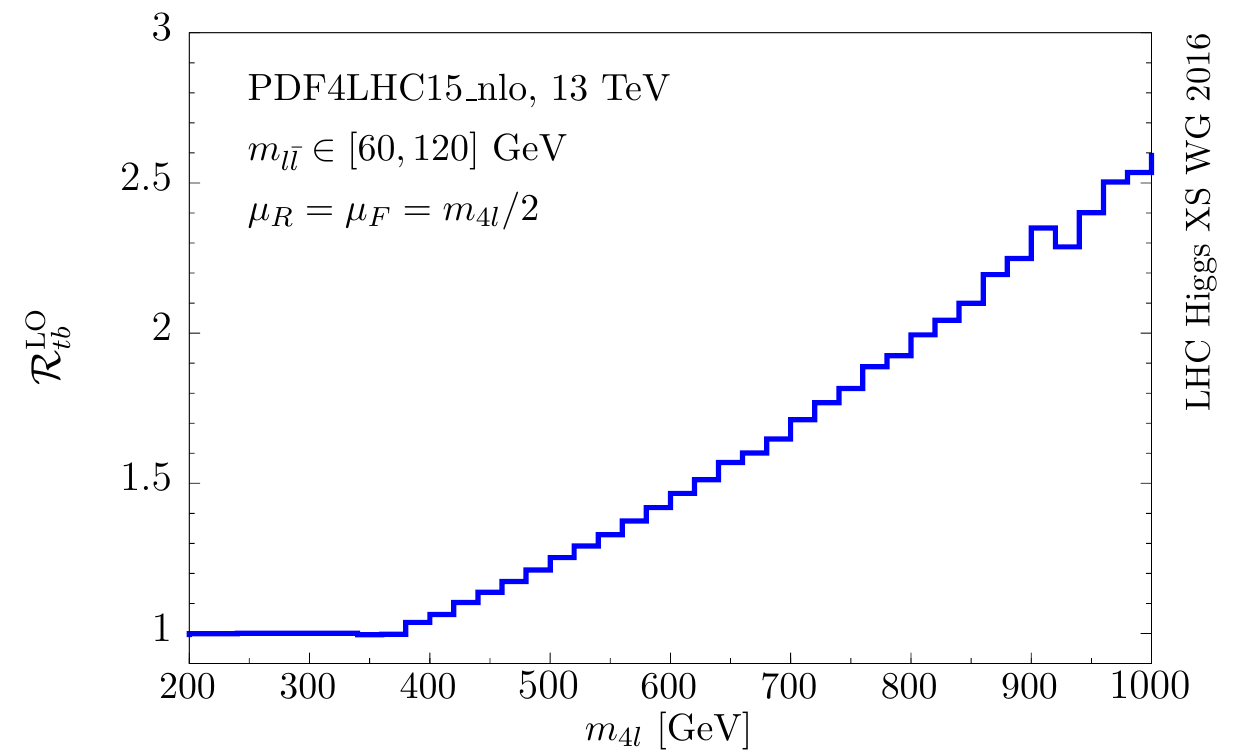}
\caption{Top quark mass contribution to $gg\to ZZ\to 4l$ at LO. Left: comparison between the exact result (blue) and the
approximation where the top quark contribution is omitted and the bottom quark is considered massless (see~\cite{Caola:2015psa} for details).
Right: ratio between the exact and approximate results for the central scale $\mu=m_{4l}/2$. See text for details.}\label{offshell_interf_ggVV_NLO:mass_fig}
\end{figure}
From the figure it is clear that below the top threshold the approximation~\cite{Caola:2015psa} is essentially
exact, while above the top quark contribution becomes rapidly important. The relative size of the top quark contribution is
quantified in the right panel of \refF{offshell_interf_ggVV_NLO:mass_fig}, where
\begin{equation}
\mathcal{R}^{\rm LO}_{tb}(m_{4l}) \equiv \left.\frac{{\rm d}\sigma^{\rm LO}_{t,b}/{\rm d}m_{4l}}{{\rm d}\sigma^{\rm LO}_{{\rm no}-t}/{\rm d}m_{4l}}
\right|_{\mu_r=\mu_f=m_{4l}/2}.\label{offshell_interf_ggVV_NLO:rtilde}
\end{equation}
For the $WW$ case, in the calculation~\cite{Caola:2015rqy} both the top and
the bottom quark contributions are omitted. At LO, top/bottom
contributions account for $\mathcal O(10\%)$ of the total $gg\to WW$
cross section.

\subsubsection{Finite top quark effects}
The effect of finite top quark mass in $gg\to ZZ$ at NLO was
investigated in~\cite{Melnikov:2015laa}. Similar to what is done in
the Higgs case, the authors performed the computation as an expansion
in the $\mt\to\infty$ limit.  The first two non trivial terms in the
expansion were kept, which allowed for a reliable description of the
top quark contribution up to invariant masses of order $m_{4l}\sim 300~{\rm GeV}$.
In this computation, only the total $gg\to ZZ$
cross-section was considered, although this should be enough to have a
rough estimate of the size of the mass effects. The result on the NLO
corrections, compared to the signal case, are shown in
\refF{offshell_interf_ggVV_NLO:mass}. For these results, the Higgs boson signal is computed in
the $\mt\to\infty$ limit as well. Also, compared to the $K$-factor
defined in~\cite{Melnikov:2015laa}, here we used NLO PDFs and $\alpha_s$
evolution for both the LO and the NLO contributions. The band represent
scale variation uncertainty, obtained from a factor of two variation around
$\mu_0=m_{4l}/2$.
\begin{figure}
\centering
\includegraphics[width=0.45\textwidth]{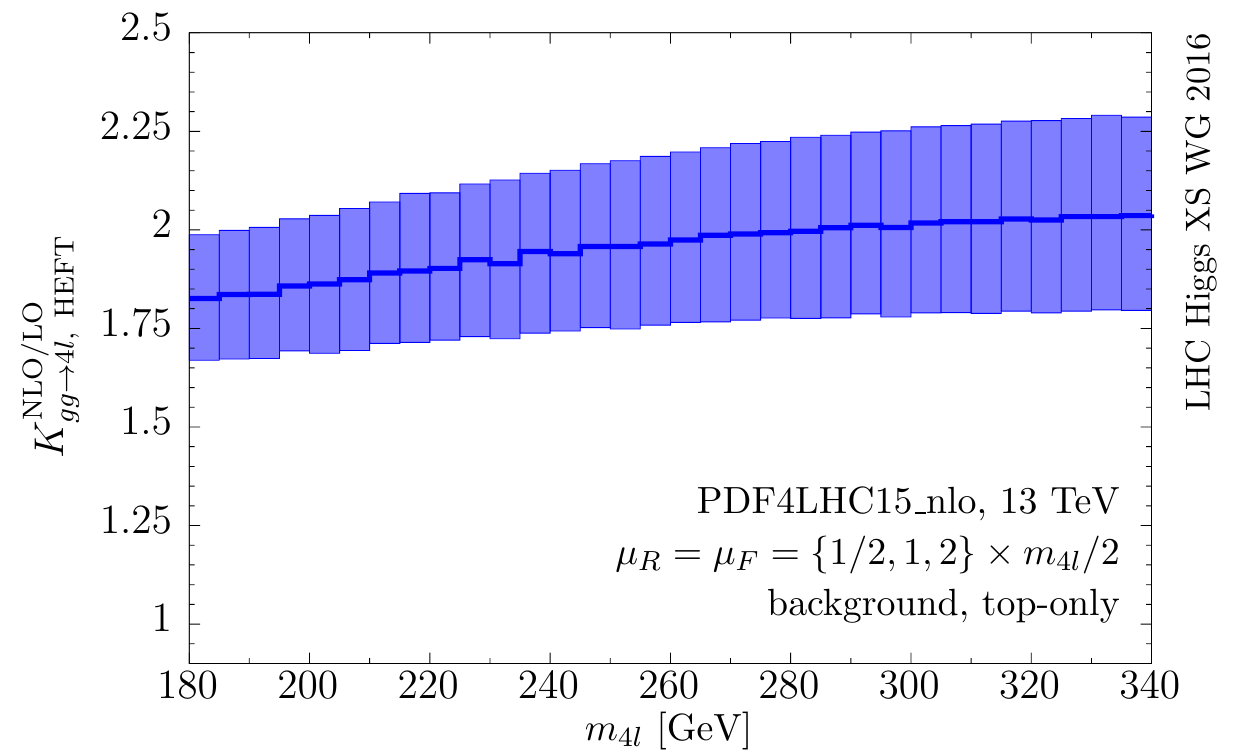}
\includegraphics[width=0.45\textwidth]{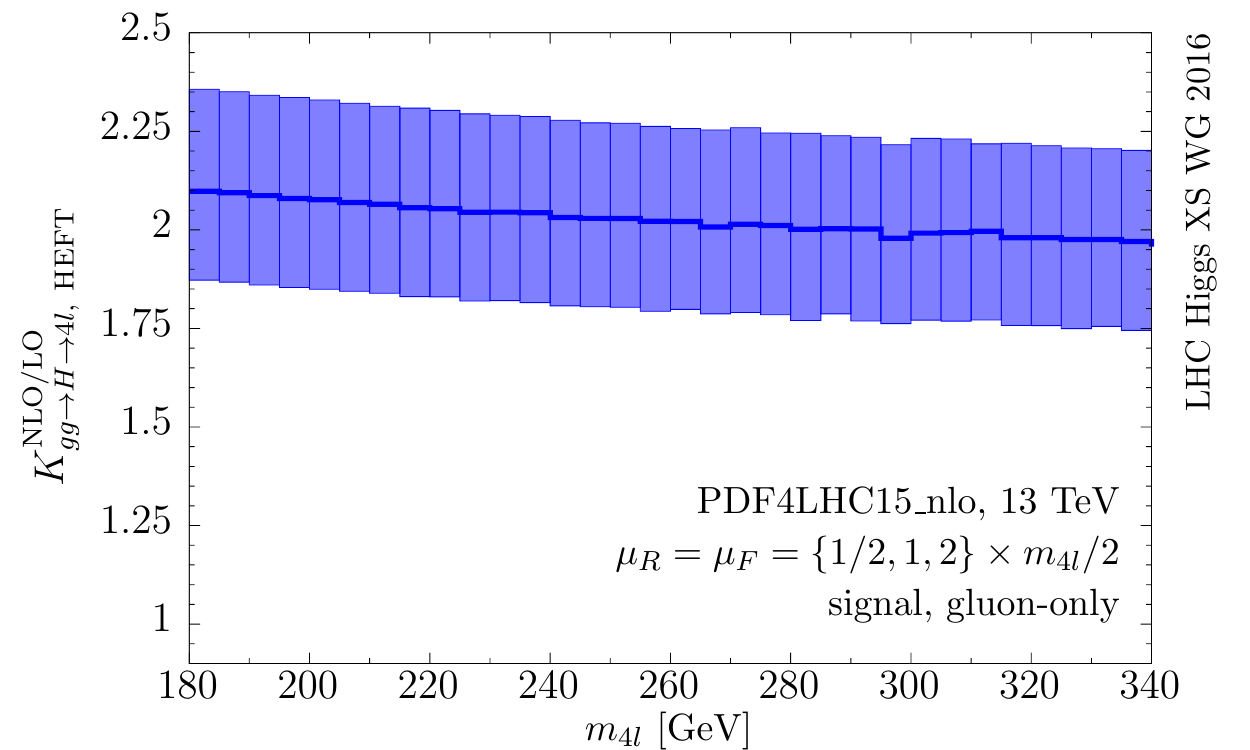}
\caption{$K-$factors for signal and background, in the heavy top expansion.
Both LO and NLO contributions are computed with NLO PDFs and $\alpha_s$. See text for details.}\label{offshell_interf_ggVV_NLO:mass}
\end{figure}

Close to the $ZZ$ threshold, the background $1/\mt$ expansion is
expected to be accurate within $\mathcal O(20\%)$~\cite{Melnikov:2015laa}.
Signal and $K-$factors are of the same order of magnitude, in
agreement with what expected from soft gluon
approximations~\cite{Bonvini:2013jha}. Below the top threshold, the
precision on the approximation~\cite{Melnikov:2015laa} can be
systematically improved by computing more terms in the $1/\mt$ expansion.
Above the top threshold $m_{4l}\sim 300~{\rm GeV}$, the
expansion~\cite{Melnikov:2015laa} alone is no longer reliable. Since
the full computation is not available, the expansion could be improved
along two directions. In principle, it could be matched against the
exact high energy behaviour. While this does not pose any conceptual
challenge, the computation of the high energy limit is technically
more involved than in the Higgs case and it is presently unknown. A
second option would be to rescale by the exact LO and hence consider
and expansion for the $K-$factor, for which the $1/\mt$ expansion
should be better behaved.

\subsection{Results and recommendation for the \texorpdfstring{$gg\ (\to H)\to ZZ$}{gg (to H) to ZZ} interference \texorpdfstring{$K$}{K}-factor
}
As explained in the previous section, exact predictions valid up to
high $\sim 1~$TeV invariant masses are only known at NLO for the
$gg\to H \to 4l$ signal and LO for the $gg\to 4l$ background. However,
several indications point towards sizeable higher order corrections,
both for signal and background. In this section we study this issue
and present a possible practical recommendation for the
signal, background and interference $K-$factors.

We start by describing the setup used for the results presented in
this section. LO and NLO results are both obtained with NLO PDFs and
$\alpha_s$. In principle, one could envision using LO PDFs (and
$\alpha_s$) for the LO results, and this would in general lead to
smaller corrections, with reduced shape dependence. However, since
PDFs fits are still dominated by DIS data, the LO gluon distribution
is almost entirely determined by DGLAP evolution. The large LO gluon
flux hence is artificially driven by the large NLO DIS $K-$factor and
it is not reliable. Until LO gluon PDFs are obtained by fitting
hadronic data, using the NLO gluon distribution is preferable, see the
PDFs section of this report for more details.
NNLO PDFs could be used as well, since the $gg\to 4l$ process
enters at NNLO in the $q\bar q \to 4l$ computation. However, here we are
mostly interested in interference effects, so for consistency with 
the Higgs case we use NLO PDFs for NLO signal, 
$gg\to 4l$ background and interference.

Regarding the scale choice, it is well known that for Higgs boson production
an optimal choice would be $\mu\sim
\mh/2$~\cite{Anastasiou:2002yz}. Theoretically, it is justified both
by all-order analysis of the $Hgg$ form factor and by the fact that
the average $p_\perp$ of the Higgs boson is $\sim \mh/2$.
Empirically, a much better convergence is observed with this scale
choice, as well as a reduced impact of resummation
effects~\cite{Anastasiou:2016cez}. For off-shell studies, this
translates into choosing as a central scale half of the virtuality of
the Higgs boson, i.e. $\mu = m_{4l}/2$. Since most of the above
consideration are only based on the colour flow of the process, they
also apply for the background and interference scale
choice. Incidentally, we note that this was also the preferred choice
for the NNLO $pp\to WW/ZZ$ computations~\cite{Gehrmann:2014fva,Cascioli:2014yka}.

In the region $m_{4l} < 2 m_t$, precise results exist for both the
signal and the background.  In more detail, NNLO results for the signal
can be obtained from~\cite{Harlander:2009my,Pak:2009dg}. For the background, NLO
contributions from massless quarks can be obtained
using~\cite{Caola:2015psa}\footnote{A numerical code for background
  predictions should be made public soon.} while
top quark contributions can be obtained from~\cite{Melnikov:2015laa}.
In principle, these results could be used to obtain a NLO prediction
for the interference. However, this calculation has not been performed
yet.
Given the similarity of signal and background $K-$factors, until
a better computation is available the interference $K-$ factor can be
obtained as the geometric average of the signal and background $K-$
factors. Scale variation uncertainties should account for missing
higher order in the perturbative expansion. Alternatively, we note
that even with our scale choice the signal still exhibits a non
negligible NNLO $K-$factor, and it is not unreasonable to expect a
similar $K-$factor also for the background~\cite{Bonvini:2013jha}.
One may then apply the signal NNLO $K-$factor to the background as
well, and take the difference between NNLO and NLO as a conservative
estimate of perturbative uncertainties.

In the high invariant mass region $m_{4l}>2 m_t$, it is not possible
at this stage to provide a full NNLO (NLO) theoretical prediction for the
signal (background), since exact heavy quark mass effects at NLO are
unknown. In the following, we investigate signal and background
$K-$factors in this region making different assumptions for missing
top quark contributions.
First, we compare in \refF{offshell_interf_ggVV_NLO:kfactor}
results for signal -- with full top and bottom mass dependence through NLO --
 and background neglecting top quark contributions, as described in the
previous sections and in~\cite{Caola:2015psa}. For
reference, we also show the effect of NNLO QCD corrections (computed
with NNLO PDFs and $\alpha_s$, and in the heavy-top approximation).
\begin{figure}
\centering
\includegraphics[width=0.45\textwidth]{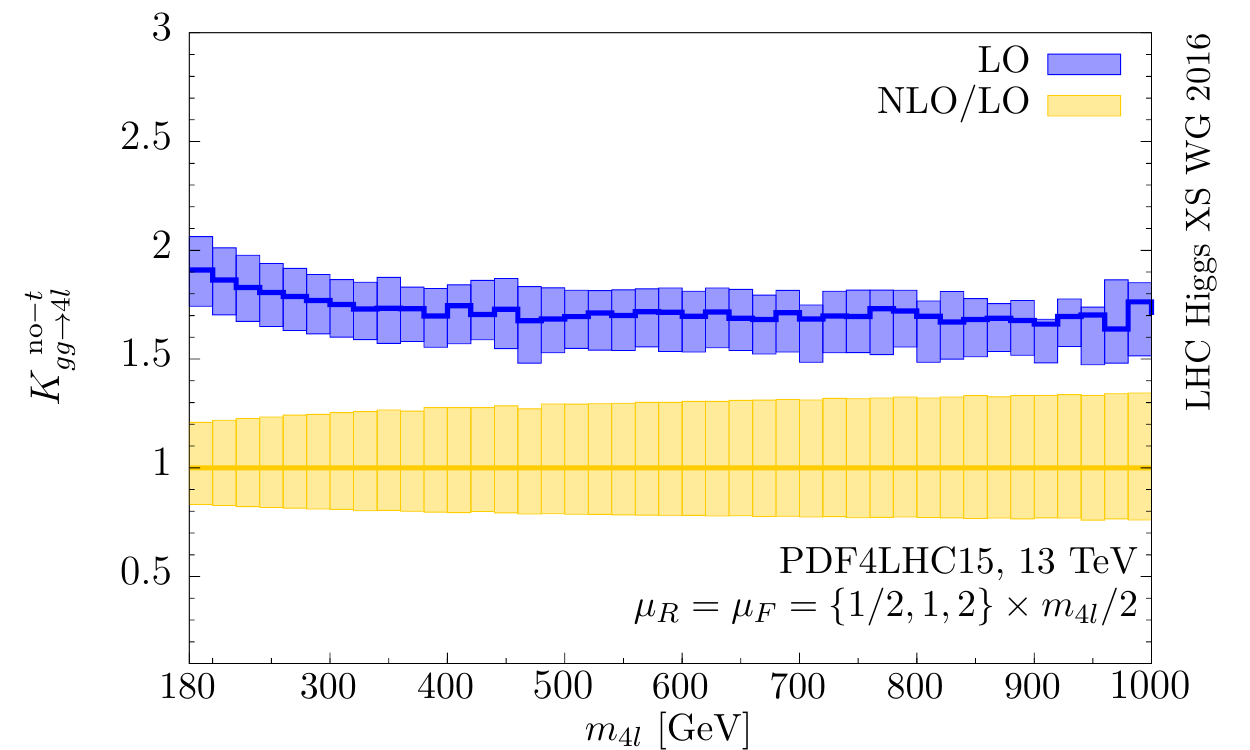}
\includegraphics[width=0.45\textwidth]{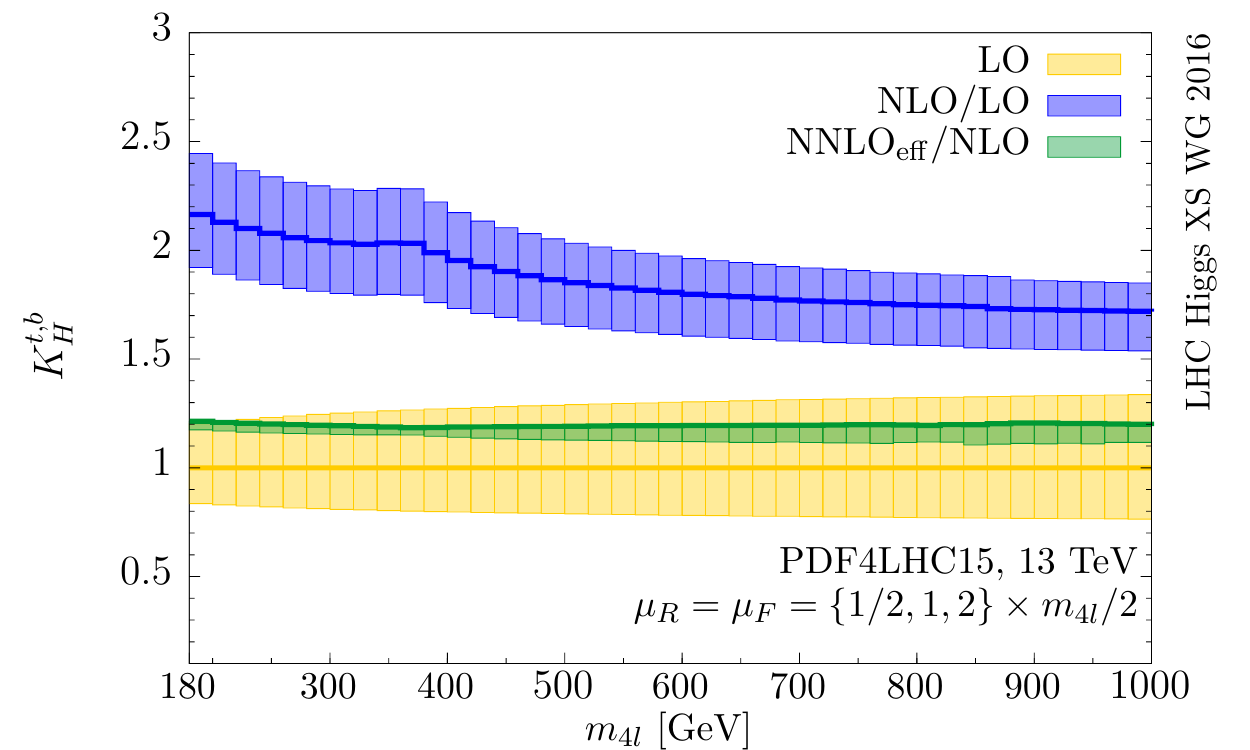}
\caption{Left: NLO $K$-factor for $gg\to 4l$ background, massless quark contribution.
Right: $K$-factor for $gg\to H \to 4l$ signal. NLO with full mass dependence, NNLO in the HEFT
approximation. See text for details.}\label{offshell_interf_ggVV_NLO:kfactor}
\end{figure}
This figure shows that signal and background $K$-factors are
similar throughout the whole invariant mass spectrum considered here.

To quantify the effect of the missing top quark contribution in the
background, we study two extreme approaches. First, we assume that the
$K-$factor for massive and massless contributions is identical. Given
their similarity in the low-mass region, we believe this assumptions
to be reasonable. This leads to the $K$-factor shown in
\refF{offshell_interf_ggVV_NLO:kfactor} (see also
Eq.~\ref{offshell_interf_ggVV_NLO:rtilde})
\begin{equation}
K_{gg\to 4l} = \frac{{\rm d}\sigma^{\rm LO}_{t,b}/{\rm d}m_{4l}+\mathcal{R}^{\rm LO}_{t,b}{\rm d}\Delta\sigma^{\rm NLO}_{{\rm no}-t}/{\rm d}m_{4l}}{{\rm d}\sigma^{\rm LO}_{t,b}/{\rm d}m_{4l}} = \frac{{\rm d}\sigma^{\rm LO}_{{\rm no}-t}/{\rm d}m_{4l}+{\rm d}\Delta\sigma^{\rm NLO}_{{\rm no}-t}/{\rm d}m_{4l}}{{\rm d}\sigma^{\rm LO}_{{\rm no}-t}/{\rm d}m_{4l}} = K_{gg\to 4l}^{{\rm no}-t}.\label{offshell_interf_ggVV_NLO:kfactor_k}
\end{equation}
Second, we use full mass dependence in the LO contribution and only
add NLO corrections for massless quarks\footnote{Note that this second
  approach is rather unrealistic, as it assumes no interference
  between LO massive amplitudes and NLO massless ones. We consider it here
  only as a way to estimate possible top quark effects in a conservative way.}

\begin{equation}
\tilde K_{gg\to 4l} = \frac{{\rm d}\sigma^{\rm LO}_{t,b}/{\rm d}m_{4l}+{\rm d}\Delta\sigma^{\rm NLO}_{{\rm no}-t}/{\rm d}m_{4l}}{{\rm d}\sigma^{\rm LO}_{t,b}/{\rm d}m_{4l}}.\label{offshell_interf_ggVV_NLO:kfactor_kt}
\end{equation}
A comparison between $K$ Eq.~\ref{offshell_interf_ggVV_NLO:kfactor_k}
and $\tilde K$ Eq.~\ref{offshell_interf_ggVV_NLO:kfactor_kt} is shown
in \refF{offshell_interf_ggVV_NLO:comparison}. Up to invariant masses
$m_{4l}\sim500~$GeV the two results are in good agreement, while they differ
significantly at higher mass. The spread of these two results is a way
to probe the uncertainty due to unknown mass effects.

\begin{figure}
\centering
\includegraphics[width=0.45\textwidth]{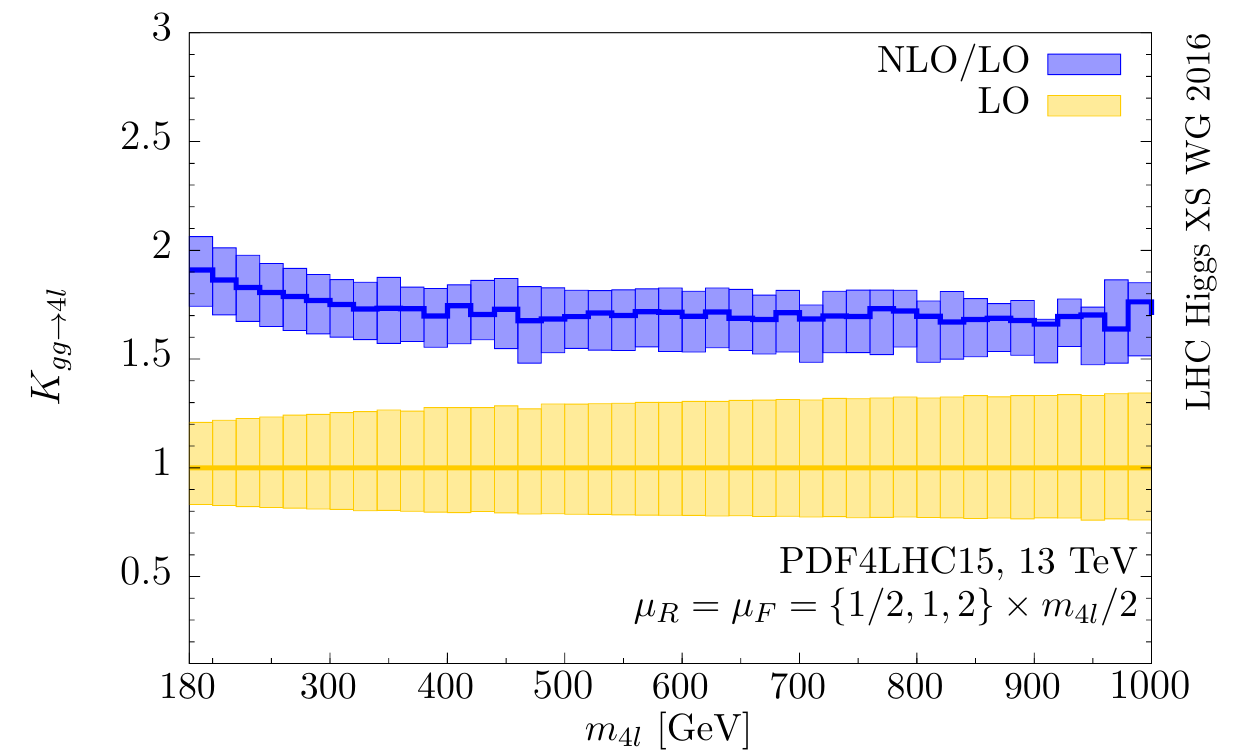}
\includegraphics[width=0.45\textwidth]{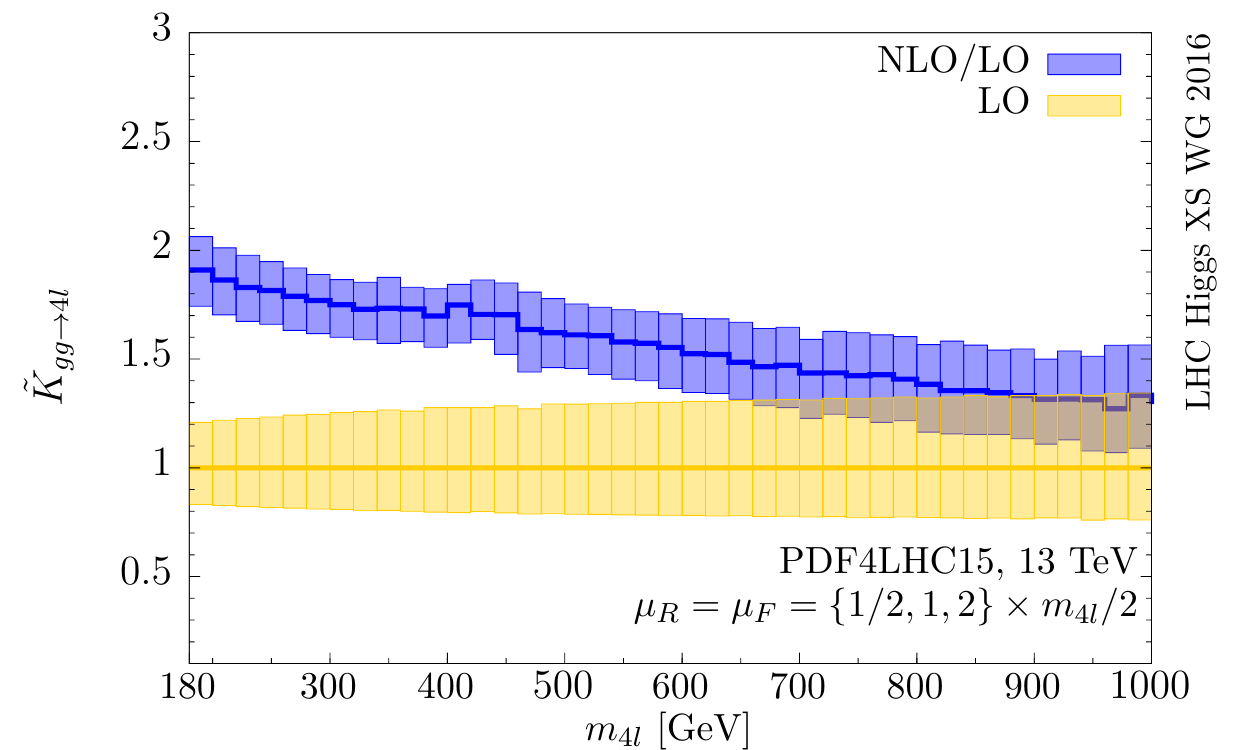}
\caption{Comparison of different ways of treating quark mass effects
  at higher orders. Left: assume identical correction to massive and
  massless contributions.  Right: assume zero corrections for massive
  contributions. See text for
  details.}\label{offshell_interf_ggVV_NLO:comparison}
\end{figure}

Summarizing, for background predictions in the high invariant mass
region we suggest to use exact LO multiplied by the massless
$K$-factor Eq.~\ref{offshell_interf_ggVV_NLO:kfactor_k}. The spread
shown in \refF{offshell_interf_ggVV_NLO:comparison} may be used as
a way to estimate the uncertainty of this procedure until a better
computation becomes available. As for low invariant mass region, the
interference $K$-factor is then determined as geometric mean of signal
and background $K$-factors.
Alternatively, given the similarity of signal and background
$K-$factors and the size of uncertainties a simpler alternative -- until
more precise theoretical predictions are available -- would be to
assume the same $K$-factor for signal and background, and assign to it
a systematic uncertainty which covers the effects described
above. Note that both these approaches lead to a smooth interference
$K-$factor over the whole $m_{4l}$ spectrum, with an uncertainty
increasing at large invariant masses to reflect the effect of unknown
top quark mass effects.
While this report was finalized, \Brefs{Campbell:2016ivq,Caola:2016trd} appeared.  The results for the NLO corrections to the signal-background interference presented there support the approach advocated in this section. 




\newcommand\f[2]{\frac{#1}{#2}}
\def\beq{\begin{equation}}
\def\eeq{\end{equation}}
\def\to{\rightarrow}
\def\nn{\nonumber}
\def\beeq{\begin{eqnarray}}
\def\eeeq{\end{eqnarray}}
\def\sh{\hat{s}}
\def\ca{{\cal A}}
\def\eqn#1{eq.~(\ref{#1})}
\def\Eqn#1{Equation~(\ref{#1})}
\def\eqns#1#2{eqs.~(\ref{#1}) and~(\ref{#2})}
\def\Eqns#1#2{Eqs.~(\ref{#1}) and~(\ref{#2})}

\section{\texorpdfstring{$H \to \gamma \gamma$}{H to gamma gamma} mode}
\label{sec:interf_2gamma}

In this section we will review the status of the theoretical and experimental
treatments of the interference term between the $gg\to H\to\gamma\gamma$ and $gg\to\gamma\gamma$.

The natural width of the Higgs boson is an important physics property that could reveal new physics in
case of disagreement between the prediction and the measured values.
Direct measurements of the Higgs boson widths are not possible, as the experimental mass resolution is significantly larger than the expected width.
The mass resolution of the $\gamma\gamma$ system is about $1.7$ GeV for $m_{\gamma\gamma} = 125$ GeV,
400 times larger than the natural width.
Measurements of coupling strengths paired with limits on the invisible branching fraction indirectly constrain the width to close to its SM value~\cite{Barger:2012hv},
but this strategy cannot take into account unobserved (but not truly invisible) decay modes.

A new method as introduced by Dixon, Li, and Martin~\cite{Dixon:2013haa,Martin:2012xc},
allows to extract an indirect limit on the Higgs boson width using the interference of the $H \to \gamma \gamma$ signal
with respect to the continuum diphoton background ($gg \to \gamma \gamma$ box diagrams). This interference has two parts.
\begin{itemize}
 \item[1.] An imaginary component reduces the total signal yield by $2-3\%$. Because this effect is degenerate with the coupling (signal strength) measurements,
it is only measurable using constraints on the production rates from other channels.
 \item[2.] The real component is odd around the Higgs boson mass and does not change the yield. However, when folded with the experimental resolution,
it engenders a negative shift in the apparent mass.
\end{itemize}
In the SM, this shift was originally estimated using a simplified resolution model to be approximately 80 MeV~\cite{Dixon:2013haa},
and for a width 20 times larger than the SM value, the shift was estimated to approximately 400 MeV.


In this section, we will review the latest developments on theoretical calculations, available
MC tools, as well as experimental analyses from ATLAS and CMS collaborations.

\subsection{Theory overview}
\label{sec:HggIntTheory}

The Higgs boson is dominantly produced by gluon fusion through a top quark loop.
Its decay to two photons, $H\to\gamma\gamma$, provides a very clean signature for probing Higgs boson properties, including its mass.
However, there is also a large continuum background to its detection in this channel.
It is important to study how much the coherent interference between the Higgs boson signal and the background could affect distributions in
diphoton observables, and possibly use it to constrain Higgs boson properties.

The interference of the resonant process $ij \to X+H ( \to \gamma \gamma )$ with the continuum QCD background $ij \to X+\gamma\gamma $
induced by quark loops can be expressed at the level of the partonic cross section as:

\begin{eqnarray}
\delta\hat{\sigma}_{ij\to X+ H\to \gamma\gamma} &=& 
-2 (\sh-m_H^2) \,\, { \text{Re} \left( \ca_{ij\to X+H} \ca_{H\to\gamma\gamma} 
                          \ca_{\rm cont}^* \right) 
        \over (\sh - m_H^2)^2 + m_H^2 \Gamma_H^2 }
\nonumber\\
&& 
-2 m_H \Gamma_H \,\, { \text{Im} \left( \ca_{ij\to X+H} \ca_{H\to\gamma\gamma} 
                          \ca_{\rm cont}^* \right)
        \over (\sh - m_H^2)^2 + m_H^2 \Gamma_H^2 } \,,
\label{intpartonic}
\end{eqnarray}
where $m_H$ and $\Gamma_H$ are the Higgs boson mass and decay width, and  $\sh$ is the partonic invariant mass.
The interference is written in two parts, proportional to the real and imaginary parts of the Higgs Breit-Wigner propagator respectively,
to which will be referred to as the real and imaginary part of the interference from now on.

The real part interference is odd in $\sh$ around the Higgs boson mass peak, and thus its effect on the total $\gamma\gamma$ rate is subdominant as pointed out in ref.~\cite{Dixon:2003yb,Dicus:1987fk}.
The imaginary part of the interference, depending on the phase difference between the signal and background amplitudes, could significantly affect the total cross section.
However, for the gluon-gluon partonic subprocess, it was found that the loop-induced background continuum amplitude has a quark mass suppression in its imaginary part for the relevant helicity combinations,
making it dominantly real, therefore bearing the same phase as the Higgs boson production and decay amplitudes \cite{Dicus:1987fk}.
As a result, the contribution of the interference to the total cross section in the gluon fusion channel is highly suppressed at leading order (LO).
The main contribution of the interference to the total rate comes from the two-loop imaginary part of the continuum amplitude $gg \to \gamma\gamma$, and only amounts to around $3\%$ of the total signal rate~\cite{Dixon:2003yb}.

Later, in ref. \cite{Martin:2012xc} it was shown that even though the real part of the interference hardly contributes to the total cross section,
it has a quantifiable effect on the position of the diphoton invariant mass peak, producing a shift of $ {\cal O}(100\,\text{MeV})$ towards a lower mass region,
once the smearing effect of the detector was taken into account.
In ref. \cite{deFlorian:2013psa}, the $qg$ and $q\bar{q}$ channels of this process were studied, completing the full ${\cal O}(\alpha_{\mathrm{S}}^2)$
computation of the interference effects between the Higgs diphoton signal and the continuum background at the LHC.
Note that the extra $qg$ and $q\bar{q}$ channels involve one QCD emission in the final states, but the corresponding background amplitudes start at tree level,
and therefore the relevant interference is of the same order as the LO $gg$ channel in which the background amplitude is induced by a quark loop.
The extra LO $qg$ interference is depicted by the top right diagram in \refF{FeynDiag}, and the $q\bar{q}$ channel is related by cross symmetry.
It was found that the contribution from the $q\bar{q}$ channel is numerically negligible due to the quark PDF suppression.

\begin{figure}
\includegraphics[width=0.8\textwidth]{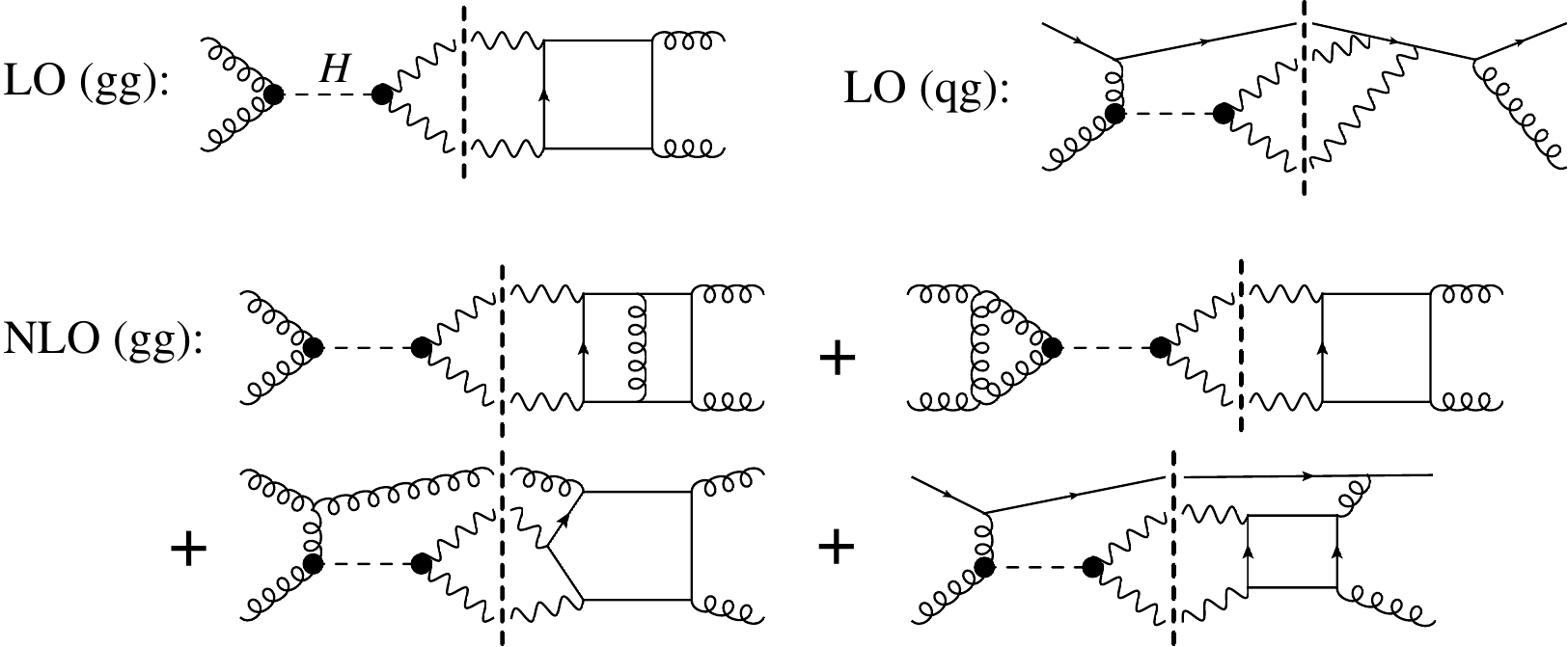}
\caption{\label{FeynDiag}
  Representative diagrams for interference between the
Higgs boson resonance and the continuum in the diphoton channel.
The dashed vertical lines separate the resonant amplitudes
from the continuum ones.}
\end{figure}

More recently, the dominant next-to-leading order (NLO) QCD corrections to the interference were calculated in ref. \cite{Dixon:2013haa}, where the dependence of the mass shift on the acceptance cuts was also studied. The left panel of \refF{Mdistr} shows the Gaussian-smeared diphoton invariant mass distribution for the pure signal at both LO and NLO in QCD. Standard acceptance cuts were applied to the photon transverse momenta, $p_{T,\gamma}^{\textrm{hard/soft}}>40/30$~GeV, and rapidities, $|\eta_{\gamma}|<2.5$. In addition, events were discarded when a jet with $p_{T,j}>3$~GeV was within $\Delta R_{\gamma j}<0.4$ of a photon.
The scale uncertainty bands were obtained by varying $m_H/2<\mu_F,\mu_R<2m_H$ independently. For NLO, an additional $qg$ process was included, where the background is induced by a quark loop as shown in the bottom right diagram of \refF{FeynDiag}; this is required as part of NLO $gg$ channel to cancel the quark to gluon splitting in PDF evolution and reduces dependence on the factorization scale $\mu_F$. As a result, the scale uncertainty bands come mostly from varying the renormalization scale $\mu_R$. 

The right panel of \refF{Mdistr} shows the corresponding Gaussian-smeared interference contributions. Each band is labelled according to \refF{FeynDiag}. The destructive interference from the imaginary part shows up at two-loop order in the gluon channel in the zero mass limit of light quarks~\cite{Dixon:2003yb}. It produces the offset of the NLO $gg$ curve from zero at $M_{\gamma\gamma} = 125$~GeV.

\begin{figure}
  \begin{center}
	\begin{tabular}{c c}
	\includegraphics[width=0.45\textwidth]{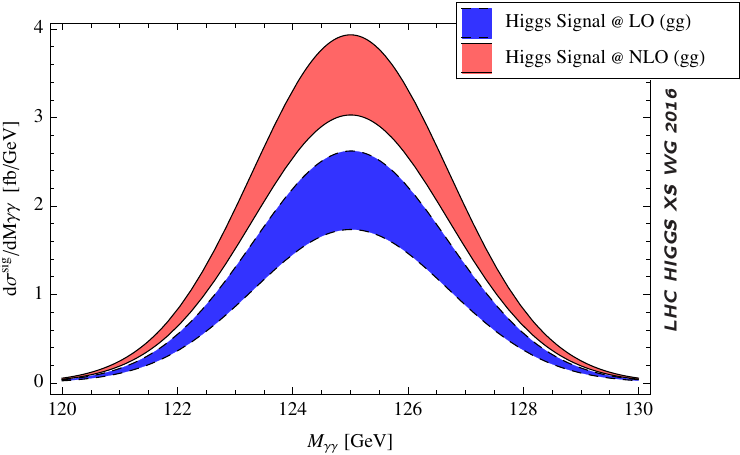} &
    \includegraphics[width=0.45\textwidth]{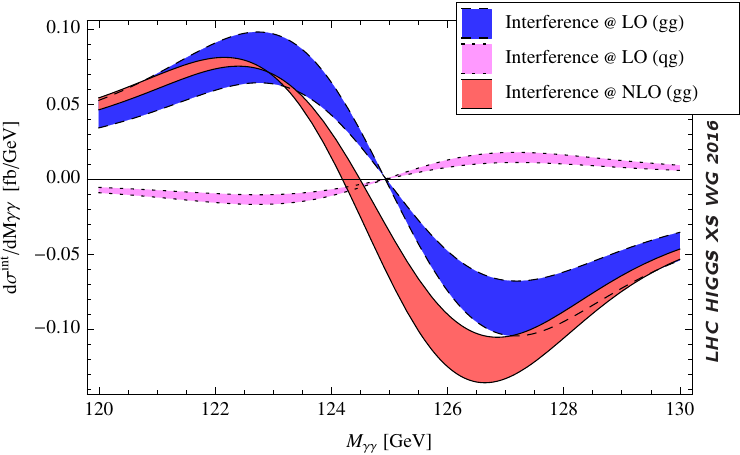}
\end{tabular}	   
    \end{center}
\vspace{-0.5cm}
  \caption{\label{Mdistr}
   Diphoton invariant mass $M_{\gamma\gamma}$ distribution for pure signal 
   (left panel) and interference term (right panel) after Gaussian smearing.}
\end{figure}

Figure \ref{dMvspTh} shows the study of the mass shift dependence on a lower cut on the Higgs boson transverse momentum $p_T > p_{T,H}$. This strong dependence could potentially be observed experimentally, completely within the 
$\gamma\gamma$ channel, without having to compare against a mass measurement using the only other high-precision channel, $ZZ^*$\footnote{The mass shift for $ZZ^*$ is much smaller than for $\gamma\gamma$, as can be inferred from Figure~17 of ref. \cite{Kauer:2012hd},
because $H \to ZZ^*$ is a tree-level decay, while the continuum background $gg \to ZZ^*$ arises at one loop, the same order
as $gg\to \gamma\gamma$.}.
Using only $\gamma\gamma$ events might lead to reduced experimental systematics associated with the absolute photon energy scale.
The $p_{T,H}$ dependence of the mass shift was first studied in ref.~\cite{Martin:2013ula}. The dotted red band includes, in addition,
the continuum process $qg\to\gamma\gamma q$ at one loop via a light quark loop, a part of the full ${\cal O}(\alpha_s^3)$ correction as explained above.
This new contribution partially cancels against the tree-level $qg$ channel, leading to a larger negative Higgs boson mass shift. The scale variation of the mass shift at finite $p_{T,H}$ is very small, because it is essentially a LO analysis; the scale variation largely cancels in the ratio between interference and signal that enters the mass shift.

Due to large logarithms, the small $p_{T,H}$ portion of \refF{dMvspTh} is less reliable than the large $p_{T,H}$ portion. In using the $p_{T,H}$ dependence of the mass shift to constrain the Higgs boson width, the theoretical accuracy will benefit from using a wide first bin in $p_T$. One could take the difference between apparent Higgs boson masses for $\gamma\gamma$ events in two bins, those 
having $p_T$ above and below, say, 40~GeV.

\begin{figure}
  \begin{center}
    \includegraphics[width=0.5\textwidth]{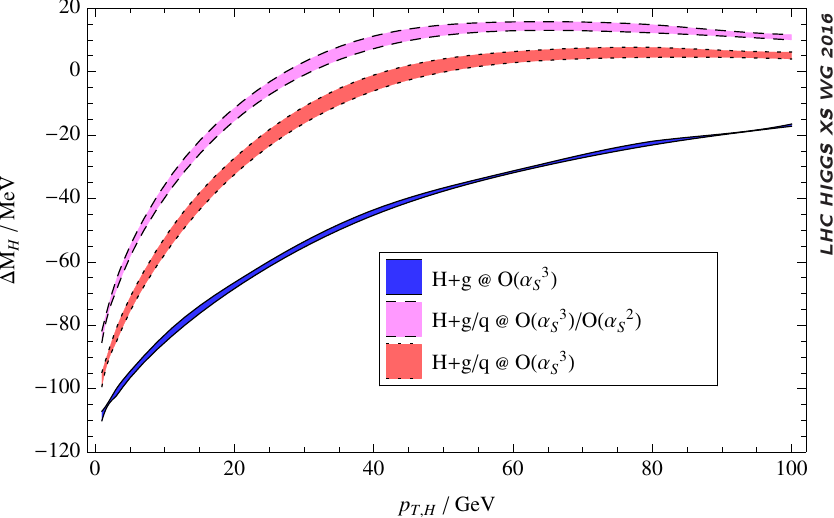}
  \end{center}
  \vspace{-0.5cm}
  \caption{\label{dMvspTh}
   Apparent mass shift for the SM Higgs boson versus the lower cut
   on the Higgs boson transverse momentum, $p_T > p_{T,H}$.}
\end{figure}

The Higgs boson width in the SM is $\Gamma_{H,\text{SM}} = 4.07$ MeV, far too narrow to observe directly at the LHC. In global analyses of various Higgs boson decay channels \cite{Dobrescu:2012td,Djouadi:2013qya,CMS:yva}, it is impossible to decouple the Higgs boson width from the couplings in experimental measurements without a further assumption, because the Higgs boson signal strength is always given by the product of squared couplings for Higgs boson production and for decay, divided by the Higgs boson total width $\Gamma_H$. Typically, the further assumption is that the Higgs boson coupling to electroweak vector bosons does not exceed the SM value. However, as was also pointed out in ref.~\cite{Dixon:2013haa}, the apparent mass shift could be used to bound the value of the Higgs boson width. This is because the interference effect has different dependence on the Higgs boson width, allowing $\Gamma_H$ to be constrained independently of assumptions about couplings or new decay modes in a lineshape model. Such a measurement would complement more direct measurements of the Higgs boson width at future colliders such as the ILC \cite{Richard:2007ru,Peskin:2012we} or a muon collider \cite{Han:2012rb,Conway:2013lca}, but could be accomplished much earlier.

Using $\mu_{\gamma\gamma}$ to denote the ratio of the experimental signal strength in $gg\to H\to \gamma\gamma$ to the SM prediction
($\sigma/\sigma^{\rm SM}$), the following equation can be set up,
\begin{equation}
\frac{c_{g\gamma}^2 S}{m_H\Gamma_H}+c_{g\gamma} I 
= \left( \frac{S}{m_H\Gamma_{H, \text{SM}}} + I \right)
 \mu_{\gamma\gamma} \,,
\label{constyield}
\end{equation}
where $c_{g\gamma} = c_g c_\gamma$  is the rescaling factor to be solved to preserve the signal yield when the Higgs boson width is varied. Once the relation between the $c_{g\gamma}$ and the Higgs boson width $\Gamma_H$ is obtained, it can be used to determine the size of the apparent mass shift as a function of $\Gamma_H$. Neglecting the interference contribution $I$ to the total rate, and assuming $\mu_{\gamma\gamma}=1$, the mass shift was found to be proportional to the square root of the Higgs boson width, $\delta m_H \propto \sqrt{\Gamma_H/\Gamma_{H,\text{SM}}}$, given that the width is much less than the detector resolution. \refF{dMvsGam} plots the mass shift with $\mu_{\gamma\gamma}=1$ and a smearing Gaussian width of $1.7$ GeV. It is indeed proportional to $\sqrt{\Gamma_H}$ up to small corrections. If new physics somehow reverses the sign of the Higgs diphoton amplitude, the interference $I$ would be constructive and the mass shift would become positive.

\begin{figure}
  \begin{center}
    \includegraphics[width=0.5\textwidth]{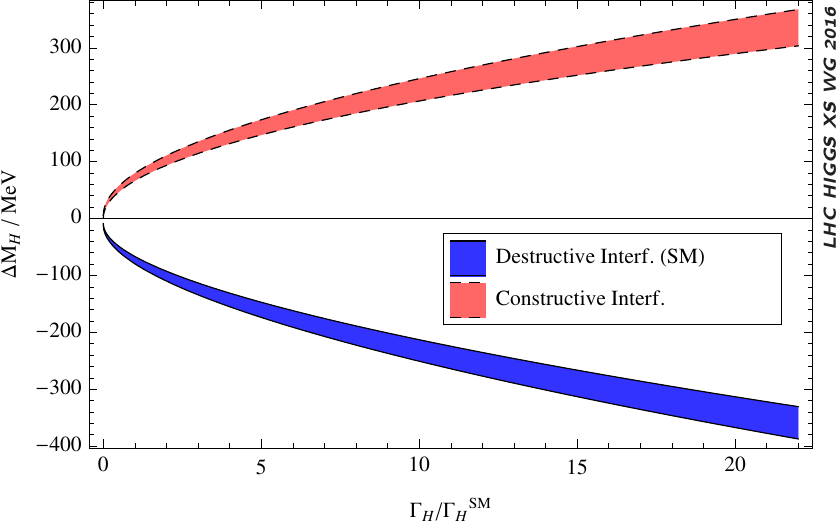}
  \end{center}
  \vspace{-0.5cm}
  \caption{\label{dMvsGam}
   Higgs boson mass shift as a function of the Higgs boson width. The coupling 
   $c_{g\gamma}$ has been adjusted to maintain a constant signal strength,
   in this case $\mu_{\gamma\gamma}=1$.}
\end{figure}


In ref. \cite{Coradeschi:2015tna} it was proposed to use another $\gamma\gamma$ sample to determine the Higgs boson resonance peak, in which the two photons were produced in association with two jets. Although this process is relatively rare, so is the background, making it possible to obtain reasonable statistical uncertainties on the position of the mass peak in this channel despite the lower number of events. The production of a Higgs boson in association with two jets is characteristic of the Vector Boson Fusion (VBF) production mechanism. While, in general terms, VBF is subdominant with respect to GF, it has a very different kinematical signature and can be selected through an appropriate choice of the experimental cuts. From a theoretical point of view, the VBF production mechanism has the additional advantage that perturbative corrections are much smaller than for GF (see e.g.~ref.~\cite{Bolzoni:2010xr}). The effect of the signal-background interference for both the GF and VBF production mechanisms were studied, and the relevant diagrams are given in \refF{VBFDiag}. There are two kinds of backgrounds amplitudes, each of QCD and EW origin. It turns out that the interferences between GF signal and EW background or VBF signal and QCD background are highly suppressed by QCD colour factors, and therefore only the remaining combinations are shown in the first two diagrams of \refF{VBFDiag}. In addition, the interference with loop-induced QCD background, as given in the third diagram of \refF{VBFDiag}, was also considered, since it is enhanced by large gluonic luminosity at the LHC.

\begin{figure}
  \begin{center}
    \includegraphics[width=0.5\textwidth]{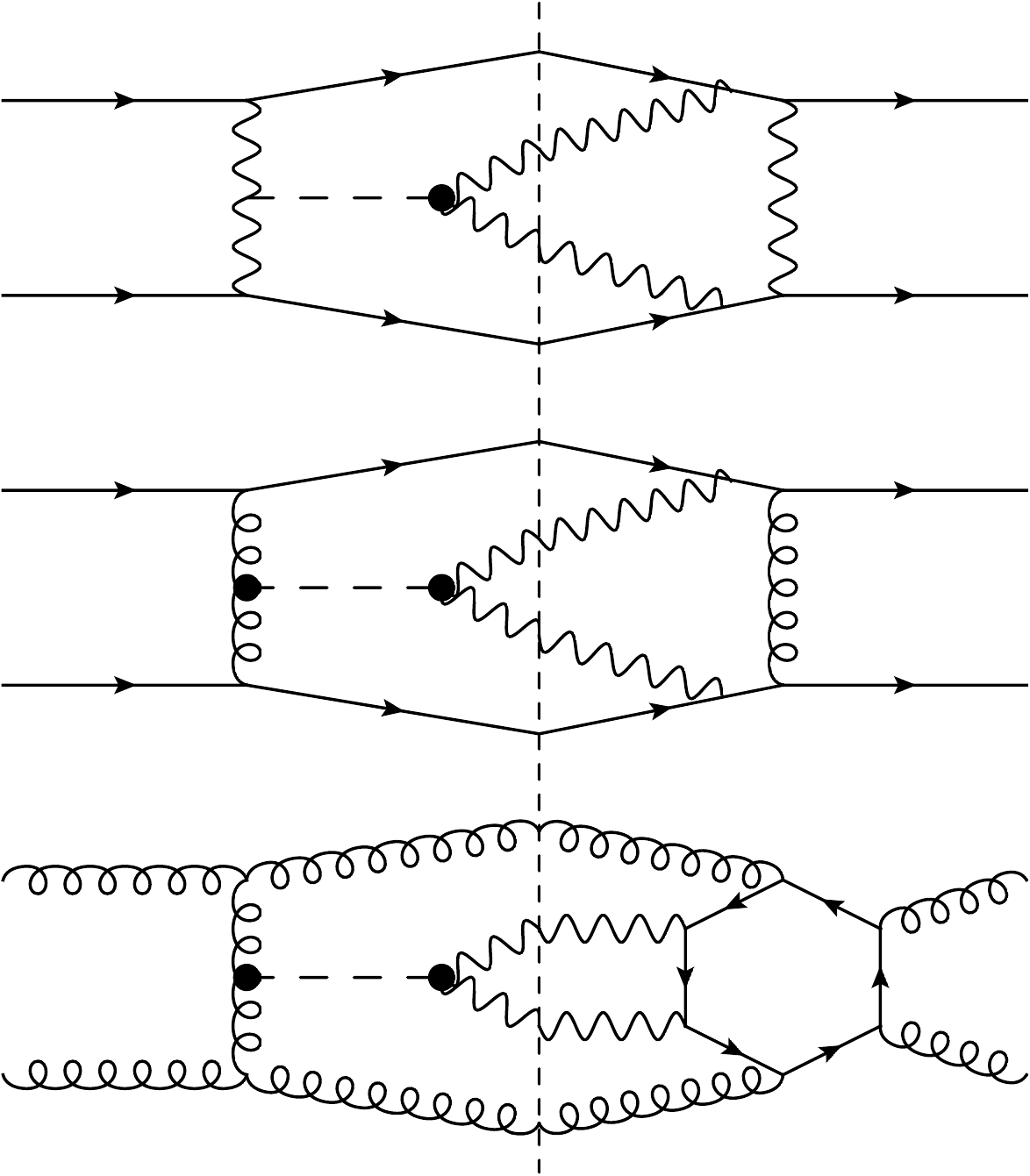}
  \end{center}
  \vspace{-0.5cm}
  \caption{\label{VBFDiag}
   Examples of the Feynman diagrams computed for
the calculation. The vertical dotted line separates signal from
background. Above, the VBF signal and EW background con\-tributions; in the middle the GF signal with tree level QCD
mediated background; below, gluon-initiated signal, with the
corresponding loop-induced LO background.}
\end{figure}

In \refF{fig:eta} the values of the apparent mass shift $\delta m_H$ obtained for different cuts on the difference in pseudorapidities between the jets $|\Delta\eta_{jj}|$ are shown. The contributions from VBF and GF are presented separately, as well as the total shift. At the bottom of the plot, the total integrated signal is shown, also separated into VBF and GF contributions for the same cuts. For this plot no cut in $p_{T,H}$ was applied, and only events with the invariant mass of the dijet system $M_{jj}>400$~GeV were considered. When no cut in $|\Delta\eta_{jj}|$ is applied, the shift in the Higgs boson invariant mass peak position produced by these two main production mechanisms is of the same magnitude, but of opposite sign; hence one observes a partial cancellation between them, with a net shift of around $-6\text{ MeV}$. As the value of $|\Delta\eta_{jj}|_{\text{min}}$ is increased, VBF becomes the dominant contribution, and GF becomes negligible, leading to a shift of around 20~MeV toward lower masses.

\begin{figure}
\begin{center}
\hskip-0.4cm \includegraphics[width=0.48\textwidth]{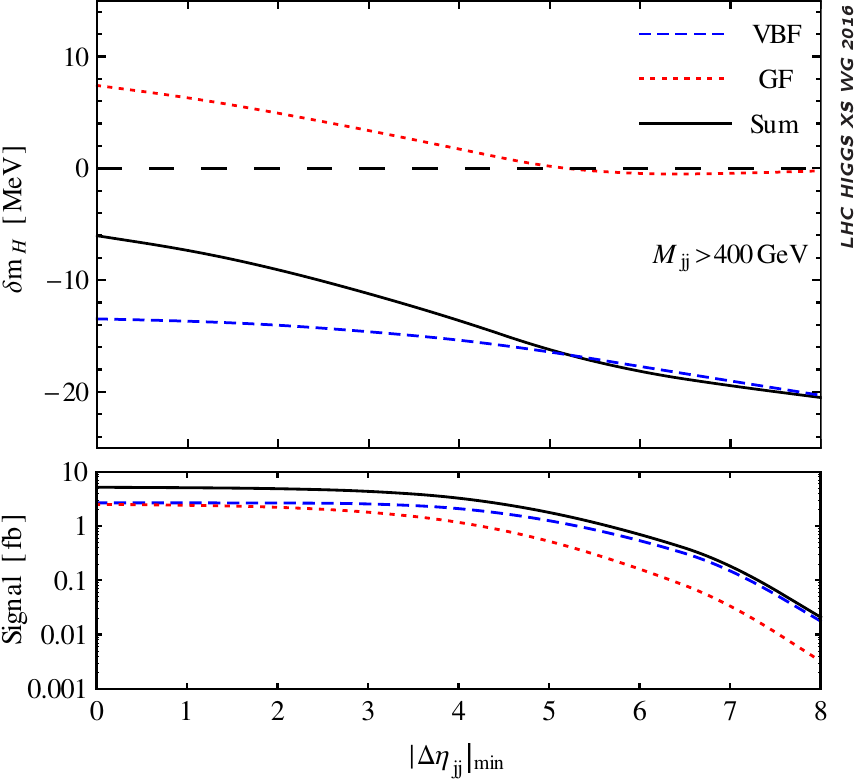}
\caption{Top: Plot of mass shift $\delta m_H$ for different values of $|\Delta \eta_{jj}|_{\text{min}}$. The dashed blue line represents the contribution from the VBF mechanism alone, the dotted red line shows GF only, and the solid black line displays the total shift of the Higgs boson invariant mass peak. Bottom: Total integrated signal cross section, also separated into VBF and GF contributions for the same cuts. No cut on $p_{T,H}^{\text{min}}$ was applied, and an additional cut was set of $M_{jj}>400$GeV.}
\label{fig:eta}
\end{center}
\end{figure}

Next, the dependence of the mass shift on $p_{T,H}^{\text{min}}$ was studied. In \refF{fig:ptH_theory} the mass shift and the signal cross section for a range of $p_{T,H}^{\text{min}}$ between 0~GeV and 160~GeV is presented. The curves are labelled in the same way as in \refF{fig:eta}. Once again, both production mechanisms contribute to the shift in invariant mass with opposite signs. For this plot, additional cuts in $M_{jj}>400$~GeV and $|\Delta\eta_{jj}|>2.8$ were applied, enhancing in this way the VBF contributions. However, at higher $p_{T,H}^{\text{min}}$, GF becomes as important as VBF.

\begin{figure}
\begin{center}
\hskip-0.4cm \includegraphics[width=0.48\textwidth]{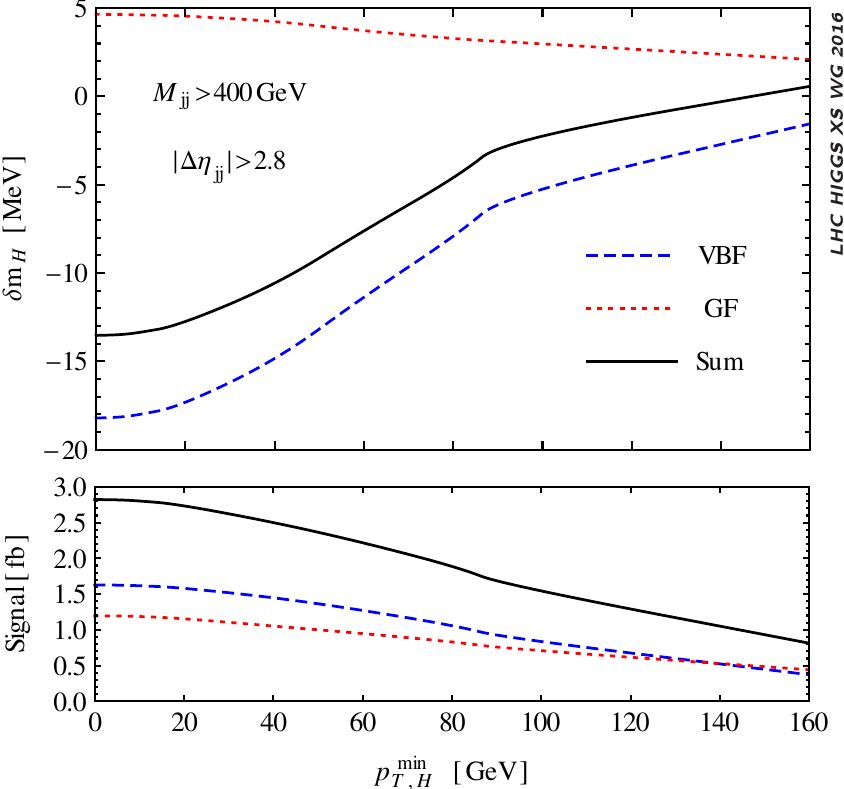}
\caption{Top: Plot of mass shift $\delta m_H$ for different values of $p_{T,H}^{\text{min}}$ for VBF, GF and total contributions.
The curves are labelled as in \refF{fig:eta}. Bottom:  Total integrated signal, also separated into VBF and GF contributions for the same cuts. The following additional cuts were applied: $M_{jj}>400$~GeV and $|\Delta\eta_{jj}|>2.8$.}
\label{fig:ptH_theory}
\end{center}
\end{figure} 

As has already been mentioned, the shift in the Higgs boson invariant mass peak in $pp\to H(\to\gamma\gamma)+2\text{ jets}\,+\,X$ is considerably smaller than in the inclusive channel $pp\to H(\to\gamma\gamma)\,+\,X$. For appropriate cuts it can be almost zero. This makes it useful as a reference mass for experimental measurement of the mass difference,

\begin{equation}
\Delta m_H^{\gamma\gamma} \equiv \delta m_H^{\gamma\gamma,\,\text{incl}} - \delta m_H^{\gamma\gamma,\,\text{VBF}} \,,
\label{BigDeltaDef}
\end{equation}

where $\delta m_H^{\gamma\gamma,\,\text{incl}}$ is the mass shift in the inclusive channel, as computed at NLO in ref. \cite{Dixon:2013haa},
and $\delta m_H^{\gamma\gamma,\,\text{VBF}}$ is the quantity computed in ref. \cite{Coradeschi:2015tna}.
In computing $\delta m_H^{\gamma\gamma,\,\text{VBF}}$ for use in \eqn{BigDeltaDef} the basic photon and jet $p_T$ and $\eta$ cuts were imposed, and also $M_{jj} > 400$~GeV,
but no additional cuts on $p_{T,H}$ or $\Delta\eta_{jj}$ were applied. This choice of cuts results in a small reference mass shift and a relatively large rate with which to measure it.

The lineshape model of ref.~\cite{Dixon:2013haa}, as introduced earlier for the $gg \to \gamma\gamma$ inclusive process, was used in ref. \cite{Coradeschi:2015tna} to
compute the mass shift for the VBF process.
It is in a way relatively independent of the new physics that may increase $\Gamma_H$ from the SM value.
The couplings of the Higgs boson to other SM particles must be modified if the Higgs boson width is varied,
in order to be consistent with the Higgs boson signal strength measurements already made by the LHC,
and prevent the total cross section from suffering large variations.
Here, the deviation from SM coupling is described by a rescaling factor $c_{V\gamma} = c_V c_\gamma$,
similar to $c_{g\gamma}$ in the $\gamma\gamma$ inclusive case, which is adjusted for different values of $\Gamma_H$ to maintain the Higgs boson signal strength near the SM value.

\begin{figure}
\begin{center}
\hskip-0.4cm \includegraphics[width=0.5\textwidth]{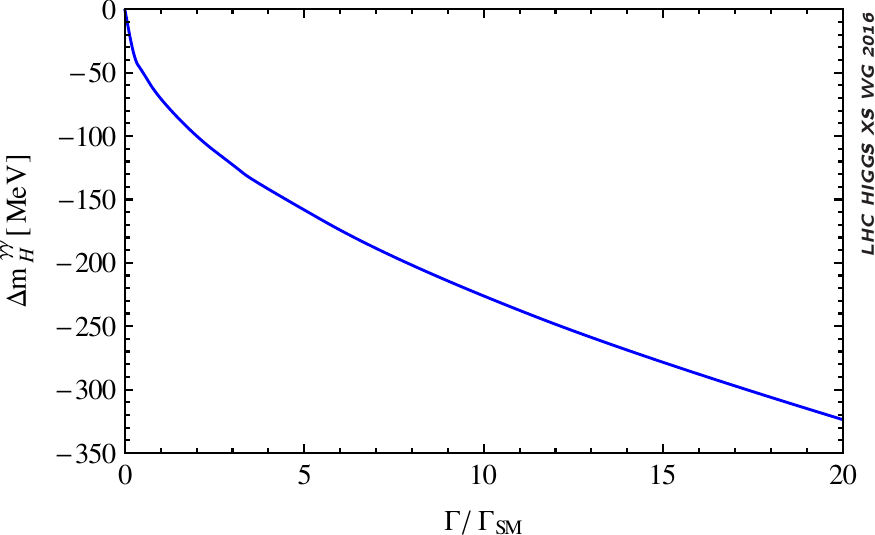}
\caption{Plot of measurable mass shift $\Delta m_H^{\gamma\gamma}$ defined in \eqn{BigDeltaDef}, as a function of $\Gamma_H / \Gamma_{H,\text{SM}}$.}
\label{fig:widths}
\end{center}
\end{figure} 
%
Figure~\ref{fig:widths} shows how the observable $\Delta m_H^{\gamma\gamma}$ depends on the value of the Higgs boson width.
The dependence is proportional to $\sqrt{\Gamma_H/\Gamma_{H,\text{SM}}}$ to a very good accuracy,
as dictated by the linearity of the produced shift in $c_{g\gamma}$ or $c_{V\gamma}$ (in the range shown).
It is dominated by the mass shift for the inclusive sample \cite{Dixon:2013haa}. 
As was stated before, the main theoretical assumption was that the couplings of the Higgs rescale by real factors,
and the same rescaling for the Higgs boson coupling to gluons as for its coupling to vector boson pairs was assumed;
this assumption could easily be relaxed, to the degree allowed by current measurements of the relative yields in different channels.
The strong dependence the shift shows on the Higgs boson width might allow LHC experiments to measure or bound the width.

\subsection{Monte Carlo interference implementations}
\label{sec:intef_2gamma_mc}

An overview of the Monte Carlo tools available to describe the 
Higgs lineshape and the 
signal-background interference is presented in this section.
A first study using these tools is also presented.

\subsubsection{Available Tools: Sherpa 2.2.0 with DIRE parton shower}
\label{sec:dire-sub}

The calculations of \cite{Dixon:2013haa,Coradeschi:2015tna} have been implemented in Sherpa 2.2.0.  Parton showers have been used for more than three decades 
to predict the dynamics of multi-particle final states in collider
experiments~\cite{Webber:1986mc,Buckley:2011ms}. Recently, a new model was 
proposed~\cite{Hoche:2015sya}, which combines the careful treatment of 
collinear configurations in parton showers with the correct resummation 
of soft logarithms in colour dipole cascades~\cite{Gustafson:1986db,Gustafson:1987rq,Lonnblad:1992tz,Kharraziha:1997dn}.
Following the basic ideas of the 
dipole formalism, the ordering variable is chosen as the transverse momentum 
in the soft limit. The evolution equations are based on the parton picture.
Colour-coherence is implemented by partial fractioning the soft eikonal 
following the approach in~\cite{Catani:1996vz}, and matching each term 
to the double logarithmically enhanced part of the DGLAP splitting functions.
Enforcing the correct collinear anomalous dimensions then determines all 
splitting kernels to leading order.

\subsubsection{Exercise with DIRE parton shower}
This sensitivity study follows the basic search strategy exploited in the past by
both the CMS and ATLAS experiments for what concerns the $H\to \gamma\gamma$ search~\cite{Khachatryan:2014ira,Aad:2012tfa}.
The study is performed only at generator level assuming only gluon fusion production mode (GGH).
The parton shower model assumed is the one described in Section~\ref{sec:dire-sub}.
Two isolated photons fulfiling loose identification criteria are selected
and required to be within the the detector acceptance of $|\eta|<2.5$
and the leading (subleading) photon must have $p_{T1}>40$ GeV and $p_{T2}>30$ GeV.
The diphoton invariant mass distribution is constructed from these photons and required to be in the $[110-150]$ GeV energy range.
Figures~\ref{fig:pt1H} and~\ref{fig:pt2H} show the transverse momentum distributions obtained for the two photons after the selection.

\begin{figure}
  \begin{center}
	\begin{tabular}{c c}
	\includegraphics[width=0.65\textwidth]{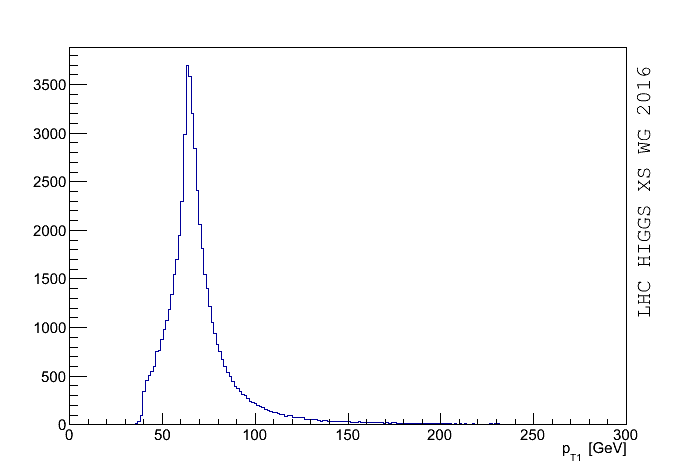} &
   \end{tabular}	   
    \end{center}
\vspace{-0.5cm}
  \caption{\label{fig:pt1H}
   Transverse momentum distribution of the leading photon of the $H\to \gamma\gamma$ process produced via gluon fusion.}
\end{figure}
\begin{figure}
  \begin{center}
	\begin{tabular}{c c}
    \includegraphics[width=0.65\textwidth]{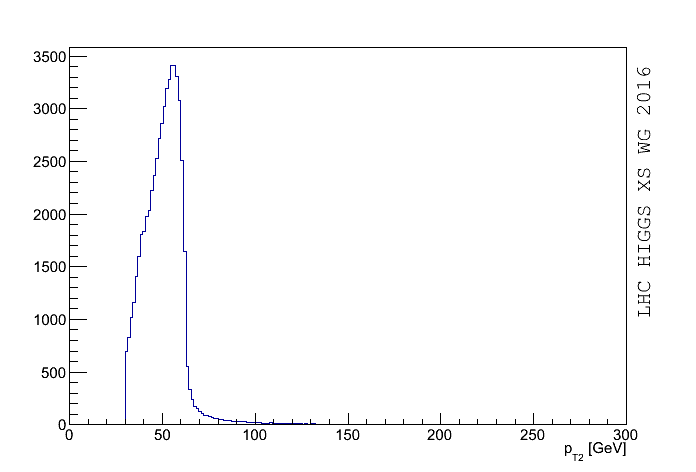}
\end{tabular}	   
    \end{center}
\vspace{-0.5cm}
  \caption{\label{fig:pt2H}
   Transverse momentum distribution of the subleading photon of the $H\to \gamma\gamma$ process produced via gluon fusion.}
\end{figure}
Figures~\ref{fig:ptH} and~\ref{fig:MH}  show the transverse momentum and the pure invariant mass of the diphoton system assuming no interference effect.
\begin{figure}
  \begin{center}
	\begin{tabular}{c c}
	\includegraphics[width=0.65\textwidth]{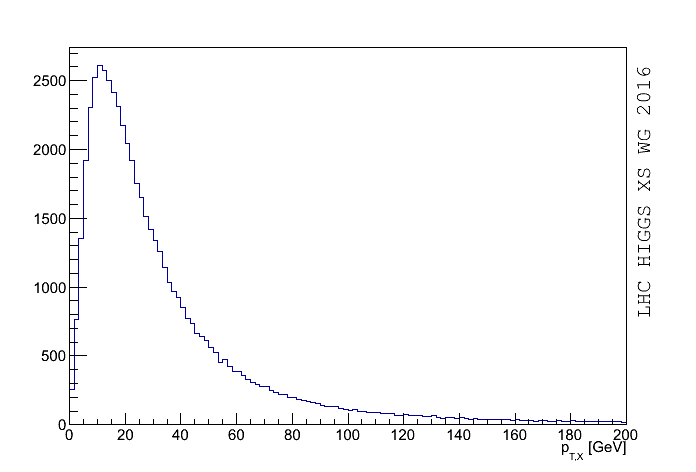} &
  \end{tabular}	   
    \end{center}
\vspace{-0.5cm}
  \caption{\label{fig:ptH}
   Diphoton transverse momentum distribution for pure $H\to \gamma\gamma$ signal produced via gluon fusion.}
\end{figure}
\begin{figure}
  \begin{center}
	\begin{tabular}{c c}
    \includegraphics[width=0.65\textwidth]{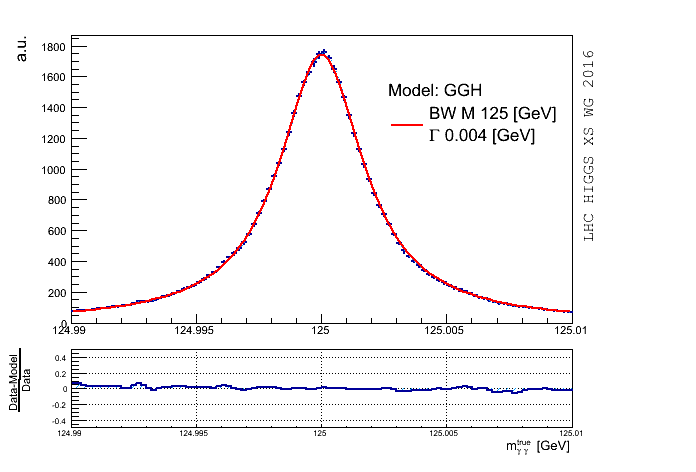}
\end{tabular}	   
    \end{center}
\vspace{-0.5cm}
  \caption{\label{fig:MH}
   Diphoton invariant mass distribution for pure  $H\to \gamma\gamma$ signal produced via gluon fusion.}
\end{figure}
Finally \refFs{fig:Int} and~\ref{fig:SigInt} show the diphoton mass shapes for only the interference term and for the signal+interference cross--section. Interference effect is considered between the $H\to \gamma\gamma$ resonant process and the non resonant diphoton production.
\begin{figure}
  \begin{center}
	\begin{tabular}{c c}
	\includegraphics[width=0.65\textwidth]{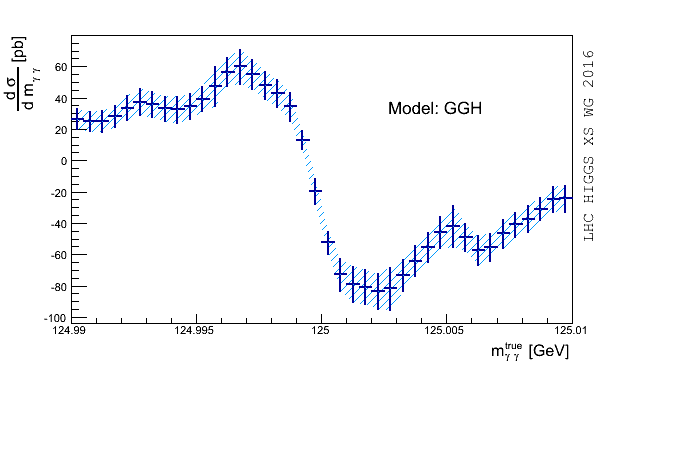} &
  \end{tabular}	   
    \end{center}
\vspace{-0.5cm}
  \caption{\label{fig:Int}
   Pure interference term of the diphoton production cross--section. }
\end{figure}
\begin{figure}
  \begin{center}
	\begin{tabular}{c c}
    \includegraphics[width=0.65\textwidth]{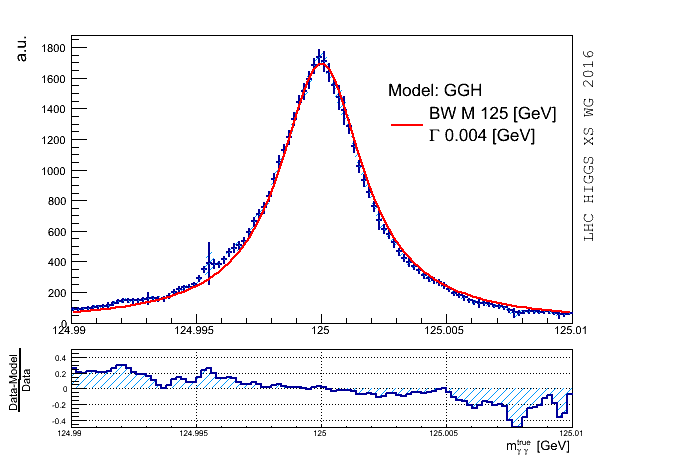}
\end{tabular}	   
    \end{center}
\vspace{-0.5cm}
  \caption{\label{fig:SigInt}
   Total cross--section (signal+interference terms) distribution of the diphoton production. Signal refers to the $H\to \gamma\gamma$ process produced via gluon fusion. }
\end{figure}
A convolution of the pure cross--section shape with a gaussian model can be applied to simulate the effects of the limited resolution of the detector in the photon energy measurement. Different values for the energy resolution (the $\sigma$ of the gaussian function) can be assumed to fold the generator shape.
Figure~\ref{fig:smear} shows the effect of the resolution smearing on the interference term assuming resolution values in the range [1.2-2.2] GeV.
\begin{figure}
  \begin{center}
	\begin{tabular}{c c}
	\includegraphics[width=0.65\textwidth]{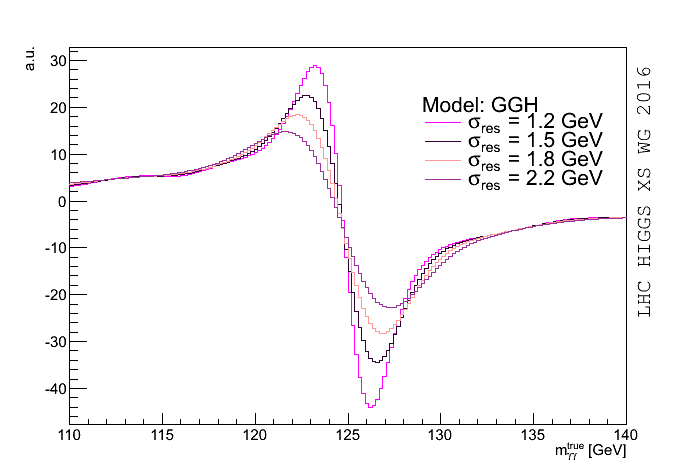} 
\end{tabular}	   
    \end{center}
\vspace{-0.5cm}
  \caption{\label{fig:smear}
   Interference cross--section term smeared assuming different values for the energy resolution in the range [1.2-2.2] GeV.}
\end{figure}
A realistic energy resolution value of 1.7 GeV is eventually assumed before comparing the shapes of the pure signal term and of the signal + interference terms in order to evaluate the relative shift introduced by the interference term itself. Figures~\ref{fig:shift} and~\ref{fig:shiftZ}  show this effect. In this case the shift is evaluated by fitting the two distributions with a gaussian function and taking the difference of the fitted mean values of the two models. The inclusive shift obtained is equal to $\Delta m = -89$ MeV. The trend of this shift varying the assumption on the value of the energy resolution is also shown in Figure~\ref{fig:smearvsres}. The uncertainties associated to the shifts come only from the statistical propagation of the errors on the fit parameters.
\begin{figure}
  \begin{center}
	\begin{tabular}{c c}
	  \includegraphics[width=0.45\textwidth]{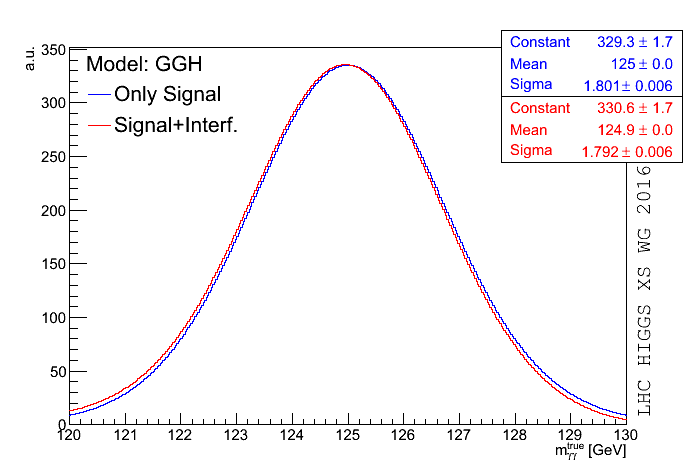} &
      \end{tabular}	   
    \end{center}
\vspace{-0.5cm}
  \caption{\label{fig:shift}
   Pure signal and signal + interference shapes after applying a gaussian energy smearing of 1.7 GeV to simulate detector resolution effects. Red distribution corresponds to the pure $H\to \gamma\gamma$ process while the blue distribution includes the interference effect. Cross--section distributions are fitted with a gaussian function. Results of the fit are shown on the plot with the corresponding colours.}
\end{figure}
\begin{figure}
  \begin{center}
	\begin{tabular}{c c}
          \includegraphics[width=0.45\textwidth]{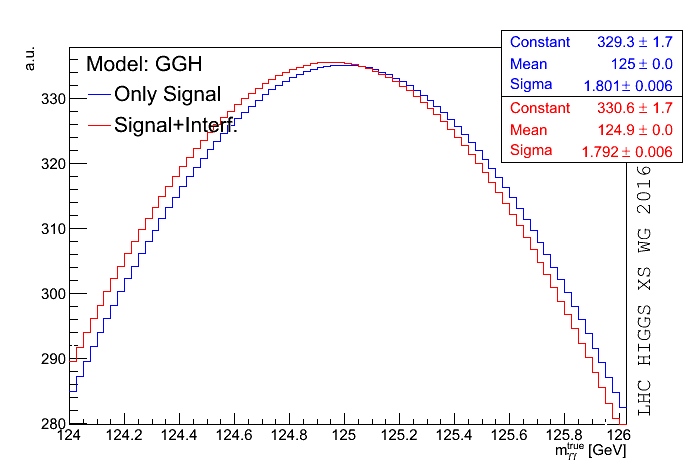}
\end{tabular}	   
    \end{center}
\vspace{-0.5cm}
  \caption{\label{fig:shiftZ}
    This figure shows the same results of \refF{fig:shift} with a zoom around the peak region, applied to better visualize the shift introduced by the interference effect. The inclusive shift obtained is equal to $\Delta m = -89$ MeV.}
\end{figure}
\begin{figure}
  \begin{center}
	\begin{tabular}{c c}
	  \includegraphics[width=0.65\textwidth]{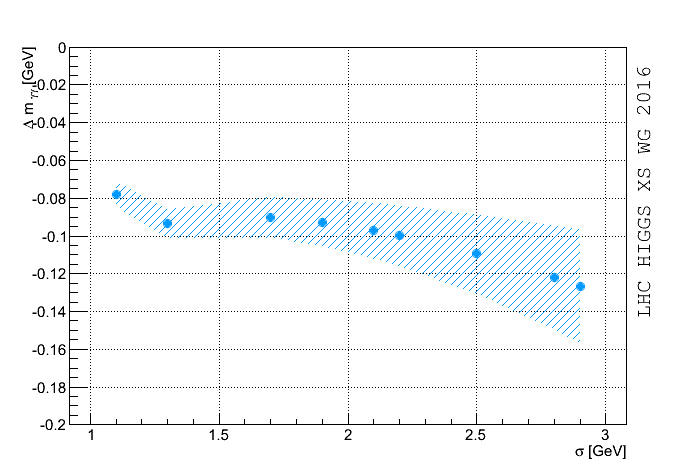} 
\end{tabular}	   
    \end{center}
\vspace{-0.5cm}
  \caption{\label{fig:smearvsres}
   Shift in the mass peak position as a function of the energy mass resolution assumed. The smearing resolution being fixed, both the signal only and the signal+interference cross--section distributions are fitted with a gaussian function. The smearing is evaluated by the difference of the mean values for
   the two gaussian functions. The uncertainties associated with the shifts comes only from the statistical propagation of the errors on the fit parameters. }
\end{figure}
As outlined in Section~\ref{sec:HggIntTheory} the effect of the shift depends strongly upon the minimum threshold applied on the transverse momentum of the diphoton system. Figure~\ref{fig:shiftvspt} reproduces the results shown in Section~\ref{sec:HggIntTheory} showing that the greater the requirement on the minimum value of the diphoton momentum, the smaller the shift in the mass peak position.
\begin{figure}
  \begin{center}
	\begin{tabular}{c c}
	  \includegraphics[width=0.65\textwidth]{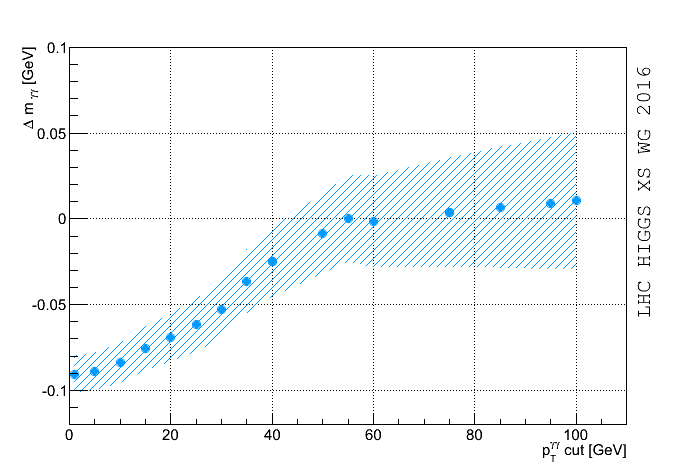} 
\end{tabular}	   
    \end{center}
\vspace{-0.5cm}
  \caption{\label{fig:shiftvspt}
   Shift in the mass peak position as a function of the minimum requirement on the diphoton transverse momentum. The smearing resolution being fixed, both the signal only and the signal+interference cross--section distributions are fitted with a gaussian function. The smearing is evaluated by the difference of the mean values for the two gaussian functions. The uncertainties associated to the shifts comes only from the statistical propagation of the errors on the fit parameters. }
\end{figure}
Additional studies are ongoing in order to evaluate the dependence of the shift upon the natural width of the Higgs.

\subsection{Studies from ATLAS}

This section documents the studies by the ATLAS collaboration~\footnote{Contact: C.~Becot, F.~Bernlochner, L.~Fayard, S.~Yuen}.

\subsubsection{Interference impact on the Higgs boson mass}

A recent study has been conducted by ATLAS \cite{ATL-PHYS-PUB-2016-009} to give a realistic estimate of the impact of the
interference term on the Higgs boson mass measured in the $h \rightarrow \gamma\gamma$ channel \cite{Aad:2014aba}.
Sherpa 2.0 is used to generate the $gg \rightarrow H \rightarrow \gamma\gamma$ signal samples
as well as samples corresponding to the interference between this signal and its irreducible background,
which is achieved using weighted events. The invariant mass spectrum of the di-photon system produced by these samples may be seen in \refF{fig:SvsIvsSplusI} for a specific category used in the ATLAS mass measurement. This generation has been done for a Higgs boson mass of $m_H = 125$ GeV
and a Higgs boson width of $\Gamma_H = 4$ MeV.
The NLO computation implemented in Sherpa 2 is matched to the CSS parton shower \cite{Schumann:2007mg},
which accounts for additional QCD radiations in the initial state.
In order to give the best description of the interference and signal $p_T$ spectra the behaviour of
the shower has been tuned so that the Higgs boson signal $p_T$ distribution generated by Sherpa matches the one generated by
HRes 2.0 \cite{deFlorian:2012mx} as well as possible.
This has been done by modifying the shower parameter CSS\_IS\_AS\_FAC which modifies the
energy at which the strong coupling constant is evaluated during the parton-shower evolution.
The best agreement between the two distributions is obtained for CSS\_IS\_AS\_FAC = 1.5.
This tuning is also applied for the generation of the interference term.
After generation, the di-photon mass is smeared according to the signal model derived in \cite{Aad:2014aba}
which is dominated by a Crystal-Ball component.
In order to reproduce the experimental efficiencies, the Monte-Carlo weights are folded by multiplicative weights
that have the values of these efficiencies.

\begin{figure}
	\centering
	\includegraphics[width=0.6\textwidth]{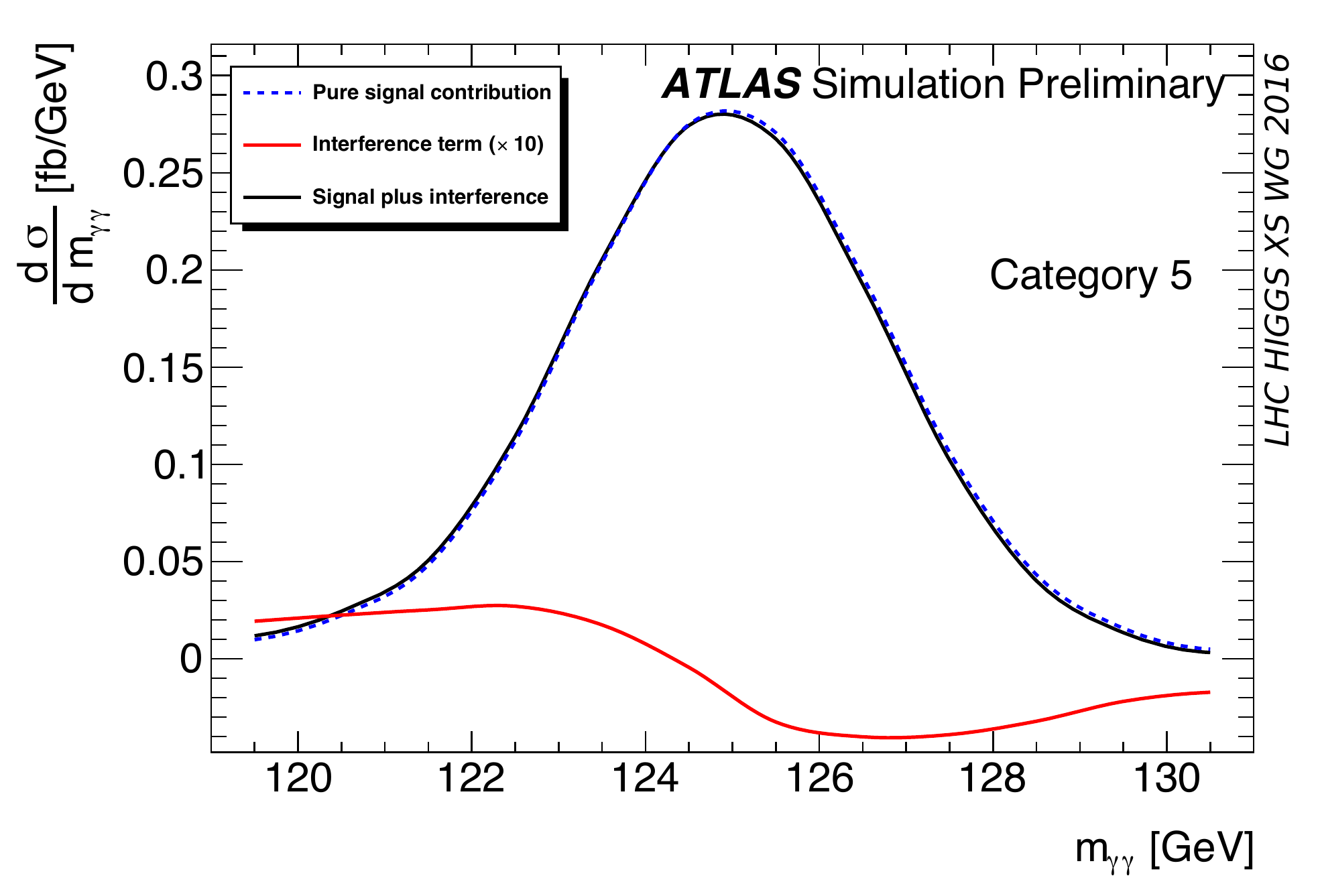}
	\caption{ Mass distribution generated by the signal and interference term, as well as their sum, for a specific category of the ATLAS mass measurement (category no. 5). For illustrative purpose the magnitude of the interference-only term has been multiplied by 10\label{fig:SvsIvsSplusI}.}
\end{figure}

The background is determined from a fit to data, as is usually done is the construction of
the 'Asimov' dataset \cite{Cowan:2010js} and is therefore not subject to consideration on the physics modelling.
In order to improve the analysis performances the mass measurement is carried out in event categories
that are afterwards combined \cite{Aad:2014aba}, and the actual shape used for this fit of the background
depends on the actual category of events and are the same than those used in \cite{Aad:2014aba}.
The additional production mechanisms with associated objects (vector-boson fusion, Higgs-strahlung and $t\bar{t}H$)
are added by a re-scaling of the cross-section of the signal samples.
As they have kinematical properties that differ from the main gluon-fusion production mechanism (and especially a different $p_T$ spectrum),
the templates determined in each category are rescaled separately using the fraction of gluon-fusion events of this particular category.

Two 'Asimov' datasets are then determined: one that contains only the signal and background templates,
and another one that contains the same contributions plus the interference template.
Each of these datasets contains one template for each of the ten categories used in \cite{Aad:2014aba}.
The best-estimate of the Higgs boson mass is obtained separately on each of these datasets with a maximum-likelihood fit that uses the
statistical model derived in \cite{Aad:2014aba}, which is based on the signal and background models described above.
The Higgs boson mass shift is then estimated as the signed difference between the two dataset $\Delta m_H = m_H^{S+B+I}-m_H^{S+B}$
and has been estimated to be of $\Delta m_H = -35$ MeV.

In order to assess that the Monte-Carlo samples are sufficiently large to give a negligible statistical uncertainty on $\Delta m_H$,
four equivalent signal and interference samples have been generated with different random seeds.
The mass shift has been determined separately on each of these, giving a variance of less than 1 MeV.
The imperfect closure of the estimate of the mass on the signal-plus-background only sample has been added as a systematic uncertainty.
The choice the actual background shape used has also been considered and added as an uncertainty.
Both of these systematic uncertainties are at the level of 3 MeV.

Theoretical uncertainties have been estimated by varying the signal and background K-factors as well as the QCD scales involved in this problem.
For the main result the signal K-factor $K_S$ was set to $K_S = 1.45$, which effectively rescales the signal prediction
from Sherpa to the NNLO+NNLL signal cross-section. This factor has been varied by $\pm 0.1$, which accounts for PDF and $\alpha_S$ uncertainties.
Ideally the background K-factor $K_B$ would rescale the background cross-section to the same order than the signal (NNLO),
however no computation of the $gg \rightarrow \gamma\gamma$ background have been performed beyond NLO so far.
A conservative uncertainty on $k_B$ has been assessed by varying it from 1 to $K_S$, using $K_B = K_S$ as a central value.
These two factors modify the interference template by rescaling it by a factor $\sqrt{K_S K_B}$.
At the end the uncertainty due to the K-factors has been taken has the biggest envelope of all these variations
and gave an error of the mass-shift of $\pm 7$ MeV.
The three QCD scales (renormalization, factorization and resummation) have been varied,
first separately then all at the same time.
In spite of having a sizeable impact on the $p_T$ spectrum, the resummation scale has almost no impact on
the overall mass-shift estimated from the combined fit to ten categories,
as most of the statistical power of this measurement is carried out by low $p_T$ categories on which this scale does not have a big impact.
The renormalization and factorization scales are varied by a factor 2, the central value being set to $m_{\gamma\gamma}$.
Although the factorization scale has the dominant effect, the scale uncertainty is estimated as the variation that gives the biggest effect,
which happens when the three scales are varied at once and gives an uncertainty of $\pm 5$ MeV on the mass-shift.

All these four uncertainties are summed quadratically, which gives an estimate of the mass-shift of $\Delta m_H = -35 \pm 9$ MeV.
This is valid only for the mass measurement carried out by ATLAS in the $h \rightarrow \gamma\gamma$ channel.

An illustration of the dependence of this shift on the analysis details is provided in \cite{ATL-PHYS-PUB-2016-009},
where an equivalent number is provided for an 'inclusive' analysis where the events are not split into categories.
In this case the shift is estimated to be of $\Delta m_H = -49$ MeV, which is sizeably larger than in the actual measurement
combining the ten different categories because of the different resolution.
Moreover as the associated production components do not suffer from such large interference effects,
their relative weights in the different categories may also give big variations of the actual mass-shift.
For instance, it was estimated in \cite{ATL-PHYS-PUB-2016-009} that for the inclusive fit
and with the associated production removed the mass-shift would be of $\Delta m_H = -54$ MeV.
It was also noted in \cite{Dixon:2013haa} that the mass-shift had a linear dependence on the invariant mass resolution of the detector.

\subsubsection{The choice of \texorpdfstring{$gg\to (H)\to \gamma\gamma$}{gg to H to gamma gamma} k-factors}

As the most precise computation of the $gg \rightarrow \gamma\gamma$ continuum background has been done at NLO \cite{Bern:2002jx},
the interference term is also limited to a NLO precision. However the signal $gg \rightarrow H \rightarrow \gamma\gamma$ is
known up to NNLO with threshold resummation up to NNLL \cite{deFlorian:2012mx},
while the computation provided in Sherpa 2 is only done at NLO.
The increase of cross-section due to higher-order effects is usually implemented,
for the signal, as a multiplicative k-factor $K_S$ that rescales the cross-section of the signal Monte-Carlo.
In this particular case this factor is of $K_S = 1.45$.
If the impact of higher orders on the background cross-section was known the same approach could be carried out,
using a factor $K_B$. As these two factors correspond to the impact of additional diagrams in the signal and background amplitudes,
they also have an impact on the interference term whose cross-section will then scale as $\sqrt{K_S K_B}$.

Although an exact value for $K_B$ cannot yet be determined, it is possible to determine an interval within which it should be.
The dominating contribution to the Higgs boson signal is carried by a loop of top-quarks while for the continuum $gg \rightarrow \gamma\gamma$
background it comes from a loop of light quarks.
At NLO, it was noticed in \cite{Bern:2002jx} that this implied larger short-distance renormalization effects for the
signal calculation than for the background, which gave a LO to NLO K-factor larger by $\approx 20\%$ for the signal than for the background.
Although no higher-order computations exist for the background yet, it is expected that the same analysis will hold for the NLO to NNLO K-factor,
and hence a reasonable interval within which $K_B$ should be is $[1,\, K_S]$.







\clearpage

\chapter{Summary}
\label{chap:WG1Summary}

In this chapter of the Report we have presented the state of the art for the SM
Higgs cross-section and branching-ratio calculations.

Here we summarize the Higgs boson production cross sections which are obtained
following the new recommendation for the choice of
parton distribution functions (PDFs) and their combined uncertainty assessment
together with the one for the strong coupling constant $\alphas$ 
(the new PDF4LHC recipe) as described in Chapter~\ref{chap:PDFs}.
Moreover, we combine this PDF + $\alphas$ uncertainty with the
theoretical uncertainty (THU).
The combination of the two theoretical uncertainties has been discussed at length and
while in the vast majority of the cases the theoretical advice is to sum them linearly,
the experiments generally sum them in quadrature and assign a gaussian distribution to their
density function.

A particular case is the uncertainty assigned to the ggF cross section calculation at N3LO
order in QCD (see Section~\ref{subsec:recommendation}).

The detailed analysis of the different calculations  suggests the following recommendation:  \\
use the F-uncertainty
\begin{equation}\label{eq:thuncf}
\Delta_{\rm th}=[-6.7,+4.6]\%
\end{equation}
which is a 100\% flat interval. If it is highly preferred to have Gaussian
uncertainties, then symmetrize the flat interval and divide it by $\sqrt{12}$, obtaining
\begin{equation}\label{eq:thuncg}
\Delta_{\rm th}=\pm3.9\% .
\end{equation}

The corresponding gluon fusion cross-sections expanded to a scan over SM
Higgs boson masses are presented in \refTs{tab:ggF_XStot_7}--\ref{tab:ggF_XStot_14}.
\refTs{tab:ggF_XScan_125} and \ref{tab:ggF_XScan_12509} summarize the Standard Model
gluon fusion cross-sections and the corresponding uncertainties
for the different proton--proton collision energies
for a Higgs boson mass $\MH=125\UGeV$ and  $\MH=125.09\UGeV$, respectively.

The following figures present the results as described in the above chapter, 
for a Higgs boson of mass ranging from $120 \UGeV$ to $130 \UGeV$ for $13 \UTeV$, \refF{fig:SMXS13TeV} and
$14 \UTeV$, \refF{fig:SMXS14TeV},
centre-of-mass-energies, with the combined parametric and theoretical uncertainties summed in quadrature
and treated as gaussian pdf, illustrated by bands.
The labels on the bands briefly indicated the type of radiative corrections that are
included in the predictions.

Figure~\ref{fig:SMXS_Escan}, presents the cross section for a Higgs boson of $125 \UGeV$ 
as a function of the centre-of-mass-energies.

The branching ratios for the SM Higgs boson are shown in \refF{fig:SMBRs}.
Tables containing explicit numbers on partial widths, branching ratios, and on the 
total width can be found in 

The results shown in this section will be regularly updated at our
webpage\footnote{\url{https://twiki.cern.ch/twiki/bin/view/LHCPhysics/LHCHXSWG}}.
f
Each experiment is recommended to use the common Standard Model input
parameters as presented in Chapter~\ref{chapter:input},
the best known N3LO/NNLO/NLO cross sections and branching ratios
reported in this Report as common basis for Higgs physics at LHC.

The SM cross section calculations has been extended to low and to high Higgs boson masses as a basis
for Beyond Standard Model analysis and calculations.
For these calculation the Narrow Width Approach has been used and Electro-Weak correction have  not been
included.

The following figures present the results a function of the Higgs boson mass for $13 \UTeV$ ( 
\refF{fig:BSM_H_13TeV}) and $14 \UTeV$ (\refF{fig:BSM_H_14TeV})
centre-of-mass-energies, with the combined parametric and theoretical uncertainties summed in quadrature,
illustrated by bands.
The labels on the bands briefly indicated the type of radiative corrections that are
included in the predictions.


\begin{figure}[H]
	\begin{center}
	\includegraphics[width=0.55\textwidth]{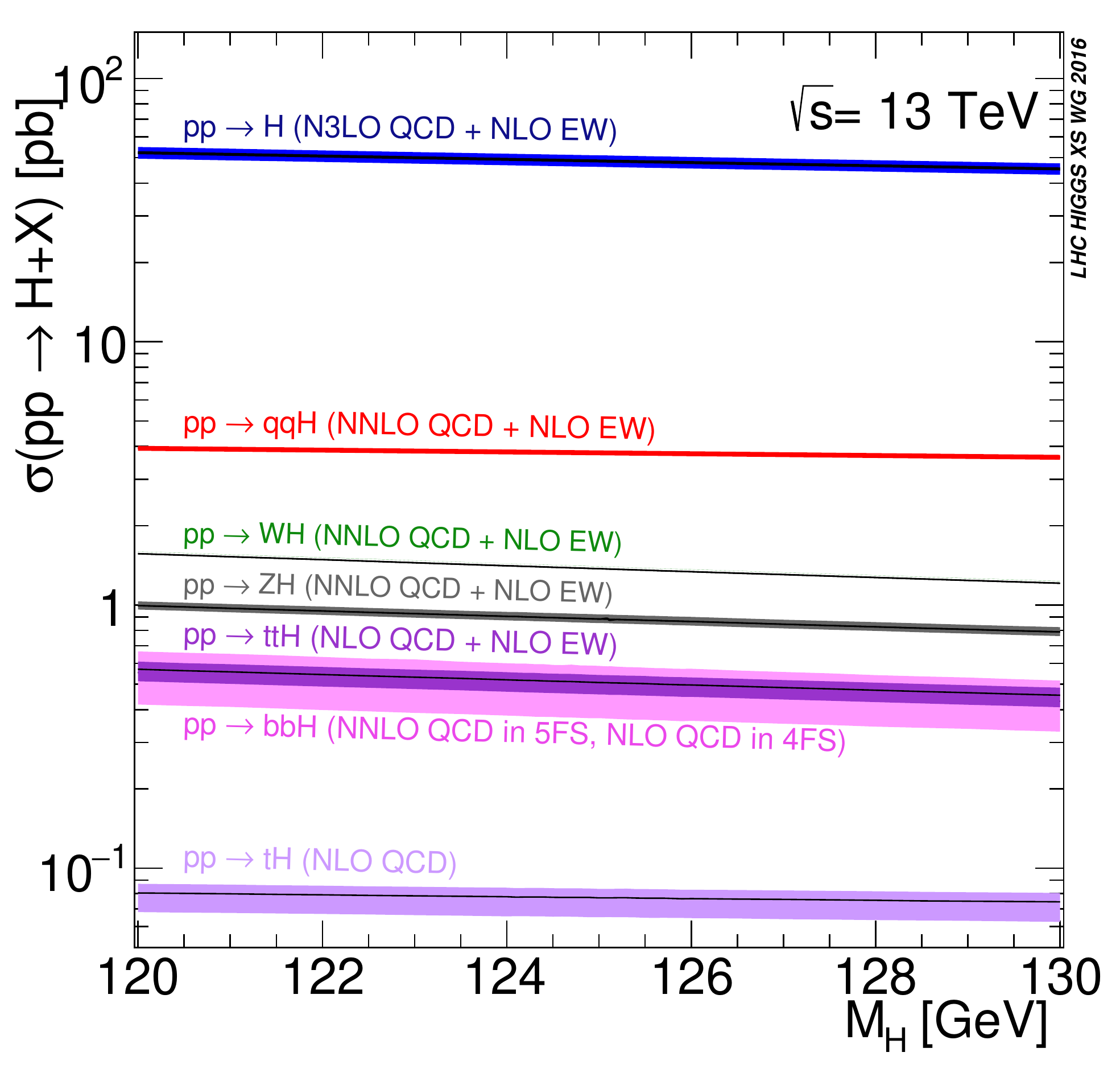}
	\caption{The SM Higgs boson production cross section at $\sqrt{s} = 13$\UTeV}
	\label{fig:SMXS13TeV}
	\end{center}
\end{figure}

\begin{figure}[H]
	\begin{center}
	\includegraphics[width=0.55\textwidth]{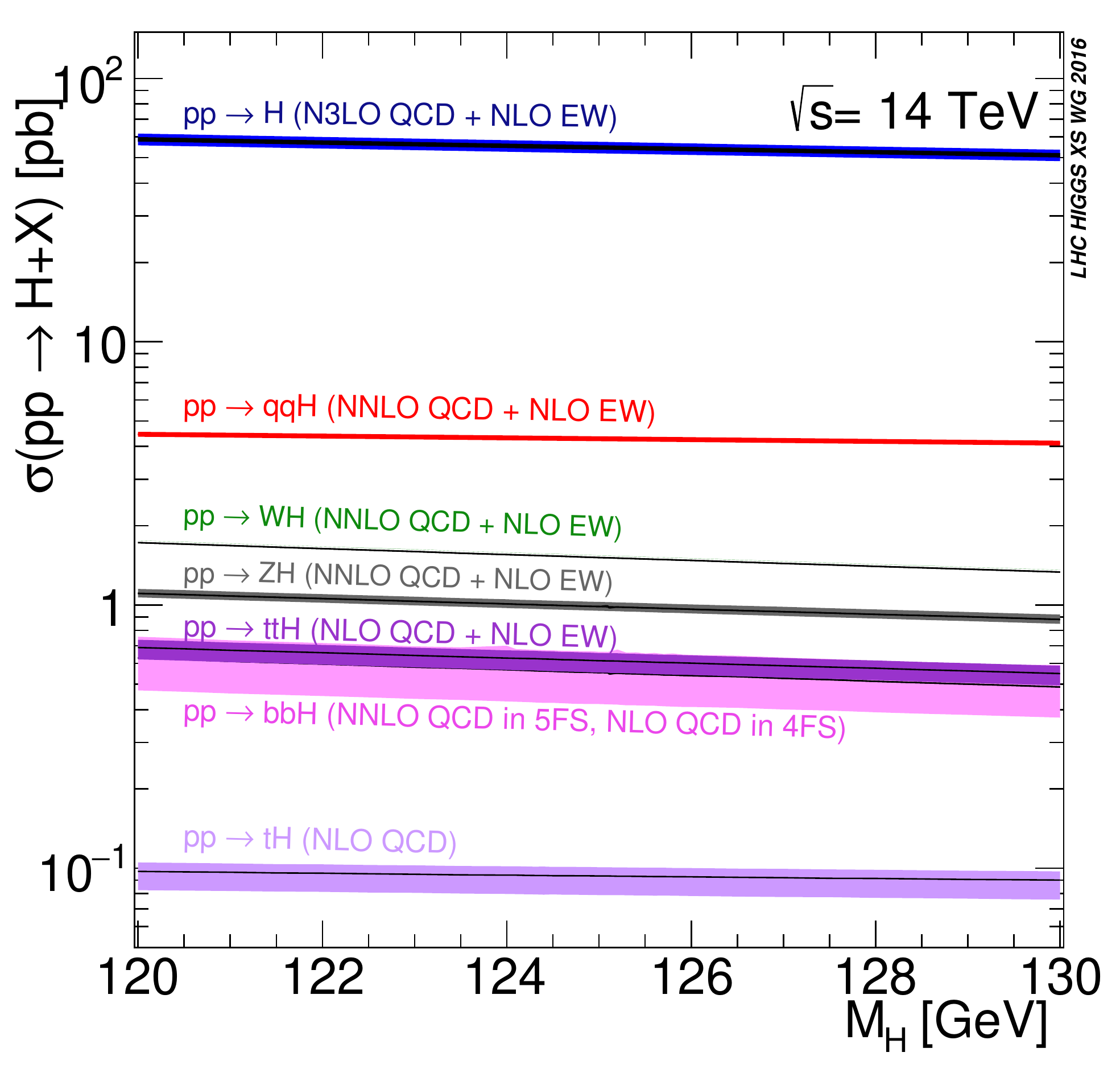}
	\caption{The SM Higgs boson production cross section at $\sqrt{s} = 14$\UTeV.}
	\label{fig:SMXS14TeV}
	\end{center}
\end{figure}

\vfill 
\newpage

\begin{figure}
	\begin{center}
	\includegraphics[width=0.55\textwidth]{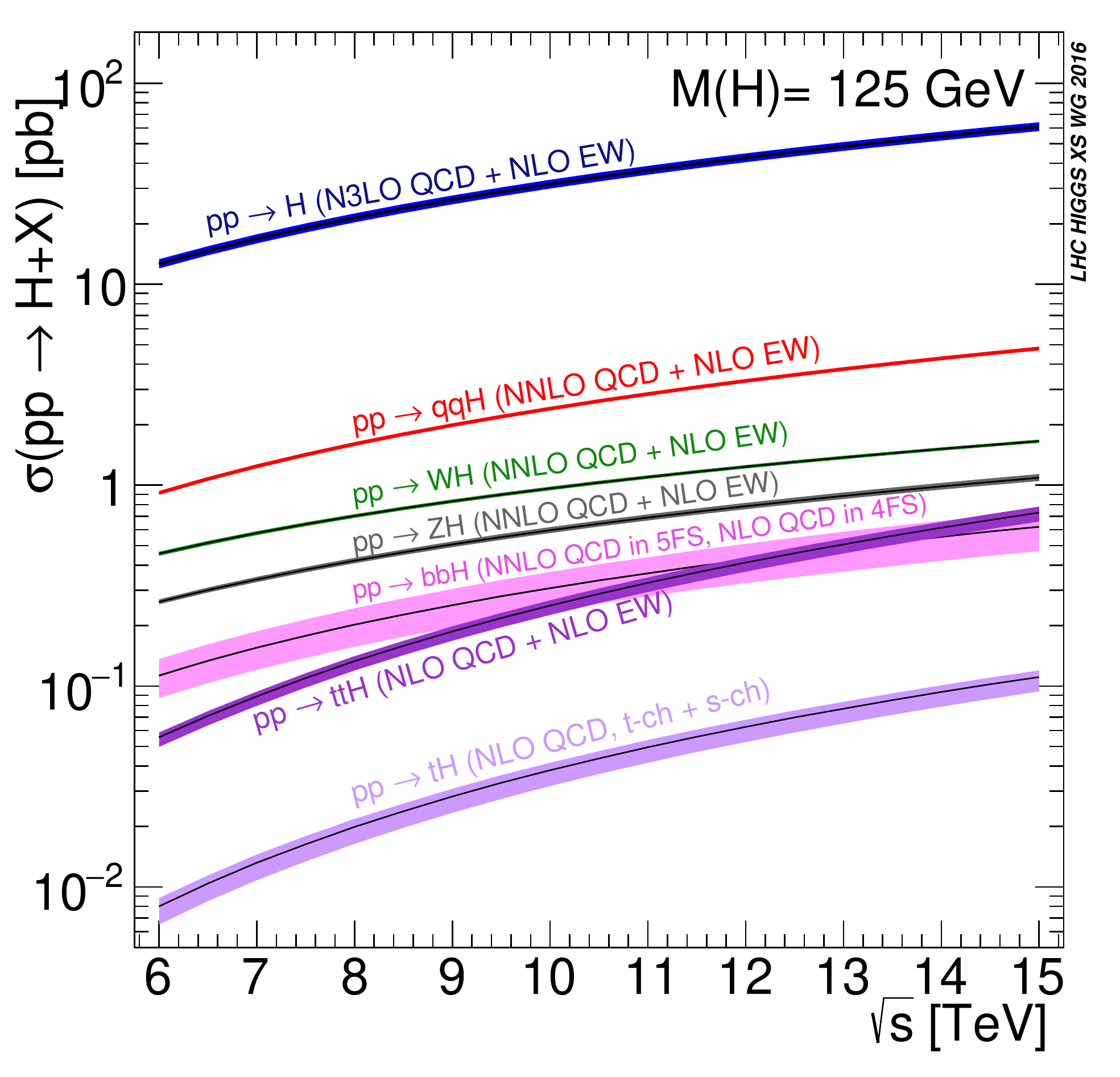}
	\caption{The SM Higgs boson production cross sections as a function of the LHC centre of mass energy.}
	\label{fig:SMXS_Escan}
	\end{center}
\end{figure}

\begin{figure}
	\begin{center}
	\includegraphics[width=0.55\textwidth]{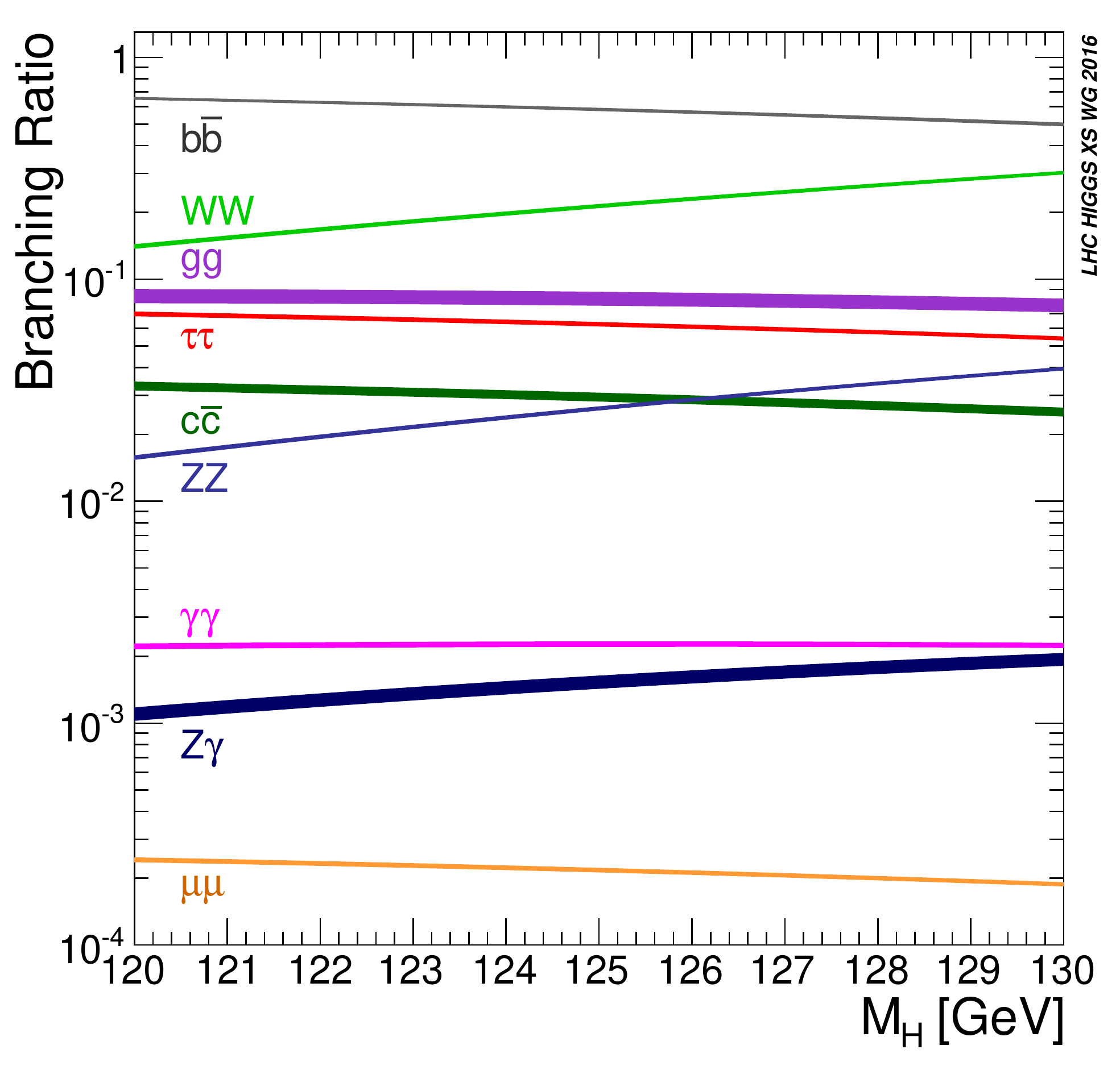}	
	\caption{The SM Higgs boson branching ratios as a function of the Higgs boson
              mass.}
	\label{fig:SMBRs}
	\end{center}
\end{figure}

\vfill
\newpage

\begin{figure}
	\begin{center}
	\includegraphics[width=0.75\textwidth]{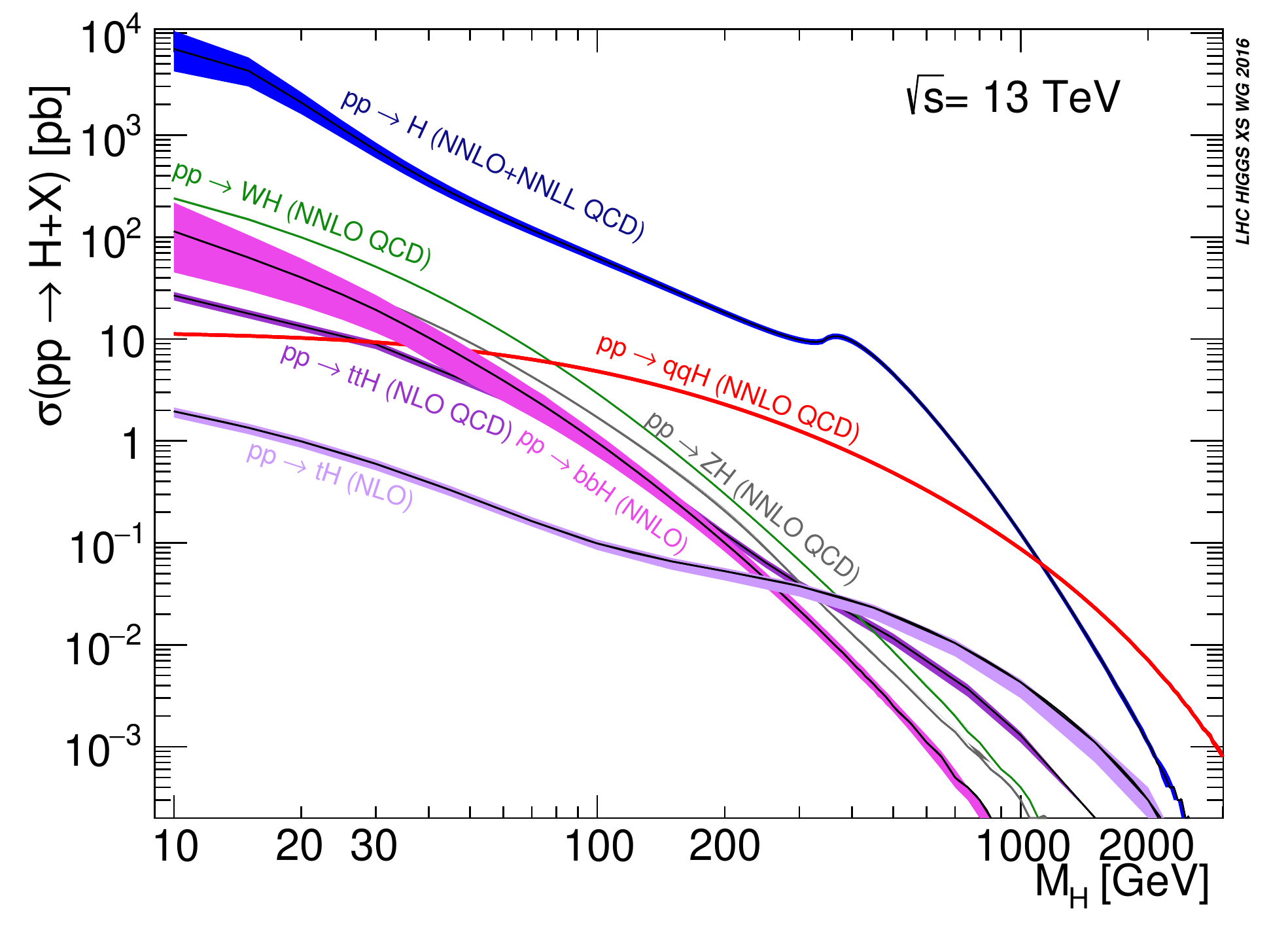}
	\caption{The SM Higgs boson production cross section as a function of the Higgs boson mass at $\sqrt{s} = 13$\UTeV}
	\label{fig:BSM_H_13TeV}
	\end{center}
\end{figure}

\begin{figure}
	\begin{center}
	\includegraphics[width=0.75\textwidth]{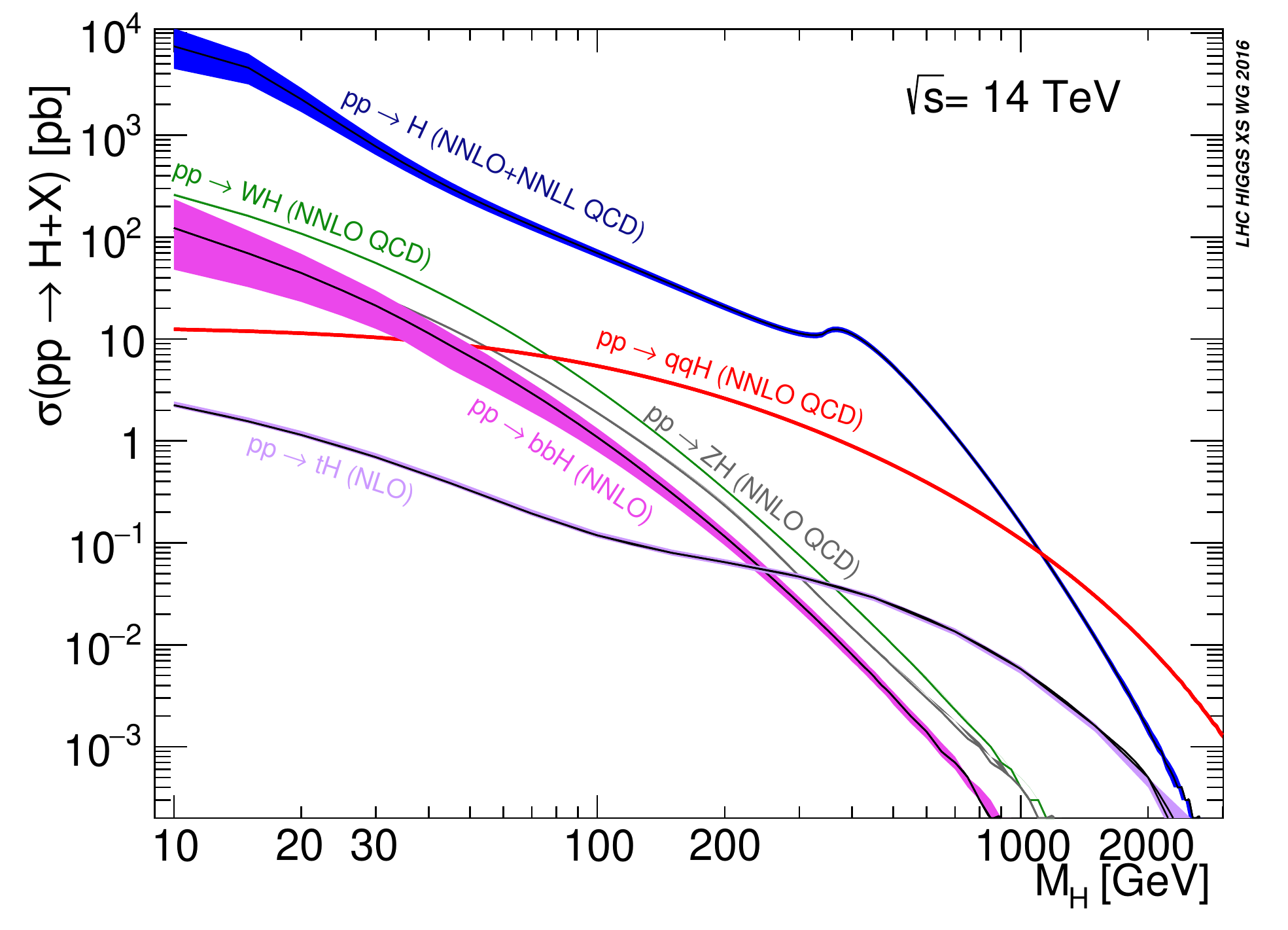}
	\caption{The SM Higgs boson production cross section as a function of the Higgs boson mass at $\sqrt{s} = 14$\UTeV.}
	\label{fig:BSM_H_14TeV}
	\end{center}
\end{figure}

\cleardoublepage

\newpage
\renewcommand*{\thefootnote}{\fnsymbol{footnote}}
\part[Effective Field Theory Predictions]{Effective Field Theory Predictions \footnote{M.~Chen, A.~David, M.~D\"uhrssen, A.~Falkowski, C.~Hays, G.~Isidori~(Eds.)}}
\label{chap:EFT}
\renewcommand*{\thefootnote}{\Roman{part}.\arabic{footnote}}
\setcounter{footnote}{0} 
\chapter[Executive Summary of Parts II and III]{Executive Summary of Parts~\ref{chap:EFT} and \ref{chap:MO}}
\ChapterAuthor{M.~Chen, A.~David, M.~D\"uhrssen, A.~Falkowski, C.~Hays, G.~Isidori}

The 2012 discovery of a new particle, subsequently shown by the
ATLAS and CMS collaborations to be a Higgs boson,
has closed a chapter in particle physics.
Not only on the experimental side, putting an end to a decades-long search,
but also, and perhaps more sharply, by completing the set of predictions by the standard model (SM) for elementary particles.
The challenge that is ahead for the LHC and future machines is now fully in the
BSM realm.

The LHC Run 2, which started in 2015, now has a qualitatively different goal
in what regards the program for measuring the properties of this Higgs boson
and the search for deviations from the SM predictions.

The WG2 contributions to this Yellow Report therefore naturally cluster around two main axes, supplemented by a third aspect:
\begin{enumerate}
  \item How to expand the palette of measurements that can be performed by the experiments.
  \item How to interpret existing measurements to set limits on and constrain new physics and characterize discoveries.
  \item Tools with which to proceed in practice.
\end{enumerate}

The two main axes are complementary to and feed off of each other:
measurements pave the way for different interpretations,
while interpretation frameworks motivate new measurements.
This being said, there are caveats to this interaction.
For instance,
while almost any framework can be used to motivate particular measurements,
the interpretation of a measurement and the definition of (pseudo)-observables
can only be consistently done in a well-defined theory framework.
In other words,
much in the same way that finding a significant deviation
with the kappa framework would clearly point to BSM physics,
its meaning and interpretation would require a well-defined theory,
which the kappa framework alone is not.

Complementing the chapters on measurements and interpretation,
there are also two chapters describing tools
that can be used in the different aspects
of the measurement and interpretation steps.

This Executive Summary provides an overview of the WG2 chapters, comprised in Part~\ref{chap:EFT} and, in collaboration with WG1, in Part~\ref{chap:MO}.
The goal is not to exhaustively review the contents,
but to offer a ``lay of the land'',
providing the reader with the most salient and distinctive features of each chapter and how the different chapters are related and connected
with each other.

\subsection*{Measurements}

In this Yellow Report,
the existing kappa framework for the search of deviations from the SM predictions
is substantially expanded in two, complementary, ways:
simplified template cross-sections (STXS) and
pseudo-observables (PO). 
This dichotomy arises naturally from the fact that at the LHC the Higgs boson interaction
with SM particles is probed at multiple energy scales.
For instance,
while $H \to 4\ell$ probes  
the amplitude coupling the Higgs to four fermions in a region of transferred momenta kinematically bounded by $m_H$,
the associated ZH production  will probe the same amplitude
(or at least part of it) at significantly higher momentum transfer,
possibly even in the multi-TeV region.
That explains the different approaches presented in this Yellow Report, with some chapters focusing on production properties and others on decay properties.

Chapter~\ref{STCS} presents a way to partition the phase-spaces of different Higgs boson production processes into simplified template cross-sections.
The goal of the STXS partitioning is two-fold:
\begin{enumerate}
 \item To separate regions of the phase-space for which theory uncertainties can evolve with time.
 \item To single out parts of the production phase-space where BSM physics predicts large deviations from the SM expectation. In this case, rare corners of SM production can be used to probe for BSM-induced deviations.
\end{enumerate}
The STXS are mostly a tool that generalizes the notion of production process into sub-processes and the result of their use is a measurement of fully-extrapolated and unfolded cross-sections that can be also expressed as signal strengths relative to the SM predictions.
This allows to recombine the measurements and update total cross section measurements ex post facto.
This is for instance the case with jet binning for gluon-fusion cross-sections: measuring the ggH plus 0-jet, 1-jet, and 2-jet sub-processes allows to avoid to commit to a single prescription for jet bin migration that is needed to extract the ggH total cross-section, allowing for the prescription to evolve and be introduced later.

The STXS can be thought of as fully extrapolated and unfolded cross sections that can be inferred differentially in the production properties.
As they are extrapolated from a simultaneous fit, this allows for advanced experimental techniques (including multi-variate observables and discriminants) to be employed in the analyses.
The use of such techniques is not possible, for instance, when measuring fiducial cross-sections, as it is very hard, if not impossible, to define the fiducial volume for a multi-variate observable.

In a completely complementary way, fiducial cross-section (FXS) measurements provide easy-to-reproduce phase-spaces.
Many practical aspects of FXS measurements are discussed in Chapter~\ref{chap:FXS}, paving the way for common extractions of more model-independent quantities that are easy to collect in persistent form, using the HepData database and the Rivet toolkit.
Attention is also paid to the interplay between signal and background processes in a given fiducial volume, as well as to unfolding of experimental effects from the measurements.

Finally, Chapter~\ref{chap:PO} discusses how on-shell Higgs boson decays and production cross sections, close to the threshold region, can be parameterized in terms of pseudo-observables. 
The PO framework builds up on the similar approach introduced for Z-pole observables at LEP. In Higgs physics the formalism is 
a bit more complicated by the multiple poles involved in Higgs boson decays into 3 and 4 bodies, as well as in Higgs boson production cross sections.
This richer kinematical structure is decomposed in terms of independent Lorentz structures, as well as resonant and non-resonant contributions, 
whose form is 
dictated by the general analytic properties of the amplitudes under the  assumption that no BSM particles appear on-shell.
The purpose and the main philosophy of Higgs PO is the same of the Z-pole (pseudo)-observables at LEP:
PO are well-defined quantities from the quantum field theory point of view,
that can be measured by experiments and then interpreted in generic BSM scenarios, 
including effective theory approaches.

\subsection*{Interpretation}

Given that the SM has been completed, the focus on extending the SM is very strong.
There are two fundamentally different ways to go about extending the SM Lagrangian:
via concrete BSM alternatives (such as SUSY), for which different predictions are provided in Part~\ref{chap:BSM}, or through an effective descriptions of sufficiently high-mass (hence partially decoupled) degrees of freedom, as discussed in Part~\ref{chap:EFT}.
The latter approach,
referred to as the effective field theory (EFT) approach to Higgs physics,
is the one discussed in the chapters contributed from WG2.
In this case, the SM Lagrangian is extended by adding higher-dimension operators written in terms of SM fields. 
Such a framework can be used to describe the effects of new heavy
particles on Higgs physics in a large class of models beyond the SM.
The Wilson coefficients of the higher-dimension operators in the EFT
encode information about masses and couplings in the UV theory that
completes the SM.

The EFT approach to Higgs physics is conceptually different from ``top-down'' EFT approaches, such as HQET, where the ultraviolet completion of the theory is known.
In the Higgs EFT case, the full theory is unknown and the working conditions are ``bottom-up''.
This means that, a priori, there is a large range for the possible values of the couplings of the higher-dimensional operators.
The latter can be restricted employing additional dynamical or symmetry assumptions about the overarching BSM model. 
This is why different EFT approaches
(based on different symmetry hypotheses,  
and order of the expansion in the various couplings) are discussed in the chapters from WG2.\\

More generally, two main themes are addressed:\\[2mm]
\hspace{.3cm}-- The definition of the theory frameworks that can be used to extend the SM.\\[2mm]
\hspace{.3cm}-- Discussion of the limitations that such effective descriptions have in describing different BSM physics scenarios (UV completions).\\

In terms of Lagrangian formulation, two avenues are explored.
In  Section~\ref{s.nleft} the  chiral Lagrangian relevant to the case  of a possible non-linear realization of electroweak symmetry breaking is presented. 
The largest effort, though, was devoted to the so-called SM EFT where, much as in the SM,  the electroweak symmetry is  realized linearly  and broken spontaneously by the VEV of the Higgs field.  
Within this framework, Sections~\ref{s.eftbasis} and \ref{s.eftnlo} discuss how the Wilson coefficients of dimension-6 operators are related to deformations of the Higgs boson couplings from the SM predictions. 
Furthermore, Chapter~\ref{s.eftbasis} proposes a parameterization of the space of dimension-6 operators, the so-called Higgs basis,  that is convenient for  calculating Higgs observables at the leading order (LO) in the SM EFT.
Section~\ref{s.eftnlo} provides NLO results in the SM EFT after performing the renormalization programme.
Given the differences between the nonlinear formulation, the SM EFT at the LO,  and the NLO formulation, it is important to note that, depending on the exact UV completion realized in nature, there may or
may not be a close correspondence between interpretations of the data in these different frameworks. 
Calculations within the LO EFT are simpler from the theoretical point of view and introduce a minimal number of parameters to describe leading deformations of Higgs observables.
However, there may be physical situations where the LO EFT does not provide an adequate description and going beyond the LO is necessary; they are discussed in Sections~ \ref{s.eftval} and \ref{s.eftnlo}.  
EFT with a non-linearly realized electroweak symmetry is less predictive than the SM EFT, but it may be relevant for certain classes of  UV completions of the SM.

Sections~\ref{s.eftval} and \ref{s.eftmodels} then explore the applicability of the SM EFT construction,
both concluding that the SM EFT enjoys a broad range of validity in the absence of new particles with masses in the hundreds of GeV.
One topic that is particularly difficult to address in interpreting measurements in terms of the underlying theory parameters,
Wilson coefficients in this case, is that of theory uncertainties.
These studies also shed some light into how operators with dimensions higher than 6 can play a role in the interpretation.

One conclusion arising from the discussion in this chapter is that there isn't
a unique EFT approach to be recommended as different EFT approaches have different validity limits.
In principle, the more general, the better, as there are fewer
implicit assumptions about the ultraviolet completion.
However, generality comes at the price of increased complexity and
some compromise between generality and simplicity may be necessary,
especially in the early phases of the LHC Run 2.
This highlights the importance for EFT interpretations of LHC data to always
be pursued in parallel to more general (although less predictive) approaches
such as FXS, STXS, and POs.

Finally, it should be noted that the SM EFT has overarching implications to electroweak physics and even in other sectors in the sense that 
in its full generality, it can be constrained not only from precise measurements of Higgs boson properties but also other measurements, ranging from multiboson production, to top quark properties.
Global fits of EFT parameters may allow for better constraints on BSM
physics and amplify the power of the data.
Indeed, the ability to combine distinct measurements within one
consistent framework is a great strength of the EFT approach and one
of its main motivations.

\subsection*{Tools}

Of course none of the concepts above can be put to practice without computational tools
that allow for the simulation of the different SM deformations,
as well as tools that simplify the practical aspects related to the
statistical inference on the parameters of interest.

Section~\ref{s.efttools} overviews the available frameworks upon which many of the interpretations previously discussed, but also the measurements, can be based on.
The effective Lagrangian formulation is widely used in order to make predictions starting
from Wilson coefficients, but it is also used without loss of generality for
encoding pseudo-observables.

In Section~\ref{s.eftmorphing} the reader can find a practical proposal to
model predictions for different processes in a multidimensional context of SM deformations.
The tools provide for continuous interpolation between parameter-space points, are based on commonly use software at the LHC, and may also find use in other applications.
By providing smooth and continuous interpolations, they allow for simple application
of likelihood ratio methods commonly used by ATLAS and CMS to determine the allowed confidence regions for parameters.

\chapter{EFT Formalism} 
\label{chap:EFTForm}
\ChapterAuthor{
N.~Belyaev, A.~Biek{\"o}tter, J.~Brehmer, I.~Brivio, G.~Buchalla, O.~Cata, A.~Celis, R.~Contino, T.~Corbett, R.L.~Delgado, A.~Dobado, C.~Englert, D.~Espriu, A.~Falkowski,  A.~Freitas, F.~Goertz, D.~Gon{\c c}alves, J.~Gonzalez-Fraile,  M.~Gorbahn, C.~Grojean,  M.~Herrero, R.~Kogler, R.~Konoplich,  C.~Krause, F.J.~Llanes-Estrada, D.~Lopez-Val, L.~Merlo, K.~Mimasu, J.M.~No, T.~Ohl,  G.~Passarino, T.~Plehn, M.~Rauch, J.~Reuter,  M.~Riembau, F.~Riva V.~Sanz, J.J.~Sanz-Cillero, M.~Spannowsky, M.~Trott}

\section[Bases for the Standard Model Effective Field Theory]{Bases for the Standard Model Effective Field Theory\SectionAuthor{N.~Belyaev, A.~Falkowski, F.~Goertz, R.~Konoplich, K.~Mimasu, T.~Ohl, J.~Reuter, M.~Riembau, F.~Riva}}
\label{s.eftbasis}
\subsection{Introduction}

For a large class of models beyond the SM, 
physics at energies below the mass scale $\Lambda$ of the new particles  can be parameterized by an effective field theory (EFT) where the SM Lagrangian is supplemented by new operators  with canonical dimensions $D$ larger than 4.   
The theory has the same field content and the same linearly realized $SU(3) \times SU(2) \times U(1)$ local symmetry as the SM.\footnote{%
The latter assumption can be relaxed, leading to an EFT with a non-linearly realized electroweak symmetry. 
This framework is discussed in Section~\ref{s.nleft}.} 
The higher-dimensional operators are organized in a systematic expansion in $D$, where each consecutive term is suppressed by a larger power  of $\Lambda$.
For a general introduction to the EFT formalism see e.g. \cite{Manohar:1996cq,Rothstein:2003mp,Kaplan:2005es,Burgess:2007pt,Weinberg:2009bg};
for recent review articles about EFT in connection with Higgs physics  see e.g. \cite{Willenbrock:2014bja,Masso:2014xra,Pomarol:2014dya,Henning:2014wua,Falkowski:2015fla,David:2015waa}.

Quite generally, the EFT Lagrangian takes the form:  
\begin{equation}
\label{eq:EFT_Lops}
{\cal L}_{\textrm{eff}}={\cal L}_{\rm SM}
+ \sum_{i}  \frac{c_{i}^{(5)}}{\Lambda}{\cal O}_{i}^{(5)} \
+ \sum_{i}  \frac{c_{i}^{(6)}}{\Lambda^2} {\cal O}_{i}^{(6)} \
+ \sum_{i}  \frac{c_{i}^{(7)}}{\Lambda^3}{\cal O}_{i}^{(7)} \
+ \sum_{i}  \frac{c_{i}^{(8)}}{\Lambda^4} {\cal O}_{i}^{(8)}  + \cdots\,,
\end{equation}
where each ${\cal O}_i^{(D)}$ is an $SU(3) \times SU(2) \times U(1)$  invariant operator of dimension $D$ and the parameters $c_i^{(D)}$ multiplying the operators in the Lagrangian are called  the {\em Wilson  coefficients}.  
This EFT is intended to parameterize observable effects of a large class of BSM theories where new particles, with mass of order $\Lambda$, 
are much heavier than the SM ones and much heavier than the energy  scale at which the experiment is performed. 
The main motivation to use this framework is that the constraints on the EFT parameters can be later re-interpreted as constraints on masses and couplings  of new particles in many BSM theories. 
In other words, translation of experimental data into a theoretical framework has to be done only once in the EFT context, rather than for each BSM model separately.

The  contribution of each $\mathcal{O}_i^{(D)}$ to amplitudes of physical processes at the energy scale of order $v$ scales\footnote{%
Apart from the scaling with $\Lambda$, the effects of higher-dimensional operators also scale with appropriate powers of couplings in the UV theory. The latter is important to assess the validity range of the EFT description, as discussed  in Section~\ref{s.eftval} and Ref.~\cite{Contino:2016jqw}.} 
as $(v/\Lambda)^{D-4}$. 
Since $v/\Lambda < 1$ by construction, the EFT in its validity regime typically describes {\em small} deviations from the SM predictions, although, under certain conditions, it may be consistent to use this framework to describe large deviations \cite{Biekoetter:2014jwa,Contino:2016jqw}.     

A complete and non-redundant set of operators that can be constructed from the SM fields  is known for $D$=5 \cite{Weinberg:1979sa}, $D$=6 \cite{Grzadkowski:2010es},  $D$=7  \cite{Lehman:2014jma,Henning:2015alf}, and $D$=8 \cite{Lehman:2015coa,Henning:2015alf}. 
All $D$=5 operators violate the lepton number \cite{Weinberg:1979sa}, while all $D$=7 operators violate $B-L$ (the latter is true for  all odd-D operators \cite{deGouvea:2014lva}). Then,  experimental constraints dictate that their Wilson coefficients must be suppressed at a level which makes them unobservable at the LHC \cite{Deppisch:2015yqa}, and for this reason $D$=5 and $7$ operators will not be discussed here. 
Consequently, the leading new physics effects are expected from operators with $D$=6  \cite{Buchmuller:1985jz},  whose contributions scale as $(v/\Lambda)^2$.
Contributions from operators with $D\geq 8$ are suppressed by at least  $(v/\Lambda)^4$, and in most of the following discussion we will assume that they can be neglected. 

In this section, we discuss in detail the  $D$=6 operators that can be constructed from the SM fields. 
We review various possible choices of these operators (the so-called {\em basis}) and their phenomenological effects.  
Only the operators that  conserve the baryon and lepton numbers are considered.  
On the other hand,  we do not impose a-priori any flavour symmetry. 
Also, we include CP violating operators in our discussion.
One purpose of this section is to propose a common EFT language and conventions that could be universally used in LHC Higgs analyses and be implemented in numerical tools.

In Section~\ref{sec:smeft} we introduce the SM Lagrangian extended by dimension-6 operators.  
Two popular bases of dimension-6 operators using the manifestly $SU(2)\times U(1)$ invariant formalism  are  introduced.  
In Section~\ref{sec:pel} we discuss the interactions of  the SM mass eigenstates that arise in the presence of dimension-6 operators beyond the SM, with the emphasis on the Higgs interactions.
We also provide a map between the couplings in that effective Lagrangian and Wilson coefficients of  dimension-6 operators  introduced in  Section~\ref{sec:smeft}. 
In Section~\ref{sec:hb} we define a new basis of $D$=6 operators, the so-called Higgs basis, which is spanned by a subset of the independent couplings of the mass eigenstate Lagrangian.

\subsection{SM EFT with dimension-6 operators}  
\label{sec:smeft}
  
We consider an EFT Lagrangian where the SM is extended by dimension-6 operators: 
\begin{equation}
\label{eq:left}
{\cal L}_{\rm EFT}  = {\cal L}_{\rm SM}  +  \sum_i \bar c_i^{(6)} O_i^{(6)}. 
\end{equation}  
In our conventions, the scale $\Lambda$ has been absorbed in the
definition of the Wilson coefficients, $\bar c_i^{(6)} =   c_i^{(6)}
v^2/\Lambda^2$,  and we divide the dimension-6 operators by $v^2$,
$O_i^{(6)} = {\cal O}_i^{(6)}/v^2$.

To fix our notation and conventions, we first write down the SM Lagrangian: 
\begin{eqnarray} 
\label{eq:EFT_lsm}
{\cal L}_{\rm SM} &=& 
  -{1 \over 4} G_{\mu \nu}^a G_{\mu \nu}^a  - {1 \over 4} W_{\mu \nu}^i W_{\mu \nu}^i  -  {1 \over 4}    B_{\mu \nu} B_{\mu \nu} 
  + D_\mu H^\dagger D_\mu H  + \mu_H^2 H^\dagger H - \lambda (H^\dagger H)^2  
\nonumber \\  &+& 
\sum_{f \in q, \ell}   i \bar f_L   \gamma_\mu D_\mu f_L  + \sum_{f \in u ,d,e}   i \bar f_R   \gamma_\mu D_\mu f_R
\nonumber \\  &-& 
 \left [  q_L \tilde H  y_u u_R +   \bar q_L H  y_d  d_R +  \bar \ell_L  H y_e \ell_R   + {\mathrm h.c.} \right ].
\end{eqnarray}
Here,  $G_\mu^a$, $W_\mu^i$, and $B_\mu$ denote the gauge fields of  the $SU(3) \times SU(2) \times U(1)$ local symmetry.
The corresponding gauge couplings are denoted by $g_s$, $g$, $g'$; 
we also define the electromagnetic coupling $e = g g'/\sqrt{g^2 + g'{}^2}$, and the Weinberg angle $s_\theta = g'/\sqrt{g^2 + g'{}^2}$.   
The field strength tensors are defined as  
$G_{\mu \nu}^a = \partial_\mu G_\nu^a  - \partial_\nu G_\mu^a + g_s f^{abc} G_\mu^b G_\nu^c $, 
$W_{\mu \nu}^i = \partial_\mu W_\nu^i  - \partial_\nu W_\mu^i + g \epsilon^{ijk}W_\mu^j W_\nu^k $, 
$B_{\mu \nu} = \partial_\mu B_\nu  - \partial_\nu B_\mu$.
The Higgs doublet is denoted as $H$, and we also define $\tilde H_i = \epsilon_{ij} H^*_j$.  
It acquires the VEV  $\langle  H^\dagger H \rangle = v^2/2$. 
In the unitary gauge we have $H = (0,(v + h)/\sqrt 2)$, where $h$ is the Higgs boson field. 
After electroweak symmetry breaking, the electroweak gauge boson mass eigenstates are defined as 
$W^\pm = (W^1 \mp i W^2)/\sqrt 2$, $Z = c_\theta W^3 -  s_\theta B$, $A =  s_\theta W^3 +  c_\theta B$, where $c_\theta = \sqrt{1 - s_\theta^2}$. 
The tree-level  masses of W and Z bosons are given by $m_W = g v/2$, $m_Z = \sqrt{g^2 + g'{}^2} v/2$. 
The left-handed Dirac fermions $q_L= (u_L, d_L)$ and $\ell_L = (\nu_L, e_L)$ are doublets of the SU(2) gauge group, and the right-handed Dirac fermions $u_R$, $d_R$, $e_R$ are SU(2) singlets. 
All fermions are 3-component vectors in the generation space, and $y_f$ are $3 \times 3$ matrices. 
The 3 electroweak  parameters $g$, $g'$, $v$  are customarily derived from the  Fermi constant $G_F$ measured in muon decays, Z boson mass $m_Z$, and the low-energy electromagnetic coupling $\alpha(0)$.  The Higgs quartic couplings $\lambda$ can then be fixed from the measured Higgs boson mass.    
The tree-level relations between the input observables and the electroweak parameters are given by: 
\begin{equation}
\label{eq:EFT_sminput}
G_F = {1 \over \sqrt 2 v^2}, \qquad  \alpha  = {g^2 g'^2 \over 4 \pi (g^2 + g'{}^2)}, \qquad m_Z = {\sqrt {g^2 + g'{}^2 } v \over 2}, \qquad m_h^2 = 2 \lambda v^2. 
\end{equation}

We demand that the dimension-6 operators $O_i^{(6)}$ in Eq.~(\ref{eq:left}) form a complete, non-redundant set -  a  so-called  {\em basis}. 
Complete means that any dimension-6 operator is either a part of the basis or can be obtained from a combination of operators  in the basis using equations of motion, integration by parts, field redefinitions, and Fierz transformations. 
Non-redundant means it is a minimal such set. 
Any complete basis leads to the same physical predictions concerning possible new physics effects.   
Several bases have been proposed in the literature, and they may be convenient for specific applications.   
Historically, a complete and non-redundant set of $D$=6 operators was first identified in Ref.~\cite{Grzadkowski:2010es}, and is usually referred to as the {\em Warsaw basis}. 
This basis is described in detail in Section~\ref{s.eftnlo}., and the relevant formulas are summarized in Appendix~A~of~Ref.~\cite{LHCHXSWG-INT-2015-001}. 
Below, we work with another basis choice commonly used in the literature: the so-called SILH basis ~\cite{Contino:2013kra}. 
Later, in Section.~\ref{sec:hb},  we propose a new basis choice that is particularly convenient for leading-order LHC Higgs analyses in the EFT framework.


\begin{table}[h]
\caption{
\label{tab:silh_bosonic}
Bosonic $D$=6 operators in the SILH basis. 
}
\begin{center}
\small
\begin{minipage}[t]{5cm}
\renewcommand{\arraystretch}{1.5}
\begin{tabular}[t]{c|c}
\multicolumn{2}{c}{Bosonic CP-even} \\
\midrule
$O_{H}$ & $ {1 \over 2 v^2} \left [ \partial_\mu (H^\dag H) \right ]^2$ \\
$O_{T}$   & ${1 \over 2 v^2}  \left (H^\dagger {\overleftrightarrow { D_\mu}} H \right)^2 $  \\
$O_{6}$       & $- {\lambda \over v^2}  (H^\dag H)^3$ \\
$O_{g}$     & $  {g_s^2 \over m_W^2} H^\dagger H\, G^a_{\mu\nu} G^a_{\mu\nu}$ \\
$O_{\gamma}$     & ${g'{}^2 \over m_W^2} H^\dagger H\, B_{\mu\nu} B_{\mu\nu}$ \\
$O_W$ &  $\frac{i g}{2 m_W^2}  \left(H^\dagger \sigma^i {\overleftrightarrow {D_\mu}} H \right) D_\nu W_{\mu\nu}^i$  \\
$O_B$ &  $\frac{ig'}{2 m_W^2} \left(H^\dagger  {\overleftrightarrow { D_\mu}} H \right) \partial_\nu B_{\mu\nu}$ \\
$O_{HW}$ &   $\frac{i g}{m_W^2}  \left(D_\mu H^\dagger \sigma^i  D_\nu H \right) W^i_{\mu\nu}$  \\ 
$O_{HB}$ &   $\frac{i g'}{m_W^2}   \left(D_\mu H ^\dagger D_\nu H \right) B_{\mu\nu} $ \\ 
$O_{2W}$  &  ${1 \over m_W^2} D_\mu W_{\mu \nu}^i  D_\rho W_{\rho \nu}^i $ \\ 
$O_{2B}$  &   ${1 \over m_W^2} \partial_\mu B_{\mu \nu}  \partial_\rho B_{\rho \nu}$ \\
$O_{2G}$ &    ${1 \over m_W^2} D_\mu G_{\mu \nu}^a  D_\rho G_{\rho \nu}^a$  \\
$O_{3W}$                & $ {g^3 \over m_W^2}  \epsilon^{ijk} W_{\mu \nu}^i W_{\nu \rho}^j W_{\rho\mu}^k$ \\
$O_{3G}$                & ${g_s^3 \over m_W^2}  f^{abc} G_{\mu \nu}^{a} G_{\nu\rho}^b G_{\rho\mu}^c $ 
\end{tabular}
\end{minipage}
\qquad
\begin{minipage}[t]{5cm}
\renewcommand{\arraystretch}{1.5}
\begin{tabular}[t]{c|c}
\multicolumn{2}{c}{Bosonic CP-odd} \\
\midrule
 & \\
&  \\
    &  \\
$\widetilde O_{g}$         & ${g_s^2 \over m_W^2}  H^\dag H\, \widetilde G^a_{\mu\nu} G^a_{\mu\nu}$ \\
$\widetilde O_{\gamma}$         & ${g'{}^2 \over m_W^2} H^\dag H\, \widetilde B_{\mu\nu} B_{\mu\nu}$ \\
\\ \\ 
$\widetilde O_{{HW}}$ &   $\frac{i g}{m_W^2}  \left(D_\mu H^\dagger \sigma^i  D_\nu H \right) \widetilde W^i_{\mu\nu}$  \\
$\widetilde O_{{HB}}$ &   $\frac{i g}{m_W^2}  \left(D_\mu H ^\dagger D_\nu H \right)  \widetilde  B_{\mu\nu}$ \\
\\ \\ \\ 
$\widetilde O_{ {3W} }$          & $ {g^3 \over m_W^2}  \epsilon^{ijk} \widetilde  W_{\mu \nu}^i W_{\nu \rho}^j W_{\rho\mu}^k$ \\
$\widetilde O_{ {3G} }$          & $ {g_s^3 \over m_W^2}  f^{abc} \widetilde  G_{\mu \nu}^{a} G_{\nu\rho}^b G_{\rho\mu}^c $ 
\end{tabular}
\end{minipage}

\end{center}
\end{table}

\begin{table}[h]
\caption{
\label{tab:silh_2f}
Two-fermion dimension-6 operators in the SILH basis. 
They are the same as in the Warsaw basis, except that the operators $[O_{H \ell}]_{11}$, $[O_{H \ell}']_{11}$  are absent by definition. 
We define $\sigma_{\mu \nu} = i [\gamma_\mu,\gamma_\nu]/2$. 
In this table,  $e,u,d$ are always right-handed  fermions, while $\ell$ and $q$ are left-handed.
For complex operators  the complex conjugate operator is implicit. 
}
\begin{center}
\small
\begin{minipage}[t]{6.5cm}
\renewcommand{\arraystretch}{1.5}
\begin{tabular}[t]{c|c}
\multicolumn{2}{c}{Vertex} \\
\midrule
$[O_{H \ell }]_{ij}$      & ${i \over v^2}  \bar \ell_i \gamma_\mu  \ell_j H^\dagger \overleftrightarrow {D_\mu} H$\\
$[O_{H \ell}']_{ij}$      & ${i \over v^2}  \bar \ell_i \sigma^k \gamma_\mu  \ell_j  H^\dagger \sigma^k \overleftrightarrow {D_\mu} H $\\
$[O_{H e}]_{ij}$            & ${i \over v^2}   \bar e_i \gamma_\mu \bar   e_j  H^\dagger  \overleftrightarrow {D_\mu} H$\\
$[O_{H q}]_{ij}$      & $ {i \over v^2}  \bar q_i  \gamma_\mu   q_j H^\dagger  \overleftrightarrow {D_\mu} H  $\\
$[O_{H q}']_{ij}$      & ${i \over v^2}  \bar q_i \sigma^k  \gamma_\mu  q_j H^\dagger \sigma^k \overleftrightarrow {D_\mu} H$\\
$[O_{H u}]_{ij}$            & $ {i \over v^2}  \bar u_i \gamma_\mu   u_j  H^\dagger  \overleftrightarrow {D_\mu} H $\\
$[O_{H d}]_{ij}$            & ${i \over v^2}   \bar d_i \gamma_\mu d_j H^\dagger  \overleftrightarrow {D_\mu} H $\\
$[O_{H u d}]_{ij}$  & ${i \over v^2}  \bar u_i \gamma_\mu  d_j  \tilde H^\dagger {D_\mu} H$ \\
\end{tabular}
\end{minipage}
\vspace{0.25cm}
\begin{minipage}[t]{5.2cm}
\renewcommand{\arraystretch}{1.5}
\begin{tabular}[t]{c|c}
\multicolumn{2}{c}{Yukawa and  Dipole} \\
\midrule
$[O_{e}]_{ij}$           & $ {\sqrt{2 m_{e_i} m_{e_j}} \over v^3} H^\dagger H \bar \ell_i H  e_j$ \\
$[O_{u}]_{ij}$          & $ {\sqrt{2 m_{u_i} m_{u_j}} \over v^3}  H^\dagger H \bar q_i  \widetilde H  u_j$ \\
$[O_{d}]_{ij}$           & $  {\sqrt{2 m_{d_i} m_{d_j}} \over v^3} H^\dagger H \bar q_i H d_j$\\
$[O_{eW}]_{ij}$      & ${g \over m_W^2} {\sqrt{2 m_{e_i} m_{e_j}} \over v} \bar \ell_i \sigma^k H \sigma_{\mu\nu} e_j  W_{\mu\nu}^k$ \\
$[O_{eB}]_{ij}$        & ${g' \over m_W^2} {\sqrt{2 m_{e_i} m_{e_j}} \over v} \bar \ell_i H \sigma_{\mu\nu} e_j B_{\mu\nu}$ \\
$[O_{uG}]_{ij}$        & ${g_s \over m_W^2}{\sqrt{2 m_{u_i} m_{u_j}} \over v} \bar q_i \tilde H  \sigma_{\mu\nu} T^a u_j  G_{\mu\nu}^a$ \\
$[O_{uW}]_{ij}$        & ${g \over m_W^2}  {\sqrt{2 m_{u_i} m_{u_j}} \over v} \bar q_i   \sigma^k \tilde H  \sigma_{\mu\nu} u_j  W_{\mu\nu}^k$ \\
$[O_{uB}]_{ij}$        & ${g' \over m_W^2} {\sqrt{2 m_{u_i} m_{u_j}} \over v} \bar q_i \tilde H  \sigma_{\mu\nu} u_j  B_{\mu\nu}$ \\
$[O_{dG}]_{ij}$        & ${g_s \over m_W^2}  {\sqrt{2 m_{d_i} m_{d_j}} \over v} \bar q_i H  \sigma_{\mu\nu} T^a d_j  G_{\mu\nu}^a$ \\
$[O_{dW}]_{ij}$         & ${g \over m_W^2}   {\sqrt{2 m_{d_i} m_{d_j}} \over v}  \bar q_i  \sigma^k H \sigma_{\mu\nu} d_j  W_{\mu\nu}^k$ \\
$[O_{dB}]_{ij}$        & ${g' \over m_W^2}  {\sqrt{2 m_{d_i} m_{d_j}} \over v} \bar q_i H \sigma_{\mu\nu} d_j   B_{\mu\nu}$ 
\end{tabular}
\end{minipage}
\end{center}
\end{table}

\begin{table}[h]
\caption{\label{tab:silh_4f}
Four-fermion operators in the SILH basis. 
They are the same as in the Warsaw basis \cite{Grzadkowski:2010es}, except that  the operators $[O_{\ell \ell}]_{1221}$, $[O_{\ell \ell }]_{1122}$,  $[O_{uu}]_{3333}$ are absent by definition. 
In this table,  $e,u,d$ are always right-handed  fermions, while $\ell$ and $q$ are left-handed. 
A flavour index is implicit for each fermion field.  
For complex operators  the complex conjugate operator is implicit. 
}
\begin{center}
\begin{minipage}[t]{5.5cm}
\renewcommand{\arraystretch}{1.5}
\begin{tabular}[t]{c|c}
\multicolumn{2}{c}{$(\bar LL)(\bar LL)$ and $(\bar LR)(\bar L R)$} \\
\midrule
$O_{\ell\ell}$        & ${1 \over v^2 } (\bar \ell \gamma_\mu \ell)(\bar \ell \gamma_\mu \ell)$ \\
$O_{qq}$  & ${1 \over v^2 } (\bar q \gamma_\mu q)(\bar q \gamma_\mu q)$ \\
$O_{qq}'$  & ${1 \over v^2 } (\bar q \gamma_\mu \sigma^i q)(\bar q \gamma_\mu \sigma^i q)$ \\
$O_{\ell q}$                & ${1 \over v^2 } (\bar \ell \gamma_\mu \ell)(\bar q \gamma_\mu q)$ \\
$O_{\ell q}'$                & ${1 \over v^2 } (\bar \ell \gamma_\mu \sigma^i \ell)(\bar q \gamma_\mu \sigma^i q)$  \\ 
$O_{quqd}$ & ${1 \over v^2 } (\bar q^j u) \epsilon_{jk} (\bar q^k d)$ \\
$O_{quqd}'$ & ${1 \over v^2 } (\bar q^j T^a u) \epsilon_{jk} (\bar q^k T^a d)$ \\
$O_{\ell equ}$ & ${1 \over v^2 } (\bar \ell^j e) \epsilon_{jk} (\bar q^k u)$ \\
$O_{\ell equ}'$ & ${1 \over v^2 } (\bar \ell^j \sigma_{\mu\nu} e) \epsilon_{jk} (\bar q^k \sigma^{\mu\nu} u)$ \\ 
$O_{\ell edq}$ & ${1 \over v^2 } (\bar \ell^j e)(\bar d q^{j})$
\end{tabular}
\end{minipage}
\begin{minipage}[t]{5.25cm}
\renewcommand{\arraystretch}{1.5}
\begin{tabular}[t]{c|c}
\multicolumn{2}{c}{$(\bar RR)(\bar RR)$} \\
\midrule
$O_{ee}$               & ${1 \over v^2 } (\bar e \gamma_\mu e)(\bar e \gamma_\mu e)$ \\
$O_{uu}$        & ${1 \over v^2 } (\bar u \gamma_\mu u)(\bar u \gamma_\mu u)$ \\
$O_{dd}$        & ${1 \over v^2 } (\bar d \gamma_\mu d)(\bar d\gamma_\mu d)$ \\
$O_{eu}$                      & ${1 \over v^2 } (\bar e \gamma_\mu e)(\bar u \gamma_\mu u)$ \\
$O_{ed}$                      & ${1 \over v^2 } (\bar e \gamma_\mu e)(\bar d \gamma_\mu d)$ \\
$O_{ud}$                & ${1 \over v^2 } (\bar u \gamma_\mu u)(\bar d \gamma_\mu d)$ \\
$O_{ud}'$                & ${1 \over v^2 } (\bar u \gamma_\mu T^a u)(\bar d \gamma_\mu T^a d)$ \\
\end{tabular}
\end{minipage}
\begin{minipage}[t]{4.75cm}
\renewcommand{\arraystretch}{1.5}
\begin{tabular}[t]{c|c}
\multicolumn{2}{c}{$(\bar LL)(\bar RR)$} \\
\midrule
$O_{\ell e}$               & ${1 \over v^2 } (\bar  \ell \gamma_\mu \ell)(\bar e \gamma_\mu e)$ \\
$O_{\ell u}$               & ${1 \over v^2 } (\bar  \ell \gamma_\mu \ell)(\bar u \gamma_\mu u)$ \\
$O_{\ell d}$               & ${1 \over v^2 } (\bar \ell \gamma_\mu \ell)(\bar d \gamma_\mu d)$ \\
$O_{qe}$               & ${1 \over v^2 } (\bar q \gamma_\mu q)(\bar e \gamma_\mu e)$ \\
$O_{qu}$         & ${1 \over v^2 } (\bar  q \gamma_\mu q)(\bar u \gamma_\mu u)$ \\ 
$O_{qu}'$         & ${1 \over v^2 } (\bar q \gamma_\mu T^a q)(\bar u \gamma_\mu T^a u)$ \\ 
$O_{qd}$ & ${1 \over v^2 } (\bar q \gamma_\mu q)(\bar d \gamma_\mu d)$ \\
$O_{qd}'$ & ${1 \over v^2 } (\bar q \gamma_\mu T^a q)(\bar d \gamma_\mu T^a d)$\\
\end{tabular}
\end{minipage}
\end{center}
\end{table}

The full set of operators in the SILH basis is given in Tables~\ref{tab:silh_bosonic}, \ref{tab:silh_2f}, and \ref{tab:silh_4f}.
We use the normalization and conventions of Ref.~\cite{Contino:2013kra}.\footnote{%
In Ref.~\cite{Contino:2013kra} it was assumed that the flavour indices of fermionic $D$=6 operators are proportional to the unit matrix.
Generalizing this to an arbitrary  flavour structure, one needs to specify flavour indices of the operators $[O_{H \ell}]$, $[O_{H \ell}']$, $[O_{\ell \ell}]$ and  $[O_{uu}']$ which are absent in the SILH basis to avoid redundancy.   
Here, for concreteness,  we made a particular though somewhat arbitrary choice of these indices.}

\subsection{Effective Lagrangian of mass eigenstates}
\label{sec:pel}

In Section.~\ref{sec:smeft} we introduced an EFT with the SM  supplemented by $D$=6 operators, using a manifestly  $SU(2)\times U(1)$ invariant notation. 
At that point, the connection between the new operators and phenomenology is not obvious. 
To relate to high-energy collider observables, it is more transparent to express the EFT Lagrangian in terms of  the mass eigenstates after electroweak symmetry breaking (Higgs boson, $W$, $Z$, photon, etc.). 
Once this step is made, only the unbroken $SU(3)_c \times U(1)_{\rm em}$ local symmetry is manifest in the Lagrangian.  
Moreover, to simplify the interaction vertices, we will make further field transformations that respect only $SU(3)_c \times U(1)_{\rm em}$.  
Since field redefinitions do not affect physical predictions, the gauge invariance of the EFT we started with ensures that observables calculated using this mass eigenstate Lagrangian are also gauge invariant.   
This is possible because the full $SU(2) \times U(1)$ electroweak symmetry is still present, albeit in a non-manifest way, in the form of  non-trivial relations between different couplings of mass eigenstates.
Finally, for the sake of calculating observables beyond the tree-level one needs to specify the gauge fixing terms.  
Again, the gauge invariance of the starting point ensures that physical observables are independent of the gauge fixing procedure. 
Below we only  present  the Lagrangian in the unitary gauge when the Goldstone bosons eaten by $W$ and $Z$  are set to zero, which is completely sufficient to calculate LHC Higgs observables at tree level;  see Appendix~C~of~Ref.~\cite{LHCHXSWG-INT-2015-001} for a generalization to the $R_\xi$ gauge.


In this section we relate the Wilson coefficients of dimension-6 operators in the SILH basis to the parameters of the tree-level effective Lagrangian describing the interactions of the mass eigenstates.
The analogous relations can be derived for any other basis; see Appendix~A~of~\cite{LHCHXSWG-INT-2015-001} for the map from the Warsaw basis.  The form of the mass eigenstate Lagrangian obtained directly by inserting  the Higgs VEV and eigenstates into Eq.~(\ref{eq:left})  is not convenient for practical applications.
However, at this point one is free to make the following redefinitions of fields and couplings in the Lagrangian: 
\begin{eqnarray}
\label{eq:redef}
G^a_\mu  &\to &  (1 + \delta_G) G^a_\mu ,\quad
W^\pm_\mu  \to   (1 + \delta_W) W^\pm_\mu ,\quad 
Z_\mu \to   (1 + \delta_{Z}) Z_\mu, \quad   
A_\mu \to   (1 + \delta_{A}) A_\mu + \delta_{AZ} Z_\mu  , 
\nonumber \\ 
v & \to &   v (1 + \delta v),  \qquad 
g_s  \to     g_s (1 + \delta g_s),  \qquad  
g  \to     g (1 + \delta g),  \qquad  
g'  \to    g' (1 + \delta g'), \nonumber \\  
\lambda &\to &  \lambda (1  + \delta \lambda),   \qquad 
h \to   (1 +  \delta_1 ) h + \delta_2 h^2/v + \delta_3 h^3/v^2 , 
\end{eqnarray} 
where the free parameters $\delta_i$ are ${\cal O}(\Lambda^{-2})$ in the EFT expansion. 
Note that the non-linear transformation of the Higgs boson field does not generate any new interaction terms at ${\cal O}(\Lambda^{-2})$  in the effective Lagrangian that cannot be generated by $D$=6 operators.\footnote{For example, applied to the $h^4$ self-interaction term in the SM Lagrangian, it generates $h^5$ and $h^6$ self-interactions  at ${\cal O}(\Lambda^{-2})$, which are also generated by the $O_6$ operator in the SILH basis. 
Rather than applying the non-linear transformation, one can equivalently use the equations of motion for the Higgs boson field.} 
In addition, one is free to add to the Lagrangian a total derivative and/or interactions terms that vanish by equations of motion.  
These redefinitions of course do not change the physical predictions or symmetries of the theory. 
However, they allow one to bring the theory to a more convenient form to perform practical calculations.\footnote{Editor footnote: 
Another point of view is expressed in Section~\ref{s.eftnlo}, where it is argued that this kind of transformations make one-loop calculations harder to develop.} 
We will use this freedom to demand that the mass eigenstate Lagrangian has the following features: 
\begin{itemize} 
\item[\#1] All kinetic and mass terms are diagonal and canonically normalized. 
In particular, higher-derivative kinetic terms are absent. 
\item[\#2] 
The non-derivative photon and  gluon interactions with fermions are the same as in the SM.  
\item[\#3] Tree-level relations between the electroweak parameters and input observables are the same as the SM ones in  Eq.~(\ref{eq:EFT_sminput}).   
\item[\#4] 
Two-derivative self-interactions of the Higgs boson (e.g. $h \partial_\mu h \partial_\mu h$) are absent. 
\item[\#5] In the Higgs boson interactions with gauge bosons, the derivative does not act on the Higgs (e.g., there is no $\partial_\mu h  V_\nu V_{\mu \nu}$ terms). 
\item[\#6] 
For each fermion pair, the coefficient of the vertex-like Higgs interaction terms 
$\left ( 2 {h \over v}  + {h^2 \over v^2} \right) V_\mu \bar f \gamma_\mu f$  is equal to the vertex correction to the respective $V_\mu \bar f \gamma_\mu f$ interaction. 
\end{itemize} 
These conditions are a choice of conventions (one among many possible ones) how to represent interactions in the mass eigenstate Lagrangian. 
It is always possible to implement this choice starting from any D=6 basis: SILH, Warsaw, or any other.  
 The condition \#1 simplifies extracting physical predictions of the EFT,  and is essential to implement the theory in existing Monte Carlo simulators. 
The conditions \#2-\#3 simplify the interpretation of the SM parameters  $g$, $g'$ and $v$. 
If the [$G_F$, $\alpha$, $m_Z$] input is used to determine them (as assumed here), their numerical values should be the same as in the SM, and the input observables are not affected by $D$=6 operators at the leading order.\footnote{If other input observables are used, for example  [$G_F$, $m_W$, $m_Z$] or   [$\alpha$, $m_W$ $m_Z$], the shift of input observables due to the presence of $D$=6 operators must be taken into account to correctly derive physical predictions of the theory. Much as in the SM,  the input observables  [$G_F$, $\alpha$, $m_Z$]  are affected by loop corrections, and this has to be taken into account if the framework is used beyond tree level.}
The conditions \#4-\#6 are conventions commonly used in the literature that allow one to fix the remaining freedom of fields and couplings redefinitions. 
These particular conventions match the ones used e.g. in the Higgs characterization framework of Ref.~\cite{Artoisenet:2013puc}. 
See Appendix~D~of~Ref.~\cite{LHCHXSWG-INT-2015-001} for physical examples showing these redefinitions do not change the S-matrix. 
Other convention choices can be made, leading to the same predictions for observables. 
For example, the features  \#3, \#4, and  \#6 are not enforced in Section~\ref{s.eftnlo}. 

In general, dimension-6 operators do induce interaction terms  that do not respect the   features \#1-\#6.
However, these features can always be achieved, {\em without any loss of generality}, 
by using  equations of motion, integrating by parts, and redefining the fields and couplings.   
Starting from  the SILH  basis, the conditions \#1-\#6 fix the free parameters in  Eq.~(\ref{eq:redef}) as 
\begin{eqnarray}
\delta_{G} & =  &  {4 g_s^2 \over g^2} \bar c_g,  \nonumber \\ 
\delta_W  & =  & \bar c_W,  \nonumber \\ 
\delta_Z  & =  & \bar c_W +  {g'{}^2 \over g^2 }\bar c_B + {4 g'{}^4 \over g^2 (g^2 + g'{}^2)} \bar c_\gamma,   \nonumber \\ 
\delta_{AZ}  & =  &  {g'{} \over g }\left (  \bar c_W  -  \bar c_B  \right )
-  {8 g'{}^3 \over g (g^2 + g'{}^2)} \bar c_\gamma,   \nonumber \\ 
\delta_A & =  &  {4 g'{}^2 \over g^2 + g'{}^2} \bar c_\gamma  \nonumber \\ 
\delta v & =  & {[\bar c'_{H \ell}]_{22} \over 2},   \nonumber \\  
 \delta g_s & =  &  - {4 g_s^2 \over g^2} \bar c_g, \nonumber \\ 
\delta g  & =  & -   {g^2  \over g^2 - g'{}^2} \left ( 
 \bar c_W  +  \bar c_{2W}  
 +  {g'{}^2 \over g^2 }\bar c_B  +  {g'{}^2 \over g^2 }\bar c_{2B}
-  {1 \over 2}  \bar c_T  +  {1 \over 2} [\bar c'_{H \ell}]_{22} \right ), 
 \nonumber \\ 
 \delta g'  & =  &  {g'{}^2  \over g^2 - g'{}^2} \left ( 
 \bar c_W  + \bar c_{2W}  + {g'{}^2 \over g^2 }\bar c_B  +  {g'{}^2 \over g^2 }\bar c_{2B}
-  {1 \over 2}  \bar c_T  + {1 \over 2} [\bar c'_{H \ell}]_{22} - 4{g^2 - g'{}^2 \over g^2} \bar c_\gamma  \right ), 
\nonumber \\ 
\delta \lambda & =  & \bar c_H - {3 \over 2} \bar c_6  - [\bar c'_{H \ell}]_{22}, 
 \nonumber \\ 
\delta_1  & =  &  - {\bar c_H \over 2}, \qquad 
\delta_2   =    - {\bar c_H \over 2},  \qquad 
\delta_3   =    - {\bar c_H \over 6}. 
\end{eqnarray}
Finally, the Higgs boson mass term in the SM Lagrangian is related by vacuum equations  to the other parameters by 
$\mu_H^2 = \lambda v^2 (  1 + \delta \lambda + 2 \delta v + 3/4 \bar c_6) $. 
One can repeat this procedure starting from any other basis than SILH, and find a unique solution to the conditions \#1-\#6 in terms of the Wilson coefficients in that basis.   

We move to  discussing the interactions in the mass eigenstate Lagrangian  once conditions   \#1-\#6 are satisfied. 
We will focus on interaction terms that are most relevant for LHC phenomenology. 
To organize the presentation, we split the Lagrangian into the following parts, 
\begin{equation}
\label{eq:PEL_left}
 {\cal L}_{\rm EFT} 
=  {\cal L}_{\rm kinetic}  +   {\cal L}_{\rm aff}  +   {\cal L}_{\rm vertex}   +  {\cal L}_{\rm dipole} 
+  {\cal L}_{\rm tgc}   
+  {\cal L}_{\rm hff}  
    +  {\cal L}_{\rm hvv}    + {\cal L}_{hvff} +  {\cal L}_{hdvff}   + L_{h,\rm self}  + {\cal L}_{h^2}    
+  {\cal L}_{\rm other}.  
\end{equation}   
Below we define each term in order of appearance. 
We also express the corrections to the SM interactions in ${\cal L}_{\rm EFT} $  in terms of linear combinations of Wilson coefficients of $D$=6 operators in the SILH basis (the analogous formulas for the Warsaw basis are given in Appendix~A~of~Ref.~\cite{LHCHXSWG-INT-2015-001}.   
These corrections start at  ${\cal O}(1/\Lambda^2)$ in the EFT expansion, and we will ignore all  ${\cal O}(1/\Lambda^4)$ and higher contributions.

\subsubsection*{Kinetic Terms}

By construction, the kinetic terms of the mass eigenstates are diagonal and canonically normalized: 
\begin{eqnarray} 
\label{eq:PEL_lkin}
{\cal L}_{\rm kinetic} &= & 
 - {1 \over 2} W_{\mu \nu}^+  W_{\mu \nu}^- -   {1 \over 4} Z_{\mu \nu} Z_{\mu \nu} 
-   {1 \over 4} A_{\mu \nu} A_{\mu \nu}   - {1 \over 4} G_{\mu \nu}^a  G_{\mu \nu}^a
\nonumber \\  & + &  
{g^2 v^2 \over 4} \left (1+ \delta m \right )^2  W_{\mu}^+  W_{\mu}^-  
+ {(g^2 + g^{\prime 2}) v^2  \over 8 } Z_\mu Z_\mu 
\nonumber \\   & + & 
 {1 \over 2} \partial_\mu h \partial_\mu h  -   \lambda v^2 h^2 
 +  \sum_{f \in q,\ell, u,d,e}  \bar f \left (  i \gamma_\mu \partial_\mu - m_f \right ) f . 
\end{eqnarray} 
Above, the parameter $\lambda$ is defined by the tree-level relation $m_h^2 = 2 \lambda v^2$. 
There is no correction to the Z boson mass terms, in accordance with the condition $\#3$. 
With this convention, the corrections to the W boson mass cannot be in general redefined away, and are parameterized by $\delta m$. 
The relation between $\delta m$ and the Wilson coefficients in the SILH basis is given by  
\begin{eqnarray}
\label{eq:PEL_dm}
\delta m &=&  - {g'{}^2 \over g^2 - g'{}^2} 
\left ( \bar c_W + \bar c_B +  \bar c_{2W} + \bar c_{2B}  - {g^2  \over 2 g'{}^2} \bar c_T + {1  \over 2} [\bar c'_{H\ell}]_{22}  \right ) . 
\end{eqnarray}  

\subsubsection*{Gauge boson interactions with fermions}
\label{sec:ewpt}

By construction (condition $\#2$), the non-derivative photon and gluon  interactions with fermions are the same as in the SM: 
\begin{eqnarray} 
\label{eq:PEL_laff}
{\cal L}_{aff} & =& 
e A_\mu \sum_{f  \in u,d,e}  \bar f \gamma_\mu Q_f  f  
+ g_s G_\mu^a \sum_{f  \in u,d} \bar f \gamma_\mu T^a  f .  
\end{eqnarray} 
The analogous interactions of the W and Z boson may in general be affected by dimension-6 operators: 
\begin{eqnarray} 
\label{eq:PEL_lvertex}
{\cal L}_{\rm vertex} &= &
{g \over \sqrt 2}  \left ( 
  W_\mu^+  \bar \nu_L \gamma_\mu \left (I_3 + \delta g^{W\ell}_L \right) e_L + 
W_\mu^+ \bar u_L \gamma_\mu  \left (I_3 +  \delta g^{Wq}_L  \right)  d_L 
 +  W_\mu^+  \bar u_R \gamma_\mu \delta g^{Wq}_R  d_R  + {\mathrm h.c.} \right )
\nonumber \\ 
&+& \sqrt{g^2 + g'{}^2} Z_\mu   \left [  
   \sum_{f \in u, d,e,\nu}  \bar f_L  \gamma_\mu \left (T^3_f - s_\theta^2 Q_f +   \delta g^{Zf}_L  \right)  f_L 
+    \sum_{f \in u, d,e}  \bar f_R  \gamma_\mu \left ( - s_\theta^2 Q_f  +   \delta g^{Zf}_R  \right)  f_R   \right ].
\nonumber \\ 
\end{eqnarray} 
Here, $I_3$ is the $3 \times 3$ identity matrix, and the {\em vertex corrections} $\delta g$ are $3 \times 3$ Hermitian matrices in the generation space, except for  $\delta g^{Wq}_R$ which  is a general $3 \times 3$ complex matrix. 
The vertex corrections  to W and Z boson couplings to fermions  are expressed by the Wilson coefficients in the SILH  basis as 
\begin{eqnarray}
\label{eq:PEL_dg}
\delta g^{Z \nu }_L & = &    {1 \over 2} \bar c'_{H\ell} - {1\over 2}  \bar c_{H\ell}   + \hat f(1/2,0),  
\nonumber \\
\delta g^{Ze}_L & = &    - {1 \over 2}  \bar c'_{H\ell} - {1\over 2}  \bar c_{H\ell}    +  \hat  f(-1/2, -1), 
\nonumber \\
\delta g^{Ze}_R & = &    - {1\over 2} \bar c_{He}   + \hat f(0, -1),  
\nonumber \\
\delta g^{Zu}_L & = &   {1 \over 2}  \bar c'_{Hq} - {1\over 2} \bar c_{Hq}   +\hat f(1/2,2/3), 
\nonumber \\
\delta g^{Zd}_L & = &    -{1 \over 2} V_{\rm CKM}^\dagger  \bar c'_{Hq}  V_{\rm CKM}   - {1\over 2}  V_{\rm CKM}^\dagger  \bar c_{Hq}  V_{\rm CKM}    + \hat f(-1/2,-1/3),
\nonumber \\
\delta g^{Zu}_R & = &    - {1\over 2} \bar c_{Hu}   + \hat f(0,2/3),  
\nonumber \\
\delta g^{Zd}_R & = &    - {1\over 2} \bar c_{Hd}  + \hat f(0,-1/3),
\nonumber \\
\delta g^{W \ell}_L & = &   \bar c'_{H \ell} + \hat f(1/2,0) -  \hat f(-1/2,-1),
\nonumber \\
\delta g^{Wq}_L & = &   \left ( \bar c'_{Hq}   + \hat f(1/2,2/3) -  \hat f(-1/2,-1/3) \right ) V_{\rm CKM},
\nonumber \\  
\delta g^{Wq}_R & = &   - {1 \over 2 } \bar c_{Hud}, 
\end{eqnarray}
where  
\begin{eqnarray}
\hat f(T^3_f,Q_f) &\equiv&  
\left [  \bar c_{2W} +  {g'{}^2 \over g^2} \bar c_{2B} + {1 \over 2}  \bar c_T - {1 \over 2}   [\bar c'_{H \ell}]_{22}  \right ] T^3_f
\nonumber \\ &-  &  {g'{}^2 \over  (g^2 - g'{}^2)} 
\left  [ { (2 g^2 - g'{}^2) \over g^2} \bar c_{2B}   +  \bar c_{2W} + \bar c_W +  \bar c_B  - {1 \over 2}  \bar c_T +  {1 \over 2}  [\bar c'_{H \ell}]_{22}   \right ] Q_f , 
\nonumber \\
\end{eqnarray}
and it is  implicit that $[\bar c'_{H \ell}]_{11}= [\bar c_{H \ell}]_{11}=0$. 

Another type of gauge boson interactions with fermions   are the so-called dipole interactions. 
These do not occur  in the tree-level  SM Lagrangian, but they in general may appear in the EFT with $D$=6 operators. 
We parameterize  them as follows: 
\begin{eqnarray}
\label{eq:PEL_ldipole}
\! \! \! \! \!
{\cal L}_{\rm dipole}  &=& 
- {1 \over 4 v}  \left [ 
g_s \sum_{f \in u,d}    {\sqrt{m_{f_i} m_{f_j}} \over v} \bar f_{L,i}  \sigma_{\mu \nu} T^a  [d_{Gf}]_{ij}  f_{R,j}  G_{\mu\nu}^a 
+ e  \sum_{f \in u,d,e}   {\sqrt{m_{f_i} m_{f_j}} \over v}  \bar f_{L,i}  \sigma_{\mu \nu}  [d_{A f}]_{ij}  f_{R,j}   A_{\mu\nu} 
\right . \nonumber \\ && \left . 
+ \sqrt{g^2 + g'{}^2}  \sum_{f \in u,d,e}   {\sqrt{m_{f_i} m_{f_j}} \over v} 
\bar f_{L,i}  \sigma_{\mu \nu}  [d_{Zf}]_{ij}  f_{R,j}   Z_{\mu\nu}
\right . \nonumber \\ && \left . 
+ \sqrt {2} g 
   {\sqrt{m_{u_i} m_{u_j}} \over v}  \bar d_{L,i}  \sigma_{\mu \nu}  [d_{Wu}]_{ij}  u_{R,j} W_{\mu\nu}^- 
+ \sqrt {2} g    {\sqrt{m_{d_i} m_{d_j}} \over v}   \bar u_{L,i}  \sigma_{\mu \nu}  [d_{Wd}]_{ij}  d_{R,j} W_{\mu\nu}^+ 
 \right . \nonumber \\ && \left . 
+ \sqrt {2} g   {\sqrt{m_{e_i} m_{e_j}} \over v}  \bar \nu_{L,i}  \sigma_{\mu \nu}  [d_{We}]_{ij}  e_{R,j}  W_{\mu\nu}^+    \right ] + {\rm h.c.} ,
  \nonumber \\   
\end{eqnarray} 
where $\sigma_{\mu\nu}  = i[\gamma_\mu, \gamma_\nu]/2$, and $d_{G f}$, $d_{A f}$,   $d_{Z f}$,  and  $d_{Wf}$ are complex $3 \times 3$ matrices.  
The field strength tensors are defined as $X_{\mu \nu} = \partial_\mu X_\nu - \partial_\nu X_\mu$, and $\tilde X_{\mu \nu}  = \epsilon_{\mu\nu\rho\sigma} \partial_\rho X_\sigma$. 
The coefficients $d_{vf}$ are related to the Wilson coefficients in the SILH  basis as 
\begin{eqnarray}
\label{eq:PEL_dipole2}
d_{Gf}  &=& - {16 \over g^2} \bar c_{fG},
\nonumber \\ 
d_{A f}   &=& - {16 \over g^2}   \left ( \eta_f \bar c_{fW} + \bar c_{fB} \right ),
\nonumber \\  
d_{Z f}  &=& - {16 \over g^2}  \left (  \eta_f  c_\theta^2 \bar c_{fW} - s_\theta^2 \bar c_{fB} \right ),
\nonumber \\ 
d_{Wf} &= & - {16 \over g^2} \bar  c_{fW},   
\end{eqnarray} 
where $\eta_u = +1$, $\eta_{d,e} = -1$.

\subsubsection*{Gauge boson self-interactions}

Gauge boson self-interactions are not directly relevant for LHC Higgs searches, 
however we include them in this presentation  because of the important synergy between the triple gauge couplings and Higgs boson couplings measurements \cite{Dumont:2013wma,Pomarol:2013zra,Corbett:2013pja,Masso:2014xra,Ellis:2014jta,Falkowski:2015jaa,Butter:2016cvz}. 
The triple gauge interactions in the effective Lagrangian are parameterized by 
\begin{eqnarray} 
\label{eq:PEL_ltgc}
 {\cal L}_{\rm tgc}  &= &  
 i  e    \left ( W_{\mu \nu}^+ W_\mu^-  -  W_{\mu \nu}^- W_\mu^+ \right ) A_\nu   
 + 
i  e  \left [  (1 + \delta \kappa_\gamma )  A_{\mu\nu}\,W_\mu^+W_\nu^-   
+ \tilde \kappa_\gamma  \tilde A_{\mu\nu}\,W_\mu^+W_\nu^-  \right ]
\nonumber \\  &  + & i g c_\theta  \left [  (1 + \delta g_{1,z})   \left ( W_{\mu \nu}^+ W_\mu^-  -  W_{\mu \nu}^- W_\mu^+ \right ) Z_\nu 
 + (1 +  \delta \kappa_z) \, Z_{\mu\nu}\,W_\mu^+W_\nu^-   +  \tilde  \kappa_z \,  \tilde Z_{\mu\nu}\,W_\mu^+W_\nu^-   \right ] 
  \nonumber \\  & + &    
 i {e  \over m_W^2 } \left [  \lambda_\gamma W_{\mu \nu}^+W_{\nu \rho}^- A_{\rho \mu}   +  \tilde  \lambda_\gamma W_{\mu \nu}^+W_{\nu \rho}^- \tilde A_{\rho \mu} \right]
  +  i {g  c_\theta \over m_W^2}  \left [ \lambda_z W_{\mu \nu}^+W_{\nu \rho}^- Z_{\rho \mu} +  \tilde \lambda_z W_{\mu \nu}^+W_{\nu \rho}^-  \tilde Z_{\rho \mu} \right ]
  \nonumber \\  & - &   
   g_s  f^{abc} \partial_\mu G_{\nu}^a G_\mu^b G_\nu^c 
 + {c_{3g} \over v^2} g_s^3 f^{abc} G_{\mu \nu}^{a} G_{\nu\rho}^b G_{\rho\mu}^c  
 +{ \tilde c_{3g}  \over v^2}  g_s^3 f^{abc} \widetilde  G_{\mu \nu}^{a} G_{\nu\rho}^b G_{\rho\mu}^c .
\end{eqnarray} 
The couplings of electroweak gauge bosons follow the customary parameterization of Ref.~\cite{Hagiwara:1993ck}.
The anomalous triple gauge couplings of electroweak gauge bosons are related to the Wilson coefficients in the SILH basis as  
\begin{eqnarray}
\label{eq:PEL_tgc}
\delta g_{1z} &=&
 - {g^2 + g'{}^2 \over g^ 2 - g'{}^2} \left [
 {g^2 - g'{}^2 \over g^2 } \bar c_{HW} + \bar c_W + \bar c_{2W} + {g'{}^2 \over g^2} \bar c_B +  {g'{}^2 \over g^2} \bar c_{2B}
  - {1 \over 2}  \bar c_T +  {1 \over 2}   [\bar c'_{H \ell}]_{22} \right ],
\nonumber \\  
\delta \kappa_\gamma  &=& 
-  \bar c_{HW}  - \bar c_{HB},
\nonumber \\ 
\delta \kappa_{z} &=&  
 - \bar c_{HW} +  {g'{}^2 \over g^2}  \bar c_{HB} 
- {g^2 + g'{}^2 \over g^ 2 - g'{}^2} \left [
 \bar c_W + \bar c_{2W} + {g'{}^2 \over g^2} \bar c_B +  {g'{}^2 \over g^2} \bar c_{2B}
  - {1 \over 2}  \bar c_T +  {1 \over 2}   [\bar c'_{H \ell}]_{22} \right ],
\nonumber \\  
\lambda_z &= &  - 6  g^2 \bar c_{3W},  \qquad \lambda_\gamma = \lambda_z, 
\nonumber \\ 
\delta \tilde \kappa_\gamma  &=& 
- \tilde c_{HW} -  \tilde c_{HB},
\nonumber \\ 
\delta \tilde \kappa_z  &=& 
{g'{}^2 \over g^2} \left [\tilde c_{HW} + \tilde c_{HB} \right ],
\nonumber \\
\tilde \lambda_z &= &  - 6 g^2 \tilde c_{3W}, \qquad \tilde \lambda_\gamma = \tilde \lambda_z ,
\nonumber \\
c_{3g} &=& {4 \over g^2} \bar c_{3 G}, \qquad  \tilde c_{3g} =  {4 \over g^2} \tilde c_{3 G} .  
\end{eqnarray} 
The tilded Wilson coefficients refer to the tilded (CP-odd) operators
in Table~\ref{tab:silh_bosonic}.

\subsubsection*{Single Higgs boson couplings}
\label{sec:singlehiggs}

In this subsection we discuss the terms in  the effective Lagrangian that involve a single Higgs boson field $h$.
This part is the most relevant one from the point of view of the LHC Higgs phenomenology.

We first define the Higgs boson couplings to a pair of fermions: 
\begin{eqnarray}
\label{eq:PEL_lhff}
 {\cal L}_{\rm hff} &= & 
-{h \over v} \sum_{f \in u,d,e} \sum_{ij}    
\sqrt{m_{f_i} m_{f_j}} \left ( \delta_{ij} + [\delta y_f]_{ij}  e^{i [\phi_f]_{ij}}   \right ) \bar f_{R,i} f_{L,j} + {\rm h.c.},
 \end{eqnarray} 
 where $ [\delta y_f]_{ij} $ and $ \phi_{ij}$ are general $3 \times 3$  matrices with real elements.  
The corrections to the SM Yukawa interactions are related to the Wilson coefficients in the SILH basis by 
\begin{eqnarray}
\label{eq:PEL_dy}
\, [\delta y_f]_{ij}  e^{i  [\phi_f]_{ij}}   & =&  
- [\bar c_f]_{ij}  -  \delta_{ij} {1 \over 2}  \left [\bar c_H   +  [\bar c'_{H \ell}]_{22} \right ]. 
\end{eqnarray}
 
Next, we define the following single Higgs boson couplings  to a pair of the SM gauge fields: 
\begin{eqnarray}
\label{eq:PEL_lhvv}
 {\cal L}_{\rm hvv} &= & {h \over v} \left [ 
  \left (1 +  \delta c_w \right )  {g^2 v^2 \over 2} W_\mu^+ W_\mu^- 
  +    \left (1 +  \delta c_z \right )  {(g^2+g^{\prime 2}) v^2 \over 4} Z_\mu Z_\mu
\right . \nonumber \\ & & \left . 
+ c_{ww}  {g^2 \over  2} W_{\mu \nu}^+  W_{\mu\nu}^-  + \tilde c_{ww}  {g^2 \over  2} W_{\mu \nu}^+   \tilde W_{\mu\nu}^- 
+ c_{w \Box} g^2 \left (W_\mu^- \partial_\nu W_{\mu \nu}^+ + {\mathrm h.c.} \right )  
\right . \nonumber \\ & & \left . 
+  c_{gg} {g_s^2 \over 4 } G_{\mu \nu}^a G_{\mu \nu}^a   + c_{\gamma \gamma} {e^2 \over 4} A_{\mu \nu} A_{\mu \nu} 
+ c_{z \gamma} {e \sqrt{g^2 + g'{}^2}  \over  2} Z_{\mu \nu} A_{\mu\nu} + c_{zz} {g^2 + g'{}^2 \over  4} Z_{\mu \nu} Z_{\mu\nu}
\right . \nonumber \\ & & \left . 
+c_{z \Box} g^2 Z_\mu \partial_\nu Z_{\mu \nu} + c_{\gamma \Box} g g' Z_\mu \partial_\nu A_{\mu \nu}
\right . \nonumber \\ & & \left . 
+  \tilde c_{gg} {g_s^2 \over 4} G_{\mu \nu}^a \tilde G_{\mu \nu}^a  
+ \tilde c_{\gamma \gamma} {e^2 \over 4} A_{\mu \nu} \tilde A_{\mu \nu} 
+ \tilde c_{z \gamma} {e \sqrt{g^2 + g'{}^2} \over  2} Z_{\mu \nu} \tilde A_{\mu\nu}
+ \tilde c_{zz}  {g^2 + g'{}^2  \over  4} Z_{\mu \nu} \tilde Z_{\mu\nu}
\right ],  
\nonumber \\ 
 \end{eqnarray} 
 where all the couplings above are real. 
The terms in the first two lines describe corrections to the SM Higgs boson couplings to W and Z, while the remaining terms introduce Higgs boson couplings to gauge bosons with a tensor structure that is absent in the SM Lagrangian. 
  Note that, using equations of motion, we could get rid of certain 2-derivative interactions between the Higgs and  gauge bosons: 
$h Z_\mu \partial_\nu Z_{\nu \mu}$,  $h Z_\mu \partial_\nu A_{\nu \mu}$,  and $h W_\mu^\pm \partial_\nu W_{\nu \mu}^\mp$.
These interactions would then be traded for contact interactions of the Higgs, gauge bosons and fermions in Eq.~(\ref{eq:PEL_lvertex}).
However, one of the defining features of our effective Lagrangian is that the coefficients of the latter couplings are equal to the corresponding vertex correction in Eq.~(\ref{eq:PEL_lvertex}).
This form can be always obtained, without any loss of generality,  starting from an arbitrary dimension-6 Lagrangian provided the 2-derivative $h V_\mu \partial_\nu V_{\nu \mu}$ are kept in the Lagrangian. 
Note that we work in the limit where the neutrinos are massless and the Higgs boson does not couple to the neutrinos. In the EFT context, the couplings to neutrinos induced by dimension-5 operators are proportional to neutrino masses, therefore they are far too small to have any relevance for LHC phenomenology. 

The shifts  of the Higgs boson couplings to W and Z bosons are related to the Wilson coefficients in the  SILH basis by 
\begin{eqnarray}
\label{eq:PEL_dcv}
\delta c_w &= & - {1 \over 2} \bar c_H -  { 1 \over g^2 - g'{}^2} \left [
4 g'{}^2 (\bar c_W +  \bar c_B +  \bar c_{2B} + c_{2W} ) 
 - 2 g^2 \bar c_T + {3 g^2 + g'{}^2 \over 2} [\bar c'_{H \ell}]_{22}  \right ],
\nonumber \\ 
\delta c_z &= &  - {1 \over 2} \bar c_H   -  {3 \over 2}  [\bar c'_{H \ell}]_{22} .
\end{eqnarray} 
The two-derivative Higgs boson couplings to gauge bosons are related to the Wilson coefficients in the SILH basis by  
\begin{eqnarray}
\label{eq:PEL_cvv}
c_{gg} &=&{16 \over g^2}  \bar c_{g}, 
\nonumber \\
c_{\gamma \gamma} &= & {16 \over g^2} \bar c_{\gamma},  
\nonumber \\  
c_{zz} &= & -{4 \over g^2 + g'{}^2} \left [   \bar c_{HW} + {g'{}^2 \over g^2} \bar c_{HB}  - 4 {g'{}^2 \over g^2 } s_\theta^2 \bar c_\gamma \right ],
\nonumber \\ 
c_{z\Box} &= &  \frac{2}{g^2} \left[ 
\bar c_W + \bar c_{HW} +  \bar c_{2W}  
+ {g'{}^2 \over g^2} ( \bar c_B + \bar c_{HB} + \bar c_{2B})  - {1 \over 2} \bar c_T  + {1 \over 2} [\bar c'_{H \ell}]_{22}  \right ],
\nonumber \\ 
c_{z\gamma} &= & {2 \over g^2}  \left ( \bar c_{HB} - \bar c_{HW}  - 8 s_\theta^2 \bar c_{\gamma} \right ), 
\nonumber \\ 
c_{\gamma \Box} &= & {2 \over g^2} \left ( \bar c_{HW} -\bar c_{HB} \right )
  + {4  \over g^2 - g'{}^2} \left [ 
  \bar c_W+  \bar c_{2W}  +  {g'{}^2 \over g^2 }  (\bar c_B + \bar c_{2B} ) - {1 \over 2} \bar c_T + {1 \over 2} [\bar c'_{H \ell}]_{22} \right ],
\nonumber \\ 
c_{ww} & = & - {4 \over g^2 } \bar c_{HW},   
\nonumber \\ 
c_{w \Box} &=&  {2 \bar c_{HW} \over g^2}    + {2  \over g^2 - g'{}^2} \left [ 
 \bar c_W+   \bar c_{2W}  +  {g'{}^2 \over g^2 } (\bar c_B + \bar c_{2B} ) - {1 \over 2} \bar c_T + {1 \over 2} [\bar c'_{H \ell}]_{22} \right ], 
\end{eqnarray} 
\begin{eqnarray}
\label{eq:PEL_cvvodd}
 \tilde c_{gg} &=&{16 \over g^2}   \tilde c_{g}, 
\nonumber \\
 \tilde c_{\gamma \gamma} &= & {16 \over g^2}  \tilde c_{\gamma},  
\nonumber \\  
 \tilde c_{zz} &= & -{4 \over g^2 + g'{}^2} \left [    \tilde c_{HW} + {g'{}^2 \over g^2}  \tilde c_{HB}  - 4 {g'{}^2 \over g^2 } s_\theta^2  \tilde c_\gamma \right ],
\nonumber \\ 
 \tilde c_{z\gamma} &= & {2 \over g^2}  \left (  \tilde c_{HB} -   \tilde c_{HW}  - 8 s_\theta^2  \tilde c_{\gamma} \right ), 
\nonumber \\ 
\tilde c_{ww} & = & - {4 \over g^2 } \tilde c_{HW}. 
\end{eqnarray} 

Next, couplings of the Higgs boson to a gauge field and two fermions (which are not present in the SM Lagrangian) can be generated by dimension-6 operators. 
The vertex-like contact interactions between the Higgs, electroweak gauge bosons, and fermions are parameterized as:    
\begin{eqnarray}
\label{eq:PEL_lhvff}
{\cal L}_{hvff} &=&  \sqrt 2  g {h \over v}   W_\mu^+   \left (
  \bar u_L \gamma_\mu \delta g^{hWq}_L  d_L
+   \bar u_R \gamma_\mu \delta g^{hWq}_R  d_R
+    \bar \nu_L \gamma_\mu \delta g^{h W \ell }_L e_L \right ) + {\mathrm h.c.} 
\nonumber \\
&+& 2 {h \over v}  \sqrt{g^2 + g'{}^2} Z_\mu 
\left [ \sum_{f = u,d,e,\nu}   \bar f_L \gamma_\mu \delta g^{hZf}_L f_L  +  
\sum_{f = u,d,e}    \bar f_R \gamma_\mu \delta g^{hZf}_R f_R  \right ] .
\end{eqnarray}
By construction (condition \#6), the coefficients of these interaction are equal to the corresponding  vertex correction in Eq.~(\ref{eq:PEL_lvertex}): 
\begin{equation}
\label{eq:contact}
 \delta g^{hZf}  =  \delta g^{Zf}, \qquad   \delta g^{hWf}  =  \delta g^{Wf}. 
 \end{equation}
 
The  dipole-type contact interactions of the Higgs boson are parameterized as:  
\begin{eqnarray}
\label{eq:PEL_lhdvff}
\! \! \! \! \!
{\cal L}_{hdvff}  &=& 
- {h \over 4 v^2}  \left [ 
g_s \sum_{f \in u,d}    {\sqrt{m_{f_i} m_{f_j}} \over v} \bar f_{L,i}  \sigma_{\mu \nu} T^a  [d_{hGf}]_{ij}  f_{R,j}  G_{\mu\nu}^a 
\right . \nonumber \\ && \left . 
+ e  \sum_{f \in u,d,e}   {\sqrt{m_{f_i} m_{f_j}} \over v}  \bar f_{L,i}  \sigma_{\mu \nu}  [d_{hA f}]_{ij}  f_{R,j}   A_{\mu\nu} 
\right . \nonumber \\ && \left . 
+ \sqrt{g^2 + g'{}^2}  \sum_{f \in u,d,e}   {\sqrt{m_{f_i} m_{f_j}} \over v} 
\bar f_{L,i}  \sigma_{\mu \nu}  [d_{hZf}]_{ij}  f_{R,j}   Z_{\mu\nu}
\right . \nonumber \\ && \left . 
+ \sqrt {2} g 
   {\sqrt{m_{u_i} m_{u_j}} \over v}  \bar d_{L,i}  \sigma_{\mu \nu}  [d_{hWu}]_{ij}  u_{R,j} W_{\mu\nu}^- 
+ \sqrt {2} g    {\sqrt{m_{d_i} m_{d_j}} \over v}   \bar u_{L,i}  \sigma_{\mu \nu}  [d_{hWd}]_{ij}  d_{R,j} W_{\mu\nu}^+ 
 \right . \nonumber \\ && \left . 
+ \sqrt {2} g   {\sqrt{m_{e_i} m_{e_j}} \over v}  \bar \nu_{L,i}  \sigma_{\mu \nu}  [d_{hWe}]_{ij}  e_{R,j}  W_{\mu\nu}^+    \right ] + {\rm h.c.} ,
\end{eqnarray} 
where $d_{hGf}$, $d_{hA f}$, $d_{hZ f}$,  and  $d_{hWf}$  are general complex $3 \times 3$ matrices.
The coefficients are simply related to the corresponding dipole interactions in Eq.~(\ref{eq:PEL_ldipole}):
\begin{equation}
d_{hVf} = d_{Vf}. 
\end{equation} 

Dimension-6 operators can also induce single Higgs boson couplings to more than 2 gauge bosons, but we do  not display them here.  

\subsubsection*{Higgs boson self-couplings and double Higgs boson couplings}

The cubic Higgs boson self-coupling and couplings of two Higgs boson fields to matter play a role in the EFT description of double Higgs boson production \cite{Contino:2010mh,Goertz:2014qta}. 
The cubic Higgs boson self-coupling is parameterized as 
\begin{equation} 
\label{eq:PEL_lhself}
{\cal L}_{h, \rm self} = - (\lambda +  \delta \lambda_3)  v h^3. 
\end{equation} 
The relation between the cubic Higgs boson coupling correction and the Wilson coefficients in the SILH basis is given by  
\begin{eqnarray} 
 \delta \lambda_3   &=& 
   \lambda  \left ( \bar c_{6} -  {3 \over 2} \bar c_H  -  {1 \over 2} [\bar c'_{H \ell}]_{22} \right ) .
\end{eqnarray}
In accordance with the condition \#4, the 2-derivative Higgs boson self-couplings have been traded for other equivalent interactions and do not occur in the mass eigenstate Lagrangian.  
Self-interactions terms with 4, 5, and 6 Higgs boson fields may also arise from dimension-6 operators,  but  we do  not display them  here.  

The  interactions between two Higgs bosons and two other SM fields  are parameterized as follows:   
\begin{eqnarray} 
\label{eq:PEL_lhh}
\hspace{-2cm}
{\cal L}_{h^2}  &= &
h^2  \left (1 + 2  \delta c_z^{(2)} \right ) {g^2 + g'{}^2 \over 4}  Z_\mu Z_\mu  
+ h^2  \left (1 + 2  \delta c_w^{(2)} \right )  {g^2 \over 2} W_\mu^+  W_\mu^-
\nonumber \\  &-& 
{h^2 \over 2 v^2} \sum_{f;ij} \sqrt{m_{f_i} m_{f_j}} 
\left [ \bar f_{i,R}  [y_f^{(2)}]_{ij}   f_{j,L} + {\mathrm h.c.} \right ]
\nonumber \\  &+& 
 {h^2 \over 8 v^2}   \left ( c_{gg}^{(2)}  g_s^2 G_{\mu \nu}^a G_{\mu \nu}^a   +   2 c_{ww}^{(2)} g^2 W_{\mu \nu}^+ W_{\mu \nu}^-   + c_{zz}^{(2)} (g^2 + g'{}^2)Z_{\mu \nu} Z_{\mu \nu}   +  2 c_{z\gamma}^{(2)} g g' Z_{\mu \nu} A_{\mu \nu}    +  c_{\gamma \gamma}^{(2)} e^2  A_{\mu \nu} A_{\mu \nu} 
 \right )
\nonumber \\  &+& 
{h^2 \over 8 v^2} \left (   \tilde c_{gg}^{(2)}  g_s^2G_{\mu \nu}^a \tilde G_{\mu \nu}^a +   2   \tilde c_{ww}^{(2)} g^2 W_{\mu \nu}^+ \tilde W_{\mu \nu}^-   +   \tilde c_{zz}^{(2)} (g^2 + g'{}^2)Z_{\mu \nu}  \tilde Z_{\mu \nu}   +  2   \tilde c_{z\gamma}^{(2)} g g' Z_{\mu \nu} \tilde A_{\mu \nu}    +    \tilde c_{\gamma \gamma}^{(2)} e^2  A_{\mu \nu} \tilde A_{\mu \nu}  \right ) 
\nonumber \\ &&  
- {h^2 \over 2 v^2} \left  ( g^2 c_{w \Box}^{(2)} (W_\mu^+ \partial_\nu W_{\nu \mu}^- + W_\mu^- \partial_\nu W_{\nu \mu}^+ ) 
+ g^2  c_{z \Box}^{(2)} Z_\mu \partial_\nu Z_{\nu \mu} + g g'  c_{\gamma \Box}^{(2)} Z_\mu \partial_\nu A_{\nu \mu}  \right ). 
\end{eqnarray}  
All double Higgs boson couplings arising from $D$=6 operators can be expressed by the single Higgs boson couplings: 
\begin{eqnarray}
\delta c_z^{(2)} &=& \delta c_z, \qquad   \delta c_w^{(2)} = \delta c_z + 3 \delta m, \quad 
\nonumber \\ 
\,  [y_f^{(2)}]_{ij}   & = &  3  [\delta y_f]_{ij} e^{i \phi_{ij}} -  \delta c_z \, \delta_{ij}, 
\nonumber \\ 
c_{vv}^{(2)}  & =& c_{vv}, \qquad \tilde c_{vv}^{(2)}  = \tilde  c_{vv}, \qquad v \in \{g,w,z,\gamma \},   
\nonumber \\
c_{v\Box}^{(2)}  & =& c_{v\Box},  \qquad v \in \{w,z,\gamma \}.   
 \end{eqnarray}
Other interaction terms with two Higgs bosons involve at least 5 fields:  e.g  the $h^2 V^3 $  or $h^2 f f V$ contact interactions, and are not displayed here. 
 
\subsubsection*{Other terms}

In this section we have written down the interaction terms of mass eigenstates in the $D$=6 EFT Lagrangian which  are most relevant  for LHC Higgs phenomenology.  
They either enter the single and double Higgs boson production at tree level, or they affect electroweak precision observables that are complementary to Higgs boson couplings measurements.     
The  remaining terms in the mass eigenstate Lagrangian, which  are not explicitly displayed in this chapter,  are contained in ${\cal L}_{\rm other}$ in Eq.~(\ref{eq:PEL_left}). 
They include 4-fermion terms, couplings of a single Higgs boson to 3 or more gauge bosons, quartic Higgs and gauge boson self-interactions, dipole-like interactions of two gauge bosons and two fermions, and interaction terms with 5 or more fields.
For a future reference, we only comment on two 4-lepton terms involving left-handed electrons and muons and the corresponding neutrinos: 
\begin{equation}
{\cal L}_{4 \ell} \supset {1\over v^2} \left [  [c_{\ell \ell}]_{1122}  (\bar \ell_1 \gamma_\mu  \ell_1)  (\bar \ell_2 \gamma_\mu  \ell_2) + [c_{\ell \ell}]_{1221}  (\bar \ell_1 \gamma_\mu  \ell_2)  (\bar \ell_2 \gamma_\mu  \ell_1)  \right ]. 
\end{equation}
The coefficients of these 4-lepton terms are related to the Wilson coefficients in the SILH basis by 
\begin{eqnarray}
\label{eq:PEL_4f}
\,  [c_{\ell \ell}]_{1122} & =&  {2 g'{}^2 \over g^2} \bar c_{2B} - 2 \bar c_{2W}, 
\nonumber \\ 
\,  [c_{\ell \ell}]_{1221} & =&  4 \bar c_{2W}. 
\end{eqnarray}
Note that the corresponding 4-fermion operators are absent in the SILH basis. 
However, in the mass eigenstate Lagrangian, these operators do appear, once the SILH operators $O_{2W}$ and $O_{2B}$ are traded for other interactions terms by using equations of motion. 
By the same token, the 4-top term $[O_{uu}]_{3333}$ does appear in the mass eigenstate Lagrangian, with the coefficient proportional to $\bar c_{2 G}$.

\subsection{Higgs basis}
\label{sec:hb}

In the previous section we related the Wilson coefficients in the SILH bases of $D$=6 operators  to the couplings of mass eigenstates in the Lagrangian. 
With this information at hand, one can proceed to calculating observables at a given order in the EFT as a function of the Wilson coefficients. 
The information provided above is enough to calculate the leading order EFT corrections to SM predictions for single and double Higgs boson production and decays in all phenomenologically relevant channels.     

There is no theoretical obstacle to present the results of LHC Higgs analyses as constraints  on the Wilson coefficients in the SILH, Warsaw, or any other basis.
However, this procedure may not be the most efficient one from the experimental point of view.
The reason is that the relation between the Wilson coefficients  in the SILH basis and the relevant couplings of the Higgs boson in the mass eigenstate  Lagrangian is somewhat complicated, c.f. Eqs~(\ref{eq:PEL_dg}),  (\ref{eq:PEL_dy}),  (\ref{eq:PEL_dcv}), (\ref{eq:PEL_cvv}).  
The situation is similar for the Warsaw basis, see Appendix~A~of~Ref.~\cite{LHCHXSWG-INT-2015-001}. 
In this section we propose another, equivalent parameterization of the EFT with $D$=6 operators. 
The idea, put forward in Ref.~\cite{Gupta:2014rxa},  
is to parameterize the space of $D$=6 operators using a subset of  couplings in a mass eigenstate Lagrangian, such as the one defined   in Eq.~(\ref{eq:PEL_left})  of Section.~\ref{sec:pel}. 
The parameterization described in this section, which differs slightly from that in Ref.~\cite{Gupta:2014rxa}, is referred to as the {\em Higgs basis}.\footnote{
Here, the Higgs basis is introduced in a different manner than how the SILH or Warsaw basis were defined in the literature.
Rather than by choosing a set  $SU(3) \times SU(2) \times U(1)$ invariant $D$=6 operators, we introduce the Higgs basis as a parameterization of the space of all possible deformations of the SM mass eigenstate Lagrangian that can arise in the presence of $D$=6 operators. 
However, both ways can be shown to be equivalent, which justifies using the term {\em basis} for our construction. 
In particular, it is possible to define the Higgs basis as a complete non-redundant set of $SU(3) \times SU(2) \times U(1)$ invariant $D$=6 operators, see Section~\ref{sec:HB_summary}.}

The salient  features  of the Higgs basis are the following. 
The goal is to parameterize the space of $D$=6 operators in a way that can be more directly connected  to observable quantities in Higgs physics. 
The variables  spanning  the Higgs basis correspond to a  subset of the couplings parameterizing interaction terms in the mass eigenstate Lagrangian  in Eq.~(\ref{eq:PEL_left}). 
Since these couplings have been expressed as linear combinations of the SILH basis Wilson coefficients, technically the Higgs basis is defined as a linear transformation from the SILH basis. 
All couplings in the subset have to be independent, in the sense that none can be expressed by the remaining ones at the level of a  general $D=6$ EFT Lagrangian. 
It is also a maximal such subset, which implies that their number is the same as the number of independent operators in the Warsaw or SILH basis. 
 We  will refer to this set as the  {\em independent couplings}.  
 They parameterize all possible deformations of the SM Lagrangian in the presence of $D$=6 operators. 
 Therefore, they can be  used on par with any other basis to describe the effects of dimension-6 operators on any physical observables (also those unrelated to Higgs physics). 
 By definition of the Higgs basis, 
the independent couplings will include single Higgs boson couplings to gauge bosons and fermions.  
Thanks to that, the parameters of the Higgs basis can be connected in a more intuitive way to LHC Higgs observables calculated at leading order in the EFT. 
Furthermore, the vertex corrections to the Z boson interactions with fermions are chosen to be among the independent couplings.  
As a consequence, combining experimental information from Higgs and electroweak precision observables is more transparent in the Higgs basis.

\subsubsection*{Independent couplings}
\label{s.HB_ic}

%

We now describe the choice of independent couplings which defines the Higgs basis. 

The first group of independent couplings parameterizes the interactions of the Higgs boson with itself and  with the SM gauge bosons and fermions: 
\begin{eqnarray} 
\label{eq:HB_ind_higgs} 
 & c_{gg}, \  \delta c_z,  \ c_{\gamma \gamma}, \ c_{z \gamma},  \ c_{zz},  \ c_{z \Box},   \  \tilde c_{gg}, \  \tilde c_{\gamma \gamma}, \  \tilde c_{z \gamma}, \ \tilde c_{zz},  \ \delta \lambda_3,  
\nonumber \\
&  [\delta y_u]_{ij}, \   [\delta y_d]_{ij},  \ [\delta y_e]_{ij}, \   [\phi_u]_{ij}, \    [\phi_d]_{ij},  \   [\phi_e]_{ij}.
\end{eqnarray} 
The parameters in the first line are defined by Eq.~(\ref{eq:PEL_lhvv}) and Eq.~(\ref{eq:PEL_lhh}), and in the second line by Eq.~(\ref{eq:PEL_lhff}). 
Overall, there is 65 independent parameters in Eq.~(\ref{eq:HB_ind_higgs}), 
and they all affect Higgs boson production and/or decay at the leading order in the EFT expansion. 
Therefore they are of crucial importance for LHC Higgs phenomenology. 
Moreover, at the leading order, they are not constrained at all by LEP-1 electroweak precision tests or low-energy precision observables.
   
The second group of independent couplings parameterizes the W boson mass and the Z and W boson couplings to fermions:  
\begin{eqnarray} 
\label{eq:HB_ind_ewpt}
& \delta m,  \   [\delta g^{Ze}_L]_{ij}, \ [\delta g^{Ze}_R]_{ij}, \ [\delta g^{W \ell}_L]_{ij}, \  [\delta g^{Zu}_L]_{ij},  \ [\delta g^{Zu}_R]_{ij},   \  [\delta g^{Zd}_L]_{ij},  \ [\delta g^{Zd}_R]_{ij},  \ [\delta g^{Wq}_R]_{ij}, 
\nonumber \\ & 
[d_{G u}]_{ij}, \ [d_{G d}]_{ij},   \ [d_{A e}]_{ij}, \  [d_{A u}]_{ij}, \ [d_{A d}]_{ij}, \  [d_{Z e}]_{ij}, \ [d_{Z u}]_{ij}, \ [d_{Z d}]_{ij}. 
\end{eqnarray} 
Here the mass correction $\delta m$ is defined in Eq.~(\ref{eq:PEL_lkin}), the vertex corrections $\delta g^i$ are defined in Eq.~(\ref{eq:PEL_lvertex}), and the dipole moments $d_i$ are defined in Eq.~(\ref{eq:PEL_ldipole}).
All these parameters also affect the Higgs boson production and/or decay at the leading order in the EFT. 
However, as opposed to the ones in Eq.~(\ref{eq:HB_ind_higgs}), they affect at the same order electroweak and/or low-energy precision observables.  

The third group of  independent couplings parameterizes the self-couplings of gauge bosons: 
\begin{equation}
\label{eq:HB_ind_tgc}  
\lambda_z, \  \tilde \lambda_z,  \ c_{3G}, \ \tilde  c_{3G}.  
\end{equation} 
They are defined in Eq.~(\ref{eq:PEL_ltgc}).  
These couplings  do not affect Higgs boson production and decay at the leading order in EFT. 

To complete the definition of the Higgs basis, one has to select the independent couplings corresponding to 4-fermion operators. 
We choose to parameterize them by the same set of Wilson coefficients as in the SILH basis, 
c.f. Table~\ref{tab:silh_4f}: 
\begin{eqnarray} 
\label{eq:HB_ind_4f}  
& c_{\ell\ell}, \  c_{qq} , \ c_{qq}' , \  c_{\ell q} , \ c_{\ell q}' , \ c_{quqd},  \ c_{quqd}' , \  c_{\ell equ} , \ c_{\ell equ}' , \  c_{\ell edq}, 
\nonumber \\ & 
 c_{\ell e} , \  c_{\ell u} , \  c_{\ell d} , \  c_{qe} , \  c_{qu}, \  c_{qu}', \  c_{qd} , \  c_{qd}',  \  c_{ee} , \  c_{uu} , \  c_{dd} , \  c_{eu} , \  c_{ed} , \  c_{ud} , \  c_{ud}' . 
\end{eqnarray}
Each parameter $c_{ff}$ has 4 flavour indices, which are not displayed here. 
The non-trivial  question of which combination of flavour indices constitutes an independent set was worked out in Ref.~\cite{Alonso:2013hga}. 
In the Higgs basis we take the same choice of independent 4-fermion couplings as in that reference, with one exception. 
As explained in the next subsection, in a $D$=6 EFT Lagrangian,  the coupling $[c_{\ell \ell}]_{1221}$ multiplying a particular 4-lepton operator can be expressed by $\delta m$ and $\delta g^i$.  
Therefore $[c_{\ell \ell} ]_{1221}$ is not among the independent couplings defining the Higgs basis.  

\subsubsection*{Dependent couplings}
\label{sec:HB_dc}

The number of parameters characterizing departure from the SM Lagrangian in  Eq.~(\ref{eq:PEL_left})  is larger than the number of Wilson coefficients in a  basis of $D$=6 operators. 
Due to this fact,  there must be relations among these parameters. 
Working in the Higgs basis, some of the parameters in the mass eigenstate Lagrangian can be expressed by the independent couplings;  we  call them the {\em dependent} couplings.
The relations between dependent and independent couplings can be inferred from the matching between the effective Lagrangian and the Warsaw or SILH basis in Section.~\ref{sec:pel}. 
These relations {\em hold at the level of the dimension-6 Lagrangian}, and they are in general not respected in the presence of dimension-8 and higher operators. 

We start with the dependent couplings in  Eq.~(\ref{eq:PEL_lhvv}) parameterizing the single Higgs boson interactions with gauge bosons. 
They can be expressed in terms of the independent couplings as\footnote{%
The relation between $c_{ww}$, $\tilde c_{ww}$ and other parameters can also be viewed as a consequence of the accidental custodial symmetry at the level of the dimension-6 operators \cite{Contino:2013kra}.}  
\begin{eqnarray}
\label{eq:HB_dep_h}
\delta  c_{w} &=&  \delta c_z + 4 \delta m , 
\nonumber \\ 
c_{ww} &=&  c_{zz} + 2 s_\theta^2 c_{z \gamma} + s_\theta^4 c_{\gamma \gamma}, 
\nonumber \\ 
\tilde c_{ww} & = &  \tilde c_{zz} + 2 s_\theta^2  \tilde c_{z \gamma} + s_\theta^4  \tilde c_{\gamma \gamma} , 
\nonumber \\
c_{w \Box}  &= & {1 \over g^2 - g'{}^2} \left [ 
g^2 c_{z \Box} + g'{}^2 c_{zz}  - e^2 s_{\theta}^2   c_{\gamma \gamma}  -(g^2 - g'{}^2) s_\theta^2  c_{z \gamma} 
\right ] ,  
\nonumber \\ 
  c_{\gamma \Box}  &= &  
  {1 \over g^2 - g'{}^2} \left [ 
2 g^2 c_{z \Box} + (g^2+ g'{}^2) c_{zz}  - e^2  c_{\gamma \gamma}  -(g^2 - g'{}^2)   c_{z \gamma} 
\right ] . 
  \end{eqnarray} 
The coefficients of W-boson dipole interactions in   Eq.~(\ref{eq:PEL_ldipole})   are related to those of the $Z$ and the photon as  
\begin{equation}
\label{eq:HB_dep_dip}
\eta_f  d_{Wf}  = d_{Z f}   + s_\theta^2 d_{Af},  
\end{equation} 
where $\eta_u = 1$ and $\eta_{d,e} = -1$. 
The coefficients of the dipole-like Higgs boson couplings in  Eq.~(\ref{eq:PEL_lhdvff})  are simply related to the corresponding dipole moments: 
\begin{equation} 
\label{eq:dep_hdip}
d_{hVf} = d_{Vf}, \quad \tilde d_{hVf} = \tilde d_{Vf}, \qquad V \in \{G,W,Z,A \} . 
\end{equation} 

Coefficients of  all interaction terms with   two Higgs bosons in   Eq.~(\ref{eq:PEL_lhh})  are dependent couplings. 
The   can be expressed in terms of  the independent couplings as:    
\begin{eqnarray}
\label{eq:HB_dep_hh}
\delta c_z^{(2)} &=& \delta c_z, \qquad   \delta c_w^{(2)} = \delta c_z + 3 \delta m, \quad 
\nonumber \\ 
\,  [y_f^{(2)}]_{ij}   & = &  3  [\delta y_f]_{ij} e^{i \phi_{ij}} -  \delta c_z \, \delta_{ij}, 
\nonumber \\ 
c_{vv}^{(2)}  & =& c_{vv}, \qquad \tilde c_{vv}^{(2)}  = \tilde  c_{vv}, \qquad v \in \{g,w,z,\gamma \},   
\nonumber \\
c_{v\Box}^{(2)}  & =& c_{v\Box},  \qquad v \in \{w,z,\gamma \}.   
 \end{eqnarray}
The dependent vertex corrections are expressed in terms of  the independent couplings as
\begin{equation}
\label{eq:HB_dep_dg}
 \delta g^{Z\nu}_L  =   \delta g^{Ze}_L   +  \delta g^{W \ell}_L , \qquad    
\delta g^{Wq}_L  =     \delta g^{Zu}_L V_{\rm CKM} -     V_{\rm CKM}  \delta g^{Zd}_L . 
\end{equation} 
All but four  triple gauge couplings in  Eq.~(\ref{eq:PEL_ltgc})  are dependent couplings expressed in terms of  the independent couplings as  
\begin{eqnarray}
\label{eq:dep_tgc}
\delta g_{1,z}  & =&  
{1 \over 2 (g^2 - g'{}^2)} \left [   c_{\gamma\gamma} e^2 g'{}^2 + c_{z \gamma} (g^2 - g'{}^2) g'{}^2  - c_{zz} (g^2 + g'{}^2) g'{}^2  - c_{z\Box} (g^2 + g'{}^2) g^2 \right ], 
 \nonumber \\
 \delta \kappa_\gamma &=& - {g^2 \over 2} \left ( c_{\gamma\gamma} {e^2 \over g^2 + g'{}^2}   + c_{z\gamma}{g^2  - g'{}^2 \over g^2 + g'{}^2} - c_{zz} \right ) , 
\nonumber \\
 \tilde \kappa_\gamma &=& - {g^2 \over 2} \left ( \tilde c_{\gamma\gamma} {e^2 \over g^2 + g'{}^2}   + \tilde c_{z\gamma}{g^2  - g'{}^2 \over g^2 + g'{}^2} - \tilde c_{zz} \right ), 
\nonumber \\ 
\delta \kappa_z &=& \delta g_{1,z} -  t_\theta^2 \delta \kappa_\gamma, \qquad \tilde \kappa_z =  -  t_\theta^2 \tilde \kappa_\gamma, 
\nonumber \\ 
\lambda_\gamma &=& \lambda_z, \quad \tilde \lambda_\gamma = \tilde \lambda_z. 
\end{eqnarray}

Finally, we discuss how the Wilson coefficient $[c_{\ell \ell}]_{1221}$ is expressed by the independent couplings. 
One defining feature of  the mass eigenstate  Lagrangian   Eq.~(\ref{eq:PEL_left})  is that the tree-level relations between the SM electroweak parameters and input observables are not affected by $D$=6 operators (condition \# 3). 
On the other hand, one of the four-fermion couplings in the Lagrangian, 
\begin{equation}
\label{eq:4F_cll}
{\cal L}_{4f}^{D=6}  \supset   [c_{\ell \ell}]_{1221} (\bar \ell_{1,L} \gamma_\rho \ell_{2,L} ) (\bar \ell_{2,L} \gamma_\rho \ell_{1,L} ),   
\end{equation} 
does affect the relation between the parameter $v$ and the muon decay width from which 
$v = (\sqrt 2 G_F)^{-2}$  is determined: 
\begin{equation}
{\Gamma(\mu \to e \nu \nu) \over \Gamma(\mu \to e \nu \nu)_{\rm SM}} \approx 
1  +  2 [\delta g_L^{We}]_{11} +  2 [\delta g_L^{We}]_{22}  -  4 \delta m   -  [c_{\ell \ell}]_{1221}. 
\end{equation}   
Therefore,  the muon decay width is unchanged with respect to the SM when $[c_{\ell \ell}]_{1221}$ is related to  $\delta m$ and $\delta  g$ as 
\begin{equation}
\label{eq:HB_dep_cll}
[c_{\ell \ell}]_{1221} =   2 \delta [g_L^{We}]_{11} + 2 [\delta g_L^{We}]_{22}  -  4 \delta m .
\end{equation}  
This relation can be verified using the expressions of these parameters in terms of the SILH Wilson coefficients in  Eqs.~(\ref{eq:PEL_dm}),  (\ref{eq:PEL_dg}), and (\ref{eq:PEL_4f}). 
In other words, due to the fact that we selected $ \delta m$ and $\delta g$  as  an independent coupling in the Higgs basis,  $[c_{\ell \ell}]_{1221}$ has to be a dependent coupling. 
Of course, one could equivalently  choose   $[c_{\ell \ell}]_{1221}$ to define a basis, and remove e.g. $\delta m$ from the list of independent couplings. 
The remaining 4-fermion parameters in Eq.~(\ref{eq:HB_ind_4f}) are independent couplings.

 \subsubsection*{Gauge invariant definition}
 \label{sec:HB_summary}

In summary, in the Higgs basis the parameters spanning the space of $D$=6 EFT operators 
 are the independent couplings in Eqs.~(\ref{eq:HB_ind_higgs}),  (\ref{eq:HB_ind_ewpt}), (\ref{eq:HB_ind_tgc}), and (\ref{eq:HB_ind_4f}).  
 In the EFT expansion, the independent couplings are formally of order ${\cal O}(\Lambda^{-2})$. 
These parameters are directly linked to deviations from the SM  interactions in the mass eigenstate Lagrangian in Eq.~(\ref{eq:PEL_left}). 
All other deviations in the mass eigenstate Lagrangian can be expressed by the independent couplings.

In this section, the Higgs basis was introduced by choosing a subset of independent couplings in the mass eigenstate Lagrangian defined in Section~\ref{sec:pel}. 
The latter is not manifestly invariant under the full  gauge symmetry of the SM, as the electroweak symmetry  $SU(2) \times U(1)$ is broken to $U(1)_{\rm em}$ at the mass eigenstate level. 
Nevertheless,  one can provide an equivalent and manifestly gauge invariant definition of the Higgs basis.
To this end, one can introduce the $SU(3) \times SU(2) \times U(1)$ invariant $D$=6 operators as follows: 
\begin{eqnarray}
\label{eq:hb_invariant}
O_{\delta \lambda_3} & = & - {1 \over  v^2} (H^\dagger H)^3, 
 \nonumber  \\  
O_{c_{gg}} & = & {g_s^2  \over 4 v^2} H^\dagger H\, G^a_{\mu\nu} G^a_{\mu\nu}
 \nonumber  \\  
O_{\delta c_z} & = & - {1 \over  v^2} \left [ \partial_\mu (H^\dagger H) \right ]^2 
+ {3 \lambda \over  v^2} (H^\dagger H)^3
+  \left ( \sum_f {\sqrt{2} m_{f_i}  \over v^3} H^\dagger H \bar f_{L,i} H  f_{R,i}  + {\rm h.c.} \right ),
\nonumber  \\ 
O_{c_{z \Box}} & = & {i g^3 \over v^2 (g^2 - g'{}^2) }\left(H^\dagger \sigma^i {\overleftrightarrow {D_\mu}} H \right) D_\nu W_{\mu\nu}^i -  {i g^2 g'  \over v^2 (g^2 - g'{}^2) } \left(H^\dagger  {\overleftrightarrow { D_\mu}} H \right) \partial_\nu B_{\mu\nu} ,
\nonumber  \\  
O_{c_{zz}}  & = &
 {i g(g^2 + g'{}^2) \over 2 v^2 (g^2 - g'{}^2) }\left(H^\dagger \sigma^i {\overleftrightarrow {D_\mu}} H \right) D_\nu W_{\mu\nu}^i 
 -  {i g' (g^2 + g'{}^2)  \over 2 v^2 (g^2 - g'{}^2) } \left(H^\dagger  {\overleftrightarrow { D_\mu}} H \right) \partial_\nu B_{\mu\nu}
 \nonumber  \\ 
 & - &  \frac{i g}{v^2}  \left(D_\mu H^\dagger \sigma^i  D_\nu H \right) W^i_{\mu\nu} 
- \frac{i g'}{ v^2}   \left(D_\mu H ^\dagger D_\nu H \right) B_{\mu\nu}, 
\nonumber  \\ 
O_{c_{z\gamma}}  & = &
 - {2 i g  g'{}^2 \over  v^2 (g^2+  g'{}^2) }    \left(D_\mu H^\dagger \sigma^i  D_\nu H \right) W^i_{\mu\nu} 
 +  {2 i  g'{} g^2  \over  v^2 (g^2 + g'{}^2) }  \left(D_\mu H ^\dagger D_\nu H \right) B_{\mu\nu}, 
\nonumber  \\ 
O_{c_{\gamma \gamma}}  & = &
 - {i g  g'{}^4 \over  2v^2 (g^4 -  g'{}^4) } \left(H^\dagger \sigma^i {\overleftrightarrow {D_\mu}} H \right) D_\nu W_{\mu\nu}^i 
 +  {i  g'{}^5   \over 2 v^2 (g^4 -  g'{}^4) } \left(H^\dagger  {\overleftrightarrow { D_\mu}} H \right) \partial_\nu B_{\mu\nu}
 \nonumber  \\ 
 & - &
  \frac{i g g'{}^4}{v^2 (g^2 + g'{}^2)^2}  \left(D_\mu H^\dagger \sigma^i  D_\nu H \right) W^i_{\mu\nu} 
+ \frac{i g'{}^3 (2 g^2 + g'{}^2)}{(g^2 + g'{}^2)^2 v^2}   \left(D_\mu H ^\dagger D_\nu H \right) B_{\mu\nu}
 \nonumber  \\ 
 & + &
{g'{}^2 \over 4 v^2} H^\dagger H\, B_{\mu\nu} B_{\mu\nu} , 
\nonumber  \\
\, [O_{\delta y_f}]_{ij}  & = & - {\sqrt{2 m_{f_i} m_{f_j}} \over v^3} H^\dagger H \bar f_{L,i} H  f_{R,j} + {\rm h.c.}  , 
\nonumber  \\ 
\dots 
\end{eqnarray}
The coefficients of the operators on the right-hand side in Eq.~(\ref{eq:hb_invariant}) are determined by the linear map relating  the SILH Wilson coefficients to those in the Higgs basis,  which can be obtained by inverting the relations between the Higgs and SILH coefficients derived earlier in this section. 
By following this algorithm,  a complete and non-redundant set of $D$=6 operators $ O_{c_i}$  defining the Higgs basis  can be constructed.  
Then the Higgs basis Lagrangian can be defined in a manifestly gauge invariant way as  ${\cal L}_{\rm EFT} = {\cal L}_{\rm SM} + \sum_i c_i O_{c_i}$. 
 
 \subsubsection*{Simplified scenarios}
 \label{sec:HB_simplified} 
  
In total, the Higgs basis, as any complete basis at the dimension-6 level,  is parameterized by 2499 independent real couplings \cite{Alonso:2013hga}. 
One should not, however, be intimidated by this number.  
The point is that a much smaller subset of the independent couplings is relevant for analyses of Higgs data at leading order in EFT. 
First of all, the coefficients of 4-fermion interactions in Eq.~(\ref{eq:HB_ind_4f}) and triple gauge interactions in   Eq.~(\ref{eq:HB_ind_tgc})  do not enter Higgs observables at the leading order.  
At that order, the parameters relevant for LHC Higgs analyses are those in Eqs.~(\ref{eq:HB_ind_higgs}) and (\ref{eq:HB_ind_ewpt}), which  already reduces the number of variables significantly. 
Furthermore, there are several motivated assumptions about the UV theory underlying the EFT  which could be used to further reduce the number of parameters: 
\begin{itemize} 
\item {\em Minimal flavour violation},  in which case the matrices $\delta y_f$, $\phi_f$, $d_{Vf}$, and $\delta g^{Vf}$,  reduce to a single number for each $f$.  
\item  {\em CP conservation}, in which case all CP-odd couplings vanish:  $\tilde c_i  =  \phi_f = {\rm Im} d_f = 0$. 
\item {\em Custodial symmetry}, in which case $\delta m = 0$.\footnote{Custodial symmetry implies several relations between Higgs boson couplings to gauge bosons: 
$\delta c_w = \delta c_z$, $c_{w \Box} = c_\theta^2 c_{z \Box} + s_\theta^2 c_{\gamma \Box}$, $c_{ww} = c_{zz} + 2 s_\theta^2 c_{z \gamma} + s_\theta^4 c_{\gamma}$, 
and $\tilde c_{ww} = \tilde  c_{zz} + 2 s_\theta^2 \tilde  c_{z \gamma} + s_\theta^4 \tilde  c_{\gamma}$.
The last three are satisfied automatically at the level of dimension-6 Lagrangian, while the first one is true for $\delta m =0$, see Eq.~(\ref{eq:HB_dep_h}). }
\end{itemize} 
We stress that independent couplings should not be arbitrarily set to zero without an underlying symmetry assumption.  
Furthermore, the relations between the dependent and independent couplings in the mass eigenstate Lagrangian  should be consistently imposed, so as to preserve the structure of the $D$=6 EFT Lagrangian. 

Finally, to reduce the number of free parameters in an analysis, one may take advantage of the fact that,  in addition to Higgs observables, other measurements are sensitive to  the parameters in Eq.~(\ref{eq:HB_ind_ewpt}). 
In particular, the parameters in the first line of Eq.~(\ref{eq:HB_ind_ewpt}) are constrained by electroweak precision tests in LEP-1.   
These are among the most stringent constraints on EFT parameters, and they have an important impact on possible signals in Higgs searches.
Assuming minimal flavour violation, all the vertex corrections in Eq.~(\ref{eq:HB_ind_ewpt})  are constrained to be smaller than $O(10^{-3})$ (for the leptonic vertex corrections and $\delta m$), or $O(10^{-2})$ (for the quark vertex corrections) \cite{Pomarol:2013zra,Falkowski:2014tna,Ellis:2014jta}.\footnote{%
These constraints may be relaxed if the leading-order $D$=6 EFT does not provide an adequate  description of electroweak precision observables~\cite{Berthier:2015gja}. 
If that is the case, the vertex-like and dipole-like Higgs boson couplings in Eqs.~(\ref{eq:PEL_lhvff}) and (\ref{eq:PEL_lhdvff}) could in principle be sizeable enough  to be relevant  for  the LHC searches without conflict with electroweak precision constraints. 
However, it is not clear whether there exist explicit BSM models where this concern is relevant. 
}
Even when the assumption of  minimal flavour violation is not imposed,  all the leptonic, bottom and charm quark vertex corrections are still constrained at the level of  $O(10^{-2})$ or better \cite{Efrati:2015eaa}.  
Similarly, many parameters in the second line of Eq.~(\ref{eq:HB_ind_ewpt})  are strongly constrained by measurements of the magnetic and electric dipole moments.
In the LHC environment, experimental sensitivity is often not sufficient to probe these parameters with a comparable accuracy.   
If that is indeed the case, it is well-motivated  to neglect the parameters in Eq.~(\ref{eq:HB_ind_ewpt})  in LHC Higgs analyses.\footnote{Editor footnote: 
Another point of view is expressed in Section~\ref{s.eftnlo} which argues against neglecting the parameters in Eq.~(\ref{eq:HB_ind_ewpt}) in EFT analyses of LHC Higgs data.} 

Once  the parameters in Eq.~(\ref{eq:HB_ind_ewpt}) are neglected, this leaves the parameters collected in Eq. (\ref{eq:HB_ind_higgs}) to describe leading order deformations of Higgs observables. 
This set  consists of 11 bosonic and  $2 \times 3 \times 3 \times 3 = 54$ fermionic couplings.
While that number is still large, it represents a significant simplification  compared to the 2499 Wilson coefficients parameterizing a complete $D$=6 basis. 
Further simplifications can be introduced by making more specific assumptions about the high-energy theory that generates $D$=6 operators in the EFT.
For example,  if the high-energy theory respects the minimal flavour violation paradigm, the flavour structure of the fermionic  parameters   in Eq.~(\ref{eq:HB_ind_higgs}) is proportional to the unit matrix: 
$[\delta y_f]_{ij} = \delta_{ij} \delta y_f $ and $[\phi_f]_{ij} = \delta_{ij} \phi_f$.  
This  reduces down to 17 (11 bosonic and 6 fermionic)  the number of parameters relevant for LHC Higgs observables. 
In the Higgs basis, these parameters are:    
\begin{eqnarray} 
\label{eq:HB_17par}  {\rm CP\text{-}even:}& 
c_{gg}, \  \delta c_z,  \ c_{\gamma \gamma}, \ c_{z \gamma},  \ c_{zz},  \ c_{z \Box},  \  \delta y_u, \   \delta y_d,  \ \delta y_e, \ \delta \lambda_3; 
&\nonumber \\  {\rm CP\text{-}odd:} & 
\tilde c_{gg},   \  \tilde c_{\gamma \gamma}, \ \tilde c_{z \gamma},  \ \tilde c_{zz},    \  \phi_u, \   \phi_d,  \phi_e. 
\end{eqnarray}
Assuming in addition CP conservation\footnote{%
The CP-odd parameters affect inclusive Higgs observables only at the quadratic level,  (${\cal O}(\Lambda^{-4})$ in the EFT expansion). Therefore they can be neglected in the leading order approximation, even without assuming CP conservation, if one restricts the analysis to inclusive measurements, such as  the Higgs boson signal strength measurements at the  LHC. } 
in the Higgs sector leaves only 10 CP-even parameters  to describe leading order EFT corrections to single and double Higgs boson production and decay.

Providing model-independent constraints on the 17 parameters in Eq.~(\ref{eq:HB_17par}), or at least the 10 CP-even ones, is a realistic target for run-2 LHC Higgs searches. 
The CP-even parameters are weakly constrained by prior precision experiments, 
with ${\cal O}(0.1)$- ${\cal O}(1)$ values allowed by current global fits to Higgs and electroweak data \cite{Falkowski:2015jaa}. 
The CP-odd parameters are even less constrained  by Higgs and electroweak  data, though they are indirectly constrained by low-energy  probes of CP violation \cite{Brod:2013cka,Altmannshofer:2015qra,Chien:2015xha,Cirigliano:2016njn}.   
Better constraints on this  reduced sets of EFT parameters  from the ensemble of LHC Higgs measurements  would  already be a valuable input for constraining a large class of theories beyond the SM.  
 
 \subsubsection*{Relation to other frameworks}
 \label{sec:HB_relation} 

The Higgs basis can be used in par with any other basis  to describe the effects of dimension-6 operators on physical observables.
Other popular SM EFT approaches  in the literature use the so-called SILH \cite{Contino:2013kra}, Warsaw \cite{Grzadkowski:2010es}, or HISZ \cite{Hagiwara:1993ck} bases of $D$=6 operators.   
At the leading order in EFT all these approaches are completely equivalent, as there exists a 1-to-1 correspondence between the parameter of the Higgs basis and Wilson coefficients  of any other $D$=6 basis.  
Therefore, the results of leading order EFT analyses can be always translated from and to the Higgs basis without any loss of generality  (see e.g. \cite{Falkowski:2015jaa} for the translation of the LHC Higgs and TGC constraints). 
Formulas necessary for translations between various bases  are provided: 
see  Section~\ref{sec:pel} for the Higgs-SILH basis translation, 
and Appendix~A~of~Ref.~\cite{LHCHXSWG-INT-2015-001} for the Higgs-Warsaw  basis translation. 
A map between the Higgs basis parameters in Eq.~(\ref{eq:HB_17par} and the HISZ basis can be found  in Appendix~B~of~Ref.~\cite{LHCHXSWG-INT-2015-001}.  
These maps are used by  the Rosetta package \cite{Falkowski:2015wza}, which provides automated translation between different bases and an interface to Monte Carlo simulations in the MadGraph~5 framework \cite{Alwall:2011uj}. 

Using the Higgs basis for leading order Higgs EFT analysis is then simply a matter of convenience.  
Its usefulness is in the fact that description of Higgs observables and electroweak precision observables at the leading EFT order  (tree-level ${\cal O}(\Lambda^{-2})$) is more transparent than in other bases.
This also implies simplification of Monte Carlo simulation of collider signals, as relevant Higgs observables typically depend on a smaller  number of parameters than in other bases.   
The advantages of the Higgs basis are especially pronounced when simplified approaches to LHC Higgs data  are employed. 
The main point of the Higgs basis is to separate parameters affecting only Higgs observables at leading order from those that also affect electroweak precision observables. 
If the latter are neglected in an analysis, a small subset of Higgs basis parameters  in Eq.~(\ref{eq:HB_17par} is adequate to describe all leading order effects of $D$=6 operators on Higgs observables.\footnote{Editor footnote: 
Another point of view is expressed in Section~\ref{s.eftnlo} which advocates using the Warsaw basis formalism as it is more readily applicable to NLO calculations.}
  
Beyond tree level,  advantages of using  the Higgs basis are yet to be demonstrated.
Indeed, one-loop corrections will introduce a dependence of the Higgs observables on a larger number of parameters, 
and the neat separation of parameters affecting precision observables is not maintained.
As of this time, no one-loop EFT calculations  using the Higgs basis formalism exists in the literature;
the existing ones are typically performed in the SILH \cite{Elias-Miro:2013gya,Elias-Miro:2013mua,Contino:2014aaa,Grober:2015cwa,Grazzini:2015gdl,Mimasu:2015nqa}
or Warsaw 
\cite{Grojean:2013kd,Jenkins:2013zja,Jenkins:2013wua,Alonso:2013hga,Ghezzi:2015vva,Hartmann:2015aia,Gauld:2015lmb}  basis. 
Note however that any constraint on a coefficient derived in a certain basis at a given order can
be straightforwardly translated to other parameterizations.

We will now comment on the relationship between the Higgs basis and other frameworks that also do not introduce new particles beyond the SM but are {\em not} equivalent to an EFT. 
The Higgs basis (and dimension-6 EFT in general) is an extension of the {\bf $\kappa$-formalism} \cite{Heinemeyer:2013tqa}. 
That formalism, widely used in LHC Run1 analyses, assumes that only the Higgs boson couplings already present in the SM receive corrections from new physics. 
This way, the kinematics of the Higgs boson production and decay in various channels is unchanged with respect to the SM, and only the signal strength is affected. 
Moreover, the standard approach  allows for new effective Higgs boson coupling to gluons and photons, as they lead to subleading modifications of  the Higgs kinematics  when one restrict experimental analyses to inclusive signal strength  observables.
Recent applications of the $\kappa$-formalism include global fits to the Higgs data with 7 independent coupling modifiers \cite{Khachatryan:2016vau}. 
This is still less general than the dimension-6 EFT, even in its restricted form with the free parameters Eq.~(\ref{eq:HB_17par}). 
In particular, the $D$=6 operators may induce Higgs boson couplings with a different Lorentz structure than that present in the SM (see e.g. Eq.~(\ref{eq:PEL_lhvv})) and thus they my violate the assumptions of the $\kappa$-formalism by modifying the Higgs kinematics. 
Therefore, the results obtained within the $\kappa$-formalism cannot be in general translated into the EFT language, whereas the translation is always possible in the opposite direction.\footnote{Note however that, in the dimension-6 EFT,  modifications of the relative Higgs boson coupling strength to $W_\mu W_\mu$ and $Z_\mu Z_\mu$ are always correlated with corrections to the W-boson mass, see Eq.~(\ref{eq:HB_dep_h}),   which is not taken into account in the $\kappa$-formalism. 
Strictly speaking, one can thus project general dim-6 EFT results onto a subset of 6 $\kappa$ parameters of Ref.~\cite{Khachatryan:2016vau}: 
$\kappa_{gZ}$, $\lambda_{Zg}$, $\lambda_{tg}$, $\lambda_{\gamma Z}$,  $\lambda_{\tau Z}$, $\lambda_{bZ}$, with $\lambda_{WZ}$ set to zero.
In  the LO EFT, these 6 $\kappa$'s are in the 1-to-1 correspondence with a subset of 6 parameters in Eq.~(\ref{eq:HB_17par}):  $c_{gg}$,  $\delta c_z$,  $c_{\gamma \gamma}$,  $\delta y_u$, $\delta y_d$,  $\delta y_e$.   
}

A framework more  general than the SM EFT  discussed in Chapter.~\ref{chap:PO} is referred to as {\bf pseudo-observables}. 
In Refs.~\cite{Gonzalez-Alonso:2014eva,Greljo:2015sla},  pseudo-observables are defined as form factors parameterizing amplitudes of physical processes  subject to constraints from Lorentz invariance.  
These form factors are expanded in powers of kinematical invariants of the process around the known poles  of SM particles, assuming poles from BSM particles are absent in the relevant energy regime.  
Such a framework is more general than SM EFT with $D$=6 operators and involves a larger number of parameters, as it does not impose relations between different form factors or between amplitudes of different processes that are predicted by $D$=6 EFT.   
Constraints on pseudo-observables can always be projected into constraints on the Higgs basis  parameters, provided the complete likelihood function (with  correlations)  is given; see Ref.~\cite{Gonzalez-Alonso:2014eva} for a map between observables relevant for $h \to 4f$ decays and EFT parameters.  
The converse is in general not true: constraints on the Higgs basis parameters cannot always be translated into constraints on pseudo-observables. 

In Section~\ref{sec:pel} we introduced the effective Lagrangian that arise when $D$=6 EFT is rewritten in terms of mass eigenstates after electroweak symmetry breaking. 
The crucial feature of this Lagrangian is that various interaction terms are not independent but are instead related by the formulas summarized in  Section~\ref{sec:HB_dc}.
These relations are required by the SM gauge symmetry realized linearly at the level of operators with $D\leq 6$.  
However, one could consider the same Lagrangian without imposing the correlations listed  in  Section~\ref{sec:HB_dc}, and treating instead all parameters as independent.  
Such a construction is referred to as the {\bf Beyond-the-Standard Model Characterization} (BSMC). 
The  BSMC Lagrangian is more general than $D$=6 EFT, and involves more parameters. 
At leading order, it can be used to parameterize new physics effects on Higgs and other observables in a manner akin to pseudo-observables.
Once the likelihood function for the parameters of the BSMC Lagrangian is provided by experiment,  it can be projected into constraints on the Higgs basis parameters by imposing the relations of Section~\ref{sec:HB_dc}.
At the same time, the BSMC likelihood can be used to constrain some more general theories that do not reduce to a SM EFT at low energies.
The BSMC Lagrangian is a part of the Rosetta package \cite{Falkowski:2015wza}. 

Another well-known framework to describe Higgs observables is the so-called {\bf Higgs Characterization} (HC) \cite{Artoisenet:2013puc}. 
In the HC Lagrangian, one describes the effective Higgs boson couplings to the SM gauge bosons and fermions using 20 new parameters.
The HC framework is distinct from the SM EFT.    
On the one hand, the relations between various 2-derivative Higgs boson couplings to  gauge bosons required by $D$=6 EFT are not imposed. 
In this aspect HC is more general than the Higgs or other $D$=6  basis, where these relations follow automatically from the structure of the EFT Lagrangian.   
On the other hand, the HC Lagrangian  does not include all possible deformations of the SM Lagrangian predicted in the presence of $D$=6 operators. 
For example, corrections to SM gauge boson couplings to fermions, dipole interactions,  or contact Higgs interactions with one gauge boson and 2 fermions are not implemented.  
In this aspect, the HC framework is less general than the SM EFT.  

Thus, it is in general not  possible to translate the constraints from the HC framework to the Higgs basis or the other way around.
However, it is possible to do so in certain situations when a simplified EFT description is employed. 
In particular, one can project constraints on the HC parameters onto the subset of the Higgs basis parameters in Eq.~(\ref{eq:HB_17par}, assuming other parameters in the Higgs basis  are not relevant for  these constraints. 
For such a special case, the relation between the HC parameters and the Higgs basis parameters is given in Appendix~B~of~Ref.~\cite{LHCHXSWG-INT-2015-001}.


%
\section[Comments on the validity of the EFT approach to physics beyond the SM]{Comments on the validity of the Effective Field Theory approach to physics beyond the Standard Model\SectionAuthor{N.~Belyaev, R.~Contino, T.~Corbett, A.~Falkowski, F.~Goertz, C.~Grojean, R.~Konoplich, T.~Ohl, J.~Reuter, F.~Riva}}
\label{s.eftval}
\subsection{Introduction}

We consider an EFT where the SM is extended by a set of higher-dimensional operators, and assume that it reproduces the low-energy limit of a 
more fundamental UV description. The theory has the same field content and the same 
linearly-realized $SU(3) \times SU(2) \times U(1)$ local symmetry as the SM.
The difference is the presence  of operators  with canonical dimension $D$ larger than 4.  
These are organized in a systematic expansion in $D$, where each consecutive term is suppressed by a larger power of a high mass scale.
Assuming baryon and lepton number conservation, the Lagrangian takes the form
\begin{equation}
\label{Lops}
{\cal L}_{\textrm{eff}}={\cal L}_{\rm SM}+ \sum_{i}  c_{i}^{(6)} {\cal O}_{i}^{(6)} \
+ \sum_{j}  c_{j}^{(8)} {\cal O}_{j}^{(8)}  + \cdots\,,
\end{equation}
where each ${\cal O}_i^{(D)}$ is a gauge-invariant operator of dimension $D$ and $c_i^{(D)}$ is the corresponding coefficient. 
Each coefficient has dimension $4-D$ and scales like a given power of the couplings of the UV theory; in particular, for an operator made of $n_i$ fields one has
\begin{equation}
\label{eq:counting}
c^{(D)}_i \sim \frac{(\text{coupling})^{n_i-2}}{(\text{high mass scale})^{D-4}}\, .
\end{equation}
This scaling holds in any UV completion which admits some perturbative expansion in its couplings~\cite{Giudice:2007fh}.
An additional suppressing factor $(\text{coupling}/4\pi)^{2L}$ may arise with respect to the naive scaling if the operator is first generated at $L$ loops
in the perturbative expansion.  If no perturbative expansion is possible in the UV theory because this is maximally strongly coupled, 
then  Eq.~(\ref{eq:counting}) gives a correct estimate of the size of the effective coefficients by replacing the numerator with $(4\pi)^{n_i-2}$ ({\it i.e.} setting $\text{coupling}\sim 4\pi$)~\cite{Manohar:1983md}.

The EFT defined by Eq.~(\ref{Lops}) is able to parameterize observable effects of a large class
of beyond the SM (BSM) theories. In fact all decoupling $B-L$ conserving BSM physics where new particles are much heavier than the SM ones and much heavier than the energy scale at which the experiment is performed can be mapped to such a Lagrangian.
The main motivation to use this framework is that the constraints on the EFT parameters can be later re-interpreted as constraints on masses and couplings of  new particles in many BSM theories. 
In other words, translation of experimental data into a theoretical framework has to be done only once in the EFT context, rather than for each BSM model separately. Moreover, the EFT can be used to establish a consistent picture of deviations from the SM by itself and thus can provide guidance for constructing a UV completion of the SM.

The EFT framework contains higher-dimensional operators (non-renormalizable in the traditional sense). As a consequence,  physical amplitudes in general grow with the energy scale of the process,  and therefore the EFT inevitably has a limited energy range of validity.
In this note we address  the question of the validity range at the quantitative level.
We will discuss the following points: 
\begin{itemize} 
\item 
Under what conditions does the EFT give  a faithful description of the low-energy phenomenology of some BSM theory? 
\item  When is it justified to truncate the EFT expansion at the level of dimension-6 operators?  To what extent can  experimental limits on dimension-6 operators 
be affected by the presence of dimension-8 or higher operators? 
\end{itemize}

It is important to realize that  addressing the above questions  
cannot be done in a completely model-independent way, but requires a number of (broad) assumptions about the new physics.
An illustrative example is that of the {\em Fermi theory}, 
which is an EFT for the SM degrees of freedom below the weak scale after the $W$ and $Z$ bosons have been integrated out. 
In this language, the weak interactions of the SM fermions are described at leading order by 4-fermion operators of $D$=6, such as: 
\beq
{\cal L}_{\rm eff} \supset 
 c^{(6)} \, (\bar e \gamma_\rho P_L \nu_e)    (\bar \nu_\mu \gamma_\rho P_L \mu) + {\rm h.c.} \, , 
 \qquad\quad c^{(6)} = -\frac{g^2/2}{m_W^2} = -\frac{2}{v^2}\, .
\eeq 
This operator captures several aspects of the low-energy phenomenology of the SM, including for example the  decay of the muon,  $\mu \to e \nu \nu$,
and the inelastic scattering  of neutrinos on electrons $\nu e \to \nu \mu$.  
It can be used to adequately describe these processes as long as the energy scale involved ({\it i.e.} the momentum transfer between the electron current and the muon current)
is well below $m_W$.
However,  the information concerning $m_W$ is {\em not} available to a low-energy observer.
Instead, only the scale $|c^{(6)}|^{-1/2}  \sim  v = 2 m_W/g$ is measurable at low energies, which is not sufficient to determine
$m_W$ without knowledge of the coupling $g$.
For example, from a bottom-up viewpoint, a  precise measurement of the muon lifetime gives indications on the energy at which some new
particle ({\it i.e.} the $W$ boson) is expected to be produced in a higher-energy process, like
the scattering $\nu e \to \nu \mu$, only after making an assumption on the strength of its coupling 
to electrons and muons. 
Weaker couplings imply lower scales: for example, the Fermi theory could have ceased to be valid  right above the muon mass scale had
the SM been very weakly coupled,  $g \approx  10^{-3}$. On the other hand, a precise measurement of the muon lifetime sets an upper bound 
on the mass of the $W$ boson, $m_W \lesssim 1.5$~TeV,   
corresponding to the limit in which the UV completion is maximally strongly coupled, $g \sim 4 \pi$. 

This example illustrates the necessity of making assumptions (in this case on the value of the coupling $g$, see also Section~\ref{app:BSMmodel}
for another BSM example) when assessing the validity range of the EFT, that is, when estimating the mass scale at which new particles appear.
 On the other hand, the very interest in the EFT stems from its model-independence, and from the possibility of deriving the results from experimental analyses 
using Eq.~(\ref{Lops}) without any reference to specific UV completions.
In this note we identify under which physical conditions Eq.~(\ref{Lops}), and in particular its truncation at the level of dimension-6 operators, can be used 
to set limits on, or determine, the value of the effective coefficients.
Doing so, we also discuss the importance that results be reported by the experimental collaborations
in a way which makes it possible to later give a quantitative assessment of the validity range of the EFT approach used in the analysis.
As we will discuss below, this entails estimating the energy scale characterizing the physical process under study.
Practical suggestions on how experimental results should be reported will be given in this note.
A more theoretical discussion of the EFT validity issues and scaling of higher-dimensional operators can be found in Ref.~\cite{Contino:2016jqw}.

\subsection{General discussion}
\label{sec:gen}

\subsection{Model-independent experimental results}

Let us first discuss how an experimental analysis can be performed  in the context of the EFT.
We start considering Eq.~(\ref{Lops}) truncated at the level of $D\!=\!6$ operators, and assume that it gives an approximate low-energy description of the UV theory. 
Below we discuss the theoretical error associated with this truncation and identify the situations where the truncation is not even possible.
Physical observables are computed from the truncated EFT Lagrangian in a perturbative expansion according to the usual rules of effective field theories 
\cite{Weinberg:1995mt}. The perturbative order to be reached depends on the experimental precision and on the aimed theoretical accuracy, as we
discuss in the following. Theoretical predictions obtained in this way depend on the coefficients $c_i^{(6)}$ and can be used to perform a fit to the experimental data.
The fit to the coefficients  $c_i^{(6)}$ should be performed by correctly including the effect of all the theoretical uncertainties (such as those from the PDFs
and missing SM loop contributions~\footnote{These latter can be estimated as usual by varying the factorization and renormalization scales.}) not originating 
from the EFT perturbative expansion. 
The errors due to the truncation at the $D\!=\!6$ level and higher-loop diagrams involving
insertions of different effective operators, on the other hand, are not quantifiable in a model-independent way and should thus be reported separately.
Below we discuss how the neglected contributions from $D\geq 8$ operators can be estimated;   the effects of EFT loops are discussed elsewhere [NLO note].  

Let us consider a situation in which no new physics effect is observed in future data (the discussion follows likewise in the case of observed deviations from the SM).
In this case, the experimental  results can be  expressed into
 the limits~\footnote{In general, the experimental constraints on different $c^{(6)}_i$ may have non-trivial correlations.   
Depending on a chosen basis, the left-hand-side of  Eq.~(\ref{eq:bounds}) may contain linear combinations of several Wilson coefficients.
If a deviation from the SM is observed, Eq.~(\ref{eq:bounds}) turns into a confidence interval, 
$\delta_i^\text{d,exp}(M_{\rm cut}) <  c^{(6)}_i < \delta_i^\text{u,exp}(M_{\rm cut})$. }
\begin{equation}
\label{eq:bounds}
c^{(6)}_i < \delta_i^\text{exp}(M_{\rm cut})\,.
\end{equation}
The functions $\delta_i^\text{exp}$ depend on the {upper} value,
here collectively denoted by $M_{\rm cut}$,  of the kinematic variables (such as transverse momenta or 
invariant masses) that set the typical energy scale characterizing the process and, in general, Eq.~(\ref{eq:bounds}) is obtained by imposing cuts on these variables and making use of the differential kinematic distributions of the process.
For example, when the EFT is applied to describe inclusive on-shell Higgs boson decays  one has $M_{\rm cut} \approx m_h$. 
Another example is  $e^+ e^- $ collisions at a fixed centre-of-mass energy $\sqrt{s}$, in which case $M_{\rm cut} \approx \sqrt{s}$. 
For certain physically important  processes these considerations are less trivial, especially in the context of hadron collider experiments.
The relevant scale for the production of two on-shell particles  in proton-proton collisions, for example, is the centre-of-mass energy of the partonic collision 
$\sqrt{\hat s}$;  this varies in each event and may not be fully reconstructed in practice.
Important examples of this kind are the vector boson  scattering ({\it e.g.} with final states $WW\to 2\ell2\nu$ and $ZZ\to 4\ell$), and Higgs  production in association with 
a vector boson ($Vh$) or a jet ($hj$). In all these processes the relevant energy is given by the invariant mass of the final pair; when this cannot be fully
reconstructed, other correlated variables such as the transverse momentum of the Higgs or a lepton, or the transverse invariant mass can be considered.\footnote{However, one needs to be aware that it is the former
which determines if one is within the validity range of the EFT.}
Since the energy scale of the process determines the range of validity of the EFT description, it is extremely important that the experimental limits $\delta_i^\text{exp}$  
are reported by the collaborations for various values of~$M_{\rm cut}$.
For processes occurring over a wide energy range (unlike Higgs boson decays or $e^+ e^- $ collisions),  knowledge of {only the limit $\delta_i^\text{exp}$
obtained by making use of all the events without any restriction on the energy  ({\it i.e.} for $M_\text{cut}\to \infty$)}
severely limits the interpretation of the EFT results in terms of constraints on specific BSM models. 
If the relevant energy of the process cannot be determined  ({\it e.g.} because the kinematics cannot be closed),  setting consistent bounds 
requires a more careful procedure, similar to the one proposed in Ref. \cite{Racco:2015dxa} in the context of DM searches.
%

\subsection{EFT validity and interpretation of the results}

Extracting bounds on {(or measuring)} the EFT coefficients can be done by experimental collaborations
 in a completely model-independent way. 
However, the {\em interpretation} of these bounds  is always model-dependent. 
In particular, whether or not  the EFT is valid in  the parameter space probed by the experiment  
depends on  further assumptions about the (unknown) UV theory. 
These assumptions {correspond, in the EFT language, to a choice of power counting},
{\it i.e.} a set of rules to estimate the  coefficients of the effective operators in terms of the couplings and mass scales of the UV dynamics.

The simplest situation is when the microscopic dynamics is characterized by a single mass scale $\Lambda$ and a single 
new coupling $g_*$ \cite{Giudice:2007fh}. 
This particular power counting  prescription smoothly interpolates between the naive dimensional analysis ($g_* \sim 4 \pi$) \cite{Manohar:1983md,Cohen:1997rt}, 
the case $g_* \sim  1$ as {\it e.g.} in the Fermi theory,
and the very weak coupling limit $g_* \muchless 1$.    
While this is not a unique prescription, it covers a large selection of popular scenarios beyond the SM.
In this class falls the Fermi theory described previously, as well as other weakly coupled models where a narrow resonance 
is integrated out. 
Moreover, despite the large number of resonances, also some theories with a  strongly-interacting BSM sector belong to this category ({\it e.g.} the holographic composite 
Higgs models \cite{Agashe:2004rs} or, more generally, theories where the strong sector has a large-N description).  
The scaling of the effective coefficients with $g_*$ is then determined by Eq.~(\ref{eq:counting}) and by symmetries and selection rules.

For a given power counting, it is relatively simple to derive limits on the theoretical parameter space that are automatically consistent with the EFT expansion, provided 
the relevant energy of the process is known. Consider the case of a single scale $\Lambda$ and a single coupling strength~$g_*$. Then the bounds~(\ref{eq:bounds}) can be recast as limits on these two parameters by using the power counting to estimate $c_i^{(6)} = \tilde c^{(6)}_i(g_*)/\Lambda^2$, 
and setting the maximum relevant energy scale to $M_\text{cut} = \kappa \Lambda$. 
Here $\tilde c_i^{(6)}(g_*)$ is a (dimensionless) polynomial of $g_*$ and of the SM couplings, while $0 < \kappa < 1$ controls the size of the tolerated error due to neglecting 
higher-derivative operators (the value of $\kappa$ can be chosen according to the sensitivity required in the analysis).
One finds
\begin{equation} \label{eq:bounds2}
\frac{\tilde c^{(6)}_i(g_*)}{\Lambda^2} < \delta_i^\text{exp}(\kappa\Lambda)   \, .
\end{equation}
These inequalities determine the region of the plane $(\Lambda, g_*)$ which is excluded consistently with the EFT expansion, with a relative error of order $\kappa^2$. 
These are a conservative bounds, since they are obtained by using only a subset of the events (effectively only those with relevant energy up to 
$M_\text{cut} = \kappa \Lambda$).
They are thus less stringent than the bounds one would obtain in the full theory with the full dataset, but they are by construction consistent with the EFT expansion.
They give a useful indication of how effective are the experimental data in constraining the class of theories under consideration ({\it i.e.} those respecting the assumed power counting).
A detailed re-analysis of experimental results based on the $M_{\rm cut}$ technique that we propose here, was performed in Ref.~\cite{Biekoetter:2014jwa} for processes with $Vh$ associated production.
The same reasoning can be applied to more complicated theories  following a different power counting than the simple $g_*$-scaling presented above.


The usefulness of power counting stems from a number of reasons. 
First of all it provides a physically motivated range in which the coefficients 
$c_i^{(D)}$ are expected to vary. 
Secondly, and very importantly, it allows one to estimate the relative importance of higher-order terms in the EFT series.
As an example,  consider a $2\to 2$ scattering process, where the SM contribution to the amplitude is at most of order $g_{SM}^2$ 
at high energy ($g_{SM}$ denotes a SM coupling). 
The correction from $D\!=\!6$ operators involving derivatives will in general grow quadratically with the energy and can be as large as $g_*^2 (E^2/\Lambda^2)$.\footnote{
Effects growing with energy can also be induced by operators without additional derivatives, if they yield new contact interactions relevant for the process, or if they disrupt cancellations between ${\cal O}(E^2)$ contribution of different SM diagrams, see {\it e.g.} \cite{Dror:2015nkp,Bylund:2016phk}. The following discussion is unchanged in these cases.} 
If the coupling strength $g_*$ is much larger than $g_{SM}$, then the BSM contribution dominates over the SM one at sufficiently high energy 
({\it i.e.} for $\Lambda > E > \Lambda\, (g_{SM}/g_*)$), while the EFT expansion is still valid. 
The largest contribution to the cross section in this case comes from the square of the $D\!=\!6$ term, rather than from its interference with the SM.
The best sensitivity to $c_i^{(6)}$ is thus expected to come from the highest value of the relevant energy scale accessible in the experiment.
In this example the contribution of $D\!=\!6$ derivative operators is 
enhanced by a factor $(g_*/g_{SM})^2$ compared to the naive expansion parameter~$(E/\Lambda)^2$; such enhancement is a consequence of the fact that 
the underlying strong coupling $g_*$ only appears at the level of $D\!=\!6$ operators, while SM  operators mediate weaker interactions.
In this example, no further enhancement exists between $D\!=\!6$ and $D\!=\!8$ operators, {\it i.e.} $D\!=\!8$ operators are subdominant and the EFT series is converging.
In other words, although the contributions to the cross section proportional to $(c_i^{(6)})^2$ and $c_i^{(8)}$ 
are both of order $1/\Lambda^4$, the latter (generated by the interference of $D\!=\!8$ operators with the SM) is smaller by a factor
$(g_{SM}/g_*)^2$  independently of the energy, and can thus be safely neglected.  
A well known process where the above situation occurs
 is the scattering of longitudinally-polarized vector bosons.
 Depending on the UV dynamics, the same can happen in other $2\to 2$ scatterings,
such as Higgs associated production with a $W$ or $Z$ boson (VH) \cite{Biekoetter:2014jwa,Biekotter:2016ecg} or dijet searches at the LHC \cite{Domenech:2012ai}.  
A simple illustrative example is discussed in the next section.
Finally, the domination of  $(c_i^{(6)})^2$ terms can also happen when $g_*$ is moderate or small but  at the same time $g_{SM}$ is even more suppressed. 
One possible example concerns flavour-changing neutral current processes which in the SM are  strongly suppressed by a loop and CKM factors, see {\it e.g.} \cite{Durieux:2014xla}. 
An even sharper example is lepton-flavour violating processes ({\it e.g.} $h \to \mu \tau$) for which $g_{\rm SM} = 0$ exactly.  

\subsection{On the importance of loop corrections}
\label{sec:loop}

So far our discussion was limited to tree-level effects of $D\!=\!6$ operators. 
The EFT can be consistently extended to an arbitrary loop order 
by computing observables perturbatively in the SM couplings.
The corresponding series is controlled by the expansion parameter $g_{SM}^2/16 \pi^2$, which adds to the two EFT parameters $\kappa_v^2= (g_*v/\Lambda)^2$ (assuming again a simple $g_*$-scaling of the effective couplings)
and $\kappa_E^2=(E/\Lambda)^2$ controlling the effects of the neglected higher-dimensional operators.
One-loop effects of $D$=6 operators are formally suppressed by $O(g_{SM}^2/16 \pi^2)$, and are thus {generally} subleading compared to the tree-level
contributions.
Including loop corrections in the EFT context is, at present, less crucial than for a pure SM calculation. 
This is because the experimental precision is typically better than the magnitude of the SM loop corrections, therefore going beyond tree level
in a SM calculation  is essential to obtain a correct description of physical processes.
In the case of the EFT, on the other hand, we are yet to observe any leading-order effect of higher-dimensional operators. 

There do  exist situations, however,  where including NLO corrections may be important for obtaining an adequate description of physical processes in the EFT  (see Refs.~\cite{Willenbrock:2014bja,Henning:2014wua} for an extended discussion). 
For example, it is well known that  NLO QCD corrections to the SM predictions of certain processes
at the LHC can be of order 1, and  large k-factors are expected to apply to the EFT corrections as well.
Another example is the one-loop Higgs corrections to electroweak precision observables.
Since deviations of the Higgs boson couplings due to $D$=6 operators can be relatively large (up to $O(10\%)$) without conflicting with current experimental data, 
the 1-loop effects, in spite of the suppression factor,   can be numerically important for observables measured with a per mille 
precision~\cite{Barbieri:2007bh,Elias-Miro:2013eta,Henning:2014gca}.
%

More generally, 1-loop corrections are important if they stem from large coefficients and correct precisely measured observables whose tree-level contribution
arises from smaller coefficients.
The tree-level contribution of a $D\!=\!6$ operator may be suppressed, for example, because its coefficient is
generated  at the 1-loop level by the UV dynamics.
In this case, both the 1-loop and tree-level
contributions from $D\!=\!6$ operators would correspond to 1-loop processes in the UV theory.
An example of this kind is the decay of the Higgs boson to two photons, $h \to \gamma \gamma$, which arises necessarily at the 1-loop level if the UV theory
is minimally coupled (see Ref.~\cite{Giudice:2007fh} and the appendix of Ref.~\cite{Liu:2016idz}) and perturbative.
The calculation of NLO effects in the context of the EFT is currently an active field of study.
As suggested by the above discussion, it is very important to identify all cases  where 1-loop effects of $D\!=\!6$ operators can be relevant.\footnote{Editor footnote: 
Another point of view is expressed in Section~\ref{s.eftnlo} where it is argued that, considering projections for the precision to be reached in LHC RunII analyses, the LO approach may not be sufficient.}

Besides one-loop effects, it is sometimes also important to include corrections from real emission processes.
In particular, including additional jets may be important when exclusive observables, {\it i.e.} quantities particularly sensitive to extra radiation,
are studied.

\subsection{An Explicit Example}\label{app:example}
\label{app:BSMmodel}

In this section we illustrate our general arguments by comparing the predictions of the EFT and of a specific BSM model which reduces to that EFT at low energies. 
To this end we discuss the $q \bar q \to V h$ process at the LHC, along the lines of Ref.~\cite{Biekoetter:2014jwa}.  
The purpose of the example presented below is to demonstrate that, as in the Fermi theory, the knowledge of the $D\!=\!6$ coefficients of an effective Lagrangian is not enough to determine  the validity range of the EFT approximation. 
Therefore, the theoretical error incurred as a result of the truncation of the EFT Lagrangian cannot be quantified in a model-independent way.

We consider the SM extended by a triplet of vector bosons $V_\mu^i$ with mass $M_V$  transforming in the adjoint representation  of the SM $SU(2)_L$ symmetry. 
Its couplings to the SM fields are described by~\cite{Low:2009di,deBlas:2012qp,Pappadopulo:2014qza}
\beq
\label{eq:lvt}
{\cal L} \supset    i g_H  V_\mu^i H^\dagger \sigma^i  \overleftrightarrow{D_\mu} H 
+   g_q  V_\mu^i  \bar q_L \gamma_\mu \sigma^i q_L,
\eeq 
where $q_L = (u_L,d_L)$ is a doublet of the 1st generation left-handed quarks. 
In this model $V_\mu^i$ couples to light quarks, the Higgs boson, and electroweak gauge bosons, 
and it contributes to the $q \bar q \to V h$ process at the LHC. 
Below the scale $M_V$,  the vector resonances can be integrated out, giving rise to an EFT where the SM is extended by $D$=6 and higher-dimensional operators. 
Thus, $M_V$ plays the role of the EFT cut-off scale $\Lambda$. 
Using the language of the Higgs basis introduced in Section~\ref{s.eftbasis}, at the $D$=6 leve the EFT l is described by the parameter  $\delta c_z$  (relative correction to the SM Higgs boson couplings to $WW$ and $ZZ$) and $\delta g_L^{Zq}$ (relative corrections to the $Z$ and $W$ boson couplings to left-handed quarks),   plus other parameters that do not affect the $q \bar q \to V h$ process at tree level. 
The relevant EFT parameters are matched to those in the UV model as 
\beq 
\label{eq:eftvsmodel}
\delta c_z  =  -  \frac{3 v^2}{2 M_V^2} g_H^2, 
 \qquad  
\, [\delta g^{Zu}_L]_{11}  = - [\delta g^{Zd}_L]_{11}  =   - \frac{v^2}{2 M_V^2}  g_H g_{q}\, . 
 \eeq 
When these parameters are non-zero, certain EFT amplitudes grow as the square of the centre-of-mass energy $s \equiv M_{Wh}^2$ of the analysed process,  ${\cal M} \sim M_{Wh}^2/M_V^2$.
Then, for a given value of the parameters, the observable effects of the parameters become larger at higher energies.
However, above a certain energy scale, the EFT may no longer approximate correctly the UV theory defined by Eq.~(\ref{eq:lvt}), and then experimental constraints on the EFT parameters do not provide any information about the UV theory. 
  
To illustrate this point, we compare the UV and EFT descriptions of $q \bar q \to W^+ h$ for three benchmark points: 
\begin{itemize} 
\item {\bf Strongly coupled:} $M_V = 7$~TeV, $g_H= -g_q = 1.75$; 
\item {\bf Moderately  coupled:} $M_V = 2$~TeV, $g_H=-g_q = 0.5$; 
\item {\bf Weakly  coupled:} $M_V = 1$~TeV, $g_H=-g_q = 0.25$; 
\end{itemize}  
Clearly, all three benchmarks lead to the same EFT parameters at the $D\!=\!6$ level.
However, because $M_V = \Lambda$ varies, these cases imply different validity ranges in the EFT. 
This is illustrated in \refF{fig:su2l}, where we show (in the left panel) the production cross section as a function of
$M_{Wh}$, for both the full model and the EFT.
While, as expected, in all cases the EFT description is valid
near the production threshold, above a certain point $M^{\rm max}_{Wh}$ the EFT is no longer a good approximation of the UV theory. 
Clearly, the value of $M^{\rm max}_{Wh}$ is different in each case. 
For the moderately coupled case,  it coincides with the energy at which the linear and quadratic EFT approximations diverge. 
From the EFT perspective, this happens because $D\!=\!8$ operators can no longer be neglected. 
However, for the strongly coupled  case, the validity range extends beyond that  point. 
In this case, it is the quadratic approximation that provides a good effective description of the UV theory.  
As discussed in the previous section, that is because, for strongly-coupled UV completions, the quadratic contribution from $D\!=\!6$ operators dominates 
over that of $D \geq$ 8 operators in an energy range below the cutoff scale.

\begin{figure}[t]
\centering
\includegraphics[width=0.48 \textwidth]{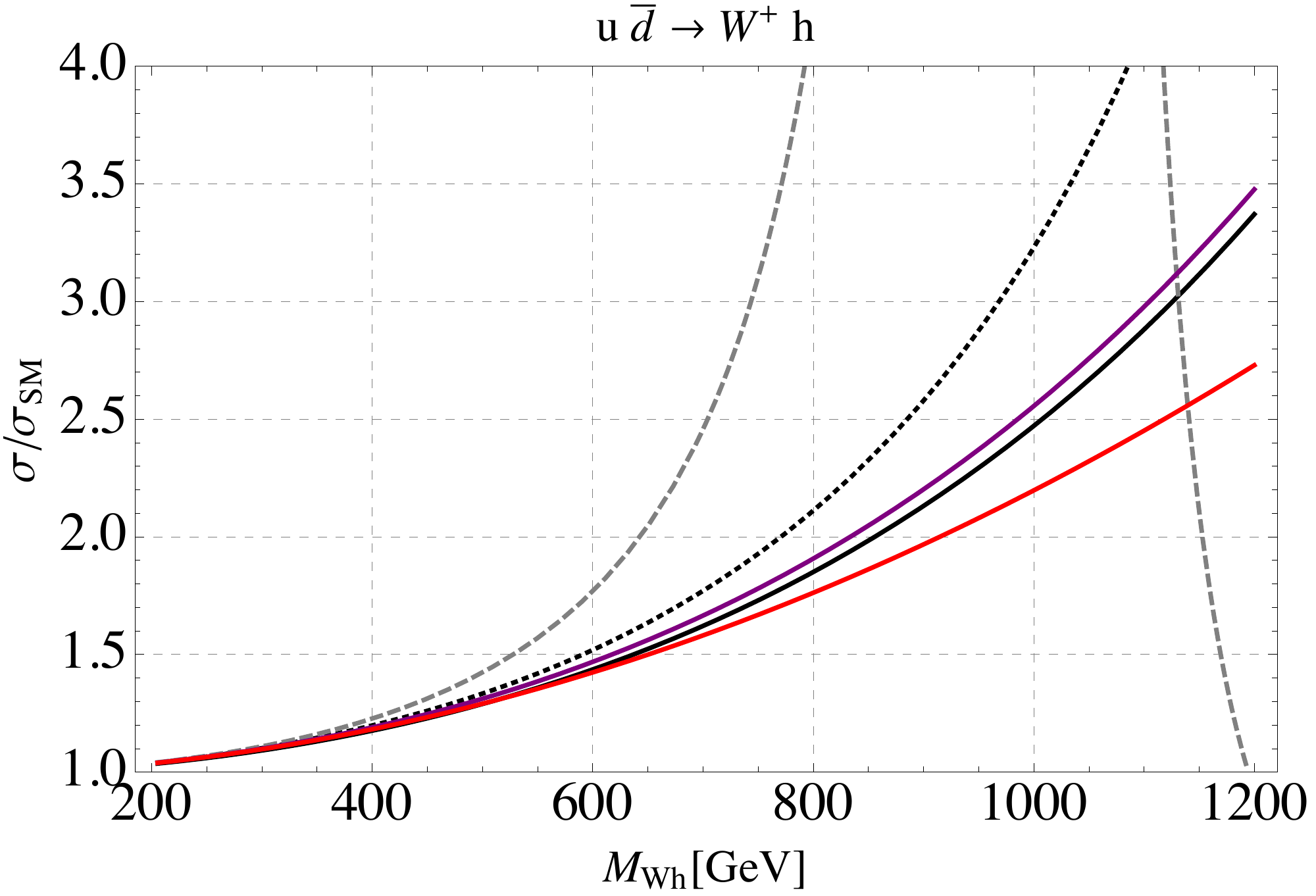}
\quad 
\includegraphics[width=0.48 \textwidth]{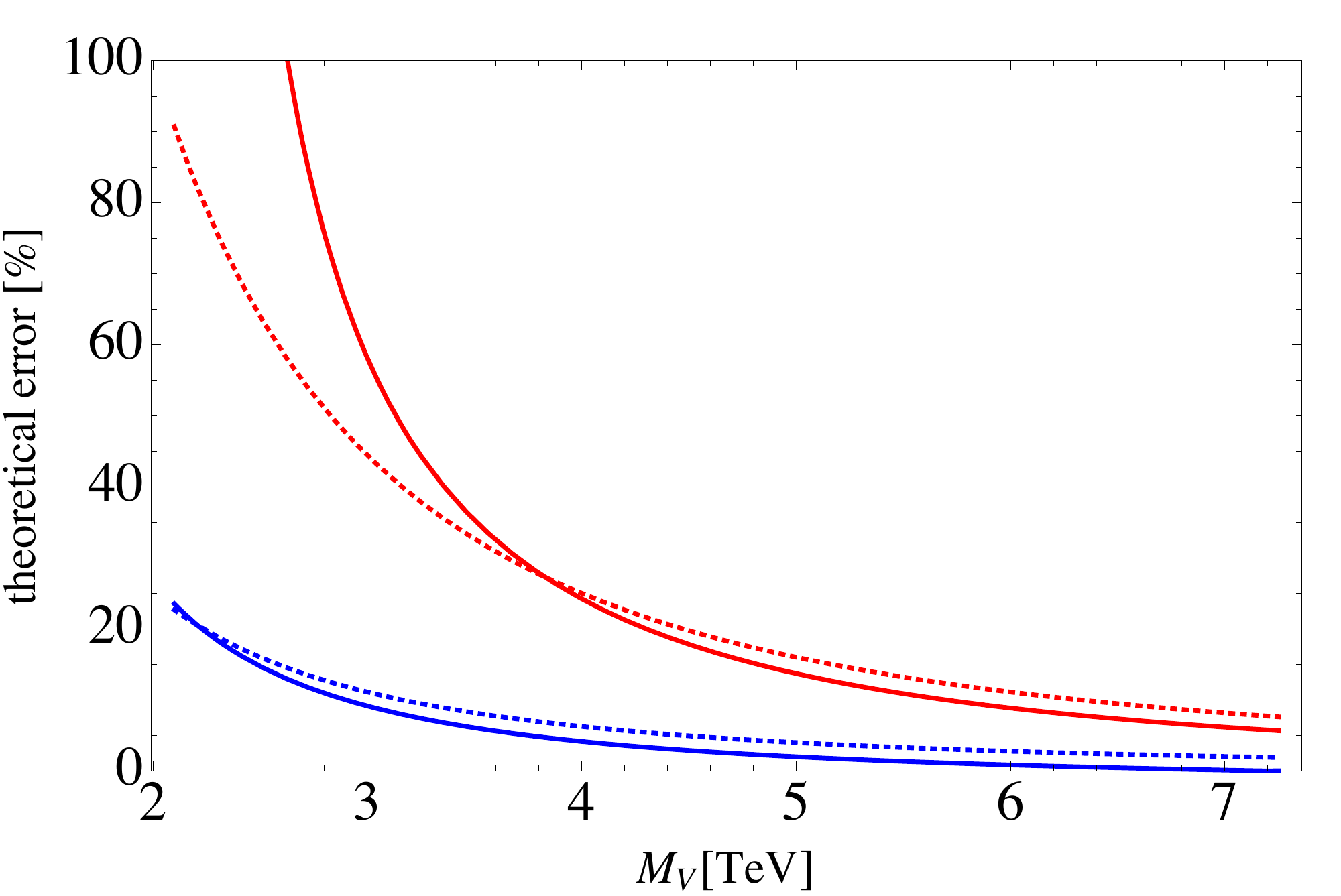} 
\caption{\emph{Left: The partonic $u \bar d \to W^+ h$ cross section as a function of  the centre-of-mass energy of the parton collision. 
The black lines correspond to the $SU(2)_L$ triplet model with  $M_V =1\,$TeV,  $g_H=-g_q = 0.25$ (dashed),   $M_V = 2\,$TeV, 
$g_H=-g_q = 0.5$ (dotted), and  $M_V = 7\,$TeV, and $g_H=-g_q = 1.75$ (solid).  The corresponding  EFT predictions are shown 
in the linear approximation (solid red), and when quadratic terms in $D\!=\!6$ parameters are included in the calculation of the cross section (solid purple). 
Right:  Theory error as a function of $M_V$ (solid line).
The error is defined to be the relative difference between the constraints on $g_*^2 \equiv g_H^2 = g_q^2$ obtained by recasting the limits 
derived in the framework of a $D\!=\!6$ EFT and those derived from the resonance model.
The limits come from re-interpreting the hypothetical experimental  constraints with $M_{\rm cut} = 3\,$TeV, as  described in the text. 
The dotted line corresponds to the naive estimate $(M_{\rm cut}/M_V)^2$.
 }}
\label{fig:su2l}
\end{figure} 

As an illustration of our discussion of setting limits on EFT parameters and estimating associated theoretical errors, consider the following example of an idealized measurement.
Suppose an experiment  makes the following measurement of the $\sigma(u \bar d \to W^+ h)$  cross section at different values of $M_{Wh}$: 
{
\begin{center}
\begin{tabular}{c|cccccc}
$M_{Wh}$[TeV] & 0.5 & 1 & 1.5  &  2 & 2.5 & 3 
 \\[0.1cm] \midrule  &&&&&& \\[-0.45cm]
$\sigma/\sigma_{\rm SM}$ & $1 \pm 1.2$ & $1 \pm 1.0$  & $1 \pm 0.8$  & $1 \pm 1.2$ & $1 \pm 1.6$ & $1 \pm 3.0$  
\end{tabular}
\end{center}
}

This is meant as a simple proxy for more realistic measurements at the LHC, for example measurements of a fiducial $\sigma(pp \to W^+ h)$ cross section in several bins of $M_{Wh}$. 
For simplicity, we assume that the errors are Gaussian and uncorrelated.  
These measurements can be recast as constraints on $D$=6 EFT parameters for different $M_{\rm cut}$ identified in this case with the maximum  $M_{Wh}$ bin included in the analysis. 
For simplicity, in this discussion we only include $\delta g^{Wq}_L  \equiv [\delta g^{Zu}_L]_{11} - [\delta g^{Zd}_L]_{11}$
and ignore other EFT parameters (in general, a likelihood function in the multi-dimensional space of EFT parameters should be quoted by experiments). 
Then the ``measured" observable is related to the EFT parameters as 
\beq
\frac{\sigma}{\sigma_{SM}} \approx \left ( 1 + 160\, \delta g^{Wq}_L 
\frac{M_{Wh}^2}{{\rm TeV}^2}\right )^2.
\eeq  
Using this formula, one can recast the measured cross sections as $95\%$~CL confidence intervals on $\delta g^{Wq}_L$: 
Combining the $M_{Wh}$ bins up to $M_{\rm cut}$, one finds the following  $95\%$ confidence intervals:  
{
\begin{center}
\begin{tabular}{c|cccccc}
$M_{\rm cut}$[TeV] & 0.5 & 1 & 1.5  &  2 & 2.5 & 3  
 \\[0.08cm]  \midrule {\vrule height 16pt depth 10pt width 0pt}
$\delta g^{Wq}_L \times 10^{3}$ & [-70, 20] & [-16,4] & [-7,1.6] & [-4.1,1.1] &  [-2.7,0.8] & [-2.2,0.7] 
\end{tabular}
\end{center}
}

Suppose these constraints are quoted by experiment. 
A theorist may try to interpret them as constraints on the vector resonance model with $g_q = - g_H \equiv g_*$ using the map in Eq.~(\ref{eq:eftvsmodel}).
This way one would obtain the constraints on $g_*$ as a function of $M_V$: 
for example, for $M_V =3$~TeV one would find  $g_* \leq 0.80(0.49)$ for $M_{\rm cut,1} = 1$~TeV  ($M_{\rm cut,2}=2$~TeV).  
Note that, for our (arbitrary) choice of data points, the limits on $g_*$ obtained from the measurement with $M_{\rm cut,2}$ are stronger from the one with $M_{\rm cut,1}$. 
However, the result for  $M_{\rm cut,1}$ is also useful for theorists. 
First, it can be used also for $M_V \approx 2$~TeV, whereas the one with $M_{\rm cut,2} = 2$~TeV does not have a meaningful interpretation in this mass range. 
Furthermore, the theory error is smaller for $M_{\rm cut,1}$ ($\sim 10\%$ for $M_V=3$~TeV) than  for $M_{\rm cut,2}$  ($\sim 40\%$  for $M_V=3$~TeV).  
Here, we define the theory error as the fractional difference between the bound on $g_*^2$ interpreted from the EFT constraints, and the true bound obtained by fitting the full resonance model to the experimentally ``measured" cross sections using the bins up to a given $M_{\rm cut}$. 
The theory errors are plotted as solid lines in the  right panel of \refF{fig:su2l}. 
By general arguments discussed in the previous section, one expects the theory error to scale as $\kappa^2$ for $M_V \gg M_{\rm cut}$, where $\kappa = M_{\rm cut }/\Lambda = M_{Wh}/M_V$, and this expectation, which is shown as dotted lines in the same plot, is confirmed.
While the limits  on $g_*$ obviously depend on the experimental central values and errors we assumed,  the theory errors as defined here are very weakly dependent on it. 

\subsection{Summary}

In this note we have discussed the validity of an EFT where the SM is extended by $D$=6 operators. 
{We have emphasized} that the validity range cannot be determined using only low-energy information. 
The reason is that, while the EFT is valid  up to energies of order of the mass $\Lambda$ of the new particles,  low-energy observables  depend on the combinations $\tilde c^{(6)}/\Lambda^2$, where the Wilson coefficients $\tilde c^{(6)}$ of $D$=6 operators  are function of the couplings of the UV theory.

The question of a theoretical uncertainty due to the truncation of the EFT at the level of $D$=6 operators depends on the impact of $D>6$ operators on the studied processes.   
The relative size of the contribution of $D$=8 operators is controlled by $\tilde c^{(8)}/\tilde c^{(6)} (E_{\rm exp}^2/\Lambda)$, where $E_{\rm exp}$ is the typical energy scale of the process. 
We have discussed  the physical assumptions that lead to a situation with $\tilde c^{(8)}\approx\tilde c^{(6)}$. In this situation the energy at which the EFT breaks down  coincides with the scale at which the contribution of $D$=8 and higher-dimensional operators is of the same order 
as that of $D$=6 operators.   
Conversely, when the EFT expansion is well convergent at the LHC energies,  the effects of $D$=8 operators can be  neglected. 
We have also shown that the power counting is  necessary to estimate the range of variation of the effective
coefficients $\tilde c_i^{(6)}$, and to identify situations in which departures from the SM can be sizeable (even bigger than the SM itself), compatibly with the EFT expansion. 
Exceptions from this rule, in the form $\tilde c^{(8)}\gg\tilde c^{(6)}$, may arise in a controlled way as a consequence of symmetries and selection rules or for certain well-defined classes of  processes. 
The concrete examples where this occurs are discussed in Ref.~\cite
{Contino:2016jqw}. 
The inclusion of $D$=8 operators in experimental analyses is justified only when dealing with these  special cases,  and would represent  an inefficient strategy in a generic situation.

We have stressed that the ratio  $\tilde c^{(8)}/\tilde c^{(6)}$, which controls the  theoretical uncertainty of the EFT predictions,  depends on the assumptions about the UV theory that generates the $D$=6 and $D$=8 operators. 
Only when a particular power counting is adopted, for example the $g_*$-scaling discussed in this note, can the  contributions from $D\!=\!6$ and  $D\!=\!8$ relative to the SM be estimated in a bottom-up approach, and the error associated with the series truncation be established. 
 For this reason we suggest to report the estimated uncertainty due to the truncation separately from the other errors, and to clearly state on which assumptions the estimate is based.

If no large deviations from the SM are observed at the LHC Run-2, stronger constraints on $D\!=\!6$ operators can be set. 
As we discussed, this will extend the EFT validity range to a larger class of UV theories ({\it i.e.} those with smaller $c^{(6)}$) and
leave less room for contributions of $D\!=\!8$ operators.
As a consequence, the internal consistency and the validity range of the LO $D\!=\!6$ EFT will increase.\footnote{Editor footnote: 
A different conclusion is presented in Section~\ref{s.eftnlo}.  The discrepancy is due to different assumptions about the underlying UV theory. For example, in a situation in which both $\Lambda$ and $c^{(6)}$ are small while $c^{(8)}$ is sizeable, the general validity discussion presented in this contribution would not be applicable.} 
The validity range can also be improved by means of a global analysis combining different measurements,  which often lifts flat directions in the parameter space~\cite{Pomarol:2013zra,Falkowski:2014tna} and leads to stronger constraints on $D\!=\!6$ effective coefficients, {see {\it e.g.}~\cite{Falkowski:2015jaa,Butter:2016cvz}}. 
{On the other hand, if a deviation from} the SM is observed, efforts to include EFT loop corrections and to estimate the effects of $D>6$ operators may be crucial to
better characterize the underlying UV theory.

Most of the discussion in this note is relevant at the level of the interpretation of the EFT results, rather than at the level of experimental measurements.  
However, there are also practical conclusions for experiments.  
We have proposed  a concrete strategy to extract bounds on (or determine) the effective coefficients of $D\!=\!6$ operators in a way which is automatically 
consistent  with the EFT expansion. This requires reporting the experimental results as functions of the upper cuts (here collectively denoted by $M_{\rm cut}$) 
on the kinematic variables, such as transverse momenta or invariant masses, that set the relevant energy scale of the process.
This is especially important for hadron collider experiments, such as those performed at the LHC, where collisions probe a wide range of energy scales.  
In general, knowledge of the experimental results as a function of $M_\text{cut}$ allows one to constrain a larger class of theories beyond the SM in a larger range 
of their parameter space.  
As a quicker (though less complete) way to get an indication on the validity range of the EFT description, it is also useful to present the experimental results both 
with and without  the contributions to the measured cross sections and decay widths that are quadratic in the effective coefficients.  
This gives an indication on whether the constraints only apply to strongly-interacting UV theories or they extend also to weakly-coupled ones.
Notice that even in situations where it makes sense to expand the cross section at linear order in the coefficients of $D\!=\!6$ operators, quadratic terms should 
always be retained in the calculation of the likelihood function.
With this way of presentation, the experimental results can be applied to constrain a larger class  of theories beyond the SM in a larger range of their parameter space.  
Other frameworks to present results, for example the template cross-sections or the pseudo-observables discussed elsewhere in this volume, 
should also be pursued in parallel, as they may address some of the special situations discussed in this note.  
Finally, given its model-dependency, we suggest to report the estimated uncertainty on the results implied by the EFT truncation separately 
from the other errors, and to clearly state on which assumptions the estimate is based.\footnote{Editor footnote:
Another point of view is expressed in Section~\ref{s.eftnlo} that puts more emphasis on the uncertainty due to the EFT truncation.  There it is argued that this uncertainty is an essential part of the theory prediction, to compare to the experimental results.}

Note that even in the case of BSM discoveries in the next LHC runs, the EFT approach and the results 
presented here remain still useful. 
For measurements with a characteristic scale $M_{\rm cut}$ considerably below
the new physics threshold, the new particle(s) can be integrated out (in analogy to the Fermi Theory) and deviations from the predicted values of the $D$=6 coefficients can be probed. 
Such an EFT approach may give a more economical description of the relevant precess, with fewer parameters (the effective coefficients) that can be directly measured from low-energy data.
For processes involving higher scales, an EFT including the BSM degrees of freedom cam be set up and all results generalize straightforwardly.

A concluding comment is in order when it comes to constrain explicit models from the bounds derived in an EFT analysis of the data.
Although EFT analyses aim at a global fit with all the operators included, it is important to ensure that the reported results are complete enough  
to later consider more specific scenarios  where one can focus on a smaller set of operators. Reporting the full likelihood function, or at the very least the correlation 
matrix, would be a way to address this issue.

\section[The Standard Model EFT and Next to Leading Order]{The Standard Model Effective Field Theory and Next to Leading Order\SectionAuthor{G.~Passarino, M.~Trott. The present content of this section reflects the initial contribution and also the editing of the WG2 conveners (A.~Falkowski, C.~Hays, G.~Isidori, M.~Chen, A.~Tinoco)}}
\label{s.eftnlo}
\subsection{Overview}

In this section we discuss how to interpret data in the Standard Model Effective Field Theory (SMEFT)
in a transparent manner at leading order (LO) and explain why a next to leading order (NLO)
interpretation of the data is important. 
The approach presented  for LO is the one we consider the most simple to enable the
ongoing development of the SMEFT to NLO. The LO approach we present
is written in terms of mass eigenstate fields and is trivially connected to Higgs observables and electroweak precision observables. It can be directly used at LO to interpret the data.

Interpreting the data using theoretical results developed beyond LO (in perturbation theory) can often be crucial to do in the SMEFT. NLO calculations should be used if they are available.
We discuss the basic issues involved in improving calculations to NLO, and review the advances in this direction that have been achieved to date.
These calculations help characterize (and reduce) theoretical errors of a LO result
and allow the consistent incorporation of precise measurements, such as the LEP pseudo-observables, in the SMEFT.
NLO interpretations of the data are particularly critical in the event that deviations from the Standard Model (SM) emerge over the course of LHC operations.
NLO results are being developed in the theoretical community and will become increasingly available over the course of RunII.
Experimental analyses can adopt approaches to LO that will allow these results to be incorporated in the future as efficiently as possible. \footnote{Editor footnote: 
Another point of view is expressed in Section~\ref{s.eftbasis} which advocates using an operator basis that at leading order simplifies/diagonalizes the relation between the observables and the EFT parameters and focuses the attention on the least constrained directions in the EFT space.}

This review provides scientific support for the above statements. The reader who is mostly interested in the LO and NLO summary conclusions can skip directly to the end of this review.

\subsection{Introduction to the SMEFT}

As exact non-perturbative solutions to
quantum field theories are rarely known approximate solutions that expand observables perturbatively in a small
coupling constant or in a ratio of scales are generally developed. Such quantum field theories can be regarded as examples of Effective Field Theory (EFT),
the treatment of which was pioneered in~\cite{Weinberg:1980wa,Coleman:1969sm,Callan:1969sn}.
The predictions of the  LO Lagrangian of any EFT are
approximations of limited applicability and precision.
Developing such predictions beyond leading order is in general extremely useful and straightforward if the LO EFT is well
defined.
The ability to improve EFTs from LO to NLO largely explains why they
have become the standard approach to interpreting data sets of constraints on the SM, as reducing theoretical errors to be below experimental errors is required for a precise interpretation of an experimental measurement.

At LHC it is of interest to treat the Standard Model itself as a general EFT.
In this section we briefly outline how the standard straightforward LO formulation of this SMEFT is defined.
We then discuss extending the SMEFT approach to NLO in order to incorporate important QCD and Electroweak
corrections.

The SMEFT assumes that $\rm SU(2)_L \times U(1)_Y$ is spontaneously
broken to $\rm U(1)_{em}$ by the vacuum expectation value of the Higgs field
($\it v$) and that the observed $J^P =0^+$ scalar is embedded in the Higgs doublet.
The Lagrangian is schematically
\begin{eqnarray}
{\cal L}_{SMEFT} = {\cal L}_{SM} + {\cal L}_{5}+ {\cal L}_{6} + {\cal L}_{7} + {\cal L}_{8} + \cdots
\end{eqnarray}
${\cal L}_{5}$ has one operator suppressed by one power of the cut off scale
($\Lambda$)~\cite{Weinberg:1979sa}. ${\cal L}_{6}$ has $76$ parameters
that preserve Baryon number~\cite{Weinberg:1979sa,Buchmuller:1985jz,Abbott:1980zj,Grzadkowski:2010es} in the $N_f = 1$ limit\footnote{Here $N_f$ counts the number of fermion generations.} and four that do not. The baryon preserving operators in ${\cal L}_{6}$ has  $2499$ parameters in the case
$N_f = 3$ \cite{Alonso:2013hga}.  ${\cal L}_{7}$ and ${\cal L}_{8}$ are now known, see Refs. \cite{Lehman:2014jma,Henning:2015alf}.
We label the Wilson coefficients of the operators in ${\cal L}_5$ as $C_i^5$, operators
in ${\cal L}_6$ as $C_i^6$ etc., and have implicitly absorbed the appropriate power of $1/\Lambda$ into the definition of the $C_i$.
When $1/\Lambda$ is made explicit, and pulled out of the Wilson coefficient we will use the tilde superscript as a notation to indicate this, for example $\tilde{C}_i/\Lambda^2$.

The SMEFT is a different theory than the SM as it has local contact operators suppressed
by powers of $1/\Lambda$. To get a feeling for the nature of the LO and NLO predictions in the SMEFT, consider a (lepton number preserving)
amplitude that can be written as
\arraycolsep 0.14em\begin{eqnarray}
{\cal A} &=& \sum_{n=\mathrm{N}}^{\infty}\,g_{SM}^n\,{\cal A}^{(4)}_n +
       \sum_{n=\mathrm{N}_6}^{\infty}\,\sum_{l=1}^n\,\sum_{k=1}^{\infty}\,
        g_{SM}^n\,\left[\frac{1}{(\sqrt{2}\,G_{F}\,\Lambda^2)^k}\right]^l\,
        {\cal A}^{(4+2\,k)}_{n\,l\,k}
\end{eqnarray}
where $g_{SM}$ is a SM coupling. $G_{F}$ is the Fermi coupling constant and $\Lambda$ is
again the cut off scale.
$l$ is an index that indicates the number of SMEFT operator insertions leading to the amplitude,
and $k$ indicates the inverse mass dimension of the Lagrangian
terms inserted. $N$ is process dependent and  indicates the order of the
coupling dependence for the leading non-vanishing term in the SM (\eg $N = 1$ for
$\PH \to \PV\PV$ \etc but $N = 3$ for $\PH \to \PGg\PGg$). $N_6 = N$ for tree
initiated processes in the SM. For processes that first occur at loop level in the SM,
$N_6 = N - 2$ as operators in the SMEFT can mediate such decays directly thought a contact
operator, for example, through a $\mathcal{L}_6$ operator for $\PH \to \PGg\PGg$. For instance,
the $\PH\PGg\PGg$ (tree) vertex is generated by
$O_{HB} = H^\dagger \, H \, B^{\mu\nu}\,B_{\mu\nu}$, by
$O_{HW}^{8}= \  H^\dagger \,B^{\mu\nu}\,B_{\mu \rho} \,\mathrm{D}^{\rho}\,\mathrm{D}_{\nu}\, H$ \etc
An example of the Feynman diagrams leading to ${\cal A}$ is given in \refF{class}.

\begin{figure}[t]
   \centering
   \includegraphics[width=0.9\textwidth, trim = 30 250 50 80, clip=true]{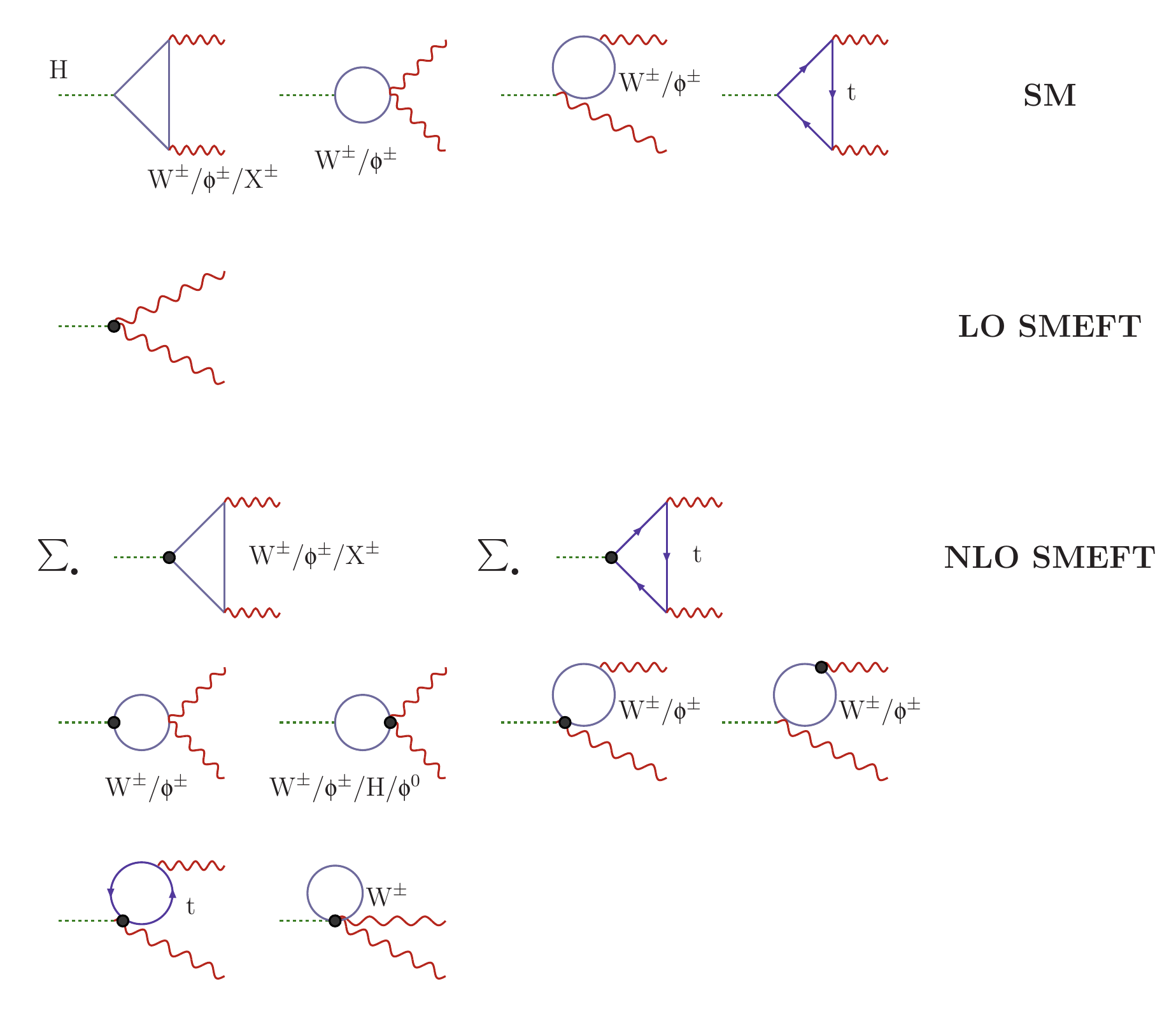}
\caption[]{
Diagrams contributing to the amplitude for
$\PH \to \PGg\PGg$ in the $\mathrm{R}_{\xi}\,$-gauge:
SM (first row), LO SMEFT (second row), and NLO SMEFT.
Black circles denote the insertion of one $\mathcal{L}_6$ operator.
$\sum_{_{\bullet}}$ implies summing over all insertions
in the diagram (vertex by vertex).
For triangles with internal charge flow
($\PQt, \PW^{\pm}, \upphi^{\pm},\HepParticle{\PX}{}{\pm}\Xspace$)
only the clockwise orientation is shown.
Non-equivalent diagrams obtained
by the exchange of the two photon lines are not shown.
Higgs and photon wave-function factors are not included.
The Fadeev-Popov ghost fields are denoted by $\PX$.}
\label{class}
\end{figure}

An example of how the SMEFT orders a double expansion in $1/\Lambda$ and the perturbative expansion in SM couplings is as follows.
Consider a tree level $2$ body decay of a single field. The double expansion of such a process
is given as the following Table~\footnote{Here we have introduced short hand notation where
$g_{4+2\,k} = 1/(\sqrt{2}\,G_{F}\,\Lambda^2)^k$, so that $g_6$ denotes a single $\mathcal{O}^{(6)}$
insertion, $g_8$ denotes a single $\mathcal{O}^{(8)}$ insertion, $g^2_6$ denotes two, distinct,
$\mathcal{O}^{(6)}$ insertions, etc..}:
\begin{eqnarray} \label{tableNLO}
\begin{array}{llll}
g_{SM}\,/\,\mathrm{D}im & \longrightarrow & & \\
\downarrow   & g_{SM}\,{\cal A}^{(4)}_1  &
               \; + \; g_{SM}\,g_6\,{\cal A}^{(6)}_{1,1,1} &
               \; + \; g_{SM}\,g_8\,{\cal A}^{(8)}_{1,1,2} \\
             & g_{SM}^3\,{\cal A}^{(4)}_3  &
               \; + \; g_{SM}^3\,g_6\,{\cal A}^{(6)}_{3,1,1} &
               \; + \; g_{SM}^3\,g^2_6\,{\cal A}^{(6)}_{3,2,1} \\
             &  \dots\dots & \dots\dots & \dots\dots
\end{array}
\end{eqnarray}
The combination of parameters $g_{SM}\,g_6\,{\cal A}^{(6)}_{1,1,1}$ defines the LO SMEFT expression
for the process, including the leading insertion of a higher dimensional operator, and is
generally well known.
$g_{SM}^3\,g_6\,{\cal A}^{(6)}_{3,1,1}$ defines the NLO SMEFT amplitude in the perturbative expansion,
and $g_{SM}\,g_8\,{\cal A}^{(8)}_{1,1,2}$ defines the NLO SMEFT Lagrangian expansion contribution to the amplitude.
We will refer to these two different NLO effects in this manner in this document.
The discussion here generalizes to cases other than
two body decays of a single field directly. Currently NLO terms in the double expansion present in the SMEFT are generally unknown, in almost every
process that is of interest phenomenologically.

The construction of the SMEFT, to all orders, is not based on assumptions on the size of the Wilson coefficients of the higher dimensional operators,
although it does assume that a valid perturbative expansion is present.
Constructing an NLO SMEFT result means including all operators at a fixed order in the power counting of the theory
or performing a complete one loop calculation for a process, including all of the operators in ${\cal L}_{6}$ that can contribute. One must add results for real emission (if present) to get a complete
description of a process at NLO in perturbation theory.\footnote{There are different uses of the phrase ``NLO" in the literature. This can refer to a fixed-order NLO
calculation including non-logarithmic terms not fixed by renormalization group evolution, only  an
approximate fixed-order NLO calculation, which includes logarithmic terms
fixed by renormalization group evolution to NLO, and a genuine leading-log calculation, which
uses exact solutions to the RG equations to actually do a resummation. In this work ``NLO in the perturbative expansion" refers to a complete perturbative correction due to SM interactions to the operators in $\mathcal{L}_6$.}

NLO corrections are a necessary consequence of the SMEFT being a well defined field theory.
The {\it numerical size of the higher order terms} depends upon the high energy (UV) scenario dictating the $\tilde{C}_i$ and $\Lambda$, which is unknown.
Restricting to a particular UV case is not an integral part of a general SMEFT treatment and various cases can be chosen once the general calculation is performed.
All explicit references to the underlying theory are introduced via the matching procedure in the standard approach to EFTs and power counting, see Refs.\cite{Weinberg:1980wa,Coleman:1969sm,Callan:1969sn,Manohar:1983md,Georgi:1994qn,Kaplan:1995uv,Manohar:1996cq,Cohen:1997rt,Luty:1997fk,Polchinski:1992ed,Rothstein:2003mp,Skiba:2010xn,Burgess:2007pt,Jenkins:2013fya,Jenkins:2013sda,Buchalla:2014eca,Buchalla:2013eza,Gavela:2016bzc} for reviews.
Below we briefly summarize the standard definitions of these terms.

\subsubsection{Power counting}
The size of corrections to SM results due to ${\cal L}_{SMEFT}$ interactions are estimated with power counting.\footnote{Differences of opinion about the size of NLO corrections exist in the theory community. Our claim is that any differences of opinion regarding NLO analyses are due to different implicit UV assumptions
and the data should be reported in a manner that maximizes its potential use in the future, including its use in NLO analyses. This means formalisms that cannot be improved to NLO should be avoided.
We return to this point below.}
A naive power counting scheme based on the mass dimensions of the operators simply normalizes an operator by the appropriate power of $1/\Lambda$.
Expansions in $(v/\Lambda)^m$ and $(p^2/\Lambda^2)^m$ are then present, where $p^2$ is a typical invariant momentum flow of a process. Both expansions are relative to the SM interactions.

The Naive Dimensional Analysis (NDA) power counting scheme
incorporates the counting of $\Lambda$ and an estimate of factors of $4 \, \pi$ in the normalization of the operators, see Refs.\cite{Manohar:1983md,Cohen:1997rt,Luty:1997fk,Gavela:2016bzc}
for details. By definition any remaining $4 \, \pi$ dependence, coupling dependence, or alternate scales present in the EFT, can be absorbed into the Wilson coefficients in the matching procedure
if the naive power counting scheme is used.

\subsubsection{Matching}

Wilson coefficients are determined by calculating  on-shell amplitudes in the UV theory and in the SMEFT and taking
the low energy limit ($E/\Lambda << 1$). The mismatch of the finite terms defines the Wilson coefficient in the matching condition.

If the value of Wilson coefficients in broad UV scenarios could be inferred in general this would be of significant scientific value.
An example of a scheme that applies to a fairly large set of UV scenarios is the  Artz-Einhorn-Wudka ``potentially-tree-generated''
(PTG) scheme~\cite{Arzt:1994gp,Einhorn:2013kja}. This approach classifies Wilson coefficients for operators in $\mathcal{L}_6$
as tree or loop level (suppressed by $g^2/16 \, \pi^2$) essentially using topological matching arguments. This classification scheme corresponds only to a subset of weakly
coupled and renormalizable UV physics cases, as the topologies considered are (effectively) limited by Lorentz invariance and renormalizability.
This scheme does not apply to scenarios where any high energy physics is strongly interacting or an EFT itself \cite{Jenkins:2013fya}.
This scheme should be only considered with caution, as it is not the result of a precise matching calculation.

One can study the Wilson coefficients using dimensional analysis, by restoring $\hbar \neq 1$ in the Lagrangian, as recently discussed in Refs.\cite{Pomarol:2014dya,Panico:2015jxa}.
{In this note we do not assume any hierarchy among the couplings as discussed in these works.}
The reasons we do not adopt these claims is that,
{in our opinion, one cannot}
unambiguously identify the powers of hypothetical UV couplings present in the $\tilde{C}_i$, as the SM couplings also carry $\hbar$ dimensions and the UV theory is not known.
Further, the matching procedure introduces order one constant terms that can be as large as, or dominant over, any such coupling dependence.
{For these reasons, we adopt an agnostic position and treat the $\tilde{C}_i$ as anything other than parameters to be constrained by experiment. In this approach, by performing the calculations without unnecessary
assumptions, it is still possible to study the effect of particular hierarchies and specific UV completions (when they are precisely defined allowing a matching calculation) a posteriori.}

\subsubsection{Operator bases for the SMEFT}

The Warsaw basis \cite{Grzadkowski:2010es} for the SMEFT is given in
Table \ref{warsaw}.
This basis is completely and precisely defined and is fully reduced by the Equations of Motion (EOM). It was the first basis of this form, building upon Ref.\cite{Buchmuller:1985jz}.
No fully reduced basis was present in the literature prior to 2010 when this result was reported.
{The Warsaw basis is one of the most (if not the most) standard SMEFT bases in use in the theoretical community.}
 This is the basis we use to define the straightforward LO approach in subsequent sections.

One can make small field redefinitions of $\mathcal{O}(1/\Lambda^2)$ to shift $\mathcal{L}_{SMEFT}$ by operators proportional to the EOM.
This procedure can be used to eliminate redundant operators from the Lagrangian to obtain a fully reduced basis or to change to a different operator basis.
Operator bases that are related by field redefinitions give equivalent results for physically measured quantities due to the equivalence theorem, see Refs.\cite{Kallosh:1972ap,'tHooft:1972ue,Politzer:1980me}
for the proof of this theorem and its conditions. Field redefinitions are a change of variables in a path integral and do not affect $S$-matrix elements although the source terms in Greens functions can get modified.
If a modification of how $\mathcal{L}_{SMEFT}$ is presented uses manipulations that are not gauge independent field redefinitions, it does not directly satisfy the conditions of the equivalence theorem.
Any LO Lagrangian construction based on intrinsically gauge dependent manipulations is
different from a gauge independent operator basis, like the Warsaw basis,
and we believe it would not be referred to as an operator basis in standard EFT literature \cite{Weinberg:1980wa,Coleman:1969sm,Callan:1969sn,Manohar:1983md,Georgi:1994qn,Kaplan:1995uv,Manohar:1996cq,Cohen:1997rt,Luty:1997fk,Polchinski:1992ed,Rothstein:2003mp,Skiba:2010xn,Burgess:2007pt,Jenkins:2013fya,Jenkins:2013sda,Buchalla:2014eca,Buchalla:2013eza,Gavela:2016bzc}.
\footnote{Editor footnote: 
Another point of view is expressed in Section~\ref{s.eftbasis} that advocates the use of field
transformations to simplify tree level calculations and separate
different classes of observables.}

{In principle, any well-defined basis (in the sense specified above) can be used. There are very few such bases defined in the literature: the Warsaw basis and various constructions that can be related to the Warsaw basis using the EOM. A notable example is the so-called
SILH basis,  a later version of which was reported in Ref.\cite{Contino:2013kra} in 2013.
The different operators that are present in this SILH basis (we adopt here the definition of the basis of Ref.~\cite{Alonso:2013hga}),
denoted $\mathcal{O}_i$,  are given by }
\begin{align}
\mathcal{O}_{HW} &=  -i \, g_2 \, (D^\mu H)^\dagger \, \tau^I \, (D^\nu H) \,  W^I_{\mu \, \nu}, &
\mathcal{O}_{HB} &=  -i \, g_1 \, (D^\mu H)^\dagger \,  (D^\nu H) \,  B_{\mu \, \nu}, \\
\mathcal{O}_{W} &=  -\frac{i \, g_2}{2} \, (H^\dagger \,  \overleftrightarrow D^I_\mu H) \,  (D^\nu W^I_{\mu \, \nu}), &
\mathcal{O}_{B} &=  -\frac{i \, g_1}{2} \, (H^\dagger \, \overleftrightarrow{D}^\mu H) \,  (D^\nu B_{\mu \, \nu}),  \\
\mathcal{O}_{T} &=  (H^\dagger \, \overleftrightarrow{D}^\mu H) \,  (H^\dagger \, \overleftrightarrow{D}^\mu H).
\label{SILHops}
\end{align}
We use $Q_i$ for the Warsaw basis operators, $\tau$ is the Pauli matrix, $g_1$ is the $U(1)_Y$ coupling and $g_2$ is the $SU(2)_L$ coupling. See Refs.\cite{Grzadkowski:2010es,Alonso:2013hga}
for more details on notation.
All other operators are the same in these bases.
The  transformation from the Warsaw basis
to the $\mathcal{O}_i$ operators is derived using the SM EOM and found\footnote{
Operator relations of this form were partially discussed in Refs \cite{SanchezColon:1998xg,Kilian:2003xt,Grojean:2006nn} previously.}  to be \cite{Alonso:2013hga}
\begin{align}
g_1 \, g_2 \, Q_{HWB} &=  4 \, \mathcal{O}_{B} - 4 \, \mathcal{O}_{HB} - 2 \,  \mathsf{y}_H \, g_1^2 \, Q_{HB},  \\
g_2^2 \, Q_{HW} &=  4 \, \mathcal{O}_{W} - 4 \, \mathcal{O}_{B} - 4 \, \mathcal{O}_{HW} + 4 \, \mathcal{O}_{HB} + 2 \,  \mathsf{y}_H \, g_1^2 \, Q_{HB},  \\
g_1^2 \, \mathsf{y}_\ell \, Q_{\substack{H l \\ tt}}^{(1)} &= 2 \, \mathcal{O}_{B} +  \mathsf{y}_H \, g_1^2\, \mathcal{O}_T -
g_1^2\left[\mathsf{y}_e Q_{\substack{H e \\ rr}} + \mathsf{y}_q Q_{\substack{H q \\ rr}}^{(1)}+\mathsf{y}_u Q_{\substack{H u \\ rr}}+ \mathsf{y}_d Q_{\substack{H d \\ rr}} \right],  \\
g_2^2  \, Q_{\substack{H l \\ tt}}^{(3)} &=  4 \, \mathcal{O}_{W} - 3 \, g_2^2 \, Q_{H \Box} + 2 \, g_2^2 m_h^2 \, (H^\dagger \, H)^2 - 8 \, g_2^2 \, \lambda \, Q_H - g_2^2 \, Q_{H q}^{(3)},  \\
&- 2 \, g_2^2\left( [Y_u^\dagger]_{rr} Q_{\substack{ uH \\ rr}} + [Y_d^\dagger]_{rr} Q_{\substack{dH \\ rr}}+ [Y_e^\dagger]_{rr} Q_{\substack{eH \\ rr}}+h.c. \right).
\label{inversetransform1}
\end{align}
Here the $t$ subscript is a flavour index and the $(1),(3)$ superscripts are operator labels, see Table 1.
In these relations only the flavour singlet component of the operators appears - given by the $tt$ subscript and
the notation $Q_{\substack{H d \\ rr}}$ for the Warsaw basis operators. It is necessary to define what flavour components of the operators are removed and retained in this procedure, as first pointed out in
Ref.\cite{Alonso:2013hga}.\footnote{{In our opinion, using the SILH basis to describe interactions of vector bosons with fermions, for example in Electroweak Precision Data (EWPD) is less transparent than using the Warsaw basis {(see Ref.~\cite{Trott:2014dma}).}}}
Note that these relationships between operators are not gauge dependent as they follow from gauge independent field redefinitions that satisfy the equivalence theorem.

{The complete renormalization program for $\mathcal{L}_6$ was only carried out in the Warsaw basis.
In the latter basis higher derivative terms are systematically removed using the EOM in favour of other operators without derivatives.}
{ This is done for a number of technical reasons which, in our opinion, were crucial to complete the renormalization program in
Refs.~\cite{Alonso:2013hga,Grojean:2013kd,Jenkins:2013zja,Jenkins:2013wua}.
 Only partial renormalization results were derived in the SILH basis. Any LO construction introducing operator normalizations,
 redefinitions of the SM parameters and EOM manipulations that are intrinsically gauge dependent will, in our opinion, make very hard the use of the results in  Refs.~\cite{Alonso:2013hga,Grojean:2013kd,Jenkins:2013zja,Jenkins:2013wua}.}
\begin{table}
\caption{\label{warsaw} The $\mathcal{L}_6$ operators built from Standard Model fields which conserve baryon number in the Warsaw basis
\cite{Grzadkowski:2010es}. The flavour labels of the form $p,r,s,t$ on the $Q$ operators are suppressed on the left hand side of the tables.}
\begin{center}
\small
\begin{minipage}[t]{4.6cm}
\renewcommand{\arraystretch}{1.5}
\begin{tabular}[t]{c|c}
\multicolumn{2}{c}{$1:X^3$} \\
\midrule
$Q_G$                & $f^{ABC} G_\mu^{A\nu} G_\nu^{B\rho} G_\rho^{C\mu} $ \\
$Q_{\widetilde G}$          & $f^{ABC} \widetilde G_\mu^{A\nu} G_\nu^{B\rho} G_\rho^{C\mu} $ \\
$Q_W$                & $\epsilon^{IJK} W_\mu^{I\nu} W_\nu^{J\rho} W_\rho^{K\mu}$ \\
$Q_{\widetilde W}$          & $\epsilon^{IJK} \widetilde W_\mu^{I\nu} W_\nu^{J\rho} W_\rho^{K\mu}$ \\
\end{tabular}
\end{minipage}
\begin{minipage}[t]{2.7cm}
\renewcommand{\arraystretch}{1.5}
\begin{tabular}[t]{c|c}
\multicolumn{2}{c}{$2:H^6$} \\
\midrule
$Q_H$       & $(H^\dag H)^3$
\end{tabular}
\end{minipage}
\begin{minipage}[t]{5.4cm}
\renewcommand{\arraystretch}{1.5}
\begin{tabular}[t]{c|c}
\multicolumn{2}{c}{$3:H^4 D^2$} \\
\midrule
$Q_{H\Box}$ & $(H^\dag H)\Box(H^\dag H)$ \\
$Q_{H D}$   & $\ \left(H^\dag D_\mu H\right)^* \left(H^\dag D_\mu H\right)$
\end{tabular}
\end{minipage}
\begin{minipage}[t]{2.7cm}

\renewcommand{\arraystretch}{1.5}
\begin{tabular}[t]{c|c}
\multicolumn{2}{c}{$5: \psi^2H^3 + \hbox{h.c.}$} \\
\midrule
$Q_{eH}$           & $(H^\dag H)(\bar l_p e_r H)$ \\
$Q_{uH}$          & $(H^\dag H)(\bar q_p u_r \widetilde H )$ \\
$Q_{dH}$           & $(H^\dag H)(\bar q_p d_r H)$\\
\end{tabular}
\end{minipage}

\vspace{0.25cm}

\begin{minipage}[t]{4.7cm}
\renewcommand{\arraystretch}{1.5}
\begin{tabular}[t]{c|c}
\multicolumn{2}{c}{$4:X^2H^2$} \\
\midrule
$Q_{H G}$     & $H^\dag H\, G^A_{\mu\nu} G^{A\mu\nu}$ \\
$Q_{H\widetilde G}$         & $H^\dag H\, \widetilde G^A_{\mu\nu} G^{A\mu\nu}$ \\
$Q_{H W}$     & $H^\dag H\, W^I_{\mu\nu} W^{I\mu\nu}$ \\
$Q_{H\widetilde W}$         & $H^\dag H\, \widetilde W^I_{\mu\nu} W^{I\mu\nu}$ \\
$Q_{H B}$     & $ H^\dag H\, B_{\mu\nu} B^{\mu\nu}$ \\
$Q_{H\widetilde B}$         & $H^\dag H\, \widetilde B_{\mu\nu} B^{\mu\nu}$ \\
$Q_{H WB}$     & $ H^\dag \tau^I H\, W^I_{\mu\nu} B^{\mu\nu}$ \\
$Q_{H\widetilde W B}$         & $H^\dag \tau^I H\, \widetilde W^I_{\mu\nu} B^{\mu\nu}$
\end{tabular}
\end{minipage}
\begin{minipage}[t]{5.2cm}
\renewcommand{\arraystretch}{1.5}
\begin{tabular}[t]{c|c}
\multicolumn{2}{c}{$6:\psi^2 XH+\hbox{h.c.}$} \\
\midrule
$Q_{eW}$      & $(\bar l_p \sigma^{\mu\nu} e_r) \tau^I H W_{\mu\nu}^I$ \\
$Q_{eB}$        & $(\bar l_p \sigma^{\mu\nu} e_r) H B_{\mu\nu}$ \\
$Q_{uG}$        & $(\bar q_p \sigma^{\mu\nu} T^A u_r) \widetilde H \, G_{\mu\nu}^A$ \\
$Q_{uW}$        & $(\bar q_p \sigma^{\mu\nu} u_r) \tau^I \widetilde H \, W_{\mu\nu}^I$ \\
$Q_{uB}$        & $(\bar q_p \sigma^{\mu\nu} u_r) \widetilde H \, B_{\mu\nu}$ \\
$Q_{dG}$        & $(\bar q_p \sigma^{\mu\nu} T^A d_r) H\, G_{\mu\nu}^A$ \\
$Q_{dW}$         & $(\bar q_p \sigma^{\mu\nu} d_r) \tau^I H\, W_{\mu\nu}^I$ \\
$Q_{dB}$        & $(\bar q_p \sigma^{\mu\nu} d_r) H\, B_{\mu\nu}$
\end{tabular}
\end{minipage}
\begin{minipage}[t]{5.4cm}
\renewcommand{\arraystretch}{1.5}
\begin{tabular}[t]{c|c}
\multicolumn{2}{c}{$7:\psi^2H^2 D$} \\
\midrule
$Q_{H l}^{(1)}$      & $(H^\dag i\overleftrightarrow{D}_\mu H)(\bar l_p \gamma^\mu l_r)$\\
$Q_{H l}^{(3)}$      & $(H^\dag i\overleftrightarrow{D}^I_\mu H)(\bar l_p \tau^I \gamma^\mu l_r)$\\
$Q_{H e}$            & $(H^\dag i\overleftrightarrow{D}_\mu H)(\bar e_p \gamma^\mu e_r)$\\
$Q_{H q}^{(1)}$      & $(H^\dag i\overleftrightarrow{D}_\mu H)(\bar q_p \gamma^\mu q_r)$\\
$Q_{H q}^{(3)}$      & $(H^\dag i\overleftrightarrow{D}^I_\mu H)(\bar q_p \tau^I \gamma^\mu q_r)$\\
$Q_{H u}$            & $(H^\dag i\overleftrightarrow{D}_\mu H)(\bar u_p \gamma^\mu u_r)$\\
$Q_{H d}$            & $(H^\dag i\overleftrightarrow{D}_\mu H)(\bar d_p \gamma^\mu d_r)$\\
$Q_{H u d}$ + h.c.   & $i(\widetilde H ^\dag D_\mu H)(\bar u_p \gamma^\mu d_r)$\\
\end{tabular}
\end{minipage}

\vspace{0.25cm}

\begin{minipage}[t]{4.95cm}
\renewcommand{\arraystretch}{1.5}
\begin{tabular}[t]{c|c}
\multicolumn{2}{c}{$8:(\bar LL)(\bar LL)$} \\
\midrule
$Q_{ll}$        & $(\bar l_p \gamma_\mu l_r)(\bar l_s \gamma^\mu l_t)$ \\
$Q_{qq}^{(1)}$  & $(\bar q_p \gamma_\mu q_r)(\bar q_s \gamma^\mu q_t)$ \\
$Q_{qq}^{(3)}$  & $(\bar q_p \gamma_\mu \tau^I q_r)(\bar q_s \gamma^\mu \tau^I q_t)$ \\
$Q_{lq}^{(1)}$                & $(\bar l_p \gamma_\mu l_r)(\bar q_s \gamma^\mu q_t)$ \\
$Q_{lq}^{(3)}$                & $(\bar l_p \gamma_\mu \tau^I l_r)(\bar q_s \gamma^\mu \tau^I q_t)$
\end{tabular}
\end{minipage}
\begin{minipage}[t]{5.45cm}
\renewcommand{\arraystretch}{1.5}
\begin{tabular}[t]{c|c}
\multicolumn{2}{c}{$8:(\bar RR)(\bar RR)$} \\
\midrule
$Q_{ee}$               & $(\bar e_p \gamma_\mu e_r)(\bar e_s \gamma^\mu e_t)$ \\
$Q_{uu}$        & $(\bar u_p \gamma_\mu u_r)(\bar u_s \gamma^\mu u_t)$ \\
$Q_{dd}$        & $(\bar d_p \gamma_\mu d_r)(\bar d_s \gamma^\mu d_t)$ \\
$Q_{eu}$                      & $(\bar e_p \gamma_\mu e_r)(\bar u_s \gamma^\mu u_t)$ \\
$Q_{ed}$                      & $(\bar e_p \gamma_\mu e_r)(\bar d_s\gamma^\mu d_t)$ \\
$Q_{ud}^{(1)}$                & $(\bar u_p \gamma_\mu u_r)(\bar d_s \gamma^\mu d_t)$ \\
$Q_{ud}^{(8)}$                & $(\bar u_p \gamma_\mu T^A u_r)(\bar d_s \gamma^\mu T^A d_t)$ \\
\end{tabular}
\end{minipage}
\begin{minipage}[t]{4.75cm}
\renewcommand{\arraystretch}{1.5}
\begin{tabular}[t]{c|c}
\multicolumn{2}{c}{$8:(\bar LL)(\bar RR)$} \\
\midrule
$Q_{le}$               & $(\bar l_p \gamma_\mu l_r)(\bar e_s \gamma^\mu e_t)$ \\
$Q_{lu}$               & $(\bar l_p \gamma_\mu l_r)(\bar u_s \gamma^\mu u_t)$ \\
$Q_{ld}$               & $(\bar l_p \gamma_\mu l_r)(\bar d_s \gamma^\mu d_t)$ \\
$Q_{qe}$               & $(\bar q_p \gamma_\mu q_r)(\bar e_s \gamma^\mu e_t)$ \\
$Q_{qu}^{(1)}$         & $(\bar q_p \gamma_\mu q_r)(\bar u_s \gamma^\mu u_t)$ \\
$Q_{qu}^{(8)}$         & $(\bar q_p \gamma_\mu T^A q_r)(\bar u_s \gamma^\mu T^A u_t)$ \\
$Q_{qd}^{(1)}$ & $(\bar q_p \gamma_\mu q_r)(\bar d_s \gamma^\mu d_t)$ \\
$Q_{qd}^{(8)}$ & $(\bar q_p \gamma_\mu T^A q_r)(\bar d_s \gamma^\mu T^A d_t)$\\
\end{tabular}
\end{minipage}

\vspace{0.25cm}

\begin{minipage}[t]{3.75cm}
\renewcommand{\arraystretch}{1.5}
\begin{tabular}[t]{c|c}
\multicolumn{2}{c}{$8:(\bar LR)(\bar RL)+\hbox{h.c.}$} \\
\midrule
$Q_{ledq}$ & $(\bar l_p^j e_r)(\bar d_s q_{tj})$
\end{tabular}
\end{minipage}
\begin{minipage}[t]{5.5cm}
\renewcommand{\arraystretch}{1.5}
\begin{tabular}[t]{c|c}
\multicolumn{2}{c}{$8:(\bar LR)(\bar L R)+\hbox{h.c.}$} \\
\midrule
$Q_{quqd}^{(1)}$ & $(\bar q_p^j u_r) \epsilon_{jk} (\bar q_s^k d_t)$ \\
$Q_{quqd}^{(8)}$ & $(\bar q_p^j T^A u_r) \epsilon_{jk} (\bar q_s^k T^A d_t)$ \\
$Q_{lequ}^{(1)}$ & $(\bar l_p^j e_r) \epsilon_{jk} (\bar q_s^k u_t)$ \\
$Q_{lequ}^{(3)}$ & $(\bar l_p^j \sigma_{\mu\nu} e_r) \epsilon_{jk} (\bar q_s^k \sigma^{\mu\nu} u_t)$
\end{tabular}
\end{minipage}
\end{center}
\end{table}

\subsubsection{Rotating to mass eigenstate fields}\label{rotation}
Expanding around the vev in unitary gauge and rotating to mass eigenstate fields, the LO modification of the SM interactions in the SMEFT come about
in a straightforward manner.\footnote{The operator basis for the SMEFT remains the Warsaw basis when the interaction terms are expanded in terms of
mass eigenstate fields in unitary gauge. Operator bases are gauge independent, satisfy the equivalence theorem, and do not change when the SMEFT is improved from LO to NLO.}
Here we list the most phenomenologically relevant terms present for mass eigenstate fields, the remaining interactions unlisted come from Class $1,3,5,6,8$ operators in Table 1.
It is not required in our approach to specify all interactions as we make no assertion that these mass eigenstate interactions listed are an operator basis.
As the theory should be canonically normalized,
we denote coupling parameters in the canonically normalized SMEFT with bar superscripts. This use of bar notation is distinct from
bar superscripts on fermion fields where $\bar{\psi} = \psi^\dagger \gamma^0$.
The following section is largely taken from Ref. \cite{Alonso:2013hga}.

\subsubsubsection{SM Lagrangian}
We define the SM Lagrangian as
\begin{align}
\mathcal{L}_{\rm SM} &= -\frac14 G_{\mu \nu}^A G^{A\mu \nu}-\frac14 W_{\mu \nu}^I W^{I \mu \nu} -\frac14 B_{\mu \nu} B^{\mu \nu}
+ (D_\mu H^\dagger)(D^\mu H)
+\sum_{\psi=q,u,d,l,e} \overline \psi\, i \slashed{D} \, \psi\nn \\
&-\lambda \left(H^\dagger H -\frac12 v^2\right)^2- \biggl[ H^{\dagger j} \overline d\, Y_d\, q_{j} + \widetilde H^{\dagger j} \overline u\, Y_u\, q_{j} + H^{\dagger j} \overline e\, Y_e\,  l_{j} + \hbox{h.c.}\biggr],
\label{sm}
\end{align}
which implicitly defines most of our notational conventions. Note $\tilde{H}^j = \epsilon_{jk} \, H^{\dagger k}$. We have suppressed reference to the
$\tilde{\theta}$ gauge dual operators of the form $\tilde{\theta} \, F^{\mu \, \nu} \, \tilde{F}_{\mu \, \nu}$. These terms are known to be experimentally small. For dual gauge fields we use the convention $\tilde{F}_{\mu \, \nu} = \epsilon_{\mu \, \nu \, \alpha \, \beta} \, F^{\alpha \, \beta}$
with $\epsilon_{0123} = + 1$. See Ref \cite{Alonso:2013hga,Jenkins:2013zja,Jenkins:2013wua} for more details on notation.

\subsubsubsection{Higgs boson mass and self-couplings}\label{sec:Hmass}

The potential in the SMEFT is
\begin{equation}
V(H) = \lambda \left(H^\dagger H -\frac12 v^2\right)^2 - C_H \left( H^\dagger H \right)^3,
\label{eq:smeftpot}
\end{equation}
yielding the new minimum
\begin{align}
\langle H^\dagger H \rangle &= \frac{v^2}{2} \left( 1+ \frac{3 C_H v^2}{4 \lambda} \right) \equiv \frac12 v_T^2.
\end{align}
The scalar field can be written in unitary gauge as
\begin{align}
H &= \frac{1}{\sqrt 2} \left(\begin{array}{c}
0 \\
 \left[ 1+ c_{H,\text{kin}} \right]  h + v_T
 \end{array}\right),
 \label{Hvev}
\end{align}
where
\begin{align}\label{chkindef}
c_{H,\text{kin}} &\equiv \left(C_{H\Box}-\frac14 C_{HD}\right)v^2, &
v_T &\equiv \left( 1+ \frac{3 C_H v^2}{8 \lambda} \right) v.
\end{align}
The coefficient of $h$ in Eq.~(\ref{Hvev}) is no longer unity, in order for the Higgs boson kinetic term to be properly normalized when the dimension-six operators are included. In what follows
we can exchange $v_T$ for $v$ when this parameter multiplies a operator in $\mathcal{L}_6$ as the difference is NLO in the lagrangian expansion.The kinetic terms
\begin{align}
{\cal L}^{(6)} &= (D_\mu H^\dagger)(D^\mu H) + C_{H \Box} \left( H^\dagger H \right) \Box \left( H^\dagger H \right) + C_{HD} \left( H^\dagger D^\mu H \right)^* \left( H^\dagger D_\mu H \right),
\end{align}
and the potential in Eq.~(\ref{eq:smeftpot}) yield
\begin{align}
{\cal L}^{(6)} &= {1 \over 2} \left( \partial_\mu h \right)^2 -c_{H,\text{kin}} \left[ \frac{h^2}{v^2} +2 \,  \frac{h}{v} \right](\partial_\mu h)^2 - \frac{m_h^2}{2} h^2 -\lambda v_T \left(1-\frac{5 C_H v^2}{2 \lambda} +3 c_{H,\text{kin}}  \right)h^3 \nn \\
&-\frac14 \lambda  \left(1-\frac{15 C_H v^2}{2 \lambda} +4 c_{H,\text{kin}} \right)h^4 +\frac34 C_H v h^5+\frac18 C_H h^6,
\end{align}
for the $h$ self-interactions. The Higgs boson mass is
\begin{align}
m_h^2 &= 2 \lambda v_T^2 \left(1-\frac{3 C_H v^2}{2 \lambda} + 2 c_{H,\text{kin}} \right)\,.
\end{align}

\subsubsubsection{Yukawa couplings}\label{sec:yuk}

The Lagrangian terms in the unbroken theory
\begin{align}
{\cal L} &= - \biggl[ H^{\dagger j} \overline d_r\, \left[Y_d \right]_{rs}\, q_{j s} + \widetilde H^{\dagger j} \overline u_r\, \left[ Y_u \right]_{rs} \, q_{j s}
+ H^{\dagger j} \overline e_r\, \left[Y_e\right]_{rs}\,  l_{j s} + \hbox{h.c.}\biggr] \\
&+ \left[ C^*_{\substack{dH \\ sr}} \left( H^\dagger H \right) H^{\dagger j}  \overline d_r q_{j s} + C^*_{\substack{uH \\ sr}} \left( H^\dagger H \right) \tilde H^{\dagger j}  \overline u_r q_{j s} + C^*_{\substack{eH \\ sr}} \left( H^\dagger H \right) H^{\dagger j}  \overline e_r l_{j s} + \hbox{h.c.} \right], \nonumber
\end{align}
yield the fermion mass matrices
\begin{align}
\left[ M_\psi \right]_{rs} &= \frac{v_T}{\sqrt 2} \left( \left[Y_\psi \right]_{rs}   - \frac12 v^2 C^*_{\substack{\psi H \\ sr}}   \right), \qquad \psi=u,d,e
\end{align}
in the broken theory.  The  coupling matrices of the $h$ boson to the fermions $\mathcal{L}=- h \, \overline u \, \mathcal{Y}\, q + h.c + \ldots$ are
\begin{align}
\left[ {\cal Y}_\psi \right]_{rs} &= \frac{1}{\sqrt 2}  \left[ Y_\psi \right]_{rs}\left[ 1+ c_{H,\text{kin}} \right]  - \frac{3}{2 \, \sqrt{2}} v^2 C^*_{\substack{\psi H \\ sr}}
\nn \\
& = \frac{1}{v_T}\left[ M_\psi \right]_{rs} \left[ 1+ c_{H,\text{kin}}  \right]   - \frac{v^2}{\sqrt{2}} C^*_{\substack{\psi H \\ sr}} ,
\qquad \psi=u,d,e.
\label{5.12}
\end{align}
The fermion fields can be rotating to diagonal mass eigenstates with $3 \times 3$ unitary matrices
$\mathcal{U}$ as
\begin{align}
\psi_L &= \mathcal{U}(\psi,L) \, \psi_L', & \psi_R &= \mathcal{U}(\psi,R) \, \psi_R',
\end{align}
where the measured masses $\hat{m}^i_\psi$ are
\begin{eqnarray}
\mathcal{U}^\dagger(\psi,R)  \, \left[ M_\psi \right] \, \mathcal{U}(\psi,L) = \delta_{ij} \, \hat{m}^i_\psi, \quad \quad
\begin{array}{c} i = \{u,c,t \}, \quad \psi =u, \\ i = \{d,s,b \}, \quad \psi =d, \\ i = \{e,u,\tau \}, \quad \psi =e. \end{array}
\end{eqnarray}
For the complex Yukawa coupling the higgs to the mass eigenstate fermion fields
\begin{eqnarray}
\mathcal{L}=- h \, \left[ {\cal Y}_\psi \right]_{rs} \, \bar{\psi}_r \, P_L \,  \psi_s + h.c.
\end{eqnarray}
where $P_L =  (1- \gamma_5)/2$ and
\begin{eqnarray}
\left[ {\cal Y}_\psi \right]_{rs} =  \delta_{rs} \, \frac{\hat{m}^r_\psi}{v_T} \,  \left[ 1+ c_{H,\text{kin}}  \right]  - \frac{v^2}{\sqrt{2}} \left[\mathcal{U}^\dagger(\psi,R) \, C^*_{\substack{\psi H}}  \, \mathcal{U}(\psi,L) \right]_{rs}.
\end{eqnarray}
The Yukawa matrices are off diagonal in general and not simply proportional to the fermion mass matrices as in the SM, as indicated by the second term.
The CKM and PMNS matrices control flavour violating interactions in the SM and are defined as
\begin{eqnarray}
V_{CKM} =  \mathcal{U}(u,L)^\dagger \, \mathcal{U}(d,L), \quad \quad U_{PMNS} = \mathcal{U}(e,L)^\dagger \, \mathcal{U}(\nu,L),
\end{eqnarray}
when the $\mathcal{U}$ matrices only rotate between the weak and mass eigenstates in the SM.
The definition of these matrices in the SMEFT is a convention choice. Here we choose to define these matrices so that the masses are taken to diagonal form including the $\mathcal{L}_6$ interactions.

\subsubsubsection{Gauge boson masses and couplings}\label{sec:Gmass}
The relevant CP even $\mathcal{L}_6$ terms are
\begin{align}
\mathcal{L}^{(6)} &=  C_{HG} H^\dagger H G_{\mu \nu}^A G^{A\mu \nu} + C_{HW} H^\dagger H W_{\mu \nu}^I W^{I \mu \nu}  + C_{HB} H^\dagger H B_{\mu \nu} B^{\mu \nu}  \nn \\
&+ C_{HWB} H^\dagger \tau^I H W^I_{\mu \nu} B^{\mu \nu} + C_G f^{ABC} G^{A \nu}_\mu G^{B \rho}_\nu G^{C \mu}_\rho + C_W \epsilon^{IJK} W^{I \nu}_\mu W^{J \rho}_\nu W^{K \mu}_\rho \ .
\end{align}
The gauge fields need to be redefined, so that the kinetic terms are properly normalized and diagonal. The first step is to redefine the gauge fields
\begin{align}
G_\mu^A &= \mathcal{G}_\mu^A \left(1 + C_{HG} v_T^2 \right), &
W^I_\mu  &=  \mathcal{W}^I_\mu \left(1 + C_{HW} v_T^2 \right), &
B_\mu  &=  \mathcal{B}_\mu \left(1 + C_{HB} v_T^2 \right).
\label{5.16a}
\end{align}
The modified coupling constants are
\begin{align}
{\overline g_{3}} &= g_3 \left(1 + C_{HG} \, v_T^2 \right), & {\overline g_{2}} &= g_2 \left(1 + C_{HW} \, v_T^2 \right), & {\overline g_{1}} &= g_1 \left(1 + C_{HB} \, v_T^2 \right),
\label{5.16b}
\end{align}
so that the products $g_3 G_\mu^A={\overline g_{3}} \mathcal{G}_\mu^A$, etc.\ are unchanged. The mass eigenstate basis is given by~\cite{Grinstein:1991cd}
\begin{align}
\left[ \begin{array}{cc}  \mathcal{W}_\mu^3 \\ \mathcal{B}_\mu \end{array} \right]
&=
\left[ \begin{array}{cc}  1   &  -  \frac{1}{2} \,  v_T^2 \,  C_{HWB} \\
- \frac{1}{2} \,  v_T^2 \,  C_{HWB} & 1 \end{array} \right] \, \left[ \begin{array}{cc} \cos {\overline \theta}  &  \sin {\overline \theta} \\
-\sin {\overline \theta} & \cos {\overline \theta} \end{array} \right] \left[ \begin{array}{cc}  \mathcal{Z}_\mu \\ \mathcal{A}_\mu \end{array} \right],
\end{align}
where the rotation angle is
\begin{align}
\tan {\overline \theta} &= \frac{{\overline g_{1}}}{{\overline g_{2}}} + \frac{v_T^2}{2}  \,  C_{HWB} \, \left[1 -  \frac{{\overline g_{1}}^2}{{\overline g_{2}}^2}\right].
\end{align}
The $W$ and $Z$ masses are
\begin{align}
\bar{M}_W^2 &= \frac{{\overline g_{2}}^2 v_T^2}{4} , \nn \\
\bar{M}_Z^2 &= \frac{v_T^2}{4}({\overline g_{1}}^2+{\overline g_{2}}^2)+\frac{1}{8}v_T^4 C_{HD} ({\overline g_{1}}^2+{\overline g_{2}}^2)+\frac{1}{2}v_T^4 {\overline g_{1}}{\overline g_{2}} C_{HWB}.
\end{align}
The covariant derivative is
\begin{align}
D_{\mu} = \partial_\mu + i \, \frac{{\overline g_{2}}}{\sqrt{2}} \left[\mathcal{W}_{\mu}^+ T^+ + \mathcal{W}_{\mu}^- T^- \right] + i \bar{g}_Z \left[T_3 - {\overline s}^2 Q \right] \mathcal{Z}_\mu +
i \, {\overline e} \, Q \, \mathcal{A}_\mu,
\end{align}
where $Q=T_3+Y$,  and the effective couplings are given by
\begin{align}
{\overline e} &= \frac{{\overline g_{1}} {\overline g_{2}}}{\sqrt{{\overline g_{2}}^2+{\overline g_{1}}^2}} \left[ 1 - \frac{{\overline g_{1}} {\overline g_{2}}}{{\overline g_{2}}^2+{\overline g_{1}}^2} v_T^2 C_{HWB} \right] = {\overline g_{2}} \, \sin {\overline \theta} - \frac{1}{2} \, \cos {\overline \theta}  \, {\overline g_{2}} \, v_T^2 \, C_{HWB}, \nn \\
{\overline g_{Z}} &= \sqrt{{\overline g_{2}}^2 + {\overline g_{1}}^2} + \frac{{\overline g_{1}} {\overline g_{2}}}{\sqrt{{\overline g_{2}}^2 + {\overline g_{1}}^2} } v_T^2  C_{HWB}=  \frac{{\overline e}}{\sin {\overline \theta} \cos {\overline \theta}}  \left[1 +  \frac{{\overline g_{1}}^2+{\overline g_{2}}^2}{2 {\overline g_{1}} {\overline g_{2}}} v_T^2C_{HWB}\right] ,\nn \\
{\overline s}^2 &= \sin^2 {\overline \theta} =  \frac{{\overline g_{1}}^2}{{{\overline g_{2}}^2 + {\overline g_{1}}^2}} + \frac{{\overline g_{1}} 
{\overline g_{2}} ({\overline g_{2}}^2-{\overline g_{1}}^2)}{({\overline g_{1}}^2+{\overline g_{2}}^2)^2}  v_T^2 C_{HWB} .
\end{align}
The relevant CP odd $\mathcal{L}^{\cancel{CP}}_6$ terms are
\begin{align}
\mathcal{L}^{\cancel{CP}}_6 &=  C_{H\tilde{G}} H^\dagger H \tilde{G}_{\mu \nu}^A G^{A\mu \nu} + C_{H\tilde{W}} H^\dagger H \tilde{W}_{\mu \nu}^I W^{I \mu \nu}  + C_{H \tilde{B}} H^\dagger H \tilde{B}_{\mu \nu} B^{\mu \nu}  \nn \\
&+ C_{H \tilde{W} B} H^\dagger \tau^I H \tilde{W}^I_{\mu \nu} B^{\mu \nu} + C_{\tilde{G}} f^{ABC} \tilde{G}^{A \nu}_\mu G^{B \rho}_\nu G^{C \mu}_\rho + C_{\tilde{W}} \epsilon^{IJK} \tilde{W}^{I \nu}_\mu W^{J \rho}_\nu W^{K \mu}_\rho \ .
\end{align}
The modified couplings and gauge fields introduced in Eqns.\ref{5.16a}, \ref{5.16b} do not cancel the new contribution  from these operators suppressed by $v^2_T/\Lambda^2$ to the $CP$ violating $\tilde{\theta}$ parameters. These extra contributions strongly indicate
that without fine tuning the Wilson coefficients $C_{H\tilde{G}},C_{H\tilde{W}},C_{H \tilde{B}},C_{H \tilde{W} B}$ are suppressed by a large $CP$ violating scale and can be neglected, similar to the treatment of $\mathcal{L}_5$.

\subsubsubsection{\texorpdfstring{$h \to WW$}{h to WW} and \texorpdfstring{$h \to ZZ$}{h to ZZ}}
The relevant $CP$-even Lagrangian terms are
\begin{align}
\mathcal{L} &= (D_\mu H)^\dagger (D^\mu H) - \frac{1}{4} \left(W^I_{\mu\nu} W^{I \mu\nu} + B_{\mu\nu} B^{\mu\nu} \right), \nn \\
&\hspace{0.4cm} + C_{HW} \, Q_{HW} + C_{HB} \, Q_{HB} + C_{HWB} Q_{HWB} + C_{HD} \, Q_{HD},
\label{gaugebosontree}
\end{align}
which lead to the interactions
\begin{align}
\mathcal{L}&=\frac12 {\overline g_{2}}^2 v_T     h \, W^+_\mu \, W^-_\mu \, \left[1 + c_{H,\text{kin}}  \right] +  C_{HW}  v_T  h  \, W^+_{\mu \, \nu} \, W^-_{\mu \, \nu}.
\end{align}
for the $W$, and
\begin{align}
\mathcal{L}&=\frac14 ({\overline g_{2}}^2+{\overline g_{1}}^2)v_T h (\mathcal{Z}_\mu)^2 \left[1 + c_{H,\text{kin}} +v_T^2 C_{HD} \right] +\frac12  {\overline g_{1}} {\overline g_{2}} v_T^3 h (\mathcal{Z}_\mu)^2 C_{HWB} \nn \\
& +  v_T  h (\mathcal{Z}_{\mu\nu})^2 \left( \frac{{\overline g_{2}}^2 C_{HW}+ {\overline g_{1}}^2 C_{HB} +{\overline g_{1}} {\overline g_{2}} C_{HWB}}{{\overline g_{2}}^2+{\overline g_{1}}^2} \right)
\label{5.42}
\end{align}
for the $Z$. Normalizing the SM $\tilde{\theta}$ operators by two powers of the appropriate gauge coupling so that Eqns.\ref{5.16a}, \ref{5.16b} do not introduce extra terms, the $\cancel{CP}$ contributions are
\begin{align}
\mathcal{L}^{\cancel{CP}}_6 =  C_{H\tilde{W}}  v_T  h  \, \tilde{W}^+_{\mu \, \nu} \, W^-_{\mu \, \nu} + v_T  h (\mathcal{\tilde{Z}}_{\mu\nu} \, \mathcal{Z}^{\mu\nu}) \left( \frac{{\overline g_{2}}^2 C_{H\tilde{W}}+ {\overline g_{1}}^2 C_{H \tilde{B}} +{\overline g_{1}} {\overline g_{2}} C_{H\tilde{W}B}}{{\overline g_{2}}^2+{\overline g_{1}}^2} \right).
\end{align}

\subsubsubsection{\texorpdfstring{$h \to \gamma \, \gamma$, $h \to \gamma \,Z$ and $h \rightarrow gg$}{h to gamma gamma, h to gamma Z, and h to gg}}

The CP even and odd couplings of $h\rightarrow \gamma \, \gamma$ and $h\rightarrow \gamma \,Z$ are given by \cite{Manohar:2006gz}
\begin{align}
\mathcal{L}&= h \, v_T \, \bar{e}^2 \, \left[C_{\gamma \, \gamma} \, \mathcal{A}^{\mu \, \nu} \, \mathcal{A}_{\mu \, \nu} + \tilde{C}_{\gamma \, \gamma} \, \tilde{\mathcal{A}}^{\mu \, \nu} \, \mathcal{A}_{\mu \, \nu}
+ C_{\gamma \, Z} \, \mathcal{A}^{\mu \, \nu} \, \mathcal{Z}_{\mu \, \nu} + \tilde{C}_{\gamma \, Z} \, \tilde{\mathcal{A}}^{\mu \, \nu} \, \mathcal{Z}_{\mu \, \nu} \right].
\end{align}
Here
\begin{align}
C_{\gamma \, \gamma} &= \frac{C_{HW}}{\bar{g}^2_2} + \frac{C_{HB}}{\bar{g}^2_1} - \frac{C_{HWB}}{\bar{g}_1 \, \bar{g}_2}, \\
\tilde{C}_{\gamma \, \gamma} &= \frac{C_{H\tilde{W}}}{\bar{g}^2_2} + \frac{C_{H\tilde{B}}}{\bar{g}^2_1} - \frac{C_{H\tilde{W}B}}{\bar{g}_1 \, \bar{g}_2}, \\
C_{\gamma \, Z} &= \frac{1}{\bar{g}_1 \, \bar{g}_2} \left( C_{HW} - C_{HB}\right) - \left(\frac{1}{2 \,\bar{g}_1^2} - \frac{1}{2 \,\bar{g}_2^2} \right) C_{HWB} , \\
\tilde{C}_{\gamma \, Z} &=\frac{1}{\bar{g}_1 \, \bar{g}_2} \left( C_{H\tilde{W}} - C_{H\tilde{B}}\right) - \left(\frac{1}{2 \,\bar{g}_1^2} - \frac{1}{2 \,\bar{g}_2^2} \right) C_{H\tilde{W}B}.
\end{align}
The CP even and odd couplings of $h\rightarrow gg$ are trivially
\begin{align}
\mathcal{L} =  h \, v_T \, \left[C_{HG} \, \mathcal{G}^{\mu \, \nu} \, \mathcal{G}_{\mu \, \nu} + C_{H\tilde{G}} \,  \mathcal{\tilde{G}}^{\mu \, \nu} \, \mathcal{G}_{\mu \, \nu} \right].
\end{align}
\subsubsubsection{Dipoles and Higgs dipole interactions}

In the broken phase the dipole interactions with neutral gauge bosons are
\begin{align}
\mathcal{L} &=\frac{\bar{e} \, (v+ h)}{\sqrt 2} \mathcal{C}_{\substack{\psi \gamma \\ rs }}\  \overline \psi_{r} \sigma^{\mu \nu} P_R \psi_{s}\, \mathcal{A}_{\mu \nu}
+ \frac{\bar{e} \, (v+ h)}{\sqrt 2} \mathcal{C}_{\substack{\psi \mathcal{Z} \\ rs }}\  \overline \psi_{r} \sigma^{\mu \nu} P_R \psi_{s}\, \mathcal{Z}_{\mu \nu}, \nn \\
&+ \frac{(v+ h) }{\sqrt 2} C_{\substack{d G \\ rs }}\  \overline d_{r} \sigma^{\mu \nu} T_A \, P_R d_{s}\, \mathcal{G}^A_{\mu \nu}
+ \frac{ (v+ h) }{\sqrt 2} C_{\substack{u G \\ rs }}\  \overline u_{r} \sigma^{\mu \nu} T_A P_R u_{s}\, \mathcal{G}^A_{\mu \nu} + h.c.
\end{align}
where $r$ and $s$ are flavour indices  and $\psi = \{e, u, d\}$ so that
\begin{align}
\mathcal{C}_{\substack{e \gamma \\ rs }} &= \left[\mathcal{U}(e,L)^\dagger\left( \frac{C_{\substack{e B}}}{g_1}  - \frac{C_{\substack{e W  }}}{g_2}\right) \mathcal{U}(e,R)\right]_{rs}&
\mathcal{C}_{\substack{e Z \\ rs }} &= - \left[\mathcal{U}(e,L)^\dagger \left(\frac{C_{\substack{e B}}}{g_2} + \frac{C_{\substack{e W}}}{g_1}\right) \mathcal{U}(e,R) \right]_{rs}  \nn \\
\mathcal{C}_{\substack{d \gamma \\ rs }} &=  \left[\mathcal{U}(d,L)^\dagger\left( \frac{C_{\substack{d B}}}{g_1}  - \frac{C_{\substack{d W  }}}{g_2}\right) \mathcal{U}(d,R)\right]_{rs}&
\mathcal{C}_{\substack{d Z \\ rs }} &=- \left[\mathcal{U}(d,L)^\dagger \left(\frac{C_{\substack{d B}}}{g_2} + \frac{C_{\substack{d W}}}{g_1}\right) \mathcal{U}(d,R) \right]_{rs}  \nn \\
\mathcal{C}_{\substack{u \gamma \\ rs }} &=   \left[\mathcal{U}(u,L)^\dagger\left( \frac{C_{\substack{u B}}}{g_1}  + \frac{C_{\substack{u W  }}}{g_2}\right) \mathcal{U}(u,R)\right]_{rs}&
\mathcal{C}_{\substack{u Z \\ rs }} &= - \left[\mathcal{U}(u,L)^\dagger \left(\frac{C_{\substack{u B}}}{g_2} - \frac{C_{\substack{u W}}}{g_1}\right) \mathcal{U}(u,R) \right]_{rs}.
\label{2.9}
\end{align}
$C_{uW}$ has the opposite sign for $u$-type quarks in Eq.~(\ref{2.9}) because of the opposite sign for $T_{3L}$. Note that $\sigma_{\mu \,\nu} = i \, \left[\gamma_\mu, \gamma_\nu \right]/2$.
The dipole interactions with charged gauge bosons are
\begin{align}
\mathcal{L} &= (v+ h) \, \left[ \overline \nu_r \, \sigma^{\mu \nu} \, P_R \, e_s \, \mathcal{W}^+_{\mu \nu} \mathcal{C}_{\substack{e \, W \\ rs }}
+ \overline u_r \, \sigma^{\mu \nu} \, P_R \, d_s \, \mathcal{W}^+_{\mu \nu} \, \mathcal{C}_{\substack{d \, W \\ rs}}
+ \overline d_r \, \sigma^{\mu \nu} \, P_R \, u_{s}\, \mathcal{W}^-_{\mu \nu} \, \mathcal{C}_{\substack{u \, W \\ rs}} \right]+ h.c.
\end{align}
where
\begin{align}
\mathcal{C}_{\substack{e \, W \\ rs }} &= \left[\mathcal{U}(\nu,L)^\dagger \, C_{\substack{e \, W}} \, \mathcal{U}(e,R)\right]_{rs}& \mathcal{C}_{\substack{d \, W \\ rs}} =  \left[\mathcal{U}(u,L)^\dagger \, C_{\substack{d \, W}} \, \mathcal{U}(d,R)\right]_{rs} \nn \\
\mathcal{C}_{\substack{u \, W \\ rs}}  &= \left[\mathcal{U}(d,L)^\dagger \, \mathcal{C}_{\substack{u \, W}} \, \mathcal{U}(u,R)\right]_{rs}.
\end{align}
\subsubsubsection{\texorpdfstring{$(h + v)^2\, V \bar{\psi} \, \psi$}{(h+v)^2 V psibar psi} interactions}

In the broken phase the interactions of the Higgs with  fermions and an associated gauge boson are given by
\begin{align}
\mathcal{L} &= \frac{\sqrt{\bar{g}_1^2+ \bar{g}_2^2}}{2} \, (h + v_T)^2 \, \mathcal{Z}_\mu \, \bar{\nu}_{r} \, \gamma^\mu P_L \nu_s  \left[\mathcal{U}(\nu,L)^\dagger \, \left(C_{\substack{H \ell }}^{(1)} - C_{\substack{H \ell}}^{(3)} \right) \mathcal{U}(\nu,L)\right]_{rs}, \nn \\
&+ \frac{\sqrt{\bar{g}_1^2+ \bar{g}_2^2}}{2} \, (h + v_T)^2 \, \mathcal{Z}_\mu \, \bar{e}_{r} \, \gamma^\mu P_L e_s  \left[\mathcal{U}(e,L)^\dagger \, \left(C_{\substack{H \ell }}^{(1)} + C_{\substack{H \ell}}^{(3)} \right) \mathcal{U}(e,L)\right]_{rs}, \nn \\
&+\frac{\sqrt{\bar{g}_1^2+ \bar{g}_2^2}}{2} \, (h + v_T)^2 \, \mathcal{Z}_\mu \, \bar{u}_{r} \, \gamma^\mu P_L u_s  \left[\mathcal{U}(u,L)^\dagger \, \left(C_{\substack{H q }}^{(1)} - C_{\substack{H q}}^{(3)} \right) \mathcal{U}(u,L)\right]_{rs}, \nn \\
&+ \frac{\sqrt{\bar{g}_1^2+ \bar{g}_2^2}}{2} \, (h + v_T)^2 \, \mathcal{Z}_\mu \, \bar{d}_{r} \, \gamma^\mu P_L d_s  \left[\mathcal{U}(d,L)^\dagger \, \left(C_{\substack{H q }}^{(1)} + C_{\substack{H q}}^{(3)} \right) \mathcal{U}(d,L)\right]_{rs}, \nn \\
& + \frac{\sqrt{\bar{g}_1^2+ \bar{g}_2^2}}{2} \, (h + v_T)^2 \, \mathcal{Z}_\mu \, \bar{\psi}_{r} \, \gamma^\mu P_R \, \psi_s \,  \left[\mathcal{U}(\psi,R)^\dagger \, C_{\substack{H \psi}} \, \mathcal{U}(\psi,R)\right]_{rs}, \nn \\
&- \frac{\bar{g}_2}{\sqrt{2}}  \, (h + v_T)^2 \,  \mathcal{W}^+_\mu \, \bar{\nu}_r \, \gamma^\mu P_L  \,  e_s \,\left[\mathcal{U}(\nu,L)^\dagger \, C_{\substack{H \ell}}^{(3)} \, \mathcal{U}(e,L)\right]_{rs} \nn \\
&- \frac{\bar{g}_2}{\sqrt{2}}  \, (h + v_T)^2 \,  \mathcal{W}^+_\mu \, \bar{u}_r \, \gamma^\mu P_L  \,  d_s \,\left[\mathcal{U}(u,L)^\dagger \, C_{\substack{H q}}^{(3)} \, \mathcal{U}(d,L)\right]_{rs},\nn \\
&+ \frac{i \, \bar{g}_2}{2}  \, (h + v_T)^2 \,  \mathcal{W}^+_\mu \, \bar{u}_r \, \gamma^\mu P_R  \,  d_s \,\left[\mathcal{U}(u,R)^\dagger \, C_{\substack{H ud}} \, \mathcal{U}(d,R)\right]_{rs} + h.c
\end{align}
where $\psi = \{u,d,e \}$.
\subsubsubsection{TGC parameters}
The off-shell Triple gauge coupling parameters are given by
\begin{align}
\left(-\mathcal{L}_{TGC} \right)/ \bar{g}_{VWW} &=i \bar{g}_{1}^{V} \left( \mathcal{W}_{\mu \nu}^{+} \mathcal{W}^{- \mu} - \mathcal{W}_{\mu \nu}^{-} \mathcal{W}^{+ \mu}\right)\mathcal{V}^{\nu} +i \bar{\kappa}_{V} \mathcal{W}^{+}_{\mu}\mathcal{W}^-_{\nu}\mathcal{V}^{\mu \nu}, \\
&+ i \frac{\bar{\lambda}_{V}}{\bar{M}^2_W}\mathcal{V}^{\mu \nu} \mathcal{W}^{+ \rho}_{\nu}\mathcal{W}^{-}_{\rho \mu} \nonumber
\end{align}

where $V = \{\mathcal{Z},\mathcal{A} \}$. In the SM $g_{\mathcal{AWW}} = e$ and $g_{\mathcal{ZWW}} = g_2 \, c_{\theta}$.
In the SMEFT the canonically normalized couplings are modified to $\bar{g}_{\mathcal{AWW}} = \bar{e}$ and $\bar{g}_{\mathcal{ZWW}} = \bar{g}_2 \, \bar{c}_{\theta}$
and the shifts compared to these normalized couplings are
\begin{align}
\delta \bar{g}_1^{\mathcal{A}}  &=- \delta \bar{\kappa}_{\mathcal{A}} = - \frac{v_T^2}{2}\frac{c_{\bar{\theta}}}{s_{\bar{\theta}}} C_{HWB},  \quad \quad \quad &\delta \bar{g}_1^{\mathcal{Z}} = -\delta \kappa_{\mathcal{Z}} = \frac{v_T^2}{2}\frac{s_{\bar{\theta}}}{c_{\bar{\theta}}} C_{HWB},
\end{align}
and
\begin{align}
\delta \bar{\lambda}_{\mathcal{A}} &=  6 \, s_{\bar{\theta}} \, C_W \, \frac{\bar{M}^2_W}{\bar{g}_{\mathcal{AWW}}},  \quad \quad \quad &\delta \bar{\lambda}_{\mathcal{Z}} =  6 \, c_{\bar{\theta}} \, C_W \frac{\bar{M}^2_W}{\bar{g}_{\mathcal{ZWW}}}.
\end{align}
An important check of gauge invariance in TGC shifts is that the relationships
\begin{eqnarray}
\bar{\kappa}_{\mathcal{Z}} = \bar{g}_1^{\mathcal{Z}} - (\bar{\kappa}_{\mathcal{A}}-1) \, t_{\bar{\theta}}^2, \quad \quad \bar{\lambda}_{\mathcal{Z}} =  \bar{\lambda}_{\mathcal{A}},
\end{eqnarray}
are respected when the shifts in the Lagrangian parameters are expressed in terms of the SM parameters.
These shifts respect these relationships.

\subsubsection{Summary of mass eigenstate interactions and symmetries}

$\mathcal{L}_6$ has $2499$ parameters in general \cite{Alonso:2013hga}. Clearly restricting to
a Minimal Flavour Violating (MFV) scenario \cite{Chivukula:1987py,D'Ambrosio:2002ex,Cirigliano:2005ck}, which imposes a $\rm U(3)^5$  flavour symmetry broken only by the SM Yukawas
is desirable. This reduces the number of parameters to $76$.
Assuming that $\rm CP$ violating effects can also be neglected, the number of parameters is restricted to $53$ for $\mathcal{L}_6$ \cite{Alonso:2013hga}. This is a reasonable symmetry based
limit to assume.\footnote{Custodial symmetry is broken by gauge interactions in the SM and the mass splitting of fermion doublet fields.
The number of parameters removed due to this strongly broken symmetry being assumed are negligible compared to the effects of the $\rm CP$ even and MFV assumptions.}
In this symmetric case $\left[ {\cal Y}_\psi \right]_{rs} \in \mathbb{R}$ and
\begin{eqnarray}
\left[ {\cal Y}_\psi \right]_{rs} =  \delta_{rs} \, \frac{\hat{m}^r_\psi}{v_T} \,  \left[ 1+ c_{H,\text{kin}}   - v^2\,  \mathcal{C}_{\substack{\psi H}} \right], \quad {\rm where} \quad \mathcal{C}_{\substack{\psi H}} [Y_\psi]_{rs} =  {\rm Re} \left[C^*_{\substack{\psi H \\ rs}}\right].
\end{eqnarray}
Further all Wilson coefficients for operators with dual fields (denoted with tilde subscripts or superscripts) are neglected. The dipole and Higgs dipole interactions are all flavour diagonal proportional to the corresponding
fermion mass and real. The $(h + v)^2\, \mathcal{Z} \bar{\psi} \, \psi$ interactions are flavour diagonal while the $(h + v)^2\, \mathcal{W}^+ \bar{\psi} \, \psi$ interactions are proportional to $V_{CKM}$ or $\mathcal{U}_{PMNS}^\dagger$. Finally, flavour violating interactions in the Class 8 operators follow an MFV pattern \cite{Chivukula:1987py,D'Ambrosio:2002ex,Cirigliano:2005ck}.

These symmetries, if assumed in $\mathcal{L}_6$, are broken at least by the SM interactions, which violate these symmetries. NLO calculations are required to define the perturbative breaking of these symmetric limits.

\subsubsection{Input parameters and defining conditions}

{Differently than in~Section~\ref{s.eftbasis}, we have not imposed the following conditions on the mass eigenstate construction: }
\begin{itemize}
\item{Tree-level relations between the electroweak parameters and a choice of input parameter set (IPS) are the same as the SM ones.}
\item{Two-derivative self-interactions of the Higgs boson are absent.}
\item{For each fermion pair, the coefficients of the $h V \bar{\psi} \, \psi$, $h^2 V \bar{\psi} \, \psi$ interaction terms are equal to the vertex correction of $V  \bar{\psi} \, \psi$.}
\end{itemize}
{These conditions are not required to interpret the data in the SMEFT and, in our view,   introduce technical complications in a LO approach that could make NLO calculations harder to develop.}

{Considering condition one above, we emphasize that an operator basis is IPS independent. If one were to modify the construction of
$\mathcal{L}_{SMEFT}$ to make some relationships to a particular IPS the same in the SM and the SMEFT with algebraic manipulations that were only defined classically (i.e.~at LO), this would make such a construction an example of a ``phenomenological effective Lagrangian'' whose advantage is limited to LO.
Claims of intuitive connections to LHC Higgs and EWPD observables in such approaches should be considered with great care. We believe it is advantageous not to tie a phenomenological Lagrangian construction to any specific IPS, for a series of reasons:}
\begin{itemize}
\item{Monte Carlo programs do not all use the same IPS. Further, the IPS $\{\alpha_{ew}, G_F, M_Z \}$ is not in common use when
automated calculations for the SM beyond LO are generated to define the SM event rate in a measurement.
Before any SMEFT implementation is used it must first be checked what IPS set or sets are used to define the SM event rate in the measurement of interest.
If a construction tied to the specific IPS $\{\alpha_{ew}, G_F, M_Z \}$ were to be used it must be confirmed that all simulation tools and SM results only use this specific IPS or inconsistent
results will be reported.}
\item{When  the IPS $\{\alpha_{ew}, G_F, M_Z \}$ is used in the analysis of ``high'' energy data it is afflicted with
hadronic uncertainties entering already at the one loop level and arising because it must be
``run up'' from low energy, crossing the hadronic resonance region.
The Fermi coupling constant, obtained from the muon lifetime, does not suffer from this
disadvantage (even in the full SM one loop hadronic effects are mass
suppressed)~\cite{vanRitbergen:1999fi}.}
\end{itemize}

The parameters $v_T$, $\bar{g}_1$, $\bar{g}_2$, $\bar{e}$, $s_{\bar{\theta}}$, $c_{\bar{\theta}}$ etc. in the Lagrangian do have to be assigned numerical values consistent with some IPS.
This is sometimes known as a ``finite renormalization". This is distinct from rotating to the mass eigenstate fields in the canonically normalized SMEFT and does not require the conditions above be imposed
in a gauge dependent manner.
We now illustrate how a straightforward LO implementation is related to the IPS $\{\alpha_{ew}, G_F, M_Z \}$ in the $\rm U(3)^5$ limit for tree level gauge boson fermion interactions.
\subsubsubsection{Input parameters measurements}
Define the local effective interaction for muon decay as
\begin{align}
\mathcal{L}_{G_F} =  -\frac{4G_{F}}{\sqrt{2}} \, \left(\bar{\nu}_\mu \, \gamma^\mu P_L \mu \right) \left(\bar{e} \, \gamma_\mu P_L \nu_e\right).
\end{align}
$G_F$ is defined as the following parameter measured in $\mu$ decay, $\mu^- \rightarrow e^- + \bar{\nu}_e + \nu_\mu$.
In the SMEFT ($e$ and $\mu$ are generation indices 1 and 2, and are not summed over)
\begin{align}
-\frac{4G_{F}}{\sqrt{2}} &=  -\frac{2}{v_T^2} +  \left(C_{\substack{ll \\ \mu ee \mu}} +  C_{\substack{ll \\ e \mu\mu e}}\right) - 2 \left(C^{(3)}_{\substack{Hl \\ ee }} +  C^{(3)}_{\substack{Hl \\ \mu\mu }}\right).
\label{gfermi}
\end{align}
The parameter $\alpha_{ew}$ is measured in the Thompson ($p^2 \rightarrow 0$) limit and discussed in Section \ref{alphaew}, and $M_Z$ is defined in the resonance pole scan of LEP measurements.

\subsubsubsection{Gauge boson couplings for the \texorpdfstring{$\alpha$}{alpha} IPS}
Our notational conventions are that shifts due to the
SMEFT are denoted as $\delta X = (X)_{SMEFT} - X_{SM}$ for a parameter $X$.\footnote{See Refs.~\cite{Grinstein:1991cd,Han:2004az,Burgess:1993vc,Bamert:1996px,Burgess:1993qk,Pomarol:2013zra} for the development of this approach and Refs. \cite{Berthier:2015gja,Berthier:2015oma} for details.}
Measured input observables or parameters directly defined by combinations of input observables are denoted with hat superscripts.
The shifts in the commonly appearing Lagrangian parameters $M_Z$, $M_W$, $G_F$, $s_\theta^2$ are
\begin{eqnarray}
\delta M_Z^2 &\equiv&  \frac{1}{2 \, \sqrt{2}} \, \frac{\hat{m}_Z^2}{\hat{G}_F} C_{HD} + \frac{2^{1/4} \sqrt{\pi} \, \sqrt{\hat{\alpha}} \, \hat{m}_Z}{\hat{G}_F^{3/2}} C_{HWB}, \\
\delta M_W^2 &=& -\hat{m}_W^2 \left(\frac{\delta s_{{\hat \theta}}^2}{s_{{\hat \theta}}^2}+\frac{c_{{\hat \theta}}}{s_{{\hat \theta}} \sqrt{2} \hat{G}_F}C_{HWB} + \sqrt{2} \delta G_F\right),\\
\delta G_F &=&  \frac{1}{\sqrt{2} \,  \hat{G}_F} \left(\sqrt{2} \, C^{(3)}_{\substack{Hl}} - \frac{C_{\substack{ll}}}{\sqrt{2}}\right), \\
\delta s_\theta^2 &=&  - \frac{s_{\hat \theta} \, c_{\hat \theta}}{2 \, \sqrt{2} \, \hat{G}_F (1 - 2 s^2_{\hat \theta})} \left[s_{\hat \theta} \, c_{\hat \theta} \, (C_{HD} + 4 \, C^{(3)}_{\substack{H l}} - 2 \, C_{\substack{ll}})
+ 2 \, C_{HWB} \right].
\end{eqnarray}
These shifts lead to modifications of the $Z$ couplings with the normalization
\begin{eqnarray}
\mathcal{L}_{Z,eff}  =  g_{Z,eff}  \,   \left(J_\mu^{Z \ell} Z^\mu + J_\mu^{Z \nu} Z^\mu + J_\mu^{Z u} Z^\mu +  J_\mu^{Z d} Z^\mu \right),
\end{eqnarray}
where $g_{Z,eff} = - \, 2 \, 2^{1/4} \, \sqrt{\hat{G}_F} \, \hat{m}_Z$, $(J_\mu^{Z x})^{pr} = \bar{x}_p \, \gamma_\mu \left[(\bar{g}^{x}_V)_{eff}^{pr}- (\bar{g}^{x}_A)_{eff}^{pr} \, \gamma_5 \right] x_r$ for $x = \{u,d,\ell,\nu \}$.
In general, these currents are matrices in flavour space. When we restrict our attention to the case of a MFV scenario $(J_\mu^{Z x})_{pr} \simeq (J_\mu^{Z x}) \delta_{pr}$.
In the Warsaw basis, the effective axial and vector couplings are modified from the SM values by a shift
\begin{eqnarray}
\delta (g^{x}_{V,A})_{pr} = (\bar{g}^{x}_{V,A})^{eff}_{pr} - (g^{x}_{V,A})^{SM}_{pr},
\end{eqnarray}
where
\begin{eqnarray}\label{higherdgvga}
\delta (g^{\ell}_V)_{pr}&=&\delta \bar{g}_Z \, (g^{\ell}_{V})^{SM}_{pr} - \frac{1}{4 \sqrt{2} \hat{G}_F} \left(C_{\substack{H e \\pr}} + C_{\substack{H l \\ pr}}^{(1)} + C_{\substack{H l \\ pr}}^{(3)} \right) - \delta s_\theta^2, \\
\delta(g^{\ell}_A)_{pr}&=&\delta \bar{g}_Z \, (g^{\ell}_{A})^{SM}_{pr} + \frac{1}{4 \, \sqrt{2} \, \hat{G}_F}
\left( C_{\substack{H e \\pr}} - C_{\substack{H l \\ pr}}^{(1)} - C_{\substack{H l \\ pr}}^{(3)} \right),  \\
\delta (g^{\nu}_V)_{pr}&=&\delta \bar{g}_Z \, (g^{\nu}_{V})^{SM}_{pr} - \frac{1}{4 \, \sqrt{2} \, \hat{G}_F} \left( C_{\substack{H l \\ pr}}^{(1)} - C_{\substack{H l \\ pr}}^{(3)} \right),
\\
\delta(g^{\nu}_A)_{pr}&=&\delta \bar{g}_Z \,(g^{\nu}_{A})^{SM}_{pr}  - \frac{1}{4 \, \sqrt{2} \, \hat{G}_F}
\left(C_{\substack{H l \\ pr}}^{(1)} - C_{\substack{H l \\ pr}}^{(3)} \right),  \\
\delta (g^{u}_V)_{pr}&=&\delta \bar{g}_Z \, (g^{u}_{V})^{SM}_{pr}  +
\frac{1}{4 \, \sqrt{2} \, \hat{G}_F} \left(- C_{\substack{H q \\ pr}}^{(1)} + \, C_{\substack{H q \\ pr}}^{(3)} -C_{\substack{H u \\ pr}} \right) + \frac{2}{3} \delta s_\theta^2,\\
\delta(g^{u}_A)_{pr}&=&\delta \bar{g}_Z \, (g^{u}_{A})^{SM}_{pr}
-\frac{1}{4 \, \sqrt{2} \, \hat{G}_F} \left( C_{\substack{H q \\ pr}}^{(1)} -  \, C_{\substack{H q \\ pr}}^{(3)} - C_{\substack{H u \\ pr}} \right), \\
\delta (g^{d}_V)_{pr}&=&\delta \bar{g}_Z \,(g^{d}_{V})^{SM}_{pr}
-\frac{1}{4 \, \sqrt{2} \, \hat{G}_F} \left( C_{\substack{H q \\ pr}}^{(1)}  +  \, C_{\substack{H q \\ pr}}^{(3)} + C_{\substack{H d \\ pr}} \right) -  \frac{1}{3} \delta s_\theta^2, \\
\delta(g^{d}_A)_{pr}&=&\delta \bar{g}_Z \,(g^{d}_{A})^{SM}_{pr}
+\frac{1}{4 \, \sqrt{2} \, \hat{G}_F} \left(-C_{\substack{H q \\ pr}}^{(1)}  -  \, C_{\substack{H q \\ pr}}^{(3)} + C_{\substack{H d \\ pr}} \right),
\end{eqnarray}
where
\begin{eqnarray}
\delta \bar{g}_Z =- \frac{\delta G_F}{\sqrt{2}} - \frac{\delta M_Z^2}{2\hat{m}_Z^2} + \frac{s_{\hat{\theta}} \, c_{\hat{\theta}}}{\sqrt{2} \hat{G}_F} \, C_{HWB},
\end{eqnarray}
and similarly the $W$ couplings are defined as
\begin{eqnarray}
\delta(g^{W_{\pm},\ell}_V)_{rr} = \delta(g^{W_{\pm},\ell}_A)_{rr}  &=&  \frac{1}{2\sqrt{2} \hat{G}_F} \left(C^{(3)}_{\substack{H l \\ rr}} + \frac{1}{2} \frac{c_{\hat{\theta}}}{ s_{\hat{\theta}}} \, C_{HWB} \right)
+ \frac{1}{4} \frac{\delta s_\theta^2}{s^2_{\hat{\theta}}},
\end{eqnarray}
\begin{eqnarray}
\delta(g^{W_{\pm},q}_V)_{rr} = \delta(g^{W_{\pm},q}_A)_{rr}  &=&  \frac{1}{2\sqrt{2} \hat{G}_F} \left(C^{(3)}_{\substack{H q \\ rr}} + \frac{1}{2} \frac{c_{\hat{\theta}}}{ s_{\hat{\theta}}} \, C_{HWB} \right)
+ \frac{1}{4} \frac{\delta s_\theta^2}{s^2_{\hat{\theta}}}.
\end{eqnarray}
Here our chosen normalization is $(g^{x}_{V})^{SM} = T_3/2 - Q^x \, s_\theta^2, (g^{x}_{A})^{SM} = T_3/2$ where $T_3 = 1/2$ for $u_i,\nu_i$ and $T_3 = -1/2$ for $d_i,\ell_i$
and $Q^x = \{-1,2/3,-1/3 \}$ for $x = \{\ell,u,d\}$.
The set of $\delta X$ parameters are not an operator basis for the SMEFT.

\subsubsection{Fitting at LO and NLO: constraints and covariance}

The mapping of an experimental constraint to the underlying $C_i$ is based on a linear expansion of a cross section or a pseudo-observable based decomposition of a cross section.
A fit at LO to mass eigenstate parameters should include a theoretical covariance matrix
and a theoretical error due to neglected higher order effects in the SMEFT \cite{Berthier:2015gja,Berthier:2015oma,David:2015waa}.
A fit in terms of the underlying weak eigenstate Wilson coefficients is straightforward and
{we believe it will in general have a much simpler theoretical covariance matrix.}

\subsubsubsection{Digression on theoretical uncertainty}
In the SM, when a particular process is calculated, a common practice is that
a theoretical error is assigned. For example, for parametric and theoretical uncertainties within the SM, see Table~1 of
\Bref{Heinemeyer:2013tqa}.  It can be subtle to assign such an error~\cite{David:2013gaa}
due to the neglect of missing higher order perturbative terms in the SM.
The need to include theoretical errors when perturbatively expanding the SMEFT is tied to the fact that different truncations of such expansions can
be constructed. Suppose that a given quantity $\mathrm{Q}(a)$ is given in perturbation theory by the following expansion:
\begin{equation}
\mathrm{Q} = a + g\,\Bigl[ a^2 + f_1(a) \Bigr] + g^2\,\Bigl[ a^3 + f_2(a) \Bigr] + \mathcal{O}(g^3)
= {\bar a} + g\,f_1(a) + \mathcal{O}(g^2),
\end{equation}
where ${\bar a} = a/(1 - g a)$. Suppose that only the $f_1$ term is actually known. It could be
decided that ${\bar a}$ is the effective expansion parameter (or that in the full expression we
change variable $a \to {\bar a}$). This is equivalent, in the truncated expansion, to introducing
\begin{equation}
\mathrm{Q} = \bar{a} + g\,f_1(a) = {\bar a} + g\,f_1({\bar a}),
\end{equation}
which gives $\Delta \mathrm{Q} = g^2\,f'_1(a)$, the difference in the two results due to neglected higher order terms is an estimate of the associated theoretical uncertainty.
A fit to observables defined in a perturbative expansion must always include an estimate of the missing higher order terms~\cite{Bardin:1999gt}, which specifies
a theoretical uncertainty.

\subsubsubsection{The importance of NLO results for theoretical uncertainty}

An excellent example of the importance of theory errors is provided by another effective field theory, NRQED, as discussed in
Refs.\cite{Bauer:2004ve,Manohar:2003xv,Caswell:1985ui,Bodwin:1994jh,Luke:1999kz,Pineda:1998kn,Grinstein:1997gv}.
The Hydrogen hyperfine splitting is measured to fourteen digits, but only computed to seven digits.
This introduces a theoretical error when using this measurement. Comparatively, the Positronium hyperfine splitting is measured
and computed to eight digits. It would simply be a mistake to give the $H$ hyperfine splitting a weight $10^6$ larger than the $P_s$ hyperfine
splitting in a global fit to the fundamental constants, and to totally ignore theory errors. A careful consideration of NLO effects can help in
avoiding similar errors when using the SMEFT formalism.
{In our opinion, neglecting such considerations may lead to incorrect conclusions. For example, in Ref.~\cite{Berthier:2015gja,Berthier:2015oma} it has been shown that the per mille level due to the LEP experiments projected into the SMEFT might not be as strong as claimed in the previous literature when SMEFT theoretical errors are taken into account.}
This should not be surprising, as in EWPD the modifications of the $\PW$ mass, the $\uprho$ parameter and the effective weak-mixing angle are
loop-induced quantities and a study of their SM deviations requires an analysis at NLO in the SMEFT.
As a result of these developments, constraints on parameters in the SMEFT (that are not symmetry based) are not robustly below LHC sensitivity.

{For this reason, we believe it is not wise}
to set parameters that contribute to EWPD to zero in LHC analyses in the SMEFT.
The experimental bound should be imposed on these parameters, with a clearly specified theory error.
As a rule of thumb when experimental bounds descend below the $10 \%$ level SMEFT theory errors should not be neglected in an EFT interpretation of the data. \footnote{Editor footnote: 
Another point of view is expressed in Section~\ref{s.eftbasis}  which advocates to focus the LHC Higgs analyses, at least in the early phase, on the parameters that are most unconstrained by previous experiments in order to maximize the sensitivity to new physics.}

\subsubsubsection{Covariance due to operator basis in \texorpdfstring{$\mathcal{L}_6$}{L_6}}\label{variancesection}

Consider two mass eigenstate interaction shifts $\delta X_1, \delta X_2$ that contribute to a particular cross section that reports an experimental bound.
Several SMEFT Wilson coefficients  generally contribute to any one observable through $\delta X_1, \delta X_2$.  All such parameters must be retained unless symmetries, or knowledge of the UV theory, allows a reduction. One can directly interpret the data at LO in terms of the underlying Wilson coefficients that are present in $\delta X_1, \delta X_2$ and defined in linear
expansions of these parameters, so long
as theoretical errors are carefully accounted for.

Alternatively fit results can be reported in terms of $\delta X_1, \delta X_2$. However in this case it is critical that a theoretical covariance matrix is included.
As the shifts $\delta X_1, \delta X_2$ are linear in the Wilson coefficients, the bi-linearity property of covariance
can be used to obtain the theoretical covariance matrix directly. Schematically the matrix can be build up for  $\delta X_1 = a C_1 + b C_2 + \cdots$ and $\delta X_2 = c C_1+ d C_3 + \cdots$
as follows
\begin{eqnarray}
Cov \left[\delta X_1,  \delta X_2\right] = a \,c \, Cov [C_1,C_1] + a \,d \, Cov [C_1,C_3] + b \,c \, Cov [C_2,C_1] + b \,d \, Cov [C_2,C_3] + \cdots \nonumber
\end{eqnarray}
Assuming that the $C_1,C_2,C_3$ are independent operators $Cov [C_1,C_1] = Var[C_1]$ and all other entries vanish.
The appropriate covariance matrix can be constructed so long as a theoretical error is included for each of the terms in the perturbative expansion of the $\delta X$.
Estimating a theoretical error for these terms to obtain the individual variances requires an estimate of neglected NLO corrections.
A NLO mapping can be carried out in the same manner. The only modification is the use of NLO formula in the expansion of the cross section and smaller theoretical errors, as we illustrate below.
Fits to mass eigenstate parameters in general have very non trivial covariance matrices (due to gauge invariance of the underlying operator basis) that have to be defined. The required
theoretical errors can only be estimated by an understanding of NLO corrections to a LO formalism.

\subsubsubsection{Fitting at LO or NLO?}

A NLO  treatment of the data is always advisable if the required theoretical results are available.
NLO analyses are required to consistently map
lower energy measurements in the SMEFT to the cut off $\mu = \Lambda$, or to consistently
combine data sets measured at different effective scales ($\mu_1 \neq \mu_2$).
Whether or not a NLO treatment of the data is {\it required} in the SMEFT is defined by three considerations:

\begin{itemize}
\item{What is the cut off scale ($\Lambda$) and what is the matching pattern of Wilson coefficients into the SMEFT?}

\item{What is the experimental precision that will be reached in a measurement?}

\item{How will a bound projected into the SMEFT formalism at LO be used?}
\end{itemize}

Considering the first question, it is interesting to consider the cases where
$1 \, {\rm TeV} \lesssim \Lambda/\sqrt{\tilde{C}_i} \lesssim 3 \, {\rm TeV}$.
In these cases, deviations in processes measured at the LHC could possibly be observable.
If deviations are seen then a NLO analysis is well motivated to learn as precisely as possible about
the underlying physics sector through the measured deviation.
Cut off scales of this form are not implausible or ruled out. On the contrary they are well motivated by the Hierarchy problem. Further model building exercises for decades have indicated
that such cut off scales are not robustly ruled out when considering EWPD. If the ratio $ \Lambda/\sqrt{\tilde{C}_i} $ lies in this interesting range,
the effect of NLO corrections are clearly not negligible
\cite{Grojean:2013kd,Berthier:2015gja,Berthier:2015oma,David:2015waa,Englert:2014cva,Hagiwara:1993ck,Hagiwara:1993qt,Alam:1997nk,Passarino:2012cb,Wells:2015cre,Gauld:2015lmb,Hartmann:2015oia,Hartmann:2015aia,Ghezzi:2015vva}.
Considering the second question, as we have stressed, when experimental precision starts to reach the $10\%$ level a NLO analysis should be pursued.
The answer to the third question differs among analyses and authors but in general NLO results will always be useful to
authors interested in LO results while the converse is not true.
\subsubsubsection{Theory errors in a LO formalism on the IPS}\label{alphaew}
As a specific example of a theory error to include in a LO analysis, any LO approach does not take into account that the scales characterizing the measurements of the input parameters $\alpha_{ew}, G_F, M_Z$ differ.
Consider the error
introduced due to the neglect of this NLO effect in the SMEFT, compared to the errors quoted on  $\alpha_{ew}$ in the SM.
This parameter is measured at low energies in the $p^2 \rightarrow 0$ limit.\footnote{$\alpha_{ew}$  is frequently extracted in the Thompson limit $p^2 \rightarrow 0$ when probing some Coulomb potential
of a charged particle, for example in a measurement of $g-2$ for the electron or muon. Recently, extractions with a competitive error budget have emerged where $\alpha_{ew}$ is extracted
from the measured ratio of $\hbar/M_{atom}$ via the recoil velocity for a stable atom, such as ${\rm Rb}^{87}$ \cite{Hanneke:2008tm} or ${\rm C_s}$ \cite{PST}. The important point is to realize that
this input parameter differs in the SM and in the SMEFT at NLO.} The value of this input parameter is given in Table 2.
In the SMEFT, the running of $\alpha_{ew}$ is modified compared to the SM as given in Ref.\cite{Jenkins:2013zja}.
As a simple approximation of the error introduced in the SMEFT, one finds that the neglected NLO SMEFT correction to $\alpha_{ew}$ is then
\begin{eqnarray}
\frac{(\Delta \alpha_{ew})_{SMEFT}}{(\Delta \alpha_{ew})_{SM}} \simeq - 250 \, \left(\frac{1 {\rm TeV}}{\Lambda} \right)^2 \tilde{C}_{HB} - 80 \, \left(\frac{1 {\rm TeV}}{\Lambda} \right)^2 \tilde{C}_{HW},
\end{eqnarray}
running from $p^2 \sim 1 \, {\rm GeV^2}$ to $m_h$.\footnote{This is only an approximation, as formally all of the SM states with masses $m^2 \gg p^2$ should be integrated out in sequence
when running down from the high scale.}
Here $(\Delta \alpha_{ew})_{SM}$ is the SM error quoted in the Table. Depending on $\tilde{C}_{HB}$ and $\tilde{C}_{HW}$ and $\Lambda$, which are unknown, the neglected
NLO SMEFT effects can lead to an error on this input parameter far larger than in the SM.
This should be completely unsurprising. Neglected NLO effects in the SMEFT in this case include corrections of order $ g_{1,2}^2 v_T^2/ (16 \, \pi^2) \, \Lambda^2$.
The theoretical errors due to such neglected effects can
obviously compete with the SM theoretical errors, introduced in a QED calculation out to {\it tenth order} in the SM.
Similarly, neglected NLO corrections on the other input parameters modify their theoretical error.
\begin{center}
\begin{table}
\caption{Current experimental best estimates of $\alpha_{ew}, G_F, M_Z$. }
\centering
\tabcolsep 8pt
\begin{tabular}{c|c|c}
\toprule
Parameter & Input Value & Ref.  \\ 
\midrule
$\hat{m}_Z$ & $91.1875 \pm 0.0021$ & \cite{ALEPH:2005ab,Agashe:2014kda,Mohr:2012tt} \\
$\hat{G}_F$ & $1.1663787(6) \times 10^{-5} $ &  \cite{Agashe:2014kda,Mohr:2012tt} \\
$\hat{\alpha}_{ew}$ & $1/137.035999074(94) $ &  \cite{Hanneke:2008tm,Agashe:2014kda,Mohr:2012tt,Bouchendira:2010es} \\
\bottomrule
\end{tabular}
\end{table}
\end{center}
\subsubsubsection{Approximating unknown SMEFT theory errors}
Various ways exist to estimate SMEFT theory errors. One can
compute the same observable with different ``options'', \eg linearization or
quadratization of the squared matrix element, resummation or expansion of the (gauge invariant)
fermion part in the wave function factor for the external legs, variation of the renormalization
scale, $G_{\mathrm{F}}$ renormalization scheme or $\alpha\,$-scheme, \etc

A conservative estimate of the associated theoretical uncertainty is obtained by taking the
envelope over all ``options''; the interpretation of the envelope is a log-normal distribution
(commonly done in the experimental community) or a flat Bayesian
prior~\cite{Cacciari:2011ze,David:2013gaa} (a solution preferred in a large part of the theoretical
community).

{In our opinion, to properly characterize the perturbative error,}
it is essential to calculate at least to one loop
order in the SMEFT, including the leading insertion of operators in ${\cal L}_6$. Until such
calculations are performed,
{we recommend}
conservative theoretical errors should be
applied to theoretical relations in the SMEFT. Further, the introduction of a ``non-perturbative''
error, due to ${\cal L}_8$ when bounding ${\cal L}_6$ should be done.
In Eqn.\ref{tableNLO}, the $g^3\,g^2_6\,{\cal A}^{(6)}_{3,2,1}$ terms can be used as estimators
of missing higher order non-perturbative terms in the SMEFT.
This approach is not particularly novel, but is simply the obvious extension of the widely
accepted approach to assigning theoretical error in the SM to the SMEFT.
\footnote{Editor footnote: 
Another point of view is expressed in Section~\ref{s.eftval}, where it is argued that the validity range of an EFT analysis cannot be determined using only low-energy information or the truncation error of the EFT Lagrangian.} 

As a specific example, a reasonable approximation of a theoretical error to introduce for an observable $i$ when fitting
to the leading parameters in ${\cal L}_6$, is given by~\cite{Berthier:2015gja,Berthier:2015oma}
\begin{eqnarray}
\Delta_{SMEFT}^i(\Lambda) &\simeq& \sum_{j}  \, x_{ij} \,  \tilde{C}^8_{ij} \,
\frac{v_T^4}{\Lambda^4} + \sum_{j} \frac{(g^{ij}_{SM})^2}{16 \, \pi^2} \, \tilde{C}^6_{ij} \, y_{ij} \,
\ln \left[\frac{\Lambda^2}{v_T^2}\right] \, \frac{v_T^2}{\Lambda^2} \,.
\end{eqnarray}
Non log dependence in the second term is also present, but is suppressed for a simplifying
approximation. Here $x_{ij},y_{ij}$ label the observable dependence and are $\mathcal{O}(1)$.
One can further define
\begin{eqnarray}
x'_i \, \sqrt{N^{i}_8}  = \sum_{j} \sqrt{x^2_{ij} \, (\tilde{C}^8_{ij})^2},  \quad
\quad y'_i \, \sqrt{N^i_6}  = \sum_{j} \sqrt{y^2_{ij} \,  (\tilde{C}^6_{ij})^2}
\end{eqnarray}
as the product of $\mathcal{O}(1)$ numbers that characterize the multiplicity of the operators
that contribute to a process ($N_{6,8}$) and the typical numerical dependence $x'_i,y'_i$.
The square root is because errors are assumed to add in quadrature.
As an alternative, a Bayesian uniform prior for the $C_i$ could be used.

Although the number of
operators is large, the relevant number of operators that contribute in a process is far less
then the full operator set;
in known examples $N_{6,8} \sim \mathcal{O}(10)$.
No complete operator basis of ${\cal L}_8$ has ever been encoded in a Monte-Carlo program
and used to fit the data, and we do not recommend that
fits should explicitly include all terms in $\mathcal{L}_8$ and vary corrections in general. Rough error estimates of this form should be sufficient for most purposes.

This error is multiplicative and the absolute
error is obtained as $\Delta_{SMEFT}^i(\Lambda)$ times the SM prediction for an observable.
For cut off scales and Wilson coefficients in the range $1 \, {\rm TeV} \lesssim \Lambda/\sqrt{\tilde{C}_i} \lesssim 3 \, {\rm TeV}$ and order one
numbers for $x_i,y_i,N_{6,8}$ the value of $\Delta_{SMEFT}^i(\Lambda)$
is in the range of few $\mathcal{O}(\%)$ to
$\mathcal{O}(0.1\%)$ \cite{Berthier:2015gja,Berthier:2015oma,David:2015waa}.
This is the reason
{we believe}
that once experimental errors descend to the $\mathcal{O}(10 \%)$ level SMEFT theory errors should
be considered to be conservative.  It is widely considered to be the case that the precision expected in LHC analyses can be
expected to approach a few per cent in well measured channels \cite{CMS:2013xfa,Flechl:2015foa}.

{We believe that a percentage error}
can be motivated for SMEFT theoretical uncertainties using these approximations and
then directly applied (and varied) when reporting a bound.

\subsubsection{NLO SMEFT loop corrections}

{We believe that including loop corrections in the SMEFT context is somehow even more crucial than for a pure SM calculation.}\footnote{Editor footnote: 
Another point of view is expressed in Section~\ref{s.eftval} where it is argued that the NLO
corrections due to $D$=6 EFT coefficients are important to include
only in particular well-defined situations.}

One loop corrections can introduce a dependence on Wilson coefficients that
do not contribute at tree level to a particular process and some of these Wilson coefficients are very poorly bounded.
This is different from the SM where all of the Lagrangian terms are extremely well known.
We will refer to the introduction of such dependence as ``non-factorizable'' corrections. Such corrections can significantly change
the interpretation of a mapping of experimental constraints at NLO in the SMEFT, as we illustrate below.
Loop corrections also introduce a perturbative rescaling of the dependence on an operator's
Wilson coefficient. These corrections help define the variance discussed Section~\ref{variancesection} for a LO analysis.

Improving the SMEFT to one loop requires
a renormalization scheme be defined, a systematic
renormalization of the SMEFT be carried out on the new parameters in $\mathcal{L}_6$, and
loop corrections be performed in a particular chosen gauge.
We now discuss each of these steps in the NLO program in more detail.

\subsubsection{SMEFT: renormalization in practice}\label{inpractice}
In this section we describe a general renormalization procedure in the SMEFT.
The results
presented have been developed in \Brefs{Passarino:2012cb,Ghezzi:2015vva}, based on the
conventional formalism widely used in the SM \cite{'tHooft:1972ue,'tHooft:1972fi,'tHooft:1978xw,Passarino:1978jh}.
To perform renormalization in an EFT it is appropriate to use a dimensionless regulator, see
Refs.~\cite{Manohar:1996cq} for a review. We work with dimensional regularization and define
\begin{equation}
\Delta_{\mathrm{UV}} = \frac{2}{4 - d} - \mathswitch \gamma - \ln \pi - \ln\frac{\muRs}{\mu^2}
\end{equation}
where $d$ is space-time dimension, the loop measure is $\mu^{4 - d}\,d^nq$
and $\muR$ is the renormalization scale; $\mathswitch \gamma$ is the Euler-Mascheroni constant. Counter-terms
for SM parameters and fields are defined by
\begin{equation}
{\mathrm{Z}}_i = 1 + \frac{g^2}{16\,\pi^2}\,\lpar d{\mathrm{Z}}^{(4)}_i + g_6\,d{\mathrm{Z}}^{(6)}_i \rpar\,\Delta_{\mathrm{UV}} \,.
\end{equation}
With field/parameter counter-terms we can make UV finite the self-energies and the
corresponding Dyson resummed propagators. However, these counterterm subtractions are not
enough to make UV finite the Green's functions with more than two legs
(at $\mathcal{O}(g^{\mathrm{N}_6} g_6)$). A mixing matrix among Wilson coefficients is needed:
\begin{equation}
C_i = \sum_j\,{\mathrm{Z}}^{\mathrm{W}}_{ij}\,C^{\ren}_j
\qquad
{\mathrm{Z}}^{\mathrm{W}}_{ij} = \delta_{ij} + \frac{g^2}{16\,\pi^2}\,d{\mathrm{Z}}^{\mathrm{W}}_{ij}\,\Delta_{\mathrm{UV}} \,.
\end{equation}


For example, in this way we can renormalize the (on-shell) $\mathrm{s}\,$-matrix for
$\PH(P) \to \PA_{\mu}(p_1)\PA_{\nu}(p_2)$ and
$\PH(P) \to \PA_{\mu}(p_1)\PZ_{\nu}(p_2)$ which have only one (transverse) Lorentz structure.
By on-shell $\mathrm{s}\,$-matrix for an arbitrary process (involving unstable particles) we mean
the corresponding (amputated) Green's function supplied with LSZ factors and sources, computed
at the (complex) poles of the external lines~\cite{Grassi:2000dz,Kniehl:2001ch,Goria:2011wa}. For
processes that involve stable particles this can be straightforwardly transformed into a
physical observable.

The connection of the $\PH\PV\PV, \PV = \PZ,\PW$ (on-shell) $\mathrm{s}\,$-matrix with the off shell
vertex $\PH \to \PV\PV$ and the full process $\Pp\Pp \to 4\,\psi$ is more complicated and
is discussed in some detail in Sect.~3 of \Bref{David:2015waa}.
The ``on-shell'' $\mathrm{s}\,$-matrix for $\PH\PV\PV$, being built with the the residue of the
$\PH{-}\PV{-}\PV$ poles in $\Pp\Pp \to 4\, \psi$ is gauge invariant by construction (it can be
proved by using Nielsen identities~\cite{Grassi:2001bz}) and represents one of the building
blocks for the full process: in other words, it is a
pseudo-observable~\cite{Passarino:2010qk,Gonzalez-Alonso:2014eva,David:2015waa}.
Technically speaking the ``on-shell'' limit for external legs should be understood
``to the complex poles'' (for a modification of the LSZ reduction formulas for unstable
particles, see \Bref{Weldon:1975gu}) but, as well known, at one loop we can use
on-shell masses (for unstable particles) without breaking the gauge parameter independence
of the result. Residues of complex poles are what matters, as far as renormalization is concerned.

The $\PH(P) \to \PZ_{\mu}(p_1)\PZ_{\nu}(p_2)$ (on-shell) matrix contains a part of the amplitude
proportional to $g^{\mu\nu}$ (referred to as $\mathcal{D}_{HZZ }$ below) and a part of the amplitude
proportional to $p^{\mu}_2\,p^{\nu}_1$ (referred to as $\mathcal{P}_{HZZ }$ below).
Both of these terms get renormalized through a mixing.

Consider now the $\PH(P) \to \PWm_{\mu}(p_1)\PWp_{\nu}(p_2)$ (on-shell) matrix:
it has the same Lorentz decomposition of $\PH \to \PZ\PZ$ and it is UV finite in the
$\mathrm{D}im = 4$ part. The $\mathcal{D}_{HWW}$ part at $\mathrm{D}im = 6$ is renormalized through a mixing;
however, there are no Wilson coefficients in $\mathcal{P}_{HWW}$ that are not also present in
$\mathcal{P}_{HZZ }$, so that the UV finiteness of this term is related by gauge symmetry to the
renormalization of $\mathcal{P}_{HZZ }$. This is the first part of the arguments
used in \Brefs{Passarino:2012cb,Ghezzi:2015vva} in proving closure of NLO SMEFT
under renormalization.

The (on-shell) decays $\PH(P) \to \PQb(p_1)\PAQb(p_2)$ and
$\PZ(P) \to \bar{\psi}(p_1)\psi(p_2)$ are more involved to improve to NLO in the SMEFT.
The SM contribution to these amplitudes are rendered finite by the SM counter-terms,
however renormalizing the contributions due to ${\cal L}_6$ requires an extensive
treatment of this operator mixing. See Ref. \cite{Gauld:2015lmb} for recent results on these decays.


Some structure present in the SM is not preserved when extending an analysis into the SMEFT.
Manifestly, processes that first appear at one loop in the SM can occur at tree level in
the SMEFT, due to the presence of local contact operators.
However, some symmetries of the SM are preserved. For example, consider the universality of the
electric charge.
In pure QED there is a Ward identity \cite{Ward:1950xp} telling us that $e$ can be renormalized in terms of
vacuum polarization (which is a way to understand the universality of the coupling), and Ward-Slavnov-Taylor (WST) identities \cite{Ward:1950xp,Slavnov:1972fg,Taylor:1971ff} allow us to generalize the argument
to the full spontaneously broken SM symmetry group.
The previous statement means that the contribution from vertices (at zero momentum transfer) in
the full SM exactly cancel those from (fermion) wave function renormalization factors. Therefore,
by directly computing the vertex $A \, \bar{\psi} \, \psi$ (at $q^2 = 0$) and the $\PZ_\psi$ wave function
factor in the SMEFT, one can directly prove (or check) that the WST identity is extended to the SMEFT at $\mathcal{L}_6$.
This is expected as the corresponding identities are the consequence of symmetries. However,
this is technically non-trivial even after the previous steps in the renormalization program
discussed above. Once (non-trivial) finiteness of this vertex is established, the finiteness
of $\Pep\Pem \to \bar{\psi} \, \psi$ (including the four-point functions in the non resonant part) follows.
This is the second part in proving closure of the NLO SMEFT under renormalization, using the arguments
of \Brefs{Passarino:2012cb,Ghezzi:2015vva}.

At NLO one first has to render all SM and SMEFT
parameters finite. Considering the arguments above, and the complete renormalization results of
all the operators in ${\cal L}_6$ reported in
Refs.~\cite{Alonso:2013hga,Grojean:2013kd,Jenkins:2013zja,Jenkins:2013wua}
in the Warsaw basis, this step in the NLO program has been accomplished. This result has not been
{derived so far  in any other basis.
We believe that translating these results to other bases may be very challenging.}

\subsubsection{Input parameter choices}
The detailed fixing of poles and residues that make up precise renormalization conditions require a lengthy discussion.
For detailed reviews in the case of the SM, see Refs.~\cite{Denner:1994xt,Denner:1991kt}. Below we summarize the
results of the finite renormalization in the relationship to the input observables.

\subsubsubsection{Using a `\texorpdfstring{$G_F$}{G_F}-scheme' with \texorpdfstring{$G_{F}$}{G_F}, \texorpdfstring{$M_W$}{M_W}, \texorpdfstring{$M_Z$}{M_Z} }
In the `$G_{F}\,$-scheme', one uses $\{G_{F}\,,\,\mw\,,\,\mz\}$ to fix terms in the Lagrangian. In this case, we write the following
equation for the $g$ finite renormalization
\begin{equation}
g_{\ren} = g_{{\mbox{\scriptsize exp}}} + \frac{g^2_{{\mbox{\scriptsize exp}}}}{16\,\pi^2}\,\lpar
d\mathcal{Z}^{(4)}_g + g_6\,d\mathcal{Z}^{(6)}_g \rpar
\label{gexp}
\end{equation}
where $g_{{\mbox{\scriptsize exp}}}$ will be expressed in terms of the Fermi coupling constant $G_{F}$.
Furthermore, ${\mathrm{c}}_{_{\theta}} = \mw/\mz$. The $\mu\,$-lifetime can be written in the form
\begin{equation}
\frac{1}{\tau_{\mu}} = \frac{M^5_{\PGm}}{192\,\pi^3}\,\frac{g^4}{32\,M^4}\,
\lpar 1 + \delta_{\mu} \rpar \,.
\end{equation}
The radiative corrections are $\delta_{\mu} = \delta^{\PW}_{\mu} + \delta_{\mathrm{G}}$
where $\delta_{\mathrm{G}}$ is the sum of vertices, boxes etc and $\delta^{\PW}_{\mu}$ is due
to the $\PW$ self-energy. The renormalization equation becomes
\begin{equation}
\frac{G_{F}}{\sqrt{2}} = \frac{g^2}{8\,M^2}\,\left\{
1 + \frac{g^2}{16\,\pi^2}\,\Bigl[ \delta_{\mathrm{G}} + \frac{1}{M^2}\,\Sigma_{\PW \PW}(0)
\Bigr]
\right \}
\label{GFscheme}
\end{equation}
where we expand the solution for $g$
\begin{equation}
g^2_{\ren} = 4\,\sqrt{2}\,G_{F}\,M^2_{\PW\,;\,{\mathrm{OS}}}\,\left\{
 1 + \frac{G_{F} M^2_{\PW\,;\,{\mathrm{OS}}}}{2\,\sqrt{2}\,\pi^2}\,
\Bigl[ \delta_{\mathrm{G}} + \frac{1}{M^2}\,\Sigma_{\PW \PW  \,;\,{\mbox{\scriptsize fin}}}(0) \Bigr]
\right \} \,.
\label{GFsol}
\end{equation}
Note that the non universal part of the corrections is given by
\begin{equation}
\delta_{\mathrm{G}} = \delta^{(4)}_{\mathrm{G}} + g_6\,\delta^{(6)}_{\mathrm{G}}
\quad
\delta^{(4)}_{\mathrm{G}} = 6 + \frac{7 - 4\,{\mathrm{S}}_{_{\theta}}^2}{2\,{\mathrm{S}}_{_{\theta}}^2}\,\ln{\mathrm{c}}_{_{\theta}}s
\end{equation}
but the contribution of $\mathcal{L}_6$ to muon decay at NLO is not available yet and
has not be included in the calculation. It is worth noting that \eqn{GFscheme} defines the finite
renormalization in the $\{G_{F}\,,\,\mw\,,\,\mz\}$ IPS.
\subsubsubsection{The `\texorpdfstring{$\alpha\;$}{alpha-}scheme', using \texorpdfstring{$\alpha, G_{\rm{F}},M_Z$}{alpha, G_F, M_Z} }
This scheme uses the fine structure constant $\alpha$ and is based on using
$\{\alpha\,,\,G_{F}\,,\,\mz\}$ as the IPS. The new finite-renormalization equation is
\begin{equation}
g^2\,{\mathrm{S}}_{_{\theta}}^2 = 4\,\pi\alpha\,\Bigl[ 1 - \frac{\alpha}{4\,\pi}\,\frac{\Pi_{{\mathrm{A} \mathrm{A} }}(0)}{{\mathrm{S}}_{_{\theta}}^2} \Bigr]
\end{equation}
where $\alpha = \alpha_{\rm{\scriptscriptstyle{QED}}}(0)$ and $\Pi_{{\mathrm{A} \mathrm{A} }}$ defines the vacuum polarization. Therefore,
in this scheme, the finite counter-terms are
\begin{equation}
g^2_{\ren} = g^2_{\mathrm{A}}\,\Bigl[ 1 + \frac{\alpha}{4\,\pi}\,d\mathcal{Z}_{g} \Bigr]
\quad
{\mathrm{c}}_{_{\theta}}r = {\hat {\mathrm{c}}}_{_{\theta}}\,\Bigl[ 1 + \frac{\alpha}{4\,\pi}\,d\mathcal{Z}_{{\mathrm{c}}_{_{\theta}}} \Bigr],
\quad
M_{\ren} = \mathswitch{M_{\PZ\,;\,{\mathrm{OS}}}}\,{\hat {\mathrm{c}}}_{_{\theta}}s\,\Bigl[ 1 + \frac{\alpha}{8\,\pi}\,d\mathcal{Z}_{\mw} \Bigr]
\label{eqIPSb}
\end{equation}
where the parameters ${\hat {\mathrm{c}}}_{_{\theta}}$ and $g_{\mathrm{A}}$ are defined by
\begin{equation}
g^2_{\mathrm{A}} = \frac{4\,\pi\,\alpha}{{\hat {\mathrm{S}}}^2_{_{\theta}}}
\qquad
{\hat {\mathrm{S}}}^2_{_{\theta}} = \frac{1}{2}\,\Bigl[ 1 - \sqrt{1 - 4\,\frac{\pi\,\alpha}{\sqrt{2}\,G_{F}\,\mathswitch{M^2_{\PZ\,;\,{\mathrm{OS}}}}}}\Bigr] \,.
\label{defscipsb}
\end{equation}
The reason for introducing this scheme is that the $\mathrm{s}, {\mathrm{T}}$ and ${\mathrm{U}}$ parameters
\Bref{Peskin:1990zt} have been originally given in the $\{\alpha\,,\,G_{F}\,,\,\mz\}$
scheme, and these input parameters are very well measured in the SM.
When calculating processes involving photons final states, this scheme can be transparent to adopt.
For other processes, the $\{G_{F}\,,\,\mw\,,\,\mz\}$ scheme can be more appropriate, and
is in wider use in the SM in higher order calculations. In the $\alpha\,$-scheme, after
requiring that $\mathswitch{M^2_{\PZ\,;\,{\mathrm{OS}}}}$ is a zero of the real part of the inverse $\PZ$ propagator, we are
left with one finite counterterm, $d\mathcal{Z}_{g}$. The latter is fixed by using $G_{F}$ and
requiring that
\begin{equation}
\frac{1}{\sqrt{2}}\,G_{F} = \frac{g^2}{8\,M^2}\,\Bigl\{ 1 + \frac{g^2}{16\,\pi^2}\,\Bigl[
                       \delta_{\mathrm{G}} + \frac{1}{M^2}\,\Delta_{\PW \PW}(0)
                     - \lpar d{\mathrm{Z}}_{\PW} + d{\mathrm{Z}}_{\mw} \rpar\,\Delta_{\mathrm{UV}} \Bigr] \Bigr\}
\end{equation}
where we use the following relations for UV and finite renormalization,
\begin{equation}
g = g_{\ren}\,\lpar 1 + \frac{g^2_{\ren}}{16\,\pi^2}\,d{\mathrm{Z}}_{g}\,\Delta_{\mathrm{UV}} \rpar
\qquad
g_{\ren} = g_{\mathrm{A}}\,\lpar 1 + \frac{\alpha}{8\,\pi}\,d\mathcal{Z}_{g} \rpar \,.
\end{equation}
Note that SM EW calculations available in literature generally use $G_{F}$ for the pure
weak part or evolve $\alpha(0) \to \alpha(M)$ and use $\alpha(M)$ as the expansion parameter
at the scale $M$. For a comprehensive discussion see Sect.~5.3 of \Bref{Binoth:2010xt}.

\subsubsection{Background field gauge}

Any well defined gauge can be used in a calculation, see Ref.~\cite{Leibbrandt:1987qv} for an
excellent review on gauge fixing.
There can be some advantage to organizing a calculation in a manner that enforces relationships
between counter terms due to gauge invariance. A technique that accomplishes
this is known as the Background Field (BF) method~\cite{DeWitt:1967ub,Abbott:1981ke}. The idea is that
fields are split into classical and quantum components and a gauge fixing term is added that
maintains the gauge invariance of the classical background fields, while breaking the gauge
invariance of the quantum fields.
Due to the resulting Ward identities, one finds the relations among the SM counter-terms~\cite{Denner:1994xt}.
The gauge fixing in the BF method can be imposed as in Ref.~\cite{Denner:1994xt,Einhorn:1988tc}.
Use of the background field method can make extending the WST
relations between counter-terms manifest and transparent, even when including the effects
of ${\cal L}_6$. It is worth noting that the WST identities have been explicitly verified in the straightforward LO approach detailed in this note.
Proving such identities in any LO approach verifies the gauge-independence of the results.

Extending any gauge fixing procedure to the case of the SMEFT is subtle, due to the order
by order redefinition of the fields that are gauged due to terms in ${\cal L}_{SMEFT}$.
Optimally resolving the technical complications that result is a challenge.
These subtleties are some of the reasons it is difficult to directly
modify computer programs that have been developed for automatic NLO calculations in the SM, to the case of the SMEFT.
The development of NLO SMEFT Monte-Carlo tools is still very much a work in progress.

\subsection{Known results in the SMEFT to NLO}
Despite all of the challenges to advancing SMEFT results to NLO, progress in this area is rapid
and steady. In this section we briefly summarize some of these theoretical developments.

\subsubsubsection{Renormalization results}
The complete renormalization of the Warsaw basis was reported in
Refs.~\cite{Alonso:2013hga,Grojean:2013kd,Jenkins:2013zja,Jenkins:2013wua}.
In the approach outlined in Section \ref{inpractice}, results for the Warsaw basis
operator renormalization were reported in \Brefs{Passarino:2012cb,Ghezzi:2015vva}.
Use of SMEFT renormalization results (including a subset of NLO finite terms) to leverage
EWPD to bound operators not contributing at tree level was reported in Ref.~\cite{deBlas:2015aea}.
Partial results for renormalizing some alternate operator sets in a so called ``SILH basis''
were given in Refs.~\cite{Elias-Miro:2013eta,Elias-Miro:2013mua}. A recent study of RGE effects on the
oblique parameters, in a subset of UV models, was reported in Ref.\cite{Wells:2015cre}.

\subsubsubsection{Advances in one loop matching techniques}
Recently, the covariant derivative expansion discussed in Refs.~\cite{Cheyette:1985ue,Cheyette:1987qz,Gaillard:1985uh} has re-emerged
in Refs.\cite{Henning:2014wua,Drozd:2015kva,Drozd:2015rsp} as a powerful technique to perform matching calculations
to underlying UV theories at one loop. The basic idea at work is that, the contribution to the effective action that results when integrating out a heavy field $\mathcal{X}$
at one loop is schematically given by
\begin{eqnarray}
\Delta S \propto i \, {\rm Tr} \, \log \left[D^2+ m_\mathcal{X}^2 + U(x) \right]
\end{eqnarray}
where $m_\mathcal{X}$ is the mass of the $\mathcal{X}$ field integrated out, $D^2 = D_\mu \, D^\mu$, $D_\mu$ is the covariant derivative,
and $U(x)$ depends on the SM field content. The covariant derivative expansion allows this functional trace to be directly
evaluated, while keeping gauge covariance manifest. This simplifies and systematizes one loop matching calculations in the SMEFT,
in many simple UV physics cases.\footnote{It is worth noting, that some questions remain about the effect of mixing between the heavy and light field content in this approach \cite{Brehmer:2015rna}. These questions were recently clarified in \cite{delAguila:2016zcb,Fuentes-Martin:2016uol}.}

\subsubsubsection{Full Lagrangian expansion results to NLO \texorpdfstring{$(\mathcal{L}_8$)}{(L_8)}}

Refs.\cite{Lehman:2014jma,Henning:2015alf,Lehman:2015coa,Lehman:2015via,Henning:2015daa}
have developed the theoretical technology (essentially advanced use of Hilbert series techniques) to characterize the number of
independent operators present at each order in the SMEFT expansion. This has lead to the complete characterization of the operator
sets in $\mathcal{L}_7$ and $\mathcal{L}_8$.

\subsubsubsection{Perturbative NLO results in the SMEFT}

Full results to NLO in the SMEFT have started to appear in the literature. The first pioneering calculations of this
form were for the process $\mu \rightarrow e \, \gamma$ in Ref.\cite{Pruna:2014asa} and for the process
$\Gamma(\PH \rightarrow \PGg \, \PGg)$ in Refs.~\cite{Hartmann:2015oia,Hartmann:2015aia,Ghezzi:2015vva}.
In \cite{Hartmann:2015aia} the full NLO perturbative SMEFT result for
this decay with no assumption
in the underlying UV scenario was reported. Ref.~\cite{Ghezzi:2015vva} also reported NLO results for
$\Gamma(\PH \rightarrow \PZ \, \PGg)$, $\PH \rightarrow \PZ \, \PZ^\star$,
$\PH \rightarrow \PW \, \PW^\star$ under the assumption of a PTG scenario and presented
results to NLO for the $\PW$ mass and other EWPD parameters. Recently Ref.~\cite{Gauld:2015lmb}
also reported NLO perturbative results for $\PH\rightarrow \PAQb\PQb$ and
$\PH \rightarrow \PGtm\PGtp$  in the general SMEFT, including finite terms, in the
large $m_{\PQt}$ limit.  NLO QCD results for a set of higher dimensional operators contributing to the Higgs boson pair production process
were given in Ref.\cite{Grober:2015cwa}, for the Higgs characterization model in Ref.\cite{Artoisenet:2013puc} and for associated Higgs boson production in
Ref.\cite{Mimasu:2015nqa}.

\subsubsection{A study of constraints}
As a particular example, we
discuss the impact of NLO corrections on inferred LO bounds, in the case of
$\Gamma(\PH \rightarrow \PGg \, \PGg)$, using the results of
Refs.~\cite{Hartmann:2015oia,Hartmann:2015aia}.
We consider the general SMEFT case, consider unknown $\tilde{C}_i \sim 1$ and vary the unknown parameters
over $0.8 \leq \Lambda \leq 3$ in ${\rm TeV}$ units.
Note that $\bar{v}_T^2/(0.8 \, {\rm TeV})^2 \sim 0.1$. Taking $\kappa_{\gamma}$ from
Ref.~\cite{Aad:2015gba} to be $0.93^{+0.36}_{-0.17}$,
and neglecting light fermion ($m_{\Pf} < m_{\PH}$) effects for simplicity, one finds
the $1 \, \sigma$ range
\begin{eqnarray}\label{masternumerics}
-0.02 \, \leq \left(\tilde{C}^{1,NP}_{\PGg\,\PGg} + \frac{\tilde{C}^{NP}_i\,f_i}{16\,\pi^2}\right)
\, \frac{\bar{v}^2_T}{\Lambda^2} \leq 0.02 \,.
\end{eqnarray}
Here, the tilde superscript denotes that the scale $1/\Lambda^2$ has been factored out of a
Wilson coefficient. The $f_i$ terms correspond to the ``nonfactorizable'' terms, and $\tilde{C}^{1,NP}_{\PGg\,\PGg}$
corresponds to the one loop improvement of the Wilson coefficient that gives this decay at tree level -- $\tilde{C}^{0,NP}_{\PGg \, \PGg}$.
The difference in the mapping of this constraint to the coefficient of $\tilde{C}^{0,NP}_{\PGg \, \PGg}$
at tree level, and at one loop, can now be characterized.

To determine this correction we determine the percentage change on the inferred
value of the bounds of $\tilde{C}^{0,NP}_{\PGg \, \PGg}$, while shifting the quoted upper and
lower experimental bounds by the NLO SMEFT perturbative correction. The
envelope of the two percentage variations on the bounds is quoted in the form $\left[ ,\right]$,
for values of $\Lambda$ varying from $\left[0.8, \, 3\right] {\rm TeV}$.
For one specific choice of signs for $C_i$, we find the following characteristic results.
The net impact of one-loop corrections (added in quadrature) due to higher dimensional
operators on the bound of the tree level Wilson coefficient  is
\begin{eqnarray}
\Delta_{\text{quad}} \,  \tilde{C}^{0,NP}_{\PGg \, \PGg} \sim \left[29, 4 \right]  \% \,.
\end{eqnarray}
Similarly, CMS reports $\kappa_{\gamma}= 0.98^{+0.17}_{-0.16}$ \cite{Khachatryan:2014jba},
which gives
\begin{eqnarray}
\Delta_{\text{quad}} \,  \tilde{C}^{0,NP}_{\PGg \, \PGg} \sim \left[52, 7 \right]  \% \,.
\end{eqnarray}
It is possible that these corrections could add up in a manner that is not in quadrature, as
this depends on the unknown $\tilde{C}_i$ values. The impact of the one-loop corrections listed
above is on {\it current} experimental bounds of $\Gamma(\PH \rightarrow \PGg \PGg)$, following from our conservative
treatment of unknown UV effects.
As the experimental precision of the measurement of $\Gamma(\PH \rightarrow \PGg \PGg)$
increases, the impact of the neglected corrections directly scales up. Repeating the exercise
above, with a chosen projected RunII value $\kappa_{\gamma} = 1 \pm 0.045$ which is consistent with projected
future bounds (CMS - scenario II~\cite{Flechl:2015foa,CMS:2013xfa})
\begin{eqnarray}
(\Delta_{\text{quad}} \, \tilde{C}^{0,NP}_{\PGg \, \PGg})^{\text{proj:RunII}} \sim
\left[167, 21 \right]  \% \,.
\end{eqnarray}
High luminosity LHC runs are further quoted to have a sensitivity between $2 \%$ and $5 \%$ in
$\kappa_{\gamma}$ \cite{Dawson:2013bba}. Choosing a value
$\kappa_{\gamma} = 1 \pm 0.03$ for this case, one finds
\begin{eqnarray}
(\Delta_{\text{quad}} \, \tilde{C}^{0,NP}_{\PGg \, \PGg})^{\text{proj:HILHC}} \sim
\left[250, 31 \right]  \% \,.
\end{eqnarray}
Neglected one loop corrections can have an important effect on the projection of an
experimental bound into the LO SMEFT formalism, when measurements become sufficiently precise and
the cut off scale is not too high. \footnote{Editor footnote: 
A different conclusion is presented in Section~\ref{s.eftval} that argues that the theoretical uncertainty of the EFT computation decreases when the bounds on the deviations from the SM predictions improve.}

\subsubsection{A study of SM-deviations}
Here the reference process is the off-shell $\Pg\Pg \to \PH$ production.
It is important to go off-shell because the correct use of the SMEFT proves that scaling couplings
on a resonance pole is not the same thing as scaling them off of a resonance pole, which has important
consequences in bounding the Higgs intrinsic width, see
\Brefs{Englert:2015bwa,Englert:2015zra,Ghezzi:2014qpa}.

In the $\upkappa\,$ approach, which was developed out of Refs.\cite{Azatov:2012bz,Espinosa:2012ir,Carmi:2012yp}, and formalized in
Ref.~\cite{LHCHiggsCrossSectionWorkingGroup:2012nn}, one writes the amplitude as
\begin{equation}
{\mathrm{A}}^{{\scriptscriptstyle{\Pg\Pg}}} = \sum_{\PQq=\PQt,\PQb}\,\upkappa^{{\scriptscriptstyle{\Pg\Pg}}}_{\PQq}\,
{\cal A}^{{\scriptscriptstyle{\Pg\Pg}}}_{\PQq} + \upkappa^{{\scriptscriptstyle{\Pg\Pg}}}_c
\label{kappafact}
\end{equation}
${\cal A}^{{\scriptscriptstyle{\Pg\Pg}}}_{\PQt}$ being the SM $\PQt\,$-loop \etc The contact term (which is the
LO SMEFT) is given by $\upkappa^{{\scriptscriptstyle{\Pg\Pg}}}_c$. Furthermore
$\upkappa^{{\scriptscriptstyle{\Pg\Pg}}}_{\PQq} = 1 + \Delta\,\upkappa^{{\scriptscriptstyle{\Pg\Pg}}}_{\PQq}$
Next we compute the following ratio
\begin{equation}
{\mathrm{R}} =
\sigma\lpar \upkappa^{{\scriptscriptstyle{\Pg\Pg}}}_{\PQq}\,,\,\upkappa^{{\scriptscriptstyle{\Pg\Pg}}}_c \rpar/\sigma_{\mySM} - 1 \quad  [\%]
\,.
\end{equation}

In LO SMEFT $\upkappa_c$ is non-zero and $\upkappa_{\PQq} = 1$. One measures a deviation and gets
a value for $\upkappa_c$.  However, at NLO $\Delta \upkappa_{\PQq}$ is non zero and one gets a
degeneracy: the interpretation in terms of $\upkappa^{\rm{\scriptscriptstyle{LO}}}_c$ or in terms of
$\{\upkappa^{\rm{\scriptscriptstyle{NLO}}}_c, \Delta \upkappa^{\rm{\scriptscriptstyle{NLO}}}_{\PQq}\}$ could be rather different
(we show an example in \refF{HAApdf}). Going interpretational we consider
\begin{equation}
{\mathrm{A}}^{{\scriptscriptstyle{\Pg\Pg}}}_{\rm{\scriptscriptstyle{SMEFT}}} =
\frac{g\,g_3}{\pi^2}\,\sum_{\PQq=\PQt,\PQb}\,\upkappa^{{\scriptscriptstyle{\Pg\Pg}}}_{\PQq}\,{\cal A}^{{\scriptscriptstyle{\Pg\Pg}}}_{\PQq}
+   2\, g_3 \,g_{_6}\,\frac{s}{\mws}\, \tilde{C}_{\PH\,\Pg}  +
    \frac{g\,g_3\,g_{_6}}{\pi^2}\,\sum_{\PQq=\PQt,\PQb}\,{\cal A}^{{\mbox{\scriptsize nfc}}\,;\,{\scriptscriptstyle{\Pg\Pg}}}_{\PQq}\,
      \tilde{C}_{\PQq \Pg}
\label{SMEFTfact}
\end{equation}
where $g_3$ is the $SU(3)$ coupling constant. Using \eqn{SMEFTfact} we adopt the Warsaw basis
and eventually work in the (PTG) scenario~\cite{Arzt:1994gp,Einhorn:2013kja}. The following options are available:
LO SMEFT: $\upkappa_{\PQq} = 1$ and $\tilde{C}_{\PH\,\Pg} $ is scaled by $1/16\,\pi^2$
being ``loop-generated'' (LG);
NLO PTG-SMEFT: $\upkappa_{\PQq} \not= 1$ but only PTG operators
inserted in loops (non-factorizable terms absent), $\tilde{C}_{\PH\,\Pg}$  scaled as above;
NLO full-SMEFT: $\upkappa_{\PQq} \not= 1$ LG/PTG operators
inserted in loops (non-factorizable terms present), LG coefficients scaled as above.
Again we note the PTG  classification scheme is not valid for all
possible UV.

It is worth noting the difference between \eqn{kappafact} and \eqn{SMEFTfact}, showing that the
original $\upkappa\,$-framework can be made consistent at the price of adding
``non-factorizable'' sub-amplitudes. At NLO, $\Delta \upkappa = g_{_6}\,\rho$ and
\arraycolsep 0.14em\begin{eqnarray}
g^{-1}_{_6} = \sqrt{2}\,G_{F}\,\Lambda^2
&\qquad&
4\,\pi\,\alphas = g_3   \\
\rho^{{\scriptscriptstyle{\Pg\Pg}}}_{\PQt} = \tilde{C}_{\PH\,\PW} + \tilde{C}_{\PQt\PH}  + 2\, \tilde{C}_{\PH\,\Box} -
\frac{1}{2}\, \tilde{C}_{\PH\,\PD} 
&\qquad&
\rho^{{\scriptscriptstyle{\Pg\Pg}}}_{\PQb} = \tilde{C}_{\PH\,\PW} - \tilde{C}_{\PQb\PH}+ 2\, \tilde{C}_{\PH\,\Box} -
\frac{1}{2}\, \tilde{C}_{\PH\,\PD} \,. \nn \\
\end{eqnarray}
Relaxing the PTG assumption introduces non-factorizable sub-amplitudes proportional to
$\tilde{C}_{\PQt\PH} , \tilde{C}_{\PQb\PH} $ with a mixing among $\tilde{C}_{\PH\,\Pg}, \tilde{C}_{\PQt \Pg},
\tilde{C}_{\PQb \Pg}$. Meanwhile, renormalization
has made one-loop SMEFT finite, \eg in the $G_{F}\,$-scheme, with a residual
$\muR\,$-dependence.

We allow each Wilson coefficient to vary in some interval ${\mathrm{I}}_n = [-n\,,\,+n]$ and fix a
value for $\Lambda$. Next we generate points from ${\mathrm{I}}_n$ for the Wilson coefficients
with uniform probability and calculate ${\mathrm{R}}$. Finally, we calculate the ${\mathrm{R}}$ probability
distribution function (pdf), as shown in Figs.~\ref{figurepdf2},\ref{figurepdf}.

As another example, a comparison between the LO pdf and NLO pdf for $\PH \to \PGg\PGg$
using the approach of this section, and the results in \cite{Ghezzi:2015vva}, is shown in \refF{HAApdf}.

\begin{figure}[t]
\includegraphics[scale=0.37]{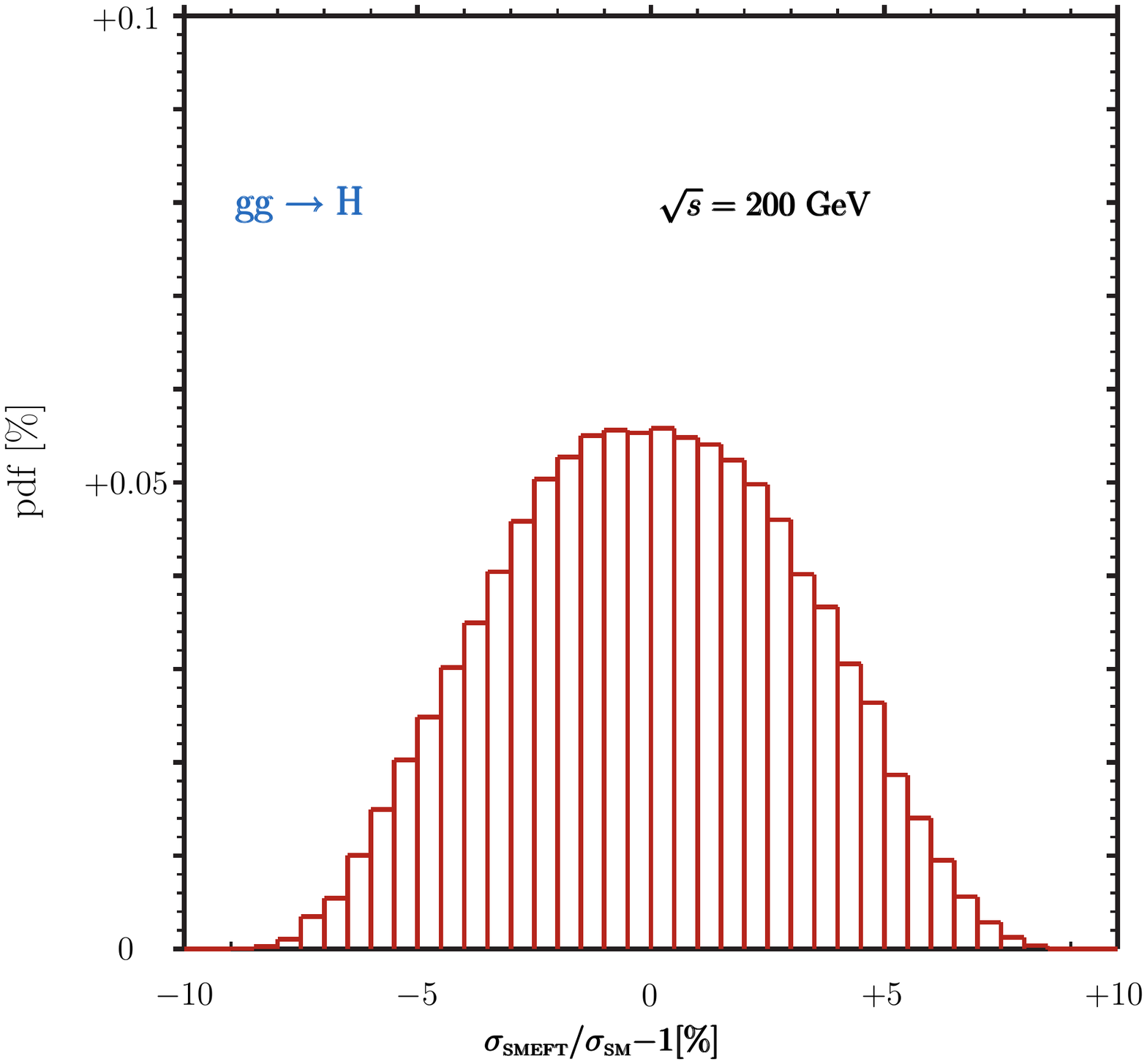}
\includegraphics[scale=0.37]{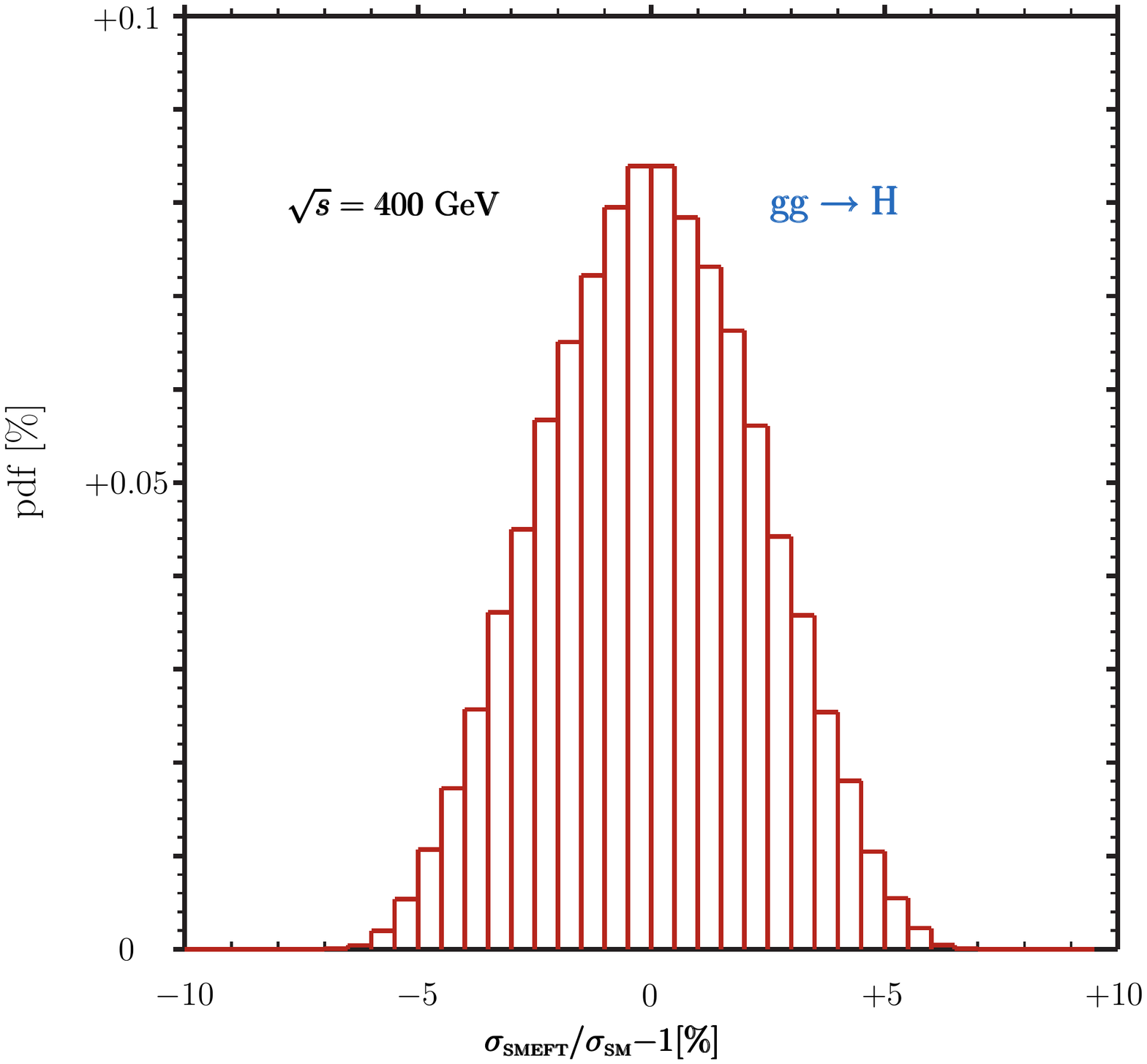}
\caption{Probability distribution function for the off-shell process $\Pg\Pg \to \PH$. Support is $C_i\;\in\;[-1\,,\,+1]$
with a uniform prior, and we have set $\Lambda = 3 \, {\rm TeV}$.}
\label{figurepdf2}
\end{figure}

\begin{figure}[t]
\includegraphics[scale=0.32,angle=90]{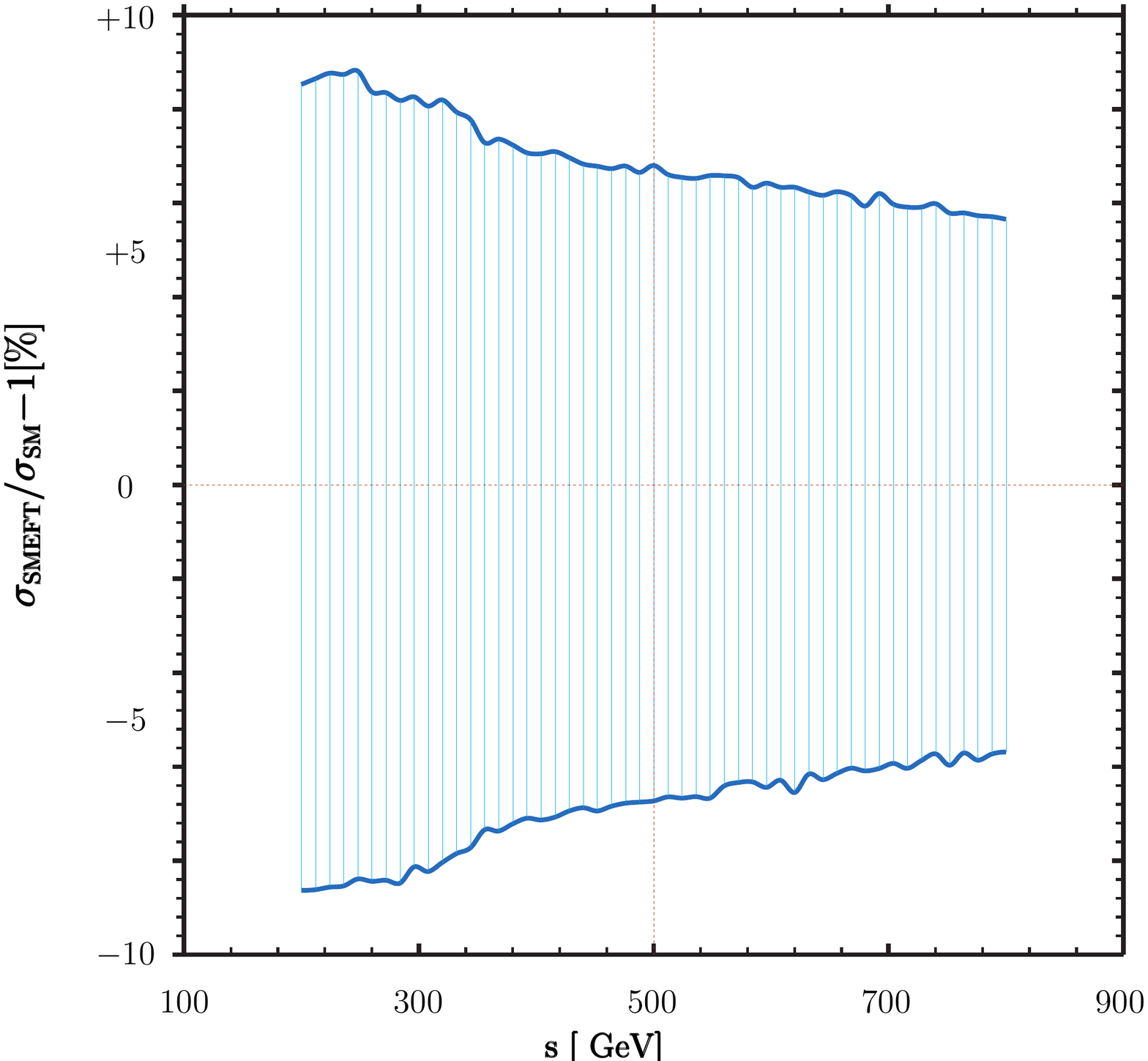}
\put(-15,-45){\includegraphics[scale=0.33]{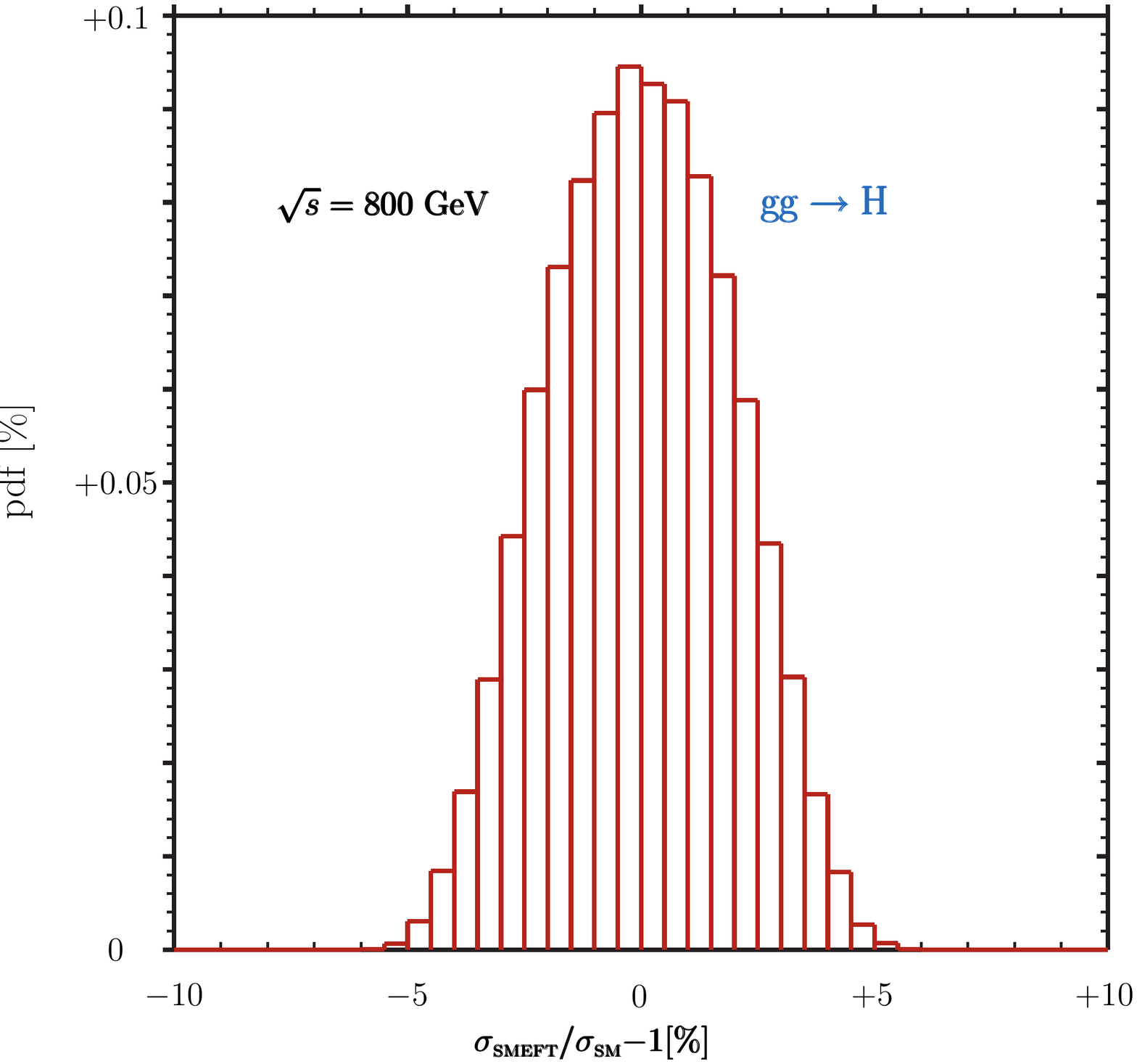}}
\caption{Probability distribution function for the off-shell process $\Pg\Pg \to \PH$. Support is $C_i\;\in\;[-1\,,\,+1]$
with a uniform prior, and we have set $\Lambda = 3 \, {\rm TeV}$.}
\label{figurepdf}
\end{figure}

\begin{figure}[t]
   \centering
   \includegraphics[width=0.7\textwidth, trim = 30 250 50 80, clip=true]{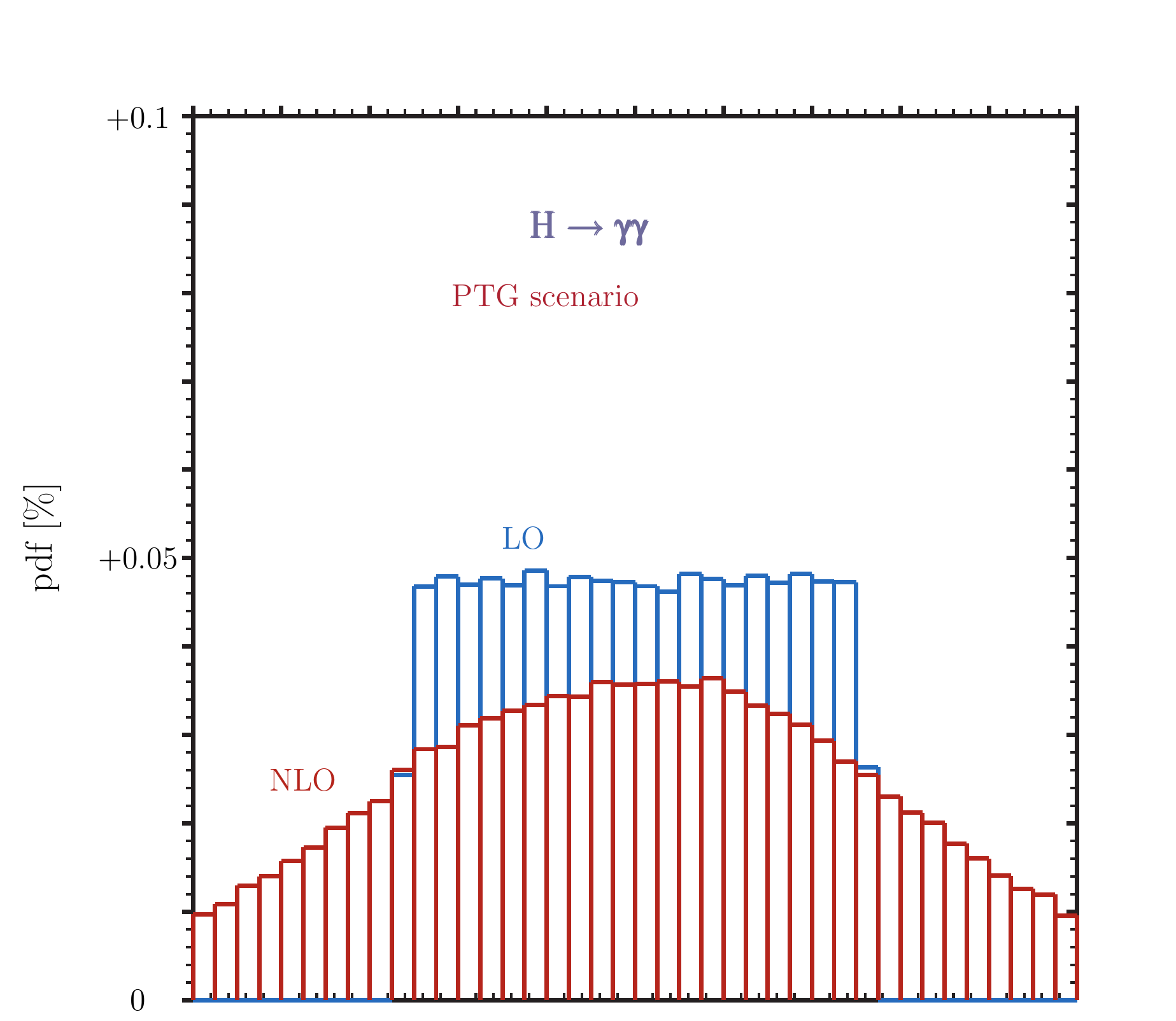}
\caption[]{Probability distribution function for the decay $\PH \to \PGg\PGg$ with a comparison
between the LO and the NLO predictions. Here $\Lambda = 3\UTeV$ and $n = 1$. X axis as in previous figures.}
\label{HAApdf}
\end{figure}

\subsubsection{Comments on Pole observables vs tails of distributions}

When analysing data near poles, scaling arguments that apply to the suppression of local contact (non resonant) four fermion operators in ${\cal L}_6$
also apply to NLO ${\cal L}_8$ corrections. This is fortunate as the very large number of parameters present in
${\cal L}_8$ and ${\cal L}_6$ are primarily present in four fermion operators. In the
case of ${\cal L}_6$ $2205$ of the $2499$ parameters present are due to four fermion operators \cite{Alonso:2013hga}.  NLO power corrections in ${\cal L}_8$, higher order in $(v/\Lambda)^m$, are suppressed compared to ${\cal L}_6$ by the
power counting parameter $v^2/\Lambda^2$,which
varies from $\sim 6\%$ to $\sim 0.6\%$ for $\Lambda/\sqrt{\tilde{C_i}} = 1,3 \, {\rm TeV}$ respectively.

The suppression of NLO terms in the Lagrangian expansion that scale as $p^2/\Lambda^2$ can be far less in the tails of distributions\footnote{See for example discussion in Ref.\cite{Ellis:2012xd,Isidori:2013cga,Biekoetter:2014jwa}.}.
Tails of distributions can also have a very large number of  SMEFT parameters contributing due to non-resonant fermion pair (and higher multi-body) production background
processes. The SMEFT
expansion breaks down when $p^2/\Lambda^2 \sim 1$, and Pseudo Observable/form factor \cite{David:2015waa,Passarino:2010qk,Gonzalez-Alonso:2014eva,Isidori:2013cla,Isidori:2013cga}
methods are required to characterize the data in this case.  In doing so, it is appropriate to bin the data in a manner that is transparent as to the momentum scale being probed.

It is also worth noting that unlike the case of pole data, NLO corrections to tails of distributions
are complicated in their analysis, as the $p^2/\Lambda^2$ terms are in general not gauge invariant alone, and need to always be combined
with the interference with non-resonant part of the SM, and SMEFT background processes. The requirement for joint analysis
including SMEFT corrections on the background that results, further complicates the analysis of non-pole data.

\subsection{Summary  and comments}

{We summarize here the main points discussed and advocated in this note.}

\begin{itemize}
\item{{NLO results have already had an important impact on the SMEFT physics program. For example, it has been shown
{that the inclusion} of these effects may relax, in some cases,
the bounds on some EFT parameters from  $\mathcal{O}(10^{-3})$ to $\mathcal{O}(10^{-2})$.  This is
why we advocate not to use LEP constraints to set to zero effective SMEFT parameters in  ${\cal L}_6$, or combinations of such parameters for vector boson couplings to fermions in LHC analyses. In general, care should be used when fixing combinations of parameters from EW constraints in LHC analyses. For example, currently, $\mathcal{O}(10^{-3})$ bounds are based on LO SMEFT analyses without any theoretical error assigned.}}

\item{ It is important to preserve the original data, not
just the interpretation results, as the estimate of the missing higher order terms can change over time,
modifying the lessons drawn from the data and projected into the SMEFT.}

\item{Overall, the neglect of NLO (perturbative EW) corrections, considering the precision of LHC RunI measurements,
is (retrospectively) justified in most channels. On the other hand, NLO QCD corrections are not negligible, even in RunI.
{However, considering projections for the precision to be reached in LHC RunII analyses,
we believe that the LO approach may not be sufficient. This may in particular be the case if the cut off scale is in the few TeV range.}}

\item{NLO results are starting to become available in the SMEFT. These results allow the
consistent interpretation of the data combining measurements at different scales, and can
robustly accommodate the precision projected to be achieved in RunII analyses, even for lower
cut off scales.}

\item{
{In a sense, the SMEFT allow the kappa-framework~\cite{LHCHiggsCrossSectionWorkingGroup:2012nn} to be extended/replaced,
and NLO results are crucial in this respect.}
The idea is that interpretations can transition to the linear SMEFT, which is a systematically improvable  EFT formalism.
NLO results more consistently include kinematic deviations from the SM,
and define higher order calculations in relation to a measured observable, in a well defined field theory.
A properly formulated SMEFT is not limited to LO and can include
QCD and EW corrections.}

\item{
The assignment of  a theoretical error for LO SMEFT analyses is, in general, important. In our opinion, this
is essential if the cut off
scale is assumed to be in the ``interesting range''
$1\, {\rm TeV} \lesssim \Lambda /\sqrt{\tilde{C_i}}  \lesssim 3\, {\rm TeV}$ and the experimental precision of analyses descends below the $10\%$ level.
The exact size of NLO corrections depends on the particular UV model, which is unknown, and also the particular channel
analysed.}

\item{ We do not advocate absorbing the effects of $\mathcal{L}_8$  corrections and/or or absorbing logarithmic NLO perturbative corrections into an ``effective'' parameter to attempt to incorporate NLO correction.
Such a redefinition cannot simultaneously be made in different measurements
generally measured at different scales. Correlating different measurements
is necessary if the SMEFT is to be used in a predictive fashion for constraints on LHC measurements.}

\item{We think that the experimental collaborations should restrict the bulk of their efforts to
defining and reporting clean measurements that can be interpreted in any well defined basis in the SMEFT.
The focus for data reporting should be on fiducial cross sections and/or pseudo-observables.
If a LO interpretation of the data in the SMEFT is reported there is no barrier to using the straightforward LO formalism of the Warsaw basis discussed in this note. This approach is convenient and well defined.}

\end{itemize}

We have supplied the outline and details of a LO implementation in Section \ref{rotation}.
{We believe that the adoption of this approach for LO fits may be advantageous as it is more readily extendable to NLO.}
We have sketched out how fits can be pursued at LO and NLO in a consistent fashion using this formalism.
The approach presented is well defined, is not intrinsically tied to a particular IPS, can be informed by theoretical errors determined at NLO and can be directly improved to NLO.
The gauge invariance of the approach presented has been checked at NLO by explicit confirmation of the WST identities.

We have stressed the standard usage of EFT terminology in this discussion, in particular the definition of an operator basis, to clarify discussion on these issues.
EFT is traditionally a very successful paradigm to use to interpret the data because it is implemented as a well defined field theory.
{Standard EFTs can be systematically improved from LO to NLO and we think that severe caution should be exercised
when considering approaches that are that are not constructed in such a standard manner.}

\section[Non-linear EFT]{Non-linear EFT\SectionAuthor{I.~Brivio, G.~Buchalla, O.~Cata, A.~Celis, R.L.~Delgado, A.~Dobado, D.~Espriu, M.~Herrero, C.~Krause, F.J.~Llanes-Estrada, L.~Merlo, J.J.~Sanz-Cillero}}
\label{s.nleft}
\subsection{Motivation and leading-order Lagrangian}

The following section describes the chiral-Lagrangian framework,
in which electroweak symmetry breaking is nonlinearly realized,
as an effective field theory (EFT) for physics beyond the Standard Model (SM).
The motivation for and the main properties of this approach, in particular
for describing anomalous Higgs boson couplings, will be reviewed.
The connection with the more common EFT based on power counting by
canonical dimension (SM + dimension-6 operators, sometimes referred to as
SMEFT) will also be discussed.
We start with a phenomenologically oriented introduction, which will be
followed by a systematic formulation of the nonlinear EFT.

A central goal of the LHC after the discovery of the Higgs boson will be
a more comprehensive investigation of its properties in order to test
the underlying dynamics of electroweak symmetry breaking.
At present, the Higgs boson couplings to gauge bosons and top quarks are
compatible with the SM, but deviations of ${\cal O}(10\%)$ are still
possible \cite{Khachatryan:2016vau}. For the couplings to other fermions, or the
triple-Higgs boson coupling, even larger effects are not excluded.
Anomalous Higgs boson couplings have the potential to give much larger
effects than new physics in electroweak gauge interactions, which is
typically constrained to the ${\cal O}(1\%)$ level by
electroweak precision measurements \cite{ALEPH:2005ab}.

It then appears natural to focus the attention, in a first step,
on the couplings of the Higgs particle. This goal is also well motivated
by the foreseeable precision at the LHC with 300 fb$^{-1}$, projected to
reach several per cent accuracy for the Higgs boson couplings to gauge bosons and
heavy fermions \cite{CMS:2013xfa}.

Following this line of reasoning, one is led to consider a generalization
of the SM, in which the gauge interactions are unchanged (at leading order),
but general
anomalous couplings are introduced for the physical Higgs boson.
To do this in a consistent, gauge-invariant way, the scalar fields have
to be decomposed into the three Goldstone fields $\varphi^a$, described by
\begin{equation}\label{udef}
U = \exp(2i\varphi^a T^a/v)
\end{equation}
where $T^a$ are the generators of $SU(2)$ with normalization
${\rm Tr}[T^a T^b]=\delta^{ab}/2$,
and the physical Higgs field $h$. This corresponds to a decomposition of
the usual Higgs doublet $\phi_i$, $\tilde\phi_i=\varepsilon_{ij}\phi^*_j$,
into polar coordinates
\begin{equation}\label{phipolar}
\sqrt{2}(\tilde\phi,\phi)\equiv (v+h)U
\end{equation}
Under electroweak gauge transformations $SU(2)_L\times U(1)_Y$
\begin{equation}
U\to g_L U g^\dagger_Y,\qquad  h\to h
\end{equation}
such that $h$ is invariant, and its couplings can be consistently
modified.\footnote{The generic name of ``nonlinear'' comes from the fact that
the scalar sector of the SM has a larger symmetry $SU(2)_L\times SU(2)_R$
(usually called chiral EW symmetry), under which the EW Goldstone bosons
$\varphi^a$ in (\ref{udef}) transform nonlinearly, in contrast to the usual
Higgs doublet field, which transforms linearly. The relevant symmetry
breaking pattern in the scalar sector is then given by
$SU(2)_L\times SU(2)_R \to SU(2)_{L+R}$, where the $SU(2)_{L+R}$
is usually called the custodial symmetry group.}

The resulting generalized Lagrangian can be written as
\begin{eqnarray}\label{l2}
{\cal L}_2 &=& -\frac{1}{2} \langle G_{\mu\nu}G^{\mu\nu}\rangle
-\frac{1}{2}\langle W_{\mu\nu}W^{\mu\nu}\rangle
-\frac{1}{4} B_{\mu\nu}B^{\mu\nu}
+\sum_{\psi=q_L,l_L,u_R,d_R,e_R}\bar \psi i\!\not\!\! D\psi
\nonumber\\
&& +\frac{v^2}{4}\ \langle D_\mu U^\dagger D^\mu U\rangle\, \left( 1+F_U(h)\right)
+\frac{1}{2} \partial_\mu h \partial^\mu h - V(h) \nonumber\\
&& - v \left[ \bar q_L \left( Y_u +
       \sum^\infty_{n=1} Y^{(n)}_u \left(\frac{h}{v}\right)^n \right) U P_+ q_R
+ \bar q_L \left( Y_d +
     \sum^\infty_{n=1} Y^{(n)}_d \left(\frac{h}{v}\right)^n \right) U P_- q_R
  \right. \nonumber\\
&& \quad\quad\left. + \bar l_L \left( Y_e +
   \sum^\infty_{n=1} Y^{(n)}_e \left(\frac{h}{v}\right)^n \right) U P_- l_R
+ {\rm h.c.}\right]
\end{eqnarray}
where
\begin{equation}\label{dcovu}
D_\mu U=\partial_\mu U+i g W_\mu U -i g' B_\mu U T_3,
\end{equation}
and  $P_\pm = 1/2\pm T_3$.
The trace of a matrix $A$ is denoted by $\langle A\rangle$. The left-handed
doublets of quarks and leptons are written as $q_L$ and $l_L$, the
right-handed singlets as $u_R$, $d_R$, $e_R$. Generation indices are omitted.
In the Yukawa terms the right-handed quark and lepton fields are collected
into $q_R=(u_R,d_R)^T$ and $l_R=(\nu_R,e_R)^T$, respectively. In general,
different flavour couplings $Y^{(n)}_{u,d,e}$ can arise at every order in
the Higgs field $h^n$, in addition to the usual Yukawa matrices $Y_{u,d,e}$.
The detailed assumptions underlying (\ref{l2}) are summarized in points
(i) -- (iii) below.

The first line in (\ref{l2}) represents the unbroken SM and the remaining
lines describe the sector of electroweak symmetry breaking.
The $h$-dependent functions, analytic near zero field, are
\begin{equation}\label{fuv}
F_U(h)=\sum^\infty_{n=1} f_{U,n} \left(\frac{h}{v}\right)^n,\qquad
V(h)=v^4\sum^\infty_{n=2} f_{V,n} \left(\frac{h}{v}\right)^n
\end{equation}
In addition to modifying the Higgs boson couplings present in the SM, new couplings
with higher powers in the field $h$ are introduced.
All these couplings may deviate, in principle, by corrections of
${\cal O}(1)$ from their (dimensionless) SM values.
For smaller deviations, the Lagrangian in (\ref{l2}) continues to describe
the leading new-physics effects, as long as the anomalous couplings in the
Higgs sector dominate over other corrections from physics beyond the SM.
(Those would be represented by operators of
chiral dimension 4 and higher, see the discussion of power counting below.)

While ${\cal L}_2$ in (\ref{l2}) is gauge invariant, it is no longer
renormalizable for general Higgs boson couplings. Renormalizability would be
recovered in the SM limit where
\begin{equation}\label{smlimit}
f_{U,1}=2,\quad f_{U,2}=1,
\quad f_{V,2}=f_{V,3}=\frac{m^2_h}{2v^2}, \quad f_{V,4}=\frac{m^2_h}{8v^2},
\quad Y^{(1)}_f=Y_f,
\end{equation}
and all other couplings $f_{U,n}$, $f_{V,n}$, $Y^{(n)}_f$ equal to zero.
In this limit (\ref{l2}) is just the SM written in somewhat unconventional
variables.
All $S$-matrix elements are of course identical to the ones obtained with the
familiar linear Lagrangian.

If the deviations of the couplings from their SM values are smaller
than unity, it is useful to parameterize them by
a quantity $\xi\equiv v^2/f^2$, where $f > v$ represents a new scale
(which could be related to a new strongly interacting dynamics).
In models of a composite, pseudo-Goldstone Higgs \cite{Kaplan:1983fs,Kaplan:1983sm,Banks:1984gj,Dugan:1984hq,Agashe:2004rs,Contino:2006qr,Contino:2010rs,Falkowski:2007hz,Carena:2014ria,Alonso:2014wta,Hierro:2015nna} $f$ corresponds to the
Goldstone-boson decay constant. Experimentally, values of $\xi={\cal O}(10\%)$
are currently still allowed.

For general Higgs boson couplings, the Lagrangian ${\cal L}_2$, nonrenormalizable
in the traditional sense, is still renormalizable in the modern sense,
order by order in a consistent expansion \cite{Weinberg:1978kz}.
It therefore continues to serve
as a fully consistent effective field theory. This EFT is known as
the {\it electroweak chiral Lagrangian\/} including a light Higgs boson.
For the case without Higgs the electroweak chiral Lagrangian has been
formulated and applied in
\cite{Appelquist:1980vg,Longhitano:1980iz,Appelquist:1984rr,Cvetic:1988ey,Dobado:1989ax,Dobado:1989ue,Dobado:1989gr,Dobado:1990jy,Feruglio:1992wf,Dobado:1995qy,Dobado:1999xb,Dobado:1990zh,Espriu:1991vm,Espriu:2000fq,Alonso:2012jc,Buchalla:2012qq,Buchalla:2013wpa}.
The generalization to include a light Higgs boson has been developed in
\cite{Feruglio:1992wf,Bagger:1993zf,Koulovassilopoulos:1993pw,Burgess:1999ha,Wang:2006im,Grinstein:2007iv,Alonso:2012px,Espriu:2013fia,Buchalla:2013rka,Brivio:2013pma,Delgado:2013hxa,Buchalla:2013eza,Gavela:2014vra}.

Having motivated the basic structure of the electroweak chiral Lagrangian,
it is useful to summarize the most important assumptions that define it as a
systematic EFT. These concern the {\it particle content} below a certain
mass gap, the relevant {\it symmetries}, and the {\it power counting}:
\begin{description}
\item[(i)]
SM particle content, where (transverse) gauge bosons and fermions
are weakly coupled to the Higgs-sector dynamics.
\item[(ii)]
SM gauge symmetries;
conservation of lepton and baryon number; conservation
{\it at lowest order\/} of custodial symmetry in the strong sector,
CP invariance in the Higgs sector and fermion flavour.
The latter symmetries are violated at some level, but this
would only affect terms at subleading order.
Generalizations may in principle be introduced if necessary.
\item[(iii)]
Power counting by chiral
dimensions \cite{Urech:1994hd,Knecht:1999ag,Nyffeler:1999ap,Hirn:2004ze},
equivalent to a loop expansion \cite{Buchalla:2013eza}, with the simple
assignment of $0$ for bosons (gauge fields $X_\mu$, Goldstones $\varphi$ and
Higgs $h$) and $1$ for each derivative, weak coupling (e.g. gauge or Yukawa),
and fermion bilinear:
\begin{equation}\label{chidim}
[X_\mu, \varphi, h]_\chi = 0\, ,\qquad
[\partial_\mu, g, y, \psi\bar\psi]_\chi = 1
\end{equation}
The loop order $L$ of a term in the Lagrangian is equivalent
to its chiral dimension (or chiral order) $2L+2$.
\end{description}

Under these assumptions the expression in (\ref{l2}) follows
as the most general Lagrangian built from
terms of chiral dimension 2 (corresponding to loop-order $L=0$).
This is the systematic
basis for the leading-order electroweak chiral Lagrangian.

Functions $F(h)$ multiplying the Higgs or the fermion
kinetic terms can be removed by field redefinitions and are therefore
omitted in (\ref{l2}) \cite{Buchalla:2013rka,Giudice:2007fh}.

Note that the Higgs potential $V(h)$, being related to the light
Higgs boson mass $\sim m^2_h$, carries chiral dimension 2.
This is explicitly realized in models where the Higgs is a
pseudo-Goldstone and its potential is generated at one loop
(proportional to two powers of weak coupling, hidden in the
coefficients $f_{V,n}$) \cite{Agashe:2004rs,Contino:2006qr,Contino:2010rs,Falkowski:2007hz,Carena:2014ria}.

Expressions of the form $(\bar\psi\psi)^2 (h/v)^n$,
$\bar\psi\sigma_{\mu\nu}\psi X^{\mu\nu}(h/v)^n$,
$X_{\mu\nu}X^{\mu\nu} (h/v)^{n+1}$, $n\geq 0$,
where $\psi$ is a fermion and $X_{\mu\nu}$ a gauge field-strength tensor, might
superficially look like terms entering the Lagrangian at chiral dimension 2.
However, they represent local interactions arising from the (weak) coupling
of $\psi$ and $X$ to the new-physics sector, according to assumption
(i) above. The weak coupling associated
with $\bar\psi\psi$ or $X_{\mu\nu}$ carries chiral dimension. The operators
above then acquire a chiral dimension of at least 4, which eliminates
them from the leading-order Lagrangian \cite{Buchalla:2013eza}.

\subsection{Renormalization of the chiral Lagrangian}

As the electroweak chiral Lagrangian defines a consistent quantum field theory,
loop corrections can be systematically included.
For the case without Higgs field this has been discussed in detail in
\cite{Herrero:1992zq,Herrero:1993nc,Herrero:1994iu,Dittmaier:1995cr,Dittmaier:1995ee}.
The one-loop divergent parts arising from the scalar sector have recently
been also obtained in the chiral Lagrangian including the light 
Higgs boson\cite{Espriu:2013fia,Delgado:2013loa,Delgado:2014jda,Gavela:2014uta,Guo:2015isa,Alonso:2015fsp}.

At one-loop order, terms up to chiral dimension 4 need to be included and
the Lagrangian can be written as
${\cal L} = {\cal L}_2 + {\cal L}_4 + {\cal L}_{\rm GF} + {\cal L}_{\rm FP}$,
including also gauge-fixing and ghost terms.
In general, the leading-order approximation is given by the tree-level
amplitudes from ${\cal L}_2$. The next-to-leading order corrections consist
of the one-loop amplitudes with vertices from ${\cal L}_2$, together with
tree-level contributions to first order in ${\cal L}_4$.
The latter comprise new interactions, not present in ${\cal L}_2$, and
act as counterterms for the one-loop divergences.
In general, they may get contributions from heavy states with masses of
order $\Lambda$ that are integrated out in the EFT
\cite{Pich:2013fea,Cata:2014fna,Pich:2015kwa}. This pattern is known
from the chiral perturbation theory of pions. It is typical for the
systematics of a nonrenormalizable EFT. Explicit examples
are discussed in Section~\ref{subsec:applications}.

The local operators in ${\cal L}_4$ have been discussed for the bosonic
sector in \cite{Alonso:2012px}, a subset of the fermionic terms has been
considered in \cite{Alonso:2012pz}. A systematic presentation of the
complete basis of local operators in ${\cal L}_4$ can be
found in \cite{Buchalla:2013rka}.
Concentrating on the electroweak bosonic sector one has
(with $2a=f_{U,1}$, $b=f_{U,2}$ in (\ref{fuv}))
\begin{eqnarray}\label{l2l4}
{\cal L}_2 &=&
-\frac{1}{2} \langle{W}_{\mu\nu}{W}^{\mu\nu}\rangle -
\frac{1}{4} {B}_{\mu\nu} {B}^{\mu\nu}\nonumber\\
&& +\frac{v^2}{4}\left[%
  1 + 2a \frac{h}{v} + b \frac{h^2}{v^2} + \dots\right]
\langle D^\mu U^\dagger D_\mu U \rangle
 + \frac{1}{2} \partial^\mu h \, \partial_\mu h + \dots
\nonumber\\
\nonumber\\
{\cal L}_{4} &=&
  a_1 g' g\langle U T_3 {B}_{\mu\nu} U^\dagger {W}^{\mu\nu}\rangle
  + i a_2 g'  \langle U T_3 {B}_{\mu\nu} U^\dagger [V^\mu, V^\nu ]\rangle
  - i a_3 g \langle {W}_{\mu\nu}[V^\mu, V^\nu]\rangle
  \nonumber \\
&&
  + a_4 \langle V_\mu V_\nu\rangle \langle V^\mu V^\nu\rangle
  + a_5 \langle V_\mu V^\mu \rangle \langle V_\nu V^\nu \rangle
+\frac{e^2}{16\pi^2}c_{\gamma\gamma} \frac{h}{v} F_{\mu\nu} F^{\mu\nu}
+\frac{g^{hh}}{v^4}(\partial_\mu h\partial^\mu h)^2
 \nonumber \\
&&
+\frac{d^{hh}}{v^2}(\partial_\mu h\partial^\mu h)
\langle D_\nu U^\dagger D^\nu U\rangle
+\frac{e^{hh}}{v^2}(\partial_\mu h\partial^\nu h)
\langle D^\mu U^\dagger D_\nu U\rangle +\dots
\end{eqnarray}
where $V_\mu\equiv (D_\mu U)U^\dagger$
and $F_{\mu\nu}$ is the photon field strength.
Here only a subset of the operators in ${\cal L}_4$ has been displayed,
corresponding to those needed in the discussion below.
All operators that need to be included as counterterms are manifestly
custodially preserving, except for the custodial breaking from $U(1)_Y$.
This is so because the initial theory is custodially invariant
when Yukawas are neglected.

%


As a simple example for renormalization, consider the oblique $S$-parameter.
The first non-vanishing contribution to $S$ appears at NLO.
One finds that the one-loop amplitude is  UV--divergent
and needs to be renormalized by means of the NLO parameter $a_1$.
In the $\overline{MS}$ scheme one
obtains~\cite{Dobado:1999xb,Delgado:2014jda,Pich:2013fea}
\begin{eqnarray}
S &=& \, -\, 16 \pi a^r_1
\,\,\, +\,\,\, \frac{(1-a^2)}{12 \pi}\, \bigg( \,\frac{5}{6}\, +
\, \ln\frac{\mu^2}{m_h^2}\,\bigg)
\label{eq.S-prediction}
\end{eqnarray}
In this expression, the oblique parameter is defined with
the reference value $m_h^{\rm Ref}$ set to the physical Higgs boson mass~\cite{Peskin:1990zt}.
Since fermionic couplings to gauge bosons receive only NLO contributions
from new physics, fermion loops do not affect this result. Their impact
would be a NNLO effect..


Renormalization leads to a scale dependence of the coefficients.
In general, the relation between a given
renormalized chiral parameter $C^r(\mu)$ and the corresponding bare
pa\-ra\-me\-ter $C^{(B)}$ from the ${\cal L}_4$ Lagrangian (e.g. $a_1$),
together with the resulting $\mu$-dependence, is given by
\begin{equation}
\frac{dC^r}{d\ln{\mu}} = -\frac{\Gamma_C}{16\pi^2},
\qquad
C^r(\mu) = C^{(B)} + \frac{\Gamma_C}{32\pi^2}  \frac{1}{\hat{\epsilon}}
\end{equation}
where an $\overline{MS}$ subtraction of the UV divergence has been performed.
Here $1/\hat{\epsilon} = \mu^{-2\epsilon} (1/\epsilon-\gamma_E +\ln{4\pi})$,
with $D=4-2\epsilon$.

\begin{table}[!t]
\begin{center}
\caption{\small
Example of the renormalization structure. Running of some NLO
coefficients \cite{Espriu:2013fia,Delgado:2013loa,Delgado:2014jda}.}
\vspace{.2cm}
\label{tab:running}
\begin{tabular}{c|c|c|c|c|c}
\toprule
 $\Gamma_{a_1-a_2+a_3}$
& $\Gamma_{c_{\gamma\gamma}}$
&  $\Gamma_{a_1}$
& $\Gamma_{a_2-a_3}$
& $\Gamma_{a_4}$
& $\Gamma_{a_5}$ \\
\midrule
$0$ & $0$ & $-\frac{1}{6}(1-a^2)$ & $-\frac{1}{6}(1-a^2)$ &
 $\frac{1}{6}(1-a^2)^2$ & $\frac{1}{8}(b-a^2)^2+\frac{1}{12}(1-a^2)^2$ \\
\bottomrule
\end{tabular}
\end{center}
\end{table}

The running of the ${\cal L}_4$ parameters
$C=a_1,\, a_2,\, a_3, \, c_{\gamma\gamma}$ \cite{Delgado:2014jda}
(relevant e.g. for $\gamma\gamma\to w^a w^b$) and of $C=a_4,\, a_5$
(contributing to
$ZZ$~and~$W^+W^-$~scattering~\cite{Espriu:2013fia,Delgado:2013loa})
is shown in Table~\ref{tab:running}.
It is apparent that the $S$-parameter in (\ref{eq.S-prediction})
is independent of the renormalization scale $\mu$.

\subsection{Connection of chiral Lagrangian
to \texorpdfstring{$\kappa$}{kappa}-formalism}

The couplings of the leading-order Lagrangian in (\ref{l2}),
which are non-standard in general, are displayed in
\refF{fig:hcouplings}.
\begin{figure*}[t]
\begin{center}
\includegraphics[width=14cm]{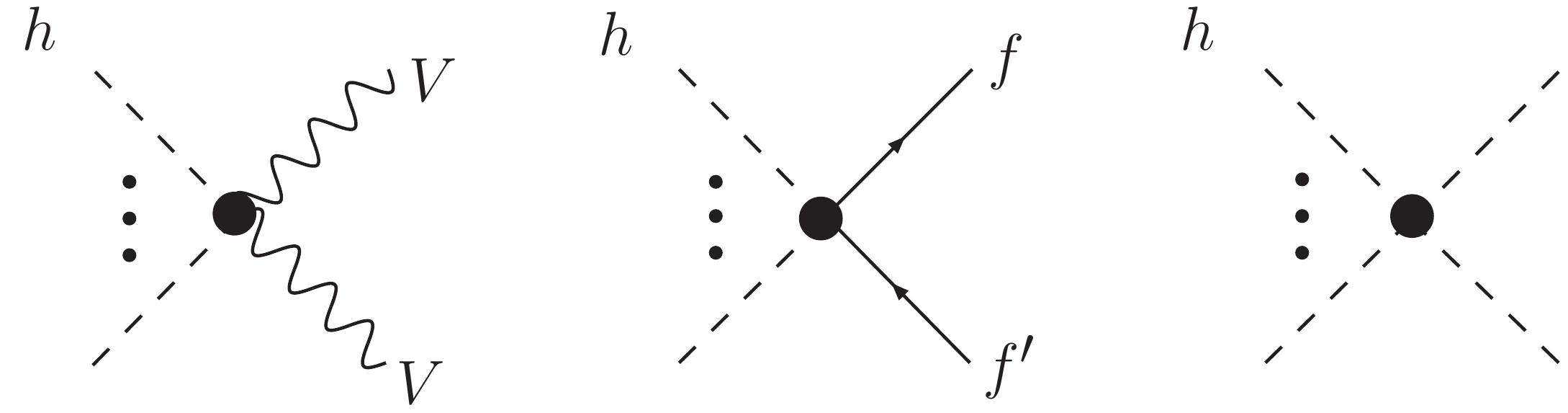}
\end{center}
\caption{The Higgs vertices from the leading-order Lagrangian ${\cal L}_2$
in unitary gauge.They are represented by a black dot and may deviate
sizably from the SM.
The pair of dashed lines with dots in between signifies any number of
Higgs lines. The massive vector bosons are denoted by $V=W,Z$.
$f=f'$ if flavour conservation is assumed to hold at leading order.
All other couplings are identical to the SM.}
\label{fig:hcouplings}
\end{figure*}
They parameterize the leading new-physics effects in tree-level processes.

A further consideration is needed for the application of the chiral
Lagrangian to processes that arise only at one-loop level in the SM.
Important examples are $h\to gg$, $h\to\gamma\gamma$ and $h\to Z\gamma$.
In this case local terms at NLO will also become
relevant, in addition to the standard loop amplitudes with modified
couplings from (\ref{l2}). The reason is that both contributions can lead to
deviations of the amplitude from the SM at the same order, $\sim \xi/16\pi^2$.
The complete list of NLO operators has first been worked out in
\cite{Buchalla:2013rka}. The terms that are relevant here are
\begin{equation}\label{hxx}
e^2F_{\mu\nu}F^{\mu\nu}h,\qquad e g' F_{\mu\nu}Z^{\mu\nu}h,\qquad
g^2_s\langle G_{\mu\nu} G^{\mu\nu}\rangle h
\end{equation}
On the other hand, the analogous terms $g'^2 Z_{\mu\nu}Z^{\mu\nu}h$
and $g^2 W^+_{\mu\nu} W^{-\mu\nu}h$ in the subleading Lagrangian
yield only subleading contributions, of ${\cal O}(\xi/16\pi^2)$,
to the tree-level amplitudes for $h\to ZZ$ and $h\to W^+W^-$, which receive
new-physics corrections of ${\cal O}(\xi)$ from (\ref{l2}).
They can thus be neglected in a first approximation
(see \cite{Buchalla:2015qju} for a discussion of generic NLO effects).

In summary, the Higgs boson couplings from NLO operators that are relevant for a
LO analysis of loop-induced processes are shown in \refF{fig:hnlo}.
\begin{figure*}[t]
\begin{center}
\includegraphics[width=14cm]{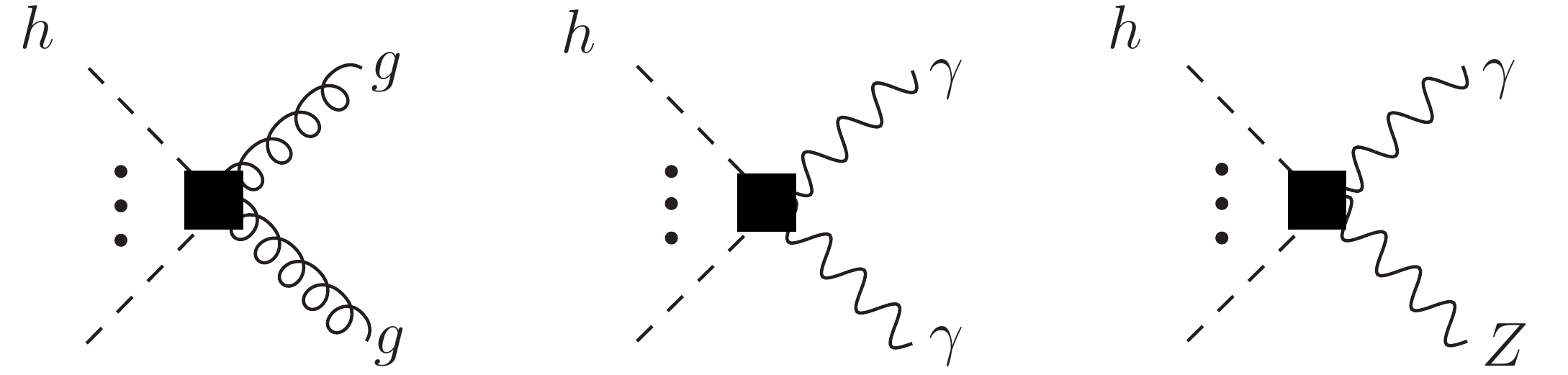}
\end{center}
\caption{Higgs vertices from the NLO Lagrangian ${\cal L}_4$, represented
by black squares, that contribute to $gg$, $\gamma\gamma$ and $Z\gamma$
amplitudes. Since the latter arise only at one-loop order from the
interactions of ${\cal L}_2$, the NLO couplings give relative corrections of
the same order in this case and have to be retained.}
\label{fig:hnlo}
\end{figure*}

\vspace*{0.4cm}

Based on the preceding discussion, one can now define anomalous Higgs boson couplings 
for specific classes of interactions, corresponding to the
leading order approximation within the chiral Lagrangian framework.

An important example are interactions involving a single Higgs field.
Focusing on these terms, and working in unitary gauge, (\ref{l2})
supplemented by the local NLO terms for $h\to\gamma\gamma$, $Z\gamma$ and
$gg$, implies  the interaction Lagrangian ($c_V\equiv a$ in (\ref{l2l4}))
\begin{equation}\label{llopar1}
\begin{array}{ll}
  \mathcal{L} &=2 c_{V} \left(m_{W}^{2}W_{\mu}^{+}W^{-\mu}
+\frac{1}{2} m^2_Z Z_{\mu}Z^{\mu}\right) \dfrac{h}{v} \vspace*{0.3cm} \\
&-\sum_{i,j} (y^{(1)}_{u,ij} \bar u_{Li}u_{Rj} + y^{(1)}_{d,ij} \bar d_{Li}d_{Rj}
  +y^{(1)}_{e,ij} \bar e_{Li}e_{Rj} + {\rm h.c.}) h \vspace*{0.3cm} \\
 &+ \dfrac{e^{2}}{16\pi^{2}} c_{\gamma\gamma} F_{\mu\nu}F^{\mu\nu} \dfrac{h}{v}
+ \dfrac{e g'}{16\pi^{2}} c_{Z\gamma} Z_{\mu\nu}F^{\mu\nu} \dfrac{h}{v}
+\dfrac{g_{s}^{2}}{16\pi^{2}} c_{gg}\langle G_{\mu\nu}G^{\mu\nu}\rangle\dfrac{h}{v}
\end{array}
\end{equation}
Neglecting flavour violation, the very small Yukawa couplings to light
fermions, and concentrating on those Higgs processes that have already become
accessible at the LHC, the parameterization reduces to a set of six anomalous
couplings, described by \cite{Buchalla:2015qju,Buchalla:2015wfa}
\begin{equation}\label{llopar2}
\begin{array}{ll}
  \mathcal{L} &=2 c_{V} \left(m_{W}^{2}W_{\mu}^{+}W^{-\mu}
+\frac{1}{2} m^2_Z Z_{\mu}Z^{\mu}\right) \dfrac{h}{v} -c_{t} y_{t} \bar{t} t h
- c_{b} y_{b} \bar{b}b h -c_{\tau} y_{\tau} \bar{\tau}\tau h  \\
 &+ \dfrac{e^{2}}{16\pi^{2}} c_{\gamma\gamma} F_{\mu\nu}F^{\mu\nu} \dfrac{h}{v}
+\dfrac{g_{s}^{2}}{16\pi^{2}} c_{gg}\langle G_{\mu\nu}G^{\mu\nu}\rangle\dfrac{h}{v}
\end{array}
\end{equation}
where $y_f=m_f/v$.
The SM at tree level is given by $c_{V}= c_{t} = c_{b}=c_{\tau} =1$ and
$c_{gg} =c_{\gamma\gamma} =0$. Deviations due to new physics are expected to
start at $\mathcal{O}(\xi)$.

A few important points should be emphasized:
\begin{description}
\item[(i)]
The parameterization of anomalous Higgs boson couplings in (\ref{llopar2})
essentially corresponds to the {\it $\kappa$-formalism\/}
\cite{Heinemeyer:2013tqa}, which is frequently used in experimental
analyses. Here, (\ref{llopar2}) has been derived from the electroweak
chiral Lagrangian.
\item[(ii)]
The minimal version in (\ref{llopar2}) can be generalized to include
more of the couplings contained in (\ref{llopar1}), such as
$h\to Z\gamma$, $h\to\mu\mu$, or the lepton-flavour violating $h\to\tau\mu$.
\item[(iii)]
The treatment can be further extended, for instance to 
double-Higgs boson production, where additional couplings with two or three $h$-fields
from (\ref{l2}) need to be considered
(see Section~\ref{subsec:higgspp} below).
\item[(iv)]
The anomalous couplings $c_i$ of the nonlinear EFT at leading order
are able to account for deviations of ${\cal O}(1)$ from the SM.
It is then consistent to retain the terms
quadratic in these couplings when computing cross sections and rates.
This is in contrast to the linear case, where a linearization in the
dimension-6 corrections has to be performed at this level of accuracy.
\item[(v)]
Eventually the computation of an observable at a given chiral order must
incorporate the loop corrections at that order if one wants to perform an
accurate determination of the Higgs parameters at the
LHC \cite{Buchalla:2015qju}.
\end{description}

\subsection{Linear vs. nonlinear EFT}

In this section the relation between the SMEFT organized by canonical
dimensions (often referred to as ``linear'' EFT) and the one organized by
chiral dimensions (usually referred to as ``nonlinear'' EFT)
will be discussed.

SMEFT is the most common approach to the SM as an EFT and starts from
the renormalizable, dimension-4 Lagrangian, adding operators of higher
canonical dimension to account for the physics at shorter distances. Assuming
conservation of baryon and lepton number, the leading corrections come from
the terms of dimension 6 \cite{Buchmuller:1985jz,Grzadkowski:2010es}.

In the case of the electroweak chiral Lagrangian
three relevant energy scales may be distinguished:
The electroweak scale $v$, the scale $f$ of the Higgs-sector dynamics,
and the cut-off scale $\Lambda=4\pi f$, where the low-energy description of
this dynamics breaks down.
These three scales allow for two independent expansion parameters,
$\xi = v^2/f^2$ and the loop factor $1/(16\pi^2)= f^2/\Lambda^2$.
In full generality, a double expansion can thus be performed on the
new-physics effects.

The resulting picture is sketched in \refF{fig:loopdim}, where
the powers of $\xi$ are plotted on the vertical and the loop order
on the horizontal axis \cite{Buchalla:2014eca}.
The dots indicate, schematically, (classes of) operators in
the effective Lagrangian or, alternatively, terms in a physical
amplitude.

Without expanding in $\xi$, the effective theory takes the form of
a loop expansion as in the usual chiral Lagrangians \cite{Weinberg:1978kz}.
This amounts to proceeding from left to right in \refF{fig:loopdim},
order by order in the loop expansion, resumming at each order all terms
along the vertical axis.

Alternatively, the expansion may be organized in powers of $\xi$,
proceeding from bottom to top of \refF{fig:loopdim} and including,
in principle, at each power of $\xi$ terms of arbitrary order in the
loop expansion. This scheme corresponds to the conventional expansion
of the effective theory in terms of the canonical dimension $d$ of
operators, where the power of $\xi$ is given by $(d-4)/2$.
Since the dimensional expansion requires only a hierarchy between $v$ and
the new-physics scale $f$, $\xi\muchless 1$, it is not restricted to
a pseudo-Goldstone Higgs scenario, typically underlying the chiral
Lagrangian.

These observations clarify the relation between
an effective theory organized by canonical dimension and
the electroweak chiral Lagrangian organized as a loop expansion:
The former is constructed row by row, the latter column by column
from the terms in \refF{fig:loopdim}.
\begin{figure*}[t]
\begin{center}
\includegraphics[width=8cm]{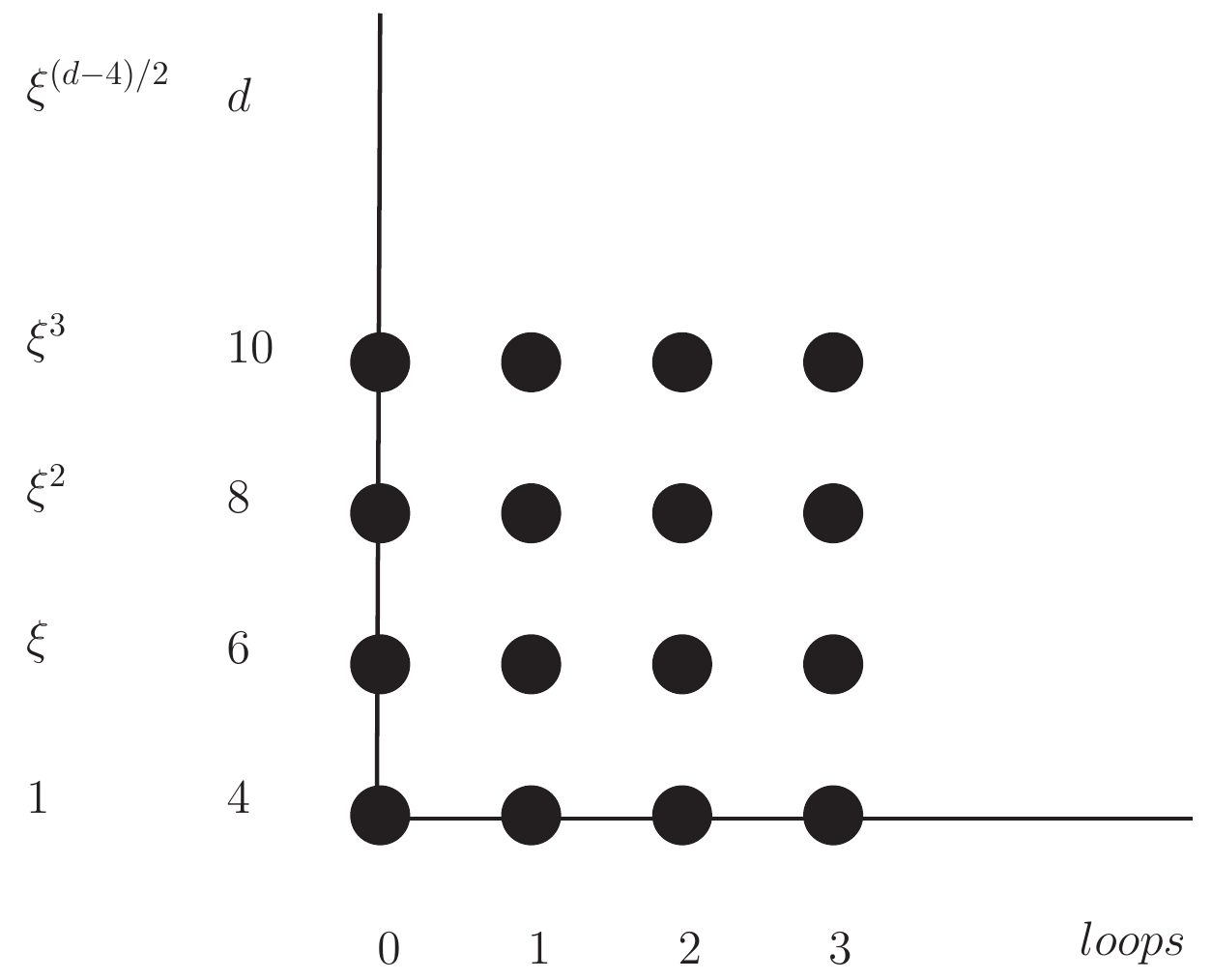}
\end{center}
\caption{Systematics of the effective theory with sizeable anomalous
couplings in the Higgs sector. The dots indicate operators in
the effective Lagrangian (or terms in a physical amplitude). In general,
they may be organized both in powers of $\xi=v^2/f^2$ (vertical axis)
and according to their order $L$ in the loop expansion (horizontal axis).
The latter is equivalent to the chiral dimension $2L+2$.}
\label{fig:loopdim}
\end{figure*}
In conclusion, both EFTs, the one based on canonical dimension and the
one based on chiral dimensions, can in principle account for low-energy
deviations from the SM and could in general cover the same correction terms.
The difference consists in the way these terms are organized or resummed.
It will ultimately depend on the pattern of new-physics effects which
of the two formulations will eventually be more appropriate.

\subsubsection*{Phenomenological implications of
linear vs. nonlinear EFT}

The essential difference between the EFT with a chiral expansion
(`nonlinear') and the one with a dimensional expansion (`linear')
consists in a reordering of terms as illustrated in \refF{fig:loopdim}.
This reordering is dictated by the different dynamics.
The chiral framework will be the
relevant one if the Higgs-coupling deviations characterized by $\xi$ are
parametrically larger than the loop factor, $\xi\gg 1/16\pi^2$, or
equivalently $f\muchless 4\pi v\approx 3\,{\rm TeV}$. This would typically be
the case for $\xi={\cal O}(10\%)$, within reach of the precision
achievable at the LHC.
The experimental program to explore such a scenario
will then be the search for anomalous Higgs boson couplings with sizeable
deviations from the SM values. Given the precision goal of the LHC in
Run 2 and 3, this search should be focussed on the leading-order
couplings contained in ${\cal L}_2$. Important targets
are the $hVV$, $ht\bar t$, $hb\bar b$, $h\tau\bar\tau$ couplings, but also
$h^3$ from the Higgs potential or $h\to gg$, $h\to \gamma\gamma$,
$h\to Z\gamma$ local contributions.
Longitudinal gauge-boson scattering, although challenging experimentally,
might also yield important
information \cite{Contino:2010mh,Espriu:2013fia,Delgado:2013loa}.
The same is true for $\gamma\gamma$ scattering and other
photon-related observables \cite{Delgado:2014jda}.

Besides the expected size of the deviations, one of the generic features
of the chiral Lagrangian is the {\it decorrelation} between Higgs boson couplings
\cite{Buchalla:2014eca,Brivio:2013pma}, which arises already at leading order.
For instance, the quark mass and
the Yukawa interaction at LO are controlled by different coefficients
(see (\ref{l2}) above). If an expansion at fixed order in $\xi$ is
performed on the Wilson coefficients of the electroweak chiral Lagrangian,
then correlations will appear. If $\xi$ is sufficiently small, these
correlations will eventually be the same as in the SMEFT
(see for instance \cite{Giudice:2007fh,Buchalla:2014eca} for a discussion).
The distinction between a linear
and nonlinear framework therefore depends crucially on the size of $\xi$.

As another example, the corrections to the oblique parameter
$a_1$ in (\ref{l2l4}), or to the triple gauge boson vertex (parameterized
by $a_2$ and $a_3$), or to longitudinal $WW$ scattering ($a_4$ and $a_5$) all
appear at chiral dimension 4 in the nonlinear EFT,
whereas in the linear realization corrections to
the triple gauge boson vertex appear at $D=6$ while anomalous contribution to
the quartic gauge boson vertex appear only at $D=8$.
In the linear case there is thus a strong hierarchy between those corrections.
In the nonlinear case  $a_1$, $a_2$, $a_3\sim\xi/16\pi^2$ while
$a_4$, $a_5\sim\xi^2/16\pi^2$, thus their hierarchy depends on the
size of $\xi$ and would disappear for $\xi={\cal O}(1)$.

\subsection{Sample applications}
\label{subsec:applications}

The following examples illustrate how Higgs-related processes are affected
by new physics as described by the electroweak chiral Lagrangian.

\subsubsection{\texorpdfstring{\boldmath $h\to Z \ell^+\ell^-$}{h to Zll}}

The process $h\to Z \ell^+\ell^-$, shown in \refF{fig:hzz}
may serve as a prototype for a tree-level decay of the Higgs boson.
\begin{figure*}[h]
\begin{center}
\includegraphics[width=5cm]{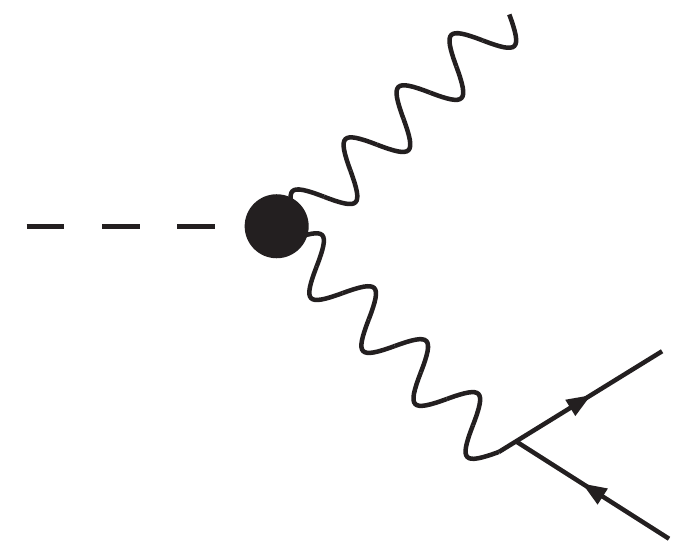}
\end{center}
\caption{$h\to Z \ell^+\ell^-$ decay at leading order
in the chiral Lagrangian. The black dot indicates the
(anomalous) $hZZ$ coupling $c_V$ from ${\cal L}_2$.
The $Z$-fermion coupling is not modified at this order.}
\label{fig:hzz}
\end{figure*}
At leading order only the rate is affected by an anomalous coupling
that modifies the $hZZ$ vertex.
The operators contributing at NLO have been listed in
\cite{Buchalla:2015qju,Buchalla:2013mpa}.
These give the dominant contributions to the angular distributions.
This hierarchy between rates and distributions
is a generic feature of the chiral Lagrangian for tree-level processes:
in the SMEFT corrections to
rates and distributions come at the same (NL) order in the expansion.

The case of $h\to WW^*$ decay is similar.
Likewise, the decay of Higgs into a pair of fermions $h\to f\bar f$
is modified multiplicatively by a leading-order anomalous coupling.

\subsubsection{\texorpdfstring{\boldmath $h\to \gamma\gamma$,
$h\to Z\gamma$}{h to gamma gamma, h to Z gamma}}

Further important examples are the decays $h\to\gamma\gamma$ and
$h\to Z\gamma$. Based on the chiral Lagrangian, the leading contributions
are displayed in \refF{fig:hgamma}.
\begin{figure*}[h]
\begin{center}
\includegraphics[width=14cm]{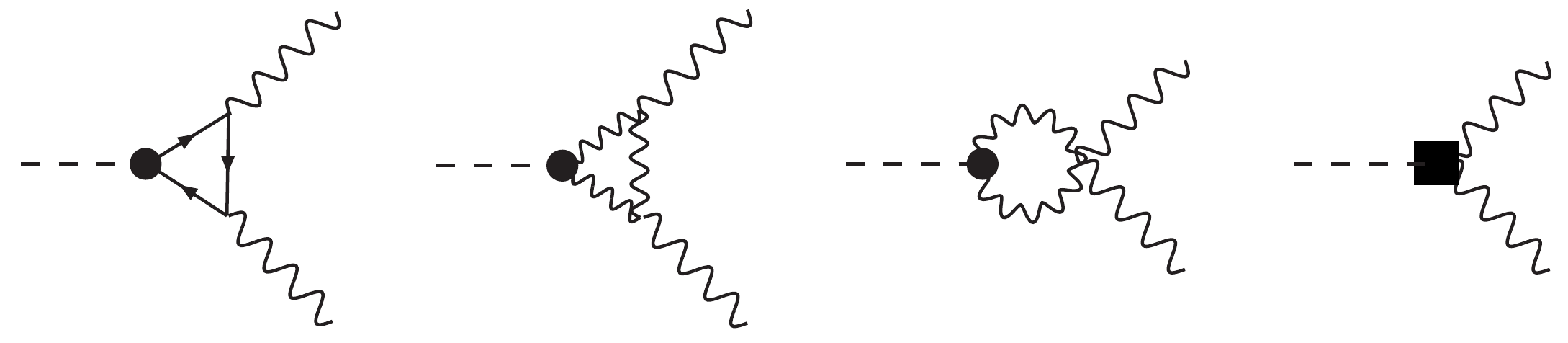}
\end{center}
\caption{$h\to\gamma\gamma$, $h\to Z\gamma$ decay at leading order
in the chiral Lagrangian. The black dots indicate vertices from ${\cal L}_2$
(couplings $c_V$, $c_t$),the black squares denote local terms from
${\cal L}_4$ (couplings $c_{\gamma\gamma}$, $c_{Z\gamma}$).}
\label{fig:hgamma}
\end{figure*}
Leading-order vertices inside loops contribute at the same level as
next-to-leading order local terms at tree level. The (non-Higgs) photon
and $Z$-boson couplings are identical to those in the SM.

\subsubsection{\texorpdfstring{\boldmath $pp\to h + jet$}{pp to h+jet}}
\label{subsec:higgsjet}

The high-$p_T$ distribution of a boosted Higgs in $pp\to h + jet$
has been proposed as a tool to disentangle the contributions
from $c_t$ and $c_{gg}$, which cannot be separated in the $gg\to h$
total rate  \cite{Grojean:2013nya}. The dependence on these couplings
in the framework of the electroweak chiral Lagrangian at leading order
is illustrated for the partonic process $gg\to gh$ in \refF{fig:hjet}.
\begin{figure*}[h]
\begin{center}
\includegraphics[width=10cm]{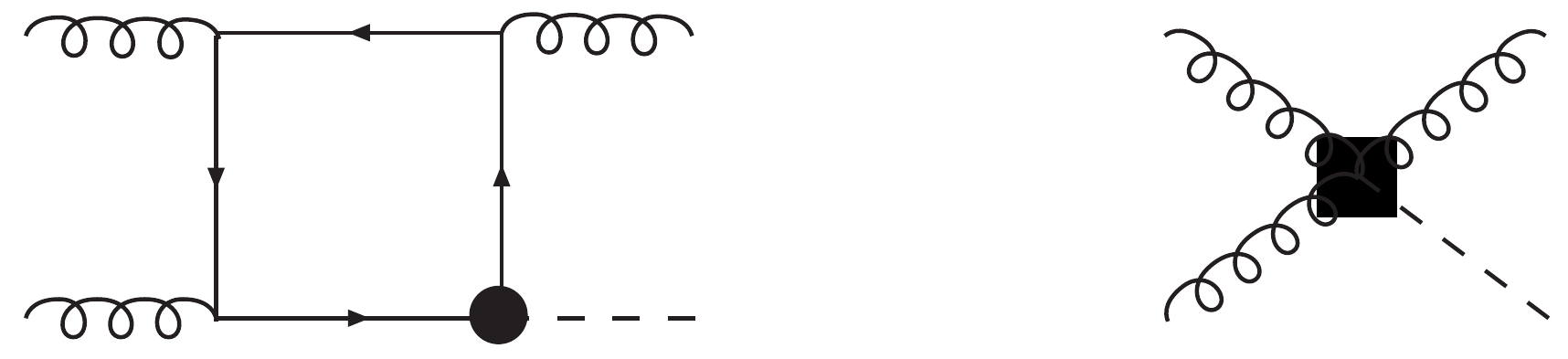}
\end{center}
\caption{Sample diagrams for $pp\to h + jet$ at leading order
in the chiral Lagrangian. The black dot denotes $c_t$,
the black square $c_{gg}$.}
\label{fig:hjet}
\end{figure*}
At high $p_T$ the loop involving $c_t$ can be distinguished
from the local interaction described by $c_{gg}$.

\subsubsection{Higgs-pair production in gluon fusion}
\label{subsec:higgspp}

Another interesting case is Higgs-pair production in gluon-gluon
fusion, which has recently been discussed
in \cite{Grober:2015cwa,Contino:2012xk}
in the framework of the chiral Lagrangian. NLO QCD effects have been
consistently included \cite{Grober:2015cwa}.
The leading-order diagrams are shown in \refF{fig:hprocess}.
\begin{figure*}[h]
\begin{center}
\includegraphics[width=14cm]{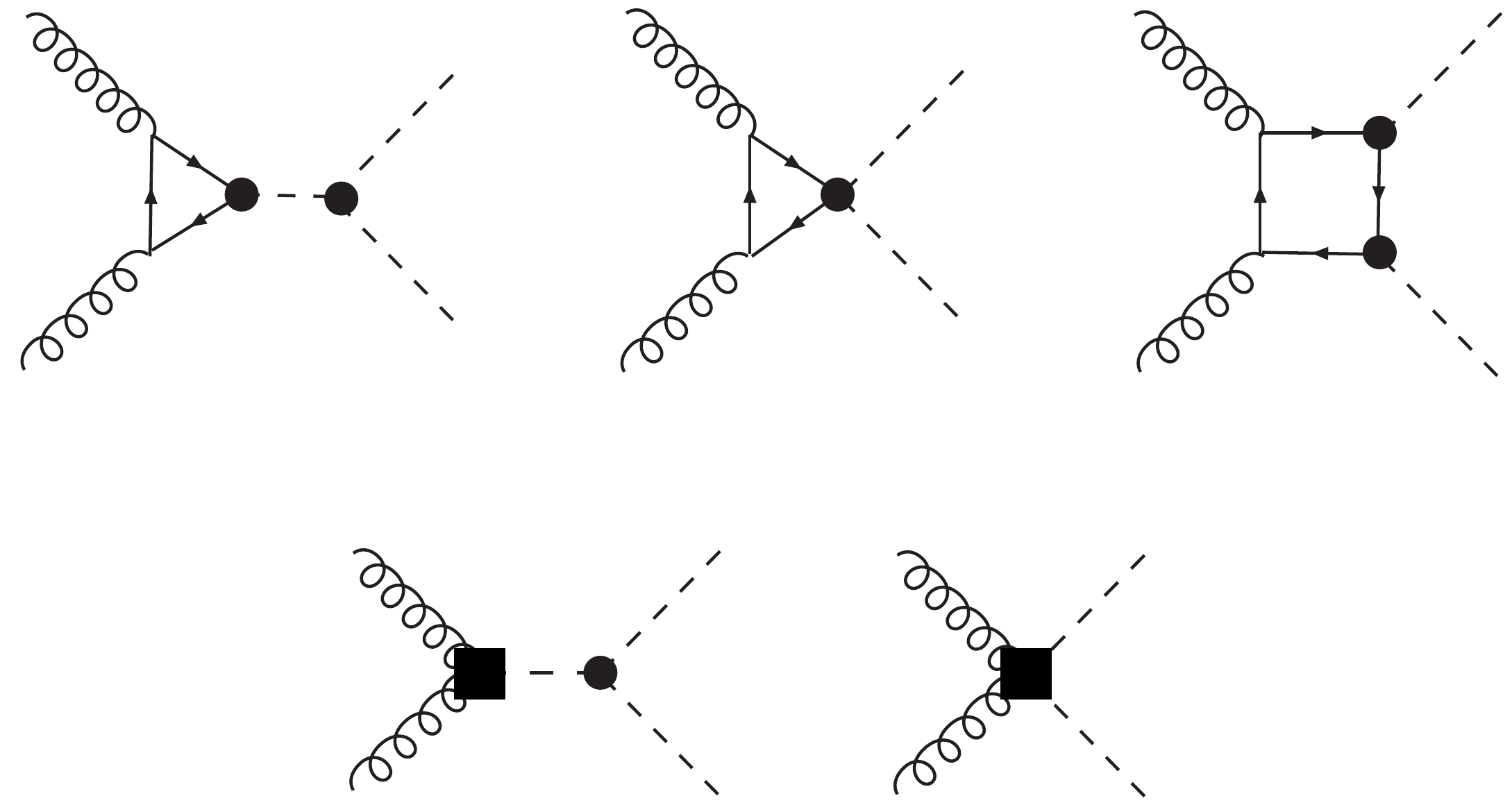}
\end{center}
\caption{Higgs-pair production in gluon fusion at leading order
in the chiral Lagrangian. The black dots indicate vertices from ${\cal L}_2$,
the black squares denote local terms from ${\cal L}_4$.}
\label{fig:hprocess}
\end{figure*}
All diagrams are at the same order in the chiral counting.
They illustrate again the interplay between leading order anomalous couplings
(black dots) within loops, and next-to-leading order terms
(black squares) at tree level. Note that five different couplings
appear, $ht\bar t$, $hht\bar t$, $hhh$ from ${\cal L}_2$,
and $hgg$, $hhgg$ from ${\cal L}_4$.


\subsubsection{Photon-photon scattering}

The scattering processes $\gamma\gamma\to W_L^+ W_L^-$ and
$\gamma\gamma \to Z_L Z_L$ have been studied in ~\cite{Delgado:2014jda} in
the framework of the electroweak chiral Lagrangian with a light Higgs to NLO
and by means of the equivalence theorem, where the relevant amplitudes are
those for the corresponding Goldstone bosons $\gamma\gamma\to w^+w^-$ and
$\gamma\gamma\to zz$, respectively. These Goldstone bosons are introduced via
the $U$-matrix in (\ref{udef}).
Also some related observables like the oblique
$S$-parameter, the electromagnetic form factor $\gamma^*\to w^+w^-$, and the
Higgs transition form factor $\gamma^*\gamma^* \to h$
have been studied in~\cite{Delgado:2014jda}.
All these photon-related observables  are sensitive to the nature and the
couplings of the Higgs boson via loops and via internal Higgs propagators, 
allowing the investigation of a possible dynamical electroweak
symmetry breaking and strongly coupled scenarios. This motivates the study
of these observables in the context of the LHC and other future colliders,
and in particular the photon-photon scattering processes have received
increased interest also in the experimental community.
The CMS Collaboration has published Run~1 results on the charged
channel~\cite{Chatrchyan:2013akv,Khachatryan:2016mud},
showing the feasibility of this type of analysis and new forward proton
detectors, CMS-TOTEM Precision Proton Spectrometer (CT-PPS)~\cite{Albrow:2006xt,Albrow:2008pn}
and ATLAS-AFP~\cite{Royon:2007ah}, will be incorporated.
The goal would be to search for exclusive or quasi-exclusive $W^+W^-$
production by photon-photon interactions in $pp \to p^{(*)} W^+W^-p^{(*)}$,
where the two intermediate photons are radiated collinearly from the protons,
which come out undetected along the beam-pipe~\cite{Chatrchyan:2013akv,Khachatryan:2016mud}. Tagging the
outgoing $p^{(*)}$ with the CT-PPS and ATLAS-AFP forward detectors
will highly increase the efficiency of this type of analyses. On the other
hand, a further study could be done if the extra jets, being produced in the
deep inelastic regime of these photon mediated processes, are also required
to be detected in the forward/backward region. This could also provide
interesting additional information on these subprocesses where the photons
are virtual.

Within the approximation of the equivalence theorem ($W_L^\pm\to w^\pm$,
$Z_L\to z$) considered in \cite{Delgado:2014jda}, the photon-photon scattering
amplitudes and the previously mentioned related observables can be described
by just a few terms in (\ref{l2l4}), parameterized by $a$ in ${\cal L}_2$ and
$a_1$, $a_2$, $a_3$, $c_{\gamma\gamma}$ in ${\cal L}_4$.
(Related work on photon-photon processes in the chiral perturbation theory
of pions is described in
\cite{Dobado:1995ka,Bijnens:1991za,Bijnens:1987dc,Burgi:1996qi}.)

The results for the scattering amplitudes are presented in terms of the two
helicity-independent scalars $A(s,t,u)$ and $B(s,t,u)$, in the form
\begin{equation}
{\cal M}(\gamma(k_1,\epsilon_1)\gamma(k_2,\epsilon_2)\to w^a(p_1) w^b(p_2))
 \,\,\,  =\,\,\, ie^2 (\epsilon_1^\mu \epsilon_2^\nu T_{\mu\nu}^{(1)}) A(s,t,u)+
ie^2 (\epsilon_1^\mu \epsilon_2^\nu T_{\mu\nu}^{(2)})B(s,t,u)
\nonumber
\end{equation}
which are written in terms of the two independent Lorentz structures
and the external photon polarizations $\epsilon_j$,
\begin{eqnarray}
(\epsilon_1^\mu\epsilon_2^\nu T^{(1)}_{\mu\nu}) &=&
\frac{s}{2} (\epsilon_1 \epsilon_2) - (\epsilon_1 k_2)(\epsilon_2 k_1),
\label{eq.T1-def}
\\
(\epsilon_1^\mu\epsilon_2^\nu T^{(2)}_{\mu\nu}) &=&
2 s (\epsilon_1 \Delta)(\epsilon_2 \Delta)
- (t-u)^2 (\epsilon_1\epsilon_2)
- 2 (t-u) [(\epsilon_1\Delta)(\epsilon_2 k_1) -
(\epsilon_1 k_2) (\epsilon_2 \Delta)],
\nonumber\\
\nonumber
\end{eqnarray}
with the Mandelstam variables defined as usual, $s=(p_1+p_2)^2$, $t=(k_1-p_1)^2$
and $u=(k_1-p_2)^2$, and the relevant momentum combination is defined as
$\Delta^\mu\equiv p_1^\mu -p_2^\mu$.

The diagrams contributing to $\gamma \gamma \to ww$ are summarized
schematically in \refF{fig:diags-gammagamma}.
\begin{figure}[!t]
\begin{center}
\includegraphics[width=10cm]{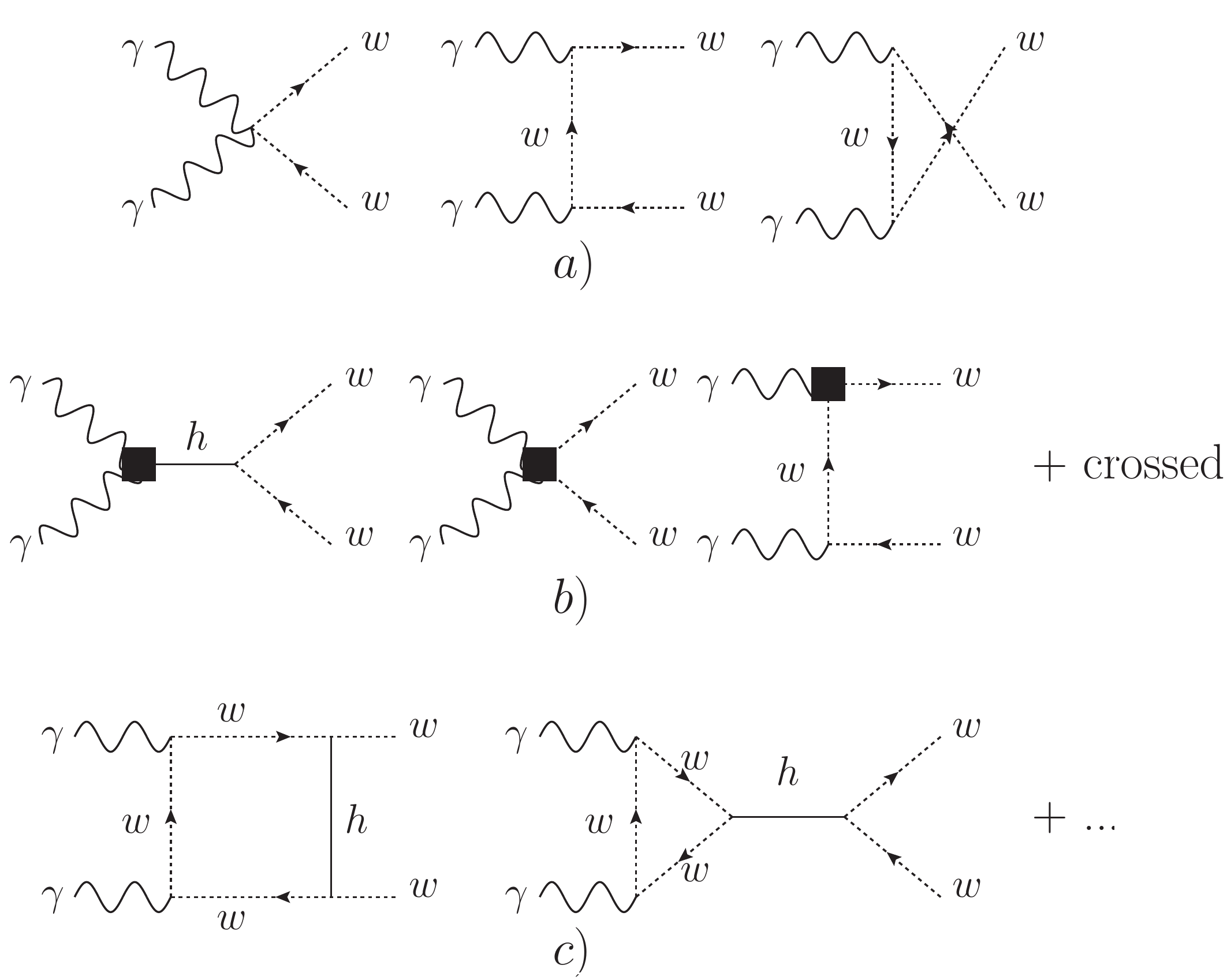}
\end{center}
\caption{Diagrams contributing to  $\gamma \gamma \to ww$ in the nonlinear
EFT up to NLO: a) from just ${\cal L}_2$ at tree level,
b) from both ${\cal L}_4$ (in black boxes) and ${\cal L}_2$ at tree level,
c) from ${\cal L}_2$ at one-loop. The full set of diagrams for the two
channels  $\gamma \gamma \to ww, zz$ can be found in ~\cite{Delgado:2014jda}.}
\label{fig:diags-gammagamma}
\end{figure}
The LO amplitudes are provided by tree-level diagrams with vertices from
${\cal L}_2$, \refF{fig:diags-gammagamma} a) for $ww$
(the diagrams for $zz$ can be found in ~\cite{Delgado:2014jda}):
\begin{eqnarray}
A(s,t,u)^{\gamma\gamma\to zz}_{\rm LO} \, =\,
B(s,t,u)^{\gamma\gamma\to zz}_{\rm LO} &=& 0 \, ,
\nonumber\\
A(s,t,u)^{\gamma\gamma\to w^+w^-}_{\rm LO} \,
=\, 2 s B(s,t,u)^{\gamma\gamma\to w^+w^-}_{\rm LO} &=&
-\frac{1}{t} -\frac{1}{u} \, .
\end{eqnarray}
At NLO, there are additional contributions from diagrams of type b) and c)
in \refF{fig:diags-gammagamma}:
\begin{eqnarray}
A(s,t,u)^{\gamma\gamma\to zz}_{\rm NLO} &=& -\frac{a c_{\gamma\gamma}^r}{4\pi^2 v^2} +
\frac{ (a^2-1)}{4\pi^2v^2},
\nonumber\\
A(s,t,u)^{\gamma\gamma\to w^+w^-}_{\rm NLO}
&=&
\frac{8(a^r_1-a^r_2+a^r_3)}{v^2} - \frac{a c_{\gamma\gamma}^{r}}{4\pi^2 v^2} +
\frac{(a^2-1)}{8\pi^2v^2},
\end{eqnarray}
and $B(s,t,u)^{\gamma\gamma\to zz}_{\rm NLO}
=B(s,t,u)^{\gamma\gamma\to w^+w^-}_{\rm NLO}=0$. Notice that the last term
above coming from the loops with Goldstone bosons cancels for the input
value of the SM, i.e for $a=1$.  Fermion loops are missing in this calculation
but as the electroweak Goldstones and the Higgs are the only 
ones coupling derivatively in the LO Lagrangian
their contribution is expected to be suppressed by $m_{t,b}^2/E^2$,
with $E^2=s,t,u$.

The above results have been expressed in terms of the renormalized
couplings $c_{\gamma\gamma}^r$, $a_1^r$, $a_2^r$ and $a_3^r$. The renormalization
of these couplings and their running were first computed
in \cite{Delgado:2014jda} and are summarized in Table \ref{tab:running}.

 \subsubsection{Electromagnetic form factor
\texorpdfstring{\boldmath $\gamma^*\to w^+w^-$}{gamma* to ww}}

There are related subprocesses that depend on different combinations of the
same effective couplings and can be potentially explored in future collider
studies. The electromagnetic transition $\gamma^*\to w^+w^-$ from a deeply
virtual photon with momentum $q^\mu=p_1^\mu+p_2^\mu$ is described through the
matrix element
\begin{eqnarray}
\langle w^+(p_1) \,w^-(p_2)|\,  J^\mu_{\rm EM} \, |0\rangle &=&
 \,   e  \,
 (p_1^\mu-p_2^\mu)\, \mathbb{F}_{\gamma^*ww}(q^2)  \, .
\end{eqnarray}
This matrix element is crucial in the production of two longitudinal weak
bosons in future $e^+e^-$ colliders. The electromagnetic vector form factor
(VFF) can be computed  with  the chiral Lagrangian up to NLO.
At high momentum-transfer squared (with $q^2=(p_1+p_2)^2$), where the
equivalence theorem applies, one finds \cite{Delgado:2014jda}
\begin{eqnarray}
\mathbb{F}_{\gamma^*ww} &=&
\underbrace{   \phantom{\frac{A}{B}}     1  \phantom{\frac{A}{B}}  }_{\rm LO}
\quad  + \quad
\underbrace{   \frac{ 2 q^2       ( a_3^r-a_2^r)         }{v^2}
+    (1-a^2)\frac{q^2 }{96\pi^2 v^2}\left(
\frac{8}{3}   -\ln\frac{-q^2}{\mu^2}\right)  }_{\rm NLO} \, .
\end{eqnarray}
The NLO chiral couplings $a_2^r$ and $a_3^r$ are
renormalized in the $\overline{MS}$ scheme at the scale $\mu$ in dimensional
regularization and provide the tree-level NLO contribution to the form factor.
They renormalize the UV--divergences that show up in the one-loop NLO
contribution, which is given by the term proportional to $(1-a^2)$.
Fermion loops are missing in this calculation~\cite{Delgado:2014jda} but
their contribution is expected to be suppressed by  $m_{t,b}^2/q^2$ for the
same reasons previously exposed for $\gamma\gamma$--scattering.

\subsubsection{Higgs transition form factor
\texorpdfstring{\boldmath $\gamma^*\gamma^* \to h$}{gamma* gamma* to h} / associated production}

An interesting observable in order to pin down the $h\gamma\gamma$ coupling
$c_{\gamma\gamma}$ is the  Higgs transition form factor (HTFF),
which describes the process $\gamma^*(k_1)\gamma^*(k_2)\to h(p)$
\cite{Watanabe:2013ria,Alloul:2013naa,Isidori:2013cga}.
This transition is given by the matrix element
\begin{eqnarray}
\int d^4x\, e^{-i k_1x} \, \langle h(p)|\, T\{ \, J_{\rm EM}^\mu(x)\,
J_{\rm EM}^\nu(0)\, \}\, |0\rangle
&=& \,i  \, e^2  \,
\,\left[\, k_1\cdot k_2 \, g^{\mu\nu}\,- \, k_2^\mu \, k_1^\nu  \, \right]
\,\,\mathbb{F}_{\gamma^*\gamma^* h}(k_1^2,k_2^2)
\nonumber\\
\end{eqnarray}

In the case where one of the photons is on-shell and the other is highly
virtual ($k^2\gg m_h^2$), one finds that the HTFF is zero at LO:
$\mathbb{F}_{\gamma^*\gamma h}(k^2, 0)_{\rm LO} =0$.
The first non-zero contribution shows up at NLO~\cite{Delgado:2014jda}:
\begin{eqnarray}
\mathbb{F}_{\gamma^*\gamma h}(k^2, 0)_{\rm NLO}  &=&
\frac{c_{\gamma\gamma}^r}{4\pi^2} \, .
\end{eqnarray}
We note that the NLO form factor comes exclusively from the renormalized
tree-level $h\gamma\gamma$ vertex with $c_{\gamma\gamma}^r$.
The one-loop diagrams vanish within the configuration of momenta
studied here.
Here again we provide the result in the equivalence theorem approximation
and neglect corrections due to boson masses.
Notice that this result does not correspond to
the same kinematical regime as $\Gamma(h\to\gamma\gamma)$,
since we are considering  $m_h^2\muchless k^2 \muchless 16\pi^2 v^2$ in this form factor.
For the same reason as in the previous photon observables,
fermion loops are expected to be suppressed by powers $m_{t,b}^2/k^2$,
as fermions do not couple derivatively in the LO chiral Lagrangian.

As it occurred for the $\gamma\gamma$--scattering, in order to pin down the
$\gamma^* \gamma\to h$ process at LHC one should look for events
where one of the protons radiates a collinear photon with low virtuality and
comes out again undetected, while the other radiates a deeply virtual photon
and gives rise to a jet. Again, tagging the outgoing collinear protons
$p^{(*)}$ with the forward detectors CT-PPS~\cite{Albrow:2006xt,Albrow:2008pn} and
ATLAS-AFP~\cite{Royon:2007ah} will increase the efficiency in LHC analyses.
Likewise, electromagnetic subprocesses of this type would be important in
future $e^+e^-$ machines or dedicated $e\gamma$
colliders~\cite{Watanabe:2013ria}.


\subsubsection{TeV-scale particle-pair production at NLO}

At the TeV scale, much simplification takes place, in that many couplings
become negligible with respect to the sought derivative vertices that grow
with $s=E^2_{\rm cm}$. It is then possible to ignore, in first approximation,
all SM masses and gauge couplings, and concentrate on the electroweak symmetry
breaking sector. The relevant effective Lagrangian follows from (\ref{l2l4})
and, even at NLO, contains only seven
parameters~\cite{Delgado:2014dxa,Delgado:2015kxa}:
the $a$ and $b$ LO parameters coupling the Higgs boson to the longitudinal
electroweak $W_L\sim w$ bosons, the classical $a_4$ and $a_5$ from the
Higgsless electroweak chiral Lagrangian and, highlighted here,
the NLO counterterms to one-loop computations of boson-boson scattering
including Higgs in the initial or final state, $g^{hh}$, $d^{hh}$ and
$e^{hh}$.

The Lagrangian (\ref{l2l4}) with these couplings can be used to compute
scattering amplitudes between the $ww$ and $hh$ two-body channels,
as explained below.
Because these amplitudes get strong in the TeV region if there are few
per cent level deviations from the SM value couplings, it should be feasible
to spot them in experimental diboson production. The reason is Watson's
final state rescattering theorem, which corrects the SM production amplitude
as implemented, {\it e.g.} in a Monte Carlo simulation,  by a strong
rescattering form factor setting the correct phase, as illustrated in
\refF{fig:treelevelprod}.
\begin{figure}[h]
\begin{center}
\includegraphics[width=4.5cm]{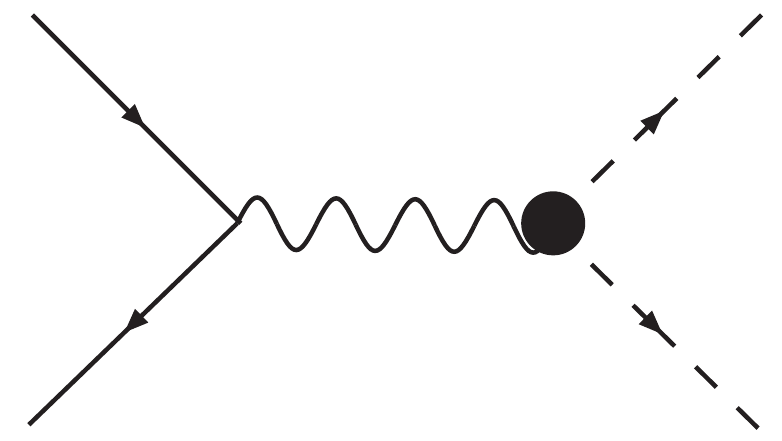}
\caption{Diboson $W_LW_L$ production with $I=1$
from a fermion pair via an intermediate $W_T$. If there are strong
interactions in the final state, the production amplitude is multiplied by
a universal rescattering form factor, here the thick round blob.}
\label{fig:treelevelprod}
\end{center}
\end{figure}
Such form factor needed for the Feynman diagram in
\refF{fig:treelevelprod} can be computed from the scattering
amplitudes presented shortly, typically as
\cite{Delgado:2014dxa,Delgado:2015kxa,Dobado:2015hha}
(using the inverse amplitude method (IAM) for unitarization)
\begin{equation} \label{FF}
F_V(s) = F_{11}(s)
      = \left[1-\frac{A_{11}^{(1)}(s)}{A_{11}^{(0)}(s)}\right]^{-1} .
\end{equation}
{\bf Diboson production.}
Since strong couplings are derivative, their ``low energy''
($E\muchless 4\pi v\sim 3$ TeV) scattering amplitude $A(s,t)$ can be very
economically represented in terms of very few partial waves, because the
expansion should quickly converge:
\begin{equation}
  A_{i\to j,I}(s,t) = 64\pi \sum_{J=0}^\infty (2J+1) P_J(x) A_{i\to j,IJ}(s),
\end{equation}
with $x=\cos\theta=1- 2t/s$ the cosine of the scattering angle for
boson-boson elastic ($i\to i$) or inelastic ($i\to j\neq i$) processes.
$P_J(x)$ are the Legendre polynomials.

If the custodial isospin is $I=1,2$, the $ww$ scattering is
elastic.~\footnote{For $I=1$ the $t\bar{t}$ channel is active, which is more
weakly coupled with an amplitude $\sim m_t/\sqrt{s}$.} 
But in the isoscalar case one confronts a $ww\to hh$
coupled-channel problem, so that it is useful to define a reaction matrix
for each partial wave,
\begin{equation}
   F_{IJ} = \begin{pmatrix}
                A_{ww\to ww, IJ} & A_{ww\to hh, IJ} \\
                A_{hh\to ww, IJ} & A_{hh\to hh, IJ}
            \end{pmatrix} %
            \equiv %
            \begin{pmatrix}
                A_{IJ}  & A'_{IJ} \\
                A'_{IJ} & A''_{IJ}
            \end{pmatrix} ,
\end{equation}
whose chiral expansion  has the generic form
\begin{subequations}\label{chLagr:PartialWaves:AIJ}
\begin{eqnarray}
  F_{IJ}^{(0)}(s) &=& K_{IJ}s \\
  F_{IJ}^{(1)}(s) &=& \left(B_{IJ}(\mu) + D_{IJ}\log\frac{s}{\mu^2} +
    E_{IJ}\log\frac{-s}{\mu^2}\right) s^2\ .
\end{eqnarray}
\end{subequations}
Here, the constants $K_{IJ}$, $B_{IJ}$, $D_{IJ}$, $E_{IJ}$ are, in general,
matrices whose elements depend on the NLO Lagrangian parameters.
For elastic $ww\to ww$ scattering in the
vector-isovector channel, $I=J=1$, they are numbers given by
\begin{eqnarray} \label{isovectorconstants}
\nonumber  K_{11}      &=& \frac{1}{96\pi v^2}(1-a^2) \\
\nonumber  B_{11}(\mu) &=& \frac{1}{110592\pi^3 v^4}
 \left[8(1-a^2)^2 - 75(a^2-b)^2 + 4608\{a_4(\mu) - 2a_5(\mu)\}\pi^2 \right] \\
           D_{11}      &=& \frac{1}{9216\pi^3v^4}
                          \left[(1-a^2)^2 + 3(a^2-b)^2\right] ,\; %
           E_{11}      = -\frac{1}{9216\pi^3v^4}(1-a^2)^2\ ,
\label{k11}
\end{eqnarray}

The form factor in Eq.~(\ref{FF}) then modifies SM production,
as
\begin{equation}\label{modifiedFF}
F_V(s)\sim 1 + \frac{A_{11}^{(1)}(s)}{A_{11}^{(0)}(s)}\ .
\end{equation}
Using the above expressions, the $ww\to ww$ amplitude
in the $I=1$, $J=1$ (vector) channel is given by
\begin{equation} \label{AV}
A_V(s)=\left[1-\frac{A_{11}^{(1)}(s)}{A_{11}^{(0)}(s)}\right]^{-1} A_{11}^{(0)}(s)\ ,
\end{equation}
from which a fit to the $a_4-2a_5$ parameter combination from
(\ref{k11}) can be attempted, for example.
The required experimental measurements are, however, rather challenging
\cite{Contino:2010mh}.
The case of resonances and unitarization is discussed in detail in
\cite{Espriu:2013fia,Delgado:2014dxa,Delgado:2015kxa,Dobado:2015hha}.

Note that formulae used in this section are derived using the equivalence
theorem (ET). By using exact formulae for the tree-level expressions and the
optical theorem itself one finds that the ET results are generally reliable,
except for the $I=2$ channel when $a>1$ where the $I=2$, $J=0$ resonance
found with the ET disappears \cite{Espriu:2013fia}.
Using ``exact'' amplitudes also implies that propagation of transverse
$W$ and $Z$ needs to be included and this modifies slightly the dependence
on the coefficients of chiral order 4 (for instance $a_3$ enters), although
corrections are small.

\subsection{Concluding remarks}

Assuming that the largest effects of new physics arise in the 
Higgs sector, and parameterizing the non-standard Higgs boson couplings in a
gauge-invariant way, automatically leads to an electroweak chiral Lagrangian
as the low-energy EFT.

This EFT represents a consistent quantum field theory framework, in which
improvements, through higher-order radiative corrections or by going
to next-to-leading order in the new-physics effects, can be systematically
included. Focussing first on the leading-order nonstandard couplings has
the advantage of reducing the new-physics parameters to a manageable set,
in a well-defined and consistent manner. In fact, the formalism essentially
corresponds to Higgs-coupling parameterizations routinely used in the
experimental analyses of the ATLAS and CMS collaborations
($\kappa$-framework).

The electroweak chiral Lagrangian can then be used to systematically
improve the $\kappa$-formalism, in particular such that differential
distributions in general Higgs processes can be studied.
However, the detection of such effects will be rather challenging at the LHC,
since they are expected to be of ${\cal{O}}(\xi/16\pi^2)$ and therefore rather
suppressed.

Consequently, the best strategy to follow at the LHC is to focus the
experimental analysis to the leading-order chiral Lagrangian, which matches
well the precision goals for Higgs boson properties anticipated for Run~2 and 3.
If a deviation is found, the next natural step is to study the differential
distributions. This sequential analysis is in contrast with what
follows from the SMEFT, where deviations in {\emph{both}} rates and
distributions are expected at NLO and should therefore
be studied simultaneously. Additional interesting information on deviations
from the Higgs boson couplings and NLO contributions could also be obtained via
the study of EW boson scattering ($WW$, $\gamma\gamma$, etc.).

\subsection*{Acknowledgements}

Comments and suggestions from Belen Gavela and Veronica Sanz are
gratefully acknowledged.

\section[Fitting EFT parameters and constraining models]{Fitting EFT parameters and constraining models\SectionAuthor{N.~Belyaev, A.~Biek{\"o}tter, J.~Brehmer, C.~Englert, A.~Freitas, D.~Gon{\c c}alves, J.~Gonzalez-Fraile, M.~Gorbahn, R.~Kogler, D.~Lopez-Val, J.M.~No, T.~Plehn, M.~Rauch, V.~Sanz, M.~Spannowsky}}
\label{s.eftmodels}



\subsection{The problem}
\label{sec:introEFT_Models}

Extending the Higgs boson couplings framework to an effective field theory,
usually truncated after including dimension-6 operators~\cite{Leung:1984ni,Buchmuller:1985jz,DeRujula:1991ufe,Hagiwara:1993qt,Hagiwara:1995vp,GonzalezGarcia:1999fq}, addresses two
short-comings of the classic Higgs boson couplings fit in the $\kappa$
framework~\cite{Duhrssen:2004cv,Lafaye:2009vr}:
\begin{enumerate}
\item on the theory side it allows us to systematically include
  loop corrections, not only in perturbative QCD but also in the weak
  coupling;
\item on the experimental side it describes modified kinematic
  distributions, like for example the transverse momentum of the
  Higgs;
\item on the theory and experimental sides allows us to combine
  measurements in the Higgs sector for example with anomalous gauge
  couplings or low-energy precision measurements.
\end{enumerate}
The number of free parameters of the dimension-6 Higgs Lagrangian is an
extended set compared to the Higgs boson couplings ansatz. This larger set
of free parameters leads to strong correlations when we extract the
dimension-6 Wilson coefficients from the usual total cross sections
measured at the LHC. If, and only if the measured kinematic
distributions can be measured and predicted with similar accuracy as
total rates, they will resolve these degeneracies. This also means
that the marginalization of the multi-dimensional parameter space will
have a sizeable impact on the allowed range for a given dimension-6
Wilson coefficient.

In general, there appear two kinds of operators in the dimension-6
Lagrangian. For example operators simply adding $\phi^\dagger
\phi/\Lambda^2$ to the Standard Model do not change the Lorentz
structure of interactions, so they do not change kinematic
distributions. New operators including a derivative will after a
Fourier transformation lead to momentum-dependent Higgs boson couplings. 
Schematically written the two kinds of operators are
\begin{align}
\mathcal{O} \propto \frac{g_\Lambda^2 v^2}{\Lambda^2}
\qquad \text{and} \qquad 
\mathcal{O} \propto \frac{g_\Lambda^2 \partial^2}{\Lambda^2} \; .
\label{eq:operators}
\end{align}
Only studying total rates at the LHC, we can safely
assume that the series of higher-dimensional operators will be ordered
by factors $g_\Lambda^2 m_h^2/\Lambda^2$, were $g_\Lambda$ is the scale of the 
coupling to new physics. For a reasonably weakly interacting
theory with tree-level modifications, an assumed LHC accuracy of $10\%$ 
directly translates into a new physics reach around
\begin{align}
\left| \frac{\sigma \times \text{BR}}{\left( \sigma \times \text{BR} \right)_\text{SM}} - 1 \right|
= \frac{g_\Lambda^2 m_h^2}{\Lambda^2} \gtrsim 10\%
\qquad \Leftrightarrow \qquad 
\Lambda < \frac{g_\Lambda \, m_h}{\sqrt{10\%} } \, < 400~\gev \; .
\label{eq:lmax}
\end{align}
For this estimate we assume $g_\Lambda < 1$, corresponding to a (reasonable) weakly interacting extension of the Standard Model. For momentum-dependent modified Lorentz
structures the picture changes.  For them the suppression will depend
on additional energy scale in LHC processes, for example $g^2
p_{T,h}^2/\Lambda^2$. Taking the above values of $g < 1$ and $\Lambda
< 400$~GeV, the experimentally accessible transverse momentum
distributions $p_{T,h} > 400$~GeV will then receive order-one
corrections through dimension-6 operators. Depending on the sign of
the interference term between the Standard Model coupling and the
dimension-6 Wilson coefficient, this leading interference correction to the
differential rate can even drive the number of predicted events
through zero and negative.


\subsection{Measuring dimension-6 Wilson coefficients}
\label{sec:eft_fits}

While there exist many experimental and theoretical challenges to an
analysis of LHC Higgs data other than in complete, renormalizable
models, the crucial question is if we can sensibly measure dimension-6
Wilson coefficients, and if these results are useful.  The aim of this
section as well as the following Section~\ref{sec:weak_completions} is
to illustrate how
\begin{itemize}
\item a fit of dimension-6 Wilson coefficients to LHC Higgs data
  can be done (and has been done for Run~I data) by non-members of
  the ATLAS and CMS collaborations and based on published results;
\item kinematic distributions can significantly improve the
  multi-dimensional parameter fit by resolving strong correlations
  induced by total rate measurements;
\item communicating the relevant information, in particular
  related to kinematic distributions, is a challenge which needs to be
  resolved in close collaboration with the fitting projects;
\item the results of a dimension-6 fit can be translated into
  weakly interacting extensions of the Standard Model, and many of the
  theoretical issues are clearly separated from experimental uncertainties;
\item the language of dimension-6 Lagrangians can intuitively be
  linked to the structure of ultraviolet completions of the Standard
  Model gauge and Higgs sectors.
\end{itemize}
Unlike the more general discussion of Section~\ref{s.eftval} we focus
on weakly interacting extensions of the Higgs and gauge sector, and
how a dimension-6 Lagrangian approach can be useful in
practice.\bigskip

\begin{figure}[t]
\centering
\includegraphics[width=0.35\textwidth]{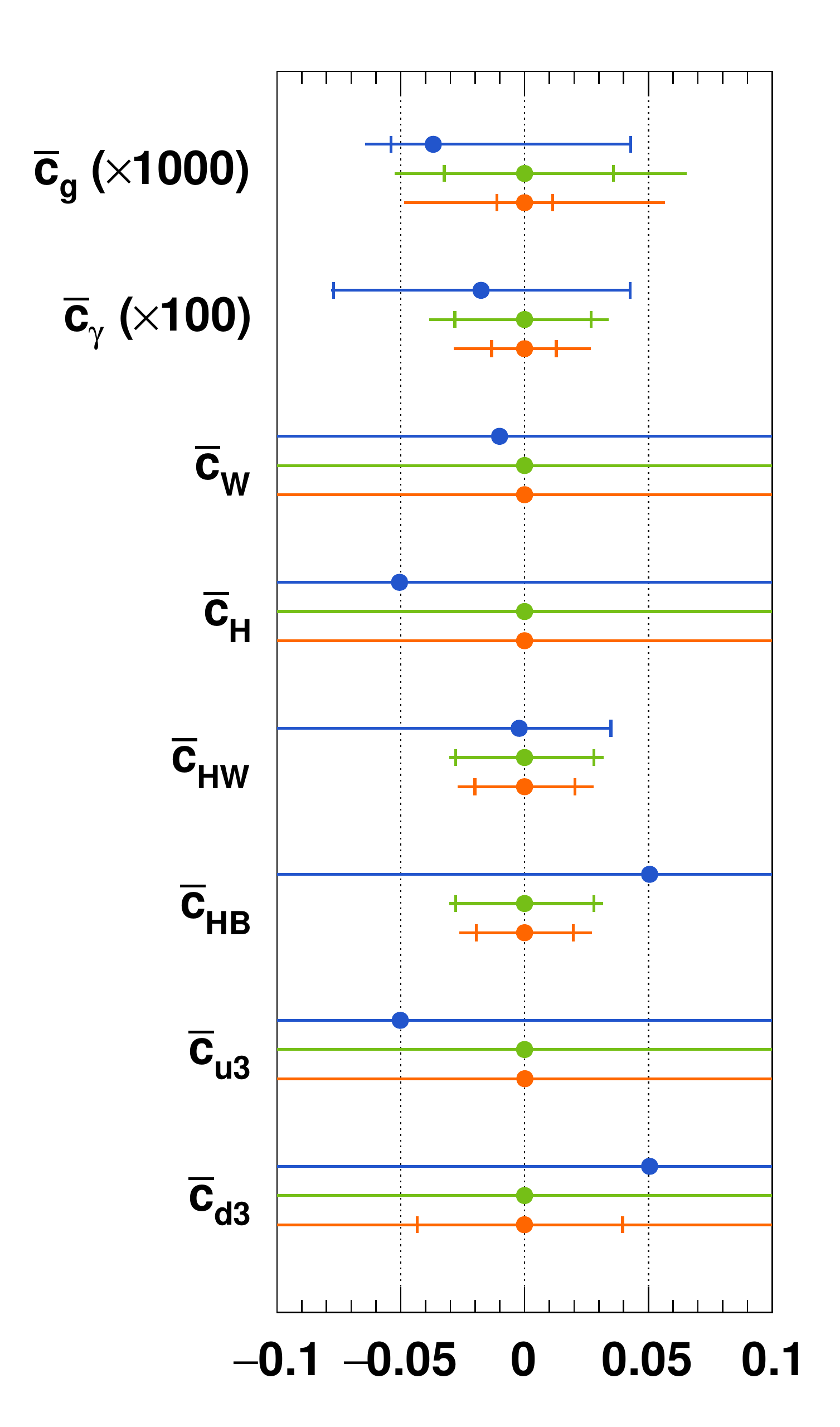}
\includegraphics[width=0.35\textwidth]{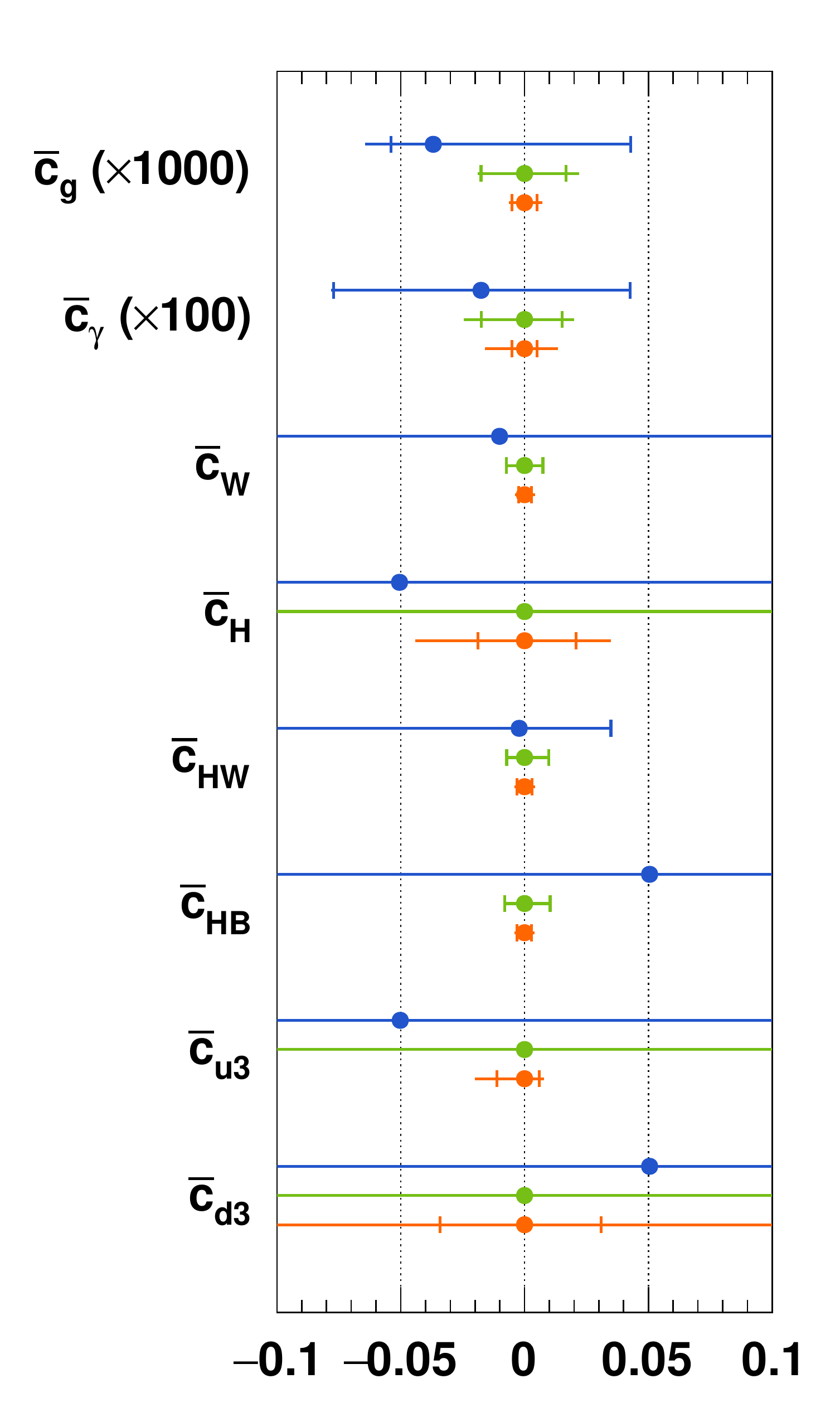}
\caption{Marginalized 95\% confidence level constraints for the
  dimension-six operator coefficients for current data (blue), the LHC
  at 14~TeV with $300~\text{fb}^{-1}$ (green), and with
  $3000~\text{fb}^{-1}$ (orange).  The expected constraints are
  centred around zero by construction.  For the left panel we only
  use signal strengths, while on the right differential $p_{T,h}$
  measurements are included.  The inner error bar depicts the
  experimental uncertainty, the outer error bar shows the total
  uncertainty. Figure from Ref.~\cite{Englert:2015hrx}.}
\label{fig:summary_hpt} 
\end{figure}

From the extraordinarily successful and well-established $\kappa$
framework we know that measurements of Higgs boson couplings at the level of
several per cent can be expected from the upcoming LHC run(s)~\cite{Corbett:2015ksa}. Towards
higher luminosity the Higgs boson couplings to weak bosons will likely be
the best-measured parameters, also because the theoretical
uncertainties linked to the corresponding LHC production cross
sections are under control. 

Already in the $\kappa$ framework, a few select kinematic
distributions for example in the gluon fusion production process can
be used to collect information on modified Higgs boson couplings.  In the
top-gluon-Higgs sector we can compare three different analysis
strategies: a modified $p_{T,h}$ spectrum of boosted Higgs boson production
in gluon fusion~\cite{Banfi:2013yoa}, off-shell Higgs boson production, and
a measurement of the gluon fusion vs $t\bar{t}h$ production rates.
Unfortunately, explicit threshold effects in boosted Higgs boson production
are too small to be observable in the near
future~\cite{Buschmann:2014twa}. We can compare the different methods
for a simple benchmark model where a 30\% reduction in the top Yukawa
coupling is compensated in the total rate through an effective
Higgs-gluon coupling.  Ignoring anything but statistical
uncertainties, boosted Higgs boson production can rule out this scenario
based on $700~\ifb$ of LHC data, while off-shell Higgs boson production will
require more than $1~\text{ab}^{-1}$ for the same
purpose~\cite{Buschmann:2014sia}. These numbers are expected to become
significantly worse once we include systematic and theoretical
uncertainties. Unfortunately, global analyses including kinematic
information in all Higgs channels cannot rely on the $\kappa$
framework, but they can be based on a Higgs EFT. Below, we will describe
the potential and the challenges in such analyses.\bigskip

When focussing on kinematic distributions,
the first choice we need to make concerns the 
the dimension-6 squared terms in our Lagrangian.  Writing the Lagrangian as $\mathscr{L}
= \mathscr{L}_\text{SM} + c/\Lambda^2 \, O_c $ we compute the
amplitude and the matrix element squared to leading order in the new
interactions,
\begin{align}
\mathcal{M} = \mathcal{M}_\text{SM} + \frac{c}{\Lambda^2} \mathcal{M}_\text{D6} 
\qquad \Rightarrow \qquad
|\mathcal{M}|^2 
= |\mathcal{M}_\text{SM}|^2 
+ \frac{2c}{\Lambda^2} \, \text{Re} \mathcal{M}_\text{D6}^\ast \mathcal{M}_\text{SM} 
\stackrel{?}{+} \frac{c^2}{\Lambda^4} \; |\mathcal{M}_\text{D6}|^2 \; .
\label{eq:d6}
\end{align}
In the following, we will discuss two analyses choosing different options concerning the dimension-6 squared terms.

\subsubsection*{Linearized EFT analysis}

Following pure power counting, the squared dimension-6 term
enters as a dimension-8 contribution. If sizeable, it can signal a
breakdown of the systematic EFT approach.  Drawing inspiration from
fixed-order QCD calculations, where negative event weights are present
and are interpreted as a shortcoming of this particular order of the
perturbative series expansion, we only include the interference term
and require that the differential distributions have positive cross
sections for each of the bins.  A prediction of negative event rates
(or a destructively interfering correction of the same order as the SM
expectation) signalizes a breakdown of the perturbative series in a
particular Wilson coefficient, so the underlying model should be
disregarded.  In other words, negative differential cross sections
can be used as an estimate of the validity range of a certain Wilson
coefficient, unless the symmetries or accidental cancellations render
the interference in Eq.\eqref{eq:d6} zero or negligibly small.\bigskip

\begin{figure}[t]
\begin{center}
\includegraphics[width=0.35\textwidth]{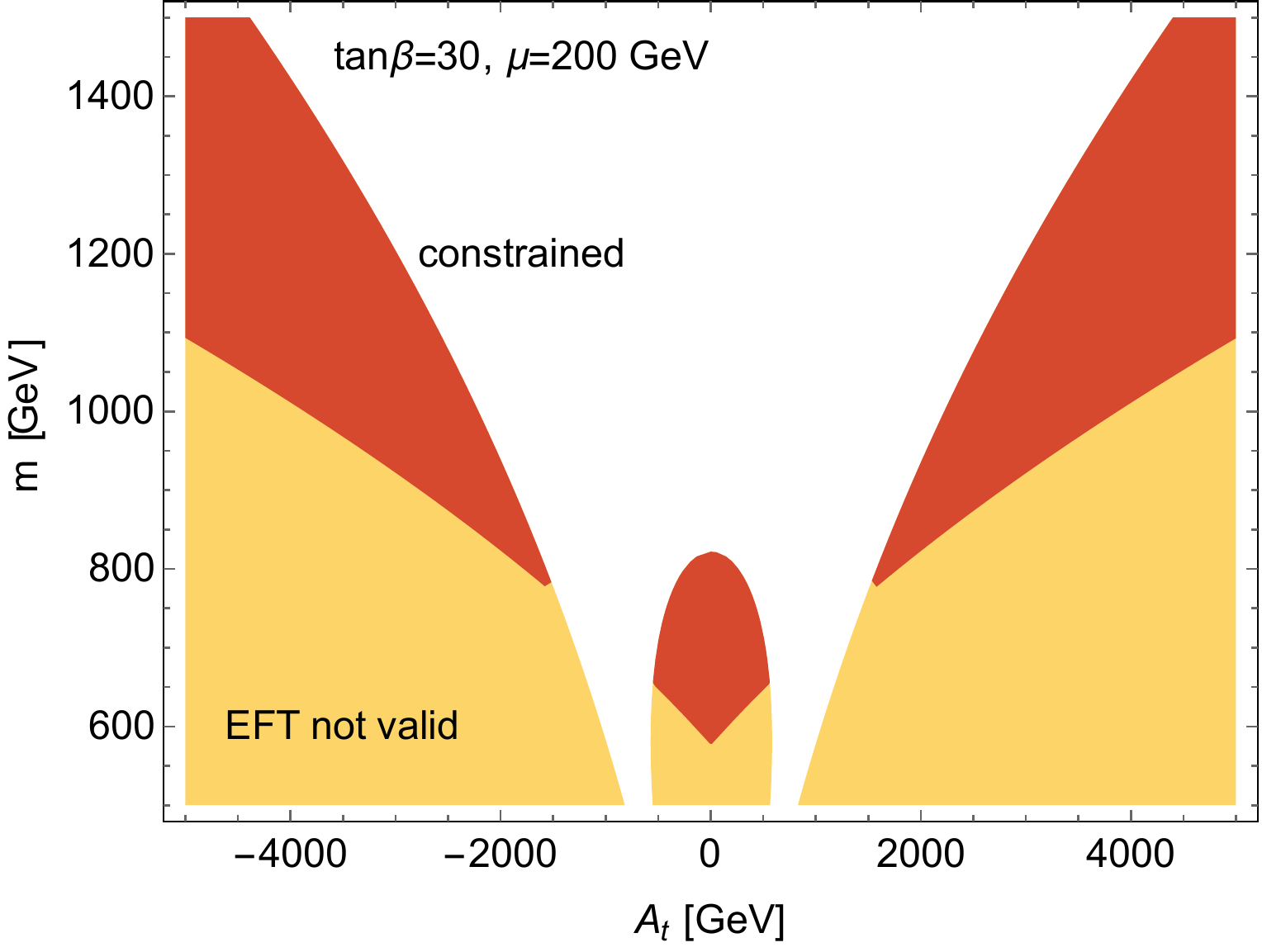}
\end{center}
\caption{Constraints appearing through $h^\dagger h \,
  G^{a\mu\nu}G^{a}_{\mu\nu}$ corresponding to MSSM stops, assuming
  that no hints for stops exist. The excluded parameter range is
  indicated by the red region, while the EFT consistency condition
  removes the orange region.  Figure from Ref.~\cite{Englert:2015hrx}.
}
\label{fig:stops} 
\end{figure}

For Run~I data as well as for different 14~TeV luminosity benchmarks
this approach has been adopted in the Higgs EFT fit of
Ref.~\cite{Englert:2015hrx}.  Its particular emphasis on the
question how differential distributions lift degeneracies in a fit to
the full set of Higgs-related Wilson coefficients. To this end,
$p_{T,h}$ distributions were added to all Higgs boson production and decay
signatures.  The results shown in \refF{fig:summary_hpt} indicate
that for Run~I the EFT approach is in poor shape without this
additional information. The EFT energy scales $\Lambda$ which can be
probed lie in the few hundred GeV range, quickly driving constraints
on actual TeV-scale models into a non-perturbative regime. Even worse,
energy scales in the same range are already resolved in the $p_{T,h}$
distributions, which strictly speaking invalidates the ideal EFT
approach.  The picture starts to change for Run~II at 14~TeV with an
assumed $300~\ifb$, in which case already the fit to total rates gives
more meaningful results. Finally, for the high-luminosity running with
$3000~\ifb$ the kinematic distributions clearly dominate the precision
of the expected limits. This conclusion rests on vastly improved
experimental systematic uncertainties and good control over
theoretical uncertainties. The experimental systematic uncertainties
are assumed to scale like statistical uncertainties with the squared
root of the number of events. For example, a 50\% uncertainty assigned
to a signature at 7-8~TeV and $25~\ifb$ can turn into a
$50\%/\sqrt{200} \approx 3.5\%$ uncertainty at 14~TeV assuming
$3000~\ifb$. The theoretical uncertainties are not reduced for the
Run~II scenarios, but nevertheless a stringent assumption is made,
that they are flat as function of $p_{T,h}$.  In that sense the
results of \refF{fig:stops} support the statement that it might
well be possible to rely on a consistent Higgs EFT approach in the
long-term of LHC running, if the systematic and theoretical
uncertainties are controlled.  Interpreting the projections for
high-luminosity running in a concrete decoupled stop scenario we
obtain \refF{fig:stops}.

\subsubsection*{Run~I analysis including distributions}

\begin{figure}[t]
  \centering
  \includegraphics[width=0.65\textwidth]{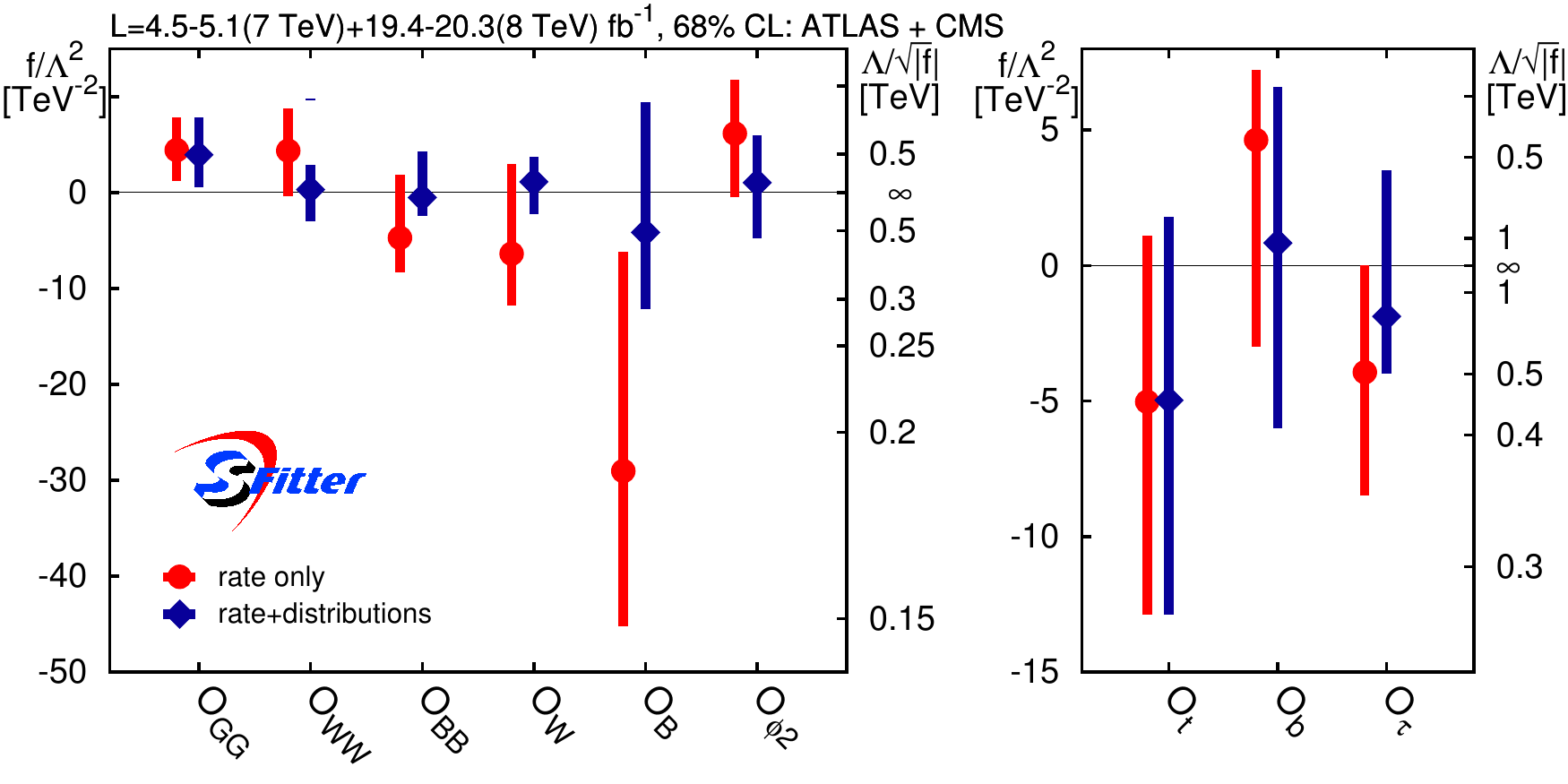}
  \caption{68\% CL error bars on the Wilson coefficients
    $f_x/\Lambda^2$ for the dimension-6 operators. For the Yukawa
    couplings as well as for $\mathcal{O}_{GG}$ we only show the SM-like
    solution. Figure from Ref.~\cite{Corbett:2015ksa}.}
\label{fig:dim6kin}
\end{figure}

\begin{figure}[t]
  \includegraphics[width=0.29\textwidth]{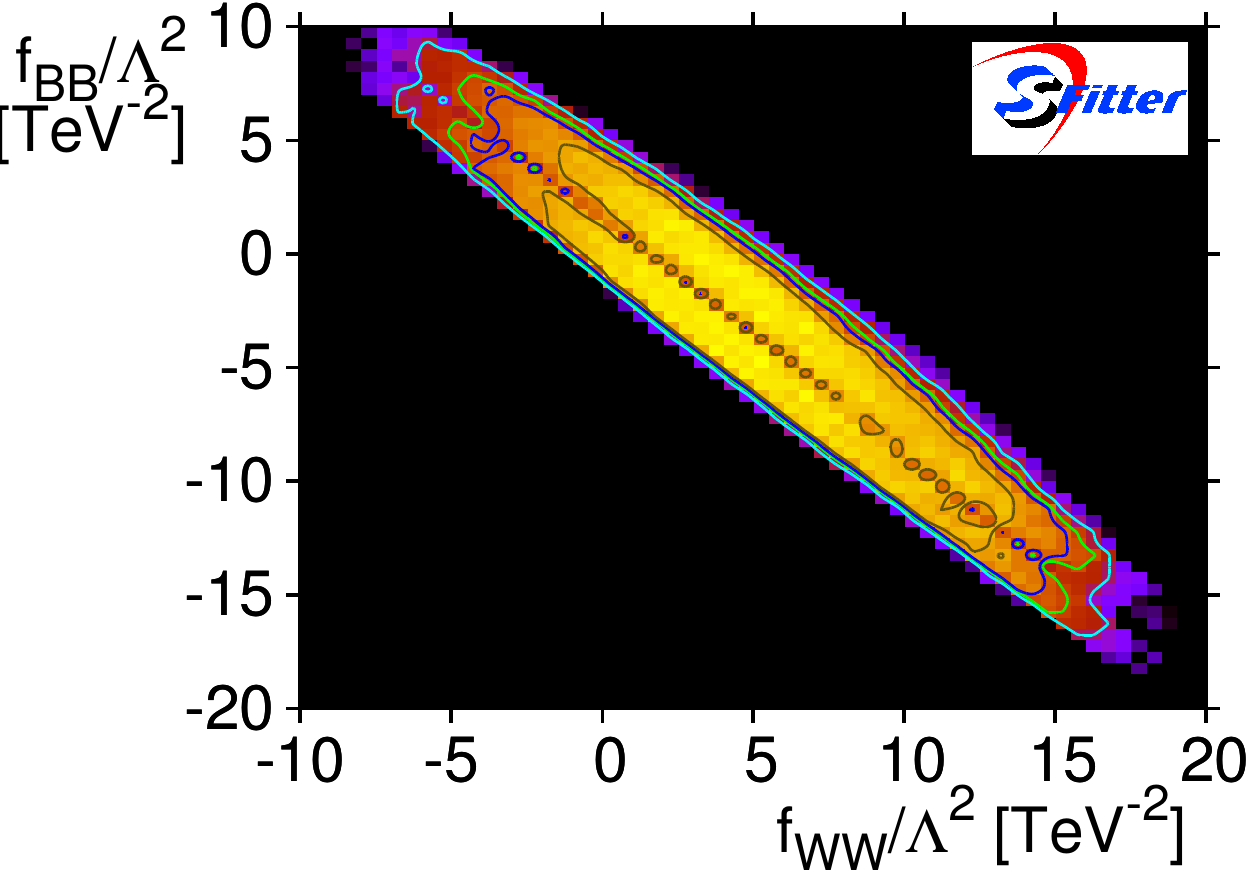} 
  \hspace*{1ex}
  \includegraphics[width=0.29\textwidth]{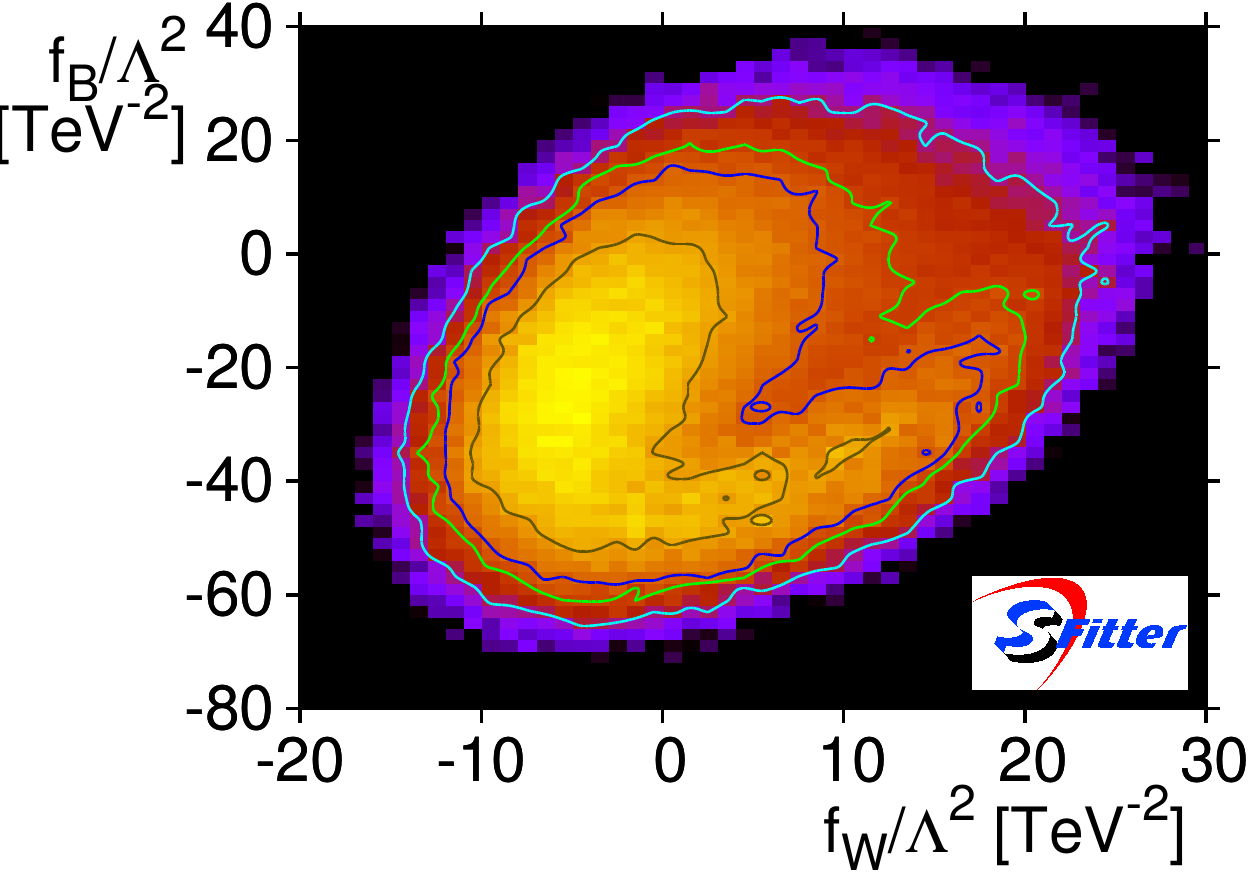} 
  \hspace*{1ex}
  \includegraphics[width=0.29\textwidth]{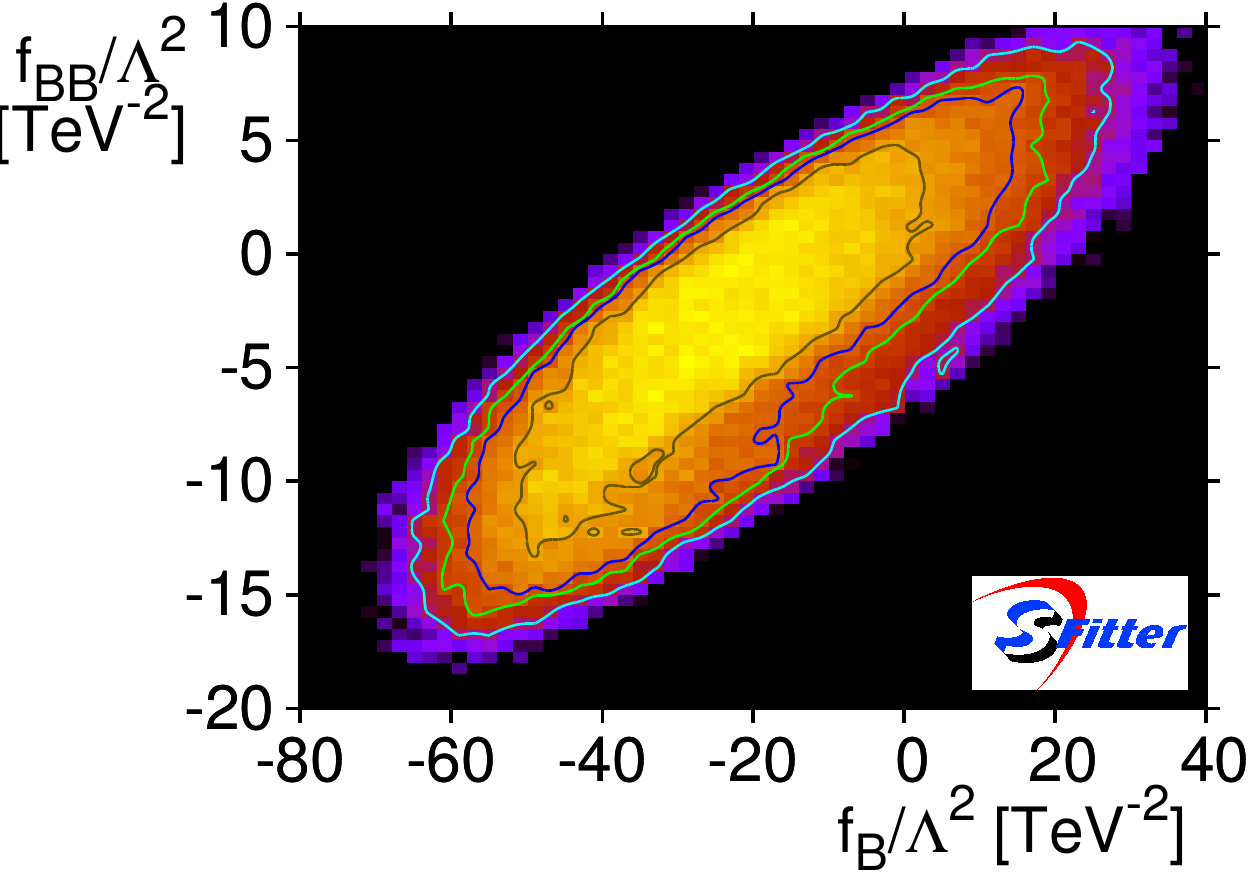}
  \hspace*{1ex}
  \raisebox{3pt}{\includegraphics[width=0.0545\textwidth]{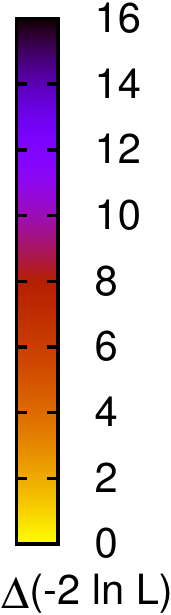}}\\[1ex]
  \includegraphics[width=0.29\textwidth]{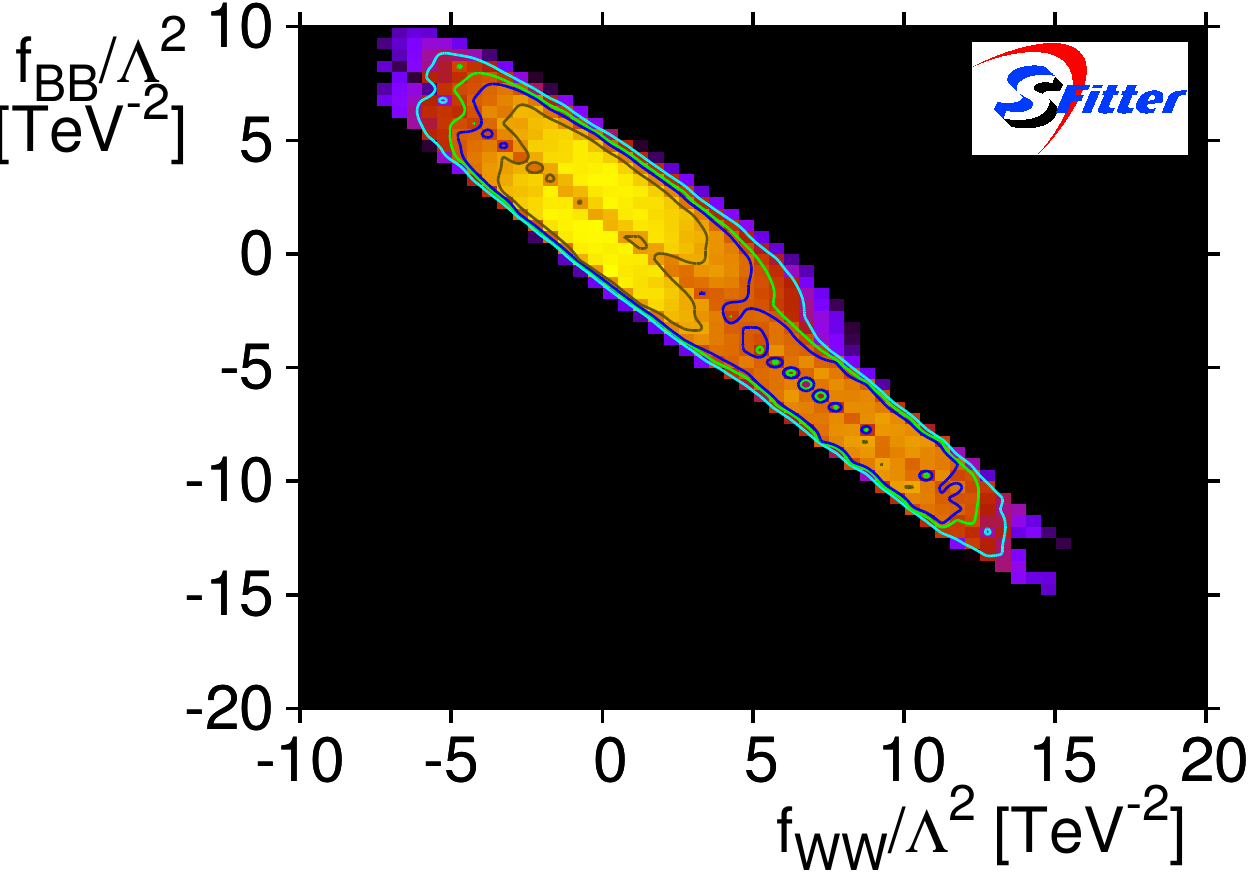} 
  \hspace*{1ex}
  \includegraphics[width=0.29\textwidth]{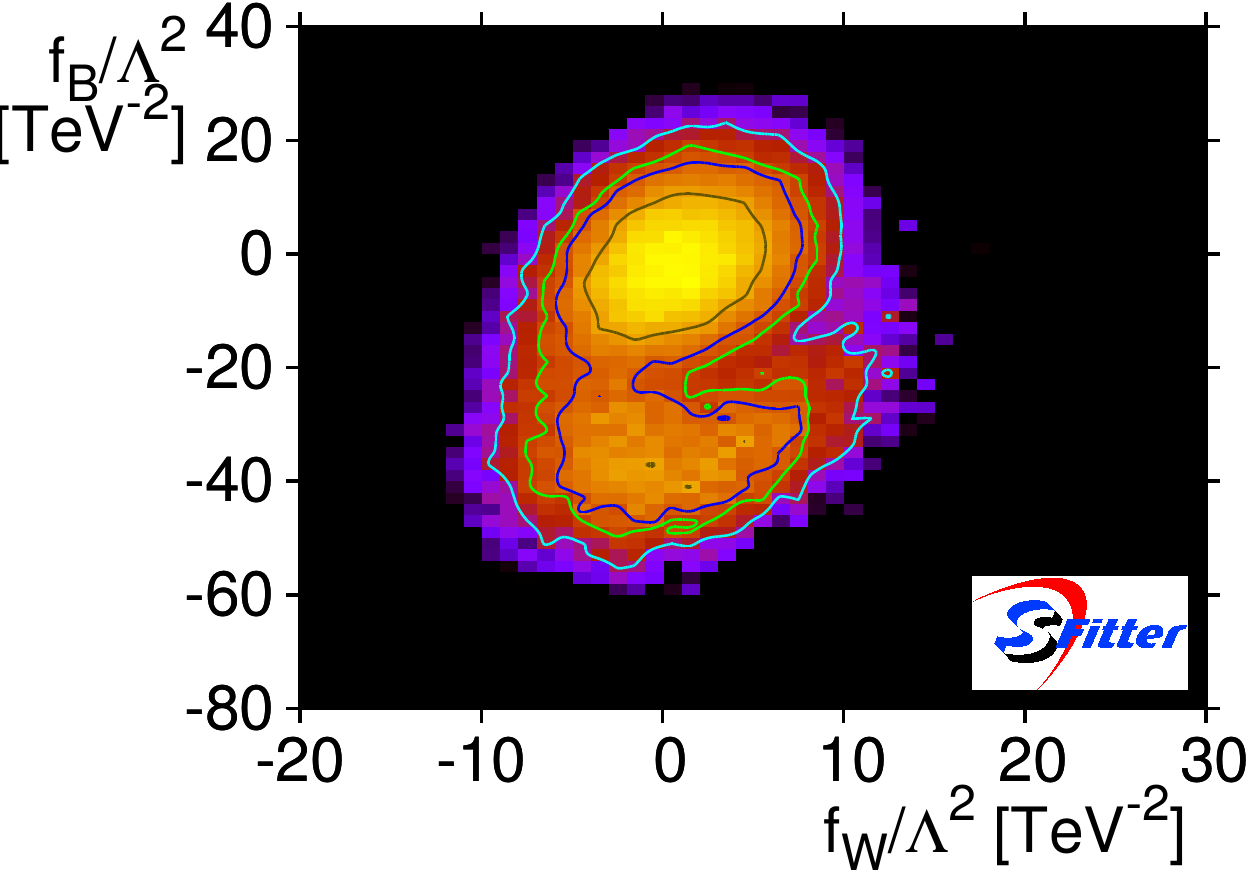} 
  \hspace*{1ex}
  \includegraphics[width=0.29\textwidth]{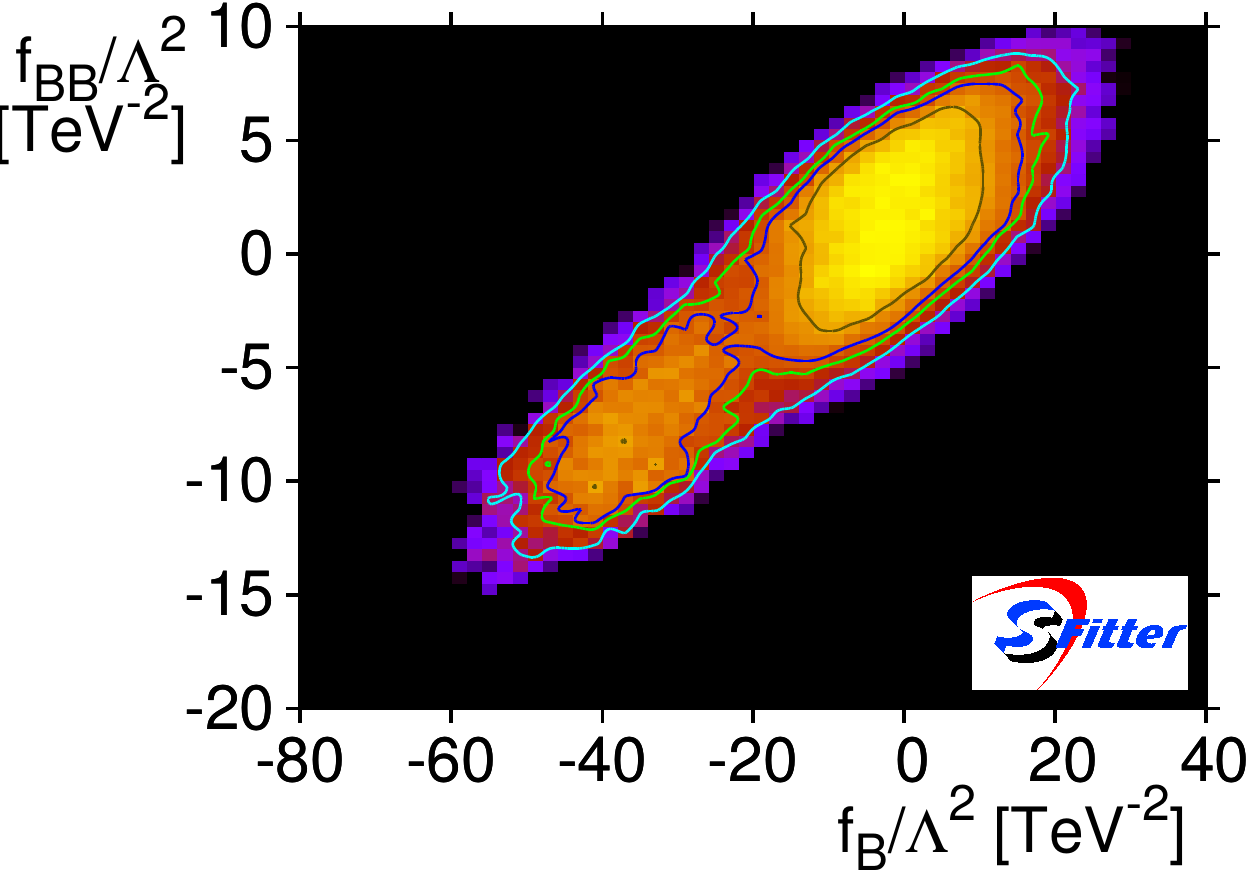}
  \phantom{\hspace*{1ex}
  \includegraphics[width=0.0545\textwidth]{WG2/WG2_6_EFT_Models/colorbox}}\\[1ex]
  \includegraphics[width=0.29\textwidth]{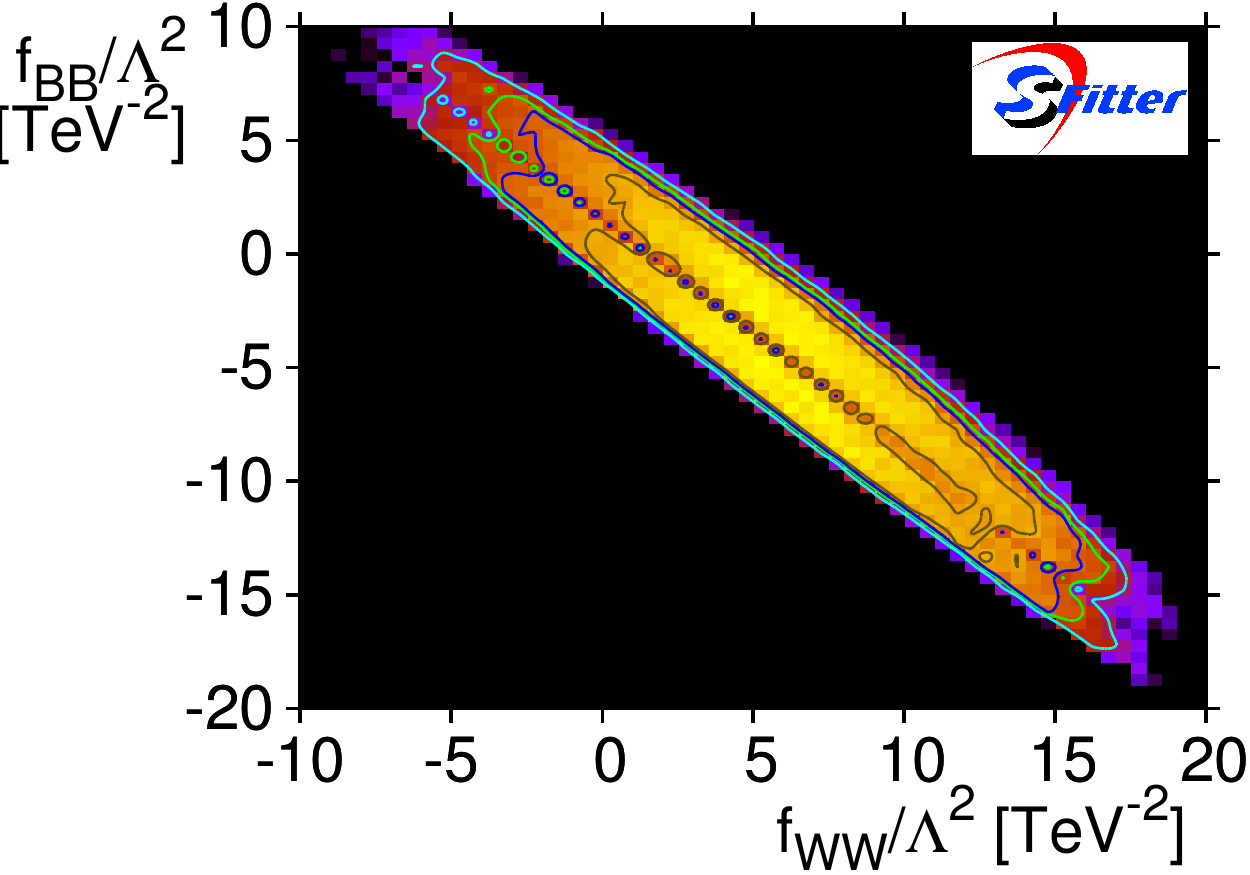}
  \hspace*{1ex}
  \includegraphics[width=0.29\textwidth]{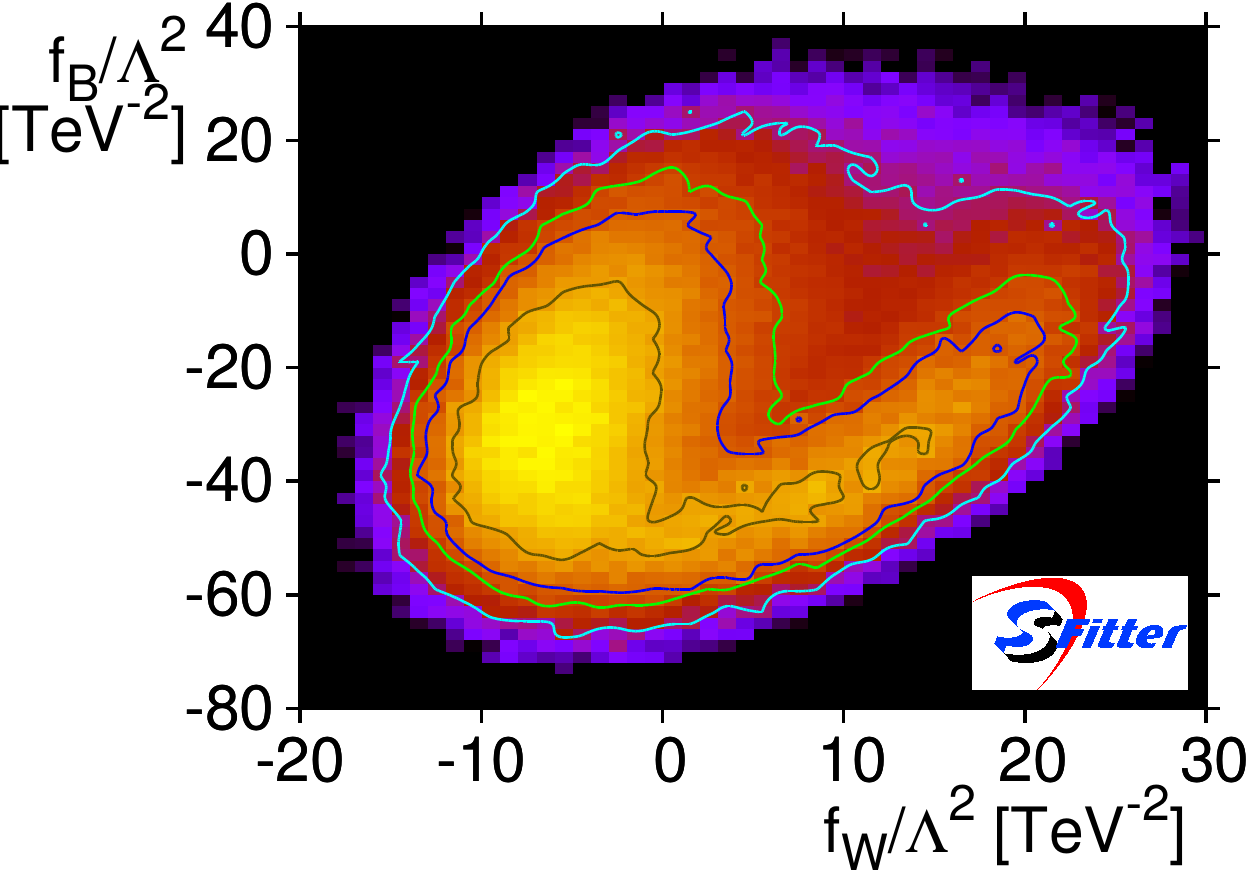}
  \hspace*{1ex}
  \includegraphics[width=0.29\textwidth]{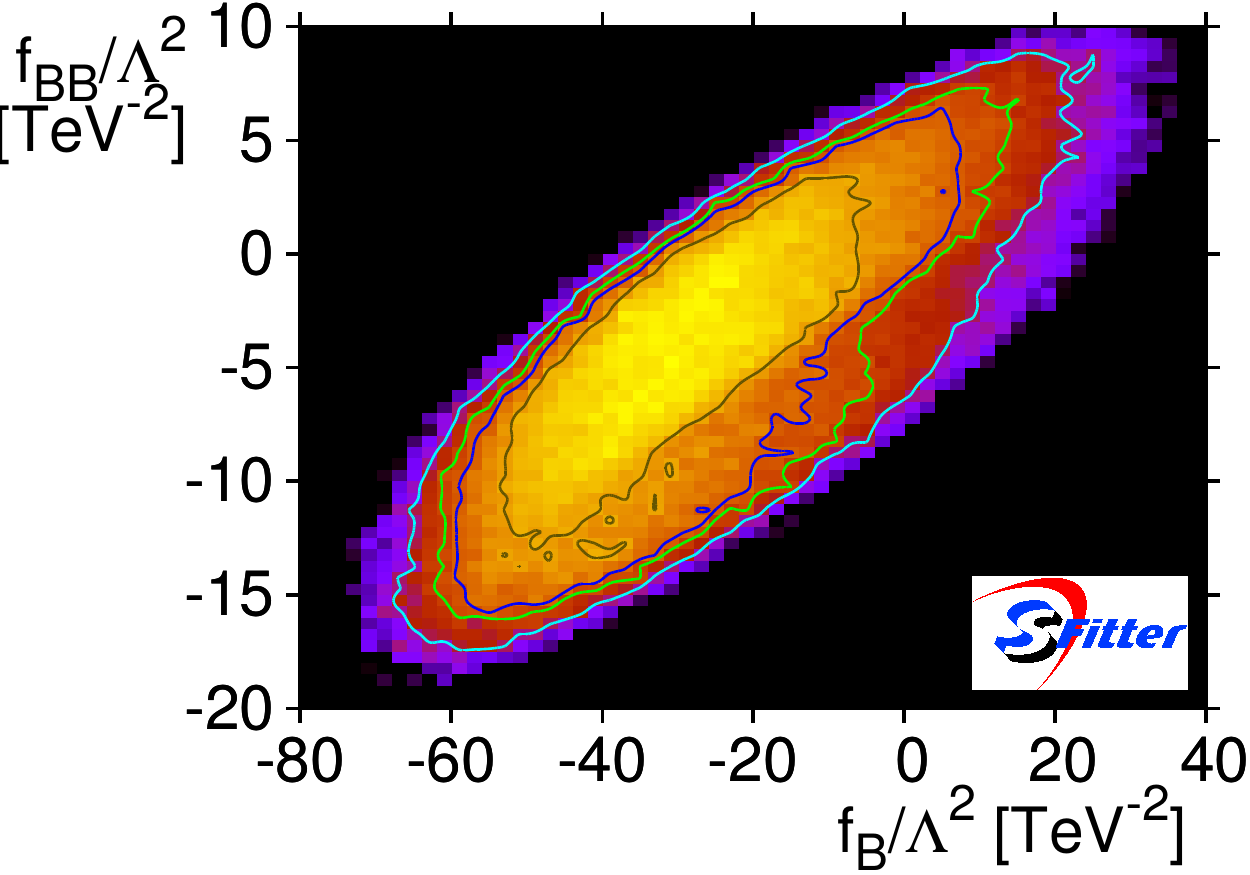}
  \phantom{\hspace*{1ex}
  \includegraphics[width=0.0545\textwidth]{WG2/WG2_6_EFT_Models/colorbox}}
  \caption{Correlations between different coefficients $f_x/\Lambda^2$
    after including kinematic distributions. In the top row we only
    add the $\Delta \phi_{jj}$ distribution; in the second row we also
    include $p_{T,V}$ from $Vh$ production; in the bottom row we then
    remove the highest bin associated with large momentum flow through
    the dimension-6 vertex. Figure from Ref.~\cite{Corbett:2015ksa}.}
\label{fig:dim6_corrkin}
\end{figure}

Also based on the Run~I Higgs measurements, \refF{fig:dim6kin} 
shows the \textsc{SFitter}
analysis~\cite{Corbett:2015ksa} of the dimension-6 Lagrangian defined in
Refs.~\cite{Hagiwara:1993ck,Corbett:2012ja}. Unlike in the previous analysis, the
dimension-6 squared terms indicated in Eq.\eqref{eq:d6} are included.
The red error bars are only based on total rate measurements, while
the blue bars includes the $p_{T,V}$ distribution in $Vh$ production
as well the the $\Delta \phi_{jj}$ distribution in weak boson fusion
with a decay $h \to \gamma \gamma$. The Higgs-fermion sector is
limited to modified Yukawa couplings of the 3rd generation, due to
scarce experimental data probing other Yukawa couplings or the
structure of the Higgs boson couplings to heavy fermions.  Without any
information from the distributions the constraints on the Wilson
coefficients can be translated into $\Lambda \gtrsim 300$~GeV for $g =
1$. Including kinematic distributions the reach increases to typically
$\Lambda \gtrsim 500$~GeV, where the entire improvement comes from the
overflow bin of the $p_{T,V}$ distribution defined as $p_{T,V} >
200$~GeV. As discussed before, this limited mass reach implies that the
dimension-6 fit should not be viewed as the leading term of an EFT
expansion in $1/\Lambda$.\bigskip

Technically, this kinematic information is included based on published
kinematic distributions. Looking at future LHC runs there are several
ways of making kinematic information available to 
an independent Higgs operator analysis:
\begin{enumerate}
\item kinematic distributions compared to different signal and
  background predictions shown bin-by-bin. In that case the event
  counts in each bin are statistically independent measurements;
\item unfolded signal distributions as discussed in Chapter~\ref{chap:FXS}, which are technically simple to
  include, but might introduce statistical correlations between bins and an additional
  systematic uncertainty from the unfolding model;
\item fiducial cross sections as discussed in Chapter~\ref{chap:FXS},
  which are also technically simple to include, but might provide less
  information than the full distributions;
\item exclusive likelihood maps, which are highly processed by the
  experiments and remove the treatment of experimental uncertainties
  from the operator analysis. Theoretical uncertainties can be
  decoupled using the procedure presented in
  Ref.\cite{Cranmer:2013hia}.
\end{enumerate}
For the Run~I analysis shown here we rely on the first and third of
these methods, \textsl{i.e.}  on published kinematic distributions
compared to signal and background predictions.  The theoretical
predictions for the distributions come from \textsc{Madgraph} and can
trivially be translated from one operator basis to another. Because
the measured total rate in a given Higgs boson production and decay channel
does not correspond to the integral under the published distributions,
we define ratios of bins such that the measured total rate is not
related to the measured distribution. To ensure that kinematic
distributions to not develop negative event counts we have to include
the dimension-6 squared contribution, even though from a technical
perspective we could also find other ways to remove problems in the
interpretation of these phase space regions.  The technical setup of our
fit to a linear representation can easily be extended to a non-linear
representation, where additional degrees of freedom appear in
the anomalous gauge sector and the fermionic operators~\cite{Corbett:2015mqf}.  All operators
in the fit are defined and evaluated at the weak scale, but can
obviously be renormalization-group evolved to other, experimentally
relevant scales of an extended fit.\bigskip

In \refF{fig:dim6_corrkin} we show how kinematic distributions
improve the dimension-6 fit because they reduce strong non-Gaussian
correlation, for example between $\mathcal{O}_B$ and $\mathcal{O}_W$ or between
$\mathcal{O}_{BB}$ and $\mathcal{O}_{WW}$ in model space.  Using a profile
likelihood analysis leads to poor limits on each of the two operators
individually. If we were to vary only one operator at a time, the limit
for example on the Wilson coefficients for $\mathcal{O}_{BB}$ or $\mathcal{O}_{WW}$
would improve by an order of magnitude. The top panels show the
results after including total rates and the $\Delta \phi_{jj}$
distribution in weak boson fusion.  The 1-dimensional profile
likelihoods from this setup largely correspond to the red bars in
\refF{fig:dim6kin}.  In the second row we show the improvement
from the $p_{T,V}$ distribution in $Vh$ production. This includes an
overflow bin, where at least the lower end of the bin is not beyond
the region of validity of the effective field theory.  The
corresponding 1-dimensional profile likelihoods are shown as blue bars
in \refF{fig:dim6kin}. The main source of improvement in the
1-dimensional results is an improved control of correlations when we
include kinematic distributions as additional observables.
It turns out that after including LHC measurements on pair-produced weak bosons these non-trivial correlations essentially vanish~\cite{Ellis:2014dva,Ellis:2014jta,Butter:2016cvz}.
 In the
bottom row of \refF{fig:dim6_corrkin} we illustrate what happens if
we remove the overflow bin of the $p_{T,V}$ distribution from our
analysis: this limited scenario is essentially equivalent to not
including the $p_{T,V}$ distribution at all.\bigskip

Finally, we can include off-shell Higgs boson production in a 
global Higgs analysis, either based on the $\kappa$ framework or based on a
dimension-6 Lagrangian~\cite{Corbett:2015ksa}. The tree-level $ZZh$
vertex will only run logarithmically with the relevant momentum scale,
and additional operator structures changing the $ZZh$ interaction will
hardly modify the $m_{4\ell}$ kinematics. What remains is a combined
effect of the gluon-Higgs boson coupling described by $\mathcal{O}_{GG}$ and
a modified top Yukawa coupling through $\mathcal{O}_t$. Hence, the
main effect of including off-shell Higgs boson production in the
\textsc{SFitter} Higgs analysis is to again reduce possible
degeneracies in the $\mathcal{O}_{GG}$ vs $\mathcal{O}_t$ plane. The
same would be the effect of including a transverse momentum
distribution in gluon fusion Higgs boson production. Technically, ATLAS and
CMS report off-shell Higgs results in terms of rate measurements in a
given phase space region. This approach is related to fiducial cross
section measurements on the above list of possible formats. In the
future, the same information might be published in the form of $m_{4
  \ell}$ distributions, which could be easily accommodated by the
global analysis tools.

\subsubsection*{Outlook}

Current studies have shown that a dimension-6 analysis of 
LHC Higgs data is entirely feasible. Only when we try to
embed the truncated dimension-6 analysis into a consistent EFT
framework we face major issues like theoretical uncertainties due to missing higher-dimensional contributions. One way to test the validity of the EFT description is the
treatment of dimension-6 squared terms in Monte Carlo simulations ---
on the one hand including these terms assures positive event numbers
all over phase space, but on the other hand large contributions from
these squared terms suggest poor convergence of the underlying
framework.  As long as we consider a global analysis only in terms of
dimension-6 operators the treatment of the dimension-6 terms is part
of the underlying hypothesis and only has to be stated clearly.  In
general, including kinematic distributions makes the EFT analysis more
vulnerable to such model assumptions. However, for increased luminosity the
hierarchy of scales will be improved at least for inclusive measurements, and the limiting theoretical
uncertainties might become smaller with time.


\subsection{Weakly interacting new physics to dimension-6}
\label{sec:weak_completions}

The discussion in the last section shows that an effective theory
approach to LHC Higgs data faces serious issues linked to 
theoretical consistency arguments, triggered by the limited 
reach of the LHC Run~I in terms of the new physics scale. 
However, these limitations do not
necessarily imply that we cannot interpret LHC Higgs data, including
kinematic distributions, in terms of a dimension-6 Lagrangian. This
has to be tested explicitly for different structures of the new
physics model. For pedagogical reviews see \textsl{e.g.} Refs~\cite{Kilian:2003pc,Englert:2014uua,Falkowski:2015fla}. We study weakly interacting modifications of the
Higgs sector and the electroweak gauge sector~\cite{Brehmer:2015rna},
\begin{itemize}
\item singlet extension of the Higgs potential, a so-called
  Higgs portal with Higgs-scalar mixing;
\item two-Higgs-doublet model, including the specific type-2 setup in
  the MSSM;
\item non-Higgs extension of the scalar sector, for example through
  scalar top partners;
\item gauge-triplet extension of the electroweak gauge sector.
\end{itemize}
For each of these models we construct and match the linear dimension-6
Lagrangian, compute LHC observables, and compare the LHC predictions
from the dimension-6 Lagrangian and from the full model. The
(dis-)agreement of these predictions determines to what degree we can
rely on a dimension-6 analysis at the LHC, without explicitly testing for
example dimension-8 operators. The point where our analysis differs
from the general consistency arguments of the last section is that for
our classes of models we can derive the structure and the size of the
dimension-6 Wilson coefficients. Moreover, a specific set of benchmark
models allows us to make quantitative statements.\bigskip

One key ingredient to our analysis is the appropriate matching of
models which, to be observable at the LHC, have to feature a low new
physics scale $M_\text{heavy}$ in the extended Lagrangian.  In an
ideal world, where the effective theory is well defined and we can
systematically neglect terms suppressed by $g^2 v^2/M_\text{heavy}^2$,
the matching of the effective field theory is uniquely defined. First,
the masses of the new particles will be of the order $m_\text{heavy} =
M_\text{heavy}$ and will serve as the natural matching scale
$\Lambda$. Second, at this scale $\Lambda = M_\text{heavy}
\gg v$ we match to the dimension-6 Lagrangian in the unbroken phase
and assume $v \to 0$ throughout. For a dimension-6 truncation this
also means that we neglect terms of the order $1/\Lambda^4$ in the
Wilson coefficients and terms of the order $1/M_\text{heavy}^4$ in the
full model, which renders the matching condition unique.

For many models we are interested in, the underlying mass scale
and the masses of new particles will be linked like
\begin{align}
m_\text{heavy} = M_\text{heavy} \pm g v + \cdots
\label{eq:v-improvement}
\end{align}
Our choice of $\Lambda$ is now driven by the phenomenological argument
that the effective theory will break down the moment we observe a new
resonance. Independent of the hierarchy of the two new physics mass
parameters we define a $v$-improved matching at the scale
\begin{align}
\Lambda = m_\text{heavy} \; .
\end{align}
Similarly, at this scale we want to incorporate as much information of
the full model in the Wilson coefficients as possible. This is why we
express them in terms of all-order model parameters like masses and
mixing angles, \textsl{i.e.} we do not truncate them in terms of $1/\Lambda$.
This $v$-improved matching leads to a significant improvement in the
comparison of the dimension-6 Lagrangian and the full model in LHC
simulations, until actual poles of new particles
appear~\cite{Brehmer:2015rna}.\bigskip

Because differential information on LHC processes is not described by one
single mass scale, we need to test each of our new physics models
based on the critical kinematic distributions.  For each of these
weakly interacting new physics models we therefore study
\begin{itemize}
\item Higgs boson decays, where the momentum flow through the vertices is
  typically smaller than $m_h$ and we can study new
  physics effects in the $m_{4 \ell}$ distribution;
\item $Vh$ production, where the momentum flow through the $hVV$
  vertex can be reconstructed as $m_{Vh}$, and where we can search for
  new gauge resonances in the $s$-channel;
\item weak boson fusion production, where the $2 \to 3$ kinematics
  with two $t$-channel propagators complicates the experimental access
  to momentum flow for example based on the leading $p_{T,j}$;
\item Higgs boson pair production, where issues arise not at large energies,
  but at threshold, and where we are likely to observe 
  new Higgs boson resonances in $m_{hh}$.
\end{itemize}
The list of models and the list of observables define a matrix which
allows us to systematically study the level of agreement between full
models and the dimension-6 Lagrangian at the LHC. More information on
such new physics models can be found in Section~\ref{s.eftmodels}. Unlike
in that section, the benchmark point described below are specifically
designed to challenge an effective theory description of the LHC
features of the full models.

\subsubsection*{Singlet and doublet extensions}

First, we extend the minimal Higgs sector of the Standard
Model by a real scalar singlet $S$,
\begin{alignat}{5}
V(\phi,S) = 
  \mu^2_1\,(\phi^\dagger\,\phi) 
+ \lambda_1\,|\phi^{\dagger}\phi|^2 
+ \mu^2_2\,S^2
+ \lambda_2\,S^4 
+ \lambda_3\,|\phi^{\dagger}\,\phi|S^2 \,,
\label{eq:singlet-potential}
\end{alignat}
where the singlet VEV $v_s \gg v$ induces a Higgs-singlet mixing. All
Higgs boson couplings, with the exception of the Higgs self-coupling, are
modified as
\begin{align}
\Delta_x 
= \frac{g_{xxh}}{g_{xxh}^\text{SM}} -1 
= \cos\alpha -1 
\approx - \frac{\alpha^2}{2} 
\approx - \frac{\lambda_3^2}{8 \lambda_2^2}\, \left( \frac{v}{v_S} \right)^2 
\qquad \text{with} \qquad 
\tan\alpha \approx \frac{\lambda_3}{2\lambda_2}\,\frac{v}{v_s} 
\ll 1 \; .
\label{eq:singlet-shift}
\end{align}
In our dimension-6 basis the singlet extension only induces the
operator $\mathcal{O}_H$ or $\mathcal{O}_{\phi,2} = \partial^\mu
(\phi^\dagger \phi) \partial_\mu (\phi^\dagger \phi)$. The matching
scale and the corresponding Wilson coefficient from $v$-improved
matching are
\begin{align}
\Lambda = m_H = \sqrt{2 \lambda_2} v_s
\qquad \text{and} \qquad 
\bar{c}_H = 2 (1 - \cos \alpha) \; .
\label{eq:singlet_ch}
\end{align}
This means that a singlet extension of the Higgs potential does not
introduce momentum-dependent Higgs boson couplings.  The coupling
modification to all Standard Model particles is universal and only
leads to modified total Higgs rates.  In
\refT{tab:singlet_benchmarks} we show the parameters of our
benchmark points as well as the modification of the Higgs boson couplings
and the Higgs rates. In spite of the fact that the mass of the new
scalar $m_h$ can be very low, we hardly find visible effect in the
Higgs boson couplings. Moreover, the dimension-6 approximation to the
modified total rates is by definition fully justified, except for the
on-shell contribution from the second scalar shown in
\refF{fig:singlet_distributions}.\bigskip

\begin{table}[t]
  \caption{Benchmarks for the singlet extension. We show the model
    parameters and the universal coupling modification for the
    complete model, as well as the matching scale $\Lambda$, the
    Wilson coefficient $\bar{c}_H$, and the universal coupling
    modification for the dimension-6 Lagrangian. $m_H$ and $\Lambda$
    are in GeV. Table from Ref.~\cite{Brehmer:2015rna}.}
  \label{tab:singlet_benchmarks}
  \renewcommand{\arraystretch}{1.2}
  \centering
    \begin{tabular}{c c rccr c cc}
      \toprule
      \multirow{2}{*}{} &\hspace*{1em}& 
      \multicolumn{4}{c}{Singlet} &\hspace*{1em} \\
      \cmidrule{3-6} \cmidrule{8-9} 
                && $m_H$ & $\sin\alpha$ & $v_s/v$ & $\Delta_x^\text{singlet}$ 
                &&  $\bar{c}_H$ & $\Delta_x^\text{D6}$ \\
      \midrule
      S1 &&  500 & 0.2 & 10 & $-0.020$ && 0.040 &  $-0.020$ \\ 
      S2 &&  350 & 0.3 & 10 & $-0.046$ && 0.092 &  $-0.046$ \\ 
      S3 &&  200 & 0.4 & 10 & $-0.083$ && 0.167 &  $-0.083$ \\ 
      S4 && 1000 & 0.4 & 10 & $-0.083$ && 0.167 & $-0.092$ \\ 
      S5 && 500 & 0.6 & 10 & $-0.200$ &&  0.400 &  $-0.200$ \\ 
      \bottomrule
    \end{tabular}
\end{table}

\begin{figure}[b!]
  \centering
  \includegraphics[width=0.43\textwidth]{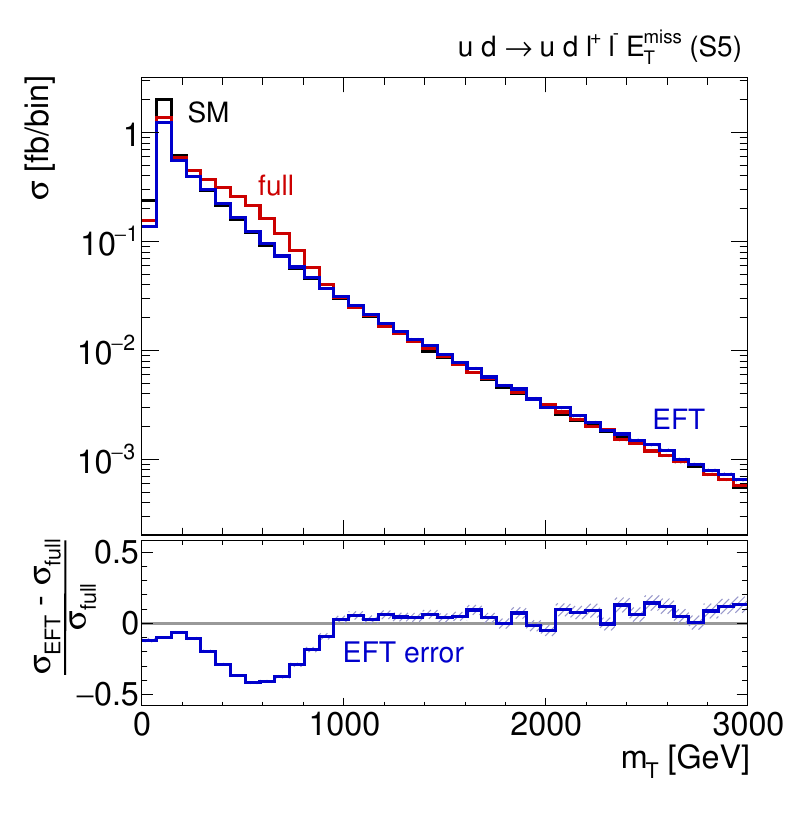}
  \hspace*{0.05\textwidth}
  \includegraphics[width=0.43\textwidth]{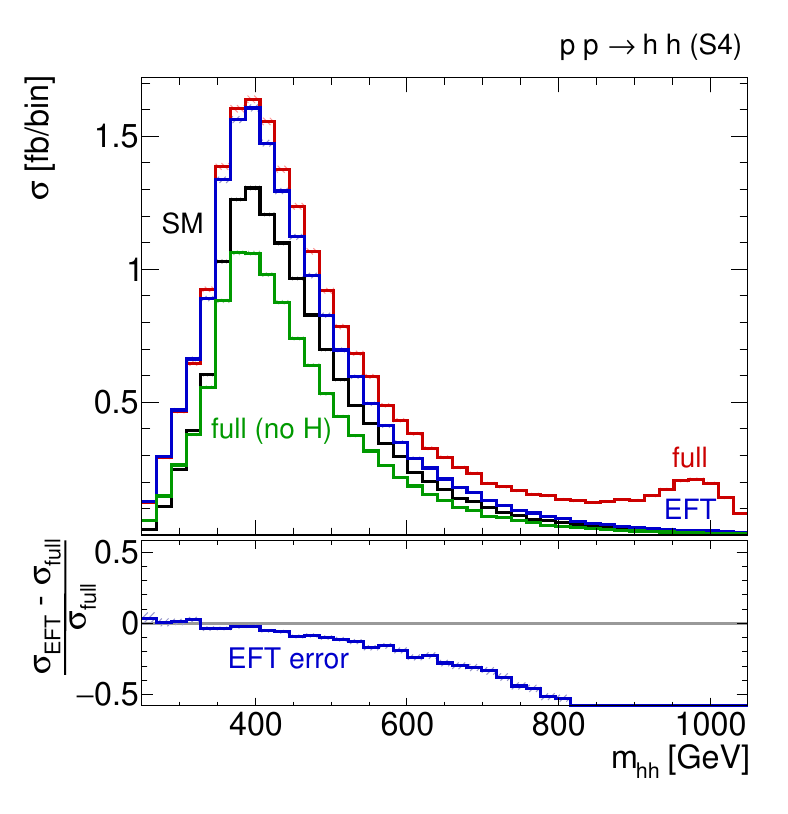}
  \caption{Kinematic distributions for the singlet extension. Left:
    transverse mass in WBF Higgs boson production. Right: invariant mass
    $m_{hh}$ linked to the new resonance. Figure from
    Ref.~\cite{Brehmer:2015rna}.}
  \label{fig:singlet_distributions}
\end{figure}

The only exception to this simple structure will be the appearance of
the new resonance $H$, for example in Higgs boson pair production $pp \to
hh$. If we do not remove the operator $\mathcal{O}_H$ from our operator basis
through equations of motion, it induces a momentum-independent and a
momentum-dependent correction to the Higgs self-coupling
\begin{align}
 \mathscr{L} \supset
 - \dfrac{m_{h}^2} {2v} 
    \left( 1 - \dfrac{1}{2} \bar{c}_H \right) 
    h^3 
   + \dfrac {g}{2 m_W} \bar{c}_H \; h \; \partial_\mu h \; \partial^\mu h \; .
\label{eq:singlet-self}
\end{align}
The shift in the SM-like Higgs self-coupling and the 2-derivative term
are driven by the same parameter in our model. If,
instead, we break the $Z_2$ symmetry of the Lagrangian in
Eq.\eqref{eq:singlet-potential}, there will be an additional
contribution to the shift in the self-coupling at tree-level. However, at the LHC we
expect the strongly interacting modification in 
Eq.\eqref{eq:singlet-self} to dominate.  In the absence of the heavy
scalar such a momentum-dependent Higgs self-coupling would indicate
the onset of a strongly interacting theory. However, in our case it
only describes the onset of a new resonance $m_H$. Both scalars
together lead to a well-defined and unitary UV-behaviour of LHC cross
sections, as shown in \refF{fig:singlet_distributions}. As
expected, the dimension-6 description breaks down once we approach the
resonance peak above $m_{hh} = 600$~GeV.\bigskip

Exactly the same analysis we can perform for two-Higgs-doublet
models with a potential of the form~\cite{Gorbahn:2015gxa}
\begin{alignat}{5}
 V(\phi_1,\phi_2) 
=& \, m^2_{11}\,\phi_1^\dagger\phi_1
 + m^2_{22}\,\phi_2^\dagger\phi_2
 + \frac{\lambda_1}{2} \, (\phi_1^\dagger\phi_1)^2
 + \frac{\lambda_2}{2} \, (\phi_2^\dagger\phi_2)^2
 + \lambda_3 \, (\phi_1^\dagger\phi_1)\,(\phi_2^\dagger\phi_2) 
 + \lambda_4 \, |\phi_1^\dagger\,\phi_2|^2 \notag \\
& + \left[ - m^2_{12}\,\phi_1^\dagger\phi_2 
        + \frac{\lambda_5}{2} \, (\phi_1^\dagger\phi_2)^2 
        + \text{h.c.} 
   \right] \,.
\label{eq:2hdmpotential}
\end{alignat}
The self-couplings $\lambda_1 \dots \lambda_5$ are bounded 
from above only if we require our model to remain perturbative. If they contribute to the light Higgs interactions
they can lead to a non-decoupling behaviour.  In the conventions of Eq.\eqref{eq:singlet_ch}
the modification of the Higgs boson couplings to weak bosons scales like
\begin{align}
\Delta_{W,Z} 
= \sin (\beta - \alpha) 
\approx \frac{\sin^2 (2\beta)}{8} \, \left( \dfrac{v}{m_{A^0}} \right)^4 \; ,
\end{align}
with a suppression in terms of the mass of the heavy pseudoscalar
$A^0$. Through $m_{12}$ it sets the mass scale of the heavy Higgs states, 
which need to have similar masses to respect custodial
symmetry. This degeneracy can be broken by large scalar couplings
$\lambda_j$ and can shift one of the neutral scalar masses with
respect to the charged scalar mass. In this situation the $v$-improved
matching at the mass of the lightest new state will be most helpful in
the numerical comparison.  At tree level the corresponding
modifications of the light Higgs rates are as unspectacular as in the
Higgs singlet extension.  Unlike the dimension-8 effect in the gauge
sector, the Yukawa couplings of the light Higgs encounter
modifications of the kind $v^2/m_{A^0}^2$, with possible additional
powers of $\tan \beta$.  For example, in type-II models we find
\begin{align}
\Delta_b 
\approx - \tan \beta \; \frac{\sin (2\beta)}{2} \, \left( \dfrac{v}{m_{A^0}} \right)^2 \,.
\label{eq:2hdm_delayed}
\end{align}
This dimension-6 effect can already for moderate values of $\tan
\beta$ significantly delay the decoupling of the heavy 2HDM states in
the Yukawa sector. Generalizing Eq.\eqref{eq:2hdm_delayed}, we can
compute the fermionic corrections which the 2HDM generates at tree
level. For the type-I and type-II setups we find
\begin{align}
  \bar{c}_u^\text{I} = \bar{c}_u^\text{II} = \bar{c}_d^\text{I\phantom{I}} = \bar{c}_\ell^\text{I\phantom{I}} &=
\phantom{-} \dfrac{\sin (2\beta) \cot\beta}{2}
\left[\frac{\lambda_1}{2} -\frac{\lambda_2}{2}
     +\left(\frac{\lambda_1}{2} +\frac{\lambda_2}{2} -\lambda_3-\lambda_4-\lambda_5\right)\cos (2\beta)
\right]
\left(\dfrac {v}{m_{A^0}} \right)^2 \,, \notag \\
\bar{c}_d^\text{II} = \bar{c}_\ell^\text{II} &=
- \dfrac{\sin (2\beta) \tan \beta}{2} 
\left[\frac {\lambda_1} 2-\frac {\lambda_2} 2+\left(\frac {\lambda_1} 2
      +\frac {\lambda_2} 2-\lambda_3-\lambda_4-\lambda_5\right)\cos (2\beta)
\right] 
\left(\dfrac{v}{m_{A^0}} \right)^2 \; .
\label{eq:2hdmmatching1}
\end{align}

Unlike for the singlet extension, a new charged Higgs $H^\pm$ induces
new features in the loop-induced coupling $g_{h \gamma
  \gamma}$,
\begin{align}
  \bar{c}_\gamma = \frac {g^2} {11\,520 \, \pi^2  } \Bigg[ 
& 30 \left(1 - [\cot \beta + \tan \beta] \frac {m_{12}^2} {m_{H^\pm}^2}  \right) \notag \\
 &\!+  \left(19 - 4  [\cot \beta + \tan \beta] \frac {m_{12}^2} {m_{H^\pm}^2} \right) \frac {m_{h^0}^2} {m_{H^\pm}^2} - 30 \cot (2 \beta)   [\cot \beta + \tan \beta]  \frac {m_{12}^2} {m_{H^\pm}^2} \; x 
   \Bigg] \,.
   \label{eq:THDM_cgamma}
\end{align}
Unfortunately, such loop-induced modifications will most
likely not be visible at the LHC. In contrast, the heavy additional
Higgs particles should appear in LHC Higgs searches as new resonances,
provided they are not too heavy. From a dimension-6 perspective the test of 
two-Higgs-doublet models at the LHC is as little of a challenge as 
the Higgs singlet extension~\cite{Gorbahn:2015gxa}.

\subsubsection*{Scalar top partners}

\begin{table}[t]
\small
\caption{Scalar top-partner Lagrangian parameters, physical
   parameters, and selected Wilson coefficient. All masses are in
   GeV. Table from Ref.~\cite{Brehmer:2015rna}.}
  \label{tab:partner-benchmarks}
\renewcommand{\arraystretch}{1.2}
\centering
\begin{tabular}{c c rrrr c rrr c rrr}
  \toprule 
  \multirow{2}{*}{}
  &\hspace*{1em}& \multicolumn{8}{c}{Top partner} 
  &\hspace*{1em}& \multicolumn{3}{c}{D6} \\
  \cmidrule{3-10}   \cmidrule{12-14} 
  && $M$ & $\kappa_{LL}$ & $\kappa_{RR}$ & $\kappa_{LR}$
                  && $m_{\tilde{t}_1}$ & $m_{\tilde{t}_2}$ & $\theta_{\tilde{t}}$ 
    && $\bar{c}_H$ & $\bar{c}_W$ &  $\bar{c}_{HW}$ \\  
  \midrule
  P1 && 500 & $-1.16$ & 2.85 & 0.147 && 500 & 580 & $-0.15$ && 0.006 & $-3.1 \cdot 10^{-7}$  & $4.0 \cdot 10^{-7}$ \\
  P2 && 350 & $-3.16$ & $-2.82$ & 0.017 && 173 & 200 & $-0.10$ && 0.018 & $-1.0 \cdot 10^{-3}$  & $1.0 \cdot 10^{-3}$\\      
  P3 && 500 & $-7.51$ & $-7.17$ & 0.012 && 173 & 200 & $-0.10$ && 0.139 & $-2.5 \cdot 10^{-3}$  & $2.5 \cdot 10^{-3}$ \\      
  \bottomrule
 \end{tabular}
 \end{table}

\begin{figure}[b!]
  \centering
  \includegraphics[width=0.43\textwidth]{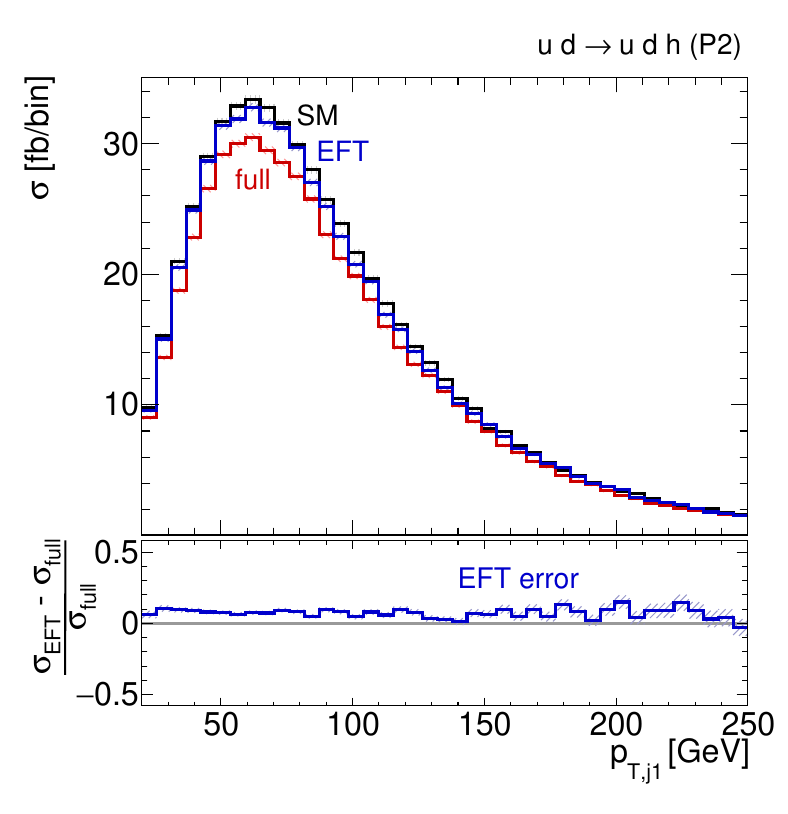}
  \hspace*{0.05\textwidth}
  \includegraphics[width=0.43\textwidth]{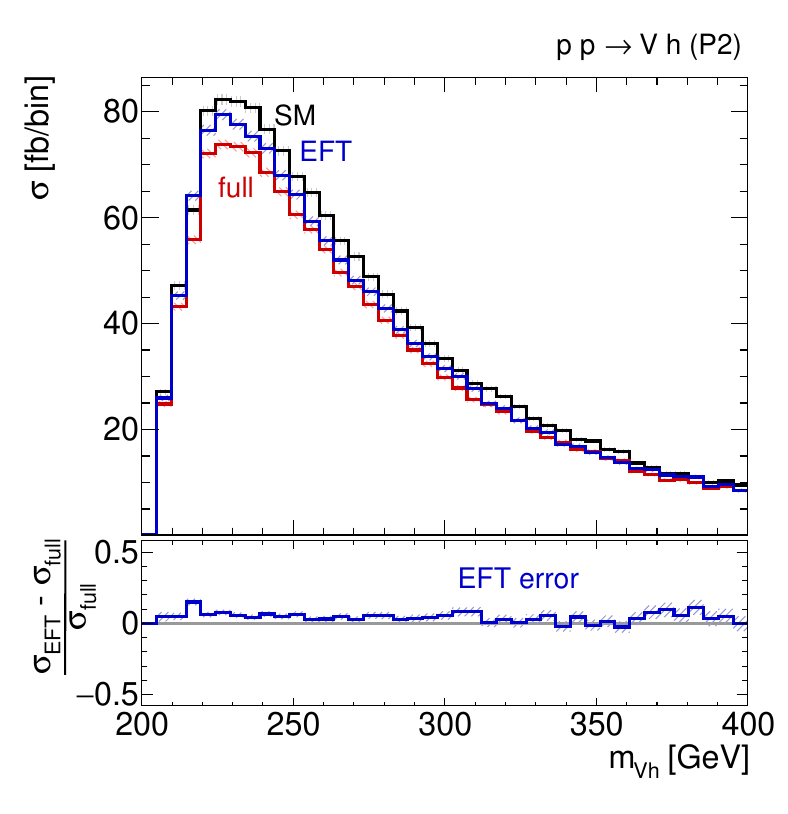}
  \caption{Kinematic distributions for the top partner model.  Left:
    tagging jet properties in WBF Higgs boson production.  Right: $m_{Vh}$
    distribution in Higgs-strahlung. Figure from
    Ref.~\cite{Brehmer:2015rna}.}
  \label{fig:partners_distributions}
\end{figure}

New scalar particles do not have to be part of the Higgs sector, in
the sense that they do not have to participate in electroweak symmetry
breaking.  We introduce a scalar top partner similar to the stop of
the MSSM.  The masses of the additional isospin doublet and singlet in
the fundamental representation of $SU(3)_c$ can be different, but for
the sake of simplicity we unify them to $M$. We consider three
relatively light new scalars $\tilde{t}_1$, $\tilde{t}_2$ and
$\tilde{b}_2= \tilde{b}_L$. The stop mass matrix reads
\begin{alignat}{5}
 \begin{pmatrix}
 \kappa_{LL}\dfrac{v^2}{2} + M^2 & \kappa_{LR}\,\dfrac{vM}{\sqrt{2}} \\
 \kappa_{LR}\,\dfrac{vM}{\sqrt{2}} & \kappa_{RR}\,\dfrac{v^2}{2}\, + M^2 
 \end{pmatrix}
  \label{eq:partner_mass}
\end{alignat}
Through loops the new scalars modify the Higgs boson couplings to gluons,
photons, and weak bosons, including new Lorentz structures in the
$hVV$ coupling. The SM-like Higgs boson coupling to weak bosons changes
in the limit of small stop mixing scales like
\begin{align}
\Delta_V \approx \frac{\kappa_{LL}^2}{16 \pi^2} \, \left( \frac{v}{m_{\tilde{t}_1}} \right)^2
\; ,
\label{eq:partner_decoup}
\end{align}
which means we expect loop effects in a dimension-6 Lagrangian.  In
the $v$-improved matching scheme we identify $\Lambda = m_{\tilde{t}_1}$.
The Wilson coefficients for the induced dimension-6 operators are then
\begin{align}
\overline{c}_{g} &=  
 \cfrac{m_W^2}{24\,(4\pi)^2\,m_{\tilde{t}_1}^2}\,\left[ \kappa_{LL} + \kappa_{RR}  - \kappa_{LR}^2\right] 
&\overline{c}_{\gamma} &=  
 \cfrac{m_W^2}{9\,(4\pi)^2\,m_{\tilde{t}_1}^2}\,\left[ \kappa_{LL} + \kappa_{RR}  - \kappa_{LR}^2\right] \notag \\
\overline{c}_{B} &=  
 -\cfrac{5m_W^2}{12\,(4\pi)^2\,m_{\tilde{t}_1}^2}\,\left[\kappa_{LL} - \cfrac{31}{50} \kappa_{LR}^2 \right] 
&\overline{c}_{W} &=  
 \cfrac{m_W^2}{4\,(4\pi)^2\,m_{\tilde{t}_1}^2}\,\left[\kappa_{LL} - \cfrac{3}{10} \kappa_{LR}^2 \right] \notag \\
\overline{c}_{HB} &=  
 \cfrac{5m_W^2}{12\,(4\pi)^2\,m_{\tilde{t}_1}^2}\,\left[\kappa_{LL} - \cfrac{14}{25} \kappa_{LR}^2 \right] 
&\overline{c}_{HW} &=  -
 \cfrac{m_W^2}{4\,(4\pi)^2\,m_{\tilde{t}_1}^2}\,\left[\kappa_{LL} - \cfrac{2}{5} \kappa_{LR}^2 \right] \notag \\
 \overline{c}_H &= 
 \cfrac{v^2}{4(4\pi)^2\,m_{\tilde{t}_1}^2}\,\Bigg[ 2\kappa_{RR}^2-\kappa_{LL}^2 - 
 \left( \kappa_{RR} - \frac{1}{2}\kappa_{LL} \right) \kappa_{LR}^2 + \cfrac{\kappa_{LR}^4}{10} \Bigg] \hspace*{-2cm}&& \notag \\
 \overline{c}_T &= \cfrac{v^2}{4(4\pi)^2\,m_{\tilde{t}_1}^2}\,\left[\kappa_{LL}^2 - \cfrac{\kappa_{LL}\,\kappa_{LR}^2}{2} + \cfrac{\kappa_{LR}^4}{10}\right] \,.
 \label{eq:c-ew}
\end{align}
We can evaluate these corrections for the most optimistic benchmark
point, \textsl{i.e.} $m_{\tilde{t}_1} \approx m_t$ and $\kappa_{ij} \gg 1$ as shown in
\refT{tab:partner-benchmarks}.  The problem is that typical
pre-factors of loop-induced Wilson coefficients are still at most
\begin{align}
 \cfrac{v^2  \kappa_{ij}^2}{4(4\pi)^2\,m_{\tilde{t}_1}^2}
< 0.16
\qquad \text{for} \quad \kappa_{ij} < 5 \; ,
\end{align}
leading to mild rate modifications at the LHC.  In
\refF{fig:partners_distributions} we show a case where rate
modifications in the 10\% range might be large enough to be
seen. However, the kinematic distributions in weak boson fusion and in
$Vh$ production are hardly modified. The matched dimension-6
Lagrangian follows the full model only for $Vh$ production, exposing
the underlying problem that we have to push the benchmarks for top
partner models aggressively towards small masses to find possibly
relevant deviations from the Standard Model.\bigskip

A more promising place to look for modification through scalar top
partners at the LHC might be Higgs plus jet production at large
$p_{T,h}$, as described in Section~\ref{sec:eft_fits} Altogether, this
means that even though they in principle generate the corresponding
momentum-dependent operators, scalar top partners are well described
by the dimension-6 Lagrangian. This is because the new structures are
loop-induced, and the corresponding suppression renders them hardly
visible at the LHC, not even talking about an agreement between the
full model and the dimension-6 Lagrangian in the relevant kinematic
distributions.

\subsubsection*{Vector triplet model}

\begin{table}[t]
  \caption{Vector triplet Lagrangian parameters, physical parameters,
    and selected Wilson coefficient. All masses are in GeV. Table
    modified from Ref.~\cite{Brehmer:2015rna}.}
  \label{tab:triplet_benchmarks}
  \renewcommand{\arraystretch}{1.2}
  \setlength{\tabcolsep}{0.3em}
  \centering
    \begin{tabular}{c c rrrrrr c rrrr}
      \toprule
      \multirow{2}{*}{} &\hspace*{1em}& \multicolumn{5}{c}{Triplet} 
      &\hspace*{1em}& \multicolumn{5}{c}{D6}\\
      \cmidrule{3-8}       \cmidrule{10-13}
      && $M_V$ & $g_V$ & $c_H$ & ${c}_{F}$ & ${c}_{VVHH}$  & $m_\xi$ 
      && $\bar{c}_W$ & $\bar{c}_H$ & $\bar{c}_6$ & $\bar{c}_f$ \\
      \midrule
      T1\;  && 591 & 3.0 & $-0.47$ & $-5.0$ & 2.0 & 1200 && $-0.011$ & 0.000 & 0.000 & 0.000 \\
      T2\;  && 946 & 3.0 & $-0.47$ & $-5.0$ & 1.0 & 1200 && $-0.011$ & 0.000 & 0.000 & 0.000 \\
      T3\;  && 941 & 3.0 & $-0.28$ & $3.0$ & 1.0 & 1200  && 0.004 & 0.046 & 0.061 & 0.015 \\
      T4\;  && 1246 & 3.0 & $-0.50$ & $3.0$ & $-0.2$ & 1200 && 0.007 & 0.111 & 0.149 & 0.037\\
      T4' && 3738 & 3.0 & $-1.50$ & $9.0$ & $-1.8$ & 3600 && 0.007 & 0.111 & 0.149 & 0.037\\
      T5\;  && 846 & 1.0 & $-0.56$ & $-1.32$ & $0.08$ & 849 && $-0.007$ & $-0.020$ & $-0.027$ & $-0.007$ \\
      T5'  && 2538 & 1.0 & $-1.68$ & $-3.96$ & $0.72$ & 2547 && $-0.007$ & $-0.020$ & $-0.027$ & $-0.007$ \\
      \bottomrule
    \end{tabular}
\end{table}

\begin{figure}[t!]
  \centering
  \includegraphics[width=0.41\textwidth]{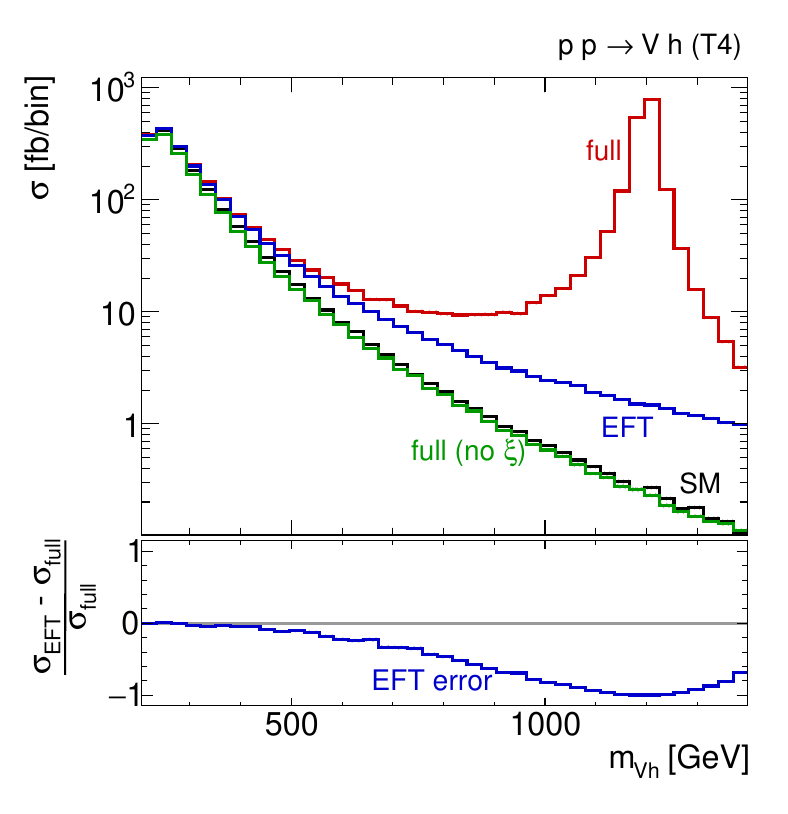} 
  \hspace*{0.05\textwidth}
  \includegraphics[width=0.41\textwidth]{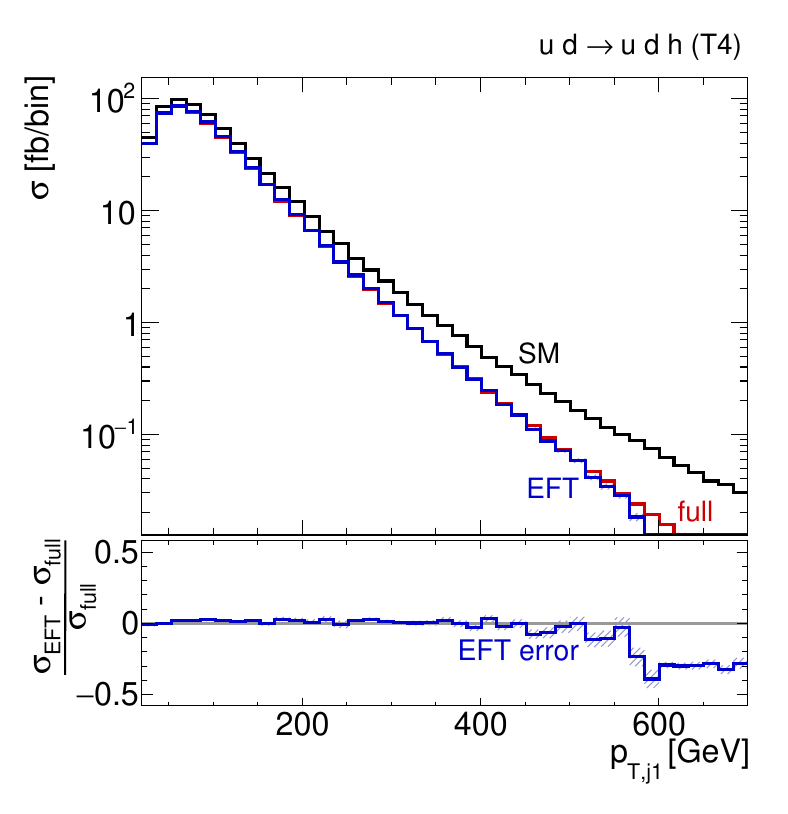} \\[-3.6mm]
  \includegraphics[width=0.41\textwidth]{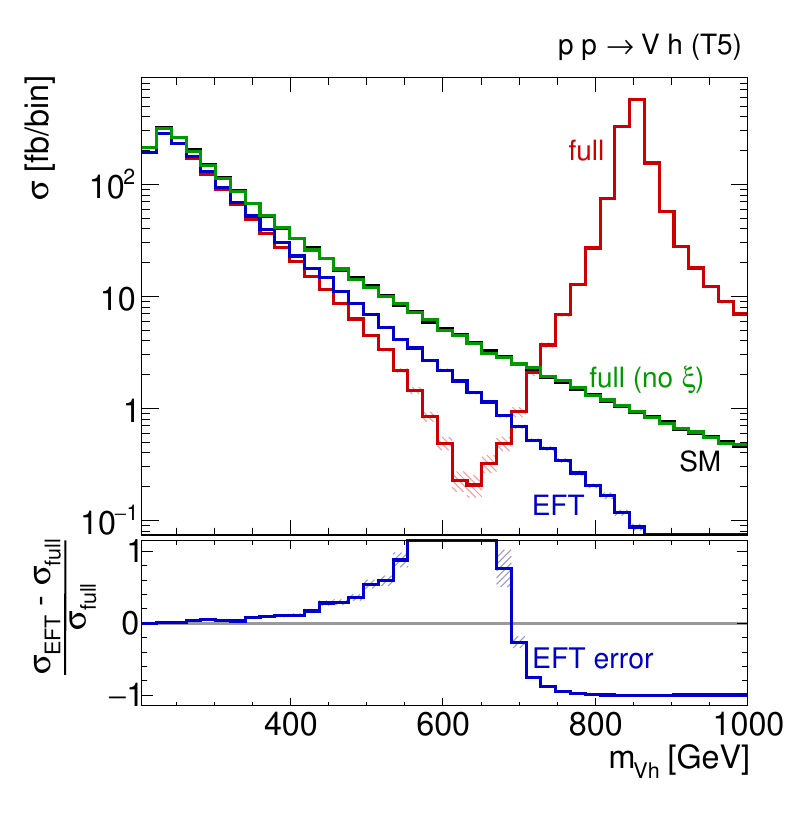} 
  \hspace*{0.05\textwidth}
  \includegraphics[width=0.41\textwidth]{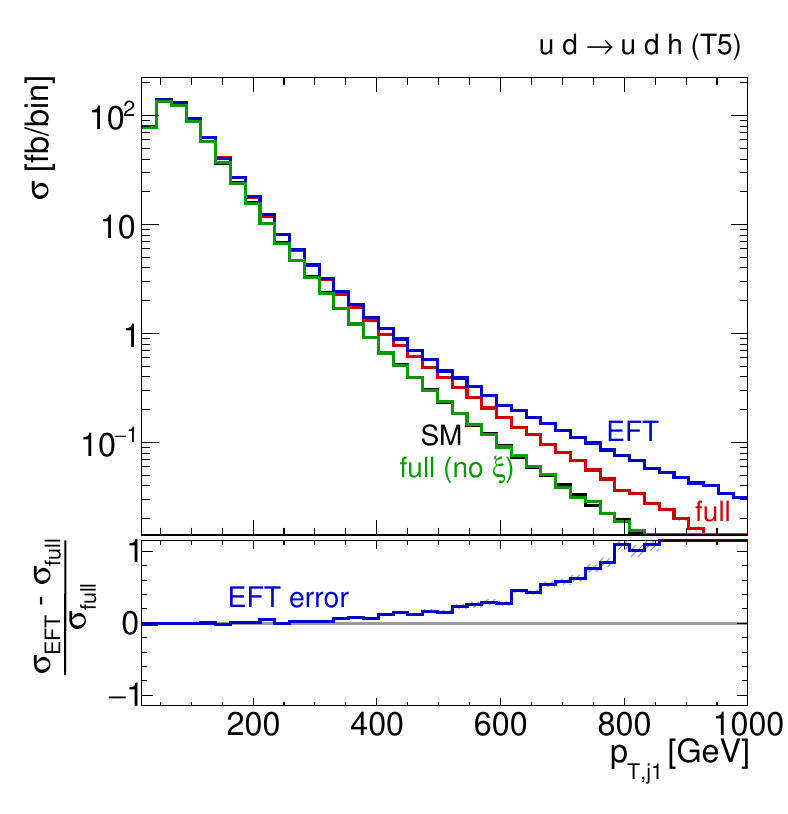} \\[-3.6mm]
  \includegraphics[width=0.41\textwidth]{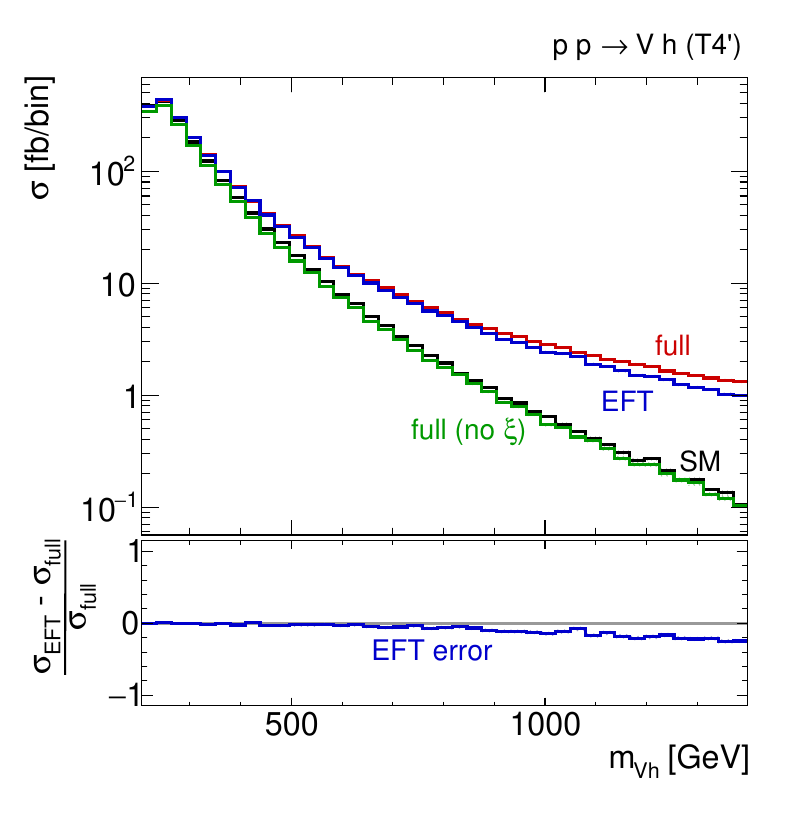} 
  \hspace*{0.05\textwidth}
  \includegraphics[width=0.41\textwidth]{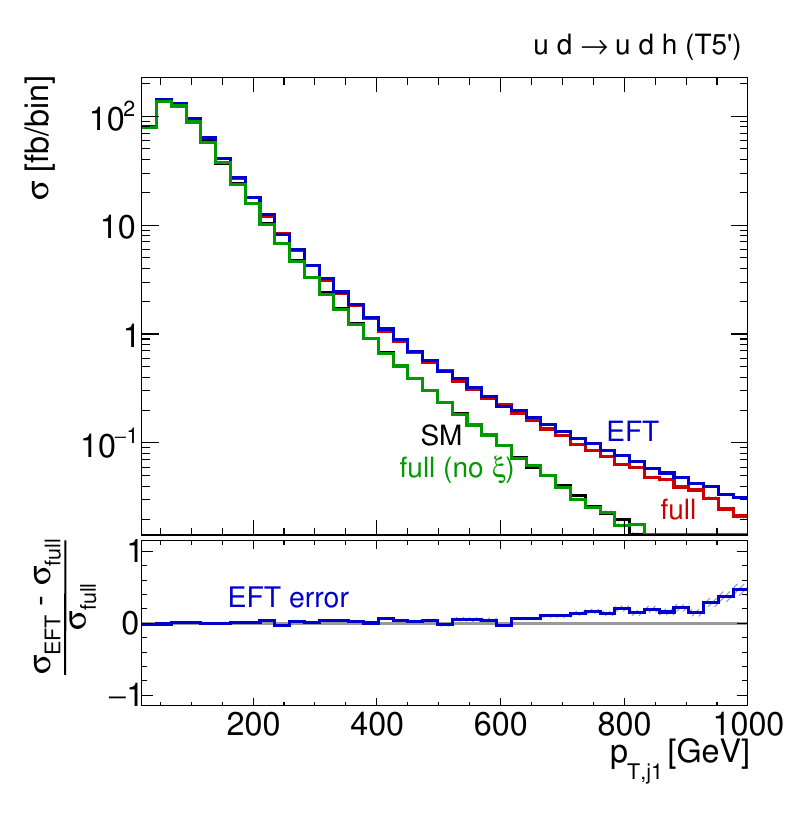} \\[-6mm]
  \caption{Distributions for WBF production and Higgs-strahlung
    in the vector triplet model. Figure modified from
    Ref.~\cite{Brehmer:2015rna}.}
  \label{fig:triplet}
\end{figure}

\begin{figure}[t!]
  \centering
  \includegraphics[width=0.43\textwidth]{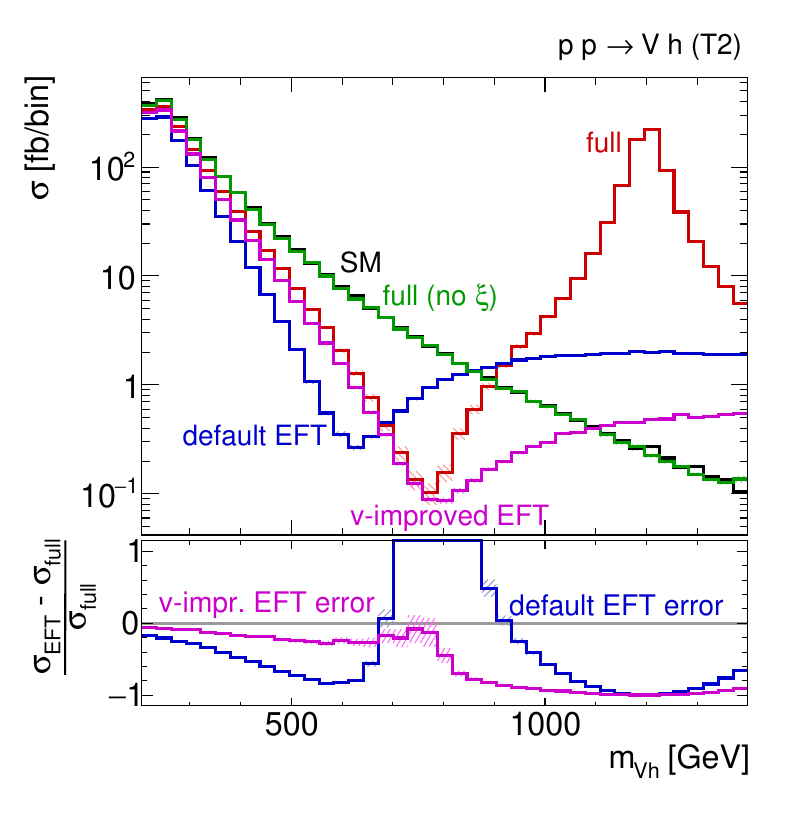} 
  \hspace*{0.05\textwidth}
  \includegraphics[width=0.43\textwidth]{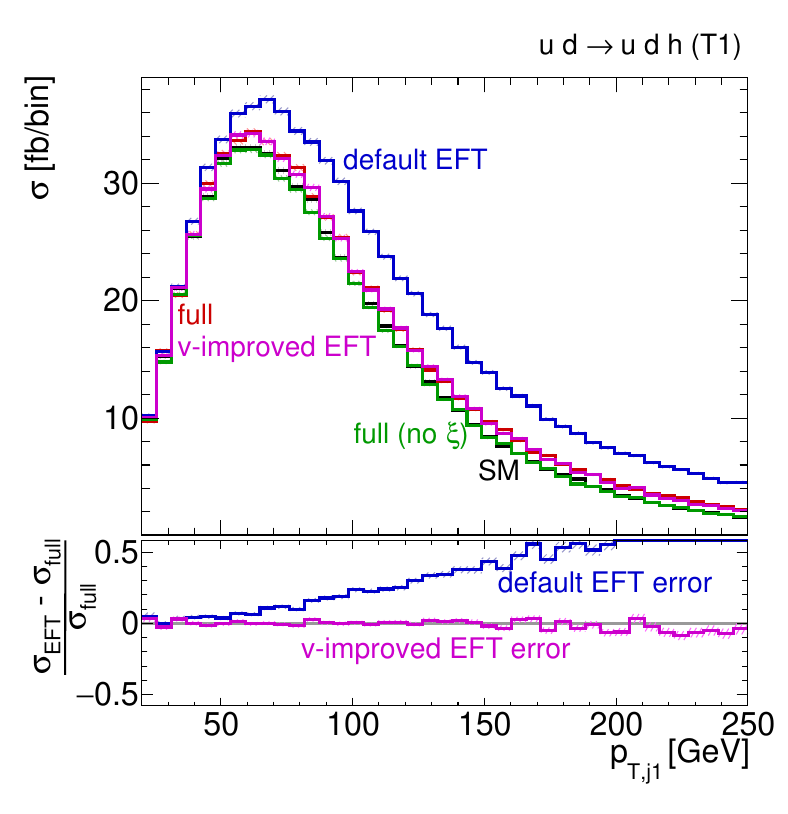} 
  \caption{Distributions for WBF Higgs boson production and Higgs-strahlung
    in the vector triplet model for both the default and $v$-improved matching conditions.
    Figure modified from Ref.~\cite{Brehmer:2015rna}.}
  \label{fig:v-improvement}
\end{figure}

\begin{figure}[t]
  \centering
  \includegraphics[width=0.43\textwidth]{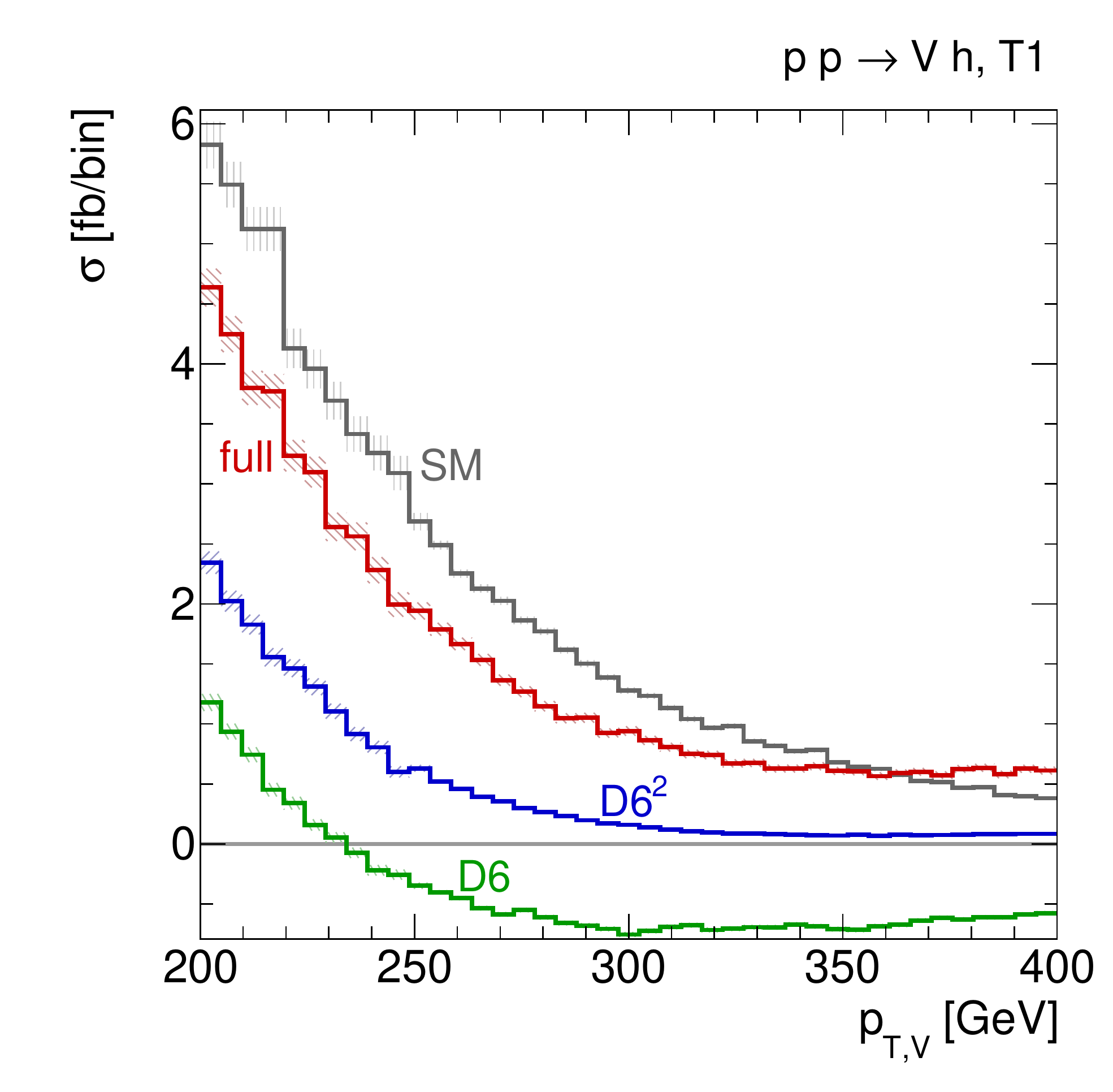} 
  \hspace*{0.05\textwidth}
  \includegraphics[width=0.43\textwidth]{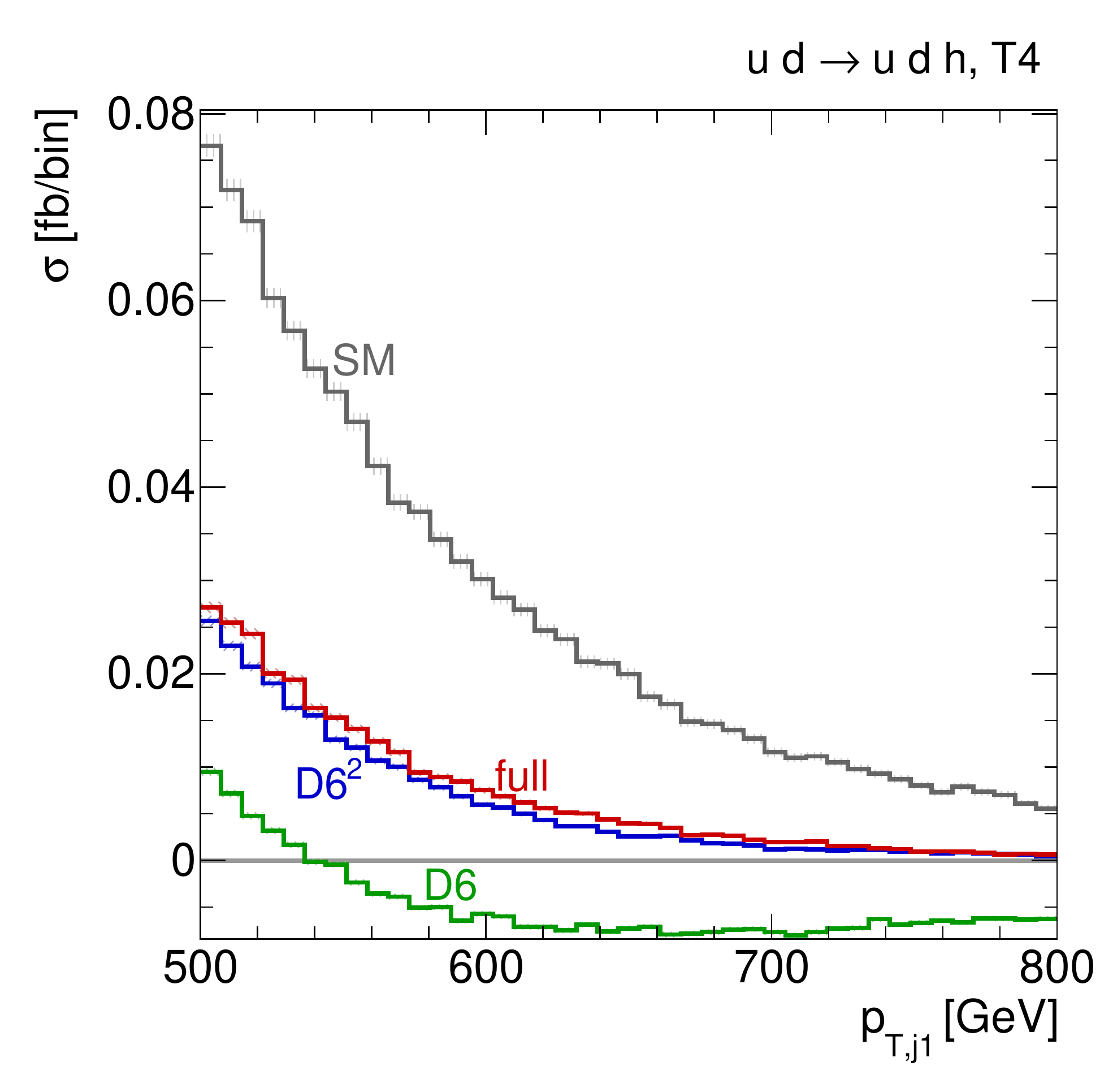}
  \caption{Distributions for WBF Higgs boson production and Higgs-strahlung
    in the vector triplet model with and without the dimension-6
    squared contribution. Figure from Ref.~\cite{Biekotter:2016ecg}.}
  \label{fig:squared}
\end{figure}

Given scalar extensions of the Standard Model do not challenge
the dimension-6 approach to LHC Higgs data we now assume a
modification of the weak gauge sector.  It consists of a massive
vector field $V^a_\mu$ forming an $SU(2)$ triplet. To allow for large,
tree-level modifications of LHC observables we assume that the new
fields mix with the weak bosons,
\begin{alignat}{5}
\mathscr{L} \supset& \,
  - \dfrac{1}{4}\,V_{\mu\nu}^a\,V^{\mu\nu\, a}
  + \dfrac{M_V^2}{2}\,V_\mu^a\,V^{\mu\,a}
 \notag \\
&  + i\,\frac{g_V}{2} \,c_H\,V_\mu^a\,\left[\phi^\dagger \sigma^a \,\overleftrightarrow{D}^\mu\,\phi\,\right]
  +\dfrac{g_w^2}{2 g_V}\,V_\mu^a\,\sum_\text{fermions}\, c_F \overline{F}_L\,\gamma^\mu\, \sigma^a \,F_L
  + g_V^2\,c_{VVHH}\,V_\mu^a\,V^{\mu a}\, \phi^\dagger \phi + \cdots
 \label{eq:lag-vectortriplet}
\end{alignat}
After mixing with the $W$ and $Z$ bosons we denote the new heavy
states as $\xi^0$ and $\xi^\pm$. The masses of these particles define
the  matching scale $\Lambda = m_\xi$ in our $v$-improved matching.
Integrating out the vector triplet leaves us with a set of dimension-6
Wilson coefficients generated at tree level
\begin{align}
\bar{c}_{H} &= \dfrac{3\,g^2\,v^2}{4\,m_\xi^2}\,\left[c_H^2\dfrac{g_V^2}{g^2}  - 2\,c_F\,c_H\right] 
&\bar{c}_{6} &= \dfrac{g^2\,v^2}{m_\xi^2}\,\left[c_H^2\dfrac{g_V^2}{g^2} - 2\,c_F\,c_H \right] \notag \\
\bar{c}_{f} &= \dfrac{g^2\,v^2}{4\,m_\xi^2}\,\left[c_H^2\dfrac{g_V^2}{g^2}  - 2\,c_F\,c_H \right] 
&\bar{c}_{W} &= - \dfrac{m_W^2}{m_\xi^2}\, c_Fc_H \; .
 \label{eq:triplet_coefficients}
\end{align}
This universal structure implies that along a line in $c_F$ vs $c_H$
all Wilson coefficients except for the operator $\mathcal{O}_W$ vanish. 
Following the argument from the scalar top partners we neglect
further, loop-induced contributions. The Wilson coefficients induced
at tree level are shown together with the model parameters for a set
of benchmarks in \refT{tab:triplet_benchmarks}. In the definition
of these benchmarks we ignore experimental constraints on those models
as well as open questions concerning their ultraviolet
completions. They are merely chosen to test the agreement between the
full model predictions and the dimension-6 Lagrangian at the
LHC. Moreover, the notion of weakly interacting new particles might
not really apply, given the size of the couplings defined in
Eq.\eqref{eq:lag-vectortriplet}.\bigskip

In \refF{fig:triplet} we show a set of sample distributions for
the vector triplet model. In the upper panel for the $Vh$ channel we
immediately see what the limitations in matching the dimension-6
Lagrangian are: the benchmark points include new states at
1.2~TeV. The invariant mass distribution $m_{Vh}$ develops a pole with
tails reaching down to around $m_{Vh} = 600$~GeV. Below this, at least
the $v$-improved dimension-6 Lagrangian follows the full model
description to around 20~\%. What is remarkable is that the
description of the full model without the heavy $\xi$ state deviates
significantly from the complete model and from the dimension-6
Lagrangian. Far below the mass shell the latter indeed describes the
effect of the $s$-channel $\xi$ exchange, while it obviously fails to
reproduce the pole region. The WBF distribution shows less dramatic
effects, because the new states occur in the $t$-channel and there is
no on-set of a resonance peak.  Moreover, for the same benchmark model
we observe a different sign of the (interference) effects; now the
full model as well as the dimension-6 approximation stay below the
Standard Model curve. Over the entire $p_{T,j_1}$ range the agreement
between the full model and the dimension-6 Lagrangian with
$v$-improved matching is excellent even in regions where deviations
from the Standard Model are due to new particle exchange.

In the second row of \refF{fig:triplet} we show another benchmark
point, where the new resonance peak in $m_{Vh}$ appears around
850~GeV, and the interference effects between the Standard Model
continuum and the developing pole above $m_{Vh} = 500$~GeV are
destructive. The full model and its dimension-6 agree well into the
range where the extended model deviates from the Standard Model. For
WBF production the new particles in the $t$-channel lead to a
significant enhancement of the $p_{T,j}$ distribution, which is
described by the dimension-6 approximation to transverse momenta
around 500~GeV.

In the lower panels of \refF{fig:triplet} we show the same
distributions for modified benchmark points where the masses of the
new particles are heavier and this decoupling effect is compensated by
larger couplings, so that the Wilson coefficients are the same as in
the scenarios shown in the upper panels. In the absence of new
resonances the agreement between the full vector triplet model and the
dimension-6 description improves to an almost perfect match,
reflecting the fact that higher-dimensional operators are still most
appropriate in describing strongly interacting models at the
LHC.\bigskip

While for the generic high-mass scenarios shown in
\refF{fig:triplet} the choice of default and $v$-improved matching
does not make a big difference, we demonstrate in
\refF{fig:v-improvement} that this question can be crucial for
other benchmarks. Here the dimension-6 description based on the
default matching fails to reproduce the full model already in the bulk
of the distributions. The large discrepancies are caused by low masses
of the new states as shown in Eq.\eqref{eq:v-improvement}.  The
$v$-improved matching, designed to include such effects, leads to
impressive agreement up to large energies.\bigskip

Finally, an open question is if in the dimension-6 approach we want to
include the squared dimension-6 term in the combination $|\mathcal{M}_\text{SM} +
\mathcal{M}_\text{D6}|^2$.  For our strictly dimension-6 approach this is an entirely
practical question, as long as it is clearly stated what is done. One
issue of the truncation without the dimension-6 squared terms is that
the matrix element squared is not guaranteed to be positive; this is
true even if the effective theory is valid and dimension-8 operators
are negligible, but the new physics effects dominate over the Standard
Model and the bulk contribution stems from the dimension-6 structures.
 
In \refF{fig:squared} we show two distributions for the vector
triplet model in an extreme benchmark point and based on $v$-improved
matching. For $Vh$ production and WBF production we choose a benchmark
point with large new physics contributions, negatively interfering
with the Standard Model contribution. For the $Vh$ process we show the
$p_{T,V}$ distribution, which is strongly correlated with $m_{Vh}$. In
spite of the fact that the new particles only occur at $m_\xi =
1.2$~TeV the predictions without the dimension-6 squared term become
negative for $p_{T,V} > 230$~GeV for $Vh$ production and for $p_{T,j}
> 550$~GeV for WBF. In particular the former is within reach of early
LHC analyses.

\subsubsection*{Outlook}

\begin{table}[b!]
 \caption{Possible sources of failure of dimension-6 Lagrangian at the
   LHC.  We use parentheses where deviations in kinematic distributions
   appear, but are unlikely to be observed in realistic scenarios.}
 \label{tab:differences}
\renewcommand{\arraystretch}{1.2}
\centering
\begin{tabular}{ll c ccc}
  \toprule
  Model & Process &\hspace*{1em}& \multicolumn{3}{c}{EFT failure} \\ 
  \cmidrule{4-6}
      & && resonance & kinematics & matching \\
  \midrule
  singlet & on-shell $h \to 4 \ell$, WBF, $Vh$, \dots && & & $\boldsymbol{\times}$  \\
      & off-shell WBF, \dots && & $(\boldsymbol{\times})$ & $\boldsymbol{\times}$ \\
      & $hh$ && $\boldsymbol{\times}$ &  $\boldsymbol{\times}$ & $\boldsymbol{\times}$ \\
  2HDM    & on-shell  $h \to 4 \ell$, WBF, $Vh$, \dots && & & $\boldsymbol{\times}$  \\
      & off-shell $h \to \gamma \gamma$, \dots && & $(\boldsymbol{\times})$ & $\boldsymbol{\times}$ \\
      & $hh$ && $\boldsymbol{\times}$ &  $\boldsymbol{\times}$ & $\boldsymbol{\times}$ \\
  top partner & WBF, $Vh$ && & & $\boldsymbol{\times}$ \\
  vector triplet & WBF  &&  & $(\boldsymbol{\times})$ & $\boldsymbol{\times}$ \\ 
      & $Vh$ && $\boldsymbol{\times}$ & $(\boldsymbol{\times})$ & $\boldsymbol{\times}$ \\       
  \bottomrule
\end{tabular}
\end{table}

In view of the limited precision of early LHC data, and the typical
size of new physics deviations, it is often assumed that a consistent
EFT framework is of little practical use in LHC Higgs analyses.  The
reason is that models with a clear hierarchy of scales will only
predict small deviations in LHC observables. For many key
observables, the situation will get better with more LHC data being
analysed, until all relevant observables become systematics-limited or
theory-limited. The crucial question then become how well an ad-hoc
truncated dimension-6 Lagrangian describes LHC observables for example
compared to a weakly interacting full model.\bigskip

\vfill
\newpage

Our first finding is that weakly interacting new physics models
usually do not generate a large number of dimension-6 operators with
non-negligible Wilson coefficients. This is particularly true for
extended scalar sectors and all operators affecting associated 
Higgs boson production and Higgs boson production in weak boson fusion. Tree-level
effects tend to be structurally simple, numerically small, and well
described by a dimension-6 Lagrangian.  Loop-induced effects for
example induced by scalar top partners will be very hard to measure,
and possible deviations between the full model and the dimension-6
approximation will not cause any problem. Effects on the gluon fusion
process can be larger, but actual model predictions are known to agree
well with the dimension-6 framework.  This line of argument becomes
stronger for more realistic new physics scenarios, compatible with all
current constraints.  If instead we want to generate large deviations
from the Standard Model we should turn to extensions of the weak gauge
sector. Here we can test how well a properly matched dimension-6
Lagrangian described the sizeable changes in the LHC kinematics.\bigskip

Second, we find very good agreement between the full models extending
the gauge and Higgs sector and the respective dimension-6
approximation for the relevant LHC observables. In cases where the
standard matching procedure reaches its limitations, a $v$-improved
matching procedure leads to excellent agreement between the full model
and the dimension-6 approximation for our set of models. Note that our
comparison between the full model and the dimension-6 approximation
does not necessarily imply that all individual dimension-8 operators
are negligible, because towers of higher-dimensional operators can
cancel due to an unknown structure or even symmetry. We illustrate our
findings in \refT{tab:differences}.\bigskip

Finally, we confirm that the dimension-6 approximation typically
breaks down when we become sensitive to new particles, often visible
as poles in the $s$-channel. Unless we start observing such new
resonances the dimension-6 approach seems valid to describe large
classes of weakly interacting new physics at the LHC. The consistent
effective field theory as well as the dimension-6 Lagrangian allow us
to describe kinematic distributions and to include electroweak quantum
corrections. The fundamental difference between the two approaches is
that a consistent effective theory allows us to assign a theoretical uncertainty
to the truncated operator basis in the model hypothesis we are testing
at the LHC. In contrast, the dimension-6 Lagrangian does not include such a
theoretical uncertainty. On the other hand, in most searches for new
physics at the LHC we do not account for theoretical uncertainties
beyond the perturbative QCD description, either. This means that the 
missing, most likely sizeable, theoretical uncertainty in the
dimension-6 approach only has to be accounted for when we translate
our findings into full weakly interacting models.
This implies that from an experimental perspective a
dimension-6 Lagrangian can be used in LHC Higgs physics as long as we
do not observe an obvious breakdown through new resonances. 



\chapter{EFT Application} 
\label{chap:EFTAppl}
\ChapterAuthor{N.~Belyaev, V.~Bortolotto,  L.~Brenner, C.D.~Burgard, F.~Campanario, B.~Chokouf{\'e} Nejad, M.~Ciuchini, R.~Contino, T.~Corbett, J.~de Blas, F.~Demartin,  A.~Denner, S.~Dittmaier, M.~D{\"u}hrssen, K.~Ecker,  A.~Falkowski,  E.~Franco, B.~Fuks, S.~Gadatsch,  M.~Ghezzi, D.~Ghosh, D.~Gray, A.~Greljo, A.~Gritsan, R.~Gr{\"o}ber, C.~Grojean, G.~Isidori, S.~Kallweit, A.~Kaluza, W.~Kilian, K.~K{\"o}neke,  R.~Konoplich, S.~Kortner, D.~Marzocca, K.~Mawatari, K.~Mimasu, S.~Mishima, A.~M{\"u}ck, M.~M{\"u}hlleitner, T.~Ohl, A.~Papaefstathiou, G.~Perez, M.~Pierini, K.~Prokofiev, M.~Rauch, L.~Reina, J.~Reuter,  F.~Riva, R.~R{\"o}ntsch, J.~Roskes, R.~Roth, V.~Sanz, U.~Sarica,  C.~Schmitt, M.~Schulze, M.~Sekulla, S.~Shim,  L.~Silvestrini, C.~Speckner,  M.~Spira, J.~Streicher, W.~Verkerke, C.~Weiss,  M.~Xiao, M.~Zaro, D.~Zeppenfeld} 

\section[High-energy physics tools for the study of the Higgs boson properties in EFT]{High-energy physics tools for the study of the Higgs boson properties in Effective Field Theories\SectionAuthor{B.~Fuks (Ed.);  F.~Campanario, B.~Chokouf{\'e} Nejad, M.~Ciuchini, R.~Contino, T.~Corbett, J.~de Blas, F.~Demartin, A.~Denner, S.~Dittmaier, A.~Falkowski, E.~Franco, M.~Ghezzi, D.~Ghosh, A.~Greljo, A.~Gritsan, R.~Gr\"ober, C.~Grojean, G.~Isidori, S.~Kallweit, W.~Kilian, D.~Marzocca, K.~Mawatari, K.~Mimasu, S.~Mishima, A.~M\"uck, M.~M\"uhlleitner, T.~Ohl, A.~Papaefstathiou, G.~Perez, M.~Pierini, M.~Rauch, L.~Reina, J.~Reuter, F.~Riva, R.~Roentsch, J.~Roskes, R.~Roth, V.~Sanz, U.~Sarica, M.~Schulze, M.~Sekulla, S.~Shim, L.~Silvestrini, C.~Speckner, M.~Spira, J.~Streicher, C.~Weiss, M.~Xiao. M.~Zaro, D.~Zeppenfeld}}
\label{s.efttools}
\subsection{Introduction}
The discovery by the ATLAS and CMS collaborations of a Higgs boson boson with a
mass of about 125~GeV~\cite{Aad:2012tfa,Chatrchyan:2012xdj} has marked an
important step forward in the study and the understanding of the electroweak
symmetry breaking mechanism. Although the currently measured properties of this
newly discovered boson seem to be compatible with the Standard Model
expectation, the recent start of the second LHC experimental run has risen new
hopes to detect phenomena beyond the Standard Model. In this context, present
and future LHC data could be interpreted in an effective field theory framework
where departures from the Standard Model are organized as a series expansion in
the new physics energy scale $\Lambda$ that is assumed to be large.

The leading effects implied by such an effective field theory description
usually consist of dimension-six operators that are supplemented to the Standard
Model Lagrangian, each of these being associated with a new interaction
strength. The number of independent coefficients is usually large, but
important classes of observables turn to only depend on a much smaller subset of
parameters. The effective field theory approach is therefore testable and the
results could be reinterpreted to constrain explicit new physics models.
Consequently, the development of high-energy physics tools able to perform
computations in the effective field theory context has been a very active field
during the last years. The recent progress described in the document addresses
total and differential cross section precision calculations for the production
of a single Higgs boson (see Section~\ref{sec:higlu} for the gluon fusion
channel and Section~\ref{sec:hawk} for the vector boson fusion and the
Higgs-strahlung channels) and of a pair of Higgs bosons (see
Section~\ref{sec:hpair}), as well as precision predictions for Higgs boson decays (see
Section~\ref{sec:ehdecay}). In addition, several machineries have been built so
that we are now able to characterize the Higgs boson properties on the basis of Monte
Carlo event generators. These include a description of the Higgs boson decays by
means of pseudo-observables (see Section~\ref{sec:PO}), as well as 
all Higgs boson properties in full generality 
within the {\tt MG}5\_a{\tt MC@NLO} (see
Section~\ref{sec:HC}) or {\tt JhuGen}/{\tt Mela} (see Section~\ref{sec:jhugen})
platforms. Monte Carlo simulations can also be performed with the
{\tt Herwig} (see  Section~\ref{sec:herwig}), {\tt Vbfnlo} (see
Section~\ref{sec:VBFNLO}) and {\tt Whizard} (see Section~\ref{sec:whizard})
packages. As all codes are using different conventions, we review the notation
in each case. Constraints on
effective operators can also be extracted by fitting the experimental results,
which can be achieved for instance via the {\tt HEPfit} package (see
Section~\ref{sec:hepfit}).

An important aspect of the effective field theory approach is the freedom in the
choice of the operator basis, so that a given effect could be modeled by several
different combinations of operators at a fixed order in the effective energy scale
expansion. This is related to the possibility of redefining the
Standard Model fields in such a way that the Standard Model Lagrangian is
unaltered (or more precisely the scattering amplitudes), while certain combinations of dimension-six operators proportional to the
Standard Model equations of motion can be eliminated up to subleading
higher-dimensional effects. Different complete operator bases have been proposed
in the past, and
although each of them yields the same predictions, they present specific and
different advantages. Existing calculations or tools are however often
bound to a given basis choice, and it is desirable to be able to reuse results
derived in the context of one basis in another basis. The {\tt Rosetta} platform
(see Section~\ref{sec:rosetta}) has been very recently released to close this
gap.

\subsection{{\tt HiGlu}: Higgs boson production via gluon fusion}
\label{sec:higlu}
The program {\tt HiGlu}~\cite{Spira:1995mt,Spira:1996if}, that is available at 
\url{http://tiger.web.psi.ch/higlu/} 
calculates the Higgs boson production cross section via gluon-fusion up to the
next-to-next-to-leading order (NNLO) accuracy in QCD in the large
top mass limit~\cite{Djouadi:1991tka,
Dawson:1990zj,Graudenz:1992pv,Spira:1995rr,Harlander:2005rq,Anastasiou:2009kn,
Catani:2001ic,Harlander:2001is,Harlander:2002wh,Ravindran:2003um,
Anastasiou:2002yz} and includes next-to-leading order (NLO) electroweak
corrections~\cite{Djouadi:1994ge,Chetyrkin:1996wr,Chetyrkin:1996ke,
Aglietti:2004nj,Aglietti:2006yd,Degrassi:2004mx,Actis:2008ug,Actis:2008ts,
Anastasiou:2008tj} within the
Standard Model, and up to the NNLO in QCD and in the large top mass
limit~\cite{Djouadi:1991tka,
Dawson:1990zj,Graudenz:1992pv,Spira:1995rr,Harlander:2005rq,Anastasiou:2009kn,
Catani:2001ic,Harlander:2001is,Harlander:2002wh,Ravindran:2003um,
Anastasiou:2002yz,Kauffman:1993nv,Djouadi:1993ji,Spira:1993bb,Dawson:1996xz,
Muhlleitner:2006wx,Bonciani:2007ex,Anastasiou:2006hc,Aglietti:2006tp,
Harlander:2002vv,Anastasiou:2002wq} for the minimal supersymmetric extension of
the Standard Model. The genuine supersymmetric corrections are not included in
the program,
although the supersymmetric QCD ones are known~\cite{Harlander:2003bb,
Harlander:2003kf,Harlander:2004tp,Harlander:2005if,Degrassi:2008zj,
Degrassi:2010eu,Degrassi:2011vq,Degrassi:2012vt,Anastasiou:2008rm,
Muhlleitner:2010nm,Muhlleitner:2010zz}.
Starting from the Standard Model Higgs results, the contributions of
dimension-six operators beyond the Standard Model are included up to the NNLO in
QCD. The latter extension is based on the effective Lagrangian (in the heavy top
limit for the Standard Model part)~\cite{Chetyrkin:1997sg,Chetyrkin:1997un,
Kramer:1996iq,Spira:1997dg},
\begin{equation}
  {\cal L}_{\rm eff} = \frac{\alpha_s}{\pi}
    \Big[ \frac{c_t}{12} (1 + \delta) + c_g \Big]\
    G_a^{\mu\nu} G^a_{\mu\nu} \frac{H}{v},
\label{eq:leffh}
\end{equation}
where $H$ denotes the Standard Model physical Higgs boson and
$v$ the vacuum expectation value of the neutral component of the Higgs field
$\Phi$. Moreover, the gluon field strength tensor is denoted by $G^{\mu\nu}_a$
and the strong coupling constant by $\alpha_s$. The contributions of dimension-six
operators are absorbed in the rescaling factor $c_t$ for the top Yukawa
coupling and the point-like coupling $c_g$. In other words, deviations of $c_t$
and $c_g$ from their Standard Model values $c_t=1$ and $c_g=0$ are understood to
originate from dimension-six operators. The contribution of the chromomagnetic
dipole operator \cite{Choudhury:2012np,Degrande:2012gr} is not included. The
Lagrangian above includes QCD corrections via the $\delta$ parameter,
\beq
  \delta = \delta_1 \frac{\alpha_s}{\pi}
    + \delta_2 \Big(\frac{\alpha_s}{\pi}\Big)^2
    + \delta_3 \Big(\frac{\alpha_s}{\pi}\Big)^3,
\eeq
whose three components read
\beq
\begin{split}
  \delta_1  = &\ \frac{11}{4}, \qquad \qquad
  \delta_2  = \frac{2777}{288} + \frac{19}{16} L_t
     + N_F \Big(\frac{L_t}{3}-\frac{67}{96} \Big),\\
  \delta_3  = & \frac{897943}{9216} \zeta_3 - \frac{2761331}{41472} +
\frac{209}{64} L_t^2 + \frac{2417}{288} L_t
     + N_F \bigg(\frac{58723}{20736} - \frac{110779}{13824} \zeta_3 +
        \frac{23}{32} L_t^2 + \frac{91}{54} L_t \bigg)\\
  &\quad
     + N_F^2 \bigg(-\frac{L_t^2}{18} + \frac{77}{1728} L_t - \frac{6865}{31104} \bigg),
\end{split}
\eeq
where $L_t = \log (\mu_R^2/m_t^2)$ with $\mu_R$ denoting the
renormalization scale, $N_F$ the number of active quark flavours and $m_t$ the
top quark pole mass.

The leading order (LO) cross section extended by the new physics contributions
induced by the above effective Lagrangian is given by
\beq
  \sigma_{LO}(pp\to H) = \sigma_0 \tau_H \frac{{\rm d}{\cal L}^{gg}}{{\rm d}\tau_H},
\label{eq:lohxs}\eeq
where ${\cal L}^{gg}$ stands for the gluon-gluon partonic luminosity and where
$\tau_H=m_h^2/S$, $m_h$ being the Higgs boson mass and $S$ the hadronic
centre-of-mass energy.
The $\sigma_0$ prefactor can be computed either in the non-linear case
($\sigma_0^{\rm NL}$) or in the SILH (strongly interacting light Higgs)
framework ($\sigma_0^{\rm SILH}$)~\cite{Giudice:2007fh},
\begin{eqnarray}
  \sigma_0^{\rm NL} &=& \frac{G_F\alpha_s^2}{288\sqrt{2}\pi}
    \bigg| \sum_Q c_Q A_{1/2}(\tau_Q) + 12 c_g \bigg|^2, \qquad \nonumber \\ 
  \sigma_0^{\rm SILH} &=&  \sigma_0^{\rm NL} - \frac{G_F\alpha_s^2}{288\sqrt{2}\pi}
    \bigg| \sum_Q (c_Q-1) A_Q(\tau_Q) + 12 c_g \bigg|^2,
\label{eq:sig0}
\end{eqnarray}
where $\tau_Q = 4m_Q^2/m_h^2$ with $m_Q$ being a generic quark mass, and where
$G_F= (\sqrt{2} v^2)^{-1}$ stands for the Fermi
constant. The LO form factors depend on the $A_{1/2}$ function that is given by
\beq
 A_{1/2}(\tau) = \frac{3}{2} \tau\Big[1+(1-\tau) f(\tau)\Big]
 \qquad{\rm with}\qquad
  f(\tau) = \left\lbrace \begin{array}{ll}
    \displaystyle \arcsin^2\frac{1}{\sqrt{\tau}} & \quad {\rm for}~\tau\geq1 \\[0.5cm]
    \displaystyle -\frac{1}{4}\bigg[\ln\frac{1+\sqrt{1-\tau}}{1-\sqrt{1-\tau}}-i\pi\bigg]^2
      & \quad {\rm for}~\tau<1\, . \end{array} \right.
\label{eq:Ahalf}\eeq
Rescaling factors $c_Q$ have been introduced in Eq.~\eqref{eq:sig0} for all
contributing quarks, \textit{i.e.}, the top ($c_t$), bottom ($c_b$) and
charm ($c_c$) quarks.

The Wilson coefficient $c_g$ does not receive QCD corrections within the
effective Lagrangian, but develops a scale dependence according to the
renormalization group equation
\beq
  \mu^2 \frac{\partial c_g(\mu^2)}{\partial \mu^2} =
   - \bigg\{\beta_1 \Big(\frac{\alpha_s(\mu^2)}{\pi}\Big)^2
     + 2 \beta_2 \Big(\frac{\alpha_s(\mu^2)}{\pi}\Big)^3\bigg\}
      c_g(\mu^2),
\eeq
with
\beq
  \beta_0 = \frac{33-2 N_F}{12},\qquad
  \beta_1 = \frac{153-19 N_F}{24}\qquad{\rm and}\qquad
  \beta_2 = \frac{1}{128} \Big( 2857 - \frac{5033}{9}N_F + \frac{325}{27} N_F^2 \Big).
\eeq
This renormalization group equation can be derived either from the
scale-invariant trace anomaly term $\beta(\alpha_s)/\alpha_s G_a^{\mu\nu}
G^a_{\mu\nu}$~\cite{Callan:1970yg,Symanzik:1970rt,Coleman:1970je,
Crewther:1972kn,Chanowitz:1972vd,Chanowitz:1972da}, or from the scale dependence
of the factor $(1+\delta)$ of the effective Lagrangian of Eq.~\eqref{eq:leffh},
since both coefficients, $c_t (1+\delta)$ and $c_g$, have to develop the same
scale dependence. The solution of the RGE for $c_g$ up to
the next-to-next-to-leading logarithmic (NNLL) level can be
cast into the form
\beq
  c_g(\mu^2) = c_g(\mu_0^2)~\frac{\beta_0 + \beta_1
    \frac{\alpha_s(\mu^2)}{\pi} + \beta_2 \Big( \frac{\alpha_s(\mu^2)}{\pi}
     \Big)^2}{\beta_0 + \beta_1 \frac{\alpha_s(\mu_0^2)}{\pi} + \beta_2
     \Big( \frac{\alpha_s(\mu_0^2)}{\pi} \Big)^2}.
\eeq
In order to compute the modified cross section up to NNLO, the mismatch of the
individual terms of the effective Lagrangian of Eq.~\eqref{eq:leffh} with
respect to the $\delta$ terms have been taken into account, the NNLL scale
dependence of the Wilson coefficient $c_g$ being properly included. In addition,
the finite NLO quark mass term effects have been added at fixed NLO to the
Standard Model contributions. This yields a consistent
determination of the gluon-fusion cross section up to NNLO including the
dimension-six operators that imply a rescaling of the top, bottom and
charm Yukawa couplings and the point-like $Hgg$ coupling parameterized by
$c_g(\mu_R^2)$.

The present version 4.34 of {\tt HiGlu} is linked to {\tt HDecay} (version
6.51)~\cite{Djouadi:1997yw,Djouadi:2006bz} and allows one to choose the usual
Standard Model Higgs input values in the separate input files {\tt higlu.in}
and {\tt hdecay.in}. In addition, the rescaling factors $c_{t,b,c}$ and the
point-like Wilson coefficient $c_g(\mu_0^2)$ can be set, together with the
corresponding input scale $\mu_0$, in the {\tt higlu.in} file. In this way,
{\tt HiGlu} provides a consistent calculation of the gluon-fusion cross
section up to NNLO QCD including dimension-six operator effects. More detailed
information about the input files {\tt higlu.in} and {\tt hdecay.in} can
be found as comment lines at the beginning of the main {\tt Fortran} files
{\tt higlu.f} and {\tt hdecay.f} shipped with the program. Finally, we note that
the pole masses of the bottom and charm quarks are computed from the
$\overline{\rm MS}$ input values $\overline{m}_b(\overline{m}_b)$ and
$\overline{m}_c(3~{\rm GeV})$ with N$^3$LO accuracy internally.

\subsection[{\tt Hawk}:  vector boson fusion and Higgs-strahlung channels]{{\tt Hawk}: Precision predictions for Higgs boson production in the vector boson fusion and the Higgs-strahlung channels}
\label{sec:hawk}
{\tt Hawk} is a parton-level Monte Carlo program providing precision predictions
for Higgs boson production in the vector-boson fusion and Higgs-strahlung modes.
For the Higgs-strahlung case, \textit{i.e.}, $VH$ production with $V=W$ or $Z$,
{\tt Hawk} includes the leptonic decays of the vector bosons, while
contributions stemming from $VH$ production with an hadronically decaying vector
boson can be included in the vector-boson-fusion calculation optionally. In the
Standard Model context, {\tt Hawk} provides fully differential predictions that
include QCD and electroweak next-to-leading-order corrections, and the results
are returned as binned distributions for important hadron-collider observables.
A detailed description of the program, that can be obtained at 
\url{http://hawk.hepforge.org}, 
can be found in Ref.~\cite{Denner:2014cla}, while details concerning the
underlying calculations are given in Refs.~\cite{Ciccolini:2007jr,
Ciccolini:2007ec,Denner:2011id}.

Concerning the inclusion of higher-dimensional operators, {\tt Hawk} supports
anomalous $HVV$ couplings which correspond to the Feynman rule
\beq
\mathrm{i}\, a_1^\mathrm{hvv}\,g_{\mu\nu}
+ \mathrm{i}\, a_2^\mathrm{hvv}\,(-k_1\cdot k_2 \, g_{\mu\nu}+k_{1\nu}k_{2\mu})
+ \mathrm{i}\, a_3^\mathrm{hvv}\,\epsilon_{\rho\sigma\mu\nu}\,k_1^\rho k_2^\sigma\ ,
\eeq
where $k_1$ and $k_2$ are the four-momenta of the gauge bosons and the $a_i$
parameters are real quantities. One can either
directly specify the coupling factors $a_1^\mathrm{hvv}$, $a_2^\mathrm{hvv}$ and
$a_3^\mathrm{hvv}$ in the input file or the coefficients of the corresponding
higher-dimensional operators, where a (modified) parameterization of the one
introduced in Ref.~\cite{Hankele:2006ma} is used. The correspondence of the
coefficients and the coupling factors is given in Ref.~\cite{Denner:2014cla}.
Anomalous coupling effects on the predictions can be calculated by including QCD
corrections. However, for the vector-boson fusion case, the dressing of the
anomalous
amplitudes with QCD corrections is restricted to the embedding of `diagonal QCD
contributions' which correspond to corrections in the $t$-channel (or DIS-like)
approximation where colour exchange between the two protons does not take place.
In the Standard Model, the remaining `non-diagonal' contributions are suppressed
to a phenomenologically irrelevant level, and the introduction of anomalous
coupling diagrams into those corrections would require an application of the
effective-field-theory approach including a renormalization of the anomalous
couplings. Moreover, the anomalous couplings related to the neutral gauge bosons
are switched off for small momentum transfer by means of a form factor,
\beq
  F_1 = |s_1| |s_2| / ( m_0^2 + |s_1| )/ ( m_0^2 + |s_2| )\ ,
\eeq
to avoid infrared singularities induced by the anomalous couplings. In this
expression, $s_1$ and $s_2$ denote the virtualities of the two intermediate
$W$ and $Z$~bosons and $m_0=1$~GeV is used. The code also allows one to use a
form factor
\beq
F_2 = \Lambda_\mathrm{hvv}^4/
(\Lambda_\mathrm{hvv}^2 \!+\! |s_1| )/ (\Lambda_\mathrm{hvv}^2 \!+\! |s_2| )\ ,
\eeq
to control the anomalous coupling effects at large momentum transfer.

The {\tt Hawk} implementation of anomalous $HVV$ has been used, \textit{e.g.},
in the ATLAS analysis~\cite{Aad:2016nal} that focuses on the
anomalous effects on vector-boson scattering via the reweighting of Standard
Model predictions.

\subsection{{\tt HPair}: Higgs boson pair production via gluon fusion}
\label{sec:hpair}
The program {\tt Hpair}, available at 
\url{http://tiger.web.psi.ch/hpair/}, 
calculates the Higgs boson pair production cross section via gluon fusion up to
the NLO in QCD~\cite{Dawson:1998py} in the limit of
heavy top quarks within the Standard Model and the quark-loop induced
contributions in the minimal supersymmetric extension of the Standard Model.
While genuine supersymmetric QCD (in the general case) and electroweak
corrections are unknown, known
subleading NLO top mass effects~\cite{Grigo:2013rya,Frederix:2014hta,
Maltoni:2014eza,Grigo:2015dia}, NNLO QCD corrections~\cite{deFlorian:2013jea,
deFlorian:2015moa,Grigo:2014jma} and NLO supersymmetric QCD corrections in the
limit of heavy superpartner masses~\cite{Agostini:2016vze} are not included in
the program. Starting from
the Standard Model Higgs result in the heavy top mass limit, the contributions of
dimension-six operators beyond the Standard Model are included up to the NLO in
QCD~\cite{Grober:2015cwa}. The latter extension is based on an effective
phenomenological Lagrangian given by
\beq
  {\cal L}_{\rm eff} = - m_t \bar{t}t \bigg[c_t \frac{H}{v} + c_{tt} \frac{H^2}{2 v^2} \bigg]
     - \frac16 c_3 \bigg[ \frac{3 m_h^2}{v} \bigg] H^3
     +  \frac{\alpha_s}{\pi} G_a^{\mu\nu} G_{\mu\nu}^a \bigg[
       c_g \frac{H}{v} + c_{gg}\frac{H^2}{2 v^2} \bigg],
\label{eq:leff}
\eeq
As in Section~\ref{sec:higlu},  the contributions of the dimension-six
operators are absorbed in the rescaling factors $c_t$ for the top Yukawa
coupling and $c_3$ for the trilinear Higgs boson coupling, \textit{i.e.}, deviations
of $c_t$ and $c_3$ from the Standard Model expectation of $c_t=c_3=1$ originate
from dimension-six operators. The remaining couplings $c_{tt}$, $c_g$ and
$c_{gg}$ are novel contributions purely arising from dimension-six operators and
not present in the Standard Model Lagrangian. Integrating out the heavy top
quark loops one arrives at the effective Lagrangian
\beq
  {\cal L}_{\rm eff} = \frac{\alpha_s}{\pi} G_a^{\mu\nu} G_{\mu\nu}^a \bigg\{
     \frac{H}{v} \bigg[
       \frac{c_t}{12} \Big( 1 + \frac{11}{4} \frac{\alpha_s}{\pi} \Big) + c_g \bigg]
   + \frac{H^2}{v^2} \bigg[
       \frac{c_{tt}-c_t^2}{24} \Big( 1 + \frac{11}{4} \frac{\alpha_s}{\pi}\Big)
       + \frac{c_{gg}}{2} \bigg]
   \bigg\},
\eeq
describing the Higgs boson couplings to gluons when only the leading QCD corrections
for the single Higgs interactions are kept (compared to Eq.~\eqref{eq:leffh}).

The partonic LO cross section extended by beyond the
Standard Model contributions induced by the above Lagrangian is given by
\begin{equation}
  \hat{\sigma}_{LO} (gg \to HH) = \int_{\hat{t}_-}^{\hat{t}_+} {\rm d} \hat{t}\,
  \frac{G_F^2 \alpha_s^2(\mu_R)}{512 (2\pi)^3} \,
  \left[ \left| C_\Delta F_1 + F_2 \right|^2   + |c_t^2 G_\Box|^2 \right],
\label{eq:cxnlo}
\end{equation}
where the Mandelstam variables are defined by
\beq
  \hat{s} = Q^2 \;, \qquad
  \hat{t} = m_h^2 - \frac{Q^2(1-\beta \cos \theta)}{2}  \qquad
  \mbox{and} \qquad
  \hat{u} = m_h^2 - \frac{Q^2(1+\beta \cos \theta)}{2},
\eeq
with $\beta = \sqrt{1-4m_h^2/Q^2}$ and with $Q$ denoting the invariant mass
of the Higgs boson pair. Moreover, the integration bounds at $\cos\theta=\pm 1$
read
\beq
  \hat{t}_\pm = m_h^2 -\frac{Q^2(1 \mp \beta)}{2}.
\eeq
The form factors $F_1$ and $F_2$ can be cast into the form
\begin{equation}
  F_1 = c_t F_\Delta + \frac{2}{3} c_\Delta \qquad \mbox{and} \qquad
  F_2 = c_t^2 F_\Box +c_{tt} F_\Delta - \frac{2}{3} c_\Box,
\end{equation}
with the explicit expressions of $F_\Delta$, $F_\Box$ and $G_\Box$ being
available in Ref.~\cite{Glover:1987nx,Plehn:1996wb}. In the limit of heavy top
quarks they simplify to
\begin{equation}
  F_1^{\rm lim} = \frac{2}{3} \Big( c_t + c_\Delta \Big) \;,
  \qquad F_2^{\rm lim} = \frac{2}{3} \Big(- c_t^2 + c_{tt} - c_\Box\Big),
\end{equation}
after introducing the abbreviations $c_\Delta$ and $c_\Box$. The latter read,
together with the $C_\Delta$ variable of Eq.~\eqref{eq:cxnlo},
\begin{equation}
C_\Delta \equiv \lambda_{hhh} \frac{m_Z^2}{Q^2-m_h^2+i m_h \Gamma_h}
\;, \qquad c_\Delta \equiv 12 c_{g} \qquad
\mbox{and} \qquad c_\Box \equiv - 12 c_{gg},
\end{equation}
with $m_Z$ denoting the $Z$-boson mass, $\Gamma_h$ the Higgs boson width, and
the trilinear Higgs boson coupling being normalized as
\begin{equation}
\lambda_{hhh} = \frac{3 m_h^2 c_3}{m_Z^2}.
\end{equation}
The partonic cross section has then to be convoluted with the gluon density in
the proton in order to obtain the hadronic cross section.

The present version 2.00 of {\tt Hpair} needs the usual 
Standard Model Higgs input values that are provided in a separate 
input file {\tt hpair.in}. In
addition, this file also allows one for choosing the anomalous dimension-six
factors $c_t$, $c_{tt}$, $c_g$, $c_{gg}$ and $c_3$. Alternatively the
dimension-six coefficients $\bar c_H$, $\bar c_u$, $\bar c_6$ and $\bar c_g$ can
be provided within the usual SILH
basis~\cite{Giudice:2007fh}, starting from the effective SILH Lagrangian
\beq
  \Delta {\cal L}_6^{\rm SILH} \supset
     \frac{\bar{c}_H}{2v^2} \partial_\mu(\Phi^\dagger \Phi) \partial^\mu (\Phi^\dagger \Phi)
   + \frac{\bar{c}_u}{v^2} y_t \Big(\Phi^\dagger \Phi \bar{q}_L \Phi^c t_R + {\rm h.c.}\Big)
   - \frac{\bar{c}_6}{6 v^2} \frac{3 m_h^2}{v^2} \Big(\Phi^\dagger \Phi\Big)^3
   + \bar{c}_g \frac{g_s^2}{m_W^2} \Phi^\dagger \Phi G_{\mu\nu}^a G_a^{\mu\nu},
\eeq
where $y_t = \sqrt{2} m_t/v$ denotes the top quark Yukawa coupling, $g_s$ the
strong coupling constant and $m_W$ the $W$-boson mass. The connection to the
Lagrangian of Eq.~\eqref{eq:leff} is given by
\begin{equation}
  c_t = 1 - \frac{\bar{c}_H}{2} - \bar{c}_u \;, \qquad
  c_{tt} = - \frac{1}{2} (\bar{c}_H + 3 \bar{c}_u ) \;, \qquad
  c_3 = 1- \frac{3}{2} \bar{c}_H + \bar{c}_6 \;, \qquad
  c_g = c_{gg} = \bar{c}_g \frac{4 \pi}{\alpha_2},
\end{equation}
with $\alpha_2 = \sqrt{2} G_F m_W^2/\pi$. Contrary to the strongly interacting
non-linear case, in the SILH case products of dimension-six coefficients are not
taken into account in accordance with a consistent expansion of the physical
observable up to linear dimension-six terms. In the SILH case, this also
requires the expansion of the LO cross section of Eq.~\eqref{eq:cxnlo} up to
linear dimension-six terms. The NLO terms are treated accordingly in the
{\tt Hpair} program that allows for both options, \textit{i.e.}, the non-linear
and SILH cases. In this way, {\tt Hpair} provides a consistent calculation of
the gluon-fusion cross section up to NLO QCD including the effects of
dimension-six operators. More detailed information about the input file
{\tt hpair.in} can be found as comment lines at the beginning of the main
{\tt Fortran} file {\tt hpair.f} shipped with the program.

\subsection{{\tt eHDecay}, Higgs boson decays in the effective Lagrangian approach}
\label{sec:ehdecay}
Ref.~\cite{Contino:2014aaa} has presented the program {\tt eHDecay}, a
{\tt Fortran} code based on a modification of the program
{\tt HDecay}~\cite{Djouadi:1997yw,Djouadi:2006bz}, and discussed the relation
between the non-linear and the linear effective Lagrangian approaches. The
program can be downloaded from 
\url{http://www.itp.kit.edu/~maggie/eHDECAY/}.
In {\tt eHDecay}, the full list of leading bosonic operators of 
the Higgs effective Lagrangian has been implemented, 
both for a linear and a non-linear
realization of the electroweak symmetry and for two benchmark 
composite Higgs models called MCHM4~\cite{Agashe:2004rs} 
and MCHM5~\cite{Contino:2006qr}. All
the relevant QCD corrections have been included and we detail in the following
the importance of higher-order QCD corrections and of mass effects on the
corrections. We also show how to consistently include electroweak
corrections whenever it is possible, focusing on the example of the Higgs boson
decay into two gluons.
As the leading part of the QCD corrections in general factorizes with respect to
the expansion in the number of fields and derivatives of the effective
Lagrangian, they can be included by taking over the results from the Standard
Model. The electroweak corrections, on the contrary, require dedicated
computations that are only partly available at present. They are consequently
only implemented up to higher orders in a $v^2/f^2$ expansion in the framework
of the linear Lagrangian, where $f=\Lambda/g_\star$ with $g_\star$ being
a typical new physics coupling strength. They are valid, for instance, in the
SILH framework~\cite{Giudice:2007fh} in which the deviations from the Standard
Model are small. On different lines, higher-order QCD corrections and mass
effects are included also in the non-linear implementation.

Denoting by $c_\psi$ the modifications of the Higgs boson couplings to fermions with
respect to the Standard Model and by $c'_{gg}$ the effective Higgs boson coupling
to gluons, the related non-linear effective Lagrangian reads
\beq
  \Delta{\cal L}_{\rm NL} =
    - \sum_{\psi=u,d,\ell} c_\psi m_\psi \bar{\psi} \psi \frac{H}{v}
    + \frac{c'_{gg}}{2} G^a_{\mu\nu} G_a^{\mu\nu} \frac{H}{v}\;.
\eeq
The decay rate into gluons implemented in {\tt eHDecay} in the framework of the
non-linear Lagrangian is then given by
\beq\begin{split}
 \Gamma_{gg}^{\rm NL}  =& \frac{G_F\alpha_s^2 m_h^3}{4\sqrt{2}\pi^3} \Bigg[
     \bigg| \sum_{q=t,b,c} \frac{c_q}{3} \, A_{1/2}(\tau_q) \bigg|^2
       c_{\rm eff}^2 \, \kappa_{\rm soft}
   + 2\, \Re\!\bigg\{\sum_{q=t,b,c} \frac{c_q}{3} \, A^*_{1/2}(\tau_q)
       \frac{2\pi c_{gg}}{\alpha_s} \bigg\}\, c_{\rm eff}\, \kappa_{\rm soft}
  \\ &
   + \bigg|\frac{2\pi c'_{gg}}{\alpha_s}\bigg|^2 \kappa_{\rm soft}
   + \frac{1}{9} \sum_{q,q'=t,b} c_q\, A^*_{1/2}(\tau_q) c_{q'}\,
       A_{1/2}(\tau_{q'}) \kappa^{\rm NLO} (\tau_q,\tau_{q'})
  \Bigg] \, .
\end{split}\label{ehdecay_label1}\eeq
On the other hand, the part of the linear Lagrangian that contributes to the
Higgs boson decay into two gluons reads, in the SILH basis,
\beq\begin{split}
  \Delta{\cal L}_6^{\rm SILH} \supset&
     \frac{\bar c_H}{2v^2} \partial^\mu(\Phi^\dagger \Phi)
        \partial_\mu (\Phi^\dagger \Phi)
   + \frac{\bar c_g g_s^2}{m_W^2} \Phi^\dagger \Phi G_{\mu\nu}^a G_a^{\mu\nu}
   + \Big( \frac{\bar c_u y_u}{v^2} \Phi^\dagger \Phi {\bar q}_L \Phi^c u_R +
           \frac{\bar c_d y_d}{v^2} \Phi^\dagger \Phi {\bar q}_L \Phi d_R +
           {\rm h.c.} \Big)\ ,
\end{split}\eeq
where $\bar c_H$, $\bar c_g$, $\bar c_u$ and $\bar c_d$ are the Wilson
coefficients of the corresponding dimension-six effective operators and
$y_{u,d}$ the Yukawa couplings of the up- and down-type quarks, respectively. In
this case, the Higgs boson decay rate into gluons implemented in {\tt eHDecay}
is given by
\beq\begin{split}
\Gamma_{gg}^{\rm SILH} =& \frac{G_F\alpha_s^2 m_h^3}{4\sqrt{2}\pi^3} \Bigg[
   \frac{1}{9} \sum_{q,q'=t,b,c}  (1-\bar{c}_H - \bar{c}_q- \bar{c}_{q'})
    A^*_{1/2}(\tau_{q'}) A_{1/2}(\tau_q) c_{\rm eff}^2\,  \kappa_{\rm soft}\\
 &
   + 2 \Re\!\bigg\{\sum_{q=t,b,c} \frac{1}{3} \, A^*_{1/2}(\tau_q)
       \frac{16\pi \bar{c}_{g}}{\alpha_2} \bigg\} \, c_{\rm eff} \, \kappa_{\rm soft}
   + \bigg|\sum_{q=t,b,c} \frac{1}{3} \, A_{1/2}(\tau_q) \bigg|^2 \,
       c_{\rm eff}^2 \, \kappa_{\rm ew} \, \kappa_{\rm soft} \\
 &
   + \frac{1}{9} \sum_{q,q'=t,b} (1-\bar{c}_H - \bar{c}_q- \bar{c}_{q'})
      A^*_{1/2}(\tau_q)  A_{1/2}(\tau_{q'}) \kappa^{\rm NLO} (\tau_q,\tau_{q'})
   \Bigg] \, .
\end{split}\label{gg-silh}\eeq
In both Eq.~\eqref{ehdecay_label1} and Eq.~\eqref{gg-silh},
$\tau_q=4m_q^2/m_h^2$ and the loop function $A_{1/2}(\tau)$ is defined as in
Eq.~\eqref{eq:Ahalf}. We use the pole masses for the top, bottom and charm quark
masses and $\alpha_s$ is computed up to N$^3$LO at a scale fixed to $m_h$ and
for $N_F=5$ active flavours. The QCD corrections have been taken into account up
to N$^3$LO in QCD in the limit of heavy loop-particle masses so that the effect
from low-energy gluon radiation, given by the coefficient $\kappa_{\rm soft}$,
factorizes in this limit. The corrections from high-energy gluon and quark
exchange (with virtuality $q^2 \gg m_t^2$) are encoded in the coefficient
$c_{\rm eff}$. Namely, for $m_h \muchless 2m_t$, the top quark can be integrated
out, leading to the effective five-flavour Lagrangian
\beq
  {\cal L}_{\rm eff} = -2^{1/4} G_F^{1/2} C_1 G^a_{\mu\nu} G_a^{\mu\nu} H \;,
\eeq
where the dependence on the top quark mass $m_t$ is included in the coefficient
function $C_1$. We then have
\beq
  \kappa_{\rm soft}=\frac{\pi}{2 m_h^4} \, \Im\big\{\Pi^{GG}(q^2=m_h^2)\big\}
  \qquad{\rm and}\qquad
  c_{\rm eff} = -\frac{12 \pi \, C_1}{\alpha_s(m_h)} \, ,
\eeq
with the vacuum polarization $\Pi^{GG} (q^2)$ being induced by the gluon
operator. The N$^3$LO expressions for $C_1$ and for $\Im\{\Pi^{GG}\}$ have been
given in Refs.~\cite{Chetyrkin:1997un,Kramer:1996iq,Schroder:2005hy,
Chetyrkin:2005ia} and Ref.~\cite{Baikov:2006ch}, respectively. In particular,
the NLO expressions for $\kappa_{\rm soft}$ and $c_{\rm eff}$
read~\cite{Inami:1982xt,Djouadi:1991tka,Chetyrkin:1997iv}
\beq
  \kappa_{\rm soft}^{\rm NLO} = 1 +
     \frac{\alpha_s}{\pi} \Big(\frac{73}{4} - \frac{7}{6} N_F \Big)
  \qquad{\rm and}\qquad
  c_{\rm eff}^{\rm NLO} = 1 + \frac{\alpha_s}{\pi} \frac{11}{4} \, ,
\eeq
in agreement with the low-energy theorem~\cite{Ellis:1975ap,Shifman:1979eb,
Kniehl:1995tn}. The additional mass effects at NLO~\cite{Spira:1995rr} in the
top and bottom quark loops are taken into account by the function
$\kappa^{\rm NLO} (\tau_q, \tau_{q^\prime})$ in the last lines of
Eqs.~\eqref{ehdecay_label1} and \eqref{gg-silh}. This function quantifies the
difference between the NLO QCD corrections for the top (bottom) contribution
taking into account finite mass effects in the loop, and the result for the top
(bottom) contribution in the limit of a large loop-particle mass.

Higher order corrections are large, with in particular the N$^3$LO QCD
corrections increasing the total width by almost up to 90\%.
The mass effects at NLO QCD are relevant for the bottom loop where they amount
to 8\%, while they are negligible for the top loop. To be consistent with the
non-linear approach, no electroweak corrections have been included in the
non-linear parameterization, as a perturbative expansion in powers of $v^2/f^2$
is not possible in the general case. In contrast, in the linear
parameterization, both QCD and electroweak corrections are included. The
treatment of the QCD corrections is in accordance with the non-linear case,
while the electroweak corrections~\cite{Djouadi:1994ge, Chetyrkin:1996wr,
Chetyrkin:1996ke, Aglietti:2004nj, Aglietti:2006yd, Degrassi:2004mx,
Actis:2008ug, Actis:2008ts} are taken into account by the factor
$\kappa_{\rm ew}$ of Eq.~\eqref{gg-silh}. In the Standard Model, they can be
factorized in the particular case of the Higgs boson decaying into two gluons.
However, this factorization is not valid in the general case. In particular,
when the contributions from the effective Lagrangian are taken into account,
modified Higgs boson couplings absent at the leading order can appear and spoil the
factorization. Hence, a consistent inclusion of the electroweak corrections is
only possible in the linear-Lagrangian case and up to higher orders in
$v^2/f^2$. More precisely, the Standard Model electroweak corrections can be
added to get a result that includes the leading ${\cal O}(v^2/f^2)$ and the
next-to-leading ${\cal O}(\alpha/4\pi)$ corrections as well as the mixed
${\cal O}[(\alpha_s/4\pi)^5 (\alpha/4\pi)]$ contributions, but that neglects terms of
${\cal O}[(\alpha/4\pi) (v^2/f^2)]$, ${\cal O}[((v^2/f^2)^2]$ and
${\cal O}[(\alpha/4\pi)^2]$,
where $\alpha$ is generically meant as the electroweak expansion parameter.

For the future it is planned to add to {\tt eHDecay} contributions from
the fermionic effective operators and those from the effective Lagrangian to the
set of input observables $m_W$, $m_Z$ and $G_F$. Moreover, in accordance with
the Higgs Cross Section Working Group Internal Note of
Ref.~\cite{LHCHXSWG-INT-2015-006}, a switch from the pole-mass scheme to the
$\overline{\rm MS}$ scheme for the fermionic masses is under consideration, as
well as the implementation of an interface for the so-called Higgs basis  
parameterization of the effective Lagrangian (see Section~\ref{s.eftbasis} and Section~\ref{sec:rosetta}).

\subsection[Higgs Pseudo-Observables in the universal {\tt FeynRules} output]{Implementation of Higgs Pseudo-Observables in the universal
  {\tt FeynRules} output}
\label{sec:PO}
With the LHC Run-II, Higgs physics is entering a precision era. This
would allow us to look for new physics effects not only in the overall signal
strengths, but also in kinematical distributions. In this perspective, the
so~called `$\kappa$-framework'~\cite{LHCHiggsCrossSectionWorkingGroup:2012nn} is
insufficient and needs to be extended. The natural extension is the so-called
Higgs pseudo-observables framework introduced in
Refs.~\cite{Gonzalez-Alonso:2014eva,Greljo:2015sla}. A detailed discussion about
this formalism is presented in Chapter~\ref{chap:PO}. Here we limit
ourself to summarize its main features in order to illustrate its implementation
via the Universal {\tt FeynRules} Output model {\tt HiggsPO} that is available
at  \url{http://www.physik.uzh.ch/data/HiggsPO}.

The pseudo-observables are a finite set of parameters that are experimentally
accessible, well-defined from the point of view of a quantum field theory,
and that characterize possible deviations from the Standard Model in processes
involving the Higgs boson in great generality. More precisely, 
the Higgs pseudo-observables are defined from a general decomposition (based on
analyticity, unitarity, and crossing symmetry) of on-shell amplitudes involving
the Higgs boson and a momentum expansion following the assumption of no new
light particles in the kinematical regime where the decomposition is assumed to
be valid. A further key assumption of the pseudo-observables formalism is that
the Higgs boson is a spin zero resonance with a narrow width, such that new
physics effects in production and decay factorize.

For the convenience of the reader, let us stress the difference between a
pseudo-observable approach, such as the one summarized here, and an effective
field theory one. On the one hand, pseudo-observables are defined from on-shell
properties of the relevant scattering amplitudes and are thus well-defined at
all orders in perturbation theory. On the other hand, effective field theory
coefficients are Lagrangian parameters and, as such, not observable quantities.
Pseudo-observables can be computed in an effective field theory approach at a
given order in perturbation theory and, doing this at the tree-level, it is
possible to derive a one-to-one correspondence between the two setups. However,
this correspondence would change, and become more complex, if the computation is
performed at higher orders: the physical meaning of the pseudo-observable would
not change while the operator coefficients would loose the direct connection to
observable quantities which was obtained at the tree-level. In other words, the
difference between pseudo-observables and effective field theory coefficients is
the same as the one between the pole mass of a particle (a pseudo-observable)
and the mass parameters in the Lagrangian.

Due to their simple kinematics, two-body Higgs boson decays (into a $\gamma \gamma$
and an $\bar{f}f$ pair) can be parameterized by only two pseudo-observables
each, one describing the $CP$-conserving amplitude and another for the
$CP$-violating one. Unless the polarization of the final states can be measured,
only a combination of the two is experimentally accessible. Three-body and
four-body Higgs boson decays (such as into $2\ell\gamma$, $4\ell$ and $2\ell 2\nu$ systems)
have a more complicated kinematics. In this case the pseudo-observables are
defined from the expansion of the on-shell amplitudes around the known physical
poles due to the propagation of intermediate Standard Model electroweak gauge
bosons. General amplitude decomposition and pseudo-observables definitions can
be found in Ref.~\cite{Gonzalez-Alonso:2014eva} for the Higgs boson decays and in
Ref.~\cite{Greljo:2015sla} for Higgs boson production in the vector-boson fusion
mode and in association with a gauge boson.

The pseudo-observables are defined directly at the amplitude level. As such, on
the one hand they can be computed from a specific Lagrangian (to a specific
order in perturbation theory), and on the other hand the pseudo-observables are
directly connected to $S$-matrix elements, providing a direct link to physical
observables~\cite{Gonzalez-Alonso:2014eva}. They are thus particularly well
suited for the analysis of experimental data with the matrix-element
method~\cite{Gainer:2013iya}. In order to use the pseudo-observable
decomposition for precision studies, it is important to account for the
long-distance contributions due to soft and collinear photon emission
(\textit{i.e.}, the leading QED radiative corrections). These represent a
universal correction factor that can be implemented, by means of appropriate
convolution functions or, equivalently, by showering algorithms in Monte Carlo
simulations, irrespective of the specific short-distance structure of the
amplitude. It has been shown that inclusion of such correction in $H\to2e2\mu$
decays recovers the complete NLO Standard Model predictions within an accuracy of
about $1\%$~\cite{Bordone:2015nqa}.

\renewcommand{\arraystretch}{1.3}
\begin{table}
 \centering
 \begin{tabular}{c | c c c c c}
   Process & $H \to  b b$ & $H \to  \tau \tau$ & $H \to  c c$ & $H \to  \mu \mu$ \\\hline
   Pseudo-observables & $\kappa_b,~ \delta^{\rm CP}_b$ & $\kappa_\tau,~ \delta^{\rm CP}_\tau$ & $\kappa_c,~ \delta^{\rm CP}_c$ & $\kappa_\mu,~ \delta^{\rm CP}_\mu$
\end{tabular}
\caption{\label{tab:POYuk} Pseudo-observables relevant for Higgs boson decays into
  two fermions.}
\begin{tabular}{c | l}
Process 			& Pseudo-observables\\[0.1cm] \hline
$H \to \gamma\gamma$ & $\kappa_{\gamma\gamma},~ \delta^{\rm CP}_{\gamma\gamma}$ \\[0.1cm]
$H \to Z\gamma$ & $\kappa_{Z\gamma},~ \delta^{\rm CP}_{Z\gamma}$ \\[0.1cm]
$H \to \gamma 2\nu$ & $\kappa_{Z\gamma},~ \delta^{\rm CP}_{Z\gamma}$ \\[0.1cm]
$H \to \gamma 2\ell$ & $\kappa_{\gamma\gamma},~ \delta^{\rm CP}_{\gamma\gamma},~ \kappa_{Z\gamma},~ \delta^{\rm CP}_{Z\gamma}$ \\[0.1cm]
$H \to Z 2\ell$ & $\kappa_{ZZ},~ \epsilon_{ZZ},~ \epsilon^{\rm CP}_{ZZ},~ \kappa_{Z\gamma},~ \delta^{\rm CP}_{Z\gamma},~ \epsilon_{Z \ell_L},~ \epsilon_{Z \ell_R}$ \\[0.1cm]
$H \to 2\ell2\ell'$ & $\kappa_{ZZ},~ \epsilon_{ZZ},~ \epsilon^{\rm CP}_{ZZ},~ \kappa_{Z\gamma},~ \delta^{\rm CP}_{Z\gamma},~ \kappa_{\gamma\gamma},~ \delta^{\rm CP}_{\gamma\gamma},~ \epsilon_{Z \ell_L},~ \epsilon_{Z \ell_R},~ \epsilon_{Z \ell'_L},~ \epsilon_{Z \ell'_R}$ \\[0.1cm]
$H \to 4\ell$ & $\kappa_{ZZ},~ \epsilon_{ZZ},~ \epsilon^{\rm CP}_{ZZ},~ \kappa_{Z\gamma},~ \delta^{\rm CP}_{Z\gamma},~ \kappa_{\gamma\gamma},~ \delta^{\rm CP}_{\gamma\gamma},~ \epsilon_{Z \ell_L},~ \epsilon_{Z \ell_R}$ \\[0.2cm]
$H \to \bar\ell\ell2\nu$ & $\left\{\!\! \begin{array}{l} \kappa_{ZZ},~ \epsilon_{ZZ},~ \epsilon^{\rm CP}_{ZZ},~ \kappa_{Z\gamma},~ \delta^{\rm CP}_{Z\gamma},~ \epsilon_{Z \ell_L},~ \epsilon_{Z \ell_R},~ \epsilon_{Z \nu} \\
       			  \kappa_{WW},~ \epsilon_{WW},~ \epsilon^{\rm CP}_{WW},~ \epsilon_{W \ell},~ \phi_{W \ell} \end{array} \right.$ \\[0.4cm]
$H \to \bar\ell\ell'2\nu$ & $\kappa_{WW},~ \epsilon_{WW},~ \epsilon^{\rm CP}_{WW},~ \epsilon_{W \ell},~ \phi_{W \ell},~ \epsilon_{W \ell'},~ \phi_{W \ell'}$
\end{tabular}
  \caption{\label{tab:POHPO} Pseudo-observables relevant for Higgs boson decays into
  a gauge-boson pair, and into a three-body or four-body system. We denote by
  $\ell$ an electron, a muon or a tau and $\nu$ indicates any of the three
  neutrino species.}
\end{table}

In order to use the pseudo-observable framework in experimental analyses, it is
convenient to have a tool capable of generating signal events for specific
values of the pseudo-observables. Such a tool has been developed in the context
of the Higgs boson decays and is publicly available on the above webpage that also
includes a detailed user manual. The implementation of pseudo-observables in
electroweak Higgs boson production including NLO QCD corrections is underway and
will be available at the same Internet link. We have implemented into
{\tt FeynRules}~\cite{Alloul:2013bka} (version 2.3.1) a {\tt HiggsPO} model,
whose source file is named {\tt HPO.fr}, by means of an effective Lagrangian generating
the corresponding vertices at the tree level. We stress that such a Lagrangian
should not be considered as a specific model or an effective field theory
description, but rather as an auxiliary tool to be used (at tree level only) in
order to reproduce the correct Higgs boson decay amplitude decomposition in terms of
pseudo-observables that is valid beyond the tree level.

The {\tt HiggsPO} {\tt FeynRules} model as implemented in the {\tt HPO.fr} file
is exported to the Universal FeynRules Output (UFO)
format~\cite{Degrande:2011ua}, which can then be used within the
{\tt MG5\_}a{\tt MC@NLO}~\cite{Alwall:2014hca} and
{\tt Sherpa}~\cite{Gleisberg:2008ta} event generators. The general idea is that
after simulating Higgs boson production with a dedicated Monte Carlo generator,
the Higgs boson can be decayed at the parton-level with the {\tt HiggsPO} model
and the partonic events can then be passed to a general purpose event generator
for subsequent parton showering and hadronization (such as
{\tt Pythia}~\cite{Sjostrand:2007gs}). This last step will automatically
account for the important radiative corrections~\cite{Bordone:2015nqa}. We note
that it is very practical to use the {\tt MadSpin}
module~\cite{Artoisenet:2012st} of {\tt MG5\_}a{\tt MC@NLO} to decay the
Higgs boson with the \textsc{HiggsPO} model on the fly. We stress again that our
{\tt FeynRules} implementation only consists of a set of effective interactions
that generate exactly the scattering amplitude of interest at the tree level and is
supposed to be used for this purpose only. It should not be used as a
Lagrangian for arbitrary process and beyond the tree level.

All the Higgs boson decay processes implemented in the
\textsc{HiggsPO} model, together with the associated Higgs pseudo-observables
accessible in the parameter card, are summarized in Table~\ref{tab:POYuk} and
Table~\ref{tab:POHPO}. The Higgs pseudo-observables relevant for
describing the Higgs boson decays into two fermions are shown in Table~\ref{tab:POYuk}.
The coupling strengths have been assigned an interaction order {\tt YUK}~$=1$.
The pseudo-observables relevant for all other Higgs boson decays are presented in
Table~\ref{tab:POHPO} and the coupling strengths have been related to an
interaction order {\tt HPO}~$=1$.
Translation of the pseudo-observable language to the Higgs basis defined in
 Section~\ref{s.eftbasis} is also available in
Refs.~\cite{Gonzalez-Alonso:2014eva,Greljo:2015sla}.

\subsection[Higgs and BSM characterization in the
{\tt MG5\_}a{\tt MC@NLO} framework]{Higgs and beyond the Standard Model characterization in the
{\tt MG5\_}a{\tt MC@NLO} framework}
\label{sec:HC}

The {\tt Higgs Characterization} (HC) implementation provides a complete
framework, based on an effective field theory description, that allows for the
study of the Higgs boson properties in a consistent, systematic and accurate
way. The HC~\cite{Artoisenet:2013puc} follows the general strategy outlined in
Ref.~\cite{Christensen:2009jx}, and has been implemented in a complete
simulation chain from Lagrangian to hadron-level events, especially in the
{\tt FeynRules/MG5\_}a{\tt MC@NLO} framework. The HC effective Lagrangian
features bosons $X(J^P)$ with various spin-parity assignments ($J^P=0^+$, $0^-$,
$1^+$, $1^-$ and $2^+$) and has been implemented in terms of
mass eigenstates into {\tt FeynRules}~\cite{Alloul:2013bka}, whose output files
are interfaced~\cite{Degrande:2011ua,deAquino:2011ub} with various event
generators. The HC model files are publicly available online, 
\url{http://feynrules.irmp.ucl.ac.be/wiki/HiggsCharacterisation}.

Later on, the HC framework has been extended as a 
{\tt Beyond} {\tt the} {\tt Standard} {\tt Model} {\tt Characterization}
(BSMC) framework~\cite{Falkowski:2015wza} where the effective
Lagrangian has been constructed as above, but starting from the Higgs basis
Lagrangian in  Section~\ref{s.eftbasis} where all possible effective
operators of dimension up to six are written in terms of mass
eigenstates. In particular, fermionic operators different from the Yukawa
interactions are now included. However, instead of imposing the realization of
the electroweak
symmetry that relates the Wilson coefficients, the latter have been kept
independent. Due to the lack of manifest $SU(2)_L\times U(1)_Y$ invariance, the
BSMC Lagrangian is associated with a larger number of independent coefficients
compared to more traditional bases such as the Warsaw~\cite{Grzadkowski:2010es}
or SILH~\cite{Giudice:2007fh,Contino:2013kra} bases. Therefore, the BSMC model
can be used in many contexts, not only in the case of linear effective field
theories but also in the non-linear case or in the pseudo-observable
parameterization (see Section~\ref{sec:PO}). As introduced in
Section~\ref{sec:rosetta}, the {\tt Rosetta} package allows for the translation
of the Wilson coefficients given in a particular basis of non-redundant
dimension-six operators (such as the Warsaw or SILH basis) to the BSMC
parameterization coefficients by including constraints that render the
Lagrangian invariant under the full electroweak symmetry group. Extensions to
other translations are foreseen. As for the HC
case, the BSMC Lagrangian has been implemented into {\tt FeynRules} and is
available online, \url{http://feynrules.irmp.ucl.ac.be/wiki/BSMCharacterisation}. 
A corresponding implementation of the dimension-six Lagrangian above the weak
scale, where $SU(2)_L \times U(1)_Y$ is an exact symmetry, has been also
achieved~\cite{Alloul:2013naa} and has overlapping as well as complementary
features with respect to the HC and BSMC Lagrangians. It is available at 
 \url{http://feynrules.irmp.ucl.ac.be/wiki/HEL}. 
In this case, the implementation has been performed in the SILH basis.

There are several advantages in having a first principle implementation in terms
of an effective Lagrangian which can be automatically interfaced to event
generators. First and most important, all relevant production and decay modes
can be studied within the same model, and the corresponding processes
automatically generated within minutes. Second, it is straightforward to modify
the model implementation to extend it further in case of need, by adding further
interactions, for example of higher dimensions in energy. Finally, higher-order
effects can be easily accounted for, by generating multi-jet merged samples or
computing NLO corrections within automatic frameworks. As accumulated data has
been increasing, the last point became more and more important. The detailed
demonstrations and analyses for all the main production modes of the Higgs boson
(gluon fusion, vector boson fusion, $VH$ associated production and $t\bar tH$
production) as well as $tH$ production at NLO accuracy in QCD have been done
recently~\cite{Maltoni:2013sma,Demartin:2014fia,Demartin:2015uha}, within the
{\tt MG5\_}a{\tt MC@NLO} program~\cite{Alwall:2014hca}. In the following,
we focus, for the sake of the example, on a spin-0 (Higgs) boson that is denoted
by $X_0$, and compute several differential distributions for some production
processes and for different benchmark scenarios.

In the HC framework, the only assumptions are that the 125~GeV resonance found
at the LHC corresponds to a spin-0 state, that no other new state coupled to
such a resonance exists below the cutoff scale $\Lambda$ and that new physics is
dominantly described by the lowest dimensional operators. We thus
include all effects stemming from the complete set of dimension-six operators
allowed by the Standard Model gauge symmetry. The effective interaction
Lagrangian is given by
\beq\begin{split}
 {\cal L}_0 =&  \bigg\{ -\sum_{f=t,b,\tau}\bar\psi_f\big(
         c_{\alpha}\kappa_{\sss Htt}g_{\sss Htt}
       +i s_{\alpha}\kappa_{\sss Att}g_{\sss Att}\, \gamma_5 \big)
      \psi_f +
  c_{\alpha}\kappa_{\rm SM}\big[\frac{1}{2}g_{\sss HZZ}\, Z_\mu Z^\mu
                                +g_{\sss HWW}\, W^+_\mu W^{-\mu}\big] \\
  & \quad -\frac{1}{4}\big[c_{\alpha}\kappa_{\sss H\gamma\gamma}
  g_{\sss H\gamma\gamma} \, A_{\mu\nu}A^{\mu\nu}
        +s_{\alpha}\kappa_{\sss A\gamma\gamma}g_{ \sss A\gamma\gamma}\,
  A_{\mu\nu}\widetilde A^{\mu\nu}
  \big]
  -\frac{1}{2}\big[c_{\alpha}\kappa_{\sss HZ\gamma}g_{\sss HZ\gamma} \,
  Z_{\mu\nu}A^{\mu\nu}
        +s_{\alpha}\kappa_{\sss AZ\gamma}g_{\sss AZ\gamma}\,Z_{\mu\nu}\widetilde A^{\mu\nu} \big] \\
  &\quad -\frac{1}{4}\big[c_{\alpha}\kappa_{\sss Hgg}g_{\sss Hgg} \, G_{\mu\nu}^aG^{a,\mu\nu}
        +s_{\alpha}\kappa_{\sss Agg}g_{\sss Agg}\,G_{\mu\nu}^a\widetilde G^{a,\mu\nu} \big]
  -\frac{1}{4}\frac{1}{\Lambda}\big[c_{\alpha}\kappa_{\sss HZZ} \, Z_{\mu\nu}Z^{\mu\nu}
        +s_{\alpha}\kappa_{\sss AZZ}\,Z_{\mu\nu}\widetilde Z^{\mu\nu} \big] \\
  &\quad -\frac{1}{2}\frac{1}{\Lambda}\big[c_{\alpha}\kappa_{\sss HWW} \, W^+_{\mu\nu}W^{-\mu\nu}
        +s_{\alpha}\kappa_{\sss AWW}\,W^+_{\mu\nu}\widetilde W^{-\mu\nu}\big] \\
  &\quad -\frac{1}{\Lambda}c_{\alpha}
    \big[ \kappa_{\sss H\partial\gamma} \, Z_{\nu}\partial_{\mu}A^{\mu\nu}
         +\kappa_{\sss H\partial Z} \, Z_{\nu}\partial_{\mu}Z^{\mu\nu}
 + \kappa_{\sss H\partial W}\big( W_{\nu}^+\partial_{\mu}W^{-\mu\nu}+h.c.\big)
 \big]
 \bigg\}\, X_0  \,,
\end{split}\label{eq:LX0}\eeq
where $A_{\mu\nu}$ ($\widetilde A_{\mu\nu}$), $Z_{\mu\nu}$
($\widetilde Z_{\mu\nu}$) and $W^\pm_{\mu\nu}$ ($\widetilde W^\pm_{\mu\nu}$)
represent the field strength tensors of the physical electroweak bosons,
and $G_{\mu\nu}^a$ and $\widetilde G^a_{\mu\nu}$ the gluon field strength
tensor and its dual. This parameterization allows for the easy recovery of the
Standard Model case by fixing the dimensionless parameters $\kappa_i$ to
appropriate values, since the dimensionful couplings $g_{\sss Xyy'}$ are set to
their Standard Model values. For instance, we enforce that
\mbox{$g_{\sss H\gamma\gamma}=47\alpha/18\pi v$} and
\mbox{$g_{\sss Hgg}=-\alpha_s/3\pi v$}, two values that can be obtained in the
limit of large $W$-boson and top-quark masses. Moreover, the model description includes
the interactions of a $0^-$ state typical of supersymmetry or of generic two
Higgs doublet models, and enables the $CP$-mixing between the $0^+$ and $0^-$
states via a mixing angle $\alpha$ whose sine and cosine are denoted by
$s_\alpha$ and $c_\alpha$. In the Standard Model, $c_\alpha=1$ and $s_\alpha=0$.

\begin{table}
\center
\begin{tabular}{l|l||l l}
 Scenario & HC parameter choice
 & Scenario & HC parameter choice \\
\hline
 $0^+$(GF, SM)  & $\kappa_{\sss Hgg/Htt}=1\ (c_{\alpha}=1)$
 &$0^+$(VBF, SM) &  $\kappa_{\sss SM}=1\ (c_{\alpha}=1)$\\
 $0^-$(GF)      & $\kappa_{\sss Agg/Att}=1\ (c_{\alpha}=0)$
 &$0^+$(VBF, HD) &  $\kappa_{\sss HZZ,HWW}=1\ (c_{\alpha}=1)$\\
 $0^{\pm}$(GF)  & $\kappa_{\sss Hgg,Agg/Htt,Att}=1\ (c_{\alpha}=1/\sqrt{2})$
 & $0^-$(VBF, HD) &  $\kappa_{\sss AZZ,AWW}=1\ (c_{\alpha}=0)$\\
 &&$0^{\pm}$(VBF, HD) & $\kappa_{\sss HZZ,HWW,AZZ,AWW}=1\ (c_{\alpha}=1/\sqrt{2})$\\
\end{tabular}
\caption{Benchmark scenarios for $X_0$ production in the gluon fusion and
  $t\bar t H$ (GF) and in the vector boson fusion (VBF) channel.}
\label{tab:HCscenarios}
\end{table}

We then make use of {\tt MG5\_}a{\tt MC@NLO} to generate both a numerical
code and events, at the NLO accuracy in QCD. In practice, for the production of
an $X_0$ state plus two jets in the gluon fusion channel, this is achieved
by issuing the following commands (with the \texttt{/ t} syntax forbidding
diagrams containing top loops),\\
\hspace*{.5cm}  \verb+> import model HC_NLO_X0-heft+\\
\hspace*{.5cm}  \verb+> generate p p > x0 j j / t [QCD]+\\
\hspace*{.5cm}  \verb+> output+\\
\hspace*{.5cm}  \verb+> launch+\\
where the {\tt -heft} suffix in the model name refers to the corresponding model
restriction. As a result, all the amplitudes featuring the Higgs--gluon
effective vertices in the heavy-top limit are generated, including corrections
up to the NLO in QCD. Analogous commands can be issued to generate events
related to the production of an $X_0$ state plus zero and one jet. After the
{\tt launch} command, one can modify the {\tt param\_card.dat} file to change
the values of the $\Lambda$, $\kappa$ and $c_\alpha$ parameters. Similarly, a
code and events can be generated in the vector boson fusion case (left, with the
\textsc{\$\$} sign forbidding diagrams with $W^{\pm}$ or $Z$ bosons
in the $s$-channel as they are included in Higgs-strahlung production) and in
the Higgs-strahlung production mode (right) by typing in\\
\hspace*{.5cm}\verb"> import model HC_NLO_X0"\\
\hspace*{.5cm}\verb"> generate p p > x0 l+ vl [QCD]"\\
\hspace*{.5cm}\verb"> generate p p > x0 j j $$ w+ w- z / a [QCD]"\\
\hspace*{.5cm}\verb"> add process p p > x0 l- vl~ [QCD]"\\
\hspace*{.5cm}\verb"> add process p p > x0 l+ l- / a [QCD]"\\
as well as in the $t \bar t H$ production mode with the command\\
\hspace*{.5cm}  \verb+> generate p p > x0 t t~ [QCD]+\\
Furthermore, the $X_0$ and the top quark decays are subsequently performed
starting from the event files generated as above with the help of the
{\tt MadSpin}~\cite{Artoisenet:2012st} package, following the procedure
described in Ref.~\cite{Frixione:2007zp} that allows one to keep spin
correlations.

\begin{figure}
 \centering
 \includegraphics[width=0.32\textwidth]{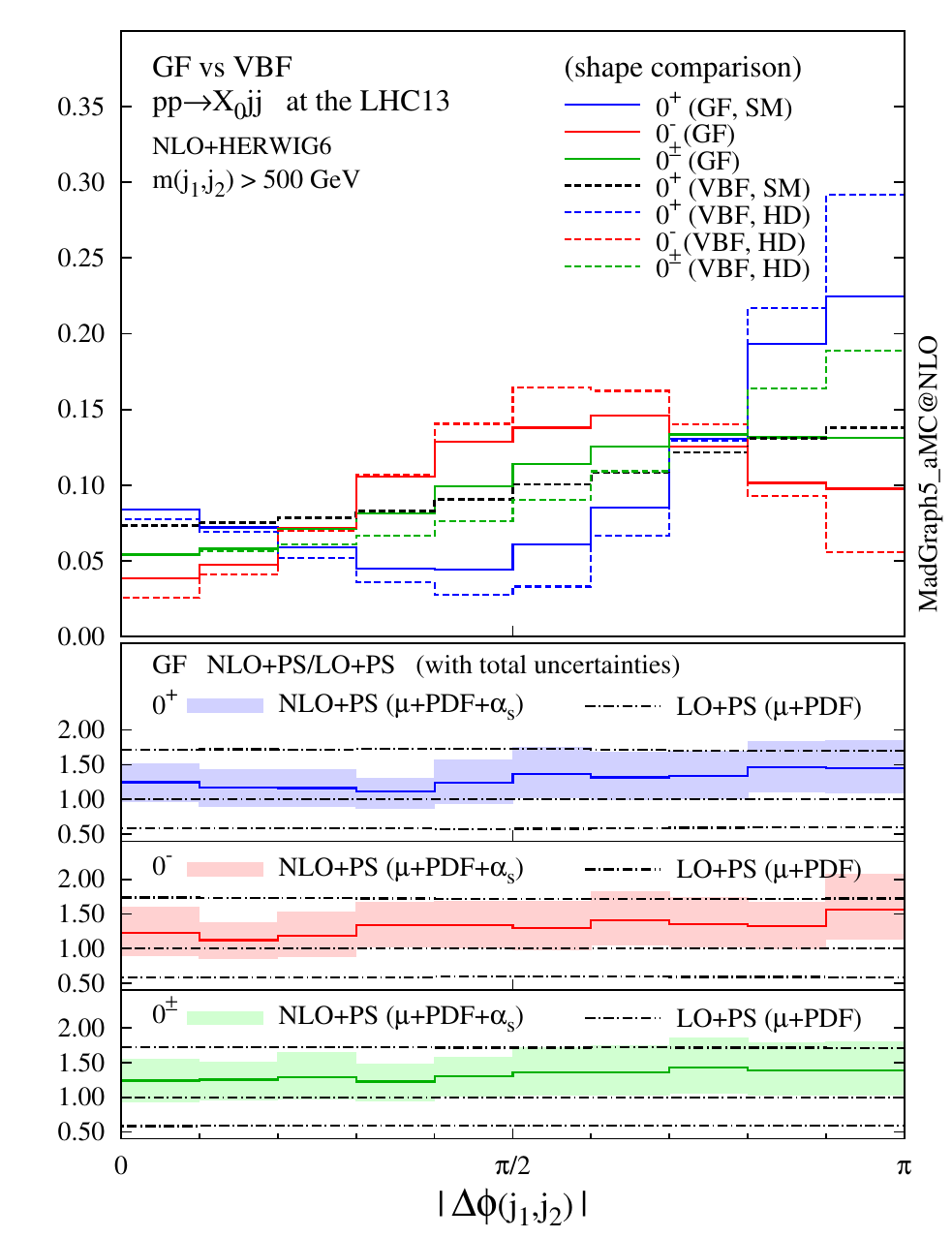}
 \includegraphics[width=0.32\textwidth]{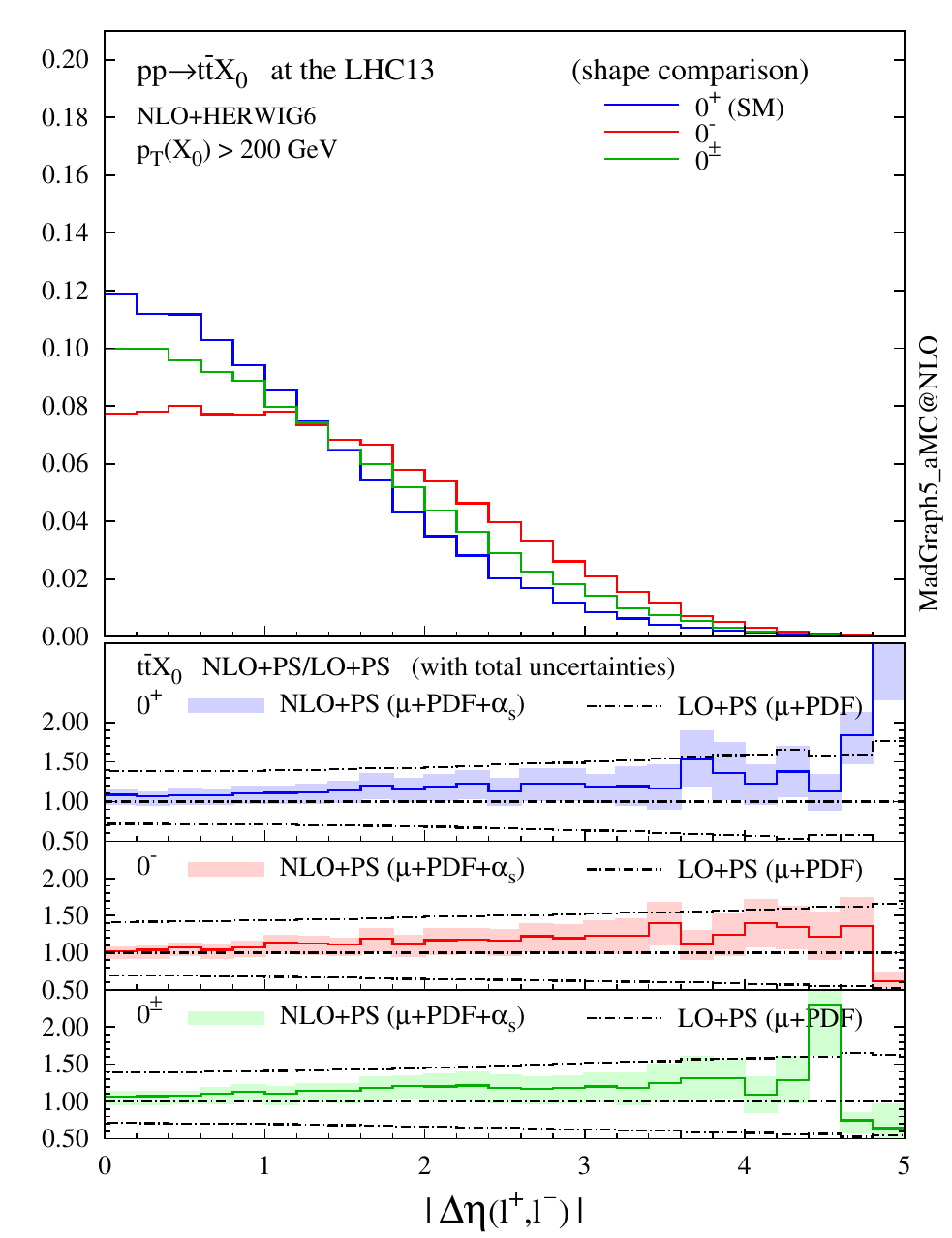}
 \includegraphics[width=0.32\textwidth]{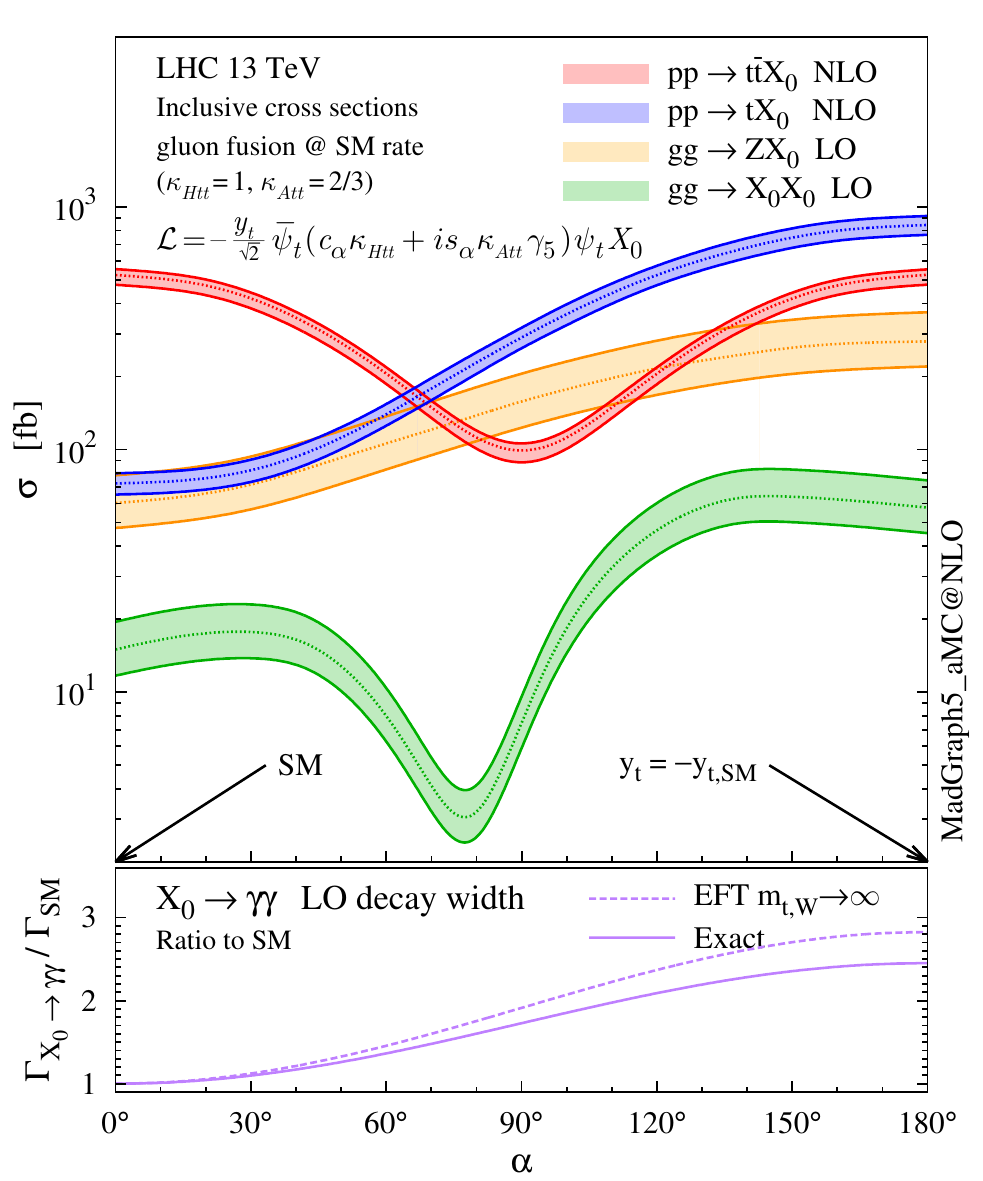}
 \caption{Normalized kinematical distributions at the LHC, running at a
   centre-of-mass energy of 13~TeV. We show the azimuthal difference between the
   two tagging jets for $pp\to X_0jj$ after imposing a $m_{jj}>500$~GeV
   selection (left) and the rapidity separation between the leptons (centre) for the
   dileptonic decay channel in $pp\to t\bar tX_0$ after enforcing a
    $p_T(X_0)>200$~GeV selection. For each
   scenario, the lower panels give the ratio of the NLO results matched to parton
   showers to the LO results matched to parton showers, together with the total
   theoretical uncertainties. In the right panel of the figure, we show NLO (or
   loop-induced LO) cross sections, presented with the associated
   scale uncertainties, for $t\bar tH$ and $t$-channel $tH$ ($ZH$ and
   $HH$) production as a function of the $CP$-mixing angle $\alpha$. The
   $\kappa_{\sss Htt}$ and $\kappa_{\sss Att}$ parameters have been set to
   reproduce the Standard Model  gluon-fusion cross section for every value of
   $\alpha$. The ratio of the $X_0\to\gamma\gamma$ partial decay width to the
   corresponding Standard Model value is also shown in the lower panel of the
   figure.}
\label{fig:dis_unc}
\end{figure}

We show in Figure~\ref{fig:dis_unc} (left and central panels) a few kinematical
distributions, including
theoretical uncertainties, for the benchmark scenarios presented in
Table~\ref{tab:HCscenarios}. The parton densities are evaluated by using the
NNPDF2.3 (LO/NLO) parameterization~\cite{Ball:2012cx} and the central value
$\mu_0$ for the renormalization ($\mu_R$) and factorization ($\mu_F$) scales
is set to $H_T/2$, $m_W$ and $\sqrt[3]{m_T(t)\,m_T(\bar t)\,m_T(X_0)}$ in the
gluon fusion, vector boson fusion and $t\bar tH$ production channel respectively.
Uncertainties have been automatically calculated within the
{\tt MG5}\_a{\tt MC@NLO} framework and consist of the linear sum of two
components
respectively related to the scale and the parton density (plus $\alpha_s$) dependence.
   Scale uncertainties have been obtained by varying independently the
   unphysical scales by a factor of two up and down with respect to the reference
   scale $\mu_0$, and parton densities and $\alpha_s$ uncertainties have been derived
   following the PDF4LHC recommendations.

Finally, in the right panel of the figure, we present the dependence of the
$t\bar tH$ and $tH$ production cross sections on the $CP$-mixing angle $\alpha$. The
nature of the top quark Yukawa coupling also affects the loop-induced Higgs
coupling to gluons and photons. In order to maintain the Standard Model
gluon-fusion production cross section, the rescaling parameters are set to
$\kappa_{\sss Htt}=1$ and $\kappa_{\sss Att}=2/3$.
The LO cross sections for the loop-induced $ZH$~\cite{Hespel:2015zea} and
$HH$~\cite{Hespel:2014sla} production processes via gluon fusion are also shown as
references.

\subsection{Higgs boson properties with the {\tt JhuGen} / {\tt Mela}
  framework} \label{sec:jhugen}
The {\tt Jhu Generator} and {\tt Mela} framework~\cite{Gao:2010qx,%
Bolognesi:2012mm,Anderson:2013afp}, available at \url{http://www.pha.jhu.edu/spin/},
is designed for the study of anomalous couplings of a resonance to vector bosons
and fermions in various processes. A wide range of production and decay
channels are supported for either spin-zero, spin-one, or spin-two resonances
and for the most general Lorentz structures of the $HVV$ and $Hff$ interaction
vertices, with the focus on the spin-zero case. These processes include the
hadronic production of the resonances in association with zero, one, or two
jets, their production via vector boson fusion, their associated production with
a vector boson ($ZH$, $WH$), and their production in association with heavy
flavour quarks (such as $t\bar{t}H$, $tH$ and $b\bar{b}H$). The supported decay
modes include $H\to ZZ$ / $Z\gamma^*$ / $\gamma^*\gamma^*\to4f$,
$H\to WW \to4f$, $H\to Z\gamma $ / $\gamma^*\gamma \to2f\gamma$,
$H\to \gamma\gamma$, $H\to \tau\tau$, and generally $H\to f\bar{f}$, with a
complete modelling of the spin correlations including the interference effects
related to identical particles. In the case of a resonance carrying a non-zero
spin or of the associated production of a spin-zero resonance, spin correlations
between the initial and final states are also fully modeled.

While the {\tt JhuGen} / {\tt Mela} framework can be used in a standalone mode,
it is also integrated with the {\tt Mcfm} Monte Carlo
package~\cite{Campbell:2010ff,Campbell:2011bn,Campbell:2013una} that allows for
the modelling of all necessary background processes and for the simulation of
off-shell Higgs boson production (including anomalous coupling effects) after
accounting for the interferences with the continuum arising from diboson
production. The simulation of the impact of an additional broad resonance is
also possible, allowing for the study of a new Higgs-like state with arbitrary
couplings and interfering with the Standard Model contributions. The program can
be interfaced to parton showers, as well as full detector simulators, through
the Les Houches Event file (LHE) format~\cite{Alwall:2006yp}. The {\tt Jhu}
generator also allows for the simulation of the decay of a spin-zero particle
when its production is taking care of by other codes (or by the {\tt Jhu}
generator itself) via an interface through LHE files. As an example, this allows
for the production of a spin-zero boson through the NLO QCD accurate
{\tt PowHeg}~\cite{Frixione:2007vw} package, and further decay this boson with
the {\tt Jhu} generator.

Additionally, the {\tt Mela} framework allows to construct various likelihood
functions in order to distinguish between different hypotheses concerning the
Lorentz structure of the $HVV$ and $Hff$ interaction vertices. These likelihood
functions are obtained from kinematic probability distributions that can be
either computed analytically or numerically. Analytical parameterizations are
currently available for the $gg$ or $q\bar{q} \to H\to VV (\to 4f)$ and
$q\bar{q}^\prime \to VH$ processes and an arbitrary spin of the $H$-boson. On
the other hand, numerical matrix element computations are provided by the
{\tt Jhu} and {\tt Mcfm} generators. Both the analytical and numerical options
are implemented as separate functions within the {\tt Mela} package and can be
accessed by an end-user directly. These matrix elements can then be used for
Monte Carlo reweighting techniques and the construction of kinematic
discriminants for an optimal analysis of the considered processes. The {\tt Jhu}
generator and {\tt Mela} package have been in this way extensively used in many
LHC analyses by both the CMS~\cite{Chatrchyan:2012xdj,Chatrchyan:2012jja,%
Chatrchyan:2013mxa,Chatrchyan:2013iaa,Khachatryan:2014iha,Khachatryan:2014ira,%
Khachatryan:2014kca,Khachatryan:2015mma,Khachatryan:2016tnr} and
ATLAS~\cite{Aad:2013xqa,%
Aad:2015mxa} collaborations, including analyses that include the
discovery~\cite{Chatrchyan:2012xdj} of the Higgs boson and the first measurement
of its spin-parity properties~\cite{Chatrchyan:2012jja}.

The implemented formalism uses equivalent formulations of the effective field
theory scattering amplitudes, in which the dependence on the virtualities of the
weak and Higgs bosons is additionally tested by means of form factors. The
general couplings of a spin-zero particle $H$ to two fermions is hence given by
the amplitude
\beq
 {A}(Hff) = - \frac{m_f}{v}
\bar f \left ( \kappa_f  + i \tilde\kappa_f  \gamma_5 \right ) f \,,
\label{eq:ampl-spin0-qq}
\eeq
where $m_f$ denotes a generic fermion mass and $f$ and $\bar f$ are the related
Dirac spinors, and where the coupling strengths $\kappa_f$ ($=1$ in the Standard
Model) and $\tilde\kappa_f$ are respectively connected to a scalar and
pseudoscalar $H$-boson. Anomalous $HVV$ couplings are described by the amplitude
\beq
A(HVV) \propto
\Biggl[ a_{1}
 - e^{i\phi_{\Lambda{1}}} \frac{\left(q_{{V}1}^2 + q_{{V}2}^2\right)}{\left(\Lambda_{1}\right)^{2}}
 - e^{i\phi_{\Lambda{Q}}} \frac{\left(q_{{V}1} + q_{{V}2}\right)^{2}}{\left(\Lambda_{Q}\right)^{2}}
\Biggr]m_{V}^2 \epsilon_{{V}1}^* \epsilon_{{V}2}^*
+ a_{2}^{}  f_{\mu \nu}^{*(1)}f^{*(2),\mu\nu}
+ a_{3}^{}   f^{*(1)}_{\mu \nu} {\tilde f}^{*(2),\mu\nu},
\label{eq:ampl-spin0}
\eeq
where $f^{(i){\mu \nu}} = \epsilon_{{V}i}^{\mu}q_{{V}i}^{\nu} -
\epsilon_{{V}i}^\nu q_{{V}i}^{\mu} $ is the field strength tensor of a gauge
boson with momentum $q_{{V}i}$ and polarization vector $\epsilon_{{V}i}$, and
${\tilde f}^{(i)}_{\mu \nu} = \frac{1}{2} \epsilon_{\mu\nu\rho\sigma} f^{(i)
\rho\sigma}$ is its dual field strength tensor. For spin-one and spin-two
resonance couplings, higher-order terms in the momentum expansion, and terms
asymmetric in $q_{V1}^2$ and $q_{V2}^2$ that are supported by the {\tt Jhu}
generator, we refer to Refs.~\cite{Gao:2010qx,Bolognesi:2012mm,Anderson:2013afp}
and the program manual for details. The $q^2$-expansion of
Eq.~\eqref{eq:ampl-spin0} can be equivalently rewritten as an effective
Lagrangian containing operators with a mass-dimension up to
five~\cite{Khachatryan:2014kca},
\beq\begin{split}
 {L}(HVV) \propto &\
a_{1}\frac{m_{\sss Z}^2}{2} H Z^{\mu}Z_{\mu}
- \frac{\kappa_1}{\left(\Lambda_{1}\right)^{2}} m_{\sss Z}^2 H  Z^{\mu} \Box Z_{\mu}
- \frac{\kappa_3}{2\left(\Lambda_{Q}\right)^{2}} m_{\sss Z}^2 \Box H  Z^{\mu} Z_{\mu}
\\
&\ 
- \frac{1}{2}a_{2} H  Z^{\mu\nu}Z_{\mu\nu}
- \frac{1}{2}a_{3} H  Z^{\mu\nu}{\tilde Z}_{\mu\nu} \\
&\ + a_{1}^{\PW\PW}{m_{\sss\PW}^2} H \PW^{+\mu} \PW^{-}_{\mu}
- \frac{1}{\left(\Lambda_{1}^{\PW\PW}\right)^{2}} m_{\sss\PW}^2 H
  \left(  \kappa_1^{\PW\PW} \PW^{-}_{\mu} \Box \PW^{+\mu} + \kappa_2^{\PW\PW} \PW^{+}_{\mu} \Box \PW^{-\mu} \right)\\
&\
 - \frac{\kappa_3^{\PW\PW}}{\left(\Lambda_{Q}\right)^{2}} m_{\sss\PW}^2 \Box H  \PW^{+\mu} \PW^{-}_{\mu}
- a_{2}^{\PW\PW} H \PW^{+\mu\nu}\PW^{-}_{\mu\nu}
- a_{3}^{\PW\PW} H \PW^{+\mu\nu}{\tilde \PW}^{-}_{\mu\nu} \\
&\
                          - a_{2}^{Z\gamma} H A^{\mu\nu} Z_{\mu\nu} - a_{3}^{Z\gamma} H  A^{\mu\nu}{\tilde Z}_{\mu\nu}
                           - \frac{1}{2}a_{2}^{\gamma\gamma} H  A^{\mu\nu}A_{\mu\nu} - \frac{1}{2}a_{3}^{\gamma\gamma}H  A^{\mu\nu}{\tilde A}_{\mu\nu}\\
&\
- \frac{1}{2}a_{2}^{\Pg\Pg} H  G^{\mu\nu}_aG^a_{\mu\nu}
- \frac{1}{2}a_{3}^{\Pg\Pg}H  G^{\mu\nu}_a{\tilde G}^a_{\mu\nu}~.
\label{eq:lagrangian}
\end{split}\eeq

Both on-shell $H$ production and off-shell $H^*$ production are considered and
there is no kinematic limit neither on $q^2_{\sss Vi}$ nor on
$(q_{\sss V1}+q_{\sss V2})^2$ other than the one due to the energy of the
colliding beams and the relevant partonc luminosities. Since the scale of
validity of the non-renormalizable higher-dimensional operators is {\it a
priori} unknown, effective cut-off scales $\Lambda_{V1,i}$, $\Lambda_{V2,i}$
and $\Lambda_{H,i}$ are introduced for each term in
Eq.~\eqref{eq:ampl-spin0} with a form factor scaling the anomalous
contribution $a_i^{\rm BSM}$ as
\beq
a_i = a_i^{\rm SM} \times \delta_{i1} +
a_i^{\rm BSM} \times \frac{\Lambda_{V1,i}^2 \Lambda_{V2,i}^2 \Lambda_{H,i}^2}
{(\Lambda_{V1,i}^2+|q^2_{\sss V1}|)(\Lambda_{V2,i}^2+|q^2_{\sss V2}|)(\Lambda_{H,i}^2+|(q_{\sss V1}+q_{\sss V2})^2|)}
\,.
\label{eq:formfact-spin0}
\eeq

In \refF{fig:lhc_angles}, representative distributions of $HVV$ observables
are shown for the $H\to VV$ decay channel (left), and for the $VH$ (centre) and
VBF (right) production modes. The anomalous coupling parameterization in terms
of the effective fractions of events follows the LHC
conventions~\cite{Khachatryan:2014kca} and is equivalent to start from
Eq.~\eqref{eq:ampl-spin0} and fix
$f_{ai}= |a_i|^2\sigma_i / \Sigma_j |a_j|^2\sigma_j$ and
$\phi_{ai}=\arg(a_i/a_1)$,
where $\sigma_i$ denotes the cross section for a given process with
$a_i=1$~\cite{Khachatryan:2014kca}.

\begin{figure}[t]
  \centering
  \includegraphics[width=0.32\textwidth]{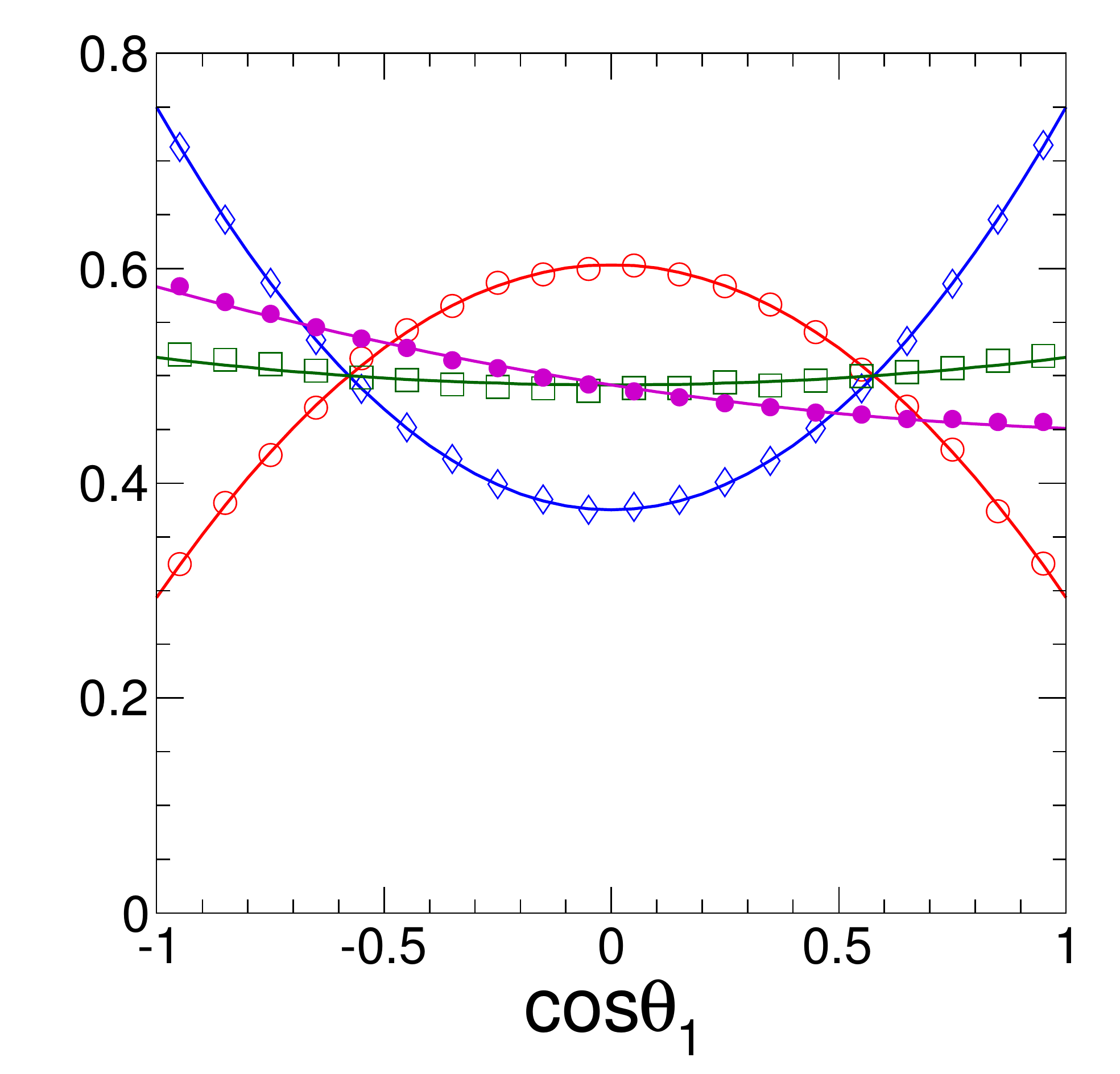}
  \includegraphics[width=0.32\textwidth]{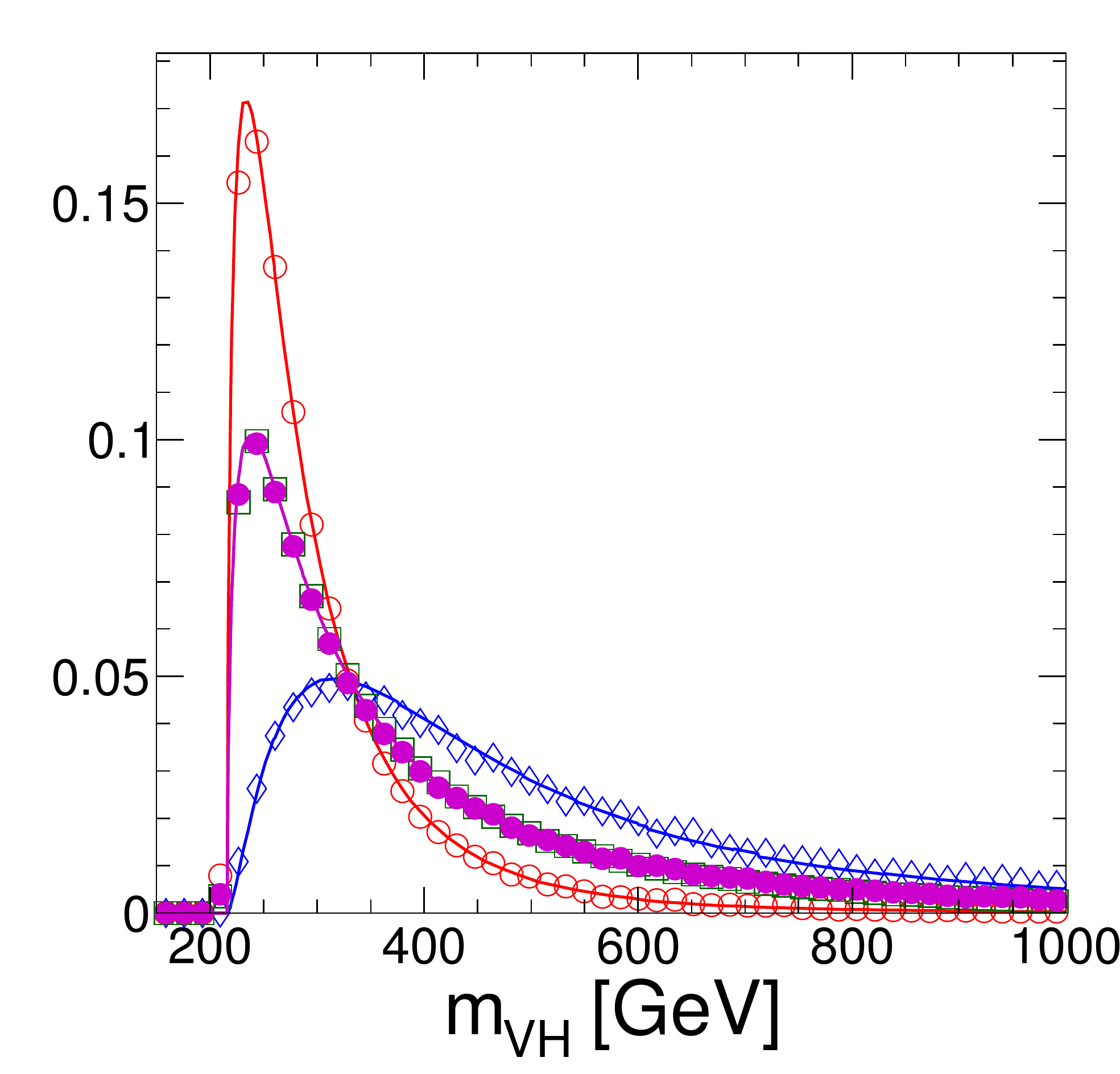}
  \includegraphics[width=0.32\textwidth]{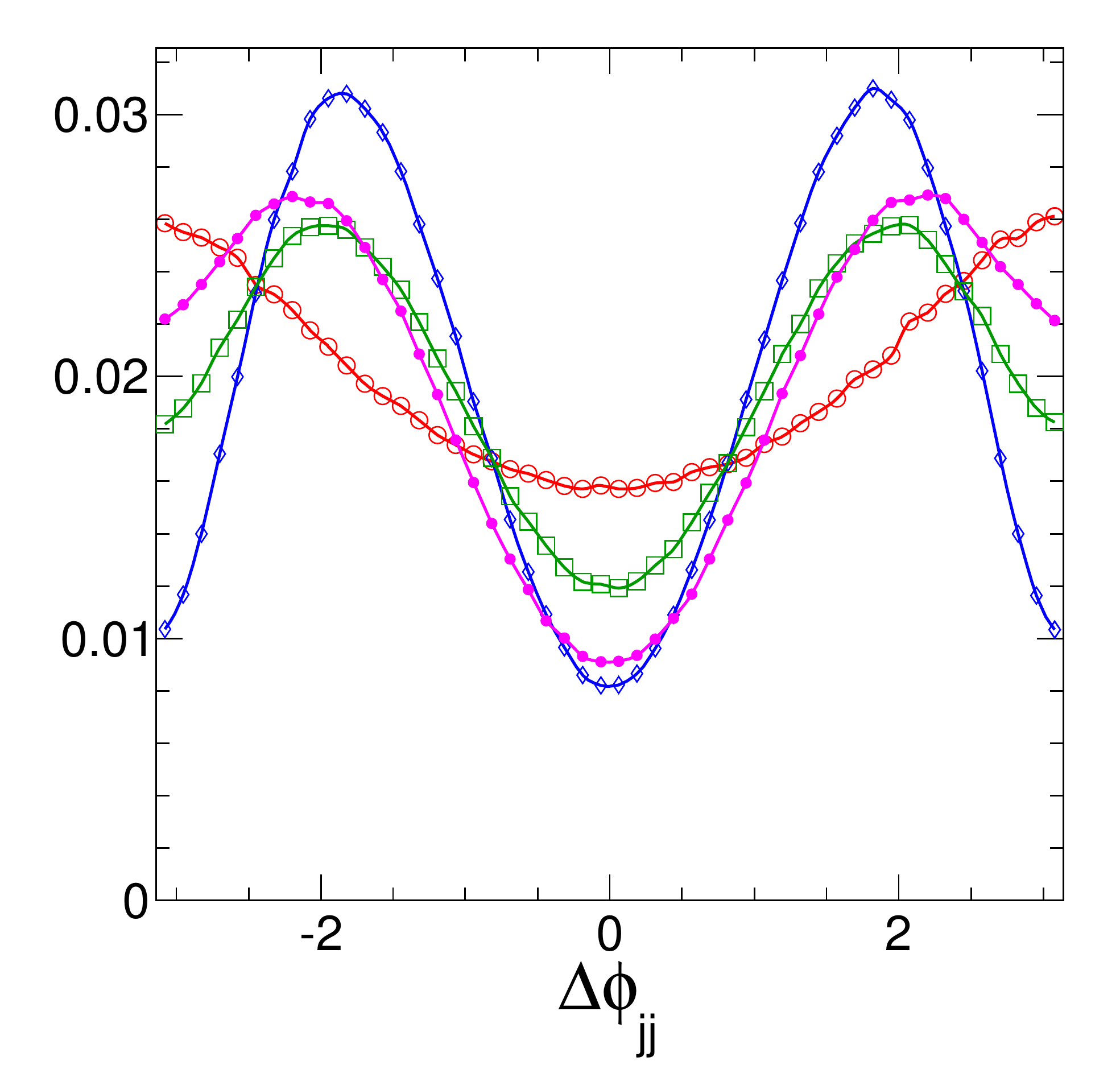}
\caption{
Representative distributions of observables depending on $HVV$ anomalous
couplings as generated with {\tt JhuGen} (dots) shown together with the
analytical results obtained with {\tt Mela} likelihood projections (smooth
curves on the left and middle panels) in the context of a Higgs boson of 125~GeV
and proton-proton collisions
at a centre-of-mass energy of 14~TeV. Left: distribution of the helicity angle
of a $Z$ boson in the decay $H\to ZZ\to 4\ell$; middle: spectrum of the $VH$
invariant-mass in the case of $q\bar{q}\to ZH$ production; right: distribution
in the azimuthal angle between the two jets in VBF production. Four scenarios
are shown: the Standard Model case ($0^+$, red open circles), a case with a
pseudoscalar boson ($0^-$, blue diamonds),
and two mixed cases corresponding to $f_{a3}=0.5$ with $\phi_{a3}=0$ (green
squares) and $\pi/2$ (magenta points).
}
\label{fig:lhc_angles}
\end{figure}

\subsection{Higgs boson pair production in \textsc{Herwig}~7}
\label{sec:herwig}
The general-purpose event generator \textsc{Herwig}~7~\cite{Bahr:2008pv,
Gieseke:2011na, Arnold:2012fq,Bellm:2013hwb, Bellm:2015jjp}, that is available
at \url{https://herwig.hepforge.org}, 
contains the \textsc{HiggsPair} and \textsc{HiggsPairOL} packages that offer the generation
of exclusive events for Higgs boson pair production via gluon fusion. The former
uses code from \textsc{HPair}~\cite{Dawson:1998py,Plehn:2005nk} (see
Section~\ref{sec:hpair}), whereas the latter uses the \textsc{OpenLoops}
one-loop generator for the matrix elements~\cite{Cascioli:2011va}.
\textsc{HiggsPair} describes leading-order Higgs boson pair
production with the option of either including an additional scalar as an
intermediate or final state particle, or including the effects of dimension-six
effective field theory operators that could extend the Standard Model.
The validation of the implementation has been performed using an
equivalent \textsc{MG5\_aMC@NLO} model~\cite{Alwall:2014hca},
implemented
in a similar (but independent) way using functions taken from the \textsc{HPair}
package. Comparisons of several distributions and the total cross section
between the two implementations have been performed, and the total cross section
output obtained by using \textsc{Herwig} was confirmed to match that obtained
using \textsf{HPair} at leading order for various parton density sets and the
scale choice $\mu=\sqrt{\hat{s}}$.
The dimension-six effective field theory extension was examined in detail
in Ref.~\cite{Goertz:2014qta}. The relevant Lagrangian terms affecting
Higgs boson pair production whose effects can be included at the time of event
generation is given, using the SILH basis conventions and mass eigenstates, by
\beq\label{eq:Lgghh}\begin{split}
  \mathcal{L}_{hh} = &
  - \frac{m_h^2}{2v} \Big[ 1 - \frac{3}{2} c_H + c_6 \Big] H^3
  - \frac{m_h^2}{8 v^2} \Big[ 1  - \frac{25}{3} c_H + 6 c_6 \Big] H^4
  + \frac{\alpha_s}{4\pi} \Big[ c_g \frac{h}{v} + c_{2g} \frac{H^2}{2v^2} \Big]
        G_{\mu\nu}^a G^{\mu\nu}_a \\
 &- \left[ \frac{m_t}{v} \left( 1 - \frac{ c_H } { 2  } + c_t   \right) \bar{t}_L t_R H +  \frac{m_b}{v} \left( 1 - \frac{ c_H } { 2  } + c_b   \right) \bar{b}_L b_R H + \textrm{h.c.} \right]
\\
&- \left[ \frac{m_t}{v^2}\left( \frac{3 c_{2t}}{2}  -  \frac{ c_H}  { 2 } \right) \bar{t}_L t_R H^2 +  \frac{m_b}{v^2}\left( \frac{3 c_{2b}}{2}  -  \frac{c_H}  { 2 } \right) \bar{b}_L b_R H^2 + \textrm{h.c.} \right] ,
\end{split}\eeq
where the Wilson coefficients $c_H$, $c_6$, $c_g$, $c_{2g}$, $c_t$,
$c_b$, $c_{2t}$ and $c_{2b}$ can be varied independently through the
input file. Note that in the case of a single Higgs boson doublet, we have the
relations  $c_g = c_{2g}$, $c_t = c_{2t}$ and $c_b = c_{2b}$.

The \textsc{HiggsPairOL} package describes Standard Model Higgs boson pair
production, with the optional use of Higgs-pair plus one jet matrix elements
merged to the parton shower via the MLM method. We refer to
Ref.~\cite{Maierhofer:2013sha} for a detailed description of this procedure.
The implementation of the effects of higher dimensional operators in this
framework is foreseen in the near future.

\subsection{Anomalous couplings in \textsc{Vbfnlo}}
\label{sec:Anomalous couplings}
\label{sec:VBFNLO}
NLO QCD predictions including anomalous coupling
effects can be studied for several processes with the flexible Monte Carlo
program \textsc{Vbfnlo}~\cite{Arnold:2008rz,Baglio:2014uba}, available at 
\url{https://www.itp.kit.edu/vbfnlo}.
The ensemble of implemented processes includes Higgs, single and double vector
boson production via vector boson fusion (VBF), $WH$ production, as well as
double and triple vector boson (plus jet) production. The Standard Model
QCD-induced background for double vector boson production in association with
two jets is also available at the NLO accuracy.
Furthermore, anomalous $HVV$ coupling effects are also included in the
gluon-induced contributions to diboson (plus jet) production as well as in the
gluon fusion processes $gg \rightarrow Hjj \rightarrow VVjj$. Although these
processes are all one-loop induced and hence computed at the leading-order
accuracy, the full top- and bottom-quark mass dependence is retained.

In the \textsc{Vbfnlo-3.0 $\beta$} release, an interface compliant with the
Binoth Les Houches Accord (BLHA) \cite{Binoth:2010xt,Alioli:2013nda} has been
added for all VBF processes including fully leptonic decays, which allows for
Monte Carlo studies at NLO in QCD including the full functionality of event
generators like {\tt Herwig}~7\cite{Bahr:2008pv,Bellm:2015jjp}.  The
$K$-matrix unitarization procedure has been implemented for the two
dimension-eight operators $\mathcal{O}_{S,0}+\mathcal{O}_{S,2}$ and
$\mathcal{O}_{S,1}$ that are defined in Eq.~\eqref{eq:vbf_dim8} below.
The strength of the anomalous triple and quartic gauge boson couplings can be
set in the file {\tt anomV.dat}. They are parameterized using an effective
Lagrangian, as described in Refs.~\cite{Buchmuller:1985jz, Hagiwara:1993ck, %
Eboli:2006wa, Degrande:2013rea},
\begin{equation}
 \mathcal{L}_{\rm eff} = \sum_i \frac{f_{i}}{\Lambda^{n}} \mathcal{O}_{i}^{n+4}\ ,
\end{equation}
where $n+4$ signifies the dimension of the operator $\mathcal{O}_{i}$.
\textsc{Vbfnlo} then defines anomalous gauge couplings in terms of the
coefficients $f_{i}/\Lambda^{n}$ of the dimension-six and dimension-eight
operators. The full list of implemented operators can be found in the
Appendix~1 of the \textsc{Vbfnlo} manual~\cite{Arnold:2011wj}.
The explicit form of the included $CP$-even dimension-six operators is
given by
\beq\begin{split}
  \mathcal{O}_{W} = (D_{\mu} \Phi)^{\dagger} \widehat{W}^{\mu \nu}(D_{\nu} \Phi), \qquad&
  \mathcal{O}_{B} = (D_{\mu} \Phi)^{\dagger} \widehat{B}^{\mu \nu}(D_{\nu} \Phi), \qquad
  \mathcal{O}_{WWW}= {\rm Tr} \Big[ \widehat{W}_{\mu\nu} \widehat{W}^{\nu\rho} \widehat{W}_\rho^{\mbox{ } \mu} \Big],\\
\mathcal{O}_{WW} = \Phi^{\dagger} \widehat{W}_{\mu\nu}\widehat{W}^{\mu\nu} \Phi,\qquad&
\mathcal{O}_{BB} = \Phi^{\dagger} \widehat{B}_{\mu\nu} \widehat{B}^{\mu\nu} \Phi.
\end{split}\eeq
The building blocks for these operators (following the notation of
Refs.~\cite{Hagiwara:1993qt,Hagiwara:1993ck}) are defined by
\beq
 \widehat{W}_{\mu\nu} = i g T_{a} W^{a}_{\mu\nu}, \qquad
 \widehat{B}_{\mu\nu} = i g' Y B_{\mu \nu}, \qquad
 D_\mu = \partial_\mu + i g T_{a} W^{a}_\mu  + i g' Y B_\mu,
\eeq
where $g$ and $g'$ are the $SU(2)_L$ and $U(1)_Y$ gauge couplings, and $T_{a}$
and $Y$ the generators of the $SU(2)$ group in the fundamental representation and
the hypercharge operator, respectively.
The $CP$-odd part of the Lagrangian is obtained replacing the field
strength tensor with the corresponding dual field strength tensors.
The dimension-eight operators that are supported are taken from
Ref.~\cite{Eboli:2006wa} although slightly different normalizations for the
field strength tensors are employed in this work,
$\widehat{W}_{\mu\nu} = T_{a} W^{a}_{\mu\nu}$ and
$\widehat{B}_{\mu\nu} = B_{\mu \nu}$. The conversion factors for the coupling
strengths $f_i$ relating Ref.~\cite{Eboli:2006wa} to our implementation can be
found in the Appendix~1 of the {\tt Vbfnlo} manual. The dimension-eight
operators that are supported can be split into three categories, namely
operators depending on the gauge-covariant derivative $D_\mu \Phi$ only,
\beq\begin{split}
   \label{eq:vbf_dim8}
   &{\cal O}_{S,0} = \Big[ (D_\mu \Phi)^\dagger D_\nu \Phi \Big]
      \Big[ ( D^\mu \Phi)^\dagger D^\nu \Phi \Big],
   \qquad
   {\cal O}_{S,1} = \Big[ ( D_\mu \Phi )^\dagger D^\mu \Phi  \Big]
      \Big[ ( D_\nu \Phi )^\dagger D^\nu \Phi \Big],\\
   &{\cal O}_{S,2} = \Big[ (D_\mu \Phi)^\dagger D_\nu \Phi \Big]
      \Big[ ( D^\nu \Phi)^\dagger D^\mu \Phi \Big],
\end{split}\eeq
operators depending on both $D_\mu \Phi$ and the electroweak field strength
tensors $\widehat{W}_{\mu \nu}$ and $\widehat{B}_{\mu \nu}$, \textit{e.g.},
\beq
  {\cal L}_{M,0} = {\rm Tr} \Big[ \widehat{W}_{\mu\nu} \widehat{W}^{\mu\nu} \Big]
      \Big[ ( D_\rho \Phi )^\dagger D^\rho \Phi \Big],
\eeq
and operators depending only on the electroweak field strength tensors
$\widehat{W}_{\mu \nu}$ and $\widehat{B}_{\mu \nu}$, such as,
\beq
 {\cal L}_{T,0} = {\rm Tr} \Big[ \widehat{W}_{\mu\nu} \widehat{W}^{\mu\nu} \Big]
      {\rm Tr}\Big[ \widehat{W}_{\rho\sigma} \widehat{W}^{\rho\sigma} \Big].
\eeq

Anomalous Higgs boson $HVV$ coupling parameters are controlled in
\textsc{Vbfnlo} via the file named {\tt anom\_HVV.dat}, where three different
parameterizations are available. The latter consist of the Wilson coefficients
associated with the subset of dimension-six operators introduced above that
contribute to the Higgs boson couplings, the parameterization used by the L3
collaboration that is defined in Ref.~\cite{Achard:2004kn}, and a
parameterization based on anomalous couplings in the mass basis that is thus
expressed in terms of the field strength and dual field
strength tensors of the $W$ and $Z$ bosons~\cite{Figy:2004pt}. The relationships
between these three parameterizations are discussed in more detail on the
\textsc{Vbfnlo} webpage.

Since the pure operators for anomalous gauge boson couplings might lead to a
violation of tree-level unitarity within the energy range of the LHC, special
care has to be taken to avoid this unphysical behaviour. Within \textsc{Vbfnlo},
we have opted for using the following form factors, all of them depending on
$\Lambda$, the characteristic scale where the form factor effects become relevant.
Equivalently, introducing these form factors is like restricting the validity
range of the effective field theory rather than implementing a sharp cutoff.
For $HVV$ vertices, two different form factors can be chosen as described in
Refs.~\cite{Figy:2004pt,Hankele:2006ma}.
\beq
F_1 = \frac{\Lambda^2}{|q_1|^2 + \Lambda^2}\ \frac{\Lambda^2}{|q_2|^2 +
  \Lambda^2}, \qquad
F_2 = -2 \,\Lambda^2 \, C_0\!\Big(q_1^2, q_2^2, (q_1+q_2)^2,
\Lambda^2\Big).
\eeq
where $q_i$ are the momenta of the vector bosons and $C_{0}$ is the scalar
one-loop three point function in the notation of
Ref.~\cite{Passarino:1978jh}.
For triple and quartic gauge couplings the form factor takes the form
\beq
 F = \Big(1 + \frac{s}{\Lambda^{2}} \Big)^{-p},
\eeq
for each phase space point, where $s$ is a universal scale identified with the
invariant mass squared of the produced bosons. Finally, for $\gamma jj$
production in VBF, the form factor reads
\beq
 F = \Big(1 + \frac{|q_1|^2}{\Lambda^{2}} + \frac{|q_2|^2}{\Lambda^{2}} + \frac{|q_3|^2}{\Lambda^{2}} \Big)^{-p},
\eeq
with $q_1^2$, $q_2^2$ and $q_3^2$ as the invariant masses squared of the three
vector bosons involved in the $VVV$ vertex.
On the \textsc{Vbfnlo} webpage, we provide the `Formfactor
Calculation Tool for aGC' which gives the maximum value of the scale
$\Lambda$ which is allowed by unitarity. The value is determined by
calculating on-shell $VV$ scattering and computing the zeroth partial wave
of the amplitude. As unitarity criterion, the absolute value of the real
part of the zeroth partial wave has to be below 0.5~\cite{Barger:1990py}.
Each channel in $VV \to VV$ scattering (with $V = W / Z / \gamma$) is
checked individually, while additionally channels with the same electrical
charge for the $VV$ system are combined~\cite{Gounaris:1993fh}. We recall
that the
definition of the partial wave expansion in Ref.~\cite{Gounaris:1993fh}
differs from ours by a factor of 2.

Finally, the $K$-matrix unitarization procedure has been implemented for
the two dimension-eight operators $\mathcal{O}_{S,0}+\mathcal{O}_{S,2}$ and
$\mathcal{O}_{S,1}$ by using the relations to
the operators $\mathcal{O}_4$ and $\mathcal{O}_5$ of the electroweak
chiral Lagrangian and the procedure worked out
in Ref.~\cite{Alboteanu:2008my}. This method guarantees the preservation
of unitarity when either of these operators is used in the study of
anomalous quartic gauge couplings. In contrast to using form factors, no
additional input parameters need to be set.
The anomalous contributions are automatically suppressed at the energy
scale where unitarity would be violated without unitarization. After
this energy scale is reached, the anomalous contributions are kept at a
finite value, representing the maximally possible anomalous
contribution. We refer to
Refs.~\cite{Alboteanu:2008my,Degrande:2013rea,Kilian:2014zja}
for details of the $K$-matrix unitarization procedure and its
implementation.

\subsection{Event generation with {\tt Whizard}}
\label{sec:whizard}
{\tt Whizard}~\cite{Kilian:2007gr} is a multipurpose event generator for hadron
and lepton colliders that can be obtained from \url{https://whizard.hepforge.org}.
It has a highly optimized internal matrix element
generator, {\tt O'Mega}, for the recursive computing of tree-level amplitudes for
almost arbitrary theories~\cite{Moretti:2001zz}. In the QCD case, it
additionally uses the colour flow formalism~\cite{Kilian:2012pz}. {\tt Whizard}
has its own parton shower implementation~\cite{Kilian:2011ka}, with both a
$p_T$-ordered shower and an analytic shower. Recently, automated FKS subtraction
and {\tt PowHeg} matching for NLO QCD corrections (using external virtual matrix
elements) have been implemented~\cite{Weiss:2015npa,Nejad:2015opa}. For external
theories beyond the Standard Model, interfaces to the packages {\tt Sarah} and
{\tt FeynRules}~\cite{Christensen:2010wz} are available, and support for the UFO
file format allowing for arbitrary Lorentz and colour structures is currently under
way.

\begin{figure}
 \centering
 \includegraphics[width=0.6\textwidth]{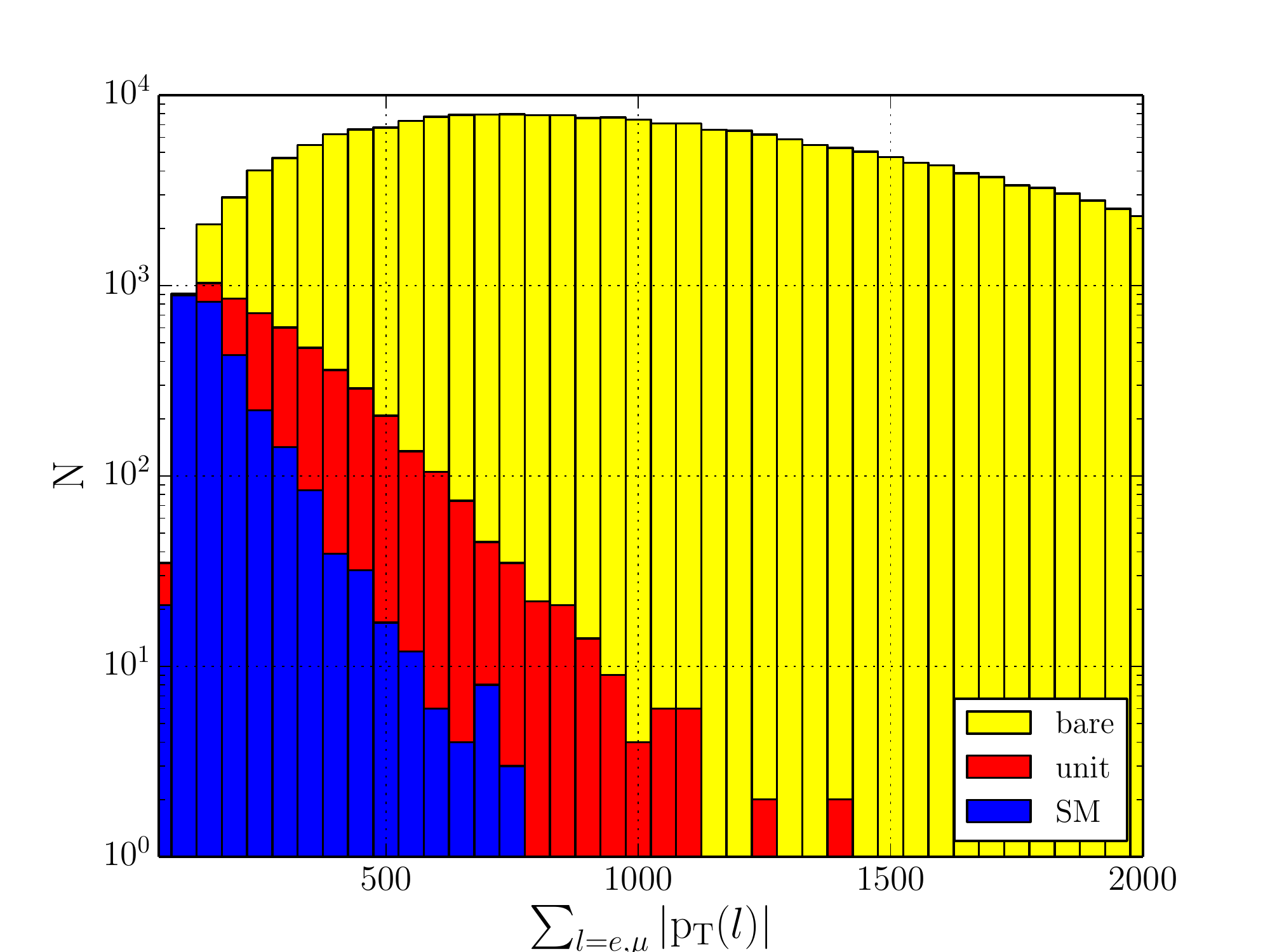}
  \caption{Events generated by {\tt Whizard} when a dimension-eight coupling
    $F_{S,0}=480$~TeV$^{-4}$ (the Wilson coefficient associated with the
    $\mathcal{O}_{S,0}$ operator of Eq.~\eqref{eq:vbf_dim8})
    is added to the Standard Model and for the
    process $pp\to e^+\mu^+\nu_e\nu_\mu jj$ at a centre-of-mass energy of 14~TeV
    and for an integrated luminosity of ${\cal L}=1000$~fb$^{-1}$. Event
    selection requires a dijet invariant mass $M_{jj} > 500$ GeV, a rapidity
    separation between the jets of $\Delta y_{jj} > 2.4$, and the two jets must
    have a transverse momentum $p_T^j > 20$~GeV and a pseudorapidity
    $|\eta_j|<4.5$. In addition, leptons are required to have a transverse
    momentum $p_T^l > 20$~GeV. Red and yellow histograms represent the naive and
    unitarized effective field theory predictions, respectively, while the pure
    Standard model expectation is shown as a blue histogram for comparison.}
  \label{fig:WHIZARDEFT}
\end{figure}
With respect to Higgs Effective Field Theories and higher-dimensional operators,
{\tt Whizard} supports the whole set of bosonic dimension-six operators in the
Warsaw basis~\cite{Grzadkowski:2010es}. For vector-boson scattering at the
LHC, usually dimension-eight operators in the coupled system of electroweak
gauge and Higgs bosons are as important as dimension-six operators as they can
be generated at tree-level in some new physics models. These electroweak
dimension-eight operators have been implemented in {\tt Whizard}, both as plain
operators and also with their interplay with new resonances in the electroweak
sector. The description of new physics contributions with a low energy effective
energy is however only valid up to an a priori unknown scale $\Lambda$. At
energies above $\Lambda$, the effective field theory will lead to unphysical
predictions. As an example, the sum of the norm of the transverse momenta of all
leptons produced in the process $pp \rightarrow e^+\mu^+\nu_e\nu_\mu$
is shown on Figure~\ref{fig:WHIZARDEFT} when the dimension-eight operator
$\mathcal{O}_{S,0}$ of Eq.~\eqref{eq:vbf_dim8} is included. The number of events
generated with a naive effective field theory description (yellow histogram)
will largely overshoot a physically possible distribution, because $S$-matrix
unitarity is violated within the experimental energy reach. A selection to avoid
energy regions where the theoretical description breaks down is not possible in
general, due to the inability to reconstruct the invariant mass for some final
states. Using the $T$-matrix scheme, a unitarization prescription for
high-energy regions of the phase space~\cite{Alboteanu:2008my,Kilian:2014zja,
Kilian:2015opv}, avoid the unphysical high number of generated events by
{\tt Whizard}. Instead, the number of events are saturated in every isospin-spin
channel to satisfy $S$-matrix unitarity (red histogram). In order to simplify
unitarization, {\tt Whizard} does not use the basis of Ref.~\cite{Eboli:2003nq}
as this basis does not respect isospin symmetry which makes calculation of the
channels in need for unitarization much easier.

On different lines, fermionic dimension-six operators are implemented for the
top quark sector as (form-factor regularized) anomalous
couplings~\cite{Bach:2012fb,Bach:2014zca}.

\subsection[Constraints on non-standard Higgs boson couplings with {\tt HEPfit}]{Model-independent constraints on non-standard Higgs boson couplings
  with {\tt HEPfit}} \label{sec:hepfit}

The {\tt HEPfit} package (formerly {\tt SusyFit}) is a general tool to combine
direct and indirect constraints on the Standard Model and its extensions and is
available under the GNU General Public License (GPL) from \url{http://hepfit.roma1.infn.it}. 
The {\tt HEPfit} code can be extended to include any observables and new physics
models (supersymmetric theories, Two-Higgs-Doublet Models, $\ldots$) which can
be added to the main core as external modules. Exploiting the Markov Chain Monte
Carlo implementation of the Bayesian Analysis Toolkit~\cite{Caldwell:2008fw},
{\tt HEPfit} can be used as a standalone program to perform Bayesian statistical
analyses.  Alternatively, it can be used in library mode to compute observables
in any implemented model, allowing for phenomenological analyses in any
statistical framework. 
The interested reader can find more details on  {\tt HEPfit}  in Refs.~\cite{hepfit}. 

In particular, {\tt HEPfit} has been used to perform statistical analyses of
electroweak precision data, including Higgs boson signal-strength measurements,
in the Standard Model and beyond.  Most importantly, these analyses have
obtained constraints on possible deviations of the Higgs boson couplings to both
gauge bosons and fermions from the Standard Model predictions. Results from the
initial stages of this project were presented in Refs.~\cite{deBlas:2014ula,%
Ciuchini:2014dea} and recently updated in Refs.~\cite{Ciuchini:2016sjh,deBlas:2016ojx} to reflect
all the most recent developments in theoretical calculations and experimental
measurements.

Within {\tt HEPfit}, new physics effects on electroweak precision observables
and on Higgs boson couplings can also be systematically studied in the context
of an effective field theory that adds to the Lagrangian of the Standard Model
new interactions of the Standard Model fields in the form of higher-dimension
(of dimension $d>4$) local operators that preserve the Standard Model gauge
symmetry, namely
\begin{equation}
\label{eq:Leff-main}
  {\mathcal L}_{\mathrm{eff}}={\mathcal L}_{\mathrm{SM}} +
    \sum_{d>4}\frac{1}{\Lambda^{d-4}}{\mathcal L}_d,
~~\mbox{with}~~{\mathcal L}_d=\sum_i C_i {\mathcal O}_i,~~\left[{\mathcal O}_i\right]=d\,.
\end{equation}
In this equation, the dependence on $\Lambda$, the scale at which direct
evidence of the new physics degrees of freedom is expected, has been made
explicit. In particular, the current public version of {\tt HEPfit} implements
as a \textit{model} the $d=6$ extension of the Standard Model Lagrangian using
the basis of $d=6$ operators proposed in Ref.~\cite{Grzadkowski:2010es},
which is quite easy to relate to electroweak precision data and Higgs
observables by means of shifts of the couplings to the Standard Model bosons.
Through a global fit of electroweak-data and Higgs boson signal-strengths
measurements {\tt HEPfit} provides constraints on the individual Wilson
coefficients $C_i$ at the electroweak scale. At the moment {\tt HEPfit} does not
include effects of operator mixing induced by renormalization-group scale
evolution, waiting for more insight on the new physics theory that determines
the initial condition of the renormalization group running. Results have been
presented in~\cite{deBlas:2014ula}, where the subset of relevant to Higgs boson
observables have been considered and constraints on the corresponding
coefficients have been derived by switching on one operator at a time, for a
fixed scale $\Lambda=1$~TeV. \textit{Vice versa}, for values of the individual
coefficients $C_i=\pm 1$, lower bounds on the scale $\Lambda$ have been found.

\subsection{{\tt Rosetta}}
\label{sec:rosetta}
Different complete and non-redundant operator bases of dimension-six effective
operators have been proposed in the literature, the most popular choices
including the Warsaw basis~\cite{Grzadkowski:2010es}, the SILH
basis~\cite{Giudice:2007fh,Contino:2013kra} and the beyond the Standard Model
primaries basis~\cite{Gupta:2014rxa,Masso:2014xra,Pomarol:2014dya}. It is
however cumbersome to express any experimental result in a basis-independent
manner. Different
bases may be convenient for particular applications, either because they
facilitate the comparison with a given class of theories or simply because
different experimental analyses look more transparent in a specific basis.
The {\tt Rosetta} package~\cite{Falkowski:2015wza} has been designed to
explicitly solve such problems by allowing for a straightforward translation
between different effective field theory languages. In addition to translating,
another important goal of the {\tt Rosetta} program is to provide a platform for
communication with Monte Carlo event generators, no matter which basis is chosen.
To achieve this, {\tt Rosetta} contains an implementation of the Higgs
basis defined in Section~\ref{s.eftbasis} and is connected to the BSMC
Lagrangian introduced in Section~\ref{sec:HC}. More precisely, the output format
of {\tt Rosetta} has been tuned so that the translation maps an effective field
theory Lagrangian given in a specific basis to the BSMC Lagrangian and generates
an output file that is compatible with the BSMC implementation into
{\tt FeynRules}~\cite{Alloul:2013bka}. As a consequence, any high-energy physics
tool that is interfaced to {\tt FeynRules} can be employed within the context of
any basis of dimension-six operators that is included in {\tt Rosetta}, as
illustrated on Figure~\ref{fig:rosettachart}.
A full description of the program including detailed example usage and
information on how to create a user-defined basis class can be found in the
manual~\cite{Falkowski:2015wza}.

\begin{figure}
 \centering
 \includegraphics[width=1.2\textwidth, trim=0 11.5cm 0 7.5cm, clip]{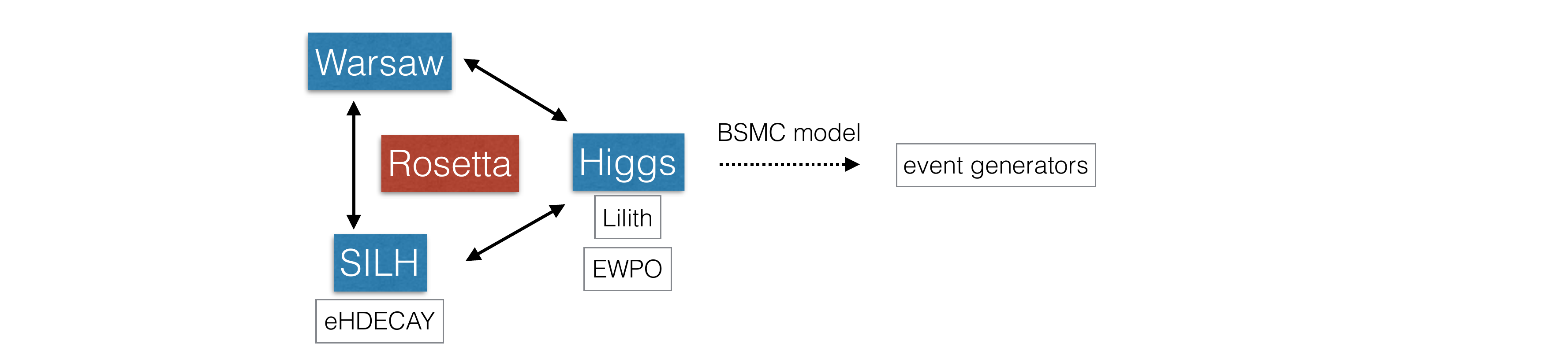}
  \caption{Description of the way the {\tt Rosetta} package works and is
    interfaced to other high-energy physics tools and connect different
    effective field theory operator basis choice. We moreover indicate tools or
    calculations that exist for a specific basis.}
  \label{fig:rosettachart}
\end{figure}

The most basic functionality of {\tt Rosetta} is to map a chosen set of input
parameters (the Wilson coefficients in a specific basis choice) onto the BSMC
coefficients such that the output can be employed within tools relying on a BSMC
Lagrangian description. In addition, the user may define his/her own map to the
BSMC coefficients (or to any other basis implementation) and proceed with event
generation using the related {\tt FeynRules} implementation. This highlights one
of the key features of {\tt Rosetta}, the possibility to easily define one's own
input basis and directly use it in the context of many programs via the
translation functionality. An example of this would be the
{\tt eHDecay} program (see Ref.~\cite{Contino:2013kra} and
Section~\ref{sec:ehdecay}) for which {\tt Rosetta} provides an interface to
calculate the Higgs boson width and branching fractions including effective field
theory effects in any basis. The strength of this approach is that it is much simpler than
developing from scratch new modules for existing tools in the context of a new
basis. Moreover, {\tt Rosetta}\ not only enables translation into the BSMC Lagrangian, but
also allows for translations into any of the other bases included in the
package that are currently the Higgs, Warsaw and SILH bases. Translations
between these three bases in any direction are possible, so that the addition of
a new basis by the user only requires the specification of translation rules to
any one of the three core bases. One is subsequently able to indirectly
translate the new basis into any of the other two bases, as well as into the
BSMC Lagrangian.

The latest release of {\tt Rosetta} can be obtained from \url{http://rosetta.hepforge.org}. 
The package contains a {\tt Python} executable named \texttt{tran\-slate}, an
information file named \texttt{README} and two directories, a first folder
(named \texttt{Cards}) collecting example input files and a second folder (named
\texttt{Rosetta}) including the source code of {\tt Rosetta}. Each basis is
implemented as a class in its own {\tt Python} module, where the coefficients
and the required inputs are declared. Moreover, the input format must respect
conventions inspired by the Supersymmetry Les Houches Accord
(SLHA)~\cite{Skands:2003cj,Allanach:2008qq} and the related SLHA block structure
is defined in the class. Translations are defined as member functions of the
class with a special syntax to identify the target basis. Several utility
functions may also be implemented to either calculate the values of parameters
declared as dependent as a function of those declared as independent or modify
the values of the Standard Model input parameters as a function of the effective
operator coefficients.

The \texttt{tran\-slate} executable takes as input an SLHA-style parameter file
with the coefficients of the dimension-six operators associated with a
particular basis. Information on the format of such an input file can be found
in the manual~\cite{Falkowski:2015wza}. Depending on the basis implementation,
some basic input quantities, such as Standard Model input constants and particle
masses may be required in addition to the effective field theory coefficients.
The execution of the \texttt{translate} script from a shell yields the
generation of an output parameter file where all parameters are this time the
coefficients of the dimension-six operators associated with a specified new
basis, the default choice being the BSMC Lagrangian. The tool can be used by
typing in\\
\hspace*{.5cm}   \verb+./translate PARAMCARD.dat OPTION+\\
where \texttt{PARAMCARD.dat} is the name of the SLHA-style input file and
\texttt{OPTIONS} stands for optional arguments. These can range from specifying
the output file name (\texttt{--output}, \texttt{-o}) or target basis into which
to translate (\texttt{--target}, \texttt{-t}) to invoking the {\tt eHDecay}
(\texttt{--ehdecay}, \texttt{-e}) interface to additionally generate an SLHA
decay block for the Higgs as part of the output file. A particularly useful
option is the \texttt{--flavour} or \texttt{-f} one which allows users to
specify the treatment of the flavour structure relevant for fermionic operators.
This can take the values \texttt{general}, \texttt{diagonal} and
\texttt{universal} depending on the desired degree of simplicity. A more
complicated example of usage might read\\
 \hspace*{.5cm}   \verb+./translate myinput.dat -t warsaw -f universal -e -o myoutput.dat+\\
which would read the \texttt{myinput.dat} file and translate it to the Warsaw
basis, assuming the universal flavour structure which is designed to map easily
to the Minimal Flavour Violation assumption~\cite{D'Ambrosio:2002ex}. The
{\tt eHDecay} interface will also be called, writing in the output file,
specified to be \texttt{myoutput.dat} an additional SLHA block containing 
Higgs boson decay information.

With the advent of NLO-accurate Monte Carlo event generation software, it is
important that {\tt Rosetta} remains flexible enough to eventually provide
compatibility with this new generation of tools. The future development plans of
the program are connected to the recent progresses that have been made
on the theory side both in implementing the linear dimension-six description in
the {\tt FeynRules} framework~\cite{Degrande:2016dqg} and in calculating the
renormalization group evolution of the full set of operators and their
mutual mixing~\cite{Jenkins:2013zja,Jenkins:2013wua,Alonso:2013hga,
Henning:2014wua,Boggia:2016asg}. In the former case, {\tt Rosetta} can simply be extended to
provide an output compatible with the eventual NLO model implementation,
analogously to the BSMC Lagrangian. The latter case of evaluating the
renormalization group running effects, while being a slightly separate issue,
highlights a key feature of {\tt Rosetta}, given that the calculation of these
effects has only been performed in the Warsaw basis and that {\tt Rosetta} could
allow for the application of these results in any desired basis.

\section[Morphing implementation]{Morphing implementation\SectionAuthor{N.~Belyaev, V.~Bortolotto, L.~Brenner, C.D.~Burgard, M.~D{\"u}hrssen, K.~Ecker, S.~Gadatsch, D.~Gray, A.~Kaluza, K.~K{\"o}neke, R.~Konoplich, S.~Kortner, K.~Prokofiev, C.~Schmitt, W.~Verkerke}}
\label{s.eftmorphing}
\label{sec:mprinc}
The properties of the newly discovered Higgs boson have been
extensively probed by the ATLAS and CMS experiments using LHC Run\,1
proton-proton collision data at $\sqrt{s}=7$ and
$8$\,TeV \cite{Aad:2015mxa,Khachatryan:2014kca,Aad:2015zhl,Khachatryan:2016vau}.
The studies of the tensor structure of the Higgs boson couplings to
gauge bosons were based on signal models including at most one or two
Beyond the Standard Model coupling parameters at a time, with all
remaining Beyond the Standard Model (BSM) parameters set to zero. For Run\,2, it is envisioned to
have signal models which depend on a larger number of coupling
parameters, in order to account for possible correlations among
them. Additional coupling parameters in the Higgs boson coupling to Standard Model (SM) particles
change the predicted cross section, as well as the shape of differential distributions.
 In this context, it is necessary to revise the existing signal
modelling methods and provide alternatives which are better suited for
 such a multidimensional parameter space.

For this purpose, a morphing method has been developed and
implemented. It provides a continuous description of arbitrary
physical signal observables such as cross sections or differential
distributions in a multidimensional space of coupling parameters. The
morphing-based signal model is a linear combination of a minimal set
of orthogonal base samples (templates) spanning the full coupling
parameter space. The weight of each template is derived from the
coupling parameters  appearing in the signal matrix element.

Morphing is more than a simple interpolation technique, in that it is
not limited to the points in the range spanned by the input
samples. In fact, the choice of the input samples is arbitrary, and
any set of input samples satisfying the required conditions to build
the morphing function will span the entire space, independent of their
precise coordinates.%

A full explanation and validation of this method is shown in reference \cite{LHCHXSWG-INT-2016-004}.

\subsection{Morphing principles}
The morphing procedure is based on the concepts of the morphing of (possibly multi-dimensional) histograms described in Ref~\cite{Baak:2014fta}. It is introduced to describe the dependence of a given physical observable $T$ on an arbitrary configuration of a set of non-SM Higgs boson couplings $\vec{g}_{\text{target}} \equiv \{ g_{\SM}, g_{{\rm BSM} ,1}, .. , g_{{\rm BSM} ,n} \}$ to known particles. This dependence is described by a morphing function
\begin{align}
  T_{\text{out}}(\vec{g}_{\text{target}}) = \sum_{i}w_{i}(\vec{g}_{\text{target}}; \vec{g}_{i})T_{\text{in}}(\vec{g}_{i}),
  \label{eq:derivation_morphingFunction}
\end{align}
which linearly combines the values or differential distributions
$T_{\text{in}}$ at a number of selected discrete coupling
configurations $\vec{g}_i=\{ \tilde {g}_{\SM,i}, \tilde{g}_{{\rm BSM} ,1}, .. , \tilde{g}_{{\rm BSM} ,n}\}$.  The
input distributions $T_{\text{in}}$ are normalized to their expected
cross sections such that $T_{\text{out}}$ includes not only the
correct shape, but also the correct cross section prediction. Here,
$g_{\SM}$ denotes the Higgs boson coupling predicted by the Standard
Model. Morphing only requires that any differential cross section can
be expressed as a polynomial in coupling parameters. For calculation at lowest order and using the narrow-width approximation
for a resonance, this yields a second order polynomial each in
production and decay.

In practice, the template distributions $T_{\text{in}}$ are obtained
from the Monte Carlo (MC) simulation of the signal process for a given
coupling configuration $\vec{g}_i$. The minimal number $N$ of Monte
Carlo samples needed to describe the signal at all possible coupling
configurations, depends on the number $n$ of studied non-SM coupling
parameters. The contribution of each sample $T_{\text{in}}$ is
weighted by a weight $w_i$ based on the assumption that the value of a
physical observable is proportional to the squared matrix element for
the studied process
\begin{align}
T \propto \left\vert \mathcal{M}\right\vert^2.
  \label{eq:derivation_ME}
\end{align}
The weights $w_i$ can therefore be expressed as functions of the coupling parameters in the matrix element $\mathcal{M}$. In this case $T$ can be anything derived from the Matrix element, for example a whole MC sample.

The described procedure allows for a continuous description in an
$n$-dimensional parameter space. A feature-complete implementation
has been developed within the RooFit package \cite{Verkerke:2003ir}, making use
of HistFactory \cite{Cranmer:2012sba}. The provided signal model can
therefore be used in commonly used RooFit workspaces in a
straightforward, black-box-like way. A visual representation of the idea for a simple case is shown in Figure~\ref{fig:morphexample}.

\begin{figure}
\includegraphics[width=\textwidth]{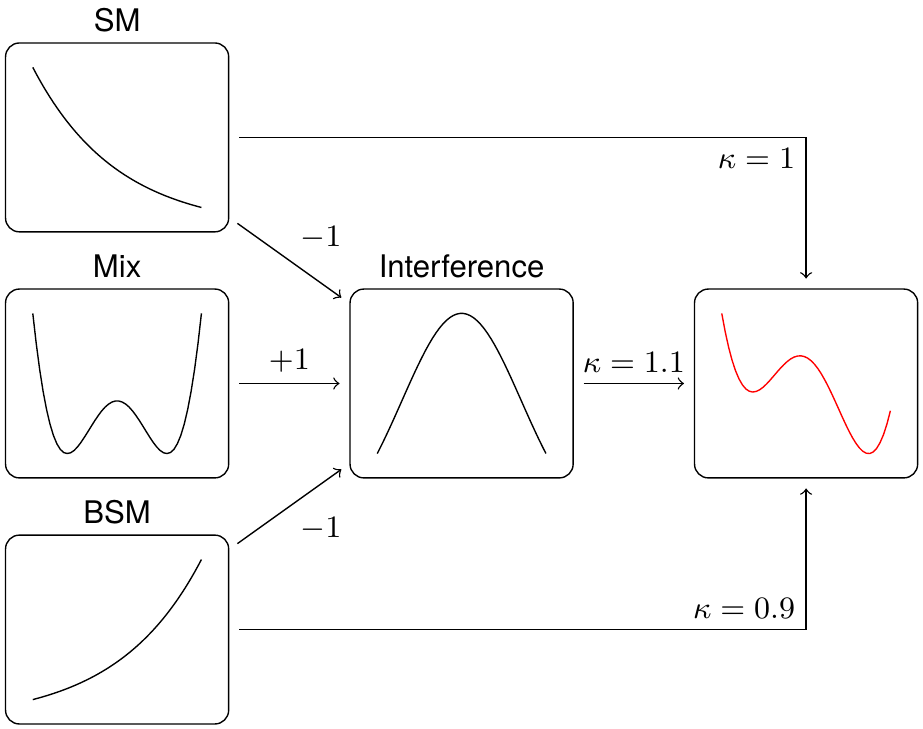}
\caption{Illustration of the morphing procedure in a simple showcase.\label{fig:morphexample}}
\end{figure}

\subsection{General procedure to construct morphing function}
\label{sec:procedure}

A step-by-step explanation on how to construct the morphing function for processes with an arbitrary number of free coupling parameters in two vertices is outlined below.
\begin{enumerate}
\item Construct a general matrix element squared
\begin{align}
\left|{\rm ME}(\vec{g})\right|^{2} = \underbrace{\left(\sum_{x\in p,s}g_{x}\mathcal{O}(g_{x})\right)^{2}}_{\text{production}}\cdot
                      \underbrace{\left(\sum_{x\in d,s}g_{x}\mathcal{O}(g_{x})\right)^{2}}_{\text{decay}},
\end{align}
denoting operators appearing only in the production vertex with $p$,
such only appearing in the decay vertex with $d$, and such shared
between both vertices with $s$, and assuming that production and decay
vertices are uncorrelated, which is the case for a scalar intermediate
particle.
\item Expand the matrix element squared to a 4th degree polynomial in the coupling parameters
\begin{align}
\left|{\rm ME}(\vec{g})\right|^{2}=\sum_{i=1}^{N}X_{i}\cdot P_{i}\left(\vec{g}\right),
\end{align}
$X_i$ is a prefactor, which will be represented by an input distribution. In the 4th degree polynomial $P_i\left(\vec{g}\right) = g_ag_bg_cg_d$ of the coupling parameters $\vec{g}$, the same coupling can occur multiple times (e.g.~$g_\SM^4$ or $g_{{\rm BSM},1}g_{{\rm BSM},2}g_{{\rm BSM},3}^2$). The number of different expressions in the polynomial $N$ is equal to the number of samples needed for the morphing.
\item Next generate input distributions at arbitrary but fixed parameter points $\vec{g}_i$
\begin{align}
T_{\text{in},i}\propto\left|{\rm ME}(\vec{g_{i}})\right|^{2}.
\end{align}
\item Construct the morphing function with an ansatz
\begin{align}
T_{\text{out}}(\vec{{g}})&=\sum_{i=1}^{N}\underbrace{\left(\sum_{j=1}^{N}A_{ij} P_{j}\left(\vec{g}\right)\right)}_{w_{i}(\vec{{g}})} T_{\text{in},i}.\\
&= \vec{P}\left(\vec{{g}}\right)\cdot A\vec{T},
\end{align}
where the second line is the first one recast in matrix notation.  The
matrix $A$ has to be calculated to obtain the full morphing function.
\item Thus, exploit that the output distribution should be equal to the input distribution at the respective input parameters
\begin{align}
T_\text{out}\left(\vec{{g}}_i\right) = T_{\text{in},i} \qquad \text{for}\qquad i=1,\dots ,N.
\end{align}
which can also be cast in matrix notation as
\begin{align}
  \begin{split}
    A \cdot \left(P_j\left(\vec{g}_i\right)\right)_{ij}
    = 1\!\!1\\
    \Leftrightarrow \qquad A\cdot G=1\!\!1.
  \end{split}
\end{align}
\item The unique solution $A=G^{-1}$ requires the input parameters to fulfil the condition $\text{det}(G)\ne 0$.

\end{enumerate}

When the aim is to perform a likelihood fit on some (pseudo-)data $T_d$, the minimization condition is
\begin{align}
  \widehat{\vec{g}}\left(T_d\right) = {\rm argmin}_{\vec{g}} -2 \ln P\left(T_d\mid \mu =
\sum_{i=1}^{N}\left(\sum_{j=1}^{N}A_{ij} P_{j}\left(\vec{g}\right)\right)T_{\text{in},i}\right).
\end{align}
From this it becomes apparent that only the polynomials $P_{j}\left(\vec{g}\right)$ need to be
recalculated during the minimization process, while the non-trivial
quantities $A_{ij}$ and $T_{\text{in},i}$ stay fixed.

The error propagation of statistical uncertainties to the output
$T_\text{out}$ is conceptually straightforward. Since the $\vec{g}_i$
are free parameters, the matrix $A$ carries no uncertainty
besides numerical fluctuations. Thus, uncertainties only propagate via
linear combinations. The question of how the input parameters
$\vec{g}_i$ need to be chosen such that the expected uncertainty of
the output is minimal, within some parameter region of interest, is
non-trivial and will be addressed in future studies.

The number $N$ of input base samples
depends on how many of coupling parameters enter the production and/or the decay vertex.
However, the general morphing principle remains the same and the method can be
generalized to a higher-dimensional coupling parameter space.

A general expression for the number of input samples $N$ with $n_p$ couplings appearing only in production, 
$n_d$ couplings appearing only in decay and $n_s$ couplings shared in production and decay is given by
\begin{align}
N&=\quad\frac{n_{p}\left(n_{p}+1\right)}{2}\cdot\frac{n_{d}\left(n_{d}+1\right)}{2}+{4+n_{s}-1 \choose 4}\label{eq:p2d2b4}\\
&+\left(n_{p}\cdot n_{s}+\frac{n_{s}\left(n_{s}+1\right)}{2}\right)\cdot\frac{n_{d}\left(n_{d}+1\right)}{2}\label{eq:pbb2d2}\\
&+\left(n_{d}\cdot n_{s}+\frac{n_{s}\left(n_{s}+1\right)}{2}\right)\cdot\frac{n_{p}\left(n_{p}+1\right)}{2}\label{eq:dbb2p2}\\
&+\frac{n_{s}\left(n_{s}+1\right)}{2}\cdot n_{p}\cdot n_{d}+\left(n_{p}+n_{d}\right){3+n_{s}-1 \choose 3}. \label{eq:b2pdb3pb3d}
\end{align}
In this expression the counting is split for (\ref{eq:p2d2b4}) terms pure in production and decay, or pure in shared, (\ref{eq:pbb2d2})
terms pure in decay and mixed in production and shared or purely shared, (\ref{eq:dbb2p2}) terms pure in production
and mixed in decay and shared or purely shared, and (\ref{eq:b2pdb3pb3d} terms mixed in both, and terms mixed in one and
purely shared in the other.

This is a general definition of the number of samples N in terms of
number of coupling parameters $n_p$, $n_d$, and $n_s$.  In case of the gluon
fusion process with subsequent decays to vector bosons, the production
and decay will have a completely disjoint set of couplings, and the
number of input samples will be given by Eq.\,\ref{eq:p2d2b4} by
setting $n_s = 0$. For the VBF Higgs boson production with subsequent
decay into vector bosons, when considering the same set of couplings
in the production and the decay vertex, the number of samples is given
by Eq.\,\ref{eq:p2d2b4} with $n_p = 0$ and $n_d = 0$.

\subsection{Conclusions}
\label{sec:conclusions}

This note describes a method for modelling signal parameters and
distributions in a multidimensional space of coupling
parameters. This method is capable of continuously morphing signal
distributions and rates based on a minimal orthogonal set of
independent base samples. Therefore it allows to directly fit for the
coupling parameters that describe the SM and possibly non-SM
interaction of the Higgs boson with fermions and bosons of the SM.

This method can be utilized to test the properties of the Higgs boson
during the LHC Run 2 data-taking period and beyond and has already
been tested successfully \cite{LHCHXSWG-INT-2016-004}.

\clearpage

\cleardoublepage

\newpage
\renewcommand*{\thefootnote}{\fnsymbol{footnote}}
\part[Measurements and Observables]{Measurements and Observables \footnote{M.~Chen, A.~David, M.~D\"uhrssen, A.~Falkowski, M.~Grazzini, R.~Harlander, C.~Hays, G.~Isidori, B.~Mellado, 
P.~Musella~(Eds.)}}
\label{chap:MO}
\renewcommand*{\thefootnote}{\Roman{part}.\arabic{footnote}}
\setcounter{footnote}{0} 

\chapter{Pseudo-observables}
\label{chap:PO}
\ChapterAuthor{A.~David, A.~Greljo, G.~Isidori, J.~Lindert, D.~Marzocca, G.~Passarino}
\renewcommand{\be}{\begin{equation}}
\renewcommand{\ee}{\end{equation}}
\newcommand{\hatkappa}{\hat\kappa}

\newcommand{\vareps}{\varepsilon}
\newcommand{\cpeps}{\epsilon^{\rm CP}}
\newcommand{\ACP}{A^{\rm CP}}
\newcommand{\lamCP}{\delta^{\rm CP}}
\newcommand{\no} {\nonumber}
\newcommand{\cL} {\mathcal L}
\newcommand{\cA} {\mathcal A}
\newcommand{\cT} {\mathcal T}
\newcommand{\cd} {{\cdot }}
\newcommand{\aem} {\alpha_{\rm em}}
\newcommand{\mZ} {m_Z}
\newcommand{\gZf} {g_Z^f}
\newcommand{\SU}{\text{SU}}
\newcommand{\LHAPDF}{{\rmfamily\scshape Lhapdf}\xspace}
\newcommand{\SherpaOpenLoops}{{\rmfamily\scshape Sherpa+OpenLoops}\xspace}
\newcommand{\pTjone}{\ensuremath{p_\mathrm{T,j_1}}\xspace}
\newcommand{\pTjtwo}{\ensuremath{p_\mathrm{T,j_2}}\xspace}
\newcommand{\HThalf}{\ensuremath{H_\mathrm{T}/2}\xspace}
\newcommand{\pTZ}{\ensuremath{p_\mathrm{T,Z}}\xspace}
\newcommand{\minvZH}{\ensuremath{m_\mathrm{HZ}}\xspace}

\newcommand{\mcA}{\mathcal{A}}
\newcommand{\mcC}{\mathcal{C}}
\newcommand{\mcD}{\mathcal{D}}
\newcommand{\mcF}{\mathcal{F}}
\newcommand{\mcP}{\mathcal{P}}
\newcommand{\mcQ}{\mathcal{Q}}
\newcommand{\mcT}{\mathcal{T}}
\newcommand{\mrc}{{\mathrm{c}}}
\newcommand{\mrs}{{\mathrm{s}}}
\newcommand{\mrv}{{\mathrm{v}}}
\newcommand{\mrA}{{\mathrm{A}}}
\newcommand{\mrD}{{\mathrm{D}}}
\newcommand{\mrF}{{\mathrm{F}}}
\newcommand{\mrG}{{\mathrm{G}}}
\newcommand{\mrH}{{\mathrm{H}}}
\newcommand{\mrI}{{\mathrm{I}}}
\newcommand{\mrL}{{\mathrm{L}}}
\newcommand{\mrN}{{\mathrm{N}}}
\newcommand{\mrP}{{\mathrm{P}}}
\newcommand{\mrQ}{{\mathrm{Q}}}
\newcommand{\mrR}{{\mathrm{R}}}
\newcommand{\mrS}{{\mathrm{S}}}
\newcommand{\mrT}{{\mathrm{T}}}
\newcommand{\mrV}{{\mathrm{V}}}
\newcommand{\mrW}{{\mathrm{W}}}
\newcommand{\TR}{\mathrm{TR}}
\newcommand{\spc}{\,,}
\newcommand{\spp}{\,.}
\newcommand{\ssI}{{\mathrm{I}}}
\newcommand{\mrdim}{\mathrm{dim}}
\newcommand{\myLO}{\mathrm{\scriptscriptstyle{LO}}}
\newcommand{\SMEFT}{\rm{\scriptscriptstyle{SMEFT}}}
\newcommand{\gds}{g_{_6}}
\newcommand{\nfact}{{\mbox{\scriptsize nfc}}}
\newcommand{\proc}{{\mbox{\scriptsize proc}}}
\newcommand{\sPF}{{\scriptscriptstyle{\PF}}}
\newcommand{\sPH}{{\scriptscriptstyle{\PH}}}
\newcommand{\sPZ}{{\scriptscriptstyle{\PZ}}}
\newcommand{\myGF}{G_{\sPF}}
\newcommand{\apD}{a_{\phi\,\scriptscriptstyle{\PD}}}
\newcommand{\apW}{a_{\phi\,\scriptscriptstyle{\PW}}}
\newcommand{\apBox}{a_{\phi\,\scriptscriptstyle{\Box}}}
\newcommand{\abp}{a_{\PQb\,\upphi}}
\newcommand{\sPHAA}{{\scriptscriptstyle{\PH\PAA}}}
\newcommand{\PAA}{\PA\PA}
\newcommand{\stWs}{\mrs_{_{\PW}}^2}
\newcommand{\stWvi}{\mrs_{_{\PW}}^6}
\newcommand{\aAA}{a_{\scriptscriptstyle{\PAA}}}
\newcommand{\atWB}{a_{\PQt\,\scriptscriptstyle{\PW\PB}}}
\newcommand{\atBW}{a_{\PQt\,\scriptscriptstyle{\PB\PW}}}
\newcommand{\srt}{\sqrt{2}}
\newcommand{\ctw}{\mrc_{_{\theta}}}
\newcommand{\ctW}{\mrc_{_{\PW}}}
\newcommand{\ctWs}{\mrc_{_{\PW}}^2}
\newcommand{\ctWc}{\mrc_{_{\PW}}^3}
\newcommand{\sPA}{{\scriptscriptstyle{\PA}}}
\newcommand{\sPZZ}{{\scriptscriptstyle{\PZZ}}}
\newcommand{\sPAZ}{{\scriptscriptstyle{\PA\PZ}}}
\newcommand{\sPAA}{{\scriptscriptstyle{\PA\PA}}}
\newcommand{\sPV}{{\scriptscriptstyle{\PV}}}
\newcommand{\sths}{{\hat \mrs}^2_{_{\theta}}}
\newcommand{\aZZ}{a_{\scriptscriptstyle{\PZZ}}}
\newcommand{\aAZ}{a_{\scriptscriptstyle{\PAZ}}}
\newcommand{\PZZ}{\PZ\PZ}
\newcommand{\PAZ}{\PA\PZ}
\newcommand{\sPB}{{\scriptscriptstyle{\PB}}}
\newcommand{\sPW}{{\scriptscriptstyle{\PW}}}
\newcommand{\stW}{\mrs_{_{\PW}}}             
\newcommand{\atp}{a_{\PQt\,\upphi}}
\newcommand{\abWB}{a_{\PQb\,\scriptscriptstyle{\PW\PB}}}
\newcommand{\sone}{s_{_1}}
\newcommand{\stwo}{s_{_2}}
\newcommand{\proA}[1]{\Delta_{\sPA}(#1)}
\newcommand{\proZ}[1]{\Delta_{\sPZ}(#1)}
\newcommand{\NR}{\mathrm{NR}}
\newcommand{\gamZ}{\mathswitch{\gamma_{\sPZ}}}
\newcommand{\muZ}{\mathswitch{\mu_{\sPZ}}}
\newcommand{\muZs}{\mathswitch{\mu^2_{\sPZ}}}
\newcommand\mybar[1]{\ensuremath{\bar{#1}}}
\newcommand{\spinu}[1]{{\mathrm{u}}_{#1}}
\newcommand{\spinub}[1]{\mybar{{\mathrm{u}}}_{#1}}
\newcommand{\spinubL}[1]{\mybar{{\mathrm{u}}}_{#1\,\scriptscriptstyle{\mrL}}}
\newcommand{\spinubR}[1]{\mybar{{\mathrm{u}}}_{#1\,\scriptscriptstyle{\mrR}}}
\newcommand{\mcV}{\mathcal{V}}
\newcommand{\spinv}[1]{{\mathrm{v}}_{#1}}
\newcommand{\spinvL}[1]{{\mathrm{v}}_{#1\,\scriptscriptstyle{\mrL}}}
\newcommand{\spinvR}[1]{{\mathrm{v}}_{#1\,\scriptscriptstyle{\mrR}}}
\newcommand{\sPHZ}{{\scriptscriptstyle{\PH\PZ}}}
\newcommand{\sPHZZ}{{\scriptscriptstyle{\PH\PZ\PZ}}}
\newcommand{\DR}{\mathrm{DR}}
\newcommand{\SR}{\mathrm{SR}}
\newcommand{\apfV}{a_{\phi\,\Pf\,\scriptscriptstyle{\PV}}}
\newcommand{\aplV}{a_{\phi\,\Pl\,\scriptscriptstyle{\PV}}}
\newcommand{\aplA}{a_{\phi\,\Pl\,\scriptscriptstyle{\PA}}}
\newcommand{\stWq}{\mrs_{_{\PW}}^4}
\newcommand{\ctWq}{\mrc_{_{\PW}}^4}
\newcommand{\aqp}{a_{\PQq\,\upphi}}
\newcommand{\gen}{{\mbox{\scriptsize gen}}}
\newcommand{\ssdCZ}{{\mathrm{d}{\mathcal{Z}}}}
\newcommand{\ssdZ}{{\mathrm{dZ}}}
\newcommand{\sPHWW}{{\scriptscriptstyle{\PH\PW\PW}}}
\newcommand{\aptV}{a_{\phi\,\PQt\,\scriptscriptstyle{\PV}}}
\newcommand{\aptA}{a_{\phi\,\PQt\,\scriptscriptstyle{\PA}}}
\newcommand{\apbV}{a_{\phi\,\PQb\,\scriptscriptstyle{\PV}}}
\newcommand{\apbA}{a_{\phi\,\PQb\,\scriptscriptstyle{\PA}}}
\newcommand{\apUD}{a_{\upphi\,\PQU\,\PQD}}
\newcommand{\abBW}{a_{\PQb\,\scriptscriptstyle{\PB\PW}}}
\newcommand{\afWB}{a_{\Pf\,\scriptscriptstyle{\PW\PB}}}
\newcommand{\apWB}{a_{\phi\,\scriptscriptstyle{\PW\PB}}}
\newcommand{\apfA}{a_{\phi\,\Pf\,\scriptscriptstyle{\PA}}}
\newcommand{\aUW}{a_{\sPQU\,\scriptscriptstyle{\PW}}}
\newcommand{\sPQD}{{\scriptscriptstyle{\PQD}}}
\newcommand{\sPQU}{{\scriptscriptstyle{\PQU}}}
\newcommand{\sPZWW}{{\scriptscriptstyle{\PZ\PW\PW}}}
\newcommand{\aDW}{a_{\sPQD\,\scriptscriptstyle{\PW}}}
\newcommand{\aQW}{a_{\scriptscriptstyle{\PQQ\PW}}}
\newcommand{\apFt}{a^{(3)}_{\phi\,\PF}}
\newcommand{\apB}{a_{\phi\,\scriptscriptstyle{\PB}}}
\newcommand{\sPHAV}{{\scriptscriptstyle{\PH\PA\PV}}}
\newcommand{\sPHAZ}{{\scriptscriptstyle{\PH\PA\PZ}}}
\newcommand{\PO}{\mathrm{\scriptscriptstyle{PO}}}
\DeclareRobustCommand{\PQQ}{\HepParticle{Q}{}{}\Xspace}
\DeclareRobustCommand{\Ppz}{\HepParticle{\upphi}{}{0}\Xspace}
\DeclareRobustCommand{\PQU}{\HepParticle{U}{}{}\Xspace}
\DeclareRobustCommand{\PQD}{\HepParticle{D}{}{}\Xspace}

\newcommand{\myPH}{h}
\newcommand{\myPV}{V}
\newcommand{\myPZ}{Z}
\newcommand{\myPA}{A}
\newcommand{\myPW}{W}
\newcommand{\myPAA}{AA}
\newcommand{\myPAZ}{AZ}
\newcommand{\myPGg}{\gamma}
\newcommand{\myPf}{f}
\newcommand{\myPAf}{f}
\newcommand{\myPep}{e^+}
\newcommand{\myPem}{e^-}
\newcommand{\myPGm}{\mu^-}
\newcommand{\myPGp}{\mu^+}
\newcommand{\myPQb}{b}
\newcommand{\myPAQb}{\bar b}

\renewcommand{\mw}{m_W}
\renewcommand{\mZ}{m_Z}
\renewcommand{\mz}{m_Z}
\renewcommand{\mws}{m^2_W}
\renewcommand{\mzs}{m^2_Z}
\renewcommand{\mhs}{m^2_h}

\newcommand{\arXhref}[1]{\href{http://arxiv.org/abs/#1}{#1}}

\definecolor{darkblue}{cmyk}{1,0.3,0,0.2}
\definecolor{violet}{cmyk}{0,1,0,0.2}

\newcommand{\xxx}[1]{{\color{red} #1}}

\section{Introduction}

The idea of PO has been formalized the first time in the context of electroweak
observables around the $Z$ pole~\cite{Bardin:1999gt}.
A generalization of this concept to describe possible deformations
from the SM in Higgs boson production and decay processes has been discussed in Refs.~~\cite{Gonzalez-Alonso:2014eva,Gonzalez-Alonso:2015bha,Ghezzi:2015vva,Passarino:2010qk,David:2015waa,Greljo:2015sla}.
The basic idea is to identify a set of quantities that are
\begin{itemize}
\item[\bf I.] experimentally accessible,
\item[\bf II.] well-defined from the point of view of QFT,
\end{itemize}
and capture all relevant New Physics (NP) effects (or all relevant deformations from the SM) without losing information and with  minimum theoretical bias.
The last point implies that changes in the underlying NP model should not require any new processing of raw experimental data.
In the same spirit, the PO should be independent from the theoretical precision (e.g. LO, NLO, ...) at which NP effects
are computed. Finally, the PO are obtained after removing (via a proper deconvolution) the effect of the soft SM radiation (both QED and QCD radiation), that is assumed to be free from NP effects. In the case of observables around the $Z$ pole, the $\Gamma(Z\to f\bar f)$ partial decay rates provide good examples of PO.

The independence from NP models can not be fulfilled in complete generality.  However, it can be fulfilled under very general assumptions. In particular, we require the PO to
\begin{itemize}
\item[\bf III.] capture all relevant NP effects in the limit of no new (non-SM) particles
propagating on-shell (in the amplitudes considered) in the kinematical range where the
decomposition is assumed to be valid.

\end{itemize}

\noindent
Under this additional hypothesis, the PO provide a bridge between the fiducial cross-section measurements and the determination of NP couplings in explicit NP frameworks.

On a more theoretical footing, the Higgs PO are defined from a
general decomposition of on-shell amplitudes involving the Higgs boson -- based
on analyticity, unitarity, and crossing symmetry -- and a momentum expansion
following from the dynamical assumption of no new light particles (hence no
unknown physical poles in the amplitudes) in the kinematical regime where the
decomposition is assumed to be valid. These conditions ensure the generality of this
approach and the possibility to match it to a wide class of explicit NP model,
including the determination of Wilson coefficients in the context of Effective Field Theories.

The old $\kappa$ framework~\cite{Heinemeyer:2013tqa,LHCHiggsCrossSectionWorkingGroup:2012nn}
satisfied the conditions I and II, but not the condition III,
since the framework was not general enough to describe modifications in $(n>2)$-body
Higgs boson decays resulting in non-SM kinematics. Similarly, the old $\kappa$ framework could not describe
modifications of the Higgs-cross sections that cannot be reabsorbed into a simple overall re-scaling with respect to the SM.

Similarly to the case of electroweak observables, it is convenient to introduce two complementary sets of Higgs PO:
\begin{itemize}
\item{} a set of {\em physical} PO, namely a set
of (idealized) partial decay rates and asymmetries;
\item{} a set of {\em effective-couplings} PO,
parameterizing the on-shell production and decay amplitudes.
\end{itemize}
The two sets are in one-to-one correspondence: by construction, the effective-couplings PO are directly related to the {\it physical} PO after properly working out the decay kinematics. The effective-couplings PO are particularly useful to build tools to simulate data, taking into account  the effect of soft QCD and QED radiation.\footnote{A first public tool for Higgs PO is available
in Ref.~\cite{HiggsPO}.}  This is why, from the practical point of view,
the effective-couplings  PO are first extracted from data in the LHC Higgs analysis, and from these the {\it physical} PO are indirectly derived. As we discuss below, the latter  provide a more intuitive and effective presentation of the measurements performed.

The note is organized as follows: the PO for Higgs boson decays are discussed in Section \ref{sec:PO-2}-\ref{sec:decomposition}, separating
two, three, and four-body decay modes. General aspects of PO in electroweak production processes
are discussed in Section~\ref{sec:PO-EW}, whereas the specific implementation for VH  and VBF  is presented in Section~\ref{sec:PO-pheno}.
The total number of PO to discuss both production and decay processes is summarized in Section~\ref{sectPO:PCSL},
where we also address the reduction of the number of independent terms under specify symmetry assumptions
(in particular CP conservation and flavour universality). Finally, a discussion about the
matching between the PO approach and the SM Effective Field Theory (SMEFT) is presented in Sections~\ref{sectPO:SMEFT}.
The latter section is not needed to discuss the PO implementation in data analyses, but it provides
a bridge between this chapter of the YR (Measurements and Observables) and the one devoted to the EFT
approaches.

\section{Two-body decay modes}
\label{sec:PO-2}

In the case of two-body Higgs boson decays into on-shell SM particles, namely $h\to   f  \bar f$ and $h\to   \gamma \gamma$, the natural {\em physical PO } for each mode are the partial decay widths, and possibly the polarization asymmetry
if the spin of the final state is accessible.

In the  $h\to   f  \bar f$ case the main issue to be addressed is the optimal definition of the partial decay width taking into account the final state QED and QCD radiation.

In the  $h\to  \gamma\gamma$ case the point to be addressed is the extrapolation to real photons of electromagnetic showers with non-vanishing invariant mass.

\subsection{\texorpdfstring{$h\to   f  \bar f$}{h to ff}  }
\label{sect:hff}

For each fermion species we can decompose the on-shell  $h\to   f  \bar f$ amplitude in terms of
two effective couplings ($y^f_{S,P}$), defined by
\be
\cA( h\to f \bar f) =   -\frac{i}{\sqrt{2}} \left( y^f_S   \bar f  f +  i y^f_P \bar f  \gamma_5  f \right)~,
\label{eq:one}
\ee
where $f$, $\bar f$ in the right hand side are spinor wave functions.
These couplings are real in the limit where we neglect re-scattering effects,
 that is an excellent approximation (also beyond the SM if we assume no new light states),
 for all the accessible $h\to f \bar f$ channels.   If $h$ is a CP-even state (as in the SM),
then $y^f_P$ is a CP-violating coupling.

In order to match our notation with the $\kappa$ framework~\cite{Heinemeyer:2013tqa}, we define the two {\em effective couplings} PO of the $h\to   f  \bar f$  decays as follows:
\be
\kappa_f = \frac{ y^f_{S} }{ y^{f, \rm SM}_S }~,  \qquad
\lamCP_f = \frac{ y^f_{P} }{ y^{f, \rm SM}_S }~.
\label{eq:two}
\ee
Here $y^{f, \rm SM}_S$ is the SM effective coupling that provides the best SM prediction in the $\kappa_f \to 1$ and $\lamCP_f \to 0$ limit.

The measurement of $\Gamma(h\to f\bar f)_{\rm (incl)}$ determines the combination $| \kappa_f |^2 + |\lamCP_f |^2$,
while the $\lamCP_f/\kappa_f$ ratio can be determined only if the fermion polarization
is  experimentally accessible.
With this notation, the inclusive decay rates, computed  assuming
a pure bremsstrahlung spectrum can be  written as
\be
\Gamma(h\to f \bar f)_{\rm (incl)} =  \left[ |\kappa_f|^2 + |\lamCP_f |^2  \right]  \Gamma(h\to f \bar f)_{\rm (incl)}^{\rm (SM)}~,
\label{eq:physPOgg}
\ee
where fermion-mass effects, of per-mil level even for the $b$ quark, have been neglected.
In experiments $\Gamma(h\to f \bar f)_{\rm (incl)} $ cannot be directly accessed,
given tight cuts on the $f \bar f$ invariant mass to suppress the background: $\Gamma(h\to f \bar f)_{\rm (incl)} $
is extrapolated from the experimentally accessible  $\Gamma(h\to f \bar f)_{\rm (cut)}$
assuming a pure bremsstrahlung spectrum, both as far as QED and as far as
QCD (for the $q\bar q$ channels only) radiation is concerned.

The SM decay width is given by
\be
	\Gamma(h\to f \bar f)_{\rm (incl)}^{\rm (SM)} = N_c^f \frac{|y^{f,\SM}_{\rm eff}|^2}{16 \pi} m_H~,
	\label{eq:hffSM}
\ee
where the colour factor $N_c^f$ is 3 for quarks and 1 for leptons.
Using the best SM prediction of the branching ratios presented in this report (Tables~\ref{tab:YRHXS4_1}--\ref{tab:YRHXS4_5}), 
for $m_H = 125.09 ~\UGeV$, and $\Gamma_H^{\rm tot} = 4.100 \times 10^{-3} ~\UGeV$ ($\pm 1.4 \%$), we extract the values of the $|y^{f,\SM}_{\rm eff}|$ couplings in 
Eq.~\eqref{eq:hffSM}:
\begin{equation*}\begin{aligned}
&\begin{array}{c | cccc}
					& \bar{b} b & \bar{\tau} \tau \\ \hline
	\mathcal{B}(h \to \bar{f} f)& 5.809 \times 10^{-1} ~(\pm 1.2 \%)& 6.256 \times 10^{-2} ~(\pm 1.6 \%) \\
	|y^{f,\SM}_{\rm eff}|	& 1.79 \times 10^{-2}  ~(\pm 0.92 \%) & 1.02 \times 10^{-2}  ~(\pm 1.1 \%)
\end{array}~, \\
&\begin{array}{c | cccc}
					& \bar{c} c & \bar{\mu} \mu \\ \hline
	\mathcal{B}(h \to \bar{f} f)& 2.884 \times 10^{-2}  ~(\pm 5.5 \%)& 2.171 \times 10^{-4} ~(\pm 1.7 \%) \\
	|y^{f,\SM}_{\rm eff}|	& 3.98 \times 10^{-3} ~(\pm 2.8 \%) & 5.98 \times 10^{-4}  ~(\pm 0.68 \%)
\end{array}~,
\end{aligned}\end{equation*}

As anticipated, the {\it physical} PO sensitive to  $\lamCP_f/\kappa_f$ necessarily involve a determination (direct or indirect)
of the fermion spins. Denoting by $\vec{k}_f$ the 3-momentum of the fermion $f$ in the Higgs centre of mass frame,
and  with $\{ \vec{s}_f, \vec{s}_{\bar f}\} $ the two fermion spins, we can define the following
CP-odd asymmetry~\cite{Bernreuther:1993df}
\be
\cA^{\rm CP}_f = \frac{1}{ |\vec{k}_f| }  \langle \vec{k}_f  \cdot (\vec{s}_f  \times  \vec{s}_{\bar f}) \rangle    =  - \frac{\lamCP_f \kappa_f }{ \kappa_f^2 +  (\lamCP_f )^2 }
\ee
As pointed out in Ref.~\cite{Berge:2008wi}, in the $h\to \tau^+\tau^- \to X_{\tau^+} X_{\tau^-}$ decay chains
asymmetries proportional to $\cA^{\rm CP}_f$ are accessible
through the measurement of the angular distribution of the $\tau^\pm$ decay products.

Note that, by construction, the effective couplings PO depend on the SM normalization.
This imply an intrinsic theoretical uncertainty in their determination related to the theory error on the SM reference value.
On the other hand, the {\it physical} PO are independent of any reference to the SM.
Indeed the (conventional) SM normalization of $\kappa_f$ cancels in
Eq.~(\ref{eq:physPOgg}).

\subsection{\texorpdfstring{$h\to \gamma\gamma$}{h to gamma gamma}}
\label{sec:hgg}

The general  decomposition for  the $h\to \gamma\gamma$ amplitude is
\bqa
\cA \left[ h \to  \gamma (q, \epsilon)  \gamma(q^\prime, \epsilon^\prime) \right]  &=&   i \frac{2 }{ v_F}    \epsilon^\prime_\mu  \epsilon_\nu
 \left[   \epsilon_{\gamma\gamma}  (  g^{\mu\nu}~ {q}\cd{q^\prime} - {q}^\mu {q}^{\prime \nu})
 +    \cpeps_{\gamma\gamma}   \vareps^{\mu\nu\rho\sigma} q_{\rho} q^\prime_{\sigma}  \right]~, \qquad
 \label{eq:h2gamma}
\eqa
where $\epsilon^{\mu\nu\rho\sigma}$ is the fully antisymmetric tensor and $\epsilon^{0123} = 1$ in our convention. From this we identify the two  effective couplings $\epsilon_{\gamma\gamma}$ and $\cpeps_{\gamma\gamma}$
that, similarly to $y^f_{S,P}$, can be assumed to be real in the limit where we assume no new light states
and small deviations from the SM limit. Here $v_F = (\sqrt{2} G_F)^{-1/2} $,
and  $G_F$ is the Fermi constant extracted from the muon decay.
We define the effective couplings  PO for this channels as
\be
\kappa_{\gamma\gamma} =   \frac{  {\rm Re}( \epsilon_{\gamma \gamma} ) }{  {\rm Re}( \epsilon_{\gamma\gamma}^{\rm SM} ) }~, \qquad
\lamCP_{\gamma\gamma} =   \frac{  {\rm Re}( \cpeps_{\gamma \gamma} ) }{  {\rm Re}( \epsilon_{\gamma\gamma}^{\rm SM} ) }~,
\label{eq:epsgg}
\ee
where $\epsilon^{\gamma\gamma}_{\rm SM}$ is the value of the PO which reproduces the best SM prediction of the decay width.
By construction, the SM expectation for the two PO  is  $\kappa^{\rm SM}_{\gamma\gamma}=1$ and
$(\lamCP_{\gamma\gamma})^{\rm SM} =0$.

If the photon polarization is not accessible, the only {\it physical} PO for this channel is $\Gamma(h\to \gamma\gamma)$.
Starting from realistic observables, where the electromagnetic showers have non-vanishing invariant mass,
$\Gamma(h\to \gamma\gamma)$ is defined as the extrapolation to the limit of zero invariant  mass for the
electromagnetic showers.
The relation between  $\Gamma(h\to \gamma\gamma)$ and the two effective couplings  PO is
\be
\Gamma(h\to \gamma\gamma) = \left[ \kappa_{\gamma\gamma}^2  +   (\lamCP_{\gamma\gamma})^2   \right]
 \Gamma(h\to \gamma\gamma)^{\rm (SM)}~,
 \label{eq:hggrate}
\ee
where
\be
\Gamma(h\to \gamma\gamma)^{\rm (SM)} = \frac{|\epsilon^{\SM, {\rm eff}}_{\gamma\gamma}|^2}{16 \pi} \frac{m_H^3}{v_F^2}.
\ee
Using the SM prediction for the branching ratios in two photons (Table~\ref{tab:YRHXS4_3}), 
for $m_H = 125.09 ~\UGeV$, and $\Gamma_H^{\rm tot} = 4.100 \times 10^{-3} ~\UGeV$ ($\pm 1.4 \%$), we obtain
\be
	\mathcal{B}(h \to \gamma\gamma)^{\SM} = 2.270 \times 10^{-3} ~(\pm 2.1 \%) \quad \rightarrow \quad
	\epsilon^{\gamma\gamma}_{\rm SM} = 3.81 \times 10^{-3}~(\pm 1.2 \%) ~.
\ee
This value corresponds to the 1-loop contribution in the SM, which also fixes the relative sign.
Similarly to the $f \bar f$ case, the  SM normalization cancels in the definition of the {\it physical} PO.

The {\it physical} PO linear in the CP-violating coupling  $\lamCP_{\gamma\gamma}$ necessarily
 involves the measurement of the photon polarization and is therefore hardly accessible at the LHC (at least in a direct way,
 see for example~\cite{Bishara:2013vya}).
Denoting by $\vec{q}_{1,2}$ the 3-momenta of the two photons in the centre of mass frame,
and  with $\vec{\epsilon}_{1,2}$ the corresponding polarization vectors,
we can define:
\be
\cA^{\rm CP}_{\gamma\gamma} = \frac{1}{ m_h }  \langle (\vec{q}_1 -\vec{q}_2)  \cdot (\vec{\epsilon}_{1}  \times  \vec{\epsilon}_{2}) \rangle    \propto    \frac{\lamCP_{\gamma\gamma}
\kappa_{\gamma\gamma} }{ \kappa_{\gamma\gamma}^2 +  (\lamCP_{\gamma\gamma} )^2 }~.
 \label{eq:ACPgg}
\ee

\section{Three-body decay modes}
\label{sec:PO-3}

The guiding principle for the definition of PO in multi-body channels is the decomposition of the  decay amplitudes
in terms of contributions associated to a specific single-particle  pole structure. In the absence of new light states,
such poles are generated only by the exchange of the SM electroweak bosons ($\gamma$, $Z$, and $W$)
or by hadronic resonances (whose contribution appears only beyond the tree level and is largely suppressed).
Since positions and residues on the poles are gauge-invariant quantities, this decomposition satisfies the general
requirements for the definitions of PO.

\subsection{\texorpdfstring{$h\to  f\bar f \gamma$}{h to ff gamma}}
\label{sec:hgll}

The general form factor decomposition for these  channels is
\bqa
 \cA \left[ h \to  f (p_1) \bar f (p_2) \gamma (q, \epsilon) \right]  &=&   i \frac{2  }{ v_F}   \sum_{f = f_L, f_R}     (\bar f  \gamma_\mu f)  \epsilon_\nu
\times \no \\
& \times & \!\!\!   \left[    F_T^{f\gamma} (p^2)  ( {p}\cd {q}~ g^{\mu\nu}-{q}^\mu {p}^\nu )
 +   F_{CP}^{f\gamma} (p^2)     \vareps^{\mu\nu\rho\sigma} q_{\rho} p_{\sigma}  \right],
 \label{eq:hZgamma}
\eqa
where $p=p_1+p_2$.
The form factors can be further decomposed as
\bqa
&& F^{f\gamma}_T (p^2) =   \epsilon_{Z\gamma}   \frac{   g_Z^f  }{ P_Z( p^2) }   +   \epsilon_{\gamma\gamma}   \frac{   e  Q_f  }{ p^2  } + \Delta_{f\gamma}^{\rm SM}(p^2)~,
 \label{eq:h2l2b}
   \\
&& F^{f\gamma}_{CP} (p^2)  =    \cpeps_{Z\gamma}     \frac{   g_Z^f  }{ P_Z( p^2) }   +   \cpeps_{\gamma\gamma}   \frac{   e  Q_f   }{ p^2 }~.
 \label{eq:h2l2}
\eqa
Here $g_Z^f$ are the effective PO describing on-shell $Z\to f\bar f$ decays\footnote{We have absorbed a factor $g/\cos(\theta_W)$ with respect to the definition of the effective $Z$ couplings adopted at LEP-1, see
Eq.~\eqref{eq:LEPZcoupl}.
}
and $P_Z(q^2) =  q^2 -m_Z^2 + i m_Z \Gamma_Z$.
In other words, we decompose the form factors identifying the physical poles associated to the $Z$ and $\gamma$ propagators.

The term $\Delta_{f\gamma}^{\rm SM}(p^2)$ denotes the remnant of the SM $h\to f\bar f \gamma$ loop function that is
regular both in the limit $p^2\to 0$ and in the limit $p^2\to m^2_Z$. This part of the amplitude is largely subdominant (being not enhanced by  a physical single-particle pole) and cannot receive non-standard contributions from operators of dimension up to 6 in the EFT approach to Higgs physics. For this reason it is fixed to its SM value.

In this channel we thus have four effective couplings PO, related to the four $\epsilon_X$ terms in  Eqs. (\ref{eq:h2l2b}) and (\ref{eq:h2l2}),
two of which are accessible also in $h\to 2 \gamma$.\footnote{~In the decomposition~(\ref{eq:hZgamma}) we have also neglected possible dipole-type (helicity-suppressed) amplitudes. The latter necessarily give a strongly suppressed
contribution to the decay rate since the interference with the leading
$h\to f\bar f \gamma$ amplitude vanishes in the limit $m_f \to 0$. There is no obstacle, in principle, to add
a corresponding set of PO for these suppressed amplitudes. However, for all light fermions they will be un-measurable
in realistic scenarios even in the high-luminosity phase of the LHC (see e.g.~\cite{Gonzalez-Alonso:2014rla} for a numerical discussion in the $h\to 2 \mu \gamma$ case).}

Similarly to the $h\to 2 \gamma$ case, it is convenient to
define the PO normalizing them the corresponding reference SM values of the amplitudes. We thus define
\be
\kappa_{Z\gamma} =   \frac{  {\rm Re}( \epsilon_{Z \gamma} ) }{  {\rm Re}( \epsilon_{Z\gamma}^{\rm SM} ) }~, \qquad
\lamCP_{Z\gamma} =   \frac{  {\rm Re}( \cpeps_{Z\gamma} ) }{  {\rm Re}( \epsilon_{Z\gamma}^{\rm SM} ) }~,
\label{eq:epsZg}
\ee
where the numerical value of the SM contribution $\epsilon_{Z\gamma}^{\rm SM}$ is obtained from the best SM prediction for the $h \to Z \gamma$ decay width.

The simplest {\it physical} PO that can be extracted from this channel is $\Gamma(h\to Z\gamma)$,
where both the $Z$ boson and the photon are on-shell.
By construction, this can be written as
\be
\Gamma(h\to Z\gamma) = \left[ \kappa_{Z\gamma}^2  +  (\lamCP_{Z\gamma})^2   \right]
 \Gamma(h\to Z\gamma)^{\rm (SM)}~,
 \label{eq:hZArate}
\ee
where
\be
\Gamma(h\to Z\gamma)^{\rm (SM)} = \frac{|\epsilon^{\SM, {\rm eff}}_{Z\gamma}|^2}{8 \pi} \frac{m_H^3}{v^2} \left(1 - \frac{m_Z^2}{m_H^2}\right)^3~.
\ee
The SM prediction for this decay rate (Table~\ref{tab:YRHXS4_4}),  
for $m_H = 125.09 ~\UGeV$, and $\Gamma_H^{\rm tot} = 4.100 \times 10^{-3} ~\UGeV$ ($\pm 1.4 \%$), provides the value of $\epsilon^{Z\gamma}_{\rm SM}$:
\be
\mathcal{B}(h\to Z\gamma)^{\rm (SM)} = 1.541 \times 10^{-3} ~(\pm 5.8\%) \quad\to\quad
\epsilon_{Z\gamma}^{\rm SM} = 6.91 \times 10^{-3} ~(\pm 3.0\%) ~.
\ee

The independent {\it physical} PO linear in the coupling $\lamCP_{Z\gamma}$ is the following CP-odd asymmetry at the $Z$ peak:
\be
\cA^{\rm CP}_{Z\gamma} = \left.  \frac{1}{ | \vec{p} | | \vec{q} | }   \langle  \vec{p} \cdot  (\vec{q} \times \vec{\epsilon_\gamma} ) \rangle \right|_{(p^2=m_Z^2)}
\propto  \frac{\lamCP_{Z\gamma}
\kappa_{Z\gamma} }{ \kappa_{Z\gamma}^2 +  (\lamCP_{Z\gamma} )^2 }~,
\ee
where all 3-momenta are defined in the Higgs centre of mass frame.

This channel is also sensitive to $\Gamma(h\to \gamma\gamma)$ and $\cA^{\rm CP}_{\gamma\gamma}$ via the effective couplings
$\kappa_{\gamma\gamma}$ (or $\epsilon_{\gamma\gamma}$)  and $\lamCP_{\gamma\gamma}$ (or $\cpeps_{\gamma\gamma}$).
Determining such couplings from a fit to the from factors in the low $p^2$ region, one can indirectly
determine $\Gamma(h\to \gamma\gamma)$ and $\cA^{\rm CP}_{\gamma\gamma}$
by means of  Eq.~(\ref{eq:hggrate}) and  Eq.~(\ref{eq:ACPgg}), respectively.

\section{Four-fermion decay modes}
\label{sec:decomposition}

Similarly to the three-body modes, also in this case the guiding principle for the definition of PO
is the decomposition of the  decay amplitudes in terms of contributions associated to a specific pole structure.
Such decomposition for the $h\to 4f$ channels has been presented in  Ref.~\cite{Gonzalez-Alonso:2014eva}.
The effective coupling PO that appear in these channels consist of four sets:
\begin{itemize}
\item{} 3 flavour-universal charged-current PO: $\{\kappa_{WW},\epsilon_{WW}, \cpeps_{WW} \}$;
\item{} 7 flavour-universal neutral-current PO,  4 of which are appearing already in $h\to \gamma\gamma$ and $h\to f\bar f\gamma$ :  $\{ \kappa_{\gamma \gamma},\lamCP_{\gamma\gamma},\kappa_{Z \gamma},\lamCP_{Z \gamma}\}$, and another
3 which are  specific for $h\to 4f$: $\{ \kappa_{ZZ},\epsilon_{ZZ},\cpeps_{ZZ}\}$;
\item{} the set of flavour non-universal charged-current PO:~$\{ \epsilon_{Wf} \}$;
\item{} the set of flavour non-universal neutral-current PO: ~$\{ \epsilon_{Zf} \}$.
\end{itemize}

While the number of  flavour-universal PO is fixed,   the number of flavour non-universal PO depend
on the fermion species we are interested in. For instance, looking only at light leptons $(\ell=e,\mu)$,
we have 4  flavour non-universal  PO contributing to  $h \to 4\ell$ modes ($\epsilon_{Zf}$, with $f=e_L,e_R, \mu_L,\mu_R$)
and 4 PO contributing to $h \to 2\ell 2\nu$ modes  ($\epsilon_{W e_L}, \epsilon_{W \mu_L},  \epsilon_{Z \nu_e}, \epsilon_{Z \nu_\mu}$).
The definition of these PO is done at the amplitude level, separating neutral-current and charged-current
contributions to the $h\to 4f$ processes, as discussed below.

\medskip

Starting from each of the effective couplings PO we can define a corresponding {\it physical} PO.
In particular, $\Gamma(h\to ZZ)$ is defined as the (ideal) rate extracted from the full $\Gamma(h\to 4f)$,
extrapolating the result in the limit $\kappa_{ZZ}\not=0$ and all the other effective couplings set to zero. Similarly $\Gamma(h\to Z f\bar f)$ is defined
from the extrapolation in the limit $\epsilon_{Zf} \not =0$ and all the other effective couplings set to zero
(see extended discussion below).

\bigskip

 \subsection{\texorpdfstring{$h\to 4 f$}{h to 4f} neutral currents}
 \label{sect:h4fneutral}

Let us consider the case of two different (light) fermion species: $h\to f\bar f + f^\prime \bar f^\prime$.
Neglecting helicity-violating terms (yielding contributions suppressed by light fermion masses
in the rates), we can decompose the  neutral-current contribution to the amplitude in the following way
\bqa
&& \cA_{n.c.} \left[ h \to  f (p_1) \bar f (p_2) f^\prime  (p_3) \bar f^\prime (p_4) \right]  =  i \frac{2 m^2_Z}{ v_F}   \sum_{f = f_L, f_R}   \sum_{f^\prime = f^\prime_L, f^\prime_R}
(\bar f  \gamma_\mu f) (\bar f^\prime  \gamma_\nu f^\prime) \cT^{\mu\nu}_{n.c.} (q_1,q_2) \nonumber \\
&&
\cT^{\mu\nu}_{n.c.} (q_1, q_2) =  \left[ F^{f f^\prime}_L  (q_1^2, q_2^2) g^{\mu\nu} +  F^{f f^\prime}_T (q_1^2, q_2^2)  \frac{ {q_1}\cd {q_2}~g^{\mu\nu} -{q_2}^\mu {q_1}^\nu }{m_Z^2}   \right. \nonumber \\
&& \left. \qquad \qquad\qquad \qquad   +   F^{f f^\prime}_{CP} (q_1^2, q_2^2)  \frac{  \vareps^{\mu\nu\rho\sigma} q_{2\rho} q_{1\sigma}   }{m_Z^2}  \right]~,
 \label{eq:h4l1}
 \eqa
where $q_1=p_1 +p_2$ and $q_2=p_3 +p_4$.
The form factor $F_L$ describes the interaction with the longitudinal part of the current, as in the SM, the $F_T$ term describes the interaction with the transverse part, while $F_{CP}$ describes the CP-violating part of the interaction (if the Higgs is assumed to be a CP-even state).

We can further expand  the form factors in full generality  around the poles,  providing the definition of the neutral-current PO~\cite{Gonzalez-Alonso:2014eva}:
\bqa
F^{f f^\prime}_L (q_1^2, q_2^2) &=& \kappa_{ZZ}  \frac{ g_Z^f  g_Z^{f^\prime}  }{P_Z(q_1^2) P_Z(q_2^2)}
  +  \frac{\epsilon_{Z f}}{m_Z^2}  \frac{ g_Z^{f^\prime}   }{  P_Z(q_2^2)} +    \frac{\epsilon_{Z f^\prime}}{m_Z^2}   \frac{ g_Z^f    }{  P_Z(q_1^2)}  +\Delta^{\rm SM}_{L} (q_1^2, q_2^2) ~,   \label{eq:h4lF1}\\
F^{f f^\prime}_T (q_1^2, q_2^2) &=&  \epsilon_{ZZ}  \frac{ g_Z^f  g_Z^{f^\prime}  }{P_Z(q_1^2) P_Z(q_2^2)}   +   \epsilon_{Z\gamma}  \left(
 \frac{  e  Q_{f^\prime} g_Z^f   }{ q_2^2  P_Z( q_1^2) }   +  \frac{ e Q_f   g_Z^{f^\prime}   }{ q_1^2  P_Z( q_2^2) }  \right) +   \epsilon_{\gamma\gamma}   \frac{   e^2 Q_f Q_{f^\prime} }{ q_1^2 q_2^2  } \no \\
 && +\Delta^{\rm SM}_{T} (q_1^2, q_2^2),   \label{eq:h4lF3}
\\
F^{f f^\prime}_{CP} (q_1^2, q_2^2) &=&  \cpeps_{ZZ}  \frac{ g_Z^f  g_Z^{f^\prime}  }{P_Z(q_1^2) P_Z(q_2^2)}   +   \cpeps_{Z\gamma}  \left(
 \frac{  e  Q_{f^\prime} g_Z^f   }{ q_2^2  P_Z( q_1^2) }   +  \frac{ e Q_f   g_Z^{f^\prime}   }{ q_1^2  P_Z( q_2^2) }  \right) +   \cpeps_{\gamma\gamma}   \frac{   e^2 Q_f Q_{f^\prime} }{ q_1^2 q_2^2  }~.
 \label{eq:h4lF4}
\eqa
Here $g_{Z}^f$ are $Z$-pole PO extracted from $Z$ decays at LEP-I, the translation to the notation used at LEP being very simple
\be
	g_{Z}^{f} = \frac{2 m_Z}{v_F} g_{f}^{\rm LEP}~, \qquad \text{and} \qquad
	(g_Z^f)_{\rm SM} = \frac{2 m_Z}{v_F} (T_3^f - Q_f s_{\theta_W}^2)~.
	\label{eq:LEPZcoupl}
\ee
As anticipated, all the parameters but $\epsilon_{Z f}$ and $g_{Z}^f$ are flavour universal, i.e.~they do not
depend on the fermion species. In fact, flavour non-universal effects in $g_{Z}^f$ have been very tightly constrained at
LEP-I \cite{ALEPH:2005ab}, however, sizeable effects in $\epsilon_{Z f}$ are possible and should be tested at the LHC. In the limit where we neglect re-scattering effects, both   $\kappa_{ZZ}$ and  $\epsilon_X$ are real.
The functions  $\Delta^{\rm SM}_{L,T} (q_1^2, q_2^2)$ denote subleading non-local contributions that are
regular both in the limit  $q_{1,2}^2\to 0$ and in the limit $q_{1,2}^2\to m^2_Z$.
As in the 3-body decay case, this part of the amplitude is largely subdominant and not affected by operators with dimension up to 6, therefore it is fixed it to its SM value.

 \subsection{\texorpdfstring{$h\to 4f$}{h to 4f} charged currents}
 \label{sect:h4fcharged}

Let us consider the $h\to \ell \bar \nu_{\ell}   \bar{\ell'} \nu_{\ell'}$ process.\footnote{The analysis of a process involving quarks is equivalent, with the only difference that the
$\epsilon_{Wf}$ coefficients are in this case non-diagonal matrices in flavour space, as the $g_{ud}^W$ effective couplings.}
Employing the same assumptions used in the neutral current case, we can decompose the amplitude in the following way:
\bqa
&& \cA_{c.c.} \left[ h \to  \ell (p_1) \bar \nu_\ell (p_2)    \nu_{\ell'} (p_3)   \bar \ell^\prime  (p_4)  \right]  = i  \frac{2 m^2_W}{ v_F}
(\bar \ell_L \gamma_\mu \nu_{\ell L} ) (\bar \nu_{\ell' L}  \gamma_\nu \ell^\prime_{L}) \cT^{\mu\nu}_{c.c.} (q_1,q_2) \nonumber \\
&&
\cT^{\mu\nu}_{c.c.} (q_1, q_2) =  \left[ G^{\ell \ell'}_L (q_1^2, q_2^2) g^{\mu\nu} +  G^{\ell \ell'}_T (q_1^2, q_2^2)  \frac{ {q_1}\cd {q_2}~g^{\mu\nu} -{q_2}^\mu {q_1}^\nu }{m^2_W}   \right. \nonumber \\
&& \left. \qquad \qquad\qquad \qquad
 +   G^{\ell \ell'}_{CP} (q_1^2, q_2^2)  \frac{  \vareps^{\mu\nu\rho\sigma} q_{2\rho} q_{1\sigma}   }{m^2_W}  \right]~,
 \label{eq:h4lCharged}
\eqa
where $q_1=p_1 +p_2$ and $q_2=p_3 +p_4$. The  decomposition of the form factors, that allows us to define the
charged-current PO, is~\cite{Gonzalez-Alonso:2014eva}
\bqa
G^{\ell \ell'}_L  (q_1^2, q_2^2) &=& \kappa_{WW}  \frac{ (g_W^{\ell})^*  g_W^{\ell'}  }{P_W(q_1^2) P_W(q_2^2)}
  +  \frac{(\epsilon_{W \ell})^*}{m_W^2}  \frac{ g_W^{\ell'}   }{  P_W(q_2^2)} + \frac{\epsilon_{W \ell'}}{m_W^2}  \frac{ (g_W^{\ell})^*   }{  P_W(q_1^2)} ~,   \\
G^{\ell \ell'}_T (q_1^2, q_2^2) &=&  \epsilon_{WW}  \frac{ (g_W^{\ell})^* g_W^{\ell'}  }{P_W(q_1^2) P_W(q_2^2)} ~, \\
G^{\ell \ell'}_{CP} (q_1^2, q_2^2) &=&  \cpeps_{WW}  \frac{ (g_W^{\ell})^*  g_W^{\ell'}   }{P_W(q_1^2) P_W(q_2^2)} ~,
 \label{eq:h4lChargedEFT}
\eqa
where $P_W(q^2)$ is the $W$ propagator defined analogously to $P_Z(q^2)$ and $g_W^f$ are the effective couplings describing on-shell $W$ decays (we have absorbed a factor of $g$ compared to standard notations).
In the SM,
\be
	(g^{ik}_{W})_{\rm SM} = \frac{g}{\sqrt{2}} V_{ik}~,
	\label{eq:Wcoupl}
\ee
where $V$ is the CKM mixing matrix.\footnote{More precisely,  $(g^{ik}_{W})_{\rm SM} =\frac{g}{\sqrt{2}} V_{ik}$ if $i$ and $k$ refers to left-handed quarks, otherwise $(g^{ik}_{W})_{\rm SM}=0$.}
In absence of rescattering effects,
the Hermiticity of the underlying effective Lagrangian implies that
$\kappa_{WW}$,  $\epsilon_{WW}$ and $\cpeps_{WW}$
 are real couplings,   while $\epsilon_{W \ell}$ can be complex.

 \subsection{\texorpdfstring{$h\to 4f$}{h to 4f} complete decomposition}
 \label{sect:h4ffull}

The complete decomposition of a generic $h\to 4f$ amplitude is obtained combining neutral- and charged-current contributions depending on the
nature of the fermions involved. For instance   $h \to 2 e 2\mu$ and $h \to \ell \bar \ell q \bar q$ decays are determined by a single neutral current amplitude,
while the case of two identical lepton pairs is obtained from Eq.~(\ref{eq:h4l1}) taking into account the proper (anti-)symmetrization of the amplitude:
\bqa
\cA \left[ h \to  \ell (p_1) \bar \ell (p_2) \ell  (p_3) \bar \ell (p_4) \right] &=&     \cA_{n.c.} \left[ h \to  f (p_1) \bar f (p_2) f^\prime  (p_3) \bar f^\prime (p_4) \right]_{f=f^\prime=\ell} \no \\
& - &   \cA_{n.c.} \left[ h \to  f (p_1) \bar f (p_4) f^\prime  (p_3) \bar f^\prime (p_2) \right]_{f=f^\prime=\ell}~. \label{eq:NCinterference}
\eqa
The $h \to  e^\pm  \mu^\mp \nu  \bar \nu $ decays receive contributions from a single charged-current amplitude,
while in the  $h \to  \ell \bar \ell  \nu \bar \nu $ case we have to sum charged and neutral-current contributions:
\bqa
\cA \left[ h \to  \ell (p_1) \bar \ell (p_2) \nu  (p_3) \bar \nu (p_4) \right] &=&     \cA_{n.c.} \left[ h \to  \ell (p_1) \bar \ell (p_2)  \nu  (p_3) \bar \nu (p_4) \right]
\no \\
& - &   \cA_{c.c.} \left[ h \to   \ell (p_1) \bar \nu (p_4)  \nu (p_3)  \bar \ell  (p_2) \right]~.
\eqa

\subsection{{\it Physical} PO for \texorpdfstring{$h\to 4 \ell$}{h to 4l}}

To define the idealized {\it physical} PO we start with the quadratic terms for each of the form factors in Eqs.~(\ref{eq:h4lF1}-\ref{eq:h4lF4}), and  compute  their contribution to the double differential decay rate for $h \rightarrow e^+ e^- \mu^+ \mu^-$ (for $\kappa_{ZZ}$, $\epsilon_{ZZ}$ and $\cpeps_{ZZ}$) and for $h \rightarrow Z \ell^+ \ell^-$ (for the contact terms $\epsilon_{Z \ell}$). It is important to stress that {\it physical} PO as defined here, are extracted from the {\it effective-coupling} PO, and represent a more intuitive presentation of the experimental measurements.

\subsubsection*{Decay channel $h\to e^+ e^- \mu^+ \mu^-$}

We choose this particular decay channel for the (conventional) definition of the {\it physical} PO because it depends on all the PO relevant for $h \rightarrow 4\ell$ and because it does not contain interference between the two fermion currents as in Eq.~\eqref{eq:NCinterference}.
The independent contributions of the three form factors to the decay rate are:
\be
\begin{split}
\frac{d\Gamma^{\rm LL}}{dm_{1}dm_{2}} &=~ \frac{\lambda_{p} \beta_{10}}{2304 \pi^5} \frac{m_{Z}^{4} m_h^3}{v_{F}^2 }m_{1} m_{2} \sum_{f,f'} \left| F_{\rm L}^{ff'} \right|^2~, \\
\frac{d\Gamma^{\rm TT}}{dm_{1}dm_{2}}&=~ \frac{\lambda_{p} \beta_4}{1152 \pi^5} \frac{m_h^3}{v_{F}^2} m_{1}^{3}m_{2}^{3} \sum_{f,f'} \left| F_{\rm T}^{ff'}  \right|^2~, \\
\frac{d\Gamma^{\rm CP}}{dm_1 dm_2}&=~ \frac{\lambda_{p} \beta_2}{1152 \pi^5} \frac{m_h^3}{v_{F}^2} m_1^3 m_2^3 \sum_{f,f'} \left| F_{\rm CP}^{ff'}  \right|^2~,
\label{eq:diff-dis-tens}
\end{split}
\ee
where $f = e_L, e_R$, $f^\prime = \mu_L, \mu_R$, $m_{1(2)} \equiv \sqrt{q_{1(2)}^2}$ and
\be
\lambda_{p}=\sqrt{1+\left(\frac{m_{1}^{2}-m_{2}^{2}}{m_{h}^{2}}\right)^{2}-2\frac{m_{1}^{2}+m_{2}^{2}}{m_{h}^{2}}}~,
\qquad
\beta_{N}=1+\frac{m_{1}^{4} + N m_{1}^2 m_{2}^2 + m_{2}^{4}}{m_{h}^{4}} - 2\frac{m_{1}^{2}+m_{2}^{2}}{m_{h}^{2}}~.
\ee
By integrating over $m_1$ and $m_2$ we obtain the partial decay rate as
\be
\Gamma(h\to 2e 2 \mu) = \Gamma(h\to 2e 2 \mu)^{\rm{SM}} \times \sum_{j \ge i} X_{ij}^{2e2\mu} \kappa_i  \kappa_j~,
\ee
where $X_{ij}^{2e2\mu}$ are the numerical coefficients reported in Ref.~\cite{Gonzalez-Alonso:2015bha}, while $\kappa_i$ are the corresponding {\it effective-coupling} PO. We define the {\it physical PO} as the specific contribution to the partial decay rate. Namely, {\it physical PO} is the partial decay rate due to a single PO, while setting other PO to zero. In particular,
\bqa
	\!\!\!\!\!\!\Gamma(h \rightarrow 2e2\mu)[\kappa_{ZZ}] & = & 4.929 \times 10^{-2} (|g_{Z e_L}|^2 + |g_{Z e_R}|^2)(|g_{Z \mu_L}|^2 + |g_{Z \mu_R}|^2) ~ |\kappa_{ZZ}|^2 ~ {\rm MeV} \no \\
	\Gamma(h \rightarrow 2e2\mu)[\epsilon_{ZZ}] &= & 4.458 \times 10^{-3} (|g_{Z e_L}|^2 + |g_{Z e_R}|^2)(|g_{Z \mu_L}|^2 + |g_{Z \mu_R}|^2) ~ |\epsilon_{ZZ}|^2 ~ {\rm MeV} \no \\
	\Gamma(h \rightarrow 2e2\mu)[\epsilon_{ZZ}^{\rm CP}] & = & 1.884 \times 10^{-3} (|g_{Z e_L}|^2 + |g_{Z e_R}|^2)(|g_{Z \mu_L}|^2 + |g_{Z \mu_R}|^2) ~ |\epsilon_{ZZ}^{\rm CP}|^2 ~ {\rm MeV}  \no \\
	\label{eq:h4lpart}
\eqa
The numerical coefficients in Eq.~(\ref{eq:h4lpart}) have been obtained neglecting QED corrections. The latter must be included
at the simulation level by appropriate QED showering programs, such as PHOTOS~\cite{Davidson:2010ew}.
As shown in Ref.~\cite{Bordone:2015nqa}: the impact of such corrections is negligible after integrating over the full phase space,
hence in the overall normalization of the partial rates in Eq.~(\ref{eq:h4lpart}),
while they can provide sizeable distortions of the spectra
in specific phase-space regions.

Since each effective coupling PO correspond to a well-defined pole contribution to the amplitude (with one or two poles of the $Z$ boson),
and a well-defined Lorentz and flavour structure, we can associate to those contributions to partial rates
an intuitive physical meaning.
In particular, we define the following \emph{physical PO}  for the $h\to 4\ell$ decays:
\be\begin{split}
	\Gamma(h \rightarrow Z_L Z_L) &\equiv~ \frac{\Gamma(h \rightarrow 2e2\mu)[\kappa_{ZZ}]}{\mathcal{B}(Z \rightarrow 2e)\mathcal{B}(Z \rightarrow 2\mu)} = 0.209 ~ |\kappa_{ZZ}|^2 ~ {\rm MeV}  \\
	\Gamma(h \rightarrow Z_T Z_T) &\equiv~ \frac{\Gamma(h \rightarrow 2e2\mu)[\epsilon_{ZZ}]}{\mathcal{B}(Z \rightarrow 2e)\mathcal{B}(Z \rightarrow 2\mu)} = 0.0189 ~ |\epsilon_{ZZ}|^2 ~ {\rm MeV} \\
	\Gamma^{\rm CPV}(h \rightarrow Z_T Z_T) &\equiv~ \frac{\Gamma(h \rightarrow 2e2\mu)[\cpeps_{ZZ}]}{\mathcal{B}(Z \rightarrow 2e)\mathcal{B}(Z \rightarrow 2\mu)} = 0.00799 ~ |\cpeps_{ZZ}|^2  ~ {\rm MeV} \\
	\label{eq:physPOh4l}
\end{split}
\ee
where, due to the double pole structure of the amplitude, we have removed the (physical) branching ratios of the $Z \rightarrow e^+ e^-$ and $Z \rightarrow \mu^+ \mu^-$ decays.
Here
\be
	\mathcal{B}(Z \rightarrow 2\ell) = \frac{\Gamma_0}{\Gamma_Z} R^\ell  \left( (g_{Z}^{ \ell_L})^2 + (g_{Z}^{ \ell_R})^2 \right) \approx 0.4856 \left( (g_{Z}^{ \ell_L})^2 + (g_{Z}^{ \ell_R})^2 \right) ~,
	\label{eq:ZBR2l}
\ee
where $\Gamma_0 = \frac{m_Z}{24 \pi}$, $\Gamma_Z$ is the total decay width and $R^\ell = \left( 1 + \frac{3}{4 \pi} \alpha(m_Z) \right)$ describes final state QED radiation. The relative uncertainty in the numerical coefficients in Eqs.~(\ref{eq:physPOh4l},\ref{eq:ZBR2l}), due to the experimental error on $\Gamma_Z$ and $m_Z$ \cite{Agashe:2014kda}, is at the $\sim 10^{-3}$ level, thus negligible even given the expected long-term sensitivity in these Higgs measurements.

\subsubsection*{Decay channel $h\to Z \ell^+ \ell^-$}

The idealized {\it physical} PO related to the contact terms can be defined directly from the on-shell decay $h \rightarrow Z \ell^+ \ell^-$, where $\ell = e_L, e_R, \mu_L, \mu_R$
and the $Z$ boson is assumed to be on-shell  (narrow width approximation). We compute this decay rate, neglecting QED corrections and light lepton masses, in presence of the contact terms $\epsilon_{Z \ell}$  only.
The Dalitz double differential rate in $s_{12} \equiv (p_{\ell^+} + p_{\ell^-})^2$ and $s_{23} \equiv (p_{\ell^-} + p_Z)^2$ is
\be
	\frac{d \Gamma}{d s_{12} d s_{23}} = \frac{1}{(2\pi)^3} \frac{1}{32 m_h^2} \frac{4 |\epsilon_{Z \ell}|^2}{v^2} \left( s_{12} + \frac{(s_{23} - m_Z^2)(m_h^2 - s_{12} - s_{23})}{m_Z^2} \right)~,
\ee
The allowed kinematical region is $0 < s_{12} < (m_h - m_Z)^2$ and, for any given value of $s_{12}$, $s_{23}^{\rm min} < s_{23} < s_{23}^{\rm Max}$ with
\be
	s_{23}^{\rm min (Max)} = (E_2^* + E_Z^*)^2 - \left(E_2^* \pm \sqrt{(E_Z^*)^2 - m_Z^2}\right)^2~,
\ee
where $E_2^* = \sqrt{s_{12}}/2$ and $E_Z^* = \frac{m_h^2 - s_{12} - m_Z^2}{2 \sqrt{s_{12}} }$. The decay rate defines the relation between the {\it physical} PO and the effective couplings PO as:
\be
	\Gamma(h \rightarrow Z \ell^+ \ell^-) = 0.0366 |\epsilon_{Z \ell}|^2 ~ {\rm MeV}~.
	\label{eq:physPOhZll}
\ee
As before, the {\it physical} PO is defined as the contribution to partial decay rate from the relevant PO, assuming others are set to zero. The only inputs in the numerical coefficient in Eq.~\eqref{eq:physPOhZll} are the $Z$ and Higgs boson masses, as well as the Higgs vev, therefore the relative uncertainty is at the $10^{-3}$ level.
Together with the {\it physical} PO already defined for  $h\rightarrow \gamma\gamma$ and $h \rightarrow Z \gamma$, we have thus
established a complete mapping between the effective couplings PO and the {\it physical} PO appearing in $h\to 4\ell$ decays.

\subsection{{\it Physical} PO for \texorpdfstring{$h\to 2 \ell 2\nu$}{h to 2l2nu}}

{\it Physical} PO for charged-current processes can be defined in a very similar way as the neutral-current ones. In particular, we use the $h \to e^+ \nu_e \mu^- \bar{\nu}_\mu$ process for the {\it physical} PO corresponding to $k_{WW}$, $\epsilon_{WW}$, and $\cpeps_{WW}$, and $h \to W^+ \ell \bar{\nu}_\ell$ for the contact terms.

\subsubsection*{Decay channel $h\to e^+ \nu_e \mu^- \bar{\nu}_\mu$}

Integrating the differential distributions analogous to Eq.~\eqref{eq:diff-dis-tens} we obtain the expression of the partial decay rate in this channel, in the limit where only one PO is turned on:
\bqa
	\!\!\!\!\!\!\Gamma(h \rightarrow e\mu2\nu)[\kappa_{WW}] & = & 2.20 \times 10^{-4} |g_{W e_L}|^2 |g_{W \mu_L}|^2 ~ |\kappa_{WW}|^2 ~ {\rm MeV} \no \\
	\Gamma(h \rightarrow e\mu2\nu)[\epsilon_{WW}] &= & 4.27 \times 10^{-5} |g_{W e_L}|^2 |g_{W \mu_L}|^2 ~ |\epsilon_{WW}|^2 ~ {\rm MeV} \no \\
	\Gamma(h \rightarrow e\mu2\nu)[\epsilon_{WW}^{\rm CP}] & = & 1.77 \times 10^{-5} |g_{W e_L}|^2 |g_{W \mu_L}|^2 ~ |\epsilon_{WW}^{\rm CP}|^2 ~ {\rm MeV}  \no \\
	\label{eq:h2l2vpart}
\eqa
As in the neutral channel, the \emph{physical PO} are defined from these quantities by factorizing the $W$ branching ratios:
\be\begin{split}
	\Gamma(h \rightarrow W_L W_L) &\equiv~ \frac{\Gamma(h \rightarrow e\mu2\nu)[\kappa_{WW}]}{\mathcal{B}(W \rightarrow e \bar{\nu}_e)\mathcal{B}(W \rightarrow \mu \bar{\nu}_\mu)} = (0.841 \pm 0.016 )~ |\kappa_{WW}|^2 ~ {\rm MeV}  \\
	\Gamma(h \rightarrow W_T W_T) &\equiv~ \frac{\Gamma(h \rightarrow e\mu2\nu)[\epsilon_{WW}]}{\mathcal{B}(W \rightarrow e \bar{\nu}_e)\mathcal{B}(W \rightarrow \mu \bar{\nu}_\mu)} = (0.1634 \pm 0.0030)~ |\epsilon_{WW}|^2 ~ {\rm MeV} \\
	\Gamma^{\rm CPV}(h \rightarrow W_T W_T) &\equiv~ \frac{\Gamma(h \rightarrow e\mu2\nu)[\cpeps_{WW}]}{\mathcal{B}(W \rightarrow e \bar{\nu}_e)\mathcal{B}(W \rightarrow \mu \bar{\nu}_\mu)} = (0.0677 \pm 0.0012) ~ |\cpeps_{WW}|^2  ~ {\rm MeV}~,
\end{split}
\ee
where the uncertainty has been obtained from the experimental error on $\Gamma_W$ \cite{Agashe:2014kda}
that, as recently pointed out in \cite{Murray:2016czw}, it has a non-neglible impact in the prediction of
$h\to 2 \ell 2\nu$ branching ratios.
The $W$ branching ratios are given by
\be
	\mathcal{B}(W \rightarrow \ell \bar{\nu}_\ell) = \frac{\Gamma_0}{\Gamma_W} (g_{W \ell_L})^2 \approx 0.511 (g_{W \ell_L})^2 ~,
\ee
where $\Gamma_0 = \frac{m_W}{24 \pi}$, $\Gamma_W$ is the total decay width.

\subsubsection*{Decay channel $h \to W^+ \ell \bar{\nu}_\ell$}

Also in this case the {\it physical} PO corresponding to the charged-current contact terms are defined in complete analogy to the neutral-current case, starting from the 3-body decay $h \to W^+ \ell \bar{\nu}_\ell$.
The partial decay width computed in the limit where only the contact term PO is switched on defines the relation between the {\it physical} PO and the effective couplings PO as:
\be
	\Gamma(h \rightarrow W^+ \ell \bar{\nu}_\ell) = 0.143  |\epsilon_{W \ell}|^2 ~ {\rm MeV}~,
\ee
where the relative uncertainty in the coefficient due to the experimental error on $m_W$ \cite{Agashe:2014kda} is below $\sim 2 \times 10^{-3}$.

\begin{table}[t]
\caption{\label{Tab:physicalPO}
Summary of the \emph{effective coupling} PO and the corresponding \emph{physical} PO. The parameter $N_c^f$ is 1 for leptons and 3 for quarks. In the case of the charged-current contact term, $f'$ is the $SU(2)_L$ partner of the fermion $f$.
See the main text for a discussion about the errors on the numerical coefficient in the table and the reference values of $\{\kappa_X, \delta_X, \epsilon_X\}$ within the SM.}
\begin{center}
\begin{tabular}{c| r l} 
\toprule
 PO  & {\it Physical} PO  & \qquad Relation to the eff. coupl. \\ 
 \midrule
 \raisebox{-2pt}[0pt][10pt]{ $\kappa_{f},~\delta^{\rm CP}_{f}$ } & \raisebox{-2pt}[0pt][10pt]{  $\Gamma(h \rightarrow f \bar{f})$ }
   &    \raisebox{-2pt}[0pt][10pt]{ $\!\! =~\Gamma(h \rightarrow f \bar{f})^{\rm (SM)} [(\kappa_{f} )^2 + (\delta^{\rm CP}_{f} )^2]$ } \\
 \raisebox{0pt}[0pt][8pt]{ $\kappa_{\gamma\gamma}, ~ \delta^{\rm CP}_{\gamma\gamma}$ }
 &  $\Gamma(h \rightarrow \gamma\gamma)$  & $=~\Gamma(h \rightarrow \gamma\gamma)^{\rm (SM)} [(\kappa_{\gamma\gamma} )^2 + (\delta^{\rm CP}_{\gamma\gamma} )^2]$  \\
 \raisebox{0pt}[0pt][8pt]{ $\kappa_{Z\gamma},~ \delta^{\rm CP}_{Z\gamma}$  }& $\Gamma(h \rightarrow Z\gamma)$ & $=~ \Gamma(h \rightarrow Z\gamma)^{\rm (SM)} [(\kappa_{Z\gamma} )^2 + (\delta^{\rm CP}_{Z\gamma} )^2]$ \\
 \raisebox{0pt}[0pt][8pt]{ $\kappa_{ZZ}$ } & $\Gamma(h \rightarrow Z_L Z_L)$ & $=~(0.209~\UMeV) \times |\kappa_{ZZ} |^2$ \\
 \raisebox{0pt}[0pt][8pt]{ $\epsilon_{ZZ}$ } & $\Gamma(h \rightarrow Z_T Z_T)$ & $=~ (1.9 \times 10^{-2}~\UMeV) \times  |\epsilon_{ZZ} |^2$ \\
 \raisebox{0pt}[0pt][8pt]{ $\cpeps_{ZZ}$ } & $\Gamma^{\rm CPV}(h \rightarrow Z_T Z_T)$ & $=~(8.0\times 10^{-3}~\UMeV)\times |\cpeps_{ZZ} |^2$ \\
 \raisebox{0pt}[0pt][8pt]{ $\epsilon_{Zf}$  }& $\Gamma(h \rightarrow Z f \bar{f})$ & $=~ (3.7\times 10^{-2}~\UMeV)\times N_c^f ~ | \epsilon_{Zf} |^2$ \\
 \raisebox{0pt}[0pt][8pt]{  $\kappa_{WW}$ } & $\Gamma(h \rightarrow W_L W_L)$ & $=~ (0.84~\UMeV)\times  |\kappa_{WW} |^2$ \\
 \raisebox{0pt}[0pt][8pt]{  $\epsilon_{WW}$ } & $\Gamma(h \rightarrow W_T W_T)$ & $=~ (0.16~\UMeV)\times   |\epsilon_{WW} |^2$ \\
 \raisebox{0pt}[0pt][8pt]{  $\cpeps_{WW}$ } & $\Gamma^{\rm CPV}(h \rightarrow W_T W_T)$ & $=~ (6.8 \times 10^{-2}~\UMeV) \times  |\cpeps_{WW} |^2$ \\
 \raisebox{0pt}[0pt][8pt]{  $\epsilon_{Wf}$ } & $\Gamma(h \rightarrow W f \bar{f'})$ & $=~ (0.14~\UMeV)\times  N_c^f ~ | \epsilon_{Wf} |^2$ \\ 
\midrule
 \raisebox{0pt}[0pt][8pt]{  $\kappa_{g}$ } & $\sigma(pp \to h)_{gg-\rm{fusion}}$ & $=~ \sigma(pp \to h)^{\rm SM}_{gg-\rm{fusion}} \kappa^2_{g}$ \\
 \raisebox{0pt}[0pt][8pt]{  $\kappa_{t}$ } & $\sigma(pp \to t \bar t h)_{\rm Yukawa}$ & $=~ \sigma(pp \to t \bar t  h)^{\rm SM}_{\rm Yukawa} \kappa^2_{t}$ \\
 \raisebox{0pt}[0pt][8pt]{  $\kappa_{H}$ } & $\Gamma_{\rm tot} (h) $ & $=~ \Gamma_{\rm tot}^{\rm SM} (h) \kappa^2_{H}$ \\ 
\bottomrule
 \end{tabular}
\end{center}
\end{table}

\section{PO in Higgs electroweak production: generalities}
\label{sec:PO-EW}

The PO decomposition of $h\to 4f$ amplitude discussed above can naturally be
generalized to describe
electroweak Higgs-production processes, namely Higgs-production via
vector-boson fusion (VBF) and  Higgs-production in association with a massive SM gauge boson
(VH).

The interest of such production processes is twofold.
On the one hand, they are closely connected to the $h\to4\ell,2\ell2\nu$ decay processes
by crossing symmetry, and by the exchange of lepton currents into quark currents.
As a result, some of the Higgs PO necessary to describe the $h\to4\ell,2\ell2\nu$ decay kinematics
appear also in the description of the VBF and VH cross sections (independently of the Higgs boson decay mode).
This facts opens the possibility of combined analyses of  production cross sections
and differential decay distributions, with a significant reduction on the experimental error on the extraction of the PO.
On the other hand, the production cross sections allow to explore different  kinematical regimes compared to the
decays. By construction, the momentum transfer appearing in the Higgs boson decay amplitudes is limited by the  Higgs boson mass, while such limitation is not present in the production amplitudes.
The higher energies probed in the production processes provide an increased sensitivity to new physics effects.
This fact also allows to test the momentum expansion that is intrinsic in the PO decomposition, as well as in any effective field theory approach to physics beyond the SM.

Despite the similarities at the fundamental level,
the phenomenological description of VBF and VH in terms of PO is significantly more challenging compared to that of Higgs boson decays.
On the one hand, QCD corrections plays a non-negligible role in the production processes.
Although technically challenging,  this fact does not represent a conceptual problem for the PO approach: the leading QCD corrections
factorize in VBF and VH, similarly to the factorization of QED corrections in $h\to 4\ell$.  This implies that NLO QCD corrections can
be incorporated in general terms with suitable modifications of the existing Montecarlo tools.
On the other hand, the relation between the kinematical variables at the basis of the PO decomposition (i.e.~the momentum transfer
of the partonic currents, $q^2$) and the kinematical variables accessible in $pp$ collisions is not straightforward,
especially in the VBF case. This problem finds a natural solution in the VBF case due to strong correlation
between $q^2$ and the $p_T$ of the VBF tagged jets, while in the VH case invariant mass of the VH system is correlated to the vector $p_T$.

\subsection{Amplitude decomposition}
\label{sect:ampdec}

Neglecting the light fermion masses, the electroweak production processes VH and VBF or,
more precisely, the electroweak partonic amplitudes $f_1  f_2  \to h +  f_3  f_4$,
can be completely described by the three-point correlation function of the Higgs boson and two (colour-less) fermion currents
\be
	\langle 0 | \cT\left\{ J_f^\mu(x), J_{f^\prime}^\nu (y), h(0) \right\}| 0\rangle ~,
	\label{eq:corr_func}
\ee
where all the states involved are on-shell. The same correlation function controls also the four-fermion Higgs boson decays discussed above.
In the  $h\to4\ell,2\ell2\nu$ case both currents are leptonic and all fermions are in the final state. In case of VH associate production one of the currents describes the initial state quarks, while the other describes the decay products of the (nearly on-shell) vector boson. Finally, in VBF production the currents are not in the $s$-channel as in the previous cases, but in the $t$-channel.
Strictly speaking, in VH and VBF  the quark states  are not on-shell; however,
their off-shellness can be neglected compared to the
electroweak scale   characterizing the process (both within and beyond the SM).

As in the $h\to 4f$ case, we can expand the correlation function in Eq.~\eqref{eq:corr_func}
around the known physical poles due to the propagation of intermediate SM electroweak gauge bosons. The PO are then
defined by the residues on the poles and by the non-resonant terms in this expansion. By construction, terms
corresponding to a double pole  structure are independent from the nature of the fermion current involved. As a result, the corresponding PO are universal and can be extracted from any of the above mention processes, both in production and in decays~\cite{Greljo:2015sla}.

\subsubsection{Vector boson fusion Higgs boson production}
\label{sec:VBFdec}

Higgs boson production via vector boson fusion (VBF) receives contribution both from neutral- and charged-current channels. Also, depending on the specific partonic process, there could be two different ways to construct the two currents, and these two terms interfere with each other. For example, for $u u \to u u h$ one has the interference between two neutral-current processes, while in $u d \to u d h$ the interference is between neutral and charged currents. In this case it is clear that one should sum the two amplitudes with the proper symmetrization, as done in the case of $h\to4e$.

We now proceed describing how each of these amplitudes can be parameterized in terms of PO.
Let us start with the neutral-current one.
The amplitude for the on-shell process $q_i(p_1) q_j(p_2) \to q_i(p_3) q_j(p_4) h(k)$ can be parameterized by
\beq
	\cA_{n.c} (q_i(p_1) q_j(p_2) \to q_i(p_3) q_j(p_4) h(k) ) = i \frac{2m_Z^2}{v} \bar{q_i}(p_3) \gamma_\mu q_i(p_1) \bar{q_j}(p_4) \gamma_\nu q_j(p_2) \cT^{\mu\nu}_{n.c.}(q_1, q_2) ,
	\label{eq:VBFnc}
\eeq
where $q_1 = p_1 - p_3$, $q_2 = p_2 - p_4$ and $\cT^{\mu\nu}_{n.c.}(q_1, q_2)$ is the same tensor structure appearing in $h \to 4f$ decays.
Indeed, proceeding as in  Eq.~(\ref{eq:h4l1}), using Lorentz invariance we decompose this tensor structure in term of three from factors:
\bqa
\cT^{\mu\nu}_{n.c.} (q_1, q_2) &=&  \left[ F^{q_i q_j}_L (q_1^2, q_2^2) g^{\mu\nu} +  F^{q_i q_j}_T (q_1^2, q_2^2)  \frac{ {q_1}\cd {q_2}~g^{\mu\nu} -{q_2}^\mu {q_1}^\nu }{m_Z^2}   \right. \nonumber \\
&& \left.
 +   F^{q_i q_j}_{CP} (q_1^2, q_2^2)  \frac{  \vareps^{\mu\nu\rho\sigma} q_{2\rho} q_{1\sigma}   }{m_Z^2}  \right]~.
\eqa
Similarly, the charged-current contribution to the amplitude for the on-shell process $u_i(p_1) d_j(p_2) \to d_k(p_3) u_l(p_4) h(k)$ can be parameterized by
\beq
	\cA_{c.c} (u_i(p_1) d_j(p_2) \to d_k(p_3) u_l(p_4) h(k) ) = i \frac{2m_W^2}{v} \bar{d_k}(p_3) \gamma_\mu u_i(p_1) \bar{u_l}(p_4) \gamma_\nu d_j(p_2) \cT^{\mu\nu}_{c.c.}(q_1, q_2) ,
	\label{eq:VBFcc}
\eeq
where, again, $\cT^{\mu\nu}_{c.c.}(q_1, q_2)$ is the same tensor structure
appearing in the charged-current $h \to 4f$ decays:
\bqa
	\cT_{c.c.}^{\mu\nu} (q_1, q_2) &=&  \left[ G^{ijkl}_L (q_1^2, q_2^2) g^{\mu\nu} +  G^{ijkl}_T (q_1^2, q_2^2)  \frac{ {q_1}\cd {q_2}~g^{\mu\nu} -{q_2}^\mu {q_1}^\nu }{m^2_W}    \right. \nonumber \\
&& \left.
 +   G^{ijkl}_{CP} (q_1^2, q_2^2)  \frac{  \vareps^{\mu\nu\rho\sigma} q_{2\rho} q_{1\sigma}   }{m^2_W}  \right]
\eqa
The amplitudes for the processes with initial anti-quarks can easily be obtained from the above ones.

The next step is to perform a momentum expansion of the form factors around the physical poles due to the propagation of SM electroweak gauge bosons ($\gamma$, $Z$ and $W^{\pm}$), and to define the PO (i.e.~the set $\{\kappa_i, \epsilon_i\}$) from the residues of such poles.
We stop this expansion neglecting terms which can be generated only by local operators with dimension higher than six.
A discussion about limitations and consistency checks of this procedure will be presented later on.
The decomposition of the form factors closely follows the procedure already introduced for the decay amplitudes
and will not be repeated here.
We report explicitly only expression of the longitudinal form factors,
where the contact terms not accessible in the leptonic decays appear:
\be
\begin{split}
F^{q_i q_j}_L (q_1^2, q_2^2) &= \kappa_{ZZ}  \frac{ g_Z^{q_i}  g_Z^{q_j}  }{P_Z(q_1^2) P_Z(q_2^2)}
  +  \frac{\epsilon_{Z q_i}}{m_Z^2}  \frac{ g_Z^{q_j}   }{  P_Z(q_2^2)} + \frac{\epsilon_{Z q_j}}{m_Z^2}   \frac{ g_Z^{q_i}   }{  P_Z(q_1^2)}  + \Delta^{\rm SM}_{L, n.c.} (q_1^2, q_2^2) ~, \\
G^{ijkl}_L (q_1^2, q_2^2) &= \kappa_{WW} \frac{ g_W^{ik}  g_W^{jl}  }{P_W(q_1^2) P_W(q_2^2)}
  +  \frac{\epsilon_{W ik}}{m_W^2}  \frac{ g_W^{jl}   }{  P_W(q_2^2)} + \frac{\epsilon_{W jl}}{m_W^2}   \frac{ g_W^{ik}   }{  P_W(q_1^2)}  + \Delta^{\rm SM}_{L, c.c} (q_1^2, q_2^2) ~.
\end{split}
\label{eq:FLGL}
\ee
Here  $P_V(q^2) = q^2 - m_V^2 + i m_V \Gamma_V$, while $g_Z^f$ and $g^{ik}_{W}$ are the PO characterizing the on-shell couplings of $Z$ and $W$ boson to a pair of fermions, see Eqs.~\eqref{eq:LEPZcoupl} and \eqref{eq:Wcoupl}.
The functions  $\Delta^{\rm SM}_{L, n.c. (c.c.)} (q_1^2, q_2^2)$ denote non-local contributions generated
at the one-loop level (and encoding multi-particle cuts) that cannot be re-absorbed in the definition of
$\kappa_i$ and $\epsilon_i$. At the level of precision we are working, taking into account also
the high-luminosity phase of the LHC, these contributions can be safely fixed to their SM values.

As anticipated, the crossing symmetry between $h\to 4 f$ and $2 f \to h \, 2 f$ amplitudes ensures that the
PO are the same in production and decay (if the same fermions species are involved). The amplitudes are
explored in different kinematical regimes in the two type of processes
(in particular the momentum-transfers, $q_{1,2}^2$, are space-like in VBF and time-like in $h\to 4f$).
However, this does not affect the definition of the PO.
This implies that  the  fermion-independent PO associated to a double pole structure,
such as $\kappa_{ZZ}$ and $\kappa_{WW}$ in Eq.~(\ref{eq:FLGL}),
are expected to be measured with higher accuracy in  $h\to 4 \ell$ and $h\to 2\ell 2\nu$
rather than in VBF. On the contrary, VBF is particularly useful to constrain
the fermion-dependent contact terms $\epsilon_{Z q_i}$ and $\epsilon_{Wu_i d_j}$,
that appear only in the longitudinal form factors.

\subsubsection{Associated vector boson plus Higgs boson production}
\label{sec:VHprod_ff}

The VH production process denote the production of a Higgs boson with a nearly
on-shell massive vector boson ($W$ or $Z$).  For simplicity, in the following we will assume
that the vector boson is on-shell and that the interference with
the VBF amplitude can be neglected. However, we stress that the PO formalism clearly allow to describe
both these effects (off-shell $V$ and interference with VBF in case of $V\to \bar q q$ decay) simply
applying the general decomposition of  neutral-  and charged-current
amplitudes as outlined above.

Similarly to VBF, Lorentz invariance allows us to decompose the amplitudes for the on-shell processes $q_i(p_1) \bar{q}_i(p_2) \to h(p) Z(k)$ and $u_i(p_1) \bar{d}_j(p_2) \to h(p) W^+(k)$ in three possible tensor structures: a longitudinal one, a transverse one, and a CP-odd one,
\beq\begin{split}
	\cA &(q_i(p_1) \bar{q}_i(p_2) \to h(p) Z(k)) = i \frac{2m_Z^2}{v} \bar{q_i}(p_2) \gamma_\nu q_i(p_1) \epsilon_\mu^{Z*}(k) \times \\
	&\quad \times \left[ F_{L}^{q_i Z}(q^2) g^{\mu\nu} + F_{T}^{q_i Z}(q^2) \frac{- (q\cdot k) g^{\mu\nu} + q^\mu k^\nu}{m_Z^2} + F_{CP}^{q_i Z}(q^2) \frac{ \epsilon^{\mu\nu\alpha\beta} q_\alpha k_\beta}{m_Z^2} \right]~,
	\label{eq:AmplqqZh}
\end{split}\eeq
\beq\begin{split}
	\cA & (u_i(p_1) \bar{d}_j(p_2) \to h(p) W^+(k)) = i \frac{2m_W^2}{v} \bar{d_j}(p_2) \gamma_\nu u_i(p_1) \epsilon_\mu^{W*}(k) \times \\
	&\quad \times \left[ G_{L}^{q_{ij} W}(q^2) g^{\mu\nu} +  G_{T}^{q_{ij} W}(q^2) \frac{- (q\cdot k) g^{\mu\nu} + q^\mu k^\nu}{m_W^2} + G_{CP}^{q_{ij} W}(q^2) \frac{ \epsilon^{\mu\nu\alpha\beta} q_\alpha k_\beta}{m_W^2} \right]~,
	\label{eq:AmplqqWh}
\end{split}\eeq
where $q = p_1 + p_2 = k + p$.  In the limit where we neglect the off-shellness of the final-state $V$,
 the form factors depend only on $q^2$. Already from this decomposition of the amplitude it is clear
the importance of providing measurements of the differential cross-section as a function of $q^2$,
as well as differential measurements in terms of the  angular variables that allow to disentangle the different tensor structures.

Performing the momentum expansion of the form factors around the physical poles, and defining
the PO as in Higgs boson decays and VBF, we find
\be\begin{array}{rcl rcl}
	F_{L}^{q_i Z}(q^2)	&=& \kappa_{ZZ} \frac{g_{Zq_i}}{P_Z(q^2)} + \frac{\epsilon_{Zq_i}}{m_Z^2}
\qquad&	G_{L}^{q_{ij} W}(q^2)	&=& \kappa_{WW} \frac{(g^{u_i d_j}_{W})^*}{P_W(q^2)} + \frac{\epsilon_{Wu_i d_j}^*}{m_W^2} \\
	F_{T}^{q_i Z}(q^2)	&=& \epsilon_{ZZ} \frac{g_{Zq_i}}{P_Z(q^2)} + \epsilon_{Z\gamma} \frac{e Q_q}{q^2}
\qquad&	G_{T}^{q_{ij} W}(q^2)	&=& \epsilon_{WW} \frac{(g^{u_i d_j}_{W})^*}{P_W(q^2)} \\
	F_{CP}^{q_i Z}(q^2)	&=& \epsilon^{\rm CP}_{ZZ} \frac{g_{Zq_i}}{P_Z(q^2)} - \epsilon^{\rm CP}_{Z\gamma} \frac{e Q_q}{q^2}
\qquad&	G_{CP}^{q_{ij} W}(q^2) &=& \epsilon^{\rm CP}_{WW} \frac{(g^{u_i d_j}_{W})^*}{P_W(q^2)}
\end{array}
	\label{eq:FFVh}
\ee
where we have omitted the indication of the (tiny) non-local terms, fixed to their corresponding SM values.
As in the VBF case, only the longitudinal form factors $F_L$ and $G_L$ contain PO not accessible in  the leptonic decays,
namely the  quark contact terms $\epsilon_{Zq_i}$ and $\epsilon_{Wu_i d_j}$.


\section{PO in Higgs electroweak production: phenomenology}
\label{sec:PO-pheno}

\subsection{Vector Boson Fusion}

\label{sect:VBF}

At the parton level (i.e.~in the $q q \to h q q$  hard scattering) the ideal observable  relevant to extract the momentum dependence of the
factor factors would be the double differential cross section $d^2 \sigma/ d q_1^2 d q_2^2$, where $q_1=p_1-p_3$ and $q_2=p_2-p_4$ are the momenta of the two fermion currents entering the process (here $p_1$, $p_2$ ($p_3$, $p_4$) are the momenta of the initial (final) state quarks).
The $q^2_{i}$ are also the key variables to test and control the momentum expansion at the basis of the PO decomposition.

A first nontrivial task is to choose the proper pairing of the incoming and outgoing quarks, given we are experimentally blind to their flavour.
For partonic processes receiving two interfering contributions when the final-state quarks are exchanged, such as $uu \to h u u$ or $ud \to h u d$, the definition of $q_{1,2}$ is even less transparent since a univocal pairing of the momenta can not be assigned, in general, even if one knew the flavour of all partons.
This problem can be simply overcome at a practical level by making use of the VBF kinematics, in particular the fact that the two jets are always very forward. This implies one can always pair the momenta of the jet going, for example, on the $+z$ direction with the initial parton going in the same direction, and vice versa. The same argument can be used to argue that the interference between different amplitudes (e.g.~neutral current and charged current) is negligible in VBF. In order to check this,
we have performed a leading order parton level simulation of the VBF Higgs boson production ($p p \to h j j$) using \MadGraph~\cite{Alwall:2014hca} (version 2.2.3) at $13$~TeV c.m. energy. We have imposed the basic set of cuts,
\begin{align}
\label{eq:vbf_cuts}
p_{\mathrm{T},\mathrm{j}_{1,2}} >30~\UGeV,\quad  |\eta_{\mathrm{j}_{1,2}}| <4.5, \quad \mathrm{and}  \quad m_{\mathrm{j}_1\mathrm{j}_2}>500~\UGeV.
 \end{align}
   In \refF{fig:colorVSkinemtics}, we show the distribution in the opening angle of the incoming and outgoing quark momenta for the two different pairings. The left plot is for the SM, while the right plot is for a specific NP benchmark point. Shown in blue is the pairing based on the leading colour connection using the colour flow variable while in red is the opposite pairing. The plot shows that the momenta of the colour connected quarks tend to form a small opening angle and the overlap between the two curves, i.e.~where the interference effects might be sizeable, is negligible. This implies that in the experimental analysis the pairing should be done based on this variable.  Importantly, the same conclusions can be drawn in the presence of new physics contributions to the contact terms.

There is a potential caveat to the above argument: the colour flow approximation ignores the interference terms that are higher order in $1/N_C$, where $N_C$ is number of colours.
Let us consider a process with two interfering amplitudes with the final state quarks exchanged, for example in $u u \to u u h$. The differential cross section receives three contributions proportional to  
$$
|F^{f f^\prime}_L (t_{13},t_{24})|^2, \quad 
|F^{f f^\prime}_L (t_{13},t_{24}) F^{f f^\prime}_L (t_{14},t_{23})|~,  \quad
{\rm and}\quad |F^{f f^\prime}_L (t_{14},t_{23})|^2~, 
$$
where $t_{ij}=(p_i-p_j)^2=-2E_i E_j (1-\cos \theta_{i j})$. 
For the validity of the momentum expansion it is important that the momentum transfers ($t_{i j}$) remain smaller than the hypothesized scale of new physics. On the other hand, imposing the VBF cuts, the interference terms turns out to depend on one small and one large momentum transfer. However, thanks to the pole structure of the form factors, these interference effects
turns out to give a very small contribution.
Therefore, we can safely state that  the momentum transfers marked with the leading colour flow are reliable control variables of the momentum expansion validity.
\begin{figure}
  \begin{center}
    \includegraphics[width=0.45\textwidth]{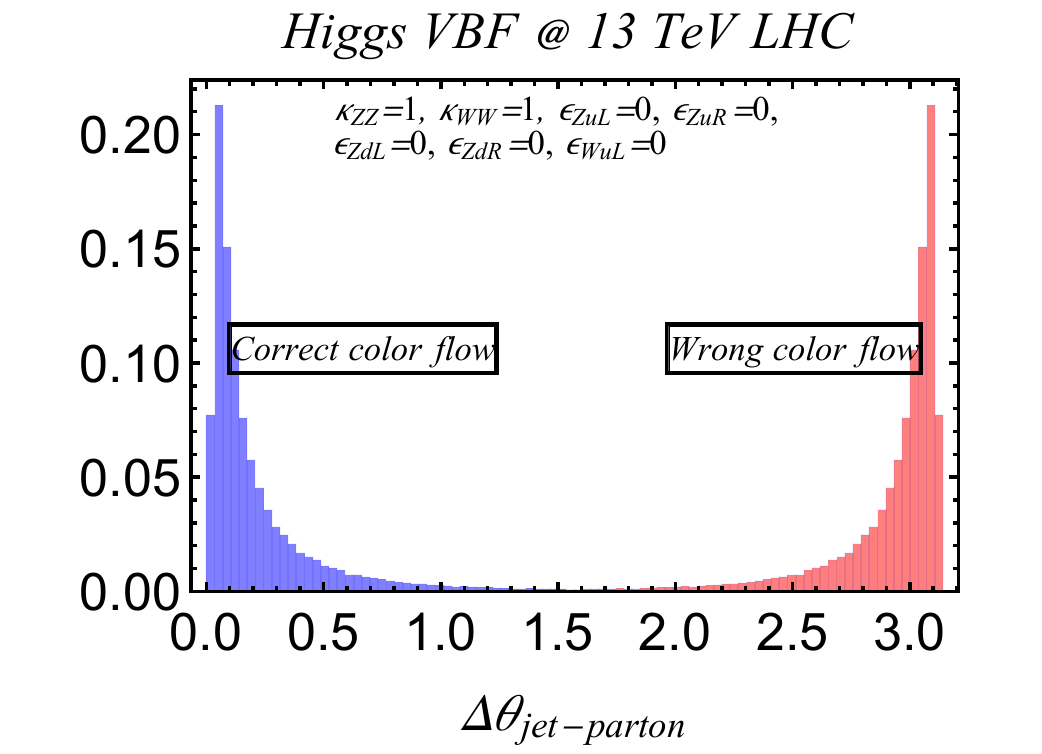}
        \includegraphics[width=0.45\textwidth]{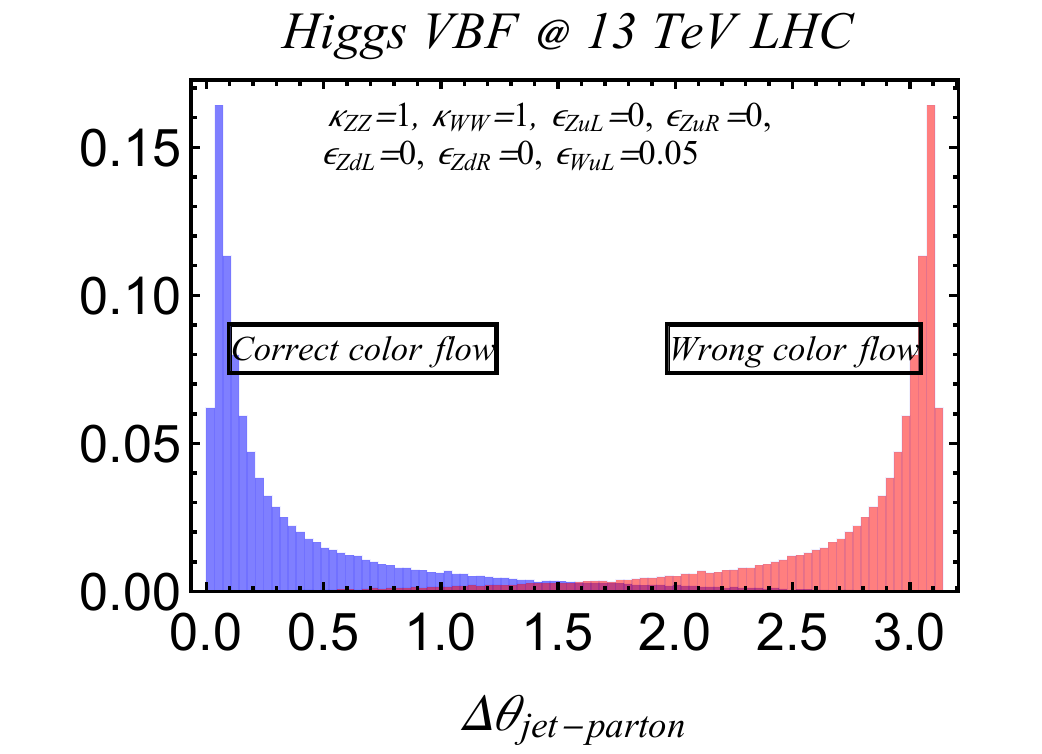}
  \end{center}
\caption{\small\label{fig:colorVSkinemtics} Leading order parton level simulation of the Higgs VBF production at $13$~TeV pp
c.m.~energy (from Ref.~\cite{Greljo:2015sla}). Show in blue is the distribution in the opening angle of the colour connected incoming and outgoing quarks $\measuredangle (\vec p_3,\vec p_1)$, while in red is the distribution for the opposite pairing, $ \angle (\vec p_3,\vec p_2)$. The left plot is for the SM, while the plot on the right is for a specific NP benchmark.
  }
\end{figure}

\begin{figure}
  \begin{center}
    \includegraphics[width=0.48\textwidth]{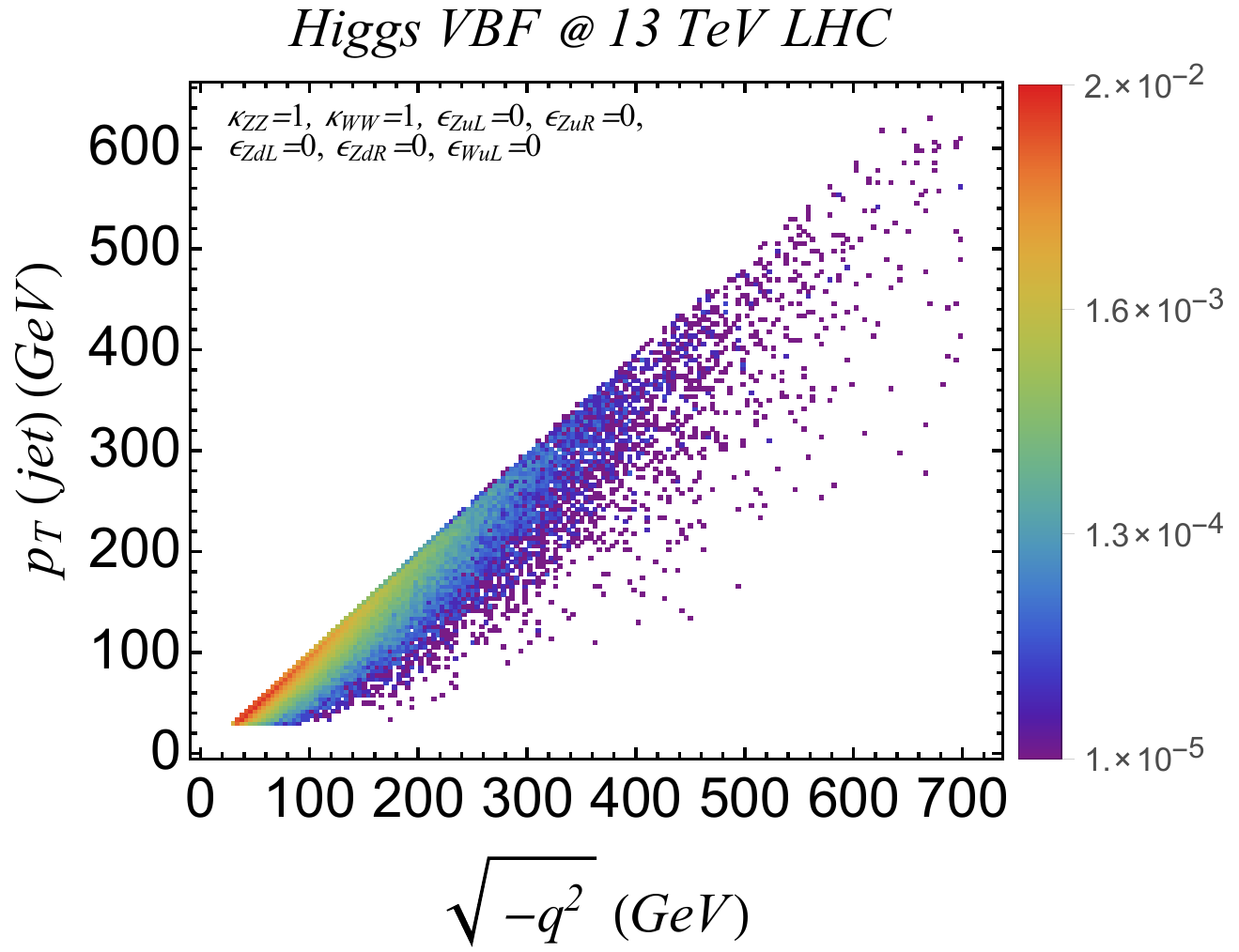}
        \includegraphics[width=0.495\textwidth]{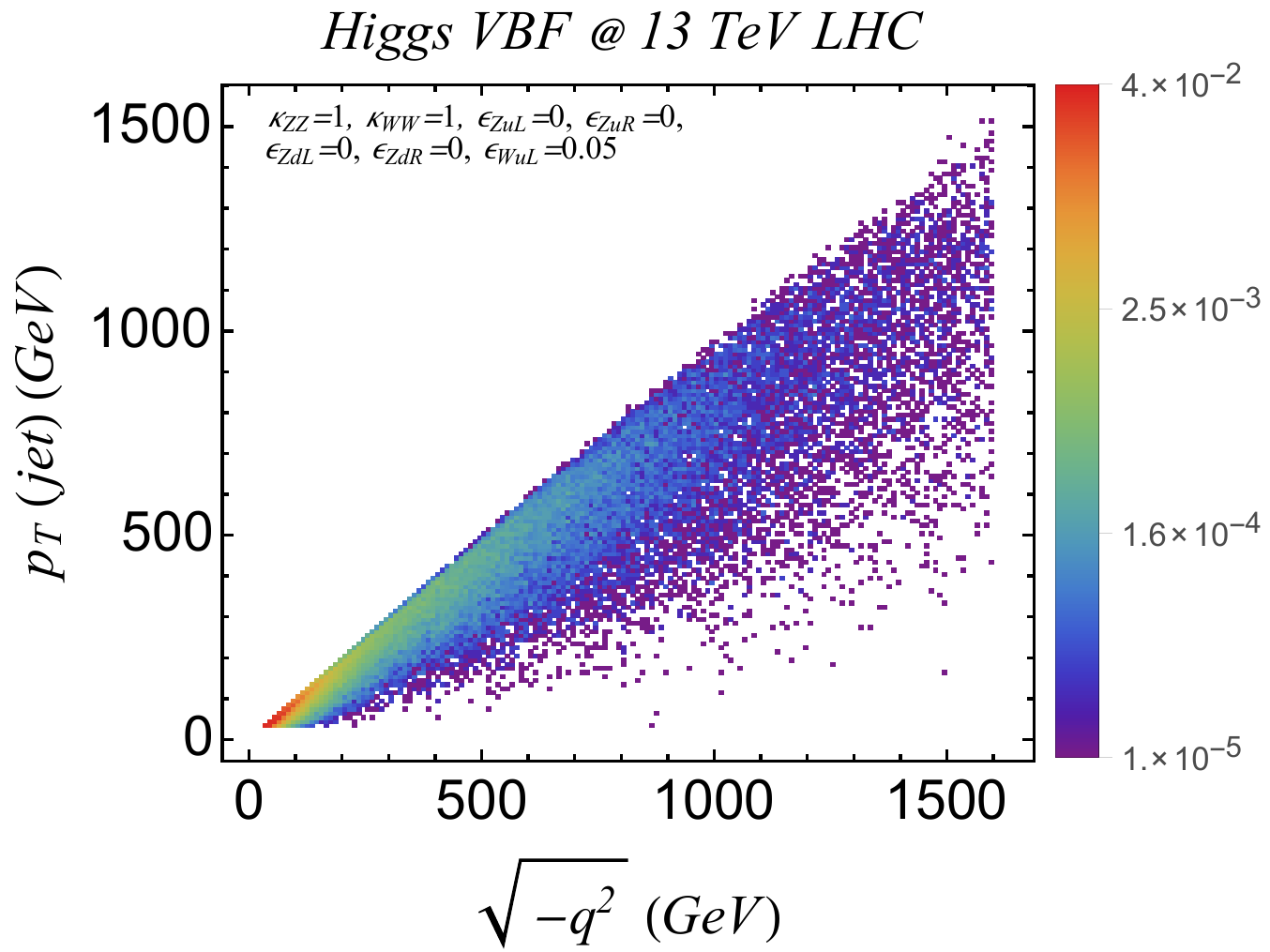}
  \end{center}
\caption{\small\label{fig:pTVSq2} Leading order parton level simulation of the Higgs VBF production at $13$~TeV pp c.m.~energy~\cite{Greljo:2015sla}. Shown here is the density histogram in two variables; the outgoing quark $p_T$ and the momentum transfer $\sqrt{- q^2}$ with the initial ``colour-connected'' quark. The left plot is for the SM, while the plot on the right is for a specific NP benchmark.     }
\end{figure}

In some realistic experimental analyses, after reconstructing the momenta of the two VBF tagged jets and the Higgs boson, one can compute the relevant momentum transfers  $q_1$ and $q_2$, adopting the pairing based on the opening angle. However, for some interesting Higgs boson decays modes, such as
$h\to 2\ell2\nu$, it is not possible to reconstruct the Higgs boson momentum. In this case, a good approximation of the momentum transfer is the jet $p_T$. This can be understood by explicitly computing the momentum transfer $q^2_{1,2}$ in the limit $|p_T|
\muchless  E_{jet}$ and for a Higgs produced close to threshold.
Let us consider the partonic momenta in c.o.m.~frame for the process:  $p_1 = (E, \vec{0}, E)$, $p_2 = (E, \vec{0}, -E)$, $p_3 = (E'_1, \vec{p}_{T1}, \sqrt{E^{\prime 2}_1-p_{T1}^2})$ and $p_4 = (E'_2, \vec{p}_{T2}, \sqrt{E^{\prime 2}_2 - p_{T2}^2})$. Conservation of energy for the whole process dictates $2E = E'_1 + E'_2 + E_h$, where $E_h^2$ is the Higgs energy, usually of order $m_h$ if the Higgs is not strongly boosted. In this case $E - E'_i = \Delta E_i  \muchless  E$ since the process is symmetric for $1 \leftrightarrow 2$. For each leg, energy and momentum conservation (along the $z$ axis) give
\be
	\left\{ \begin{array}{l} q^z_i = E - \sqrt{E^{\prime 2}_i - p_{Ti}^2} \\ q^0_i = E - E'_i \end{array} \right. \quad \to \quad
	\left\{ \begin{array}{l} q^0_i - q^z_i = \sqrt{E^{\prime 2}_i - p_{Ti}^2} - E'_i \approx - \frac{p_{Ti}^2}{2 E'_i} \\
	q^0_i + q^z_i \approx 2 \Delta E_i + \frac{p_{Ti}^2}{2 E'_i} \end{array}\right. ~.
\ee
Putting together these two relations one gets
\be
	q^2_i = (q^0_i)^2 - p_{Ti}^2 - (q^z_i)^2 =  - p_{Ti}^2 + (q^0_i - q^z_i)(q^0_i + q^z_i) \approx - p_{Ti}^2 - \frac{p_{Ti}^2 \Delta E_i}{2 E'_i} + \mathcal{O}(p_{Ti}^4 / E'^2)~.
\ee
We can thus conclude that, for a Higgs produced near threshold ($\Delta E_i <<  E'$),  $q^2 \approx - p_T^2$.

To illustrate the above conclusion, 
in \refF{fig:pTVSq2} we show a density histogram in two variables: the outgoing quark $p_T$ and the momentum transfer 
$\sqrt{-q^2}$ obtained from the correct colour flow pairing (the left and the right plots are for the SM and for a specific NP benchmark, respectively). The plots indicate the strong correlation of the jet $p_T$ with the momentum transfer $\sqrt{-q^2}$ associated with the correct colour pairing. 
We stress that this conclusion holds both within and beyond the SM.

Given the strong $q^2 \leftrightarrow p_T^2$ correlation, we strongly encourage the experimental collaborations to report the unfolded measurement of the double differential distributions in the two VBF tagged jet $p_T$'s: $\tilde F(p_{T j_1}, p_{T j_2})$.
This measurable distribution is  closely related to the form factor entering the amplitude decomposition, $F_L(q_1^2, q_2^2)$, and encode (in a model-independent way) the dynamical information about the high-energy behaviour of the process.
Moreover, the extraction of the PO in VBF
must be done preserving the validity of the momentum expansion: the latter can be checked and enforced setting
appropriate  upper cuts on the $p_T$ distribution. As an example, in \refF{fig:smpTpT}, we show the prediction in the SM (left plot) and in the specific NP benchmark (right plot) of the normalized $p_T$-ordered double differential distribution.

\begin{figure}
  \begin{center}
   \includegraphics[width=0.47\textwidth]{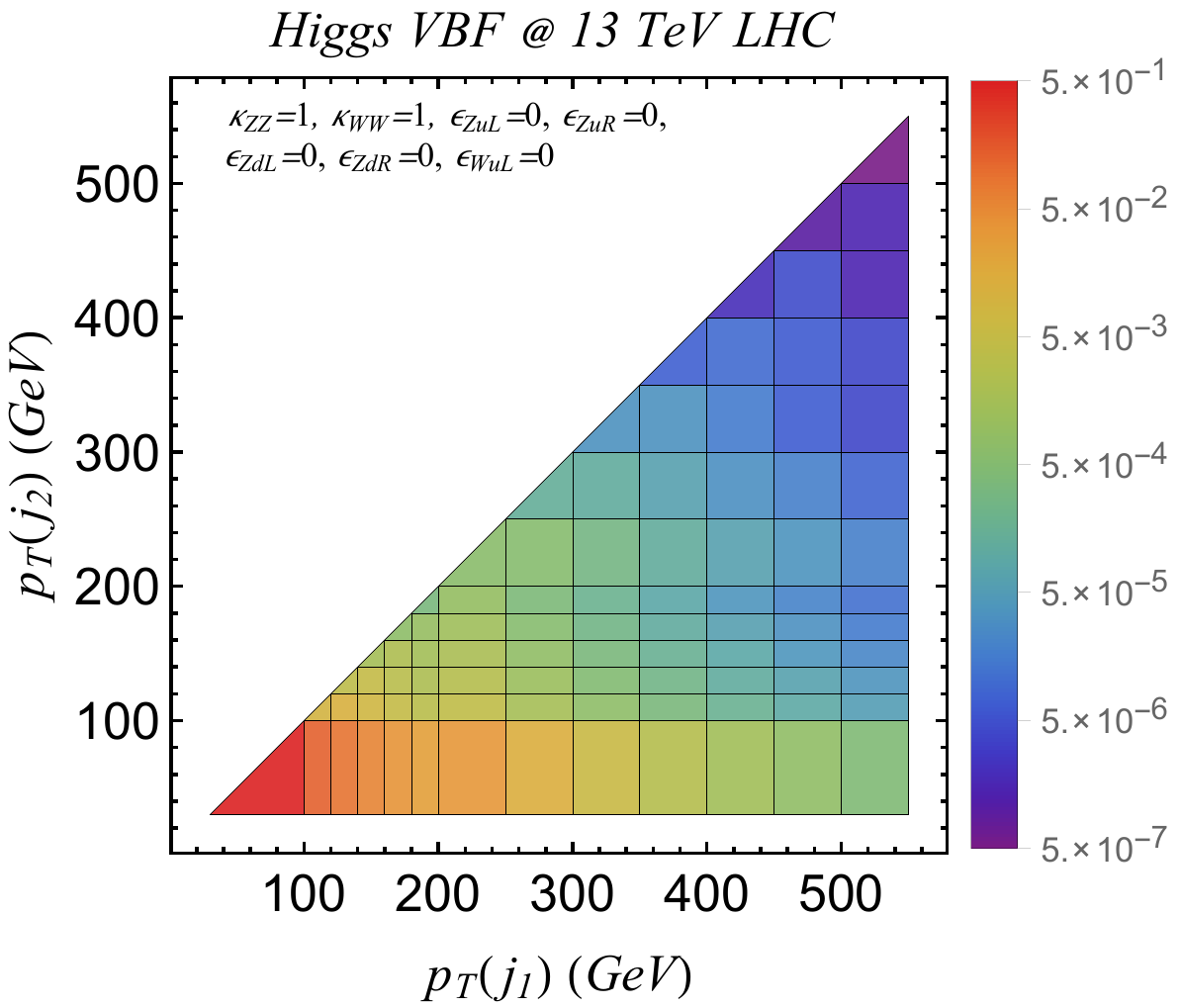}
   \includegraphics[width=0.47\textwidth]{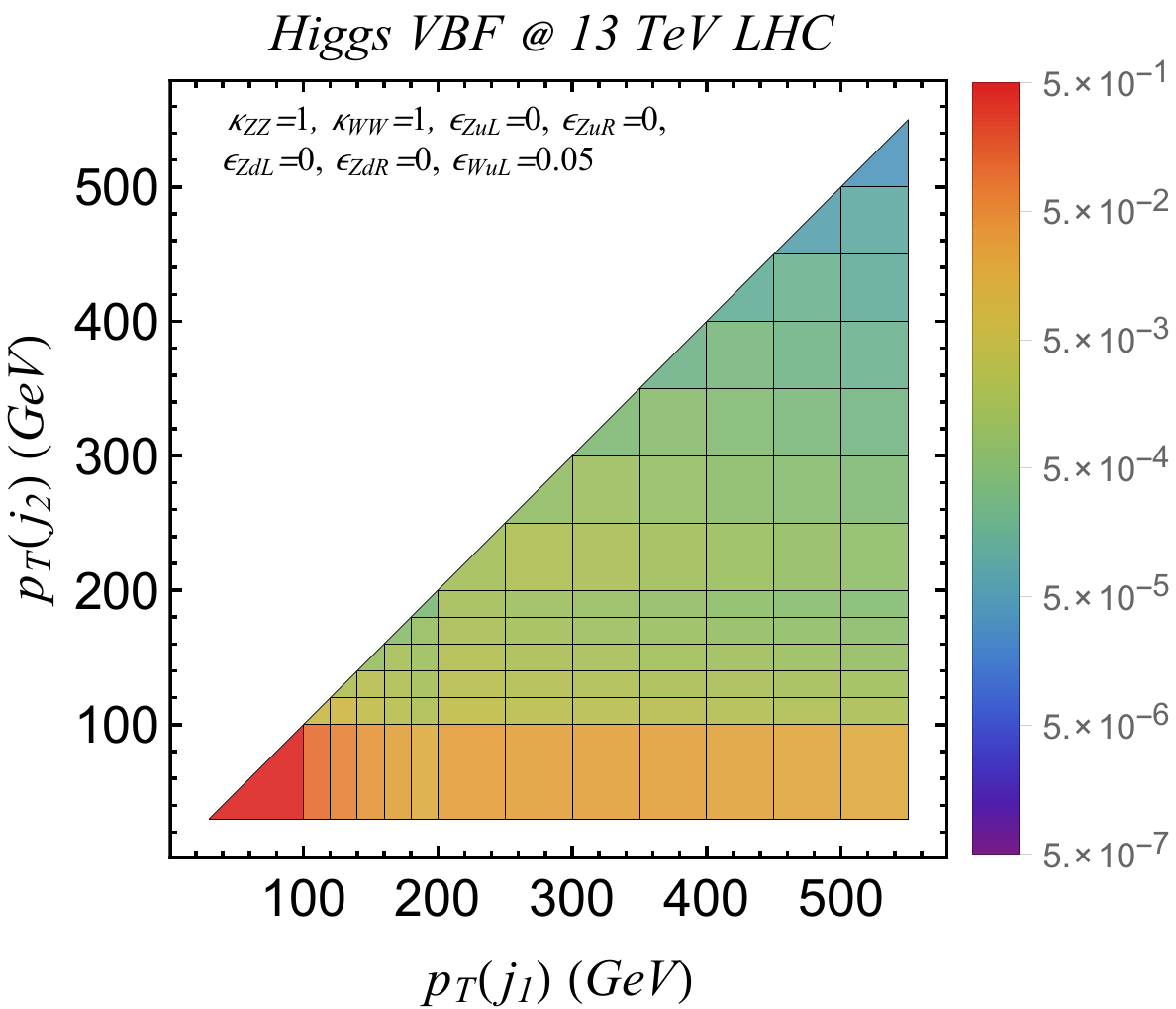}
  \end{center}
\caption{\small\label{fig:smpTpT} Double differential distribution in the two VBF-tagged jet $p_T$ for VBF Higgs boson production at 13 TeV LHC~\cite{Greljo:2015sla}. The distribution is normalized such that the total sum of events in all bins is 1. (Left) Prediction in the SM. (Right) Prediction for NP in $\epsilon_{W u_L} =0.05$.}
\end{figure}

\subsection{Associated vector boson plus Higgs boson production}
\label{sect:VH}

Higgs boson production in association with a $W$ or $Z$ boson are respectively the third and fourth Higgs boson production processes in the SM, by total cross section. Combined with VBF studies, they offer other important handles to disentangle the various Higgs PO. Due to the  lower cross section, this process is mainly studied in the highest-rate Higgs boson decay channels, such as $h\to b\bar{b}$  and $WW^*$. The drawback of these channels is the background, which is overwhelming in the $b\bar{b}$ case and of the same order as the signal in the $WW^*$ channels. Nonetheless, kinematical cuts, such as the Higgs boson $p_T$ in the $b\bar{b}$ case, and the use of multivariate analysis allow the experiments to precisely extract the the signal rates from these measurements.

\begin{figure}
  \begin{center}
    \includegraphics[width=0.45\textwidth]{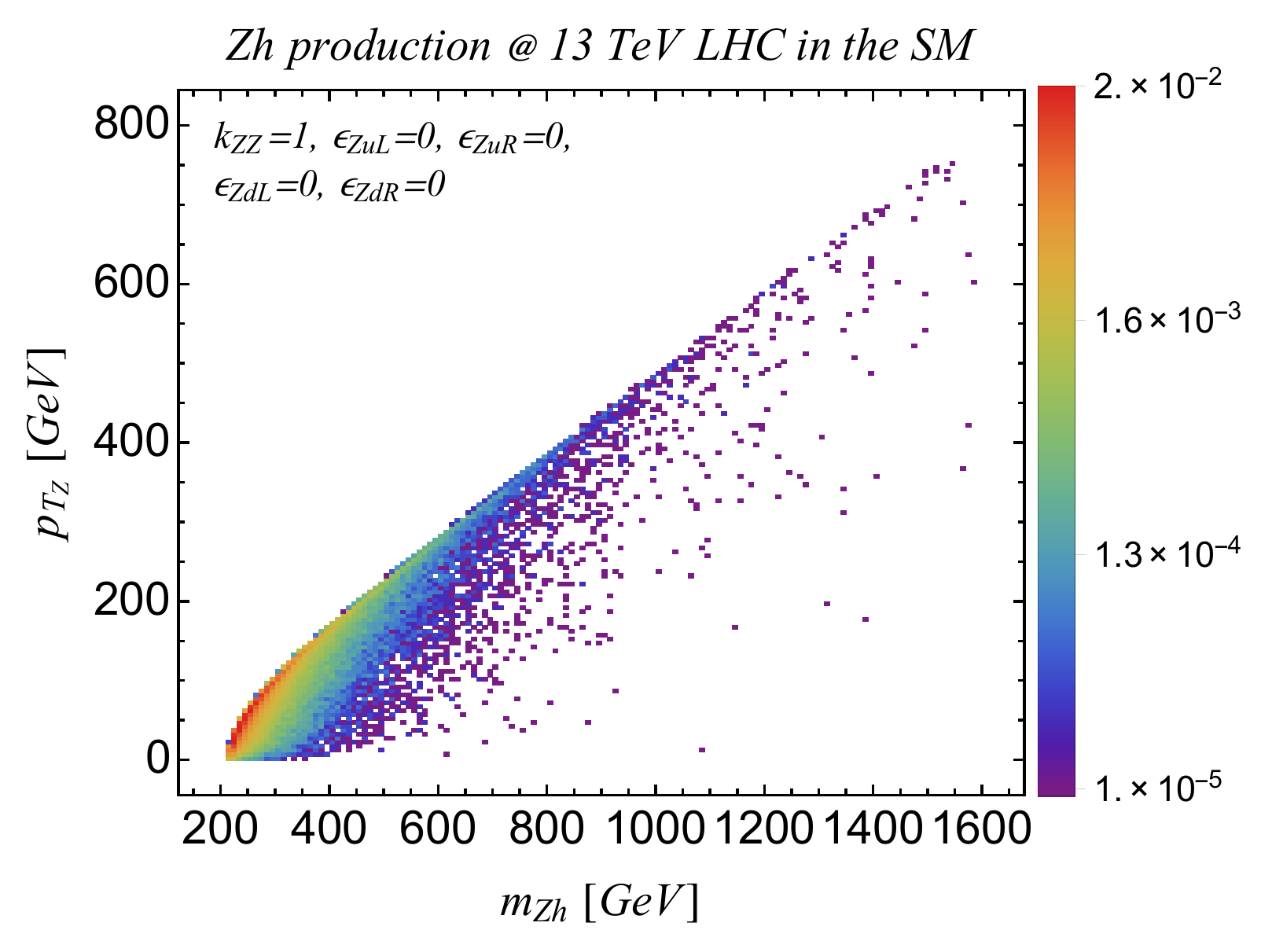} \quad
    \includegraphics[width=0.45\textwidth]{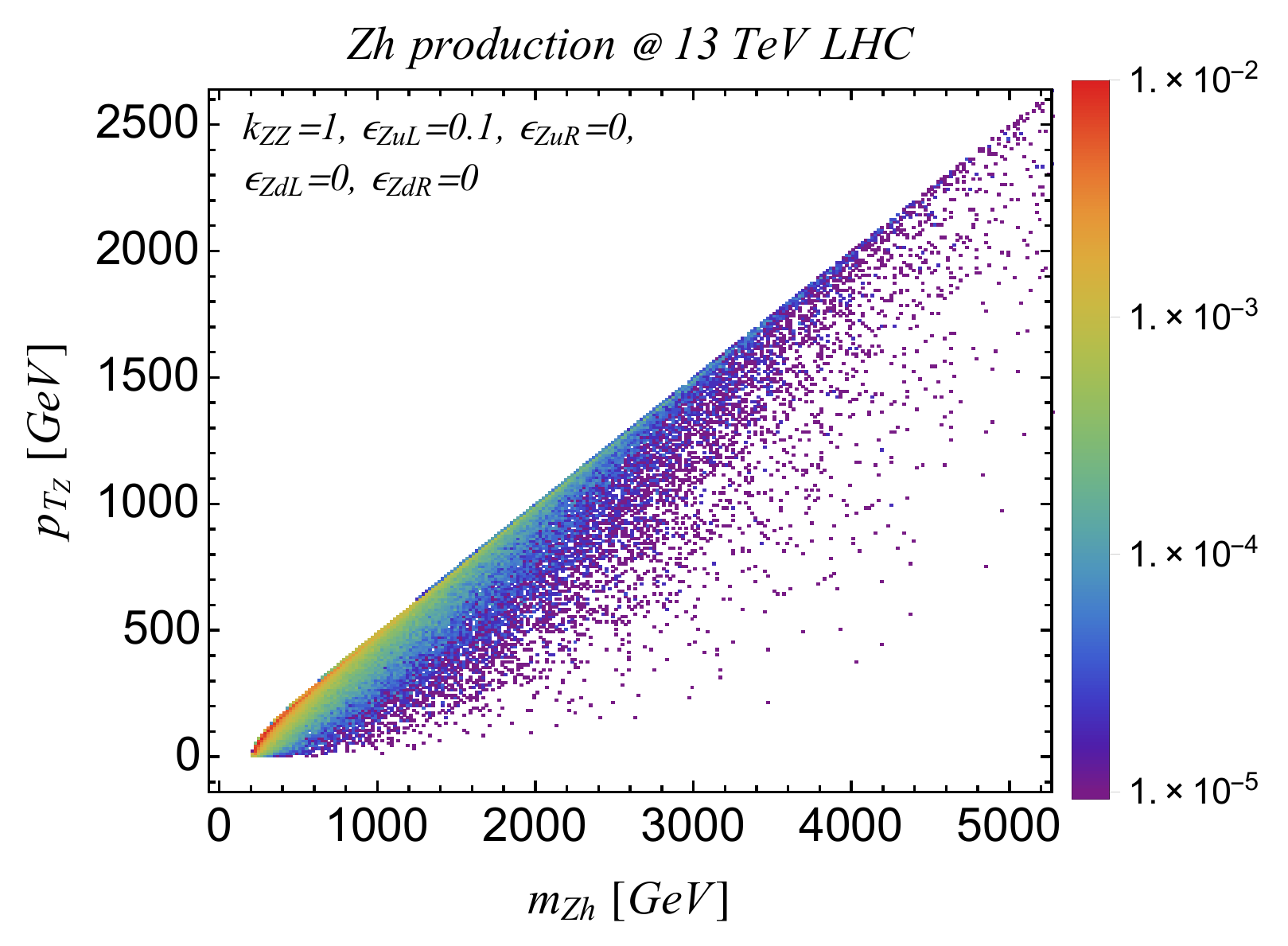}
  \end{center}
\caption{\small\label{fig:pTZ_qSQ_SM} The correlation between the $Zh$ invariant mass and the $p_T$ of the $Z$ boson in $Zh$ associate production at the 13TeV LHC in the SM (left plot) and for a BSM point $\kappa_{ZZ} = 1$, $\epsilon_{Z u_L} = 0.1$ (right plot)~\cite{Greljo:2015sla}. A very similar correlation is present in the $Wh$ channel.}
\end{figure}

An important improvement for future studies of these channels with the much higher luminosity which will be available, is to study differential distributions in some specific kinematical variables. In Section~\ref{sec:VHprod_ff} we showed that the invariant mass of the $Vh$ system is the most important observable in this process, since the form factors directly depend on it.
In those channels where the $Vh$  invariant mass can not be reconstructed due to the presence of neutrinos, another observable which shows some correlation with the $q^2$ is the $p_T$ of the vector boson, or equivalently of the Higgs, as can be seen in the \refF{fig:pTZ_qSQ_SM}. Even though this correlation is not as good as the one between the jet $p_T$ and the momentum transfer in the VBF channel, a measurement of the vector boson (or Higgs) $p_T$ spectrum, i.e. of some form factor $\tilde F^{Vh}(p_{TV})$ would still offer important information on the underlying structure of the form factors appearing in Eq.~\eqref{eq:FFVh}, $F_L^{q_i Z}(q^2)$ or $G_L^{q_{ij} W}(q^2)$.
The invariant mass of the $Vh$ system is given by $m_{Vh}^2 = q^2 = (p_V + p_h)^2 = m_V^2 + m_h^2 + 2 p_V\cdot p_h$. Going in the  centre of mass frame, we have $p_V = (E_V, \vec{p}_T, p_z)$ and $p_h = (E_h, - \vec{p}_T, - p_z)$, where $E_i = \sqrt{m_i^2 + p_T^2 + p_z^2}$ ($i=V,h$). Computing $m_{Vh}^2$ explicitly:
\be
	m_{Vh}^2 = m_V^2 + m_h^2 + 2 p_T^2 + 2 p_z^2 + 2\sqrt{m_V^2 + p_T^2 + p_z^2}\sqrt{m_h^2 + p_T^2 + p_z^2} \, \stackrel{|p_T| \to \infty}{\longrightarrow} \, 4 p_T^2~.
\ee
For $p_z = 0$ this equation gives the minimum $q^2$ for a given $p_T$, which can be seen as the left edge of the distributions in the \refF{fig:pTZ_qSQ_SM}. This is already a valuable information, for example the boosted Higgs regime used in some $b\bar{b}$ analysis implies a lower cut on the $q^2$: a bin with $p_T > 300$ GeV implies $\sqrt{q^2} \gtrsim 630$ GeV, which could be a problem for the validity of the momentum expansion.

In the $Wh$ process, if the $W$ decays leptonically its $p_T$ can not be reconstructed independently of the Higgs boson decay channel. One could think that the $p_T$ of the charged lepton from the $W$ decay would be correlated with the $Wh$ invariant mass, but we checked that there is no significant correlation between the two observables.

\subsection{Validity of the momentum expansion}

In order to control the momentum expansion at the basis of the PO composition, it is necessary to
set an upper cut on appropriate kinematical variables. These are the  $p_T$  of the leading VBF-tagged jet in VBF,
and the $Vh$ invariant mass (or the $p_T$  of the massive gauge boson) in VH.

The momentum expansion of the form factors in Eq.~\eqref{eq:FLGL} makes sense only if the higher order terms in $q_{1,2}^2$ are suppressed. Comparing the first two terms in this expansion, this
 leads to the consistency condition,
\be
\epsilon_{X_f} ~ q_{\rm{max}}^2 \lesssim m_Z^2 ~ g_X^{f}~,
\label{eq:consistency}
\ee
where $q_{\rm{max}}^2$ is the largest momentum transfer in the process. A priori we don't know which is the size of the $\epsilon_{X_f}$
or, equivalently, the effective scale of new physics. However, a posteriori we can verify by means of  Eq.~(\ref{eq:consistency})
if we are allowed to truncate the momentum expansion to the first non-trivial terms. In VBF,
setting a cut-off on  $p_T$ we implicitly define a value of $q_{\rm{max}}$. Extracting the $\epsilon_{X_f}$ for
 $p_{T}^j < (p_{T}^j)^{\rm max} \approx q_{\rm{max}}$ we can
check if Eq.~(\ref{eq:consistency}) is satisfied. Ideally, the experimental collaborations should perform the extraction of the $\epsilon_{X_f}$
for different values of $(p_{T}^j)^{\rm max}$ optimizing the range according to the results obtained.
 The issue is completely analog in VH, where the $q_{\rm{max}}^2$ is controlled by $m^2_{Vh}$.

It must be stressed that if data indicate non-vanishing values for the $\epsilon_{X_f}$, and the condition~(\ref{eq:consistency}) is falsified, this does not necessarily imply a break down of the
momentum expansion.\footnote{~The break-down of the momentum expansion occurs only when we approach,
with a given kinematical variable, a new pole in the amplitude   due to the exchange of a NP particle.
On the other hand, the dominance of the contact terms leading to a violation of the
condition~(\ref{eq:consistency}) could also occur far from the NP poles in case of strongly interacting
theories (see e.g.~Ref.~\cite{Contino:2016jqw}).}
 However, in such case it is important to check with data the size of
additional terms in this  expansion. From this point of view, it is very
important to complement the PO approach with the differential measurements of the cross-section
as function of  $p_{T}^j$   that could be  achieved via the so-called {\em template-cross-section} method.
In such distributions a possible break-down of the momentum expansion  at the basis of the PO decomposition
could indeed be seen (or excluded) directly by data.

A further check to assess the validity of the momentum expansion is obtained comparing the fit performed
including the full quadratic dependence of  the distributions, as function  on the PO, with the fit in which
such distributions are linearized in $\delta \kappa_X \equiv \kappa_X-\kappa_X^{\rm{SM}}$ and $\epsilon_X$. The idea behind this procedure is that the quadratic corrections to physical observable in $\delta \kappa_X$ and $\epsilon_X$ are  formally of the same order as the interference of the first neglected term in Eq.~\eqref{eq:FLGL} with the leading SM contribution.
If the two fits yields significantly different results, the difference can be used as an estimate of the uncertainty due to the neglected
higher-order terms in the momentum expansion. However, as discussed in detail in Ref.~\cite{Greljo:2015sla},  such procedure naturally leads to a large overestimate of the uncertainty. This is because in the linearized fit only a few linear combinations of the PO enter the observables, and thus the number of independent constraints derived from data is effectively reduced.
On general grounds, the fit obtained with the full quadratic dependence should be considered as the most reliable result,
provided that the obtained PO satisfy the consistency condition in Eq.~(\ref{eq:consistency}).

\subsection{Illustration of NLO QCD effects}

\begin{figure}[t]
   \centering
    \includegraphics[width=0.40\textwidth]{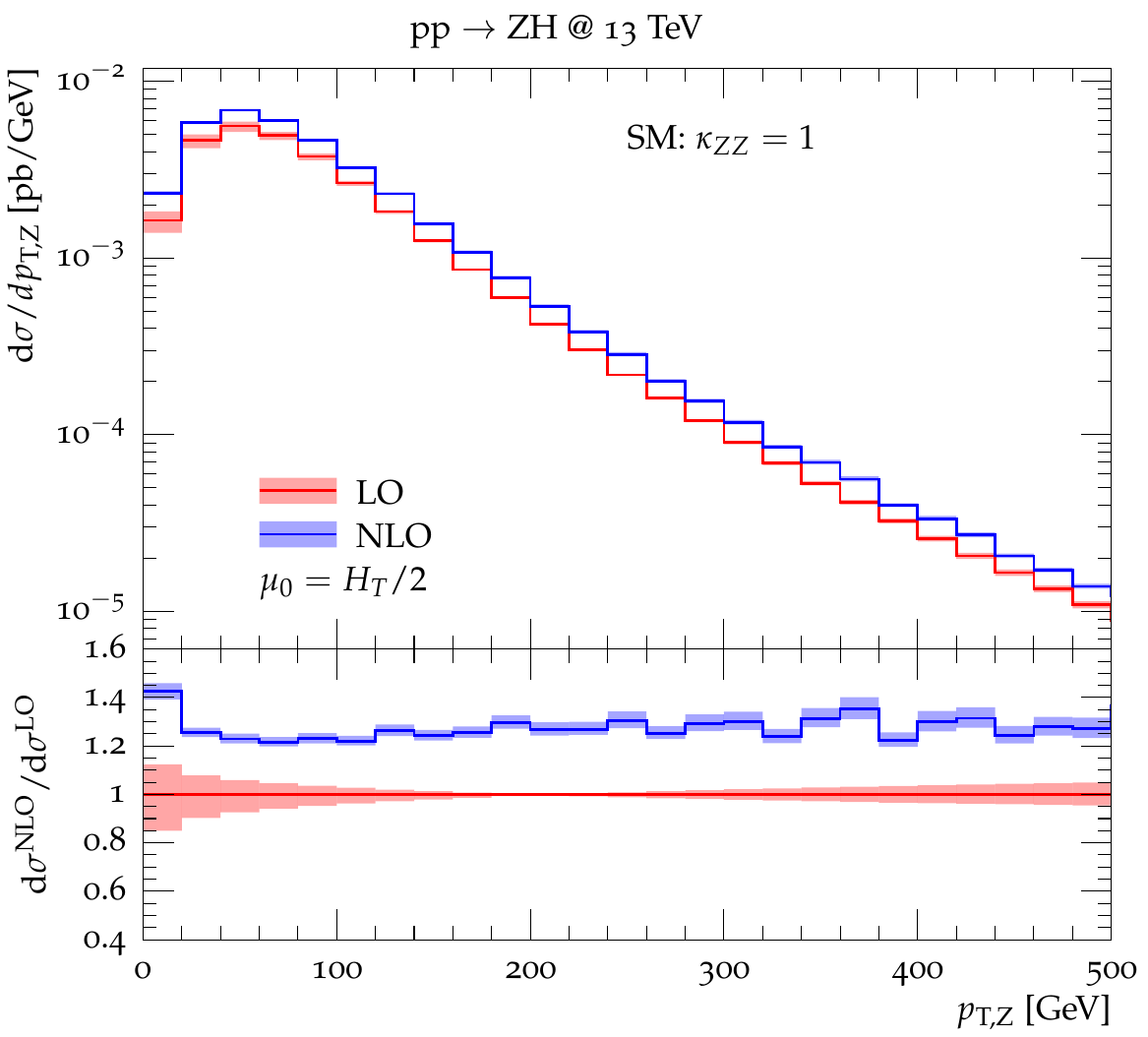}\quad
    \includegraphics[width=0.40\textwidth]{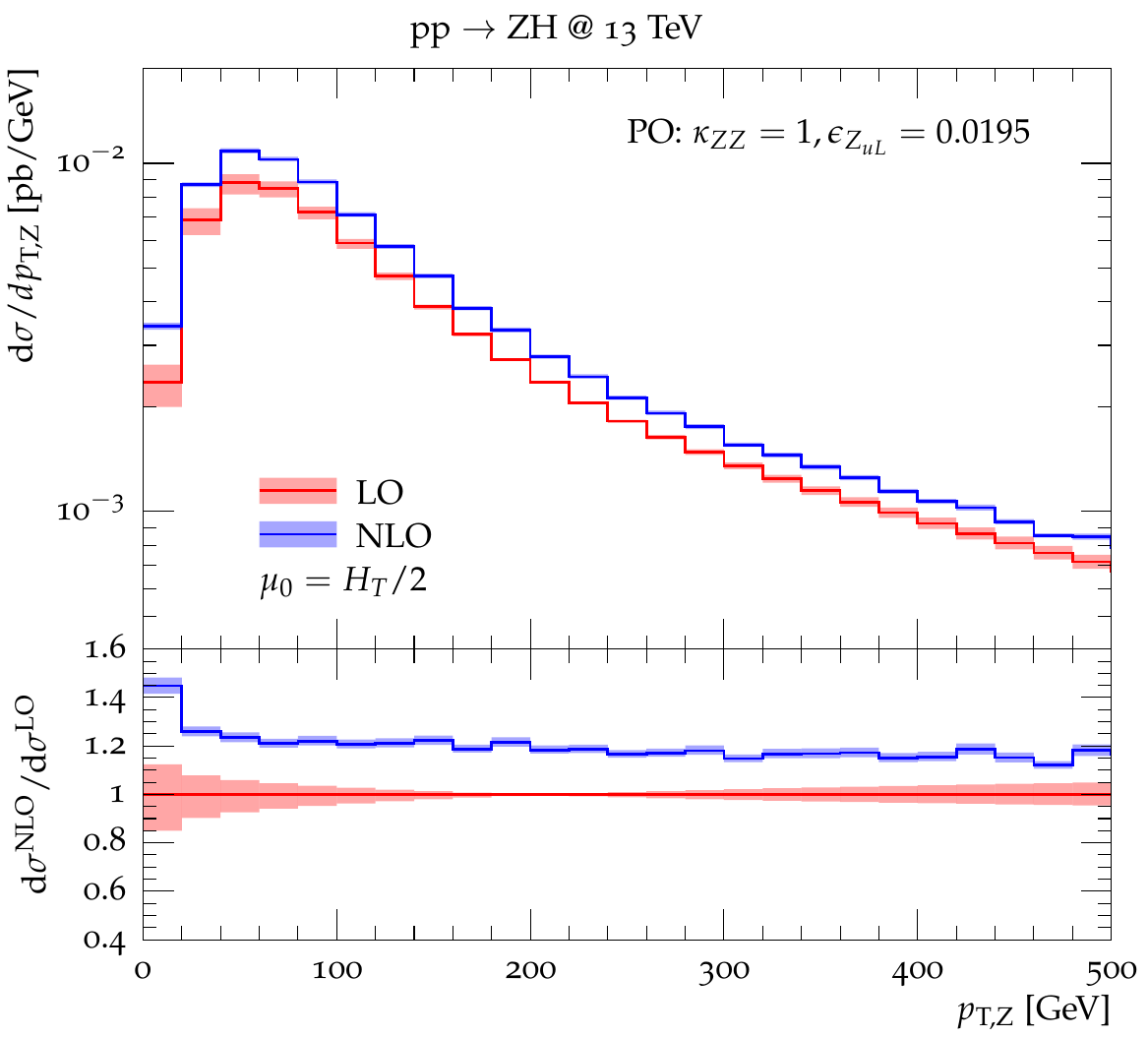}
\caption{Differential $p_{\mathrm{T},Z}$ distributions in the process $pp \to ZH$ in the SM (left) and including an example of NP within the PO framework with $\kappa_{ZZ}=1, \epsilon_{Z{uL}}=0.0195$ (right). Shown are LO (red) and NLO (blue) predictions and corresponding (7-pt) scale variations employing a central scale $\mu_0 = H_{\mathrm{T}}/2$.
}
\label{fig:nlo_ptZ}
\end{figure}

Higgs boson production via Higgs-strahlung and VBF are very stable with respect to NLO QCD corrections. At the inclusive level NLO scale uncertainties are as small as a few per cent \cite{Han:1992hr,Figy:2003nv,Dittmaier:2011ti,Dittmaier:2012vm,Heinemeyer:2013tqa} and further reduced at NNLO~\cite{Ferrera:2011bk,Ferrera:2013yga,Ferrera:2014lca,Bolzoni:2010xr,Bolzoni:2011cu,Cacciari:2015jma}.
Given an appropriate scale choice similar conclusion also hold at the differential level with residual NLO scale uncertainties at the 10\% level.

As already discussed above, similar to the factorization in QED, the dominant QCD corrections are universal and factorize from any new physics effects in EW Higgs boson production. Consequently, with respect to possible small deformations from the SM, parameterized via effective form factor contributions in the PO framework, we expect a very limited sensitivity to QCD effects assuming a similar stabilization of higher order corrections as observed for the SM.
In order to verify this assumption (and to make a corresponding tool available), the PO framework has been implemented in the \texttt{Sherpa + OpenLoops} \cite{Gleisberg:2007md,Gleisberg:2008ta,Cascioli:2011va,OLhepforge} framework, where the implementation within \texttt{Sherpa} is based on a model independent \texttt{UFO} interface~\cite{Hoche:2014kca}.
As in illustration, in \refF{fig:nlo_ptZ} we compare the NLO QCD corrections to the $p_T$ distribution of the $Z$-boson in $pp\to ZH$ between the SM and a NP point with $\kappa_{ZZ}=1, \epsilon_{Z{uL}}=0.0195$.  Despite the very different shape of the two distributions, higher order QCD corrections are very similar and do only show a very mild shape dependence. Detailed studies for VBF and VH including parton shower matching are under way.

After the inclusion of NLO QCD corrections the dominant theoretical uncertainties to Higgs observables in VBF and VH  are of EW type and dominated by large EW Sudakov logarithms at large energies~\cite{Denner:2011id,Denner:2014cla,Ciccolini:2007ec}. The dominant NLO EW effects are
factorizable corrections which can be reabsorbed into a future redefinition of the PO.

\section{Parameter counting and symmetry limits}
\label{sectPO:PCSL}

We are now ready to identify the number of independent pseudo-observables necessary to describe various sets
of Higgs boson decay amplitudes and productions cross sections. We list them below separating four set of observables:
\begin{enumerate}
\item[i)] the Yukawa decay modes ($h\to f\bar f$);
\item[ii)] the EW decays ($h\to \gamma\gamma,  f \bar f  \gamma, 4f$);
\item[iii)] the EW production production cross sections (VBF and VH);
\item[iv)] the non-EW production cross sections (gluon fusion and $t\bar t H$) and the total Higgs boson decay width.
\end{enumerate}
We list the PO needed for a completely general analysis, and the reduction
of the number of independent PO  obtained under well-defined symmetry hypotheses, such as CP
invariance or flavour universality. The latter can be more efficiently tested considering specific sub-sets of observables.

\subsection{Yukawa modes}

As discussed in Section~\ref{sect:hff} the  $h\to f\bar f$ amplitudes
are characterized by two independent PO ($\kappa_f$ and $\lamCP_f$)
for each fermion species. Considering only the decay channels relevant for LHC, the full set of 8 parameters is:
\be
\kappa_b, \kappa_c, \kappa_\tau, \kappa_\mu, \lamCP_{b}, \lamCP_{c}, \lamCP_{\tau}, \lamCP_{\mu}~.
\ee
Assuming $CP$ conservation (that implies $\lamCP_f=0$ for each $f$) the number of PO is reduced to 4.
This is also the number of independent PO effectively measurable if the spin polarization of the final-state fermions is
not accessible. The corresponding {\it physical} PO are the  $\Gamma(h \rightarrow f \bar{f})$
partial widths (see \refT{Tab:physicalPO}).

\subsection{Higgs EW decays}
The category of EW decays  includes a long list of  channels; however, not all of them are accessible at the LHC.
The clean neutral current processes $h \rightarrow e^+ e^- \mu^+ \mu^-$, $h \rightarrow e^+ e^-  e^+ e^-  $ and $h \rightarrow \mu^+ \mu^- \mu^+ \mu^-$,
together with the  photon channels $h \to \gamma \gamma$ and $h \to \ell^+ \ell^- \gamma$,  can be described in terms of 11 real parameters:
\be
	\kappa_{ZZ},  \kappa_{Z\gamma},  \kappa_{\gamma \gamma},  \epsilon_{ZZ},  \epsilon_{ZZ}^{CP},  \delta_{Z\gamma}^{CP}, \delta_{\gamma \gamma}^{CP}, \epsilon_{Z e_L}, \epsilon_{Z e_R}, \epsilon_{Z \mu_L}, \epsilon_{Z \mu_R}
\ee
(of which only the subset  $\{ \kappa_{\gamma \gamma}, \kappa_{Z\gamma}, \delta_{\gamma \gamma}^{CP}, \delta_{Z\gamma}^{CP}  \}$
is necessary to describe $h \to \gamma \gamma$ and $h \to \ell^+ \ell^- \gamma$).
The charged-current process $h \rightarrow \bar{\nu}_e e \bar{\mu} \nu_\mu$ needs 7 further independent real parameters to be completely specified:
\be
	\kappa_{WW}, \epsilon_{WW}, \epsilon_{WW}^{CP}~ ({\rm real}) \quad  + \quad \epsilon_{W e_L},  \epsilon_{W \mu_L}~({\rm complex})~.
\ee
Finally, the mixed processes $h \rightarrow e^+ e^- \nu\bar{\nu}$ and $h \rightarrow \mu^+ \mu^- \nu\bar{\nu}$ can be described by a subset of the coefficients already introduced plus 2 further real contact interactions coefficients:
\be
	\epsilon_{Z \nu_{e}}~, \epsilon_{Z \nu_{\mu}}~.
\ee
This brings the total number of (real) parameters to 20  for all the (EW) decays involving muons, electrons, and photons.

The extension to discuss $h\to 4f$ or $h\to f\bar f \gamma$
decays with one or two pairs of tau leptons is straightforward:  it requires the introduction of  the corresponding set of contact terms
($\epsilon_{Z \tau_L}, \epsilon_{Z \tau_R}$, $\epsilon_{W \tau_L}, \epsilon_{Z_{\nu_\tau}}$). Similarly, quark contact terms need to
be introduced if one or two lepton pairs are replaced by  a quark pair.

A first simple restriction in the number of parameters is obtained by assuming flavour universality.
This hypothesis imply that the contact terms are the same for all flavours. In particular,
for muon and electron modes, this implies
\be
	\epsilon_{Z e_L} = \epsilon_{Z \mu_L}~,\qquad
	\epsilon_{Z e_R} = \epsilon_{Z \mu_R}~,\qquad
	\epsilon_{Z \nu_{e}} = \epsilon_{Z \nu_{\mu}}~,\qquad
	\epsilon_{W e_L} = \epsilon_{W \mu_L}~.\qquad
	\label{eq:flavour}
\ee
Technically, this correspond to assume an underlying $U(N_\ell)^2$ flavour symmetry,
for the $N_\ell$ generations of leptons considered (namely
the maximal flavour symmetry compatible with the SM gauge group).

Since the $\epsilon_{W \ell_L}$ parameters are complex in general,
the relations (\ref{eq:flavour}) allow to reduce the total number of parameters to 15.
This assumption can be tested directly from data by comparing the extraction of the contact terms from
$h \rightarrow 2e2\mu$, $h \rightarrow 4e$ and $h \rightarrow 4\mu$ modes.

The assumption that CP is a good approximate symmetry of the BSM sector and that the Higgs is a CP-even state,
allows us to set to zero six independent (real) coefficients:
\be
	\epsilon_{ZZ}^{CP} = \delta_{Z\gamma}^{CP} = \delta_{\gamma\gamma}^{CP} = \epsilon_{WW}^{CP} = \text{Im} \epsilon_{W e_L} = \text{Im} \epsilon_{W \mu_L} = 0~.
\ee
Assuming, at the same time, flavour universality, the number of free real parameters reduces to 10.

The various cases are prorated in the upper panel of \refT{tab:POtable}: in the second column
we list the 10 PO needed assuming both CP invariance and flavour universality, while in the third and fourth
column we list the additional PO needed if these hypotheses are relaxed (for the clean modes involving only
muons and electrons).
The corresponding {\it physical} PO are the partial widths reported in \refT{Tab:physicalPO}.

\begin{table}[p]
 \caption{\label{tab:POtable} Summary of the effective couplings PO appearing in EW Higgs boson decays and in the VBF and VH production  cross-sections (see main text). The terms between square brakes in the middle table are the PO present both in
 production and decays. The last table denote the PO needed  to describe
 both production and decays under the assumption of custodial symmetry. }
\begin{center}
\begin{tabular}{c||c|c|c}
\multicolumn{4}{c}{Higgs (EW) decay amplitudes} \\ 
\toprule
\rule{0pt}{1.2em}%
Amplitudes  &  Flavour  + CP  & Flavour Non Univ. & CPV    \\
 \midrule
  $h \to \gamma\gamma,  2e \gamma, 2\mu\gamma$  &  $\kappa_{ZZ},  \kappa_{Z\gamma},  \kappa_{\gamma \gamma}, \epsilon_{ZZ} $    &
  \raisebox{-6pt}[0pt][0pt]{$\epsilon_{Z \mu_L}, \epsilon_{Z \mu_R}$}  &
  \raisebox{-6pt}[0pt][0pt]{$\epsilon_{ZZ}^{CP},  \delta_{Z\gamma}^{CP}, \delta_{\gamma \gamma}^{CP} $}
  \\
  $ 4e, 4\mu, 2e2\mu$ & $ \epsilon_{Z e_L}, \epsilon_{Z e_R}$ &   &
  \\ \midrule
   \raisebox{-10pt}[0pt][0pt]{ $h \to   2e 2\nu, 2\mu 2\nu, e\nu \mu \nu$} &
  \raisebox{-3pt}[0pt][0pt]{ $\kappa_{WW}, \epsilon_{WW}$ } & \raisebox{-4pt}[0pt][8pt]{ $\epsilon_{Z \nu_{\mu}}$, Re($\epsilon_{W \mu_L}$) } &  \raisebox{-4pt}[0pt][8pt]{$\epsilon_{WW}^{CP}$, Im($\epsilon_{W e_L}$)}
    \\
   &  $\epsilon_{Z \nu_{e}}$, Re($\epsilon_{W e_L}$)
     &   \multicolumn{2}{c}{  \raisebox{-2pt}[4pt][8pt]{
 Im($\epsilon_{W \mu_L}$)  }}  \\ 
 \bottomrule
 
 \multicolumn{4}{c}{ } \\
\multicolumn{4}{c}{Higgs (EW) production amplitudes} \\ 
\toprule
\rule{0pt}{1.2em}%
Amplitudes &   Flavour + CP  & Flavour Non Univ. & CPV    \\
 \midrule
  \raisebox{-2pt}[0pt][7pt]{ VBF neutral curr.} &  \raisebox{-2pt}[0pt][7pt]{  $\left[~ \kappa_{ZZ},  \kappa_{Z\gamma}, \kappa_{\gamma\gamma},  \epsilon_{ZZ}~ \right]$ }
   &   $\epsilon_{Z c_L}, \epsilon_{Z c_R}$    &
   \raisebox{-6pt}[0pt][0pt]{ $\left[ ~\epsilon_{ZZ}^{CP},  \delta_{Z\gamma}^{CP},  \delta_{\gamma\gamma}^{CP} ~\right]$  }
   \\
   \raisebox{0pt}[0pt][7pt]{ and $Zh$ }  & $\epsilon_{Z u_L}, \epsilon_{Z u_R}, \epsilon_{Z d_L}, \epsilon_{Z d_R}$ &   $\epsilon_{Z s_L}, \epsilon_{Z s_R}$    &
   \\ 
   \midrule
   \raisebox{-2pt}[0pt][7pt]{VBF charged  curr.}    &
    \raisebox{-1pt}[0pt][0pt]{ $\left[~\kappa_{WW}, \epsilon_{WW}~\right]$  }   & \raisebox{-4pt}[0pt][7pt]{ Re($\epsilon_{Wc_L}$)  } &
   \raisebox{-4pt}[0pt][7pt]{$[\epsilon^{CP}_{WW}]$, Im($\epsilon_{W u_L}$)}
   \\
    and $Wh$    &    Re($\epsilon_{Wu_L}$)    &   \multicolumn{2}{ c}{  \raisebox{-2pt}[4pt][7pt]{ $\qquad$  Im($\epsilon_{W c_L}$)  }}
   \\ 
   \bottomrule
   
  \multicolumn{4}{c}{ } \\
\multicolumn{4}{c}{EW production and decay modes, with custodial symmetry } \\ 
\toprule
\rule{0pt}{1.2em}%
Amplitudes  &   Flavour + CP & Flavour Non Univ. & CPV    \\
 \midrule
   \raisebox{-4pt}[0pt][0pt]{   } & & &
  \\
\raisebox{8pt}[0pt][0pt]{ production \& decays }
 &  \raisebox{8pt}[0pt][0pt]{  $\kappa_{ZZ},   \kappa_{Z\gamma},\kappa_{\gamma \gamma},  \epsilon_{ZZ}$}  &
 &  \raisebox{8pt}[0pt][0pt]{$\epsilon_{ZZ}^{CP}, \delta_{Z\gamma}^{CP} ,  \delta_{\gamma \gamma}^{CP}$}
 \\
  \raisebox{-8pt}[0pt][7pt]{ VBF and VH only  } &   \raisebox{-8pt}[0pt][7pt]{$\epsilon_{Z u_L}, \epsilon_{Z u_R}, \epsilon_{Z d_L}, \epsilon_{Z d_R}$ }    &
  $\epsilon_{Z c_L}, \epsilon_{Z c_R}$    &
 \\
 &    &    \raisebox{2pt}[0pt][7pt]{ $\epsilon_{Z s_L}, \epsilon_{Z s_R}$ }  &  \\

    \raisebox{-4pt}[0pt][0pt]{   } & & &  \\
 \raisebox{8pt}[0pt][0pt]{ decays only }
        &   \raisebox{8pt}[0pt][0pt]{  $\epsilon_{Z e_L}, \epsilon_{Z e_R}$,   Re($\epsilon_{W e_L}$)} &
         \raisebox{8pt}[0pt][0pt]{  $\epsilon_{Z \mu_L}, \epsilon_{Z \mu_R}$  }  &
          \raisebox{8pt}[0pt][0pt]{    } \\
       \bottomrule
 \end{tabular}
\end{center}
\end{table}

\subsection{EW production processes}

The fermion-independent PO present in Higgs boson decays appear also in EW production processes.
The additional PO appearing only in production (assuming Higgs boson decays to quark are not detected)
are  the
contact terms for the light quarks. In a four-flavour scheme, in absence of any symmetry assumption,
 the number of independent parameters for the neutral currents contact terms is 16
 ($\epsilon_{Z q^{i j}}$, where $q=u_L, u_R, d_L, d_R$, and $i,j=1,2$):
  8 real parameters for flavour diagonal terms
  and 4 complex flavour-violating parameters. Similarly, there are 16 independent parameters in charged currents, namely the
  8 complex  terms $\epsilon_{W u^{i}_{L} d^{j}_{L}}$ and  $\epsilon_{W u^{i}_{R} d^{j}_{R}}$.
 However, we can safely reduce the number of independent PO under neglecting the terms that violates the
 $U(1)_f$ flavour symmetry acting on each of the light fermion species, $u_R$, $d_R$, $s_R$, $c_R$, $q_L^{(d)}$, and $q_L^{(s)}$,
 where $q_L^{(d,s)}$ denotes the two quark doublets in the basis where down quarks are diagonal. This symmetry
is an exact symmetry of the SM in the limit where we neglect light quark masses. Enforcing it at the PO level is equivalent to neglecting
terms that do not interfere with SM amplitudes in the limit of vanishing light quark masses.
Under this (rather conservative) assumption,
the number of independent neutral currents contact terms reduces to 8 real parameters,
\be
\epsilon_{Z u_R},~ \epsilon_{Z c_R},~ \epsilon_{Z d_R},~ \epsilon_{Z s_R},~ \epsilon_{Z d_L},~ \epsilon_{Z s_L},~
~ \epsilon_{Z u_L},~ \epsilon_{Z c_L},
\ee
and only 2 complex parameters appear in the charged-current case:
\be
\epsilon_{W u^{i}_{L} d^{j}_{L}}\equiv V_{i j} \epsilon_{W u^{j}_{L}},~\qquad \epsilon_{W u^{i}_{R} d^{j}_{R}}=0~.
\ee

Similarly to the decays, a further interesting reduction of the number of parameters is obtained assuming
flavour universality or, more precisely, under the assumption of an $U(2)^3$ symmetry acting on the
first two generations of quarks. The latter is the the maximal flavour symmetry for the light quarks compatible with
the SM gauge group. In this case  the independent parameters in this case reduces to six:
\be
\epsilon_{Z u_L},~\epsilon_{Z u_R},~ \epsilon_{Z d_L},~ \epsilon_{Z d_R},~ \epsilon_{W u_L}~,
\label{eq:MFV_contterms}
\ee
where $\epsilon_{W u_L}$ is complex, or five if we further neglect CP-violating contributions (in such case $\epsilon_{W u_L}$ is real). This case is listed in the second column of \refT{tab:POtable} (middle panel), where the terms
between brackets denote the PO appearing also in decays.

\subsubsection*{Custodial symmetry and the combination of EW production and decay modes}
Assuming flavour universality and CP conservation, the number of independent PO necessary to describe all EW decays
and production cross sections is 15. These are the terms listed in the second column of the first two panels of
\refT{tab:POtable}.

A further reduction of the number of independent PO is obtained under the hypothesis of custodial symmetry,
that relates charged and neutral current modes. The complete list of custodial symmetry relations can be found
in Refs.~\cite{Gonzalez-Alonso:2014eva,Greljo:2015sla}.
Here we only mention the one between $\kappa_{WW}$ and $\kappa_{ZZ}$, noting that the presence of  contact terms modify it with respect to the one known in the context of the kappa-framework:
\be
	\kappa_{WW} - \kappa_{ZZ} = -\frac{2}{g} \left( \sqrt{2} \epsilon_{W \ell_L} + 2 \frac{m_W}{m_Z} \epsilon_{Z \ell_L} \right)~,
\ee
where $\ell = e, \mu$.
After imposing flavour and CP conservation, custodial symmetry allow a reduction of the number of
independent PO from 15 down to 11, as shown in the lower panel of  \refT{tab:POtable}.

\subsection{Additional PO}
The remaining PO needed for a complete description of Higgs physics at the LHC
are those related to the non-EW production processes (gluon fusion and $t\bar t H$) and to the total Higgs boson decay width
(i.e.~NP effects in invisible or undetected decay modes).

A detailed formalism, similar to the one developed for EW production and decay process, has not been
developed yet for gluon fusion and $t\bar t H$ production process.
However, it should be stressed that the latter  are on a very different footing compared to EW processes since
they involve a significantly smaller number of observables. Moreover,
a smaller degrees of modellization is required in order to analyse the corresponding
data in generic NP frameworks.
As a result, the combination of PO for the total cross sections, and template-cross-section
analyses of the kinematical distributions, provide an efficient way to report data in a sufficiently general
and unbiased way.

More precisely, for the time being we suggest to introduce the following two PO
\be
\kappa^2_{g} = \left. \frac{ \sigma(pp \to h) }{ \sigma^{\rm SM}(pp \to h) }\right|_{gg-\rm{fusion}}~,
\qquad
\kappa^2_{t} = \left. \frac{ \sigma(pp \to t \bar t h) }{ \sigma^{\rm SM}(pp \to t\bar t h) }\right|_{\rm Yukawa}~,
\ee
in close analogy to what it is presently done within the $\kappa$ formalism.
As far as the gluon fusion is concerned, it is well known that the Higgs boson $p_T$ distribution
carries additional dynamical information about the underlying process.
However, such distribution can be efficiently reported via the template-cross-section method.
Moreover, the steep fall of the $p_T$ spectrum (that is a general consequence of the infrared structure of QCD)
implies that the determination of $\kappa_{gg}$ is practically unaffected
by possible NP effects in this distribution.

Finally, as far as  the Higgs boson width is concerned, we need to  introduce a single effective {\it physical} PO
to account for all the invisible or undetected Higgs boson decay modes.
This additional partial width must be added to the various visible partial widths in order to determine the total Higgs boson width. 
Alternatively, it is possible to define an effective coupling PO as the ratio
\be
\kappa_{H}^2 =   \frac{ \Gamma_{\rm tot} (h) }{ \Gamma_{\rm tot}^{\rm SM} (h) }~.
\ee


\section{PO meet SMEFT \label{POSMEFT}}
\label{sectPO:SMEFT}

One of the main goals of the LHC is to perform high-precision studies of
possible deviations from the SM. Ideally, this would require the following four steps:
i)~for each process write down some (QFT-compatible) amplitude allowing for SM-deviations,
both for the main signal analysed (e.g.~a given Higgs cross-section, close to the resonance)
as well as for the background (non-resonant signal);
ii)~compute fiducial observables;
iii)~fit the signal (SM+NP) via an appropriate set of  conventionally-defined PO,
without subtracting the SM background;
iv)~using the PO thus obtained to derive information on the Wilson coefficients
of an appropriate Lagrangian allowing for deviations from the SM.

%
%

In the previous sections we have discussed a convenient choice for the definition of the PO relevant
to resonant Higgs physics (steps i and iii). In this section we outline how to address the last step
in the case of the so-called  SM Effective Field Theory (SMEFT), i.e.~how to extract the Wilson coefficients of the
SMEFT from the measured PO.

Before starting, it is worth stressing that PO are {\em not} Wilson coefficients, despite one can
derive a linear relation between the two sets of parameters when working at the lowest-order
(LO) in a given Lagrangian framework.
The distinction between PO and Wilson coefficients is quite clear from their different
``status'' in QFT: the PO provide a general parameterization a given
set of on-shell scattering amplitudes and are not Lagrangian parameters.
Once a PO is observed to deviate from its SM value we cannot, without further theoretical assumptions,
 predict deviations in other amplitudes. The latter can be obtained only using a given Lagrangian and
after extracting from data (or better from PO) the corresponding set of Wilson coefficients.
Conversely, Wilson coefficients are scale and scheme dependent  parameters that
require specific theoretical prescriptions to be extracted from physical observables.
This is why the PO can be measured including only the SM THU\footnote{By THU we mean
theoretical uncertainty which has two components, parametric (PU) and missing higher order uncertainties (MHOU)},
while the extraction of SMEFT Wilson coefficients require also an estimate of the
corresponding SMEFT THU\footnote{Although SMEFT converges to SM in the limit
of zero Wilson coefficients, SMEFT and SM are different theories in the UV.}.

There is a line of thought where the Wilson coefficients in any LO EFT approach to physics
beyond the SM are not actual Wilson coefficients, but parameters encoding deformation possibilities.
According to this line of though, PO and and Wilson coefficients are somehow the same object.
But this way of proceeding has a limited applicability, especially if a deviations from the SM is found.
Proceeding along this line one could write an ad-hoc effective Lagrangian, do some calculations at LO
(deviation parameters at tree-level), interpret the data, and limit the considerations to
answer the question ``are there deviations from the SM?". If we want to go a step further,
viz.~answering the question ``What do the deviations from the SM mean?" then it is important to
separate the role of PO and Wilson coefficients. Indeed after extracting the PO, two possibilities appear:
i)~{\em top-down}, namely employ a specific UV model, compute the PO and
try to figure out if it matches or not with the observed deviation;
in such case there will be an uncertainty in projecting down the UV model to the parameters and in the choice
of the input parameter set (IPS); ii)~{\em bottom-up}, namely
do a SMEFT analysis to extract from the PO conclusions on the actual Wilson coefficients;
here there will be an uncertainty from the order at which the calculation is done, as well as a
parametric uncertainty. In the following we illustrate the basic strategy of for the latter (bottom-up)
approach.



\subsection{SMEFT summary \label{summa}}
To establish our notations we observe that in the SMEFT a (lepton number preserving)
amplitude can be written as
\bqa
\mcA &=& \sum_{n=\mrN}^{\infty}\,g^n\,\mcA^{(4)}_n +
       \sum_{n=\mrN_6}^{\infty}\,\sum_{l=1}^n\,\sum_{k=1}^{\infty}\,
        g^n\,\left[\frac{1}{(\sqrt{2}\,G_{\mrF}\,\Lambda^2)^k}\right]^l\,
        \mcA^{(4+2\,k)}_{n\,l\,k} \spc
\label{noclass}
\eqa
where $g$ is a SM coupling. $G_{\mrF}$ is the Fermi coupling constant and $\Lambda$ is
the cut off scale.
$l$ is an index that indicates the number of SMEFT operator insertions leading to the amplitude,
and $k$ indicates the inverse mass dimension of the Lagrangian
terms inserted. $N$ is a label for each individual process, that indicates the order of the
coupling dependence for the leading non vanishing term in the SM (\eg $N = 1$ for
$\myPH \to \myPV\myPV$ \etc but $N = 3$ for $\myPH \to \myPGg\myPGg$). $N_6 = N$ for tree
initiated processes in the SM. For processes that first occur at loop level in the SM,
$N_6 = N - 2$ when operators in the SMEFT can mediate such decays directly thought a contact
operator, for example, through a $\mrdim = 6$ operator for $\myPH \to \myPGg\myPGg$. For instance,
the $\myPH\myPGg\myPGg$ (tree) vertex is generated by
$O_{HB} = \Phi^\dagger \, \Phi \, B^{\mu\nu}\,B_{\mu\nu}$, by
$O_{HW}^{8}= \  \Phi^\dagger \,B^{\mu\nu}\,B_{\mu \rho} \,\mrD^{\rho}\,
\mrD_{\nu}\, \Phi$ \etc
Therefore, SMEFT is a double expansion: in $g$ and $g_{_6} = \mrv^2_{\mrF}/\Lambda^2$ for
pole observables and in $g, g_{_6}\,E^2/\mrv^2_{\mrF}$ for off-shell ones;
furthermore, the combination of parameters $g\,g_6\,\mcA^{(6)}_{1,1,1}$ defines the LO SMEFT
expression for the process while $g^3\,g_6\,\mcA^{(6)}_{3,1,1}$ defines the NLO SMEFT amplitude
in the perturbative expansion.

To summarize, LO SMEFT refers to $\mrdim = 6$ operators in tree diagrams, sometimes called
``contact terms'' while NLO SMEFT refers to one loop diagrams with a single insertion
of $\mrdim = 6$ operators. One can make additional assumptions by introducing classification
schemes in SMEFT. One example of a classification scheme is the Artz-Einhorn-Wudka
``potentially-tree-generated'' (PTG) scenario~\cite{Arzt:1994gp,Einhorn:2013kja}. In this scheme,
it is argued that classes of Wilson coefficients for operators of $\mrdim = 6$ can be argued to
be tree level, or loop level (suppressed by $g^2/16 \, \pi^2$)\footnote{This classification
scheme corresponds only to a subset of weakly coupled and renormalizable UV physics cases.}.
In these cases the expansion in \eqn{noclass} is reorganized in terms of TG (we assume a
BSM model where PTG is actually TG) and LG insertions, \ie LG contact terms and one loop TG
insertions, one loop LG insertions and two loop SM \etc
It is clear that LG contact terms alone do not suffice.

Strictly speaking we are considering here the virtual part of SMEFT, under the
assumptions that LHC PO are defined \`a la LEP, \ie when QED and QCD corrections are deconvoluted.
Otherwise, the real (emission) part of SMEFT should be included and it can be shown that
the infrared/collinear part of the one-loop virtual corrections and of the real ones respect
factorization: the total = virtual ${+}$ real is IR/collinear finite at $\mcO(g^4\,g_{_6})$.

It is worth nothing that SMEFT has limitations, obviously the scale should be such that
$E \muchless \Lambda$.
Understanding SM deviations in tails of distributions requires using SMEFT, but only up to the
point where it stops to be valid, or using the kappa--BSM-parameters connection, \ie replace
SMEFT with BSM models, optimally matching to SMEFT at lower scales.

In any process, the residues of the poles corresponding to unstable particles (starting from
maximal degree) are numbers while the non-resonant part is a multivariate function that requires
some basis, \ie a less model independent, underlying, theory of SM deviations.
That is to say, residue of the poles can be PO by themselves, expressing them in terms
of Wilson coefficients is an operation the can be eventually postponed.
The very end of the chain, the non-resonant part, may require model dependent BSM interpretation.
Numerically speaking, it depends on the impact of the non-resonant part which is small in
gluon-fusion but not in Vector Boson Scattering. Therefore, the focus for data reporting should
always be on real observables, fiducial cross sections and pseudo-observables.

To explain SMEFT in a nutshell consider a process described by some SM amplitude
\bq
\mcA_{\mySM} = \sum_{i=1,n}\,\mcA^{(i)}_{\mySM} \spc
\eq
where $i$ labels gauge-invariant sub-amplitudes. In general, the same process
is given by a contact term or a collection of contact terms of $\mrdim = 6$; for instance,
direct coupling of $\myPH$ to $\myPV\myPV (\myPV= \myPGg, \myPZ, \myPW)$.
In order to construct the theory one has to select a set of higher-dimensional operators and
to start the complete procedure of renormalization. Of course, different sets of operators
can be interchangeable as long as they are closed under renormalization. It is a matter of
fact that renormalization is best performed when using the so-called Warsaw basis, see
\Bref{Grzadkowski:2010es}. Moving from SM to SMEFT we obtain
\bq
\mcA^{\myLO}_{\SMEFT} = \sum_{i=1,n}\,\mcA^{(i)}_{\mySM} + i\,\gds\,\hatkappa_{\mrc} \spc
\qquad
\mcA^{\myNLO}_{\SMEFT} = \sum_{i=1,n}\,\hatkappa_i\,\mcA^{(i)}_{\mySM} +
                i\,\gds\,\hatkappa_{\mrc} +
                \gds\,\sum_{i=1,\ssN}\,a_i\,\mcA^{(i)}_{\nfact} \spc
\label{nloeft}
\eq
where $g^{-1}_{_6} = \sqrt{2}\,\myGF\,\Lambda^2$. The last term in \eqn{nloeft} collects all
loop contributions that do not factorize and the coefficients $a_i$ are Wilson coefficients.
The $\hatkappa_i$ are linear combinations of Wilson coefficients.\footnote{We denote
these combinations of Wilson coefficients  $\hatkappa_i$,  rather than   $\kappa_i$,
in order to distinguish them from the PO defined in the  previous sections.}
We conclude that \eqn{nloeft} gives a consistent and convenient
generalization of the original
$\kappa$-framework at the price of introducing additional, non-factorizable, terms in
the amplitude.

There are several reasons why loops should not be neglected in SMEFT, one is as follows:
consider the ``off-shell''$\Pg\Pg \to \myPH$
fusion~\cite{Kauer:2012hd,Campbell:2014gha,Englert:2015bwa,Ghezzi:2014qpa},
the ``contact'' term is real while the SM amplitude crosses normal thresholds, \eg at
$s = 4\,m^2_{\PQt}$, where $s$ is the Higgs virtuality. Therefore, in the interference one
misses the large effect induced by the SM imaginary part while this effect (of the order of
$5\%$ above the $\PAQt\PQt\,$-threshold) is properly taken into account by the inclusion of
SMEFT loops, also developing an imaginary part after crossing the same normal threshold.
To summarize, the LO part (contact term) alone shows large deviations from the SM around the
$\PAQt\PQt\,$-threshold while the one-loop part reproduces, with the due rescaling, the SM
lineshape in a case where there is no reason to neglect the insertion of PTG operators in loops.
Only the formulation including loops gives an accurate result, with deviations of
$\mathcal{O}(5\%)$ wrt tree (uncritical as long as experimental precision is $\gg 10\%$ but
experiments are getting close).
\subsection{Theoretical uncertainty \label{MHOU}}
A theoretical uncertainty arises when the value of the Wilson coefficients in the PO scenario
is inferred.
A fit defined in a perturbative expansion must always include an estimate of the missing higher
order terms (MHOU)~\cite{Bardin:1999gt}. Various ways exist to estimate this uncertainty, at
any order in perturbation theory. One can compute the same observable with different ``options'',
\eg linearization or quadratization of the squared matrix element, resummation or expansion
of the (gauge invariant) fermion part in the wave function factor for the external legs (does
not apply at tree level), variation of the renormalization scale, $G_{\mathrm{F}}$ renormalization
scheme or $\alpha\,$-scheme, \etc

A conservative estimate of the associated theoretical uncertainty is obtained by taking the
envelope over all ``options''; the interpretation of the envelope is a log-normal distribution
(this is the solution preferred in the experimental community) or a flat Bayesian
prior~\cite{Cacciari:2011ze,David:2013gaa} (a solution preferred in a large part of the
theoretical community). It is clear that MHOU for the SM should always be included.

The notion of MHOU has a long history but it is worth noting that there is no statistical
foundation and that it cannot be derived from a set of consistent (incomplete) principles.
Ideally, calculations should be repeated using a well defined (and definable) set of
options, results from different calculations should be compared and their MHOU assumptions
subjectable to falsification.
Therefore, no estimate of the theoretical errors is general enough and it is clear that there
are several ways to approach the problem with conceptual differences between the
bottom-up and the top-down scenarios.
\subsection{Examples \label{exa}}
In this section we provide a number of examples connecting PO to Wilson coefficients;
results are based on the work of \Brefs{Ghezzi:2015vva,Passarino:2012cb}, and of
\Brefs{Berthier:2015oma,Trott:2014dma,Hartmann:2015oia,Hartmann:2015aia}. For simplicity we
will confine the presentation to CP-even couplings.

\begin{itemize}
\item{\bf{\underline{$\myPH \to \myPAQb\myPQb$}}} At tree level this amplitude is given by
\end{itemize}

\bq
\mcA_{\sPH\myPAQb\PQb} = - \frac{1}{2}\,g\,\frac{m_{\PQb}}{\mw}\,\Bigl[
\mrA^{\mySM}_{\sPH\myPAQb\PQb} +
g_{_6}\,\lpar \apW - \frac{1}{4}\,\apD + \apBox - \abp \rpar \Bigr] \spc
\eq
giving the following connection with Eq.~(\ref{eq:one}) (note that all deviations are real):
\bq
y^{\PQb}_{\mrS} = - \frac{i}{\sqrt{2}}\,\,g\,\frac{m_{\PQb}}{\mw}\,\Bigl[
\mrA^{\mySM}_{\sPH\myPAQb\PQb} +
g_{_6}\,\lpar \apW - \frac{1}{4}\,\apD + \apBox - \abp \rpar \Bigr] \spp
\eq

\begin{itemize}
\item{\bf{\underline{$\myPH \to \myPGg\myPGg$}}}
The amplitude for the process $\myPH(P) \to \myPGg_{\mu}(p_1)\myPGg_{\nu}(p_2)$ can be written as
\end{itemize}

\bq
\mrA^{\mu\nu}_{\sPHAA} = i\,\mcT_{\sPHAA}\,\mrT^{\mu\nu} \spc
\quad
\mhs\,\mrT^{\mu\nu} = p^{\mu}_2\,p^{\nu}_1 - \spro{p_1}{p_2}\,g^{\mu\nu} \spp
\label{HAAamp}
\eq
The $S\,$-matrix element follows from \eqn{HAAamp} when we multiply the amplitude by the
photon polarizations $e_{\mu}(p_1)\,e_{\nu}(p_2)$; in writing \eqn{HAAamp} we have
used $p\,\cdot\,e(p)= 0$. A convenient way for writing the amplitude is the following:
after renormalization we neglect all fermion masses but $m_{\PQt}, m_{\PQb}$ and write
\bq
\mcT_{\sPHAA} =
\frac{g^3_{\mrF}\,\stWs}{8\,\pi^2}\,
\sum_{\ssI=\myPW,\PQt,\PQb}\,\uprho^{\sPHAA}_{\ssI}\,\mcT^{\ssI}_{\sPHAA\,;\,\myLO} +
g_{\mrF}\,\gds\,\frac{\mhs}{\mw}\,\aAA +
\frac{g^3_{\mrF}\,\gds}{\pi^2}\,\mcT^{\nfact}_{\sPHAA} \spc
\label{linkappaHAA}
\eq
where $g^2_{\mrF} = 4\,\srt\,\myGF\,\mws$ and $\ctW= \mw/\mz$. Note that, at this point we have
selected the $\{ G_{\mrF}\,,\,\mz\,,\,\mw \}$ IPS, alternatively one could use the
$\{ \alpha\,,\,G_{\mrF}\,,\,\mz \}$ IPS where
\bq
g^2_{\sPA} = \frac{4\,\pi\,\alpha}{\sths}
\qquad
\sths = \frac{1}{2}\,\Bigl[ 1 - \sqrt{1 - 4\,\frac{\pi\,\alpha}{\srt\,G_{\mrF}\,\mzs}}\Bigr] \spc
\label{defscipsbPO}
\eq
with a numerical difference that enters the MHOU. Referring to \eqn{nloeft} we have
\bq
\hatkappa^{\sPHAA}_{\ssI} = \frac{g^3_{\mrF}\,\stWs}{8\,\pi^2}\,\uprho^{\sPHAA}_{\ssI} \spc
\qquad
\hatkappa^{\sPHAA}_{\mrc} = g_{\mrF}\,\frac{\mhs}{\mw}\,\aAA \spp
\eq
In writing deviations in terms of Wilson coefficients we introduce the following combinations:
\bqa
&\aZZ = \stWs\,a_{\phi\,\sPB} + \ctWs\,a_{\phi\,\sPW} - \stW\,\ctW\,a_{\phi\,\sPW\sPB} \spc
\qquad
\aAA = \ctWs\,a_{\phi\,\sPB} + \stWs\,a_{\phi\,\sPW} + \stW\,\ctW\,a_{\phi\,\sPW\sPB}& \spc
\nl
&\aAZ = 2\,\ctW\,\stW\,\lpar a_{\phi\,\sPW} - a_{\phi\,\sPB} \rpar +
             \lpar 2\,\ctWs - 1 \rpar\,a_{\phi\,\sPW\sPB}& \spp
\eqa
The process dependent $\rho\,$-factors are given by
\bq
\rho^{\proc}_{\ssI} = 1 + \gds\,\Delta\rho^{\proc}_{\ssI} \spc
\label{duk}
\eq
and there are additional, non-factorizable, contributions. For $\myPH \to \myPGg\myPGg$ the
$\Delta\rho$ factors are as follows:
\bqa
\Delta \rho^{\sPHAA}_{\PQt} &=&
            \frac{3}{16}\,\frac{\mhs}{\stW\,\mws}\,\atWB
            + ( 2 - \stWs )\,\frac{\ctW}{\stW}\,\aAZ
            + ( 6 - \stWs )\,\aAA
\nl
{}&-& \frac{1}{2}\,\Bigl[
            \apD + 2\,\stWs\,( \ctWs\,\aZZ - \atp  - 2\,\apBox ) \Bigr]\,\frac{1}{\stWs} \spc
\nl
\Delta \rho^{\sPHAA}_{\PQb} &=&
          - \frac{3}{8}\,\frac{\mhs}{\stW\,\mws}\,\abWB
          + ( 2 - \stWs )\,\frac{\ctW}{\stW}\,\aAZ
          + ( 6 - \stWs )\,\aAA
\nl
{}&-& \frac{1}{2}\,\Bigl[
          \apD + 2\,\stWs\,( \ctWs\,\aZZ + \abp - 2\,\apBox) \Bigr]\,\frac{1}{\stWs} \spc
\nl
\Delta \rho^{\sPHAA}_{\myPW} &=&
           ( 2 + \stWs )\,\frac{\ctW}{\stW}\,\aAZ
          + ( 6 + \stWs )\,\aAA
          - \frac{1}{2}\,\Bigl[
          \apD - 2\,\stWs\,( 2\,\apBox + \ctWs\,\aZZ) \Bigr]\,\frac{1}{\stWs} \spp
\label{kappaHAA}
\eqa
In the PTG scenario we only keep $\atp, \abp, \apD$ and $\apBox$ in \eqn{kappaHAA}.
The advantage of \eqn{linkappaHAA} is to establish a link between the EFT and the $\kappa$-framework,
which has a validity restricted to LO. As a matter of fact \eqn{linkappaHAA} tells us that
appropriate $\hatkappa\,$-factors can be introduced also at the loop level; they are combinations of Wilson
coefficients but we have to extend the scheme with the inclusion of process dependent,
non-factorizable, contributions.

We also derive the following result for the non-factorizable part of the amplitude:
\bq
\mcT^{\nfact}_{\sPHAA} = \mw\,\sum_{a\,\in\,\{A\}}\,\mcT^{\nfact}_{\sPHAA}(a)\,a \spc
\quad \{ A \} = \{ \atWB, \abWB, \aAA, \aAZ, \aZZ \} \spp
\eq
In the PTG scenario all non-factorizable amplitudes for $\myPH \to \myPGg\myPGg$ vanish.
Comparing with Eq.~(\ref{eq:h2gamma}) we obtain
\bq
\epsilon_{\myPGg\myPGg} = - \frac{1}{2}\,\frac{\mrv_{\mrF}}{\mhs}\,\mcT_{\sPHAA} \spc
\quad
\mcT^{\myLO}_{\sPHAA} = \mcT^{\mySM}_{\sPHAA} + g_{\mrF}\,\gds\,\frac{\mhs}{\mw}\,\aAA \spp
\eq

\begin{itemize}
\item{\bf{\underline{$\myPH \to 4\,\myPf$}}}
\end{itemize}

Few additional definitions are needed: by on-shell $\mrS\,$-matrix for an arbitrary process
(involving external unstable particles) we mean the corresponding (amputated) Green's function
supplied with LSZ factors and sources, computed at the (complex) poles of the external
lines~\cite{Grassi:2000dz,Goria:2011wa}. For processes that involve stable
particles this can be straightforwardly transformed into a {\it physical} PO.

The connection of the $\myPH\myPV\myPV, \myPV = \myPZ,\myPW$ (on-shell) $\mrS\,$-matrix with the off shell
vertex $\myPH \to \myPV\myPV$ and the full process $\Pp\Pp \to 4\,\psi$ is more complicated and
is discussed in some detail in Sect.~3 of \Bref{David:2015waa}.
The ``on-shell'' $\mrS\,$-matrix for $\myPH\myPV\myPV$, being built with the the residue of the
$\myPH{-}\myPV{-}\myPV$ poles in $\Pp\Pp \to 4\, \psi$ is gauge invariant by construction (it can be
proved by using Nielsen identities) and represents one of the building blocks for the full
process: in other words, it is a PO.
Technically speaking the ``on-shell'' limit for external legs should be understood
``to the complex poles'' (for a modification of the LSZ reduction formulas for unstable
particles see \Bref{Weldon:1975gu}) but, as well known, at one loop we can use on-shell masses
(for unstable particles) without breaking the gauge parameter independence of the result.
In order to understand the connection with Eqs.~(\ref{eq:h4lF1})--(\ref{eq:h4lF4}), defining neutral current PO
we consider the process
\bq
\myPH(P) \to \myPem(p_1) + \myPep(p_2) + \myPf(p_3) + \myPAf(p_4)
\label{prco}
\eq
where $\myPf \not= \Pe,\PGne$, and introduce the following invariants:
$s_{\sPH}= P^2$, $\sone = q^2_1 = \lpar p_1 + p_2 \rpar^2$ and
$\stwo = q^2_2 = \lpar p_3 + p_4 \rpar^2$,
while $s_i, i=3,\dots,5$ denote the remaining invariants describing the process.
We also introduce $s_{\sPZ}$, the $\myPZ$ complex pole. Propagators are
\bq
\proA{i} = \frac{1}{s_i} \spc
\qquad
\proZ{i} = \mrP^{-1}_{\myPZ}(s_i) = \frac{1}{s_i - s_{\sPZ}} \spp
\label{defprop}
\eq
The total amplitude for process \eqn{prco} is given by the sum of different contributions,
doubly $\myPZ$ resonant \etc
\bqa
\mcA\lpar \myPH \to \myPem\myPep\myPf\myPAf\rpar &=&
\mrA_{\sPZZ}\lpar \cph\,,\,\sone\,,\,\stwo \rpar\,\proZ{\sone}\,\proZ{\stwo}
\nl
{}&+& \mrA_{\sPAA}\lpar \cph\,,\,\sone\,,\,\stwo \rpar\,\proA{\sone}\,\proA{\stwo} +
      \mrA_{\sPAZ}\lpar \cph\,,\,\sone\,,\,\stwo \rpar\,\proA{\sone}\,\proZ{\stwo}
\nl
{}&+& \mrA_{\sPAZ}\lpar \cph\,,\,\stwo\,,\,\sone \rpar\,\proZ{\sone}\,\proA{\stwo} +
      \mrA_{\sPZ}\lpar \cph\,,\,\sone\,,\,\stwo \rpar\,\proZ{\sone}
\nl
{}&+& \mrA_{\sPZ}\lpar \cph\,,\,\stwo\,,\,\sone \rpar\,\proZ{\stwo} +
      \mrA_{\sPA}\lpar \cph\,,\,\sone\,,\,\stwo \rpar\,\proA{\sone}
\nl
{}&+& \mrA_{\sPA}\lpar \cph\,,\,\stwo\,,\,\sone \rpar\,\proA{\stwo} +
      \mrA_{\NR} \spp
\label{deco}
\eqa
To describe in details the various terms in \eqn{deco} we introduce fermion currents defined by
\bqa
J^{\mu}_{\myPZ\,\myPf}\lpar p\,;\,q,k\rpar &=& {\spinub{\myPf}}(q)\,\gamma^{\mu}\,
                \Bigl[ \mcV_{\myPf}(p^2) + \mcA_{\myPf}(p^2)\,\gamma^5 \Bigr]\,\spinv{\myPf}(k)
\nl
{}&=&
 \mcV^{+}_{\myPf}(p^2)\,{\spinubL{\myPf}}(q)\,\gamma^{\mu}\,\,\spinvL{\myPf}(k) +
 \mcV^{-}_{\myPf}(p^2)\,{\spinubR{\myPf}}(q)\,\gamma^{\mu}\,\,\spinvR{\myPf}(k) \spc
\nl
J^{\mu}_{\myPA\,\myPf}\lpar p\,;\,q,k\rpar &=& \mcQ_{\myPf}(p^2)\,
                {\spinub{\myPf}}(q)\,\gamma^{\mu}\,\spinv{\myPf}(k) \spc
\eqa
where $p = q + k$. At tree level we have
\bq
\mcV^{+}_{\myPf} = \frac{g}{\ctW}\,\lpar \mrI^{(3)}_{\myPf} - \mrQ_{\myPf}\,\stWs \rpar \spc
\qquad
\mcV^{-}_{\myPf} = - g\,\mrQ_{\myPf}\,\frac{\stWs}{\ctW} \spc
\qquad
\mcQ_{\myPf} = g\,\mrQ_{\myPf}\,\stW \spp
\eq
The amplitude for $\myPH(P) \to \myPGg(q_1) + \myPGg(q_2)$ is
\bq
\mrA_{\sPAA}\lpar \cph\,,\,\sone\,,\,\stwo \rpar =
\mcT_{\sPHAA}\lpar \cph\,,\,\sone\,,\,\stwo \rpar\,\mrT_{\mu\nu}\lpar q_1\,,\,q_2 \rpar\,
J^{\mu}_{\myPA\,\Pe}\lpar q_1\,;\,p_1,p_2\rpar\,
J^{\nu}_{\myPA\,\myPf}\lpar q_2\,;\,p_3,p_4\rpar \spc
\eq
with $q^2_1= \sone$ and $q^2_2= \stwo$. Similarly, the amplitude for $\myPH(P) \to
\myPZ(q_1) + \myPZ(q_2)$ is
\bqa
\mrA_{\sPZZ}\lpar \cph\,,\,\sone\,,\,\stwo \rpar &=& \Bigl[
\mcP_{\sPHZZ}\lpar \cph\,,\,\sone\,,\,\stwo \rpar\,q_{2\mu}\,q_{1\nu} -
\mcD_{\sPHZZ}\lpar \cph\,,\,\sone\,,\,\stwo \rpar\,g_{\mu\nu}\Bigr]
\nl
{}&\times& J^{\mu}_{\myPZ\,\Pe}\lpar q_1\,;\,p_1,p_2\rpar\,
J^{\nu}_{\myPZ\,\myPf}\lpar q_2\,;\,p_3,p_4\rpar \spp
\eqa
The $\myPZ\myPZ$, $\myPAA$ or $\myPAZ$, doubly-resonant parts of the amplitude are shown in \refF{fig:crabVV}
while the singly, $\myPZ$ or $\myPA$, parts are shown in \refF{fig:acrab}.
\begin{figure}[t]
 \vspace{-8.cm}
 \centering
 \includegraphics[width=0.9\textwidth, trim = 30 250 50 80, clip=true]{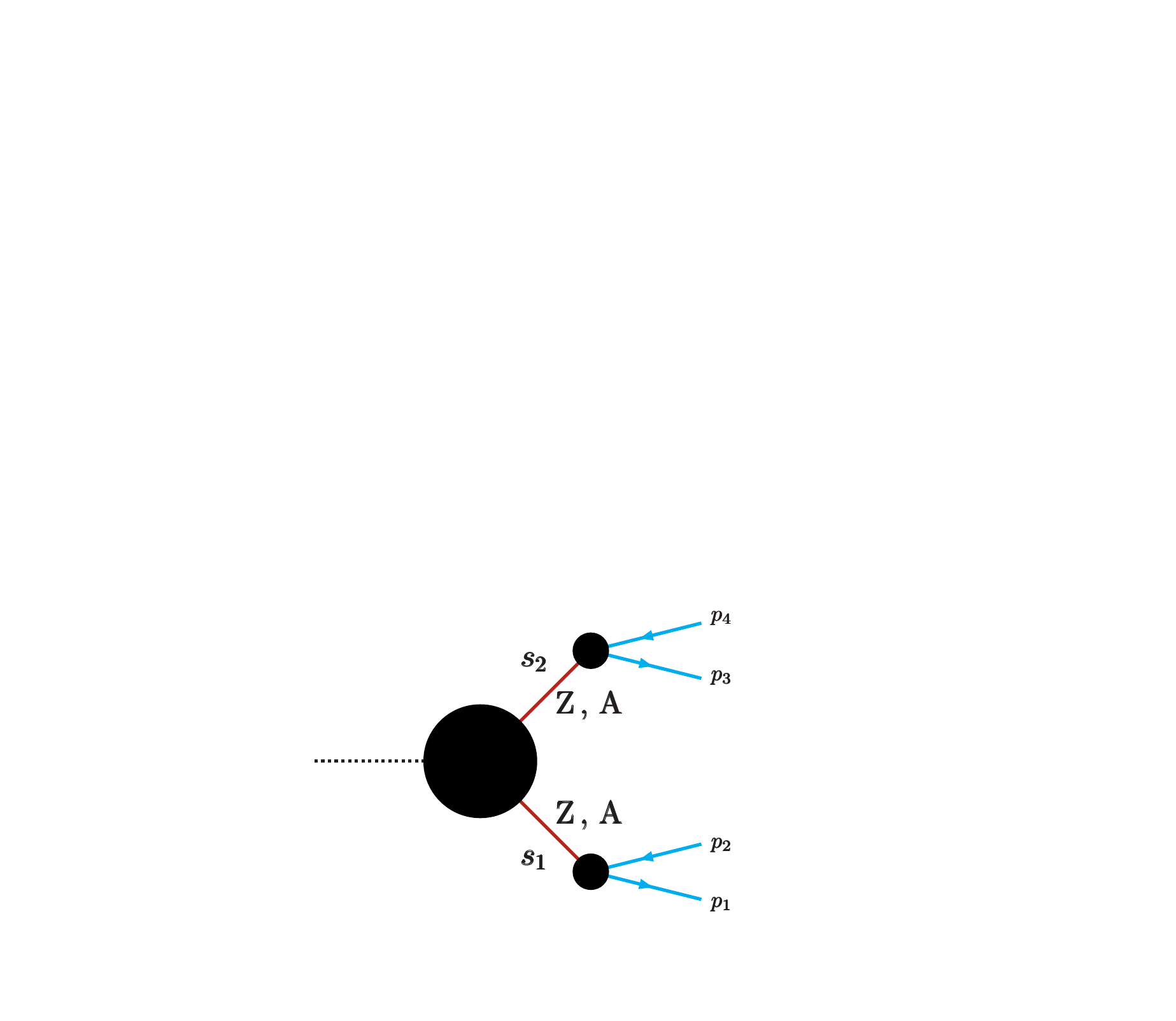}
 \vspace{-1.cm}
 \caption{
Doubly-resonant ($\myPZ\myPZ$, $\myPAA$ or $\myPAZ$) part of the amplitude for the process of
\eqn{prco}.
\label{fig:crabVV}
         }
\end{figure}
\begin{figure}[t]
 \centering
  \vspace{-8.cm}
 \includegraphics[width=0.9\textwidth, trim = 30 250 50 80, clip=true]{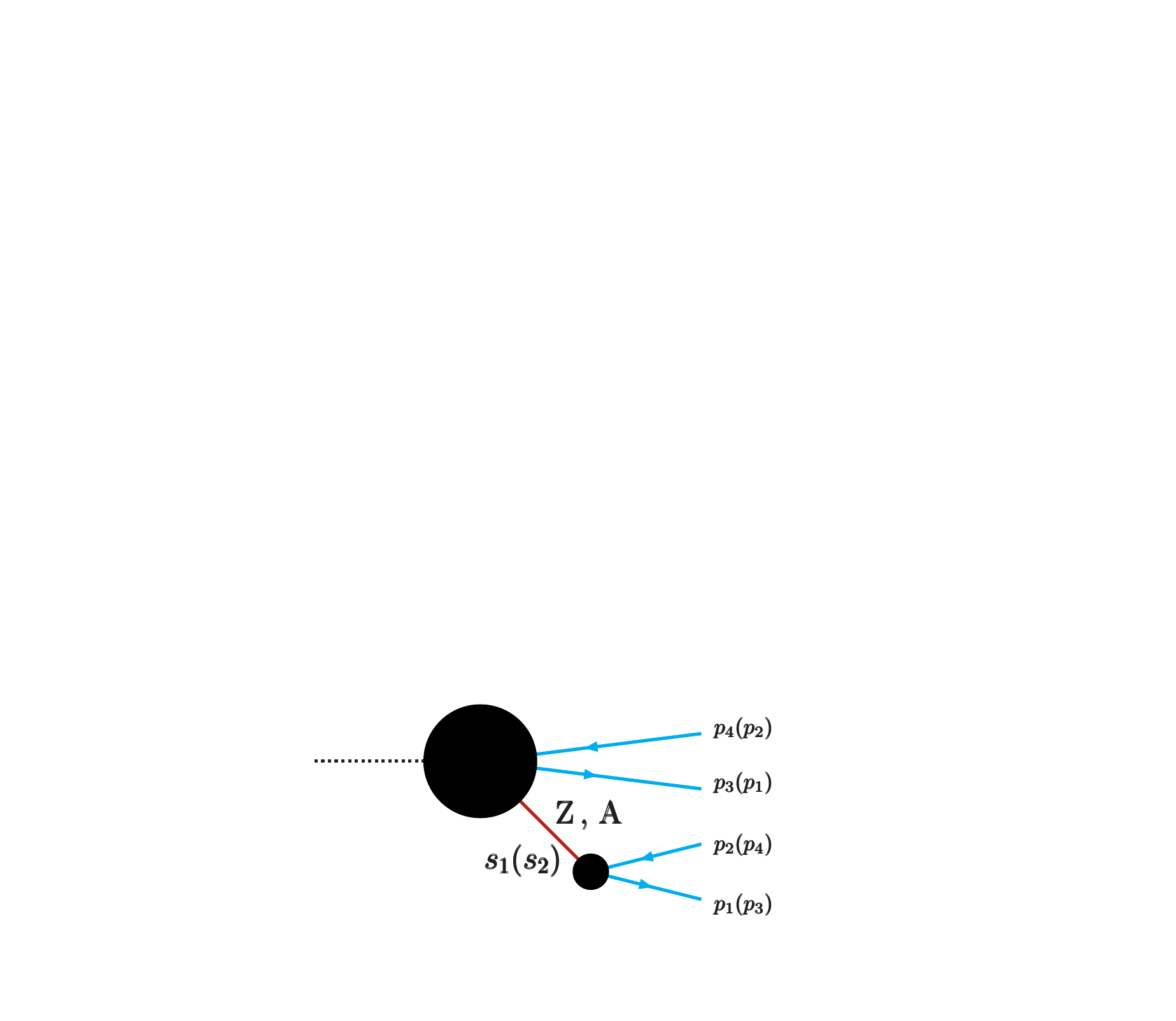}
 \vspace{-1.cm}
 \caption{
Singly-resonant ($\myPZ$ or $\myPA$) part of the amplitude for the process of
\eqn{prco}.
\label{fig:acrab}
         }
\end{figure}
For the singly-resonant amplitudes we write
\bq
\mrA_{\sPZ}\lpar \cph\,,\,\sone\,,\,\stwo \rpar =
\spinub{\myPf}(p_3)\,\mcF_{\sPHZ\,\mu}\lpar \cph\,,\,\sone\,,\,\stwo \rpar \spinv{\myPf}(p_4)\,
J^{\mu}_{\myPZ,\Pe}\lpar q_1\,;\,p_1,p_2\rpar \spc
\eq
where the form factor $\mcF$ is again decomposed as follows:
\bqa
&\spinub{\myPf}(p_3)\,\mcF_{\sPHZ\,\mu}\,\spinv{\myPf}(p_4) =
\sum_i\,\mcF^i_{\sPHZ}\,\mcC_{i\,\mu}& \spc
\nl
&\mcC_{i\,\mu} = \{\,
\spinub{\myPf}(p_3)\,\gamma_{\mu}\,\spinv{\myPf}(p_4) \;,\;
\spinub{\myPf}(p_3)\,\gamma_{\mu}\,\gamma^5\,\spinv{\myPf}(p_4) \;,\;
\dots \,\}& \spp
\eqa
Having the full amplitude we start expanding, \eg
\bqa
\mrA_{\sPZZ}\lpar \cph\,,\,\sone\,,\,\stwo \rpar &=&
   \mrA_{\sPZZ}\lpar \cph\,,\,\cpz\,,\,\stwo \rpar +
   \lpar \sone - \cpz \rpar\,
   \mrA^{(1)}_{\sPZZ}\lpar \cph\,,\,\sone\,,\,\stwo \rpar \spc
\nl
\mrA^{(1)}_{\sPZZ}\lpar \cph\,,\,\sone\,,\,\stwo \rpar &=&
   \mrA^{(1)}_{\sPZZ}\lpar \cph\,,\,\sone\,,\,\cpz \rpar +
   \lpar \stwo - \cpz \rpar\,
   \mrA^{(12)}_{\sPZZ}\lpar \cph\,,\,\sone\,,\,\stwo \rpar \spc
\eqa
\etc The total amplitude of \eqn{deco} can be split into several components,
$\myPZ$ doubly resonant (DR) $\;\dots\;$ $\myPZ$ singly resonant (SR) $\;\dots\;$
non resonant (NR). Note that NR includes multi-leg functions, up to pentagons:
\bqa
\mcA_{\DR\,;\,\sPZZ} &=&
     \mrA_{\sPZZ}\lpar \cph\,,\,\cpz\,,\,\cpz \rpar\,\proZ{\sone}\,\proZ{\stwo} \spc
\nl
\mcA_{\DR\,;\,\sPAA} &=&
     \mrA_{\sPAA}\lpar \cph\,,\,0\,,\,0 \rpar\,\proA{\sone}\,\proA{\stwo} \spc
\nl
\mcA_{\SR\,;\,\sPZ} &=& \Bigl[
     \mrA_{\sPZ}\lpar \cph\,,\,\cpz\,,\,\stwo \rpar +
     \mrA^{(2)}_{\sPAZ}\lpar \cph\,,\,\stwo\,,\,\cpz \rpar +
     \mrA^{(2)}_{\sPZZ}\lpar \cph\,,\,\cpz\,,\,\stwo \rpar \Bigr]\,\proZ{\sone} \spc
\nl
{}&+& \Bigl[
     \mrA_{\sPZ}\lpar \cph\,,\,\cpz\,,\,\sone \rpar +
     \mrA^{(1)}_{\sPAZ}\lpar \cph\,,\,\sone\,,\,\cpz \rpar +
     \mrA^{(1)}_{\sPZZ}\lpar \cph\,,\,\sone\,,\,\cpz \rpar \Bigr]\,\proZ{\stwo} \spc
\nl
\mcA_{\SR\,;\,\sPA} &=& \Bigl[
     \mrA_{\sPA}\lpar \cph\,,\,0\,,\,\stwo \rpar +
     \mrA^{(2)}_{\sPAA}\lpar \cph\,,\,0\,,\,\stwo \rpar +
     \mrA^{(2)}_{\sPAZ}\lpar \cph\,,\,0\,,\,\stwo \rpar ]\,\proA{\sone}
\nl
{}&+& \Bigl[
     \mrA_{\sPA}\lpar \cph\,,\,0\,,\,\sone \rpar +
     \mrA^{(1)}_{\sPAA}\lpar \cph\,,\,\sone\,,\,0 \rpar +
     \mrA^{(1)}_{\sPAZ}\lpar \cph\,,\,0\,,\,\sone \rpar  ]\,\proA{\stwo}
\nl
\mcA_{\DR\,;\,\sPAZ} &=&
     \mrA_{\sPAZ}\lpar \cph\,,\,\cpz\,,\,\sone \rpar\,
     \Bigl[ \proZ{\sone}\,\proA{\stwo} + \proA{\sone}\,\proZ{\stwo} \Bigr]
\nl
\mcA_{\NR} &=&
     \mrA^{(1,2)}_{\DR\,;\,\sPZZ}\lpar \cph\,,\,\sone\,,\,\stwo \rpar +
     \mrA^{(2,1)}_{\DR\,;\,\sPZZ}\lpar \cph\,,\,\sone\,,\,\stwo \rpar +
     \mrA^{(1,2)}_{\DR\,;\,\sPAZ}\lpar \cph\,,\,\stwo\,,\,\sone \rpar
\nl
{}&\quad +& \mrA^{(2,1)}_{\DR\,;\,\sPAZ}\lpar \cph\,,\,\sone\,,\,\stwo \rpar +
            \mrA^{(1,2)}_{\DR\,;\,\sPAA}\lpar \cph\,,\,\sone\,,\,\stwo \rpar +
            \mrA^{(2,1)}_{\DR\,;\,\sPAA}\lpar \cph\,,\,\sone\,,\,\stwo \rpar
\nl
{}&\quad +& \mrA^{(1)}_{\SR\,;\,\sPZ}\lpar \cph\,,\,\sone\,,\,\stwo \rpar +
            \mrA^{(2)}_{\SR\,;\,\sPZ}\lpar \cph\,,\,\stwo\,,\,\sone \rpar +
            \mrA^{(1)}_{\SR\,;\,\sPA}\lpar \cph\,,\,\sone\,,\,\stwo \rpar
\nl
{}&\quad +& \mrA^{(2)}_{\SR\,;\,\sPA}\lpar \cph\,,\,\stwo\,,\,\sone \rpar +
            \mrA_{\NR} \spp
\eqa
Each $\mcA$ amplitude is gauge parameter independent. Let us consider SMEFT at tree level, so that
\bq
\mrA_{\sPAA}\lpar \cph\,,\,\sone\,,\,\stwo \rpar =
\mrA^{\mySM}_{\sPAA}\lpar \cph\,,\,\sone\,,\,\stwo \rpar +
\Delta\mrA_{\sPAA}\lpar \cph\,,\,\sone\,,\,\stwo \rpar \spc
\eq
\etc. In $\Delta\mrA$ we do not include loops with $\mrdim = 6$ insertions. Taking $\myPf = \myPGm$
and neglecting fermion masses we obtain the following result
\bqa
{}&{}& \Delta\mcA\lpar \myPH \to \myPem\myPep\myPGm\myPGp\rpar =
 -\,i\,g^3\,g_{_6}\,
 J^{\mu}_{\mrL}\lpar q_1\,;\,p_1,p_2\rpar\,
 J_{\mu\,\mrL}\lpar q_2\,;\,p_3,p_4\rpar\,\Delta\mcA_{\mrL}
\nl
{}&\qquad\qquad -&
 i\,\frac{g^3\,g_{_6}}{\mw}\,\lpar q_{2\mu}\,q_{1\nu} - \spro{q_1}{q_2}\,g_{\mu\nu} \rpar\,
 J^{\mu}_{\mrL}\lpar q_1\,;\,p_1,p_2\rpar\,
 J^{\nu}_{\mrL}\lpar q_2\,;\,p_3,p_4\rpar \Bigr]\,\Delta\mcA_{\mrT} \spc
%
\eqa
where $J^{\mu}_{\mrL}$ is the left-handed fermion current (fermion masses are neglected)
and $\Delta\mcA_{\mrL\,,\,\mrT}$ are the longitudinal and transverse parts of the LO SMEFT
deviations. We obtain
\bqa
\Delta\mcA_{\mrT} &=&
       2\,\stWs\,\aAA\,\proA{\sone}\,\proA{\stwo} +
      \frac{1}{2}\,\mrv_{\Pl}\,\frac{\stW}{\ctW}\,\aAZ\,
      \Bigl[ \proA{\sone}\,\proZ{\stwo} + \proA{\stwo}\,\proZ{\sone} \Bigr]
\nl
{}&-&
      \frac{1}{2}\,\mrv^2_{\Pl}\,\frac{\aZZ}{\ctWs}\,\proZ{\sone}\,\proZ{\stwo} \spc
\label{Tdev}
\eqa
\bqa
\Delta\mcA_{\mrL} &=&
     \frac{1}{4}\,\frac{\mrv_{\Pl}}{\ctWs}\,\frac{1}{\mw}\,
         \lpar \aplV + \aplA \rpar\,\Bigl[  \proZ{\sone} +\proZ{\stwo} \Bigr]
\nl
{}&-& \frac{1}{16}\,\Bigl[
          \lpar 7 - 20\,\stWs + 12\,\stWq \rpar\,\frac{\apD}{\ctWq}
          + 4\,\mrv^2_{\Pl}\,\frac{\apBox}{\ctWq}
          + 8\,\frac{\mrv_{\Pl}}{\ctWq}\,\lpar \aplV + \aplA \rpar
\nl
{}&+& 4\,\lpar 3 - 7\,\stWs + 4\,\stWvi\rpar\,\frac{\aZZ}{\ctWq}
          - 4\,\lpar 5 - 12\,\stWs + 4\,\stWq\rpar\,\frac{\stW}{\ctWq}\,
            \lpar \stW\,\aAA + \ctW\,\aAZ \rpar
          \Bigr]
\nl
{}&\times& \mw\,\proZ{\sone}\,\proZ{\stwo} \spc
\label{Ldev}
\eqa
where $\mrv_{\Pl} = 1 - 2\,\stWs$. With the help of \eqns{Tdev}{Ldev}  it is straightforward to
establish the relation between the PO of Sect.~4 and the SMEFT Wilson coefficients
(when the complex poles are identified with
on-shell masses). It is worth noting that we have not included dipole operators.

It is worth noting that there are subtleties when the $\myPH$ is off-shell, they are described in
Appendix~C.1 of \Bref{Goria:2011wa}. Briefly, there is a difference between performing an
analytical continuation ($\myPH$ virtuality $\to$ $\myPH$ on-shell mass) in the off-shell decay
width and using leading-pole approximation (LPA) of \Bref{Denner:2005es}, \ie the DR part, where
the matrix element (squared) is projected but not the phase-space. Analytical continuation is a
unique, gauge invariant procedure, the advantage of LPA is that it allows for a straightforward
implementation of cuts.

In order to extend the SMEFT-PO connection to loop-level SMEFT we have to consider various
ingredients separately.

\begin{itemize}
\item{\bf{\underline{$\myPH \to \myPZ\myPZ$}}}
The amplitude for $\myPH(P) \to \myPZ_{\mu}(p_1)\myPZ_{\nu}(p_2)$ can be written as
\end{itemize}

\bq
\mrA^{\mu\nu}_{\sPHZZ} = i\,\lpar
\mcP^{11}_{\sPHZZ}\,p^{\mu}_1\,p^{\nu}_1 +
\mcP^{12}_{\sPHZZ}\,p^{\mu}_1\,p^{\nu}_2 +
\mcP^{21}_{\sPHZZ}\,p^{\mu}_2\,p^{\nu}_1 +
\mcP^{22}_{\sPHZZ}\,p^{\mu}_2\,p^{\nu}_2 -
\mcD_{\sPHZZ}\,g^{\mu\nu} \rpar \spp
\label{HZZamp}
\eq
The result in \eqn{HZZamp} is fully general and can be used to prove Ward-Slavnov-Taylor
identities (WSTI). As far as the partial decay width is concerned only $\mcP^{21}_{\sPHZZ}
\equiv \mcP_{\sPHZZ}$ will be relevant, due to $p\,\cdot\,e(p) = 0$ where $e$ is the polarization
vector. Note that computing WSTI requires additional amplitudes, \ie $\myPH \to \Ppz\myPGg$
and $\myPH \to \Ppz\Ppz$. The result can be written as follows:
\bqa
\mcD_{\sPHZZ} &=& - g_{\ssF}\,\frac{\mw}{\ctWs}\,\rho^{\sPHZZ}_{\ssD\,;\,\myLO}
   + \frac{g^3_{\ssF}}{\pi^2}\,\Bigl[ \sum_{\PI{=} \PQt,\PQb,\myPW}\,
   \rho^{\sPHZZ}_{\PI\,;\,\ssD\,;\,\myNLO}\,\mcD^{\PI}_{\sPHZZ\,;\,\myNLO} +
   \mcD^{(4)\,;\,\nfact}_{\sPHZZ}
 + \gds\,\sum_{\{a\}}\,\mcD^{(6)\,;\,\nfact}_{\sPHZZ}(a) \Bigr] \spc
\nl\nl
\mcP_{\sPHZZ} &=& 2\,g_{\ssF}\,\gds\,\frac{\aZZ}{\mw}
   + \frac{g^3_{\ssF}}{\pi^2}\,\Bigl[ \sum_{\PI{=} \PQt,\PQb,\myPW}\,
   \rho^{\sPHZZ}_{\PI\,;\,\ssP\,;\,\myNLO}\,\mcP^{\PI}_{\sPHZZ\,;\,\myNLO} +
   \gds\,\sum_{\{a\}}\,\mcP^{(6)\,;\,\nfact}_{\sPHZZ}(a) \Bigr] \spp
\eqa
\bq
\Delta \rho^{\sPHZZ}_{\ssD\,;\,\myLO} =
    \stWs\,\aAA + \Bigl[ 4 + \ctWs\,\lpar 1 - \frac{\mhs}{\mws} \rpar \Bigr]\,\aZZ
    + \ctW\,\stW\,\aAZ + 2\,\apBox \spc
\eq
\bqa
\Delta \rho^{\sPHZZ}_{\PQq\,;\,\ssD\,;\,\myNLO} &=&
\Delta \rho^{\sPHZZ}_{\PQq\,;\,\ssP\,;\,\myNLO} =
 2\,\mrI^{(3)}_{\PQq}\,\aqp + 2\,\apBox - \frac{1}{2}\,\apD + 2\,\aZZ + \stWs\,\aAA \spc
\nl
\Delta \rho^{\sPHZZ}_{\myPW\,;\,\ssD\,;\,\myNLO} &=&
          \frac{1}{12}\,\lpar 4 + \frac{1}{\ctWs} \rpar\,\apD
          + 2\,\apBox
          + \stWs\,\aAA
\nl
{}&+& \stWs\,\lpar 3\,\ctW + \frac{5}{3}\,\frac{1}{\ctW} \rpar\,\aAZ
          + \lpar 4 + \ctWs \rpar\,\aZZ \spc
\nl
\Delta \rho^{\sPHZZ}_{\myPW\,;\,\ssP\,;\,\myNLO} &=&
    4\,\apBox + \frac{5}{2}\,\apD + 12\,\aZZ + 3\,\stWs\,\aAA
\eqa
It is convenient to define sub-amplitudes; however, to respect a factorization into
$\PQt, \PQb$ and bosonic components, we have to introduce the following quantities:
\bq
\begin{array}{ll}
\mrW_{\myPH} = \mrW_{\myPH\,\;\,\myPW} + \mrW_{\myPH\,\;\,\PQt}  + \mrW_{\myPH\,\;\,\PQb} \quad & \quad
\mrW_{\myPZ} = \mrW_{\myPZ\,\;\,\myPW} + \mrW_{\myPZ\,\;\,\PQt}  + \mrW_{\myPZ\,\;\,\PQb} +
               {\overline{\sum}}_{\gen}\,\mrW_{\myPZ\,;\,\myPf} \\
\ssdCZ_{g} = \ssdCZ_{g\,;\,\myPW} + \sum_{\gen}\,\ssdCZ_{g\,;\,\myPf} \quad & \quad
\ssdCZ_{\ctW} = \ssdCZ_{\ctW\,;\,\myPW}  + \ssdCZ_{\ctW\,;\,\PQt}  + \ssdCZ_{\ctW\,;\,\PQb}
               + {\overline{\sum}}_{\gen}\,\ssdCZ_{\ctW\,;\,\myPf} \\
\ssdCZ_{\mw} = \ssdCZ_{\mw\,;\,\myPW} + \sum_{\gen}\,\ssdCZ_{\mw\,;\,\myPf} & \\
\end{array}
\eq
where $\mrW_{\phi\,;\,\phi}$ denotes the $\phi$ component of the $\phi$ (LSZ)
wave-function factor \etc
By $\ssdCZ_{\mathrm{par}}$ we denote the (UV finite) counterterm that is needed in
connecting the renormalized parameters to an input parameter set (IPS). In the actual
calculation we use IPS = $\{G_{\mrF}\,,\,\mz\,,\,\mw\}$.
Furthermore, $\sum_{\gen}$ implies summing over all fermions and all generations, while
${\overline{\sum}}_{\gen}$ excludes $\PQt$ and $\PQb$ from the sum.
\bq
\mcD^{(4)\,;\,\nfact}_{\sPHZZ} = \frac{1}{32}\,\frac{\mw}{\ctWs}\,\lpar
   2\,{\overline{\sum}}_{\gen}\,\mrW_{\myPZ\,;\,\myPf} - \sum_{\gen}\,\ssdZ_{\mw\,;\,\myPf}
   + 4\,{\overline{\sum}}_{\gen}\,\ssdZ_{\ctW\,;\,\myPf}
- 2\,\sum_{\gen}\,\ssdZ_{g\,;\,\myPf} \rpar \spp
\eq
The connection to Eqs.~(20)--(24) is given by
\bq
\mrv_{\mrF}\,g^f_{\myPZ}\,g^{f'}_{\myPZ}\,\mcP_{\sPHZZ} = - 2\,\epsilon_{\myPZ\myPZ} \spc
\qquad
\mrv_{\mrF}\,g^f_{\myPZ}\,g^{f'}_{\myPZ}\,
\lpar \spro{p_1}{p_2}\,\mcP_{\sPHZZ} - \mcD_{\sPHZZ} \rpar = 2\,\mzs\,\kappa_{\myPZ\myPZ} \spp
\eq

\begin{itemize}
\item{\bf{\underline{$\myPH \to \myPW\myPW$}}}
The derivation of the amplitude for $\myPH \to \myPW\myPW$ follows closely the one for
$\myPH \to \myPZ\myPZ$.
\end{itemize}
\bqa
\mcD_{\sPHWW} &=& - g_{\ssF}\,\mw\,\rho^{\sPHWW}_{\ssD\,;\,\myLO}
   + \frac{g^3_{\ssF}}{\pi^2}\,\Bigl[ \sum_{\PI{=} \PQq,\myPW}\,
   \rho^{\sPHWW}_{\PI\,;\,\ssD\,;\,\myNLO}\,\mcD^{\PI}_{\sPHWW\,;\,\myNLO} +
   \mcD^{(4)\,;\,\nfact}_{\sPHWW}
 + \gds\,\sum_{\{a\}}\,\mcD^{(6)\,;\,\nfact}_{\sPHWW}(a) \Bigr] \spc
\nl\nl
\mcP_{\sPHWW} &=& 2\,g_{\ssF}\,\gds\,\frac{1}{\mw}\,\apW
 + \frac{g^3_{\ssF}}{\pi^2}\,\Bigl[ \sum_{\PI{=} \PQq}\,
   \rho^{\sPHWW}_{\PI\,;\,\ssP\,;\,\myNLO}\,\mcP^{\PI}_{\sPHWW\,;\,\myNLO} +
   \gds\,\sum_{\{a\}}\,\mcP^{(6)\,;\,\nfact}_{\sPHWW}(a) \Bigr] \spc
\eqa
where we have introduced
\bq
\mcD^{(4)\,;\,\nfact}_{\sPHWW} = \frac{1}{32}\,\mw\,{\overline{\sum}}_{\gen}\,\lpar
   2\,\mrW_{\myPW\,;\,\myPf}
   - \ssdCZ_{\mw\,;\,\myPf}
   - 2\,\ssdCZ_{g\,;\,\myPf} \rpar \spp
\eq
\bq
\Delta \rho^{\sPHWW}_{\ssD\,;\,\myLO} =
 \stWs\,\lpar \frac{\mhs}{\mw} - 5\,\mw \rpar\,\lpar \aAA + \aAZ + \aZZ \rpar
+ \frac{1}{2}\,\mw\,\apD - 2\,\mw\,\apBox \spc
\eq
\bqa
\Delta \rho^{\sPHWW}_{\PQq\,;\,\ssD\,;\,\myNLO} &=&
\Delta \rho^{\sPHWW}_{\PQq\,;\,\ssP\,;\,\myNLO} =
           \aptV + \aptA + \apbV + \apbA
           - \abp + 2\,\apBox - \frac{1}{2}\,\apD
\nl
{}&+& \stW\,\abWB + \ctW\,\abBW + 5\,\stWs\,\aAA + 5\,\ctW\,\stW\,\aAZ + 5\,\ctWs\,\aZZ \spc
\nl
\Delta \rho^{\sPHWW}_{\myPW\,;\,\ssD\,;\,\myNLO} &=&
           \frac{1}{96}\,\frac{\stWs}{\ctWs}\,\apD
          + \frac{23}{12}\,\apBox
          - \frac{35}{96}\,\apD
\nl
{}&+& 4\,\,\stWs\,\aAA
          + \frac{1}{12}\,\stW\,\lpar 3\,\frac{1}{\ctW} + 49\,\ctW \rpar\,\aAZ
          + \frac{1}{2}\,\lpar 9\,\ctWs + \stWs \rpar\,\aZZ \spc
\nl
\Delta \rho^{\sPHWW}_{\myPW\,;\,\ssP\,;\,\myNLO} &=&
           7\,\apBox - 2\,\apD  + 5\,\stWs\,\aAA
          + 5\,\ctW\,\stW\,\aAZ + 5\,\ctWs\,\aZZ \spp
\eqa
These results allow us to write $\epsilon_{\myPW\myPW}$ and $\kappa_{\myPW\myPW}$ of
Eqs.~(25)--(27) in terms of Wilson coefficients.

\begin{itemize}
\item{\bf{\underline{$\myPZ \to \myPAf\myPf$}}}
Let us consider the $\myPZ\myPAf\PF$ vertex, entering the process $\myPH \to 4\,\myPf$:
\end{itemize}

\bq
J^{\mu}_{\myPZ\,\myPf}\lpar p\,;\,q,k\rpar =
{\spinub{\myPf}}(q)\,\Bigl\{
\gamma^{\mu}\,\Bigl[ \mcV_{\myPf}(p^2) + \mcA_{\myPf}(p^2)\,\gamma^5 \Bigr] +
\mcT_{\myPf}(p^2)\,\sigma^{\mu\nu}\,p_{\nu}\,
\spinv{\myPf}(k) \Bigr\} \spp
\eq
At lowest order we have deviations defined by
\bq
\mcV_{\myPf} = i\,\mrI^{3)}_{\myPf}\,\frac{g}{\ctW}\,
             \lpar \mrv_{\myPf} + g_{_6}\,\Delta\mcV_{\myPf} \rpar \spc
\quad
\mcA_{\myPf} = i\,\mrI^{(3)}_{\myPf}\,\frac{g}{\ctW}\,
             \lpar \frac{1}{2} + g_{_6}\,\Delta\mcA_{\myPf} \rpar \spc
\quad
\mcT_{\myPf} = - \frac{g}{2}\,g_{_6}\,\frac{m_{\myPf}}{\mw}\,\afWB \spc
\eq
\bqa
\Delta\mcV_{\myPf} &=&
           \apfV +
           \Bigl[ \mrv_{\myPf} + \ctWs\,\lpar \mrv_{\myPf} - 1 \rpar \Bigr]\,
           \lpar  \aAA + \frac{\ctW}{\stW}\,\aAZ - \frac{1}{4\,\stWs}\,\apD \rpar +
           \lpar 1 - \mrv_{\myPf} \rpar\,\ctWs\,\aZZ \spc
\nl
\Delta\mcA_{\myPf} &=&
           \apfA + \frac{1}{2}\,\lpar \apW - \frac{1}{4}\,\apD \rpar \spp
\eqa
where the vector couplings are
$\mrv_{\PQu} = 1/2 - 2\,Q_{\PQu}\,\stWs$,
$\mrv_{\PQd} = 1/2 + 2\,Q_{\PQd}\,\stWs $
and $\PQu, \PQd$ are generic up, down fermions.
When loops are included the decomposition in gauge invariant sub-amplitudes is not as simple
as in the previous case, fermion loops and boson loops. Here the decomposition is
given in terms of abelian and non-abelian ($\myPZ$ and $\myPW$) parts, $\mrQ\,$-components
(those proportional to $\gamma^{\mu}$) and $\mrL\,$-parts (those proportional to
$\gamma^{\mu}\,\gamma_{+}$). Details can be found in Sect.~6.15 of \Bref{Bardin:1999ak}.
The general expression in SMEFT will not be reported here. It is worth noting that
\bq
\mrA_{\myPZ\myPAf\myPf}(P^2) = \mrA^{\inv}_{\myPZ\myPAf\myPf}(\mzs) + \lpar P^2 - \mzs \rpar\,
                         \mrA^{\xi}_{\myPZ\myPAf\myPf}(P^2) \spc
\eq
where $\xi$ denotes the collection of gauge parameters.

\begin{itemize}
\item{\bf{\underline{$\myPW \to \PQu\PQd$}}}
Similarly to the $\myPZ$ vertex we obtain
\end{itemize}

\bq
i\,\frac{g}{2\,\sqrt{2}}\,\gamma^{\mu}\,\Bigl[
    \mrV^{(+)}_{\PF}\,\gamma^+ +
    \mrV^{(-)}_{\PF}\,\gamma^- \Bigr] +
  \frac{g}{4}\,g_{_6}\,\lpar \frac{m^2_{\PQU}}{\mw}\,\aUW - \frac{m^2_{\PQD}}{\mw}\,\aDW \rpar\,
   \sigma^{\nu\mu}\,p_{\nu} \spc
\eq
\bq
\mrV^{(+)}_{\PF} = 1 + \sqrt{2}\,g_{_6}\,\lpar \apFt + \frac{1}{2}\,\apW \rpar \spc
\qquad
\mrV^{(-)}_{\PF} = \sqrt{2}\,g_{_6}\,\apUD \spp
\eq
Here $\PF$ is a generic doublet of components $\PQU = \PQu$ or $\PGnl$ and
$\PQD = \PQd$ or $\Pl$. Note that $\apUD = 0$ for leptons. The general expression in SMEFT will
not be reported here.

\begin{itemize}
\item{\bf{\underline{$\myPZ \to \myPW\myPW$}}}
Triple gauge boson couplings are dscribed by the following deviations (all momenta flowing
inwards):
\end{itemize}

\bqa
{}&{}& \mrV^{\mu\nu\rho}_{\sPZWW}\lpar p_1\,,\,p_2\,,\,p_3 \rpar =
     g\,\ctW\,\mrF^{\mu\nu\rho}\lpar p_1\,,\,p_2\,,\,p_3 \rpar +
    \frac{3}{2}\,g\,g_{_6}\,\mrH^{\mu\nu\rho}\lpar p_1\,,\,p_2\,,\,p_3 \rpar\,
     \frac{\aQW}{\mws}
\nl
{}&\qquad +&  g\,g_{_6}\ctW\,\mrF^{\mu\nu\rho}\lpar p_1\,,\,p_2\,,\,p_3 \rpar\,
     \Bigl[ \lpar 1 - 2\,\stWs \rpar\,\lpar \frac{\stW}{\ctW}\,\apWB - \apW \rpar +
            2\,\stWs\,\apB + \frac{1}{4}\,\apD \Bigr]
\nl
{}&\qquad +&  g\,g_{_6}\ctW\,\stW\,\mrG^{\mu\nu\rho}\lpar p_1\,,\,p_2\,,\,p_3 \rpar\,
     \Bigl[ \lpar 1 - 2\,\stWs \rpar\,\,\frac{\stW}{\ctW} + 2\,\ctW \Bigl]\,\apWB \spc
\eqa
\bqa
{}&{}& \mrF^{\mu\nu\rho}\lpar p_1\,,\,p_2\,,\,p_3 \rpar =
         p^{\rho}_1\,p^{\mu}_2\,p^{\nu}_3
       - p^{\nu}_1\,p^{\rho}_2\,p^{\mu}_3
       + g^{\nu\rho}\,\lpar p^{\mu}_3\,\spro{p_1}{p_2} - p^{\mu}_2\,\spro{p_1}{p_3} \rpar
\nl
{}&\qquad +& g^{\nu\mu}\,\lpar p^{\rho}_2\,\spro{p_1}{p_3} - p^{\rho}_1\,\spro{p_2}{p_3} \rpar
       + g^{\rho\mu}\,\lpar p^{\nu}_1\,\spro{p_2}{p_3} - p^{\nu}_3\,\spro{p_1}{p_2} \rpar \spc
\nl
{}&{}& \mrG^{\mu\nu\rho}\lpar p_1\,,\,p_2\,,\,p_3 \rpar =
         g^{\nu\rho}\,\lpar p^{\mu}_2 - p^{\mu}_3 \rpar
       + g^{\nu\mu}\,\lpar p^{\rho}_1 - p^{\rho}_2 \rpar
       + g^{\rho\mu}\,\lpar p^{\nu}_3 - p^{\nu}_1 \rpar \spc
\nl
{}&{}& \mrH^{\mu\nu\rho}\lpar p_1\,,\,p_2\,,\,p_3 \rpar =
         g^{\nu\rho}\,\lpar p^{\mu}_3 - p^{\mu}_2 \rpar
       + g^{\nu\mu}\,\lpar p^{\rho}_2 - p^{\nu}_3 \rpar \spp
\eqa

\begin{itemize}
\item{{\underline{VBF}}}
The process that we want to consider is
\end{itemize}

\bq
\PQu(p_1) + \PQu(p_2) \to \PQu(p_3) + \myPem(p_4) + \myPep(p_5) + \myPGm(p_6) + \myPGp(p_7) + \PQu(p_8)
\spp
\eq
At LO SMEFT we introduce the triply-resonant (TR) part of the amplitude ($t\,$-channel
propagators are never resonant):
\bq
J^{\mu}_{\pm}(p_i\,,\,p_j) = \spinub{p_i}\,\gamma^{\mu}\,\gamma_{\pm}\,\spinu{p_j} \spc
\qquad
\Delta^{-1}_{\Phi}(p) = s - M^2_{\Phi}  \spc
\quad s= p^2 \spc
\eq
\bqa
\mcA^{\TR}_{\myLO} &=&
\Bigl[ J^{\mu}_{-}(p_4\,,\,p_5)\,(1 - v_{\Pl}) + J^{\mu}_{+}(p_4\,,\,p_5)\,(1 + v_{\Pl}) \Bigr]\,
\Bigl[ J^{-}_{\mu}(p_6\,,\,p_7)\,(1 - v_{\Pl}) + J^{+}_{\mu}(p_6\,,\,p_7)\,(1 + v_{\Pl}) \Bigr]
\nl &\times&
\Bigl[ J^{\nu}_{-}(p_3\,,\,p_2)\,(1 - v_{\PQu}) + J^{\nu}_{+}(p_3\,,\,p_2)\,(1 + v_{\PQu}) \Bigr]\,
\Bigl[ J^{-}_{\nu}(p_8\,,\,p_1)\,(1 - v_{\PQu}) + J^{+}_{\nu}(p_8\,,\,p_1)\,(1 + v_{\PQu}) \Bigr]
\spc
\nl
\eqa
\bqa
\mcA^{\TR}_{\SMEFT} &=& -\,\frac{g^6}{4096}\,
\Delta_{\myPH}(q_1+q_2)\,\prod_{i=1,4}\,\Delta_{\myPZ}(q_i)\,\frac{\mws}{c^8_{\theta}}\,
\rho_{\myLO}\,\mcA^{\TR}_{\myLO} + g^6\,g_{_6}\,\mcA^{\TR\,;\,\nfact}_{\SMEFT}
\nl
\Delta \rho_{\myLO} &=& 2\,\apBox
- \frac{2\,\mzs - 2\,\mhs + \spro{q_1}{q_2} + \spro{q_2}{q_2}}{\mws}\,\ctWs\,\aZZ \spc
\eqa
where $q_1 = p_8 - p_1,\; q_2 = p_3 - p_2$ are the incoming momenta in VBF and
$q_3 = p_4 + p_5,\; q_4 = p_6 + p_7$ are the outgoing ones. Furthermore,
$\gamma_{\pm} = 1 \pm \gamma^5$.
\subsection{SMEFT and {\it physical} PO \label{physPO}}
In this section we describe the connection between a possible realization of {\it physical} PO
and SMEFT.

Multi pole expansion (MPE) has a dual role: as we mentioned, poles and their residues are
intimately related to the gauge invariant splitting of the amplitude (Nielsen identities);
residues of poles (after squaring the amplitude and after integration over residual variables)
can be interpreted as {\it physical} PO, which requires factorization into subprocesses.
However, gauge invariant splitting is not the same as ``factorization'' of the process into
sub-processes, indeed phase space factorization requires the pole to be inside the physical
region. For all technical details we refer to the work in Sect.~3 of \Bref{David:2015waa}
which is based on the following decomposition of the square of a propagator
\bq
\Delta = \frac{1}{\lpar s - M^2\rpar^2 + \Gamma^2\,M^2} =
\frac{\pi}{M\,\Gamma}\,\delta\lpar s - M^2\rpar +
\mathrm{PV}\,\left[ \frac{1}{\lpar s - M^2\rpar^2}\right] \spp
\label{deltap}
\eq
and on the n-body decay phase space
\bq
d\Phi_n\lpar P, p_1 \dots p_n\rpar =
\frac{1}{2\,\pi}\,dQ^2\,d\Phi_{n-j+1}\lpar P, Q, p_{j+1} \dots p_n\rpar\,
d\Phi_j\lpar Q, p_1 \dots p_j \rpar \spp
\label{fact}
\eq
To ``complete'' the decay ($d\Phi_j$) we need the $\delta\,$-function in \eqn{fact}.
We can say that the $\delta\,$-part of the resonant (squared) propagator opens the corresponding
line allowing us to define {\it physical} PO ($t\,$-channel propagators cannot be cut). Consider the
process $\PQq\PQq \to \myPAf_1\myPf_1\myPAf_2\myPf_2 jj$, according to the structure of the resonant
poles we have different options in extracting {\it physical} PO, \eg
\bqa
\sigma(\PQq\PQq \to \myPAf_1\myPf_1\myPAf_2\myPf_2 jj) &\stackrel{PO}{\longmapsto}&
\sigma(\PQq\PQq \to \myPH jj)\,\mathrm{Br}(\myPH \to \myPZ\myPAf_1\myPf_1)\,
\mathrm{Br}(\myPZ \to \myPAf_2\myPf_2) \spc
\nl
\sigma(\PQq\PQq \to \myPAf_1\myPf_1\myPAf_2\myPf_2 jj) &\stackrel{PO}{\longmapsto}&
\sigma(\PQq\PQq \to \myPZ\myPZ jj)\,\mathrm{Br}(\myPZ \to \myPAf_1\myPf_1)\,
\mathrm{Br}(\myPZ \to \myPAf_2\myPf_2) \spp
\eqa
There are fine points when factorizing a process into ``physical'' sub-processes (PO):
extracting the $\delta$ from the (squared) propagator, \eqn{deltap}, does not necessarily
factorize the phase space; if cuts are not introduced, the interference terms among different
helicities oscillate over the phase space and drop out, \ie we achieve factorization,
see \Brefs{Uhlemann:2008pm}.
Furthermore, MPE should be understood as ``asymptotic expansion'', see
\Brefs{Nekrasov:2007ta,Tkachov:1999qb}, not as Narrow-Width-Approximation (NWA). The phase space
decomposition obtains by using the two parts in the propagator expansion of \eqn{fact}: the
$\delta\,$-term is what we need to reconstruct PO, the PV-term (understood as a
distribution~\cite{Nekrasov:2007ta}) gives the remainder and PO are extracted without making
any approximation. It is worth noting that, in extracting PO, analytic continuation (on-shell
masses into complex poles) is performed only after integrating over residual
variables~\cite{Goria:2011wa}.

We can illustrate the SMEFT - MPE - PO connection by using a simple but non-trivial example:
Dalitz decay of the Higgs boson, see \Brefs{Passarino:2013nka,David:2015waa}. Consider the process
\bq
\myPH(P) \to \myPAf(p_1) + \myPf(p_2) + \myPGg(p_3) \spc
\label{prct}
\eq
and introduce invariants $s_{\sPH}= - P^2$, $s = - \lpar p_1 + p_2 \rpar^2$
and propagators, $\proA{i} = 1/s_i, \; \proZ{i} = 1/(s_i - s_{\sPZ})$.
With $s_{\sPH} = \muhs - i\,\muh\,\gh$ we denote the $\myPH$ complex pole \etc
In the limit $m_{\myPf} \to 0$ the total amplitude for process \eqn{prct} is given by the sum of
three contributions, $\myPZ,\myPA\,$-resonant and non-resonant:
\bq
\mcA\lpar \myPH \to \myPAf\myPf\myPGg\rpar =
 \Bigl[ \mrA^{\mu}_{\sPZ}\lpar \cph\,,\,s \rpar\,\proZ{s} +
        \mrA^{\mu}_{\sPA}\lpar \cph\,,\,s \rpar\,\proA{s} \Bigr]\,e_{\mu}\lpar p_3,l\rpar +
 \mrA_{\NR} \spc
\label{dect}
\eq
where $e_{\mu}$ is the photon polarization vector. The two resonant components are given by
\bq
\mrA^{\mu}_{\sPV}\lpar \cph\,,\,s \rpar =
\mcT_{\sPHAV}\lpar \cph\,,\,s \rpar\,\mrT^{\mu}_{\nu}\lpar q\,,\,p_3 \rpar\,
J^{\nu}_{\myPV\,\myPf}\lpar q\,;\,p_1,p_2\rpar \spc
\eq
where $J^{\mu}_{\myPV\,\myPf}$ is the $\myPV$ fermion ($\myPf$) current, $\myPV = \myPA,\myPZ$, $q = p_1 + p_2$
and
$
\mrT^{\mu\nu}\lpar k_1\,,\,k_2 \rpar =  \spro{k_1}{k_2}\,g^{\mu\nu} - k^{\nu}_1\,k^{\mu}_2
$.
Having the full amplitude we start expanding (MPE) according to
\bq
\mcT_{\sPHAZ}\lpar \cph\,,\,s \rpar =
\mcT_{\sPHAZ}\lpar \cph\,,\,\cpz \rpar +
\lpar s - \cpz \rpar\,\mcT^{(1)}_{\sPHAZ}\lpar \cph\,,\,s \rpar
\qquad \mbox{\etc}
\label{DMPE}
\eq
Derivation continues till we define physical PO:
\bqa
\Gamma_{\PO}\lpar \myPH \to \myPZ\myPGg \rpar &=&
\frac{1}{16\,\pi}\,\frac{1}{\mh}\,\lpar 1 - \frac{\muZs}{\mhs} \rpar\,
\mrF_{\myPH \to \myPZ\myPGg}\lpar \cpz\,,\,\muZs\rpar \spc
\nl
\Gamma_{\PO}\lpar \myPZ \to \myPAf\myPf \rpar &=&
\frac{1}{48\,\pi}\,\frac{1}{\muZ}\,
\mrF_{\myPZ \to \myPAf\myPf}\lpar \cpz\,,\,\muZs\rpar \spp
\nl
\Gamma_{\SR}\lpar \myPH \to \myPAf\myPf\myPGg \rpar &=&
\frac{1}{2}\,
\Gamma_{\PO}\lpar \myPH \to \myPZ\myPGg \rpar\,
\frac{1}{\gamZ}\,
\Gamma_{\PO}\lpar \myPZ \to \myPAf\myPf \rpar + \;\; \mbox{remainder} \spp
\label{physPOdef}
\eqa
In the NWA the remainder is neglected while we keep it in our formulation
where the goal is extracting PO without making approximations.
The interpretation in terms of SMEFT is based on $\mcT_{\sPHAZ}\lpar \cph\,,\,\cpz \rpar$.
A convenient way for writing $\mcT_{\sPHAZ}$ is the following:
\bq
\mcT_{\sPHAZ} =
\frac{g^3_{\mrF}}{\pi^2 \mz}\,
\sum_{\ssI=\myPW,\PQt,\PQb}\,\rho^{\sPHAZ}_{\ssI}\,\mcT^{\ssI}_{\sPHAZ\,;\,\myLO} +
g_{\mrF}\,\gds\,\frac{\mhs}{M}\,\aAZ +
\frac{g^3_{\mrF}\,\gds}{\pi^2}\,\mcT^{\nfact}_{\sPHAZ} \spp
\label{linkappaHAZ}
\eq
The factorizable part is defined in terms of $\rho\,$-factors
\bqa
\Delta\rho^{\sPHAZ}_{\PQq} &=&
 \lpar 2\,\mrI^{(3)}_{\PQq}\,\aqp + 2\,\apBox - \frac{1}{2}\,\apD + 3\,\aAA + 2\,\aZZ \rpar \spc
\nl
\Delta\rho^{\sPHAZ}_{\myPW} &=&
            \frac{1 + 6\,\ctWs}{\ctWs}\,\apBox
          - \frac{1}{4}\,\frac{1 + 4\,\ctWs}{\ctWs}\,\apD
          - \frac{1}{2}\,\frac{1 + \ctWs - 24\,\ctWq}{\ctWs}\,\aAA \spc
\nl
{}&+& \frac{1}{4}\,\lpar 1 + 12\,\ctWs - 48\,\ctWq \rpar\,\frac{\stW}{\ctWc}\,\aAZ
          + \frac{1}{2}\,\frac{1 + 15\,\ctWs - 24\,\ctWq}{\ctWs}\,\aZZ \spp
\label{kappaHAZ}
\eqa
In the PTG scenario we only keep $\atp, \abp, \apD$ and $\apBox$ in \eqn{kappaHAZ}.
We also derive the following result for the non-factorizable part of the amplitude:
\bq
\mcT^{\nfact}_{\sPHAZ} = \sum_{a\,\in\,\{A\}}\,\mcT^{\nfact}_{\sPHAZ}(a)\,a \spc
\eq
where $\{ A \} = \{ \aptV, \atBW, \atWB, \apbV, \abWB, \abBW, \apD, \aAZ, \aAA, \aZZ \}$.
In the PTG scenario there are only $3$ non-factorizable amplitudes for $\myPH \to \myPGg\myPZ$, those
proportional to $\aptV, \apbV$ and $\apD$.
\subsection{Summary on the PO-SMEFT matching\label{sum}}

As we have shown, there are different layers of measurable parameters that can be extracted from LHC data.
An external layer, where the kinematics is kept exact, is represented by {\em physical PO}
such as $\Gamma_{\SR}\lpar \myPH \to \myPAf\myPf\myPGg \rpar$ of \eqn{physPOdef}: these are similar
to the $\sigma^{\peak}_{\myPf}$ measured at LEP and, similarly to the LEP case, can be
extracted from data via a non-trivial NWA. A first intermediate inner layer is represented by the
{\em effective-couplings PO} introduced in Sections~\ref{sec:PO-2}--{sec:PO-phen} (and summarized in Section~\ref{sectPO:PCSL}): these are
similar to effective $\myPZ$-boson couplings ($g^{\Pe}_{\sPV\,\sPA}$)
measured at LEP and control the parameterization of on-shell amplitudes.
A further internal layer is represented by the $\hatkappa_i$ introduced in this section, that are
appropriate combinations of  Wilson coefficients in the SMEFT.
Finally,  the innermost layer is represented by the Wilson coefficients (or the Lagrangian couplings)
of the specific EFT (or explicit NP model) employed to analyse the data.
When moving to the innermost layer in the SMEFT context we still have the option of performing
the tree-level translation, which is well defined and should be integrated with the corresponding estimate
of MHOU, or we can go to SMEFT at the loop level, again with its own MHOU.


\section{Conclusions}

The experimental precision on the kinematical distributions of  Higgs boson decays
and production cross sections
is expected to significantly improve in the next few years. This will allow us to  investigate in depth
a wide class of possible extensions of the SM. To reach this goal, an accurate and sufficiently general
parameterization of possible NP effects in such distributions is needed.

The Higgs PO presented in this note are conceived exactly to fulfil this goal: they provide
a general decomposition of on-shell amplitudes involving the Higgs boson, based
on analyticity, unitarity, and crossing symmetry. A further key assumption is the
absence of new light particles in the kinematical regime of interest, or better no
unknown physical poles in theses amplitudes. These conditions ensure the generality of this
approach and the possibility to match it to a wide class of explicit NP models,
including the determination of Wilson coefficients in the context of Effective Field Theories.

As we have shown, the PO can be organized in two complementary sets:
the so-called physical PO, that are nothing but a series of idealized Higgs boson partials decay widths,
and the effective-couplings PO, that are particularly useful for the developments of
simulation tools.
The two sets are in one-to-one correspondence,
and their relation is summarized in Table~\ref{Tab:physicalPO}.
The complete set of  effective-couplings PO that can be realistically accessed
in Higgs-related measurements  at the LHC, both in production and in decays,
is summarized  in Section~\ref{sectPO:PCSL}. The reduction of independent PO
obtained under specific symmetry assumptions (in particular flavour universality
and CP invariance) is also discussed in Section~\ref{sectPO:PCSL}.
In two-body processes the effective-couplings PO
are in one-to-one correspondence with the parameters of the original $\kappa$ framework.
A substantial difference arises in more complicated processes,  such as $h\to 4f$ or VBF and VH
production. Here, in order to take into account possible kinematical distortions
in the decay distributions and/or in the production cross-sections,
the PO framework requires the introduction of a series of  additional terms.
These terms encode generic NP effects in the $hVf\bar f$ amplitudes and
their complete list is summarized in Table~\ref{tab:POtable}.

The PO framework  can be systematically improved to include the effect of higher-order
QCD and QED corrections, recovering the best up-to-date SM predictions in absence
of new physics. The effective-couplings PO
should not be confused with EFT Wilson coefficients. However, their measurement
can facilitate the extraction of Wilson coefficients in any EFT approach to Higgs physics, 
as briefly illustrated in Section~\ref{sectPO:SMEFT} in the context of the so-called SMEFT.
The physical and the effective-couplings
PO can be considered as the most general and external layers
in the characterization of physics beyond the SM,
whose innermost layer is represented by the couplings
of some explicit NP model.

\chapter{Simplified Template Cross Sections}
\label{STCS}
\ChapterAuthor{M.~D{\"u}hrssen, P.~Francavilla, F.J.~Tackmann, K.~Tackmann}
We acknowledge discussions in Les Houches 2015 and WG2 and contributions and feedback from Aaron Armbruster, Josh Bendavid, Fawzi Boudjema, Andr\'e David, Marco Delmastro, Dag Gillberg, Admir Greljo, Thibault Guillemin, Chris Hays, Gino Isidori, Sabine Kraml, Kirtimaan Mohan, James Lacey, Carlo Pandini, Elisabetta Pianori, Tilman Plehn, Michael Rauch, Chris White, and many others.

\section{Overview}
\label{PO_STCS:STCS_Overview}

After the successful Higgs boson coupling measurements during the LHC Run1, which had
as their main results measured signal strength and multiplicative coupling modifiers,
it is important to discuss in which way the experiments should present and perform
Higgs boson coupling measurements in the future.
Simplified template cross sections were developed to provide a natural way to
evolve the signal strength measurements used during Run1. Compared to the Run1
measurements, the simplified template cross section framework allows one to
reduce in a systematic fashion the theory dependences that must be directly folded into
the measurements. This includes both the dependence on the theoretical uncertainties in the SM
predictions as well as the dependence on the underlying physics model (i.e.\ the SM or BSM models).
In addition, they provide more finely-grained measurements (and hence more information for theoretical
interpretations), while at the same time allowing and benefitting from the global combination of
the measurements in all decay channels.

The primary goals of the simplified template cross section framework are to maximize the sensitivity
of the measurements while at the same time to minimize their theory dependence. This means in
particular
\begin{itemize}
\item combination of all decay channels
\item measurement of cross sections instead of signal strengths, in mutually exclusive regions of
  phase space
\item cross sections are measured for specific production modes (with the SM production serving as kinematic template)
\item measurements are performed in abstracted/simplified fiducial volumes
\item allow the use of advanced analysis techniques such as event categorization, multivariate techniques, etc.
\end{itemize}

The measured exclusive regions of phase space, called ``bins'' for simplicity, are specific to
the different production modes. Their definitions are motivated by
\begin{itemize}
\item minimizing the dependence on theoretical uncertainties that are directly folded into the measurements
\item maximizing experimental sensitivity
\item isolation of possible BSM effects
\item minimizing the number of bins without loss of experimental sensitivity
\end{itemize}
These will of course be competing requirements in some cases and some compromise has to be achieved.
The implementation of these basic design principles is discussed in more detail below.

\begin{figure}
\begin{center}
\includegraphics[width=\textwidth]{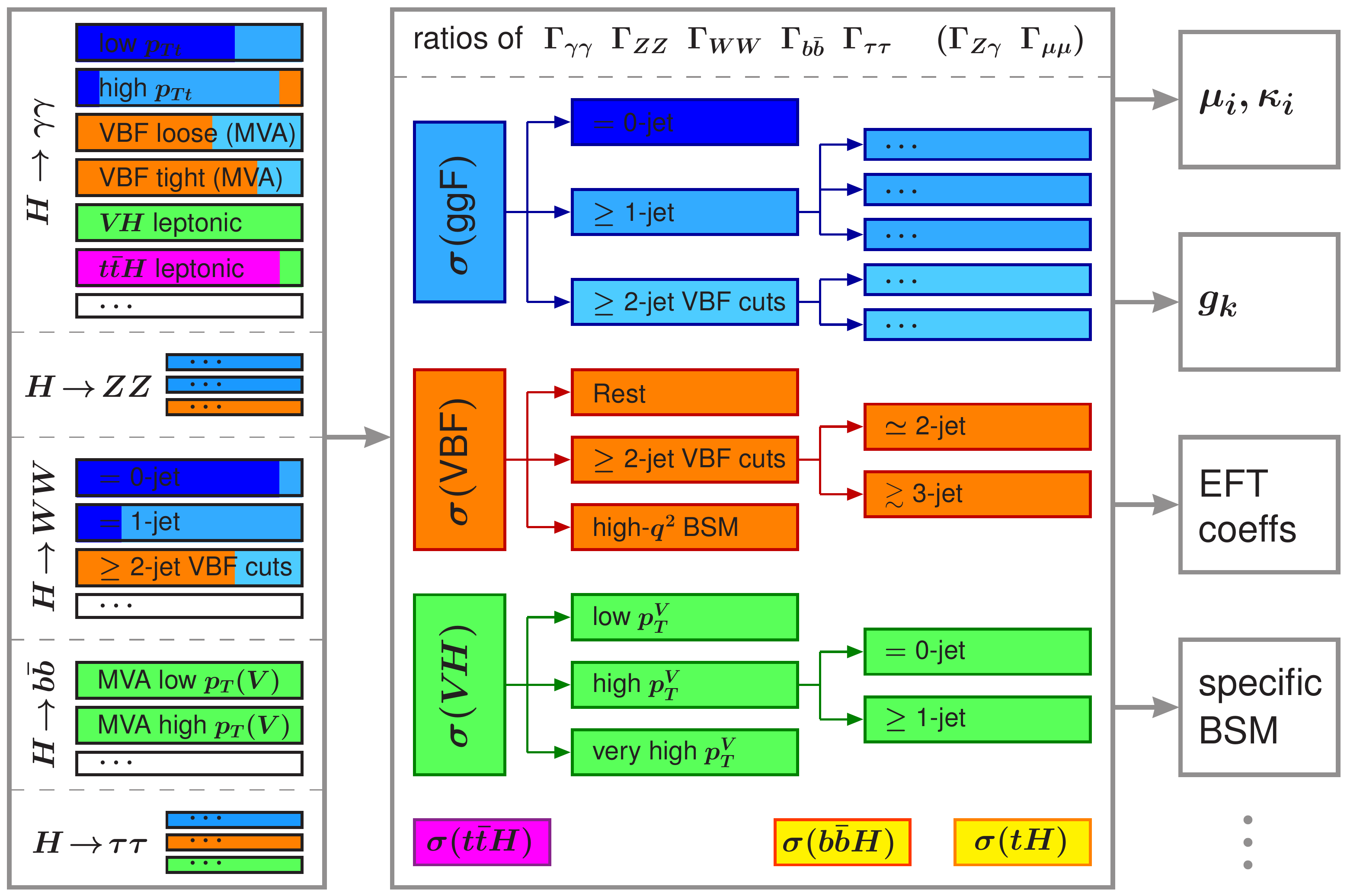}
\end{center}
\caption{Schematic overview of the simplified template cross section framework.}
\label{fig:STCS:sketch}
\end{figure}

A schematic overview of the simplified template cross section framework is shown in \refF{fig:STCS:sketch}.
The experimental analyses shown on the left are very similar to the Run1 coupling measurements. For each decay channel, the
events are categorized in the analyses, and there are several motivations
for the precise form of the categorization. Typically, a subset of the experimental event categories
is designed to enrich events of a given Higgs boson production mode, usually making use of specific
event topologies. This is what eventually allows the splitting of the production modes in the global fit. Another subset
of event categories is defined to increase the sensitivity of the analysis by splitting events
according to their expected signal-to-background ratio and/or invariant-mass resolution. In other cases,
the categories are motivated by the analysis itself, e.g. as a consequence of the backgrounds being estimated
specifically for certain classes of events. While these are some of the primary motivations, in the future the details of the event
categorization can also be optimized in order to give good sensitivity to the
simplified template cross sections to be measured.

The centre of \refF{fig:STCS:sketch} shows a sketch of the simplified template cross sections, which are determined from
the experimental categories by a global fit that combines all decay channels and which represent the main results of
the experimental measurements. They are cross sections per production mode, split into mutually exclusive kinematic bins for each of
the main production modes.
In addition, the different Higgs boson decays are treated by fitting the partial decay widths.
Note that as usual, without additional assumptions on the total width, only ratios of partial widths
and ratios of simplified template cross sections are experimentally accessible.

The measured simplified template cross sections together with the partial decay widths then serve as input for
subsequent interpretations, as illustrated on the right of \refF{fig:STCS:sketch}.
Such interpretations could for example be the determination of signal strength modifiers or
coupling scale factors $\kappa$ (providing compatibility with earlier results), EFT coefficients, tests
of specific BSM models, and so forth. For this purpose, the experimental results should quote the full covariance among the different bins.
By aiming to minimize the theory dependence that is folded into the first step of determining the simplified
template cross sections from the event categories, this theory dependence is shifted into the second interpretation step, making the measurements more long-term useful. For example, the treatment of theoretical uncertainties can be decoupled from the measurements and can be dealt with at the interpretation stage. In this way, propagating improvements in theoretical predictions and their uncertainties into the measurements itself, which is a very time-consuming procedure and unlikely to be feasible for older datasets, becomes much
less important. Propagating future theoretical advances into the interpretation, on the other hand, is generally much easier.

To increase the sensitivity to BSM effects, the simplified template cross sections can be
interpreted together with e.g. POs in Higgs boson decays. To make this possible, the experimental
and theoretical correlations between the simplified template cross sections and the decay POs would need
to be evaluated and taken into account in the interpretation. This point will not be expanded on further
in this section, but would be interesting to investigate in the future.

While the simplified template cross section bins have some similarity to a differential cross section measurement,
they aim to combine the advantages of the signal strength measurements and fiducial
and differential measurements. In particular, they are complementary to full-fledged fiducial and
differential measurements and are neither designed nor meant to replace these. Fully fiducial differential
measurements are of course essential but can only be carried out in a subset of decay channels in
the foreseeable future. They are explicitly optimized for maximal theory independence. In practice, this
means that in the measurements acceptance corrections are minimized, typically,
simple selection cuts are used, and the measurements are unfolded to a fiducial volume that is as close as possible to the fiducial
volume measured for a particular Higgs boson decay channel. In contrast,
simplified template cross sections are optimized for sensitivity while reducing the dominant theory
dependence in the measurement. In practice, this means that simplified fiducial volumes are used and larger acceptance corrections are allowed in order to maximally
benefit from the use of event categories and multivariate techniques. They are also
inclusive in the Higgs boson decay to allow for the combination of the different decay channels. The
fiducial and differential measurements are designed to be agnostic to the production modes as much
as possible. On the other hand, the separation into the production modes is an essential aspect of
the simplified template cross sections to reduce their model dependence.

\section{Guiding principles in the definition of simplified template cross section bins}
\label{PO_STCS:STCS_GuidingPrinciple}

As outlined above, several considerations have been taken into account in the definition
of the simplified template cross section bins.

One important design goal is to reduce the dependence of the measurements on theoretical uncertainties
in SM predictions. This has several aspects. First, this requires
avoiding that the measurements have to extrapolate from a certain region in phase space to the full
(or a larger region of) phase space whenever this extrapolation carries nontrivial or sizeable theoretical
uncertainties. A example is the case where an event category selects an exclusive
region of phase space, such as an exclusive jet bin. In this case, the associated theoretical uncertainties
can be largely avoided in the measurement by defining a corresponding truth jet bin.
The definition of the bins is preferably in terms of quantities that are directly measured by the experiments
to reduce the needed extrapolation.

There will of course always be residual theoretical uncertainties due to the experimental acceptances for each
truth bin. Reducing the theory dependence thus also requires to avoid cases with large variation in the
experimental acceptance within one truth bin, as this would introduce a direct dependence on the
underlying theoretical distribution in the simulation. If this becomes an issue, the bin can be further split into two
or more smaller bins, which reduces this dependence in the measurement and moves it to the interpretation step.

To maximize the experimental sensitivity, the analyses should continue to use event categories primarily optimized for sensitivity, while
the definition of the truth bins should take into consideration the experimental requirements. However, in cases where multivariate analyses are used in the analyses, it has to be carefully checked and balanced against the requirement to not introduce theory dependence, e.g., by selecting specific regions of phase space.

Another design goal is to isolate regions of phase space, typically at large kinematic scales, where BSM effects could be potentially large and visible above the SM background. Explicitly separating these also reduces the dependence of the measurements on the assumed SM kinematic distribution.

In addition, the experimental sensitivity is maximized by allowing the combination of all decay channels, which requires the framework
to be used by all analyses. To facilitate the experimental implementation, the bins should be mutually exclusive to avoid introducing statistical correlations between different bins. In addition, the number of bins should be kept
minimal to avoid technical complications in the individual analyses as well as the global fit, e.g. in the evaluation of the full covariance matrix. For example, each bin should typically have some sensitivity from at least one event category in order to avoid the need to statistically combine many poorly constrained or unconstrained measurements. On the other hand, in BSM sensitive bins experimental limits are already very useful for the theoretical interpretation.

\subsection{Splitting of production modes}
\label{PO_STCS:STCS_GuidingPrincipleSplitting}

The definition of the production modes has some notable differences compared to Run1 to deal
with the fact that the naive distinction between the $q\bar{q}\to VH$ and VBF processes, and
similarly between $gg\to VH$ and gluon-fusion production, becomes ambiguous at higher order
when the $V$ decays hadronically. For this reason, the $VH$ production mode is explicitly
defined as Higgs boson production in association with a leptonically decaying $V$ boson.
The $q\bar{q}\to VH$ process with a hadronically decaying $V$ boson is considered to be part of
what is called ``VBF production'', which is defined as electroweak $qqH$ production. Similarly,
the $gg\to ZH$ process with hadronically decaying $Z$ boson is included in what is called
``gluon-fusion production''.

In principle, also the separation of $ZH$ production with a leptonic $Z$ into $q\bar{q}$ or $gg$ initial states
becomes ambiguous at higher order. For present practical purposes, on the experimental side the split
can be defined according to the separate MC samples for $q\bar{q}\to ZH$ and $gg\to ZH$ used in the analyses.

\subsection{Staging}
\label{PO_STCS:STCS_GuidingPrincipleStaging}

In practice, it will be impossible to define a set of bins that satisfies all of the above requirements for every analysis.
Some analyses will only be able to constrain a subset of all bins or only constrain the sum of a set of bins.
In addition, the number of bins that will be possible to measure increases with increasing amount of
available data. For this reason, several stages with an increasing number of bins are defined.
The evolution from one stage to the next can take place independently for each production mode.

\begin{figure}
\begin{center}
\includegraphics[width=\textwidth]{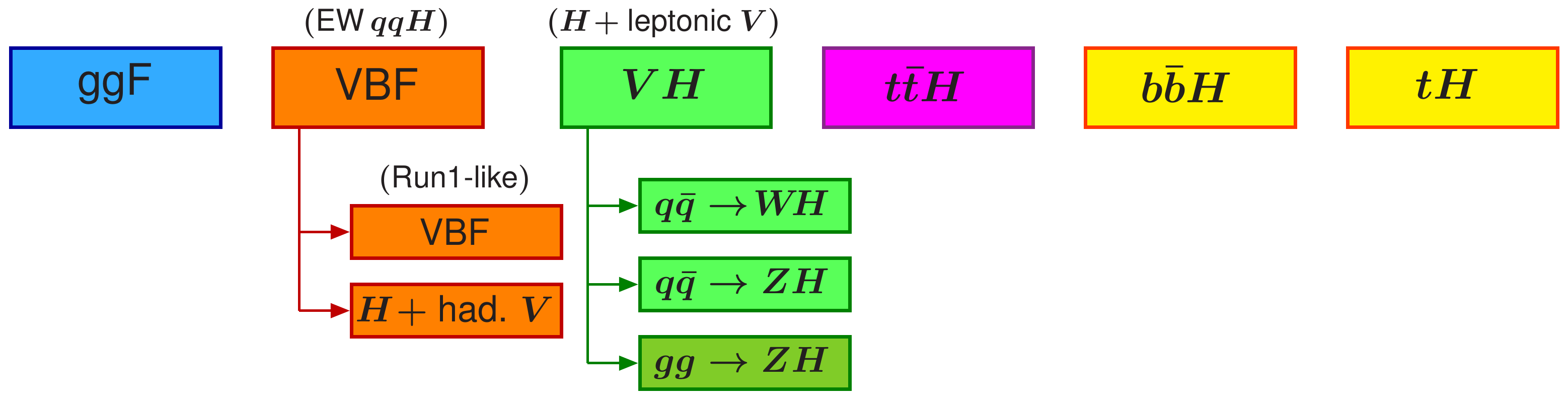}
\end{center}
\caption{Stage 0 bins.}
\label{fig:STCS:stage0}
\end{figure}

\subsubsection{Stage 0} Stage 0 is summarized in \refF{fig:STCS:stage0} and corresponds most closely to the
measurement of the production mode $\mu$ in Run1. At this stage, each main production mode has a single
inclusive bin,
with associated Higgs boson production separated into $q\bar q\to WH$, $q\bar q\to ZH$ and $gg\to ZH$ channels.

As discussed in Section~\ref{PO_STCS:STCS_GuidingPrincipleSplitting}, VBF production is
defined as electroweak $qqH$ production. For better compatibility with Run1 measurements,
the VBF production is split into a Run1-like VBF and
Run1-like $V(\to jj)H$ bin, where the splitting is defined by the conventional Feynman diagrams included in the
simulations. In practice, most decay channels will only provide a measurement for
the Run1-like VBF bin.

\subsubsection{Stage 1} Stage 1 defines a binning that is targeted to be used by all analyses on an intermediate
time scale. In principle, all analyses should aim to eventually implement the full stage 1 binning.
If necessary, intermediate stages to reach the full stage 1 binning can be implemented by a given analysis by
merging bins that cannot be split. In this case, the analysis should ensure that the merged bins have similar acceptances,
such that the individual bins can still be determined in an unbiased way in the global combination of all channels.
In the diagrams presented below, the possibilities for merging bins are indicated by ``(+)''.

\subsubsection{Stage 2} Defining the stage 2 binning in full detail is very difficult before having gained experience with the
practical implementation of the framework with the stage 1 binning. Therefore, instead of giving a detailed
proposal for the stage 2 binning, we only give indications of interesting further separation of bins that should be considered
for the stage 2 binning.

\section{Definition of leptons and jets}
\label{PO_STCS:STCS_ObjectsAndJets}

The measured event categories in all decay channels are unfolded by the global fit to the
simplified template cross sections bins. For this purpose, and for the
comparison between the measured bins and theoretical predictions from either analytic
calculations or MC simulations, the truth final state particles need to be defined unambiguously.
The definition of the final state particles, leptons, jets, and in particular also the Higgs boson 
are explicitly kept simpler and more idealized than in the fiducial cross section measurements.
Treating the Higgs boson as a final state particle is what allows the combination of the
different decay channels.

For the moment, the definitions are adapted to the current scope of the measurements. Once a
finer binning is introduced for $t\bar{t}H$ or processes such as VBF with the emission of a hard
photon are added, some of the definitions below might have to be adapted or refined.

\subsection{Higgs boson}
\label{PO_STCS:STCS_ObjectsAndJets_Higgs}

The simplified template cross sections are defined for the production of an on-shell Higgs boson,
and the unfolding should be done accordingly. A global cut on the Higgs boson rapidity at $|Y_H| < 2.5$
is included in all bins. As the current measurements have no sensitivity beyond this rapidity range, this
part of phase space would only be extrapolated by the MC simulation. On the other
hand, it is in principle possible to use forward electrons (up to $|\eta|$ of 4.9) in
$H\to ZZ^*\to 4\ell$ and extend the accessible rapidity range. For this purpose, an
additional otherwise
inclusive bin for $|Y_H| > 2.5$ can be included.

\subsection{Leptons}
\label{PO_STCS:STCS_ObjectsAndJets_leptons}

Electrons and muons from decays of signal vector bosons, e.g. from $VH$ production, are defined
as dressed, i.e. FSR photons should be added back to the
electron or muon. $\tau$ leptons are defined from the sum of their decay products (for any $\tau$ decay
mode). There should be
no restriction on the transverse momentum or the rapidity of the leptons.
That is, for a leptonically decaying vector boson
the full decay phase space is included.

\subsection{Jets}
\label{PO_STCS:STCS_ObjectsAndJets_Jets}

Truth jets are defined as anti-$k_t$ jets with a jet radius of $R = 0.4$, and are built from all
stable particles (exceptions are given below), including neutrinos, photons and leptons
from hadron decays or produced in the shower.
Stable particles here have the usual definition, having a lifetime greater than 10~ps, i.e. those
particles that are passed to GEANT in the experimental simulation chain. All decay products from
the Higgs boson decay are removed as they are accounted for by the truth Higgs boson.
Similarly, leptons (as defined above) and neutrinos from decays of the signal $V$ bosons are
removed as they are
treated separately, while decay products from hadronically decaying signal $V$ bosons are included in
the inputs to the truth jet building.

By default, truth jets are defined without restriction on their rapidity. A possible rapidity cut can
be included
in the bin definition. A common $p_\mathrm{T}$ threshold for jets should be used for all truth jets.
A lower threshold would in principle have the advantage to split the events more evenly between the
different jet bins.
Experimentally, a higher threshold at 30~GeV is favored due to pile up and is therefore used for the
jet definition to limit the amount of phase-space extrapolation in the measurements.

\section{Bin definitions for the different production modes}
\label{PO_STCS:STCS_bins}

In the following, the bin definitions for the different production modes in each stage are given.
The bins are easily visualized through cut flow diagrams. In the diagrams, the bins on each branch
are defined to be mutually exclusive and sum up to the preceding parent bin. For simplicity, sometimes
not all cuts are explicitly written out in the diagrams, in which case the complete set of cuts are specified in the text.
In case of ambiguities, a more specific bin is excluded from a more generic bin. As already mentioned, for the stage 1 binning
the allowed possibilities for merging bins at intermediate stages are indicated by a ``(+)'' between two bins.

\subsection{Bins for \texorpdfstring{$gg \to H$}{gg to H} production}
\label{PO_STCS:STCS_ggH}

\subsubsection{Stage 0} Inclusive gluon fusion cross section within $|Y_H|<2.5$. Should the
measurements start to have acceptance beyond 2.5, an additional bin for $|Y_H| > 2.5$ can be
included.

\begin{figure}
\begin{center}
\includegraphics[width=\textwidth]{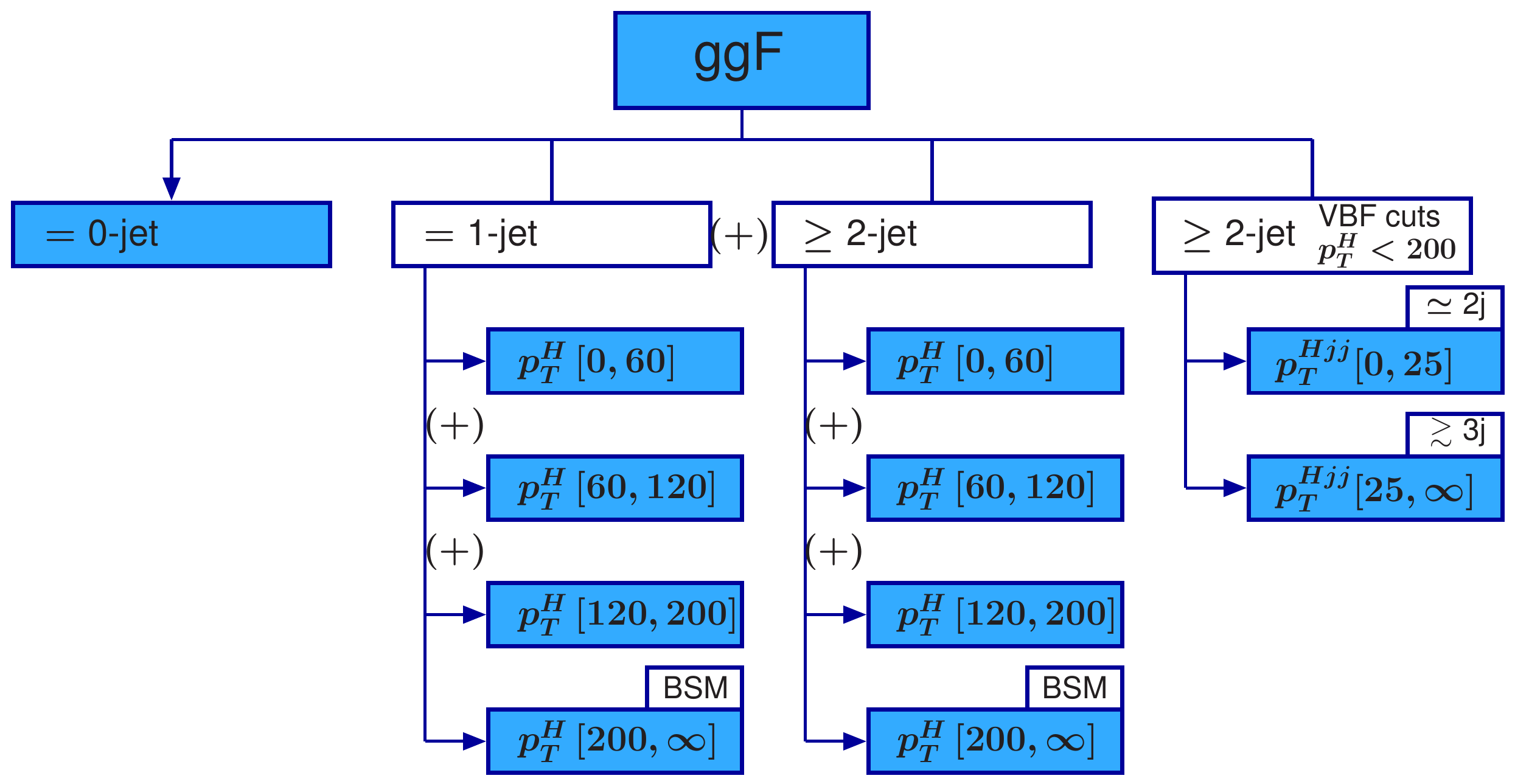}
\end{center}
\caption{Stage 1 binning for gluon fusion production.}
\label{fig:STCS:ggH}
\end{figure}

\subsubsection{Stage 1}

Stage 1 refines the binning for $|Y_H|<2.5$.
The stage 1 binning is depicted in \refF{fig:STCS:ggH} and summarized as follows:
\begin{itemize}
\item Split into jet bins: $N_j=0$, $N_j=1$, $N_j\geq2$, $N_j\geq2$ with VBF topology cuts
  (defined with the same cuts as the corresponding bin in VBF production). For the
  $N_j\geq2$ with VBF topology cuts, $p_\mathrm{T}^H <$ 200~GeV is required, which gives priority
  to the $p_\mathrm{T}^H >$ 200~GeV bin for $N_j\geq2$. Otherwise, the $N_j \geq 2$ with VBF topology cuts
  is excluded from the $N_j \geq 2$ bins. The jet bins are
  motivated by the use of jet bins in the experimental analyses. Introducing them also for the
  simplified template cross sections avoids folding the associated theoretical uncertainties into
  the measurement. The separation of the $N_j\geq 2$ with VBF topology cuts is motivated by the wish to
  separately measure the gluon fusion contamination in the VBF selection. If the fit has no
  sensitivity to determine the gluon fusion and the VBF contributions with this topology, the sum
 of the two contributions can be quoted as result.
\item The $N_j\geq2$ with VBF topology bin is split further into an exclusive $2$-jet-like and inclusive $3$-jet-like
  bin. The split is implemented by a cut on $p_\mathrm{T}^{Hjj} = |\vec p_\mathrm{T}^H + \vec p_\mathrm{T}^{j1} + \vec p_\mathrm{T}^{j2}|$ at 25~GeV.
  See the corresponding discussion for VBF for more details.
  This split is explicitly included here since it induces
  nontrivial theory uncertainties in the gluon-fusion contribution.
\item The $N_j=1$ and $N_j\geq2$ bins are further split into $p_\mathrm{T}^H$ bins.
  \begin{itemize}
  \item 0~GeV $< p_\mathrm{T}^H<$ 60~GeV: The boson channels have most sensitivity in the low
    $p_\mathrm{T}^H$ region. The upper cut is chosen as low as possible to give a more even split
    of events but at the same time high enough that no resummation effects are expected. The cut
    should also be sufficiently high that the jet $p_{\rm T}$ cut introduces a negligible bias.
  \item 60~GeV $< p_\mathrm{T}^H<$ 120~GeV: This is the resulting intermediate bin between the low and high
    $p_\mathrm{T}^H$ regions. The lower cut here is high enough that this bin can be safely treated
    as a hard $H+j$ system in the theoretical description.
  \item 120~GeV $< p_\mathrm{T}^H<$ 200~GeV: The boosted selection in $H\to\tau\tau$ contributes to
    the high $p_\mathrm{T}^H$ region. Defining a separate bin avoids large extrapolations for the
    $H\to\tau\tau$ contribution. For $N_j = 2$, this bin likely provides a substantial part of the gluon-fusion
    contribution in the hadronic $VH$ selection.
  \item $p_\mathrm{T}^H>$ 200~GeV: Beyond the top-quark mass, the top-quark loop gets resolved and top-quark mass effects become relevant.
    Splitting off the high-$p_\mathrm{T}^H$ region ensures the usability of the heavy-top expansion
    for the lower-$p_\mathrm{T}^H$ bins. At the same time, the high $p_\mathrm{T}^H$ bin in principle
    offers the possibility to distinguish a pointlike $ggH$ vertex induced by heavier BSM particles in the loop from
    the resolved top-quark loop.
\end{itemize}
\end{itemize}

At intermediate stages, all lower three $p_\mathrm{T}^H$ bins, or any two adjacent bins, can be merged. Alternatively
or in addition the $N_j=1$ and $N_j\geq2$ bins can be merged by individual analyses as needed, and potentially
also when the combination is performed at an intermediate stage.

\subsubsection{Stage 2}

In stage 2, the high $p_\mathrm{T}^H$ bin should be split further, in particular if evidence for new heavy particles arises.
In addition, the low $p_\mathrm{T}^H$ region can be split further to reduce any theory dependence there. If desired by
the analyses, another possible option is to further split the $N_j \geq 2$ bin into $N_j=2$ and $N_j \geq 3$.

\subsection{Bins for VBF production}
\label{PO_STCS:STCS_VBF}

At higher order, VBF production and $VH$ production with hadronically decaying $V$ become ambiguous.
Hence, what we refer to as VBF in this section, is defined as as electroweak $qq'H$ production,
which includes both VBF and $VH$ with hadronic $V$ decays.

\subsubsection{Stage 0} Inclusive vector boson fusion cross section within $|Y_H|<2.5$. Should the
measurements start to have acceptance beyond 2.5, an additional bin for $|Y_H| > 2.5$ can be
included.

\begin{figure}
\begin{center}
\includegraphics[width=\textwidth]{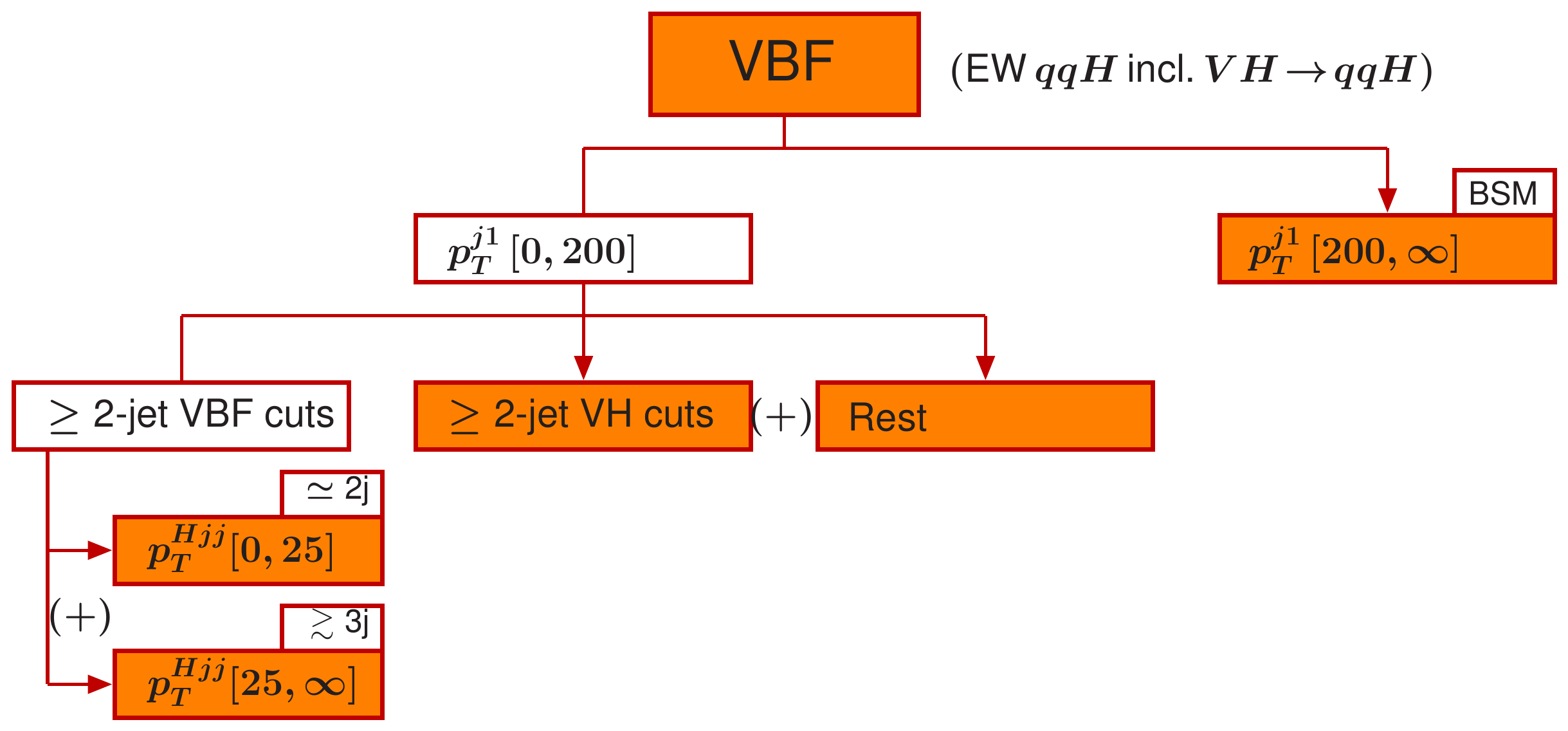}
\end{center}
\caption{Stage 1 binning for vector boson fusion production.}
\label{fig:STCS:VBF}
\end{figure}

\subsubsection{Stage 1}

Stage 1 refines the binning for $|Y_H|<2.5$.
The stage 1 binning is depicted in \refF{fig:STCS:VBF} and summarized as follows:
\begin{itemize}
\item VBF events are split by $p_\mathrm{T}^{j1}$, the transverse momentum of the highest-$p_\mathrm{T}$
jet. The lower $p_\mathrm{T}^{j1}$ region is expected to be dominated by SM-like events, while the
high-$p_\mathrm{T}^{j1}$ region is sensitive to potential BSM contributions, including events with
typical VBF topology as well as boosted $V(\to jj)H$ events where the $V$ is reconstructed as one jet.
The suggested cut is
at 200~GeV, to keep the fraction of SM events in the BSM bin small. Note that events with $N_j = 0$,
corresponding to $p_\mathrm{T}^{j1}<$ 30~GeV, is included in the $p_\mathrm{T}^{j1}<$ 200~GeV bin.
\item The $p_\mathrm{T}^{j1}<$ 200~GeV bin is split further:
  \begin{itemize}
  \item Typical VBF topology: The adopted VBF topology cuts are $m_{jj} >$ 400~GeV,
    $\Delta\eta_{jj} > 2.8$ (and without any additional rapidity cuts on the signal jets).
    This should provide a good intermediate compromise among the various VBF selection cuts
    employed by different channels.
  \begin{itemize}
  \item The bin with typical VBF topology is split into an exclusive $2$-jet-like and inclusive $3$-jet-like
  bin using a cut on $p_\mathrm{T}^{Hjj}$ at 25~GeV, where the cut value is a compromise between providing
  a good separation of gluon fusion and VBF and the selections used in the measurements.
  $p_\mathrm{T}^{Hjj}$ as quantity to define this split is chosen as a compromise between
  the different kinematic
  variables used by different channels to enrich VBF production. (In particular the kinematic
  variables $\Delta\phi_{H-jj}$
  and $p_\mathrm{T}^{j3}$ are both correlated with $p_\mathrm{T}^{Hjj}$).
\end{itemize}
\item Typical $V(\to jj)H$ topology: events with at least two jets and 60~GeV $< m_{jj} <$ 120~GeV.
\item  Rest: all remaining events, including events with zero or one jet.
  The ``rest'' bin can be sensitive to certain BSM
  contributions that do not follow the typical SM VBF signature with two forward jets.

\end{itemize}
\end{itemize}

\begin{figure}
\begin{center}
\includegraphics[width=\textwidth]{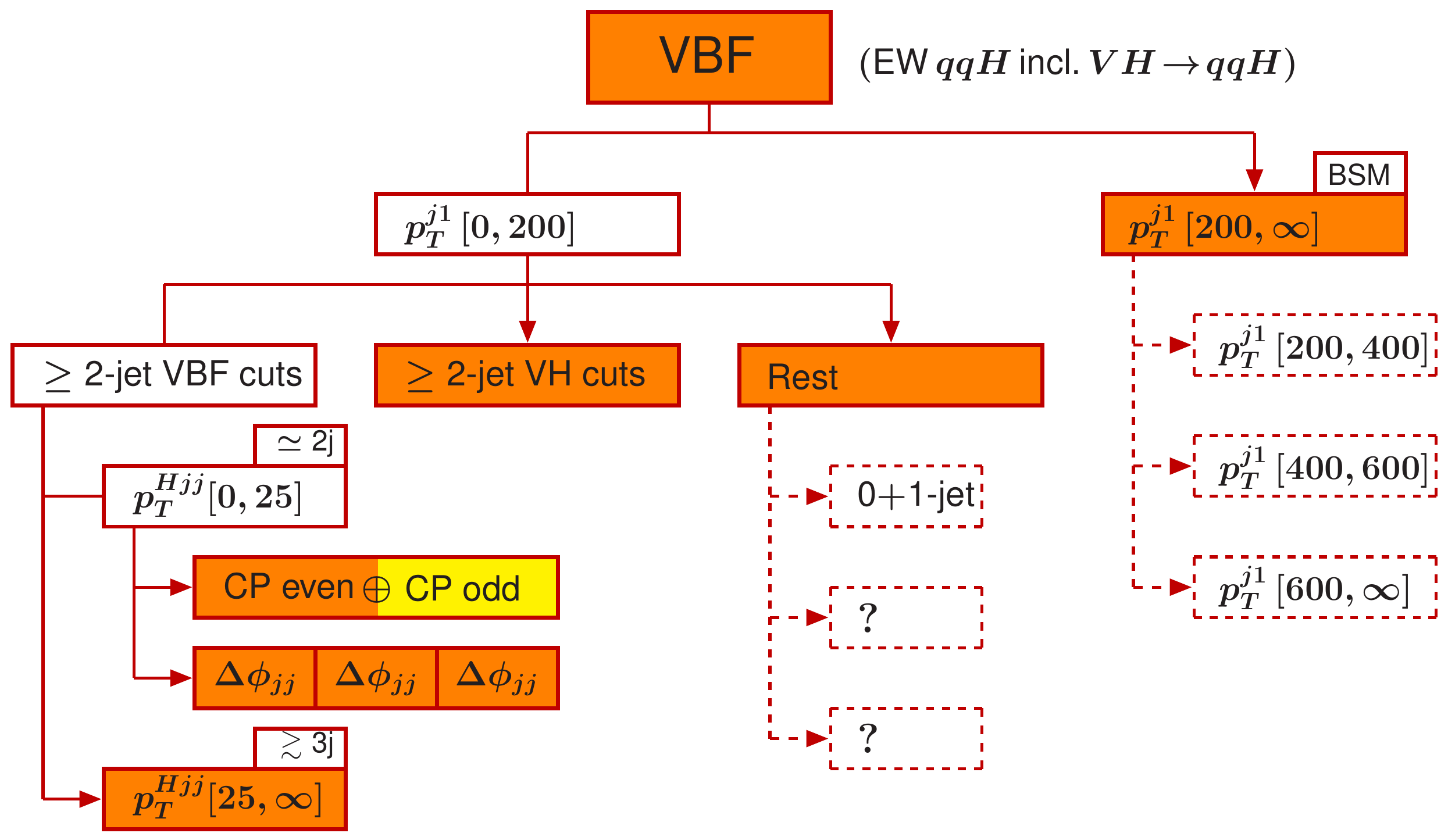}
\end{center}
\caption{Possible stage 2 binning for vector boson fusion production.}
\label{fig:STCS:VBF2}
\end{figure}

\subsubsection{Stage 2}

More splits are introduced at stage 2 as illustrated in \refF{fig:STCS:VBF2}. While the details require more discussion and
cannot be finalized at the present, this could include

\begin{itemize}
\item The high-$p_\mathrm{T}^{j1}$ bin can be split further by separating out very high-$p_\mathrm{T}^{j1}$
  events for example with additional cuts at 400~GeV and 600~GeV.
\item The ``rest'' bin can be split further, e.g., by explicitly separating out a looser VBF selection,
  and/or by separating out events with zero or one jets.
\item The $N_j\simeq 2$ VBF topology bin can be split further to gain sensitivity to CP odd
  contributions, e.g. by splitting it into subbins of $\Delta\phi_{jj}$ or alternatively by measuring a
  continuous parameter.
\end{itemize}

\subsection{Bins for \texorpdfstring{$VH$}{VH} production}
\label{PO_STCS:STCS_VH}

In this section, $VH$ is defined as Higgs boson production in association with a leptonically
decaying $V$ boson.

Note that $q\bar{q}\to VH$ production with a hadronically decaying $V$ boson is considered part of
VBF production. Similarly, $gg\to VH$ production with hadronically decaying $V$ boson is
considered part of gluon fusion production.

\subsubsection{Stage 0} Inclusive associated production with vector bosons cross section within
 $|Y_H|<2.5$. Should the
measurements start to have acceptance beyond 2.5, an additional bin for $|Y_H| > 2.5$ can be
included.

\begin{figure}
\begin{center}
\includegraphics[width=\textwidth]{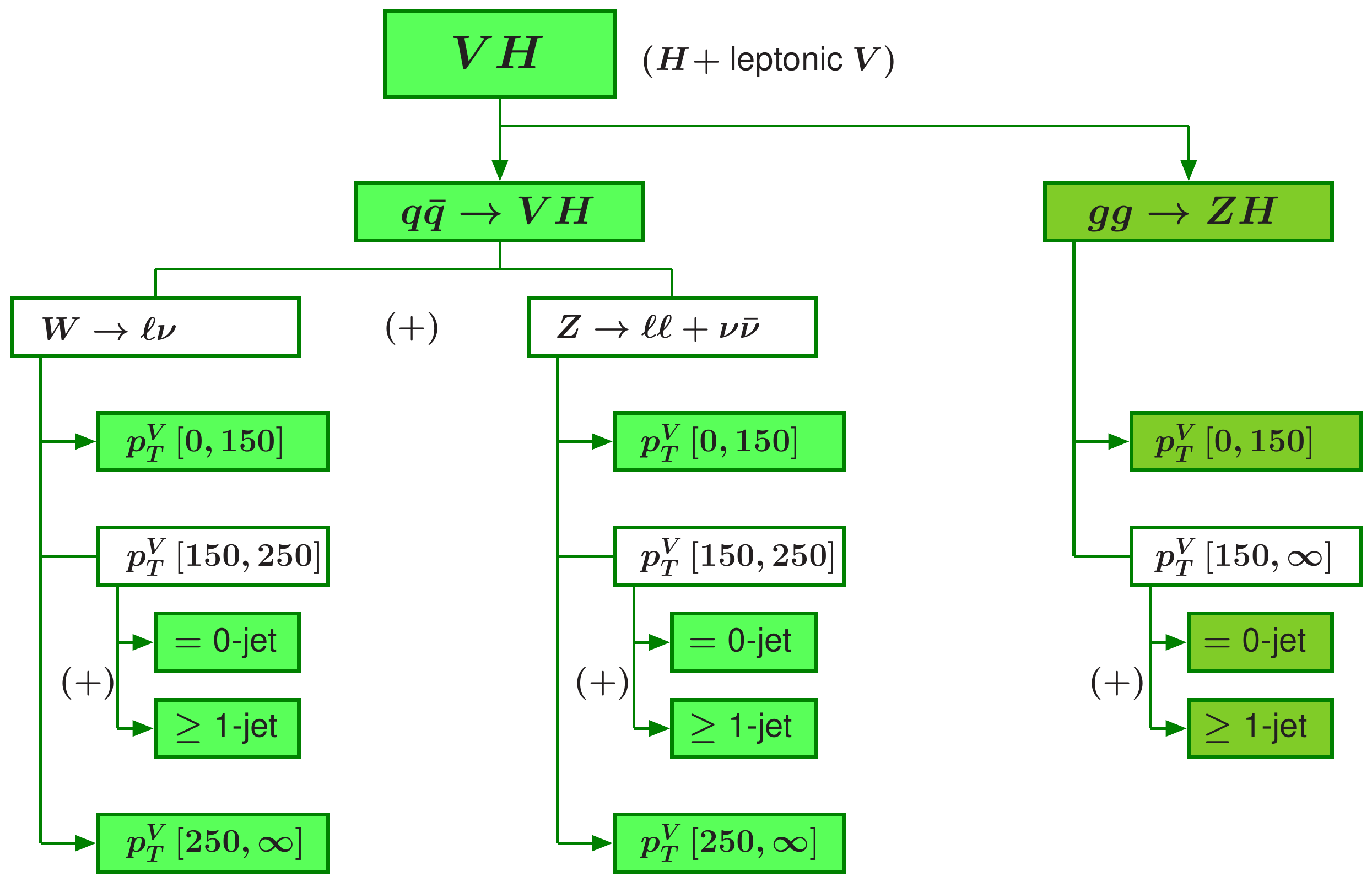}
\end{center}
\caption{Stage 1 binning for associated production with vector bosons.}
\label{fig:STCS:VH}
\end{figure}

\subsubsection{Stage 1}

Stage 1 refines the binning for $|Y_H|<2.5$.
The stage 1 binning is depicted in \refF{fig:STCS:VH} and summarized as follows:
\begin{itemize}
\item $VH$ production is first split into the production via a $q\bar{q}$ or $gg$ initial state.
This split becomes ambiguous at higher order. For practical
purposes, on the experimental side the split can be defined according to the MC samples used in
the analyses, which are split by $q\bar{q}$ and $gg$.
\begin{itemize}
\item The production via $q\bar{q}\to VH$ is split according to the vector boson: $W\to\ell\nu$ and
  $Z\to\ell\ell + \nu\bar{\nu}$.
\item $W\to\ell\nu$ and $Z\to\ell\ell + \nu\bar{\nu}$ are split further into bins of $p_\mathrm{T}^V$,
  aligned with the quantity used in the $H\to b\bar{b}$ analysis, which is one of the main
  contributors to the $VH$ bins.
  \begin{itemize}
  \item  $p_\mathrm{T}^V<$ 150~GeV receives contributions from the bosonic decay channels and from
    $H\to b\bar{b}$ with $W\to\ell\nu$ and $Z\to\ell\ell$, which do not rely on $E_\mathrm{T}^\mathrm{miss}$
    triggers.
  \item 150~GeV $<p_\mathrm{T}^V<$ 250~GeV receives contributions from $H\to b\bar{b}$ with
    $Z\to\nu\bar{\nu}$ due to the high threshold of the $E_\mathrm{T}^\mathrm{miss}$ trigger, as
    well as from $H\to b\bar{b}$ with $W\to\ell\nu$ and $Z\to\ell\ell$.
    \begin{itemize}
    \item This bin is split further into a $N_j=0$ and a $N_j \geq 1$ bin, reflecting the different
      experimental sensitivity and to avoid the corresponding theory dependence.
    \end{itemize}
  \item $p_\mathrm{T}^V>$ 250~GeV is sensitive to BSM contributions.
  \end{itemize}
\item The production via $gg \to ZH$ is split in analogy to production from the $q\bar{q}$
  initial state, apart from the $p_\mathrm{T}^V>$ 250~GeV bin, which is not split out.
\end{itemize}
\end{itemize}

\begin{figure}
\begin{center}
\includegraphics[width=\textwidth]{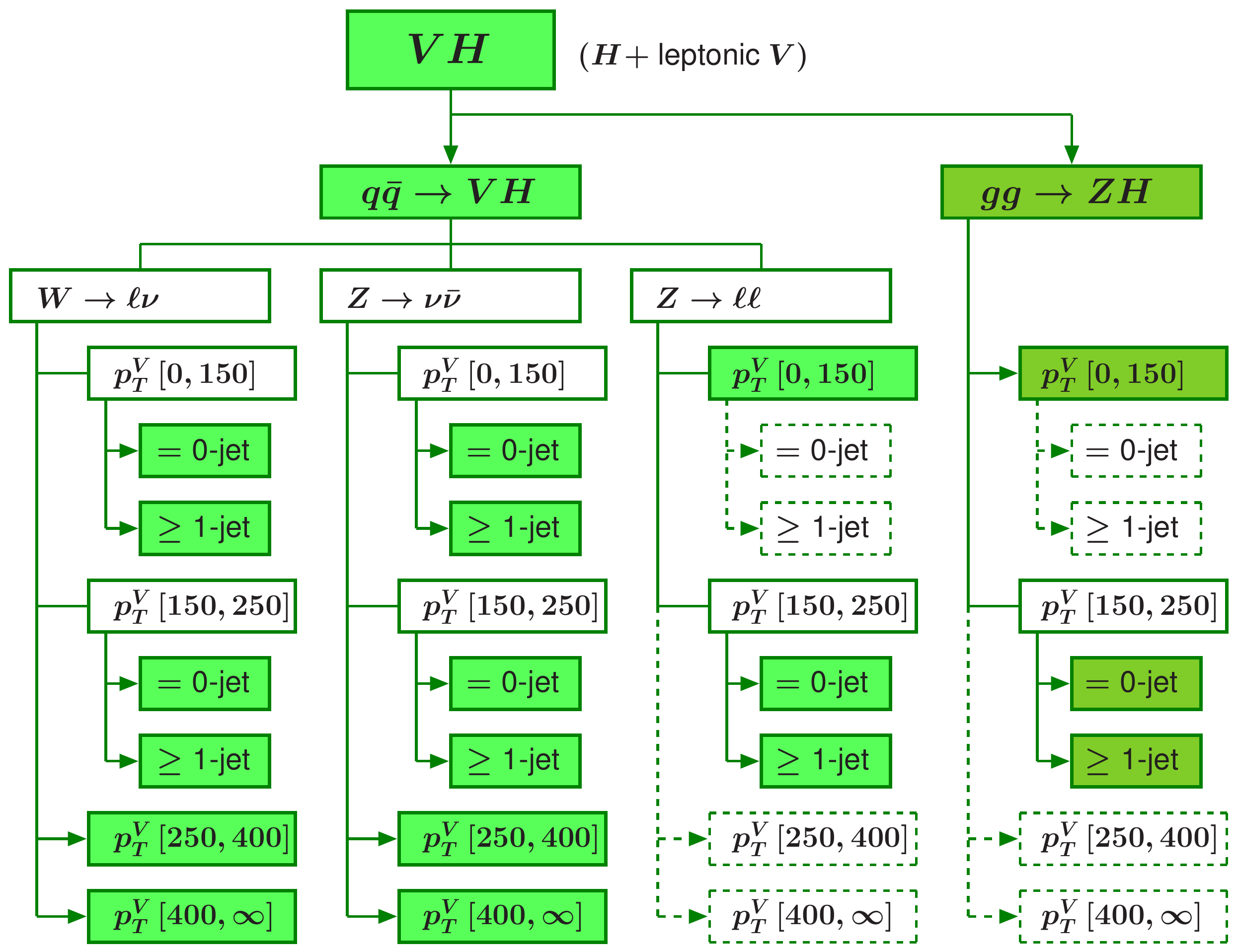}
\end{center}
\caption{Possible Stage 2 binning for associated production with vector bosons.}
\label{fig:STCS:VH2}
\end{figure}

\subsubsection{Stage 2}

More splits are introduced at stage 2 as illustrated in \refF{fig:STCS:VBF2}. While the details need more discussion, this could include

\begin{itemize}
\item Split of the $Z\to\ell\ell + \nu\bar{\nu}$ into $Z\to\ell\ell$ and $Z\to\nu\bar{\nu}$.
\item Split of the $p_\mathrm{T}^V<$ 150~GeV into a $N_j=0$ and a $N_j \geq 1$ bin, except maybe for
  the $Z\to\ell\ell$ channel, which will suffer from the low $Z\to\ell\ell$ branching ratio.
\item Split of the $p_\mathrm{T}^V>$ 250~GeV bin into $p_\mathrm{T}^V<$ 400~GeV and $p_\mathrm{T}^V>$ 400~GeV,
  to increase the sensitivity to BSM contributions with very high $p_\mathrm{T}^V$, potentially apart
  from the $Z\to\ell\ell$.
\item Potentially analogous splits for $gg\to ZH$ production.
\end{itemize}

\subsection{Treatment of \texorpdfstring{$t\bar tH$}{ttH} production}
\label{PO_STCS:STCS_ttH}

\subsubsection{Stage 0} Inclusive $t\bar t H$ production with $|Y_H|<2.5$. Should the
measurements start to have acceptance beyond 2.5, an additional bin for $|Y_H| > 2.5$ can be
included.

\subsubsection{Stage 1}

Currently no additional splits beyond stage 0 are foreseen. One option might be to
separate different top decay channels for $|Y_H|<2.5$.

\subsubsection{Stage 2}

In the long term it could be useful to split into bins with 0 and $\geq 1$ additional jets or
one or more bins tailored for BSM sensitivity.

\subsection{Treatment of \texorpdfstring{$b\bar b H$}{bbH} and \texorpdfstring{$tH$}{tH} production}
\label{PO_STCS:STCS_bbH}

In the foreseeable future, there will only be one inclusive bin for $b\bar{b}H$ production and
only one inclusive bin for $tH$ production for $|Y_H|<2.5$. Should the
measurements start to have acceptance beyond 2.5, an additional bin for $|Y_H| > 2.5$ can be
included.

\section{Practical considerations}
\label{PO_STCS:practicalities}

To facilitate the combination of the results from ATLAS and CMS, the same bin definitions
need to be used by the two collaborations.
As for the Run1 Higgs boson coupling measurements, a combination of results from ATLAS and CMS will
also require that the two collaborations estimate systematic and residual theoretical uncertainties
in a compatible way.
This might be facilitated for example by the use of the same Monte Carlo generators in the measurements.

After first experience with the measurement and
interpretation has been collected, the stage 1 bin definitions should be reviewed. This
should in particular include the definition of the VBF topology cuts as well as the $p_\mathrm{T}^{Hjj}$ split.
In cases where the bin definitions are clearly inadequate, they should be improved for future
measurements. The stage 2 bins will be defined in detail taking into account the
experience gained during the measurements based on the stage 1 definitions.

A reference implementation of the bin definitions in Rivet has been
developed~\cite{STXSRivet}.
This will facilitate a consistent treatment in both experiments as well as
in theoretical studies.

\section{Summary}
\label{PO_STCS:summary}

Simplified template cross sections provide a way to evolve the signal strength measurements that were
performed during LHC Run1, by reducing the theoretical uncertainties that are directly folded into
the measurements and by providing more finely-grained measurements, while at the same time
allowing and benefitting from the combination of measurements in many decay channels. Several stages
are proposed: stage 0 essentially corresponds to the production mode measurements of Run1 and stage 1
defines a first complete setup, with indications for potential bin merging when a given channel
cannot yet afford the full stage 1 granularity. A complete proposal for the stage 2 binning will need to be based on
experience of using the simplified template cross section framework in real life, but some indications
of what could be interesting are already given here.

\chapter{Higgs Fiducial Cross Sections}
\label{chap:FXS}
\ChapterAuthor{F.U.~Bernlochner, S.~Kraml, P.~Milenovic, P.F.~Monni~(Eds.);  M.~Ahmad, F.~Caola, N.~Chanon, D.~de~Florian, G.~Ferrera, F.F.~Freitas, M.~Grazzini, D.~Gon{\c c}alves, J.~Huston, S.~Kallweit, P.~Lenzi, A.C.~Marini, K.~Melnikov,  S.~Menary, C.~Meyer, A.~Pilkington, T.~Plehn, M.~Queitsch-Maitland, D.~Rathlev, V.~Sanz, H.~Sargsyan, M.~Sch\"onherr,  M.~Schulze, D.M.~Sperka, D.~Tommasini, L.~Viliani}
\providecommand{\slash}[1]{\ooalign{$\hfil/\hfil$\crcr${#1}$}}
\providecommand{\ltap}{\raisebox{-.6ex}{\rlap{$\,\sim\,$}} \raisebox{.4ex}{$\,<\,$}} 
\providecommand{\gtap}{\raisebox{-.6ex}{\rlap{$\,\sim\,$}} \raisebox{.4ex}{$\,>\,$}} 
\providecommand{\lra}{\leftrightarrow} 
\providecommand{\naive}{na\"{\i}ve} 
\providecommand{\bom}[1]{{\mbox{\boldmath ${#1}$}}} 
\providecommand{\to}{\rightarrow}
\providecommand{\ito}{\leftarrow} 
\providecommand{\nn}{\nonumber} 
\providecommand{\arrowlimit}[1]{\mathrel{\mathop{\longrightarrow}\limits_{#1}}} 
\providecommand{\ptmin}{p_{T{\rm min}}}
\providecommand{\ptmax}{p_{T{\rm max}}}
\providecommand{\ep}{\epsilon}
\providecommand{\ms}{${\overline {\rm MS}}$}
\providecommand{\mH}{m_H}
\providecommand{\mb}{m_b}
\providecommand{\mt}{m_t}
\providecommand{\pT}{p_T}
\providecommand{\tL}{{\widetilde L}}

\providecommand{\HRule}{\rule{\linewidth}{0.5mm}}
\providecommand{\href}[2]{#2}
\providecommand\as{\alpha_{\mathrm{S}}} 
\providecommand\f[2]{\frac{#1}{#2}} 
\providecommand\Matrix{{\sc Matrix}}
\providecommand\Munich{{\sc Munich}}
\providecommand\OpenLoops{{\sc OpenLoops}}
\providecommand\Collier{{\sc Collier}}
\providecommand{\CutTools}{{\sc CutTools}}
\providecommand{\OneLOop}{{\sc OneLOop}}

\providecommand{\eqn}[1]{Eq.\,(\ref{#1})}
\providecommand{\eqns}[2]{Eqs.\,(\ref{#1}) -- (\ref{#2})}
\providecommand{\refF}[1]{Figure~\ref{#1}}
\providecommand{\refFs}[2]{Figures~\ref{#1} -- \ref{#2}}
\providecommand{\refT}[1]{Table~\ref{#1}}
\providecommand{\refTs}[2]{Tables~\ref{#1} -- \ref{#2}}
\providecommand{\refS}[1]{Section~\ref{#1}}
\providecommand{\refSs}[2]{Sections~\ref{#1} -- \ref{#2}}
\providecommand{\refC}[1]{Chapter~\ref{#1}}
\providecommand{\refCs}[2]{Chapters~\ref{#1} -- \ref{#2}}
\providecommand{\refA}[1]{Appendix~\ref{#1}}
\providecommand{\refAs}[2]{Appendices~\ref{#1} -- \ref{#2}}


\providecommand{\Mfourl}{\mathswitch {m_{4\Pl}}}
\providecommand{\MVH}{\mathswitch {m_{\PV\PH}}}
\providecommand{\PV}{\HepParticle{V}{}{}\Xspace} 
\providecommand\pTH{\ensuremath{p_{\mathrm{T}}^{\PH}}} 
\providecommand\pTV{\ensuremath{p_{\mathrm{T}}^{\PV}}} 
\providecommand\pTj{\ensuremath{p_{\mathrm{T},j}}} 
\providecommand{\ttH}{\ttbar\PH} 
\providecommand{\ggH}{\Pg\Pg \to \PH}

\providecommand{\ltap}{\raisebox{-.6ex}{\rlap{$\,\sim\,$}} \raisebox{.4ex}{$\,<\,$}} 
\providecommand{\gtap}{\raisebox{-.6ex}{\rlap{$\,\sim\,$}} \raisebox{.4ex}{$\,>\,$}} 


\section{Introduction}\label{sec:introFXS} 
Over the past years fiducial measurements, both differential and total, became standard practice for characterizing Standard Model (SM) processes at the LHC. The advantage of quoting such results over inclusive cross section measurements lies in the almost complete factorization of experimental and theoretical uncertainty sources: whereas for inclusive cross section measurements one extrapolates back to account for acceptance and phase-space regions not measured by the detector or removed by the analysis, fiducial cross sections are defined within an experiment dependent fiducial volume of acceptance and phase space. The large extrapolation from the measured subset to the entirety of phase space is reduced to account solely for reconstruction efficiencies and to revert resolution effects and migrations inside and outside the fiducial region. If one now wishes to confront the measured fiducial cross section with a new SM or beyond the SM (BSM) theory, one can readily do so once one accounted for the acceptance and no repetition of the analysis of the original data is needed. This helps preserve the measured results and allow for a comparison even in a scenario of a new theory being developed years after an original measurement has been carried out.

The measurement of differential or total fiducial cross sections offers an alternative approach to study the properties of the Higgs boson: a range of physical phenomena are accessible, such as the theoretical modelling of different Higgs boson production mechanism or BSM contributions, by measuring kinematic distributions constructed from the Higgs boson decay products or other objects produced in association with the Higgs boson. The fiducial or differential cross sections may be split into four categories based on what underlying physical aspect they are probing: 

\begin{itemize}
\item[1.] \underline{Higgs boson kinematics:} In Higgs boson decay final states with no undetectable particles, such as neutrinos, the transverse momentum of the Higgs boson, $\pTH$, and the absolute rapidity, $\left| y^{\PH} \right|$, can be measured with good experimental precision. Inclusive Higgs boson production is dominated by gluon fusion for which the transverse momentum is largely balanced by the emission of QCD radiation. Measuring $p_T^H$ probes the perturbative QCD modelling of this production mechanism. The rapidity distribution of the Higgs boson is sensitive to the gluon fusion production mechanism as well as to the parton distribution functions (PDFs) of the colliding protons.
 \item[]
 \item[2.] \underline{Jet activity:} The jet multiplicity, $N_{\rm jets}$ and the transverse momentum and rapidity distributions of the leading and subleading jets are sensitive to the  theoretical modelling and relative contributions of different Higgs boson production mechanisms.  In the SM events with zero and one jet are dominated by gluon fusion production of the Higgs boson and the transverse momentum and rapidity of the leading jet probes the theoretical modelling of hard quark and gluon radiation in this process. The contribution from vector boson fusion and associated production with vector bosons becomes more important for two-jet events. The contribution from Higgs boson production in association with top-antitop production becomes more relevant in the highest jet multiplicities. 
 \item[]
\item[3.] \underline{Spin and CP quantum numbers:} Angular observables, such as the angle between the Higgs boson decay products and the beam axis or the azimuthal angle between the two leading jets in events containing two or more jets are sensitive to the spin and charge conjugation and parity properties of the Higgs boson, respectively. 
\item[]
\item[4. ] \underline{Higgs boson production mechanisms}: Specific fiducial regions may be constructed which target certain Higgs boson production mechanism. For instance in events with two or more jets, one may look at the subset of events with jets with a large rapidity separation between the two leading jets and a large dijet mass. This selects events with suppressed colour flow between the two jets, which is expected for Higgs boson production from the vector boson fusion process, but not from gluon fusion. Using similar criteria Higgs boson production associated with vector bosons or top-antitop pairs can be targeted and fiducial cross sections enriched with such can be studied. 
\end{itemize}

For BSM contributions some regions are more interesting than others: in some scenarios large contributions near kinematic corners of phase space are expected. The purpose of this chapter is to review the state of the art of predicting fiducial cross sections inside the SM and provide a summary of interesting observables and fiducial regions that can be used to search for physics beyond the SM. In addition a summary of the experimental aspects of fiducial measurements are reviewed and a range of suggestions for future measurements is given. The rest of this chapter is organized as follows: 
Section~\ref{sec:run1} provides a brief summary of fiducial measurements carried out in Run~1 of the LHC. 
Section~\ref{sec:SM} reviews the state-of-the-art of predicting fiducial cross sections for the SM 
and discusses the current limitations. These studies are carried out using template fiducial regions with acceptance and kinematic cuts close to the expected definition for Run~2 for the ATLAS and CMS experiments. 
Section~\ref{sec:BSM} reviews the effects that can arise from new physics contributions. In particular, a range of suggestions is presented which kinematic regions might be probed in future measurements. Section~\ref{sec:EXP} provides a review of the necessary experimental methods and prerequisites to carry out fiducial measurements. Among the aspects discussed is a brief review of what points should be considered when reverting migrations inside the fiducial region or in and outside the fiducial region. The combination of several measurements, as well as the implications of different treatments of the Higgs boson mass is discussed as well. The chapter concludes with Section~\ref{sec:future} with a list of recommendations on how future measurements should be presented. 

\section{Review of Run~1 and early Run~2 results} \label{sec:run1} 
In the first running period of the LHC, several measurements of fiducial cross sections have been carried out by the ATLAS and CMS experiments. These measurements mark the transition from the discovery of the Higgs boson, and first measurement characterizing its quantum numbers, to less statistical powerful but more model independent statements about its properties. Such measurements were first carried out in the channels with high mass resolution, $\PH \to \PGg\PGg$~\cite{Aad:2014lwa,Khachatryan:2015rxa} and $\PH \to \PZ\PZ^*\to 4\Pl$~\cite{Aad:2014tca,Khachatryan:2015yvw}, 
More recently, the first measurements of the differential fiducial cross sections in $H \to WW$ decay channel also appeared~\cite{Khachatryan:2016vnn}.
First combinations of total and differential information from $\PH \to \PGg\PGg$ and $\PH \to \PZ\PZ^*\to 4\Pl$ also were carried out, extrapolating into the inclusive phase space and correcting for the difference in branching fraction, cf. Ref.~\cite{Aad:2015lha}. A variety of results were reported, ranging from inclusive and exclusive jet multiplicities to differential information characterizing the Higgs boson, its decay products, or objects produced in association with it: the Higgs boson $p_T$ distribution and absolute rapidity has been measured, the number of jets as well as the $p_T$ and rapidity distributions of the leading and sub-leading jets. Angular observables from jets and jets plus the Higgs boson decay products have been reported. Figure~\ref{fig:run1:results} shows a summary of the fiducial regions measured by ATLAS in the $\PH \to \PGg\PGg$  channel and the measured Higgs boson $p_T$ spectrum reported by CMS using $\PH \to \PZ\PZ^*\to 4\Pl$ events. The precision of the probed fiducial regions and differential observables are all statistically limited at this point.

\begin{figure}
\includegraphics[width=0.54\textwidth]{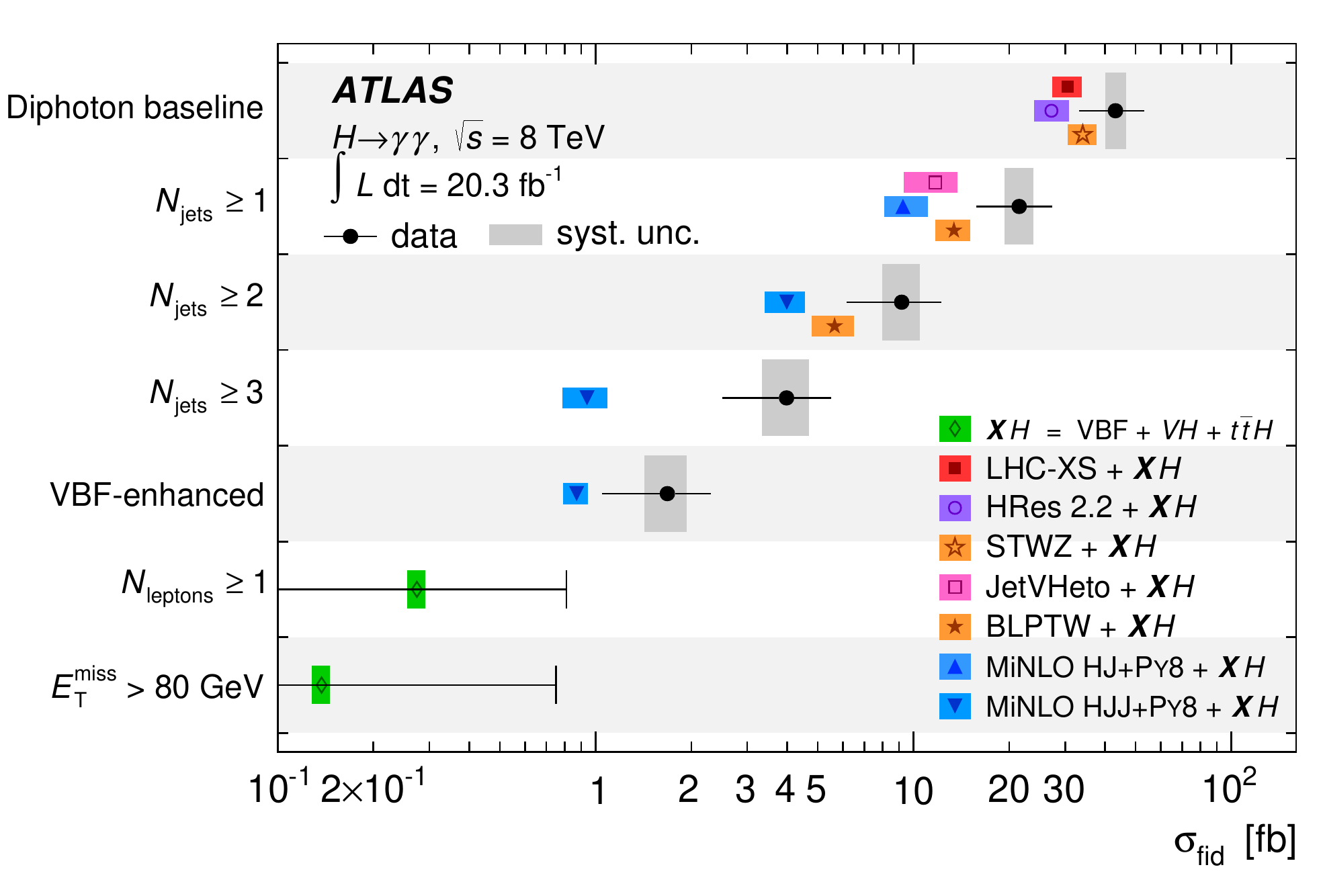}
\includegraphics[width=0.38\textwidth]{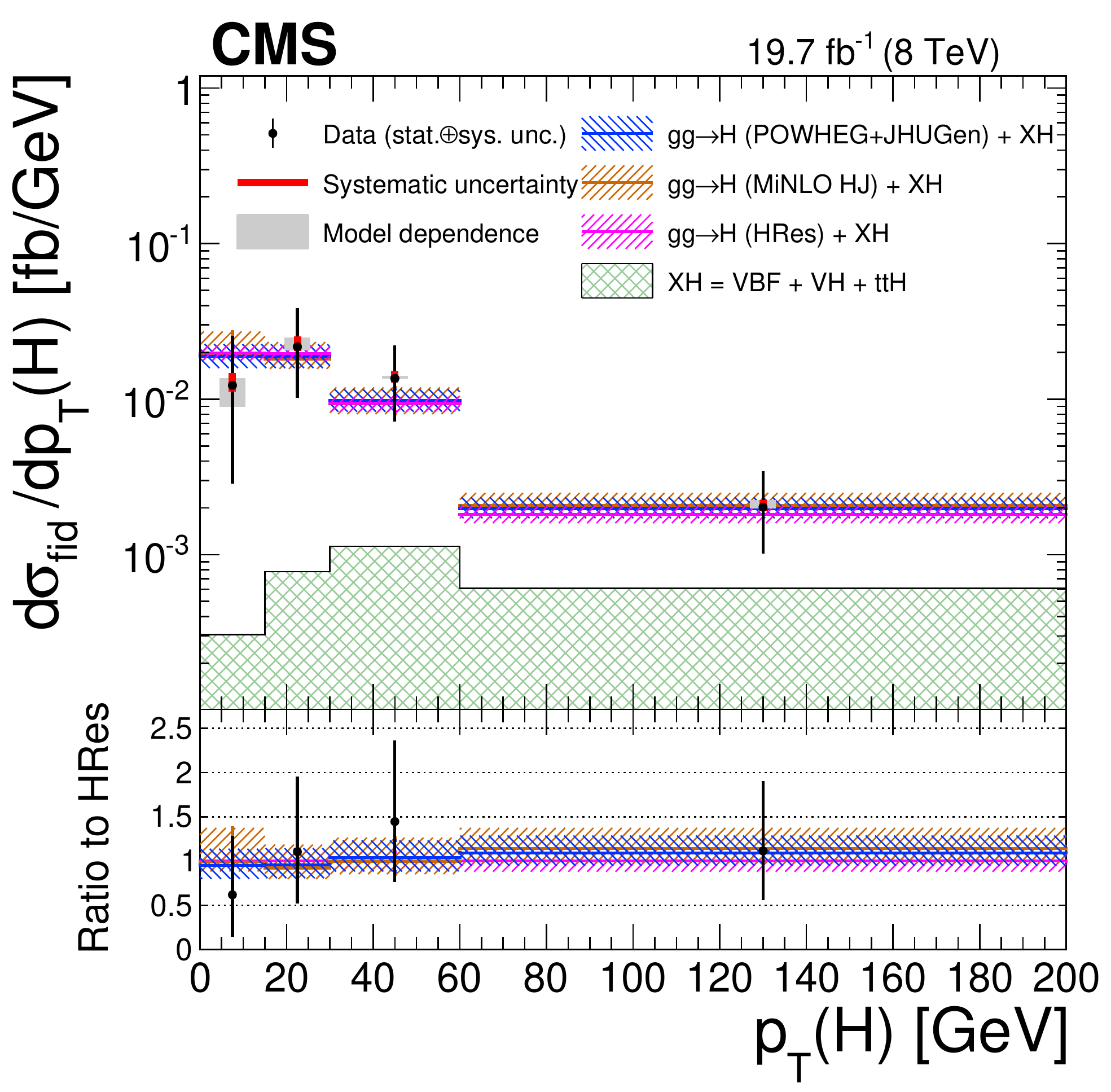}
\caption{
(left) Various fiducial regions measured by Ref.~\cite{Aad:2014lwa} using $\PH \to \PGg\PGg$ decays: the data points show measured cross sections of total, inclusive jet multiplicities as well as VBF and $\PV\PH$ enhanced regions. The coloured bands show several theory predictions for the different fiducial regions for gluon fusion and other SM Higgs boson production. (right) The measured differential Higgs boson $\pT$ spectrum is shown from Ref.~\cite{Khachatryan:2015yvw} using $\PH \to \PZ\PZ^*\to 4\Pl$ events. The data points show the measured cross section and the different shaded bands theory predictions from gluon fusion and other SM Higgs boson production. 
}
\label{fig:run1:results}
\end{figure}

The reported cross sections are unfolded to the particle level, defined by particles that have lifetimes such that $c \tau > 10$ mm. The reported fiducial volumes differ for each final state and between experiments. The defining criteria of the fiducial volumes were chosen to be very similar to the criteria applied at detector level to ensure minimal model dependence in the final measurements. In both experiments, leptons are identified using an isolation criterion by summing over energetic clusters in a cone around the charged track trajectory. This can be mimicked at particle level by requiring a similar isolation in a cone around the particle-level lepton. For photons a similar requirement can be imposed to closely match the experimental selection. Other requirements that enter the fiducial acceptance is the overall detector acceptance, trigger threshold energies, and analysis selection cuts. Once defined, selection related efficiencies can be reverted to convert fitted yields into fiducial cross sections. Migrations inside the fiducial volumes from finite resolution are reverted as well, what allows direct comparison with theory predictions. Two different methods to accomplish this are used right now: the ATLAS $\PH \to \PGg\PGg$ and $\PH \to \PZ\PZ^*\to 4\Pl$ results chose to use correction factors, while the CMS $\PH \to \PZ\PZ^*\to 4\Pl$ and the CMS $\PH \to \PGg\PGg$ results directly inverted the migration matrix. 
Several results~\cite{Aad:2014lwa,Aad:2014tca} are already published on HEPDATA~\cite{hepdata} along with example RIVET routines~\cite{rivet} to apply the corresponding particle level fiducial selection of each analysis, while the others are expected to follow~\cite{Khachatryan:2015rxa, Khachatryan:2015yvw, Khachatryan:2016vnn}. 

The ATLAS experiment reported in a follow up publication to Ref.~\cite{Aad:2014lwa} also the statistical correlations between five measured differential distributions in Ref.~\cite{Aad:2015tna}. This allows the simultaneous analysis of several differential distributions and the result of a proof-of-concept analysis constraining BSM physics is shown in Figure~\ref{fig:run1:results2} along with the determined statistical correlations between exclusive jet bins and the Higgs boson $\pT$. Finally, ATLAS reported fiducial cross sections of $\Pp\Pp \to \PZ\PZ^{(*)}\to 4\Pl$ in which the Higgs boson contribution from $\PH \to \PZ\PZ^*\to 4\Pl$ was included as part of the signal definition~\cite{Aad:2015rka}.

\begin{figure}
\includegraphics[width=0.56\textwidth]{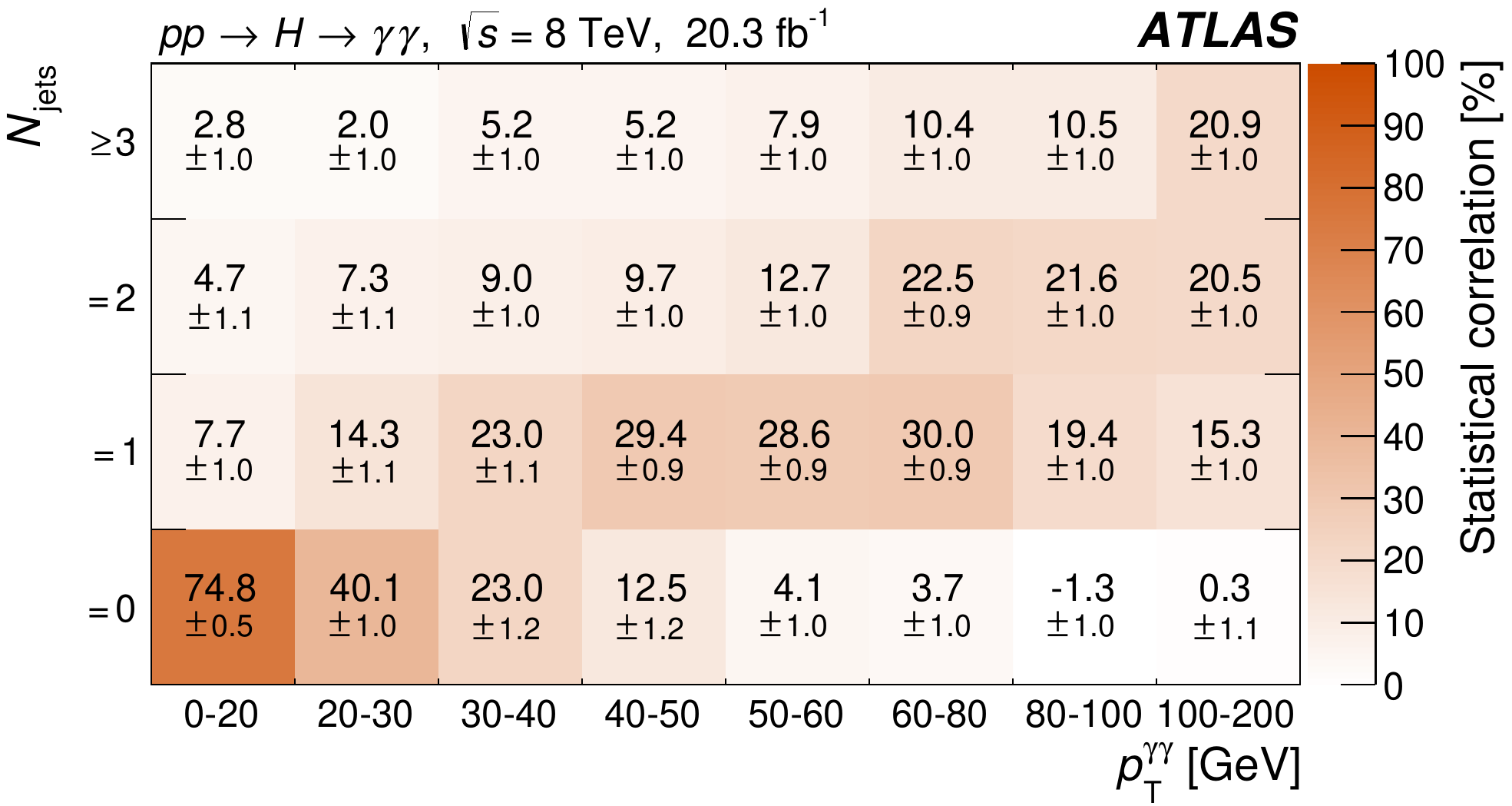}
\includegraphics[width=0.42\textwidth]{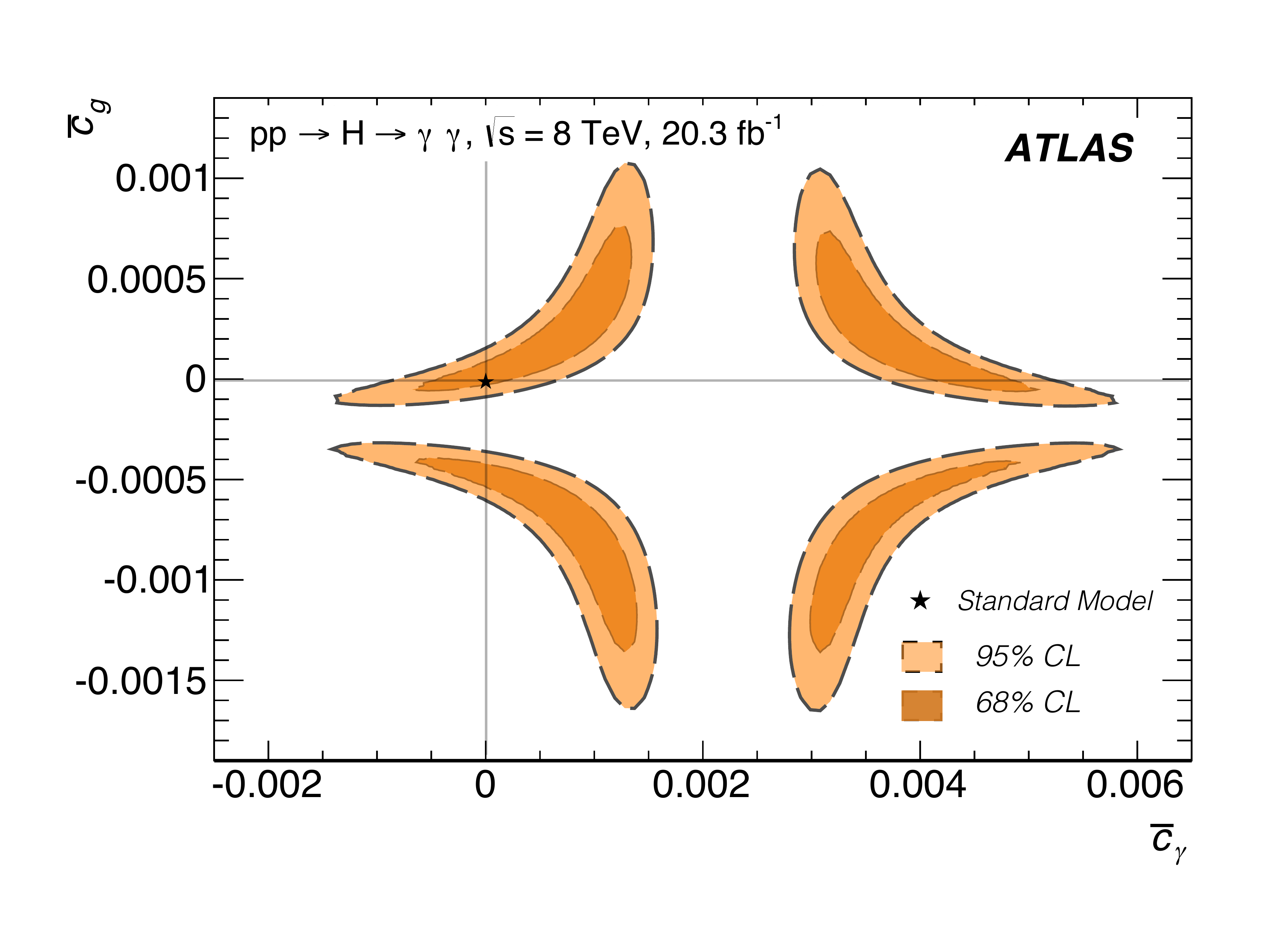}
\caption{
(left) Statistical correlations between exclusive jet bins and Higgs boson $p_T$. (right) Proof-of-concept analysis using five differential distributions: allowed 95\% and 68\% CL for two Wilson coefficients which add additional point-like interactions for Higgs boson production via gluon fusion ($c_g$) and Higgs boson decay into two photons ($c_\gamma$) are shown.
}
\label{fig:run1:results2}
\end{figure}

Fiducial cross sections results using early Run~2 data were reported by both ATLAS and CMS for the $\PH \to \PGg\PGg$ and $\PH \to \PZ\PZ^*\to 4\Pl$ channels. \refF{fig:run1:results3} shows preliminary results of the total fiducial cross section for centre-of-mass energies at $7$, $8$ and $13$ \UTeV\ from Refs.~\cite{ATLAS-CONF-2015-060,CMS:2016rqf}.

\begin{figure}
\includegraphics[width=0.53\textwidth]{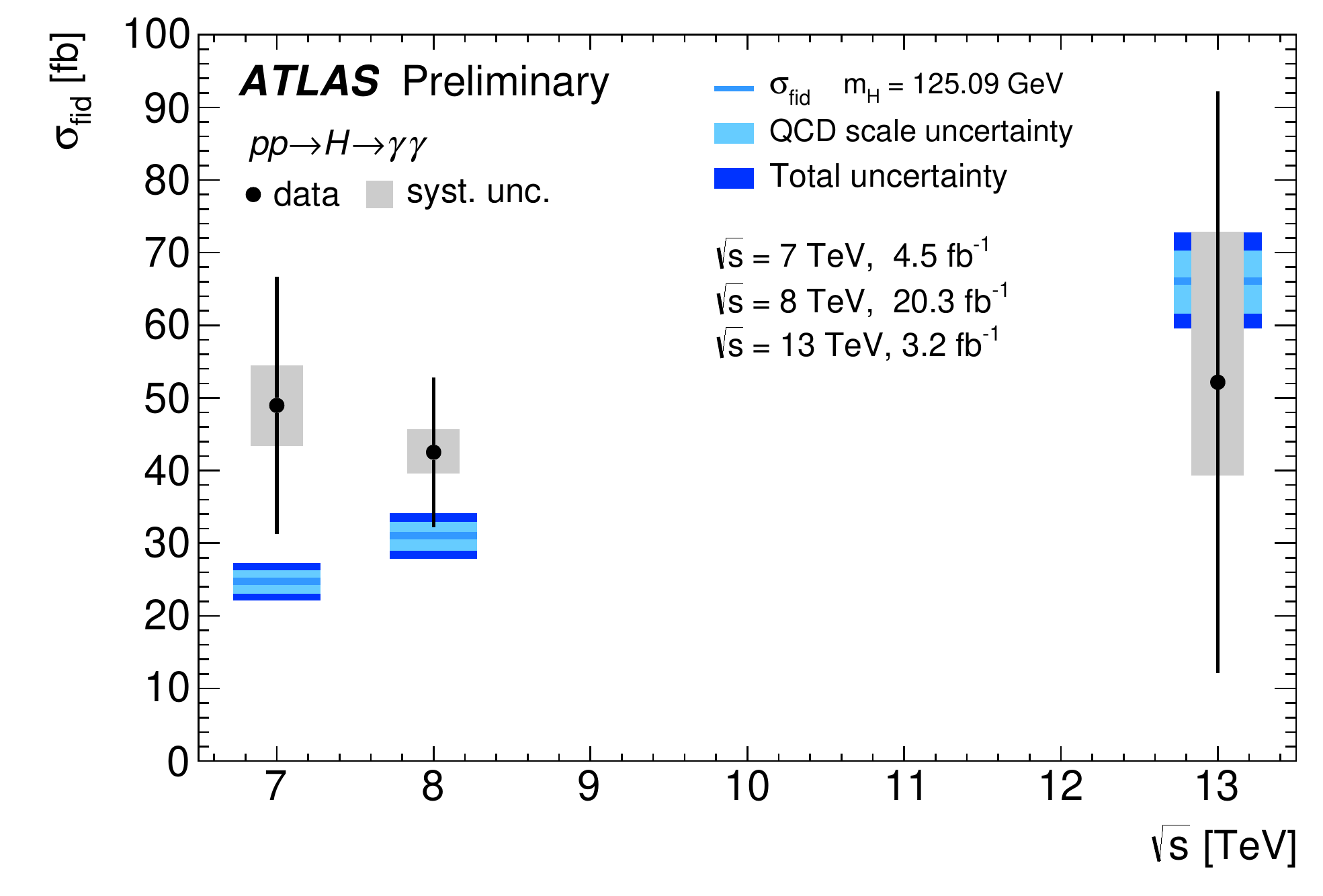}
\includegraphics[width=0.47\textwidth]{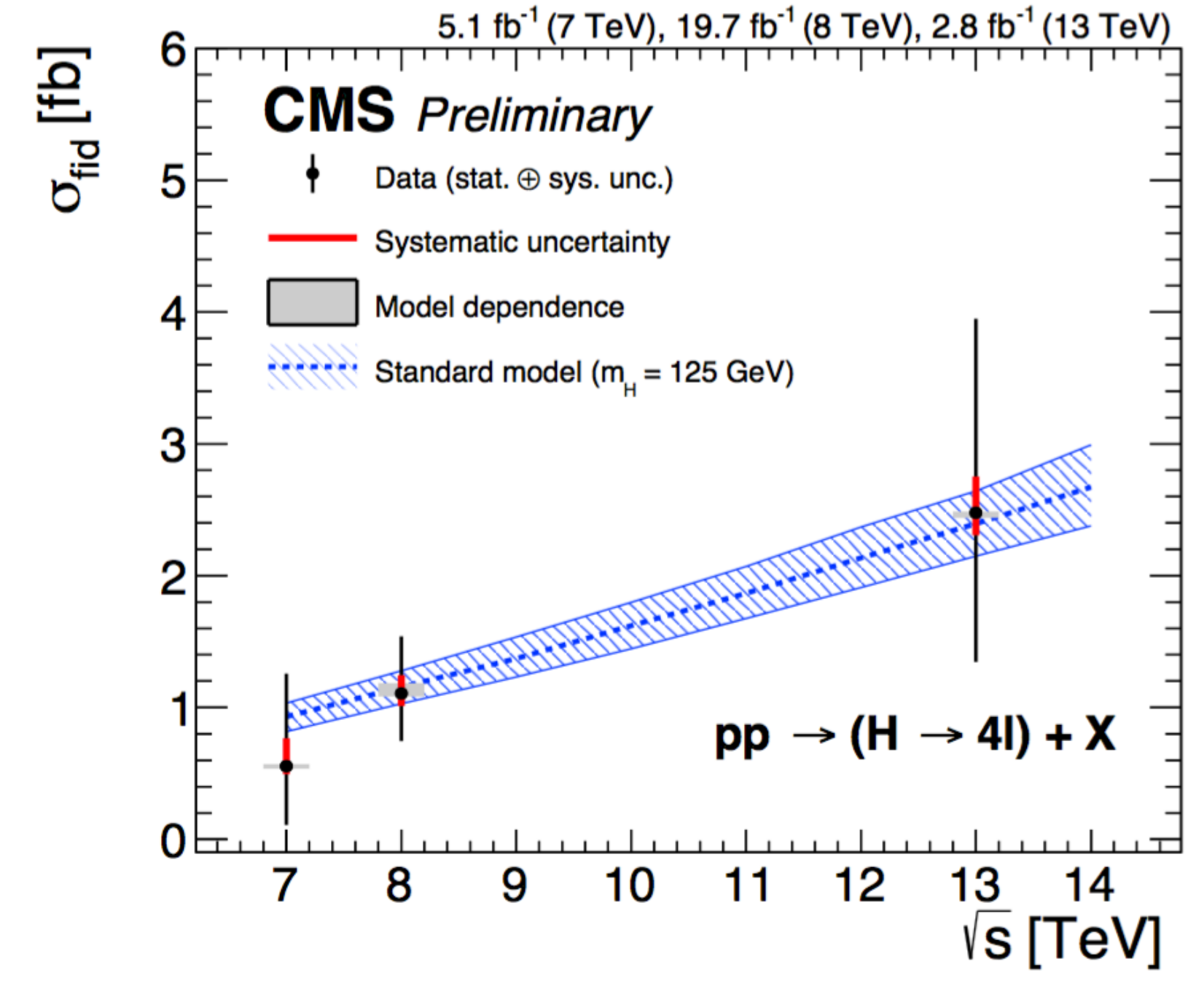}
\caption{
(left) Fiducial cross sections from $\PH \to \PGg\PGg$ for $\sqrt{s} = 7$, $8$ and $13$\UTeV. The fiducial volumes were extrapolated so that there is an identical definition between all centre-of-mass energies. The hatched theory band shows the prediction from the SM. (right) Fiducial cross sections from $\PH \to \PZ\PZ^*\to 4\Pl$ for $\sqrt{s} = 7$, $8$ and $13$\UTeV also with matching fiducial volume definitions. 
}
\label{fig:run1:results3}
\end{figure}

\clearpage
\section{State-of-the-art Standard Model predictions}\label{sec:SM}

Sometimes in the experimental analyses the fiducial cross sections are
extrapolated onto the full phase space by simulating the relevant
geometric acceptances by means of various Monte Carlo generators in
order to quote a total cross section for the underlying process
(cf. Ref.~\cite{Aad:2015lha} and Section~\ref{sec:exp:xsec_comb}).
In view of the high precision expected in Run 2 fiducial measurements,
it is of primary relevance to provide the experiments with accurate
theory predictions for two reasons.
On the one hand, it is necessary to perform an comparison to data at
the fiducial level, before extrapolating to the inclusive phase space,
and to quote the relative measurement publicly. 
On the other hand, it is important to study how the tools used in the
comparison and in the extrapolation behave in the presence of the
fiducial cuts, in order to avoid the propagation of unwanted
generator-dependent effects into the quoted total cross sections.
In order to have a robust control over the extrapolation procedure, it
is in fact necessary to assess precisely the performance of the event
generators as well as the theoretical uncertainties associated with
such simulations, both at the perturbative and non-perturbative level.
The ultimate and future goal of this section of the Task Force is to
perform a comprehensive comparison and validation of the predictions
obtained with the event generators currently used in experimental
analyses to the best available results.

The development of several computational techniques in the past few
years considerably improved the known perturbative accuracy of many of
the relevant signal and background processes. In the rest of this
section we provide templates for the fiducial volumes relative to the
three final states $\gamma\gamma$, $\PW\PW^{*}\to 2\Pl 2\PGn$, $\PZ\PZ^{*}\to
4\Pl$, followed by the state-of-the-art predictions for the signal and,
whenever possible, background processes.
As far as the signal is concerned, since the template fiducial volumes
that will be given below are dominated by the gluon fusion production
mode, we limit ourselves to considering the latter in this section. A
comprehensive review of the available predictions and public tools can
be found in Chapter~\ref{chap:ggF} of this volume.
The contribution of additional production modes, notably VBF and VH
associate production, should be also taken into account when
considering the fiducial template volumes presented here. A review of
the available results for both QCD and EW effects can be found in the
Chapter~\ref{chap:VBF+VH} of the present volume. However, considering
the moderate size of the latter production channels in comparison to
the gluon-fusion mode, it is necessary to define different sets of
fiducial cuts which enhance their contribution.
This study is addressed in Chapter~\ref{chap:VBF+VH} of this report,
and it will be considered in the future by this Task Force.
Analogously, the precise study of subdominant contributions to a given
fiducial volume may require the definition of multiple specific
fiducial sub-categories, with the goal of enhancing different
kinematic regimes and increasing the experimental
sensitivity. Examples of these regions are the tails of the
differential distributions, or the off-shell production regime.

In the rest of this section, we classify the signal predictions into
two categories, according to their jet multiplicity: totally inclusive
in the number of QCD jets, and with at least one jet. For such
processes NNLO QCD predictions for the fiducial cross sections and
distributions will be reported in
Sections~\ref{sec:hres-YR4},~\ref{sec:hjet} below.
Since some analyses use the theory calculations to perform the
 background subtraction, the same type of validation is necessary for
 the relevant background processes. As far as the irreducible
 background reactions are concerned, while in $\gamma\gamma$
 production a data-driven fit is used to estimate the background, in
 the $\PZ\PZ^*$ and $\PW\PW^*$ final states a precise theoretical calculation
 is necessary. Section~\ref{sec:ZZ-YR4} reports NNLO QCD predictions for
 $\PZ\PZ^*\to 4\Pl$ production, while the $\PW\PW^*$ case is left for future
 work.

A similar comprehensive benchmarking in the presence of exclusive cuts
on the QCD activity accompanying Higgs boson production is being conducted
within the context of the Les Houches 2015 Workshop~\cite{Badger:2016bpw}. We
conclude this section with a summary of the detailed comparisons which
have been carried out. The future validation of tools in the presence
of realistic fiducial cuts should benefit from the interaction and
coordination of activities between the two working groups.

\def\ltap{\raisebox{-.6ex}{\rlap{$\,\sim\,$}} \raisebox{.4ex}{$\,<\,$}}
\def\gtap{\raisebox{-.6ex}{\rlap{$\,\sim\,$}} \raisebox{.4ex}{$\,>\,$}}
\subsection{Template fiducial regions for benchmark }
\label{sec:fiducialvolumes}
In this section we report the definitions of the template fiducial
volumes for Higgs boson production.
These templates do not serve as a reference for the fiducial
definitions that will be used in Run 2 analyses, but rather as a
plausible set of cuts that will be used in the benchmarking
process. Kinematic cuts are listed according to the Higgs boson decay
mode, for the three final states $\PH \to \PZ\PZ^*\to 4\Pl$,
$\PH \to \PGg \PGg$, and $\PH \to \PW\PW^*\to 2\Pl 2 \PGn$.
As mentioned in the introduction to this chapter, the definition of
the fiducial regions should have as little impact as possible on the
phase space available for BSM searches in Higgs signatures, and should
be defined mainly in order to minimize the model dependence.
Therefore, fiducial regions will be largely defined by the detector
coverage and by trigger performances, rather than by theoretical
prejudice.

\subsubsection{Fiducial volume for \texorpdfstring{$\PH \to \PZ\PZ^*\to 4\Pl$}{H to ZZ* to 4l}}
Muons (electrons) are required to have a transverse momentum larger
than $5$\,\UGeV ( $7$\,\UGeV) and rapidity $|\eta|\leq 2.5$.
The leading lepton pair is defined as the
same-flavoured-opposite-signed (SFOS) lepton pair with the smallest
$|m_{\PZ} - m_{\Pl\Pl}|$. The leading-lepton-pair invariant mass is denoted by
$m_{12}$. The subleading lepton pair is the remaining pair of SFOS
leptons with smallest $|m_{\PZ} - m_{\Pl\Pl}|$. Its invariant mass is denoted
by $m_{34}$.
Jets are reconstructed using the anti-$k_t$
algorithm~\cite{Cacciari:2008gp} with a radius parameter of 0.4. All
jets with a transverse momentum larger than $30$\,\UGeV and rapidity
$|\eta|\leq 4.4$ are used in the selection criteria. Neutrinos are not
considered in the jet definition.

\noindent The fiducial volume for the $\PH \to \PZ\PZ^*\to 4\Pl$ channel is reported
in Table~\ref{tab:fiducialZZ}
\begin{table}[htp]
\caption{Template fiducial cuts for the $\PH \to \PZ\PZ^*\to 4\Pl$ channel.}
\label{tab:fiducialZZ}
\centering
\footnotesize
\begin{tabular}{l}
  \toprule
  Template fiducial region for $\PH \to \PZ\PZ^*\to 4\Pl$\\
  \midrule
  Leading lepton:  $p_{t}>20$\,\UGeV\\\\
  1$^{st}$ subleading lepton: $p_{t}>10$\,\UGeV\\\\
  2$^{nd}$ subleading lepton: $p_{t}>7\,(5)$\,\UGeV for electrons
  (muons)\\\\
  3$^{rd}$ subleading lepton: $p_{t}>7\,(5)$\,\UGeV for electrons
  (muons)\\\\
  All leptons are required to be isolated:\\
  ratio of the sum of $p_t$'s of all charged particles within $\Delta R =  [ \left( \Delta \phi \right)^2 + \left( \Delta \eta \right)^2 ]^{1/2} <
  0.4$\\
  from the lepton to the lepton's $p_t$ must be smaller than $0.4$  \\\\
  Mass requirements: $40\,{\rm GeV} \leq m_{12} \leq 120\,\UGeV$;
  $12\,\UGeV \leq m_{34} \leq 120\,\UGeV$\\\\
  Lepton separation: $\Delta R(i,j) > 0.1$ for all leptons $i$, $j$\\\\
  $J/\Psi$ invariant mass veto: $m_{ij} > 4$\,\UGeV for all SFOS leptons
  $i$, $j$\\\\
  Invariant mass cut: $120\,\UGeV \leq m_{4\Pl} \leq 130\,\UGeV$\\
  \\ \bottomrule
\end{tabular}
\end{table}

\subsubsection{Fiducial volume for \texorpdfstring{$\PH \to \PGg\PGg$}{H to gamma gamma}}
Photons are requested to have a transverse momentum larger than
$25$\,\UGeV and rapidity $|\eta|\leq 2.5$.
The photon pair with the largest transverse momentum is denoted as the
leading photon pair. Its invariant mass is denoted by
$m_{\PGg\PGg}$.
Jets are reconstructed using the anti-$k_t$
algorithm~\cite{Cacciari:2008gp} with a radius parameter of 0.4. All
jets with a transverse momentum larger than $30$\,\UGeV and rapidity
$|\eta|\leq 4.4$ are used in the selection criteria. Neutrinos are not
considered in the jet definition.

\noindent The fiducial volume for the $\PH \to \PGg\PGg$ channel is reported
in Table~\ref{tab:fiducialyy}
\begin{table}[htp]
\caption{Template fiducial cuts for the $\PH \to \PGg\PGg$ channel.}
\label{tab:fiducialyy}
\centering
\footnotesize
\begin{tabular}{l}
  \toprule
  Fiducial region for $\PH \to \PGg\PGg$\\
  \midrule
  Leading photon:  $p_{t}/m_{\PGg\PGg}>0.35$\\\\
  Subleading photon:  $p_{t}/m_{\PGg\PGg}>0.25$\\\\
  All photons are required to be isolated:\\
  ratio of the sum of $E_t$'s of all charged particles within $\Delta R =  [ \left( \Delta \phi \right)^2 + \left( \Delta \eta \right)^2 ]^{1/2} <
  0.2$\\
  from the photon to the photon's $E_t$ must be smaller than $0.2$  \\ \\
  Invariant mass cut: $105\,\UGeV \leq m_{\PGg\PGg} \leq 160\,\UGeV$\\
  \\ 
  \bottomrule
\end{tabular}
\end{table}

\subsubsection{Fiducial volume for \texorpdfstring{$\PH\to W^+W^-\to 2\Pl 2\PGn$}{H to WW to 2l2nu}}
The leading (subleading) lepton is required to have a transverse
momentum larger than $20$\,\UGeV ($10$\,\UGeV), and rapidity
$|\eta|\leq 2.5$.
Jets are reconstructed using the anti-$k_t$
algorithm~\cite{Cacciari:2008gp} with $R=0.4$. All jets with a
transverse momentum larger than $30$\,\UGeV and rapidity
$|\eta|\leq 4.4$ are used in the selection criteria. Neutrinos are not
considered in the jet definition.

\noindent The fiducial volume for the $\PH\to W^+W^-\to 2\Pl 2\PGn$
channel is reported in Table~\ref{tab:fiducialWW}
\begin{table}[htp]
\caption{Template fiducial cuts for the $\PH\to W^+W^-\to 2\Pl 2\PGn$ channel.}
\label{tab:fiducialWW}
\centering
\footnotesize
\begin{tabular}{l}
  \toprule
  Fiducial region for $\PH\to W^+W^-\to 2\Pl 2\PGn$\\
  \midrule
  Lepton invariant mass:     $m(ll) > 12$\,\UGeV \\\\
  Lepton transverse momentum:     $p_{t, \Pl\Pl} > 30$\,\UGeV \\\\
  All leptons are required to be isolated:\\
  ratio of the sum of $p_t$'s of all charged particles within $\Delta R =  [ \left( \Delta \phi \right)^2 + \left( \Delta \eta \right)^2 ]^{1/2} <
  0.4$\\
  from the lepton to the lepton's $p_t$ must be smaller than $0.4$  \\\\
  Lepton transverse mass:     $m_{T}(ll,\PGn\PGn) > 50$\,\UGeV \\\\
  where  $m_{T}(ll,\PGn\PGn) = \sqrt{(E_{t,\Pl\Pl} + p_{t,\PGn\PGn})^2 - |\vec{p}_{t,\Pl\Pl} +
  \vec{p}_{t,\PGn\PGn}|^2}$, and $E_{t,\Pl\Pl}=\sqrt{p_{t,\Pl\Pl}^2+m_{\Pl\Pl}^2}$ \\\\
  Missing transverse energy:     $E^{\rm miss}_{T} > 15$\,\UGeV \\
  \\
  \bottomrule
\end{tabular}
\end{table}

\def\ltap{\raisebox{-.6ex}{\rlap{$\,\sim\,$}} \raisebox{.4ex}{$\,<\,$}}
\def\gtap{\raisebox{-.6ex}{\rlap{$\,\sim\,$}} \raisebox{.4ex}{$\,>\,$}}

\subsection{Fiducial cross sections for Higgs boson production in   association with \texorpdfstring{$n_{\rm jet} \geq 1$}{n_jet >= 0} jets}
\label{sec:hres-YR4}

In this section we present results at $13$\,TeV for the fiducial cross
sections and some kinematic distributions for Higgs boson production,
inclusive in the number of QCD jets.
The calculation is performed with the {\ttfamily
  HRes}\,\cite{deFlorian:2012mx,Grazzini:2013mca} program, which
computes the cross section for SM Higgs boson ($\PH$) production by
gluon-gluon fusion. {\ttfamily HRes}\footnote{The program can be
  downloaded from Ref.\,\cite{hres}, together with some accompanying
  notes.} combines the fixed-order calculation of the cross section up
to NNLO with the resummation of the logarithmically-enhanced
contributions at small transverse momentum ($q_T$) up to NNLL accuracy
in QCD. The method that is used to perform the resummation is
presented in Ref.~\cite{Catani:2000vq}.

The produced Higgs boson subsequently decays into the three final
states $\PH\rightarrow \PGg\PGg$,
$\PH\rightarrow \PW\PW^* \rightarrow 2\Pl 2\PGn$ or
$\PH \rightarrow Z Z^* \rightarrow 4\Pl$.  In the case of
$\PH\rightarrow \PW\PW^*$ and $\PH\rightarrow \PZ\PZ^*$ decays, finite-width
effects of the vector bosons and the appropriate interference
contributions are also included.
The results below are generally obtained in the large-$m_t$
approximation, and the effects of top- and bottom-quarks in the
production are included following ref.~\cite{Grazzini:2013mca}.
The calculation retains the full kinematics of the Higgs boson and of
its decay products, allowing the user to apply arbitrary cuts on these
final-state kinematical variables, and to plot the corresponding
distributions in the form of bin histograms. Given that the
Higgs boson transverse momentum resummation is fully inclusive in the
QCD initial-state radiation, we consider the sets of fiducial volumes
presented in the previous subsection, ignoring the cuts on QCD
jets. This in particular affects the isolation requirements on photons
and charged leptons.


The parton distribution functions are chosen according to the
PDF4LHC15 recommendation~\cite{Butterworth:2015oua}, with densities
and $\alpha_S$ evaluated at each corresponding perturbative order.  As
for the electroweak parameters, we follow the LHCHXSWG
recommendations~\cite{Heinemeyer:2013tqa}, in particular we set the Higgs boson
mass at $\MH = 125$~\UGeV .  The renormalization and factorization
scales are fixed to the value $\mu_R=\mu_F=\MH$ while the resummation
scale is fixed at the value $Q=\MH/2$.

We start by considering the diphoton decay channel
$\PH\rightarrow\PGg\PGg$.  The cuts on final state photons are
reported in \refT{tab:fiducialyy}, and QCD jets are not considered
in the isolation criterion.
The corresponding fiducial cross sections are shown in
\refT{tab}, which reports the resummed results at NLL+NLO and
NNLL+NNLO level, and we compare them with the NNLO fixed order
predictions obtained with the {\ttfamily HNNLO}
code.\,\cite{Catani:2007vq}.

\begin{table}
  \caption{
    Fiducial cross sections for $p p\to \PH+ X \to \PGg\PGg+ X$, $p p\to \PH+ X \to W^+W^-  + X\to 2\Pl2\PGn + X$ and
    $p p\to \PH+ X \to \PZ\PZ  + X\to 4\Pl + X$ at the LHC ($\sqrt{s}=13$~TeV): fixed-order
    results at NNLO and corresponding resummed results at NLL+NLO and
    NNLL+NNLO. The result in the $W^+W^-$ channel refers to a single
    lepton family.
    The uncertainties are obtained by varying the scales $\mu_R=\mu_F$
    by a factor of two around the central value. For the resummed
    calculations, in addition, the resummation scale $Q$ is varied by
    a factor of two in either direction by keeping $\mu_R=\mu_F=2Q$.}
\label{tab}
\begin{center}
\begin{tabular}{lcccc}
\toprule
Cross section [fb]        &  NLL+NLO          &  NNLL+NNLO       &   NNLO             \\
\midrule
$\PH\to \PGg\PGg$    &  41.63$^{+9\%}_{-8\%}$& 54.2$^{+9\%}_{-8\%}$  &    54.6$^{+9\%}_{-8\%}$  \\
$\PH\to W^+W^- \to 2\Pl2\PGn$   & 34.99$^{+9\%}_{-8\%}$ & 45.4$^{+9\%}_{-8\%}$  &  46.0 $^{+9\%}_{-8\%}$  \\
$\PH\to \PZ\PZ \to e^+e^-\PGm^+\PGm^-$  & 0.792$^{+9\%}_{-8\%}$ &  1.042$^{+9\%}_{-8\%}$   &  1.042$^{+9\%}_{-8\%}$ \\
$\PH\to \PZ\PZ \to e^+e^-e^+e^-$    &   0.441$^{+9\%}_{-8\%}$  &  0.581$^{+9\%}_{-8\%}$   &    0.583$^{+9\%}_{-8\%}$  \\
$\PH\to \PZ\PZ \to \PGm^+\PGm^-\PGm^+\PGm^-$    &   0.521$^{+9\%}_{-8\%}$  &  0.685$^{+9\%}_{-8\%}$   &    0.687$^{+9\%}_{-8\%}$  \\
\bottomrule
\end{tabular}
\end{center}
\end{table}

We recall that at large values of $q_T$ the resummed calculation
implements a perturbative unitarity constraint\,\cite{Catani:2000vq},
that guarantees to exactly reproduce the NNLO value of the total cross
section for the Higgs boson production after integration over $q_T$.
However, kinematic cuts affects in a different way fixed order and
resummed predictions, leading to small differences in the fiducial
cross sections, as shown in \refT{tab}.
%
%

Figures~\ref{fig:pTgg} and~\ref{fig:pTglead} show some relevant
differential distributions. The left plot of \refF{fig:pTgg}
shows the diphoton transverse-momentum $p_T^{\PGg\PGg}$ distribution,
and the right plot shows the $p_T^{t,\PGg\PGg}$ distribution defined
as the magnitude of the transverse momentum of the diphoton system
perpendicular to the diphoton thrust axis $\hat t$.  The transverse
momentum distributions for the leading and subleading photons are
shown in \refF{fig:pTglead}. By inspecting these plots we see
that the resummed calculations are essential to restore the
predictivity of perturbation theory in the small-transverse-momentum
region. Moreover, NNLL resummation gives a sizeable effect, with
respect to the NNLO calculations, in a wide intermediate region up to
the chosen resummation scale $q_T \ltap \MH/2$. Finally in the
large-$q_T$ region ($q_T\sim \MH$), where the resummation does not
improve the accuracy of the fixed-order expansion, the NNLL+NNLO
predictions obtained with {\ttfamily HRes} show perfect agreement with
the NNLO ones.

\begin{figure} \centering
\includegraphics[width=0.485\textwidth]{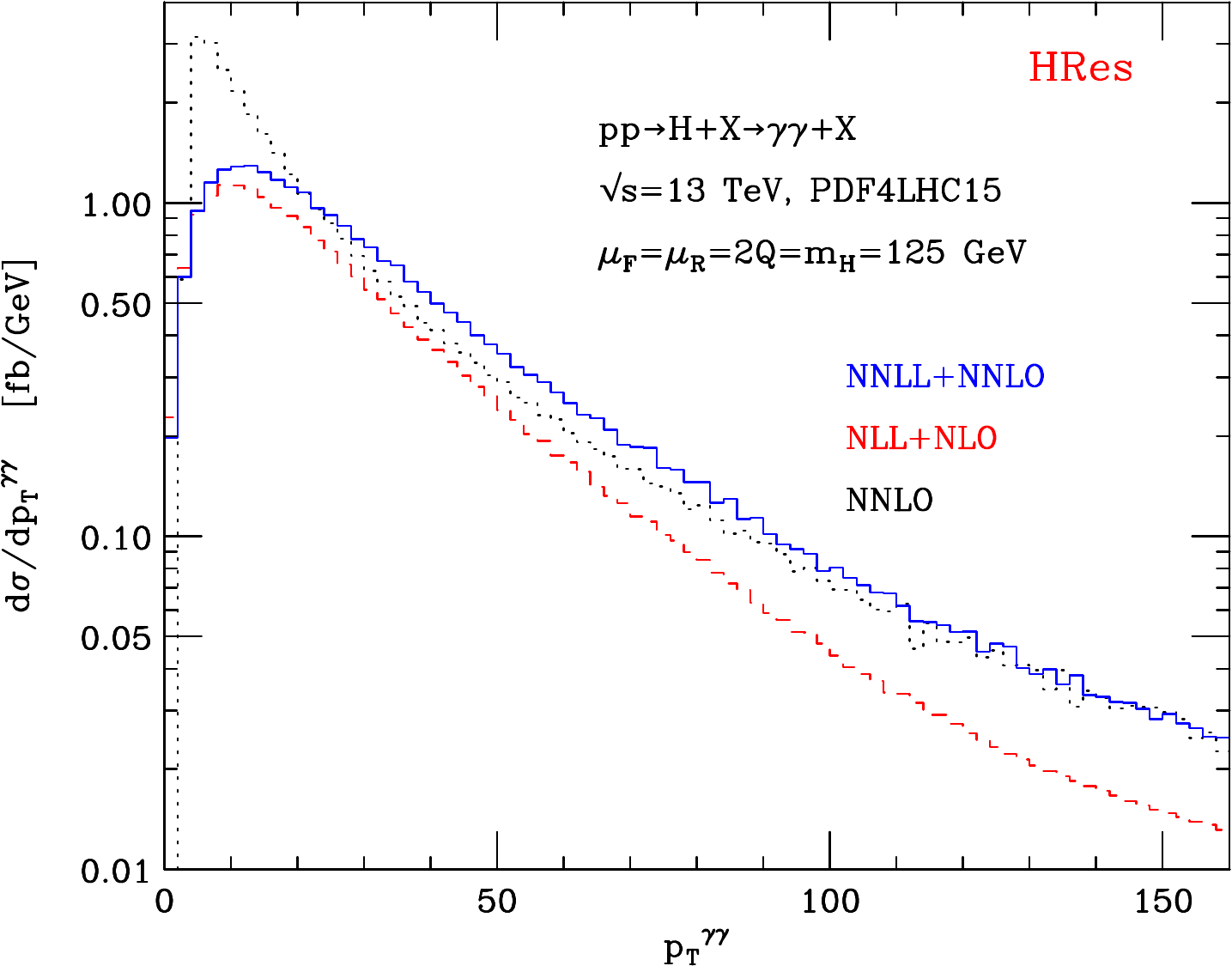}
\includegraphics[width=0.485\textwidth]{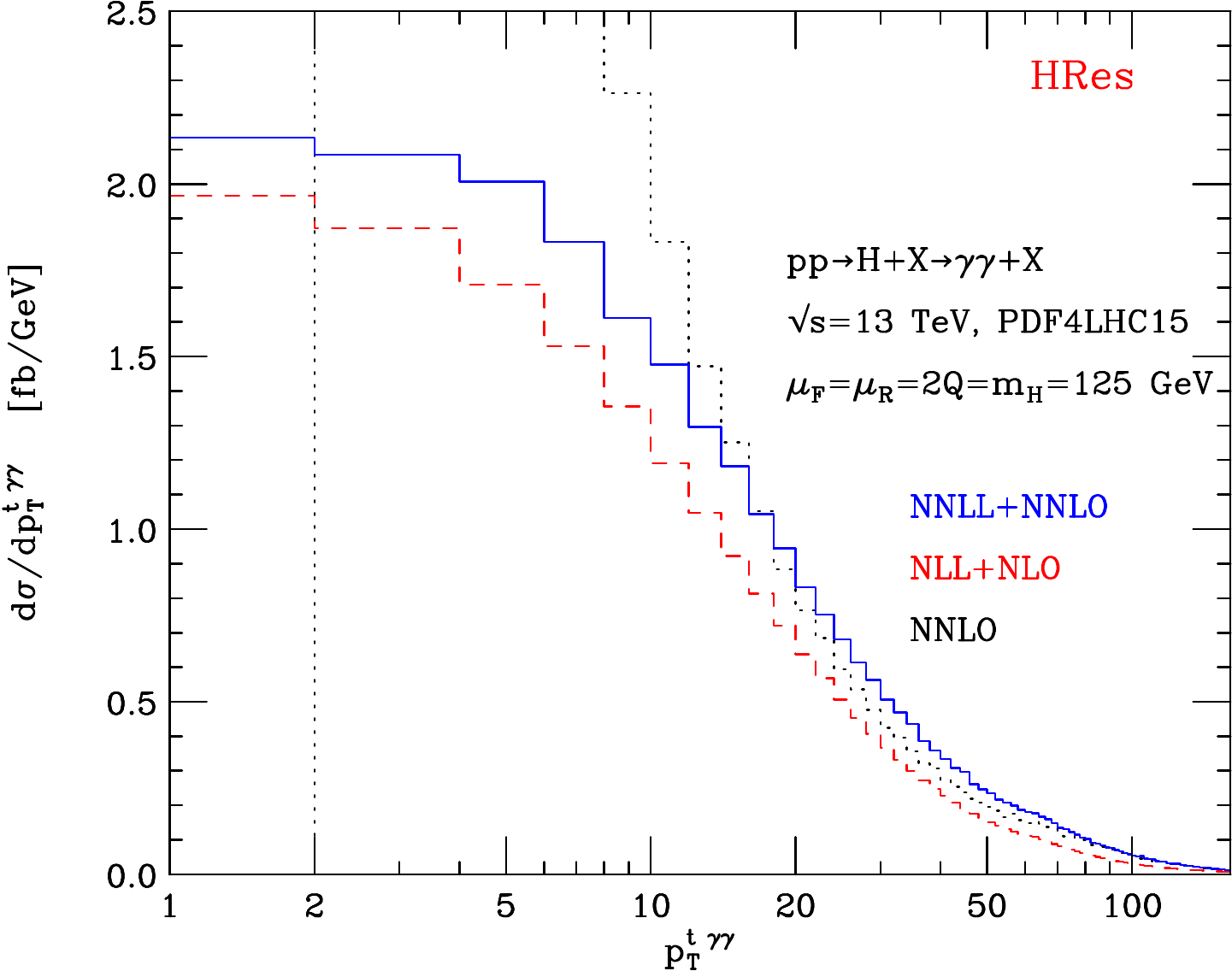}
\caption{\label{fig:pTgg}{ Higgs boson production and diphoton decay
at the LHC.  Transverse-momentum distribution $p_T^{\PGg\PGg}$ (left)
and $p_T^{t,\PGg\PGg}$ distribution (right) obtained by resummed
NLL+NLO and NNLL+NNLO and fixed-order NNLO calculations.  Selection
cuts on the final-state photons are described in the text.  }}
\end{figure}

\begin{figure}
\centering
\includegraphics[width=0.485\textwidth]{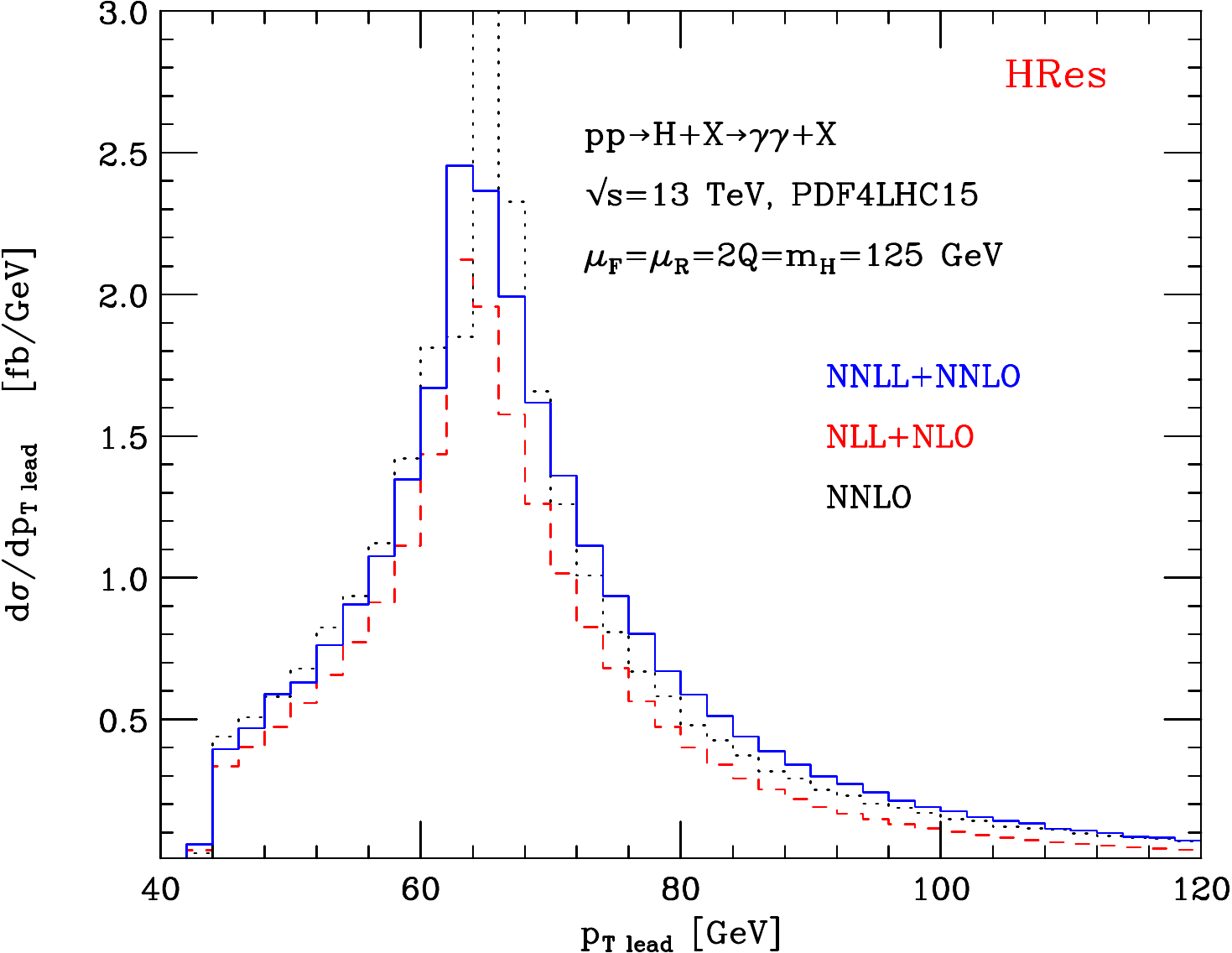}
\includegraphics[width=0.485\textwidth]{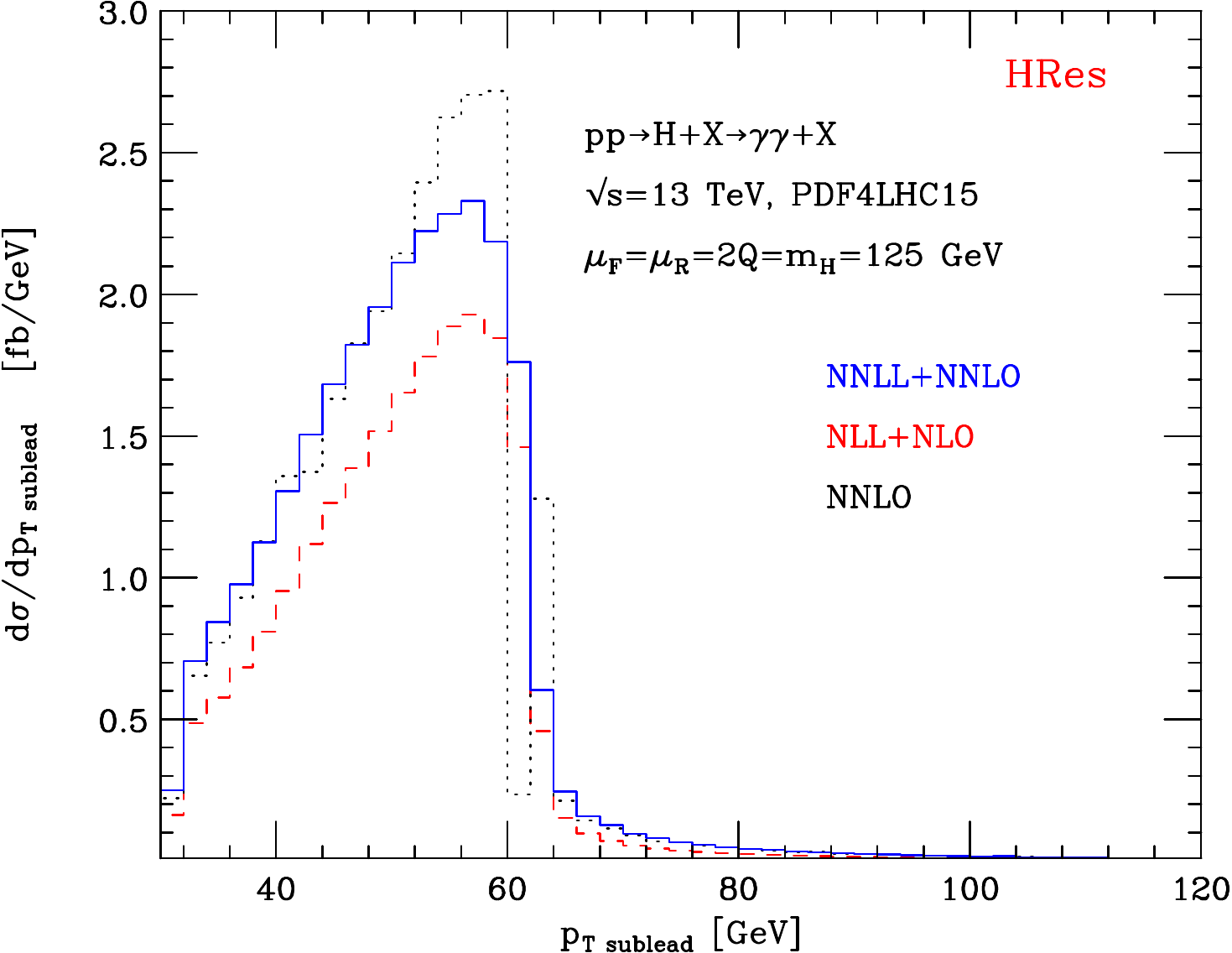}
\caption{\label{fig:pTglead}{ Higgs boson production and diphoton
    decay at the LHC.  Transverse-momentum distributions for the
    leading $p_{ T,{\rm lead}}$ (left) and subleading
    $p_{ T,{\rm sublead}}$ (right) photon obtained by resummed NLL+NLO
    and NNLL+NNLO and fixed-order NNLO calculations.  Selection cuts
    on the final-state photons are described in the text.  }}
\end{figure}

Next we consider the decay channels
$\PH \rightarrow Z Z \rightarrow 4\Pl$ and
$\PH \rightarrow W^+W^- \rightarrow 2\Pl 2\PGn$ with the cuts on the
final state leptons as in
tabs.~\ref{tab:fiducialZZ},~\ref{tab:fiducialWW}.
The corresponding fiducial cross sections at NLL+NLO, NNLL+NNLO and
NNLO are reported in \refT{tab}.  In the left plot of
\refF{fig:pTVV} we show the transverse-momentum spectrum of the
$e^+e^-\mu^+\mu^-$ system relative to the
$\PH \rightarrow Z Z \rightarrow 4\Pl$ channel.
Regarding the $\PH \rightarrow W^+W^- \rightarrow 2\Pl 2\PGn$ final
state, the right plot of \refF{fig:pTVV} shows the transverse
momentum of the $2\Pl2\PGn$ (right plot) system, while the leading-
and subleading-lepton transverse momentum distributions are displayed
in \refF{fig:pTllead}.

\begin{figure}
\centering
\includegraphics[width=0.485\textwidth]{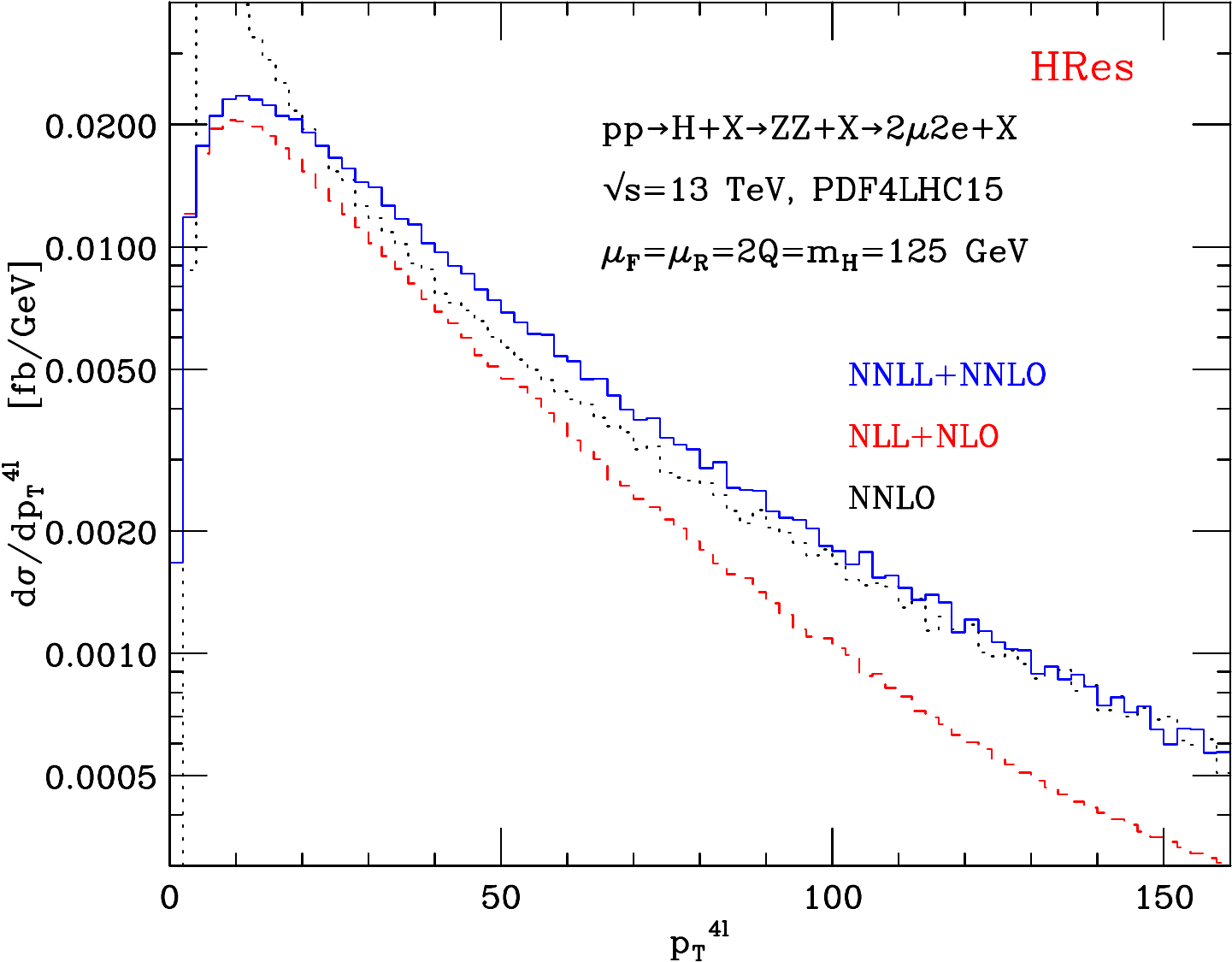}
\includegraphics[width=0.485\textwidth]{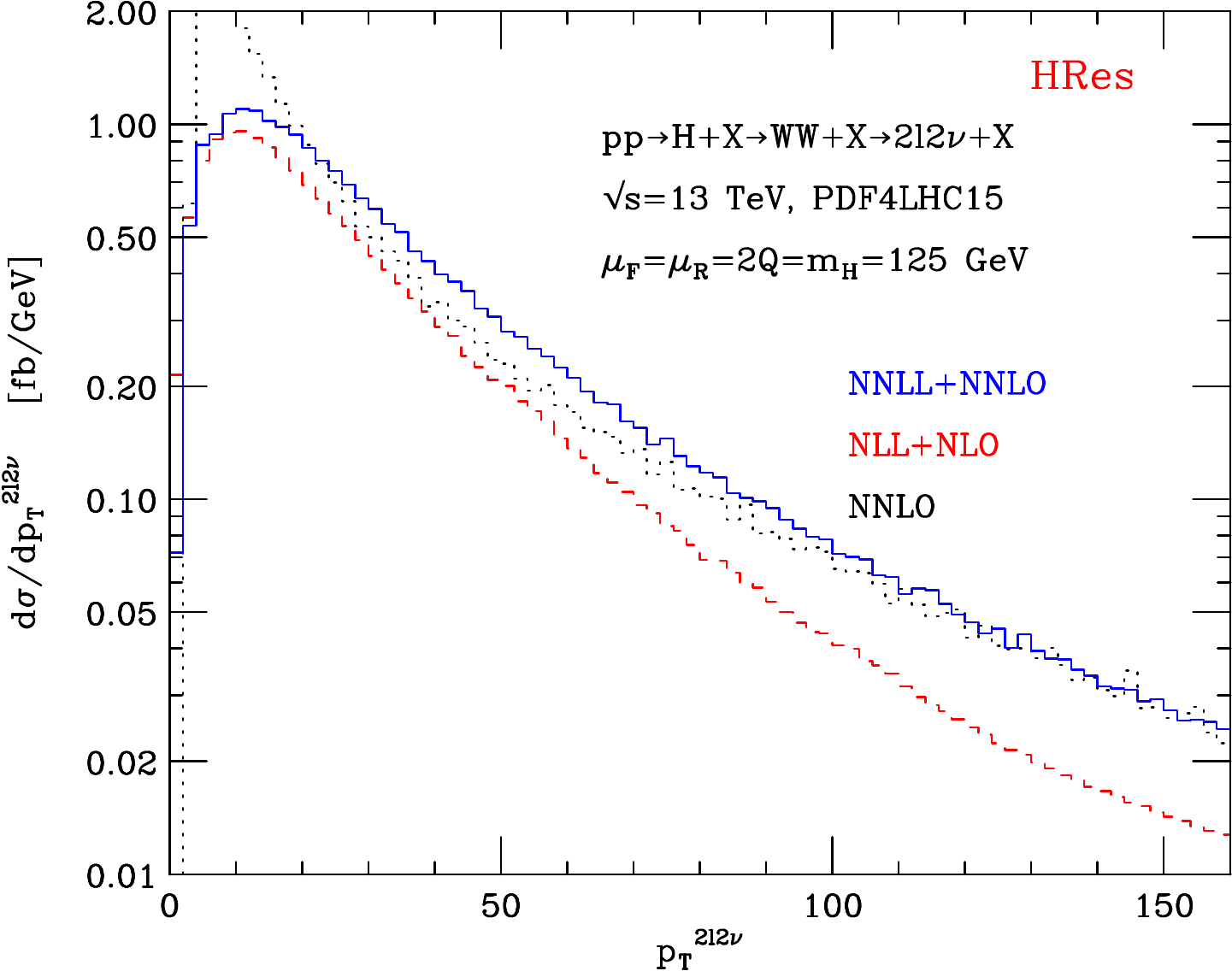}
\caption{\label{fig:pTVV}{Transverse-momentum distributions
    $p_T^{4\Pl}$  for the
$\PH\to \PZ\PZ \to e^+e^-\mu^+\mu^-$ signal (left) and $p_T^{2\Pl2\PGn}$ for
the  $\PH\to W^+W^- \to 2\Pl2\PGn$ signal (right) at the LHC.
Prediction for NNLL+NNLO, NLL+NLO and NNLO  calculations.
Selection cuts on the final-state leptons are described in the text.}}
\end{figure}

\begin{figure}
\centering
\includegraphics[width=0.485\textwidth]{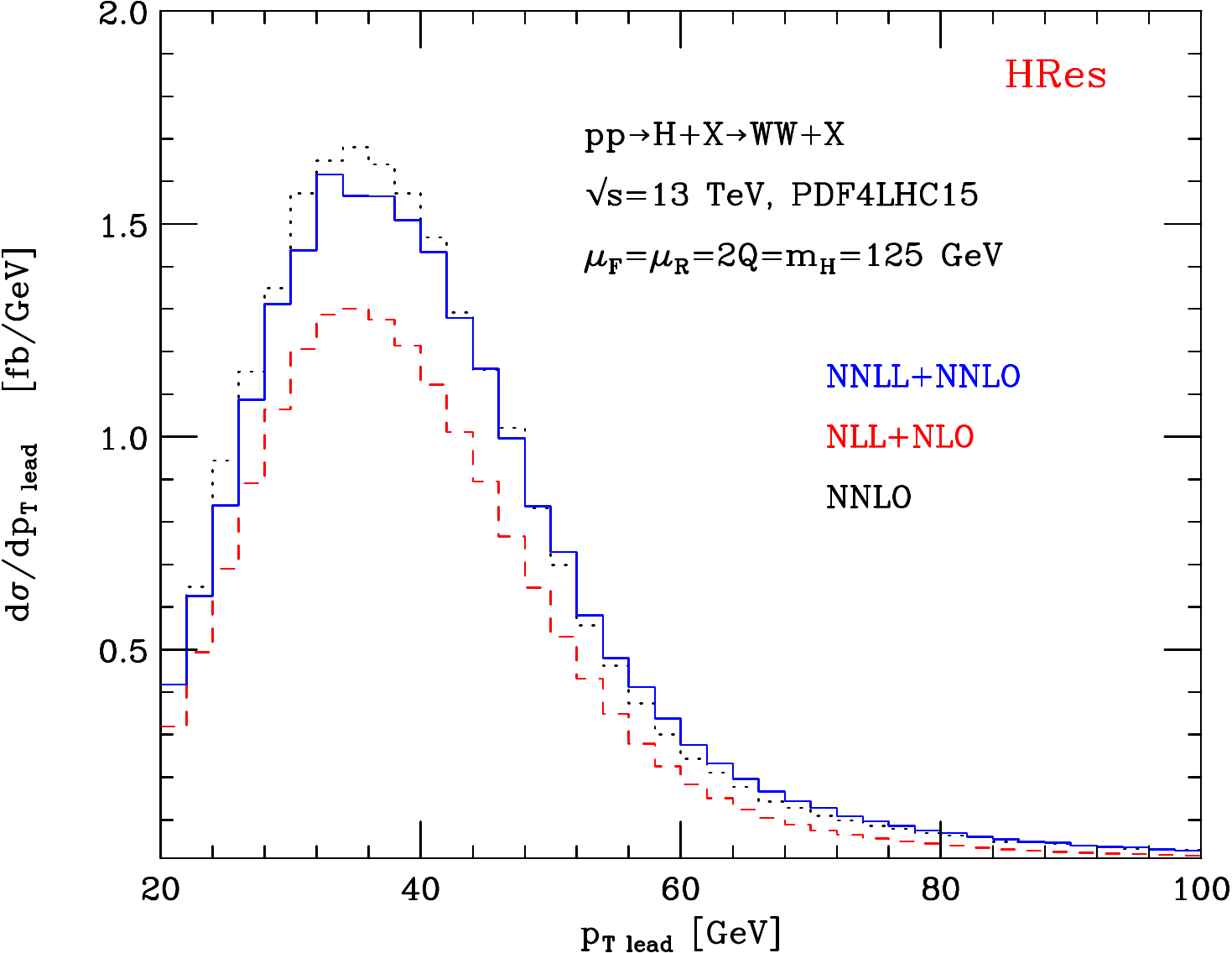}
\includegraphics[width=0.485\textwidth]{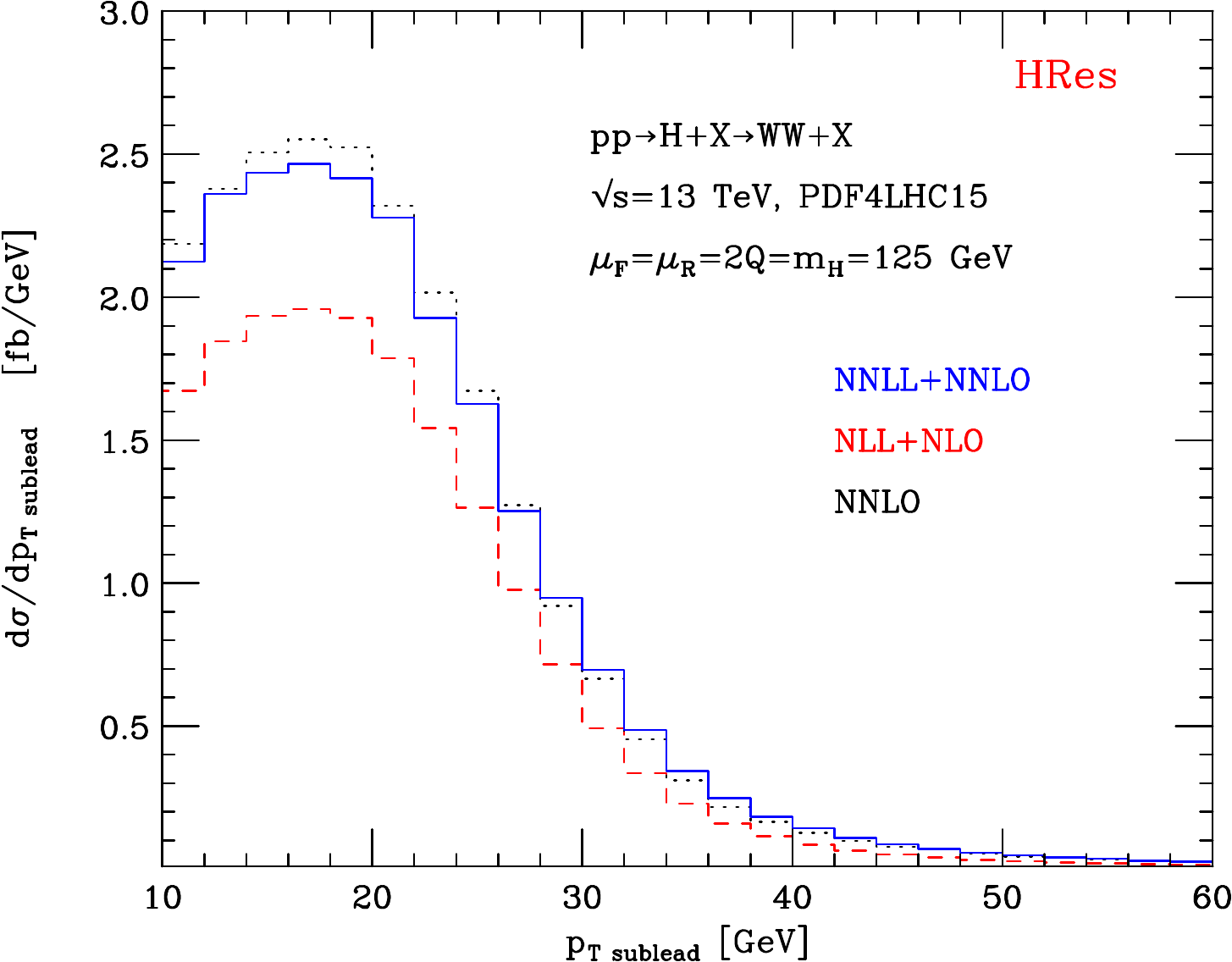}
\caption{\label{fig:pTllead}{Transverse-momentum distributions for the
    leading $p_{ T,{\rm lead}}$ (left) and subleading
    $p_{ T,{\rm sublead}}$ (right) charged lepton obtained by resummed
    NLL+NLO and NNLL+NNLO and fixed-order NNLO calculations in the
    $\PH\to W^+W^- \to 2\Pl2\PGn$ channel.  Selection cuts on the
    final-state photons are described in the text.  }}
\end{figure}

From the above results we observe that resummation effects on top of
the exclusive NNLO prediction tend to be very sizeable in specific
kinematic configurations. It is therefore important to validate the
Monte-Carlo event generators used in experimental analyses for a
complete set of kinematic distributions in the presence of realistic
fiducial cuts. This validation should be carried out over the whole
spectrum - both in the multijet region and in the regime where the
radiation is mainly unresolved - by comparing with the
state-of-the-art perturbative predictions.

\subsection{Fiducial cross sections for Higgs boson production in association with \texorpdfstring{$n_{\rm jet} \geq 1$}{n_jet >= 1} jets}
\label{sec:hjet}

In this section we present results for fiducial cross section and
differential observables obtained from the NNLO QCD computation of
Higgs boson production in association with one hard jet at the LHC
\cite{Caola:2015wna}.  First, we briefly describe the setup of the
computation. We work in a Higgs Effective Theory where the massive top
quark is integrated out. This approximation is parametric in
$\frac{q^2}{m_t}$ where $q$ is a typical scale of the process, and it
is known to work well up to $q \sim m_t$ \cite{Harlander:2012hf}.  In
principle, the exact top quark mass dependence is known at LO
\cite{Ellis:1987xu,Baur:1989cm} and could be included in our
predictions.  However, for simplicity we refrain from including them
in the results presented here. In our computation we include all
partonic channels at NLO. For the NNLO corrections we include the $gg$
and $qg$ channels which are the only ones relevant for phenomenology
within the $p_T$ range considered here.  The effect of the missing
channels is expected to be much smaller than the residual scale
uncertainty and probably comparable to finite top quark mass
corrections.  We treat the Higgs boson on-shell and include all
relevant decays to $\gamma\gamma$ and four lepton final states.  We
neglect interference effects in the case of identical fermions and
$W-Z$ interference for $2\ell \, 2\nu$ final states. All these effects
are expected to be small in the fiducial region considered in this
section, with the exception of off-shell production in the $WW$
channel.  Indeed, the fiducial region studied in this section does not
include any cut on the $WW$ transverse mass, $m_T$. As it is well
known~\cite{Kauer:2012hd}, this leads to a sizeable off-shell and
signal/background effects on the Higgs cross-section\footnote{Note
  that in the the $ZZ$ fiducial region off-shell effects are
  negligible thanks to the $m_{4l}$.}. One should carefully take this
into account when performing any study in the fiducial region defined
in this section. As we said, our numbers are for on-shell Higgs and do
not include $pp\to 2l 2\nu$ signal/background interference, so by
construction they don't account for such (potentially large)
effects. As a consequence, they lead to reliable predictions only in a
region where off-shell effects are negligible.
Finally, we use $\mu=\mu_\mathrm{fact}=\mu_\mathrm{ren}=m_H$ as the
central scale and vary by a factor of two upwards and downwards.  We
believe that this choice gives a more conservative estimate of
residual theoretical uncertainty while not changing much (at NNLO) the
central value compared to the more traditional choice $\mu=m_H/2$.
Also, we note that the stability w.r.t. scale variation of the NNLO
results presented in this section makes the use of dynamical scales
unnecessary at this order.

Our results show that NNLO corrections lead to a stabilization of
fiducial cross sections and shapes of differential distributions.  The
unphysical scale dependence of the NNLO result is reduced by a factor
of more than two with respect to NLO\footnote{Our results do not
  include PDFs uncertainties which are estimated to be at the level of
  5\%~\cite{Boughezal:2015dra}.}.  In general, we find that applying
fiducial cuts does not spoil the convergence of the perturbative
expansion.  Acceptances are found to be stable when moving from NLO to
NNLO, in agreement to our results in Ref.~\cite{Caola:2015wna}. To
illustrate this, and to explore the validity range of pure fixed order
computations, in \refF{fig:hjetcml}, we present the cumulative
leading jet $p_\perp$ distribution as a function of the lower cut on
$p_\perp$, for the $\gamma\gamma$ channel in the fiducial region.  In
the lower pane we plot the NLO and NNLO $K$-factors and observe very
stable corrections to $p_\mathrm{cut}$ as low as 30~GeV\footnote{In
  this section, both NLO and the NNLO results are computed with
  NNLO PDFs and $\alpha_s$.}.  There is no indication of a breakdown of
perturbation theory for these values of transverse momentum.  Hence,
it appears, that also in the fiducial region the NNLO result already
captures the dominant logarithmic enhancements and additional
resummation effects are small~\cite{Banfi:2015pju} for $p_\perp \ge
30$~GeV.  This is in contrast to NLO predictions where missing higher
logarithmic terms still lead to sizeable corrections. A similar
behaviour is observed also in the $WW$ and $ZZ$ channels.

\begin{figure*}[ht]
\centering
\includegraphics[width=0.45\textwidth]{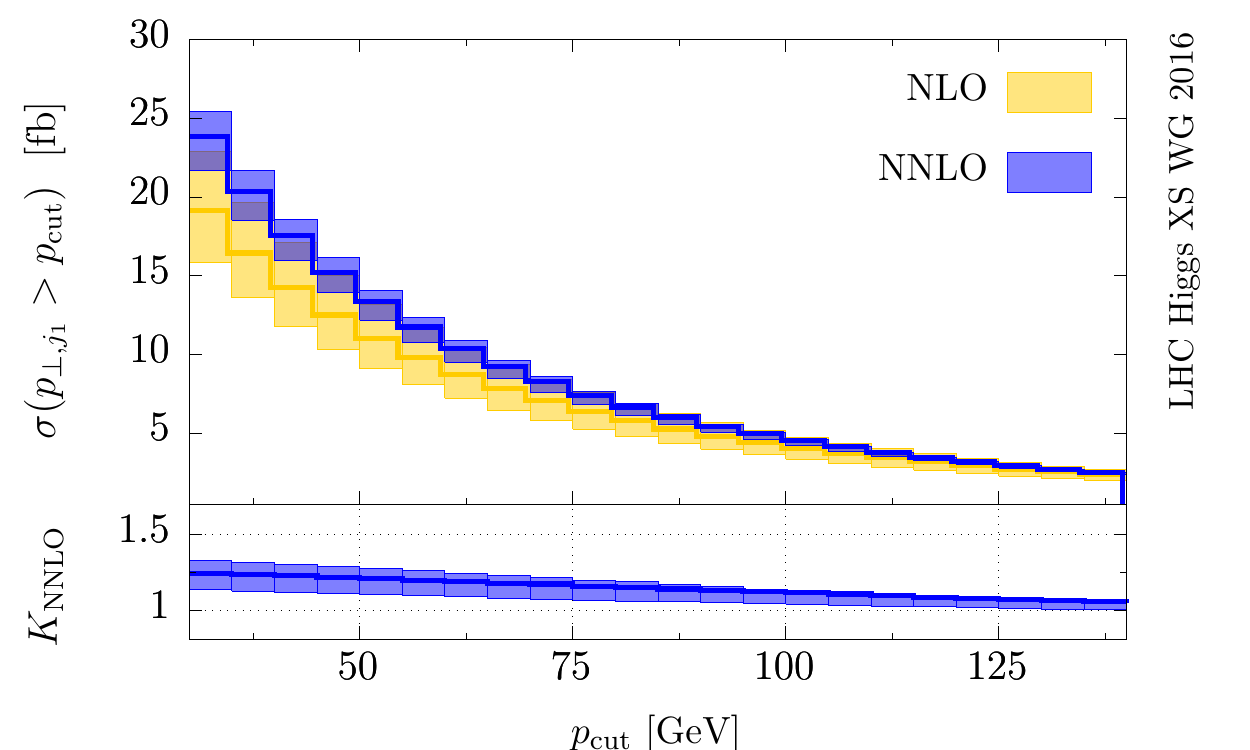}
\caption{ \label{fig:hjetcml}
Cumulative jet $p_\perp$
Distribution in the fiducial volume at NLO (yellow) and NNLO (blue) for $H\to \gamma\gamma$. Both NLO and NNLO
curves obtained with NNLO PDFs and $\alpha_s$. Solid line: value for $\mu_r=\mu_f=m_H$. Filled band: scale uncertainty.
In the lower pane the ratio of NNLO to the NLO central $\mu_r=\mu_f$ value is shown. See text for details.
}
\end{figure*}

Next, we show in \refT{tab:hjetfid} the NLO and NNLO cross-sections in the fiducial region for
the $\gamma\gamma$, $WW$ and $ZZ$ channel. Note that since all leptons are massless in our 
computation, there is no distinction between $2e2\nu$ and $2\mu 2\nu$ predictions for the fiducial
region considered here. Our result show that the $K-$factor is similar for all the three channels, and
similar to the inclusive $K-$factor.

\begin{table*}[ht]
\caption{The NLO and NNLO cross-sections in the fiducial region for the $\gamma\gamma$, $WW$ and $ZZ$ channel.}
\label{tab:hjetfid}
\centering
\renewcommand{\arraystretch}{1.5}
\begin{tabular}{c|c|c|ccc}
\toprule
     & $\gamma \gamma$ & $WW\to e\mu \nu\nu$ & \multicolumn{3}{c}{$ZZ\to 4l$} \\
     &                 &                     & $4\mu$ & $2e2\mu$ & $4e$ \\
\midrule
NLO (fb)  & $    19.1^{+     3.8}_{-     3.2}$ & $14.9^{+     3.0}_{-     2.5}$     & $0.336^{+   0.065}_{-   0.057}$ & $0.611^{+   0.120}_{-   0.103}$ & $   0.282^{+   0.055}_{-   0.048}$         \\
NNLO (fb) & $    22.7^{+     1.4}_{-     1.9}$ & $17.9^{+     1.0}_{-     1.6}$     & $   0.388^{+   0.010}_{-   0.030}$ & $   0.707^{+   0.020}_{-   0.055}$       & $   0.327^{+   0.008}_{-   0.026}$         \\
\bottomrule
\end{tabular}
\end{table*}

Finally, we present result for selected differential distributions in
the fiducial region, for the $\gamma\gamma$
(Figs.~\ref{fig:hjetaa01},\ref{fig:hjetaa02},\ref{fig:hjetaa03},\ref{fig:hjetaa04}),
$WW$ (Figs.~\ref{fig:hjetww01},\ref{fig:hjetww02},\ref{fig:hjetww03}) and $ZZ$
(Figs.~\ref{fig:hjetzz01},\ref{fig:hjetzz02}).
Note that for the $ZZ$ channel we only show results for the $4\mu$ sub-channel and
for lepton observables.
Results for other sub-channels are very similar to these ones, and results for
jet observables are very similar to the $WW$ case already shown.

\begin{figure*}[ht]
\centering
\includegraphics[width=0.45\textwidth]{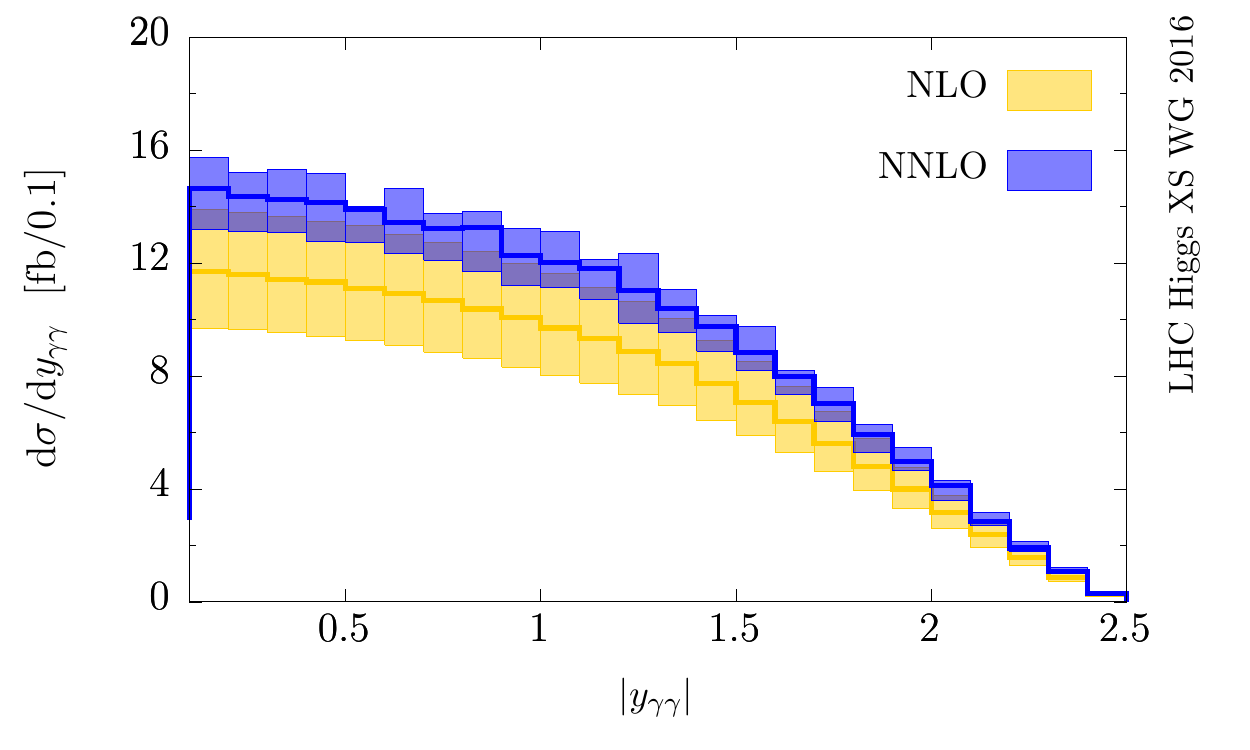}
\hfill
\includegraphics[width=0.45\textwidth]{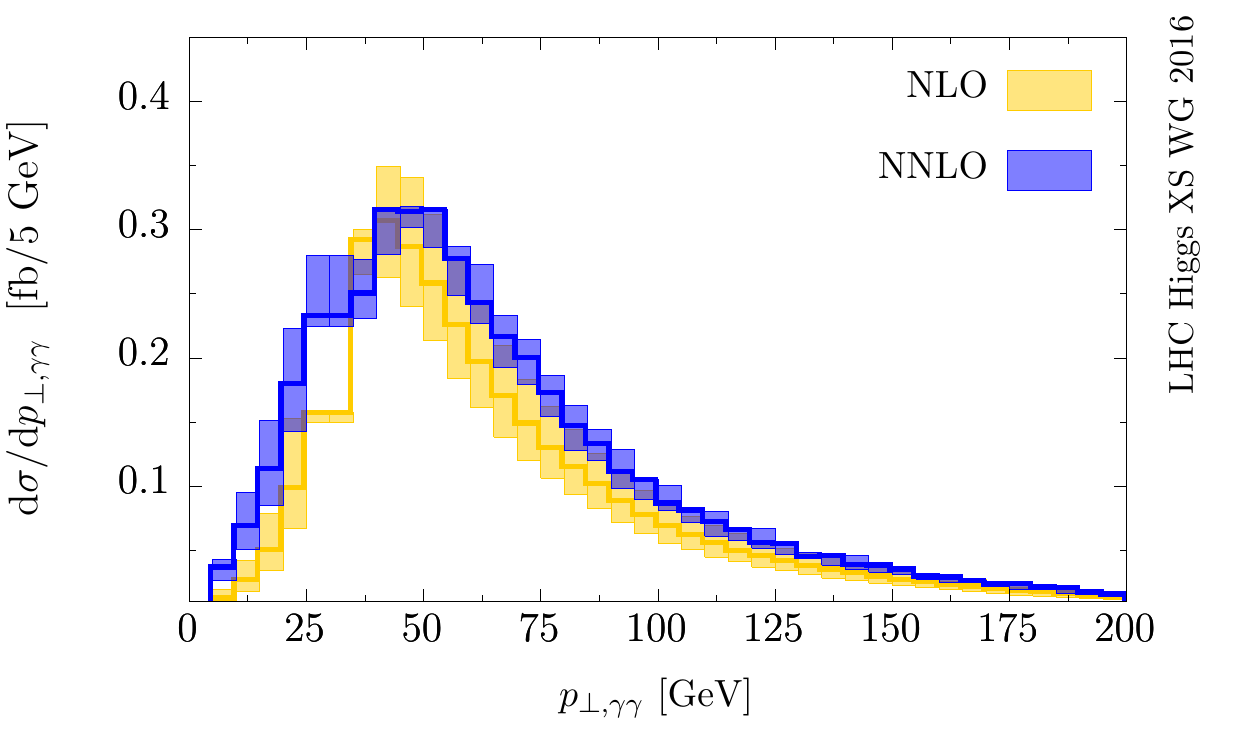}
\caption{ \label{fig:hjetaa01} Distribution in the fiducial volume at
  NLO (yellow) and NNLO (blue) for $H\to \gamma\gamma$. Left:
  di-photon rapidity. Right: di-photon $p_\perp$. Both NLO and NNLO
  curves obtained with NNLO PDFs and $\alpha_s$. Solid line: value for
  $\mu_r=\mu_f=m_H$. Filled band: scale uncertainty. See text for
  details.   }
\end{figure*}

\begin{figure*}[ht]
\centering
\includegraphics[width=0.45\textwidth]{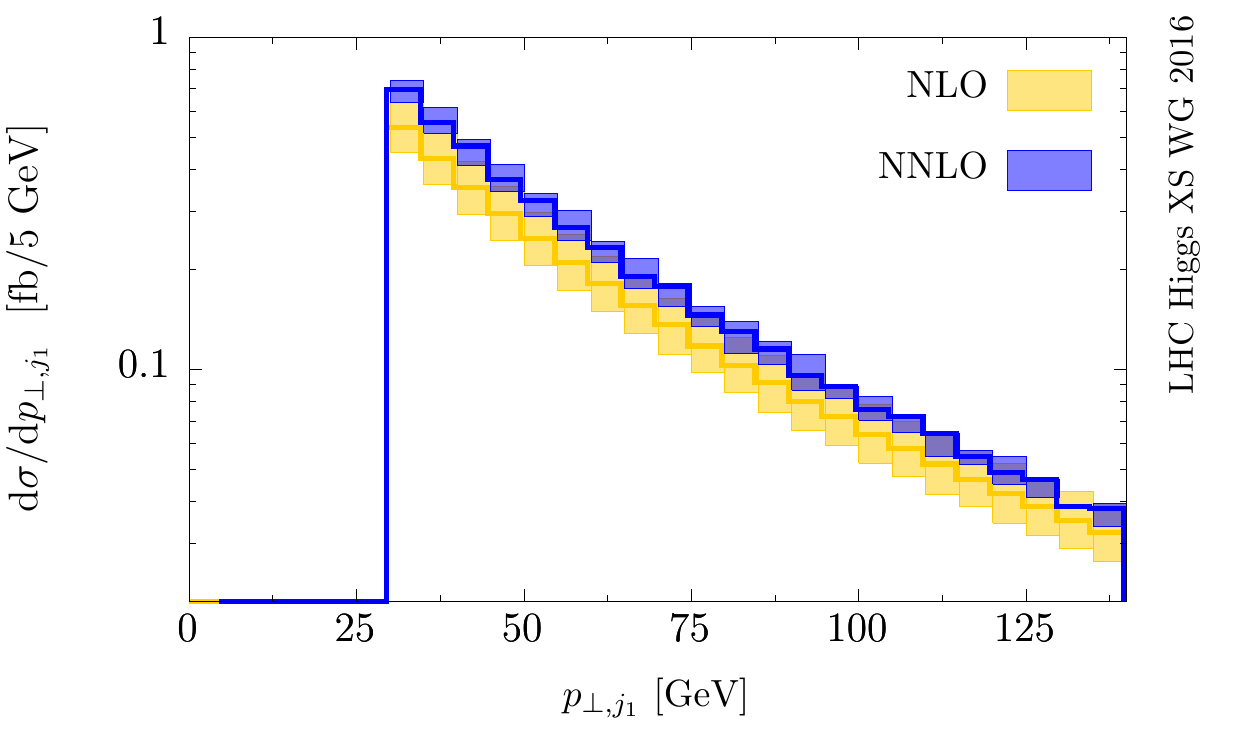}
\hfill
\includegraphics[width=0.45\textwidth]{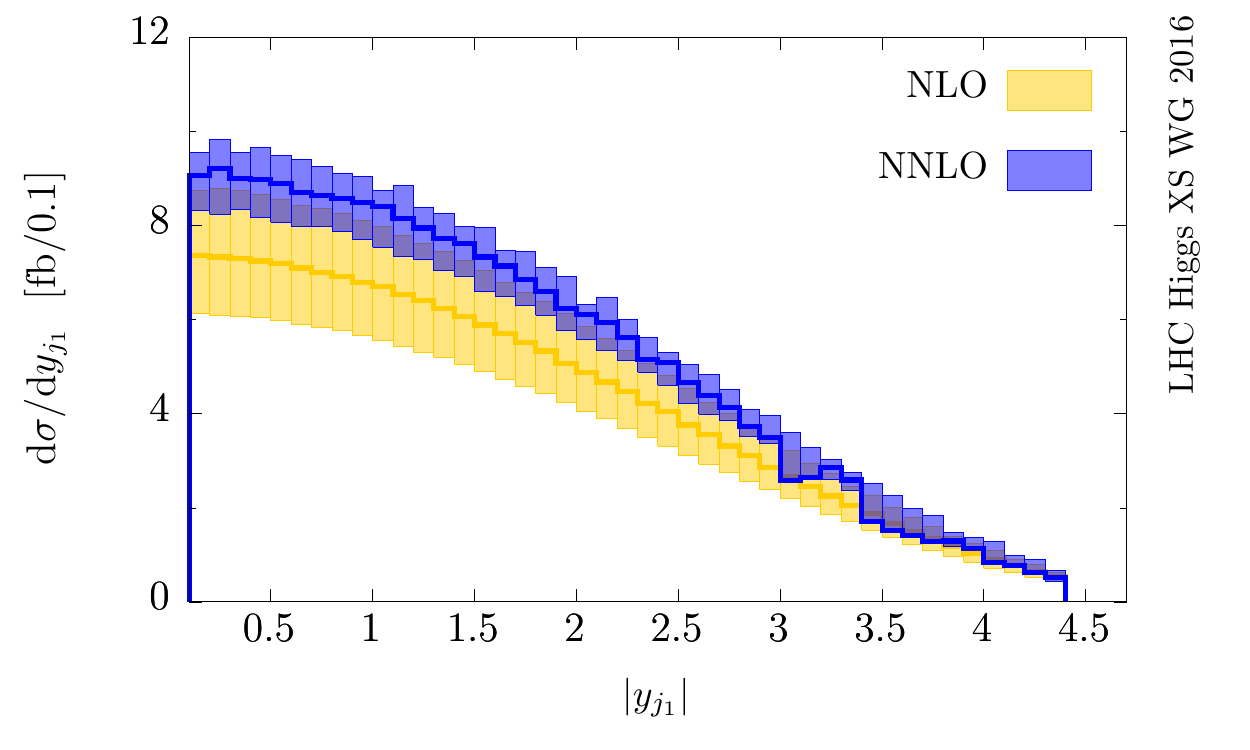}
\caption{ \label{fig:hjetaa02} Distribution in the fiducial volume at
  NLO (yellow) and NNLO (blue) for $H\to \gamma\gamma$. Left: leading
  jet $p_\perp$.  Right: leading jet rapidity. Both NLO and NNLO
  curves obtained with NNLO PDFs and $\alpha_s$. Solid line: value for
  $\mu_r=\mu_f=m_H$. Filled band: scale uncertainty. See text for
  details.   }
\end{figure*}

\begin{figure*}[ht]
\centering
\includegraphics[width=0.45\textwidth]{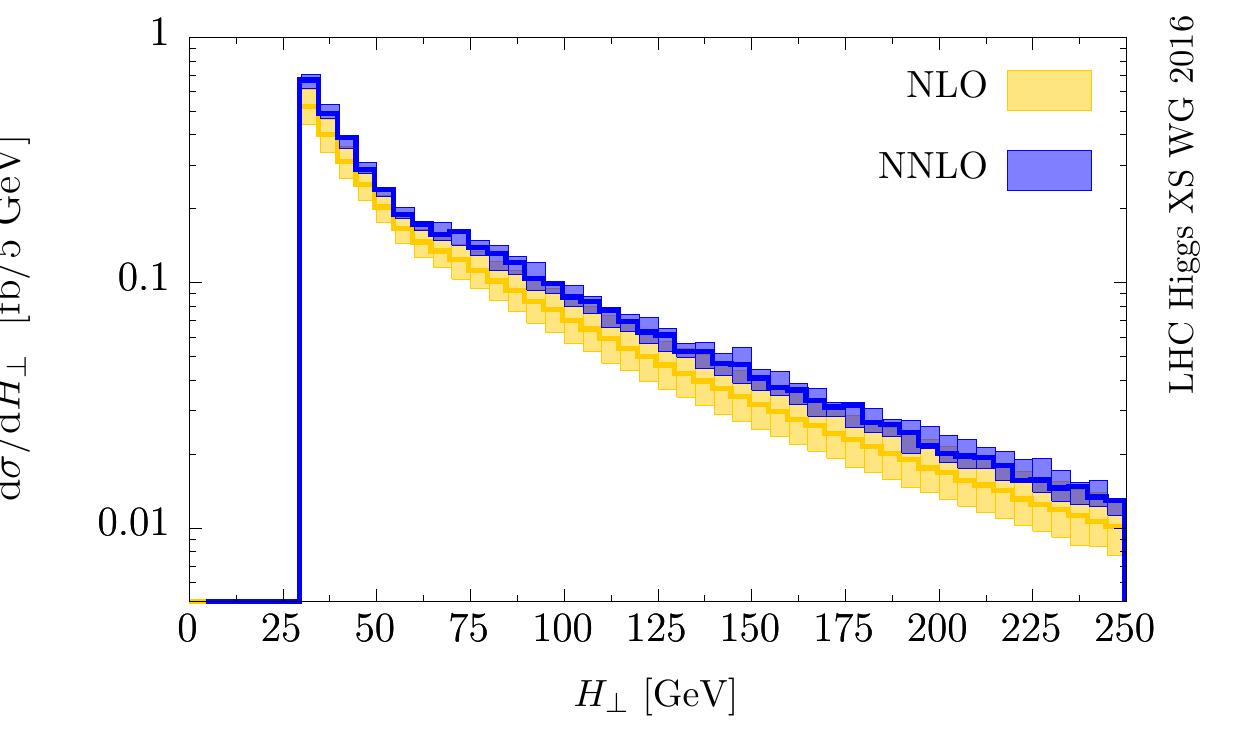}
\hfill
\includegraphics[width=0.45\textwidth]{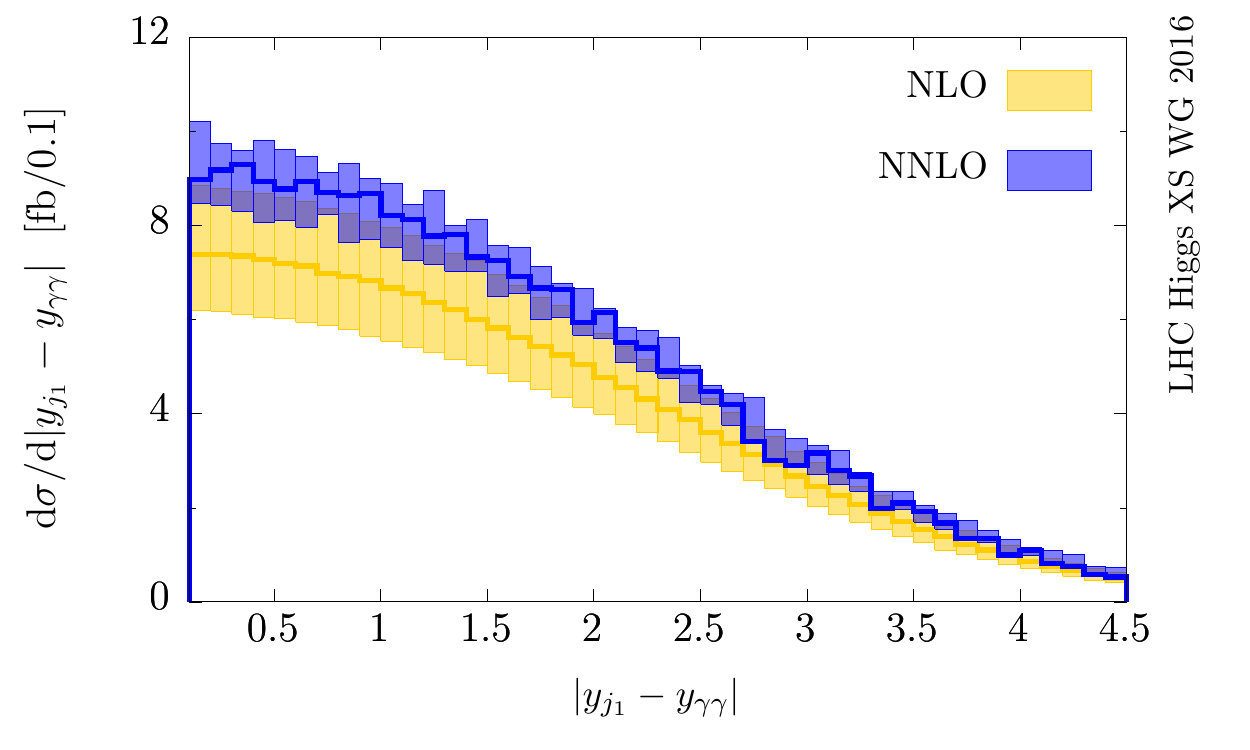}
\caption{ \label{fig:hjetaa03} Distribution in the fiducial volume at
  NLO (yellow) and NNLO (blue) for $H\to \gamma\gamma$. Left:
  $H_\perp$. Right: rapidity difference between the di-photon system
  and the leading jet. Both NLO and NNLO curves obtained with NNLO
  PDFs and $\alpha_s$. Solid line: value for $\mu_r=\mu_f=m_H$. Filled
  band: scale uncertainty. See text for details.}
\end{figure*}

\begin{figure*}[ht]
\centering
\includegraphics[width=0.45\textwidth]{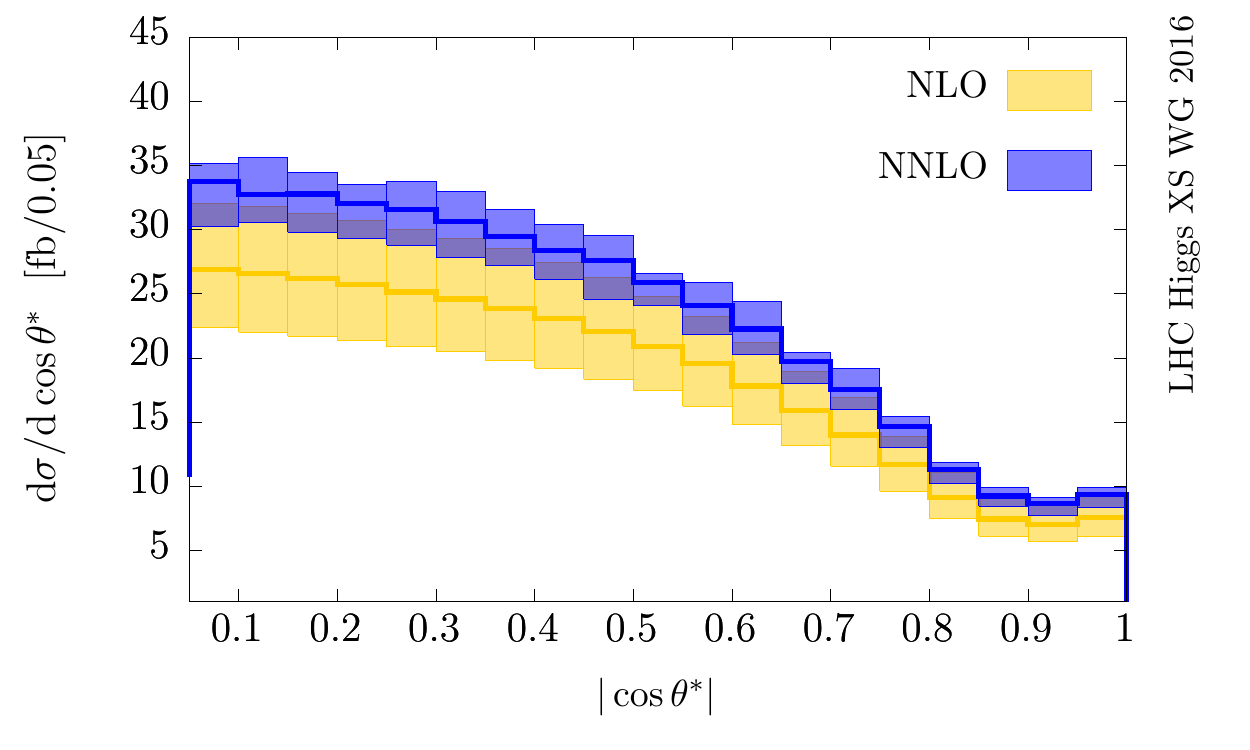}
\caption{ \label{fig:hjetaa04} $\cos\theta^*$ distribution in the
  fiducial volume at NLO (yellow) and NNLO (blue) for $H\to
  \gamma\gamma$. Both NLO and NNLO curves obtained with NNLO PDFs and
  $\alpha_s$. Solid line: value for $\mu_r=\mu_f=m_H$. Filled band:
  scale uncertainty. See text for details.}
\end{figure*}

\begin{figure*}[ht]
\centering
\includegraphics[width=0.45\textwidth]{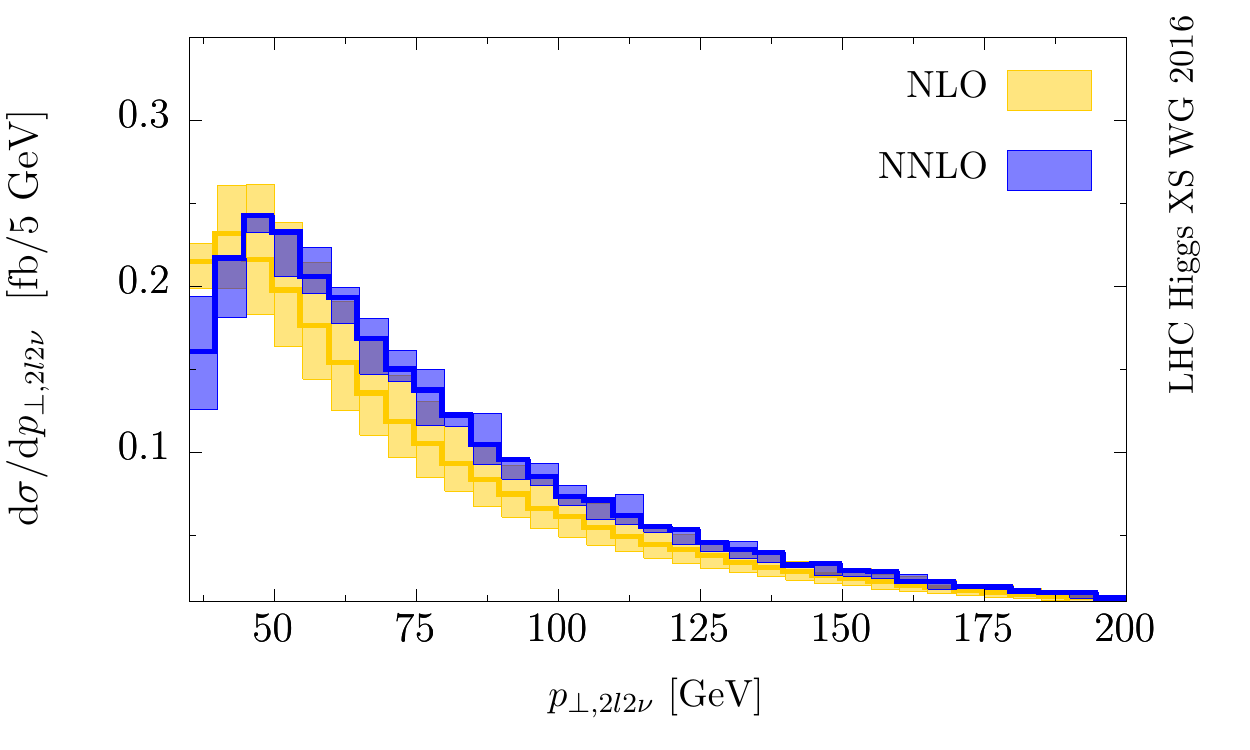}
\hfill
\includegraphics[width=0.45\textwidth]{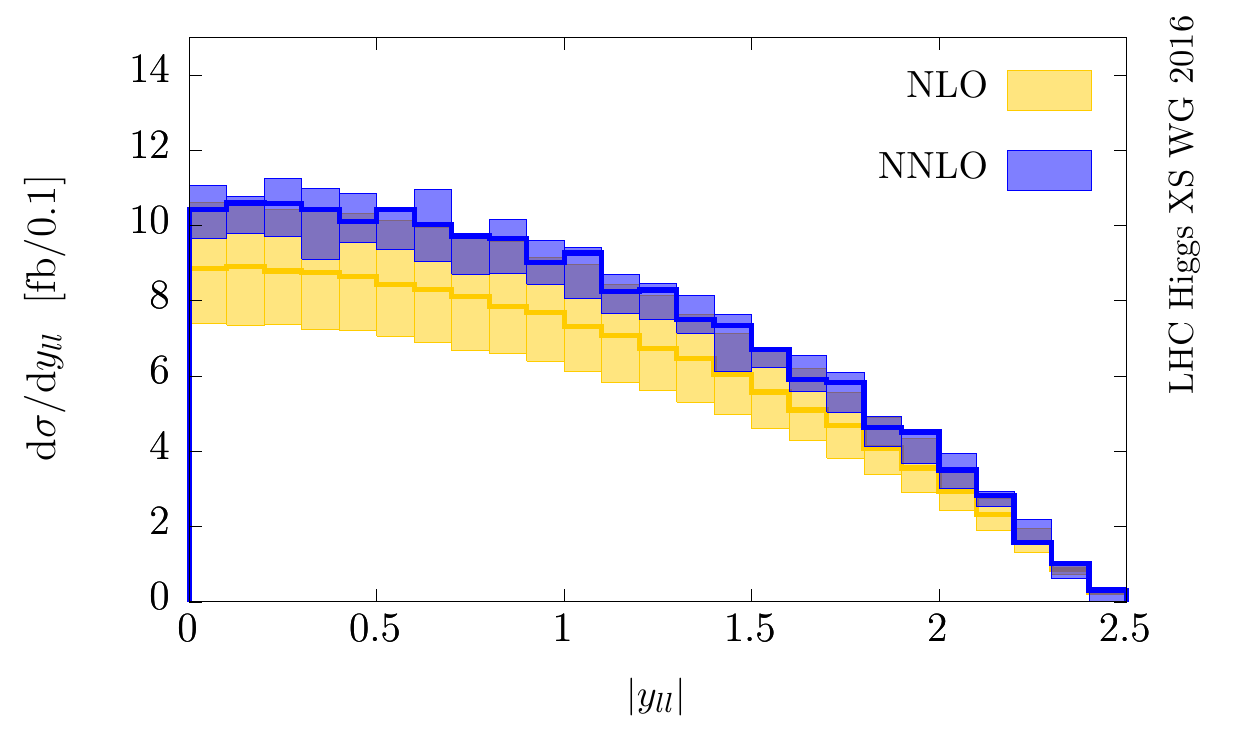}
\caption{ \label{fig:hjetww01} Distribution in the fiducial volume at
  NLO (yellow) and NNLO (blue) for $H\to WW \to e\mu \nu\nu$. Left:
  Higgs boson $p_\perp$.  Right: di-lepton rapidity. Both NLO and NNLO
  curves obtained with NNLO PDFs and $\alpha_s$. Solid line: value for
  $\mu_r=\mu_f=m_H$. Filled band: scale uncertainty. See text for
  details.   }
\end{figure*}

\begin{figure*}[ht]
\centering
\includegraphics[width=0.45\textwidth]{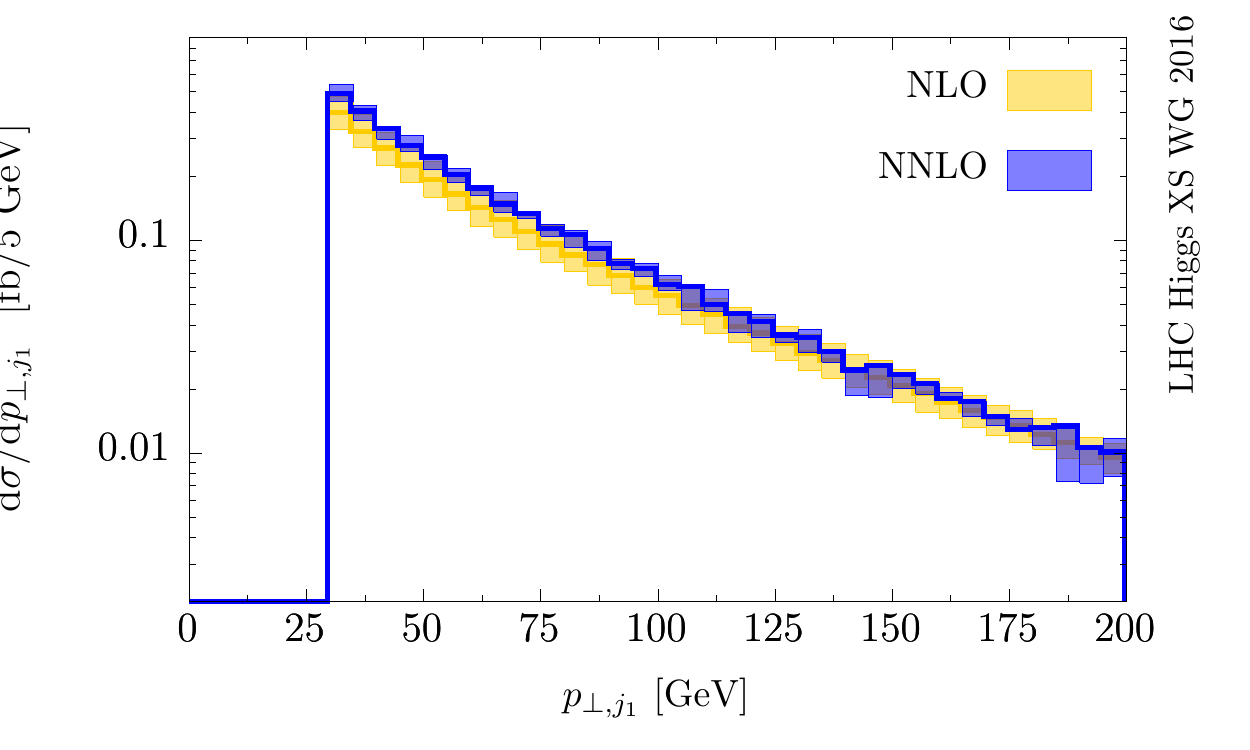}
\hfill
\includegraphics[width=0.45\textwidth]{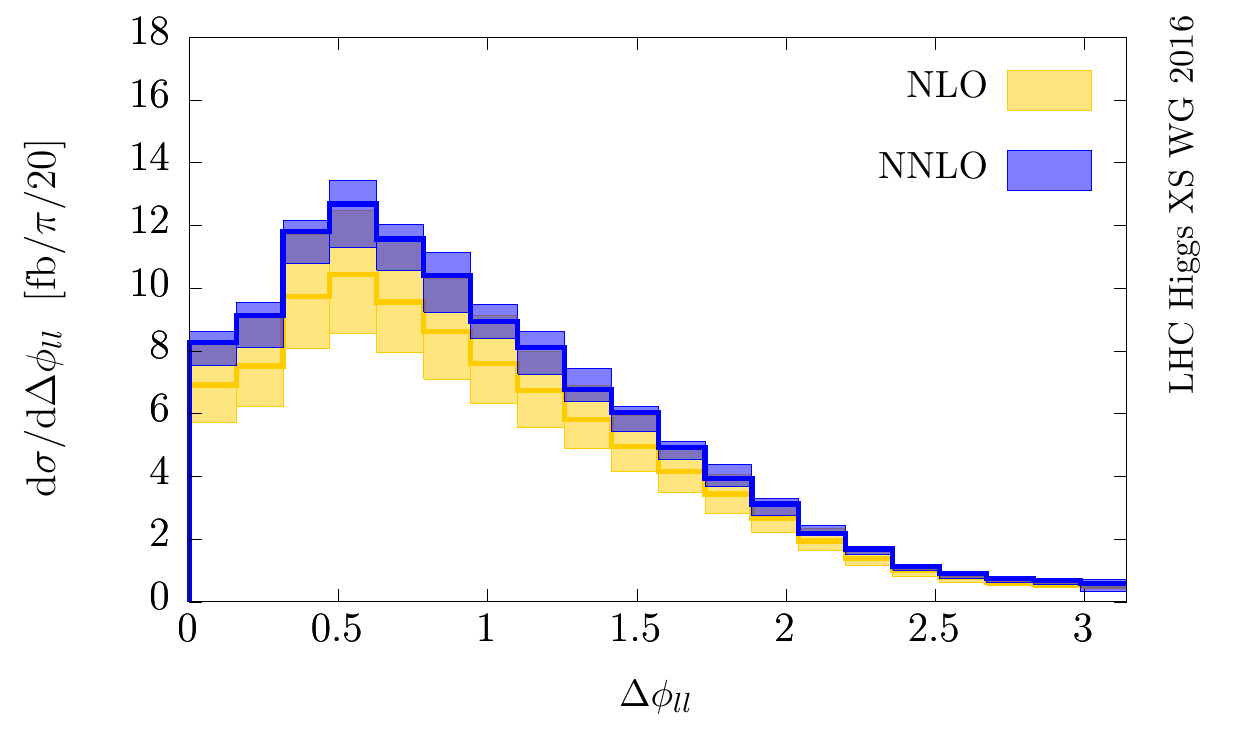}
\caption{ \label{fig:hjetww02}
Distribution in the fiducial volume at NLO (yellow) and NNLO (blue) for $H\to WW \to e\mu \nu\nu$.
Left: leading jet $p_\perp$. Right: di-lepton azimuthal separation. Both NLO and NNLO
curves obtained with NNLO PDFs and $\alpha_s$. Solid line: value for $\mu_r=\mu_f=m_H$. Filled band: scale uncertainty. See text for details.
}
\end{figure*}

\begin{figure*}[ht]
\centering
\includegraphics[width=0.45\textwidth]{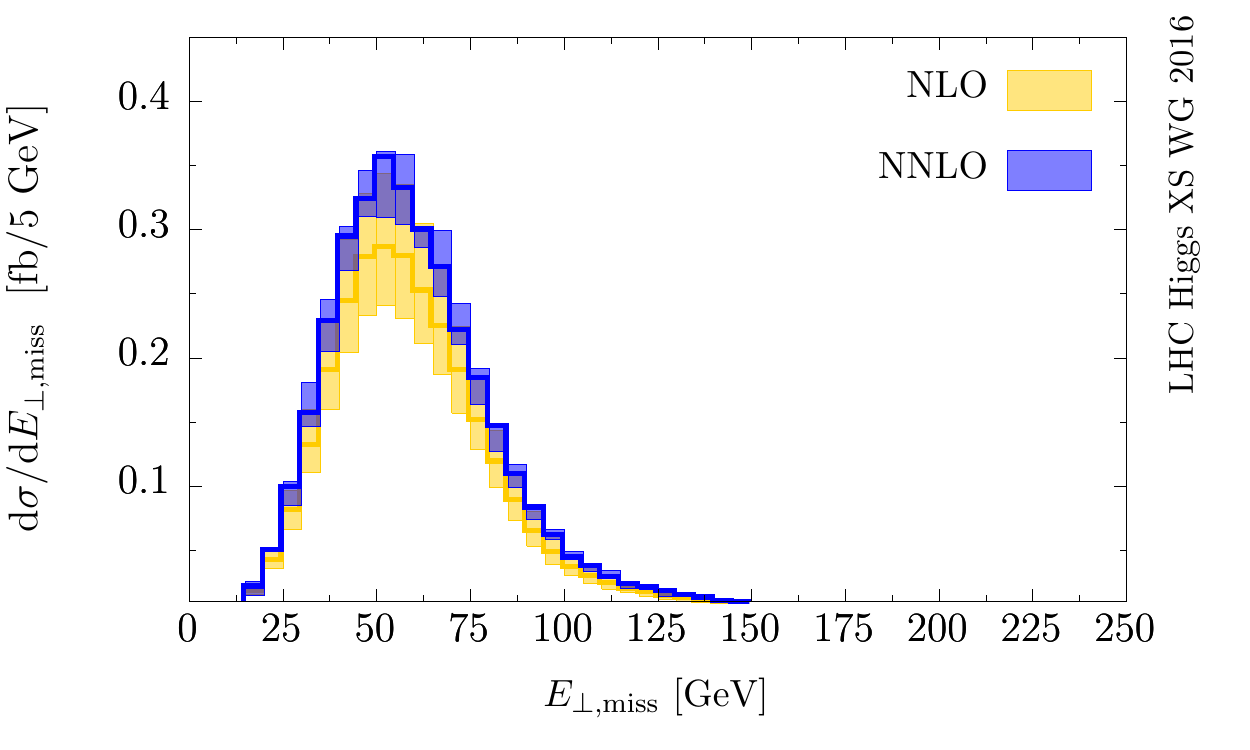}
\hfill
\includegraphics[width=0.45\textwidth]{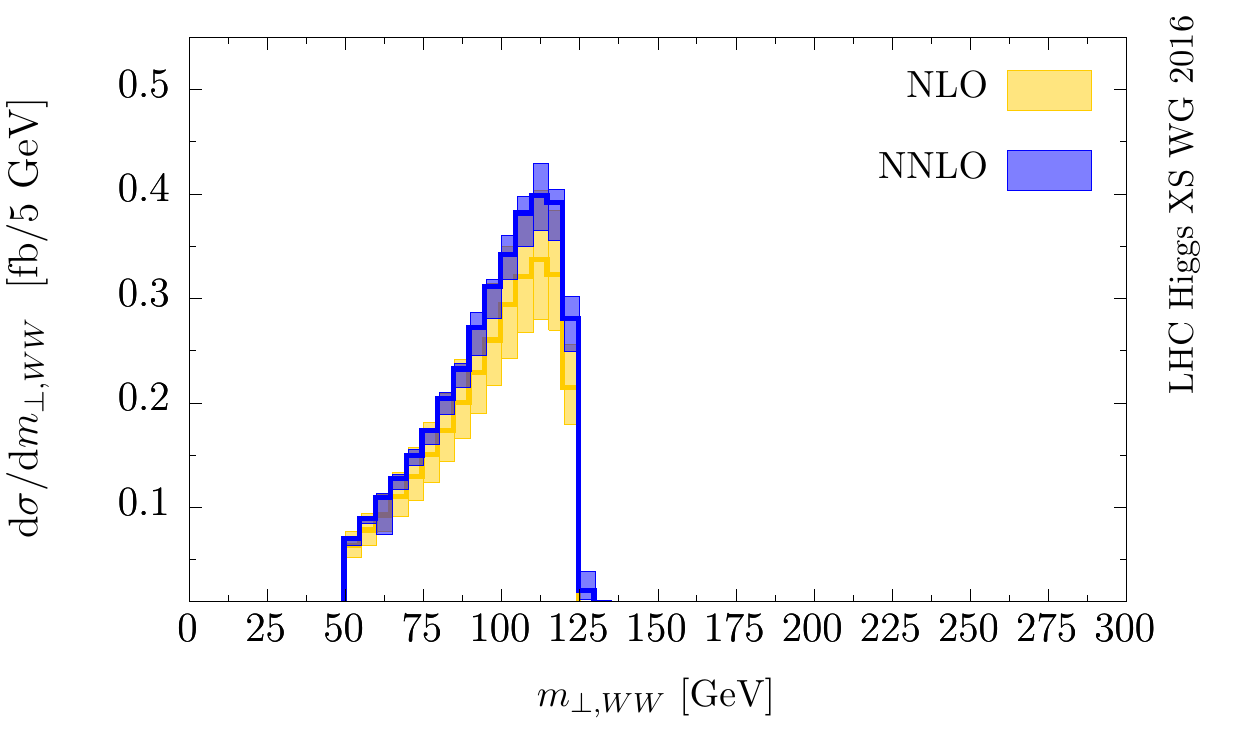}
\caption{ \label{fig:hjetww03} Distribution in the fiducial volume at
  NLO (yellow) and NNLO (blue) for $H\to WW \to e\mu \nu\nu$. Left:
  $E_{\perp,\mathrm{miss}}$. Right: $WW$ transverse mass
  $m_\perp$. Both NLO and NNLO curves obtained with NNLO PDFs and
  $\alpha_s$. Solid line: value for $\mu_r=\mu_f=m_H$. Filled band:
  scale uncertainty. See text for details.}
\end{figure*}

\begin{figure*}[ht]
\centering
\includegraphics[width=0.45\textwidth]{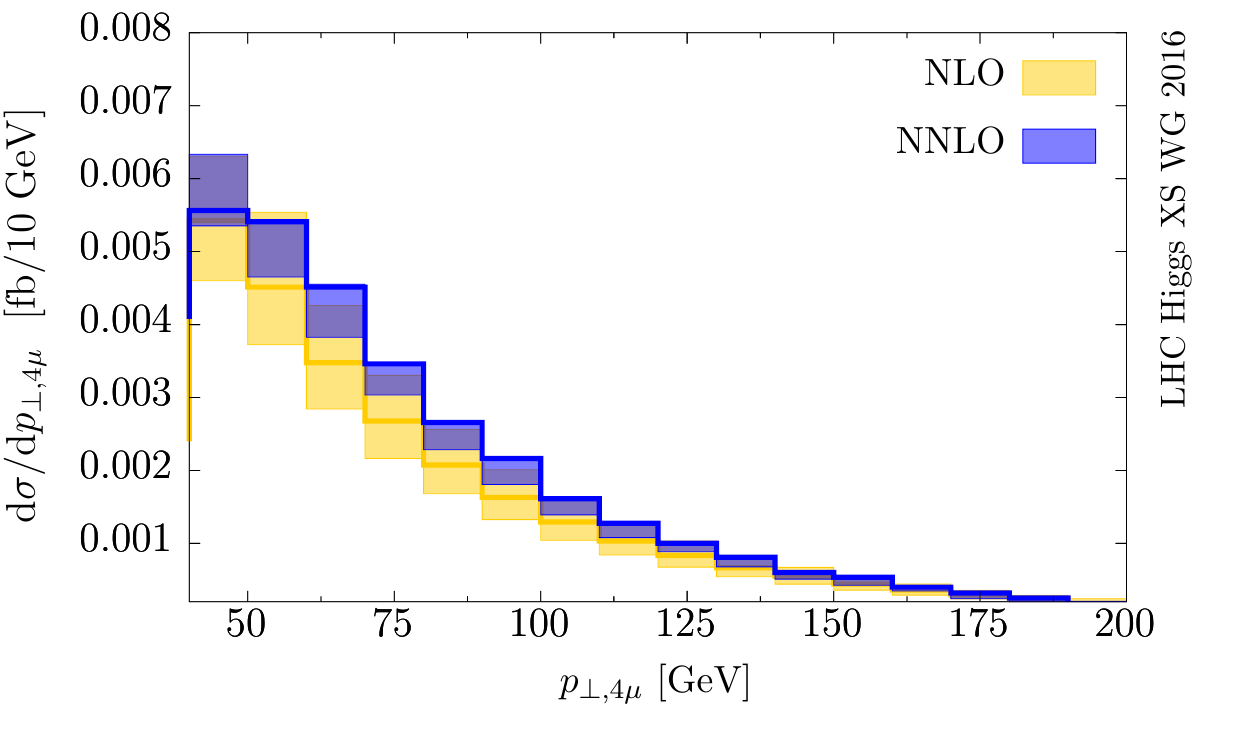}
\hfill
\includegraphics[width=0.45\textwidth]{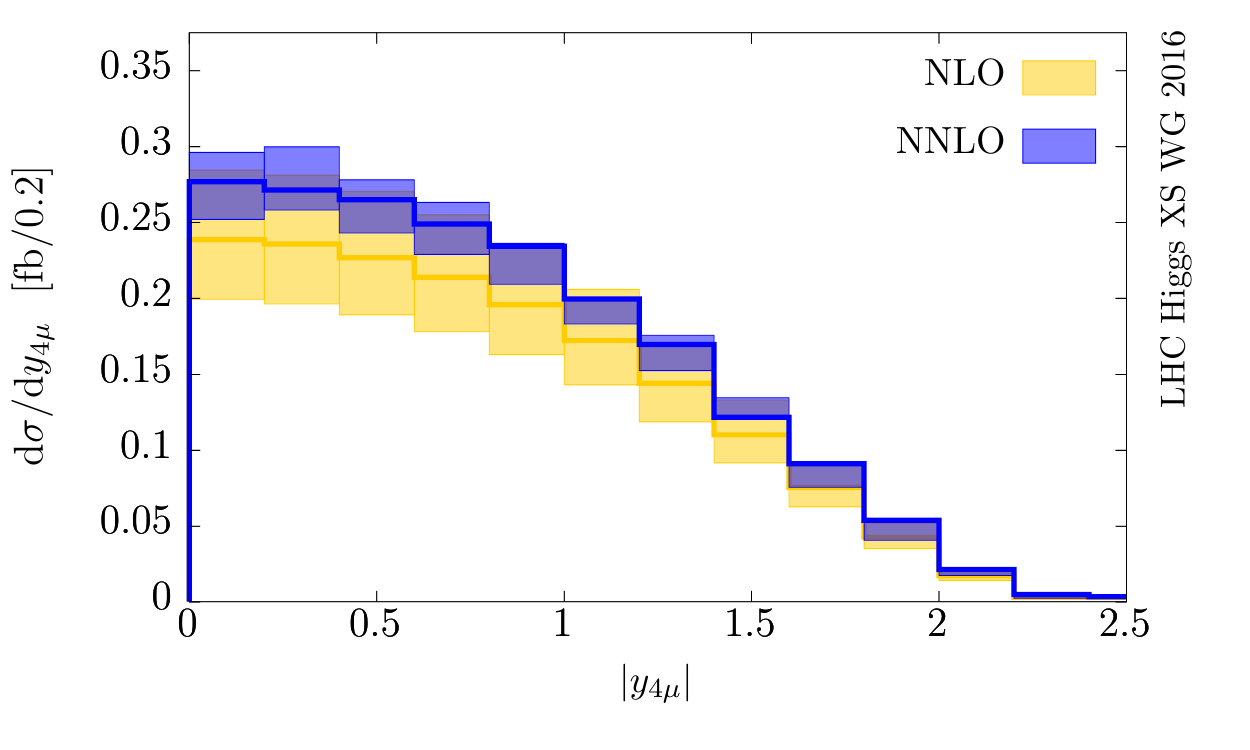}
\caption{ \label{fig:hjetzz01} Distribution in the fiducial volume at
  NLO (yellow) and NNLO (blue) for $H\to ZZ \to 4\mu$. Left: Higgs boson $p_\perp$.  
  Right: Higgs boson rapidity. Both NLO and NNLO curves obtained
  with NNLO PDFs and $\alpha_s$. Solid line: value for
  $\mu_r=\mu_f=m_H$. Filled band: scale uncertainty. See text for
  details.   }
\end{figure*}

\begin{figure*}[ht]
\centering
\includegraphics[width=0.45\textwidth]{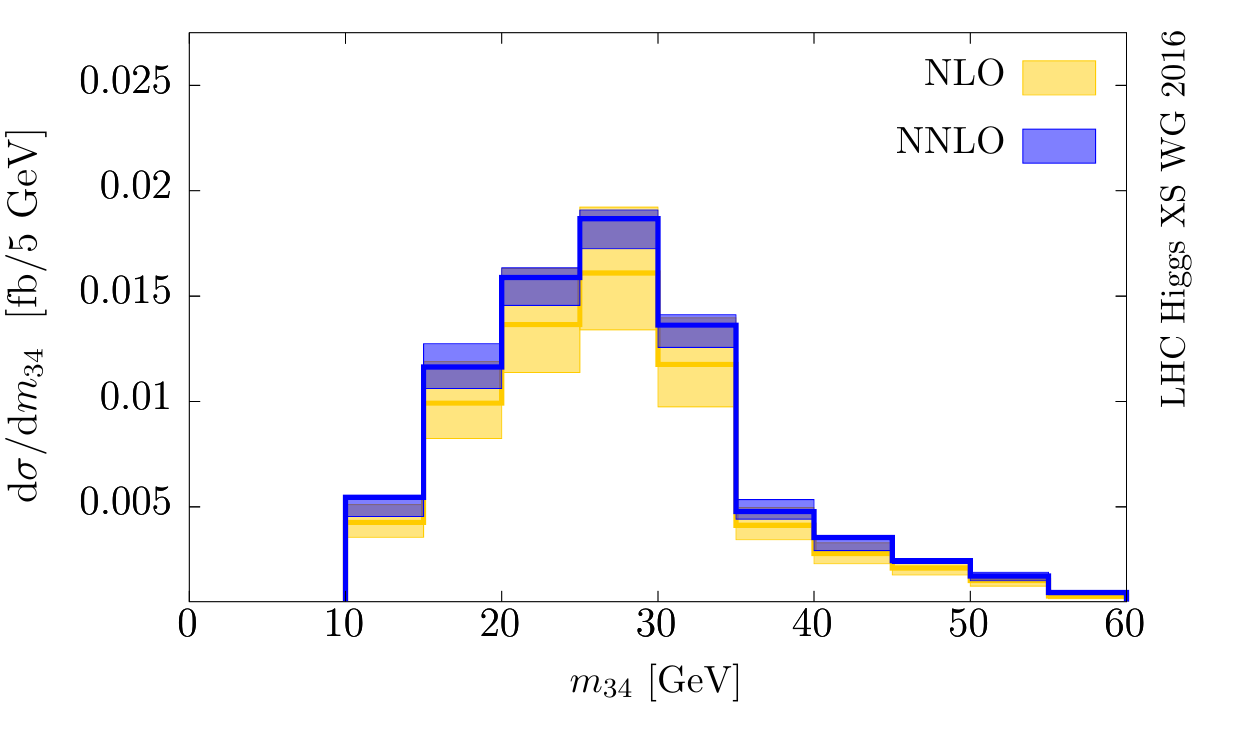}
\hfill
\includegraphics[width=0.45\textwidth]{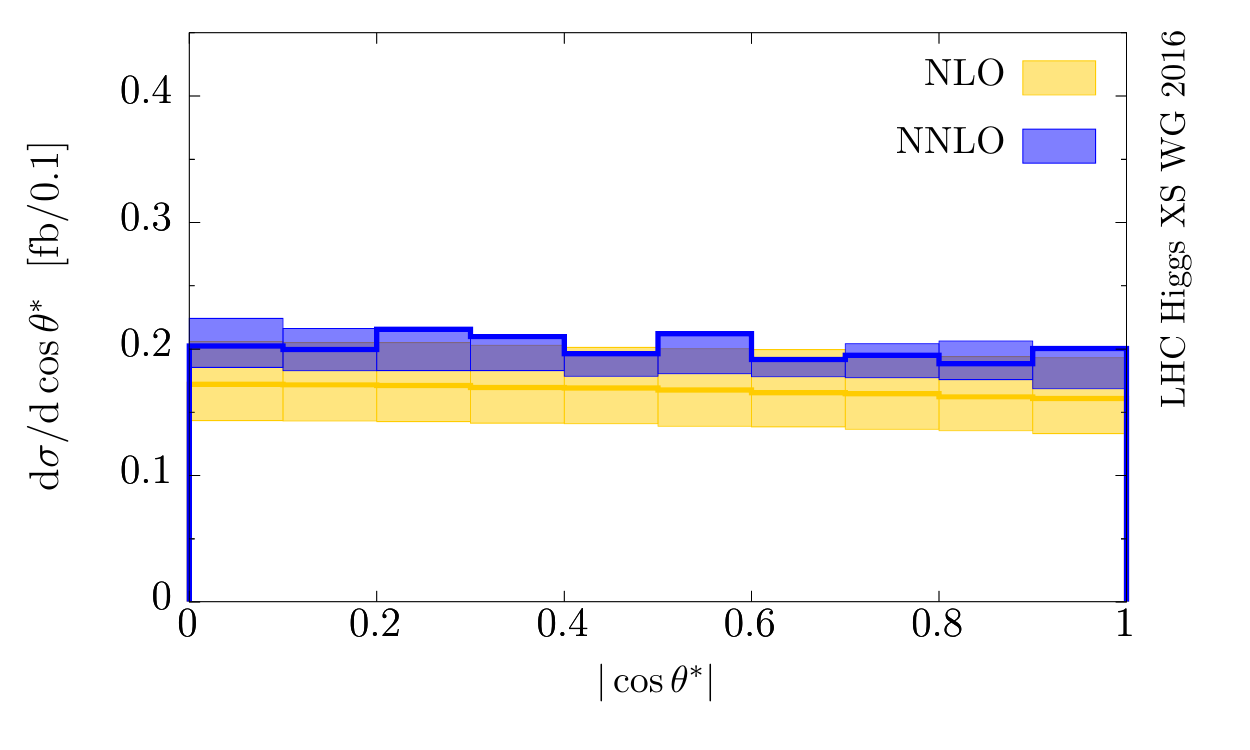}
\caption{ \label{fig:hjetzz02} Distribution in the fiducial volume at
  NLO (yellow) and NNLO (blue) for $H\to ZZ \to 4\mu$. Left: sub-leading
  di-lepton invariant mass. Right: $\cos\theta^*$. Both NLO and
  NNLO curves obtained with NNLO PDFs and $\alpha_s$. Solid line:
  value for $\mu_r=\mu_f=m_H$. Filled band: scale uncertainty. See
  text for details.   }
\end{figure*}



\subsection{Fiducial cross section and distribution for the irreducible background}
\label{sec:ZZ-YR4}

In this contribution
we present NNLO predictions for four-lepton production in the Higgs background region.
The calculation includes the leptonic decays of the vector bosons 
together with
spin correlations and off-shell effects. 
Contributions from $Z\gamma^*$ and $\gamma^*\gamma^*$ production as well as from
$pp\to Z/\gamma^*\to4\textrm{ leptons}$ topologies are also
consistently included with all interference terms.
The corresponding results for the $ZZ\to 4l$ signal region have been presented in \cite{Grazzini:2015hta}.

The calculation is performed with the numerical program
\Matrix\footnote{\Matrix{} is the abbreviation of 
``\Munich{} Automates qT subtraction and Resummation
to Integrate X-sections'', by M.~Grazzini, S.~Kallweit, D.~Rathlev, M.~Wiesemann. 
In preparation.}, which combines the $q_T$-subtraction~\cite{Catani:2007vq} and 
-resummation~\cite{Bozzi:2005wk} formalisms with the 
\Munich{} Monte Carlo framework~\cite{Kallweit:Munich}.
\Munich{} provides a fully automated implementation of the Catani--Seymour dipole 
subtraction method~\cite{Catani:1996jh,Catani:1996vz},
an efficient phase-space integration,
as well as an interface to the one-loop generator \OpenLoops{}~\cite{Cascioli:2011va} 
to obtain all required (spin- and colour-correlated) 
tree-level and one-loop amplitudes. The two-loop helicity amplitudes for this process
have been computed in Refs.~\cite{Caola:2014iua,Gehrmann:2015ora}.

Our calculation allows us to apply
arbitrary cuts on the final-state leptons and
the associated QCD radiation. The central values for the factorization and renormalization scales are fixed
to the invariant mass of the four-lepton system, i.e. $\mu_F=\mu_R=\mu_0=m_{ZZ}$. Perturbative uncertainties are estimated as usual by varying $\mu_F$ and $\mu_R$ in
the range $0.5 \mu_0\leq \mu_F,\mu_R\leq 2\mu_0$ with the constraint $0.5\leq \mu_F/\mu_R\leq 2$.

In Table \ref{tab:zz} we report the fiducial cross sections at LO, NLO and NNLO in the three decay channels.
We also compare our result with the approximation in which only the loop-induced $gg$ contribution is included.


\begin{table}[ht]
\caption{\label{tab:zz} Fiducial cross sections and scale uncertainties at LO, NLO and NNLO in the three channels. The NLO+gg result is also shown for comparison.}
\begin{center}
\renewcommand{\baselinestretch}{1.5}
\begin{tabular}{c| c c c c}
\toprule
Channel & $\sigma_{LO}$ (fb) & $\sigma_{NLO}$ (fb) & $\sigma_{NLO+gg}$ (fb) & $\sigma_{NNLO}$ (fb)  \\ [0.5ex]
\midrule
$e^+e^-e^+e^-$ & 0.1347(1)$^{+10\%}_{-11\%}$ & 0.1485(2)$^{+2.4\%}_{-3.6\%}$ & 0.1584(2)$^{+2.4\%}_{-3.6\%}$ & 0.159(1)$^{+0.7\%}_{-0.9\%}$ \\
$\mu^+\mu^-\mu^+\mu^-$ & 0.1946(2)$^{+10\%}_{-11\%}$ & 0.2150(2)$^{+2.4\%}_{-3.6\%}$ & 0.2291(2)$^{+2.4\%}_{-3.6\%}$ & 0.230(1)$^{+0.9\%}_{-0.8\%}$ \\
 $e^+e^-\mu^+\mu^-$ & 0.3165(3)$^{+10\%}_{-11\%}$ & 0.3457(3)$^{+2.4\%}_{-3.6\%}$ & 0.3677(2)$^{+2.3\%}_{-3.5\%}$ & 0.3690(6)$^{+0.5\%}_{-0.8\%}$ \\
\bottomrule
\end{tabular}
\end{center}
\end{table}
The NNLO effect increases the NLO result by about 7\%.
By using NNLO PDFs throughout,
the loop-induced gluon-fusion contribution provides
about $83\%-85\%$ of the full NNLO result.
The impact of the gluon fusion contribution is thus higher than what was found in Ref.~\cite{Grazzini:2015hta} in the case of the $ZZ$ analysis, in which the $ZZ$ bosons are essentially produced on shell. This is due to the selection cuts, which suppress the impact of genuine radiative corrections to the $q{\bar q}$ channel. This effect is visible already at NLO, where the impact of radiative corrections is reduced from $+23\%$ in the case of inclusive
on shell $ZZ$ production to $+9\%-10\%$.

In \refF{fig:m4ly4l} we present our LO, NLO and NNLO predictions for the invariant mass (left) and the rapidity (right) of the four leptons. The lower panels show the NNLO results normalized to the NLO predictions.
The NLO+gg prediction is also shown. We see that the impact of NNLO corrections is rather stable for these distributions.

\begin{figure}[htpb]
  \centering
    \subfigure[]{\includegraphics[width=0.49\textwidth]{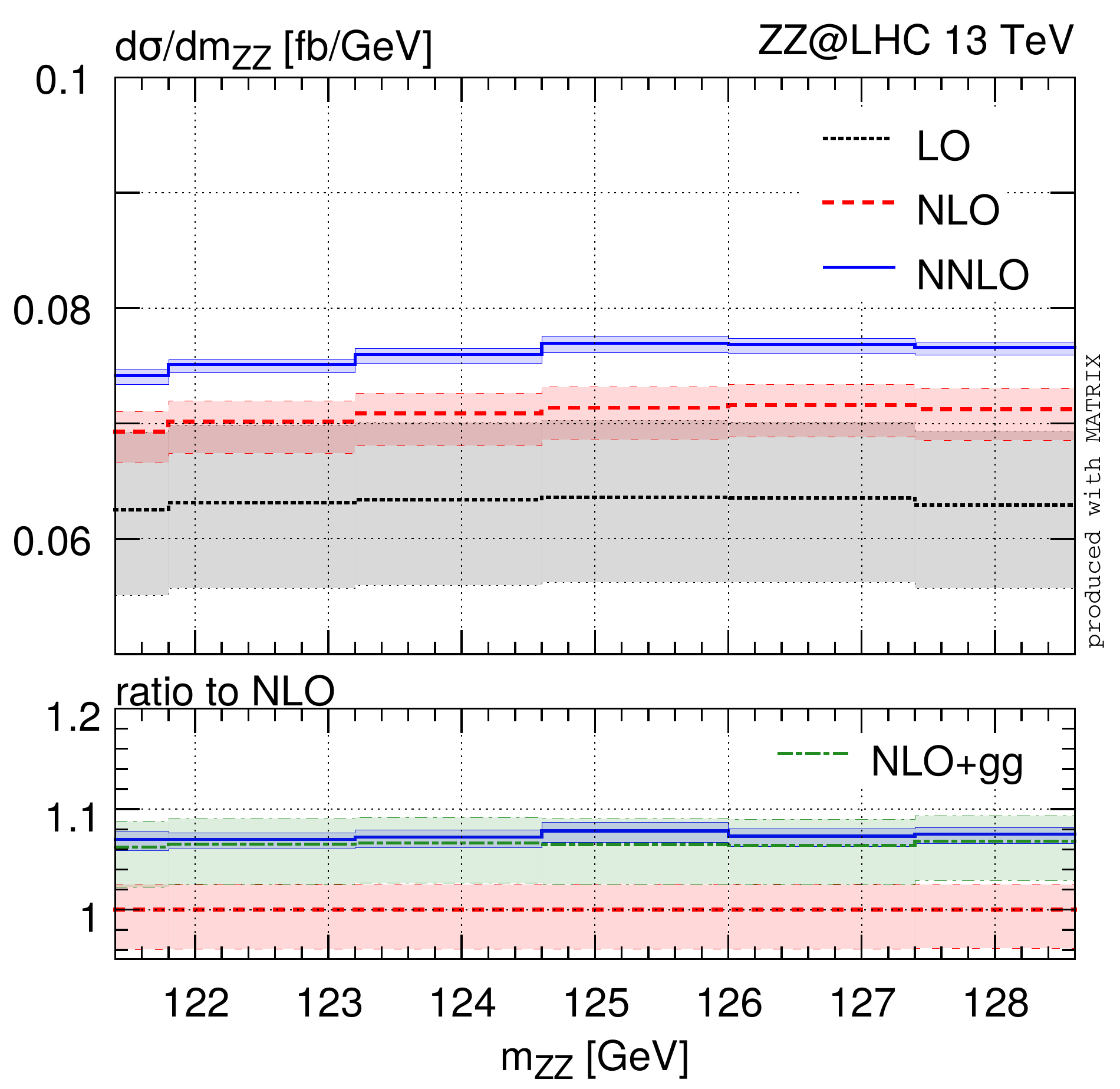}}
    \subfigure[]{\includegraphics[width=0.49\textwidth]{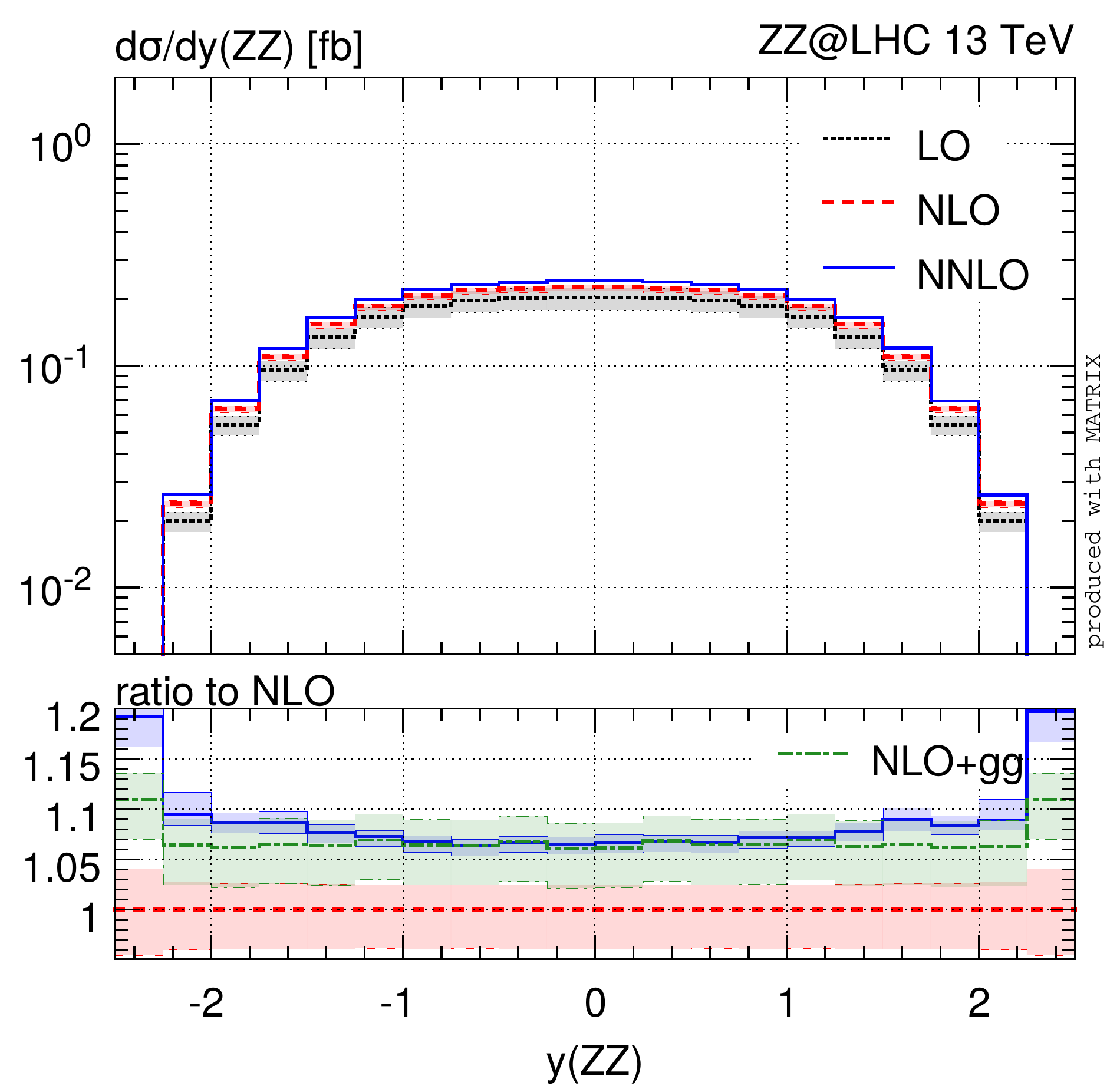}}
    \caption{The invariant mass (left) and rapidity (right) distributions of the four-lepton system at LO, NLO and NNLO. The lower panel shows the NNLO result normalized to NLO. The NLO+gg result is also shown for comparison.}
    \label{fig:m4ly4l}
\end{figure}

In \refF{fig:m34cos12} we show the distributions in the
invariant mass of the subleading lepton pair (left) and in $\cos\theta_{12}$ (right).
Here we notice that the NNLO corrections are larger at large $m_{34}$ and small $\cos\theta_{12}$.

\begin{figure}[htpb]
    \centering
    \subfigure[]{\includegraphics[width=0.49\textwidth]{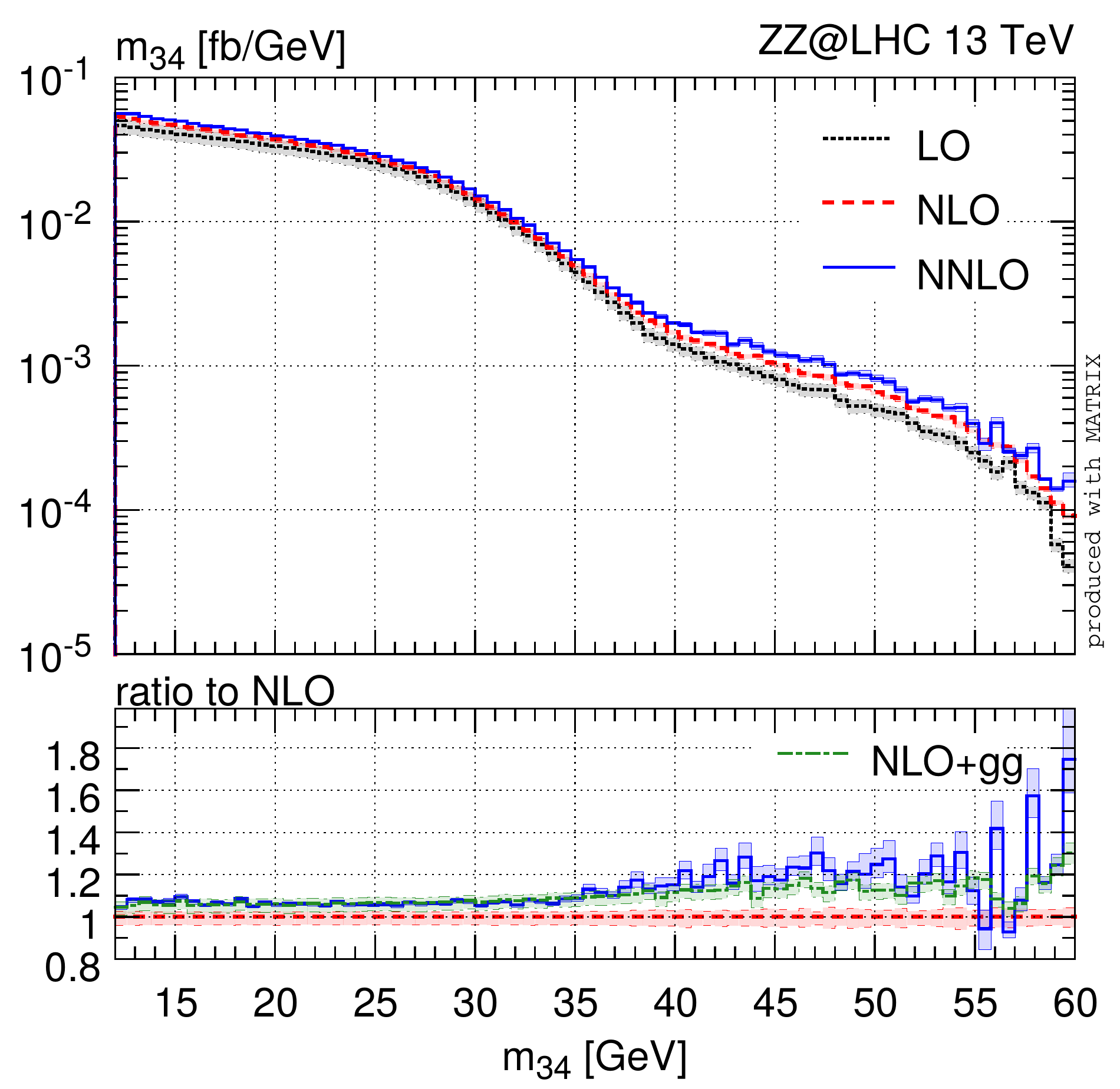}}
    \subfigure[]{\includegraphics[width=0.49\textwidth]{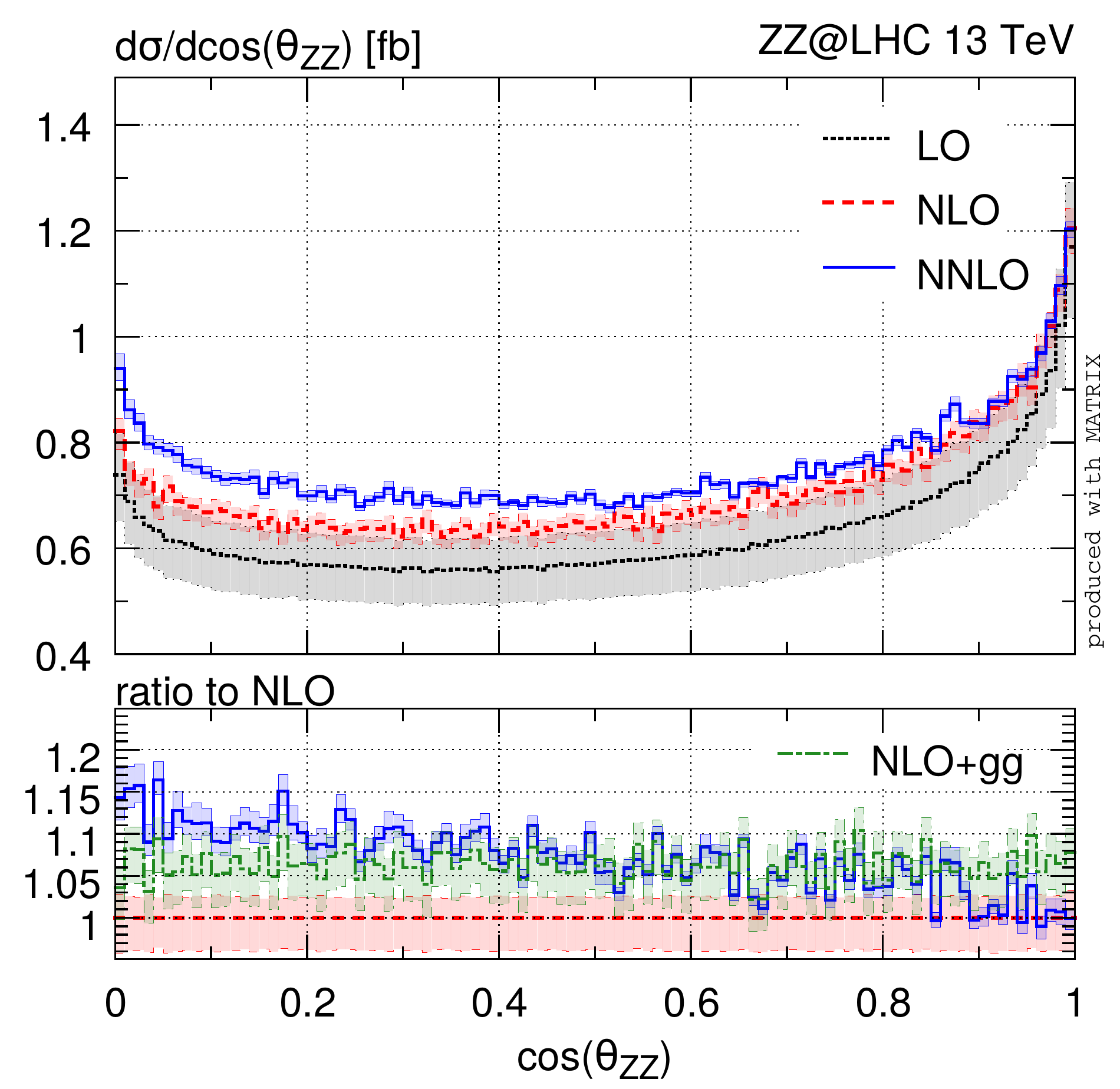}}
    \caption{As in \refF{fig:m4ly4l}, but for the invariant mass of the subleading lepton pair (left) and the $\cos\theta_{12}$ distributions.}
    \label{fig:m34cos12}
\end{figure}

In \refF{fig:pT4lj1} we show the $p_T$ distributions of the four-lepton system (left) and of the leading jet (right). These distributions are identical at NLO, when only one parton recoils against the four-lepton system, and show significant corrections at NNLO. This is not unexpected since the NNLO calculation is effectively NLO in this case.

\begin{figure}[htpb]
    \centering
    \subfigure[]{\includegraphics[width=0.49\textwidth]{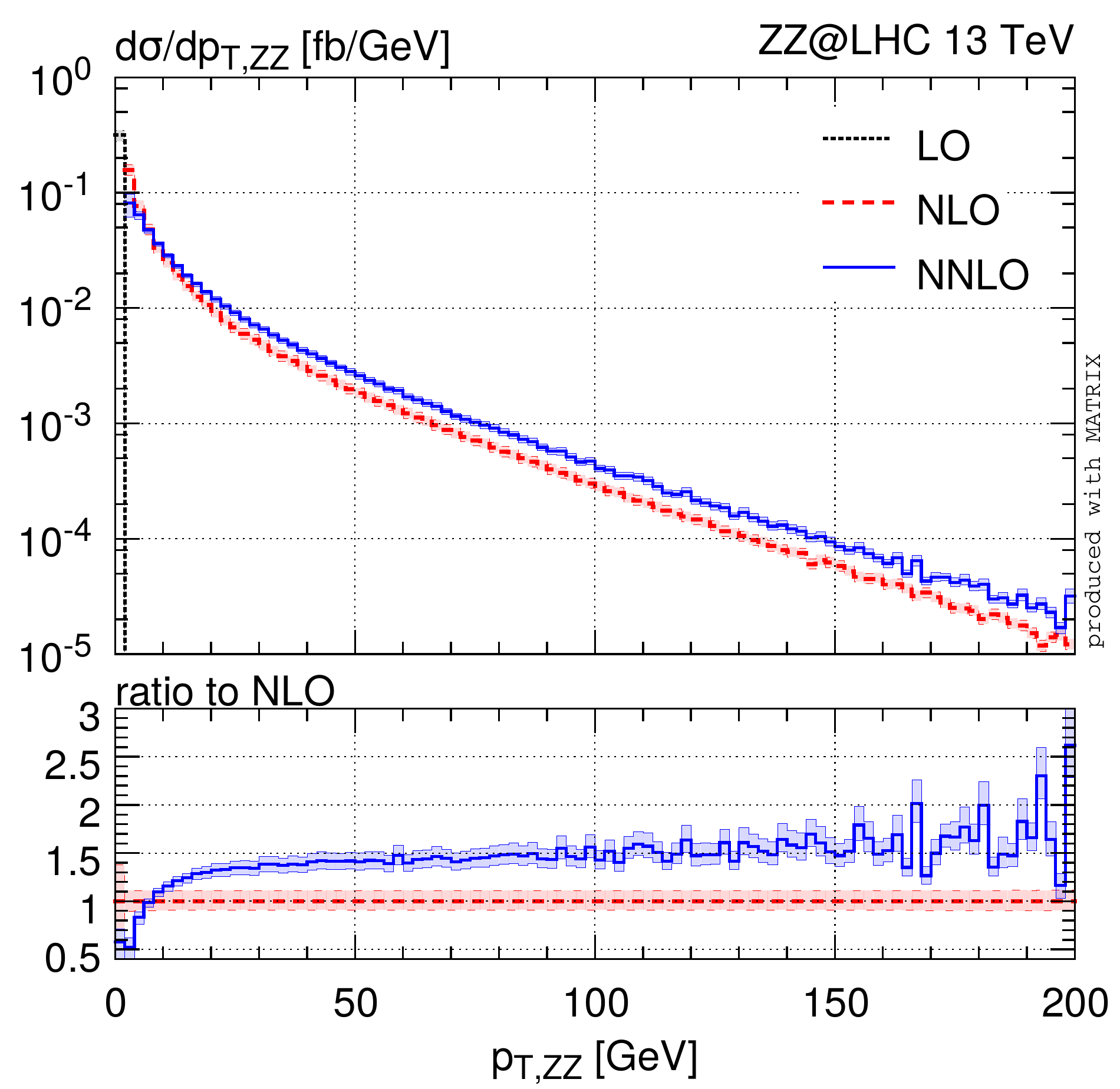}}
    \subfigure[]{\includegraphics[width=0.49\textwidth]{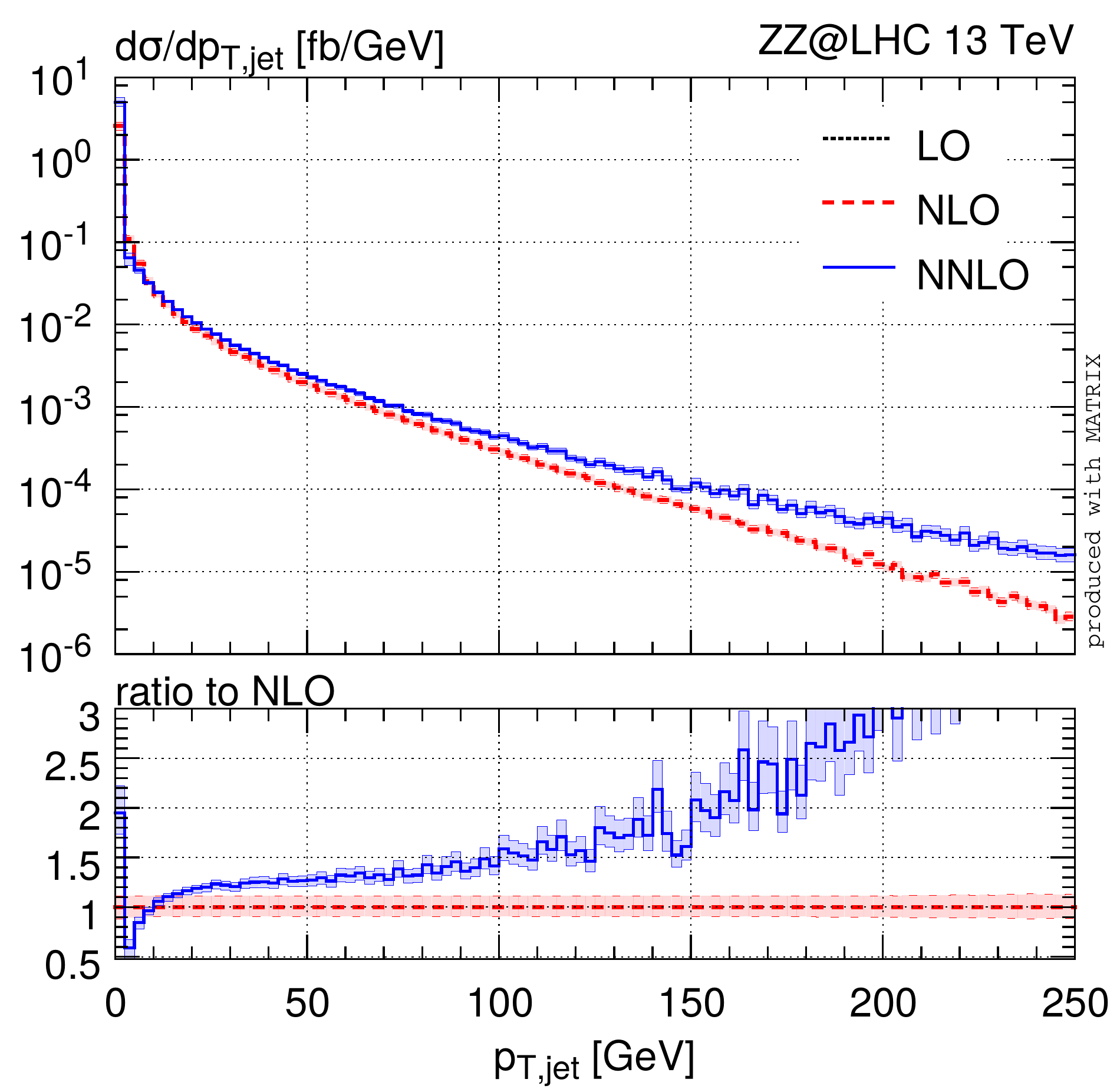}}
    \caption{As in \refF{fig:m4ly4l}, but for the $p_T$ distributions of the four-lepton system (left) and of the leading jet (right).}
    \label{fig:pT4lj1}
\end{figure}

Analogous comments apply to \refF{fig:yj1dy4lj}, which shows our results for
the rapidity distribution of the leading jet (left) and for the rapidity difference between the leading jet and the
four-lepton system (right).

\begin{figure}[htpb]
    \centering
    \subfigure[]{\includegraphics[width=0.49\textwidth]{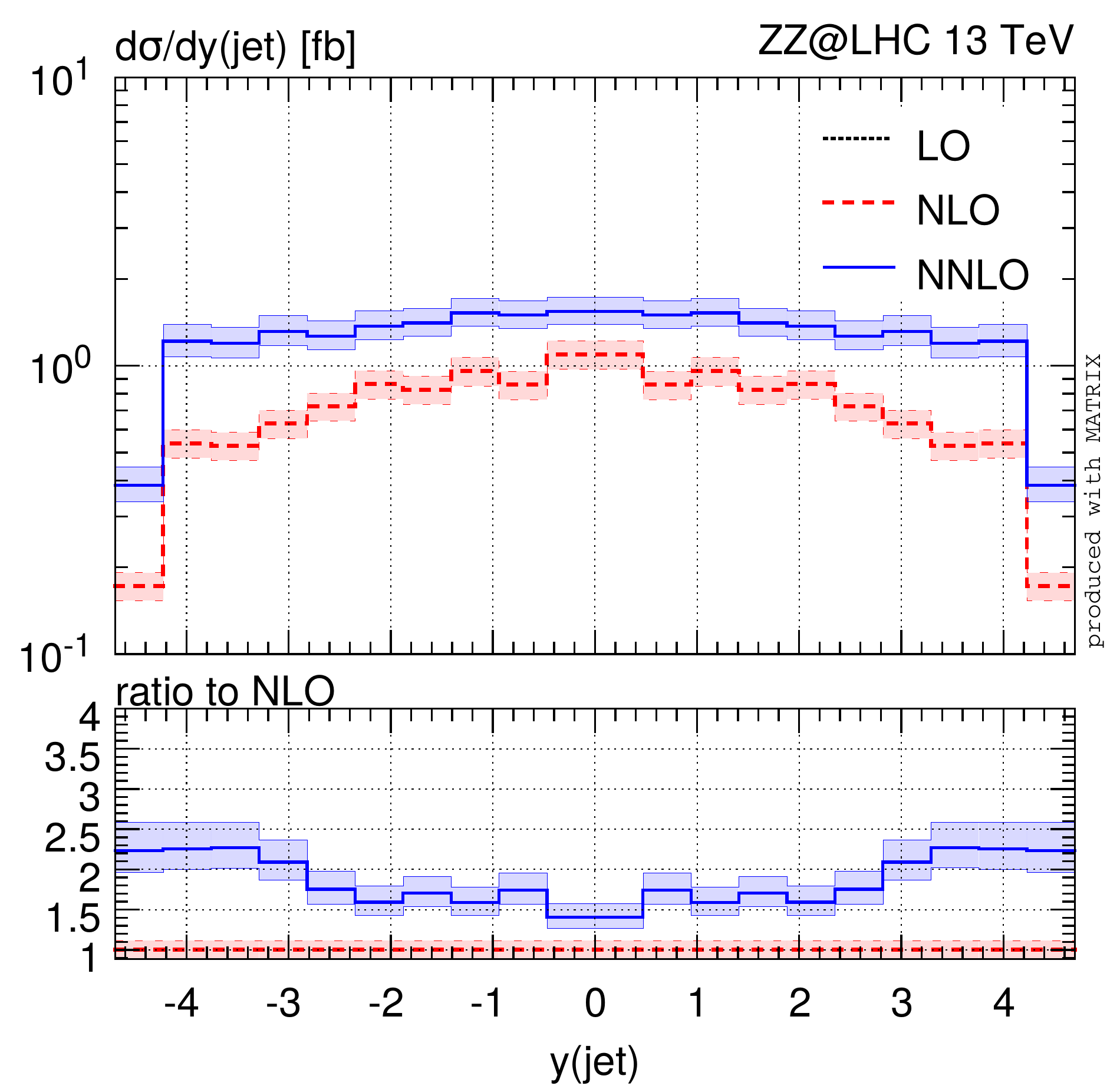}}
    \subfigure[]{\includegraphics[width=0.49\textwidth]{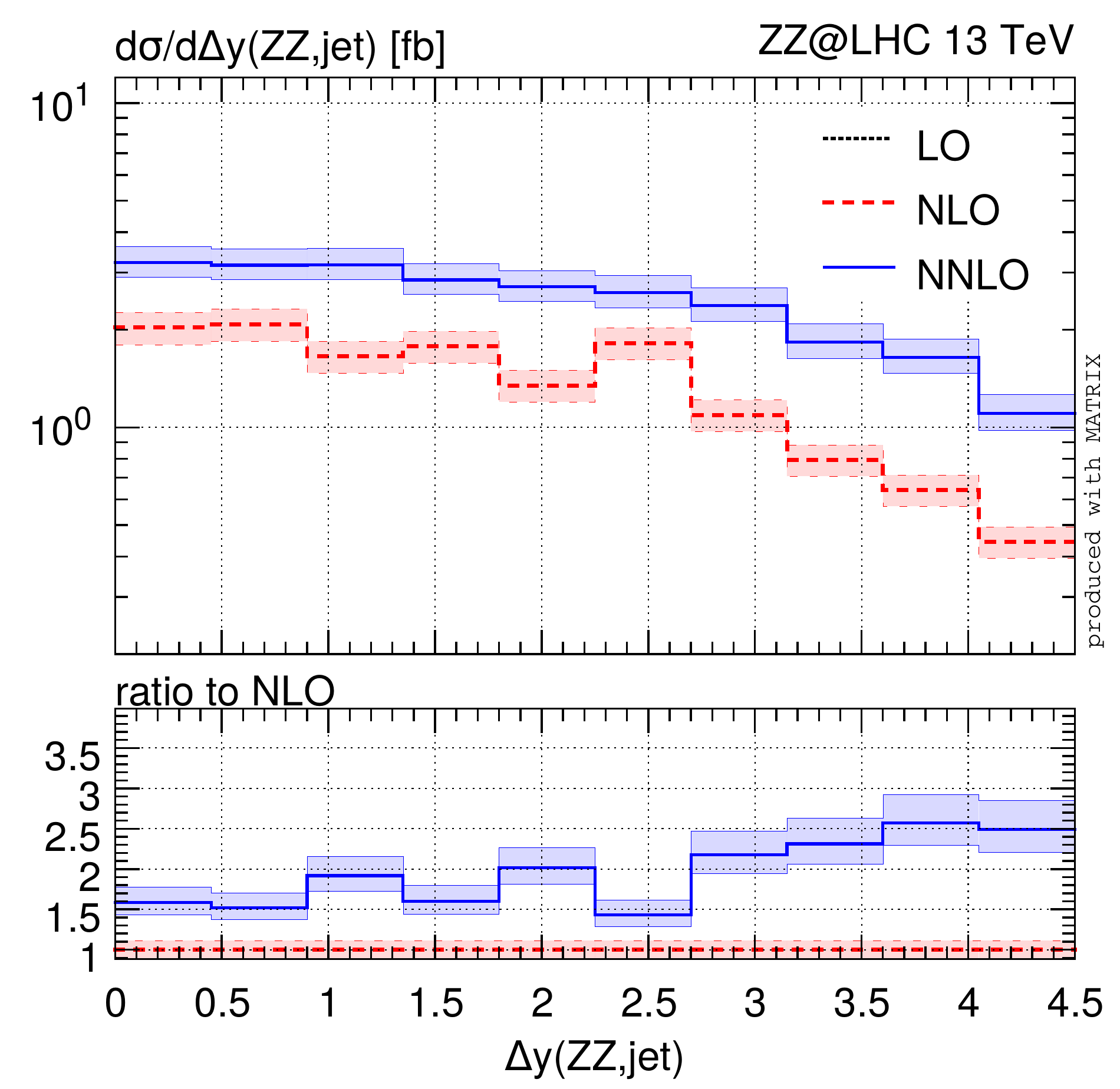}}
    \caption{As in \refF{fig:m4ly4l}, but for the rapidity distribution of the leading jet (left) and the rapidity difference of the leading jet and the four-lepton system (right).}
    \label{fig:yj1dy4lj}
\end{figure}

\subsection[QCD activity associated with Higgs  production in gluon fusion]{Comparison of the description of QCD activity associated with Higgs  production in gluon fusion} 
\label{sec:lh}

In this section we present a brief summary of the extensive comparison of 
the description of QCD activity in association with Higgs boson production in 
gluon fusion, undertaken within the context of the Les Houches 2015 
workshop~\cite{Badger:2016bpw}. It aims at comparing the description of the QCD 
activity accompanying the production of a Higgs boson in gluon fusion. 
To this end contributions obtained from from a multitude of authors 
applying different approximations and calculational schemes were subjected 
to a comprehensive list of inclusive and successively exclusive observables. 
To facilitate the comparison a common setup was adopted. 
Beside working in the pure Higgs effective theory (HEFT) in the 
$m_t\to\infty$ limit, scales reducing to $\tfrac{1}{2}\,\MH$ in the 
zero jet limit were adopted along with the common PDF sets 
\texttt{MMHT2014nlo68clas0118}/\texttt{MMHT2014nnlo68cl} \cite{Harland-Lang:2014zoa}, 
as appropriate, with $\alpha_s(\MZ)=0.118$ were used where possible. 
It is important to stress that for many tools this does not necessarily 
constitute their respective best setup, but was adopted for the sake of 
the comparison.

\noindent
The contributions comprise the analytical resummations of 
\begin{itemize}
  \item \textsc{HqT} \cite{Bozzi:2005wk,deFlorian:2011xf} for the Higgs boson 
transverse momentum, 
  \item STWZ \cite{Stewart:2013faa} for the jet veto cross sections, the leading jet 
        transverse momentum, the inclusive cross section and the exclusive 
        zero jet cross section, 
  \item \textsc{ResBos}~2 \cite{Wang:2012xs,Sun:2016kkh} for inclusive 
        zero and one jet observables, 
\end{itemize}
the fixed-order computations of 
\begin{itemize}
  \item \textsc{Sherpa Nnlo} \cite{Gleisberg:2008ta,Hoche:2014dla} calculation 
        of $pp\to h+X$, 
  \item BFGLP \cite{Boughezal:2015aha,Boughezal:2015dra} NNLO  calculation 
        of $pp\to h+j+X$, 
  \item \textsc{GoSam+Sherpa} \cite{Cullen:2011ac,Cullen:2014yla,
          Gleisberg:2008ta,Gleisberg:2008fv} NLO calculation of 
        $pp\to h+1,2,3j+X$ \cite{Cullen:2013saa,Greiner:2015jha}, 
        here also a \textsc{MiNlo} \cite{Hamilton:2012np} and 
        \textsc{LoopSim} \cite{Rubin:2010xp} (labelled nNLO) calculation 
        are available
\end{itemize}
the \textsc{NnloPs} matched computations of 
\begin{itemize}
  \item \texttt{Powheg} \textsc{NnloPS} \cite{Hamilton:2013fea}, showered 
        with \textsc{Pythia~8.253} \cite{Sjostrand:2014zea},
  \item \textsc{Sherpa NnloPs} \cite{Gleisberg:2008ta,Hoche:2014dla},
\end{itemize}
the \textsc{Nlo} multijet merged computations of 
\begin{itemize}
  \item \texttt{MG5\_aMC@NLO} in the FxFx scheme \cite{Alwall:2014hca,Frederix:2012ps}, 
  showered with \textsc{Pythia 8.210} \cite{Sjostrand:2014zea}
  \item \textsc{Sherpa} in the \textsc{MePS@Nlo} scheme \cite{Gleisberg:2008ta,
          Hoeche:2012yf,Hoeche:2014lxa,Hoeche:2011fd,Hoche:2012wh} using 
          one-loop matrix elements from \textsc{GoSam} \cite{Cullen:2011ac,
          Cullen:2014yla,Cullen:2013saa,Greiner:2015jha},
  \item \textsc{Herwig 7.1} in the unitarized merging scheme \cite{Bellm:2015jjp,
          Platzer:2012bs,Lonnblad:2012ix} using its dipole shower 
          \cite{Platzer:2009jq} and matrix elements from 
          \texttt{MG5\_aMC@NLO} \cite{Alwall:2014hca} and 
          \textsc{OpenLoops} \cite{Cascioli:2011va},
\end{itemize}
and the BFKL resummation of 
\begin{itemize}
  \item \textsc{Hej} \cite{Andersen:2009nu,Andersen:2011hs,Andersen:2008ue} 
        describing $pp\to h+2j+X$,
\end{itemize}
and therefore cover a large space of calculations available. Uncertainties 
are determined varying the appropriate scales, cf.\ \cite{Badger:2016bpw} for details. 

\begin{figure}
  \begin{minipage}{0.32\textwidth}
    \includegraphics[width=\textwidth]{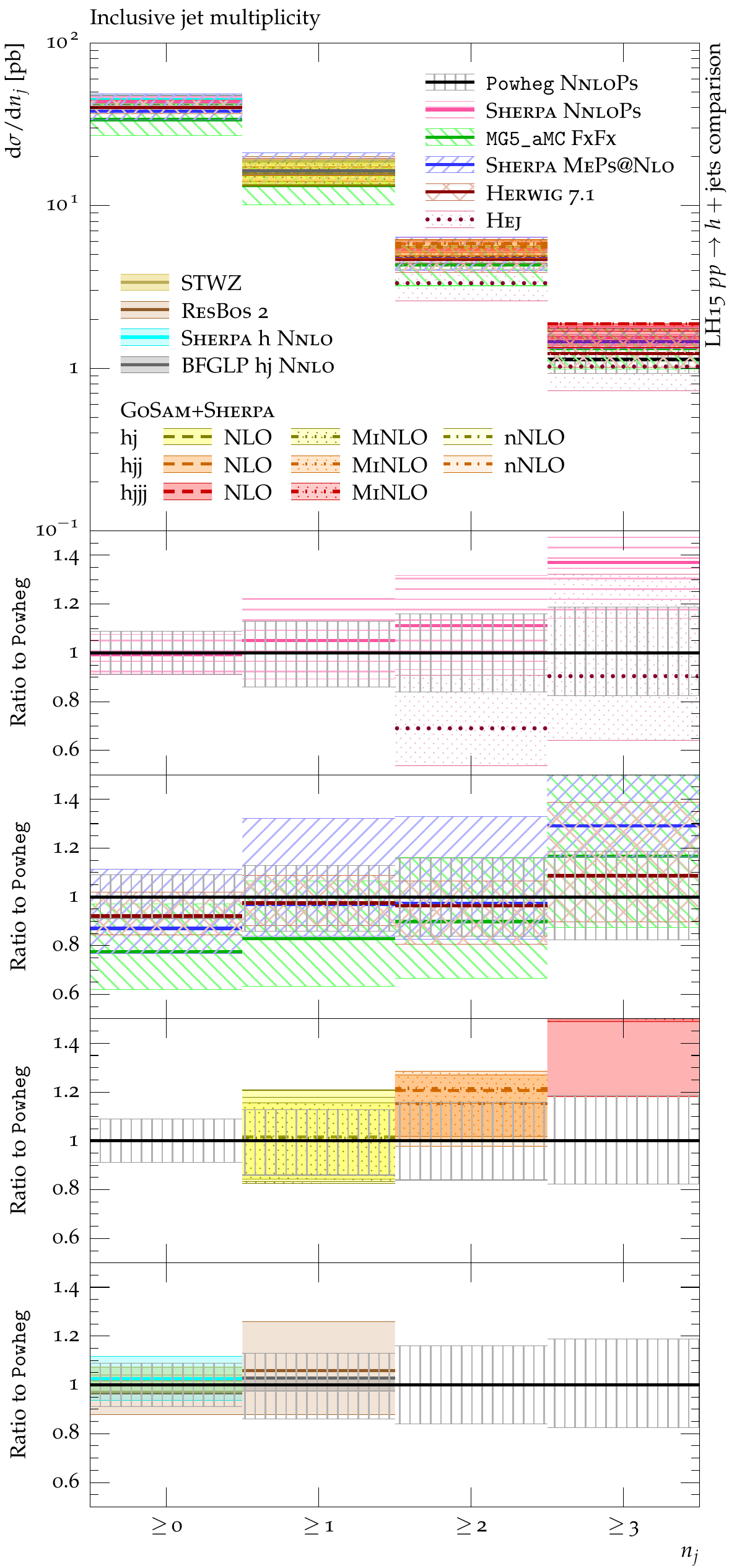}
  \end{minipage}
  \begin{minipage}{0.32\textwidth}
    \includegraphics[width=\textwidth]{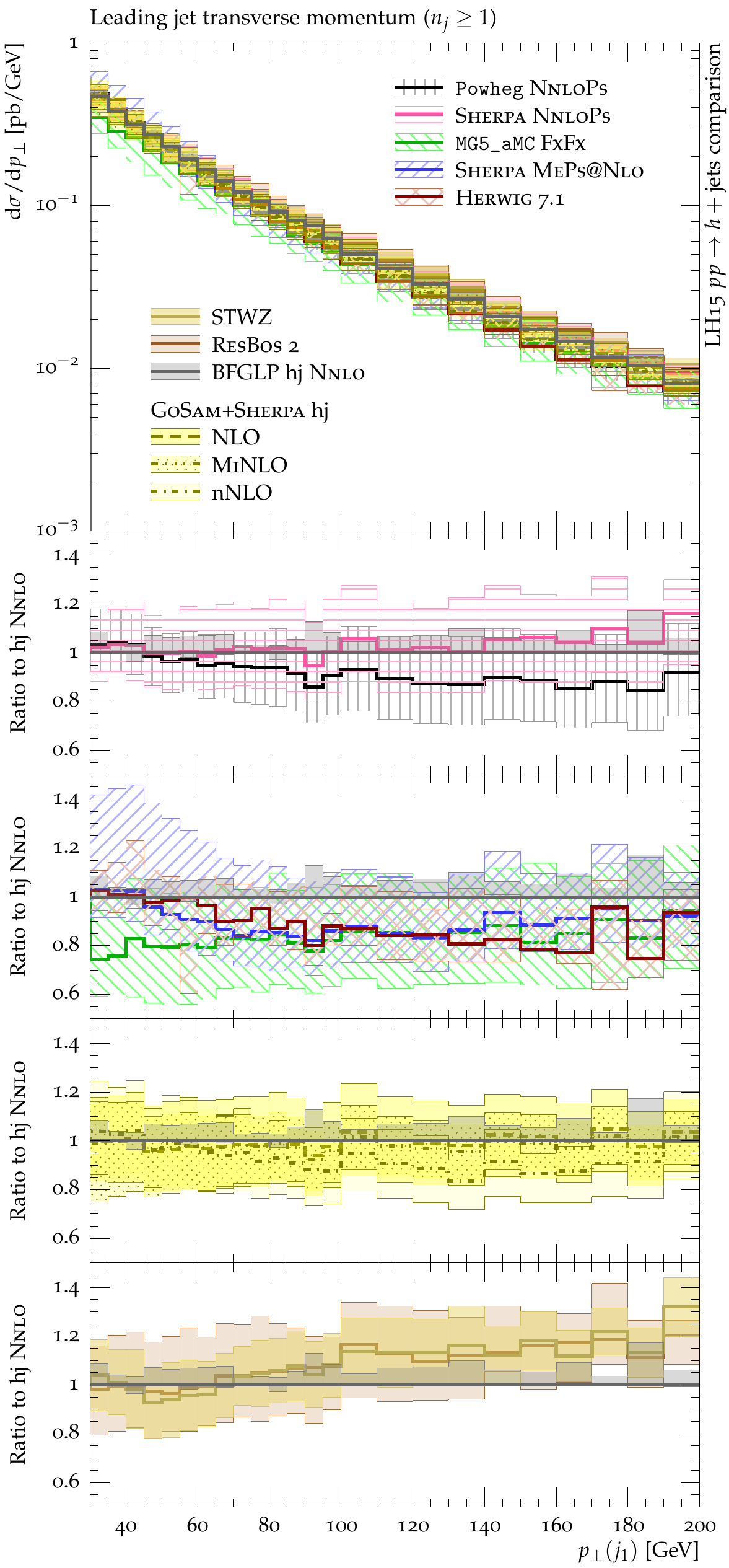}
  \end{minipage}
  \begin{minipage}{0.32\textwidth}
    \lineskip-1.35pt
    \includegraphics[width=\textwidth]{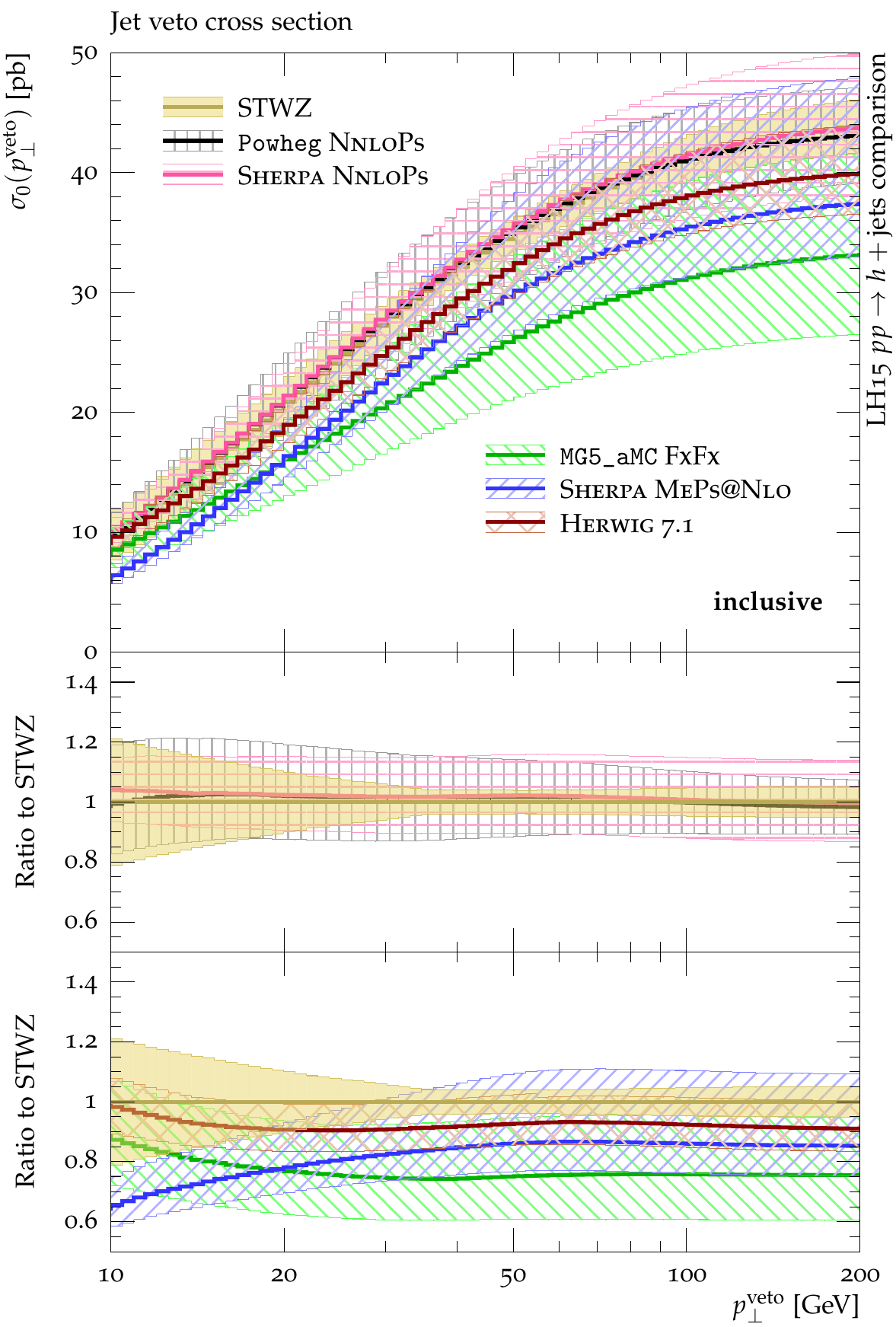}\\
    \includegraphics[width=\textwidth]{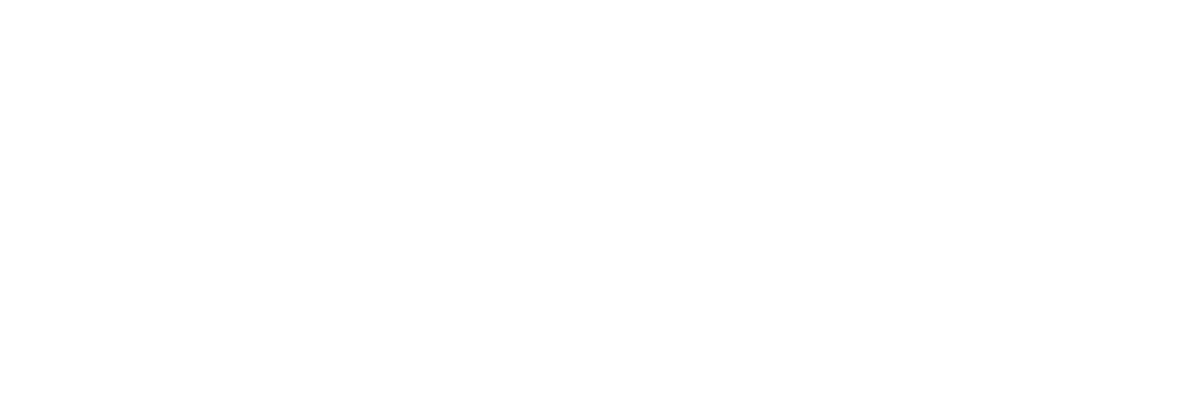}\\
    \includegraphics[width=\textwidth]{./WGx/FXS/figs/ratiopanelplaceholder}
  \end{minipage}
  \caption{
    Example comparisons of the description of QCD activity accompanying 
    Higgs boson production in gluon fusion, taken from~\cite{Badger:2016bpw}. 
    The inclusive jet multiplicity (top left), the leading jet 
    transverse momentum (right) and the jet vetoed inclusive cross 
    section (bottom left) are shown. The ratio panels compare to the 
    appropriate reference as indicated. See text for details.
    \label{fig:lh_hjets_comp:njet_ptj1_jvxs}
  }
\end{figure}

To facilitate comparisons with as diverse calculations as possible we 
consider inclusive Higgs boson production with no restriction on its decay 
products. Jets are identified using the anti-$k_{\mathrm{T}}$ algorithm 
\cite{Cacciari:2008gp} with $R=0.4$ and are required to have $\pT>30\,\mathrm{GeV}$ 
and $|\eta|<4.4$. An implementation in \textsc{Rivet} \cite{Buckley:2010ar} exists. 
\refF{fig:lh_hjets_comp:njet_ptj1_jvxs} displays exemplary results 
of this comparison for three observables of interest out of the 79 observables 
considered: the inclusive jet multiplicity, the leading jet transverse 
momentum, and the jet vetoed inclusive cross section. All plots show 
a main plot accompanied by multiple ratio plots, grouping the individual 
contributions to ease the comparison. Noteworthy in the observables 
exhibited here are the agreement 
in the inclusive $n_j\ge 1$ cross section between all tools considered 
which possess at least NLO accuracy in this observable. While for $n_j\ge 2$ 
all predictions including parton showering agree well, for $n_j\ge 3$ the 
spread is larger. However, it has to be kept in mind that the \textsc{NnloPs} 
matched calculations revert leading order accuracy for $n_j\ge 2$ and to 
pure parton shower accuracy for $n_j\ge 3$ while 
\texttt{MG5\_aMC@NLO} and \textsc{Herwig 7.1} are NLO 
and LO accurate there, respectively. Only, the \textsc{GoSam+Sherpa} and the 
\textsc{Sherpa MePs@Nlo} prediction possess NLO accuracy in both cases. 
\textsc{Hej}, being LO accurate for $n_j\ge 2,3$, predicts slightly 
different inclusive jet multiplicities. 
The leading jet transverse momentum spectrum shows a consensus between 
(almost) all parton shower matched and/or multijet merged calculations 
exhibiting NLO accuracy for this observable: \texttt{Powheg} \textsc{NnloPs}, 
\textsc{Sherpa Nnlops}, \texttt{MG5\_aMC@NLO}, 
\textsc{Sherpa MePs@Nlo} and \textsc{Herwig 7.1}. The parton level calculations 
deviate mostly due to the individual scales set. However, the pure fixed 
order NLO and NNLO calculation, employing the same scale, agree very well, 
indicating a $K$-factor very close to unity. Finally, the 
degree of congruence in the jet vetoed inclusive cross section between the 
\textsc{NnloPs} matched calculations and the STWZ dedicated resummation is 
remarkable. Here, the NLO multijet merged tools primarily suffer from their 
NLO normalization in the $\pT^{\mathrm{veto}}\to\infty$ limit.
Generally, a remarkable level of agreement is found between the individual 
calculations throughout most observables.

\section{Beyond the Standard Model effects} 
\label{sec:BSM}
New physics beyond the SM could affect Higgs physics in total rates
(including differences in efficiencies of selection cuts) and
differential distributions.  While one can attempt to isolate effects
in either production or decay processes, a full BSM scenario typically
affects both simultaneously.  (Differential) fiducial cross sections
are an appropriate complementary tool for scrutinizing the Lagrangian
structure of the Higgs boson interactions, for instance through tests for
new tensorial couplings, non-standard production modes, and effective
form factors. In addition to the measurement of specific fiducial
regions, the combined analysis of all available fiducial measurements
in a global fit seems a promising approach.

Below we address the effects that can be expected in a variety of BSM
scenarios and we discuss interesting distributions and fiducial
regions that can be used to target them. Unless otherwise specified,
we will not consider any generic fiducial cuts on the results shown
below. The eventual feasibility to measure such regions in a model
independent way needs though to be scrutinized by experimentalists,
since in some cases the poor resolution for some of the observable
used to define the fiducial volumes could lead to non-negligible
migration effects, cf.~\refS{sec:exp:fid}.
We also note that a parallel effort is required from the theory
community, in obtaining predictions with adequate precision, and a
robust determination of the associated uncertainties, and their
correlations, within the theoretical framework used in the comparison
to the measurements, be it specific BSM models or effective Lagrangian
descriptions. The status of the presently existing tools is discussed 
in the various other sections of this report. 

\subsection{Higgs boson production in gluon fusion}

Higgs boson production in gluon fusion has one specific feature which allows
us to test to what degree the observed Higgs is described by the
SM: its production amplitude at one loop is mediated by 
virtual top quarks, which means that any new, strongly interacting particle can
lead to order-one corrections to the effective Higgs-gluon
coupling. The relevant Lagrangian describing this channel is 
\begin{align}
\mathcal{L}
= \mathcal{L} \Bigg|_{\kappa_j =0} + 
\left[  \kappa_{\PQt} \; g_{\Pg\Pg\PH} + \kappa_{\Pg}\frac{\alpha_s}{12 \pi}  \right] \; 
\frac{\PH}{v} \; 
G_{\mu\nu}G^{\mu\nu}
- \kappa_{\PQt} \; 
\frac{\Mt}{v} \PH \left( \PAQt_R \PQt_L + \mathrm{h.c.} \right)
\; . 
\label{eq:fxs-bsm-lagrangian}
\end{align}
In combination with a second measurement of the top Yukawa
coupling, for example in $\ttbar\PH$ production, already a total rate
measurement in this largest Higgs boson production channel can constrain
particles like light top partners.

While the Lagrangian of \eqn{eq:fxs-bsm-lagrangian} only features shifts
in SM-like Higgs boson couplings, the interplay between the renormalizable
top Yukawa coupling and the dimension-6 Higgs-gluon coupling can
affect kinematic distributions.  For example, the reach for new
particles contributing to the effective Higgs-gluon coupling can be
enhanced by adding off-shell Higgs boson production to the set of
measurements~\cite{Campbell:2011cu,Kauer:2012hd,Caola:2013yja}. Strictly speaking, off-shell Higgs
production with a subsequent decay $\PH \to 4\Pl$ is best described by
a shape analysis of the $\Mfourl$ distribution.  Based on the
Lagrangian of \eqn{eq:fxs-bsm-lagrangian} we can write the complete
gluon-induced amplitude $\Pg\Pg\to \PZ\PZ$ as
\begin{alignat}{5}
\mathcal{M}_{\PZ\PZ} & =
\kappa_{\PQt} \mathcal{M}_{\PQt} + \kappa_{\Pg}\mathcal{M}_{\Pg} + \mathcal{M}_c  \; ,
\label{eq:fxs-bsm-amplitude_offshell}
\end{alignat}
where the last term arises from the Higgs-independent continuum
diagram. Numerically, the interference between the Higgs and continuum diagrams 
is one of the key features in the measurement of off-shell Higgs
effects at the LHC.  
While this phase space region is not included in the template fiducial regions in \refS{sec:fiducialvolumes}, 
considering it is extremely beneficial~\cite{Aad:2015rka}.%
\footnote{Off-shell Higgs boson production and Higgs interference are discussed in detail in \refC{chap:offshell_interf} of this report.}
The longitudinal components to the different
contributions feature different dependences on $\Mfourl$, namely
\begin{alignat}{5}
\mathcal{M}_{\Pg}^{++00}
&\approx
-\frac{\Mfourl^2}{2\MZ^2} \qquad \qquad  
&&\mathrm{with} \;\; \Mt \gg \Mfourl \gg \MH,\MZ \notag \\
\mathcal{M}_{\PQt}^{++00}
&\approx
+\frac{\Mt^2}{2\MZ^2}
\log^2 \frac{\Mfourl^2}{\Mt^2}\qquad \qquad 
&&\mathrm{with} \;\; \Mfourl \gg \Mt \gtrsim \MH,\MZ \notag \\
\mathcal{M}_c^{++00}
&\approx
-\frac{\Mt^2}
{2\MZ^2}\log^2 \frac{\Mfourl^2}{\Mt^2} 
&&\mathrm{with} \;\; \Mfourl \gg \Mt \gtrsim \MZ \; .
\label{eq:fxs-bsm-m4l_tgc}
\end{alignat}
In the proper limit a logarithmic dependence on $\Mfourl/\Mt$ develops
far above the Higgs boson mass shell.  The ultraviolet logarithm cancels
between the correct Higgs amplitude and the continuum, ensuring the
proper ultraviolet behaviour of the full amplitude. The sensitivity due
to this logarithmic dependence on $\Mfourl/\Mt$, as shown in the left
panel of \refF{fig:fxs-bsm-m4l}, should allow us to extract the top
mass dependence of the observed signal from the $\Mfourl$
distribution. This means Higgs boson production is a good example of a
process where it is not clear where we can expect to best find BSM
effects: on the one hand, a precise measurement of the total rates of
inclusive Higgs boson production and $\PQt\PAQt\PH$ production can be
expected to constrain the Lagrangian if
\eqn{eq:fxs-bsm-lagrangian}. On the other hand, off-shell Higgs
production, or better the $\Mfourl$ distribution might well benefit
from its known dependence on the ratio $\Mfourl/\Mt$ in the Standard
Model, in that case relying on the tails of a momentum-related
kinematic distribution.


\begin{figure}\centering
 \includegraphics[width=0.44\textwidth]{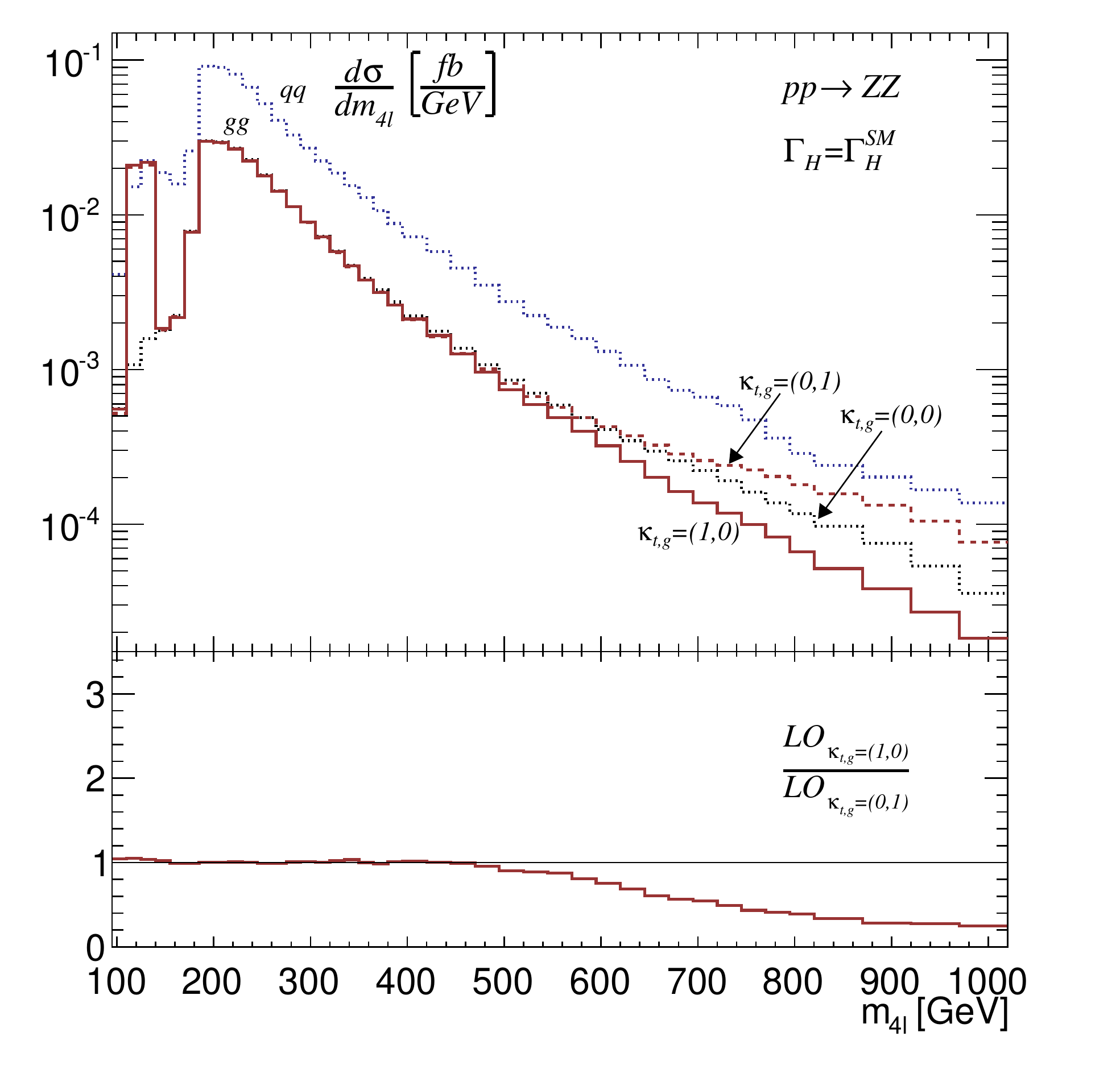} \quad
 \raisebox{-4mm}{\includegraphics[width=0.47\textwidth]{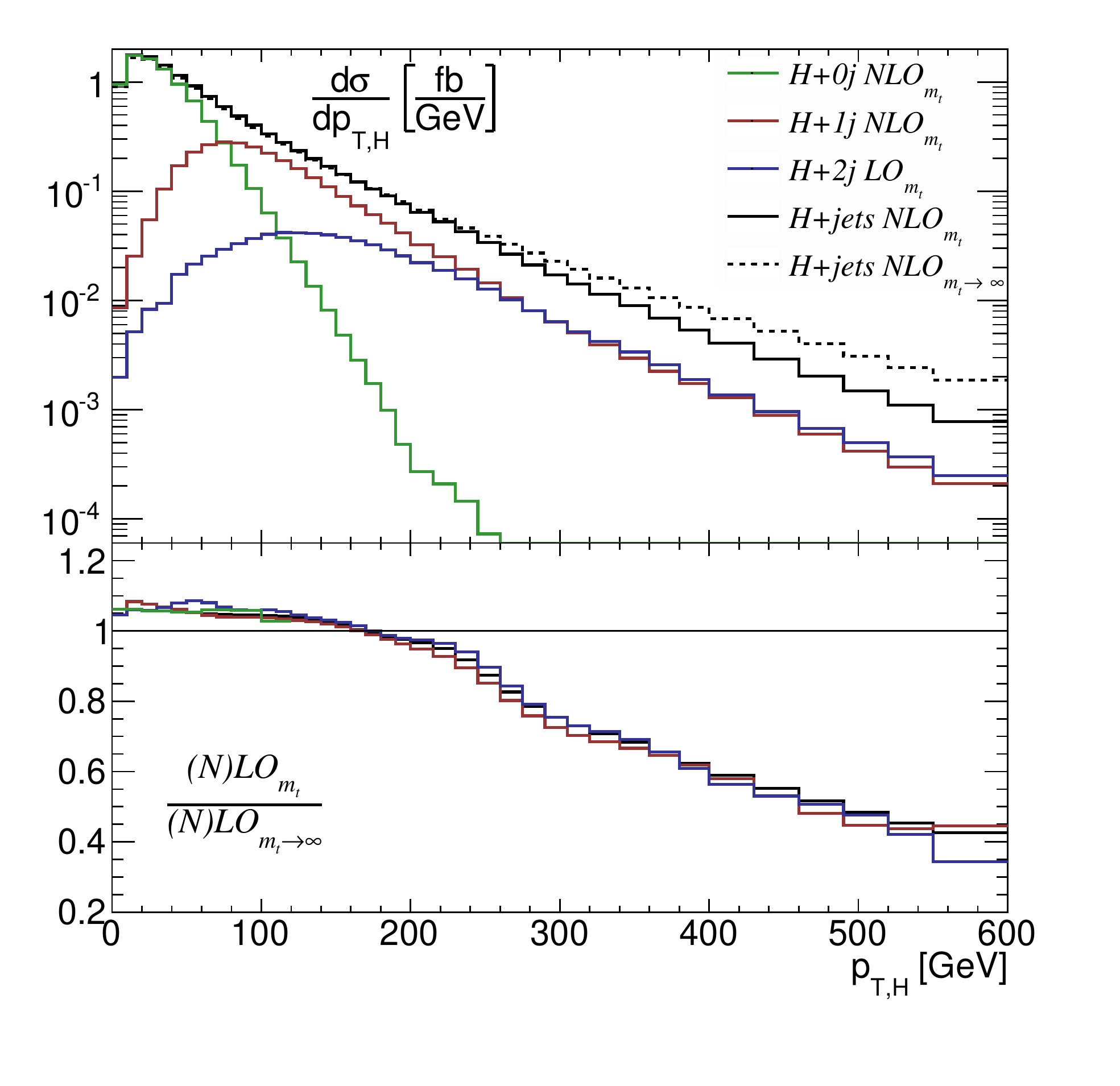}}
\caption{(left) Invariant $\Mfourl$ distributions for the process
  $\qqbar (\Pg\Pg)\rightarrow \PZ\PZ$ at $13\UTeV$, obtained with MCFM~\cite{Campbell:2016jau} (right) $\pTH$
  including jets recoiling against the on-shell Higgs. We merge 0-jet and 1-jet production to NLO with
  the full top mass dependence, 2-jet production to LO with the full top mass dependence, and parton
  shower effects. Figure from Ref.~\cite{Buschmann:2014sia}. }
\label{fig:fxs-bsm-m4l}
\end{figure}

\subsection{Boosted Higgs boson production in gluon fusion}

An alternative way to search for BSM effects in gluon-fusion Higgs boson production 
is to require a large boost of the Higgs to generate a
large momentum flow through its production vertex. The leading
partonic signal process is $\Pg\Pg \to \PH\Pg$, where the $2 \to 2$
kinematics defines the momentum flow through the Higgs vertex for
example in terms of $\pTH$ or, equivalently, $\pTj$. When a second jet
is considered, the $\pTH$ spectrum provides the information about this
production mode. Following the Lagrangian given in
\eqn{eq:fxs-bsm-lagrangian} we can again compute the leading
${\cal O}(\alpha_s)$ dependence on the ratio
$\pTH/\Mt$~\cite{Buschmann:2014twa,Buschmann:2014sia},
\begin{alignat}{5}
|\mathcal{M}_{\PH j(j)}|^2 
\propto 
\Mt^4 \; \log^4 \frac{(\pTH)^2}{\Mt^2} \; .
\label{eq:fxs-bsm-pt_log}
\end{alignat}
The effect of the different partonic sub-processes is shown in the
right panel of \refF{fig:fxs-bsm-m4l}. In this observable, sizeable top mass effects already appear for $\pTH > 250\UGeV$. In the tail of the distributions, where for $\pTH > 500\UGeV$ the SM expectations have dropped below a per mille of all events, the corrections from the new dimension-6 Higgs-gluon operator dominate the distribution. The sensitivity of off-shell Higgs boson production and boosted Higgs boson production in terms of the modified Lagrangian of \eqn{eq:fxs-bsm-lagrangian} can be compared. Because both methods rely on a small number of events in the tail of the kinematic distribution, large luminosities will be needed to detect sizeable deviations from the SM.

Going beyond the description in terms of dimension-6 operators, benchmarks for searches in this channel are coloured partners to the top, which participate in gluon fusion but whose contribution cancels at leading order. In this case, the $\PH+j$ channel provides the best handle on BSM. Examples of this situation-- cancellation of effects at LO Higgs boson production-- are fermionic top-partners in Composite Higgs models~\cite{Banfi:2013yoa,Azatov:2013xha} and stops in the so-called {\it funnel region}~\cite{Espinosa:2012in,Grojean:2013nya}. 

In the right plot of \refF{fig:fxs-bsm-VH_Hj} we show the $\pTj$
distribution in the SM case (blue), and the effect of introducing
top-partners of different masses ($M_T$) and mixing angles
($\sin \theta$). This distribution has been obtained using a modified
version of MCFM~\cite{Campbell:2016jau} including the effect of top partners, see
Ref.~\cite{Banfi:2013yoa} for details. Note that the lowest order
effect in gluon fusion of these top-partners is exactly cancelled by
the top, due to a low-energy
theorem~\cite{Ellis:1975ap,Shifman:1979eb,Kniehl:1995tn}, hence
information on these new particles is delegated to the $\PH$+jets
channel, or parametrically small finite-mass effects.

\subsection{VH associated production}

In this channel, the Higgs recoils against a vector boson, and thus it
has an inherent boost. This boost enhances the momentum-dependent
effects of New Physics with respect to the SM production, where the
$\PV\PV\PH$ interaction has no momentum dependence. Because the signal
process has a simple $2 \to 2$ kinematic structure, the description of
these effects in terms of $\pTH$ or $\pTV$, or $\MVH$ is, from a
theory perspective, essentially equivalent. The relevant question is
for which observable the experimental boundary conditions allow for
the best coverage of phase space.

Generally speaking, the higher the momentum of the Higgs and vector
boson one has access to, the higher the sensitivity to BSM
effects~\cite{Ellis:2012xd,Ellis:2013ywa}. In the left plot of
\refF{fig:fxs-bsm-VH_Hj} we show this dependence in the context of an
EFT approach to BSM performed at NLO QCD accuracy in {\sc POWHEG} and
showered through {\sc PYTHIA}v8~\cite{Mimasu:2015nqa}.  The selection
criteria applied to produce this plot are jets using $k_T$ algorithm
of $\Delta R=0.4$, $\pT > 25\UGeV$ and $|\eta_j| < 2.5$, 2 $\PQb$-jets
with $\pT > 25\UGeV$ and $|\eta_{\PQb}| < 2.5$, and 1 lepton
($\Pl=\Pe$ or $\PGm$) with $\pT > 25\UGeV$, and $|\eta_{\Pl}| <
2.5$.
Needless to say, in this channel there is a strong correlation between
the distributions in $\pTH$ (or $\pTV$) and the invariant mass of the
system, $\MVH$ for resolved final states or $m_{T}$ in the channels
with missing energy, see Figure~10 in~\cite{Ellis:2014dva} and Figure~3
in~\cite{Biekotter:2016ecg}.

\begin{figure}
   \hspace*{-5mm}
   \includegraphics[width=.54\textwidth]{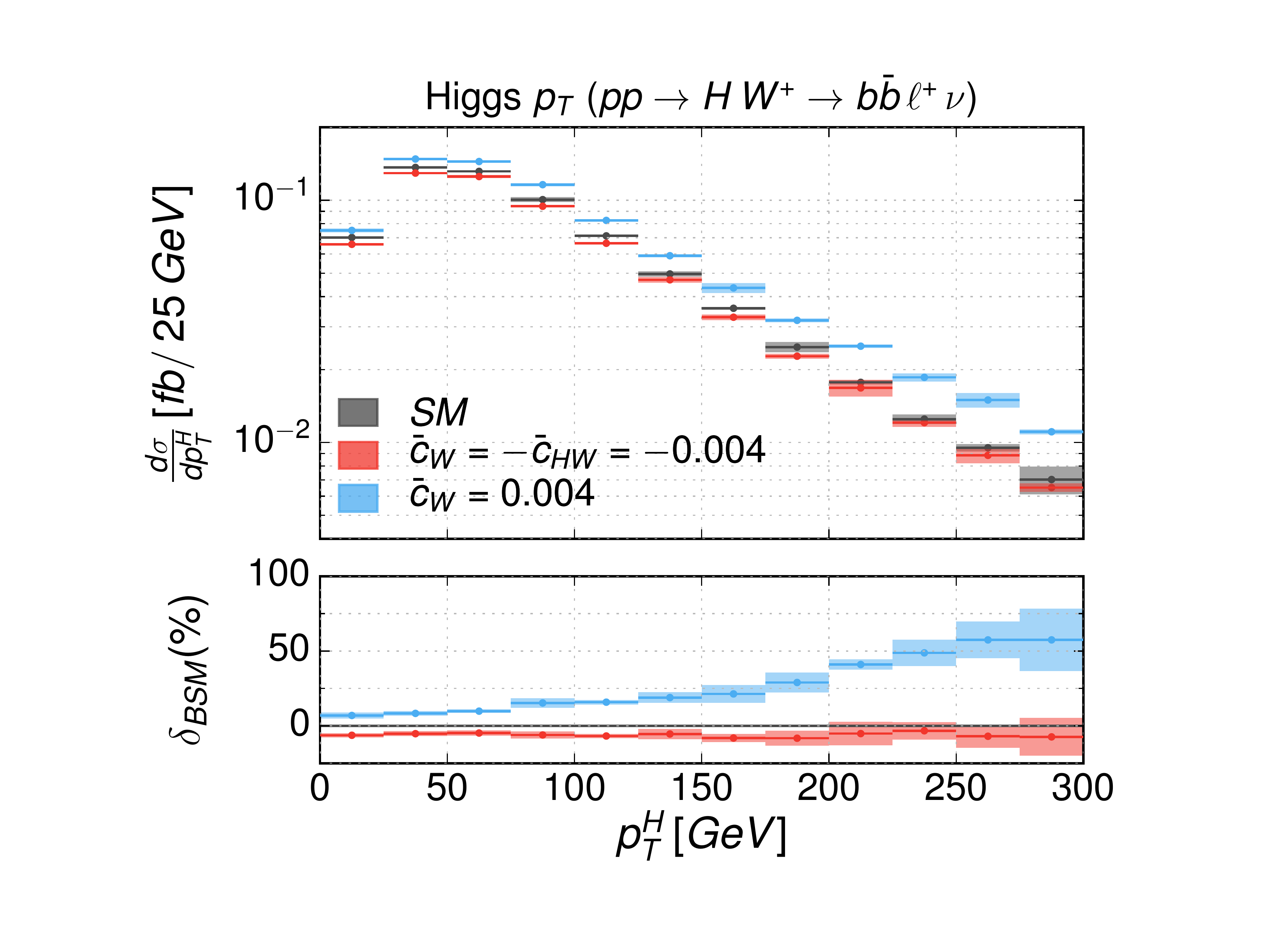}\hspace*{-4mm}%
   \includegraphics[width=.5\textwidth]{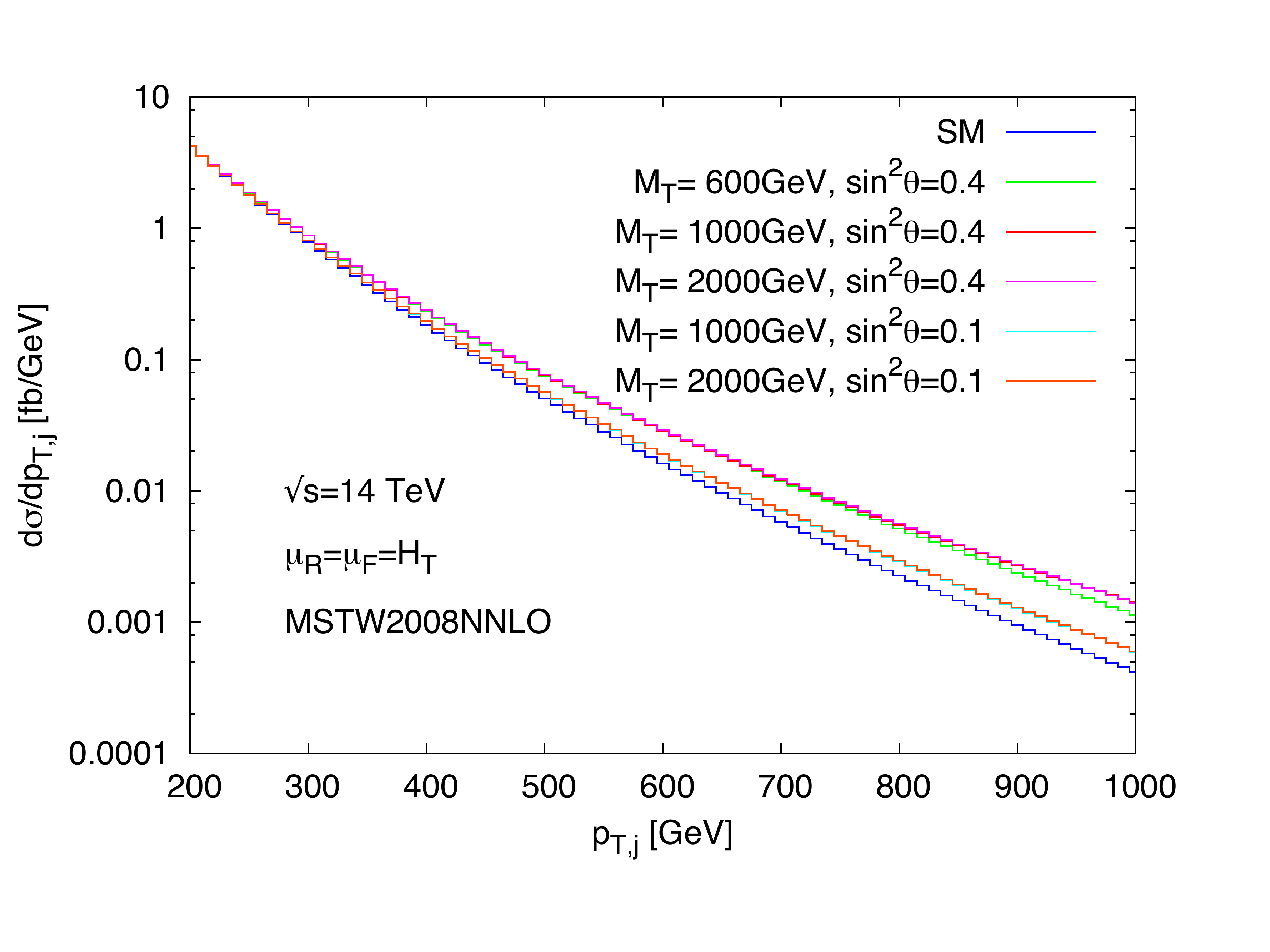}	
   \caption{ (left) Comparison of differential distribution of
       the Higgs $\pT$ in the SM and the two EFT benchmarks of $\bar
       c_W=0.004$ and $\bar c_W=-\bar c_{HW}=-0.004$ using {\sc
         Powheg} + {\sc Pythia8} in the process $\Pp\Pp\to \PWp \PH \to
       \Plp \PGn \bbbar$. The lower panel show the percentage
       deviation of the EFT benchmarks from the SM prediction,
       $\delta_{\mathrm{BSM}}$. Figure taken from
       Ref.~\cite{Mimasu:2015nqa}. (right) Jet $\pT$ differential
       distribution in the channel $\PH+j$, with the SM case in blue,
       and the effect of introducing top-partners of different masses
       ($M_T$) and mixing angles ($\sin \theta$) shown in other
       colours. Figure from Ref.~\cite{Banfi:2013yoa}. }
 \label{fig:fxs-bsm-VH_Hj}
 \end{figure}

In Run~1, the best limit on BSM phenomena using this channel  was obtained by looking at the last reported bin, the overflow bin, which typically contains a low number of events. Despite this, one can use the estimate of SM background to set limits on new physics~\cite{Ellis:2014dva,Corbett:2015ksa}. This procedure is analogous to searches for anomalous trilinear gauge couplings in $\PW\PW$ by looking at the overflow bin in the leading lepton $p_T$ distribution performed at LEP and now at the LHC~\cite{Chatrchyan:2013yaa,Chatrchyan:2013oev,Aad:2016wpd}. The combination of both sets of measurements would in fact be useful to enhance the sensitivity to anomalous trilinear couplings. 

The sensitivity to BSM obtained via the last bin raises questions on the validity of the EFT approach at high-momentum transfer, as in this region one {\it could} be able to resolve new physics effects. This question is discussed in Sections~\ref{s.eftval}~and~\ref{s.eftmodels}.
Tools incorporating BSM (incl.\ EFT) effects at higher order in precision are discussed in Section~\ref{s.efttools}.

\subsection{Vector boson fusion}

In several ways, vector boson fusion is the most prolific of the usual
Higgs boson production channels when it comes to measuring the properties of
the Higgs boson. The first reason is that its $2 \to 3$ kinematics
allows us to test a sizeable number of observables, including pure
tagging jet correlations; second, we can test modifications of the
gauge sector as well as modifications of the scalar or Higgs sector as
long as they affect the central $\PV\PV\PH$ coupling; third, it allows
us to separate the very specific Lorentz structure of the $\PV\PV\PH$
coupling in the SM from many modified structures induced by BSM
physics; and finally, as an electroweak process with further
suppressed QCD corrections we expect theoretical uncertainties to be
under better control than in the gluon-fusion process.

A prime example for a general test of the SM nature of the Higgs boson
is the direct test of the Lorentz structure of the $\PV\PV\PH$
coupling which can be performed via the measurement of the azimuthal
angle between the tagging jets~\cite{Plehn:2001nj}. While the SM
predicts a rather flat distributions, the typical CP-even and CP-odd
structures could be identified with essentially zero events at 90
degrees or for back-to-back configurations, respectively. Similarly,
the transverse momentum spectra of the tagging jets in the SM show a
strong peak below $\pTj \sim \MW/2$. Generally, the longitudinal or
transverse polarization of the gauge bosons affects this
spectrum~\cite{Brehmer:2014pka}, as originally computed in the
effective $\PW$ approximation~\cite{Dawson:1984gx}. Finally, the
rapidity difference of the two tagging jets is particularly large for
SM Higgs boson production, and the tagging jets should become significantly
more central if we change the $\PV\PV\PH$ coupling in any
way~\cite{Hagiwara:2009wt,Englert:2012xt,Djouadi:2013yb}.  The
important aspect of all these measurements (illustrated in
\refF{fig:fxs-bsm-jj}) is that they do not require a reconstructed
Higgs 4-momentum, provided that global cuts ensure that we are working
with a Higgs-rich event sample.  This implies that it is secondary how
we modify the $VVH$ coupling, \textsl{i.e.}  through a modification of
the Higgs quantum numbers or through higher-dimensional operators for
a SM-like Higgs.  On the other hand, adding information from the Higgs boson decay 
will of course enhance the power of these measurements, for
example re-formulating all questions originally asked in the framework
of $\PW\PW \to \PW\PW$ scattering at high
energies~\cite{Han:2009em,Kilian:2014zja,Brehmer:2014pka}.

\begin{figure}
 \includegraphics[width=0.24\textwidth]{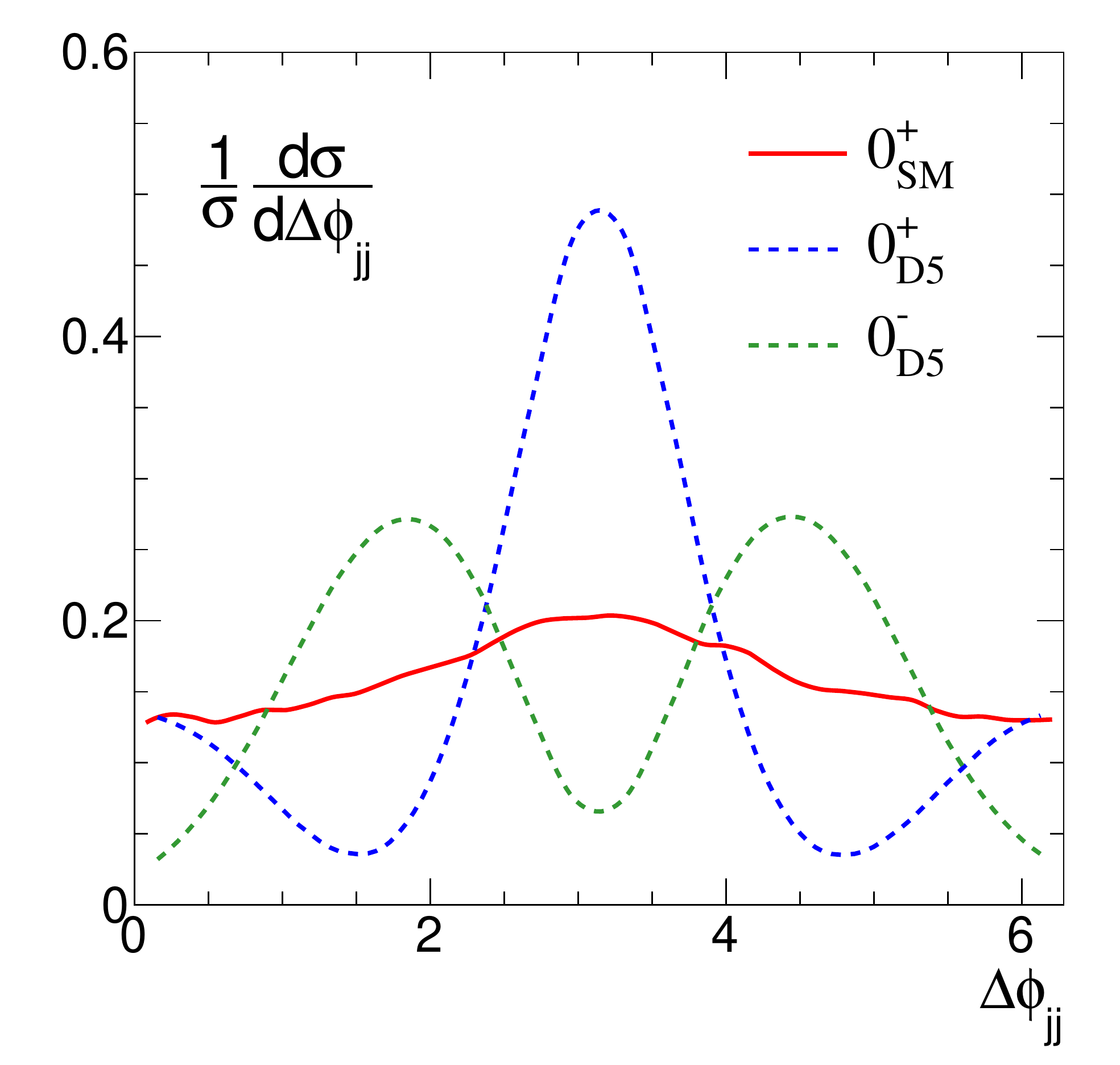}
 \hfill
 \includegraphics[width=0.24\textwidth]{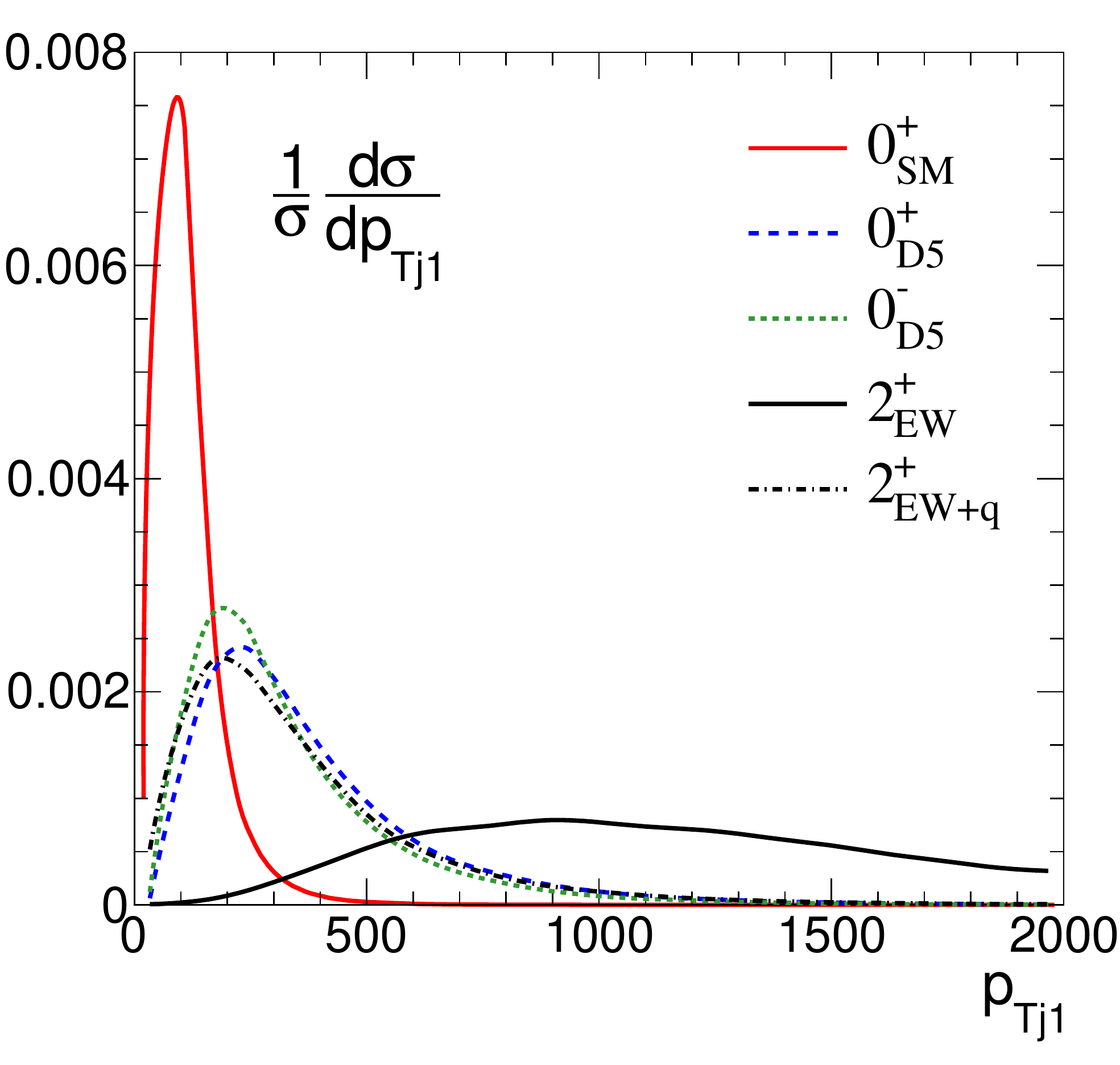}
 \hfill
 \includegraphics[width=0.24\textwidth]{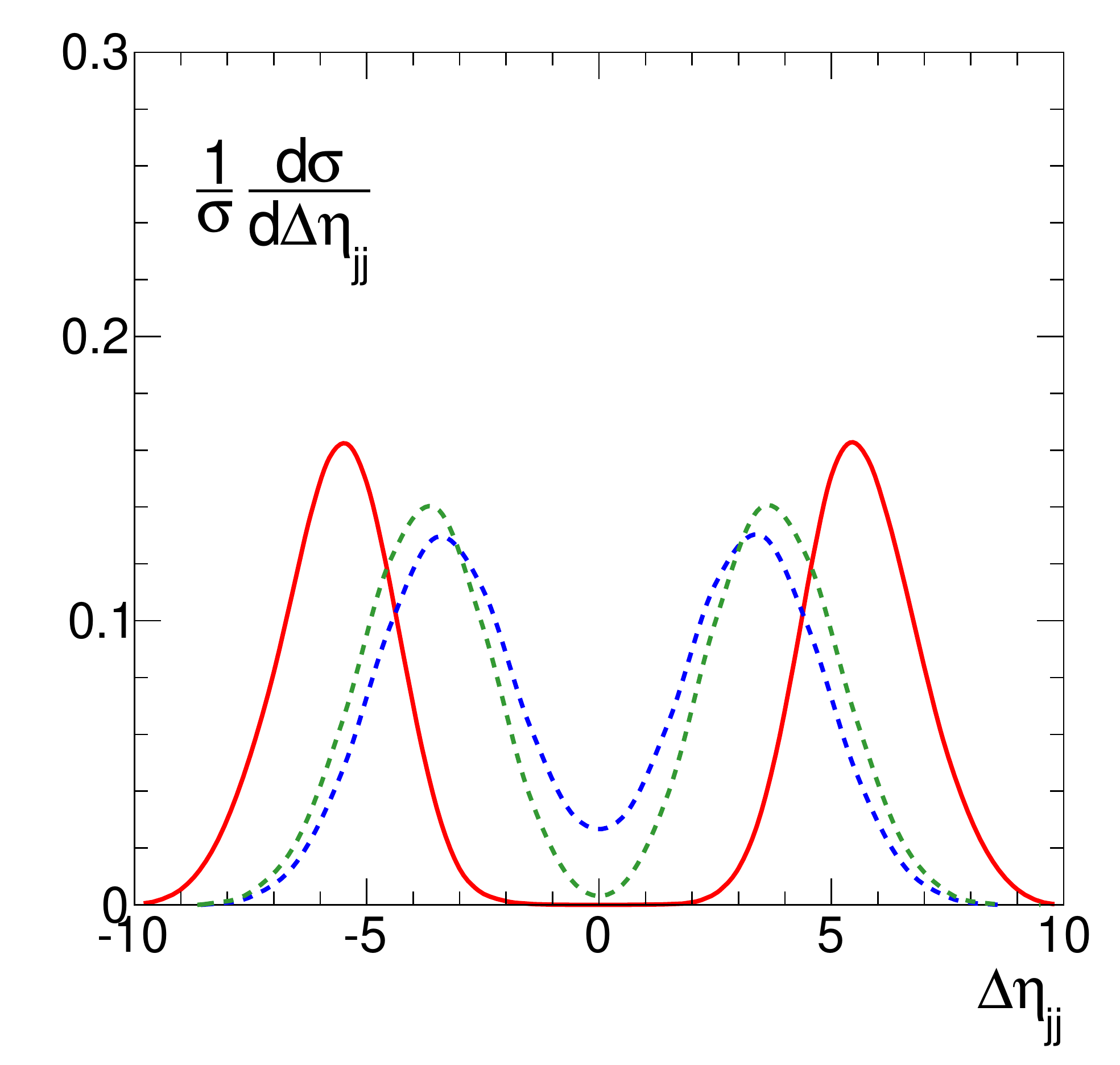}
 \hfill
 \includegraphics[width=0.24\textwidth]{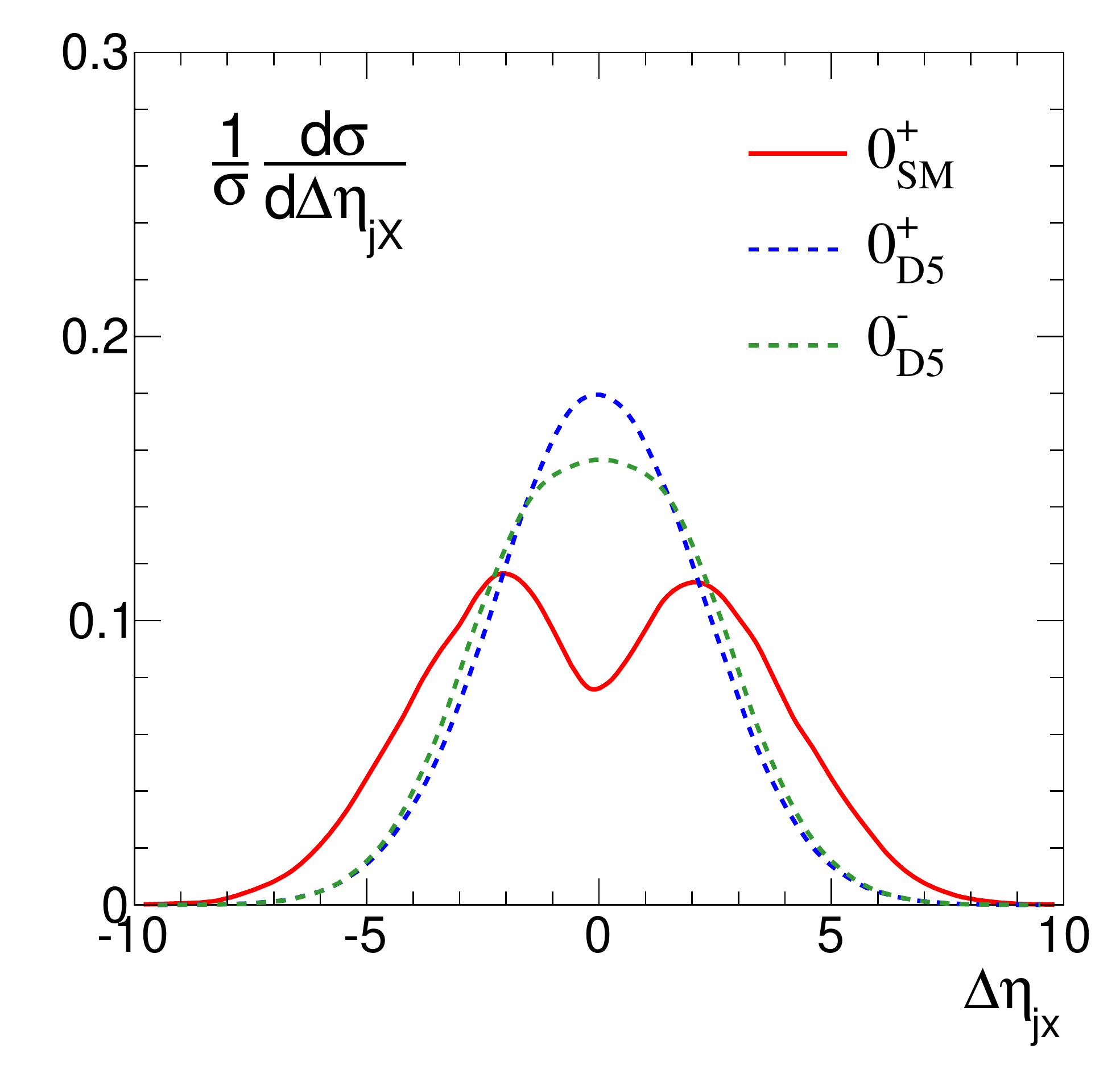}
 \caption{Normalized correlations between the two tagging jets in VBF
   production of a SM-like Higgs and a CP-even or CP-odd scalar
   coupled through a higher-dimensional operator. We show the
   difference in the azimuthal angle $\Delta \phi_{jj}$, the tagging
   jet $p_T$, the rapidity difference between the tagging jets $\Delta
   \eta_{jj}$, and the rapidity difference between the tagging jet and
   the Higgs-like resonance $X$. Figure from
   Ref.~\cite{Englert:2012xt}.}
\label{fig:fxs-bsm-jj}
\end{figure}

In the spirit of the discussion of gluon fusion Higgs boson production and
VH associated production we focus on experimental tests enhanced by
the momentum flow through the Higgs boson production vertex.  We can link
the transverse momenta of the two tagging jets or the Higgs to the
virtuality of the weak bosons. Obviously, all of them are strongly
correlated, so it becomes a theoretical as well as experimental
question which of these observables to include in an
analysis~\cite{Biekotter:2016ecg}. The only key requirement is that we
do not bias these distributions for example by cutting on the tagging
jet correlations discussed before. In \refF{fig:fxs-bsm-VBF_monoH} we
present the $p_T$ of the Higgs, as a function of CP-violating
operators in an EFT approach to BSM. 
Basic cuts applied to these events are $m_{jj} > 400\UGeV$, $\Delta \eta_{jj}>$ 2.8 and $|\eta_j|< 4.5$.

\subsection{Invisible Higgs boson decays}

Once it is possible to experimentally target a specific Higgs
production mechanism in terms of fiducial cross sections, we can focus
on specific Higgs boson decay modes in this production mechanism. Arguably
the hardest Higgs boson decay mode to search for the LHC are invisible Higgs boson decays. 
The SM predicts a very small invisible Higgs boson branching ratio
through $\PH \to \PZ \PZ^* \to 4\PGn$, but for example in Higgs portal
models~\cite{Djouadi:2011aa} or in supersymmetry~\cite{Butter:2015fqa}
this decay can be an observable effect of weakly interacting dark
matter.  To date, searches for invisible Higgs boson decays in weak boson
fusion, \textit{i.e.} two tagging jets combined with missing
transverse momentum~\cite{Eboli:2000ze}, appear to be the most
promising strategy.  The key feature of this signature are two tagging
jets with exactly the same kinematics as in other VBF Higgs boson production
channels. This means that fiducial volumes are related to other VBF
studies by replacing the central Higgs boson decay products by missing
transverse momentum.

Current projections for different LHC luminosities are shown in
\refT{tab:fxs-bsm-reach}. The main background is $\PZ$+jets
production, with an invisible $\PZ$ decay. Two production mechanisms
contribute to the background, one at the order $\alpha_s^2 \alpha$ and
one at the order $\alpha_s \alpha^2$. The QCD-like channel can be
strongly reduced by a central jet veto, while weak boson fusion
$\PZ$-production is essentially irreducible, with some kinematic
differences for example in the azimuthal correlation of the tagging
jets reflecting the Lorentz structures of the $\PZ$ and $\PH$ production
vertices. For a success of this channel it is crucial to understand
the central jet activity, so the right columns of
\refT{tab:fxs-bsm-reach} should be considered a challenge to the
experimental performance of jet and particle-flow-like algorithms.  In
general, weak boson fusion signatures might also be extracted just
based on one tagging jet~\cite{Mellado:2004tj}, a channel which has
not (yet) been studied for invisible Higgs boson decays.

\begin{table}
\caption{Exclusion reach in $\textrm{BR}_{\textrm{inv}} = \Gamma_{\textrm{inv}}/\Gamma_H$ at 95\% CLs
  to an invisible Higgs boson width at various luminosities and different
  combinations of cuts and multivariate analyses. Here, $\Gamma_H$ is defined to be 
  the width of the Higgs boson in the SM without the additional invisible
  component due to new physics. Table from Ref.~\cite{Bernaciak:2014pna}.}
\label{tab:fxs-bsm-reach}
\centering
\begin{tabular}{@{\extracolsep{\fill}}r|cccc|cc} 
\toprule
& \multicolumn{4}{c|}{$\pTj > 20\UGeV$} & \multicolumn{2}{c}{$\pTj > 10\UGeV$} \\ 
$\mathcal{L} [\text{fb}^{-1}]$   & VBF cuts & + jet veto  &  + $\Delta\phi_{jj}$ & BDT 2-jets & BDT 2-jets  & + BDT 3-jets \\ 
\midrule
10           & 1.02                & 0.49                       & 0.47                  & 0.28  & 0.18  & 0.16 \\
100          & 0.49                & 0.20                       & 0.18                  & 0.10  & 0.07  & 0.061 \\
3000         & 0.25                & 0.094                      & 0.069                 & 0.035 & 0.025 & 0.021 \\
 \bottomrule
\end{tabular}
\end{table}

\subsection{Mono-Higgs signatures}

An alternative way to probe the connection between the Higgs and Dark
Matter are channels where the Higgs recoils against missing energy,
\textit{i.e.} mono-Higgs signatures. Again, fiducial measurements are
closely linked to a SM signatures, $\Pp\Pp \to \PGn\PAGn\PH$ 
arising in the VH topology. Unlike for invisible Higgs boson decays, the
kinematical structure of BSM mono-Higgs events will not resemble that
in VH production. Instead, we expect significant deviations for
example in the distributions of the reconstructed Higgs. Nevertheless,
we would be able to utilize fiducial volumes defined for associated VH
production with only a transverse reconstruction of the gauge boson.

Studies at Run~1 on the mono-Higgs signature have been done in the context of the Higgs portal~\cite{Aad:2015yga}, and other extensions of the SM are now being considered. Particularly interesting distributions  are the transverse mass of the system or the $\pT$ distribution of the Higgs. In \refF{fig:fxs-bsm-VBF_monoH}, we show the Higgs $\pT$ distribution in events where the Higgs is produced in association with a pair of Dark Matter particles of mass 500 GeV at LHC13. The labels correspond to different assumptions of the coupling of Dark Matter to the Higgs sector with the blacks line the benchmark of standard Higgs portal coupling $\lambda_{hs} h^2 S^2$. Note that to produce this distribution, no cuts at generation level have been applied. 

\begin{figure}
   \includegraphics[width=.48\textwidth]{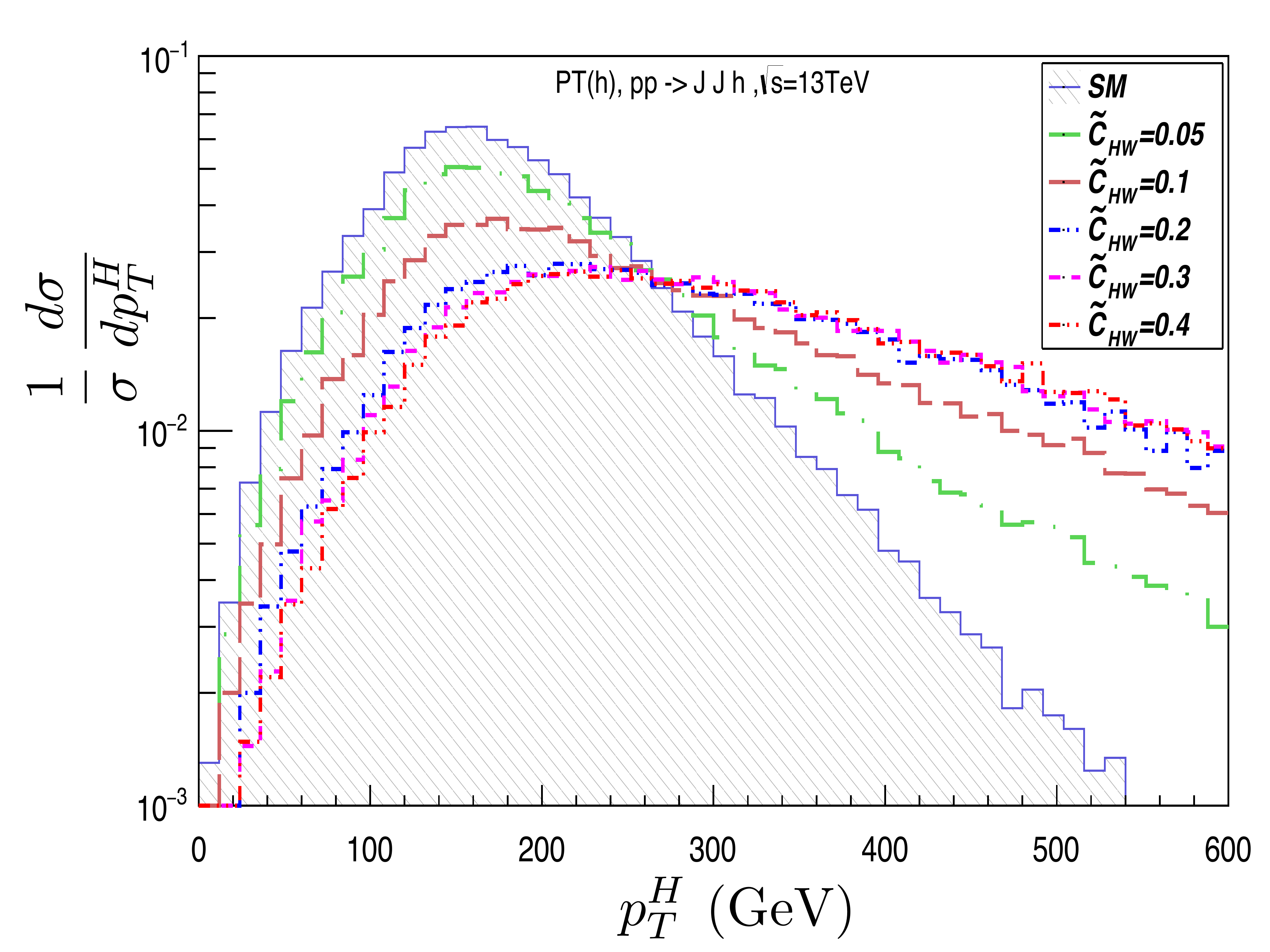} \quad %
   \raisebox{3mm}{\includegraphics[width=.47\textwidth]{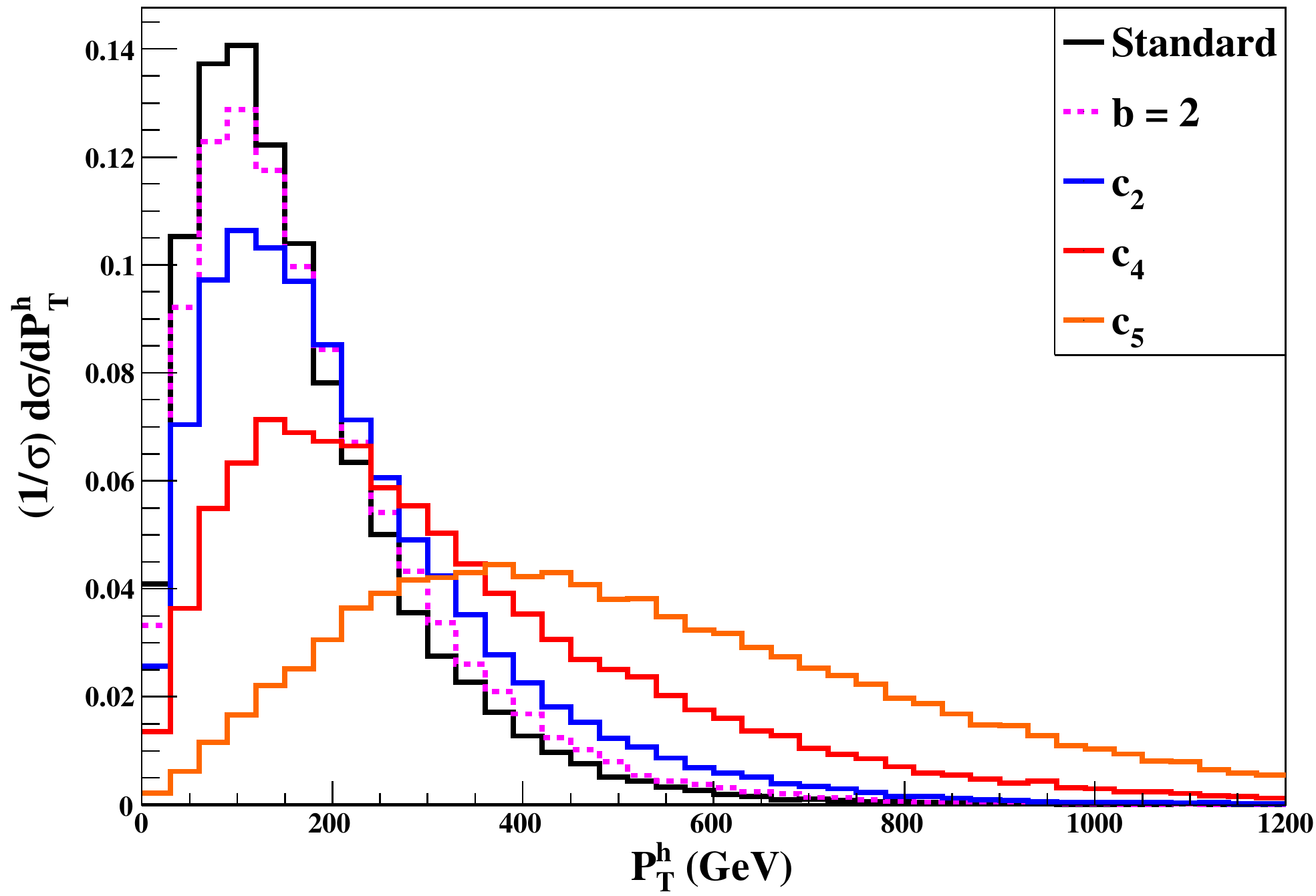}}	
   \vspace*{1mm}
     \caption{Distributions of the Higgs $\pT$ in two channels. (left) Vector boson fusion, with the SM case represented by the solid distribution, and the SM plus additional modifications due to different values for the possible CP-violating operator $\tilde c_{HW}$~\cite{Alloul:2013naa} are given by the coloured lines. (right) The production of the Higgs boson in association with a pair of Dark Matter scalar particles of mass 500 GeV~\cite{Brivio:2015kia} at LHC13. The black line corresponds to the standard portal coupling $h^2 S^2$, and the other lines represent different contributions due to a non-linear nature of electroweak symmetry breaking. 
     }
 \label{fig:fxs-bsm-VBF_monoH}
 \end{figure}

\section{Experimental aspects}\label{sec:EXP}

In this section a brief overview of the experimental aspects important for the fiducial cross section measurements is given. It includes the criteria for the particle level fiducial volume definition, a brief review of the unfolding procedures, estimation of the remaining model dependence, treatment of the Higgs boson mass parameter, aspects related to the statistical combination of the fiducial measurements between different processes and/or different experiments, and also recommends the points that should be carefully studied when designing the fiducial measurements. It primarily focuses on the measurement that are deemed feasible in the short and medium term of the LHC running.

\subsection{Definition of the fiducial phase space}
\label{sec:exp:fid}

The acceptance and selection efficiency for the particular Higgs boson decay channel can vary significantly between different Higgs boson production mechanisms and different exotic models of Higgs boson properties. In processes with large jet activity such as the $\ttH$ production or those with the kinematics of the decay products very different from the SM prediction (such as in case of the exotic Higgs-like spin-one models), the acceptance of signal events within a certain part of the phase space can significantly differ from the acceptance for the SM Higgs boson decays. In order to minimize the dependence of the measurement on the specific theoretical model assumption, the fiducial phase space for the Higgs boson cross section measurements should be defined to match as closely as possible the experimental acceptance in terms of the kinematics of the decay products and topological reconstruction-level event selection.

The fiducial phase space is typically defined using the stable particles or more complex objects built out of them (leptons, photons, jets, missing transverse momentum, etc.) at the hard scattering level, before their interaction with the detector material. 
In order to minimize the model dependence, fiducial-level particles and objects are typically defined to be as close as possible to the particles and objects used at the reconstruction level. 
In case of the leptons, it is typical that fiducial-level leptons are defined as the leptons ``dressed'' with the photons from the final state radiation (the photons that are within certain distance $\Delta R$ from the lepton), as at the reconstruction level those photons are typically recovered by the experimental methods.
In case of differential measurements as a function of jet-related observables, it is recommended that jets are reconstructed from the individual stable particles, excluding neutrinos, using the anti-$k_t$ clustering algorithm with a distance parameter identical to the one used at the reconstruction level.

\subsubsection{Isolation requirement in the definition of the fiducial volume\SectionAuthor{S.~Menary, A.~Pilkington}}

The inclusion of isolation of photons and leptons can be important in the fiducial phase space definition whenever object isolation is used at the reconstruction level, as it can reduce the differences in signal selection efficiency between different models. It has been verified in simulation that this difference can be significant if the lepton isolation requirement is included at the reconstruction but not at the fiducial level~\cite{Aad:2014tca,Khachatryan:2015yvw}. This can be especially pronounced in case of large associated jet activity such as in the $\ttH$ production mode. Exclusion of neutrinos from the computation of the isolation sum typically brings the definition of the fiducial phase space closer to the reconstruction level selection, and can additionally improve the model independence of the signal selection efficiency. It is recommended that these effects are studied in each particular analysis separately.

The experimental analyses measuring fiducial cross sections in the $H \to \gamma\gamma$ channel typically require two isolated photons with a $p_T$ above a certain threshold. In order to minimize the extrapolation when correcting experimental yields to particle level fiducial cross sections, it is useful to impose a similar criterion which duplicates a similar requirement using stable particles. Imposing such drastically reduces the underlying model dependence: this can be readily understood if one compares for example Higgs boson production with gluon fusion versus in association with a top quark pair: $H \to \gamma\gamma$ photons from the latter fail more often the isolation criterion due to the large hadronic activity and thus have a lower reconstruction efficiency. Figure~\ref{fig:exp:isolation} shows the correction factors mapping reconstructed yields into fiducial cross sections with and without imposing a similar particle level isolation cut. A particle level isolation criterion can be imposed by summing around a fixed cone the energies of all stable particles and events are similarly rejected when the isolation energy is larger than a certain threshold. The exact cut can be tuned such that the model dependence becomes minimal, i.e. that the efficiency difference between rejecting a reconstructed event and a true event is very similar: In Figure~\ref{fig:exp:isolation} the correlation between true and reco isolation is shown, and an illustrative reconstruction cut is mapped to a given true value using a profile of both observables. The effect of imposing this criterion is illustrated as well, resulting in near matching correction factors for all Higgs boson production processes. 

\begin{figure}
\includegraphics[width=0.55\textwidth]{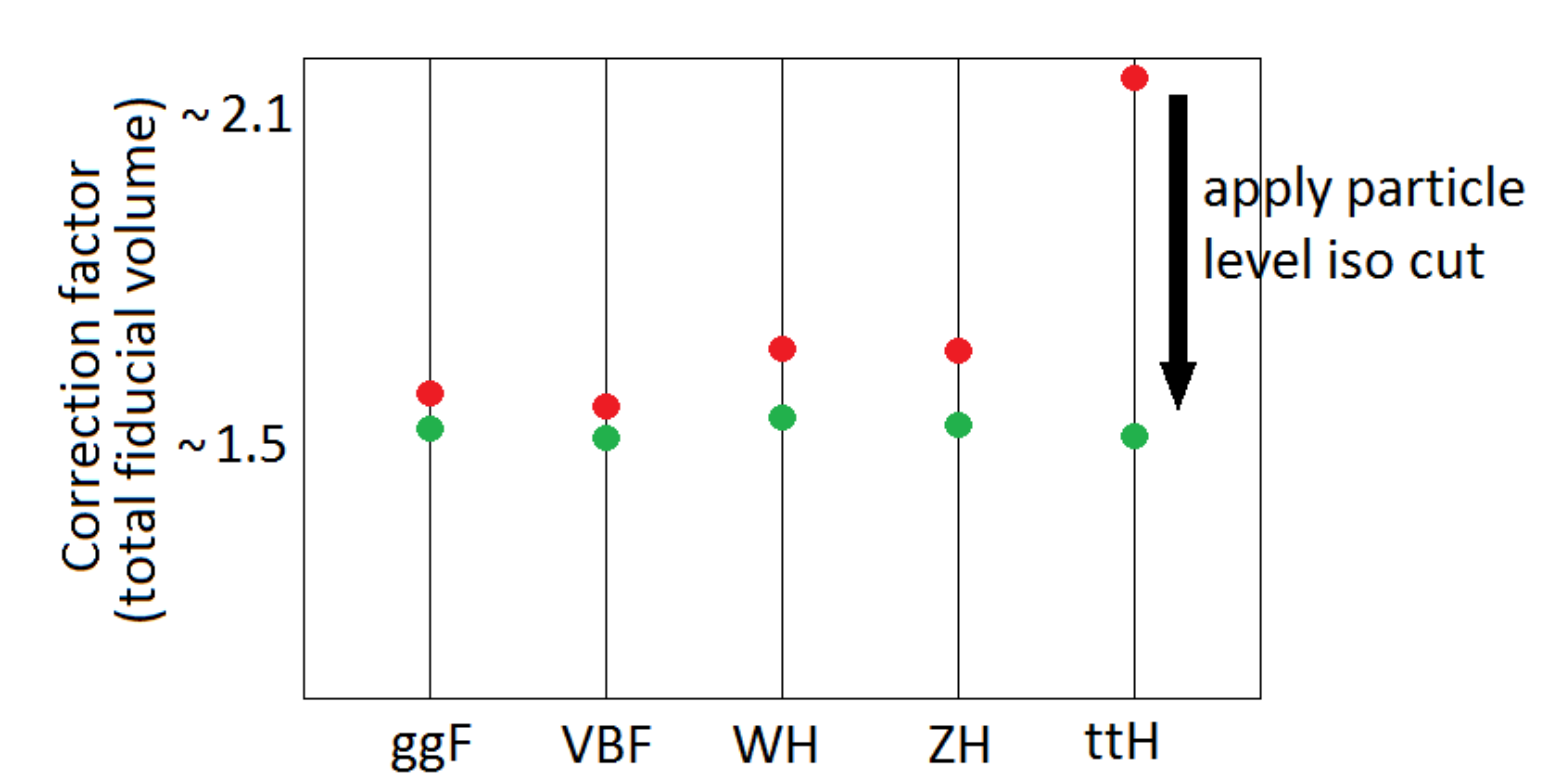}
\includegraphics[width=0.45\textwidth]{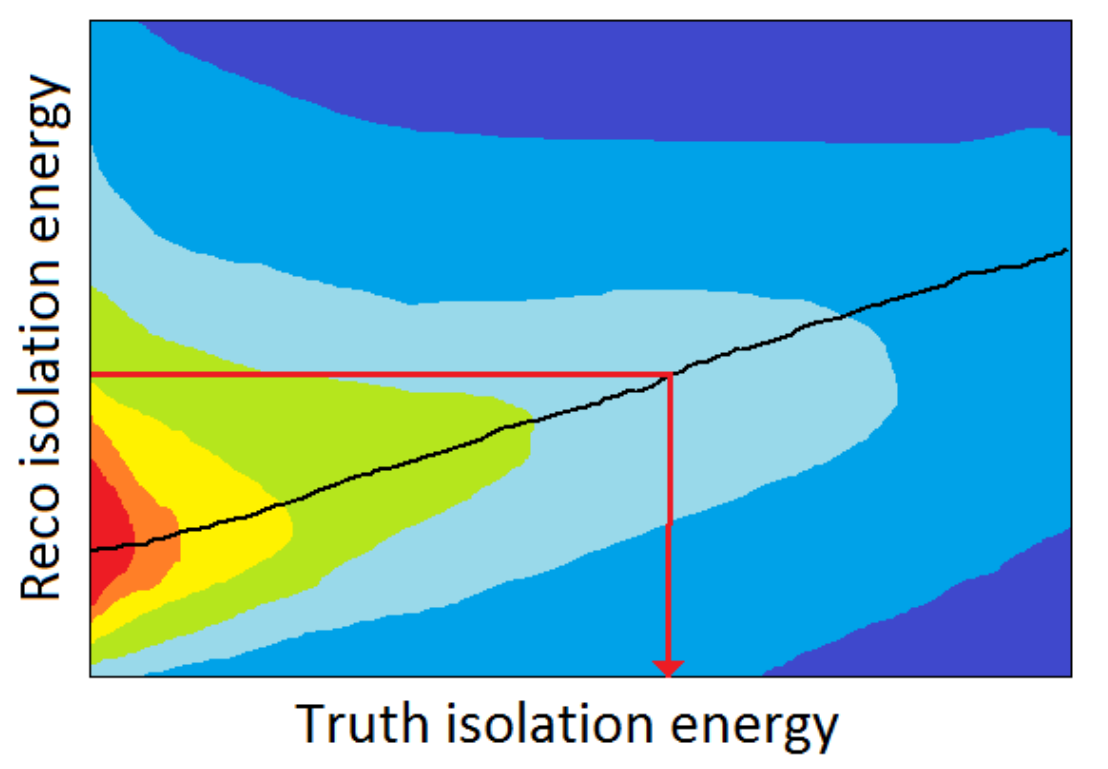}
\caption{
(left) Illustration for correction factor which maps reconstructed yields to fiducial cross sections without (red) and with (green) imposing a particle level isolation criterion are shown. Imposing particle level isolation significantly reduces the differences between different Higgs boson production modes which minimizes the model dependence. (right) The procedure to map a reconstructed isolation criterion to a particle level isolation criterion using profiles is illustrated.
}
\label{fig:exp:isolation}
\end{figure}

\subsubsection{Signal contributions from outside of the fiducial phase space}

At the reconstruction level, additional signal contribution from events that do not originate from the fiducial phase space can arise due to detector resolution effects that cause differences between the quantities used for the fiducial phase space definition (such as the lepton or photon isolation, jet transverse momentum, missing transverse momentum etc.) and the analogous quantities used for the event selection. This contribution should be treated as background and subtracted before the unfolding procedure is applied. Hereafter we refer to this contribution as the ``nonfiducial signal'' contribution. It has been shown in simulation that the shape of these events is typically very similar to the shape of the fiducial signal.
In order to minimize the model dependence of the measurement - it should be studied how to optimize fiducial phase space definition to minimize the effect that arises from nonfiducial signal' contribution, and how to experimentally treat this contribution in the measurement. Studies in simulation have shown that this component can vary from just few per cent e.g. for the $\ggH$ production mode to several per cent for the $\ttH$ production mode~\cite{Aad:2014tca,Khachatryan:2015yvw}. The variation of this fraction between different signal models can be included in the model dependence estimation.

The nonfiducial signal contribution deserves special attention when the observables used to define the signal region have poor experimental resolution (such as missing transverse energy, transverse momentum of jets, etc.). In those cases effects of migration of the signal events can be large, and it might be worth studying if the measurement can benefit (in terms of the overall model dependence) from relaxing the requirements on such observables at the fiducial level with respect to the reconstruction level. These effects have been discussed in the light of the fiducial measurements of the Higgs boson transverse momentum in the $H \to WW$ decay channel~\cite{Khachatryan:2016vnn}.

The fraction of signal events within the fiducial phase space $\mathcal{A}_{\rm fid}$, the reconstruction efficiency $\epsilon$ for signal events within the fiducial phase space for individual SM production modes and exotic signal models, as well as the fraction of signal events outside of the fiducial phase space $f_{\rm nonfid}$ are listed in Table~\ref{tab:Accept_Eff_fOut}. Values are given for characteristic signal models assuming $m_{\rm H} = 125.0 \UGeV$, $\sqrt{s}=8\,\UTeV$, and the overall picture is similar in case of the pp collision at $\sqrt{s}=13\,\UTeV$.

\begin{table}[!h!tb]
\begin{center}
\small
\caption{
The fraction of signal events within the fiducial phase space (acceptance $\mathcal{A}_{\rm fid}$), reconstruction efficiency ($\epsilon$) for signal events from within the fiducial phase space, and ratio of reconstructed events which are from outside the fiducial phase space to reconstructed events which are from within the fiducial phase space ($f_{\rm nonfid}$). 
Values are given for characteristic signal models assuming $m_{\rm H} = 125.0\,\UGeV$, $\sqrt{s}=8\,\UTeV$, and the uncertainties include only the statistical uncertainties due to the finite number of events in MC simulation.
\label{tab:Accept_Eff_fOut}
}
\begin{tabular}{l|cccc} 
\toprule
\textbf{Signal process} & $\mathcal{A}_{\rm fid}$ & $\epsilon$ & $f_{\rm nonfid}$  & $(1+f_{\rm nonfid})\epsilon$ \\ 
\midrule
\multicolumn{5}{c}{Individual Higgs boson production modes} \\
\midrule
$\ggH$ ({\sc Powheg+JHUGen})   & 0.422 $\pm$ 0.001 & 0.647 $\pm$ 0.002 & 0.053 $\pm$ 0.001  & 0.681 $\pm$ 0.002 \\ 
VBF ({\sc Powheg})                      & 0.476 $\pm$ 0.003 & 0.652 $\pm$ 0.005 & 0.040 $\pm$ 0.002  & 0.678 $\pm$ 0.005 \\ 
WH ({\sc Pythia})                          & 0.342 $\pm$ 0.002 & 0.627 $\pm$ 0.003 & 0.072 $\pm$ 0.002  & 0.672 $\pm$ 0.003 \\ 
ZH ({\sc Pythia})                          & 0.348 $\pm$ 0.003 & 0.634 $\pm$ 0.004 & 0.072 $\pm$ 0.003  & 0.679 $\pm$ 0.005 \\ 
$\ttH$ ({\sc Pythia})                      & 0.250 $\pm$ 0.003 & 0.601 $\pm$ 0.008 & 0.139 $\pm$ 0.008  & 0.685 $\pm$ 0.010 \\ 
\midrule
\multicolumn{5}{c}{Some characteristic models of a Higgs-like boson with exotic decays and properties} \\
\midrule
${\rm q\bar{q}} \to {\rm H}(J^{CP}=1^{-})$ ({\sc JHUGen})         & 0.238 $\pm$ 0.001 & 0.609 $\pm$ 0.002 & 0.054 $\pm$ 0.001  & 0.642 $\pm$ 0.002 \\ 
${\rm q\bar{q}} \to {\rm H}(J^{CP}=1^{+})$ ({\sc JHUGen})        & 0.283 $\pm$ 0.001 & 0.619 $\pm$ 0.002 & 0.051 $\pm$ 0.001  & 0.651 $\pm$ 0.002 \\ 
${\rm gg} \to {\rm H} \to {\rm Z}\gamma^{*}$ ({\sc JHUGen})         & 0.156 $\pm$ 0.001 & 0.622 $\pm$ 0.002 & 0.073 $\pm$ 0.001  & 0.667 $\pm$ 0.002 \\ 
${\rm gg} \to {\rm H} \to \gamma^{*}\gamma^{*}$ ({\sc JHUGen}) & 0.188 $\pm$ 0.001 & 0.629 $\pm$ 0.002 & 0.066 $\pm$ 0.001  & 0.671 $\pm$ 0.002 \\ 
\bottomrule
\end{tabular}
\normalsize
\end{center}
\end{table}

\subsubsection{Signal definition}

The requirement on the invariant masses of the Higgs boson decay products is also important as the off-shell production cross section in the dominant gluon fusion production mode can be sizeable and can amount up to a few per cent of the total cross section~\cite{Kauer:2012hd}.

\subsection{Unfolding of experimental data}
\label{sec:exp:unfold}

Dealing with experimental resolutions will become a major aspect of differential fiducial measurements in Run 2: the large increase in integrated luminosity will mark the transition of the total uncertainties  being statistically dominated to becoming systematically limited. This will offer new challenges to reverting experimental resolutions. A summary about the caveats and various approaches of unfolding can be found for instance in Ref.~\cite{Cowan:2002in}, which this overview follows. 

\subsubsection{Introduction}

In measurements of differential cross sections, one often faces the problem of non-negligible migrations due to the finite resolution of the experimental apparatus. The reversion of such resolution migrations is typically called 'unfolding' or 'deconvolution' and compromises an essential ingredient that allows the easy comparison of theory predictions with measured fiducial cross sections. Mathematically the problem can be formulated in finding an inversion to the function
\begin{align}\label{eq:exp:unfold:master}
 f_{\rm meas} (x) = \int \, R(x | y| f_{\rm true}(y) \, \text{d} y \, ,
\end{align}
where $ f_{\rm meas} (x) $ is the PDF of the measured values $x$, and   $ f_{\rm true} (y) $ the PDF of true (but unknown) values $y$, smeared out by a detector response $R(x|y)$. In practice measurements are carried out in bins of observables, reducing Eq.~\ref{eq:exp:unfold:master}, to a matrix multiplication
\begin{align}\label{eq:exp:unfold:master2}
 x_i = \sum_{j=1}^N R_{ij} y_j \, .
\end{align}
and the response matrix $R_{ij}$ can be interpreted as a conditional probability
\begin{align}\label{eq:exp:unfold:matrix}
 R_{ij} = \mathcal{P}( {\rm reconstructed \, in \, bin \,} i \, | \, {\rm true \, value \, in \, bin \,} j)
\end{align}
with the sum 
\begin{align}
\sum_{i=1}^N R_{ij} = \mathcal{P}( {\rm observed \, anywhere} \, | \, { \rm true \, value \, in \, bin \,} j) = \epsilon_j \, ,
\end{align}
resulting in the the reconstruction efficiency. The task of unfolding is now to revert Eq.~\ref{eq:exp:unfold:master2} to convert measured values to true values. There exist several approaches for this, each with different strengths and caveats. 

\subsubsection{Inverting the response matrix and correction factors}

The most straightforward approach of unfolding involves the inversion of the matrix Eq.~\ref{eq:exp:unfold:matrix} and construct $R_{ij}^{-1}$.
This is generally often possible, but has some drawbacks: if the response matrix has large off-diagonal elements, e.g. if the chosen bin size is too small compared to the measurement resolution or one tries to measure an observable with an intrinsic poor resolution, such as jet multiplicities, the resulting expression for the true value
\begin{align}
y = R_{ij}^{-1} \, x
\end{align}
can have extremely large variances and strong negative correlations between neighbouring bins. If the measured values $x$ themselves are affected by large statistical fluctuations, these get amplified as one tries to revert migrations in a given bin using the estimated bin content of neighbouring bins with large variances themselves. In scenarios with small measured variances the resulting variances also can get amplified, if a high degree of fine structure is present, cf. Ref.~\cite{Cowan:2002in}. Technically the inversion of $R_{ij}$ can be implemented using least square estimators and in case the inversion is not possible, a pseudo-inverse may be constructed. The advantage of this inversion approach is that the resulting values for $y$, albeit in generally affected by large variances are in fact unbiased and the variance itself has the smallest possible value for any unbiased estimator. Thus any other method that aims to reduce the variance will necessarily introduce a bias. Thus the strategy that is followed is to accept a small bias in exchange for a large reduction in variance, i.e. trading statistical for systematic errors. 

\subsubsection{Correction factor method}

A relative simple method, often used in low statistics situation, is based on multiplicative correction factors derived from Monte Carlo simulations. The estimator for $y$ in a given bin $i$ is constructed as
\begin{align}
 y_i = C_i x_i
\end{align}
where the correction factor $C_i$ is
\begin{align}
 C_i = \frac{y_i^{\rm MC}}{x_i^{\rm MC}} \, ,
\end{align}
where $y_i^{\rm MC}$ and $x_i^{\rm MC}$ are the expected true and reco yields from the simulation. This inversion has a much smaller variance than the inversion of the migration matrix, but has a potential bias of the size
\begin{align}\label{eq:exp:bias}
 b_i = \left(  \frac{y_i^{\rm MC}}{x_i^{\rm MC}} - \frac{y_i^{\rm true}}{x_i^{\rm true}} \right) x_i^{\rm obs}
\end{align}
which has to be carefully estimated. In Eq.~\ref{eq:exp:bias} the quantities $y_i^{\rm true}$ and $x_i^{\rm true}$ are the true underlying mean population of the bin. The bias is zero in case the model is correct, which is not something that can be inferred prior a measurement. Typically the size of the bias is estimated with respect to some baseline scenario (e.g. the SM) and a maximal deviation one expects (e.g. a certain amount of new physics). One is satisfied when $b_i$ is small with respect to the variance and the size of the estimated bias is added to the systematic error of the measurement. This maximal deviation can also be composed by an entire of ensemble of scenarios. 

\subsubsection{Regularized unfolding}

Regularized unfolding tries to find some middle ground between these two approaches: inverting the migration matrix assumes that all the relevant information on how to revert migrations comes from the neighbouring bins of a measured distributions. The correction factor method excludes all the information from neighbouring bins to revert migrations. In regularized unfolding the information from both is used, and the weighting of either piece of information is typically controlled by one or many regularization parameters. There exist a range of different methods following different philosophies, often used are for example Refs.~\cite{Hocker:1995kb,D'Agostini:1994zf}. It is important to note, that regularized unfolding also has to carefully control and estimate the size of a potential bias. Here, as for the correction factor method, the found bias using a baseline scenario and a scenario for the largest to be expected deviation is added to the systematic error of the unfolded spectrum.  In addition special attention has to be payed to not 'over-regularize' the unfolded distributions, that is to impose a too strong dampening of statistical fluctuations. This is usually achieved by careful tuning of the regularization parameter(s).

A priori any method for unfolding is fine to use, as long as the bias is properly estimated. Even the combination of information gained via different unfolding methods is not a priori a problem, as long as the bias is negligible with respect to the statistical precision and is included into the error budget properly. With the expected large data sets of Run 2 of the LHC, it is expected that experiments will shift away from simpler methods and follow suit what is standard practice in SM measurements.

\subsection{Model dependence}
\label{sec:exp:mod}

The underlying assumptions on the signal model used to extract the fiducial cross sections unavoidably introduce some remaining systematic effects on the final measurement. The size of these effects can be estimated by extracting the fiducial cross sections from data assuming a range of alternative signal models, and by comparing them to the fiducial cross sections obtained assuming the SM Higgs boson. The range of alternative models to consider can include models with an arbitrary fraction of different SM Higgs boson production modes, models of Higgs-like resonances with anomalous interactions with a pair of neutral gauge bosons, models of Higgs-like resonances with some exotic decays to the final state of interest, and similar.

After the comparison is performed, the largest deviation between the fiducial cross section measured assuming any of the models from a particular range of alternative signal models, and the fiducial cross section measured under the SM Higgs boson assumption, should be reported as the systematic effect associated with the model dependence.

Given the fact that a wide range of exotic signal models has been excluded using the LHC Run 1 data at the 95\% C.L. or better [Ref.], it is recommended to impose those existing experimental constraints to narrow the range of the considered exotic signal models, and to report the model dependence computed using this reduced range of models. Naturally, in cases when it is considered useful, analyses are welcome to report also the model dependence computed from some wider range of alternative models (e.g. obtained by completely neglecting the existing experimental constraints).

\subsection{Treatment of the Higgs boson mass in fiducial and differential measurements}
\label{sec:xsec_mH}

All fiducial and differential cross section measurements rely on knowledge of the Higgs boson mass and analyses typically face two choices:
\begin{itemize}
 \item[1.] To extract the Higgs boson mass simultaneously along with the desired cross section parameters.
 \item[2.] To fix the Higgs boson mass to the current world average and treat it as an external parameter.
\end{itemize}

A simultaneous extraction of the Higgs boson mass and cross section is not always possible due to poor mass resolution, as for instance in $H \to WW \to 2 \ell 2 \nu$ cross section measurements. Both approaches are justifiable and have a number of advantages and disadvantages.

\subsubsection{Simultaneous extraction of Higgs boson mass and cross section}

In channels with good mass resolution, the simultaneous extraction of the Higgs boson mass and the fiducial cross sections is possible. Carrying out such a simultaneous analysis has the benefit that one avoids using information twice, once in the fiducial measurement and once from the Higgs boson mass measurement or average it might enter. In addition, one could argue that such measurements are less prone to possible systematic biases due to an accidental miscalibration of object energy scales, which relate the measured invariant mass with the physical Higgs boson mass. The clear disadvantage of the approach is that knowledge on the Higgs boson mass from complementary channels is completely discarded.

\subsubsection{Treating the Higgs boson mass as an external parameter }

Treating the Higgs boson mass as an external parameter allows one to incorporate such external knowledge from other centre-of-mass energies or channels. The apparent complication that enters the analyses though is the double use of data events in high resolution channels, such as $H \to \gamma \gamma$ and $H \to 4\ell$, which might have been used to determine the Higgs boson mass world averages in the first place. In practice, this is not an issue as the extracted quantities of interest, the Higgs boson mass and the reconstructed cross section yields, are uncorrelated entities and as such can be extracted individually from the same dataset without introducing a bias or error under-coverage.

\subsubsubsection{Updating the Higgs boson mass}

The knowledge of the Higgs boson mass is bound to improve in the future. In order to update existing and future measurements which treat the Higgs boson mass as an external parameter, the impact of shifting the Higgs boson mass within its experimental error on the measured fiducial and differential cross sections should be determined and quoted along with other systematic shifts.

\subsection{Combination of inclusive cross sections for Higgs boson production}
\label{sec:exp:xsec_comb}

\subsubsection{Introduction}
\label{sec:xsec_comb:intro}

This section discusses the combination of measured Higgs boson cross sections across multiple decay channels.
The combination is only possible in the inclusive phase space, where the effects of the Higgs boson decay products have been removed.
This introduces some model dependence: namely, the correction for the acceptance of the fiducial selection, and the assumption of Standard Model decay branching fractions for each decay channel.
Although measurements of cross sections within a fiducial phase space offer a more model-independent way of probing the properties of the Higgs boson, measurements at the LHC are currently statistically limited~\cite{Aad:2014lwa,Aad:2014tca,Khachatryan:2015yvw,Khachatryan:2015rxa,Khachatryan:2016vnn}.
It is therefore beneficial to combine the data across several channels, and maximize its potential.
This has been performed in Run-1 by the ATLAS Collaboration in the diphoton and four-lepton decay channels~\cite{Aad:2015lha}.

Both total and differential cross sections can be combined, provided that there is a consistent definition of the observable of interest between the different decay channels.
If only the shape of the differential distribution is of interest, the uncertainties from the combination can be reduced further as the assumptions of the SM decay branching fractions for each decay channel can be neglected.

\subsubsection{Combination method}
\subsubsubsection{Extrapolation to the total phase space}
\label{sec:xsec_comb:extrap}

The inclusive phase space must be carefully defined such that all observables of interest are independent of the Higgs boson decay products.
It is preferable to retain the philosophy of the fiducial cross-section measurements and keep the theoretical predictions and measurement as disentangled as possible, to maximize the longevity of the data.
The data should therefore ideally be corrected to the particle level in the inclusive phase space.
In order to compare theoretical predictions at the parton level to measurements, non-perturbative corrections accounting for the impact of underlying event, multi-parton interactions and hadronization can be provided separately.
The central values and uncertainties should be evaluated using a range of generators with different tunes for showering, hadronization and underlying event, such as in Refs.~\cite{ATLAS:2011krm,ATLAS:2011gmi,ATLAS:2011zja}.

\subsubsubsection{Jets in the inclusive phase space}

When measuring differential cross sections, it is important that all observables of interest are independent of the decay products of the Higgs boson, in particular when considering jets.
Generally the inputs to jet finding at the particle level are all final state particles with lifetimes $c\tau > 10$\ mm, preferably excluding neutrinos, electrons, and muons that do not originate from hadronic decays (as suggested in Ref.~\cite{ATL-PHYS-PUB-2015-013}).
Using this method, some of the resulting particle-level jets will contain decay products of the Higgs boson.
It is then possible for jets to veto -- or be vetoed by -- nearby decay products, due to overlap removal between physics objects.
This creates an inconsistency in the definition of a jet between different decay channels, such as diphoton (two decay products in the final state) and four-lepton (four decay products in the final state).

The effect of the Higgs boson decay can be effectively removed by reconstructing jets at the particle level and explicitly excluding the decay products of the Higgs boson.
This is possible in generators that retain the history of the Higgs boson decay on the event record.
The effects of final-state radiation from the Higgs boson decay products can also be removed.
For example, photons can be excluded from jet finding if they lie inside a cone of radius $\Delta\mathrm{R} < 0.1$ of an electron or muon, where neither the photon nor lepton originate from a hadron decay.

Optimally, a similar definition can be used when reconstructing detector-level jets.
This can be done using the so-called particle-flow method, which reconstructs individual leptons and photons before using them as inputs to the jet finding algorithm.

\subsubsubsection{Binning of differential observables}

In order to carry out a statistical combination of differential observables, matching bin boundaries are helpful. This reduces the combination in the inclusive phase space to a statistical problem of statistically combining coarse cross sections which hold information about the sum of more fine cross section sums. In case no matching bin boundaries are present, a combination requires additional theory input to approximate such matching bin boundaries. For the Run 1 results, a range of criteria have been used to justify the binning of a given observable: binning choices aiming to have equal (expected) statistical significance in each bin or identical expected purities for example, offer a first rough guideline but no unique choice. In analyses which rely on extracting signal yields by subtracting non-resonant production, a certain given binning choice cannot introduce a bias. Thus future measurements should be encouraged to use matching bin boundaries wherever possible and practical to facilitate the possibility of a later inclusive combination. 

\subsubsection{Treatment of uncertainties}
\label{sec:xsec_comb:uncert}

Care must be taken to appropriately correlate shared uncertainty sources (both experimental and theoretical) between the different decay channels.
In particular, the uncertainties on the corrections for acceptance and branching fractions need to be assessed appropriately.

The branching fraction uncertainties for Run-1 are described in the LHC HXSWG YR3~\cite{Heinemeyer:2013tqa}.
For example, in Ref.~\cite{Aad:2015lha} five nuisance parameters were used to describe the branching fraction uncertainties; three fully correlated and two uncorrelated between the diphoton and four-lepton decay channels.

Since the acceptance corrections are largely driven by the Higgs boson rapidity distribution, which is shaped by the phase space of the PDF, the choice of PDF set is one of the largest contributions to the acceptance factor uncertainty.
In Run-1 this uncertainty was assessed by following the PDF4LHC recommendations~\cite{Botje:2011sn}.
Uncertainties associated with missing higher-order corrections are evaluated by varying the renormalization and factorization scales, also following the PDF4LHC recommendations.

The acceptance may also be sensitive to the choice of the assumed mass of the Higgs boson, if for example a mass window around the peak is chosen~\cite{Aad:2014tca}.
In this case the mass of the Higgs boson in the Monte Carlo samples used to calculate the fiducial acceptance should be varied within the current uncertainties.
In order to cover a range of different event topologies and modest deviations from the SM couplings, the composition of the SM signal should also be varied within current experimental constraints.
Finally, uncertainties in the MC simulation of underlying event, multi-parton interactions and hadronization should be included.
This can be done using different MC tunes, preferably the ``systematic variation'' eigentunes which give more accurate variations of the perturbative effects~\cite{ATLAS:2011krm}.

\subsubsubsection{Statistical procedure}

Provided that the uncertainties are normal distributed, a weighted average between the two measurements may be performed.
However, because channels are often statistically limited the use of Poisson statistics is frequently necessary.
In this case a combined likelihood fit that accounts for common theoretical and experimental uncertainties can be used.

\subsubsection{Summary}
\label{sec:xsec_comb:summary}

Combining different Higgs boson decay channels to reduce the large statistical uncertainty will continue to be important until a larger data sample is available.
For example, the combination of the diphoton and four-lepton final states reduced the total uncertainty on the ATLAS 8 TeV inclusive measurement by on average $25{-}30\%$.
While special care must be taken to unify the object definitions and to account for appropriate correlations between decay channels, the recommendations given here should provide a useful starting point for future combined measurements.

\section{Summary and recommendations for future measurements}\label{sec:future}
Fiducial measurements will play an important role in characterizing the Higgs boson in Run 2 and beyond: differential and total fiducial cross sections allow the study of a wide range of physical phenomena, such as for example the accuracy of the theoretical model of different Higgs boson production mechanisms or the presence of beyond the SM contributions. To take full opportunity of these measurements further advancements in certain theory and experimental questions should be prioritized. In the following a list of such prioritized areas is given, which is based on the contributions of this report and the discussions inside our working group:

\begin{itemize}
\item[1.] \underline{Fiducial definitions:} The particle-level fiducial phase space should be designed with the goal to minimize the underlying model dependence. Often this leads to particle-level fiducial phase space definitions which closely follow the reconstruction level definition. In Run 1 such model errors contributed only negligible ot the total error budget for most observables, but with an expected integrated luminosity of 100/fb in Run 2 such aspects will become more important. Section~\ref{sec:EXP} discussed several experimental aspects on how such model dependence can be minimized. Among the most important ones is the use of particle-level isolation, which does mimic the reconstruction level requirements of isolated objects. If tuned right the ratio of reconstruction level and particle level fiducial efficiency can be brought close to be near production process independent. Both experiments should continue such studies and explore alternative fiducial regions and definitions, as there is no need to stay in sync with past definitions or other experiments.
\item[]
\item[2.] \underline{Unfolding of detector effects:} Reverting migrations induced by the finite resolution of the detectors is an important aspect of how measured differential cross sections are presented. Without such an unfolding a direct comparison of the measured fiducial cross sections with particle-level differential theory predictions cannot be made. In Run 1 both experiments followed a slightly different philosophy in unfolding, briefly summarized in Section~\ref{sec:EXP}. In principle any method of unfolding is fine to use and combine as long as underlying model biases are properly estimated and included in the error budget. Besides making the unfolded results available (cf. recommendation 3), it became clear that it would be useful to also publish the migration matrices and the folded yield. One core aspect of fiducial measurements is the preservation of results, such that measured differential information can be confronted with new theory ideas even many years after the original measurement was carried out. As it is difficult to forsee the needs of future analyses, we recommend to provide the full array of information which lead to unfolded results: that is the original yields, the migration and efficiency matrix, as well as the full experimental on either such that the published results can be reproduced.
\item[]
\item[3.] \underline{Preservation of data:} An important trend, which is already very much being prioritized for SM measurements, is the publication of experimental information on HEPDATA~\cite{hepdata}, an online archive which allows easy access of many experimental results. The remaining Run 1 measurements which are not published there, should follow suit and for Run 2 a full summary of the experimental result of a paper should be made available there. In addition, experiments should provide validated RIVET routines~\cite{rivet} to apply the corresponding particle level fiducial selection of each analysis, as done in some cases in Run 1 already. This will allow phenomenologists or other interested parties to make maximal use of the measurements and minimize the possibility for errors. This is particular important if non-trivial fiducial definitions are used (cf. recommendation 1). 
\item[]
\item[4.] \underline{Higgs boson mass shifts:} In Run 1 the experiments followed two different approaches when dealing with the Higgs boson mass: some either assumed that this is external knowledge and fixed the mass at the current best known world average. Although this implied a double use of the same data, the near complete decorrelation of cross section yields and resonance mass value, justified such an approach, as then both properties can be measured independently from each other. The alternative approach was to determine the Higgs boson mass using the available data and neglecting any other information from other channels. Both approaches are fine and could be used in the future. As the knowledge on the Higgs boson mass will improve with Run 2, we recommend that measurements which fix the Higgs boson mass at its world average value provide the necessary information to allow updating their cross sections if our knowledge there improves. This information is contained completely in the shifts in cross sections within the current experimental bounds of the Higgs boson mass. 
\item[]
\item[5.] \underline{Additional object definitions and bin boundaries:} Albeit not mandatory, the harmonization of object definitions (such as jets or of additional particles in the event) and bin boundaries would simplify comparisons between the experiments and theory considerably. In addition, it would allow both experiments to provide inclusive averages of their results. Such are more model independent, but will provide interesting future input to probe SM properties such as the modelling of hard and soft quark and gluon radiation or help constraining PDFs. In Run 1 such a combination between channels was carried out in Ref.~\cite{Aad:2015lha} and experiments are encouraged to discuss a full combination of all experimental information to provide differential information of inclusive Higgs boson production. A review of some crucial points which need to be taken into account is provided in Section~\ref{sec:exp:xsec_comb}.
\item[]
\item[6.] \underline{Continued benchmarking of fiducial acceptance:} An important aspect of fiducial measurements as well as for a combination between several channels outlined in point 5, is the understanding of the fiducial acceptance. In Section~\ref{sec:SM} a first set of public results of differential Higgs boson $p_T$ spectra are provided for a benchmark fiducial region. Such studies should continue to better understand the accuracy of the range of tools used to calculate such factors in an experimental context. A first contribution in this direction on the object level was carried out by Ref.~\cite{Badger:2016bpw}, but future studies will be needed. 
\item[]
\item[7.a] \underline{Signal definitions:} In Ref.~\cite{Aad:2015rka} a first fiducial cross section measurement of $\Pp\Pp \to \PZ\PZ^{(*)}\to 4\Pl$ was reported which absorbs the Higgs boson signal into its cross section definition. Such measurements are very interesting and should be pursuit also in Run 2, as they allow for a simultaneous analysis of resonant and non-resonant production: if for instance new physics would alter the coupling of the Higgs boson to the Weak bosons, the non-resonant background also would be modified what in turn would change the extracted yield. In addition, the maximal sensitivity to probe for such scenarios would involve the simultaneous analysis of resonant and non-resonant $\Pp\Pp \to \PZ\PZ^{(*)}\to 4\Pl$ production.
\item[]
\item[7.b] \underline{Signal definitions:} In Ref.~\cite{Aad:2016lvc} a value for the gluon fusion fiducial cross section is provided from $\PH \to \PW\PW^{(*)}\to 2\Pl 2 \nu$ decays. Albeit in terms of this measurement it is natural to subtract non-gluon Higgs boson production, we advocate a fiducial cross section definition which is inclusive over the Higgs boson production mechanism. Albeit this complicates the eventual analysis (as migration patterns are more complicated), such results allow for a wider range of interpretations. In particular, they do not make weaker assumptions of non-gluon fusion Higgs boson production, which is then subtracted. By providing inclusive measurements in all channels, the information is complementary to the information provided in the  $\PH \to \PZ\PZ^*\to 4\Pl$ and $\PH \to \PGg\PGg$ channels, where no such subtractions are carried out and is left to an additional interpretation step. 
\item[]
\item[8.] \underline{Global beyond the SM analyses:} A combined global analysis of several fiducial observables offers a unique probe to constrain beyond the SM physics. Such an analysis can either happen inside a given BSM model or in an effective field theory framework (cf. Section~\ref{sec:BSM} and the EFT review inside this report). This approach is complementary and more model independent to the simplified template cross sections, but also more limited as only channels can be included which allow for fiducial measurements. To maximize the sensitivity of such an analysis, the experimental and theory communities should both work together to create the necessary tools and identity the necessary measurements that maximize the sensitivity. This for example requires theory predictions which match the experimental sensitivity and the determination of statistical correlations between observables. A proof-of-concept global analysis which could serve as a template for future global analyses was carried out in Ref.~\cite{Aad:2014lwa} using Run 1 data, where Wilson coefficients of dimension six operators were constrained using the differential information of five differential observables and their experimental correlations. 
\item[]
\item[9.] \underline{Targeted beyond the SM analyses:} Complementary to global analyses, which use a broad collection of differential or fiducial regions to constrain beyond the SM physics, is the Ansatz of targeting regions with a specific sensitivity for a given model. If for example one wishes to constrain a dark matter scenario in association with a Higgs boson, that produces a lepton and missing transverse energy, most sensitivity will come from a matching fiducial region. Theorists are encouraged to enter a dialogue with their experimental colleagues if they have specific suggestions. Section~\ref{sec:BSM} provides a first discussion and specific regions targeting VBF, VH, and dark matter associated production seem among the most promising suggestions. 
\end{itemize}


\cleardoublepage

\newpage
\renewcommand*{\thefootnote}{\fnsymbol{footnote}}
\part[Beyond the Standard Model Predictions]{Beyond the Standard Model Predictions \footnote{I.~Low, M.~M\"uhlleitner, M.~Pelliccioni, N.~Rompotis, R.~Wolf~(Eds.)}}
\label{chap:BSM}
\renewcommand*{\thefootnote}{\Roman{part}.\arabic{footnote}}
\setcounter{footnote}{0} 
\chapter{Neutral MSSM}
\label{chap:NeutralMSSM}
\ChapterAuthor{R.~Lane, S.~Liebler, A.~McCarn, P.~Slavich, M.~Spira, D.~Winterbottom (Eds.)
E.~Bagnaschi, F.~Frensch, R.~Harlander, S.~Heinemeyer,  G.~Lee, H.~Mantler, M.~M{\"u}hlleitner, J.~Quevillon, N.~Rompotis, A.~Vicini, C.~Wagner, M.~Wiesemann, R.~Wolf} 








\providecommand{\ltbh}{``low-tb-high''}
\providecommand{\mHpm}{M_{\PSHpm}}
\providecommand{\mh}{m_{\Ph}}
\providecommand{\mhiggs}{M_{\phi}}
\providecommand{\mphi}{M_\phi}
\providecommand{\mSUSY}{M_{\scriptscriptstyle{\rm SUSY}}}
\providecommand{\mstone}{M_{\PSQtDo}}
\providecommand{\msttwo}{M_{\PSQtDt}}
\providecommand{\msbar}{\overline{\rm MS}}
\providecommand{\smallmsbar}{{\scriptscriptstyle \msbar}}
\providecommand{\mbmb}{\Mb^\smallmsbar(\Mb)}
\providecommand{\mcmc}{\Mc^\smallmsbar(\Mc)}
\providecommand{\tb}{\tan\beta}
\providecommand{\abbrev}{\scalefont{.9}}
\providecommand{\hmw}{{\abbrev HMW}}
\providecommand{\bv}{{\abbrev BV}}
\providecommand{\AR}{{\abbrev AR}}
\providecommand{\mcnlo}{{\sc MC@NLO}}
\providecommand{\powheg}{{\sc POWHEG}}
\providecommand{\prophecy}{{\sc Prophecy4f}}
\providecommand{\feynhiggs}{{\sc FeynHiggs}}
\providecommand{\hdecay}{{\sc HDECAY}}
\providecommand{\suspect}{{\sc SuSpect}}
\providecommand{\sushi}{{\sc SusHi}}
\providecommand{\moresushi}{{\sc MoRe-SusHi}}
\providecommand{\rootf}{{\sc ROOT}}
\providecommand{\thdm}{2HDM}
\providecommand{\nlo}{{\abbrev NLO}}
\providecommand{\fnlo}{f\nlo}
\providecommand{\lo}{{\abbrev LO}}
\providecommand{\pt}{\ensuremath{p_T}}
\providecommand{\pth}{p^{\phi}_T}
\providecommand{\pthtwo}{p^{\phi~2}_T}
\providecommand{\dd}{\mathrm{d}}
\providecommand{\qrest}{Q_t}
\providecommand{\qresb}{Q_b}
\providecommand{\qresint}{Q_\text{int}}
\providecommand{\wrest}{w_t}
\providecommand{\wresb}{w_b}
\providecommand{\wresint}{w_\text{int}}
\providecommand{\qres}{\ensuremath{Q_{\text{res}}}}
\providecommand{\qmatch}{\ensuremath{Q_{\text{match}}}}
\providecommand{\qsh}{\ensuremath{Q_\text{sh}}}
\providecommand{\largeb}{large-$b$}
\providecommand{\largebot}{\largeb{}}
\providecommand{\largetop}{large-$t$}
\providecommand{\order}[1]{{\cal O}(#1)}
\providecommand{\sm}{{\abbrev SM}}


\section{Introduction}
\label{sec:mssm-intro}

In contrast to the SM, the \emph{Minimal Supersymmetric Standard
  Model} (MSSM) requires the introduction of two complex Higgs doublets, $H_{u}$ and $H_{d}$, to provide masses for up- and down-type
fermions via the spontaneous breaking of the $SU(2)_{L}\times
U(1)_{Y}$ gauge symmetry. If the MSSM Lagrangian does not contain new
sources of $CP$ violation, the presence of two complex Higgs doublets
implies the existence of two charged Higgs bosons, $\PSHpm$, and three
neutral Higgs bosons: a $CP$-odd (i.e., pseudoscalar) state $\PA$, and
two $CP$-even (i.e., scalar) states, $\Ph$ and $\PH$, with $\Mh <
\MH$.  At the tree level in the MSSM, the masses of these five Higgs bosons and their mixing can be expressed in terms of the gauge-boson
masses $\MW$ and $\MZ$ plus two additional parameters, which can be
chosen as the pseudoscalar mass $\MA$ and the ratio of the vacuum
expectation values of the neutral components of the two Higgs doublets, $\tan\beta \equiv v_u/v_d$. The tree-level mass of the
charged states is given by $\mHpm^2 = \MA^2 + \MW^2$, and the
tree-level masses of the neutral $CP$-even states are:
\begin{equation}
\label{treemasses}
  M_{\Ph,\, \PH}^{2}
~=~ \frac{1}{2}\left(\MA^2 + \MZ^2\, \mp\,
\sqrt{\left(\MA^2 + \MZ^2\right)^{2} - 4 \,\MA^2\,\MZ^2\,\cos^{2}2\beta}\right)~.
\end{equation}

In the MSSM the role of the SM Higgs boson is shared between the
scalars $\Ph$ and $\PH$. In particular, the couplings of the neutral
scalars to pairs of massive vector bosons (${\rm VV}$) and of SM fermions
($uu$, $dd$ and $\ell\ell$), relative to the corresponding SM
couplings, are:

\vspace*{2mm}
\begin{equation}
\label{tab:couplingsMSSM}
    \begin{tabular}{lccc}
       & ${g}_{{\rm VV}}$ & ${g}_{uu}$ & ${g}_{dd,\ell\ell}$ \\
      $\PA$ &  $0$                   &  $\cot\beta$     & $\tan\beta$              \\[2mm]
      $\PH$ &  $\cos(\beta-\alpha)$  &  $\sin\alpha / \sin\beta$  & $\;\;\cos\alpha / \cos\beta$  \\[2mm]
      $\Ph$ &  $\sin(\beta-\alpha)$  &  $\cos\alpha / \sin\beta$  & $   -\sin\alpha / \cos\beta$  \\
    \end{tabular}
\end{equation}
\vspace*{2mm}

\noindent
where $\alpha$ is the angle that diagonalizes the $CP$-even mass matrix,
given at tree level by:
\begin{equation}
\label{alpha}
  \tan\alpha~=~ \frac{-(\MA^2+\MZ^2)\,\sin2\beta}
{(\MZ^2-\MA^2)\cos2\beta + \sqrt{\left(\MA^2 + \MZ^2\right)^{2}
- 4 \,\MA^2\,\MZ^2\,\cos^{2}2\beta}}~.
\end{equation}
\noindent
In addition, there are non-SM couplings of the neutral scalars to $\PZ
\PA$ and to $\PWmp \PSHpm\,$. These are proportional to
$\cos(\beta-\alpha)$ in the case of $\Ph$ and to $\sin(\beta-\alpha)$ in
the case of $\PH$, while the $\PZ\PA\PA$ coupling vanishes and the $\PWmp
\PSHpm \PA$ coupling does not depend on $\alpha$ or $\beta$. Also
relevant for the discussion in \refS{sec:mssm-lowtanB} is the trilinear
coupling of one heavy scalar to two light scalars, whose tree-level
value reads, in units of $\MZ^2/v\,$ where $v \equiv
(v_u^2+v_d^2)^{1/2} \,\approx 246$\UGeV,
\begin{equation}
\label{lamHhh}
\lambda_{\PH\Ph\Ph,\,{\rm tree}} ~=~ 2 \,\sin2\alpha\,\sin(\beta+\alpha)
\,-\, \cos2\alpha\,\cos(\beta+\alpha)~.
\end{equation}

In the {\em decoupling limit}, $\MA \gg \MZ\,$, the mixing angle in
the $CP$-even sector simplifies to $\alpha \approx \beta - \pi/2$. As
a result, the tree-level mass of the light neutral scalar $\Ph$
becomes $\Mh \approx \MZ \,|\cos 2\beta|$, and its couplings to gauge
bosons, quarks and leptons in \Eq~(\ref{tab:couplingsMSSM}) become
SM-like. The masses of $\PH$ and $\PSHpm$ become approximately
degenerate with $\MA\,$, the couplings of $\PH$ to two massive gauge
bosons vanish, the couplings of $\PH$ to two up-type (down-type) SM
fermions are suppressed (enhanced) for large $\tan\beta$, and the
coupling of $\PH$ to two light neutral scalars is suppressed for large
$\tan\beta$. Therefore, in this limit, the Higgs sector of the MSSM
reduces to an SM-like Higgs boson with tree-level mass $\Mh < \MZ\,$,
and a heavy and mass-degenerate multiplet $(\PH,\PA,\PSHpm)$ with
vanishing couplings to two massive gauge bosons. In contrast, for low
values of $\MA$ there is a crossing point where $\PH$ and $\Ph$ swap
their roles, i.e.~the heavy neutral scalar is the one whose mass is
independent of $\MA$ and whose couplings approach SM strength. For
$\tan\beta\gtrsim 10$, the decoupling behaviour of the tree-level
scalar masses is rather sharp, with a clear crossing point around $\MA
\approx \MZ$. For lower values of $\tan\beta$, the onset of the
decoupling behaviour at $\MA \gg \MZ$ (or $\MA \muchless \MZ$) is
delayed to larger (or smaller) values of $\MA$. Indeed, for
$\tan\beta=3$ a heavy scalar $\PH$ of mass around 300\UGeV\ can still
have non-negligible couplings to two massive gauge bosons (as well as
to two light scalars, due to the reduced $\tan\beta$
suppression). However, for low $\tan\beta$ the upper bound on the
tree-level mass of the light scalar can be considerably lower than
$\MZ$, with $\tan\beta=1$ corresponding to a vanishing tree-level
mass.

As is well known, the tree-level predictions for the masses of the
MSSM Higgs bosons are subject to substantial radiative corrections,
which can lift the lightest-scalar mass well above the tree-level
bound and introduce a dependence on several other parameters of the
MSSM.  The dominant one-loop contribution to the lightest-scalar mass
arises from loops of top quarks and their scalar superpartners, the
top squarks (stops), and in the decoupling limit takes the
approximate form
\begin{equation}
 \left(\Delta \Mh^{2}\right)_{1\ell}^{t/\tilde t} ~\approx~
\frac{3\, \Mt^{4}}{2\,\pi^{2}\,v^{2}}\left(\log \frac{\mSUSY^2}{\Mt^2} +
\frac{X_t^2}{\mSUSY^{2}} - \frac{X_t^4}{12\,\mSUSY^{4}}\right)~,
 \label{eq:1loop-correction-mh}
\end{equation}
where $\mSUSY = \sqrt{\mstone\,\msttwo}$ is an average scale for the
stop masses, and $X_t = A_t-\mu\cot\beta$ is the stop mixing term,
where $A_t$ is the soft SUSY-breaking Higgs-stop coupling and $\mu$ is
the higgsino mass parameter. The one-loop top/stop contribution to
$\Mh$ is maximized for large values of $\mSUSY$ (due to the
logarithmic term) and for the {\em maximal mixing} condition $|X_t| =
\sqrt{6}\,\mSUSY$.  A smaller negative contribution from sbottom
loops, not shown in the equation above, can be relevant only for large
values of $\tan\beta$.

Due to the crucial role of radiative corrections in pushing the
prediction for the lightest-scalar mass above the tree-level bound, an
impressive theoretical effort has been devoted over a quarter-century
to the precise determination of the Higgs sector of the
MSSM.\footnote{\,We focus here on the MSSM with real
  parameters. Significant efforts have also been devoted to the
  Higgs boson mass calculation in the presence of $CP$-violating phases, as
  well as in non-minimal supersymmetric extensions of the SM.} After
the first computations of the one-loop top/stop
contributions~\cite{Okada:1990vk, Okada:1990gg, Ellis:1990nz,
  Ellis:1991zd, Brignole:1991pq, Haber:1990aw}, full one-loop
computations of the MSSM Higgs boson masses have become
available~\cite{Chankowski:1991md, Chankowski:1992er, Brignole:1991wp,
  Brignole:1992uf, Dabelstein:1994hb, Pierce:1996zz}, leading
logarithmic corrections beyond one loop have been included via
renormalization-group (RG) methods~\cite{Barbieri:1990ja,
  Espinosa:1991fc, Haber:1993an, Casas:1994us, Carena:1995bx,
  Carena:1995wu, Haber:1996fp, Espinosa:2001mm}, and the genuine
two-loop corrections have been evaluated in the limit of vanishing
external momentum in the Higgs self-energies~\cite{Hempfling:1993qq,
  Heinemeyer:1998jw, Heinemeyer:1998kz, Heinemeyer:1998np,
  Carena:2000dp, Heinemeyer:2004xw, Zhang:1998bm, Espinosa:1999zm,
  Espinosa:2000df, Degrassi:2001yf, Brignole:2001jy, Brignole:2002bz,
  Dedes:2003km, Martin:2002iu, Martin:2002wn}. The external-momentum
dependence of the dominant two-loop corrections involving the strong
gauge coupling or the third-family Yukawa couplings has subsequently
been computed~\cite{Martin:2004kr, Borowka:2014wla, Borowka:2015ura,
  Degrassi:2014pfa}, and some of the dominant three-loop corrections
to the lightest-scalar mass have also been obtained, both via RG
methods~\cite{Martin:2007pg, Hahn:2013ria, Draper:2013oza} and by
explicit calculation of the Higgs self-energy at zero external
momentum~\cite{Harlander:2008ju, Kant:2010tf}.

As in the SM, one of the most important production mechanisms for the
neutral Higgs bosons in the MSSM is gluon fusion~\cite{Georgi:1977gs},
mediated by loops involving the top and bottom quarks and their
superpartners. However, for intermediate to large values of $\tb$
bottom-quark annihilation can become the dominant production mechanism
for the neutral Higgs bosons that have enhanced couplings to down-type
fermions.
For what concerns gluon fusion, the knowledge of the contributions of
diagrams involving only quarks and gluons includes: NLO-QCD
contributions~\cite{Dawson:1990zj, Djouadi:1991tka, Graudenz:1992pv,
  Spira:1993bb, Spira:1995rr, Harlander:2005rq, Anastasiou:2006hc,
  Aglietti:2006tp, Bonciani:2007ex} with full dependence on the Higgs boson and quark masses; NNLO-QCD contributions in the heavy-top
limit~\cite{Kramer:1996iq, Spira:1997dg, Chetyrkin:1997sg,
  Chetyrkin:1997un, Chetyrkin:1998mw, Harlander:2000mg, Catani:2001ic,
  Harlander:2001is, Harlander:2002wh, Harlander:2002vv,
  Anastasiou:2002yz, Anastasiou:2002wq, Ravindran:2003um} and
including finite top-mass effects~\cite{Marzani:2008az,
  Harlander:2009bw, Harlander:2009mq, Harlander:2009my, Pak:2009bx,
  Pak:2009dg, Pak:2011hs, Caola:2011wq}; partial NNNLO-QCD
contributions~\cite{Moch:2005ky, Ravindran:2006cg, Ball:2013bra,
  Buehler:2013fha, Anastasiou:2014vaa, Anastasiou:2014lda}; soft-gluon
resummation effects~\cite{Kramer:1996iq, Catani:2003zt, Idilbi:2005ni,
  Idilbi:2006dg, Ahrens:2008nc}.
The NLO-QCD contributions of diagrams involving only squarks and
gluons are fully known~\cite{Dawson:1996xz, Anastasiou:2006hc,
  Muhlleitner:2006wx, Bonciani:2007ex}. For the NLO-QCD contributions
of diagrams involving quarks, squarks and gluinos, approximate
analytic results assuming different hierarchies between the Higgs,
quark and superparticle masses are available~\cite{Harlander:2003bb,
  Harlander:2004tp, Harlander:2005if, Harlander:2010wr,
  Degrassi:2008zj, Degrassi:2010eu, Degrassi:2011vq, Degrassi:2012vt},
and exact results relying on a combination of analytic and numerical
methods have been presented~\cite{Anastasiou:2008rm,
  Muhlleitner:2010nm}. Approximate results, in the limit of vanishing
Higgs boson mass, also exist for the NNLO-QCD contributions of diagrams
involving superparticles~\cite{Harlander:2003kf, Pak:2010cu,
  Pak:2012xr}. Finally, the effects of non-decoupling, $\tb$-enhanced
corrections to the Higgs-bottom coupling can be taken into account via
an effective-Lagrangian approach~\cite{Hempfling:1993kv, Hall:1993gn,
  Carena:1994bv, Carena:1999py, Guasch:2003cv, Noth:2008tw,
  Noth:2010jy, Mihaila:2010mp}, and the subset of electroweak (EW)
contributions involving loops of light quarks and gauge bosons can be
adapted to the MSSM from the corresponding SM
calculation~\cite{Aglietti:2004nj, Bonciani:2010ms}.
For Higgs boson production in bottom-quark annihilation, the cross section
is known in the four-flavour scheme at NLO-QCD~\cite{Dittmaier:2003ej,
  Dawson:2003kb} and in the five-flavour scheme up to
NNLO-QCD~\cite{Dicus:1998hs, Maltoni:2003pn, Harlander:2003ai}.  As in
the case of gluon fusion, the $\tb$-enhanced contributions from
diagrams involving superpartners can be taken into account by means of
an effective Higgs-bottom coupling. The remaining one-loop
contributions from superpartners have been found to be
small~\cite{Dittmaier:2006cz, Dawson:2011pe,Dittmaier:2014sva}.

The kinematic distributions of the Higgs bosons of the MSSM can
exhibit different properties with respect to the corresponding ones of
an SM-like Higgs boson of equal mass~\cite{Langenegger:2006wu,
  Brein:2007da, Bagnaschi:2011tu, Grazzini:2013mca, Harlander:2013oja,
  Banfi:2013yoa, Azatov:2013xha, Grojean:2013nya, Harlander:2014uea,
  Dawson:2014ora, Bagnaschi:2015qta, Langenegger:2015lra}.  In the
case of gluon fusion, these effects are due to the modified relative
importance of the top and bottom amplitudes and to the additional
contributions from SUSY particles that mediate the interaction between
the Higgs and the gluons.  To understand the size of these effects and
to estimate the discriminating power in distinguishing them from the
SM behaviour, it is important to compute precise predictions where all
theoretical uncertainties are under control~\cite{Bagnaschi:2014zla}.

The decays of the MSSM Higgs bosons should also be calculated
including higher-order corrections.
%
Decays to SM fermions have been evaluated at the full one-loop level
in the MSSM~\cite{Dabelstein:1995js, Williams:2011bu} (see also
\Bref{Heinemeyer:2000fa}), where higher-order SUSY corrections can be
included via resummation~\cite{Hempfling:1993kv, Hall:1993gn,
  Carena:1994bv, Carena:1999py, Guasch:2003cv, Noth:2008tw,
  Noth:2010jy, Mihaila:2010mp}.  Decays to (lighter) Higgs bosons have
been evaluated at the full one-loop level in the MSSM in
\Brefs{Brignole:1992zv,Williams:2011bu} (see
also~\Brefs{Barger:1991ed,Heinemeyer:1996tg}).  Decays to SM gauge
bosons have been evaluated at the full one-loop level in
\Bref{Hollik:2011xd}, but no corresponding computer code is available.
In an approximative way they can be evaluated using the full SM
one-loop result~\cite{Bredenstein:2006rh,Bredenstein:2006ha} rescaled
with the appropriate effective LO coupling factors. The latter are
available in the form of radiatively corrected effective couplings and
$Z$~factors up to the two-loop level~\cite{Frank:2006yh} and ensure,
also in the other decay modes, the on-shell properties of the decaying
Higgs boson.
The (heavy) MSSM Higgs bosons can (if kinematically allowed) also decay
to SUSY particles, i.e.~to charginos, neutralinos and scalar fermions,
where these modes can dominate the heavy Higgs boson decays.  The
lightest neutral Higgs boson, on the other hand, can have a substantial
branching ratio into the lightest neutralino, i.e.~the Dark Matter
candidate in the MSSM~\cite{Goldberg:1983nd,Ellis:1983ew}.
Higher-order contributions to the decays of the MSSM Higgs bosons to
scalar fermions have been evaluated at the full one-loop level in
\Brefs{Weber:2003tw,Heinemeyer:2014yya}. In \Bref{Accomando:2011jy} the
${\cal O}(\alpha_s)$ corrections to Higgs boson decays to scalar
quarks were re-analysed. Full one-loop corrections to the decays to
charginos and neutralinos can be found in
\Brefs{Bharucha:2012nx,Eberl:2004ic,Heinemeyer:2015pfa}.

\bigskip

The discovery of an approximately SM-like Higgs boson with mass around
$125$\UGeV\ -- and the non-observation of SUSY particles or of
additional (neutral or charged) Higgs bosons -- in the first few years
of operation of the LHC have led to strong constraints on the allowed
parameter space of the MSSM.
For low values of the parameter $\tan\beta$, the direct bounds on the
masses of the additional Higgs bosons are still relatively weak, but
very heavy top squarks are required to reproduce the observed mass of
the SM-like Higgs boson. In \refS{sec:mssm-lowtanB} we discuss two
approaches, proposed in
\Brefs{Djouadi:2013vqa,Maiani:2013nga,Djouadi:2013uqa,
  Djouadi:2015jea} and \cite{Svennote}, for predicting the properties
of the Higgs bosons in the region with low $\tan\beta$ and heavy SUSY
particles. We also compare the predictions of the two approaches with
those of a recent calculation based on the effective-field-theory
(EFT) method~\cite{Lee:2015uza}.

In \refS{sec:mssm-mssmrootfiles} we discuss the content and structure of
the \rootf\ files which are provided on the webpages of the subgroup
for various MSSM benchmark scenarios.  They contain crucial
information about Higgs boson masses, cross sections and branching ratios
relevant for the experimental analysis. We also describe a comparison
between the \rootf\ files for the two above-mentioned MSSM scenarios
with low $\tb$.

In \refS{sec:mssm-pt} we turn our attention to the Higgs
transverse-momentum distribution.  In the context of the matching
between resummed and fixed-order computations, we investigate the main
sources of matching ambiguities by means of a twofold comparison.  On
the one hand, we present a comparison of two recently introduced
algorithms for determining the matching
scale~\cite{Harlander:2014uea,Bagnaschi:2015qta}, an auxiliary
parameter of these matched-resummed computations, whose variation can
be used as a probe of the theoretical uncertainties akin to what is
usually done with the renormalization and factorization scales in the
case of fixed-order results.  On the other hand, we compare the
predictions of two NLO+PS frameworks, \mcnlo~\cite{Frixione:2002ik}
(in its implementation {\sc aMCSusHi}~\cite{Mantler:2015vba,
  amcsushiHP, Alwall:2014hca, Harlander:2012pb}) and
\powheg~\cite{Nason:2004rx} (using the {\sc
  POWHEG-BOX}~\cite{Alioli:2010xd, powheg-box, Bagnaschi:2011tu}), and
of the NLO+NLL code \moresushi~\cite{Mantler:2012bj, moresushiHP}
based on collinear analytic resummation~\cite{Collins:1984kg,
  Bozzi:2005wk}. We consider for simplicity the process of heavy Higgs boson production in a type-II Two-Higgs-Doublet Model (\thdm).

We remark that \refS{sec:mssm-lowtanB} and part of
\refS{sec:mssm-mssmrootfiles} summarize the content of a recent
LHC-HXSWG public note, \Bref{Bagnaschi:2015hka}. \refS{sec:mssm-pt}
summarizes another recent study, \Bref{Bagnaschi:2015bop}.

\section{Benchmark scenarios for low \texorpdfstring{${\tan\beta}$}{tan beta} in the MSSM}
\label{sec:mssm-lowtanB}

The Run 1 data taken at the LHC in 2011 and 2012 have led to strong
constraints on the allowed parameter space of the
MSSM.~\footnote{Additional constraints on the MSSM parameter space
  arise, e.g., from cold dark matter density, $(g-2)_\mu$ and
  $B$-physics observables. In particular, the latter can exclude
  regions of the $(\MA,\tan\beta)$ plane in scenarios where all SUSY
  contributions decouple.  However, such indirect constraints are
  independent of -- and complementary to -- those arising from Higgs phenomenology at the LHC, and will not be discussed further here.}
These constraints are imposed by (i) the discovery of a scalar
particle with a mass of $125.09\pm0.24$\UGeV
\cite{Aad:2012tfa,Chatrchyan:2012xdj,Aad:2015zhl} and couplings
compatible with the predictions for the SM Higgs boson within an
experimental accuracy of $\pm(10\!-\!20)
\%$~\cite{Khachatryan:2014jba,Aad:2015gba,
  Khachatryan:2016vau}; (ii) the non-observation so far of additional
neutral or charged Higgs bosons in direct
searches~\cite{Khachatryan:2014wca,Aad:2014vgg,
  CMS:2014cdp,Aad:2014kga}; and (iii) the non-observation so
far of SUSY particles.

Within the MSSM, the newly discovered particle is usually interpreted
as the light neutral scalar $\Ph$, while the interpretation as the heavy
neutral scalar $\PH$ is disfavored by the
data~\cite{CMS:2014cdp,Aad:2014kga}. For the set-up and testing
of benchmark scenarios, the light-scalar mass is usually treated as a
constraint on the unknown SUSY parameters, with the requirement
\begin{equation}
 \Mh=125\pm 3\UGeV~,
 \label{eq:experimental-mass-constraint}
\end{equation}
where the $\pm 3$\UGeV\ variation corresponds to an approximate estimate of the
theoretical uncertainty of the MSSM prediction for $\Mh$, due to the
unknown effect of higher-order
corrections~\cite{Degrassi:2002fi,Allanach:2004rh}.

For $\tan\beta \gtrsim 10$ and $\MA$ in the decoupling region, the
tree-level mass of the light scalar saturates the bound $\Mh < \MZ\,$;
values of $\mSUSY$ around $1$\UTeV\ are then necessary to
reproduce the observed $\Mh$ in the maximal mixing case, whereas
multi-TeV stop masses are necessary for smaller $|X_t|$.
However, the $\tan\beta$-enhancement of the couplings of the heavy
Higgs bosons to bottom quarks and to $\PGt$ leptons leads to
significant constraints on the $(\MA,\tan\beta)$ plane from direct
searches by ATLAS and CMS~\cite{Khachatryan:2014wca, Aad:2014vgg,
  CMS:2014cdp, Aad:2014kga}. For example, in the benchmark MSSM
scenario $\mh^{\rm mod\,+}$, described in \Bref{Carena:2013ytb},
values of $\tan\beta\gtrsim 10\;(20)$ are directly excluded for
$\MA\lesssim 300\;(500)$\UGeV. For Run 2 of the LHC the allowed
parameter space is expected to shrink further, unless a discovery is
made.

For lower values of $\tan\beta$, heavy Higgs bosons with masses as low
as 200\UGeV\ are not yet excluded by direct searches at the
LHC. Moreover, thanks to the delayed approach to the decoupling limit
and to the reduced $\tan\beta$-suppression of the three-scalar
coupling, the decays
\begin{equation}
\label{interestingdecays}
\PH\,\rightarrow\,\PW\PW~,~~~~~~~~
\PH\,\rightarrow\,\PZ\PZ~,~~~~~~~~
\PH\,\rightarrow\,\Ph\Ph~,~~~~~~~~
\PA\,\rightarrow\,\PZ\Ph~,
\end{equation}
may still have significant branching ratios, especially below the
threshold for the decay to a top-quark pair (see, e.g.,
\Brefs{Arbey:2013jla,Djouadi:2013vqa}). However, as mentioned above,
lower values of $\tan\beta$ imply a reduced tree-level mass for the
lightest scalar, and hence require larger values of $\mSUSY$ entering
the radiative corrections to satisfy the mass constraint in
\Eq~(\ref{eq:experimental-mass-constraint}). For $\tan\beta$ in the
low single digits, the required hierarchy between $\mSUSY$ and $\Mt$
is so large that a fixed-order result such as the one in
\Eq~(\ref{eq:1loop-correction-mh}) would be inadequate even if
extended to two- or three-loop order, because the unknown higher-order
corrections contain higher powers of the large logarithm of
$\mSUSY/\Mt$. In this case, the large logarithmic corrections to the
Higgs boson masses should be {\em resummed} to all orders via an EFT
approach: the heavy SUSY particles are integrated out at the scale
$\mSUSY$, where appropriate boundary conditions, free of logarithmic
enhancements, are imposed on the quartic Higgs boson couplings; the latter
are evolved down to the weak scale with the corresponding
renormalization group equations (RGE); finally, the Higgs boson masses are
computed from the quartic Higgs boson couplings, including the radiative
corrections due to the contributions of the remaining light particles
at the weak scale.

An EFT calculation of the MSSM Higgs boson masses in scenarios where all the
SUSY particles (except possibly charginos and neutralinos) are far
above the TeV scale, while all Higgs bosons are below it, has recently
been completed~\cite{Lee:2015uza}, extending the earlier work in
\Bref{Draper:2013oza} to the case of two light Higgs doublets. The
boundary conditions on the quartic couplings are computed at two
loops, and the RG evolution is performed at two or three loops (the
latter only in the region where the relevant effective theory is the
SM). However, no public code implementing the results of such a
calculation is currently available. For the analysis of
low-$\tan\beta$ scenarios by ATLAS and CMS, this limitation has been
circumvented in two ways: (i) in the phenomenological {\tt hMSSM}
approach of \Brefs{Djouadi:2013vqa,Maiani:2013nga,Djouadi:2013uqa,
  Djouadi:2015jea}, which will be briefly described in
\refS{sec:mssm-hMSSM}, the experimental knowledge of $\Mh$ can be
traded -- under certain assumptions -- with the calculation of the
radiative corrections, and used to predict the remaining masses and
couplings of the MSSM Higgs bosons; (ii) in an alternative
approach~\cite{Svennote}, the accurate fixed-order calculation of the
MSSM Higgs boson masses provided by the code
\feynhiggs~\cite{Heinemeyer:1998yj, Heinemeyer:1998np,
  Degrassi:2002fi, Frank:2006yh} has been supplemented with a partial
resummation of the large logarithmic corrections~\cite{Hahn:2013ria},
and used to produce a new benchmark scenario with $\MA \leq 500$\UGeV,
$\tan\beta \leq 10\,$ and sufficiently heavy SUSY particles, whose
predictions for $\Mh$ are compatible with the requirement of
\Eq~(\ref{eq:experimental-mass-constraint}).  This scenario, referred
to as {\tt low-tb-high}, will be briefly described in
\refS{sec:mssm-low-tanb-high}.

\subsection{The hMSSM approach}
\label{sec:mssm-hMSSM}

In the hMSSM
approach~\cite{Djouadi:2013vqa,Maiani:2013nga,Djouadi:2013uqa,
  Djouadi:2015jea}, the Higgs sector of the MSSM is described in terms
of just the parameters entering the tree-level expressions for masses
and mixing, \Eqs~(\ref{treemasses}) and (\ref{alpha}), plus the
experimentally known value of $\Mh$. In this sense, the hMSSM approach
can be considered ``model independent'', because the predictions for
the properties of the MSSM Higgs bosons do not depend -- with some
caveats which will be discussed below -- on the details of the
unobserved SUSY sector.

The mass matrix for the neutral $CP$-even states can be decomposed
into a tree-level part and radiative corrections as
\begin{equation}
\label{hmatrixcorr}
{\cal M}_\Phi^2 ~=~
\left(\begin{array}{cc}
\MA^2\, \sin^2\beta + \MZ^2 \, \cos^2\beta&
-(\MA^2+\MZ^2)\,\sin\beta\cos\beta\\[2mm]
-(\MA^2+\MZ^2)\,\sin\beta\cos\beta &
\MA^2\, \cos^2\beta + \MZ^2\, \sin^2\beta
\end{array}\right)~+~
\left(\begin{array}{cc}
\Delta {\cal M}^2_{11} & \Delta {\cal M}^2_{12}\\[2mm]
\Delta {\cal M}^2_{12} & \Delta {\cal M}^2_{22}
\end{array}\right)~.
\end{equation}
The hMSSM approach is based on the following assumptions: (i) the
observed Higgs boson is the light scalar $\Ph$; (ii) of the radiative
corrections in \Eq~(\ref{hmatrixcorr}), only the element
$\Delta\mathcal{M}^2_{22}$, which contains the leading logarithmic
terms arising from top and stop loops, needs to be taken into account;
(iii) all SUSY particles are heavy enough to escape detection at the
LHC, and their effects on the Higgs sector other than those on the
mass matrix, e.g.~via direct loop corrections to the Higgs boson
couplings or via modifications of the total decay widths, can be
neglected.

With these assumptions $\Delta\mathcal{M}^2_{22}$ can be traded for
the known value of $\Mh$, inverting the relation that gives the
lightest eigenvalue of the mass matrix in \Eq~(\ref{hmatrixcorr}):
\begin{equation}
\label{DM22}
 \Delta\mathcal{M}^2_{22} =
\frac{\Mh^{2}\,(\MA^2+\MZ^{2}-\Mh^{2}) - \MA^{2}\,\MZ^{2}\,\cos^{2}2
 \beta}{\MZ^{2}\cos^{2}\beta + \MA^{2}\sin^{2}\beta - \Mh^{2}}~,
\end{equation}
which leads to the following expressions for the heavy-scalar mass
and for the mixing angle
\begin{eqnarray}
\label{hmH}
  \MH^{2} &=&
\frac{(\MA^{2}+\MZ^{2}-\Mh^{2})(\MZ^{2}\cos^{2}\beta+\MA^{2}
  \sin^{2}\beta)-\MA^{2}\,\MZ^{2}\cos^{2}2\beta}
{\MZ^{2}\cos^{2}\beta+\MA^{2}\sin^{2}\beta- \Mh^{2}}~, \\[2mm]
  \tan\alpha &=& - \frac{(\MZ^{2}+\MA^{2})\cos\beta\sin\beta}{\MZ^{2}
  \cos^{2}\beta+\MA^{2}\sin^{2}\beta-\Mh^{2}}~.
\label{halpha}
\end{eqnarray}

The mass of the charged scalars coincides with the tree-level value
$\mHpm^2 = \MA^2 + \MW^2$ in this approximation. The couplings of
the neutral Higgs bosons to fermions and to gauge bosons are fixed to
their tree-level form as in \Eq~(\ref{tab:couplingsMSSM}), but they are
expressed in terms of the effective (i.e., loop-corrected) angle
$\alpha$ obtained in \Eq~(\ref{halpha}). In contrast, the triple and
quartic Higgs self-couplings receive additional contributions. In
particular, the effective $\PH\Ph\Ph$ coupling in the hMSSM reads
\begin{equation}
\label{hlamHhh}
\lambda_{\PH\Ph\Ph}~=~\lambda_{\PH\Ph\Ph,\,{\rm tree}} \,+~
3\,\frac{\Delta{\cal M}_{22}^2}{\MZ^2}\,\frac{\sin\alpha}{\sin\beta}
\,\cos^2\alpha~,
\end{equation}
where the tree-level coupling, see \Eq~(\ref{lamHhh}), is also
expressed in terms of the effective $\alpha$, and the correction
$\Delta{\cal M}_{22}^2$ is given in \Eq~(\ref{DM22}). Under the
assumptions that characterize the hMSSM, the information encoded in
\Eqs~(\ref{hmH})--(\ref{hlamHhh}) is sufficient to determine the
production cross sections and the decay branching ratios of all the
MSSM Higgs bosons, as function of only $\MA$ and $\tan\beta$ for a
fixed value of the light-scalar mass (which we can take as
$\Mh=125$\UGeV). The precise calculation of these observables will be
described in \refS{sec:mssm-mssmrootfiles} below.

It should be noted that the hMSSM approach is well defined only in the
region of the $(\MA,\,\tan\beta)$ plane where the denominator in
\Eqs~(\ref{DM22})--(\ref{halpha}) is greater than zero (indeed, as the
denominator approaches zero $\Delta\mathcal{M}_{22}^{2}$ diverges, and
we get $\alpha\rightarrow-\pi/2$ and $\MH\rightarrow\infty$).  In
other words, for any given value of $\tan\beta$ there is a minimum
value $\MA^{\rm min}$ below which it is not possible to reproduce the
desired $\Mh$ with only a correction to the $(2,2)$ element of the
Higgs boson mass matrix. For large $\tan\beta$ one has $\MA^{\rm min}
\approx \Mh$, while for decreasing $\tan\beta$ the minimum value of
$\MA$ increases, up to $\MA^{\rm min} = (2 \,\Mh^2-\MZ^2)^{1/2}$ for
$\tan\beta=1$ (for $\Mh=125$\UGeV, this corresponds to $\MA^{\rm
  min}\approx 151$\UGeV). However, in \Bref{Djouadi:2015jea} it is
argued that the region where the hMSSM approach breaks down is already
excluded, both by direct searches for $\PSHpm$ and $\PA$ at the LHC and
by the requirement that the couplings of $\Ph$ be approximately SM-like.

The validity of the assumption (ii), that $\Delta\mathcal{M}_{11}^{2}$
and $\Delta\mathcal{M}_{12}^{2}$ can be neglected, is also discussed
in \Brefs{Djouadi:2013uqa, Djouadi:2015jea}. Direct inspection of the
dominant one-loop contributions from top/stop loops shows that the
corrections to the $(1,1)$ and $(1,2)$ elements of the Higgs boson mass
matrix are proportional to powers of the ratio $\mu\,
X_t/\mSUSY^2$. Since the sbottom contributions to those matrix
elements are not enhanced at the moderate $\tan\beta$ values of
interest here, the assumption (ii) is satisfied as soon as $\mu\,
X_t/\mSUSY^2$ is suppressed. In MSSM scenarios with $\mSUSY$ up to a
few TeV, the inclusion of the full one-loop contributions and of the
known two-loop contributions does not alter this picture. This was
shown in \Brefs{Djouadi:2013uqa, Djouadi:2015jea} via numerical
comparisons between the predictions for $\MH$ and $\alpha$ obtained
with the codes \suspect~\cite{Djouadi:2002ze} and
\feynhiggs~\cite{Heinemeyer:1998yj, Heinemeyer:1998np,
  Degrassi:2002fi, Frank:2006yh, Hahn:2013ria} and those obtained with
the hMSSM approximations, \Eqs~(\ref{hmH}) and (\ref{halpha}), using
the values of $\Mh$ produced by the codes as input. To extend this
check to the very large values of $\mSUSY$ required to obtain the
observed value of $\Mh$ at low $\tan\beta$, a comparison against the
proper EFT calculation becomes necessary. The studies in
\Bref{Lee:2015uza} indicate that, even in such heavy-SUSY scenarios,
the predictions of \Eqs~(\ref{hmH})--(\ref{hlamHhh}) agree within a few
per cent with the exact results for $\MH$, $\alpha$ and
$\lambda_{\PH\Ph\Ph}\,$, as long as $\mu\, X_t/\mSUSY^2 \lesssim 1$.

Concerning the assumption (iii), i.e.~the absence of direct SUSY
corrections to the Higgs boson couplings, we recall that the couplings to
bottom quarks are subject to potentially large, $\tan\beta$-enhanced
SUSY corrections -- often called $\Delta_b$ corrections -- which do
not decouple in the limit of heavy superparticles. However, those
corrections are not particularly relevant at the values of $\tan\beta$
considered here, and in addition they scale like $\mu/\mSUSY$,
i.e.~they could be suppressed by the same choices of SUSY parameters
that guarantee the validity of the assumption (ii).

\subsection{The \texorpdfstring{\ltbh\ }{low tanb high }scenario}
\label{sec:mssm-low-tanb-high}

\begin{figure}[t]
  \begin{center}
    \setlength{\unitlength}{\textwidth}
    \includegraphics[width=0.57\textwidth]{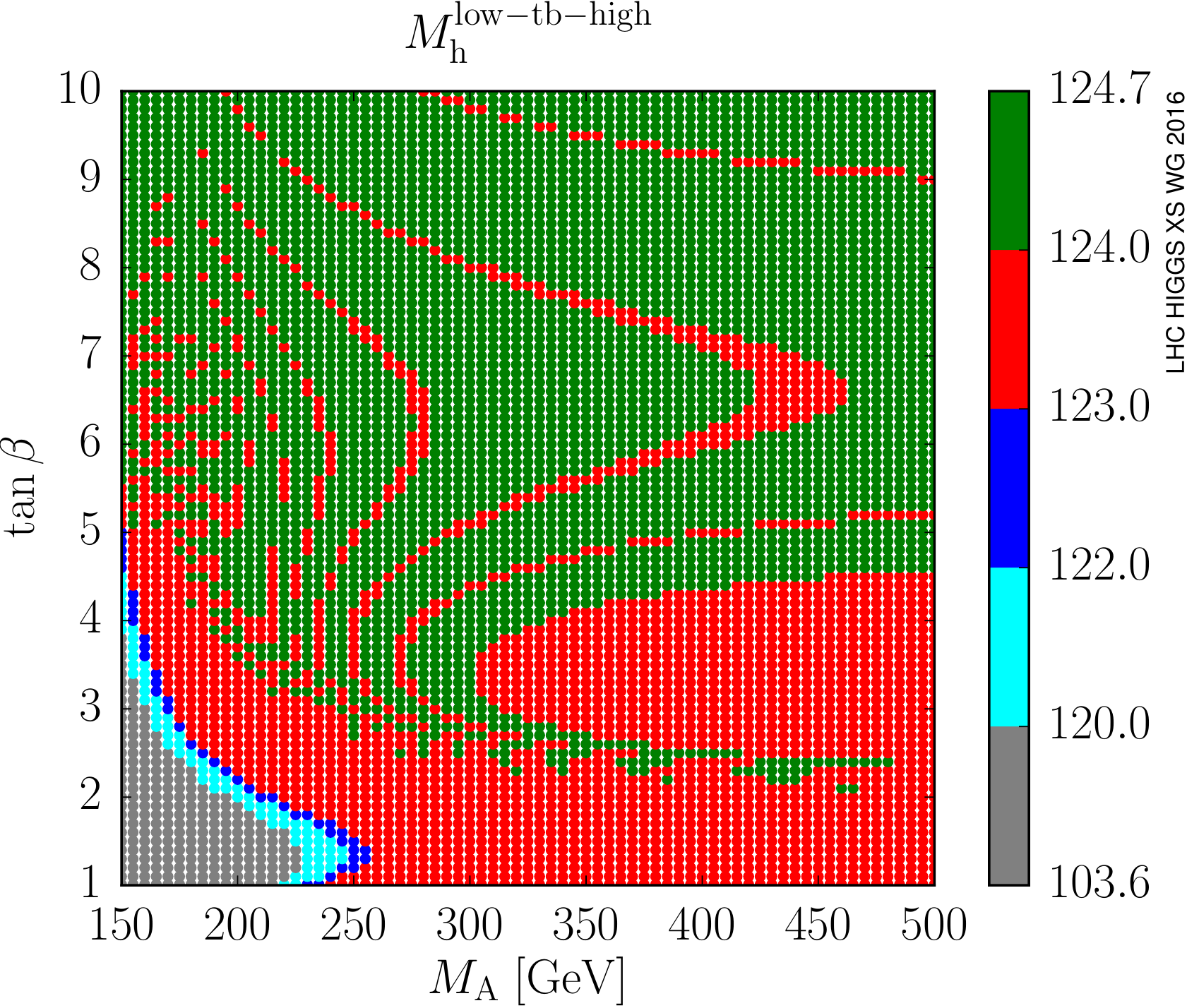}
    \caption {Mass of the light scalar $\Ph$ as computed by
      \feynhiggs~{\sc 2.10.4} in the \ltbh\ scenario, as a function of
      $\MA$ and $\tan\beta$.}
    \label{fig:low-tanb-high-mh}
  \end{center}
\end{figure}

The second approach~\cite{Svennote} to the study of low-$\tan\beta$
scenarios in the MSSM is essentially orthogonal to the one outlined in
the previous section. Instead of treating $\Mh$ as an input, and using
it to obtain a simple but approximate description of the Higgs sector
which is largely independent of the underlying SUSY parameters, one
looks for choices of SUSY parameters that, using a high-precision
calculation of the Higgs boson masses and mixing, allow to obtain the
desired value of $\Mh$ in most of the $(\MA,\,\tan\beta)$ plane.

As discussed in \refS{sec:mssm-intro}, for low $\tan\beta$ the values
of $\mSUSY$ required to obtain $\Mh \approx 125$\UGeV\ are so large
that a fixed-order calculation of the Higgs boson masses becomes inadequate,
and a resummation of the large logarithmic corrections is
unavoidable. Starting from version {\sc 2.10.0}, the public code
\feynhiggs~\cite{Heinemeyer:1998yj, Heinemeyer:1998np,
  Degrassi:2002fi, Frank:2006yh} does include such
resummation~\cite{Hahn:2013ria}, with some limitations that will be
discussed below. The so-called \ltbh\ scenario is defined for
$0.5\leq\tan\beta\leq10$ and $150\UGeV\leq \MA\leq 500\UGeV$, and the
masses and mixing of all the MSSM Higgs bosons are computed with
version {\sc 2.10.4} of \feynhiggs.

To obtain values of $\Mh$ in the desired range, the SUSY parameters --
in the on-shell scheme adopted by \feynhiggs\ -- are chosen as
follows:
(i) all soft SUSY-breaking masses for the sfermions (both squarks and
sleptons) as well as the gluino mass are set equal to $\mSUSY$;
(ii) $\mSUSY$ is varied between few TeV (for large values of
$\MA$ or $\tan\beta$) and up to $100$\UTeV\ (for small values of $\MA$
or $\tan\beta$), keeping the following relations between $X_{t}$,
$\mSUSY$ and $\tan\beta$:
\begin{equation*}
 \begin{split}
  \tan\beta\leq2     \hspace{0.6cm} &: \qquad X_{t}/\mSUSY = 2 ; \\
  2<\tan\beta\leq8.6 \hspace{0.3cm} &: \qquad X_{t}/\mSUSY
= 0.0375\,\tan^{2}\beta-0.7\,\tan\beta+3.25; \\
  8.6<\tan\beta      \hspace{1.3cm} &: \qquad X_{t}/\mSUSY = 0 ; \\
 \end{split}
 \label{eq:low-tanb-high-variaton}
\end{equation*}
(iii) for what concerns the remaining SUSY parameters, all
Higgs-sfermion trilinear couplings other than $A_t$ are set to
$2$\UTeV, $\mu$ is set to = $1.5$\UTeV\ and the $SU(2)$ gaugino mass
$M_{2}$ is set to $2$\UTeV\ (this fixes also the $U(1)$ gaugino mass
$M_{1}$ via the GUT relation $M_1/M_2 = 5/3\,\tan^2\theta_W$). With
these choices of SUSY parameters, the prediction of \feynhiggs\ for
the light-scalar mass $\Mh$ is shown in \refF{fig:low-tanb-high-mh} as
a function of $\MA$ and $\tan\beta$. As can be seen, the requirement
of \Eq~(\ref{eq:experimental-mass-constraint}) can be met over most of
the parameter space, with the exception of the lower-left corner
corresponding to very low values of both $\MA$ and
$\tan\beta$. However, it has not been tested whether the predictions
for the production cross section and the branching ratios of the light
scalar are in full agreement with the latest results of the ATLAS and
CMS collaborations~\cite{Khachatryan:2014jba,Aad:2015gba},
which indicate an SM-like Higgs boson with uncertainties in the
$(10\!-\!20)\%$ range.
For the heavy Higgs bosons, the chosen values of $\mu$ and $M_2$
ensure that all decays to charginos and neutralinos (henceforth,
electroweakinos or EW-inos) are kinematically forbidden, thus maximizing
the branching ratios for the decays in \Eq~(\ref{interestingdecays}).

A limitation of the \ltbh\ scenario should be taken into account.  The
resummation procedure currently implemented in \feynhiggs\ -- which
accounts only for the leading and next-to-leading logarithmic
corrections to the Higgs boson masses controlled by the strong gauge
coupling and by the top Yukawa coupling -- relies on the assumption
that all SUSY masses as well as the heavy-Higgs boson masses are of the
order of $\mSUSY$. However, in the \ltbh\ scenario the parameters
$\mu$ and $M_{1,2}$ are fixed to ${\cal O}$(TeV), and $\MA$ is below
$500$\UGeV. To assess the adequacy of \feynhiggs\ in an MSSM scenario
with heavy sfermions and gluinos but relatively light EW-inos and
additional Higgs bosons, a comparison with a proper EFT calculation --
where the effective theory below $\mSUSY$ is a \thdm\ augmented with
EW-inos -- would be necessary.

The studies in \Bref{Lee:2015uza} indicate that, with the choices of
SUSY parameters of the \ltbh\ scenario, the EFT predictions for $\Mh$
can be considerably lower than those of the current \feynhiggs\
implementation.
In particular, for $\tan\beta >5.5$ the EFT calculation yields values
of $\Mh$ that are about $2$\UGeV\ lower than those obtained by
\feynhiggs. For lower values of $\tan\beta$, the disagreement is more
severe: the difference is greater than $5$\UGeV\ ($10$\UGeV) for
$\tan\beta <3.5$ ($\tan\beta <2$). When one looks at the minimal value
of $\mSUSY$ required to obtain $\Mh$ in the desired range, the
logarithmic dependence of $\Mh$ on $\mSUSY$ amplifies the discrepancy.
For $\tan\beta=2$, the EFT calculation requires $\mSUSY >
10^{8}$\UTeV\ ($\mSUSY> 200$\UTeV) to obtain $\Mh > 122$\UGeV\ with
$\MA = 200$\UGeV\ ($\MA = 500$\UGeV). This should be compared with
the maximal value $\mSUSY= 100$\UTeV\ adopted in the \ltbh\ scenario
for the lowest values of $\MA$ and $\tan\beta$.
Discrepancies up to $(10\!-\!12)\%$ between \feynhiggs\ and the EFT
calculation can also be found in the predictions for $\MH$ and
$\alpha$ at very small values of $\MA$ and $\tan\beta$.
Further investigation will be required to ascertain how these
discrepancies are related to the presence of a light \thdm, to the
presence of light EW-inos, and to other aspects of the calculation
such as the determination of the top Yukawa coupling.

On the other hand, the fact that $\mu \muchless \mSUSY$ over the whole
parameter space ensures that the dominant top/stop corrections to the
elements other than $(2,2)$ of the $CP$-even Higgs boson mass matrix are
suppressed. Therefore, a meaningful comparison with the results
obtained in the hMSSM approach can be performed, as will be discussed
in \refS{sec:mssm-comparison}.


\section{\texorpdfstring{\rootf\ }{ROOT }files for cross sections and branching ratios}
\label{sec:mssm-mssmrootfiles}

For Run 1 of the LHC, benchmark scenarios for the MSSM were
delivered by the LHC-HXSWG in the form of \rootf\ files, which
contain masses, inclusive production cross sections and branching
ratios of the Higgs bosons. The files can be downloaded from the
webpages of the MSSM subgroup~\cite{mssmsubgroupwebpage}.  Their
structure and content were significantly updated for Run 2 of the LHC,
and in this section we describe the changes in some detail.  We first
focus on the content of the files, including a description of how the
relevant quantities were calculated.  Afterwards the structure of the
files and the possible ways to access the included data are
explained. Finally we describe a comparison of the \rootf\ files for
the hMSSM and the \ltbh\ scenario.

\subsection{Content of the \texorpdfstring{\rootf\ }{ROOT }files}

\begin{table}
\label{tab:mssmbenchmarkscenarios}
\caption{Benchmark scenarios provided on the webpage of the MSSM
  subgroup of the LHC-HXSWG~\cite{mssmsubgroupwebpage}
  in the form of new \rootf\ files for Run 1 and Run 2 of the LHC.
  We give information about the ranges of $\MA$ and $\tan\beta$,
  and the centre-of-mass energies $\sqrt{s}$. The $\mh^{\rm mod+}$ scenario
  is delivered for different values of $\mu\in
  \mu^{\rm val}=\left\{ -1000,-500,-200,200,500,1000\right\} $\UGeV\
  with the restriction $\tan\beta<40$ for $\mu=-1000$\UGeV. In the
  light-stop scenario the gaugino and higgsino mass parameters are fixed
  as $M_1 = 350$\UGeV, $M_2=\mu=400$\UGeV\ to avoid the bounds from direct
  stop searches.}
\begin{center}
    \begin{tabular}{lccc}
      \toprule
       scenario & $\MA$\,[GeV] & $\tan\beta$ & $\sqrt{s}$~[TeV] \\
      \midrule
      \ltbh\ \cite{Bagnaschi:2015hka}                               &  $150-500$   &  $0.5-10$  & $8,13$  \\[2mm]
      hMSSM~\cite{Djouadi:2013vqa,Maiani:2013nga,Djouadi:2013uqa,Djouadi:2015jea}   &  $130-1000$  &  $1-60$    & $8,13$  \\[2mm]
      $\mh^{\rm max}$~\cite{Carena:2013ytb}                         &  $90-2000$   &  $0.5-60$  & $13,14$ \\[2mm]
      $\mh^{\rm mod+}$~\cite{Carena:2013ytb}, $\mu\in \mu^{\rm val}$&  $90-2000$   &  $0.5-60$  & $8,13,14$ \\[2mm]
      $\mh^{\rm mod-}$~\cite{Carena:2013ytb}                        &  $90-2000$   &  $0.5-60$  & $13,14$ \\[2mm]
      light stau~\cite{Carena:2013ytb}                              &  $90-2000$   &  $0.5-60$  & $13,14$ \\[2mm]
      light stop~\cite{Carena:2013ytb}                              &  $90-650$  &  $0.5-60$  & $13,14$ \\[2mm]
      $\PGt$-phobic~\cite{Carena:2013ytb}                           &  $90-2000$   &  $0.5-50$  & $13,14$ \\[2mm]
      \bottomrule
    \end{tabular}
\end{center}
\end{table}

All \rootf\ files contain the Higgs boson masses, cross sections (including
estimates of the theoretical uncertainties via scale variations) and
branching ratios for a grid of $\MA$ and $\tan\beta$
values. \refT{tab:mssmbenchmarkscenarios} provides an overview of the
delivered benchmark scenarios, including the ranges of $\MA$ and
$\tan\beta$ and the centre-of-mass energies for which the cross
sections were produced. In the following we describe how the
quantities included in the \rootf\ files were obtained. We detail the
calculations of the Higgs boson masses, cross sections, and decay widths and
branching ratios, and then we describe the SM input parameters used by
the codes.

For the benchmark scenarios already presented in the third report of
the LHC-HXSWG~\cite{Heinemeyer:2013tqa}, which are based on
\Bref{Carena:2013ytb}, the neutral-scalar masses, the Higgs mixing
angle $\alpha$ and the charged Higgs boson mass are computed with
\feynhiggs~{\sc 2.10.2}. In contrast, for the \ltbh\ scenario the
masses and mixing are computed with \feynhiggs~{\sc 2.10.4},
activating the option to resum large logarithms due to the heavy SUSY
scale.  Finally, in the \rootf\ files for the hMSSM the mass of the
light neutral scalar is fixed to $\Mh=125$\UGeV, $\MH$ and $\alpha$
are computed as function of $\MA$ and $\tan\beta$ using
\Eqs~(\ref{hmH}) and (\ref{halpha}), and the charged Higgs boson mass is
obtained from the tree-level relation $\mHpm^2 = \MA^2+\MW^2$.

We use \sushi~{\sc 1.5.0}~\cite{Harlander:2012pb} to obtain the
inclusive cross sections for the production of the neutral Higgs bosons $\phi \equiv (\Ph, \PH, \PA)$ via gluon fusion,
$\Pg\Pg\rightarrow \phi$, and bottom annihilation, $\PAQb \PQb
\rightarrow \phi$.  In preparation for Run 2 of the LHC the \rootf\
files are produced for centre-of-mass energies of $13$\UTeV\ and
$14$\UTeV, with the exception of the scenarios for the low-$\tan\beta$
studies, available for $8$ and $13$\UTeV, and the $\mh^{\rm mod+}$
scenario with different values of $\mu$, available for $8$, $13$ and
$14$\UTeV. In the calculation of the gluon-fusion cross section,
\sushi\ implements the NLO-QCD top and bottom contributions from
\Brefs{Spira:1995rr, Harlander:2005rq}.  Moreover, NNLO-QCD top
contributions from \Brefs{Harlander:2002wh, Harlander:2002vv} are
taken into account in the heavy-top limit, and the electroweak
contributions by light quarks from \Brefs{Aglietti:2004nj,
  Bonciani:2010ms} are added.  In the benchmark scenarios of
\Bref{Carena:2013ytb}, the squark contributions are taken into account
in different limits: For the heavy scalar and the pseudoscalar,
\sushi\ employs the NLO virtual corrections in the limit of heavy SUSY
masses from \Brefs{Degrassi:2010eu, Degrassi:2011vq, Degrassi:2012vt}.
For the light scalar, \sushi\ employs the results implemented in {\sc
  evalcsusy}~\cite{Harlander:2003bb, Harlander:2004tp,
  Harlander:2005if}, which were obtained in the limit of vanishing
Higgs boson mass, for the NLO corrections involving stops, and the results
of \Bref{Degrassi:2010eu} for those involving sbottoms. In addition,
\sushi\ takes into account the resummation of $\tan\beta$-enhanced
SUSY corrections in the Higgs-bottom coupling, with a correction
factor $\Delta_b$ computed by \feynhiggs.  In the \ltbh\ scenario,
characterized by very heavy squarks, the squark-loop contributions to
gluon fusion are negligible and are omitted, but the non-decoupling
$\Delta_b$ corrections are included. Finally, in the case of the hMSSM
\sushi\ is Run with a \thdm\ input file providing the Higgs boson masses and
the Higgs mixing angle, thus no SUSY contributions to the cross
section are included.

For bottom annihilation, \sushi\ calculates the inclusive cross
section at NNLO-QCD in the five-flavour scheme (5F), based on
\Bref{Harlander:2003ai}. For each neutral Higgs boson $\phi$, the
amplitude for the production of a SM Higgs boson with mass $\mphi$ is
reweighted with the effective Higgs-bottom coupling. As for gluon
fusion, the $\Delta_b$ corrections are included in all scenarios
except the hMSSM.  The \rootf\ files also include reweighted
four-flavour scheme (4F) cross sections for bottom-quark associated
production, $\Pg\Pg\rightarrow \PQb\PAQb\, \phi$, based on
\Brefs{Dittmaier:2003ej,Dawson:2003kb}.  These 4F cross sections now
also include a mixed top- and bottom-quark induced contribution, which
is reweighted with effective up- and down-type quark couplings, taken
from \sushi.  Finally, the 4F and 5F descriptions can be combined into
the ``Santander matched'' cross sections~\cite{Harlander:2011aa} by
the output routines delivered for the \rootf\ files.

The \rootf\ files also include estimates for the theoretical
uncertainties of the cross sections.  The central values of the
renormalization scale $\mu_{\scriptscriptstyle R}$ and the
factorization scale $\mu_{\scriptscriptstyle F}$ are set to
$\mu_{\scriptscriptstyle R}=\mu_{\scriptscriptstyle F}=\mphi/2$ for
gluon fusion and to $\mu_{\scriptscriptstyle
  R}=4\,\mu_{\scriptscriptstyle F}= \mphi$ for bottom
annihilation. Scale uncertainties are then determined from the
envelope of seven independent variations of $\mu_{\scriptscriptstyle
  R}$ and $\mu_{\scriptscriptstyle F}$ by factors of $2$, with the
additional constraint $1/2\leq\mu_{\scriptscriptstyle
  R}/\mu_{\scriptscriptstyle F}\leq2$ for gluon fusion and
$2\leq\mu_{\scriptscriptstyle R}/\mu_{\scriptscriptstyle F}\leq8$ for
bottom annihilation. For the parton distribution functions the {\sc
  MSTW2008}~\cite{Martin:2009iq} set has been used, and the residual
uncertainties on the parton distribution functions and on the strong
coupling constant, $\alpha_{s}$, are obtained from the corresponding
relative uncertainties for a SM Higgs boson of mass $\mphi$,
evaluated as proposed in \Bref{pdf-uncerts}. Note however that
PDF and $\alpha_s$ uncertainties are only available for gluon fusion
and bottom annihilation~(5F), not for bottom associated
production~(4F).

For the calculation of branching ratios and their inclusion in the
\rootf\ files there is a substantial difference between standard MSSM
benchmark scenarios and the hMSSM.  For the standard MSSM benchmark
scenarios, including the \ltbh\ scenario, the branching ratios are
computed as recommended by the
LHC-HXSWG~\cite{Heinemeyer:2013tqa}. They are thus a combination of
the results of \hdecay\ for the decays to quark pairs with the results
of \feynhiggs\ for the remaining decays. In particular, for the decays
to massive gauge bosons \feynhiggs\ approximates the MSSM results by
reweighting the SM results from the code
\prophecy~\cite{Bredenstein:2006rh, Bredenstein:2006ha} with the
appropriate Higgs-gauge boson coupling.  For the decays to Higgs bosons, \feynhiggs\ implements a full one-loop calculation within the
(complex) MSSM~\cite{Williams:2011bu}, improved -- starting from
version {\sc 2.10.4} -- with the resummation of potentially large
logarithmic corrections to the decay $\PH\rightarrow \Ph\Ph$ that are
relevant for the \ltbh\ scenario. Finally, the \rootf\ files for the
standard MSSM benchmark scenarios also include the branching ratios
for the decays of the charged Higgs boson, as well as the branching
ratio for the decay $\PQt\rightarrow \PSHp \PQb$ (relevant for low
masses of the charged Higgs boson), all obtained from \feynhiggs.
In contrast, in the \rootf\ files for the hMSSM the branching ratios
for the decays of the neutral and charged Higgs bosons, as well as for
$\PQt\rightarrow \PSHp \PQb$, are solely computed with the code
\hdecay~\cite{Djouadi:1997yw, Spira:1997dg, Djouadi:2006bz}, which
starting from version {\sc 6.40} allows for hMSSM input.  \hdecay\
implements N$^4$LO-QCD corrections to the decays to quark
pairs~\cite{Braaten:1980yq, Sakai:1980fa, Inami:1980qp, Drees:1989du,
  Drees:1990dq, Gorishnii:1990zu, Gorishnii:1991zr, Kataev:1993be,
  Gorishnii:1983cu, Surguladze:1994gc, Larin:1995sq, Chetyrkin:1995pd,
  Chetyrkin:1996sr, Baikov:2005rw}; LO results for the decays to
lepton pairs and for the decays involving massive gauge bosons, both
on-shell and off-shell; a LO calculation of the decays to Higgs boson
pairs, both on-shell and off-shell, using effective hMSSM couplings.

The SM parameters used as input by the codes that compute the Higgs boson masses, production and decays mostly coincide with the recommendations
of the LHC-HXSWG, see \Bref{LHCHXSWG-INT-2015-006} and Chapter~\ref{chapter:input},  with the following exceptions: for the calculation of Higgs boson
masses in \feynhiggs\ and the calculation of branching ratios in
\feynhiggs\ and \hdecay\ $\alpha_s(\MZ)=0.119$ is used. The cross
sections are calculated by \sushi\ with the strong coupling constant
$\alpha_s$ taken from the PDF set employed.  In both \feynhiggs\ and
\sushi, the $\msbar$ masses of the bottom and charm quarks are set to
$\mbmb = 4.16$\UGeV\ and $\mcmc=1.28$\UGeV, respectively, and the
$\PW$~boson mass is set to $\MW=80.398$\UGeV. In addition, \sushi\
requires the pole bottom mass, set to $\Mb^{\rm pole}=4.75$\UGeV,
and \feynhiggs\ requires the widths of the massive gauge bosons, set
to $\GW=2.118$\UGeV\ and $\GZ=2.4952$\UGeV, and the $\tau$ lepton
mass, set to $\Mtau=1777.03$\UMeV.  In contrast, \hdecay\ employs
$\Mb^{\rm pole}=4.49$\UGeV, $\Mc^{\rm pole}=1.42$\UGeV,
$\MZ=91.15349$\UGeV, $\MW=80.36951$\UGeV, $\GZ=2.49581$\UGeV\ and
$\GW=2.08856$\UGeV, corresponding to the on-shell parameters in the
complex-mass scheme.

\subsection{Technical details and data access}

The setup to combine Higgs boson masses, cross sections and branching ratios
into two-dimensional \rootf\ histograms for MSSM benchmark scenarios
was reorganized in 2015.  We will now summarize the procedure to
access the data stored in the \rootf\ files. The user can download the
{\tt C++} class {\tt mssm\_xs\_tools.C} and the header file {\tt
  mssm\_xs\_tools.h} from the webpage of the MSSM subgroup. Together
with a scenario, here named ${\tt ``scenario.root"}$, the class is
initialized within \rootf\ through
\begin{equation*}
{\tt mssm\_xs\_tools \quad mssm(``scenario.root",INT,0).}
\end{equation*}
Setting {\tt INT} to {\tt true} enables linear
interpolation between the grid points in $\MA$ and $\tan\beta$,
providing results for intermediate values of $\MA$ or $\tan\beta$.
The last argument with default setting {\tt 0} controls the printout
level for debugging. After the initialization, data can be
accessed either through the predefined routines listed in the header
file or through the commands
\begin{equation*}
{\tt mssm.COMMAND(STRING,}\,\MA,\tan\beta{\tt )}\qquad,
\end{equation*}
where {\tt COMMAND} can be one of the member functions {\tt
  mass,\,width,\,br} or {\tt xsec} for the Higgs boson masses, the total
decay widths, branching ratios or cross sections,
respectively. Examples for {\tt STRING} are
\begin{equation*}
\begin{split}
&{\tt mssm.mass(``H",}\,\MA,\tan\beta{\tt )}\\
&{\tt mssm.br(``A\to tautau",}\,\MA,\tan\beta{\tt )}\\
&{\tt mssm.xsec(``gg\to H::scaleUp",}\,\MA,\tan\beta{\tt )}\quad,
\end{split}
\end{equation*}
where the latter provides the upper bound of the uncertainty of the
gluon fusion cross section $\Pg\Pg\rightarrow \PH$ obtained from the
variation of the renormalization and factorization scales.  A complete
list of possible strings can again be deduced from the class
description in the header file.  Alternatively, a {\sc python} wrapper
is available on the webpage.  An example for its usage is included at
the end of the {\sc python} file.

\subsection{Comparison of benchmark scenarios for low \texorpdfstring{$\tan\beta$}{tan beta}}
\label{sec:mssm-comparison}

As mentioned in \refS{sec:mssm-low-tanb-high}, the choices of SUSY
parameters in the \ltbh\ scenario satisfy all of the assumptions that
underlie the hMSSM, thus inviting a comparison between the predictions
for the Higgs boson properties obtained within the two approaches.
However, a direct comparison between the two sets of \rootf\ files is
hindered by the fact that the light-scalar mass is fixed as
$\Mh=125$\UGeV\ in the hMSSM files, whereas it varies with $\MA$ and
$\tan\beta$ in the \ltbh\ files.  To circumvent this problem, the
predictions of the \ltbh\ scenario for $\MH$, $\alpha$ and the
branching ratios were compared with the corresponding results obtained
in the hMSSM approach taking as input the values of $\Mh$ from the
\ltbh\ scenario and computing all branching ratios with \hdecay.

In the left and right panels of \refF{fig:hMSSM-vs-ltbh_mH-alpha} we
show the relative differences between the predictions of the \ltbh\
scenario and those of the hMSSM for $\MH$ and $\alpha$, respectively,
on the $(\MA,\,\tan\beta)$ plane with $150\UGeV\leq \MA \leq500\UGeV$
and $1\leq\tan\beta\leq10$. The figure shows that, for the SUSY
parameters that characterize the \ltbh\ scenario, the results of
\feynhiggs\ for $\MH$ and $\alpha$ and the approximate results
obtained via \Eqs~(\ref{hmH}) and (\ref{halpha}) differ by less than
$1\%$ over most of the parameter space. Larger discrepancies, up to a
few per cent, occur only in the lower-left corner at very low $\MA$ and
$\tan\beta$.  In view of this good accord, we can expect any
significant discrepancy in the predictions for cross sections and
branching ratios to be due to differences in the calculation of the
physical observables themselves, rather than to the approximation in
\Eqs~(\ref{hmH}) and (\ref{halpha}). While the production cross
sections are computed with \sushi\ in both cases, discrepancies can
arise in the widths for the decays $\PH\rightarrow
\PW\PW,\PZ\PZ,\Ph\Ph$ and $\PA\rightarrow \PZ\Ph$, which in the \ltbh\
and hMSSM files are computed with \feynhiggs+\prophecy\ and with
\hdecay, respectively.

\begin{figure}[t]
  \begin{center}
    \setlength{\unitlength}{\textwidth}
    \includegraphics[width=0.47\textwidth]{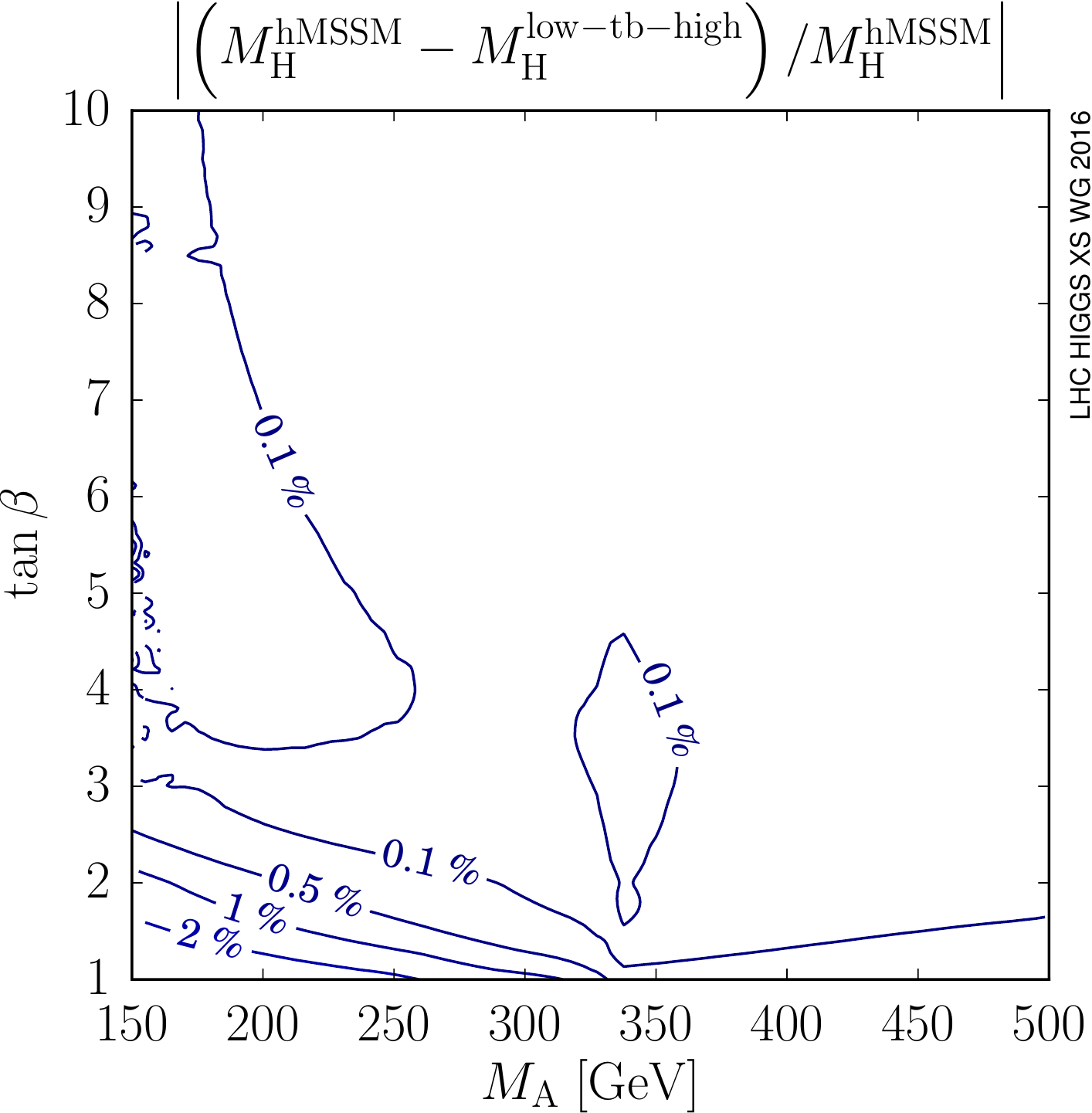}~~~~~~~
    \includegraphics[width=0.47\textwidth]{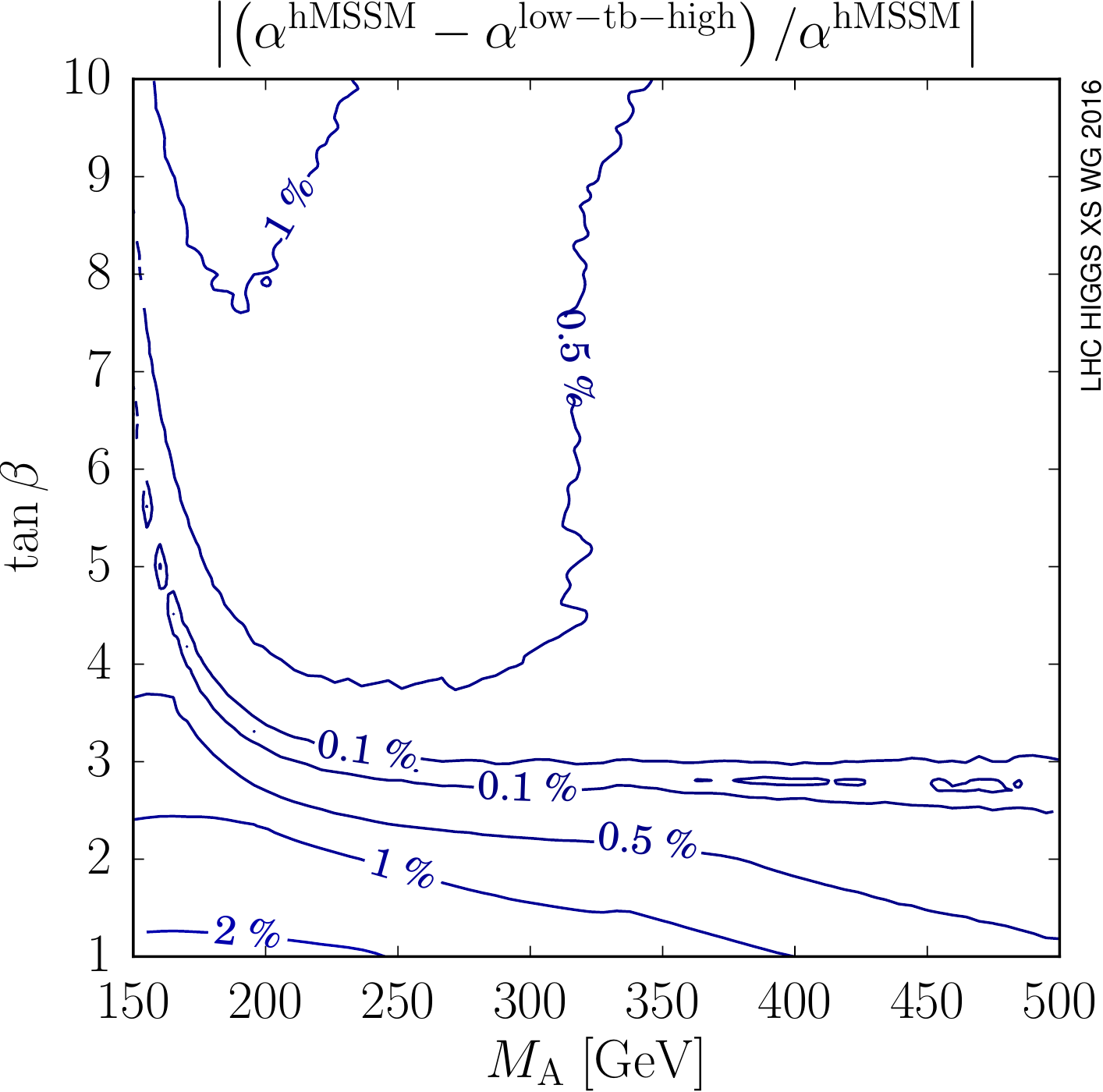}
    \caption{Relative differences in $\MH$ (left) and $\alpha$ (right)
      between the predictions of \feynhiggs\ for the \ltbh\ scenario
      and the corresponding predictions obtained in the hMSSM approach
      via \Eqs~(\ref{hmH}) and (\ref{halpha}), starting from the values
      of $\Mh$ computed by \feynhiggs.}
    \label{fig:hMSSM-vs-ltbh_mH-alpha}
  \end{center}
\vspace*{-2.5mm}
\end{figure}

\begin{figure}[p]
  \begin{center}
    \setlength{\unitlength}{\textwidth}
    \includegraphics[width=0.47\textwidth]{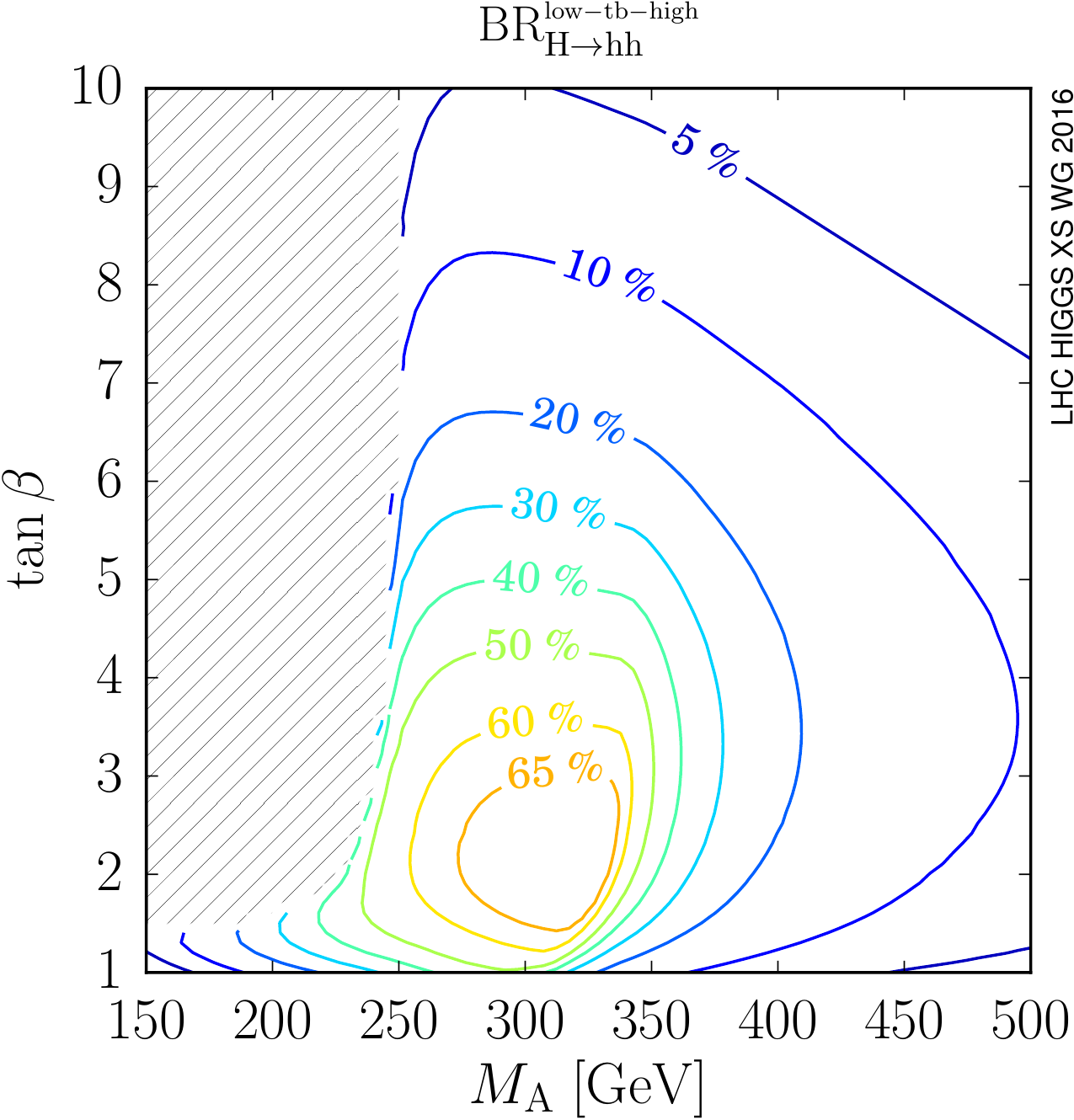}~~~~~~~
    \includegraphics[width=0.47\textwidth]{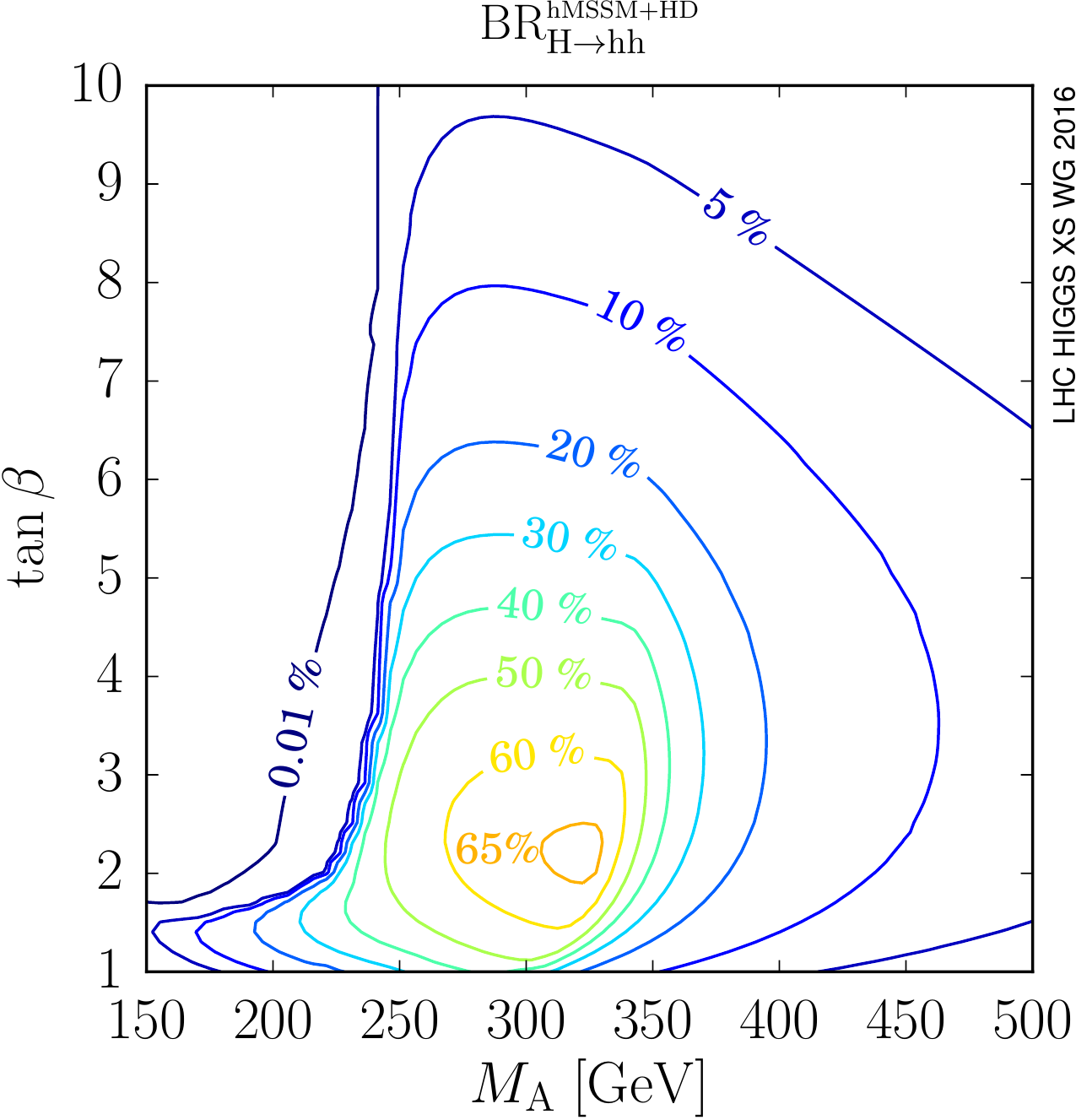}
    \caption{Left: Branching ratio for the decay $\PH\rightarrow \Ph\Ph$ as
      computed in the \ltbh\ scenario following the LHC-HXSWG
      recommendations for the decay widths (in particular,
      $\Gamma(\PH\rightarrow \Ph\Ph)$ is computed with \feynhiggs). Right:
      The same branching ratio obtained with the hMSSM+\hdecay\
      combination -- namely, starting from the values of $\Mh$
      computed by \feynhiggs\ in the \ltbh\ scenario, then computing
      the branching ratio with \hdecay, which obtains $\MH$, $\alpha$
      and $\lambda_{\PH\Ph\Ph}$ from the hMSSM prescriptions in
      \Eqs~(\ref{hmH})--(\ref{hlamHhh}).}
    \label{fig:hMSSM-and-ltbh_HH_BRs}
  \end{center}
\end{figure}

\begin{figure}[p]
  \begin{center}
    \setlength{\unitlength}{\textwidth}
    \includegraphics[width=0.47\textwidth]{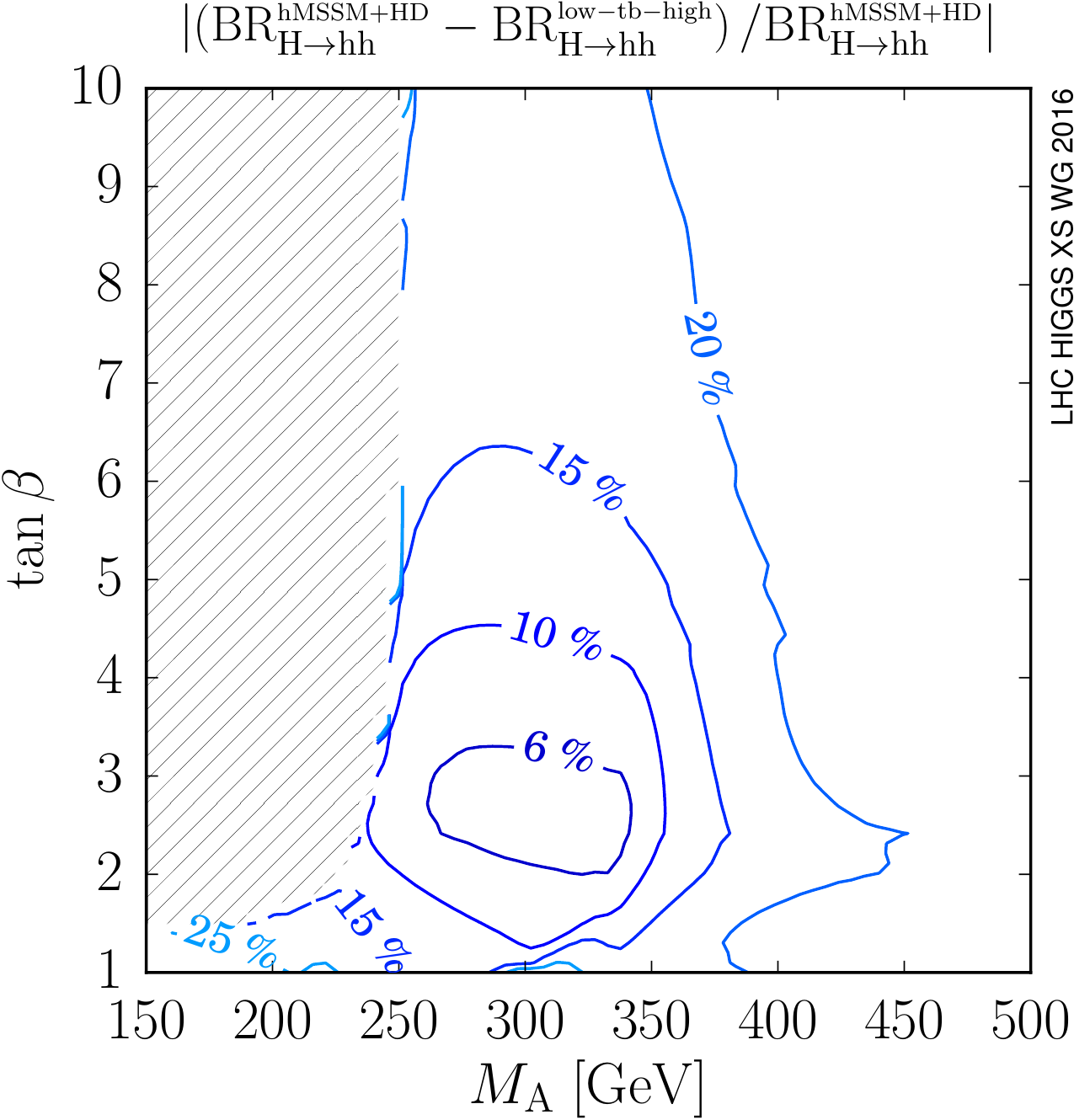}~~~~~~~
    \includegraphics[width=0.47\textwidth]{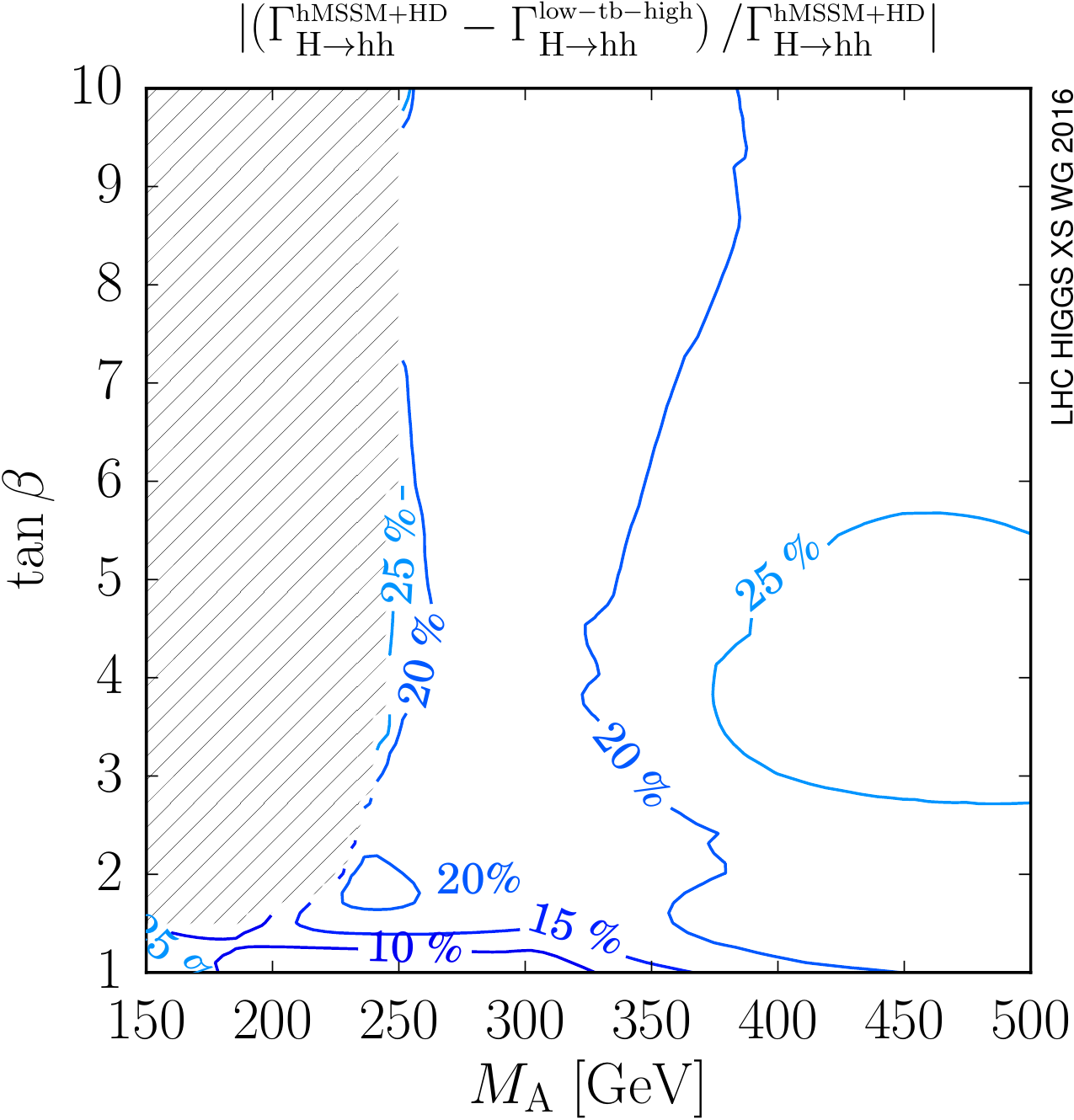}
    \caption{Relative differences in ${\rm BR}(\PH\rightarrow \Ph\Ph)$
      (left) and $\Gamma(\PH\rightarrow \Ph\Ph)$ (right) between the
      predictions of the \ltbh\ scenario (where $\Gamma(\PH\rightarrow
      \Ph\Ph)$ is computed with \feynhiggs) and the corresponding
      predictions obtained with the hMSSM+\hdecay
      combination. For the latter we start from the values of $\Mh$
      computed by \feynhiggs\ in the \ltbh\ scenario, then we
      compute width and branching ratio with \hdecay, which
      obtains $\MH$, $\alpha$ and $\lambda_{\PH\Ph\Ph}$ from the hMSSM
      prescriptions in \Eqs~(\ref{hmH})--(\ref{hlamHhh}).}
    \label{fig:hMSSM-vs-ltbh_HH_BRs}
  \end{center}
\end{figure}

The left and right panels in \refF{fig:hMSSM-and-ltbh_HH_BRs} show the
branching ratio for the decay $\PH\rightarrow \Ph\Ph$ in the \ltbh\
scenario and in the hMSSM+\hdecay\ combination, respectively. Again,
in the hMSSM plot the mass $\Mh$ used to compute $\MH$ and $\alpha$
via \Eqs~(\ref{hmH}) and (\ref{halpha}) in a given point of the
$(\MA,\,\tan\beta)$ plane has been adjusted to the value computed by
\feynhiggs\ in the corresponding point of the \ltbh\ scenario. In the
hatched region on the left plot the decay is below threshold, and the
corresponding width is set to zero by {\sc FeynHiggs} (in contrast, in
the right plot \hdecay\ computes also the small width to off-shell
scalars). The plots show that, in this scenario, ${\rm
  BR}(\PH\rightarrow \Ph\Ph)$ can be larger than $50\%$ for
$\tan\beta\lesssim 4$ and for values of $\MA$ such that $\MH$ sits
between the kinematic threshold for the decay to a light-scalar pair
and the one for the decay to a top-quark pair. A visual comparison of
the left and right plots also shows that the qualitative dependence of
${\rm BR}(\PH\rightarrow \Ph\Ph)$ on $\MA$ and $\tan\beta$ is the same
in both approaches, but the branching ratio takes on somewhat larger
values in the \ltbh\ plot than it does in the hMSSM plot.

To quantify the previous statement, the left plot in
\refF{fig:hMSSM-vs-ltbh_HH_BRs} shows the relative difference between
the values of ${\rm BR}(\PH\rightarrow \Ph\Ph)$ computed in the \ltbh\
scenario and those computed in the hMSSM with \hdecay.  The plot shows
that the discrepancy in the branching ratio is less than $10\%$ in the
region where the decay $\PH\rightarrow \Ph\Ph$ is dominant, and
exceeds $20\%$ for larger values of $\MA$ and, hence, $\MH$. However,
in the region where a decay channel is dominant a comparison between
branching ratios can mask the true extent of a discrepancy.  The right
plot of \refF{fig:hMSSM-vs-ltbh_HH_BRs} shows instead the relative
difference between the corresponding values of the decay width
$\Gamma(\PH\rightarrow \Ph\Ph)$. It appears that, at the level of the
decay width, the discrepancy between the results obtained in the two
approaches is above $15\%$ in most of the relevant parameter space,
and exceeds $25\%$ for large $\MA$ and intermediate $\tan\beta$.  The
size of the discrepancy can be understood in view of the different
accuracy of the $\Gamma(\PH\rightarrow \Ph\Ph)$ calculation in the two
approaches. Indeed, while in the hMSSM the effect of the top/stop
contributions is included via the effective coupling in
\Eq~(\ref{hlamHhh}), \feynhiggs\ implements a full calculation of the
one-loop corrections to the decay width, supplemented with the
resummation of large logarithmic terms. Thus, the \feynhiggs\ result
accounts for potentially large threshold effects in diagrams with
loops of SM particles, which are not captured by using an effective
coupling alone.

\begin{figure}[p]
  \begin{center}
    \setlength{\unitlength}{\textwidth}
    \includegraphics[width=0.47\textwidth]{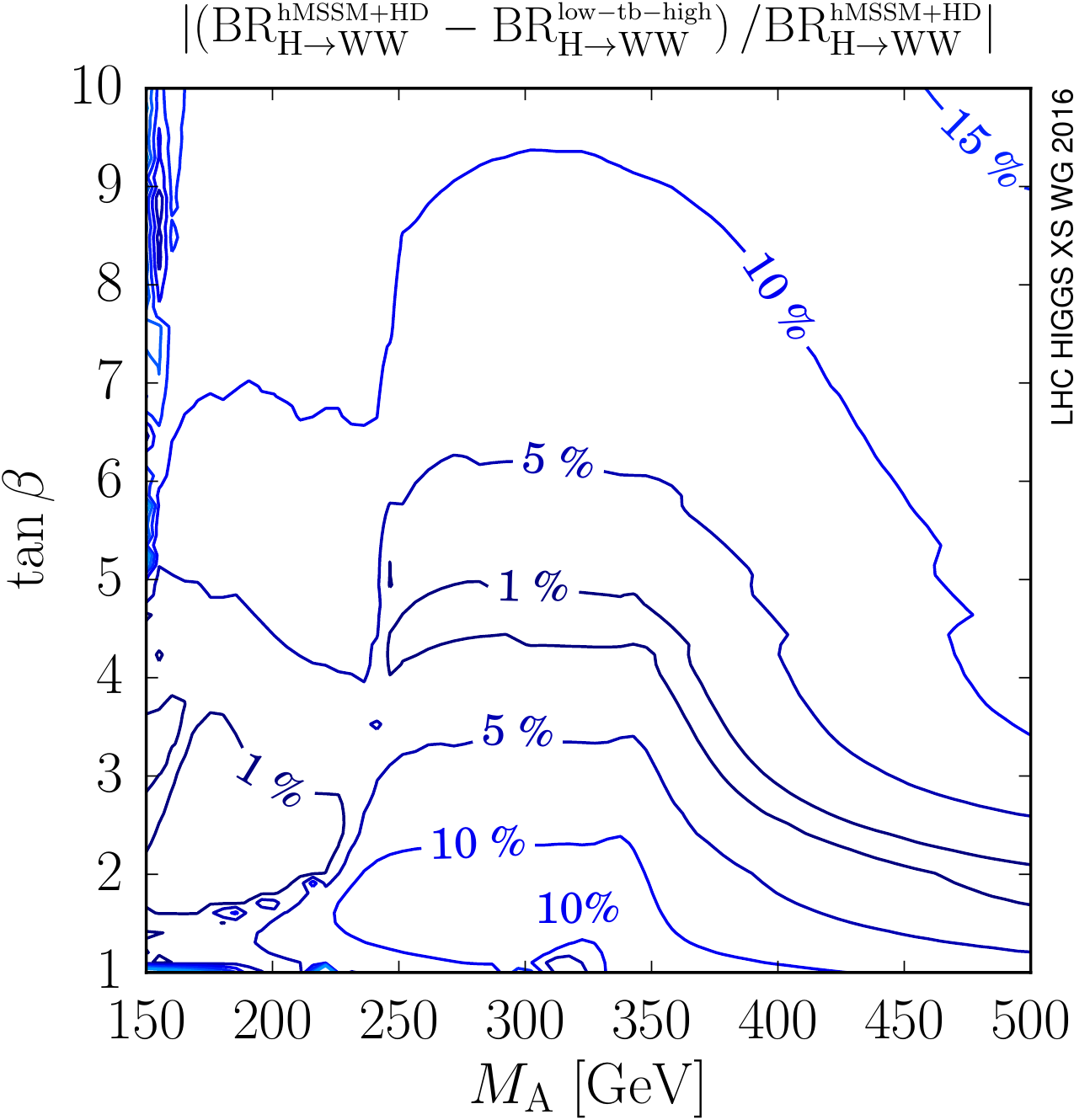}~~~~~~~
    \includegraphics[width=0.47\textwidth]{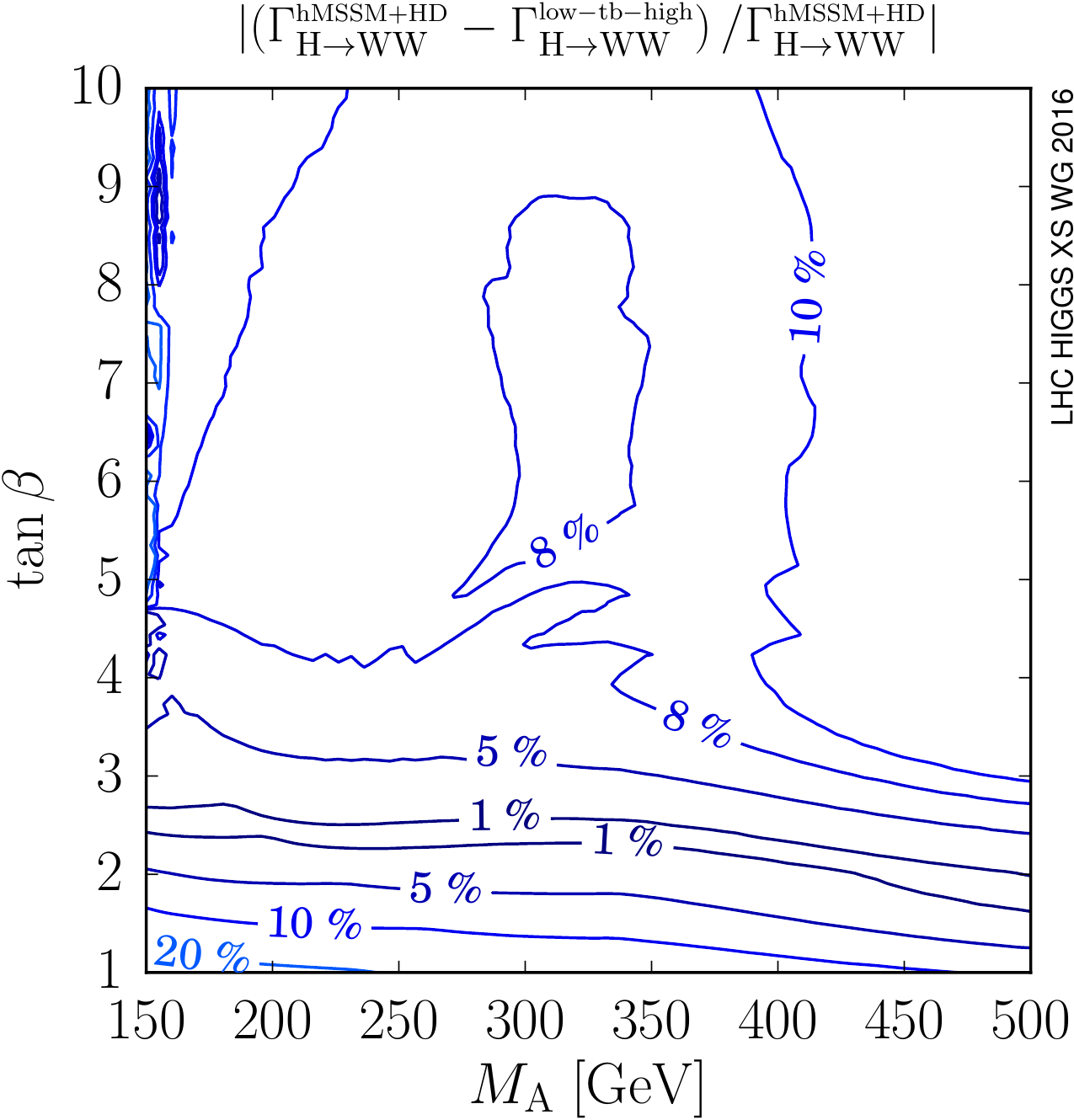}
    \caption{Relative differences in ${\rm BR}(\PH\rightarrow \PW\PW)$
      (left) and $\Gamma(\PH\rightarrow \PW\PW)$ (right) between the
      predictions of the \ltbh\ scenario (where $\Gamma(\PH\rightarrow
      \PW\PW)$ is computed with \feynhiggs+\prophecy) and the
      corresponding predictions obtained with the hMSSM+\hdecay
      combination. For the latter we start from the values of $\Mh$
      computed by \feynhiggs\ in the \ltbh\ scenario, then we compute
      width and branching ratio with \hdecay, which obtains $\MH$,
      $\alpha$ and $\lambda_{\PH\Ph\Ph}$ from the hMSSM prescriptions
      in \Eqs~(\ref{hmH})--(\ref{hlamHhh}).}
    \label{fig:hMSSM-vs-ltbh_WW_BRs}
  \end{center}
\end{figure}
\begin{figure}[p]
  \begin{center}
    \setlength{\unitlength}{\textwidth}
    \includegraphics[width=0.47\textwidth]{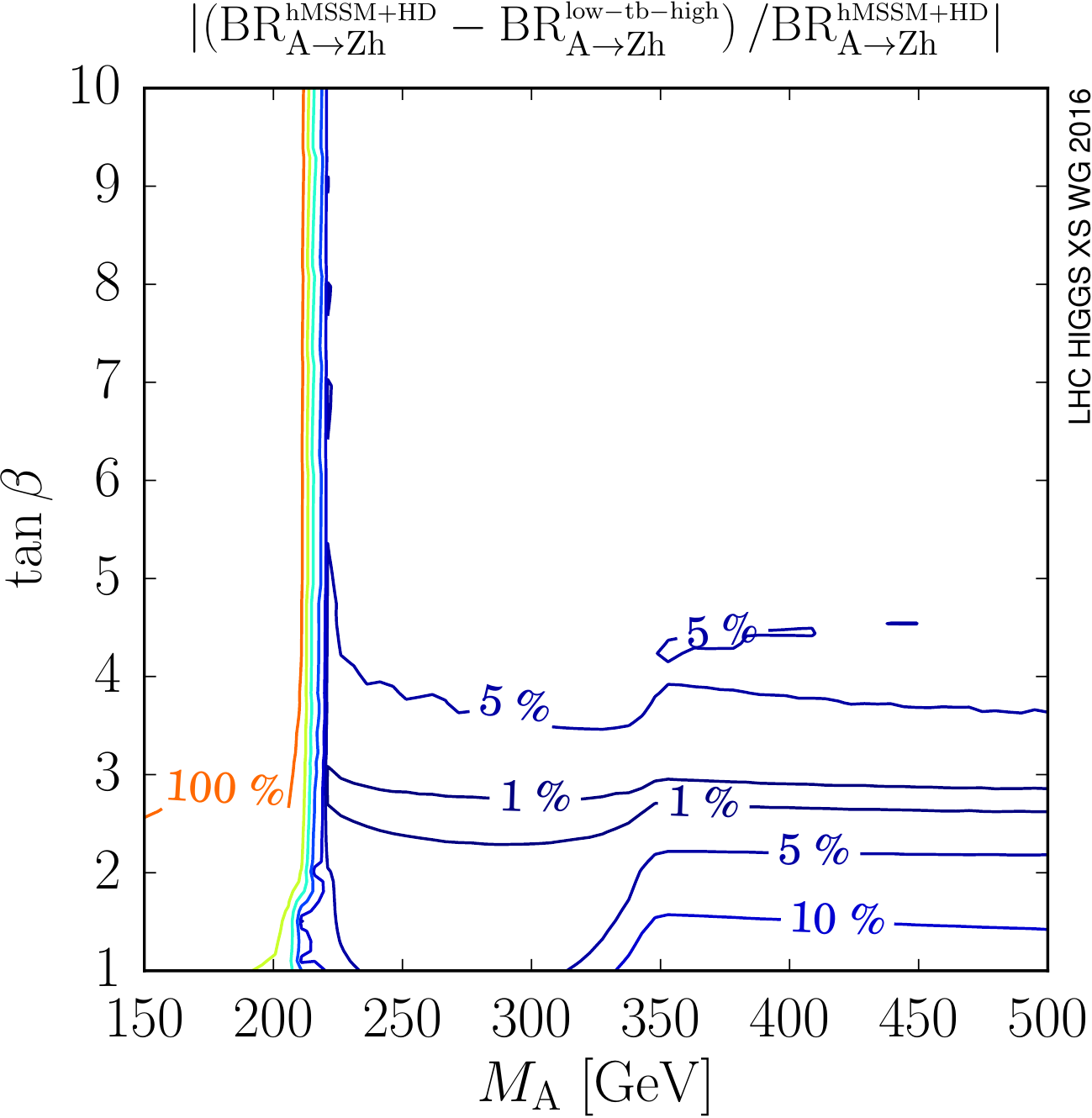}~~~~~~~
    \includegraphics[width=0.47\textwidth]{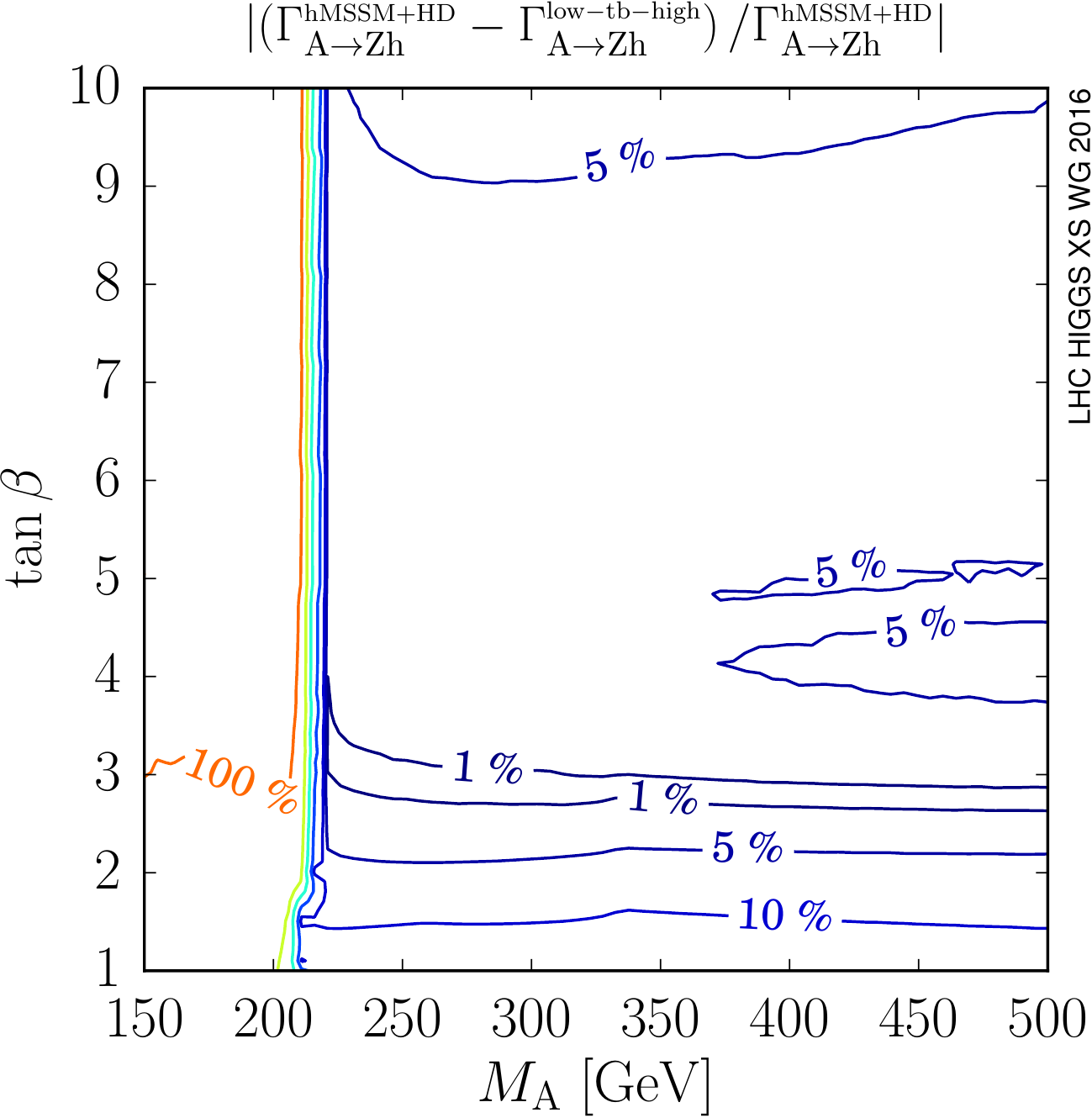}
    \caption{Relative differences in ${\rm BR}(\PA\rightarrow \PZ\Ph)$
      (left) and $\Gamma(\PA\rightarrow \PZ\Ph)$ (right) between the
      predictions of the \ltbh\ scenario (where $\Gamma(\PA\rightarrow
      \PZ\Ph)$ is computed with \feynhiggs) and the corresponding
      predictions obtained with the hMSSM+\hdecay combination. For the
      latter we start from the values of $\Mh$ computed by \feynhiggs\
      in the \ltbh\ scenario, then we compute width and branching
      ratio with \hdecay, which obtains $\MH$, $\alpha$ and
      $\lambda_{\PH\Ph\Ph}$ from the hMSSM prescriptions in
      \Eqs~(\ref{hmH})--(\ref{hlamHhh}).}
    \label{fig:hMSSM-vs-ltbh_Zh_BRs}
  \end{center}
\end{figure}

In \refFs{fig:hMSSM-vs-ltbh_WW_BRs} and \ref{fig:hMSSM-vs-ltbh_Zh_BRs}
we show plots analogous to those in \refF{fig:hMSSM-vs-ltbh_HH_BRs}
for the decays $\PH\rightarrow \PW\PW$ and $\PA\rightarrow
\PZ\Ph$. For what concerns the decays of $\PH$ to $\PW$-boson pairs,
the relative differences between the two calculations of the widths
are -- over most of the $(\MA, \tan\beta)$ plane -- smaller than
$15\%$. In the case of the decays to $\PZ$-boson pairs, not shown
here, we find differences smaller than $10\%$. Again, such
discrepancies can be explained by the fact that, in the \ltbh\ files,
the MSSM results for the $\PH\rightarrow {\rm VV}$ decay widths are
approximated by reweighting the state-of-the-art SM results of
\prophecy\ with the appropriate Higgs-gauge boson couplings, whereas
in the hMSSM files those widths are computed at LO with \hdecay.  For
the decay $\PA\rightarrow \PZ\Ph$, \refF{fig:hMSSM-vs-ltbh_Zh_BRs}
shows that -- in the region with $\MA > \MZ+\Mh$ where the decay is
kinematically open -- the relative differences between the two
calculations of the widths are smaller than $10\%$ unless $\tan\beta$
is very close to 1. For lower values of $\MA$, where the pseudoscalar
must decay to off-shell bosons, large discrepancies appear, due to
differences in both the implementation of the calculations and the
input value of $\alpha$. However, the decay width is extremely
suppressed in that region, and the process is not relevant to the
low-$\tan\beta$ analysis.

Finally, we performed analogous comparisons for all the remaining
decay channels of $\PH$ and $\PA$, but we discuss here only the decays
to pairs of third-family SM fermions, which can reach sizeable
branching ratios in the considered scenario. The widths for the decays
to top and bottom pairs are computed with \hdecay\ in both approaches,
therefore any discrepancy must be due to different input values for
$\MH$ and $\alpha$. For the decays to top quarks, we find
discrepancies of ${\cal O}(1\%)$ in the region where the relevant
Higgs boson mass is above the threshold for the production of a real-top
pair and the decay is unsuppressed. For the decays to bottom quarks,
the discrepancies for $\tan\beta \gtrsim 3\,$, where the branching
ratio becomes significant, are also of ${\cal O}(1\%)$. In contrast,
the decays to tau leptons are computed at LO with \hdecay\ in the
hMSSM files, and at one loop with \feynhiggs\ in the \ltbh\ files. In
this case, the discrepancies for $\tan\beta \gtrsim 3$ are smaller
than $5\%$ for $\Gamma(\PH\rightarrow\PGt\PGt)$, and smaller than $8\%$
for $\Gamma(\PA\rightarrow\PGt\PGt)$.

In summary, this comparison shows that -- in an MSSM scenario where
its underlying assumptions are satisfied -- the hMSSM approach
provides a good approximation to the results of a direct calculation
of the Higgs boson properties. The observed discrepancies of order
$(10\!-\!20)\%$ in the decays of \Eq~(\ref{interestingdecays})
originate from the different accuracy in the calculations of the decay
widths, and could be reduced by including in \hdecay\ the effect of EW
corrections from loops involving SM particles or additional Higgs bosons. However, we stress again that a direct comparison between the
\rootf\ files for the hMSSM and those for the \ltbh\ scenario would
yield larger discrepancies than those shown in
\refFs{fig:hMSSM-and-ltbh_HH_BRs}--\ref{fig:hMSSM-vs-ltbh_Zh_BRs}, due
to the different values of $\Mh$ used in the two sets of files.

\section{Description of the transverse momentum of the Higgs boson in gluon fusion}
\label{sec:mssm-pt}

The transverse momentum distribution of the Higgs boson at fixed order
exhibits a logarithmic divergence in the limit of $\pth \to 0$. To
obtain meaningful predictions, it is necessary to resum the
logarithmically enhanced terms to all orders in $\alpha_s$. The
resummed result is then matched to the fixed-order result by avoiding
double counting. Various matching approaches have been proposed;
common to all of them is the introduction of an auxiliary momentum
scale (from now on generically referred to as ``matching scale''),
which indicates the transverse-momentum region of the transition from
the resummed to the fixed-order result.  The dependence of the
distribution on this matching scale is of higher logarithmic
order. Variations of the theoretical prediction with the matching
scale may be used as estimates of the residual uncertainty due to the
resummation/matching procedure.

Assuming that the dominant contributions to Higgs boson production are dominated by
 top- and bottom-quark mediated Higgs-gluon coupling, the process
is characterized by three scales, namely $\mphi$, $\Mt$ and $\Mb$. The
wide separation of these scales prohibits an intuitive choice of the
resummation scale, in particular in BSM models (see, e.g.,
\Brefs{Mantler:2012bj,Bagnaschi:2011tu,Grazzini:2013mca}).  In the
following sections we will try to address this issue (for a more
detailed discussion, see \Bref{Bagnaschi:2015bop}).  At first we will
compare two possible determinations of the matching parameter and
then, using these values, we will compare the predictions of three
different resummation frameworks.

\subsection{Determination of the matching scale}

In this section we describe and compare two recently proposed
algorithms to determine the matching scales, defined in
\Bref{Harlander:2014uea} and \Bref{Bagnaschi:2015qta} and referred to
as \hmw\ and \bv, respectively, in what follows.  In both approaches,
the matching scale $\mu_i$ ($i=\PQt,\PQb,\text{int}$) is determined
separately for the component of the cross section involving only the
top- or the bottom-quark loop ($\mu_{\PQt}$, $\mu_{\PQb}$), and for the
top-bottom interference contribution ($\mu_{\mathrm{int}}$).  The
resummed results for each of these terms are then added in order to
yield the best prediction for the $\pth$ distribution:
\begin{align}
\frac{\dd\sigma}{\dd\pth} = \frac{\dd \sigma_{\PQt}}{\dd \pth}\bigg|_{\mu_{\PQt}} + \frac{\dd \sigma_{\PQb}}{\dd \pth}\bigg|_{\mu_{\PQb}} + \frac{\dd \sigma_{\mathrm{int}}}{\dd \pth}\bigg|_{\mu_{\mathrm{int}}}\,.
\label{eq:tbsplit}
\end{align}
The interference term, at variance with the first two, is not positive
definite; in particular, it may vanish for a specific value of the Higgs boson mass.
Note that, due to the fact that the scales are determined separately
for each component, it is possible to use them in any model with
arbitrary relative strength of the couplings of the Higgs boson to the
top and bottom quarks.

\subsubsection{Matching scale determination \`a la HMW and BV}

In the HMW method, one first defines, for each different contribution
(top, bottom and interference), $\qres^\text{max}$ as the maximum
value of $\qres$ for which the analytically resummed
$\pth$-distribution stays within the interval
\mbox{$\left[0,\!2\right]\!\cdot\!
  [\dd\sigma\!/\dd\pthtwo]_\text{f.o.}$} of the fixed-order
distribution, for $\pth\ge\mphi$.  The default matching scale $Q$ is
then defined to be half of that maximum value.  As it turns out, for
the results based on analytic resummation this choice of the central
matching scale indeed leads to a behaviour of the matched result in the
large $\pth$ region that is very close to the fixed-order result.

In the \bv{} approach, the exact squared matrix elements of the
subprocesses $\Pg\Pg\to \Pg\PH$ and $\PQq\Pg\to \PQq\PH$ are compared
to their collinear approximation, again separately for each different
contribution.  A deviation by more than 10\% from the exact result is
interpreted as a breakdown of the latter. The upper limit $w$ of the
range of Higgs boson transverse momenta where the collinear approximation is
accurate is chosen as the value for the matching scale. The two
partonic subprocesses initiated by $\Pg\Pg$ and by $\PQq\Pg$ have a
different collinear behaviour, which leads to two different scales
$w^{\Pg\Pg}$ and $w^{\PQq\Pg}$; the final scale $w$ is computed as the
average of the two previous values, weighted differentially by their
relative importance to the transverse momentum distribution of the
Higgs, in the $\pth$ range between $w^{\Pg\Pg}$ and $w^{\PQq\Pg}$.

\begin{figure}[t]
\begin{center}
\includegraphics[width=0.85\textwidth]{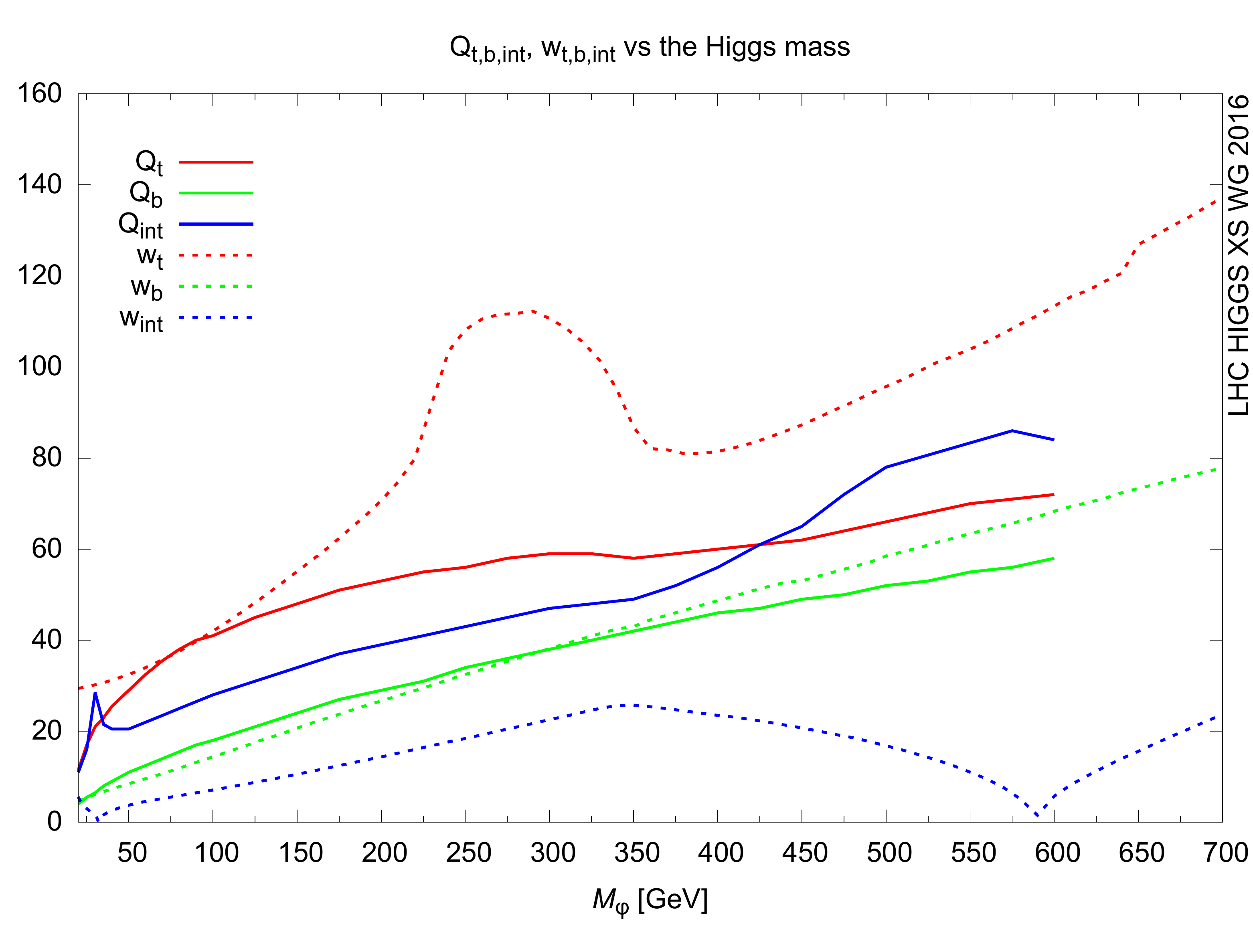}
\caption{\label{fig:scales-compare}Comparison of the matching scales
  in the \bv{} and the \hmw{} approach for a scalar Higgs. Solid
  (dashed) curves correspond to the \hmw{} (\bv{}) scales. The scale
  corresponding to the top (bottom) quark squared matrix element is
  shown in red (green), while the values to be used for the
  interference term are in blue.}
\end{center}
\end{figure}

\subsubsection{Comparison of the two approaches}

Since the matching scale is unphysical, its choice is formally
arbitrary, and any prescription for its determination is necessarily
heuristic.  The \bv{} and the \hmw{} approach are complementary in at
least two aspects.  While \bv{} works at the {\it partonic} level and
considers the {\it low-$\pth$} region, the \hmw{} approach uses the
{\it large-$\pth$} region of the {\it hadronic} distribution in order
to choose a value for the matching scale.  It is thus not surprising
that the numerical values of the resulting scales are different. The
spread of the results is likely to cover in a quite conservative way
the ambiguities of this scale determination.

In \refF{fig:scales-compare} we compare the BV scales $w_i$ and the
HMW scales $Q_i$ $(i=t,b,\text{int})$.  The numerical results of
\refF{fig:scales-compare} exhibit a moderate agreement between the BV
and the HMW scale for the top contribution, and a very good agreement
for the bottom contribution.  Concerning the former, we notice that
the largest deviation between the approaches appears to be close to
the $t\bar{t}$ threshold, around which the BV scales exhibit a
non-trivial structure, while it has no effect on the HMW scale
determination. The difference between the scales of \bv{} and \hmw{}
is largest in cases where the \lo{} term is much smaller than the
\nlo{} term.  This only happens for the interference contribution
which is not required to be positive definite.  The collinear
approximation is proportional to the \lo{} term, which is why the
\bv{} scale will be very small in these cases.  On the contrary, since
the matched curve becomes almost identical to the fixed-order one, and
since the \hmw{} algorithm looks for the largest scale that fulfils
the \hmw{} criteria, the resulting matching scale will tend to be very
large. This explains the respective behaviours of the two blue curves
around $M_{\mathrm{\phi}} \simeq 30$ and $M_{\mathrm{\phi}} \simeq
590$ GeV.

\subsection{Resummation frameworks}

We consider the following three representative theoretical approaches
for the resummation of the collinear logarithms, indicating the specific
software implementations (``codes'') used in this study\footnote{Note
  that all approaches feature NLO accuracy (up to $\alpha_s^3$) on the
  total Higgs boson production cross section, which implies, however, a
  formally LO accurate prediction at large $\pth$.}
\begin{itemize}
\item analytic resummation (\AR) as formulated
in \Brefs{Collins:1984kg,Bozzi:2005wk}\\
(code: \moresushi{}\cite{Mantler:2012bj,Harlander:2014uea,moresushiHP});
\item the \powheg{} method, described in
  \Brefs{Nason:2004rx,Frixione:2007vw}\\
(code: {\sc
    gg\_H\_quark-mass-effects} and {\sc gg\_H\_2HDM}
  \cite{Bagnaschi:2011tu} of the {\sc POWHEG-BOX}
  \cite{Alioli:2010xd,powheg-box});
\item the \mcnlo{} method of \Bref{Frixione:2002ik}\\
  (code: {\sc aMCSusHi} \cite{Mantler:2015vba,amcsushiHP}).
\end{itemize}
All codes work at NLO-QCD accuracy in the prediction of the Higgs boson production total cross section, i.e., $\mathcal{O}(\alpha_s^3)$.  The
differences in the $\pth$ distribution are formally
subleading,\footnote{Note that the meaning of ``subleading terms'' is
  somewhat different for \AR\ and the MC generators. AR consistently
  resums NLL terms to all orders, while the PS in the Monte Carlo
  approaches strictly includes only the leading logarithms, but resums
  also some logarithms beyond the leading ones.}  but can be
numerically sizeable, as we will see later on.  In order to assess the
impact of these differences, we compare their numerical results using
the same values of the matching scales for all of them. On the other
hand, we compare the results of a single code for the two different
strategies of setting the matching scale proposed in
\Brefs{Harlander:2014uea,Bagnaschi:2015qta}.  As we will see, both the
intrinsic difference in the formulation of the codes as well as the
dependence on their matching scales are a source of sizeable
ambiguities in the theoretical prediction of the Higgs $\pth$
distribution, in particular at intermediate and large $\pth$.

\subsection{Phenomenological analysis in the \thdm}
\label{sec:mssm-largeb}
\begin{table}
  \caption{\label{tab:scenariospth}Mixing angle values specific of the
    \largetop\ and \largebot{} \thdm{} scenarios considered in
    \refS{sec:mssm-largeb}. The cross sections for the production of a heavy
    scalar and a pseudoscalar Higgs boson have been obtained with \sushi{}
    (the integration error at \nlo{} is of the order of $0.1\%$,
    and negligible at \lo{}).}
\begin{center}
\begin{tabular}{c|cc|c|rr|rr|rr}
\toprule
\multirow{2}{*}{scenario} &
\multirow{2}{*}{$\tan\beta$} &
\multirow{2}{*}{$\sin(\beta-\alpha)$} &
\multirow{2}{*}{$\phi$}
&\multicolumn{2}{c|}{$\sigma_{\PQt}$/pb} &
\multicolumn{2}{c|}{$\sigma_{\PQb}$/pb} &
\multicolumn{2}{c}{$-\sigma_\text{int}$/pb}\\
&&&& \lo{} & \nlo{} & \lo{} & \nlo{} & \lo{} & \nlo{}\\
\midrule
\multirow{2}{*}{\largetop{}} & \multirow{2}{*}{1.0} & \multirow{2}{*}{0.999} &
$\PH$ & 3.715 &6.788& 0.002&0.003 & $-0.132$&$-0.168$\\
&&& $\PA$ & 12.844 & 23.832 & 0.004 &0.005&0.334&0.428\\
\midrule
\multirow{2}{*}{\largebot{}} & \multirow{2}{*}{50} & \multirow{2}{*}{0.999} &
$\PH$ & 0.002 & 0.005 & 5.085 & 7.089 & $0.163$&$0.199$\\
&&& $\PA$& 0.005 & 0.010 & 9.984 & 13.408 & 0.334 & 0.412\\
\bottomrule
\end{tabular}
\end{center}
\end{table}
In order to compare the predictions of the three codes, we compute the
uncertainty band obtained varying only the resummation scale in a
range $[1/2,2]$ times the central value (chosen either with the BV or
HMW procedure), while keeping the renormalization and factorization
scales fixed.  Moreover, specifically for AR, which consistently
matches the fixed-order results at large transverse momenta, we follow
\Bref{Harlander:2014uea} and apply a suppression factor to the error
band which damps it towards large values of $\pth$.  For our
phenomenological study, we construct two theoretical scenarios (i.e.,
not necessarily compatible with current experimental constraints),
defined on the basis of a type-II \thdm{}, with parameters and \nlo{}
total cross sections reported in \refT{tab:scenariospth}; we discuss
in these two examples the production of the heavy neutral Higgs boson
with mass $\MH=300$\UGeV\ (other scenarios have been considered in
\Bref{Bagnaschi:2015bop}).
\begin{figure}[p]
\centering
\includegraphics[width=0.45\textwidth]{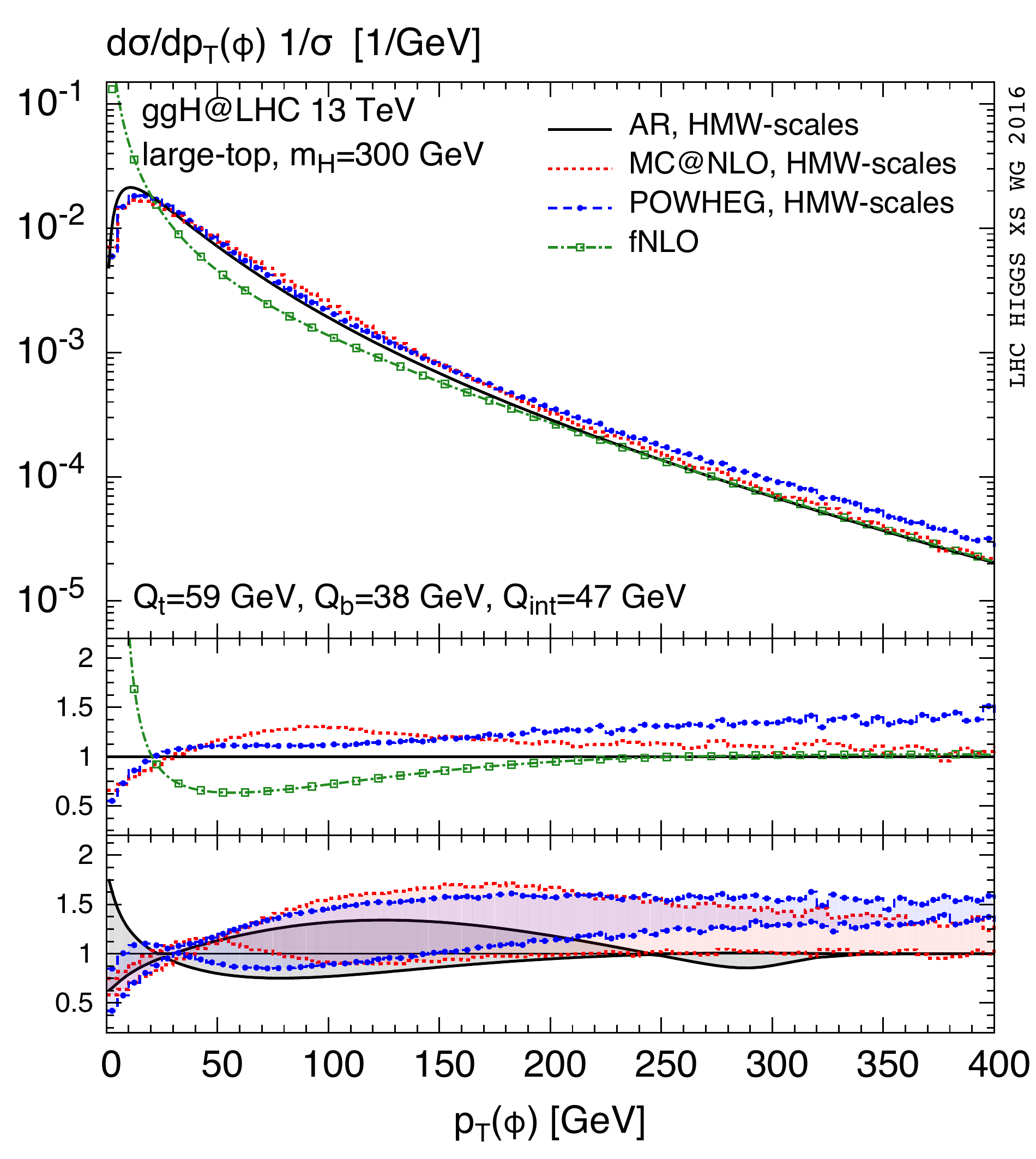}
\includegraphics[width=0.45\textwidth]{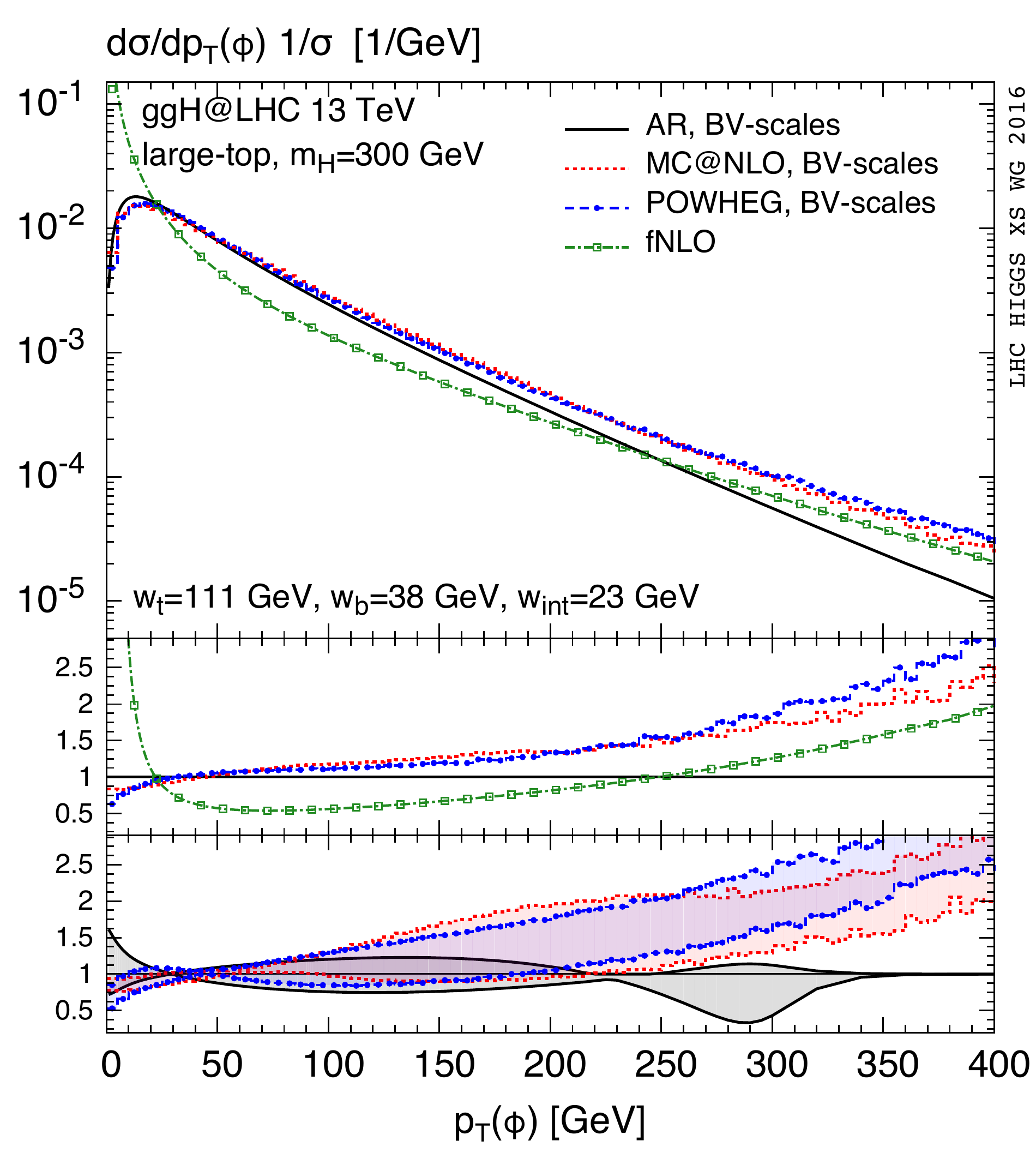}
\caption{ Shapes of the transverse-momentum distributions (i.e.,
  normalized such that the integral yields one) for the heavy Higgs boson with $\MH=300$\UGeV\ in the \largetop{} scenario. The
  distributions are computed with \AR{} (black, solid), \mcnlo{} (red,
  dotted), and \powheg{} (blue, dashed overlaid by points), by setting
  the matching scales to the \hmw{} values (left) and the \bv{} values
  (right).  For reference, we also show the fixed-order \nlo{} (\fnlo{})
  prediction (green, dash-dotted with open boxes).  The main frame shows
  the absolute distributions, the first inset the shape-ratio of the
  central values to the \AR{} distribution, and the second inset the
  uncertainty bands, normalized again to the central \AR~value.
\label{fig:results-larget-hH}
}
\end{figure}

In the first scenario, dubbed {\it \largetop{}} in what follows, the
production cross section is top-loop dominated.  The results for the
$\pth$ distribution are shown in \refF{fig:results-larget-hH}.  Within
uncertainties (which are based solely on resummation scale variation),
the three predictions are compatible with each other for $\pth\lesssim
200$\UGeV. Using \hmw{} scales, the central value of \AR{} merges into
the fixed-order \nlo{} prediction (\fnlo{}) at about $\pth=200$\UGeV;
for \mcnlo{}, this transition is a bit slower, while \powheg{}'s
asymptotic value towards large $\pth$ appears to be about $40 \!-\!
50$\% above \fnlo{}. This overshooting of \fnlo{} at large $\pth$ is a
general feature of the default \powheg{} matching; its origin will be
discussed in more detail below.

Using \bv{} scales, one notices that \AR{} deviates quite
significantly ($\sim 50$\%) from the \fnlo{} result already at $\pth=
400$\UGeV; this deviation tends to further increase towards larger
$\pth$ values. This is not unexpected since the \hmw{} scales are
designed to guarantee similarity between the resummed and the \fnlo{}
curve at large $\pth$.  Scale choices larger than the values
determined by \hmw{} will therefore necessarily lead to a deviation
from the \fnlo{} predictions in that region.  The agreement between
the two Monte Carlos turns out to be excellent, at least up to $\pth$
values as large as the Higgs boson mass.  Despite the large deviations of
\AR{} in the tail and the much softer \AR{} spectrum, all approaches
are compatible within uncertainties at small to intermediate
transverse momenta ($\pth\lesssim 200$\UGeV).  It should be noted that
this is partly due to the fact that the uncertainty bands are
significantly larger (almost by a factor of two) than in the \sm{}.

In the second scenario, dubbed {\it \largebot{}} in what follows, the
production cross section for the heavy Higgs is bottom-loop dominated.
Since the associated matching scales $\wresb$ and $\qresb$ are very
close to each other, any difference in the $\pth$ distributions are
due to the conceptional variants of the matching in the three codes
under consideration.

\begin{figure}[p]
\centering
\includegraphics[width=0.43\textwidth]{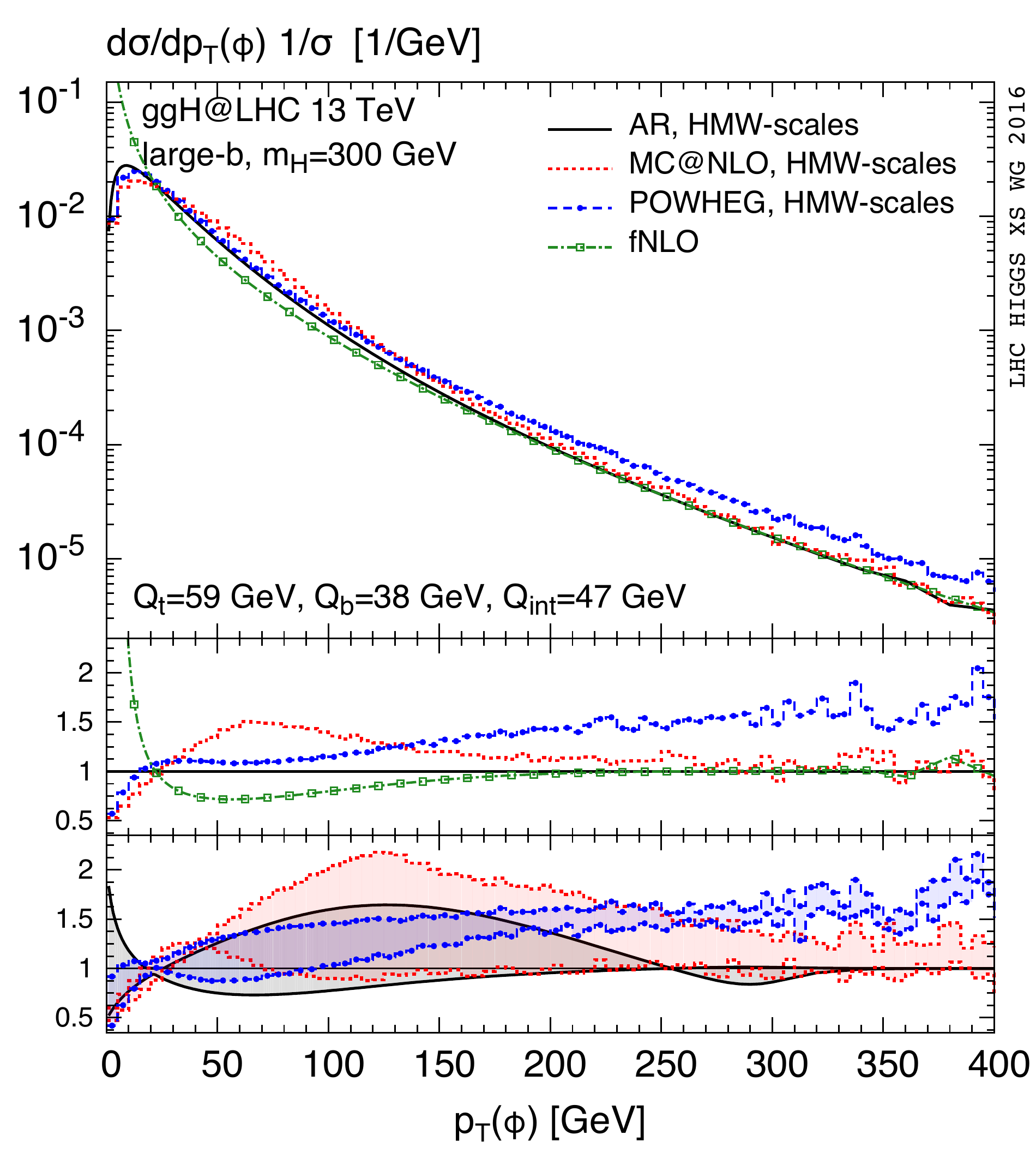}
\includegraphics[width=0.43\textwidth]{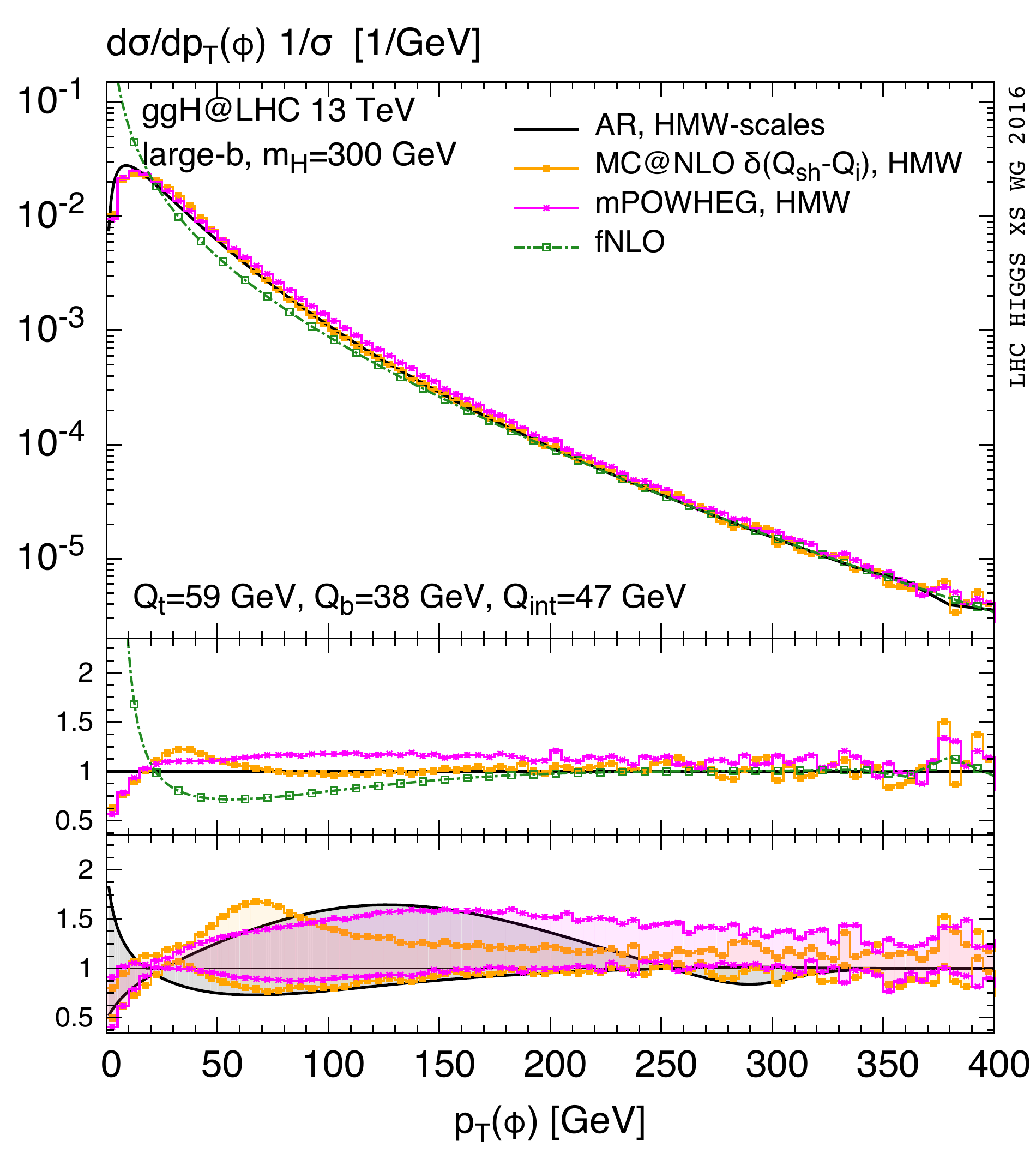}
\caption{ Left: Same as left plot in \refF{fig:results-larget-hH},
  but for the \largebot{} scenario. Right: Same, but using a fixed
  shower scale for the soft events in \mcnlo{}, and m\powheg{} (see main
  text). (\hmw{} scales are assumed here; the plots for \bv{} scales are
  identical for all practical purposes.)
\label{fig:results-scenB-hH}
}
\end{figure}

Let us first discuss the left plot of \refF{fig:results-scenB-hH}. The
large-$\pth$ behaviour is similar to the \largeb{} scenario, when using
\hmw{} scales, discussed above.  Apparently, the specific matching
procedure of \powheg{} has a significant impact on the large-$\pth$
region, where the Parton Shower, based on the soft/collinear
approximation, is outside its region of validity.  Note that the size
of the error bands is very different in the two Monte Carlo
approaches: the \mcnlo{} band blows up to $\order{100\%}$ around
$\pth\sim 125$\UGeV; the \powheg{} band remains very narrow over the
whole range.

It turns out that both the enhanced high-$\pth$ tail of the \powheg{}
curve and its small uncertainty band can be tackled by the same
modification of the matching procedure. In the original \powheg{}
approach, the starting scale of the shower ($t_1$) for each event is
identified with the transverse momentum of the first emission.  If the
latter is very large, the shower will act up to scales which are way
beyond the validity range of the underlying approximations.  In the
m\powheg{} modification\,\cite{Bagnaschi:2015bop}, on the other hand, $t_1$
is defined to remain below the matching scale for all ``remnant
events'' (i.e., events which are described by the pure fixed-order
real emission matrix element; this restriction ensures that the formal
accuracy of the original \powheg{} approach remains unaffected). This
results in the magenta, solid curve in the right plot of
\refF{fig:results-scenB-hH}, which exhibits a \fnlo{}-like high-$\pth$
behaviour; also the uncertainty band appears to describe the matching
uncertainty more realistically.

In the \mcnlo{} implementation the choice of the shower scale
$Q_\text{sh}$ of soft events follows a distribution peaking at the
value of the matching scale $Q_i$.  Restricting the range of that
distribution has a significant effect on the central \mcnlo{}
prediction: in particular, in the limit where this distribution turns
into a delta function $\delta(Q_\text{sh} - Q_i)$, also the size of
the uncertainty band is strongly reduced, as can be observed from the
orange curve in the right plot of Figure~\ref{fig:results-scenB-hH}.

This study simply shows that the predictions in bottom-quark dominated
scenarios, both in the \powheg{} and in the \mcnlo{} approaches,
strongly depend on the details of the matching procedure; this feature
is reflected in the large associated uncertainties.

The distributions for the pseudo-scalar Higgs in the \largebot{}
scenario largely resemble the ones of the heavy Higgs shown in
\refF{fig:results-scenB-hH}, and we do not need to discuss them
separately.

\vfill
\newpage


\chapter{Neutral Higgs Boson Production in Association with Bottom Quarks}
\ChapterAuthor{M.~Beckingham, A.~Nikitenko, M.~Spira, M.~Wiesemann (Eds.)
M.~Bonvini, S.~Forte, H.B.~Hartanto, B.~J{\"a}ger, S.~Liebler, D.~Napoletano, A.~Papanastasiou, F.J.~Tackmann, M.~Ubiali}
\label{chap:bbH}
\section{Introduction}

Higgs boson production modes that feature a $\bbH$ vertex at 
tree level are a viable alternative to determine the Higgs-bottom Yukawa 
coupling ($\yb$), since the $H\to b\bb$ decay is problematic from an experimental 
viewpoint: it suffers from the huge $b$-quark QCD background; 
the absolute value of the total width is extremely small;
and the $H\to b\bb$ branching ratio is large,
which renders the determination of the relative partial decay widths at a 
certain accuracy difficult. Besides loop-induced Higgs boson production through 
gluon fusion, that receives a contribution from both top- and bottom-quark 
loops, the direct production of a Higgs boson in association with bottom quarks 
(i.e., tree-level processes that contain a $b$-quark radiating a Higgs boson) 
gives access to the $\bbH$ coupling. 

The associated production of a Higgs boson with bottom quarks ($\bbH$ production) 
is suppressed in the \sm{} by almost two orders of magnitude with respect 
to the gluon-fusion process. 
Furthermore, this inclusive rate strongly decreases when requiring 
realistic $b$-tagging (i.e., minimal transverse momentum and 
centrality requirements on the $b$ jets) in order to render it 
distinguishable from other production mechanisms. However, in theories with 
extended Higgs sector, such as a generic \thdm{} or  
the MSSM, the Higgs boson coupling to bottom quarks can be significantly 
enhanced and the $\bbH$ process can, in fact, become the dominant production mode.
Since experimentally a scalar sector richer than the one of the SM has 
not been ruled out so far, this constitutes a strong motivation for 
precision computations of the total 
rate (see \refS{sec:bbHtotal}), a proper modelling of the $\bbH$ signal 
in Monte Carlo (MC) generators (see \refS{sec:bbHMC}) and the study
of uncertainties related to the experimental acceptance 
(see \refS{sec:bbHacceptance}). We will further report total inclusive cross sections for the 
$c\bar{c}\phi$ production mode (see \refS{sec:ccH}) which may become relevant in specific models with enhanced charm 
Yukawa coupling.
Although all presented predictions are in the SM, they 
are directly applicable to all neutral Higgs bosons ($\phi=h,H,A$) 
in a \thdm{}, by a proper rescaling of the bottom Yukawa; for the MSSM 
this has been shown~\cite{Dittmaier:2006cz,Dawson:2011pe,Dittmaier:2014sva} to be a good 
approximation of the full result.

\begin{figure}
\begin{center}
    \includegraphics[height=2cm]{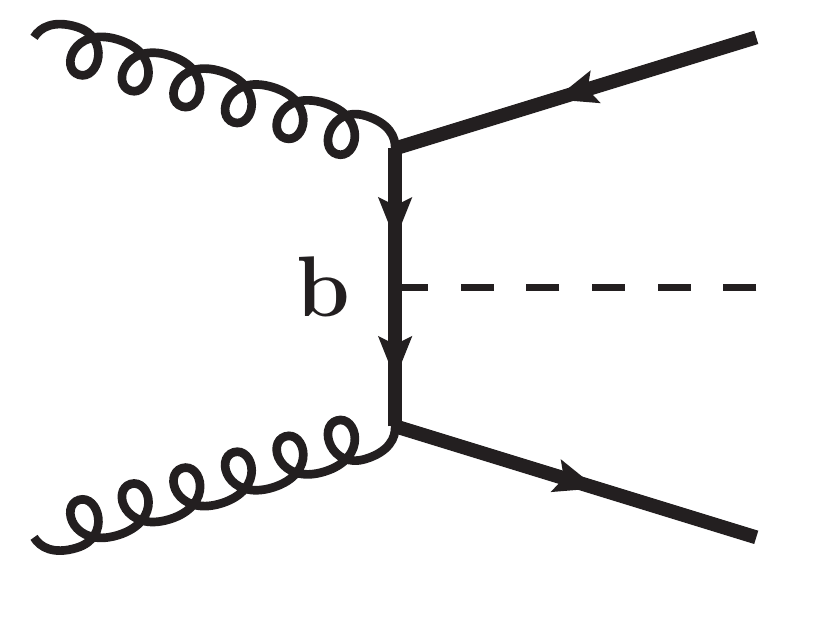}\hspace*{2cm}
    \includegraphics[height=2cm]{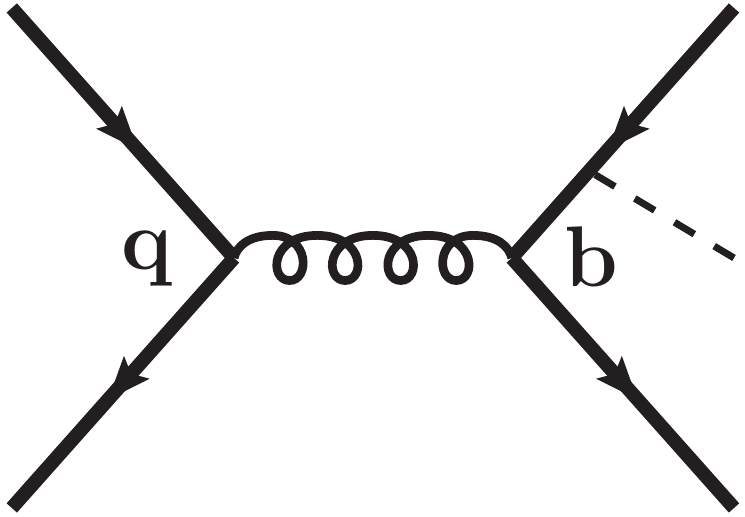}\hspace*{2cm}
    \includegraphics[height=2cm]{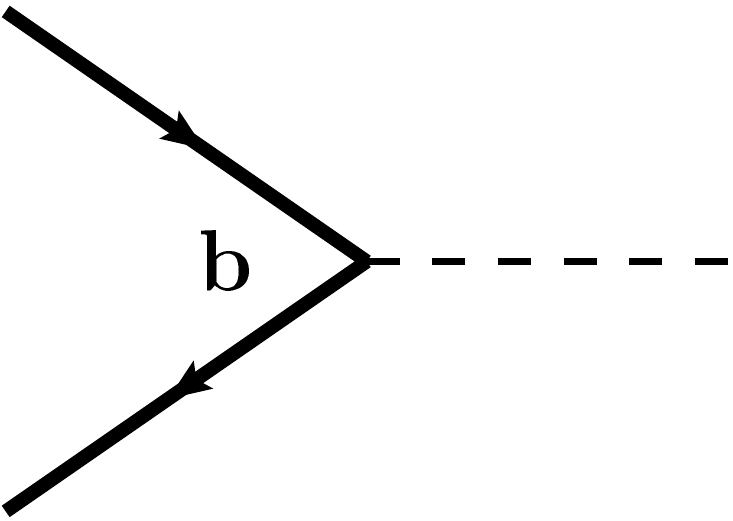}
	\vspace{0.2cm}
  \caption{Sample of LO Feynman diagrams for $\bbH$ production in
the four-flavour scheme (left, centre) and the five-flavour scheme (right).}
  \label{fig:LO}
\end{center}
\end{figure}

As for all processes that feature $b$ quarks at the level of the hard-scattering
process, there are two viable approaches to compute the $\bbH$ cross section. 
In the four-flavour scheme (4FS), bottom quarks are treated as massive 
particles, hence no bottom quarks can appear in the initial state of the 
partonic scattering process. This is relevant for those cases where
the physical mass of the $b$ quark is considered as a hard scale and implies
that observables with tagged final-state $b$ quarks are well defined (and thus can
be computed) at fixed $\as$ order in perturbation theory. The leading-order 
(LO) partonic processes in the 4FS are (see left and centre diagrams in \refF{fig:LO})
\beq
gg\to b\bb H\,,\;\;\;\;\;\;\;\;
q\bar{q}\to b\bb H\,,
\label{LO4FS}
\eeq
where $q$ denotes a light quark.

At any order in perturbation theory the 4FS involves terms $\sim\as^k\log^k(\mb/Q)$,
where $Q$ is the characteristic scale of the $g\rightarrow b\bar{b}$ splitting. 
These logarithmically 
enhanced terms remain small as long as $Q\sim\mb$, but can spoil the perturbative 
convergence when $Q\gg\mb$. Such terms are generally dealt with by re-organizing 
the perturbative series while resumming them to all orders in $\as$. This is 
precisely achieved by working in second viable approaches to compute the 
$\bbH$ cross section, the five-flavour scheme (5FS), which is particularly 
important when the characteristic of an observable is that of being 
dominated by such logarithms. In this scheme, one assumes massless $b$ quarks 
($\mb\equiv 0$) at the level of the short-distance cross section, which are therefore 
treated at equal footing as the other light quarks and may appear 
as initial state particles; the potentially large logarithms are 
effectively resummed through the DGLAP evolution of the $b$-quark PDFs.
Hence, the LO cross section in the 5FS is simply given by (see right diagram of \refF{fig:LO})
\beq
b\bb\to H\,.
\label{LO5FS}
\eeq

It is clear from the discussion above that 4FS computations do not account for
logarithmic terms beyond the first few, while 5FS results lack 
power-suppressed terms $(\mb/Q)^n$. If either of these properties is 
important the other scheme must be preferred. Being highly observable 
dependent\footnote{This receives an additional complication from the fact that
an observable may be 
associated with different powers of $\as$ in the four- and five-flavour schemes.}, 
at least for inclusive quantities neither resummation nor mass effects 
are dominant and the two approaches lead to generally similar results.
One must bear in mind, however, that reasonable agreement is found only 
by judicious choices of hard scales, i.e., the resummation and factorization 
scales, which must be chosen significantly smaller than 
$\mH$~\cite{Maltoni:2003pn,Boos:2003yi,Harlander:2003ai,Maltoni:2012pa}---the 
hardness one would naively associate with $\bbH$ production.
\footnote{This is also supported by arguments based on collinear dominance
\cite{Maltoni:2012pa} and the small upper $p_T$ limit of the factorizing
part in the $p_T$ distribution \cite{Rainwater:2002hm}.}

For inclusive observables the 5FS process in Eq.\,\eqref{LO5FS} has the 
technical advantage of being much simpler ($2\to 1$ at the LO), which 
renders feasible radiative corrections beyond the NLO (and even beyond the 
NNLO with current technology), while the $2\to 3$ Born-level processes 
of Eq.\,\eqref{LO4FS} in the 4FS
limit perturbative computations in this scheme to NLO.

Considering more exclusive observables, in particular regarding the final-state 
$b$ quarks, which are relevant for a realistic description of the $\bbH$ signal, 
the 5FS loses its advantage mentioned above: The process in Eq.\,\eqref{LO5FS} 
has much more limited information on the final-state kinematics than 
the one in Eq.~\eqref{LO4FS}. Only higher orders in the 5FS recover such 
information, e.g., $1$-$b$ and $2$-$b$ tag observables can be described in the 
5FS only starting from NLO and NNLO, respectively, while the 4FS tree-level process 
in Eq.~\eqref{LO4FS} describes observables involving 
$0$-, $1$-, or $2$-$b$ tags formally already with leading perturbative accuracy.
Furthermore, $b$-tagged objects in the 5FS can not be consistently defined 
at any order in perturbation theory, because the corresponding cross section 
becomes infinite beyond the LO, when the massless $b$ quarks are not considered as
(and clustered into) jets or integrated over. The massive $b$ quarks in the 4FS, 
on the other hand, can be associated with physical objects, allowing for 
realistic $b$ tagging with arbitrary selection cuts on the $b$ quark kinematics\footnote{Note that 
fragmentation effects when turning the $b$ quarks into $B$ hadrons are 
assumed to be moderate and neglected in general at fixed order.}, whereas too small 
$\pt$ cuts directly lead to a divergence in the 5FS. The problems related to a 
consistent definition of $b$-tagged objects in the 5FS can be alleviated by matching 
the fixed-order computation to parton showers (PS). Due to the backward evolution of the initial-state 
$b$ quarks the shower will generate $b$-flavoured hadrons already at the LO, 
rendering realistic any $b$-tagging requirements. However, one must
not forget that the backward evolution in the Monte Carlos is not trivial and 
has only leading logarithmic (\llog{}) accuracy. Furthermore, the kinematic 
reshuffling of massless into massive $b$ quarks can have sizeable effects 
on the $B$ hadron kinematics. Both come at the price of an additional 
uncertainty. As far as the 4FS is concerned, the PS matching particularly 
improves the Sudakov-suppressed small-$\pt$ initial-state radiation, although the
impact of the PS is less crucial than in the 5FS. In both schemes, 
the PS introduces 
additional power-suppressed contributions due to long-distance phenomena.

Before considering phenomenological predictions, let us discuss 
the general coupling structure in the two schemes. Being a $2\to 3$ 
process the lowest order cross section in the 4FS starts at 
${\cal O}(\as^2)$. In the 5FS each bottom PDF can be considered 
as being of ${\cal O}(\as)$ with respect to the gluon and hence the 
LO features no power of $\as$. In both schemes the LO 
cross section is proportional to $\yb^2$, since the Higgs boson is always 
coupled to a $b$ quark. Considering higher order corrections, the 
coupling structure becomes more involved, in particular in the 4FS.
In this case virtual diagrams with a $\bbH$ final state 
may involve a top quark circulating in the loop, which couples 
to the Higgs boson (e.g., see \refF{fig:4FSyt}), and thus are 
proportional to the Higgs-top coupling ($\yt$). Such diagrams 
are generally attributed to the gluon-fusion Higgs boson production mode
(their square enters the NNLO gluon-fusion cross section), but their 
interference with diagrams proportional to $\yb$ must be carefully 
accounted for in the cross section of the $\bbH$ production mode. 
Such contributions will be 
generically referred to as $\yb\yt$ terms. Including such interference effects, 
but neglecting all contributions that already appear in the gluon-fusion process, 
we may thus express the $\bbH$ cross section in the two schemes as follows:
\begin{align}
\sigma_{\bbh}^{\text{4FS}}&=
\underbrace{\as^2\,\yb^2\,\Delta^{(0)}_{\yb^2}+
\as^3\,\Big(\yb^2\,\Delta^{(1)}_{\yb^2}}_{\equiv \sigma_{\yb^2}}+
\underbrace{\yb\,\yt\,\Delta^{(1)}_{\yb\yt}}_{\equiv \sigma_{\yb\yt}/\as^3}\Big)+
\as^4\left(\yb^2\,\Delta^{(2)}_{\yb^2} + \yb\,\yt\,\Delta^{(2)}_{\yb\yt}\right)+\mathcal{O}(\alpha_s^5)\,.
\label{sig4FS}\\
\sigma_{\bbh}^{\text{5FS}}&=
\yb^2\,\left(\Delta^{(0)}_{\yb^2}+
\as\,\Delta^{(1)}_{\yb^2}+
\as^2\,\Delta^{(2)}_{\yb^2} 
+\as^3 \Delta^{(3)}_{\yb^2} + \mathcal{O}(\alpha_s^4)\right).
\label{sig5FS}
\end{align}
The 4FS cross section at NLO (being the current state-of-the-art) can 
be decomposed in terms proportional to $\yb^2$ ($\sigma_{\yb^2}$) and 
$\yb\yt$ ($\sigma_{\yb\yt}$). Any component with a $\bbh$ final state, but 
proportional to $\yt^2$, must originate from a squared gluon-fusion amplitude 
(i.e., with a Higgs radiated from a closed top-quark loop) and can be 
incoherently added to the $\bbh$ cross section above.\footnote{Note that 
starting from the NNLO in both schemes such squared gluon-fusion diagrams 
contribute also to the $\yb^2$ and $\yb\yt$ components of the $\bbh{}$ 
cross section, which, as stated above and being common practice, are not 
attributed to the $\bbh$ cross section, but to the gluon-fusion one.}
In the 5FS, on the other hand, interference terms (proportional to $\yb\yt$) 
between the gluon-fusion and $\bbh$ processes exactly vanish order by 
order in perturbative QCD, since they involve a helicity flip of the
bottom quarks that leads to a vanishing interference term with the
generic 5FS amplitudes in the massless limit.

\begin{figure}
\begin{center}
    \includegraphics[height=2.15cm]{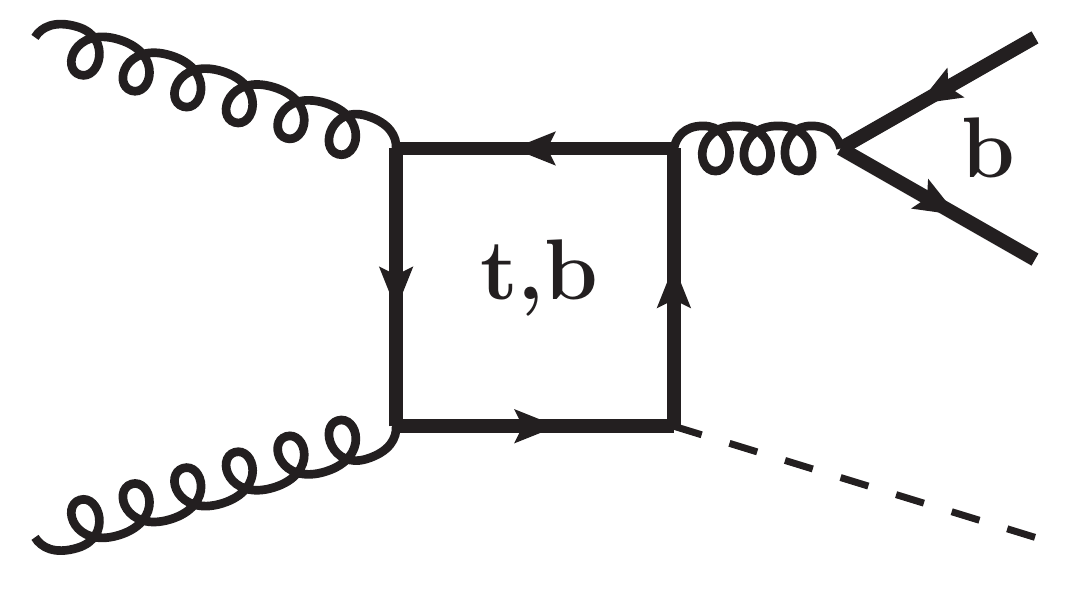}\hspace{1.2cm}
    \includegraphics[height=2.15cm]{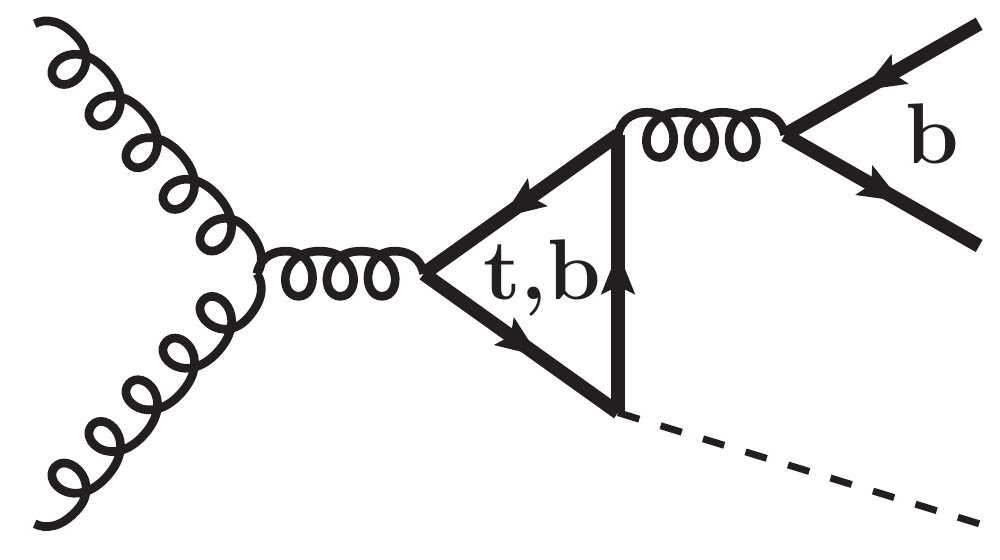}\hspace{1.2cm}
    \includegraphics[height=2.15cm]{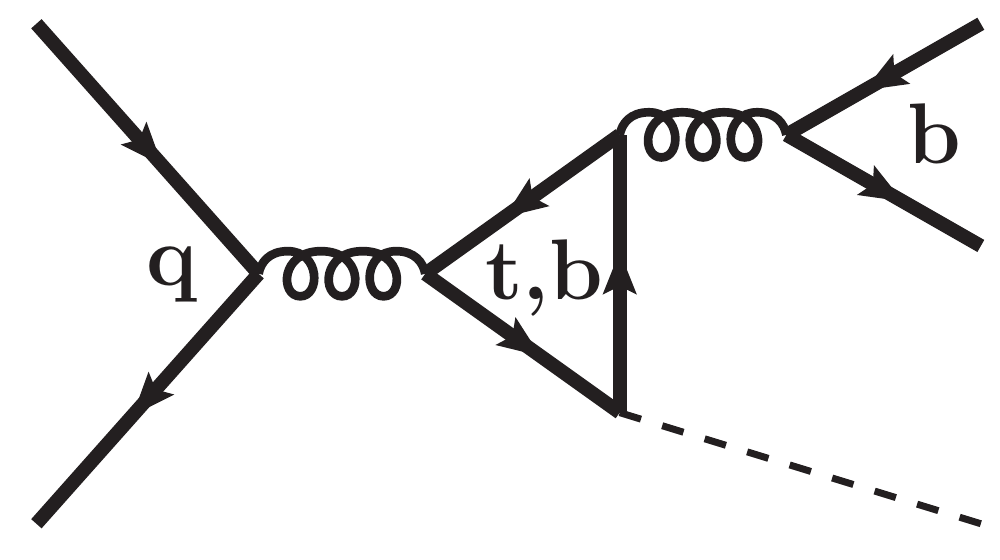}
  \caption{Sample of one-loop Feynman diagrams for $\bbH$ production in the
	four-flavour scheme featuring a $\yt$ coupling.}
  \label{fig:4FSyt}
\end{center}
\end{figure}
%
%
%
%
%
%
%

\section{Total inclusive cross section \label{bbH-inclusive-cross-section}}
\label{sec:bbHtotal}
In this section, we study and compare the state-of-the-art predictions for the
total inclusive $\bbH$ cross section within the 4FS, within the 5FS and matched 
predictions of the two schemes.

Before introducing the features of the different matched computations 
under consideration, let us summarize the available total cross section computations 
for $\bbH$-production in the literature. The 4FS cross section is known through 
NLO QCD in the SM~\cite{Dittmaier:2003ej,Dawson:2003kb,Wiesemann:2014ioa};
MSSM-type couplings have been studied in \Bref{Dawson:2005vi}; and the
SUSY-QCD corrections in the MSSM were computed in 
\Brefs{Liu:2012qu,Dittmaier:2014sva}. Owing to its technically simpler 
process structure, radiative corrections in the 5FS are known through 
NLO~\cite{Dicus:1998hs,Balazs:1998sb} and NNLO QCD \cite{Harlander:2003ai}. Even the relevant matrix elements for the full N$^3$LO prediction are 
already available \cite{Ahmed:2014pka,Gehrmann:2014vha} (their combination 
being far from trivial though).

Matched four- and five-flavour scheme computations will be discussed and 
their phenomenological results compared in the following sections. The 
Santander matching~\cite{Harlander:2011aa} has been introduced 
as an empirical approach to combine 4FS and 5FS predictions by a 
weighted average with respect to the Higgs boson mass. More recently, two 
consistent flavour-scheme matching procedures have been applied to the 
$\bbH$ production process. In \Bref{Bonvini:2015pxa} the 
formally NLO+NLL accurate (NLO in the 4FS, and NLL refers to the logarithms 
resummed through the 5FS) cross section is presented, 
while \Bref{Forte:2015hba} works in the FONLL approach, but includes only 
the LO 4FS cross section.

\subsection{Choice of bottom PDFs in the 5FS}                                                      
\label{sec:bbh-pdfs}

The resummation of potentially large collinear logarithms due to initial-state gluons splitting into $b$ quark pairs
is achieved through their factorization into the definition of the PDFs, and consequently through DGLAP evolution.
Once this is done, the partonic coefficients do not contain collinear $b$-quark logarithms,\footnote
{In the massless 5FS, the partonic coefficients do not contain any bottom mass dependence,
while in the matched computations only power suppressed mass contribution are present.}
so the detail of their resummation resides entirely in the PDFs.
Therefore, the choice and usage of the PDF set in resummed computations must be regarded as
a (fundamental) part of the computation itself.
It is the purpose of this section to briefly review how the resummation in PDFs works
and to describe the choice of PDFs adopted in this Chapter for  $\bbH$ production.

Since splittings involving bottom quarks are finite (the bottom mass regulates the collinear limit),
factorizing or not the collinear logarithms generated by these splittings gives rise to
two factorization schemes, denoted by 5FS and 4FS respectively.
In the 5FS, a bottom PDF exists and the DGLAP evolution involves 5 active flavours,
while in the 4FS there is no bottom PDF and DGLAP evolution involves only 4 active flavours.
The PDFs in the two schemes can be related by matching conditions,
obtained imposing scheme invariance, possibly up to power corrections in the bottom mass
(for a recent review, see Ref.~\cite{Ball:2015dpa}).
The matching conditions at the generic scale $\mu_b$ read
\begin{equation}
  \label{eq:PDFmatching}
  f^{[5]}_i(\mu_b) = \sum_{j=g,u,\bar u,d,\bar d,s,\bar s,c,\bar
c} \mathcal{K}_{ij}(m_b,\mu_b) \otimes f^{[4]}_j(\mu_b),
  \qquad i= g,u,\bar u,d,\bar d,s,\bar s,c,\bar c, b, \bar b,
\end{equation}
($\otimes$ denotes Mellin convolution) where $\mathcal{K}_{ij}(m_b,\mu_b)$ are matching functions
fully known to $O(\as^2)$~\cite{Buza:1996wv,Bierenbaum:2009zt}
and partially to $O(\as^3)$~\cite{Ablinger:2010ty,Ablinger:2014lka,Ablinger:2014vwa,Ablinger:2014nga}.
Since resumming bottom mass logarithms is useful only at large scales, where phase space constraints
allow gluon splittings into $b$ quark pairs, the factorization scheme adopted in standard PDF sets
changes dynamically: for scales below the ``bottom threshold'' $\mu_b$ the 4FS is used, at the threshold scale $\mu_b$
5FS PDFs are generated from the 4FS ones through Eq.~\eqref{eq:PDFmatching}, and for scales higher than $\mu_b$
5FS evolution is used.

The value of $\mu_b$ is in principle arbitrary, however its value should be chosen to be close to the bottom 
mass $m_b$, so that the collinear logarithms of $\mu_b/m_b$
factorized in $\mathcal{K}_{ij}(m_b,\mu_b)$ are not large and a
fixed-order expression for $\mathcal{K}_{ij}$ is reliable.
The standard choice adopted in PDF sets is $\mu_b=m_b$. However, it is useful to let $\mu_b$ vary
to estimate the theoretical uncertainty due to missing higher orders in the matching procedure.
The remaining potentially large logarithms of $\mu_b/\mu_F$, where $\mu_F\sim m_H$ is the typical hard scale of the process,
are then resummed via DGLAP evolution with 5 active flavours from the scale $\mu_b$ to the factorization scale $\mu_F$.

The values of the bottom mass, $m_b$, and bottom threshold, $\mu_b$, as well as the mass-renormalization scheme
are therefore entirely contained in the PDF set one uses for resummed computation.
Therefore, when computing a cross section, one must either set the input parameters to be consistent with 
the PDF set used, or refit the PDFs using the desired input parameters.
For instance, in matched computations, it is important to ensure that
the value of the mass adopted in the computation of power-suppressed contributions is equal
(perhaps modulo mass-renormalization scheme conversion) to the one present in the PDF set.

In practice, the effect of the choice of the bottom mass value on PDF fits is very mild 
for all PDFs, barring the bottom-quark PDF~\cite{Harland-Lang:2015qea}.
Hence, one can expect that refitting the PDFs with different values of the bottom mass and threshold
should not significantly affect light PDFs at scales below the bottom threshold.
Therefore, taking any PDF set with a given bottom mass, and re-evolving it from a low scale ($<\mu_b$) using 
a different bottom mass, threshold and even renormalization scheme should yield virtually the same resulting 
PDF set as refitting with the same setting.
This ``approximate'' approach is very useful because it opens up the possibility of using any desired
setting with any input PDF.

We now discuss the settings used for the resummed calculations (5FS and matched) presented in this report.
Since the bottom mass is most precisely determined in the $\overline{\text{MS}}$ scheme,
it seems most correct to use this scheme for the mass renormalization in the PDF matching.
However, current PDF fits have been performed with pole-scheme masses,
and, to date, no fits with $\overline{\text{MS}}$-scheme masses are publicly available.
Furthermore, existing computations of the $\bbH$ cross section assumed pole-scheme masses in the PDFs~\cite{Forte:2015hba,Bonvini:2015pxa}.
In order to be consistent with existing fits and computations it is convenient at this stage to use the pole scheme.
However, since the bottom pole mass is not very well determined and is additionally affected by the renormalon ambiguity \cite{Beneke:1994sw,Beneke:1998ui},
the choice of the value of the pole mass is a delicate point.
We argue that the pole mass which can better reproduce a hypothetical NNLO $\overline{\text{MS}}$ fit
with $\overline{m_b}(\overline{m_b})=4.18$~GeV is obtained from a 1-loop conversion, which gives
\begin{equation}
  \label{eq:bbH-mbpole}
  m_b^{\text{pole}} = 4.58~\text{GeV}. 
\end{equation}
The reason for this is that the bottom mass first enters the matching
functions $\mathcal{K}_{ij}$ at $O(\as)$. 
This means that an NNLO fit, which employs the $\mathcal{K}_{ij}$ at $O(\as^2)$, is only affected by the 1-loop 
pole to $\overline{\text{MS}}$ conversion.
We have verified that evolving the same initial PDF set using either the pole scheme with $m_b^{\text{pole}} = 4.58$~GeV
or the $\overline{\text{MS}}$ scheme with $\overline{m_b}(\overline{m_b})=4.18$~GeV gives very similar results.

Taking Eq.~\eqref{eq:bbH-mbpole}, together with $\mu_b=m_b^\text{pole}$,
as the central choice for the resummed $\bbH$ computations, we can now discuss how to compute theoretical uncertainties.
As already mentioned, the uncertainty due to missing higher order terms in the matching procedure can
be probed by varying $\mu_b$. Following Ref.~\cite{Bonvini:2015pxa}, we propose to use a canonical
factor of 2 around the central value $\mu_b=m_b^\text{pole}$.
In addition to this, one can consider a parametric uncertainty on the value of the bottom mass itself.
A reasonable variation, which should be sufficient to cover the uncertainties on both the determination
of $\overline{m_b}(\overline{m_b})$ and in the conversion to the pole scheme, is $\pm0.14$~GeV.
Therefore, we recommend that uncertainties are estimated according to
\begin{equation}
  \label{eq:bbH-mbmubvariations}
  4.44~\text{GeV} \leq m_b^{\text{pole}} \leq 4.72~\text{GeV},
  \qquad
  2.29~\text{GeV} \leq \mu_b \leq 9.16~\text{GeV}.
\end{equation}
For $m_b^{\text{pole}}$ variations, the bottom threshold is kept fixed at $\mu_b=4.58$~GeV.

As far as the initial PDF set is concerned, we use the \texttt{PDF4LHC15\_nnlo\_mc} set.
All the 101 PDF members are re-evolved from an initial scale of 2~GeV, using all the possible variations
of settings mentioned above. This is done using the latest version ($\geq 2.8$) of \texttt{APFEL}~\cite{Bertone:2013vaa},
which can now handle heavy quark thresholds that are different from the mass.
The new re-evolved PDF sets\footnote
{We acknowledge V.~Bertone who cross-checked the sets and S.~Liebler who tested them against the original PDF4LHC15 set.}
are publicly available for download from \url{http://www.ge.infn.it/~bonvini/bbh}. 

The settings used here differ from the LHCHXSWG default
\cite{Butterworth:2015oua},
which in particular includes PDF4LHC15 PDFs. Specifically, the bottom pole mass
value used in the PDF sets included in the PDF4LHC15 combination
differ from each other.
This is supposed to give a rough estimate of the
PDF uncertainty associated to the choice of $m_b$ in the PDF fit, but
it contrasts with the needs of the more detailed estimate of
the uncertainties related to the bottom mass in our calculation, as
discussed above. A fully consistent treatment of the bottom mass, such
as the one adopted here, is of crucial importance for processes such as $\bbH$ production,
which are very sensitive to the bottom PDF:
as stated in the PDF4LHC15 recommendation, such cases require special care.

\subsection{Santander matching}
\label{sec:santander}

The 5FS result is obtained with the help of \sushi{}
\cite{Harlander:2012pb} which includes the NNLO QCD corrections to the
process $\bbH$ \cite{Harlander:2003ai}.  The bottom Yukawa coupling is
renormalized in the $\overline{\rm MS}$ scheme. It is evaluated
according to the prescription of Chapter~\ref{chapter:input} by running
$\mb(\mu)$ from $\mu=4.18$\,GeV to $\mH$ at 4-loop level, which
corresponds to our central choice of the renormalization scale.  The
central factorization scale is $\mu_F=\mH/4$
\cite{Rainwater:2002hm,Boos:2003yi,Maltoni:2003pn,Harlander:2003ai}. The
scale uncertainties are determined by taking the minimum and the maximum
of the seven values for the cross section, corresponding to the choices
$\mu_R,4\,\mu_F\in\{\mH/2,\mH,2\,\mH\}$ while $1/2\leq 4\mu_F/\mu_R <
2$. The Yukawa coupling at $\mu_R$ is obtained from its previously
determined value at $\mH$ by 3-loop evolution.

The 4FS result uses the implementation of the $b\bar b\phi$ processes by
means of \aNLO{} at NLO \cite{Wiesemann:2014ioa}.\footnote{See 
also \refS{sec:bbHMG5} for details on the dedicated $b\bar b\phi$ 
implementation in \aNLO{}.}
 The bottom Yukawa coupling has been
adopted according to the 5FS, while for the PDFs 4-flavour PDF4LHC15
densities are used consistently with the appropriate choice of a
4-flavour coupling $\alpha_s$. The central renormalization and
factorization scales have been chosen as $\mu_R=\mu_F=(M_H+2m_b)/4$. The
renormalization scales of $\alpha_s$ and the bottom Yukawa coupling have
been identified. This implementation has been cross checked against the
published 4FS results of
Refs.~\cite{Dittmaier:2003ej,Dawson:2003kb,Wiesemann:2014ioa}.

In the asymptotic limits $\mH/\mb \to 1$ and $\mH/\mb \to \infty$, the
4FS and 5FS results tend to provide the unique description of the inclusive
$\bbh$ cross section, respectively. In the Santander matching
prescription \cite{Harlander:2011aa}, the 4FS and 5FS are thus combined
in such a way that they are given variable weight, depending on the
value of the Higgs boson mass. Since the difference between the two
approaches is formally logarithmic, the dependence of their relative
importance on the Higgs boson mass should be controlled by a logarithmic
term, i.e.
\begin{equation}
\begin{split}
\sigma^\text{matched}= \frac{\sigma^\text{4FS} +
  w\,\sigma^\text{5FS}}{1+w}\,,
\end{split}
\end{equation}
with the weight $w$ defined as 
\begin{equation}
\begin{split}
w = \ln\frac{\mH}{\mb}  - 2\,.
\label{eq::t}
\end{split}
\end{equation}
The theoretical errors are combined according to
\begin{equation}
\begin{split}
\Delta\sigma_\pm = \frac{\Delta\sigma_\pm^\text{4FS}
  + w\,\Delta\sigma_\pm^\text{5FS}}{1+w}\,,
\end{split}
\label{eq::error}
\end{equation}
where $\Delta\sigma_\pm^\text{4FS}$ and
$\Delta\sigma_\pm^\text{5FS}$ are the upper/lower uncertainty limits
of the 4FS and the 5FS, respectively.

The corresponding $\bbH$ cross-sections expanded to a scan over SM
Higgs boson masses are presented in \refTs{tab:bbH_XStot_7}--\ref{tab:bbH_XStot_14}.
\refTs{tab:bbH_XStot_125} and \ref{tab:bbH_XStot_12509} summarize the Standard Model
$\bbH$  cross-sections and the corresponding uncertainties
for the different proton--proton collision energies
for a Higgs boson mass $\MH=125\UGeV$ and  $\MH=125.09\UGeV$, respectively.

\subsection{NLO+NLL matching}

This section summarizes the approach of Ref.~\cite{Bonvini:2015pxa}
which combines the virtues of the 4F and 5F schemes into a fully
consistent fixed-order + resummation matched result, which is valid for
any parametric scale hierarchy between $m_b$ and $m_H$. The updated
results for LHC 13 TeV together with the estimate for the perturbative
and parametric uncertainties  are discussed in detail in
Ref.~\cite{Bonvini:2016fgf}. The code for the NLO+NLL matched
predictions will be available at
\href{http://www.ge.infn.it/~bonvini/bbh}{\texttt{http://www.ge.infn.it/$\sim$bonvini/bbh}}.

To describe the matched result, it is instructive to first briefly recall the 4FS and 5FS results.
The 4FS calculation of the $bbH$ cross section is formally valid in the $m_b \sim m_H$ limit.
In this scheme, bottom quarks do not appear in the initial-state, but rather via gluon splitting
into $b$-quark pairs.
The 4FS cross section includes both power corrections in the bottom-quark mass, $O(m_b^2/m_H^2)$,
and logarithmic terms $\sim\ln(m_b^2/m_H^2)$ at fixed order in $\alpha_s$ that arise from collinear
gluon splittings into $b$-quark pairs.

The 4FS (fixed-order) result is given by the Mellin convolution of coefficient functions\break
$D_{ij}(m_H,m_b,\mu_F)$, which depend explicitly on $m_b$, and evolved 4FS PDFs, 
\begin{align}
  \label{eq:bbH4FSexplicit}
  \sigma^{\rm FO} &= \sum_{i,j=g,q,\bar q} D_{ij}(m_H,m_b,\mu_F) \otimes
                    f^{[4]}_i(\mu_F) \otimes f^{[4]}_j(\mu_F) \nonumber\\
                  &= \sum_{i,j,k,l=g,q,\bar q}  D_{ij}(m_H,m_b,\mu_F) \otimes
                    \left[U^{[4]}_{ik}(\mu_F,\mu_0) \otimes f^{[4]}_k(\mu_0)\right] \otimes
                    \left[U^{[4]}_{jl}(\mu_F,\mu_0) \otimes f^{[4]}_l(\mu_0)\right].
\end{align}
The sum is restricted to gluons and light quarks and antiquarks ($q=d,u,s,c$),
$\mu_F$ is the hard (factorization) scale of the process, $\mu_0$ is the scale
at which PDFs are fitted,%
\footnote{We consider the charm quark as fitted light-quark PDF; in most PDF fits, the charm PDF is generated perturbatively
similar to the bottom PDF, but this is not of relevance for the present discussion.}
and $U^{[4]}$ are DGLAP evolution factors with $n_f = 4$ active quark flavours.
The 4FS predictions for the inclusive cross section are most accurate at small values of the Higgs boson masses 
(parametrically $m_H \sim m_b$), where power corrections are important and the logarithms are small.

The 5FS is formally valid in the limit $m_b \muchless m_H$. In this scheme, bottom quarks do appear in the initial state.
The large collinear logarithms are resummed through DGLAP evolution by the introduction of a bottom-quark PDF, as 
described in Sect.~\ref{sec:bbh-pdfs}. In this scheme, the bottom quark mass is set to zero in the partonic cross section and 
therefore the power-corrections $O(m_b^2/m_H^2)$ are not included. 
The cross section is given by the convolution of coefficient functions $C_{ij}(m_H,\mu_F)$, which contain no $m_b$ dependence,
with evolved 5FS PDFs,
\begin{align}
  \label{eq:bbH5FSexplicit}
  \sigma^{\rm Resum} &= \sum_{i,j=g,q,\bar q,b,\bar b} C_{ij}(m_H,\mu_F) \otimes
                    f^{[5]}_i(m_b,\mu_F) \otimes f^{[5]}_j(m_b,\mu_F) \nonumber\\
  &= \sum_{i,j=g,q,\bar q,b,\bar b} C_{ij}(m_H,\mu_F)
  \otimes \Bigg[\sum_{\substack{k=g,q,\bar q,b,\bar b\\l,p=g,q,\bar q}} U^{[5]}_{ik}(\mu_F,\mu_b) \otimes \mathcal{K}_{kl}(m_b,\mu_b)
    \otimes U^{[4]}_{lp}(\mu_b,\mu_0) \otimes f^{[4]}_p(\mu_0)\Bigg] \nonumber\\
  &\hspace{8em}\otimes \Bigg[\sum_{\substack{k=g,q,\bar q,b,\bar b\\l,p=g,q,\bar q}} U^{[5]}_{jk}(\mu_F,\mu_b) \otimes \mathcal{K}_{kl}(m_b,\mu_b)
    \otimes U^{[4]}_{lp}(\mu_b,\mu_0) \otimes f^{[4]}_p(\mu_0)\Bigg].
\end{align}
In the second equation we explicitly wrote out how the 5FS PDFs at $\mu=\mu_F$ are perturbatively constructed from the fitted 4FS PDFs at the
initial scale $\mu_0$, the 4-flavour evolution from $\mu_0$ to the bottom matching scale $\mu_b$,
the matching Eq.~\eqref{eq:PDFmatching} at $\mu_b$ and the 5-flavour evolution from $\mu_b$ to $\mu_F$.

To facilitate the matching of the two results, it is advantageous to bring the 5FS result into a form that is
consistent with the logarithms present in the fixed-order result. 
To do so, the perturbative counting assigned to the resummed cross section is revisited.
Usually, the PDFs are treated as external $O(1)$ objects and one assigns a perturbative counting to the cross section based
on a perturbative counting of the coefficient functions alone. 
For the $b$-quark PDF, this would be justified in the limit where the off-diagonal evolution factor $U^{[5]}_{bg}(\mu_F,\mu_b) \sim 1$.
However, $U^{[5]}_{bg}$ is $\alpha_s \ln(\mu_F/\mu_b)$-suppressed relative to the diagonal evolution factors, and therefore this only holds for scales $\mu_F \ggg \mu_b$. Numerically, for $\mu_b\sim O(m_b)$ this is only attained for scales $\mu_F \gtrsim 1$~TeV. Hence, for the relevant scales of interest here it is more appropriate to count $U^{[5]}_{bg}$ as $O(\alpha_s)$, and therefore also
count $f_b^{[5]}$ as $O(\alpha_s)$. We refer to the 5FS result reorganized with this counting as the resummed result. For a more detailed discussion we refer to Ref.~\cite{Bonvini:2015pxa}.

This counting has the important property that it provides a perturbative treatment of the $b$-quark PDF that is consistent between the fixed-order (4FS) and resummed results. The 5FS cross section in Eq.~\eqref{eq:bbH5FSexplicit} is expanded in $\alpha_s$ by counting $U_{bg}^{[5]}\sim\alpha_s$ and expanding the coefficient functions $C_{ij}$ \emph{together} with the $b$-quark matching coefficients $\mathcal{K}$. The key feature is that order by order in $\alpha_s$ the limit $\mu_b \to \mu_F$ in the resummed cross section then exactly reproduces \emph{all} the logarithmic terms (and nothing more) that are present in the $m_b\to 0$ limit of the fixed-order cross section. (Or in other words, the reexpansion of the resummed result to fixed order is simply given by setting $\mu_b \to \mu_F$.)
This in turn means that for $\mu_b < \mu_F$ the evolution factors $U^{[5]}$ in this expansion precisely resum the singular logarithms
present in the fixed-order result. 

With this re-arrangement of the resummed cross section, all that is missing in the latter compared to the fixed-order result are
purely nonsingular terms (i.e. terms that do vanish in the limit $m_b \to 0$). Therefore, matching the two results becomes completely
straightforward. All that is required is to add to the purely resummed result the remaining nonsingular fixed-order terms.
The matched cross section is given by
\begin{align}
\sigma^{\rm FO+Resum} = \sigma^{\rm Resum} + \big( \sigma^{\rm FO} - \sigma^{\rm Resum}\big|_{\mu_b=\mu_F} \big).
\end{align} 
By construction, it satisfies $\sigma^{\rm FO+Resum} \to \sigma^{\rm FO}$
in the limit $\mu_b \to \mu_F$, as required for a consistently matched prediction. 
The terms in brackets are precisely the nonsingular terms. For the practical implementation, they can be conveniently absorbed into modified gluon and light-quark coefficient functions, $\bar{C}_{ij}(m_H,m_b,\mu_F)$, which now carry an explicit dependence on $m_b$, convolved with 5F PDFs.%
\footnote{Moving the nonsingular corrections underneath the 5F resummation corresponds to including some resummation effects for power-suppressed terms, which is beyond the order one is working in either the 4FS or 5FS.}
The final matched result is then written as
\begin{align}\label{eq:bbHMatchedexplicit}
\sigma^{\rm FO+Resum}
&= \sum_{i,j=b,\bar b} C_{ij}(m_H,\mu_F) \otimes f^{[5]}_i(m_b,\mu_F)
\otimes
f^{[5]}_j(m_b,\mu_F) \nonumber\\
&+ \sum_{\substack{i=b,\bar b\\ j=g,q,\bar q}}
\Big[C_{ij}(m_H,\mu_F) \otimes f^{[5]}_i(m_b,\mu_F) \otimes
f^{[5]}_j(m_b,\mu_F)
+ (i \leftrightarrow j) 
 \Big]
\nonumber\\
&+ \sum_{i,j=g,q,\bar q} \bar C_{ij}(m_H,m_b,\mu_F) \otimes
f^{[5]}_i(m_b,\mu_F) \otimes
f^{[5]}_j(m_b,\mu_F),
\end{align}
where $f^{[5]}_{i,b}$ are perturbative objects, and an expansion of $C_{ij}$ and $\bar{C}_{ij}$ against $f^{[5]}_{i,b}$ as discussed above is implicit. It was explicitly checked that for values of $m_b/m_H \lesssim 0.1$ a strict expansion of the square brackets in Eq.~\eqref{eq:bbH5FSexplicit} 
against the expansion of $C_{ij}$ gives practically the same numerical results as expanding $C_{ij}$ but keeping the square brackets 
unexpanded, as in the first line of Eq.~\eqref{eq:bbH5FSexplicit}, under the condition that one counts $f^{[5]}_b \sim \alpha_s$.
This allows for a significant simplification in the implementation of the matched cross section, as it avoids
having to split up the $b$-quark PDF into its pieces but allows working with a common $b$-quark PDF.%
\footnote{This applies to all Higgs boson mass values considered in this report except for $m_H=25$~GeV, for which the strict expansion would be necessary.}

With this simplification, expanding the all-orders result for the matched cross section in powers of $\alpha_s=\alpha_s(\mu_F)$, the following
perturbative expansion is obtained
\begin{align}
 \text{LO+LL} && \sigma &= \quad \alpha_s^2 \bar C^{(2)}_{ij}\otimes
f^{[5]}_{i}\otimes f^{[5]}_{j} + \alpha_s 4 C^{(1)}_{bg}\otimes
f^{[5]}_b \otimes f^{[5]}_g + \phantom{\alpha_s}2C^{(0)}_{b\bar
b}\otimes f^{[5]}_b \otimes f^{[5]}_b \nonumber\\
 \text{NLO+NLL} &&  & \quad + \alpha_s^3 \bar C^{(3)}_{ij}\otimes
f^{[5]}_{i}\otimes f^{[5]}_{j} +\alpha_s^2 4 C^{(2)}_{bk}\otimes
f^{[5]}_b \otimes f^{[5]}_k +\alpha_s 2C^{(1)}_{b\bar b}\otimes
f^{[5]}_b \otimes f^{[5]}_b \nonumber\\
 \text{NNLO+NNLL} &&  & \quad + \alpha_s^4 \bar C^{(4)}_{ij}\otimes
f^{[5]}_{i}\otimes f^{[5]}_{j} +\alpha_s^3 4 C^{(3)}_{bk}\otimes
f^{[5]}_b \otimes f^{[5]}_k +\alpha_s^2 (2C^{(2)}_{b\bar
b}+2C^{(2)}_{bb}) \otimes f^{[5]}_b \otimes f^{[5]}_b \nonumber\\
 && & \quad + \ldots. \label{eq:bbhCountingRes}
\end{align}
The factors of two and four account for the exchange of partons among
the two protons and (to a first approximation) the equality
$f^{[5]}_b=f^{[5]}_{\bar b}$. A sum over \emph{light} quark and gluons is implicitly assumed when latin indices $i,j,k,...$ are repeated.
The superscripts on the coefficient functions indicate the order in $\alpha_s$ to which these are computed.
In \refF{fig:bbh-diagrams} the counting of Eq.~\eqref{eq:bbhCountingRes} is pictorially summarized and compared
to the 4FS and 5FS counting.
\begin{figure}
\centering
\includegraphics[width=0.7\textwidth]{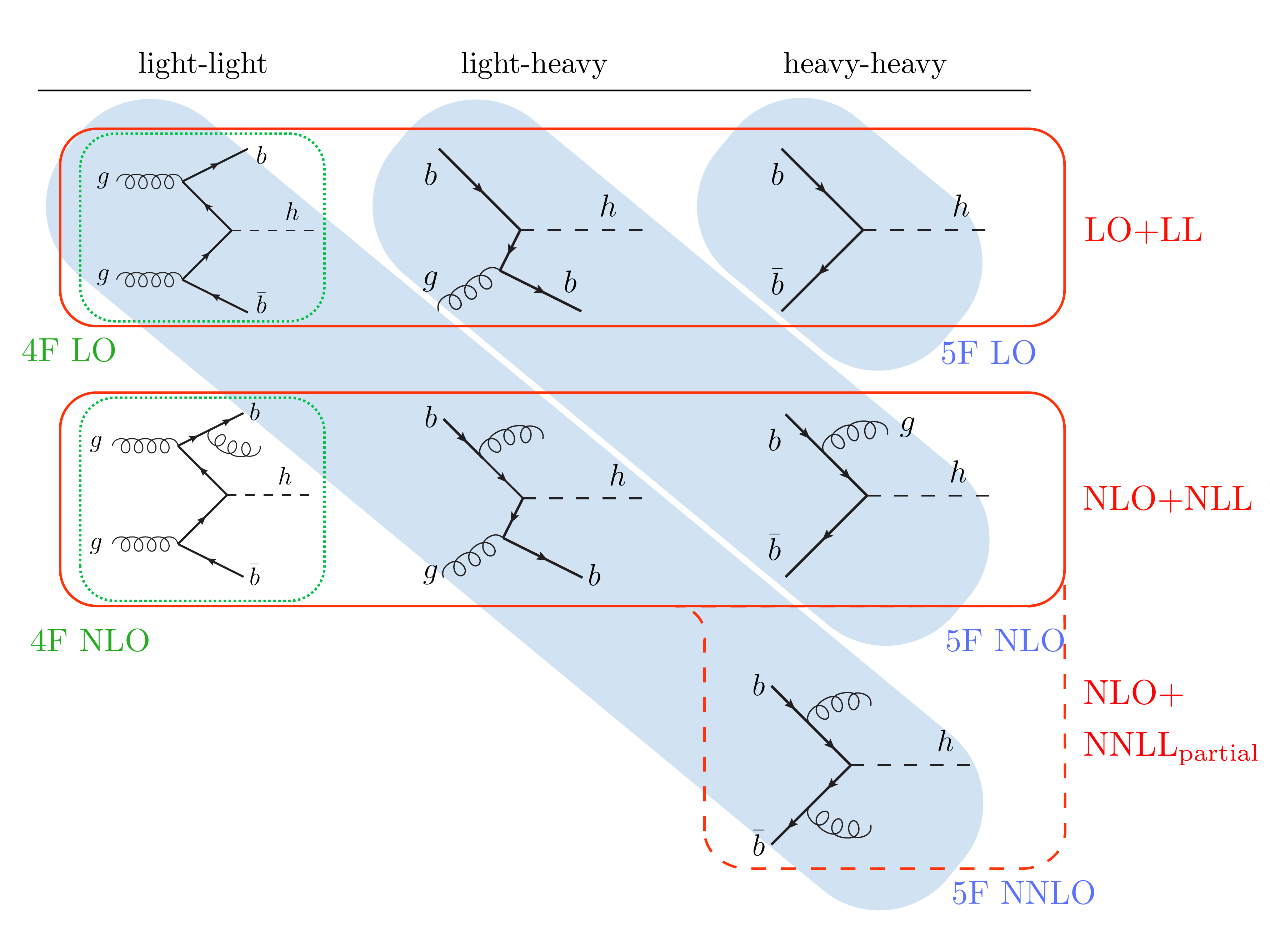}
\caption{Pictorial representation with sample diagrams appearing in the computation of the $bbH$ cross section,
  grouped according to the different perturbative countings adopted in the 4FS (green), 5FS (blue areas)
  and the matched resummed result of Ref.~\cite{Bonvini:2015pxa} (red).}
\label{fig:bbh-diagrams}
\end{figure}

At this point, it is important to point out that the construction of the
coefficient functions $\bar{C}_{ij}$ is formally the same as the
corresponding construction in the the FONLL
approach~\cite{Forte:2015hba} (and to a hypothetical S-ACOT
construction).  There are however, two main differences between these
approaches. Firstly, the matched NLO+NLL result of
Refs.~\cite{Bonvini:2015pxa,Bonvini:2016fgf} counts the effective
$b$-quark PDF as an $O(\alpha_s)$ perturbative object, which follows
from including all perturbative ingredients in the perturbative
counting, as explained above. Secondly, the results of
Refs.~\cite{Bonvini:2015pxa,Bonvini:2016fgf} and those provided here,
explicitly include an estimate of the resummation uncertainty associated
with the 5F resummation by varying the (in principle arbitrary) matching
scale $\mu_b$.

At present, all coefficient functions in Eq.~\eqref{eq:bbhCountingRes}
except $\bar C^{(4)}_{ij}$ and $C^{(3)}_{bk}$ are
known~\cite{Harlander:2003ai,Buehler:2012cu,Dittmaier:2003ej,Dawson:2003kb}.
Therefore, the highest possible accuracy that can be currently achieved
is NLO+NLL and it is not yet possible to produce full NNLO+NNLL results,
which would require knowledge of the full NNLO 4FS result.  However, the
two-loop coefficients $C^{(2)}_{b\bar b}$ and $C^{(2)}_{bb}$ are
known~\cite{Harlander:2003ai}, and can in principle be added to the
NLO+NLL result. In our counting this provides a partial NNLL result,
denoted as NLO+NNLL$_{\rm partial}$ in \refF{fig:bbh-diagrams}.%
\footnote{In Ref.~\cite{Bonvini:2015pxa}, this result was called
NLO+NLL+$C^{(2)}_{b\bar b}$.} Including these higher-order terms at
low/intermediate Higgs boson masses would include part of the resummation of
some NNLL logarithms that are not present in the NLO fixed-order result
(i.e.\ in the matched result the resummation is no longer consistent
with the fixed order).  This could potentially bias the result and/or
lead to underestimating the perturbative uncertainty. For this reason,
we do not recommend this as the default result.  Nevertheless, we also
provide it here, as it gives a good indication of the size of the
next-order correction, thus providing a useful additional cross check on
the estimated perturbative uncertainties of the NLO+NLL result, which
indeed fully cover the NLO+NNLL$_{\rm partial}$ result.  In the limit of
very large $m_H$, including these terms may be beneficial, once the size
of the resummed logarithms grows, $\alpha_s\log(\mu_F^2/\mu_b^2)\sim 1$,
and the original strict 5FS counting applies.

Regarding the nonsingular terms, at LO there is only a contribution
proportional to $\yb^2$. At NLO there are terms proportional to $\yb^2$
as well as terms $\sim \yb \yt$ due to the interference of the
Born-level diagrams diagrams with diagrams involving a top-quark loop,
see \eqn{sig4FS}. These interference terms can lead to a noticeable
correction. For the scale setup chosen, the $\yb \yt$ terms reduce the
cross section for $m_H\lesssim300$~GeV and increase it for
$m_H\gtrsim300$~GeV. They were not yet included in
Ref.~\cite{Bonvini:2015pxa} but are straightforward to add as they
correspond to purely nonsingular corrections, and their effect was
studied in Ref.~\cite{Bonvini:2016fgf}. Here, we provide the results
both with and without the interference terms included.

Next, we turn to the discussion of how to estimate the theoretical
uncertainties.  The approach taken in Ref.~\cite{Bonvini:2015pxa}, and
summarized in Sect.~\ref{sec:bbh-pdfs} is that perturbative
uncertainties are most realistically estimated by varying both the hard
scale $\mu_F$ as well as the scale $\mu_b$ that resides in the PDFs.  In
this report we further explore the dependence on the renormalization
scale $\mu_R$ at which $\alpha_s$ is computed, and on the
renormalization scale $\mu_Y$ at which the Yukawa coupling is evaluated.
For the parametric uncertainty on the bottom mass, the prescription of
Sect.~\ref{sec:bbh-pdfs}, Eq.~\eqref{eq:bbH-mbmubvariations}, is used.
It is important of course that we consistently change the value of the
pole mass in the coefficient functions $\bar C_{ij}$ when varying its
value in the PDFs.

The Yukawa coupling is computed according to the prescription of
Sect.~\ref{chapter:input}, namely evolving from
$\overline{m}_b(\overline{m}_b)=4.18$~GeV to the central Yukawa scale
$\mu_Y$ with 4-loop evolution, while $\mu_Y$ variations are computed
using 2-loop evolution.  The strong coupling $\alpha_s$ in the
coefficient functions is evaluated at the renormalization scale $\mu_R$.
While both $\mu_R$ and $\mu_Y$ are renormalization scales, they do not
necessarily need to be the same; it is always possible to evolve
$\alpha_s$ and the Yukawa coupling to different scales using their own
renormalization group evolution, compensating in the partonic
coefficients with the appropriate fixed-order logarithms.  We have found
that varying $\mu_R$ and $\mu_Y$ together gives the maximal uncertainty
and we therefore set $\mu_Y=\mu_R$ in all our results and variations.

The factorization scale $\mu_F$ is taken to be $\mu_F=(m_H+2m_b)/4$, and
varied by a factor of two up and down.  The choice is motivated by the
well-known fact that in $bbH$ such a small factorization scale leads to
improved perturbative convergence.  This choice is also consistent with
the scale adopted in the 4FS and 5FS computations, but we emphasize that
the matched NLO+NLL result is significantly less sensitive to the
central value of $\mu_F$~\cite{Bonvini:2015pxa}.  The value of $m_b$ in
the definition of $\mu_F$ is taken to be the central pole mass value
$m_b=4.58$~GeV, and is kept fixed under $m_b$ variations.  While varying
$\mu_F$, the threshold scale $\mu_b$ is varied by the same factor of two
up and down.  As discussed in Ref.~\cite{Bonvini:2015pxa}, this enables
us to interpret the $\mu_F$ variation as an overall fixed-order
uncertainty. In addition, we estimate a resummation uncertainty by
separately varying $\mu_b$ by a factor of two while keeping all other
scales and $m_b$ fixed.

In Ref.~\cite{Bonvini:2015pxa}, the central value for $\mu_R$ ($=\mu_Y$)
was taken to be the same as $\mu_F$, and all scales were varied
together. This choice leads to an excellent perturbative convergence.
Despite this, here an even more conservative approach is taken whereby
$\mu_R$ and $\mu_F$ are varied independently.  In addition, we use a
larger central value for the renormalization scale, namely
$\mu_R=m_H/2$.  This is motivated by the fact that the primary reason
for a small scale $\sim m_H/4$ is related to the collinear factorization
($\mu_F$) and not the renormalization ($\mu_R$).  On the other hand,
choosing $\mu_R=m_H$ produces somewhat artificial leftover $\ln 4$ terms
in the cross section, so $\mu_R=m_H/2$ seems a sensible compromise and
also lies between the 4FS and the 5FS choices for $\mu_R$.  This choice
has the additional advantage that the NLO+NNLL$_{\rm partial}$ turns out
to be a tiny correction over the NLO+NLL.

For each of the fixed-order (envelope of individual $\mu_F$ and $\mu_R$
variations omitting cases where the product of the variations from their
respective central values exceeds a factor of two), resummation
($\mu_b$), and parametric $m_b$ uncertainties we take the maximum
deviation from the central result as the symmetrized uncertainty. The
total perturbative uncertainty is obtained by adding the $\mu_F+\mu_R$
and the $\mu_b$ uncertainties in quadrature. We have checked that this
includes the envelope of the individual variations of the three scales.
The total parametric uncertainty emerges from the quadratic sum of the
$m_b$ and the asymmetric PDF+$\alpha_s$ (computed according to the
PDF4LHC15 prescription) uncertainties. The full uncertainty band is then
obtained by adding the perturbative and parametric uncertainties
linearly. For a detailed discussion of the different uncertainties in
the cross section see Ref.~\cite{Bonvini:2016fgf}.

\subsection{FONLL matching}

\newcommand{\dD}{\mathrm{d}}

The FONLL method is based on the observation that
the four- and the five-flavour scheme perturbative
series are made up, although formally at different orders, 
by the same type of terms with the difference that perturbative
coefficients in the four-flavour expansion exhibit full 
mass dependence.  The idea is therefore to replace
in the five-flavour scheme expansion those terms, 
that are known in the four-flavour scheme, 
with their counterpart calculated in the four-flavour scheme.
This procedure can thus be seen as 
including fixed-order (of the order of the four-flavour 
scheme calculation) mass effects to a resummed calculation
(of the order of the five-flavour scheme).

In the case of Higgs boson production in bottom-quark fusion
the five-flavour scheme total cross section is known up
to NNLL, $\order{\as^2}$, while the four-flavour one
is known up to NLO, $\order{\alpha_s^3}$. The FONLL
matching can thus be performed in two different ways:
one could either match the NNLL five-flavour with the 
LO four-flavour which we refer to as FONLL-A \cite{Forte:2015hba},
or the NNLL five-flavour to the NLO four-flavour
scheme calculation, which we will refer to as 
FONLL-B \cite{Forte:2016sja}.

In order to perform the matching procedure in either
of the mentioned schemes, we need to have comparable
perturbative series. This means in particular that both
the four- and the five-flavour scheme have to be expressed
in terms of the same scheme couplings and PDFs.
To achieve this, one needs to compute matching conditions
in both schemes.
In the four-flavour scheme, one has to express the value
of $\alpha_S(\mu_R^2)$ calculated with four active flavour
to that evolved with five active flavour up to the desired 
order. The same must be done for the PDFs evolution 
kernel as well.
This can be done by using the following relations
\begin{align}
\alpha_S^{4F}(\mu^2) &= \alpha_S(\mu^2) \left[1+ \sum_{p=1}^\infty c_i\left(\log\frac{\mu^2}{m_b^2}\right) \, \alpha_S^p(\mu^2)\right]\, ; \\
f_i^{4F}(x,\mu^2) &= \sum_{j=q,\bar{q},g}\int_{x}^{1}\frac{\dD y}{y}\left[\delta_{ij}\,\delta(1-y) + \sum_{p=1}^\infty \alpha_S^p(\mu^2)\,K_{ij}^{(p)}(y,\mu^2)\right] f_j\left(\frac{x}{y},\mu^2\right)\,.
\end{align}
Quantities were the flavour number scheme is not explicit 
are intended as to be computed in the five-flavour scheme.
In order to perform the FONLL-A type matching we don't
need any of these formulae, however we need the $\order{\alpha_S}$
expression for both in the FONLL-B case.

Likewise, in the five-flavour scheme one also has to compute
some matching conditions. In particular, to make the
two scheme comparable, one needs to re-express the
bottom-quark PDF in terms of the light-quarks and gluon PDFs.
In both cases only the terms of the needed order have to be 
kept into account.
Thus we need respectively the $\order{\alpha_S}$  and 
$\order{\alpha_S^2}$ of:
\begin{equation}
f_b(x,\mu^2) = \sum_{j = q,\bar{q},g}\int_x^1 \frac{\dD y}{y}\left[ \sum_{p=1}^{\infty}\alpha_S^p(\mu^2) \mathcal{A}_{bj}^{(p)}(y,\mu^2) f_j\left(\frac{x}{y},\mu^2\right)\right]
\end{equation}
for the FONLL-A and -B matching.

We need then to subtract from the five-flavour scheme
those terms which are computed with exact mass dependence
in the four-flavour scheme. The terms that we need for this
purpose are those terms which are either logarithmic
or constant in $m_b$, {\it i.e} all non-vanishing terms
in the four-flavour scheme when we take the limit
of $m_b \rightarrow 0$.
Clearly those terms must also be present in the five-flavour
scheme and can be computed from the five-flavour scheme
as well.
Critically, we choose this latter method to compute
the massless limit up to $\order{\alpha_S^2}$ and 
$\order{\alpha_S^3}$ respectively for the 
FONLL-A and -B scheme.
This enables us to perform the
subtraction in the five-flavour scheme of those terms that are known, 
in the four-flavour scheme, with full mass dependence up to the desired
order in $\alpha_S$.

To obtain numerical results we use the bbh@nnlo code for both
the five-flavour scheme partonic cross-section and those
terms needed to compute the massless limit of the 
four-flavour scheme. The NLO four-flavour scheme 
is obtained with a modified version of the 
SHERPA code such as to accommodate for the described
change of scheme, and NLO matrix elements are obtained
from the OpenLoops matrix element generator.
As we stated, the four-flavour scheme has to be computed
with a five-flavour scheme $\alpha_S$, bottom Yukawa 
coupling $y_b$ and PDFs (plus all the corrections
needed to express four-flavour scheme quantities in the 
five-flavour scheme). The running coupling constant,
as well as parton distribution functions, 
are obtain from the NNLO, five-flavour, PDF4LHC
combined PDF set through the LHAPDF interface. The running
of the bottom Yukawa coupling is computed at two-loop in the 
five-flavour scheme with $m_b(m_b) = 4.18$~GeV and
a pole mass $m_b=4.58$~GeV.
The four-flavour scheme calculation includes
both $y_b^2$ proportional terms and $y_b y_t$ ones.
Scale variations bands are obtained by varying $\muR$
and $\muF$ separately in the range $K_\mu \in [1/2, 2]$
excluding the two extrema $(\mu_R=1/2,\mu_F=2)$ and ($\mu_R=2$, $\mu_F=1/2$).
The central scale is chosen to be $\mu = (m_H + 2\,m_b)/4$.

\subsection{Comparison of different matching approaches}

\begin{figure}
\begin{center}
        \vspace{-1.0cm}
    \includegraphics[height=0.85\textwidth]{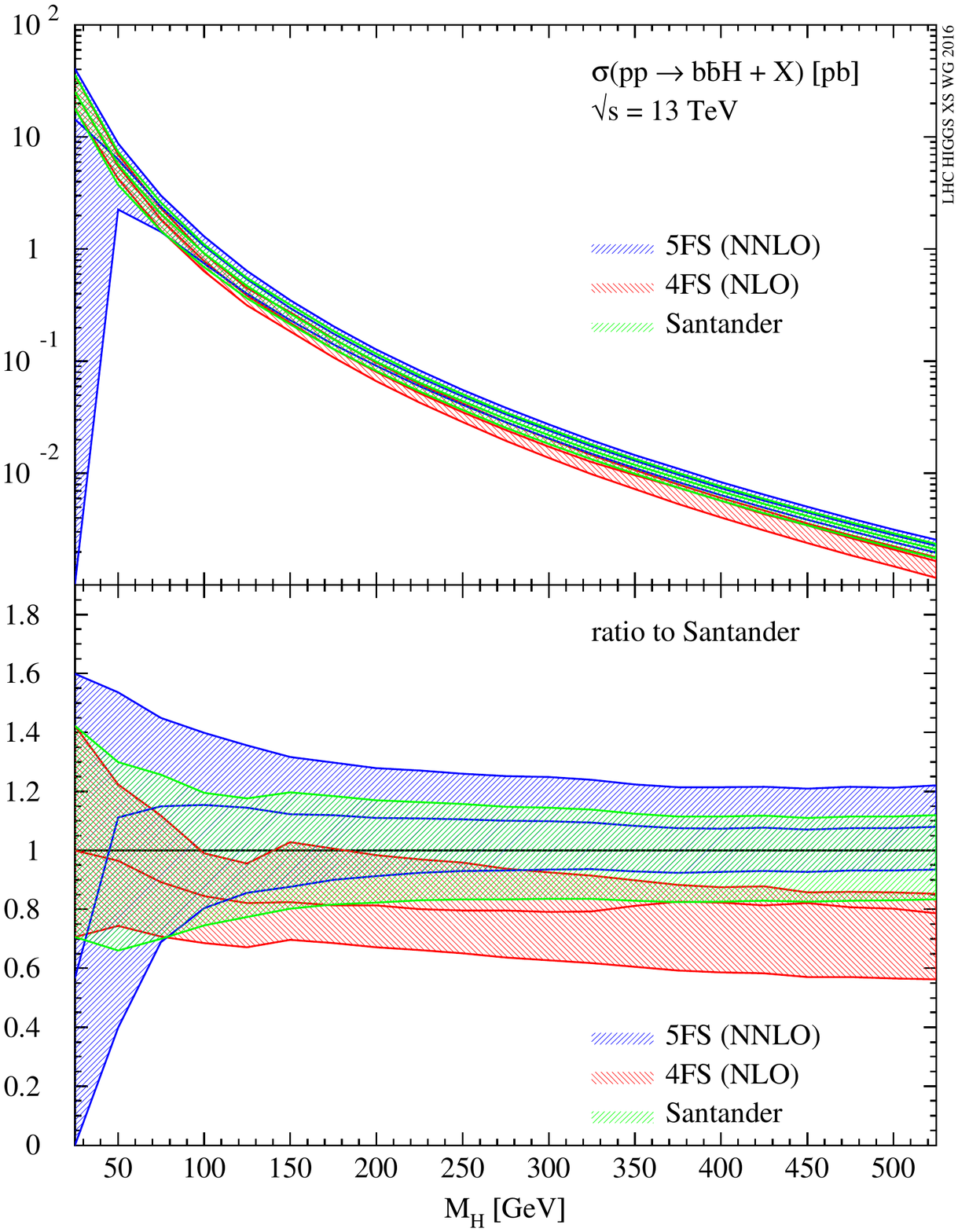}
        \vspace{-2.0cm}
  \caption{Comparison of the inclusive $\bbH$ production cross sections
in the 4FS and 5FS with the Santander-matched result. The lower plot
displays the ratios to the central Santander-matched prediction.}
  \label{fig:santander}
\end{center}
\end{figure}

First we compare the results of the 4FS, 5FS and the Santander
matching in \refF{fig:santander} including the total uncertainty
bands. The blue band displays the 5FS, the red one the 4FS and the green
band the Santander-matched result. For small Higgs boson masses the 4FS and
5FS overlap considerably and the Santander-matched result overlaps with
both of them by construction. Only for small Higgs boson masses below 50 GeV
the 5FS develops an ill-defined uncertainty due to the fact that the
central scale choices determined in terms of the Higgs boson mass become too
small to give sense to a sophisticated definition of the bottom PDFs.
For the Santander-matched result this is also reflected by a negative
value of the weight $w$ of Eq.~\ref{eq::t}. Thus for Higgs boson masses
below 50 GeV the Santander-matched result has been defined as the 4FS
one. This procedure guarantees that for small Higgs boson masses the 4FS is
singled out for the matched result, while for large Higgs boson masses the
Santander matching follows the 5FS. From the lower plot of
\refF{fig:santander} one can infer that the deviations between all
central predictions are less than about 30\% for larger Higgs boson masses.

\begin{figure}
\begin{center}
        \vspace{-1.0cm}
    \includegraphics[height=0.85\textwidth]{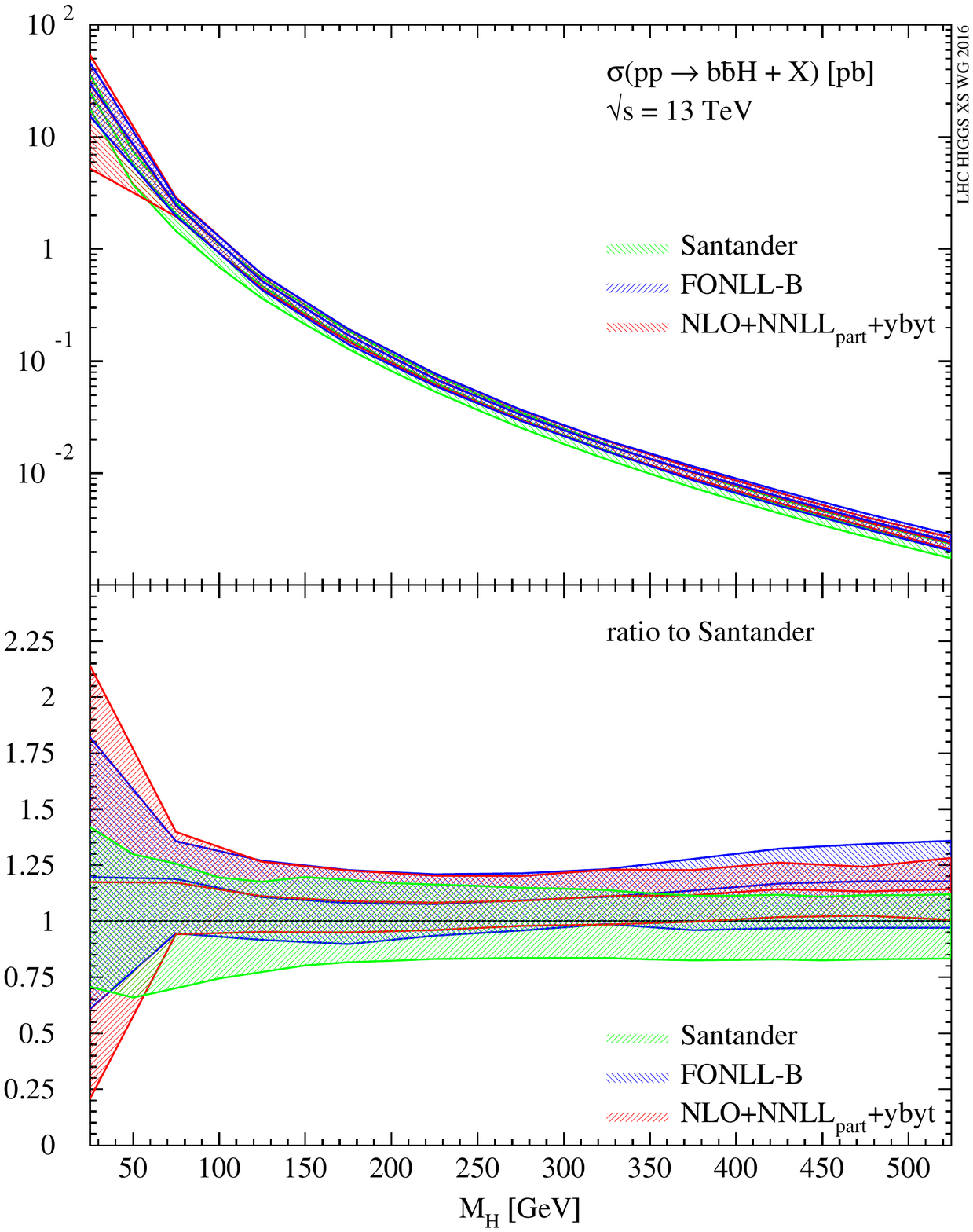}
        \vspace{-2.0cm}
  \caption{Comparison of the inclusive $\bbH$ production cross sections
for the different matching procedures, i.e. Santander matching, FONLL-B
and NLO+NNLL$_{\rm partial}$+ybyt. The lower plot displays the ratios to
the central Santander-matched prediction.}
  \label{fig:matching}
\end{center}
\end{figure}

In \refF{fig:matching} we compare the Santander matching with the
two consistent matching procedures FONLL-B and NLO+NNLL$_{\rm partial}$+ybyt
as discussed in the previous sections. The green band represents
the uncertainty band of the Santander matching, the blue one the FONLL-B
matching and the red band the NLO+NNLL$_{\rm partial}$+ybyt approach. It is
clearly visible that all three approaches overlap over the full Higgs boson mass 
range which means that also the systematic resummation approach
tends to favor the 4FS results for small Higgs boson masses and the 5FS one
for large Higgs boson masses. It should be noted that for larger Higgs boson masses
both approaches, FONLL-B and NLO+NNLL$_{\rm partial}$+ybyt, show that
resummation effects are indeed important, while finite bottom mass
effects on top of resummation effects turn out to be tiny. 
The two systematic matching procedure provide comparable
results provided the renormalization and factorization scales are set to the same values.
The residual differences in the plot are due to the different choice of renormalization
scale ($M_H/2$ in NLO+NNLL$_{\rm partial}$ and $(M_H+2m_b)/4$ in FONLL-B)
and to a different way of computing the error associated to the final prediction.
In particular, in the FONLL-B matching procedure the threshold $\mu_b$ is kept fixed to
$m_b$, while it is varied independently of $m_b$ in the NLO+NNLL$_{\rm
partial}$ approach and the scale variation is
symmetrized in the latter, while it is kept asymmetric in the former.
The matched results instead are systematically away from the Santander matched
results. The matched results are about 15\% higher than the Santander-matched
ones for all values of $M_H$, although mostly within its uncertainty band.
For large Higgs boson masses
the FONLL-B and NLO+NNLL$_{\rm partial}$+ybyt results tend towards the 5FS
as expected. The uncertainty bands of the matched calculations increase
for small Higgs boson masses due to the problems to define reasonable bottom
PDFs for small scales.

In total the consistent NLO+NNLL$_{\rm partial}$+ybyt and FONLL-B
matched results follow the expected pattern, i.e.~a clear tendency
towards the 5FS for larger Higgs boson masses. For a unique procedure to
provide a single prediction for the inclusive $b\bar b\phi$ cross
section we recommend to use the envelope of both predictions for the
uncertainty band and the central values inside as the central
prediction. 
The consistently matched results should be used when available and can
be seen as an important cross-check of the empirical Santander central
values in case a matched result is not available.

\section{Differential Monte-Carlo predictions}
\label{sec:bbHMC}
In this section, we present Monte-Carlo predictions at the 13~TeV LHC 
for both total rates with and without cuts, and several differential distributions, reconstructed 
from the final-state momenta in $\bbH$ production,
and compare them among the different generators under consideration.

Before introducing the different Monte Carlo generators under 
consideration, let us summarize the available computations for differential
$\bbH$-production in the literature. The first distributions
in the 4FS cross section have been presented (to a very limited extent though)
through NLO QCD in \Brefs{Dittmaier:2003ej,Dawson:2003kb}. A more 
comprehensive study of differential quantities in the 4FS and the first matching of the
NLO cross section to the PS has later been performed in \Bref{Wiesemann:2014ioa} 
within the MC@NLO approach \cite{Frixione:2002ik}. Recently, the corresponding computation 
in the POWHEG framework \cite{Nason:2004rx,Frixione:2007vw} was done in \Bref{Jager:2015hka}. The number of differential 
computations in the 5FS is considerably larger. At parton-level 
NLO corrections are known for the $H+b$ and $H+$jet processes
~\cite{Campbell:2002zm,Harlander:2010cz}, as well as NNLO corrections for 
the jet-vetoed rate~\cite{Harlander:2011fx} and the fully-differential cross 
section~\cite{Buehler:2012cu}. The transverse momentum distribution of the 
Higgs boson was studied analytically at NNLO \cite{Ozeren:2010qp}
and  including resummation at NLO+NLL~\cite{Belyaev:2005bs} and 
NNLO+NNLL~\cite{Harlander:2014hya}. 
NLO+PS predictions in the 5FS were computed in \Bref{Wiesemann:2014ioa}.

\subsection{\texorpdfstring{$b\bar{b}\phi$}{bb phi} in {\tt\small MG5\_aMC@NLO}}
\label{sec:bbHMG5}
The first NLO simulation matched to parton showers of the $\bbH$ signal in the
4FS has been performed in \Bref{Wiesemann:2014ioa}. This computation has been treated as 
a special case in the automated framework of \aNLO{} due to the necessity of 
a $\msbar{}$ renormalization of the bottom-quark Yukawa coupling, which 
can not be handled by the public version so far. Therefore, dedicated 
process folders have been provided\footnote{See 
\url{https://cp3.irmp.ucl.ac.be/projects/madgraph/wiki/bbH}.}, which were also used for 
all NLO+PS simulations with \aNLO{} throughout this chapter.

\aNLO{} allows for the computation of LO and NLO cross sections both with and
without matching to parton showers. NLO results not matched to parton showers
are obtained by adopting the FKS method~\cite{Frixione:1995ms,Frixione:1997np}
automated in the module \MadFKS~\cite{Frederix:2009yq}, and the OPP
integral-reduction procedure~\cite{Ossola:2006us} for the computation 
of the one-loop matrix elements (automated in the module 
\MadLoop~\cite{Hirschi:2011pa}, which makes use of
\CutTools~\cite{Ossola:2007ax} and of an in-house implementation 
of the optimizations proposed in \Bref{Cascioli:2011va} (\OpenLoops)).
Matching with parton showers is achieved
by means of the MC@NLO formalism~\cite{Frixione:2002ik}.
\aNLO\ is maximally automated.

The default treatment of Yukawa couplings in the \aNLO{} code, however, 
is that of an on-shell scheme 
renormalization, which is not optimal in the case of $\bbH$ production, 
where the $\msbar{}$ scheme has to be preferred~\cite{Braaten:1980yq}.
The advantage of an $\msbar{}$ renormalized bottom Yukawa $\overline{y}_b(\muR)$ 
is the resummation of potentially large logarithms
of \mbox{$\mH/\mb$}, when \mbox{$\muR\sim\mH$} is chosen.
An additional complication emerges from the fact that
$\yb$ enters at different powers in the $\sigma_{\yb^2}$ and 
$\sigma_{\yb\yt}$ terms introduced in eq.~(\ref{sig4FS}). 
At the moment the implementation does not warrant a completely 
general and automated solution. However, since such complication 
is recurrent in the mixed-coupling expansion as for EW corrections, 
we have included a general $\overline{y}_b(\muR)$ implementation applicable 
not only to $\bbH$ production in this context, which will be made available in the future.
This will allow any user to generate the $\bbH$ process in the general \aNLO{} 
interface and replace the current necessity of using the dedicated 
process folders.

The inputs are coordinated to be the same as the ones for the \POWHEGBOX{} 
described in the upcoming section
to warrant a consistent comparison of the results, starting from a reasonable 
agreement in the normalization, i.e., the total rate. We set the
renormalization and factorization scales to the sum of the transverse masses
of all Born-level particles, divided by a factor of four, which in the soft/collinear limit is in 
keeping with the scales used for the total 4FS rate (see Sect.\,\ref{sec:santander}):
\beq
\muR=\muF=\mu_0=\frac{\Ht^{\rm Born}}{4}\equiv 
\frac{1}{4}\sum_{i\in \{b,\bar{b},\phi\}}\sqrt{m_i^2+\pt^2(i)}\,.
\label{scref}
\eeq
The respective scale uncertainties are estimated by the independent variation
$0.5\,\mu_0\le\muR,\muF\le 2\,\mu_0$ with the constraint $0.5\le\muR/\muF\le 2$.

The additional factor of $1/4$ in the scale settings of the 4FS reflects the
fact that the optimal values for the hard scales that enter the $\bbH$ calculation
appear to be significantly smaller than the hardness of the process
would suggest. As pointed out in \Bref{Wiesemann:2014ioa} another 
hard scale is affected by this choice, when considering simulations matched to parton showers, 
namely the shower scale $\Qshow$, which loosely 
speaking can be identified with the largest hardness accessible to the shower.
It is the MC that determines, event-by-event, the value of $\Qshow$, by choosing it
so as to maximize the kinematic population of the phase-space due to
shower radiation, without overstretching the approximations upon which 
the MC is based. 

In \aNLO\ one is given the possibility of
setting the upper value\footnote{If the MC-determined $\Qshow$ value 
is lower than that set by the user, the latter is ignored. Also
bear in mind that the physical meaning of $\Qshow$ depends on the
specific MC employed -- see \Bref{Alwall:2014hca}.}
of $\Qshow$; this value is actually picked up at random in a 
user-defined range:
\beq
\alpha f_1\sqrt{s_0}\le\Qshowmax\le\alpha f_2\sqrt{s_0}\,,
\label{murange}
\eeq
so as to avoid possible numerical inaccuracies due to the presence 
of sharp thresholds.\footnote{More details can be found in Sect.~2.4.4 of
\Bref{Alwall:2014hca} (see in particular Eq.~(2.113) and the
related discussion).} $s_0$ is the Born-level
partonic centre of mass energy squared, and $\alpha$, $f_1$, and $f_2$ are numerical
constants whose default inputs are $1$, $0.1$, and $1$, respectively. 
The way, in which $\Qshowmax$ is generated, results in a 
distribution peaked at values slightly larger than 
\mbox{$\alpha(f_1+f_2)\sqrt{\langle s_0\rangle}/2$}.
As argued in \Bref{Wiesemann:2014ioa} the typical shower 
scales in the default setup are rather large as compared to the 
factorization scale, even though their origin is quite similar being both 
based on the soft/collinear approximation. By setting
$\alpha=1/4$ the two scales become significantly closer.
\Bref{Wiesemann:2014ioa} further studied the distribution of the 
Born-level ``system'' ($\ptsyst$), which showed a strongly improved 
matching behaviour of the NLO+PS curve with the NLO curve in the 
high-$\ptsyst$ tail for smaller values of $\alpha$. 
In conclusion a reduced shower scale by setting 
$\alpha=1/4$ has to be preferred over the default choice in \aNLO\ and will
be the default choice fo all numerical results produced with \aNLO\
throughout this section.

Additionally, we assign a theoretical uncertainty to the shower scale choice 
by varying $\alpha\in[1/(4\sqrt{2}),\sqrt{2}/4]$, which is added linearly to 
the $\muR$-$\muF$ scale uncertainties.

\subsection{\texorpdfstring{$b\bar{b}\phi$}{bb phi} in the \POWHEGBOX{}}
\label{sec:powheg-box}
In Ref.~\cite{Jager:2015hka} an implementation of the NLO-QCD calculation of Ref.~\cite{Dawson:2003kb} in the framework of the \POWHEGBOX{}~\cite{Nason:2004rx,Frixione:2007vw,Alioli:2010xd} has been presented. 
While the virtual corrections to the $pp \to \bbH$ process have been extracted from the fixed-order calculation of Ref.~\cite{Dawson:2003kb}, the tree-level amplitudes were generated with a tool based on {\tt MadGrap~4}~\cite{Stelzer:1994ta,Alwall:2007st}. The process-independent ingredients of the implementation are provided internally by the \POWHEGBOX{}.  All building blocks have been implemented in the 4FS, i.e.\ no contributions from incoming bottom quarks have been taken into account, and the bottom-quark was always assumed to be massive.

In the virtual corrections, not only diagrams including a $\bbh$ coupling emerge, but also loop diagrams with a $t\bar t \phi$ coupling. Both contributions are fully taken into account in the implementation. However, a switch is provided that allows the user to deactivate the contributions involving a $t\bar t \phi$ coupling. 

By default, the renormalization of the bottom-quark Yukawa coupling is defined in the $\msbar$
renormalization scheme~\cite{Dittmaier:2003ej,Dawson:2003kb,Dawson:2005vi}. In addition, an option for defining the bottom-quark Yukawa coupling in the on-shell renormalization scheme is provided. 

In its default version, the \POWHEGBOX{} code for $pp\to \bbH$ provides three different options for the renormalization and factorization scales: a fixed scale, $\mu_0 = (m_H+2 m_b)/2$, as used for total rates, a dynamical scale defined via the transverse masses $m_T(f)=(m_f^2+p_{T,f}^2)^{1/2}$ of the born-level final state particles $f$ in an event, $\mu_0 = m_T(H)+m_T(b)+m_T(\bar b)$, or the geometrical mean of the transverse masses, $\mu_0 = \left(m_T(H)\, m_T(b)\, m_T(\bar b)\right)^{1/3}$; the latter two being particularly relevant as far as tails of kinematical distributions are concerned. Scaling factors, $\xi_R$ and  $\xi_F$, can be set individually for the factorization and renormalization scales, $\mu_R$ and $\mu_F$, such that the relevant scales are given by $\mu_R = \xi_R \mu_0$ and  $\mu_F = \xi_F \mu_0$. Throughout this section we use the second scale setting in the list above with 
$\xi_R=\xi_F=1/4$, in keeping with the settings quoted for \aNLO{} in the previous section.

In order to assess the intrinsic uncertainties associated with the matching of the NLO-QCD calculation with parton shower programs, the \POWHEGBOX{} offers the possibility to vary the so-called {\tt hdamp} parameter, defined as \cite{Alioli:2010xd}
\begin{equation}
D_h= \frac{h^2}{h^2+p_T^2}\,. 
\end{equation}
Here, $p_T$ denotes the transverse momentum of the hardest parton in the real emission contributions, and $h$ is a parameter that can be chosen by the user. If no explicit choice is made, the  {\tt hdamp} parameter is set to one. In general, this parameter determines the separation of the cross section in a part at low transverse momentum of the extra
emission, generated mainly with the Sudakov form factor, and a part at
high transverse momentum, generated mainly with the real-emission
diagrams only. The uncertainty of observables simulated at NLO+PS level associated with a variation of the $h$~parameter provides an estimate of the intrinsic matching uncertainty of the  \POWHEGBOX{}. We choose $h=1/4\,(m_H+2\,m_b)$, consistent with the shower scale setting proposed in \Bref{Wiesemann:2014ioa} and used in \aNLO{} throughout.

Residual uncertainties are estimated by performing a seven-point variation of the renormalization and factorization scales by 
a factor of two with respect to their central values. We then combine these uncertainties linearly with the variation of $D_h$ in the same range.

\subsection{\texorpdfstring{$b\bar{b}\phi$}{bb phi} in \protect\Sherpa}
\newcommand{\CSS}{C\protect\scalebox{0.8}{SS}\xspace}
\newcommand{\Comix}{C\protect\scalebox{0.8}{OMIX}\xspace}
\newcommand{\Amegic}{A\protect\scalebox{0.8}{MEGIC++}\xspace}
\newcommand{\MCatLO}{M\protect\scalebox{0.8}{C}@L\protect\scalebox{0.8}{O}\xspace}

In this section, the setup of the \Sherpa{} event generation
framework~\cite{Gleisberg:2008ta} is presented. Two classes of 
results are considered for \Sherpa:
\begin{itemize}
\item {4FS using the \MCatNLO{} matching:}\\
  One set of results is in the 4FS, and based on the
  \MCatNLO technique~\cite{Frixione:2002ik}, as implemented in
  \Sherpa~\cite{Hoeche:2011fd}.  At tree-level the simulation thus starts
  from processes such as $gg\to b\bar{b}H$ and $q\bar{q}\to b\bar{b}H$,
  where no specific cuts are applied on the $b$ quarks.  Their finite
  mass regulates collinear divergences that would appear in
  the massless case.  In most cases, therefore, a $b$ jet actually
  originates from the parton shower evolution and hadronization of a
  $b$ quark. At the NLO, our computation involves contributions proportional 
  to $\yb^2$ and $\yb\yt$, see Eq.~\eqref{sig4FS}.
\item {5FS using the \MEPSatNLO{} multi-jet merging:}\\
  Alternatively, we employ the 5FS with massless  $b$ quarks.  
  In order to account for bins with zero and one $b$ jets, multi-jet
  merging is being employed.  In \Sherpa, the well-established mechanism for
  combining into one inclusive sample towers of matrix elements with
  increasing jet multiplicity at tree level~\cite{Catani:2001cc} has
  recently been extended to next-to leading order matrix elements, in
  a technique dubbed \MEPSatNLO~\cite{Hoeche:2012yf}.  This is the technique
  chosen here.  Merging rests on a jet criterion, applied to the matrix
  elements.  As a result, jets are being produced by the fixed-order
  matrix elements and further evolved by the parton shower.  As a consequence,
  the jet criterion separating the two regimes is typically chosen such
  that the jets produced by the matrix elements are softer than the jets
  entering the analysis.  This is realized here by a cut-off of
  $\mu_{\rm jet}\,=\, 20 $ GeV.  In the \MEPSatNLO{} simulation matrix elements
  for $\bbh$ production in the 5FS up to 2 jets at NLO accuracy have been 
  included, i.e.\ final states for $\phi$, $\phi+j$, and $\phi+jj$ with $\phi$ emitted 
  from a massless bottom-quark line are calculated using the \MCatNLO{} technique, 
  while $\phi+jjj$ matrix elements are accounted for only at the LO, 
  where the jets can be light jets or $b$ jets. 
  For the former, it is of course always possible that a
  light jet originating from, e.g., a gluon, can turn into a $b$ jet through
  a $g\to b\bar{b}$ splitting during the parton shower.  
  We note that only contributions proportional to $\yb^2$ do not vanish in 
  the 5FS computation, see Eq.~\eqref{sig5FS}.
\end{itemize}
In \Sherpa, tree-level cross sections are provided by two matrix
element generators, \Amegic~\cite{Krauss:2001iv} and
\Comix~\cite{Gleisberg:2008fv}, which also implement the automated
infrared subtraction~\cite{Gleisberg:2007md} through the Catani-Seymour
scheme~\cite{Catani:1996vz,Catani:2002hc}.  For parton showering, the
implementation of~\cite{Schumann:2007mg} is employed with the difference
that for $g\to b\bar{b}$ splitting the invariant mass instead of the
transverse momentum are being used as scale.  One-loop matrix elements
are instead obtained from \OpenLoops~\cite{Cascioli:2011va,
Cascioli:2014wya}.

Our central scales, both of perturbative origin ($\mu_R$, $\mu_F$) and relevant 
to the shower ($\mu_Q$), are computed according to the so-called reverse clustering 
algorithm; for further details we refer the reader to \Bref{Hoeche:2012yf}. The residual uncertainties 
are computed through variations by a factor of two from the central scales, where we combine the 
seven-point variation of the renormalization and factorization scales linearly with the uncertainties 
related to $\mu_Q$. We adopt these scale settings and variations for both our 4FS and 5FS setups, 
introduced above.

\subsection{Comparison of the Monte-Carlo tools}
We compare predictions of the different Monte Carlo generators for the simulation of a $\bbh$ signal 
introduced in the preceding section in four- and five-flavour schemes which are at least 
NLO accurate and matched to PS.
Higgs boson decays are not considered. In the \aNLO{} and \POWHEGBOX{} computations 
we employ \PYe~\cite{Sjostrand:2007gs} for the parton-shower matching.
Throughout this section we consider Higgs boson production in association with bottom 
quarks for a SM Higgs boson with mass $\mphi=125$\,GeV at the $13$\,TeV LHC. 
We use a top-quark pole mass of $m_t=172.5$\,GeV relevant to the $\yb\yt$
contribution, see Eq.~(\ref{sig4FS}). The internal bottom-quark mass is set to its pole value of
$m_b = 4.92$\,GeV, while the bottom-quark mass in the Yukawa is renormalized in the 
$\msbar{}$ scheme and set to $m_b(\mu_R)$. The central value $m_b(\mu_R\equiv \mu_0)$ 
is evaluated with $n_f=4$ ($n_f=5$) four-loop running from $m_b(m_b)=4.18$\,GeV in the 4FS (5FS); 
scale variations are done from that central value with two-loop accuracy (which is consistent with the NLO order 
of the computations). Finally, we use the {\tt PDF4LHC15} sets of parton 
distribution functions in its four and five flavour versions where applicable.

In Table\,\ref{tab:brates} and Table\,\ref{tab:jrates} we report predictions of the 
various tools (\aNLO{}, \POWHEGBOX{}, \Sherpa 4FS, \Sherpa 5FS) for total 
rates with requirements on the final-state $b$-jets and jets, respectively. In 
Table\,\ref{tab:brates} we also give the total inclusive cross section. The three 
4FS predictions for the inclusive rate are in rather good agreement at the 1-2\% 
level, which is roughly of the order of the numerical accuracy. Also the uncertainties 
are similarly large, bearing in mind that the central scale for the Sherpa prediction differs 
from the other two 4FS results. We recall that all uncertainties quoted for the predictions 
of each code are given by a $7$-point $\muR$-$\muF$ variation combined linearly with 
the uncertainties coming from the scale related to the respective shower matching procedure
of each code. The latter of course should not have any impact on the total inclusive cross 
section, although unexpectedly the bulk of the Sherpa 4FS uncertainties on this quantity 
originates from variations of the shower starting scale. This issue could not be resolved 
in the course of this comparison.
The 5FS prediction is significantly larger than the 4FS ones and quite far beyond the quoted uncertainties
owing to the positive effect due to the merging of higher multiplicities.
In a full 5FS NNLO computation the cross section is usually  
reduced by the additional two-loop contribution, leading to a far better agreement at the level of the total 
inclusive cross section. 
In any case, the focus throughout this section is on the kinematics and distributions
of the final-state particles, which is why we will rescale the 5FS result, once we 
consider distributions below.

For the total rates with requirements on the $b$ jets, we define a $b$ jets 
as any jet that contains a $B$ hadron, using the anti-$k_T$ algorithm \cite{Cacciari:2008gp} with $R=0.4$, a minimal 
transverse momentum of $25$\,GeV and a rapidity
of $|y|<2.5$. We consider the cross section with a $b$-jet veto ($0j_b$),
one or more $b$ jets ($\ge 1 j_b$), two or more $b$ jets ($\ge 2 j_b$), exactly one $b$ jet ($1 j_b$),
and exactly two $b$ jets ($2 j_b$). To allow a comparison with the 5FS prediction we also 
show the respective acceptances for each code, i.e., the ratio of the prediction within cuts with 
respect to the total rate. The general conclusions that can be drawn from the table are the following:

\begin{itemize}
\item The cross section without any $b$-tagged object in the final state ($b$-jet veto) has the smallest value with the \aNLO{} generator, being 
roughly 20\% below \powheg{} and 10\% below the 4FS result of Sherpa. Given the well-known fact that $\bbH$ production in the 4FS comes 
with rather large ($\sim$20\%) uncertainties even at NLO, due to the logarithmic structure, 
the mutual agreement among the codes is 
still well within scale uncertainties for the jet-vetoed rate.
\item As pointed out before the 5FS is significantly larger due to the different normalization.
Looking at the acceptances, however, we see that the 5FS result is just in between the 4FS predictions by \powheg{} and Sherpa. Due to the very 
similar total rates the conclusions among the 4FS results for the acceptances are, by construction, identical to the ones for the absolute cross sections.
Overall, a large fraction (of the order of $70$\%) of the events have no $b$-tagged objects in the final state.
\item Since the cross section with the requirement of one or more $b$-jets is fully determined by the total and the 
jet-vetoed rate, the general conclusions are identical to the ones for the latter. The uncertainties are, however, larger 
(except for \powheg{}), in particular for the 5FS prediction. Let us point out again that tagging at least one of the $b$ jets
reduces the total rate by about a factor of $3$--$4$, which is rather large. Requiring a second $b$ jet further reduces the
cross section by one order of magnitude. This is consistent with the findings of \Bref{Wiesemann:2014ioa}.
\item The various predictions become increasingly different for higher $b$-jet multiplicities starting at two $b$ jets. While 
\powheg{} and Sherpa 4FS are still quite close to each other, the 5FS results and the \aNLO{} prediction are rather different; 
the former predicting generally a smaller jet activity and the latter a larger one. Still, there is by and large agreement within 
the respective uncertainties, in particular because the 5FS uncertainties are quite sizeable.
\end{itemize}

Considering total rates with requirements on the jets in Table\,\ref{tab:jrates}, we define a jet 
with the anti-$k_T$ algorithm \cite{Cacciari:2008gp} with $R=0.4$ and a minimal transverse momentum of $25$\,GeV.
We consider the same types of rates as in the case of $b$ jets for jets without 
any flavour tagging, and also report the corresponding acceptances in the table.
The general conclusions are very similar as compared to the $b$-jet case and can by summarized as follows:

\begin{itemize}
\item The $0$-jet rates in the 4FS are quite different. In particular the \powheg{} prediction is rather large, which is 25\% 
larger than Sherpa and 45\% than \aNLO{}, having only barely overlapping uncertainties with the latter. As expected, 
the 5FS rate is significantly larger, but the acceptance is quite similar to the one predicted in by Sherpa in the 4FS.
\item The $1$-jet exclusive bin agrees very well among the three 4FS codes and also the acceptances are very close 
among all four predictions.
\item The biggest difference emerges again from higher jet multiplicities (two and more), which is largest in 
the case of \aNLO{} and smallest for \powheg{}. The uncertainties become quite sizeable for the high multiplicities though.
\end{itemize}

We turn now to kinematical distribution of the final-state particles both generated already at 
the hard-matrix element level and by the shower. The figures are all organized according to 
the same pattern: In the main frame the relevant predictions for the different codes are shown 
as cross section per bin (namely, the sum of the contents of the bins is equal to the total cross section, 
possibly within cuts), with 
\aNLO{}+\PYe{} (black, solid), \powheg{}+\PYe{} (red, dotted), \Sherpa{} 4FS (blue, dashed with dots) 
and \Sherpa{} 5FS (green, dash-dotted with open boxes). In the first inset we display the bin-by-bin ratio 
of all the histograms which appear in the main frame over the black solid curve, chosen as a reference.
Finally, in a second inset the bands that represent the fractional scale dependence are given by taking 
the bin-by-bin ratios of the maximum and the minimum of a given simulation over the same central prediction 
that has been used as reference for the ratios of the first inset.
We recall that the \Sherpa{} 5FS normalization is generally much larger than 4FS results and that therefore its curve 
is rescaled to have the identical normalization as the \aNLO{}+\PYe{} result, since we use it as reference also in the ratios.

In \refF{fig:pTH_bbH} we consider the 
transverse-momentum distribution of the Higgs boson $p_T(\phi)$ with different requirements on the 
$b$ jets: inclusive (upper left), no $b$-jets (upper right), one or more $b$ jets 
(lower left) and two or more $b$ jets (lower right). Considering the inclusive 
case first, it is clear that overall there is a reasonable agreement among the predictions of the 
different codes. This conclusion can be drawn from the nicely overlapping uncertainty bands
in the second inset. The size of the uncertainties are also very similar with \powheg{} having 
a slightly smaller scale dependence than the other results. Considering only the shape of the curves 
while ignoring the bands for the moment, as shown in the first inset, some differences emerge 
even though they are not too severe. In the low-$p_T(\phi)$ region the agreement of the two Sherpa 
predictions and \powheg{} is quite remarkable in terms of shape, with \aNLO{} being a bit 
harder than the other predictions. Around $p_T(\phi)\sim 100$\,GeV the 5FS \Sherpa{} prediction 
departs from the other two predictions and gets closer to the \aNLO{} curve, while \powheg{} 
and the 4FS \Sherpa{} result staying rather close to each other over the whole $p_T(\phi)$ range 
that is displayed. One should note, however, that apart from $p_T(\phi)\lesssim 50$\,GeV 
all three 4FS results are very similar in terms of shape.

The $p_T(\phi)$ distribution with a veto on $b$ jets in the upper right panel of \refF{fig:pTH_bbH}, 
shows a quite similar pattern as in the inclusive case, with the relative size of the 
differences among the curves being amplified though. Except for the first bin the \powheg{} and \Sherpa{} 
4FS results are again in very good agreement, while the \aNLO{} prediction is quite much harder 
and up to $\sim 40\%$ away for $p_T(\phi)\lesssim 100$\,GeV. In that region also the 
\Sherpa{} 5FS prediction is very close to the \powheg{} and \Sherpa{} 4FS results, but gets 
significantly harder than all the 4FS predictions in the tail of the distribution. Due to the 
considerably increased size of the uncertainty bands, however, the predictions agree by and large 
within their respective uncertainties.

The picture essentially reverses when considering one or more $b$-tagged objects in the final state: 
the \aNLO{} result now featuring the softest spectrum and \Sherpa{} 5FS the hardest. In this case,
from $p_T(\phi)\gtrsim 30$\,GeV all four predictions (in particular the 4FS ones) are in excellent agreement 
in terms of both normalization and shape, with entirely overlapping uncertainty bands.

Finally, in the case with two or more observed $b$ jets, the biggest difference comes from the fact, that 
the overall normalization, i.e. the rate, is larger
for \aNLO{} as already pointed out in the discussion of Table\,\ref{tab:brates}.
Still, the predictions agree largely within the given uncertainty bands. Considering only the shape, 
the \aNLO{} result is in very good agreement with the \Sherpa{} 5FS curve and also quite similar to 
the one of \powheg{}. The \Sherpa{} 4FS shape in this case is a bit harder than all the other curves.

In conclusion, we find a decent agreement among the various predictions for the Higgs boson transverse-momentum 
spectrum with and without cuts, once the respective uncertainties are taken into consideration.

Let us move now to distributions relevant to the associated jets and $b$ jets 
shown in \refF{fig:b1j1_bbH}. We start by discussing the transverse-momentum 
$p_T(b_1)$  (left panel) and rapidity distributions $y(b_1)$ (right panel) 
of the hardest $b$ jet in the upper panel of \refF{fig:b1j1_bbH}. We first notice 
that the mutual agreement of the central predictions for $p_T(b_1)$ among 
all the codes is very good; none of the curves differs by more 
than $\sim 20\%$ from the others. 
Hence, we observe well overlapping uncertainty bands, which, however, turn 
out to be rather large in the case of \aNLO{} and the \Sherpa{} 5FS prediction.
The agreement is even better in terms of shape. In particular the shapes 
of \powheg{}, \Sherpa{} 5FS and \aNLO{} are essentially identical up to statistical 
fluctuations. For the rapidity $y(b_1)$ the situation is pretty much the same. 
In this case, however, all predictions are in perfect agreement apart from the 
slightly different normalization of \powheg{} and \Sherpa{} 5FS being roughly 
25\% lower than \aNLO{} as already pointed out in the discussion of 
Table\,\ref{tab:brates}. Furthermore, the \aNLO{} uncertainty is much lower 
in case of the $y(b_1)$ distribution and quite similar to the \Sherpa{} 4FS band,
while \powheg{} still has the smallest uncertainty band.

Finally, we can draw similar conclusion for the hardest jet distributions 
in the lower panel of \refF{fig:b1j1_bbH}: The residual uncertainties are rather 
large in case of the $p_T(j_1)$ distribution 
and all predictions agree well within uncertainties. The central 
predictions are quite quite similar in terms of their shape, where again the 
\Sherpa{} 5FS and the \aNLO{} result are essentially identical up to statistical 
fluctuations. For the rapidity of the hardest jet the shapes are quite similar to 
each other except for the one of \aNLO{} which is slightly enhanced in the 
forward region. Nevertheless, we find agreement within the quoted uncertainties 
for the predictions of all codes.


\begin{table}
\caption{\label{tab:brates}
Predictions for the total rates (in pb) of the various Monte-Carlo tools under consideration inclusive and within cuts on the final-state $b$ jets. For comparison, also the 
respective acceptances are given.
}
\begin{center}
\scriptsize
\begin{tabular}{cl|ccccccc}
\toprule
\multicolumn{2}{c|}{\quad} & inclusive & $0j_b$ & $\ge 1j_b$& $\ge 2j_b$ & $1j_b$& $2j_b$\\
\midrule
\multirow{4}{*}{$\sigma$[pb]} & 
{\sc MG5\_aMC}&
$0.369^{+ 19.7 \% }_{-18.8\%}$ &
$0.243^{+ 22.5\%}_{-23.0\%}$ &
$0.126^{+ 32.5\%}_{-28.3\%}$ & 
$0.0160^{+ 47.2\%}_{-39.8\%}$ &
$0.110^{+ 30.4\%}_{-26.7\%}$ & 
$0.0154^{+ 46.0\%}_{-38.9\%}$ \\
&{\sc POWHEG} &
$0.375^{+ 20.3 \% }_{-17.9\%}$ &
$0.281^{+ 21.8\%}_{-18.6\%}$ & 
$0.0943^{+ 16.6\%}_{-16.5\%}$ &
$0.00761^{+ 15.0\%}_{14.8\%}$ & 
$0.0867^{+ 16.8\%}_{-16.7\%}$ &
$0.00754^{+ 14.0\%}_{-14.9\%}$ \\
&{\sc Sherpa 4FS} &
$0.370^{+ 15.4 \% }_{-26.8\%}$ &
$0.264^{+ 11.8\%}_{-26.0\%}$ & 
$0.105^{+ 26.9\%}_{-28.8\%}$ &
$0.00955^{+ 74.9\%}_{-45.4\%}$ & 
$0.0952^{+ 22.2\%}_{-28.6\%}$ &
$0.00934^{+ 70.9\%}_{-44.1\%}$ \\
&{\sc Sherpa 5FS} &
$0.586^{+ 30.4 \% }_{-22.7\%}$ &
$0.423^{+ 20.6\%}_{-15.7\%}$ & 
$0.162^{+ 56.1\%}_{-40.7\%}$ &
$0.00773^{+ 68.9\%}_{-59.7\%}$ & 
$0.155^{+ 55.5\%}_{-40.4\%}$ &
$0.00746^{+ 68.7\%}_{-59.0\%}$ \\
\midrule
\multirow{4}{*}{acceptance} & 
{\sc MG5\_aMC}&
$1$ &
$0.659$ &
$0.342$ & 
$0.0432$ & 
$0.298$ &
$0.0417$ \\
&{\sc POWHEG} &
$1$ &
$0.749$ & 
$0.251$ &
$0.0203$ & 
$0.231$ &
$0.0201$ \\
&{\sc Sherpa 4FS} &
$1$ &
$0.717$ & 
$0.283$ &
$0.0258$ & 
$0.258$ &
$0.0253$ \\
&{\sc Sherpa 5FS} &
$1$ &
$0.723$ & 
$0.277$ &
$0.0132$ & 
$0.264$ &
$0.0127$ \\
\bottomrule
\end{tabular}
\end{center}
\end{table}

\begin{table}
\caption{\label{tab:jrates}
Predictions for the total rates (in pb) of the various Monte-Carlo tools under consideration within cuts on the final-state jets. For comparison, also the 
respective acceptances are given.
}
\begin{center}
\scriptsize
\begin{tabular}{cl|cccccc}
\toprule
\multicolumn{2}{c|}{\quad} & $0j$ & $\ge 1j$& $\ge 2j$ & $1j$& $2j$\\
\midrule
\multirow{4}{*}{$\sigma$[pb]} & 
{\sc MG5\_aMC}&
$0.163^{+ 27.5\%}_{-25.8\%}$ &
$0.206^{+ 31.0\%}_{-30.5\%}$ &
$0.102^{+ 44.4\%}_{-41.3\%}$ &
$0.104^{+ 18.5\%}_{-19.9\%}$ &
$0.0613^{+ 37.9\%}_{-36.5\%}$ \\
&{\sc POWHEG} &
$0.239^{+ 21.6\%}_{-19.3\%}$ &
$0.136^{+ 23.0\%}_{-20.2\%}$ &
$0.0347^{+ 41.7\%}_{-26.7\%}$ &
$0.101^{+ 17.3\%}_{-18.2\%}$ &
$0.0299^{+ 38.7\%}_{-26.2\%}$ \\
&{\sc Sherpa 4FS} &
$0.192^{+ 11.7\%}_{-15.5\%}$ &
$0.177^{+ 32.6\%}_{-48.6\%}$ &
$0.0637^{+ 73.2\%}_{-77.2\%}$ &
$0.113^{+ 9.9\%}_{-38.3\%}$ &
$0.0428^{+ 41.6\%}_{-68.4\%}$ \\
&{\sc Sherpa 5FS} &
$0.328^{+ 34.6\%}_{-28.6\%}$ &
$0.258^{+ 94.2\%}_{-69.4\%}$ &
$0.0851^{+93.2\%}_{-68.6\%}$ &
$0.173^{+ 94.7\%}_{-69.7\%}$ &
$0.068^{+ 90.0\%}_{-65.3\%}$ \\
\midrule
\multirow{4}{*}{acceptance} & 
{\sc MG5\_aMC}&
$0.442$ &
$0.558$ &
$0.276$ &
$0.283$ &
$0.166$  \\
&{\sc POWHEG} &
$0.637$ &
$0.363$ &
$0.0927$ &
$0.270$ &
$0.0798$  \\
&{\sc Sherpa 4FS} &
$0.521$ &
$0.479$ &
$0.172$ &
$0.307$ &
$0.116$  \\
&{\sc Sherpa 5FS} &
$0.559$ &
$0.440$ &
$0.145$ &
$0.295$ &
$0.116$  \\
\bottomrule
\end{tabular}
\end{center}
\end{table}


\begin{figure}
\centering
\hspace{-0.2cm}\includegraphics[width=0.46\textwidth]{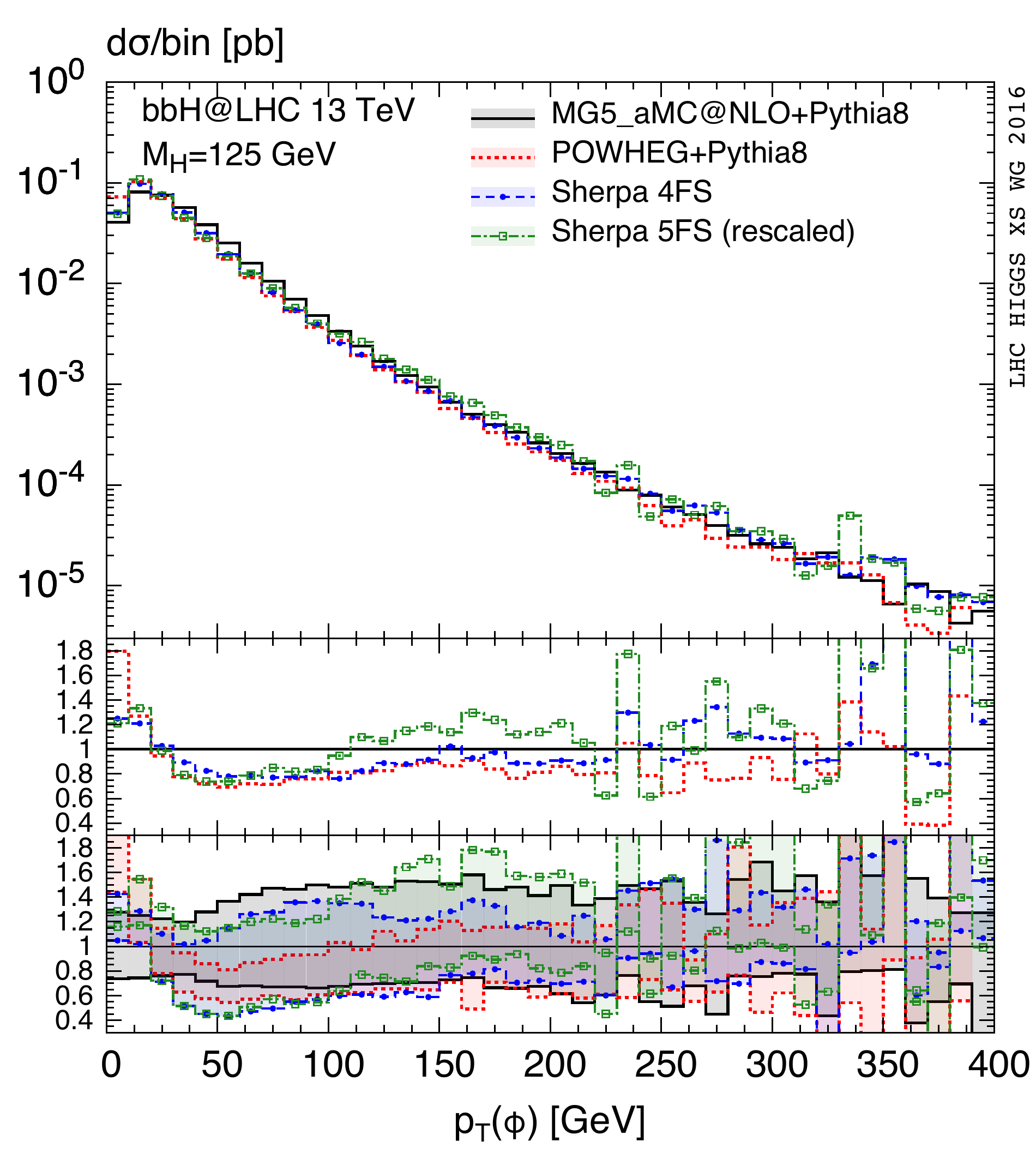}\hspace{1cm}
\includegraphics[width=0.46\textwidth]{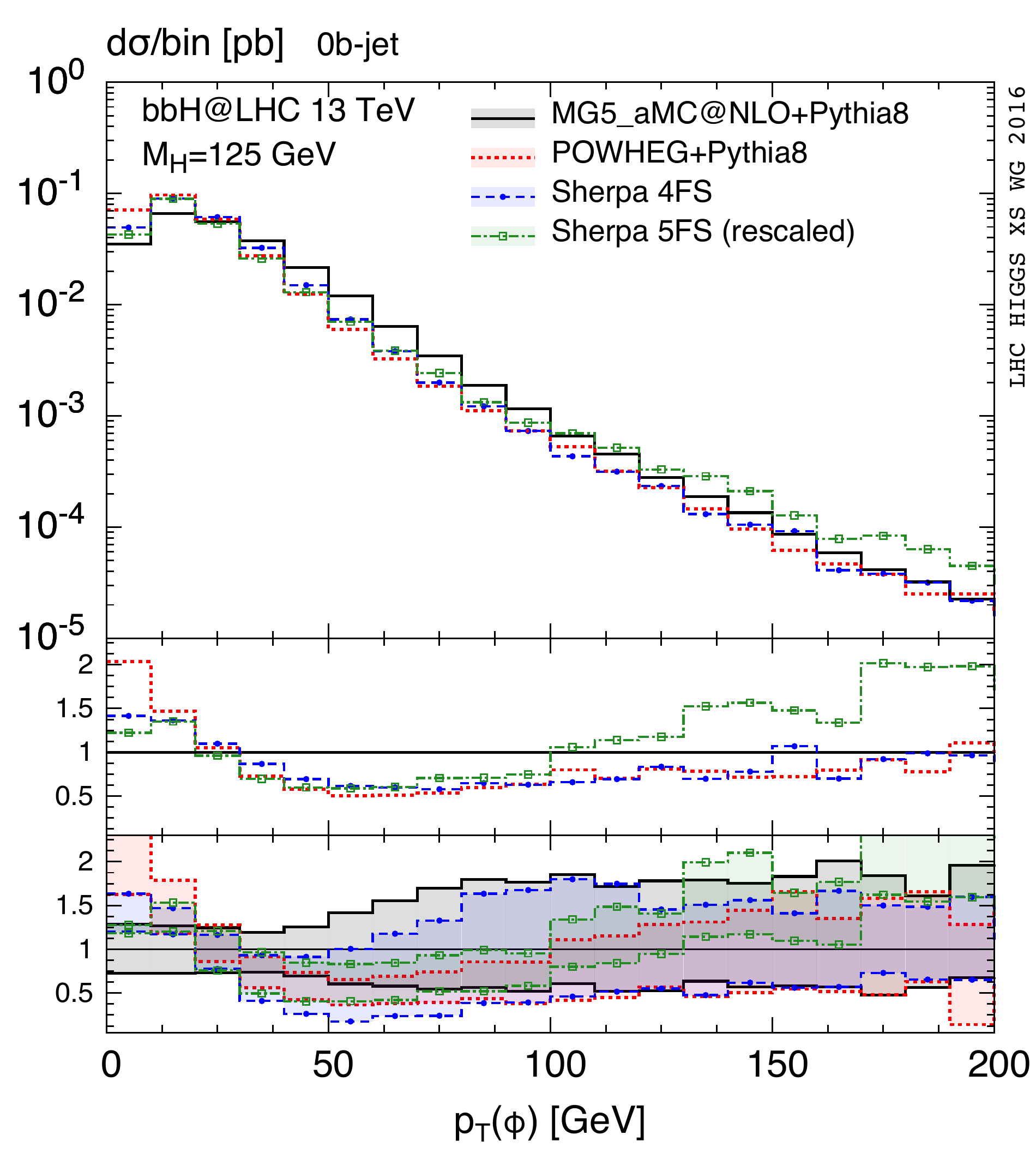}\\
\hspace{-0.2cm}\includegraphics[width=0.46\textwidth]{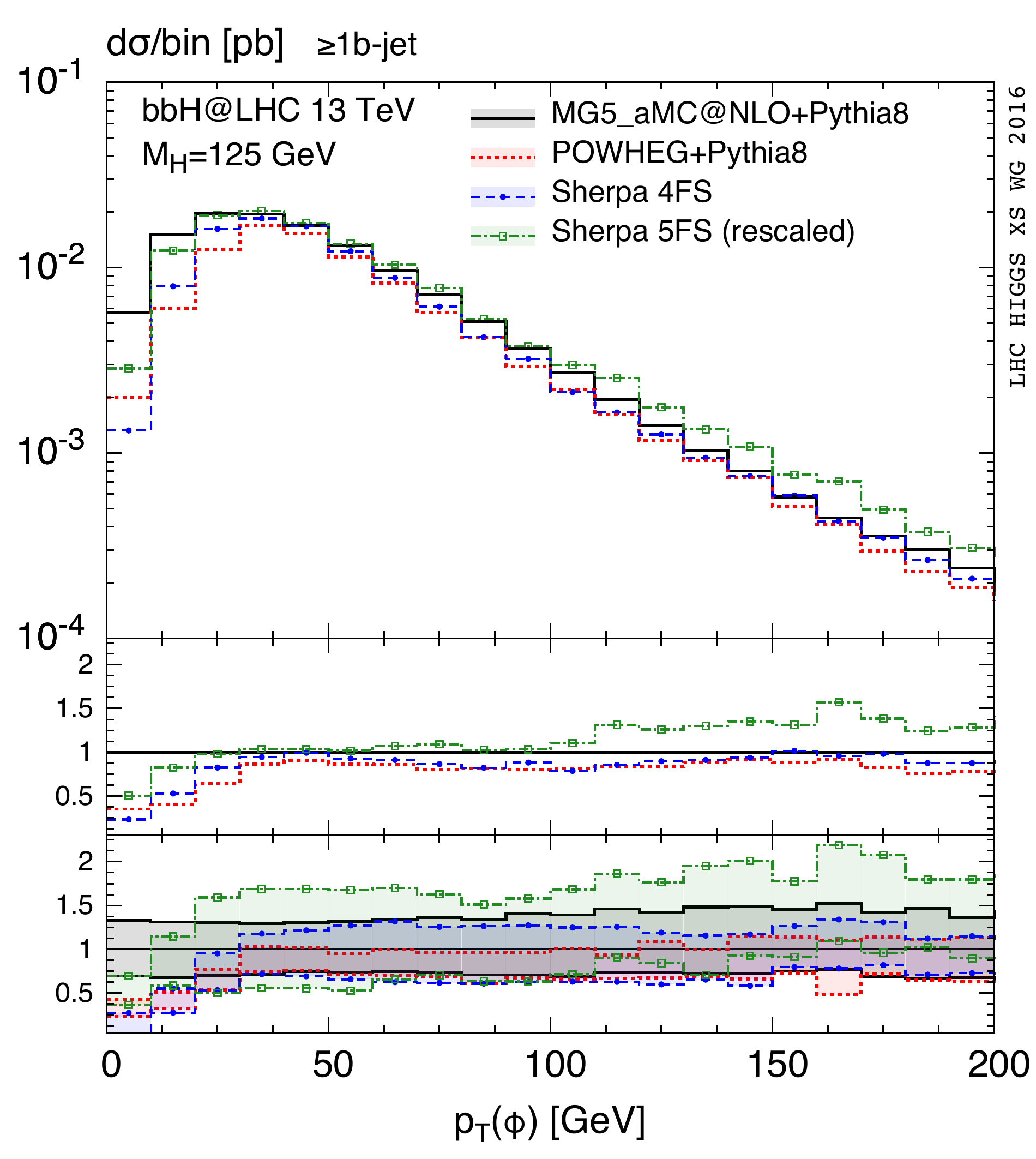}\hspace{1cm}
\includegraphics[width=0.46\textwidth]{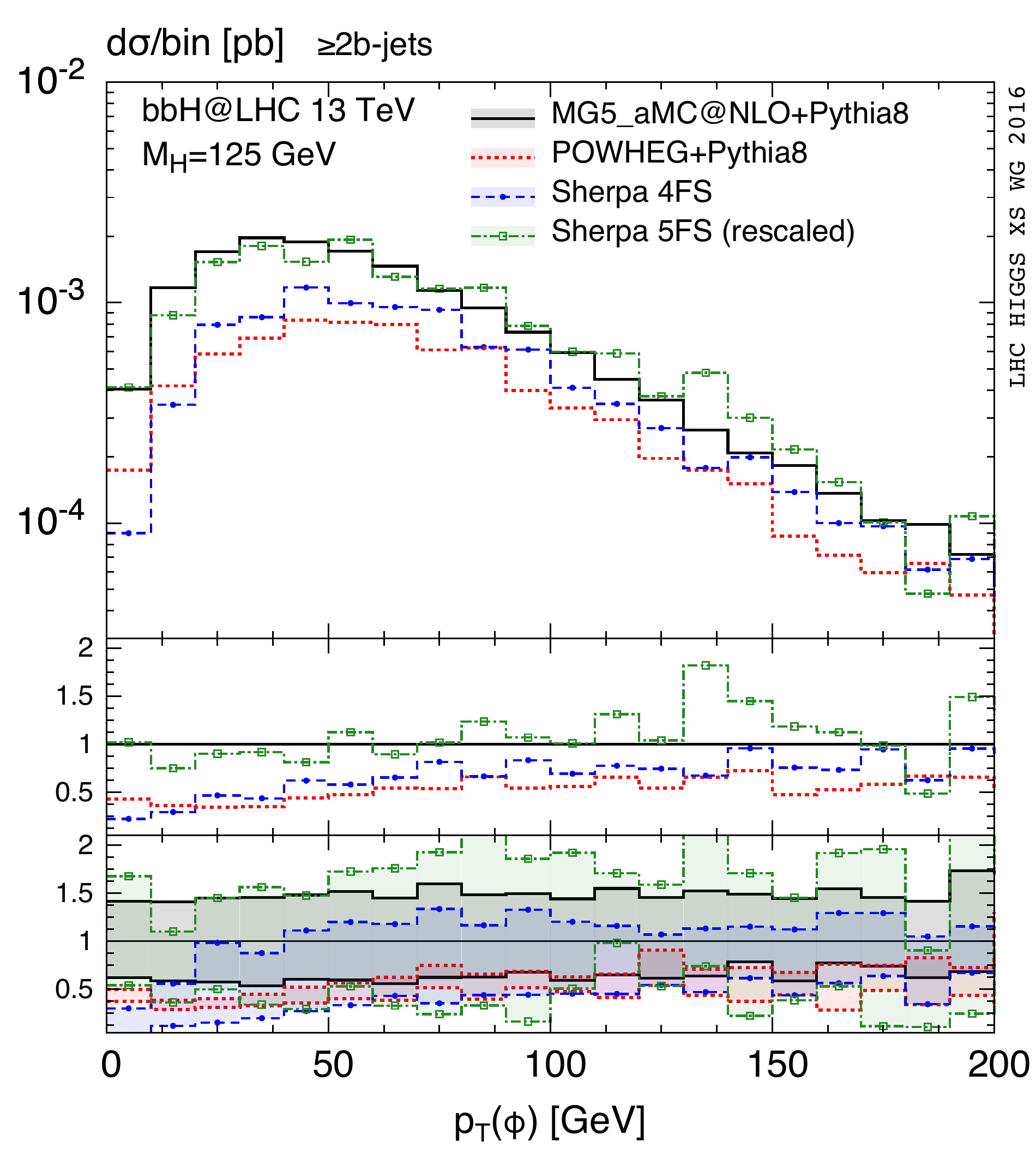}
\caption{Transverse-momentum distribution of the Higgs boson as predicted by the various codes with requirements on the 
final-state $b$ jets; upper left panel: inclusive; 
upper right panel: with a veto on $b$-tagged jets; lower left panel: with one or more $b$ jets; lower right panel: with at least two observed $b$ jets; see text for details.
\label{fig:pTH_bbH}
}
\end{figure}

\begin{figure}
\centering
\hspace{-0.2cm}\includegraphics[width=0.46\textwidth]{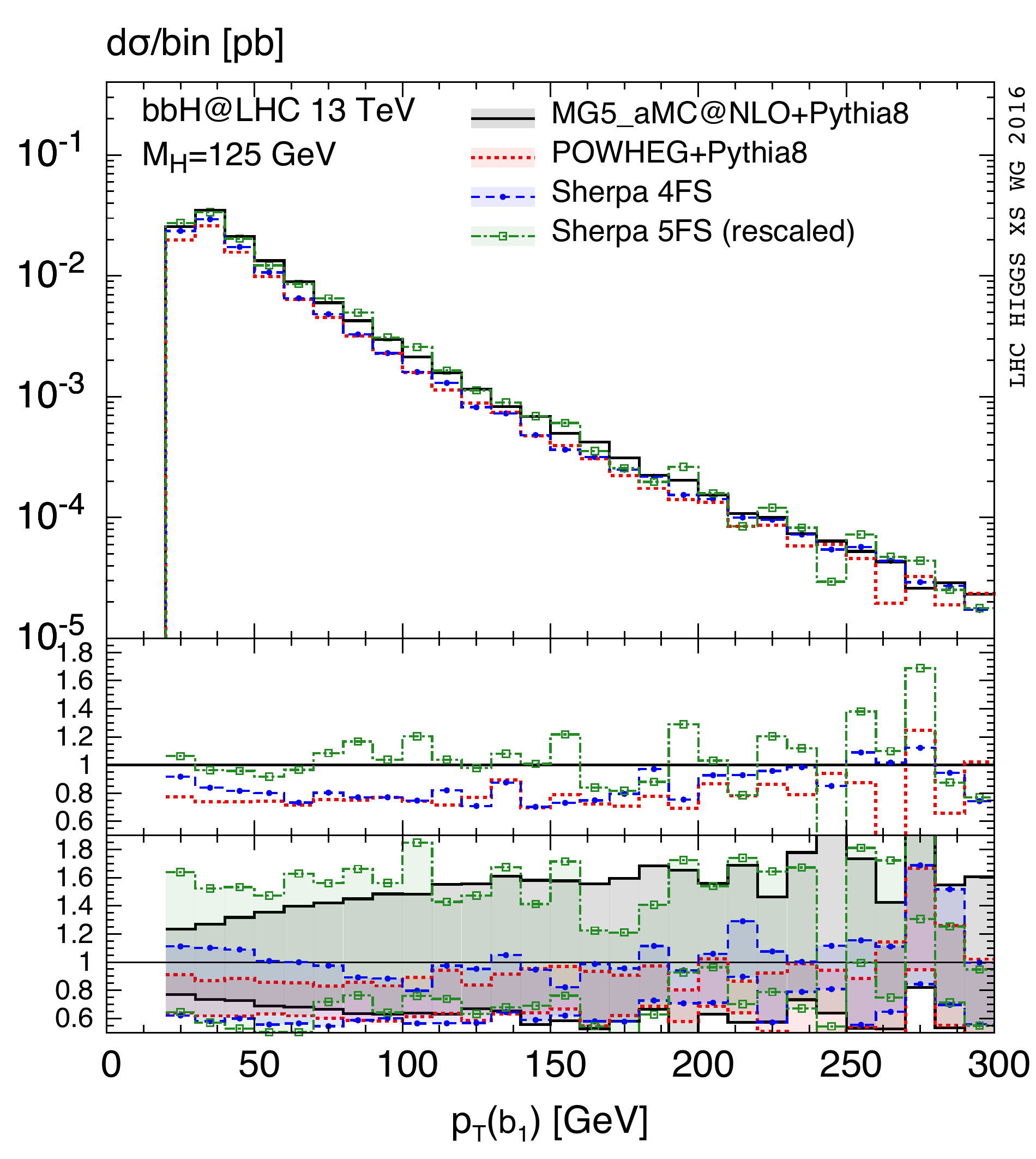}\hspace{1cm}
\includegraphics[width=0.46\textwidth]{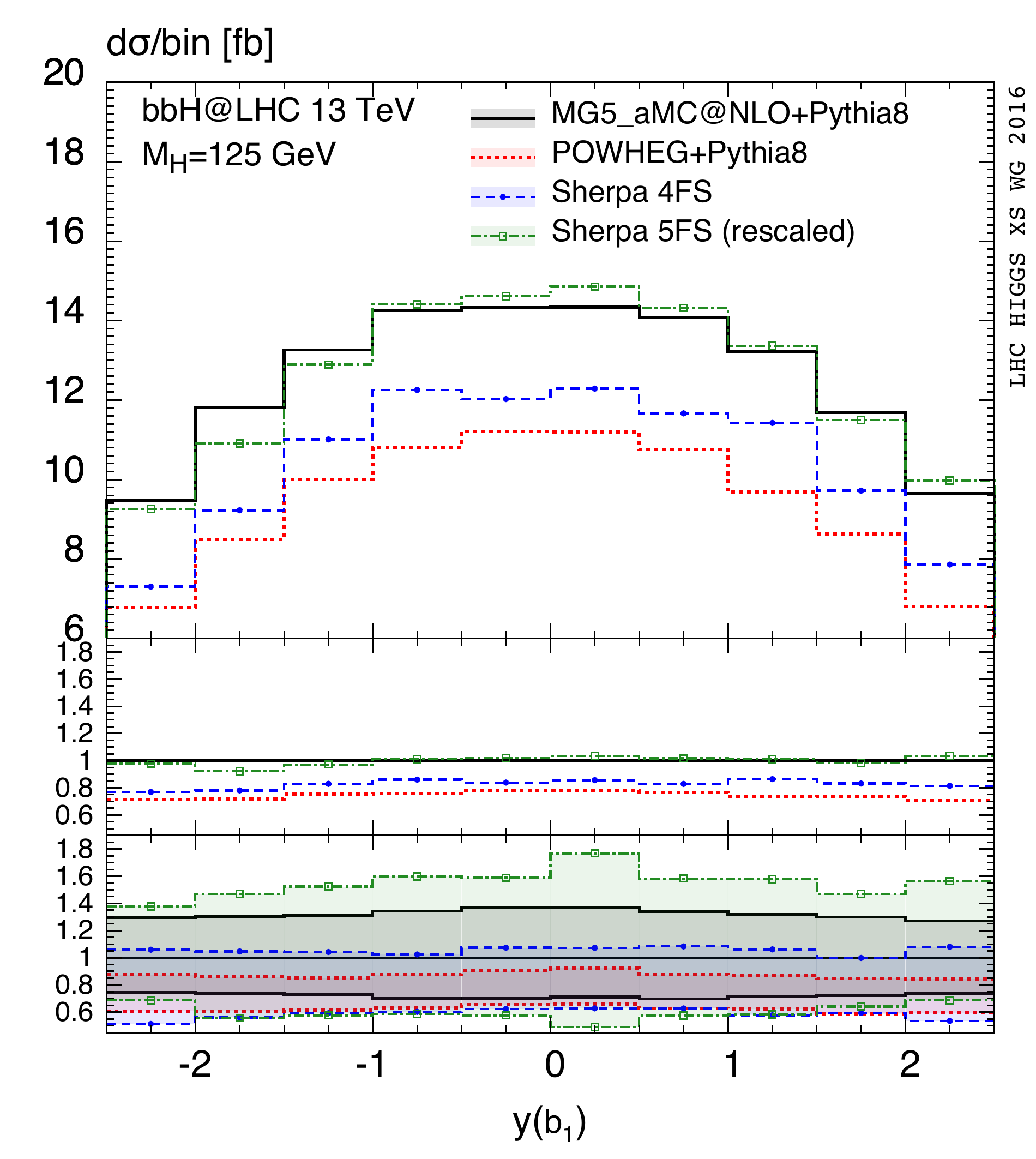}


\centering
\hspace{-0.2cm}\includegraphics[width=0.46\textwidth]{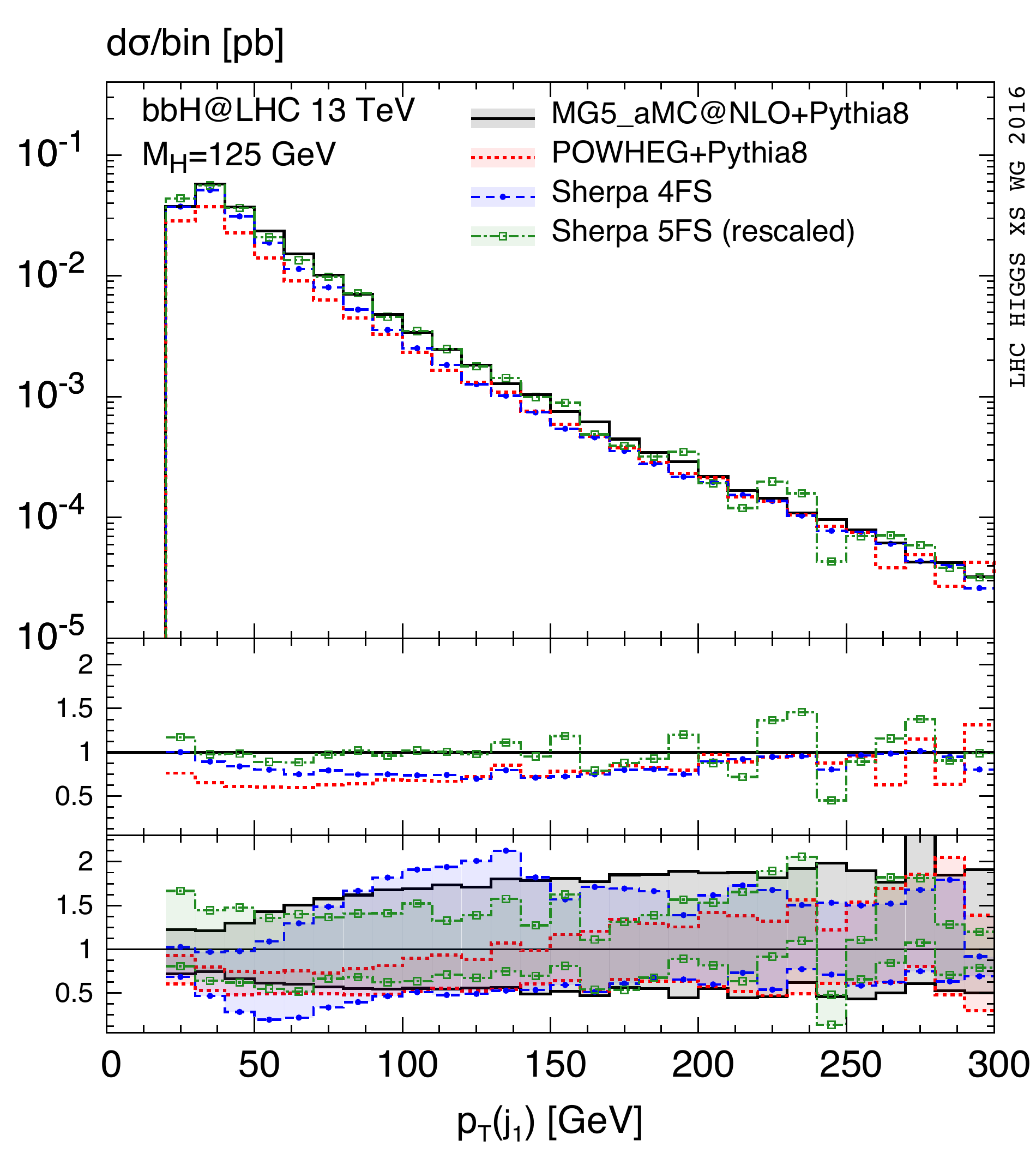}\hspace{1cm}
\includegraphics[width=0.46\textwidth]{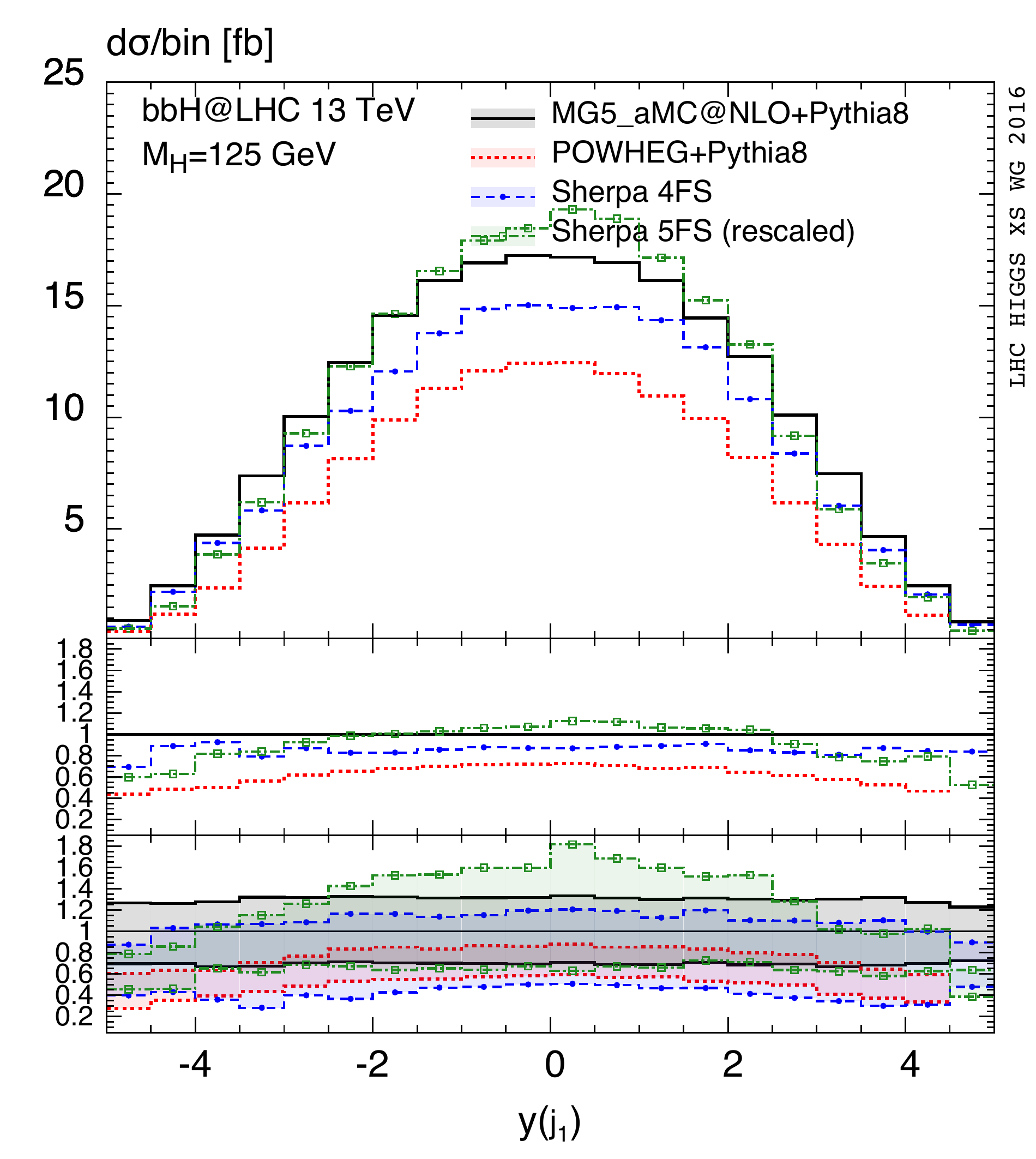}
\caption{Transverse-momentum (left) and rapidity distributions (right) of the hardest $b$ jet (upper) and 
hardest jet without requirements on its flavour (lower); see text for details.
\label{fig:b1j1_bbH}
}
\end{figure}


Due to the limited statistics of the samples under consideration, we refrain from showing 
results relevant to the second hardest jets or $b$ jets. The general conclusions  in these cases 
should be rather similar though to what has been observed so far.

\subsection{Recommendations for \texorpdfstring{$\bbh$}{bb phi} signal simulation}
\label{mc-generation-recommendations}
Let us conclude this section by formulating recommendations for the 
simulation of Higgs boson production in association with bottom quarks at the LHC Run 2.
In summary of the preceding section, all the Monte-Carlo tools under consideration 
provide decent predictions for the simulation of a $\bbh$ signal, with the agreement 
among the different codes being reasonably well within their respective uncertainties.
As pointed out in the introduction, the 5FS has the disadvantage of being less accurate 
in the perturbative prescription of observables exclusive in the degrees of freedom of 
the final-state bottom quarks, which are most relevant in the context of the $\bbh$ process. 
This is alleviated to some extent by the merging of higher multiplicities, but the prescription 
of the bottom kinematics does still rely, in part, on the poorly described $g\rightarrow b\bar{b}$ 
splittings in the backward evolution of the shower. This fact is also reflected in the very large 
uncertainties of the merged 5FS results, see Table\,\ref{tab:brates} and the related discussion.
Nevertheless, the agreement of the 5FS with the 4FS predictions is overall quite satisfactory
within their respective uncertainties and the 5FS computation serves as a crucial consistency 
check.

In conclusion, we recommend the use of fully-differential 4FS predictions for any realistic $\bbh$
signal simulation in experimental searches. We further point out that in NLO calculations matched 
to parton showers precaution must be taken for the choice of a suitable matching scale. As pointed out 
in \Bref{Wiesemann:2014ioa} such a scale must be chosen of the order of the other unphysical 
scales (in particular the factorization scale), which assume generally much lower 
values than $m_\phi$, the scale naively associated with the 
hardness of the $\bbh$ process. As far as theoretical systematics are concerned, the variation 
of all unphysical scales must be taken into account as well as PDF+$\as$ uncertainties.

As pointed out before, all three 4FS Monte-Carlo codes provide valid predictions. At least two of 
them should be used in order to assess the systematic differences among them.
The \Sherpa{} $\bbh$ simulation develops a dependence of the total inclusive cross section
on the shower starting scale, while its factorization- and renormalization-scale uncertainties 
is four time smaller than expected from from other computations of that quantity; a cross check 
of the \Sherpa{} with fixed scales $\muR=\muF=(m_H+2\,m_b)/4$
showed an unexpected decrease of the total cross section with respect to the default dynamical 
scale settings; these issues still have to be resolved. 
The use of two different Monte Carlos comes with the potential advantage of using two 
complementary matching schemes, i.e., MC@NLO and POWHEG. 


\section{Acceptance uncertainties}
\label{sec:bbHacceptance}
The $b \bar{b} \phi,~\phi\rightarrow \tau\tau$ process is the most sensitive
one in the searches for the Higgs bosons, $\phi=\rm h,H,A$ in the MSSM at
large values of tan$\beta$. The final state with two $\tau$--jets ($\tau_{\rm
h}\tau_{\rm h}$) has the largest sensitivity compared to $\tau_{\ell}\tau_{\rm
h}$ or $\tau_{\ell}\tau_{\ell}$ in the searches for the heavy Higgs bosons, A
or H with $m_{\rm A/H}\geq400$ GeV (ATLAS 13 TeV reference).
 
The parton level acceptance and its uncertainties of the $\tau_{\rm h}
\tau_{\rm h}$ final state have been evaluated with {\tt\small
MG5\_aMC@NLO} using the selection criteria of the recent ATLAS and CMS
analyses with 13 TeV data. We take a Higgs boson mass of $m_{\rm H}=700$ GeV
and $\sqrt{s}=\rm 13~TeV$ at the LHC. The selection criteria for the
$\tau_{\rm h} \tau_{\rm h}$ final state in the CMS analysis are:

\begin{itemize}
\item two $\tau$--jets with $p_{\rm T}^{\tau_{\rm h}}>40$ GeV, $|\eta^{\tau_{\rm h}}|<2.1$
\item at least one b--jet with $p_{\rm T}>20$ GeV, $|\eta|<2.4$
\item no more than one jet with $p_{\rm T}>30$ GeV, $|\eta|<4.7$
\end{itemize}

In the ATLAS analysis the $\tau_{\rm h}\tau_{\rm h}$ final state is selected
requiring two $\tau$--jets with the leading jet $p_{\rm T}^{\tau_{\rm
h1}}>135$ GeV and the sub--leading jet $p_{\rm T}^{\tau_{\rm h2}}>55$ GeV in
the pseudo--rapidity region of $|\eta^{\tau_{\rm h}}|<2.5$. The $\tau$--jets
are required to be back-to-back in the transverse plane of the detector,
$\Delta\phi(\tau_{\rm h1},\tau_{\rm h2})>2.7$.

The parton jets are reconstructed from the gluons and quarks after {\tt\small
PYTHIA8} showering using the anti--$k_{T}$ algorithm with $R=0.4$. The parton
jet is identified as a b--jet if it has a b--quark as a jet constituent. The
$p_{\rm T}$ of $\tau_{\rm h}$ is a vector sum $p_{\rm T}$ of the hadronic
$\tau$ decay products. 

The acceptance uncertainties are separated into the uncertainties due to the QCD scale choice,
due to the PDF uncertainty and due to the shower scale, $Q_{\rm sh}$ choice in
{\tt\small MG5\_aMC@NLO} (see Section~\ref{mc-generation-recommendations}). 
The $\rm PDF4LHC15\_nlo\_nf4\_30$ set is used.

The QCD scale is varied as $0.5\mu_{0}\leq\mu_{\rm R},\mu_{\rm F}\leq 2\mu_{0}$ with the
constraint $0.5 \leq \mu_{\rm R} / \mu_{\rm F} \leq 2$. The shower scale, $Q_{\rm sh}$ is varied
by varying the parameter $\alpha \in [1/(4\sqrt{2}),\sqrt{2}/4]$ as described in
Section~\ref{mc-generation-recommendations}. 

Table~\ref{tab:bbHacceptance_aMCNLO} shows
the parton level acceptance and its uncertainties for ATLAS and CMS analysis selections evaluated
with {\tt\small MG5\_aMC@NLO}. One can see that the uncertainty is dominated by the choice
of the shower scale. 


\begin{table} 
\caption{The parton level acceptance of ATLAS and CMS $\tau_{\rm h}\tau_{\rm h}$ analysis selections
         and its uncertainties evaluated with {\tt\small MG5\_aMC@NLO} for $m_{\rm H}=700$ GeV
         and $\sqrt{s}=\rm~13 TeV$ LHC.}
\label{tab:bbHacceptance_aMCNLO}
\begin{center}
\begin{tabular}{c|c|c}
\toprule
                          & ATLAS                                & CMS \\
\midrule
selections                &     \multicolumn{2}{c}{acceptance}  \\
\cmidrule{1-3}
  $\tau_{\rm h}$ kinematics                  &          0.671                       &    0.816         \\
    jet selections                           &    no jet selections                 &    0.161         \\
\midrule
source of uncertainty     & \multicolumn{2}{c}{acceptance uncertainty in \%} \\
\cmidrule{1-3}
QCD scale                 &         +2.6,~--1.4                  &    +3.0,~-2.0     \\
PDF                       &         +0.2,~-0.9                   &    +0.4,~--0.9    \\
shower scale, $Q_{\rm sh}$ &        +1.0,~-7.2                   &    +4.4,~-10.9    \\
\bottomrule
\end{tabular}
\end{center}
\end{table}


\section{Total cross sections for \texorpdfstring{$c\bar{c}\phi$}{cc phi} production}
\label{sec:ccH}
\newcommand{\cch}{c\bar c\phi}

As shown in Ref.\ \cite{Harlander:2015xur}, the partonic 5FS results for
$\bbh$ production allow for the calculation of other quark-initiated
cross sections by simply changing the parton density flavour.  Amplitudes
with different Yukawa couplings do not interfere with each other if
kinematical quark masses and terms which involve a loop-induced
Higgs-gluon coupling are neglected.

Of particular interest in certain extended theories may be the $\cch$
cross section. Table \ref{tab:cch} shows the total inclusive $\cch$
cross section through NNLO. To obtain these numbers, the parton
{\tt PDF4LHC15} parton densities have been used. The Yukawa coupling has
been set to its SM value, with the charm quark mass $m_c(\mu_R)$ derived
from $m_c(3~{\rm GeV})=0.986$\,GeV by running it at four-loop order to
$m_c(M_H)$, and from there at three-loop order to $m_c(\mu_R)$.

These numbers have been obtained with version 1.6 of the program {\tt
  SusHi} \cite{Harlander:2012pb}.

\begin{table}
\caption{\label{tab:cch} Total inclusive cross section for $\cch$
  production at $\sqrt{s}=13$\,TeV.}
\begin{center}
  \renewcommand{\arraystretch}{1.2}
  \tabcolsep5pt
\begin{tabular}{cccc}
  \toprule
  $\MH$[GeV] & $\sigma_{\cch}$[pb] & $\Delta_\mathrm{scale}$[\%] & $\Delta_\mathrm{PDF}$[\%]\\
\midrule
$25$ & $  8.120\cdot10^{0}$ & $_{-44.1}^{+ 22.6}$ & $\pm  8.0$ \\
$45$ & $  1.890\cdot10^{0}$ & $_{-21.9}^{+ 10.5}$ & $\pm  5.7$ \\
$65$ & $  6.657\cdot10^{-1}$ & $_{-14.3}^{+  6.8}$ & $\pm  5.1$ \\
$85$ & $  2.944\cdot10^{-1}$ & $_{-10.4}^{+  5.0}$ & $\pm  5.1$ \\
$105$ & $  1.500\cdot10^{-1}$ & $_{ -8.1}^{+  4.0}$ & $\pm  4.9$ \\
$125$ & $  8.429\cdot10^{-2}$ & $_{ -6.5}^{+  3.3}$ & $\pm  4.9$ \\
$145$ & $  5.082\cdot10^{-2}$ & $_{ -5.4}^{+  2.8}$ & $\pm  4.9$ \\
$165$ & $  3.234\cdot10^{-2}$ & $_{ -4.6}^{+  2.4}$ & $\pm  4.9$ \\
$185$ & $  2.148\cdot10^{-2}$ & $_{ -3.9}^{+  2.1}$ & $\pm  5.1$ \\
$205$ & $  1.476\cdot10^{-2}$ & $_{ -3.4}^{+  1.9}$ & $\pm  5.2$ \\
$225$ & $  1.044\cdot10^{-2}$ & $_{ -3.0}^{+  1.7}$ & $\pm  5.2$ \\
$245$ & $  7.562\cdot10^{-3}$ & $_{ -2.7}^{+  1.6}$ & $\pm  5.1$ \\
$265$ & $  5.591\cdot10^{-3}$ & $_{ -2.4}^{+  1.4}$ & $\pm  5.0$ \\
$285$ & $  4.207\cdot10^{-3}$ & $_{ -2.1}^{+  1.3}$ & $\pm  5.2$ \\
$305$ & $  3.215\cdot10^{-3}$ & $_{ -1.9}^{+  1.2}$ & $\pm  5.3$ \\
$325$ & $  2.491\cdot10^{-3}$ & $_{ -1.7}^{+  1.2}$ & $\pm  5.4$ \\
$345$ & $  1.953\cdot10^{-3}$ & $_{ -1.6}^{+  1.1}$ & $\pm  5.6$ \\
$365$ & $  1.549\cdot10^{-3}$ & $_{ -1.4}^{+  1.0}$ & $\pm  5.6$ \\
$385$ & $  1.240\cdot10^{-3}$ & $_{ -1.3}^{+  1.0}$ & $\pm  5.6$ \\
$405$ & $  1.001\cdot10^{-3}$ & $_{ -1.2}^{+  0.9}$ & $\pm  5.6$ \\
$425$ & $  8.150\cdot10^{-4}$ & $_{ -1.1}^{+  0.9}$ & $\pm  5.7$ \\
$445$ & $  6.684\cdot10^{-4}$ & $_{ -1.0}^{+  0.8}$ & $\pm  5.7$ \\
$465$ & $  5.518\cdot10^{-4}$ & $_{ -0.9}^{+  0.8}$ & $\pm  5.7$ \\
$485$ & $  4.585\cdot10^{-4}$ & $_{ -0.8}^{+  0.7}$ & $\pm  5.8$ \\
$505$ & $  3.831\cdot10^{-4}$ & $_{ -0.8}^{+  0.7}$ & $\pm  6.0$ \\
$525$ & $  3.218\cdot10^{-4}$ & $_{ -0.8}^{+  0.7}$ & $\pm  6.0$ \\
\bottomrule
\end{tabular}
\end{center}
\end{table}

\chapter{Charged Higgs Bosons}
\ChapterAuthor{M.~Flechl, S.~Sekula, M.~Ubiali, M.~Zaro (Eds.)
C.~Degrande, H.E.~Haber, M.~Spira, M.~Wiesemann}
\label{chap:ChargedHiggs}
\providecommand{\mhpm}{m_{\PSHpm}}

\section{Introduction}
Charged Higgs bosons $\PSHpm$ appear in many extensions of the Standard Model, in particular when adding
additional doublets or triplets to its scalar sector. Here, the focus is 
on charged Higgs bosons in 2-Higgs-doublet models (2HDM) including the special case of the Higgs sector 
of the minimal supersymmetric extension of the standard model (MSSM). The dominant production mode
for a charged Higgs boson depends on its mass. In particular, for masses below the top quark mass, the charged Higgs boson
is dominantly produced in top-quark decays. Therefore, the production cross section corresponds to the top pair production times 
the branching ratio $\Pt \to \PH^+ \Pb$. For values of the mass close to the top quark mass ($160-180\UGeV$), both contributions with resonant and 
non-resonant top quarks are equally important, and the full $ \PWm \PH^+ \Pb \bar\Pb$ has to be simulated. Finally, heavy charged Higgs bosons
are dominantly produced in association with a top quark. Most of the parameter space for a light charged Higgs boson has already been excluded
at the LHC Run 1~\cite{Aad:2014kga, Khachatryan:2015qxa}. For what concerns the intermediate mass region $\mhpm \sim m_{\Pt}$, no search has been performed to date due to the 
lack of accurate predictions for the signal. In fact, NLO predictions for the total cross section have been made available only recently~\cite{Degrande:2016hyf}. Further theoretical developments in this direction are encouraged. In this chapter we will focus on the heavy mass
range, up to masses of $2\UTeV$, which is being probed at the Run 2.\\
In the following, we present updated NLO predictions for heavy charged Higgs boson production in a type-II 
2HDM. These cross sections can also be translated into predictions 
for a type-I, type-III or type-IV 2HDM according to the recipe in Ref.~\cite{Flechl:2014wfa}. We continue by showing differential cross sections for this production 
process with an emphasis on the comparisons of the 4-flavour scheme (4FS) 
and 5-flavour scheme (5FS) predictions. We conclude the chapter providing some recommendations for the signal simulation in experimental searches.

\section{Inclusive production cross sections}
The dominant charged Higgs boson production mode for $\mhpm > m_t$ in a 2HDM is via the process
\begin{equation*} 
\Pp\Pp\, \rightarrow\, \Pt\PSHpm + X.
\end{equation*}
The cross section for associated $\Pt\PSHpm$ production can be
computed in the 4FS or the 5FS.  In the 4FS there are no $\Pb$ quarks in the initial
state, hence the lowest-order QCD production processes are
gluon-gluon fusion and quark-antiquark annihilation, $\Pg\Pg
\rightarrow \Pt\Pb\PSHpm$ and $\Pq\bar{\Pq} \rightarrow
\Pt\Pb\PSHpm$, respectively. Potentially large logarithms of the
ratio between the hard scale of the process and the mass of the bottom
quark, which arise from the splitting of incoming gluons into nearly
collinear $\Pb\bar{\Pb}$ pairs, can be summed to all orders in
perturbation theory by introducing bottom parton densities.  This
defines the five-flavour scheme (5FS). The use of bottom quark distribution
functions is based on the approximation that the outgoing $\Pb$ quark
is at small transverse momentum and massless, and the virtual $\Pb$
quark has a vanishing virtuality ($m \approx 0$). In this scheme, the LO process for the
inclusive top-quark-associated production cross section is gluon-bottom fusion,
$\Pg\Pb \rightarrow \Pt\PSHpm$. The NLO cross section in the 5FS
scheme includes $\mathcal{O}(\alphas)$ corrections to $\Pg\Pb
\rightarrow \Pt\PSHpm$, including the tree-level processes $\Pg\Pg
\rightarrow \Pt\Pb\PSHpm$ and $\Pq\bar{\Pq} \rightarrow
\Pt\Pb\PSHpm$.  To all orders in perturbation theory the two schemes
are identical, but the way of ordering the perturbative expansion is
different, and the results do not match exactly at finite order. 

Here, we present cross-section predictions for this process by following the methodology described in more detail 
in Refs.~\cite{Flechl:2014wfa,Heinemeyer:2013tqa}. The main differences are the usage of the most recent combination of PDF sets provided by
PDF4LHC15, the centre-of-mass energy of $\sqrt{s}=13 \UTeV$, and the parameters which are set according to the conventions adopted in this report. 
In addition,  we have significantly extended the mass range up to $\mhpm=2$ TeV and we explicitly calculate
the $\tan \beta$ dependence, by computing separately the contributions to the cross section proportional to $y_b^2$
and $y_t^2$ and rescaling each of them by the corresponding overall $\tan \beta$ factor. 
In the previous analysis~\cite{Flechl:2014wfa} the central value of the cross section was computed for all points 
in the $\tan \beta$ scan, the approximation was made that the size of the relative theoretical uncertainty is 
independent of the value of $\tan \beta$. The present analysis goes beyond this approximation by taking into account the theoretical uncertainty
associated to the running of the bottom Yukawa coupling up to the renormalization
scale: the uncertainties for all considered values of $\tan \beta$ are computed explicitly.

For a type-II 2HDM, the $\Pt\bar{\Pb}\PSHm$ coupling is given by $\sqrt{2}\, \left( y_{\Pt} \,P_R\cot\beta + y_{\Pb} \,P_L\tan\beta \right)$. We separately 
evaluate the $y_{\Pt}^2$ and $y_{\Pb}^2$ terms. 
The size of the interference term, $y_{\Pt} y_{\Pb}$, 
is proportional to $m_{\Pb}$ and will be neglected in the following. In the 5FS, this contribution is exactly zero, 
while in the 4FS, it has been shown~\cite{Degrande:2015vpa} that neglecting this term leads to an overestimate of the cross section 
by at most 5\% for $\mhpm=200$ GeV and less than 1\% for $\mhpm>600$ GeV. 
This estimate refers to $\tan \beta=8$; for all other values of $\tan \beta$, the size of this contribution is further suppressed by $\tan^2 \beta$ ($1/\tan^2 \beta$) for 
large (small) $\tan \beta$ values. In 
all cases, the impact of this term remains much smaller than the size of the theoretical uncertainties.
For what concerns supersymmetric corrections, effects due to virtual supersymmetric particles in the loop have to be taken into account. Such corrections
are finite and can be simply added to the total cross section. Among
these corrections the dominant ones are those that modify the relation between the $\Pb$ quark mass and its Yukawa coupling. This class of corrections are 
enhanced at large $\tan \beta$ and can be summed up to all orders through a modification of the $\Pb$ quark Yukawa 
coupling~\cite{Carena:1993bs, Hempfling:1993kv, Pierce:1996zz, Guasch:2003cv, Dittmaier:2009np, Noth:2008tw, Noth:2010jy, Mihaila:2010mp}. The remaining
SUSY-QCD effects are negligible at large $\tan \beta$ but can be of order $10\%$ at small $\tan \beta$. \\
We present results for the 4FS and 5FS schemes,
including the theoretical uncertainty, and combine the two schemes
according to the Santander matching~\cite{Harlander:2011aa}. Fully-matched computations have been presented for bottom-fusion 
initiated Higgs boson production in this
report~\cite{Forte:2015hba,Bonvini:2015pxa}, but are not available for charged Higgs boson production.
Throughout this report we present results
for the $\Pt\PH^-$ final state. The charge-conjugated final state can be included by simply multiplying the results by a factor two.

To estimate the theoretical uncertainty due to missing higher-order
contributions, we vary the renormalization scale $\mu_R$, the 
factorization scale $\mu_F$ and the scale $\mu_{\Pb}$ (which determines the 
running bottom quark mass in the Yukawa coupling and is set to $\mu_R$) by
a factor two about their central values. 
In addition to the scale uncertainties, we have computed
the PDF and $\alphas$ uncertainties following the
PDF4LHC15 recommendation. We stress that the PDF uncertainty computed with the PDF4LHC15 set
also accounts for the parametric uncertainty associated to the value of $\Mb$ used in PDF fits. 
PDF uncertainties are given at 68\% confidence level (CL). 

The results for heavy charged Higgs boson production within the 4FS 
are based on the calculation presented in
Ref.\,\cite{Dittmaier:2009np} and implemented in 
{\tt MG5\_aMC@NLO}~\cite{Alwall:2014hca, Degrande:2015vpa}, interfaced to the
LHAPDF library~\cite{Bourilkov:2006cj,Buckley:2014ana}. 
The renormalization and factorization scales are
set to $\mu$ = $(\mhpm+\Mt+\Mb)/3$. The resulting 4FS cross section and 
uncertainties are shown in \refF{fig:hplus-4fs}.

\begin{figure}[ht]
  \begin{center}
    \setlength{\unitlength}{\textwidth}
    \includegraphics[width=0.48\textwidth]{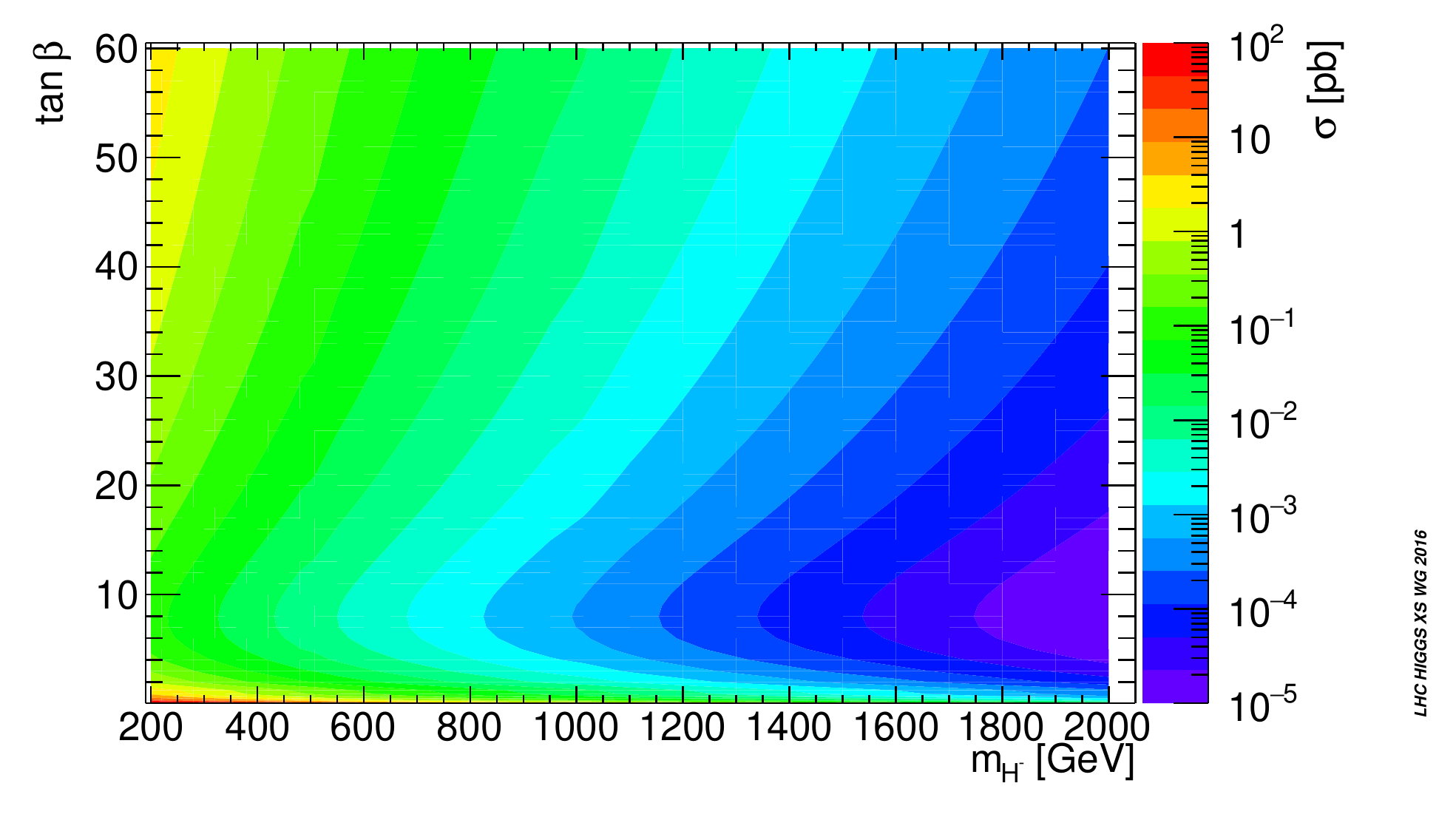}~~~~~~~
    \includegraphics[width=0.48\textwidth]{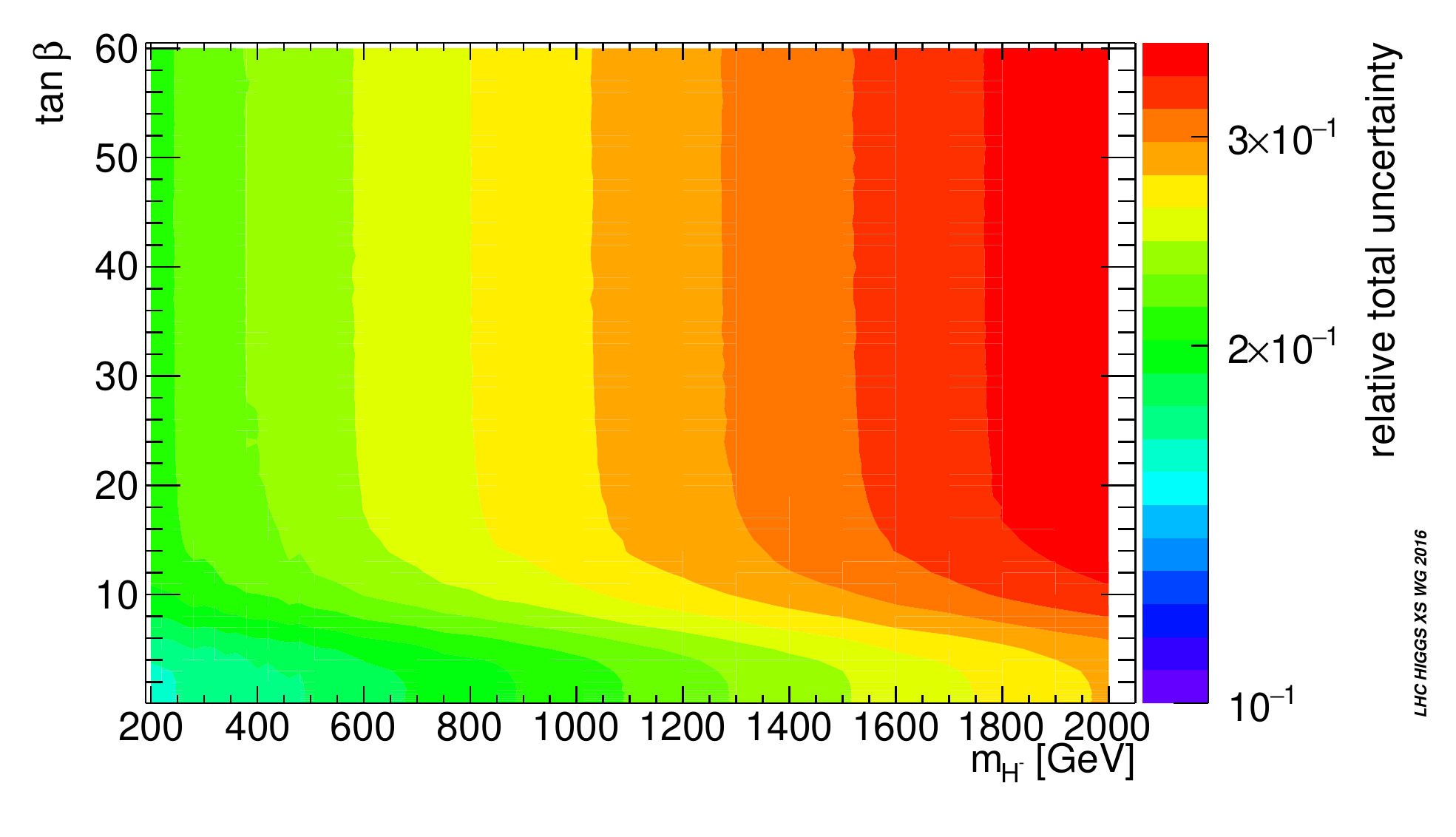}
    \caption{Cross section (left) and average relative uncertainty (right) for $\Pt\PSHpm + X$ production in the 4FS, as a function of $\mhpm$ and $\tan \beta$.}
    \label{fig:hplus-4fs}
  \end{center}
\end{figure}

\begin{figure}
  \begin{center}
    \setlength{\unitlength}{\textwidth}
    \includegraphics[width=0.48\textwidth]{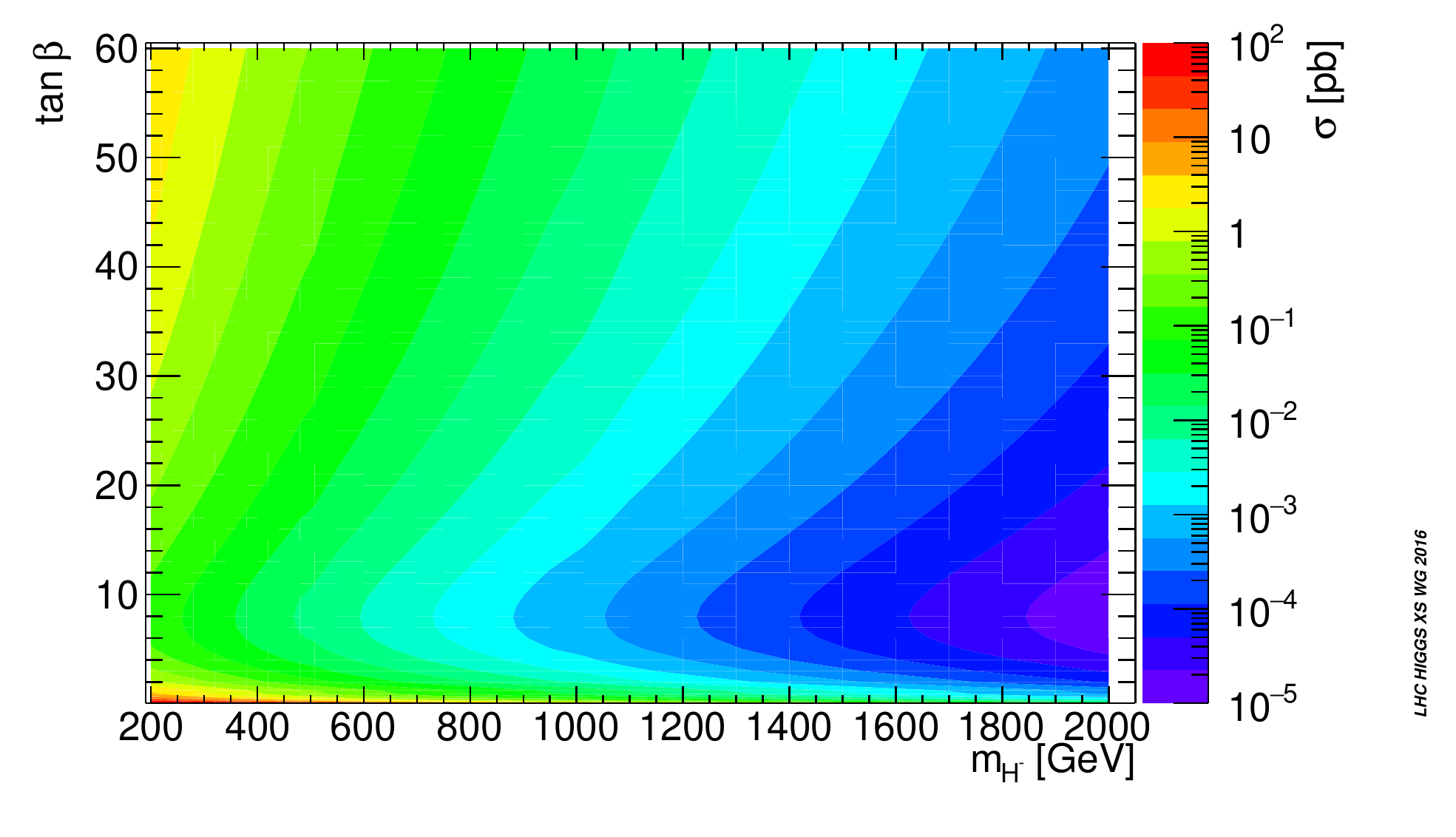}~~~~~~~
    \includegraphics[width=0.48\textwidth]{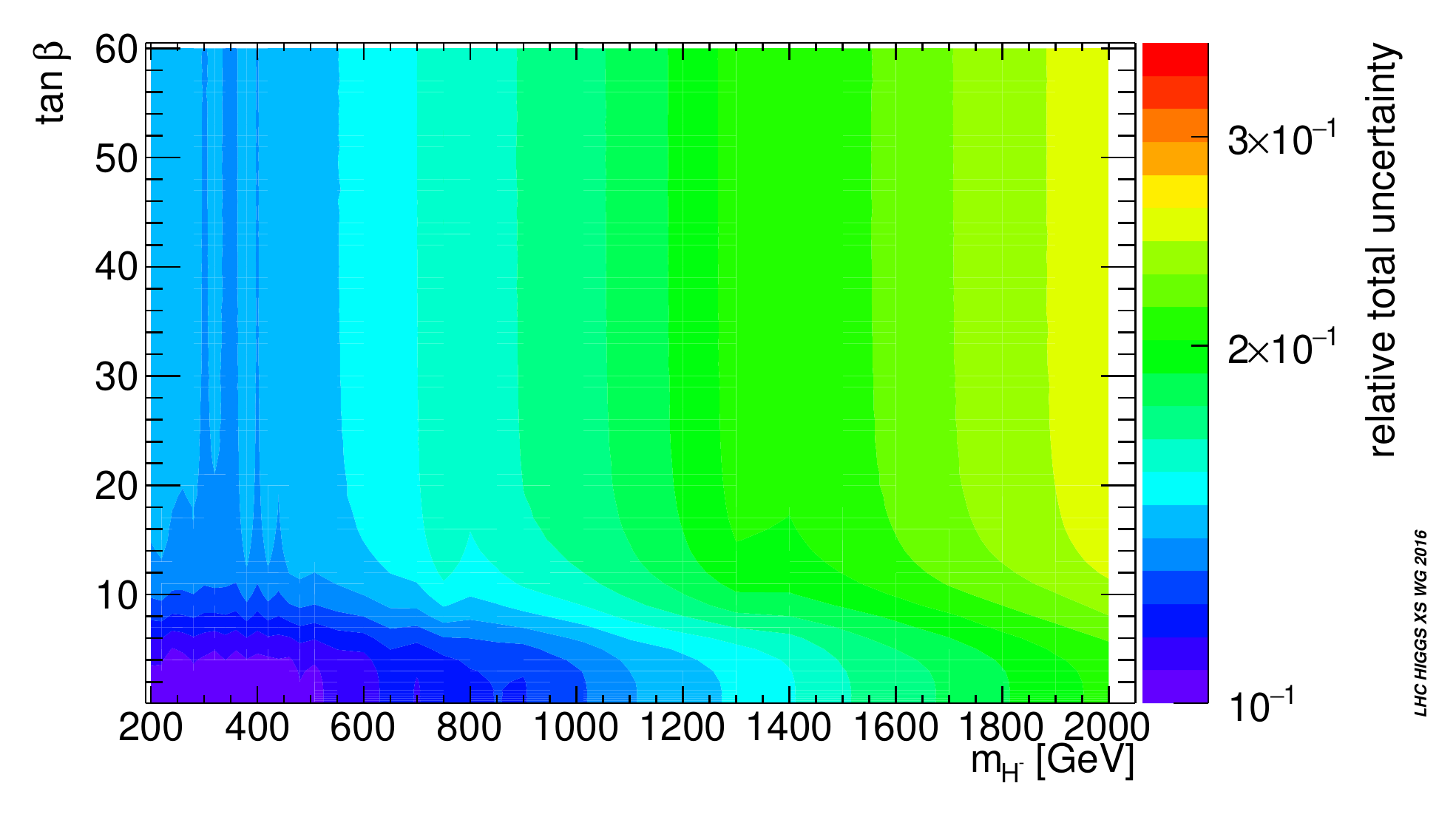}
    \caption{Cross section (left) and average relative uncertainty (right) for $\Pt\PSHpm + X$ production in the 5FS, as a function of $\mhpm$ and $\tan \beta$.}
    \label{fig:hplus-5fs}
  \end{center}
\end{figure}

For the calculation in the 5FS, the program
Prospino~\cite{Plehn:2002vy} has been employed, interfaced to the
LHAPDF library~\cite{Bourilkov:2006cj}.  The renormalization scale is
set to $\mu_{\rm R}$ = $(\mhpm+\Mt)/2$, while the factorization
scale $\mu_{\rm F}=\tilde{\mu}$ is chosen according to the method
proposed in~\cite{Maltoni:2012pa}. The effective factorization scale
entering the initial state logarithms is proportional to the hard
scale, but modified by a phase space factor which tends to reduce the
size of the logarithms for processes at hadron colliders. A table with $\tilde \mu$ values for
the various charged Higgs boson masses is provided on the Twiki 
page \url{https://twiki.cern.ch/twiki/bin/view/LHCPhysics/LHCHXSWGMSSMCharged}.
The  resulting 5FS cross section and uncertainties are shown in \refF{fig:hplus-5fs}.
Before presenting matched predictions, we would like to make some comments on the numbers, see \refT{tab:hp_xsec_summary}. 
The central values of the four- and five-flavour predictions are compatible within uncertainties, as observed also in 
Ref.~\cite{Flechl:2014wfa}, although the agreement is worse. This is mostly due to the different way of computing the bottom Yukawa coupling
compared to previous predictions. As far
as uncertainties are concerned, the 4FS numbers are affected by a total uncertainty which is about 50\% larger than the
one of the 5FS. Furthermore, in the 5FS the 
largest contribution to the total uncertainty comes from PDFs, in particular
for heavy Higgs bosons $\mhpm > 1 TeV$, while scale uncertainties remain around or below 10\% also
at higher values of the masses.  
In the 4FS, the situation is reversed, as scale uncertainties are dominant and reach up to 20\%
for $\mhpm \sim 1 TeV$. PDF uncertainties on the other hand are smaller, about half the
PDF uncertainty of the 5FS calculation. The smaller scale uncertainty of the 5FS calculation 
is also observed in the case of $\bbh$: in particular, in Ref.~\cite{Forte:2015hba} 
it has been suggested that this fact may be associated with
theoretical uncertainties coming from mass terms included only in the 4FS powers of $\Mb$.
Overall the total theoretical uncertainty is smaller in the 5FS calculation by about 30\%.\\

\begin{table}
\setlength{\tabcolsep}{4pt}
\caption{
Comparison of $pp \to tH^+ + X$ cross sections (in units of pb) and percentage uncertainties for the 4FS, the 5FS, and when matching both.
}
\label{tab:hp_xsec_summary}
\small
\begin{center}
    \begin{tabular}{rr|llll|llll|ll}
      \toprule                                                                                                                 
      $\mhpm$   & $\tan\beta$     & \multicolumn{4}{c|} {4FS}        & \multicolumn{4}{c|} {5FS}       & \multicolumn{2}{c} {matched} \\
      $[$GeV$]$ &           & $\sigma$      & $\Delta \sigma^\mathrm{scale}$ & $\Delta \sigma^\mathrm{pdf}$ & $\Delta \sigma^\mathrm{tot}$     
                            & $\sigma$      & $\Delta \sigma^\mathrm{scale}$ & $\Delta \sigma^\mathrm{pdf}$ & $\Delta \sigma^\mathrm{tot}$
                            & $\sigma$      & $\Delta \sigma^\mathrm{tot}$\\
\midrule
      200 &  1 &     2.90     & 13.1 & 3.1 & 16.6 &   3.63     & 4.1 & 6.6 & 11.8 &     3.36     & 12.5 \\ 
       200 &  8 &     0.0961   & 15.7 & 3.2 & 18.9 &   0.1194   & 6.4 & 5.9 & 13.8 &     0.1109   & 14.4 \\ 
       200 & 30 &     0.718    & 18.0 & 3.2 & 21.2 &   0.886    & 8.5 & 5.6 & 15.7 &     0.825    & 16.2 \\ 
       600 &  1 &     0.143    & 13.3 & 4.9 & 18.9 &   0.186    & 2.7 & 8.1 & 12.9 &     0.175    & 12.6 \\ 
       600 &  8 &     0.00461  & 16.7 & 5.0 & 21.9 &   0.00602  & 5.1 & 7.8 & 15.1 &     0.00566  & 14.8 \\ 
       600 & 30 &     0.0336   & 19.6 & 5.1 & 24.7 &   0.0440   & 7.3 & 7.7 & 17.0 &     0.0413   & 16.9 \\ 
      1000 &  1 &     0.0162   & 14.2 & 6.8 & 21.0 &   0.0217   & 2.3 & 10.6 & 14.7 &     0.0204   & 14.2 \\ 
      1000 &  8 &     0.000516 & 17.4 & 7.0 & 24.4 &   0.000697 & 5.2 & 10.0 & 16.6 &     0.000655 & 16.8 \\ 
      1000 & 30 &     0.00371  & 20.8 & 7.0 & 27.8 &   0.00506  & 7.9 & 9.7 & 18.8 &     0.00475  & 19.4 \\ 
    \bottomrule                                                                                                              
    \end{tabular}
\end{center}
\end{table}

To provide a final prediction for heavy charged Higgs boson production we
combine the NLO 4FS and 5FS cross sections according to Santander
matching\,\cite{Harlander:2011aa}. We note that the 4FS and 5FS calculations
provide the unique description of the cross section in the asymptotic
limits $M_\phi/\Mb \to 1$ and $M_\phi/\Mb \to \infty$, respectively (here
and in the following $M_\phi$ denotes a generic Higgs boson mass). The
4FS and 5FS are thus combined in such a way that they are given
variable weight, depending on the value of the Higgs boson mass. The
difference between the two approaches is formally logarithmic.
Therefore, the dependence of their relative importance on the
Higgs boson mass should be controlled by a logarithmic term, i.e.\
\begin{equation}
  \sigma^{\rm matched} = \frac{\sigma^{\rm 4FS} + w\,\sigma^{\rm 5FS}}{1 +
    w} \quad \mbox{with} \quad w= \ln \frac{\mathrm{M}_\phi}{\Mb} - 2\,.
\end{equation}
The theoretical uncertainties are combined according to 
\begin{equation}
  \Delta\sigma^{\rm matched}_\pm = \frac{\Delta\sigma^{\rm 4FS}_\pm + w\Delta\sigma^{\rm 5FS}_\pm}{1 +
    w}
\end{equation}
where $\Delta\sigma^{\rm 4FS}_\pm$ and $\Delta\sigma^{\rm 5FS}_\pm$
are the upper/lower uncertainty limits of the 4FS and the 5FS,
respectively.
\begin{figure}[t]
  \begin{center}
    \setlength{\unitlength}{\textwidth}
    \includegraphics[width=0.47\textwidth]{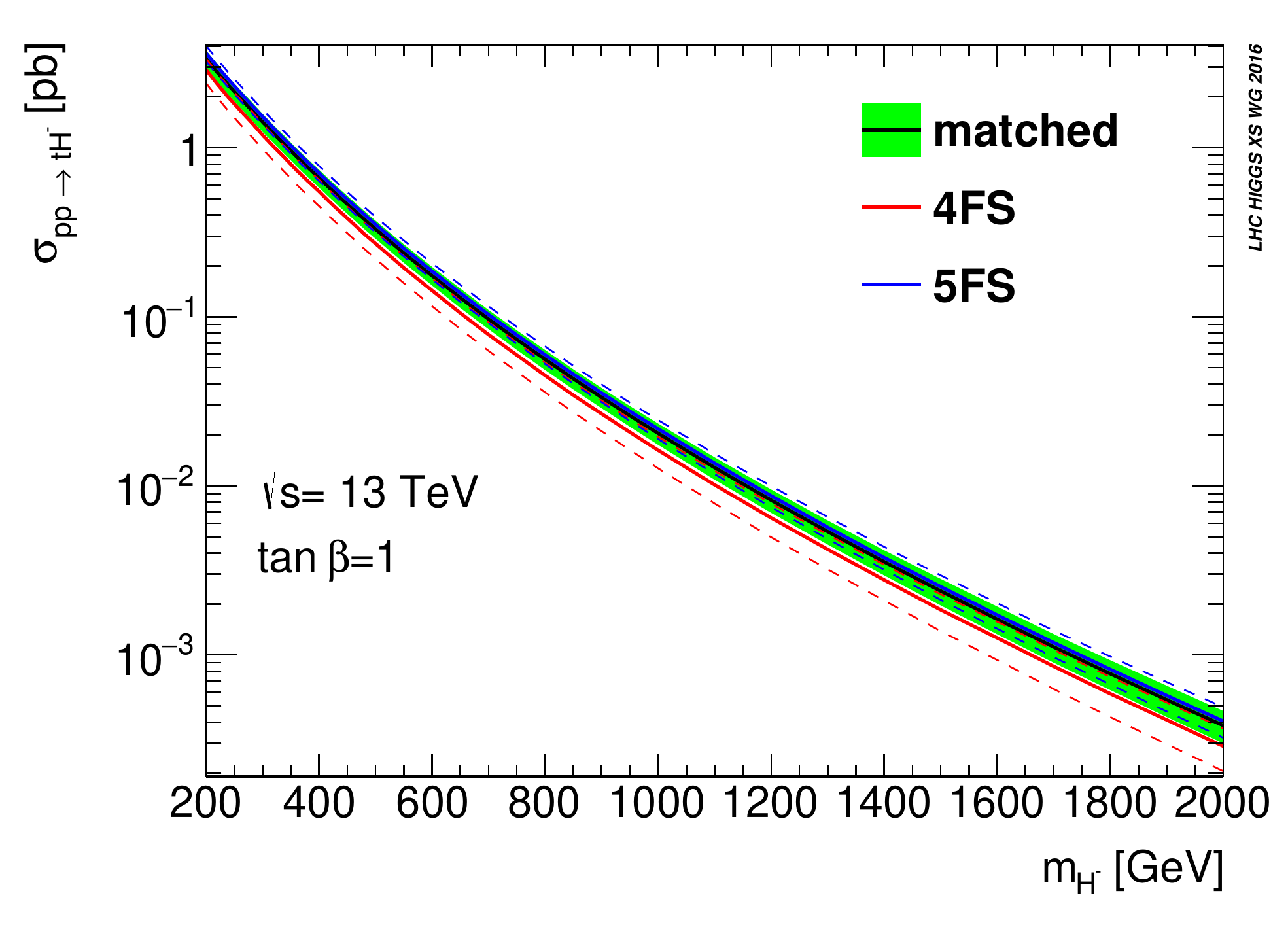}~~~~~~~
    \includegraphics[width=0.47\textwidth]{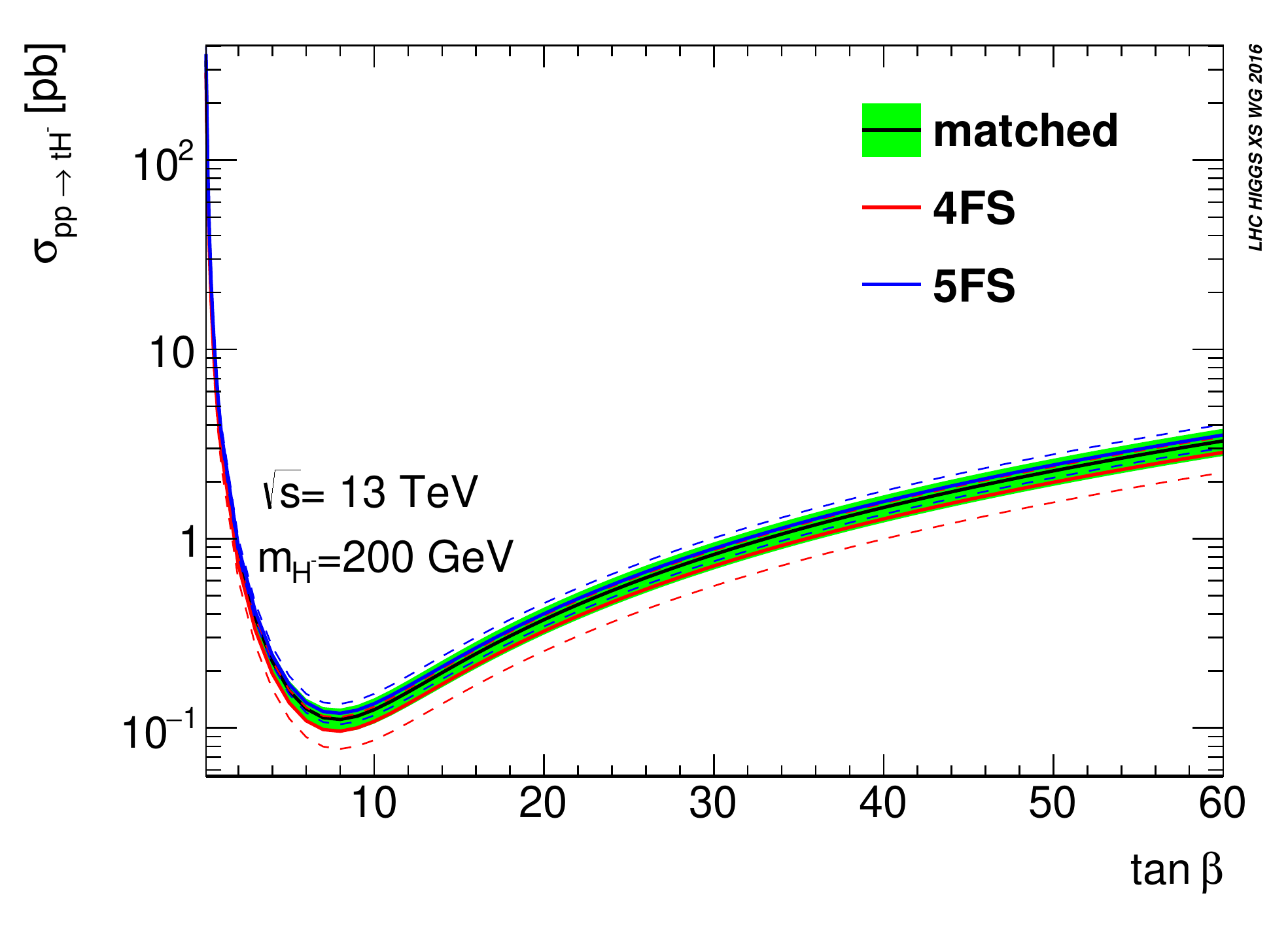}
    \includegraphics[width=0.47\textwidth]{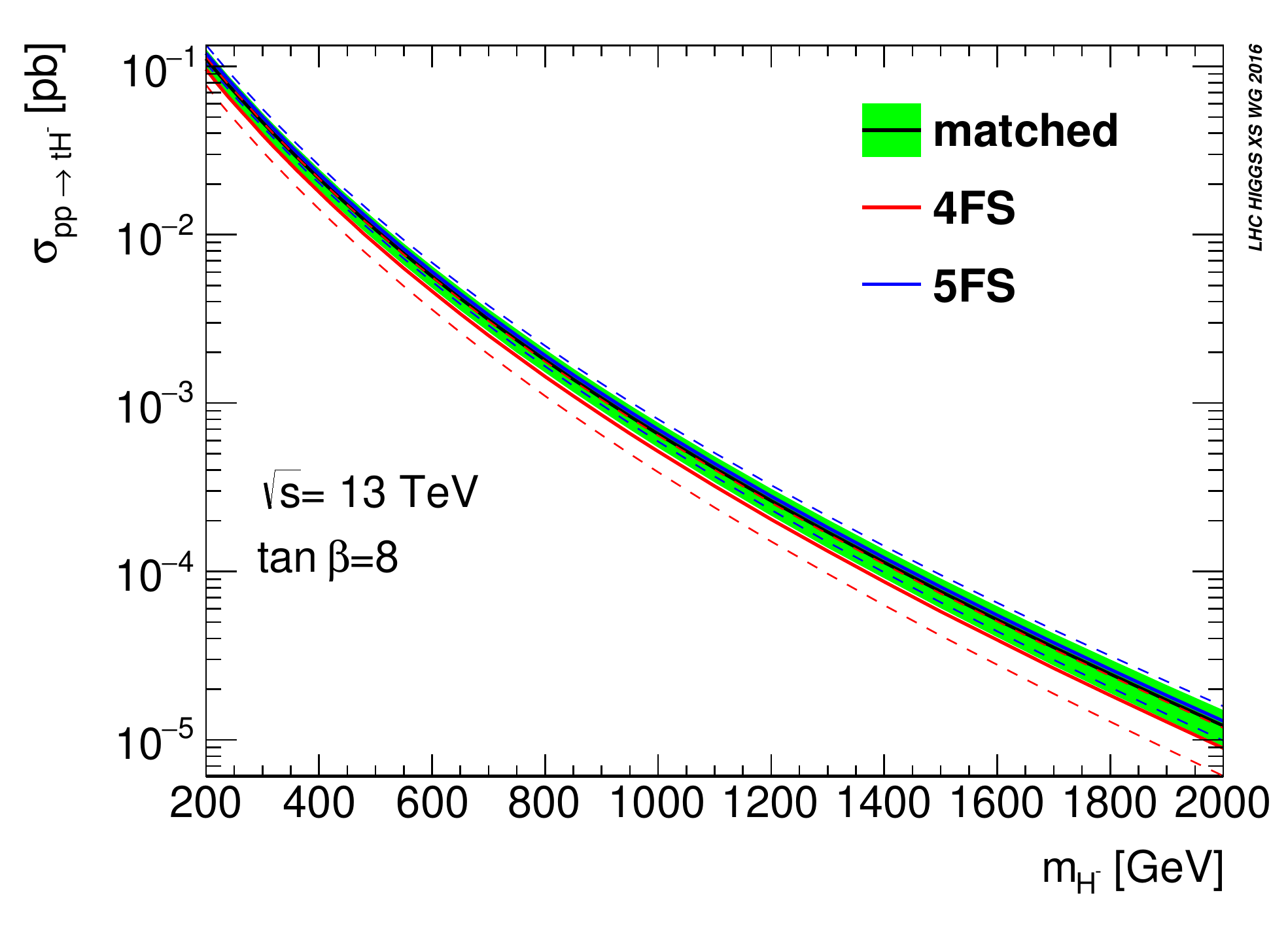}~~~~~~~
    \includegraphics[width=0.47\textwidth]{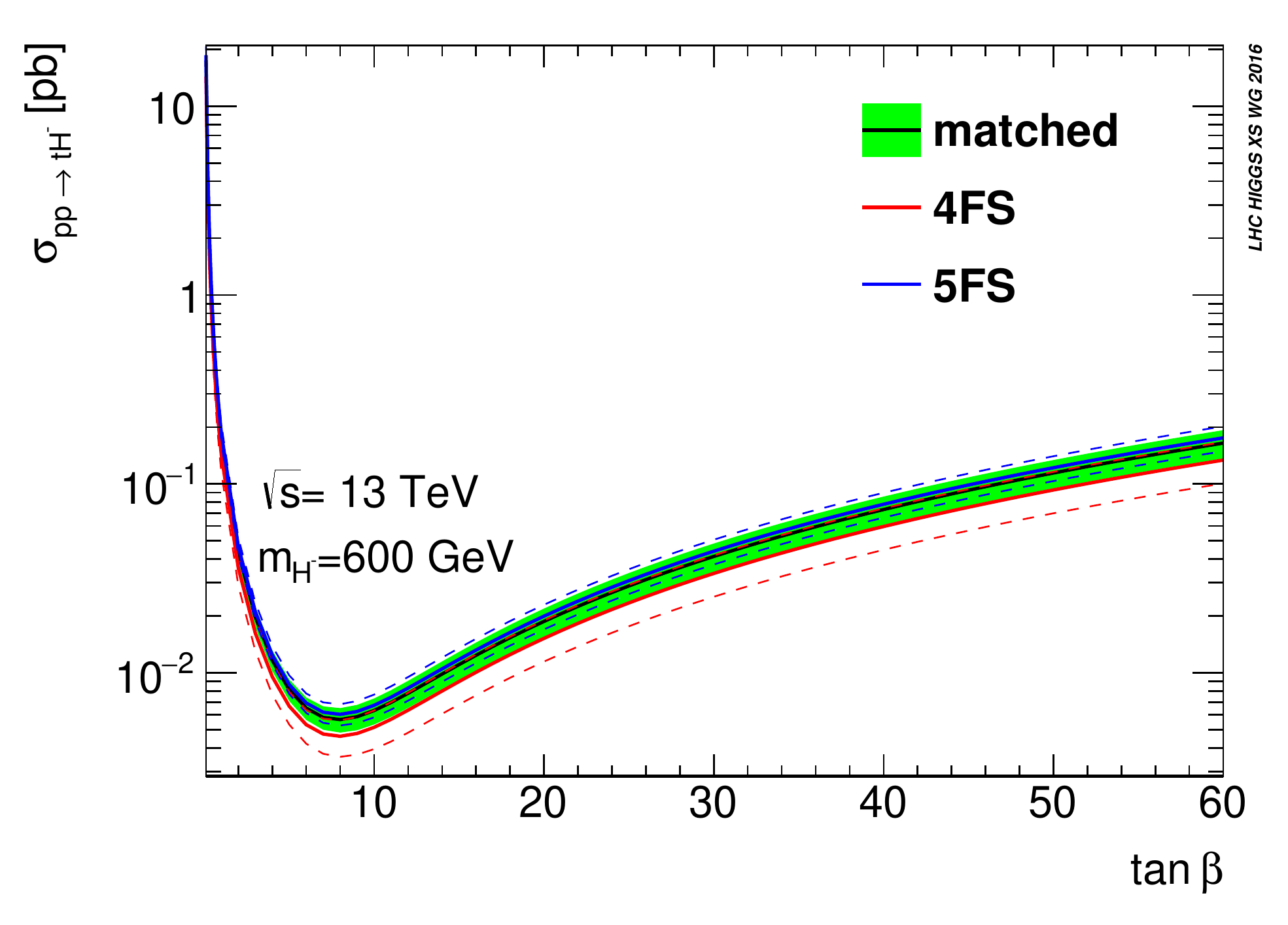}
    \includegraphics[width=0.47\textwidth]{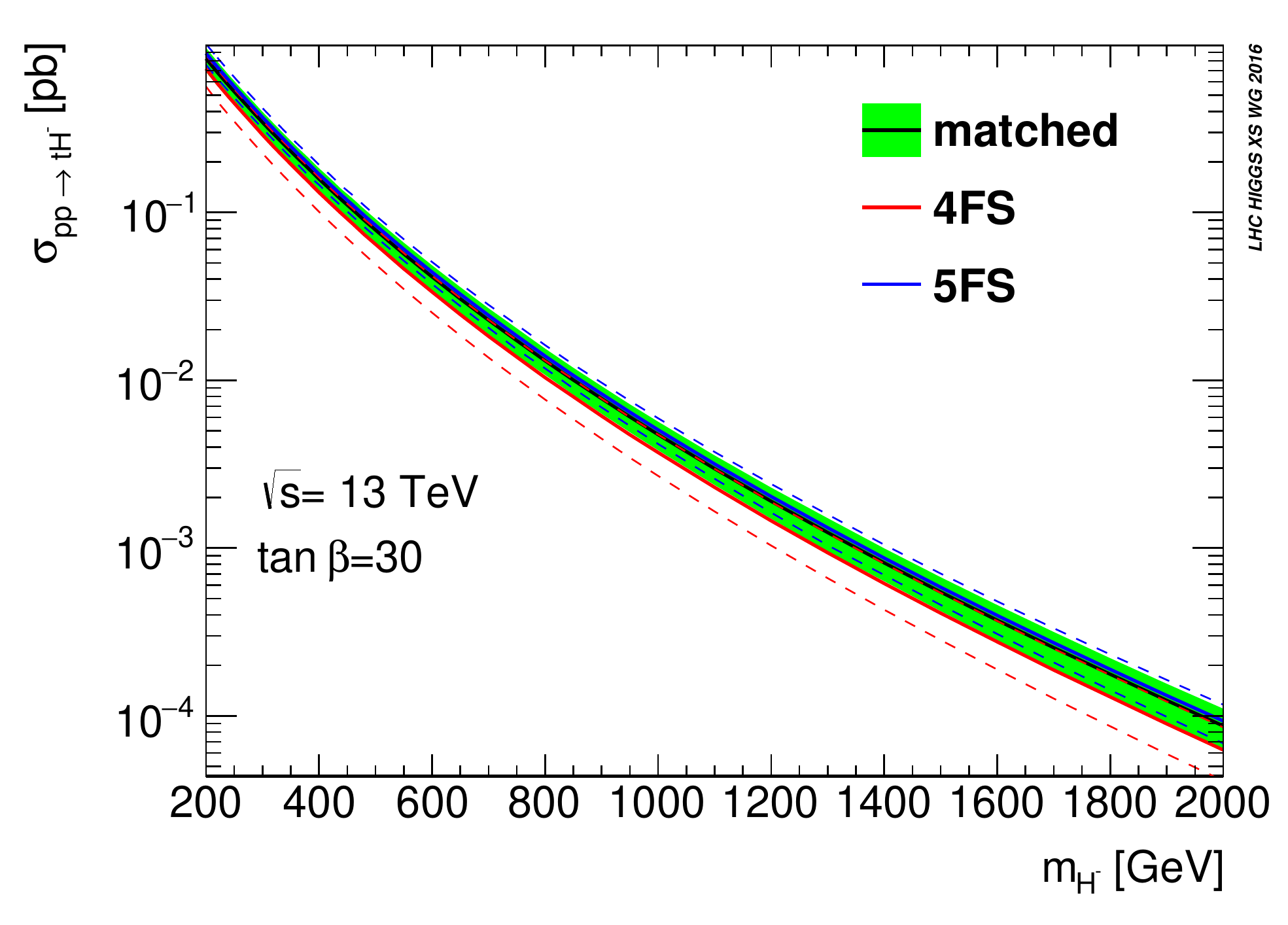}~~~~~~~
    \includegraphics[width=0.47\textwidth]{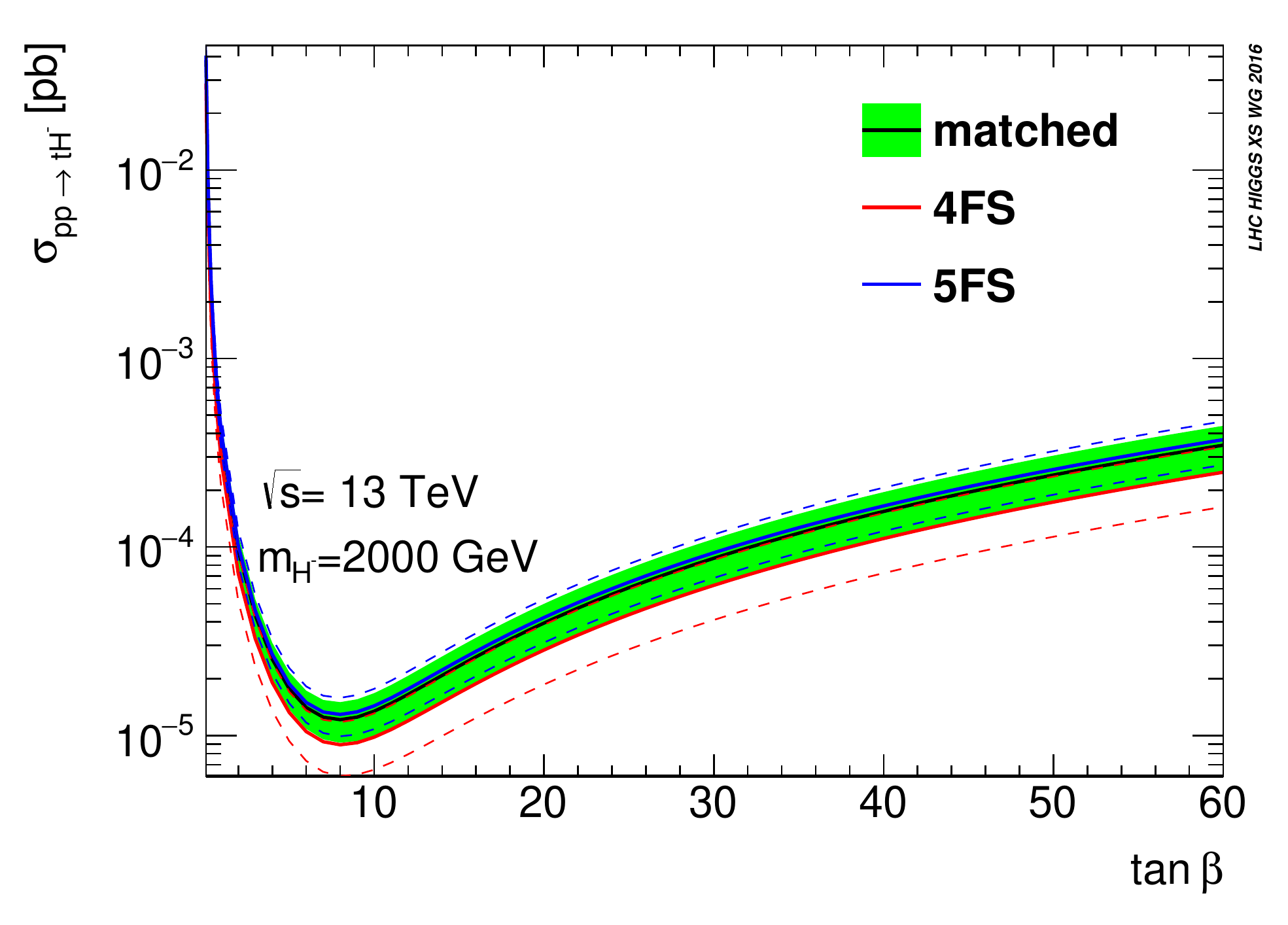}
    \caption{Cross section for $\Pt\PSHpm + X$ production, after matching the 4FS and 5FS results. The result is given for three different 
values of $\tan \beta$ (left) and of $\mhpm$ (right).}
    \label{fig:hplus-matched}
  \end{center}
\end{figure}

The resulting matched cross section is shown in \refF{fig:hplus-matched}.
We observe that the NLO 4FS and 5FS predictions are in fair mutual
agreement, with differences of the central values of roughly $20\%$. 
The dynamical choice
for $\mu_{\rm F}$ in the 5FS used here improves the matching of the
predictions in the two schemes. The overall theoretical uncertainty of
the matched NLO prediction is about 10\%.
Cross sections and uncertainties for a two-dimensional grid, 
$\mhpm=200 \UGeV - 2000 \UGeV$ and $\tan \beta=0.1-60$, can be retrieved 
online\footnote{\url{https://twiki.cern.ch/twiki/bin/view/LHCPhysics/LHCHXSWGMSSMCharged}}. 

In contrast to the type-II 2HDM, for type-I the bottom Yukawa coupling is not enhanced by $\tan\beta$,
so that $g_{\Pt\bar{\Pb}\PSHm}|_{\rm type-I} = \sqrt{2}\, \Mt/ v \,P_R\cot\beta + {\cal O}(\Mb/\Mt)$. Up to corrections suppressed by ${\cal O}(\Mb/\Mt)$, the cross section for heavy charged Higgs boson production in the
type-I 2HDM, $\sigma |_{\rm type-I} \propto g^2_{\Pt\bar{\Pb}\PSHm}|_{\rm type-I} \propto 2 (\Mt/v)^2 \cot^2\beta + {\cal O}(\Mb/\Mt)$, can thus be obtained
from the type-II cross section, $\sigma |_{\rm type-II, \tan\beta = 1} \propto g^2_{\Pt\bar{\Pb}\PSHm}|_{\rm type-II, \tan\beta = 1} \propto 2 (\Mt/v)^2 + {\cal O}(\Mb/\Mt)$, evaluated at $\tan\beta = 1$ and rescaled by $\cot^2\beta$. 
This relation is correct to all orders in QCD, but \textit{not} to all orders in the electroweak corrections. Given the overall theoretical uncertainty of the cross section prediction it is, 
however, an excellent approximation and sufficient for all practical purposes. Note that the charged Higgs boson cross section predictions for the type-I and type-II 2HDMs also hold for the so-called lepton-specific 
and flipped 2HDMs, respectively, see e.g. Ref.~\cite{Branco:2011iw}.

\section{Differential production cross sections}

\begin{figure}[t]
  \begin{center}
    \setlength{\unitlength}{\textwidth}
    \includegraphics[width=0.50\textwidth]{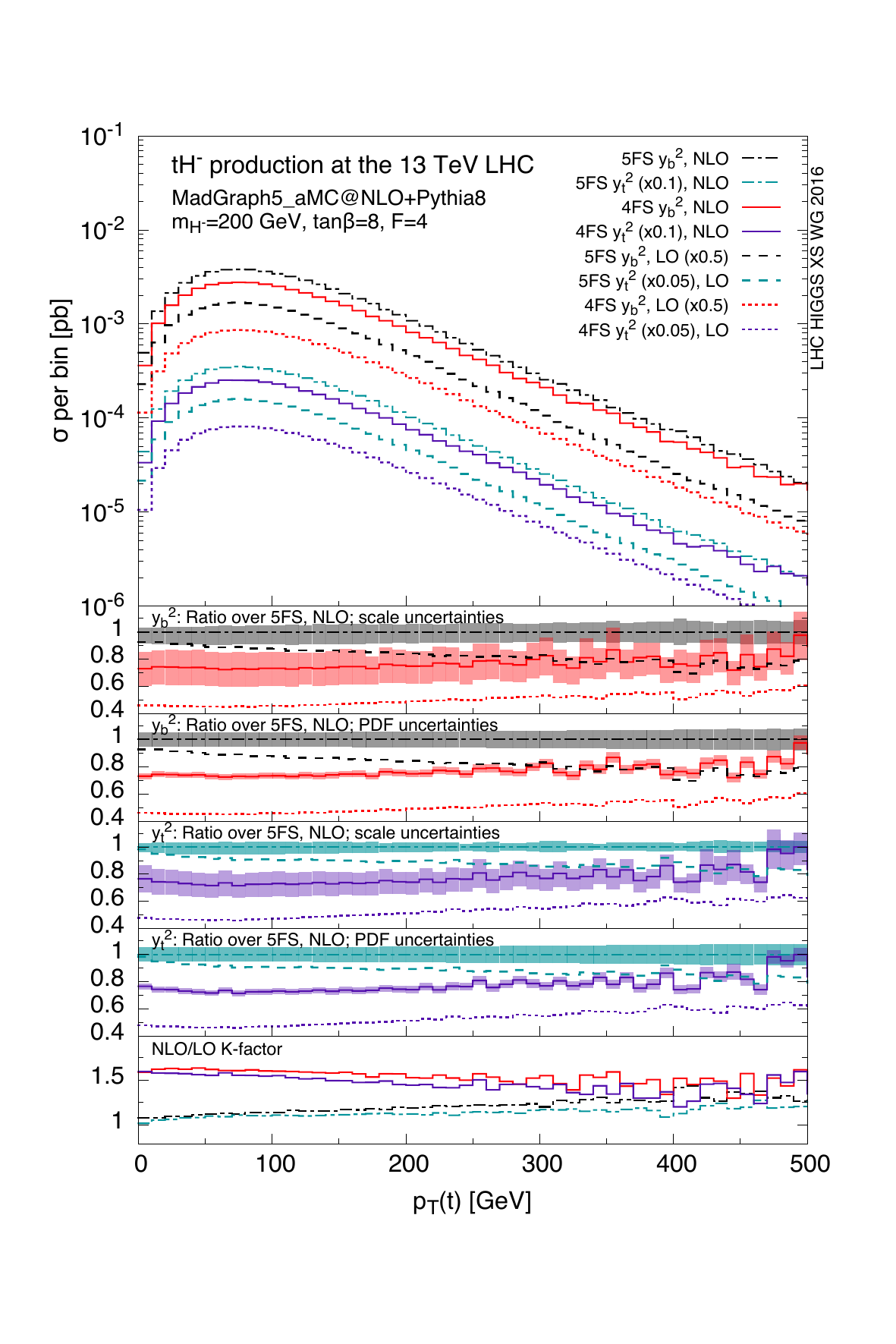}~~~~~~~
    \includegraphics[width=0.50\textwidth]{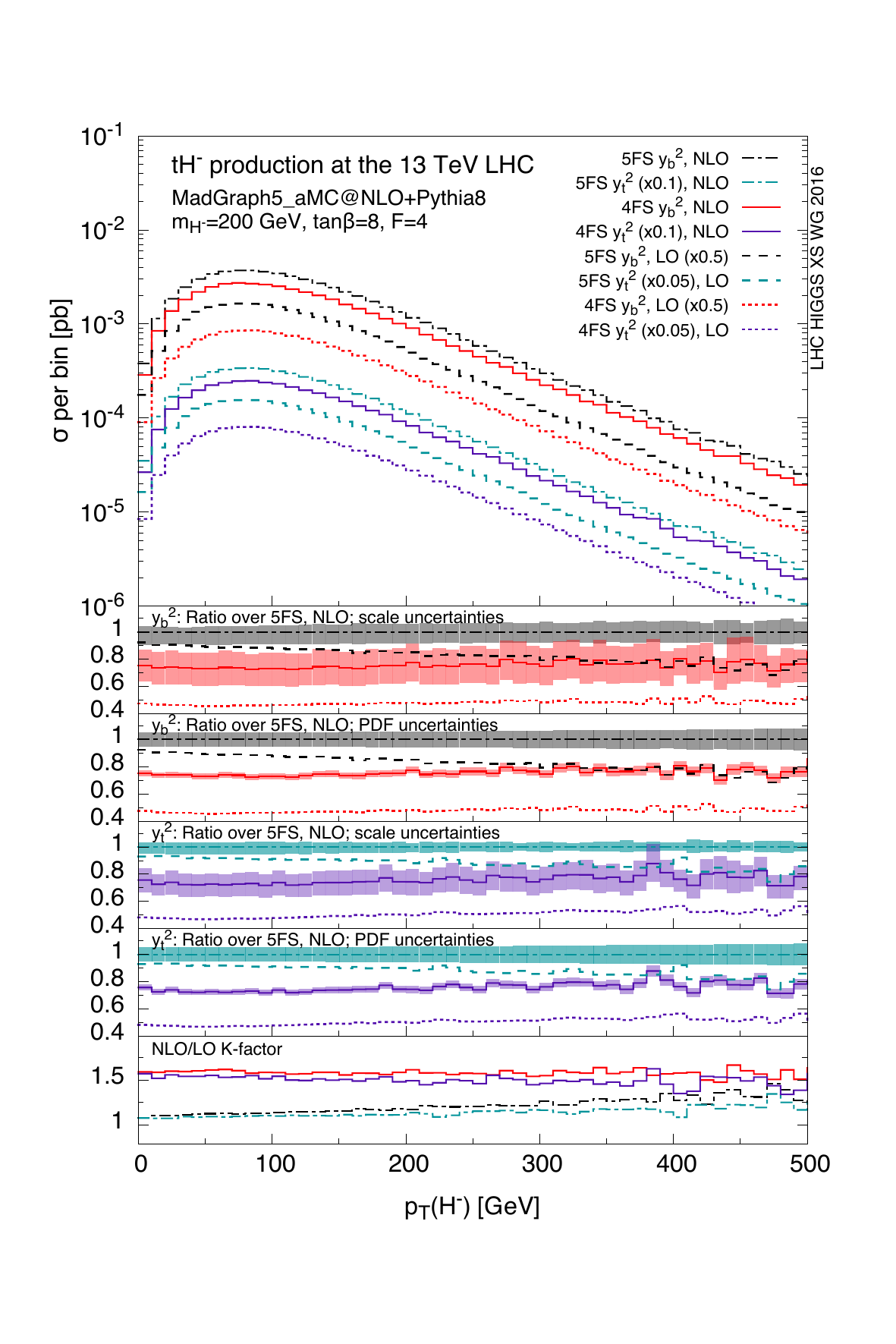}
    \vspace{-0.5cm}
    \caption{LO and NLO predictions matched with {\tt Pythia8} in the 4FS and 5FS, separately for the the $y_{\Pb}^2$ and 
$y_{\Pt}^2$ terms, for the transverse momentum of the top
quark (left) and of the charged Higgs boson (right). Rescaling factors are introduced in the main
frame for better visibility. The first four smaller frames at the bottom show the ratio over the NLO prediction in
the 5FS for the $y_b^2$ and $y_t^2$ terms, and the scale and PDF uncertainty bands for the NLO curves.
The bottom frame shows the differential $K$ factor (NLO/LO) for the four predictions. A charged Higgs boson mass of $\mhpm = 200$ GeV is assumed.}
    \label{fig:hdiff-200-ptht}
  \end{center}
\end{figure}

We now present differential distributions for the production of a heavy charged Higgs boson
in association with a top quark in a type-II 2HDM. We present results in the 4FS and
5FS up to NLO accuracy and including matching to parton shower Monte Carlos. Fully differential results in the 5FS have 
been available for some years~\cite{Weydert:2009vr, Klasen:2012wq}, while 4FS results have been presented only recently~\cite{Degrande:2015vpa}. In this
chapter, we follow
the methodology presented in Ref.~\cite{Degrande:2015vpa}, where fully-differential results in the 4FS were presented for
the first time using
{\tt MG5\_aMC@NLO}~\cite{Alwall:2014hca} together with {\tt Herwig++}~\cite{Bahr:2008pv} 
or {\tt Pythia8}~\cite{Sjostrand:2007gs}. 
In particular, we use a reduced shower scale (generated in the range $0.025 \sqrt s_0 < \mu^2_{sh} < \times 0.25 \sqrt s_0$, $s_0$ being
the born-level partonic centre of mass energy) with respect to 
the default one in {\tt MG5\_aMC@NLO}\ which
improves the matching between NLO+PS and NLO predictions at large transverse momentum. 
In the results shown here, we adapt 
relevant input parameters to match 
the recommendations followed throughout this report. In particular, differences in the setup with respect to 
Ref.~\cite{Degrande:2015vpa} include the running of the bottom Yukawa up to the renormalization scale using 
four loops  for the central predictions and two loops for the renormalization scale variations and the usage of the PDF4LHC15 parton 
distributions~\cite{Butterworth:2015oua}. We employ 4FS and 5FS PDFs consistently with the flavour scheme of the computation. In both 
cases (and also for LO predictions) PDFs are evolved at NLO. As in Ref.~\cite{Degrande:2015vpa}, we assume that the top quark decays leptonically 
while the charged Higgs remains stable. Therefore, $b$ jets in the final state will typically come from the top quark and from the matrix element.\\

\begin{figure}[t]
  \begin{center}
    \setlength{\unitlength}{\textwidth}
    \includegraphics[width=0.50\textwidth]{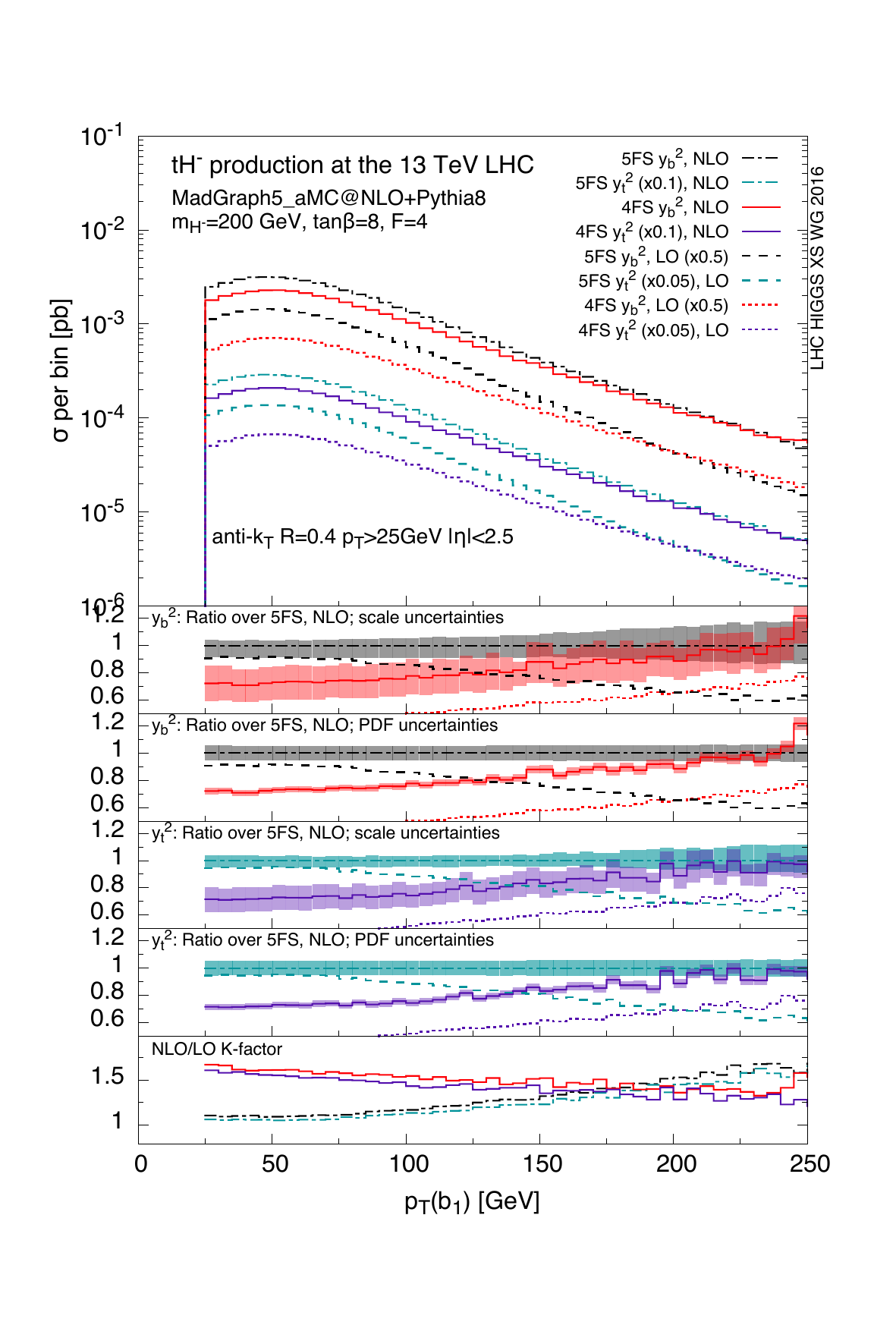}~~~~~~~
    \includegraphics[width=0.50\textwidth]{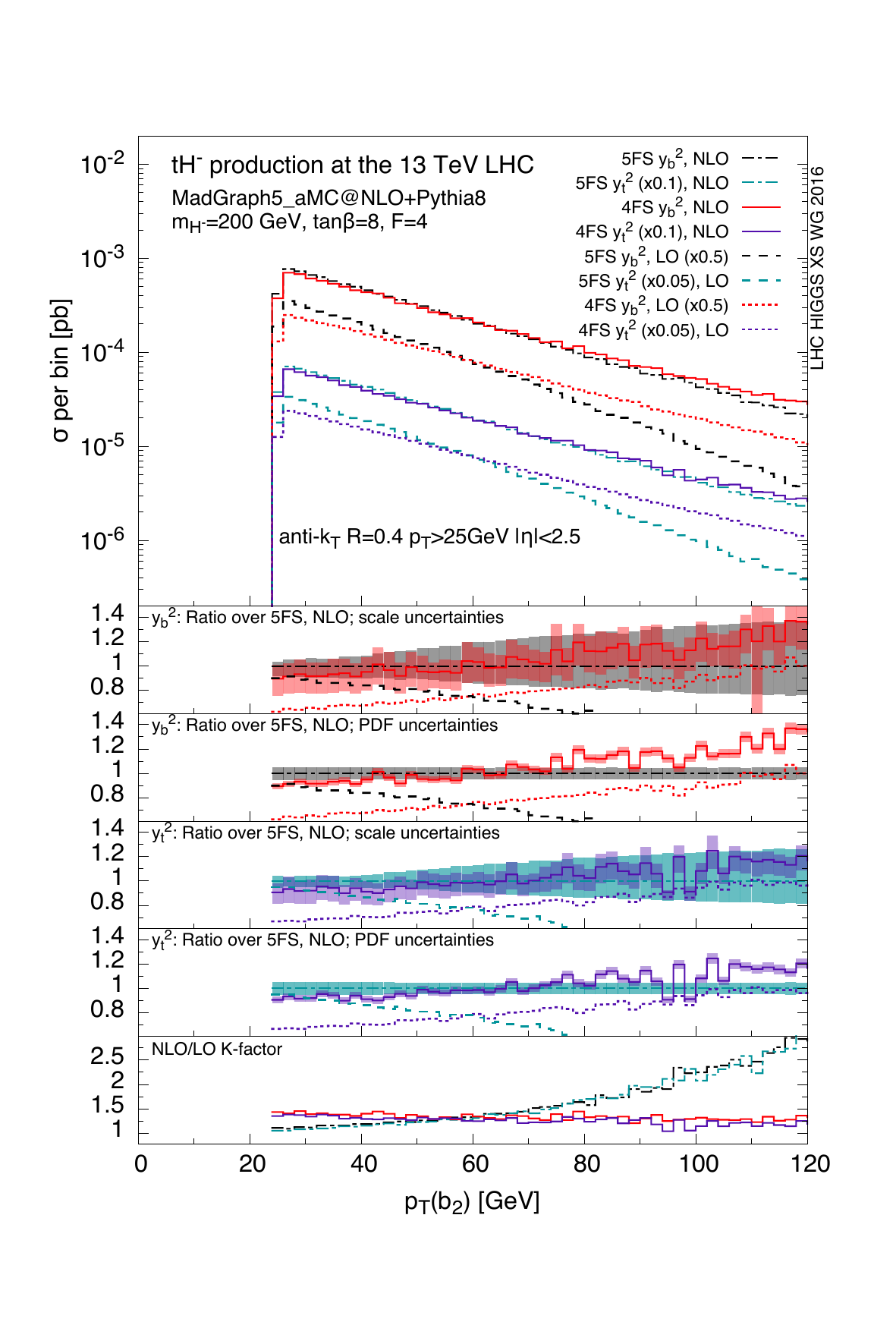}
    \vspace{-0.5cm}
    \caption{LO and NLO predictions matched with {\tt Pythia8} in the 4FS and 5FS, separately for the the $y_{\Pb}^2$ and 
$y_{\Pt}^2$ terms, for the transverse momentum of the hardest (left) and 
    second hardest $\Pb$ jet (right). Rescaling factors are introduced in the main
frame for better visibility. The first four smaller frames at the bottom show the ratio over the NLO prediction in
the 5FS for the $y_b^2$ and $y_t^2$ terms, and the scale and PDF uncertainty bands for the NLO curves.
The bottom frame shows the differential $K$ factor (NLO/LO) for the four predictions. A charged Higgs boson mass of $\mhpm = 200$ GeV is assumed.}
    \label{fig:hdiff-200-ptbj}
  \end{center}
\end{figure}

\begin{figure}[t]
  \begin{center}
    \setlength{\unitlength}{\textwidth}
    \includegraphics[width=0.50\textwidth]{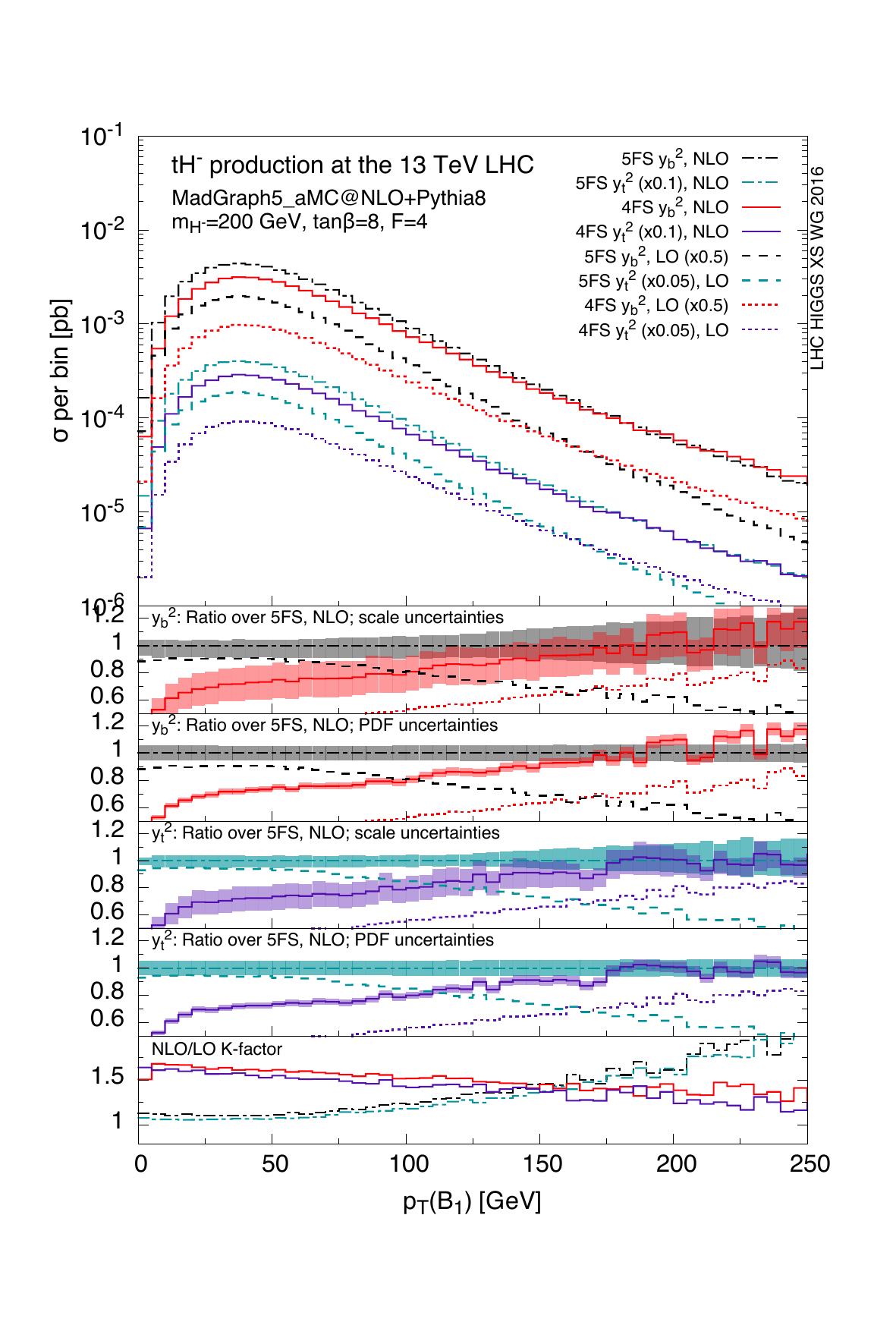}~~~~~~~
    \includegraphics[width=0.50\textwidth]{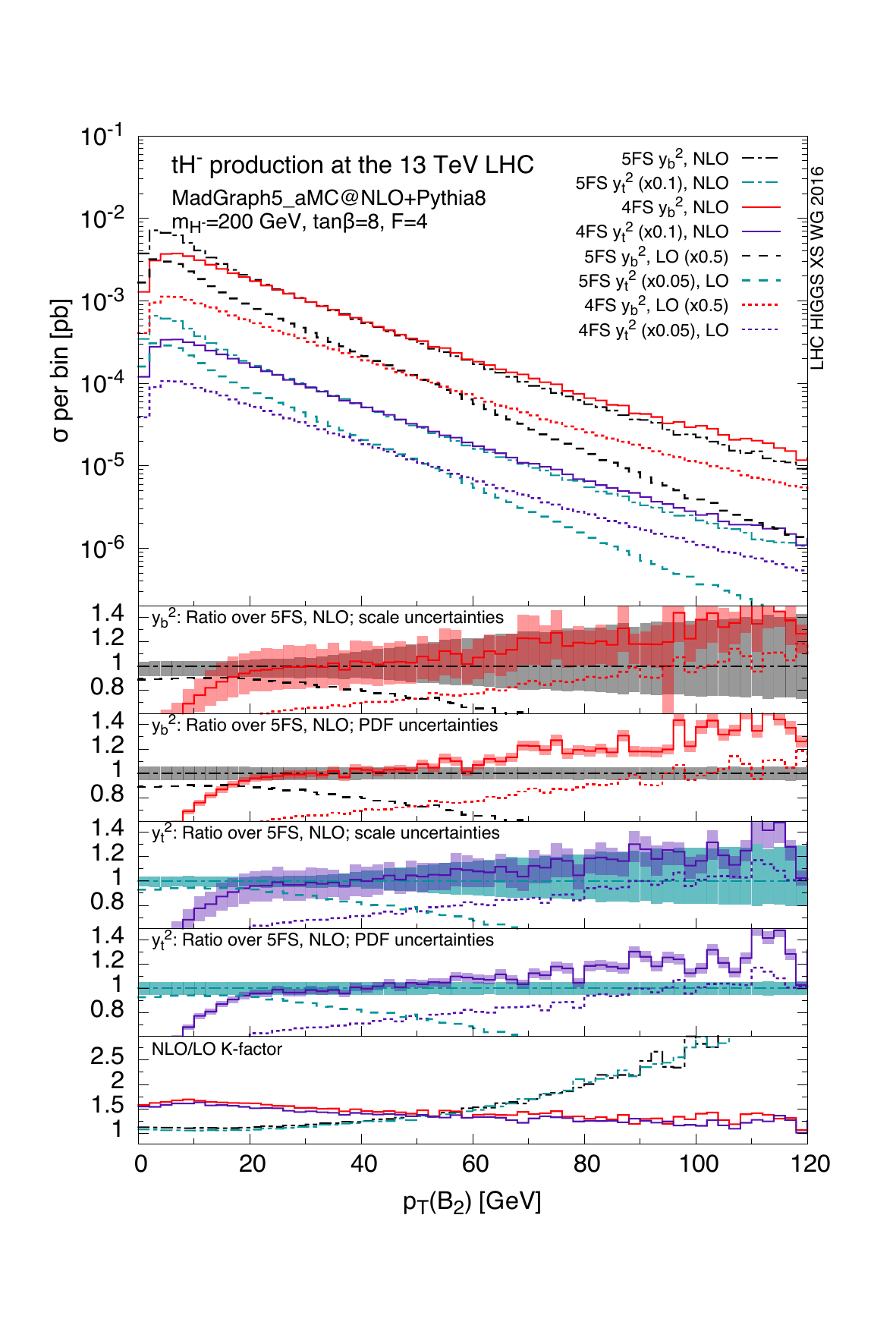}
    \vspace{-0.5cm}
    \caption{LO and NLO predictions matched with {\tt Pythia8} in the 4FS and 5FS, separately for the the $y_{\Pb}^2$ and 
$y_{\Pt}^2$ terms, for the transverse momentum of the hardest (left) and 
    second hardest $\Pb$ hadron (right). Rescaling factors are introduced in the main
frame for better visibility. The first four smaller frames at the bottom show the ratio over the NLO prediction in
the 5FS for the $y_b^2$ and $y_t^2$ terms, and the scale and PDF uncertainty bands for the NLO curves.
The bottom frame shows the differential $K$ factor (NLO/LO) for the four predictions. A charged Higgs boson mass of $\mhpm = 200$ GeV is assumed.}
     \label{fig:hdiff-200-ptbh}
  \end{center}
\end{figure}

In Figs.~\ref{fig:hdiff-200-ptht}-\ref{fig:hdiff-200-njet}, 
we present a comparison between the two schemes at LO and NLO matched with {\tt Pythia8}, for several differential observables. 
All figures refer to the case of a $\mhpm = 200\UGeV$ Higgs boson and $\tan\beta=8$. For the sake of generality, 
the  $y_{\Pt}^2$ and $y_{\Pb}^2$ contributions to the cross section are
shown separately, omitting the negligible interference term. Predictions for different values of $\tan \beta$ can be obtained
by a trivial rescaling of the shown histograms. All figures have the same layout, namely: a main frame with the absolute predictions
in the two schemes, at LO and NLO, and five smaller frames below. In the first four of these frames, the ratio of histograms in the main frame over the 5FS NLO prediction is shown,
together with scale (first and third frames, respectively for the $y_{\Pb}^2$ and $y_{\Pt}^2$ contributions) and PDF uncertainties 
(second and fourth frames). The last frame shows the differential $K$-factors (NLO/LO).

Before looking at the various observables, we outline some general features: the first one is that, as expected, the inclusion of NLO corrections brings 
predictions in the two schemes much closer than at LO. The second is about the size of uncertainties at NLO, which follows the same pattern as
the inclusive cross section described in the previous section: for observables which are described
with the same accuracy in the two schemes (e.g. top and Higgs boson $p_T$, $\Pb$-jet rates for zero and one jet), scale uncertainties in the 4FS are usually larger
than in the 5FS ($\pm10-12\%$ vs$\pm4-6\%$). PDF uncertainties display instead an opposite behaviour (at least for this value of the charged Higgs boson mass): they 
are larger in the 5FS, with a similar size as scale uncertainties, and smaller in the 4FS, where they are negligible with respect to scale variations.
Finally, due to the additional running of the bottom Yukawa, the $y_{\Pb}^2$ contribution
has a broader scale uncertainty band than the $y_{\Pt}^2$ one. 

We now turn to compare the two schemes for a number of differential observables: in \refF{fig:hdiff-200-ptht} we observe that for the transverse momentum of the top quark (reconstructed using Monte Carlo truth information)
and the Higgs boson the difference between the two schemes can be compensated by a simple overall 
rescaling of the total rates at NLO (similar to the one observed in the previous section) 
while LO predictions in the two schemes have quite different
shapes (in particular for the top quark). The same level of agreement is expected to be found also for observables related to the 
decay products of the top quark (and of the charged Higgs boson). 
Indeed, the $p_T$ spectrum of the hardest $\Pb$ jet (left plot in \refF{fig:hdiff-200-ptbj})
displays a flat ratio between the 4FS and 5FS at NLO up to $\approx 120$ GeV. While below 120 GeV the hardest $\Pb$ jet essentially 
coincides with the $\Pb$ jet from the top quark, above 120 GeV secondary $\Pg\to \Pb\bar{\Pb}$
splitting from hard gluons becomes relevant. This fact is reflected
in the growth of the 5FS scale uncertainty band and of the $k$ factor. Larger differences between the two schemes 
appear for the second-hardest $\Pb$ jet, see right plot in \refF{fig:hdiff-200-ptbj}. 
This distribution is expected to be poorly described in the 5FS. 
In particular, its kinematics in the 5FS at LO is determined
by the shower, while at NLO it is driven by a tree-level matrix element (therefore being
formally only LO accurate). As expected, the 5FS develops larger $k$ factors.
The 4FS calculation thus describes these observables significantly better, both because of its more robust 
perturbative behaviour and because of the proper modelling of the final-state $\Pb$ jets.\\
The effect of the different treatment of the bottom quark in the two schemes is even more
visible for the transverse momentum of the hardest and second hardest $\PB$ hadron 
(left and right plot in \refF{fig:hdiff-200-ptbh}). At medium
and large $\pT$ of the hardest $\PB$ hadron similar effects as for the hardest $\Pb$ jet are observed. 
At low momentum, the 4FS prediction is suppressed with respect to the 5FS . 
This is most likely due to mass effects as these kinematical regions correspond to one
$\Pb$ quark being collinear to the beam. In the 5FS these configurations are enhanced because of
the collinear singularities, while in the 4FS such singularities are screened by the $\Pb$ quark mass.
Therefore, even after the parton shower, the 5FS is reminiscent of the collinear enhancement. In the case of
the second-hardest $\PB$ such effects are further enhanced.

\begin{figure}[t]
  \begin{center}
    \setlength{\unitlength}{\textwidth}
    \includegraphics[width=0.50\textwidth]{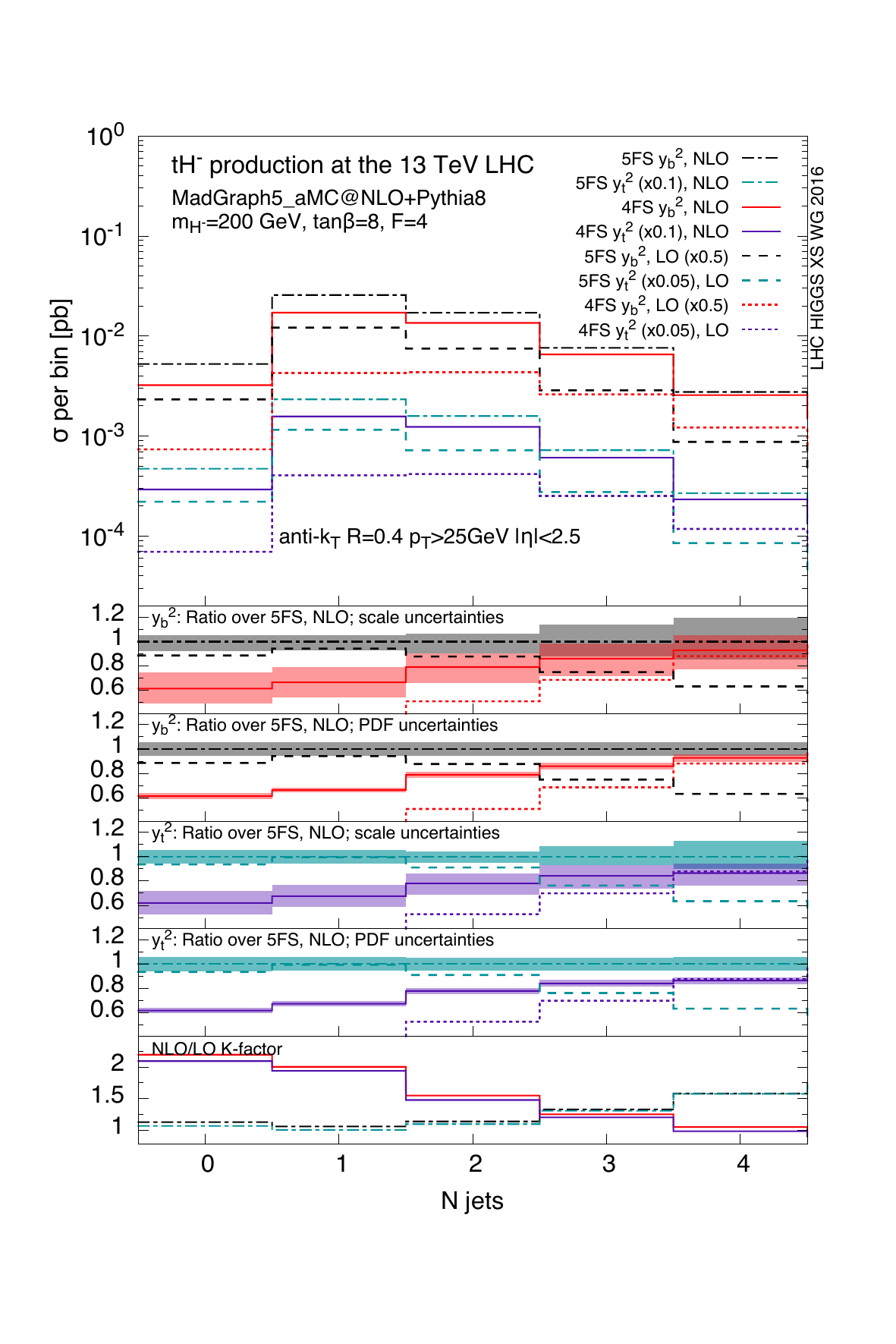}~~~~~~~
    \includegraphics[width=0.50\textwidth]{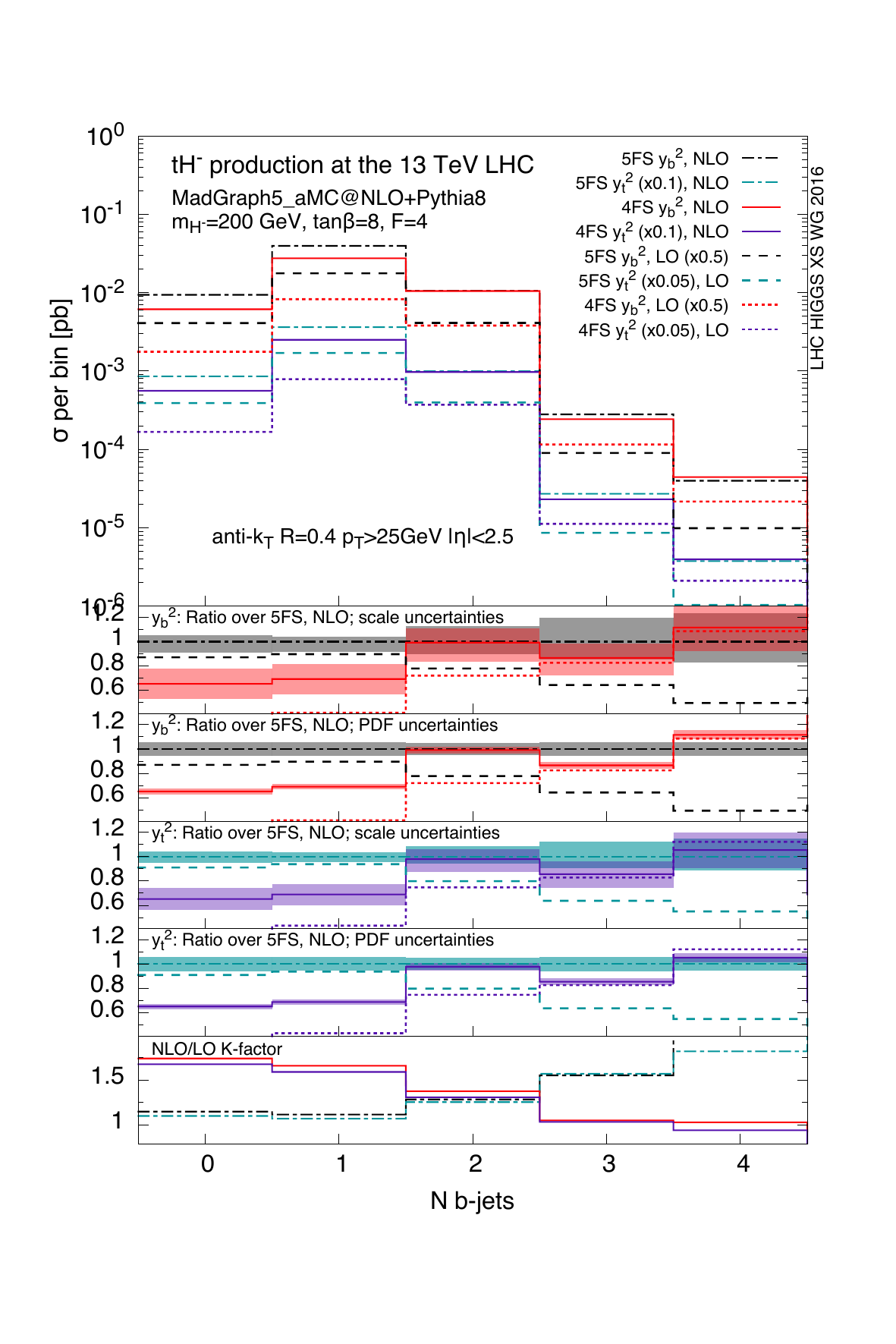}
    \vspace{-0.5cm}
    \caption{LO and NLO predictions matched with {\tt Pythia8} in the 4FS and 5FS, separately for the the $y_{\Pb}^2$ and 
$y_{\Pt}^2$ terms, for for jet (left) and $\Pb$ jet (right) multiplicity. Rescaling factors are introduced in the main
frame for better visibility. The first four smaller frames at the bottom show the ratio over the NLO prediction in
the 5FS for the $y_b^2$ and $y_t^2$ terms, and the scale and PDF uncertainty bands for the NLO curves.
The bottom frame shows the differential $K$ factor (NLO/LO) for the four predictions. A charged Higgs boson mass of $\mhpm = 200$ GeV is assumed.}
    \label{fig:hdiff-200-njet}
  \end{center}
\end{figure}
Finally, looking at jet (\refF{fig:hdiff-200-njet} left) 
and $\Pb$-jet multiplicities (\refF{fig:hdiff-200-njet} right), we observe again
that the effect of NLO corrections is very different in the two schemes: while in the 5FS NLO corrections
make the jet spectrum moderately harder, in the 4FS they tend to make it softer, with greater enhancements in the low multiplicity bins. Despite this fact,
at NLO the 4FS still shows a slightly harder spectrum than the 5FS.
Irrespective of whether $\Pb$-tagged jets are required or not, the overall effect of NLO QCD corrections is to bring the two schemes in much 
better agreement in the zero- and one-jet bin. The two-jet bin is only described at LO accuracy by the 5FS NLO prediction, therefore for this
bin the 4FS prediction is expected to be more reliable. Higher multiplicities are described with rather poor accuracy (LO or even LL) by both schemes.

All in all, the global behaviour of predictions closely follows what has been observed in 
Ref.~\cite{Degrande:2015vpa}. The interested reader can find more details 
in that paper, in particular for what concerns the comparison of matched and fixed-order computations and the usage of different parton showers.

\section{Recommendations for signal simulation}
We conclude this chapter by providing some recommendations for the simulation of heavy charged Higgs boson production at the LHC Run 2.
The use of 4FS fully-differential predictions is recommended for any realistic fully-differential signal simulation for experimental
searches. This recommendation is backed by two sets of evidences: first, for a large number of observables, 
the 4FS prediction provides a better description of the
final state kinematics; second, it reduces the systematic error related to the usage of a given parton
shower. Moreover, when matching the NLO calculation to the shower, we recommend to use
a lower shower scale by reducing the current default value from {\tt MG5\_aMC@NLO}
by a factor four. This corresponds to a shower scale generated in the range $0.025\sqrt s_0 < \mu^2_{sh} < \times 0.25 \sqrt s_0$, $s_0$ being
the born-level partonic centre of mass energy. 
This choice provides a better matching
to the fixed-order computation at large transverse momenta, slightly reduces the parton shower
dependence and also improves the agreement of four- and five-flavour scheme computations.\\
For what concerns the normalization of the signal total cross-section, the Santander-matched prediction provided in this section should 
be employed as central value.

As far as the estimate of theoretical uncertainties is concerned, the most obvious choice would be to use the scale and PDF uncertainties
coming from the 4FS fully-differential computation. 
For observables inclusive in the $\Pb$ kinematics,
this choice results into a slightly larger theoretical uncertainty than the one associated with the matched prediction (suggested for the normalization). On 
the other hand, for more exclusive observables, the 4FS uncertainty is smaller and more reliable than the 5FS one.\\

\chapter{Extended Scalar Sector}
\label{chap:ExtScalar}
\ChapterAuthor{R.~Gerosa, H.~Logan, R.~Santos, O.~St{\r a}l, S.~Su, X.~Sun (Eds.)
R.~Aggleton, D.~Barducci, F.J.~Botella, G.C.~Branco, R.~Costa, D.~Fontes, G.F.~Giudice, H.E.~Haber, S.~Huber, A.~Ilnicka, F.~Kling, M.~Krawczyk, D.~Lopez-Val, K.~Mimasu, S.~Moretti, M.~Nebot, J.M.~No, M.~Pelliccioni, M.N.~Rebelo, T.~Robens, J.C.~Rom{\~a}o, N.~Rompotis, M.O.P.~Sampaio, C.H.~Shepherd-Themistocleous, J.P.~Silva, T.~Stefaniak, M.~Zaro}
\providecommand\MSHpm{\mathswitch {M_{\PSHpm}}}
\providecommand\Mone{\mathswitch {M_{1}}}
\providecommand\Mtwo{\mathswitch {M_{2}}}
\providecommand\Mthree{\mathswitch {M_{3}}}
\providecommand\Phone{\HepParticle{h}{\!\ 1}{}\Xspace}
\providecommand\Phtwo{\HepParticle{h}{\!\ 2}{}\Xspace}
\providecommand\Phthree{\HepParticle{h}{\!\ 3}{}\Xspace}
\providecommand\Phtwothree{\HepParticle{h}{\!\ 2,3}{}\Xspace}
\providecommand\Phindexi{\HepParticle{h}{\!\ i}{}\Xspace}
\providecommand\Phindexj{\HepParticle{h}{\!\ j}{}\Xspace}
\providecommand\PQQ{\HepParticle{Q}{}{}\Xspace}
\providecommand\PSHmp{\HepParticle{\PSH}{}{\mp}\Xspace}
\providecommand\PQf{\HepParticle{f}{}{}\Xspace}
\providecommand\PAQf{\HepAntiParticle{f}{}{}\Xspace}
\providecommand\PV{\HepParticle{V}{}{}\Xspace}
\providecommand\PVV{\HepParticle{VV}{}{}\Xspace}
\providecommand\Pxi{\HepParticle{\xi}{}{}\Xspace}
\providecommand\PX{\HepParticle{X}{}{}\Xspace}
\providecommand\Pchi{\HepParticle{\chi}{}{}\Xspace}
\providecommand\PS{\HepParticle{S}{}{}\Xspace}
\providecommand\PG{\HepParticle{G}{}{}\Xspace}
\providecommand\Ps{\HepParticle{s}{}{}\Xspace}
\DeclareRobustCommand{\PSHpmFive}{\HepParticle{\PSH}{5}{\pm}\Xspace}
\DeclareRobustCommand{\PSHpmpmFive}{\HepParticle{\PSH}{5}{\pm\pm}\Xspace}
\DeclareRobustCommand{\PSHpFive}{\HepParticle{\PSH}{5}{+}\Xspace}
\DeclareRobustCommand{\PSHppFive}{\HepParticle{\PSH}{5}{++}\Xspace}
\DeclareRobustCommand{\PSHmFive}{\HepParticle{\PSH}{5}{-}\Xspace}
\DeclareRobustCommand{\PSHmmFive}{\HepParticle{\PSH}{5}{--}\Xspace}
\DeclareRobustCommand{\PHzFive}{\HepParticle{\PH}{5}{0}\Xspace}
\DeclareRobustCommand{\PHFive}{\HepParticle{\PH}{5}{}\Xspace}
\DeclareRobustCommand{\PHThree}{\HepParticle{\PH}{3}{}\Xspace}
\DeclareRobustCommand{\PSHpThree}{\HepParticle{\PSH}{3}{+}\Xspace}
\DeclareRobustCommand{\PSHmThree}{\HepParticle{\PSH}{3}{-}\Xspace}
\DeclareRobustCommand{\PSHpmThree}{\HepParticle{\PSH}{3}{\pm}\Xspace}
\DeclareRobustCommand{\PHzThree}{\HepParticle{\PH}{3}{0}\Xspace}
\DeclareRobustCommand{\PHThree}{\HepParticle{\PH}{3}{}\Xspace}
\section{\label{sec:introExtSca} Introduction}

The two-Higgs doublet model (2HDM) is one of the simplest extensions of the Standard Model (SM)
and is build by simply adding one more scalar doublet to the SM field content while keeping
the Lagrangian invariant under the same symmetries. It was first
proposed by T.D.~Lee~\cite{Lee:1973iz} as a means to provide an extra source of CP-violation thus helping to
explain the observed matter anti-matter asymmetry of the Universe. In recent years the 2HDM has been
used as a benchmark model helping to identify possible directions to minimal changes with respect to the SM. These
changes can lead to the inclusion of a dark matter candidate, new sources of CP-violation,
baryogenesis and even a more elaborate flavour structure. Even in its minimal $\mathbb{Z}_2$
symmetric and CP-conserving version, it has been used by the ATLAS and CMS collaborations,
to perform phenomenological studies involving extra scalars
production and decay. In its CP-conserving versions the model has three neutral states
- two CP-even ($\Ph$ and $\PH$) and one CP-odd ($\PA$) - and two charged states ($\PSHpm$) while
the remaining three degrees of freedom are the longitudinal components of the gauge bosons.
In its CP-violating versions the three neutral states have no definite CP numbers and are usually
denoted by $\Phone$, $\Phtwo$ and $\Phthree$.
A thorough description of the different versions of
two-Higgs doublet models can be found in~\cite{Gunion:1989we, Branco:2011iw}.

In order to avoid flavour changing neutral currents (FCNC) at tree-level, a $\mathbb{Z}_2$ symmetry
is imposed on the fields ($\Phi_1 \to \Phi_1$ and
$\Phi_2 \to - \Phi_2$). The Higgs potential invariant under $\mathbb{Z}_2$,
softly broken by a dimension-two term,
can be written as
\begin{eqnarray} \label{pot}
\mathcal{V} &=& m^2_{11}\Phi_1^\dagger\Phi_1+m^2_{22}\Phi_2^\dagger\Phi_2
  -\left(m^2_{12}\Phi_1^\dagger\Phi_2+{\rm h.c.}\right)
  +\half\lambda_1\left(\Phi_1^\dagger\Phi_1\right)^2
  +\half\lambda_2\left(\Phi_2^\dagger\Phi_2\right)^2 \nonumber \\
&& \qquad +\lambda_3\Phi_1^\dagger\Phi_1\Phi_2^\dagger\Phi_2
  +\lambda_4\Phi_1^\dagger\Phi_2\Phi_2^\dagger\Phi_1
  +\left[\half\lambda_5\left(\Phi_1^\dagger\Phi_2\right)^2+{\rm h.c.}\right]\,.
\end{eqnarray}
Since $\mathcal{V}$ has to be hermitian, $m_{12}^2$ and $\lambda_5$ can be complex
while all other parameters have to be real. Choosing the two vacuum expectation values (VEVs)
to be real we end up with a model
that is either CP-conserving or explicitly CP-violating. This particular version
of the CP-violating 2HDM, where both $m_{12}^2$ and $\lambda_5$ are complex and
the VEVs are real, was first analysed in~\cite{Ginzburg:2002wt}. In the following, the CP-conserving model is denoted by
``2HDM'' while the CP-violating model is denoted by ``C2HDM''.

Since the VEVs are real we can define $\tan \beta = v_2/v_1$ for both models.
The remaining independent parameters for the 2HDM are the
four masses $\Mh $, $\MH $, $\MA $ and $\MSHpm$, the rotation
angle $\alpha$ that diagonalizes the CP-even mass matrix, and $m_{12}^2$.
For the C2HDM the remaining free parameters are the two lighter neutral states
masses $\Mone$ and $\Mtwo$, $\MSHpm$, the three rotation angles that diagonalize
the neutral mass matrix $\alpha_1$, $\alpha_2$ and $\alpha_3$ and $\rm{Re}[m_{12}^2]$.
If $\sin \alpha_2 =0$, $h_1$ is a pure scalar and if $\sin \alpha_2 =1$ $h_1$
is a pure pseudoscalar. The 2HDM Higgs boson couplings to massive gauge bosons are given by
\begin{equation}
g_{{\rm 2HDM}}^{\Ph \PVV}  =  \sin (\beta - \alpha) \, g_{{\rm SM}}^{\Ph \PVV}
\qquad \qquad
g_{{\rm 2HDM}}^{\PH \PVV}  =  \cos (\beta - \alpha) g_{{\rm SM}}^{\Ph \PVV}
  \label{gvv}
\end{equation}
while the lightest Higgs boson couplings in the CP-violating case can be written as
\begin{equation}
g_{{\rm C2HDM}}^{\Phone \PVV}  =  \cos \alpha_2 \, \, g_{{\rm 2HDM}}^{\Ph \PVV} .
  \label{gvv2}
\end{equation}

In order to avoid FCNCs at tree level, the simple solution of coupling fermions of a given
electric charge to no more than one Higgs doublet~\cite{Glashow:1976nt, Paschos:1976ay} is used. In practice,
a $\mathbb{Z}_2$ symmetry is imposed to all fields and invariance
of the Lagrangian under that symmetry is enforced.
The $\mathbb{Z}_2$ charge assignments ($\Phi_1$ is even and
$\Phi_2$ is odd) lead to four independent
combinations~\cite{Barger:1989fj}: only $\Phi_2$ couples to all fermions (type I);
$\Phi_2$ couples to up-type quarks while $\Phi_1$ couples to charged leptons and down-type quarks
(type II); $\Phi_2$ couples to charged leptons and up-type quarks while $\Phi_1$ couples to down-type quarks
(type Flipped or Y); $\Phi_2$ couples to quarks while $\Phi_1$ couples to charged leptons (type Lepton Specific
or X). The Yukawa couplings for the CP-conserving model, relative to the SM ones,
are shown in Table~\ref{Tab:MixFactor}.

\begin{table}[h]
\caption{Yukawa couplings to the scalars $\Ph$, $\PH$ and $\PA$ normalized to the
 SM Higgs Yukawa couplings. Notation is $c(s)_\alpha =\cos (\sin) \alpha$, $c(s)_\beta =\cos (\sin) \beta$.}
\label{Tab:MixFactor}
\begin{center}
\renewcommand{\arraystretch}{1.3}
\begin{tabular}{c|c|c|c|c|c|c|c|c|c} 
\toprule
& $y_{\Ph}^{\PQu}$ & $y_{\Ph}^{\PQd}$ & $y_{\Ph}^\ell$
& $y_{\PH}^{\PQu}$ & $y_{\PH}^{\PQd}$ & $y_{\PH}^\ell$
& $y_{\PA}^{\PQu}$ & $y_{\PA}^{\PQd}$ & $y_{\PA}^\ell$ \\ 
\midrule
Type-I
& {$\frac{c_\alpha}{s_\beta}$} & $\frac{c_\alpha}{s_\beta}$ & $\frac{c_\alpha}{s_\beta}$
& {$\frac{s_\alpha}{s_\beta}$} & $\frac{s_\alpha}{s_\beta}$ & $\frac{s_\alpha}{s_\beta}$
& $\cot\beta$ & $-\cot\beta$ & $-\cot\beta$ \\
Type-II
& $\frac{s_\alpha}{s_\beta}$ & $-\frac{s_\alpha}{c_\beta}$ &
$-\frac{s_\alpha}{c_\beta}$
& {$\frac{s_\alpha}{s_\beta}$} & $\frac{c_\alpha}{c_\beta}$ & $\frac{c_\alpha}{c_\beta}$
& $\cot\beta$ & $\tan\beta$ & $\tan\beta$ \\
Flipped (Y)
& $\frac{c_\alpha}{s_\beta}$ & $-\frac{s_\alpha}{c_\beta}$ & $\frac{c_\alpha}{s_\beta}$
& {$\frac{s_\alpha}{s_\beta}$} & $\frac{s_\alpha}{s_\beta}$ & $\frac{s_\alpha}{s_\beta}$
& $\cot\beta$ & $\tan\beta$ & $-\cot\beta$ \\
Lepton Specific (X)
& $\frac{c_\alpha}{s_\beta}$ & $\frac{c_\alpha}{s_\beta}$ & $-\frac{s_\alpha}{c_\beta}$
& {$\frac{s_\alpha}{s_\beta}$} & $\frac{c_\alpha}{c_\beta}$ & $\frac{s_\alpha}{s_\beta}$
& $\cot\beta$ & $-\cot\beta$ & $\tan\beta$ \\
\bottomrule
\end{tabular}
\end{center}
\end{table}

The Yukawa couplings for the the lightest Higgs boson in the C2HDM can be obtained from the respective
2HDM $\Ph$ couplings via
\begin{equation}
y_{\Phone}^{{\rm C2HDM}}= \cos \alpha_2 \, y_{\Ph}^{{\rm 2HDM}} \pm \,  \sin \alpha_2 \, F(\tan \beta) \, \gamma_5
\end{equation}
where the sign of the pseudoscalar term and the function $F(\tan \beta)$ (that can be either $\tan \beta$
or $\cot \beta$) depend on the model type considered (see details in~\cite{Fontes:2015mea}).

We will now briefly discuss some properties of the models.
When $\cos (\beta -\alpha) = 0$ there is no mixing
of the two-Higgs-doublet fields in the Higgs basis which is the basis
where the Goldstone bosons are all in one of the doublets. For this reason it is called
the alignment limit and h has SM-like couplings to the fermions and to the
gauge bosons~\footnote{If $\sin (\beta -\alpha) = 0$ it is the heaviest Higgs boson H
that acquires SM-like couplings.}. The decoupling limit~\cite{Gunion:2002zf} is defined by the previous
condition plus having all scalar masses well above the electroweak scale.
Finally the wrong sign Yukawa coupling~\cite{Ginzburg:2001ss,Ferreira:2014naa,Ferreira:2014dya}
 regime is defined as the region of 2HDM parameter space in which at least one
of the couplings of \Ph\ to down-type and up-type fermion pairs is opposite in sign to the corresponding coupling of \Ph\
to V V. In the convention followed here ($|\alpha|\leq \pi/2$), the limit appears only in Type II and Flipped
versions of the 2HDM
for the down-type quarks and is obtained by setting $\sin (\beta + \alpha) = 1$. Other wrong sign limits
are also possible as discussed in~\cite{Ferreira:2014naa,Ferreira:2014dya}.

There are two particular cases of the 2HDM that will also be discussed. These are the
{\it Inert Model} and a { \it Fermiophobic Model} which are particular cases
of a type I 2HDM. The potential of the Inert Model~\cite{Deshpande:1977rw,
Cao:2007rm,Barbieri:2006dq,LopezHonorez:2006gr,Ilnicka:2015jba,Arhrib:2013ela,
Dolle:2009ft,Dolle:2009fn,Gustafsson:2012aj,Ginzburg:2010wa}
is obtained from the one in equation~(\ref{pot})
by setting $m_{12}^2=0$ and by taking the VEV of the second doublet to be zero. In the Yukawa
Lagrangian, only the doublet that generates a VEV, $\Phi_1$, couples to all fermions.
$\Phi_2$ is usually called the dark doublet, since it contains the dark matter candidate.
The scalars from the dark doublet couple to gauge bosons via the covariant derivative term,
but because the $\mathbb{Z}_2$ symmetry is exact, they always come in pairs in their interactions,
that is the $\mathbb{Z}_2$ charge is conserved.
The Fermiophobic Model~\cite{Barger:1992ty, Diaz:1994pk, Brucher:1999tx}
is obtained form the usual type I 2HDM. By setting $\cos \alpha = 0$, the lightest
CP-even scalar decouples from all fermions and becomes fermiophobic. The discovery of the Higgs boson
and subsequent measurement of Higgs rates has excluded the possibility of a 125~\UGeV\ fermiophobic
Higgs boson by more than 3$\sigma$~\cite{Santos:2015gaa} at the end of Run~1. There is
however the possibility of having a heavier fermiophobic CP-even Higgs boson, which implies
setting $\sin \alpha = 0$. In this scenario the lightest CP-even scalar is the SM-like
Higgs boson while the heavy CP-even scalar decouples from the fermions.

There are other versions of 2HDMs where the $\mathbb{Z}_2$ symmetry is not imposed to the scalar doublets.
These models either have tree-level FCNCs or are not stable under the renormalization group~\cite{Ferreira:2010xe}
(see also~\cite{Branco:2011iw}).
However, there are ways to force the neutral flavour changing currents to be small
at tree-level. One such example is a class of models known as BGL~\cite{Branco:1996bq} models,
where the tree-level FCNC couplings are proportional to the elements of
the CKM matrix. Therefore, the off-diagonal CKM elements naturally suppress the neutral scalars
flavour changing couplings.
BGL models were first proposed for the quark sector in \cite{Branco:1996bq} and later generalized in~\cite{Botella:2009pq}. The extension to the leptonic sector was presented in \cite{Botella:2011ne}.

There are several versions of the BGL models and the first
was obtained by imposing the following symmetry on the
quark and scalar sector of the Lagrangian,
\begin{equation}
\PQQ_{Lj}^{0}\rightarrow \exp {(i\tau )}\ \PQQ_{Lj}^{0}\ ,\qquad
\PQu_{Rj}^{0}\rightarrow \exp {(i2\tau )}\PQu_{Rj}^{0}\ ,\qquad \Phi
_{2}\rightarrow \exp {(i\tau )}\Phi_{2}\ ,  \label{S symmetry up quarks}
\end{equation}
where $\tau \neq 0, \pi$, with all other quark fields transforming
trivially under the symmetry. The index $j$ can be fixed as either 1,
2 or 3. Alternatively the symmetry may be chosen as
\begin{equation}
\PQQ_{Lj}^{0}\rightarrow \exp {(i\tau )}\ \PQQ_{Lj}^{0}\ ,\qquad
\PQd_{Rj}^{0}\rightarrow \exp {(i2\tau )}\PQd_{Rj}^{0}\ ,\quad \Phi
_{2}\rightarrow \exp {(- i \tau)}\Phi_{2}\ .  \label{S symmetry down quarks}
\end{equation}
The symmetry given by Eq.~(\ref{S symmetry up quarks}) leads to Higgs FCNC 
in the down sector and the models are known as BGL up-type models, whereas
the symmetry specified by  Eq.~(\ref{S symmetry down  quarks}) leads to Higgs FCNC in the up
sector and to the BGL down-type models. These two alternative choices of
symmetries combined with the three possible ways of fixing  the index $j$ give rise to six
different realizations of the model with the flavour structure, in the
quark sector, controlled by the CKM matrix.
In the leptonic sector, with Dirac type neutrinos, there is a perfect
analogy with the quark sector.
There are thirty six different  models corresponding to the
combinations of the six possible different implementations in each
sector.

This symmetry constrains the Higgs potential in such a way that it does
not violate CP neither explicitly nor spontaneously. As a result, once in 
the Higgs basis, the fields $\PSHp$ and $\PA$ are already physical and only $\PHz$ and $R$  (real
and imaginary neutral fields respectively) are
allowed to mix such that they are combined into the two CP-even states $\Ph$ and $\PH$ via the rotation
\begin{equation}
\Ph = s_{\beta - \alpha} \PH + c_{\beta - \alpha} R, \qquad \PH = c_{\beta - \alpha} \PHz -  s_{\beta - \alpha} R
\end{equation}
In the case of $s_{\beta - \alpha} \neq 1$ all neutral scalars mediate flavour changing neutral currents.

For a general 2HDM the Yukawa
couplings in terms of quark mass eigenstates for $\PSHp$, $\PHz$,  $R$ and $\PA$ are given by
\begin{eqnarray}
{\mathcal L}_Y (\mbox{quark, Higgs} ) & = &
 - \frac{\sqrt{2} \PSHpm}{v} \PAQu \left(
V N_{\PQd} \gamma_R - N^\dagger_{\PQu} \ V \gamma_L \right) \PQd +  \mbox{h.c.} -
\frac{H^0}{v} \left(  \bar{u} D_u u + \bar{d} D_d \ d \right)  \nonumber \\
&& -  \frac{R}{v} \left[\PAQu (N_{\PQu} \gamma_R + N^\dagger_{\PQu} \gamma_L) \PQu +
\PAQd ( N_{\PQd} \gamma_R + N^\dagger_{\PQd} \gamma_L)\ \PQd \right]   \nonumber \\
&& +  i  \frac{\PA}{v}  \left[\PAQu(N_{\PQu} \gamma_R - N^\dagger_{\PQu} \gamma_L)\PQu-
\PAQd (N_{\PQd} \gamma_R - N^\dagger_{\PQd} \gamma_L)\ \PQd \right] \, .
\end{eqnarray}

In BGL up-type models  the matrices $N_{\PQd}$ and $ N_{\PQu}$ have the following simple form
\begin{equation}
\left(  N_{\PQd}^{(\PQu_j)} \right)_{rs} = \left[ t_\beta \delta_{rs} - ( t_\beta + t_\beta^{-1}) V^\ast_{jr}
 V_{js}  \right]  (D_{\PQd})_{ss}
 \label{24}
\end{equation}
where no sum in $j$ is implied and $t_\beta$ stands for $\tan \beta$. The upper index  $(\PQu_j)$ indicates that
the BGL is of the up-type form, with index $j$, thus leading to FCNC in the down-sector.
Note that all FCNC are proportional to the factor $ ( t_\beta + t_\beta^{-1}) $
and to a product of elements of one single row of the $V_{\rm CKM}$. The corresponding $N_{\PQu}$
matrix is given by
\begin{equation}
\left(  N_{\PQu}^{(\PQu_j)} \right)_{rs} = \left[ t_\beta  - (t_\beta + t_\beta^{-1}) \delta_{rj}  \right]  (D_{\PQu})_{ss} \delta_{rs} \, .
\label{25}
\end{equation}
$N_{\PQu}$ is a diagonal matrix, the $t_\beta$ dependence is not the same for all diagonal entries.
It is  proportional to $(- t_\beta^{-1})$  for the $(jj)$ element and to $t_\beta$ for all other elements.
The index $j$ fixes the row of  the  $V_{\rm CKM}$ matrix which suppresses the flavour
changing neutral currents. Since for each up-type  BGL model a single row of  $V_{\rm CKM}$
participates in these couplings, one may choose a phase convention
where all elements of $N_{\PQu}$ and $N_{\PQd}$ are real.

In BGL down-type models the matrices $N_{\PQd}$ and $N_{\PQu}$  exchange r\^ ole,
\begin{equation}
\left(  N_{\PQu}^{(\PQd_j)} \right)_{rs}  =  \left[ t_\beta \delta_{rs} - (t_\beta + t_\beta^{-1}) V_{rj}
 V^\ast_{sj}  \right]  (D_{\PQu})_{ss}
\label{ees}
\end{equation}
\begin{equation}
\left(  N_{\PQd}^{(\PQd_j)} \right)_{rs} =   \left[ t_\beta  - (t_\beta + t_\beta^{-1}) \delta_{rj}  \right]  (D_{\PQd})_{ss} \delta_{rs}
\label{EQ:Nd:dj:01}
\end{equation}
In down-type models the flavour changing neutral currents are suppressed
by the columns of the $V_{\rm CKM}$ matrix. Since the flavour structure of
these scalar currents results from a symmetry of the Lagrangian, each BGL
model is natural and stable under the renormalization group.
Furthermore,  the resulting number of free parameters is very small and therefore BGL models
are very predictive.

\subsection{Input parameters}

\begin{itemize}
\item 2HDM

The set of input parameters are the four masses $\Mh $, $\MH $, $\MA $, $\MSHpm$,
$\tan \beta$, $\alpha$ and $m_{12}^2$.

\item Inert Model

The input parameters are the four masses $\Mh $, $\MH $, $\MA $, $\MSHpm$, $\lambda_2$
and $\lambda_{345}=\lambda_3 + \lambda_4 + \lambda_5$. The $\Ph$ boson is taken to be
the SM-like Higgs boson discovered at the LHC.

\item Fermiophobic Model

The input parameters are the four masses $\Mh $, $\MH $, $\MA $, $\MSHpm$, $\tan \beta$
and $m_{12}^2$. The fermiophobic limit is attained with the condition $\sin \alpha =0$.
The fermiophobic scalar is denoted by $\PH$
and the charged Higgs boson mass is chosen such that $\MA  = \MSHpm$. The free parameters
are then $\MH $, $\Delta M = \MH -\MA $ and $\tan \beta$.

\item C2HDM

The input parameters are three masses $\Mone$, $\Mtwo$, $\MSHpm$, $\tan \beta$, $Re[m_{12}^2]$
and the three rotation angles of the neutral sector $\alpha_1$, $\alpha_2$ $\alpha_3$. With this
choice the heavier state has a mass defined by
\begin{equation}
\Mthree ^2 = \frac{\Mone ^2\, R_{13} (R_{12} \tan{\beta} - R_{11})
+ \Mtwo ^2\ R_{23} (R_{22} \tan{\beta} - R_{21})}{R_{33} (R_{31} - R_{32} \tan{\beta})}.
\label{m3_derived}
\end{equation}
and the parameter space will be restricted to values which obey $\Mthree  > \Mtwo $.

\item BGL

The set of input parameters are the four masses $\Mh $, $\MH $, $\MA $, $\MSHpm$,
$\tan \beta$ and $\cos (\beta - \alpha)$.

\end{itemize}

\section{\label{sec:tools} Tools and constraints}
At present there are only two public codes that allows one to perform scans in the 2HDM parameter space:
2HDMC~\cite{Eriksson:2009ws, Eriksson:2010zzb} - 2-Higgs Doublet Model Calculator
and ScannerS~\cite{Coimbra:2013qq, Ferreira:2014dya}. These codes can only be used for CP-conserving
versions of the 2HDM. Several codes have been compared~\cite{Harlander:2013qxa}
regarding the neutral scalar production in gluon fusion and the decay of all scalars
in the 2HDM.

\subsection{Vacuum stability and theoretical constraints}

The CP-conserving minimum of any 2HDM is stable against tunnelling to both CP-violating and
charge breaking minima~\cite{Ferreira:2004yd,Barroso:2005sm}. However, the potential
invariant under $\mathbb{Z}_2$ and softly broken by the term $m_{12}^2$, can still have two
simultaneous CP-conserving minima~\cite{Barroso:2007rr, Ivanov:2006yq, Ivanov:2007de, Maniatis:2006fs}.
In this case, choosing the global minimum is easily achieved by imposing a simple condition on the
parameters of the potential~\cite{Barroso:2013awa, Ivanov:2008cxa}.

The two codes 2HDMC and ScannerS have in-built the following theoretical constraints:
tree-level vacuum stability~\cite{Deshpande:1977rw, Ivanov:2006yq}, that is, the potential is
forced to be bounded from below at tree-level, and perturbative unitarity~\cite{Kanemura:1993hm, Akeroyd:2000wc}
is enforced  to the quartic couplings of the potential as proposed by
Lee, Quigg and Thacker~\cite{Lee:1977eg} for the SM. In
2HDMC a further perturbativity constraint is imposed on the quartic couplings of the mass eigenstates fields
($|\lambda_{ijkl}| \leq 4\pi$) where the indices run over all allowed quartic vertices. In ScannerS
one can opt for having a global minimum at tree-level by imposing the condition proposed
in~\cite{Barroso:2013awa, Ivanov:2008cxa} to the parameters of the potential.
For all other benchmarks proposed, the points produced comply to the respective
vacuum stability and perturbative unitarity (or/and perturbativity) constraints.

\subsection{Experimental constraints}

The parameter space of the models that is used to produce benchmarks satisfies
a number of experimental constraints described in this section. If for a given
benchmark one or more constraints are not taken into account it will be explicitly stated.
This section is intended to give an overview of the most relevant experimental
constraints for all benchmark points presented. It should be noted that, for instance,
the constraints for the charged Higgs boson mass from LEP
do not apply to the Inert Model as it does not couple to fermions.
Conversely, dark matter constraints only apply to the Inert Model.

The main experimental constraints are

\begin{itemize}

\item
The parameter space complies with S and T parameters~\cite{Peskin:1991sw, Froggatt:1991qw, Grimus:2008nb,
Haber:2010bw, Baak:2011ze}
as derived from electroweak precision observables~\cite{ALEPH:2010aa}.

\item
Collider constraints are taken into account. In most cases and in particular in 2HDMC and 
ScannerS these bounds are considered via an interface with the {\tt HiggsBounds}~\cite{Bechtle:2013wla} and
{\tt HiggsSignals} \cite{Bechtle:2013xfa} codes that include all LHC data published so far
by the ATLAS and CMS collaboration and that are updated regularly. The codes also include
the LEP bounds and in particular the only bound on the charged Higgs boson mass that only assumes
that $BR(\PSHpm \to \PQc\PQs) + BR(\PSHpm \to \PGt \PGn) + BR(\PSHpm \to \PA\PWpm) = 1$~\cite{Abbiendi:2013hk}.
This leads us to roughly consider for the four type of 2HDMs (and C2HDMs) $\MSHpm \geq 90 \UGeV$.

\item
There are indirect constraints on the $(\MSHpm, \tan \beta)$ plane
stemming mainly from loop processes involving charged Higgs bosons but also
from direct measurements at the LHC. Indirect bounds come mainly from
$B$-physics observables~\cite{WahabElKaffas:2007xd, Aoki:2009ha, Su:2009fz,
Mahmoudi:2009zx, Deschamps:2009rh} and $\Rbb = \Gamma (\PZ \to \PQb \PAQb)/\Gamma (\PZ \to \text{hadrons})/$
~\cite{Denner:1991ie, Haber:1999zh, Freitas:2012sy}.
When considering all experimental results a rough bound of $\tan \beta \geq 1$ is obtained.
Of particular importance is the bound coming from the charged Higgs loop contribution
to $\PQb \to \PQs \PGg$ that applies only to Type II and Flipped (Y) and is at
present $\MSHpm \geq 480$~\UGeV~\cite{Misiak:2015xwa}. Direct constraints
in the $(\MSHpm, \tan \beta)$ plane were obtained during the LHC Run~1
in the process $\Pp \Pp \to \PQt \PAQt (\PSHp  \PWm \PQb \PAQb)$~\cite{Chatrchyan:2012vca,
Aad:2012tj}.

Except for $\PB \to \PGtpm \PGnGt$ and $\PAB\to \PD^{(*)} \PGtm \PAGnGt$
the 2HDM contributions to $\PB$-physics arise via one-loop radiative corrections.
For such observables, cancellations could occur in the loops if other sources
of new physics are considered. This would lead to
cancellations in the loop from other sources of new physics and a consequent
relaxation of the respective bounds.

\item
The CP-violating phase of the C2HDM is also constrained by electric dipole moment (EDM)
measurements. The most stringent bound~\cite{Inoue:2014nva} comes from the ACME~\cite{Baron:2013eja}
results on the ThO molecule EDM. Points are rejected if the calculated EDMs, with
Barr-Zee diagrams with fermions in the loop, are not of the
order of magnitude of ACME result. The ACME limit can only be evaded by either going
to the limit of the CP-conserving 2HDM or in scenarios where
cancellations~\cite{Jung:2013hka, Shu:2013uua} among the neutral scalars occur.
It should be noted that ref.~\cite{Jung:2013hka}
argues that the extraction of the electron EDM from the data is filled with uncertainties
and an order of magnitude larger EDM than that claimed by ACME should be allowed
for.

\item
Regarding astrophysical constraints on the dark matter part of the Inert Doublet Model,
relic density limits are required to respect the results obtained by the Planck
experiment~\cite{Ade:2015xua}:
\begin{equation}
\Omega_c\,h^2\,=\,0.1197\,\pm\,0.0022,
\end{equation}
\label{eq:planck}
where an upper limit of
\begin{equation}
\Omega_c\,h^2\,\leq\, 0.1241,
\label{eq:planck_up}
\end{equation}
is imposed, which corresponds to not over closing the universe at $\sim\,95\,\%$ confidence level.  Direct detection limits are taken into account by using an approximation function (cf. \cite{Ilnicka:2015jba} for details) to compare with limits from the LUX experiment  \cite{Akerib:2013tjd}, where multicomponent dark matter scenarios are discarded. The dark matter properties are calculated by processing through MicrOmegas (version 4.2.3) \cite{Belanger:2013oya}, where \cite{Belanger:2008sj} provides explicit details about the calculation of nucleon-dark matter cross sections.

\end{itemize}

\subsection{Calculation of cross sections and decay widths}

A detailed study of the various tools and their
accuracy has been performed in~\cite{Harlander:2013qxa}. For the CP-conserving case
it includes both production via gluon and $\PQb \PAQb$ fusion and decay of the 2HDM
scalars. For the production the tools compared were
SUSHI~\cite{Harlander:2012pb} and HIGLU~\cite{Spira:1995mt}.
Regarding the scalar decays the two codes compared were
2HDMC~\cite{Eriksson:2009ws, Eriksson:2010zzb} and
HDECAY~\cite{Djouadi:1997yw, Djouadi:2006bz} with a very good agreement.
The remaining cross sections were just rescaled from the SM ones.

In the particular cases of the Inert and Fermiophobic models, the cross
sections cannot be obtained from the SM ones and were calculated at LO
with MadGraph 5~\cite{Alwall:2011uj}. The new decays in the BGL were
calculated at LO.

\section{Benchmark points}

In this section benchmark points for the 2HDM are presented. For each point, plane or
scenario physical motivation is given together with their main features, cross sections
and branching ratios. In same cases only scenarios were proposed.

A quick summary of the points before their detailed presentation is shown in the following:
\begin{longtable}{cl}

BP$1_{\text{A}}$ &  2HDM,  non-alignment, $\Ph$ approximately SM-like;  plane: $1 < \tan\beta < 60$, $150 < \MH  < 600$~\UGeV \\
               & signatures: type-I $\PH\to \Ph\Ph$, $\PQt \PAQt$, $\PW\PW$, $\PZ\PZ$; type-II: $\PH\to \PQb\PAQb$, $\PQt \PAQt$ \\
BP$1_{\text{B}}$ & 2HDM, $\PH$ is SM-like; line of $65 < \Mh  < 120$~\UGeV;  signatures: $\Ph\to \PQb \PAQb$, $\PGt\PGt$ \\
BP$1_{\text{C}}$ & 2HDM, $\Mh  =\MA  = 125$~\UGeV; line $1 < \tan\beta < 10$ for $\MH  = \MSHpm = 300$~\UGeV \\
               & large deviations from the SM value of $\PGt\PGt $ for the 125~\UGeV\ Higgs boson\\
BP$1_{\text{D}}$ & 2HDM short cascades, $\Mh  = 125$~\UGeV, exact alignment; line $250 < \MH  < 500$~\UGeV \\
               & signatures: $\PH \to \PZ\PA$, $\PWpm \PSHpm$, $\PSHpm \PSHmp$, $\PA\PA$ \\
BP$1_{\text{E}}$ & 2HDM long cascades,  $\Mh  = 125$~\UGeV, exact alignment; line $200 < \MH  < 300$~\UGeV \\
               & signatures:  $\PSHpm\to \PWpm \PA \to \PWpm \PZ\PH$, $\PA \to \PWpm\PSHmp \to \PWpm \PWmp \PH$\\
BP$1_{\text{F}}$ & 2HDM, $\Mh  = 125$~\UGeV, opposite sign coupling to down-type fermions; \\
               & plane: $5 < \tan\beta < 50$, $150 < \MH  < 600$~\UGeV;  signatures: $\PH\to \PW\PW$, $\PZ\PZ$ \\
BP$1_{\text{G}}$ & 2HDM, $\Mh  = 125$~\UGeV, with an MSSM-like Higgs sector; \\
               & signatures $\PH \to \Ph\Ph$ and $\PH,\PA \to \PQt \PAQt$ or $\PH,\PA \to \PGt\PGt$ for large $\tan\beta$ \\[5pt]

BP2 & Exotic decays in the 2HDM alignment limit; planes provided \\[5pt]
BP3 & 2HDM, $\Mh  = 125$~\UGeV, large mass splitting between $\PH$ and $\PA$ for electroweak baryogenesis\\
   & signatures: $\PA\to \PZ\PH $ with $\PH\to \PW\PW$, $\PQb \PAQb$, $\PQt \PAQt$ \\[5pt]
BP4 & 2HDM, $\Mh  = 125$~\UGeV, $\PA$ is very light ($\MA  \lesssim \MZ $);  signatures: $\Ph\to \PZ\PA$ with $\PA\to \PQb \PAQb$, $\PGt\PGt$, $\PGm\PGm$ \\[5pt]
BP5 & Inert 2HDM, $\Mh  = 125$~\UGeV\ and SM-like, $\PH$ dark matter candidate, $\MH  > 45$~\UGeV \\
   & signatures: $\PA\to \PZ\PH$, $\PSHpm\to \PWpm \PH$; $\PH$ will give missing transverse energy in the event \\[5pt]
BP6 & Fermiophobic 2HDM, $\Mh  = 125$~\UGeV\ and SM-like, $\PH$ fermiophobic; various planes are suggested\\
   & signatures: $\Pp\Pp\to \PH\PA $, $\PH\PSHpm$, and has large branching ratios to $\PH \to \PW\PW$, $\PZ\PZ$ \\[5pt]
BP7 & C2HDM,  $\Mh  = 125$~\UGeV\ and SM-like, CP-violation detected by the simultaneous existence of \\
   & 3 decay channels signatures: e.g. $\Phthree \to \Phtwo \PZ$, $\Phtwo\to \Phone \PZ$, $\Phthree \to \Phone \PZ$ simultaneously \\[5pt]
BP8 & BGL models: Higgs bosons with flavour changing decays; plane: $\tan\beta$ versus $\cos(\beta - \alpha)$ \\
   & signatures: $\PQt\to \Ph\PQc$, $\Ph\to \PGt\PGt$, $\PQb \PAQb$, $\PQb\PQs$, $\PGm\PGt$ 
\end{longtable}
A detailed description of the benchmark points follows in the next sections.


\subsection{Benchmark points \texorpdfstring{$BP1$}{BP1}}

\begin{center}
\small
\begin{longtable}{c|l@{}m{0pt}@{}}
\toprule

\multicolumn{2}{c}  {\bf Scenario $BP1_A$:  2HDM non-alignment}\cr
\midrule
\multicolumn{2}{c}  {Howard E. Haber and Oscar St{\aa}l}\\
\multicolumn{2}{c}  {from Eur.~Phys.~J.~C {\bf 75} (2015) 491, arXiv:1507.04281 [hep-ph]} \cr
\midrule
Main Features & Departures from the alignment limit with $\Ph$ SM-like. \cr
& Scan over values of $1<\tan\beta<50$.
\cr
\midrule
\multicolumn{2}{c}  {Type-I Scenarios $BP1_{A1.1}$  
{\small with $\cos(\beta-\alpha)=0.1$, and $BP1_{A1.2}$ with $\cos(\beta-\alpha)=0.1\times (150~\UGeV/\MH )^2$} \hspace*{\fill}}\cr
\midrule
Spectrum & $\Mh =125$~\UGeV, $150\UGeV <\MH <600$~\UGeV, $\MH <\MSHpm =\MA $  \cr
\midrule
\multicolumn{2}{c}  {Production cross sections and branching fractions}\cr
\midrule
$\Ph$ & SM-like cross-sections and decays with small deviations from SM predictions.   \cr
\midrule
$\PH$ &   $\Pg\Pg$ fusion cross sections are largest at low $\tan\beta$ values.  At low $\tan\beta$, $\PH\to \Ph\Ph$   \cr
&  dominant for $2\Mh <\MH <2m_{\PQt}$ and $\PH\to \PQt\PAQt$ dominates for $\MH >2m_{\PQt}$.\cr
& At higher $\tan\beta$ values, $\PQt\PAQt$ is suppressed and $\Ph\Ph$ is dominant for $\MH >2\Mh $  \cr
& and the $\PW\PW$ and $\PZ\PZ$ modes are more relevant.\cr
\midrule
$\PA$ and $\PSHpm$ & Assumed to be very heavy with little impact on LHC Higgs phenomenology \cr
\midrule
\multicolumn{2}{c}  {Type-II Scenarios $BP1_{A2.1}$ 
{\small with $\cos(\beta-\alpha)=0.1$, and $BP1_{A2.2}$ with $\cos(\beta-\alpha)=0.1\times (150 \UGeV/\MH )^2$} \hspace*{\fill}}\cr
\midrule
Spectrum & $\Mh =125$~GeV, $150 \UGeV <\MH <600$~\UGeV, $\MH < \MSHpm =\MA $  \cr
\midrule
\multicolumn{2}{c}  {Production cross sections and branching fractions}\cr
\midrule
$\Ph$ & SM-like cross-sections and decays with small deviations from SM predictions.   \cr
\midrule
$\PH$ &   $\Pg\Pg$ fusion cross sections are largest either at low $\tan\beta$ values or at high $\tan\beta$ values \cr
& for moderate values of $\MH $.  $\PQb\PAQb$ fusion cross sections relevant at larger $\tan\beta$.  At low \cr
& $\tan\beta$, $\PH\to \PQb\PAQb$ dominant
for $\MH <2m_{\PQt}$ and $\PH\to \PQt\PAQt$ dominates for $\MH >2m_{\PQt}$.  At \cr
& higher $\tan\beta$, $\PQt\PAQt$ is somewhat suppressed above threshold
and $\PQb\PAQb$ is dominant.   \cr
\midrule
$\PA$ and $\PSHpm$ & Assumed to be very heavy with little impact on LHC Higgs phenomenology \cr
\bottomrule
\toprule

\multicolumn{2}{c}  {\bf Scenario $BP1_B$:  2HDM with a SM-like $\PH$}\cr
\midrule
\multicolumn{2}{c}  {Howard E. Haber and Oscar St{\aa}l}\\
\multicolumn{2}{c}  {from Eur.~Phys.~J.~C {\bf 75} (2015) 491, arXiv:1507.04281 [hep-ph]} \cr
\midrule
Main Features & The heavier of the two CP-even Higgs bosons, $\PH$ is SM-like. \cr
& Scan over values of $\frac{1}{2}\MH <\Mh <\MH $, with $\tan\beta=1.5$.
\cr
\midrule
\multicolumn{2}{c}  {Type-I Scenarios $BP1_{B1.1}$ with $\cos(\beta-\alpha)=1.0$, and $BP1_{B1.2}$ with $\cos(\beta-\alpha)=0.9$ \hspace*{\fill}}\cr
\midrule
Spectrum & $\MH =125$~\UGeV, $65 \UGeV <\Mh <120$~\UGeV, $\MH <\MSHpm =\MA $  \cr
\midrule
\multicolumn{2}{c}  {Production cross sections and branching fractions}\cr
\midrule
$\PH$ & SM-like cross-sections and decays with small deviations from SM predictions.   \cr
\midrule
$\Ph$ & ${\rm BR}(\Ph\to \PQb\PAQb)\sim 75{-}80\%$ and ${\rm BR}(\Ph\to \PGt\PGt)\sim 8\%$.  \cr
\midrule
$\PA$ and $\PSHpm$ & very heavy and difficult to detect at the LHC \cr
\midrule
\multicolumn{2}{c}  {Type-II Scenario  $BP1_{B2}$ with $\cos(\beta-\alpha)=1.0$ \hspace*{\fill}}\cr
\midrule
Spectrum & $\Mh =125$~\UGeV, $150 \UGeV <\MH <600$~\UGeV, $\MH <\MSHpm=\MA $  \cr
\midrule
\multicolumn{2}{c}  {Production cross sections and branching fractions}\cr
\midrule
$\PH$ & SM cross-sections and decays with no deviations from SM predictions.   \cr
\midrule
$\Ph$ & ${\rm BR}(\Ph\to \PQb \PAQb )\sim 75{-}80\%$ and ${\rm BR}(\Ph\to \PGt\PGt)\sim 8\%$.  \cr
\midrule
$\PA$ and $\PSHpm$ & Assumed to be very heavy with little impact on LHC Higgs phenomenology \cr
\bottomrule
\toprule

\multicolumn{2}{c}  {\bf Scenario $BP1_C$:  2HDM with degenerate $h$ and $A$}\cr
\midrule
\multicolumn{2}{c}  {Howard E. Haber and Oscar St{\aa}l}\\
\multicolumn{2}{c}  {from Eur.~Phys.~J.~C {\bf 75} (2015) 491, arXiv:1507.04281 [hep-ph]} \cr
\midrule
Main Features &  The CP-even $\Ph$ and the CP-odd $\PA$ are roughly mass-degenerate and both  \cr
& contribute to the observed scalar at 125~\UGeV.     $\PH$ and $\PSHpm$ are assumed heavy.  \cr
& Exact alignment limit of $\cos(\beta-\alpha)=0$ taken.  Scan over values of $1<\tan\beta<10$.
\cr
\midrule
\multicolumn{2}{c}  {Type-I Scenario  $BP1_{C1}$ \hspace*{\fill}}\cr
\midrule
Spectrum & $\Mh =\MA =125$~\UGeV,  $\MH =\MSHpm=300$~\UGeV  \cr
\midrule
\multicolumn{2}{c}  {Production cross sections and branching fractions}\cr
\midrule
$\Ph,\ \PA$ & Combined $\sigma\times{\rm BR}$ of $\Ph,\ \PA \to \PGt\PGt$ deviates significantly from SM for $\tan\beta<2$.   \cr
& At large $\tan\beta$, the combined signal approaches that of the SM. \cr
\midrule
$\PH$ and $\PSHpm$ & Assumed to be very heavy with little impact on LHC Higgs phenomenology \cr
\midrule
\multicolumn{2}{c}  {Type-II Scenario  $BP1_{C2}$ \hspace*{\fill}}\cr
\midrule
Spectrum &  $\MH =\MA =125$~\UGeV,  $\MH =\MSHpm=300$~\UGeV \cr
\midrule
\multicolumn{2}{c}  {Production cross sections and branching fractions}\cr
\midrule
$\Ph,\ \PA$ & Combined $\sigma\times{\rm BR}$ of $\Ph,\ \PA \to \PGt\PGt$ deviates significantly from SM over the entire    \cr
& $\tan\beta$ range (with a minimum enhancement of about 1.5 at at $\tan\beta=3.5$. \cr
\midrule
$\PH$ and $\PSHpm$ & Assumed to be very heavy with little impact on LHC Higgs phenomenology \cr
\bottomrule
\toprule

\multicolumn{2}{c}  {\bf Scenario $BP1_D$: - short cascade of Higgs-to-Higgs boson decay }\cr
\midrule
\multicolumn{2}{c}  {Howard E. Haber and Oscar St{\aa}l}\\
\multicolumn{2}{c}  {from Eur.~Phys.~J.~C {\bf 75} (2015) 491, arXiv:1507.04281 [hep-ph]} \cr
\midrule
Main Features &  $\Ph$ has SM couplings in the exact alignment limit, $\cos(\beta-\alpha)=0$. The Higgs boson mass\cr
&  spectrum is chosen such that either one or both decay modes, $\PH\to \PZ\PA$ and/or \cr
& $\PH\to \PWpm \PSHmp$ are open, resulting in a short cascade of Higgs-to-Higgs boson decay.\cr
& Scan over $250 \UGeV <\MH <500$~\UGeV.
\cr
\midrule
\multicolumn{2}{c}  {Type-I Scenario $BP1_{D1.1}$ or Type-II Scenario $BP1_{D1.2}$ with $\tan\beta=2$ \hspace*{\fill}}\cr
\midrule
Spectrum & $\Mh =125$~\UGeV\ and $\MH ^2=\MSHpm= \MA^2+v^2$, where $v\equiv 246$~\UGeV.  \cr
\midrule
\multicolumn{2}{c}  {Production cross sections and branching fractions}\cr
\midrule
$\Mh$ & Cross sections and branching ratios are SM-like.  Small corrections to $\Mh\to\PGg\PGg$ \cr
& due to $\PSHpm$-loop.  \cr
\midrule
$\PH$ & $\PH \to \PZ\PA$ kinematically allowed and dominant below $\PQt\PAQt$ threshold.   \cr
& $\PH\to \PA\PA$ can be significant if
kinematically allowed. \cr
\midrule
$\PSHpm$ & $\PSHpm\to \PWpm \PA$ is kinematically allowed, but may have a small BR. \cr
\midrule
\multicolumn{2}{c}  {Type-I Scenario $BP1_{D2.1}$  or Type-II Scenario $BP1_{D2.2}$ with $\tan\beta=2$\hspace*{\fill}}\cr
\midrule
Spectrum & $\Mh =125$~\UGeV\ and $\MH ^2=\MA^2=\MSHpm^2+v^2$, where $v\equiv 246$~\UGeV.  \cr
\midrule
\multicolumn{2}{c}  {Production cross sections and branching fractions}\cr
\midrule
$\Ph$ & Cross sections and branching ratios are SM-like.  Small corrections to $\Ph\to\PGg\PGg$ \cr
& due to $\PSHpm$-loop.  \cr
\midrule
$\PH$ & $\PH \to \PWpm \PSHmp$ kinematically allowed and dominant below $\PQt\PAQt$ threshold.   \cr
& $\PH\to \PSHp  \PSHm$ can be significant if
kinematically allowed. \cr
\midrule
$\PA$ & $\PA\to \PWpm \PSHmp$  is kinematically allowed, but may have a small BR. \cr
\midrule
\multicolumn{2}{c}  {Type-I Scenario $BP1_{D3.1}$ or Type-II Scenario $BP1_{D3.2}$ with $\tan\beta=2$ \hspace*{\fill}}\cr
\midrule
Spectrum & $\Mh =125$~\UGeV\ and $\MA^2=\MSHpm ^2=\MH ^2-v^2$, where $v\equiv 246$~\UGeV.  \cr
\midrule
\multicolumn{2}{c}  {Production cross sections and branching fractions}\cr
\midrule
$\Ph$ & Cross sections and branching ratios are SM-like.  Small corrections to $\Ph\to\PGg\PGg$ \cr
& due to $\PSHpm$-loop.  \cr
\midrule
$\PH$ & $\PH \to \PZ\PA$ and $H\to \PWpm \PSHmp$ are both kinematically allowed and significant below $\PQt\PAQt$    \cr
& threshold.  $\PH\to \PA\PA$ and/or $\PH\to \PSHp \PSHm$ can be significant if
kinematically allowed. \cr
\bottomrule
\toprule

\multicolumn{2}{c}  {\bf Scenario $BP1_E$:  Long cascade of Higgs-to-Higgs boson decays }\cr
\midrule
\multicolumn{2}{c}  {Howard E. Haber and Oscar St{\aa}l}\\
\multicolumn{2}{c}  {from Eur.~Phys.~J.~C {\bf 75} (2015) 491, arXiv:1507.04281 [hep-ph]} \cr
\midrule
Main Features &  $\Ph$ has SM couplings in the exact alignment limit, $\cos(\beta-\alpha)=0$. The Higgs boson mass\cr
&  spectrum is chosen such that a long cascade of Higgs-to-Higgs boson decays,  \cr
& $\PSHpm\to \PWpm \PA\to \PWpm \PZ\PH$  or $\PA\to \PWpm \PSHmp\to \PWpm \PWmp \PH$, are kinematically allowed.  \cr
& The former will compete with $\PSHpm\to \PWpm \PH$ and the latter will compete with  \cr
& $\PA\to \PZ\PH$. Scan over $200 \UGeV <\MH <300$~\UGeV.
\cr
\midrule
\multicolumn{2}{c}  {Type-I Scenario $BP1_{E1}$  with $\tan\beta=2$ \hspace*{\fill}}\cr
\midrule
Spectrum & $\Mh =125$~\UGeV\ and $\MH <\MA <\MSHpm$  \cr
\midrule
\multicolumn{2}{c}  {Production cross sections and branching fractions}\cr
\midrule
$\Ph$ & Cross sections and branching ratios are SM-like.  \cr
\midrule
$\PH$ & Decays dominantly into the heaviest kinematically accessible fermion pairs. \cr
\midrule
$\PSHpm$ & BR($\PSHpm\to \PWpm \PH)\sim 0.74$--$0.79$. BR for the long cascade of order a few per cent. \cr
\midrule
$\PA$ & BR($\PA\to \PZ \PH)\sim 0.39$--$0.62$.  \cr
\midrule
\multicolumn{2}{c}  {Type-I Scenario $BP1_{E2}$ with $\tan\beta=2$\hspace*{\fill}}\cr
\midrule
Spectrum & $\Mh =125$~\UGeV\ and $\MH <\MSHpm<\MA $.  \cr
\midrule
\multicolumn{2}{c}  {Production cross sections and branching fractions}\cr
\midrule
$\Ph$ & Cross sections and branching ratios are SM-like.    \cr
\midrule
$\PH$ & Decays dominantly into the heaviest kinematically accessible fermion pairs.   \cr
\midrule
$\PA$ & BR($\PA\to \PZ \PH)\sim 0.50$--$0.56$. BR for the long cascade of order a few per cent. \cr
 \midrule
$\PSHpm$ & BR($\PSHpm\to \PWpm \PH)\sim 0.03$--0.27.\cr
\bottomrule
\toprule

\multicolumn{2}{c}  {\bf Scenario $BP1_{F1}$: Type-II 2HDM with the opposite sign Higgs boson coupling to down-type fermions.}\cr
\midrule
\multicolumn{2}{c}  {Howard E. Haber and Oscar St{\aa}l}\\
\multicolumn{2}{c}  {from Eur.~Phys.~J.~C {\bf 75} (2015) 491, arXiv:1507.04281 [hep-ph]} \cr
\midrule
Main Features &  The couplings of $\Ph$ to vector boson pairs ($\PWpm \PWm$ and $\PZ\PZ $) and to up-type fermions \cr
& are SM-like (close to the alignment limit).  The magnitude of the coupling of $\Ph$ to \cr
& down-type fermions
is also SM-like but the sign of this coupling is flipped.  \cr
& The latter occurs in the Type-II 2HDM when $\cos(\beta-\alpha)=\sin 2\beta$.  \cr
& Scan over $150 \UGeV <\MH <600$~\UGeV\ and $5<\tan\beta <50$.
\cr
\midrule
\multicolumn{2}{c}  {Type-II Scenario $BP1_{F2}$ \hspace*{\fill}}\cr
\midrule
Spectrum & $\Mh =125$~\UGeV, $150 \UGeV <\MH <600$~\UGeV, $\MH <\MSHpm=\MA $  \cr
\midrule
\multicolumn{2}{c}  {Production cross sections and branching fractions}\cr
\midrule
$\Ph$ & Cross sections and branching ratios are mostly SM-like.  The $\Pg\Pg\Ph$ and $\PGg\PGg \Mh$ effective  \cr
& couplings exhibit small modifications due to the change of sign of the $\PQb$-quark loop \cr
& contribution.\cr
\midrule
$\PH$ & BR into vector boson pairs can be sizeable over a large part of the parameter space,   \cr
& since the departure from the alignment limit of $\cos(\beta-\alpha)=0$ can be significant.
\cr
\midrule
$\PA$ and $\PSHpm$ & Assumed to be very heavy with little impact on LHC Higgs phenomenology \cr
\bottomrule
\toprule

\multicolumn{2}{c}  {\bf Scenario $BP1_{G1}$: 2HDM with an MSSM-like Higgs sector} \cr
\midrule
\multicolumn{2}{c}  {Howard E. Haber and Oscar St{\aa}l}\\
\multicolumn{2}{c}  {from Eur.~Phys.~J.~C {\bf 75} (2015) 491, arXiv:1507.04281 [hep-ph]} \cr
\midrule
Main Features &  The Higgs scalar potential of the MSSM is employed, with one modification.  \cr
& The coefficient $\lambda_2$ of the scalar potential is modified,
$\lambda_2=\frac{1}{4}(g^2+g^{\prime\,2})+\delta\lambda_2$, \cr
& to accommodate the observed Higgs boson at 125~\UGeV\ in mass.  \cr
& Scan over $90 \UGeV <\MA <1000$~\UGeV\
and $1<\tan\beta<60$.
\cr
\midrule
\multicolumn{2}{c}  {Type-II Scenario $BP1_{G2}$ \hspace*{\fill}}\cr
\midrule
Spectrum & $\Mh =125$~\UGeV; tree-level MSSM relations $\MH ^2=\MA ^2+\MZ ^2-\Mh ^2$ and  \cr
& $\MSHpm^2=\MA ^2+\MW^2$ hold approximately. \cr
\midrule
\multicolumn{2}{c}  {Production cross sections and branching fractions}\cr
\midrule
$\Ph$ & Cross sections and branching ratios are very SM-like in the decoupling regime, \cr
& where $\PA$ is heavy.  Present precision Higgs data requires that $\MA >360$~\UGeV, \cr
& almost independently of the value of $\tan\beta$.\cr
\midrule
$\PH$, $\PA$ and $\PSHpm$ & These states are heavy in the decoupling regime, presenting a challenge for \cr
& LHC Higgs phenomenology.  For low values of $\tan\beta$, $\PH\to \Ph\Ph$ may provide a viable \cr
& signal.  Otherwise, one may have to rely on $\PH$, $\PA\to \PQt\PAQt$.  At very large $\tan\beta$,\cr
& $\PH$, $\PA\to\PGt\PGt$ and $\PSHpm\to\PGtpm \PGn$ provide the most useful final states for discovery.\cr
\bottomrule

\end{longtable}
\end{center}



\subsection{Benchmark points \texorpdfstring{$BP2$}{BP2}}

\begin{center}
\small
\begin{longtable}{c|l@{}m{0pt}@{}}
\toprule
\multicolumn{2}{c}  {\bf BP2} & \\
\midrule
\multicolumn{2}{c}  {F. Kling, J. M. No and S. Su \cite{Kling:2016opi}}\cr
\midrule
\multicolumn{2}{c}  {$BP2_1$: Exotic Decays in the Alignment Limit \hspace*{\fill}}\cr
\midrule
Fixed Param.
  & $\Mh =125$~\UGeV, $c_{\beta-\alpha}=0$, $\lambda_6=\lambda_7=0$ \cr
\midrule
\multicolumn{2}{c}  {Benchmark Planes}\cr
\midrule
$BP2_{1A}$
 & Mass Hierarchy: $\MH =\MSHpm<\MA $, \cr
 & $m_{12}^2=\MH ^2 s_\beta c_\beta \text{ for }  t_\beta=1.5,7,30 $ and $m_{12}^2=0  \text{ for }  t_\beta=1.5$. \cr
 & Open Decays: $\PA \to \PH\PZ/\PSHpm \PW$. \cr
 & See \refF{BPX:1A}  for $\tan\beta=1.5$ and $m_{12}^2=\MH ^2 s_\beta c_\beta$.\cr
\midrule
Example BP:
 & $\MH =\PSHpm=200$~\UGeV, $\MA =500$~\UGeV, $t_\beta=1.5$, $m_{12}= 135$~\UGeV, \cr
 & $\sigma(\Pg\Pg\to \PA)=3.7$ pb, BR$(\PA\to \PH\PZ)=28\%$, BR$(\PA\to \PSHpm \PW)=58\%$, \cr
 & BR$(\Ph\to \PQb\PQb)=83\%$, BR$(\PH\to \PGt\PGt)=9\%$, BR$(\PSHpm\to \PQt\PQb)=99\%$. \cr
\midrule
$BP2_{1B}$
 & Mass Hierarchy: $\MH <\MA =\MSHpm$, \cr
 & $m_{12}^2=\MH ^2 s_\beta c_\beta \text{ for }  t_\beta=1.5,7,30 $ and $m_{12}^2=0  \text{ for }  t_\beta=1.5$. \cr
 & Open Decays: $\PA \to \PH\PZ, \PSHpm \to \PH\PW$. \cr
 & See \refF{BPX:1B} for $\tan\beta=1.5$ and $m_{12}^2=\MH ^2 s_\beta c_\beta$.\cr
\midrule
Example BP:
 & $\MH =200$~\UGeV, $\MA =\MSHpm=500$~\UGeV, $t_\beta=1.5$, $m_{12}= 135$~\UGeV, \cr
 & $\sigma(\Pg\Pg\to \PA)=3.7$ pb, $\sigma(\Pg\Pg\to \PSHpm tb)=0.2$ pb, BR$(\PA\to \PH\PZ)=66\%$, \cr
 & BR$( \PSHpm \to \PH\PW)=70\%$, BR$(\PH\to \PQb\PQb)=83\%$, BR$(\PH\to \PGt\PGt)=9\%$. \cr
\midrule
$BP2_{1C}$
 & Mass Hierarchy: $\MA =\MSHpm<\MH $, \cr
 & $m_{12}^2=0  \text{ for }  t_\beta=1.5$. \cr
  & Open Decays: $\PH \to \PA\PZ/\PSHpm \PW/\PA\PA/\PSHp\PSHm$.\cr
\midrule
Example BP:
  & $\MH =400$~\UGeV, $\MA =\MSHpm=225$~\UGeV, $t_\beta=1.5$, $m_{12}= 0$~\UGeV, \cr
 & $\sigma(\Pg\Pg\to \PH)=4.2$ pb, BR$(\PH\to \PA\PZ)=27\%$, BR$(\PH\to \PSHpm \PW)=60\%$, \cr
 & BR$(\PA\to \PQb\PQb)=75\%$, BR$(\PA\to \PGt\PGt)=8\%$, BR$(\PSHpm\to \PQt\PQb)=99\%$. \cr
\midrule
$BP2_{1D}$
 & Mass Hierarchy: $\MA <\MH =\MSHpm$, \cr
 & $m_{12}^2=0  \text{ for }  t_\beta=1.5$. \cr
 & Open Decays: $\PH \to \PA\PZ/\PA\PA, \PSHpm \to \PA\PW$.   \cr
 & See \refF{BPX:1D}  for $\tan\beta=1.5$ and $m_{12}^2=0$.\cr
\midrule
Example BP:
 & $\MH =\MSHpm=400$~\UGeV, $\MA =100$~\UGeV, $t_\beta=1.5$, $m_{12}= 0$~\UGeV, \cr
 & $\sigma(\Pg\Pg\to \PH)=4.2$ pb, $\sigma(\Pg\Pg\to \PSHpm \PQt\PQb)=0.4$ pb, \cr
 & BR$(\PH\to \PA\PA)=28\%$, BR$(\PH\to \PA\PZ)=63\%$, BR$( \PSHpm \to \PA\PW)=70\%$.\cr
\midrule
\multicolumn{2}{c}  {$BP2_2$: Exotic Decays for Non-Alignment \hspace*{\fill}}\cr
\midrule
Fixed Param.
  & $\Mh =125$~\UGeV, $\lambda_6=\lambda_7=0$ \cr
\midrule
\multicolumn{2}{c}  {Benchmark Planes}\cr
\midrule
$BP2_{2A}$
 & Mass Hierarchy: $\Mh <\MH =\MA =\MSHpm$, $c_{\beta-\alpha} \in (-1,1)$,  \cr
 & $m_{12}^2=\MH ^2 s_\beta c_\beta \text{ for }  t_\beta=1.5,7,30 $ and $m_{12}^2=0  \text{ for }  t_\beta=1.5$. \cr
 & Open Decays: $\PH \to \Ph\Ph,\ \PA \to \Ph\PZ,\ \PSHpm  \to \Ph\PW$.  \cr
 &  See \refF{BPX:2A}  for $\tan\beta=7$ and $m_{12}^2=\MH ^2 s_\beta c_\beta$\cr
\midrule
Example BP:
 & $\MH =\MSHpm=\MA =500$~\UGeV, $s_{\beta-\alpha}=0.96$, $t_\beta=7$, $m_{12}= 187$~\UGeV, \cr
 & $\sigma(\Pg\Pg\to \PH)=97$ fb, $\sigma(\Pg\Pg\to \PA)=205$ fb, $\sigma(\Pg\Pg\to \PSHpm \PQt\PQb)=10$ fb, \cr
 & BR$(\PH\to \Ph\Ph)=4\%$, BR$(\PA \to \Ph\PZ)=65\%$, BR$( \PSHpm \to \Ph\PW)=70\%$.\cr
\bottomrule
\end{longtable}
\end{center}

\begin{figure}
 \centering
	\includegraphics[width=0.48\textwidth]{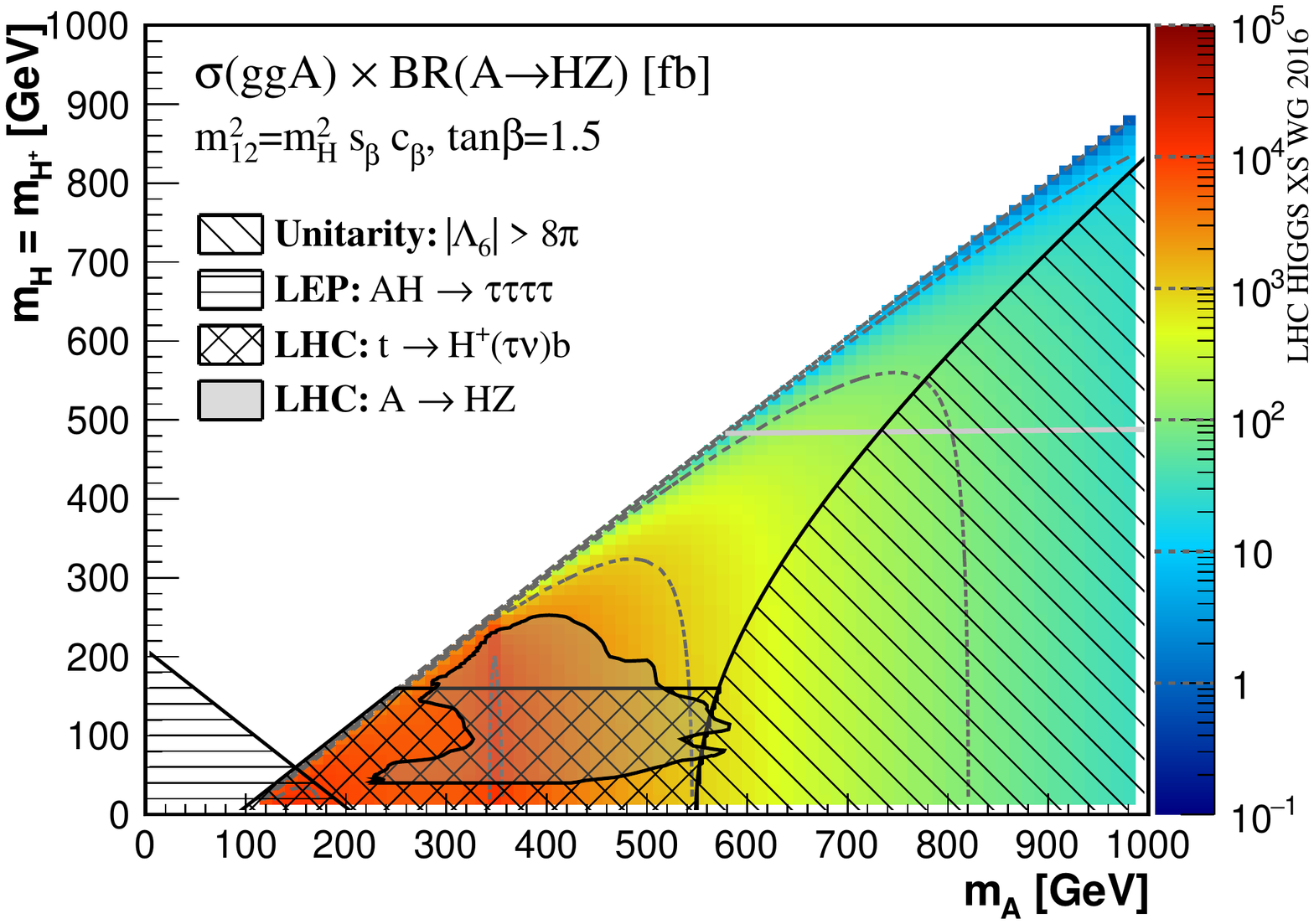}
	\includegraphics[width=0.48\textwidth]{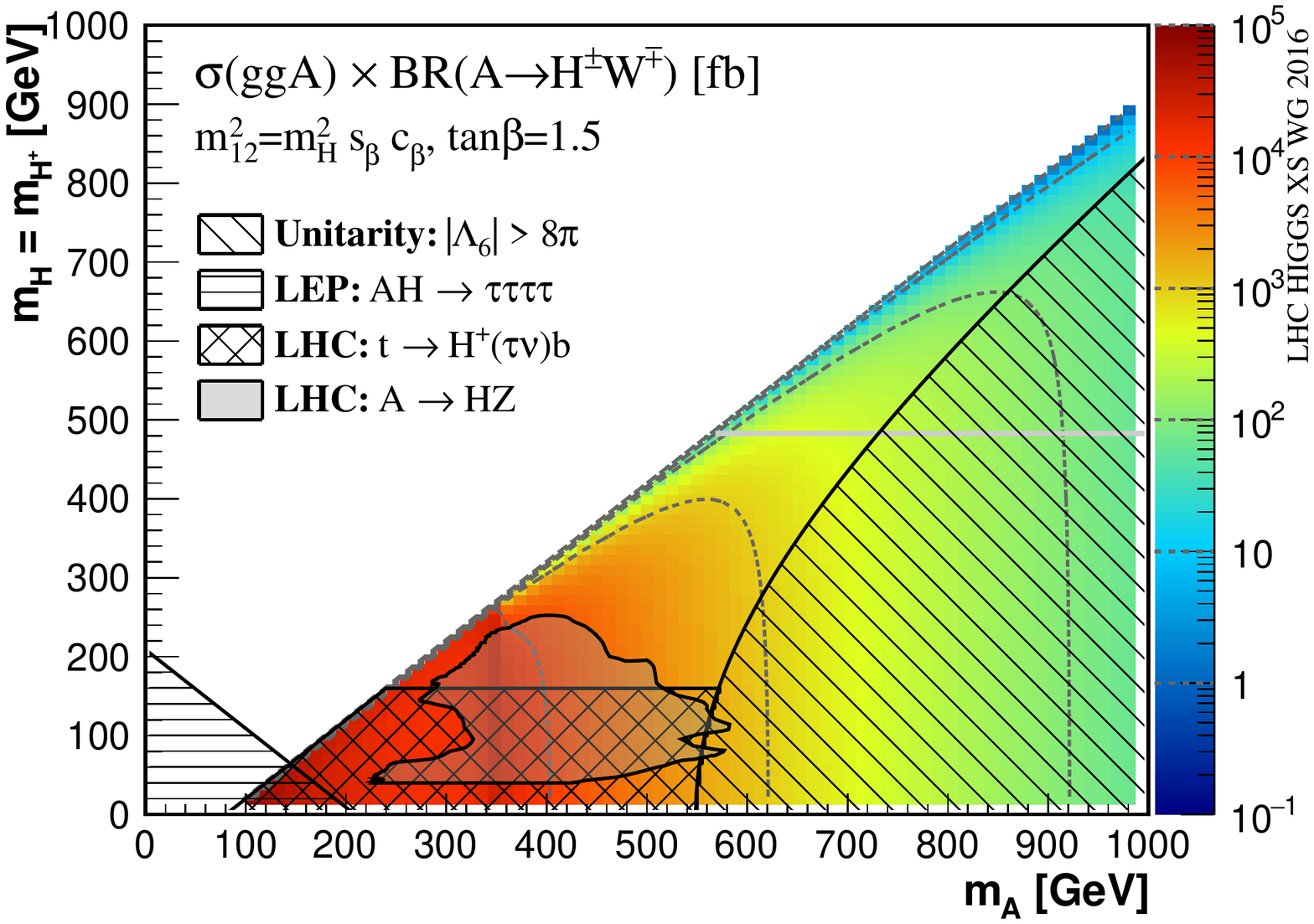}
\caption{$\sigma \times \text{BR}$ for $\Pg\Pg\to \PA \to \PH\PZ$ (left) and $\Pg\Pg\to \PA \to \PSHpm \PWmp$ (right) for the mass hierarchy $\MH =\MSHpm<\MA $. We consider the alignment limit at $\tan\beta=1.5$ and $m_{12}^2=\MH ^2s_\beta s_\beta$. We show contours in $\sigma \times \text{BR}$ (dashed lines) and excluded regions (hatched and shaded area). The solid horizontal or vertical grey line indicates the flavour constraints.}
\label{BPX:1A}
\end{figure}
\begin{figure}
 \centering
	\includegraphics[width=0.48\textwidth]{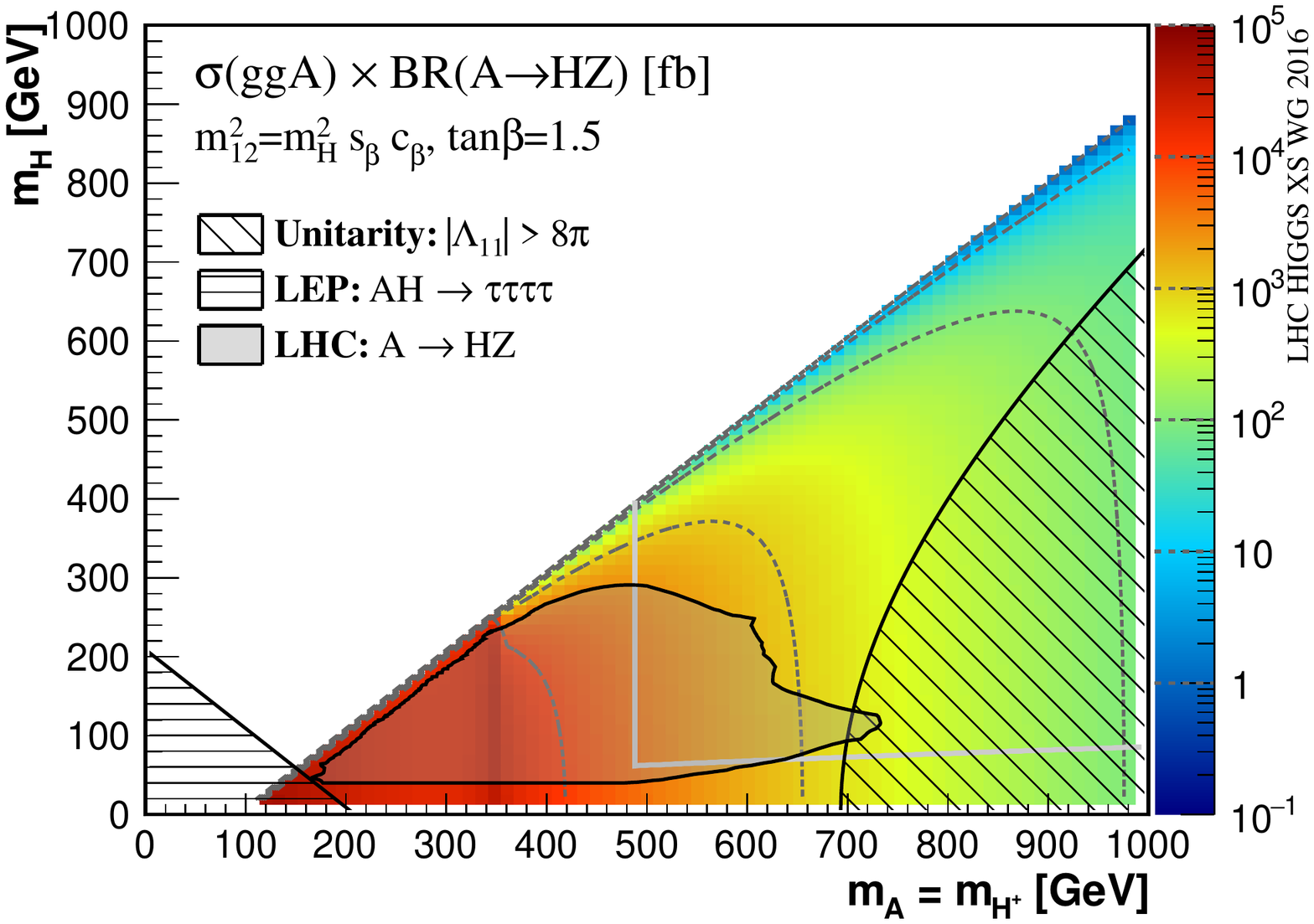}
	\includegraphics[width=0.48\textwidth]{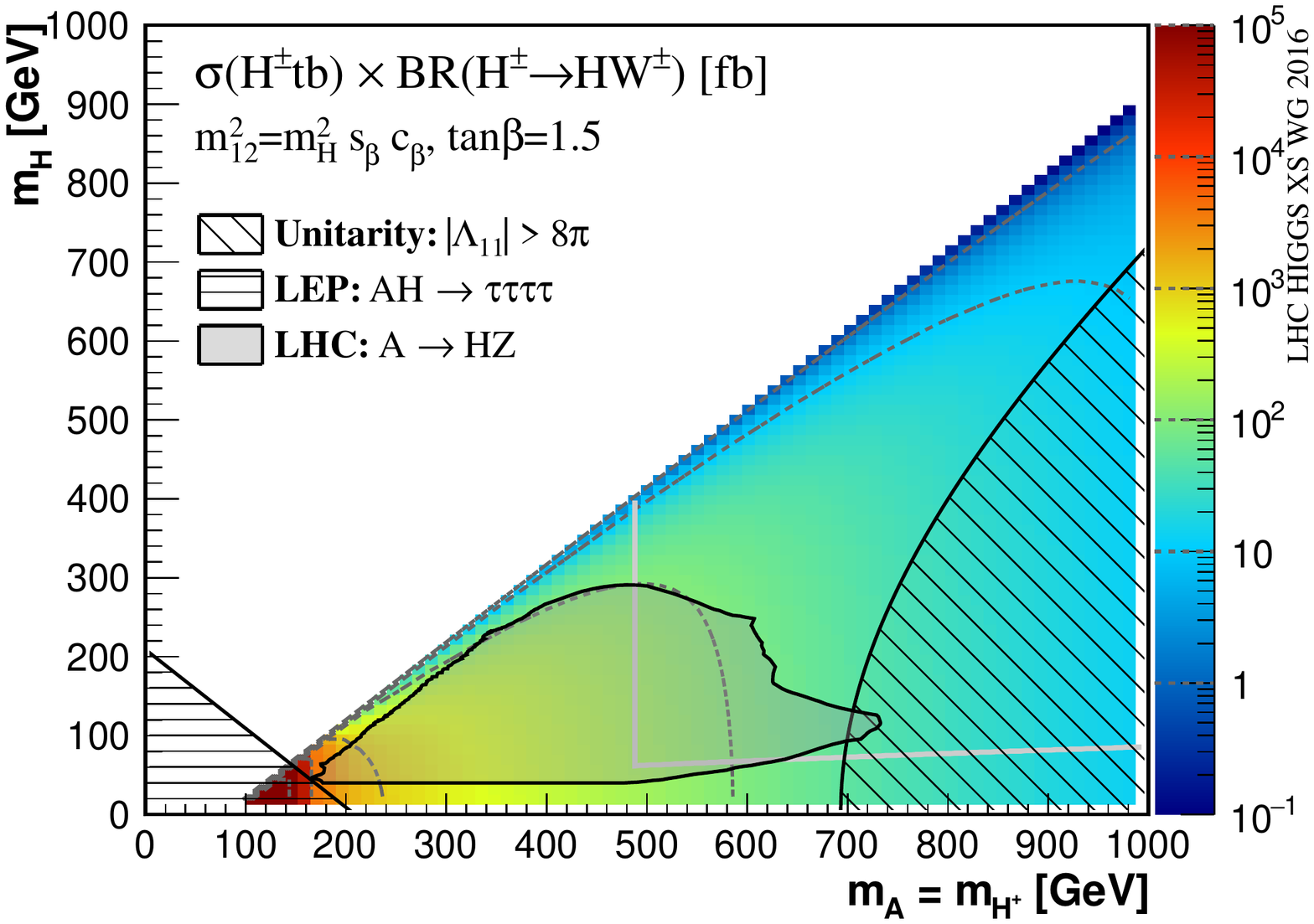}
\caption{$\sigma \times \text{BR}$ for $\Pg\Pg\to \PA \to \PH\PZ$ (left) and $\Pg\Pg\to \PSHp \PQt\PQb \to \PH\PW\PQt\PQb$ (right) for the mass hierarchy $\MH <\MA =\MSHpm$. We consider the alignment limit at $\tan\beta=1.5$ and $m_{12}^2=\MH ^2s_\beta s_\beta$.  Lines and hatched areas are the same as in \refF{BPX:1A}.}
\label{BPX:1B}
\end{figure}

\begin{figure}
\begin{tabular}{cc}
	\includegraphics[width=0.48\textwidth]{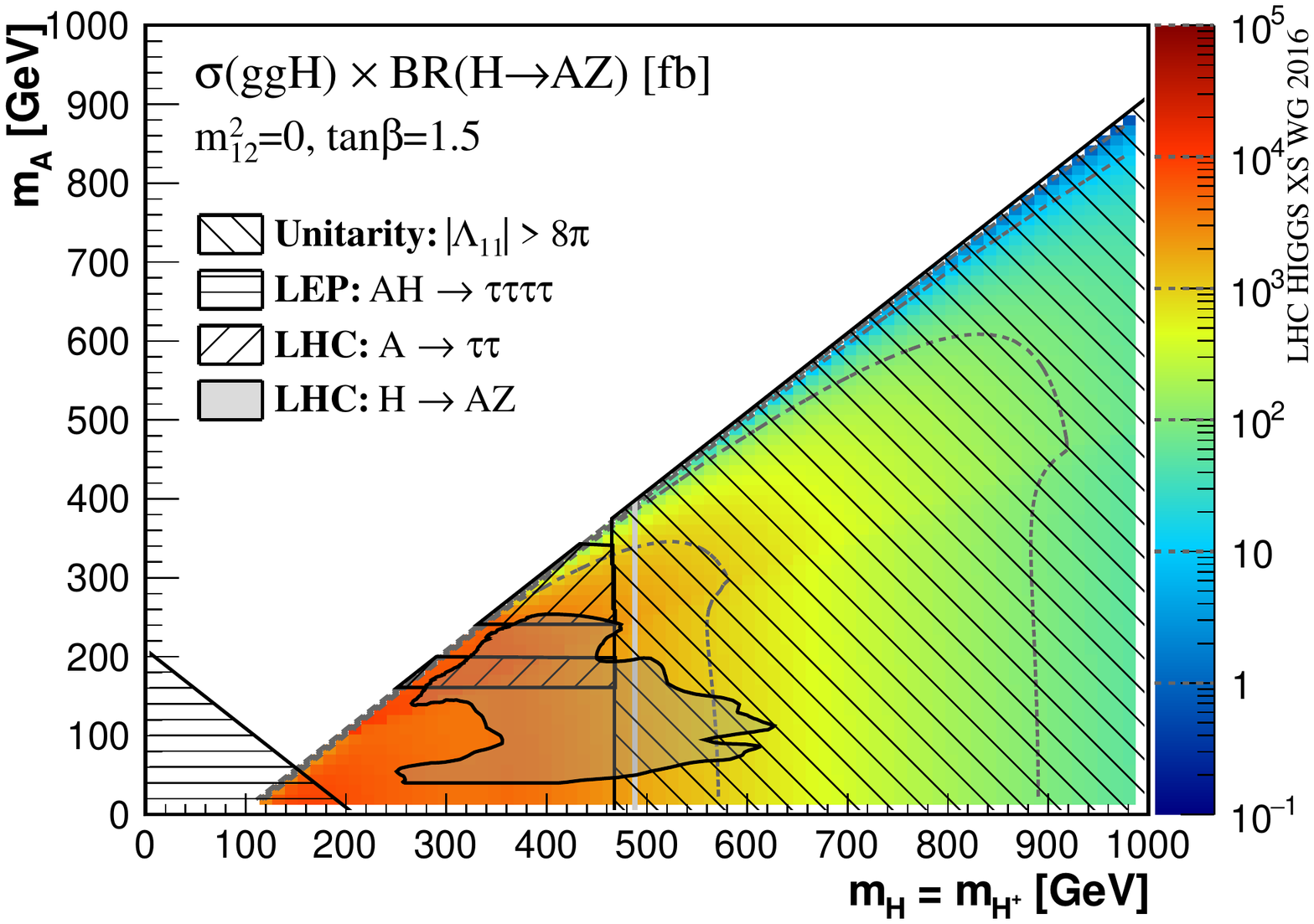} &
	\includegraphics[width=0.48\textwidth]{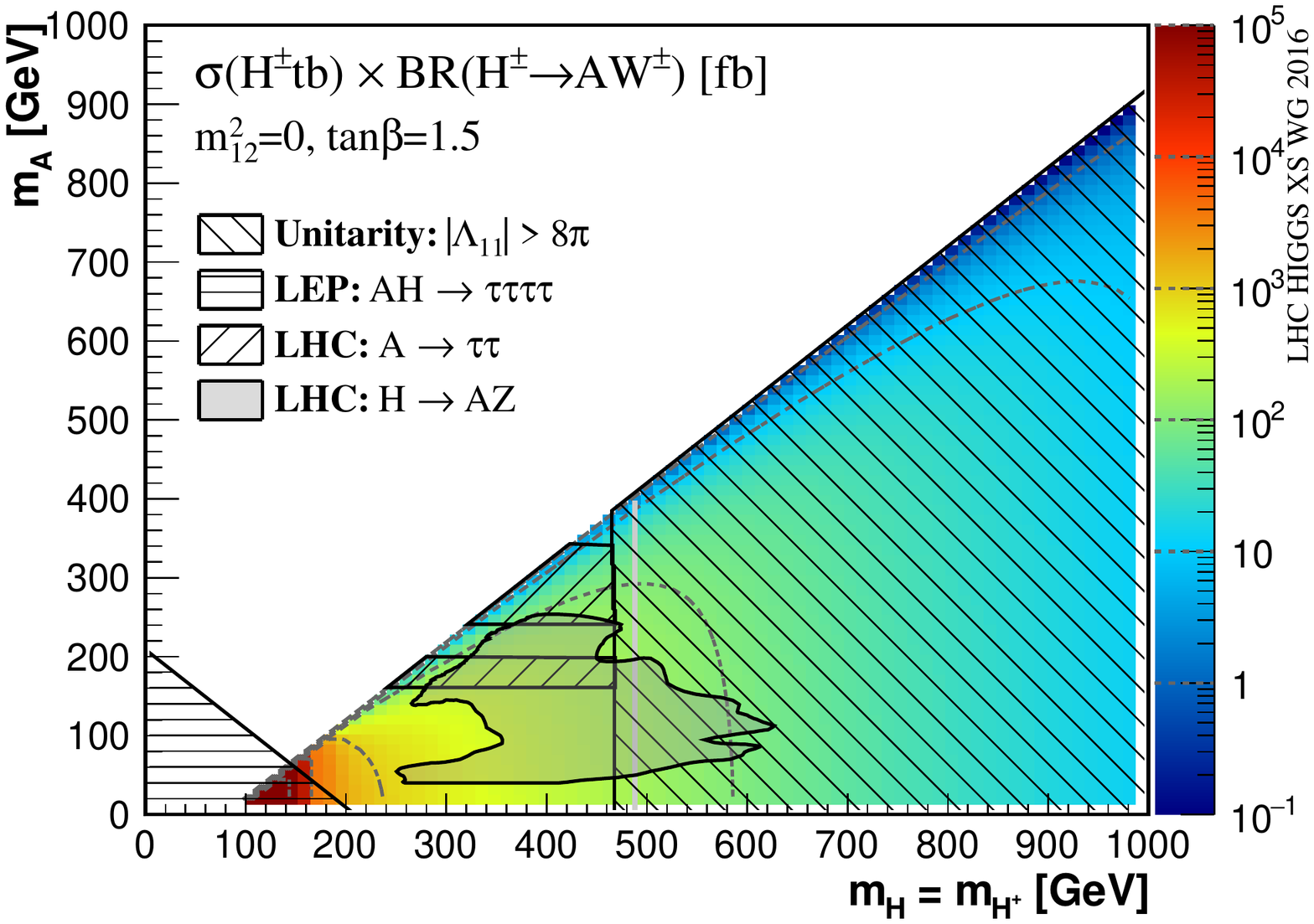} \\[5mm]
	\multicolumn{2}{c}{\includegraphics[width=0.48\textwidth]{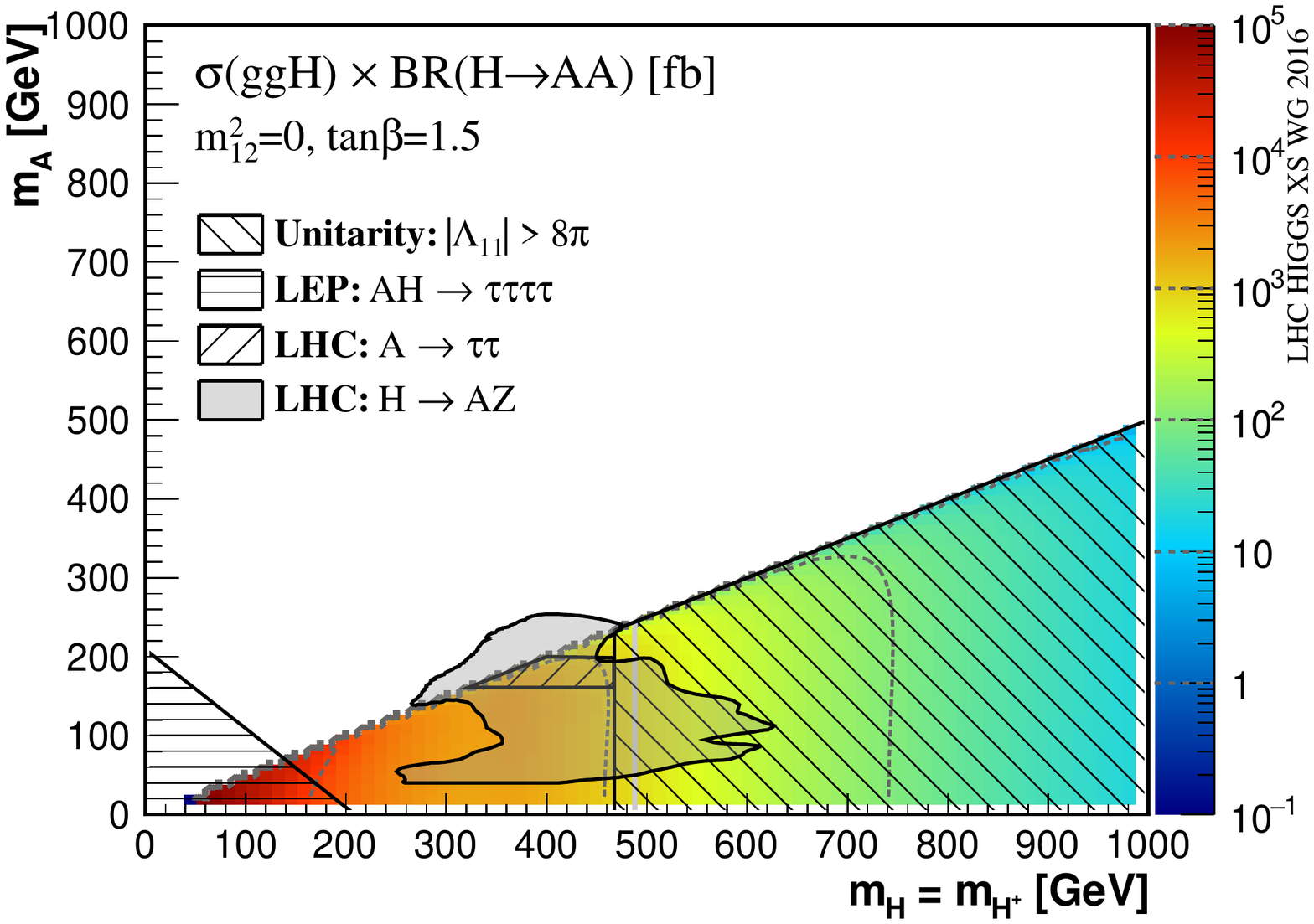}}
\end{tabular}
\caption{$\sigma \times \text{BR}$ for $\Pg\Pg\to \PH \to \PA\PZ$ (top-left), $\Pg\Pg\to \PSHp \PQt\PQb \to \PA\PW\PQt\PQb$ (top-right) and $\Pg\Pg\to \PH \to \PA\PA$ (bottom-left) for the mass hierarchy $\MA <\MH =\MSHpm$. We consider the alignment limit at $\tan\beta=1.5$ and $m_{12}^2=0$. Lines and hatched areas are the same as in \refF{BPX:1A}.}
\label{BPX:1D}
\end{figure}
\begin{figure}
\begin{tabular}{cc}
	\includegraphics[width=0.48\textwidth]{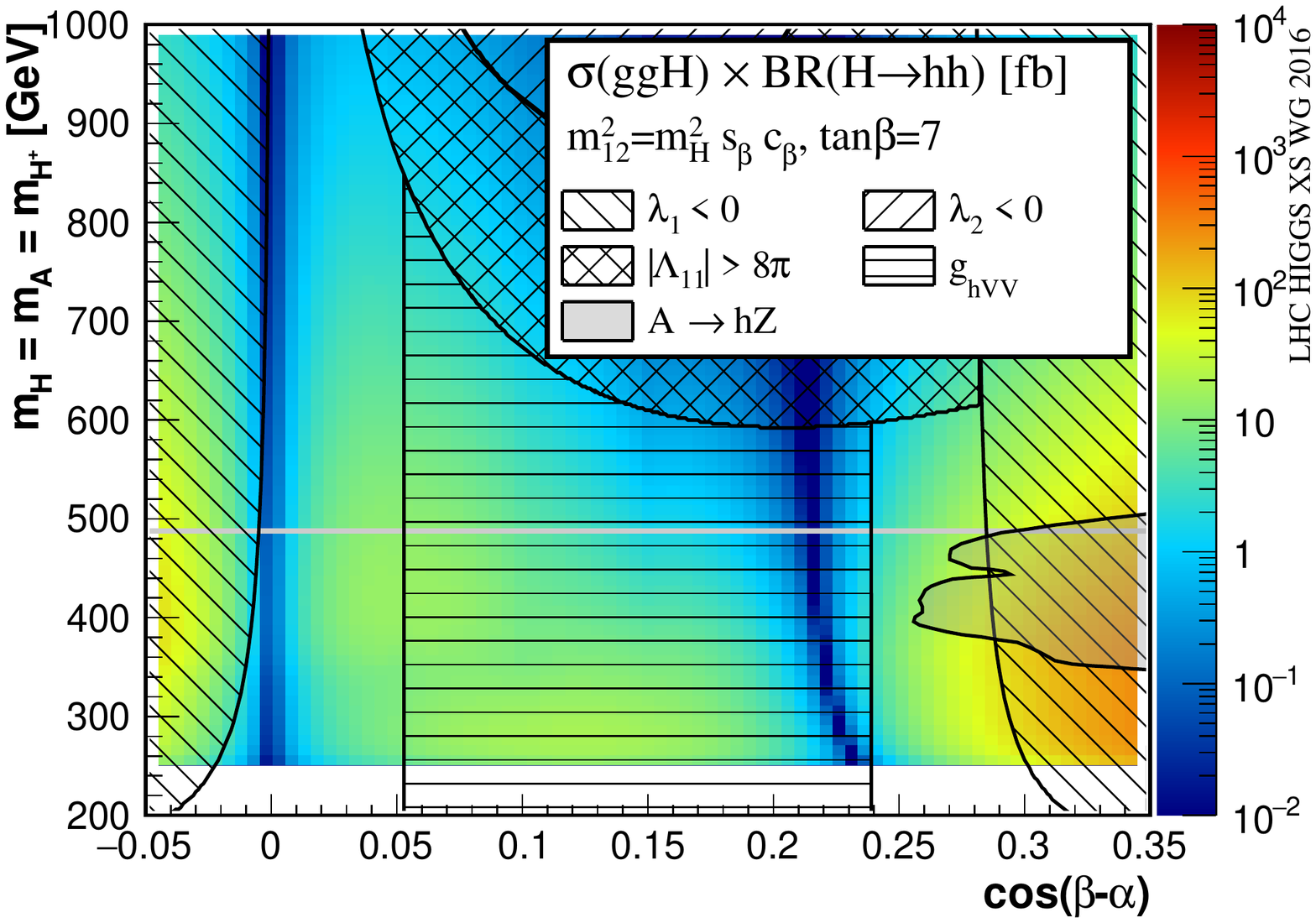} &
	\includegraphics[width=0.48\textwidth]{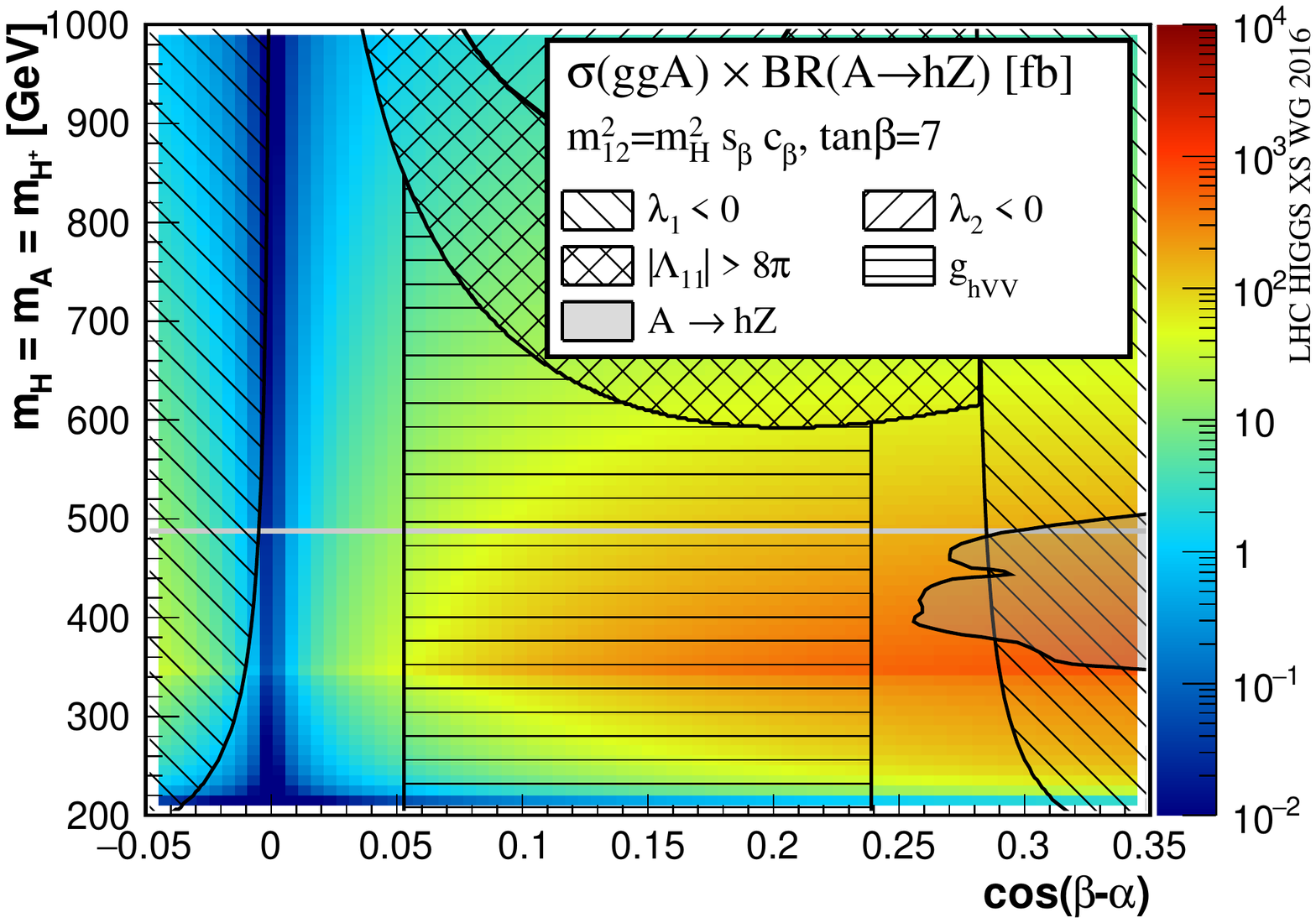} \\[5mm]
	\multicolumn{2}{c}{\includegraphics[width=0.48\textwidth]{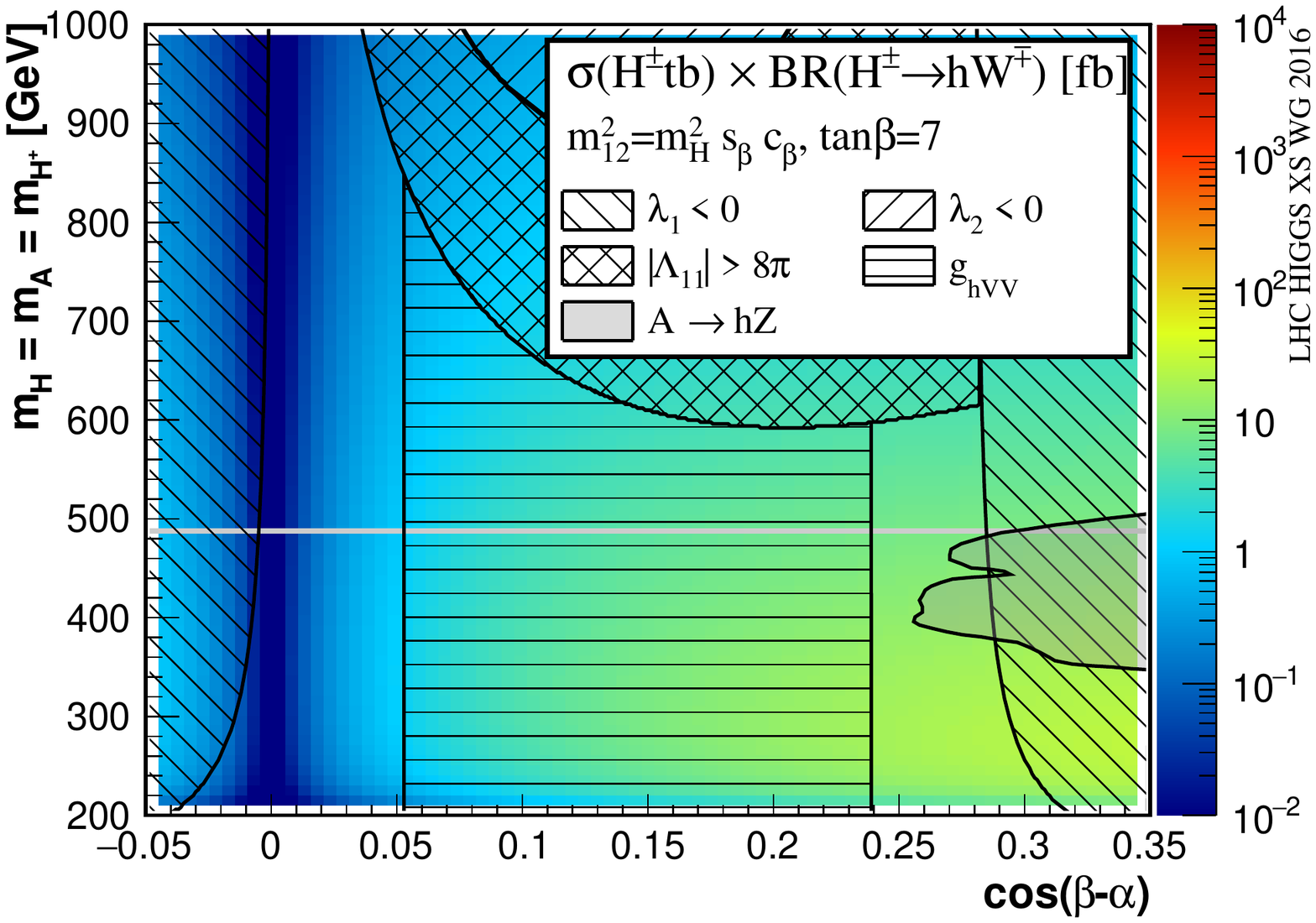}}
\end{tabular}
\caption{$\sigma \times \text{BR}$ for $\Pg\Pg\to \PH \to \Ph\Ph$ (top-left), $\Pg\Pg\to \PA\to \Ph\PZ$ (top-right) and $\Pg\Pg\to \PSHp \PQt\PQb \to \Ph\PW\PQt\PQb$ (bottom-left) for the mass hierarchy $\Mh <\MH =\MA =\MSHpm$. We consider the non-alignment case at $\tan\beta=1.5$ and $m_{12}^2=\MH ^2s_\beta s_\beta$. Lines and hatched areas are the same as in \refF{BPX:1A}.}
\label{BPX:2A}
\end{figure}



\subsection{Benchmark points \texorpdfstring{$BP3$}{BP3}}\label{tab:wg3_extscalars_BP3}
\begin{center}
\small
\begin{longtable}{c|l@{}m{0pt}@{}}
\toprule
\multicolumn{2}{c}  {\bf $BP3$: EW Cosmology Benchmarks for $\PA \to \PH\, \PZ$ Decays}& \\
\midrule
\multicolumn{2}{c}  {G. C. Dorsch, S. J. Huber, K. Mimasu and J. M. No \cite{Dorsch:2014qja}}\\
\midrule
\textsl{Physical} & $\MA - \MH > \MZ $ is a primary signature of a strong EW Phase Transition\cr
\textsl{Motivation} & in the 2HDM, potentially leading to successful EW Baryogenesis\cr
\midrule
\multicolumn{2}{c}  {{\bf $BP3_A$} \hspace*{\fill}}\cr
\midrule
Main & Type I and II 2HDM, alignment limit $c_{\beta - \alpha} = 0$\cr
Features & $\lambda_6 = \lambda_7 = 0$, no impact from $\mu_{12}^2$ \cr
\midrule
Spectrum & $\MH + \MZ  < \MA  = \MSHpm$ \cr
\midrule
Particular & $\PA \to \PZ \PH$, $\PH \to \PQb \PAQb$ (for $\MH < 340$~\UGeV); $\PH \to \PQt\PAQt$ (for $\MH > 340$~\UGeV)\cr
Signatures & See \refF{BPX:A_GHMN}(a) for Type I 2HDM with $t_{\beta} = 3$\cr
\midrule
Example: & $\MA = \MSHpm = 420$~\UGeV, $\MH = 180$~\UGeV, $t_{\beta} = 3$, $\mu = 100$~\UGeV \cr
$BP3_{A1}$ & Type I: $\sigma(\Pg\Pg\to \PA) = 2.369$ pb, $\mathrm{BR}(\PA \to \PZ \PH)$ = 0.843, $\mathrm{BR}(\PH \to \PQb\PAQb)$ = 0.711 \cr
& Type II: $\sigma(\Pg\Pg\to \PA) = 2.405$ pb, $\mathrm{BR}(\PA \to \PZ \PH)$ = 0.838, $\mathrm{BR}(\PH \to \PQb \PAQb)$ = 0.899 \cr
\midrule
Example: & $\MA = \MSHpm = 550$~\UGeV, $\MH = 400$~\UGeV, $t_{\beta} = 3$, $\mu = 210$~\UGeV \cr
$BP3_{A2}$ & Type I: $\sigma(\Pg\Pg\to \PA) = 0.548$ pb, $\mathrm{BR}(\PA \to \PZ \PH)$ = 0.498,  $\mathrm{BR}(\PH \to \PQt\PAQt)$ = 0.992 \cr
& Type II: $\sigma(\Pg\Pg\to \PA) = 0.570$ pb, $\mathrm{BR}(\PA \to \PZ \PH)$ = 0.486,  $\mathrm{BR}(\PH \to \PQt\PAQt)$ = 0.866 \cr
\midrule
\multicolumn{2}{c}  {{\bf $BP3_B$} \hspace*{\fill}}\cr
\midrule
Main & Type I 2HDM, non-alignment $\left| c_{\beta - \alpha}\right| > 0.1$\cr
Features & Type II 2HDM, non-alignment $s_{\beta + \alpha} \sim 1$ ($t_{\beta} \geq 3$)\cr
& $\lambda_6 = \lambda_7 = 0$, $\mu^2 = \MH ^2 s_{\beta} c_{\beta}$\cr
\midrule
Spectrum & $\MH + \MZ  < \MA = \MSHpm$ \cr
\midrule
Particular & $\PA \to \PZ \PH$, $\PH \to \PWp\PWm$\cr
Signatures & See \refF{BPX:A_GHMN}(b) for Type II 2HDM with $t_{\beta} = 3$\cr
\midrule
Example: & $\MA = \MSHpm = 420$~\UGeV, $\MH = 200$~\UGeV, $t_{\beta} = 3$, $c_{\beta-\alpha} = 0.3$, $\mu = 110$~\UGeV \cr
$BP3_{B1}$ & Type I: $\sigma(\Pg\Pg\to \PA) = 2.369$ pb \cr
&  \hspace{13mm}    $\mathrm{BR}(\PA \to \PZ \PH)$ = 0.697, $\mathrm{BR}(\PH \to \PWp\PWm)$ = 0.742   \cr
\midrule
Example: & $m_{A_0} = m_{H^{\pm}} = 420$~\UGeV, $m_{\PH} = 200$~\UGeV, $t_{\beta} = 3$, $c_{\beta-\alpha} = 0.5$, $\mu = 110$~\UGeV\ \cr
$BP3_{B2}$ & Type II: $\sigma(\Pg\Pg\to A_0) = 2.405$ pb \cr
&  \hspace{15mm}    $\mathrm{BR}(\PA \to \PZ \PH)$ = 0.517, $\mathrm{BR}(\PH \to \PWp\PWm)$ = 0.662   \cr
\bottomrule
\end{longtable}
\end{center}


\begin{figure}[h!]
\centering
\includegraphics[width=0.45\textwidth]{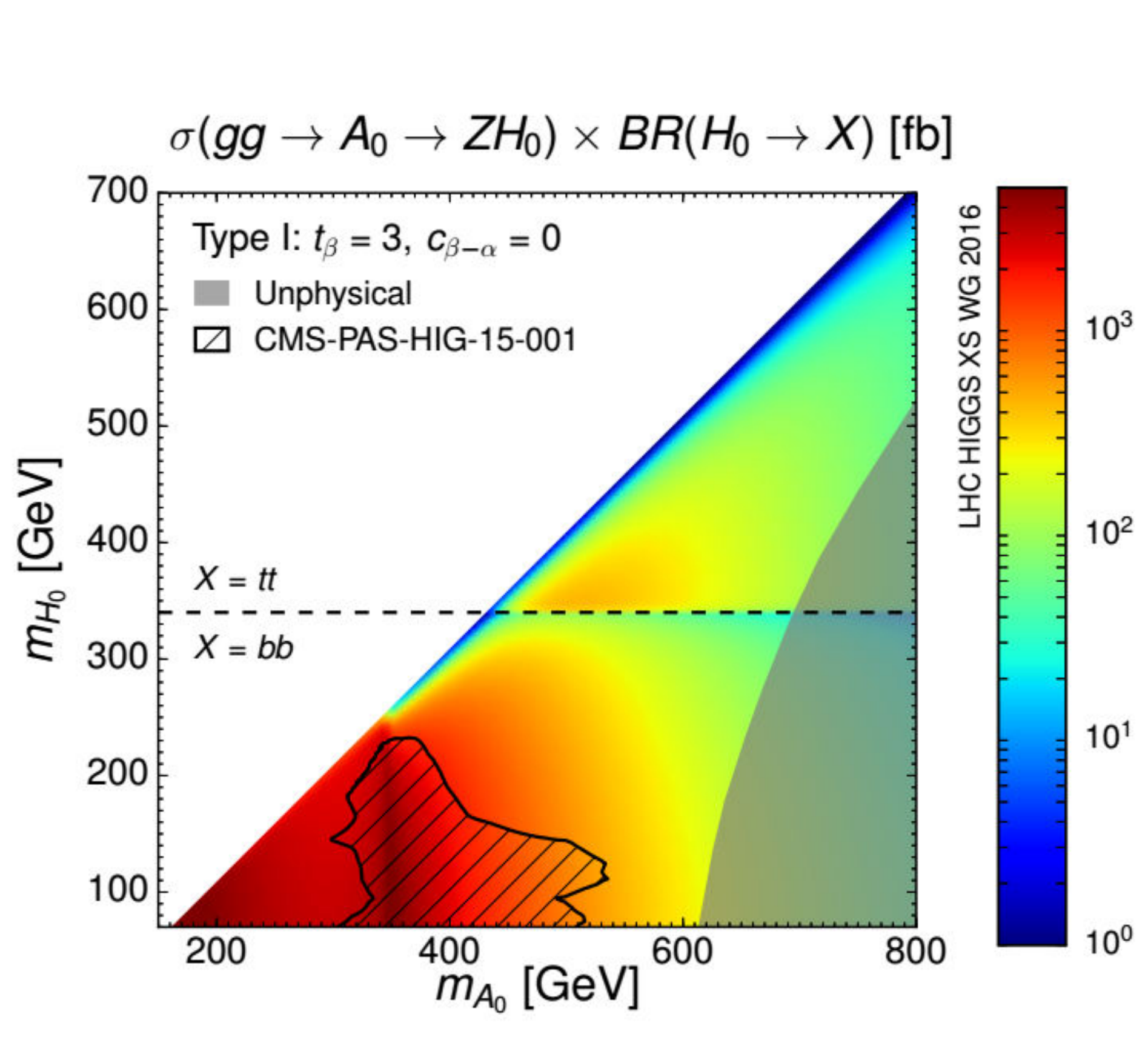}
\includegraphics[width=0.45\textwidth]{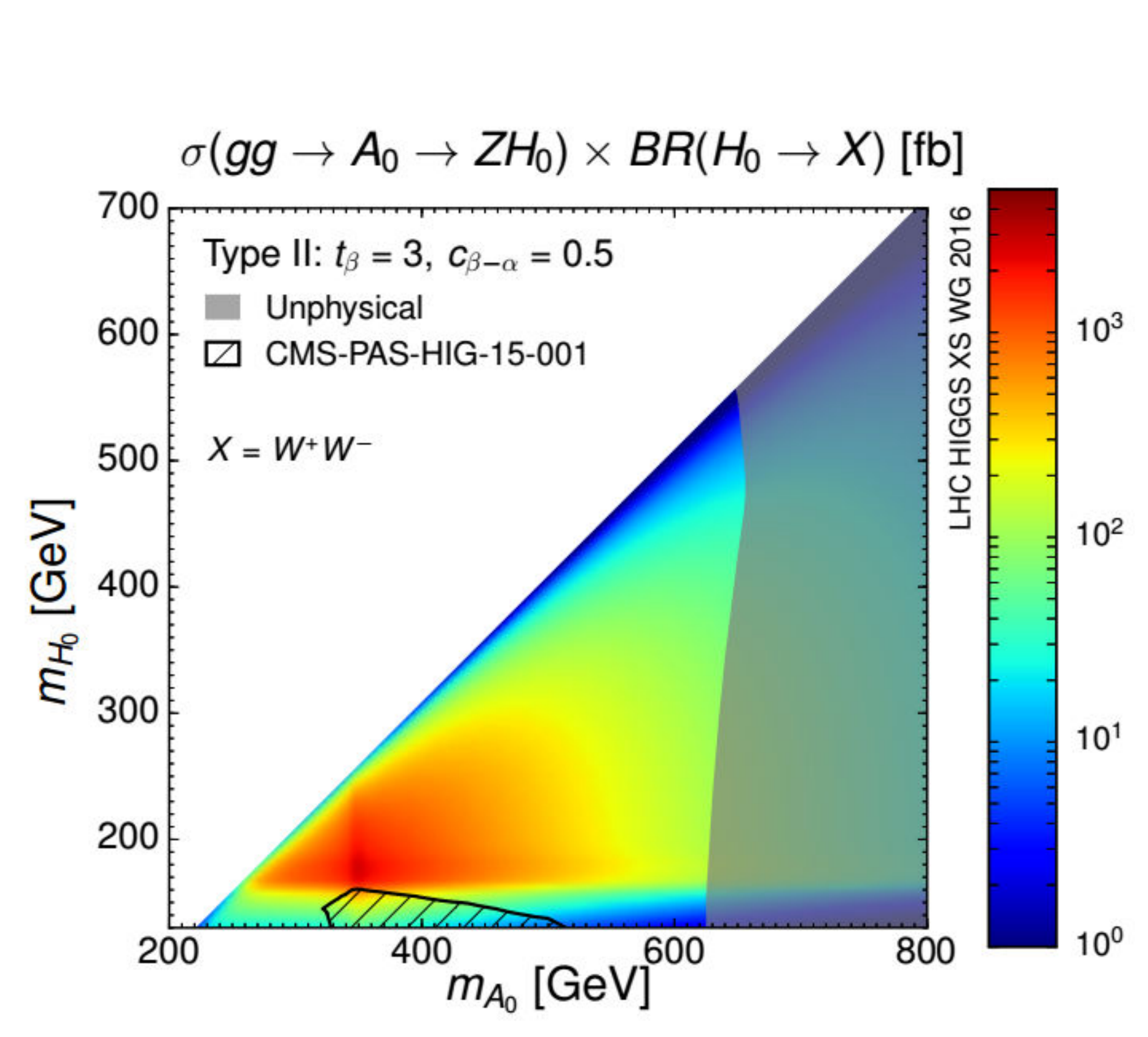}
\caption{Example planes for BP3. (a) (left)
$\sigma \times$ BR for $\Pg\Pg \to \PA  \to\PZ \PH$ ($\PH \to \PQf\PAQf$) in 2HDM Type I for $t_{\beta} = 3$ and the alignment limit $c_{\beta-\alpha} = 0$. For $\MH < 340$~\UGeV\ ($> 340$~\UGeV), $\PQf\PAQf = \PQb\PAQb$ ($ = \PQt\PAQt$). The black-hatched region corresponds to the exclusion from CMS-PAS-HIG-15-001 \cite{Khachatryan:2016are}. (b) (right) $\sigma \times$ BR for $\Pg\Pg \to \PA  \to\PZ \PH$ ($\PH \to \PWp \PWm$) in 2HDM Type II for $t_{\beta} = 3$ away from the alignment limit ($s_{\beta+\alpha} \sim 1$).
See \refT{tab:wg3_extscalars_BP3} for more details.}
\label{BPX:A_GHMN}


\end{figure}



\subsection{Benchmark points \texorpdfstring{$BP4$}{BP4}}
\begin{center}
\small
\begin{longtable}{c|l@{}m{0pt}@{}}
\toprule
\multicolumn{2}{c}  {\bf  $BP4$: Light Pseudoscalars} & \\
\midrule
\multicolumn{2}{c}  {R. Aggleton, D. Barducci, S. Moretti, A. Nikitenko and C. Shepherd-Themistocleous}\\
\midrule
\midrule
\multicolumn{2}{c}  {$BP4_1$ \hspace*{\fill}}\cr
\midrule
Main Features & 2HDM Type I: light pseudoscalar ($\sim20$~\UGeV) and substantial $\Ph\to \PZ\PA$ decay rate\cr
\midrule
Spectrum & $\Mh \approx 126$~\UGeV,  $\MA \approx  20$~\UGeV,
$\MH \approx 165$~\UGeV,\cr
& $\MSHpm \approx  444$~\UGeV\ \cr
\midrule
\multicolumn{2}{c}  {Production cross sections and branching fractions}\cr
\midrule
$\Ph \to\PZ \PA$ & $\Pg\Pg$F$(\Ph) \approx$ 38 pb (at 13~\UTeV), $BR(\Ph \to\PZ \Pa) \approx 10\%$\cr
& BR$(\PA \to \PQb\PQb) \approx 85\%$, BR$(\PA \to \PGt\PGt) \approx 6\%$,  BR$(\PA \to \PGm\PGm) \approx 0.02\%$  \cr
\midrule
Particular & On shell $\PZ$ from $\Ph\to \PZ\PA$ decay
\cr
signatures &
\cr
\midrule
Model & $\Mh =126.0$~\UGeV, $\MH =165.5$~\UGeV, $\MA =20.2$~\UGeV, $\MSHpm=444.7$~\UGeV
\cr
Parameters & $\tan{\beta}=1.9$, $\lambda_{6,7}=0$, $m_{12}^2=3891.5$~\UGeV$^2$, $\sin(\beta-\alpha)=-0.99$
\cr
\midrule
\midrule
\multicolumn{2}{c}  {$BP4_2$ \hspace*{\fill}}\cr
\midrule
Main Features & 2HDM Type I: light pseudoscalar ($\sim60$~\UGeV) and substantial $\Ph\to \PZ\PA$ decay rate\cr
\midrule
Spectrum & $\Mh \approx 126$~\UGeV,  $\MA \approx  63$~\UGeV,
$\MH \approx 153$~\UGeV,\cr
& $\MSHpm \approx  258$~\UGeV\cr
\midrule
\multicolumn{2}{c}  {Production cross sections and branching fractions}\cr
\midrule
$\Ph \to\PZ \PA$ & $\Pg\Pg$F$(\Ph) \approx$ 26 pb (at 13~\UTeV), BR$(\Ph \to\PZ \Pa) \approx 3\%$\cr
& BR$(\PA \to \PQb\PQb) \approx 79\%$, BR$(\PA \to \PGt\PGt) \approx 7\%$,  BR$(\PA \to \PGm\PGm) \approx 0.02\%$  \cr
\midrule
Particular & Off shell $Z$ from $\Ph\to \PZ\PA$ decay
\cr
signatures &
\cr
\midrule
Model & $\Mh =126.0$~\UGeV, $\MH =153.4$~\UGeV, $\MA =63.4$~\UGeV, $\MSHpm=257.7$~\UGeV\
\cr
Parameters & $\tan{\beta}=6.2$, $\lambda_{6,7}=0$, $m_{12}^2=2793.3$~\UGeV$^2$, $\sin(\beta-\alpha)=-0.85$
\cr
\midrule
\midrule
\multicolumn{2}{c}  {$BP4_3$ \hspace*{\fill}}\cr
\midrule
Main Features & 2HDM Type II: light pseudoscalar ($\sim6$~\UGeV) and substantial $\Ph\to \PZ\PA$ decay rate\cr
\midrule
Spectrum & $\Mh \approx 126$~\UGeV,  $\MA \approx  6$~\UGeV,
$\MH \approx 264$~\UGeV,\cr
& $\MSHpm \approx  308$~\UGeV\ \cr
\midrule
\multicolumn{2}{c}  {Production cross sections and branching fractions}\cr
\midrule
$\Ph \to\PZ \PA$ & $\Pg\Pg$F$(\Ph) \approx$ 51 pb (at 13~\UTeV), BR$(\Ph \to\PZ \Pa) \approx 31\%$\cr
& BR$(\PA \to \PQb\PQb) = 0\%$, BR$(\PA \to \PGt\PGt) \approx 78\%$,  BR$(\PA \to \PGm\PGm) \approx 0.3\%$  \cr
\midrule
Particular & On shell $\PZ$ from $\Ph\to \PZ\PA$ decay.
\cr
signatures & Extremely light $\PA$, decay products will be very boosted
\cr
\midrule
Model & $\Mh =126.0$~\UGeV, $\MH =263.7$~\UGeV, $\MA =6.3$~\UGeV, $\MSHpm=308.3$~\UGeV\
\cr
Parameters & $\tan{\beta}=1.9$, $\lambda_{6,7}=0$, $m_{12}^2=2737.4$~\UGeV$^2$, $\sin(\beta-\alpha)=0.99$
\cr
\midrule
\midrule
\multicolumn{2}{c}  {$BP4_4$ \hspace*{\fill}}\cr
\midrule
Main Features & 2HDM Type II: light pseudoscalar ($\sim 25$~\UGeV) and substantial $\Ph\to \PZ\PA$ decay rate\cr
\midrule
Spectrum & $\Mh \approx 126$~\UGeV,  $\MA \approx  25$~\UGeV,
$\MH \approx 227$~\UGeV,\cr
& $\MSHpm \approx  227$~\UGeV\ \cr
\midrule
\multicolumn{2}{c}  {Production cross sections and branching fractions}\cr
\midrule
$\Ph \to\PZ \PA$ & $\Pg\Pg$F$(\Ph) \approx$ 52 pb (at 13~\UTeV), BR$(\Ph \to\PZ \Pa) \approx 15\%$\cr
& BR$(\PA \to \PQb\PQb) \approx 92\%$, BR$(\PA \to \PGt\PGt) \approx 6\%$,  BR$(\PA \to \PGm\PGm) \approx 0.02\%$  \cr
\midrule
Particular & On shell $\PZ$ from $\Ph\to \PZ\PA$ decay.
\cr
signatures &
\cr
\midrule
Model & $\Mh =126.2$~\UGeV, $\MH =227.1$~\UGeV, $\MA =24.7$~\UGeV, $\MSHpm=226.8$~\UGeV
\cr
Parameters & $\tan{\beta}=1.8$, $\lambda_{6,7}=0$, $m_{12}^2=3406.8$~\UGeV$^2$, $\sin(\beta-\alpha)=0.99$
\cr
\midrule
\midrule
\multicolumn{2}{c}  {$BP4_5$ \hspace*{\fill}}\cr
\midrule
Main Features & 2HDM Type II: light pseudoscalar ($\sim 63$~\UGeV) and substantial $\Ph\to \PZ\PA$ decay rate\cr
\midrule
Spectrum & $\Mh \approx 126$~\UGeV,  $\MA \approx  63$~\UGeV,
$\MH \approx 210$~\UGeV,\cr
& $\MSHpm  \approx  333$~\UGeV\ \cr
\midrule
\multicolumn{2}{c}  {Production cross sections and branching fractions}\cr
\midrule
$\Ph \to\PZ \PA$ & $\Pg\Pg$F$(\Ph) \approx$ 57~pb (at 13~\UTeV), BR$(\Ph \to\PZ \Pa) \approx 4\%$\cr
& BR$(\PA \to \PQb\PQb) \approx 92\%$, BR$(\PA \to \PGt\PGt) \approx 7\%$,  BR$(\PA \to \PGm\PGm) \approx 0.03\%$  \cr
\midrule
Particular & Off shell $\PZ$ from $\Ph\to \PZ\PA$ decay.
\cr
signatures &
\cr
\midrule
Model & $\Mh =125.2$~\UGeV, $\MH =210.2$~\UGeV, $\MA =63.06$~\UGeV, $\MSHpm=333.5$~\UGeV\
\cr
Parameters & $\tan{\beta}=2.4$, $\lambda_{6,7}=0$, $m_{12}^2=4791.9$~\UGeV$^2$, $\sin(\beta-\alpha)=0.7$
\cr
\bottomrule
\end{longtable}
\end{center}



\subsection{Benchmark points \texorpdfstring{$BP5$}{BP5}}

\begin{center}
\small
\begin{longtable}{c|l@{}m{0pt}@{}}
\toprule
\multicolumn{2}{c}  {\bf $BP5$: Benchmarks for the Inert Doublet Model} & \\
\midrule
\multicolumn{2}{c}  {Agnieszka Ilnicka, Maria Krawczyk, Tania Robens \cite{Ilnicka:2015jba}}\cr
\midrule
Main Features & IDM, two ${\rm SU}(2)\times\,{\rm U}(1)$ doublet model \cr
&  with SM-like Higgs boson h  and dark matter candidate H \cr
\midrule
Floating parameters & masses of scalars $\MH$, $\MA$, $\MSHpm$ \cr
& $\MH$ > {45} \UGeV, mass degeneracy \cr
\midrule
Fixed parameters        & $\Mh $ = 125.1~\UGeV \cr
Irrelevant parameters & $\lambda_2$; $\lambda_{345}$ (if kept within allowed ranges); {$\lambda_2\,\in\,[0;4.2]$} \cr
\midrule
BR ($\PA \rightarrow \PZ\PH$) & 1 \cr
BR ($\PSHpm \rightarrow \PWpm \PH$) & dominant \cr
\midrule
comment & {dark scalars ($\PH,\ \PA,\ \PSHpm$) have to be produced in pairs} \cr
& signature: always  \MET\ from \PH\PH\ in final states \cr
\midrule
\multicolumn{2}{c}  {Production cross sections and branching fractions}\cr
\midrule
\midrule
\multicolumn{2}{c}  {$BP5_A$ \hspace*{\fill}}\cr
\midrule
Main Features & Low mass H [$\MH < \Mh /2$] \cr
Spectrum & $\MH$=57.5~\UGeV, $\MA$=113.0~\UGeV, $\MSHpm$=123.0~\UGeV, $\| \lambda_{345} \| \in [0.002,0.015]$ \cr
$\sigma(\Pp\Pp \rightarrow \PH\PA)$ & 0.371(4) [pb] \cr
$\sigma(\Pp\Pp \rightarrow \PSHp \PH)$ & 0.3071(4) [pb] \cr
$\sigma(\Pp\Pp \rightarrow \PSHp \PA)$ & 0.1267(1) [pb] \cr
$\sigma(\Pp\Pp \rightarrow \PSHp \PSHm)$ & 0.097(1) [pb] \cr

BR($\PSHp \rightarrow \PWp \PH$) & $>$0.99 \cr
BR($\PSHp \rightarrow \PWp \PA$) & $<$0.01 \cr
\midrule
\midrule
\multicolumn{2}{c}  {$BP5_B$ \hspace*{\fill}}\cr
\midrule
Main Features & Low mass H [$\Mh /2 < \MH < \Mh $] \cr
Spectrum & $\MH$=85.5~\UGeV, $\MA$=111.0~\UGeV, $\MSHpm$=140.0~\UGeV, $\| \lambda_{345} \| < 0.015 $ \cr
$\sigma(\Pp\Pp \rightarrow \PH\PA)$ & 0.226(2) [pb] \cr
$\sigma(\Pp\Pp \rightarrow \PSHp \PH)$ & 0.1439(2) [pb] \cr
$\sigma(\Pp\Pp \rightarrow \PSHp \PA)$ & 0.1008(1) [pb] \cr
$\sigma(\Pp\Pp \rightarrow \PSHp \PSHm)$ & 0.0605(9) [pb] \cr
BR($\PSHp  \rightarrow \PWp \PH$) & 0.96 \cr
BR($\PSHp \rightarrow \PWp \PA$) & 0.04 \cr
\midrule
\midrule
\multicolumn{2}{c}  {$BP5_C$ \hspace*{\fill}}\cr
\midrule
Main Features & Low mass H [$\MH \sim \Mh $] \cr
Spectrum & $\MH$=128.0~\UGeV, $\MA$=134.0~\UGeV, $\MSHpm$=176.0~\UGeV, $\| \lambda_{345} \| < 0.05$ \cr
$\sigma(\Pp\Pp \rightarrow \PH\PA)$ & 0.0765(7) [pb] \cr
$\sigma(\Pp\Pp \rightarrow \PSHp \PH)$ & 0.04985(5) [pb] \cr
$\sigma(\Pp\Pp \rightarrow \PSHp \PA)$ & 0.04653(5) [pb] \cr
$\sigma(\Pp\Pp \rightarrow \PSHp  \PSHm)$ & 0.0259(3) [pb] \cr
BR($\PSHp  \rightarrow \PWp\PH$) & 0.66 \cr
BR($\PSHp \rightarrow \PWp\PA$) & 0.34 \cr
\midrule
\midrule
\multicolumn{2}{c}  {$BP5_D$ \hspace*{\fill}}\cr
\midrule
Main Features & High mass H [$\MH > \Mh $]; degeneracy \cr
Spectrum & $\MH$=363.0~\UGeV, $\MA$=374.0~\UGeV, $\MSHpm$=374.0~\UGeV, $\| \lambda_{345} \| < 0.25$ \cr
$\sigma(\Pp\Pp \rightarrow \PH\PA)$ & 0.00122(1) [pb] \cr
$\sigma(\Pp\Pp \rightarrow \PSHp \PH)$ & 0.001617(2) [pb] \cr
$\sigma(\Pp\Pp \rightarrow \PSHp \PA)$ & 0.001518(2) [pb] \cr
$\sigma(\Pp\Pp \rightarrow \PSHp \PSHm)$ & 0.00124(1) [pb] \cr
BR($\PSHp  \rightarrow \PWp \PH$) & 1 \cr
\midrule
\midrule
\multicolumn{2}{c}  {$BP5_E$ \hspace*{\fill}}\cr
\midrule
Main Features & High mass H [$\MH >  \Mh $] \cr
Spectrum & $\MH$=311.0~\UGeV, $\MA$=415.0~\UGeV, $\MSHpm$=447.0~\UGeV, $\| \lambda_{345} \| < 0.19$ \cr
$\sigma(\Pp\Pp \rightarrow \PH\PA)$ & 0.00129(1) [pb] \cr
$\sigma(\Pp\Pp \rightarrow \PSHp \PH)$ & 0.001402(2) [pb] \cr
$\sigma(\Pp\Pp \rightarrow \PSHp \PA)$ & 0.0008185(8) [pb] \cr
$\sigma(\Pp\Pp \rightarrow \PSHp  \PSHm)$ & 0.000553(7) [pb] \cr
BR($\PSHp  \rightarrow \PWp \PH$) & $>$0.99 \cr
BR($\PSHp \rightarrow \PWp \PA$) & $<$0.01 \cr
\bottomrule
\end{longtable}
\end{center}

\clearpage


\subsection{Benchmark points \texorpdfstring{$BP6$}{BP6}}
\begin{center}
\small
\begin{longtable}{c|l@{}m{0pt}@{}}
\toprule
\multicolumn{2}{c}  {\bf $BP6$: Fermiophobic heavy Higgs} & \\
\midrule
\multicolumn{2}{c}  {D. L\'opez-Val \cite{Hespel:2014sla}}\\
\midrule
\multicolumn{2}{c}  {\textbf{Benchmark setup}}\cr \midrule
Main Features & $\bullet$\,SM-like light Higgs boson \cr
& $\bullet$\,Fermiophobic heavy neutral Higgs boson \cr
& $\bullet$\,$\Rightarrow$ Relatively light, yet very elusive Higgs companion \cr
& $\bullet$\,$\Rightarrow$ Warning sign: lack of signal should not rule out the model too early \cr
\midrule
Spectrum & $\Mh $= 125~\UGeV, $\MH  = 200$~\UGeV,  $\MA  = 500$~\UGeV, $\MSHpm  = 500$~\UGeV \cr
Model parameters (physical basis) & Type I Yukawas,  $\sin\alpha=0$, $\tan\beta = 20$, $m^2_{12} = 2000$~\UGeV$^2$ \cr
\midrule
\multicolumn{2}{c}  {\textbf{Production cross sections and branching fractions}}\cr
\midrule
leading signatures &
Light Higgs phenomenology essentially unaffected \cr
& Heavy Higgs sharp resonance into $\PW\PW/\PZ\PZ $ \cr
 & $\sigma(\Pp\Pp \to \PH\PA) \simeq 1.91$~fb (13~\UTeV) \cr
 & $\sigma(\Pp\Pp \to \PH \PSHpm) \simeq 0.88$~fb  (13~\UTeV) \cr
\midrule
Heavy Higgs boson total width & $\Gamma(\PH) = 3.39 \times 10^{-3}$~\UGeV \cr \midrule
Heavy Higgs boson branching fractions & BR($\PH \to \PZ\PZ $) $=$ 0.742 \cr
 & BR($\PH \to \PW\PW$) $=$ 0.258 \cr
 & BR($\PH \to \PGg\PGg$) <  $10^{-4}$ \cr
 & BR($\PH \to\PZ \gamma$) <  $10^{-4}$ \cr
 & BR($\PH \to \PQf\PAQf$) $=$ 0 \cr
\midrule
\multicolumn{2}{c}  {\textbf{Benchmark planes}}\cr \midrule
Floating parameters & $\MH $, mass splitting $\Delta M$ \cr & \cr
Fixed parameters & $\sin\alpha = 0$ (by construction) \cr
& $\MA  = \MH  + \Delta M$, $\MSHpm  = \MH  + \Delta M$, $m_{12}^2 = \MH ^2/\tan\beta$ \cr & \cr
Fiducial $\tan\beta$ choices
& $\tan\beta = 40$ (mild departure from alignment, coupling shifts of $\mathcal{O}(2\%)$) \cr
& $\tan\beta = 20$ (moderate departure, coupling shifts of $\mathcal{O}(5\%)$) \cr
& $\tan\beta = 10$ (large departure, coupling shifts of $\mathcal{O}(10\%)$) \cr
\bottomrule
\end{longtable}
\end{center}



\subsection{Benchmark points \texorpdfstring{$BP7$}{BP7}}

As proposed in \cite{Branco:1999fs, Fontes:2015xva},
CP-violation in the scalar sector can be found in the interactions with
gauge bosons in a very simple way.
Assuming CP is conserved, any decay $\Phindexi \rightarrow \Phindexj \PZ$ would imply opposite
CP parities for $\Phindexi $ and $\Phindexj $.
Moreover,
assuming only Lagrangian terms up to dimension four, any scalar $\Phindexi $
decaying into $\PZ\PZ $ would be
CP even~\footnote{There are CP conserving terms of
dimension higher than four that can mediate the decay of a
pseudoscalar into two vector bosons. A calculation performed
in the framework of the 2HDM has shown~\cite{Arhrib:2006rx} that the
loop mediated decays of the type $\Phindexi \to \PZ\PZ $ are several orders of magnitude smaller
than the tree-level ones.}.

In Table~\ref{tab:classes} we define five classes of CP-violation
with the respective decays. Classes $C_1$-$C_4$ represent CP-violation,
regardless of the origin of the neutral scalars. Class $C_5$ does not represent
necessarily CP-violation in models other than the 2HDM.
\begin{table}[h]
 \caption{\label{tab:classes}Classes of combined measurements guaranteed
 to probe CP-violation in 2HDMs.}
 \centering
\begin{tabular}{c| ccccc}
\toprule
Classes & $C_1$ & $C_2$ & $C_3$ & $C_4$ & $C_5$ \\
\midrule
        & $\Phthree \to \Phtwo  \PZ$ & $\Phtwo \to \Phone  \PZ$ & $\Phthree \to \Phone  \PZ$ & $\Phthree \to \Phtwo  \PZ$
        & $\Phthree \to\PZ \PZ$ \\
Decays  & $\Phtwo \to \Phone  \PZ$ & $\Phone \to\PZ \PZ$   & $\Phone \to\PZ \PZ$   & $\Phtwo \to\PZ \PZ$
        & $\Phtwo \to\PZ \PZ$ \\
        & $\Phthree \to \Phone  \PZ$ & $\Phtwo \to\PZ \PZ$   & $\Phthree \to\PZ \PZ$   & $\Phthree \to\PZ \PZ$
        & $\Phone \to\PZ \PZ$ \\
 \bottomrule
 \end{tabular}
 \end{table}
There are other classes of decays that constitute a sign of CP-violation with
at least one process where a scalar decays to two other scalars.
Some of them involve the decay $\Phthree  \to \Phtwo  \Phone $ which is not present in the CP-conserving
version of the 2HDM. The two remaining classes are presented in table~\ref{tab:classes2}.

\begin{table}[h]
\caption{\label{tab:classes2}Classes of combined measurements guaranteed
 to probe CP-violation.}
 \centering
\begin{tabular}{c| cc}
\toprule
Classes & $C_6$ & $C_7$ \\
\midrule
        & $\Phthree \to \Phtwo  \Phone $  & $\Phtwothree \to \Phone  \Phone $ \\
Decays  & $\Phthree \to \Phtwo  \PZ$  & $\Phtwothree \to \Phone  \PZ$ \\
        & $\Phone \to\PZ \PZ$ & $\Phone \to\PZ \PZ$ \\
 \bottomrule
 \end{tabular}
 \end{table}

Finally, other combinations like $\Phthree  \to \Phone  \Phone  (\Phtwo  \Phtwo )$, $\Phtwo  \to \Phone  \Phone $ and
$\Phone  \to \PZ\PZ $ are not possible in a CP-conserving 2HDM but are possible in the C2HDM
and can also serve to determine the CP-quantum numbers of other extensions of the
scalar sector.
%

\begin{center}
\small
\begin{longtable}{c|ll@{}m{0pt}@{}}
\toprule
\multicolumn{3}{c}  {\bf $BP7_1$ Benchmarks for the Complex Two Higgs Doublet Model} & \\
\midrule
\multicolumn{3}{c}  {D. Fontes, J.C. Rom\~{a}o, R. Santos, Jo\~{a}o
 P. Silva \cite{Fontes:2014xva}}\\
\midrule \midrule
\multicolumn{3}{c}  {\textbf{Benchmark setup}}\cr \midrule
Main Features &\multicolumn{2}{l}{ $\bullet$\, lightest Higgs boson
at 125~\UGeV} \cr
&\multicolumn{2}{l}{ $\bullet$\, CP violation through
  scalar-pseudoscalar mixing }\cr
& \multicolumn{2}{l}{$\bullet$\, CP violation detected by the
  simultaneous existence }\cr
& \multicolumn{2}{l}{of three decay channels
(see classes $C_1$-$C_7$ in the text) }\cr
\bottomrule
\toprule
\multicolumn{3}{c}  {\textbf{Benchmark point $1_\textrm{C2HDM}$}}\cr
\bottomrule
\end{longtable}
\end{center}



\subsection{Benchmark points \texorpdfstring{$BP8$}{BP8}}

\begin{center}
\small
\begin{longtable}{c|l@{}m{0pt}@{}}
\toprule
\multicolumn{2}{c}  {\bf $BP8$: Controlled Higgs flavour changing couplings}\cr
\midrule
\multicolumn{2}{c}  {F. Botella, G.C. Branco, M. Nebot \& M.N. Rebelo \cite{Botella:2015hoa}}\cr
\midrule
Main Features &  Higgs flavour changing decays, \cr
& can be probed at the 14~\UTeV\ LHC \cr
\midrule\midrule
\multicolumn{2}{l}{{\bf  $BP8_1$} \quad BGL model $(\PQb,\PGt)$}\cr
\midrule
Spectrum: & $\Mh = 125$~\UGeV, $\MH\approx \MA \approx \MSHpm \gsim  600$~\UGeV\cr
$\cos(\alpha-\beta)$ & $\leq 0.17$\cr
$\tan\beta$ & $25-100$\cr
\bottomrule
\end{longtable}
\end{center}



\section{Georgi-Machacek model}

Extensions of the Standard Model (SM) Higgs sector that contain scalars in triplet or larger isospin representations and that preserve the custodial symmetry in the scalar sector generically contain fermiophobic scalars that transform as a fiveplet under the custodial symmetry.  These custodial-fiveplet scalars play an essential role in the unitarity of vector boson scattering amplitudes~\cite{Falkowski:2012vh,Grinstein:2013fia,Bellazzini:2014waa}.  The coupling strength of the custodial-fiveplet scalars to $\PVV$ is proportional to the vacuum expectation value (vev) carried by the higher-isospin representation(s).  While many scenarios with scalars in larger isospin representations are severely constrained by the electroweak $\rho$ parameter, preservation of the custodial symmetry renders these models viable.

The experiments at the CERN Large Hadron Collider (LHC) have sensitivity to the production and decay of the fermiophobic custodial-fiveplet states,
\begin{equation}
	\PSHppFive , \PSHpFive , \PHzFive , \PSHmFive , \PSHmmFive ,
\end{equation}
through their tree-level couplings to $\PW$ and $\PZ$ boson pairs.
This states offer several interesting features, including a tree-level $\PSHpmFive \PWmp  \PZ$ interaction, $\PSHpmpmFive$ decays to like-sign $\PW$ bosons, and $\PHzFive $ decays to $\PWp \PWm$ and $\PZ\PZ $ in a different ratio than appears in the SM.  The production of $\PSHpmFive$ via vector boson fusion (VBF) followed by decays to $\PWpm  \PZ$ has already been studied by ATLAS in Run~1 of the LHC~\cite{Aad:2015nfa}.

As a prototype model containing a custodial fiveplet of scalars, we consider the Georgi-Machacek (GM) model~\cite{Georgi:1985nv,Chanowitz:1985ug}, in which the SM Higgs sector is extended by two isospin triplets while preserving custodial SU(2) symmetry.
The phenomenology of these scalars is dramatically different from that of the additional Higgs bosons found in two Higgs doublet models (2HDMs) or singlet extensions of the SM.  They do not couple to fermions and hence cannot be produced in gluon fusion or in association with a top quark.  The singly-charged scalars $\PSHpmFive$ couple to $\PWpm  \PZ$ at tree level, in contrast to the charged scalar of the 2HDM for which this coupling appears only at one loop.  The relative coupling strengths of $\PHzFive $ to $\PWp \PWm$ and $\PZ\PZ $ are different than those of the SM Higgs boson, leading to $\Gamma(\PHzFive  \to \PW\PW)/\Gamma(\PHzFive  \to \PZ\PZ ) \to 1/2$ in the high-mass limit in contrast to the SM relation $\Gamma(\Ph_{\rm SM} \to \PW\PW)/\Gamma(\Ph_{\rm SM} \to \PZ\PZ ) \to 2$.  The presence of a doubly-charged Higgs boson with decays to like-sign $\PW$ bosons is a dramatic indication of isospin representations larger than doublets that contribute to electroweak symmetry breaking.
A comprehensive tree-level phenomenological study of the production of $\PHzFive $, $\PSHpmFive$, and $\PSHpmpmFive$ at the LHC via VBF with decays to $\PWp \PWm/\PZ\PZ $, $\PWpm  \PZ$, and $\PWpm \PWpm $ respectively, with the gauge bosons decaying leptonically, was performed in~\cite{Godfrey:2010qb}.

In this section we define the H5plane benchmark for the GM model, in which the two free parameters most relevant for $\PHzFive$ searches can be varied.  The benchmark plane specification is designed to be compatible with the spectrum calculator {\tt GMCALC}~\cite{Hartling:2014xma}.
We also provide tables of cross sections for $\PHzFive $, $\PSHpmFive$, and $\PSHpmpmFive$ production in VBF for the 13~\UTeV\  LHC, and well as their decay widths to vector boson pairs.  The cross section recommendations include QCD corrections at next-to-next-to-leading order (NNLO) and were generated using the {\tt VBF@NNLO} code~\cite{Bolzoni:2010xr,Bolzoni:2011cu}.  The decay widths were calculated at tree level using the code {\tt GMCALC~1.2.0}~\cite{Hartling:2014xma}, and include doubly-offshell effects.  We concentrate on $\PHzFive$ masses in the range 200--2000~\UGeV.  For $\PHzFive$ masses below 200~\UGeV, decays to off-shell vector bosons and other final states need to be considered, which changes the experimental analysis.

A fully differential study including next-to-leading order (NLO) QCD matrix elements interfaced to the parton shower has recently been undertaken in~\cite{Degrande:2015xnm} using the {\tt MG5\_aMC@NLO} technology~\cite{Alwall:2014hca,Degrande:2014vpa}, and the corresponding model file made public.
The largest remaining theoretical uncertainties in the cross section and decay width predictions are due to NLO electroweak corrections.  A calculation of these corrections would require the full one-loop electroweak renormalization of the Georgi-Machacek model.  This is unlikely to be undertaken by hand in the few-years timescale, but might become feasible using automated NLO technology~\cite{Degrande:2014vpa}.

In the next subsection we give a brief summary of the GM model.  We specify the H5plane benchmark in Section~\ref{GM:sec:benchmark}.  The VBF production cross sections are tabulated in Section~\ref{GM:sec:xsec} and the total widths of the $\PHzFive$ states are tabulated in Section~\ref{GM:sec:widths}.

\subsection{Model parameterization}
\label{GM:sec:model}

The scalar sector of the GM model~\cite{Georgi:1985nv,Chanowitz:1985ug} consists of the usual complex isospin doublet $(\phi^+,\phi^0)$ with hypercharge\footnote{We normalize the hypercharge operator such that $Q = T^3 + Y/2$.} $Y = 1$, a real triplet $(\Pxi^+,\Pxi^0,\Pxi^-)$ with $Y = 0$, and  a complex triplet $(\Pchi^{++},\Pchi^+,\Pchi^0)$ with $Y=2$.  The doublet is responsible for the fermion masses as in the SM.

The scalar potential is chosen by hand to preserve a global SU(2)$_L \times$SU(2)$_R$ symmetry.  This ensures $\rho = 1$ at tree level.  In order to make the global SU(2)$_L \times$SU(2)$_R$ symmetry explicit, we write the doublet in the form of a bidoublet $\Phi$ and combine the triplets to form a bitriplet $X$:
\begin{equation}
	\Phi = \left( \begin{array}{cc}
	\phi^{0*} &\phi^+  \\
	-\phi^{+*} & \phi^0  \end{array} \right), \qquad
	X =
	\left(
	\begin{array}{ccc}
	\Pchi^{0*} & \Pxi^+ & \Pchi^{++} \\
	 -\Pchi^{+*} & \Pxi^{0} & \Pchi^+ \\
	 \Pchi^{++*} & -\Pxi^{+*} & \Pchi^0
	\end{array}
	\right).
\end{equation}
The vacuum expectation values (vevs) are defined by $\langle \Phi  \rangle = \frac{ v_{\phi}}{\sqrt{2}} {\bf I}_{2\times2}$  and $\langle X \rangle = v_{\Pchi} {\bf I}_{3 \times 3}$, where ${\bf I}$ is the unit matrix.  The Fermi constant $G_F$ fixes the combination of vevs,
\begin{equation}
	v_{\phi}^2 + 8 v_{\Pchi}^2 \equiv v^2 = \frac{1}{\sqrt{2} G_F} \approx (246 \UGeV)^2.
\end{equation}

The most general gauge-invariant scalar potential involving these fields that conserves custodial SU(2) is given, in the conventions of Ref.~\cite{Hartling:2014zca}, by
\begin{eqnarray}
	V(\Phi,X) &= & \frac{\mu_2^2}{2}  {\rm Tr}(\Phi^\dagger \Phi)
	+  \frac{\mu_3^2}{2}  {\rm Tr}(X^\dagger X)
	+ \lambda_1 [{\rm Tr}(\Phi^\dagger \Phi)]^2
	+ \lambda_2 {\rm Tr}(\Phi^\dagger \Phi) {\rm Tr}(X^\dagger X)   \nonumber \\
          & & + \lambda_3 {\rm Tr}(X^\dagger X X^\dagger X)
          + \lambda_4 [{\rm Tr}(X^\dagger X)]^2
           - \lambda_5 {\rm Tr}( \Phi^\dagger \tau^a \Phi \tau^b) {\rm Tr}( X^\dagger t^a X t^b)
           \nonumber \\
           & & - M_1 {\rm Tr}(\Phi^\dagger \tau^a \Phi \tau^b)(U X U^\dagger)_{ab}
           -  M_2 {\rm Tr}(X^\dagger t^a X t^b)(U X U^\dagger)_{ab}.
           \label{GM:eq:potential}
\end{eqnarray}
Here the SU(2) generators for the doublet representation are $\tau^a = \sigma^a/2$ with $\sigma^a$ being the Pauli matrices, the generators for the triplet representation are
\begin{equation}
	t^1= \frac{1}{\sqrt{2}} \left( \begin{array}{ccc}
	 0 & 1  & 0  \\
	  1 & 0  & 1  \\
	  0 & 1  & 0 \end{array} \right), \qquad
	  t^2= \frac{1}{\sqrt{2}} \left( \begin{array}{ccc}
	 0 & -i  & 0  \\
	  i & 0  & -i  \\
	  0 & i  & 0 \end{array} \right), \qquad
	t^3= \left( \begin{array}{ccc}
	 1 & 0  & 0  \\
	  0 & 0  & 0  \\
	  0 & 0 & -1 \end{array} \right),
\end{equation}
and the matrix $U$, which rotates $X$ into the Cartesian basis, is given by~\cite{Aoki:2007ah}
\begin{equation}
	 U = \left( \begin{array}{ccc}
	- \frac{1}{\sqrt{2}} & 0 &  \frac{1}{\sqrt{2}} \\
	 - \frac{i}{\sqrt{2}} & 0  &   - \frac{i}{\sqrt{2}} \\
	   0 & 1 & 0 \end{array} \right).
\end{equation}
We decompose the neutral fields into real and imaginary parts according to
\begin{equation}
	\phi^0 \to \frac{v_{\phi}}{\sqrt{2}} + \frac{\phi^{0,r} + i \phi^{0,i}}{\sqrt{2}},
	\qquad
	\Pchi^0 \to v_{\Pchi} + \frac{\Pchi^{0,r} + i \Pchi^{0,i}}{\sqrt{2}},
	\qquad
	\Pxi^0 \to v_{\Pchi} + \Pxi^0.
\end{equation}

The physical fields can then be organized by their transformation properties under the custodial SU(2) symmetry into a fiveplet, a triplet, and two singlets.  The custodial-fiveplet states are given by
\begin{equation}
	\PSHppFive  = \Pchi^{++}, \qquad
	\PSHpFive  = \frac{\left(\Pchi^+ - \Pxi^+\right)}{\sqrt{2}}, \qquad
	\PHzFive  = \sqrt{\frac{2}{3}} \Pxi^0 - \sqrt{\frac{1}{3}} \Pchi^{0,r}.
\end{equation}
Because the states in the custodial fiveplet contain no doublet field content, they do not couple to fermions.

The custodial-triplet states are given by
\begin{equation}
	\PSHpThree = - s_{\PH}\phi^+ + c_{\PH}\frac{\left(\Pchi^++\Pxi^+\right)}{\sqrt{2}}, \qquad
	\PHzThree = - s_{\PH}\phi^{0,i} + c_{\PH}\Pchi^{0,i},
\end{equation}
where the vevs are parameterized by
\begin{equation}
	c_{\PH}\equiv \cos\theta_{\PH}= \frac{v_{\phi}}{v}, \qquad
	s_{\PH}\equiv \sin\theta_{\PH}= \frac{2\sqrt{2}\,v_{\Pchi}}{v}.
\end{equation}
The quantity $s_{\PH}^2$ represents the fraction of the gauge boson masses-squared $\MW^2$ and $\MZ^2$ that is generated by the vev of the triplets, while $c_{\PH}^2$ represents the fraction generated by the usual Higgs doublet.

The states of the custodial fiveplet $(\PSHpmpmFive,\ \PSHpmFive,\ \PHzFive )$ have a common mass $m_5$ and the states of the custodial triplet $(\PSHpmThree,\ \PHzThree)$ have a common mass $m_3$.
These masses can be written (after eliminating $\mu_2^2$ and $\mu_3^2$ in favor of the vevs) as\footnote{Note that the ratio $M_1/v_{\Pchi}$ can be written using the minimization condition $\partial V/ \partial v_{\Pchi} = 0$ as
\begin{equation}
	\frac{M_1}{v_{\Pchi}} = \frac{4}{v_{\phi}^2}
	\left[ \mu_3^2 + (2 \lambda_2 - \lambda_5) v_{\phi}^2
	+ 4(\lambda_3 + 3 \lambda_4) v_{\Pchi}^2 - 6 M_2 v_{\Pchi} \right],
\end{equation}
which is finite in the limit $v_{\Pchi} \to 0$.}
\begin{eqnarray}
	m_5^2 &=& \frac{M_1}{4 v_{\Pchi}} v_\phi^2 + 12 M_2 v_{\Pchi}
	+ \frac{3}{2} \lambda_5 v_{\phi}^2 + 8 \lambda_3 v_{\Pchi}^2, \\
	m_3^2 &=&  \frac{M_1}{4 v_{\Pchi}} (v_\phi^2 + 8 v_{\Pchi}^2)
	+ \frac{\lambda_5}{2} (v_{\phi}^2 + 8 v_{\Pchi}^2)
	= \left(  \frac{M_1}{4 v_{\Pchi}} + \frac{\lambda_5}{2} \right) v^2.
\end{eqnarray}

The two custodial-singlet mass eigenstates are given by
\begin{equation}
	\Ph = \cos \alpha \, \phi^{0,r} - \sin \alpha \, \PH_1^{0\prime},  \qquad
	\PH = \sin \alpha \, \phi^{0,r} + \cos \alpha \, \PH_1^{0\prime},
\end{equation}
where
\begin{equation}
	\PH_1^{0 \prime} = \sqrt{\frac{1}{3}} \Pxi^0 + \sqrt{\frac{2}{3}} \Pchi^{0,r}.
\end{equation}
The mixing angle and masses are given by
\begin{eqnarray}
	&&\sin 2 \alpha =  \frac{2 \mathcal{M}^2_{12}}{\MH ^2 - \Mh ^2},    \qquad
	\cos 2 \alpha =  \frac{ \mathcal{M}^2_{22} - \mathcal{M}^2_{11}  }{\MH ^2 - \Mh ^2},  \\
	&&m^2_{\Ph,\PH} = \frac{1}{2} \left[ \mathcal{M}_{11}^2 + \mathcal{M}_{22}^2
	\mp \sqrt{\left( \mathcal{M}_{11}^2 - \mathcal{M}_{22}^2 \right)^2
	+ 4 \left( \mathcal{M}_{12}^2 \right)^2} \right],
\end{eqnarray}
where we choose $\Mh  < \MH $, and
\begin{eqnarray}
	\mathcal{M}_{11}^2 &=& 8 \lambda_1 v_{\phi}^2, \nonumber \\
	\mathcal{M}_{12}^2 &=& \frac{\sqrt{3}}{2} v_{\phi}
	\left[ - M_1 + 4 \left(2 \lambda_2 - \lambda_5 \right) v_{\Pchi} \right], \nonumber \\
	\mathcal{M}_{22}^2 &=& \frac{M_1 v_{\phi}^2}{4 v_{\Pchi}} - 6 M_2 v_{\Pchi}
	+ 8 \left( \lambda_3 + 3 \lambda_4 \right) v_{\Pchi}^2.
\end{eqnarray}

The fiveplet states couple to vector bosons according to the following Feynman rules~\cite{Gunion:1989we,Godfrey:2010qb,Hartling:2014zca}:
\begin{eqnarray}
	\PHzFive  \PWp _{\mu} \PWm_{\nu} : && \!\!\!\!\!
	\sqrt{\frac{2}{3}} i g^2 v_{\Pchi} g_{\mu\nu}
	= 2 i \frac{\MW^2}{v} \left( \frac{s_{\PH}}{\sqrt{3}} \right) g_{\mu\nu}
	= 2 (\sqrt{2} G_F)^{1/2} \MW^2 \left( -\frac{s_{\PH}}{\sqrt{3}} \right) (-i g_{\mu\nu}),
	\nonumber \\
	\PHzFive  \PZ_{\mu} \PZ_{\nu} : && \!\!\!\!\!
	- \sqrt{\frac{8}{3}} i \frac{g^2 v_{\Pchi}}{c_W^2} g_{\mu\nu}
	= 2 i \frac{\MZ^2}{v} \left( -\frac{2 s_{\PH}}{\sqrt{3}} \right) g_{\mu\nu}
	= 2 (\sqrt{2} G_F)^{1/2} \MZ^2 \left( \frac{2 s_{\PH}}{\sqrt{3}} \right) (-i g_{\mu\nu}),
        \nonumber \\
	\PSHpFive  \PWm_{\mu} \PZ_{\nu} : && \!\!\!\!\!
	 -\sqrt{2} i \frac{g^2 v_{\Pchi}}{c_W} g_{\mu\nu}
		= 2 i \frac{\MW \MZ}{v} \left( -s_{\PH}\right) g_{\mu\nu}
		= 2 ( \sqrt{2} G_F )^{1/2} \MW \MZ \left( s_{\PH}\right) (-i g_{\mu\nu}),
	\nonumber \\
	\PSHppFive  \PWm_{\mu} \PWm_{\nu} :&& \!\!\!\!\!
	2 i g^2 v_{\Pchi} g_{\PGm\PGn}
	= 2 i \frac{\MW^2}{v} \left( \sqrt{2} s_{\PH}\right) g_{\mu\nu}
	= 2 (\sqrt{2} G_F)^{1/2} \MW^2 \left( -\sqrt{2} s_{\PH}\right) (-i g_{\mu\nu}),
\label{GM:eq:hwzvert}
\end{eqnarray}
where we write the coupling in multiple forms to make contact with the notation of Refs.~\cite{Godfrey:2010qb,Bolzoni:2011cu}.  The triplet vev $v_{\Pchi}$ is called $v^{\prime}$ in~\cite{Godfrey:2010qb}, and the
factors $F_{\PVV}$ in Eq.~(5.2) of Ref.~\cite{Bolzoni:2011cu} correspond in this model to
\begin{eqnarray}
    F_{\PWp \PWm} &=& -\frac{s_{\PH}}{\sqrt 3} \qquad \ \ (\PHzFive \textrm{ production}), \nonumber \\
    F_{\PZ\PZ } &=& \frac{2 s_{\PH}}{\sqrt 3} \qquad \quad (\PHzFive \textrm{ production}), \nonumber \\
    F_{\PWpm \PZ} &=& s_{\PH}\qquad \qquad (\PSHpmFive \textrm{ production}), \nonumber \\
    F_{\PWpm \PWpm} &=& -\sqrt{2} s_{\PH}\qquad (\PSHpmpmFive\textrm{ production}).
\end{eqnarray}
Note in particular that, for $\PHzFive $, one cannot simply rescale the vector boson fusion cross section of the SM Higgs boson because the ratio of $\PW\PW$ to $\PZ\PZ $ couplings is different than in the SM (for SM Higgs boson production, $F_{\PWp \PWm} = F_{\PZ\PZ } = -1$).

The VBF production cross sections for a single $\PHzFive$ state depend only on the two parameters $m_5$ and $s_{\PH}$.  If the spectrum is such that decays of $\PHzFive$ to $\PHThree \PV$ or $\PHThree\PHThree $ are kinematically inaccessible, the total decay widths of the $\PHzFive$ states to vector boson pairs also depend only on $m_5$ and $s_{\PH}$.

\subsection{H5plane benchmark}
\label{GM:sec:benchmark}

The purpose of the H5plane benchmark is to facilitate searches for the $\PHFive $ states over the $m_5$--$s_{\PH}$ plane.  The other parameters are chosen so that $m_3 > m_5$ (thereby forbidding decays of $\PHFive  \to \PHThree \PV$ or $\PHThree \PHThree$) and so that the largest possible parameter region is allowed by theoretical constraints for $200 \UGeV < m_5 < 3000 \UGeV$.

The benchmark is defined as follows.  The nine parameters of the GM model scalar potential in Eq.~(\ref{GM:eq:potential}) are fixed in terms of the input parameters $G_F$, $\Mh $, $m_5$, $s_{\PH}$, $\lambda_2$, $\lambda_3$, $\lambda_4$, $M_1$, and $M_2$, i.e., the input parameters of {\sc INPUTSET = 4} in {\tt GMCALC}.  The values are given in Table~\ref{GM:tab:H5plane}.

\begin{table}
\caption{Specification of the H5plane benchmark for the Georgi-Machacek model.  These input parameters correspond to {\sc INPUTSET = 4} in {\tt GMCALC}~\cite{Hartling:2014xma}.}
\label{GM:tab:H5plane}
\begin{center}
\begin{tabular}{l |l |l}
\toprule
Fixed parameters & Variable parameters & Dependent parameters \\
\midrule
$G_F = 1.1663787 \times 10^{-5} \UGeV^{-2}$ & $m_5 \in [200, 3000] \UGeV$ & $\lambda_2 = 0.4 (m_5/1000 \UGeV)$ \\
$\Mh  = 125 \UGeV$ & $s_{\PH}\in (0,1)$ & $M_1 = \sqrt{2} s_{\PH}(m_5^2 + v^2)/v$ \\
$\lambda_3 = -0.1$ & & $M_2 = M_1/6$ \\
$\lambda_4 = 0.2$ & & \\
\bottomrule
\end{tabular}
\end{center}
\end{table}

Perturbative unitarity of $\PVV \to \PVV$ scattering amplitudes constrains~\cite{Logan:2015xpa}
\begin{equation}
	s_{\PH}^2 < \frac{3}{5} \frac{(16 \pi v^2 - 5 \Mh ^2)}{(4 m_5^2 + 5 \Mh ^2)},
	\qquad {\rm or} \qquad
	s_{\PH}^2 \lesssim \left(\frac{667 \UGeV}{m_5} \right)^2,
	\label{GM:eq:sHbound}
\end{equation}
for $m_5 \gg \MW$, $\MZ$, $\Mh$.  This implies that $\Gamma(\PHFive \to \PVV) \lesssim 0.15 \, m_5$ for each of the $\PHFive $ states.  The full set of perturbative unitarity constraints~\cite{Aoki:2007ah,Hartling:2014zca} constrain the model a little more tightly, leading to $\Gamma(\PHFive \to \PVV) \lesssim 0.10 \, m_5$.  These constraints are implemented in {\tt GMCALC}, which will return an error message if they are violated.

The H5plane benchmark has the following features:
\begin{itemize}
\item It populates nearly all of the theoretically-allowed region of the $m_5$--$s_{\PH}$ plane for $m_5 \in [200,3000]$ \UGeV, except for a small corner at low $m_5$ and high $s_{\PH}$ which is already excluded by the cross section for like-sign $\PW$ boson pair production in VBF~\cite{Chiang:2014bia} (this limits the maximum allowed production cross section for VBF$\to \PSHpmpmFive \to \PWpm  \PWpm $, and hence sets an upper bound on $s_{\PH}$ as a function of $m_5$).
\item Constraints from $\PQb \to \PQs \PGg$ (see Ref.~\cite{Hartling:2014aga}) eliminate only points that are already excluded by the cross section for like-sign $\PW$ boson pair production in VBF~\cite{Chiang:2014bia}.
\item The benchmark is not unreasonably constrained by coupling measurements of the 125~\UGeV\ Higgs boson: the region of the $m_5$--$s_{\PH}$ plane in which $| \kappa_i^{\Ph} - 1| < 0.1$, with $i = \PQf, \PV, \PGg$, is essentially the same in the H5plane benchmark as in a full parameter scan.
\item It has $m_3 \gtrsim m_5 + 10$~\UGeV\ over the whole benchmark plane, so that the Higgs-to-Higgs boson decays $\PHFive \to \PHThree \PHThree$ and $\PHFive \to \PHThree \PV$ are kinematically forbidden, leaving only the decays $\PHFive \to \PVV$ at tree level; i.e., to a very good approximation,
\begin{equation}
	{\rm BR}(\PHzFive  \to \PWp \PWm, \PZ\PZ )
	= {\rm BR}(\PSHpmFive \to \PWpm  \PZ)
	= {\rm BR}(\PSHpmpmFive \to \PWpm  \PWpm )
	= 1.
\end{equation}
\item It has $\MH  \gtrsim m_5 + 12$~\UGeV\ over the whole benchmark plane, except for a few points at $s_{\PH}> 0.7$ which are already excluded by the cross section for like-sign $\PW$ boson pair production in VBF~\cite{Chiang:2014bia}.  However, there is a large region of parameter space covering $m_5 \gtrsim 600$~\UGeV\ and $0.07 \lesssim s_{\PH}\lesssim 0.6$ in which the total decay widths of $\PHzFive $ and $\PH$ are larger than the mass splitting between these two states.  In this region, a dedicated study of the lineshape and interference effects of the two resonances in VBF$\to (\PHzFive,\ \PH) \to \PW\PW,\PZ\PZ $ will be required.
\end{itemize}

\subsection{Vector boson fusion production cross sections of the \texorpdfstring{$\HepParticle{\PH}{5}{}$}{H_5} states}
\label{GM:sec:xsec}

The total cross sections for production of $\PHzFive $, $\PSHpmFive$, and $\PSHpmpmFive$ in VBF can be computed up to NNLO accuracy using the
{\tt VBF@NNLO} code~\cite{Bolzoni:2010xr,Zaro:2010fc,Bolzoni:2011cu}, via
the structure-function approach. This approach~\cite{Han:1992hr} consists in considering the VBF process as a double
deep-inelastic scattering (DIS) attached to the colourless pure electroweak vector-boson fusion into a
Higgs boson. According to this approach one can include next-to-leading order (NLO) QCD corrections to the VBF process
employing the standard DIS structure functions $F_i(x, Q^2)$;  $i = 1, 2, 3$ at NLO~\cite{Bardeen:1978yd} or similarly the
corresponding structure functions at NNLO~\cite{Kazakov:1990fu,Zijlstra:1992kj,Zijlstra:1992qd,Moch:1999eb}.

Although the effective factorization underlying the structure-function approach holds to a very good approximation up to NNLO,
it formally does not include all types
of contributions. At leading order (LO) an additional contribution arises from the interference between identical final-state
quarks (e.g., $\PQu\PQu \to \PH\PQu\PQu$) or between processes where either a $\PW$ or a $\PZ$ boson can be exchanged (e.g.,
$\PQu\PQd \to \PH\PQu\PQd$). These LO contributions are known to be extremely small (less than 0.1\% of the total cross-section). Apart from such contributions, the structure-function approach is exact up to NLO. At
NNLO, however, several types of diagrams violate the underlying factorization. Their impact on the total
rate has been computed or estimated in~\cite{Bolzoni:2011cu} and found to be negligible. Some of them are colour
and kinematically suppressed~\cite{vanNeerven:1984ak,Blumlein:1992eh,Figy:2007kv}, and others have been shown in~\cite{Harlander:2008xn} to
be small enough not to produce a significant deterioration of the VBF signal.

NLO electroweak corrections are known for SM Higgs boson production 
in VBF~\cite{Ciccolini:2007jr,Ciccolini:2007ec}, but not for any beyond-the-SM scenario, and therefore are not included in the numbers shown here.

To produce the numbers shown in this chapter, we have used the following electroweak parameters: 
\begin{eqnarray}
    G_F=1.1663787 \cdot 10^{-5} \UGeV^{-2}, \quad &\MW = 80.385 \UGeV, \quad &\MZ = 91.1876 \UGeV,\nonumber\\
    &\GW = 2.085 \UGeV, \quad &\GZ = 2.4952 \UGeV.
    \label{eq:inputs}
\end{eqnarray}
The $\PHFive \PVV$ vertices have the form given in Eq.~(\ref{GM:eq:hwzvert}), and we have set $s_{\PH}= 1$.
The production cross sections for other values of $s_{\PH}$ are conveniently obtained using the relation
\begin{equation}
	\sigma^{\rm NNLO}({\rm VBF} \to \PHFive) = s_{\PH}^2 \sigma_1^{\rm NNLO}({\rm VBF} \to \PHFive),
\end{equation}
where $\sigma_1^{\rm NNLO}$ represents the NNLO cross section for $s_{\PH}= 1$. The values of $\sigma_1^{\rm NNLO}$ computed for the 13~\UTeV\ LHC for $\PHzFive $, $\PSHpFive $, and $\PSHmFive $ are shown in Tables~\ref{GM:tab:xsec}--\ref{GM:tab:xsec2}, and for $\PSHppFive $ and $\PSHmmFive $ are shown in Tables~\ref{GM:tab:xsec3}--\ref{GM:tab:xsec4}.

We have employed the PDF4LHC NNLO parton distribution function~\cite{Butterworth:2015oua} with 30 sets (Hessian error estimate) plus 2 sets to estimate the $\alpha_s$ systematic uncertainties.  For the NNLO PDF set, $\alpha_s(\MZ) = 0.118$.  As is the case of the SM, the systematic uncertainty from $\alpha_s$ is rather small for VBF.
The renormalization and factorization scales have been set to $\MW$.  Scale uncertainties have been computed by varying the two scales independently by a factor in the range $[1/2, 2]$.

As for SM Higgs boson production in VBF, the impact of QCD corrections is well under control: with our setup, and using PDF sets with QCD evolution consistent with
the perturbative order of the cross-section, NLO QCD corrections increase the LO cross-section by $6{-}7\%$ and NNLO corrections contribute at most another $1\%$ to
the cross-section. The inclusion of NNLO corrections reduces the QCD scale uncertainties to the 1\% level or below, while PDF uncertainties are at the level of $2\%$ of
the cross-section.

For the uncertainty due to uncalculated NLO electroweak corrections, we suggest to adopt a fractional uncertainty of $\pm 7\%$.  This encompasses the size of the NLO electroweak correction to the SM Higgs VBF cross section~\cite{Heinemeyer:2013tqa} for SM Higgs boson masses below 700~\UGeV, where tree-level perturbative unitarity constraints in $2 \to 2$ gauge and Higgs boson scattering are satisfied.  This same tree-level perturbativity requirement results in the upper bound on $s_{\PH}$ given in Eq.~(\ref{GM:eq:sHbound}).

\begin{table}
\vspace{-0.7cm}
\caption{VBF production cross sections for $\PHzFive $, $\PSHpFive $ and $\PSHmFive $ in the GM model, computed for $s_{\PH}= 1$ at the $\sqrt s = 13~\UTeV$ LHC.  The first (asymmetric) uncertainties are the QCD scale uncertainty, the second is the PDF uncertainty, and the third is the $\alpha_s$ uncertainty.  The uncertainty from uncalculated NLO electroweak corrections should be taken as $\pm 7\%$.  The relative Monte Carlo numerical integration error is below $5 \times 10^{-4}$ in all cases.}
\label{GM:tab:xsec}
\centering
\small
 \begin{tabular}{c| c| c| c}
 \toprule
 $m_5$ [\UGeV] & $\sigma_1^{\rm NNLO}(\PHzFive )$ [fb] &  $\sigma_1^{\rm NNLO}(\PSHpFive )$ [fb] &  $\sigma_1^{\rm NNLO}(\PSHmFive )$ [fb] \\
 \midrule
$ 200.$ & $      1375.    ^{+ 0.35\%}_{- 0.20\%} \pm  1.8\% \pm  0.51\%$ & $      1770.    ^{+ 0.30\%}_{- 0.18\%} \pm  1.6\% \pm  0.46\%$ & $      1148.    ^{+ 0.36\%}_{- 0.21\%} \pm  2.2\% \pm  0.54\%$ \\
$ 210.$ & $      1288.    ^{+ 0.33\%}_{- 0.19\%} \pm  1.8\% \pm  0.49\%$ & $      1662.    ^{+ 0.28\%}_{- 0.17\%} \pm  1.7\% \pm  0.45\%$ & $      1073.    ^{+ 0.34\%}_{- 0.21\%} \pm  2.2\% \pm  0.53\%$ \\
$ 220.$ & $      1209.    ^{+ 0.30\%}_{- 0.18\%} \pm  1.8\% \pm  0.48\%$ & $      1564.    ^{+ 0.26\%}_{- 0.17\%} \pm  1.7\% \pm  0.44\%$ & $      1004.    ^{+ 0.32\%}_{- 0.20\%} \pm  2.2\% \pm  0.52\%$ \\
$ 230.$ & $      1136.    ^{+ 0.28\%}_{- 0.17\%} \pm  1.8\% \pm  0.47\%$ & $      1473.    ^{+ 0.25\%}_{- 0.16\%} \pm  1.7\% \pm  0.43\%$ & $      940.9    ^{+ 0.31\%}_{- 0.19\%} \pm  2.2\% \pm  0.51\%$ \\
$ 240.$ & $      1069.    ^{+ 0.26\%}_{- 0.17\%} \pm  1.8\% \pm  0.46\%$ & $      1388.    ^{+ 0.25\%}_{- 0.15\%} \pm  1.7\% \pm  0.42\%$ & $      883.0    ^{+ 0.29\%}_{- 0.18\%} \pm  2.3\% \pm  0.50\%$ \\
$ 250.$ & $      1006.    ^{+ 0.27\%}_{- 0.16\%} \pm  1.8\% \pm  0.46\%$ & $      1311.    ^{+ 0.25\%}_{- 0.14\%} \pm  1.7\% \pm  0.41\%$ & $      829.6    ^{+ 0.27\%}_{- 0.17\%} \pm  2.3\% \pm  0.49\%$ \\
$ 260.$ & $      948.9    ^{+ 0.27\%}_{- 0.15\%} \pm  1.8\% \pm  0.45\%$ & $      1239.    ^{+ 0.25\%}_{- 0.14\%} \pm  1.7\% \pm  0.40\%$ & $      780.4    ^{+ 0.27\%}_{- 0.17\%} \pm  2.3\% \pm  0.48\%$ \\
$ 270.$ & $      895.7    ^{+ 0.27\%}_{- 0.15\%} \pm  1.8\% \pm  0.44\%$ & $      1172.    ^{+ 0.25\%}_{- 0.13\%} \pm  1.7\% \pm  0.39\%$ & $      734.9    ^{+ 0.27\%}_{- 0.16\%} \pm  2.3\% \pm  0.48\%$ \\
$ 280.$ & $      846.3    ^{+ 0.27\%}_{- 0.14\%} \pm  1.8\% \pm  0.43\%$ & $      1110.    ^{+ 0.25\%}_{- 0.13\%} \pm  1.7\% \pm  0.38\%$ & $      692.8    ^{+ 0.28\%}_{- 0.15\%} \pm  2.3\% \pm  0.47\%$ \\
$ 290.$ & $      800.5    ^{+ 0.27\%}_{- 0.14\%} \pm  1.8\% \pm  0.42\%$ & $      1052.    ^{+ 0.26\%}_{- 0.12\%} \pm  1.7\% \pm  0.37\%$ & $      653.8    ^{+ 0.28\%}_{- 0.14\%} \pm  2.3\% \pm  0.46\%$ \\
$ 300.$ & $      757.8    ^{+ 0.27\%}_{- 0.13\%} \pm  1.8\% \pm  0.41\%$ & $      997.7    ^{+ 0.26\%}_{- 0.11\%} \pm  1.7\% \pm  0.37\%$ & $      617.5    ^{+ 0.28\%}_{- 0.14\%} \pm  2.3\% \pm  0.45\%$ \\
$ 310.$ & $      718.0    ^{+ 0.28\%}_{- 0.12\%} \pm  1.8\% \pm  0.40\%$ & $      947.3    ^{+ 0.26\%}_{- 0.10\%} \pm  1.7\% \pm  0.36\%$ & $      583.9    ^{+ 0.28\%}_{- 0.13\%} \pm  2.4\% \pm  0.45\%$ \\
$ 320.$ & $      680.9    ^{+ 0.28\%}_{- 0.12\%} \pm  1.8\% \pm  0.40\%$ & $      900.3    ^{+ 0.26\%}_{- 0.10\%} \pm  1.7\% \pm  0.35\%$ & $      552.6    ^{+ 0.28\%}_{- 0.13\%} \pm  2.4\% \pm  0.44\%$ \\
$ 330.$ & $      646.3    ^{+ 0.28\%}_{- 0.11\%} \pm  1.8\% \pm  0.39\%$ & $      856.2    ^{+ 0.27\%}_{- 0.09\%} \pm  1.7\% \pm  0.34\%$ & $      523.4    ^{+ 0.28\%}_{- 0.13\%} \pm  2.4\% \pm  0.43\%$ \\
$ 340.$ & $      614.0    ^{+ 0.28\%}_{- 0.11\%} \pm  1.9\% \pm  0.38\%$ & $      815.0    ^{+ 0.27\%}_{- 0.09\%} \pm  1.7\% \pm  0.33\%$ & $      496.1    ^{+ 0.28\%}_{- 0.12\%} \pm  2.4\% \pm  0.42\%$ \\
$ 350.$ & $      583.7    ^{+ 0.28\%}_{- 0.10\%} \pm  1.9\% \pm  0.37\%$ & $      776.3    ^{+ 0.27\%}_{- 0.08\%} \pm  1.7\% \pm  0.32\%$ & $      470.7    ^{+ 0.28\%}_{- 0.12\%} \pm  2.4\% \pm  0.42\%$ \\
$ 360.$ & $      555.2    ^{+ 0.28\%}_{- 0.10\%} \pm  1.9\% \pm  0.37\%$ & $      739.9    ^{+ 0.27\%}_{- 0.08\%} \pm  1.7\% \pm  0.31\%$ & $      446.9    ^{+ 0.28\%}_{- 0.11\%} \pm  2.4\% \pm  0.41\%$ \\
$ 370.$ & $      528.6    ^{+ 0.28\%}_{- 0.09\%} \pm  1.9\% \pm  0.36\%$ & $      705.8    ^{+ 0.27\%}_{- 0.08\%} \pm  1.7\% \pm  0.31\%$ & $      424.6    ^{+ 0.28\%}_{- 0.10\%} \pm  2.5\% \pm  0.41\%$ \\
$ 380.$ & $      503.6    ^{+ 0.28\%}_{- 0.09\%} \pm  1.9\% \pm  0.35\%$ & $      673.7    ^{+ 0.27\%}_{- 0.07\%} \pm  1.7\% \pm  0.30\%$ & $      403.7    ^{+ 0.28\%}_{- 0.10\%} \pm  2.5\% \pm  0.40\%$ \\
$ 390.$ & $      480.0    ^{+ 0.28\%}_{- 0.08\%} \pm  1.9\% \pm  0.34\%$ & $      643.4    ^{+ 0.27\%}_{- 0.06\%} \pm  1.7\% \pm  0.29\%$ & $      384.1    ^{+ 0.28\%}_{- 0.09\%} \pm  2.5\% \pm  0.39\%$ \\
$ 400.$ & $      457.9    ^{+ 0.28\%}_{- 0.07\%} \pm  1.9\% \pm  0.34\%$ & $      614.9    ^{+ 0.27\%}_{- 0.06\%} \pm  1.7\% \pm  0.28\%$ & $      365.7    ^{+ 0.28\%}_{- 0.09\%} \pm  2.5\% \pm  0.39\%$ \\
$ 410.$ & $      437.1    ^{+ 0.28\%}_{- 0.07\%} \pm  1.9\% \pm  0.33\%$ & $      588.0    ^{+ 0.27\%}_{- 0.05\%} \pm  1.7\% \pm  0.28\%$ & $      348.4    ^{+ 0.28\%}_{- 0.08\%} \pm  2.5\% \pm  0.38\%$ \\
$ 420.$ & $      417.4    ^{+ 0.28\%}_{- 0.06\%} \pm  1.9\% \pm  0.32\%$ & $      562.6    ^{+ 0.27\%}_{- 0.05\%} \pm  1.7\% \pm  0.27\%$ & $      332.1    ^{+ 0.28\%}_{- 0.07\%} \pm  2.5\% \pm  0.38\%$ \\
$ 430.$ & $      398.9    ^{+ 0.28\%}_{- 0.06\%} \pm  1.9\% \pm  0.32\%$ & $      538.5    ^{+ 0.27\%}_{- 0.04\%} \pm  1.7\% \pm  0.26\%$ & $      316.8    ^{+ 0.29\%}_{- 0.06\%} \pm  2.5\% \pm  0.37\%$ \\
$ 440.$ & $      381.4    ^{+ 0.28\%}_{- 0.06\%} \pm  1.9\% \pm  0.31\%$ & $      515.8    ^{+ 0.27\%}_{- 0.06\%} \pm  1.7\% \pm  0.25\%$ & $      302.3    ^{+ 0.29\%}_{- 0.06\%} \pm  2.6\% \pm  0.36\%$ \\
$ 450.$ & $      364.9    ^{+ 0.28\%}_{- 0.05\%} \pm  1.9\% \pm  0.30\%$ & $      494.3    ^{+ 0.27\%}_{- 0.07\%} \pm  1.7\% \pm  0.24\%$ & $      288.7    ^{+ 0.28\%}_{- 0.06\%} \pm  2.6\% \pm  0.36\%$ \\
$ 460.$ & $      349.2    ^{+ 0.28\%}_{- 0.05\%} \pm  1.9\% \pm  0.30\%$ & $      473.9    ^{+ 0.27\%}_{- 0.08\%} \pm  1.7\% \pm  0.24\%$ & $      275.9    ^{+ 0.28\%}_{- 0.06\%} \pm  2.6\% \pm  0.35\%$ \\
$ 470.$ & $      334.4    ^{+ 0.28\%}_{- 0.06\%} \pm  1.9\% \pm  0.29\%$ & $      454.6    ^{+ 0.27\%}_{- 0.09\%} \pm  1.7\% \pm  0.23\%$ & $      263.7    ^{+ 0.28\%}_{- 0.06\%} \pm  2.6\% \pm  0.35\%$ \\
$ 480.$ & $      320.4    ^{+ 0.28\%}_{- 0.07\%} \pm  1.9\% \pm  0.28\%$ & $      436.3    ^{+ 0.28\%}_{- 0.10\%} \pm  1.7\% \pm  0.22\%$ & $      252.2    ^{+ 0.28\%}_{- 0.07\%} \pm  2.6\% \pm  0.34\%$ \\
$ 490.$ & $      307.1    ^{+ 0.28\%}_{- 0.08\%} \pm  1.9\% \pm  0.28\%$ & $      418.9    ^{+ 0.28\%}_{- 0.12\%} \pm  1.7\% \pm  0.22\%$ & $      241.4    ^{+ 0.28\%}_{- 0.08\%} \pm  2.6\% \pm  0.34\%$ \\
$ 500.$ & $      294.5    ^{+ 0.28\%}_{- 0.10\%} \pm  2.0\% \pm  0.27\%$ & $      402.4    ^{+ 0.28\%}_{- 0.13\%} \pm  1.7\% \pm  0.21\%$ & $      231.1    ^{+ 0.28\%}_{- 0.09\%} \pm  2.7\% \pm  0.33\%$ \\
$ 550.$ & $      240.4    ^{+ 0.28\%}_{- 0.15\%} \pm  2.0\% \pm  0.24\%$ & $      331.0    ^{+ 0.28\%}_{- 0.18\%} \pm  1.8\% \pm  0.18\%$ & $      187.0    ^{+ 0.28\%}_{- 0.15\%} \pm  2.7\% \pm  0.31\%$ \\
$ 600.$ & $      198.0    ^{+ 0.28\%}_{- 0.20\%} \pm  2.0\% \pm  0.21\%$ & $      274.8    ^{+ 0.28\%}_{- 0.24\%} \pm  1.8\% \pm  0.14\%$ & $      152.9    ^{+ 0.28\%}_{- 0.21\%} \pm  2.8\% \pm  0.29\%$ \\
$ 650.$ & $      164.5    ^{+ 0.28\%}_{- 0.26\%} \pm  2.1\% \pm  0.19\%$ & $      230.0    ^{+ 0.28\%}_{- 0.29\%} \pm  1.8\% \pm  0.11\%$ & $      126.1    ^{+ 0.28\%}_{- 0.26\%} \pm  2.9\% \pm  0.27\%$ \\
$ 700.$ & $      137.7    ^{+ 0.29\%}_{- 0.32\%} \pm  2.1\% \pm  0.16\%$ & $      193.8    ^{+ 0.28\%}_{- 0.34\%} \pm  1.8\% \pm  0.08\%$ & $      104.8    ^{+ 0.28\%}_{- 0.32\%} \pm  3.0\% \pm  0.25\%$ \\
$ 750.$ & $      115.9    ^{+ 0.29\%}_{- 0.36\%} \pm  2.1\% \pm  0.14\%$ & $      164.3    ^{+ 0.29\%}_{- 0.39\%} \pm  1.8\% \pm  0.05\%$ & $      87.64    ^{+ 0.28\%}_{- 0.37\%} \pm  3.1\% \pm  0.23\%$ \\
$ 800.$ & $      98.20    ^{+ 0.29\%}_{- 0.41\%} \pm  2.2\% \pm  0.11\%$ & $      140.1    ^{+ 0.29\%}_{- 0.43\%} \pm  1.8\% \pm  0.02\%$ & $      73.75    ^{+ 0.29\%}_{- 0.42\%} \pm  3.2\% \pm  0.21\%$ \\
$ 850.$ & $      83.60    ^{+ 0.29\%}_{- 0.46\%} \pm  2.2\% \pm  0.09\%$ & $      120.0    ^{+ 0.29\%}_{- 0.48\%} \pm  1.8\% \pm  0.00\%$ & $      62.39    ^{+ 0.29\%}_{- 0.47\%} \pm  3.2\% \pm  0.20\%$ \\
$ 900.$ & $      71.50    ^{+ 0.29\%}_{- 0.51\%} \pm  2.2\% \pm  0.07\%$ & $      103.3    ^{+ 0.29\%}_{- 0.53\%} \pm  1.9\% \pm  0.03\%$ & $      53.03    ^{+ 0.29\%}_{- 0.52\%} \pm  3.3\% \pm  0.18\%$ \\
 \bottomrule
 \end{tabular}
\end{table}

\begin{table}
\vspace{-0.4cm}
\caption{Continuation of Table~\ref{GM:tab:xsec}.}
\label{GM:tab:xsec2}
\centering
\small
 \begin{tabular}{c| c| c| c}
 \toprule
 $m_5$ [\UGeV] & $\sigma_1^{\rm NNLO}(\PHzFive )$ [fb] &  $\sigma_1^{\rm NNLO}(\PSHpFive )$ [fb] &  $\sigma_1^{\rm NNLO}(\PSHmFive )$ [fb] \\
 \midrule
$ 950.$ & $      61.41    ^{+ 0.29\%}_{- 0.55\%} \pm  2.3\% \pm  0.05\%$ & $      89.21    ^{+ 0.29\%}_{- 0.57\%} \pm  1.9\% \pm  0.06\%$ & $      45.27    ^{+ 0.29\%}_{- 0.57\%} \pm  3.4\% \pm  0.17\%$ \\
$1000.$ & $      52.94    ^{+ 0.30\%}_{- 0.60\%} \pm  2.3\% \pm  0.03\%$ & $      77.35    ^{+ 0.29\%}_{- 0.62\%} \pm  1.9\% \pm  0.08\%$ & $      38.80    ^{+ 0.29\%}_{- 0.62\%} \pm  3.5\% \pm  0.16\%$ \\
$1050.$ & $      45.79    ^{+ 0.30\%}_{- 0.64\%} \pm  2.4\% \pm  0.01\%$ & $      67.28    ^{+ 0.30\%}_{- 0.66\%} \pm  1.9\% \pm  0.11\%$ & $      33.38    ^{+ 0.29\%}_{- 0.67\%} \pm  3.6\% \pm  0.15\%$ \\
$1100.$ & $      39.74    ^{+ 0.30\%}_{- 0.69\%} \pm  2.4\% \pm  0.00\%$ & $      58.70    ^{+ 0.30\%}_{- 0.71\%} \pm  1.9\% \pm  0.13\%$ & $      28.81    ^{+ 0.30\%}_{- 0.72\%} \pm  3.7\% \pm  0.14\%$ \\
$1150.$ & $      34.58    ^{+ 0.31\%}_{- 0.74\%} \pm  2.4\% \pm  0.02\%$ & $      51.34    ^{+ 0.30\%}_{- 0.75\%} \pm  1.9\% \pm  0.15\%$ & $      24.93    ^{+ 0.30\%}_{- 0.77\%} \pm  3.8\% \pm  0.13\%$ \\
$1200.$ & $      30.17    ^{+ 0.30\%}_{- 0.79\%} \pm  2.5\% \pm  0.04\%$ & $      45.03    ^{+ 0.30\%}_{- 0.80\%} \pm  2.0\% \pm  0.17\%$ & $      21.64    ^{+ 0.30\%}_{- 0.81\%} \pm  3.9\% \pm  0.12\%$ \\
$1250.$ & $      26.39    ^{+ 0.31\%}_{- 0.83\%} \pm  2.5\% \pm  0.05\%$ & $      39.58    ^{+ 0.33\%}_{- 0.84\%} \pm  2.0\% \pm  0.20\%$ & $      18.83    ^{+ 0.32\%}_{- 0.86\%} \pm  4.0\% \pm  0.11\%$ \\
$1300.$ & $      23.13    ^{+ 0.34\%}_{- 0.87\%} \pm  2.6\% \pm  0.07\%$ & $      34.86    ^{+ 0.35\%}_{- 0.88\%} \pm  2.0\% \pm  0.22\%$ & $      16.43    ^{+ 0.33\%}_{- 0.91\%} \pm  4.1\% \pm  0.11\%$ \\
$1350.$ & $      20.32    ^{+ 0.36\%}_{- 0.92\%} \pm  2.6\% \pm  0.08\%$ & $      30.77    ^{+ 0.37\%}_{- 0.92\%} \pm  2.0\% \pm  0.24\%$ & $      14.36    ^{+ 0.36\%}_{- 0.95\%} \pm  4.2\% \pm  0.10\%$ \\
$1400.$ & $      17.88    ^{+ 0.38\%}_{- 0.96\%} \pm  2.7\% \pm  0.09\%$ & $      27.20    ^{+ 0.39\%}_{- 0.97\%} \pm  2.0\% \pm  0.26\%$ & $      12.58    ^{+ 0.38\%}_{- 1.00\%} \pm  4.3\% \pm  0.10\%$ \\
$1450.$ & $      15.77    ^{+ 0.40\%}_{- 1.00\%} \pm  2.7\% \pm  0.11\%$ & $      24.09    ^{+ 0.41\%}_{- 1.01\%} \pm  2.1\% \pm  0.28\%$ & $      11.04    ^{+ 0.40\%}_{- 1.04\%} \pm  4.4\% \pm  0.10\%$ \\
$1500.$ & $      13.92    ^{+ 0.43\%}_{- 1.05\%} \pm  2.7\% \pm  0.12\%$ & $      21.37    ^{+ 0.43\%}_{- 1.05\%} \pm  2.1\% \pm  0.30\%$ & $      9.704    ^{+ 0.43\%}_{- 1.09\%} \pm  4.5\% \pm  0.09\%$ \\
$1550.$ & $      12.32    ^{+ 0.45\%}_{- 1.09\%} \pm  2.8\% \pm  0.13\%$ & $      18.98    ^{+ 0.45\%}_{- 1.10\%} \pm  2.1\% \pm  0.32\%$ & $      8.545    ^{+ 0.46\%}_{- 1.14\%} \pm  4.6\% \pm  0.09\%$ \\
$1600.$ & $      10.91    ^{+ 0.47\%}_{- 1.14\%} \pm  2.8\% \pm  0.14\%$ & $      16.89    ^{+ 0.47\%}_{- 1.14\%} \pm  2.1\% \pm  0.34\%$ & $      7.536    ^{+ 0.48\%}_{- 1.19\%} \pm  4.7\% \pm  0.09\%$ \\
$1650.$ & $      9.677    ^{+ 0.50\%}_{- 1.18\%} \pm  2.9\% \pm  0.15\%$ & $      15.04    ^{+ 0.50\%}_{- 1.19\%} \pm  2.1\% \pm  0.36\%$ & $      6.656    ^{+ 0.50\%}_{- 1.23\%} \pm  4.8\% \pm  0.09\%$ \\
$1700.$ & $      8.594    ^{+ 0.51\%}_{- 1.23\%} \pm  2.9\% \pm  0.16\%$ & $      13.41    ^{+ 0.52\%}_{- 1.23\%} \pm  2.2\% \pm  0.37\%$ & $      5.886    ^{+ 0.53\%}_{- 1.28\%} \pm  4.9\% \pm  0.09\%$ \\
$1750.$ & $      7.641    ^{+ 0.54\%}_{- 1.28\%} \pm  3.0\% \pm  0.17\%$ & $      11.97    ^{+ 0.54\%}_{- 1.28\%} \pm  2.2\% \pm  0.39\%$ & $      5.211    ^{+ 0.55\%}_{- 1.33\%} \pm  5.0\% \pm  0.09\%$ \\
$1800.$ & $      6.802    ^{+ 0.56\%}_{- 1.33\%} \pm  3.0\% \pm  0.18\%$ & $      10.70    ^{+ 0.55\%}_{- 1.33\%} \pm  2.2\% \pm  0.41\%$ & $      4.620    ^{+ 0.57\%}_{- 1.38\%} \pm  5.1\% \pm  0.09\%$ \\
$1850.$ & $      6.061    ^{+ 0.58\%}_{- 1.38\%} \pm  3.1\% \pm  0.19\%$ & $      9.571    ^{+ 0.58\%}_{- 1.37\%} \pm  2.2\% \pm  0.43\%$ & $      4.100    ^{+ 0.60\%}_{- 1.43\%} \pm  5.3\% \pm  0.09\%$ \\
$1900.$ & $      5.405    ^{+ 0.61\%}_{- 1.41\%} \pm  3.1\% \pm  0.20\%$ & $      8.568    ^{+ 0.60\%}_{- 1.42\%} \pm  2.2\% \pm  0.44\%$ & $      3.642    ^{+ 0.62\%}_{- 1.48\%} \pm  5.4\% \pm  0.10\%$ \\
$1950.$ & $      4.826    ^{+ 0.64\%}_{- 1.46\%} \pm  3.2\% \pm  0.20\%$ & $      7.678    ^{+ 0.63\%}_{- 1.47\%} \pm  2.2\% \pm  0.46\%$ & $      3.239    ^{+ 0.65\%}_{- 1.53\%} \pm  5.5\% \pm  0.10\%$ \\
$2000.$ & $      4.312    ^{+ 0.67\%}_{- 1.51\%} \pm  3.3\% \pm  0.21\%$ & $      6.885    ^{+ 0.66\%}_{- 1.51\%} \pm  2.3\% \pm  0.47\%$ & $      2.883    ^{+ 0.67\%}_{- 1.58\%} \pm  5.7\% \pm  0.11\%$ \\
$2050.$ & $      3.856    ^{+ 0.69\%}_{- 1.55\%} \pm  3.3\% \pm  0.22\%$ & $      6.179    ^{+ 0.68\%}_{- 1.56\%} \pm  2.3\% \pm  0.49\%$ & $      2.568    ^{+ 0.71\%}_{- 1.63\%} \pm  5.8\% \pm  0.11\%$ \\
$2100.$ & $      3.450    ^{+ 0.71\%}_{- 1.60\%} \pm  3.4\% \pm  0.22\%$ & $      5.549    ^{+ 0.70\%}_{- 1.61\%} \pm  2.3\% \pm  0.50\%$ & $      2.289    ^{+ 0.73\%}_{- 1.68\%} \pm  5.9\% \pm  0.12\%$ \\
$2150.$ & $      3.090    ^{+ 0.73\%}_{- 1.65\%} \pm  3.4\% \pm  0.23\%$ & $      4.986    ^{+ 0.73\%}_{- 1.65\%} \pm  2.3\% \pm  0.52\%$ & $      2.043    ^{+ 0.75\%}_{- 1.72\%} \pm  6.1\% \pm  0.13\%$ \\
$2200.$ & $      2.769    ^{+ 0.76\%}_{- 1.70\%} \pm  3.5\% \pm  0.23\%$ & $      4.484    ^{+ 0.76\%}_{- 1.71\%} \pm  2.3\% \pm  0.53\%$ & $      1.824    ^{+ 0.78\%}_{- 1.78\%} \pm  6.2\% \pm  0.13\%$ \\
$2250.$ & $      2.483    ^{+ 0.79\%}_{- 1.75\%} \pm  3.6\% \pm  0.24\%$ & $      4.034    ^{+ 0.78\%}_{- 1.76\%} \pm  2.4\% \pm  0.55\%$ & $      1.629    ^{+ 0.80\%}_{- 1.83\%} \pm  6.4\% \pm  0.14\%$ \\
$2300.$ & $      2.228    ^{+ 0.82\%}_{- 1.80\%} \pm  3.6\% \pm  0.24\%$ & $      3.632    ^{+ 0.81\%}_{- 1.80\%} \pm  2.4\% \pm  0.56\%$ & $      1.457    ^{+ 0.83\%}_{- 1.88\%} \pm  6.5\% \pm  0.15\%$ \\
$2350.$ & $      2.000    ^{+ 0.85\%}_{- 1.85\%} \pm  3.7\% \pm  0.24\%$ & $      3.271    ^{+ 0.83\%}_{- 1.86\%} \pm  2.4\% \pm  0.58\%$ & $      1.303    ^{+ 0.86\%}_{- 1.92\%} \pm  6.7\% \pm  0.16\%$ \\
$2400.$ & $      1.796    ^{+ 0.87\%}_{- 1.90\%} \pm  3.8\% \pm  0.25\%$ & $      2.947    ^{+ 0.86\%}_{- 1.90\%} \pm  2.4\% \pm  0.59\%$ & $      1.166    ^{+ 0.88\%}_{- 1.98\%} \pm  6.9\% \pm  0.17\%$ \\
$2450.$ & $      1.614    ^{+ 0.90\%}_{- 1.95\%} \pm  3.8\% \pm  0.25\%$ & $      2.656    ^{+ 0.88\%}_{- 1.94\%} \pm  2.5\% \pm  0.60\%$ & $      1.044    ^{+ 0.91\%}_{- 2.03\%} \pm  7.0\% \pm  0.18\%$ \\
$2500.$ & $      1.451    ^{+ 0.92\%}_{- 2.00\%} \pm  3.9\% \pm  0.25\%$ & $      2.395    ^{+ 0.91\%}_{- 1.99\%} \pm  2.5\% \pm  0.62\%$ & $     0.9357    ^{+ 0.94\%}_{- 2.08\%} \pm  7.2\% \pm  0.20\%$ \\
$2550.$ & $      1.305    ^{+ 0.95\%}_{- 2.05\%} \pm  4.0\% \pm  0.25\%$ & $      2.161    ^{+ 0.93\%}_{- 2.03\%} \pm  2.5\% \pm  0.63\%$ & $     0.8387    ^{+ 0.97\%}_{- 2.13\%} \pm  7.4\% \pm  0.21\%$ \\
$2600.$ & $      1.174    ^{+ 0.97\%}_{- 2.10\%} \pm  4.1\% \pm  0.25\%$ & $      1.950    ^{+ 0.95\%}_{- 2.09\%} \pm  2.5\% \pm  0.64\%$ & $     0.7522    ^{+ 1.00\%}_{- 2.18\%} \pm  7.6\% \pm  0.22\%$ \\
$2650.$ & $      1.057    ^{+ 1.00\%}_{- 2.15\%} \pm  4.2\% \pm  0.25\%$ & $      1.760    ^{+ 0.98\%}_{- 2.14\%} \pm  2.5\% \pm  0.66\%$ & $     0.6748    ^{+ 1.03\%}_{- 2.24\%} \pm  7.8\% \pm  0.24\%$ \\
$2700.$ & $     0.9512    ^{+ 1.03\%}_{- 2.19\%} \pm  4.2\% \pm  0.25\%$ & $      1.590    ^{+ 1.00\%}_{- 2.18\%} \pm  2.6\% \pm  0.67\%$ & $     0.6057    ^{+ 1.05\%}_{- 2.29\%} \pm  8.0\% \pm  0.25\%$ \\
$2750.$ & $     0.8566    ^{+ 1.06\%}_{- 2.24\%} \pm  4.3\% \pm  0.25\%$ & $      1.436    ^{+ 1.03\%}_{- 2.24\%} \pm  2.6\% \pm  0.68\%$ & $     0.5437    ^{+ 1.09\%}_{- 2.33\%} \pm  8.2\% \pm  0.27\%$ \\
$2800.$ & $     0.7718    ^{+ 1.08\%}_{- 2.30\%} \pm  4.4\% \pm  0.25\%$ & $      1.297    ^{+ 1.06\%}_{- 2.29\%} \pm  2.6\% \pm  0.69\%$ & $     0.4883    ^{+ 1.12\%}_{- 2.38\%} \pm  8.4\% \pm  0.29\%$ \\
$2850.$ & $     0.6955    ^{+ 1.11\%}_{- 2.35\%} \pm  4.5\% \pm  0.25\%$ & $      1.172    ^{+ 1.10\%}_{- 2.34\%} \pm  2.6\% \pm  0.70\%$ & $     0.4387    ^{+ 1.14\%}_{- 2.45\%} \pm  8.6\% \pm  0.30\%$ \\
$2900.$ & $     0.6268    ^{+ 1.14\%}_{- 2.39\%} \pm  4.6\% \pm  0.25\%$ & $      1.059    ^{+ 1.13\%}_{- 2.39\%} \pm  2.7\% \pm  0.71\%$ & $     0.3943    ^{+ 1.18\%}_{- 2.49\%} \pm  8.8\% \pm  0.32\%$ \\
$2950.$ & $     0.5652    ^{+ 1.17\%}_{- 2.45\%} \pm  4.7\% \pm  0.25\%$ & $     0.9579    ^{+ 1.14\%}_{- 2.45\%} \pm  2.7\% \pm  0.73\%$ & $     0.3543    ^{+ 1.21\%}_{- 2.54\%} \pm  9.0\% \pm  0.34\%$ \\
$3000.$ & $     0.5096    ^{+ 1.19\%}_{- 2.50\%} \pm  4.8\% \pm  0.25\%$ & $     0.8662    ^{+ 1.18\%}_{- 2.51\%} \pm  2.7\% \pm  0.74\%$ & $     0.3186    ^{+ 1.22\%}_{- 2.59\%} \pm  9.3\% \pm  0.36\%$ \\
 \bottomrule
 \end{tabular}
\end{table}

\begin{table}
\vspace{-0.7cm}
\caption{VBF production cross sections for $\PSHppFive $ and $\PSHmmFive $ in the GM model, computed for $s_{\PH}= 1$ at the $\sqrt s = 13~\UTeV$ LHC.  The first (asymmetric) uncertainties are the QCD scale uncertainty, the second is the PDF uncertainty, and the third is the $\alpha_s$ uncertainty.  The uncertainty from uncalculated NLO electroweak corrections should be taken as $\pm 7\%$.  The relative Monte Carlo numerical integration error is below $5 \times 10^{-4}$ in all cases.}
\label{GM:tab:xsec3}
\centering
\small
 \begin{tabular}{c| c| c}
 \toprule
 $m_5$ [\UGeV] &  $\sigma_1^{\rm NNLO}(\PSHppFive )$ [fb] &  $\sigma_1^{\rm NNLO}(\PSHmmFive )$ [fb] \\
 \midrule
$ 200.$ & $      2511.    ^{+ 0.24\%}_{- 0.14\%} \pm  1.9\% \pm  0.40\%$ & $      1070.    ^{+ 0.33\%}_{- 0.21\%} \pm  2.9\% \pm  0.54\%$ \\
$ 210.$ & $      2364.    ^{+ 0.24\%}_{- 0.14\%} \pm  1.9\% \pm  0.39\%$ & $      997.0    ^{+ 0.31\%}_{- 0.20\%} \pm  2.9\% \pm  0.53\%$ \\
$ 220.$ & $      2229.    ^{+ 0.23\%}_{- 0.13\%} \pm  1.9\% \pm  0.38\%$ & $      930.3    ^{+ 0.29\%}_{- 0.19\%} \pm  3.0\% \pm  0.52\%$ \\
$ 230.$ & $      2104.    ^{+ 0.24\%}_{- 0.13\%} \pm  1.9\% \pm  0.37\%$ & $      869.2    ^{+ 0.27\%}_{- 0.19\%} \pm  3.0\% \pm  0.51\%$ \\
$ 240.$ & $      1988.    ^{+ 0.24\%}_{- 0.12\%} \pm  1.9\% \pm  0.35\%$ & $      813.3    ^{+ 0.25\%}_{- 0.18\%} \pm  3.0\% \pm  0.51\%$ \\
$ 250.$ & $      1881.    ^{+ 0.24\%}_{- 0.11\%} \pm  1.9\% \pm  0.34\%$ & $      762.0    ^{+ 0.25\%}_{- 0.18\%} \pm  3.1\% \pm  0.50\%$ \\
$ 260.$ & $      1781.    ^{+ 0.24\%}_{- 0.10\%} \pm  1.9\% \pm  0.33\%$ & $      714.8    ^{+ 0.25\%}_{- 0.18\%} \pm  3.1\% \pm  0.49\%$ \\
$ 270.$ & $      1689.    ^{+ 0.25\%}_{- 0.09\%} \pm  1.9\% \pm  0.32\%$ & $      671.3    ^{+ 0.25\%}_{- 0.17\%} \pm  3.1\% \pm  0.49\%$ \\
$ 280.$ & $      1602.    ^{+ 0.25\%}_{- 0.09\%} \pm  1.9\% \pm  0.31\%$ & $      631.2    ^{+ 0.25\%}_{- 0.16\%} \pm  3.1\% \pm  0.48\%$ \\
$ 290.$ & $      1522.    ^{+ 0.24\%}_{- 0.09\%} \pm  1.9\% \pm  0.30\%$ & $      594.1    ^{+ 0.26\%}_{- 0.15\%} \pm  3.2\% \pm  0.47\%$ \\
$ 300.$ & $      1447.    ^{+ 0.25\%}_{- 0.08\%} \pm  1.9\% \pm  0.29\%$ & $      559.8    ^{+ 0.26\%}_{- 0.14\%} \pm  3.2\% \pm  0.47\%$ \\
$ 310.$ & $      1377.    ^{+ 0.25\%}_{- 0.07\%} \pm  1.9\% \pm  0.28\%$ & $      527.9    ^{+ 0.26\%}_{- 0.14\%} \pm  3.2\% \pm  0.46\%$ \\
$ 320.$ & $      1311.    ^{+ 0.25\%}_{- 0.06\%} \pm  1.9\% \pm  0.28\%$ & $      498.4    ^{+ 0.26\%}_{- 0.13\%} \pm  3.3\% \pm  0.45\%$ \\
$ 330.$ & $      1249.    ^{+ 0.25\%}_{- 0.06\%} \pm  1.9\% \pm  0.27\%$ & $      471.0    ^{+ 0.26\%}_{- 0.13\%} \pm  3.3\% \pm  0.45\%$ \\
$ 340.$ & $      1192.    ^{+ 0.25\%}_{- 0.06\%} \pm  1.9\% \pm  0.26\%$ & $      445.4    ^{+ 0.26\%}_{- 0.12\%} \pm  3.3\% \pm  0.44\%$ \\
$ 350.$ & $      1137.    ^{+ 0.25\%}_{- 0.05\%} \pm  1.9\% \pm  0.25\%$ & $      421.6    ^{+ 0.26\%}_{- 0.12\%} \pm  3.3\% \pm  0.44\%$ \\
$ 360.$ & $      1086.    ^{+ 0.25\%}_{- 0.05\%} \pm  1.9\% \pm  0.24\%$ & $      399.4    ^{+ 0.26\%}_{- 0.11\%} \pm  3.4\% \pm  0.43\%$ \\
$ 370.$ & $      1038.    ^{+ 0.25\%}_{- 0.07\%} \pm  1.9\% \pm  0.23\%$ & $      378.7    ^{+ 0.26\%}_{- 0.10\%} \pm  3.4\% \pm  0.43\%$ \\
$ 380.$ & $      992.6    ^{+ 0.25\%}_{- 0.08\%} \pm  2.0\% \pm  0.22\%$ & $      359.3    ^{+ 0.26\%}_{- 0.10\%} \pm  3.4\% \pm  0.42\%$ \\
$ 390.$ & $      949.8    ^{+ 0.25\%}_{- 0.09\%} \pm  2.0\% \pm  0.21\%$ & $      341.1    ^{+ 0.25\%}_{- 0.10\%} \pm  3.5\% \pm  0.42\%$ \\
$ 400.$ & $      909.3    ^{+ 0.25\%}_{- 0.11\%} \pm  2.0\% \pm  0.21\%$ & $      324.1    ^{+ 0.26\%}_{- 0.09\%} \pm  3.5\% \pm  0.41\%$ \\
$ 410.$ & $      871.1    ^{+ 0.25\%}_{- 0.12\%} \pm  2.0\% \pm  0.20\%$ & $      308.1    ^{+ 0.26\%}_{- 0.09\%} \pm  3.5\% \pm  0.41\%$ \\
$ 420.$ & $      835.0    ^{+ 0.25\%}_{- 0.13\%} \pm  2.0\% \pm  0.19\%$ & $      293.1    ^{+ 0.26\%}_{- 0.08\%} \pm  3.6\% \pm  0.41\%$ \\
$ 430.$ & $      800.8    ^{+ 0.25\%}_{- 0.14\%} \pm  2.0\% \pm  0.18\%$ & $      279.0    ^{+ 0.26\%}_{- 0.07\%} \pm  3.6\% \pm  0.40\%$ \\
$ 440.$ & $      768.4    ^{+ 0.26\%}_{- 0.16\%} \pm  2.0\% \pm  0.17\%$ & $      265.8    ^{+ 0.26\%}_{- 0.07\%} \pm  3.6\% \pm  0.40\%$ \\
$ 450.$ & $      737.7    ^{+ 0.26\%}_{- 0.17\%} \pm  2.0\% \pm  0.16\%$ & $      253.3    ^{+ 0.26\%}_{- 0.08\%} \pm  3.6\% \pm  0.39\%$ \\
$ 460.$ & $      708.5    ^{+ 0.26\%}_{- 0.18\%} \pm  2.0\% \pm  0.16\%$ & $      241.5    ^{+ 0.27\%}_{- 0.10\%} \pm  3.7\% \pm  0.39\%$ \\
$ 470.$ & $      680.9    ^{+ 0.26\%}_{- 0.19\%} \pm  2.0\% \pm  0.15\%$ & $      230.5    ^{+ 0.27\%}_{- 0.11\%} \pm  3.7\% \pm  0.39\%$ \\
$ 480.$ & $      654.5    ^{+ 0.26\%}_{- 0.20\%} \pm  2.0\% \pm  0.14\%$ & $      220.0    ^{+ 0.27\%}_{- 0.13\%} \pm  3.7\% \pm  0.38\%$ \\
$ 490.$ & $      629.5    ^{+ 0.26\%}_{- 0.21\%} \pm  2.0\% \pm  0.13\%$ & $      210.2    ^{+ 0.26\%}_{- 0.14\%} \pm  3.8\% \pm  0.38\%$ \\
$ 500.$ & $      605.7    ^{+ 0.26\%}_{- 0.22\%} \pm  2.0\% \pm  0.13\%$ & $      200.8    ^{+ 0.27\%}_{- 0.15\%} \pm  3.8\% \pm  0.38\%$ \\
$ 550.$ & $      502.4    ^{+ 0.26\%}_{- 0.27\%} \pm  2.0\% \pm  0.09\%$ & $      161.1    ^{+ 0.27\%}_{- 0.21\%} \pm  3.9\% \pm  0.36\%$ \\
$ 600.$ & $      420.3    ^{+ 0.26\%}_{- 0.33\%} \pm  2.0\% \pm  0.05\%$ & $      130.6    ^{+ 0.27\%}_{- 0.28\%} \pm  4.1\% \pm  0.35\%$ \\
$ 650.$ & $      354.3    ^{+ 0.27\%}_{- 0.38\%} \pm  2.0\% \pm  0.02\%$ & $      106.9    ^{+ 0.26\%}_{- 0.33\%} \pm  4.3\% \pm  0.34\%$ \\
$ 700.$ & $      300.7    ^{+ 0.27\%}_{- 0.43\%} \pm  2.1\% \pm  0.01\%$ & $      88.12    ^{+ 0.26\%}_{- 0.39\%} \pm  4.4\% \pm  0.33\%$ \\
$ 750.$ & $      256.7    ^{+ 0.27\%}_{- 0.48\%} \pm  2.1\% \pm  0.05\%$ & $      73.17    ^{+ 0.27\%}_{- 0.45\%} \pm  4.6\% \pm  0.32\%$ \\
$ 800.$ & $      220.3    ^{+ 0.27\%}_{- 0.53\%} \pm  2.1\% \pm  0.08\%$ & $      61.13    ^{+ 0.27\%}_{- 0.50\%} \pm  4.7\% \pm  0.31\%$ \\
$ 850.$ & $      189.9    ^{+ 0.27\%}_{- 0.57\%} \pm  2.1\% \pm  0.11\%$ & $      51.36    ^{+ 0.27\%}_{- 0.56\%} \pm  4.9\% \pm  0.31\%$ \\
$ 900.$ & $      164.4    ^{+ 0.28\%}_{- 0.62\%} \pm  2.1\% \pm  0.14\%$ & $      43.37    ^{+ 0.27\%}_{- 0.61\%} \pm  5.0\% \pm  0.30\%$ \\
\bottomrule
 \end{tabular}
\end{table}

\begin{table}
\vspace{-0.4cm}
\caption{Continuation of Table~\ref{GM:tab:xsec3}.}
\label{GM:tab:xsec4}
\centering
\small
 \begin{tabular}{c| c| c}
 \toprule
 $m_5$ [\UGeV] &  $\sigma_1^{\rm NNLO}(\PSHppFive )$ [fb] &  $\sigma_1^{\rm NNLO}(\PSHmmFive )$ [fb] \\
 \midrule
$ 950.$ & $      142.8    ^{+ 0.28\%}_{- 0.67\%} \pm  2.2\% \pm  0.17\%$ & $      36.79    ^{+ 0.28\%}_{- 0.66\%} \pm  5.2\% \pm  0.30\%$ \\
$1000.$ & $      124.5    ^{+ 0.28\%}_{- 0.71\%} \pm  2.2\% \pm  0.20\%$ & $      31.33    ^{+ 0.28\%}_{- 0.72\%} \pm  5.4\% \pm  0.30\%$ \\
$1050.$ & $      108.9    ^{+ 0.28\%}_{- 0.76\%} \pm  2.2\% \pm  0.23\%$ & $      26.79    ^{+ 0.29\%}_{- 0.78\%} \pm  5.5\% \pm  0.30\%$ \\
$1100.$ & $      95.49    ^{+ 0.30\%}_{- 0.80\%} \pm  2.2\% \pm  0.25\%$ & $      22.98    ^{+ 0.27\%}_{- 0.84\%} \pm  5.7\% \pm  0.30\%$ \\
$1150.$ & $      83.95    ^{+ 0.32\%}_{- 0.84\%} \pm  2.3\% \pm  0.28\%$ & $      19.78    ^{+ 0.30\%}_{- 0.88\%} \pm  5.9\% \pm  0.30\%$ \\
$1200.$ & $      73.98    ^{+ 0.34\%}_{- 0.89\%} \pm  2.3\% \pm  0.31\%$ & $      17.07    ^{+ 0.33\%}_{- 0.93\%} \pm  6.0\% \pm  0.31\%$ \\
$1250.$ & $      65.33    ^{+ 0.36\%}_{- 0.94\%} \pm  2.3\% \pm  0.34\%$ & $      14.77    ^{+ 0.34\%}_{- 0.98\%} \pm  6.2\% \pm  0.31\%$ \\
$1300.$ & $      57.81    ^{+ 0.38\%}_{- 0.98\%} \pm  2.3\% \pm  0.36\%$ & $      12.81    ^{+ 0.37\%}_{- 1.04\%} \pm  6.4\% \pm  0.32\%$ \\
$1350.$ & $      51.24    ^{+ 0.39\%}_{- 1.02\%} \pm  2.4\% \pm  0.39\%$ & $      11.14    ^{+ 0.40\%}_{- 1.09\%} \pm  6.6\% \pm  0.32\%$ \\
$1400.$ & $      45.50    ^{+ 0.42\%}_{- 1.07\%} \pm  2.4\% \pm  0.41\%$ & $      9.706    ^{+ 0.43\%}_{- 1.14\%} \pm  6.8\% \pm  0.33\%$ \\
$1450.$ & $      40.46    ^{+ 0.44\%}_{- 1.11\%} \pm  2.4\% \pm  0.44\%$ & $      8.474    ^{+ 0.45\%}_{- 1.19\%} \pm  6.9\% \pm  0.34\%$ \\
$1500.$ & $      36.04    ^{+ 0.46\%}_{- 1.16\%} \pm  2.4\% \pm  0.46\%$ & $      7.412    ^{+ 0.48\%}_{- 1.24\%} \pm  7.1\% \pm  0.35\%$ \\
$1550.$ & $      32.14    ^{+ 0.48\%}_{- 1.21\%} \pm  2.5\% \pm  0.48\%$ & $      6.493    ^{+ 0.50\%}_{- 1.29\%} \pm  7.3\% \pm  0.36\%$ \\
$1600.$ & $      28.70    ^{+ 0.50\%}_{- 1.25\%} \pm  2.5\% \pm  0.51\%$ & $      5.698    ^{+ 0.53\%}_{- 1.34\%} \pm  7.5\% \pm  0.38\%$ \\
$1650.$ & $      25.66    ^{+ 0.52\%}_{- 1.30\%} \pm  2.5\% \pm  0.53\%$ & $      5.008    ^{+ 0.55\%}_{- 1.40\%} \pm  7.7\% \pm  0.39\%$ \\
$1700.$ & $      22.97    ^{+ 0.54\%}_{- 1.34\%} \pm  2.5\% \pm  0.56\%$ & $      4.408    ^{+ 0.57\%}_{- 1.45\%} \pm  8.0\% \pm  0.41\%$ \\
$1750.$ & $      20.57    ^{+ 0.57\%}_{- 1.39\%} \pm  2.6\% \pm  0.58\%$ & $      3.885    ^{+ 0.60\%}_{- 1.50\%} \pm  8.2\% \pm  0.42\%$ \\
$1800.$ & $      18.45    ^{+ 0.59\%}_{- 1.43\%} \pm  2.6\% \pm  0.60\%$ & $      3.428    ^{+ 0.62\%}_{- 1.56\%} \pm  8.4\% \pm  0.44\%$ \\
$1850.$ & $      16.56    ^{+ 0.62\%}_{- 1.48\%} \pm  2.6\% \pm  0.62\%$ & $      3.028    ^{+ 0.65\%}_{- 1.61\%} \pm  8.6\% \pm  0.46\%$ \\
$1900.$ & $      14.88    ^{+ 0.64\%}_{- 1.53\%} \pm  2.7\% \pm  0.65\%$ & $      2.678    ^{+ 0.68\%}_{- 1.66\%} \pm  8.9\% \pm  0.48\%$ \\
$1950.$ & $      13.37    ^{+ 0.66\%}_{- 1.58\%} \pm  2.7\% \pm  0.67\%$ & $      2.371    ^{+ 0.70\%}_{- 1.72\%} \pm  9.1\% \pm  0.50\%$ \\
$2000.$ & $      12.03    ^{+ 0.69\%}_{- 1.63\%} \pm  2.7\% \pm  0.69\%$ & $      2.101    ^{+ 0.73\%}_{- 1.77\%} \pm  9.4\% \pm  0.52\%$ \\
$2050.$ & $      10.83    ^{+ 0.71\%}_{- 1.67\%} \pm  2.8\% \pm  0.71\%$ & $      1.863    ^{+ 0.76\%}_{- 1.83\%} \pm  9.6\% \pm  0.54\%$ \\
$2100.$ & $      9.756    ^{+ 0.74\%}_{- 1.72\%} \pm  2.8\% \pm  0.73\%$ & $      1.654    ^{+ 0.79\%}_{- 1.87\%} \pm  9.9\% \pm  0.57\%$ \\
$2150.$ & $      8.793    ^{+ 0.76\%}_{- 1.77\%} \pm  2.8\% \pm  0.75\%$ & $      1.469    ^{+ 0.82\%}_{- 1.93\%} \pm 10.2\% \pm  0.59\%$ \\
$2200.$ & $      7.930    ^{+ 0.78\%}_{- 1.82\%} \pm  2.9\% \pm  0.77\%$ & $      1.306    ^{+ 0.85\%}_{- 1.99\%} \pm 10.4\% \pm  0.62\%$ \\
$2250.$ & $      7.155    ^{+ 0.80\%}_{- 1.88\%} \pm  2.9\% \pm  0.80\%$ & $      1.162    ^{+ 0.88\%}_{- 2.04\%} \pm 10.7\% \pm  0.64\%$ \\
$2300.$ & $      6.459    ^{+ 0.83\%}_{- 1.92\%} \pm  2.9\% \pm  0.82\%$ & $      1.035    ^{+ 0.90\%}_{- 2.09\%} \pm 11.0\% \pm  0.67\%$ \\
$2350.$ & $      5.833    ^{+ 0.86\%}_{- 1.98\%} \pm  3.0\% \pm  0.84\%$ & $     0.9221    ^{+ 0.94\%}_{- 2.15\%} \pm 11.3\% \pm  0.70\%$ \\
$2400.$ & $      5.270    ^{+ 0.89\%}_{- 2.03\%} \pm  3.0\% \pm  0.86\%$ & $     0.8222    ^{+ 0.96\%}_{- 2.21\%} \pm 11.6\% \pm  0.73\%$ \\
$2450.$ & $      4.763    ^{+ 0.92\%}_{- 2.08\%} \pm  3.1\% \pm  0.88\%$ & $     0.7334    ^{+ 0.99\%}_{- 2.26\%} \pm 11.9\% \pm  0.76\%$ \\
$2500.$ & $      4.306    ^{+ 0.95\%}_{- 2.12\%} \pm  3.1\% \pm  0.90\%$ & $     0.6545    ^{+ 1.02\%}_{- 2.32\%} \pm 12.3\% \pm  0.80\%$ \\
$2550.$ & $      3.895    ^{+ 0.96\%}_{- 2.18\%} \pm  3.1\% \pm  0.92\%$ & $     0.5844    ^{+ 1.05\%}_{- 2.37\%} \pm 12.6\% \pm  0.83\%$ \\
$2600.$ & $      3.524    ^{+ 1.00\%}_{- 2.23\%} \pm  3.2\% \pm  0.94\%$ & $     0.5221    ^{+ 1.08\%}_{- 2.43\%} \pm 12.9\% \pm  0.86\%$ \\
$2650.$ & $      3.189    ^{+ 1.02\%}_{- 2.28\%} \pm  3.2\% \pm  0.96\%$ & $     0.4666    ^{+ 1.12\%}_{- 2.47\%} \pm 13.3\% \pm  0.90\%$ \\
$2700.$ & $      2.886    ^{+ 1.05\%}_{- 2.33\%} \pm  3.3\% \pm  0.98\%$ & $     0.4172    ^{+ 1.14\%}_{- 2.53\%} \pm 13.6\% \pm  0.94\%$ \\
$2750.$ & $      2.613    ^{+ 1.07\%}_{- 2.38\%} \pm  3.3\% \pm  1.00\%$ & $     0.3732    ^{+ 1.18\%}_{- 2.57\%} \pm 14.0\% \pm  0.97\%$ \\
$2800.$ & $      2.367    ^{+ 1.10\%}_{- 2.44\%} \pm  3.3\% \pm  1.02\%$ & $     0.3340    ^{+ 1.22\%}_{- 2.63\%} \pm 14.4\% \pm  1.01\%$ \\
$2850.$ & $      2.144    ^{+ 1.14\%}_{- 2.49\%} \pm  3.4\% \pm  1.04\%$ & $     0.2990    ^{+ 1.24\%}_{- 2.69\%} \pm 14.8\% \pm  1.05\%$ \\
$2900.$ & $      1.942    ^{+ 1.17\%}_{- 2.54\%} \pm  3.4\% \pm  1.06\%$ & $     0.2677    ^{+ 1.27\%}_{- 2.74\%} \pm 15.2\% \pm  1.09\%$ \\
$2950.$ & $      1.760    ^{+ 1.19\%}_{- 2.61\%} \pm  3.5\% \pm  1.07\%$ & $     0.2398    ^{+ 1.30\%}_{- 2.80\%} \pm 15.6\% \pm  1.14\%$ \\
$3000.$ & $      1.594    ^{+ 1.21\%}_{- 2.65\%} \pm  3.5\% \pm  1.09\%$ & $     0.2149    ^{+ 1.33\%}_{- 2.86\%} \pm 16.0\% \pm  1.18\%$ \\
 \bottomrule
 \end{tabular}
\end{table}

\subsection{Decay widths of the \texorpdfstring{$\HepParticle{\PH}{5}{}$}{H_5} states}
\label{GM:sec:widths}

We computed the tree-level decay partial widths $\Gamma(\PSHpmpmFive \to \PWpm  \PWpm )$, $\Gamma(\PSHpm \to \PWpm  \PZ)$, $\Gamma(\PHzFive  \to \PWp \PWm)$, and $\Gamma(\PHzFive  \to \PZ\PZ )$ using {\tt GMCALC~1.2.0}~\cite{Hartling:2014xma}.  In the H5plane benchmark, ${\rm BR}(\PHFive \to \PVV) = 1$, so that these partial widths correspond to the total widths of the $\PHFive $ scalars.  The decay calculation includes the effects of both of the final-state gauge bosons off-shell.  The numerical calculation has been benchmarked against {\tt HDECAY~6.42}~\cite{Djouadi:1997yw} with agreement to within 1\%.  We use the electroweak input parameters given in Eq.~(\ref{eq:inputs}) and scale all decay widths to $s_{\PH}= 1$.
For $\PSHpmpmFive$ and $\PSHpmFive$, the decay width is the same for the charge-conjugate process.  The total widths of the $\PHFive $ states are given in Tables~\ref{GM:tab:widths}--\ref{GM:tab:widths2} for $\PSHpmpmFive$, $\PSHpmFive$ and $\PHzFive $ in the H5plane benchmark, for which ${\rm BR}(\PHFive \to \PVV) = 1$.  We also give ${\rm BR}(\PHzFive  \to \PWp \PWm)$.  The branching ratio ${\rm BR}(\PHzFive  \to \PZ\PZ )$ is then equal to $1 - {\rm BR}(\PHzFive  \to \PWp \PWm)$.  So long as ${\rm BR}(\PHFive \to \PVV) = 1$, these decay widths depend only on $m_5$ and $s_{\PH}$.  The branching ratios of $\PHzFive $ to $\PW\PW$ and $\PZ\PZ $ depend only on $m_5$.

\begin{table}
\vspace{-0.7cm}
\caption{Tree-level total decay widths for $\PSHpmpmFive$, $\PSHpmFive$, and $\PHzFive $ in the GM model, rescaled to $s_{\PH}= 1$ and assuming that ${\rm BR}(\PHFive \to \PVV) = 1$.  The uncertainty on the total widths from uncalculated NLO electroweak corrections should be taken as $\pm 12\%$.  We also give ${\rm BR}(\PHzFive  \to \PWp \PWm)$, assuming that ${\rm BR}(\PHzFive  \to \PWp  \PWm) + {\rm BR}(\PHzFive  \to \PZ\PZ ) = 1$, and its uncertainty from the uncalculated NLO electroweak corrections.}
\label{GM:tab:widths}
\centering
\small
 \begin{tabular}{c| c|c|c|c}
 \toprule
 $m_5$ [\UGeV] & $\Gamma_1^{\rm tot}(\PSHpmpmFive)$ [\UGeV] & $\Gamma_1^{\rm tot}(\PSHpmFive)$ [\UGeV] & $\Gamma_1^{\rm tot}(\PHzFive )$ [\UGeV] & ${\rm BR}(\PHzFive  \to \PWp \PWm)$ \\
 \midrule
 200. &       1.006     &      0.8608     &      0.8008     & $      0.4187     ^{+14.\%} _{-14.\%} $ \\
 210. &       1.275     &       1.118     &       1.071     & $      0.3969     ^{+15.\%} _{-14.\%} $ \\
 220. &       1.578     &       1.410     &       1.362     & $      0.3863     ^{+15.\%} _{-14.\%} $ \\
 230. &       1.921     &       1.737     &       1.686     & $      0.3799     ^{+15.\%} _{-14.\%} $ \\
 240. &       2.307     &       2.105     &       2.051     & $      0.3749     ^{+15.\%} _{-15.\%} $ \\
 250. &       2.739     &       2.516     &       2.459     & $      0.3714     ^{+16.\%} _{-15.\%} $ \\
 260. &       3.219     &       2.975     &       2.912     & $      0.3685     ^{+16.\%} _{-15.\%} $ \\
 270. &       3.750     &       3.484     &       3.414     & $      0.3661     ^{+16.\%} _{-15.\%} $ \\
 280. &       4.333     &       4.045     &       3.968     & $      0.3640     ^{+16.\%} _{-15.\%} $ \\
 290. &       4.972     &       4.660     &       4.577     & $      0.3621     ^{+16.\%} _{-15.\%} $ \\
 300. &       5.666     &       5.332     &       5.241     & $      0.3604     ^{+16.\%} _{-15.\%} $ \\
 310. &       6.420     &       6.063     &       5.965     & $      0.3588     ^{+16.\%} _{-15.\%} $ \\
 320. &       7.235     &       6.854     &       6.748     & $      0.3574     ^{+16.\%} _{-15.\%} $ \\
 330. &       8.112     &       7.708     &       7.595     & $      0.3560     ^{+16.\%} _{-15.\%} $ \\
 340. &       9.054     &       8.627     &       8.506     & $      0.3548     ^{+16.\%} _{-15.\%} $ \\
 350. &       10.06     &       9.612     &       9.483     & $      0.3537     ^{+16.\%} _{-15.\%} $ \\
 360. &       11.14     &       10.67     &       10.53     & $      0.3526     ^{+16.\%} _{-15.\%} $ \\
 370. &       12.29     &       11.79     &       11.65     & $      0.3517     ^{+16.\%} _{-15.\%} $ \\
 380. &       13.51     &       12.99     &       12.83     & $      0.3508     ^{+16.\%} _{-15.\%} $ \\
 390. &       14.80     &       14.26     &       14.10     & $      0.3499     ^{+16.\%} _{-15.\%} $ \\
 400. &       16.17     &       15.60     &       15.44     & $      0.3491     ^{+16.\%} _{-15.\%} $ \\
 410. &       17.62     &       17.03     &       16.85     & $      0.3484     ^{+16.\%} _{-15.\%} $ \\
 420. &       19.14     &       18.53     &       18.35     & $      0.3477     ^{+16.\%} _{-15.\%} $ \\
 430. &       20.75     &       20.12     &       19.93     & $      0.3471     ^{+16.\%} _{-15.\%} $ \\
 440. &       22.45     &       21.79     &       21.59     & $      0.3465     ^{+16.\%} _{-15.\%} $ \\
 450. &       24.23     &       23.55     &       23.34     & $      0.3459     ^{+16.\%} _{-15.\%} $ \\
 460. &       26.09     &       25.39     &       25.18     & $      0.3454     ^{+16.\%} _{-15.\%} $ \\
 470. &       28.05     &       27.33     &       27.10     & $      0.3449     ^{+16.\%} _{-15.\%} $ \\
 480. &       30.09     &       29.35     &       29.12     & $      0.3445     ^{+16.\%} _{-15.\%} $ \\
 490. &       32.24     &       31.47     &       31.23     & $      0.3440     ^{+16.\%} _{-15.\%} $ \\
 500. &       34.47     &       33.68     &       33.44     & $      0.3436     ^{+16.\%} _{-15.\%} $ \\
 550. &       47.15     &       46.25     &       45.97     & $      0.3419     ^{+16.\%} _{-15.\%} $ \\
 600. &       62.49     &       61.48     &       61.16     & $      0.3406     ^{+16.\%} _{-15.\%} $ \\
 650. &       80.74     &       79.63     &       79.27     & $      0.3395     ^{+16.\%} _{-15.\%} $ \\
 700. &       102.1     &       100.9     &       100.5     & $      0.3387     ^{+17.\%} _{-15.\%} $ \\
 750. &       126.9     &       125.6     &       125.2     & $      0.3380     ^{+17.\%} _{-15.\%} $ \\
 800. &       155.4     &       153.9     &       153.5     & $      0.3375     ^{+17.\%} _{-15.\%} $ \\
 850. &       187.7     &       186.1     &       185.6     & $      0.3370     ^{+17.\%} _{-15.\%} $ \\
 900. &       224.1     &       222.4     &       221.9     & $      0.3367     ^{+17.\%} _{-15.\%} $ \\
 \bottomrule
 \end{tabular}
\end{table}

\begin{table}
\vspace{-0.4cm}
\caption{Continuation of Table~\ref{GM:tab:widths}.}
\label{GM:tab:widths2}
\centering
\small
 \begin{tabular}{c| c|c|c|c}
 \toprule
 $m_5$ [\UGeV] & $\Gamma_1^{\rm tot}(\PSHpmpmFive)$ [\UGeV] & $\Gamma_1^{\rm tot}(\PSHpmFive)$ [\UGeV] & $\Gamma_1^{\rm tot}(\PHzFive )$ [\UGeV] & ${\rm BR}(\PHzFive  \to \PWp \PWm)$ \\
 \midrule
 950. &       264.9     &       263.1     &       262.5     & $      0.3363     ^{+17.\%} _{-15.\%} $ \\
1000. &       310.3     &       308.4     &       307.8     & $      0.3361     ^{+17.\%} _{-15.\%} $ \\
1050. &       360.5     &       358.5     &       357.8     & $      0.3358     ^{+17.\%} _{-15.\%} $ \\
1100. &       415.8     &       413.7     &       413.0     & $      0.3356     ^{+17.\%} _{-15.\%} $ \\
1150. &       476.5     &       474.2     &       473.4     & $      0.3355     ^{+17.\%} _{-15.\%} $ \\
1200. &       542.7     &       540.3     &       539.5     & $      0.3353     ^{+17.\%} _{-15.\%} $ \\
1250. &       614.7     &       612.2     &       611.3     & $      0.3352     ^{+17.\%} _{-15.\%} $ \\
1300. &       692.7     &       690.1     &       689.2     & $      0.3350     ^{+17.\%} _{-15.\%} $ \\
1350. &       777.1     &       774.3     &       773.4     & $      0.3349     ^{+17.\%} _{-15.\%} $ \\
1400. &       868.0     &       865.1     &       864.1     & $      0.3348     ^{+17.\%} _{-15.\%} $ \\
1450. &       965.7     &       962.6     &       961.6     & $      0.3347     ^{+17.\%} _{-15.\%} $ \\
1500. &       1070.     &       1067.     &       1066.     & $      0.3347     ^{+17.\%} _{-15.\%} $ \\
1550. &       1182.     &       1179.     &       1178.     & $      0.3346     ^{+17.\%} _{-15.\%} $ \\
1600. &       1302.     &       1298.     &       1297.     & $      0.3345     ^{+17.\%} _{-15.\%} $ \\
1650. &       1429.     &       1425.     &       1424.     & $      0.3345     ^{+17.\%} _{-15.\%} $ \\
1700. &       1564.     &       1560.     &       1559.     & $      0.3344     ^{+17.\%} _{-15.\%} $ \\
1750. &       1708.     &       1704.     &       1702.     & $      0.3344     ^{+17.\%} _{-15.\%} $ \\
1800. &       1860.     &       1855.     &       1854.     & $      0.3343     ^{+17.\%} _{-15.\%} $ \\
1850. &       2020.     &       2016.     &       2014.     & $      0.3343     ^{+17.\%} _{-15.\%} $ \\
1900. &       2190.     &       2185.     &       2184.     & $      0.3342     ^{+17.\%} _{-15.\%} $ \\
1950. &       2369.     &       2364.     &       2362.     & $      0.3342     ^{+17.\%} _{-15.\%} $ \\
2000. &       2557.     &       2552.     &       2550.     & $      0.3342     ^{+17.\%} _{-15.\%} $ \\
2050. &       2755.     &       2750.     &       2748.     & $      0.3341     ^{+17.\%} _{-15.\%} $ \\
2100. &       2962.     &       2957.     &       2956.     & $      0.3341     ^{+17.\%} _{-15.\%} $ \\
2150. &       3180.     &       3175.     &       3173.     & $      0.3341     ^{+17.\%} _{-15.\%} $ \\
2200. &       3409.     &       3403.     &       3401.     & $      0.3341     ^{+17.\%} _{-15.\%} $ \\
2250. &       3648.     &       3642.     &       3640.     & $      0.3340     ^{+17.\%} _{-15.\%} $ \\
2300. &       3898.     &       3892.     &       3890.     & $      0.3340     ^{+17.\%} _{-15.\%} $ \\
2350. &       4159.     &       4153.     &       4151.     & $      0.3340     ^{+17.\%} _{-15.\%} $ \\
2400. &       4431.     &       4425.     &       4423.     & $      0.3340     ^{+17.\%} _{-15.\%} $ \\
2450. &       4716.     &       4709.     &       4707.     & $      0.3340     ^{+17.\%} _{-15.\%} $ \\
2500. &       5011.     &       5005.     &       5002.     & $      0.3339     ^{+17.\%} _{-15.\%} $ \\
2550. &       5319.     &       5312.     &       5310.     & $      0.3339     ^{+17.\%} _{-15.\%} $ \\
2600. &       5640.     &       5633.     &       5630.     & $      0.3339     ^{+17.\%} _{-15.\%} $ \\
2650. &       5973.     &       5965.     &       5963.     & $      0.3339     ^{+17.\%} _{-15.\%} $ \\
2700. &       6319.     &       6311.     &       6308.     & $      0.3339     ^{+17.\%} _{-15.\%} $ \\
2750. &       6678.     &       6669.     &       6667.     & $      0.3339     ^{+17.\%} _{-15.\%} $ \\
2800. &       7050.     &       7041.     &       7039.     & $      0.3339     ^{+17.\%} _{-15.\%} $ \\
2850. &       7435.     &       7427.     &       7424.     & $      0.3338     ^{+17.\%} _{-15.\%} $ \\
2900. &       7835.     &       7826.     &       7823.     & $      0.3338     ^{+17.\%} _{-15.\%} $ \\
2950. &       8249.     &       8239.     &       8236.     & $      0.3338     ^{+17.\%} _{-15.\%} $ \\
3000. &       8676.     &       8667.     &       8664.     & $      0.3338     ^{+17.\%} _{-15.\%} $ \\
 \bottomrule
 \end{tabular}
\end{table}

The decay widths for other values of $s_{\PH}$ are conveniently obtained using the relation
\begin{equation}
	\Gamma(\PHFive \to \PVV) = s_{\PH}^2 \Gamma_1(\PHFive \to \PVV),
\end{equation}
where $\Gamma_1$ represents the decay width for $s_{\PH}= 1$ as given in Tables~\ref{GM:tab:widths}--\ref{GM:tab:widths2}.

The widths given here were computed at tree level.  For the uncertainty due to uncalculated NLO electroweak corrections, we suggest to adopt a fractional uncertainty on each partial width of $\pm 12\%$.  This encompasses the size of the NLO electroweak correction to the SM Higgs boson decay partial widths to $\PW\PW$ and $\PZ\PZ $~\cite{Bredenstein:2006rh} for SM Higgs boson masses below 700~\UGeV, where tree-level perturbative unitarity constraints in $2 \to 2$ gauge and Higgs boson scattering are satisfied.  This same tree-level perturbativity requirement results in the upper bound on $s_{\PH}$ given in Eq.~(\ref{GM:eq:sHbound}).

For $\PHzFive $ we take the electroweak uncertainty on the total width to be also $\pm 12\%$, corresponding to fully correlated variations of $\Gamma(\PHzFive  \to \PWp \PWm)$ and $\Gamma(\PHzFive  \to \PZ\PZ )$.  We compute the electroweak uncertainty on ${\rm BR}(\PHzFive  \to \PWp \PWm)$ by assuming fully anticorrelated variations of $\Gamma(\PHzFive  \to \PWp \PWm)$ and $\Gamma(\PHzFive  \to \PZ\PZ )$; we give these latter uncertainties in Tables~\ref{GM:tab:widths}--\ref{GM:tab:widths2}.  This is the maximally conservative approach to combining the theory uncertainties on the two contributing decay widths.

%
%
%




\section{Singlet}
\subsection{\label{sec:introExtSca2} Introduction}

The simplest extension of the SM where a resonant di-Higgs final state would be detectable
is the one where an extra real or complex spin-zero gauge singlet is added to its field content.
This is also the minimal model for dark matter~\cite{Silveira:1985rk, McDonald:1993ex, Burgess:2000yq, Bento:2000ah,
Davoudiasl:2004be, Kusenko:2006rh, vanderBij:2006ne, He:2008qm, Gonderinger:2009jp, Mambrini:2011ik, He:2011gc}
and for electroweak baryogenesis  by allowing a strong first-order
phase transition during the era of EWSB~\cite{Menon:2004wv, Huber:2006wf, Profumo:2007wc,
Barger:2011vm, Espinosa:2011ax}. Although minimal, this extension provides a rich collider
phenomenology leading to some distinctive signatures that
can be tested at the LHC~\cite{Datta:1997fx, Schabinger:2005ei, BahatTreidel:2006kx,
Barger:2006sk, Barger:2007im, Barger:2008jx, O'Connell:2006wi,Gupta:2011gd}.
There is also some discussion about the singlet model in Chapter~\ref{chap:HH} of this document.

In this section we present the real (RxSM)
and the complex (CxSM) singlet extensions of the SM. The minimal
versions of these models that we will discuss can have at least two phases. A
$\mathbb{Z}_2$-symmetric phase, with one dark matter candidate, and a
broken phase, where the singlet component(s) mix with the neutral Higgs field fluctuation of the SM Higgs doublet. Both models can be used as simple
benchmarks for resonant double Higgs boson production $\Pp\Pp \to \PH \to
\Ph\Ph$, with $\MH  > 2 \Mh $. Furthermore, the CxSM is the simplest extension that also provides a scenario where a heavy scalar $\Phthree$ may decay into two other scalars
$\Phone$, $\Phtwo $ with different masses, $\Phthree \to \Phtwo  \Phone$.


The CxSM is an extension of the SM where a complex singlet field
\begin{eqnarray}
\mathbb{S}=\PS+i \PA \;,
\end{eqnarray}
with hypercharge zero, is added to the SM field content. The most
general renormalizable scalar potential with a
$U(1)$ global symmetry that is softly broken by terms with mass dimension up to two, is given by
\begin{multline}
V_{\rm CxSM}=\dfrac{m^2}{2}\PH^\dagger \PH+\dfrac{\lambda}{4}(\PH^\dagger \PH)^2+\dfrac{\delta_2}{2}\PH^\dagger \PH |\mathbb{S}|^2+\dfrac{b_2}{2}|\mathbb{S}|^2+
\dfrac{d_2}{4}|\mathbb{S}|^4+\left(\dfrac{b_1}{4}\mathbb{S}^2+a_1\mathbb{S}+c.c.\right)
\, .  \label{eq:V_CxSM}
\end{multline}
The soft breaking terms in Eq.~\eqref{eq:V_CxSM}  are shown in parenthesis. The doublet and the complex singlet are
\begin{equation}\label{eq:vacua_CxSM}
\PH=\dfrac{1}{\sqrt{2}}\left(\begin{array}{c} \PG^+ \\
    v+\Ph+i\PG^0\end{array}\right) \quad \mbox{and} \quad
\mathbb{S}=\dfrac{1}{\sqrt{2}}\left[v_{\PS}+\Ps+i(v_{\PA}+ \Pa)\right] \;,
\end{equation}
respectively. Here $v\approx 246 \UGeV$ and $v_{\PS}$ and $v_{\PA}$
are the real and imaginary parts of the complex singlet field VEV, respectively.

The various phases of the model were discussed in~\cite{Coimbra:2013qq}.
A number of models can be obtained with the same field content by imposing extra symmetries
on the potential. For example the exact $U(1)$-symmetric potential has $a_1=b_1=0$, leading
to either one or to two dark matter candidates depending on the pattern of symmetry breaking.
Here we will focus on the version of the model where the potential is symmetric under
$\mathbb{S}\rightarrow \mathbb{S}^*$, or, equivalently symmetric under a $\mathbb{Z}_2$ symmetry for the imaginary component $\PA$. In this case
both the soft breaking terms, $\{a_1,b_1\}$, and the other parameters,
$\{m, \lambda, \delta_2, b_2,d_2\}$, have to be real. Under these conditions
there are two phases, namely,
\begin{itemize}
\item $v_{\PA}=0$ and $v_{\PS}\neq 0$ - mixing between the doublet
field $\Ph$ and the real component $\Ps$ of the singlet;
the  imaginary component $\PA\equiv \Pa$ becomes a dark matter candidate.
We call it the symmetric or \textit{dark matter phase}.

\item $v_{\PS}\neq 0~{\rm and}~ v_{\PA}\neq 0$ - no dark matter
candidate and mixing among all scalars.
We call it the \textit{broken phase}.

\end{itemize}
The model phases are summarized in table~\ref{tab:phases}.
\begin{table}[h!]
\caption{Phase classification for the version of the CxSM
with the $\mathbb{S}\rightarrow \mathbb{S}^*$ symmetry.}
\label{tab:phases}
\begin{center}
\begin{tabular}{c | c  c}
\toprule
  Phase & Scalar content & VEVs at global minimum   \\
\midrule
Symmetric (dark) &   2 mixed + 1  dark &  $\left<S\right>\neq 0$ and
$\left<A\right>=0$ \\

Broken ($\cancel{\mathbb{Z}}_2$ ) & 3 mixed &
$\left<S\right>\neq 0$ and
$\left<A\right>\neq 0$\\
\bottomrule
\end{tabular}
\end{center}
\end{table}

In order to obtain the couplings of the scalars to the SM particles one first defines the mass eigenstates
$\Phindexi$ (i$=1,2,3$). These are obtained from the gauge eigenstates $\Ph$, $\Ps$ and $\Pa$ through the mixing matrix~$R$
\begin{equation}
\left(
\begin{array}{c}
\Phone\\
\Phtwo \\
\Phthree
\end{array}
\right)
= R
\left(
\begin{array}{c}
\Ph\\
\Ps\\
\Pa
\end{array}
\right) \, ,
\label{h_as_eta}
\end{equation}
with
\begin{equation}
R\, {\cal M}^2\, R^T = \textrm{diag} \left(m_1^2, m_2^2, m_3^2 \right)\;.
\end{equation}
Here $m_1 \leq m_2 \leq m_3$ are the masses of the neutral Higgs particles.
The mixing matrix $R$ is parameterized as
\begin{equation}
R =
\left(
\begin{array}{ccc}
c_1 c_2 & s_1 c_2 & s_2\\
-(c_1 s_2 s_3 + s_1 c_3) & c_1 c_3 - s_1 s_2 s_3  & c_2 s_3\\
- c_1 s_2 c_3 + s_1 s_3 & -(c_1 s_3 + s_1 s_2 c_3) & c_2 c_3
\end{array}
\right)
\label{matrixR}
\end{equation}
with $s_i \equiv \sin{\alpha_i}$ and
$c_i \equiv \cos{\alpha_i}$ ($i = 1, 2, 3$) and
\begin{equation}
- \pi/2 < \alpha_i \leq \pi/2 \;.
\label{range_alpha}
\end{equation}

All scalar couplings to the SM particles are modified by the
same matrix element $R_{i1}$, which is independent of the SM particle. This means that for any SM coupling
$\lambda_{\Ph_{SM}}^{(p)}$, where $p$ runs over all SM fermions and
gauge bosons, the corresponding coupling in the singlet model, for the
scalar $h_i$, is given by
\begin{equation}\label{eq:couplings_SM}
\lambda_{i}^{(p)}=R_{i1}\lambda_{h_{SM}}^{(p)}\; .
\end{equation}
The self interactions are presented in~\cite{Costa:2015llh} and in the code~\texttt{sHDECAY}.
In the particular case of the dark matter phase, $R_{i1} =
(R_{11} , R_{21}, 0)$. The state $i=3$ (i.e. $\PA$) then corresponds to the dark
matter candidate which does not couple to any of the remaining SM particles.
The model has seven free parameters which are chosen to be $\{\alpha_1,\alpha_2,\alpha_3,v,v_{\PS},m_1,m_3\}$
for the broken phase and $\{\alpha_1,v,v_{\PS},a_1,m_1,m_2, m_3\equiv \MA \}$ for the dark matter phase.
\vspace{2mm}


The real singlet model, RxSM, is obtained with the addition of a real singlet $\PS$ with a
$\mathbb{Z}_2$  symmetry ($\PS\rightarrow -\PS$) to the SM. The most
general renormalizable potential then reads
\begin{equation}
V_{\rm RxSM}=\dfrac{m^2}{2}\PH^\dagger \PH+\dfrac{\lambda}{4}(\PH^\dagger
\PH)^2+\dfrac{\lambda_{\PH\PS}}{2}\PH^\dagger \PH \PS^2 +\dfrac{m^2_{\PS}}{2} \PS^2 +
\dfrac{\lambda_{\PS}}{4!}\PS^4 \, ,  \label{eq:V_RxSM}
\end{equation}
 where $m,\ \lambda,\ \lambda_{\PH\PS},\ m_{\PS}$ and $\lambda_{\PS}$ are real and
\begin{equation}\label{eq:vacua:RxSM}
\PH=\dfrac{1}{\sqrt{2}}\left(\begin{array}{c} \PG^+ \\
    v+\Ph+i\PG^0\end{array}\right) \quad \mbox{and}  \quad \PS =v_{\PS} + \Ps \;.
\end{equation}
Here, again, $v \approx 246 \UGeV$ is the SM Higgs VEV, and $v_{\PS}$ is
the singlet VEV. The benchmarks presented for this model are for the
broken phase, $v_{\PS}\neq 0$, where $m_1$ and $m_2$ are the mass-ordered
scalar states. The mixing matrix has the same form as the sub-block responsible for the mixing in the dark phase of the CxSM, i.e. when $\alpha_1\equiv
\alpha$ and $\alpha_2=\alpha_3=0$. In the symmetric phase of the RxSM, which we will not
discuss here, one of the scalars is the SM-like Higgs boson while the other
is the dark matter candidate.

The model has five independent parameters. Two different sets of parameters
were proposed. Common to both sets are the particle masses, $\{m_1,m_2\}$, the angle $\alpha_1 \equiv \alpha$
(or $\sin \alpha$) and the SM VEV $v$, which is determined from the
Fermi constant $G_F$. The remaining independent parameter is chosen to be $v_{\PS}$
or $\tan \beta = v/v_{\PS}$.

The renormalization of the RxSM has recently been addressed
in~\cite{Kanemura:2015fra, Bojarski:2015kra}.
It was found~\cite{Kanemura:2015fra} that the electroweak corrections to Higgs bosons decays to gauge bosons and fermions
are at most of the order of 1\%
and this maximal value is attained in the limit where the theory
becomes  indistinguishable from the SM.
In~\cite{Bojarski:2015kra} it was found that the corrections to the triple scalar vertex ($Hhh$)
are small, typically of a few per cent, once all theoretical and
experimental constraints
are taken into account.

The issue of interference in beyond the SM scenarios has been raised in~\cite{Maina:2015ela} for
the particular case of a real singlet extension of the SM.
Although these interference effects can be large, it was shown in~\cite{Kauer:2015hia} for the process
$\Pg\Pg \to \Ph^*, \PH^{(*)} \to \PZ\PZ  \to 4\ell$
at 8~\UTeV, that judicious kinematical cuts can be used in the
analysis to reduce the interference  effects to ${\cal O}$(10\%).
Interference effects were also studied for
$\Pg\Pg \to \Ph^*, \PH^{(*)} \to \Ph\Ph$ at next-to-leading order (NLO) in
QCD~\cite{Dawson:2015haa}. It was found that the interference effects
distort the double Higgs boson invariant mass distributions. Depending on
the heavier Higgs boson mass value, they can either decrease by 30\% or increase
by 20\%. Furthermore, it was shown that the NLO QCD corrections are
large and can significantly distort kinematic distributions near
the resonance peak. This means that any experimental analysis to be performed in
the future should take these effects into account.

\subsubsection{Constraints Applied to the Singlet Models}\label{sec:Cons}

To restrict the parameter space of the singlet models  various
theoretical and phenomenological constraints have been applied.
These have been described in detail in~\cite{Pruna:2013bma,
Lopez-Val:2014jva,Robens:2015gla,Robens:2016xkb}
for the first
set of benchmarks and in~\cite{Coimbra:2013qq, Costa:2014qga}
for the second set. Here we will only discuss them briefly.

\subsubsubsection*{Theoretical constraints}
The following theoretical constraints are applied both to the CxSM and to the RxSM:
: i) the potential must be bounded from below; ii) the vacuum
is the global minimum and iii) perturbative unitarity holds.

The first set of benchmarks also demands perturbativity of the couplings.
These are imposed using one-loop RGEs~\cite{Lerner:2009xg}. The perturbativity
and minimization conditions i), ii) are required to hold at a scale
of $\sim\,10^{10}$~\UGeV.

For the second set of benchmarks, the conditions that the potential is bounded from below and that the couplings remain perturbative were required to hold at higher scales with the two-loop RGE evolution of the couplings, as discussed in~\cite{Costa:2014qga}.

\subsubsubsection*{Dark matter constraints}
In the dark phase of the CxSM, the relic density $\Omega_A
h^2$ is computed using {\tt micrOMEGAS}~\cite{Belanger:2014hqa} and points in
parameter space are excluded if $\Omega_A h^2$ is larger
than $\Omega_ch^2+3\sigma$. Here $\Omega_{c}h^2=0.1199\pm 0.0027$
is the combined result from the WMAP and Planck
satellites~\cite{Ade:2013zuv,Hinshaw:2012aka} and $\sigma$ is the
standard deviation.
Direct detection bounds are imposed with the spin-independent scattering cross section
of weakly interacting massive particles (WIMPs) on nucleons also with
{\tt micrOMEGAS} with the procedure described in~\cite{Coimbra:2013qq}.
Points are rejected if the cross section is larger than the upper bound
obtained by the \textsc{LUX2013} collaboration~\cite{Akerib:2013tjd}.

\subsubsubsection*{Electroweak precision observables}
A $95\%$ exclusion limit is applied from the electroweak precision
observables $S,T,U$~\cite{Maksymyk:1993zm,Peskin:1991sw}.
The first set of benchmarks also takes explicit limits from a higher
order calculation of the $\PW$-boson mass \cite{Lopez-Val:2014jva} into account.

\subsubsubsection*{Collider constraints}

The strongest phenomenological constraints come from collider data
and in particular from the the LHC data. Whichever the model or phase
under discussion, one of the scalars has to match the observed signal
for a Higgs boson with a mass of $\simeq 125$~\UGeV. The remaining scalars
must be compatible with the exclusion limits set by the Tevatron, LEP
and LHC searches. $95\%$ C.L.~exclusion limits were applied using
{\tt HiggsBounds}~\cite{Bechtle:2013wla}. As for consistency with
the Higgs boson signal measurements, we test the global signal strength of
the $125$~\UGeV\ Higgs boson with the latest combination of the ATLAS
and the CMS LHC Run~1 datasets, i.e.~$\mu_{125}=1.09\pm
0.11$~\cite{Khachatryan:2016vau}.

In the first set of benchmark points {\tt HiggsSignals} \cite{Bechtle:2013xfa} is
used to explore the Higgs boson signal rate constraints in the mass range
$100{-}150 \UGeV$. This way a potential signal overlap of
the two Higgs states in the LHC signal rate measurement can
approximately be taken into account.

{\tt HiggsBounds}~\cite{Bechtle:2008jh, Bechtle:2011sb, Bechtle:2015pma}
computes internally various experimental quantities
such as the signal rates
\begin{equation}
\mu_{\Phindexi} =\dfrac{\sigma_{\rm New}(\Phindexi){\rm BR}_{\rm
    New}\left(\Phindexi\rightarrow \PX_{\rm SM}\right)}{\sigma_{\rm
    SM}(\Phindexi){\rm BR}_{\rm SM} \left(\Phindexi\rightarrow \PX_{\rm
      SM}\right)} \, \,  \; .
\label{mu}
\end{equation}
Here $\sigma_{\rm New}$ denotes the production cross section
  of the Higgs boson $\Phindexi$ in the new model under consideration and
  $\sigma_{\rm SM}$ the SM production cross section of a Higgs boson with
  the same mass. Similarly, ${\rm BR}_{\rm New}$ is the
  branching ratio in the new model for $\Phindexi$ to decay into a final state with SM particles $\PX_{\rm SM}$. ${\rm BR}_{\rm SM}$ is
  the corresponding SM quantity for a Higgs boson with the same mass.
In singlet models, at leading order (and also at higher order in QCD), the
cross section ratios are all simply given by the suppression factor squared,
$R_{i1}^2$, see Eq.~\eqref{eq:couplings_SM}.

\subsection{Tools}

Cross sections for Higgs boson production via gluon fusion were calculated with the programs
{\tt SusHi}~\cite{Harlander:2012pb} and {\tt HIGLU}~\cite{Spira:1995mt}
at next-to-next-to-leading order (NNLO) QCD and with higher order
electroweak corrections consistently turned off. All other cross sections
were rescaled from the SM ones by \texttt{HiggsBounds} using its internal tables.

The points for the first set are all obtained from the rescaled SM ones except
for the width for $\PH \to \Ph\Ph$ which was calculated at LO.

The points for the second set of benchmarks were obtained with the {\tt ScannerS}
code~\cite{ScannerS,Coimbra:2013qq} where all the constraints previously described
are either in-built or interfaced with other codes. The branching ratios for the new scalars
were calculated with the new implementation of the CxSM and RxSM in
HDECAY~\cite{Djouadi:1997yw, Djouadi:2006bz}
called {\tt sHDECAY}\footnote{The code {\tt sHDECAY} is available at
\url{http://www.itp.kit.edu/~maggie/sHDECAY/}.}~\cite{Costa:2015llh}.

\subsection{Benchmarks}

In this section benchmark points for the real and complex singlet extensions
are presented. For each set of points, plane or scenario a physical motivation is given
together with the main features of the benchmark.

\subsubsection{Benchmark Points and Planes for the RxSM}
\subsubsubsection{Benchmark Points \texorpdfstring{$BP1$}{BP1}}
See Tables~\ref{tab:BMH2}-\ref{tab:BHL}.

\subsubsubsection{Benchmark planes}
See Tables \ref{tab:highm1} and \ref{tab:highm2ExtSca}.

As input parameters, we have three independent parameters,
\begin{equation*}
{m\equiv m_{H/h}},\,\sin\alpha,\,\tan\beta
\end{equation*}
where the latter affects the collider phenomenology only via the additional decay channel $\PH\to \Ph\Ph$. \\
Note that from a collider perspective, for cases where the decay mode $\PH\,\rightarrow\,\Ph\Ph$ is kinematically allowed, the {\sl third} input variable could be replaced by either the total width of the heavier state, the branching ratio $\text{BR} \left( \PH\,\rightarrow\,\Ph\Ph \right)$, or the partial decay width into this channel respectively, such that
\begin{equation*}
\left\{ m\equiv M_{\PH/\Ph},\,\sin\alpha,\,\tan\beta\right\};\,\left\{m\equiv M_{\PH/\Ph},\,\sin\alpha,\,\Gamma(\PH)\right\};\,\left\{m\equiv M_{\PH/\Ph},\,\sin\alpha,\,\text{BR} (\PH\rightarrow\,\Ph\Ph)\right\}
\end{equation*}
are all viable parameter choices.

\begin{table}[h]
\caption{\label{tab:BMH2} Benchmarks for fixed masses and $|\sin\alpha|$, floating $\tan\beta$ (between scenarios a and b)}
\begin{center}
\small{
\begin{tabular}{l|l}
\toprule
\multicolumn{2}{c}  {\bf Benchmarks for the Real Singlet}\cr
\midrule
\multicolumn{2}{c}  { Tania Robens, Tim Stefaniak}\cr
\multicolumn{2}{c}  {Reference: Eur.Phys.J. C76 (2016) no.5, 268 \cite{Robens:2016xkb}} \cr
\midrule
Main Features & real singlet extension, with two VEVs and no hidden sector interaction \cr
&  with heavy Higgs boson H and light Higgs boson h \cr
\midrule
Fixed parameters	& $\Mh $ = 125.1~\UGeV\ or $\MH$ = 125.1~\UGeV\  \cr
Irrelevant parameters & $\tan\beta$ when channel $\PH\,\rightarrow\,\Ph\Ph$ not accessible\cr
& (LO, factorized production and decay) \cr
additional comments & a,b signify maximal and minimal BR for $\PH\,\rightarrow\,\Ph\Ph$ decay; for b, $\sin\alpha\,<\,0$.\cr
& any values  for $\tan\beta$ between scenario a and b are allowed \cr
&$\text{BR} \left( \PH\,\rightarrow\,SM \right)$ needs to be rescaled by $1-\text{BR}\left( \PH\,\rightarrow \Ph \Ph \right)$ \cr
&for SM final states. \cr
\toprule
\multicolumn{2}{c}  {Production cross sections at 14~\UTeV\ [pb]  and branching fractions}\cr
\midrule
\multicolumn{2}{c}  {BHM300 a,b \hspace*{\fill}}\cr
\midrule
Spectrum & $\MH$=300~\UGeV, $|\sin\alpha|=0.31,\,\tan\beta(a)=0.79, \,\tan\beta(b)=0.79 $ \cr
$\sigma(\Pg\Pg \rightarrow \Ph)$ & 44.91  \cr
$\sigma(\Pg\Pg \rightarrow \PH)$ &  1.09  \cr

BR($\PH \rightarrow \Ph\Ph$)  & 0.41 (a), 0.17 (b) \cr
BR($\PH \rightarrow \PW \PW$) & 0.41 (a), 0.57 (b) \cr
BR($\PH \rightarrow\PZ \PZ$)  & 0.18 (a), 0.25 (b) \cr

\midrule
\multicolumn{2}{c}  {BHM400a,b \hspace*{\fill}}\cr
\midrule
Spectrum & $\MH$=400~\UGeV, $|\sin\alpha|=0.26,\,\tan\beta (a)=0.58,\,\tan\beta(b)=0.59$ \cr
$\sigma(\Pg\Pg  \rightarrow h)$ &  46.32 \cr
$\sigma(\Pg\Pg  \rightarrow H)$ &  0.76 \cr

BR($\PH \rightarrow \Ph\Ph$)    & 0.32 (a), 0.20 (b)\cr
BR($\PH \rightarrow \PW \PW$)   & 0.40 (a), 0.47 (b) \cr
BR($\PH \rightarrow\PZ \PZ$)    & 0.18 (a), 0.22 (b) \cr
BR($\PH \rightarrow \PQt\PAQt$) & 0.10 (a), 0.12 (b) \cr

\midrule
\multicolumn{2}{c}  {BHM500a,b \hspace*{\fill}}\cr
\midrule
Spectrum & $\MH$=500~\UGeV, $|\sin\alpha|=0.24,\,\tan\beta (a)=0.46,\,\tan\beta (b)=0.47$ \cr
$\sigma(\Pg\Pg  \rightarrow h)$ & 46.82\cr
$\sigma(\Pg\Pg  \rightarrow H)$ & 0.31 \cr

BR($\PH \rightarrow \Ph\Ph$)    & 0.26 (a), 0.19 (b) \cr
BR($\PH \rightarrow \PW \PW$)   & 0.41 (a), 0.44 (b) \cr
BR($\PH \rightarrow\PZ \PZ$)    & 0.19 (a), 0.21 (b) \cr
BR($\PH \rightarrow \PQt\PAQt$) & 0.14 (a), 0.16 (b) \cr

\bottomrule
\end{tabular}}
\end{center}
\end{table}

\begin{table}
\caption{\label{tab:BHM1} Benchmarks for fixed masses and $|\sin\alpha|$, floating $\tan\beta$ (between scenarios a and b)}
\begin{center}
\small{
\begin{tabular}{l|l}
\toprule
\multicolumn{2}{c}  {BHM600a,b \hspace*{\fill}}\cr
\midrule
Spectrum & $\MH$=600~\UGeV, $|\sin\alpha|=0.22,\,\tan\beta (a) =0.38,\,\tan\beta(b)=0.38 $ \cr
$\sigma(\Pg\Pg  \rightarrow \Ph)$ &  47.28 \cr
$\sigma(\Pg\Pg  \rightarrow \PH)$ &   0.12 \cr

BR($\PH \rightarrow \Ph\Ph$)      &  0.25 (a), 0.19 (b) \cr
BR($\PH \rightarrow \PW \PW$)     &  0.41 (a), 0.45 (b) \cr
BR($\PH \rightarrow\PZ \PZ$)      &  0.21 (a), 0.22 (b)\cr
BR($\PH \rightarrow \PQt\PAQt$)   &  0.13 (a), 0.14 (b) \cr

\midrule
\multicolumn{2}{c}  {BHM700a,b \hspace*{\fill}}\cr
\midrule
Spectrum & $\MH$=700~\UGeV, $|\sin\alpha|=0.21,\,\tan\beta(a)=0.31,\,\tan\beta(b)=0.32$ \cr
$\sigma(\Pg\Pg  \rightarrow \Ph)$ & 47.49\cr
$\sigma(\Pg\Pg  \rightarrow \PH)$ & 0.050\cr

BR($\PH \rightarrow \Ph\Ph$)    & 0.24 (a), 0.19 (b) \cr
BR($\PH \rightarrow \PW \PW$)   & 0.44 (a), 0.47 (b) \cr
BR($\PH \rightarrow\PZ \PZ$)    & 0.22 (a), 0.23 (b)\cr
BR($\PH \rightarrow \PQt\PAQt$) & 0.10 (a), 0.11 (b) \cr

\midrule
\multicolumn{2}{c}  {BHM800a,b \hspace*{\fill}}\cr
\midrule
Spectrum & $\MH$=800~\UGeV, $|\sin\alpha|=0.2,\,\tan\beta (a)=0.25, \tan\beta(b)=0.27 $ \cr
$\sigma(\Pg\Pg  \rightarrow \Ph)$ & 47.46 \cr
$\sigma(\Pg\Pg  \rightarrow \PH)$ & 0.022 \cr

BR($\PH \rightarrow \Ph\Ph$)    & 0.23 (a), 0.19 (b) \cr
BR($\PH \rightarrow \PW \PW$)   & 0.46 (a), 0.48 (b) \cr
BR($\PH \rightarrow\PZ \PZ$)    & 0.23 (a), 0.24 (b) \cr
BR($\PH \rightarrow \PQt\PAQt$) & 0.08 (a), 0.09 (b) \cr

\midrule
\multicolumn{2}{c}  {BHM200 \hspace*{\fill}}\cr
\midrule
Spectrum & $\MH$=200~\UGeV, $|\sin\alpha|=0.29,\,\tan\beta = 1.19 $ \cr
$\sigma(\Pg\Pg  \rightarrow \Ph)$ & 45.50 \cr
$\sigma(\Pg\Pg  \rightarrow \PH)$ & 1.74 \cr
BR($\PH \rightarrow \text{SM}$)   & as a 200~\UGeV\ SM Higgs boson \cr

\bottomrule
\end{tabular}}
\end{center}
\end{table}

\begin{table}
\caption{\label{tab:BHL} Benchmarks for fixed masses and $|\sin\alpha|$, floating $\tan\beta$ (between scenarios a and b). In scenario b $\tan\beta\,=\,-\cot\,\alpha$. Low mass cross sections are courtesy of M. Grazzini}
\begin{center}
\small{
\begin{tabular}{l|l}
\toprule
\multicolumn{2}{c}  {BHM60a,b \hspace*{\fill}}\cr
\midrule
Spectrum & $\Mh $=60~\UGeV, $|\sin\alpha|=0.9997,\,\tan\beta (a) =3.48,\,\tan\beta(b)=0.025 $ \cr
$\sigma(\Pg\Pg  \rightarrow \Ph)$ &  0.10 \cr
$\sigma(\Pg\Pg  \rightarrow \PH)$ & 49.65 \cr
BR($\PH \rightarrow \Ph\Ph$) & 0.26 (a), 0. (b) \cr
BR($\PH \rightarrow \text{SM}$) & rescaled by 0.74 (a), as in SM (b) \cr

\midrule
\multicolumn{2}{c}  {BHM50a,b \hspace*{\fill}}\cr
\midrule
Spectrum & $\Mh $=50~\UGeV, $|\sin\alpha|=0.9998,\,\tan\beta(a)=3.25,\,\tan\beta(b)=0.020$ \cr
$\sigma(\Pg\Pg  \rightarrow \Ph)$ & 0.098\cr

BR($\PH \rightarrow \Ph\Ph$) & 0.26 (a), 0. (b) \cr
BR($\PH \rightarrow \text{SM}$) & rescaled by 0.74 (a), as in SM (b) \cr

\midrule
\multicolumn{2}{c}  {BHM40a,b \hspace*{\fill}}\cr
\midrule
Spectrum & $\Mh $=40~\UGeV, $|\sin\alpha|=0.9998,\,\tan\beta (a)=3.13, \tan\beta(b)=0.020 $ \cr
$\sigma(\Pg\Pg  \rightarrow \Ph)$ & 0.16 \cr

BR($\PH \rightarrow \Ph\Ph$) & 0.26 (a), 0. (b) \cr
BR($\PH \rightarrow \text{SM}$) & rescaled by 0.74 (a), as in SM (b) \cr

\midrule
\multicolumn{2}{c}  {BHM30a,b \hspace*{\fill}}\cr
\midrule
Spectrum & $\Mh $=30 GeV, $|\sin\alpha|=0.9998,\,\tan\beta (a)=3.16, \tan\beta(b)=0.020 $ \cr
$\sigma(\Pg\Pg  \rightarrow \Ph)$ & 0.31 \cr
BR($\PH \rightarrow \Ph\Ph$) & 0.26 (a), 0. (b) \cr
BR($\PH \rightarrow \text{SM}$) & rescaled by 0.74 (a), as in SM (b) \cr

\midrule
\multicolumn{2}{c}  {BHM20a,b \hspace*{\fill}}\cr
\midrule
Spectrum & $\Mh $=20~\UGeV, $|\sin\alpha|=0.9998,\,\tan\beta (a)=3.23, \tan\beta(b)=0.020 $ \cr
$\sigma(\Pg\Pg  \rightarrow \Ph)$ & 0.90 \cr
BR($\PH \rightarrow \Ph\Ph$) & 0.26 (a), 0. (b) \cr
BR($\PH \rightarrow \text{SM}$) & rescaled by 0.74 (a), as in SM (b) \cr

\midrule
\multicolumn{2}{c}  {BHM10a,b \hspace*{\fill}}\cr
\midrule
Spectrum & $\Mh $=10~\UGeV, $|\sin\alpha|=0.9998,\,\tan\beta (a)=3.29, \tan\beta(b)=0.020 $ \cr
$\sigma(\Pg\Pg  \rightarrow \Ph)$ & 2.98 \cr
BR($\PH \rightarrow \Ph\Ph$) & 0.26 (a), 0. (b) \cr
BR($\PH \rightarrow \text{SM}$) & rescaled by 0.74 (a), as in SM (b) \cr

\bottomrule
\end{tabular}}
\end{center}
\end{table}

\begin{table}
\caption{\label{tab:highm1} Benchmark points for mass ranges where the onshell decay $\PH\,\rightarrow\,\Ph\Ph $ is kinematically forbidden. Maximal values of $\tan\beta$ were calculated at the maximal mixing angle, and should be applied for consistency reasons. $\PH$ decays the same way a SM-like Higgs boson of the same mass would decay, and the production cross sections need to be rescaled by $\sin^2\alpha$ with respect to SM predictions for the heavy and $\cos^2\alpha$ for the light Higgs boson.  Production cross sections range from 8.21 pb (for $\MH =130$~\UGeV) to 1.8 pb (for $\MH =245$~\UGeV).}
\begin{center}
\small{
\begin{tabular}{ccc|ccc}
\toprule
$\MH $~[\UGeV]&$|\sin\alpha|_\text{max}$&$\tan\beta_\text{max}$ &$\MH [\UGeV]$&$|\sin\alpha|_\text{max}$&$\tan\beta_\text{max}$ \\ \midrule
130&	0.42& 1.79 & 195 & 0.28 &1.22\\
135&	0.38& 1.73 & 200 & 0.29 &1.19\\
140&	0.36& 1.69 & 210 & 0.28 &1.14\\
145&	0.35& 1.62 & 215 & 0.33 &1.12\\
150&	0.34& 1.57 & 220 & 0.34 &1.10\\
160&    0.36& 1.49 & 230 & 0.35 &1.05\\
180&    0.30& 1.32 & 235 & 0.34 &1.03\\
185&    0.27& 1.28 & 240 & 0.31 &1.00\\
190&    0.29& 1.26 & 245 & 0.28 &0.98\\
\bottomrule
\end{tabular}}
\end{center}
\end{table}

\begin{table}
\caption{\label{tab:highm2ExtSca} Maximal and minimal allowed branching ratios, taken at the maximal allowed value of $|\sin\alpha|$. Note that minimal values for the BR stem from $\sin\alpha\,\leq\,0$. Decay branching ratios correspond to the BRs of a SM Higgs boson of the same mass, rescaled by $1-\text{BR}(H\,\rightarrow\,\Ph\Ph)$. Production cross sections range from 1.71 pb (for $\MH =250$~\UGeV) to 0.004 pb (for $\MH =1$~\UTeV). }
\begin{center}
\small{
\begin{tabular}{cccc}
\toprule
 $\MH [\UGeV]$ & $|\sin\alpha|_\text{max}$ & BR$(\PH\to\Ph\Ph)_\text{min}$ & BR$(\PH\to\Ph\Ph)_\text{max}$ \\
\midrule
255& 0.31 & 0.09 & 0.27  \\
260& 0.34 & 0.11 & 0.33  \\
265& 0.33 & 0.13 & 0.36  \\
280& 0.32 & 0.17 & 0.40  \\
290& 0.31 & 0.18 & 0.40  \\
305& 0.30 & 0.20 & 0.40  \\
325& 0.29 & 0.21 & 0.40  \\
345& 0.28 & 0.22 & 0.39  \\
365& 0.27 & 0.21 & 0.36  \\
395& 0.26 & 0.20 & 0.32  \\
 430 & 0.25 & 0.19 & 0.30  \\
 470 &	0.24 & 0.19 & 0.28 \\
 520 &	0.23 & 0.19 & 0.26 \\
 590 &	0.22 & 0.19 & 0.25 \\
 665 &	0.21 & 0.19 & 0.24 \\
 770 &	0.20 & 0.19 & 0.23 \\
 875 &	0.19 & 0.19 & 0.22 \\
 920 & 0.18 & 0.19 & 0.22  \\
 975 & 0.17 & 0.19 & 0.21  \\
1000 & 0.17 & 0.19 & 0.21  \\
\bottomrule
\end{tabular}}
\end{center}
\end{table}

\subsection{Benchmark points for the CxSM and RxSM}
\label{Sec:LHC_Run2_benchs}

In the numbers presented we have set the SM-like
Higgs boson mass to $125.1$~\UGeV\ which is the central value\footnote{The
  reported value with the experimental errors is  $\Mh=125.09\pm
  0.21 {\rm(stat)} \pm 0.11 {\rm(syst)}$~\UGeV.} of the ATLAS/CMS
combination reported in~\cite{Aad:2015zhl}.

\subsubsection{Benchmark Points for the CxSM and RxSM}

The benchmark points to be presented were chosen as to cover
different physical situations. The first goal is to maximize
the number of scalars being produced while preserving
consistency with the LHC Run~1 measurements. 
Hence we require to be consistent with the global signal
strength obtained by the combination of the ATLAS and CMS data from
the LHC Run~1, within at most $3\sigma$. 
In most scenarios we find points satisfying the required properties within $2\sigma$.

Besides phenomenological requirements, whenever possible, we choose points
for which the model remains stable up to a large cutoff scale
$\mu$, where the theory reaches a Landau pole or the scalar potential develops a
runaway direction (a detailed two-loop analysis can be found in~\cite{Costa:2014qga}).
In the dark matter phase, we require that the dark matter relic
density predicted by the model, $\Omega_A h^2$, is within 3$\sigma$ of the central value for the WMAP and Planck combination quoted in Section~\ref{sec:Cons}.

\subsubsection{CxSM Broken Phase}

 In Tables~\ref{tabTh:CxSM_Broken} and~\ref{tabPh:CxSM_Broken} we show
a sample of various kinematically allowed set-ups for the three mixing
scalars of the broken phase of the CxSM. The first,
Table~\ref{tabTh:CxSM_Broken}, contains the parameters which define
the chosen benchmark points and the production rates of the lightest
and next-to-lightest Higgs bosons $\Phone $ and $\Phtwo $ in the various final
states. The corresponding values for $\Phthree $ are listed in
Table~\ref{tabPh:CxSM_Broken}. For $\Phtwo $ and $\Phthree $ the tables contain, in
particular, the Higgs-to-Higgs boson decay rates. We also give
the signal rates $M_{\Phindexi}$ (i$=1,2,3$) as defined in
Eq.~\eqref{mu}. An interesting feature is that there are points which stabilize the model up to a large cutoff scale, $\log_{10}(\mu/\UGeV)$, as shown for example for CxSM.B3 and CxSM.B4 in Table~\ref{tabPh:CxSM_Broken}. Most points are such that the cross sections for the indirect
decay channels of the new scalars can compete with the direct
decays. In particular, in most cases, we have tried to maximize
$\Phthree \rightarrow \Phtwo  \Phone $ where all three scalars could be observed at
once.  We have furthermore chosen points with large cross sections for the
new scalars, so that they can also be detected directly in their
decays.

\begin{table}[t!]
\vspace{-0.8cm}
\caption{{\em Benchmark points for the CxSM broken phase:}  The
  parameters of the theory that we take as input values are denoted
  with a star ($\star$). The cross sections are for $\sqrt{s}\equiv 13$~\UTeV. }
\label{tabTh:CxSM_Broken}
\centering
\footnotesize
\begin{tabular}{c|c|c|c|c|c}
\toprule
\multicolumn{6}{c}  {\bf Benchmarks for the CxSM -- Broken phase}\cr
\midrule
\multicolumn{6}{c}  {Raul Costa, Margarete M\"uhlleitner,  Marco O. P.Sampaio, Rui Santos}\cr
\multicolumn{6}{c}  {Reference:~\cite{Costa:2015llh} (see also~\cite{Costa:2014qga})} \\[3pt]
\toprule
 	 & CxSM.B1 	 & CxSM.B2 	 & CxSM.B3 	 & CxSM.B4 	 & CxSM.B5\\ 
\midrule	 
$\star$ $M_{h1}$ (\UGeV) 	 & $125.1$ 	 & $125.1$ 	 & $57.83$ 	 & $86.79$ 	 & $33.17$\\
$\phantom{\star}$ $M_{\Phtwo }$ (\UGeV) 	 & $260.6$ 	 & $228$ 	 & $125.1$ 	 & $125.1$ 	 & $64.99$\\
$\star$ $M_{\Phthree }$ (\UGeV) 	 & $449.6$ 	 & $311.3$ 	 & $299$ 	 & $291.8$ 	 & $125.1$\\
$\star$ $\alpha_1$ 	 & $-0.04375$ 	 & $0.05125$ 	 & $-1.102$ 	 & $-1.075$ 	 & $1.211$\\
$\star$ $\alpha_2$  	 & $0.4151$ 	 & $-0.4969$ 	 & $1.136$ 	 & $0.8628$ 	 & $-1.319$\\
$\star$ $\alpha_3$ 	 & $-0.6983$ 	 & $-0.5059$ 	 & $-0.02393$ 	 & $-0.0184$ 	 & $1.118$\\
$\star$ $v_{\PS}$ (\UGeV) 	 & $185.3$ 	 & $52.3$ 	 & $376.9$ 	 & $241.9$ 	 & $483.2$\\
$\phantom{\star}$ $v_{\PA}$ (\UGeV) 	 & $371.3$ 	 & $201.6$ 	 & $236.3$ 	 & $286.1$ 	 & $857.8$\\
$\lambda$ 	 & $1.148$ 	 & $1.018$ 	 & $0.869$ 	 & $0.764$ 	 & $0.5086$\\
$\delta_2$ 	 & $-0.9988$ 	 & $1.158$ 	 & $-0.4875$ 	 & $-0.4971$ 	 & $0.01418$\\
$d_2$ 	 & $1.819$ 	 & $3.46$ 	 & $0.6656$ 	 & $0.9855$ 	 & $0.003885$\\
$m^2$ ($\UGeV^2$) 	 & $5.118\times 10^{4}$ 	 & $-5.597\times 10^{4}$ 	 & $2.189\times 10^{4}$ 	 & $1.173\times 10^{4}$ 	 & $-2.229\times 10^{4}$\\
$b_2$ ($\UGeV^2$) 	 & $-3.193\times 10^{4}$ 	 & $-5.147\times 10^{4}$ 	 & $-3.484\times 10^{4}$ 	 & $-3.811\times 10^{4}$ 	 & $1362$\\
$b_1$ ($\UGeV^2$) 	 & $9.434\times 10^{4}$ 	 & $5.864\times 10^{4}$ 	 & $1.623\times 10^{4}$ 	 & $1.599\times 10^{4}$ 	 & $3674$\\
$a_1$ ($\UGeV^3$) 	 & $-1.236\times 10^{7}$ 	 & $-2.169\times 10^{6}$ 	 & $-4.325\times 10^{6}$ 	 & $-2.735\times 10^{6}$ 	 & $-1.255\times 10^{6}$\\[3pt]
\toprule
$\mu^{C}_{\Phone }/\mu^T_{\Phone }$ 	 & $0.0127$ 	 & $0.0407$ 	 & $0.365$ 	 & $0.117$ 	 & $0.687$\\ 
\midrule
$\mu_{\Phone }$ 	 & $0.836$ 	 & $0.771$ 	 & $0.0362$ 	 & $0.0958$ 	 & $0.00767$\\ 
\midrule
$\sigma_1\equiv \sigma(\Pg \Pg\rightarrow \Phone )$ 	 & $36.1$ [pb] 	 & $33.3$ [pb] 	 & $6.42$ [pb] 	 & $8.03$ [pb] 	 & $4.61$ [pb]\\
$\sigma_1\times {\rm BR}(\Phone \rightarrow \PW\PW)$ 	 & $7.55$ [pb] 	 & $6.96$ [pb] 	 & $0.345$ [fb] 	 & $10.3$ [fb] 	 & $<0.01$ [fb]\\
$\sigma_1\times {\rm BR}(\Phone \rightarrow \PZ\PZ )$ 	 & $944$ [fb] 	 & $871$ [fb] 	 & $0.106$ [fb] 	 & $2.44$ [fb] 	 & $<0.01$ [fb]\\
$\sigma_1\times {\rm BR}(\Phone \rightarrow \PQb\PQb)$ 	 & $21.3$ [pb] 	 & $19.6$ [pb] 	 & $5.48$ [pb] 	 & $6.6$ [pb] 	 & $4.01$ [pb]\\
$\sigma_1\times {\rm BR}(\Phone \rightarrow \PGt\PGt)$ 	 & $2.29$ [pb] 	 & $2.11$ [pb] 	 & $501$ [fb] 	 & $659$ [fb] 	 & $323$ [fb]\\
$\sigma_1\times {\rm BR}(\Phone \rightarrow \PGg\PGg)$ 	 & $83.7$ [fb] 	 & $77.2$ [fb] 	 & $2.87$ [fb] 	 & $9.13$ [fb] 	 & $0.617$ [fb]\\[3pt]
\toprule
$\mu^{C}_{\Phtwo }/\mu^T_{\Phtwo }$ 	 & $0.0958$ 	 & $0$ 	 & $0.0128$ 	 & $0.0104$ 	 & $0.353$\\
\midrule
$\mu_{\Phtwo }$ 	 & $0.0752$ 	 & $0.0759$ 	 & $0.782$ 	 & $0.785$ 	 & $0.0106$\\
\midrule
${\rm BR}(\Phtwo \rightarrow \PX_{\rm SM})$ \% 	 & $87.9$ 	 & $100$ 	 & $96.2$ 	 & $100$ 	 & $100$\\
\midrule
$\sigma_2 \equiv \sigma(\Pg\Pg\rightarrow \Phtwo )$ 	 & $1.01$ [pb] 	 & $1.11$ [pb] 	 & $35.1$ [pb] 	 & $33.9$ [pb] 	 & $1.51$ [pb]\\
$\sigma_2\times {\rm BR}(\Phtwo \rightarrow \PW\PW)$ 	 & $618$ [fb] 	 & $784$ [fb] 	 & $7.06$ [pb] 	 & $7.09$ [pb] 	 & $0.185$ [fb]\\
$\sigma_2\times {\rm BR}(\Phtwo \rightarrow \PZ\PZ )$ 	 & $265$ [fb] 	 & $319$ [fb] 	 & $883$ [fb] 	 & $887$ [fb] 	 & $0.0553$ [fb]\\
$\sigma_2\times {\rm BR}(\Phtwo \rightarrow \PQb\PQb)$ 	 & $0.83$ [fb] 	 & $1.66$ [fb] 	 & $19.9$ [pb] 	 & $20$ [pb] 	 & $1.27$ [pb]\\
$\sigma_2\times {\rm BR}(\Phtwo \rightarrow \PGt\PGt)$ 	 & $0.103$ [fb] 	 & $0.201$ [fb] 	 & $2.14$ [pb] 	 & $2.15$ [pb] 	 & $120$ [fb]\\
$\sigma_2\times {\rm BR}(\Phtwo \rightarrow \PGg\PGg)$ 	 & $0.0189$ [fb] 	 & $0.0373$ [fb] 	 & $78.3$ [fb] 	 & $78.6$ [fb] 	 & $0.873$ [fb]\\
\midrule
${\rm BR}(\Phtwo \rightarrow \Phone \Phone )$ \% 	 & $12.1$ 	 & $0$ 	 & $3.82$ 	 & $0$ 	 & $0$\\
\midrule
$\sigma_2\times {\rm BR}(\Phtwo \rightarrow \Phone  \Phone )$ 	 & $122$ [fb] 	 & $0$ 	 & $1.34$ [pb] 	 & $0$ 	 & $0$\\
$\sigma_2\times {\rm BR}(\Phtwo \rightarrow \Phone  \Phone \rightarrow \PQb\PQb\PQb\PQb)$ 	 & $42.5$ [fb] 	 & $0$ 	 & $977$ [fb] 	 & $0$ 	 & $0$\\
$\sigma_2\times {\rm BR}(\Phtwo \rightarrow \Phone  \Phone \rightarrow \PQb\PQb\PGt\PGt)$ 	 & $9.13$ [fb] 	 & $0$ 	 & $179$ [fb] 	 & $0$ 	 & $0$\\
$\sigma_2\times {\rm BR}(\Phtwo \rightarrow \Phone  \Phone \rightarrow \PQb\PQb\PW\PW)$ 	 & $30.1$ [fb] 	 & $0$ 	 & $0.123$ [fb] 	 & $0$ 	 & $0$\\
$\sigma_2\times {\rm BR}(\Phtwo \rightarrow \Phone  \Phone \rightarrow \PQb\PQb\PGg\PGg)$ 	 & $0.334$ [fb] 	 & $0$ 	 & $1.02$ [fb] 	 & $0$ 	 & $0$\\
$\sigma_2\times {\rm BR}(\Phtwo \rightarrow \Phone  \Phone \rightarrow \PGt\PGt\PGt\PGt)$ 	 & $0.491$ [fb] 	 & $0$ 	 & $8.16$ [fb] 	 & $0$ 	 & $0$\\
\bottomrule
\end{tabular}
\end{table}

\begin{table}[t!]
\caption{{\em CxSM broken phase benchmarks (continuation of
    \refT{tabTh:CxSM_Broken})} \label{tabPh:CxSM_Broken}}
\begin{center}
\footnotesize
\begin{tabular}{c|c|c|c|c|c}
\toprule
 	 & CxSM.B1 	 & CxSM.B2 	 & CxSM.B3 	 & CxSM.B4 	 & CxSM.B5\\[3pt]
\toprule	 
$\mu_{\Phthree }$ 	 & $0.0558$ 	 & $0.0791$ 	 & $0.0788$ 	 & $0.0491$ 	 & $0.855$\\
\midrule	 
${\rm BR}(\Phthree \rightarrow \PX_{\rm SM})$ \% 	 & $71$ 	 & $51.6$ 	 & $52.2$ 	 & $41.2$ 	 & $87.1$\\
\midrule	 
$\sigma_3 \equiv \sigma(\Pg\Pg \rightarrow \Phthree )$ 	 & $520$ [fb] 	 & $1.46$ [pb] 	 & $1.48$ [pb] 	 & $1.2$ [pb] 	 & $42.4$ [pb]\\
$\sigma_3\times {\rm BR}(\Phthree \rightarrow \PW\PW)$ 	 & $201$ [fb] 	 & $519$ [fb] 	 & $536$ [fb] 	 & $344$ [fb] 	 & $7.72$ [pb]\\
$\sigma_3\times {\rm BR}(\Phthree \rightarrow \PZ\PZ )$ 	 & $95$ [fb] 	 & $232$ [fb] 	 & $238$ [fb] 	 & $152$ [fb] 	 & $966$ [fb]\\
$\sigma_3\times {\rm BR}(\Phthree \rightarrow \PQb\PQb)$ 	 & $0.0569$ [fb] 	 & $0.401$ [fb] 	 & $0.468$ [fb] 	 & $0.323$ [fb] 	 & $21.8$ [pb]\\
$\sigma_3\times {\rm BR}(\Phthree \rightarrow \PGt\PGt)$ 	 & $<0.01$ [fb] 	 & $0.0513$ [fb] 	 & $0.0594$ [fb] 	 & $0.0408$ [fb] 	 & $2.34$ [pb]\\
$\sigma_3\times {\rm BR}(\Phthree \rightarrow \PGg\PGg )$ 	 & $<0.01$ [fb] 	 & $<0.01$ [fb] 	 & $0.0105$ [fb] 	 & $<0.01$ [fb] 	 & $85.6$ [fb]\\
\midrule	 
${\rm BR}(\Phthree \rightarrow \Phone \Phone )$ \% 	 & $8.53$ 	 & $48.4$ 	 & $29.5$ 	 & $35.4$ 	 & $11.0$\\
\midrule	 
$\sigma_3\times {\rm BR}(\Phthree \rightarrow \Phone  \Phone )$ 	 & $44.3$ [fb] 	 & $706$ [fb] 	 & $438$ [fb] 	 & $426$ [fb] 	 & $4.66$ [pb]\\
$\sigma_3\times {\rm BR}(\Phthree \rightarrow \Phone  \Phone \rightarrow \PQb\PQb\PQb\PQb)$ 	 & $15.4$ [fb] 	 & $246$ [fb] 	 & $319$ [fb] 	 & $289$ [fb] 	 & $3.52$ [pb]\\
$\sigma_3\times {\rm BR}(\Phthree \rightarrow \Phone  \Phone \rightarrow \PQb\PQb\PGt\PGt)$ 	 & $3.32$ [fb] 	 & $52.8$ [fb] 	 & $58.2$ [fb] 	 & $57.6$ [fb] 	 & $567$ [fb]\\
$\sigma_3\times {\rm BR}(\Phthree \rightarrow \Phone  \Phone \rightarrow \PQb\PQb\PW\PW)$ 	 & $10.9$ [fb] 	 & $174$ [fb] 	 & $0.0401$ [fb] 	 & $0.897$ [fb] 	 & $0.011$ [fb]\\
$\sigma_3\times {\rm BR}(\Phthree \rightarrow \Phone  \Phone \rightarrow \PQb\PQb\PGg\PGg)$ 	 & $0.121$ [fb] 	 & $1.93$ [fb] 	 & $0.334$ [fb] 	 & $0.798$ [fb] 	 & $1.08$ [fb]\\
$\sigma_3\times {\rm BR}(\Phthree \rightarrow \Phone  \Phone \rightarrow \PGt\PGt\PGt\PGt)$ 	 & $0.178$ [fb] 	 & $2.84$ [fb] 	 & $2.66$ [fb] 	 & $2.88$ [fb] 	 & $22.9$ [fb]\\
\midrule	 
${\rm BR}(\Phthree \rightarrow \Phone \Phtwo )$ \% 	 & $20.5$ 	 & $0$ 	 & $5.98$ 	 & $17.2$ 	 & $1.93$\\
\midrule	 
$\sigma_3\times {\rm BR}(\Phthree \rightarrow \Phone  \Phtwo )$ 	 & $107$ [fb] 	 & $0$ 	 & $88.8$ [fb] 	 & $207$ [fb] 	 & $820$ [fb]\\
$\sigma_3\times {\rm BR}(\Phthree \rightarrow \Phone  \Phtwo \rightarrow \PQb\PQb\PQb\PQb)$ 	 & $0.0518$ [fb] 	 & $0$ 	 & $43$ [fb] 	 & $100$ [fb] 	 & $603$ [fb]\\
$\sigma_3\times {\rm BR}(\Phthree \rightarrow \Phone  \Phtwo \rightarrow \PQb\PQb\PGt\PGt)$ 	 & $0.012$ [fb] 	 & $0$ 	 & $8.55$ [fb] 	 & $20.8$ [fb] 	 & $105$ [fb]\\
$\sigma_3\times {\rm BR}(\Phthree \rightarrow \Phone  \Phtwo \rightarrow \PQb\PQb\PW\PW)$ 	 & $38.6$ [fb] 	 & $0$ 	 & $15.2$ [fb] 	 & $35.8$ [fb] 	 & $0.0883$ [fb]\\
$\sigma_3\times {\rm BR}(\Phthree \rightarrow \Phone  \Phtwo \rightarrow \PQb\PQb\PGg\PGg)$ 	 & $<0.01$ [fb] 	 & $0$ 	 & $0.191$ [fb] 	 & $0.534$ [fb] 	 & $0.506$ [fb]\\
$\sigma_3\times {\rm BR}(\Phthree \rightarrow \Phone  \Phtwo \rightarrow \PGt\PGt\PGt\PGt)$ 	 & $<0.01$ [fb] 	 & $0$ 	 & $0.422$ [fb] 	 & $1.08$ [fb] 	 & $4.56$ [fb]\\
\midrule	 
${\rm BR}(\Phthree \rightarrow \Phtwo \Phtwo )$ \% 	 & $0$ 	 & $0$ 	 & $12.3$ 	 & $6.24$ 	 & $0$\\
\midrule	 
$\sigma_3\times {\rm BR}(\Phthree \rightarrow \Phtwo  \Phtwo )$ 	 & $0$ 	 & $0$ 	 & $182$ [fb] 	 & $75.2$ [fb] 	 & $0$\\
$\sigma_3\times {\rm BR}(\Phthree \rightarrow \Phtwo  \Phtwo \rightarrow \PQb\PQb\PQb\PQb)$ 	 & $0$ 	 & $0$ 	 & $58.7$ [fb] 	 & $26.2$ [fb] 	 & $0$\\
$\sigma_3\times {\rm BR}(\Phthree \rightarrow \Phtwo  \Phtwo \rightarrow \PQb\PQb\PGt\PGt)$ 	 & $0$ 	 & $0$ 	 & $12.6$ [fb] 	 & $5.63$ [fb] 	 & $0$\\
$\sigma_3\times {\rm BR}(\Phthree \rightarrow \Phtwo  \Phtwo \rightarrow \PQb\PQb\PW\PW)$ 	 & $0$ 	 & $0$ 	 & $41.6$ [fb] 	 & $18.5$ [fb] 	 & $0$\\
$\sigma_3\times {\rm BR}(\Phthree \rightarrow \Phtwo  \Phtwo \rightarrow \PQb\PQb\PGg\PGg)$ 	 & $0$ 	 & $0$ 	 & $0.462$ [fb] 	 & $0.206$ [fb] 	 & $0$\\
$\sigma_3\times {\rm BR}(\Phthree \rightarrow \Phtwo  \Phtwo \rightarrow \PGt\PGt\PGt\PGt)$ 	 & $0$ 	 & $0$ 	 & $0.679$ [fb] 	 & $0.303$ [fb] 	 & $0$\\[3pt]
\toprule
$\log_{10}\left(\frac{\mu}\UGeV\right)$ 	 & $9.40$ 	 & $6.05$ 	 & $19.3$ 	 & $15.7$ 	 & $6.64$\\
\bottomrule
\end{tabular}
\end{center}\vspace{-4mm}
\end{table}

Regarding the various kinematical possibilities, we have chosen:
\begin{itemize}
 \item {\em Two points where the SM-like Higgs boson is the lightest scalar (CxSM.B1 and CxSM.B2)}: For CxSM.B1 all Higgs-to-Higgs boson decay channels are open apart from
$\Phthree \rightarrow \Phtwo  \Phtwo $.\footnote{In
  this scenario we do not present a case with all channels open
  because the spectrum would be even heavier and more difficult to be
  tested.} and it has a large ratio $\mu_{\Ph_{125}}^C/\mu_{\Ph_{125}}^T$. The latter measures the importance of the production of the SM-like Higgs boson in the chain decay of a heavier scalar, $\mu_{\Ph_{125}}^C$ , compared to the total rate $\mu_{\Ph_{125}}^T$ -- see also~\cite{Costa:2015llh}.
The additional Higgs-to-Higgs boson decays, e.g.~$\Phtwothree \to \Phone  \Phone  \to \PQb\PQb \PGt\PGt$, have rates of a few fb, suggesting that all new scalars are expected to be observed. On the other hand, CxSM.B2 has a lighter Higgs boson mass spectrum since only the $\Phthree \rightarrow \Ph_{125} \Ph_{125}$  Higgs-to-Higgs boson decay is open. This has a large branching ratio of $\sim 48\%$ and allows for the largest fraction of chain decays, $\mu_{\Ph_{125}}^C/\mu_{\Ph_{125}}^T\simeq 4\%$, of all the five benchmark points. In addition to the direct decays to SM particles (mostly into massive vector bosons), the scalar $\Phthree $ should also be accessible in the chain decays into a pair of $\Ph_{125}$ bosons ($4\PQb$, $2\PQb 2\PGt$ or $2\PQb 2\PW$ final states), whereas $\Phtwo$ would be visible in its direct decays (also mostly into massive vector bosons).

\item {\em Two points where the SM-like Higgs boson is the next-to-lightest scalar (CxSM.B3 and CxSM.B4)}: For CxSM.B3 all kinematic situations for the scalar decays are available while the spectrum remains light.  Both the Higgs-to-Higgs boson decays for $\Phtwothree$ and the direct decays of $\Phone $ have been maximized so that all scalars may be discovered either in chain decays (where the cross-sections can be of the order of $ 10{-}100$~[fb]) or through their direct decays. The most significant difference in CxSM.B4 is the larger $\Phone $ mass so that the channel $\Phtwo  \to \Phone  \Phone $ is kinematically closed.


\item {\em One point where the SM-like Higgs boson is the heaviest scalar (CxSM.B5)}: This does not allow for SM-like Higgs boson production through chain decays and, since $\Phthree  \equiv \Ph_{125}$, the overall spectrum is very light. This point was chosen to have large branching fractions
for the Higgs-to-Higgs boson decays $\Phthree  \to \Phone \Phone $ or $\Phthree  \to \Phone  \Phtwo $ ($\Phtwo  \to \Phone  \Phone $ is closed), which can reach up to $\sim 500$~fb and $\sim 100$~fb respectively, in the $\PQb\PQb \PGt\PGt$ final state.  The lightest Higgs boson $\Phone $ can also be observed directly, for example in decays to~$\PQb\PQb$ or $\PGt\PGt$, and so can $\Phtwo $ (though with smaller rates).

\end{itemize}

\subsubsection{CxSM Dark Phase}

For the first two points, CxSM.D1 and CxSM.D2, the lightest of the two
visible scalars is the SM-like Higgs boson and, except for $\Phone $ in CxSM.D2, all visible scalars have large invisible decay branching ratios. The
branching ratios for the Higgs-to-Higgs boson decays $\Phtwo  \to \Phone  \Phone $ are
large with 18.4\% for CxSM.D1 and 28\% for CxSM.D2, and the cross sections for the direct production of $\Phtwo $  are also large, so that it can be discovered in its direct decays into SM particles.
An attractive feature of these two points is that the new heavy
scalar $\Phtwo $ can stabilize the theory up to a high scale as can be inferred from $\log_{10}(\mu/\UGeV)$ in the last row of the table.
\begin{table}[t!]
\vspace{-0.8cm}
\caption{{\em Benchmark points for the CxSM dark phase:} The parameters of the theory that we take as input values are denoted with a star ($\star$). The cross-sections are for $\sqrt{s}\equiv 13$~\UTeV.
\label{tab:CxSM_Dark}}
\centering
\footnotesize
\begin{tabular}{c|c|c|c|c}
\toprule
\multicolumn{5}{c}  {\bf Benchmarks for the CxSM -- Dark phase}\cr
\midrule
\multicolumn{5}{c}  {Raul Costa, Margarete M\"uhlleitner,  Marco O. P.Sampaio, Rui Santos, Reference:~\cite{Costa:2015llh} (see also~\cite{Costa:2014qga})} \\[3pt]
\toprule
 	 & CxSM.D1 	 & CxSM.D2 	 & CxSM.D3 & CxSM.D4 \\[3pt]
\toprule	 
$\star$ $M_{\Phone }$ (\UGeV) 	 & $125.1$ 	 & $125.1$ 	 & $56.12$ 	 & $121.2$\\
$\star$ $M_{\Phtwo }$ (\UGeV) 	 & $335.2$ 	 & $341.4$ 	 & $125.1$ 	 & $125.1$\\
$\star$ $\MA $ (\UGeV) 	 & $52.46$ 	 & $93.97$ 	 & $139.3$ 	 & $51.96$\\
$\star$ $\alpha$ 	 & $0.4587$ 	 & $-0.4156$ 	 & $1.507$ 	 & $1.358$\\
$\star$ $v_{\PS}$ (\UGeV) 	 & $812.5$ 	 & $987.5$ 	 & $177.9$ 	 & $909.7$\\
$\lambda$ 	 & $1.142$ 	 & $1.059$ 	 & $0.5146$ 	 & $0.5149$\\
$\delta_2$ 	 & $-0.3839$ 	 & $0.3066$ 	 & $-0.0362$ 	 & $-0.001764$\\
$d_2$ 	 & $0.2669$ 	 & $0.164$ 	 & $0.1653$ 	 & $0.03508$\\
$m^2$ ($\UGeV^2$) 	 & $9.21\times 10^{4}$ 	 & $-1.816\times 10^{5}$ 	 & $-1.503\times 10^{4}$ 	 & $-1.488\times 10^{4}$\\
$b_2$ ($\UGeV^2$) 	 & $-6.838\times 10^{4}$ 	 & $-6.027\times 10^{4}$ 	 & $1.848\times 10^{4}$ 	 & $-1.154\times 10^{4}$\\
$b_1$ ($\UGeV^2$) 	 & $2570$ 	 & $1.132\times 10^{4}$ 	 & $-1.883\times 10^{4}$ 	 & $-2479$\\
$\star$ $a_1$ ($\UGeV^3$) 	 & $-3.057\times 10^{6}$ 	 & $-1.407\times 10^{7}$ 	 & $-7.362\times 10^{4}$ 	 & $-1.418\times 10^{5}$\\[3pt]
\toprule
$\mu^{C}_{\Phone }/\mu^T_{\Phone }$ 	 & $0.019$ 	 & $0.0235$ 	 & $0.97$ 	 & $0$\\
\midrule
$\mu_{\Phone }$ 	 & $0.804$ 	 & $0.837$ 	 & $0.00404$ 	 & $0.0444$\\
\midrule
$ {\rm BR}(\Phone \rightarrow \PX_{\rm SM})$ \% 	 & $70.5$ 	 & $100$ 	 & $100$ 	 & $1.56$\\
\midrule
$\sigma_1\equiv \sigma(\Pg\Pg \rightarrow \Phone )$ 	 & $34.7$ [pb] 	 & $36.2$ [pb] 	 & $759$ [fb] 	 & $2.03$ [pb]\\
$\sigma_1\times {\rm BR}(\Phone \rightarrow \PW\PW)$ 	 & $5.12$ [pb] 	 & $7.56$ [pb] 	 & $0.0331$ [fb] 	 & $4.81$ [fb]\\
$\sigma_1\times {\rm BR}(\Phone \rightarrow \PZ\PZ )$ 	 & $640$ [fb] 	 & $945$ [fb] 	 & $0.0103$ [fb] 	 & $0.561$ [fb]\\
$\sigma_1\times {\rm BR}(\Phone \rightarrow \PQb\PQb)$ 	 & $14.4$ [pb] 	 & $21.3$ [pb] 	 & $649$ [fb] 	 & $20.4$ [fb]\\
$\sigma_1\times {\rm BR}(\Phone \rightarrow \PGt\PGt)$ 	 & $1.55$ [pb] 	 & $2.29$ [pb] 	 & $58.9$ [fb] 	 & $2.18$ [fb]\\
$\sigma_1\times {\rm BR}(\Phone \rightarrow \PGg\PGg )$ 	 & $56.8$ [fb] 	 & $83.8$ [fb] 	 & $0.317$ [fb] 	 & $0.0723$ [fb]\\
\midrule
$\sigma_1\times {\rm BR}(\Phone \rightarrow A A)$ 	 & $10.2$ [pb] 	 & $0$ 	 & $0$ 	 & $2.00$ [pb]\\[3pt]
\toprule
$\mu_{\Phtwo }$ 	 & $0.138$ 	 & $0.108$ 	 & $0.710$ 	 & $0.834$\\
\midrule
$ {\rm BR}(\Phtwo \rightarrow \PX_{\rm SM})$ \% 	 & $70.3$ 	 & $66.1$ 	 & $71.3$ 	 & $87.3$\\
\midrule
$\sigma_2 \equiv \sigma(\Pg\Pg \rightarrow \Phtwo )$ 	 & $1.83$ [pb] 	 & $1.55$ [pb] 	 & $43$ [pb] 	 & $41.3$ [pb]\\
$\sigma_2\times {\rm BR}(\Phtwo \rightarrow \PW\PW)$ 	 & $886$ [fb] 	 & $704$ [fb] 	 & $6.41$ [pb] 	 & $7.54$ [pb]\\
$\sigma_2\times {\rm BR}(\Phtwo \rightarrow \PZ\PZ )$ 	 & $402$ [fb] 	 & $320$ [fb] 	 & $802$ [fb] 	 & $943$ [fb]\\
$\sigma_2\times {\rm BR}(\Phtwo \rightarrow \PQb\PQb)$ 	 & $0.553$ [fb] 	 & $0.417$ [fb] 	 & $18.1$ [pb] 	 & $21.3$ [pb]\\
$\sigma_2\times {\rm BR}(\Phtwo \rightarrow \PGt\PGt)$ 	 & $0.0717$ [fb] 	 & $0.0542$ [fb] 	 & $1.95$ [pb] 	 & $2.29$ [pb]\\
$\sigma_2\times {\rm BR}(\Phtwo \rightarrow \PGg\PGg )$ 	 & $0.012$ [fb] 	 & $<0.01$ [fb] 	 & $71.1$ [fb] 	 & $83.6$ [fb]\\
\midrule
${\rm BR}(\Phtwo \rightarrow \Phone \Phone )$ \% 	 & $18.4$ 	 & $28$ 	 & $28.7$ 	 & $0$\\
\midrule
$\sigma_2\times {\rm BR}(\Phtwo \rightarrow \Phone  \Phone )$ 	 & $337$ [fb] 	 & $436$ [fb] 	 & $12.3$ [pb] 	 & $0$\\
$\sigma_2\times {\rm BR}(\Phtwo \rightarrow \Phone  \Phone \rightarrow \PQb\PQb\PQb\PQb)$ 	 & $58.3$ [fb] 	 & $152$ [fb] 	 & $9.02$ [pb] 	 & $0$\\
$\sigma_2\times {\rm BR}(\Phtwo \rightarrow \Phone  \Phone \rightarrow \PQb\PQb\PGt\PGt)$ 	 & $12.5$ [fb] 	 & $32.6$ [fb] 	 & $1.64$ [pb] 	 & $0$\\
$\sigma_2\times {\rm BR}(\Phtwo \rightarrow \Phone  \Phone \rightarrow \PQb\PQb\PW\PW)$ 	 & $41.3$ [fb] 	 & $107$ [fb] 	 & $0.92$ [fb] 	 & $0$\\
$\sigma_2\times {\rm BR}(\Phtwo \rightarrow \Phone  \Phone \rightarrow \PQb\PQb\PGg\PGg)$ 	 & $0.458$ [fb] 	 & $1.19$ [fb] 	 & $8.81$ [fb] 	 & $0$\\
$\sigma_2\times {\rm BR}(\Phtwo \rightarrow \Phone  \Phone \rightarrow \PGt\PGt\PGt\PGt)$ 	 & $0.675$ [fb] 	 & $1.75$ [fb] 	 & $74.3$ [fb] 	 & $0$\\
\midrule
$\sigma_2\times {\rm BR}(\Phtwo \rightarrow \PA \PA)$ 	 & $207$ [fb] 	 & $91.3$ [fb] 	 & $0$ 	 & $5.23$ [pb]\\[3pt]
\toprule
$\Omega_A h^2$ 	 & $0.118$ 	 & $0.123$ 	 & $0.116$ 	 & $0.125$\\[3pt]
\toprule
$\log_{10}\left(\frac{\mu}\UGeV\right)$ 	 & $14.9$ 	 & $17.1$ 	 & $6.69$ 	 & $6.69$\\
\bottomrule
\end{tabular}
\end{table}

In the scenarios CxSM.D3 and CxSM.D4 the SM-like Higgs boson is
the heaviest of the two visible Higgs bosons. The overall
spectrum is lighter and the theory must have a UV
completion above $\sim 10^3$~\UTeV. Point CxSM.D3 represents a case with
no invisible decays, and in CxSM.D4
decays of the SM-like Higgs boson $\Phtwo $ into a lighter Higgs boson pair are
forbidden while allowing for a large
invisible decay into the dark matter state $\PA$.  In CxSM.D3 $\mu_{\Phtwo }$ is at the edge of compatibility with the LHC data, while allowing for the light
Higgs state $\Phone $ to be discovered in the chain
decay of the SM-like Higgs boson $\Phtwo $ into an $\Phone $ pair or through its direct production (compare for example the $4b$ and the $\PQb\PQb\PGt\PGt$ final state with the direct production).


CxSM.D4, represents the challenging case where the non-SM-like light $\Phone $ has a mass close to the SM-like Higgs boson and a very large invisible decay.  This point is only accessible in its direct decays into SM particles with largest rates in the $\PQb\PQb$ and $\PGt\PGt$ final states.

\subsubsection{RxSM Broken Phase}
Table~\ref{tab:benchmarks_RxSM_Broken} contains four benchmark points
for the two possible kinematic configurations in this model.

For the points RxSM.B1 and RxSM.B2 the SM-like Higgs boson is the lightest of the two scalar states. Benchmark
RxSM.B1 allows a relatively large $\sigma_2 \times {\rm BR}(\Phtwo \to \Phone  \Phone )$, comparable to the direct $\Phtwo $ production cross section, and in particular the $\PQb\PQb\PGt\PGt$ final state reaches 72~fb. For RxSM.B2 the decay into scalars is kinematically closed, but instead various direct decay channels of
$\Phtwo $ are enhanced compared to RxSM.B1, most notably the $\PW\PW$ final state but also $\PQb\PQb$, $\PGt\PGt$ and $\PGg\PGg$.
\begin{table}[t!]
\caption{{\em Benchmark points for the RxSM broken phase:} The
  parameters of the theory that we take as input values are denoted
  with a star ($\star$). The cross sections are for $\sqrt{s}\equiv
  13$~\UTeV.
\label{tab:benchmarks_RxSM_Broken}}
\begin{center}
\footnotesize
\begin{tabular}{c|c|c|c|c}
\toprule
\multicolumn{5}{c}  {\bf Benchmarks for the RxSM -- Broken phase}\cr
\midrule
\multicolumn{5}{c}  {Raul Costa, Margarete M\"uhlleitner,  Marco O. P.Sampaio, Rui Santos}\cr
\multicolumn{5}{c}  {Reference:~\cite{Costa:2015llh}} \\[3pt]
\toprule
	 & RxSM.B1 	 & RxSM.B2 	 & RxSM.B3 	 & RxSM.B4\\[3pt]
\toprule	 
$\star$ $M_{\Phone }$ (\UGeV) 	 & $125.1$ 	 & $125.1$ 	 & $55.26$ 	 & $92.44$\\
$\star$ $M_{\Phtwo }$ (\UGeV) 	 & $265.3$ 	 & $172.5$ 	 & $125.1$ 	 & $125.1$\\
$\star$ $\alpha$ 	 & $-0.4284$ 	 & $-0.4239$ 	 & $1.376$ 	 & $1.156$\\
$\star$ $v_{\PS}$ (\UGeV) 	 & $140.3$ 	 & $94.74$ 	 & $591$ 	 & $686.1$\\
$\lambda$ 	 & $0.828$ 	 & $0.595$ 	 & $0.5007$ 	 & $0.4782$\\
$\lambda_{\PH\PS}$ 	 & $0.599$ 	 & $0.2268$ 	 & $-0.01646$ 	 & $-0.01552$\\
$\lambda_{\PS}$ 	 & $9.294$ 	 & $9.149$ 	 & $0.03029$ 	 & $0.06182$\\
$m^2$ ($\UGeV^2$) 	 & $-3.688\times 10^{4}$ 	 & $-2.007\times 10^{4}$ 	 & $-9426$ 	 & $-7190$\\
$m^2_{\PS}$ ($\UGeV^2$) 	 & $-4.863\times 10^{4}$ 	 & $-2.056\times 10^{4}$ 	 & $-1265$ 	 & $-4380$\\[3pt]
\toprule
$\mu^{C}_{\Phone }/\mu^T_{\Phone }$ 	 & $0.051$ 	 & $0$ 	 & $0.557$ 	 & $0$\\
\midrule
$\mu_{\Phone }$ 	 & $0.827$ 	 & $0.831$ 	 & $0.0376$ 	 & $0.163$\\
$\sigma_1\equiv \sigma(\Pg\Pg \rightarrow \Phone )$ 	 & $35.7$ [pb] 	 & $35.9$ [pb] 	 & $7.26$ [pb] 	 & $12.2$ [pb]\\
$\sigma_1\times {\rm BR}(\Phone \rightarrow \PW\PW)$ 	 & $7.47$ [pb] 	 & $7.5$ [pb] 	 & $0.285$ [fb] 	 & $35.4$ [fb]\\
$\sigma_1\times {\rm BR}(\Phone \rightarrow \PZ\PZ )$ 	 & $935$ [fb] 	 & $938$ [fb] 	 & $0.0887$ [fb] 	 & $6.17$ [fb]\\
$\sigma_1\times {\rm BR}(\Phone \rightarrow \PQb\PQb)$ 	 & $21.1$ [pb] 	 & $21.2$ [pb] 	 & $6.21$ [pb] 	 & $9.9$ [pb]\\
$\sigma_1\times {\rm BR}(\Phone \rightarrow \PGt\PGt)$ 	 & $2.27$ [pb] 	 & $2.28$ [pb] 	 & $562$ [fb] 	 & $1$ [pb]\\
$\sigma_1\times {\rm BR}(\Phone \rightarrow \PGg\PGg )$ 	 & $82.8$ [fb] 	 & $83.2$ [fb] 	 & $2.93$ [fb] 	 & $16.1$ [fb]\\[3pt]
\toprule
$\mu_{\Phtwo }$ 	 & $0.0887$ 	 & $0.169$ 	 & $0.857$ 	 & $0.837$\\
$\sigma_2 \equiv \sigma(\Pg\Pg \rightarrow \Phtwo )$ 	 & $1.97$ [pb] 	 & $4.06$ [pb] 	 & $41.6$ [pb] 	 & $36.2$ [pb]\\
$\sigma_2\times {\rm BR}(\Phtwo \rightarrow \PW\PW)$ 	 & $708$ [fb] 	 & $3.9$ [pb] 	 & $7.73$ [pb] 	 & $7.56$ [pb]\\
$\sigma_2\times {\rm BR}(\Phtwo \rightarrow \PZ\PZ )$ 	 & $305$ [fb] 	 & $112$ [fb] 	 & $967$ [fb] 	 & $946$ [fb]\\
$\sigma_2\times {\rm BR}(\Phtwo \rightarrow \PQb\PQb)$ 	 & $0.897$ [fb] 	 & $30.5$ [fb] 	 & $21.8$ [pb] 	 & $21.3$ [pb]\\
$\sigma_2\times {\rm BR}(\Phtwo \rightarrow \PGt\PGt)$ 	 & $0.111$ [fb] 	 & $3.48$ [fb] 	 & $2.35$ [pb] 	 & $2.29$ [pb]\\
$\sigma_2\times {\rm BR}(\Phtwo \rightarrow \PGg\PGg )$ 	 & $0.0204$ [fb] 	 & $0.582$ [fb] 	 & $85.8$ [fb] 	 & $83.9$ [fb]\\
\midrule
${\rm BR}(\Phtwo \rightarrow \Phone \Phone )$ \% 	 & $48.6$ 	 & $0$ 	 & $11$ 	 & $0$\\
\midrule
$\sigma_2\times {\rm BR}(\Phtwo \rightarrow \Phone  \Phone )$ 	 & $960$ [fb] 	 & $0$ 	 & $4.57$ [pb] 	 & $0$\\
$\sigma_2\times {\rm BR}(\Phtwo \rightarrow \Phone  \Phone \rightarrow \PQb\PQb\PQb\PQb)$ 	 & $334$ [fb] 	 & $0$ 	 & $3.35$ [pb] 	 & $0$\\
$\sigma_2\times {\rm BR}(\Phtwo \rightarrow \Phone  \Phone \rightarrow \PQb\PQb\PGt\PGt)$ 	 & $71.8$ [fb] 	 & $0$ 	 & $605$ [fb] 	 & $0$\\
$\sigma_2\times {\rm BR}(\Phtwo \rightarrow \Phone  \Phone \rightarrow \PQb\PQb\PW\PW)$ 	 & $237$ [fb] 	 & $0$ 	 & $0.307$ [fb] 	 & $0$\\
$\sigma_2\times {\rm BR}(\Phtwo \rightarrow \Phone  \Phone \rightarrow \PQb\PQb\PGg\PGg)$ 	 & $2.62$ [fb] 	 & $0$ 	 & $3.16$ [fb] 	 & $0$\\
$\sigma_2\times {\rm BR}(\Phtwo \rightarrow \Phone  \Phone \rightarrow \PGt\PGt\PGt\PGt)$ 	 & $3.86$ [fb] 	 & $0$ 	 & $27.4$ [fb] 	 & $0$\\
\bottomrule
\end{tabular}
\end{center}
\end{table}

The points RxSM.B3 and RxSM.B4 feature a SM-like Higgs boson
which is the heaviest of the two scalars. RxSM.B3 is such
that the non-SM-like Higgs boson $\Phone $ can be found directly or in the decay
$\Phtwo \rightarrow \Phone  \Phone $. One should note the large rates
for the direct $\Phone $ production and decay into the $\PQb\PQb$ and also $\PGt
\PGt$ final states and compare with the indirect processes
$\Phtwo \rightarrow \Phone  \Phone \rightarrow \PQb\PQb+\PQb\PQb$ and $\PQb\PQb+\PGt\PGt$, where the
magnitude of the latter two is comparable to the former two. RxSM.B4 represents the situation where the indirect channel is closed but the cross section for direct $\Phone $ production is larger. While this allows for larger rates into the $\PQb\PQb$ and $\PGt\PGt$ final states its discovery would be challenging since its mass is very close to the $\PZ$ boson resonance.

\chapter{NMSSM}
\label{chap:manual}
\ChapterAuthor{U.~Ellwanger, M.~M\"uhlleitner, F.~Staub, D.~Strom, R. Yohay (Eds.)
R.~Aggleton, B.~Allanach, M.~Badziak, D. Barducci, G.~B{\'e}langer, C.~Beskidt, N.E.~Bomark, N.~Christensen, W.~de Boer,  H.E.~Haber, C.~Han, T.~Han, C.~Hugonie, D.~Kazakov, S.~Liebler, Z.~Liu, S.~Moretti,  S.~Munir, R.~Nevzorov, C.T.~Potter, A.~Pukhov, C.H.~Shepherd-Themistocleous, A.M.~Teixeira, S.~Wayand, G.~Weiglein,  R.~Ziegler}
\providecommand\SARAH{{\tt SARAH}\xspace}
\providecommand\SPheno{{\tt SPheno}\xspace}
\providecommand\FlexibleSUSY{{\tt FlexibleSUSY}\xspace}
\providecommand\SoftSUSY{{\tt SoftSUSY}\xspace}
\providecommand\NMSSMCalc{{\tt NMSSMCALC}\xspace}
\providecommand\NMSSMTools{{\tt NMSSMTools}\xspace}
\providecommand\Mathematica{{\tt Mathematica}\xspace}
\providecommand\Vevacious{{\tt Vevacious}\xspace}
\providecommand\NMSPEC{{\tt NMSPEC}\xspace}
\providecommand\NMHDECAY{{\tt NMHDECAY}\xspace}
\providecommand\NMSDECAY{{\tt NMSDECAY}\xspace}
\providecommand\SDECAY{{\tt SDECAY}\xspace}
\providecommand\NMGMSB{{\tt NMGMSB}\xspace}
\providecommand\MO{{\tt MicrOmegas}\xspace}
\providecommand\HDECAY{{\tt HDECAY}\xspace}

\providecommand\PHSM{\ensuremath{H_{125}}}
\providecommand\PHS{\ensuremath{H_S}}
\providecommand\PAS{\ensuremath{A_S}}
\providecommand\PA{\ensuremath{A}}

\providecommand\alphat{\ensuremath{\alpha_{\mathrm{t}}}}
\providecommand\alphab{\ensuremath{\alpha_{\mathrm{b}}}}
\providecommand\alphatau{\ensuremath{\alpha_{\mathrm{\tau}}}}
\providecommand\ZT{\ensuremath{{\mathbb{Z}}_3}}

\section{Introduction}
The Next-to-Minimal Supersymmetric Standard Model (NMSSM) \cite{Fayet:1974pd,Dine:1981rt,%
Nilles:1982dy,Derendinger:1983bz,Frere:1983ag,Veselov:1985gd,Ellis:1988er,Ellwanger:1993xa,%
King:1995vk,Ellwanger:1996gw,Franke:1995tc,Ananthanarayan:1996zv,Ellwanger:1999ji,%
Ellwanger:2009dp,Maniatis:2009re}
shares the benefits of supersymmetric extensions of the Standard
Model (SM) with the MSSM: the hierarchy problem can be strongly reduced,
the presence of dark matter can be explained, and the running gauge
couplings are automatically consistent with a Grand Unified Theory (GUT).

In addition, the \ZT-invariant version of the NMSSM (with a
scale invariant superpotential) solves the $\mu$-problem of the MSSM~\cite{Kim:1983dt}.
Both, the general and the \ZT-invariant versions of the NMSSM,
render more natural the mass of $\sim 125~\UGeV$ of the SM-like Higgs boson
\cite{BasteroGil:2000bw,Dermisek:2005gg,Ellwanger:2011mu,Ross:2011xv,Ross:2012nr,Gherghetta:2012gb,%
Perelstein:2012qg,King:2012tr,Kim:2013uxa,Kaminska:2014wia,Binjonaid:2014oga} 
and the non-observation of sparticles like squarks and gluinos at
the Run~1 of the LHC \cite{Ellwanger:2014hia}. For these reasons the NMSSM has become more
and more appealing in the recent years.

The field content of the NMSSM differs from the MSSM by an additional
gauge singlet superfield $\hat S$ which contains a Majorana fermion
(the singlino), a CP-even, and a CP-odd scalar. The couplings of
the components of $\hat S$ to the MSSM-like Higgs fields $H_u$, $H_d$ 
and sparticles are proportional to a dimensionless coupling $\lambda$,
and the self couplings of the components of $\hat S$ are proportional
to a dimensionless coupling $\kappa$. With these conventions, 
the general, renormalizable and $R$-parity conserving superpotential for the model
reads
\begin{equation}
\label{eq:GNMSSM}
 \mathcal{W}_S = \mathcal{W}_\mathrm{Yukawa}  + \frac{1}{3}\kappa \hat S^3+ 
 (\mu + \lambda \hat S) \hat H_u \hat H_d +  \frac{1}{2} \mu_s \hat
 S^2 + t_s \hat S  \;. 
\end{equation}
Here, $\mathcal{W}_\mathrm{Yukawa}$ contains the Yukawa part which is
the same as  
in the MSSM, and the superfields containing the Higgs doublets are called $\hat H_u$, $\hat H_d$. In the scale invariant version 
of the model, $\mu=\mu_s=t_s=0$ holds.
Assuming the absence of a Landau
singularity for the running coupling $\lambda$ up to the GUT scale
imposes $\sqrt{\kappa^2 + \lambda^2} \lesssim 0.7$
at the weak (SUSY) scale, and $\kappa$
satisfies typically $\kappa \lesssim \lambda$.
The soft SUSY breaking terms in the Higgs and singlet sector are 
\begin{align}
 V_\mathrm{soft} = &  m_s^2 |S|^2 + m_{H_u}^2 |H_u|^2+ m_{H_d}^2 |H_d|^2 + \nonumber \\ 
  & \hspace{2cm} \left(B_\mu H_u H_d +  \frac12 B_s S S + \chi_s S +  A_\lambda \lambda S H_u H_d + \frac{1}{3} A_\kappa \kappa S^3  + h.c.\right) \;,
\label{soft}
\end{align}
where $S$ is the scalar component of $\hat S$, and $B_\mu =B_s = \chi_s=0$
in the scale invariant version.

After electroweak symmetry breaking the components of $S$
mix with the neutral components of $H_u$, $H_d$. If CP is conserved, the CP-even 
sector of the NMSSM contains three states $\PH_i$, $i=1,2,3$, which are ordered in
mass. Typically, two of them have the properties of the MSSM-like states
$\PSh$ (mostly SM-like) and $\PH$, and a third state $\PHS$ is mostly singlet-like.
The CP-odd sector of the NMSSM contains two states $\PA_i$, $i=1,2$,
ordered in mass, one of which has typically the properties of the
MSSM-like state $\PA$, and a second state $\PAS$ is mostly
singlet-like. Past and present searches for Higgs bosons 
at LEP, the Tevatron and the LHC do not exclude masses of $\PHS$
and/or $\PAS$ below 125~GeV. These masses depend
  on unknown parameters, like the NMSSM couplings $\lambda$ and
  $\kappa$, soft SUSY breaking mass terms, the trilinear couplings $A_\lambda$ and
$A_\kappa$ and $\tan\beta$, the ratio of the two vacuum expectation
values (VEVs) of the Higgs doublets, as well as on the VEV of the singlet,
\cite{Ellwanger:2009dp}. A charged Higgs boson
$\PSHpm$ remains present as in the MSSM, but with slightly modified
relations among the masses of $\PSHpm$, $\PH$ and $\PA$.

The singlino Majorana fermion mixes after electroweak symmetry breaking
with the four neutralinos (bino, neutral wino and two Higgsinos) of the
MSSM. Like in the scalar sector, the mixing angle is proportional to the
coupling $\lambda$. The mostly singlet-like neutralino can well be the
lightest supersymmetric particle (LSP) and a good candidate for dark
matter with a relic density consistent with present results from Planck
and WMAP, with a mass from below a GeV to hundreds of GeV depending on
the unknown parameters.

The NMSSM with an enlarged Higgs sector and an additional
singlino-like neutralino leads to a plethora 
of interesting signatures. On the one hand these can be more
challenging for the discovery of physics beyond the SM (BSM). This is
because the components of $\hat S$ have no couplings to
the SM so that all states in the Higgs and neutralino sectors
which mix with components of $\hat S$ have reduced production cross
sections compared to the MSSM. On the other hand there may be exotic signatures
like Higgs-to-Higgs cascade decays e.g.~with striking BSM signatures
of multi-fermion and/or photon final states
\cite{Dreiner:2012ec,King:2014xwa}. It is 
the target of this document to point out, which signatures allow to test
which regions of the parameter space of the NMSSM in future runs of
the LHC, which ones are clear signs of BSM physics and which may serve
to distinguish between different new physics models. In fact, 
in the decoupling limit of the NMSSM $\lambda \to 0$ (with a fixed ratio
$\kappa/\lambda \lesssim 1$), the phenomenology of the NMSSM turns into
the phenomenology of the MSSM up to a possible additional singlino-like
LSP. The solution of the $\mu$-problem can be maintained in the decoupling
limit, but not the naturalness of the mass of the SM-like Higgs boson.

In order to test the NMSSM, in a first step 
its parameters have to be translated into physical quantities: the masses, 
couplings, production cross sections, and decay 
branching fractions of all states.
Including the presently known radiative corrections, this task can be
undertaken only with the help of numerical codes. In Section~\ref{sec:tools-NMSSM} we
present the currently publicly available codes (tools) which allow
to compute the physical quantities in terms of the parameters in the
Lagrangian for different versions of the NMSSM: the
\ZT-invariant NMSSM, the general NMSSM, the NMSSM without
or with explicit CP violation, the NMSSM with general soft SUSY 
breaking terms at the SUSY scale, or high-scale scenarios as minimal supergravity (mSUGRA)
or gauge mediated SUSY breaking (GMSB). 
Some
of these tools allow, in addition, to verify present constraints on
the Higgs, heavy flavour and/or dark matter sectors.

With these tools at hand NMSSM-specific signatures can be searched for
that are in accordance with present data and can be tested during future
runs of the LHC. Ideally, in the 
case of a discovery, these should also allow to distinguish the NMSSM
from the MSSM (outside the decoupling limit). In Section~\ref{NMSSM-BP} we present
such signatures and a
list of benchmark points in the parameter space of the NMSSM.
Some of these points lead to NMSSM mass spectra, that could be discovered via 
different processes. Practically, all of these processes involve an
NMSSM specific scalar or pseudoscalar. Production cross sections and
branching fractions are provided, while more details can be found in the indicated
original publications. In some cases these references
include proposals for cuts
and estimates of SM backgrounds. Of course, the masses, production cross
sections and branching fractions of the involved particles can vary
 sometimes over wide ranges, but the benchmark points represent useful
targets for desirable sensitivities to BSM signatures.


\section{Tools for the NMSSM}
\label{sec:tools-NMSSM}

\subsection{Calculation of the spectrum and of the branching fractions}
\label{sec:SG}
A precise determination of the full SUSY and Higgs spectrum in the NMSSM is a difficult task.
In particular, for the Higgs state which is associated with the SM-like
 boson with a mass of about 125~GeV, it is 
very well known that radiative corrections are crucial. Today, full
one-loop corrections  
of the Higgs boson masses including the contributions from all states and
the momentum dependence are known in the $\overline{\mathrm{DR}}$
\cite{Degrassi:2009yq,Staub:2010ty} and on-shell (OS) scheme
\cite{Ender:2011qh,Graf:2012hh,Drechsel:2016jdg}. At the two-loop level the dominant
corrections involving the strong coupling constant are known: For the
real case \cite{Degrassi:2009yq} results for the corrections
${\cal O}(\alphas(\alphab+\alphat))$ exist , while for complex parameters
the corrections ${\cal O}(\alphas \alphat)$ have been calculated 
 \cite{Muhlleitner:2014vsa}. 
Other two-loop corrections involving only superpotential couplings such 
as Yukawa and singlet interactions have been derived in \Bref{Goodsell:2014pla}.
 
There are several numerical codes available which calculate the mass spectrum
for different versions of the NMSSM making use of these results. 
A detailed comparison of the calculations of the 
  NMSSM Higgs boson masses in  the $\overline{\mathrm{DR}}$ scheme is given
in \Bref{Staub:2015aea}. We summarize in the following  
the main features of the codes in the order in which the tools became available\footnote{A public version of {\tt FeynHiggs} for the NMSSM, based on \Bref{Drechsel:2016jdg}, is in preparation.}.

\subsubsection{\NMSSMTools} 
\NMSSMTools is a collection of the codes 
\NMHDECAY~\cite{Ellwanger:2004xm,Ellwanger:2005dv},
\NMSPEC~\cite{Ellwanger:2006rn} and \NMGMSB and can be downloaded from 
\url{http://www.th.u-psud.fr/NMHDECAY/nmssmtools.html}. 
\NMSSMTools allows to study the NMSSM with or without $\mathbb{Z}_3$ symmetry,
without or with Grand Unification of its gauge couplings and soft SUSY
breaking terms. So far \NMSSMTools is restricted to the real NMSSM, 
but a version to support also CP violation at one-loop level 
 is under construction~\cite{Domingo:2015qaa}. 

\NMSSMTools allows to
 calculate the Higgs boson masses for a parameter point with three different
 options. The most precise calculation makes use of the NMSSM corrections
  of \Bref{Degrassi:2009yq}. This provides a full one-loop 
  calculation of the Higgs boson masses including all
  contributions and the momentum dependence. At two-loop the 
  corrections ${\cal O}(\alphas (\alphab + \alphat))$ 
and the corrections known from the MSSM at 
${\cal O}((\alphat+\alphab)^2 + \alphatau (\alphatau + \alphab))$ 
\cite{Brignole:2001jy,Degrassi:2001yf,Brignole:2002bz,Dedes:2002dy,Dedes:2003km}
are included.

Other calculations performed by \NMSSMTools are the SUSY spectrum at the 
one-loop level, Higgs and sparticle decay branching fractions (\cite{Das:2011dg}, 
based on \HDECAY \cite{Djouadi:1997yw} and \SDECAY \cite{Muhlleitner:2003vg}, 
respectively). $B$-physics 
observables and the muon anomalous magnetic moment $(g-2)_{\PGm}$ are computed following
\Brefs{Domingo:2007dx,Domingo:2008bb}. An NMSSM-version of MicrOmegas \cite{Belanger:2005kh} is
included which allows to determine the dark matter relic density, direct
and indirect detection rates. All these can be compared to present constraints.
Bounds on the Higgs boson couplings from
LEP~\cite{Schael:2006cr} and on the signal rates of the SM-like Higgs boson from the LHC Run~1 (from \Bref{Bernon:2014vta}) are implemented.

\subsubsection{\SPheno and \SARAH} 
A \SPheno \cite{Porod:2003um,Porod:2011nf} version for the NMSSM can
be generated by the \Mathematica\ code \SARAH 
\cite{Staub:2008uz,Staub:2009bi,Staub:2010jh,Staub:2012pb,Staub:2013tta}. Both
tools are available at {\tt hepforge}: 
\url{http://spheno.hepforge.org/} and at \url{http://sarah.hepforge.org/}. 
By default, \SPheno has included a GUT scenario based on minimal supergravity,
 but other SUSY breaking
mechanism can be implemented via the \SARAH interface. 

\SPheno calculates the full one-loop corrections to all Higgs and SUSY
masses, including
the entire momentum dependence \cite{Staub:2010ty}. At two-loop all
corrections in the gaugeless limit with vanishing external momenta are
calculated for the real and complex NMSSM
\cite{Goodsell:2014bna,Goodsell:2015ira}, including the
NMSSM-specific corrections of ${\cal O}(\alpha_\lambda
(\alpha_\lambda+\alpha_\kappa+\alphat))$ even in the case of CP
violation \cite{Goodsell:2014pla,Goodsell:2016udb}

\SPheno also computes the most important quark-flavour violating observables 
($B$ and $K$ decays,
$\Delta M_{B_{d,s}}$/$\Delta M_K$, $\epsilon_K$) at one-loop using the {\tt FlavourKit}
functionality \cite{Porod:2014xia}, and calculations for $(g-2)_l$, $\delta\rho$ as well as
electromagnetic dipole moments are included. Moreover, all two- and
three-body decays of SUSY particles, and two-body decays of the Higgs scalars are calculated. The sparticle decays are purely tree-level,
while for the Higgs bosons the next-to-leading order (NLO) QCD
corrections to decays in two 
quarks, photons and gluons are included. Also decays in virtual vector
bosons are taken into account. \SPheno  writes all necessary
input files to test points with {\tt HiggsBounds}
\cite{Bechtle:2008jh,Bechtle:2011sb,Bechtle:2013wla} and {\tt
  HiggsSignals} \cite{Bechtle:2013xfa}.

\subsubsection{\NMSSMCalc}
\label{sec:nmssmcalc}
The Fortran code \NMSSMCalc  \cite{Baglio:2013iia} allows the
computation of the Higgs boson masses and branching fractions both in the
CP-conserving and CP-violating NMSSM. It can be downloaded from 
\url{http://www.itp.kit.edu/~maggie/NMSSMCALC/} 
and comes together with an NMSSM extension of \HDECAY
\cite{Djouadi:1997yw,Djouadi:2006bz,Butterworth:2010ym} for the Higgs boson decays.

\NMSSMCalc makes use of mixed $\overline{\mathrm{DR}}$--OS
renormalization conditions for the computation of the Higgs boson masses. The Higgs boson mass calculation at one-loop 
level is performed including the full momentum dependence and all
possible contributions \cite{Ender:2011qh,Graf:2012hh}. At the
two-loop level the ${\cal O}(\alphas \alphat)$ corrections are included
\cite{Muhlleitner:2014vsa}. They include the ${\cal O}(\alphas \alphat)$
part relating the vacuum expectation value to physical observables,
which is missing so far in the other spectrum generators. 

Decays are calculated in the CP-conserving and
CP-violating NMSSM including the dominant higher-order
QCD, SUSY-QCD and SUSY-electroweak (EW) corrections:
The neutral Higgs boson decays into quarks
include the fully massive NLO corrections near threshold \cite{Braaten:1980yq,Sakai:1980fa,Inami:1980qp,Drees:1989du,Drees:1990dq}
and massless ${\cal O}(\alphas^4)$ corrections far above threshold
\cite{Gorishnii:1990zu,Gorishnii:1991zr,Kataev:1993be,Gorishnii:1983cu,Surguladze:1994gc,Larin:1995sq,Chetyrkin:1995pd,Chetyrkin:1996sr,Baikov:2005rw}. Large
logarithms are resummed by taking into account the running of the
quark masses and the strong coupling constant. The charged Higgs boson
decay into a heavy quark pair includes the QCD corrections given in
\cite{Mendez:1990jr,Li:1990ag,Djouadi:1994gf} and taken over to the
NMSSM case. By adapting the results from the MSSM, the decays of the
neutral Higgs bosons into a bottom pair 
include SUSY-QCD \cite{Hempfling:1993kv,Hall:1993gn,Carena:1994bv,Pierce:1996zz,Carena:1998gk,Carena:1999py,Carena:2002bb,Guasch:2003cv}
and the approximate SUSY-EW corrections
\cite{Noth:2008tw,Noth:2010jy,Mihaila:2010mp}. 
For the decays into gluons the QCD corrections to quark loops have
been included up to N$^3$LO in the limit of heavy quark masses \cite{Inami:1982xt,Djouadi:1991tka,Spira:1993bb,Spira:1995rr,Kramer:1996iq,Chetyrkin:1997iv,Chetyrkin:1997un,Schroder:2005hy,Chetyrkin:2005ia,Baikov:2006ch,Baikov:2006ch} and 
NLO QCD corrections to the squark loops in the heavy squark mass
limit \cite{Dawson:1996xz,Djouadi:1996pb}. 
In the photon decays the QCD corrections to quark
\cite{Spira:1995rr,Zheng:1990qa,Djouadi:1990aj,Dawson:1992cy,Djouadi:1993ji,Melnikov:1993tj,Inoue:1994jq,Muhlleitner:2006wx}
and squark loops have been taken into account including the full mass
dependence both for the quarks and the squarks. All two-body decays into
SUSY particles have been implemented
\cite{Djouadi:1996pj,Djouadi:1996mj}. For the CP-conserving case the 
decays into stop and sbottom pairs come with the SUSY-QCD corrections \cite{Bartl:1997yd,Arhrib:1997nf,Eberl:1999he,Accomando:2011jy,Baglio:2015noa}. Finally,
all relevant off-shell decays into two massive gauge boson final
states \cite{Cahn:1988ru}, into gauge and Higgs boson final states \cite{Djouadi:1995gv,Moretti:1994ds,Spira:1997dg,Djouadi:2005gj}, into Higgs boson pairs as
well as into heavy quark pairs are included \cite{Djouadi:1995gv,Moretti:1994ds}. 

Recently, the electric dipole moments (EDMs) with NMSSM contributions \cite{Manohar:1983md,Arnowitt:1990eh,Ellis:1996dg,Weinberg:1989dx,Dicus:1989va,Dai:1990xh,Khriplovich:1997ga,Pospelov:1999ha,Pospelov:1999mv,Pospelov:2000bw,Demir:2002gg,Demir:2003js,Pospelov:2005pr,Olive:2005ru,Giudice:2005rz,Ellis:2008zy,Li:2008kz,Ellis:2010xm,Cheung:2011wn}
have been implemented in {\tt NMSSMCALC} \cite{King:2015oxa} allowing
to check for the compatibility of CP-violating phases with the
experimental constraints on the EDMs
\cite{Regan:2002ta,Baker:2006ts,Griffith:2009zz,Baron:2013eja}. The
output file contains furthermore the necessary couplings to test
points with {\tt HiggsBounds}. 

\subsubsection{\SoftSUSY}
 The NMSSM version of \SoftSUSY includes different high-scale scenarios like 
mSUGRA-inspired semi-constrained NMSSM and a general high-scale
boundary condition which allows one to set all soft parameters independently.  
The users may also easily create their own boundary conditions.
For the determination of the parameters at the SUSY/weak scale,
\SoftSUSY uses the full three-family one- and two-loop RGEs for the NMSSM.

The homepage of \SoftSUSY is \url{http://softsusy.hepforge.org/}.

The one-loop self-energies and tadpole corrections for the Higgs boson masses were extended to the NMSSM using the expressions in
\Brefs{Degrassi:2009yq,Staub:2010ty}. The NMSSM extension also
includes the two-loop corrections at order ${\cal O}(\alphas(\alphat
+ \alphab))$ from \Bref{Degrassi:2009yq} and uses
the MSSM results from \Brefs{Brignole:2001jy,Degrassi:2001yf,%
Brignole:2002bz,Dedes:2002dy,Dedes:2003km}.

Higgs and sparticle decays may be obtained by interfacing SoftSUSY
with \NMHDECAY \cite{Ellwanger:2004xm,Ellwanger:2005dv} and \NMSDECAY
\cite{Das:2011dg}, which is based upon \Bref{Muhlleitner:2003vg}.
These are both distributed as part of the \NMSSMTools package, and
\SoftSUSY provides a script to do the interface with this package
automatically.

\subsubsection{\FlexibleSUSY}
\FlexibleSUSY\cite{Athron:2014yba} is a Mathematica and C++ package, which also makes use
of \SARAH to obtain model-dependent details. It can be downloaded from \url{http://flexiblesusy.hepforge.org/}.
The boundary conditions on the model parameters at
different scales, as well as spectrum-generator specific configuration
details can be specified in a \FlexibleSUSY model file.

\FlexibleSUSY computes the Higgs spectrum employing the full two-loop RGEs
and one-loop corrections.  At the two-loop
level it makes use of Higgs boson mass corrections available in the
literature. In the case of the NMSSM \FlexibleSUSY uses the
$\alphas(\alphat + \alphab)$ corrections given in
\Bref{Degrassi:2009yq}. In addition, the MSSM two-loop
corrections can be used as \SoftSUSY does.

\subsection{Check for the vacuum stability}
A valid spectrum obtained with one of the mentioned spectrum generators
necessarily has a local minimum of
the scalar potential where the electroweak 
symmetry is broken such that the electroweak data like the $\PZ$ boson mass
are fulfilled. However, none of the above
  described codes performs an exhaustive check if this is
also the global minimum. Possible scenarios where deeper minima arise
are given in the following situations:
(i) In the case of light stops and a large mixing to accommodate the Higgs boson mass as in the MSSM, it can happen that at the global minimum charge and
colour are broken by VEVs for the stops
\cite{Camargo-Molina:2013sta,Blinov:2013fta, 
Chowdhury:2013dka,Camargo-Molina:2014pwa,Chattopadhyay:2014gfa}. (ii) In the
case of large $A_\lambda$ and $A_\kappa$ deeper minima with different
numerical values for the singlet and the two doublet VEVs can exist
\cite{Ellwanger:1999bv,Kanehata:2011ei,Kobayashi:2012xv}. Both
possibilities are disastrous 
and the corresponding parameter points are ruled out if the local minimum
with correct EWSB is short-lived on cosmological times scales. \\
The software package \Vevacious \cite{Camargo-Molina:2013qva} addresses
these issues and performs a numerical check of the vacuum stability.
Given a model file and a parameter point in the SLHA format, \Vevacious
checks for the global minimum of the one-loop effective potential
including also finite temperature effects. If necessary, it calculates
the life-time for the input minimum using {\tt Cosmotransitions}
\cite{Wainwright:2011kj}. \Vevacious includes model files not only to
check the possibility of stop, but also of stau VEVs in addition to
the Higgs and singlet VEVs. Model files for other scenarios can be
generated with \SARAH.

\subsection{Calculation of the neutral Higgs boson production cross
  sections} 
The essential features of Higgs boson production in the NMSSM can
directly be adapted from the ones of the MSSM which were broadly
analysed in the previous
reports~\cite{Dittmaier:2011ti,Dittmaier:2012vm,Heinemeyer:2013tqa}. 
Thus, when we subsequently discuss the production of the five neutral Higgs bosons we mainly elaborate on the most prominent
differences with respect to the MSSM. 

The dominant production mechanism for neutral Higgs bosons is
gluon fusion
(ggF)~\cite{Georgi:1977gs,Djouadi:1991tka,Dawson:1990zj,Spira:1995rr} 
mediated through loops involving the third generation quarks,
top and bottom, as in the SM. In SUSY also the
squarks of the third generation, stops and sbottoms, contribute, but are
typically suppressed by the ratio $m_{\PZ}^2/m_{\tilde{q}}^2$ at leading
order (LO) QCD. The NLO QCD contributions are important and have been
provided with full quark mass
dependence~\cite{Spira:1995rr,Harlander:2005rq}. 
The next-to-next-to leading order (NNLO) QCD corrections to the top
quark loops have been calculated 
in the heavy top quark mass limit~\cite{Harlander:2002wh,Anastasiou:2002yz,Ravindran:2003um,
Harlander:2002vv,Anastasiou:2002wq,Pak:2011hs} and with finite top-quark mass
effects~\cite{Marzani:2008az,Harlander:2009bw,Harlander:2009mq,
Harlander:2009my,Pak:2009bx,Pak:2009dg,Pak:2011hs,Caola:2011wq}.
Partial NNNLO QCD corrections to the top-quark contribution
were provided in~\Brefs{Moch:2005ky,Ravindran:2006cg,Ball:2013bra,
Buehler:2013fha,Anastasiou:2014vaa,Anastasiou:2014lda}.
The QCD corrections to top- and bottom-quark loops
can be taken over to the NMSSM case. Moreover, even though $\tan\beta$ is mostly
chosen small in the NMSSM, non-decoupling $\Delta_b$
terms~\cite{Carena:1999py,Guasch:2003cv,Noth:2008tw,Noth:2010jy,Mihaila:2010mp},
i.e.~the $\tan\beta$-enhanced resummed SUSY corrections to the
Higgs-bottom coupling,  are known
in the NMSSM~\cite{Baglio:2013iia} and can be included
in the NMSSM Higgs Yukawa coupling to the bottom quarks. The code {\tt
  SusHi}~\cite{Harlander:2012pb}, starting with version 1.5.0, allows
for the inclusion of the above mentioned higher order quark contributions 
for the NMSSM Higgs bosons as well~\cite{Liebler:2015bka},
but additionally takes into account genuine SUSY contributions,
namely stop and sbottom contributions up to NLO QCD
in the expansion of heavy SUSY
masses~\cite{Harlander:2005if,Degrassi:2011vq,Degrassi:2012vt}. 
Furthermore, for the Higgs bosons of the NMSSM there exists a private
version of the code {\tt
  HIGLU}~\cite{Spira:1995mt,King:2012tr,King:2014xwa}.
It takes into account the NLO QCD corrections to the quark and squark
loops including the full mass dependence, see \Brefs{Spira:1995rr,Harlander:2005rq}
and \Brefs{Aglietti:2006tp,Anastasiou:2006hc,Muhlleitner:2006wx}, respectively.

Electroweak contributions mediated by light
quarks~\cite{Aglietti:2004nj,Bonciani:2010ms} can be added in {\tt SusHi}
for the CP-even Higgs bosons of the NMSSM, while they are absent
for the CP-odd Higgs bosons.
They are known to capture the dominant fraction of the
full SM electroweak correction factor~\cite{Djouadi:1994ge,Aglietti:2004nj,
Degrassi:2004mx,Actis:2008ug,Bonciani:2010ms}
for the SM-like Higgs boson with a mass of $\sim 125$\,\UGeV.

The discussion of theory uncertainties for the gluon fusion process
applies to large extent to the case of the MSSM.
We therefore refer to the detailed discussion of theory uncertainties
in \Brefs{Heinemeyer:2013tqa,Bagnaschi:2014zla}.
In order to mimic next-to-next-to-leading-log (NNLL)
resummation~\cite{Catani:2003zt,Ravindran:2005vv,Moch:2005ky} 
of large logarithms for the top-quark induced contributions,
central scales of $m_\phi/2$ for the renormalization and
factorization scale are advisable, where
$m_\phi$ generically denotes the mass of the Higgs boson under consideration.
The scale uncertainties can then be obtained through a combination
of five different scale choices, as presented in
\Brefs{Bagnaschi:2014zla,Liebler:2015bka}. PDF$+\alphas$ uncertainties
were found to be mostly a function of the mass $m_\phi$ of the Higgs boson
involved~\cite{Bagnaschi:2014zla,Liebler:2015bka}.
Thus, the relative PDF$+\alphas$ uncertainties
can be taken from the ggF cross section
of a SM Higgs boson with mass $m_\phi$. Since $\tan\beta$ is usually chosen small
in the NMSSM, the uncertainties which stem from missing contributions
to $\Delta_b$ and from the renormalization prescription taken for the
bottom-quark Yukawa coupling are not as relevant as in the MSSM.
In contrast, the negligence of NNLO stop contributions to Higgs boson production
can be important for pseudo-scalars, where the
couplings to quarks can vanish for a CP-odd singlet-like
state~\cite{Liebler:2015bka}. 

The other production mechanisms can be treated in a
similar fashion as in the MSSM. Thus, vector-boson fusion
(VBF)~\cite{Cahn:1983ip,Hikasa:1985ee,Altarelli:1987ue} 
and top-quark associated production ($\PQt \PAQt \PH$)~\cite{Raitio:1978pt,Ng:1983jm,Kunszt:1984ri,Gunion:1991kg,
Marciano:1991qq}
can be reweighted with the effective couplings of the Higgs boson under consideration
to the heavy gauge bosons $\PZ,\PW$ and the top quark, respectively.
In the MSSM SUSY-QCD corrections are known to be small for
vector-boson fusion~\cite{Djouadi:1999ht,Hollik:2008xn}
and moderate for top-quark associated
production~\cite{Peng:2005ti,Hollik:2006vn,Hafliger:2006zz,Walser:2008zz},
and thus - as a first approximation - are neglected in the NMSSM.
Similarly, bottom-quark associated production ($\PQb \PAQb \PH$), which is relevant
for large values of $\tan\beta$ and can either be described
in the four-flavour scheme~\cite{Dittmaier:2003ej,Dawson:2003kb} at
NLO QCD accuracy or the five-flavour
scheme~\cite{Dicus:1998hs,Balazs:1998sb,Harlander:2003ai} 
at NNLO QCD accuracy, can be adjusted to
the NMSSM by reweighting the SM predictions with the effective coupling
of the Higgs boson under consideration to the bottom quark.
Again non-decoupling $\Delta_b$ terms should be added to the
effective Higgs-bottom coupling. {\tt SusHi} provides
the effective coupling to bottom quarks and also the 
bottom-quark annihilation cross sections (5FS) directly.
As a first approximation Higgs-strahlung
(${\rm V}\PH$)~\cite{Glashow:1978ab,Kunszt:1991xk} can be
adjusted from the SM to the NMSSM through 
a proper reweighting with the Higgs to gauge boson couplings.
However gluon induced contributions to Higgs-strahlung, which
contribute $\mathcal{O}(10\%)$~\cite{Altenkamp:2012sx} to the
inclusive cross section 
and are mediated through top-quark or bottom-quark loops, need
to be treated differently.
Moreover $s$-channel Higgs induced contributions
can arise in Higgs sectors with more than one Higgs doublet.
SUSY-QCD corrections to Higgs-strahlung in the MSSM are known to be
small~\cite{Djouadi:1999ht}. A code, which reweights
the individual contributions to ${\rm V}\PH$ according
to the quark and gauge boson couplings in the NMSSM and which adds
$s$-channel Higgs induced contributions, is desirable.

The relative couplings of the Higgs boson under consideration
to quarks, the heavy gauge bosons, gluons and photons with respect
to an SM Higgs boson of identical mass can also be obtained with
the different spectrum generators presented in \refS{sec:SG}:
In \NMSSMTools the relative coupling to a pair of bottom quarks includes 
$\Delta_b$ corrections and the relative coupling to gluons and
photons take into account the LO one-loop induced contributions
with the full NMSSM particle spectrum. \SPheno includes in addition the NLO QCD 
corrections in the coupling of the scalars to a pair of gluons.

Of large relevance for the phenomenology of Higgs bosons are their
transverse momentum ($p_T$) distributions in gluon fusion. We refer to
the MSSM section, see \refS{sec:mssm-pt}, for a detailed discussion
of their knowledge in the SM and beyond. Neglecting squark contributions,
the top-quark, bottom-quark and the top-bottom-interference contributions
can be reweighted from the SM to the NMSSM as depicted in the MSSM section
for the example of a 2HDM. For this purpose
the 2HDM/MSSM implementations~\cite{Bagnaschi:2011tu} of the gluon fusion
process in the {\tt POWHEG-BOX}~\cite{Alioli:2010xd}
or the two codes {\tt MoRe-SusHi}~\cite{Mantler:2012bj,Harlander:2014uea}
and {\tt aMCSusHi}~\cite{Mantler:2015vba} can be used,
which all allow to extract the three mentioned quark contributions individually
and reweight them with the corresponding Yukawa couplings
of the NMSSM Higgs boson under consideration.
An extension of the codes {\tt MoRe-SusHi} and {\tt aMCSusHi}
from the MSSM to the NMSSM to directly obtain $p_T$~distributions 
for the five neutral Higgs bosons in gluon fusion is planned.

\section{NMSSM benchmark points}
\label{NMSSM-BP}

\subsection{NMSSM specific processes}

The benchmark points presented here cover various possible NMSSM
specific processes to be searched for during the ongoing and future
runs of the LHC. The focus is on possible discovery scenarios, as adequate at the
present and foreseen LHC energies and luminosities. Additionally, 
the benchmark points can also be exploited to test specific
features of the NMSSM Higgs bosons and be used to
distinguish the NMSSM from other SUSY extensions like e.g.~the MSSM.

The production mechanisms for NMSSM Higgs bosons are given by the same
mechanisms as for an SM-like Higgs boson, i.e.~ggF, VBF,
${\rm V}\PH$ and
associated production $\PQb \PAQb \PH$ or $\PQt\PAQt \PH$ with, however,
reduced cross sections\footnote{With the exception of
    production processes where $\tan\beta$ enhanced couplings to
    $\PQb$-quarks are involved.}. Alternatively, NMSSM specific Higgs states can 
appear in decay cascades of heavier (MSSM-like or NMSSM-like) Higgs states, or in
sparticle decay cascades.
Likewise, NMSSM specific Higgs states can decay into the same final states
as an SM-like Higgs boson and additionally, if
  kinematically allowed, into pairs of other Higgs states, other Higgs states and a $\PZ$ or $\PW$ boson, or pairs of SUSY particles.

With the here presented benchmark points designed to be potential 
search and discovery modes for the NMSSM Higgs states, it
is useful to start with their classification according to production mechanisms
and decay chains. This is summarized in \refTs{tab:1}--\ref{tab:5}
below, together with the corresponding combinations that appear for
the various benchmark points described in the next subsection.
We choose the following notation: the mostly
SM-like Higgs boson is denoted by $\PHSM$, the dominantly MSSM-like
CP-even and CP-odd scalars are given by $\PH$ and $\PA$, respectively,
and the mostly NMSSM-like CP-even and CP-odd scalars by $\PHS$ and
$\PAS$ each. Note that the latter two can be lighter or heavier than all other
scalars. Neutralinos are indicated by $\PSGcz_i$ ($i=1,...,5$) where,
in most considered cases, $\PSGczDo$ is mostly singlino-like and can
have a very small mass. The c.m.~energy $\sqrt{s}$
  used by the authors of the various benchmark points has been set to either 13 or
  14 TeV, namely we have $\sqrt{s}=13$~TeV in BP1, BP2 ,BP3, BP4, BP5,
  BP7 and $\sqrt{s}=14$~TeV in BP6, BP8, BP9. Note that the gluon fusion
  cross section increases, depending on the Higgs boson mass value, by
  $\sim$10-20\% when increasing the c.m.~energy from 13 to 14 TeV.

\subsubsection*{Direct $\PHSM$ Production and Decays}
The set I of signatures given in \refT{tab:1} summarizes
scenarios that feature a directly produced SM-like Higgs boson
$\PHSM$ that decays into lighter singlet-like Higgs boson pair or
neutralino final states. The branching fractions corresponding to the
NMSSM-specific $\PHSM$
decays listed in \refT{tab:1} are (or will be) limited by the
presently (or prospectively) measured signal rates of $\PHSM$ decays
into SM channels. These indirect 
constraints on NMSSM-specific $\PHSM$ decays can be stronger than
the limits from direct searches of the corresponding final states.
Some benchmark points lead also to MSSM-like decays 
$\PHSM\to \text{invisible}$. Here and in the following
MSSM-like decays have been omitted for simplicity.

{\renewcommand{\arraystretch}{1.2}
\begin{table}
\caption{Summary of NMSSM-specific $\PHSM$ decays and their
  signatures. The last column indicates the benchmark points in which
  the scenarios are realized.} 
\label{tab:1}
\begin{center}
\begin{tabular}{c|l|l|l|l}
\toprule
I &
\multicolumn{4}{c}  {Direct $\PHSM$ production and decays} \cr
\midrule
 &Process & Signatures  & Comments & BM points  \cr
\midrule
a &$\PHSM\to \PHS+\PHS$ or &  Combinations of decays &Notably $\PAS$ can
 & BP2\_1, BP4\_1,2, \cr 
 &$\PHSM\to \PAS+\PAS$ \phantom{or$\,$} & into $\PQb \PAQb$, $\PGtp \PGtm$,
   $\PGmp \PGmm$, \PGg \PGg&
be very light &  BP9\_1 \cr
\midrule
b &$\PHSM\to \PHS+\PHS$ & Combinations of decays &$\PAS$ can be very & BP4\_2  \cr 
 &  $\to 4 \PAS$ & into $\PQb \PAQb$, $\PGtp \PGtm$,
   $\PGmp \PGmm$, \PGg \PGg&
 light & \cr
\midrule
c &$\PHSM\to \PSGczDo+\PSGczDt$, & $\PHS$ decay
products $+ \MET$ &   & Not necessary \cr 
 & $\PSGczDt \to \PSGczDo+\PHS$ & &  &  \cr
\bottomrule
\end{tabular}
\end{center}
\end{table}
}

\subsubsection*{Direct Light $\PHS/\PAS$ Production and Decays}
In set II, {\it cf.}~\refT{tab:2}, we collect
  signatures stemming from direct production of 
singlet-like CP-even Higgs boson $\PHS$ and $\PAS$. These may decay
into SM particle final states, into lighter NMSSM Higgs boson pairs, among
which also the $\PHSM$ is possible, into a gauge boson and an NMSSM
Higgs boson or into a lightest neutralino pair.
The production cross sections for $\PHS/\PAS$ in \refT{tab:2} are proportional
to the singlet-doublet mixing angles (squared) and can be very small.

{\renewcommand{\arraystretch}{1.2}
\begin{table}
\caption{NMSSM-specific $\PHS/\PAS$ production and decays as well as the corresponding collider signatures. The last column 
indicates the benchmark points in which the scenarios are realized.}
\label{tab:2}
\begin{center}
\begin{tabular}{c|l|l|l|l}
\toprule
II &
\multicolumn{4}{c}  {Direct $\PHS/\PAS$ production and decays} \cr
\midrule
 &Process & Signatures  & Comments & BM points  \cr
\midrule
a &ggF$(\PHS/\PAS)$ &  $\PQb \PAQb$, $\PGtp \PGtm$, $\PGmp \PGmm$, \PGg \PGg& &
BP1\_1,2, BP4\_1,2,\cr 
 & & & &  BP7\_1,2, BP8\_1,2,  \cr
 & & & &  BP9\_1,2   \cr
\midrule
b &ggF$(\PHS)\to \PAS \PAS$ & Combinations of decays &$\PAS$ can be  &
BP2\_1,2, BP3,  BP4\_2, \cr 
 & & into $\PQb \PAQb$, $\PGtp \PGtm$, $\PGmp \PGmm$, \PGg \PGg&
 very light & BP7\_2, BP9\_2 \cr
\midrule
c &ggF$(\PHS)\to \PHSM \PHSM$ & &  & BP9\_2 \cr 
\midrule
d &ggF$(\PAS)\to \PZ \PHS$ & $\PZ+\PQb \PAQb$ &  & BP8\_1,2 \cr 
\midrule
e &ggF$(\PHS/\PAS)\to \PSGczDo \PSGczDo$ &  &  & BP4\_1,2, BP8\_2, BP9\_2 \cr 
\bottomrule
\end{tabular}
\end{center}
\end{table}
}

\subsubsection*{Direct $\PH/A$ Production and Decays}
The production cross sections of set III for direct MSSM-like Higgs boson, $\PH$, $\PA$, production in \refT{tab:3} are dominated
by ggF for low values of $\tan\beta$, which are typical for the NMSSM. 
Some benchmark points lead also to MSSM-like decays $\PH/A\to \PHSM+
\PHSM$, $\PH/A\to \text{invisible}$, $\PH/A\to \PQt\PAQt$, $\PH/A\to
\PSGcp\PSGcm$, $\PH/A\to \PSGcz_i\PSGcz_j$, which have been omitted.
The decays IIIe), IIIf) including $\PZ$ bosons occur only if
$\PHS/\PAS$ have doublet components through mixing, and if other
channels are suppressed. 

{\renewcommand{\arraystretch}{1.2}
\begin{table}
\caption{NMSSM-specific $\PH/A$ production and decays as well as the
  corresponding collider signatures. The last column indicates the
  benchmark points in which the scenarios are realized.} 
\label{tab:3}
\begin{center}
\begin{tabular}{c|l|l|l|l}
\toprule
III &
\multicolumn{4}{c}  {Direct $\PH/A$ production and decays} \cr
\midrule
 &Process & Signatures  & Comments & BM points  \cr
\midrule
a &ggF$(H) \to \PHS+\PHS$ &  $\PQb \PAQb+\PQb \PAQb$, & $\PAS$ can be & BP7\_1,2,  \cr
 &ggF$(H) \to \PAS+\PAS$ &  $\PQb \PAQb+\PGtp \PGtm$, & very light & BP8\_2  \cr
 &ggF$(A) \to \PHS+\PAS$ & $\PQb \PAQb+\PGg\PGg$, $4\PGg$ &  &  \cr
\midrule 
b &ggF$(H)\to \PHSM+\PHS$ & $\PQb \PAQb+\PQb \PAQb$, $\PQb \PAQb+\PGtp \PGtm$,
  &$\PAS$ can be  & BP7\_1,2,\cr 
 &ggF$(A)\to \PHSM+\PAS$ & $\PQb \PAQb+\PGg\PGg$, $4\PGg$& very light & BP8\_1,2  \cr
\midrule
c &ggF$(H)\to \PHSM+\PHSM$ & $\PQb \PAQb+\PQb \PAQb$, $\PQb
\PAQb+\PGg\PGg$, $\PGtp \PGtm+\PGg\PGg$ &  & BP7\_2\cr 
\midrule
d &ggF$(H)\to \PHSM+\PHS$ & $\PQb \PAQb+\PQb \PAQb+\PQb \PAQb$, $\PQb \PAQb+\PQb \PAQb+\PGtp \PGtm$,
&  & BP7\_2 \cr 
&\hfill $\to \PHSM+\PAS+\PAS$  &$\PQb \PAQb+\PQb \PAQb+\PGg\PGg$,
 $\PQb \PAQb+\PGtp \PGtm+\PGg\PGg$,
& & \cr
 &ggF$(A)\to \PAS+\PHS$ & $\PQb \PAQb+4\PGg$, $\PGtp \PGtm+4\PGg$
&  &  \cr
&\hfill $\to \PAS+\PAS+\PAS$  & & & \cr
\midrule
e &ggF$(H)\to \PZ+\PAS$ & $\Plp\Plm+\PQb \PAQb$, $\Plp\Plm+\PGtp \PGtm$,
 & $\Plp\Plm$ from  & BP2\_2, \cr 
 &ggF$(A)\to \PZ+\PHS$ & $\Plp\Plm+\PGg\PGg$ & $\PZ$ decays &  BP7\_1, \cr
 & & & &  BP8\_1,2\cr
\midrule
f &ggF$(A)\to \PZ+\PHS$ & $\Plp\Plm+\PQb \PAQb+\PQb \PAQb$&$\Plp\Plm$ from  &  \cr 
 &\hfill $\to \PZ+\PAS+\PAS$ &  & $\PZ$ decays & BP7\_2 \cr
\bottomrule
\end{tabular}
\end{center}
\end{table}
}

\subsubsection*{Higgs Bosons in Squark/Chargino/Neutralino
  decays, Singlino-Like LSP}
Set IV is dedicated to NMSSM Higgs boson production through
squark/chargino/neutralino decay cascades. They are given in 
\refT{tab:4} and  the NMSSM-specific features consist in the
appearance of the 5th singlino-like lightest SUSY particle (LSP),
which couples  weakly to all other sparticles. Hence sparticle decay
cascades end up  provisionally in the next-to-lightest SUSY particle
(NLSP), e.g. the one that is mostly bino-like, which finally decays 
into the LSP plus $\PHSM$ or $\PHS$, depending on $M_{\PHS}$ and the
available phase space. If the LSP is light and the phase space is
narrow, the LSP carries little energy and the $\MET$ can
become very small.

{\renewcommand{\arraystretch}{1.2}
\begin{table}
\caption{NMSSM-specific squark/chargino/neutralino decay
  cascades as well as the corresponding collider signatures. The last column 
indicates the benchmark points in which the scenarios are realized.}
\label{tab:4}
\begin{center}
\begin{tabular}{c|l|l|l|l}
\toprule
IV &
\multicolumn{4}{c}  {Higgs bosons in squark/chargino/neutralino decays,
singlino-like LSP} \cr
\midrule
 &Process & Signatures  & Comments & BM points  \cr
\midrule
a & $\PSGczDt \to \PSGczDo + \PHSM$ &  Jets + $\PHSM+ \PHSM$ + $\MET$ & &
 BP1\_2  \cr 
 & &$\PHSM \to \PQb \PAQb$, $\PGtp \PGtm$ or \PGg \PGg & &  \cr
\midrule
b & $\PSGczDt \to \PSGczDo + \PHSM$ &  Jets + $\PHSM+ \PHSM$ & $\PSGczDo$ very light, &
 BP5\_1,2  \cr 
 & &$\PHSM \to \PQb \PAQb$, $\PGtp \PGtm$ or \PGg \PGg  & little $\MET$ &  \cr
\midrule 
c &$\PSGczDt \to \PSGczDo + \PHS$ &  Jets + $\PHS+ \PHS$ + $\MET$ & &BP1\_2  \cr 
 & &  $\PHS \to \PQb \PAQb,\ \PGtp \PGtm$ or \PGg \PGg & &  \cr 
\midrule
d & $\PSGczDT \to \PSGczDo + \PHS$, &  Jets +
$\MET$& & BP3 \cr 
 & $\PHS \to \PAS+\PAS$ & + up to $4\PAS$ & &  \cr 
\midrule
e & $\PSGcz_i \to \PSGczDo + \PAS$ &  Trileptons + $\MET$ &
 $\PAS$ very light & BP6 \cr 
\bottomrule
\end{tabular}
\end{center}
\end{table}
}

\subsubsection*{Displaced vertices}

Displaced vertices as given by set V in \refT{tab:5} can occur as in the MSSM in
models with light Gravitinos, like in gravity mediated SUSY breaking
(GMSB) for instance, but also for very weakly coupled singlino-like LSPs.

{\renewcommand{\arraystretch}{1.2}
\begin{table}
\caption{Processes which can cause displaced vertices, their collider signatures
and corresponding benchmark points}
\label{tab:5}
\begin{center}
\begin{tabular}{c|l|l|l}
\toprule
V &
\multicolumn{3}{c}  {Displaced vertices} \cr
\midrule
 &Process & Signatures  & BM points  \cr
\midrule
a & Squark/gluino production &  Jets + displaced vertices &BP1\_1,2  \cr 
b & chargino/slepton production & Leptons  + displaced vertices&BP1\_1,2  \cr
\bottomrule
\end{tabular}
\end{center}
\end{table}
}

\subsection{Benchmark points}

In this section we present (pairs of) benchmark
points which are aimed to cover most of the NMSSM-specific processes
and final states, that have been presented in \refTs{tab:1}--\ref{tab:5}. 
The benchmark points BP1--9\_1,2 have been 
provided by different author groups and are classified accordingly.

For better readability and easy identification of the striking
features and signatures, we give only the main features,
i.e.~the relevant parts of the spectra and the signatures. For more
information we refer the reader to the publications, on which the
points are based and which are given in the following tables, as well
as to the TWiki page: \url{https://twiki.cern.ch/twiki/bin/view/LHCPhysics/LHCHXSWGNMSSM}. 
Here detailed information on the benchmark spectra, the production
cross sections, the branching ratios and the rates to be expected can
be found. Furthermore, the program codes with which the benchmark
points have been generated are specified, so that they can be
reproduced. Since, in contrast to other BSM theories like the MSSM or
the 2HDM, there is no official recommendation for NMSSM tools yet, the
various authors of the benchmark points chose different tools. 
Note, that the numbers presented here and on the TWiki page sometimes
differ from the ones of the originally proposed points in the given
references. This is due to the following reasons: (i) The SM input
parameters have been unified according to the  official proposal of
\Bref{LHCHXSWG-INT-2015-006}. (ii) Some benchmarks use \NMSSMTools to
calculate the Higgs boson mass spectrum. If not done before, the highest
precision available in \NMSSMTools is now used throughout. 
(iii) Parameters were tweaked to resurrect scenarios ruled out by previous LHC runs.

The corresponding publications
contain typically more benchmark points with similar features
and sometimes proposals for search strategies including cuts and estimates
of the background.

All benchmark points have been chosen such that they are not in conflict with
previous searches at the Run~1 of the LHC, and use 
the SM parameters recommended by the LHC-HXSWG, see \Bref{LHCHXSWG-INT-2015-006} and \refC{chapter:input}.


\begin{center}
\small
\begin{longtable}{c|l@{}m{0pt}@{}}
\toprule
\multicolumn{2}{c}  {\bf BP1: GMSB combined with \ZT-invariant NMSSM} & \\
\midrule
\multicolumn{2}{c}  {B. Allanach, M. Badziak, C. Hugonie and R. Ziegler}\\ 
\multicolumn{2}{c}  {from Phys. Rev. D92 (2015) 015006, arXiv:1502.05836 \cite{Allanach:2015cia}} \cr
\midrule
Main Features & GMSB combined with
 \ZT-invariant NMSSM (gravitino $\tilde{G}$ LSP)\cr
 & $\PSGczDo$: singlino-like NLSP\cr
\midrule
\multicolumn{2}{c}  {BP1\_1 \hspace*{\fill}}\cr
\midrule
Spectrum & $M_{\PHSM} \approx 123~\UGeV$, $M_{\PAS} \approx  26~\UGeV$, $M_{\PHS} \approx 93~\UGeV$,
  $M_H \approx  895~\UGeV$, \cr 
 & $M_{A} \approx 895~\UGeV$, $M_{\PSGczDo} \approx 103~\UGeV$,
 $M_{\PSGtmDo} \approx 332~\UGeV$ (NNLSP)\cr
\midrule
\multicolumn{2}{c}  {Production cross sections (at $13~\UTeV$) and branching fractions}\cr
\midrule 
$\PHS$ & via ggF $\approx$ 16 \Upb, $BR(\PHS \to \PQb \PAQb) \approx 83\%,$
$BR(\PHS \to \PGtp \PGtm) \approx 8\%$    \cr 
\midrule
$\PAS$ & via ggF $\approx$ 11 \Ufb, $BR(\PAS \to \PQb \PAQb) \approx 91\%,$
$BR(\PAS \to \PGtp \PGtm) \approx 8\%$    \cr 
\midrule
$\PSGczDo$ & $BR(\PSGczDo \to \tilde{G} + \PAS ) =100\%$,
displaced vertex (mostly inside the detector)\cr
\midrule
$\PSGtmDo$ & $BR(\PSGtmDo \to \PGtm + \PSGczDo)=100\%$ \cr
\midrule
Particular & $\PAS$ appears at the end of every sparticle decay
chain \cr
signatures & $\PHS$ potentially observable via direct production\cr
\midrule
\multicolumn{2}{c}  {BP1\_2 \hspace*{\fill}}\cr
\midrule
Spectrum & $M_{\PHSM} \approx 124~\UGeV$, $M_{\PAS} \approx  32~\UGeV$,
 $M_{\PHS} \approx 94~\UGeV$,  $M_H \approx  1.4~\UTeV$, \cr 
 & $M_{A} \approx 1.4~\UTeV$, $M_{\PSGczDo} \approx 105~\UGeV$,
 $M_{\PSGczDt} \approx 397~\UGeV$ (NNLSP)\cr
\midrule
\multicolumn{2}{c}  {Production cross sections (at $13~\UTeV$) and branching fractions}\cr
\midrule 
$\PHS$ & via ggF $\approx$ 17\Upb, $BR(\PHS \to \PQb \PAQb) \approx 83\%,$
$BR(\PHS \to \PGtp \PGtm) \approx 8\%$    \cr 
\midrule
$\PAS$ & via ggF $\approx$ 0.1 \Ufb, $BR(\PAS \to \PQb \PAQb) \approx 91\%,$
$BR(\PAS \to \PGtp \PGtm) \approx 9\%$    \cr 
\midrule
$\PSGczDo$ & $BR(\PSGczDo \to \tilde{G} + \PAS ) =100\%$,
displaced vertex (mostly outside the detector)\cr
\midrule
$\PSGczDt$ & $BR(\PSGczDt\to \PSGczDo + \PHS)\approx 24\%$,
$BR(\PSGczDt\to \PSGczDo + \PHSM)\approx 76\%$ \cr
\midrule
Particular & $\PHS$ or $\PHSM$ appear at the end of every sparticle decay
chain \cr
signatures & $\PHS$ potentially observable via direct production\cr
\bottomrule
\toprule

\multicolumn{2}{c}  {\bf BP2: Light Pseudoscalars} & \\
\midrule
\multicolumn{2}{c}  {R. Aggleton, D. Barducci, N-E. Bomark, S. Moretti,}\\ 
\multicolumn{2}{c}  {A. Nikitenko,
C. Shepherd-Themistocleous, L. Roszkowski}\\ 
\multicolumn{2}{c}  {from JHEP 1502 (2015) 044, arXiv:1409.8393 \cite{Bomark:2014gya}, 
and J. Phys. G43 (2016) 105003, arXiv:1503.04228 \cite{Bomark:2015fga}} \cr
\midrule
\multicolumn{2}{c}  {BP2\_1 \hspace*{\fill}}\cr
\midrule
Main Features & A light pseudoscalar with $M_{\PAS} \approx  9~\UGeV$\cr
\midrule
Spectrum & $M_{\PHSM} \approx 123.3~\UGeV$,  $M_{\PAS} \approx  8.6~\UGeV$,
 $M_{\PHS} \approx 480~\UGeV$,\cr 
 & $M_H \approx  2254~\UGeV$, $M_{A} \approx 2255~\UGeV$\cr
\midrule
\multicolumn{2}{c}  {Production cross sections (at $13~\UTeV$) and branching fractions}\cr
\midrule 
$\PHS \to \PAS \PAS$ & ggF$(\PHS) \approx$ 43.2 \Upb, $BR(\PHS \to \PAS+\PAS) \approx 9.7\%$\cr 
 & $BR(\PAS \to \PGtp \PGtm) \approx 88.4\%$,  $BR(\PAS \to \PGmp \PGmm) \approx 0.34\%$  \cr 
 & ggF$(\PHS) \to \PAS+\PAS \to 4\PGtpr \approx 3.27$~\Upb,\cr
 & ggF$(\PHS) \to \PAS+\PAS \to 2\PGtpr+2\PGm \approx 0.0254$~\Upb \cr
\midrule
Particular & Considerable cross-section for very light pseudoscalar boson production
 \cr
signatures & with 4$\PGtpr$ final signature
state, with possibility for $2\PGtpr+2\PGm$ final state, \cr
 & free from Upsilon contamination \cr
\midrule
\multicolumn{2}{c}  {BP2\_2 \hspace*{\fill}}\cr
\midrule
Main Features & GMSB, lightest scalar SM-like, light pseudoscalar
just above $M_{\PHSM}/2$. \cr
\midrule
Spectrum & $M_{\PHSM} \approx 125.9~\UGeV$, $M_{\PHS} \approx 201~\UGeV$,
 $M_{\PAS} \approx  65~\UGeV$,\cr 
 & $M_H \approx  448~\UGeV$, $M_{A} \approx 440~\UGeV$\cr
\midrule
\multicolumn{2}{c}  {Production cross sections (at $13~\UTeV$) and branching fractions}\cr
\midrule 
$\PHS \to \PAS \PAS$ & ggF$(\PHS) \approx$ 0.86~\Upb, $BR(\PHS \to \PAS+\PAS) \approx 91\%$\cr 
 & $BR(\PAS \to \PQb \PAQb) \approx 91\%$, $BR(\PAS \to \PGtp \PGtm) \approx 8.8\%$\cr
 & ggF$(\PHS) \to \PAS+\PAS \to 4\PQb \approx 0.641$~\Upb \cr
 & ggF$(\PHS) \to \PAS+\PAS \to 2\PQb+2\PGtpr \approx 0.124$~\Upb \cr
 & ggF$(\PHS) \to \PAS+\PAS \to 4\PGtpr \approx 6$~\Ufb \cr
\midrule
$\PH \to \PZ + \PAS$ & ggF$(H) \approx$ 1.254~\Upb (at 13~\UTeV), $BR(H \to \PZ + \PAS) \approx 3.8\%$ \cr
 & ggF$(H) \to \PZ+\PAS \to
\Plp\Plm + \PQb \PAQb \approx 2.88$~\Ufb\cr 
\midrule
Particular & $\PH \to \PZ+\PAS$ of particular interest, \cr
signatures & could potentially use fat $\PQb$-jet techniques for $\PAS \to \PQb \PAQb$ \cr
\bottomrule
\toprule
%

\multicolumn{2}{c}  {\bf BP3: Discovery Through Stop to Electroweakinos} & \\
\midrule
\multicolumn{2}{c}  {C.T. Potter }\cr
\multicolumn{2}{c} {from Eur. Phys. J. C76 (2016) 44,  arXiv:1505.05554 \cite{Potter:2015wsa}}\cr
\midrule
Main Features & Natural NMSSM with discovery potential at LHC13 via stop $\PSQtDo$\cr
 & to electroweakinos: light $\PHS$, $\PAS$, singlino-like $\PSGczDo$, wino-like $\PSGczDT$  \cr
\midrule
Spectrum & $M_{\PHSM} \approx 123~\UGeV$, $M_{\PHS} \approx 55.7~\UGeV$, $M_{\PAS} \approx 10.0~\UGeV$,\cr 
 & $M_H \approx 1062~\UGeV$, $M_A \approx 1061~\UGeV$\cr 
 & $M_{\PSQtDo} \approx 336~\UGeV$,
 $M_{\PSGczDT} \approx 122~\UGeV$ \cr
\midrule
\multicolumn{2}{c}  {Production cross sections (at $13~\UTeV$) and branching fractions}\cr
\midrule 
Stop pair & $\Pp\Pp \to  \PASQtDo \PSQtDo \approx$ 5~\Upb, 
$BR(\PSQtDo \to \PSGcpDt + \PQb) \approx 75\%$, $BR(\PSGcpDt \to \PSGczDT + \PWp) \approx 38\%$ \cr 
production & $BR(\PSQtDo \to \PSGczDT + \PQt) \approx 15\%$,
 $BR(\PSGczDT \to \PSGczDo + \PHS) \approx 80\%$\cr
 & $BR(\PHS \to \PAS+\PAS) \approx 72\%$ \cr
 & $BR(\PAS \to \PGtp \PGtm) \approx 80\%$, $BR(\PAS \to \PGmp \PGmm) \approx 0.3\%$ \cr
\midrule
Particular & At least two tops and up to 4$\PAS$ as final states  \cr
signatures & \cr
\bottomrule
\toprule


\multicolumn{2}{c}  {\bf BP4: nMSSM Scenarios} & \\
\midrule
\multicolumn{2}{c}  {D.~Barducci, G.~Belanger, C.~Hugonie and
  A.~Pukhov }\cr
\multicolumn{2}{c}  {from JHEP 1601 (2016) 050, arXiv:1510.00246 \cite{Barducci:2015zna}}\cr
\midrule
Main Features & nMSSM ($\kappa = 0$, not \ZT-invariant), very light
singlino-like LSP \cr
\midrule
\multicolumn{2}{c}  {BP4\_1 \hspace*{\fill}}\cr
\midrule
Spectrum & $M_{\PHSM} \approx 122~\UGeV$, $M_{\PHS} \approx 37~\UGeV$,  $M_{\PAS} \approx  7~\UGeV$,\cr 
 & $M_{H} \approx 2.1~\UTeV$, $M_A \approx 2.1~\UTeV$, $M_{\PSGczDo} \approx 3~\UGeV$\cr 
\midrule
\multicolumn{2}{c}  {Production cross sections (at $13~\UTeV$) and branching fractions}\cr
\end{tabular}

\multicolumn{2}{c}  {\bf BP5: Light Singlino LSPs} & \\
\midrule
\multicolumn{2}{c}  {U. Ellwanger and A. M. Teixeira}\cr
\multicolumn{2}{c}  {from JHEP 1410 (2014) 113, arXiv:1406.7221 \cite{Ellwanger:2014hia} and
JHEP 1504 (2015) 172, arXiv:1412.6394 \cite{Ellwanger:2014hca} }\cr
\midrule
\multicolumn{2}{c}  {BP5\_1 \hspace*{\fill}}\cr
\midrule
Main Features & Light singlino-like LSP, all sparticle decay cascades end
with \cr
 &  $\PSGczDt \to \PSGczDo + \PHSM$, the LSP carries little $\MET$  \cr
\midrule
Spectrum & $M_{\PHSM} \approx 125~\UGeV$, $M_{\PHS} \approx 91.8~\UGeV$, \cr 
 & $M_{\PSGczDo} \approx 3.3~\UGeV$, $M_{\PSGczDt} \approx 130~\UGeV$,\cr
\end{tabular}

\multicolumn{2}{c}  {\bf BP6: Light Singlino LSP and Singlet-like Pseudoscalar, good Relic Density} & \\
\midrule
\multicolumn{2}{c}  {C. Han, D. Kim, S. Munir and M. Park }\cr
\multicolumn{2}{c} {from JHEP 1507 (2015) 002, arXiv:1504.05085 \cite{Han:2015zba} } \cr
\midrule
Main Features &  Good relic density, light singlino-like LSP, singlet-like
pseudoscalar, \cr
 & can be probed at the 14~\UTeV LHC in the di-muon decay channel \cr
\midrule
Spectrum & $M_{\PHSM} \approx 123.7~\UGeV$, $M_{\PHS} \approx 14.8~\UGeV$, $M_{\PAS}
 \approx 3.0~\UGeV$,\cr 
 & $M_{H} \approx 1504~\UGeV$, $M_{A} \approx 1504~\UGeV$, $M_{\PSGczDo}\approx 1.34~\UGeV$,\cr
 & $M_{\PSGczDt}\approx 131.7~\UGeV$,  $M_{\PSGczDT}\approx 166.8~\UGeV$, 
 $M_{\PSGcpmDo}\approx 147.2~\UGeV$\cr
\midrule
\multicolumn{2}{c}  {Production cross sections (at $14~\UTeV$) and branching fractions}\cr
\midrule
$\PSGcz_{2,3} + \PSGcpmDo$ & $\Pp\Pp \to \PSGcz_{2,3} + \PSGcpmDo  \to \PAS+  \PSGczDo +\PWpm + \PSGczDo$\cr 
production & \hspace*{2.5cm} $ \to \PGmp \PGmm + \PSGczDo + \Plpm+\PGn_l +\PSGczDo$: 3.16~\Ufb\cr 
 &   After cuts: 0.126 \Ufb, 38 events/300 \Ufb$^{-1}$,
 S/B $\approx 8$\cr 
\midrule
Particular & significance via $\PAS$ much larger than in standard trilepton channel   \cr
signatures &\cr
\bottomrule
\toprule

\multicolumn{2}{c}  {\bf BP7: Natural NMSSM and Cascade Higgs-to-Higgs   Decays} & \\
\midrule
\multicolumn{2}{c} {S.F. King, M. Muhlleitner, R. Nevzorov and K. Walz}\cr
\multicolumn{2}{c} {from Nucl. Phys. B870 (2013) 323,  arXiv:1211.5074 \cite{King:2012tr}}\cr
\multicolumn{2}{c} {and Phys. Rev. D90 (2014) 095014, arXiv:1408.1120 \cite{King:2014xwa}}\cr
\midrule
\multicolumn{2}{c}  {BP7\_1 \hspace*{\fill}}\cr
\midrule
Main Features & natural NMSSM: overall light Higgs spectrum testable at LHC13 \cr
\midrule
Spectrum & $M_{\PHSM} \approx 124.4~\UGeV$, $M_{\PHS} \approx 95.6~\UGeV$, $M_{\PAS} \approx  108~\UGeV$,\cr
 & $M_H \approx  299~\UGeV$, $M_A \approx 298~\UGeV$\cr
\end{tabular}

\multicolumn{2}{c}  {\bf BP8: Light Higgs Spectrum, Higgs-to-Higgs Decays} & \\
\midrule
\multicolumn{2}{c}  {C. Beskidt, W. de Boer, D. Kazakov and S. Wayand }\cr
\multicolumn{2}{c} {from Phys. Lett. B759 (2016) 141-148, arXiv:1602.08707 \cite{Beskidt:2016egy}}\cr
\midrule
\multicolumn{2}{c}  {BP8\_1 \hspace*{\fill}}\cr
\midrule
Main Features &  Light Higgs spectrum with heavier Higgs bosons just above 
  $\PQt\PAQt$\cr
 &  threshold, so main decay into $\PQt\PAQt$ (absent in CMSSM) \cr
\midrule
Spectrum & $M_{\PHSM} \approx 125.2~\UGeV$, $M_{\PHS} \approx
100.0~\UGeV$, $M_{\PAS} \approx 300.0~\UGeV$,\cr 
 & $M_{H} \approx 450.0~\UGeV$, $M_{A} \approx 444.9~\UGeV$\cr
\end{tabular}

\multicolumn{2}{c}  {\bf BP9: Singlino-Like and Bino-Like LSP Scenario} & \\
\midrule
\multicolumn{2}{c}  {N. Christensen, T. Han, Z. Liu and S. Su}\cr
\multicolumn{2}{c}  {from JHEP 1308 (2013) 019, arXiv:1303.2113 \cite{Christensen:2013dra} and JHEP
  1408 (2014) 093, arXiv:1406.1181 \cite{Han:2014nba}}\cr
\midrule
\multicolumn{2}{c}  {BP9\_1 \hspace*{\fill}}\cr
\midrule
Main Features & $\approx 30~\UGeV$ singlino-like LSP (good DM candidate), 
\cr
 &  two singlet-like Higgs states below 100 GeV decay mainly into $\PQb \PAQb$, \cr
 &   $\PHSM$ can decay into $\PHS \PHS$ with $\gsim 10\%$ BR\cr
\midrule
Spectrum & $M_{\PHSM} \approx 126~\UGeV$, $M_{\PHS} \approx 19.1~\UGeV$, $M_{\PAS}
 \approx 73.2~\UGeV$,\cr
\end{tabular}
\end{longtable}
\end{center}

\vspace{0.5cm}
\noindent
{\bf Acknowledgements}

The editors are grateful to the following people who have provided  input for the definition of the various benchmark points discussed in this chapter: A.~Nikitenko and L.~Roszkowski (BP2), D.~Kim and M.~Park (BP6), S.F.~King and W.~Walz (BP7), S.~Su (BP9).

\chapter{Exotic Higgs Decays}
\label{chap:ExoticDecay}
\ChapterAuthor{S.~Bressler, S.~Gori, A.~Mohammadi, J.~Shelton (Eds.)
F.~Bishara, L.~Caminada, R.~Caminal Armadans, D.~Curtin, G.~Isidori, Z.~Liu, V.I.~Martinez Outschoorn, M.~Neubert, K.~Nikolopoulos, T.~Orimoto, M.~Pelliccioni, F.~Petriello, M.J.~Strassler, R.~Teixeira de Lima, M.~Trott, J.~Zupan}
\section{Introduction and motivation}

As the program to characterize the properties of the observed Standard
Model (SM)-like Higgs boson advances, one of the major new discovery
opportunities it offers is potential new physics (NP) revealed 
in the Higgs boson's rare and exotic decays.
For any newly-discovered particle, a
comprehensive characterization of its decay modes is imperative; rare
decays of SM particles are prime places to search for signs of new
physics, and the Higgs boson is no exception.  It is worth
emphasizing, however, that among the SM particles the Higgs is
unique in its sensitivity to new physics.  The tiny SM width of the
Higgs, $\Gamma(h) = 4.08\, {\rm{MeV}}\,\pm 3.9\%$
\cite{Heinemeyer:2013tqa} for a $m_h=125.09$ GeV Higgs boson
\cite{Khachatryan:2016vau}, combined with the ease with which the Higgs can couple to physics
beyond the SM (BSM), make exotic decays of the SM Higgs a natural and
often leading signature of a broad class of theories of physics beyond
the SM.  Within the SM, the observation of rare exclusive decay modes
involving mesons would provide either confirmation or disproof
of the SM origin of mass for light quarks, which would otherwise
remain out of reach at the LHC.  This chapter discusses both these
{\em rare} decays to exclusive mesonic final states as well as {\em
  exotic} decays involving on-shell BSM particles.

Run 1 measurements of the discovered Higgs boson properties limit the exotic
branching fraction of the Higgs boson to be $\br(\hsm\to\mathrm{BSM})<
34\%$ at 95\% confidence level~\cite{Khachatryan:2016vau} assuming
that $\kappa_V\leq 1$, but allowing for the potential influence of new
physics on the Higgs boson couplings with gluons and photons, $\kappa_g$,
$\kappa_\gamma $. Thus substantial potential branching fractions of
the Higgs to new BSM particles are allowed by current data.  The
anticipated LHC data set, 3000 fb$^{-1}$ at 13 TeV, will contain
$\mathcal{O}(10^8)$ Higgs bosons, allowing branching ratios as small as
$\lesssim 10^{-7}$ to be probed, given a sufficient detection
efficiency as well as sufficient separation between the signal and the
standard model background.  For exotic decays, this enormous sample of
Higgs bosons translates into potential sensitivity to very small
Higgs-BSM couplings: e.g., for a fermionic state, an effective Yukawa
coupling of order $10^{-3}$ smaller than the bottom Yukawa, and for a
vector boson, a loop-induced coupling of order $10^{-2}$ times the
effective coupling of the Higgs to gluon pairs. For
rare SM Higgs boson decays, this data set offers the prospect of measuring
many exclusive decay modes such as $\hsm\to J/\Psi\gamma$, $\hsm\to
\rho\gamma $, and $h\to \phi\gamma$, for which the predicted SM
branching fractions are $\mathcal O(10^{-5}-10^{-6})$.  This program builds
on the pioneering Run 1 searches for $h\to J/\Psi\gamma$
\cite{Aad:2015sda,Khachatryan:2015lga}, which were for the first time
able to exclude couplings of the Higgs to charm quarks at the level of
220 times  the SM charm Yukawa coupling \cite{Perez:2015aoa}, and
extends it to yield insights into the strange, up, and down Yukawas,
which have remained almost entirely untested to date.

Exotic Higgs boson decays are a generic prediction of many well-motivated
theories of physics beyond the SM.  They occur frequently in theories
with extended Higgs sectors, as demanded by (e.g.) the NMSSM
\cite{Dobrescu:2000yn,Ellwanger:2003jt,Dermisek:2005ar,Chang:2008cw,Morrissey:2008gm,Belyaev:2010ka}
or theories with a first-order electroweak phase transition
\cite{Profumo:2007wc,Blinov:2015sna}; in models of dark matter
\cite{Silveira:1985rk, Pospelov:2007mp,
  Draper:2010ew,Ipek:2014gua,Martin:2014sxa}; in
theories of neutral naturalness \cite{Burdman:2006tz, Craig:2015pha,
  Curtin:2015fna}; and, more broadly, represent a generic signature of
physics beyond the Standard Model \cite{Strassler:2006ri,
  Curtin:2013fra}.
  In many cases, e.g. \cite{Strassler:2006ri,
  Pospelov:2007mp,Martin:2014sxa, Craig:2015pha, Curtin:2015fna},
exotic Higgs boson decays together with $\mathcal{O}(5\%)$ deviations in
Higgs boson properties may be the {\em only} observable signal of new
physics at the LHC.

In this chapter, we first cover rare exclusive mesonic
decays of the Higgs boson in  Section~\ref{sec:rare}. We give recommendations for predicted SM
branching ratios for these modes, present the predictions of a broad set of NP models
that give rise to enhanced mesonic branching ratios, and discuss
experimental prospects.
Next, we present our overarching
recommendations for a successful search program for 
exotic Higgs boson decays at the LHC in Section~\ref{sec:recommendations}. In
Section~\ref{sec:4b}, we discuss the decay topology $h\to XX\to 2Y 2Y'$
where $X$ is a NP particle and $Y,Y'$ are SM particles. In particular,
we study the kinematics of this final state as realized in the
prototypical decay $h\to aa\to 4b$, compare predictions from different
Monte Carlo generators, and highlight the relatively low $p_T$ objects
in the final state; these studies provide a guide for trigger and
analysis strategies for decays with this overall topology. In
Section~\ref{sec:2g}, we focus on prompt Higgs boson decays containing missing
energy, and perform a sensitivity study for Higgs boson decays into two
(resonant or non-resonant) photons plus missing energy.  This study
provides a careful examination of the interplay of different possible
trigger and reconstruction strategies for decays of the Higgs to
electroweak objects in combination with missing energy.  Finally, in
Section~\ref{sec:displaced}, we discuss theoretical motivations and
experimental prospects for searches for displaced Higgs boson decays,
together with recommendations for presenting results in such
searches.

\section{Exclusive mesonic and flavour-violating Higgs boson decays}\label{sec:rare}

Rare exclusive decays of the SM-like Higgs boson
to mesonic final states provide a unique window onto light quark
Yukawa couplings.  We discuss the SM predictions for the branching
ratio for Higgs boson decays into a meson plus a photon in Section~\ref{SMphoton}
and for Higgs boson decay to a meson plus a massive gauge boson in
Section~\ref{SMmassivegauge}.  In Section~\ref{NP} we summarize the impact
of different frameworks for physics beyond the SM on these exclusive
branching ratios, and in Section~\ref{ExperimentalProspects} we discuss
prospects for their detection at the LHC.

\subsection[Theoretical predictions: photon plus a meson]{Theoretical predictions: photon plus a meson\SectionAuthor{M.~Neubert, F.~Petriello}}
\label{SMphoton}

The SM predictions that the Higgs boson couplings to heavy gauge bosons and fermions are given by $2m_{W,Z}^2/v$ and $m_f/v$, where $v\approx 246$\,GeV is the Higgs vacuum expectation value, have been confirmed within experimental uncertainties for the $W$ and $Z$ bosons and for the third-generation fermions. However, no direct measurements of the Higgs boson couplings to the light fermions of the first two generations are available at present. As discussed in Section~\ref{NP} of this report, in several BSM models these couplings can deviate significantly from those predicted in the SM. Indeed, this is a generic prediction in many models trying to explain the hierarchies seen in the spectrum of fermion masses and mixing angles. Probing the Higgs boson couplings to light fermions is thus of paramount importance. This includes both flavour-diagonal and flavour-changing interactions. 

The measurement of the rare exclusive decays $h\to M\gamma$, where $M$ denotes a vector meson, would allow a unique probe of the Higgs boson coupling to light quarks at the LHC. While the absolute value of the bottom-quark Yukawa coupling can be accessed by measuring $b$-tagged jets in the associated production of the Higgs boson with a $W$ or $Z$ boson, this method becomes progressively more difficult for the lighter-quark couplings.  Advanced charm-tagging techniques may allow some access to the charm-quark Yukawa coupling~\cite{Perez:2015lra}, but no other way of directly measuring even lighter-quark couplings is currently known. The tiny branching ratios for these exclusive decays renders them inaccessible at future $e^+ e^-$ colliders. The program of measuring these decay modes is therefore only a possibility for the LHC and future hadron-collider facilities.  

The possibility of measuring rare exclusive Higgs boson decays was first pointed out in \cite{Bander:1978br,Keung:1983ac} and in more modern discussions in \cite{Bodwin:2013gca,Kagan:2014ila}, and the theoretical framework for their prediction was further developed in~\cite{Bodwin:2014bpa,Grossmann:2015lea,Koenig:2015pha}. Our discussion follows closely the techniques introduced in Refs.~\cite{Bodwin:2013gca,Kagan:2014ila,Bodwin:2014bpa,Grossmann:2015lea,Koenig:2015pha}, and we only summarize the salient features here. We begin our discussion of the theoretical predictions for these modes by introducing the effective Yukawa Lagrangian
\begin{equation}
   {\cal L} = - \sum_q\,\kappa_q\,\frac{m_q}{v}\,H\,\bar{q}_L q_R 
    - \sum_{q\ne q'}\,\frac{y_{qq'}}{\sqrt2}\,H\,\bar{q}_L q'_R + h.c. \,,
\end{equation}
where in the SM $\kappa_q=1$ while the flavour-changing Yukawa couplings $y_{qq'}$ vanish. The effective Lagrangian leads to two categories of exclusive Higgs boson decays: flavour-conserving decays involving the $\kappa_q$ couplings, where $M=\rho,\omega,\phi,J/\psi,\Upsilon(nS)$, and flavour-violating decays involving the $y_{qq'}$ couplings, where $M=B^{*0}_s,B^{*0}_d,K^{*0},D^{*0}$. In view of the very strong indirect bounds on flavour off-diagonal Higgs boson couplings to light quarks~\cite{Harnik:2012pb}, the flavour-violating decays $h\to M\gamma$ are bound to be very strongly suppressed. We will therefore restrict our discussion here to flavour-conserving processes.

The exclusive decays $H\to M\gamma$ are mediated by two distinct mechanisms, which interfere destructively.
\begin{itemize}
\item 
In the {\em indirect process}, the Higgs boson decays (primarily through loops involving heavy top quarks or weak gauge bosons) to a real photon $\gamma$ and a virtual $\gamma^*$ or $Z^*$ boson, which then converts into the vector meson $M$. This contribution only occurs for the flavour-conserving decay modes. The effect of the off-shellness of the photon and the contribution involving the $h\gamma Z^*$ coupling are suppressed by $m_M^2/m_h^2$, with $m_M$ the mass of the meson, and hence are very small~\cite{Koenig:2015pha}.
\item 
In the {\em direct process}, the Higgs boson decays into a quark and an antiquark, one of which radiates off a photon. This process introduces the dependence of the decay amplitude on the $\kappa_q$ parameters. The formation of the vector meson out of the quark-antiquark pair involves some non-trivial hadronic dynamics.
\end{itemize}
The relevant lowest-order Feynman diagrams contributing to the direct and indirect processes are shown in \refF{figure:diagrams} (left-middle and right panel, respectively).

\begin{figure}[thb]
\includegraphics[width=0.33\textwidth]{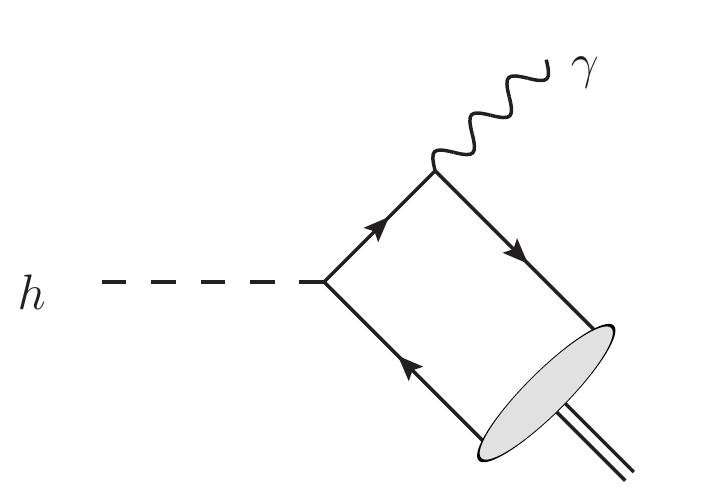}
\includegraphics[width=0.33\textwidth]{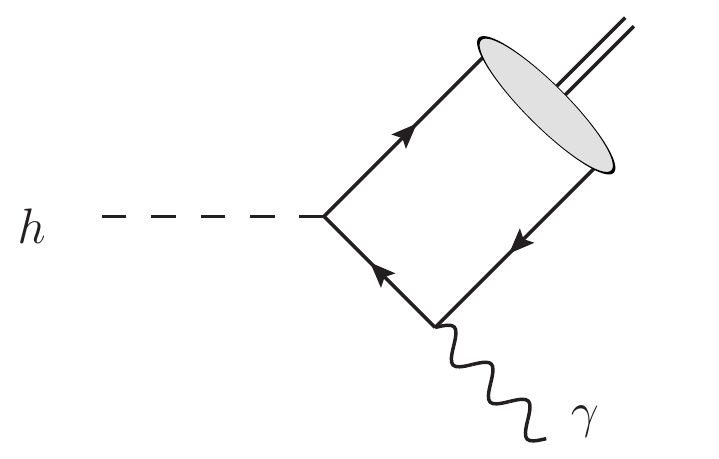}
\includegraphics[width=0.29\textwidth]{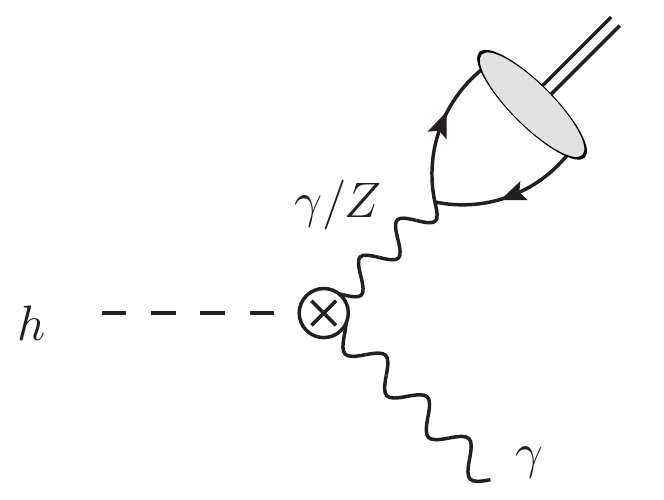}
\caption{\label{figure:diagrams}
Direct (left and centre) and indirect (right) contributions to the $h\to M\gamma$ decay amplitude. The blob represents the non-perturbative meson wave function. The crossed circle in the third diagram denotes the off-shell $h\to\gamma\gamma^*$ and $h\to\gamma Z^*$ amplitudes, which in the SM arise first at one-loop order.}
\end{figure}

We begin by outlining the calculation of the indirect amplitude. The virtual photon or $Z$ boson couples to
the vector meson through the matrix element of a local current, which can be parameterized in terms of a single hadronic parameter: the vector-meson decay constant $f_M$. This quantity can be obtained directly from experimental data. In particular, the leptonic decay rate of the vector meson can be written as
\begin{equation}
   \Gamma(M\to l^+l^-) = \frac{4\pi Q_M^2 f_M^2}{3 m_M} \alpha^2(m_M) \,,
\end{equation}
where $Q_M$ is the relevant combination of quark electric charges. The effective $h\gamma\gamma^{*}$ and $H\gamma Z^*$ vertices, which appear in the indirect amplitude, can be calculated with high accuracy in the SM. The by far dominant contributions involve loop diagrams containing heavy top quarks or $W$ bosons. The two-loop electroweak and QCD corrections to this amplitude are known, and when combined shift the leading one-loop expression by less than 1\% for the measured value of the Higgs boson mass~\cite{Degrassi:2005mc}. However, physics beyond the SM could affect these couplings in a non-trivial way, either through modifications of the $h t\bar t$ and $hW^+W^-$ couplings or by means of loops containing new heavy particles. The measurement of the light-quark couplings to the Higgs should therefore be considered together with the extraction of the effective $h\gamma\gamma$ coupling. As pointed out in~\cite{Koenig:2015pha}, by taking the ratio of the $h\to M\gamma$ and $h\to\gamma\gamma$ branching fractions one can remove this sensitivity to unknown new contributions to the $h\gamma\gamma$ coupling.

We now consider the theoretical prediction for the direct amplitude. This quantity cannot be directly related to data, unlike the indirect amplitude. Two theoretical approaches have been used to calculate this contribution. The hierarchy $m_h\gg m_M$ implies that the vector meson is emitted at very high energy $E_M\gg m_M$ in the Higgs boson rest frame. The partons making up the vector meson can thus be described by energetic particles moving collinear to the direction of $M$. This kinematic hierarchy allows the QCD factorization approach~\cite{Lepage:1980fj,Efremov:1979qk} to be utilized. Up to corrections of order $(\Lambda_{\rm QCD}/m_h)^2$ for light mesons, and of order $(m_M/m_h)^2$ for heavy vector mesons, this method can be used to express the direct contribution to the $h\to M\gamma$ decay amplitude as a perturbatively calculable hard-scattering coefficient convoluted with the leading-twist light-cone distribution amplitude (LCDA) of the vector meson. This approach was pursued in~\cite{Koenig:2015pha}, where the full next-to-leading order (NLO) QCD corrections were calculated and large logarithms of the form $[\alpha_s\ln(m_h/m_M)]^n$ were resummed at NLO, and in~\cite{Kagan:2014ila}, where an initial LO analysis was performed. The dominant theoretical uncertainties remaining after this calculation are parametric uncertainties associated with the non-perturbative LCDAs of the vector mesons. Thanks to the high value $\mu\sim m_h$ of the factorization scale, however, the LCDAs are close to the asymptotic form $\phi_M(x,\mu)=6x(1-x)$ attained for $\mu\to\infty$, and hence the sensitivity to not yet well-known hadronic parameters turns out to be mild. For the heavy vector mesons $M=J/\psi,\Upsilon(nS)$, the quark and antiquark which form the meson are slow-moving in the $M$ rest frame. This allows the non-relativistic QCD framework (NRQCD)~\cite{Bodwin:1994jh} to be employed to facilitate the calculation of the direct amplitude. This approach was pursued in~\cite{Bodwin:2014bpa}, where the NLO corrections in the velocity $v$ of the quarks in the $M$ rest frame, the next-to-leading order corrections in $\alpha_s$, and the leading-logarithmic resummation of collinear logarithms were incorporated into the theoretical predictions. The dominant theoretical uncertainties affecting the results for $h\to J/\psi\,\gamma$ and $h\to\Upsilon(nS)\,\gamma$ after the inclusion of these corrections are the uncalculated ${\cal O}(v^4)$ and ${\cal O}(\alpha_s v^2)$ terms in the NRQCD expansion.

\begin{table}
{\caption{\label{table:SM_BRs} 
Theoretical predictions for the $h\to M\gamma$ branching ratios in the SM, obtained using different theoretical approaches.}}
\begin{center}
\begin{tabular}{c|ccc}
\toprule
Mode & \multicolumn{3}{c}{Branching Fraction [$10^{-6}$]} \\
Method &  ~~NRQCD~\cite{Bodwin:2014bpa}~~ & ~~LCDA LO~\cite{Kagan:2014ila}~~
 & ~~LCDA NLO~\cite{Koenig:2015pha}~~ \\
\midrule
$\br(h\to\rho\gamma)$ & -- & $19.0\pm 1.5$ & $16.8\pm 0.8$ \\
$\br(h\to\omega\gamma)$ & -- & $1.60\pm 0.17$ & $1.48\pm 0.08$ \\
$\br(h\to\phi\gamma)$ & -- & $3.00\pm 0.13$ & $2.31\pm 0.11$ \\
$\br(h\to J/\psi\,\gamma)$ & -- & $2.79\,_{-0.15}^{+0.16}$ & $2.95\pm 0.17$ \\
$\br(h\to\Upsilon(1S)\,\gamma)$ & $(0.61\,_{-0.61}^{+1.74})\cdot 10^{-3}$ & -- 
 & $(4.61\,_{-\,1.23}^{+\,1.76})\cdot 10^{-3}$ \\
$\br(h\to\Upsilon(2S)\,\gamma)$ & $(2.02\,_{-1.28}^{+1.86})\cdot 10^{-3}$ & -- 
 & $(2.34\,_{-\,1.00}^{+\,0.76})\cdot 10^{-3}$ \\
$\br(h\to\Upsilon(3S)\,\gamma)$ & $(2.44\,_{-1.30}^{+1.75})\cdot 10^{-3}$ & -- 
 & $(2.13\,_{-\,1.13}^{+\,0.76})\cdot 10^{-3}$ \\
\bottomrule
\end{tabular}
\end{center}
\end{table} 

Table~\ref{table:SM_BRs} collects theoretical predictions for the various $h\to M\gamma$ branching fractions in the SM. The inclusion of NLO QCD corrections and resummation help to reduce the theoretical uncertainties. There is in general good agreement between the results obtained by different groups. The $h\to\phi\gamma$ branching ratio obtained in~\cite{Koenig:2015pha} is lower than that found in~\cite{Kagan:2014ila} because of an update of the $\phi$-meson decay constant performed in the former work. Also, in~\cite{Koenig:2015pha} the effects of $\rho$--$\omega$--$\phi$ mixing are taken into account. One observes that the $h\to M\gamma$ branching fractions are typically of order few times $10^{-6}$, which makes them very challenging to observe. The most striking feature of the results shown in the table concerns the $h\to\Upsilon(nS)\,\gamma$ modes, whose branching fractions are very strongly suppressed. This suppression results from an accidental and almost perfect cancellation between the direct and indirect amplitudes. In the case of $h\to\Upsilon(1S)\,\gamma$ the cancellation is so perfect that the small imaginary part of the direct contribution induced by one-loop QCD corrections gives the leading contribution to the decay amplitude. The fact that this imaginary part was neglected in~\cite{Bodwin:2014bpa} explains why a too small branching fraction for this mode was obtained there.

\begin{figure}[thb]
\includegraphics[width=0.45\textwidth]{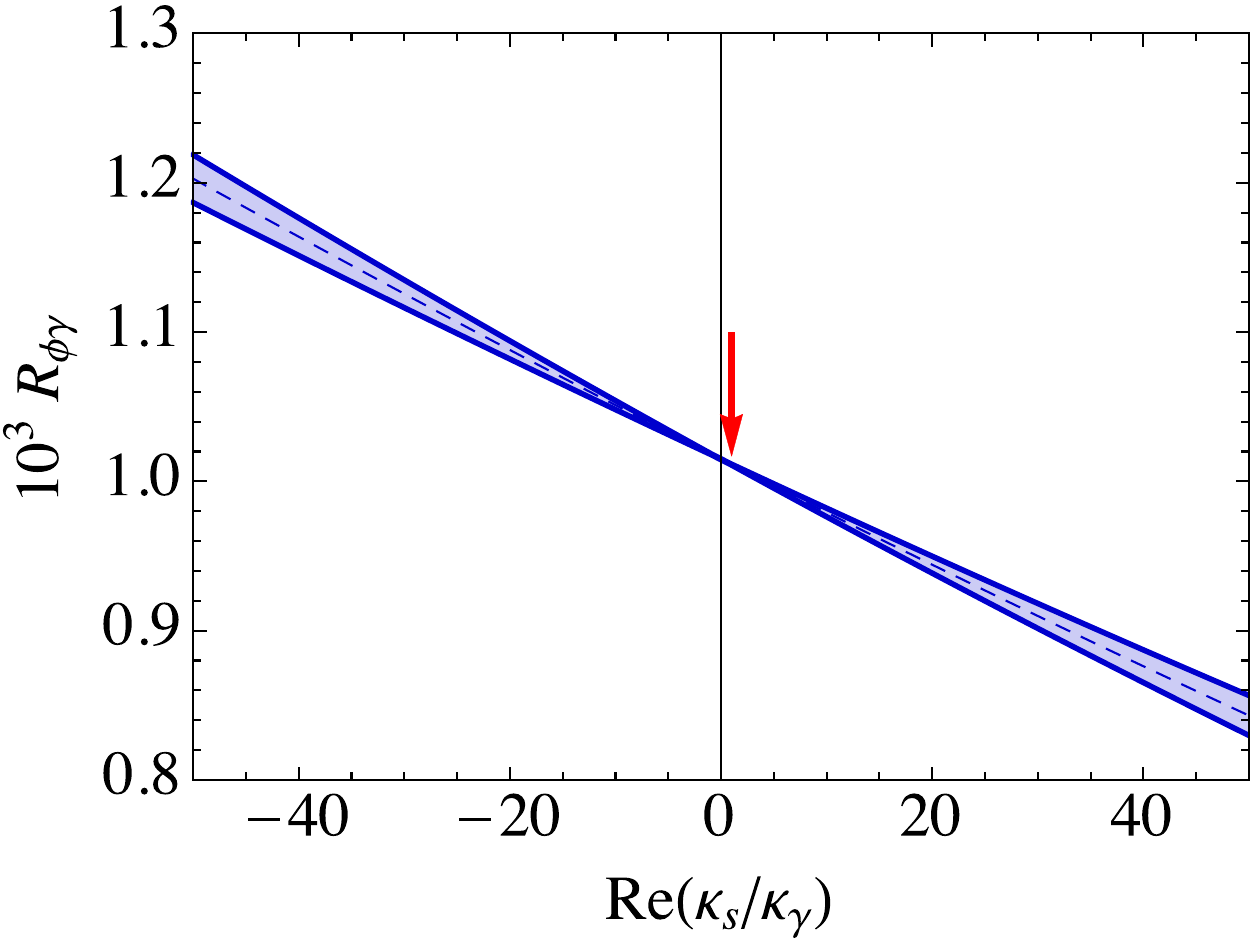} \quad
\includegraphics[width=0.45\textwidth]{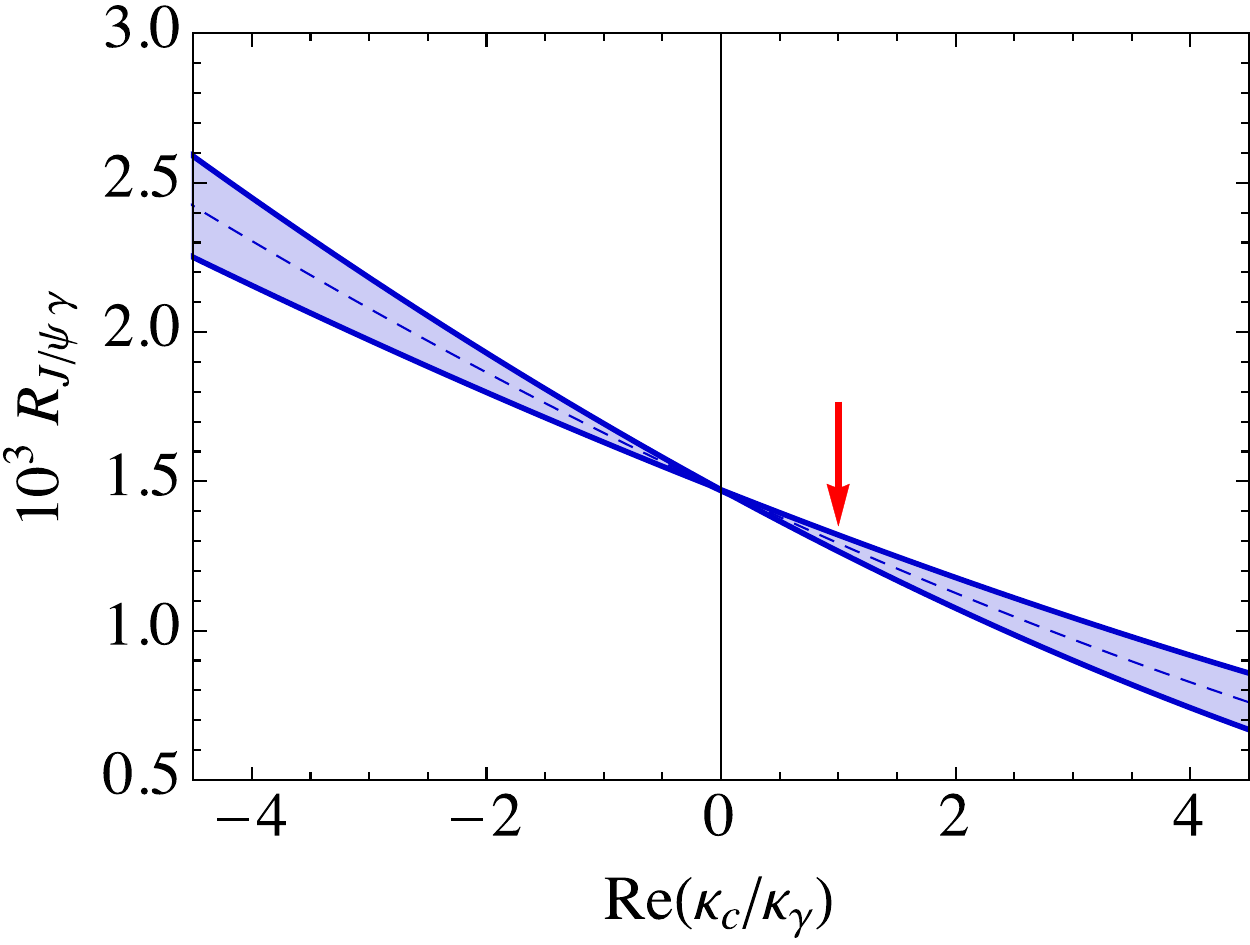}
\caption{\label{figure:phi_Jpsi}
$h\to\phi\gamma$ and $H\to J/\psi\,\gamma$ branching ratios, normalized to the $h\to\gamma\gamma$ branching fraction, as functions of the real part of $\kappa_{s,c}/\kappa_\gamma$. The SM values are indicated by the red arrows.}
\end{figure}

The main purpose of searching for the decays $h\to M\gamma$ is to use them for probing the light-quark Yukawa couplings. In order to eliminate possible new physics effects in the $h\to\gamma\gamma$ rate, it is of advantage to consider the ratio $R_{M\gamma}=\br(h\to M\gamma)/\br(h\to\gamma\gamma)$~\cite{Koenig:2015pha}, where in the SM $\mbox{BR}(h\to\gamma\gamma)=(2.28\pm 0.11)\cdot 10^{-3}$~\cite{Heinemeyer:2013tqa}. In the limit where the CP-violating contributions to the $h\to\gamma\gamma$ amplitude are neglected (the dominant such contributions would likely arise from the top-quark loop, but Electric Dipole Moment (EDM) constraints limit the imaginary part of $\kappa_t$ to be less than 1\%~\cite{Brod:2013cka}), one finds~\cite{Koenig:2015pha}
\begin{equation}\label{wonderful_ratio}
   R_{M\gamma} = \frac{8\pi\alpha^2(m_M)}{\alpha}\,\frac{Q_M^2 f_M^2}{m_M^2}
    \left( 1 - \frac{m_M^2}{m_h^2} \right)^2 
    \left( \big|1-\Delta_M\big|^2 + \big|\tilde\Delta_M\big|^2 \right) .
\end{equation}
The parameters $\Delta_M$ ($\tilde\Delta_M$) are proportional to the real (imaginary) part of the relevant $\kappa_q$ parameter and can be calculated using the QCD factorization approach, as described earlier. For all mesons other than the $\Upsilon(nS)$ states the interference of the direct amplitude with the dominant indirect one is a small effect, and hence the ratio $R_{M\gamma}$ is to excellent approximation a linear function of the real part of the ratio $\kappa_q/\kappa_\gamma$, where $\kappa_\gamma$ is the new physics modification of the entire $h\to\gamma\gamma$ matrix element. This quantity is known to be close to its SM value~1. \refF{figure:phi_Jpsi} shows theoretical predictions for the ratios $R_{\phi\gamma}$ and $R_{J/\psi\,\gamma}$ (times $10^3$) obtained in~\cite{Koenig:2015pha}. The width of the bands reflects the theoretical uncertainties. The corresponding predictions for the lighter mesons $\rho$ and $\omega$ suffer from significant hadronic uncertainties due to $\rho$--$\omega$--$\phi$ mixing.

\begin{figure}[thb]
\includegraphics[width=0.45\textwidth]{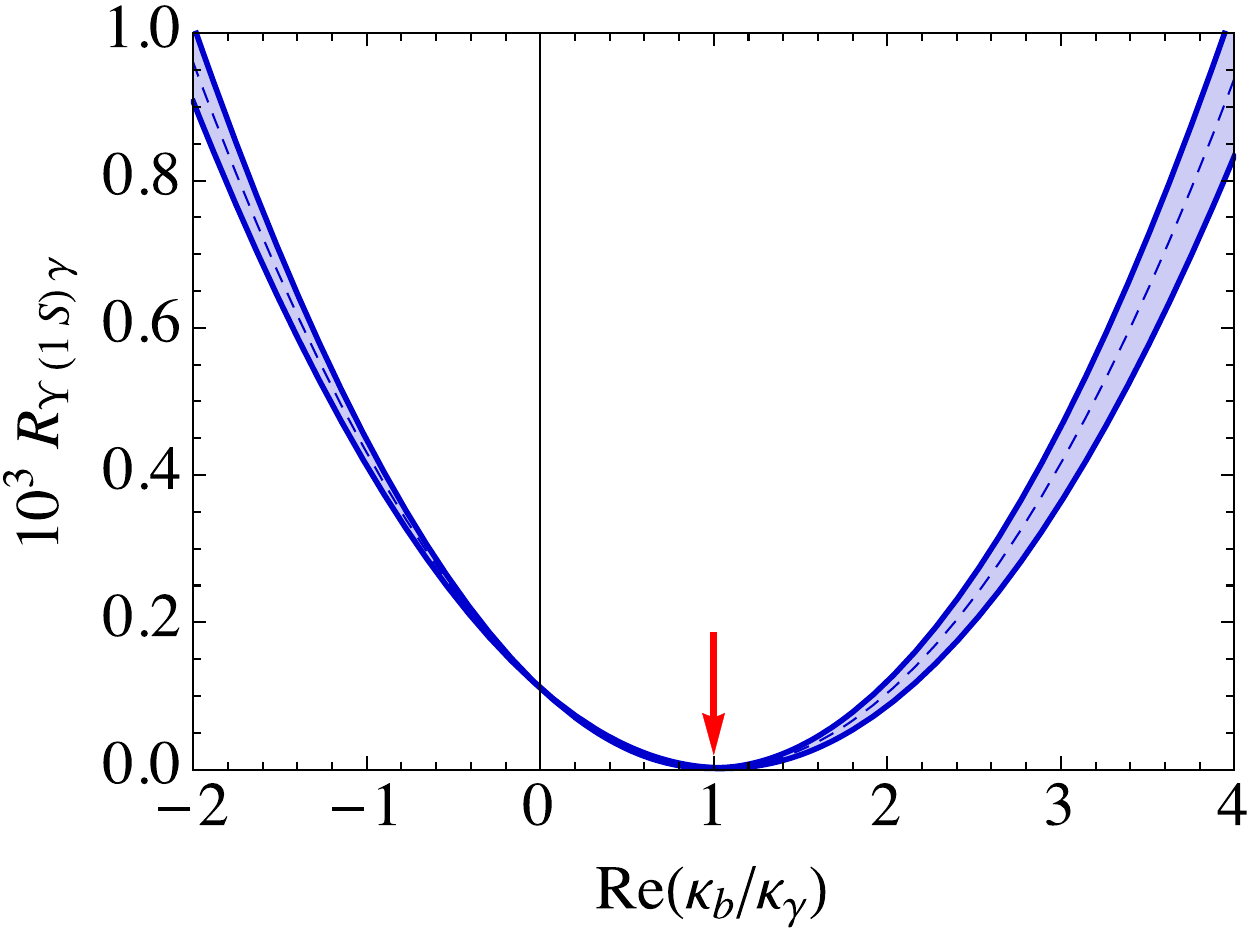} \quad
\includegraphics[width=0.45\textwidth]{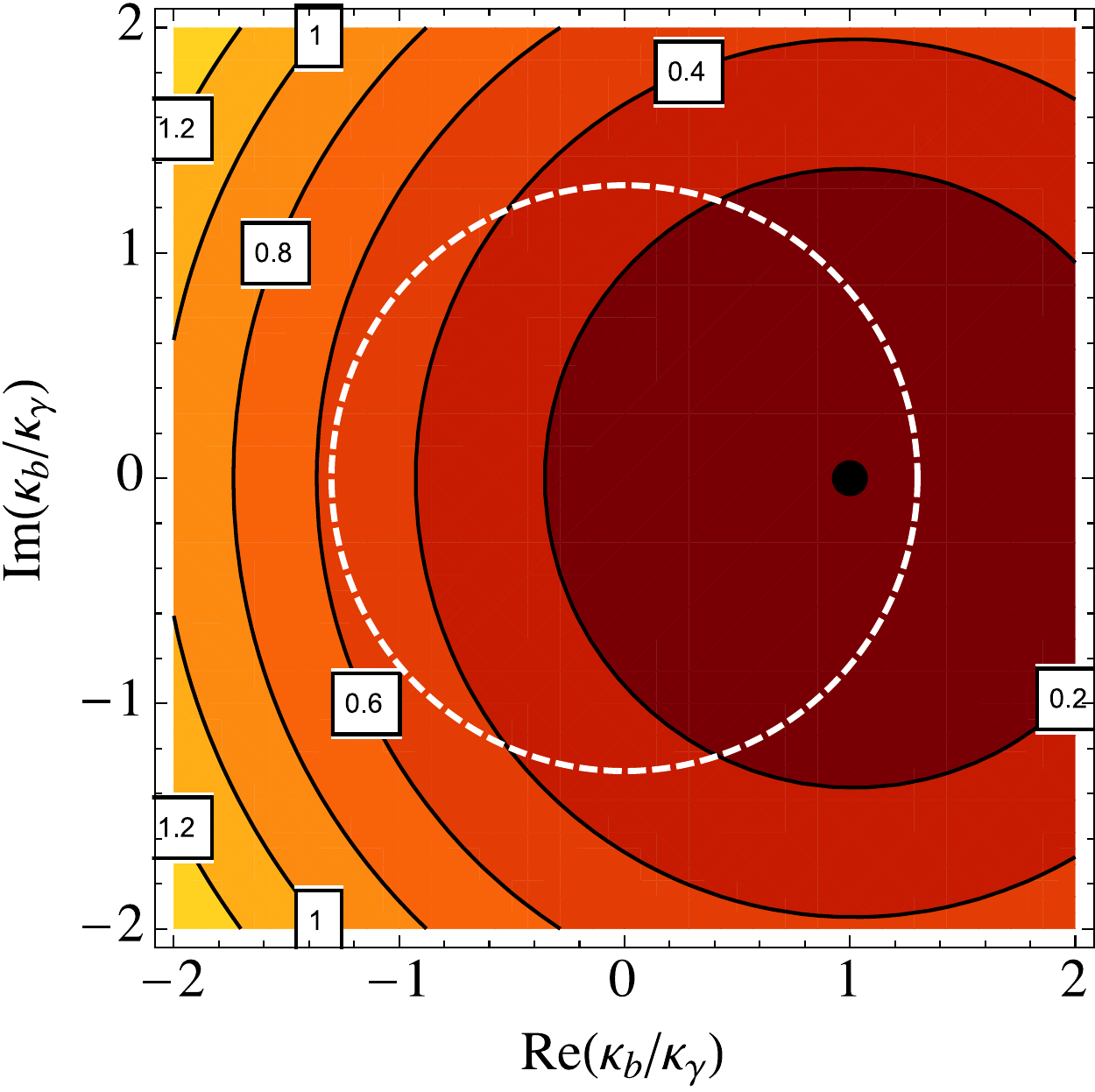}
\caption{\label{figure:contours}
Ratio $R_{\Upsilon(1S)\gamma}$ as a function of the real and imaginary parts of $\kappa_b/\kappa_\gamma$. In the left plot the imaginary part is set to zero. The right plot shows contour lines of $10^3\,R_{\Upsilon(1S)\gamma}$ in the complex $\kappa_b/\kappa_\gamma$ plane. The black dot and the arrow indicate the SM values. Coupling parameters inside the dashed white circle are preferred by the current LHC data on $h\to b\bar b$.}
\end{figure}

In the case of the $h\to\Upsilon(nS)\,\gamma$ decay modes the SM branching ratios are so small that a discovery at the LHC (or any other conceivable collider) is all but elusive. The direct contributions are no longer a small correction, and hence the quadratic terms in $\kappa_b$ are important. On the other hand, the almost perfect cancellation between the direct and indirect amplitudes no longer holds in the presence of new physics. The left plot in \refF{figure:contours} shows the dependence of the ratio $R_{\Upsilon(1S)\gamma}$ on the real part of $\kappa_b/\kappa_\gamma$, assuming that the CP-violating imaginary part vanishes. It is evident that the SM value accidentally coincides with the minimum of the curve, while significantly larger branching fractions are possible when new physics alters the value of $\kappa_b$. As an interesting benchmark for LHC experiments, we consider the case where $\kappa_b=-1$, while $\kappa_\gamma$ takes its SM value of 1. This benchmark is and will be in great agreement with LHC Higgs boson coupling fits, since Higgs boson coupling measurements cannot probe the sign of $\kappa_b$. We then obtain the branching fractions
\begin{equation}
\begin{aligned}
   \br(h\to\Upsilon(1S)\,\gamma) 
   &= (0.98\pm 0.06) \cdot 10^{-6} \,, \\
   \br(h\to\Upsilon(2S)\,\gamma) 
   &= (0.45\pm 0.03) \cdot 10^{-6} \,, \qquad (\kappa_b=-1) \\
   \br(h\to\Upsilon(3S)\,\gamma) 
   &= (0.33\pm 0.03) \cdot 10^{-6} \,,
\end{aligned}
\end{equation}
more than two orders of magnitude larger than in the SM. The right plot in \refF{figure:contours} shows contours of $10^3\,R_{\Upsilon(1S)\gamma}$ in the complex $\kappa_b/\kappa_\gamma$ plane. The dashed white circle indicates the current upper bound on the combination $\lambda_{b\gamma}=|\kappa_b/\kappa_\gamma|$, which to an excellent approximation measures the deviation of the ratio $\br(h\to b\bar b)/\br(h\to\gamma\gamma)$ from its SM value. The Higgs bosons must be produced via the same production mechanism in both cases, so that possible new physics effects in Higgs boson production cancel out. Since the $h\to b\bar b$ mode is measured at the LHC in the rare $VH$ and $t\bar t H$ associated-production channels, at present no accurate direct measurements of $\lambda_{b\gamma}$ are available. However, from the model-independent global analyses of Higgs boson couplings, in which all couplings to SM particles (including the effective couplings to photons and gluons) are rescaled by corresponding $\kappa_i$ parameters and also invisible Higgs boson decays are allowed, one obtains $\lambda_{b\gamma}=0.63\pm 0.27$ for CMS~\cite{Khachatryan:2014jba} and $\lambda_{b\gamma}=0.67\pm 0.32$ for ATLAS~\cite{Aad:2015gba}. At 95\% CL this implies $\lambda_{b\gamma}<1.3$. Within this allowed region, the $h\to\Upsilon(1S)$ branching ratio varies by more than two orders of magnitude and can take values as large as $1.3\cdot 10^{-6}$, which may be within reach of the high-luminosity run at the LHC.

The decays $h\to\Upsilon(nS)\,\gamma$ provide a golden opportunity to probe new-physics effects on the bottom-quark Yukawa couplings. Any measurement of such a decay would be a clear signal of new physics. A combined measurement of the two ratios $\br(h\to\Upsilon(nS)\,\gamma)/\br(h\to\gamma\gamma)$ and $\mbox{BR}(h\to b\bar b)/\br(h\to\gamma\gamma)$ can provide complementary information on the real and imaginary parts of the $b$-quark Yukawa coupling. We can think of no other way in which one can probe the magnitudes and signs of the real and imaginary parts of $\kappa_b$ separately.

\subsection[\texorpdfstring{$h \rightarrow V P$}{h to VP} and \texorpdfstring{$h \rightarrow V P^\star$}{H to VP*}]{\texorpdfstring{$h \rightarrow V P$}{h to VP} and \texorpdfstring{$h \rightarrow V P^\star$}{H to VP*}\SectionAuthor{G.~Isidori, M.~Trott}}
\label{SMmassivegauge}


In this section, we discuss decays of the form $h \rightarrow V M$
where $V$ is a massive on-shell SM vector boson ($V=Z,W$) and $M$ is
an associated meson (vector or pseudoscalar) produced in the decay of
the $h$ particle.

We will focus on the decay of the Higgs, assuming a narrow width
approximation to factorize the decay and production mechanisms.  The
decays we will discuss are very rare decays in the SM, with extremely
small branching ratios. Despite these small rates, it is still
important to search for such rare decays, to learn experimentally
about the properties of the discovered $h$ state robustly. It was
pointed out in Ref.~\cite{Isidori:2013cla} that such rare exclusive
decays of the $h$ particle with an associated massive vector boson
offer complementary information about the properties of this state,
and how it couples to the SM fields. This information is complementary
to what can be determined experimentally from more inclusive $h$ decay
modes.
Reaching the experimental sensitivity required to observe such
extremely rare decays, with {\it any} associated $V = \{\gamma, Z, W
\}$, is extremely challenging but worth the effort.

Given the strong suppression of such exclusive decay modes,
 the theoretical predictions of the corresponding decay rates
are subject to an irreducible uncertainty due to the limited knowledge  of the $h$ dominant decay modes, both within and, especially, beyond the SM.
In particular, the total uncertainty in the width of the $h$ particle ($\Gamma_h$) feeds into the uncertainty in the predicted $\br(h \rightarrow i)$ as
\bea
\delta BR (h \rightarrow i) = {\rm BR (h \rightarrow i)} \,  \left(\frac{\delta \Gamma_{h \rightarrow i}}{\Gamma_{h \rightarrow i}} + \frac{\delta \Gamma_h}{\Gamma_h}
-2  \frac{{\Delta}_{i \, \Gamma_h}}{\Gamma_{h \rightarrow i} \Gamma_{h}} \right)^{1/2},
\eea
using simple Gaussian error propagation, where $\delta$ indicates a $1 \sigma$ error. Here $\Delta_{i \, \Gamma_h}$ is the covariance of the total width and the decay channel $h \rightarrow i$.
Although possible tests of the decay width---within the SM---have been proposed in the literature~\cite{Kauer:2012hd,Campbell:2013una,Caola:2013yja} and carried out by the experimental collaborations~\cite{Khachatryan:2014iha,Aad:2015xua},
the corresponding constraint on the decay width does not hold in an EFT generalization of the SM~\cite{Englert:2014aca}.
Indeed, as emphasized in Ref.~\cite{Englert:2014aca}, the uncertainty in $\mu_{ggF}$ and $\mu_{ZZ}$ directly feeds into such a measurement.
When we discuss theoretical uncertainties for the rare modes in Tables \ref{tab:pseudo} and \ref{tab:vector}, we report theoretical uncertainties of the form $ \delta \Gamma_{h \rightarrow i}/\Gamma_{h \rightarrow i}$ for the SM,
added in quadrature to the total theoretical width defined as $\Gamma_{h} =4.08~{\rm MeV}$ with a $\pm 3.9\%$ relative error.
We neglect the unknown $\Delta_{i \, \Gamma_h}$ in this estimate.

\subsubsection{SM predictions, dominant electroweak dependence}

 The SM prediction for the decay proceeds dominantly through the diagrams shown in \refF{fig1:exclusivedecay}.
Defining the SM currents coupling to the massive vector bosons as
\bea
\label{currentdefn}
\mathcal{L}_{J} = \frac{e }{ \sqrt{2}\, \sin \theta_W}  J_\mu^{\pm} W^\mu_{\pm}   + \frac{e}{\sin \theta_W \cos \theta_W}  J_\mu^0 Z^\mu,
\eea
and the pseudoscalar ($P$) and vector meson ($P^\star$) decay constants with the following normalizations
\bea
\langle P(q) | J_\mu(q) | 0 \rangle 0 = \frac{1}{2} \, F_P \, q_\mu, \quad \quad \langle P^\star (q) | J_\mu(q) | 0 \rangle  = \frac{1}{2} \, F_P^\star \, m_p^\star \, \epsilon_\mu,
\eea
then the tree-level (\refF{fig1:exclusivedecay}a) SM contribution is \cite{Isidori:2013cla},
\begin{align}
\cB^{\rm SM}(h\to VP)= \frac{m_h^3 \, G_F^2}{8 \, \pi} \frac{\left| C_V F_P \right|^2}{(\Gamma_h)_{SM}} \lambda^3(1, \rho, \hat{q}^2) ,
\label{eq:hVPSM}
\end{align}
where $C_V = \{1/\sqrt{2},1 \}$ for the cases $W$ and $Z$ respectively.
Here $\rho = m_V^2/m_h^2$ and $\hat{q}^2 = m_P^2/m_h^2$ and $\lambda(a, b, c) = \sqrt{a^2 + b^2 + c^2 - 2 (ab + ac + bc)}$. $G_F$ is the fermi constant
and $(\Gamma_h)_{SM}$ is the SM Higgs boson decay width. The case of a decay to a vector meson
through a $Z^\star$, for example $h \rightarrow J/\psi Z$, through \refF{fig1:exclusivedecay}a gives a branching ratio \cite{Gonzalez-Alonso:2014rla}
\begin{align}
\cB^{\rm SM} (h\to VJ/\Psi)=\frac{m_h^3 \, G^2_F}{8 \, \pi} \frac{\left| F_{J/\Psi}^\star \right|^2}{(\Gamma_h)_{SM}} \frac{\sqrt{\lambda(1, \rho, \hat{q}^2)}}{(1 - \hat{q}^2/\rho)^2} \left[(1-\rho)^2\left(1- \frac{\hat{q}^2}{1-\rho}\right)^2 + 8 \,  \hat{q}^2 \, \rho \right].
\end{align}
The tree-level contribution in \refF{fig1:exclusivedecay}a is usually largely dominant but for the case of charmonium vector resonances
($J/\psi$, $\psi'$, $\ldots$). In the latter case the accidental suppression of the $Z$-boson vector coupling to charm makes the
formally subleading (one-loop induced) $h \rightarrow Z \, \gamma$ amplitude (\refF{fig1:exclusivedecay}b) compete with the tree-level one~\cite{Bodwin:2013gca,Gao:2014xlv}.
The full SM expression including this contribution and the interference term can be found in Ref.~\cite{Gao:2014xlv}\footnote{See Ref.~\cite{Alte:2016yuw} for the latest computation using the QCD factorization approach. A future LHC HXSWG note will contain updated predictions.}.
The SM predictions of pseudoscalar decays are given in \refT{tab:pseudo}. We use the value $ C_{Z\gamma}^{SM} = -5.540$ for the Wilson coefficient of this operator, consistent with the normalization of the operator defined in Ref.~\cite{Gao:2014xlv,Korchin:2013ifa}.

\begin{figure}[thb]
\begin{tikzpicture}
\draw[dashed] (0,0) -- (2.0,0) ;
\draw[decorate,decoration=snake] (2.0,0) -- (3.5,1.5) ;
\draw[decorate,decoration=snake] (2.0,0) -- (3,-1.5) ;
\draw [->-] (3,-1.5) -- (4,-1.0);
\draw [-<-] (3,-1.5) -- (4,-2.0);
\draw (4.1,-1.5) ellipse (0.1cm and 0.7cm);
\node [above][ultra thick] at (2.2,0.8) {$Z (W)$};
\node [above][ultra thick] at (1.6,-1.3) {$Z^\star (W^\star)$};
\node [above][ultra thick] at (2.5,-2.8) {$(a)$};
\end{tikzpicture}
\begin{tikzpicture}
\draw[dashed] (0.5,0) -- (2.5,0) ;
\filldraw (2.5,0) circle (0.095);
\draw[decorate,decoration=snake] (2.5,0) -- (4,1.5) ;
\draw[decorate,decoration=snake] (2.5,0) -- (3.5,-1.5) ;
\draw [->-] (3.5,-1.5) -- (4.5,-1.0);
\draw [-<-] (3.5,-1.5) -- (4.5,-2.0);
\draw (4.6,-1.5) ellipse (0.1cm and 0.7cm);
\node [above][ultra thick] at (3,0.8) {$Z$};
\node [above][ultra thick] at (2.5,-1.3) {$\gamma^\star$};
\node [above][ultra thick] at (3,-2.8) {$(b)$};
\end{tikzpicture}
\begin{tikzpicture}
\draw[dashed] (0,0)  (1,2.6) ;
\draw[dashed] (1,2.6)  -- (3,2.6) ;
\draw [->-] (3,2.6) -- (4.9,3.8) ;
\draw [-<-] (3,2.6) -- (3.5,2) ;
\draw [-<-] (3.5,2)  -- (4.9,3.2);
\draw[decorate,decoration=snake] (3.5,2)-- (4.85,1.2) ;
\draw (5,3.5) ellipse (0.1cm and 0.6cm);
\node [above][ultra thick] at (5.5,0.8) {$Z (W)$};
\node [above][ultra thick] at (3,0) {$(c)$};
\end{tikzpicture}
\caption{\label{fig1:exclusivedecay}
Direct contributions to exclusive decay modes of the form $h \rightarrow V P$ and $h \rightarrow V P^\star$. Diagram
(a) is generally the dominant contribution in the SM, while diagram $(b)$ can also contribute significantly for narrow light vector mesons.
Diagram (c) is generally negligible in the SM, but can be significantly enhanced in beyond the SM scenarios.}
\end{figure}

\subsubsection{SM predictions, subdominant Yukawa dependence}

In the SM branching ratios reported in Tables \ref{tab:pseudo} and \ref{tab:vector} the contribution
from \refF{fig1:exclusivedecay}c is neglected, being suppressed by a small Yukawa
coupling.  This is always a good approximation in the SM if the
associated vector meson is a $Z$ or a $W$, i.e.~when there is a
tree-level contribution not suppressed by small Yukawa couplings.

The Yukawa amplitude is not necessarily negligible in the radiative
modes ($V=\gamma$), when the tree-level amplitude is absent.  The
possibility of determining this contribution in the modes $V=\gamma$
has been already discussed in. Section~\ref{SMphoton}.
However, as we briefly illustrate below, this goal is
extremely challenging  for the SM values of $\kappa_q=1$, given that the Yukawa contribution is typically
subleading.

New physics in the Higgs sector at some high scale can induce
large deviations of the light-quark Yukawa couplings from their SM
values even when
the cut-off scale is parametrically greater than the electroweak scale.
Indeed the corresponding amplitudes scale as
\bea
A(\bar{\psi} \, \psi \rightarrow W_L \, W_L) = \frac{m_\psi \sqrt{s}}{v^2} (1 - \kappa_\psi \, \kappa_W).
\eea
Due to the presence of a small fermion ($\psi$) mass scale, large
deviations in $\kappa_\psi$ still allow the cut-off
scale of the theory to remain parametrically separated from the scale
$v$.  While such enhancements would generally occur along with
 other deviations in Higgs phenomenology,
including deviations in the electroweak couplings present in \refF{fig1:exclusivedecay}a
and \refF{fig1:exclusivedecay}b, some classes of BSM models can predict  parametrically large enhancements to $\kappa_\psi$  while leaving other Higgs boson couplings largely unaffected, as discussed in Section\ref{NP}.  It is important to bear in mind that in BSM models which do yield large deviations in other Higgs boson couplings, the extraction of any
information on the sub-leading Yukawa contribution could be out of reach.

As a specific example, consider the results quoted in Ref.~\cite{Kagan:2014ila}, and discussed in Section~\ref{SMphoton}, for the $\gamma$-tagged decays:
\bea
\frac{\br(h \rightarrow \phi \, \gamma)}{\br(h \rightarrow b \bar{b})}  &\simeq&  \frac{\kappa_\gamma \left[(3.0 \pm 0.13) \, \kappa_\gamma  - 0.02 \, \kappa_s\right]}{0.57 \, \kappa_b^2} \, \times 10^{-6}, \nn \\
\frac{\br(h \rightarrow \rho \, \gamma)}{{\br}(h \rightarrow b \bar{b})} &\simeq&  \frac{\kappa_\gamma \left[(1.9 \pm 0.15) \, \kappa_\gamma  - 1 \times 10^{-4} \, \kappa_u  - 1 \times 10^{-4} \, \kappa_d \right]}{0.57 \, \kappa_b^2} \, \times 10^{-5}, \nn \\
\frac{{\br}(h \rightarrow \omega \, \gamma)}{{\br}(h \rightarrow b \bar{b})} &\simeq&  \frac{\kappa_\gamma \left[(1.6 \pm 0.17) \, \kappa_\gamma  - 3 \times 10^{-4} \, \kappa_u  - 3 \times 10^{-4} \kappa_d \right]}{0.57 \, \kappa_b^2} \, \times 10^{-6}.
\label{eq:rarebruncertainty}
\eea
The only modification here compared to Ref.~\cite{Kagan:2014ila} is to
utilize $\kappa_q$ factors normalized to the light quark masses,
consistent with standard usage.  Recall that currently $\kappa_\gamma
= 0.92^{+0.12}_{-0.11}$ when $\br_{BSM}$ is assumed to be $0$ in the
ATLAS/CMS coupling fit combination \cite{Khachatryan:2016vau}.
While enhancements of $\kappa_{s,u,d}$ by $\mathcal{O}(100)$ may thus be
 measurable (see Section~\ref{SMphoton}), Eq.~\ref{eq:rarebruncertainty}
 demonstrates that
a very large reduction in the uncertainties of the
tree-level couplings of the $h$ state, and the effective one-loop
couplings of this state (i.e. $\kappa_\gamma$), would be required for
these rare decays to be
sensitive to light quark Yukawa couplings of $\mathcal O(10)$ their SM values or smaller.

\subsubsection{SM Theory errors}
The dominant theoretical errors in the branching ratios are due to the
lattice (or experimental) errors on the meson decay constants $F_P$ (ranging
between few $\%$ and $10\%$), which are combined in quadrature with
the uncertainties on the elements of the Cabibbo-Kobayashi-Maskawa
mixing matrix ($V_{\rm CKM}$).  This uncertainty is given in column
three of the branching ratio tables. This theoretical error, combined
with the theoretical error on the Higgs boson total width ($\sim 4\%$
\cite{Beringer:1900zz}) dictates the error given in the fifth column
of the tables.

\begin{table}
\caption{SM branching ratios $\cB^{\rm SM}$  for selected  $h\to VP$ decays. The decay constants are defined as $F_{\pi^0} = f_\pi/\sqrt{2}$, $F_{K^\pm} = V_{us} \, f_K$
$F_{\pi^{\pm}} = V_{ud} f_\pi$, $F_{D^\pm}= V_{cd} f_D$, $F_{D_s} = V_{cs} f_{D_s}$, $F_{B^\pm}= V_{ub} f_B$, $F_{B_c^\pm}= V_{cb} f_B$,  and $F_{\eta_c} = f_{\eta_c}/2$,
where $f_P$ are the standard meson decay constants reported in \cite{Beringer:1900zz,Colangelo:2010et,Davies:2010ip} for $N_f = 2+1$ (when available)
and the CKM parameters are PDG values \cite{Beringer:1900zz}. The theoretical errors quoted are $1 \sigma$ values.}
\label{tab:pseudo}
\centering
\tabcolsep 8pt
\begin{tabular}{c|c|c|c|c}
\toprule
$VP$~mode & $P$ mass & $F_P$ & $\cB^{\rm SM}$  & Th. Error  \\ \hline
 $W^- \pi^+$  & 139.57018 $\pm$ 0.00035 {\rm MeV}  & 126.6 $\pm$ 1.4 {\rm MeV} &  $0.42 \times 10^{-5}$  & $\pm 5 \%$ \\
 $W^- K^+$  & 493.677 $\pm$ 0.016 {\rm MeV}  & 35.2 $\pm$ 0.3 {\rm MeV}  &    $0.33 \times 10^{-6}$ &   $\pm 4 \%$ \\
 $W^- D^+_s$  & 1968.30 $\pm$ 0.11 {\rm MeV}  & 248.6 $\pm$ 2.4 {\rm MeV}  &   $1.6 \times 10^{-5}$ & $\pm 4 \%$  \\
$W^- D^+$  & 1869.61 $\pm$ 0.09 {\rm MeV}  & 47.07 $\pm$ 2.4 {\rm MeV} &  $0.58 \times 10^{-6}$  &  $\pm 11 \%$\\
$W^- B^+$  &  5279.29 $\pm$ 0.15  {\rm MeV}  & 0.79 $\pm$ 0.10 {\rm MeV} &  $1.6 \times 10^{-10}$  & $\pm 26 \%$ \\
$W^- B_c^+$  & 6275.1 $\pm$ 1.0 {\rm MeV}  & 7.82 $\pm$ 0.42 {\rm MeV} &  $1.6 \times 10^{-8}$  & $\pm 11 \%$\\
\midrule
 $Z \pi^0$  &  134.9766 $\pm$ 0.0006 {\rm MeV} & 92.1 $\pm$ 1.0 {\rm MeV}  &  $0.23 \times 10^{-5}$  & $\pm 5 \%$ \\
$Z \eta_c$  & 2984.3 $\pm$ 0.84 {\rm MeV}  & 197.4 $\pm$ 0.30 {\rm MeV}  &  $1.0 \times 10^{-5}$ & $\pm 5 \%$   \\
\bottomrule
\end{tabular}
\end{table}

\begin{table}
\caption{SM branching ratios for selected  $h\to VP^\star$ decays. The normalizations are defined as in the case of the
pseudoscalar mesons. We use $F_B^\star/F_B = 1.02 \pm 0.08$  \cite{Gelhausen:2013wia}. The theoretical errors quoted are $1 \sigma$ values.
Total errors quoted at $\gtrsim 4 \%$  do not have a decay constant theoretical error assigned.}
\label{tab:vector}
\centering
\tabcolsep 8pt
\begin{tabular}{c|c|c|c|c}
\toprule
$VP^\star$~mode & $P^\star$ mass & $F_P^\star/2$ & $\cB^{\rm SM}$  & Th. Error \\ 
\midrule
 $W^- \rho^+$  &  775.26 $\pm$ 0.25 {\rm MeV}  & 210 $\pm$ 5.5 {\rm MeV} \cite{Chen:2015tpa}&  $1.5 \times 10^{-5}$  & $\pm 6 \%$ \\
 $W^- K^{\star +}$  & 891.66 $\pm$ 0.026 {\rm MeV}  & 35.8 $\pm$ 0.3 {\rm MeV}  &   $4.3 \times 10^{-7}$   &  $\pm 4 \%$ \\
 $W^- D^+$  &  2010.26 $\pm$ 0.07 {\rm MeV}  & 61.1 $\pm$ 0.6 {\rm MeV}  &   $1.3 \times 10^{-6}$ & $\pm 6 \%$ \\
$W^- D^{\star \, +}_s$  & 2112.1  $\pm$  0.4 {\rm MeV}  & 320.5 $\pm$ 3.1 {\rm MeV} &  $3.5 \times 10^{-5}$ &  $\pm 6 \%$ \\
$W^- B^{\star +}$  &  5325.2 $\pm$0.4  {\rm MeV}  & 194.3 $\pm$ 15.8 {\rm MeV}  \cite{Gelhausen:2013wia} & $1.3 \times 10^{-5}$ & $\pm 17 \%$  \\
\midrule
$Z J/\Psi(1S)$  &  3096.916 $\pm$ 0.011 {\rm MeV} & 405 $\pm$ - {\rm MeV}  &  $3.2 \times 10^{-6}$ &  $\gtrsim 4 \%$ \\
$Z J/\Psi(2S)$  & 3686.109 $\pm$ 0.013 {\rm MeV}  & 290 $\pm$ - {\rm MeV}  &  $1.5 \times 10^{-6}$  & $\gtrsim 4 \%$ \\
\midrule
$Z \Upsilon(1S)$  & 9460.30 $\pm$ 0.26 {\rm MeV}  & 680 $\pm$ - {\rm MeV}  &  $1.7 \times 10^{-5}$  & $\gtrsim 4 \%$ \\
$Z \Upsilon(2S)$  &10023.26 $\pm$ 0.31 {\rm MeV}  & 485 $\pm$ - {\rm MeV}  &  $8.9 \times 10^{-6}$  & $\gtrsim 4 \%$ \\
$Z \Upsilon(3S)$  & 10355.2 $\pm$ 0.5 {\rm MeV}  & 420 $\pm$ - {\rm MeV}  &  $6.7 \times 10^{-6}$  & $\gtrsim 4 \%$ \\
\midrule
$Z \rho^0$  &  775.26 $\pm$ 0.25 {\rm MeV} & 216 $\pm$ 5.5 {\rm MeV} \cite{Chen:2015tpa} &  $1.4 \times 10^{-5}$ &  $\pm 6 \%$ \\
$Z \omega^0$  & 782.65 $\pm$ 0.12 {\rm MeV} & 216 $\pm$ 5.5 {\rm MeV} &  $1.6 \times 10^{-6}$ &  $\pm 6 \%$ \\
$Z \phi^0$  &  1019.461 $\pm$ 0.019 {\rm MeV} & 233 $\pm$ 5 {\rm MeV} \cite{Becirevic:2003pn} &  $4.2 \times 10^{-6}$ &  $\pm 6 \%$ \\
\bottomrule
\end{tabular}
\end{table}

\subsection[NP benchmarks for enhanced branching ratios]{NP benchmarks for enhanced branching ratios\SectionAuthor{F.~Bishara, J.~Zupan}}
\label{NP}

\def\Slash#1{{#1\!\!\!\!\!\slash}}
\def\SlashD{\,\slash\negthickspace \negmedspace\negmedspace D}

The Higgs boson couplings to the SM fermions, $f$, can differ from their SM
values due to NP. We describe the size of the modification using a
generalized $\kappa$ framework,
\begin{equation}\label{eq:Lh:CPV2}
\begin{split}
	\mathcal{L}_{\rm eff} &=- \kappa_f\frac{m_f}{v}\bar{f}fh
        - i\tilde \kappa_f\frac{m_f}{v}\bar{f}\gamma_5 f h  
	- \Big[\big(\kappa_{ff'}+i \tilde \kappa_{ff'}\big) \bar{f}_L
          f_R'h +{\rm h.c.}\Big], 
\end{split}
\end{equation}
where a sum over $f=t,b,c,s,d,u,\tau,\mu,e$ is understood. The first
two terms are flavour-diagonal with the first term CP-conserving and
the second CP-violating.  The terms in square brackets are flavour
violating. The real (imaginary) part of the coefficient is CP
conserving (violating). In the SM, we have $\kappa_f=1$ while $\tilde
\kappa_f=\kappa_{ff'}=\tilde \kappa_{ff'}=0$.

The Higgs boson production and decay strengths measured at the LHC constrain the flavour-diagonal CP-conserving Yukawa couplings to be \cite{Khachatryan:2016vau,Perez:2015aoa,Kagan:2014ila} (for future prospects see also \cite{Perez:2015lra,Brivio:2015fxa,Koenig:2015pha,Aad:2015sda,Bodwin:2014bpa,Bodwin:2013gca})
\begin{align}
\kappa_t&=1.43\pm0.23, & \kappa_b&=0.60\pm 0.18,&\kappa_c&\lesssim 6.2,\\
\kappa_s&< 65,  &\kappa_d&<1.4 \cdot 10^3, &\kappa_u&<3.0 \cdot 10^{3},\\
 \kappa_\tau&=0.88\pm0.13,  &\kappa_\mu&=0.2^{+1.2}_{-0.2},  &\kappa_e& \lesssim 630.
\end{align}
Here, $\kappa_{t,b,c,s,d,u,\tau}$ constraints have been obtained by
allowing BSM particles to modify the $h\to gg$ and $h\to \gamma\gamma$
couplings, i.e. $\delta\kappa_{g,\gamma}$ were floated, while assuming
that there are no new decay channels, $BR_{\rm BSM}=0$. The
$\kappa_{\mu,e}$ were required to be non-negative and, in addition,
when obtaining the respective bounds, $\delta\kappa_{g,\gamma}$ were
set to zero.  The upper bounds on $\kappa_{c,s,d,u}$ roughly
correspond to the size of the SM bottom Yukawa coupling and are thus
much bigger than the corresponding SM Yukawa couplings. The upper bounds can be
saturated only if one allows for large cancellations between the
contribution to fermion masses from the Higgs vev and an equally large
but opposite in sign contribution from NP. We will show that in models
of NP motivated by the hierarchy problem, the effects of NP are
generically well below these bounds.

The CP-violating flavour-diagonal Yukawa couplings, $\tilde \kappa_f$,
are well constrained from bounds on the electric dipole moments (EDMs)
\cite{Brod:2013cka,Chien:2015xha,Altmannshofer:2015qra} under the assumption of no other contribution to EDMs beyond the Higgs contributions. The flavour
violating Yukawa couplings are well constrained by the low-energy
flavour-changing neutral current measurements
\cite{Harnik:2012pb,Blankenburg:2012ex,Gorbahn:2014sha}. A notable
exception are the flavour-violating couplings involving a tau lepton. The
strongest constraints on $\kappa_{\tau\mu}, \kappa_{\mu\tau},
\kappa_{\tau e}, \kappa_{e \tau}$ are thus from direct searches of flavour-violating Higgs boson decays at
the LHC \cite{Khachatryan:2016rke, Aad:2015gha}. This is especially interesting in light of a potential hint
of a signal in $h\to \tau \mu$ \cite{Khachatryan:2015kon,Aad:2015gha}.

In the rest of the section we review the expected sizes of $\kappa_i$
in popular models of weak scale NP, some of them motivated by the
hierarchy problem.  At the end of the section we also discuss the
implications of a potential nonzero $\br(h\to \tau \mu)$ close to the
present experimental upper bound \cite{Khachatryan:2016rke}.


\subsubsection{Modified Yukawa couplings and electroweak New Physics} 
\label{ModifiedYukawas}
Tables~\ref{tab:upyukawa}, ~\ref{tab:downyukawa},
and~\ref{tab:leptyukawa}, adapted from
\cite{Bishara:2015cha,Dery:2014kxa,Dery:2013aba,Dery:2013rta,Bauer:2015kzy},
summarize the predictions for the effective Yukawa couplings,
$\kappa_f$, in the Standard Model, multi-Higgs-doublet models
(MHDM) with natural flavour conservation (NFC)~\cite{Glashow:1976nt,
  Paschos:1976ay}, the MSSM at tree level, a single Higgs doublet with
a Froggat-Nielsen mechanism (FN)~\cite{Froggatt:1978nt}, the
Giudice-Lebedev model of quark masses modified to 2HDM
(GL2)~\cite{Giudice:2008uua}, NP models with minimal flavour violation
(MFV)~\cite{D'Ambrosio:2002ex}, Randall-Sundrum models
(RS)~\cite{Randall:1999ee}, and models with a composite Higgs where
Higgs is a pseudo-Nambu-Goldstone boson (pNGB)~\cite{Dugan:1984hq,
  Georgi:1984ef, Kaplan:1983sm, Kaplan:1983fs}. The flavour-violating
couplings in the above set of NP models are collected in Tables
\ref{tab:upFVyukawa} and \ref{tab:downFVyukawa}. Next, we briefly
discuss each of the above models, and show that the effects are either
suppressed by $1/\Lambda^2$, where $\Lambda$ is the NP scale, or are
proportional to the mixing angles with the extra scalars.

\begin{table}[t]
\caption{Predictions for the flavour-diagonal up-type Yukawa couplings
  in a sample of NP models (see text for details).
}
\label{tab:upyukawa}
\begin{center}
\begin{tabular}{l  c  c  c c  }
\toprule[0.1em]
Model	& $\kappa_t$ & $\kappa_{c (u)}/\kappa_t$  & $\tilde \kappa_t/\kappa_t$ & $\tilde \kappa_{c (u)}/\kappa_t$ \\ \midrule[0.05em]
SM	& 1	& 1 & 0 & 0 \\
MFV &$1+\frac{\Re(a_uv^2+2b_u m_t^2)}{\Lambda^2}$
&$1-\frac{2\Re(b_u)m_t^2}{\Lambda^2}$
&$\frac{\Im(a_uv^2+2b_u m_t^2)}{\Lambda^2}$ & $\frac{\Im(a_u
  v^2)}{\Lambda^2} $ \\
NFC & $V_{hu}\,v/v_u$	& 1 &  0 &0 \\
MSSM	& $\cos\alpha/\sin\beta$	&1 &0  &0\\
FN & $1+\mcO\left(\frac{v^2}{\Lambda^2}\right)$ &
	$1+\mcO\left(\frac{v^2}{\Lambda^2}\right)$ &
	$\mcO\left(\frac{v^2}{\Lambda^2}\right)$ &
	$\mcO\left(\frac{v^2}{\Lambda^2}\right)$ \\
GL2 	& $\cos\alpha/\sin\beta$& $\simeq 3(7)$ & 0 & 0 \\
RS &$1-{\mathcal O}\Big(\frac{ v^2}{m_{KK}^2}\bar Y^2\Big)$&$1+{\mathcal O}\Big(\frac{ v^2}{m_{KK}^2}\bar Y^2\Big)$ &${\mathcal O}\Big(\frac{ v^2}{m_{KK}^2}\bar Y^2\Big)$ &${\mathcal O}\Big(\frac{ v^2}{m_{KK}^2}\bar Y^2\Big)$ \\
pNGB & $1+{\mathcal O}\Big(\frac{ v^2}{f^2}\Big)+{\mathcal O}\Big(y_*^2 \lambda^2 \frac{ v^2}{M_*^2}\Big)$ & $1+{\mathcal O}\Big(y_*^2 \lambda^2 \frac{ v^2}{M_*^2}\Big)$ & ${\mathcal O}\Big(y_*^2 \lambda^2 \frac{ v^2}{M_*^2}\Big)$ & ${\mathcal O}\Big(y_*^2 \lambda^2 \frac{ v^2}{M_*^2}\Big)$ \\
\bottomrule[0.1em]
\end{tabular}
\end{center}
\end{table}

\begin{table}[h]
\caption{Same as Table \ref{tab:upyukawa} but for down-type Yukawa
  couplings. 
}
\label{tab:downyukawa}
\begin{center}
\begin{tabular}{ l   c  c c c}
\toprule[0.1em]
Model	& $\kappa_b$ & $\kappa_{s(d)}/\kappa_b$ & $\tilde \kappa_b/\kappa_b$ & $\tilde \kappa_{s(d)}/\kappa_b$ \\ \midrule[0.05em]
SM	& 1 & 1 &0 &0\\
MFV & $1+\frac{\Re(a_d v^2 +2 c_d m_t^2)}{\Lambda^2}$&$1-\frac{2\Re(c_d)m_t^2}{\Lambda^2}$&$ \frac{\Im(a_d v^2+2 c_d m_t^2)}{\Lambda^2}$&$ \frac{\Im(a_d v^2+2 c_d |V_{ts(td)}|^2 m_t^2)}{\Lambda^2}$ \\
NFC & $V_{hd}\,v/v_d$	& 1 &0 &0\\
MSSM	 & $-\sin\alpha/\cos\beta$	&1 &0 &0\\
FN & $1+\mcO\left(\frac{v^2}{\Lambda^2}\right)$ &
	$1+\mcO\left(\frac{v^2}{\Lambda^2}\right)$ &
	$\mcO\left(\frac{v^2}{\Lambda^2}\right)$ &
	$\mcO\left(\frac{v^2}{\Lambda^2}\right)$ \\
GL2	& $-\sin\alpha/\cos\beta$	& $\simeq 3(5)$ & 0 & 0 \\
RS &$1-{\mathcal O}\Big(\frac{ v^2}{m_{KK}^2}\bar Y^2\Big)$&$1+{\mathcal O}\Big(\frac{ v^2}{m_{KK}^2}\bar Y^2\Big)$ &${\mathcal O}\Big(\frac{ v^2}{m_{KK}^2}\bar Y^2\Big)$ &${\mathcal O}\Big(\frac{ v^2}{m_{KK}^2}\bar Y^2\Big)$ \\
pNGB & $1+{\mathcal O}\Big(\frac{ v^2}{f^2}\Big)+{\mathcal O}\Big(y_*^2 \lambda^2 \frac{ v^2}{M_*^2}\Big)$ & $1+{\mathcal O}\Big(y_*^2 \lambda^2 \frac{ v^2}{M_*^2}\Big)$ & ${\mathcal O}\Big(y_*^2 \lambda^2 \frac{ v^2}{M_*^2}\Big)$ & ${\mathcal O}\Big(y_*^2 \lambda^2 \frac{ v^2}{M_*^2}\Big)$\\
\bottomrule[0.1em]
\end{tabular}
\end{center}
\end{table}

\begin{table}[t]\centering
\caption{Same as Table \ref{tab:upyukawa} but for lepton Yukawa
  couplings. NP effects in the pNGB model are negligible and therefore we do not report them here.
}
\label{tab:leptyukawa}
\begin{tabular}{ l   c  c c c}\toprule[0.1em]
Model	& $\kappa_\tau$ & $\kappa_{\mu(e)}/\kappa_\tau$ & $\tilde \kappa_\tau/\kappa_\tau$ & $\tilde \kappa_{\mu(e)}/\kappa_\tau$ \\ \midrule[0.05em]
SM & 1 & 1 & 0 & 0\\
MFV & $1+\frac{\Re\left(a_\ell\right) v^2}{\Lambda^2}$ &
	$1-\frac{2\Re{\left(b_\ell\right)}m_\tau^2}{\Lambda^2}$ &
	$\frac{\Im\left(a_\ell\right) v^2}{\Lambda^2}$ &
	$\frac{\Im\left(a_\ell\right) v^2}{\Lambda^2}$\\
NFC & $V_{h\ell}\,v/v_\ell$	& 1 &0 &0\\
MSSM & $-\sin\alpha/\cos\beta$ & 1 & 0 & 0 \\
FN & $1+\mathcal{O}\left(\frac{v^2}{\Lambda^2}\right)$ & $1+\mathcal{O}\left(\frac{v^2}{\Lambda^2}\right)$ & $\mathcal{O}\left(\frac{v^2}{\Lambda^2}\right)$  & $\mathcal{O}\left(\frac{v^2}{\Lambda^2}\right)$ \\
GL2 &  $-\sin\alpha/\cos\beta$ & $\simeq 3(5)$ & 0 & 0 \\
RS & $1+\mcO\left(\yvmkk\right)$ &
	$1+\mcO\left(\yvmkk\right)$ &
	$\mcO\left(\yvmkk\right)$ &
	$\mcO\left(\yvmkk\right)$ \\
\bottomrule[0.1em]
\end{tabular}
\end{table}

\begin{table}[t]
\caption{Same as Table \ref{tab:upyukawa} but for flavour-violating up-type Yukawa couplings. In the SM,
  NFC and the tree-level MSSM the Higgs Yukawa couplings are flavour
  diagonal. The CP-violating $\tilde \kappa_{ff'}$ are obtained by replacing the real part, ${\Re}$, with the imaginary part, ${\Im}$. All the other models predict a zero contribution to these flavour changing couplings.
}
\label{tab:upFVyukawa}
\begin{center}
\begin{tabular}{l  c  c  c }
\toprule[0.1em]
Model	& $\kappa_{ct (tc)}/\kappa_t$ & $\kappa_{ut (tu)}/\kappa_t$  & $\kappa_{uc (cu)}/\kappa_t$ \\ \midrule[0.05em]\vspace{0.15cm}
MFV &$ \frac{\Re\big( c_u m_b^2 V_{cb}^{(*)}\big)}{\Lambda^2}\frac{\sqrt2 m_{t(c)}}{v} $~&~ $ \frac{\Re\big( c_u m_b^2 V_{ub}^{(*)}\big)}{\Lambda^2} \frac{\sqrt2 m_{t(u)}}{v}$~&~ $ \frac{\Re\big( c_u m_b^2 V_{ub(cb)}V_{cb(ub)}^{*}\big)}{\Lambda^2} \frac{\sqrt2 m_{c(u)}}{v}$\\\vspace{0.15cm}
FN &  $\mathcal{O}\left(\frac{v m_{t(c)}}{\Lambda^2} |V_{cb}|^{\pm 1}\right)$ &
	$\mathcal{O}\left(\frac{v m_{t(u)}}{\Lambda^2} |V_{ub}|^{\pm 1}\right)$ &
	$\mathcal{O}\left(\frac{v m_{c(u)}}{\Lambda^2} |V_{us}|^{\pm 1}\right)$\\\vspace{0.15cm}
GL2	& $\epsilon (\epsilon^2)$ & $\epsilon (\epsilon^2)$ & $\epsilon^3$ \\\vspace{0.15cm}
RS & $\sim \lambda^{(-)2} \frac{m_{t(c)}}{v} \bar Y^2\frac{v^2}{m_{KK}^2} $&$\sim \lambda^{(-)3} \frac{m_{t(u)}}{v} \bar Y^2\frac{v^2}{m_{KK}^2} $&$\sim \lambda^{(-)1} \frac{m_{c(u)}}{v} \bar Y^2\frac{v^2}{m_{KK}^2} $ \\\vspace{0.15cm}
pNGB & ${\mathcal O}(y_*^2 \frac{m_t}{v}\frac{\lambda_{L (R),2} \lambda_{L(R),3}m_W^2}{M_*^2})$ & ${\mathcal O}(y_*^2 \frac{m_t}{v}\frac{\lambda_{L (R),1} \lambda_{L(R),3}m_W^2}{M_*^2})$  & ${\mathcal O}(y_*^2 \frac{m_c}{v}\frac{\lambda_{L (R),1} \lambda_{L(R),2}m_W^2}{M_*^2})$ \\
\bottomrule[0.1em]
\end{tabular}
\end{center}
\end{table}

\begin{table}[t]
\caption{Same as Table \ref{tab:upFVyukawa} but for flavour-violating down-type Yukawa couplings. 
}
\label{tab:downFVyukawa}
\begin{center}
\begin{tabular}{l  c  c  c }
\toprule[0.1em]
Model	&   $\kappa_{bs (sb)}/\kappa_b$ & $\kappa_{bd (db)}/\kappa_b$  & $\kappa_{sd (ds)}/\kappa_b$ \\ \midrule[0.05em]\vspace{0.15cm}
MFV &$\frac{\Re\big(c_d m_t^2 V_{ts}^{(*)}\big)}{\Lambda^2} \frac{\sqrt2m_{s(b)}}{v}$~&~$\frac{\Re\big(c_d m_t^2 V_{td}^{(*)}\big)}{\Lambda^2} \frac{\sqrt2 m_{d(b)}}{v}$~&~$\frac{\Re\big(c_d m_t^2 V_{ts(td)}^*V_{td(ts)}\big)}{\Lambda^2} \frac{\sqrt2 m_{s(d)}}{v}$ \\\vspace{0.15cm}
FN &  $\mathcal{O}\left(\frac{v m_{b(s)}}{\Lambda^2} |V_{cb}|^{\pm 1}\right)$ &
	$\mathcal{O}\left(\frac{v m_{b(d)}}{\Lambda^2} |V_{ub}|^{\pm 1}\right)$ &
	$\mathcal{O}\left(\frac{v m_{s(d)}}{\Lambda^2} |V_{us}|^{\pm 1}\right)$\\\vspace{0.15cm}
GL2	&$\epsilon^2 (\epsilon)$ & $\epsilon$ & $\epsilon^2(\epsilon^3)$\\\vspace{0.15cm}
RS & $\sim \lambda^{(-)2} \frac{m_{b(s)}}{v} \bar Y^2\frac{v^2}{m_{KK}^2} $&$\sim \lambda^{(-)3} \frac{m_{b(d)}}{v} \bar Y^2\frac{v^2}{m_{KK}^2} $&$\sim \lambda^{(-)1} \frac{m_{s(d)}}{v} \bar Y^2\frac{v^2}{m_{KK}^2} $ \\\vspace{0.15cm}
pNGB & ${\mathcal O}(y_*^2 \frac{m_b}{v}\frac{\lambda_{L (R),2} \lambda_{L(R),3}m_W^2}{M_*^2})$ & ${\mathcal O}(y_*^2 \frac{m_b}{v}\frac{\lambda_{L (R),1} \lambda_{L(R),3}m_W^2}{M_*^2})$  & ${\mathcal O}(y_*^2 \frac{m_s}{v}\frac{\lambda_{L (R),1} \lambda_{L(R),2}m_W^2}{M_*^2})$\\
\bottomrule[0.1em]
\end{tabular}
\end{center}
\end{table}

\begin{table}[t]
\caption{Same as Table  \ref{tab:upFVyukawa} but for flavour-violating lepton Yukawa couplings. 
}
\label{tab:leptFVyukawa}
\begin{center}
\begin{tabular}{l  c  c  c }
	\toprule[0.1em]
	Model	&   $\kappa_{\tau\mu (\mu\tau)}/\kappa_\tau$ & $\kappa_{\tau e (e\tau)}/\kappa_\tau$  & $\kappa_{\mu e (e\mu)}/\kappa_\tau$ \\ \midrule[0.05em]\vspace{0.15cm}
		FN & $\mathcal{O}\left(\frac{v m_{\mu(\tau)}}{\Lambda^2} |U_{23}|^{\mp1}\right)$ & $\mathcal{O}\left(\frac{v m_{e(\tau)}}{\Lambda^2}|U_{13}|^{\mp1}\right)$ & $\mathcal{O}\left(\frac{v m_{e(\mu)}}{\Lambda^2}|U_{12}|^{\mp1}\right)$ \\\vspace{0.15cm}
GL2	&$\epsilon^2 (\epsilon)$ & $\epsilon$ & $\epsilon^2(\epsilon^3)$\\
	RS & $\sim\sqrt{\frac{m_{\mu(\tau)}}{m_{\tau(\mu)}}}\,\yvmkk$ &
		$\sim\sqrt{\frac{m_{e(\tau)}}{m_{\tau(e)}}}\,\yvmkk$&
		$\sim\sqrt{\frac{m_{e(\mu)}}{m_{\mu(e)}}}\,\yvmkk$\\
	\bottomrule[0.1em]
\end{tabular}
\end{center}
\end{table}

\underline{\it Dimension-Six Operators with Minimal Flavour Violation (MFV).}
We first assume that there is a mass gap between the SM and NP. Integrating out the NP states leads to dimension six operators (after absorbing the modifications of kinetic terms using equations of motion~\cite{AguilarSaavedra:2009mx}), 
\begin{equation}
\begin{split}\label{eq:EFT:MFV}
	\mathcal{L}_{\rm EFT} &=  \frac{Y_u^\prime}{\Lambda^2}\bar{Q}_L H^c u_R
        (H^\dagger H)+ \frac{Y_d^\prime}{\Lambda^2} \bar{Q}_L H d_R
        (H^\dagger H)+\frac{Y_\ell^\prime}{\Lambda^2} \bar{L}_L H \ell_R
        (H^\dagger H)+\text{h.c.}\,, 
\end{split}
\end{equation}
which correct the SM Yukawa interactions, $Y_u \bar{Q}_L H^c u_R + Y_d \bar{Q}_L
        H d_R +Y_\ell \bar{L}_L
        H \ell_R$. Here $\Lambda$ is the NP scale and $H^c =
i\sigma_2H^\ast$. The fermion mass matrices and
Yukawa couplings after EWSB are 
\begin{equation}
M_f=\frac{v}{\sqrt2}\Big(Y_f +Y_f' \frac{v^2}{2 \Lambda^2}\Big)\,,
\qquad y_f=Y_f +3 Y_f' \frac{v^2}{2 \Lambda^2}\,, \qquad \quad f=u,d, \ell\,. 
\end{equation}
Because $Y_f$ and $Y_{f}'$ appear in two different combinations in
$M_f$ and in the physical Higgs Yukawa couplings, $y_f$, the two, in general, cannot be made diagonal in the
same basis and will lead to flavour-violating Higgs boson couplings.

In Tables \ref{tab:upyukawa}-\ref{tab:upFVyukawa} we show the resulting $\kappa_f$ assuming MFV, i.e.,  that the flavour breaking in the NP sector is only due to the SM
Yukawas \cite{D'Ambrosio:2002ex, Chivukula:1987py,
  Gabrielli:1994ff, Ali:1999we, Buras:2000dm, Buras:2003jf,
  Kagan:2009bn}. This gives  $Y_u^{\prime}= a_uY_u +
        b^{\phantom\dagger}_uY^{\phantom\dagger}_uY_u^\dagger
        Y^{\phantom\dagger}_u + c^{\phantom\dagger}_u
        Y^{\phantom\dagger}_dY_d^\dagger
        Y^{\phantom\dagger}_u+\cdots\,,$ and similarly for $Y_d'$ with $u\leftrightarrow d$, while
        $a_q, b_q, c_q\sim {\mathcal O}(1)$ and are in general complex. For leptons we follow \cite{Dery:2013rta} and assume that the SM $Y_{\ell}$ is the only flavour-breaking spurion even for the neutrino mass matrix (see also \cite{Cirigliano:2005ck}). Then $Y_\ell'$ and $Y_{\ell}$ are diagonal in the same basis and there are no flavour-violating couplings. The flavour-diagonal $\kappa_\ell$ are given in Table \ref{tab:leptyukawa}.
        

\underline{\it Multi-Higgs-doublet model with natural flavour conservation (NFC).}
Natural flavour conservation in multi-Higgs-doublet models is an assumption that only one doublet, $H_u$, couples to the up-type quarks, only one Higgs doublet, $H_d$, couples to the down-type quarks, and only one doublet, $H_\ell$ couples to leptons (it is possible that any of these coincide, as in the SM where $H=H_u=H_d=H_\ell$)~\cite{Glashow:1976nt, Paschos:1976ay}. The neutral scalar components of $H_i$ are $(v_i+h_i)/\sqrt2$, where
$v^2=\sum_i v_i^2$. The dynamical fields $h_{i}$ are a linear combination of the neutral
Higgs boson mass eigenstates (and include $h_u$ and $h_d$). We thus have
$h_i=V_{hi} h + \ldots$, where $V_{hi}$ are elements of the unitary
matrix $V$ that diagonalizes the neutral-Higgs boson mass terms and we only
write down the contribution of the lightest Higgs, $h$. NFC means that there are no tree-level Flavour Changing Neutral Currents (FCNCs) and no $CP$
violation in the Yukawa interactions $\kappa_{qq'}=\tilde \kappa_{qq'}=0\,, \tilde \kappa_q=0$. 

There is a universal shift in all up-quark Yukawa couplings, $\kappa_u = \kappa_c = \kappa_t = V_{hu}{v}/{v_u}$. Similarly there is a (different) universal shift in all down-quark Yukawa couplings and in all lepton Yukawa couplings, see Tables \ref{tab:upyukawa} - \ref{tab:leptyukawa}.

%


\underline{\it Higgs sector of the MSSM at tree level.}
The MSSM tree-level Higgs potential and the couplings to quarks are
the same as in the type-II two-Higgs-doublet model, see, e.g.,
\cite{Haber:1984rc}. This is an example of a 2HDM with natural flavour
conservation in which $v_u=\sin\beta\, v$, $v_d=\cos\beta\,
v$. The mixing of $h_{u,d}$ into the Higgs boson mass-eigenstates $h$ and
$H$ is given by $h_u=\cos \alpha h+\sin\alpha H$, $h_d = -\sin\alpha h +\cos\alpha H$,
where $h$ is the observed SM-like Higgs. The up-quark Yukawa couplings are rescaled universally, 
$	\kappa_u = \kappa_c = \kappa_t=\cos\alpha/\sin\beta$, and similarly the down-quark Yukawas, 
	$\kappa_d = \kappa_s = \kappa_b = -\sin\alpha/\cos\beta$.
The flavour-violating and CP-violating Yukawas are zero\footnote{Note that beyond the tree level, in fine-tuned regions of parameter space the loops of sfermions and gauginos can lead to substantial corrections to these expressions \cite{Aloni:2015wvn}.}. In Tables \ref{tab:upyukawa}-\ref{tab:leptyukawa} we limit ourselves to the tree-level expectations, which are a good approximation for a large part of the MSSM parameter space.

In the alignment limit, $\beta-\alpha=\pi/2$ \cite{Gunion:2002zf,Carena:2013ooa,Dev:2014yca,Carena:2014nza,Dev:2015bta,Haber:2015pua,Carena:2015moc}, the Yukawa couplings tend toward their
SM value, $\kappa_i=1$.  The global fits to Higgs data in type-II 2HDM
already constrain $\beta-\alpha$ to be not to far from $\pi/2$~\cite{Carmi:2012in,
  Falkowski:2013dza, Grinstein:2013npa} so that the couplings of the light Higgs are also constrained to be close to their SM values. 
  Note that the decoupling limit of the 2HDM, where the heavy Higgs bosons become much heavier than the SM Higgs, implies the alignment limit while the reverse is not necessarily true \cite{Carena:2013ooa}.


\underline{\it A single Higgs doublet with Froggatt-Nielsen mechanism (FN).}
The Froggatt-Nielsen~\cite{Froggatt:1978nt} mechanism provides a simple explanation of the size and hierarchy of the SM Yukawa couplings. In the simplest realization this is achieved by a $U(1)_H$ horizontal symmetry under which different generations of fermions carry different charges. The $U(1)_H$ is broken by a spurion, $\epsilon_H$.
The entries of the SM Yukawa matrix are then parametrically suppressed by powers of $\epsilon_H$ as, for example, in the lepton sector
\begin{equation}
\big(Y_\ell\big)_{ij}\sim \epsilon_H^{H(L_i)-H(e_j)},
\end{equation}
where $H(e,L)$ are the FN charges of the right- and left-handed charged lepton, respectively. The dimension 6 operators in \eqref{eq:EFT:MFV} due to electroweak NP have similar flavour suppression, $\big(Y_\ell'\big)_{ij}\sim \epsilon_H^{H(e_j)-H(L_i)} v^2/\Lambda^2$ \cite{Dery:2013rta,Dery:2014kxa}. After rotating to the mass eigenbasis, the lepton masses and mixing angles are then given by~\cite{Leurer:1993gy,Grossman:1995hk}
\begin{equation}
m_{\ell_i}/v \sim \epsilon_H^{|H(L_i)-H(e_i)|},\quad
	|U_{ij}|\sim \epsilon_H^{|H(L_i)-H(L_j)|},
\end{equation}
giving the Higgs Yukawa couplings in Tables~\ref{tab:leptyukawa}~and~\ref{tab:leptFVyukawa} in the row labelled `FN' \cite{Dery:2014kxa}.
Similarly for the quarks, after rotating to the mass eigenbasis, the masses and the mixings are given by~\cite{Leurer:1993gy} 
\begin{equation}
m_{u_i(d_i)}/v \sim \epsilon_H^{|H(Q_i)-H(u_i(d_i))|},\quad
|V_{ij}|\sim \epsilon_H^{|H(Q_i)-H(Q_j)|},
\end{equation}
where $V$ is the Cabibbo-Kobayashi-Maskawa (CKM) mixing matrix and $H(u,d,Q)$ are the FN charges of the right-handed up and down and the left-handed quark fields, respectively.

\underline{\it Higgs-dependent Yukawa couplings (GL2)} In the model of
quark masses introduced by Giudice and Lebedev~\cite{Giudice:2008uua}, the quark masses, apart from the top
mass, are small because they arise from higher dimensional
operators. The original GL proposal is ruled out by data, while the
straightforward modification to a 2HDM (GL2) is
\begin{equation}\label{eq:Higgs-dep}
\begin{split}
{\cal L}_{f}&=c_{ij}^{u} \bigg( \frac{H_1^\dagger H_1}{M^2}
  \bigg)^{n_{ij}^{u}} \, \bar Q_{L,i} u_{R,j} H_1 + c_{ij}^{d} \bigg( \frac{H_1^\dagger H_1}{M^2}
  \bigg)^{n_{ij}^{d}} \, \bar Q_{L,i} d_{R,j} H_2 + \phantom{c}\\
  &\qquad c_{ij}^\ell \bigg( \frac{H_1^\dagger H_1}{M^2}
  \bigg)^{n_{ij}^{\ell}} \, \bar L_{L,i} e_{R,j} H_2 +\text{h.c.}\,,
\end{split}
\end{equation}
where $M$ is the mass scale of the mediators.  In the original GL
model $H_2$ is identified with the SM Higgs, $H_2=H$, while
$H_1=H^c$. Taking $c_{ij}^{u,d}\sim {\mathcal
  O}(1)$, the ansatz $n_{ij}^{u,d}=a_i +b_j^{u,d}$ with $a=(1,1,0)$,
$b^d=(2,1,1)$, and $b^u=(2,0,0)$ then reproduces the hierarchies of
the observed quark masses and mixing angles for $\epsilon \equiv
v^2/M^2 \approx 1/60$. The Yukawa couplings are of the form $
y_{ij}^{u,d} = (2n_{ij}^{u,d} + 1) (y_{ij}^{u,d})_\text{SM}$. The SM
Yukawas are diagonal in the same basis as the quark masses,
while the $y_{ij}^{u,d}$ are not.  Because the bottom Yukawa is
largely enhanced, $\kappa_b \simeq 3$, this simplest version of the GL
model is already excluded by the Higgs data. Its modification, GL2, is
still viable, though \cite{Bishara:2015cha}. For
$v_1/v_2=\tan\beta\sim 1/\epsilon$ one can use the same ansatz for
$n_{ij}^{u,d}$ as before, modifying only $b^d$, so that $b^d=(1,0,0)$,
with the results shown in Tables
\ref{tab:upyukawa}-\ref{tab:leptFVyukawa}. For leptons we use the same
scalings as for right-handed quarks. Note that the $H_1^\dagger H_1$
is both a gauge singlet and a flavour singlet. From symmetry point of
view it is easier to build flavour models, if $H_1 H_2$ acts as a
spurion in \eqref{eq:Higgs-dep}, instead of $H_1^\dagger H_1$. This
possibility is severely constrained phenomenologically, though
\cite{Bauer:2015fxa,Bauer:2015kzy}.

\underline{\it Randall-Sundrum models (RS).}
The Randall-Sundrum warped extra-dimensional model
has been proposed to address the hierarchy
problem and simultaneously explain the hierarchy of the SM fermion
masses 
~\cite{Randall:1999ee, Gherghetta:2000qt, Grossman:1999ra,
Huber:2000ie, Huber:2003tu}. 
Integrating out the
Kaluza-Klein (KK) modes of mass $m_{KK}$, and working in the limit of a brane-localized
Higgs, keeping only terms of leading order in
$v^2/m_{KK}^2$, the SM quark mass matrices are given by~\cite{Azatov:2009na} (see
also~\cite{Casagrande:2008hr, Bauer:2009cf, Malm:2013jia,
  Archer:2014jca, Blanke:2008zb, Blanke:2008yr, Albrecht:2009xr,
  Agashe:2006wa, Agashe:2014jca}, and Ref.~\cite{Dillon:2014zea} for a
bulk Higgs scenario)
\begin{equation}\label{eq:RS:Mdu}
M^{d(u)}_{ij}=\big[F_q Y_{1(2)}^{5D}F_{d(u)}\big]_{ij} v\,.
\end{equation}
The $F_{q,u,d}$ are $3\times 3$ matrices of fermion wave-function
overlaps with the Higgs and are diagonal and hierarchical.
Assuming
flavour anarchy, the 5D Yukawa matrices,
$Y_{1,2}^{5D}$, are general $3\times 3$ complex matrices with
$\bar Y\sim {\mathcal O}(1)$ entries, but usually  
$\bar Y \lesssim 4$, see, e.g., \cite{Archer:2014jca}.  
At leading order in $v^2/m_{KK}^2$ the Higgs Yukawas are aligned
with the quark masses, i.e.,
$M_{u,d}=y_{u,d} v/{\sqrt2}+{\mathcal O}(v^2/m_{KK}^2)$.
The misalignments 
are generated by tree-level KK quark exchanges, giving
\begin{equation}\label{eq:misalignment}
\big[y_{u(d)}\big]_{ij}-\frac{\sqrt2}{v}\big[M_{u,d}\big]_{ij}\sim -
\frac{1}{3}F_{q_i} \bar Y^3 F_{u_j(d_j)}\frac{v^2}{m_{KK}^2}\,.
\end{equation}

For the charged leptons, there are two choices for generating the hierarchy in the masses~\cite{Azatov:2009na}. If left- and right-handed fermion profiles are both hierarchical (and taken to be similar) then the misalignment between the masses and Yukawas is $\sim \sqrt{{m_i m_j}/{v^2}} \times \mcO\big(\bar Y^2 v^2/m_{KK}^2\big)$.
If only the right-handed profiles are hierarchical the misalignment is given by (see also Tables~\ref{tab:leptyukawa}~and~\ref{tab:leptFVyukawa})
\begin{equation}\label{eq:lept-misalignment}
\big[y_\ell\big]_{ij}-\frac{\sqrt2}{v}\big[M_\ell\big]_{ij}\sim -
\frac{1}{3}\yvmkk\,\frac{m_j^\ell}{v}\,.
\end{equation}

 The Higgs
mediated FCNCs are suppressed by the same zero-mode wave-function
overlaps that also suppress the quark masses, \eqref{eq:RS:Mdu}, giving rise to the RS
GIM mechanism~\cite{Cacciapaglia:2007fw, Agashe:2004cp,
  Agashe:2004ay}. Using the fact that the CKM matrix elements are given by $V_{ij}\sim
F_{q_i}/F_{q_j}$ for $i<j$, Eq.~\eqref{eq:misalignment}, one can rewrite 
the $\kappa_i$ as in
Tables~\ref{tab:upyukawa}-\ref{tab:downFVyukawa}. The numerical
analysis of Ref.~\cite{Azatov:2009na} found that for diagonal Yukawas typically $\kappa_i<1$, with deviations in
$\kappa_{t(b)}$ up to $30\%(15\%)$, and in $\kappa_{s,c (u,d)}$ up
to $\sim 5\%(1\%)$.
For the charged leptons one obtains deviations in $\kappa_{\tau\mu(\mu\tau)}\sim 1(5)\times 10^{-5}$~\cite{Azatov:2009na}.
These estimates were obtained fixing the mass of the first KK gluon excitation to
$3.7$~TeV, above the present ATLAS bound \cite{Aad:2015fna}.

\underline{\it Composite pseudo-Goldstone Higgs (pNGB).}
Finally, we assume that the Higgs is a
pseudo-Goldstone boson arising from the spontaneous breaking of a
global symmetry in a strongly coupled sector, and  couples to the composite
sector with a typical coupling $y_*$ \cite{Dugan:1984hq,
  Georgi:1984ef, Kaplan:1983sm, Kaplan:1983fs} (for a review, see~\cite{Panico:2015jxa}).
Assuming partial compositeness, the SM fermions couple linearly to composite operators
$O_{L,R}$,
$\lambda_{L,i}^q \bar Q_{L,i} O_R^i+\lambda_{R,j}^u \bar u_{R,j}
O_L^j+h.c. \,,$
where $i,j$ are flavour indices~\cite{Kaplan:1991dc}. This is the 4D dual of fermion
mass generation in 5D RS models.  The SM masses and Yukawa
couplings arise from expanding the two-point functions of the
$O_{L,R}$ operators in powers of the Higgs field~\cite{Agashe:2009di}.

The new ingredient compared to the EFT analysis in \eqref{eq:EFT:MFV}
is that the shift symmetry due to the pNGB nature of the Higgs dictates 
the form of the higher-dimensional operators. The flavour
structure and the composite Higgs coset structure completely factorize
if the SM fields couple to only one composite operator. The general
decomposition of Higgs boson couplings then becomes \cite{Agashe:2009di}
(see also \cite{Gillioz:2012se,Delaunay:2013iia,Azatov:2014lha})
\begin{equation}\label{eq:comp}
	Y_u \bar{Q}_L H u_R + Y_u^\prime\bar{Q}_L H u_R
        \frac{(H^\dagger H)}{\Lambda^2}+\ldots \quad \to \quad
        c_{ij}^u \, P(h/f) \, \bar{Q}_L^i H u_R^j \,,
\end{equation}
and similarly for the down quarks. Here $f\gtrsim v$ is the equivalent
of the pion decay constant, while $P(h/f)=a_0+a_2 (H^\dagger H/f^2)+\ldots
$ is an analytic function whose form is fixed by the  pattern of the
spontaneous breaking and the embedding of the SM fields in the global
symmetry of the strongly coupled sector. In \eqref{eq:comp} the flavour
structure of $Y_u$ and $Y_u'$ 
is the same.
The resulting 
corrections to the quark Yukawa couplings 
are
therefore strictly diagonal, 
\begin{equation}\label{eq:kappaq:estimate}
\kappa_q\sim 1+{\mathcal O}\big({v^2}/{f^2}\big).
\end{equation}
For example, for the models based on the breaking of $SO(5)$ to
$SO(4)$, the diagonal Yukawa couplings can be written
as
$\kappa_q=(1+2m-(1+2m+n)(v/f)^2)/{\sqrt{1-(v/f)^2}}$,
where $n,m$ are positive integers~\cite{Pomarol:2012qf}. The MCHM4 model corresponds to
$m=n=0$, while MCHM5 is given by $m=0,n=1$.

The flavour-violating contributions to the quark Yukawa couplings
arise only from corrections to the quark kinetic terms~\cite{Agashe:2009di},
\begin{equation}
\bar q_L i \SlashD q_L \frac{H^\dagger H}{\Lambda^2}, \,\,
\bar u_R i \SlashD u_R \frac{H^\dagger H}{\Lambda^2},
\dots\,,
\end{equation}
due to the exchanges of composite vector resonances
with typical mass $M_* \sim \Lambda$. After using the equations of
motion 
these give (neglecting relative ${\mathcal O}(1)$ contributions
in the sum)~\cite{Agashe:2009di, Azatov:2014lha, Delaunay:2013pja},
\begin{equation}
\kappa_{ij}^u\sim 2 y_*^2 \frac{v^2}{M_*^2}
\Big(\lambda_{L,i}^q\lambda_{L,j}^q \frac{m_{u_j}}{v}
+\lambda_{R,i}^u\lambda_{R,j}^u \frac{m_{u_i}}{v}\Big)\,,
\end{equation}
and similarly for the down quarks. If the strong sector is CP
violating, then $\tilde \kappa_{ij}^{u,d}\sim \kappa_{ij}^{u,d}$.

The exchange of composite vector resonances also contributes to the
flavour-diagonal Yukawa couplings, shifting the
estimate~\eqref{eq:kappaq:estimate} by 
$\Delta \kappa_{q_i}\sim 2 y_*^2 \frac{v^2}{M_*^2}
\Big[\big(\lambda_{L,i}^q\big)^2+ \big(\lambda_{R,i}^u\big)^2\Big] \,$.
This shift can be large for the quarks with a large composite
component if the Higgs is strongly coupled to the vector resonances,
$y_*\sim 4\pi$, and these resonances are relatively light, $M_*\sim
4\pi v\sim 3$ TeV. The left-handed top and bottom, as well as the
right-handed top, are expected to be composite, explaining the large
top mass (i.e., $\lambda_{L,3}^q\sim \lambda_{R,3}^u\sim 1$). In the
anarchic flavour scenario, one expects the remaining quarks to be
mostly elementary (so the remaining $\lambda_i \muchless 1$). 
If
there is some underlying flavour alignment, it is also possible that
the light quarks are composite. This is most easily achieved in the
right-handed sector~\cite{Redi:2011zi, Redi:2012uj,
  Delaunay:2013iia}.

In the case of the lepton sector, if we assume that there are no
hierarchies in the composite sector~\cite{Redi:2013pga} (see also~\cite{Csaki:2008qq,delAguila:2010vg,Hagedorn:2011un,Hagedorn:2011pw}), then the NP
effects in the flavour diagonal and off-diagonal Yukawas are
negligible. For this reason, we do not report them in Tabs. \ref{tab:leptyukawa} and \ref{tab:leptFVyukawa}.

\subsubsection{Models with large flavour-violating Higgs boson decays}
In Section~\ref{ModifiedYukawas} we explored the modifications of
Higgs Yukawa couplings in a number of popular NP models, some of which
are motivated by the hierarchy problem. The deviations from the SM
predictions share several common features. If the scale of NP is well
above the weak scale, $\Lambda\gg v$, the deviations from the SM
expectations become increasingly small.

Flavour-violating Higgs Yukawa couplings to quarks are significantly
constrained by meson mixing constraints
\cite{Harnik:2012pb,Blankenburg:2012ex}. If the tree-level Higgs exchange 
is the dominant NP contribution, the constraints from $D-\bar
D$, $B_d-\bar B_d$, $B_s-\bar B_s$ and $K-\bar K$ mixing translate to
$\br(h\to c\bar u +u \bar c)< 3.7 \times 10^{-6}$, $\br(h\to b\bar d
+d \bar b)< 1.7 \times 10^{-5}$, $\br(h\to b\bar s +s \bar b)< 1.3
\times 10^{-3}$, and $\br(h\to s\bar d +d \bar s)< 4.2 \times 10^{-7}$
at 95\% C.L., respectively. These branching ratios are too small to be
experimentally searched for, with the possible exception of $h\to \bar
b s+ b\bar s$. 
 The indirect bounds can be relaxed to per-cent-level branching ratios
only if there is substantial cancellation between the flavour-violating
Higgs exchange and other NP contributions to the mixing amplitude.

The flavour-violating couplings of the Higgs involving top quarks,
$\kappa_{tc,tu}$ and $\kappa_{ct,ut}$, are more loosely constrained
experimentally. The most important constraints come from direct
searches for the flavour-violating top decays at the LHC, $t\to h c,
hu$, giving $\sqrt{\kappa_{tc}^2+\kappa_{ct}^2}<0.13$ and
$\sqrt{\kappa_{tu}^2+\kappa_{ut}^2}<0.13$
\cite{Aad:2015pja,CMS:2015qhe,Khachatryan:2016atv}.
The $D-\bar D$ mixing also constrains combinations of the
couplings, $|\kappa_{ut}\kappa_{ct}|, |\kappa_{tu}\kappa_{tc}|< 7.6
\times 10^{-3}$, $|\kappa_{tu}\kappa_{ct}|, |\kappa_{ut}\kappa_{tc}|<
2.2 \times 10^{-3}$,
$|\kappa_{ut}\kappa_{ct}\kappa_{tu}\kappa_{tc}|^{1/2}< 0.9 \times
10^{-3}$ \cite{Harnik:2012pb}.

Similarly, the flavour-violating Yukawa couplings involving the $\tau$
lepton are relatively loosely
constrained. 
 The most stringent
constraints come from direct searches for $h\to \tau\mu,\tau e$ at the LHC
\cite{Khachatryan:2015kon,Aad:2015gha,Khachatryan:2016rke}, giving
$\sqrt{\kappa_{\tau\mu}^2+\kappa_{\mu\tau}^2}<3.6\times 10^{-3}$
\cite{Khachatryan:2015kon}, $\sqrt{\kappa_{\tau e}^2+\kappa_{e
    \tau}^2}<2.4\times 10^{-3}$ \cite{Khachatryan:2016rke}. These numbers should be compared to the indirect constraints from searches for $\tau\to \mu \gamma$ and $\tau\to e\gamma$: $\sqrt{\kappa_{\tau\mu}^2+\kappa_{\mu\tau}^2}<0.016$, $\sqrt{\kappa_{\tau e}^2+\kappa_{e
    \tau}^2}<0.014$ \cite{Harnik:2012pb,Blankenburg:2012ex}.
    
In contrast, the most stringent constraint on $\kappa_{\mu e}$ is due
to $\mu\to e \gamma$, giving $\sqrt{\kappa_{\mu e}^2+\kappa_{e
    \mu}^2}<3.6\times 10^{-6}$ \cite{Harnik:2012pb} to be compared
with the bound from the direct search $h\to e\mu$ which is two orders
of magnitude less stringent, $\sqrt{\kappa_{\tau e}^2+\kappa_{e
    \tau}^2}<5.4\times 10^{-4}$ \cite{Khachatryan:2016rke}.

The bounds from indirect constraints also limit the relative sizes of
$h\to \tau \mu$ and $h\to \tau e$ branching ratios. Since $\mu\to
e\gamma$ and $\mu\to e$ conversion provide complementary information
on the relevant Yukawa couplings, it is possible to write a relation
\cite{Dorsner:2015mja}
\beq\label{eq:relations:htaumu:taue}
\br(h\to \tau\mu) \times \br(h\to \tau e)=7.95 \times 10^{-10}\Big(\frac{\br(\mu\to e\gamma)}{10^{-13}}\Big)+3.15 \times 10^{-4} \Big(\frac{\br(\mu\to e)_{\rm Au}}{10^{-13}}\Big),
\eeq
where the present experimental limits are $\br(\mu \to e \gamma)<5.7
\times 10^{-13}$ \cite{Adam:2013mnn} and $\br(\mu \to e)_{\rm Au}<7
\times 10^{-13}$ \cite{Bertl:2006up}. This relation will become
phenomenologically interesting once the sensitivity of the $\mu\to e$
conversion experiments is improved.  Using the central value of the
experimental hint for nonzero $h\to \tau \mu$, $\Br(h\to \tau \mu)=
(0.84 \pm0.38)\%$ \cite{Khachatryan:2015kon},
Eq.~\eqref{eq:relations:htaumu:taue} gives at present only a very
loose bound, $\br(h\to \tau e)<26\%$ compared to the direct constraint
$\Br(h\to \tau e)< 0.69\%$.

An interesting question is what kind of models could explain the
relatively large $\Br(h\to \tau \mu)= (0.84 \pm0.38)\%$, if this hint
were to become statistically significant. In terms of Yukawa couplings
this hint is $\sqrt{|\kappa_{\tau
    \mu}|^2+|\kappa_{\mu\tau}|^2}=(2.6\pm 0.6 )\cdot 10^{-3}$. For
$\kappa_{\mu\tau}\sim \kappa_{\tau\mu}$ this measurement would imply
that the product of off-diagonal Yukawa couplings is of the same order
of magnitude as the product of the diagonal ones,
$\kappa_{\mu\tau}\sim \sqrt{m_\tau m_\mu}/v$. In particular, the
off-diagonal couplings should not be additionally suppressed by powers
of $v^2/\Lambda^2$, unlike what we found for the models considered in
Section \ref{ModifiedYukawas}.

A potential obstacle to any viable model of large $\kappa_{\tau\mu}$
couplings is that the same type of NP that generates $\kappa_{\tau
  \mu, \mu\tau}$ will also generate $\tau\to \mu \gamma$. Let us take for
instance a model for which $\kappa_{\tau\mu}$ is generated at one-loop, with NP
particles running in the loop. Attaching a photon anywhere in the loop
will then produce the $\tau\to \mu\gamma$ transition. It is
instructive to attempt a naive dimensional analysis (NDA) estimate of
this transition, assuming that the Higgs is the only source of
electroweak symmetry breaking generating the charged lepton masses,
i.e. that in the limit $v\to 0$ also $m_\tau, m_\mu\to 0$. The NDA
then gives a $\br(\tau\to \mu\gamma)$ that is four orders above the
experimental bound, if one is to explain the present central value of
$\br(h\to \tau \mu)$ \cite{Altmannshofer:2015esa}. One is therefore
led to conclude that an observation of $h\to\tau\mu$ at the present level
would strongly suggest that there is an additional source of charged
lepton masses (another option is that the NDA estimate is badly violated,
for instance by large slepton mass hierarchies in the MSSM
\cite{Aloni:2015wvn} or by ad-hoc cancellations
\cite{Dorsner:2015mja}).

A well-motivated possibility is that the Higgs is predominantly responsible for the masses of the third generation charged fermions, while the masses of the first two generations fermions are generated from a new source of electroweak symmetry breaking \cite{Altmannshofer:2015esa,Ghosh:2015gpa}. This second contribution is a subleading correction, explaining the approximate $U(2)$ symmetry of the charged fermion masses (i.e. that $m_{\mu,e}\muchless m_\tau$) \cite{Ghosh:2015gpa}. The lepton mass matrix is then of the form
\beq
{\cal M}^\ell={\cal M}_0^{\ell}+\Delta {\cal M}^\ell, \label{eq:M:decomp}
\eeq
where a rank 1 matrix ${\cal M}^\ell_0$ is due to the vev of a scalar doublet $\Phi_1$ (the primary component of the Higgs), and accounts for the bulk of the third generation mass, $m_\tau$. The matrix $\Delta {\cal M}_\ell$ is due to an additional source of EWSB, can be rank 2 or 3, and accounts for first and second generation masses, $m_{e}$ and $m_\mu$. Note that the above mass matrix ${\cal M}^\ell$ in general does not imply anything about the texture of neutrino masses, which can still come from the see-saw at or close to the GUT scale. 

The simplest example is a 2HDM where ${\cal M}_0^{\ell}$ is due to
$\Phi_1$ and $\Delta {\cal M}^\ell$ due to an extra doublet $\Phi_2$
\cite{Altmannshofer:2015esa,Ghosh:2015gpa} (for other considerations
in the context of 2HDM see
\cite{Crivellin:2015lwa,Crivellin:2015mga,Campos:2014zaa,Varzielas:2015iva}). This
version of 2HDM is quite different from type II 2HDM covered in
Section \ref{ModifiedYukawas}.

Note that the off-diagonal entries in $\Delta {\cal
  M}^\ell$ can lead to the current central value for $\Br(h\to \tau\mu)$ without violating
the $\tau\to \mu\gamma$ bound (the above mentioned NDA scaling fails
since one always pays an extra charged lepton Yukawa insertion). Scans
over a reasonable range of $\Delta {\cal M}^\ell$ entries predict that
$|\kappa_\mu/\kappa_\tau|<1$ is preferred, with $\kappa_\tau-1\sim
{\mathcal O}({\rm few}~10\%)$ typical \cite{Altmannshofer:2015esa}.

Another realization of (\ref{eq:M:decomp}) are models in which $\Delta {\cal M}^\ell$ comes
from strong dynamics, and is due to a condensate of a new set of
strongly interacting fermions that also carry electroweak quantum
numbers \cite{Altmannshofer:2015esa,Ghosh:2015gpa}. In that case the
$\tau\to \mu\gamma$ constraint is avoided if the NP scale is $\gtrsim
8$~TeV. In both examples it is straightforward to construct the correct
texture of the mass matrix using flavour models, e.g. using the FN
mechanism, so that a large enough $\br(h\to \tau\mu)$ is obtained
\cite{Altmannshofer:2015esa}. It is also possible to apply the same
principle to the quark sector, so that the bottom and top quark masses
are due to the Higgs, while the light quark masses are due to a new
source of EWSB. In that case a deviation in $B_s\to \mu\mu$ , as well
as signals in $\Br(B_s\to \tau\mu),\, \Br(B\to K^{(*)}\tau \mu)\sim
{\mathcal O}(10^{-7})$ are expected \cite{Altmannshofer:2015esa}.

\subsection[Experimental status and prospects]{Experimental status and prospects\SectionAuthor{L.~Caminada, K.~Nikolopoulos}}
\label{ExperimentalProspects}

In the SM, the charged fermions obtain mass ($m_f$) via a direct
Yukawa coupling to the Higgs field, which, following electroweak
symmetry breaking, results in fermion mass terms and a fermion-Higgs boson 
interaction with a strength proportional to $m_f$. This is the
most economic way to generate fermion masses, but it is not imposed by any fundamental symmetry principle. Several
viable models of physics beyond the SM predict modifications of these
couplings, as discussed in Section~\ref{NP}.

To date, the only direct experimental evidence of the Yukawa mechanism
for fermion mass generation is the Higgs boson coupling to third
generation fermions. The ATLAS and CMS Collaborations have reported
evidence for the observation of the Higgs boson decays to a pair of
$\tau$-leptons, $h\to\tau^+\tau^-$, in line with the SM
expectation~\cite{Aad:2015vsa,Chatrchyan:2014nva}.
For Higgs boson decays to b-quark pairs, $h\to b\bar{b}$, the Tevatron
experiments have reported evidence~\cite{Aaltonen:2012qt},
complemented by indications at the
LHC~\cite{Aad:2014xzb,Chatrchyan:2013zna}.
CMS has also reported the strong evidence for the direct coupling of the 125 GeV Higgs boson to down-type fermions (combining Higgs boson decays to pair of tau leptons and b-quark)~\cite{Chatrchyan:2014vua}.
Higgs boson decays to top-quark pairs are kinematically forbidden,
thus the associated production of a Higgs boson with a top-quark pair,
$t\bar{t}h$, is exploited: several final states have been
analysed~\cite{Aad:2015gba,Khachatryan:2014qaa}, and the sensitivity
begins to approach the SM expectation.
The muon Yukawa coupling is probed through the $h\to\mu^+\mu^-$ decay,
yielding 95\% confidence level upper limits of 7-10 times the
predicted SM rate~\cite{Aad:2014xva,Khachatryan:2014aep}. Higgs boson
decays to an electron-positron pair have been also searched
for~\cite{Khachatryan:2014aep}, although the resulting bound is many
orders of magnitude larger than the predicted SM value, which is
anticipated to remain out of reach even at the HL-LHC
\cite{Altmannshofer:2015qra}.
Flavour-violating Higgs boson interactions have been searched for in
top-quark decays~\cite{Aad:2015pja,Khachatryan:2014jya} and via $h\to e^\pm\mu^\mp$, $h\to \tau^\pm e^\mp$
$h\to\tau^\pm\mu^\mp$~\cite{Aad:2015gha,Khachatryan:2015kon}, yielding
an intriguing excess in the latter.

The Yukawa couplings of the first and second generation
quarks are among the most challenging of the SM couplings to test experimentally.
Indirect
constraints are sparse, with some specifically derived for
flavour-violating interactions via meson-anti-meson
mixing~\cite{Harnik:2012pb,Blankenburg:2012ex}.
These Yukawa couplings have been, generally, considered beyond the reach of
the LHC experiments.  Recently, the possibility to probe these
couplings using rare exclusive decays of the Higgs boson to a vector
meson and a photon has resurfaced\footnote{Such rare decays were
considered as a discovery channel for a 
light Higgs boson~\cite{Bander:1978br} but then abandoned.}, initially for the
charm quark, and subsequently for
all the light quarks, including
flavour-violation (see Section~\ref{SMphoton} for details).  
 The rare exclusive decays
of a Higgs boson to a meson and massive vector boson, $h\to MV$, and their
potential in clarifying the nature of the Higgs boson have also been
considered (see Section~\ref{SMmassivegauge}).
These developments prompted significant interest, with theoretical
investigations towards precise estimates of the relevant SM
predictions, as detailed in
Sections~\ref{SMphoton}--\ref{SMmassivegauge}, and phenomenological
sensitivity studies, e.g. Ref.~\cite{Perez:2015lra}.

Currently, these exclusive decays constitute the only available method
to probe the Higgs boson interactions with first and second generation
quarks. Recently, charm-quark specific suggestions regarding the
feasibility of the $h\to c\bar{c}$ channel~\cite{Delaunay:2013pja},
using charm-tagging techniques~\cite{Aad:2015gna}, and of the
associated production of a Higgs boson with a charm
quark~\cite{Brivio:2015fxa}, were made and are under investigation.

Similar exclusive decays of the $W^\pm$ and $Z$
bosons have also attracted
interest~\cite{Grossmann:2015lea,Huang:2014cxa,Achasov:2012zzb},
offering a physics programme in precision quantum chromodynamics
(QCD), electroweak physics, and physics beyond the SM.

Using 20.3~fb$^{-1}$ of 8\,TeV proton-proton collision data, the ATLAS
Collaboration has performed a search for Higgs and $Z$ boson decays to
$J/\psi\,\gamma$ or and $\Upsilon(nS)\,\gamma$
($n=1,2,3$)~\cite{Aad:2015sda}.
No significant excess has been observed and 95\% confidence level upper
limits where placed on the respective branching ratios.  In the
$J/\psi\,\gamma$ final state the limits are $1.5\times10^{-3}$ and
$2.6\times10^{-6}$ for the Higgs and $Z$ boson decays, respectively,
while in the $\Upsilon(1S,2S,3S)\,\gamma$ final states the limits are
$(1.3,1.9,1.3)\times10^{-3}$ and $(3.4,6.5,5.4)\times10^{-6}$,
respectively. The CMS Collaboration has placed a 95\% C.L. upper limit
of $1.5\times10^{-3}$ on the $h\to J/\psi\,\gamma$ branching
ratio~\cite{Khachatryan:2015lga}. 
Using 2.7\,fb$^{-1}$ of 13\UTeV\ proton-proton collision data, the ATLAS Collaboration 
has recently performed a search for Higgs and Z boson decays to $\phi\,\gamma$, 
that yielded 95\% confidence level upper limits of $1.4 \times 10^{-3}$ and $8.3 \times 10^{-6}$ 
on the respective branching ratios~\cite{Aaboud:2016rug}. 
In all cases, an SM production rate for the observed Higgs boson is assumed. 
Currently, no other direct experimental constraint on these decays is available.

Looking to the future, the ATLAS Collaboration estimated the expected
sensitivity for Higgs and $Z$ boson decays to a $J/\psi$ and a photon,
assuming up to 3000~fb$^{-1}$ of data collected with the ATLAS
detector at the centre-of-mass energy of
14\,TeV~\cite{ATL-PHYS-PUB-2015-043}, during the operation of the High
Luminosity LHC. The expected sensitivity for the $h\to J/\psi\,\gamma$
branching ratio, assuming 300 and 3000~fb$^{-1}$ at 14 TeV, is
$153\times 10^{-6}$ and $44\times 10^{-6}$,
respectively~\cite{ATL-PHYS-PUB-2015-043}. The corresponding
sensitivities for the $Z\to J/\psi\,\gamma$ branching ratios are
$7\times 10^{-7}$ and $4.4\times 10^{-7}$,
respectively~\cite{ATL-PHYS-PUB-2015-043}.  In this analysis, the same
overall detector performance as in LHC Run~1 is assumed, while an
analysis optimization has been performed and a multivariate
discriminant using the same kinematic information as the published
analysis~\cite{Aad:2015sda} has been introduced.  As the search
sensitivity approaches the SM expectation for the $h\to
J/\psi\,\gamma$ branching ratio, the contribution from $h\to
\mu\mu\gamma$ decays, with a non-resonant dimuon pair, needs to be
included. These can be separated statistically from the $h\to
J/\psi\,\gamma$ signal using dimuon mass information.

\begin{wrapfigure}{R}{0.5\textwidth}
 \centering
\vspace{-0.7cm}
 \includegraphics[width=0.48\textwidth]{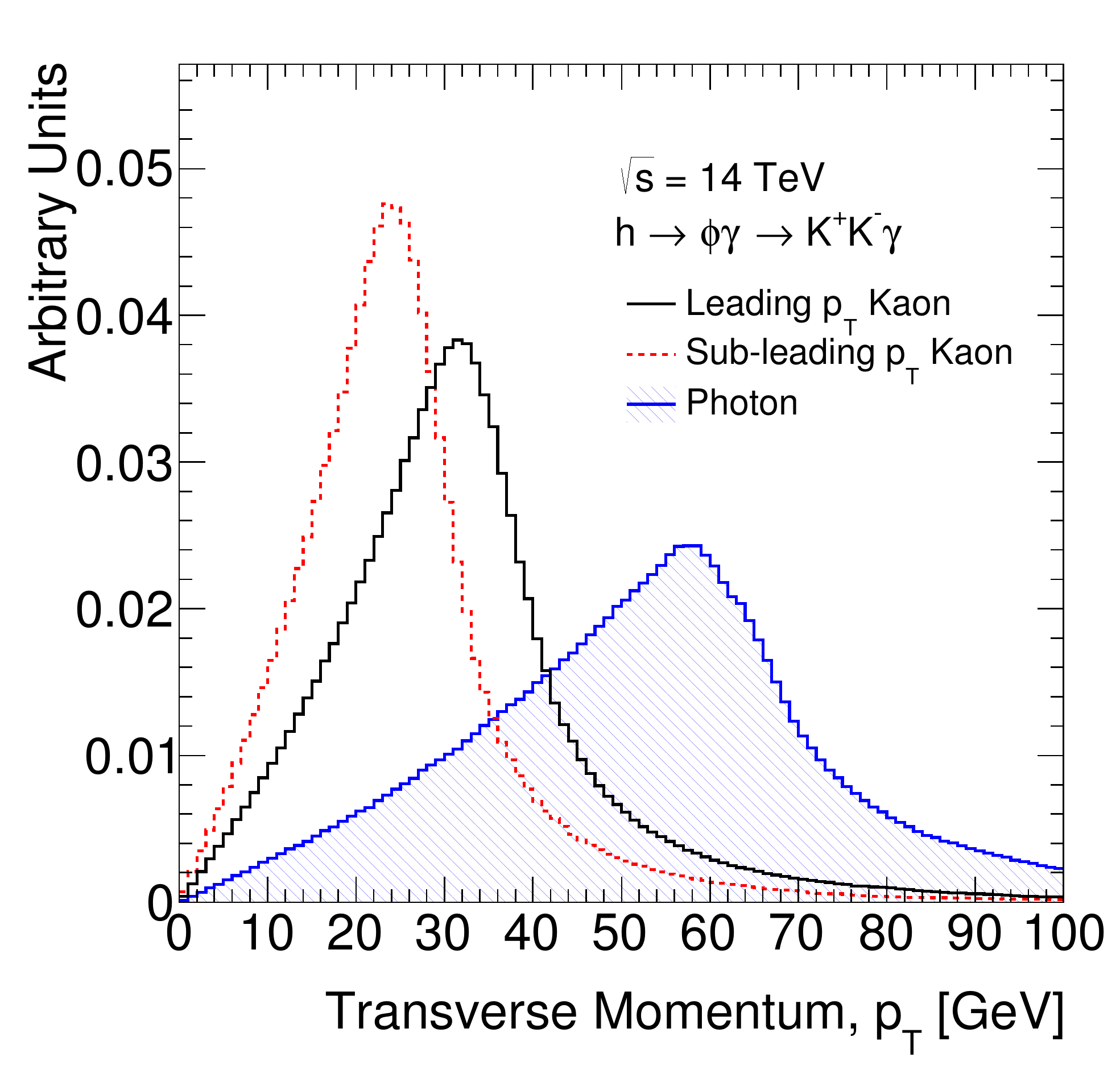}
\vspace*{-0.5cm}
\caption{Transverse momentum distribution of decay products in $h\to\phi\gamma\to K^+K^-\gamma$ decays~\cite{phigamma}.\label{fig:KKpt}}
\end{wrapfigure}
Regarding the Higgs boson coupling to the strange-quark, the
$h\to\phi\gamma$ decay is a potential probe. The
subsequent $\phi\to K^+K^-$ decay features a large branching ratio of
about 49\% and gives access to a simple final state of a hard photon
recoiling against two collimated high transverse momentum tracks, as
can be seen in \refF{fig:KKpt}. With the SM branching ratio prediction presented in \refT{table:SM_BRs},  about 6.5
events are expected to be produced with 100 fb$^{-1}$ at 14\,TeV.  For
the first generation quarks, the $h\to\omega\gamma$ and
$h\to\rho\gamma$ are being considered, followed by the
$\omega\to\pi^+\pi^-\pi^0$ and $\rho\to\pi^+\pi^-$ decays, both with
large branching ratios of about 89\% and 100\%, respectively.
The corresponding expected number of events, assuming the SM branching
ratios for these decays, are about 7.6 and 96, respectively.  The
experimental acceptance for these decays, assuming reasonable
geometrical acceptance and transverse momentum requirements, is
expected to range between 40 and 70\%~\cite{phigamma}.

These rare decays to a vector meson and a photon feature very
interesting and experimentally challenging boosted topologies. The
signature is distinct, but the QCD backgrounds require careful
consideration. A primary challenge arises from the trigger
availability to collect the required datasets. In the considered
cases, the decay signature is a photon of large transverse momentum
that is isolated from hadronic activity, recoiling against a narrow
hadronic jet. Triggering on such signatures, especially under pile-up,
will benefit from the upcoming detector upgrades, such as the ATLAS
Fast TracKer (FTK)~\cite{Shochet:1552953} that will provide rapid
track finding and reconstruction in the inner detector for every event
that passes the level-1 trigger.  The search for the final
states $\omega\gamma$ and $\rho\gamma$ is complicated further due to
the large natural width of the $\rho$ meson and the $\omega$-$\rho$
interference.

In the following, as an example of a potentially interesting target
rare decay for the high-statistics proton-proton collision data at a
centre-of-mass energy of 14\,TeV, the search for the Higgs boson decay
to $WD$ is presented in some detail. This is an interesting
experimental target due to its clean experimental signature.
The decay signature includes a high transverse momentum lepton from the $W$-boson decay, which gives the trigger for the
event, and a displaced vertex from the charmed meson decay. Since the
branching ratio of the Higgs boson decay to $WD_s^{(*)}$ is more than
an order of magnitude larger than the branching ratio to $WD^{(*)}$ (see \refT{tab:pseudo}),
the discussion here focuses on the former.

The search for this rare Higgs boson decay utilizes the dominant Higgs boson production 
mechanisms at the LHC, namely gluon-gluon fusion (ggF) and
vector boson fusion (VBF). The cross section for VBF Higgs boson production
is about an order of magnitude smaller compared to ggF, but features 
distinct event kinematics, with two jets in the forward regions
of the detector that can be used to tag the event. The main challenge
in the search for $H\rightarrow WD_s^{(*)}$ is to suppress the large
background from $W$ bosons produced in association with charm quarks
fragmenting into $D_s$ mesons. In order to estimate the sensitivity of
the search signal and background events have been produced using the
PYTHIA event generator. The $W$-boson is required to decay
leptonically to either an electron or a muon. Based on the detector
acceptance and trigger requirements, the fiducial region is defined by
lepton $p_{\rm T}>30$\,GeV, $|\eta|<2.5$ and neutrino $p_{\rm
  T}>25$\,GeV. Since the transverse momentum of the $D_s$ meson in
signal events peaks at higher values compared to background events, a
requirement of $D_s$ $p_{\rm T}>30$\,GeV, $|\eta|<2.1$ is applied. The
acceptance of these requirements is 18\% and 22\% for ggF and VBF
Higgs boson production, respectively, while 0.6\% of the background events
fulfil these requirements.

The $D^{(*)}_s$ meson is identified by reconstructing displaced
vertices from its hadronic decay to charged pions or kaon, in
particular $D_s\rightarrow K^{+}K^{-}\pi^{+}(\pi_0)$. The excited
$D^*_s$ state decays almost exclusively to $D_s+\gamma/\pi_0$ and is
tagged by the subsequent $D_s$ decay. The measurements of the
SM $W+c$ production at the LHC~\cite{Aad:2014xca,
  Chatrchyan:2013uja} demonstrate that hadronic charm decay signatures
can be reconstructed in the detector with reasonable efficiency of 30\%
to 40\%. Combinatorial background in the charm reconstruction is
largely rejected by exploiting the charge correlation between the $W$
boson and the charmed meson. However, the major background from $W+c$
production also predominantly yields opposite sign signatures and therefore is
not reduced by this requirement.

The main discriminating variable against the non-resonant $W+D_s$
background is the Higgs boson mass as reconstructed from the $W$-boson and
$D_s$ meson four-momenta. Since an excellent mass resolution of about
1\% is achieved for the $D_s$ meson, the Higgs boson mass resolution
is dominated by the measurement of the missing transverse energy in
the detector which has a resolution of about 10\% to 30\% in the
kinematic region relevant for this study. Requiring the reconstructed
mass to be within a window of 20\,GeV centred around the Higgs boson mass
reduces the background by about a factor of 6.

A further distinct characteristic of the signal events is the
isolation of the $D_s$ meson. In background events the $D_s$ meson
originates from charm-quark fragmentation and is thus seen within
a particle jet in the detector. By applying isolation criteria in
the reconstruction of the $D_s$ meson about 80\% of the background
events are rejected. In order to tag VBF events, two jets with an
invariant mass of $m_{jj}>500$\,GeV and separated by $\Delta \eta>3$
are required.

About 12 signal events and 16000 background events can be expected in
the ggF channel assuming an integrated luminosity of
$\mathcal{L}=3000\,\rm{fb^{-1}}$ and the Standard Model branching
ratio for $H\rightarrow WD_s^{(*)}$. For the VBF channel predicted
numbers of 1 signal event and 120 background events are obtained. Since
the branching ratio of this decay can be significantly enhanced in
many scenarios beyond the Standard Model (see Section~\ref{NP} of this report for a review), setting upper limits on the
decay branching ratio is of considerable interest in exploring the
phase space of these models. Based on these results an upper limit of
$7\times 10^{-4}$ on the branching ratio of $H\rightarrow WD_s^{(*)}$
can be achieved.



\section{Recommendations for searches for exotic Higgs boson decays}
\label{sec:recommendations}

In Run 2 of the LHC, the programmatic search for exotic Higgs boson decays will
 increasingly become an important topic of study.  To help guide this experimental 
program, in this section we provide a set of recommendations for searches for
the production of exotic particles in the decays of the Higgs boson.

\paragraph*{\textbf{A signature-based search program for exotic decays:}}
We
recommend that the search program for exotic Higgs boson decays take a
signature-based approach, targeting individual signatures rather than
specific BSM models.
A model that gives rise to exotic decays of the Higgs boson
will typically lead to many different possible final states.  For
instance, a model containing a light pseudo-scalar with
Yukawa-weighted decays will yield, in addition to the dominant $h\to
aa\to 4b$ decay, the much rarer but cleaner $h\to aa\to bb\mu\mu$
decay \cite{Curtin:2014pda}.  Combining the results of searches for
both exotic modes can substantially boost overall sensitivity to the
overall exotic branching fraction $h\to aa$ \cite{Curtin:2013fra}.  On
the other hand, a single given final state, such as $h\to 4b$, may be
predicted by a wide range of different theories; for a detailed
discussion of this point, see Ref.~\cite{Curtin:2013fra}.

\paragraph*{\textbf{Presentation of results:}}
We recommend that searches for specific signatures quote their results
in terms of $\sigma\times\br$.  This model-independent prescription
allows for more flexible interpretation of results in a broad range of
theories, and facilitates the ultimate combination of results obtained
in different final states to achieve the best sensitivity to a
specific model.

\paragraph*{\textbf{Production cross-sections:}}  
The SM Higgs boson width is dominated by the small $b$-quark Yukawa coupling
and is accidentally small, which means that even small couplings of
the Higgs to BSM physics can generate exotic branching fractions of
order the current upper bound, ${\rm{BR}}(h\to\mathrm{BSM})\sim 30\%
$.  However, Higgs boson production cross-sections are controlled by
the much larger electroweak and top couplings, and are thus
substantially less affected by small couplings of the Higgs to new
physics.  As a simple example, consider the case of a real scalar $s$
which mixes with the SM Higgs boson through a small Higgs portal
interaction, $\mathcal {L}_{int} = -\epsilon/2\, |H|^2s^2$, after both
$H$ and $s$ acquire vacuum expectation values, $v$ and $\langle
s\rangle$, respectively.  The Higgs boson production cross-section then
receives corrections at the order $\mathcal{O}(\theta^2)$, where
$\theta = \epsilon \langle s\rangle v/(m_h^2-m_s^2)$ is the $s$-$h$
mixing angle.  However, the values of $\theta$ that induce $\br(\hsm
\to ss) \sim 20\%$ are $\mathcal{O}(10^{-2})$ \footnote{In this
model, a given exotic branching fraction does not uniquely determine the
mixing angle; these numbers are obtained in the generic regime
$\langle s\rangle\sim m_s$.}.
Thus, {\em corrections to Higgs boson production cross-sections are
subleading} in theories yielding exotic Higgs boson decays consistent with
Run 1 data.  As these corrections are model-dependent, and as many
different models can yield the same Higgs boson decay final states, to
maximize the flexibility and utility of experimental searches for
exotic decays we recommend that searches for exotic Higgs boson decays
assume SM Higgs boson production cross-sections.  In the event that an
exotic decay mode is discovered, it will then be of high interest to
consider how possible effects of specific related new physics models
would affect Higgs boson properties and production cross-sections.  However,
to institute and support a broad program of searches for
direct production of exotic particles in Higgs boson decays, we recommend
that SM Higgs boson production be used as the baseline for generating signal
events and performing searches.

\paragraph*{\textbf{Signal event generation:} }
Apart from the special cases of decays with detector-stable or
highly-displaced objects, exotic Higgs boson decays result in at least three
objects in the final state. 
Since the Higgs itself is not heavy with respect to the LHC's
centre-of-mass (CM) energy, one generic feature of exotic Higgs boson decays
is thus that the particles in the final state tend to be
soft. 
The spectrum of the Higgs boson decay products is a major factor in
determining signal acceptance.  Consequently, in carrying out searches
for exotic Higgs boson decays, it is important to model Higgs boson production and
in particular the Higgs boson $p_T$ spectrum with some degree of care.  On
the other hand, the ease and flexibility with which BSM models can be
implemented in {\tt MadGraph} is also of practical importance to
support a broad search program capable of covering the vast number of
possible BSM decays.

Our baseline recommendation for signal event generation in searches
for BSM particles produced in exotic Higgs boson decays is to use {\tt
MadGraph5}~\cite{Alwall:2011uj}, followed by showering and
hadronization in {\tt Pythia}~\cite{Sjostrand:2007gs}.  This
recommendation is supported by the studies in the following section,
where we compare kinematic distributions for the Higgs and its decay
products in the prototypical decay $h\to aa\to 4b$ as predicted by
{\tt MadGraph}$+${\tt Pythia} to those predicted by {\tt
POWHEG}~\cite{Nason:2004rx,Frixione:2007vw,Alioli:2010xd}$+${\tt
Pythia}, demonstrating good agreement overall.  For gluon fusion
production in {\tt MadGraph}, Higgs boson production should be matched to at
least one jet; see Section~\ref{sec:4b} for further relevant settings.
For VBF and VH production, Higgs boson production is well-modeled in
{\tt MadGraph} without matching. In all cases the inclusive Higgs boson production
cross-section should be normalized to the predictions from~\ref{subsec:ADDFGHLM}.
For searches that
dominantly rely on gluon fusion production, the Higgs boson $p_T$ spectrum
should be reweighted according to the (N)NLO predictions following the
recommendations of~\ref{subsec:ADDFGHLM}.

\paragraph*{\textbf{Mass ranges:} }
While the primary focus of searches for exotic Higgs boson decays is and
should remain the decays of the discovered 125 GeV SM-like Higgs boson, 
analyses should keep in mind the possibility of extending the
search to cover other possible masses for the parent particle,
particularly but not exclusively lower masses.  This extends the
sensitivity of searches to include cases where the originating 125 GeV
Higgs boson decay includes small amounts of missing energy
\cite{Curtin:2013fra} as well as the potential direct production of
BSM Higgs bosons.

\section[Partonic distributions for the prompt decay topology \texorpdfstring{$h\to XX\to  2Y2Y'$}{h to XX to 2Y2Yprime}]{Partonic distributions for the prompt decay topology \texorpdfstring{$h\to XX\to  2Y2Y'$}{h to XX to 2Y2Yprime}\SectionAuthor{R.~Caminal Armadans, Z.~Liu, V.~Martinez Outschoorn}}
\label{sec:4b}
\subsection{Introduction}

In this section we study the decay topology where the Higgs boson decays to
two exotic particles of the same mass, each of which then undergoes a
prompt two-body decay to visible SM particles $Y,\,Y'$: $h\to XX\to 2Y2Y'$.
This decay topology is naturally realized in many well-motivated BSM
frameworks.  In particular, extensions of the Higgs sector by an
additional, possibly complex, singlet scalar can naturally have Higgs boson 
decays to (pseudo-)scalars as one of their leading signatures.
These (pseudo-)scalars decay mainly to fermions, with preference for
heavy flavour. Signals of this class of models, SM+S and 2HDM+S, are
described in detail in \cite{Curtin:2013fra}. The NMSSM is one of the
best-studied examples of this type of extended Higgs sector. It has a
large portion of parameter space where an approximate $R$-symmetry
yields a SM-like Higgs boson with appreciable branching ratio into a
pair of light pseudo-scalars $a$~\cite{Dobrescu:2000yn,Ellwanger:2003jt,Dermisek:2005ar,Morrissey:2008gm,Belyaev:2010ka}. Other motivations for
singlet-extended Higgs sectors include models of first-order
electroweak phase transitions \cite{Profumo:2007wc,Blinov:2015sna} and thermal dark matter
\cite{Silveira:1985rk, Pospelov:2007mp, Ipek:2014gua, Martin:2014sxa}. In composite
Higgs models, a symmetry-protected light pseudo-scalar in the spectrum
may be fermiophobic, with dominant decays to gluon or photon pairs,
thus yielding the exotic decay modes $h\to aa \to 4g, 2\gamma 2g,
4\gamma$~\cite{Dobrescu:2000jt, Chang:2006bw, Bellazzini:2009xt,
  Chen:2010wk, Falkowski:2010hi}.  Another
well-studied extension of the SM is a Higgsed dark $U(1)$ that
kinetically mixes with SM hypercharge
\cite{Holdom:1985ag,Galison:1983pa,Dienes:1996zr}, in which case the
Higgs boson decay $h\to Z_D Z_D\to 2f 2f'$ yields final states weighted by
gauge couplings, rather than Yukawa couplings, giving relatively
leptophilic signatures
\cite{Gopalakrishna:2008dv,Martin:2011pd,Chang:2013lfa,Davoudiasl:2013aya,Curtin:2014cca}.
All these exotic decays have very similar parton-level kinematics
\footnote{Angular correlations present in the decay $h\to Z_D Z_D\to
  4f$ will modify the final state fermion distributions relative to the
  decay $h\to ss (aa)\to 4f$, but the (pseudo-)scalar decays are still
  a good guide to the overall kinematics.}.

As discussed in Section~\ref{sec:recommendations}, exotic Higgs boson
decays are characterized by low $p_T$ objects in the final state.
Thus, object acceptance becomes one of the main limiting factors in
recording and reconstructing many exotic Higgs boson decays.  Consequently,
a good understanding of the parent Higgs boson and resulting decay
product kinematics is necessary to assess realistic triggering
opportunities and analysis strategies.

Here we study the (parton-level) kinematics of the final state particles in the
prototypical exotic decay $h\to aa\to 4b$ in depth, considering gluon
fusion, VBF, and WH associated production modes.  In addition, we
compare the predictions of LO and NLO generators, showing results for
Higgs boson production using both {\tt MadGraph}~\cite{Alwall:2011uj} and
{\tt POWHEG}~\cite{Nason:2004rx,Frixione:2007vw,Alioli:2010xd}.

\subsection{Signal model and event generation}

We consider augmenting the SM with a singlet pseudo-scalar $a$, which
obtains interactions with SM fermions through mixing with the
pseudo-scalar state $A^0$ in a 2HDM.  After electroweak symmetry
breaking, the relevant interaction terms are
\beq
\mathcal{L}^{\rm BSM}\supset i y^a_b a \bar b \gamma_5 b + \frac 1 2
\lambda_{aH} h a^2 + \frac 1 2 m_a^2 a^2,
\eeq
where $m_a<m_h/2$ is the mass of the pseudo-scalar, the effective
Yukawa coupling $y^a_b$ controls the singlet's decay into $b\bar b$
pairs, and the trilinear coupling $\lambda_{aH}$ determines the
partial width for the Higgs boson decay into pairs of pseudo-scalars.  A
discussion of how these parameters depend on the couplings in the full
2HDM+S Lagrangian can be found in \cite{Curtin:2013fra}.  The
trilinear coupling and the pseudo-scalar mass can be independently
adjusted, making the pseudo-scalar mass and the Higgs boson branching fraction $\br
(\hsm\to aa)$ independent parameters.  We assume that $y^a_b$ is large
enough to yield prompt decays; displaced decays are discussed in
Section~\ref{sec:displaced}.

The exotic Higgs boson decay mode $\hsm \to a a (ss) \to b \bar b b
\bar b $ will be the leading  signature of a broad class of SM+S, 
2HDM+S theories.  Here we explore the
kinematics of the final state $b$-partons in this decay for a range of
pseudo-scalar masses $m_a = 20, 30, 40, 50$ and $60$~GeV. 
The kinematics for a scalar
field $s$ decaying to bottom quark pairs are identical.  We consider
three Higgs boson production channels: gluon fusion (ggF); $W$-boson
associated (WH) production; and weak vector-boson-fusion (VBF)
production.  In each production channel, we compare the differential
predictions from events generated at LO ({\tt Madgraph5}$+$ {\tt
  Pythia}) to those generated at NLO ({\tt Powheg}$+${\tt Pythia}).
We find that predictions for the $b$-parton kinematics from {\tt
  Madgraph} and {\tt Powheg} event generators agree very well 
overall, justifying the use of LO signal event generation for BSM
signal models.

We use the {\tt CTEQ6L1}~\cite{Stump:2003yu} PDF set for {\tt
  MadGraph 5}~\cite{Alwall:2014hca} signal samples, with the
factorization and renormalization scales set to {\tt MadGraph}
default.  The signal model is implemented using a modification of {\tt
  MadGraph}'s {\tt heft} model to include an additional
pseudo-scalar.  For ggF production we match events to one jet using
the MLM matching scheme \cite{Alwall:2007fs} with matching parameters
${\tt xqcut} = 15$ GeV and ${\tt QCUT} = 20$ GeV; {\tt
  Pythia6}~\cite{Sjostrand:2006za} is used for showering.  In all
three production modes, the final state is generated inclusively
except for the forward tagging jets in VBF, where cuts of $|\eta|< 5$
and $p_T>20$ GeV are applied.

In {\tt Powheg-Box}
v2~\cite{Frixione:2007nw,Nason:2004rx,Frixione:2007vw,Alioli:2010xd},
signal samples for all three production modes
are generated using the CT10 PDF set~\cite{Lai:2010vv}.  The events
are interfaced with {\tt Pythia} 8.186 which is used to decay the
Higgs boson into a pair of pseudo-scalars $a$, which are themselves
decayed to a pair of $b$-quarks with $\br(a \rightarrow b \bar{b})$ =
1.  Event generation is fully inclusive for all three production
modes.

\subsection{Results}

\begin{figure}[thb]
\centering
\subfigure{
\centering
\includegraphics[width=0.45\textwidth]{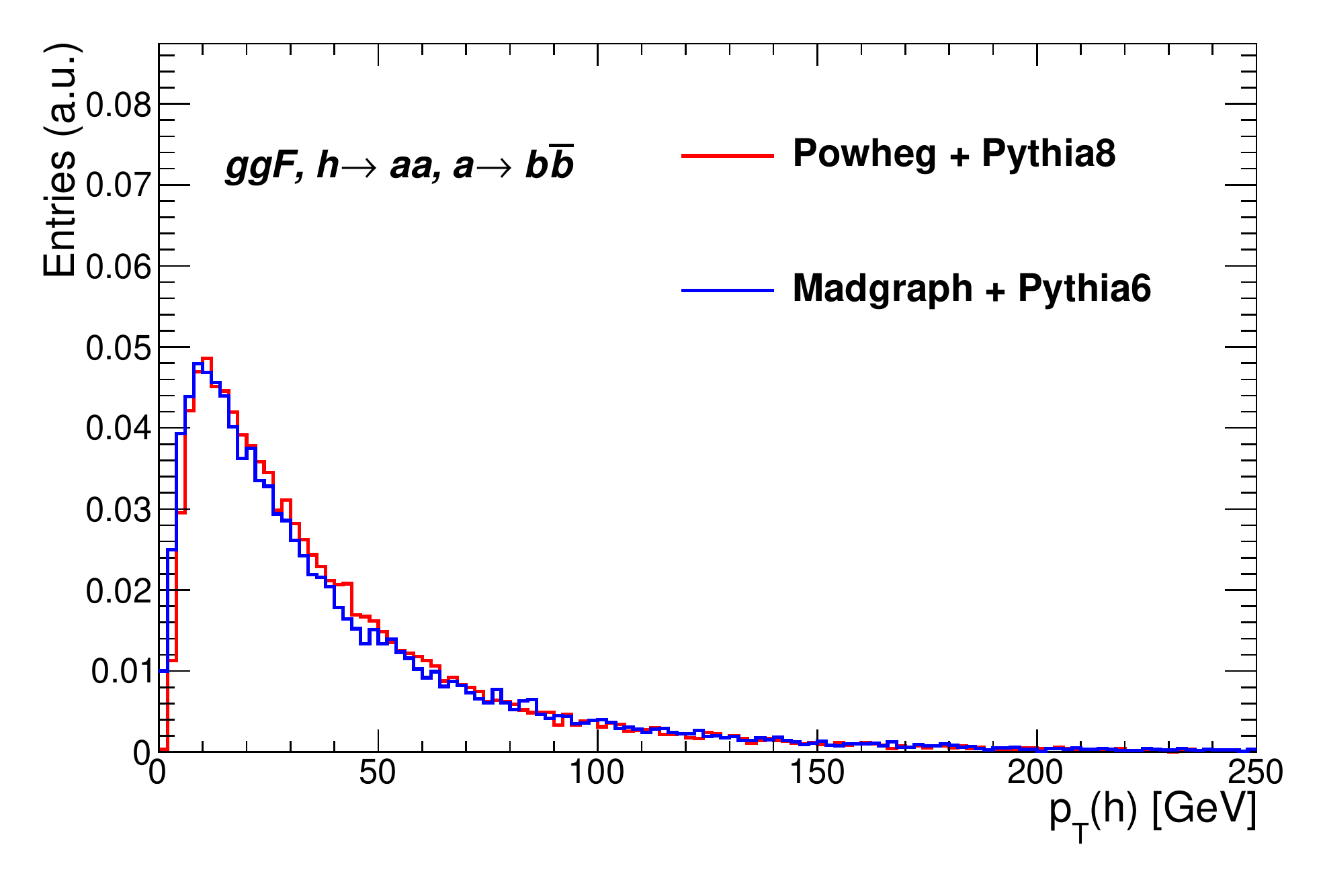} }
\subfigure{
\centering
\includegraphics[width=0.45\textwidth]{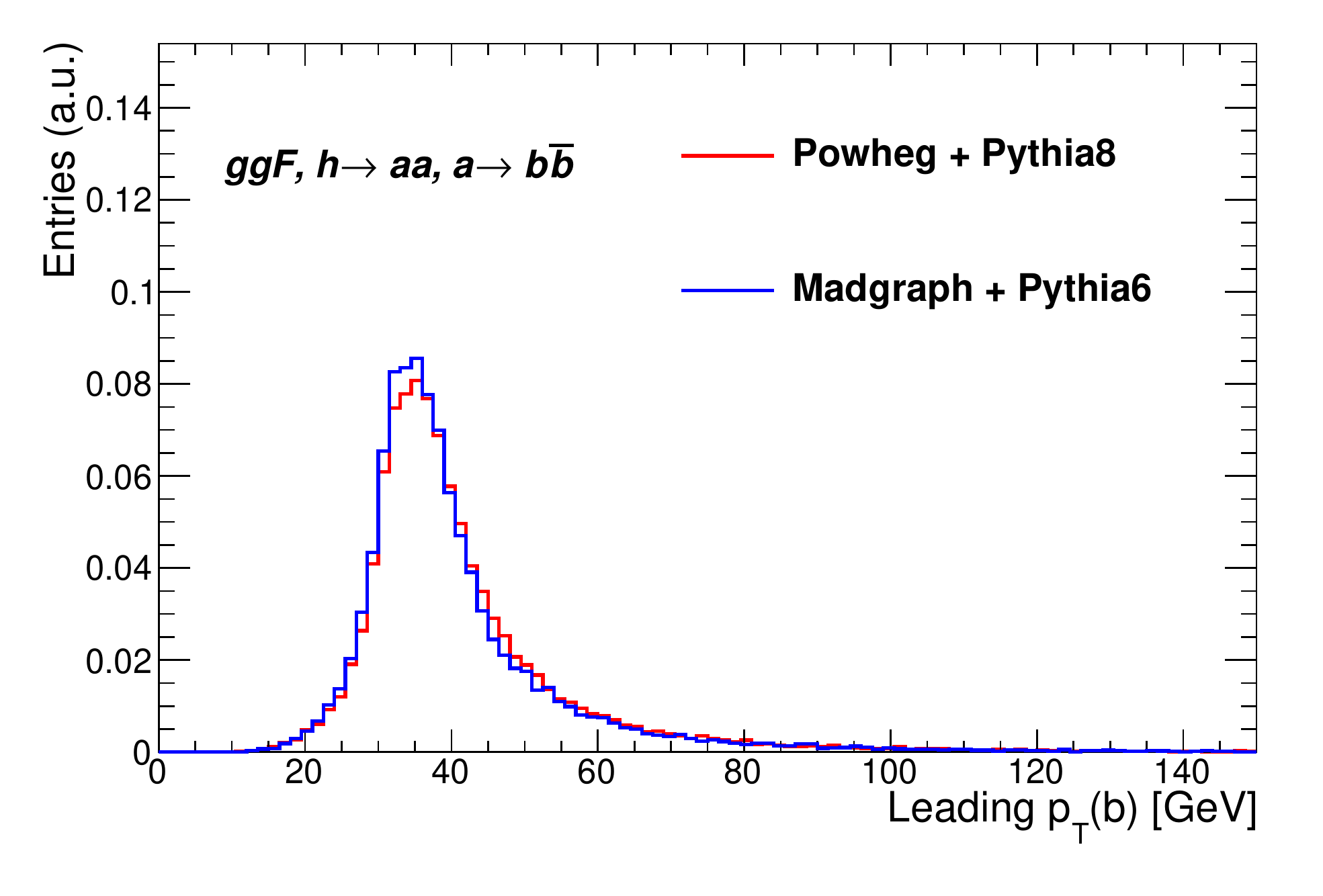} }
\subfigure{
\centering
\includegraphics[width=0.45\textwidth]{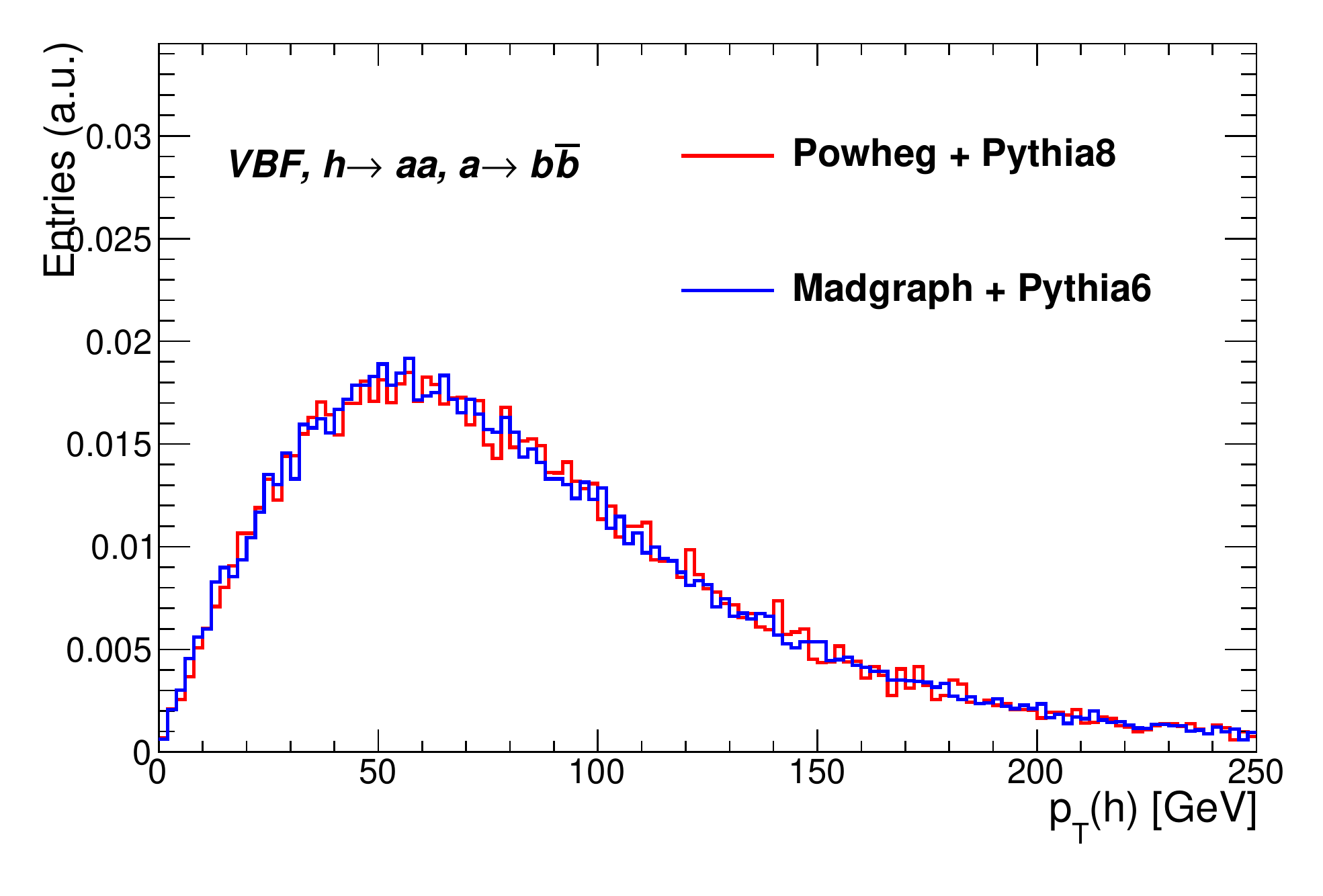} }
\subfigure{
\centering
\includegraphics[width=0.45\textwidth]{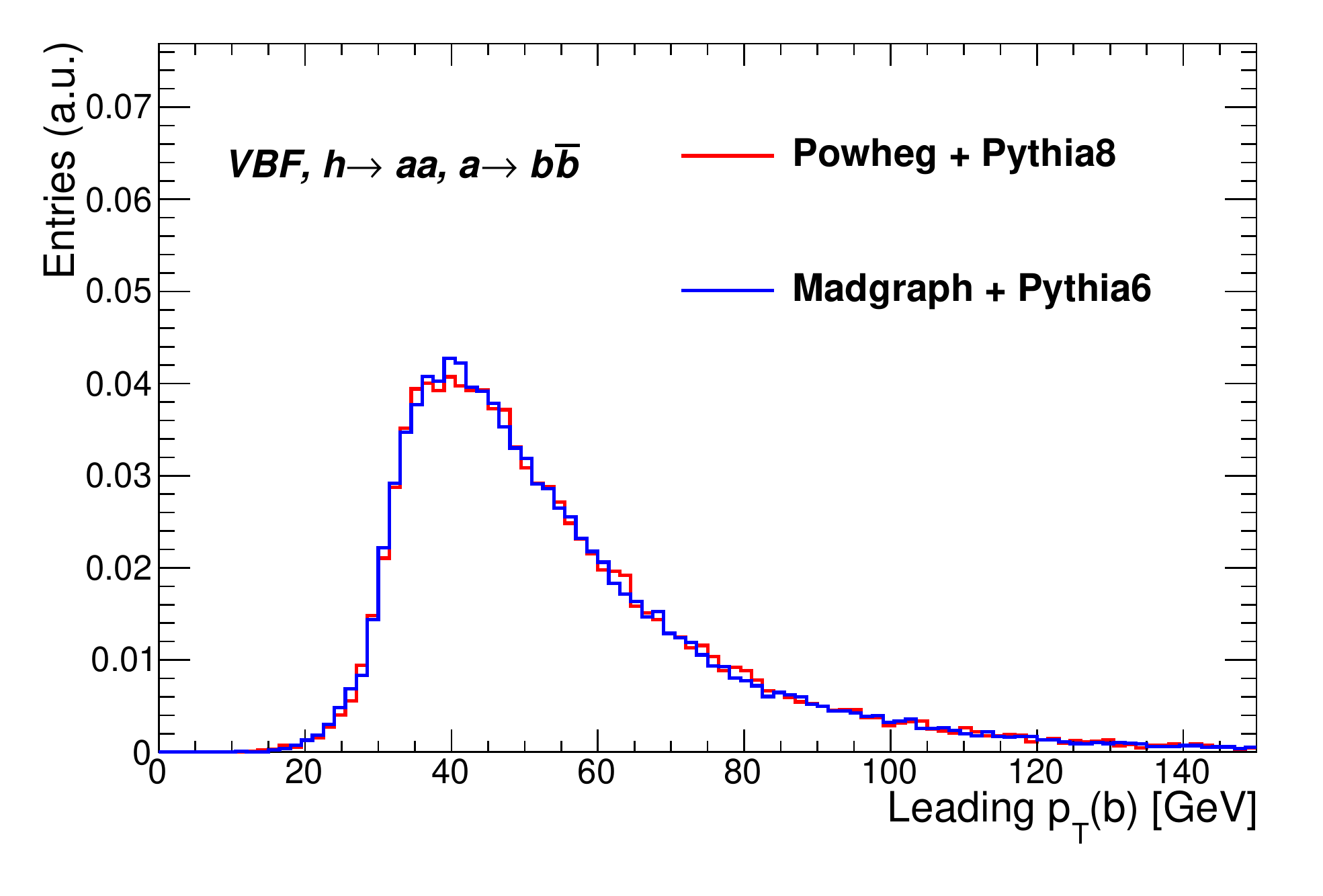} }
\subfigure{
\centering
\includegraphics[width=0.45\textwidth]{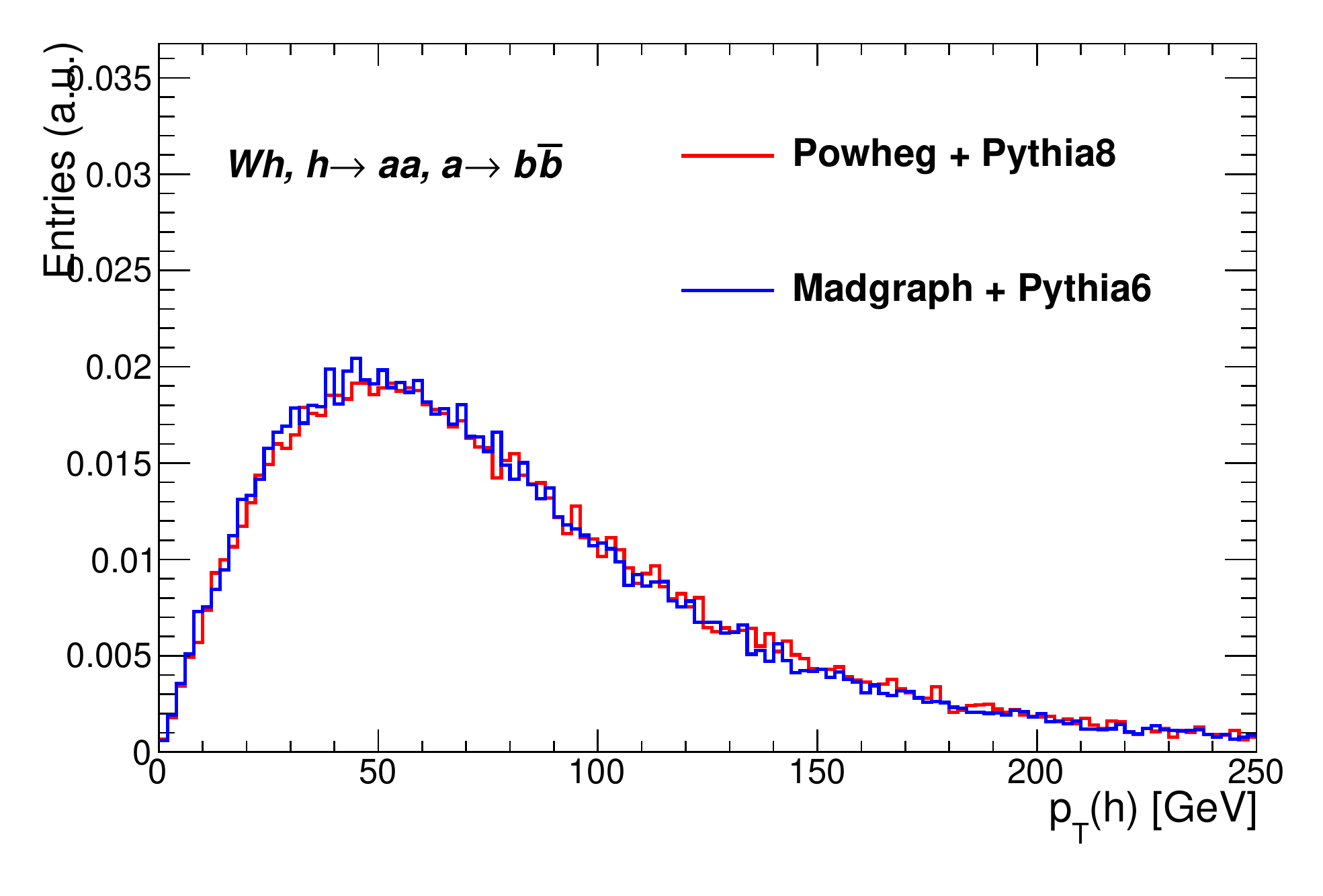} }
\subfigure{
\centering
\includegraphics[width=0.45\textwidth]{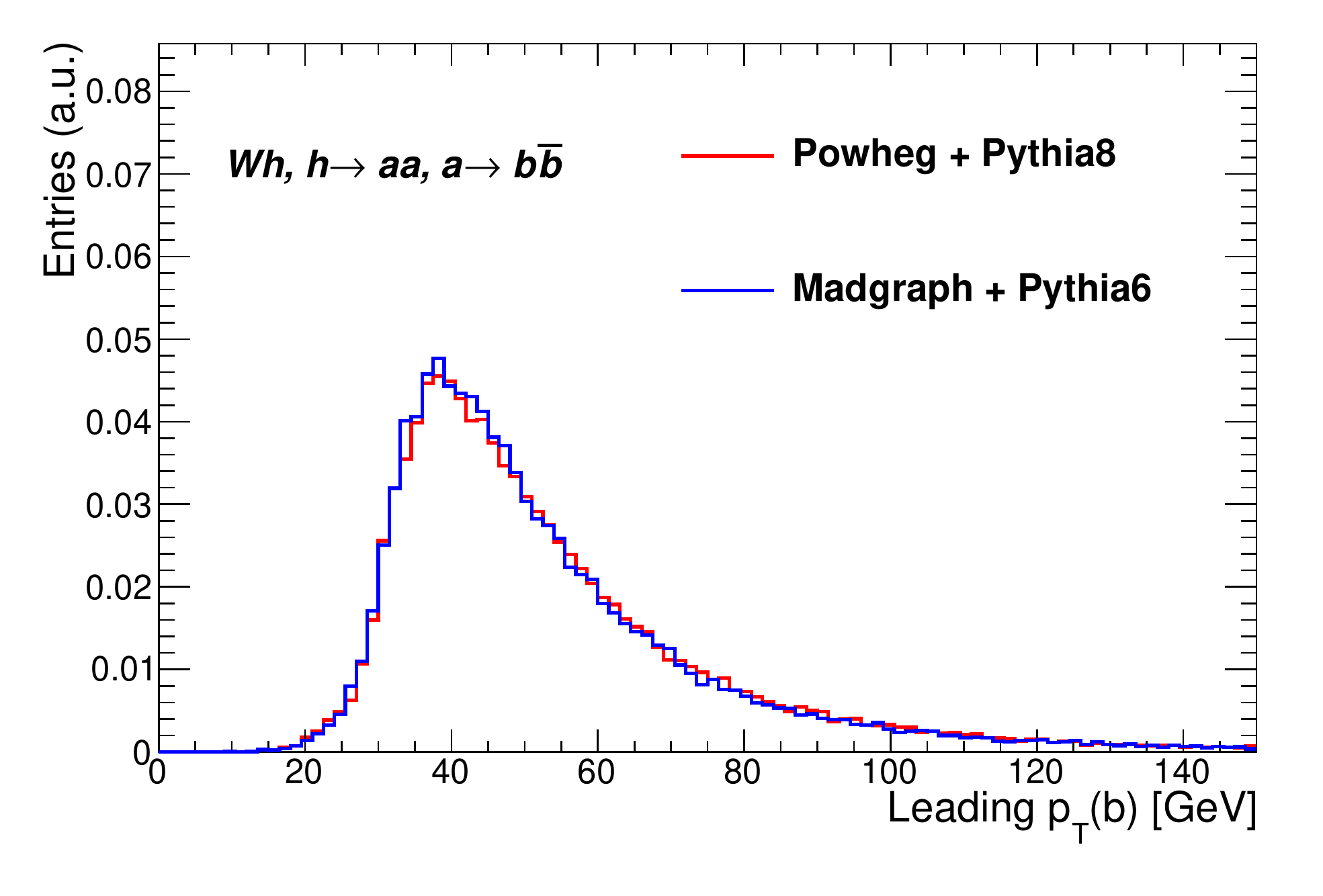} }
\caption[]{Left panels: the $p_T$ distribution of the Higgs boson at
  LHC 13 TeV from different production modes: ggF (top), VBF (centre)
  and WH (bottom) using {\tt MadGraph 5} (blue) and {\tt Powheg}
  (red).  Right panels: same, but for the $p_T$ distribution of the
  leading $b$-parton, for the pseudo-scalar mass $m_a=60$
  GeV.}\label{fig:hpt}
\end{figure}

We begin with the Higgs boson $p_T$ distribution for ggF, WH and VBF
Higgs boson production at the 13 TeV LHC, shown in the left panels of
\refF{fig:hpt}. For VBF production, we show the Higgs boson $p_T$
distribution in events containing two jets with $p_T>20$ GeV and
$|\eta|<5$.  The Higgs boson $p_T$ spectrum is a key quantity that affects
all subsequent decay product distributions. Overall we find good
agreement between the predictions of the two event generators, {\tt
  MadGraph} and {\tt Powheg}. For VBF and WH, the $p_T$ spectra from
the two generators are in excellent agreement. The greatest
differences are seen in ggF, where {\tt Powheg} predicts a slightly
harder spectrum than does {\tt MadGraph}; however, even here the $p_T$
distributions are quite similar, with good agreement in the tail and
in the peak.  As discussed in Section~\ref{sec:recommendations}, we
recommend that searches that rely on ggF for the bulk of their
sensitivity reweight the Higgs boson $p_T$ distribution according to the
recommendations of WG1.
In the right panels of \refF{fig:hpt}, we
compare predictions for the $p_T$ spectrum of the leading $b$-parton
in the same set of events, for the case $m_a=60$ GeV. Good agreement
is shown for VBF and WH production modes, while in ggF the spectrum
predicted by {\tt MadGraph} is slightly softer than the spectrum
predicted by {\tt Powheg}.

We further show various parton-level differential distributions, as
generated in {\tt Powheg}, for masses of the pseudo-scalar $a$ ranging
between 20 to 60 GeV in steps of 10 GeV. In these plots no cuts are
applied to the final state particles; in particular, no cuts on VBF
jets have been imposed.
\begin{figure}[thb]
\centering
\subfigure{
\centering
\includegraphics[width=0.45\textwidth]{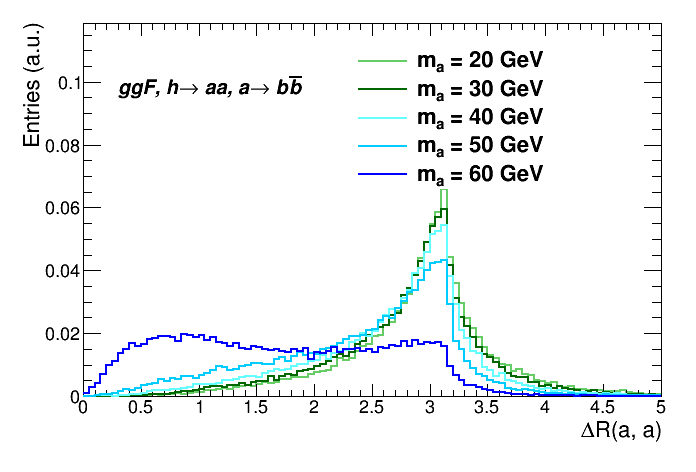} }
\subfigure{
\centering
\includegraphics[width=0.45\textwidth]{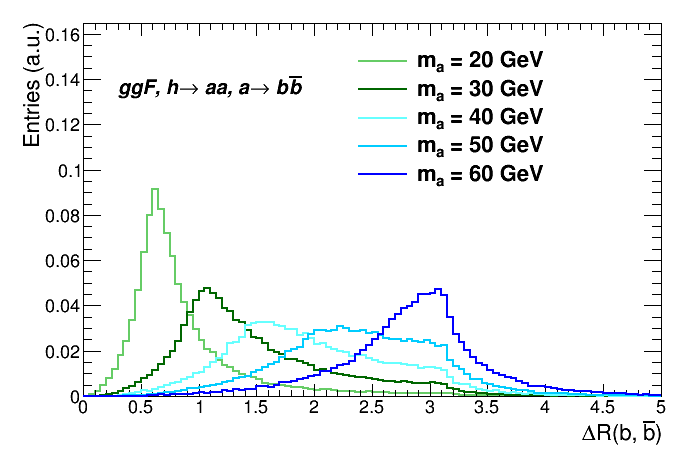} }

\subfigure{
\centering
\includegraphics[width=0.45\textwidth]{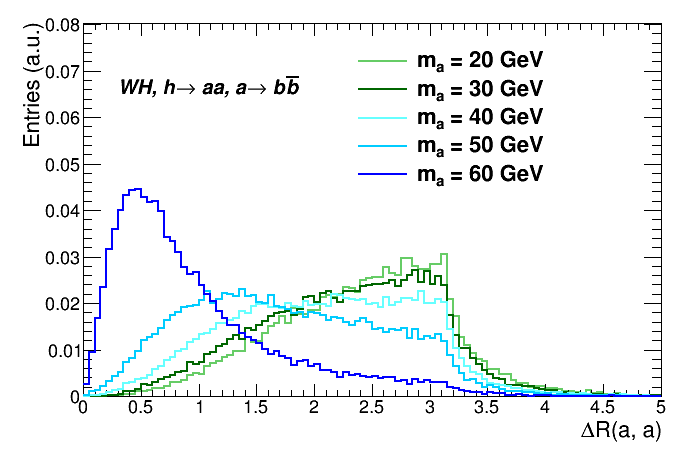} }
\subfigure{
\centering
\includegraphics[width=0.45\textwidth]{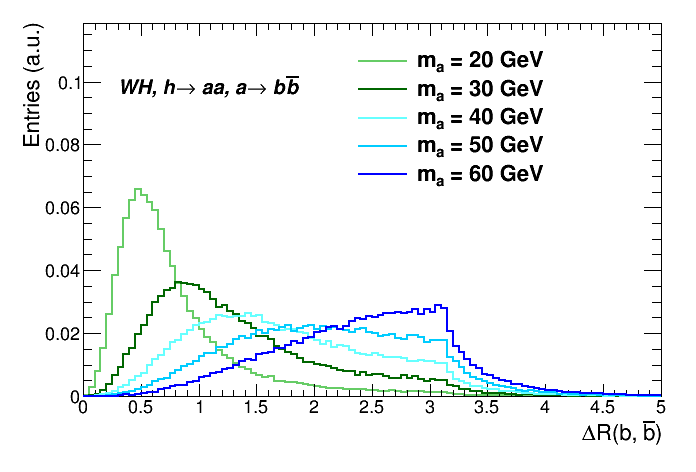} }

\subfigure{
\centering
\includegraphics[width=0.45\textwidth]{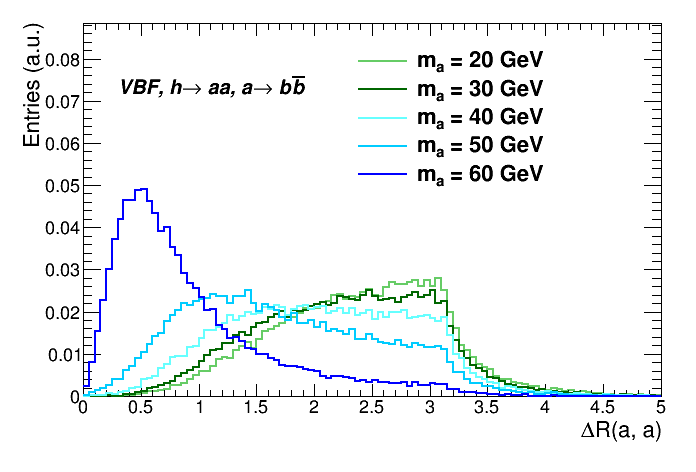} }
\subfigure{
\centering
\includegraphics[width=0.45\textwidth]{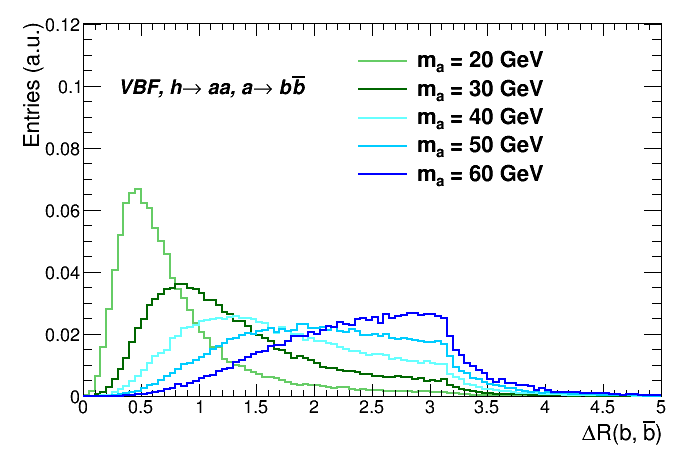} }
\caption[]{The angular separation $\Delta R$ between the two
  pseudo-scalars from Higgs boson decay (left panels), and between the $b\bar b$ pairs from a 
  pseudo-scalar $a$
  decay (right panels) as computed by {\tt Powheg} at LHC 13 TeV, normalized to unity. The upper,
  middle and lower panels correspond to Higgs boson production modes of ggF,
  WH and VBF, respectively.  }\label{fig:ssdR}
\end{figure}
In \refF{fig:ssdR} we show the angular separation $\Delta R$
between the two pseudo-scalars $a$ and between the $b\bar b$ pair
originating from the same pseudo-scalar $a$ decay for various Higgs boson production 
modes.  From the left panels of this figure, we can see
that heavy $a$'s are less separated than lighter $a$'s, as they are less
boosted in the Higgs rest frame. Similarly, from the right panels we
can see that the $b\bar b$ pairs from heavier $a$'s are more
back-to-back.  The higher average Higgs boson $p_T$s in the VBF and WH
production channels result in more collimated decay products.  We have
checked that the {\tt MadGraph} samples agree well with the {\tt
  Powheg} samples in modelling these distributions.
\begin{figure}[thb]
\centering
\subfigure{
\centering
\includegraphics[width=0.42\textwidth]{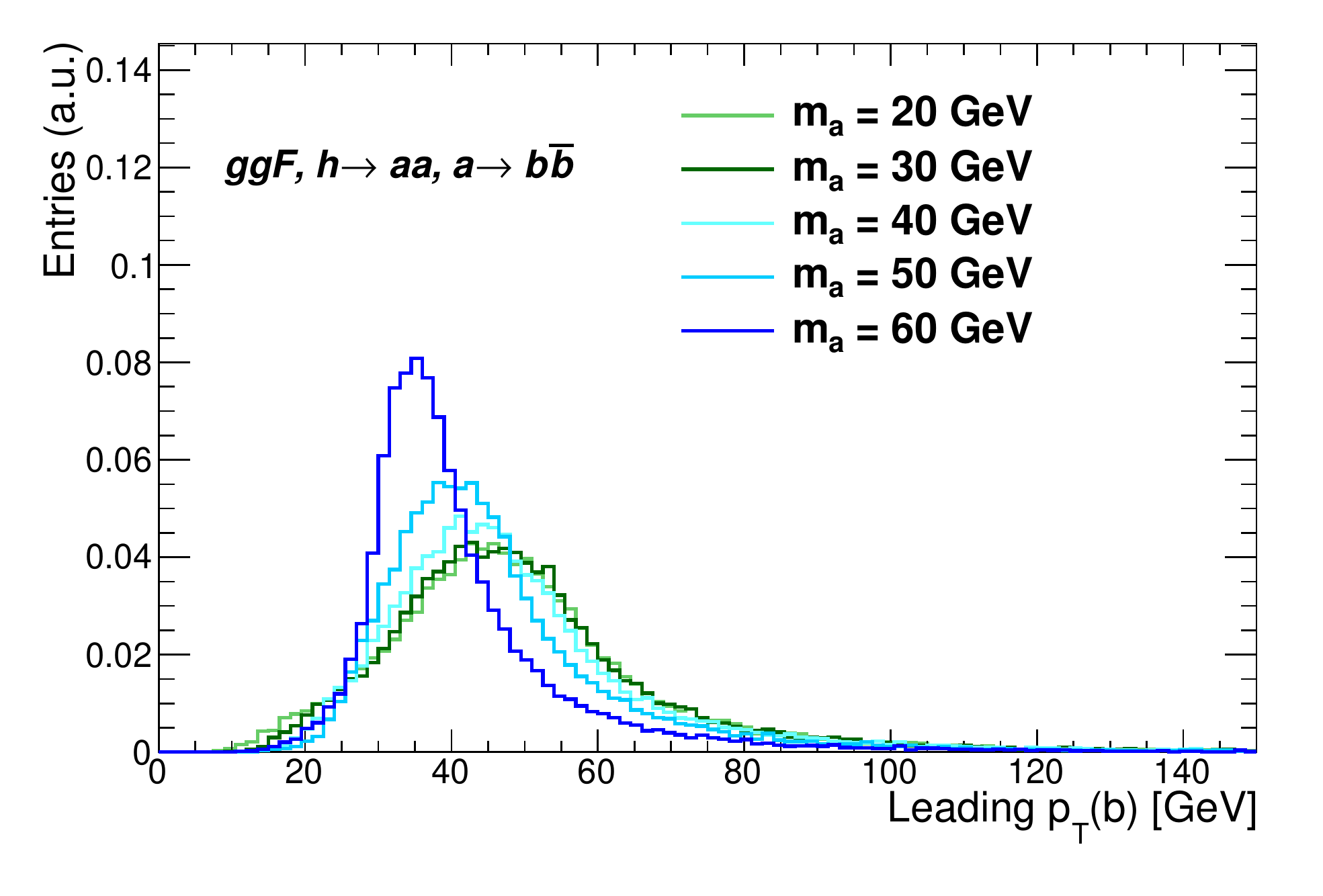} }
\subfigure{
\centering
\includegraphics[width=0.42\textwidth]{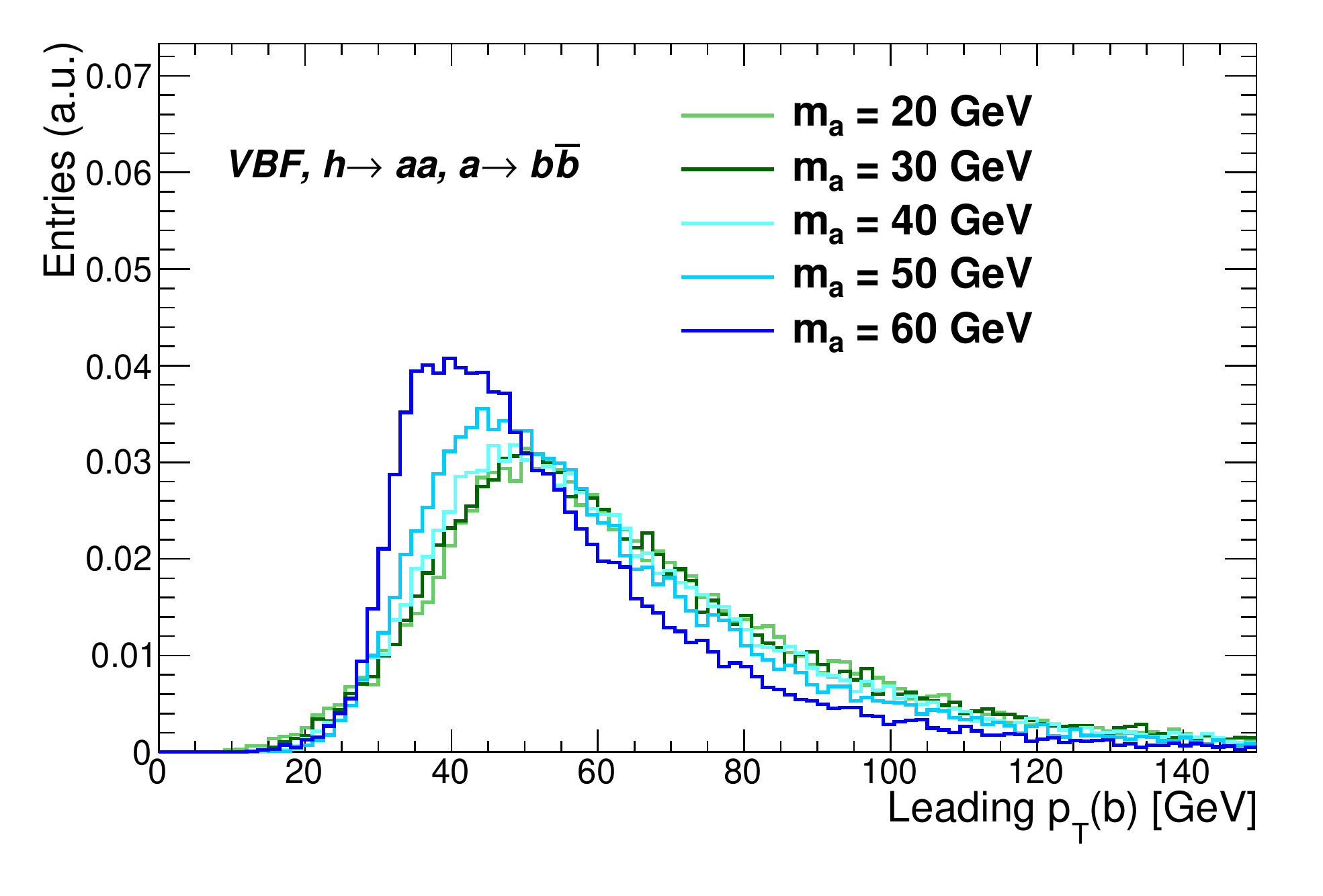} }
\subfigure{
\centering
\includegraphics[width=0.42\textwidth]{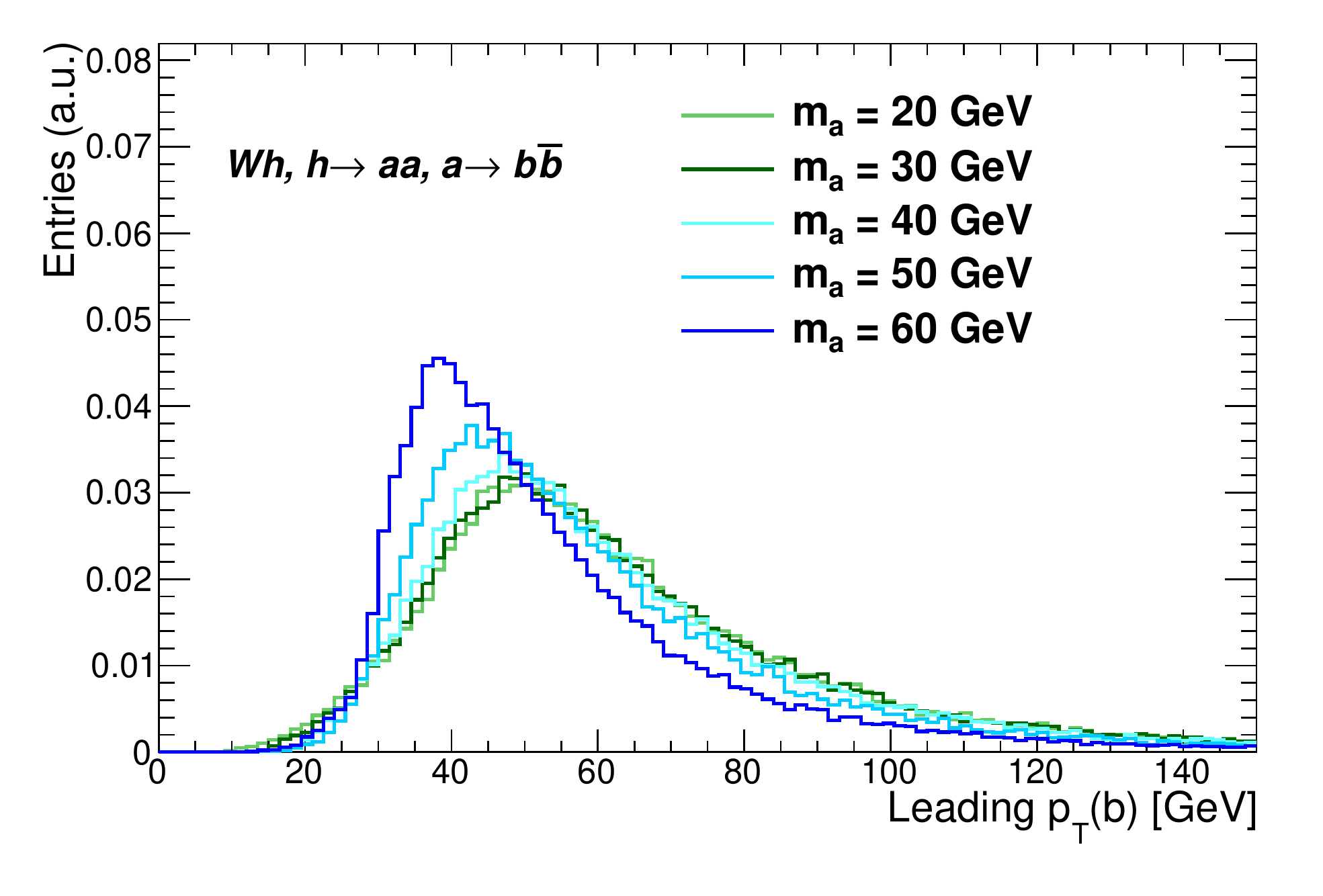} }
\caption[]{The $p_T$ spectrum of the leading $b$-parton, as computed by {\tt Powheg} at LHC 13 TeV, normalized to unity, for ggF (left, upper panel), VBF (right upper panel), and WH (lower panel).}
\label{fig:ptb1}
\end{figure}
\refF{fig:ptb1} shows the $p_T$ spectrum of the leading $b$-parton
for various Higgs boson production modes. 
The broader distributions for lighter $a$ masses reflect the bigger
boost of those $a$'s in the Higgs rest frame.  Again, we have checked
that the {\tt MadGraph} samples agree well with the {\tt Powheg}
samples in modelling these distributions in VBF and WH production
modes, while in ggF there are minor differences comparable to those in
\refF{fig:hpt}.
\begin{figure}[thb]
\centering
\includegraphics[width=0.98\textwidth]{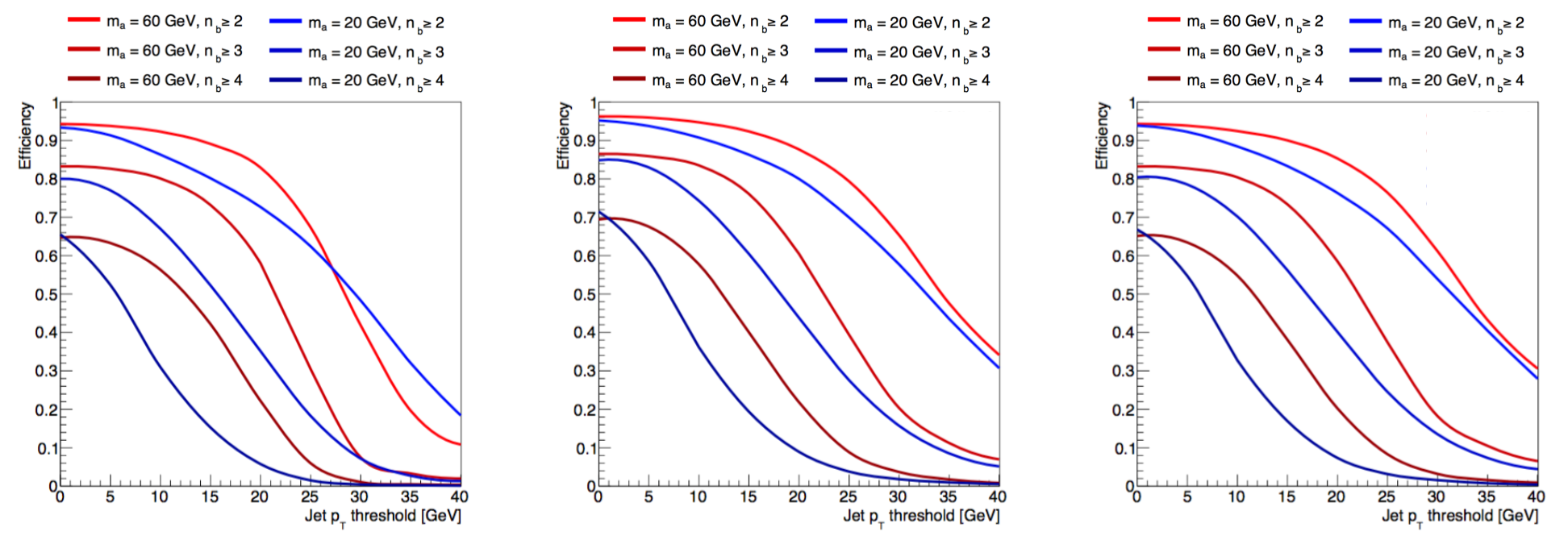}
\caption[]{Fraction of events having $N_b = 2, 3, 4$ partons above a given $p_T$ threshhold and $|\eta|<2.5$, as computed by {\tt Powheg} at LHC 13 TeV, for ggF (left), VBF (centre), and WH (right).  No other cuts are applied.
Results are shown for both $m_a=20$ GeV and $m_a=60$ GeV. }
\label{fig:ptcuts}
\end{figure}

In \refF{fig:ptcuts} we show the fraction of events which have
$N_b = 2, 3, 4$ partons above a given $p_T$ for models with heavy
($m_a = 60$ GeV) and light ($m_a = 20$ GeV) pseudoscalars.  Here the
$b$-quarks are required to have $|\eta|<2.5$, but no further cuts are
applied.  These plots quantify the overall softness of the final state
particles.  In all three production modes, the efficiency drops
quickly with the $p_T$ threshold.  For the light (heavy) pseudoscalar,
the efficiency reaches the level of 30\% around a threshold of $\sim
10$ (17) GeV for $N_b=4$, $\sim 20 (25) $ GeV for $N_b=3$, and $35
(35)$ GeV for $N_b=2$. Again, ggF gives rise to an overall softer $b$
spectrum.  The broader $p_T$ distribution produced by the lighter
pseudo-scalar results in a more rapid falloff of efficiency with
increasing $p_T$.

\afterpage{\clearpage}

\section[Prospects for prompt decays with MET: \texorpdfstring{$h\to 2\gamma+\met$}{h to 2gamma + MET} test case]{Prospects for prompt decays with MET: \texorpdfstring{$h\to 2\gamma+\met$}{h to 2gamma + MET} test case\SectionAuthor{T.~Orimoto, R.~Teixeira De Lima}}
\label{sec:2g}
\subsection{Introduction}

In this section, we discuss exotic Higgs boson decays to two photons
together with missing transverse energy (\MET).  This decay is an
example of an interesting class of {\it semi-invisible} decays, where
visible objects in the final state are accompanied by one or more
detector-stable particles.  Decays featuring multiple electroweak
objects together with \MET\ generally have good prospects at the
LHC~\cite{Curtin:2013fra}, and present an obvious target for Run~2.
However, this class of signatures poses some questions for analysis
design.  First, the multiple possible topologies that can contribute
to any specific final state raise the question of how to design an
analysis strategy capable of providing good coverage to more than one
signal model.  Second, such decays offer several possible trigger
strategies. The presence of electroweak objects in the final state can
potentially make it possible to trigger on the large population of
Higgs bosons produced through gluon fusion, significantly enhancing
the statistical reach.  On the other hand, the relative softness of
these electroweak objects can mean that weak vector boson-associated
or vector boson fusion production modes offer better sensitivity given
realistic trigger thresholds.  Which trigger strategy offers the best
sensitivity to a given decay mode is generally not immediately
obvious, and will depend in detail on the mass spectrum of the BSM
particles produced in the Higgs boson decay.

Here, to illustrate these general points and to begin to answer these
questions, we consider two simplified models yielding Higgs boson decays to
a final state consisting of two photons and \MET.
In the first, non-resonant, case, the photons arise from opposite
sides of an initial two-body decay: $h\to XX, X\to\gamma Y$, where $Y$
is a stable neutral particle. Such a decay can occur for instance
within gauge-mediated supersymmetry breaking (GMSB) SUSY models, in
which the $X$ corresponds to a neutralino next-to-lightest
supersymmetric particle (NLSP) with mass less than half the 
Higgs boson mass, and the $Y$ corresponds to a gravitino LSP~\cite{Djouadi:1997gw,
  Mason:2009qh, Petersson:2012dp}.
In the second, resonant, case, the photons are produced through an
intermediate resonance: $h\to S_1 S_2$, with $S_1 \to\gamma\gamma$ on
one side of the decay, while $S_2$ escapes detection, appearing as
\MET\hspace{0.03cm} in the detector. This signal can arise in
e.g. hidden valley scenarios \cite{Strassler:2006ri}.  The resonant
signal benefits from a peak in the diphoton invariant mass spectrum.
The Feynman diagrams for the non-resonant and resonant decays can be
seen in Figure~\ref{fig:FEYN_SIG}.

Previous searches for the $\gamma\gamma+\MET$ final state in the low
energy regime include searches for the non-resonant Higgs boson decay in the
supersymmetric scenario described above. 
The current bounds on this decay mode come from  searches using both gluon fusion and ZH production for CMS \cite{Khachatryan:2015vta}, and using vector boson fusion for ATLAS  \cite{ATLAS:2015bra}. Both the CMS and ATLAS analyses directly search for the decay $\hsm \to \gamma + \met$, which can be sensitive to the decay $\hsm \to 2\gamma+\met$ when one of the photons is not reconstructed. The CMS search sets a 95\% CL limit on branching ratios larger than 8\%-10\% on this decay, with the neutralino mass ranging between 1 and 60 GeV, and assuming SM Higgs boson production and depending on the assumed topology of the decay; ATLAS sets a  95\% CL limit of 20\%-30\%  under the same assumptions.

In this study, we devise a search strategy for the $\gamma\gamma+\MET$
final state, motivated by the exotic decays of the Higgs described
above. We estimate the sensitivity of this search for $100$ fb$^{-1}$
of $\sqrt{s}=14$ TeV $pp$ data from the LHC.

\begin{figure}[thbp]
\centering
\includegraphics[scale=0.2]{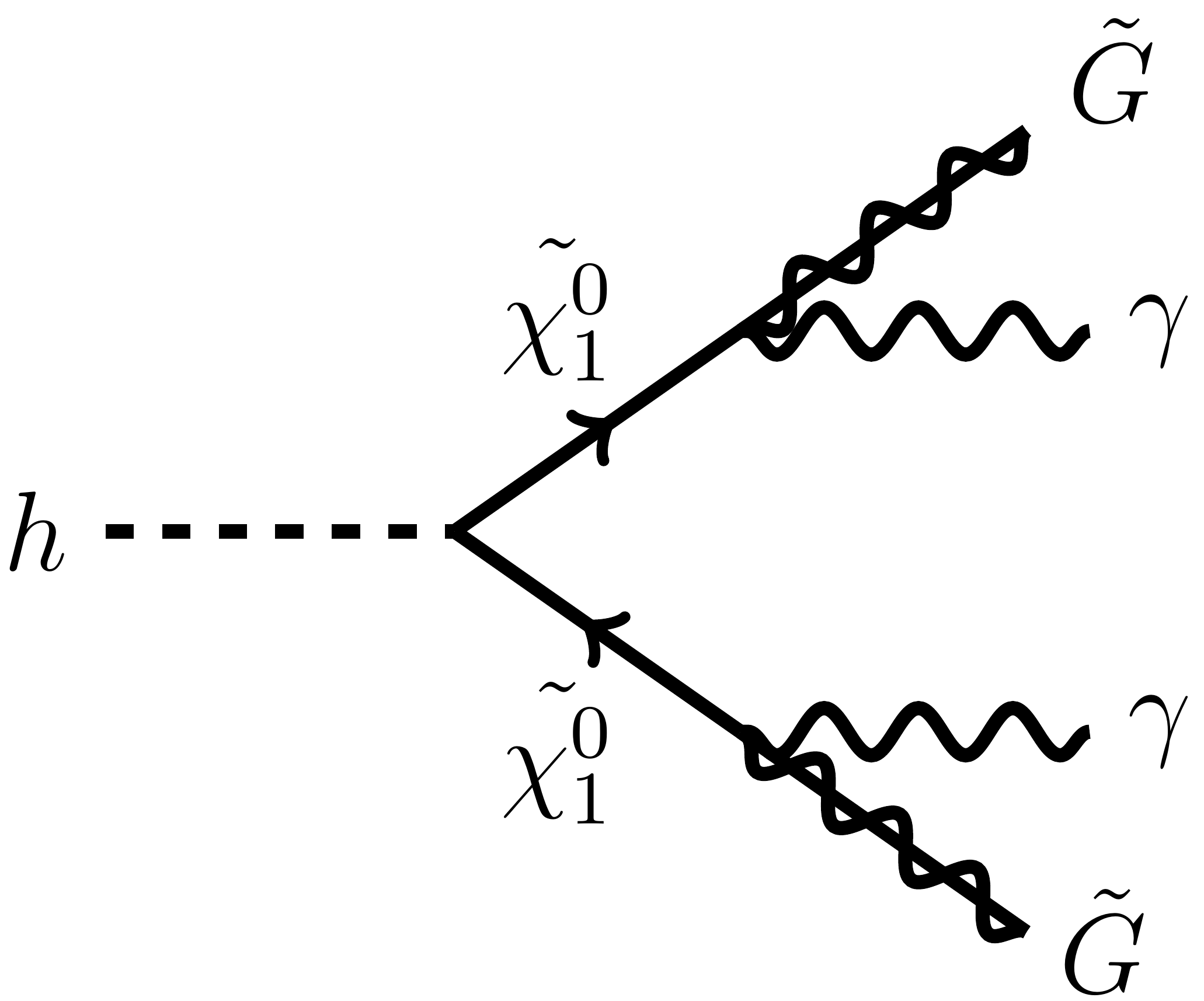}
\hspace{1cm}
\includegraphics[scale=0.2]{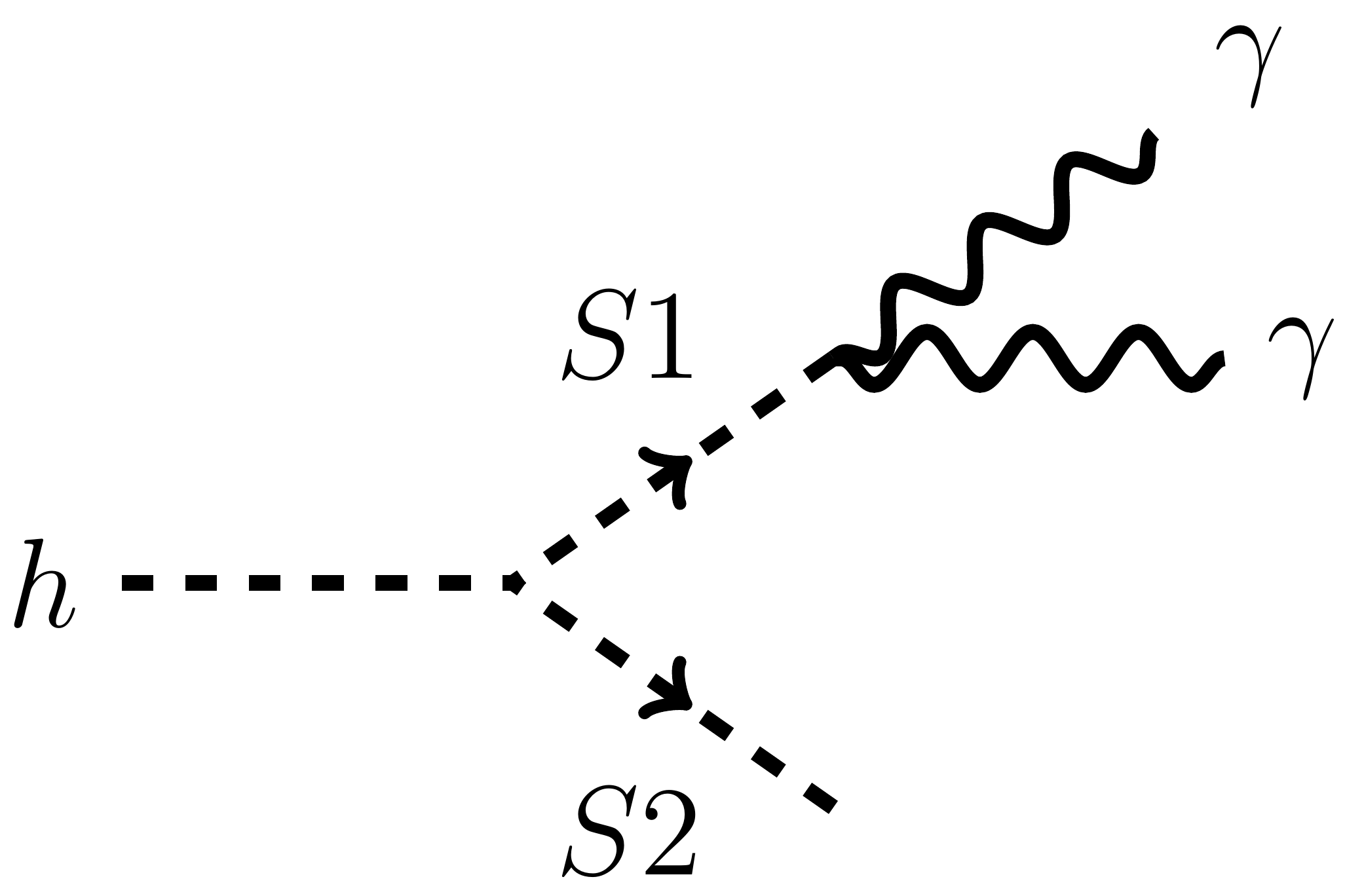}
\caption{Feynman diagrams for (left) the non-resonant and (right) the resonant signal scenarios.}
\label{fig:FEYN_SIG}
\end{figure}

\subsection{Methodology}
\subsubsection{Simulation Samples \label{sub:simulation}}

Signal and background Monte Carlo (MC) samples were generated with
{\tt MadGraph 5}~\cite{Alwall:2011uj} and hadronized with {\tt Pythia
  8}~\cite{Sjostrand:2007gs}, with the detector simulation provided by
{\tt DELPHES 3}~\cite{deFavereau:2013fsa}.
The samples were produced at $\sqrt{s} = 14$ TeV.  The object
reconstruction and identification are performed with {\tt DELPHES},
according to the information provided in the detector configuration
card. For photon reconstruction and identification, we assume an
efficiency of $95\%$ in the electromagnetic calorimeter barrel
($|\eta| < 1.5$) and $85\%$ in the endcap ($1.5 < |\eta| < 2.5$). We
also impose an isolation cut on the photons by requiring all tracks, neutral hadrons and photons reconstructed by {\tt DELPHES}
 within a cone of $\Delta R < 0.3$ of the photon candidate to
have an energy ratio less than 0.1 with respect to the photon
candidate. For muons, we assume an efficiency of $95\%$ for the whole
detector acceptance ($|\eta| < 2.5$). An isolation cut similar to the
photons is also applied. Jets are reconstructed with the anti-$k_T$
algorithm with jet radius $R = 0.4$.

\medskip
\noindent {\bf Signal Monte Carlo.}
\noindent The signal for the non-resonant case was based on the
supersymmetric cascade decay of the Higgs boson into two neutralinos,
which subsequently decay into two gravitinos and two photons (Figure
\ref{fig:FEYN_SIG}, left). This class of models has been implemented
in {\tt FeynRules} \cite{Christensen:2013aua} and generated using {\tt
  MadGraph}. We assume a gravitino mass close to zero, which is
consistent with gauge mediated low-scale SUSY breaking models with
$\sqrt{f} \approx$ TeV \cite{Petersson:2012dp}. We simulate neutralino
masses in the range $10\,\mathrm{GeV}\leq m_\chi\leq 60$ GeV in steps
of $5$ GeV, with 100,000 events per mass point.

For the resonant case, we assume the Higgs boson decays into two
scalar particles, $S_1$ and $S_2$ (Figure \ref{fig:FEYN_SIG},
right). One of the scalars then decays into two photons, while the
other escapes detection. For this study, we assume the masses of these
two particles are the same; this choice was made for simplicity, but
for detailed studies, more combinations should be investigated. We
generate samples with $M_{1} = M_{2}$ in the range $10\,\mathrm{GeV}
\leq M_1\leq 60$ GeV, in steps of $5$ GeV, with 100,000 events per
mass point.

We investigate the production of the Higgs boson through both gluon
fusion (ggF) and associated production with a $Z$ boson (ZH), with
the $Z$ boson decaying to two muons. The inclusion of the di-electron
decay of the $Z$ can also be considered for future studies.
A branching ratio of $\br(\hsm\to\gamma\gamma+\MET) = 10\%$ is assumed
for the signal. This value of the branching ratio was chosen to be
within the current bounds on the Higgs boson width, yet close to the 8
TeV limits from the search for Higgs boson decays to the monophoton final
state ($h\to\gamma+\MET$)~\cite{Khachatryan:2015vta}.

\medskip
\noindent {\bf Background Monte Carlo.}
\noindent Although this analysis is not guaranteed to be entirely free
from QCD multi-jet backgrounds, it has been shown in similar analyses
primarily targeting $\hsm \to \gamma+\MET$ (such as
\cite{Khachatryan:2015vta}) that it is possible to reduce QCD
backgrounds to a sub-dominant contribution. As we require two photons
for most aspects of this analysis, we expect that multi-jet
backgrounds will typically be less important than in
Ref.~\cite{Khachatryan:2015vta}.  For this reason, no pure QCD sample
was produced for this study. As such, the remaining backgrounds for
this analysis arise predominantly from single boson ($\gamma$/$Z$/$W$)
plus jets and diboson processes.

Backgrounds were modeled using the Snowmass LHE simulation
samples~\cite{Anderson:2013kxz}. These consist of single boson samples
($\gamma/Z/W$) with at least one jet and inclusive diboson
($\gamma\gamma/Z\gamma/W\gamma/$ $WW/ZZ/WZ$) samples. The samples
include both hadronic and leptonic decays of the $W$ and $Z$
bosons. The cross sections used for normalizing the single boson
samples were estimated with {\tt MCFM}~\cite{Campbell:2010ff},
assuming an efficiency of $15\%$ for the one jet requirement (as
obtained with {\tt MadGraph}). For the diboson samples, the cross
sections used were estimated from Ref.~\cite{Campbell:2011bn}. The
cross sections and number of events in the samples are shown in Table
\ref{tab:sel_eff}.

\subsubsection{Event Selection}

\medskip
\noindent {\bf Trigger Projections.}

For the ZH channel, the trigger strategy is expected to be
straightforward and can be based on the decay of the $Z$ to two muons.
On the other hand, triggering is one of the main challenges for the
ggF channel, since the final state consists of two soft
photons plus (a relatively small amount of) missing energy. The
standard triggers used for $h\rightarrow\gamma\gamma$ analyses in CMS
typically have a diphoton invariant mass cut which makes it
incompatible with the low energy spectrum of this
analysis. However, we have
identified three possible trigger strategies for this channel, based
on unprescaled triggers used by the CMS experiment in Run~2:
\begin{itemize}
\item Asymmetric Diphoton Trigger: This trigger requires two photons
  with different $E_{T}$ and trigger-level identification
  requirements, plus a diphoton invariant mass cut. This type of
  trigger usually has a non-negligible turn-on curve in the leading
  and subleading photon $E_{T}$.
\item Symmetric Diphoton Trigger: This trigger requires two photons
  with the same $E_{T}$ requirement, without any extra requirements.
\item $\gamma+\MET$ Trigger: This trigger requires only one barrel
  photon passing identification requirements and a $E_{T}$ requirement
  that is usually higher than the previous two triggers. In addition,
  there is a calorimetric \MET\, requirement. We expect non-negligible
  turn-on curves with respect to both photon and \MET\, for this
  trigger.
\end{itemize}
The three triggers described here represent different selection
strategies that were investigated and will be described below.

\medskip
\noindent {\bf Offline Selection.}
\noindent In the ggF analysis, events are triggered based on the
properties of the photons, and the selection cuts must reflect the
chosen trigger strategy, while maintaining a good signal efficiency.
The ZH-produced signal events are tagged through the decay of the
$Z$ boson to muons, minimizing the largest backgrounds.  The photon
selection is chosen to maximize the signal acceptance in the ZH
case, with $E_{T}$ thresholds as low as possible.  The final event
selection requirements for the ggF and ZH channels are
summarized in Table~\ref{tab:sel_all}. In this table, we use the
following definitions for transverse mass:
\begin{align}
M_{T}(\gamma\gamma,\MET) &= \sqrt{2E_{T}(\gamma\gamma)\MET(1-\cos(\Delta\phi(\gamma\gamma,\MET))}, \\
M_{T}(\gamma\gamma+\MET, \mu\mu) &= \sqrt{2E_{T}(\gamma\gamma+\MET)p_{T}(\mu\mu)(1-\cos(\Delta\phi(\gamma\gamma+\MET,\mu\mu))}.
\end{align}


\begin{table}
\caption{Analysis selection for the ggF channel (for each trigger scenario) and the ZH channel.}
\label{tab:sel_all}
\centering
\begin{tabular}{r | l | l | l| l}
\toprule
 & \multicolumn{3}{c|}{ggF} & \multicolumn{1}{c}{ZH} \\ 
\midrule
Variable & Asymmetric $\gamma\gamma$ & Symmetric $\gamma\gamma$ & $\gamma+\MET$ & \\ \hline
Number of photons       & $> 1$         & $> 1$         & $> 1$         & $> 1$\\
$p_{T}(\gamma_{1})$     & $ > 45$ GeV   & $ > 40$ GeV   & $ > 55$ GeV   & $ > 20$ GeV\\
$|\eta(\gamma_{1})|$    & $ < 2.5$      & $ < 2.5$      & $ < 1.4$      & $< 2.5$ \\
$p_{T}(\gamma_{2})$     & $ > 30$ GeV   & $ > 40$ GeV   & $ > 20$ GeV   & $ > 20$ GeV\\
$|\eta(\gamma_{2})|$    & $ < 2.5$      & $ < 2.5$      & $ < 2.5$      & $ < 2.5$ \\
$M(\gamma\gamma)$       & $\in [15, 100]$ GeV & $< 100$ GeV & $< 100$ GeV & $< 100$ GeV \\
$\MET$                  & $> 90$ GeV    & $> 90$ GeV    & $> 90$ GeV    & $> 60$ GeV \\
$M_{T}(\gamma\gamma,\MET)$                    & $< 140$ GeV   & $< 140$ GeV   & $< 140$ GeV   & $< 140$ GeV \\
$\Delta\phi(\gamma\gamma,\MET) $& $< 1.5$ & $< 1.5$ & $< 1.5$ & $< 1.5$ \\
Number of leptons       & $< 1$         & $< 1$         & $< 1$         & 2 muons \\
\midrule
$p_{T}(\mu_{1,2})$ & -  & - & - & $ > 20$ GeV \\
$|\eta(\mu_{1,2})|$ & - & - & - & $ < 2.5$ \\
$M(\mu\mu)$ & - & - & - & $\in [75,115]$ GeV\\
$M_{T}(\gamma\gamma+\MET, \mu\mu)$ & - & - & - & $> 400$ GeV\\
\bottomrule
\end{tabular}
\end{table}

To exploit the topology of the resonant signature, we apply an
additional requirement of a $\pm10$ GeV mass window, in the diphoton
invariant mass distribution ($M(\gamma\gamma)$), around the signal
mass ($M_1$). The efficiencies for each individual process and the
different searches, after the full selection (without the
$M(\gamma\gamma)$ mass window requirement), are shown in Table
\ref{tab:sel_eff}.

For the ZH case, we also explore the strategy performed by CMS in
their Run~1 result \cite{Khachatryan:2015vta}, in which one or more
photons are required in the event, instead of two or more. In this
case, we gain back the efficiency that is lost due to the inefficiency
in reconstructing the subleading photon, which can have very low
$E_T$. The selection is similar to what is described in Table
\ref{tab:sel_all}, but without the $M(\gamma\gamma)$ cut or the mass
window requirement for the non-resonant topology. The other variables
that use the diphoton information are instead reconstructed using only
the leading photon in the event. Plots of some of the discriminating
variables are available at 
\url{https://twiki.cern.ch/twiki/bin/view/LHCPhysics/LHCHXSWGExoticDecayYR4ExtraMaterials}.

\begin{table}
\caption{Cross-sections, numbers of events generated per process, and selection efficiencies for background processes and signal points, for ggF and ZH production mechanisms.  Signal cross-sections are quoted for a 10\% branching ratio.}
\label{tab:sel_eff}
\centering
\resizebox{\textwidth}{!}{
\begin{tabular}{c | c | c | c | c | c| c}
\toprule
 \multirow{ 2}{*}{Process} &  \multirow{ 2}{*}{$\sigma$ (pb)} & \multirow{ 2}{*}{$N_\text{Generated}$} & \multicolumn{3}{c|}{ggF} & \multirow{ 2}{*}{ZH} \\ 
 & & & Asymmetric $\gamma\gamma$ & Symmetric $\gamma\gamma$ & $\gamma+\MET$ & \\ 
 \midrule
\multicolumn{7}{c}{Backgrounds} \\
\midrule
$\gamma$ + Jets    & $1.0\times10^{5}$  &    5425448    &    $1.9\times10^{-6}$  &   $4.7\times10^{-7}$  &   $8.9\times10^{-7}$  &   $\approx 0$   \\
$Z$ + Jets         & $0.94\times10^{4}$   &    1888446    &  $5.6\times10^{-4}$    & $1.5\times10^{-4}$    & $5.0\times10^{-5}$    & $\approx 0$     \\
$W$ + Jets         & $2.96\times10^{4}$  &    5263872    &   $6.2\times10^{-4}$   &  $1.9\times10^{-4}$   &  $2.7\times10^{-5}$   &  $\approx 0$    \\
$\gamma\gamma$     & $10.8\times10^{1}$  &    4268781    &   $3.1\times10^{-5}$   &  $1.0\times10^{-5}$   &  $1.1\times10^{-5}$   &  $\approx 0$             \\
$Z\gamma$          & $6.30\times10^{2}$  &    3406151    &   $4.3\times10^{-4}$   &  $1.4\times10^{-4}$   &  $5.7\times10^{-5}$   &  $\approx 0$    \\
$W\gamma$          & $1.03\times10^{3}$   &    5258034    &  $1.4\times10^{-4}$    & $4.6\times10^{-5}$    & $5.4\times10^{-5}$    & $\approx 0$     \\
$WW$               & $1.24\times10^{2}$  &    8059829    &   $2.6\times10^{-1}$   &  $8.4\times10^{-2}$   &  $9.8\times10^{-5}$   &  $8.2\times10^{-8}$    \\
$ZZ$               & $1.8\times10^{1}$ &    1101611    &     $1.4\times10^{-2}$ &    $4.7\times10^{-3}$ &    $6.7\times10^{-4}$ &    $7.3\times10^{-6}$ \\
$WZ$               & $5.1\times10^{1}$ &    3319770    &     $3.6\times10^{-1}$ &    $1.2\times10^{-1}$ &    $2.5\times10^{-4}$ &    $2.9\times10^{-6}$  \\
\midrule
\multicolumn{7}{c}{Signals} \\
\midrule
Res., M = 10 GeV     &  $10.8\times10^{1}$   &    4268781    &  $2.5\times10^{-4}$   &  $2.2\times10^{-4}$   &  $1.7\times10^{-4}$   &  $5.7\times10^{-4}$       \\
Res., M = 40 GeV     &  $6.30\times10^{2}$   &    3406151    &  $8.7\times10^{-3}$   &  $5.7\times10^{-3}$   &  $5.0\times10^{-3}$   &  $6.9\times10^{-3}$      \\
Res., M = 60 GeV     &  $1.03\times10^{3}$    &    5258034    & $1.6\times10^{-2}$    & $1.1\times10^{-2}$    & $1.1\times10^{-2}$    & $9.2\times10^{-3}$         \\
Non-Res., M = 10 GeV &  $1.24\times10^{2}$   &    8059829    &  $1.5\times10^{-3}$   &  $9.5\times10^{-4}$   &  $1.1\times10^{-3}$   &  $1.1\times10^{-3}$       \\
Non-Res., M = 40 GeV &  $1.8\times10^{1}$  &    1101611    &    $8.0\times10^{-3}$ &    $5.5\times10^{-3}$ &    $5.2\times10^{-3}$ &    $6.3\times10^{-3}$     \\
Non-Res., M = 60 GeV &  $5.1\times10^{1}$  &    3319770    &    $1.1\times10^{-2}$ &    $7.6\times10^{-3}$ &    $6.9\times10^{-3}$ &    $8.1\times10^{-3}$     \\
\bottomrule
\end{tabular}
}
\end{table}

\afterpage{\clearpage}

\subsubsection{Background Estimation for Misidentified Photons}

Background processes with misidentified (or ``fake'') photons, such as
jets and electrons erroneously reconstructed as photons, that pass the final
selection generally have very low efficiency at the LHC. Nonetheless,
such backgrounds may be non-negligible since the production
cross-sections can be large. Such mis-identification rates are
typically measured with data-driven methods at the LHC.
Although this study was limited by MC statistics in measuring fake
photon backgrounds, a method was developed to mitigate this problem,
which we describe below.

The object reconstruction and selection is done at {\tt DELPHES}
level, where, given the photon identification requirements described
in Section \ref{sub:simulation}, we obtain an associated fake
rate. These fake rates are accounted for in the overall efficiencies
in Table \ref{tab:sel_eff}. In order to bypass the efficiency loss due
to the small fake rates, we select jets and electrons to be
redesignated as fake photon candidates. For the background processes
with one prompt photon ($\gamma$+jets, $W\gamma$ and $Z\gamma$), we
select one fake photon candidate. For the processes with no prompt
photons ($W/Z$+jets, $WW$, $WZ$ and $ZZ$), we select two fake photon
candidates. No fake photon selection is done for the
$\gamma\gamma$+jets sample.

With the assumption of a flat fake rate for both jets and electrons,
the fake photon candidates are randomly selected from the jets and
electrons that passed the photon acceptance requirements. One extra
assumption is that the electron-to-photon fake rate is set to be one
order of magnitude larger than the jets-to-photon fake
rate. Therefore, electrons are set to have a probability of being
selected to be redesignated as a photon that is ten times higher than
for jets.

After the choice of fake photon candidates, we calculate weights for
the individual samples based on the $E_{T}$ spectrum of the selected
photons (prompt and fake) to match the spectrum found with the photon
candidates reconstructed directly from {\tt DELPHES}. This
reweighting is done on the sum of $E_{T}$ of the two leading photons
for samples with at least one prompt photon, and on the $E_{T}$ of the
leading photon for samples with no prompt photon. An independent
reweighting is also done in $\eta$. Both reweightings reflect the
different reconstruction efficiencies and energy resolutions of
objects that are not reconstructed as photons (i.e., electrons and
jets). After applying the weights, we observe good agreement between
the kinematic distributions of interest arising from photons
reconstructed by {\tt DELPHES}, and in particular from events with two photons
reconstructed by {\tt DELPHES} in samples containing only one prompt
photon at truth level, and from our fake photon candidates.

\subsection{Results}

We present the expected sensitivity of this search in terms of the
necessary $h\rightarrow\gamma\gamma+\MET$ branching ratio to reach a
$5\sigma$ sensitivity for an assumed integrated luminosity of $100$
fb$^{-1}$ at $\sqrt{s} = 14$ TeV, with the sensitivity defined as:
\begin{equation}
\mathcal{S} = \frac{N_\text{Signal}}{\sqrt{N_\text{Background}}}.
\end{equation}

In Figure \ref{fig:triggers}, we show the sensitivity plot for the
different trigger scenarios of the ggF case.  This plot shows
that, after the full selection, the performance of the different
trigger strategies is comparable. Although it is safe to assume that a
diphoton trigger with a low $M(\gamma\gamma)$ cut will be present in
the future trigger menus of CMS and ATLAS, we choose to perform the
analysis in the $\gamma+\MET$ case. We make this choice as an effort
to make the case for the existence of such a trigger strategy for the
future LHC runs. While the diphoton triggers are designed with
specific usages that are already well established, the
$h\rightarrow\gamma\gamma+\MET$ analysis could be viewed as a
benchmark for the $\gamma+\MET$ trigger for three reasons:
\begin{itemize}
\item It is a trigger that is already present at the LHC experiments,
  but can be retuned with a specific analysis as benchmark;
\item A dedicated trigger for this analysis requiring two photons
  might not be as efficient at trigger level, given the soft spectrum
  of the second photon;
\item This trigger can also be used for other exotic searches, such as
  the extension to low energies of the dark matter searches in the
  monophoton channel.
\end{itemize}

\begin{figure}[thbp]
\centering
\includegraphics[height=3.8in]{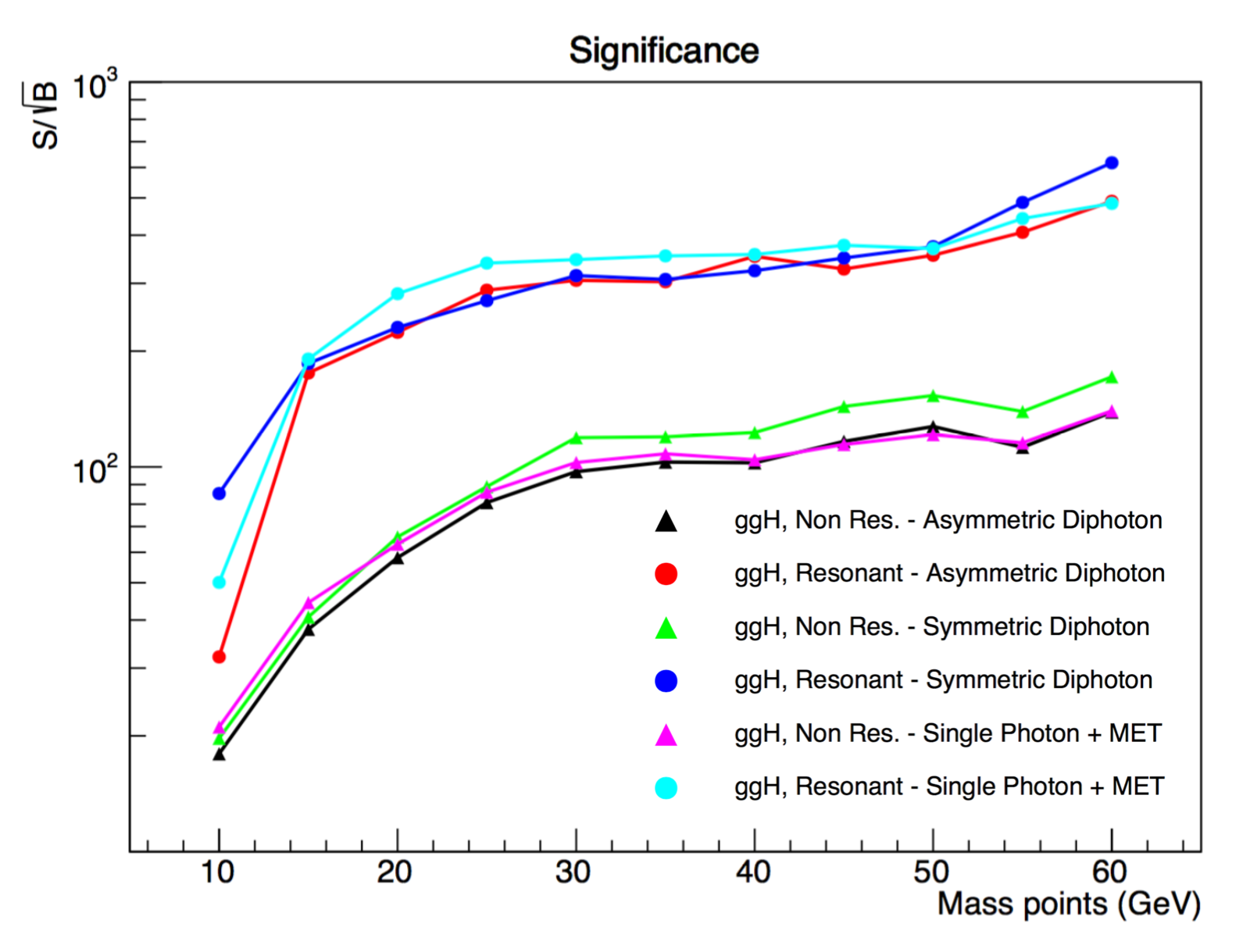}\caption{Statistical
  significance corresponding to different trigger scenarios in the gluon
  fusion analysis, for a reference signal branching ratio of
  $\br(\hsm\to\gamma\gamma+\MET) = 10\%$.}
\label{fig:triggers}
\end{figure}

In Figure \ref{fig:branching_5sigma}, on the left, we show the
branching ratio of $h\rightarrow\gamma\gamma+\MET$ needed for a
significance of $5\sigma$, assuming the Standard Model Higgs cross
section, for the ggF analysis (assuming the $\gamma+\MET$
trigger strategy and selection). On the right, we show the branching
ratio $h\rightarrow\gamma\gamma+\MET$ needed for a significance of
$2\sigma$, which represents the $95\%$ confidence level for exclusion,
assuming SM ZH production. For the ZH case, we show the results
for the strategies requiring at least one ($N_\gamma \geq 1$) and at
least two ($N_\gamma \geq 2$)
photons.

\subsubsection{Systematic Uncertainties}

While the uncertainties in the ZH channel are expected to be
dominated by statistics, the ggF channel is very sensitive to
the systematic uncertainties associated with the background
predictions. We estimate the effect of these uncertainties by
parameterizing the sensitivity as:
\begin{equation}
\mathcal{S}_{sys} = \frac{N_\text{Signal}}{\sqrt{N_\text{ Background}}+\sigma_{sys}\times N_\text{Background}},
\label{eqn:syst}
\end{equation}
with $\sigma_{sys}$ representing a source of uncertainty that does not scale with the amount of statistics.
Figure \ref{fig:branching_5sigma} shows the effect on the $5\sigma$ branching ratios due to the addition of a $10\%$ systematic uncertainty according to Eq. (\ref{eqn:syst}).

\begin{figure}[thbp]
\centering
\includegraphics[height=2.1in]{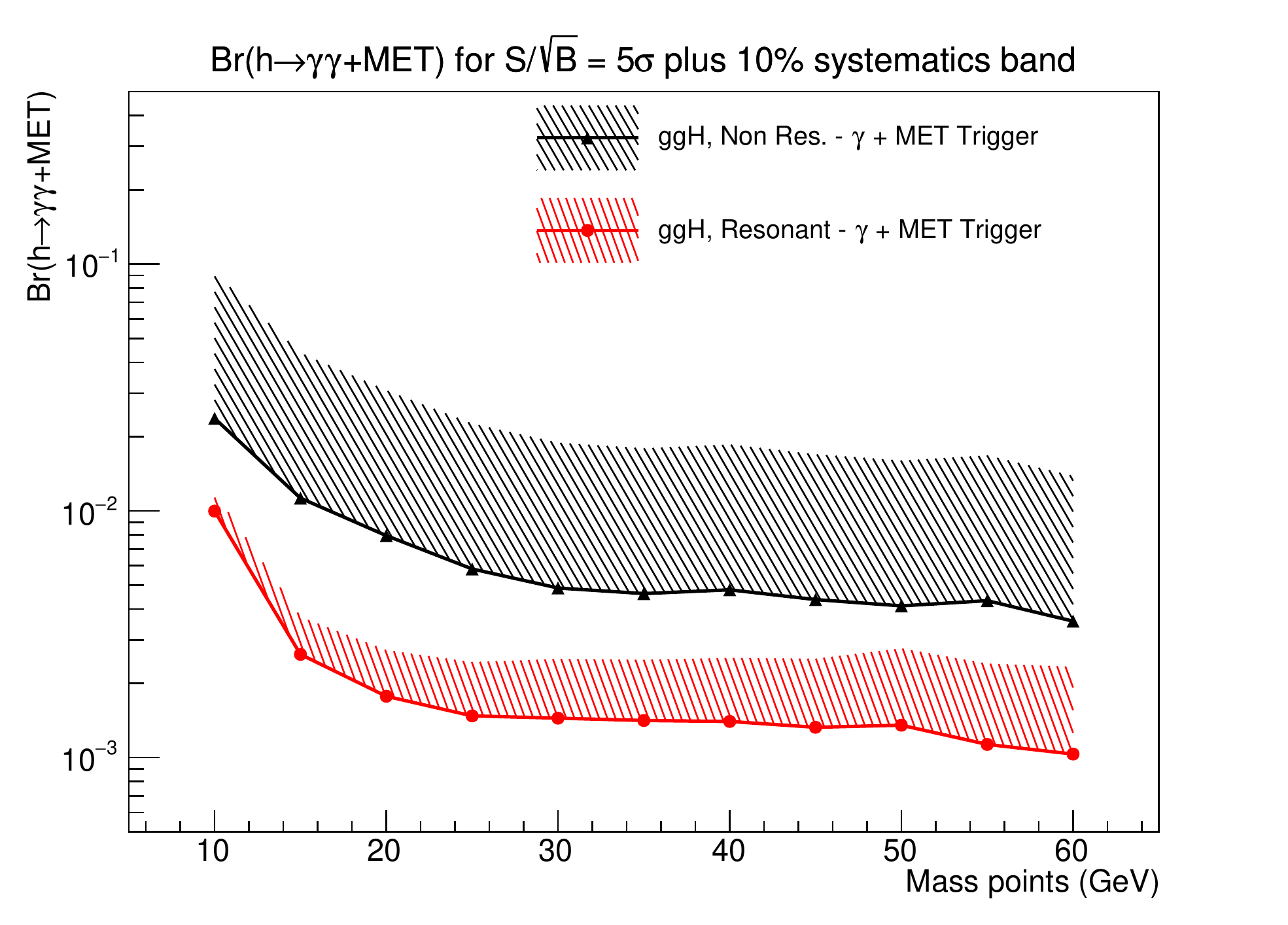}
\includegraphics[height=2.1in]{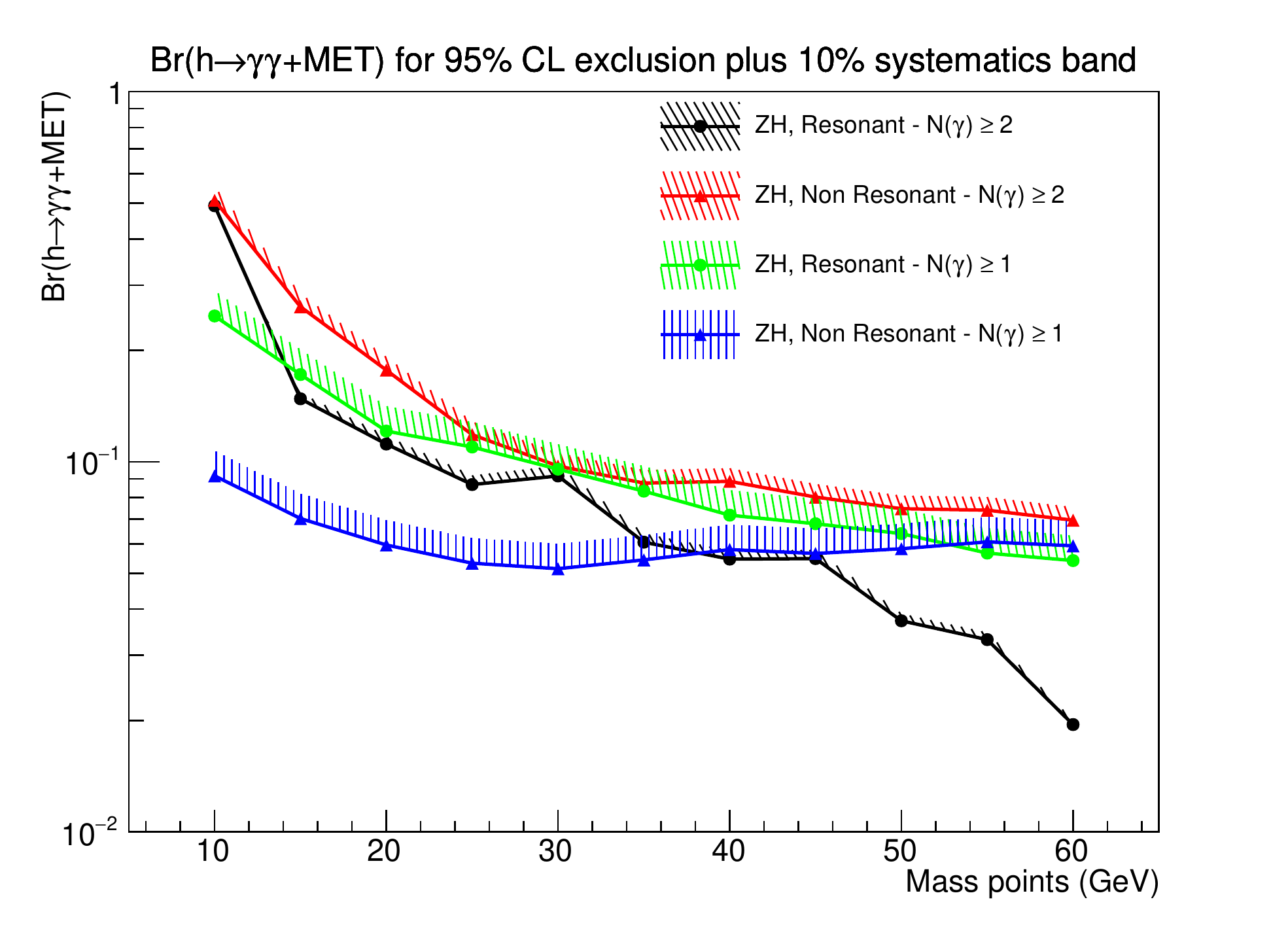}
\caption{(Left) $5\sigma$ branching ratios for the ggF
  channel, for resonant (in red) and non-resonant (in black) final
  states, using the $\gamma+\MET$ trigger. (Right) Branching ratios
  for 95$\%$ confidence level exclusion in the ZH case, resonant and
  non-resonant topologies, requiring at least one photon ($N_\gamma
  \geq 1$, in green and blue, respectively) and at least two photons
  ($N_\gamma \geq 2$ in black and red, respectively). The shaded areas
  correspond to a variation in systematic uncertainties up to $10\%$. }
\label{fig:branching_5sigma}
\end{figure}

\afterpage{\clearpage}

\section[Long lived particles from Higgs boson decays]{Long lived particles from Higgs boson decays\SectionAuthor{D.~Curtin, M.~Strassler}}
\label{sec:displaced}
\subsection{Overview and motivation}
\label{ss.overview}

Long-lived particles (LLPs), specifically meta-stable particles with
proper lifetimes $c \tau \gtrsim \mu$m, arise in a large variety of
BSM scenarios. Such particles, once produced at the LHC or other
colliders, can decay within the detector volume with measurable
displacement from the interaction point. For experimental searches,
this represents both a challenge and an opportunity. On the one hand,
the ATLAS and CMS detectors were not specifically optimized for
displaced decays, which can make triggering and reconstruction
challenging. On the other hand, events with displaced decays are
spectacular and relatively background-free. This makes LLP searches
enticing discovery avenues for new physics, especially in light of
null results from prompt BSM searches at the LHC Run~1.

LLP's often arise as a part of hidden sectors, which make exotic 
Higgs boson decays a very promising production mechanism
\cite{Curtin:2013fra}. One simple toy model that realizes this
possibility involves the Higgs boson coupling to two (pseudo)scalars $X$
which decay back to the SM, resulting in the decay chain $h \to XX \to
\mathrm{SM}$. The $X$ couplings to SM particles may be inherited from
mixing with the SM-like Higgs (and with any additional doublets, if
present), which gives them Yukawa-weighted branching fractions that
prefer third-generation fermion final states like $X \to \bar b b$ or
$\tau^+ \tau^-$. Due to this preference for heavy-flavour final states,
we will call this scenario ``\HXXHF.'' For prompt $X$-decays this toy
model arises as part of the NMSSM \cite{Dermisek:2005ar} or more
generally the SM+S or 2HDM+S models described in
\cite{Curtin:2014pda}.  This toy model is also commonly relevant in
Hidden Valley models \cite{Strassler:2006im, Strassler:2006ri,
  Strassler:2006qa, Han:2007ae}.  Perhaps the most compelling
motivation for this type of decay is the connection to \emph{Neutral
  Naturalness}, where models like the Fraternal Twin Higgs \cite{Craig:2015pha} 
or Folded SUSY \cite{Burdman:2006tz} give rise to
this decay (as well as others) through the structure of their \mirror
sectors.

Although we will not address most of them here, there are many other
possible scenarios for LLPs arising from exotic Higgs boson decays.  For
example, the Higgs might decay to spin-one bosons with displaced
decays \cite{Strassler:2006im}. Ref.~\cite{Curtin:2014cca} studied the
well-motivated scenario where the Higgs boson decays to two dark photons via
mixing with a dark Higgs. The dark photons can then decay with
long-lifetimes via a small kinetic mixing with SM hypercharge. This
also realizes a $h \to X X \to \mathrm{SM}$ signal model, but now with
\emph{gauge-ordered} branching fractions of $X$, yielding lepton-rich
final states. This scenario (\HXXGO) may be even easier to discover,
and is already more constrained \cite{CMS:2014hka, Aad:2014yea} than
is \HXXHF.\footnote{There is some overlap of the issues that displaced
  dilepton searches \cite{CMS:2014hka, Aad:2014yea} and searches for
  the \HXXHF~scenario have to contend with. However, since triggering
  on displaced leptons, especially muons, is generally much easier
  than triggering on displaced jets, we do not discuss the case of
  displaced dileptons further.}
Other examples can arise naturally within weak-scale extensions of the
SM.  For instance, models with weak-scale right-handed (RH) neutrinos
can feature Higgs boson decays into RH neutrino pairs, with the RH neutrinos
generically long-lived and decaying via $W^{(*)}\ell$ or $Z^{(*)}\nu$
\cite{Graesser:2007yj,Graesser:2007pc,Keung:2011zc,Maiezza:2015lza}.
The MSSM offers many ways to obtain LLPs, which can result in
displaced Higgs boson decays.  For instance, the Higgs can decay to a pair
of bino-like neutralinos with a displaced R-parity-violating decay
\cite{Carpenter:2006hs, AristizabalSierra:2008ye}.  Another example,
which we study in Section~\ref{ss.DP}, occurs in models with
gauge-mediated SUSY-breaking, where the initial Higgs boson decay to
neutralinos is followed by the displaced decay of the neutralino to a
gravitino and a photon \cite{Morrissey:2008gm,Mason:2009qh}.

The simplicity and motivation of the \HXXHF\ scenario makes it an
obvious starting point to explore this class of signals.  In fact,
several experimental searches looking for \HXXHF\ have already been
conducted \cite{Aad:2015asa, Aad:2015uaa}, and these kinds of analyses
will have significant power to probe scenarios like Neutral
Naturalness at LHC Run~2. The success of these studies is highly
encouraging, and prompts us to examine how to improve and broaden
their reach for new physics.

In the course of this discussion, we find it useful to carefully define the following:
\begin{itemize}
\item \emph{Associated Object} (AO): Any conventional detector object,
  such as leptons, VBF jets, or a hard initial state radiation (ISR)
  jet, that is produced in the same event as the LLP(s).
\item \emph{Displaced Object} (DO): An LLP decaying into visible SM
  particles in the detector with a \emph{potentially measurable
    displacement} from the primary interaction point.  Importantly, a
  DO is a ``truth-level'' definition, and does not necessarily imply
  the object can be detected or reconstructed.
\item \emph{Triggerable Displaced Object} (tDO): A DO which can, in
  principle, \emph{be triggered upon}, either by itself or in
  conjunction with an AO.
\item \emph{Displaced Vertex} (DV): an \emph{off-line reconstructed}
  displaced vertex in the tracker or the ATLAS muon system.
 \end{itemize}

 The most obvious strategy for improving the experimental reach is to
 include searches which only require a \emph{single} DO, but have a
 low trigger threshold. In this context, background reduction and
 triggering usually require the presence of an AO arising from 
Higgs boson production. Compared to what is possible in current searches, these
 searches would give access to long-lived Higgs daughters that have
 shorter lifetimes (and thus decay only in the tracker) or lower
 masses (where the LLP's decay products are insufficient to pass
 trigger thresholds).  They may also in some cases increase trigger
 efficiencies. Crucially, such searches are also sensitive to more
 general classes of signals, e.g. $h \to X X'$ where only $X$ decays
 in the detector. Timing may also be an important search strategy for
 heavier $X$.

 Below, in \ssref{DV} we consider displaced objects arising in the
 \HXXHF\ scenario from exotic Higgs boson decays.  In \ssref{DP} we consider
 signals that give rise to displaced photons.

\subsection{Displaced objects}
\label{ss.DV}

We now consider DOs produced in Higgs boson decays.  For brevity we consider
only the \HXXHF\ scenario.  In \sssref{simpmodel} we define a
simplified model for this scenario, and show how to simulate this
signal in {\tt Madgraph}. \sssref{nn} gives a review of Neutral Naturalness
and how it generates the \HXXHF\ scenario. An overview of present
experimental searches is given in \sssref{exp}. Finally, we suggest
new searches for the future years of the LHC in \sssref{suggestions},
and supply some benchmark points to aid in the design of DO searches
that cover the most theoretically motivated parameter space of the
\HXXHF\ scenario.

\subsubsection{A simplified model for the \HXXHF\ scenario}
\label{sss.simpmodel}

The easiest way to parameterize the \HXXHF\ scenario is by introducing a small mixing between a scalar $X$ and the Higgs. (For the purpose of signal event generation the distinction between scalar and pseudoscalar $X$ is immaterial).
However, the coupling which controls the $h \to X X$ decay does not arise from this mixing,  so that the exotic Higgs boson decay branching fraction $\br(h\to XX)$ and the $X$ decay length $c\tau_X$ are independently adjustable. A simple Lagrangian to realize this possibility after electroweak symmetry breaking is the following:
\begin{equation}
\mathcal{L} \supset \frac{1}{2} (\partial_\mu X)^2 - \frac{1}{2} m_X^2 X^2
- g_X v h X X - \epsilon_v v^2 \, h X\ ,
\end{equation}
The three important parameters are 
\begin{itemize}
\item[(a)] the mass $m_X$ of the long-lived particle, 
\item[(b)] $\br(h\to X X)$, which is controlled by the effective coupling $g_X$, and 
\item[(c)] the decay length $c \tau_X = 1/\Gamma_X$, which is controlled by the small mixing parameter $\epsilon_v$. 
(Displaced decays of $X$ require $\epsilon_v \lesssim 10^{-4}$, and so $\epsilon_v$ has a negligible effect on $m_h$ and on $h$ branching fractions to SM particles.)
\end{itemize}

\subsubsection*{Madgraph implementation}

The \HXXHF\ scenario can be realized by repurposing the SM + dark vector + dark Higgs {\tt MadGraph} model of~\cite{Curtin:2014cca} available at 
\url{http://insti.physics.sunysb.edu/~curtin/hahm_mg.html}. 
In this model, $X$ is identified with the dominantly singlet scalar state $\texttt{hs}$, with $m_X$ corresponding to the model parameter $\texttt{MHSinput}$. The singlet $\texttt{hs}$ decays to SM fermions via its mixing with the SM-like Higgs $\texttt{h}$, which is controlled by the model parameter \texttt{kap}.
This {\tt MadGraph} model includes the couplings between the Higgs and SM gauge bosons, including the gluons via an effective operator. $X$ therefore inherits the same couplings, and decays like $X \to V^{(*)} V^{(*)}$ can also be generated.

For a given $\br(h\to X X)$, $\Gamma_X = 1/(c \tau_X)$ and $m_X$, the procedure for generating $h \to XX \to \mathrm{SM}$ events is the following:
\begin{enumerate}
\item Switch off dark photon effects (by setting the \texttt{epsilon} and \texttt{mZDinput} parameters to, say, \texttt{10e-09} and \texttt{1000e+00} respectively.)
\item Set the parameter \texttt{MHSinput} to the desired $m_X$.

\item 
Event generation will depend on how LLPs are handled, specifically whether $X$ is decayed in {\tt MadGraph} or {\it e.g.} in {\tt Pythia}. In the former case, one could generate the processes\\
\texttt{p p $>$ hs hs, hs $>$ b b\~}\\
\texttt{p p $>$ hs hs, hs $>$ ta+ ta-}\\
\texttt{p p $>$ hs hs, hs $>$ ta+ ta-, hs $>$ b b\~}\\
separately (for the most important $b \bar b$ and $\tau^+ \tau^-$ final states), then manually displace the decays. 
In the latter case, $X$ can be left undecayed in {\tt MadGraph}. The lifetime can then be written to the LHE file before running through {\tt Pythia}. 

Another possibility is to produce events with the $X$ decay implemented directly in {\tt Pythia}. In that case the lifetime can be written to the SLHA file. 

\item  Each sample can then be rescaled to the desired $\sigma_h \times \br(h\to XX) \times \mathrm{BR}(XX \to f \bar f f' \bar f ')$. Here $\sigma_h$ is the inclusive Higgs boson production cross section, and $\mathrm{BR}(X \to f \bar f)$ can be computed for a SM-like Higgs boson of mass $m_X$ using HDECAY 6.42 \cite{Djouadi:1997yw}. This method ensures that important NLO QCD and threshold effects are handled accurately, which is not guaranteed if using LO branching fractions generated internally by {\tt  Madgraph} or {\tt Pythia}. Higher-order differential effects in Higgs boson production can be taken into account by reweighing events using Higgs boson $p_T$ spectra according to the recommendations of Secs.~\ref{sec:recommendations} and \ref{sec:4b}. 
\end{enumerate}

\subsubsection{Neutral naturalness}
\label{sss.nn}

Perturbative solutions to the hierarchy problem introduce top partners that cancel the quadratically divergent one-loop Higgs boson mass contribution of the top quark. In most theories, this top partner is related to the top quark by a continuous symmetry like supersymmetry, and carries SM colour charge. In models of Neutral Naturalness the symmetry relating the top to its partner includes a discrete group like $\mathbb{Z}_2$, and does not commute with SM colour. This leads to the possibility of colour-neutral top partners.

Moreover, a \mirror QCD gauge group is usually required in Neutral Naturalness theories. Without it, the top partner's coupling to the Higgs and the SM top Yukawa coupling will run differently, ruining the cancellation between the top loop and the top partner loop.
Neutral Naturalness therefore realizes a Hidden Valley scenario, where \mirror gluons couple to the Higgs via top partner loops. This allows \mirror hadrons to be produced in exotic Higgs boson decays, as shown in \refF{f.mirrrorhadronproduction}. The same coupling then allows \mirror glueballs and \mirror quarkonia to decay back to the SM via an off-shell Higgs, producing DO signatures. Neutral Naturalness is therefore one of the best motivated scenarios producing the $h\to X X \to \mathrm{SM}$ displaced vertex signature \cite{Craig:2015pha}.

Here we describe two archetypal examples of Neutral Naturalness. The first is \emph{Folded SUSY} (FSUSY) \cite{Burdman:2006tz} which features a \mirror sector of sparticles carrying SM electroweak quantum numbers but charged under the \mirror QCD.  The \mirror QCD confines at a (few--10) GeV, and since LEP limits generically forbid  EW-charged particles below $\sim$ 100 GeV, the lightest new particles are always \mirror glueballs.

The second example is the \emph{Twin Higgs} \cite{Chacko:2005pe} featuring SM-singlet fermionic top partners which are part of a \mirror sector containing copies of all SM particles and gauge forces. The original mirror Twin Higgs model has several cosmological problems due to an  abundance of light invisible \mirror states. A simple modification which satisfies all cosmological constraints is the Fraternal Twin Higgs (FTH) model \cite{Craig:2015pha}, which only duplicates the third generation in the \mirror sector.  In that case, the hadrons of \mirror QCD can be made up of \mirror glueballs, \mirror quarkonia, or a mixture of both. 

We now discuss Higgs boson decays to \mirror hadrons, following~\cite{Craig:2015pha} and~\cite{Curtin:2015fna}, with additional results from  \cite{Csaki:2015fba}.

\begin{figure}[thb]
\begin{center}
\includegraphics[width=9cm]{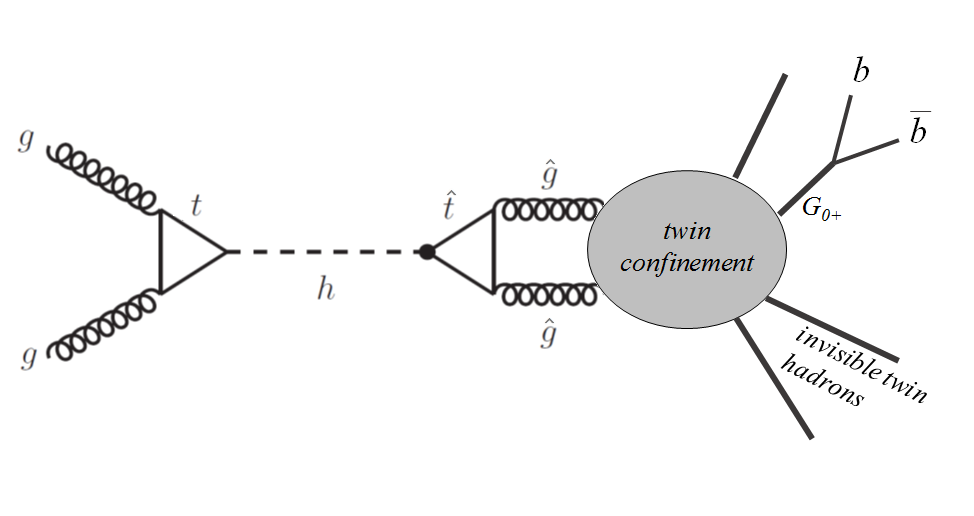}
\end{center}
\caption{Production of \mirror hadrons in exotic Higgs boson decays, and their decay back to the SM, in the Fraternal Twin Higgs model. Figure from \cite{Craig:2015pha}.}
\label{f.mirrrorhadronproduction}
\end{figure}

\subsubsection*{\Mirror Glueballs}

In the absence of light \mirror matter, the lightest states in the \mirror sector are glueballs. A pure $SU(3)$ gauge theory has $\sim$12 stable low-energy states \cite{Morningstar:1999rf}, which can decay on detector timescales when Higgs portal interactions \cite{Juknevich:2009gg} are present. The lightest state is the $0^{++}$ state, $G_0$, with mass $m_0 \approx 7 \Lambda_\mathrm{QCD}$; the heaviest has order twice this mass.  We concentrate on $G_0$ since it has a potentially detectable lifetime and is kinematically the easiest to produce.\footnote{There is a second $0^{++}$ state, heavier and shorter-lived, which may be phenomenologically relevant. In Folded SUSY, the $0^{-+}$ state can also decay through mixing with the SM $Z$-boson; the lifetime and branching fractions of this state are still under study \cite{curtinkatz}. Note also that all formulas in this section have order-one uncertainties due to RG effects, lattice uncertainties, etc.} One can show using RG arguments \cite{Craig:2015pha,Curtin:2015fna} that FSUSY and FTH prefer glueballs in the $\sim (10 - 60) \gev$ mass range. This motivates the study of hidden glueball production in exotic Higgs boson decays.

\Mirror gluons couple to $|H|^2$ via a dimension-6 operator, allowing $G_0$ to decay via its mixing with $h$. For $m_0 \gtrsim 2 m_b$ and top partner mass $m_T$, the decay length is approximately
\begin{eqnarray}
\label{e.glueballlifetime}
c \tau \ &\approx& \  \left( \frac{m_T}{400~\gev}\right)^4  \ \left(\frac{20~\gev}{m_0}\right)^7 \times
\left\{ 
\begin{array}{lll}
\displaystyle 
(35 \mathrm{cm}) & \ \ \ \ \ & \mbox{[FSUSY]}\\
\displaystyle (8.8 \mathrm{cm}) & \ \ \ \ \ & \mbox{[FTH]}
\end{array}
\right.
\end{eqnarray}
where we assume $m_T \gg m_t/2$ for FTH and degenerate unmixed stops for FSUSY. As shown in \refF{f.glueballlifetimegenericestimate} (left) for FSUSY, these decay lengths can range from $\mathcal{O}(10 \mu\mathrm{m})$ to $\mathcal{O}(\rm{km})$ and more, motivating searches for displaced vertices in all detector subsystems.

\begin{figure}[thb]
\begin{center}
\hspace*{-1cm}
\begin{tabular}{ccc}
\includegraphics[height=7cm]{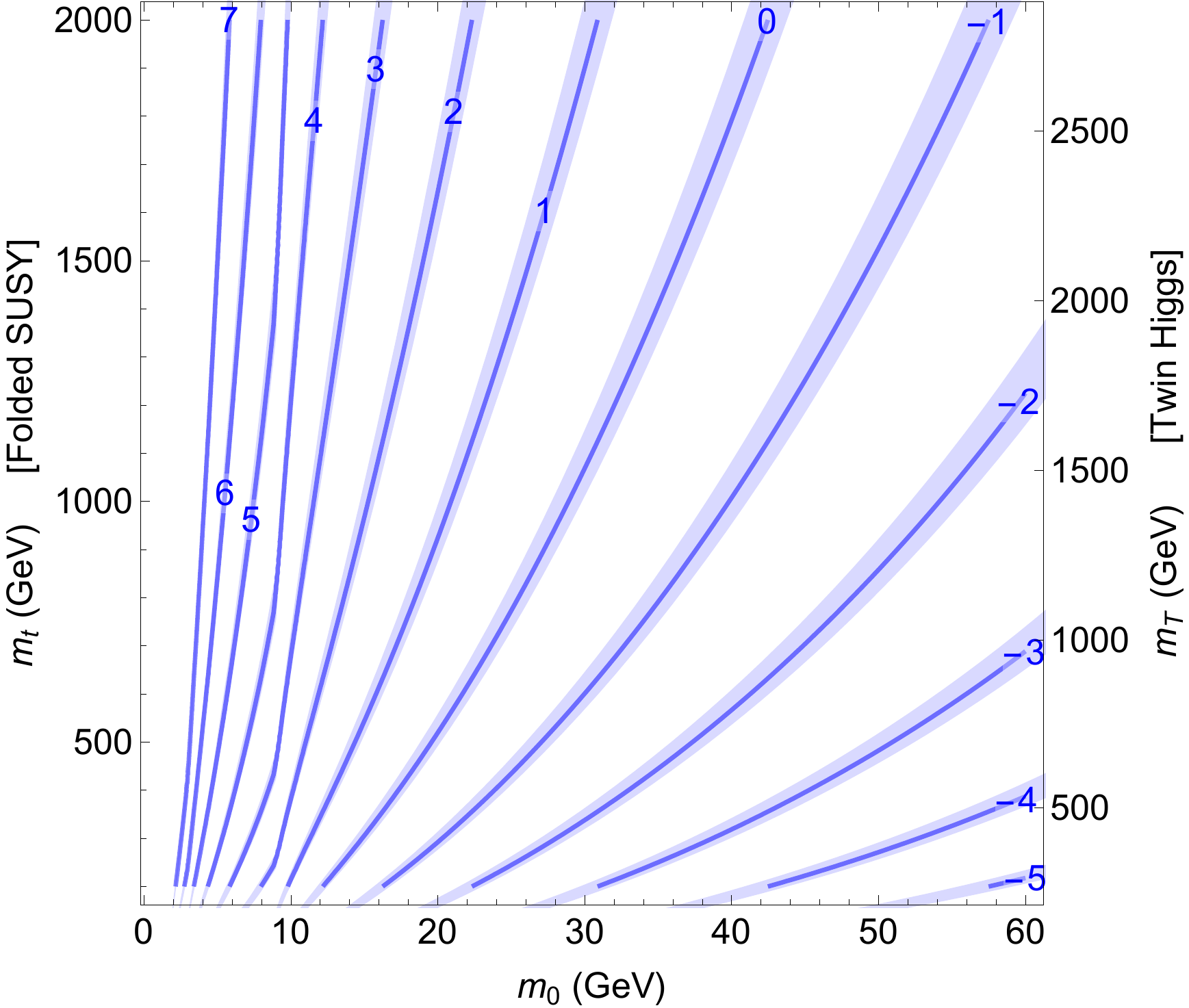}
&&
\includegraphics[height=7cm]
{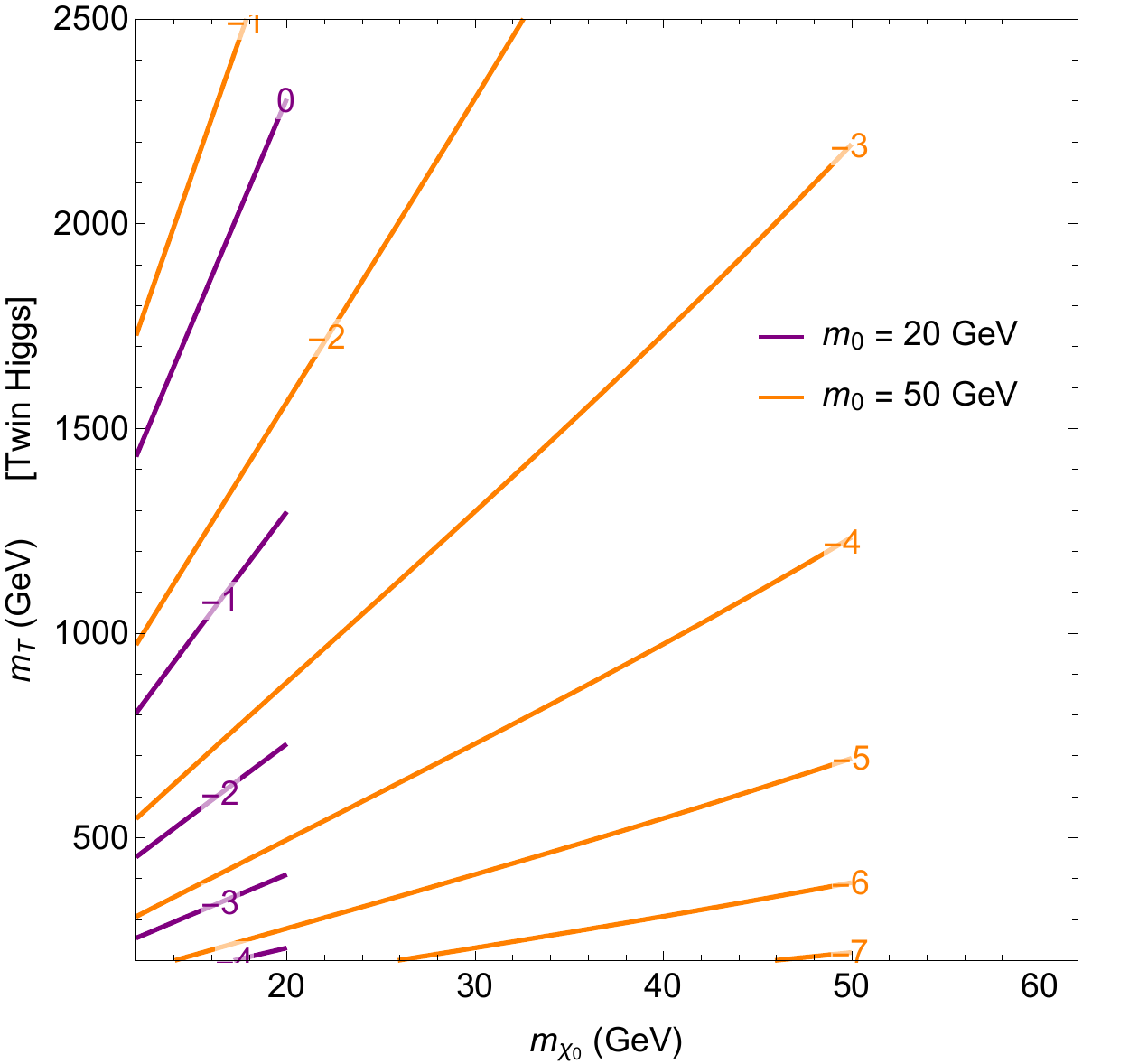}
\\
(a) & & (b)
\end{tabular}
\vspace*{-8mm}
\end{center}
\caption{ (a) Contours show  $\log_{10} (c \tau/ 1\rm{m})$, where $c\tau$ is the proper decay length of the lightest \mirror glueball state $G_0$. The blue bands correspond to the shift of the contours resulting from the $25\%$ uncertainty in the total $G_0$ width \cite{Curtin:2015fna}.
(b) Estimate of the decay length of the \mirror bottomonium $\chi_0$ state, \eref{GammaBottomonium}, for $m_0 = 20$ (in purple) and $50 \gev$ (in orange).  Uncertainties are not shown.
}
\label{f.glueballlifetimegenericestimate}
\end{figure}

The rate for \emph{inclusive} production of \mirror glueballs  from exotic Higgs boson decays can be estimated by rescaling the $\br(h \to \mathrm{gluons})$ which is of the order of $\sim 8\%$ in the SM 
\begin{equation}
\label{e.Brhgginclusive}
\br(h \to \mathrm{\mirror\ glue}) \  \approx \ 10^{-3} \ \left(\frac{400~\gev}{m_T}\right)^4 \  \times  \ 
\left\{ 
\begin{array}{lll}
\displaystyle 
1 & \ \ \ \ \ & \mbox{[FSUSY]}\\
\displaystyle 4 & \ \ \ \ \ & \mbox{[FTH]}
\end{array}
\right.
\end{equation}
Nonperturbative effects could in some regimes reduce or enhance this by a factor of order one \cite{Craig:2015pha}; meanwhile RG effects on the \mirror QCD coupling at scale $m_h$ can increase the branching ratio by a factor of two \cite{Curtin:2015fna}.

In the FTH model this branching fraction can be greatly enhanced if the \mirror bottom quark $B$ has a mass in the range $m_0 \lesssim  m_B < m_h/2$.  In this case $h\to \bar B B$ is possible, but any \mirror bottomonium states annihilate to \mirror glueballs.  The effect is to
enhance the inclusive twin glueball rate to
\begin{equation}
\label{e.BrhBB}
\br(h \to \mathrm{\mirror\ glue})  \approx (h \to \bar B B) \  \approx  \ 0.15 \  \left( \frac{m_{B}}{12~\gev}\right)^2 \ \left( \frac{400~\gev}{m_T}\right)^4  
\ \ \ \   , \ \ (m_B<m_h/2) \
\end{equation}
which can be as large as current limits on the branching ratio into exotics.  From \eref{Brhgginclusive} and \eref{BrhBB},  we expect at least $10^4$ and up to $10^6$ such decays at Run~2 with 300 fb$^{-1}$ data, giving the LHC experiments an attractive target.

Since only a fraction of \mirror glueballs may be $G_0$ states, explicit signal estimates for DO searches require an estimate of \emph{exclusive} $G_0$ production, and thus of $G_0$ production in \mirror hadronization. A conservative assumption is that $G_0$ are produced only in two-body decays $h \to G_0G_0$, which is the \HXXHF\ scenario. This probably underestimates the signal, since light glueballs (which are very long-lived) are likely produced at greater multiplicities and lower boost, effects which increase the number of observable DOs.
Introducing $\kappa$ as a nuisance parameter that encapsulates our uncertainty of \mirror hadronization, we can write the exclusive branching ratio as
\begin{equation}
\label{e.brhgg}
\br(h \to G_0G_0) \ = \  \br(h \to \mathrm{\mirror\ glue})  \ \cdot \ \kappa \ \cdot \sqrt{1 - \frac{4 m_0^2}{m_h^2}} \ ,
\end{equation}
where the phase space factor ensures the branching ratio goes to zero for $m_0 \to m_h/2$. We adopt two benchmark values for $\kappa$ which bracket the range of likely physical outcomes. The most optimistic estimate is that all glueballs produced are $G_0$ pairs, giving $\kappa_\mathrm{max} = 1$. Conversely, a reasonable lower bound on $\kappa$ is to assume democratic production of the light $C$-even glueballs $0^{++}, 2^{++}, 0^{-+}, 2^{-+}$ with masses  $m_i$, $i = 1, \ldots, 4$. In terms of phase space factors $\mathcal{P}(m_i) \equiv \sqrt{1 - 4 m_i^2/m_h^2}$, we then define $\kappa_\mathrm{min} = \mathcal{P}(m_0)/\sum_i n_i \mathcal{P}(m_i)$ with $n_i = 1, 4$ for spin 0, 2 glueballs, which ranges from $\kappa_\mathrm{min}\approx 1/12$ for light glueballs to 1 for heavier glueballs.

The authors of \cite{Curtin:2015fna} performed signal estimates for displaced glueballs from exotic Higgs boson decay for the two benchmark values $\kappa=1,\frac{1}{12}$, which are reproduced in \sssref{suggestions}. The LHC can probe $G_0$ lifetimes corresponding to values of $m_T$ up to the TeV scale with a variety of DO searches.

We should emphasize that there are certain non-perturbative effects and special regimes where this minimal parameterization may not be sufficient, or where $1>\kappa>1/12$ may not be broad enough.  For the near term, however, we recommend these subtleties be ignored in designing searches.

Additional complications may arise when many DO's are clustered in the same region of the detector, or are not isolated from prompt  objects. That being said, these difficulties are unlikely to be prohibitive for DO's decaying in the tracker.

\subsubsection*{\Mirror Bottomonia}

In the FTH, if $m_B < m_0/2$, then $G_0$ (and all other \mirror glueballs) will decay to \mirror bottomonium. The \mirror bottomonium spectrum also contains a $0^{++}$ state, $\chi_0$, which can decay via mixing with the SM-like Higgs. The lifetime of this  state is 
\begin{eqnarray}
\label{e.GammaBottomonium}
\Gamma_{\chi_0\to YY} & \sim & 2\times 10^{-3} \left( \frac{v}{f}\right)^4 \frac{m_{\chi_0}^{11/3} m_0^{10/3}}{v^2 m_h (m_h^2 - m_{\chi_0}^2)^2} \Gamma_{h \to YY}(m_h) \ ,
\end{eqnarray}
assuming there are no light twin neutrinos which could short-circuit this decay mode.
The corresponding proper lifetime for $m_0 = 20$ and 50\,GeV is shown in the $(m_{\chi_0},m_T)$ - plane in \refF{f.glueballlifetimegenericestimate}  (b). The phenomenology of exotic Higgs boson decays and resulting search strategies are broadly similar to the glueball case discussed above, with two notable differences:
\begin{enumerate}
\item The rate of \mirror bottomonium production is approximately given by the exotic Higgs boson decay rate to $\bar B B$, see \eref{BrhBB}, and can be much larger than the direct Higgs boson decay to glueballs; however, $\chi_0$ itself may not be commonly produced in the subsequent hadronization of the mirror bottom quark pair.
\item \Mirror bottomonia can have much shorter decay lengths at low masses $m_{\chi_0} \muchless m_h/2$ than glueballs of the same mass. (In some cases the decay can be prompt, which makes the decay harder to detect.) This motivates searches for low mass DOs at small displacement.
\end{enumerate}

\subsubsection{Experimental analyses}
\label{sss.exp}

At the time of writing, two experimental searches by the ATLAS collaboration have significant sensitivity to the \HXXHF\ scenario discussed here. Ref.~\cite{Aad:2015uaa} used a dedicated trigger sensitive to tDOs in the muon system (MS). In addition to the triggered decay, an additional DV in \emph{either} the muon system or the inner tracker (IT) was required. This stringent requirement made the search effectively background-free, giving it good sensitivity for $X$'s which live long enough to reach the MS.  An earlier search using the HCAL utilized a similar strategy \cite{Aad:2015asa} but is not as sensitive as the search in the MS. 

CMS conducted a search \cite{CMS:2014wda} for heavy particles decaying to at least one long-lived daughter, as arises in e.g. SUSY with R-parity violation (RPV). A $H_T > 300$\,GeV preselection requirement makes this search inefficient to the DOs arising from decays of the relatively light Higgs.  A recast of this search \cite{Csaki:2015fba} estimates \HXXHF\ limits on $\br(h \to X X)$ of order 10\%.  This is considerably better than indirect constraints from coupling fits.  However, it is only marginally sensitive to Neutral Naturalness, see \eref{Brhgginclusive}, though it does encroach upon the parameter space for the Higgs boson decaying to mirror bottomonia, see \eref{BrhBB}. As we discuss below, a different version of this search, with a trigger on VBF jets plus a displaced or trackless jet, could potentially be much more sensitive.

\subsubsection{Suggested searches and benchmarks}
\label{sss.suggestions}

\begin{figure}[thb]
\begin{center}
\begin{tabular}{ccc}
\includegraphics[height=6.6cm]{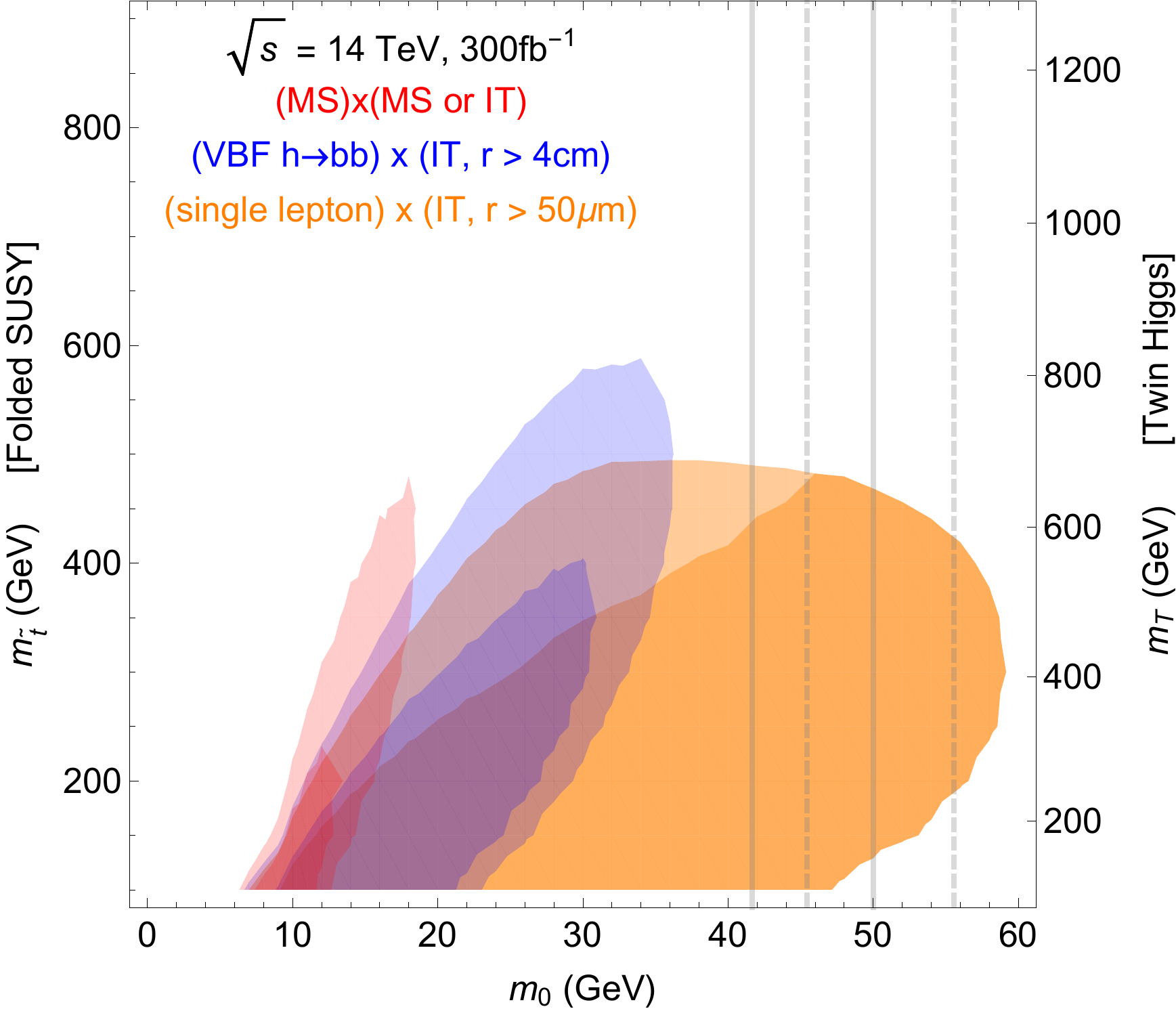} 
& &
\includegraphics[height=6.6cm]{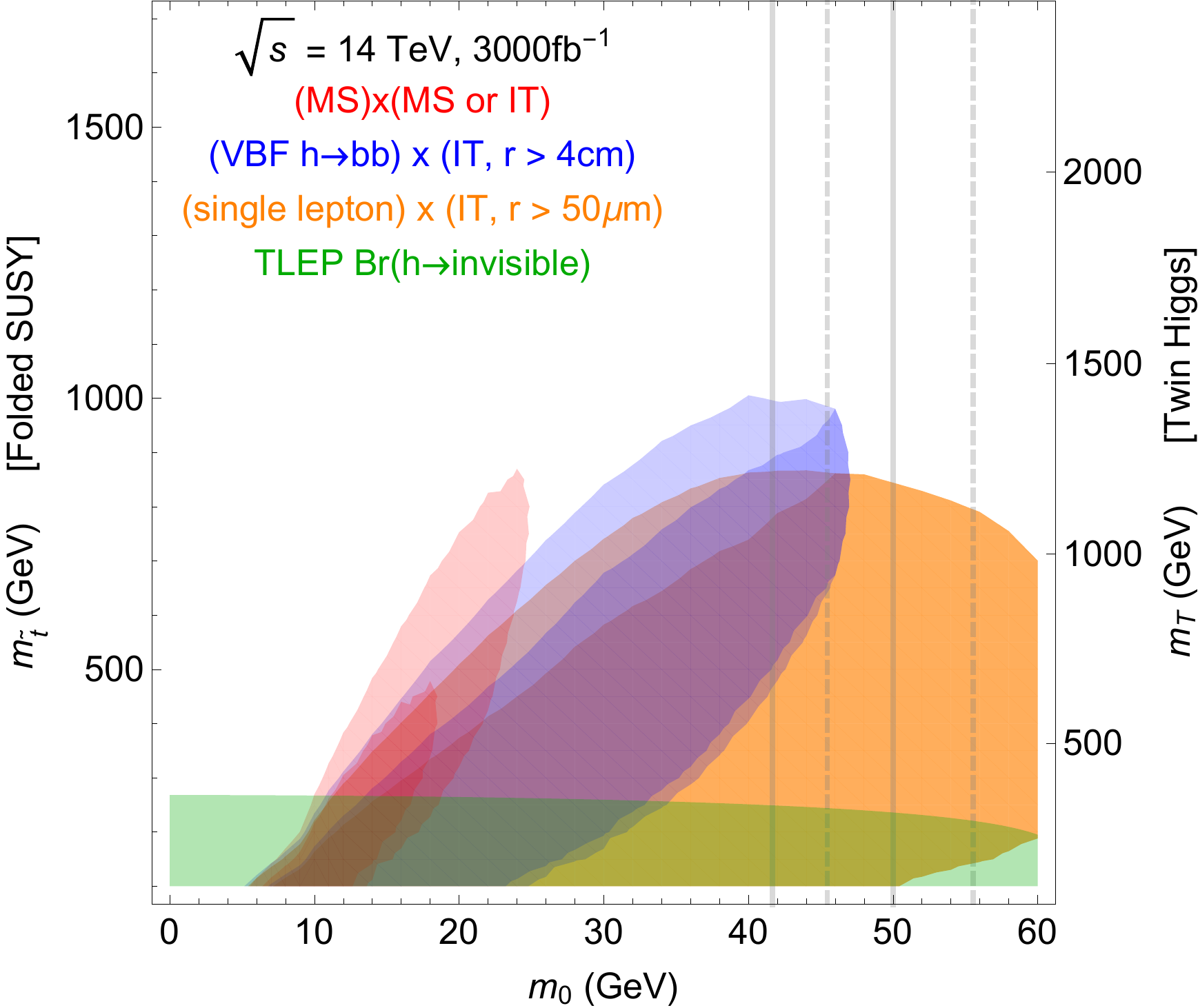} 
\end{tabular}
\end{center}
\caption{
Summary of discovery potential at LHC14 with 300$\ifb$ (left panel) and HL-LHC (right panel) from looking for (i) one DV in the muon system and one additional DV in either the MS or the inner tracker, (ii) one DV at least 4 cm from beam line and VBF jets (blue) and (iii) one DV with at least $50\,\mu$m from beam line and a single lepton (orange). Assuming negligible backgrounds and 10 events for discovery, which is likely more realistic for the MS search with two DVs than the IT searches with one DV. Electroweak top partners (Folded SUSY, Quirky Little Higgs) motivate glueball masses in the 10--60\,\UGeV\ range. See \cite{Curtin:2015fna} for details.
Note different scaling of vertical axes. 
For comparison, the inclusive TLEP $h \to$ invisible limit, as applied to the perturbative prediction for $\br(h\to \mathrm{all\ glueballs})$, is shown for future searches as well, which serves as a pessimistic estimate of TLEP sensitivity. 
Lighter and darker shading correspond to the optimistic (pessimistic) signal estimates $\kappa = \kappa_\mathrm{max}$, ($\kappa_\mathrm{min}$), under the assumption that $h$ decays dominantly to two glueballs, see \eref{brhgg}.
The effect of glueball lifetime uncertainty is small and not shown.
$m_0$ is the mass of the lightest glueball $G_0$; the vertical axes correspond to \mirror stop mass in FSUSY and \mirror top mass in FTH and Quirky Little Higgs.
Vertical solid (dashed) lines show where $\kappa$ might be enhanced (suppressed) due to non-perturbative mixing effects, see \cite{Curtin:2015fna} for details and \cite{Craig:2015pha} for additional discussion. 
}
\label{f.summaryplot}
\end{figure}

\begin{figure}[thb]
\begin{center}
\begin{tabular}{ccc}
\includegraphics[height=5cm]{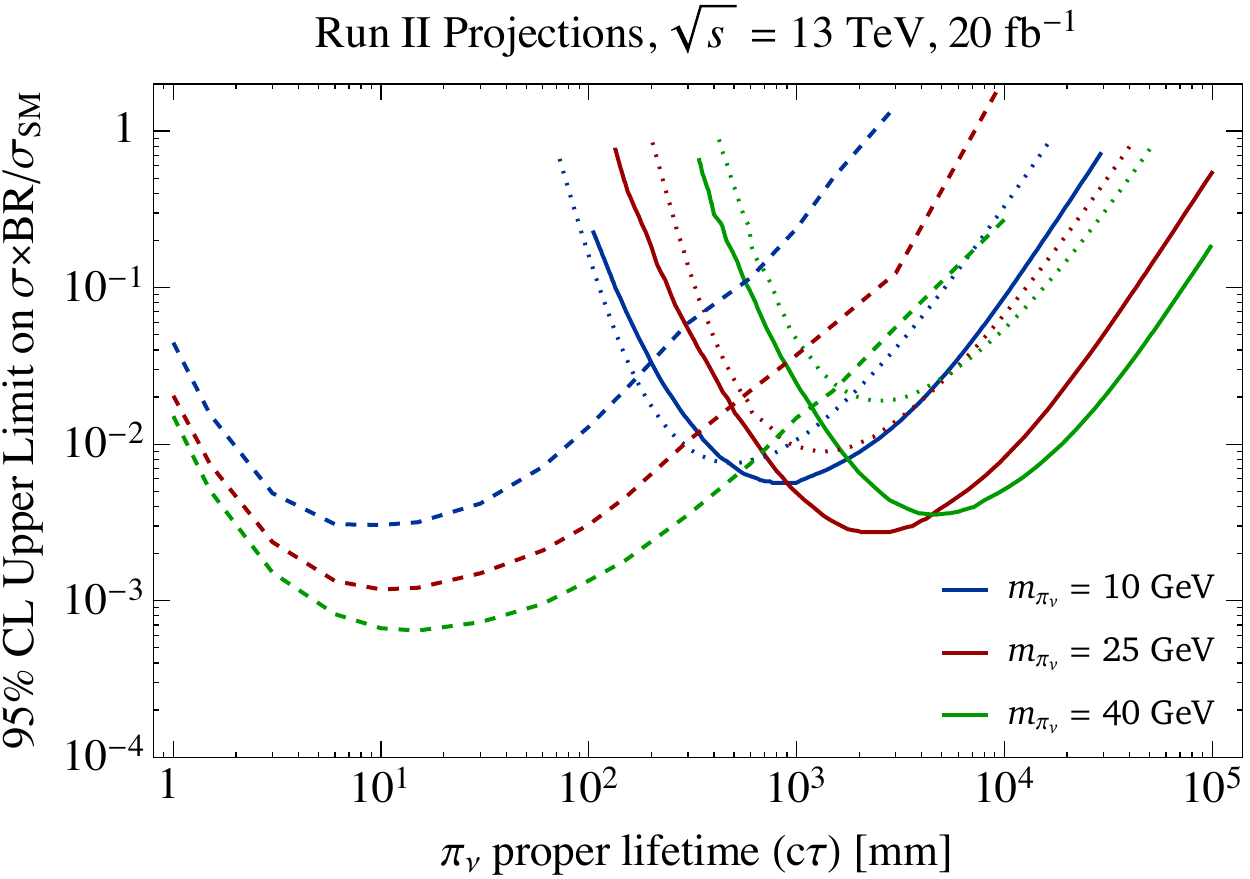} 
&&
\includegraphics[height=5cm]{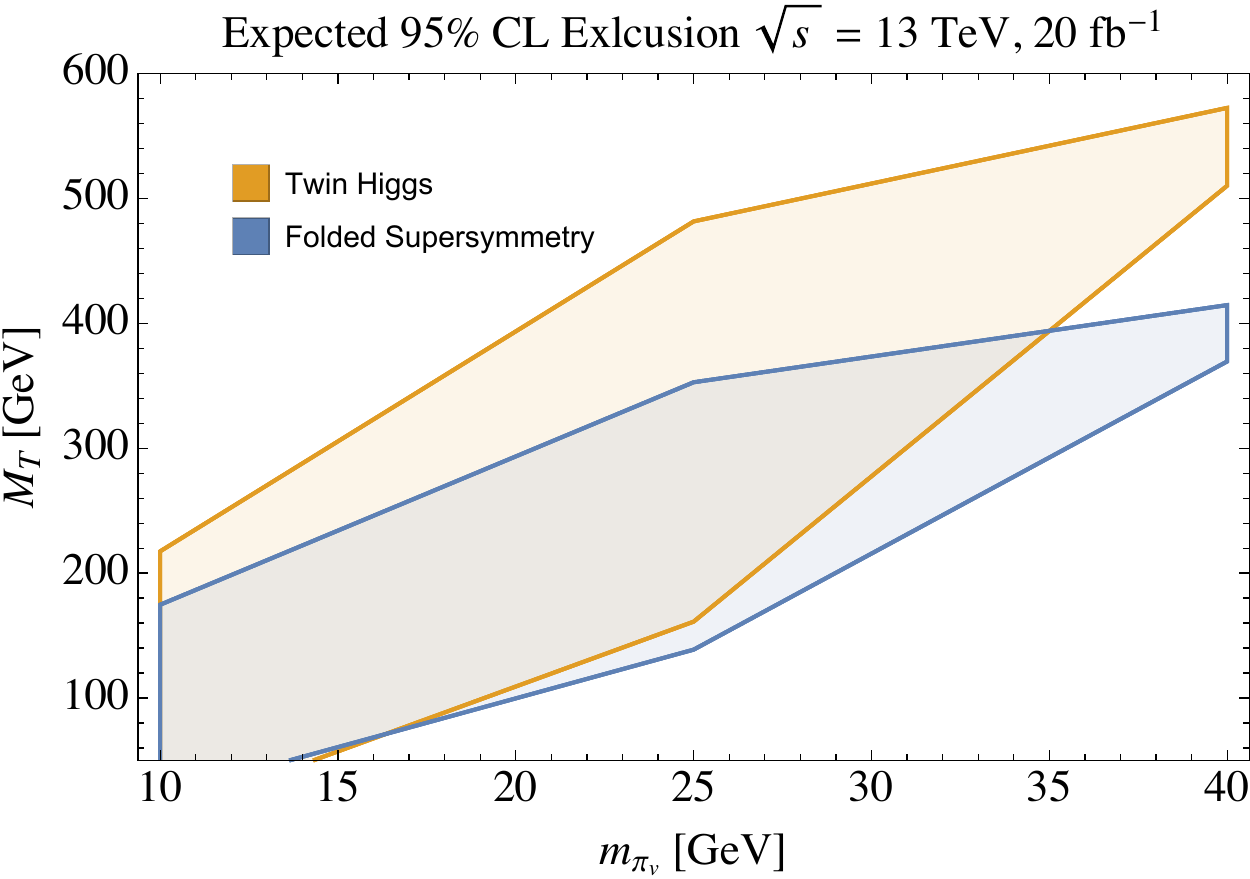} \\
(a) &&(b)
\end{tabular}
\end{center}
\caption{
(a) Projected bounds for the 13 TeV LHC with 20 $\ifb$ of data, from \cite{Csaki:2015fba}. The dashed lines represent CMS searches for single displaced vertices in the tracker, where different search and trigger strategies were separately considered. The best sensitivities are achieved using VBF or (VBF + displaced jet) triggers. The solid lines are projections of the ATLAS searches for displaced decays in the HCAL and muon systems \cite{Aad:2015uaa, Aad:2015asa}.
(b) Application of those bounds to Neutral Naturalness theories, shown in the same top partner mass $(m_T)$ vs. lightest glueball mass $m_0 = m_{\pi_v}$ as in \refF{f.summaryplot}. This assumes $\kappa = \kappa_\mathrm{max} = 1$, i.e. all exotic Higgs boson decays to \mirror glue give $h \to  G_0G_0$, and so may not be accurate at low $m_0$. }
\label{f.summaryplotcms}
\end{figure}

\subsubsection*{Suggested searches}

In the context of this discussion we carefully use the terms DO, tDO and DV as defined at the end of \ssref{overview}.

The above existing searches can be categorized into two classes. The ATLAS searches require a tDO in the Muon System or HCAL for triggering, as well as an additional DV or tDO in offline reconstruction. The CMS search triggers on a high-threshold associated object (AO) at L1 ($H_T$) and on a (tDO + high-threshold AO) at higher trigger level, where the tDO refers to the displaced jets in the tracker. 

Our central recommendation is that, in addition to existing methods, searches be added that require only a \emph{single} DO across all lifetimes, with thresholds that maintain sensitivity to production of LLPs in exotic Higgs boson decays. 

For longer lifetimes, recent progress has been made in demonstrating how existing ATLAS triggers could be used to search for exotic Higgs boson decay events with a single DO in the Muon System \cite{Coccaro:2016lnz}. The data-driven methods used in that work to control backgrounds in searches featuring a \emph{single} DO are general and may be of use for other searches as well. For shorter lifetimes, triggering and background suppression may require the presence of an AO that accompanies the Higgs, namely VBF jets or an associated $Z/W$. This allows the probing of shorter lifetimes and lower masses for $X$, while making the searches more inclusive to other final states outside the \HXXHF\  framework. In the short term, progress can be achieved by some very simple extensions of current analyses.




Specifically, we suggest searches which at off-line reconstruction level look for AO + tDO or AO + DV (since the tracker is crucial to detect shorter lifetimes). The most useful AO's for Higgs boson production are VBF jets, ISR jets, ISR jet plus MET, and leptons. These searches can utilize the following trigger strategies:
\begin{itemize}
\item a pure AO trigger, in particular VBF jets, leptons, or jets + MET (where MET comes from an undecayed or partially invisibly decaying LLP, and the jet arises from ISR); 
\item a tDO + AO trigger, such as the VBF + displaced dijet trigger explored for CMS in \cite{Csaki:2015fba};
\item a pure tDO trigger, which though not currently possible for decays in the tracker, has been used by ATLAS for decays in the HCAL and muon system. 
\end{itemize}
Furthermore, all existing searches focus on lifetimes longer than about a mm. In many scenarios, especially for relatively heavy LLP's, the lifetime can be much shorter. Therefore, in order to cover the entire naturally motivated parameter space, we also suggest study of
\begin{itemize}
\item DVs at very short displacements from the interaction point, as low as $\sim 50\mu$m if possible. 
\end{itemize}

Implementing these search strategies may give much greater coverage of Neutral Naturalness scenarios. In \cite{Curtin:2015fna} it was demonstrated that such a search program, even with pessimistically extrapolated signal yields using current ATLAS DV reconstruction efficiencies, allows discovery for colourless top partners with mass of $\mathcal{O}(\mathrm{TeV})$: see \refF{f.summaryplot} for the LHC14 ATLAS reach with 300 and 3000 $\ifb$ of data.\footnote{This study used a slightly different definition of $\kappa_\mathrm{min}$ than used here, but this does not significantly affect the results.} These projections are consistent with recasts of \cite{CMS:2014wda}, which suggest that reach can be improved  by including a VBF + displaced dijet trigger \cite{Csaki:2015fba}. The projected sensitivities with just $20\ifb$ of 13 TeV data are shown in \refF{f.summaryplotcms}, and cover uncoloured top partner masses of several hundred GeV. These searches will have even greater reach if the Higgs boson decays directly to mirror bottomonia, which increases the exotic branching fraction as shown in \eref{BrhBB}.

We now elaborate on the reasoning behind these suggestions. First, triggering on an AO is relatively independent of the Higgs boson decay final state, and certain tDO trigger thresholds can be lowered if an AO is also present in the trigger path, in compensation for events lost because they lack an AO.

An important motivation for DO + AO searches is their sensitivity to regions of the \HXXHF\ model, and to other models, in which the typical exotic $h$ decay has only one observable DO.  Single DO events may be common because it is rare for two LLPs to be produced together (in some regimes of Neutral Naturalness models, the probability to produce one $G_0$ in hadronization may be much larger than to produce two).  They may also be common because the LLPs are very long-lived, and so it is rare for two of them to decay before exiting the detector. This particularly motivates the use of an ISR jet + MET as the AO. 

Another important motivation is that low-mass DO's in the tracker can not currently be triggered on. This is unfortunate because offline reconstruction of tracker DVs can be efficient.
By triggering on an AO (and perhaps the DO as well), the overall efficiency for signal events will increase.  Backgrounds may also be controllable, since one may employ a standard ``matrix'' or ``ABCD'' method, examining events with and without an AO, and with and without a DV.

Finally, of course, requiring two reconstructed DOs reduces signal efficiency sharply.  Efficient searches that require only one should have better reach, as long as backgrounds are under control and the price of requiring an AO is not too high.

No matter what search strategy is adopted, there are also significant challenges to be overcome on the level of reconstructing any single DV. 
\begin{itemize}
\item Existing searches focus on $m_X \gtrsim 15 \UGeV$. It is important to understand how light $X$ can be while still being efficiently reconstructed as a DV in different detector systems. Small opening angles and small numbers of tracks from the DV can make reconstruction challenging, though \cite{Aad:2014yea} was able to study displaced photons with masses of a few 100 MeV. (An extreme case was studied in~\cite{Ilten:2015hya}, where long-lived dark photons with masses below 100 MeV were considered.) 
DV or AO reconstruction criteria could be loosened for $m_X \lesssim 15 \UGeV$ to maintain efficiency, but backgrounds would have to be carefully studied. In principle, there is no qualitative obstacle to the reconstruction of lower-mass DVs \cite{Aad:2015uaa, Aad:2015asa, Csaki:2015fba}.

\item When $m_X$ is fairly heavy (close to half the Higgs boson mass, or more for asymmetric Higgs boson decays), the slow-moving $X$ may have to be reconstructed out-of-time with the rest of the event. This can make analyses more complicated, but could also provide a handle to reject backgrounds.
\item Independent of triggering issues, what is the shortest lifetime that could be reconstructed and distinguished from prompt background? Going below cm or mm decay lengths opens up very well-motivated parameter space in theories of Neutral Naturalness. Maximizing the rejection of $B$-meson related backgrounds will be a high priority for such a search. 
\end{itemize}

\begin{table}
\caption{
 Displaced $h \to X X \to \mathrm{SM}$ simplified model benchmarks to give efficient coverage of Neutral Naturalness. 
  Recall that $X$ decays via its mixing with the SM-like Higgs, and thus dominantly to $\bar b b$ and $\tau^+ \tau^-$.
}
\label{t.benchmarks}
\begin{center}
\begin{tabular}{c | c |c|c|c}
\cmidrule{3-5}
\multicolumn{2}{c}{}&
\multicolumn{3}{c}{ $c \tau_X$ $(m)$}\\
\cmidrule{3-5}
\multicolumn{2}{c|}{} & $5\ \cdot \ 10^{-5}$ & $10^{-1}$ & 10 \phantom{$^0$} \\
\midrule
\multirow{5}{*}{$m_X$ (GeV)} & 7 &
 \texttt{short7} & \texttt{medium7}  &  \texttt{long7}  \\
\cmidrule{2-5}
& 15 &
\texttt{short15} & \texttt{medium15}  &  \texttt{long15}  \\
\cmidrule{2-5}
& 40 &
\texttt{short40} & \texttt{medium40}  &  \texttt{long40}  \\
\cmidrule{2-5}
& 55 &
\texttt{short55} & \texttt{medium55}  &  \texttt{long55}  \\
\bottomrule
\end{tabular}
\end{center}
\end{table}

\subsubsection*{Benchmarks}

In order to help develop these searches with full coverage we suggest 12 benchmark points, given in \tref{benchmarks}. 
They span the range of well-motivated $X$ masses that yield the $\bar b b$ and $\tau^+ \tau^-$ final states, and the range of lifetimes that can be realized in Neutral Naturalness and potentially reconstructed as a displaced decay.
Of the light benchmarks,  \texttt{long7} is most motivated for \mirror glueballs, since their lifetime increases sharply above the $b \bar b$ threshold;  \texttt{short7} and\texttt{ medium7} can be realized for \mirror bottomonia, which have potentially much shorter lifetimes. 
Each lifetime requires a different search strategy to optimally constrain $\br(h \to X X)$.
 Dealing with time-of-flight issues may be particularly challenging (or fruitful) for the $\texttt{long55}$ benchmark point, while short decay length reconstruction is yet to be demonstrated for the \texttt{short} benchmarks. The sensitivity of LHCb to these decays also deserves future study -- in some cases it might be superior to ATLAS or CMS, despite the reduced luminosity available for analysis, because of its special triggering and reconstruction capabilities.

 \subsubsection*{Presentation of limits}

Conventionally, in searches for DOs from exotic Higgs boson decays, results are represented by plotting excluded $\sigma_h \times \br/\sigma_\mathrm{SM}$, as a function of decay length $c \tau_X$, for different discrete benchmark values of the long-lived particle mass $m_X$. This is shown, for example, in \refF{f.summaryplotcms} (a).  We strongly recommend continuation of this model-independent presentation. 

We also recommend an additional means of presenting results for the \HXXHF\ signal model. This method of presentation would allow constraints to be directly applied to the Neutral Naturalness scenario where the Higgs can decay to mirror glueballs, which decay back to the SM via Higgs mixing. This is the case for FSUSY, and can occur in Twin Higgs models. 

Consider a search that is sensitive to LLPs $X$ produced in the decay $h \to X X$, setting bounds on $\br(h \to X X)$ as a function of $m_X$ and $c \tau_X$. In the above scenario where $X$ is the glueball $G_0$, its lifetime is almost uniquely determined in the $(m_0, m_T)$ plane, where $m_0, m_T$ are the $G_0$ and top partner masses. Therefore, the search gives a constraint on $\br(h \to G_0 G_0)$ as a function of $m_0$ and $m_T$.\footnote{Similar reasoning applies to searches for one LLP that set bounds on $\br(h\to X + \ldots)$.}

We can now make use of the known \emph{inclusive} exotic Higgs boson decay branching fraction $\br(h \to \mathrm{\mirror\ glue})$, see \eref{Brhgginclusive}, even if the \emph{exclusive} branching fraction to a given number of $G_0$'s cannot be computed. 
Clearly, if the limit on $\br(h \to G_0 G_0)$ is much larger than $\br(h \to \mathrm{\mirror\ glue})$ then the search has no sensitivity to this Neutral Naturalness scenario. Similarly, since we expect a sizeable fraction of the produced glueballs to be the lightest $G_0$, if the limit on $\br(h \to G_0 G_0)$ is orders of magnitude smaller than $\br(h \to \mathrm{\mirror\ glue})$, then this scenario can effectively be excluded for a given $(m_0, m_T)$. 
This reasoning can be most easily articulated by defining the parameter
\begin{equation}
\kappa^{G_0 G_0}(m_0, m_T) \equiv \frac{\sigma_h \times \br(h \to G_0 G_0)}{\sigma_\mathrm{SM} \times \br(h \to \mathrm{\mirror\ glue})}.
\end{equation}
The denominator is completely determined in the $(m_0, m_T)$ plane, while the numerator can be constrained by the search. Bounds on $\kappa^{G_0 G_0}$ can therefore be shown in the $(m_0, m_T)$ plane and constrain the fraction of glueballs that are in the $G_0$ state. Regions with $\kappa^{G_0G_0}< 0.1$ $(> 1)$ are likely (not) excluded. In regions with intermediate values of $\kappa^\mathrm{ex}$, the bounds are sensitive to pure-glue hadronization assumptions.

\subsection[Higgs boson decays to displaced photons and missing energy from Supersymmetry]{Higgs boson decays to displaced photons and missing energy from Supersymmetry\SectionAuthor{S.~Heinemeyer}}
\label{ss.DP}

Here we discuss one realization of displaced Higgs boson decays within
the MSSM, consistent with current experimental constraints.
In models of gauge-mediated supersymmetry breaking (GMSB), the next-to-lightest
Standard Model superpartner (NLSP) will decay to a gravitino LSP, which is
stable in the presence of R-parity, together with one or more SM
states.  This decay becomes displaced as the
supersymmetry-breaking scale is raised. 
In particular, in generalized GMSB
models, for gauge messenger masses of $\mathcal O(100)$~TeV, it is
possible to have the decay $h\to\tilde\chi_1\tilde\chi_1\to
\gamma\tilde G \; \gamma\tilde G$ with a long-lived Bino-like lightest
neutralino, $\tilde\chi_1$.  While the first stage of this decay is
prompt with a possibly sizeable branching ratio, 
the second decay occurs with a branching ratio of one, and the
lifetime of the neutralino can be evaluated as~\cite{Mason:2009qh}
\begin{equation}
\label{neu1lifetime}
c\tau = 48\,\pi\frac{m_{3/2}^2 M_{\rm Pl}^2}{m_{\tilde\chi_1^5}}
               \frac{1}{|P_{1\gamma}|^2}
\end{equation}
with $P_{1\gamma} = N_{11}\cos\theta_W + N_{12}\sin\theta_W$ given in
terms of the neutralino mixing matrix $N$.  Here $N_{11}$ and $N_{12}$
are the Bino and Wino $\tilde\chi_1$ components, respectively. The
gravitino mass $m_{3/2}$ is at or below the level of~eV, and $M_{\rm
  Pl}$ is the Planck mass. Thus neutralino decay lengths of up to one
meter or even more can be obtained, depending on the choice of
parameters and in particular on the neutralino mass
$m_{\tilde\chi_1}$.  As an illustrative example, in
\refF{f.photondisplaced}, we present the branching ratio for this
process, i.e.\ BR($h\to\tilde\chi_1\tilde\chi_1\to \gamma\gamma+\tilde
G\tilde G$) and the lifetime of the neutralino as a function of the Bino mass $M_1$ and $\tan\beta$, having
fixed the $\mu$ parameter, as well as the Wino mass, $M_2$, to
400~GeV. The other parameters are chosen as in the $m_h^{\rm mod+}$
scenario~\cite{Carena:2013ytb}. We note that this benchmark has not been probed by multi-lepton searches for electroweak production of Higgsinos and Winos at Run~1 LHC (see for example \cite{Aad:2014nua,Aad:2015eda,Khachatryan:2015pot}).
 One can see that the branching ratio
depends mainly on $\tan\beta$, and can reach $\sim \mathcal O(10\%)$
or larger at small values of $\tan\beta$. The lifetime, on the other
hand, depends mainly on $M_1$. The smallest values of $M_1 \lesssim
10$~GeV lead to decay lengths of 1~m or more, while values larger than
$\sim 25$~GeV lead to decay lengths below 1~cm. This shows that this
particular scenario, depending on choice of parameters, offers the
full spectrum of collider-relevant decay lengths and the corresponding
phenomenological opportunities and challenges.  

\begin{figure}[thb]
\begin{center}
\includegraphics[width=0.42\textwidth]{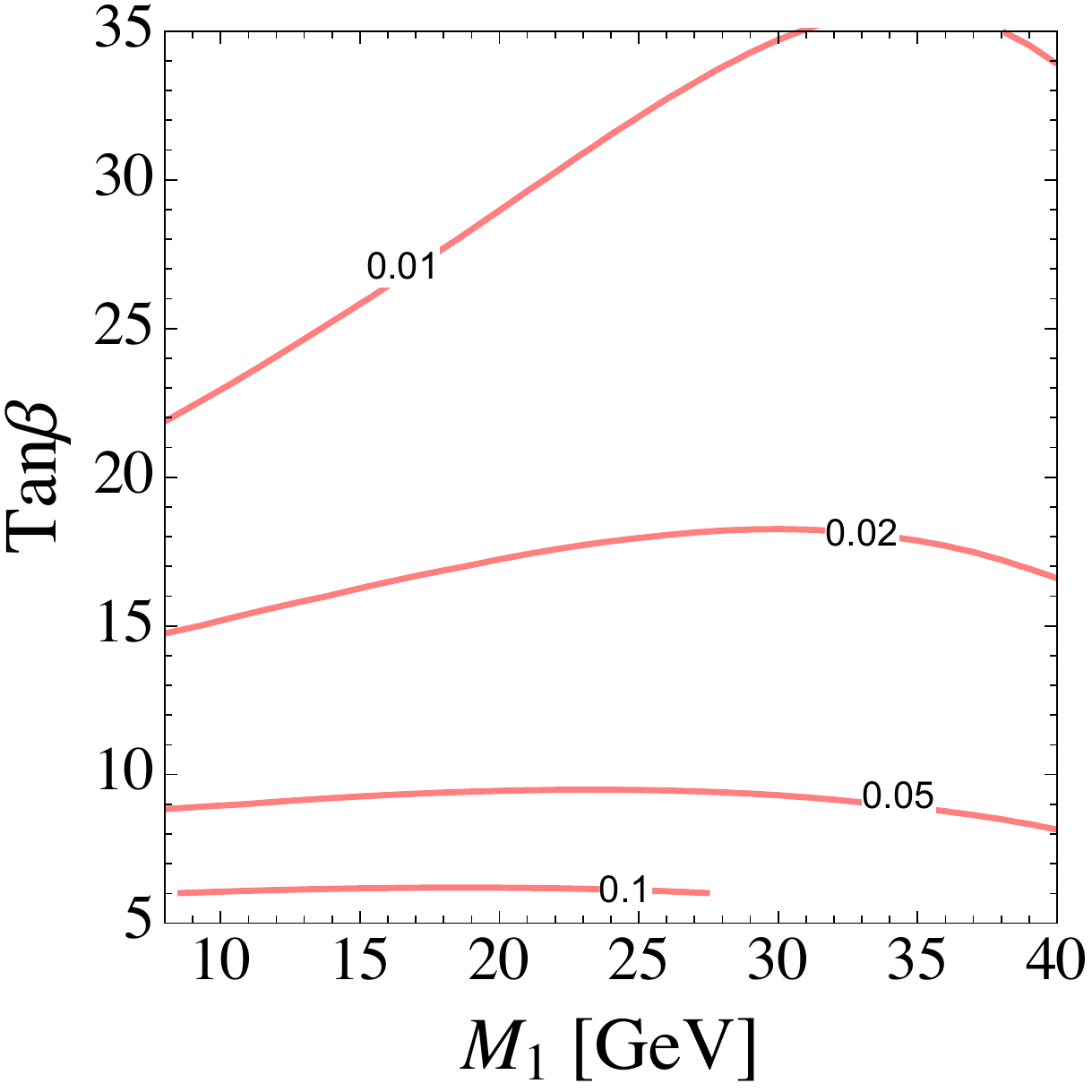}
\includegraphics[width=0.42\textwidth]{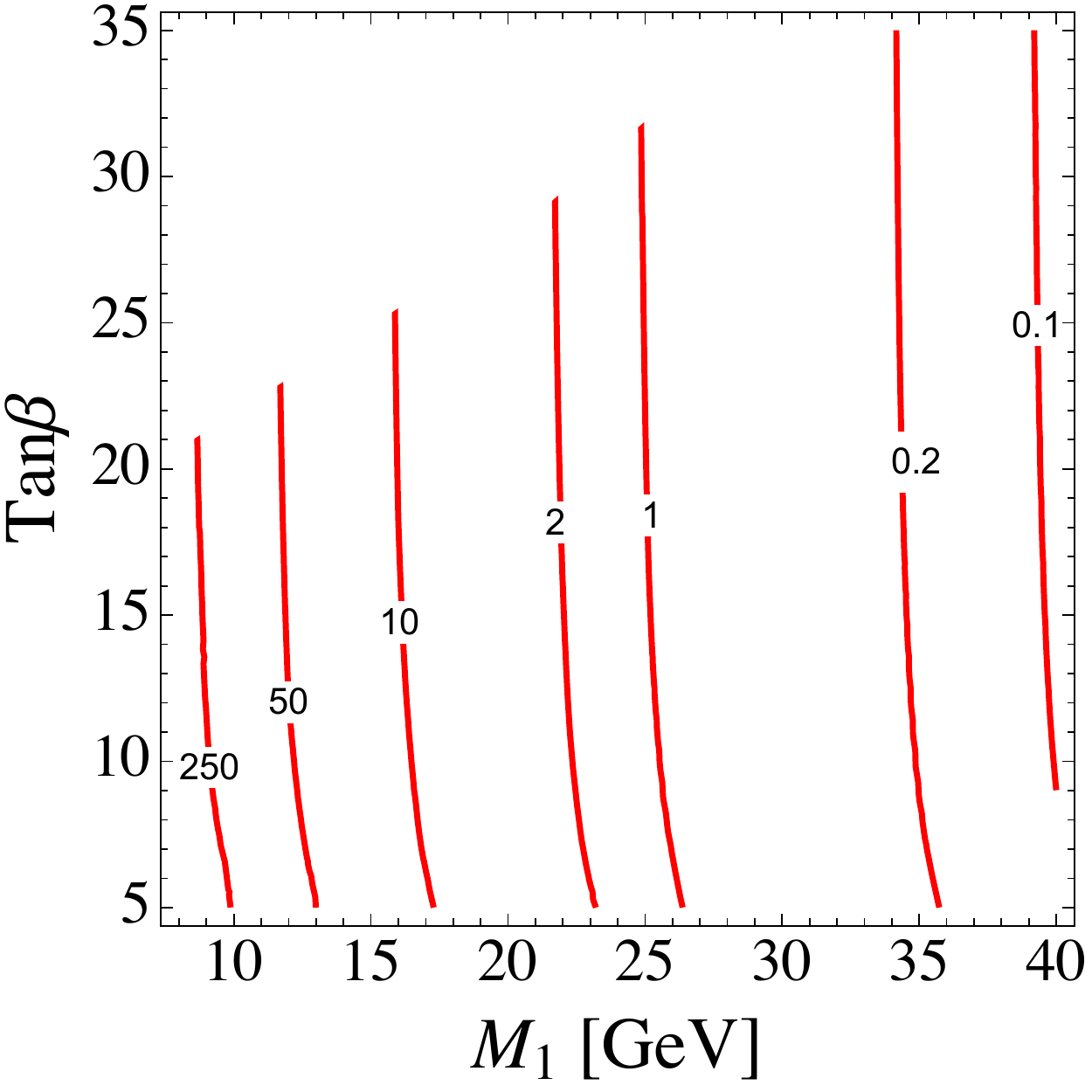}
\end{center}
\caption{Branching ratio for $h\to\tilde\chi_1\tilde\chi_1\to\tilde
  G\tilde G\gamma\gamma$ (left) and neutralino $\tilde\chi_1$ lifetime
  in cm (right), as a function of the Bino mass parameter $M_1$ and
  $\tan\beta$. The benchmark model used is the $m_h^{\rm mod+}$
  scenario~\cite{Carena:2013ytb}, with $\mu = M_2 = 400$~GeV.}
\label{f.photondisplaced}
\end{figure}



\afterpage{\clearpage}



\cleardoublepage

%

\cleardoublepage
\phantomsection
\addcontentsline{toc}{part}{Acknowledgements} 

\chapter*{Acknowledgements\markboth{Acknowledgements}{Acknowledgements}}

\indent We are obliged to CERN, in particular to the IT Department and to the
Theory Unit and the LPCC for the support with logistics and technical assistance.


\indent Fermilab is operated by Fermi Research Alliance, LLC under grant DE-AC02-07CH11359 with the US Department of Energy. \\
\indent R.~Aggleton, F.~Bishara and C.H.~Shepherd-Themistocleous are supported by the Science and Technology Facilities Council (STFC). \\
\indent B.~Allanach has been partially supported by STFC grant ST/L000385/1. \\
\indent W.~Astill, W.~Bizon, S.~Carrazza, B.~Mistlberger, E.~Re and G.~Zanderighi  are partly supported by  the ERC Consolidator grant ``HICCUP" (614577). W.~Astill and W.~Bizon thank CERN and the Mainz Institute of Theoretical Physics (MITP) for their hospitality while part of this work was carried out. \\
\indent S.~Badger is supported by an STFC Rutherford Fellowship ST/L004925/1. \\
\indent M.~Badziak has been supported in part by the Polish Ministry of Science and Higher Education (decision 1266/MOB/IV/2015/0), the US Department of Energy under Contract DE-AC02-05CH11231, the National Science Foundation under grant PHY-1316783 and by the Foundation for Polish Science through its programme HOMING PLUS. \\
\indent J.~Baglio is supported in part by the Institutional Strategy of the University of T{\"u}bingen (DFG, ZUK~63) and by the DFG grant JA~1954/1. \\
\indent D.~Barducci, G.~B\'elanger, S.~Forte, E.W.N.~Glover, M.~Grazzini, G.~Heinrich, A.~Ilnicka, S.P.~Jones, Z.~Kassabov, F.~Krauss, E.~Maina, C.~Mariotti, D.~Napoletano, G.~Passarino, P.~Slavich, M.~Spannowsky, M.~Spira, U.~Ellwanger, A.~Vicini and G.~Weiglein are partly supported by the 7th Framework Programme of the European Commission through the Initial Training Network ``HiggsTools" (PITN-GA-2012-316704). \\ D.~Barducci, G.~B{\'e}langer and S.~Kraml are partly supported by the ``Investissements d'avenir, Labex ENIGMASS", by the French ANR Project ``DMAstroLHC" (ANR-12-BS05-006). \\
\indent J.~Bellm, S.~Gieseke, A.~Papaefstathiou, P.~Schichtel and E.~Vryonidou  are supported in part by the 7th Framework Programme of the European Commission through the Initial Training Network ``MCnetITN" (PITN-GA-2012-315877). \\
\indent N.~Belyaev's work was performed  within the framework of the Center for Fundamental Research and Particle Physics  supported by MEPhI Academic Excellence Project (contract 02.a03.21.0005). \\
\indent A.~Biek{\"o}tter acknowledges support from the German Research Foundation (DFG) through the Forschergruppe ``New Physics at the LHC" (FOR 2239).. \\
\indent N.E.~Bomark is grateful to NCBJ for access to the CIS computer cluster. \\
\indent M~Bonvini and J.~Rojo have been partly supported by the ERC Starting grant ``PDF4BSM". \\
\indent S.~Borowka and E.W.N.~Glover acknowledge financial support by the ERC Advanced grant \linebreak ``MC@NNLO" (340983). \\
\indent V.~Bortolotto and K.~Prokofiev are partially supported by a grant from the Research Grant Council of the Hong Kong Special Administrative Region, China (project CUHK4/CRF/13G). \\
\indent S.~Boselli, G.~Montagna, O.~Nicrosin and F.~Piccinini were partially supported by the Research Executive Agency (REA) of the European Union under the Grant Agreement number PITN-GA-2010-264564 (LHCPhenoNet) and by the Italian Ministry of University and Research under the PRIN project 2010YJ2NYW. \\
\indent F.J.~Botella is supported by Spanish MINECO under grants FPA2015-68318-R and SEV-2014-0398, and by Generalitat Valenciana under grant GVPROMETEOII 2014-049. \\
\indent R.~Boughezal is supported by the DOE grant DE-AC02-06CH11357. Her research used resources of: the National Energy Research Scientific Computing Center, a DOE Office of Science User Facility supported by the DOE under contract DE-AC02-05CH11231; the Argonne Leadership Computing Facility, which is supported under DOE contract DE-AC02-06CH11357. R.~Boughezal and F.~Petriello thank the KITP, Santa Barbara, supported by the National Science Foundation under grant NSF PHY11-25915 for hospitality. \\
\indent G.C.~Branco and M.N.~Rebelo were partially supported by Funda{\c c}{\~a}o para a Ci{\^e}ncia e a Tecnologia (FCT, Portugal) through the projects CERN/FIS/NUC/0010/2015 and CFTP-FCT Unit 777 (UID/FIS/00777/2013) which are partially funded through POCTI (FEDER), COMPETE, QREN and EU. \\
\indent J.~Brehmer is supported by the German Research Foundation (DFG) through the research training group GRK 1940. \\
\indent S.~Bressler is supported by the I-CORE Program of the Planning and Budgeting Committee and The Israel Science Foundation (grant 1937/12). S.~Bressler, S.~Gori, A.~Mohammadi and J.~Shelton thank Fermilab for its hospitality and support. \\
\indent I.~Brivio, M.~Herrero, U.~Ellwanger and L.~Merlo  were supported by the EU networks FP7 ITN INVISIBLES (PITN-GA-2011-289442), FP10 ITN ELUSIVES (H2020-MSCA-ITN-2015-674896) and INVISIBLES-PLUS (H2020- MSCA-RISE-2015-690575). I.~Brivio, M.~Herrero, L.~Merlo and J.J.~Sanz-Cillero acknowledge partial support of the Spanish MINECO's ``Centro de Excelencia Severo Ochoa" Programme under grant SEV-2012-0249. \\
\indent H.~Brun thanks the FRS-FNRS IISN for financial support. \\
\indent G.~Buchalla and O.~Cata acknowledge support by the ERC Advanced grant ``FLAVOUR" (267104) and by the DFG Cluster of Excellence ``Origin and Structure of the Universe" (EXC 153). \\
\indent F.~Campanario thanks support to the Spanish Government and ERDF funds from the European Commission (FPA2014-53631-C2-1-P and FPA2014-54459-P). \\
\indent C.M.~Carloni Calame and A.~Ballestrero have been supported by the MIUR-PRIN project 2010YJ2NYW. \\
\indent A.~Carvalho is supported by MIUR-FIRB RBFR12H1MW grant. \\
\indent M.~Casolino and A.~Juste are supported in part by the Spanish Ministerio de Econom{\'\i}a y Competitividad under projects FPA2012-38713 and Centro de Excelencia Severo Ochoa SEV-2012-0234. \\
\indent A.~Celis is supported by the Alexander von Humboldt Foundation. \\
\indent M.~Chen is supported by the Hundred Talents Program of Chinese Academy of Sciences. \\
\indent X.~Chen thanks the IPPP at the University of Durham for hospitality, and T.~Morgan, A.~Huss, A.~Gehrmann-De Ridder, J.~Pires, J.~Currie and J.~Niehues for useful discussions and their many contributions to the NNLOJET code. \\
\indent R.~Contino and L.~Silvestrini have been partly supported by the ERC Advanced grant ``DaMeSyFla" (267985). \\
\indent T.~Corbett was supported in part by the Australian Research Council. \\
\indent R.~Costa was partially supported by FCT under contract PTDC/FIS/117951/2010.\\
\indent D.~Curtin is supported by National Science Foundation grant PHY-1315155 and the Maryland Center for Fundamental Physics. He thanks C.~Cs{\'a}ki, E.~Kuflik, S.~Lombardo, O.~Slone for helpful conversation and collaboration. \\
\indent M.~Dall'Osso is supported by grant CPDR155582 of Padua University. \\
\indent S.~Dawson is supported by the United States Department of Energy under grant DE-SC0012704. \\
\indent J.~de~Blas, D.~Ghosh and L.~Silvestrini  have been partly supported by the ERC grant ``NPFlavour" (279972). \\
\indent W.~de~Boer warmly acknowledges support from the Heisenberg-Landau program and the Deutsche Forschungsgemeinschaft (DFG grant BO 1604/3-1). \\
\indent D.~de~Florian and J.~Mazzitelli are partially supported by CONICET and ANPCyT (Argentina). J.~Mazzitelli also acknowledges support by UBACYT. \\
\indent C.~Degrande is a Durham International Junior Research Fellow. \\
\indent R.L.~Delgado, A.~Dobado and F.J.~Llanes-Estrada have been supported by grants BES-2012-056054, UCM:910309 and MINECO:FPA2014-53375-C2-1-P and by the Spanish Excellence Network on Hadronic Physics FIS2014-57026-REDT. \\
\indent F.~Demartin is supported by the IISN ``MadGraph" convention 4.4511.10 and the IISN ``Fundamental interactions" convention 4.4517.08. \\
\indent A.~Denner, R.~Feger and T.~Ohl  acknowledge support by the Bundesministerium f\"ur Bildung und Forschung (BMBF) under contract 05H12WWE. \\
\indent S.~Dittmaier is supported by the Research Training Group GRK 2044 of the German Science Foundation (DFG). \\
\indent F.A.~Dreyer, P.~Francavilla, P.~Slavich, M.~Zaro and R.~Ziegler are supported by the ILP LABEX (ANR-10-LABX-63) supported by French state funds managed by the ANR within the Investissements d'Avenir programme under reference ANR-11-IDEX-0004-02. \\
\indent C.~Duhr is supported by the ERC Starting Grant ``MathAm". \\
\indent F.~Dulat acknowledges the support of the US Department of Energy under grant DE-AC02-76SF00515. \\
\indent U.~Ellwanger, G.P.~Salam and P.~Slavich are supported in part by ERC Advanced grant ``Higgs@LHC" (321133). U.~Ellwanger and S.~Moretti are supproted in part by he EU project ``NonMinimalHiggs" (H2020-MSCA-RISE-2014 645722). U.~Ellwanger also acknowledges support from the D{\'e}fi InPhyNiTi project N2P2M-SF . \\
\indent C.~Englert is supported in part by the IPPP Associateship scheme. \\
\indent D.~Espriu is supported by the grants FPA2013-46570 and 2014-SGR-104. \\
\indent A.~Ferroglia is supported in part by the National Science Foundation grant PHY-1417354. \\
\indent T.~Figy used the Extreme Science and Engineering Discovery Environment (XSEDE), which is supported by National Science Foundation grant ACI-1053575. He thanks Mats Rynge for his assistance with implementing workflows on the Open Science Grid, which was made possible through the XSEDE Extended Collaborative Support Service (ECSS) program. \\
\indent M.~Flechl is supported by the Austrian Science Fund (FWF), project P 28857-N36. \\
\indent S.~Forte thanks M. Bonvini, A. De Roeck, S. Marzani, C. Muselli and P. Nason for discussions. He is supported in part by an Italian PRIN2010 grant, by a European Investment Bank EIBURS grant. \\
\indent P.~Francavilla is supported by the ILP LABEX (under reference ANR-10-LABX-63 and ANR-11-IDEX-0004-02). \\
\indent R.~Frederix is supported by the Alexander von Humboldt Foundation, in the framework of the Sofja Kovaleskaja Award Project ``Event Simulation for the Large Hadron Collider at High Precision". \\
\indent F.F.~Freitas work is supported by the Brazilian program ``Ciencia sem fronteras". \\
\indent B.~Fuks, S.~Kraml and K.~Mawatari acknowledges partial support by the Th\'eorie-LHC France initiative of the CNRS (INP.IN2P3). \\
\indent M.V.~Garzelli, R.~Kogler, S.~Liebler, J.~Reuter, K.~Tackmann and G.~Weiglein are supported by the German Research Foundation (DFG) in the Collaborative Research Centre (SFB) 676 ``Particles, Strings and the Early Universe". \\
\indent E.W.N.~Glover is supported in part by the UK Science and Technology Facilities Council through grant ST/G000905/1. \\
\indent F.~Goertz is supported by a Marie Curie Intra European Fellowship within the 7th European Community Framework Programme (PIEF-GA-2013-628224).. \\
\indent D.~Gon{\c c}alves is supported by STFC through the IPPP grant. \\
\indent M.~Gorbahn is supported by STFC under grant ST/L000431/1. \\
\indent S.~Gori thanks the Galileo Galilei Institute for Theoretical Physics for the hospitality and the INFN for partial support during some part of this work. \\
\indent D.~Gray and R.~Konoplich are partially supported by the US National Science Foundation under grant PHY-1402964. \\
\indent M.~Grazzini was supported in part by the Swiss National Science Foundation (SNF) under contracts CRSII2-141847, 200021-156585. \\
\indent N.~Greiner was supported by the Swiss National Science Foundation under contract PZ00P2-154829. \\
\indent A.~Greljo, G.~Isidori and D.~Marzocca are supported by the Swiss National Science Foundation under contract 200021-159720. \\
\indent A.V.~Gritsan, J.~Roskes, U. Sarica and M.~Xiao are partially supported by US NSF under grant PHY-1404302 while their calculations were performed on the Maryland Advanced Research Computing Center (MARCC). \\
\indent C.~Grojean is supported by the European Commission through the Marie Curie Career Integration Grant 631962, by the Helmholtz Association and in part by the Spanish Ministerio de Econom{\'\i}a y Competitividad under projects FPA2014-55613-P and Centro de Excelencia Severo Ochoa SEV-2012-0234 and by the Generalitat de Catalunya grant 2014-SGR-1450. \\
\indent S.~Guindon is supported by the National Science Foundation. \\
\indent H.E.~Haber is supported in part by the US Department of Energy under grant DE-FG02-04ER41286. He gratefully acknowledges the hospitality and support of S.~Heinemeyer during the Higgs Days meeting in Santander, Spain where the work on 2HDM benchmarks was initiated, and to the Theory Group at CERN, where this work was completed. \\
\indent C.~Han is supported by World Premier International Research Center Initiative (WPI), MEXT, Japan. \\
\indent T.~Han was supported in part by the US DOE and in part by the PITT PACC. \\
\indent R.~Harlander, K.~K{\"o}neke, M.~Moreno Ll{\'a}cer, C.~Schmidtt  and E.~Shabalina  are supported by Bundesministerium f\"ur Bildung und Forschung (BMBF) under contract 05H15PACC1. \\
\indent M.A.~Harrendorf's work was funded by the DFG Research Training Group 1694 and by the German BMBF under the contract  05H15VKCCA. \\
\indent H.B.~Hartanto is supported by the German Research Foundation (DFG). \\
\indent S.~Heinemeyer is supported in part by CICYT (grant FPA 2013-40715-P) and by the Spanish MICINN's Consolider-Ingenio 2010 Program under grant MultiDark CSD2009-00064. \\
\indent M.~Herrero is supported by CICYT through Grant No. FPA2012-31880. \\
\indent F.~Herzog is supported by the ERC grant ``HEPGAME" (320651). \\
\indent B.~Hespel is supported by the National Fund for Scientific Research (FRS-FNRS Belgium) under a FRIA grant. \\
\indent V.~Hirschi is supported by the Swiss National Science Foundation grant PBELP2-146525. He thanks the CP3 IT team for the computational resources put at his disposal. \\
\indent S.~Hoeche was supported by the US Department of Energy under grant DE-AC02-76SF00515. \\
\indent S.~Honeywell, L.~Reina, and C.~Reuschle are supported in part by the US Department of Energy under grant DE-FG02-13ER41942. \\
\indent S.~Huberand K.~Mimasu is supported by STFC under grant ST/L000504/1. \\
\indent C.~Hugonie acknowledges the support of France Grilles for providing cloud computing resources on the French National Grid Infrastructure. \\
\indent J.~Huston is supported by the National Science Foundation under  grant PHY-1410972. \\
\indent B.~J{\"a}ger was supported in part by the Institutional Strategy of the University of T{\"u}bingen (DFG, ZUK 63) and in part by the German Federal Ministry for Education and Research (BMBF) under contract number 05H2015. \\
\indent A.~Kardos gratefully acknowledges financial support from the Post Doctoral Fellowship programme of the Hungarian Academy of Sciences and the Research Funding Program ARISTEIA, HOCTools (co-financed by the European Union (European Social Fund ESF). \\
\indent A.~Karlberg is supported by STFC and by the Buckee Scholarship at Merton College. \\
\indent N.~Kauer and C.~O'Brien are supported by the STFC grants ST/J000485/1, ST/J005010/1 and ST/L000512/1. \\
\indent F.~Kling thanks the Munich Institute for Astro- and Particle Physics (MIAPP) of the DFG cluster of excellence ``Origin and Structure of the Universe" for hospitality. He is supported by US Department of Energy under grant DE-FG02-04ER-41298 also acknowledges support from the Fermilab Graduate Student Research Program in Theoretical Physics operated by Fermi Research Alliance, LLC under contract DE-AC02-07CH11359 with the US Department of Energy. \\
\indent C.~Krause is supported by the DFG grant BU 1391/2-1 and in part by the DFG cluster of excellence ``Origin and Structure of the Universe". \\
\indent F. Krauss is supported by the 7th Framework Programme of the European Commission through the Initial Training Network ``MCnetITN" (PITN-GA-2012-315877). \\
\indent M.~Krawczyk is partly supported by the Polish National Science Centre within an OPUS research project under  grant  2012/05/B/ST2/03306. \\
\indent A.~Kulesza is supported by the German Research Foundation DFG grant KU 3103/1. \\
\indent G.~Lee is supported by the Israel Science Foundation (grant 720/15), by the United-States-Israel Binational Science Foundation (BSF) (grant 2014397), and by the ICORE Program of the Israel Planning and Budgeting Committee (grant 1937/12). \\
\indent H.E.~Logan and P.~Savard are supported by the Natural Sciences and Engineering Research Council of Canada. \\
\indent D.~Lopez-Val acknowledges the support of the FRS-FNRS ``Fonds de la Recherche Scientifique''. \\
\indent I.~Low is supported in part by the US Department of Energy under grant DE-SC0010143. \\
\indent E.~Maina has been supported by MIUR (Italy) under contract 2010YJ2NYW006, by the Compagnia di San Paolo under contract ORTO11TPXK. \\
\indent F.~Maltoni is supported by the Fonds de la Recherche Scientifique (FNRS). \\
\indent S.~Marzani and V.~Theeuwes are supported in by the US National Science Foundation under grant PHY-0969510 ``LHC Theory Initiative". \\
\indent A.~McCarn is supported by the US Department of Energy under grant DE-SC0007859. \\
\indent L.~Merlo acknowledges partial support of CiCYT through the project FPA2012-31880. \\
\indent N.~Moretti and S.~Pozzorini are supported by the Swiss National Science Foundation under contracts PP00P2-153027 and BSCGI0157722. \\
\indent S.~Moretti is financed in part through the NExT Institute. \\
\indent L.~Motyka gratefully acknowledges support of the Polish National Science Centre grant DEC-2014/13/B/ST2/02486. \\
\indent S.~Munir was in part supported by the Swedish Research Council under contracts 2007-4071 and 621-2011-5107. \\
\indent P.~Musella is supported by the Swiss National Science Foundation. \\
\indent P.~Nadolsky is supported by the US Department of Energy under grant DE-SC0013681. \\
\indent M.~Nebot acknowledges support from Spanish MINECO under grants FPA2015-68318-R and SEV-2014-0398, and from Generalitat Valenciana under grant PROMETEOII/2013/017. \\
\indent M.~Neubert is supported by the ERC Advanced grant ``EFT4LHC", the DFG Cluster of Excellence ``Precision Physics, Fundamental Interactions and Structure of Matter" (PRISMA, EXC 1098) and grant 05H12UME of the German Federal Ministry for Education and Research  (BMBF). \\
\indent R.~Nevzorov is supported by the University of Adelaide and the Australian Research Council through the ARC Center of Excellence in Particle Physics at the Terascale. \\
\indent J.~Nielsen is supported by the US Department of Energy under grant DE-SC0010107. \\
\indent K.~Nikolopoulos is supported by the European Union's 7th Framework Programme for research, technological development and demonstration under grant agreement 334034 ``EWSB". \\
\indent J.M.~No is supported by a Marie Curie Intra European Fellowship within the 7th European Community Framework Programme (PIEF-GA-2013-625809). \\
\indent T.~Orimoto is supported by the US Department of Energy. \\
\indent D.~Pagani and H.S.~Shao are supported by the ERC Advanced grant ``LHCtheory" (291377). \\
\indent C.E.~Pandini is supported by the PhD program of the UPMC and Universit{\'e} Paris-Diderot (Paris, France). \\
\indent A.~Papaefstathiou  is supported by a Marie Curie Intra European Fellowship within the 7th European Community Framework Programme  (PIEF-GA-2013-622071). \\
\indent A.S.~Papanastasiou is supported by STFC under grant ST/L002760/1. \\
\indent M.~Pelliccioni is supported by the Istituto Nazionale di Fisica Nucleare. \\
\indent G.~Perez, R.~Podskubka and R.~Roth are supported by ``Graduiertenkolleg" of Karlsruhe Institute of Technology. \\
\indent F.~Petriello is supported by the DOE grants DE-FG02-91ER40684 and DE-AC02-06CH11357. His research used resources of: the National Energy Research Scientific Computing Center, a DOE Office of Science User Facility supported by the DOE under contract DE-AC02-05CH11231; the Argonne Leadership Computing Facility, which is supported under DOE contract DE-AC02-06CH11357. \\
\indent A.~Pilkington is supported in the UK by the Royal Society. \\
\indent S.~Pl{\"a}tzer  is supported by a Marie Curie Intra European Fellowship within the 7th European Community Framework Programme (PIEF-GA-2013-628739). \\
\indent C.T.~Potter thanks the Alder Institute for High Energy Physics (AIHEP) for financial support. \\
\indent A.~Pukhov thanks LAPTH for his hospitality. \\
\indent I.~Puljak is supported by Croatian Science Foundation under project 7118. \\
\indent J.~Quevillon is supported by  STFC under grant ST/L000326/1. \\
\indent T.~Robens thanks CERN-TH and the University of Warsaw for their hospitality while parts of this work were completed, and G.~Chalons, D.~Lopez-Val, and G.M.~Pruna for fruitful collaboration on related work. \\
\indent J.~Rojo is supported by an STFC Rutherford Fellowship and grant ST/K005227/1 ST/M003787. \\
\indent J.C.~Rom{\~a}o was partially supported by Funda{\c c}{\~a}o para a Ci{\^e}ncia e a Tecnologia (FCT, Portugal) through the projects CERN/FP/123580/2011 and CFTP-FCT Unit 777 (UID/FIS/00777/2013) which are partially funded through POCTI (FEDER), COMPETE, QREN and EU. \\
\indent M.O.P.~Sampaio is funded through the grant BPD/UI97/5528/2017 and through the CIDMA project UID/MAT/04106/2013.\\
\indent V.~Sanz is supported by STFC under grant ST/J000477/1. \\
\indent J.J.~Sanz-Cillero is partially supported by the Spanish Ministry MINECO under grant FPA2013-44773-P. \\
\indent M.~Sch{\"o}nherr acknowledges support by the Swiss National Science Foundation under contract PP00P2-128552. \\
\indent U.~Schubert is supported by the Alexander von Humboldt Foundation, in the framework of the Sofja Kovalevskaja Award 2010, ``Advanced Mathematical Methods for Particle Physics". \\
\indent S.~Sekula is supported by the US Department of Energy under grant DE-SC0010129. He gratefully acknowledges SMU's Center for Scientific Computation for their support and for the use of the SMU ManeFrame Tier 3 ATLAS System. \\
\indent J.~Shelton is partially supported by US Department of Energy under grant DE-SC0015655. \\
\indent F.~Siegert is supported by the German Research Foundation (DFG) under grant  SI 2009/1-1. \\
\indent J.P.~Silva is supported in part by the Portuguese Fundacao para a Ciencia e Tecnologia under contract UID/FIS/00777/2013. \\
\indent M.~Sjodahl is supported by the Swedish Research Council under contract 621-2012-2744. \\
\indent P.~Slavich is supported in part by the French ANR Young Researchers project ``HiggsAutomator'' (ANR-15-CE31-0002-01). \\
\indent M.~Slawinska is supported by the research programme of the Foundation for Fundamental Research on Matter (FOM), which is part of the Netherlands Organisation for Scientific Research (NWO). \\
\indent T.~Stebel acknowledges support of Marian Smoluchowski Research Consortium Matter Energy Future from KNOW funding and Polish National Science Centre grant DEC-2014/13/B/ST2/02486. \\
\indent T.~Stefaniak is supported by US Department of Energy grant number DESC0010107 and a Feodor-Lynen research fellowship sponsored by the Alexander von Humboldt foundation. \\
\indent I.W.~Stewart is supported by the US Department of Energy under grant DE-SC0011090 and by the Simons Foundation through the Investigator grant 327942. \\
\indent M.J.~Strassler thanks Harvard University's theory group for its hospitality. \\
\indent S.~Su is supported by US Department of Energy under grant DE-FG02-04ER-41298. \\
\indent X.~Sun is supported by NSFC China. \\
\indent F.J.~Tackmann is supported by the German Science Foundation (DFG) through the Emmy-Noether Grant No. TA 867/1-1. \\
\indent R.~Teixeira De Lima is supported by the US Department of Energy. \\
\indent R.S.~Thorne is supported by STFC under grant  ST/L000377/1 and and he thanks the members of the MMHT group and PDF4LHC working group for many helpful discussions. \\
\indent P.~Torrielli has received funding from the European Union 7th Framework programme for research and innovation under the Marie Curie grant agreement 609402-2020 researchers: Train to Move (T2M). \\
\indent M.~Tosi thanks Josh Bandavid. \\
\indent F.~Tramontano is partially supported by the INFN Iniziativa Specifica PhenoLNF. \\
\indent Z.~Tr{\'o}cs{\'a}nyi is supported by the Hungarian Scientific Research Fund grant K-101482 and the Swiss National Science Foundation SCOPES-JRP grant IZ73Z0-152601. \\
\indent M.~Trott thanks the Villum Foundation for support. \\
\indent I.~Tsinikos is supported by the FRS-FNRS ``Fonds de la Recherche Scientifique'' and in part by the Belgian Federal Science Policy Office through the Interuniversity Attraction Pole P7/37. \\
\indent A.~Vicini was supported in part by an Italian PRIN2010 grant, by a European Investment Bank EIBURS grant. \\
\indent D.~Wackeroth is supported in part by the US National Science Foundation under grant PHY-1417317. \\
\indent C.E.M.~Wagner is supported by the US Department of Energy under grant DE-FG02-13ER41958. His work at ANL is supported in part by the US Department of Energy under grant DE-AC02-06CH11357. \\
\indent M.~Wiesemann is supported in part by the Swiss National Science Foundation under contract 200021-156585. \\
\indent C.~Williams acknowledges support provided by the Center for Computational Research at the University at Buffalo. \\
\indent L.L.~Yang is supported in part by the National Natural Science Foundation of China under grant 11575004. \\
\indent M.~Zaro is supported by the European Union's Horizon 2020 research and innovation programme under the Marie Sklodovska-Curie grant agreement No 660171. \\
\indent J.~Zupan is supported in part by the US National Science Foundation under CAREER grant PHY-1151392. \\

\clearpage


\cleardoublepage
\phantomsection
\begin{appendices}
\renewcommand{\appendixname}{Appendix}


\chapter{Tables of branching ratios}
\label{BRappendix}

In this appendix we complete the listing of the branching fractions of the Standard Model 
Higgs boson discussed in \refS{sec:br-strategy}

\begin{table}\footnotesize
\setlength{\tabcolsep}{1.0ex}
\caption{SM Higgs boson branching ratios for $\Hbb$ and $\Htautau$,   
corresponding theoretical uncertainties (THU) and parametric
uncertainties from the quark masses (PU($m_q$)) and the strong coupling
(PU($\alphas$)) (expressed in percentage).  
Mass range around the Higgs boson resonance.
}
\label{tab:YRHXS4_1}
\begin{center}
\renewcommand{\arraystretch}{1.2}

\begin{tabular}{l|cccc|cccc}
\toprule
$\MH$ [GeV]  &  $\Hbb$ & {THU [\%]} & {PU($m_q$) [\%]} & {PU($\alphas$) [\%]}  & $\Htautau$ &  {THU [\%]} & {PU($m_q$) [\%]} & {PU($\alphas$) [\%]}\\
\midrule
120.0 & 6.536${\cdot 10^{-1}}$ & +0.58 -0.58  &  +0.60 -0.63  &  +0.67 -0.68  &6.981${\cdot 10^{-2}}$ & +1.16 -1.16  &  +1.10 -1.09  &  +0.71 -0.69  \\
120.5 & 6.472${\cdot 10^{-1}}$ & +0.59 -0.59  &  +0.60 -0.64  &  +0.68 -0.70  &6.918${\cdot 10^{-2}}$ & +1.17 -1.16  &  +1.09 -1.09  &  +0.71 -0.68  \\
121.0 & 6.406${\cdot 10^{-1}}$ & +0.59 -0.59  &  +0.62 -0.65  &  +0.69 -0.71  &6.854${\cdot 10^{-2}}$ & +1.17 -1.16  &  +1.07 -1.09  &  +0.68 -0.68  \\
121.5 & 6.338${\cdot 10^{-1}}$ & +0.60 -0.60  &  +0.63 -0.66  &  +0.70 -0.72  &6.787${\cdot 10^{-2}}$ & +1.17 -1.16  &  +1.06 -1.07  &  +0.68 -0.67  \\
122.0 & 6.269${\cdot 10^{-1}}$ & +0.61 -0.61  &  +0.64 -0.67  &  +0.71 -0.73  &6.718${\cdot 10^{-2}}$ & +1.17 -1.16  &  +1.07 -1.06  &  +0.68 -0.65  \\
122.5 & 6.198${\cdot 10^{-1}}$ & +0.61 -0.61  &  +0.66 -0.68  &  +0.72 -0.74  &6.648${\cdot 10^{-2}}$ & +1.17 -1.16  &  +1.04 -1.06  &  +0.66 -0.65  \\
123.0 & 6.126${\cdot 10^{-1}}$ & +0.62 -0.62  &  +0.67 -0.69  &  +0.73 -0.75  &6.576${\cdot 10^{-2}}$ & +1.17 -1.16  &  +1.02 -1.04  &  +0.66 -0.64  \\
\bottomrule
\end{tabular}\end{center}
\end{table}

\begin{table}\footnotesize
\setlength{\tabcolsep}{1.0ex}
\caption{SM Higgs boson branching ratios for $\Hmumu$ and $\Hcc$,
corresponding theoretical uncertainties (THU) and parametric
uncertainties from the quark masses (PU($m_q$)) and the strong coupling
(PU($\alphas$)) (expressed in percentage).  
Mass range around the Higgs boson resonance.
}
\label{tab:YRHXS4_2}
\begin{center}
\renewcommand{\arraystretch}{1.2}

\begin{tabular}{l|cccc|cccc}
\toprule
$\MH$ [GeV]  &  $\Hmumu$ & {THU [\%]} & {PU($m_q$) [\%]} & {PU($\alphas$) [\%]}  & $\Hcc$ &  {THU [\%]} & {PU($m_q$) [\%]} & {PU($\alphas$) [\%]}\\
\midrule
120.0 & 2.423${\cdot 10^{-4}}$ & +1.23 -1.23  &  +1.07 -1.09  &  +0.69 -0.70 &  3.244${\cdot 10^{-2}}$ & +1.20 -1.20  &  +5.27 -1.08  &  +1.14 -1.18  \\         
120.5 & 2.401${\cdot 10^{-4}}$ & +1.23 -1.23  &  +1.09 -1.10  &  +0.71 -0.70 &  3.212${\cdot 10^{-2}}$ & +1.20 -1.20  &  +5.26 -1.06  &  +1.17 -1.19  \\         
121.0 & 2.378${\cdot 10^{-4}}$ & +1.23 -1.23  &  +1.10 -1.07  &  +0.71 -0.67 &  3.180${\cdot 10^{-2}}$ & +1.20 -1.20  &  +5.25 -1.05  &  +1.17 -1.20  \\         
121.5 & 2.355${\cdot 10^{-4}}$ & +1.23 -1.23  &  +1.06 -1.08  &  +0.68 -0.68 &  3.147${\cdot 10^{-2}}$ & +1.20 -1.20  &  +5.23 -1.06  &  +1.15 -1.21  \\         
122.0 & 2.331${\cdot 10^{-4}}$ & +1.23 -1.23  &  +1.08 -1.04  &  +0.68 -0.64 &  3.112${\cdot 10^{-2}}$ & +1.20 -1.20  &  +5.26 -1.03  &  +1.20 -1.20  \\         
122.5 & 2.307${\cdot 10^{-4}}$ & +1.23 -1.23  &  +1.05 -1.06  &  +0.64 -0.69 &  3.077${\cdot 10^{-2}}$ & +1.20 -1.20  &  +5.26 -1.01  &  +1.18 -1.21  \\         
123.0 & 2.282${\cdot 10^{-4}}$ & +1.23 -1.23  &  +1.01 -1.07  &  +0.66 -0.65 &  3.041${\cdot 10^{-2}}$ & +1.20 -1.20  &  +5.28 -0.99  &  +1.23 -1.22  \\
\bottomrule
\end{tabular}\end{center}
\end{table}

\begin{table}\footnotesize
\setlength{\tabcolsep}{1.0ex}
\caption{SM Higgs boson branching ratios for $\Hgg$ and $\Hgaga$,
corresponding theoretical uncertainties (THU) and parametric
uncertainties from the quark masses (PU($m_q$)) and the strong coupling
(PU($\alphas$)) (expressed in percentage).  
Mass range around the Higgs boson resonance.
}
\label{tab:YRHXS4_3}
\begin{center}
\renewcommand{\arraystretch}{1.2}

\begin{tabular}{l|cccc|cccc}
\toprule
$\MH$ [GeV]  &  $\Hgg$ & {THU [\%]} & {PU($m_q$) [\%]} & {PU($\alphas$) [\%]}  & $\Hgaga$ &  {THU [\%]} & {PU($m_q$) [\%]} & {PU($\alphas$) [\%]}\\
\midrule
120.0 & 8.409${\cdot 10^{-2}}$ & +3.39 -3.40  &  +1.24 -1.26  &  +3.80 -3.72  &2.218${\cdot 10^{-3}}$ & +1.73 -1.73  &  +1.08 -1.10  &  +0.71 -0.72  \\
120.5 & 8.401${\cdot 10^{-2}}$ & +3.39 -3.40  &  +1.23 -1.24  &  +3.80 -3.70  &2.226${\cdot 10^{-3}}$ & +1.73 -1.73  &  +1.08 -1.05  &  +0.72 -0.71  \\
121.0 & 8.390${\cdot 10^{-2}}$ & +3.39 -3.40  &  +1.23 -1.23  &  +3.81 -3.69  &2.235${\cdot 10^{-3}}$ & +1.73 -1.73  &  +1.03 -1.05  &  +0.71 -0.72  \\
121.5 & 8.376${\cdot 10^{-2}}$ & +3.39 -3.40  &  +1.21 -1.22  &  +3.79 -3.68  &2.242${\cdot 10^{-3}}$ & +1.73 -1.73  &  +1.03 -1.05  &  +0.71 -0.71  \\
122.0 & 8.358${\cdot 10^{-2}}$ & +3.39 -3.40  &  +1.22 -1.19  &  +3.78 -3.66  &2.248${\cdot 10^{-3}}$ & +1.73 -1.73  &  +1.03 -1.04  &  +0.71 -0.71  \\
122.5 & 8.339${\cdot 10^{-2}}$ & +3.39 -3.41  &  +1.18 -1.20  &  +3.75 -3.66  &2.253${\cdot 10^{-3}}$ & +1.73 -1.73  &  +1.03 -0.99  &  +0.70 -0.66  \\
123.0 & 8.315${\cdot 10^{-2}}$ & +3.39 -3.41  &  +1.18 -1.19  &  +3.75 -3.64  &2.258${\cdot 10^{-3}}$ & +1.73 -1.73  &  +1.02 -0.99  &  +0.67 -0.65  \\
\bottomrule
\end{tabular}\end{center}
\end{table}

\begin{table}\footnotesize
\setlength{\tabcolsep}{1.0ex}
\caption{SM Higgs boson branching ratios for $\HZga$ and $\HWW$,
corresponding theoretical uncertainties (THU) and parametric
uncertainties from the quark masses (PU($m_q$)) and the strong coupling
(PU($\alphas$)) (expressed in percentage).  
Mass range around the Higgs boson resonance.
}
\label{tab:YRHXS4_4}
\begin{center}
\renewcommand{\arraystretch}{1.2}

\begin{tabular}{l|cccc|cccc}
\toprule
$\MH$ [GeV]  &  $\HZga$ & {THU [\%]} & {PU($m_q$) [\%]} & {PU($\alphas$) [\%]}  & $\HWW$ &  {THU [\%]} & {PU($m_q$) [\%]} & {PU($\alphas$) [\%]}\\
\midrule
120.0 & 1.100${\cdot 10^{-3}}$ & +5.72 -5.72  &  +1.09 -1.01  &  +0.72 -0.64  &1.405${\cdot 10^{-1}}$ & +1.08 -1.08  &  +1.11 -1.11  &  +0.73 -0.73  \\
120.5 & 1.143${\cdot 10^{-3}}$ & +5.72 -5.72  &  +1.05 -1.07  &  +0.70 -0.69  &1.470${\cdot 10^{-1}}$ & +1.07 -1.07  &  +1.10 -1.09  &  +0.73 -0.71  \\
121.0 & 1.186${\cdot 10^{-3}}$ & +5.72 -5.72  &  +1.10 -1.04  &  +0.76 -0.67  &1.537${\cdot 10^{-1}}$ & +1.06 -1.06  &  +1.09 -1.08  &  +0.72 -0.71  \\
121.5 & 1.229${\cdot 10^{-3}}$ & +5.72 -5.72  &  +1.06 -1.08  &  +0.73 -0.65  &1.606${\cdot 10^{-1}}$ & +1.05 -1.05  &  +1.07 -1.07  &  +0.70 -0.70  \\
122.0 & 1.272${\cdot 10^{-3}}$ & +5.72 -5.72  &  +1.11 -1.03  &  +0.70 -0.63  &1.677${\cdot 10^{-1}}$ & +1.05 -1.04  &  +1.04 -1.07  &  +0.69 -0.69  \\
122.5 & 1.316${\cdot 10^{-3}}$ & +5.72 -5.72  &  +0.99 -1.10  &  +0.60 -0.68  &1.750${\cdot 10^{-1}}$ & +1.04 -1.04  &  +1.04 -1.06  &  +0.69 -0.68  \\
123.0 & 1.359${\cdot 10^{-3}}$ & +5.72 -5.72  &  +1.03 -1.04  &  +0.67 -0.65  &1.824${\cdot 10^{-1}}$ & +1.03 -1.03  &  +1.03 -1.05  &  +0.69 -0.69  \\
\bottomrule
\end{tabular}\end{center}
\end{table}

\begin{table}\footnotesize
\setlength{\tabcolsep}{1.0ex}
\caption{SM Higgs boson branching ratios for $\HZZ$ and Higgs boson total width
  $\Gamma_{\PH}$,
corresponding theoretical uncertainties (THU) and parametric
uncertainties from the quark masses (PU($m_q$)) and the strong coupling
(PU($\alphas$)) (expressed in percentage).  
Mass range around the Higgs boson resonance.
}
\label{tab:YRHXS4_5}
\begin{center}
\renewcommand{\arraystretch}{1.2}

\begin{tabular}{l|cccc|cccc}
\toprule
$\MH$ [GeV]  &  $\HZZ$ & {THU [\%]} & {PU($m_q$) [\%]} & {PU($\alphas$) [\%]}  & $\Gamma_{\PH}$ [GeV] &  {THU [\%]} & {PU($m_q$) [\%]} & {PU($\alphas$) [\%]}\\
\midrule
120.0 & 1.572${\cdot 10^{-2}}$ & +1.08 -1.08  &  +1.11 -1.11  &  +0.73 -0.73  &3.527${\cdot 10^{-3}}$ & +0.73 -0.73  &  +1.12 -1.10  &  +0.71 -0.70  \\
120.5 & 1.662${\cdot 10^{-2}}$ & +1.07 -1.07  &  +1.10 -1.09  &  +0.73 -0.71  &3.574${\cdot 10^{-3}}$ & +0.73 -0.73  &  +1.10 -1.09  &  +0.69 -0.70  \\
121.0 & 1.755${\cdot 10^{-2}}$ & +1.06 -1.06  &  +1.09 -1.08  &  +0.72 -0.71  &3.622${\cdot 10^{-3}}$ & +0.73 -0.73  &  +1.09 -1.07  &  +0.69 -0.69  \\
121.5 & 1.851${\cdot 10^{-2}}$ & +1.05 -1.05  &  +1.07 -1.07  &  +0.71 -0.70  &3.673${\cdot 10^{-3}}$ & +0.73 -0.73  &  +1.08 -1.06  &  +0.67 -0.68  \\
122.0 & 1.951${\cdot 10^{-2}}$ & +1.05 -1.04  &  +1.04 -1.07  &  +0.69 -0.69  &3.725${\cdot 10^{-3}}$ & +0.73 -0.73  &  +1.09 -1.03  &  +0.67 -0.66  \\
122.5 & 2.055${\cdot 10^{-2}}$ & +1.04 -1.04  &  +1.04 -1.06  &  +0.68 -0.67  &3.779${\cdot 10^{-3}}$ & +0.73 -0.73  &  +1.08 -1.03  &  +0.65 -0.65  \\
123.0 & 2.161${\cdot 10^{-2}}$ & +1.03 -1.03  &  +1.03 -1.05  &  +0.68 -0.69  &3.836${\cdot 10^{-3}}$ & +0.73 -0.73  &  +1.06 -1.02  &  +0.66 -0.65  \\
\bottomrule
\end{tabular}\end{center}
\end{table}

\begin{table}\footnotesize
\setlength{\tabcolsep}{1.0ex}
\caption{SM Higgs boson branching ratios for $\PH\to4\Pl$ and 
corresponding total uncertainty (expressed in percentage).  
Mass range around the Higgs boson resonance.
}
\label{tab:YRHXS4_4f_1}
\begin{center}
\renewcommand{\arraystretch}{1.2}

\begin{tabular}{l|ccccc}
\toprule
$\MH$ [GeV]  &  $\PH \to \Plp\Plm\Plp\Plm$ &  $\PH \to \Plp\Plm\Plp\Plm$ & $\PH \to \Pep \Pem \Pep \Pem$ & $\PH \to \Pep \Pem \PGmp \PGmm$  & $\Delta$ BR[\%]\\
 & $\Pl = \Pe, \PGm, \PGt$ & $\Pl = \Pe, \PGm$ &  & &  \\
\midrule
120.0 & 1.659${\cdot 10^{-4}}$ & 7.524${\cdot 10^{-5}}$ & 1.995${\cdot 10^{-5}}$ & 3.533${\cdot 10^{-5}}$ & $\pm 2.41$  \\
120.5 & 1.753${\cdot 10^{-4}}$ & 7.950${\cdot 10^{-5}}$ & 2.107${\cdot 10^{-5}}$ & 3.735${\cdot 10^{-5}}$ & $\pm 2.40$  \\
121.0 & 1.850${\cdot 10^{-4}}$ & 8.387${\cdot 10^{-5}}$ & 2.221${\cdot 10^{-5}}$ & 3.946${\cdot 10^{-5}}$ & $\pm 2.37$  \\
121.5 & 1.950${\cdot 10^{-4}}$ & 8.837${\cdot 10^{-5}}$ & 2.337${\cdot 10^{-5}}$ & 4.162${\cdot 10^{-5}}$ & $\pm 2.34$  \\
122.0 & 2.054${\cdot 10^{-4}}$ & 9.304${\cdot 10^{-5}}$ & 2.458${\cdot 10^{-5}}$ & 4.388${\cdot 10^{-5}}$ & $\pm 2.32$  \\
122.5 & 2.161${\cdot 10^{-4}}$ & 9.785${\cdot 10^{-5}}$ & 2.581${\cdot 10^{-5}}$ & 4.622${\cdot 10^{-5}}$ & $\pm 2.30$  \\
\bottomrule
\end{tabular}\end{center}
\end{table}

\begin{table}\footnotesize
\setlength{\tabcolsep}{1.0ex}
\caption{SM Higgs boson branching ratios for $\PH\to2\Pl2\PGn$ and 
corresponding total uncertainty (expressed in percentage).  
Mass range around the Higgs boson resonance.
}
\label{tab:YRHXS4_4f_2}
\begin{center}
\renewcommand{\arraystretch}{1.2}

\begin{tabular}{l|ccccc}
\toprule
$\MH$ [GeV]  &  $\PH \to \Plp\Plm \PGn\PAGn$ &  $\PH \to \Plp\Plm \PGn \PAGn$ & $\PH \to \Pep\PGne \Pem\PAGne$ & $\PH \to \Pep\PGne \PGmm\PAGnGm$  & $\Delta$ BR[\%]\\
 & $\Pl = \Pe, \PGm, \PGt$ & $\Pl = \Pe, \PGm$ &  & &  \\
 & $\PGn= $any & $\PGn =$ any&  & &  \\

\midrule
120.0 &  1.526${\cdot 10^{-2}}$ & 6.864${\cdot 10^{-3}}$ & 1.634${\cdot 10^{-3}}$ & 1.656${\cdot 10^{-3}}$ & $\pm 2.41$  \\
120.5 &  1.598${\cdot 10^{-2}}$ & 7.190${\cdot 10^{-3}}$ & 1.712${\cdot 10^{-3}}$ & 1.733${\cdot 10^{-3}}$ & $\pm 2.40$  \\
121.0 &  1.673${\cdot 10^{-2}}$ & 7.528${\cdot 10^{-3}}$ & 1.793${\cdot 10^{-3}}$ & 1.812${\cdot 10^{-3}}$ & $\pm 2.37$  \\
121.5 &  1.749${\cdot 10^{-2}}$ & 7.874${\cdot 10^{-3}}$ & 1.876${\cdot 10^{-3}}$ & 1.894${\cdot 10^{-3}}$ & $\pm 2.34$  \\
122.0 &  1.828${\cdot 10^{-2}}$ & 8.230${\cdot 10^{-3}}$ & 1.962${\cdot 10^{-3}}$ & 1.977${\cdot 10^{-3}}$ & $\pm 2.32$  \\
\bottomrule
\end{tabular}\end{center}
\end{table}

\begin{table}\footnotesize
\setlength{\tabcolsep}{1.0ex}
\caption{SM Higgs boson branching ratios for $\PH\to2\Pl2\PQq$, $\Pl\PGn2\PQq$, $2\PGn2\PQq$, 
$4\PQq,4\Pf$ and corresponding total uncertainty (expressed in percentage).  
Mass range around the Higgs boson resonance.
}
\label{tab:YRHXS4_4f_3}
\begin{center}
\renewcommand{\arraystretch}{1.2}

\begin{tabular}{l|ccccccc}
\toprule
$\MH$ [GeV]  &  $\PH \to \Plp\Plm\PQq\PAQq$ &  $\PH \to \Plp\Plm\PQq\PAQq$ & $\PH \to \Plp \PAGnl\PQq\PAQq$ & $\PH \to \PGn \PGn\PQq\PAQq$  & $\PH \to 4\PQq$ & $\PH \to 4\Pf$ &$\Delta$ BR[\%]\\
 & $\Pl = \Pe, \PGm, \PGt$ & $\Pl = \Pe, \PGm$ & $\Pl = \Pe, \PGm$ & $\PGn =$ any  & & f=any fermion & \\
 & $\PQq=\PQu,\PQd,\PQc,\PQs,\PQb$   
 & $\PQq=\PQu,\PQd,\PQc,\PQs,\PQb$   
 & $\PQq=\PQu,\PQd,\PQc,\PQs,\PQb$   
 & $\PQq=\PQu,\PQd,\PQc,\PQs,\PQb$   
 & $\PQq=\PQu,\PQd,\PQc,\PQs,\PQb$   
&  & \\
\midrule

120.0 & 2.200${\cdot 10^{-3}}$ & 1.466${\cdot 10^{-3}}$ & 2.060${\cdot 10^{-2}}$ & 4.390${\cdot 10^{-3}}$ & 7.076${\cdot 10^{-2}}$ & 1.553${\cdot 10^{-1}}$ & $\pm 2.41$  \\
120.5 & 2.325${\cdot 10^{-3}}$ & 1.550${\cdot 10^{-3}}$ & 2.155${\cdot 10^{-2}}$ & 4.641${\cdot 10^{-3}}$ & 7.413${\cdot 10^{-2}}$ & 1.626${\cdot 10^{-1}}$ & $\pm 2.40$  \\
121.0 & 2.456${\cdot 10^{-3}}$ & 1.637${\cdot 10^{-3}}$ & 2.254${\cdot 10^{-2}}$ & 4.902${\cdot 10^{-3}}$ & 7.762${\cdot 10^{-2}}$ & 1.703${\cdot 10^{-1}}$ & $\pm 2.37$  \\
121.5 & 2.591${\cdot 10^{-3}}$ & 1.727${\cdot 10^{-3}}$ & 2.355${\cdot 10^{-2}}$ & 5.172${\cdot 10^{-3}}$ & 8.119${\cdot 10^{-2}}$ & 1.781${\cdot 10^{-1}}$ & $\pm 2.34$  \\
122.0 & 2.732${\cdot 10^{-3}}$ & 1.821${\cdot 10^{-3}}$ & 2.459${\cdot 10^{-2}}$ & 5.452${\cdot 10^{-3}}$ & 8.488${\cdot 10^{-2}}$ & 1.862${\cdot 10^{-1}}$ & $\pm 2.32$  \\
\bottomrule
\end{tabular}\end{center}
\end{table}

\begin{table}\footnotesize
\tabcolsep5pt
\caption{SM Higgs boson partials widths and their relative parametric (PU) and
  theoretical (THU) uncertainties for a selection of Higgs boson masses. 
  For PU, all the single contributions are shown. For these
  four columns, the upper percentage value (with its sign) refers 
  to the positive variation of the parameter, while the lower one
  refers to the negative variation of the parameter. 
}
\label{tab:br-correlation}
\begin{center}
\renewcommand{\arraystretch}{1.7}
\begin{tabular}{lccrrrrr}
\toprule
Channel & $\MH$ [GeV]  &  $\Gamma$ [MeV] &  $\Delta \alphas$ & $\Delta \Mb$ & $\Delta \Mc$ & $\Delta \Mt$ & THU \\
\midrule
& 124  &2.37\zehoze           &$^{-1.4\%}_{+1.4\%}$ & $ ^{+1.7\%}_{-1.7\%}$ & $ ^{+0.0\%}_{-0.0\%}$ & $ ^{+0.0\%}_{-0.0\%}$& $ ^{+0.5\%}_{-0.5\%}$\\ $\Hbb$
& 125  &2.38\zehoze           &$^{-1.4\%}_{+1.4\%}$ & $ ^{+1.7\%}_{-1.7\%}$ & $ ^{+0.0\%}_{-0.0\%}$ & $ ^{+0.0\%}_{-0.0\%}$& $ ^{+0.5\%}_{-0.5\%}$\\
& 126  &2.40\zehoze           &$^{-1.4\%}_{+1.4\%}$ & $ ^{+1.7\%}_{-1.7\%}$ & $ ^{+0.0\%}_{-0.0\%}$ & $ ^{+0.0\%}_{-0.0\%}$& $ ^{+0.5\%}_{-0.5\%}$\\ 
\midrule
& 124  &2.54\zehomi{1}        &$^{+0.0\%}_{-0.0\%}$ & $ ^{+0.0\%}_{-0.0\%}$ & $ ^{+0.0\%}_{-0.0\%}$ & $ ^{+0.0\%}_{-0.0\%}$& $ ^{+0.5\%}_{-0.5\%}$\\ $\Htautau$
& 125  &2.56\zehomi{1}        &$^{+0.0\%}_{-0.0\%}$ & $ ^{+0.0\%}_{-0.0\%}$ & $ ^{+0.0\%}_{-0.0\%}$ & $ ^{+0.0\%}_{-0.0\%}$& $ ^{+0.5\%}_{-0.5\%}$\\
& 126  &2.58\zehomi{1}        &$^{+0.0\%}_{-0.0\%}$ & $ ^{+0.0\%}_{-0.0\%}$ & $ ^{+0.0\%}_{-0.0\%}$ & $ ^{+0.0\%}_{-0.0\%}$& $ ^{+0.5\%}_{-0.5\%}$\\ 
\midrule
& 124  &8.83\zehomi{4}        &$^{+0.0\%}_{-0.0\%}$ & $ ^{+0.0\%}_{-0.0\%}$ & $ ^{+0.0\%}_{-0.0\%}$ & $ ^{+0.0\%}_{-0.0\%}$& $ ^{+0.5\%}_{-0.5\%}$\\
\bottomrule

\end{tabular}\end{center}
\end{table}

\begin{table}\footnotesize
\tabcolsep5pt
\caption{SM Higgs boson partials widths (in GeV) for 2-fermion decay channels, predictions without electroweak corrections. Low and intermediate mass range.
}
\label{tab:YRHXS4_noEW_1}
\begin{center}
\renewcommand{\arraystretch}{1.2}

\begin{tabular}{l|cccc}
\toprule
$\MH$ [GeV]  & $\Gamma_{\Hbb}$ & $\Gamma_{\Htautau}$ &$\Gamma_{\Hmumu}$ & $\Gamma_{\Hcc}$  \\
\midrule
20 & 5.244${\cdot 10^{-4}}$ & 3.949${\cdot 10^{-5}}$ & 1.465${\cdot 10^{-7}}$ & 2.936${\cdot 10^{-5}}$  \\
25 & 6.568${\cdot 10^{-4}}$ & 5.024${\cdot 10^{-5}}$ & 1.831${\cdot 10^{-7}}$ & 3.437${\cdot 10^{-5}}$  \\
30 & 7.709${\cdot 10^{-4}}$ & 6.086${\cdot 10^{-5}}$ & 2.198${\cdot 10^{-7}}$ & 3.921${\cdot 10^{-5}}$  \\
35 & 8.754${\cdot 10^{-4}}$ & 7.139${\cdot 10^{-5}}$ & 2.563${\cdot 10^{-7}}$ & 4.390${\cdot 10^{-5}}$  \\
40 & 9.752${\cdot 10^{-4}}$ & 8.193${\cdot 10^{-5}}$ & 2.932${\cdot 10^{-7}}$ & 4.851${\cdot 10^{-5}}$  \\
45 & 1.070${\cdot 10^{-3}}$ & 9.234${\cdot 10^{-5}}$ & 3.296${\cdot 10^{-7}}$ & 5.297${\cdot 10^{-5}}$  \\
50 & 1.163${\cdot 10^{-3}}$ & 1.028${\cdot 10^{-4}}$ & 3.663${\cdot 10^{-7}}$ & 5.738${\cdot 10^{-5}}$  \\
\bottomrule
\end{tabular}\end{center}
\end{table}

\begin{table}\footnotesize
\tabcolsep5pt
\caption{SM Higgs boson partials widths (in GeV) for 2-fermion decay channels, predictions without electroweak corrections. High mass range.
}
\label{tab:YRHXS4_noEW_2}
\begin{center}
\renewcommand{\arraystretch}{1.2}

\begin{tabular}{l|ccccc}
\toprule
$\MH$ [GeV]  & $\Gamma_{\Hbb}$ & $\Gamma_{\Htautau}$ &$\Gamma_{\Hmumu}$ & $\Gamma_{\Hcc}$ & $\Gamma_{\Htt}$ \\
\midrule
260 & 4.351${\cdot 10^{-3}}$ & 5.384${\cdot 10^{-4}}$ & 1.905${\cdot 10^{-6}}$ & 2.129${\cdot 10^{-4}}$ & 1.946${\cdot 10^{-7}}$  \\
270 & 4.487${\cdot 10^{-3}}$ & 5.592${\cdot 10^{-4}}$ & 1.978${\cdot 10^{-6}}$ & 2.195${\cdot 10^{-4}}$ & 1.235${\cdot 10^{-5}}$  \\
280 & 4.689${\cdot 10^{-3}}$ & 5.885${\cdot 10^{-4}}$ & 2.081${\cdot 10^{-6}}$ & 2.294${\cdot 10^{-4}}$ & 7.015${\cdot 10^{-5}}$  \\
290 & 4.756${\cdot 10^{-3}}$ & 6.008${\cdot 10^{-4}}$ & 2.125${\cdot 10^{-6}}$ & 2.328${\cdot 10^{-4}}$ & 2.247${\cdot 10^{-4}}$  \\
300 & 4.890${\cdot 10^{-3}}$ & 6.215${\cdot 10^{-4}}$ & 2.198${\cdot 10^{-6}}$ & 2.393${\cdot 10^{-4}}$ & 5.767${\cdot 10^{-4}}$  \\
310 & 5.023${\cdot 10^{-3}}$ & 6.423${\cdot 10^{-4}}$ & 2.271${\cdot 10^{-6}}$ & 2.458${\cdot 10^{-4}}$ & 1.317${\cdot 10^{-3}}$  \\
320 & 5.155${\cdot 10^{-3}}$ & 6.629${\cdot 10^{-4}}$ & 2.345${\cdot 10^{-6}}$ & 2.522${\cdot 10^{-4}}$ & 2.862${\cdot 10^{-3}}$  \\
\bottomrule
\end{tabular}\end{center}
\end{table}

\begin{table}\footnotesize
\tabcolsep5pt
\caption{SM Higgs boson partials widths (in GeV) for 2-boson decay channels, predictions without electroweak corrections. Low and intermediate mass range.
}
\label{tab:YRHXS4_noEW_3}
\begin{center}
\renewcommand{\arraystretch}{1.2}

\begin{tabular}{l|ccccc}
\toprule
$\MH$ [GeV]  & $\Gamma_{\Hgg}$ & $\Gamma_{\Hgaga}$ &$\Gamma_{\HZga}$ & $\Gamma_{\HWW}$ & $\Gamma_{\HZZ}$ \\
\midrule
20 & 1.895${\cdot 10^{-5}}$ & 2.861${\cdot 10^{-8}}$ & 0.000${\cdot 10^{0}}$ & 3.560${\cdot 10^{-11}}$ & 3.100${\cdot 10^{-12}}$   \\
25 & 1.604${\cdot 10^{-5}}$ & 5.592${\cdot 10^{-8}}$ & 0.000${\cdot 10^{0}}$ & 1.753${\cdot 10^{-10}}$ & 3.760${\cdot 10^{-11}}$   \\
30 & 1.522${\cdot 10^{-5}}$ & 9.586${\cdot 10^{-8}}$ & 0.000${\cdot 10^{0}}$ & 6.432${\cdot 10^{-10}}$ & 1.586${\cdot 10^{-10}}$   \\
35 & 1.647${\cdot 10^{-5}}$ & 1.512${\cdot 10^{-7}}$ & 0.000${\cdot 10^{0}}$ & 1.956${\cdot 10^{-9}}$ & 5.108${\cdot 10^{-10}}$   \\
40 & 1.939${\cdot 10^{-5}}$ & 2.246${\cdot 10^{-7}}$ & 0.000${\cdot 10^{0}}$ & 5.186${\cdot 10^{-9}}$ & 1.407${\cdot 10^{-9}}$   \\
45 & 2.355${\cdot 10^{-5}}$ & 3.191${\cdot 10^{-7}}$ & 0.000${\cdot 10^{0}}$ & 1.236${\cdot 10^{-8}}$ & 3.456${\cdot 10^{-9}}$   \\
50 & 2.879${\cdot 10^{-5}}$ & 4.380${\cdot 10^{-7}}$ & 0.000${\cdot 10^{0}}$ & 2.720${\cdot 10^{-8}}$ & 7.734${\cdot 10^{-9}}$   \\
\bottomrule
\end{tabular}\end{center}
\end{table}

\begin{table}\footnotesize
\tabcolsep5pt
\caption{SM Higgs boson partials widths (in GeV) for 2-boson decay channels, predictions without electroweak corrections. High mass range.
}
\label{tab:YRHXS4_noEW_4}
\begin{center}
\renewcommand{\arraystretch}{1.2}

\begin{tabular}{l|ccccc}
\toprule
$\MH$ [GeV]  & $\Gamma_{\Hgg}$ & $\Gamma_{\Hgaga}$ &$\Gamma_{\HZga}$ & $\Gamma_{\HWW}$ & $\Gamma_{\HZZ}$ \\
\midrule
260 & 3.215${\cdot 10^{-3}}$ & 9.890${\cdot 10^{-5}}$ & 4.072${\cdot 10^{-4}}$ & 3.232${\cdot 10^{0}}$ & 1.386${\cdot 10^{0}}$   \\
270 & 3.699${\cdot 10^{-3}}$ & 1.018${\cdot 10^{-4}}$ & 4.282${\cdot 10^{-4}}$ & 3.765${\cdot 10^{0}}$ & 1.632${\cdot 10^{0}}$   \\
280 & 4.315${\cdot 10^{-3}}$ & 1.061${\cdot 10^{-4}}$ & 4.545${\cdot 10^{-4}}$ & 4.351${\cdot 10^{0}}$ & 1.904${\cdot 10^{0}}$   \\
290 & 4.894${\cdot 10^{-3}}$ & 1.071${\cdot 10^{-4}}$ & 4.665${\cdot 10^{-4}}$ & 4.992${\cdot 10^{0}}$ & 2.202${\cdot 10^{0}}$   \\
300 & 5.644${\cdot 10^{-3}}$ & 1.094${\cdot 10^{-4}}$ & 4.840${\cdot 10^{-4}}$ & 5.690${\cdot 10^{0}}$ & 2.529${\cdot 10^{0}}$   \\
310 & 6.529${\cdot 10^{-3}}$ & 1.116${\cdot 10^{-4}}$ & 5.004${\cdot 10^{-4}}$ & 6.446${\cdot 10^{0}}$ & 2.885${\cdot 10^{0}}$   \\
320 & 7.597${\cdot 10^{-3}}$ & 1.137${\cdot 10^{-4}}$ & 5.157${\cdot 10^{-4}}$ & 7.264${\cdot 10^{0}}$ & 3.271${\cdot 10^{0}}$   \\
\bottomrule
\end{tabular}\end{center}
\end{table}

\clearpage



\chapter{SM gluon-gluon-fusion cross sections}
\label{ggFappendix}
In this appendix the recommended gluon-gluon fusion cross-sections  are presented.

\begin{table}
\caption{Inclusive ggF cross sections for a LHC CM energy of $\sqrt{s}=7\UTeV$, at N$^3$LO QCD,  
 together with their uncertainties. The $\mathrm{TH}$ uncertainty is interpreted as a flat $100\%$ confidence level. 
 $\Delta_{\mathrm{TH}}^{\mathrm{Gaussian}}$ uncertainty is interpreted as a one-sigma range.}
\label{tab:ggF_XStot_7}
\renewcommand{\arraystretch}{1.2}%
\begin{center}%
\begin{small}%
\tabcolsep5pt
\begin{tabular}{ccccccc}%
\toprule
$\MH$[GeV] & $\sigma^{ggF}$[pb] & $\Delta_{\mathrm{TH}}$[\%] &
$\Delta_{\mathrm{TH}}^{\mathrm{Gaussian}}$[\%] & $\Delta_{\mathrm{PDF\oplus\alpha_s}}$[\%] & $\Delta_{\mathrm{PDF}}$[\%] & $\Delta_{\mathrm{\alphas}}$[\%] 
\\
\midrule
$	120.0	$	&	$	18.31	$	&	$^{+	4.5	}_{	-7.1	}$	&	$\pm	4.1	$	&	$\pm	3.3	$	&	$\pm	1.9	$	&	$\pm	2.7	$	\\
$	120.5	$	&	$	18.16	$	&	$^{+	4.5	}_{	-7.1	}$	&	$\pm	4.1	$	&	$\pm	3.3	$	&	$\pm	1.9	$	&	$\pm	2.7	$	\\
$	121.0	$	&	$	18.00	$	&	$^{+	4.5	}_{	-7.1	}$	&	$\pm	4.1	$	&	$\pm	3.3	$	&	$\pm	1.9	$	&	$\pm	2.7	$	\\
$	121.5	$	&	$	17.85	$	&	$^{+	4.5	}_{	-7.1	}$	&	$\pm	4.1	$	&	$\pm	3.3	$	&	$\pm	1.9	$	&	$\pm	2.7	$	\\
$	122.0	$	&	$	17.71	$	&	$^{+	4.5	}_{	-7.1	}$	&	$\pm	4.1	$	&	$\pm	3.3	$	&	$\pm	1.9	$	&	$\pm	2.7	$	\\
$	122.5	$	&	$	17.56	$	&	$^{+	4.5	}_{	-7.0	}$	&	$\pm	4.1	$	&	$\pm	3.3	$	&	$\pm	1.9	$	&	$\pm	2.7	$	\\
$	123.0	$	&	$	17.41	$	&	$^{+	4.5	}_{	-7.0	}$	&	$\pm	4.1	$	&	$\pm	3.3	$	&	$\pm	1.9	$	&	$\pm	2.7	$	\\
\bottomrule
\end{tabular}%
\end{small}%
\end{center}%
\end{table}

\begin{table}
\caption{Inclusive ggF cross sections for a LHC CM energy of $\sqrt{s}=8\UTeV$, at N$^3$LO QCD,  
 together with their uncertainties. The $\mathrm{TH}$ uncertainty is interpreted as a flat $100\%$ confidence level. 
 $\Delta_{\mathrm{TH}}^{\mathrm{Gaussian}}$ uncertainty is interpreted as a one-sigma range.}
\label{tab:ggF_XStot_8}
\renewcommand{\arraystretch}{1.2}%
\begin{center}%
\begin{small}%
\tabcolsep5pt
\begin{tabular}{ccccccc}%
\toprule
$\MH$[GeV] & $\sigma^{ggF}$[pb] & $\Delta_{\mathrm{TH}}$[\%] &
$\Delta_{\mathrm{TH}}^{\mathrm{Gaussian}}$[\%] & $\Delta_{\mathrm{PDF\oplus\alpha_s}}$[\%] & $\Delta_{\mathrm{PDF}}$[\%] & $\Delta_{\mathrm{\alphas}}$[\%] 
\\
\midrule
$	120.0	$	&	$	23.22	$	&	$^{+	4.5	}_{	-7.1	}$	&	$\pm	4.1	$	&	$\pm	3.3	$	&	$\pm	1.9	$	&	$\pm	2.7	$	\\
$	120.5	$	&	$	23.03	$	&	$^{+	4.5	}_{	-7.0	}$	&	$\pm	4.1	$	&	$\pm	3.2	$	&	$\pm	1.9	$	&	$\pm	2.7	$	\\
$	121.0	$	&	$	22.85	$	&	$^{+	4.5	}_{	-7.0	}$	&	$\pm	4.1	$	&	$\pm	3.2	$	&	$\pm	1.9	$	&	$\pm	2.7	$	\\
$	121.5	$	&	$	22.66	$	&	$^{+	4.5	}_{	-7.0	}$	&	$\pm	4.0	$	&	$\pm	3.2	$	&	$\pm	1.9	$	&	$\pm	2.6	$	\\
$	122.0	$	&	$	22.48	$	&	$^{+	4.5	}_{	-7.0	}$	&	$\pm	4.0	$	&	$\pm	3.2	$	&	$\pm	1.9	$	&	$\pm	2.6	$	\\
$	122.5	$	&	$	22.30	$	&	$^{+	4.5	}_{	-7.0	}$	&	$\pm	4.0	$	&	$\pm	3.2	$	&	$\pm	1.9	$	&	$\pm	2.6	$	\\
$	123.0	$	&	$	22.12	$	&	$^{+	4.5	}_{	-7.0	}$	&	$\pm	4.0	$	&	$\pm	3.2	$	&	$\pm	1.9	$	&	$\pm	2.6	$	\\
\bottomrule
\end{tabular}%
\end{small}%
\end{center}%
\end{table}

\begin{table}
\caption{Inclusive ggF cross sections for a LHC CM energy of $\sqrt{s}=13\UTeV$, at N$^3$LO QCD,  
 together with their uncertainties. The $\mathrm{TH}$ uncertainty is interpreted as a flat $100\%$ confidence level. 
 $\Delta_{\mathrm{TH}}^{\mathrm{Gaussian}}$ uncertainty is interpreted as a one-sigma range.}
\label{tab:ggF_XStot_13}
\renewcommand{\arraystretch}{1.2}%
\begin{center}%
\begin{small}%
\tabcolsep5pt
\begin{tabular}{ccccccc}%
\toprule
$\MH$[GeV] & $\sigma^{ggF}$[pb] & $\Delta_{\mathrm{TH}}$[\%] &
$\Delta_{\mathrm{TH}}^{\mathrm{Gaussian}}$[\%] & $\Delta_{\mathrm{PDF\oplus\alpha_s}}$[\%] & $\Delta_{\mathrm{PDF}}$[\%] & $\Delta_{\mathrm{\alphas}}$[\%] 
\\
\midrule
$	120.0	$	&	$	52.22	$	&	$^{+	4.7	}_{-	6.9	}$	&	$\pm	4.0	$	&	$\pm	3.2	$	&	$\pm	1.9	$	&	$\pm	2.6	$	\\
$	120.5	$	&	$	51.84	$	&	$^{+	4.7	}_{-	6.9	}$	&	$\pm	4.0	$	&	$\pm	3.2	$	&	$\pm	1.9	$	&	$\pm	2.6	$	\\
$	121.0	$	&	$	51.46	$	&	$^{+	4.6	}_{-	6.9	}$	&	$\pm	4.0	$	&	$\pm	3.2	$	&	$\pm	1.9	$	&	$\pm	2.6	$	\\
$	121.5	$	&	$	51.08	$	&	$^{+	4.6	}_{-	6.8	}$	&	$\pm	3.9	$	&	$\pm	3.2	$	&	$\pm	1.9	$	&	$\pm	2.6	$	\\
$	122.0	$	&	$	50.71	$	&	$^{+	4.6	}_{-	6.8	}$	&	$\pm	3.9	$	&	$\pm	3.2	$	&	$\pm	1.9	$	&	$\pm	2.6	$	\\
$	122.5	$	&	$	50.35	$	&	$^{+	4.6	}_{-	6.8	}$	&	$\pm	3.9	$	&	$\pm	3.2	$	&	$\pm	1.9	$	&	$\pm	2.6	$	\\
$	123.0	$	&	$	49.98	$	&	$^{+	4.6	}_{-	6.8	}$	&	$\pm	3.9	$	&	$\pm	3.2	$	&	$\pm	1.9	$	&	$\pm	2.6	$	\\
\bottomrule
\end{tabular}%
\end{small}%
\end{center}%
\end{table}

\begin{table}
\caption{Inclusive ggF cross sections for a LHC CM energy of $\sqrt{s}=14\UTeV$, at N$^3$LO QCD,  
 together with their uncertainties. The $\mathrm{TH}$ uncertainty is interpreted as a flat $100\%$ confidence level. 
 $\Delta_{\mathrm{TH}}^{\mathrm{Gaussian}}$ uncertainty is interpreted as a one-sigma range.}
\label{tab:ggF_XStot_14}
\renewcommand{\arraystretch}{1.2}%
\begin{center}%
\begin{small}%
\tabcolsep5pt
\begin{tabular}{ccccccc}%
\toprule
$\MH$[GeV] & $\sigma^{ggF}$[pb] & $\Delta_{\mathrm{TH}}$[\%] &
$\Delta_{\mathrm{TH}}^{\mathrm{Gaussian}}$[\%] & $\Delta_{\mathrm{PDF\oplus\alpha_s}}$[\%] & $\Delta_{\mathrm{PDF}}$[\%] & $\Delta_{\mathrm{\alphas}}$[\%] 
\\
\midrule
$	120.0	$	&	$	58.71	$	&	$^{+	4.7	}_{-	6.9	}$	&	$\pm	4.0	$	&	$\pm	3.2	$	&	$\pm	1.9	$	&	$\pm	2.6	$	\\
$	120.5	$	&	$	58.29	$	&	$^{+	4.7	}_{-	6.9	}$	&	$\pm	4.0	$	&	$\pm	3.2	$	&	$\pm	1.9	$	&	$\pm	2.6	$	\\
$	121.0	$	&	$	57.87	$	&	$^{+	4.7	}_{-	6.8	}$	&	$\pm	3.9	$	&	$\pm	3.2	$	&	$\pm	1.9	$	&	$\pm	2.6	$	\\
$	121.5	$	&	$	57.45	$	&	$^{+	4.7	}_{-	6.8	}$	&	$\pm	3.9	$	&	$\pm	3.2	$	&	$\pm	1.9	$	&	$\pm	2.6	$	\\
$	122.0	$	&	$	57.04	$	&	$^{+	4.7	}_{-	6.8	}$	&	$\pm	3.9	$	&	$\pm	3.2	$	&	$\pm	1.9	$	&	$\pm	2.6	$	\\
$	122.5	$	&	$	56.64	$	&	$^{+	4.6	}_{-	6.8	}$	&	$\pm	3.9	$	&	$\pm	3.2	$	&	$\pm	1.9	$	&	$\pm	2.6	$	\\
$	123.0	$	&	$	56.24	$	&	$^{+	4.6	}_{-	6.8	}$	&	$\pm	3.9	$	&	$\pm	3.2	$	&	$\pm	1.9	$	&	$\pm	2.6	$	\\
\bottomrule
\end{tabular}%
\end{small}%
\end{center}%
\end{table}

\begin{table}
\caption{Energy scan of Inclusive ggF cross sections for a Higgs boson mass of $125\UGeV$, at N$^3$LO QCD,  
 together with their uncertainties. The $\mathrm{TH}$ uncertainty is interpreted as a flat $100\%$ confidence level. 
 $\Delta_{\mathrm{TH}}^{\mathrm{Gaussian}}$ uncertainty is interpreted as a one-sigma range.}
\label{tab:ggF_XScan_125}
\renewcommand{\arraystretch}{1.2}%
\begin{center}%
\begin{small}%
\tabcolsep5pt
\begin{tabular}{ccccccc}%
\toprule
$\sqrt{s}$\,[TeV] & $\sigma^{ggF}$[pb] & $\Delta_{\mathrm{TH}}$[\%] &
$\Delta_{\mathrm{TH}}^{\mathrm{Gaussian}}$[\%] & $\Delta_{\mathrm{PDF\oplus\alpha_s}}$[\%] & $\Delta_{\mathrm{PDF}}$[\%] & $\Delta_{\mathrm{\alphas}}$[\%] 
\\
\midrule
$	6.0	$&	$	12.65	$	&	$^{+	4.4	}_{-	7.1	}$	&	$\pm	4.1	$	&	$\pm	3.3	$	&	$\pm	1.9	$	&	$\pm	2.7	$	\\
$	6.5	$&	$	14.70	$	&	$^{+	4.4	}_{-	7.0	}$	&	$\pm	4.0	$	&	$\pm	3.3	$	&	$\pm	1.9	$	&	$\pm	2.7	$	\\
$	7.0	$&	$	16.85	$	&	$^{+	4.4	}_{-	7.0	}$	&	$\pm	4.0	$	&	$\pm	3.3	$	&	$\pm	1.9	$	&	$\pm	2.7	$	\\
$	7.5	$&	$	19.09	$	&	$^{+	4.4	}_{-	6.9	}$	&	$\pm	4.0	$	&	$\pm	3.3	$	&	$\pm	1.9	$	&	$\pm	2.7	$	\\
$	8.0	$&	$	21.42	$	&	$^{+	4.4	}_{-	6.9	}$	&	$\pm	4.0	$	&	$\pm	3.2	$	&	$\pm	1.9	$	&	$\pm	2.7	$	\\
$	8.5	$&	$	23.84	$	&	$^{+	4.5	}_{-	6.9	}$	&	$\pm	4.0	$	&	$\pm	3.2	$	&	$\pm	1.9	$	&	$\pm	2.6	$	\\
$	9.0	$&	$	26.33	$	&	$^{+	4.5	}_{-	6.8	}$	&	$\pm	3.9	$	&	$\pm	3.2	$	&	$\pm	1.9	$	&	$\pm	2.6	$	\\
\bottomrule
\end{tabular}%
\end{small}%
\end{center}%
\end{table}

\begin{table}
\caption{Energy scan of Inclusive ggF cross sections for a Higgs boson mass of $125.09\UGeV$, at N$^3$LO QCD,  
 together with their uncertainties. The $\mathrm{TH}$ uncertainty is interpreted as a flat $100\%$ confidence level. 
 $\Delta_{\mathrm{TH}}^{\mathrm{Gaussian}}$ uncertainty is interpreted as a one-sigma range.}
\label{tab:ggF_XScan_12509}
\renewcommand{\arraystretch}{1.2}%
\begin{center}%
\begin{small}%
\tabcolsep5pt
\begin{tabular}{ccccccc}%
\toprule
$\sqrt{s}$\,[TeV] & $\sigma^{ggF}$[pb] & $\Delta_{\mathrm{TH}}$[\%] &
$\Delta_{\mathrm{TH}}^{\mathrm{Gaussian}}$[\%] & $\Delta_{\mathrm{PDF\oplus\alpha_s}}$[\%] & $\Delta_{\mathrm{PDF}}$[\%] & $\Delta_{\mathrm{\alphas}}$[\%] 
\\
\midrule
$	6.0	$&	$	12.63	$	&	$^{+	4.4	}_{-	7.1	}$	&	$\pm	4.1	$	&	$\pm	3.3	$	&	$\pm	1.9	$	&	$\pm	2.7	$	\\
$	6.5	$&	$	14.68	$	&	$^{+	4.4	}_{-	7.0	}$	&	$\pm	4.0	$	&	$\pm	3.3	$	&	$\pm	1.9	$	&	$\pm	2.7	$	\\
$	7.0	$&	$	16.83	$	&	$^{+	4.4	}_{-	7.0	}$	&	$\pm	4.0	$	&	$\pm	3.3	$	&	$\pm	1.9	$	&	$\pm	2.7	$	\\
$	7.5	$&	$	19.07	$	&	$^{+	4.4	}_{-	6.9	}$	&	$\pm	4.0	$	&	$\pm	3.3	$	&	$\pm	1.9	$	&	$\pm	2.7	$	\\
$	8.0	$&	$	21.39	$	&	$^{+	4.4	}_{-	6.9	}$	&	$\pm	4.0	$	&	$\pm	3.2	$	&	$\pm	1.9	$	&	$\pm	2.7	$	\\
$	8.5	$&	$	23.80	$	&	$^{+	4.5	}_{-	6.9	}$	&	$\pm	4.0	$	&	$\pm	3.2	$	&	$\pm	1.9	$	&	$\pm	2.6	$	\\
$	9.0	$&	$	26.29	$	&	$^{+	4.5	}_{-	6.8	}$	&	$\pm	3.9	$	&	$\pm	3.2	$	&	$\pm	1.9	$	&	$\pm	2.6	$	\\
\bottomrule
\end{tabular}%
\end{small}%
\end{center}%
\end{table}

\clearpage


\chapter{SM vector-boson-fusion cross sections}
\label{VBFappendix}

In this appendix the cross-section  \refTs{tab:vbf_XStot} and \ref{tab:vbf_XSfiducial} for the SM VBF
 cross sections shown in \refS{subsec:VBF-XS} are expanded to a scan over SM Higgs boson masses.
\clearpage

\begin{table}
\caption{Total VBF cross sections in the SM for a LHC CM energy of $\sqrt{s}=7\UTeV$, including QCD and EW corrections
and their uncertainties for different Higgs boson masses $\MH$. For more details see
\refS{subsec:VBF-XS}.}
\label{tab:vbf_XStot_7}
\begin{center}%
\begin{small}%
\tabcolsep5pt
\renewcommand{\arraystretch}{1.2}%
\begin{tabular}{ccccccc|c}%
\toprule
$\MH$[GeV] & $\sigma^{\VBF}$[fb] & $\Delta_{\mathrm{scale}}$[\%] & $\Delta_{\mathrm{PDF}/\alphas/\mathrm{PDF\oplus\alphas}}$[\%] &
$\sigma_{\NNLO \QCD}^{\DIS}$[fb] & $\delta_{\ELWK}$[\%] & $\sigma_{\gamma}$[fb] & $\sigma_{\mbox{\scriptsize $s$-channel}}$[fb]
\\
\midrule
$120.0$ & $1301.6(2)$ & $^{+0.20}_{-0.22}$ & $\pm 2.1/\pm 0.4/\pm2.2$ & $1344.0(2)$ & $-4.5$ & $  17.6$ & $ 668.7(2)$ \\
$120.5$ & $1295.5(2)$ & $^{+0.20}_{-0.22}$ & $\pm 2.1/\pm 0.4/\pm2.2$ & $1337.6(2)$ & $-4.5$ & $  17.5$ & $ 659.7(2)$ \\
$121.0$ & $1289.3(2)$ & $^{+0.20}_{-0.22}$ & $\pm 2.1/\pm 0.4/\pm2.2$ & $1331.2(2)$ & $-4.5$ & $  17.5$ & $ 650.9(2)$ \\
$121.5$ & $1283.2(2)$ & $^{+0.20}_{-0.22}$ & $\pm 2.1/\pm 0.4/\pm2.2$ & $1324.8(2)$ & $-4.5$ & $  17.5$ & $ 642.1(2)$ \\
$122.0$ & $1277.2(2)$ & $^{+0.20}_{-0.22}$ & $\pm 2.1/\pm 0.4/\pm2.2$ & $1318.5(2)$ & $-4.5$ & $  17.4$ & $ 633.5(2)$ \\
$122.5$ & $1271.1(2)$ & $^{+0.20}_{-0.22}$ & $\pm 2.1/\pm 0.4/\pm2.2$ & $1312.2(2)$ & $-4.4$ & $  17.4$ & $ 624.9(2)$ \\
$123.0$ & $1265.1(2)$ & $^{+0.19}_{-0.21}$ & $\pm 2.1/\pm 0.4/\pm2.2$ & $1305.9(2)$ & $-4.4$ & $  17.3$ & $ 616.7(2)$ \\
$123.5$ & $1259.2(2)$ & $^{+0.19}_{-0.21}$ & $\pm 2.1/\pm 0.4/\pm2.2$ & $1299.7(2)$ & $-4.4$ & $  17.3$ & $ 608.4(2)$ \\
$124.0$ & $1253.2(2)$ & $^{+0.19}_{-0.21}$ & $\pm 2.1/\pm 0.4/\pm2.2$ & $1293.5(2)$ & $-4.4$ & $  17.2$ & $ 600.1(2)$ \\
$124.1$ & $1252.1(2)$ & $^{+0.19}_{-0.21}$ & $\pm 2.1/\pm 0.4/\pm2.2$ & $1292.3(2)$ & $-4.4$ & $  17.2$ & $ 598.6(2)$ \\
$124.2$ & $1250.9(2)$ & $^{+0.19}_{-0.21}$ & $\pm 2.1/\pm 0.4/\pm2.2$ & $1291.0(2)$ & $-4.4$ & $  17.2$ & $ 597.0(2)$ \\
$124.3$ & $1249.7(2)$ & $^{+0.19}_{-0.21}$ & $\pm 2.1/\pm 0.4/\pm2.2$ & $1289.8(2)$ & $-4.4$ & $  17.2$ & $ 595.4(2)$ \\
$124.4$ & $1248.5(1)$ & $^{+0.19}_{-0.21}$ & $\pm 2.1/\pm 0.4/\pm2.2$ & $1288.6(2)$ & $-4.4$ & $  17.2$ & $ 593.8(2)$ \\
$124.5$ & $1247.3(1)$ & $^{+0.19}_{-0.21}$ & $\pm 2.1/\pm 0.4/\pm2.2$ & $1287.3(2)$ & $-4.4$ & $  17.2$ & $ 592.2(2)$ \\
$124.6$ & $1246.2(1)$ & $^{+0.19}_{-0.21}$ & $\pm 2.1/\pm 0.4/\pm2.2$ & $1286.1(2)$ & $-4.4$ & $  17.2$ & $ 590.8(2)$ \\
$124.7$ & $1245.0(1)$ & $^{+0.19}_{-0.21}$ & $\pm 2.1/\pm 0.4/\pm2.2$ & $1284.9(2)$ & $-4.4$ & $  17.2$ & $ 589.0(2)$ \\
$124.8$ & $1243.8(1)$ & $^{+0.19}_{-0.21}$ & $\pm 2.1/\pm 0.4/\pm2.2$ & $1283.7(2)$ & $-4.4$ & $  17.1$ & $ 587.6(2)$ \\
$124.9$ & $1242.6(1)$ & $^{+0.19}_{-0.21}$ & $\pm 2.1/\pm 0.4/\pm2.2$ & $1282.5(2)$ & $-4.4$ & $  17.1$ & $ 586.0(2)$ \\
$125.0$ & $1241.5(1)$ & $^{+0.19}_{-0.21}$ & $\pm 2.1/\pm 0.4/\pm2.2$ & $1281.2(2)$ & $-4.4$ & $  17.1$ & $ 584.5(2)$ \\
$125.09$ & $1240.3(1)$ & $^{+0.19}_{-0.21}$ & $\pm 2.1/\pm 0.4/\pm2.2$ & $1280.0(2)$ & $-4.4$ & $  17.1$ & $ 582.8(2)$ \\
$125.1$ & $1240.3(1)$ & $^{+0.19}_{-0.21}$ & $\pm 2.1/\pm 0.4/\pm2.2$ & $1280.0(2)$ & $-4.4$ & $  17.1$ & $ 582.9(2)$ \\
$125.2$ & $1239.1(1)$ & $^{+0.19}_{-0.21}$ & $\pm 2.1/\pm 0.4/\pm2.2$ & $1278.8(2)$ & $-4.4$ & $  17.1$ & $ 581.2(2)$ \\
$125.3$ & $1238.0(1)$ & $^{+0.19}_{-0.21}$ & $\pm 2.1/\pm 0.4/\pm2.2$ & $1277.6(2)$ & $-4.4$ & $  17.1$ & $ 579.7(2)$ \\
$125.4$ & $1236.8(1)$ & $^{+0.19}_{-0.21}$ & $\pm 2.1/\pm 0.4/\pm2.2$ & $1276.4(2)$ & $-4.4$ & $  17.1$ & $ 578.1(2)$ \\
$125.5$ & $1235.7(1)$ & $^{+0.19}_{-0.21}$ & $\pm 2.1/\pm 0.4/\pm2.2$ & $1275.2(2)$ & $-4.4$ & $  17.1$ & $ 576.6(2)$ \\
$125.6$ & $1234.5(1)$ & $^{+0.19}_{-0.21}$ & $\pm 2.1/\pm 0.4/\pm2.2$ & $1273.9(2)$ & $-4.4$ & $  17.1$ & $ 575.2(2)$ \\
$125.7$ & $1233.3(1)$ & $^{+0.19}_{-0.21}$ & $\pm 2.1/\pm 0.4/\pm2.2$ & $1272.7(2)$ & $-4.4$ & $  17.1$ & $ 573.7(2)$ \\
$125.8$ & $1232.2(1)$ & $^{+0.19}_{-0.21}$ & $\pm 2.1/\pm 0.4/\pm2.2$ & $1271.5(2)$ & $-4.4$ & $  17.1$ & $ 572.1(2)$ \\
$125.9$ & $1231.0(1)$ & $^{+0.19}_{-0.21}$ & $\pm 2.1/\pm 0.4/\pm2.2$ & $1270.3(2)$ & $-4.4$ & $  17.0$ & $ 570.5(2)$ \\
$126.0$ & $1229.9(1)$ & $^{+0.19}_{-0.21}$ & $\pm 2.1/\pm 0.4/\pm2.2$ & $1269.1(2)$ & $-4.4$ & $  17.0$ & $ 569.1(2)$ \\
$126.5$ & $1224.1(1)$ & $^{+0.18}_{-0.21}$ & $\pm 2.1/\pm 0.4/\pm2.2$ & $1263.1(2)$ & $-4.4$ & $  17.0$ & $ 561.5(2)$ \\
$127.0$ & $1218.4(1)$ & $^{+0.18}_{-0.21}$ & $\pm 2.1/\pm 0.4/\pm2.2$ & $1257.2(2)$ & $-4.4$ & $  16.9$ & $ 554.2(2)$ \\
$127.5$ & $1212.7(1)$ & $^{+0.18}_{-0.21}$ & $\pm 2.1/\pm 0.4/\pm2.2$ & $1251.2(2)$ & $-4.4$ & $  16.9$ & $ 546.8(2)$ \\
$128.0$ & $1207.1(1)$ & $^{+0.18}_{-0.21}$ & $\pm 2.1/\pm 0.4/\pm2.2$ & $1245.3(2)$ & $-4.4$ & $  16.9$ & $ 539.9(2)$ \\
$128.5$ & $1201.4(1)$ & $^{+0.18}_{-0.20}$ & $\pm 2.1/\pm 0.4/\pm2.2$ & $1239.5(2)$ & $-4.4$ & $  16.8$ & $ 532.9(2)$ \\
$129.0$ & $1195.9(1)$ & $^{+0.18}_{-0.20}$ & $\pm 2.1/\pm 0.4/\pm2.2$ & $1233.7(2)$ & $-4.4$ & $  16.8$ & $ 526.0(2)$ \\
$129.5$ & $1190.3(1)$ & $^{+0.17}_{-0.20}$ & $\pm 2.1/\pm 0.4/\pm2.2$ & $1227.9(1)$ & $-4.4$ & $  16.7$ & $ 519.1(2)$ \\
$130.0$ & $1184.8(1)$ & $^{+0.17}_{-0.20}$ & $\pm 2.1/\pm 0.4/\pm2.2$ & $1222.1(1)$ & $-4.4$ & $  16.7$ & $ 512.2(2)$ \\
\bottomrule
\end{tabular}%
\end{small}%
\end{center}%
\end{table}

\begin{table}
\caption{Total VBF cross sections in the SM for a LHC CM energy of $\sqrt{s}=8\UTeV$, including QCD and EW corrections
and their uncertainties for different Higgs boson masses $\MH$. For more details see
\refS{subsec:VBF-XS}.}
\label{tab:vbf_XStot_8}
\begin{center}%
\begin{small}%
\tabcolsep5pt
\renewcommand{\arraystretch}{1.2}%
\begin{tabular}{ccccccc|c}%
\toprule
$\MH$[GeV] & $\sigma^{\VBF}$[fb] & $\Delta_{\mathrm{scale}}$[\%] &$\Delta_{\mathrm{PDF}/\alphas/\mathrm{PDF\oplus\alphas}}$[\%] &
$\sigma_{\NNLO \QCD}^{\DIS}$[fb] & $\delta_{\ELWK}$[\%] & $\sigma_{\gamma}$[fb] & $\sigma_{\mbox{\scriptsize $s$-channel}}$[fb]
\\
\midrule
$120.0$ & $1675.7(2)$ & $^{+0.26}_{-0.25}$ & $\pm 2.1/\pm 0.4/\pm2.2$ & $1733.7(2)$ & $-4.7$ & $  22.7$ & $ 811.7(3)$ \\
$120.5$ & $1668.1(2)$ & $^{+0.26}_{-0.25}$ & $\pm 2.1/\pm 0.4/\pm2.2$ & $1725.8(2)$ & $-4.7$ & $  22.6$ & $ 800.5(3)$ \\
$121.0$ & $1660.5(2)$ & $^{+0.26}_{-0.25}$ & $\pm 2.1/\pm 0.4/\pm2.2$ & $1717.8(2)$ & $-4.7$ & $  22.6$ & $ 790.0(3)$ \\
$121.5$ & $1652.9(2)$ & $^{+0.26}_{-0.25}$ & $\pm 2.1/\pm 0.4/\pm2.2$ & $1709.9(2)$ & $-4.6$ & $  22.5$ & $ 779.3(3)$ \\
$122.0$ & $1645.4(2)$ & $^{+0.26}_{-0.25}$ & $\pm 2.1/\pm 0.4/\pm2.2$ & $1702.1(2)$ & $-4.6$ & $  22.5$ & $ 768.7(3)$ \\
$122.5$ & $1637.9(2)$ & $^{+0.26}_{-0.24}$ & $\pm 2.1/\pm 0.4/\pm2.2$ & $1694.3(2)$ & $-4.6$ & $  22.4$ & $ 759.0(3)$ \\
$123.0$ & $1630.5(2)$ & $^{+0.25}_{-0.24}$ & $\pm 2.1/\pm 0.4/\pm2.2$ & $1686.5(2)$ & $-4.6$ & $  22.3$ & $ 748.9(2)$ \\
$123.5$ & $1623.2(2)$ & $^{+0.25}_{-0.24}$ & $\pm 2.1/\pm 0.4/\pm2.2$ & $1678.8(2)$ & $-4.6$ & $  22.3$ & $ 739.2(2)$ \\
$124.0$ & $1615.8(2)$ & $^{+0.25}_{-0.24}$ & $\pm 2.1/\pm 0.4/\pm2.2$ & $1671.1(2)$ & $-4.6$ & $  22.2$ & $ 729.3(3)$ \\
$124.1$ & $1614.4(2)$ & $^{+0.25}_{-0.24}$ & $\pm 2.1/\pm 0.4/\pm2.2$ & $1669.6(2)$ & $-4.6$ & $  22.2$ & $ 727.3(3)$ \\
$124.2$ & $1612.9(2)$ & $^{+0.25}_{-0.24}$ & $\pm 2.1/\pm 0.4/\pm2.2$ & $1668.1(2)$ & $-4.6$ & $  22.2$ & $ 725.5(3)$ \\
$124.3$ & $1611.4(2)$ & $^{+0.25}_{-0.24}$ & $\pm 2.1/\pm 0.4/\pm2.2$ & $1666.6(2)$ & $-4.6$ & $  22.2$ & $ 723.5(2)$ \\
$124.4$ & $1610.0(2)$ & $^{+0.25}_{-0.24}$ & $\pm 2.1/\pm 0.4/\pm2.2$ & $1665.0(2)$ & $-4.6$ & $  22.2$ & $ 721.8(3)$ \\
$124.5$ & $1608.5(2)$ & $^{+0.25}_{-0.24}$ & $\pm 2.1/\pm 0.4/\pm2.2$ & $1663.5(2)$ & $-4.6$ & $  22.2$ & $ 719.9(3)$ \\
$124.6$ & $1607.1(2)$ & $^{+0.25}_{-0.24}$ & $\pm 2.1/\pm 0.4/\pm2.2$ & $1662.0(2)$ & $-4.6$ & $  22.2$ & $ 717.9(3)$ \\
$124.7$ & $1605.6(2)$ & $^{+0.25}_{-0.24}$ & $\pm 2.1/\pm 0.4/\pm2.2$ & $1660.5(2)$ & $-4.6$ & $  22.1$ & $ 716.1(2)$ \\
$124.8$ & $1604.2(2)$ & $^{+0.25}_{-0.24}$ & $\pm 2.1/\pm 0.4/\pm2.2$ & $1659.0(2)$ & $-4.6$ & $  22.1$ & $ 714.2(3)$ \\
$124.9$ & $1602.8(2)$ & $^{+0.25}_{-0.24}$ & $\pm 2.1/\pm 0.4/\pm2.2$ & $1657.5(2)$ & $-4.6$ & $  22.1$ & $ 712.4(3)$ \\
$125.0$ & $1601.3(2)$ & $^{+0.25}_{-0.24}$ & $\pm 2.1/\pm 0.4/\pm2.2$ & $1656.0(2)$ & $-4.6$ & $  22.1$ & $ 710.4(3)$ \\
$125.09$ & $1599.8(2)$ & $^{+0.25}_{-0.24}$ & $\pm 2.1/\pm 0.4/\pm2.2$ & $1654.4(2)$ & $-4.6$ & $  22.1$ & $ 708.7(3)$ \\
$125.1$ & $1599.8(2)$ & $^{+0.25}_{-0.24}$ & $\pm 2.1/\pm 0.4/\pm2.2$ & $1654.4(2)$ & $-4.6$ & $  22.1$ & $ 708.7(3)$ \\
$125.2$ & $1598.4(2)$ & $^{+0.25}_{-0.24}$ & $\pm 2.1/\pm 0.4/\pm2.2$ & $1652.9(2)$ & $-4.6$ & $  22.1$ & $ 706.5(2)$ \\
$125.3$ & $1597.0(2)$ & $^{+0.25}_{-0.24}$ & $\pm 2.1/\pm 0.4/\pm2.2$ & $1651.4(2)$ & $-4.6$ & $  22.1$ & $ 704.8(3)$ \\
$125.4$ & $1595.5(2)$ & $^{+0.25}_{-0.24}$ & $\pm 2.1/\pm 0.4/\pm2.2$ & $1649.9(2)$ & $-4.6$ & $  22.1$ & $ 703.0(3)$ \\
$125.5$ & $1594.1(2)$ & $^{+0.25}_{-0.24}$ & $\pm 2.1/\pm 0.4/\pm2.2$ & $1648.4(2)$ & $-4.6$ & $  22.1$ & $ 701.2(3)$ \\
$125.6$ & $1592.7(2)$ & $^{+0.25}_{-0.24}$ & $\pm 2.1/\pm 0.4/\pm2.2$ & $1646.9(2)$ & $-4.6$ & $  22.0$ & $ 699.3(3)$ \\
$125.7$ & $1591.2(2)$ & $^{+0.25}_{-0.24}$ & $\pm 2.1/\pm 0.4/\pm2.2$ & $1645.4(2)$ & $-4.6$ & $  22.0$ & $ 697.5(3)$ \\
$125.8$ & $1589.8(2)$ & $^{+0.25}_{-0.24}$ & $\pm 2.1/\pm 0.4/\pm2.2$ & $1643.9(2)$ & $-4.6$ & $  22.0$ & $ 695.6(3)$ \\
$125.9$ & $1588.4(2)$ & $^{+0.25}_{-0.24}$ & $\pm 2.1/\pm 0.4/\pm2.2$ & $1642.4(2)$ & $-4.6$ & $  22.0$ & $ 693.7(3)$ \\
$126.0$ & $1587.0(2)$ & $^{+0.24}_{-0.24}$ & $\pm 2.1/\pm 0.4/\pm2.2$ & $1640.9(2)$ & $-4.6$ & $  22.0$ & $ 692.0(3)$ \\
$126.5$ & $1579.8(2)$ & $^{+0.24}_{-0.24}$ & $\pm 2.1/\pm 0.4/\pm2.2$ & $1633.5(2)$ & $-4.6$ & $  21.9$ & $ 683.1(3)$ \\
$127.0$ & $1572.8(2)$ & $^{+0.24}_{-0.24}$ & $\pm 2.1/\pm 0.4/\pm2.2$ & $1626.1(2)$ & $-4.6$ & $  21.9$ & $ 674.2(3)$ \\
$127.5$ & $1565.7(2)$ & $^{+0.24}_{-0.24}$ & $\pm 2.1/\pm 0.4/\pm2.2$ & $1618.7(2)$ & $-4.6$ & $  21.8$ & $ 665.4(2)$ \\
$128.0$ & $1558.7(2)$ & $^{+0.24}_{-0.24}$ & $\pm 2.1/\pm 0.4/\pm2.2$ & $1611.4(2)$ & $-4.6$ & $  21.8$ & $ 656.9(2)$ \\
$128.5$ & $1551.7(2)$ & $^{+0.24}_{-0.24}$ & $\pm 2.1/\pm 0.4/\pm2.2$ & $1604.2(2)$ & $-4.6$ & $  21.7$ & $ 648.5(2)$ \\
$129.0$ & $1544.8(2)$ & $^{+0.24}_{-0.23}$ & $\pm 2.1/\pm 0.4/\pm2.2$ & $1596.9(2)$ & $-4.6$ & $  21.7$ & $ 640.2(2)$ \\
$129.5$ & $1537.9(2)$ & $^{+0.24}_{-0.23}$ & $\pm 2.1/\pm 0.4/\pm2.2$ & $1589.8(2)$ & $-4.6$ & $  21.6$ & $ 631.9(2)$ \\
$130.0$ & $1531.1(2)$ & $^{+0.23}_{-0.23}$ & $\pm 2.1/\pm 0.4/\pm2.2$ & $1582.6(2)$ & $-4.6$ & $  21.5$ & $ 623.7(2)$ \\
\bottomrule
\end{tabular}%
\end{small}%
\end{center}%
\end{table}

\begin{table}
\caption{Total VBF cross sections in the SM for a LHC CM energy of $\sqrt{s}=13\UTeV$, including QCD and EW corrections
and their uncertainties for different Higgs boson masses $\MH$. For more details see
\refS{subsec:VBF-XS}.}
\label{tab:vbf_XStot_13}
\begin{center}%
\begin{small}%
\tabcolsep5pt
\renewcommand{\arraystretch}{1.2}%
\begin{tabular}{ccccccc|c}%
\toprule
$\MH$[GeV] & $\sigma^{\VBF}$[fb] & $\Delta_{\mathrm{scale}}$[\%] & $\Delta_{\mathrm{PDF}/\alphas/\mathrm{PDF\oplus\alphas}}$[\%] &
$\sigma_{\NNLO \QCD}^{\DIS}$[fb] & $\delta_{\ELWK}$[\%] & $\sigma_{\gamma}$[fb] & $\sigma_{\mbox{\scriptsize $s$-channel}}$[fb]
\\
\midrule
$120.0$ & $3935.2(7)$ & $^{+0.44}_{-0.33}$ & $\pm 2.1/\pm 0.5/\pm2.1$ & $4100.8(7)$ & $-5.3$ & $  53.0$ & $1567.0(6)$ \\
$120.5$ & $3919.4(7)$ & $^{+0.44}_{-0.33}$ & $\pm 2.1/\pm 0.5/\pm2.1$ & $4084.2(7)$ & $-5.3$ & $  52.9$ & $1546.0(7)$ \\
$121.0$ & $3903.9(7)$ & $^{+0.44}_{-0.33}$ & $\pm 2.1/\pm 0.5/\pm2.1$ & $4067.8(7)$ & $-5.3$ & $  52.8$ & $1525.7(6)$ \\
$121.5$ & $3888.3(7)$ & $^{+0.44}_{-0.33}$ & $\pm 2.1/\pm 0.5/\pm2.1$ & $4051.5(7)$ & $-5.3$ & $  52.7$ & $1506.7(6)$ \\
$122.0$ & $3873.0(7)$ & $^{+0.44}_{-0.33}$ & $\pm 2.1/\pm 0.5/\pm2.1$ & $4035.2(7)$ & $-5.3$ & $  52.5$ & $1487.6(6)$ \\
$122.5$ & $3857.6(7)$ & $^{+0.43}_{-0.33}$ & $\pm 2.1/\pm 0.5/\pm2.1$ & $4019.0(7)$ & $-5.3$ & $  52.4$ & $1468.2(6)$ \\
$123.0$ & $3842.3(7)$ & $^{+0.43}_{-0.33}$ & $\pm 2.1/\pm 0.5/\pm2.1$ & $4003.0(7)$ & $-5.3$ & $  52.3$ & $1449.3(5)$ \\
$123.5$ & $3827.0(7)$ & $^{+0.43}_{-0.33}$ & $\pm 2.1/\pm 0.5/\pm2.1$ & $3987.0(7)$ & $-5.3$ & $  52.2$ & $1430.8(5)$ \\
$124.0$ & $3811.9(7)$ & $^{+0.43}_{-0.33}$ & $\pm 2.1/\pm 0.5/\pm2.1$ & $3971.1(7)$ & $-5.3$ & $  52.1$ & $1412.9(5)$ \\
$124.1$ & $3808.9(7)$ & $^{+0.43}_{-0.33}$ & $\pm 2.1/\pm 0.5/\pm2.1$ & $3967.9(7)$ & $-5.3$ & $  52.1$ & $1409.4(5)$ \\
$124.2$ & $3805.9(7)$ & $^{+0.43}_{-0.33}$ & $\pm 2.1/\pm 0.5/\pm2.1$ & $3964.7(7)$ & $-5.3$ & $  52.0$ & $1405.8(5)$ \\
$124.3$ & $3802.9(7)$ & $^{+0.43}_{-0.33}$ & $\pm 2.1/\pm 0.5/\pm2.1$ & $3961.6(7)$ & $-5.3$ & $  52.0$ & $1401.9(5)$ \\
$124.4$ & $3799.9(7)$ & $^{+0.43}_{-0.33}$ & $\pm 2.1/\pm 0.5/\pm2.1$ & $3958.4(7)$ & $-5.3$ & $  52.0$ & $1398.4(5)$ \\
$124.5$ & $3796.9(7)$ & $^{+0.43}_{-0.33}$ & $\pm 2.1/\pm 0.5/\pm2.1$ & $3955.2(7)$ & $-5.3$ & $  52.0$ & $1395.1(5)$ \\
$124.6$ & $3793.9(7)$ & $^{+0.43}_{-0.33}$ & $\pm 2.1/\pm 0.5/\pm2.1$ & $3952.1(7)$ & $-5.3$ & $  51.9$ & $1391.5(5)$ \\
$124.7$ & $3790.9(7)$ & $^{+0.43}_{-0.33}$ & $\pm 2.1/\pm 0.5/\pm2.1$ & $3948.9(7)$ & $-5.3$ & $  51.9$ & $1388.0(5)$ \\
$124.8$ & $3788.0(6)$ & $^{+0.43}_{-0.33}$ & $\pm 2.1/\pm 0.5/\pm2.1$ & $3945.8(7)$ & $-5.3$ & $  51.9$ & $1384.9(5)$ \\
$124.9$ & $3785.0(6)$ & $^{+0.43}_{-0.33}$ & $\pm 2.1/\pm 0.5/\pm2.1$ & $3942.7(7)$ & $-5.3$ & $  51.9$ & $1381.5(5)$ \\
$125.0$ & $3782.0(6)$ & $^{+0.43}_{-0.33}$ & $\pm 2.1/\pm 0.5/\pm2.1$ & $3939.5(7)$ & $-5.3$ & $  51.9$ & $1378.1(5)$ \\
$125.09$ & $3779.0(6)$ & $^{+0.43}_{-0.33}$ & $\pm 2.1/\pm 0.5/\pm2.1$ & $3936.4(7)$ & $-5.3$ & $  51.8$ & $1374.5(5)$ \\
$125.1$ & $3779.1(6)$ & $^{+0.43}_{-0.33}$ & $\pm 2.1/\pm 0.5/\pm2.1$ & $3936.4(7)$ & $-5.3$ & $  51.8$ & $1373.9(5)$ \\
$125.2$ & $3775.9(6)$ & $^{+0.43}_{-0.33}$ & $\pm 2.1/\pm 0.5/\pm2.1$ & $3933.2(7)$ & $-5.3$ & $  51.8$ & $1370.5(5)$ \\
$125.3$ & $3773.0(6)$ & $^{+0.43}_{-0.33}$ & $\pm 2.1/\pm 0.5/\pm2.1$ & $3930.1(7)$ & $-5.3$ & $  51.8$ & $1367.2(5)$ \\
$125.4$ & $3769.9(6)$ & $^{+0.43}_{-0.33}$ & $\pm 2.1/\pm 0.5/\pm2.1$ & $3927.0(7)$ & $-5.3$ & $  51.8$ & $1364.0(5)$ \\
$125.5$ & $3767.0(6)$ & $^{+0.43}_{-0.33}$ & $\pm 2.1/\pm 0.5/\pm2.1$ & $3923.9(7)$ & $-5.3$ & $  51.7$ & $1360.0(5)$ \\
$125.6$ & $3764.2(6)$ & $^{+0.43}_{-0.33}$ & $\pm 2.1/\pm 0.5/\pm2.1$ & $3920.7(7)$ & $-5.3$ & $  51.7$ & $1356.7(5)$ \\
$125.7$ & $3761.1(6)$ & $^{+0.43}_{-0.33}$ & $\pm 2.1/\pm 0.5/\pm2.1$ & $3917.6(7)$ & $-5.3$ & $  51.7$ & $1353.3(5)$ \\
$125.8$ & $3758.1(6)$ & $^{+0.43}_{-0.33}$ & $\pm 2.1/\pm 0.5/\pm2.1$ & $3914.5(7)$ & $-5.3$ & $  51.7$ & $1349.8(5)$ \\
$125.9$ & $3755.1(6)$ & $^{+0.43}_{-0.32}$ & $\pm 2.1/\pm 0.5/\pm2.1$ & $3911.4(7)$ & $-5.3$ & $  51.6$ & $1346.5(5)$ \\
$126.0$ & $3752.2(6)$ & $^{+0.43}_{-0.32}$ & $\pm 2.1/\pm 0.5/\pm2.1$ & $3908.3(7)$ & $-5.3$ & $  51.6$ & $1343.2(5)$ \\
$126.5$ & $3737.5(6)$ & $^{+0.42}_{-0.32}$ & $\pm 2.1/\pm 0.5/\pm2.1$ & $3892.8(7)$ & $-5.3$ & $  51.5$ & $1326.4(5)$ \\
$127.0$ & $3723.0(6)$ & $^{+0.42}_{-0.32}$ & $\pm 2.1/\pm 0.5/\pm2.1$ & $3877.4(7)$ & $-5.3$ & $  51.4$ & $1310.2(5)$ \\
$127.5$ & $3708.4(6)$ & $^{+0.42}_{-0.32}$ & $\pm 2.1/\pm 0.5/\pm2.1$ & $3862.1(7)$ & $-5.3$ & $  51.3$ & $1293.4(5)$ \\
$128.0$ & $3693.9(6)$ & $^{+0.42}_{-0.32}$ & $\pm 2.1/\pm 0.5/\pm2.1$ & $3846.8(7)$ & $-5.3$ & $  51.2$ & $1277.6(5)$ \\
$128.5$ & $3679.5(6)$ & $^{+0.42}_{-0.32}$ & $\pm 2.1/\pm 0.5/\pm2.1$ & $3831.7(7)$ & $-5.3$ & $  51.0$ & $1261.8(5)$ \\
$129.0$ & $3665.1(6)$ & $^{+0.42}_{-0.32}$ & $\pm 2.1/\pm 0.5/\pm2.1$ & $3816.6(7)$ & $-5.3$ & $  50.9$ & $1246.4(5)$ \\
$129.5$ & $3650.8(6)$ & $^{+0.42}_{-0.32}$ & $\pm 2.1/\pm 0.5/\pm2.1$ & $3801.6(7)$ & $-5.3$ & $  50.8$ & $1231.0(5)$ \\
$130.0$ & $3636.7(6)$ & $^{+0.42}_{-0.32}$ & $\pm 2.1/\pm 0.5/\pm2.1$ & $3786.7(7)$ & $-5.3$ & $  50.7$ & $1216.1(5)$ \\
\bottomrule
\end{tabular}%
\end{small}%
\end{center}%
\end{table}

\begin{table}
\caption{Total VBF cross sections in the SM for a LHC CM energy of $\sqrt{s}=14\UTeV$, including QCD and EW corrections
and their uncertainties for different Higgs boson masses $\MH$. For more details see
\refS{subsec:VBF-XS}.}
\label{tab:vbf_XStot_14}
\begin{center}%
\begin{small}%
\tabcolsep5pt
\renewcommand{\arraystretch}{1.2}%
\begin{tabular}{ccccccc|c}%
\toprule
$\MH$[GeV] & $\sigma^{\VBF}$[fb] & $\Delta_{\mathrm{scale}}$[\%] & $\Delta_{\mathrm{PDF}/\alphas/\mathrm{PDF\oplus\alphas}}$[\%] &
$\sigma_{\NNLO \QCD}^{\DIS}$[fb] & $\delta_{\ELWK}$[\%] & $\sigma_{\gamma}$[fb] & $\sigma_{\mbox{\scriptsize $s$-channel}}$[fb]
\\
\midrule
$120.0$ & $4448.4(8)$ & $^{+0.46}_{-0.34}$ & $\pm 2.1/\pm 0.5/\pm2.1$ & $4640.4(8)$ & $-5.4$ & $  59.8$ & $1722.3(6)$ \\
$120.5$ & $4430.8(8)$ & $^{+0.46}_{-0.34}$ & $\pm 2.1/\pm 0.5/\pm2.1$ & $4622.1(8)$ & $-5.4$ & $  59.7$ & $1700.5(7)$ \\
$121.0$ & $4413.6(8)$ & $^{+0.46}_{-0.34}$ & $\pm 2.1/\pm 0.5/\pm2.1$ & $4603.8(8)$ & $-5.4$ & $  59.5$ & $1678.6(6)$ \\
$121.5$ & $4396.3(8)$ & $^{+0.46}_{-0.34}$ & $\pm 2.1/\pm 0.5/\pm2.1$ & $4585.6(8)$ & $-5.4$ & $  59.4$ & $1657.0(6)$ \\
$122.0$ & $4379.0(8)$ & $^{+0.46}_{-0.34}$ & $\pm 2.1/\pm 0.5/\pm2.1$ & $4567.6(8)$ & $-5.4$ & $  59.3$ & $1636.0(6)$ \\
$122.5$ & $4362.0(8)$ & $^{+0.46}_{-0.34}$ & $\pm 2.1/\pm 0.5/\pm2.1$ & $4549.6(8)$ & $-5.4$ & $  59.2$ & $1615.1(6)$ \\
$123.0$ & $4345.0(8)$ & $^{+0.46}_{-0.34}$ & $\pm 2.1/\pm 0.5/\pm2.1$ & $4531.8(8)$ & $-5.4$ & $  59.0$ & $1594.4(6)$ \\
$123.5$ & $4328.3(8)$ & $^{+0.46}_{-0.34}$ & $\pm 2.1/\pm 0.5/\pm2.1$ & $4514.0(8)$ & $-5.4$ & $  58.9$ & $1574.4(6)$ \\
$124.0$ & $4311.4(8)$ & $^{+0.45}_{-0.34}$ & $\pm 2.1/\pm 0.5/\pm2.1$ & $4496.3(8)$ & $-5.4$ & $  58.8$ & $1554.3(6)$ \\
$124.1$ & $4308.0(8)$ & $^{+0.45}_{-0.34}$ & $\pm 2.1/\pm 0.5/\pm2.1$ & $4492.8(8)$ & $-5.4$ & $  58.7$ & $1550.3(6)$ \\
$124.2$ & $4304.8(8)$ & $^{+0.45}_{-0.34}$ & $\pm 2.1/\pm 0.5/\pm2.1$ & $4489.3(8)$ & $-5.4$ & $  58.7$ & $1546.8(6)$ \\
$124.3$ & $4301.4(8)$ & $^{+0.45}_{-0.34}$ & $\pm 2.1/\pm 0.5/\pm2.1$ & $4485.8(8)$ & $-5.4$ & $  58.7$ & $1542.9(6)$ \\
$124.4$ & $4298.2(8)$ & $^{+0.45}_{-0.34}$ & $\pm 2.1/\pm 0.5/\pm2.1$ & $4482.2(8)$ & $-5.4$ & $  58.7$ & $1538.9(6)$ \\
$124.5$ & $4294.8(8)$ & $^{+0.45}_{-0.34}$ & $\pm 2.1/\pm 0.5/\pm2.1$ & $4478.7(8)$ & $-5.4$ & $  58.6$ & $1535.2(6)$ \\
$124.6$ & $4291.4(8)$ & $^{+0.45}_{-0.34}$ & $\pm 2.1/\pm 0.5/\pm2.1$ & $4475.2(8)$ & $-5.4$ & $  58.6$ & $1531.4(6)$ \\
$124.7$ & $4288.2(8)$ & $^{+0.45}_{-0.34}$ & $\pm 2.1/\pm 0.5/\pm2.1$ & $4471.7(8)$ & $-5.4$ & $  58.6$ & $1527.4(6)$ \\
$124.8$ & $4284.9(8)$ & $^{+0.45}_{-0.34}$ & $\pm 2.1/\pm 0.5/\pm2.1$ & $4468.2(8)$ & $-5.4$ & $  58.6$ & $1524.0(6)$ \\
$124.9$ & $4281.4(8)$ & $^{+0.45}_{-0.34}$ & $\pm 2.1/\pm 0.5/\pm2.1$ & $4464.7(8)$ & $-5.4$ & $  58.5$ & $1519.7(6)$ \\
$125.0$ & $4278.0(8)$ & $^{+0.45}_{-0.34}$ & $\pm 2.1/\pm 0.5/\pm2.1$ & $4461.2(8)$ & $-5.4$ & $  58.5$ & $1515.9(6)$ \\
$125.09$ & $4274.8(8)$ & $^{+0.45}_{-0.34}$ & $\pm 2.1/\pm 0.5/\pm2.1$ & $4457.8(8)$ & $-5.4$ & $  58.5$ & $1512.5(6)$ \\
$125.1$ & $4274.9(8)$ & $^{+0.45}_{-0.34}$ & $\pm 2.1/\pm 0.5/\pm2.1$ & $4457.8(8)$ & $-5.4$ & $  58.5$ & $1512.2(6)$ \\
$125.2$ & $4271.5(8)$ & $^{+0.45}_{-0.34}$ & $\pm 2.1/\pm 0.5/\pm2.1$ & $4454.3(8)$ & $-5.4$ & $  58.5$ & $1508.0(6)$ \\
$125.3$ & $4268.2(8)$ & $^{+0.45}_{-0.34}$ & $\pm 2.1/\pm 0.5/\pm2.1$ & $4450.8(8)$ & $-5.4$ & $  58.4$ & $1504.3(6)$ \\
$125.4$ & $4264.9(8)$ & $^{+0.45}_{-0.34}$ & $\pm 2.1/\pm 0.5/\pm2.1$ & $4447.3(8)$ & $-5.4$ & $  58.4$ & $1500.8(6)$ \\
$125.5$ & $4261.6(7)$ & $^{+0.45}_{-0.34}$ & $\pm 2.1/\pm 0.5/\pm2.1$ & $4443.8(8)$ & $-5.4$ & $  58.4$ & $1497.0(6)$ \\
$125.6$ & $4258.2(7)$ & $^{+0.45}_{-0.34}$ & $\pm 2.1/\pm 0.5/\pm2.1$ & $4440.4(8)$ & $-5.4$ & $  58.4$ & $1493.7(6)$ \\
$125.7$ & $4255.0(7)$ & $^{+0.45}_{-0.34}$ & $\pm 2.1/\pm 0.5/\pm2.1$ & $4436.9(8)$ & $-5.4$ & $  58.3$ & $1489.5(6)$ \\
$125.8$ & $4251.6(7)$ & $^{+0.45}_{-0.34}$ & $\pm 2.1/\pm 0.5/\pm2.1$ & $4433.4(8)$ & $-5.4$ & $  58.3$ & $1485.8(6)$ \\
$125.9$ & $4248.5(7)$ & $^{+0.45}_{-0.34}$ & $\pm 2.1/\pm 0.5/\pm2.1$ & $4430.0(8)$ & $-5.4$ & $  58.3$ & $1481.7(5)$ \\
$126.0$ & $4245.1(7)$ & $^{+0.45}_{-0.34}$ & $\pm 2.1/\pm 0.5/\pm2.1$ & $4426.5(8)$ & $-5.4$ & $  58.3$ & $1478.5(6)$ \\
$126.5$ & $4228.8(7)$ & $^{+0.45}_{-0.33}$ & $\pm 2.1/\pm 0.5/\pm2.1$ & $4409.3(8)$ & $-5.4$ & $  58.1$ & $1459.8(6)$ \\
$127.0$ & $4212.6(7)$ & $^{+0.45}_{-0.33}$ & $\pm 2.1/\pm 0.5/\pm2.1$ & $4392.2(8)$ & $-5.4$ & $  58.0$ & $1441.8(6)$ \\
$127.5$ & $4196.4(7)$ & $^{+0.45}_{-0.33}$ & $\pm 2.1/\pm 0.5/\pm2.1$ & $4375.2(8)$ & $-5.4$ & $  57.9$ & $1424.0(6)$ \\
$128.0$ & $4180.4(7)$ & $^{+0.44}_{-0.33}$ & $\pm 2.1/\pm 0.5/\pm2.1$ & $4358.2(8)$ & $-5.4$ & $  57.8$ & $1406.6(5)$ \\
$128.5$ & $4164.4(7)$ & $^{+0.44}_{-0.33}$ & $\pm 2.1/\pm 0.5/\pm2.1$ & $4341.3(8)$ & $-5.4$ & $  57.6$ & $1389.1(6)$ \\
$129.0$ & $4148.4(7)$ & $^{+0.44}_{-0.33}$ & $\pm 2.1/\pm 0.5/\pm2.1$ & $4324.6(8)$ & $-5.4$ & $  57.5$ & $1371.7(6)$ \\
$129.5$ & $4132.5(7)$ & $^{+0.44}_{-0.33}$ & $\pm 2.1/\pm 0.5/\pm2.1$ & $4307.9(8)$ & $-5.4$ & $  57.4$ & $1355.1(5)$ \\
$130.0$ & $4116.8(7)$ & $^{+0.44}_{-0.33}$ & $\pm 2.1/\pm 0.5/\pm2.1$ & $4291.3(8)$ & $-5.4$ & $  57.3$ & $1338.8(6)$ \\
\bottomrule
\end{tabular}%
\end{small}%
\end{center}%
\end{table}

\begin{table}
\caption{Fiducial VBF cross sections in the SM for a LHC CM energy of $\sqrt{s}=7\UTeV$, including QCD and EW corrections
and their uncertainties for different Higgs boson masses $\MH$. For more details see
\refS{subsec:VBF-XS}. The numbers in the $\sigma_{\NNLO \QCD}^{\DIS}$ column have been obtained from a linear interpolation. The interpolation was performed by fitting the cross section for $\MH = \{120.0,\, 122.5,\,  125.0, \, 127.5, \, 130.0\}$ GeV to a linear function. The scale uncertainty in the column $\Delta_{\mathrm{scale}}$ was only computed for $\MH=125.0$ GeV.}   
\label{tab:vbf_XSfiducial_7}
\begin{center}%
\begin{small}%
\tabcolsep5pt
\renewcommand{\arraystretch}{1.2}%
\begin{tabular}{ccccccc|c}%
\toprule
$\MH$[GeV] & $\sigma^{\VBF}$[fb] & $\Delta_{\mathrm{scale}}$[\%] & $\Delta_{\mathrm{PDF}/\alphas/\mathrm{PDF\oplus\alphas}}$[\%] &
$\sigma_{\NNLO \QCD}^{\DIS}$[fb] & $\delta_{\ELWK}$[\%] & $\sigma_{\gamma}$[fb] & $\sigma_{\mbox{\scriptsize $s$-chan}}$[fb]
\\
\midrule
$120.0$ & $ 625.8( 9)$ & $^{+1.3}_{-1.6}$ &  $\pm 2.3/\pm 0.3/\pm2.3$ & $ 655.7(10)$ & $-6.1$ & $  10.1$ & $   9.3 $ \\
$120.5$ & $ 623.5(8)$ & $^{+1.3}_{-1.6}$ &  $\pm 2.3/\pm 0.3/\pm2.3$ & $ 653.2( 9)$ & $-6.1$ & $  10.1$ & $   9.2 $ \\
$121.0$ & $ 621.2(8)$ & $^{+1.3}_{-1.6}$ &  $\pm 2.3/\pm 0.3/\pm2.3$ & $ 650.7(8)$ & $-6.1$ & $  10.1$ & $   9.1 $ \\
$121.5$ & $ 618.8(7)$ & $^{+1.3}_{-1.6}$ &  $\pm 2.3/\pm 0.3/\pm2.3$ & $ 648.2(8)$ & $-6.1$ & $  10.1$ & $   9.0 $ \\
$122.0$ & $ 616.4(7)$ & $^{+1.3}_{-1.6}$ &  $\pm 2.3/\pm 0.3/\pm2.3$ & $ 645.7(7)$ & $-6.1$ & $  10.0$ & $   8.8 $ \\
$122.5$ & $ 614.1(6)$ & $^{+1.3}_{-1.6}$ &  $\pm 2.3/\pm 0.3/\pm2.3$ & $ 643.2(6)$ & $-6.1$ & $  10.0$ & $   8.7 $ \\
$123.0$ & $ 611.7(6)$ & $^{+1.3}_{-1.6}$ &  $\pm 2.3/\pm 0.3/\pm2.3$ & $ 640.7(6)$ & $-6.1$ & $  10.0$ & $   8.6 $ \\
$123.5$ & $ 609.4(5)$ & $^{+1.3}_{-1.6}$ &  $\pm 2.3/\pm 0.3/\pm2.3$ & $ 638.3(6)$ & $-6.1$ & $  10.0$ & $   8.5 $ \\
$124.0$ & $ 607.1(5)$ & $^{+1.3}_{-1.6}$ &  $\pm 2.3/\pm 0.3/\pm2.3$ & $ 635.8(5)$ & $-6.1$ & $   9.9$ & $   8.4 $ \\
$124.1$ & $ 606.6(5)$ & $^{+1.3}_{-1.6}$ &  $\pm 2.3/\pm 0.3/\pm2.3$ & $ 635.3(5)$ & $-6.1$ & $   9.9$ & $   8.4 $ \\
$124.2$ & $ 606.2(5)$ & $^{+1.3}_{-1.6}$ &  $\pm 2.3/\pm 0.3/\pm2.3$ & $ 634.8(5)$ & $-6.1$ & $   9.9$ & $   8.4 $ \\
$124.3$ & $ 605.7(5)$ & $^{+1.3}_{-1.6}$ &  $\pm 2.3/\pm 0.3/\pm2.3$ & $ 634.3(5)$ & $-6.1$ & $   9.9$ & $   8.4 $ \\
$124.4$ & $ 605.2(5)$ & $^{+1.3}_{-1.6}$ &  $\pm 2.3/\pm 0.3/\pm2.3$ & $ 633.8(5)$ & $-6.1$ & $   9.9$ & $   8.3 $ \\
$124.5$ & $ 604.8(5)$ & $^{+1.3}_{-1.6}$ &  $\pm 2.3/\pm 0.3/\pm2.3$ & $ 633.3(5)$ & $-6.1$ & $   9.9$ & $   8.3 $ \\
$124.6$ & $ 604.3(5)$ & $^{+1.3}_{-1.6}$ &  $\pm 2.3/\pm 0.3/\pm2.3$ & $ 632.8(5)$ & $-6.1$ & $   9.9$ & $   8.3 $ \\
$124.7$ & $ 603.8(5)$ & $^{+1.3}_{-1.6}$ &  $\pm 2.3/\pm 0.3/\pm2.3$ & $ 632.3(5)$ & $-6.1$ & $   9.9$ & $   8.3 $ \\
$124.8$ & $ 603.4(5)$ & $^{+1.3}_{-1.6}$ &  $\pm 2.3/\pm 0.3/\pm2.3$ & $ 631.8(5)$ & $-6.1$ & $   9.9$ & $   8.2 $ \\
$124.9$ & $ 602.9(5)$ & $^{+1.3}_{-1.6}$ &  $\pm 2.3/\pm 0.3/\pm2.3$ & $ 631.3(5)$ & $-6.1$ & $   9.9$ & $   8.2 $ \\
$125.0$ & $ 602.4(5)$ & $^{+1.3}_{-1.6}$ &  $\pm 2.3/\pm 0.3/\pm2.3$ & $ 630.8(5)$ & $-6.1$ & $   9.9$ & $   8.2 $ \\
$125.09$ & $ 602.0(5)$ & $^{+1.3}_{-1.6}$ &  $\pm 2.3/\pm 0.3/\pm2.3$ & $ 630.3(5)$ & $-6.1$ & $   9.9$ & $   8.2 $ \\
$125.1$ & $ 602.0(5)$ & $^{+1.3}_{-1.6}$ &  $\pm 2.3/\pm 0.3/\pm2.3$ & $ 630.3(5)$ & $-6.1$ & $   9.9$ & $   8.2 $ \\
$125.2$ & $ 601.5(5)$ & $^{+1.3}_{-1.6}$ &  $\pm 2.3/\pm 0.3/\pm2.3$ & $ 629.8(5)$ & $-6.1$ & $   9.9$ & $   8.2 $ \\
$125.3$ & $ 601.1(5)$ & $^{+1.3}_{-1.6}$ &  $\pm 2.3/\pm 0.3/\pm2.3$ & $ 629.3(5)$ & $-6.1$ & $   9.9$ & $   8.2 $ \\
$125.4$ & $ 600.5(5)$ & $^{+1.3}_{-1.6}$ &  $\pm 2.3/\pm 0.3/\pm2.3$ & $ 628.8(5)$ & $-6.1$ & $   9.9$ & $   8.1 $ \\
$125.5$ & $ 600.1(5)$ & $^{+1.3}_{-1.6}$ &  $\pm 2.3/\pm 0.3/\pm2.3$ & $ 628.3(5)$ & $-6.1$ & $   9.9$ & $   8.1 $ \\
$125.6$ & $ 599.7(5)$ & $^{+1.3}_{-1.6}$ &  $\pm 2.3/\pm 0.3/\pm2.3$ & $ 627.8(5)$ & $-6.1$ & $   9.9$ & $   8.1 $ \\
$125.7$ & $ 599.2(5)$ & $^{+1.3}_{-1.6}$ &  $\pm 2.3/\pm 0.3/\pm2.3$ & $ 627.3(5)$ & $-6.1$ & $   9.9$ & $   8.1 $ \\
$125.8$ & $ 598.7(5)$ & $^{+1.3}_{-1.6}$ &  $\pm 2.3/\pm 0.3/\pm2.3$ & $ 626.8(5)$ & $-6.1$ & $   9.9$ & $   8.1 $ \\
$125.9$ & $ 598.2(5)$ & $^{+1.3}_{-1.6}$ &  $\pm 2.3/\pm 0.3/\pm2.3$ & $ 626.3(5)$ & $-6.1$ & $   9.9$ & $   8.0 $ \\
$126.0$ & $ 597.8(5)$ & $^{+1.3}_{-1.6}$ &  $\pm 2.3/\pm 0.3/\pm2.3$ & $ 625.8(5)$ & $-6.1$ & $   9.9$ & $   8.0 $ \\
$126.5$ & $ 595.4(5)$ & $^{+1.3}_{-1.6}$ &  $\pm 2.3/\pm 0.3/\pm2.3$ & $ 623.3(6)$ & $-6.1$ & $   9.8$ & $   7.9 $ \\
$127.0$ & $ 593.1(6)$ & $^{+1.3}_{-1.6}$ &  $\pm 2.3/\pm 0.3/\pm2.3$ & $ 620.8(6)$ & $-6.0$ & $   9.8$ & $   7.8 $ \\
$127.5$ & $ 590.8(6)$ & $^{+1.3}_{-1.6}$ &  $\pm 2.3/\pm 0.3/\pm2.3$ & $ 618.4(6)$ & $-6.0$ & $   9.8$ & $   7.7 $ \\
$128.0$ & $ 588.5(7)$ & $^{+1.3}_{-1.6}$ &  $\pm 2.3/\pm 0.3/\pm2.3$ & $ 615.9(7)$ & $-6.0$ & $   9.8$ & $   7.6 $ \\
$128.5$ & $ 586.1(7)$ & $^{+1.3}_{-1.6}$ &  $\pm 2.3/\pm 0.3/\pm2.3$ & $ 613.4(8)$ & $-6.0$ & $   9.7$ & $   7.5 $ \\
$129.0$ & $ 583.8(8)$ & $^{+1.3}_{-1.6}$ &  $\pm 2.3/\pm 0.3/\pm2.3$ & $ 610.9(8)$ & $-6.0$ & $   9.7$ & $   7.5 $ \\
$129.5$ & $ 581.5(8)$ & $^{+1.3}_{-1.6}$ &  $\pm 2.3/\pm 0.3/\pm2.3$ & $ 608.4(9)$ & $-6.0$ & $   9.7$ & $   7.4 $ \\
$130.0$ & $ 579.1(9)$ & $^{+1.3}_{-1.6}$ &  $\pm 2.3/\pm 0.3/\pm2.3$ & $ 605.9(10)$ & $-6.0$ & $   9.7$ & $   7.3 $ \\
\bottomrule
\end{tabular}%
\end{small}%
\end{center}%
\end{table}

\begin{table}
\caption{Fiducial VBF cross sections in the SM for a LHC CM energy of $\sqrt{s}=8\UTeV$, including QCD and EW corrections
and their uncertainties for different Higgs boson masses $\MH$. For more details see
\refS{subsec:VBF-XS}. The numbers in the $\sigma_{\NNLO \QCD}^{\DIS}$ column have been obtained from a linear interpolation. The interpolation was performed by fitting the cross section for $\MH = \{120.0,\, 122.5,\,  125.0, \, 127.5, \, 130.0\}$ GeV to a linear function. The scale uncertainty in the column $\Delta_{\mathrm{scale}}$ was only computed for $\MH=125.0$ GeV.}
\label{tab:vbf_XSfiducial_8}
\begin{center}%
\begin{small}%
\tabcolsep5pt
\renewcommand{\arraystretch}{1.2}%
\begin{tabular}{ccccccc|c}%
\toprule
$\MH$[GeV] & $\sigma^{\VBF}$[fb] & $\Delta_{\mathrm{scale}}$[\%] & $\Delta_{\mathrm{PDF}/\alphas/\mathrm{PDF\oplus\alphas}}$[\%] &
$\sigma_{\NNLO \QCD}^{\DIS}$[fb] & $\delta_{\ELWK}$[\%] & $\sigma_{\gamma}$[fb] & $\sigma_{\mbox{\scriptsize $s$-chan}}$[fb]
\\
\midrule
$120.0$ & $ 824.9(11)$ & $^{+1.3}_{-1.5}$ & $\pm 2.3/\pm 0.3/\pm2.3$ & $ 865.7(12)$ & $-6.3$ & $  13.4$ & $  12.4 $ \\
$120.5$ & $ 821.9(11)$ & $^{+1.3}_{-1.5}$ & $\pm 2.3/\pm 0.3/\pm2.3$ & $ 862.6(11)$ & $-6.3$ & $  13.3$ & $  12.3 $ \\
$121.0$ & $ 819.1(10)$ & $^{+1.3}_{-1.5}$ & $\pm 2.3/\pm 0.3/\pm2.3$ & $ 859.5(11)$ & $-6.3$ & $  13.3$ & $  12.2 $ \\
$121.5$ & $ 816.2( 9)$ & $^{+1.3}_{-1.5}$ & $\pm 2.3/\pm 0.3/\pm2.3$ & $ 856.4(10)$ & $-6.2$ & $  13.3$ & $  12.0 $ \\
$122.0$ & $ 813.3(8)$ & $^{+1.3}_{-1.5}$ & $\pm 2.3/\pm 0.3/\pm2.3$ & $ 853.3( 9)$ & $-6.2$ & $  13.2$ & $  11.9 $ \\
$122.5$ & $ 810.4(8)$ & $^{+1.3}_{-1.5}$ & $\pm 2.3/\pm 0.3/\pm2.3$ & $ 850.2(8)$ & $-6.2$ & $  13.2$ & $  11.7 $ \\
$123.0$ & $ 807.5(7)$ & $^{+1.3}_{-1.5}$ & $\pm 2.3/\pm 0.3/\pm2.3$ & $ 847.1(8)$ & $-6.2$ & $  13.2$ & $  11.6 $ \\
$123.5$ & $ 804.7(7)$ & $^{+1.3}_{-1.5}$ & $\pm 2.3/\pm 0.3/\pm2.3$ & $ 844.1(7)$ & $-6.2$ & $  13.1$ & $  11.4 $ \\
$124.0$ & $ 801.6(6)$ & $^{+1.3}_{-1.5}$ & $\pm 2.3/\pm 0.3/\pm2.3$ & $ 841.0(7)$ & $-6.2$ & $  13.1$ & $  11.3 $ \\
$124.1$ & $ 801.1(6)$ & $^{+1.3}_{-1.5}$ & $\pm 2.3/\pm 0.3/\pm2.3$ & $ 840.3(7)$ & $-6.2$ & $  13.1$ & $  11.3 $ \\
$124.2$ & $ 800.6(6)$ & $^{+1.3}_{-1.5}$ & $\pm 2.3/\pm 0.3/\pm2.3$ & $ 839.7(7)$ & $-6.2$ & $  13.1$ & $  11.2 $ \\
$124.3$ & $ 799.9(6)$ & $^{+1.3}_{-1.5}$ & $\pm 2.3/\pm 0.3/\pm2.3$ & $ 839.1(7)$ & $-6.2$ & $  13.1$ & $  11.2 $ \\
$124.4$ & $ 799.4(6)$ & $^{+1.3}_{-1.5}$ & $\pm 2.3/\pm 0.3/\pm2.3$ & $ 838.5(7)$ & $-6.2$ & $  13.1$ & $  11.2 $ \\
$124.5$ & $ 798.9(6)$ & $^{+1.3}_{-1.5}$ & $\pm 2.3/\pm 0.3/\pm2.3$ & $ 837.9(7)$ & $-6.2$ & $  13.1$ & $  11.2 $ \\
$124.6$ & $ 798.2(6)$ & $^{+1.3}_{-1.5}$ & $\pm 2.3/\pm 0.3/\pm2.3$ & $ 837.2(7)$ & $-6.2$ & $  13.1$ & $  11.1 $ \\
$124.7$ & $ 797.6(6)$ & $^{+1.3}_{-1.5}$ & $\pm 2.3/\pm 0.3/\pm2.3$ & $ 836.6(7)$ & $-6.2$ & $  13.1$ & $  11.1 $ \\
$124.8$ & $ 797.1(6)$ & $^{+1.3}_{-1.5}$ & $\pm 2.3/\pm 0.3/\pm2.3$ & $ 836.0(7)$ & $-6.2$ & $  13.1$ & $  11.1 $ \\
$124.9$ & $ 796.6(6)$ & $^{+1.3}_{-1.5}$ & $\pm 2.3/\pm 0.3/\pm2.3$ & $ 835.4(7)$ & $-6.2$ & $  13.1$ & $  11.1 $ \\
$125.0$ & $ 795.9(6)$ & $^{+1.3}_{-1.5}$ & $\pm 2.3/\pm 0.3/\pm2.3$ & $ 834.8(7)$ & $-6.2$ & $  13.1$ & $  11.1 $ \\
$125.09$ & $ 795.4(6)$ & $^{+1.3}_{-1.5}$ & $\pm 2.3/\pm 0.3/\pm2.3$ & $ 834.1(7)$ & $-6.2$ & $  13.1$ & $  11.0 $ \\
$125.1$ & $ 795.3(6)$ & $^{+1.3}_{-1.5}$ & $\pm 2.3/\pm 0.3/\pm2.3$ & $ 834.1(7)$ & $-6.2$ & $  13.1$ & $  11.0 $ \\
$125.2$ & $ 794.8(6)$ & $^{+1.3}_{-1.5}$ & $\pm 2.3/\pm 0.3/\pm2.3$ & $ 833.5(7)$ & $-6.2$ & $  13.0$ & $  11.0 $ \\
$125.3$ & $ 794.3(6)$ & $^{+1.3}_{-1.5}$ & $\pm 2.3/\pm 0.3/\pm2.3$ & $ 832.9(7)$ & $-6.2$ & $  13.0$ & $  11.0 $ \\
$125.4$ & $ 793.6(6)$ & $^{+1.3}_{-1.5}$ & $\pm 2.3/\pm 0.3/\pm2.3$ & $ 832.3(7)$ & $-6.2$ & $  13.0$ & $  10.9 $ \\
$125.5$ & $ 793.0(6)$ & $^{+1.3}_{-1.5}$ & $\pm 2.3/\pm 0.3/\pm2.3$ & $ 831.7(7)$ & $-6.2$ & $  13.0$ & $  10.9 $ \\
$125.6$ & $ 792.4(6)$ & $^{+1.3}_{-1.5}$ & $\pm 2.3/\pm 0.3/\pm2.3$ & $ 831.1(7)$ & $-6.2$ & $  13.0$ & $  10.9 $ \\
$125.7$ & $ 791.8(6)$ & $^{+1.3}_{-1.5}$ & $\pm 2.3/\pm 0.3/\pm2.3$ & $ 830.4(7)$ & $-6.2$ & $  13.0$ & $  10.8 $ \\
$125.8$ & $ 791.2(6)$ & $^{+1.3}_{-1.5}$ & $\pm 2.3/\pm 0.3/\pm2.3$ & $ 829.8(7)$ & $-6.2$ & $  13.0$ & $  10.8 $ \\
$125.9$ & $ 790.7(6)$ & $^{+1.3}_{-1.5}$ & $\pm 2.3/\pm 0.3/\pm2.3$ & $ 829.2(7)$ & $-6.2$ & $  13.0$ & $  10.8 $ \\
$126.0$ & $ 790.1(6)$ & $^{+1.3}_{-1.5}$ & $\pm 2.3/\pm 0.3/\pm2.3$ & $ 828.6(7)$ & $-6.2$ & $  13.0$ & $  10.8 $ \\
$126.5$ & $ 787.1(7)$ & $^{+1.3}_{-1.5}$ & $\pm 2.3/\pm 0.3/\pm2.3$ & $ 825.5(7)$ & $-6.2$ & $  13.0$ & $  10.7 $ \\
$127.0$ & $ 784.3(7)$ & $^{+1.3}_{-1.5}$ & $\pm 2.3/\pm 0.3/\pm2.3$ & $ 822.4(8)$ & $-6.2$ & $  12.9$ & $  10.5 $ \\
$127.5$ & $ 781.5(8)$ & $^{+1.3}_{-1.5}$ & $\pm 2.3/\pm 0.3/\pm2.3$ & $ 819.3(8)$ & $-6.2$ & $  12.9$ & $  10.4 $ \\
$128.0$ & $ 778.5(8)$ & $^{+1.3}_{-1.5}$ & $\pm 2.3/\pm 0.3/\pm2.3$ & $ 816.2(9)$ & $-6.2$ & $  12.9$ & $  10.3 $ \\
$128.5$ & $ 775.7(9)$ & $^{+1.3}_{-1.5}$ & $\pm 2.3/\pm 0.3/\pm2.3$ & $ 813.1(10)$ & $-6.2$ & $  12.8$ & $  10.2 $ \\
$129.0$ & $ 772.8(10)$ & $^{+1.3}_{-1.5}$ & $\pm 2.3/\pm 0.3/\pm2.3$ & $ 810.0(11)$ & $-6.2$ & $  12.8$ & $  10.0 $ \\
$129.5$ & $ 769.8(11)$ & $^{+1.3}_{-1.5}$ & $\pm 2.3/\pm 0.3/\pm2.3$ & $ 806.9(11)$ & $-6.2$ & $  12.8$ & $   9.9 $ \\
$130.0$ & $ 766.9(11)$ & $^{+1.3}_{-1.5}$ & $\pm 2.3/\pm 0.3/\pm2.3$ & $ 803.8(12)$ & $-6.2$ & $  12.8$ & $   9.8 $ \\
\bottomrule
\end{tabular}%
\end{small}%
\end{center}%
\end{table}

\begin{table}
\caption{Fiducial VBF cross sections in the SM for a LHC CM energy of $\sqrt{s}=13\UTeV$, including QCD and EW corrections
and their uncertainties for different Higgs boson masses $\MH$. For more details see
\refS{subsec:VBF-XS}.  The numbers in the $\sigma_{\NNLO \QCD}^{\DIS}$ column have been obtained from a linear interpolation. The interpolation was performed by fitting the cross section for $\MH = \{120.0,\, 122.5,\,  125.0, \, 127.5, \, 130.0\}$ GeV to a linear function. The scale uncertainty in the column $\Delta_{\mathrm{scale}}$ was only computed for $\MH=125.0$ GeV.}
\label{tab:vbf_XSfiducial_13}
\begin{center}%
\begin{small}%
\tabcolsep5pt
\renewcommand{\arraystretch}{1.2}%
\begin{tabular}{ccccccc|c}%
\toprule
$\MH$[GeV] & $\sigma^{\VBF}$[fb] & $\Delta_{\mathrm{scale}}$[\%] & $\Delta_{\mathrm{PDF}/\alphas/\mathrm{PDF\oplus\alphas}}$[\%] &
$\sigma_{\NNLO \QCD}^{\DIS}$[fb] & $\delta_{\ELWK}$[\%] & $\sigma_{\gamma}$[fb] & $\sigma_{\mbox{\scriptsize $s$-chan}}$[fb]
\\
\midrule
$120.0$ & $2038.9(45)$ & $^{+1.3}_{-1.2}$ & $\pm 2.1/\pm 0.4/\pm2.2$ & $2152.8(48)$ & $-6.8$ & $  33.0$ & $  32.4 $ \\
$120.5$ & $2032.3(41)$ & $^{+1.3}_{-1.2}$ & $\pm 2.1/\pm 0.4/\pm2.2$ & $2145.9(44)$ & $-6.8$ & $  32.9$ & $  32.1 $ \\
$121.0$ & $2026.3(36)$ & $^{+1.3}_{-1.2}$ & $\pm 2.1/\pm 0.4/\pm2.2$ & $2139.1(39)$ & $-6.8$ & $  32.8$ & $  31.7 $ \\
$121.5$ & $2020.0(32)$ & $^{+1.3}_{-1.2}$ & $\pm 2.1/\pm 0.4/\pm2.2$ & $2132.2(35)$ & $-6.8$ & $  32.8$ & $  31.3 $ \\
$122.0$ & $2013.4(28)$ & $^{+1.3}_{-1.2}$ & $\pm 2.1/\pm 0.4/\pm2.2$ & $2125.4(30)$ & $-6.8$ & $  32.7$ & $  30.9 $ \\
$122.5$ & $2007.1(24)$ & $^{+1.3}_{-1.2}$ & $\pm 2.1/\pm 0.4/\pm2.2$ & $2118.5(26)$ & $-6.8$ & $  32.6$ & $  30.6 $ \\
$123.0$ & $2000.5(20)$ & $^{+1.3}_{-1.2}$ & $\pm 2.1/\pm 0.4/\pm2.2$ & $2111.7(21)$ & $-6.8$ & $  32.6$ & $  30.2 $ \\
$123.5$ & $1994.5(16)$ & $^{+1.3}_{-1.2}$ & $\pm 2.1/\pm 0.4/\pm2.2$ & $2104.8(17)$ & $-6.8$ & $  32.5$ & $  29.9 $ \\
$124.0$ & $1987.9(13)$ & $^{+1.3}_{-1.2}$ & $\pm 2.1/\pm 0.4/\pm2.2$ & $2097.9(14)$ & $-6.8$ & $  32.4$ & $  29.6 $ \\
$124.1$ & $1986.9(12)$ & $^{+1.3}_{-1.2}$ & $\pm 2.1/\pm 0.4/\pm2.2$ & $2096.6(13)$ & $-6.8$ & $  32.4$ & $  29.5 $ \\
$124.2$ & $1985.3(12)$ & $^{+1.3}_{-1.2}$ & $\pm 2.1/\pm 0.4/\pm2.2$ & $2095.2(12)$ & $-6.8$ & $  32.4$ & $  29.5 $ \\
$124.3$ & $1984.2(11)$ & $^{+1.3}_{-1.2}$ & $\pm 2.1/\pm 0.4/\pm2.2$ & $2093.8(12)$ & $-6.8$ & $  32.4$ & $  29.4 $ \\
$124.4$ & $1982.7(11)$ & $^{+1.3}_{-1.2}$ & $\pm 2.1/\pm 0.4/\pm2.2$ & $2092.5(11)$ & $-6.8$ & $  32.4$ & $  29.3 $ \\
$124.5$ & $1981.4(10)$ & $^{+1.3}_{-1.2}$ & $\pm 2.1/\pm 0.4/\pm2.2$ & $2091.1(11)$ & $-6.8$ & $  32.4$ & $  29.2 $ \\
$124.6$ & $1980.4(10)$ & $^{+1.3}_{-1.2}$ & $\pm 2.1/\pm 0.4/\pm2.2$ & $2089.7(11)$ & $-6.8$ & $  32.4$ & $  29.2 $ \\
$124.7$ & $1979.2(10)$ & $^{+1.3}_{-1.2}$ & $\pm 2.1/\pm 0.4/\pm2.2$ & $2088.4(10)$ & $-6.8$ & $  32.4$ & $  29.1 $ \\
$124.8$ & $1977.6( 9)$ & $^{+1.3}_{-1.2}$ & $\pm 2.1/\pm 0.4/\pm2.2$ & $2087.0(10)$ & $-6.8$ & $  32.3$ & $  29.0 $ \\
$124.9$ & $1976.4( 9)$ & $^{+1.3}_{-1.2}$ & $\pm 2.1/\pm 0.4/\pm2.2$ & $2085.6(10)$ & $-6.8$ & $  32.3$ & $  29.0 $ \\
$125.0$ & $1975.4( 9)$ & $^{+1.3}_{-1.2}$ & $\pm 2.1/\pm 0.4/\pm2.2$ & $2084.2(10)$ & $-6.8$ & $  32.3$ & $  29.0 $ \\
$125.09$ & $1974.0( 9)$ & $^{+1.3}_{-1.2}$ & $\pm 2.1/\pm 0.4/\pm2.2$ & $2082.9(10)$ & $-6.8$ & $  32.3$ & $  28.9 $ \\
$125.1$ & $1974.1( 9)$ & $^{+1.3}_{-1.2}$ & $\pm 2.1/\pm 0.4/\pm2.2$ & $2082.9(10)$ & $-6.8$ & $  32.3$ & $  28.9 $ \\
$125.2$ & $1972.8( 9)$ & $^{+1.3}_{-1.2}$ & $\pm 2.1/\pm 0.4/\pm2.2$ & $2081.5(10)$ & $-6.8$ & $  32.3$ & $  28.8 $ \\
$125.3$ & $1971.4(10)$ & $^{+1.3}_{-1.2}$ & $\pm 2.1/\pm 0.4/\pm2.2$ & $2080.1(10)$ & $-6.8$ & $  32.3$ & $  28.7 $ \\
$125.4$ & $1970.1(10)$ & $^{+1.3}_{-1.2}$ & $\pm 2.1/\pm 0.4/\pm2.2$ & $2078.8(11)$ & $-6.8$ & $  32.3$ & $  28.7 $ \\
$125.5$ & $1968.8(10)$ & $^{+1.3}_{-1.2}$ & $\pm 2.1/\pm 0.4/\pm2.2$ & $2077.4(11)$ & $-6.8$ & $  32.2$ & $  28.6 $ \\
$125.6$ & $1967.6(11)$ & $^{+1.3}_{-1.2}$ & $\pm 2.1/\pm 0.4/\pm2.2$ & $2076.0(11)$ & $-6.8$ & $  32.2$ & $  28.5 $ \\
$125.7$ & $1966.2(11)$ & $^{+1.3}_{-1.2}$ & $\pm 2.1/\pm 0.4/\pm2.2$ & $2074.6(12)$ & $-6.8$ & $  32.2$ & $  28.5 $ \\
$125.8$ & $1965.1(12)$ & $^{+1.3}_{-1.2}$ & $\pm 2.1/\pm 0.4/\pm2.2$ & $2073.3(12)$ & $-6.8$ & $  32.2$ & $  28.4 $ \\
$125.9$ & $1963.6(12)$ & $^{+1.3}_{-1.2}$ & $\pm 2.1/\pm 0.4/\pm2.2$ & $2071.9(13)$ & $-6.8$ & $  32.2$ & $  28.4 $ \\
$126.0$ & $1962.6(13)$ & $^{+1.3}_{-1.2}$ & $\pm 2.1/\pm 0.4/\pm2.2$ & $2070.5(14)$ & $-6.8$ & $  32.2$ & $  28.3 $ \\
$126.5$ & $1956.2(16)$ & $^{+1.3}_{-1.2}$ & $\pm 2.1/\pm 0.4/\pm2.2$ & $2063.7(17)$ & $-6.8$ & $  32.1$ & $  27.9 $ \\
$127.0$ & $1949.7(20)$ & $^{+1.3}_{-1.2}$ & $\pm 2.1/\pm 0.4/\pm2.2$ & $2056.8(21)$ & $-6.8$ & $  32.1$ & $  27.6 $ \\
$127.5$ & $1943.5(24)$ & $^{+1.3}_{-1.2}$ & $\pm 2.1/\pm 0.4/\pm2.2$ & $2050.0(26)$ & $-6.8$ & $  32.0$ & $  27.4 $ \\
$128.0$ & $1937.2(28)$ & $^{+1.3}_{-1.2}$ & $\pm 2.1/\pm 0.4/\pm2.2$ & $2043.1(30)$ & $-6.7$ & $  31.9$ & $  27.1 $ \\
$128.5$ & $1930.7(32)$ & $^{+1.3}_{-1.2}$ & $\pm 2.1/\pm 0.4/\pm2.2$ & $2036.3(35)$ & $-6.7$ & $  31.9$ & $  26.8 $ \\
$129.0$ & $1924.4(36)$ & $^{+1.3}_{-1.2}$ & $\pm 2.1/\pm 0.4/\pm2.2$ & $2029.4(39)$ & $-6.7$ & $  31.8$ & $  26.5 $ \\
$129.5$ & $1917.7(41)$ & $^{+1.3}_{-1.2}$ & $\pm 2.1/\pm 0.4/\pm2.2$ & $2022.6(44)$ & $-6.8$ & $  31.7$ & $  26.2 $ \\
$130.0$ & $1911.5(45)$ & $^{+1.3}_{-1.2}$ & $\pm 2.1/\pm 0.4/\pm2.2$ & $2015.7(48)$ & $-6.7$ & $  31.7$ & $  25.9 $ \\
\bottomrule
\end{tabular}%
\end{small}%
\end{center}%
\end{table}

\begin{table}
\caption{Fiducial VBF cross sections in the SM for a LHC CM energy of $\sqrt{s}=14\UTeV$, including QCD and EW corrections
and their uncertainties for different Higgs boson masses $\MH$. For more details see
\refS{subsec:VBF-XS}. The numbers in the $\sigma_{\NNLO \QCD}^{\DIS}$ column have been obtained from a linear interpolation. The interpolation was performed by fitting the cross section for $\MH = \{120.0,\, 122.5,\,  125.0, \, 127.5, \, 130.0\}$ GeV to a linear function. The scale uncertainty in the column $\Delta_{\mathrm{scale}}$ was only computed for $\MH=125.0$ GeV.}
\label{tab:vbf_XSfiducial_14}
\begin{center}%
\begin{small}%
\tabcolsep5pt
\renewcommand{\arraystretch}{1.2}%
\begin{tabular}{ccccccc|c}%
\toprule
$\MH$[GeV] & $\sigma^{\VBF}$[fb] & $\Delta_{\mathrm{scale}}$[\%] & $\Delta_{\mathrm{PDF}/\alphas/\mathrm{PDF\oplus\alphas}}$[\%] &
$\sigma_{\NNLO \QCD}^{\DIS}$[fb] & $\delta_{\ELWK}$[\%] & $\sigma_{\gamma}$[fb] & $\sigma_{\mbox{\scriptsize $s$-chan}}$[fb]
\\
\midrule
$120.0$ & $2301.6(48)$ & $^{+1.5}_{-1.3}$ & $\pm 2.1/\pm 0.4/\pm2.1$ & $2432.4(51)$ & $-6.9$ & $  37.4$ & $  37.0 $ \\
$120.5$ & $2294.9(45)$ & $^{+1.5}_{-1.3}$ & $\pm 2.1/\pm 0.4/\pm2.1$ & $2425.4(48)$ & $-6.9$ & $  37.3$ & $  36.6 $ \\
$121.0$ & $2288.6(41)$ & $^{+1.5}_{-1.3}$ & $\pm 2.1/\pm 0.4/\pm2.1$ & $2418.3(44)$ & $-6.9$ & $  37.2$ & $  36.2 $ \\
$121.5$ & $2282.0(38)$ & $^{+1.5}_{-1.3}$ & $\pm 2.1/\pm 0.4/\pm2.1$ & $2411.3(41)$ & $-6.9$ & $  37.1$ & $  35.8 $ \\
$122.0$ & $2275.5(35)$ & $^{+1.5}_{-1.3}$ & $\pm 2.1/\pm 0.4/\pm2.1$ & $2404.3(38)$ & $-6.9$ & $  37.1$ & $  35.4 $ \\
$122.5$ & $2269.1(33)$ & $^{+1.5}_{-1.3}$ & $\pm 2.1/\pm 0.4/\pm2.1$ & $2397.3(35)$ & $-6.9$ & $  37.0$ & $  35.0 $ \\
$123.0$ & $2262.8(31)$ & $^{+1.5}_{-1.3}$ & $\pm 2.1/\pm 0.4/\pm2.1$ & $2390.3(33)$ & $-6.9$ & $  36.9$ & $  34.6 $ \\
\bottomrule
\end{tabular}%
\end{small}%
\end{center}%
\end{table}

\clearpage

\begin{table}
\caption{Inclusive VBF cross sections for a LHC CM energy of $\sqrt{s}=7\UTeV$, at NNLO QCD, with an on shell Higgs or
with off-shell effects included in the complex-pole scheme, together with their uncertainties. For more details see
\refS{sec:VBFNNLO-sub-sub}.}
\label{tab:vbfnnlo_XStot_7}
\begin{center}%
\begin{small}%
\tabcolsep5pt
\renewcommand{\arraystretch}{1.2}%
\begin{tabular}{cccccccc}%
\toprule
$\MH$[GeV] & $\sigma^{\VBF}_{\rm{OS}}$[fb] & $\sigma^{\VBF}_{\rm{CPS}}$[fb] & $\delta_{\rm{CPS}} $ & 
$\Delta_{\mathrm{scale}}$[\%] & $\Delta_{\mathrm{PDF}}$[\%] & $\Delta_{\mathrm{\alphas}}$[\%] & $\Delta_{\mathrm{PDF\oplus\alphas}}$[\%] 
\\
\midrule
 $120.0$  &  $ 1339$  &  $ 1353$  &  $1.0$  &  $^{+0.2} _{-0.2}$  &  $\pm1.9$  &  $\pm0.4$  &  $\pm1.9$ \\
 $120.5$  &  $ 1332$  &  $ 1346$  &  $1.1$  &  $^{+0.2} _{-0.2}$  &  $\pm1.9$  &  $\pm0.4$  &  $\pm1.9$ \\
 $121.0$  &  $ 1326$  &  $ 1340$  &  $1.1$  &  $^{+0.2} _{-0.2}$  &  $\pm1.9$  &  $\pm0.4$  &  $\pm1.9$ \\
 $121.5$  &  $ 1320$  &  $ 1334$  &  $1.1$  &  $^{+0.2} _{-0.2}$  &  $\pm1.9$  &  $\pm0.4$  &  $\pm1.9$ \\
 $122.0$  &  $ 1313$  &  $ 1327$  &  $1.1$  &  $^{+0.2} _{-0.2}$  &  $\pm1.9$  &  $\pm0.4$  &  $\pm1.9$ \\
 $122.5$  &  $ 1307$  &  $ 1321$  &  $1.1$  &  $^{+0.2} _{-0.2}$  &  $\pm1.9$  &  $\pm0.4$  &  $\pm1.9$ \\
 $123.0$  &  $ 1301$  &  $ 1315$  &  $1.1$  &  $^{+0.2} _{-0.2}$  &  $\pm1.9$  &  $\pm0.4$  &  $\pm1.9$ \\
\bottomrule
\end{tabular}%
\end{small}%
\end{center}%
\end{table}

\begin{table}
\caption{Inclusive VBF cross sections for a LHC CM energy of $\sqrt{s}=8\UTeV$, at NNLO QCD, with an on shell Higgs or
with off-shell effects included in the complex-pole scheme, together with their uncertainties. For more details see
\refS{sec:VBFNNLO-sub-sub}.}
\label{tab:vbfnnlo_XStot_8}
\begin{center}%
\begin{small}%
\tabcolsep5pt
\renewcommand{\arraystretch}{1.2}%
\begin{tabular}{cccccccc}%
\toprule
$\MH$[GeV] & $\sigma^{\VBF}_{\rm{OS}}$[fb] & $\sigma^{\VBF}_{\rm{CPS}}$[fb] & $\delta_{\rm{CPS}} $ & 
$\Delta_{\mathrm{scale}}$[\%] & $\Delta_{\mathrm{PDF}}$[\%] & $\Delta_{\mathrm{\alphas}}$[\%] & $\Delta_{\mathrm{PDF\oplus\alphas}}$[\%] 
\\
\midrule
 $120.0$  &  $ 1727$  &  $ 1748$  &  $1.2$  &  $^{+0.3} _{-0.2}$  &  $\pm1.9$  &  $\pm0.5$  &  $\pm1.9$ \\
 $120.5$  &  $ 1719$  &  $ 1741$  &  $1.3$  &  $^{+0.2} _{-0.2}$  &  $\pm1.9$  &  $\pm0.5$  &  $\pm1.9$ \\
 $121.0$  &  $ 1711$  &  $ 1733$  &  $1.3$  &  $^{+0.3} _{-0.2}$  &  $\pm1.9$  &  $\pm0.5$  &  $\pm1.9$ \\
 $121.5$  &  $ 1704$  &  $ 1725$  &  $1.2$  &  $^{+0.2} _{-0.2}$  &  $\pm1.9$  &  $\pm0.5$  &  $\pm1.9$ \\
 $122.0$  &  $ 1695$  &  $ 1717$  &  $1.3$  &  $^{+0.3} _{-0.2}$  &  $\pm1.9$  &  $\pm0.5$  &  $\pm1.9$ \\
 $122.5$  &  $ 1688$  &  $ 1710$  &  $1.3$  &  $^{+0.2} _{-0.2}$  &  $\pm1.9$  &  $\pm0.5$  &  $\pm1.9$ \\
 $123.0$  &  $ 1680$  &  $ 1702$  &  $1.3$  &  $^{+0.3} _{-0.2}$  &  $\pm1.9$  &  $\pm0.5$  &  $\pm1.9$ \\
\bottomrule
\end{tabular}%
\end{small}%
\end{center}%
\end{table}

\begin{table}
\caption{Inclusive VBF cross sections for a LHC CM energy of $\sqrt{s}=13\UTeV$, at NNLO QCD, with an on shell Higgs or
with off-shell effects included in the complex-pole scheme, together with their uncertainties. For more details see
\refS{sec:VBFNNLO-sub-sub}.}
\label{tab:vbfnnlo_XStot_13}
\begin{center}%
\begin{small}%
\tabcolsep5pt
\renewcommand{\arraystretch}{1.2}%
\begin{tabular}{cccccccc}%
\toprule
$\MH$[GeV] & $\sigma^{\VBF}_{\rm{OS}}$[fb] & $\sigma^{\VBF}_{\rm{CPS}}$[fb] & $\delta_{\rm{CPS}} $ & 
$\Delta_{\mathrm{scale}}$[\%] & $\Delta_{\mathrm{PDF}}$[\%] & $\Delta_{\mathrm{\alphas}}$[\%] & $\Delta_{\mathrm{PDF\oplus\alphas}}$[\%] 
\\
\midrule
  $120.0$  &  $ 4086$  &  $ 4162$  &  $1.9$  &  $^{+0.5} _{-0.2}$  &  $\pm1.8$  &  $\pm0.6$  &  $\pm1.9$ \\
  $120.5$  &  $ 4072$  &  $ 4148$  &  $1.9$  &  $^{+0.4} _{-0.3}$  &  $\pm1.8$  &  $\pm0.6$  &  $\pm1.9$ \\
  $121.0$  &  $ 4053$  &  $ 4130$  &  $1.9$  &  $^{+0.5} _{-0.2}$  &  $\pm1.8$  &  $\pm0.6$  &  $\pm1.9$ \\
  $121.5$  &  $ 4039$  &  $ 4110$  &  $1.8$  &  $^{+0.4} _{-0.3}$  &  $\pm1.8$  &  $\pm0.6$  &  $\pm1.9$ \\
  $122.0$  &  $ 4020$  &  $ 4098$  &  $1.9$  &  $^{+0.5} _{-0.2}$  &  $\pm1.8$  &  $\pm0.6$  &  $\pm1.9$ \\
  $122.5$  &  $ 4007$  &  $ 4082$  &  $1.9$  &  $^{+0.4} _{-0.3}$  &  $\pm1.8$  &  $\pm0.6$  &  $\pm1.9$ \\
  $123.0$  &  $ 3988$  &  $ 4068$  &  $2.0$  &  $^{+0.5} _{-0.2}$  &  $\pm1.8$  &  $\pm0.6$  &  $\pm1.9$ \\
\bottomrule
\end{tabular}%
\end{small}%
\end{center}%
\end{table}

\clearpage

\begin{table}
\caption{Inclusive VBF cross sections for a LHC CM energy of $\sqrt{s}=14\UTeV$, at NNLO QCD, with an on shell Higgs or
with off-shell effects included in the complex-pole scheme, together with their uncertainties. For more details see
\refS{sec:VBFNNLO-sub-sub}.}
\label{tab:vbfnnlo_XStot_14}
\begin{center}%
\begin{small}%
\tabcolsep5pt
\renewcommand{\arraystretch}{1.2}%
\begin{tabular}{cccccccc}%
\toprule
$\MH$[GeV] & $\sigma^{\VBF}_{\rm{OS}}$[fb] & $\sigma^{\VBF}_{\rm{CPS}}$[fb] & $\delta_{\rm{CPS}} $ & 
$\Delta_{\mathrm{scale}}$[\%] & $\Delta_{\mathrm{PDF}}$[\%] & $\Delta_{\mathrm{\alphas}}$[\%] & $\Delta_{\mathrm{PDF\oplus\alphas}}$[\%] 
\\
\midrule
 $120.0$  &  $ 4623$  &  $ 4718$  &  $2.1$  &  $^{+0.5} _{-0.2}$  &  $\pm1.8$  &  $\pm0.6$  &  $\pm1.9$ \\
 $120.5$  &  $ 4608$  &  $ 4700$  &  $2.0$  &  $^{+0.4} _{-0.3}$  &  $\pm1.8$  &  $\pm0.6$  &  $\pm1.9$ \\
 $121.0$  &  $ 4587$  &  $ 4687$  &  $2.2$  &  $^{+0.5} _{-0.2}$  &  $\pm1.8$  &  $\pm0.6$  &  $\pm1.9$ \\
 $121.5$  &  $ 4572$  &  $ 4673$  &  $2.2$  &  $^{+0.4} _{-0.3}$  &  $\pm1.8$  &  $\pm0.6$  &  $\pm1.9$ \\
 $122.0$  &  $ 4551$  &  $ 4651$  &  $2.2$  &  $^{+0.5} _{-0.2}$  &  $\pm1.8$  &  $\pm0.6$  &  $\pm1.9$ \\
 $122.5$  &  $ 4536$  &  $ 4634$  &  $2.2$  &  $^{+0.4} _{-0.3}$  &  $\pm1.8$  &  $\pm0.6$  &  $\pm1.9$ \\
 $123.0$  &  $ 4515$  &  $ 4605$  &  $2.0$  &  $^{+0.5} _{-0.2}$  &  $\pm1.8$  &  $\pm0.6$  &  $\pm1.9$ \\
\bottomrule
\end{tabular}%
\end{small}%
\end{center}%
\end{table}

\begin{table}
\caption{Inclusive VBF cross sections for a Higgs boson mass of $\MH=125\UGeV$, at NNLO QCD, with an on shell Higgs or
with off-shell effects included in the complex-pole scheme, together with their uncertainties. For more details see
\refS{sec:VBFNNLO-sub-sub}.}
\label{tab:table_vbf_nnlo_escan_125}
\begin{center}%
\begin{small}%
\tabcolsep5pt
\renewcommand{\arraystretch}{1.2}%
\begin{tabular}{cccccccc}%
\toprule
$\MH$[GeV] & $\sigma^{\VBF}_{\rm{OS}}$[fb] & $\sigma^{\VBF}_{\rm{CPS}}$[fb] & $\delta_{\rm{CPS}} $ & 
$\Delta_{\mathrm{scale}}$[\%] & $\Delta_{\mathrm{PDF}}$[\%] & $\Delta_{\mathrm{\alphas}}$[\%] & $\Delta_{\mathrm{PDF\oplus\alphas}}$[\%] 
\\
\midrule
 $   6.0$  &  $  938$  &  $  947$  &  $0.9$  &  $^{+0.2} _{-0.1}$  &  $\pm1.9$  &  $\pm0.4$  &  $\pm2.0$ \\
 $   6.5$  &  $ 1102$  &  $ 1114$  &  $1.1$  &  $^{+0.2} _{-0.1}$  &  $\pm1.9$  &  $\pm0.4$  &  $\pm1.9$ \\
 $   7.0$  &  $ 1276$  &  $ 1291$  &  $1.2$  &  $^{+0.2} _{-0.2}$  &  $\pm1.9$  &  $\pm0.4$  &  $\pm1.9$ \\
 $   7.5$  &  $ 1459$  &  $ 1477$  &  $1.2$  &  $^{+0.2} _{-0.2}$  &  $\pm1.9$  &  $\pm0.4$  &  $\pm1.9$ \\
 $   8.0$  &  $ 1650$  &  $ 1672$  &  $1.3$  &  $^{+0.3} _{-0.2}$  &  $\pm1.9$  &  $\pm0.5$  &  $\pm1.9$ \\
 $   8.5$  &  $ 1848$  &  $ 1873$  &  $1.4$  &  $^{+0.3} _{-0.2}$  &  $\pm1.9$  &  $\pm0.5$  &  $\pm1.9$ \\
 $   9.0$  &  $ 2054$  &  $ 2083$  &  $1.4$  &  $^{+0.3} _{-0.2}$  &  $\pm1.8$  &  $\pm0.5$  &  $\pm1.9$ \\
\bottomrule
\end{tabular}%
\end{small}%
\end{center}%
\end{table}

\begin{table}
\caption{Inclusive VBF cross sections for a Higgs boson mass of $\MH=125.09\UGeV$, at NNLO QCD, with an on shell Higgs or
with off-shell effects included in the complex-pole scheme, together with their uncertainties. For more details see
\refS{sec:VBFNNLO-sub-sub}.}
\label{tab:table_vbf_nnlo_escan_12509}
\begin{center}%
\begin{small}%
\tabcolsep5pt
\renewcommand{\arraystretch}{1.2}%
\begin{tabular}{cccccccc}%
\toprule
$\MH$[GeV] & $\sigma^{\VBF}_{\rm{OS}}$[fb] & $\sigma^{\VBF}_{\rm{CPS}}$[fb] & $\delta_{\rm{CPS}} $ & 
$\Delta_{\mathrm{scale}}$[\%] & $\Delta_{\mathrm{PDF}}$[\%] & $\Delta_{\mathrm{\alphas}}$[\%] & $\Delta_{\mathrm{PDF\oplus\alphas}}$[\%] 
\\
\midrule
 $   6.0$  &  $  937$  &  $  946$  &  $0.9$  &  $^{+0.2} _{-0.1}$  &  $\pm1.9$  &  $\pm0.4$  &  $\pm2.0$ \\
 $   6.5$  &  $ 1101$  &  $ 1113$  &  $1.1$  &  $^{+0.2} _{-0.1}$  &  $\pm1.9$  &  $\pm0.4$  &  $\pm1.9$ \\
 $   7.0$  &  $ 1275$  &  $ 1290$  &  $1.2$  &  $^{+0.2} _{-0.2}$  &  $\pm1.9$  &  $\pm0.4$  &  $\pm1.9$ \\
 $   7.5$  &  $ 1457$  &  $ 1476$  &  $1.3$  &  $^{+0.2} _{-0.2}$  &  $\pm1.9$  &  $\pm0.4$  &  $\pm1.9$ \\
 $   8.0$  &  $ 1648$  &  $ 1670$  &  $1.3$  &  $^{+0.3} _{-0.2}$  &  $\pm1.9$  &  $\pm0.5$  &  $\pm1.9$ \\
 $   8.5$  &  $ 1847$  &  $ 1872$  &  $1.4$  &  $^{+0.3} _{-0.2}$  &  $\pm1.9$  &  $\pm0.5$  &  $\pm1.9$ \\
 $   9.0$  &  $ 2052$  &  $ 2081$  &  $1.4$  &  $^{+0.3} _{-0.2}$  &  $\pm1.8$  &  $\pm0.5$  &  $\pm1.9$ \\
\bottomrule
\end{tabular}%
\end{small}%
\end{center}%
\end{table}

\clearpage

\chapter{SM Higgs-strahlung cross sections}
\label{VHappendix}

Here we expand \refTs{tab:wph_XStot}--\ref{tab:znnh_XStot},
which contain predictions for total and fiducial Higgs-strahlung cross sections
in the SM for $\MH=125\UGeV$, to a scan over SM Higgs boson masses.

\begin{table}
\caption{Total $\PWp({\to}\Pl^+\nu_{\Pl})\PH$ cross sections including QCD and EW corrections
and their uncertainties for a proton--proton collision energy $\sqrt{s}=7\UTeV$.}
\label{tab:wph_XStot_7}
\begin{center}%
\begin{small}%
\tabcolsep5pt
\renewcommand{\arraystretch}{1.2}%
\begin{tabular}{cccccccc}%
\toprule
$\MH$& $\sigma$ & $\Delta_{\mathrm{scale}}$ & 
$\Delta_{\mathrm{PDF}/\alphas/\mathrm{PDF\oplus\alphas}}$&
$\sigma_{\NNLO \QCD}^{\DY}$ & $\sigma_{\Pt\mbox{\scriptsize -loop}}$ &
$\delta_{\ELWK}$ & $\sigma_{\gamma}$
\vspace{.1cm} \\
~[GeV] & [fb] & [\%]  & [\%] & [fb]  & [fb] &  [\%]  & [fb] \\
\midrule
$120.0$ & $  46.85$ & ${}_{-0.9}^{+ 0.6}$ & $\pm 1.8/\pm 0.6/\pm 1.9$ & $  48.90$ & $   0.47$ & $-7.0$ & $  0.91_{ -0.11}^{+  1.13}$ \\
$120.5$ & $  46.22$ & ${}_{-1.0}^{+ 0.6}$ & $\pm 1.8/\pm 0.6/\pm 1.9$ & $  48.24$ & $   0.46$ & $-7.0$ & $  0.90_{ -0.11}^{+  1.12}$ \\
$121.0$ & $  45.57$ & ${}_{-0.9}^{+ 0.7}$ & $\pm 1.9/\pm 0.6/\pm 1.9$ & $  47.56$ & $   0.46$ & $-7.0$ & $  0.90_{ -0.10}^{+  1.12}$ \\
$121.5$ & $  44.95$ & ${}_{-0.9}^{+ 0.7}$ & $\pm 1.9/\pm 0.6/\pm 2.0$ & $  46.92$ & $   0.45$ & $-7.1$ & $  0.90_{ -0.10}^{+  1.12}$ \\
$122.0$ & $  44.38$ & ${}_{-0.9}^{+ 0.7}$ & $\pm 1.9/\pm 0.6/\pm 2.0$ & $  46.32$ & $   0.45$ & $-7.1$ & $  0.90_{ -0.10}^{+  1.12}$ \\
$122.5$ & $  43.80$ & ${}_{-0.9}^{+ 0.6}$ & $\pm 1.9/\pm 0.6/\pm 2.0$ & $  45.72$ & $   0.44$ & $-7.1$ & $  0.89_{ -0.10}^{+  1.11}$ \\
\bottomrule
\end{tabular}%
\end{small}%
\end{center}%
\end{table}

\begin{table}
\caption{Total $\PWp({\to}\Pl^+\nu_{\Pl})\PH$ cross sections including QCD and EW corrections
and their uncertainties for a proton--proton collision energy $\sqrt{s}=8\UTeV$.}
\label{tab:wph_XStot_8}
\begin{center}%
\begin{small}%
\tabcolsep5pt
\renewcommand{\arraystretch}{1.2}%
\begin{tabular}{cccccccc}%
\toprule
$\MH$& $\sigma$ & $\Delta_{\mathrm{scale}}$ & 
$\Delta_{\mathrm{PDF}/\alphas/\mathrm{PDF\oplus\alphas}}$&
$\sigma_{\NNLO \QCD}^{\DY}$ & $\sigma_{\Pt\mbox{\scriptsize -loop}}$ &
$\delta_{\ELWK}$ & $\sigma_{\gamma}$
\vspace{.1cm} \\
~[GeV] & [fb] & [\%]  & [\%] & [fb]  & [fb] &  [\%]  & [fb] \\
\midrule
$120.0$ & $  56.46$ & ${}_{-0.8}^{+ 0.6}$ & $\pm 1.8/\pm 0.7/\pm 2.0$ & $  58.79$ & $   0.59$ & $-7.0$ & $  1.21_{ -0.15}^{+  1.44}$ \\
$120.5$ & $  55.73$ & ${}_{-0.9}^{+ 0.5}$ & $\pm 1.8/\pm 0.7/\pm 2.0$ & $  58.04$ & $   0.59$ & $-7.1$ & $  1.21_{ -0.15}^{+  1.43}$ \\
$121.0$ & $  54.97$ & ${}_{-0.8}^{+ 0.6}$ & $\pm 1.8/\pm 0.7/\pm 2.0$ & $  57.24$ & $   0.58$ & $-7.1$ & $  1.21_{ -0.15}^{+  1.43}$ \\
$121.5$ & $  54.24$ & ${}_{-0.8}^{+ 0.6}$ & $\pm 1.8/\pm 0.7/\pm 2.0$ & $  56.48$ & $   0.58$ & $-7.1$ & $  1.20_{ -0.14}^{+  1.42}$ \\
$122.0$ & $  53.52$ & ${}_{-0.8}^{+ 0.6}$ & $\pm 1.8/\pm 0.7/\pm 2.0$ & $  55.73$ & $   0.56$ & $-7.1$ & $  1.20_{ -0.14}^{+  1.41}$ \\
$122.5$ & $  52.84$ & ${}_{-0.8}^{+ 0.6}$ & $\pm 1.8/\pm 0.7/\pm 2.0$ & $  55.02$ & $   0.56$ & $-7.2$ & $  1.20_{ -0.14}^{+  1.41}$ \\
\bottomrule
\end{tabular}%
\end{small}%
\end{center}%
\end{table}

\begin{table}
\caption{Total $\PWp({\to}\Pl^+\nu_{\Pl})\PH$ cross sections including QCD and EW corrections
and their uncertainties for a proton--proton collision energy $\sqrt{s}=13\UTeV$.}
\label{tab:wph_XStot_13}
\begin{center}%
\begin{small}%
\tabcolsep5pt
\renewcommand{\arraystretch}{1.2}%
\begin{tabular}{cccccccc}%
\toprule
$\MH$& $\sigma$ & $\Delta_{\mathrm{scale}}$ & 
$\Delta_{\mathrm{PDF}/\alphas/\mathrm{PDF\oplus\alphas}}$&
$\sigma_{\NNLO \QCD}^{\DY}$ & $\sigma_{\Pt\mbox{\scriptsize -loop}}$ &
$\delta_{\ELWK}$ & $\sigma_{\gamma}$
\vspace{.1cm} \\
~[GeV] & [fb] & [\%]  & [\%] & [fb]  & [fb] &  [\%]  & [fb] \\
\midrule
$120.0$ & $ 106.94$ & ${}_{-0.6}^{+ 0.6}$ & $\pm 1.6/\pm 0.9/\pm 1.8$ & $ 110.37$ & $   1.33$ & $-7.2$ & $  3.16_{ -0.38}^{+  3.39}$ \\
$120.5$ & $ 105.60$ & ${}_{-0.7}^{+ 0.5}$ & $\pm 1.6/\pm 0.9/\pm 1.8$ & $ 108.99$ & $   1.32$ & $-7.2$ & $  3.15_{ -0.38}^{+  3.39}$ \\
$121.0$ & $ 104.30$ & ${}_{-0.7}^{+ 0.5}$ & $\pm 1.6/\pm 0.9/\pm 1.8$ & $ 107.63$ & $   1.30$ & $-7.2$ & $  3.15_{ -0.38}^{+  3.38}$ \\
$121.5$ & $ 103.03$ & ${}_{-0.8}^{+ 0.4}$ & $\pm 1.6/\pm 0.9/\pm 1.8$ & $ 106.32$ & $   1.29$ & $-7.3$ & $  3.14_{ -0.38}^{+  3.37}$ \\
$122.0$ & $ 101.77$ & ${}_{-0.8}^{+ 0.3}$ & $\pm 1.6/\pm 0.9/\pm 1.8$ & $ 105.00$ & $   1.27$ & $-7.3$ & $  3.13_{ -0.37}^{+  3.37}$ \\
$122.5$ & $ 100.48$ & ${}_{-0.8}^{+ 0.5}$ & $\pm 1.6/\pm 0.9/\pm 1.8$ & $ 103.65$ & $   1.26$ & $-7.3$ & $  3.13_{ -0.37}^{+  3.36}$ \\
\bottomrule
\end{tabular}%
\end{small}%
\end{center}%
\end{table}

\begin{table}
\caption{Total $\PWp({\to}\Pl^+\nu_{\Pl})\PH$ cross sections including QCD and EW corrections
and their uncertainties for a proton--proton collision energy $\sqrt{s}=14\UTeV$.}
\label{tab:wph_XStot_14}
\begin{center}%
\begin{small}%
\tabcolsep5pt
\renewcommand{\arraystretch}{1.2}%
\begin{tabular}{cccccccc}%
\toprule
$\MH$& $\sigma$ & $\Delta_{\mathrm{scale}}$ & 
$\Delta_{\mathrm{PDF}/\alphas/\mathrm{PDF\oplus\alphas}}$&
$\sigma_{\NNLO \QCD}^{\DY}$ & $\sigma_{\Pt\mbox{\scriptsize -loop}}$ &
$\delta_{\ELWK}$ & $\sigma_{\gamma}$
\vspace{.1cm} \\
~[GeV] & [fb] & [\%]  & [\%] & [fb]  & [fb] &  [\%]  & [fb] \\
\midrule
$120.0$ & $ 117.37$ & ${}_{-0.6}^{+ 0.5}$ & $\pm 1.5/\pm 0.9/\pm 1.8$ & $ 120.96$ & $   1.48$ & $-7.2$ & $  3.63_{ -0.43}^{+  3.80}$ \\
$120.5$ & $ 115.95$ & ${}_{-0.7}^{+ 0.4}$ & $\pm 1.5/\pm 0.9/\pm 1.8$ & $ 119.46$ & $   1.48$ & $-7.2$ & $  3.62_{ -0.43}^{+  3.79}$ \\
$121.0$ & $ 114.51$ & ${}_{-0.7}^{+ 0.5}$ & $\pm 1.5/\pm 0.9/\pm 1.8$ & $ 117.96$ & $   1.47$ & $-7.2$ & $  3.61_{ -0.43}^{+  3.78}$ \\
$121.5$ & $ 112.91$ & ${}_{-0.6}^{+ 0.5}$ & $\pm 1.5/\pm 0.9/\pm 1.8$ & $ 116.32$ & $   1.44$ & $-7.3$ & $  3.60_{ -0.43}^{+  3.77}$ \\
$122.0$ & $ 111.56$ & ${}_{-0.6}^{+ 0.4}$ & $\pm 1.5/\pm 0.9/\pm 1.8$ & $ 114.90$ & $   1.44$ & $-7.3$ & $  3.60_{ -0.43}^{+  3.77}$ \\
$122.5$ & $ 110.21$ & ${}_{-0.8}^{+ 0.4}$ & $\pm 1.5/\pm 0.9/\pm 1.8$ & $ 113.50$ & $   1.42$ & $-7.3$ & $  3.59_{ -0.43}^{+  3.76}$ \\
\bottomrule
\end{tabular}%
\end{small}%
\end{center}%
\end{table}

\begin{table}
\caption{Total $\PWm({\to}\Pl^-\bar\nu_{\Pl})\PH$ cross sections including QCD and EW corrections
and their uncertainties for a proton--proton collision energy $\sqrt{s}=7\UTeV$.}
\label{tab:wmh_XStot_7}
\begin{center}%
\begin{small}%
\tabcolsep5pt
\renewcommand{\arraystretch}{1.2}%
\begin{tabular}{cccccccc}%
\toprule
$\MH$& $\sigma$ & $\Delta_{\mathrm{scale}}$ & 
$\Delta_{\mathrm{PDF}/\alphas/\mathrm{PDF\oplus\alphas}}$&
$\sigma_{\NNLO \QCD}^{\DY}$ & $\sigma_{\Pt\mbox{\scriptsize -loop}}$ &
$\delta_{\ELWK}$ & $\sigma_{\gamma}$
\vspace{.1cm} \\
~[GeV] & [fb] & [\%]  & [\%] & [fb]  & [fb] &  [\%]  & [fb] \\
\midrule
$120.0$ & $  26.48$ & ${}_{-0.8}^{+ 0.6}$ & $\pm 2.2/\pm 0.6/\pm 2.3$ & $  27.56$ & $   0.27$ & $-6.8$ & $  0.52_{ -0.05}^{+  0.72}$ \\
$120.5$ & $  26.10$ & ${}_{-0.8}^{+ 0.6}$ & $\pm 2.2/\pm 0.6/\pm 2.3$ & $  27.17$ & $   0.27$ & $-6.8$ & $  0.52_{ -0.05}^{+  0.72}$ \\
$121.0$ & $  25.74$ & ${}_{-0.8}^{+ 0.6}$ & $\pm 2.2/\pm 0.6/\pm 2.3$ & $  26.80$ & $   0.26$ & $-6.9$ & $  0.52_{ -0.05}^{+  0.72}$ \\
$121.5$ & $  25.38$ & ${}_{-0.9}^{+ 0.6}$ & $\pm 2.2/\pm 0.6/\pm 2.3$ & $  26.42$ & $   0.26$ & $-6.9$ & $  0.52_{ -0.05}^{+  0.71}$ \\
$122.0$ & $  25.03$ & ${}_{-0.9}^{+ 0.7}$ & $\pm 2.2/\pm 0.6/\pm 2.3$ & $  26.06$ & $   0.26$ & $-6.9$ & $  0.52_{ -0.05}^{+  0.71}$ \\
$122.5$ & $  24.68$ & ${}_{-0.8}^{+ 0.6}$ & $\pm 2.2/\pm 0.6/\pm 2.3$ & $  25.69$ & $   0.25$ & $-6.9$ & $  0.52_{ -0.05}^{+  0.71}$ \\
\bottomrule
\end{tabular}%
\end{small}%
\end{center}%
\end{table}

\begin{table}
\caption{Total $\PWm({\to}\Pl^-\bar\nu_{\Pl})\PH$ cross sections including QCD and EW corrections
and their uncertainties for a proton--proton collision energy $\sqrt{s}=8\UTeV$.}
\label{tab:wmh_XStot_8}
\begin{center}%
\begin{small}%
\tabcolsep5pt
\renewcommand{\arraystretch}{1.2}%
\begin{tabular}{cccccccc}%
\toprule
$\MH$& $\sigma$ & $\Delta_{\mathrm{scale}}$ & 
$\Delta_{\mathrm{PDF}/\alphas/\mathrm{PDF\oplus\alphas}}$&
$\sigma_{\NNLO \QCD}^{\DY}$ & $\sigma_{\Pt\mbox{\scriptsize -loop}}$ &
$\delta_{\ELWK}$ & $\sigma_{\gamma}$
\vspace{.1cm} \\
~[GeV] & [fb] & [\%]  & [\%] & [fb]  & [fb] &  [\%]  & [fb] \\
\midrule
$120.0$ & $  32.88$ & ${}_{-0.9}^{+ 0.5}$ & $\pm 2.1/\pm 0.6/\pm 2.2$ & $  34.15$ & $   0.35$ & $-6.9$ & $  0.72_{ -0.07}^{+  0.98}$ \\
$120.5$ & $  32.41$ & ${}_{-0.8}^{+ 0.5}$ & $\pm 2.1/\pm 0.6/\pm 2.2$ & $  33.66$ & $   0.35$ & $-6.9$ & $  0.72_{ -0.07}^{+  0.98}$ \\
$121.0$ & $  31.96$ & ${}_{-0.8}^{+ 0.6}$ & $\pm 2.1/\pm 0.6/\pm 2.2$ & $  33.20$ & $   0.34$ & $-6.9$ & $  0.72_{ -0.07}^{+  0.97}$ \\
$121.5$ & $  31.51$ & ${}_{-0.8}^{+ 0.5}$ & $\pm 2.1/\pm 0.6/\pm 2.2$ & $  32.73$ & $   0.34$ & $-6.9$ & $  0.72_{ -0.07}^{+  0.97}$ \\
$122.0$ & $  31.09$ & ${}_{-0.8}^{+ 0.5}$ & $\pm 2.1/\pm 0.6/\pm 2.2$ & $  32.29$ & $   0.34$ & $-7.0$ & $  0.71_{ -0.07}^{+  0.97}$ \\
$122.5$ & $  30.67$ & ${}_{-0.8}^{+ 0.5}$ & $\pm 2.1/\pm 0.6/\pm 2.2$ & $  31.85$ & $   0.33$ & $-7.0$ & $  0.71_{ -0.07}^{+  0.96}$ \\
\bottomrule
\end{tabular}%
\end{small}%
\end{center}%
\end{table}

\begin{table}
\caption{Total $\PWm({\to}\Pl^-\bar\nu_{\Pl})\PH$ cross sections including QCD and EW corrections
and their uncertainties for a proton--proton collision energy $\sqrt{s}=13\UTeV$.}
\label{tab:wmh_XStot_13}
\begin{center}%
\begin{small}%
\tabcolsep5pt
\renewcommand{\arraystretch}{1.2}%
\begin{tabular}{cccccccc}%
\toprule
$\MH$& $\sigma$ & $\Delta_{\mathrm{scale}}$ & 
$\Delta_{\mathrm{PDF}/\alphas/\mathrm{PDF\oplus\alphas}}$&
$\sigma_{\NNLO \QCD}^{\DY}$ & $\sigma_{\Pt\mbox{\scriptsize -loop}}$ &
$\delta_{\ELWK}$ & $\sigma_{\gamma}$
\vspace{.1cm} \\
~[GeV] & [fb] & [\%]  & [\%] & [fb]  & [fb] &  [\%]  & [fb] \\
\midrule
$120.0$ & $  68.17$ & ${}_{-0.6}^{+ 0.4}$ & $\pm 1.7/\pm 0.9/\pm 1.9$ & $  70.18$ & $   0.86$ & $-7.0$ & $  2.05_{ -0.23}^{+  2.38}$ \\
$120.5$ & $  67.25$ & ${}_{-0.6}^{+ 0.5}$ & $\pm 1.7/\pm 0.8/\pm 1.9$ & $  69.22$ & $   0.85$ & $-7.0$ & $  2.05_{ -0.23}^{+  2.37}$ \\
$121.0$ & $  66.35$ & ${}_{-0.6}^{+ 0.5}$ & $\pm 1.8/\pm 0.8/\pm 1.9$ & $  68.29$ & $   0.85$ & $-7.1$ & $  2.04_{ -0.23}^{+  2.37}$ \\
$121.5$ & $  65.48$ & ${}_{-0.6}^{+ 0.5}$ & $\pm 1.8/\pm 0.8/\pm 1.9$ & $  67.39$ & $   0.83$ & $-7.1$ & $  2.04_{ -0.23}^{+  2.37}$ \\
$122.0$ & $  64.60$ & ${}_{-0.6}^{+ 0.6}$ & $\pm 1.8/\pm 0.8/\pm 1.9$ & $  66.46$ & $   0.83$ & $-7.1$ & $  2.03_{ -0.23}^{+  2.36}$ \\
$122.5$ & $  63.82$ & ${}_{-0.7}^{+ 0.4}$ & $\pm 1.8/\pm 0.8/\pm 2.0$ & $  65.66$ & $   0.82$ & $-7.1$ & $  2.03_{ -0.23}^{+  2.36}$ \\
\bottomrule
\end{tabular}%
\end{small}%
\end{center}%
\end{table}

\begin{table}
\caption{Total $\PWm({\to}\Pl^-\bar\nu_{\Pl})\PH$ cross sections including QCD and EW corrections
and their uncertainties for a proton--proton collision energy $\sqrt{s}=14\UTeV$.}
\label{tab:wmh_XStot_14}
\begin{center}%
\begin{small}%
\tabcolsep5pt
\renewcommand{\arraystretch}{1.2}%
\begin{tabular}{cccccccc}%
\toprule
$\MH$& $\sigma$ & $\Delta_{\mathrm{scale}}$ & 
$\Delta_{\mathrm{PDF}/\alphas/\mathrm{PDF\oplus\alphas}}$&
$\sigma_{\NNLO \QCD}^{\DY}$ & $\sigma_{\Pt\mbox{\scriptsize -loop}}$ &
$\delta_{\ELWK}$ & $\sigma_{\gamma}$
\vspace{.1cm} \\
~[GeV] & [fb] & [\%]  & [\%] & [fb]  & [fb] &  [\%]  & [fb] \\
\midrule
$120.0$ & $  75.77$ & ${}_{-0.6}^{+ 0.4}$ & $\pm 1.7/\pm 0.8/\pm 1.9$ & $  77.89$ & $   0.97$ & $-7.0$ & $  2.39_{ -0.26}^{+  2.74}$ \\
$120.5$ & $  74.76$ & ${}_{-0.6}^{+ 0.4}$ & $\pm 1.7/\pm 0.8/\pm 1.9$ & $  76.82$ & $   0.98$ & $-7.1$ & $  2.38_{ -0.26}^{+  2.73}$ \\
$121.0$ & $  73.78$ & ${}_{-0.6}^{+ 0.4}$ & $\pm 1.7/\pm 0.8/\pm 1.9$ & $  75.82$ & $   0.96$ & $-7.1$ & $  2.37_{ -0.26}^{+  2.72}$ \\
$121.5$ & $  72.89$ & ${}_{-0.7}^{+ 0.3}$ & $\pm 1.7/\pm 0.8/\pm 1.9$ & $  74.91$ & $   0.95$ & $-7.1$ & $  2.37_{ -0.26}^{+  2.71}$ \\
$122.0$ & $  71.93$ & ${}_{-0.7}^{+ 0.4}$ & $\pm 1.7/\pm 0.8/\pm 1.9$ & $  73.90$ & $   0.94$ & $-7.1$ & $  2.36_{ -0.26}^{+  2.70}$ \\
$122.5$ & $  71.01$ & ${}_{-0.7}^{+ 0.5}$ & $\pm 1.7/\pm 0.8/\pm 1.9$ & $  72.94$ & $   0.93$ & $-7.1$ & $  2.35_{ -0.26}^{+  2.69}$ \\
\bottomrule
\end{tabular}%
\end{small}%
\end{center}%
\end{table}

\begin{table}
\caption{Total $\PZ({\to}\Pl^+\Pl^-)\PH$ cross sections including QCD and EW corrections
and their uncertainties for a proton--proton collision energy $\sqrt{s}=7\UTeV$.}
\label{tab:zllh_XStot_7}
\begin{center}%
\begin{small}%
\tabcolsep5pt
\renewcommand{\arraystretch}{1.2}%
%
%
\end{small}%
\end{center}%
\end{table}

\begin{table}
\caption{Total $\PZ({\to}\Pl^+\Pl^-)\PH$ cross sections including QCD and EW corrections
and their uncertainties for a proton--proton collision energy $\sqrt{s}=8\UTeV$.}
\label{tab:zllh_XStot_8}
\begin{center}%
\begin{small}%
\tabcolsep5pt
\renewcommand{\arraystretch}{1.2}%
%
%
\end{small}%
\end{center}%
\end{table}

\begin{table}
\caption{Total $\PZ({\to}\Pl^+\Pl^-)\PH$ cross sections including QCD and EW corrections
and their uncertainties for a proton--proton collision energy $\sqrt{s}=13\UTeV$.}
\label{tab:zllh_XStot_13}
\begin{center}%
\begin{small}%
\tabcolsep5pt
\renewcommand{\arraystretch}{1.2}%
%
%
\end{small}%
\end{center}%
\end{table}

\renewcommand{\arraystretch}{1.2}%

\begin{table}
\caption{Total $\PZ({\to}\nu\bar\nu)\PH$ cross sections including QCD and EW corrections
and their uncertainties for a proton--proton collision energy $\sqrt{s}=7\UTeV$.}
\label{tab:znnh_XStot_7}
\begin{center}%
\begin{small}%
\tabcolsep5pt
\renewcommand{\arraystretch}{1.2}%
%
%
\end{small}%
\end{center}%
\end{table}

\begin{table}
\caption{Total $\PZ({\to}\nu\bar\nu)\PH$ cross sections including QCD and EW corrections
and their uncertainties for a proton--proton collision energy $\sqrt{s}=8\UTeV$.}
\label{tab:znnh_XStot_8}
\begin{center}%
\begin{small}%
\tabcolsep5pt
\renewcommand{\arraystretch}{1.2}%
%
%
\end{small}%
\end{center}%
\end{table}

\begin{table}
\caption{Total $\PZ({\to}\nu\bar\nu)\PH$ cross sections including QCD and EW corrections
and their uncertainties for a proton--proton collision energy $\sqrt{s}=13\UTeV$.}
\label{tab:znnh_XStot_13}
\begin{center}%
\begin{small}%
\tabcolsep5pt
\renewcommand{\arraystretch}{1.2}%
%
%
\end{small}%
\end{center}%
\end{table}

\begin{table}
\caption{Total $\PZ({\to}\nu\bar\nu)\PH$ cross sections including QCD and EW corrections
and their uncertainties for a proton--proton collision energy $\sqrt{s}=14\UTeV$.}
\label{tab:znnh_XStot_14}
\begin{center}%
\begin{small}%
\tabcolsep5pt
\renewcommand{\arraystretch}{1.2}%
%
%
\end{small}%
\end{center}%
\end{table}

%

\begin{table}
\caption{Energy scan of Inclusive $WH$ cross sections including QCD and EW corrections for a Higgs boson mass of $125\UGeV$,   
 together with their corresponding uncertainties.}
\label{tab:WH_XStot_125}
\renewcommand{\arraystretch}{1.2}%
\begin{center}%
\begin{small}%
\tabcolsep5pt
%
%
\end{small}%
\end{center}%
\end{table}

\clearpage

\begin{table}
\caption{Energy scan of Inclusive $WH$ cross sections including QCD and EW corrections for a Higgs boson mass of $125.09\UGeV$,   
 together with their corresponding uncertainties.}
\label{tab:WH_XStot_12509}
\renewcommand{\arraystretch}{1.2}%
\begin{center}%
\begin{small}%
\tabcolsep5pt
%
%
\end{small}%
\end{center}%
\end{table}
\begin{table}
\caption{Energy scan of Inclusive $ZH$ cross sections including QCD and EW corrections for a Higgs boson mass of $125\UGeV$,   
 together with their corresponding uncertainties.}
\label{tab:ZH_XStot_125}
\renewcommand{\arraystretch}{1.2}%
\begin{center}%
\begin{small}%
\tabcolsep5pt
%
%
\end{small}%
\end{center}%
\end{table}

\begin{table}
\caption{Energy scan of Inclusive $ZH$ cross sections including QCD and EW corrections for a Higgs boson mass of $125.09\UGeV$,   
 together with their corresponding uncertainties.}
\label{tab:ZH_XStot_12509}
\renewcommand{\arraystretch}{1.2}%
\begin{center}%
\begin{small}%
\tabcolsep5pt
%
%
\end{small}%
\end{center}%
\end{table}


\clearpage
\def\utthxstabhead{
\toprule
$\mh$[\!\UGeV]  &
$\sigmaNLOqcd$[\!\Ufb]&
$\sigmaNLOqcdew$[\!\Ufb] &
$K_{\mathrm{QCD}}$ &
$\deltaEW$[\%] &
Scale[\%] &
$\als$[\%] &
PDF[\%] &
PDF+$\als$[\%]
\\   \midrule}
\newcommand{\eextthxs}[7]{$#1$ & $#2$ &$#3$ & #4 & #5 & $#6$ & #7  \\}
\newcommand{\eexthxs}[7]{$#1$ & $#2$ & $#3$ & #4 & #5 & $#6$ & $#7$ \\}



\chapter{\texorpdfstring{$\tth$ and $tH$}{ttH and tH} cross sections}
\label{ttHappendix}

In this appendix the cross-sections for  $\tth$ and $tH$ productions
are expanded to a scan over SM Higgs boson masses.
\clearpage



\begin{table}
\caption{\label{tab:sm7tev}SM-like $\tth$ cross section at the LHC
  with $\sqrt{s}=7$~\UTeV.
}
\vspace{1cm}
\tabcolsep5pt
\centering
\begin{small}%
\renewcommand{\arraystretch}{1.2}%

\end{small}%
\end{table}

\begin{table}
    \caption{\label{tab:sm8tev}SM-like $\tth$ cross section at the LHC
      with $\sqrt{s}=8$~\UTeV.
} \vspace{1cm}
\tabcolsep5pt
    \centering
\begin{small}%
\renewcommand{\arraystretch}{1.2}%
%
\end{small}
\end{table}

\begin{table}
\caption{\label{tab:sm13tev}SM-like $\tth$ cross section at the LHC
  with $\sqrt s=13$~\UTeV.
} \vspace{1cm}
\tabcolsep5pt
    \centering
\begin{small}%
\renewcommand{\arraystretch}{1.2}%
%
\end{small}
\end{table}

\begin{table}
\caption{\label{tab:sm14tev}SM-like $\tth$ cross section at the LHC
  with $\sqrt s=14$~\UTeV.
} \vspace{1cm}
\tabcolsep5pt
    \centering
\begin{small}%
\renewcommand{\arraystretch}{1.2}%
%
\end{small}
\end{table}

\begin{table}
    \caption{\label{tab:smescan125}$\tth$  cross section for
      $\mh=125$\UGeV \, and different collider energies.
} \vspace{1cm}
\tabcolsep5pt
    \centering
\begin{small}%
\renewcommand{\arraystretch}{1.2}%
    \begin{tabular}{ccccccccc}
  \toprule
    $\sqrt{s}$\,[\!\UTeV] &
    $\sigma_{\mathrm{QCD}}$ [\!\Ufb]&
    $\sigma_{\mathrm{REF}}$ [\!\Ufb]&
    $K_{\mathrm{QCD}}$ &
    $\delta_{\mathrm{EW}}$[\%] &
    Scale[\%] &
    $\als$[\%] &
    PDF &
    PDF+$\als$[\%] \\
\bottomrule
    \end{tabular}
\end{small}
\end{table}

\begin{table}
    \caption{\label{tab:smescan12509}$\tth$  cross section for
      $\mh=125.09$\UGeV \, and different collider energies.
} \vspace{1cm}
\tabcolsep5pt
    \centering
\begin{small}%
\renewcommand{\arraystretch}{1.2}%
    \begin{tabular}{ccccccccc}
  \toprule
    $\sqrt{s}$\,[\!\UTeV] &
    $\sigma_{\mathrm{QCD}}$[\!\Ufb] &
    $\sigma_{\mathrm{REF}}$[\!\Ufb] &
    $K_{\mathrm{QCD}}$ &
    $\delta_{\mathrm{EW}}$[\%] &
    Scale[\%] &
    $\als$[\%] &
    PDF &
    PDF+$\als$[\%] \\
\bottomrule
    \end{tabular}
\end{small}
\end{table}



\begin{table}
\caption{Cross sections for $t$-channel $tH$ and $\bar t H$ production at the 7\UTeV~LHC.}
\vspace{1cm}
\tabcolsep5pt
\centering
\begin{small}%
\renewcommand{\arraystretch}{1.2}%
%
\end{small}
\label{tab:tH_tcha_SM_lhc7}
\end{table}

\begin{table}
\caption{Cross sections for $t$-channel $tH$ and $\bar t H$ production at the 8\UTeV~LHC.}
\vspace{1cm}
\tabcolsep5pt
\centering
\begin{small}%
\renewcommand{\arraystretch}{1.2}%
%
\end{small}
\label{tab:tH_tcha_SM_lhc8}
\end{table}

\begin{table}
\caption{Cross sections for $t$-channel $tH$ and $\bar t H$ production at the 13\UTeV~LHC.}
\vspace{1cm}
\tabcolsep5pt
\centering
\begin{small}%
\renewcommand{\arraystretch}{1.2}%
%
\end{small}
\label{tab:tH_tcha_SM_lhc13}
\end{table}

\begin{table}
\caption{Cross sections for $t$-channel $tH$ and $\bar t H$ production at the 14\UTeV~LHC.}
\vspace{1cm}
\tabcolsep5pt
\centering
\begin{small}%
\renewcommand{\arraystretch}{1.2}%
%
\end{small}
\label{tab:tH_tcha_SM_lhc14}
\end{table}

\begin{table}
\caption{Cross sections for $t$-channel $tH$ and $\bar t H$ production for a Higgs boson mass of $125\UGeV$ ~and $6\UTeV\le\sqrt{s}\le 15\UTeV$.}
\vspace{1cm}
\tabcolsep5pt
\centering
\begin{small}%
\renewcommand{\arraystretch}{1.2}%
%
\end{small}
\label{tab:tH_tcha_SM_lhc6to15}
\end{table}


\begin{table}
\caption{Cross sections for $s$-channel $tH$ and $\bar t H$ production at the 7\UTeV~LHC.}
\vspace{1cm}
\tabcolsep5pt
\centering
\begin{small}%
\renewcommand{\arraystretch}{1.2}%
%
\end{small}
\label{tab:tH_scha_SM_lhc7}
\end{table}

\begin{table}
\caption{Cross sections for $s$-channel $tH$ and $\bar t H$ production at the 8\UTeV~LHC.}
\vspace{1cm}
\tabcolsep5pt
\centering
\begin{small}%
\renewcommand{\arraystretch}{1.2}%
%
\end{small}
\label{tab:tH_scha_SM_lhc8}
\end{table}

\begin{table}
\caption{Cross sections for $s$-channel $tH$ and $\bar t H$ production at the 13\UTeV~LHC.}
\vspace{1cm}
\tabcolsep5pt
\centering
\begin{small}%
\renewcommand{\arraystretch}{1.2}%
%
\end{small}
\label{tab:tH_scha_SM_lhc13}
\end{table}

\begin{table}
\caption{Cross sections for $s$-channel $tH$ and $\bar t H$ production at the 14\UTeV~LHC.}
\vspace{1cm}
\tabcolsep5pt
\centering
\begin{small}%
\renewcommand{\arraystretch}{1.2}%
%
\end{small}
\label{tab:tH_scha_SM_lhc14}
\end{table}

\begin{table}
\caption{Cross sections for $s$-channel $tH$ and $\bar t H$ production for a Higgs boson mass of $125\UGeV$ ~and  $6\UTeV\le\sqrt{s}\le 15\UTeV$.}
\vspace{1cm}
\tabcolsep5pt
\centering
\begin{small}%
\renewcommand{\arraystretch}{1.2}%
%
\end{small}
\label{tab:tH_scha_SM_lhc6to15}
\end{table}

\clearpage



\chapter{\texorpdfstring{$b\bar{b}H$}{bbH} cross sections}
\label{bbHappendix}

In this appendix the cross-sections for  $b\bar{b}H$   production are expanded to a scan over SM
Higgs boson masses as discussed in \refS{sec:bbHtotal}.

\begin{table}
\caption{Inclusive bbH cross sections for a LHC CM energy of $\sqrt{s}=7\UTeV$, obtained using Santander Matching with 5FS (NNLO) and 4FS (NLO),
 together with their total uncertainties.}
\label{tab:bbH_XStot_7}
\renewcommand{\arraystretch}{1.2}
\begin{center}%
\begin{small}%
\tabcolsep5pt
\begin{tabular}{cccc}%
\toprule
$\MH$[GeV] & $\sigma^{bbH}$[fb]  & $+{\mathrm{QCD\, scale} \oplus\mathrm{PDF\oplus\alpha_s}}$[\%] & $-{\mathrm{QCD\, scale} \oplus\mathrm{PDF\oplus\alpha_s}}$[\%]
\\
\midrule
$	120.0	$	&	$	178.20	$	&$+	21.0	$&$	-22.8	$\\
$	120.5	$	&	$	176.20	$	&$+	21.0	$&$	-22.8	$\\
$	121.0	$	&	$	173.50	$	&$+	20.9	$&$	-22.7	$\\
$	121.5	$	&	$	171.00	$	&$+	20.9	$&$	-22.7	$\\
$	122.0	$	&	$	168.80	$	&$+	20.9	$&$	-22.7	$\\
$	122.5	$	&	$	166.30	$	&$+	20.8	$&$	-22.6	$\\
$	123.0	$	&	$	163.40	$	&$+	20.7	$&$	-22.5	$\\
$	123.5	$	&	$	161.90	$	&$+	20.7	$&$	-22.5	$\\
$	124.0	$	&	$	159.70	$	&$+	20.8	$&$	-22.6	$\\
$	124.1	$	&	$	159.10	$	&$+	20.8	$&$	-22.6	$\\
$	124.2	$	&	$	158.60	$	&$+	20.8	$&$	-22.6	$\\
$	124.3	$	&	$	158.30	$	&$+	20.8	$&$	-22.7	$\\
$	124.4	$	&	$	157.90	$	&$+	20.9	$&$	-22.7	$\\
$	124.5	$	&	$	157.30	$	&$+	20.7	$&$	-22.5	$\\
$	124.6	$	&	$	157.20	$	&$+	20.8	$&$	-22.5	$\\
$	124.7	$	&	$	156.40	$	&$+	20.8	$&$	-22.5	$\\
$	124.8	$	&	$	155.70	$	&$+	20.8	$&$	-22.6	$\\
$	124.9	$	&	$	155.50	$	&$+	20.8	$&$	-22.6	$\\
$	125.0	$	&	$	155.20	$	&$+	20.7	$&$	-22.4	$\\
$	125.1	$	&	$	154.50	$	&$+	20.7	$&$	-22.5	$\\
$	125.09	$	&	$	154.70	$	&$+	20.8	$&$	-22.5	$\\
$	125.2	$	&	$	154.30	$	&$+	20.8	$&$	-22.5	$\\
$	125.3	$	&	$	153.80	$	&$+	20.8	$&$	-22.5	$\\
$	125.4	$	&	$	153.20	$	&$+	20.8	$&$	-22.6	$\\
$	125.5	$	&	$	152.70	$	&$+	20.7	$&$	-22.4	$\\
$	125.6	$	&	$	152.40	$	&$+	20.6	$&$	-22.4	$\\
$	125.7	$	&	$	152.10	$	&$+	20.7	$&$	-22.5	$\\
$	125.8	$	&	$	151.40	$	&$+	20.7	$&$	-22.5	$\\
$	125.9	$	&	$	151.00	$	&$+	20.8	$&$	-22.5	$\\
$	126.0	$	&	$	150.80	$	&$+	20.7	$&$	-22.3	$\\
$	126.5	$	&	$	148.60	$	&$+	20.6	$&$	-22.3	$\\
$	127.0	$	&	$	146.70	$	&$+	20.6	$&$	-22.3	$\\
$	127.5	$	&	$	144.60	$	&$+	20.5	$&$	-22.2	$\\
$	128.0	$	&	$	142.50	$	&$+	20.4	$&$	-22.2	$\\
$	128.5	$	&	$	140.60	$	&$+	20.4	$&$	-22.2	$\\
$	129.0	$	&	$	138.90	$	&$+	20.4	$&$	-22.1	$\\
$	129.5	$	&	$	137.00	$	&$+	20.4	$&$	-22.1	$\\
$	130.0	$	&	$	135.10	$	&$+	20.3	$&$	-22.1	$\\
\bottomrule
\end{tabular}%
\end{small}%
\end{center}%
\end{table}

\begin{table}
\caption{Inclusive bbH cross sections for a LHC CM energy of $\sqrt{s}=8\UTeV$, obtained using Santander Matching with 5FS (NNLO) and 4FS (NLO),
 together with their total uncertainties.}
\label{tab:bbH_XStot_8}
\renewcommand{\arraystretch}{1.2}
\begin{center}%
\begin{small}%
\tabcolsep5pt
\begin{tabular}{cccc}%
\toprule
$\MH$[GeV] & $\sigma^{bbH}$[fb]  & $+{\mathrm{QCD\, scale} \oplus\mathrm{PDF\oplus\alpha_s}}$[\%] & $-{\mathrm{QCD\, scale} \oplus\mathrm{PDF\oplus\alpha_s}}$[\%]
\\
\midrule
$	120.0	$	&	$	231.50	$	&$+	20.8	$&$	-22.8	$\\
$	120.5	$	&	$	228.90	$	&$+	20.8	$&$	-22.8	$\\
$	121.0	$	&	$	225.00	$	&$+	20.7	$&$	-22.7	$\\
$	121.5	$	&	$	222.40	$	&$+	20.8	$&$	-22.7	$\\
$	122.0	$	&	$	219.00	$	&$+	20.6	$&$	-22.6	$\\
$	122.5	$	&	$	216.10	$	&$+	20.7	$&$	-22.5	$\\
$	123.0	$	&	$	213.20	$	&$+	20.7	$&$	-22.5	$\\
$	123.5	$	&	$	210.40	$	&$+	20.7	$&$	-22.4	$\\
$	124.0	$	&	$	207.10	$	&$+	20.6	$&$	-22.4	$\\
$	124.1	$	&	$	206.90	$	&$+	20.7	$&$	-22.4	$\\
$	124.2	$	&	$	206.20	$	&$+	20.6	$&$	-22.5	$\\
$	124.3	$	&	$	205.60	$	&$+	20.7	$&$	-22.6	$\\
$	124.4	$	&	$	205.20	$	&$+	20.8	$&$	-22.6	$\\
$	124.5	$	&	$	204.50	$	&$+	20.6	$&$	-22.4	$\\
$	124.6	$	&	$	204.30	$	&$+	20.7	$&$	-22.5	$\\
$	124.7	$	&	$	203.60	$	&$+	20.7	$&$	-22.5	$\\
$	124.8	$	&	$	202.80	$	&$+	20.6	$&$	-22.5	$\\
$	124.9	$	&	$	202.40	$	&$+	20.7	$&$	-22.5	$\\
$	125.0	$	&	$	202.10	$	&$+	20.7	$&$	-22.3	$\\
$	125.1	$	&	$	201.50	$	&$+	20.6	$&$	-22.4	$\\
$	125.09	$	&	$	200.80	$	&$+	20.6	$&$	-22.3	$\\
$	125.2	$	&	$	200.80	$	&$+	20.7	$&$	-22.4	$\\
$	125.3	$	&	$	200.50	$	&$+	20.7	$&$	-22.4	$\\
$	125.4	$	&	$	199.80	$	&$+	20.7	$&$	-22.5	$\\
$	125.5	$	&	$	199.30	$	&$+	20.6	$&$	-22.3	$\\
$	125.6	$	&	$	199.30	$	&$+	20.6	$&$	-22.3	$\\
$	125.7	$	&	$	198.80	$	&$+	20.7	$&$	-22.4	$\\
$	125.8	$	&	$	198.20	$	&$+	20.7	$&$	-22.4	$\\
$	125.9	$	&	$	197.40	$	&$+	20.8	$&$	-22.5	$\\
$	126.0	$	&	$	196.80	$	&$+	20.6	$&$	-22.3	$\\
$	126.5	$	&	$	193.60	$	&$+	20.6	$&$	-22.1	$\\
$	127.0	$	&	$	191.40	$	&$+	20.5	$&$	-22.2	$\\
$	127.5	$	&	$	188.20	$	&$+	20.4	$&$	-22.1	$\\
$	128.0	$	&	$	186.40	$	&$+	20.4	$&$	-22.1	$\\
$	128.5	$	&	$	183.30	$	&$+	20.3	$&$	-22.0	$\\
$	129.0	$	&	$	181.10	$	&$+	20.3	$&$	-22.0	$\\
$	129.5	$	&	$	178.60	$	&$+	20.4	$&$	-22.0	$\\
$	130.0	$	&	$	176.30	$	&$+	20.3	$&$	-21.9	$\\
\bottomrule
\end{tabular}%
\end{small}%
\end{center}%
\end{table}

\begin{table}
\caption{Inclusive bbH cross sections for a LHC CM energy of $\sqrt{s}=13\UTeV$, obtained using Santander Matching with 5FS (NNLO) and 4FS (NLO),
 together with their total uncertainties.}
\label{tab:bbH_XStot_13}
\renewcommand{\arraystretch}{1.2}
\begin{center}%
\begin{small}%
\tabcolsep5pt
\begin{tabular}{cccc}%
\toprule
$\MH$[GeV] & $\sigma^{bbH}$[fb]  & $+{\mathrm{QCD\, scale} \oplus\mathrm{PDF\oplus\alpha_s}}$[\%] & $-{\mathrm{QCD\, scale} \oplus\mathrm{PDF\oplus\alpha_s}}$[\%]
\\
\midrule
$	120.0	$	&	$	553.40	$	&$+	20.3	$&$	-24.4	$\\
$	120.5	$	&	$	547.10	$	&$+	20.3	$&$	-24.4	$\\
$	121.0	$	&	$	541.40	$	&$+	20.4	$&$	-24.3	$\\
$	121.5	$	&	$	534.30	$	&$+	20.5	$&$	-24.3	$\\
$	122.0	$	&	$	526.80	$	&$+	20.4	$&$	-24.3	$\\
$	122.5	$	&	$	520.00	$	&$+	20.2	$&$	-24.2	$\\
$	123.0	$	&	$	512.40	$	&$+	21.2	$&$	-24.1	$\\
$	123.5	$	&	$	506.60	$	&$+	20.1	$&$	-24.0	$\\
$	124.0	$	&	$	499.90	$	&$+	20.1	$&$	-24.0	$\\
$	124.1	$	&	$	498.50	$	&$+	20.1	$&$	-24.0	$\\
$	124.2	$	&	$	497.60	$	&$+	20.3	$&$	-24.1	$\\
$	124.3	$	&	$	495.50	$	&$+	20.2	$&$	-24.1	$\\
$	124.4	$	&	$	495.00	$	&$+	20.3	$&$	-24.1	$\\
$	124.5	$	&	$	493.00	$	&$+	20.0	$&$	-23.9	$\\
$	124.6	$	&	$	491.80	$	&$+	20.1	$&$	-24.0	$\\
$	124.7	$	&	$	491.90	$	&$+	20.6	$&$	-24.1	$\\
$	124.8	$	&	$	489.90	$	&$+	20.2	$&$	-24.0	$\\
$	124.9	$	&	$	489.10	$	&$+	20.3	$&$	-24.1	$\\
$	125.0	$	&	$	488.00	$	&$+	20.2	$&$	-23.9	$\\
$	125.1	$	&	$	486.30	$	&$+	20.1	$&$	-23.9	$\\
$	125.09	$	&	$	486.00	$	&$+	20.5	$&$	-23.9	$\\
$	125.2	$	&	$	485.00	$	&$+	20.1	$&$	-24.0	$\\
$	125.3	$	&	$	483.30	$	&$+	20.1	$&$	-24.0	$\\
$	125.4	$	&	$	482.20	$	&$+	20.2	$&$	-24.0	$\\
$	125.5	$	&	$	480.90	$	&$+	20.1	$&$	-23.8	$\\
$	125.6	$	&	$	479.20	$	&$+	20.1	$&$	-23.9	$\\
$	125.7	$	&	$	478.90	$	&$+	20.1	$&$	-23.9	$\\
$	125.8	$	&	$	476.80	$	&$+	20.1	$&$	-23.9	$\\
$	125.9	$	&	$	475.70	$	&$+	20.2	$&$	-24.0	$\\
$	126.0	$	&	$	476.00	$	&$+	20.2	$&$	-23.8	$\\
$	126.5	$	&	$	469.50	$	&$+	20.0	$&$	-23.7	$\\
$	127.0	$	&	$	463.50	$	&$+	20.0	$&$	-23.7	$\\
$	127.5	$	&	$	458.00	$	&$+	20.0	$&$	-23.6	$\\
$	128.0	$	&	$	451.30	$	&$+	19.9	$&$	-23.6	$\\
$	128.5	$	&	$	446.30	$	&$+	19.9	$&$	-23.5	$\\
$	129.0	$	&	$	441.60	$	&$+	20.0	$&$	-23.4	$\\
$	129.5	$	&	$	435.80	$	&$+	19.9	$&$	-23.4	$\\
$	130.0	$	&	$	430.40	$	&$+	19.9	$&$	-23.3	$\\
\bottomrule
\end{tabular}%
\end{small}%
\end{center}%

\end{table}
\begin{table}
\caption{Inclusive bbH cross sections for a LHC CM energy of $\sqrt{s}=14\UTeV$, obtained using Santander Matching with 5FS (NNLO) and 4FS (NLO),
 together with their total uncertainties.}
\label{tab:bbH_XStot_14}
\renewcommand{\arraystretch}{1.2}
\begin{center}%
\begin{small}%
\tabcolsep5pt
\begin{tabular}{cccc}%
\toprule
$\MH$[GeV] & $\sigma^{bbH}$[fb]  & $+{\mathrm{QCD\, scale} \oplus\mathrm{PDF\oplus\alpha_s}}$[\%] & $-{\mathrm{QCD\, scale} \oplus\mathrm{PDF\oplus\alpha_s}}$[\%]
\\
\midrule
$	120.0	$	&	$	628.80	$	&$+	20.4	$&$	-24.7	$\\
$	120.5	$	&	$	620.40	$	&$+	20.3	$&$	-24.6	$\\
$	121.0	$	&	$	610.50	$	&$+	20.2	$&$	-24.5	$\\
$	121.5	$	&	$	603.20	$	&$+	20.2	$&$	-24.4	$\\
$	122.0	$	&	$	596.10	$	&$+	20.2	$&$	-24.4	$\\
$	122.5	$	&	$	588.40	$	&$+	20.2	$&$	-24.3	$\\
$	123.0	$	&	$	580.10	$	&$+	20.0	$&$	-24.2	$\\
$	123.5	$	&	$	574.50	$	&$+	20.1	$&$	-24.2	$\\
$	124.0	$	&	$	567.20	$	&$+	23.4	$&$	-24.2	$\\
$	124.1	$	&	$	565.60	$	&$+	20.1	$&$	-24.2	$\\
$	124.2	$	&	$	564.20	$	&$+	20.1	$&$	-24.2	$\\
$	124.3	$	&	$	562.10	$	&$+	20.1	$&$	-24.2	$\\
$	124.4	$	&	$	560.80	$	&$+	20.1	$&$	-24.3	$\\
$	124.5	$	&	$	559.60	$	&$+	20.1	$&$	-24.1	$\\
$	124.6	$	&	$	557.70	$	&$+	20.0	$&$	-24.1	$\\
$	124.7	$	&	$	556.30	$	&$+	20.1	$&$	-24.2	$\\
$	124.8	$	&	$	555.10	$	&$+	20.2	$&$	-24.2	$\\
$	124.9	$	&	$	554.50	$	&$+	20.3	$&$	-24.2	$\\
$	125.0	$	&	$	552.90	$	&$+	20.0	$&$	-24.0	$\\
$	125.1	$	&	$	552.20	$	&$+	20.1	$&$	-24.1	$\\
$	125.09	$	&	$	552.10	$	&$+	20.2	$&$	-24.1	$\\
$	125.2	$	&	$	550.10	$	&$+	23.0	$&$	-24.0	$\\
$	125.3	$	&	$	548.10	$	&$+	20.1	$&$	-24.2	$\\
$	125.4	$	&	$	547.40	$	&$+	20.6	$&$	-24.1	$\\
$	125.5	$	&	$	545.30	$	&$+	19.9	$&$	-24.0	$\\
$	125.6	$	&	$	544.80	$	&$+	20.6	$&$	-24.0	$\\
$	125.7	$	&	$	542.20	$	&$+	20.0	$&$	-24.0	$\\
$	125.8	$	&	$	541.50	$	&$+	20.8	$&$	-24.1	$\\
$	125.9	$	&	$	539.80	$	&$+	20.0	$&$	-24.1	$\\
$	126.0	$	&	$	538.40	$	&$+	19.9	$&$	-23.9	$\\
$	126.5	$	&	$	534.20	$	&$+	20.0	$&$	-23.9	$\\
$	127.0	$	&	$	526.00	$	&$+	19.9	$&$	-23.8	$\\
$	127.5	$	&	$	520.00	$	&$+	19.8	$&$	-23.7	$\\
$	128.0	$	&	$	512.40	$	&$+	19.8	$&$	-23.7	$\\
$	128.5	$	&	$	506.20	$	&$+	19.8	$&$	-23.7	$\\
$	129.0	$	&	$	500.20	$	&$+	19.7	$&$	-23.6	$\\
$	129.5	$	&	$	494.30	$	&$+	19.7	$&$	-23.6	$\\
$	130.0	$	&	$	487.80	$	&$+	21.4	$&$	-23.5	$\\
\bottomrule
\end{tabular}%
\end{small}%
\end{center}%
\end{table}

\begin{table}
\caption{Energy scan of Inclusive bbH cross sections for a Higgs boson mass of $125\UGeV$, obtained using Santander Matching with 5FS (NNLO) and 4FS (NLO),
 together with their total uncertainties.}
\label{tab:bbH_XStot_125}
\renewcommand{\arraystretch}{1.2}
\begin{center}%
\begin{small}%
\tabcolsep5pt
\begin{tabular}{cccc}%
\toprule
$\sqrt{s}$\,[TeV] & $\sigma^{bbH}$[fb]  & $+{\mathrm{QCD\, scale} \oplus\mathrm{PDF\oplus\alpha_s}}$[\%] & $-{\mathrm{QCD\, scale} \oplus\mathrm{PDF\oplus\alpha_s}}$[\%]
\\
\midrule
$	6.0	$	&	$	112.70	$	&$+	20.9	$&$	-22.5	$\\
$	6.5	$	&	$	133.50	$	&$+	20.8	$&$	-22.5	$\\
$	7.0	$	&	$	155.20	$	&$+	20.7	$&$	-22.4	$\\
$	7.5	$	&	$	177.90	$	&$+	20.6	$&$	-22.4	$\\
$	8.0	$	&	$	202.10	$	&$+	20.7	$&$	-22.3	$\\
$	8.5	$	&	$	226.70	$	&$+	20.5	$&$	-22.3	$\\
$	9.0	$	&	$	252.50	$	&$+	20.4	$&$	-22.5	$\\
$	9.5	$	&	$	280.00	$	&$+	20.5	$&$	-22.7	$\\
$	10.0	$	&	$	306.90	$	&$+	20.4	$&$	-22.8	$\\
$	10.5	$	&	$	335.60	$	&$+	20.4	$&$	-23.0	$\\
$	11.0	$	&	$	363.90	$	&$+	20.2	$&$	-23.1	$\\
$	11.5	$	&	$	394.40	$	&$+	20.3	$&$	-23.2	$\\
$	12.0	$	&	$	424.30	$	&$+	20.2	$&$	-23.4	$\\
$	12.5	$	&	$	455.70	$	&$+	20.1	$&$	-23.6	$\\
$	13.0	$	&	$	488.00	$	&$+	20.2	$&$	-23.9	$\\
$	13.5	$	&	$	520.00	$	&$+	20.1	$&$	-24.0	$\\
$	14.0	$	&	$	552.90	$	&$+	20.0	$&$	-24.0	$\\
$	14.5	$	&	$	588.00	$	&$+	20.1	$&$	-24.2	$\\
$	15.0	$	&	$	620.30	$	&$+	20.0	$&$	-24.4	$\\
\bottomrule
\end{tabular}%
\end{small}%
\end{center}%
\end{table}


\begin{table}
\caption{Energy scan of Inclusive bbH cross sections for a Higgs boson mass of $125.09\UGeV$, obtained using Santander Matching with 5FS (NNLO) and 4FS (NLO),
 together with their total uncertainties.}
\label{tab:bbH_XStot_12509}
\renewcommand{\arraystretch}{1.2}
\begin{center}%
\begin{small}%
\tabcolsep5pt
\begin{tabular}{cccc}%
\toprule
$\sqrt{s}$\,[TeV] & $\sigma^{bbH}$[fb]  & $+{\mathrm{QCD\, scale} \oplus\mathrm{PDF\oplus\alpha_s}}$[\%] & $-{\mathrm{QCD\, scale} \oplus\mathrm{PDF\oplus\alpha_s}}$[\%]
\\
\midrule
$	6.0	$	&	$	112.50	$	&$+	20.8	$&$	-22.5	$\\
$	6.5	$	&	$	133.40	$	&$+	20.8	$&$	-22.5	$\\
$	7.0	$	&	$	154.50	$	&$+	20.7	$&$	-22.5	$\\
$	7.5	$	&	$	177.40	$	&$+	20.6	$&$	-22.4	$\\
$	8.0	$	&	$	201.50	$	&$+	20.6	$&$	-22.4	$\\
$	8.5	$	&	$	226.30	$	&$+	20.5	$&$	-22.4	$\\
$	9.0	$	&	$	252.00	$	&$+	20.4	$&$	-22.5	$\\
$	9.5	$	&	$	279.20	$	&$+	20.5	$&$	-22.7	$\\
$	10.0	$	&	$	306.60	$	&$+	20.4	$&$	-22.8	$\\
$	10.5	$	&	$	334.60	$	&$+	20.4	$&$	-23.0	$\\
$	11.0	$	&	$	362.80	$	&$+	20.3	$&$	-23.1	$\\
$	11.5	$	&	$	393.00	$	&$+	20.3	$&$	-23.2	$\\
$	12.0	$	&	$	423.50	$	&$+	20.2	$&$	-23.4	$\\
$	12.5	$	&	$	454.30	$	&$+	20.1	$&$	-23.5	$\\
$	13.0	$	&	$	486.30	$	&$+	20.1	$&$	-23.9	$\\
$	13.5	$	&	$	518.30	$	&$+	20.1	$&$	-24.0	$\\
$	14.0	$	&	$	552.20	$	&$+	20.1	$&$	-24.1	$\\
$	14.5	$	&	$	584.80	$	&$+	20.0	$&$	-24.2	$\\
$	15.0	$	&	$	618.60	$	&$+	19.8	$&$	-24.4	$\\
\bottomrule
\end{tabular}%
\end{small}%
\end{center}%
\end{table}

\end{appendices}
\clearpage

\renewcommand\leftmark{References}
\renewcommand\rightmark{References}

\bibliographystyle{./StyleFilesMacros/atlasnote}
\cleardoublepage
\phantomsection
\addcontentsline{toc}{part}{References}
\bibliography{YRHXS4}


\end{document}